%% file: archiv.tex
\documentclass[reprint,natbib209,
 amsmath,amssymb,
 aps,aa
floatfix,
]{revtex4-1}
\usepackage{graphicx}% Include figure files
\usepackage{dcolumn}% Align table columns on decimal point
\usepackage{bm}% bold math
\usepackage{hyperref}% add hypertext capabilities
\begin{document}

\preprint{Fermi GeV-Excess}

\title{Molecular Clouds as  the Origin of the Fermi Gamma-Ray GeV-Excess
}% Force line breaks with \\

\author{Wim de Boer}
\email{Wim.de.Boer@kit.edu}
\author{L\'eo Bosse}
\email{leo.bosse@student.kit.edu}
\author{Iris Gebauer}%
 \email{Iris.Gebauer@kit.edu}
\author{Alexander Neumann}%
\email{alexander.neumann2@student.kit.edu}
\affiliation{Dept. of Phys., Karlsruhe Inst. for Technology KIT, Karlsruhe, Germany
}%

\author{Peter L. Biermann}
\email{plbiermann@mpifr-bonn.mpg.de}
\affiliation{MPI for Radioastronomy, Bonn, Germany
}%
\affiliation{Dept. of Phys., Karlsruhe Inst. for Technology KIT, Karlsruhe, Germany
}%
\affiliation{Dept. of Phys. \& Astr., Univ. of Alabama, Tuscaloosa, AL, USA}
\affiliation{Dept. of Phys. \& Astron., Univ. Bonn, Bonn, Germany}

\date{\today}% It is always \today, today,
\begin{abstract}

The so-called  ``GeV-excess'' of the diffuse Galactic gamma-ray emission, as observed by the Fermi-LAT satellite, is studied with a spectral template fit based on  energy spectra   for each relevant process of gamma-ray emission.  This has the advantage over ``conventional''œ analysis that one includes the spectral knowledge of physical processes  into the fit, which allows to determine 
simultaneously the standard background processes and contributions from non-standard processes, like the Fermi-
Bubbles or the ``GeV-excess'', in each sky direction. The spectral templates can be obtained in a data-driven way from the gamma-ray data, which avoids the  use of emissivity models to subtract the standard
background processes from the data. Instead, one can  determine these backgrounds simultaneously with any ``signals'' in any sky
direction, including the Galactic disk and the Galactic center.

Using the spectral template fit two hypothesis of the ``GeV-excess'' were tested: the dark matter (DM)  hypothesis assuming the excess is caused by DM annihilation and the molecular cloud (MC)  hypothesis assuming the ``GeV-excess''  is related to  a depletion of gamma-rays below 2 GeV, as is directly observed in the Central Molecular Zone (CMZ).  The origin of the depletion below 2 GeV is not important, but is most likely caused by a magnetic cutoff of cosmic rays approaching MCs, as will be discussed later. 

Both hypotheses provide acceptable fits, if one considers a limited field-of-view centered within 20$^\circ$   around the Galactic center and applies cuts on the energy range and/or excludes low latitudes,  cuts typically applied by the proponents of the DM hypothesis. However, if one considers the whole gamma-ray sky and includes gamma-ray energies up to 100 GeV we find that the MC hypothesis is  preferred over the DM hypothesis   for several reasons: i) The MC hypothesis provides significantly better fits; ii) The morphology of the  ``GeV-excess'' follows the morphology of the CO-maps, a tracer of MCs, i.e. there exists a strong  ``GeV-excess'' in the  Galactic disk also at large longitudes; iii) The  massive CMZ  with a rectangular field-of-view of $l \, \times \, b \; \simeq \; 3.5^{\circ} \, \times \, 0.5^{\circ}$ shows the maximum of the energy flux per log bin  in the diffuse gamma-ray spectrum at 2 GeV, i.e. the ``GeV-excess'', already in the raw data without any analysis. The rectangular profile contradicts the spherical morphology expected for DM annihilation.

\end{abstract}

%\pacs{95.85.Pw, 98.70.Rz, 95.35.+d;}% PACS, the Physics and Astronomy
\maketitle

%\tableofcontents
\onecolumngrid
\section{Introduction}\label{intro}
An apparent ``GeV-excess'' of diffuse gamma-rays in the data from the Fermi-LAT satellite around energies of 2 GeV towards the Galactic center  has been  studied by many  groups.  \cite{Goodenough:2009gk,Hooper:2010mq,Boyarsky:2010dr,Morselli:2010ty,Vitale:2011zz,Wharton:2011dv,Hooper:2012sr,YusefZadeh:2012nh,
Abazajian:2012pn,Hooper:2013rwa,Mirabal:2013rba,Huang:2013pda,Huang:2013apa,Gordon:2013vta,Macias:2013vya,Daylan:2014rsa,
Macias:2014sta,Lee:2014mza,Abazajian:2014fta,Abazajian:2014hsa,Calore:2014xka,Calore:2014nla,Cholis:2014lta,
Bartels:2015aea,Lacroix:2015wfx,TheFermi-LAT:2015kwa,Lee:2015fea,Cholis:2015dea,Hooper:2015jlu,DeBoer:2015yjh,Gaggero:2015nsa,Huang:2015rlu,Carlson:2016iis,Choquette:2016xsw,Yang:2016duy,TheFermi-LAT:2017vmf,Cuoco:2017rxb,Ploeg:2017vai} The ``GeV-excess''    is usually assumed to originate from the Galactic center with the most exciting interpretations being the contributions from dark matter (DM) annihilation \cite{Daylan:2014rsa}  and/or unresolved sources, like millisecond pulsars, see e.g. Refs. \cite{FaucherGiguere:2009df,Lee:2014mza,Lee:2015fea,Bartels:2015aea,Hooper:2015jlu} and references therein. 

 The  ``conventional'' approach to search for excesses is the use of  {\it spatial} templates for the gas and interstellar radiation field, which are the targets for the ``standard'' background processes: $\pi^0$ production by propagated cosmic rays (PCR), Bremsstrahlung (BR) and inverse Compton (IC) scattering. The diffuse gamma-ray emission is assumed to follow these spatial templates; the emissivity can either be calculated with  
 propagation models, as studied e.g. in Ref. \cite{Calore:2014xka} or one uses a diffuse model from Fermi, as done e.g. in Ref. \cite{Daylan:2014rsa}. The problem is that neither approach provides a good description of the gamma-ray emissivity in the  inner Galaxy and the Galactic disk, since the fitting of  spatial templates implies fitting over extended regions, thus averaging over rapidly varying emissivities, like the emissivity from molecular clouds (MCs) or unresolved sources. In these regions (MCs or unresolved sources) the gamma-ray emissivity varies  in intensity and as function of energy.  Inside MCs the maximum of the energy flux per log bin  in the diffuse gamma-ray spectrum is shifted from 0.7 GeV to up to 2 GeV, as is apparent from the spectrum of the Central Molecular Zone,  a dense assembly of MCs in the Galactic center with a total mass of $\rm 5\cdot10^7 ~M_\odot$ in the  solid angle limited by $-1.5^\circ\, <\, l\, <\, 2^\circ$ and $|b|\, <\, 0.5^\circ$ \cite{Tsuboi:1999,Jones:2011bv}. The density of molecules inside the CMZ is as high as $\rm 10^4/cm^3$. Further details on the CMZ can be found in a recent review \cite{2017arXiv170505332M} and references therein. The mass of the CMZ represents about 5\% of the total molecular mass of the Milky Way, so it is not surprising that  in the tiny solid angle of the  CMZ   the gamma-ray flux is dominated by the MCs. The origin of the shift in the maximum of the emissivity of MCs is not important, but it is most likely caused by a magnetic cutoff of cosmic rays inside MCs (MCRs) enhanced eventually with energy losses, as will be discussed later.
 
 Cosmic rays inside unresolved sources (SCRs) lead to a hard spectrum by the $\pi^0$ production in the shocked gas, as was first discussed in detail in Ref. \cite{Berezhko:2000vy}, which leads to a high energy tail in the gamma-ray emissivity. This high tail  was  investigated  in  Refs. \cite{deBoer:2014bra,DeBoer:2015yjh} and its correlation  with unresolved sources was apparent from the spatial correlation  with the 1.8 MeV line from $^{26}$Al, which traces sources. \cite{Prantzos1996}
 
 In the latest diffuse model from the Fermi collaboration \cite{Acero:2016qlg} the deficiency of low energy gamma-rays as well as the excess of high energy gamma-rays were taken into account ad hoc by an HI correction template with negative emission and an increase of emissivity above 50 GeV, but they did not realize the correlation with MCRs and SCRs.

In order to have a high spatial resolution, which can capture changes in emissivity from molecular clouds and unresolved sources we follow an approach orthogonal to the ``conventional'' approaches: instead of {\it spatial} templates we use {\it spectral} templates, one for each physical process describing the gamma-ray spectrum for that specific process. The reason for high and uncorrelated spatial resolutions is simple: including the spectral knowledge of all processes  leads to an over-constrained fit for each field-of-view in a certain sky direction (called cones in the following), since  the observed gamma-ray spectrum for a certain cone has  21 data points (= 21 energy bins) and  5  free normalization parameters for 5 physical processes, namely gamma-ray production by   PCRs, IC, BR, MCRs and SCRs.   With the fitting of spectral templates one can observe in each direction if there is an excess independent of neighboring directions. As a result one obtains an uncorrelated and  spatially highly resolved distribution of the ``GeV-excess''. We will show that the  shift in the maximum  of the energy flux per log bin  in the gamma-ray spectrum, or equivalently the ``GeV-excess'', is observed  in all directions, where MCs are present; these directions are available from the high resolution all-sky CO maps from the Planck satellite \cite{ThePlanck:2013dge}, which agree with previous CO sky maps. \cite{Dame:2000sp}

The  spectral templates for each physical process can be obtained in a data-driven way  from the gamma-ray data  \cite{deBoer:2014bra,DeBoer:2015yjh,Huang:2015rlu}, see Sect. \ref{templates}. In particular, the initial spectra for SCRs can be obtained from the high energy  tail in regions towards unresolved sources, as traced by the  $^{26}$Al line, while the inital spectral template for MCRs can be obtained from the CMZ. Initial spectral templates  for other background processes can be calculated from the locally observed cosmic ray spectra. The initial spectra for all templates are then optimized by the fit  to the data in an iterative procedure, so one obtains the spectral templates without having to rely on poorly fitting  diffuse models or propagation models.

How can one distinguish the two hypothesis for the  ``GeV-excess'': is it an excess  provided by   DM annihilation or a depletion  of low energy photons in MCs, as observed in the CMZ?  Both rely on the same observation, namely a shift in the maximum of the energy flux per log bin  in the gamma-ray spectrum  towards higher energies. Note that  the  ``GeV-excess'' is only observed along  lines-of-sight, so the spatial origin is not clear: it can originate either in the Galactic center, as expected for DM, or in the Galactic disk, as expected for MCs. This lines-of-sight argument also solves the problem that one observes the excess up to  large latitudes, although the MCs are located in the disk: the CO sky map shows a column density of MCs up to large latitudes as well with a  steeply falling latitude distribution,  as expected from the lines-of-sight crossing smaller parts of the disk with increasing latitude.  The latitude distribution from MCs resembles the latitude distribution from a DM annihilation signal with an NFW-like DM profile \cite{Navarro:1996gj}, as will be shown later.

A comparison of both hypotheses is the main goal of this paper. Is there a preference for one or the other? The result is:  if one only considers a limited field-of-view around the Galactic center and if one considers only a limited energy range of the gamma-ray spectrum (up to 10 GeV), both interpretations lead  to acceptable  fits of the data. However, if one considers as field-of-view the whole sky - and especially the whole disk, where the MCs reside - and the whole gamma-ray spectrum extending from 0.1 to 100 GeV it is clear that  the interpretation of MCs provides not only better fits, but the ``GeV-excess'' has a morphology following the MCs.

The observation that a DM template does not describe the spectrum of the ``GeV-excess'' for any WIMP mass was observed recently  by the Fermi Collaboration \cite{TheFermi-LAT:2017vmf} as well, who studied the spectrum up to 1 TeV. 
 They also observed the excess in the Galactic disk, but did not realize the correlation with MCs. The fact that the ``GeV-excess'' is so clearly observed in  the non-spherical CMZ - even in the raw data on diffuse gamma-rays without any analysis -  provides already evidence against the DM hypothesis, which predicts a centered and almost spherical morphology of the ``GeV-excess'' in the Galactic center. 

The paper is organized as follows: In Sect. \ref{anal}  the analysis procedure is described including the determination of the energy templates from the data; In Sect. \ref{results} a comparison of the two hypothesis (DM or MC) for the  ``GeV-excess'' are compared. The  comparison excludes that DM annihilation is the dominant source of the ``GeV-excess'',  as has been summarized in Sect. \ref{conclusion}. Spectral fits to all 797 uncorrelated cones   are given in the Appendices for both hypotheses (MC or DM).
%which are provided as  Online Supplemental Material (OSM). \cite{osm}

\section{Analysis  }\label{anal}
\subsection{Fermi data}
The data selection is as in our previous papers. \cite{deBoer:2014bra,DeBoer:2015yjh}
 We use  gamma-rays in the energy range between 0.1 and 100 GeV using the diffuse class of the public  P7REP\_SOURCE\_V15 data collected from August, 2008 till July 2014 (72 months) by the Fermi Space Telescope. \cite{Atwood:2009ez} The data were analyzed  with the recommended selections for the diffuse class using the  Fermi Science Tools (FST)  software. \cite{FST}  This included the  zenith angle cut of 100$^\circ$ to reduce Earth limb events and energy bins wider  than the LAT energy resolution (10 bins per decade), so no energy dispersion corrections are needed in this energy range.
 Gamma-rays converted in the front  and  back end of the detector were included. The residual hadronic background was included in the isotropic template. The point sources  from the second Fermi point source catalog \cite{Fermi-LAT:2011iqa} were subtracted using the {\it gtsrc} routine in the FST. 
 
The sky maps were binned in longitude and latitude in 0.5$^\circ \,\times\, 0.5^\circ$ bins, which were combined to form a total of 797 cones covering the whole sky. In and around the Galactic disk the cones were  one degree in latitude with a longitude size adapted to the structures, like the CMZ and the Fermi Bubbles. In the halo the cone size was increased in regions without structure, i.e. outside the Fermi Bubbles, typically to 18.5$^\circ$(10$^\circ$) for latitudes above 55$^\circ$(5$^\circ$), while the cone size in longitude was increased similarly. The precise cone sizes and fit results for each of the 797 cones are given in the fits in the Online Supplemental Material or can be estimated from the sky maps discussed later. From Fig. \ref{f7}(b) it can be seen that the binning is adequate to resolve the structure of MCs as function of longitude. 

The morphology of the ``GeV-excess'' is hardly smeared by the limited angular resolution of the Fermi-LAT instrument, given by the point-spread-function (PSF), since with our energy template fit the whole spectrum is fitted at once, so the energy dependence of the PSF is marginalized over implying that we are not sensitive to the larger PSFs at  the lower energies. The insensitivity to the PSF was checked by fitting the inner few degrees of the Galaxy, where the statistical error is small, with 0.5$^\circ \,\times\, 0.5^\circ$ bins. This did not change significantly the morphology. Hence, the morphology of the ``GeV-excess''  was not corrected for the smearing by the PSF. 

The gamma-ray flux is proportional to the product of the cosmic ray densities, the `target densities'' (gas or gamma-rays in the interstellar radiation field) and the cross sections. A template fit combines the product of these three factors into a single normalization factor  for each gamma-ray component $k$, thus eliminating the need to know them individually.

The total flux in a given direction can be described by a linear combination of the gamma-ray fluxes from various processes:
 \begin{eqnarray} |\Phi_{tot}>&=&n_1|\Phi_{PCR}>\  +\  n_2|\Phi_{BR}>\   +\  n_3|\Phi_{IC}> +\nonumber \\ &&  n_4|\Phi_{SCR}>\  +\ n_5|\Phi_{MCR}>\  +\ n_6|\Phi_{ISO}>, \label{e1}\end{eqnarray} where the normalization factors $n_i$ determine the fraction of the total flux for a given process:  PCR from the $\pi_0$ production by propagated cosmic rays, BR from Bremsstrahlung, IC from inverse Compton, SCR from the $\pi_0$ production by SCRs, MCR from the $\pi_0$ production  inside MCs and   ISO for the isotropic background.  In case one tests the DM hypothesis the MCR template is replaced by the DM template.

\begin{figure}
\centering
\includegraphics[width=0.3\textwidth,height=0.32\textwidth,clip]{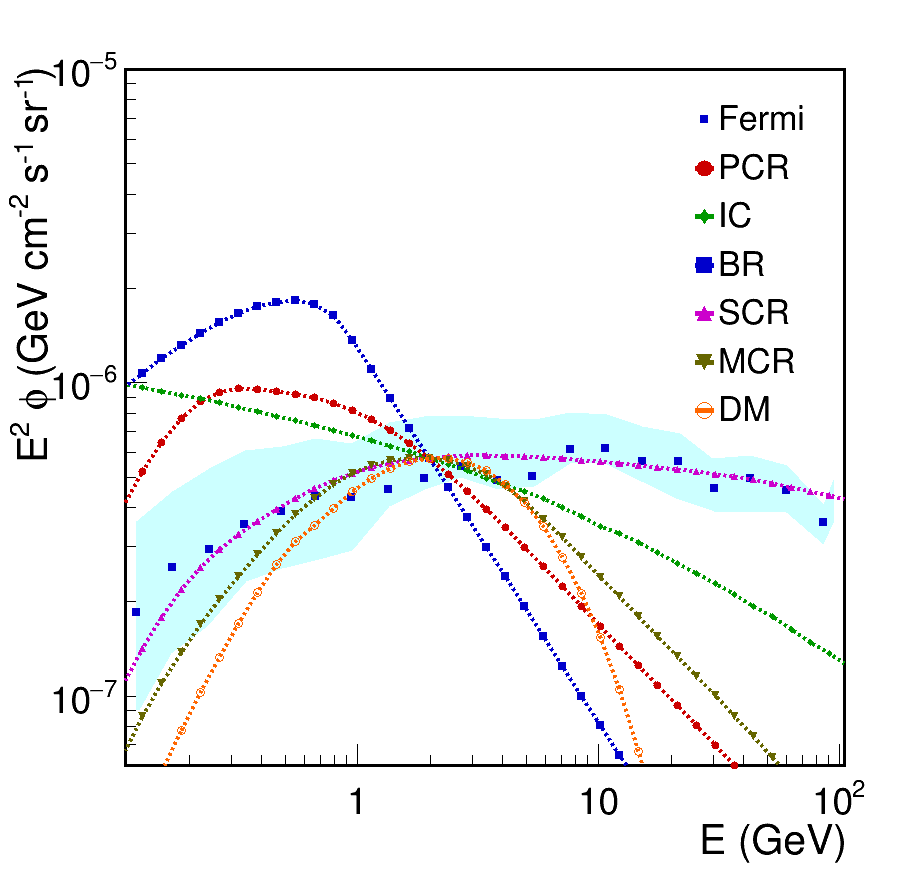}
\includegraphics[width=0.32\textwidth,height=0.32\textwidth,clip]{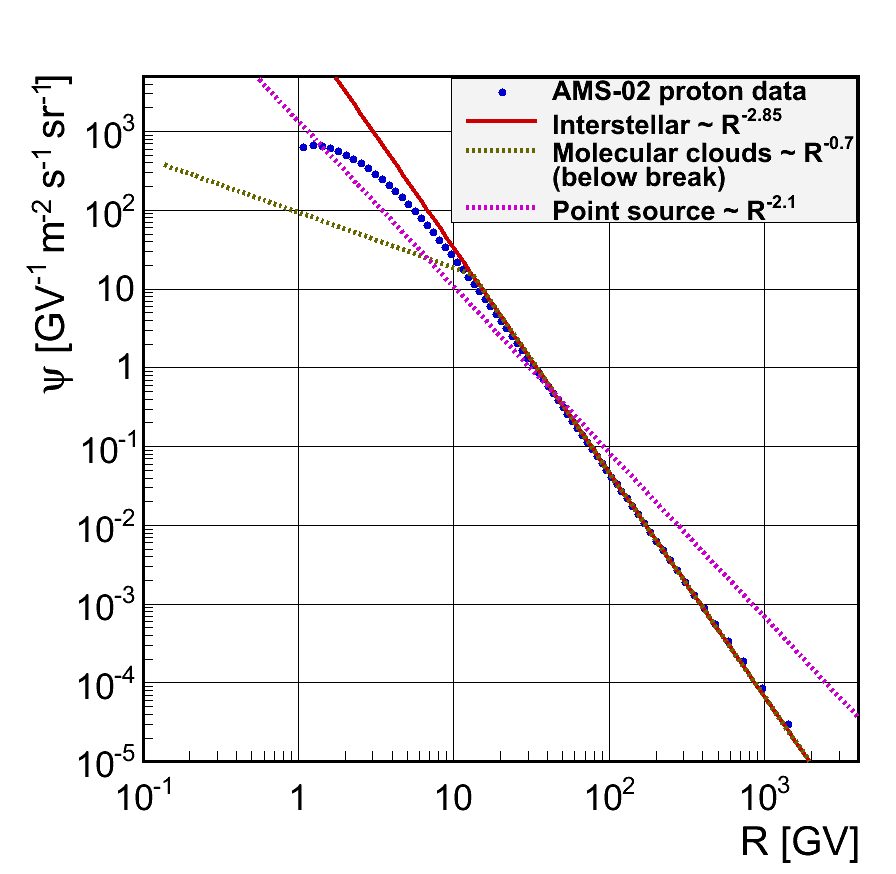}
\includegraphics[width=0.32\textwidth,height=0.32\textwidth,clip]{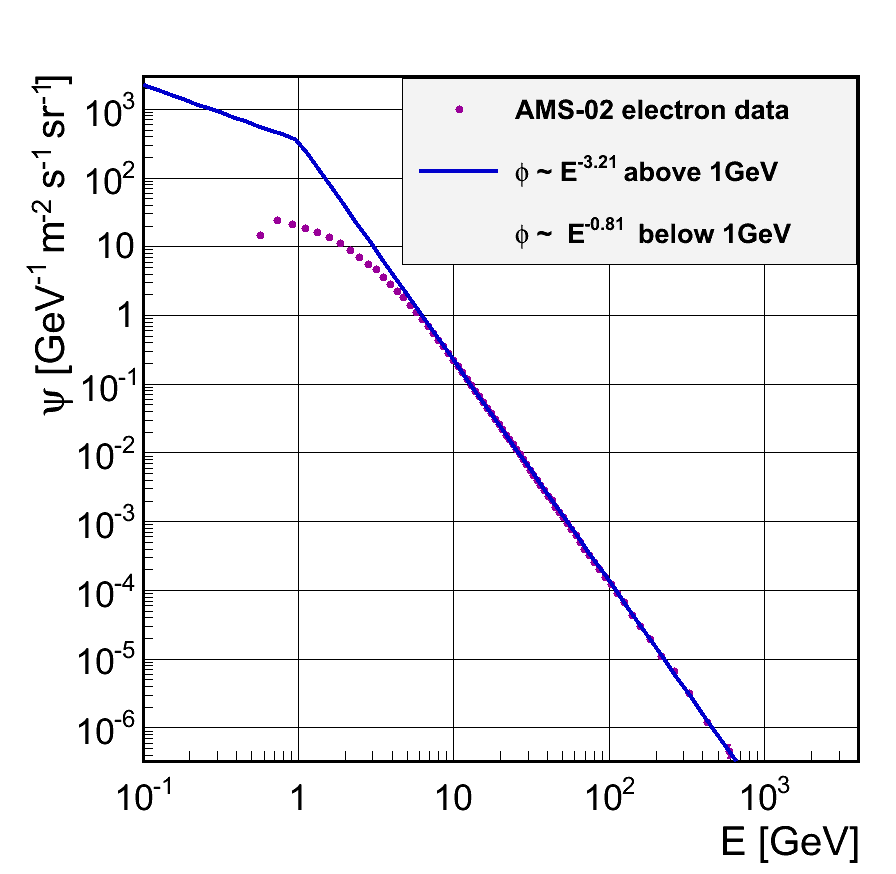}\\
(a) \hspace*{0.3\textwidth}(b)\hspace*{0.3\textwidth}(c)\\
\caption[]{(a)  Diffuse gamma-ray spectral templates fror   Bremsstrahlung (BR),  inverse Compton scattering (IC) and $\pi^0$ productions by propagated cosmic rays (PCRs), cosmic rays inside unresolved point sources (SCRs) and cosmic rays inside  MCs (MCRs). The templates are normalized around 2 GeV.  The blue band shows the allowed region for the gamma-ray spectra from the Fermi Bubbles \cite{Fermi-LAT:2014sfa}, which are consistent with the SCR template. Also shown is the gamma-ray template expected for a DM candidate with a mass of 45 GeV annihilating into $b\bar{b}$ quark pairs.
(b) Power law proton cosmic ray spectra describing the PCR, MCR, and SCR gamma-ray templates in (a). (c) A power law electron cosmic ray spectrum with a break at 1 GeV describing the IC and BR gamma-ray templates in (a). At high energies the power law cosmic ray spectra have spectral indices compatible with the ones from the locally observed electron and proton data from AMS-02 \cite{Aguilar:2014mma,Aguilar:2015ooa}, which   are shown as well. The spectra are normalized  at 70 GV.
  \label{f1}}
\end{figure}

 The factors $n_i$ can be found from a $\chi^2$ fit, which adjusts  the intensity $n_i$ of the gamma-ray fluxes from the spectral templates, one for each physical process, to best describe the data.  The spectrum of a each cone  has 21 energy bins  with only $n_i\le 6$  free parameters, so the fit is over-constrained. Furthermore, the templates  have quite  different shapes (see Fig. \ref{f1}(a)),  which allows a  determination of the gamma-ray flux for each process inside each cone.

\subsection{Test Statistic}\label{chi}
As test statistic we use the  $\chi^2$ function defined as
 \begin{equation} \chi^2=\sum_{i=1}^{N} \sum_{j=1}^{21}\left[\frac{\langle data(i,j)- \sum_{k=1}^{5} n(i,k) \times tem(i,j,k)\rangle^2}{\sigma(i,j)^2}\right], \label{e2}\end{equation}
where the sum is taken over the N=797 cones   in different sky directions $i$, $data(i,j)$ represents the total Fermi-LAT gamma-ray flux in direction $i$ for energy bin $j$, $tem(i,j,k)$ the template contribution with normalization $n(i,k)$ for template $k$  and  $\sigma(i,j)$ is the total error on $data(i,j)$, obtained by adding the statistical and systematic errors  in quadrature.

The recommended systematic errors  in the Fermi Software on the total gamma-ray flux  are 10\% for gamma-ray energies below 100 MeV, 5\% at 562 MeV, and 20\% above 10 GeV. We used  a linear interpolation for energies in between. With these large systematic errors the fit usually leads to too high probabilities resulting in a reduced $\chi^2/dof$ well below 1. Therefore, we followed the usual procedure of rescaling these errors in order to obtain $\chi^2/dof \approx 1$. This rescaling (by a factor 0.25) hardly affects
 the normalization of each template; it merely increases the $\chi^2/dof$ to $ \approx 1$. The systematic errors between the bins are correlated, which implies that all data points are allowed to move simultaneously up or down by an amount given by the correlated part of the systematic error. However,  a template fit with free normalizations for each template allows to move the fit up and down as well, which compensates a common shift in the data. So adding a correlated error in the data can slightly change the overall gamma-ray flux, but hardly affects the relative contributions of the various templates, as was verified by explicitly adding a covariance matrix to Eq. \ref{e2} with a common positive correlation between all bins, which was varied between 10\% and 70\% of the total systematic error. We did not vary the size of the correlation as function of energy, which would change the shape of the template. But the shapes are optimized from the data in an iterative way, as will be discussed in the next section.

\subsection{Determination of Spectral Templates}\label{templates}
The spectral templates for the various processes in Eq. \ref{e1} can be obtained from the gamma-ray spectra  in the following way: we assume that the leptons and nuclei follow a power law spectrum in the interstellar space with at most one break at a certain rigidity.
For the spectral index above the break we take as a first estimate the spectral index of the locally observed cosmic rays above 20 GV, a region which is not influenced strongly by solar modulation. \cite{Gleeson:1968zza} For given cosmic ray spectra the gamma-ray templates can be calculated by using e.g. the gamma-ray codes from standard propagation models, like Galprop \cite{Moskalenko:1998id,Vladimirov:2010aq} or  Dragon \cite{Evoli:2008dv}. These codes need as input the energy spectra of cosmic rays, which can either be obtained from a propagation model or one can simply provide cosmic ray power law spectra as input. In the latter case the gamma-ray codes are independent of the propagation model parameters. In our template fit we obtain the power laws of the interstellar cosmic ray spectra   from a fit to the gamma-ray data in an iterative way: we start with an initial cosmic ray spectrum parametrized with a broken power law, calculate the gamma-ray spectra, perform a fit, modify the cosmic ray spectra and fit again. This procedure is iterated until the best fit to the gamma-ray data is obtained. How to obtain the  initial cosmic ray spectra for each template will be discussed in the following sections. The resulting  templates and corresponding cosmic ray power law spectra have been summarized in Fig. \ref{f1}. 
\begin{figure}[]
\centering
\includegraphics[width=0.3\textwidth,height=0.27\textwidth,clip]{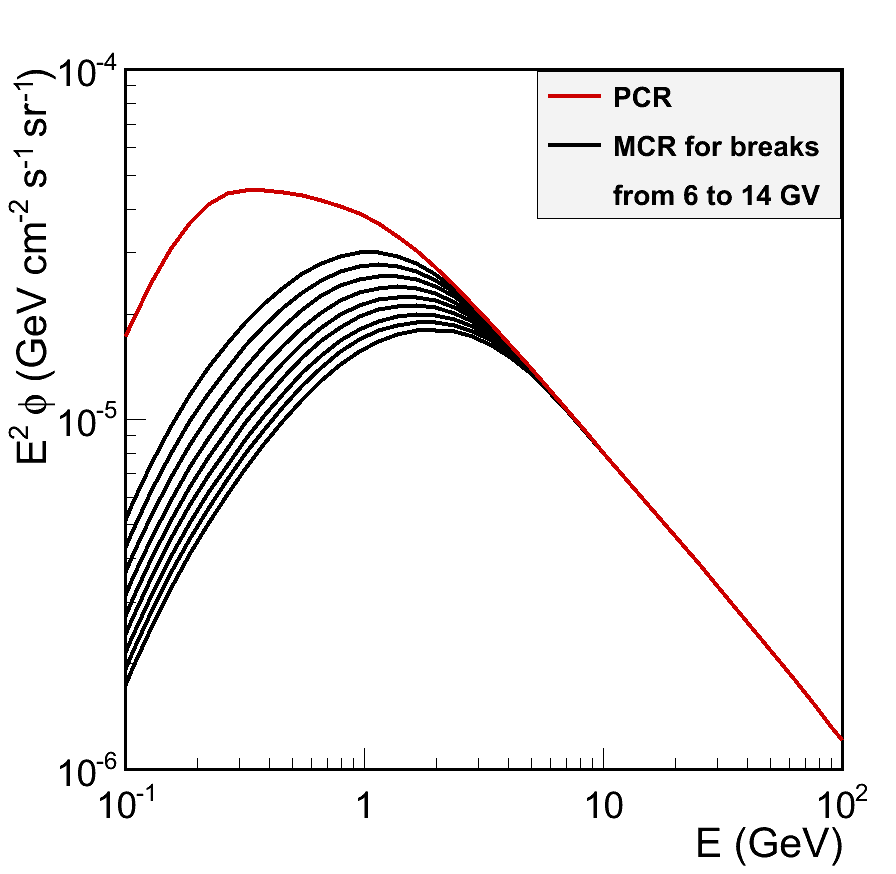}
\includegraphics[width=0.3\textwidth,height=0.27\textwidth,clip]{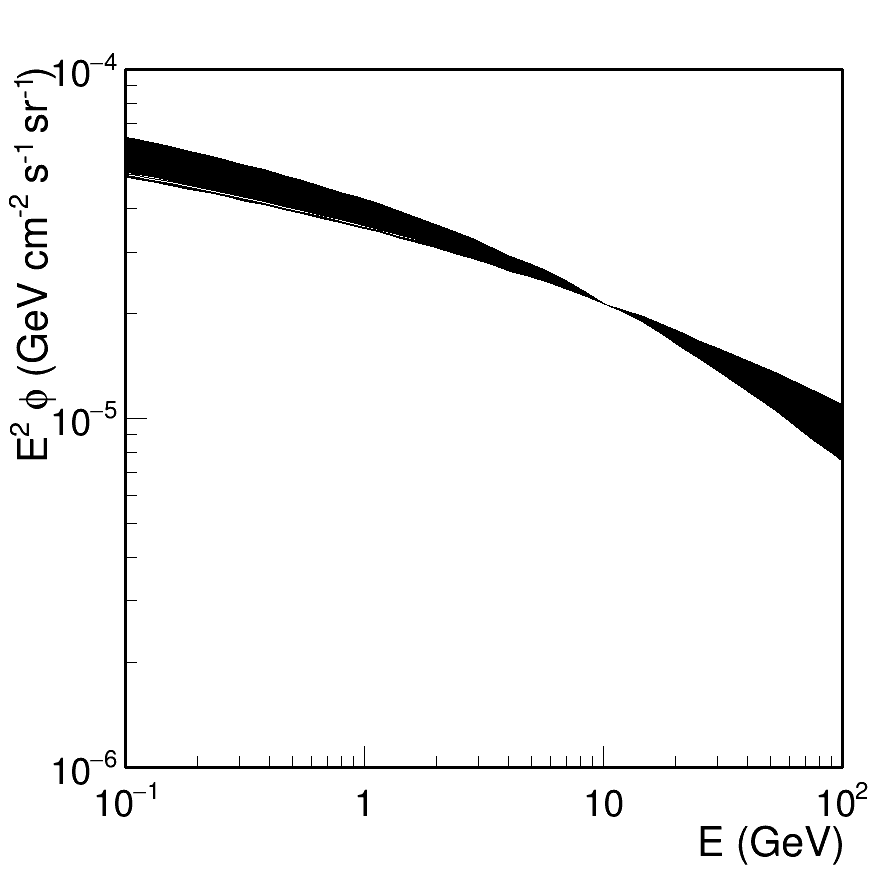}
\includegraphics[width=0.3\textwidth,height=0.27\textwidth,clip]{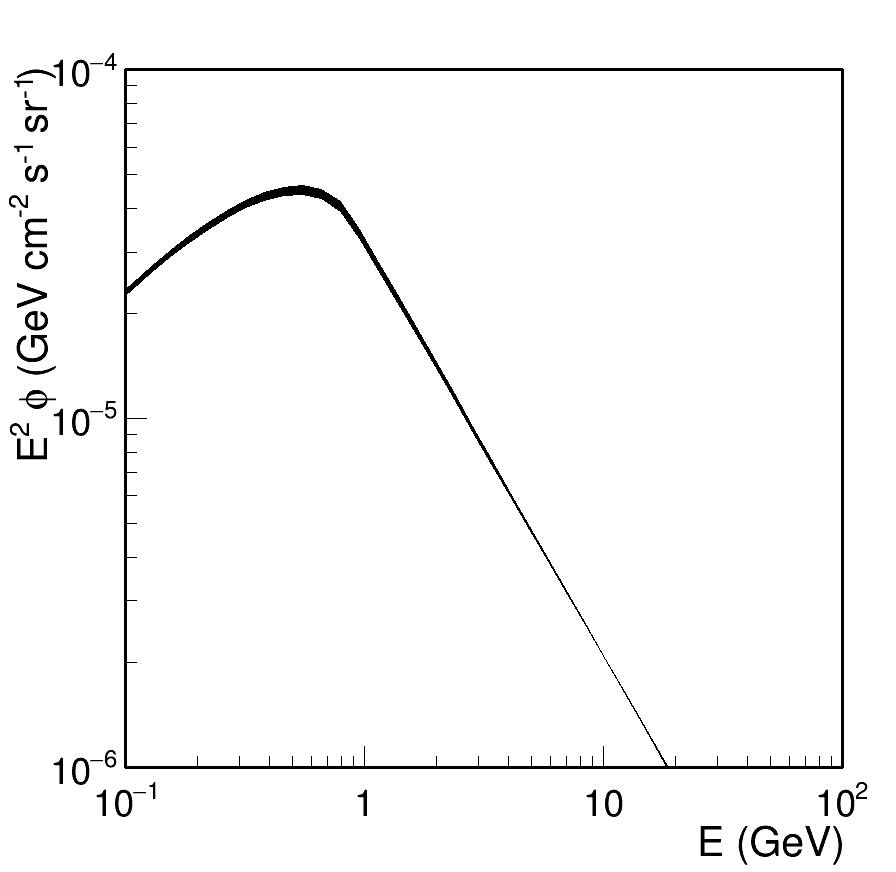}
	\caption[]{(a) MCR templates with different breaks in the proton spectra between 14 GV (maximum at the highest energy) and 6 GV in steps of 1 GV; for comparison the PCR template from Fig. \ref{f1}a is shown as well; (b) and (c):
 IC  and BR templates superimposed for all cones.
}
\label{f2}
\end{figure}

 \subsubsection{Details on the PCR  Template}
 The initial proton spectrum for the PCR template is obtained from the locally observed proton data from AMS-02 \cite{Aguilar:2015ooa}, which can  be approximated by an unbroken power law ($R^{-\alpha}$) with a spectral index ($\alpha$) of 2.85 at rigidities above 45 GV.
 At lower rigidities the data are below the power law because of solar modulation \cite{Gleeson:1968zza}, as can be seen from  Fig. \ref{f1}(b), where the AMS-02 data are plotted as well. To find the best parametrization a set of broken power laws with a grid of breaks and spectral indices above and below the break was constructed and the optimal parametrization was found by interpolation between the fits with the best test statistic.  The gamma-ray data are  well described by an unbroken power law for the protons with a spectral index ($\alpha$) of 2.85 at all rigidities.
  \subsubsection{Details on the SCR  Template}
 The proton spectra for the SCR template can be described by an unbroken power law  with a spectral index  of  2.1,  as obtained from the best fit.   The index 2.1 for the SCR template agrees with the data from the Fermi Bubbles, shown by the data points inside the shaded band in Fig. \ref{f1}(a); the index 2.1 is expected from diffuse shock wave acceleration. \cite{Hillas:2005cs,Biermann:2010qn} The fact that the Fermi Bubbles and the cosmic rays inside sources have the same spectrum strongly suggests that they are connected by point sources providing advective outflows of gas in the Galactic center. \cite{deBoer:2014bra} 
  \subsubsection{Details on the MCR Template}
 The decreasing gamma-ray emissivity from MCs below 2 GeV could be parametrized by a break in the power law of the corresponding proton spectrum. Above the break the optimal spectral index of 2.85 was found to be the same as for the PCR spectrum, as expected if the high energy propagated protons are above a certain magnetic cutoff.  But below the break, which varies according to the fit  from 13 to 6 GV for the different clouds, the optimal spectral index is 0.7, thus providing a significant suppression of protons below the break, as can be seen from Fig. \ref{f1}(b). Energy losses alone cannot reproduce such a  suppression of the proton spectrum below the break,  but magnetic cutoffs are able to do so.  Such a cutoff is well known from cosmic rays  entering the Earth's magnetic field:   particles below typically 20 GV  entering near the magnetic equator  do not reach the Earth, but are repelled into outer space  by the geomagnetic cutoff. \cite{Herbst:2013hr} The rigidity cutoff of 20 GV is proportional to the magnetic moment. Although the magnetic field near the Earth (0.5 G) is  orders  of magnitude higher than the typical magnetic fields in dense MCs \cite{2012ARAA..50...29C}, the much larger sizes of MCs - or its  substructure of filaments and cloudlets \cite{2000prpl.conf...97W} - yield magnetic moments  of the same order of magnitude as the Earth's magnetic moment, so similar magnetic cutoffs are plausible.   Variations in the magnetic cutoff in MCs are expected from the variations in size and in magnetic field; the latter increases with MC density. \cite{2012ARAA..50...29C} The variations  of the break in the proton spectrum between 13 and 6 GV varies the maximum of the gamma-ray spectrum from 2 to 1 GeV, as shown in Fig. \ref{f2}(a).  The fit prefers a constant spectral  index below the break  for all sky directions. Such a constant spectral index is plausible with regular magnetic fields  oriented in the disk \cite{Heiles:2005hr,Jansson:2012pc} and the ``cloudlets'' inside MCs \cite{2000prpl.conf...97W} form magnetic dipole fields. Then the maximum  cutoff occurs for cosmic rays entering  from the halo perpendicular into the cloud for any  orientation of the magnetic dipole.  For a given entrance angle the cutoff would provide a sharp break, but for an isotropic distribution of entrance angles the break points are smeared. A distribution of break points will provide a slope below the maximum break determined largely by the isotropic distribution of the entrance angles into the disk. Since this distribution is the same for all MCs the slopes below the break will be similar for all MCs, even if the maximum break (= maximum magnetic cutoff) varies.
 \subsubsection{Details on the BR and IC Templates}
The interstellar electron spectra needed a break around 1 GeV with a spectral index of 3.21 above the break, which is compatible with the locally observed electron spectrum (see Fig. \ref{f1}(c)); below the break the optimal  spectral index is 0.81, which implies a suppression of electrons. The break point might be related to the fact that around 1 GeV electrons have the smallest energy losses, since above this energy synchrotron, BR and IC dominate the energy losses, while below this energy  ionization losses become strong, thus depleting the electron spectrum below 1 GeV.  A similar break in the electron spectrum was needed in the Fermi diffuse model. \cite{Acero:2016qlg}

The targets for the production of gamma-rays are the interstellar gas and the interstellar radiation field. The latter consists of photons from the cosmic microwave background, the infrared radiation from hot matter, like dust and the star light, so the photon composition varies with sky direction. Hence, for the IC templates we have to calculate the templates for each sky direction. The variation over the sky is about $\pm$10\%, as shown  in  Fig. \ref{f2}(b); for the BR template the spread is considerably smaller, as shown  in  Fig. \ref{f2}(c). Note that the intensity of photons in the interstellar radiation field nor the gas density play any role in a template analysis, since the intensity of each contribution to the gamma-ray sky is determined by the fitted normalization factors in Eq. \ref{e1}.
\begin{figure}[]
\centering
\includegraphics[width=0.45\textwidth,height=0.35\textwidth,clip]{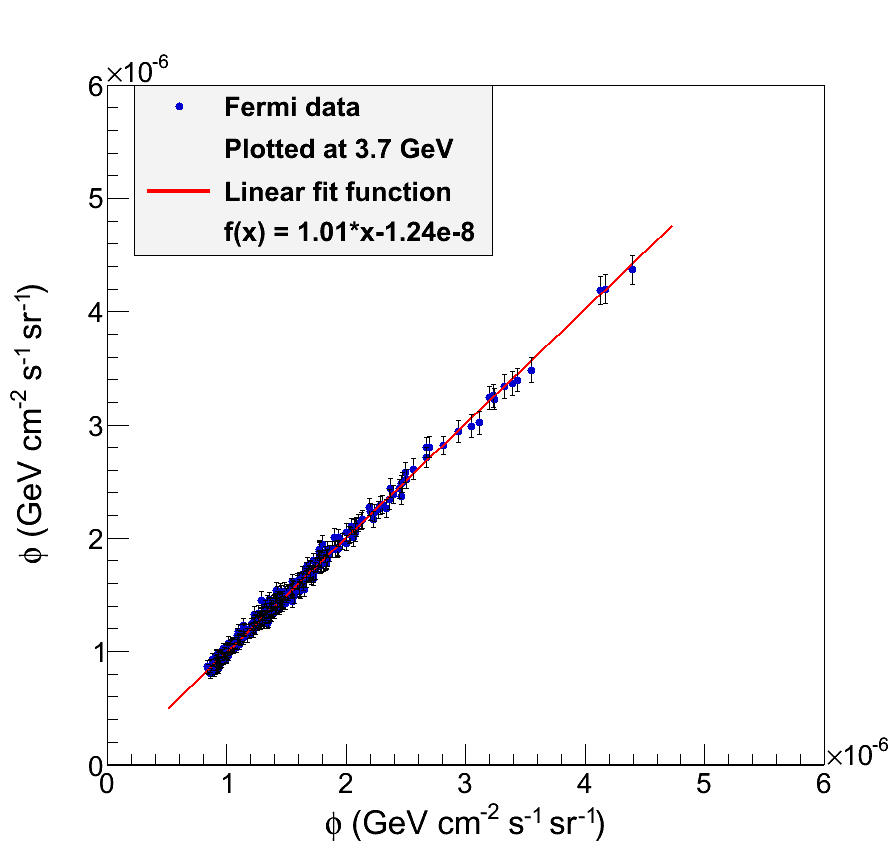}
\includegraphics[width=0.45\textwidth,height=0.35\textwidth,clip]{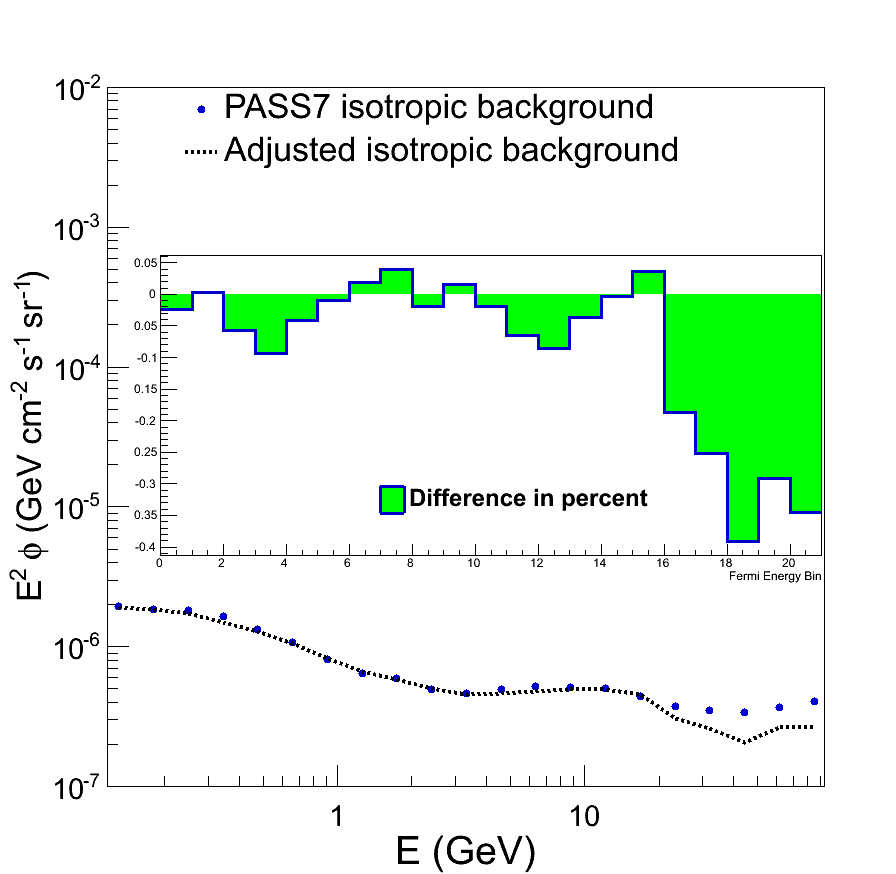}\\
(a) \hspace*{0.46\textwidth}(b)
	\caption[]{(a): The observed data versus the fitted data in various sky regions (i.e. for various gamma-ray fluxes) for a given energy, here for the energy bin between 3.7-5.2 GeV. The offset of a linear fit at the vertical axis represents the isotropic component in the data, which moves the data in all bins upwards by this amount. 
	A similar fit is repeated for all energy bins, so one determines the offset for each energy, which yields the isotropic template. (b) A comparison of the isotropic template used in our analysis and the isotropic template given in the Fermi software with the insert showing the relative difference.
}
\label{f3}
\end{figure}
 \subsubsection{Details on the Isotropic Template}
The isotropic template represents the contribution from the isotropic extragalactic background and hadron misidentification. Its spectral shape and absolute normalization are provided within the Fermi software. \cite{FST}  The isotropic template was redetermined for our analysis in the following way. We fit the data in regions outside the Bubbles and Galactic disk using  the isotropic template from the Fermi software as an initial estimate  in the fit. If one plots  the total observed gamma-ray flux versus the fitted flux in the various cones in a certain energy bin, one expects a linear relation crossing the origin, if the isotropic flux is estimated correctly. However, if there is a missing or too high isotropic contribution, this leads to an offset at the origin of the linear curve, since the isotropic component is by definition the same for all cones, so it shifts the whole curve up and down for each energy bin.   An example of such a fit is shown in Fig. \ref{f3}(a) for an energy bin between 3.7-5.2 GeV. The offset can be determined for each energy bin, which yields the  spectral template of the isotropic component. The final spectral template is obtained by iteration untill zero offset at the origin is reached. The resulting template in our analysis has  deviations from the Fermi template up to 35\% above 2 GeV, as shown in the insert of Fig. \ref{f3}(b). 

 \subsubsection{Details on the DM Template}
DM particles  are expected to annihilate and just like in electron-positron annihilation the annihilation energy of roughly twice the WIMP mass will lead to the production of hadrons, thus producing copiously gamma-rays from $\pi^0$ decays. A smaller fraction of WIMP annihilation is expected to lead to tau lepton pairs, which can lead to $\pi^0$ production in the hadronic tau decays. This contribution is expected to be small and is neglected. The DM template can be calculated with DarkSusy. \cite{Gondolo:2004sc,Gondolo:2005we} The annihilation signal peaking at 2-3 GeV requires a WIMP mass around 45 GeV, as shown in Fig. \ref{f1}(a) as well. The difference to the MCR template is the cutoff at twice the WIMP mass, which is absent in the MCR template.
\begin{figure}
\centering
\includegraphics[width=0.45\textwidth,height=0.38\textwidth,clip]{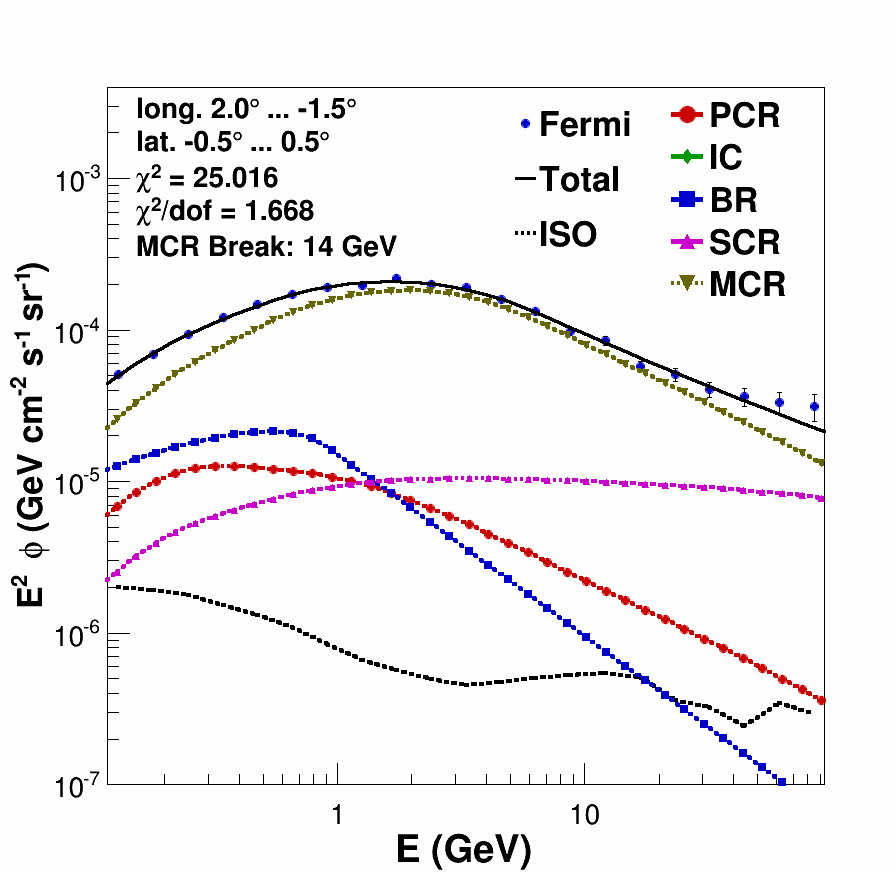}\hspace*{0.5mm}
\includegraphics[width=0.45\textwidth,height=0.38\textwidth,clip]{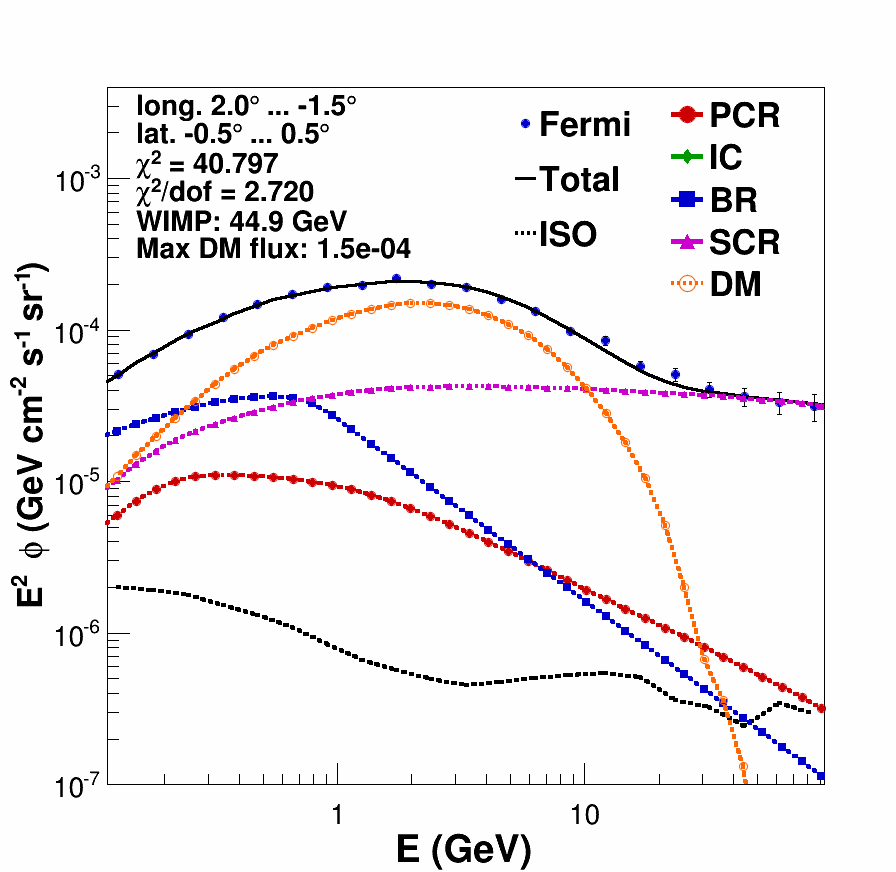}\\
(a) \hspace*{0.46\textwidth}(b)\\
\includegraphics[width=0.45\textwidth,height=0.38\textwidth,clip]{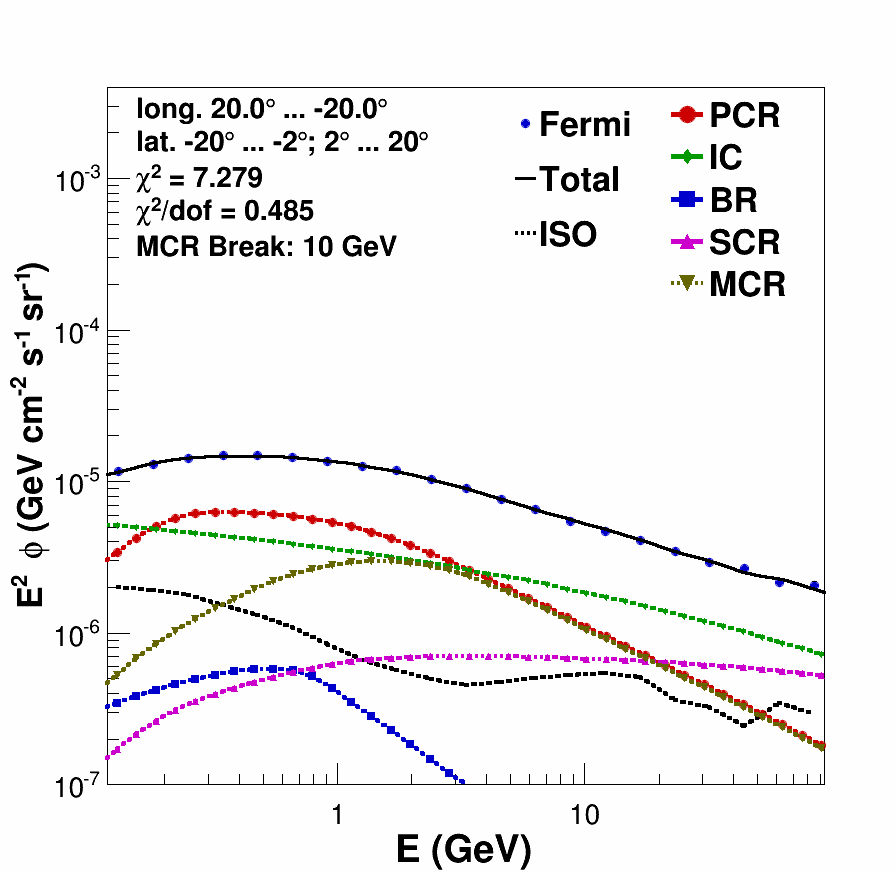}
\includegraphics[width=0.45\textwidth,height=0.38\textwidth,clip]{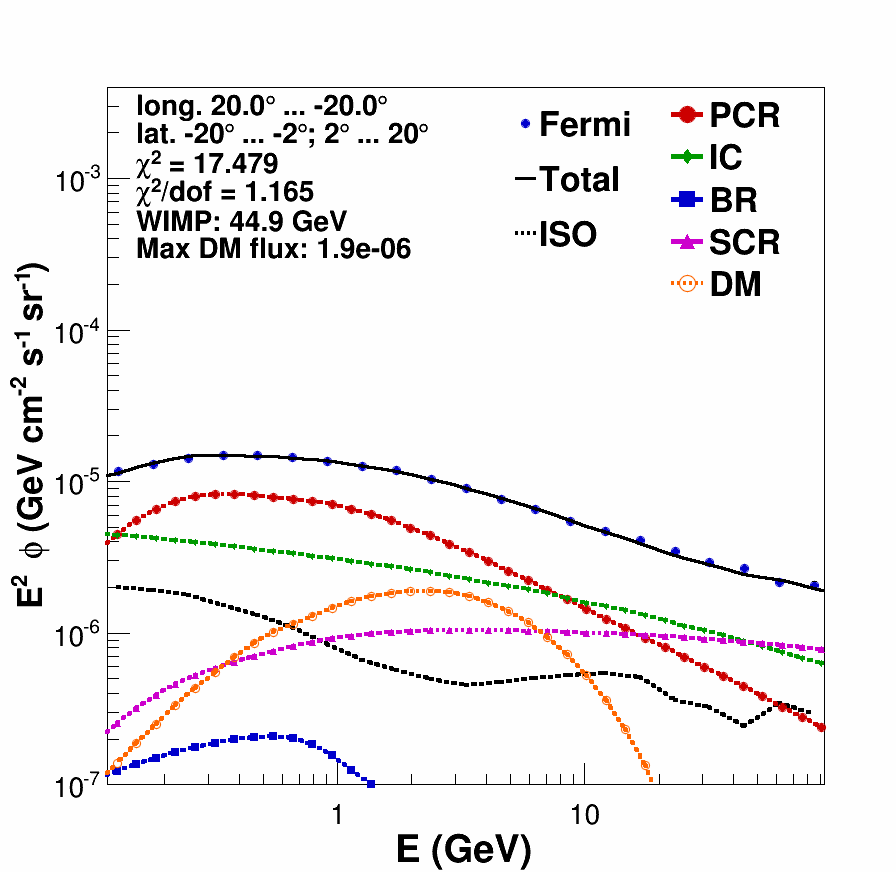}\\
(c) \hspace*{0.46\textwidth}(d)\\
\includegraphics[width=0.45\textwidth,height=0.38\textwidth,clip]{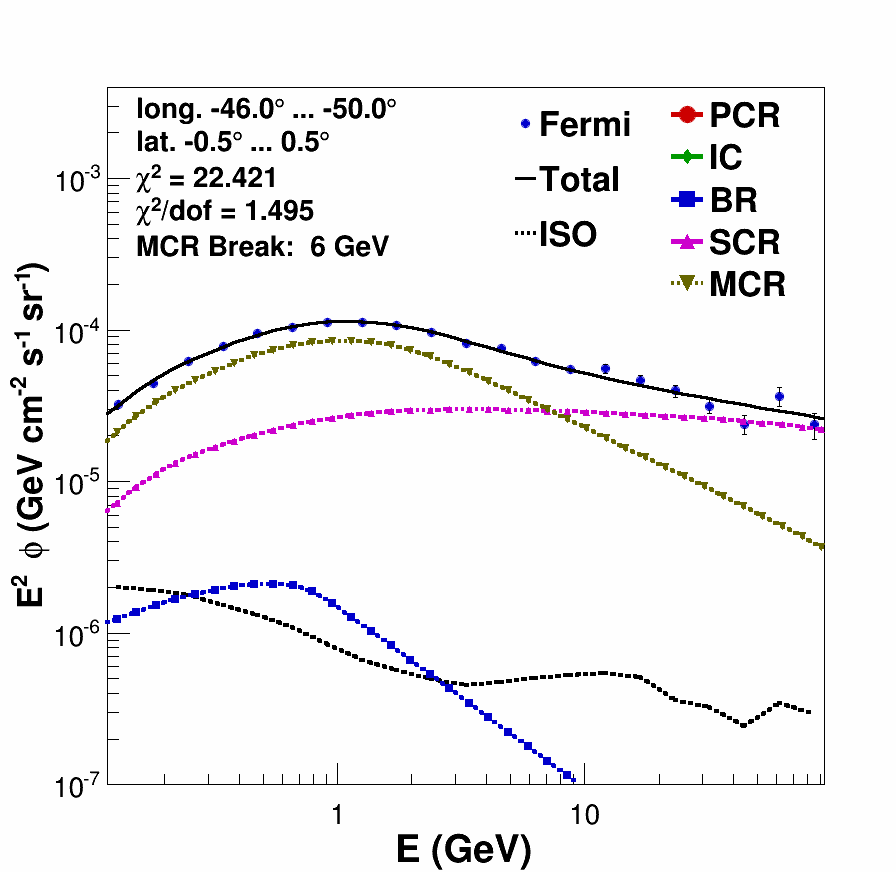}
\includegraphics[width=0.45\textwidth,height=0.38\textwidth,clip]{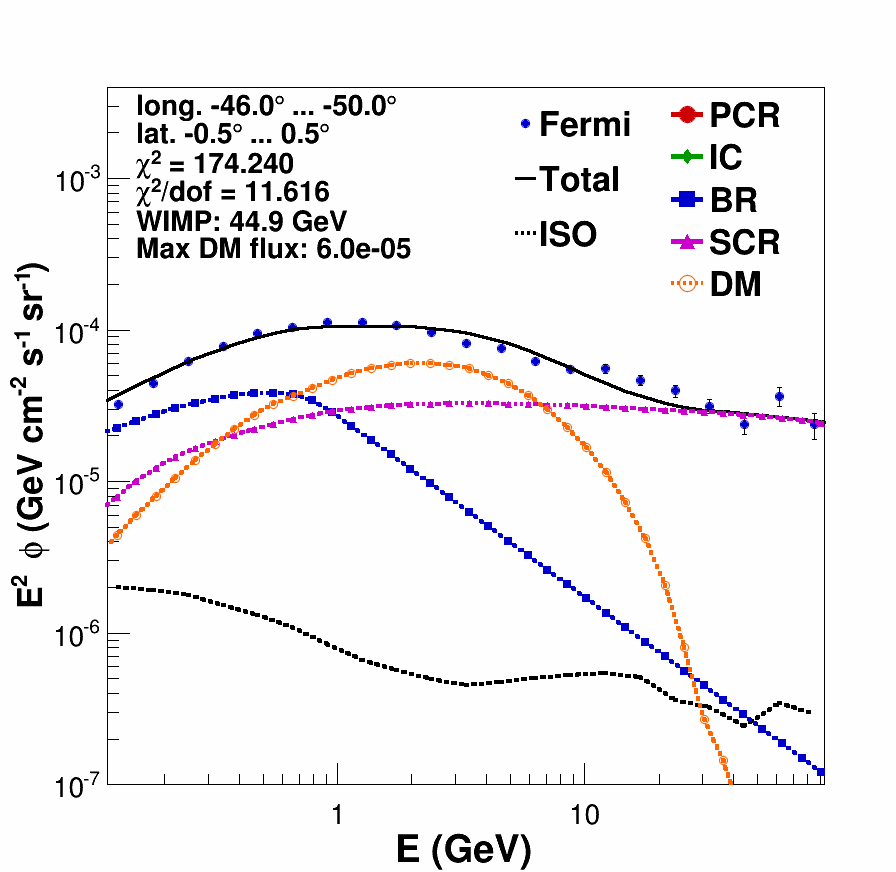}\\
(e) \hspace*{0.46\textwidth}(f)
\caption[]{Spectral template fits to the Fermi  diffuse gamma-ray data for the following regions of interest:  towards the CMZ (top row), halo (middle row) and the nearby tangent point of the Scutum-Centaurus arm (bottom row) using either the MCR template (left) or DM template (right). 
 \label{f4}
}
\end{figure}
\section{A comparison of the Fit Results with the MCR and DM Templates}\label{results}
As mentioned before, the ``GeV-excess'' can be explained by an excess at energies around a few GeV from DM annihilation or by a depletion of gamma-rays below 2 GeV from the gamma-ray emissivity from MCs. The first process would correspond to a process in the Galactic center, the second one to a process in the Galactic disk. But since we observe the excess along the lines-of-sight, these explanations are at first sight indistinguishable. However, there are two important differences: i) the DM and MCR templates differ significantly in shape above 50 GeV, so the test statistic might distiguish between them; ii) the MCs are distributed in the disk, the DM is distributed approximately spherically around the Galactic center.
So to distinguish between the two hypothesis one can either perform the fit with an MCR template or alternatively with a DM template including of course the other ``background'' templates (PCR,SCR,IC,BR, ISO) and perform a fit over the whole sky and all gamma-ray energies.
If the ``GeV-excess'' is dominated by DM annihilation one expects to see the cutoff at twice the WIMP mass in the data and a spherical distribution of the intensity of ``GeV-excess'', characterized by the normalization of the DM template in the fit. If the ``GeV-excess'' originates from the emissivity inside MCs, one expects a strong ``GeV-excess'' inside the disk  and no hint for a cutoff in the  high energy gamma-ray data.

The templates of all physical processes are allowed in the fit for all cones.
The fit is supposed to find out if the expected backgrounds from the PCR, BR, IC and ISO templates fit the data or if the maximum of the spectrum is shifted (a feature recognized by the DM or MCR template) or if the data has a high energy tail above the expectations from the known backgrounds (a feature recognized by the SCR template). 
\begin{figure}
\centering
\includegraphics[width=0.46\textwidth,height=0.46\textwidth,clip]{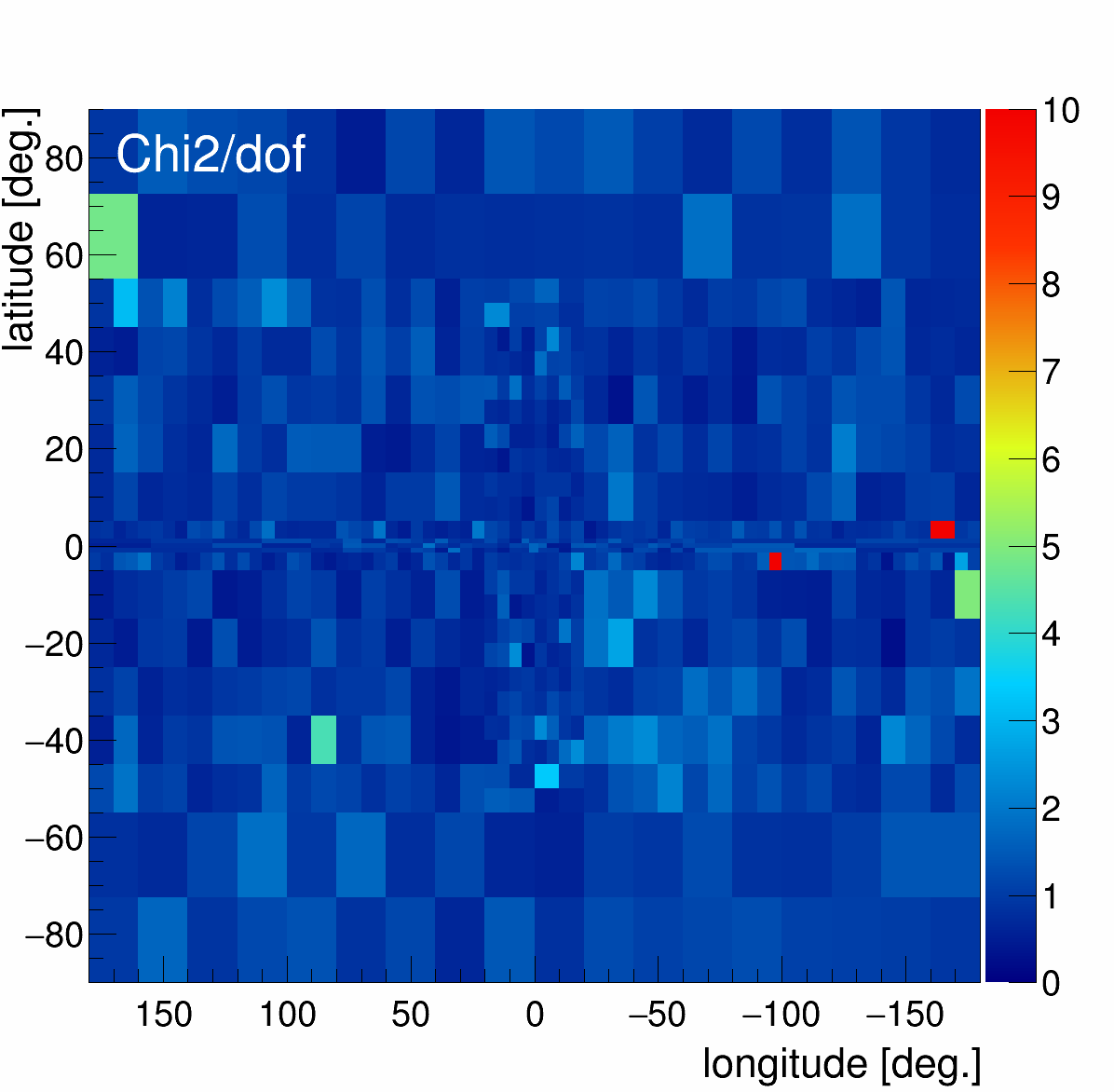}
\includegraphics[width=0.46\textwidth,height=0.46\textwidth,clip]{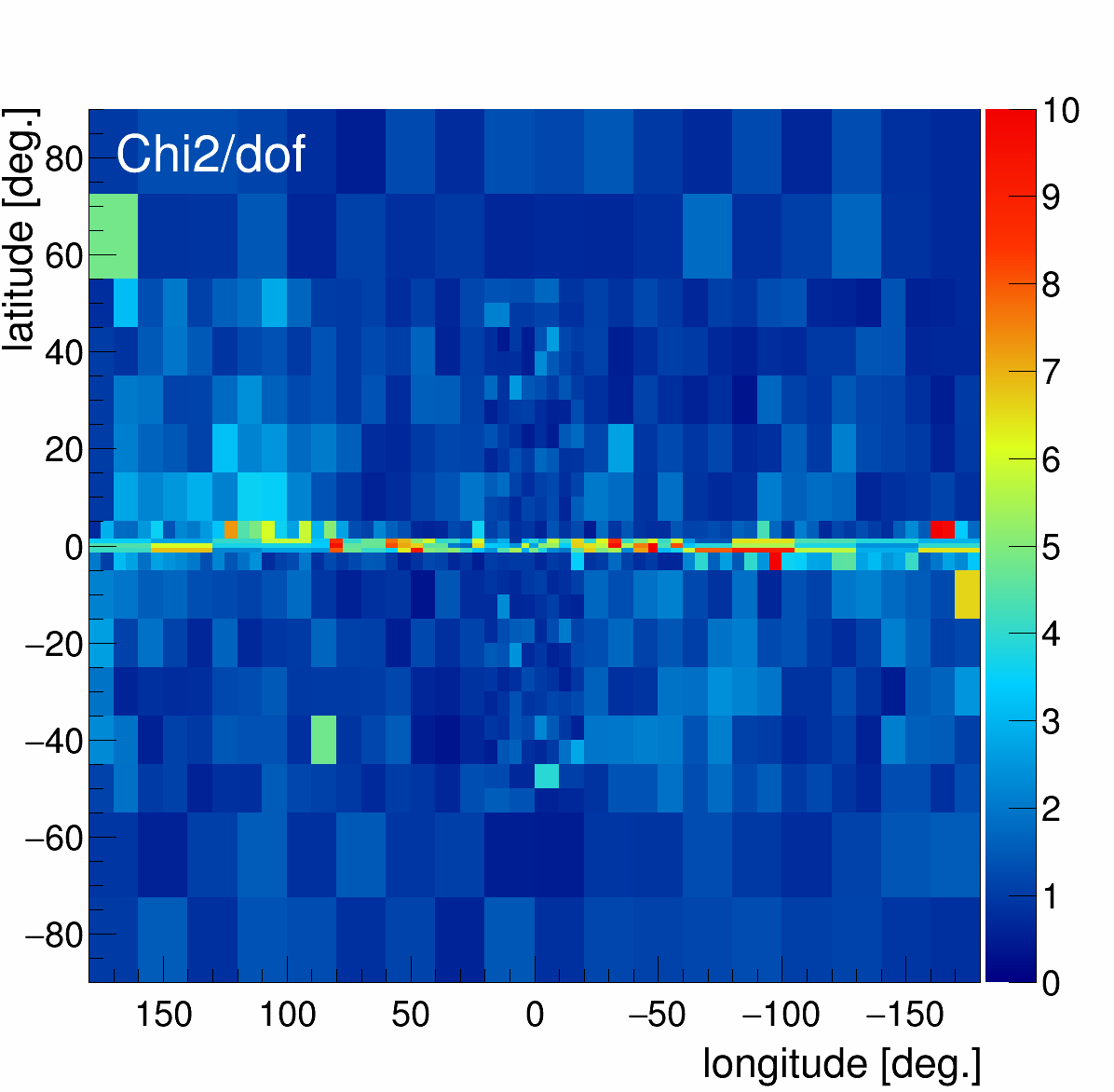}
\hspace*{0.04\textwidth}(a)\hspace*{0.46\textwidth} (b)\\ \vspace*{-1mm}
\caption[]{$\chi^2/dof$ values of fits in all 797 cones with either the MCR template (a) or a DM template for a DM candidate with a mass of 44.9 GeV annihillating into $b\bar{b}$  quark pairs (b). 
}
\label{f5}
\end{figure}
\begin{figure}
\centering
\includegraphics[width=0.46\textwidth,height=0.46\textwidth,clip]{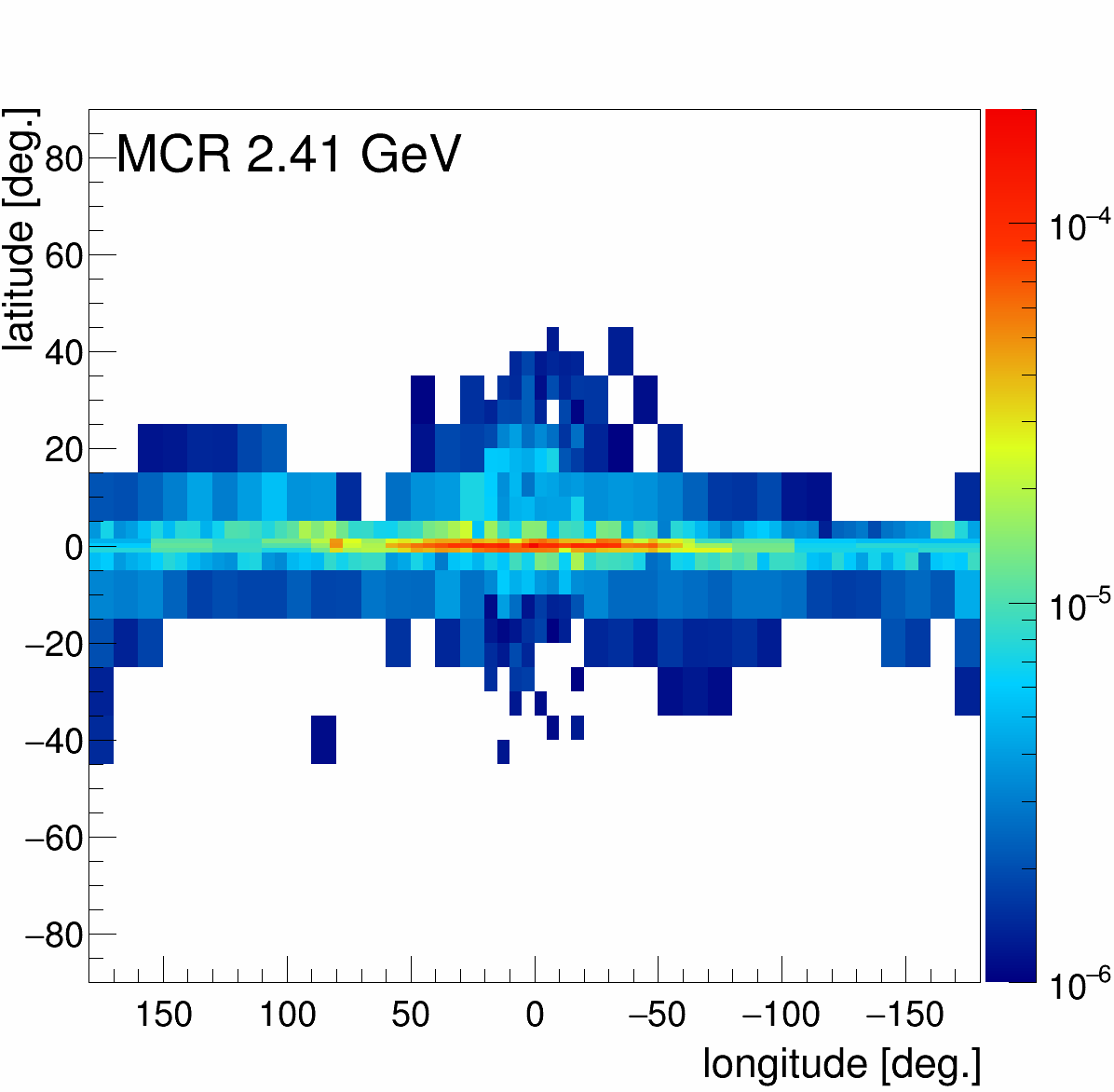}
\includegraphics[width=0.46\textwidth,height=0.46\textwidth,clip]{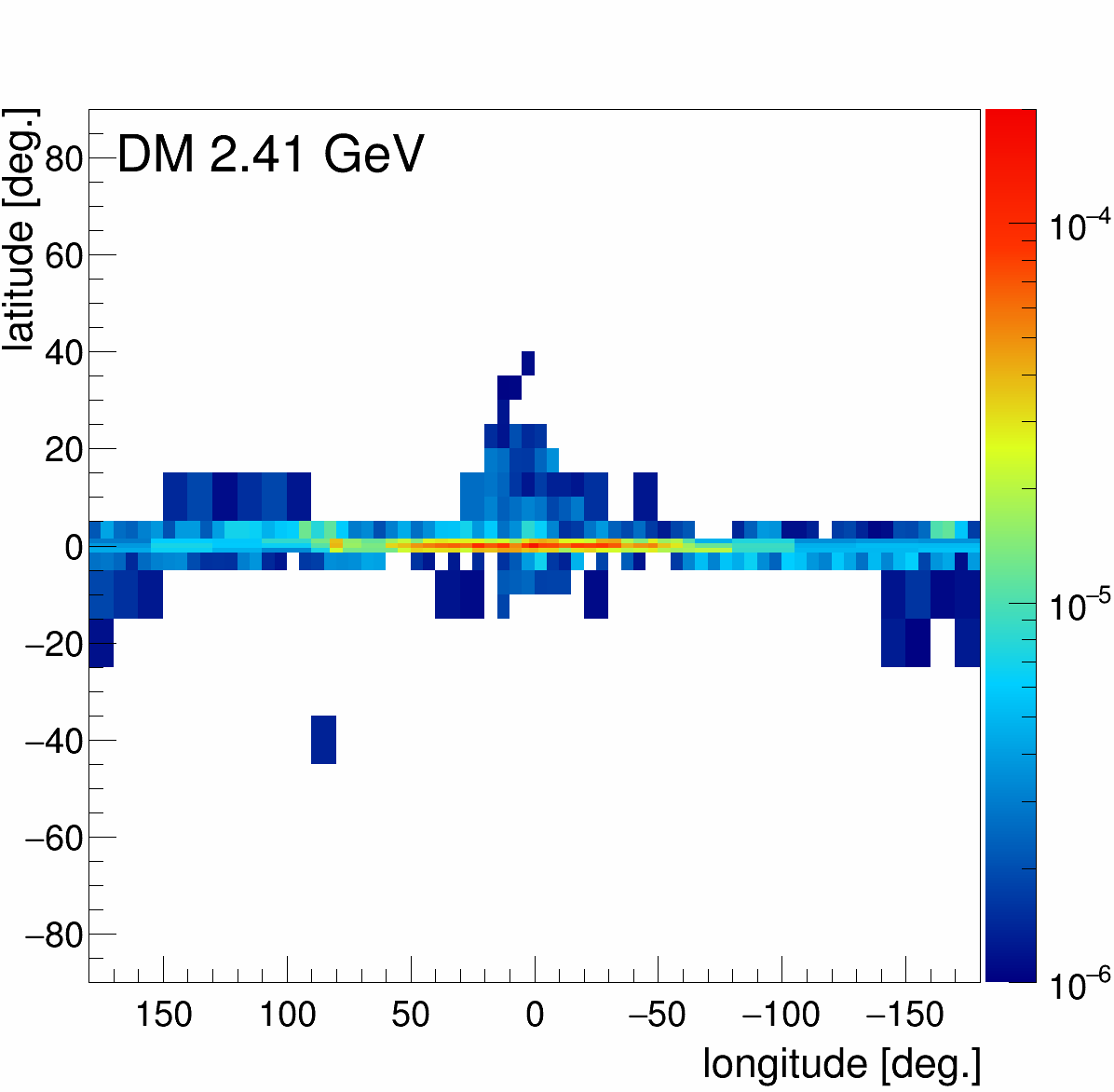}
\hspace*{0.04\textwidth}(a)\hspace*{0.46\textwidth} (b)\\ \vspace*{-1mm}
\caption[]{Sky maps of the fluxes of the MCR (a)  and DM for a DM candidate with a mass of 44.9 GeV annihiilating into $b\bar{b}$  quark pairs (b).  The fluxes are  in units of $\rm GeV cm^{-1} s^{-1} sr^{-1}$ at an energy of 2.41 GeV.
}
\label{f6}
\end{figure}

\begin{figure}
\centering
%\vspace*{-8cm}
\begin{minipage}[]{0.45\textwidth}\centering
\includegraphics[width=0.9\textwidth,height=1.4\textwidth,clip]{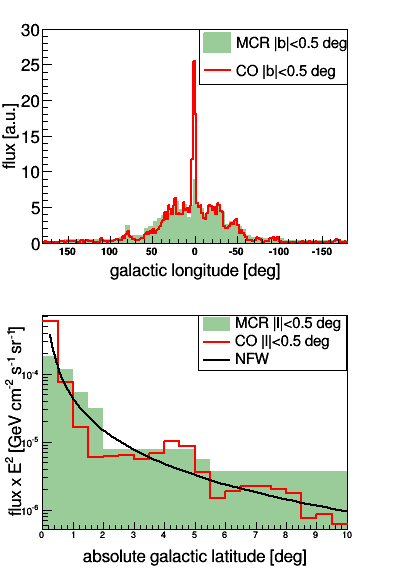}\\
\center{(a)}
\end{minipage}
\begin{minipage}[]{0.45\textwidth}\centering%\vspace*{5mm}
\includegraphics[width=0.9\textwidth,height=1.40\textwidth,clip]{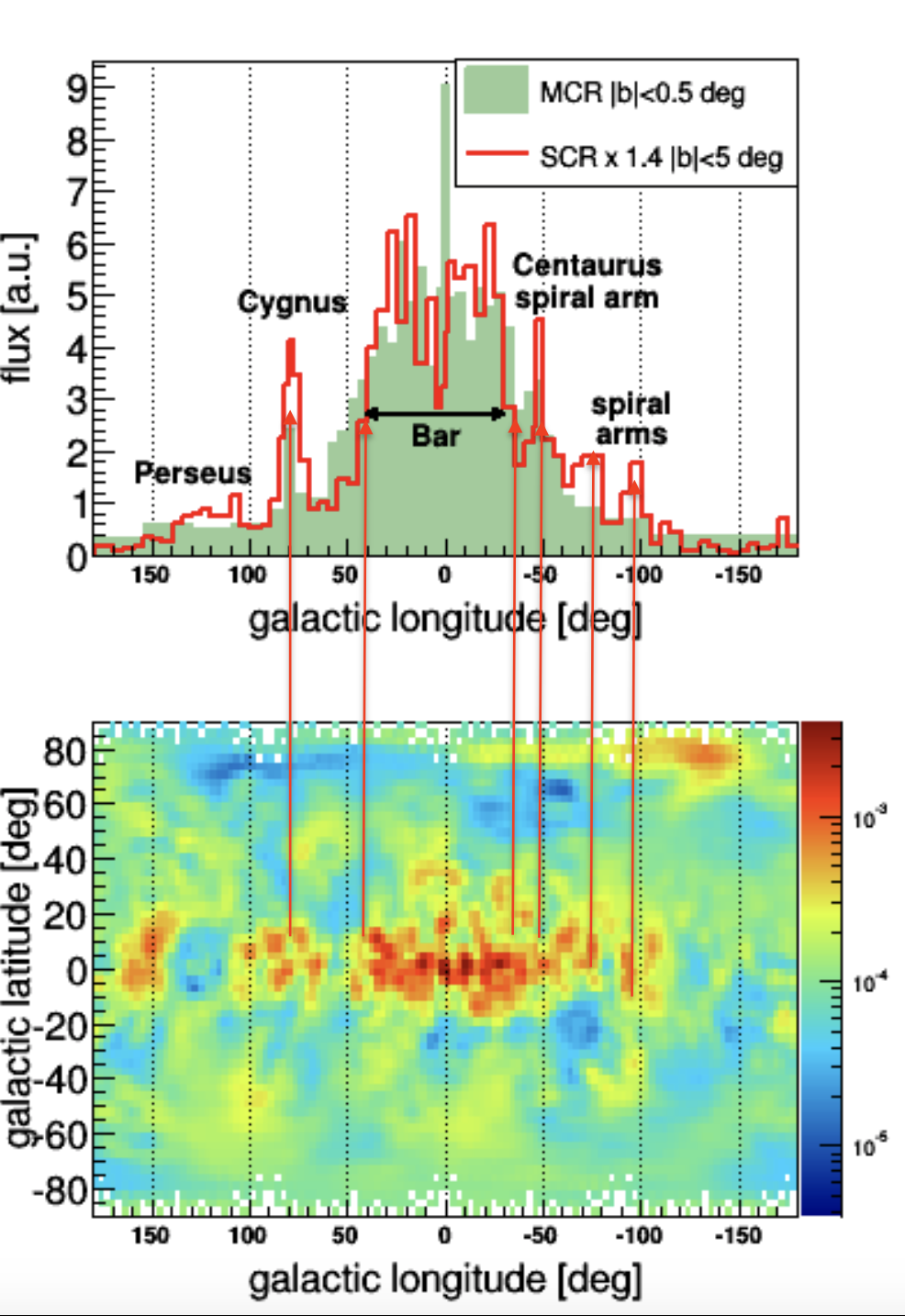}\\
\center{(b)}
\end{minipage}
	\caption[]{A summary of the fourfold correlation between the ``GeV-excess'' (= MCR), the CO maps, the $\pi^0$ production in unresolved point sources (= SCR) and the point source distribution traced by the $^{26}$Al line: (a) Longitude and latitude of   MCR (green histogram, this analysis) and CO (red line). \cite{Planck} The latitude distibution of the MCR flux at an energy of 2.41 GeV was integrated over a longitude range of $|l|\, <\, 0.5^\circ$. The black line in the bottom panel corresponds to the NFW template from the ``conventional''œ analysis, adapted from Fig. 1 from Ref.  \cite{Calore:2014nla}. One observes that the CO latitude distribution from the Planck satellite resembles  an average NFW profile with some clumpiness. (b):  the
 longitude distribution of the fluxes from the SCR and MCR templates, which have a similar morphology,  as expected since  both are connected to MCs. The lower panel  shows the  $^{26}$Al  sky map \cite{Bouchet:2015rxa,Spi}, which is correlated with the   top panel, as indicated by the vertical arrows. This correlation is expected, since both, the $^{26}$Al flux and the SCR flux, are tracers of cosmic ray sources. 
}
\label{f7}
\end{figure}
%%%
\begin{figure}
\centering
\includegraphics[width=0.7\textwidth,height=0.6\textwidth,clip]{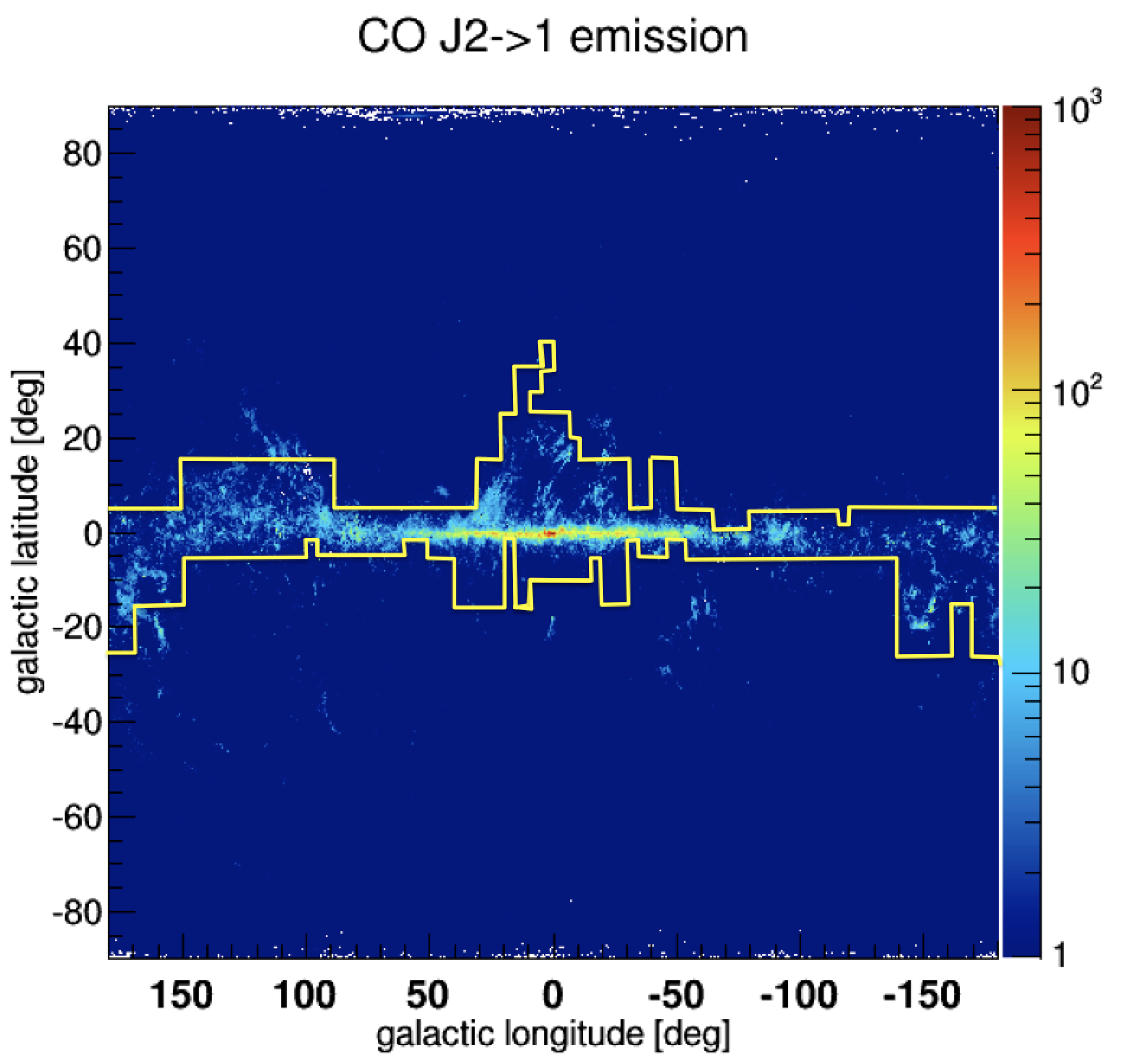}
\caption[]{Sky map of the CO rotation line as measured with the Planck satellite\cite{Planck}. The yellow (light) contour   is the DM contour from Fig. \ref{f6}(b).
}
\label{f8}
\end{figure}

Fits in three cones are shown as examples in Fig. \ref{f4}: one for the Galactic center, where the size of the CMZ has been selected ($-1.5^\circ\, <\, l\, <\, 2^\circ$ and $|b|\, <\, 0.5^\circ$), one in the halo with $-20^\circ\, <\, l\, <\, 20^\circ$ and $2^\circ\, <\, |b|\, <\, 20^\circ$ and one along the nearby tangent point of the Scutum-Centaurus spiral arm with $-50^\circ\, <l\,  <\, -46^\circ$ and $-0.5^\circ\, <\, |b|<\, 0.5^\circ$. The longitude and latitude ranges are indicated in the panels.
The left (right)  panels show the fits  with the MCR (DM) template. For the MCR fits all MCR templates in Fig. \ref{f2}(a) with varying breaks between 6 and 14 GV were tried in the fit.
 The best fitted break in the proton spectrum for the MCR template is indicated in the fits on the left, while the maximum of the DM flux is indicated  for the DM fits on the right. The DM fits all use a WIMP mass of 44.9 GeV, which is the optimal mass for the Galactic center.   Both, the WIMP mass and the DM flux from our analysis, are compatible with ``conventional'' analysis. \cite{Daylan:2014rsa,Calore:2014nla} 

The field-of-view of the CMZ ($l \, \times \, b \; \simeq \; 3.5^{\circ} \, \times \, 0.5^{\circ}$) is known from the CS or CO maps.  \cite{Tsuboi:1999,ThePlanck:2013dge} The rotation lines of both molecules are good tracers of MCs. In the top row of Fig. \ref{f4} one observes that the spectrum towards the CMZ is dominated by the ``GeV-excess'', which is either proportional to the contribution of the MCR template (left panel) or DM template (right panel). We checked that the ``GeV-excess''  is maximal inside the rectangular field-of-view of the CMZ by repeating the fit with a sliding window of constant size. If the window moved out of the field-of-view of the CMZ the flux of the ``GeV-excess'' decreased. The fact that the ``GeV-excess'' has a longitudinally elongated morphology in the inner few degrees of the Galactic center shows that the  ``GeV-excess'' cannot be dominated by DM annihilation, which would correspond to a spherical instead of a rectangular spatial morphology.

The SCR template in the middle row has  contributions from both, the $\pi^0$ production inside point sources and in the Fermi Bubbles, since they both contribute in this field-of-view up to latitudes of 20$\circ$, but they have the same hard SCR template, see Fig. \ref{f1}.  Note that our analysis does not need the spatial template for the Fermi Bubbles, since its contribution is determined by the energy template for each cone. From the contribution in each cone we find the well-known shape of the Fermi-Bubbles.

 The last row in Fig. \ref{f4} shows the fits towards the tangent point of the Scutum-Centaurus arm. Here the data are again dominated by the  ``GeV-excess'', but this region cannot be described by the DM template nor by the background templates alone (PCR, IC, BR, ISO), which is again a   strong case for the MC hypothesis of the  ``GeV-excess'', since the DM template is not expected to dominate so far from the Galactic center.  The MCR breaks can  shift the maximum of the MCR template between 1 and 2 GeV (see Fig. \ref{f2}(a)), which is not allowed for DM templates because of the requirement of the same WIMP mass in all sky directions.
 The fits for all 797 cones are shown in the Appendices, both, for the MCR and DM template fits. %These appendices are added as Online Supplemental Material (OSM). \cite{osm}
 
From the 797 fits to all cones it is clear that the $\chi^2$ of the MCR fit is better than the DM fit in regions where a strong ``GeV-excess'' is observed, as shown  in Fig. \ref{f5} and the $\chi^2$ values in the panels of Fig. \ref{f4}. The DM fit leads to $\chi^2/dof$ of typically 3 or higher inside the disk, while the fits including the MCR templates provide a good $\chi^2/dof$ over the whole gamma-ray sky.

The sky maps of the various contributions can be directly obtained by plotting the fitted normalization constants in Eq. \ref{e1}. The  sky map of the ``GeV-excess'' corresponds to the sky maps of either the MCR or DM template, which are shown in Fig. \ref{f6}.  One observes in both cases a strong component along the whole disk with rapidly decreasing latitude contributions up to 15-20$^\circ$.  This does not look like a DM profile,  but resembles the morphology of MCs.
This can be checked by comparing the MCR sky map with the CO sky map, which is a tracer of MCs and was precisely measured by the Planck satellite. \cite{ThePlanck:2013dge} These data are publicly available. \cite{Planck}
Since the agreement is difficult to visualize with color coded sky maps the longitude and latitude profile of the MCR sky map are histogrammed (green) in Fig. \ref{f7}(a)  together with the MC column density, as obtained from the Planck sky maps for the CO rotation lines (red line). The fluxes  from the MCR templates and CO sky maps are normalized.  Both show a strong contribution in the Galactic center from the CMZ. The longitude distribution decreases rapidly outside the Galactic bar region ($-50^\circ\, <\, l\, <45^\circ$) \cite{deBoer:2014bra}, both for the CO flux and the MCR flux which implies a high density of MCs in the Galactic bar. 
This similarity  in morphology between the MCR and CO fluxes points to a strong correlation, which is not expected to be exact, since the gamma-ray flux from the MCR template is determined by the molecular gas density convolved with the cosmic ray density along the lines-of-sight, while the CO maps are proportional to the MC column density only.  The latitude distribution in Fig. \ref{f7}(a) shows also the expectation from a generalized NFW profile, which was taken from the ``conventional''œ analysis, as presented in Fig. 1 from Ref.  \cite{Calore:2014nla}. From the bottom panel in Fig. \ref{f7}(a) one observes that the latitude distribution from the CO map of Planck resembles  a DM profile (compare black and red lines). In addition, this panel proves that our template fit (green histogram) is in reasonable agreement  with  the flux from the ``conventional''œ  analysis, represented by the black line. In longitude  the ``GeV-excess'' does not follow an NFW profile, as shown in the top panel of Fig. \ref{f7}(a), but follows closely the structure from the Galactic bar with the CMZ at the center. Within the first two degrees in longitude, i.e. within the CMZ,  the flux of the ``GeV-excess'' does not fall rapidly in contrast to the expectation for the DM hypothesis.
%The uncertainty  on the flux of the black line  is typically a factor two given the uncertainties   in the gas - and cosmic ray density  because of the  CMZ with a high star formation rate, see e.g. Refs.  \cite{Morris:1996th,Figer:2003tu}.

Since sources are expected to reside inside MCs one expects  a strong correlation in the spatial distributions of the SCR fluxes and the MCR fluxes, i.e. if there is a long tail in the gamma-ray spectra above 30 GeV, one expects a simultaneous shift in the maximum of the spectrum from 0.7 up to 2 GeV.
The strong correlation between the SCR and MCR fluxes is indeed observed, as shown in the top panel of Fig. \ref{f7}(b). 
Here the gamma-ray fluxes from the SCR (MCR) templates  are integrated over a latitude range of $|b|\, <\, 5^\circ(0.5^\circ)$, respectively. The larger latitude range for the SCR component is just to increase the statistics of the SCR fluxes, since the sources can have outflows towards higher latitudes, as suggested by the broad latitude distribution of $^{26}$Al in the bottom panel of Fig. \ref{f7}(b).   
 The radioactive $^{26}$Al isotope  is synthesized by proton capture of $^{25}$Mg in heavy, magnesium rich sources \cite{Prantzos1996} and can be traced by the 1.8 MeV gamma-line emitted in its decay. This line has been studied  by the Integral/Spi satellite \cite{Bouchet:2015rxa} and is publicly available as sky map. \cite{Spi}  The strong correlation between the SCR fluxes, MCR fluxes and $^{26}$Al fluxes is emphasized by the vertical arrows in Fig. \ref{f7}(b) between the sky map of the $^{26}$Al line (bottom panel) and the longitude distribution of the  SCR and MCR fluxes (top panel). 
 The sky map of the DM template in Fig. \ref{f6}(b) has the morphology of  a CO sky map instead of a spherical halo profile expected for DM. This is demonstrated in Fig. \ref{f8}.

 It was noticed recently \cite{Yang:2016jda} that the gamma-rays have a much harder spectrum towards the Galactic center than in the opposite direction. However,  the reason was not understood, but is provided by the strong SCR contribution in the bar region (see top panel in Fig. \ref{f7}(b)), which is largely absent in the opposite direction (longitude $\approx 180^\circ$).

%CHECK AGAIN: 
%SCRs are unpropagated cosmic rays  inside sources, while  MCRs are  propagated cosmic rays inside MCs. This might explain the difference in intensity in the central bin  in the top panel of  Fig. \ref{f7}(c) between the gamma-ray fluxes from the SCR and MCR templates: the SCR density is  reduced by driving the outflow into the Bubbles \cite{Everett:2007dw,Breitschwerdt:2008na}, thus reducing the density of SCRs with respect to the MCRs.
 
The sky map of the ``GeV-excess'' in Fig. \ref{f6} shows  some clumpiness, as expected from the discrete nature of MCs or its filamentary substructure. In Ref.  \cite{Lee:2015fea} some deviation from smooth sky maps for the ``GeV-excess''  was interpreted as evidence for unresolved point sources, a feature  used to support the millisecond pulsar interpretation of the ``GeV-excess''. But the nature of the clumpiness is unknown and could be related to MCs as well. \cite{Lee:2014mza}

\section{Conclusion}\label{conclusion}
We have compared two hypothesis for the ``GeV-excess'': an excess of gamma-rays peaking around 2 GeV from DM annihilation (DM hypothesis) or a depletion of gamma-rays below 2 GeV as observed in the gamma-ray emissivity of MCs (MC hypothesis). The DM hypothesis leads to an excess  falling rapidly with distance from the Galactic center, as expected for a typical DM profile. The MC hypothesis leads to a ``GeV-excess'' falling rapidly with distance from the Galactic center as well, because of the decreasing column densities of MCs along the lines-of-sight away from the Galactic center. This decrease happens to resemble a DM profile, as  is  known from the  sky maps  of the CO rotation lines, a tracer of MCs.

We find that the MC hypothesis is preferred over the DM hypothesis for the following reasons:

i) the MC hypothesis provides a significantly better fit, especially if one considers the gamma-ray energies up to 100 GeV; the  groups proposing the DM hypothesis \cite{Daylan:2014rsa} excluded data above 10 GeV for the fits towards the Galactic center, but the DM template does not describe  the ``GeV-excess'', if higher energies are included, as shown in this paper and observed recently by the Fermi Collaboration as well. \cite{TheFermi-LAT:2017vmf}
ii) the ``GeV-excess'' has in latitude for both hypotheses the morphology  from a generalized  NFW profile   (see bottom panel of Fig. \ref{f7}(a) for the MCR template), but the excess is strong  in all directions  towards MC regions in the Galactic disk, as could be proven by the spatial correlation with the CO maps from the Planck satellite. \cite{ThePlanck:2013dge} Especially, the DM sky map does not resemble the expected spherical DM halo profile if the whole gamma-ray sky is considered, but has a morphology similar to the CO sky map, as demonstrated in Fig. \ref{f8}.
iii) The single, most convincing evidence, which leads us to believe that DM cannot be the dominant source of the  ``GeV-excess'' is provided by the strong ``GeV-excess'' in the longitudinally extended field-of-view with the rectangular shape of  the CMZ, the dense MC conglomerate encircling the Galactic center. Here the ``GeV-excess'', observable as a shift in the maximum of the energy flux per log bin  in the gamma-ray spectrum to 2 GeV, is obvious already from the raw diffuse gamma-ray data without any analysis.  Such a rectangular shape is  {\it  not} compatible with the expected  spherical morphology of a DM annihilation signal.

\acknowledgments
Financial support from the Deutsche Forschungsgemeinschaft  (DFG, Grant BO 1604/3-1)  is warmly  acknowledged.  We are grateful to the Fermi scientists, engineers and technicians for collecting the Fermi data and the Fermi Science Support Center for providing the software and strong support for  guest investigators.

\input{appendix}

\clearpage
\providecommand{\href}[2]{#2}\begingroup\raggedright\endgroup
\end{document}

%% file: appendix.tex
\appendix

%\subsection{Spectral Fits for all Cones}
\begin{figure}
\centering
{\bf Appendix A: Figs. \ref{F11}-\ref{F31} show the template fits in each of the 797 cones using the MCR template to describe the ``GeV-excess''.  The figures start with the highest latitudes and in each figure the longitude varies for a given stripe in latitude, as indicated in the legends. }\vspace*{0,3cm}\\
\includegraphics[width=0.16\textwidth,height=0.16\textwidth,clip]{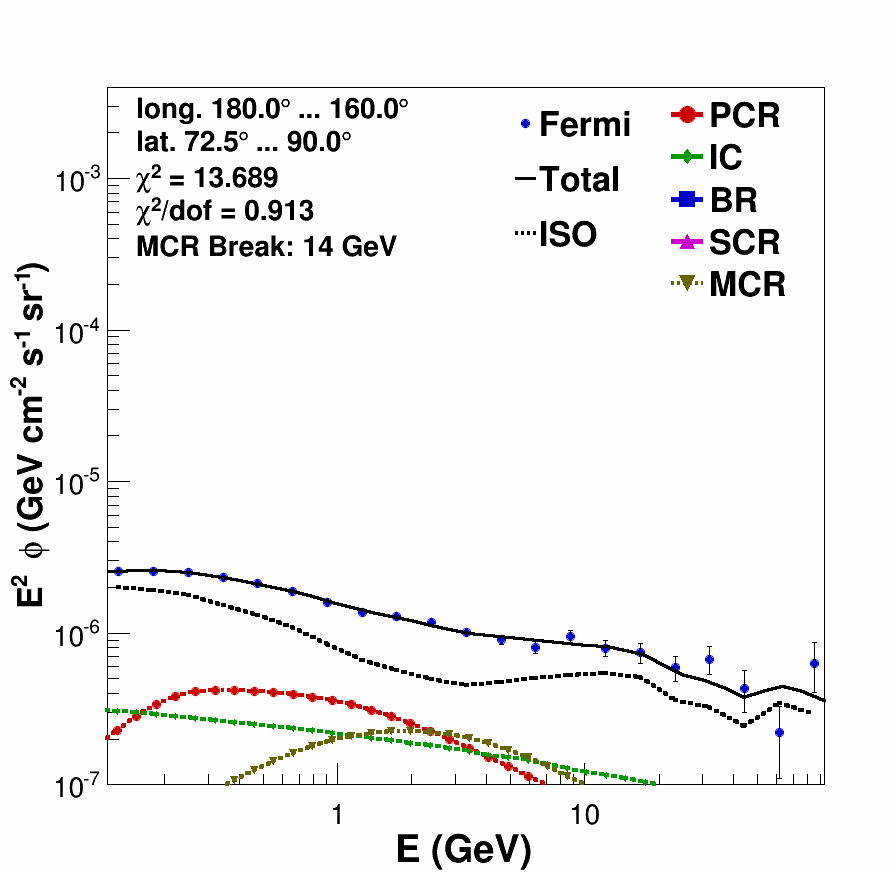}
\includegraphics[width=0.16\textwidth,height=0.16\textwidth,clip]{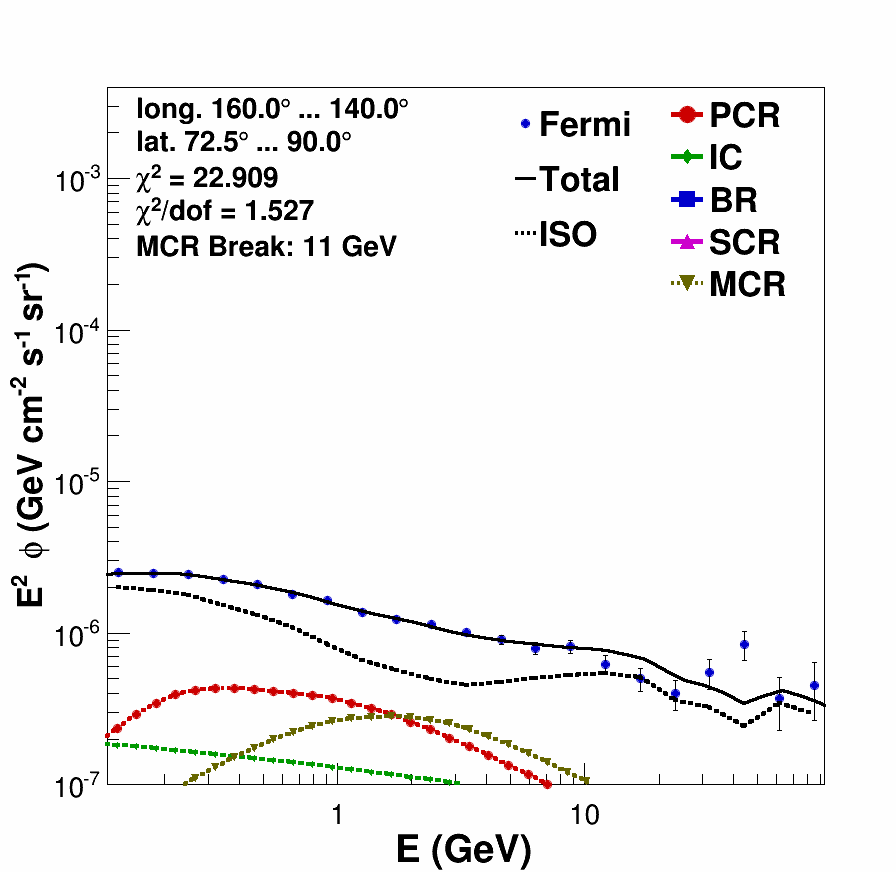}
\includegraphics[width=0.16\textwidth,height=0.16\textwidth,clip]{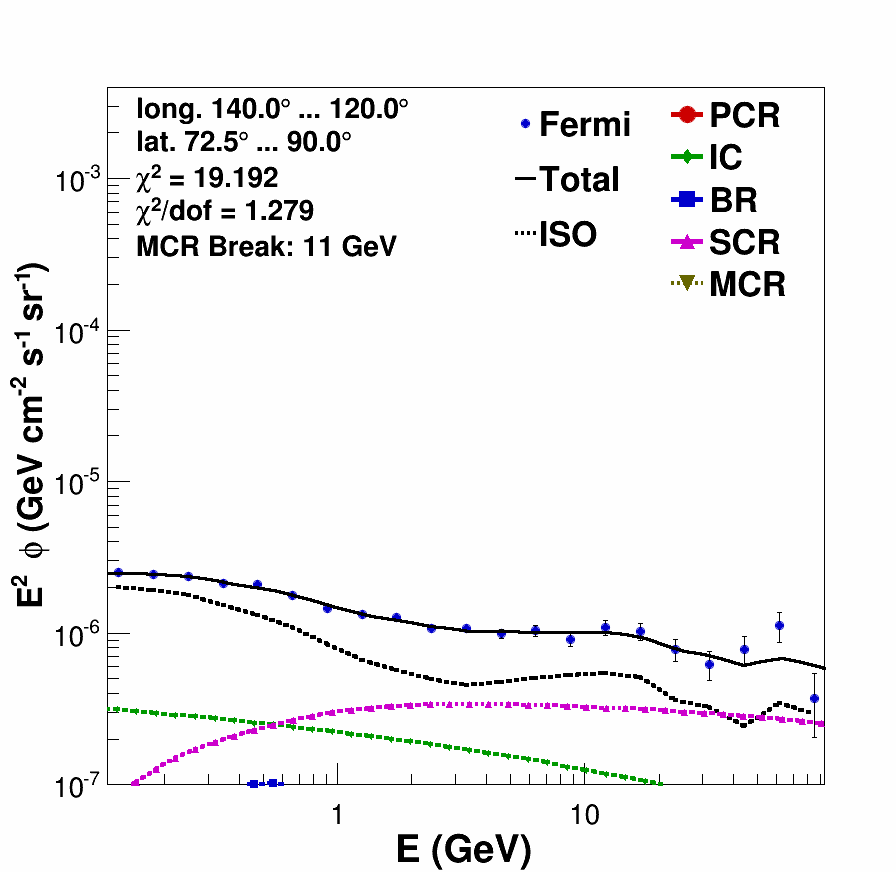}
\includegraphics[width=0.16\textwidth,height=0.16\textwidth,clip]{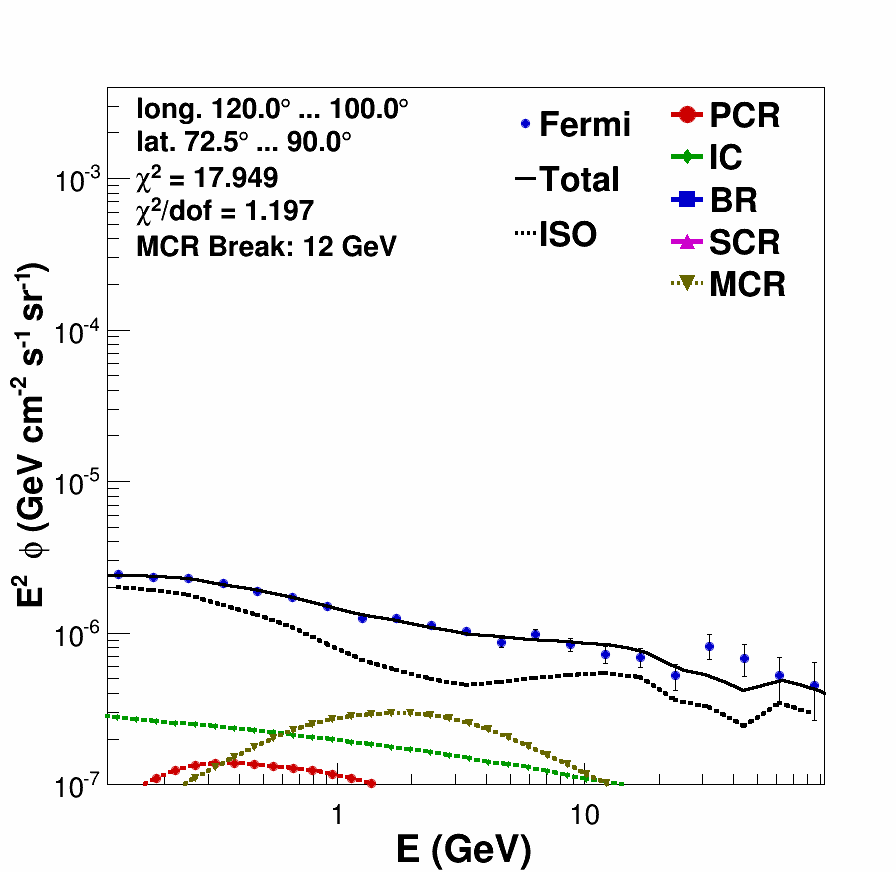}
\includegraphics[width=0.16\textwidth,height=0.16\textwidth,clip]{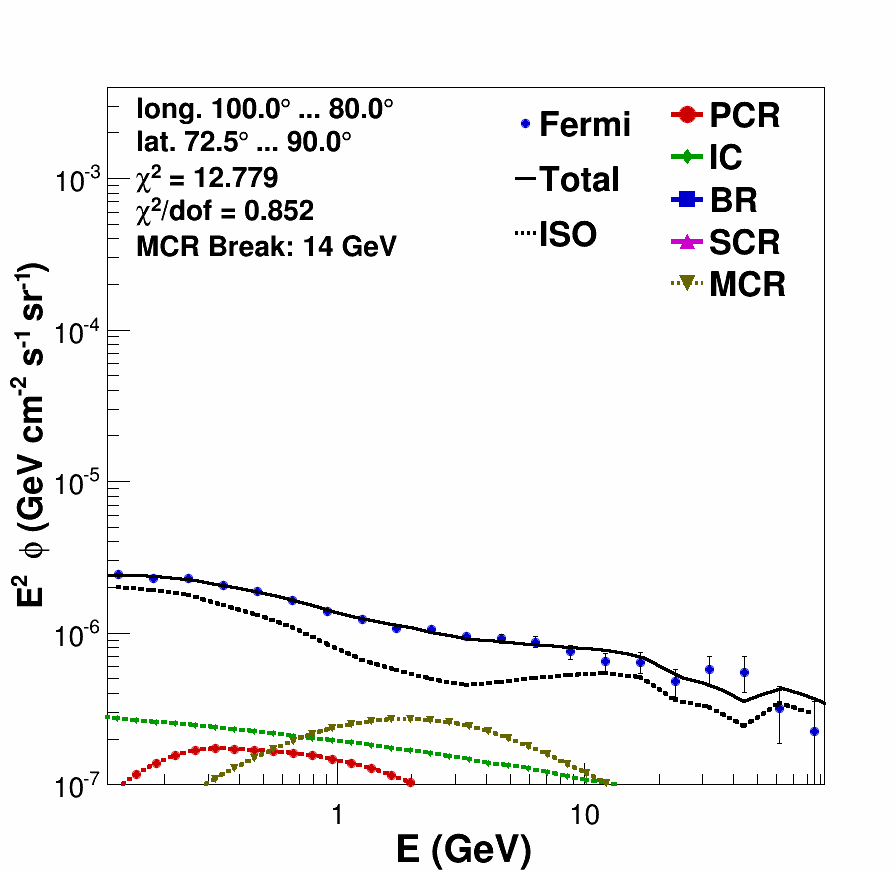}
\includegraphics[width=0.16\textwidth,height=0.16\textwidth,clip]{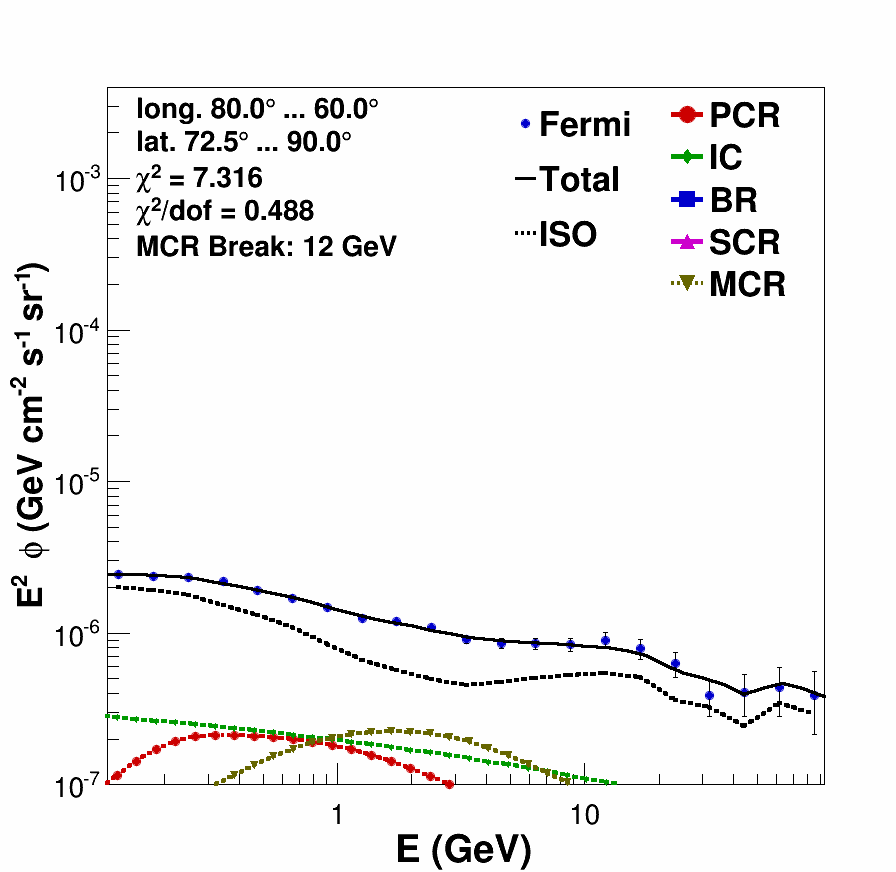}
\includegraphics[width=0.16\textwidth,height=0.16\textwidth,clip]{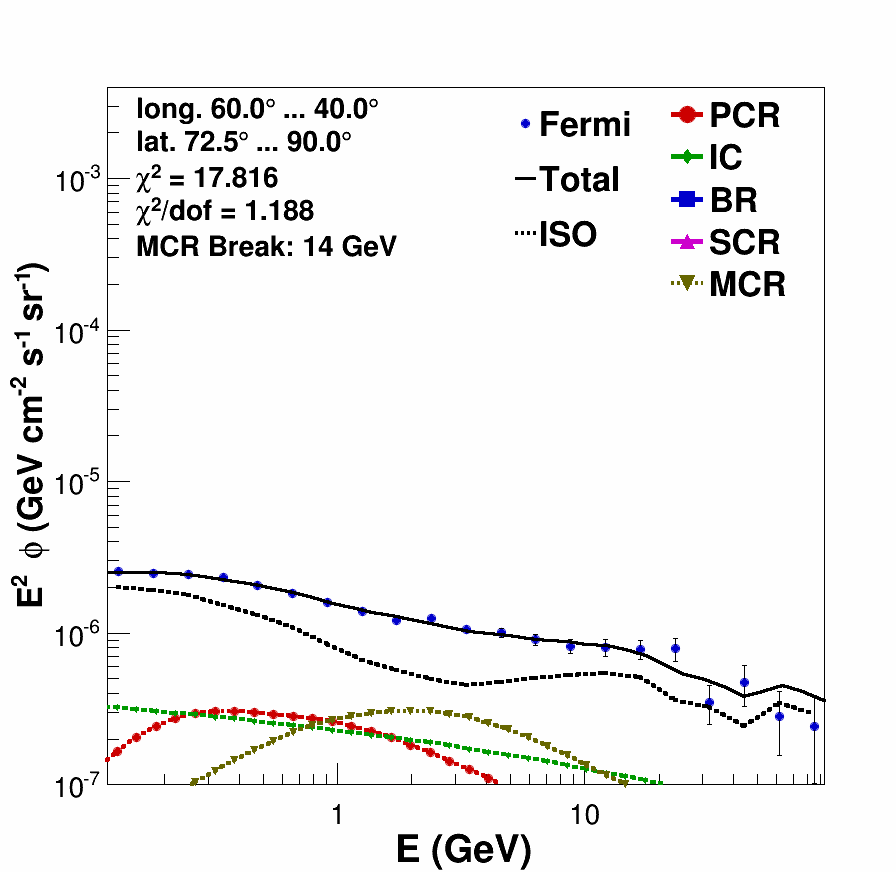}
\includegraphics[width=0.16\textwidth,height=0.16\textwidth,clip]{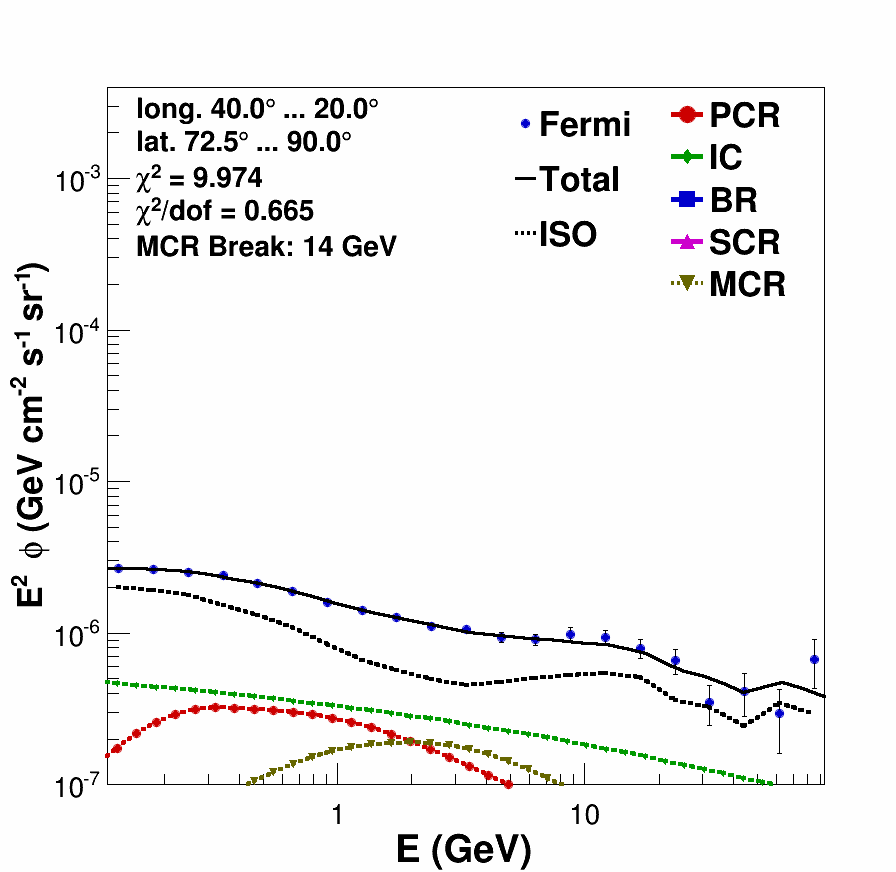}
\includegraphics[width=0.16\textwidth,height=0.16\textwidth,clip]{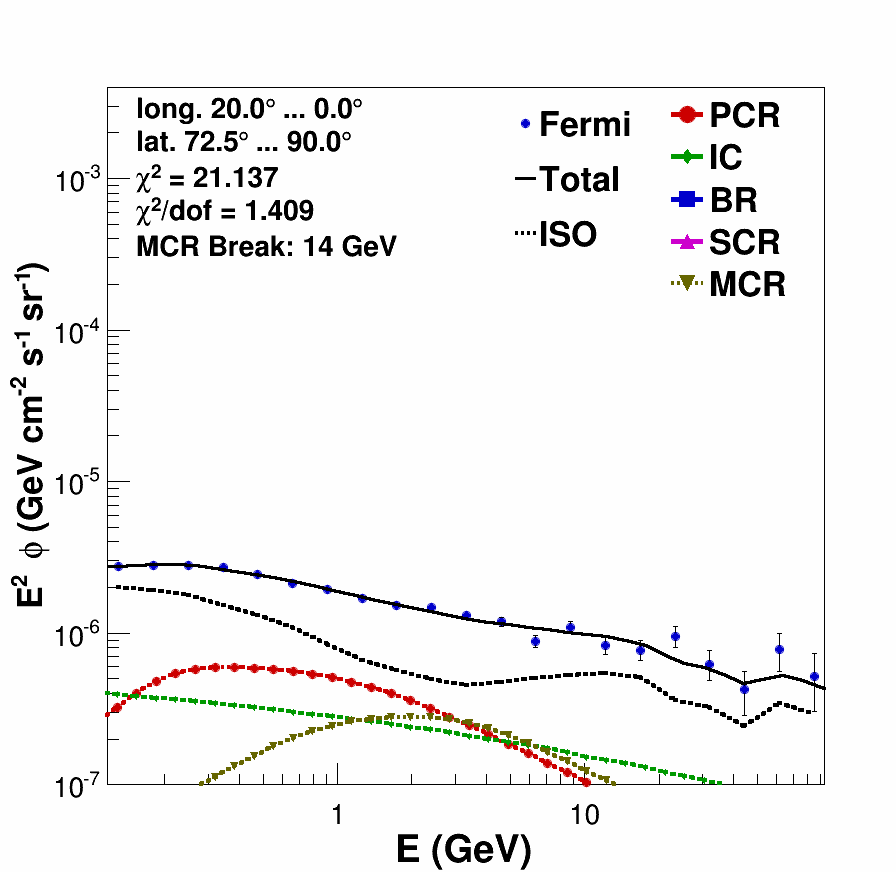}
\includegraphics[width=0.16\textwidth,height=0.16\textwidth,clip]{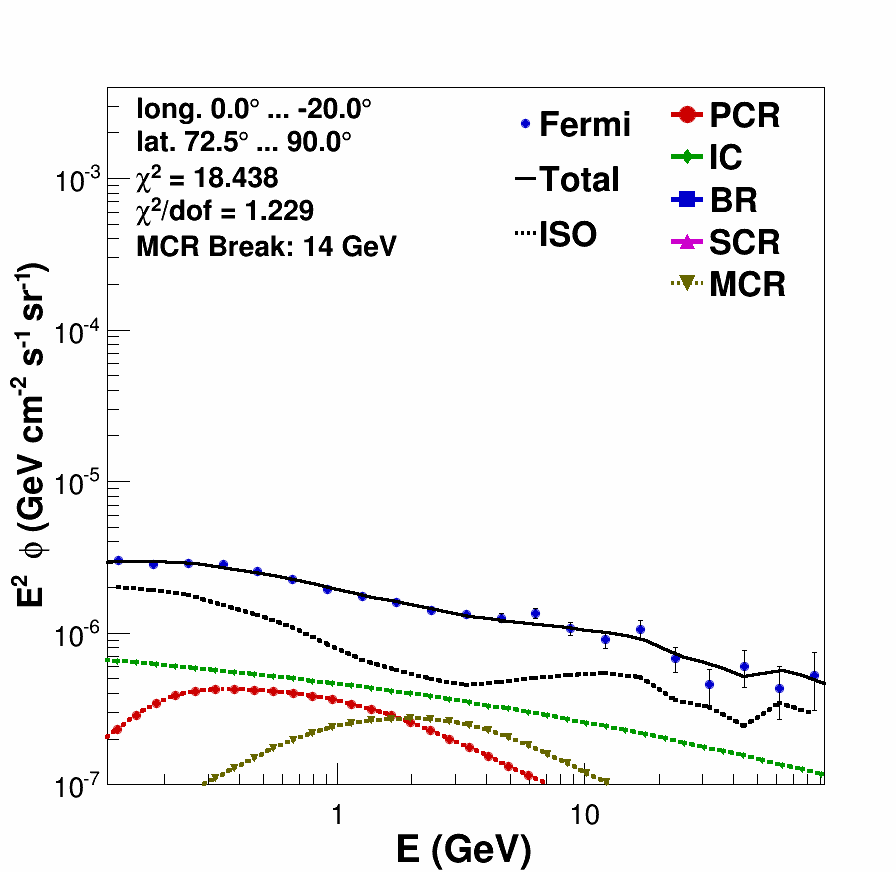}
\includegraphics[width=0.16\textwidth,height=0.16\textwidth,clip]{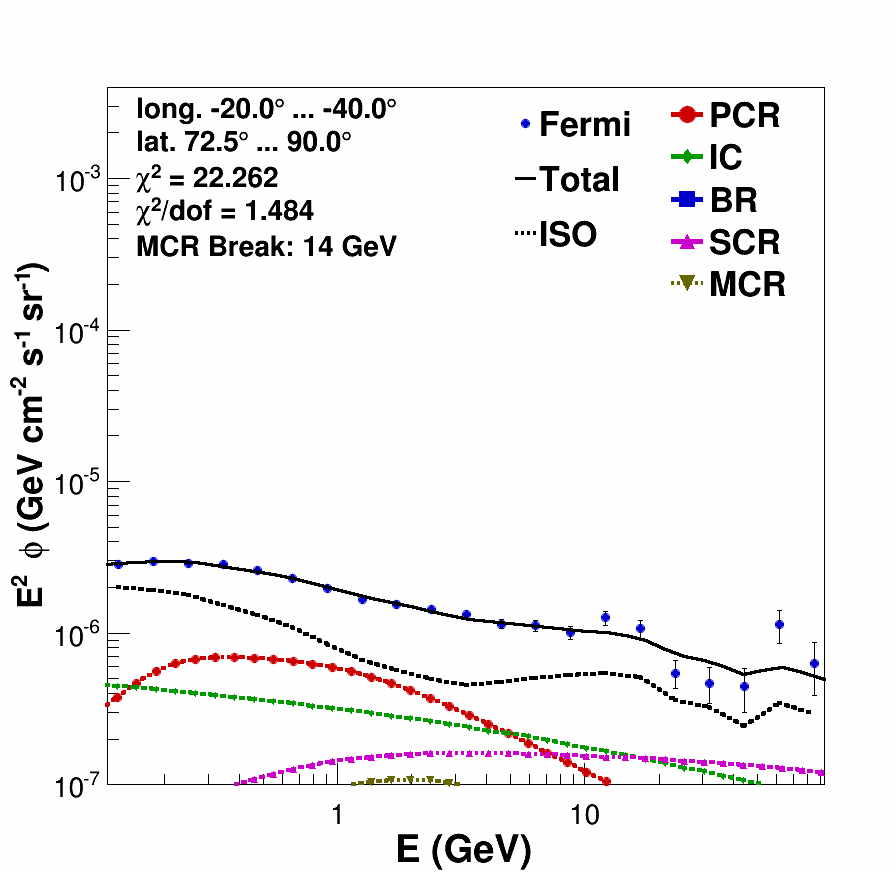}
\includegraphics[width=0.16\textwidth,height=0.16\textwidth,clip]{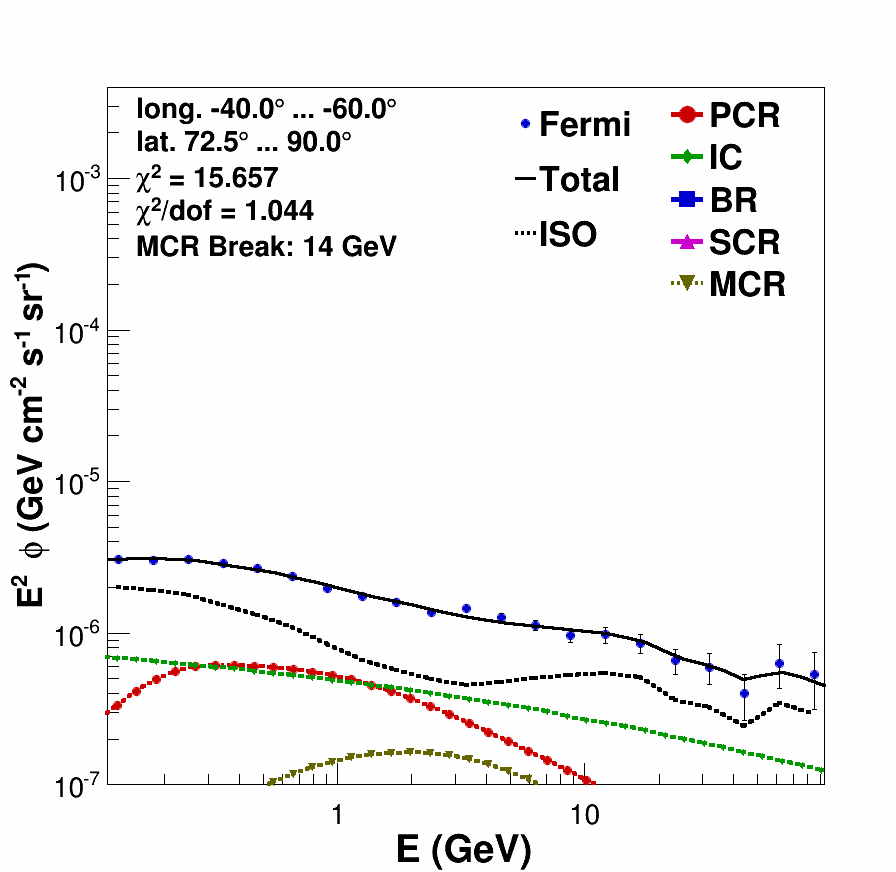}
\includegraphics[width=0.16\textwidth,height=0.16\textwidth,clip]{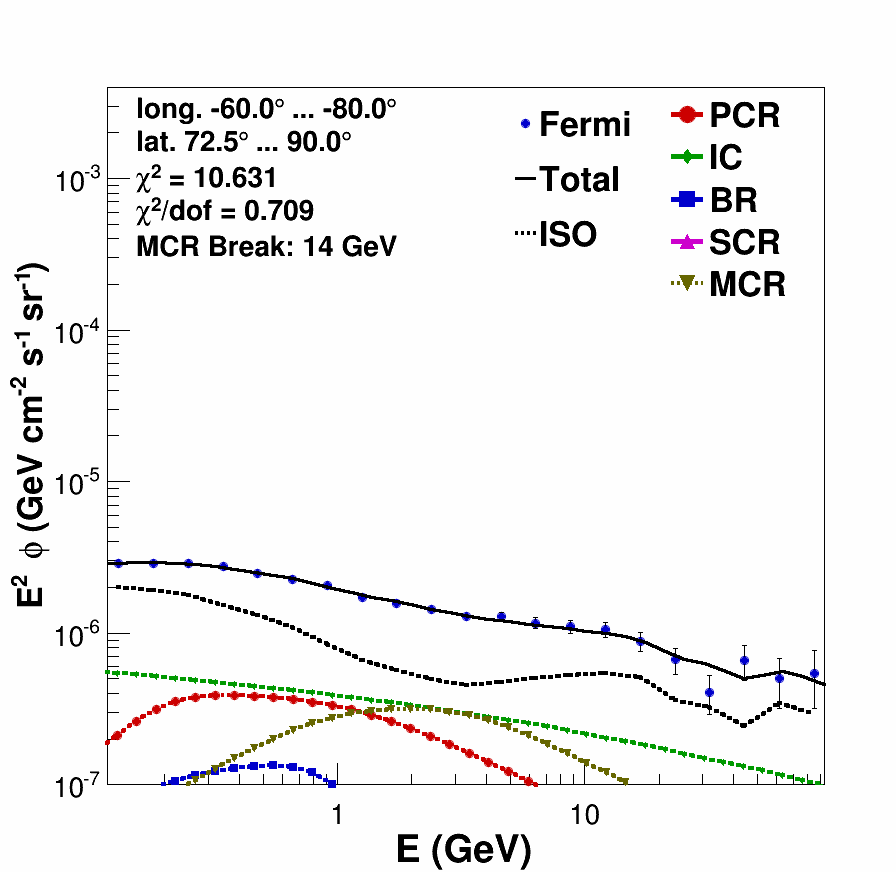}
\includegraphics[width=0.16\textwidth,height=0.16\textwidth,clip]{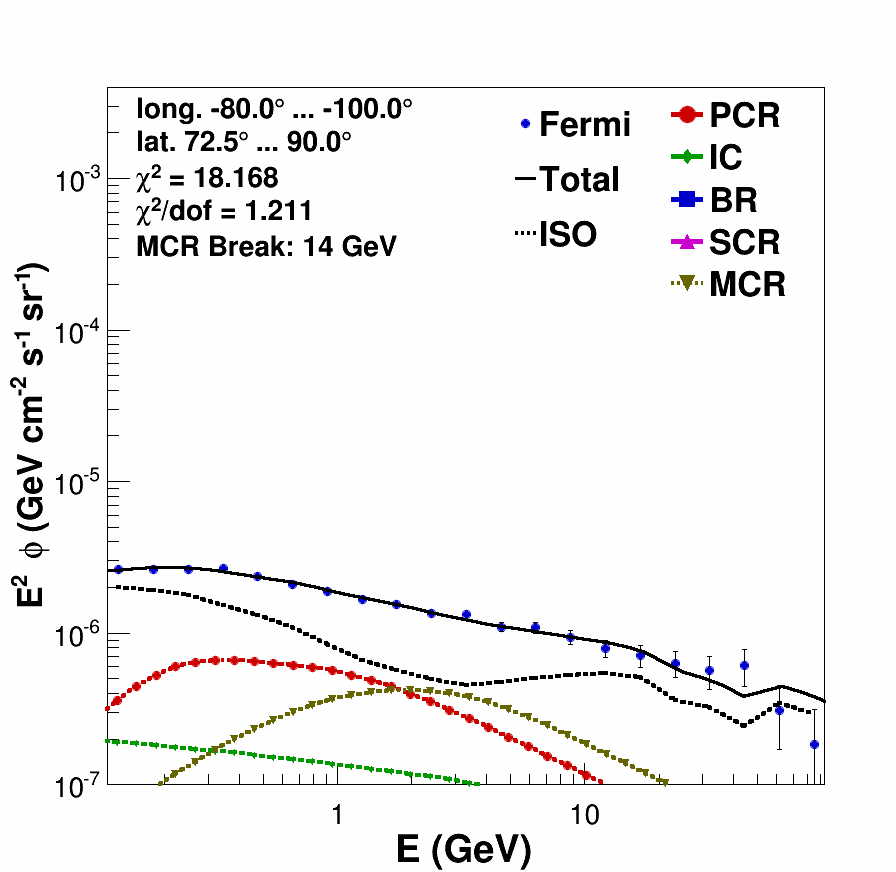}
\includegraphics[width=0.16\textwidth,height=0.16\textwidth,clip]{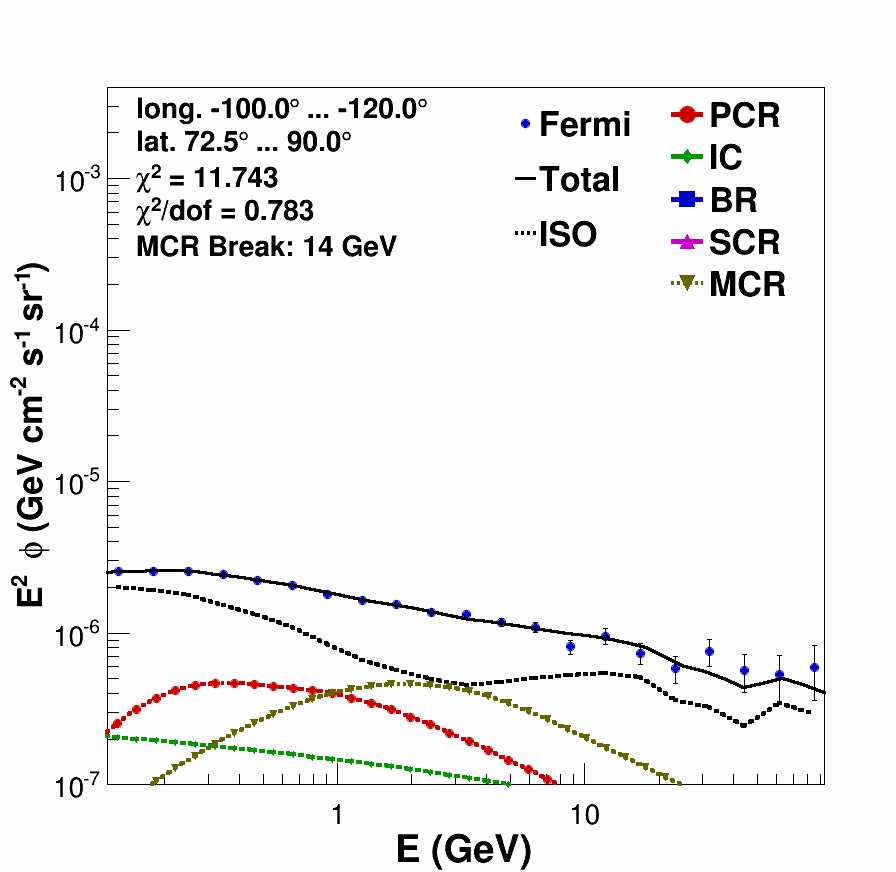}
\includegraphics[width=0.16\textwidth,height=0.16\textwidth,clip]{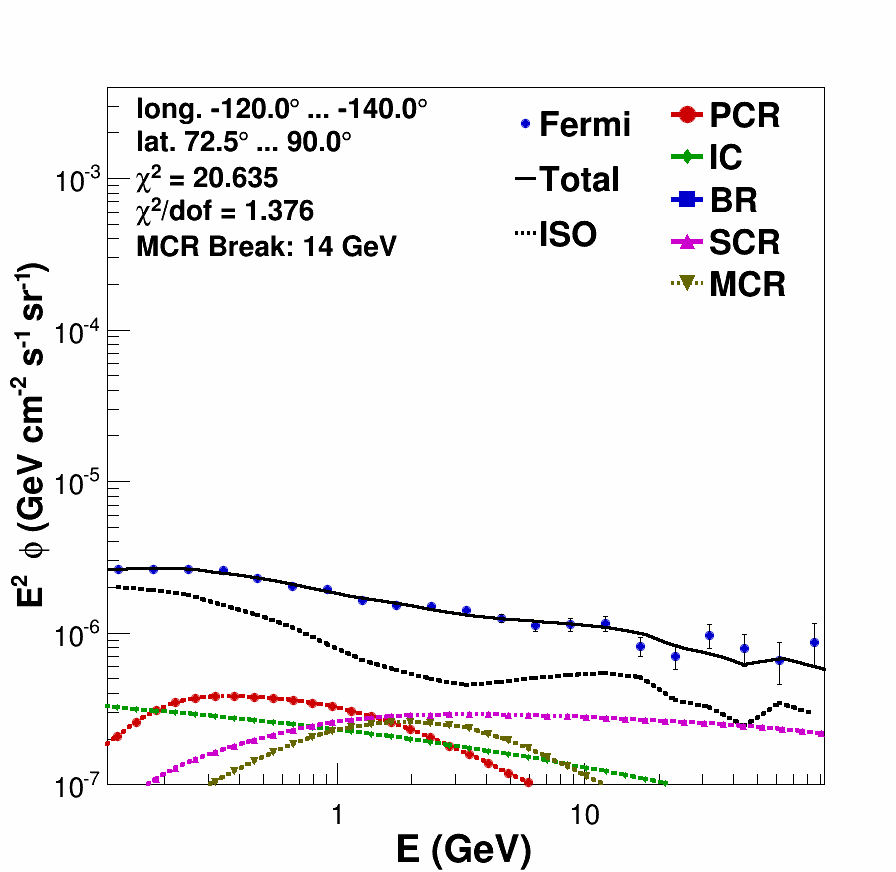}
\includegraphics[width=0.16\textwidth,height=0.16\textwidth,clip]{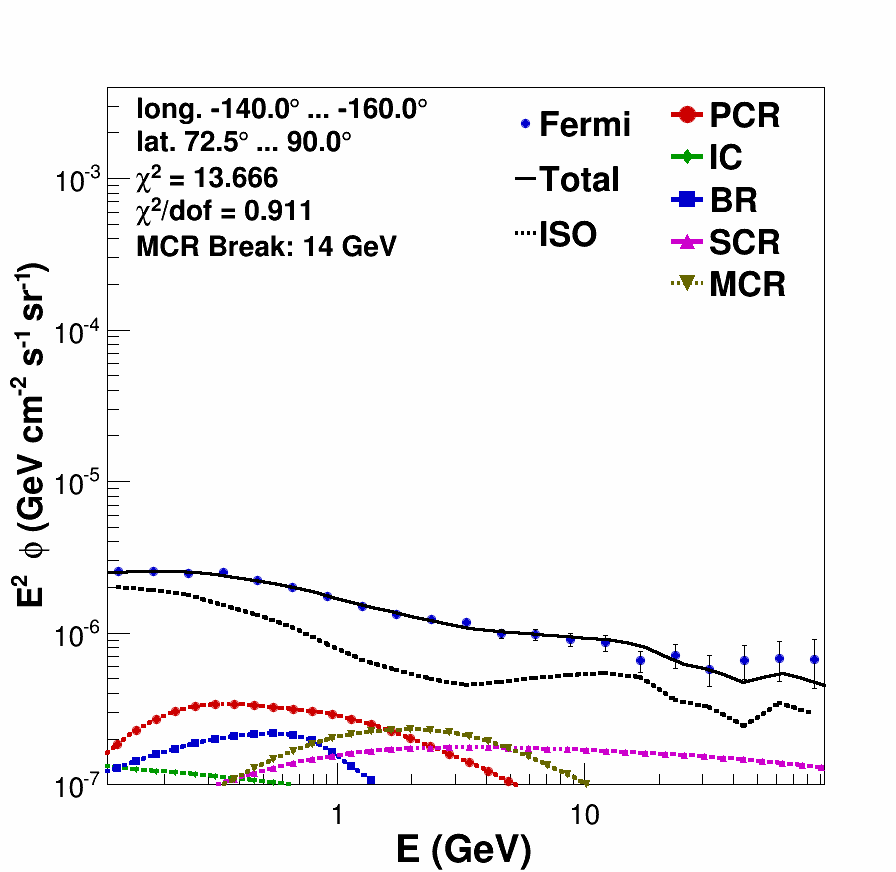}
\includegraphics[width=0.16\textwidth,height=0.16\textwidth,clip]{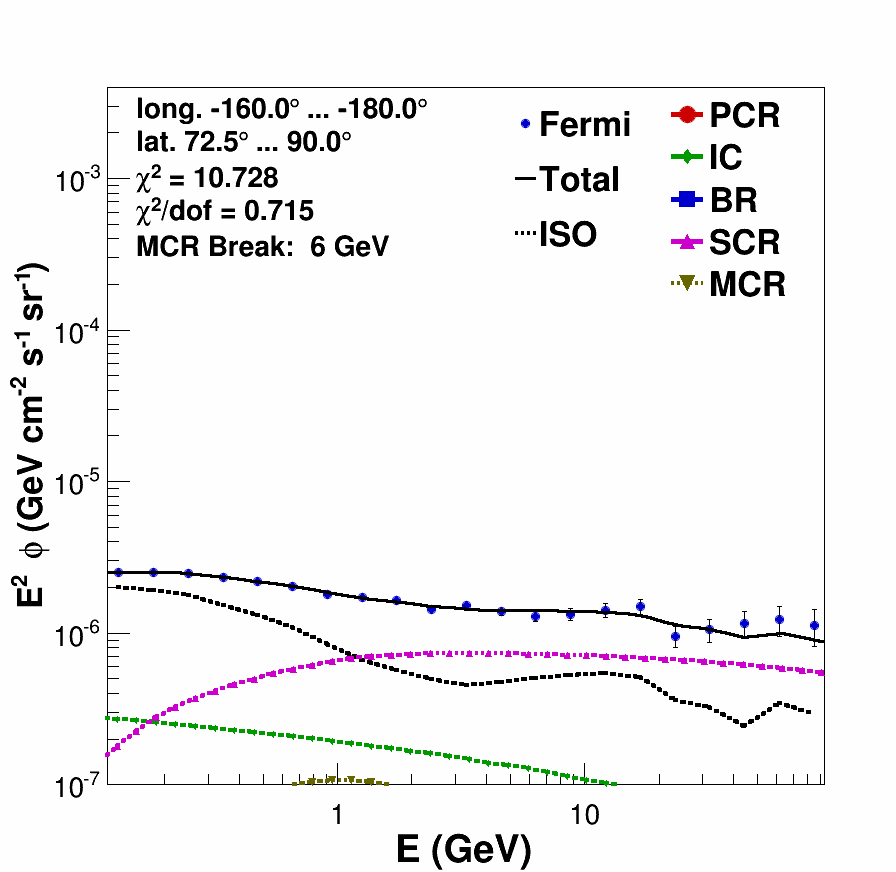}    %\\[2cm]%%%%%% first row
\caption[]{Template fits for latitudes  with $72.5^\circ<b<90.0^\circ$ and longitudes decreasing from 180$^\circ$ to -180$^\circ$.} 
\label{F11}
\end{figure}
\begin{figure}
\includegraphics[width=0.16\textwidth,height=0.16\textwidth,clip]{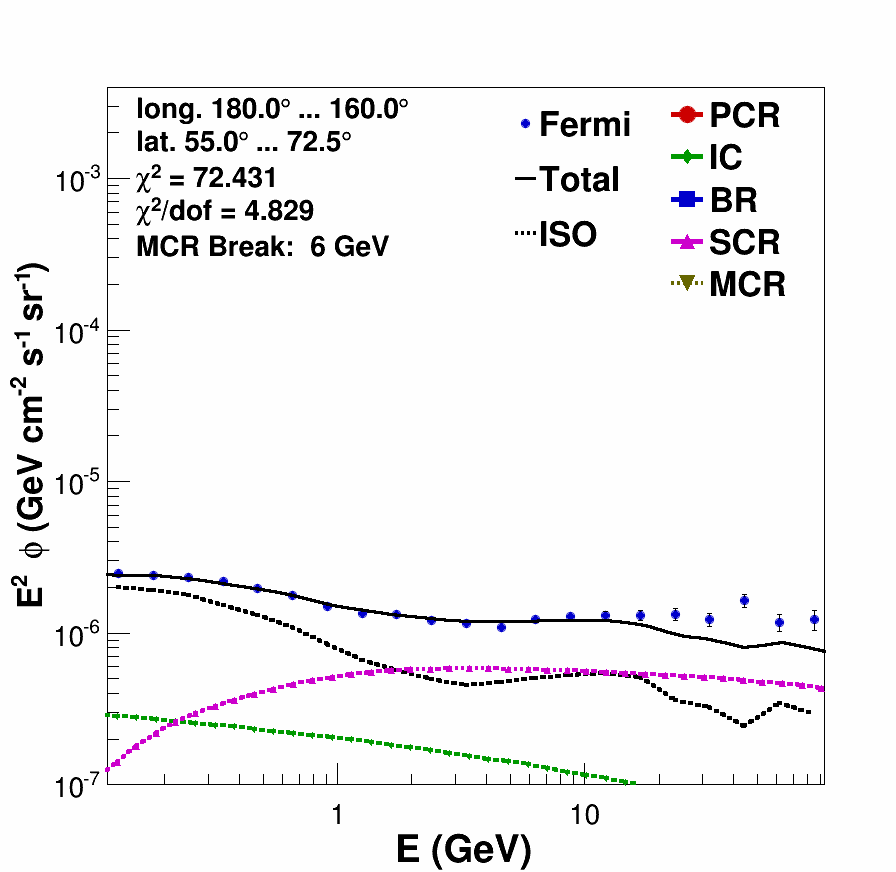}
\includegraphics[width=0.16\textwidth,height=0.16\textwidth,clip]{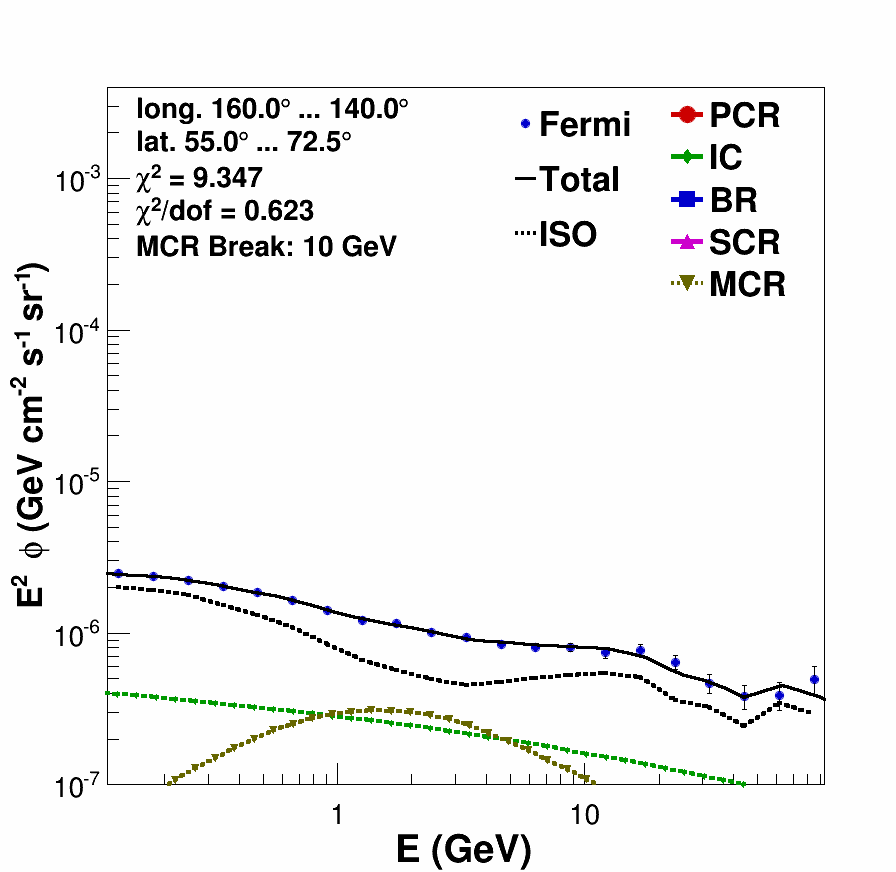}
\includegraphics[width=0.16\textwidth,height=0.16\textwidth,clip]{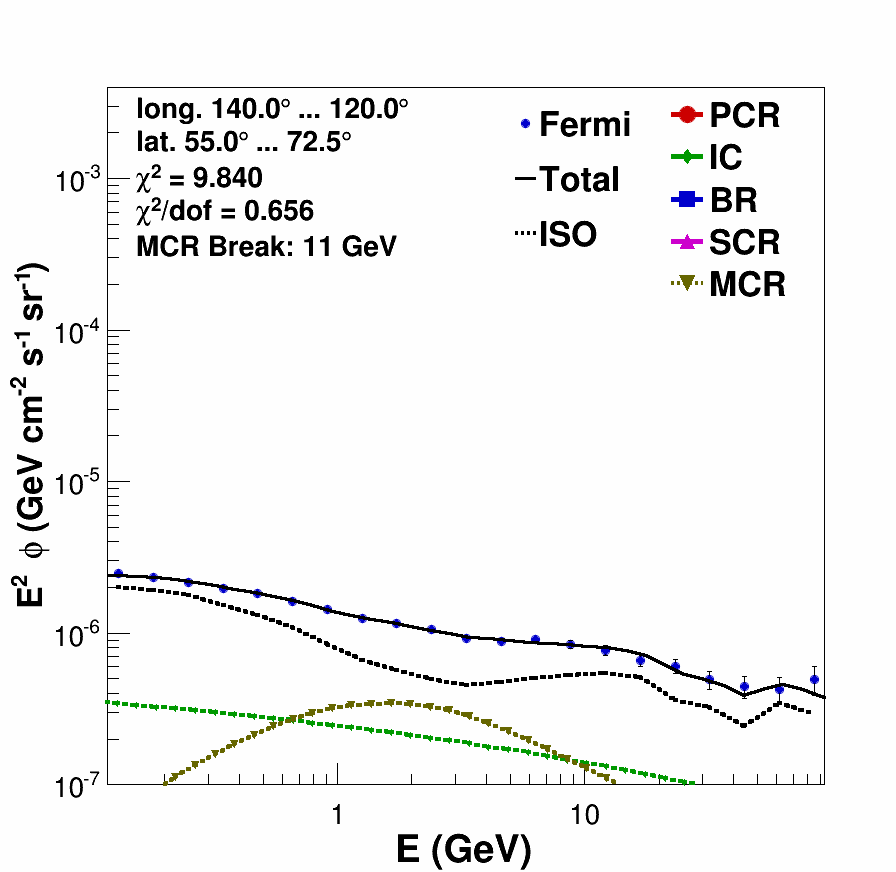}
\includegraphics[width=0.16\textwidth,height=0.16\textwidth,clip]{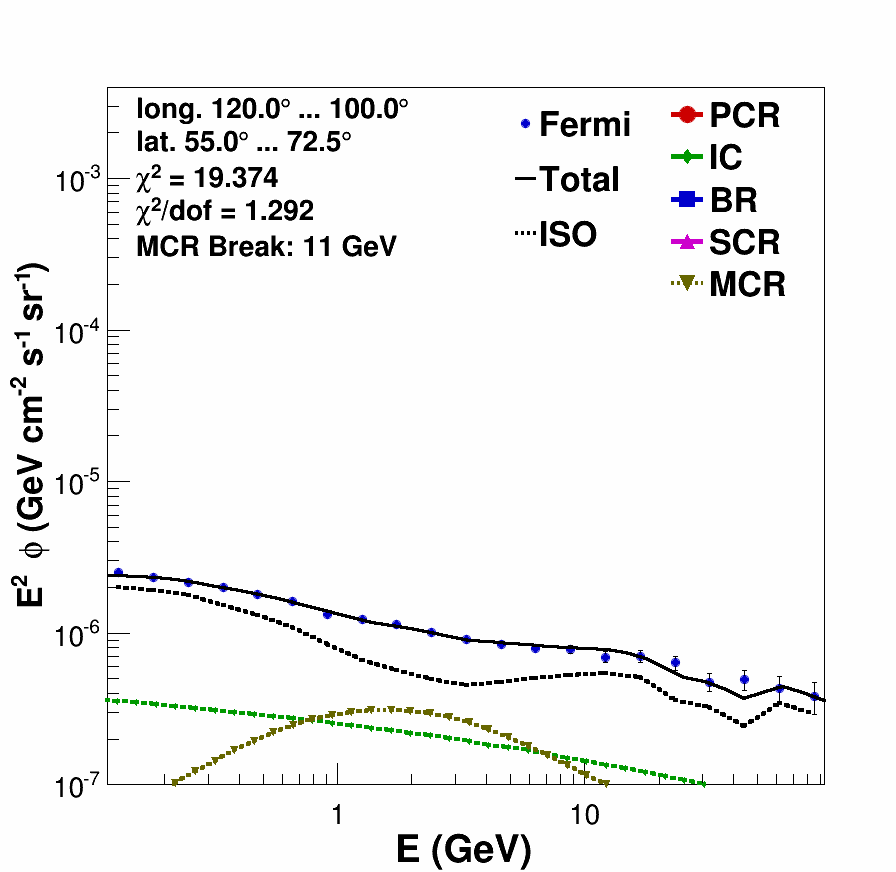}
\includegraphics[width=0.16\textwidth,height=0.16\textwidth,clip]{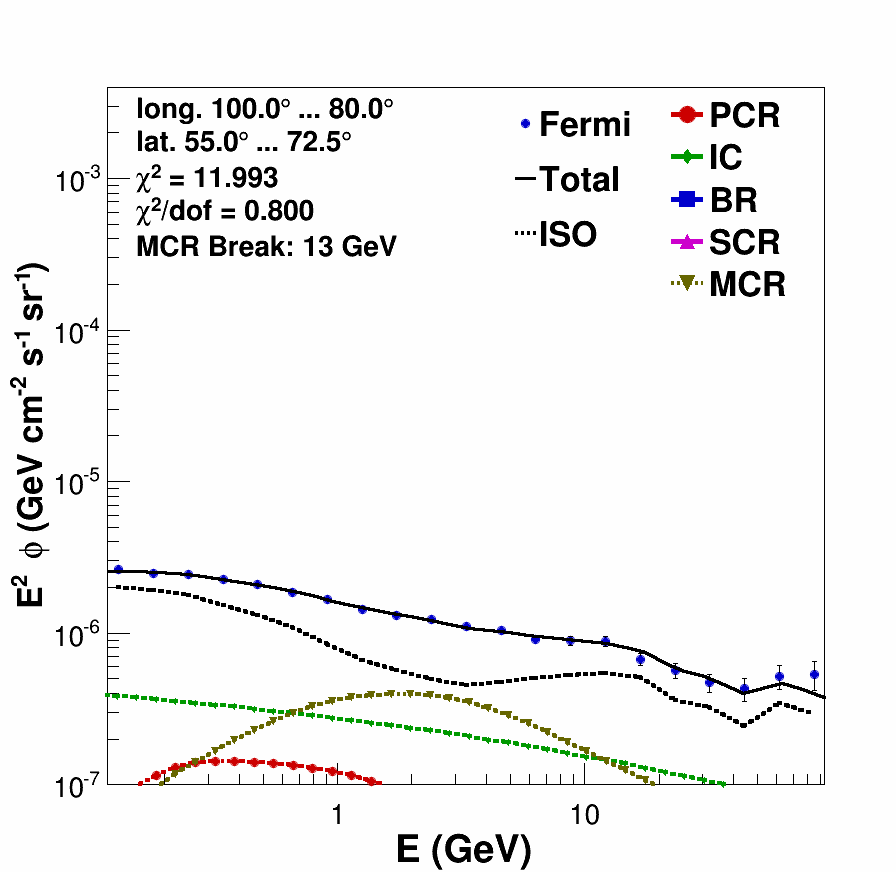}
\includegraphics[width=0.16\textwidth,height=0.16\textwidth,clip]{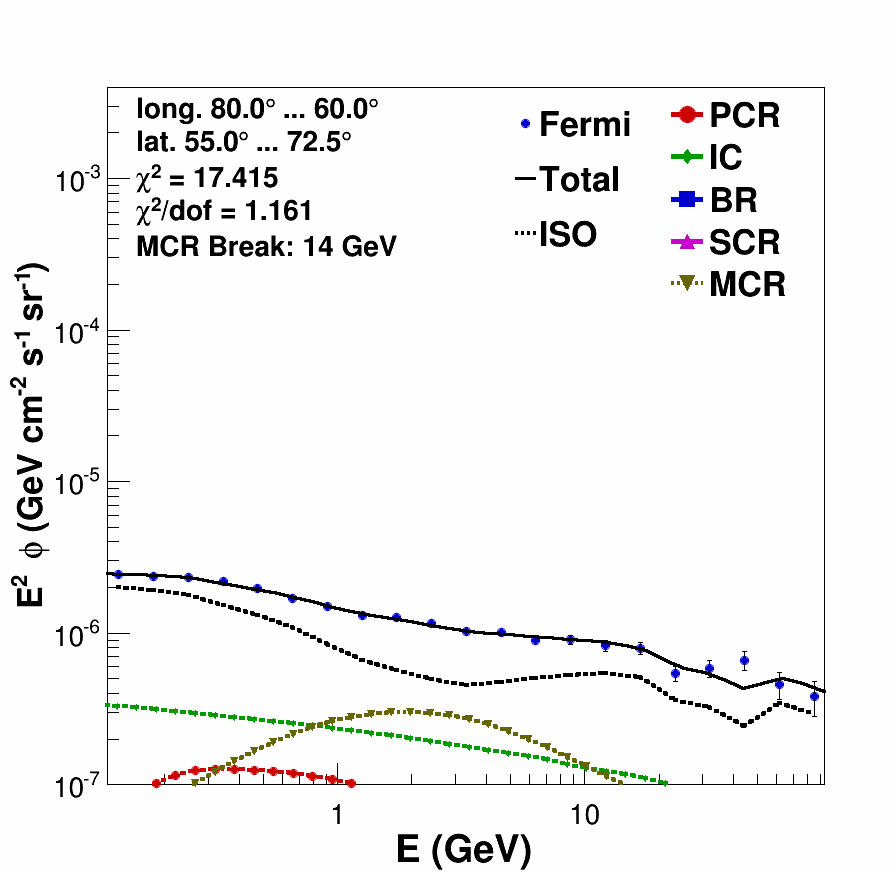}
\includegraphics[width=0.16\textwidth,height=0.16\textwidth,clip]{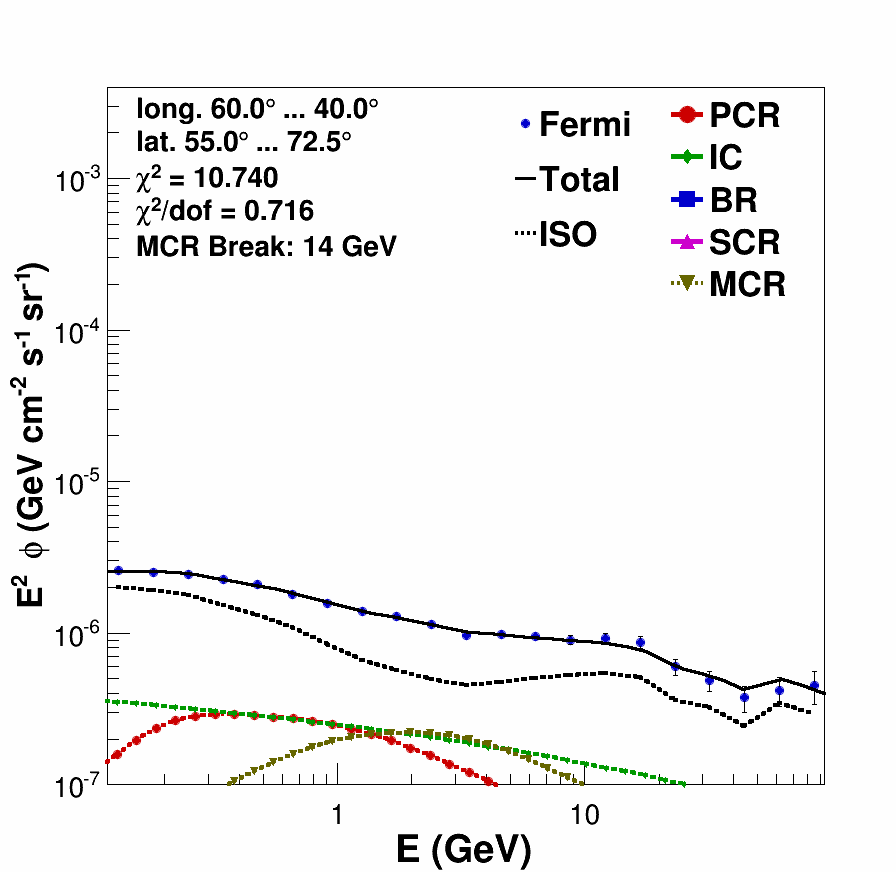}
\includegraphics[width=0.16\textwidth,height=0.16\textwidth,clip]{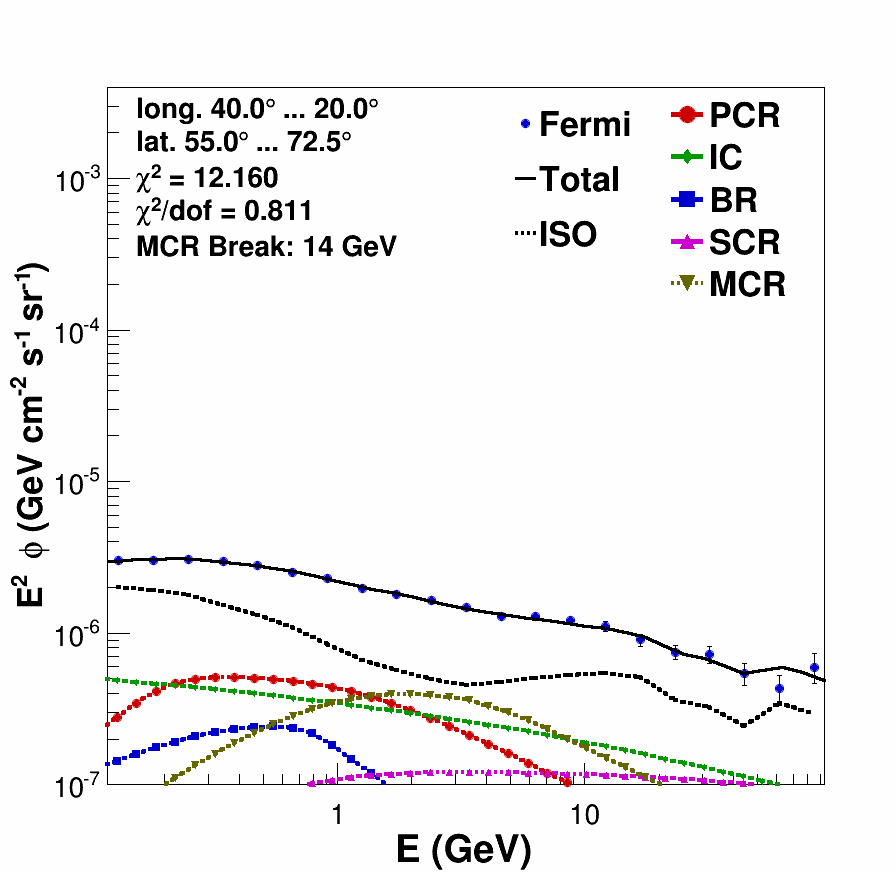}
\includegraphics[width=0.16\textwidth,height=0.16\textwidth,clip]{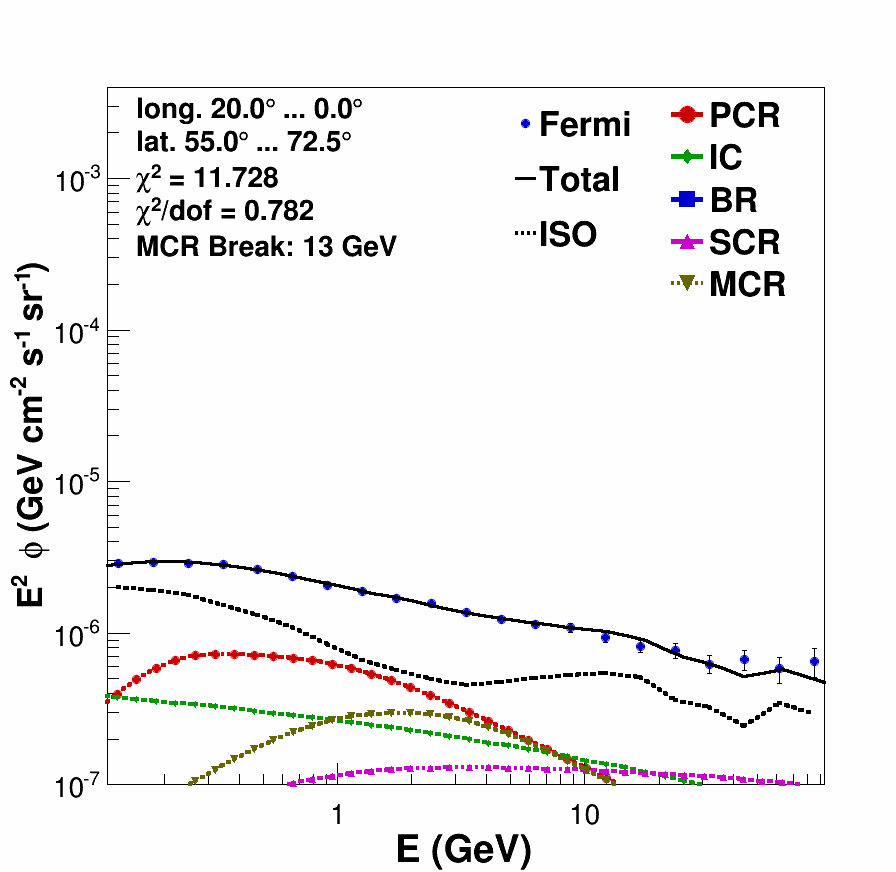}
\includegraphics[width=0.16\textwidth,height=0.16\textwidth,clip]{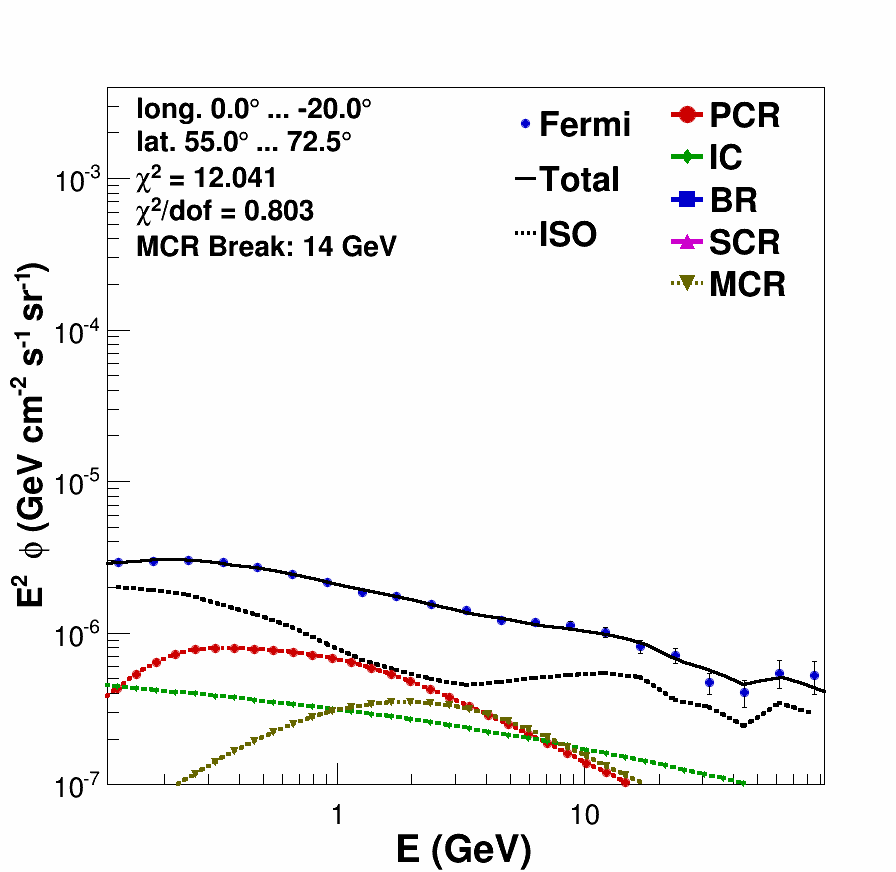}
\includegraphics[width=0.16\textwidth,height=0.16\textwidth,clip]{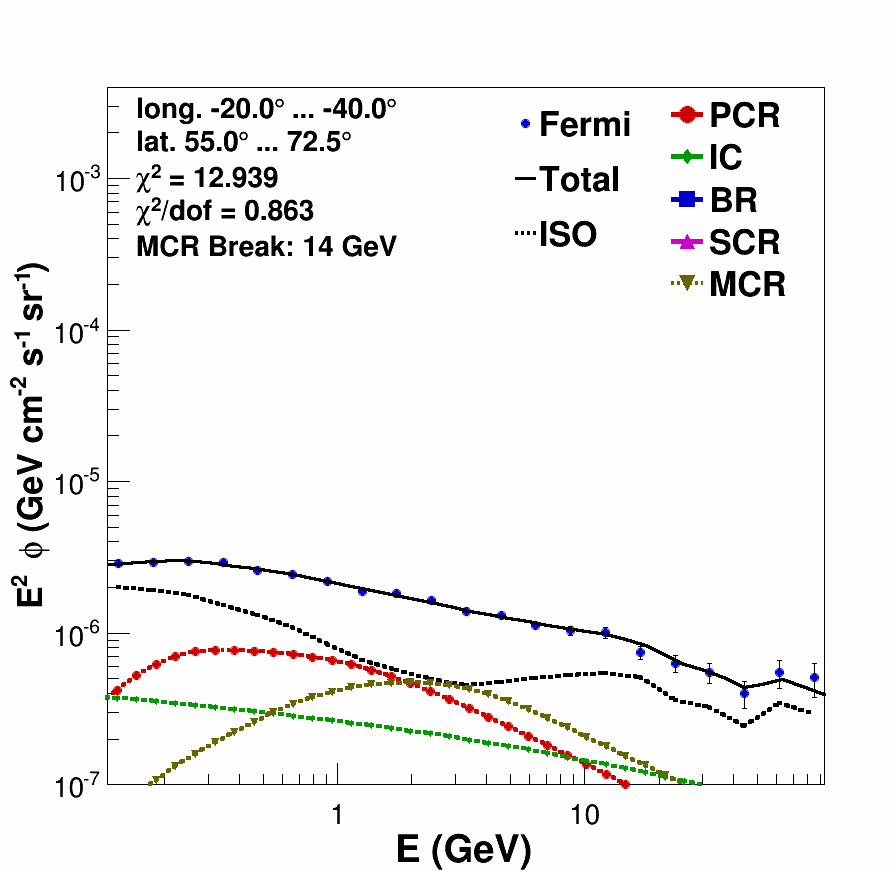}
\includegraphics[width=0.16\textwidth,height=0.16\textwidth,clip]{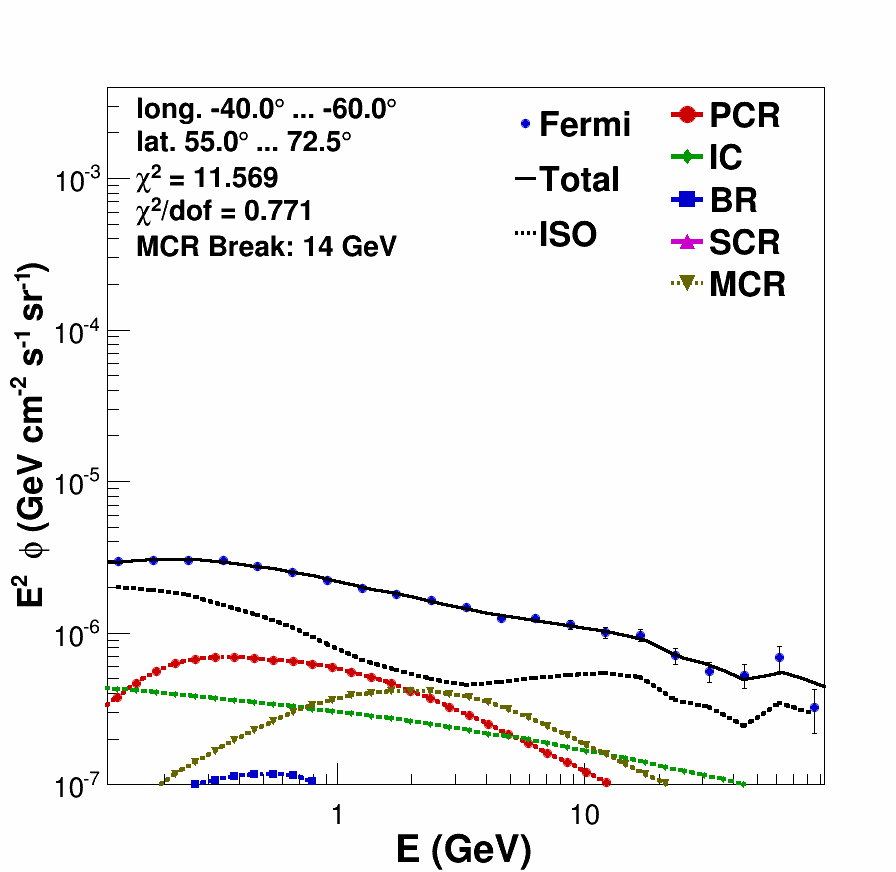}
\includegraphics[width=0.16\textwidth,height=0.16\textwidth,clip]{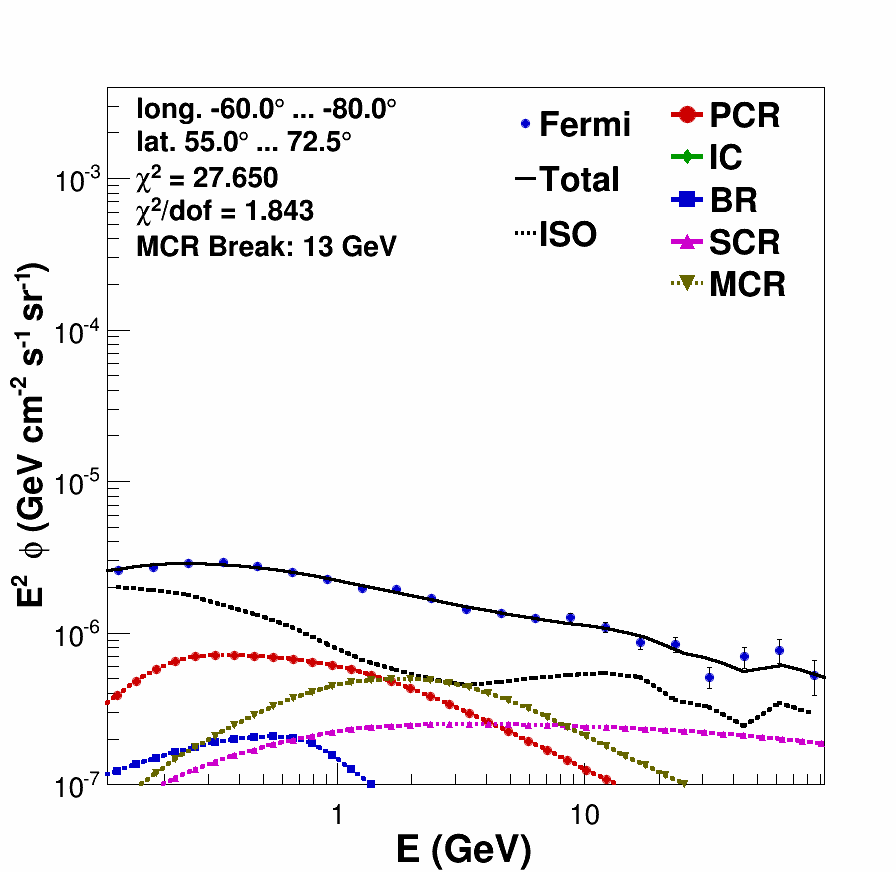}
\includegraphics[width=0.16\textwidth,height=0.16\textwidth,clip]{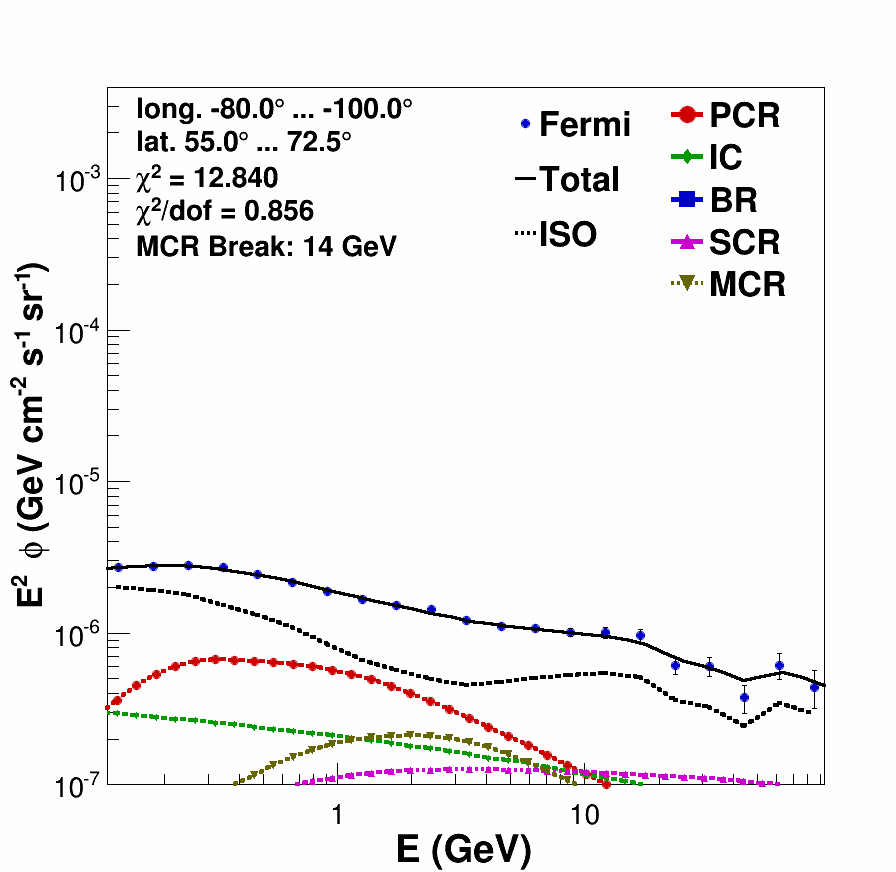}
\includegraphics[width=0.16\textwidth,height=0.16\textwidth,clip]{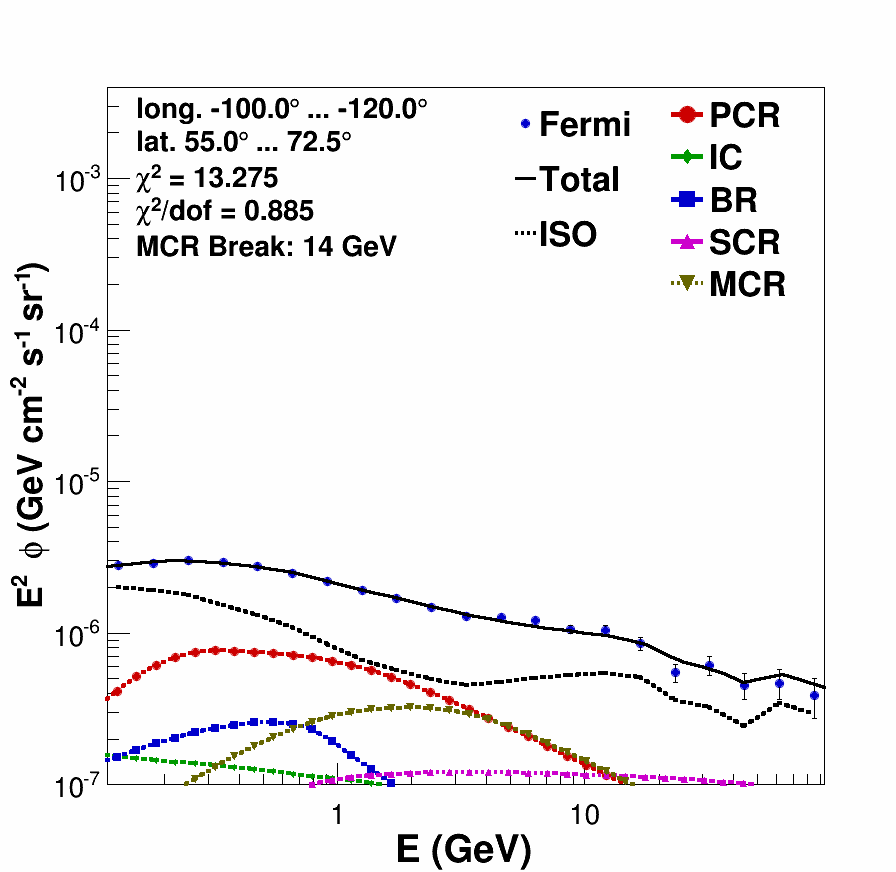}
\includegraphics[width=0.16\textwidth,height=0.16\textwidth,clip]{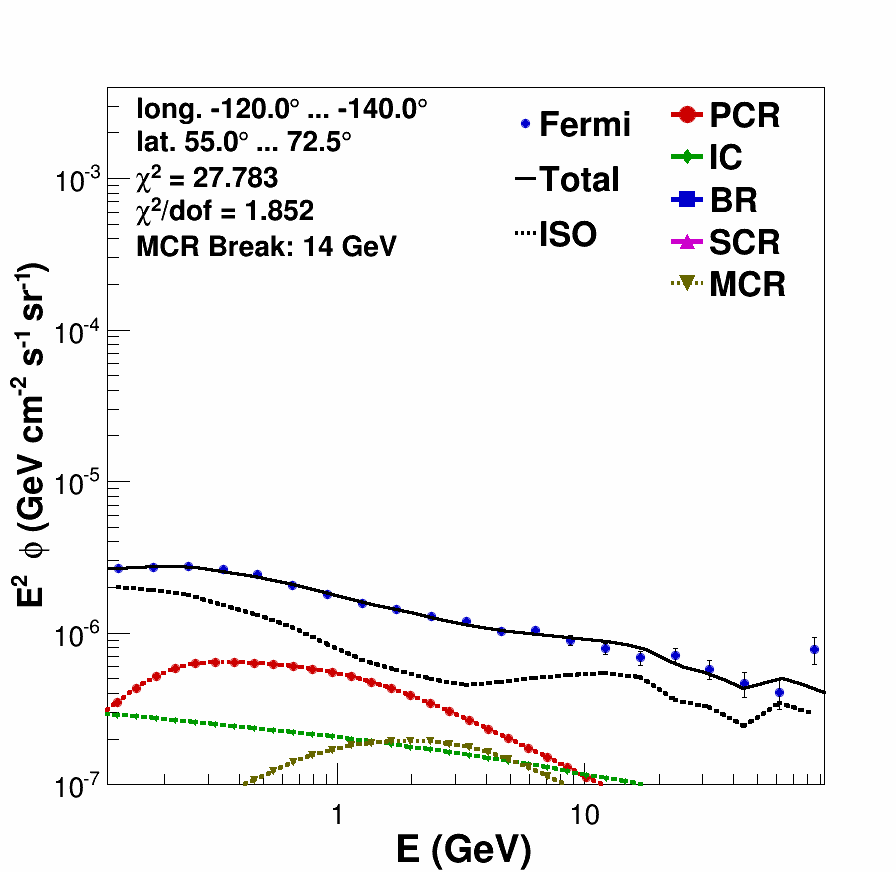}
\includegraphics[width=0.16\textwidth,height=0.16\textwidth,clip]{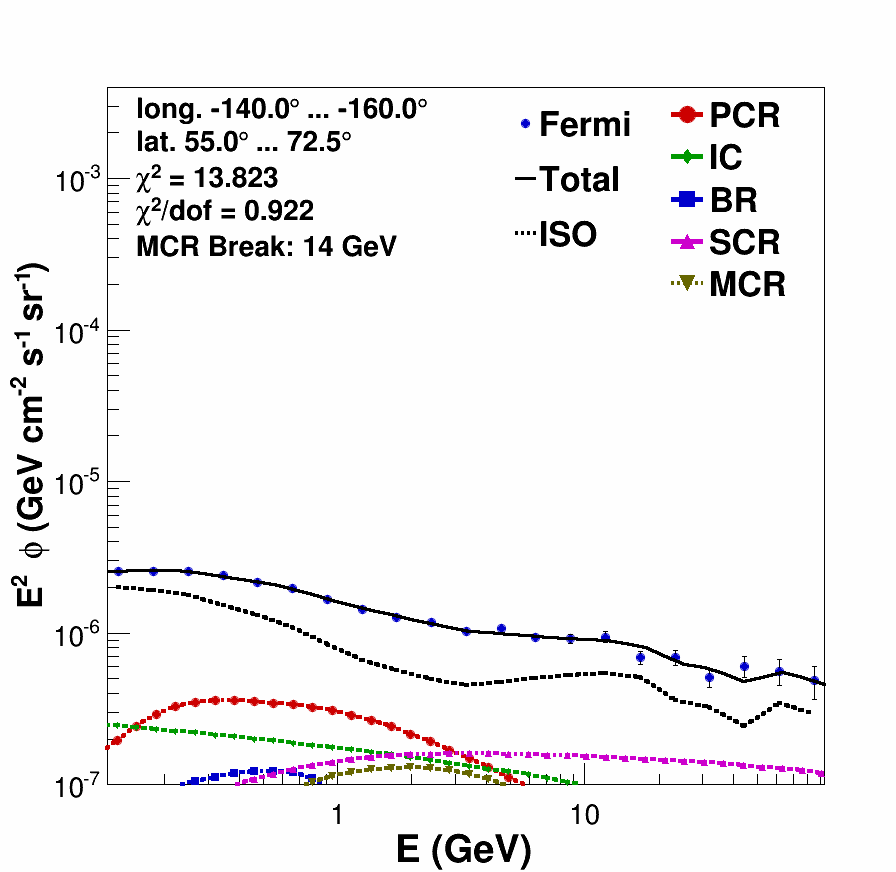}
\includegraphics[width=0.16\textwidth,height=0.16\textwidth,clip]{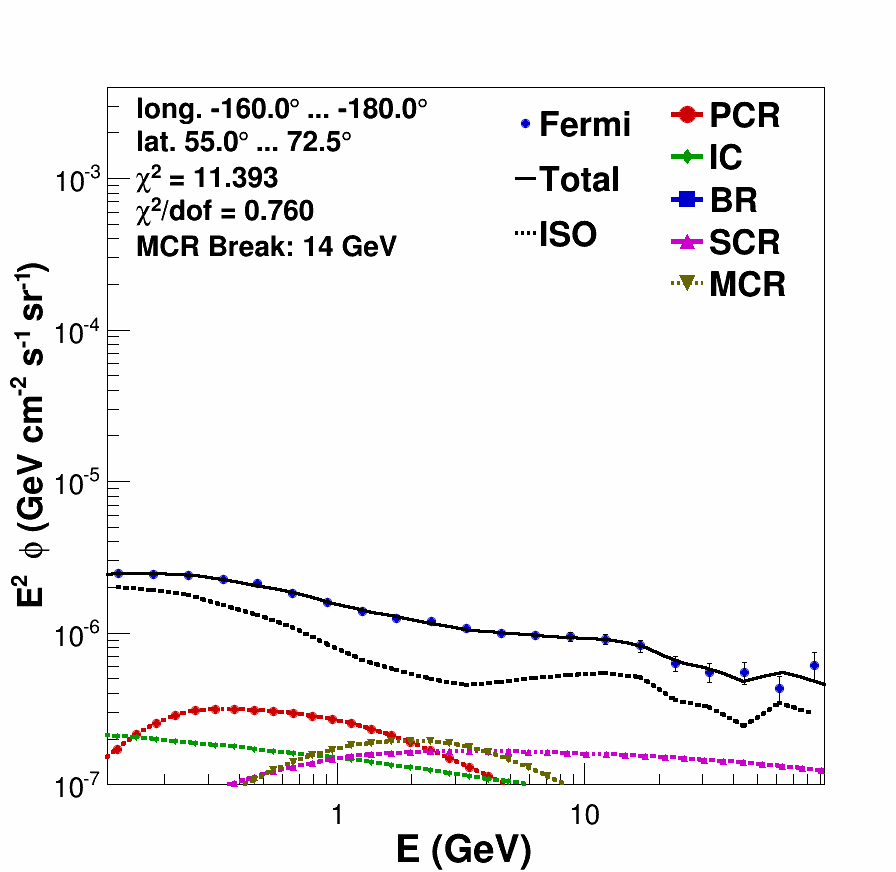}%\\%%%%%r2
\caption[]{Template fits for latitudes  with $55.0^\circ<b<72.5^\circ$ and longitudes decreasing from 180$^\circ$ to -180$^\circ$.} 
\label{F12}
\end{figure}
\begin{figure}
\centering
\includegraphics[width=0.16\textwidth,height=0.16\textwidth,clip]{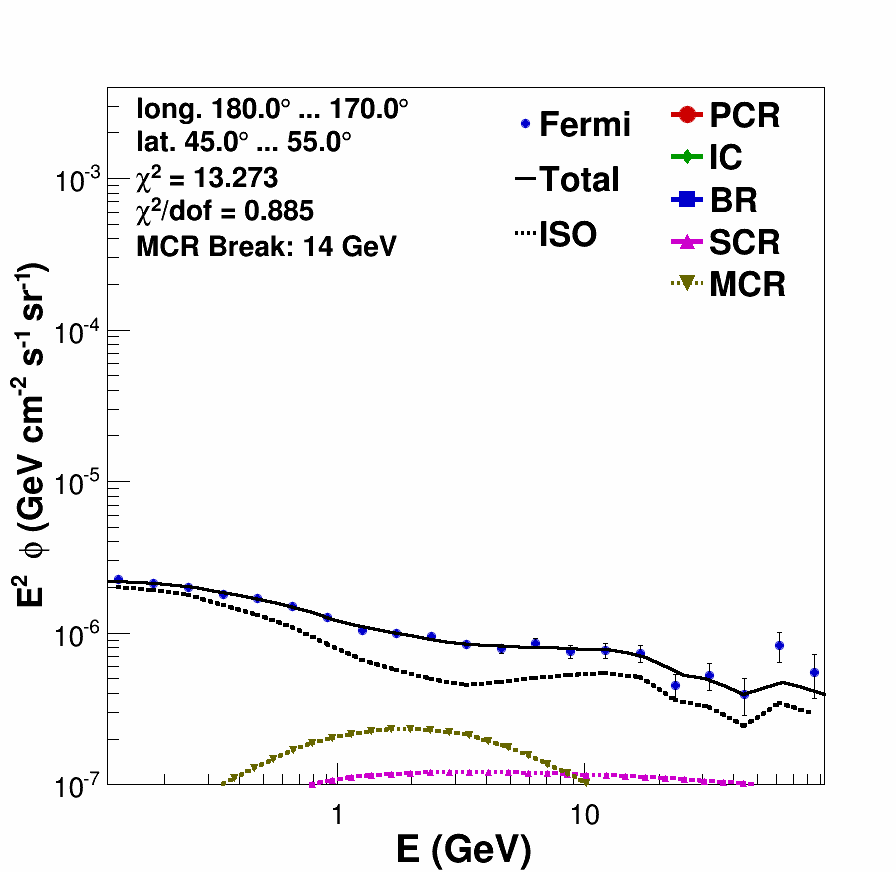}
\includegraphics[width=0.16\textwidth,height=0.16\textwidth,clip]{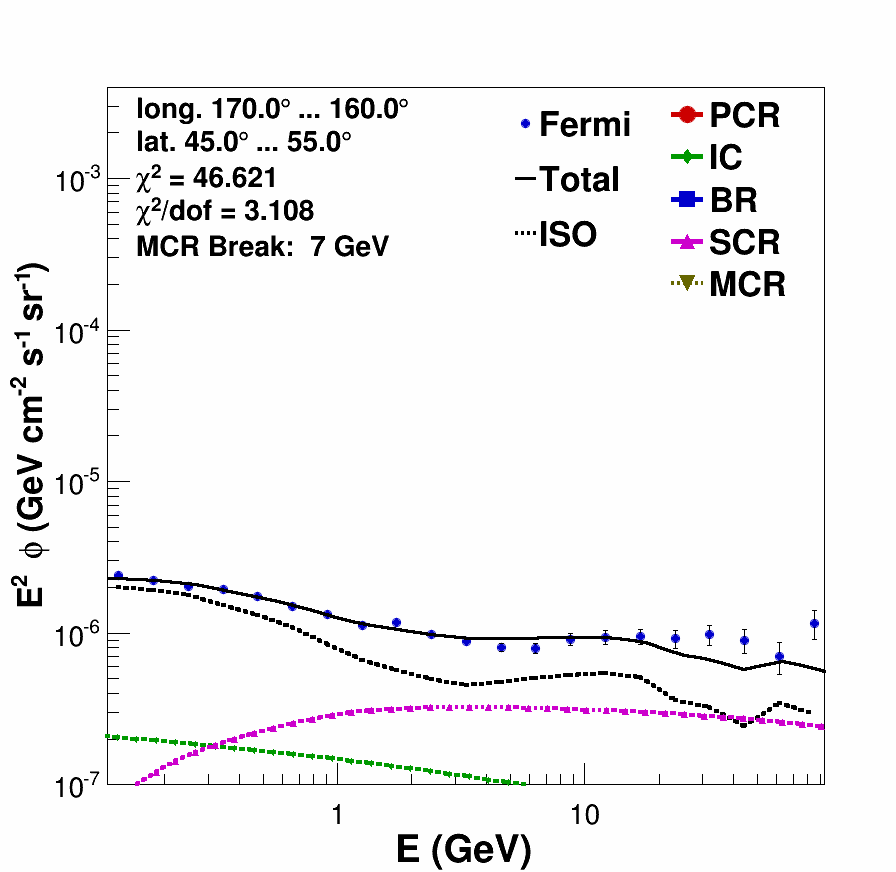}
\includegraphics[width=0.16\textwidth,height=0.16\textwidth,clip]{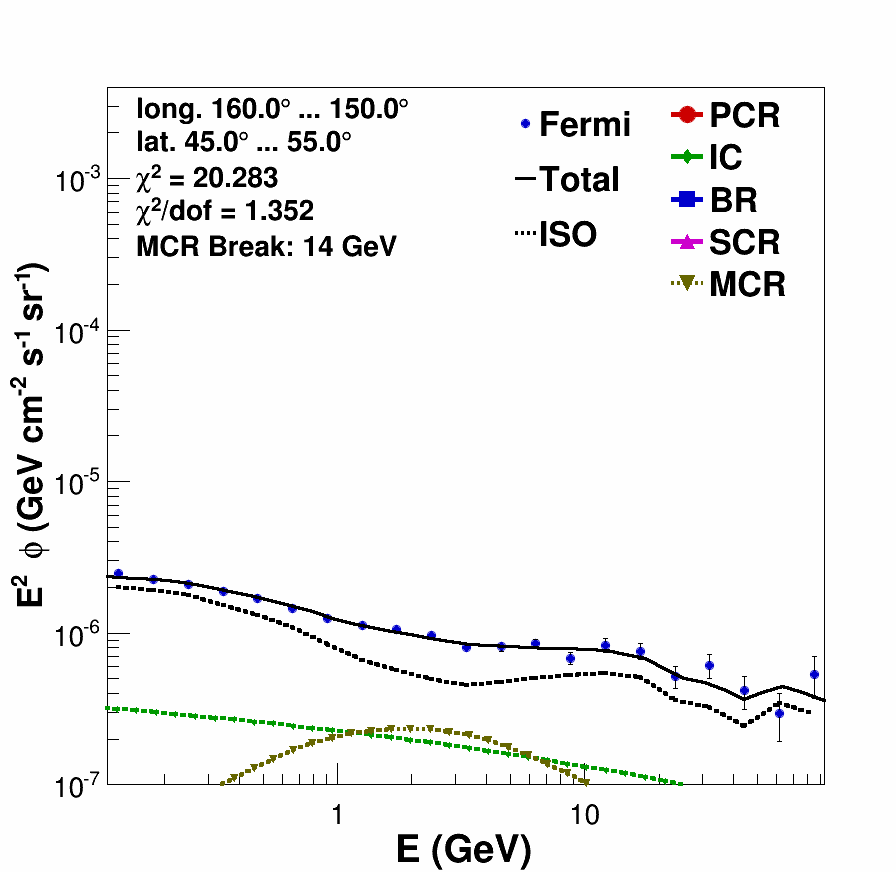}
\includegraphics[width=0.16\textwidth,height=0.16\textwidth,clip]{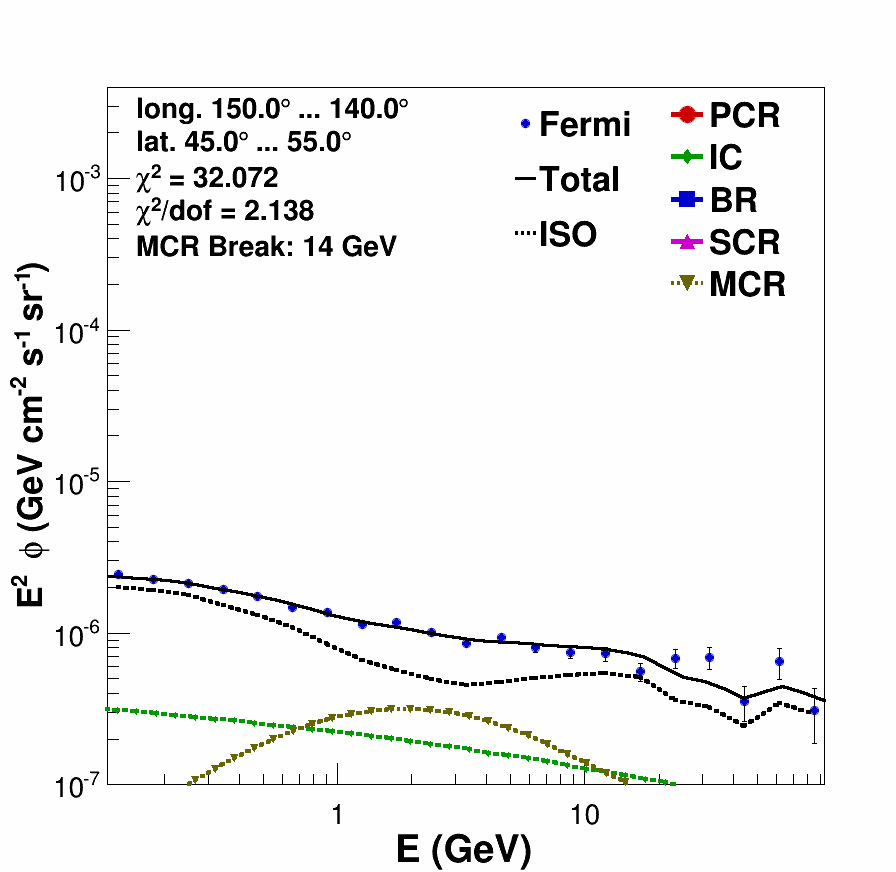}
\includegraphics[width=0.16\textwidth,height=0.16\textwidth,clip]{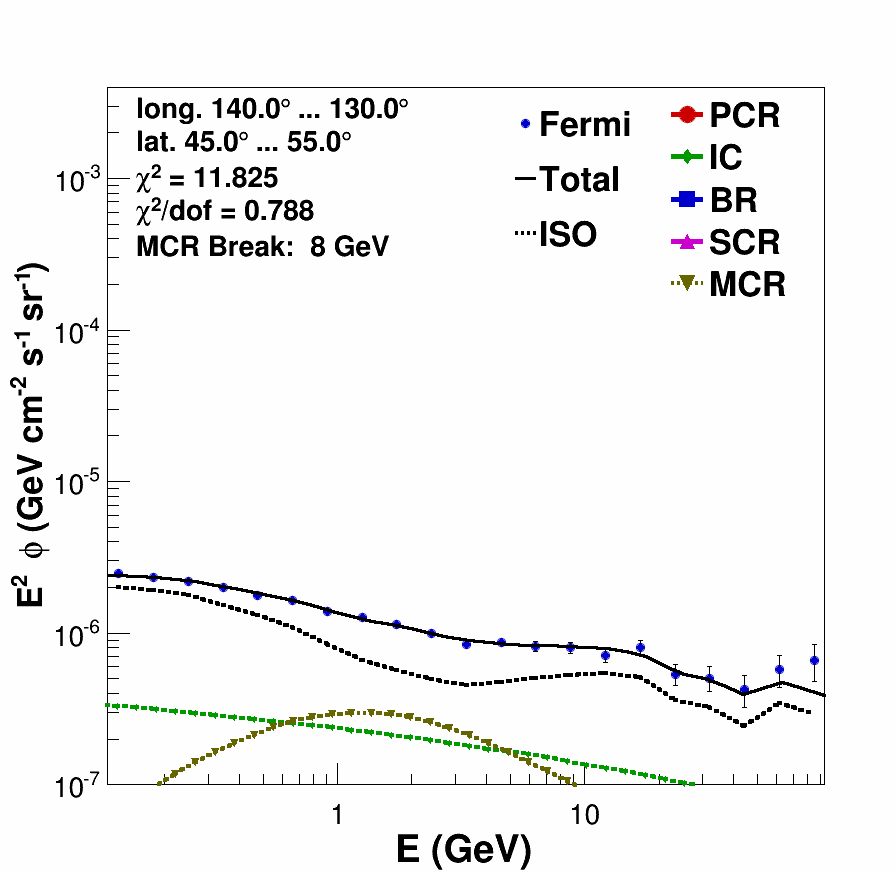}
\includegraphics[width=0.16\textwidth,height=0.16\textwidth,clip]{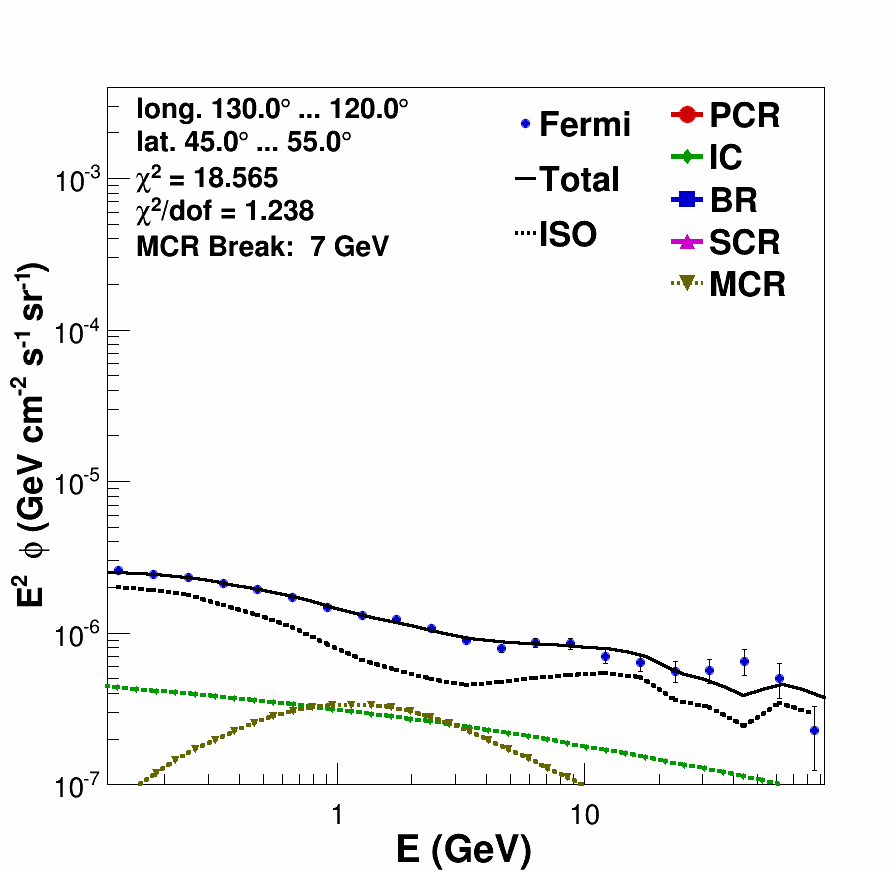}
\includegraphics[width=0.16\textwidth,height=0.16\textwidth,clip]{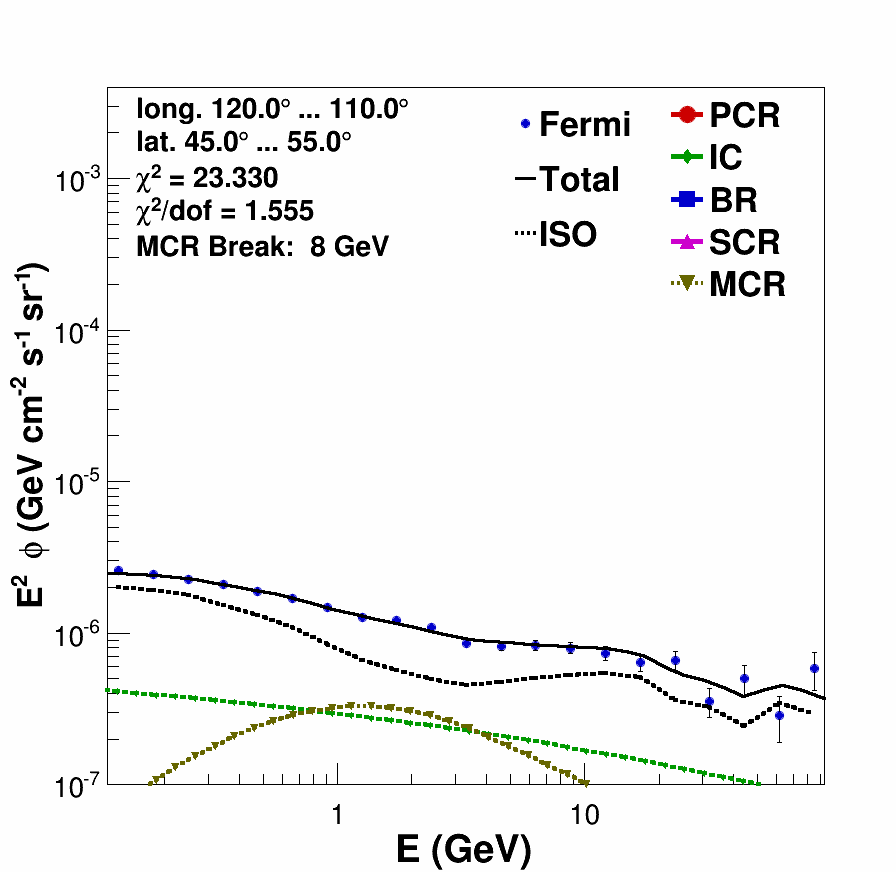}
\includegraphics[width=0.16\textwidth,height=0.16\textwidth,clip]{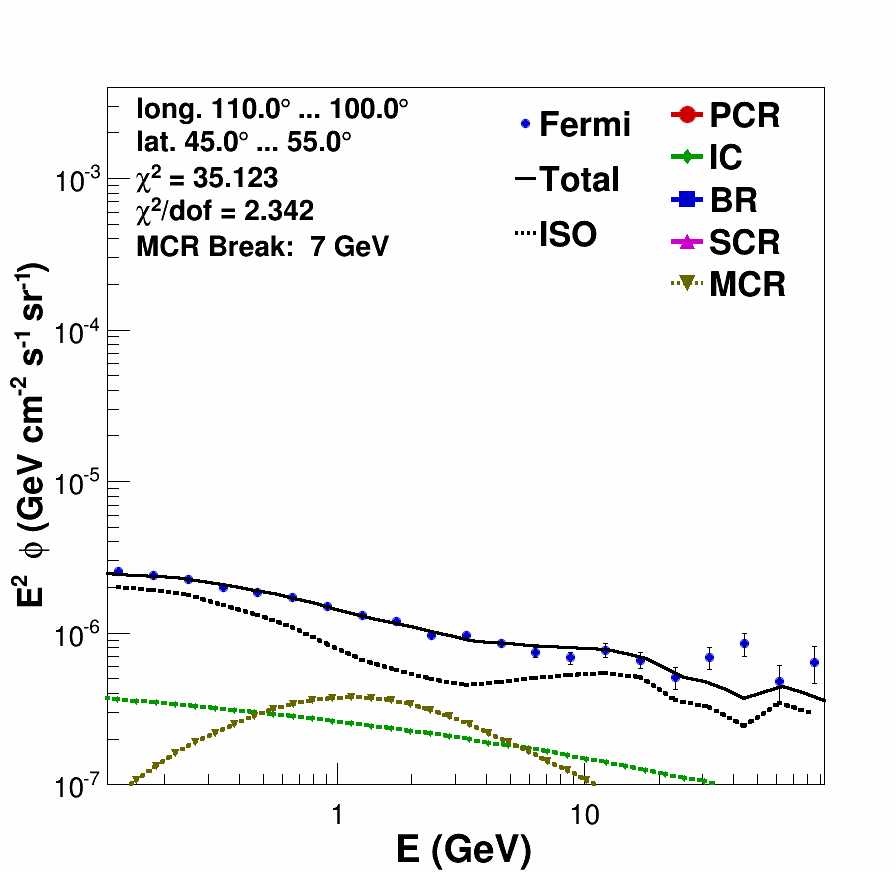}
\includegraphics[width=0.16\textwidth,height=0.16\textwidth,clip]{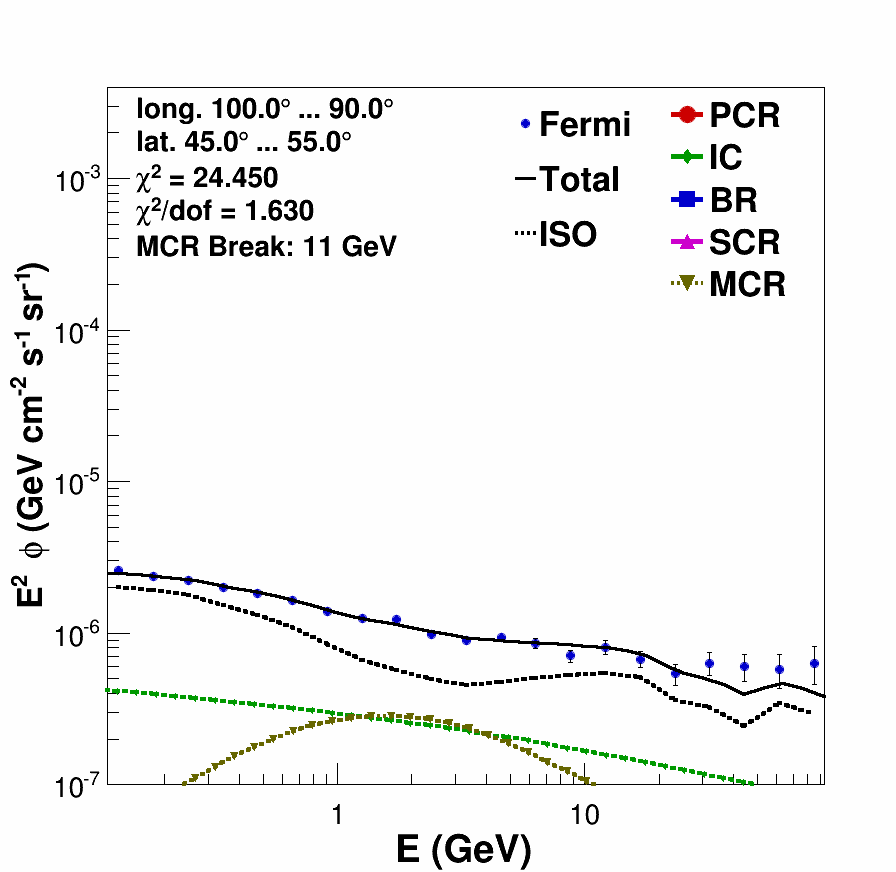}
\includegraphics[width=0.16\textwidth,height=0.16\textwidth,clip]{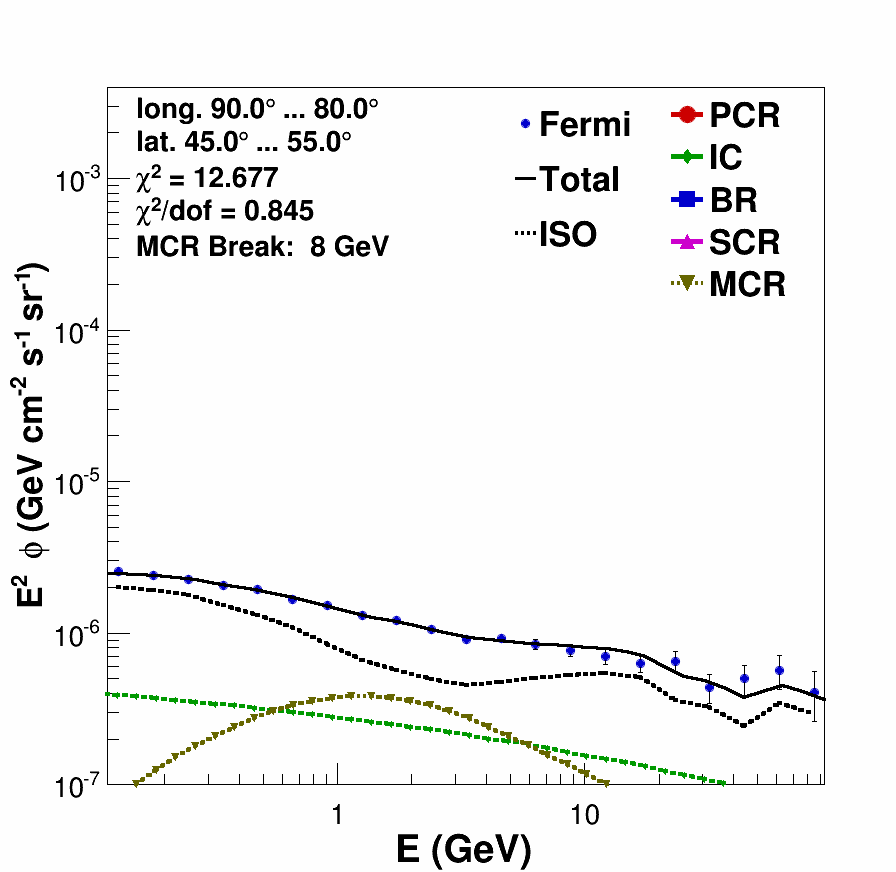}
\includegraphics[width=0.16\textwidth,height=0.16\textwidth,clip]{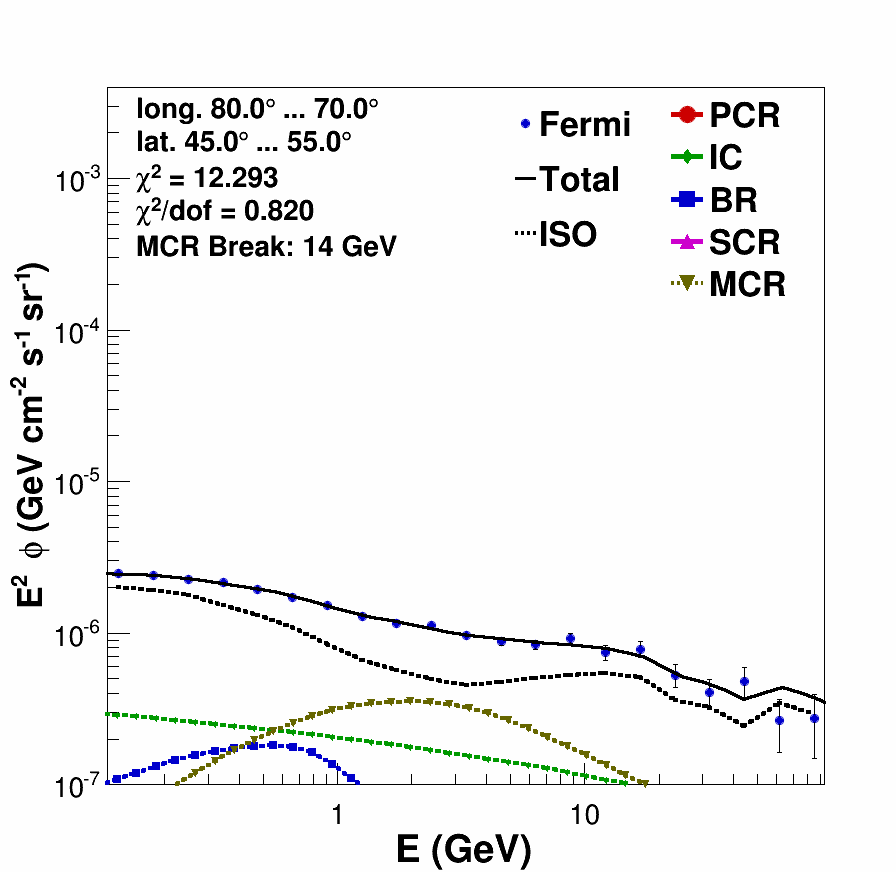}
\includegraphics[width=0.16\textwidth,height=0.16\textwidth,clip]{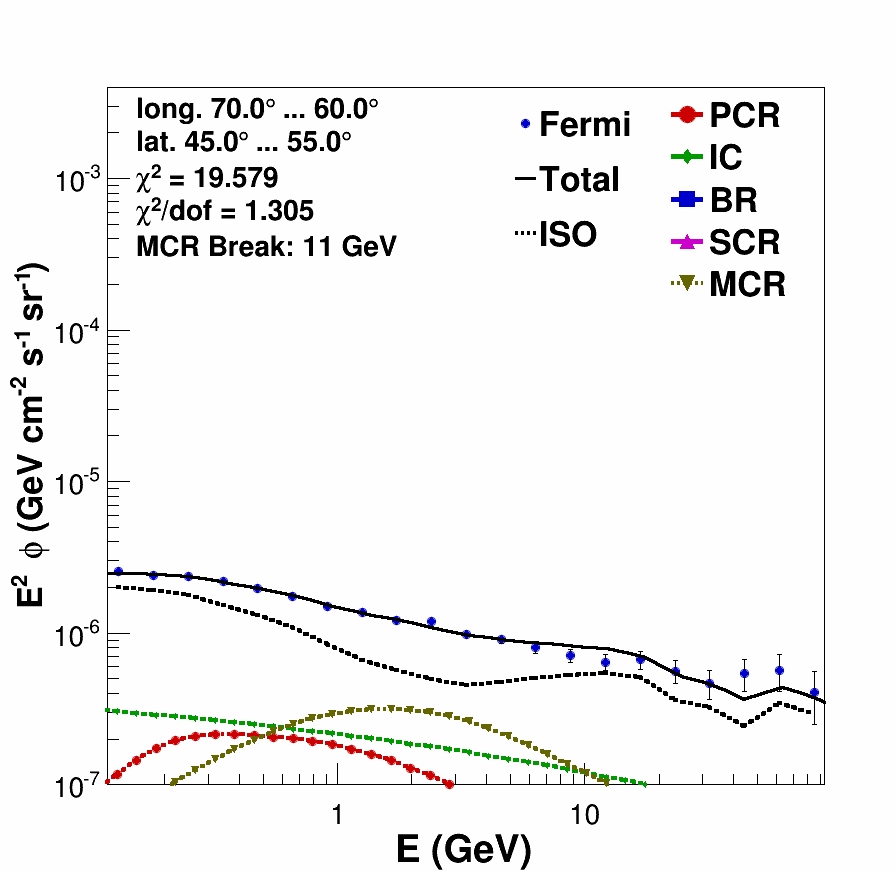}
\includegraphics[width=0.16\textwidth,height=0.16\textwidth,clip]{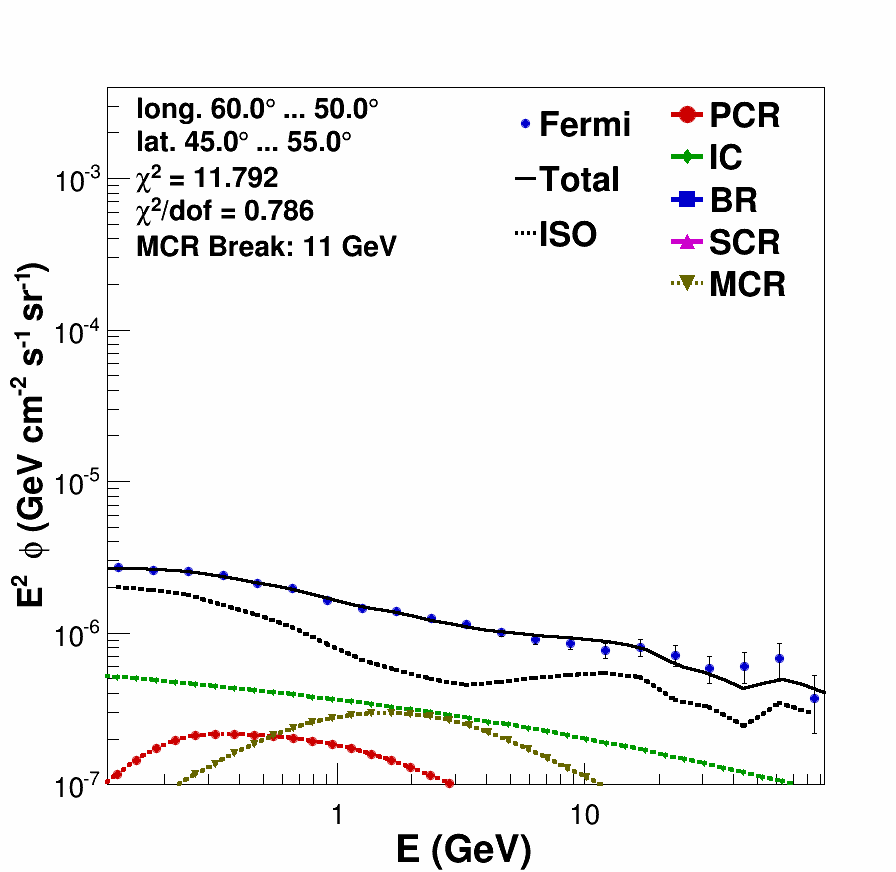}
\includegraphics[width=0.16\textwidth,height=0.16\textwidth,clip]{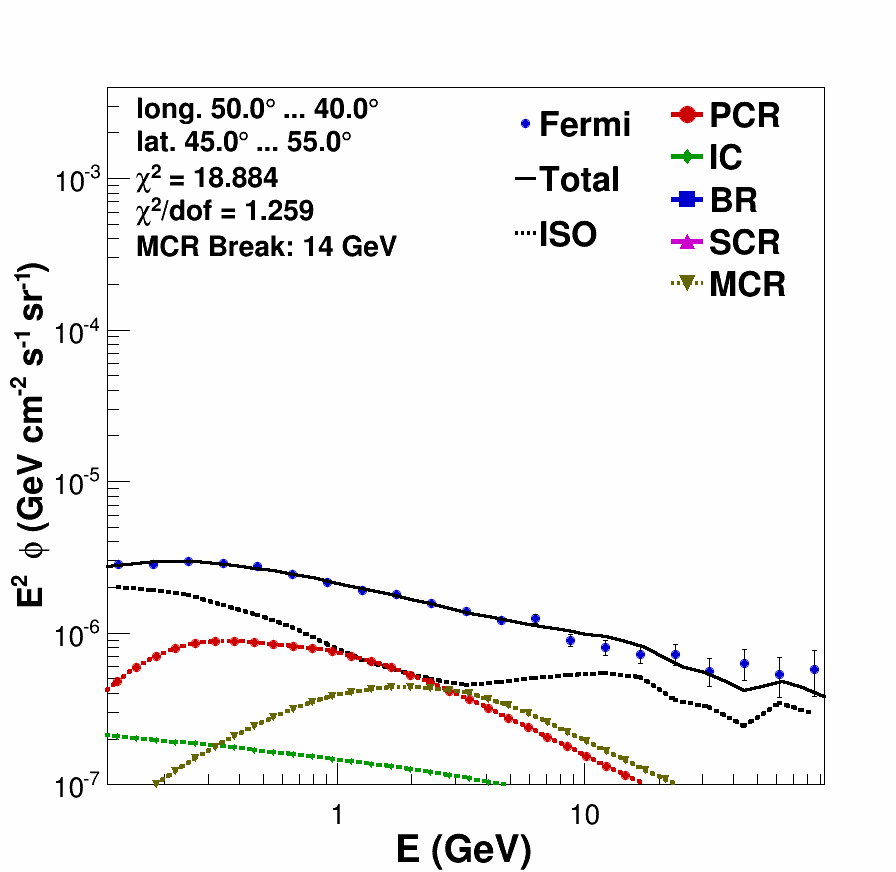}
\includegraphics[width=0.16\textwidth,height=0.16\textwidth,clip]{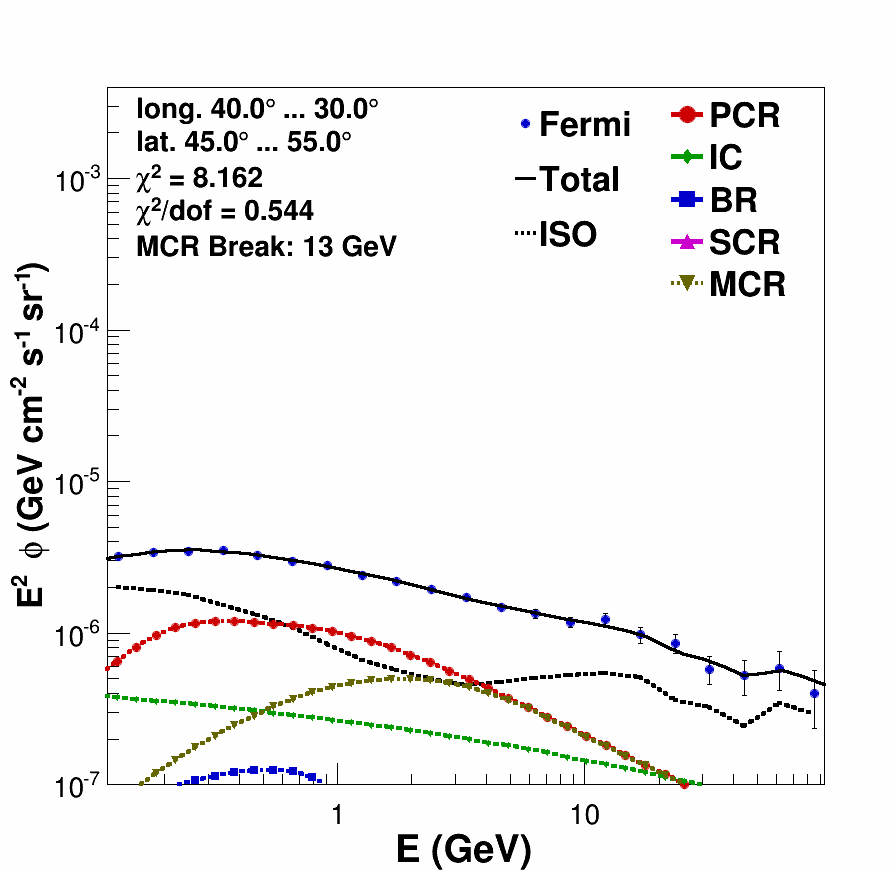}
\includegraphics[width=0.16\textwidth,height=0.16\textwidth,clip]{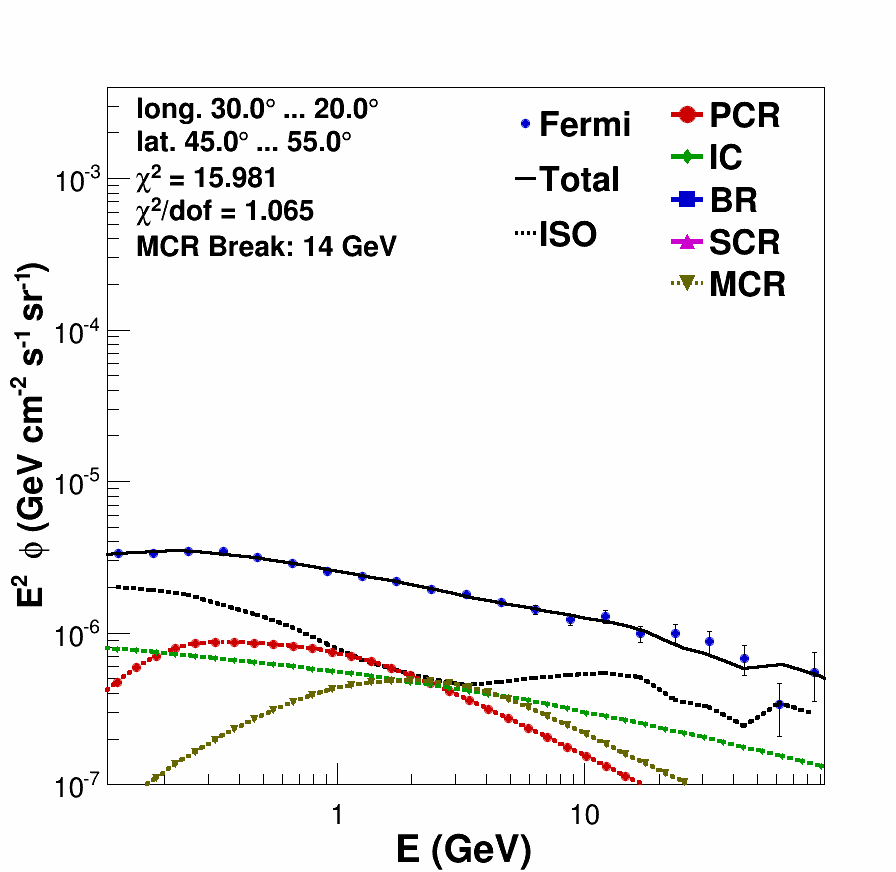}
\includegraphics[width=0.16\textwidth,height=0.16\textwidth,clip]{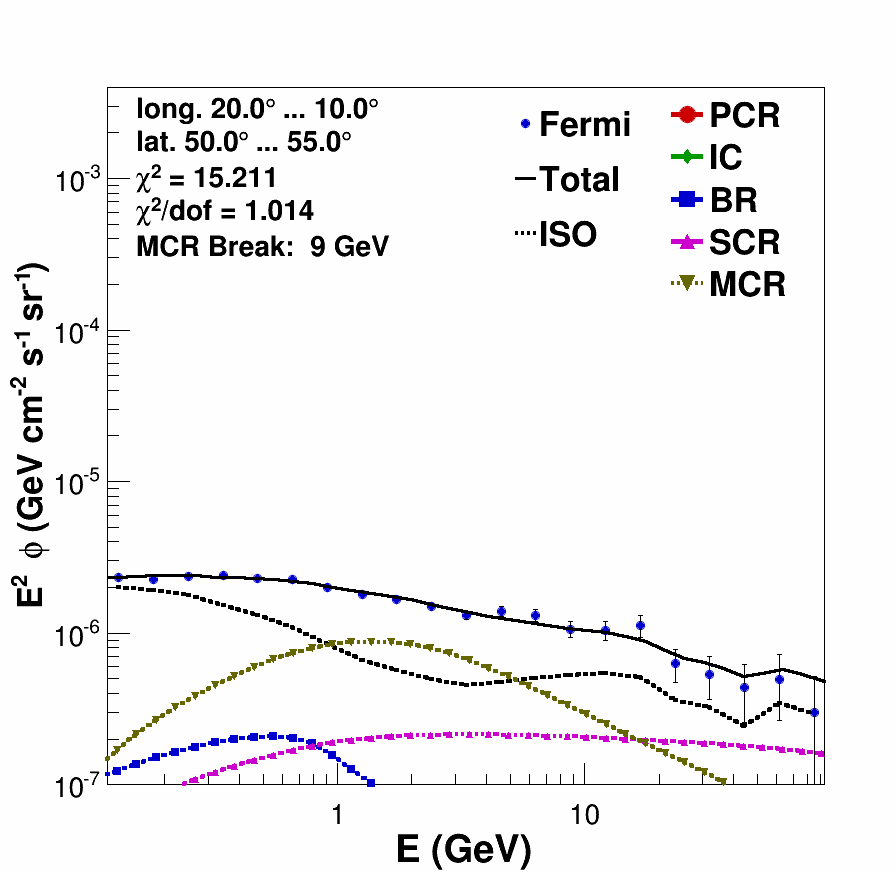}
\includegraphics[width=0.16\textwidth,height=0.16\textwidth,clip]{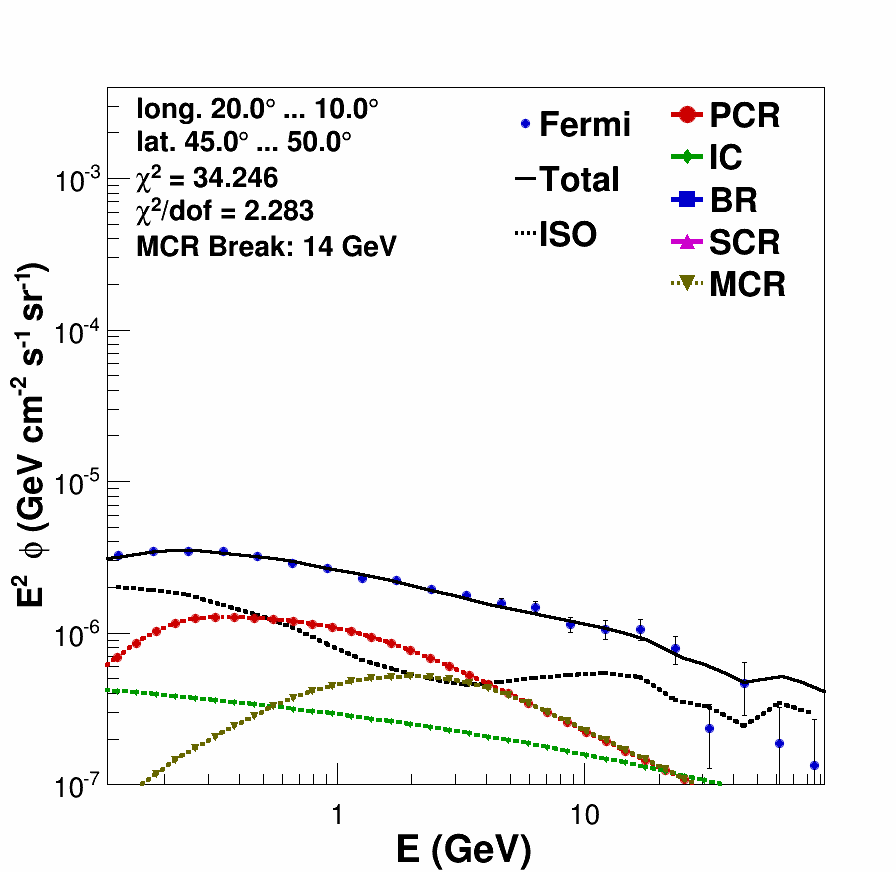}
\includegraphics[width=0.16\textwidth,height=0.16\textwidth,clip]{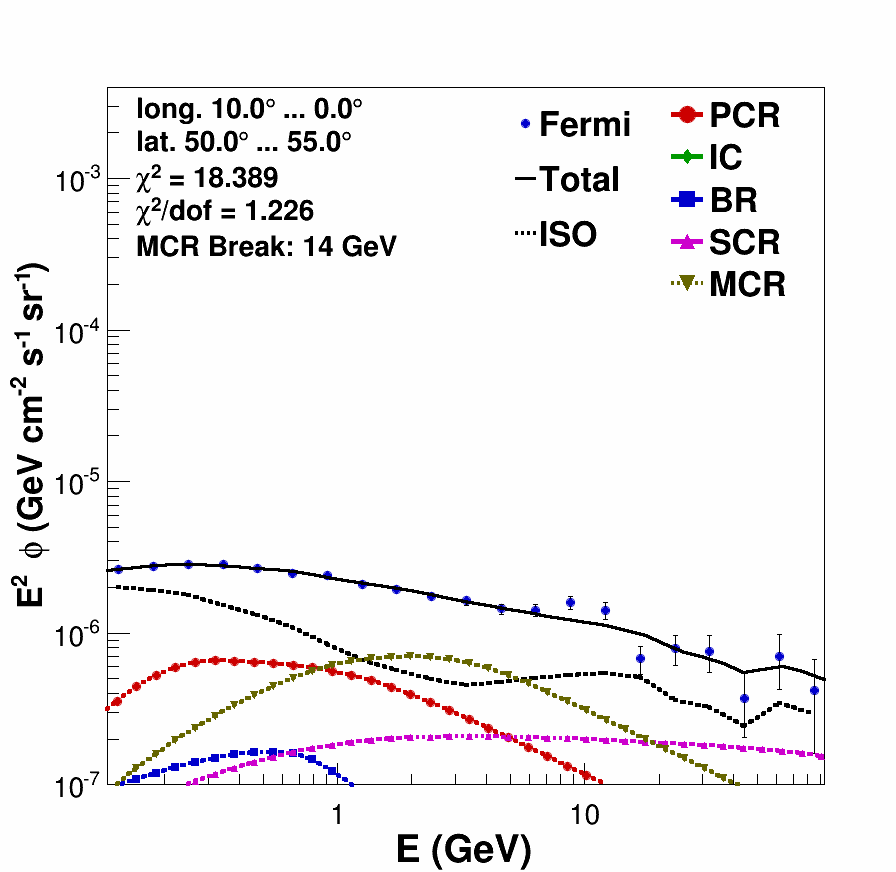}
\includegraphics[width=0.16\textwidth,height=0.16\textwidth,clip]{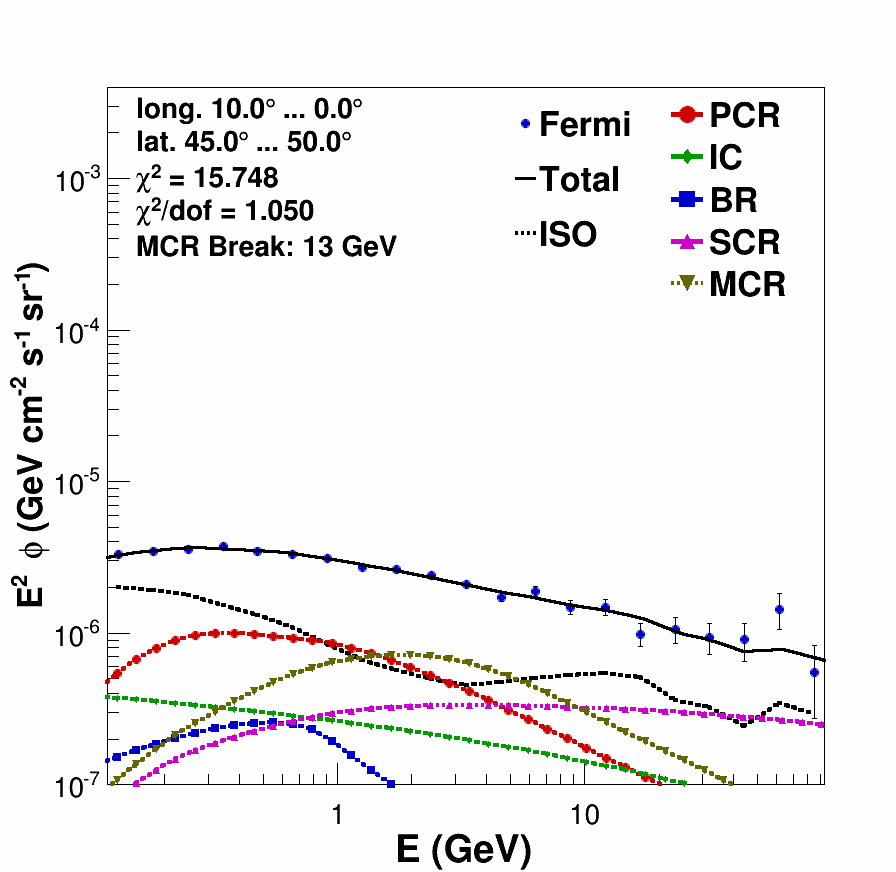}
\includegraphics[width=0.16\textwidth,height=0.16\textwidth,clip]{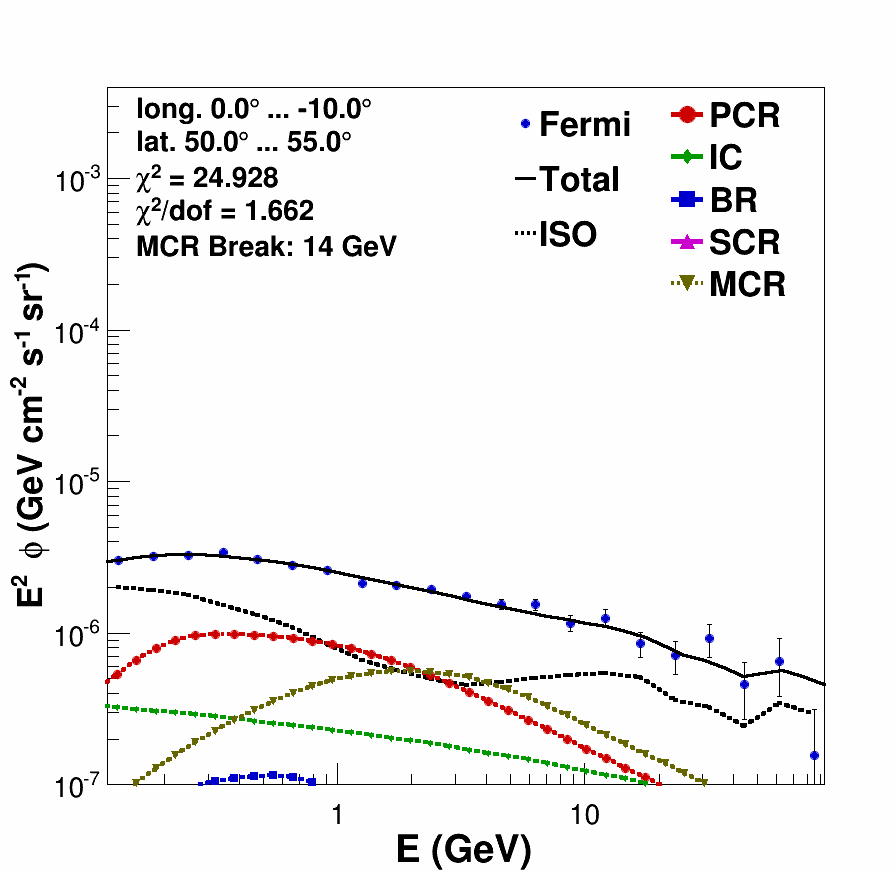}
\includegraphics[width=0.16\textwidth,height=0.16\textwidth,clip]{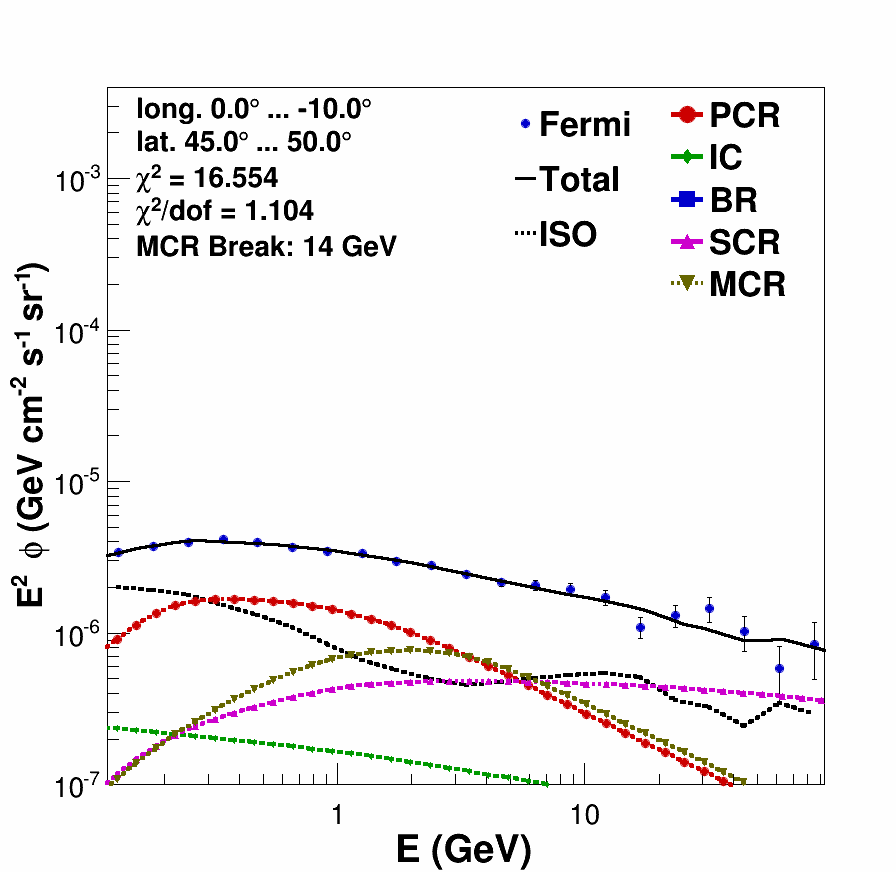}
\includegraphics[width=0.16\textwidth,height=0.16\textwidth,clip]{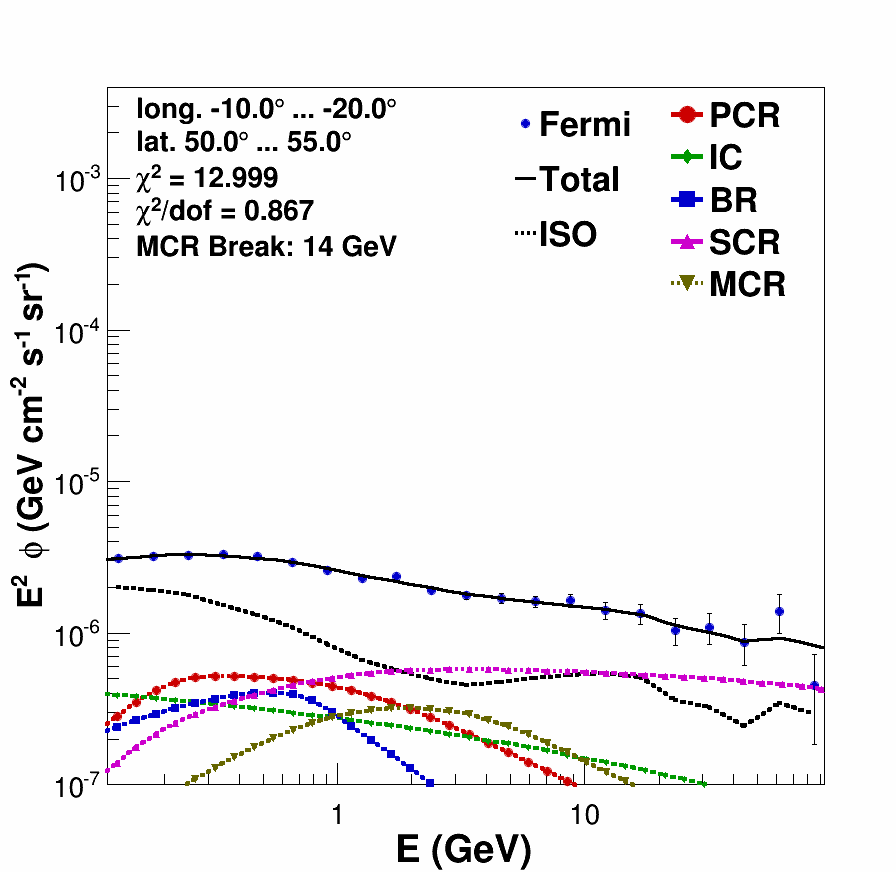}
\includegraphics[width=0.16\textwidth,height=0.16\textwidth,clip]{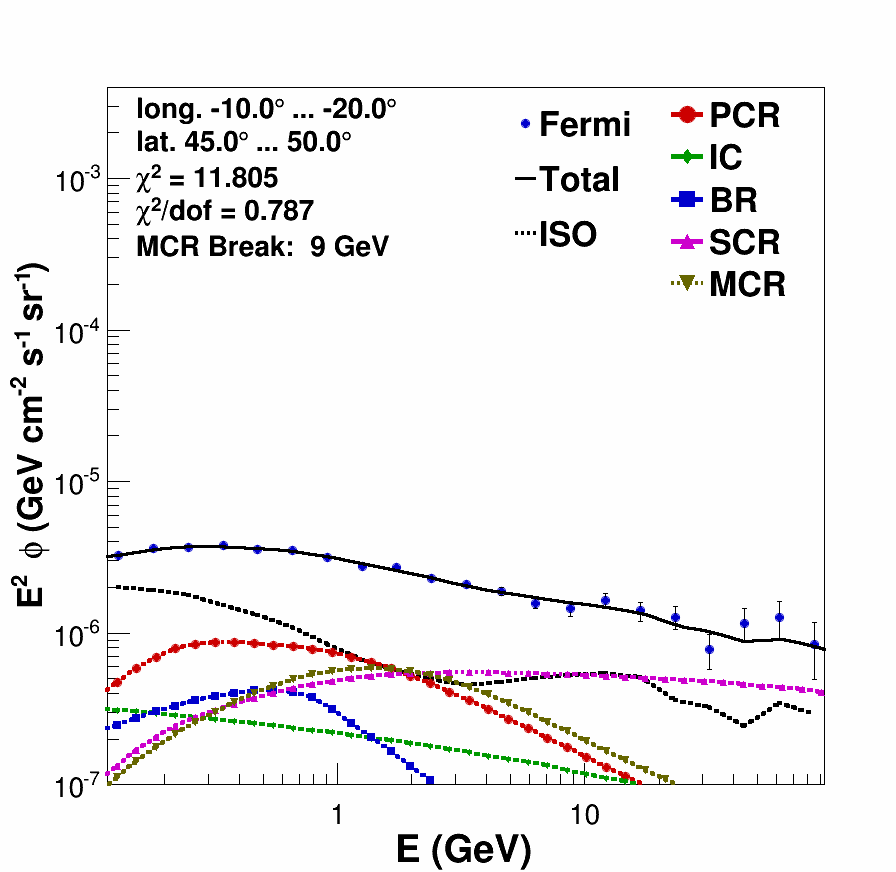}
\includegraphics[width=0.16\textwidth,height=0.16\textwidth,clip]{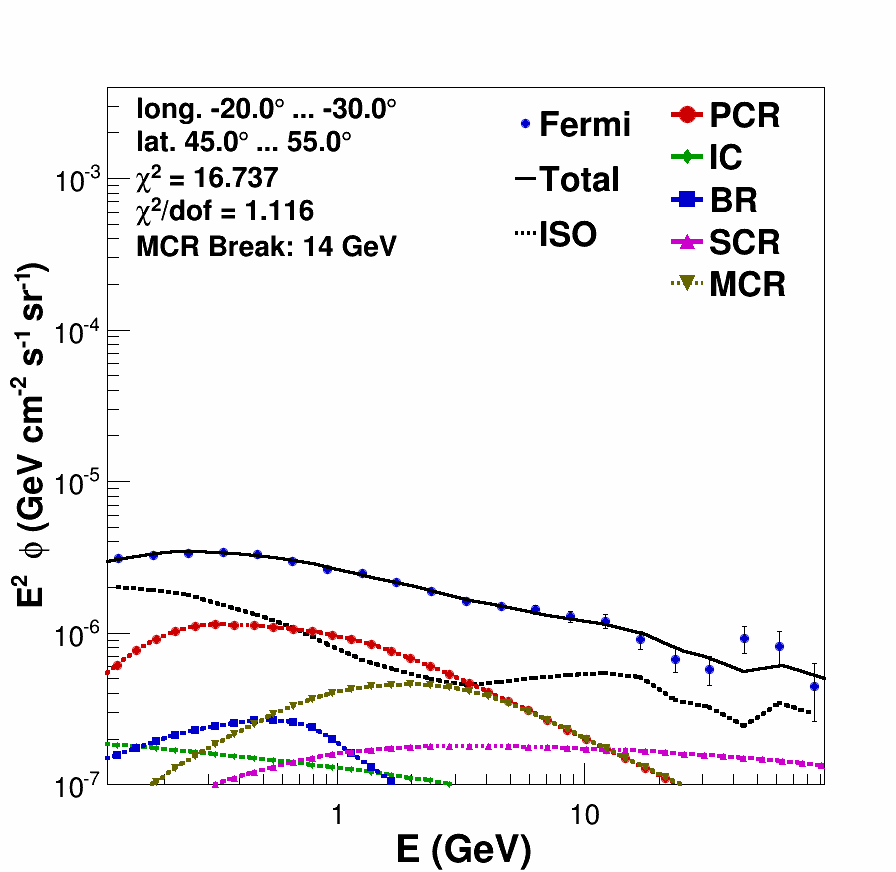}
\includegraphics[width=0.16\textwidth,height=0.16\textwidth,clip]{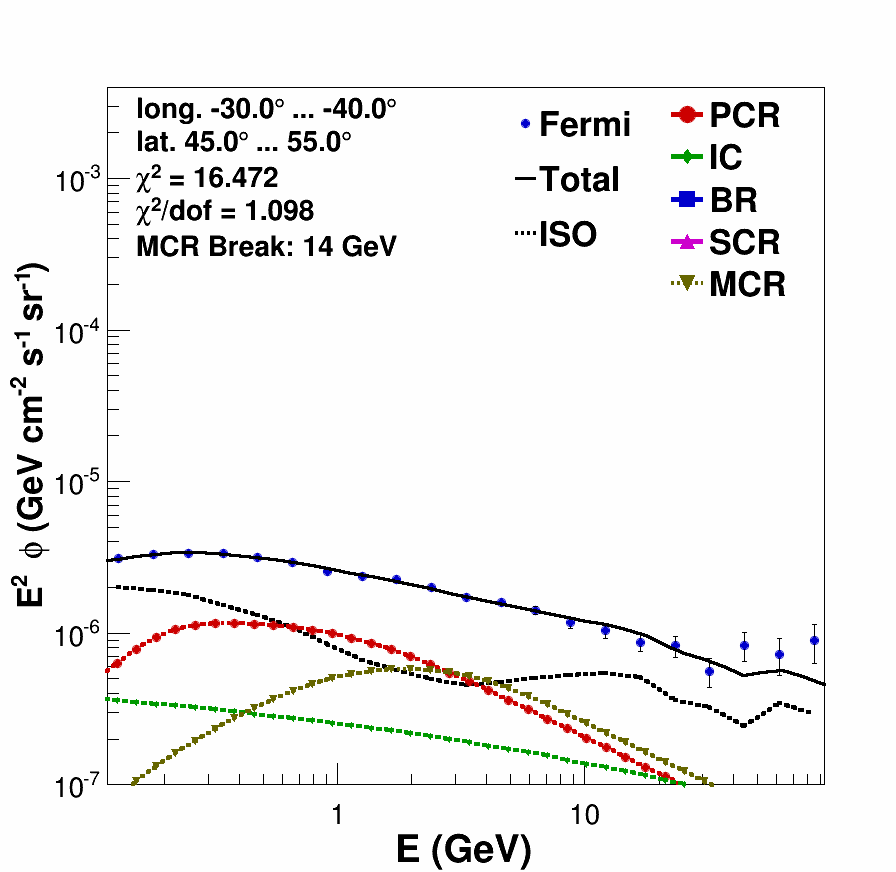}
\includegraphics[width=0.16\textwidth,height=0.16\textwidth,clip]{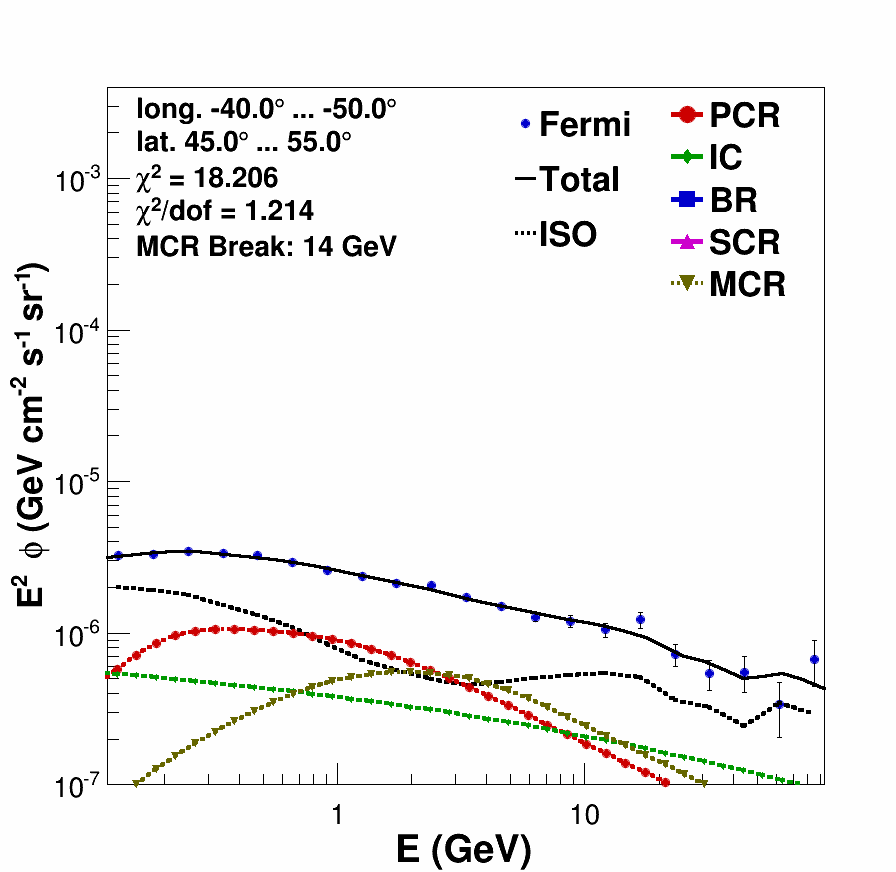}
\includegraphics[width=0.16\textwidth,height=0.16\textwidth,clip]{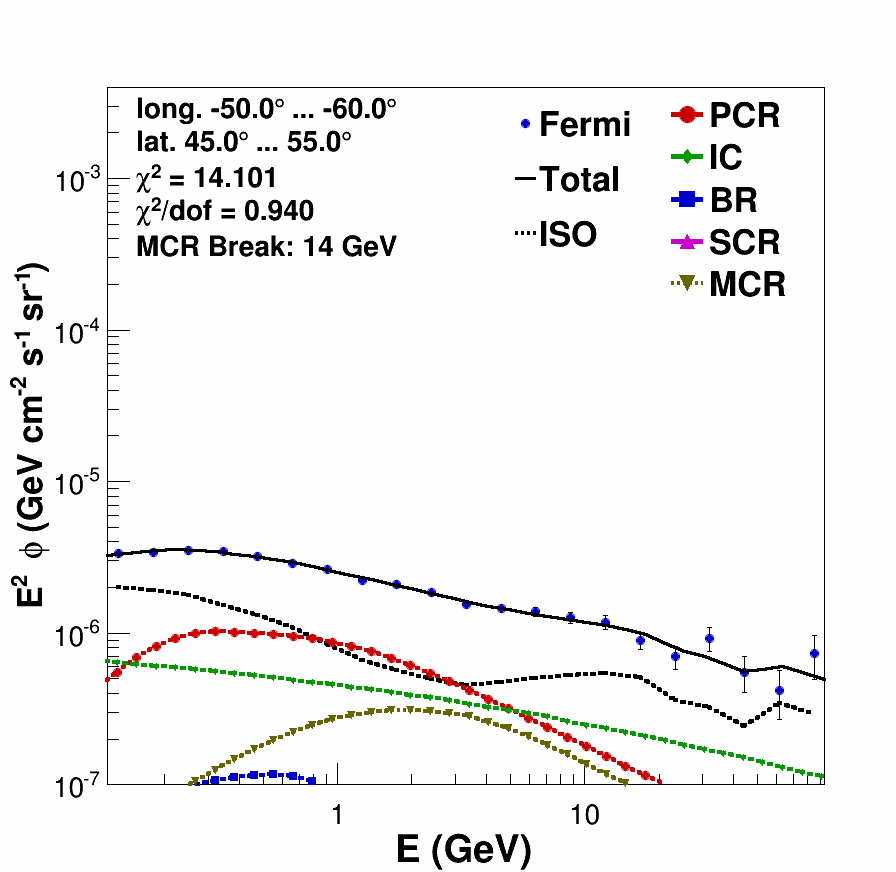}
\includegraphics[width=0.16\textwidth,height=0.16\textwidth,clip]{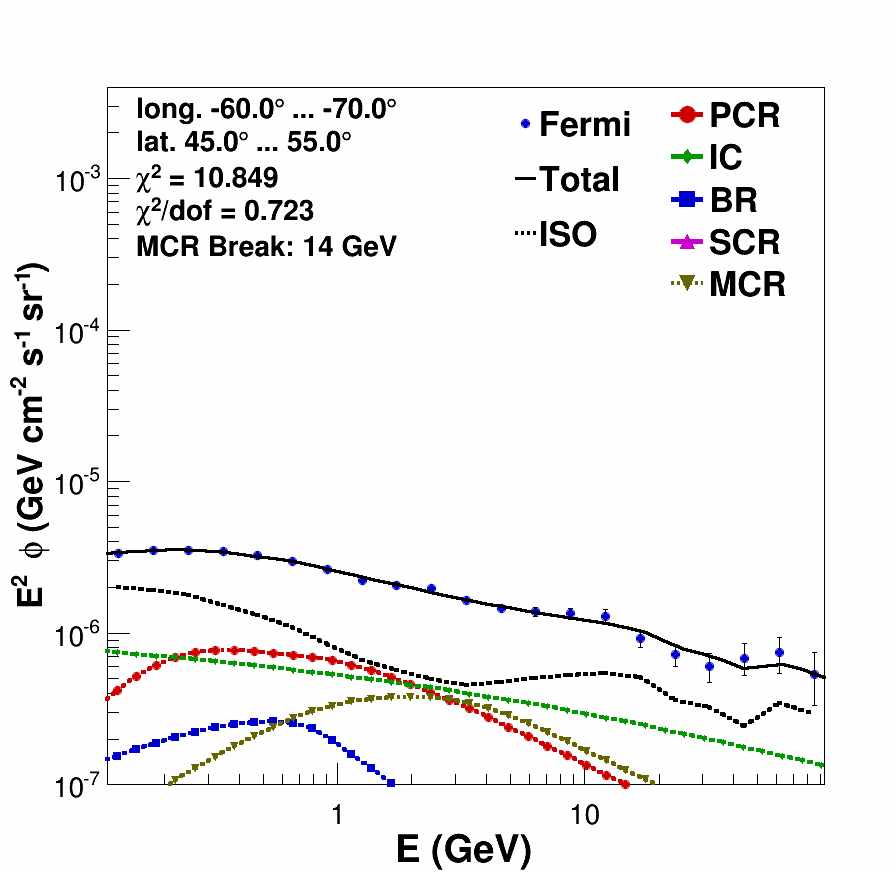}
\includegraphics[width=0.16\textwidth,height=0.16\textwidth,clip]{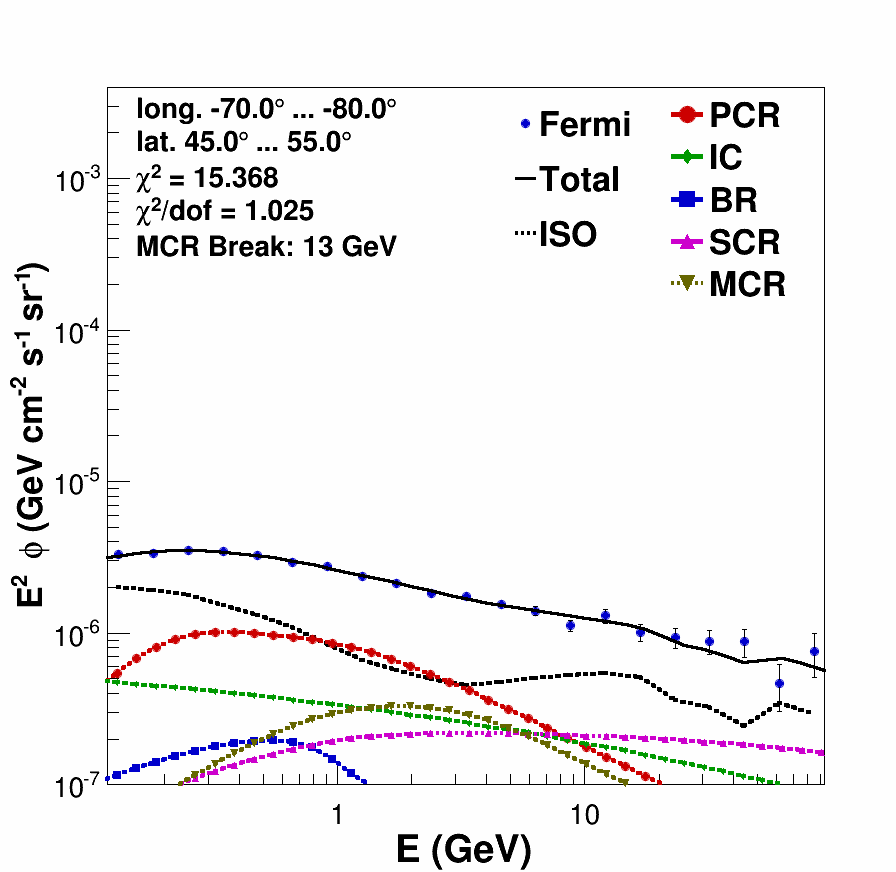}
\includegraphics[width=0.16\textwidth,height=0.16\textwidth,clip]{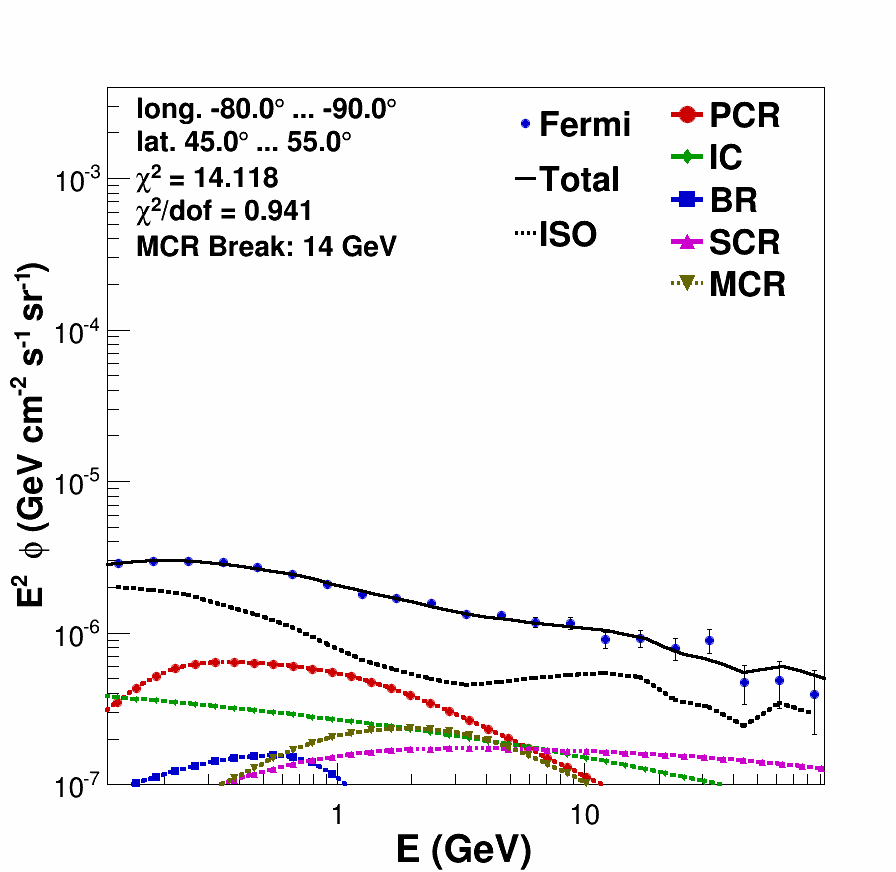}
\includegraphics[width=0.16\textwidth,height=0.16\textwidth,clip]{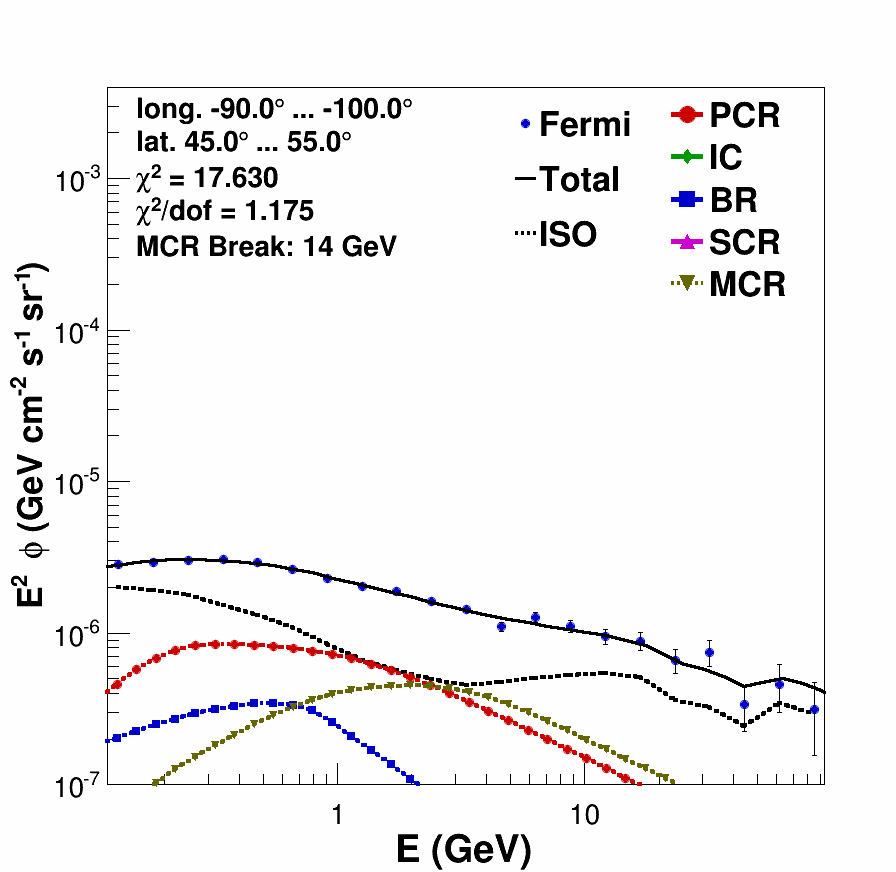}
\includegraphics[width=0.16\textwidth,height=0.16\textwidth,clip]{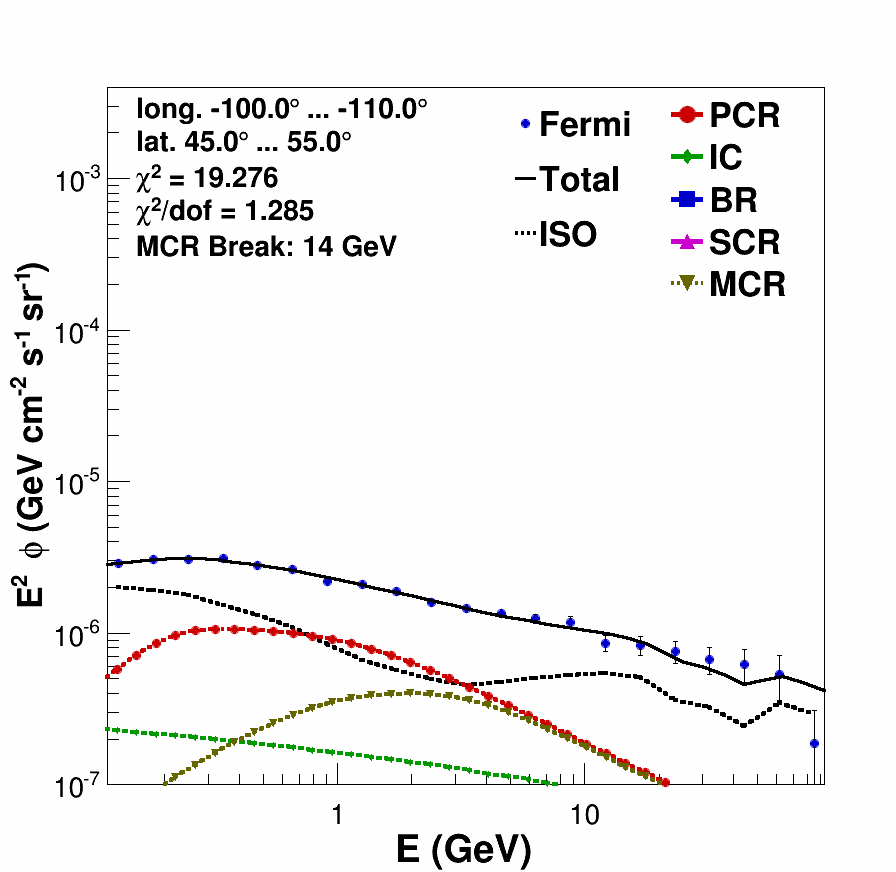}
\includegraphics[width=0.16\textwidth,height=0.16\textwidth,clip]{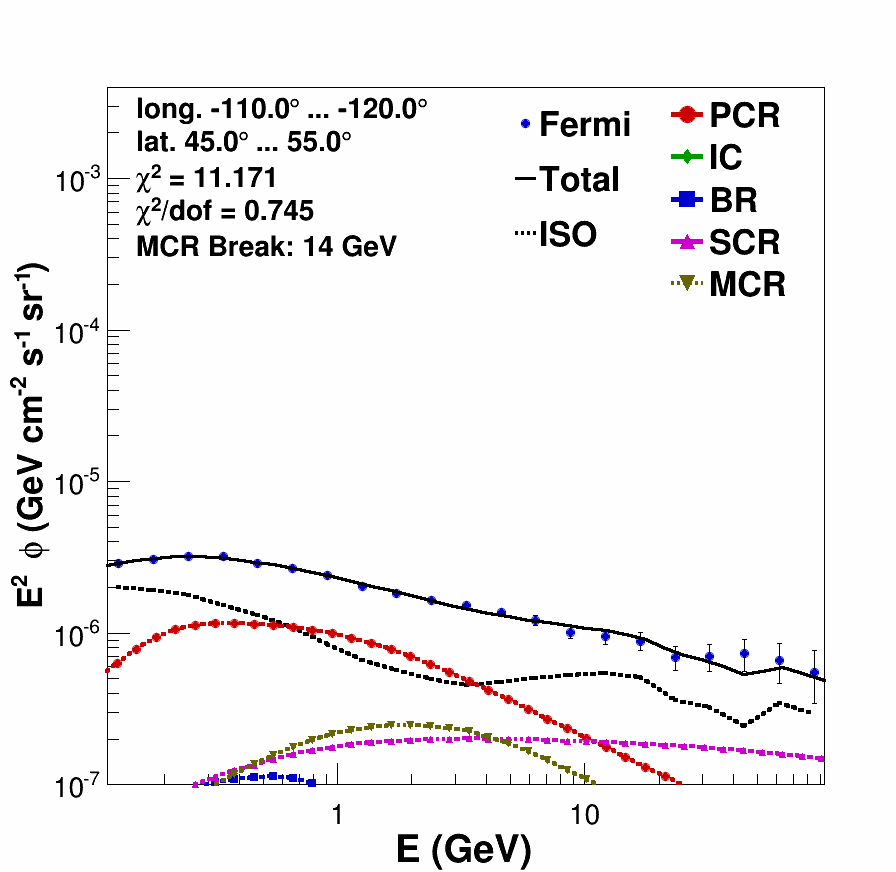}
\includegraphics[width=0.16\textwidth,height=0.16\textwidth,clip]{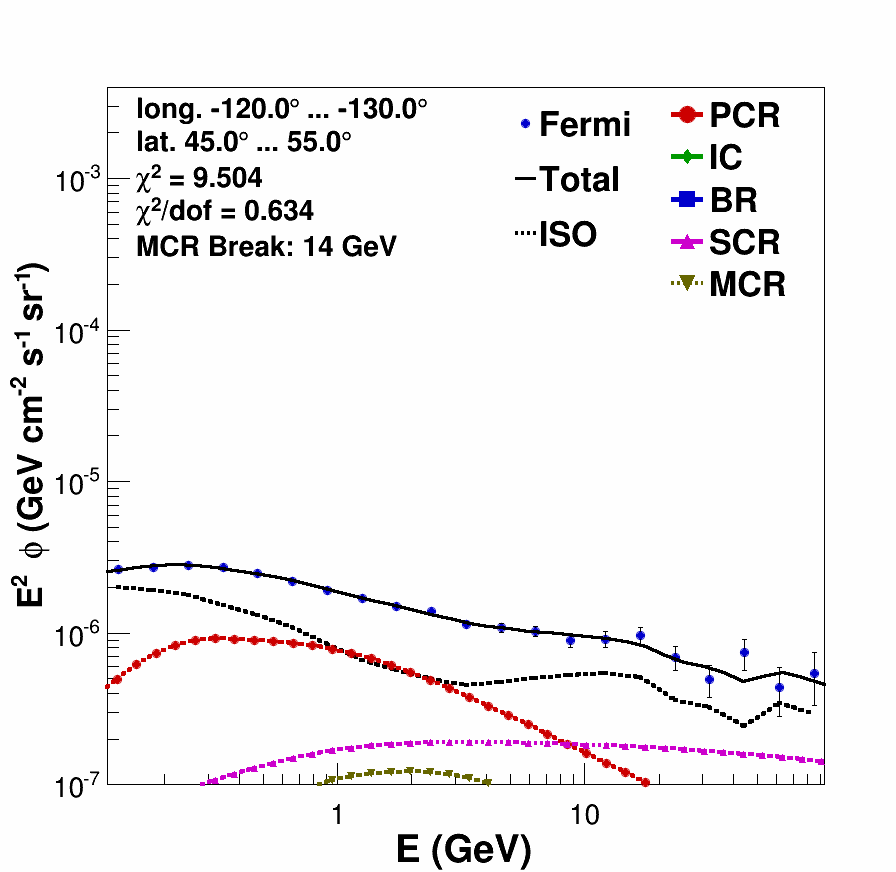}
\includegraphics[width=0.16\textwidth,height=0.16\textwidth,clip]{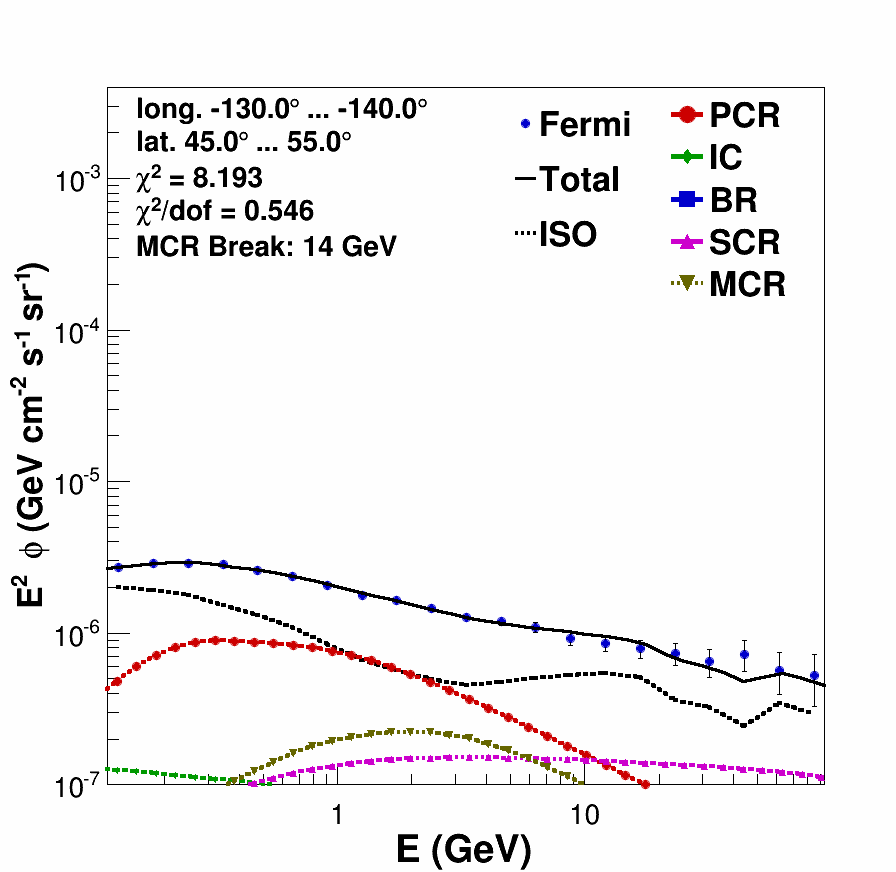}
\includegraphics[width=0.16\textwidth,height=0.16\textwidth,clip]{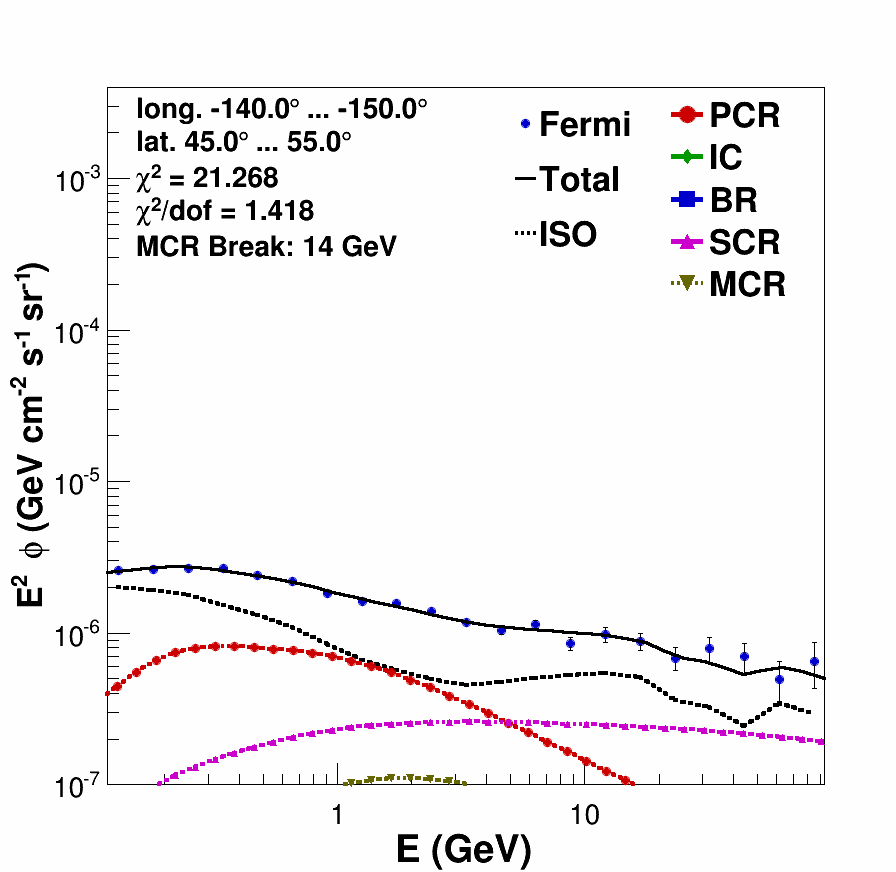}
\includegraphics[width=0.16\textwidth,height=0.16\textwidth,clip]{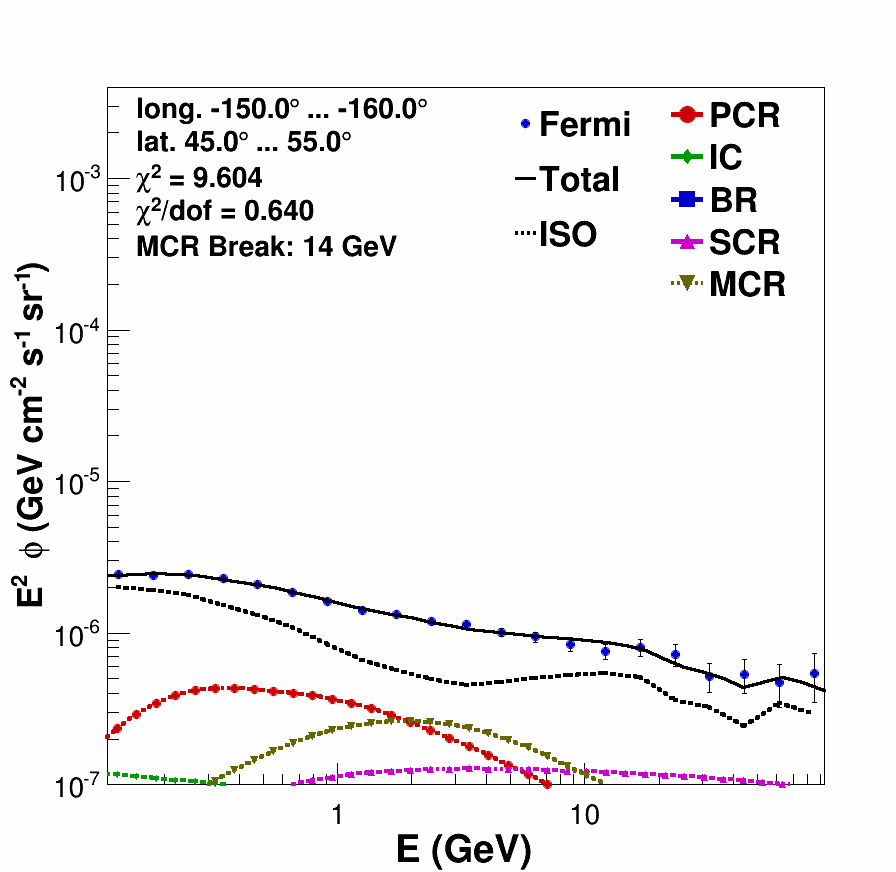}
\includegraphics[width=0.16\textwidth,height=0.16\textwidth,clip]{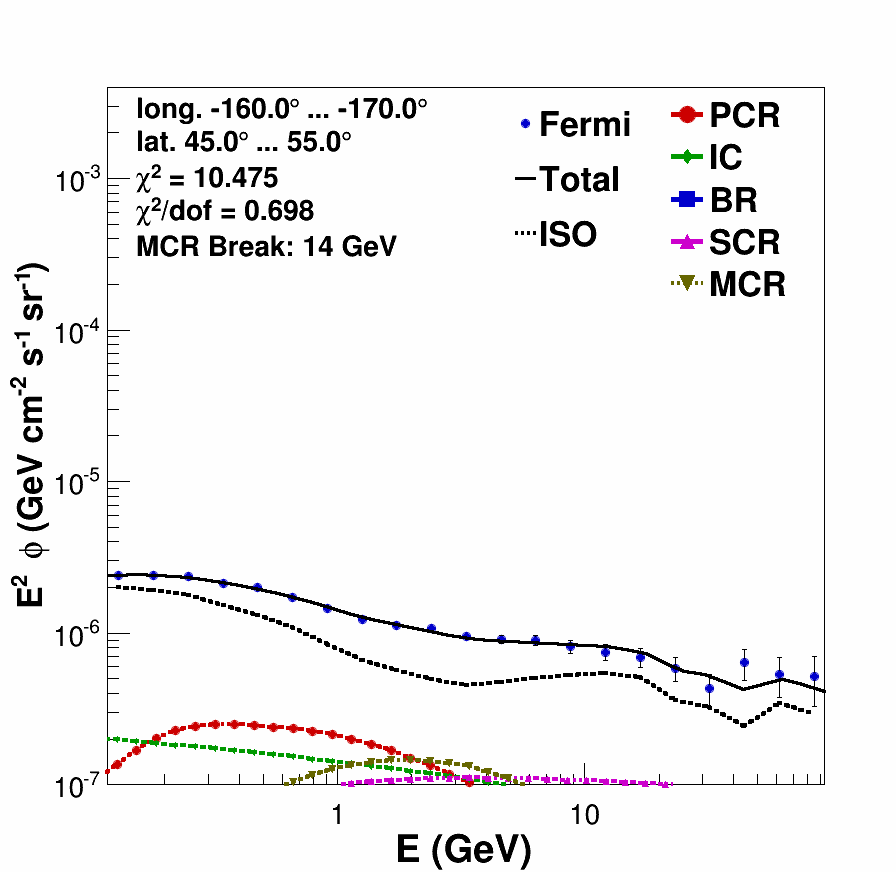}
\includegraphics[width=0.16\textwidth,height=0.16\textwidth,clip]{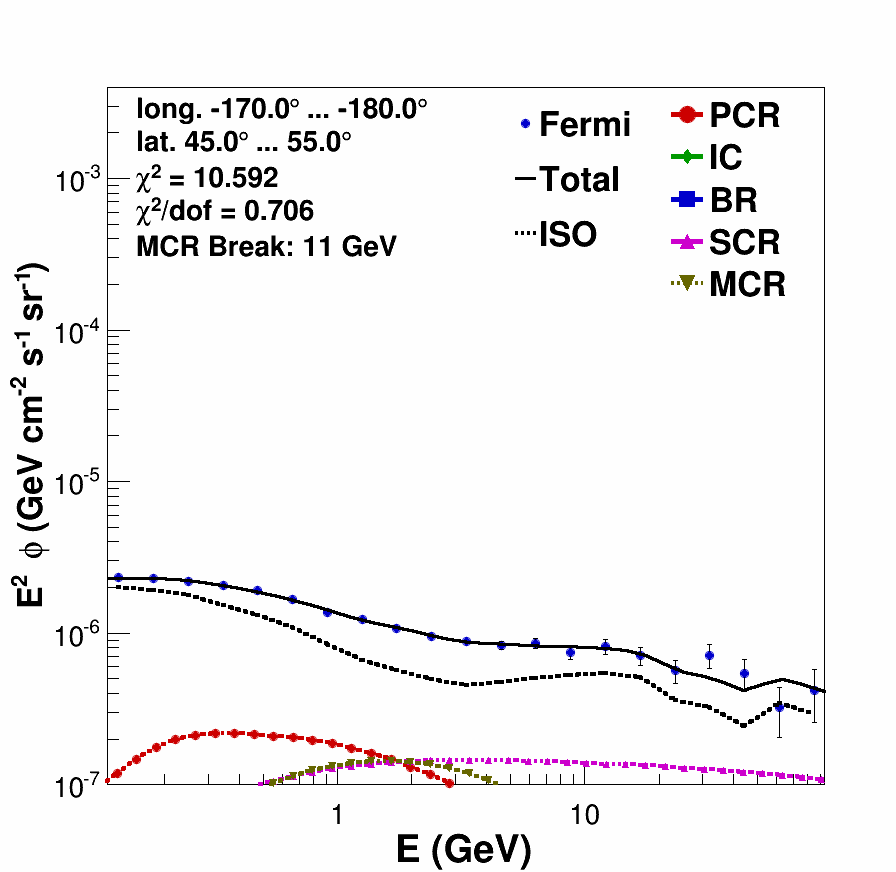}%%%%%%%r3
\caption[]{Template fits for latitudes  with $45.0^\circ<b<55.0^\circ$ and longitudes decreasing from 180$^\circ$ to -180$^\circ$. \label{F13}
}
\end{figure}
\begin{figure}
\centering
\includegraphics[width=0.16\textwidth,height=0.16\textwidth,clip]{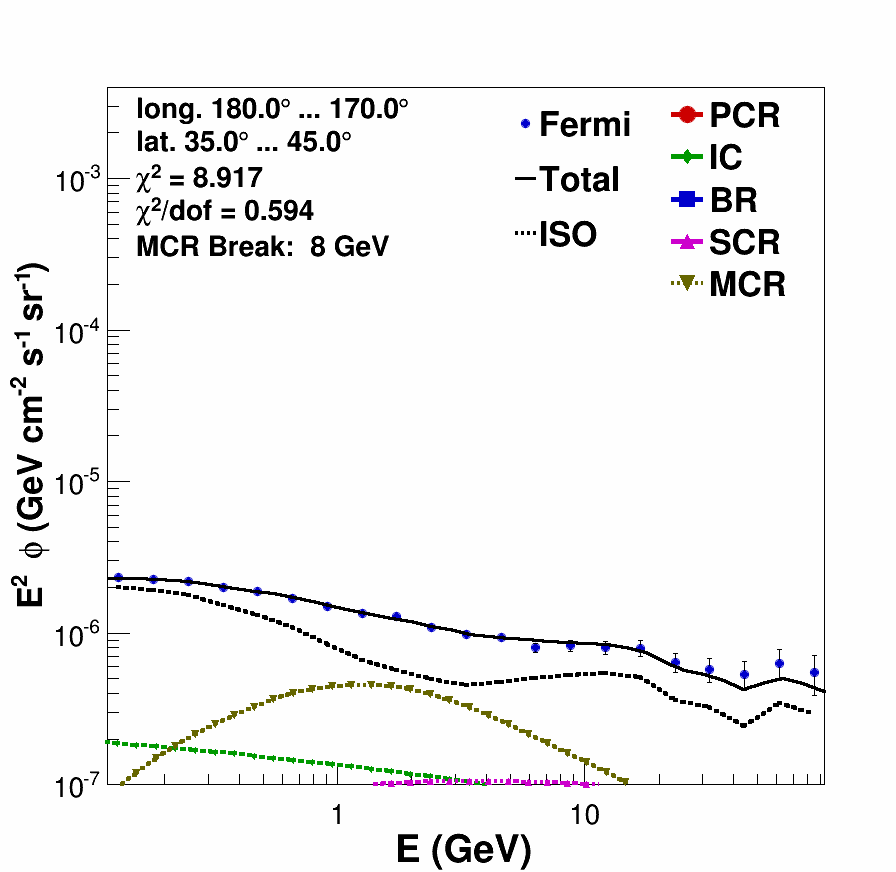}
\includegraphics[width=0.16\textwidth,height=0.16\textwidth,clip]{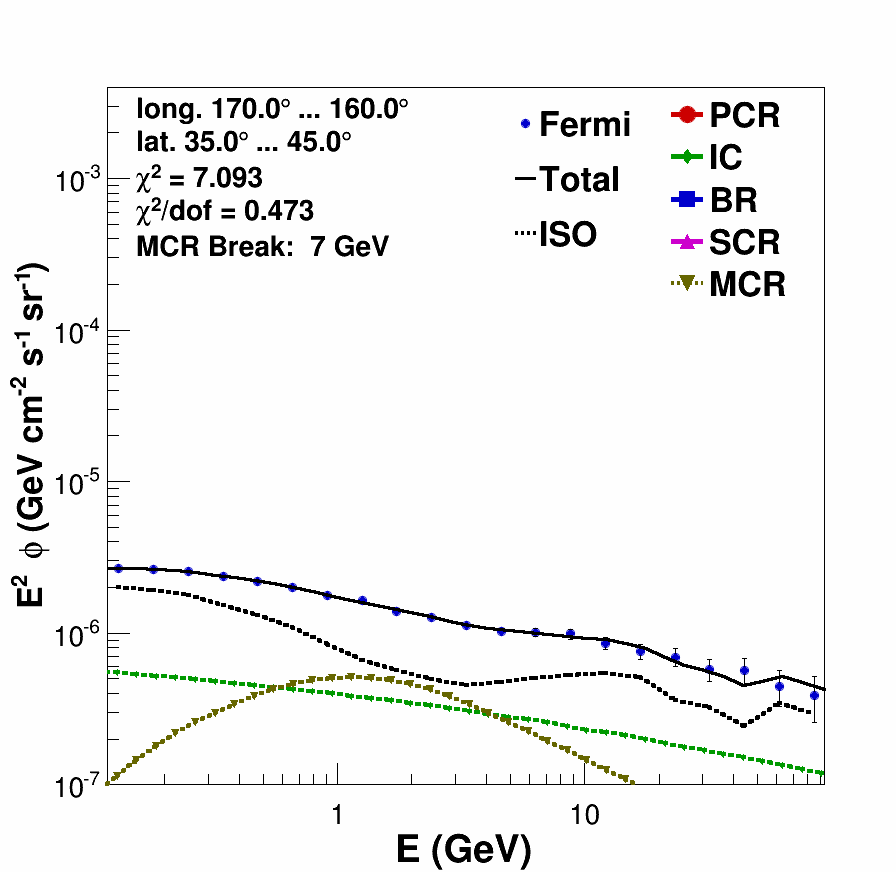}
\includegraphics[width=0.16\textwidth,height=0.16\textwidth,clip]{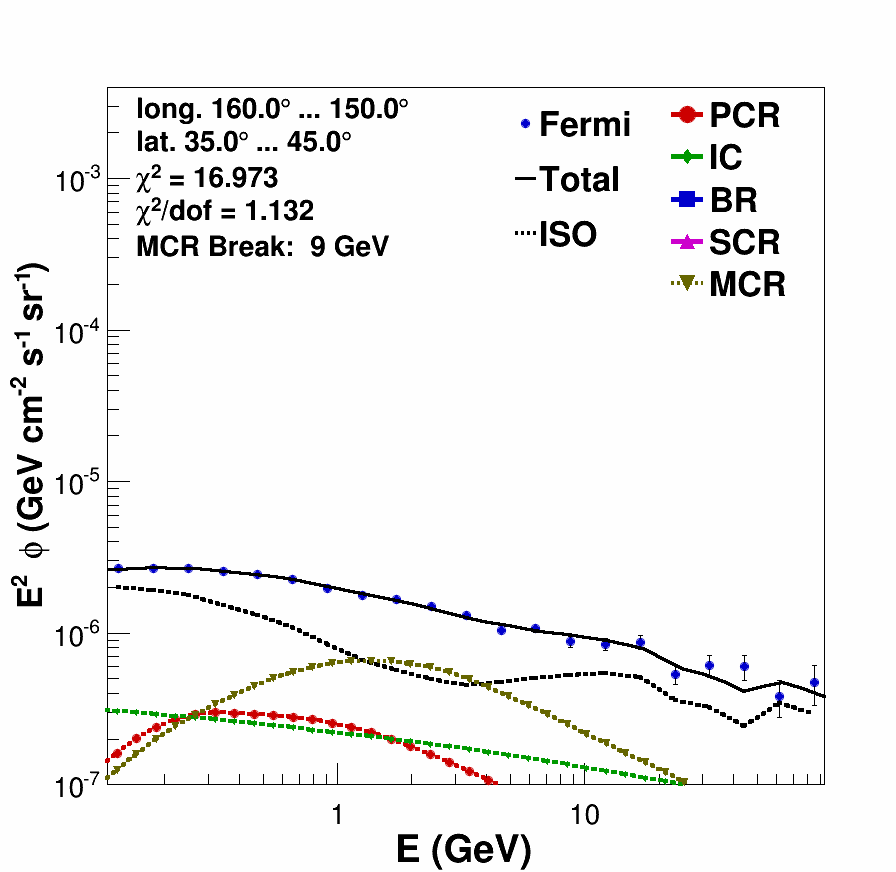}
\includegraphics[width=0.16\textwidth,height=0.16\textwidth,clip]{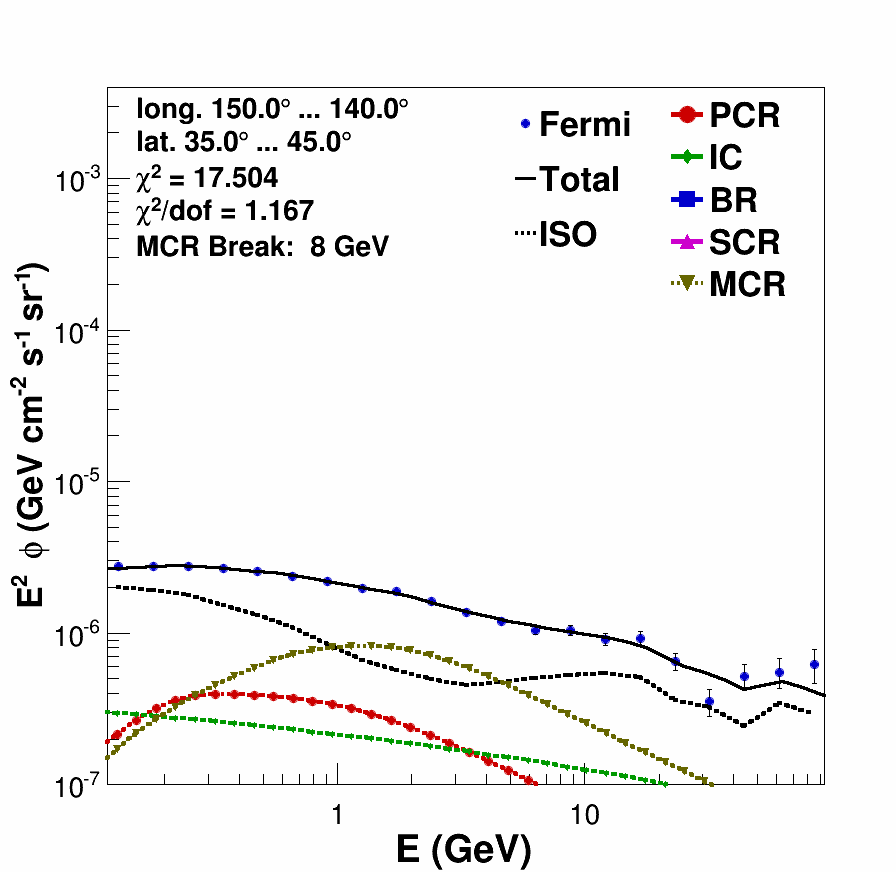}
\includegraphics[width=0.16\textwidth,height=0.16\textwidth,clip]{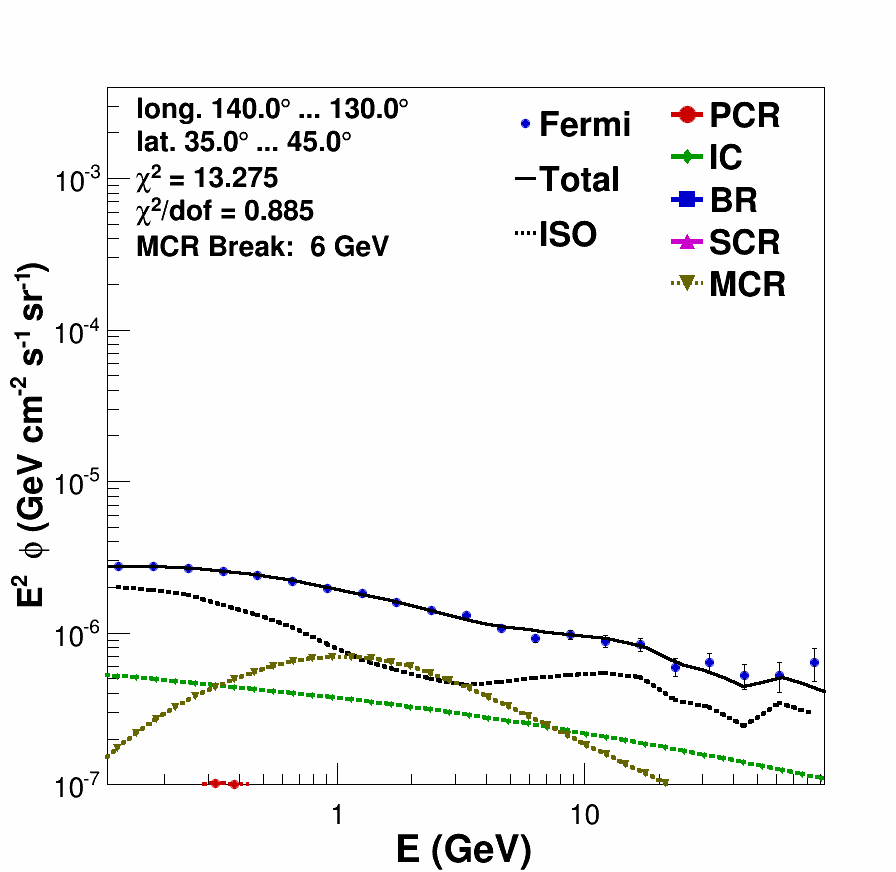}
\includegraphics[width=0.16\textwidth,height=0.16\textwidth,clip]{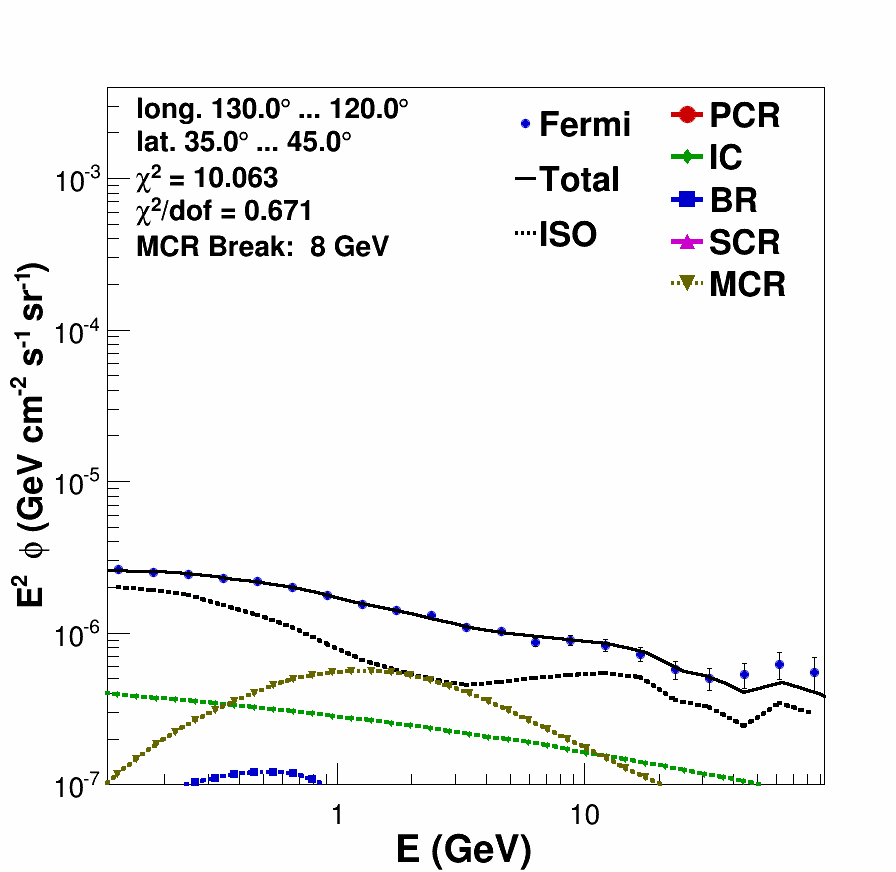}
\includegraphics[width=0.16\textwidth,height=0.16\textwidth,clip]{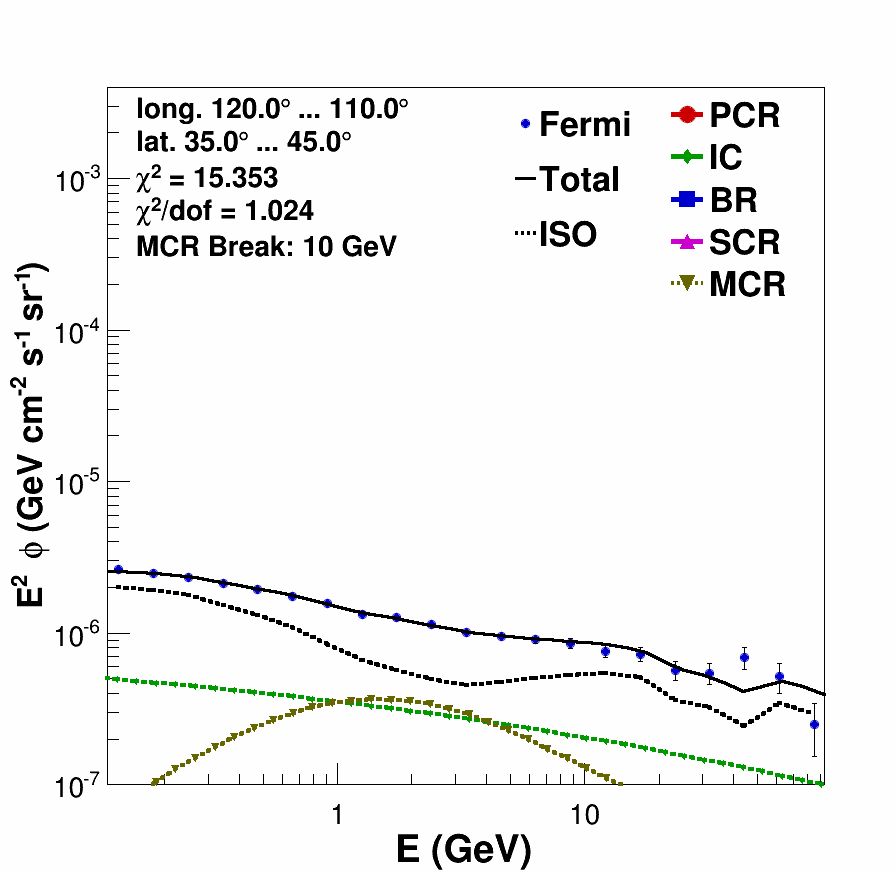}
\includegraphics[width=0.16\textwidth,height=0.16\textwidth,clip]{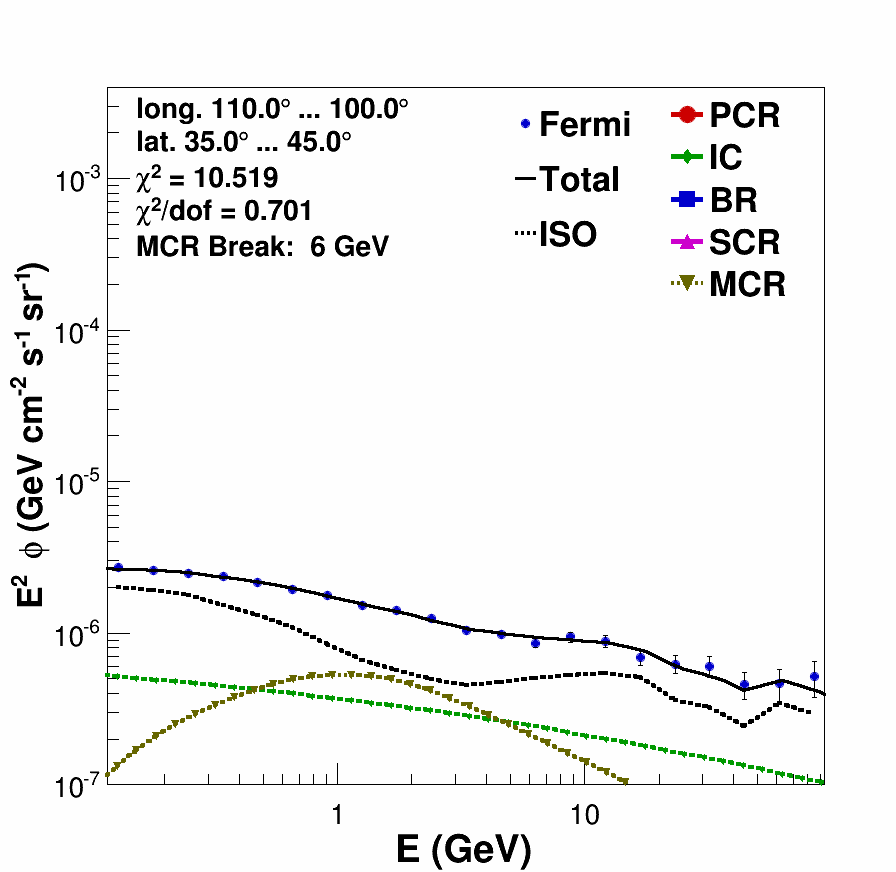}
\includegraphics[width=0.16\textwidth,height=0.16\textwidth,clip]{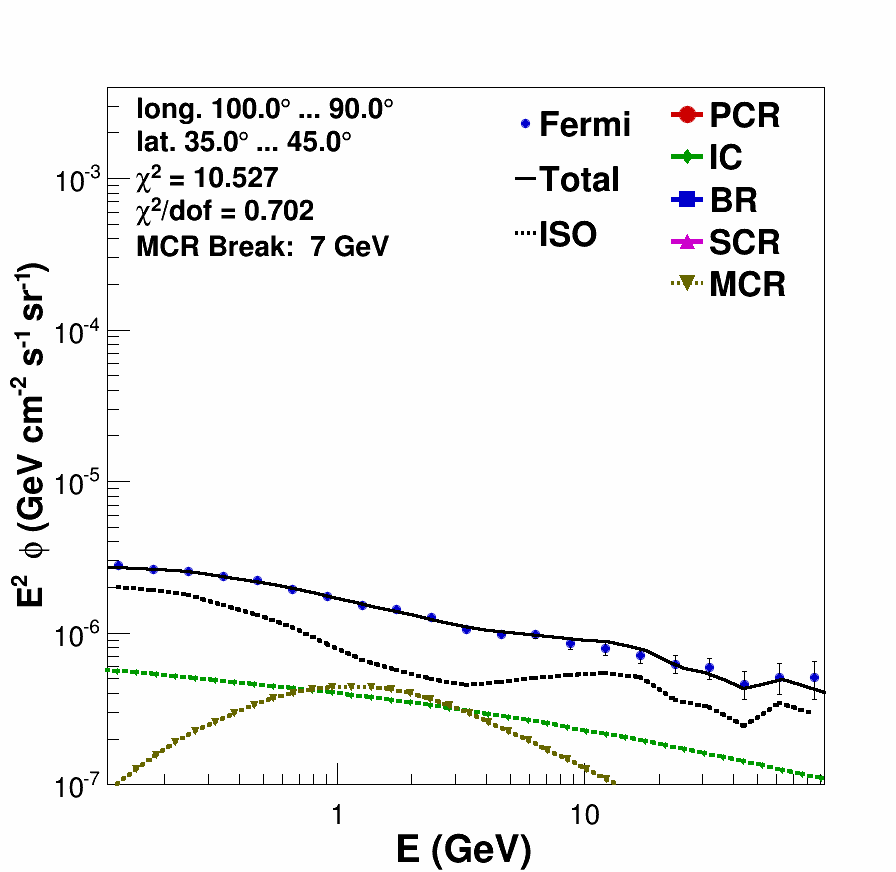}
\includegraphics[width=0.16\textwidth,height=0.16\textwidth,clip]{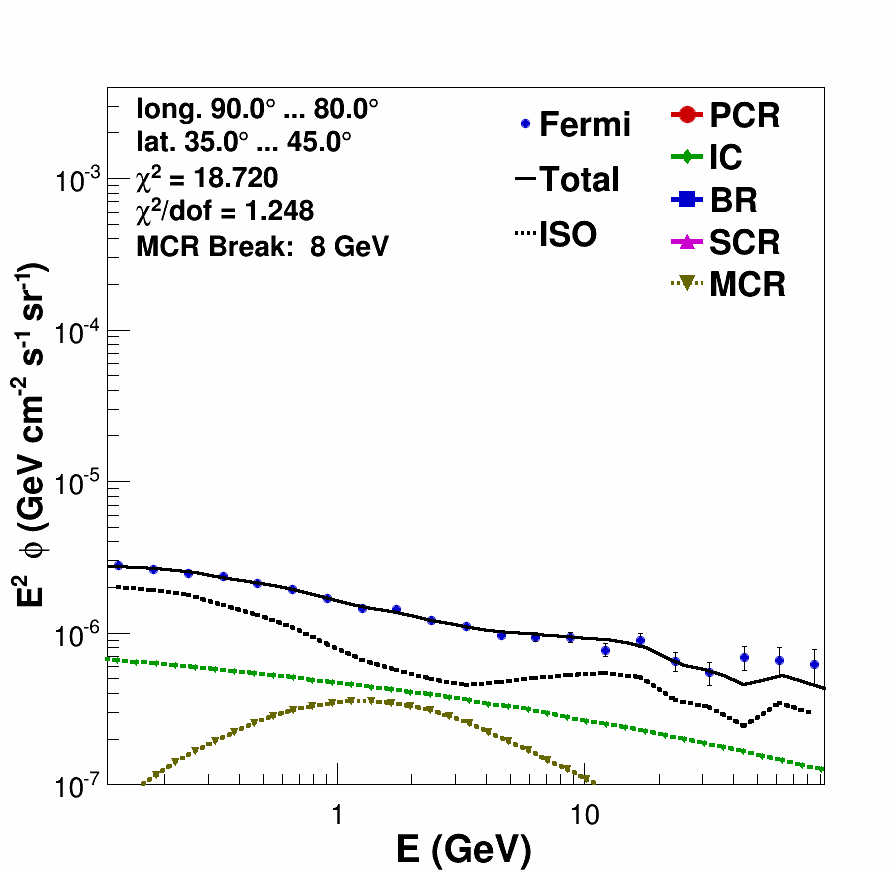}
\includegraphics[width=0.16\textwidth,height=0.16\textwidth,clip]{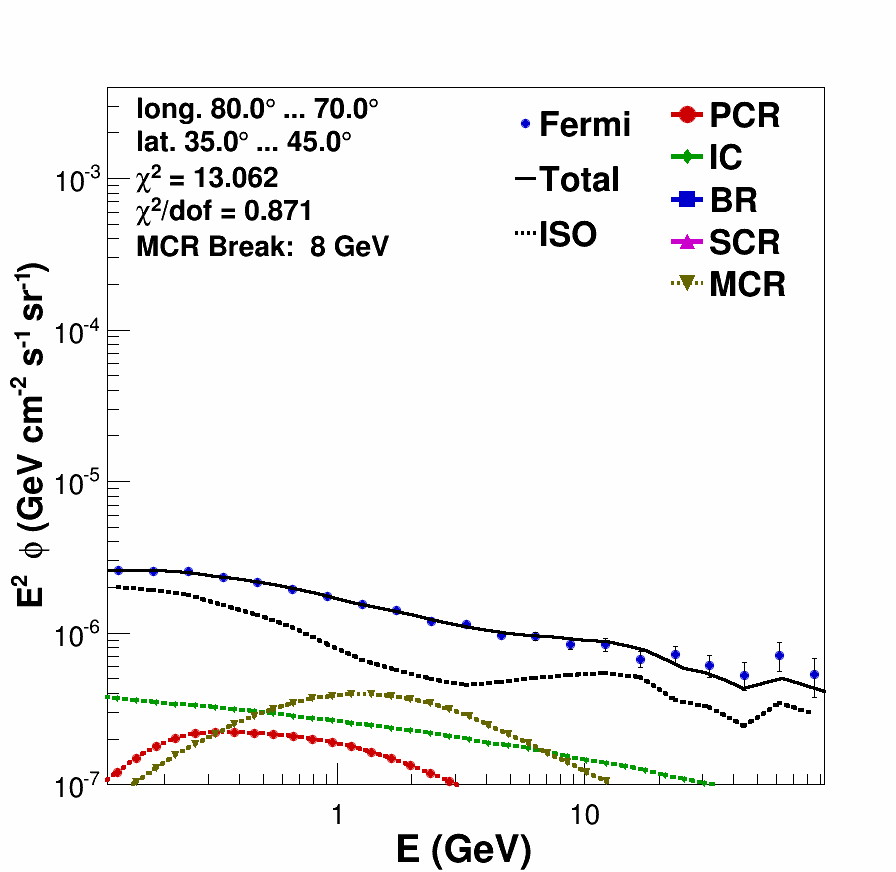}
\includegraphics[width=0.16\textwidth,height=0.16\textwidth,clip]{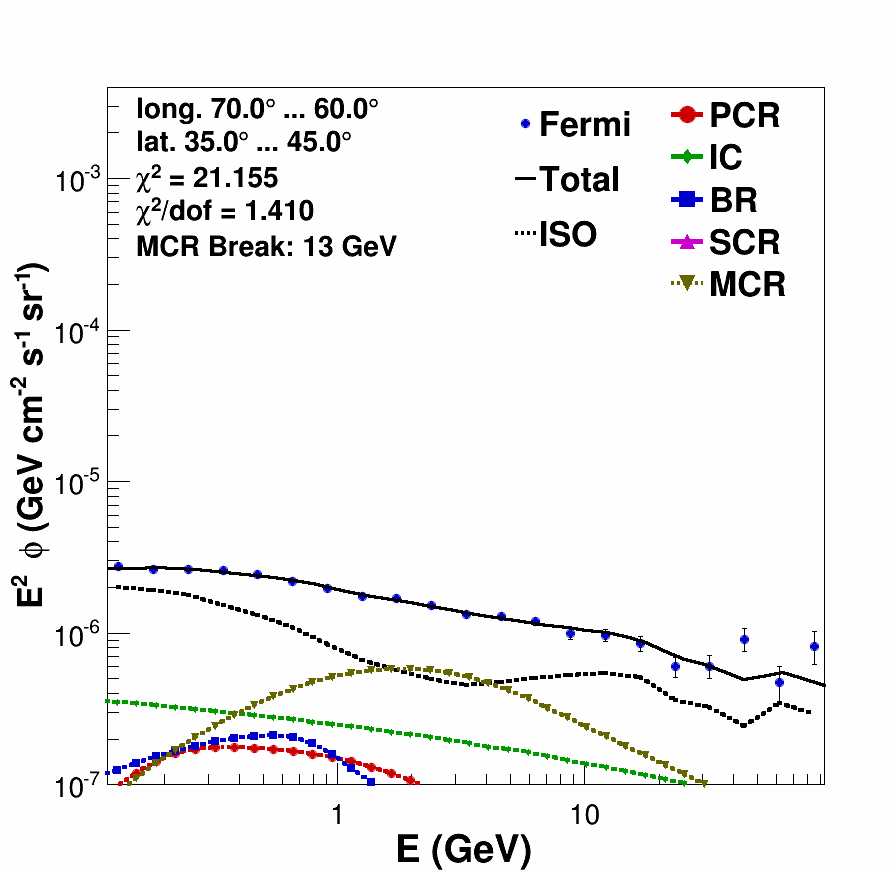}
\includegraphics[width=0.16\textwidth,height=0.16\textwidth,clip]{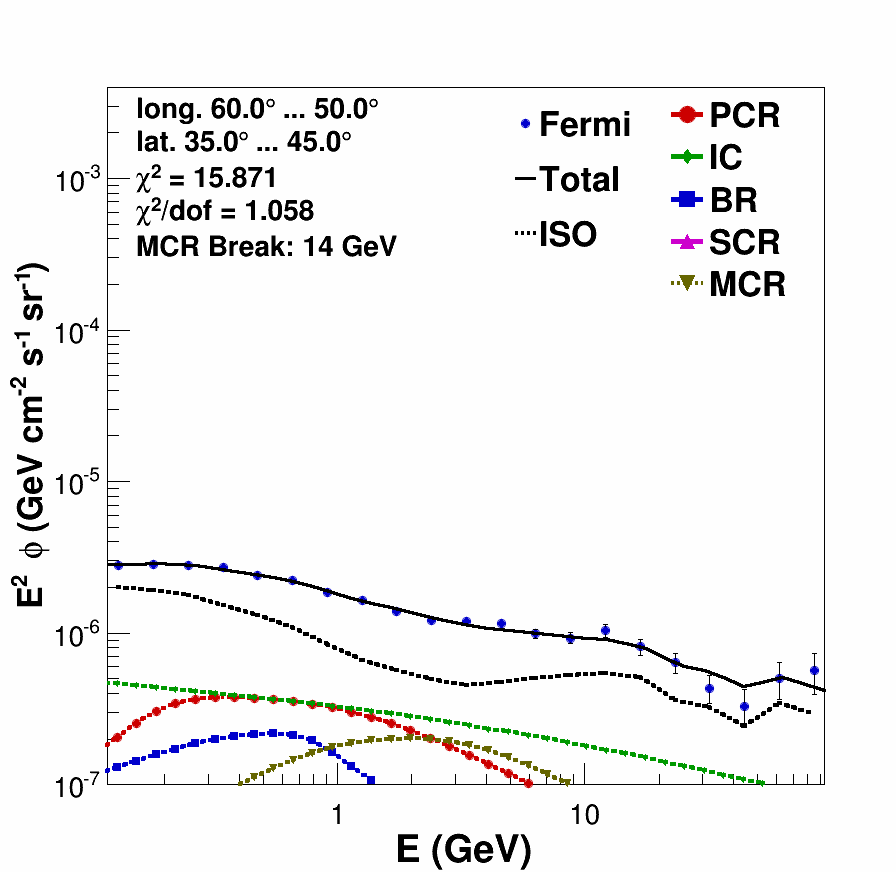}
\includegraphics[width=0.16\textwidth,height=0.16\textwidth,clip]{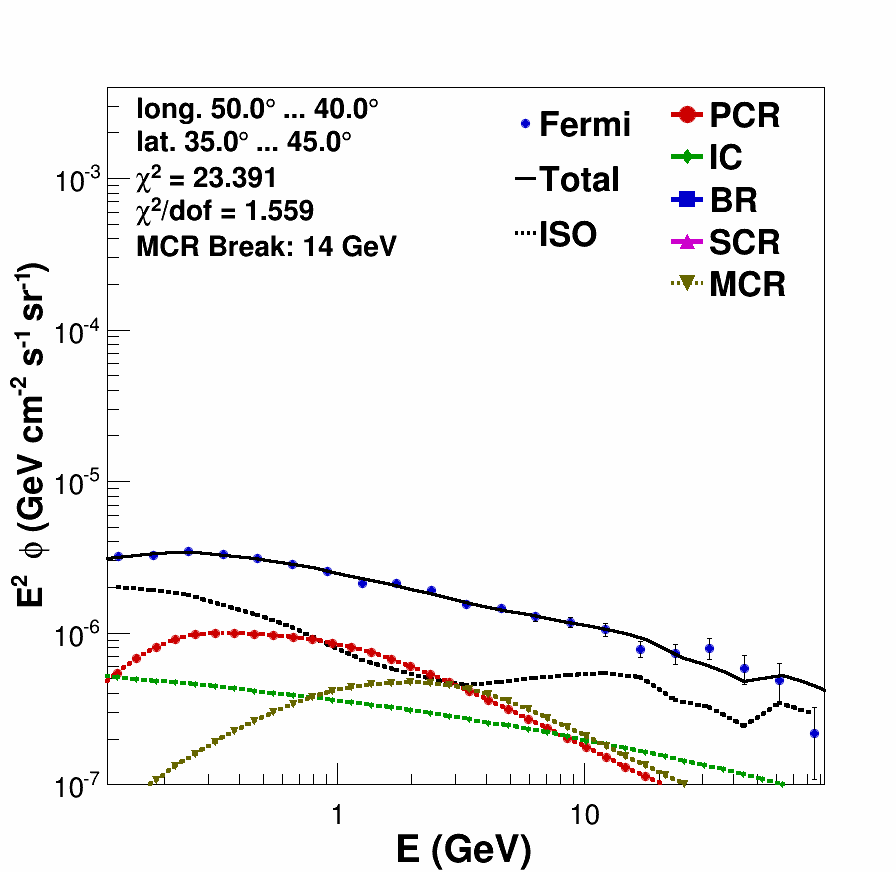}
\includegraphics[width=0.16\textwidth,height=0.16\textwidth,clip]{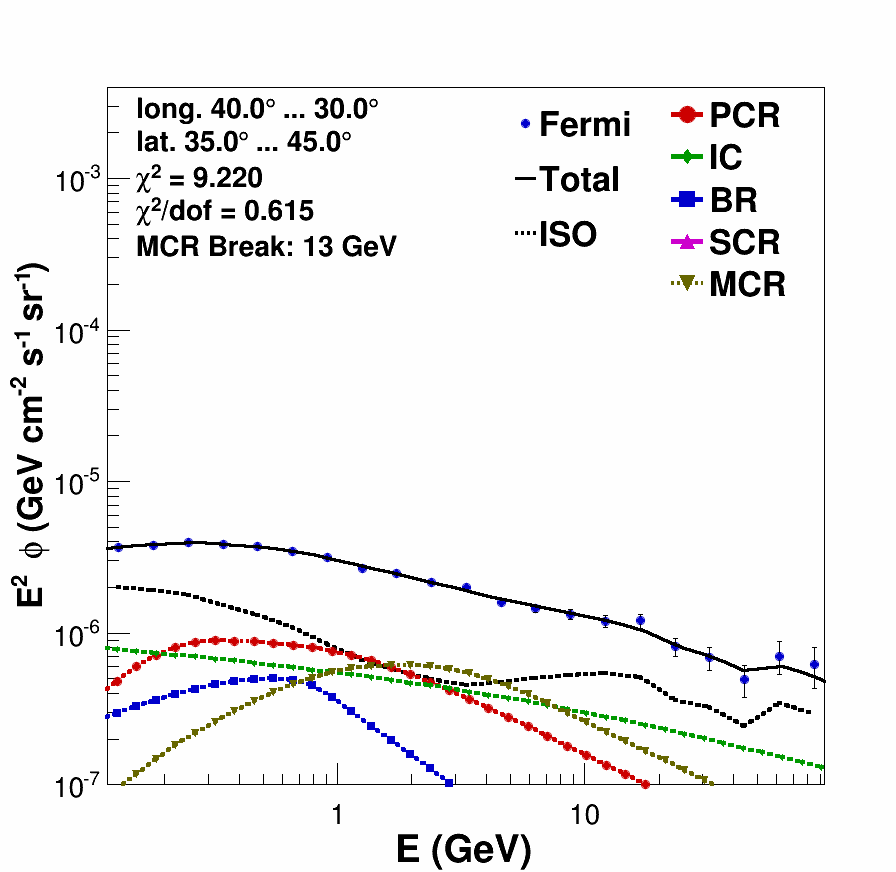}
\includegraphics[width=0.16\textwidth,height=0.16\textwidth,clip]{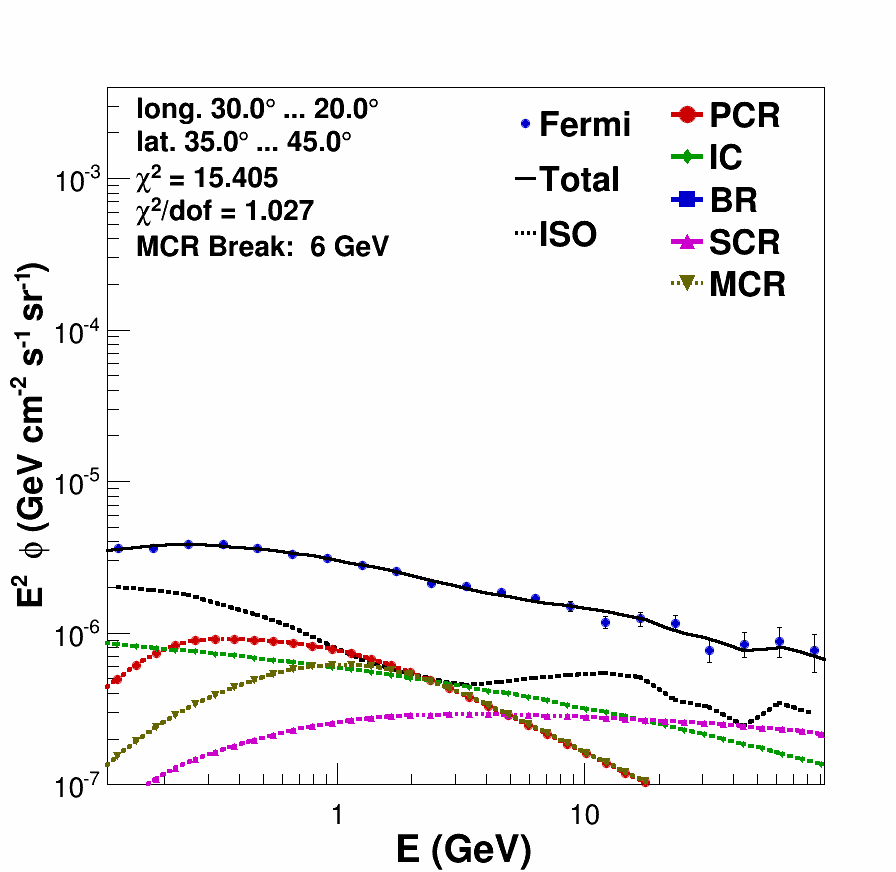}
\includegraphics[width=0.16\textwidth,height=0.16\textwidth,clip]{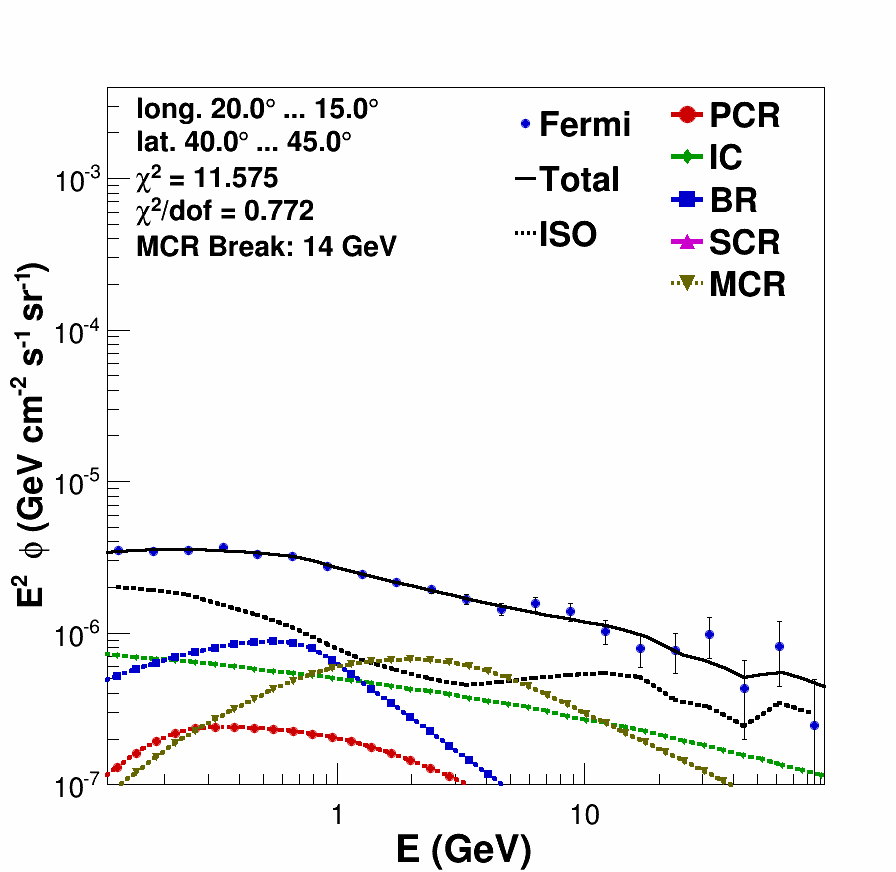}
\includegraphics[width=0.16\textwidth,height=0.16\textwidth,clip]{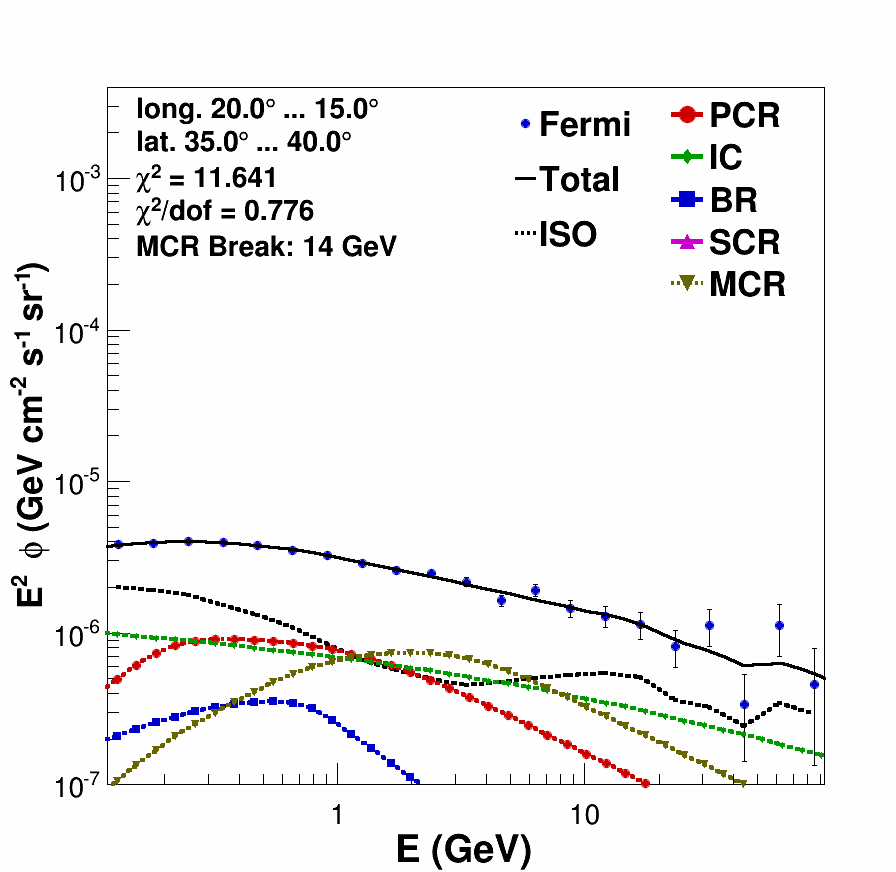}
\includegraphics[width=0.16\textwidth,height=0.16\textwidth,clip]{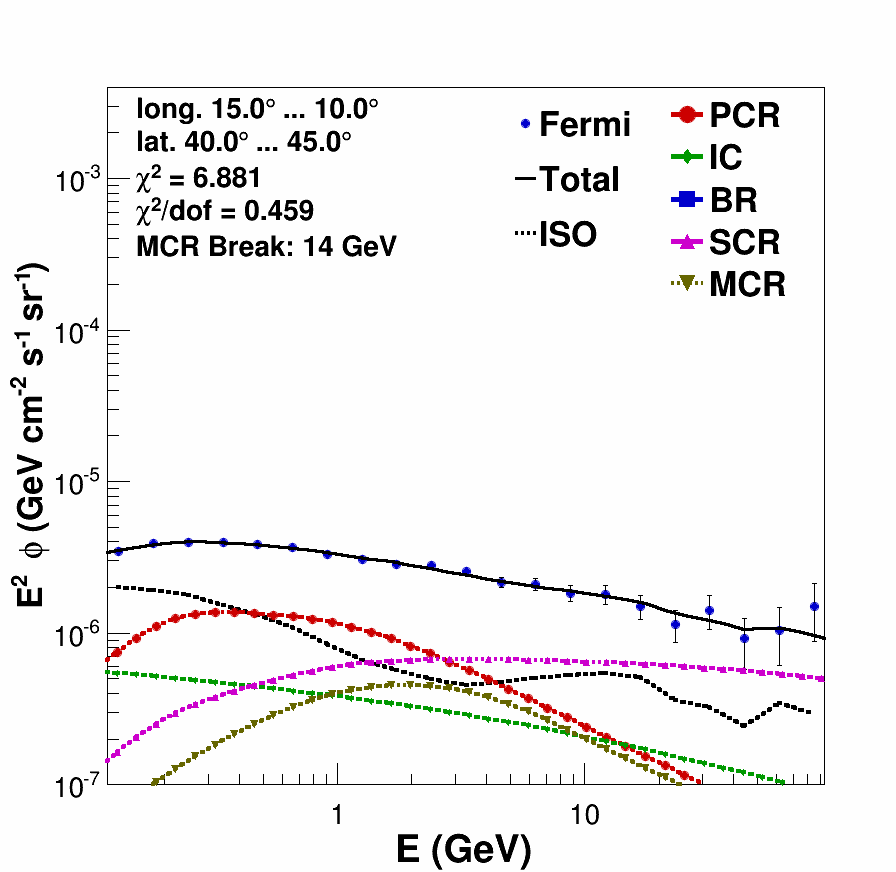}
\includegraphics[width=0.16\textwidth,height=0.16\textwidth,clip]{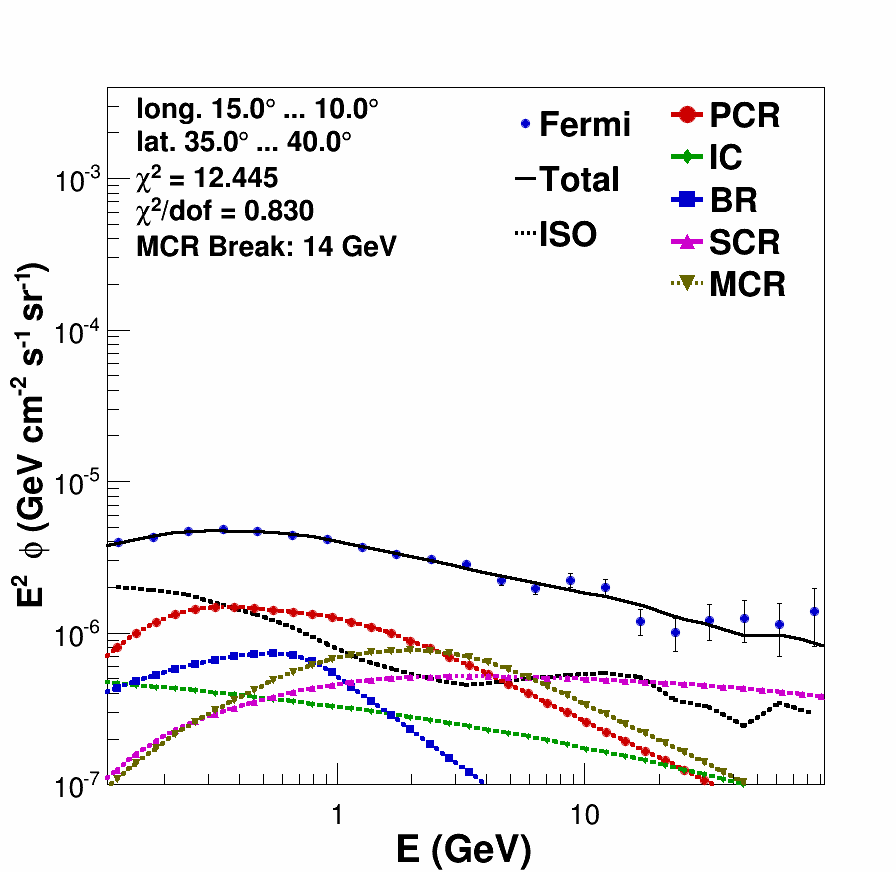}
\includegraphics[width=0.16\textwidth,height=0.16\textwidth,clip]{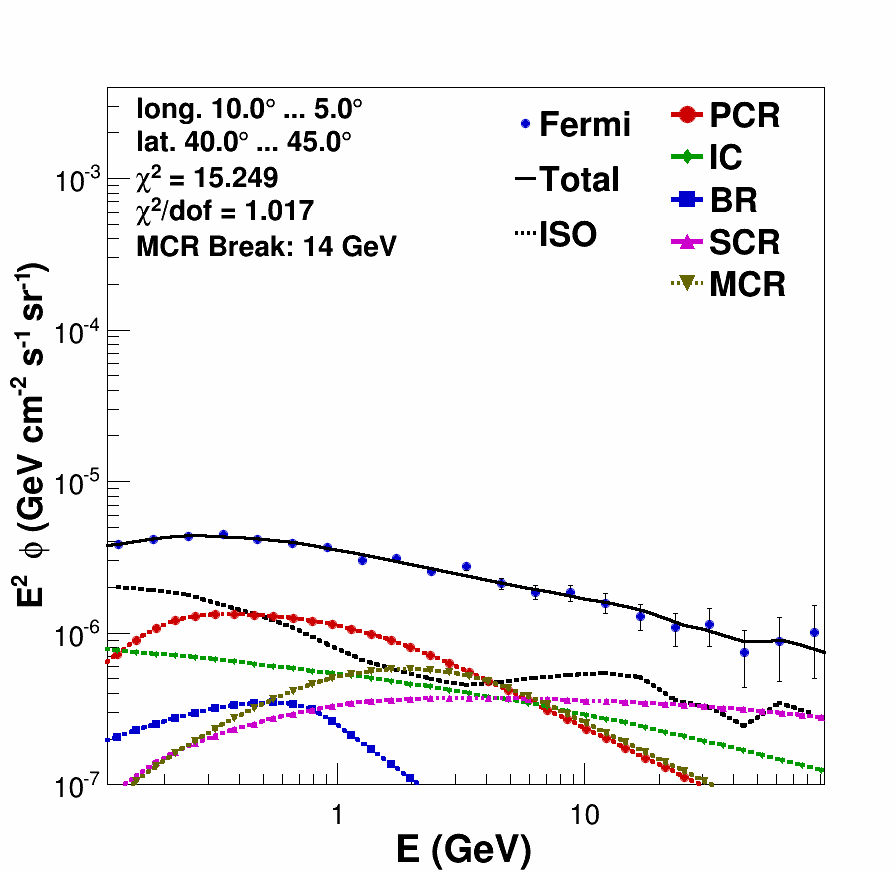}
\includegraphics[width=0.16\textwidth,height=0.16\textwidth,clip]{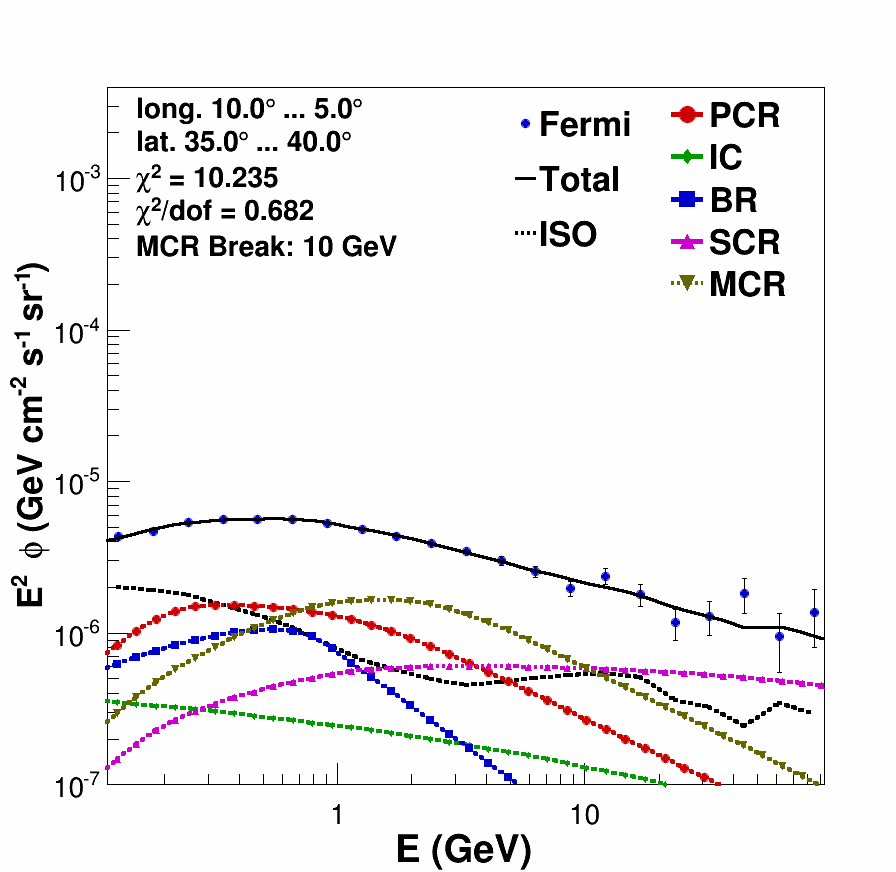}
\includegraphics[width=0.16\textwidth,height=0.16\textwidth,clip]{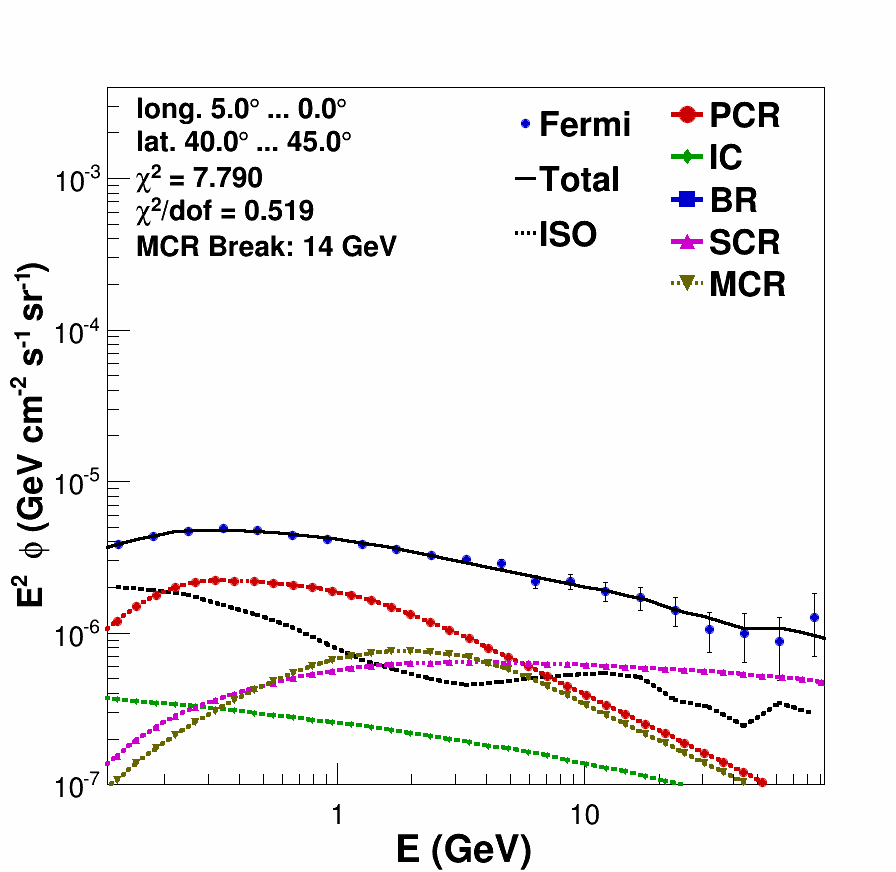}
\includegraphics[width=0.16\textwidth,height=0.16\textwidth,clip]{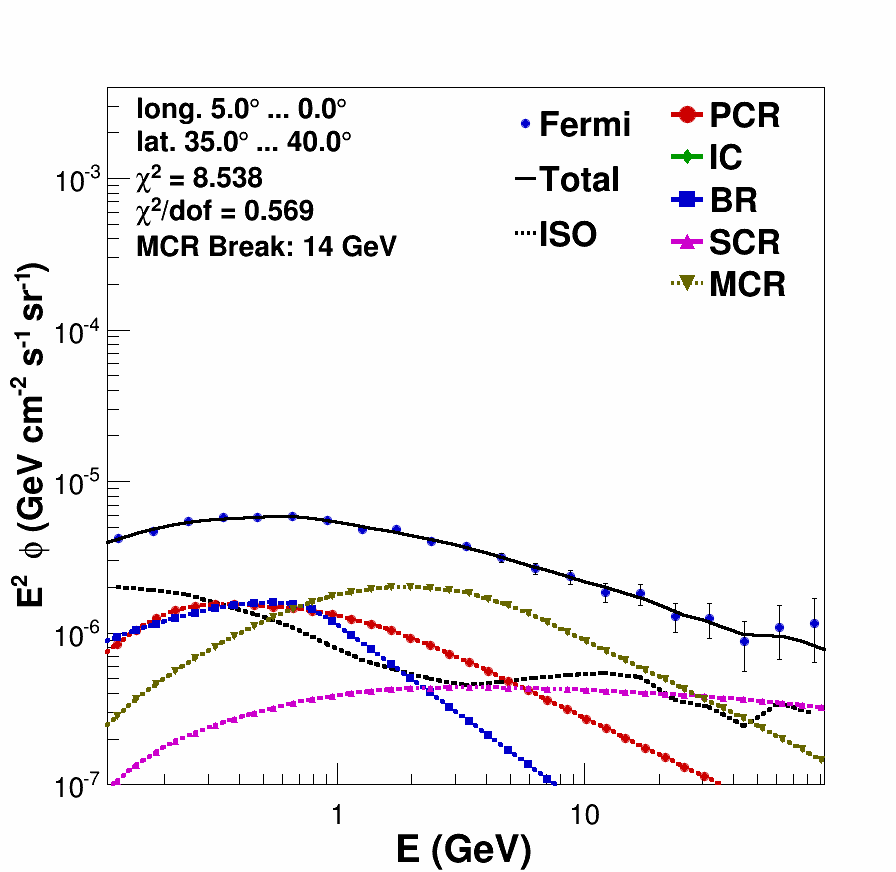}
\includegraphics[width=0.16\textwidth,height=0.16\textwidth,clip]{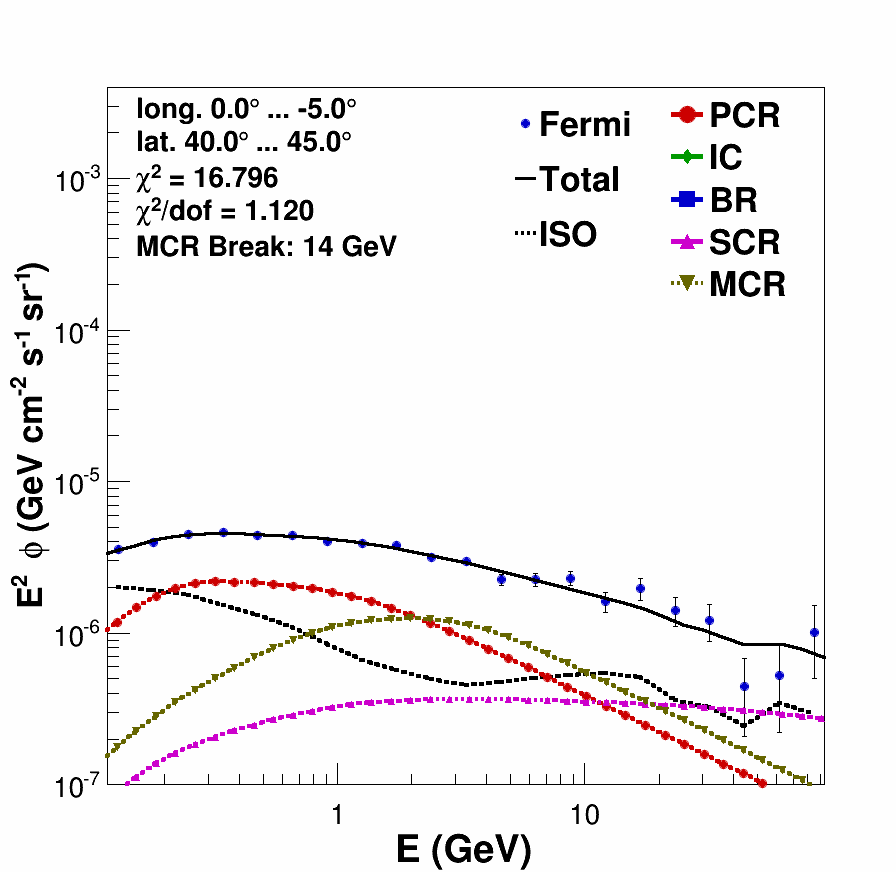}
\includegraphics[width=0.16\textwidth,height=0.16\textwidth,clip]{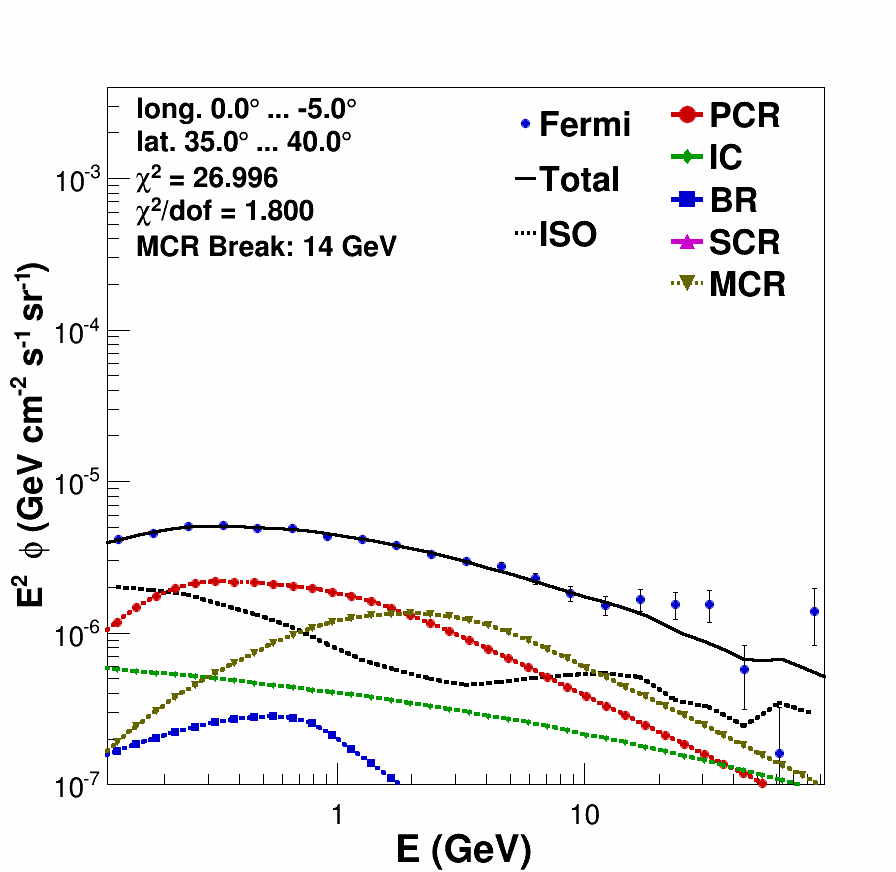}
\includegraphics[width=0.16\textwidth,height=0.16\textwidth,clip]{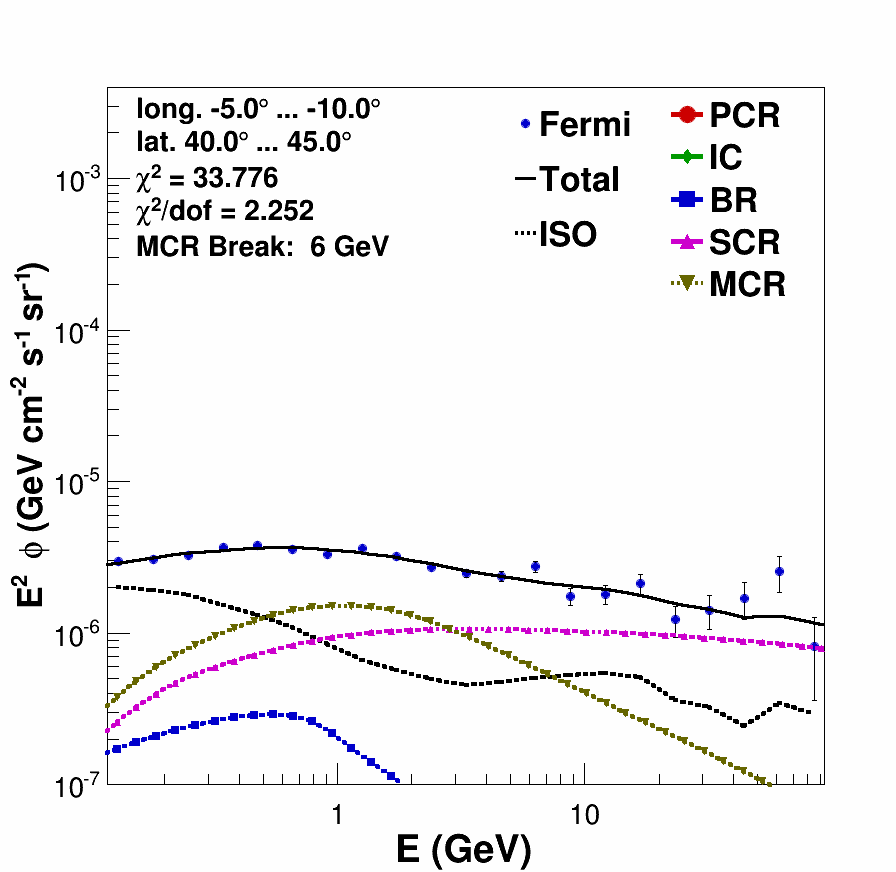}
\includegraphics[width=0.16\textwidth,height=0.16\textwidth,clip]{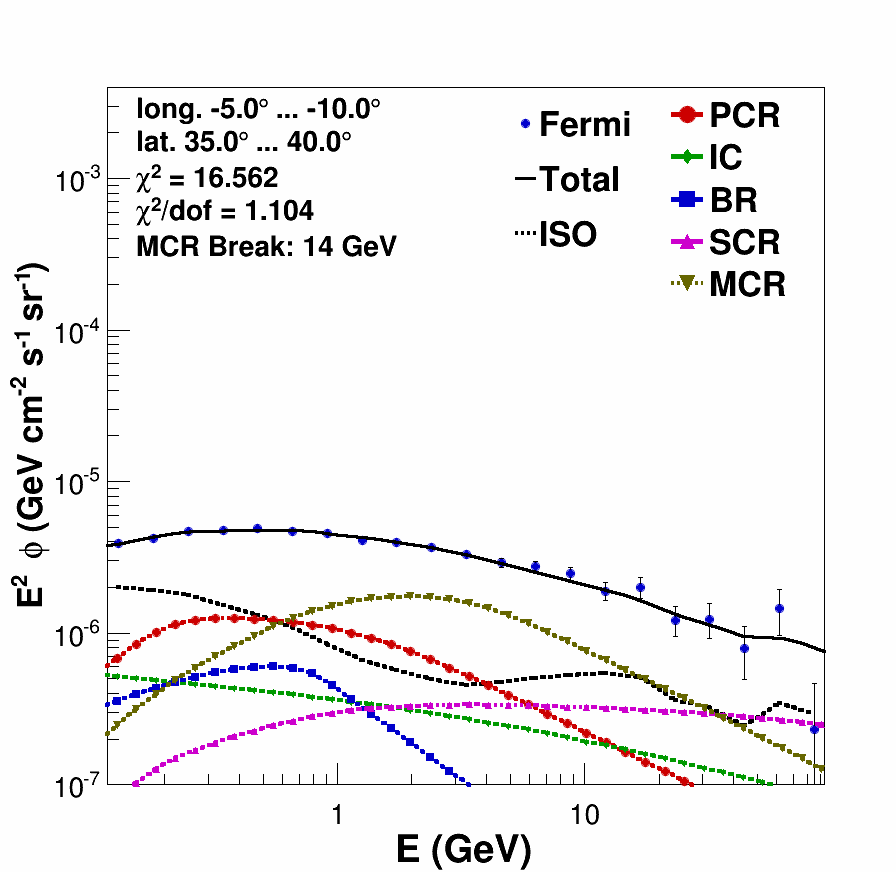}
\includegraphics[width=0.16\textwidth,height=0.16\textwidth,clip]{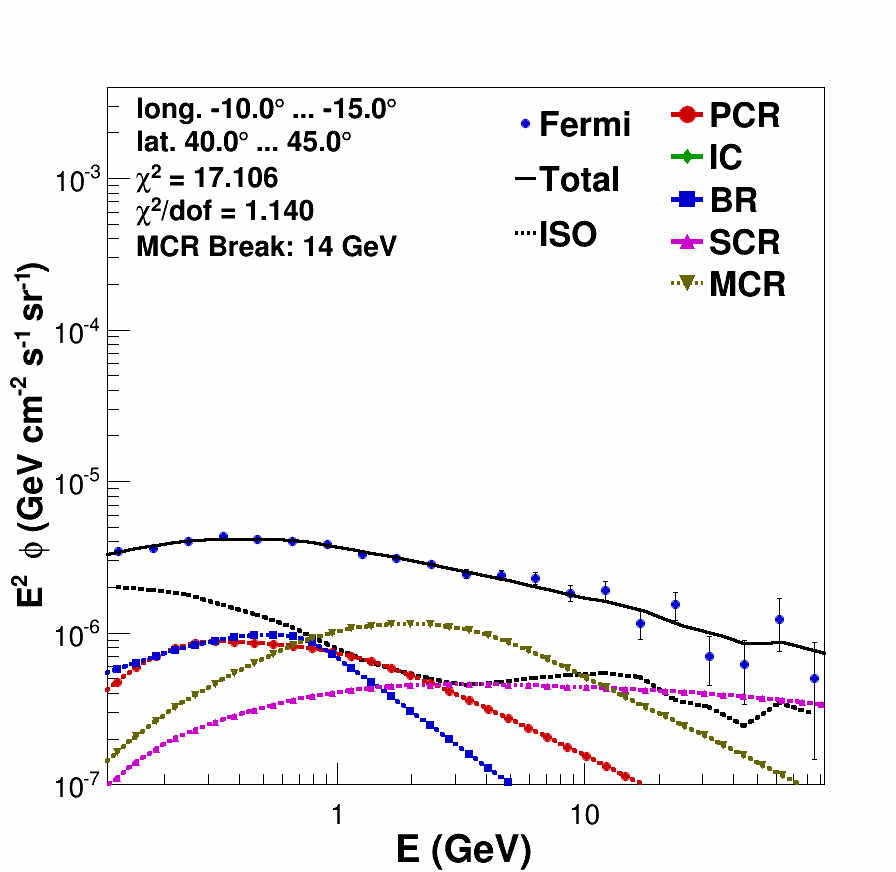}
\includegraphics[width=0.16\textwidth,height=0.16\textwidth,clip]{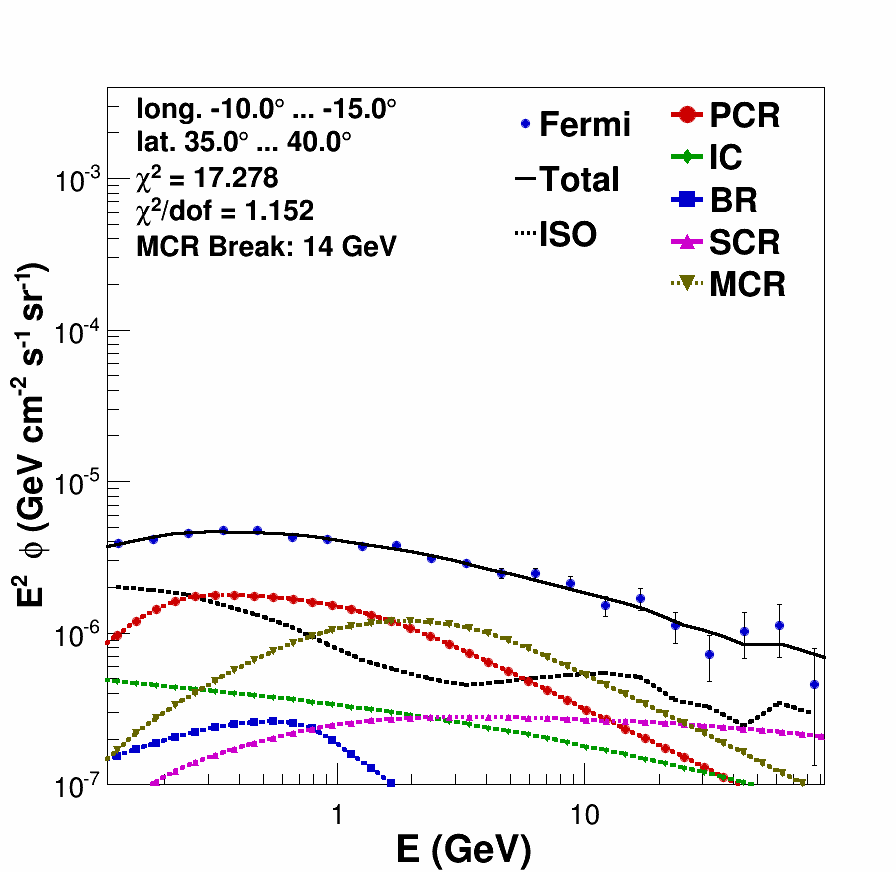}
\includegraphics[width=0.16\textwidth,height=0.16\textwidth,clip]{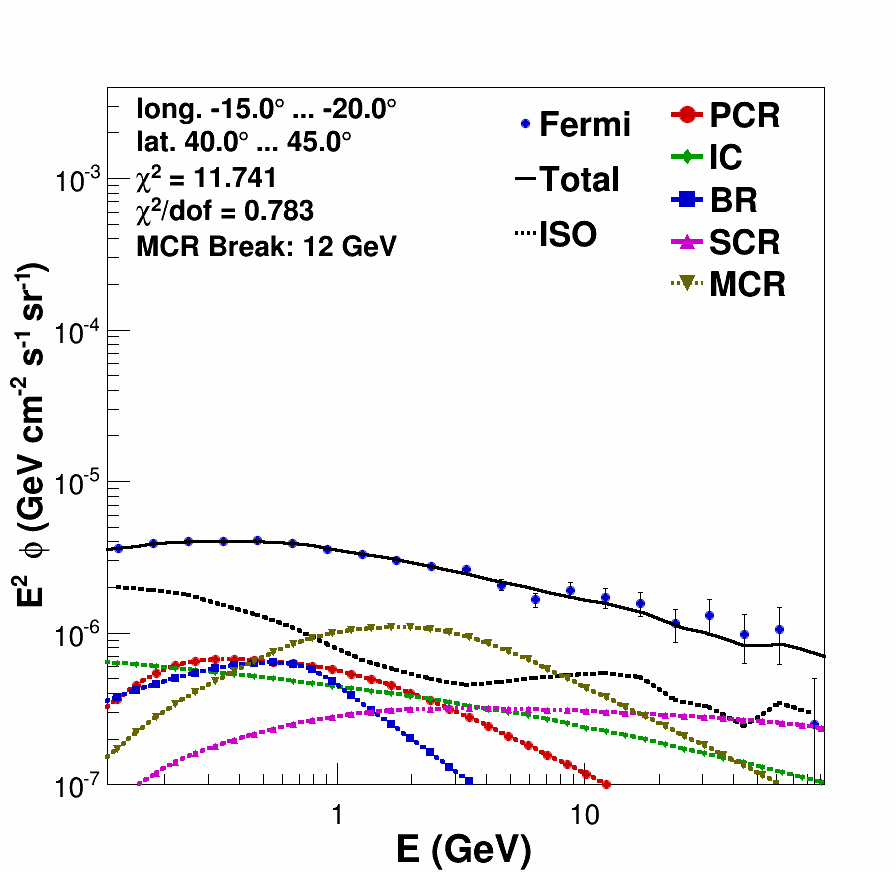}
\includegraphics[width=0.16\textwidth,height=0.16\textwidth,clip]{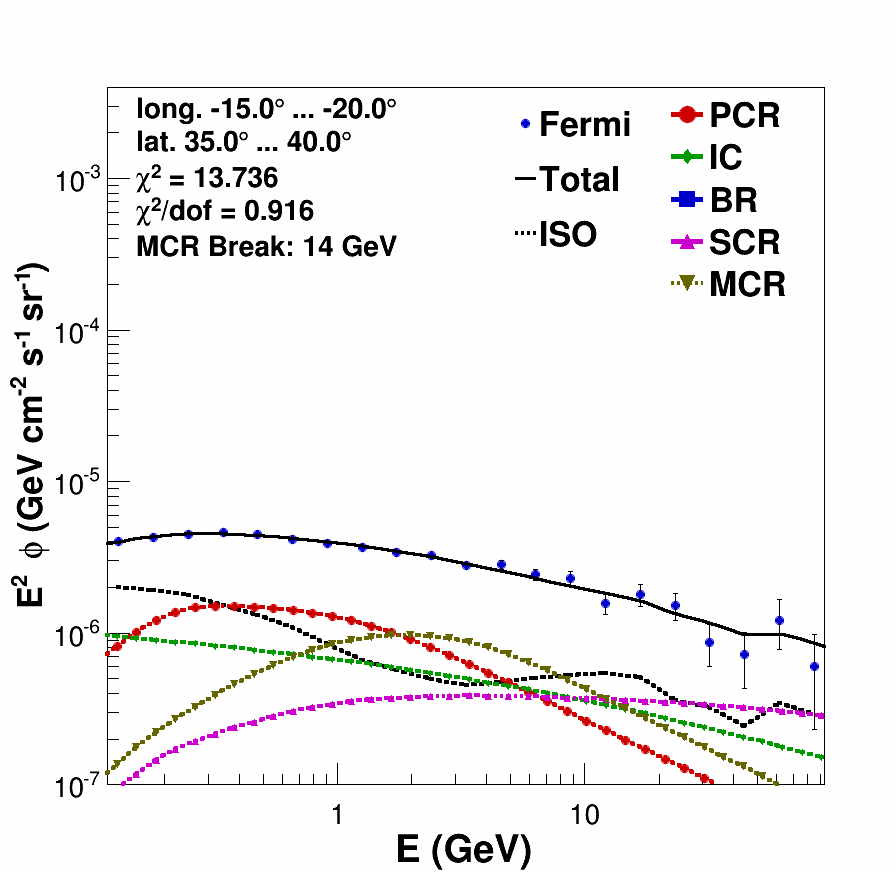}
\includegraphics[width=0.16\textwidth,height=0.16\textwidth,clip]{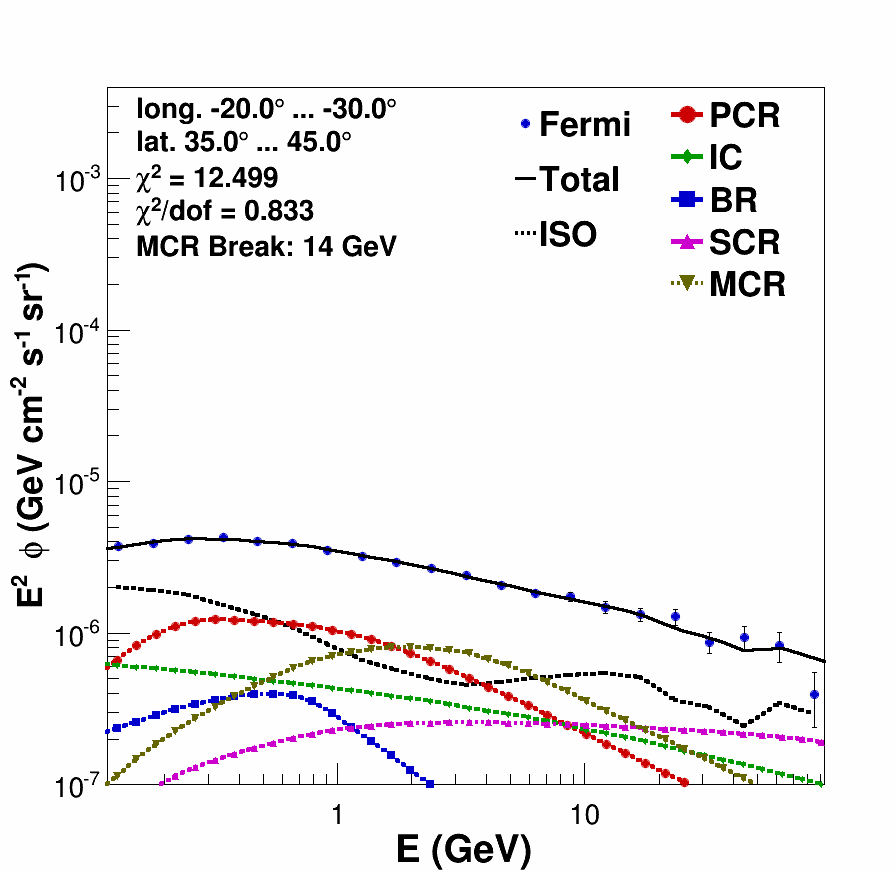}
\includegraphics[width=0.16\textwidth,height=0.16\textwidth,clip]{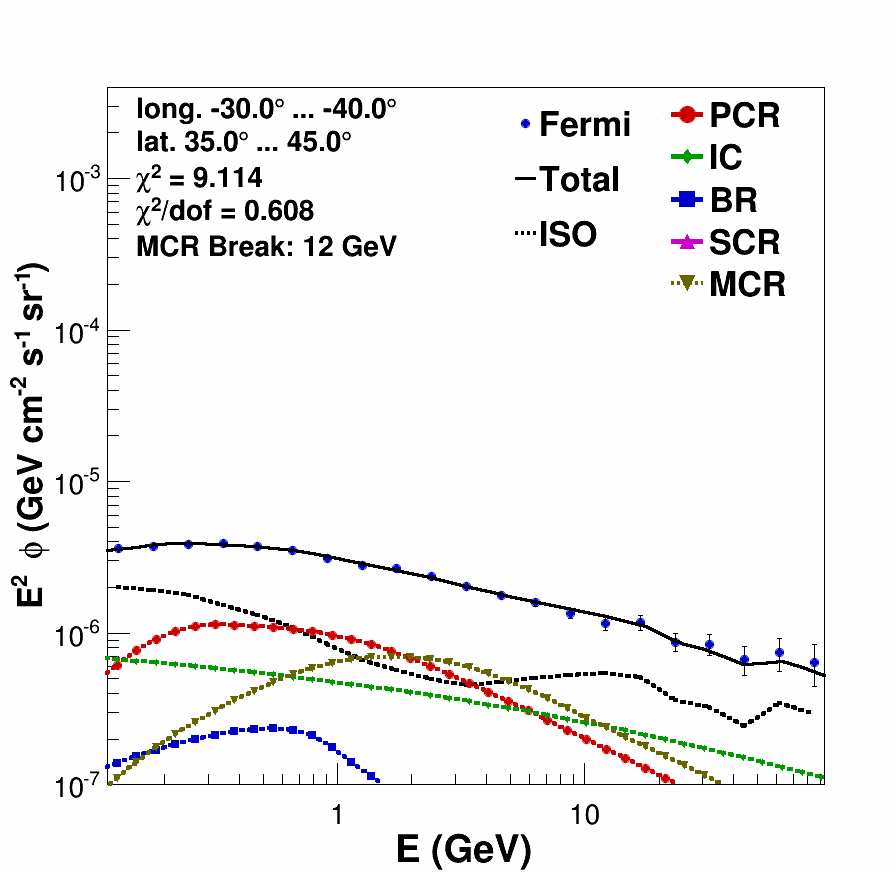}
\includegraphics[width=0.16\textwidth,height=0.16\textwidth,clip]{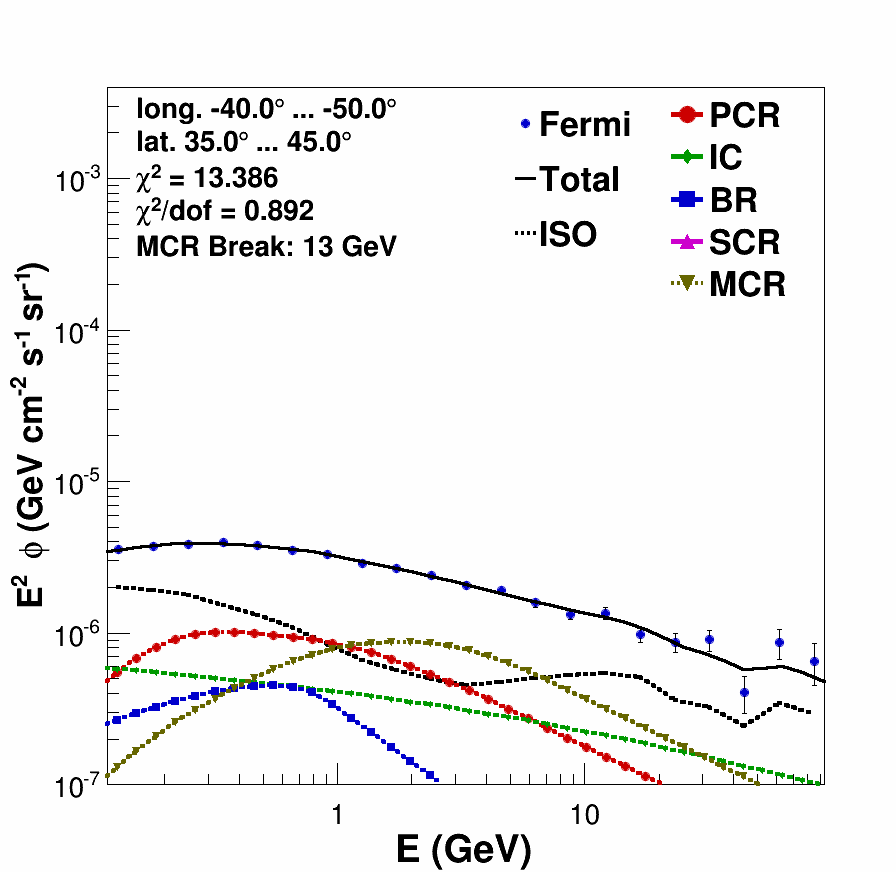}
\includegraphics[width=0.16\textwidth,height=0.16\textwidth,clip]{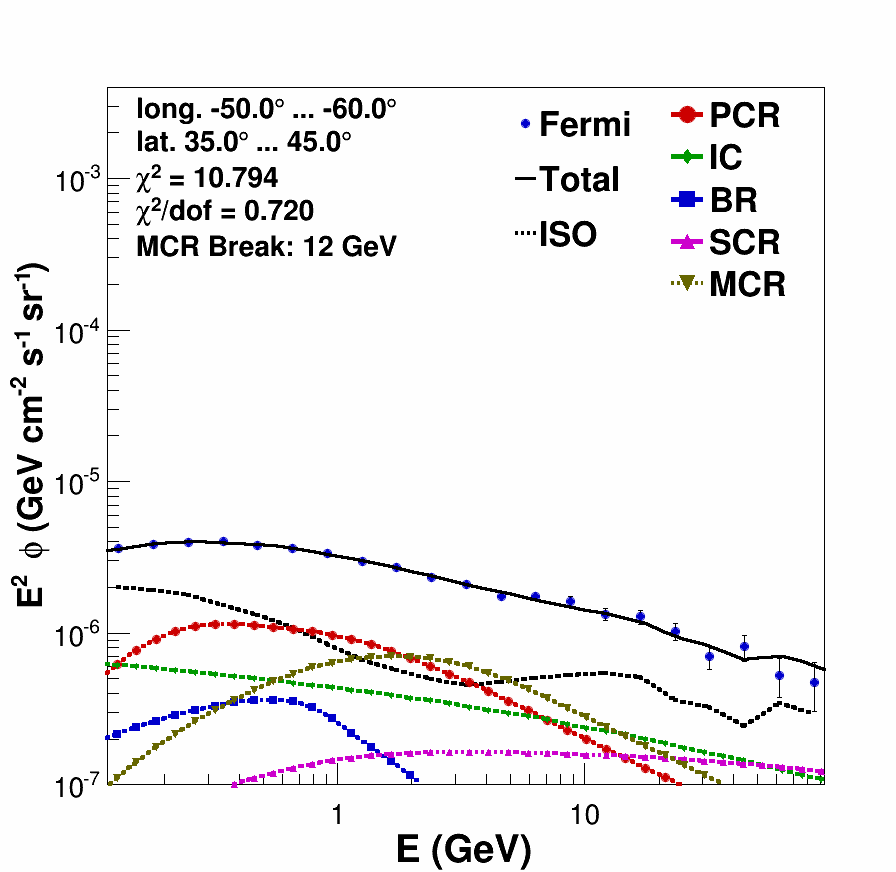}
\includegraphics[width=0.16\textwidth,height=0.16\textwidth,clip]{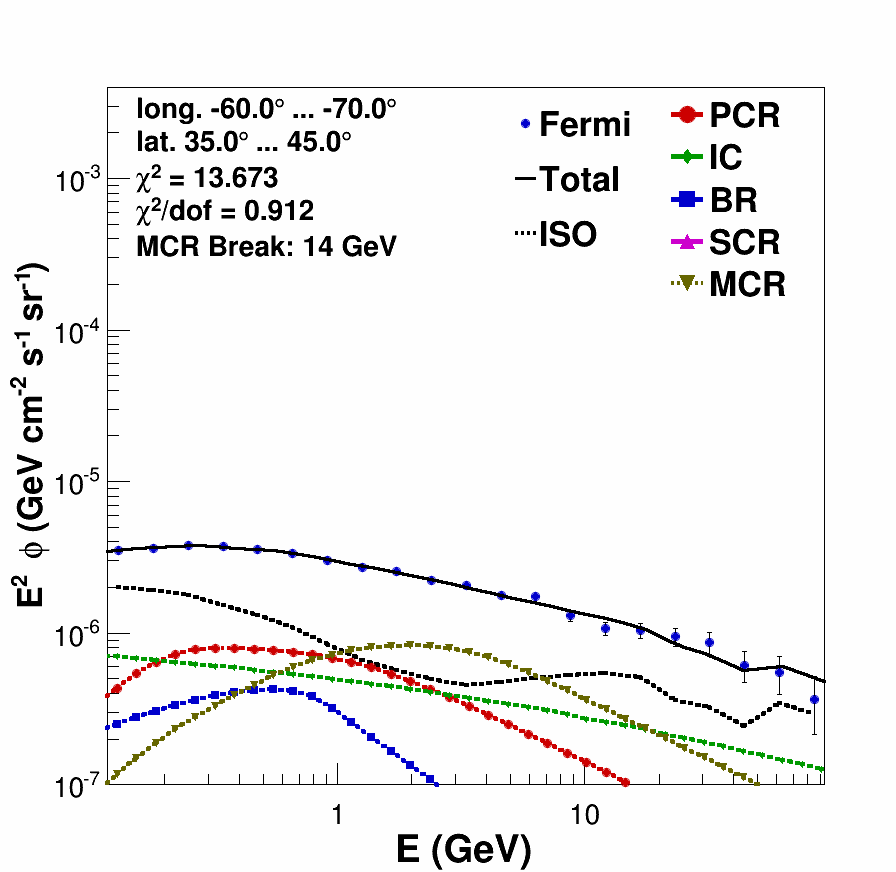}
\includegraphics[width=0.16\textwidth,height=0.16\textwidth,clip]{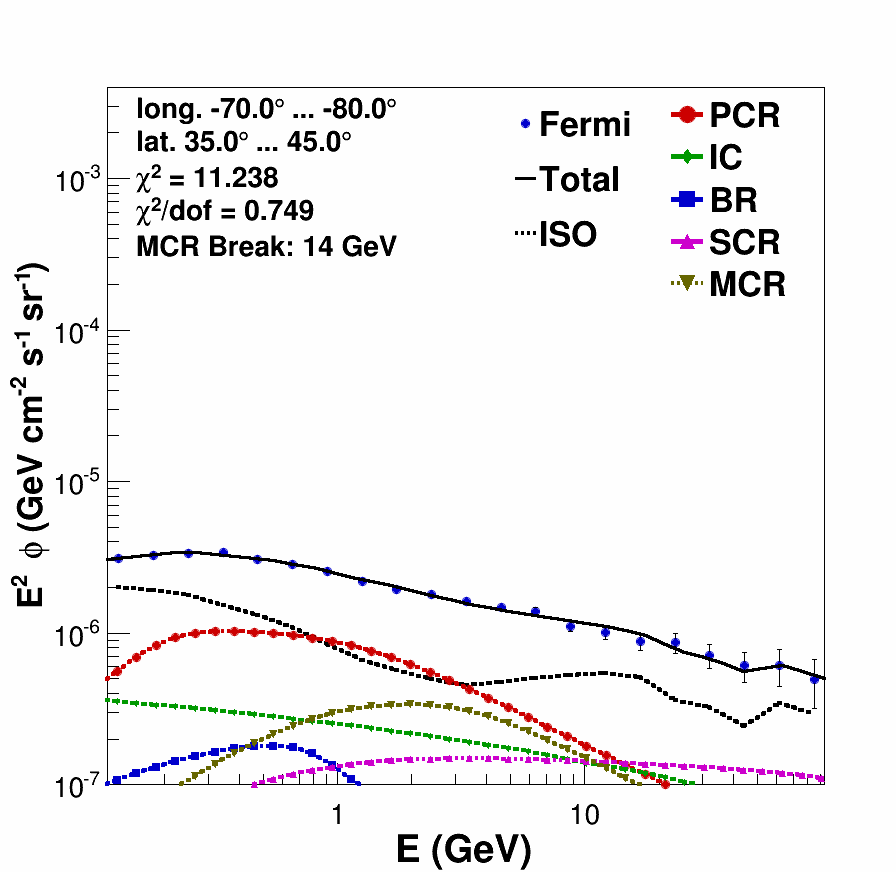}
\includegraphics[width=0.16\textwidth,height=0.16\textwidth,clip]{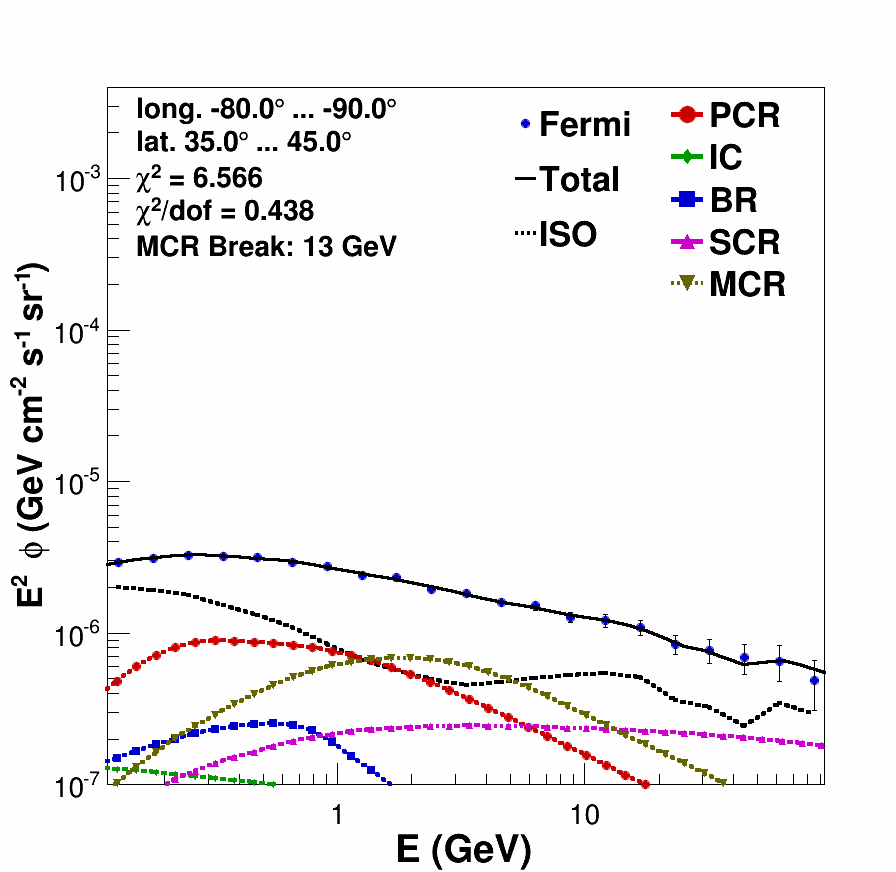}
\includegraphics[width=0.16\textwidth,height=0.16\textwidth,clip]{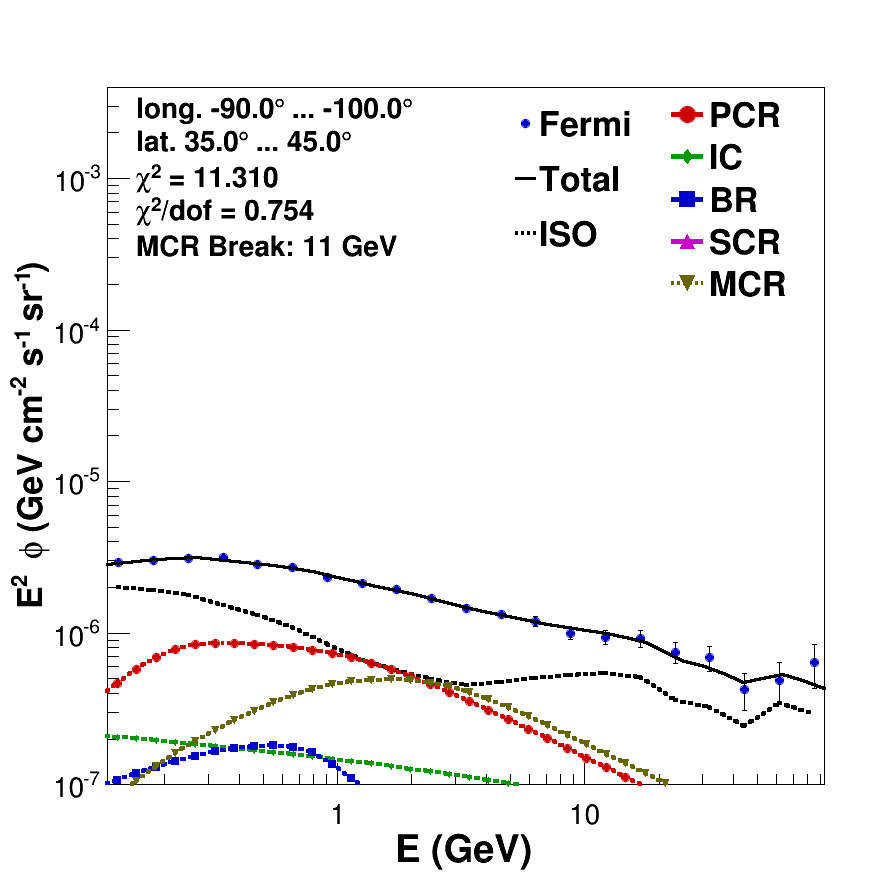}
\includegraphics[width=0.16\textwidth,height=0.16\textwidth,clip]{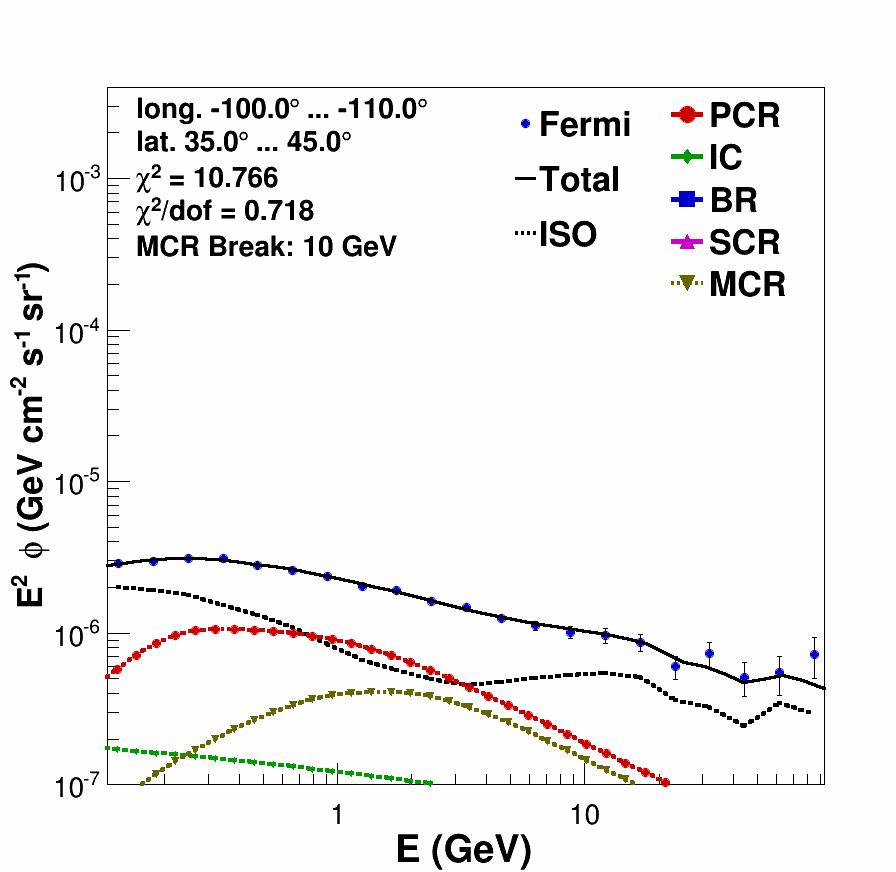}
\includegraphics[width=0.16\textwidth,height=0.16\textwidth,clip]{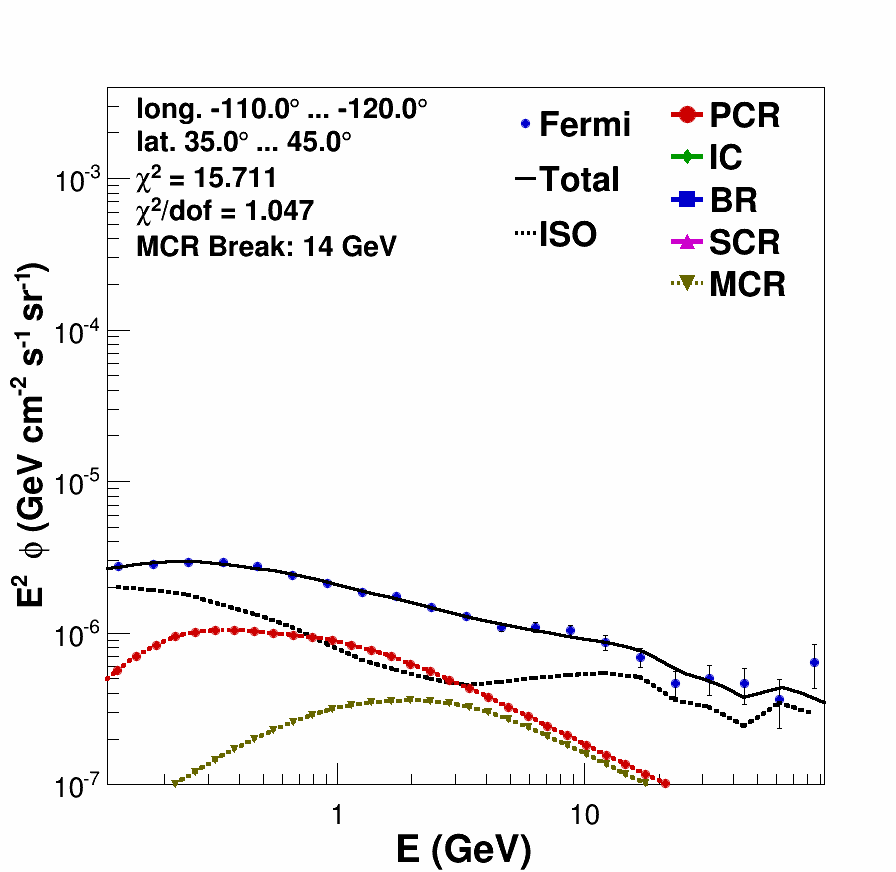}
\includegraphics[width=0.16\textwidth,height=0.16\textwidth,clip]{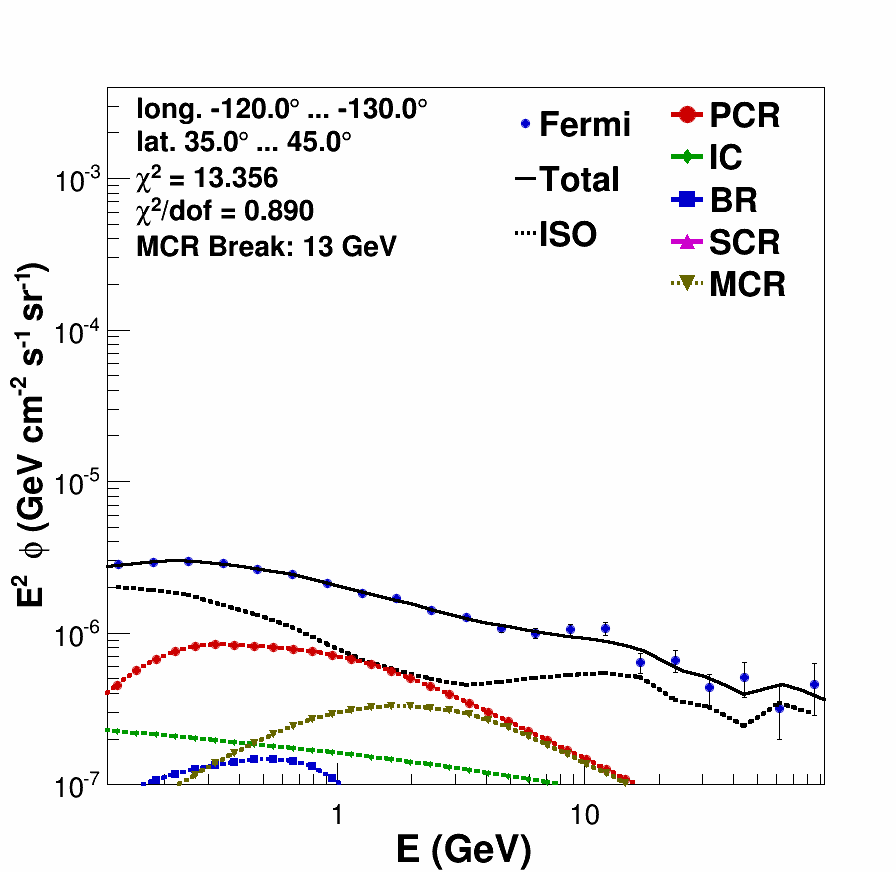}
\includegraphics[width=0.16\textwidth,height=0.16\textwidth,clip]{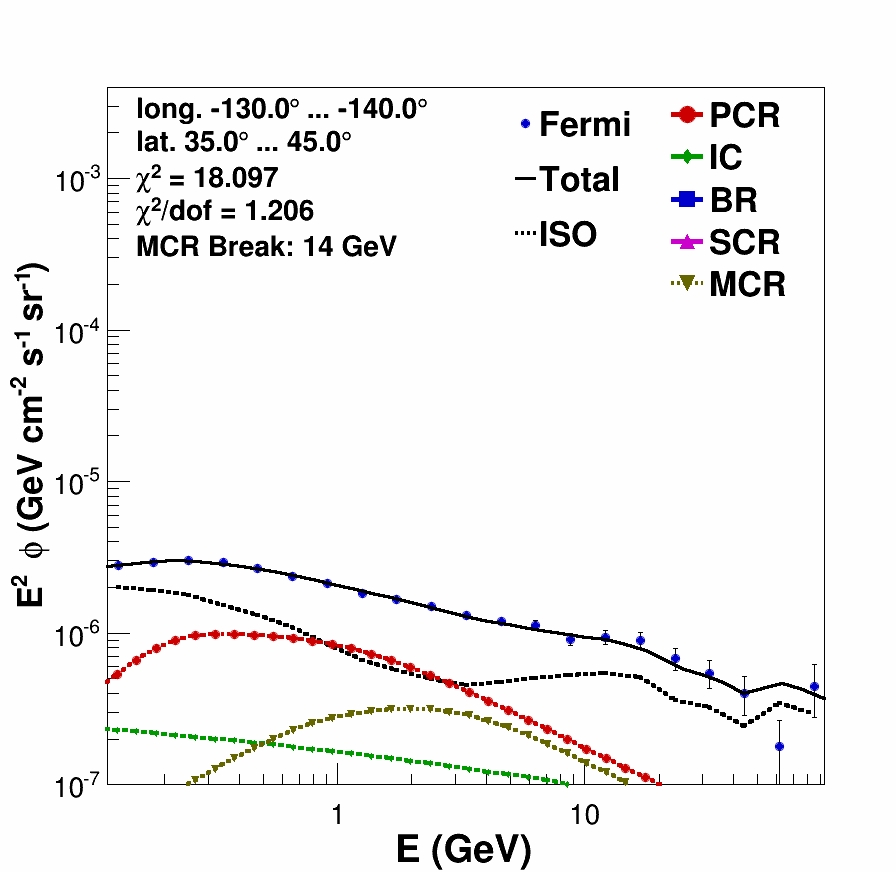}
\includegraphics[width=0.16\textwidth,height=0.16\textwidth,clip]{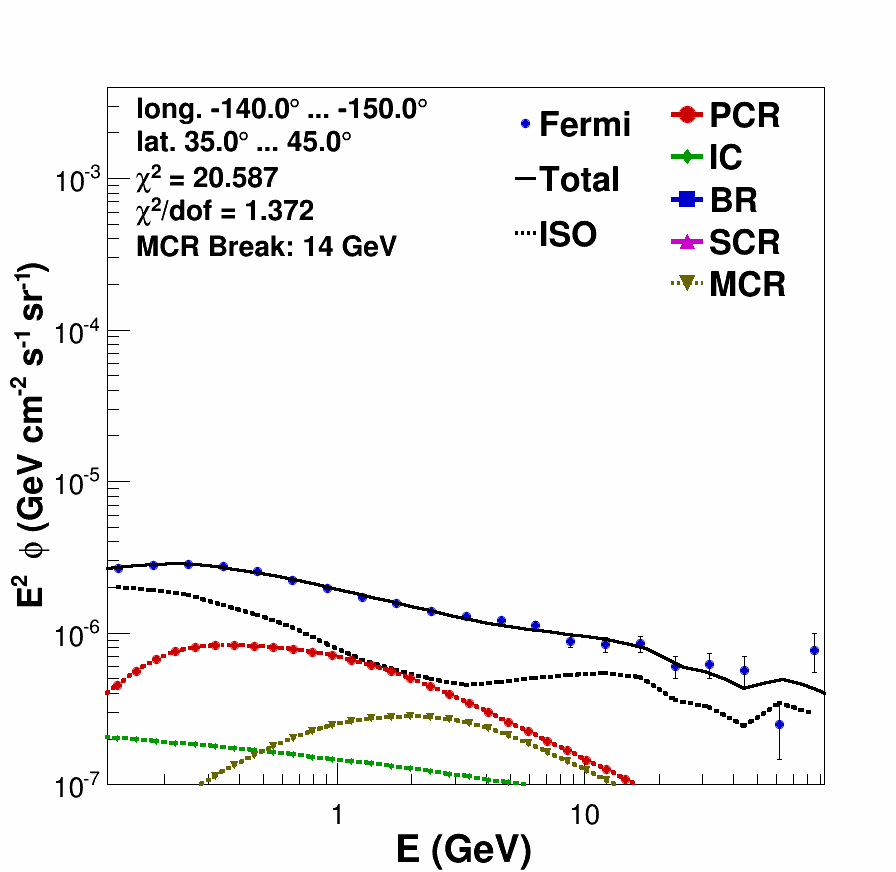}
\includegraphics[width=0.16\textwidth,height=0.16\textwidth,clip]{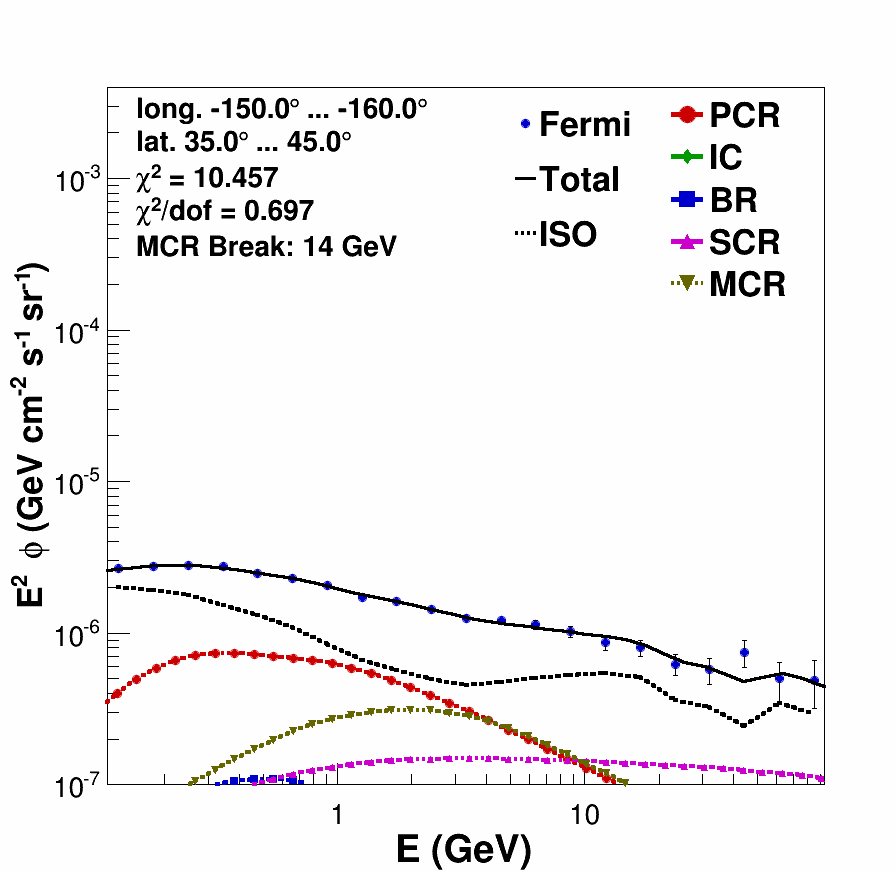}
\includegraphics[width=0.16\textwidth,height=0.16\textwidth,clip]{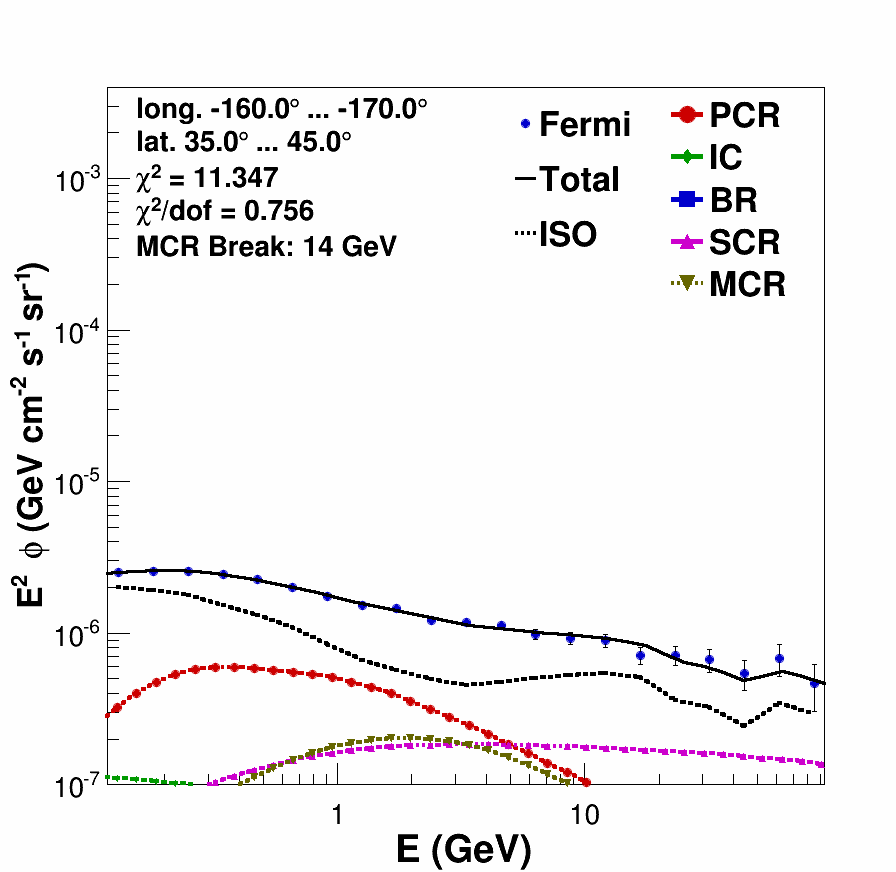}
\includegraphics[width=0.16\textwidth,height=0.16\textwidth,clip]{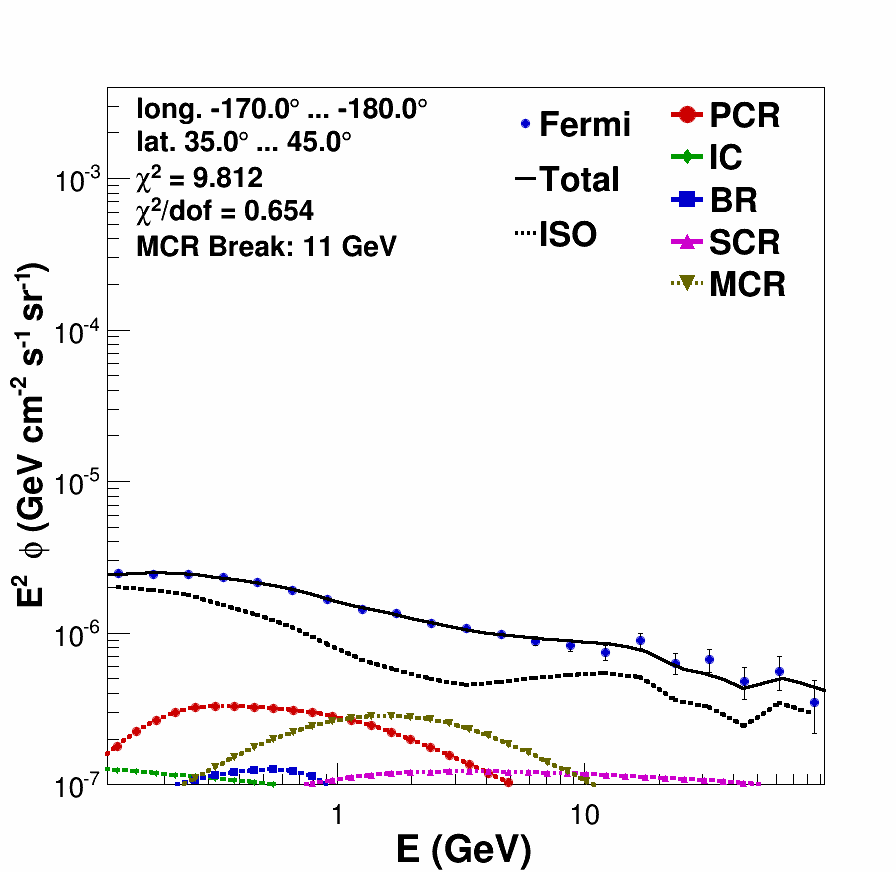}%%%%%r4
\caption[]{Template fits for latitudes  with $35.0^\circ<b<45.0^\circ$ and longitudes decreasing from 180$^\circ$ to -180$^\circ$. \label{F14}
}
\end{figure}
\clearpage
\begin{figure}
\centering
\includegraphics[width=0.16\textwidth,height=0.16\textwidth,clip]{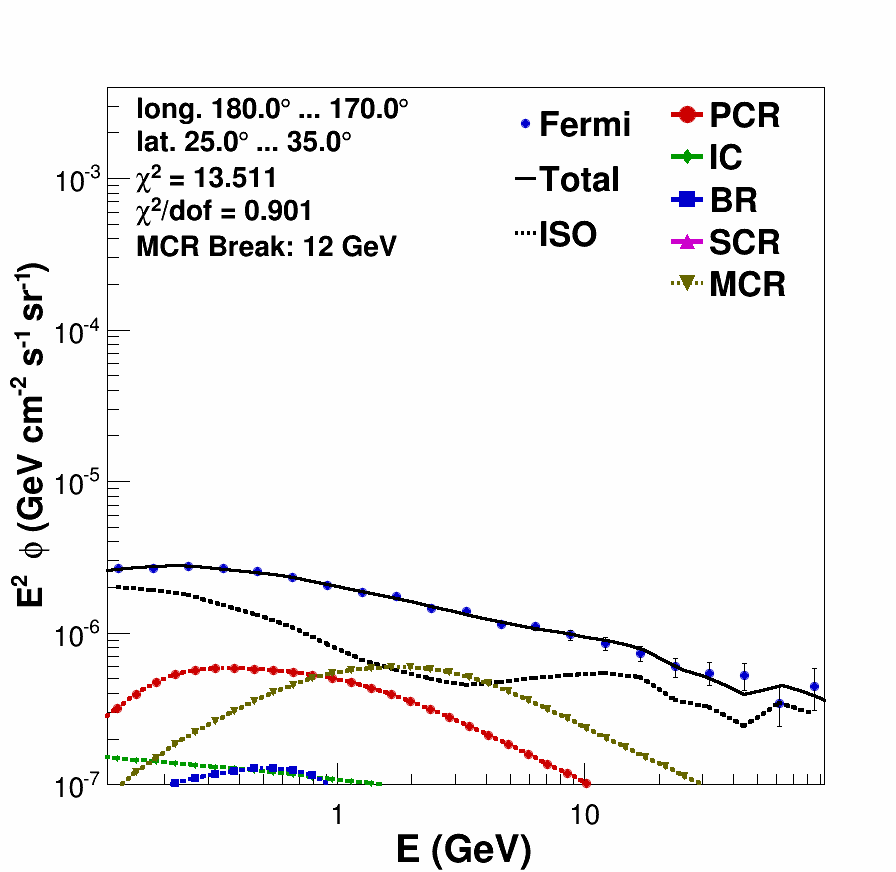}
\includegraphics[width=0.16\textwidth,height=0.16\textwidth,clip]{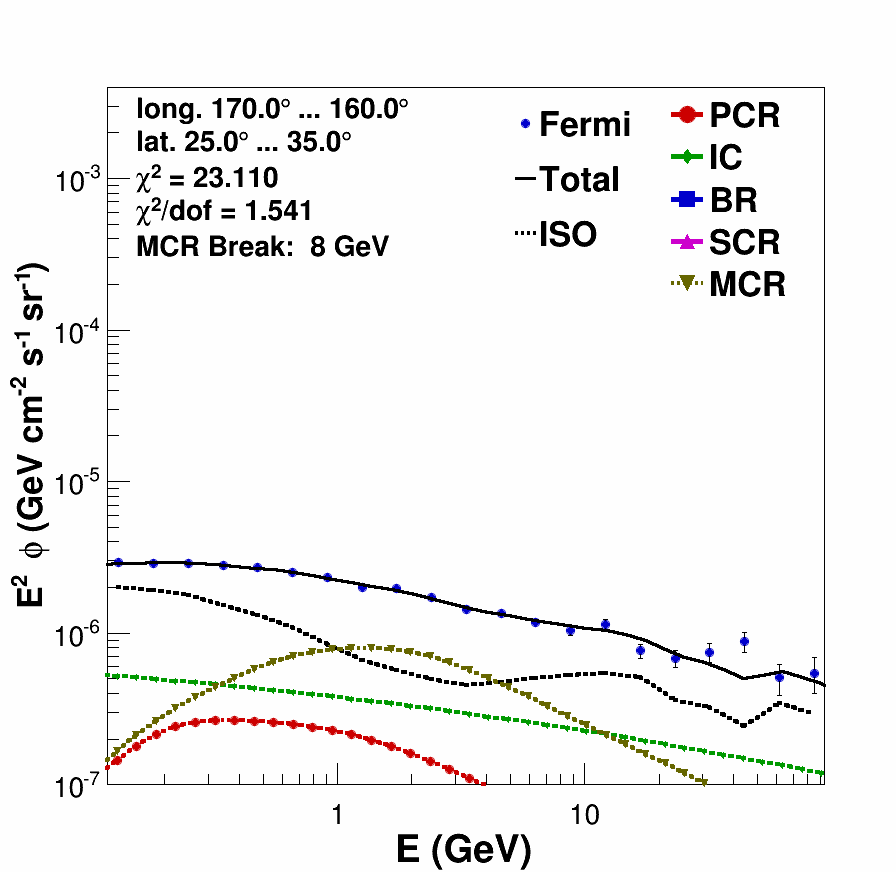}
\includegraphics[width=0.16\textwidth,height=0.16\textwidth,clip]{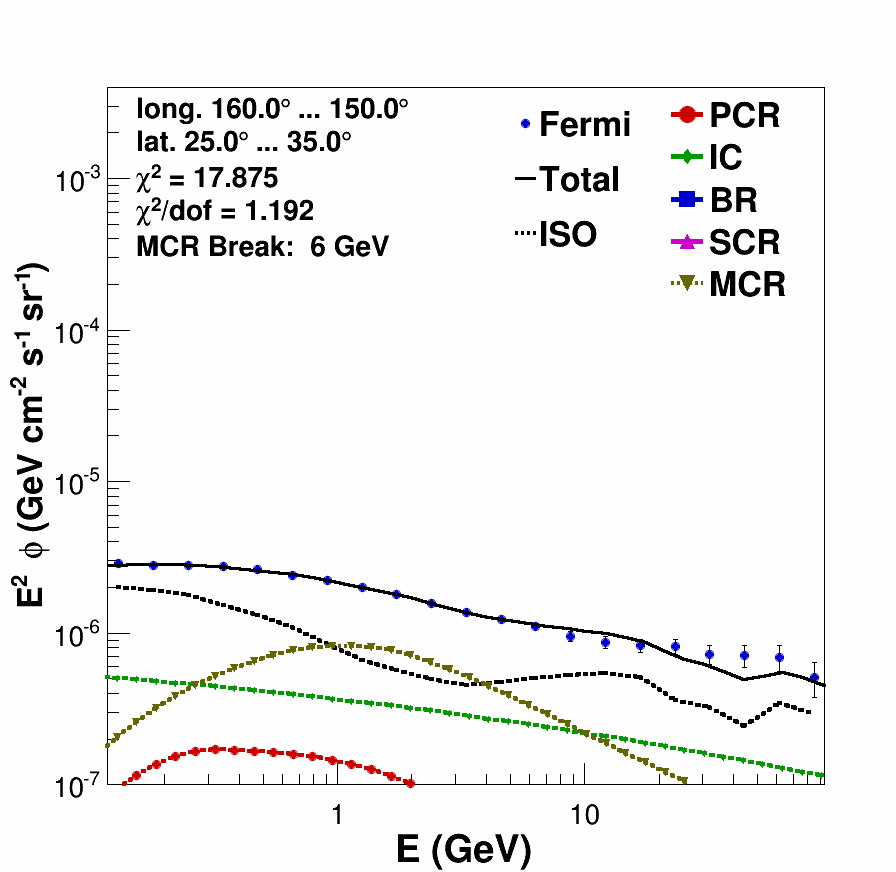}
\includegraphics[width=0.16\textwidth,height=0.16\textwidth,clip]{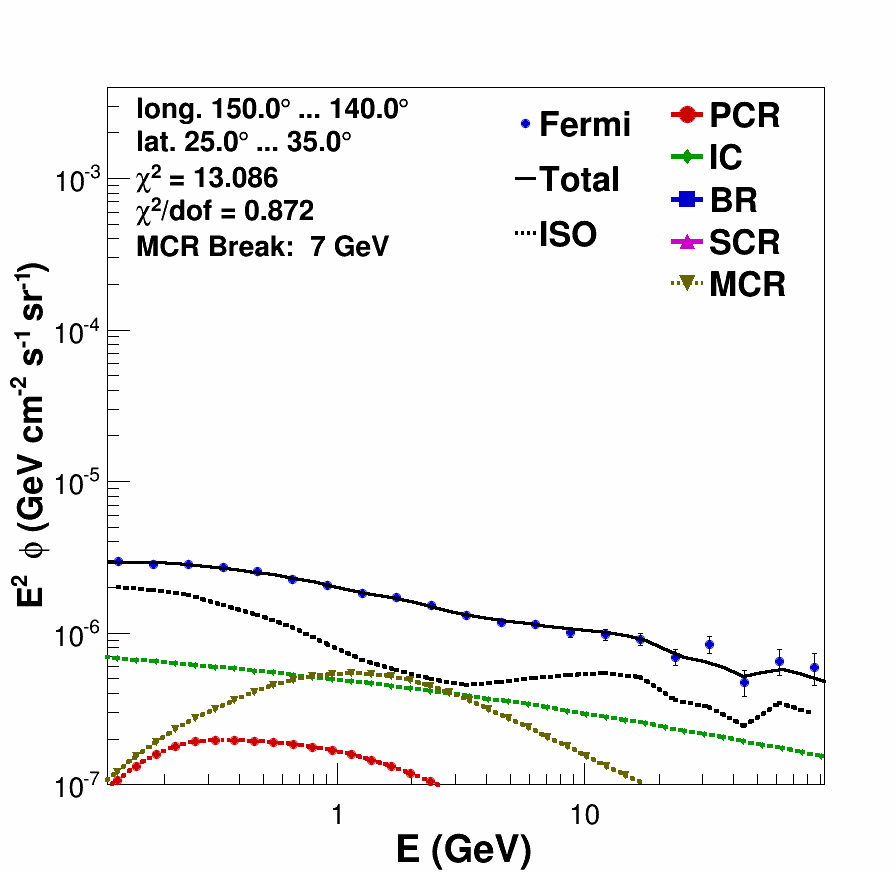}
\includegraphics[width=0.16\textwidth,height=0.16\textwidth,clip]{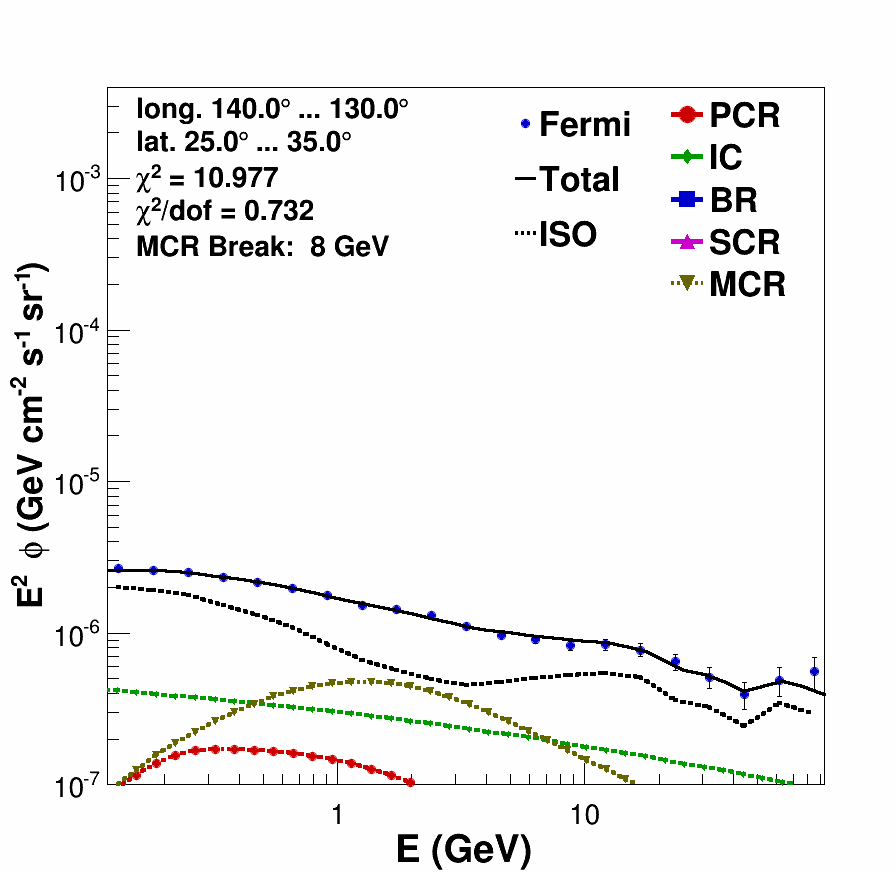}
\includegraphics[width=0.16\textwidth,height=0.16\textwidth,clip]{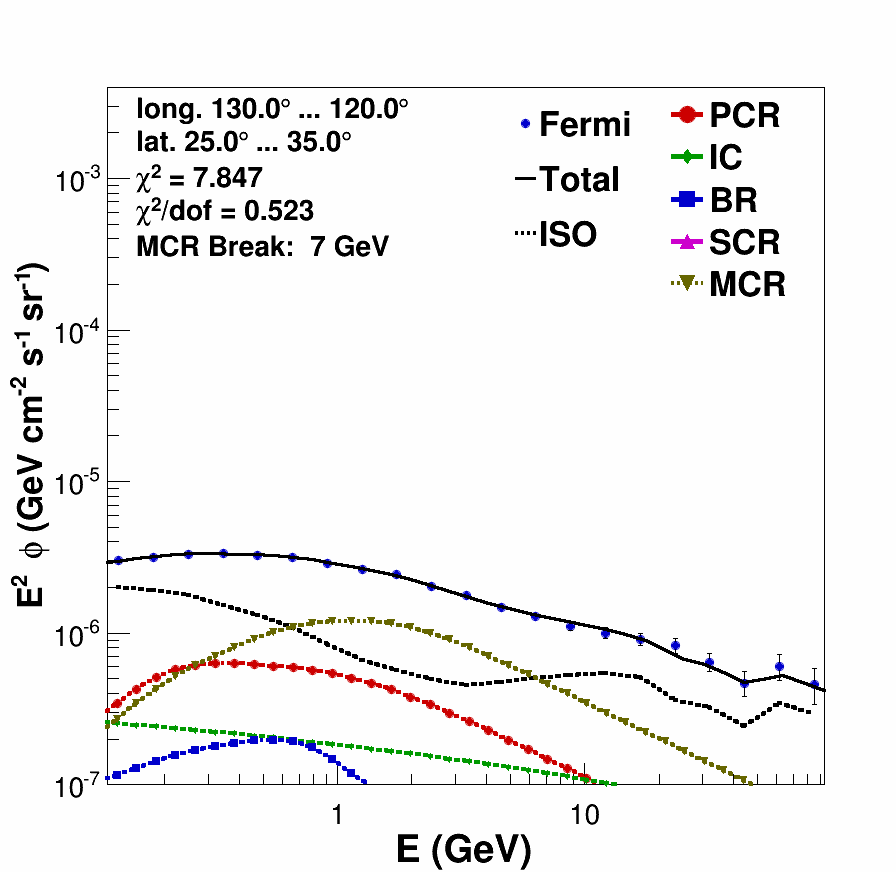}
\includegraphics[width=0.16\textwidth,height=0.16\textwidth,clip]{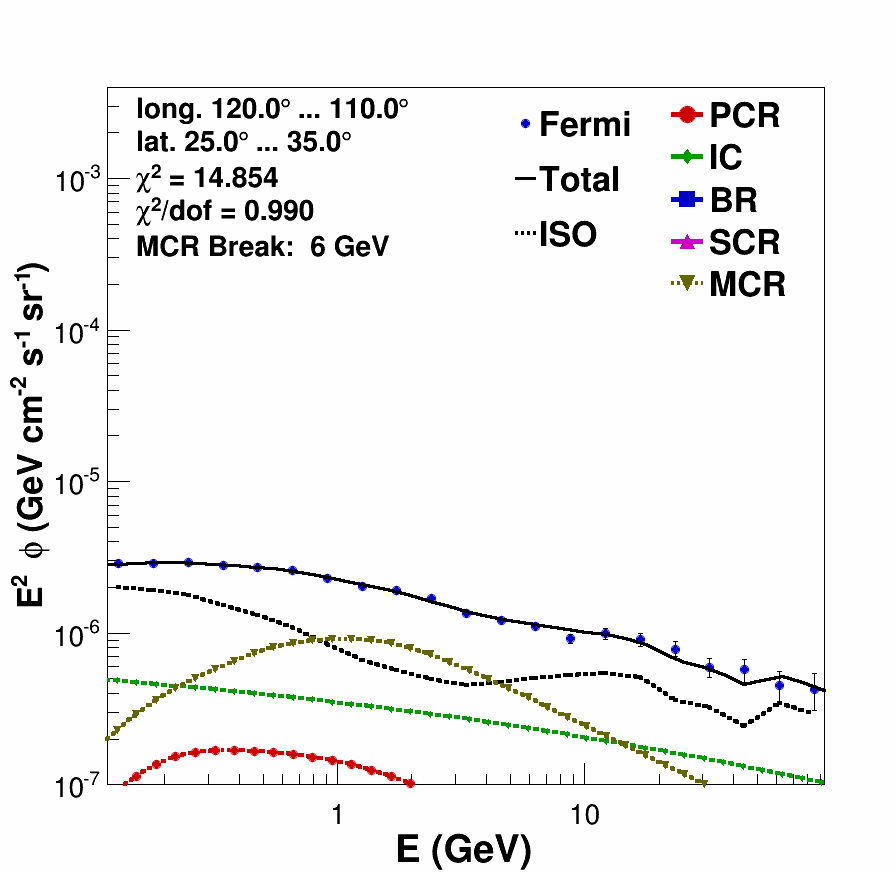}
\includegraphics[width=0.16\textwidth,height=0.16\textwidth,clip]{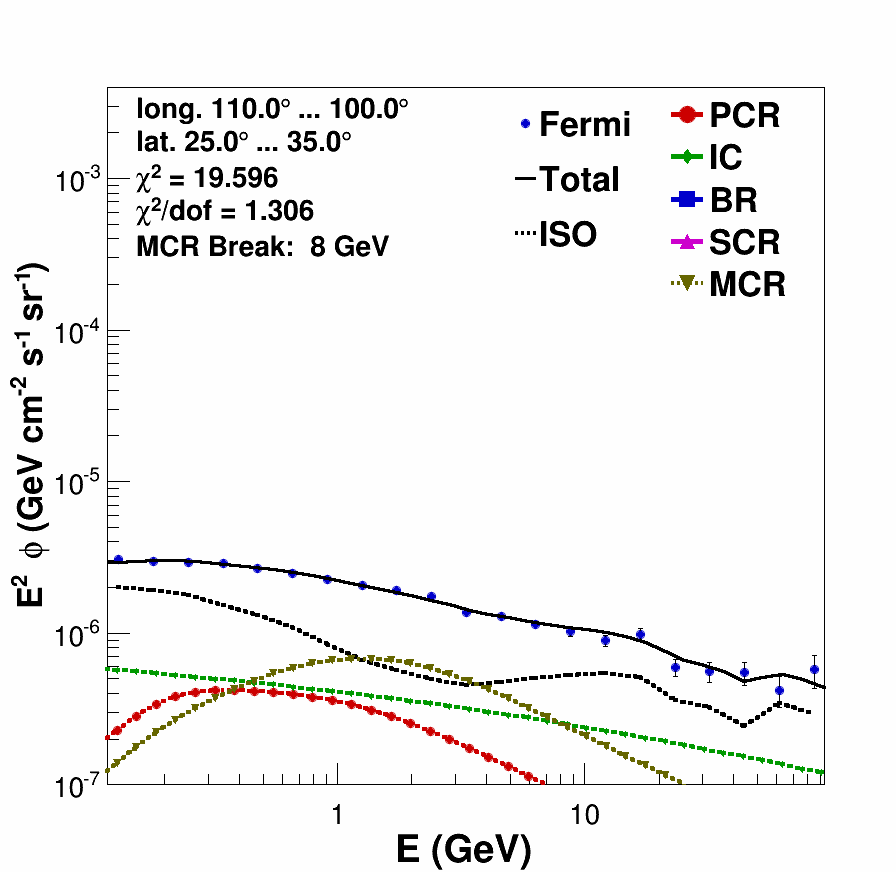}
\includegraphics[width=0.16\textwidth,height=0.16\textwidth,clip]{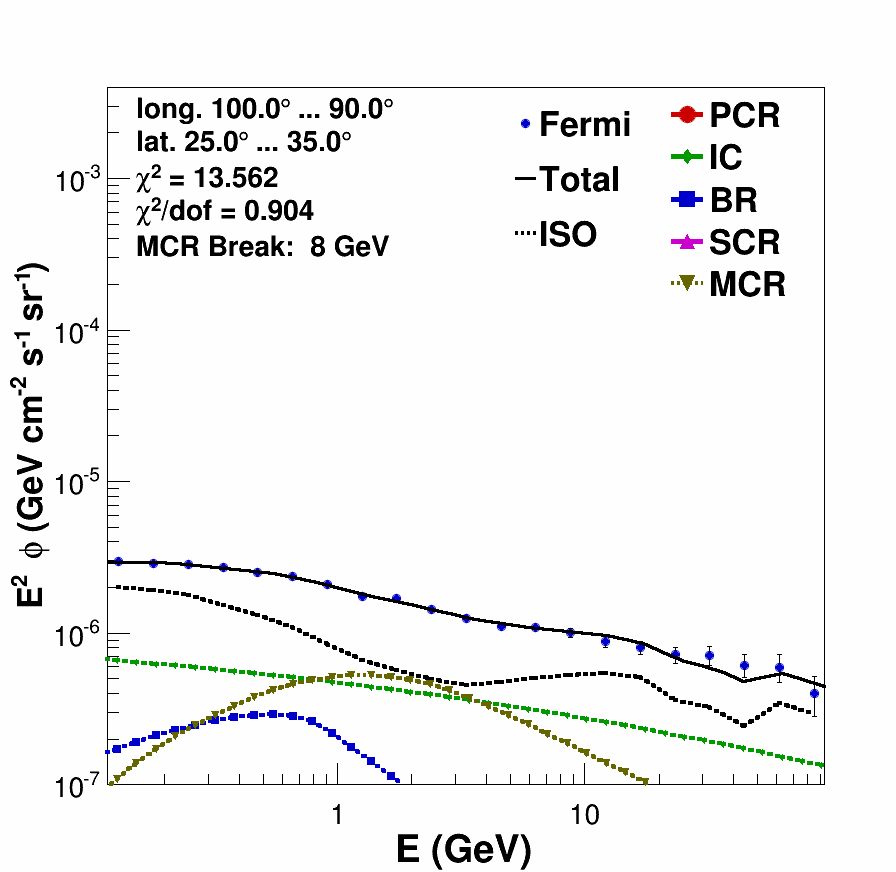}
\includegraphics[width=0.16\textwidth,height=0.16\textwidth,clip]{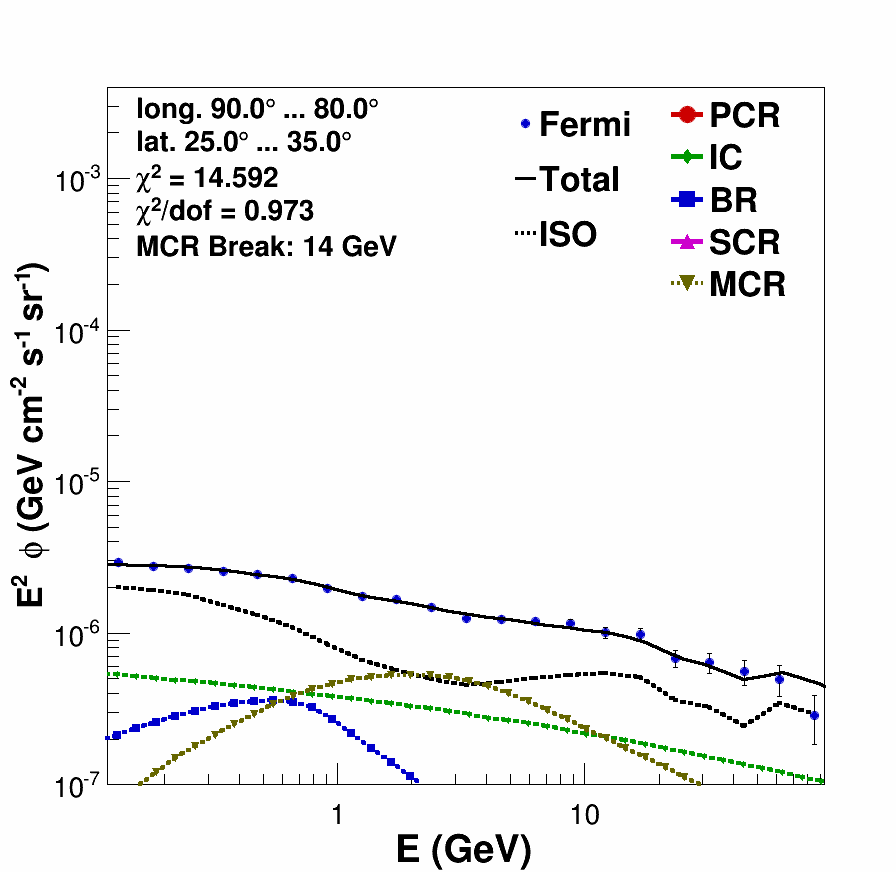}
\includegraphics[width=0.16\textwidth,height=0.16\textwidth,clip]{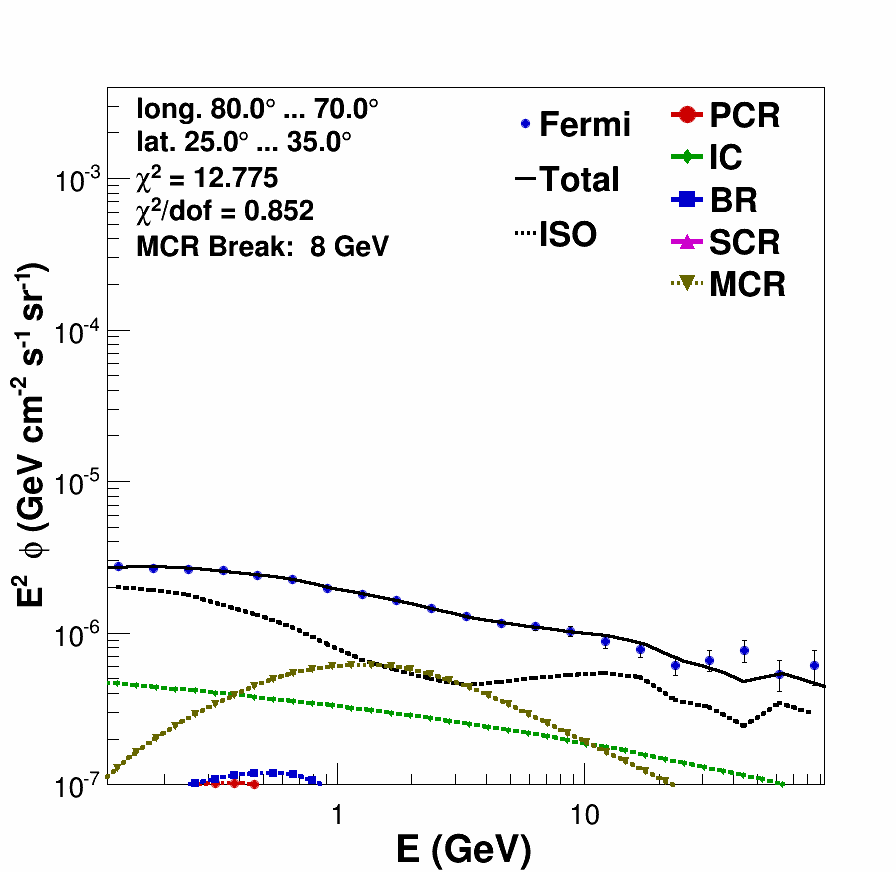}
\includegraphics[width=0.16\textwidth,height=0.16\textwidth,clip]{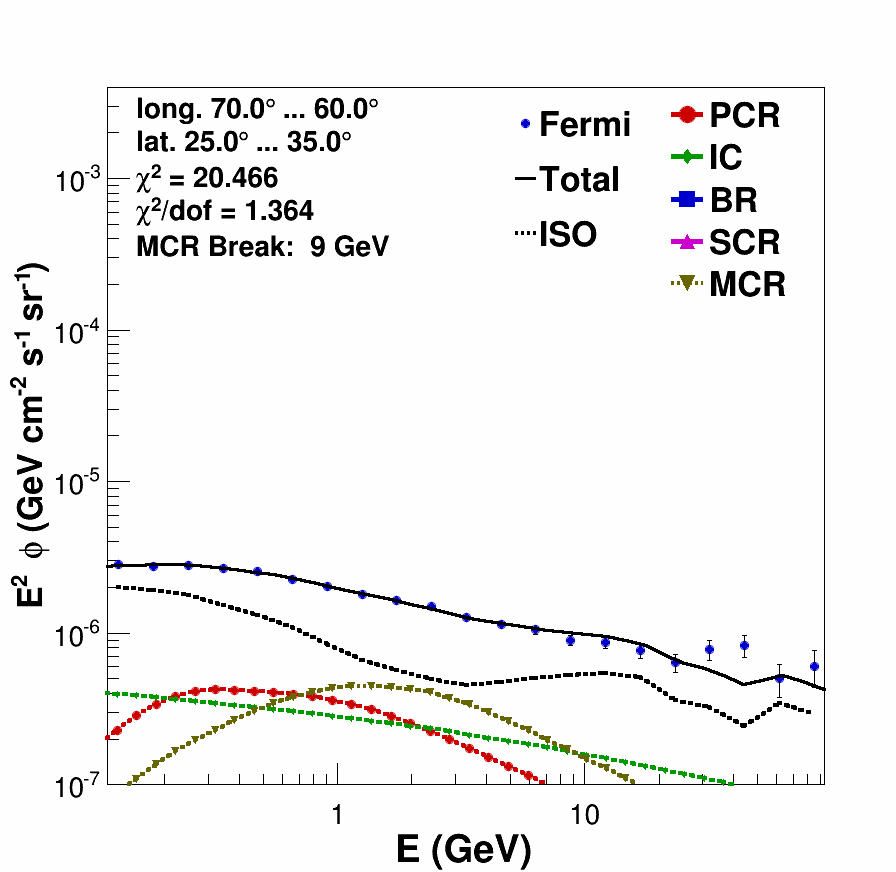}
\includegraphics[width=0.16\textwidth,height=0.16\textwidth,clip]{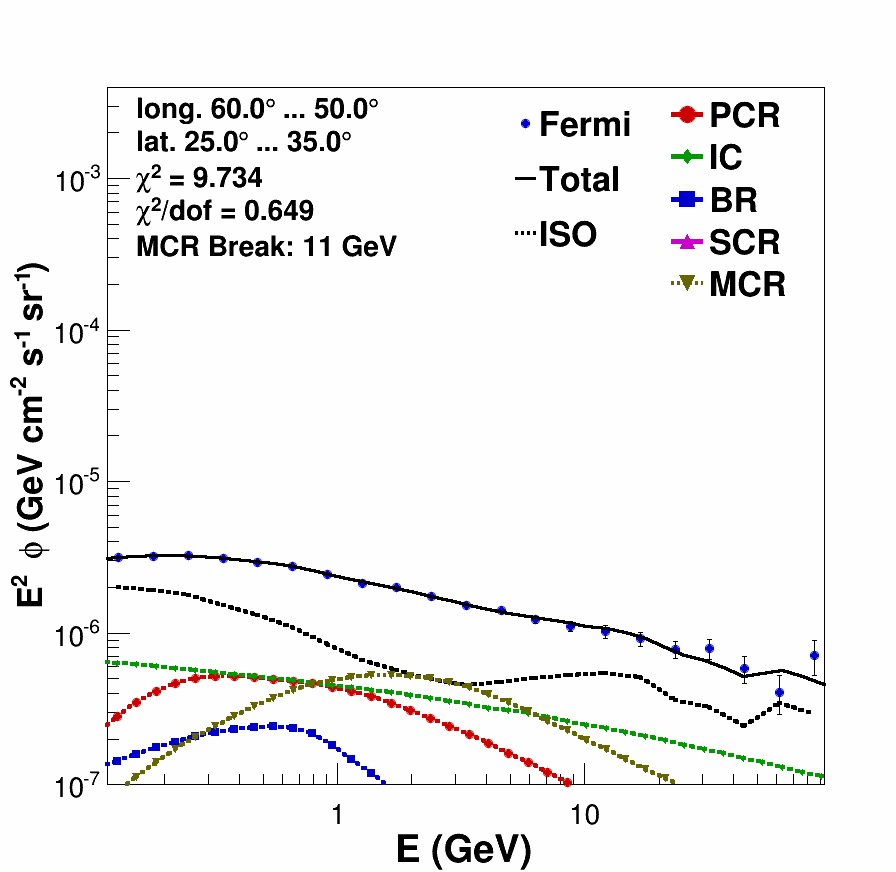}
\includegraphics[width=0.16\textwidth,height=0.16\textwidth,clip]{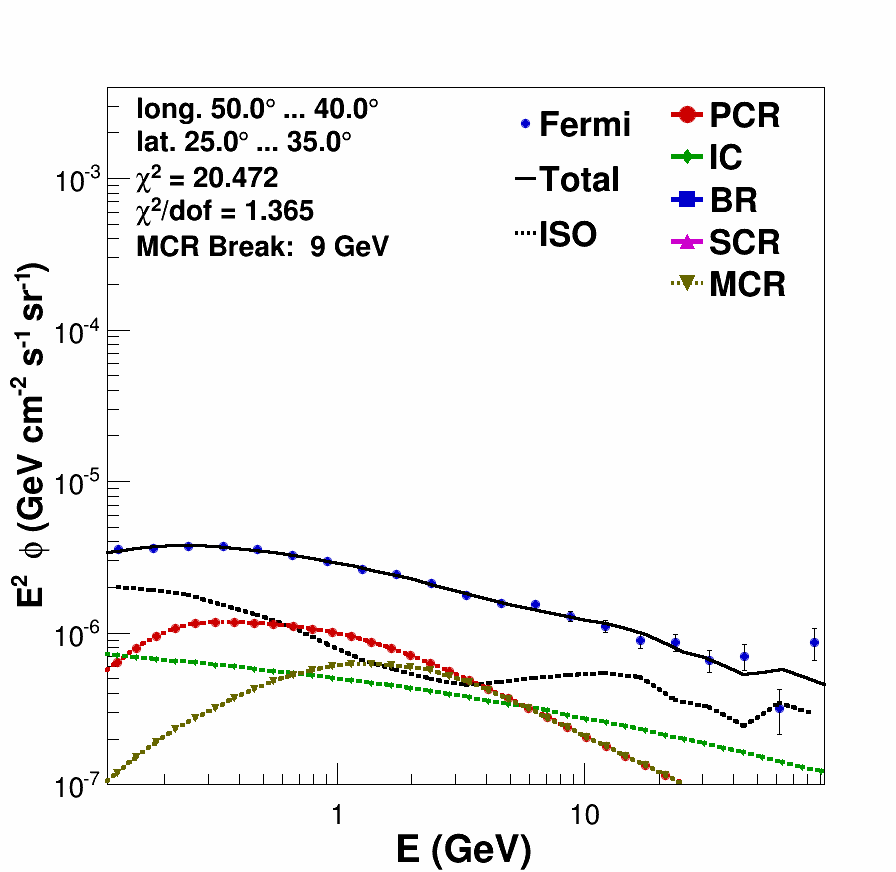}
\includegraphics[width=0.16\textwidth,height=0.16\textwidth,clip]{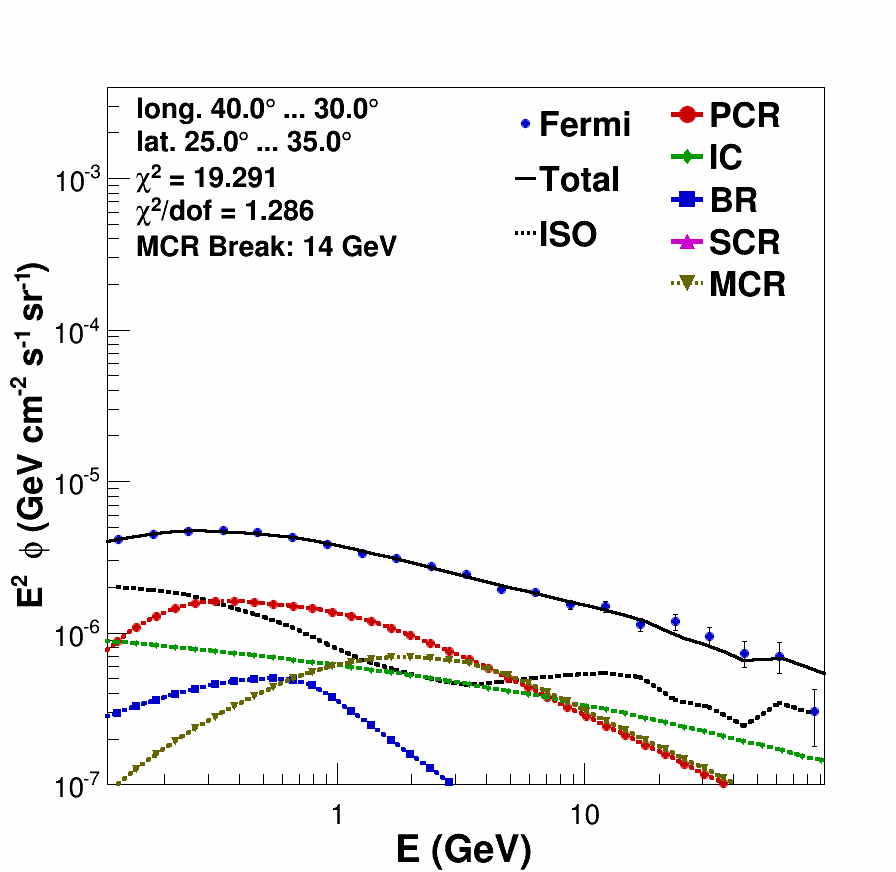}
\includegraphics[width=0.16\textwidth,height=0.16\textwidth,clip]{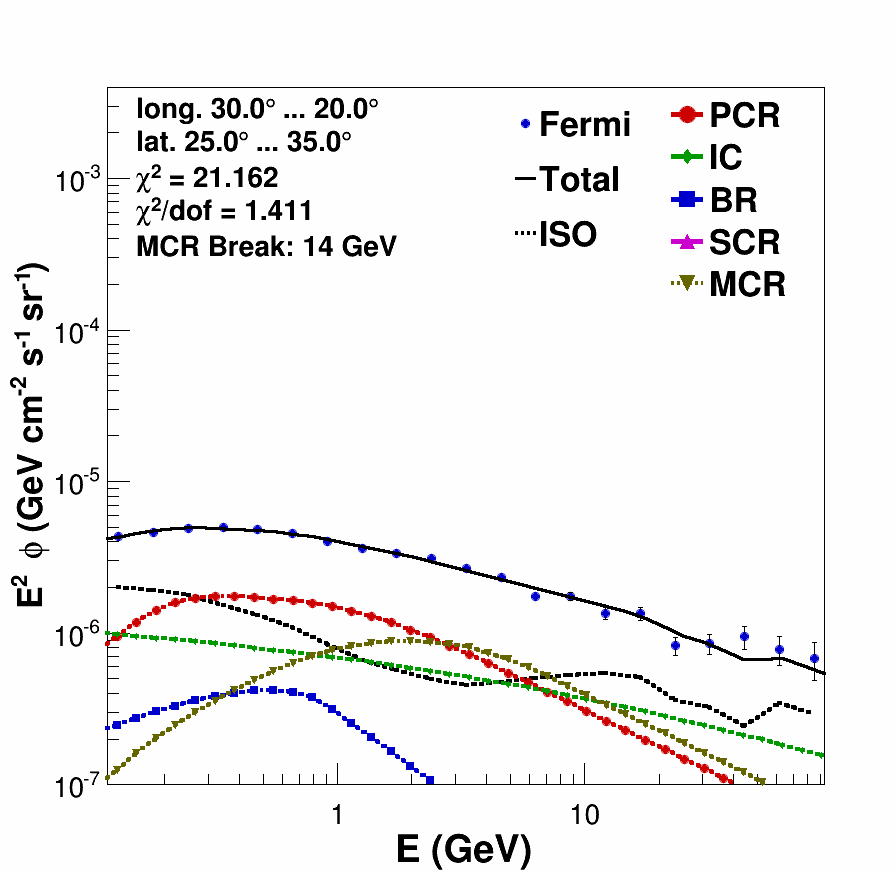}
\includegraphics[width=0.16\textwidth,height=0.16\textwidth,clip]{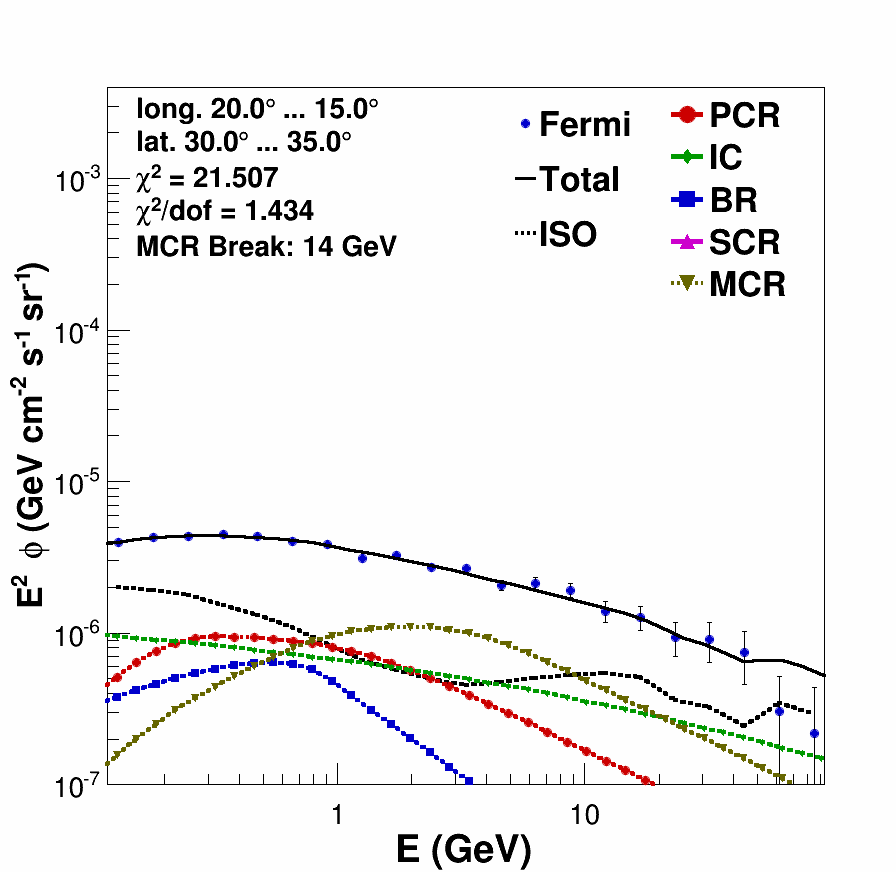}
\includegraphics[width=0.16\textwidth,height=0.16\textwidth,clip]{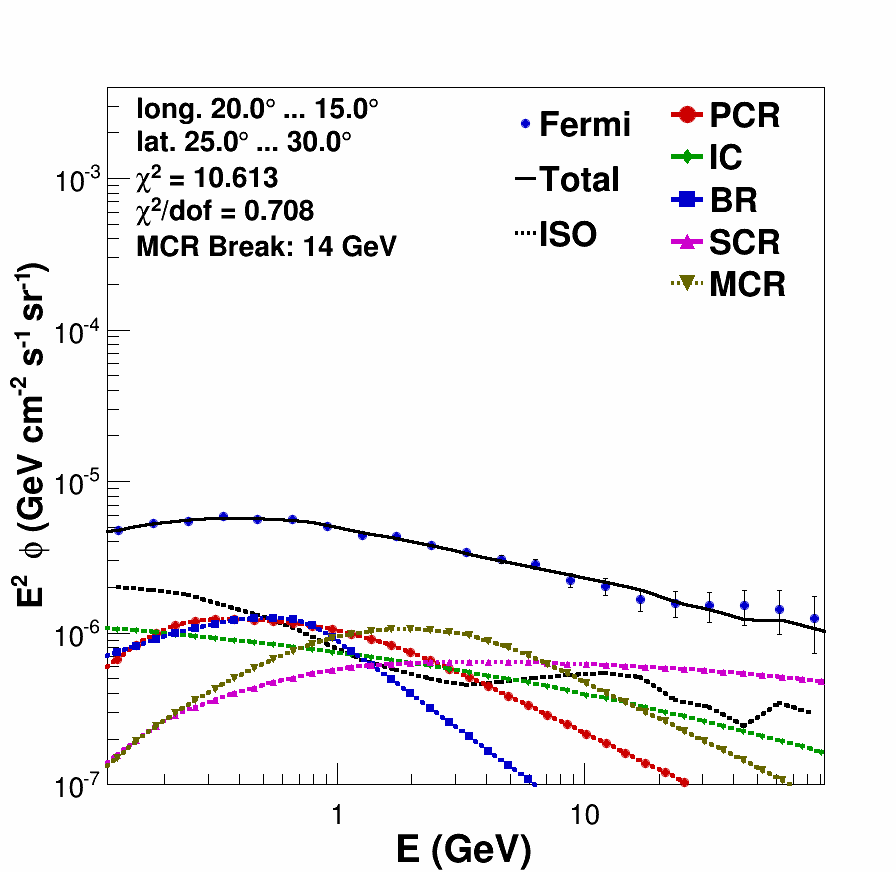}
\includegraphics[width=0.16\textwidth,height=0.16\textwidth,clip]{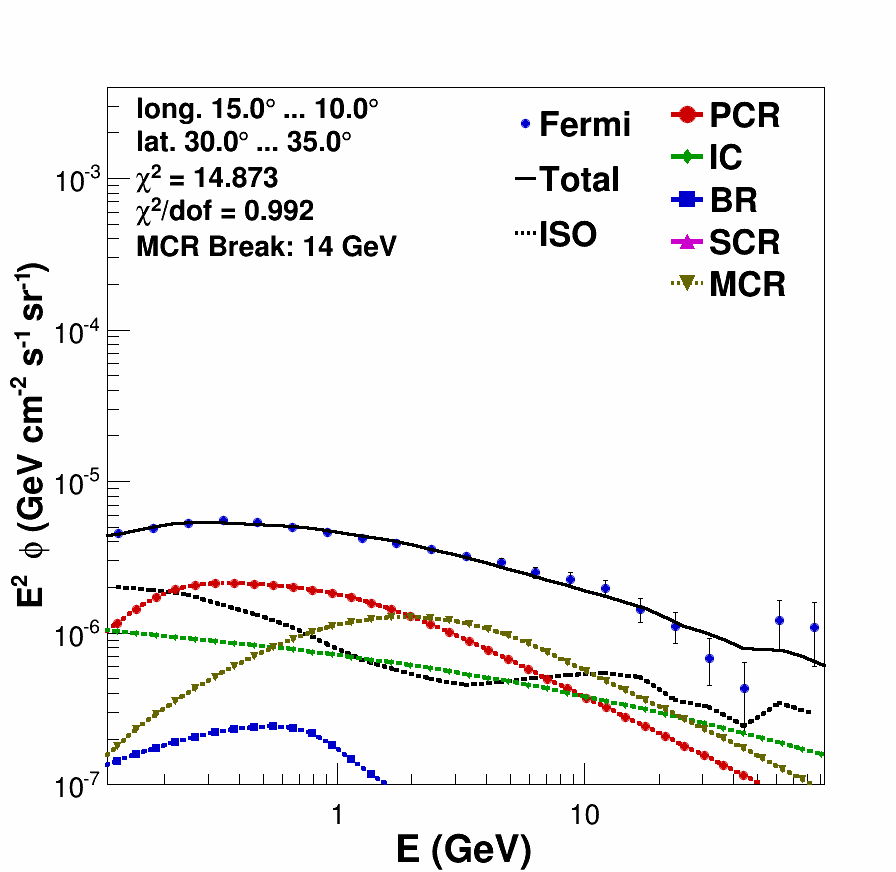}
\includegraphics[width=0.16\textwidth,height=0.16\textwidth,clip]{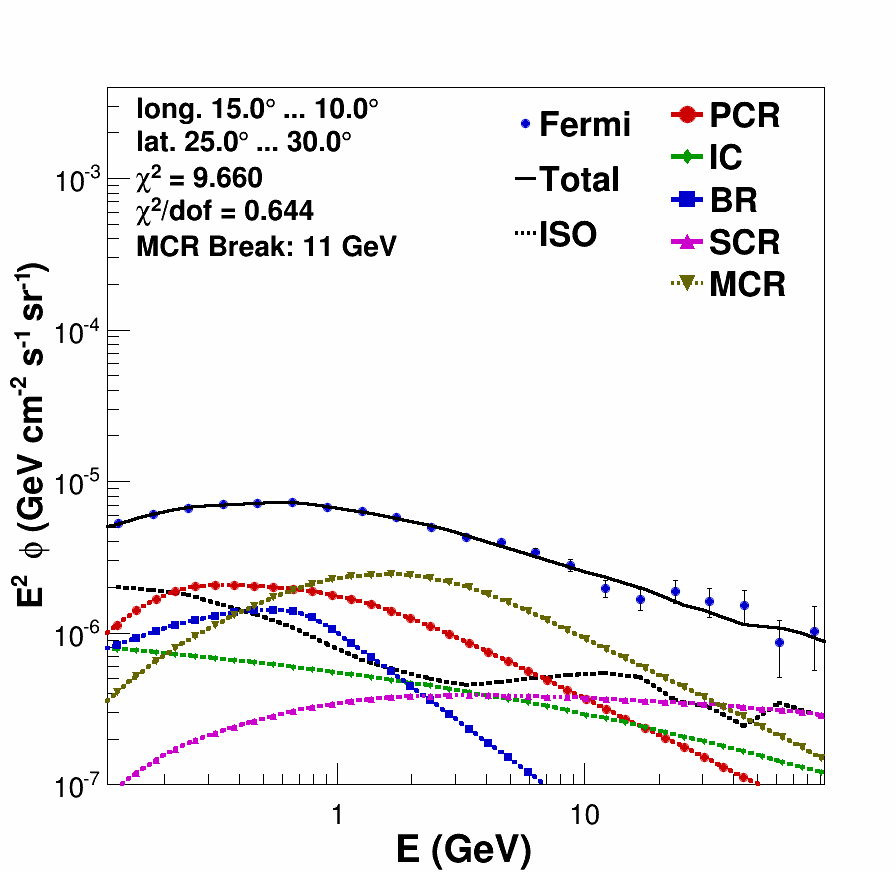}
\includegraphics[width=0.16\textwidth,height=0.16\textwidth,clip]{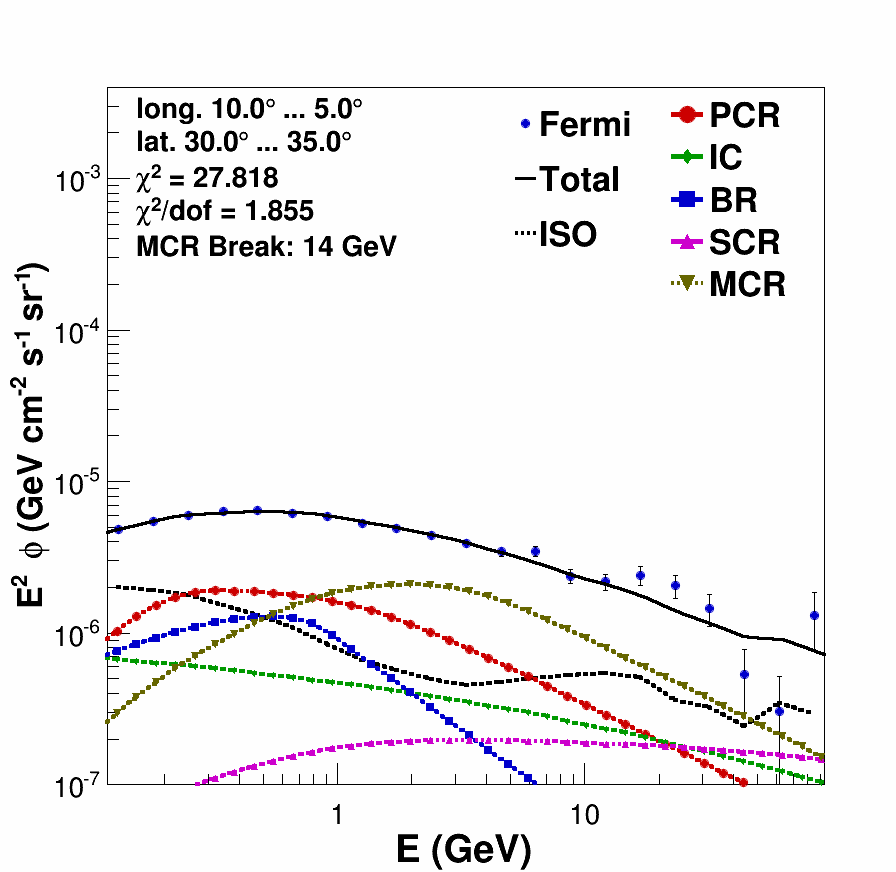}
\includegraphics[width=0.16\textwidth,height=0.16\textwidth,clip]{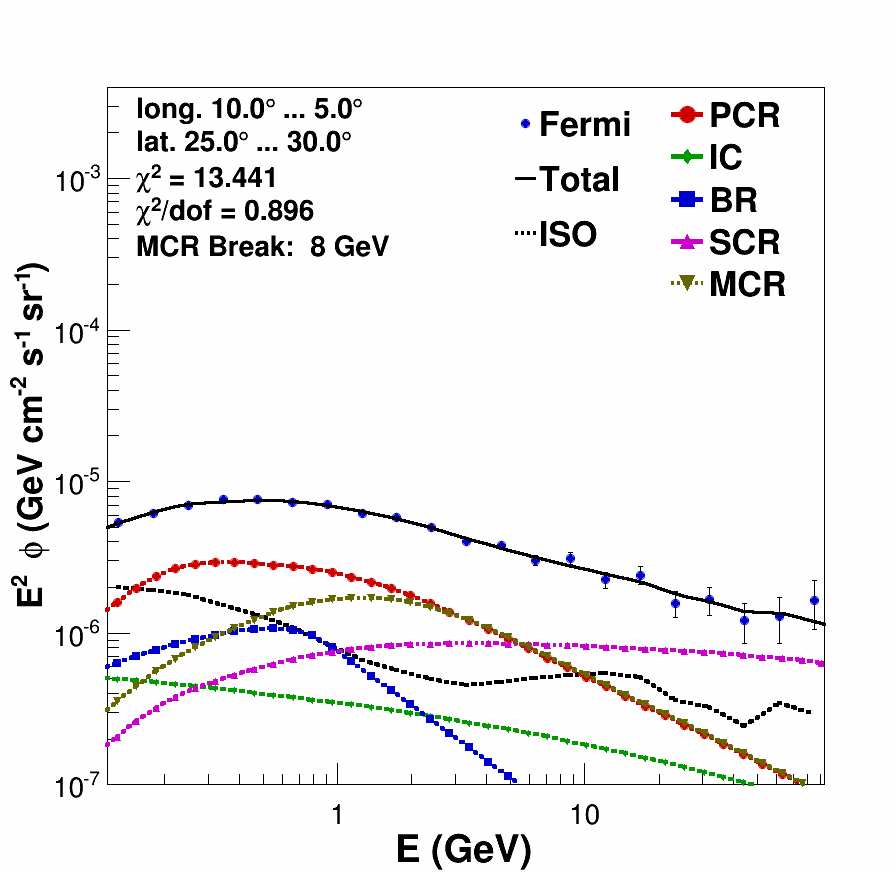}
\includegraphics[width=0.16\textwidth,height=0.16\textwidth,clip]{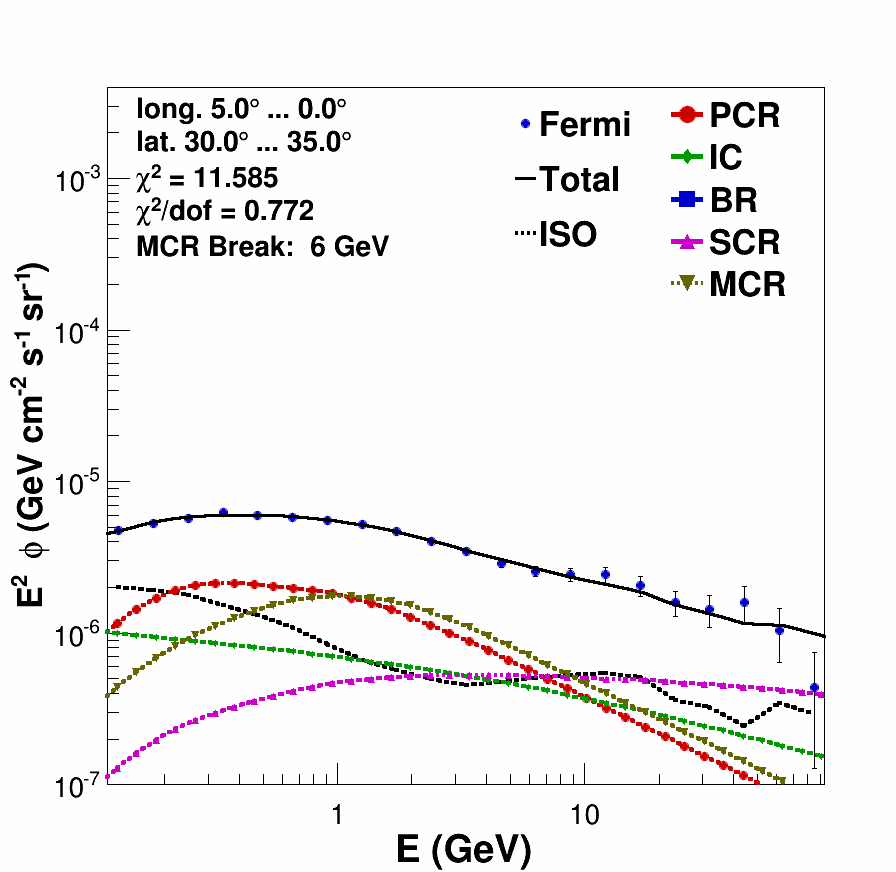}
\includegraphics[width=0.16\textwidth,height=0.16\textwidth,clip]{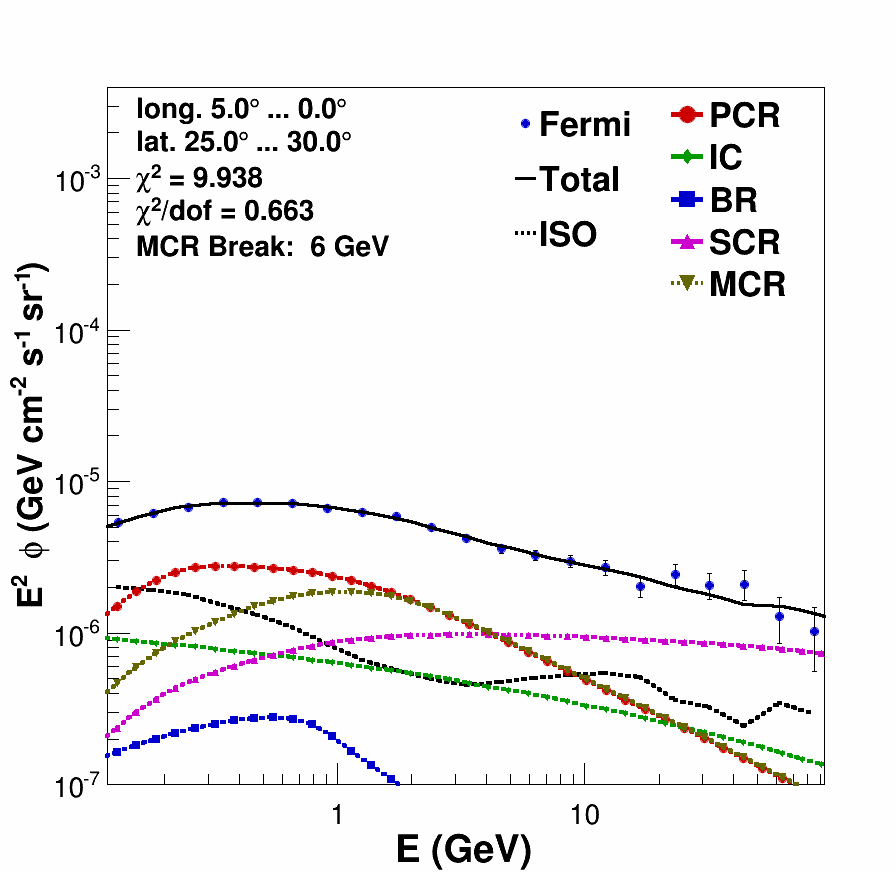}
\includegraphics[width=0.16\textwidth,height=0.16\textwidth,clip]{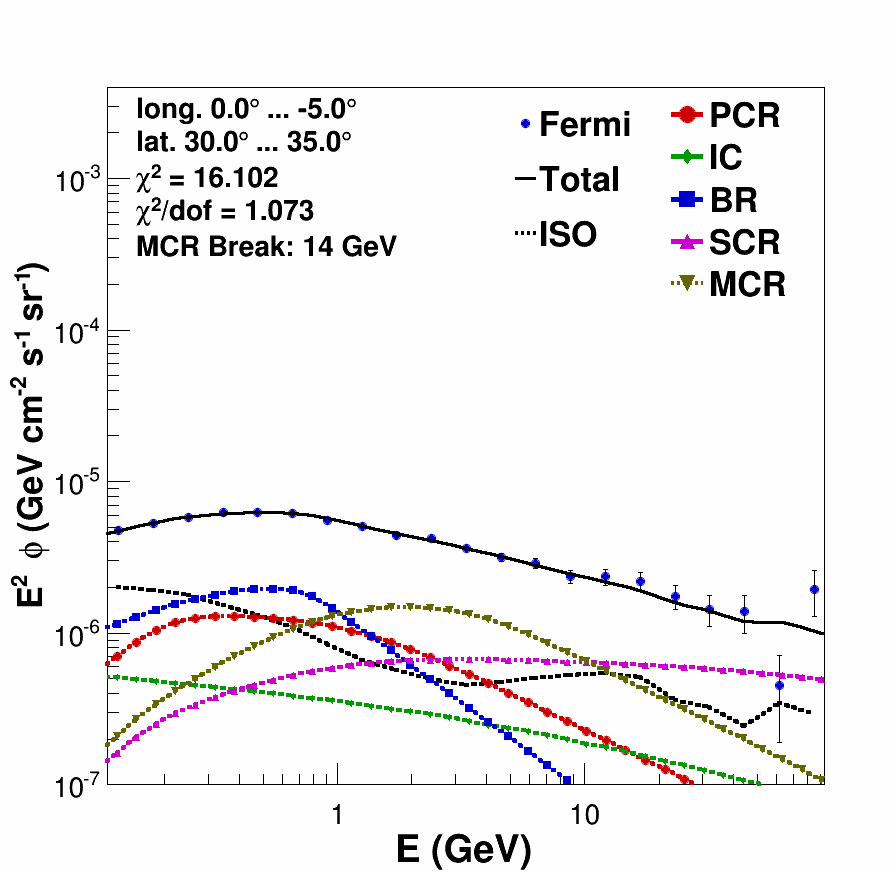}
\includegraphics[width=0.16\textwidth,height=0.16\textwidth,clip]{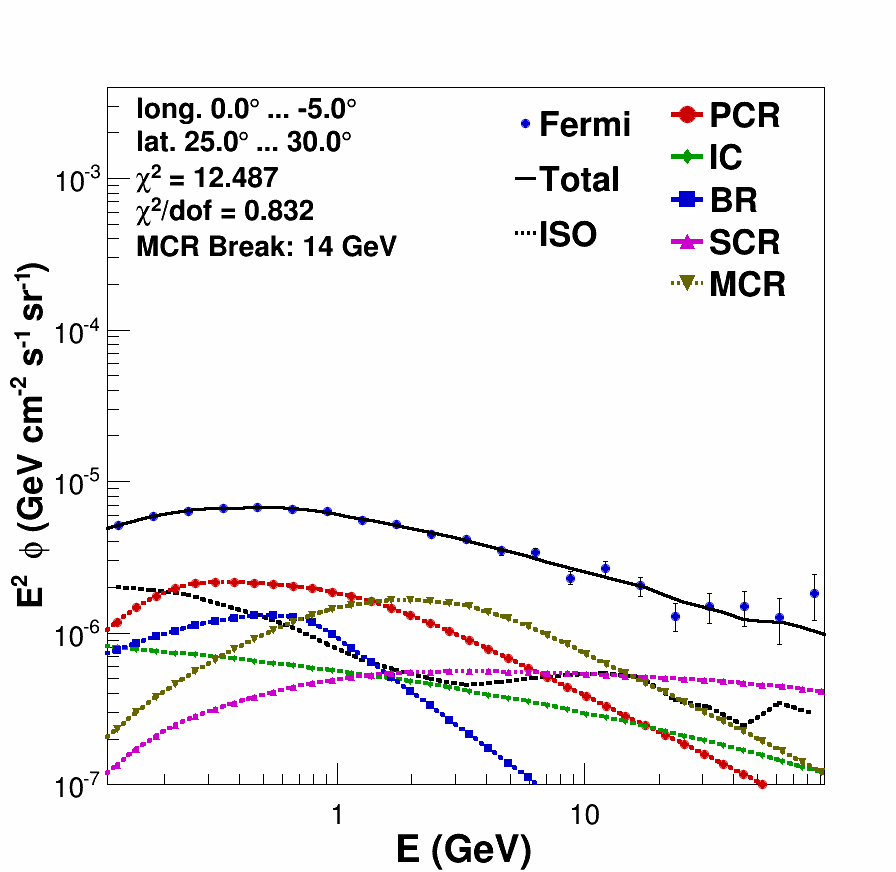}
\includegraphics[width=0.16\textwidth,height=0.16\textwidth,clip]{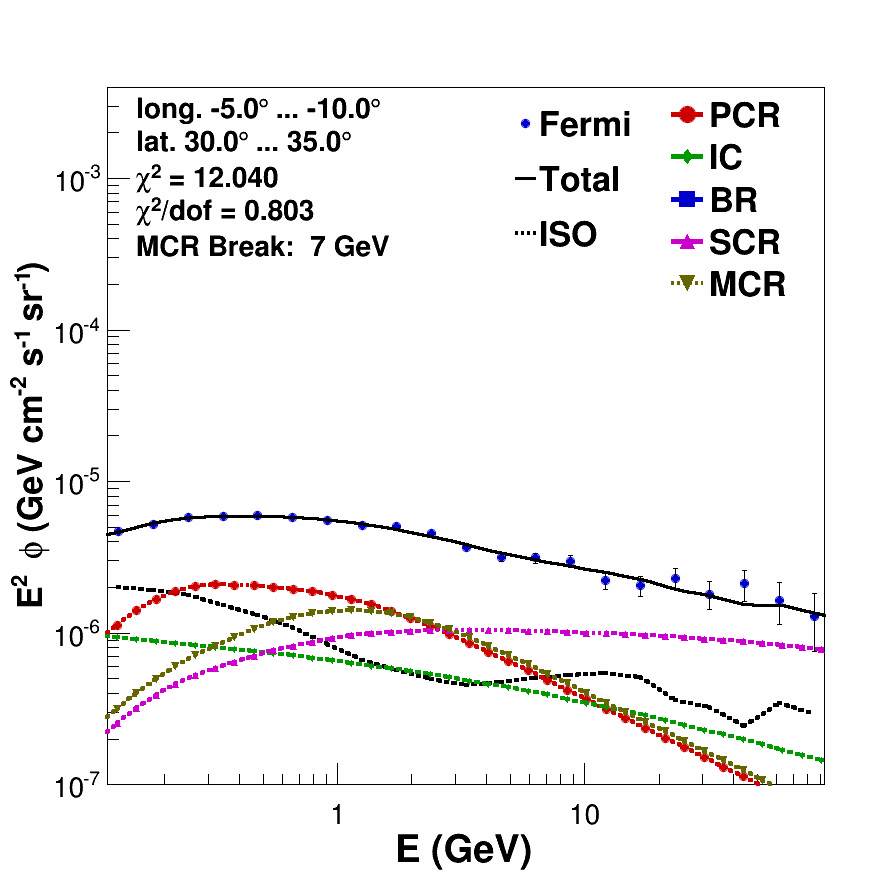}
\includegraphics[width=0.16\textwidth,height=0.16\textwidth,clip]{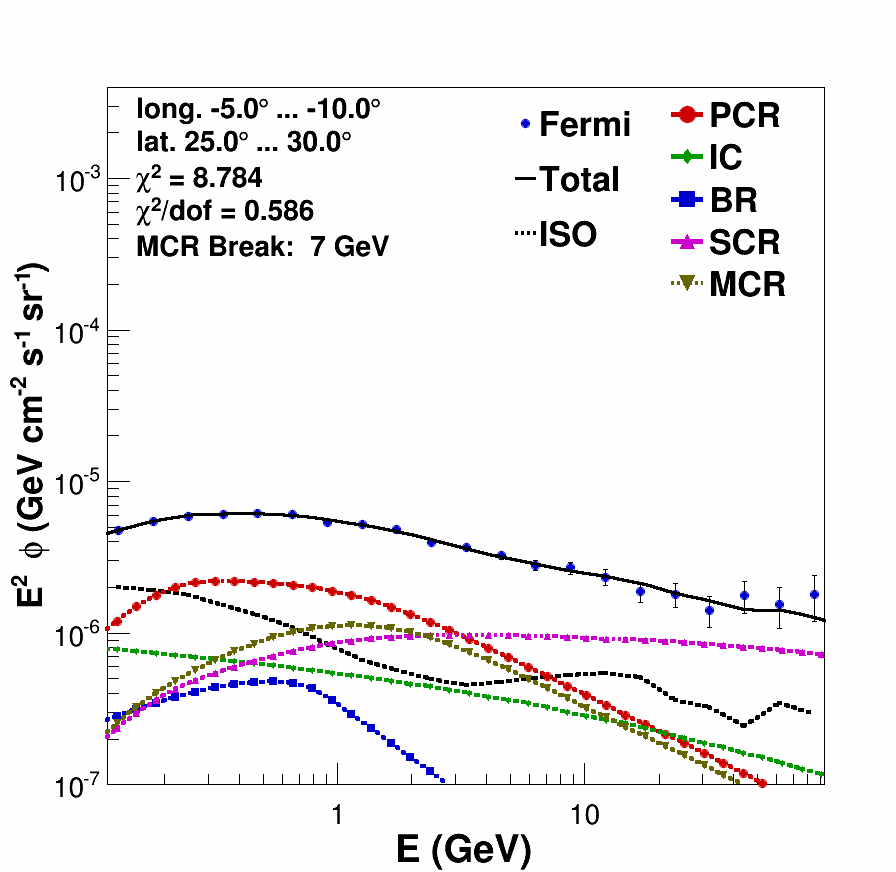}
\includegraphics[width=0.16\textwidth,height=0.16\textwidth,clip]{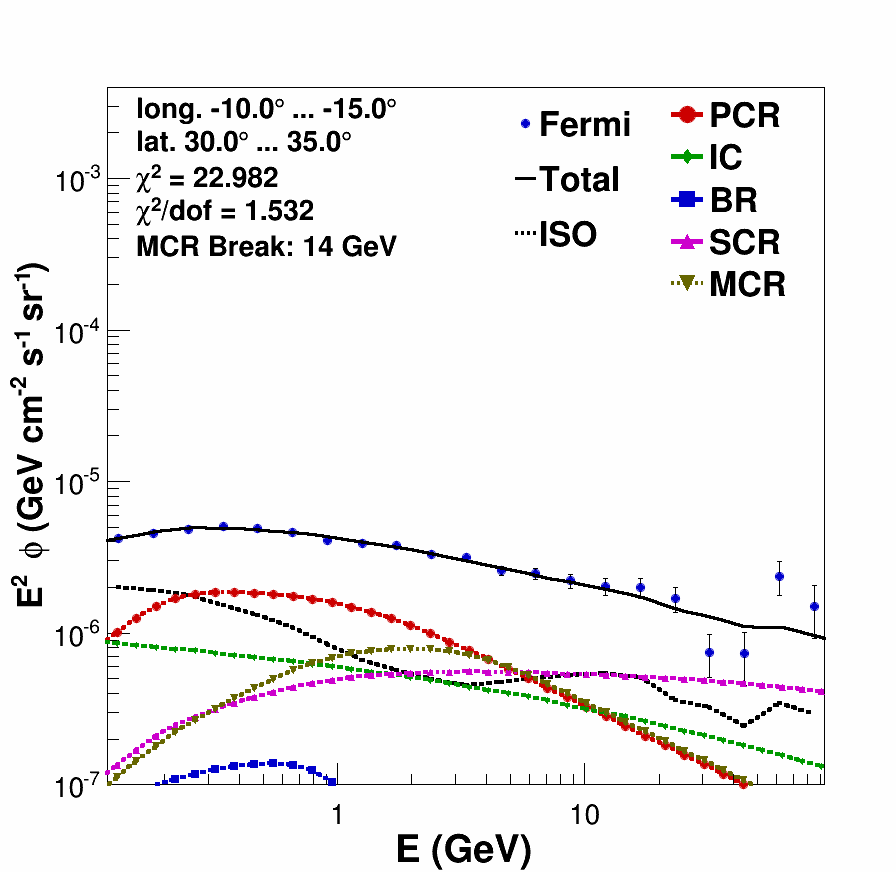}
\includegraphics[width=0.16\textwidth,height=0.16\textwidth,clip]{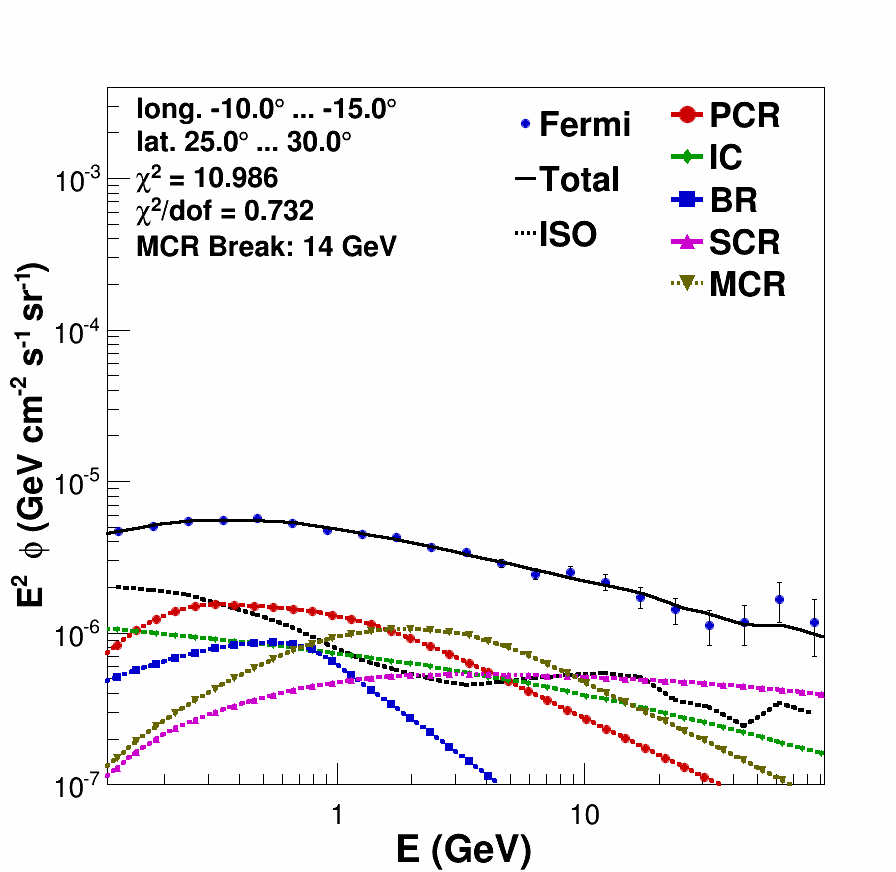}
\includegraphics[width=0.16\textwidth,height=0.16\textwidth,clip]{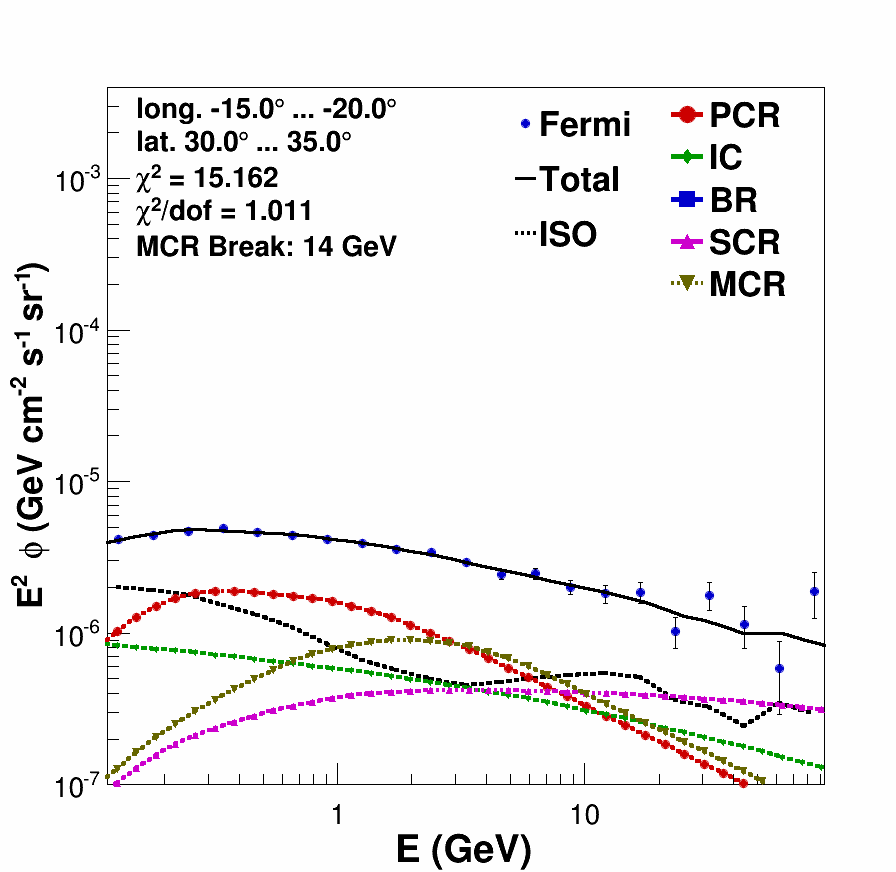}
\includegraphics[width=0.16\textwidth,height=0.16\textwidth,clip]{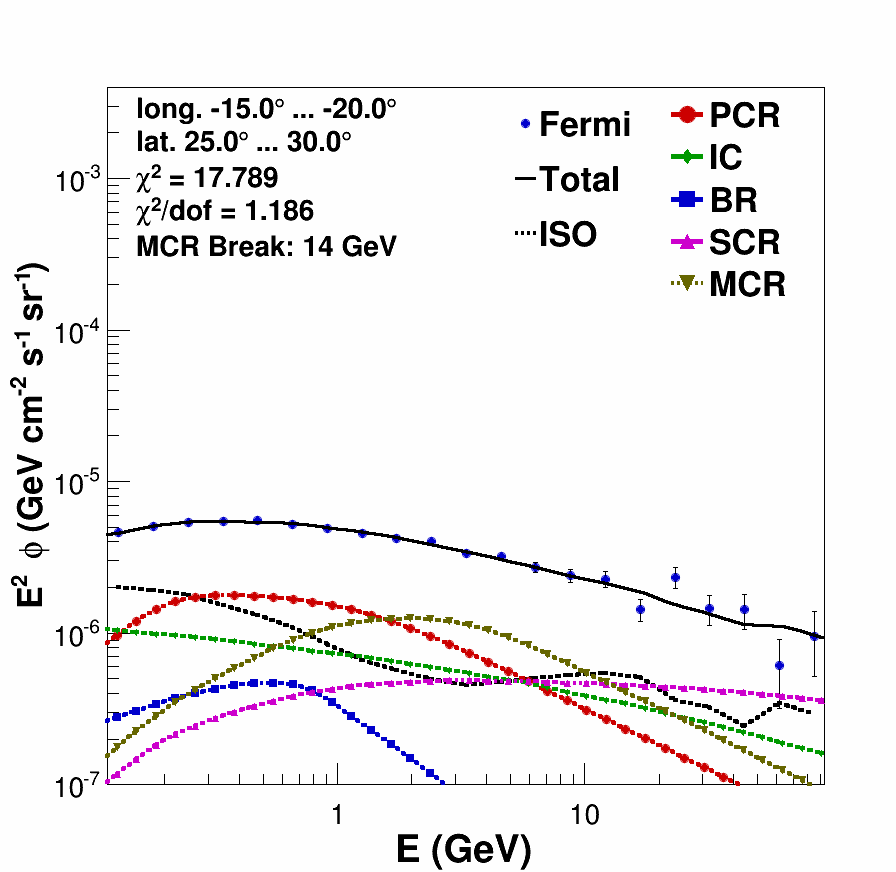}
\includegraphics[width=0.16\textwidth,height=0.16\textwidth,clip]{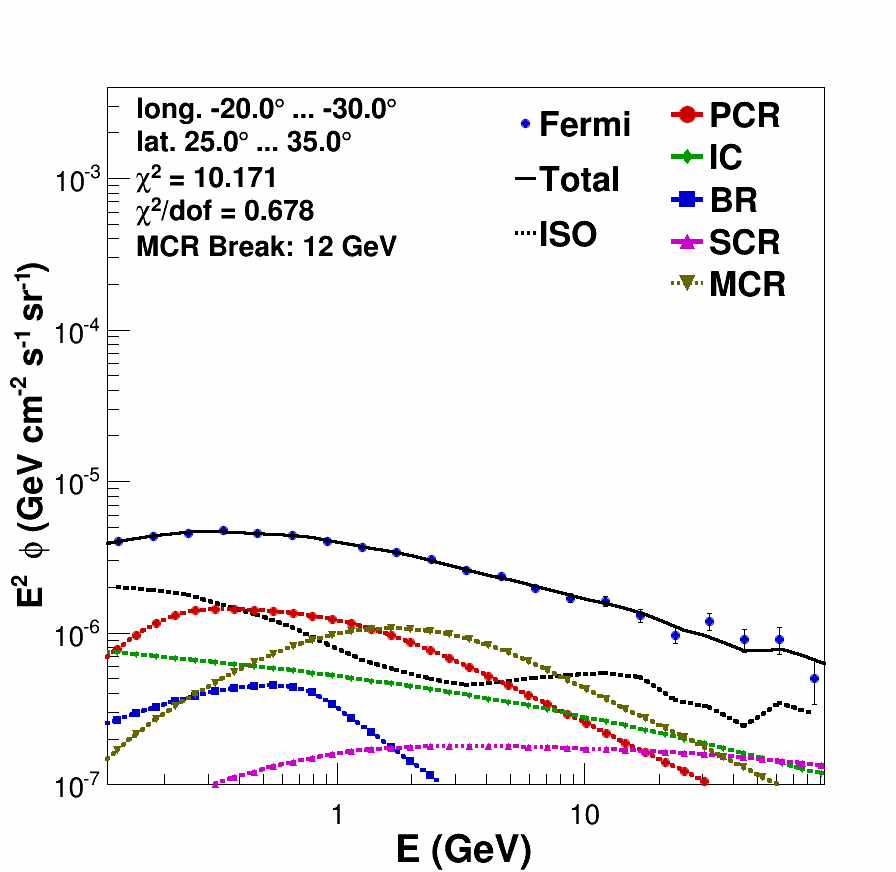}
\includegraphics[width=0.16\textwidth,height=0.16\textwidth,clip]{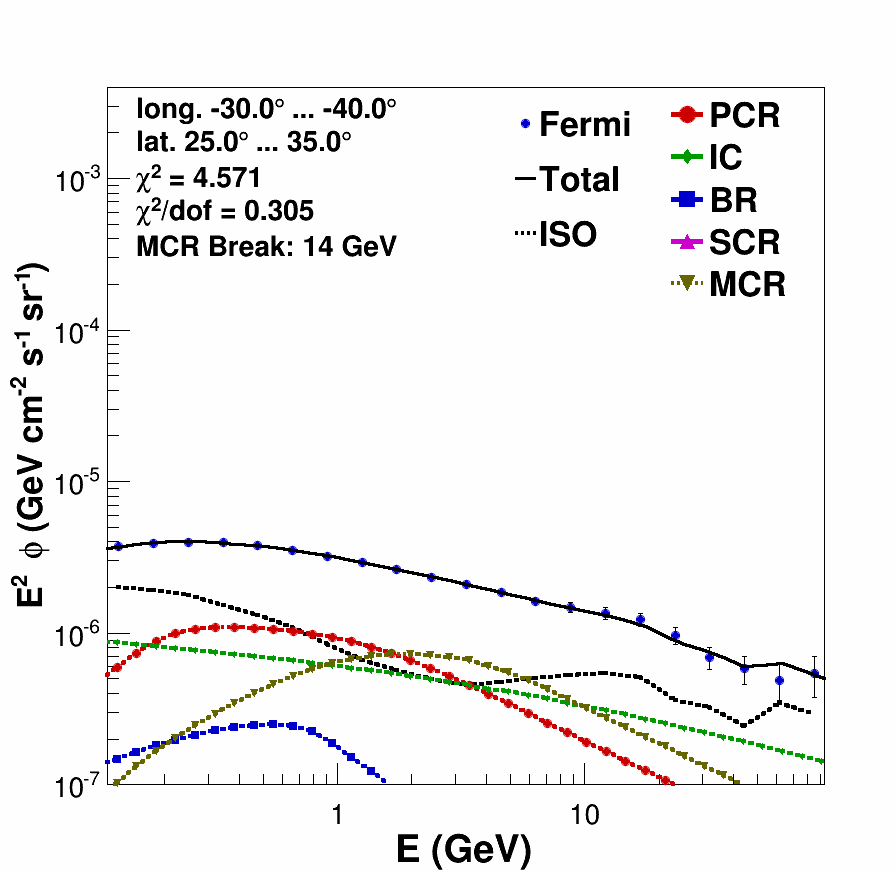}
\includegraphics[width=0.16\textwidth,height=0.16\textwidth,clip]{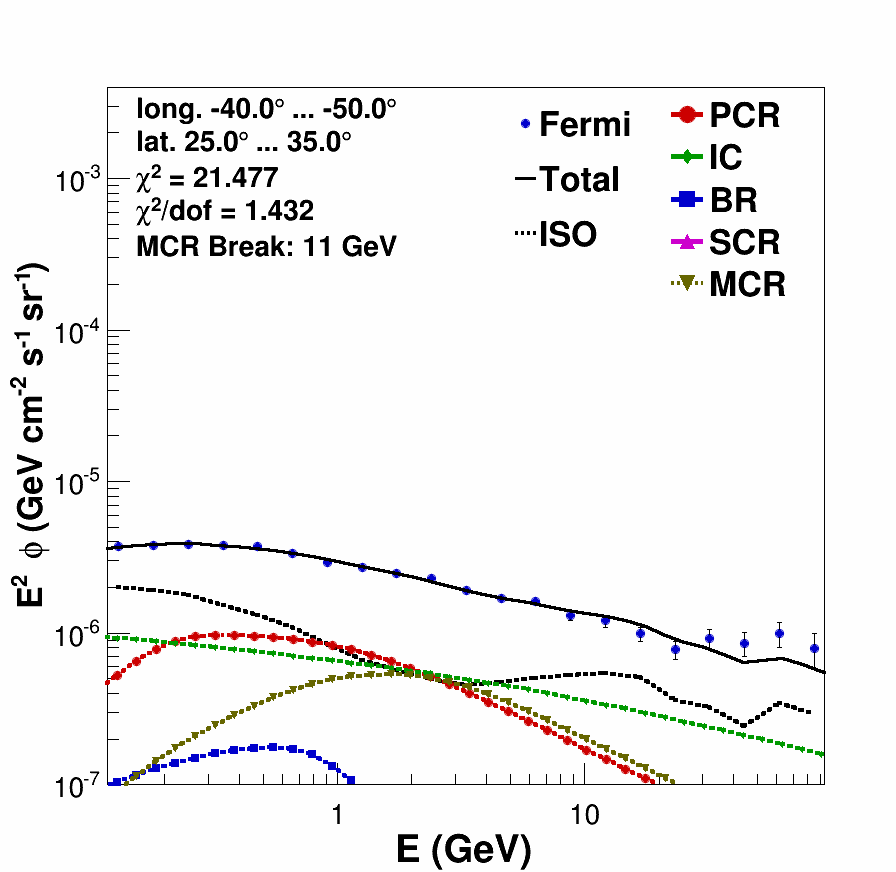}
\includegraphics[width=0.16\textwidth,height=0.16\textwidth,clip]{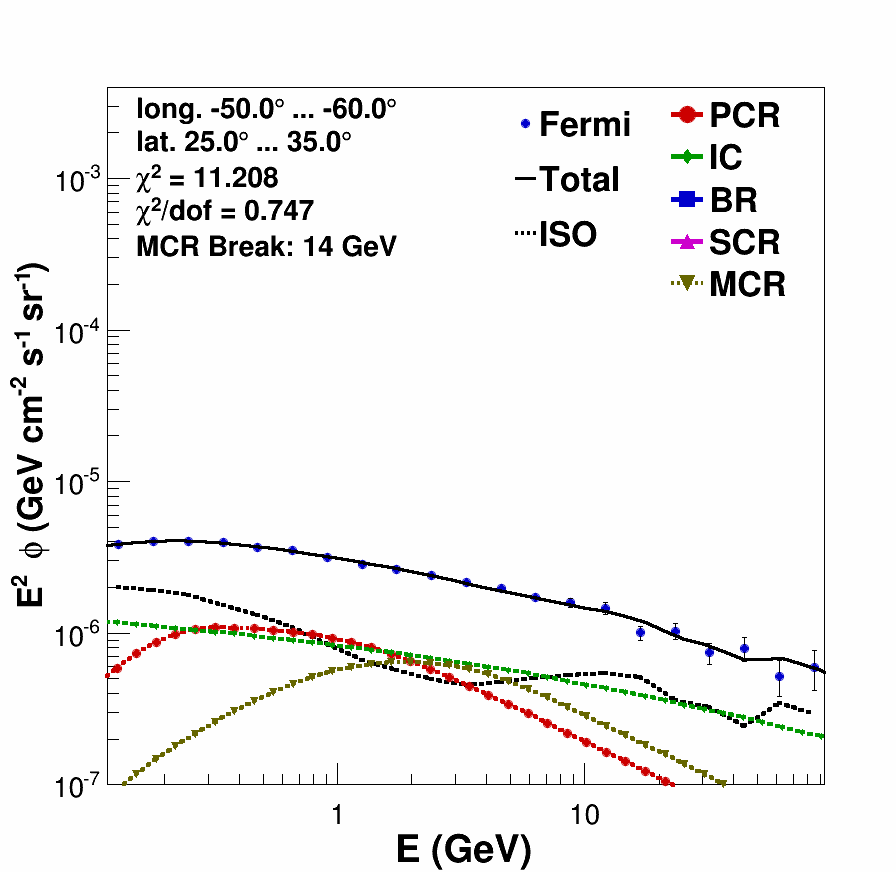}
\includegraphics[width=0.16\textwidth,height=0.16\textwidth,clip]{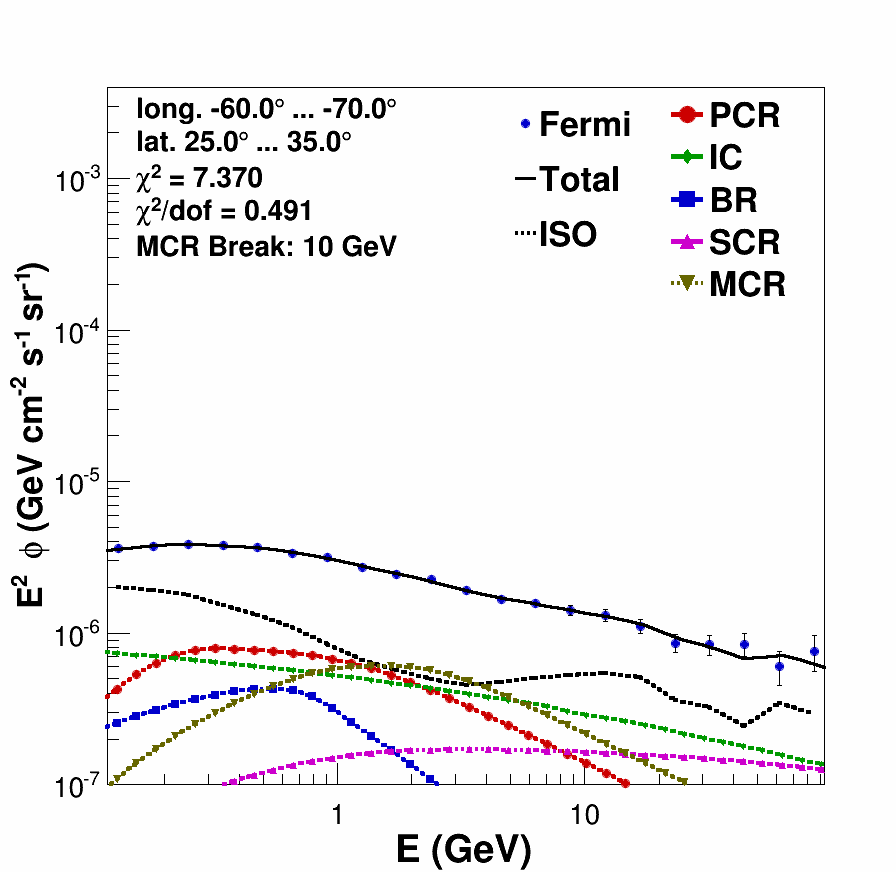}
\includegraphics[width=0.16\textwidth,height=0.16\textwidth,clip]{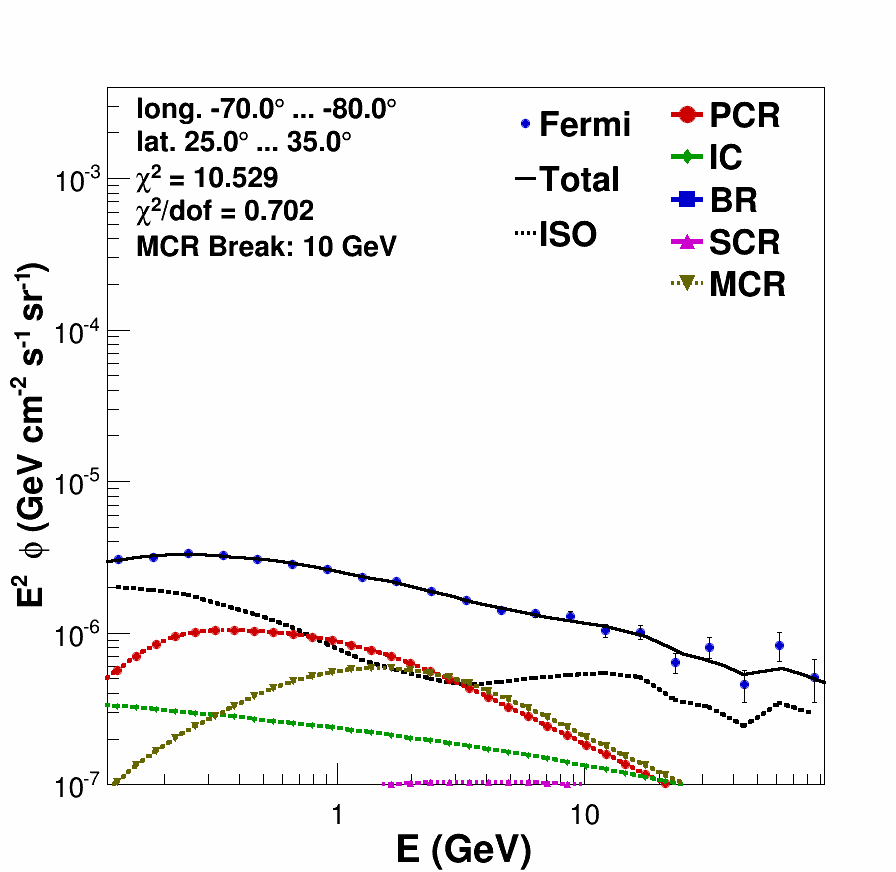}
\includegraphics[width=0.16\textwidth,height=0.16\textwidth,clip]{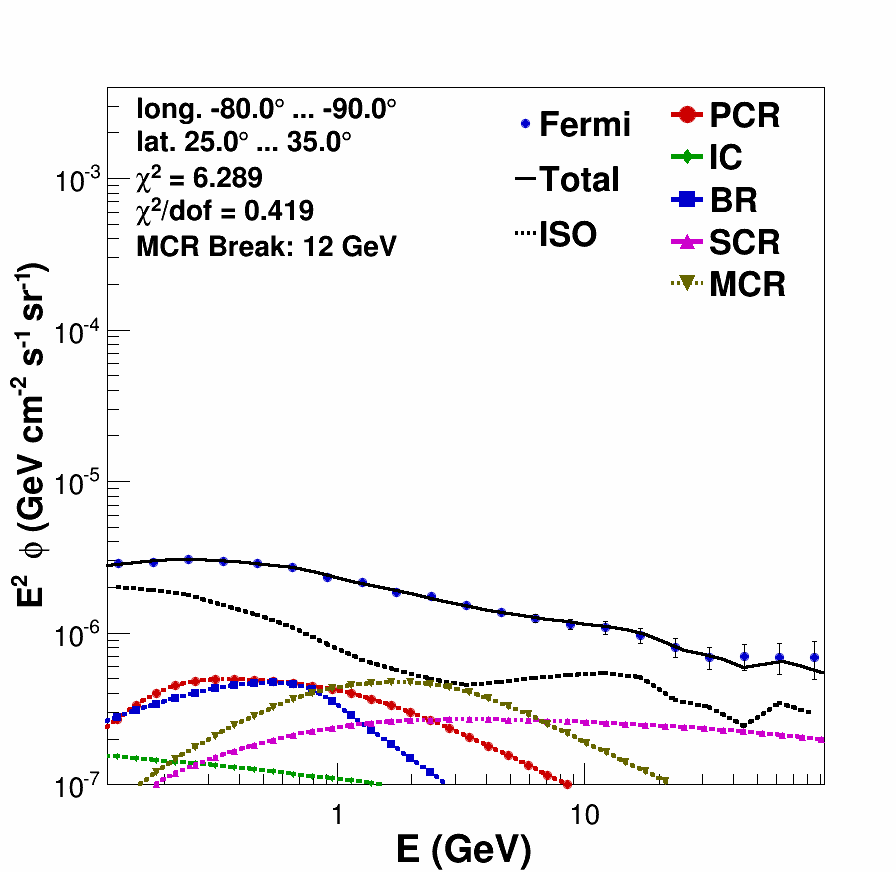}
\includegraphics[width=0.16\textwidth,height=0.16\textwidth,clip]{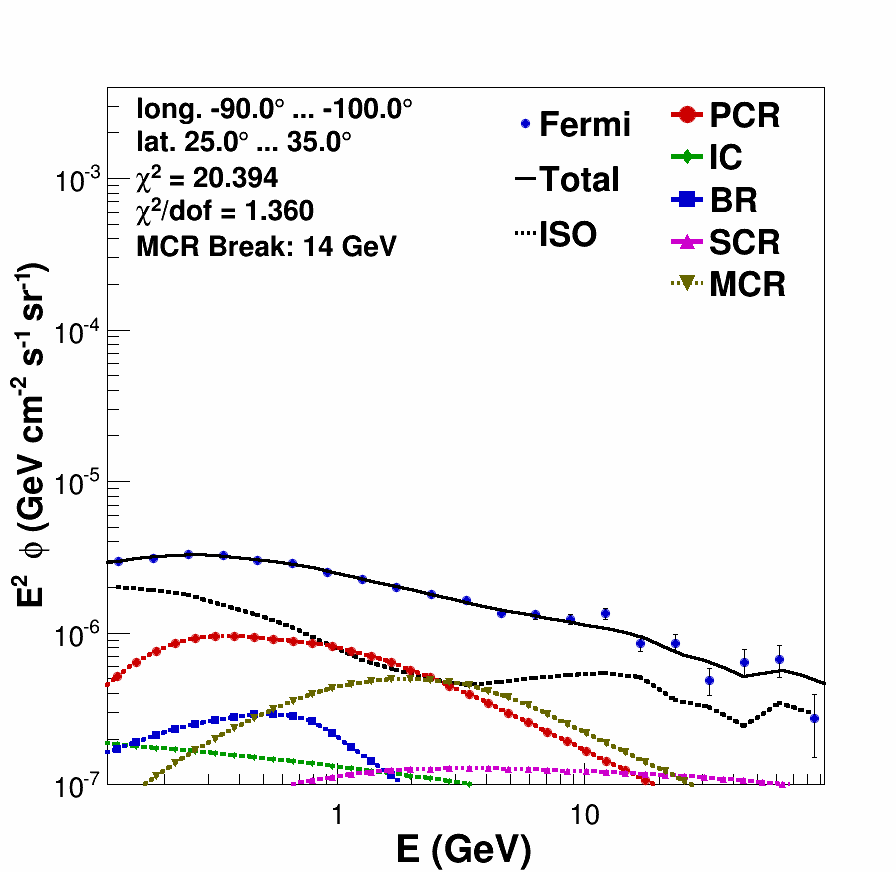}
\includegraphics[width=0.16\textwidth,height=0.16\textwidth,clip]{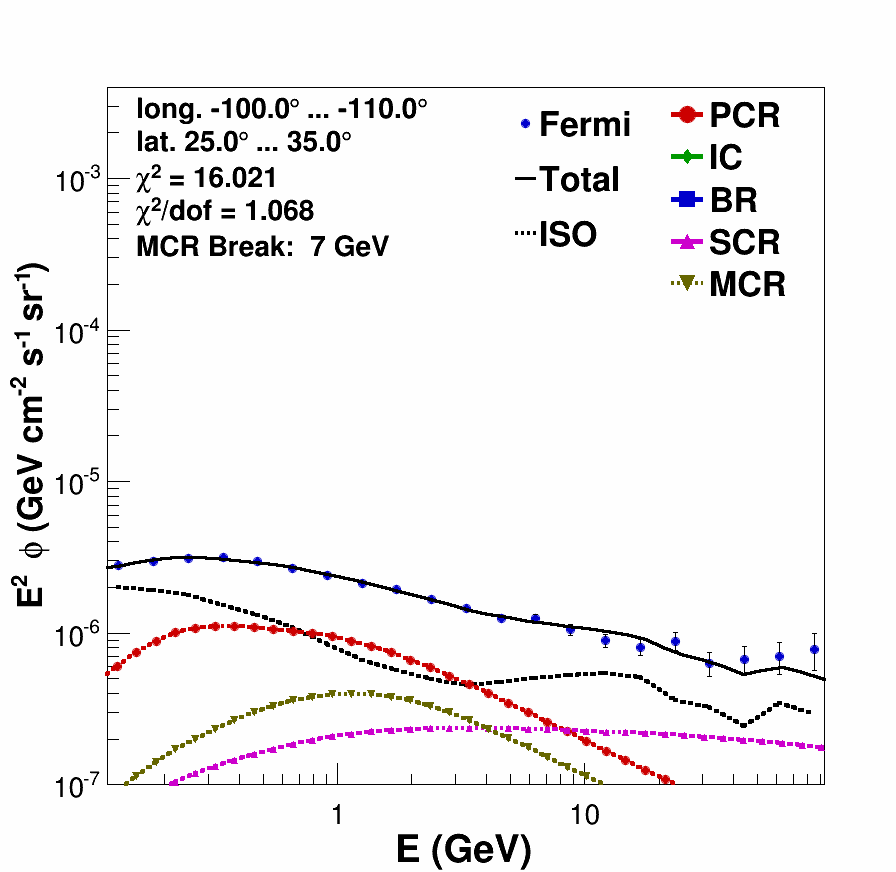}
\includegraphics[width=0.16\textwidth,height=0.16\textwidth,clip]{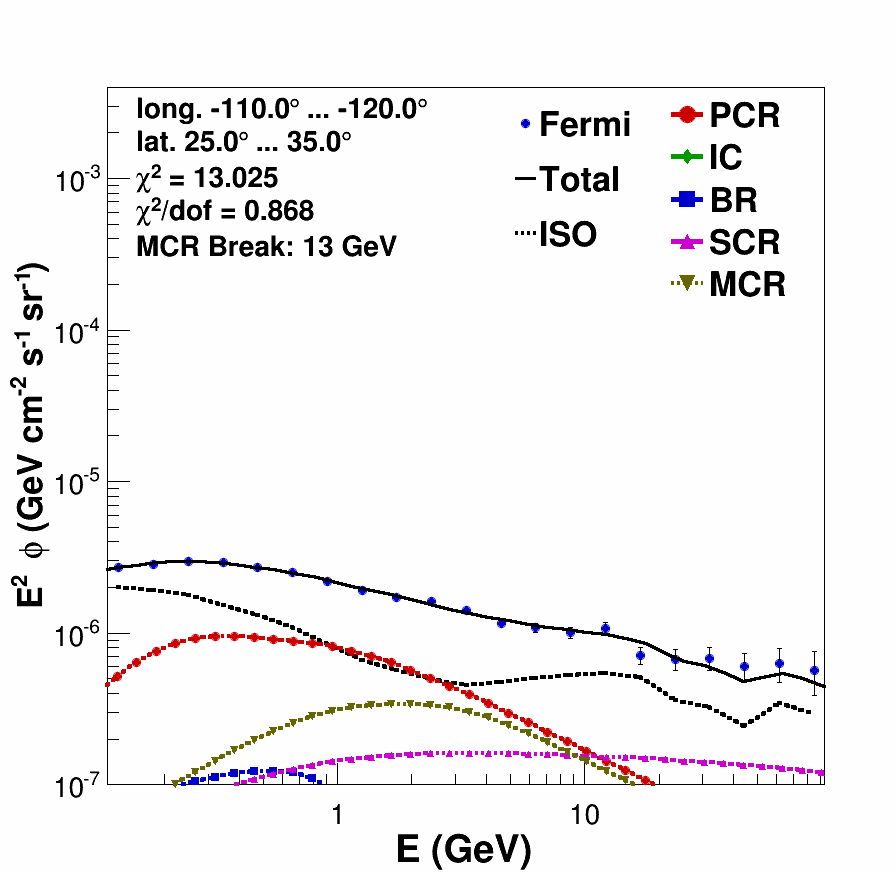}
\includegraphics[width=0.16\textwidth,height=0.16\textwidth,clip]{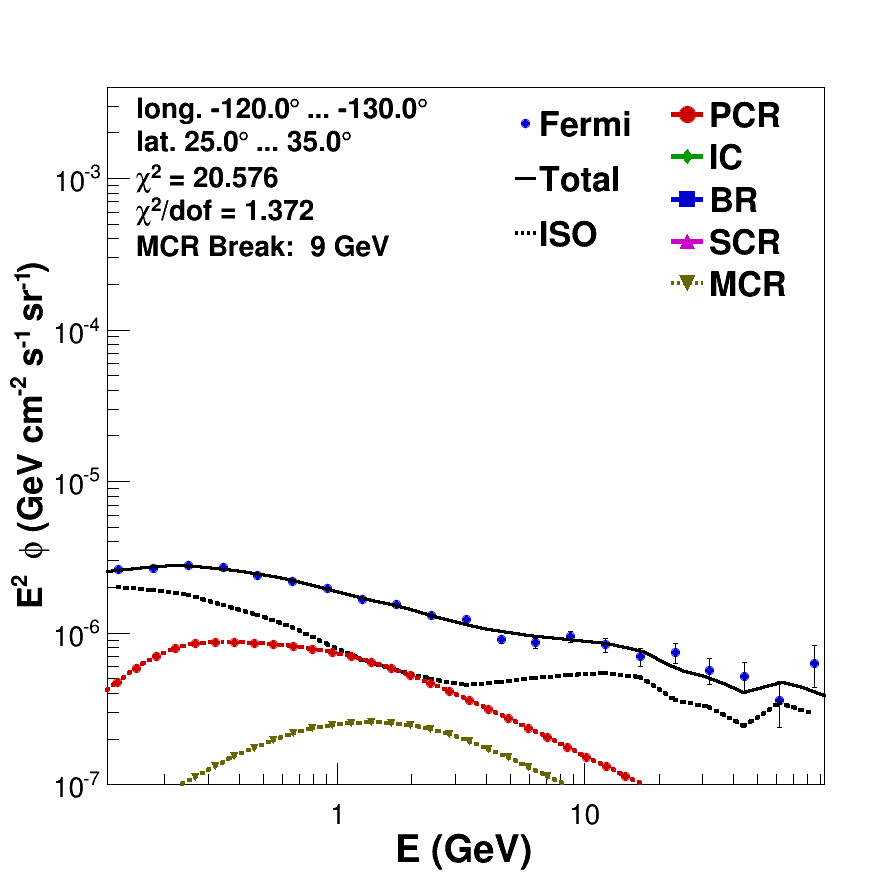}
\includegraphics[width=0.16\textwidth,height=0.16\textwidth,clip]{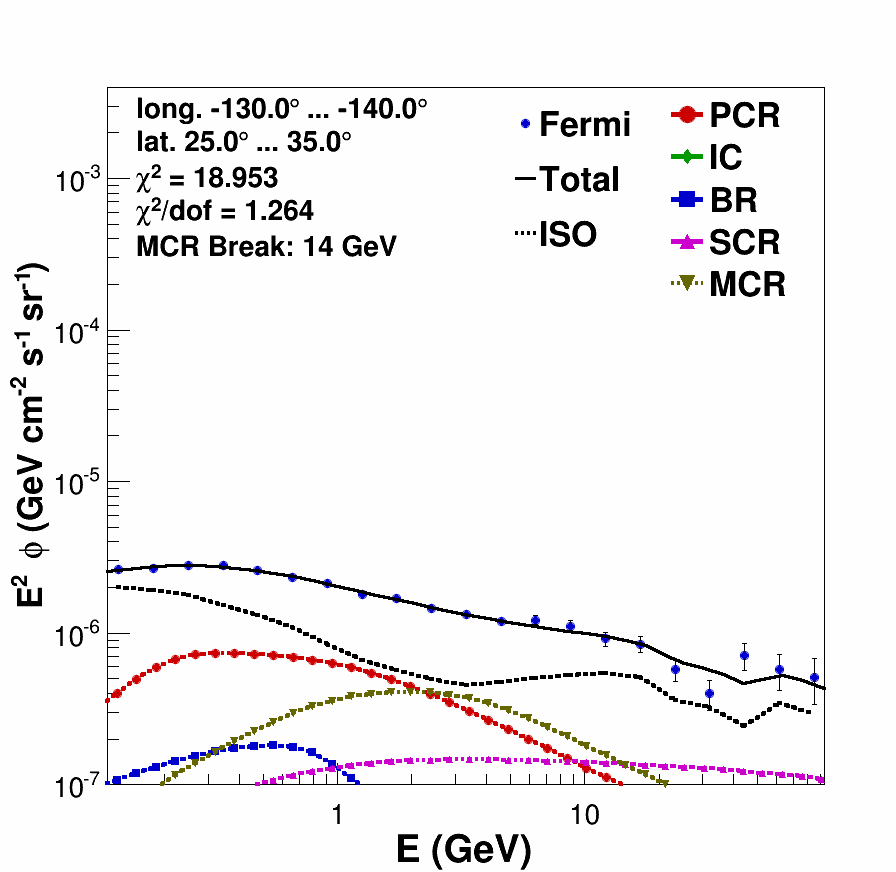}
\includegraphics[width=0.16\textwidth,height=0.16\textwidth,clip]{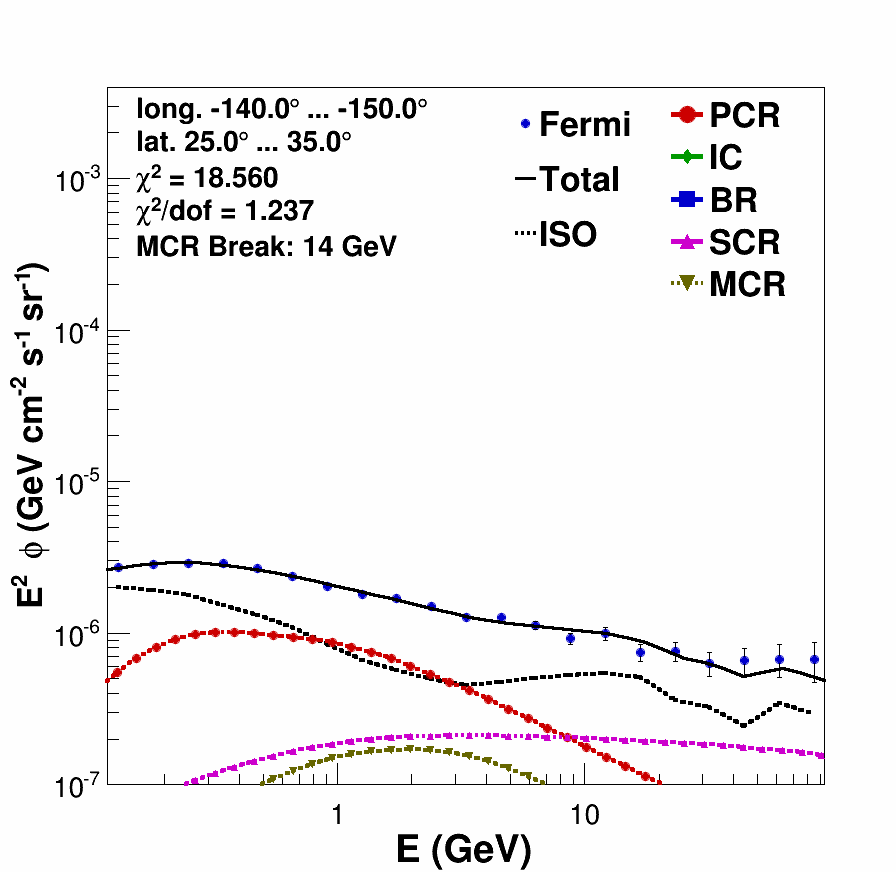}
\includegraphics[width=0.16\textwidth,height=0.16\textwidth,clip]{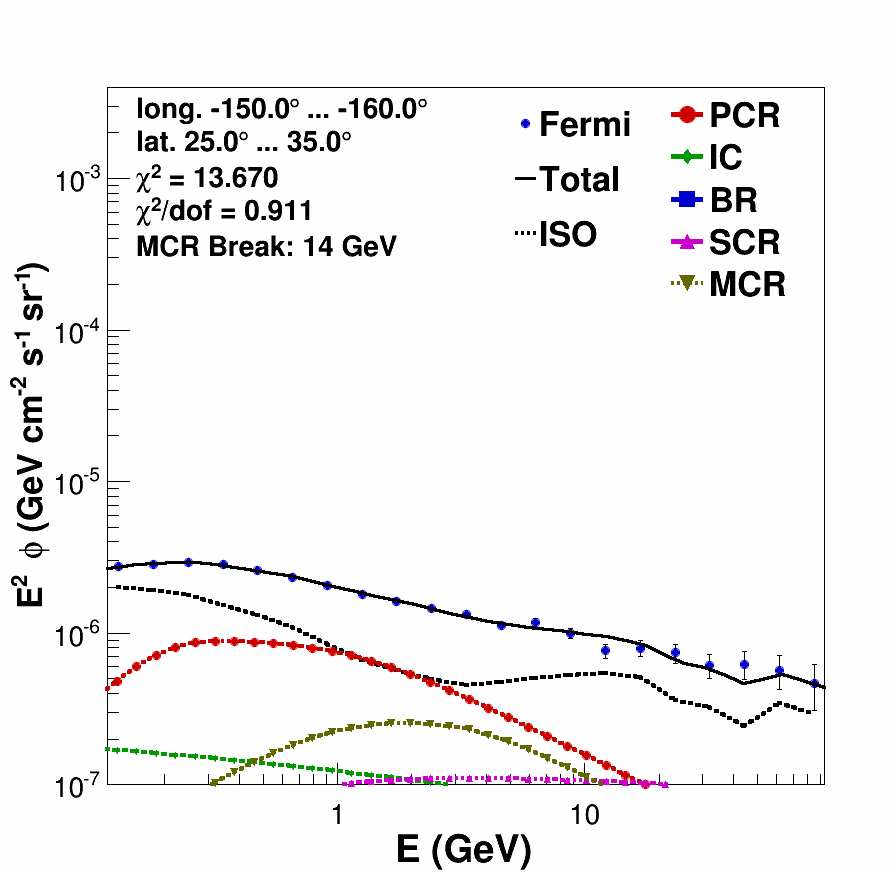}
\includegraphics[width=0.16\textwidth,height=0.16\textwidth,clip]{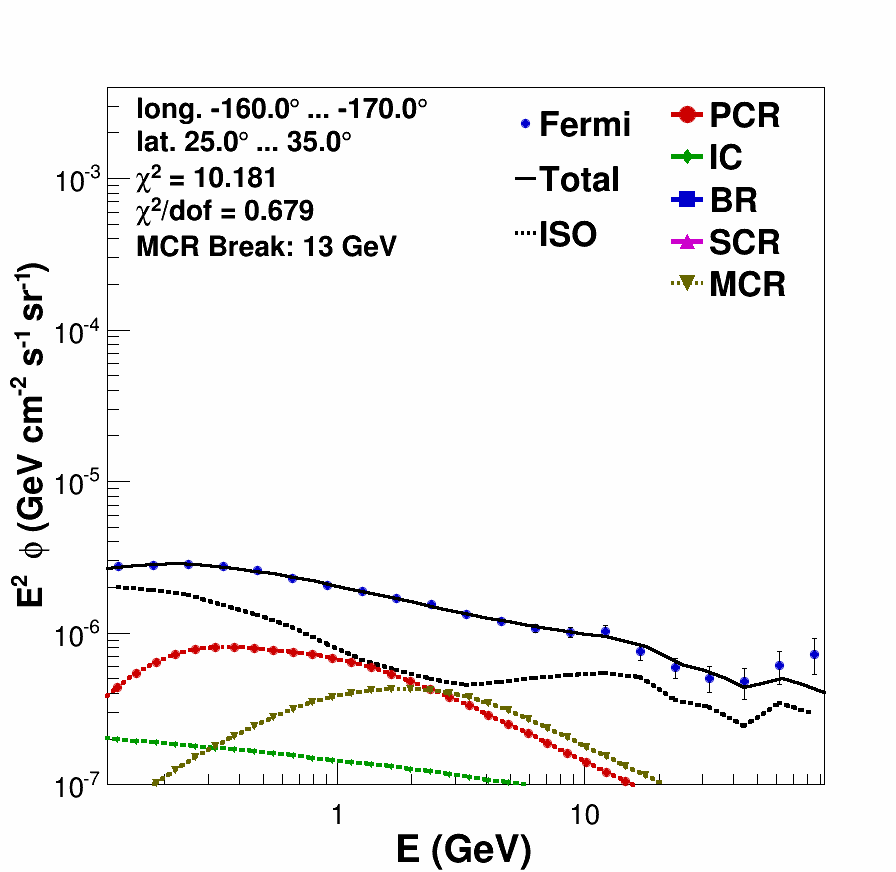}
\includegraphics[width=0.16\textwidth,height=0.16\textwidth,clip]{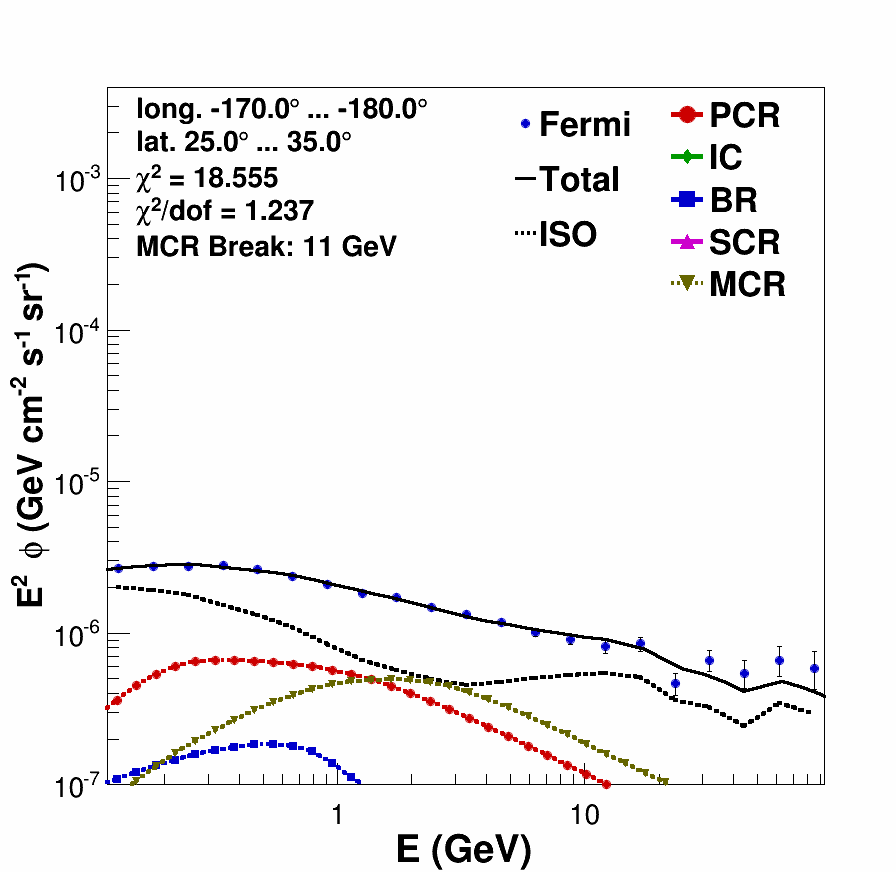}%%%%%r5
\caption[]{Template fits for latitudes  with $25.0^\circ<b<35.0^\circ$ and longitudes decreasing from 180$^\circ$ to -180$^\circ$. \label{F15}
}
\end{figure}
\begin{figure}
\centering
\includegraphics[width=0.16\textwidth,height=0.16\textwidth,clip]{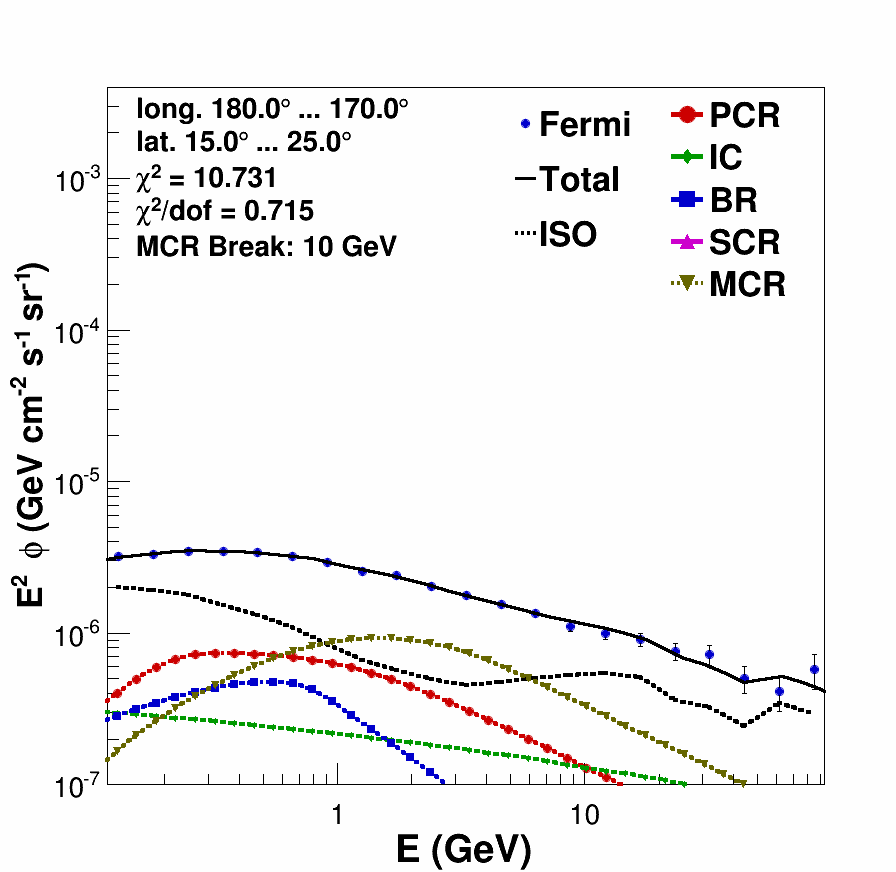}
\includegraphics[width=0.16\textwidth,height=0.16\textwidth,clip]{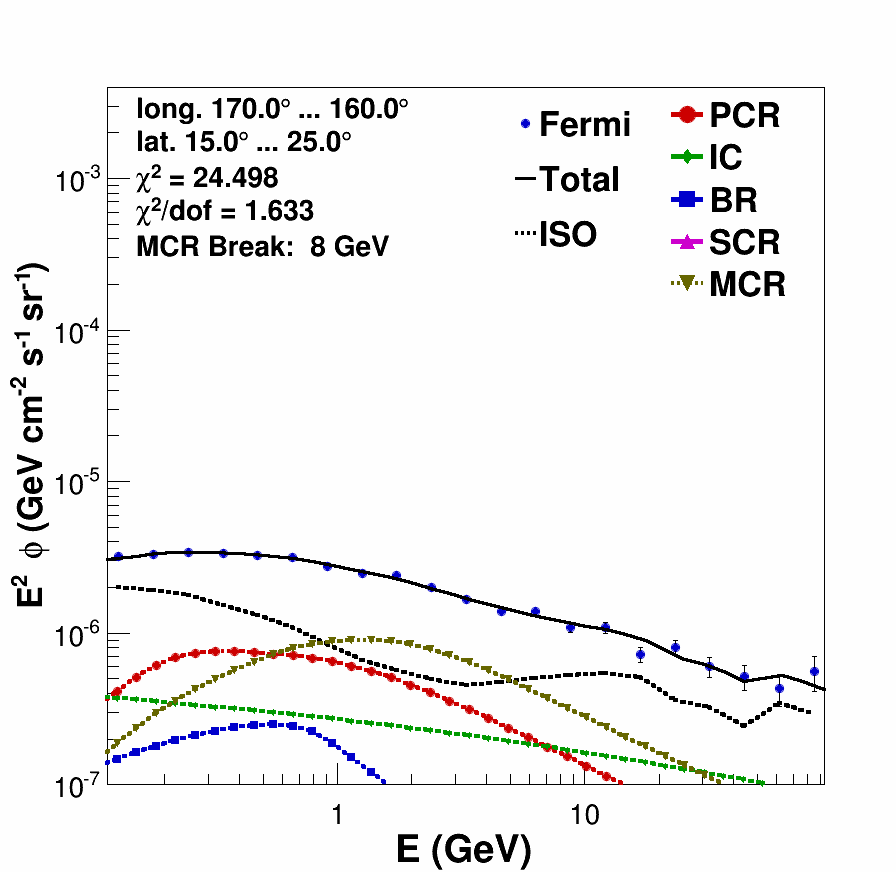}
\includegraphics[width=0.16\textwidth,height=0.16\textwidth,clip]{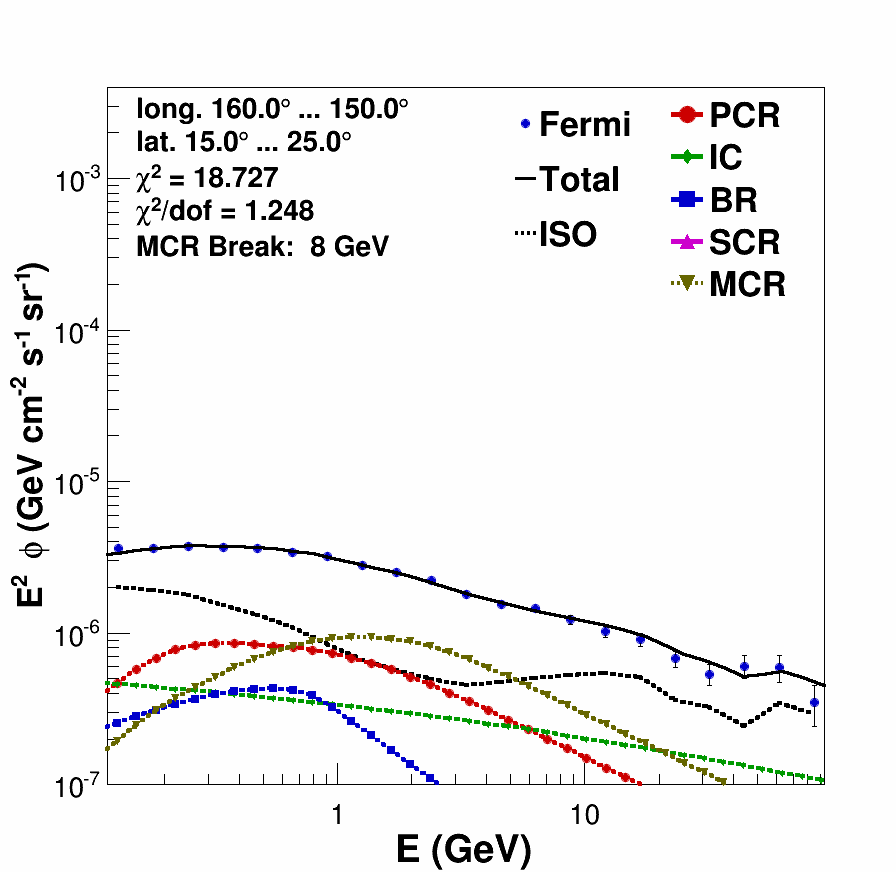}
\includegraphics[width=0.16\textwidth,height=0.16\textwidth,clip]{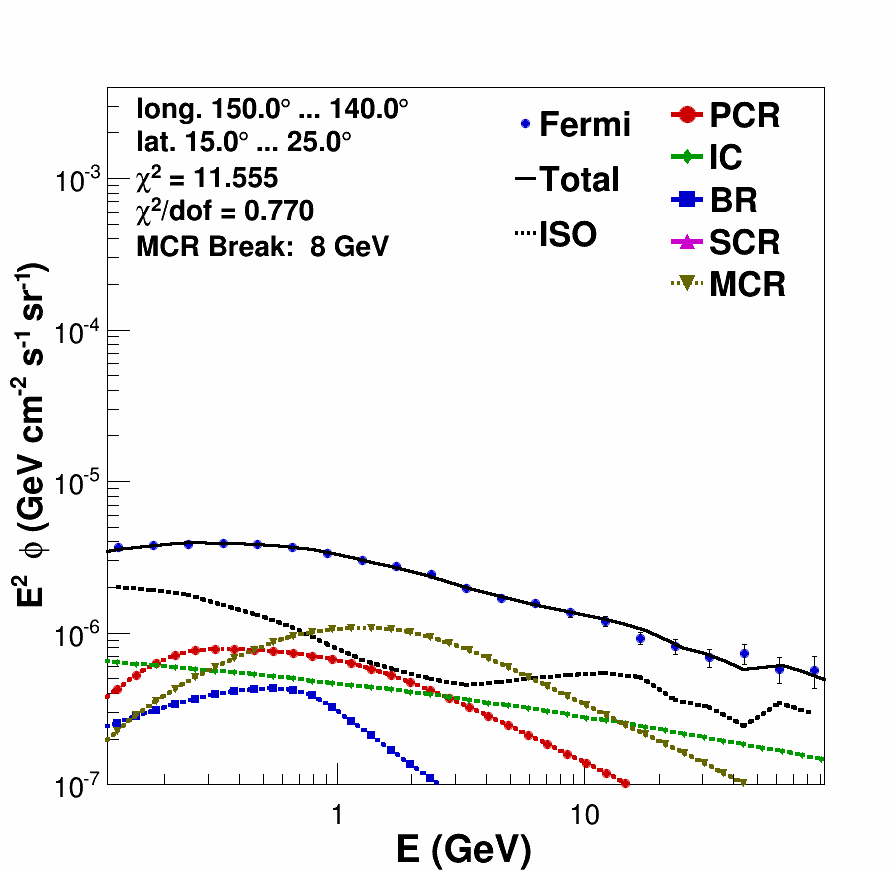}
\includegraphics[width=0.16\textwidth,height=0.16\textwidth,clip]{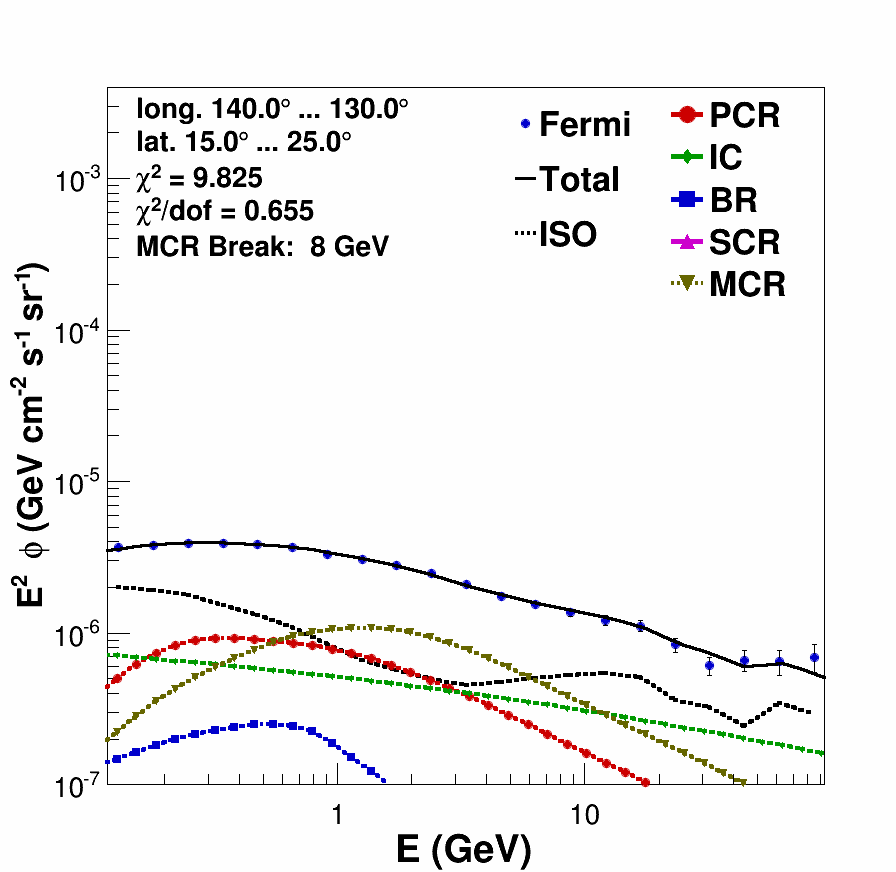}
\includegraphics[width=0.16\textwidth,height=0.16\textwidth,clip]{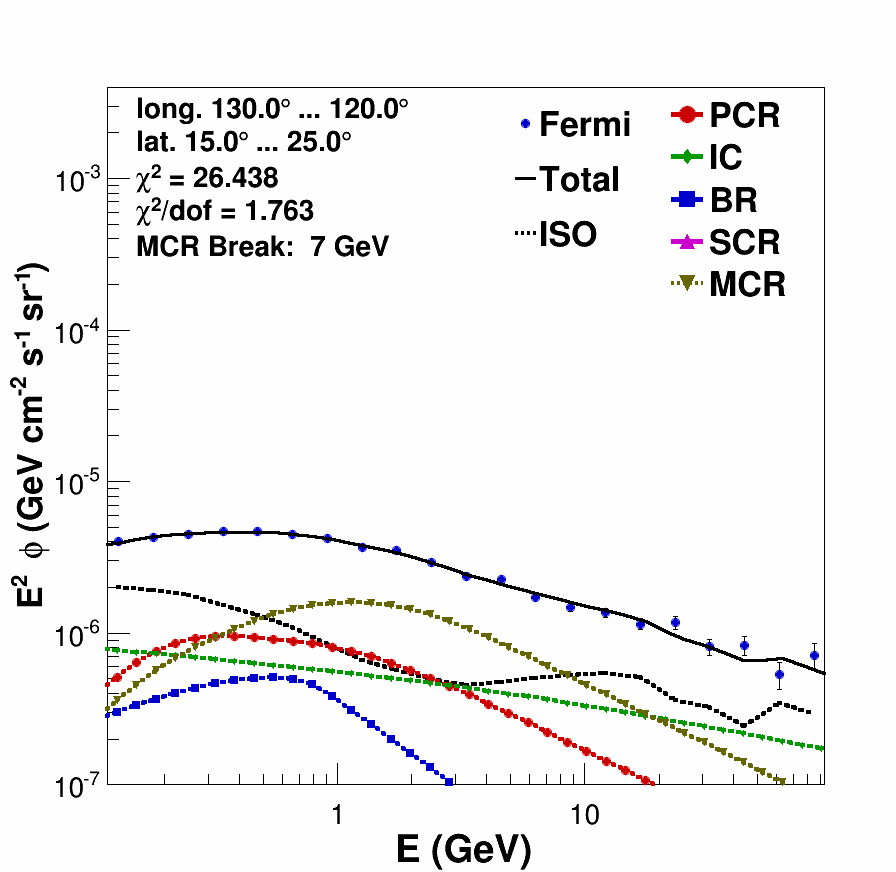}
\includegraphics[width=0.16\textwidth,height=0.16\textwidth,clip]{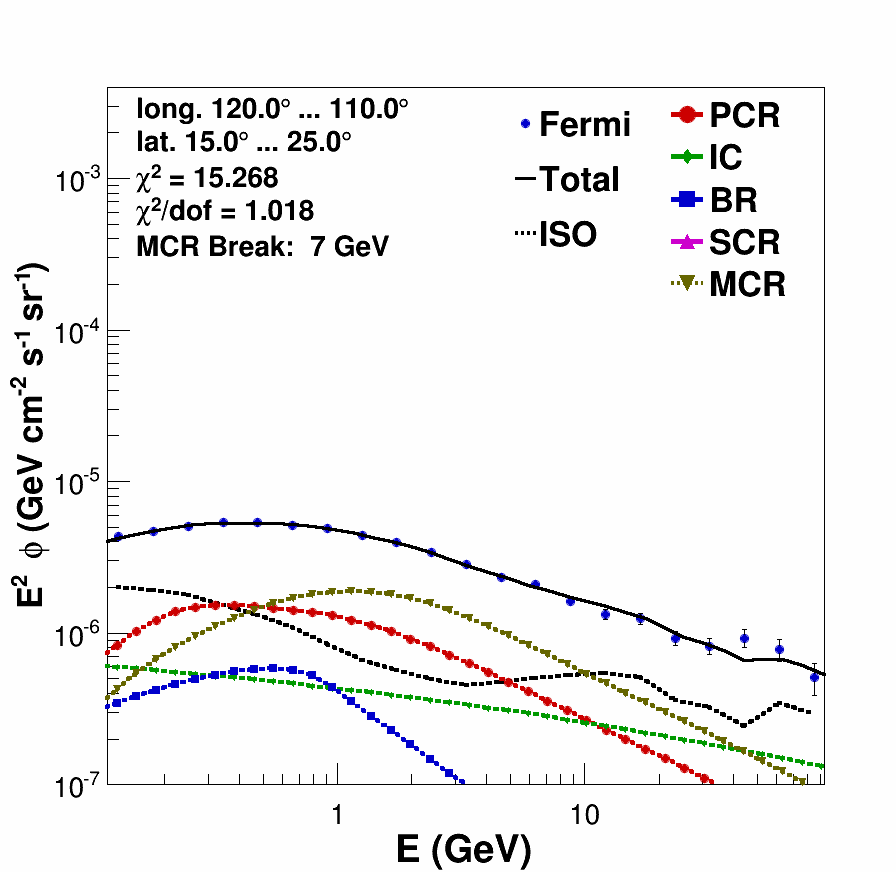}
\includegraphics[width=0.16\textwidth,height=0.16\textwidth,clip]{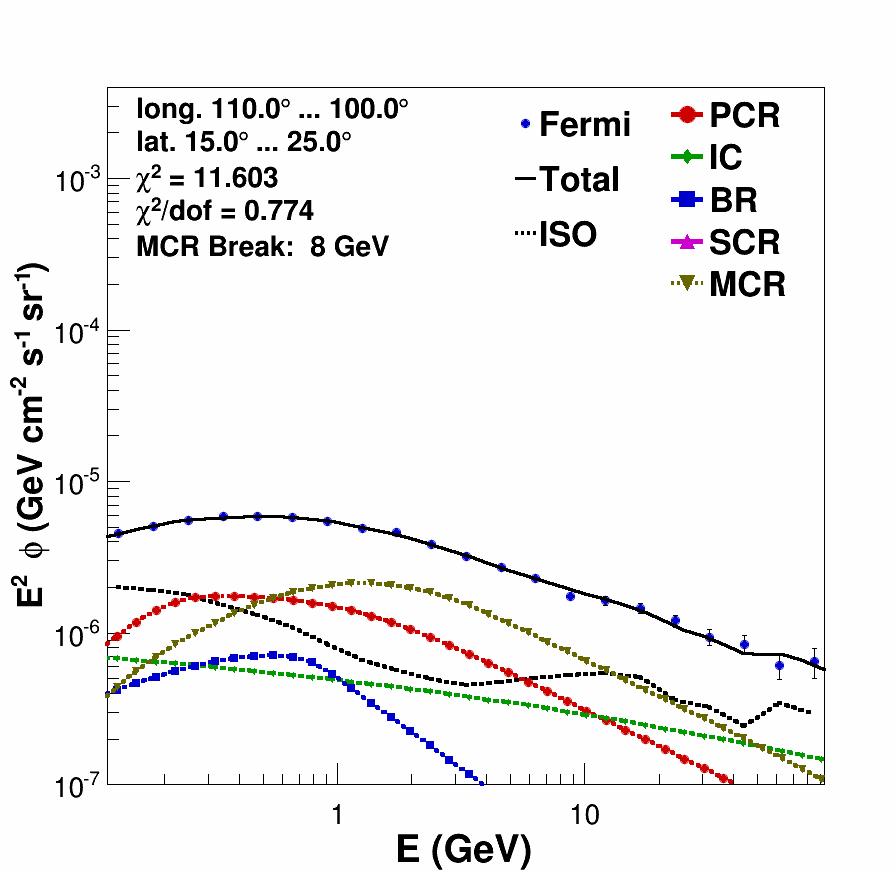}
\includegraphics[width=0.16\textwidth,height=0.16\textwidth,clip]{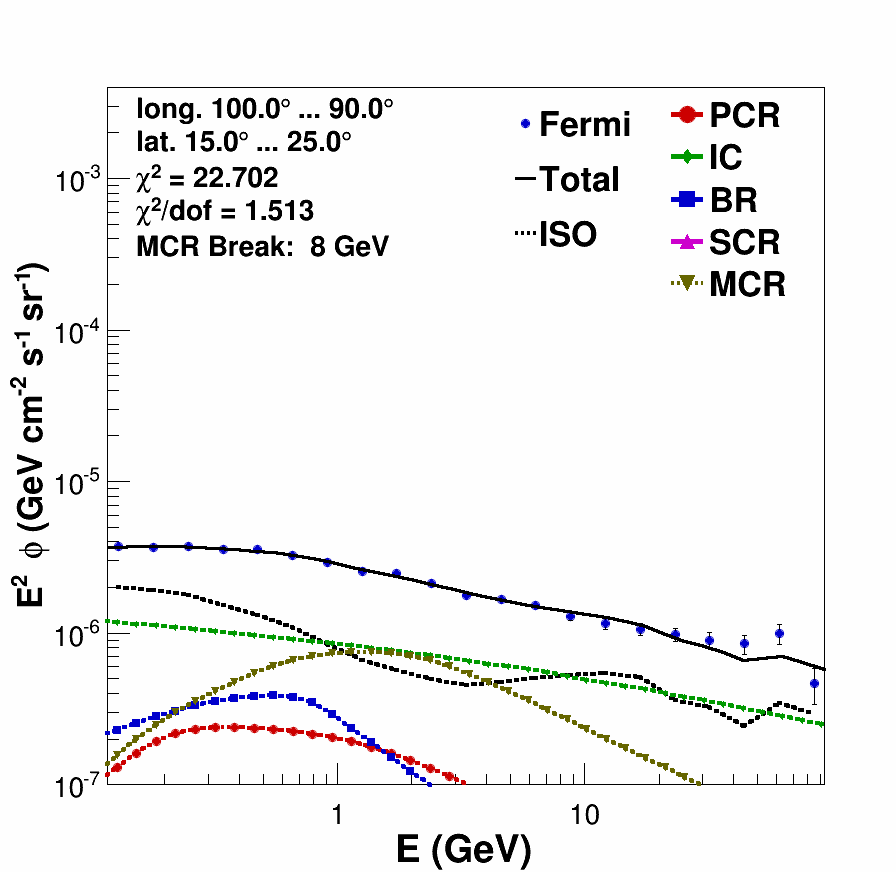}
\includegraphics[width=0.16\textwidth,height=0.16\textwidth,clip]{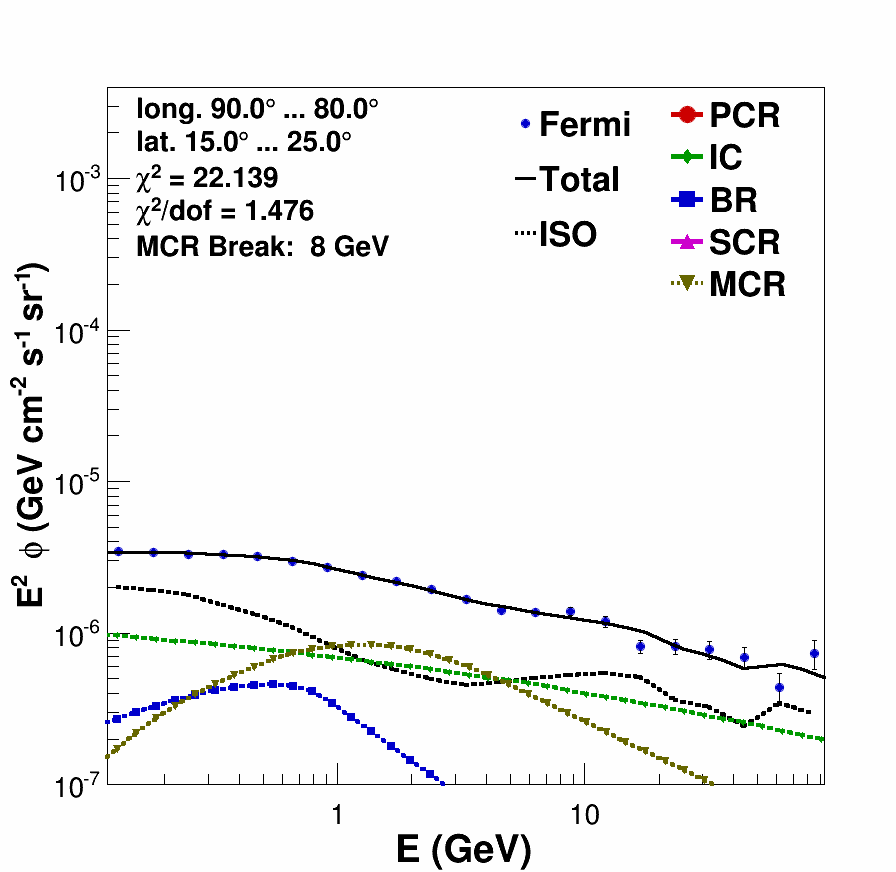}
\includegraphics[width=0.16\textwidth,height=0.16\textwidth,clip]{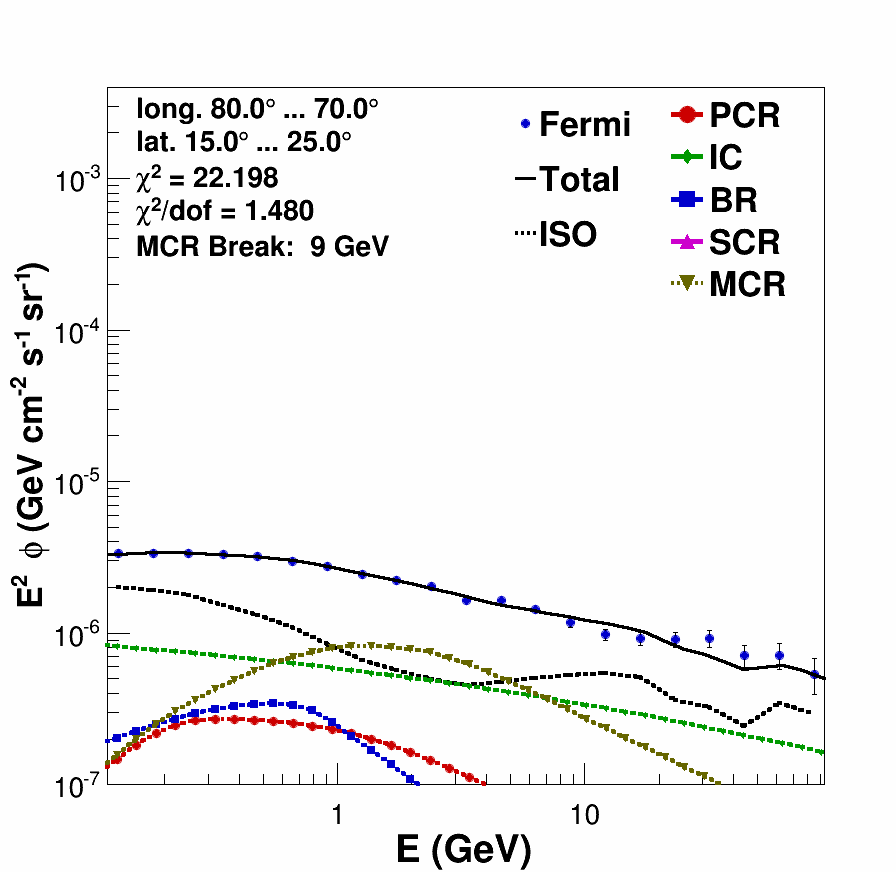}
\includegraphics[width=0.16\textwidth,height=0.16\textwidth,clip]{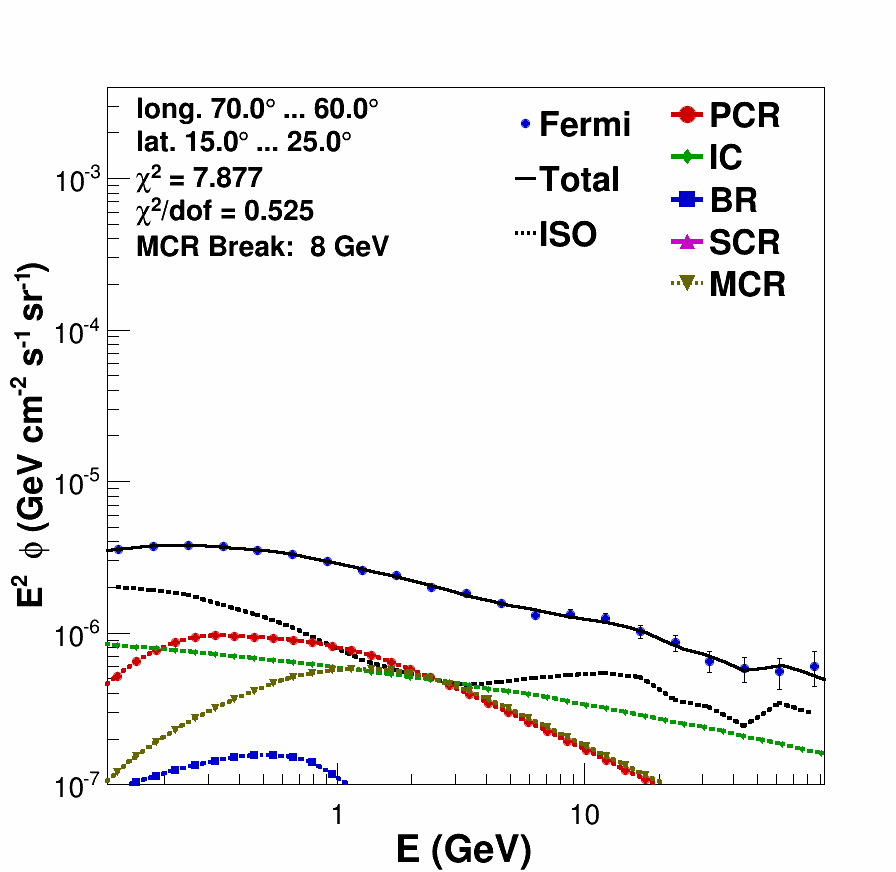}
\includegraphics[width=0.16\textwidth,height=0.16\textwidth,clip]{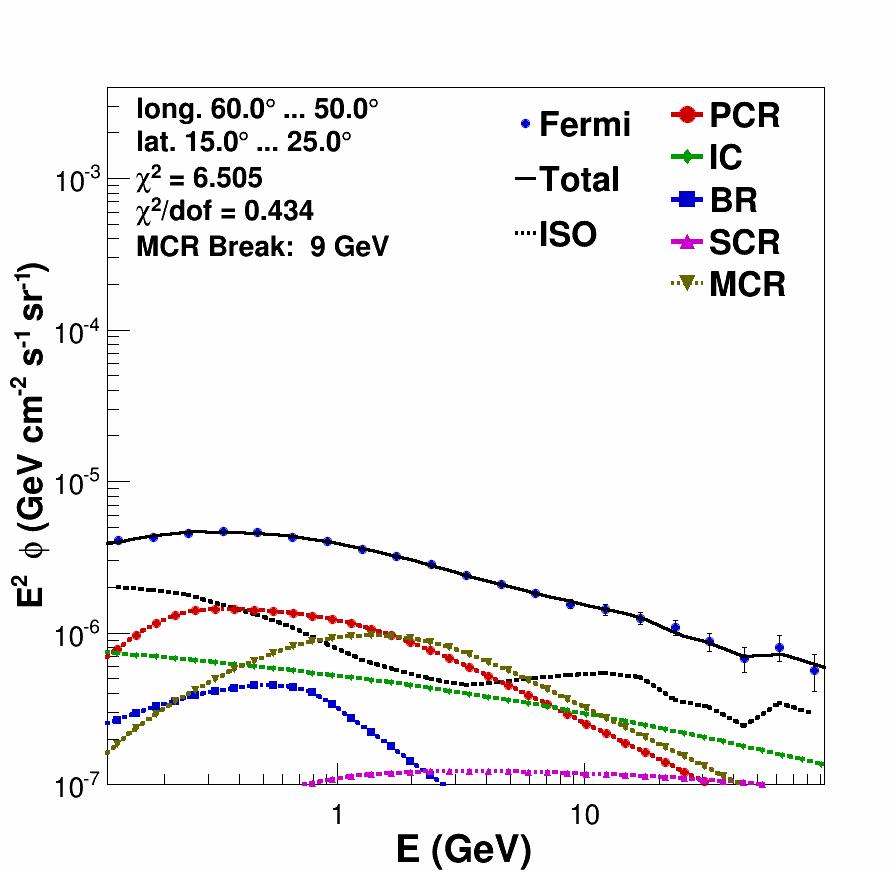}
\includegraphics[width=0.16\textwidth,height=0.16\textwidth,clip]{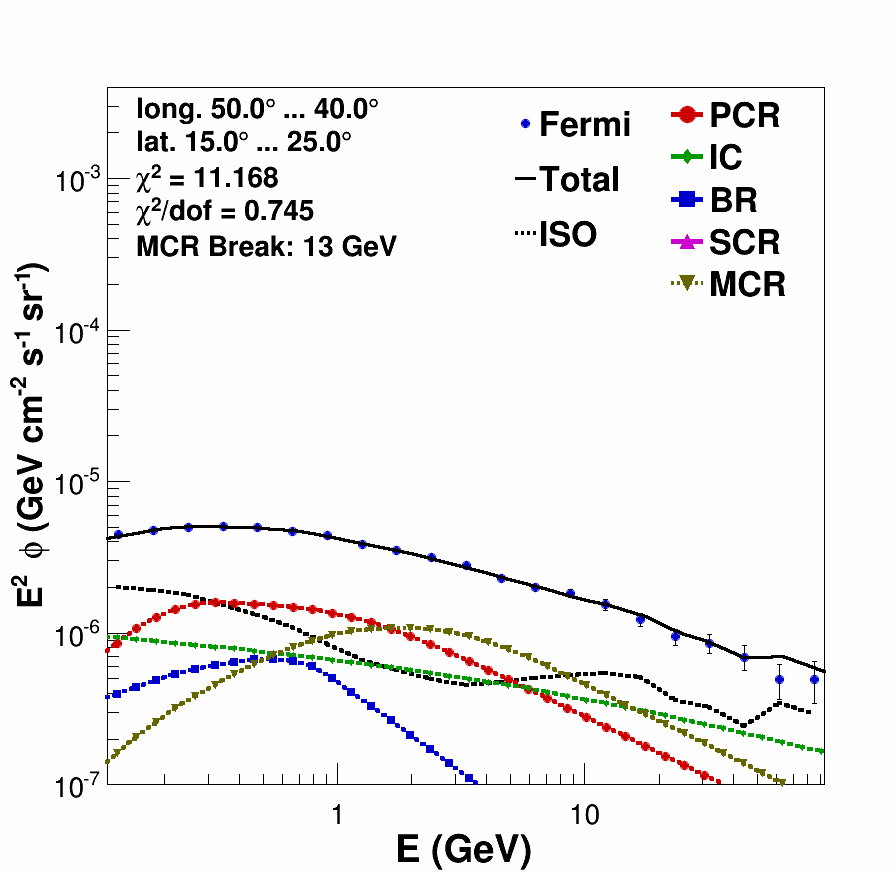}
\includegraphics[width=0.16\textwidth,height=0.16\textwidth,clip]{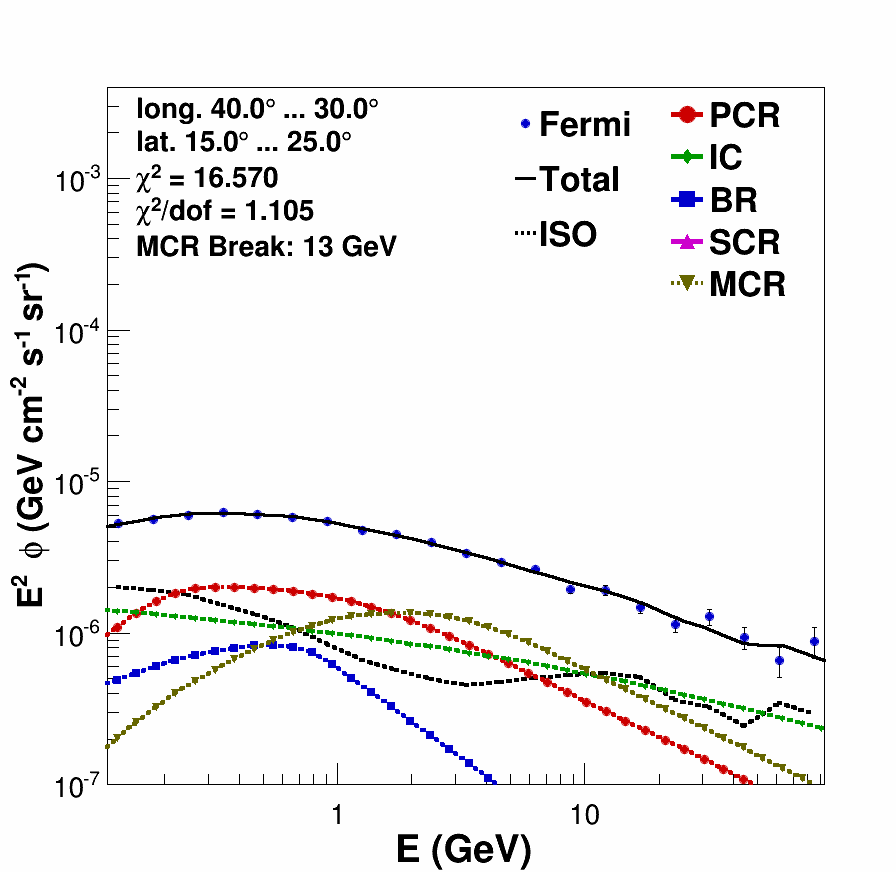}
\includegraphics[width=0.16\textwidth,height=0.16\textwidth,clip]{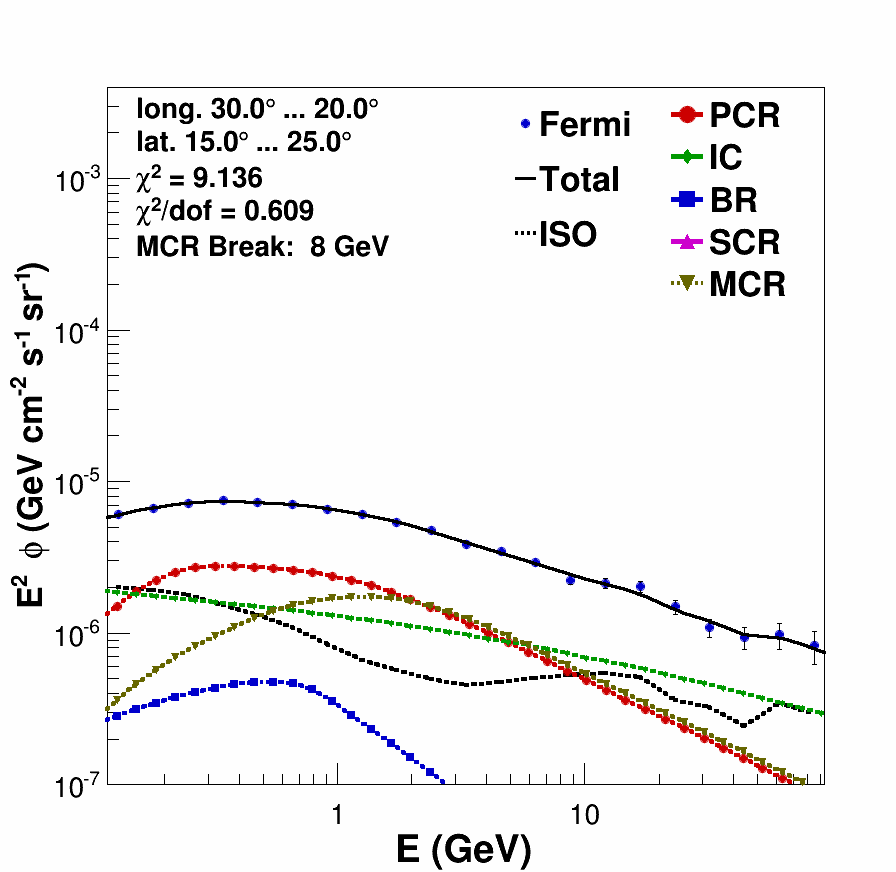}
\includegraphics[width=0.16\textwidth,height=0.16\textwidth,clip]{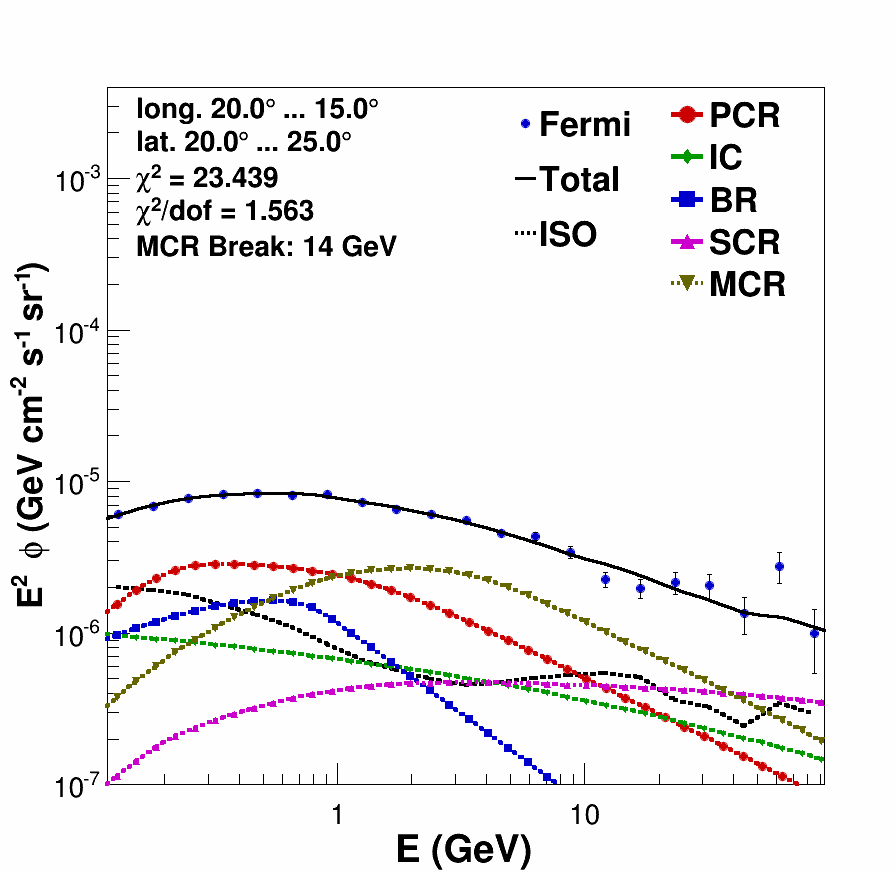}
\includegraphics[width=0.16\textwidth,height=0.16\textwidth,clip]{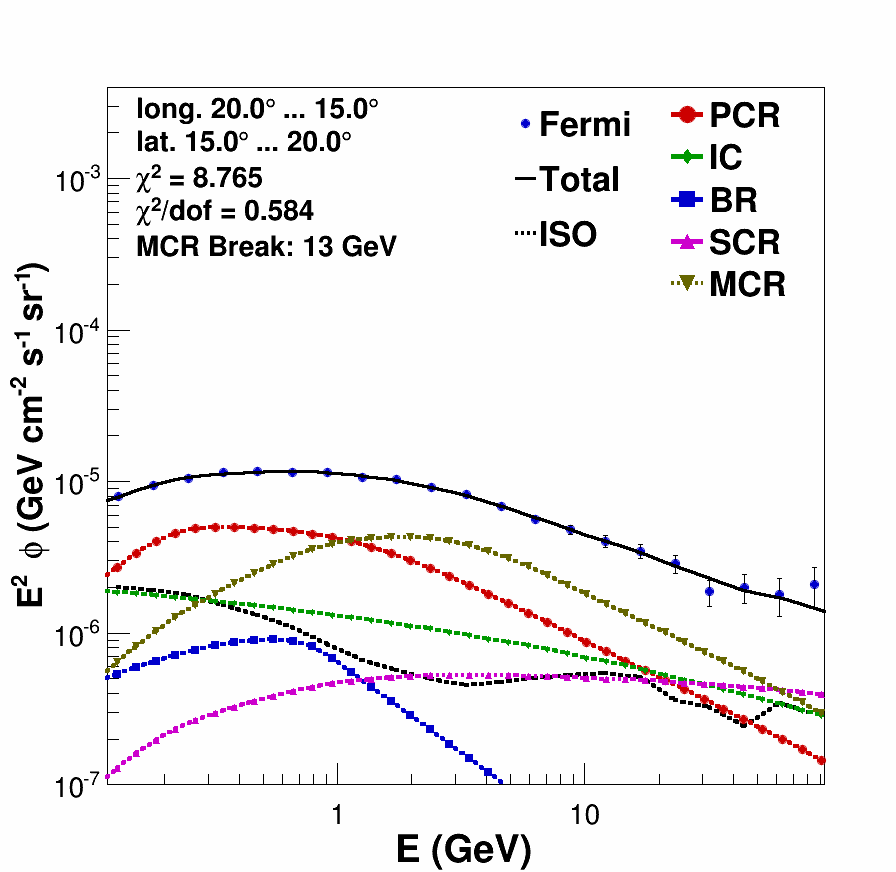}
\includegraphics[width=0.16\textwidth,height=0.16\textwidth,clip]{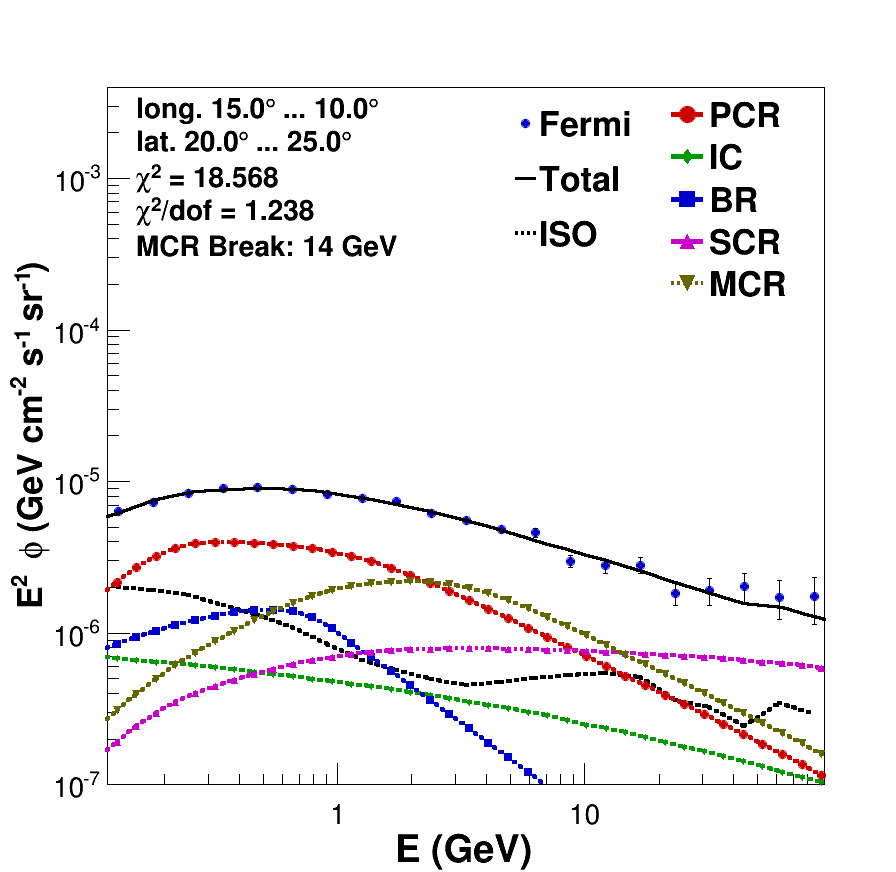}
\includegraphics[width=0.16\textwidth,height=0.16\textwidth,clip]{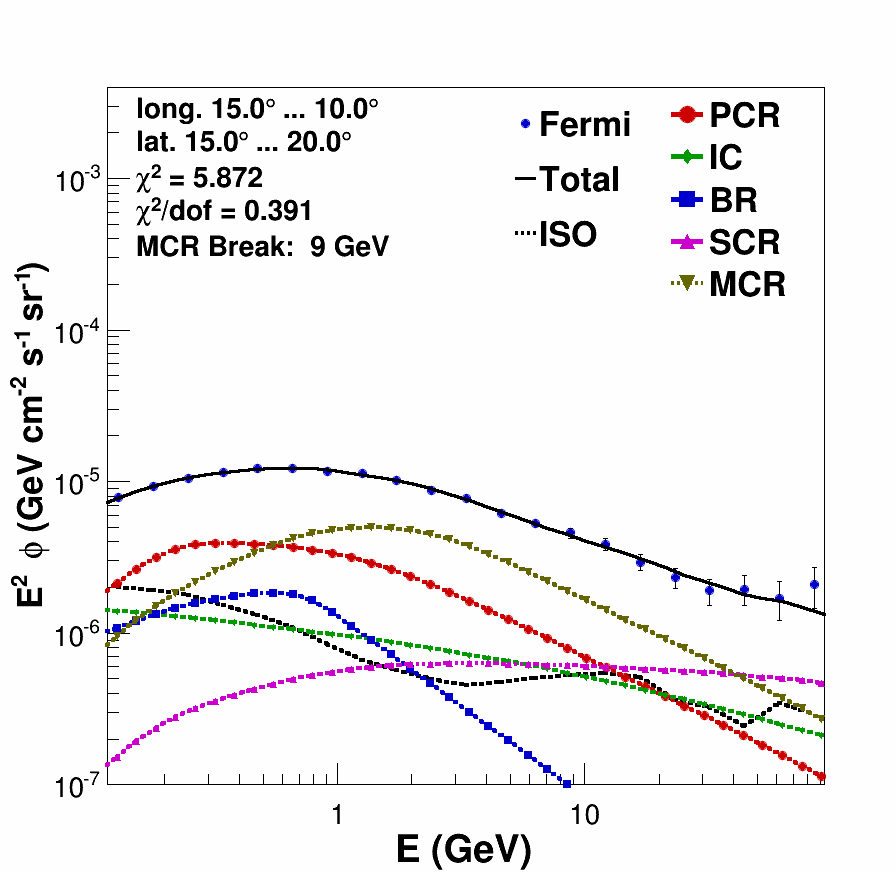}
\includegraphics[width=0.16\textwidth,height=0.16\textwidth,clip]{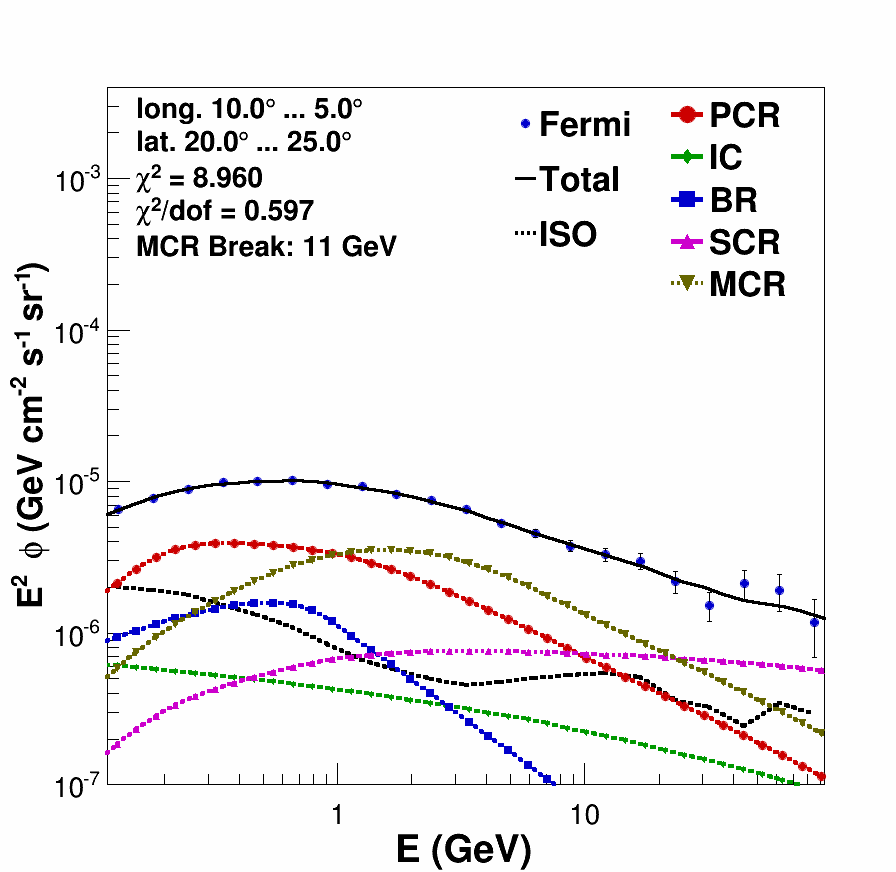}
\includegraphics[width=0.16\textwidth,height=0.16\textwidth,clip]{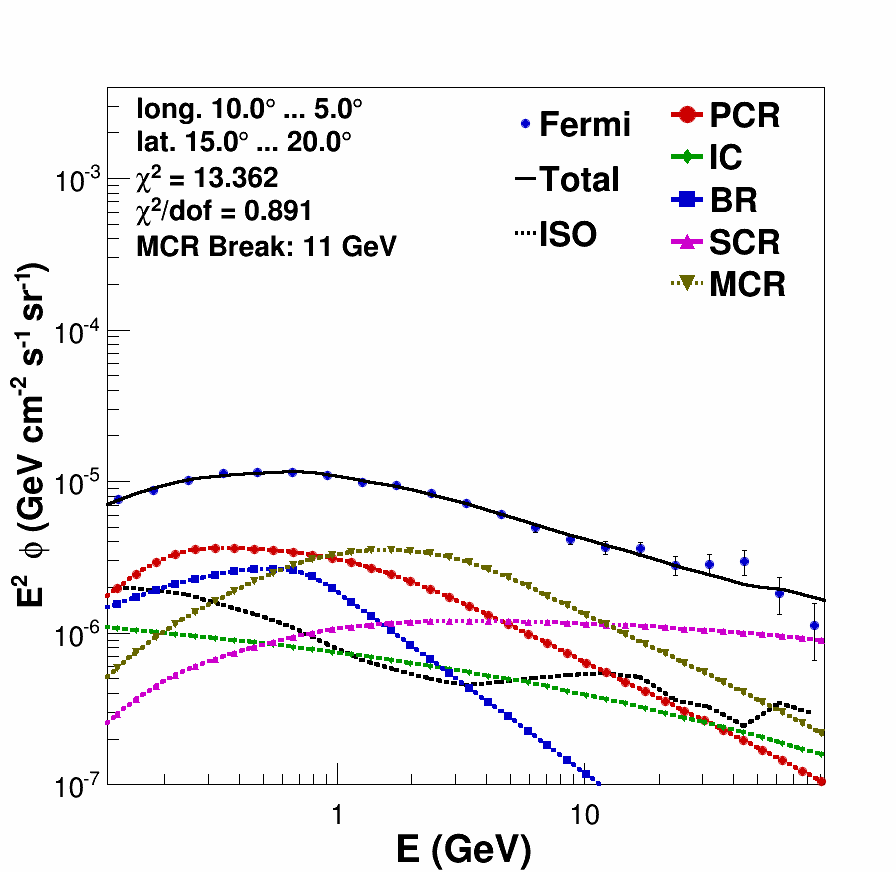}
\includegraphics[width=0.16\textwidth,height=0.16\textwidth,clip]{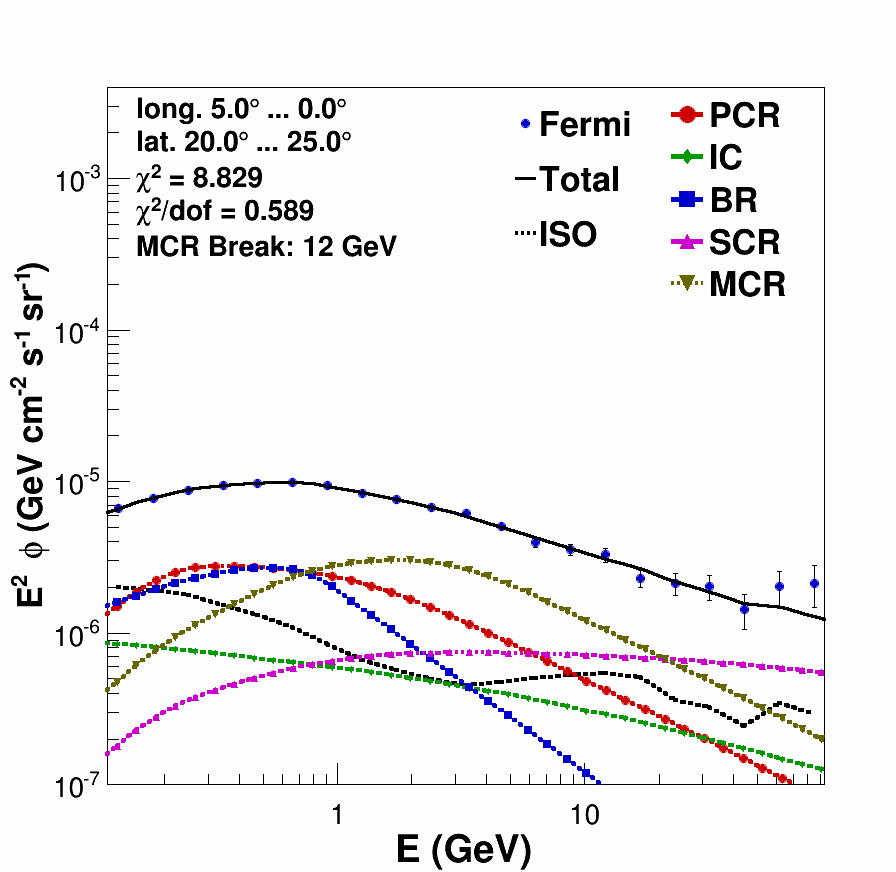}
\includegraphics[width=0.16\textwidth,height=0.16\textwidth,clip]{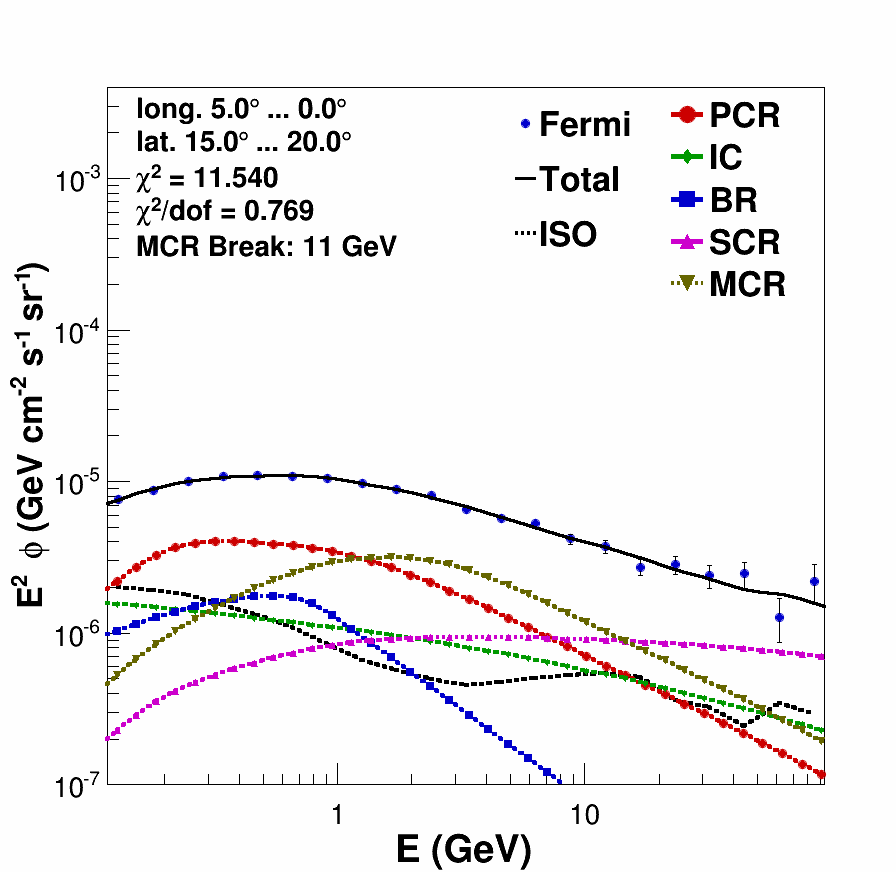}
\includegraphics[width=0.16\textwidth,height=0.16\textwidth,clip]{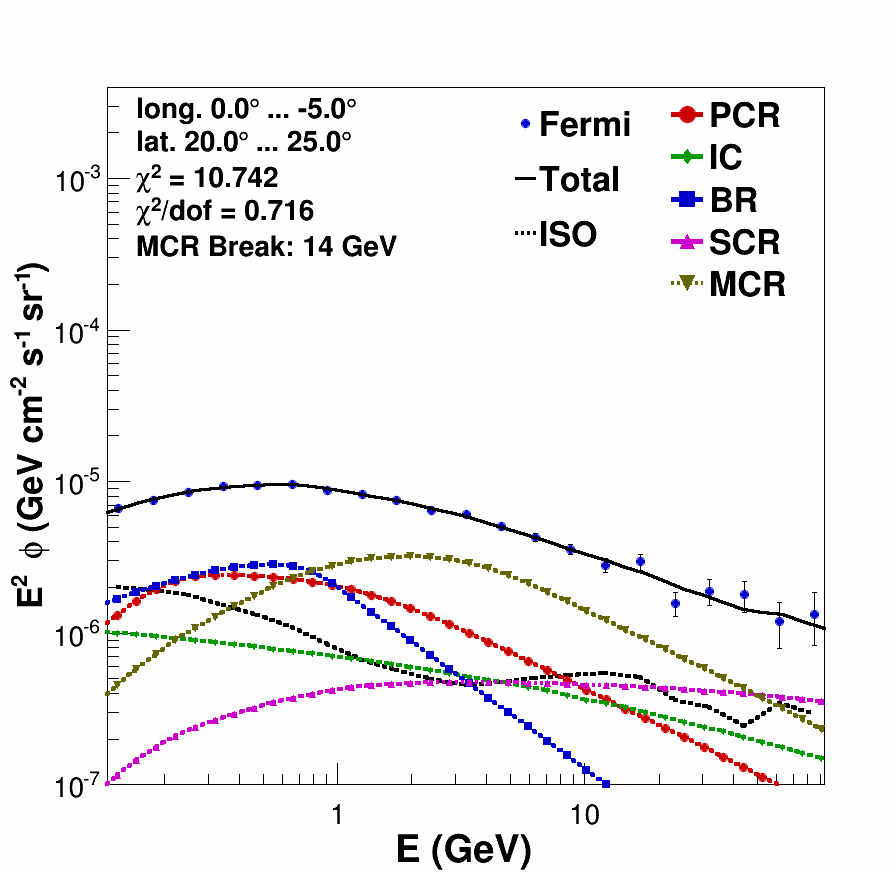}
\includegraphics[width=0.16\textwidth,height=0.16\textwidth,clip]{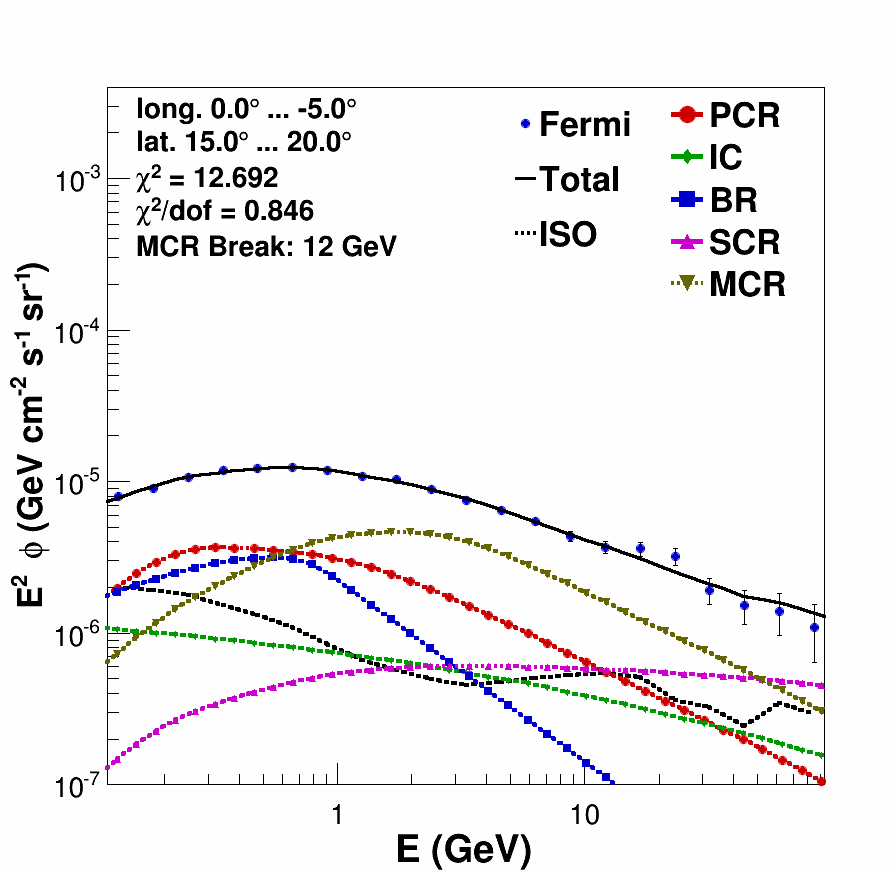}
\includegraphics[width=0.16\textwidth,height=0.16\textwidth,clip]{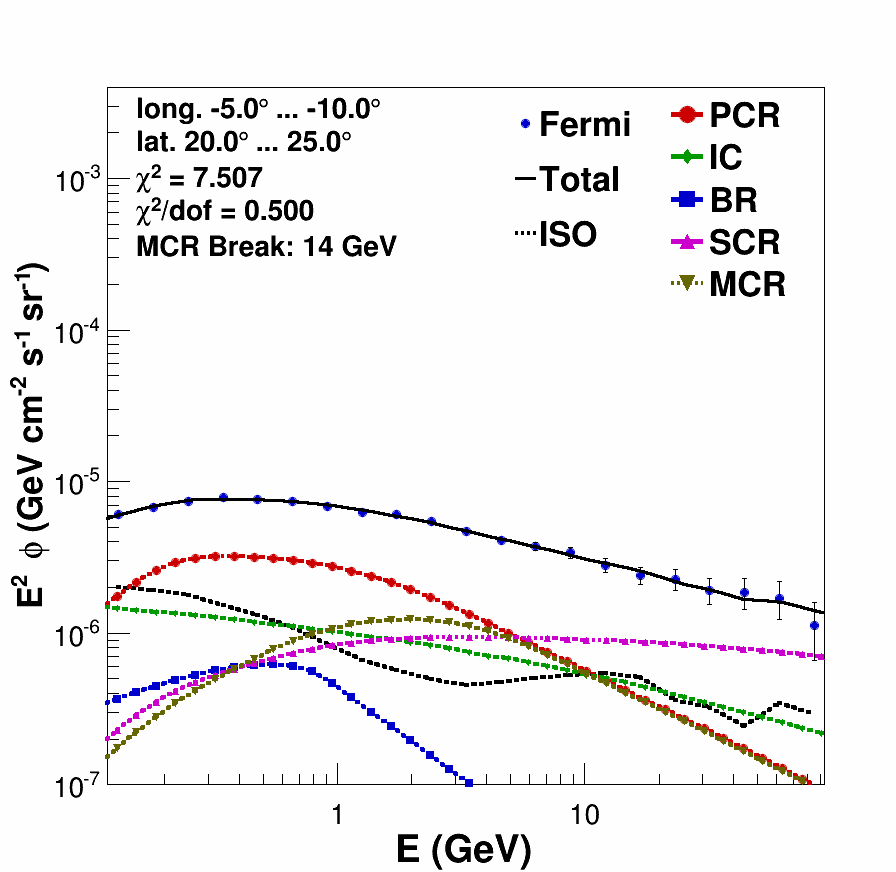}
\includegraphics[width=0.16\textwidth,height=0.16\textwidth,clip]{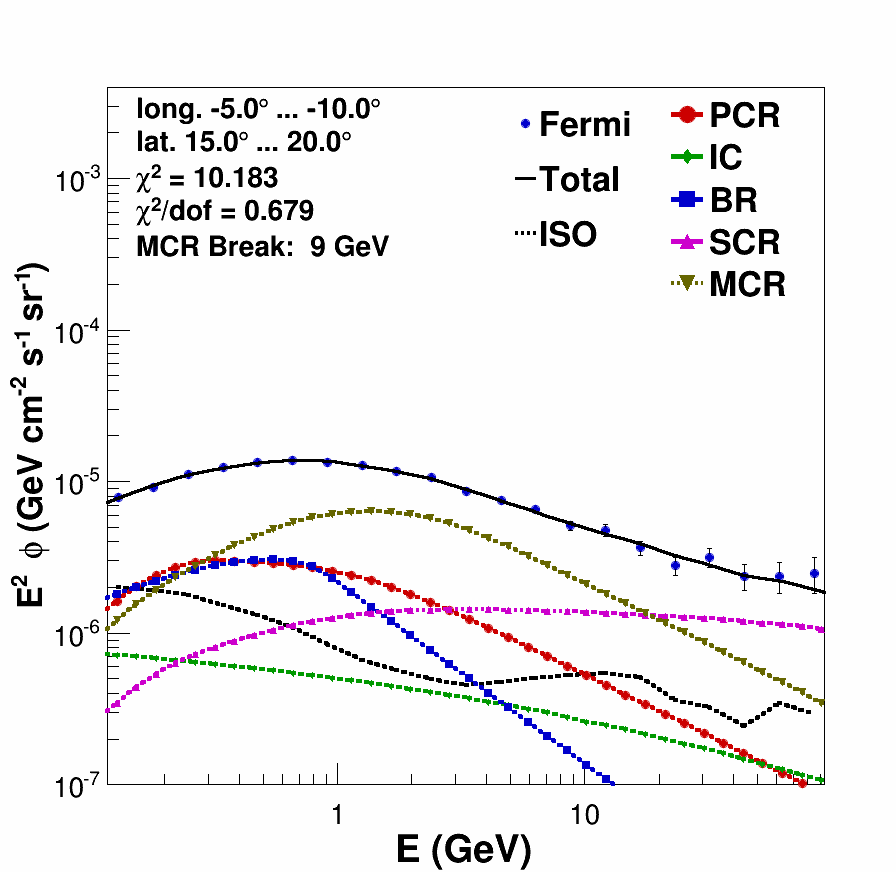}
\includegraphics[width=0.16\textwidth,height=0.16\textwidth,clip]{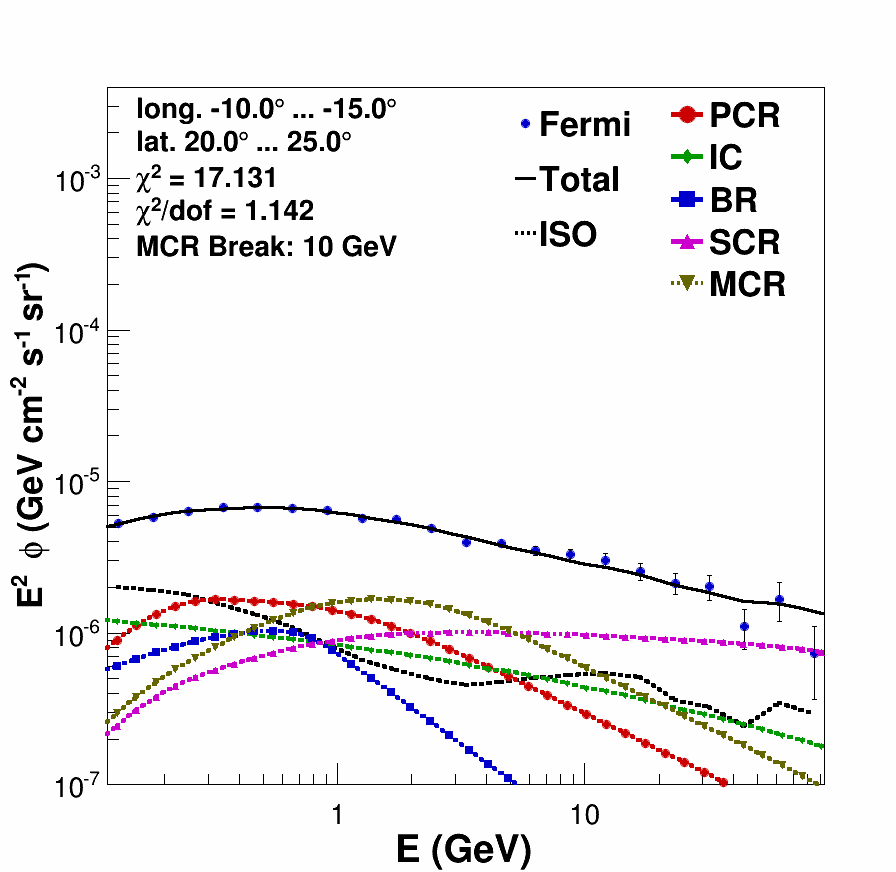}
\includegraphics[width=0.16\textwidth,height=0.16\textwidth,clip]{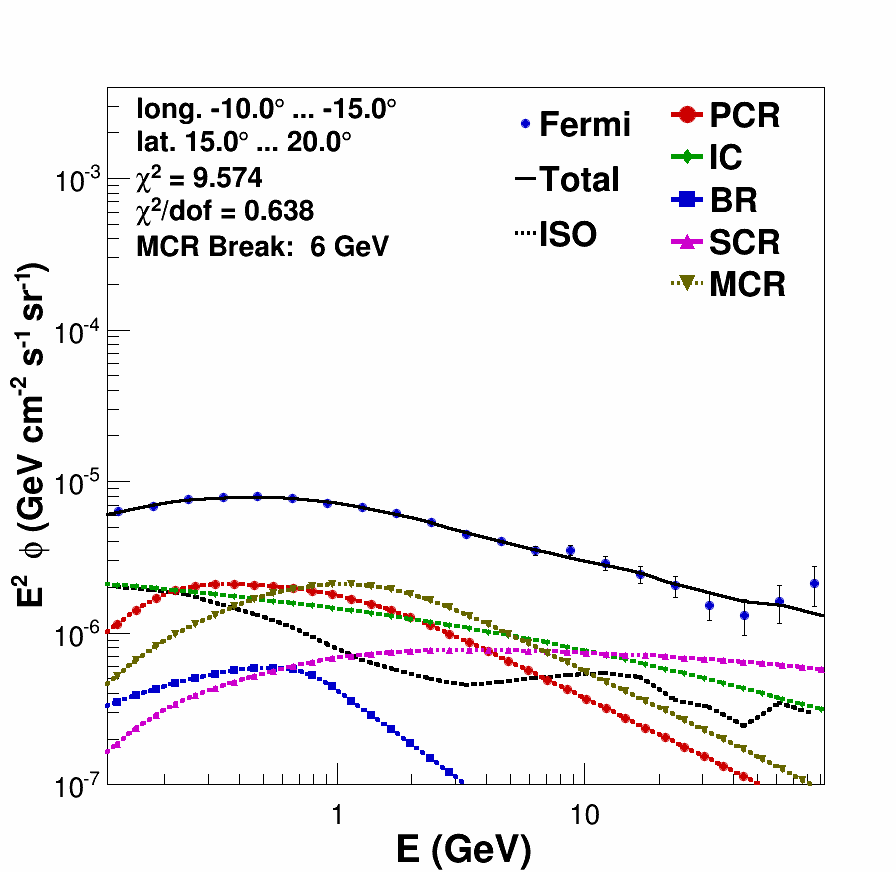}
\includegraphics[width=0.16\textwidth,height=0.16\textwidth,clip]{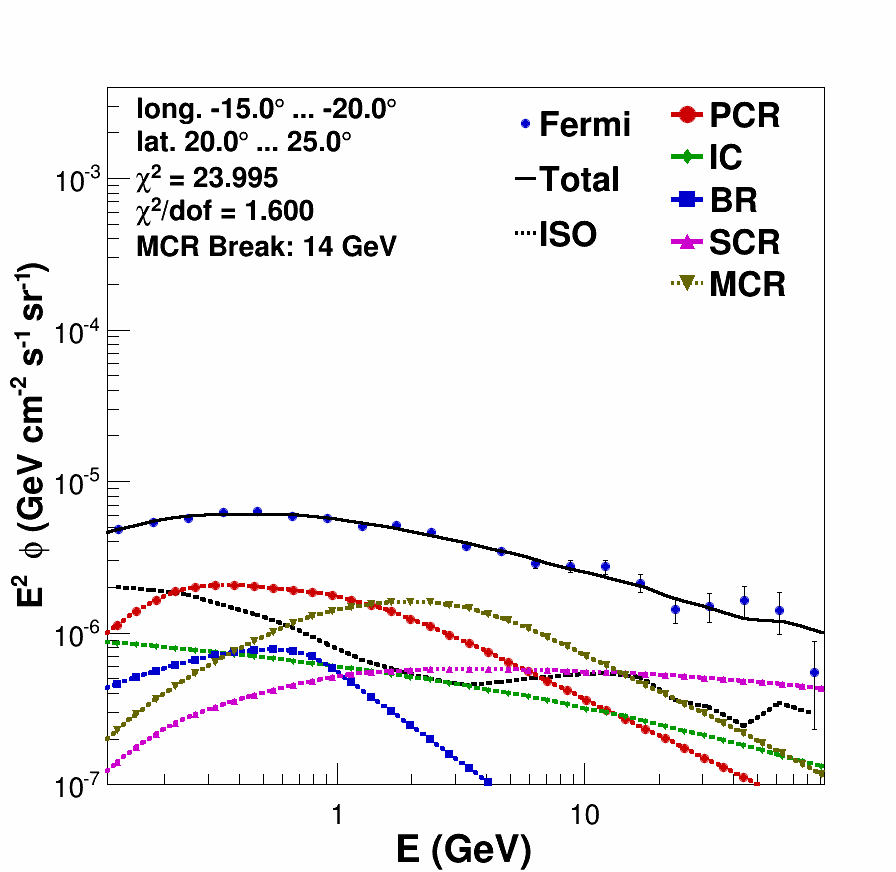}
\includegraphics[width=0.16\textwidth,height=0.16\textwidth,clip]{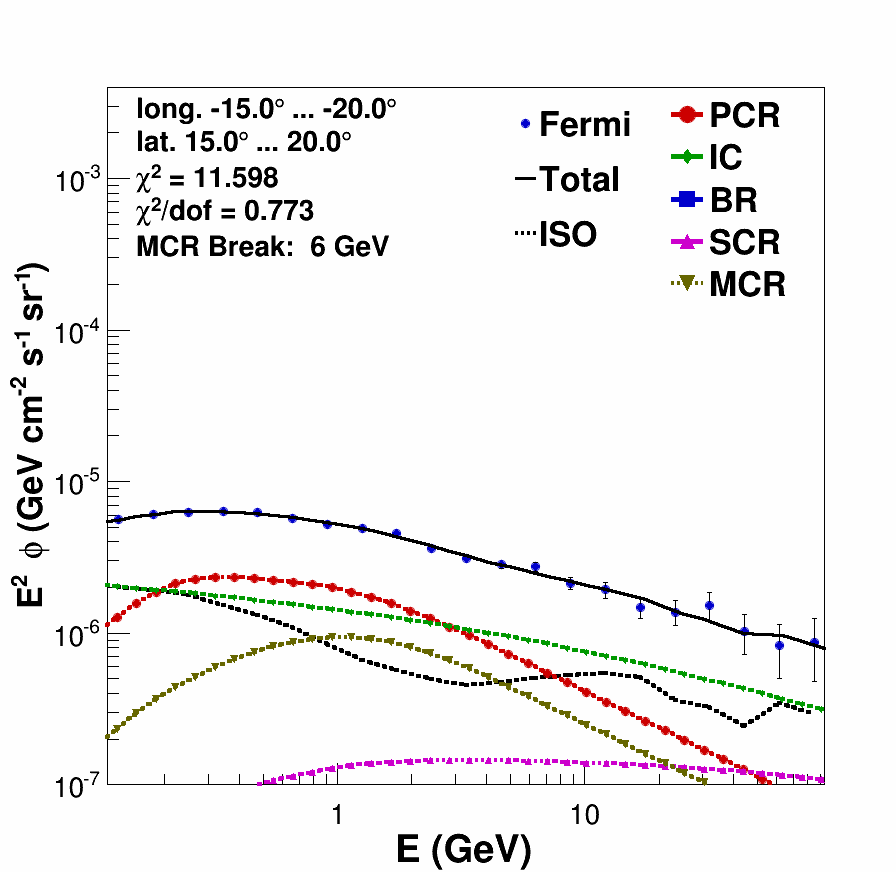}
\includegraphics[width=0.16\textwidth,height=0.16\textwidth,clip]{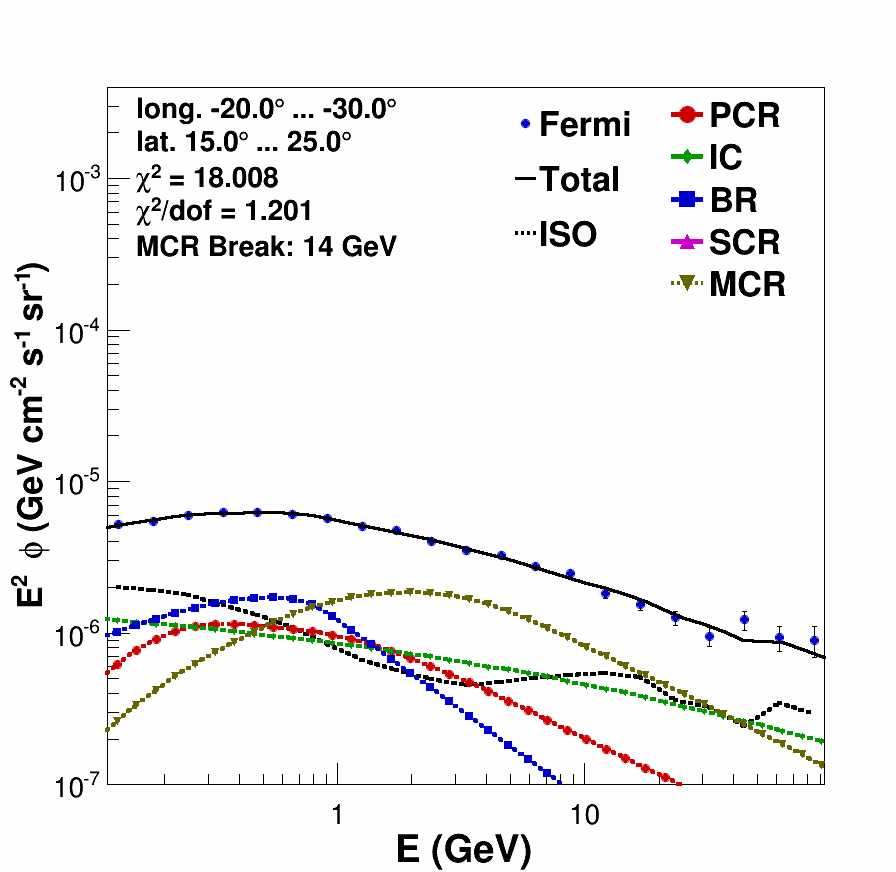}
\includegraphics[width=0.16\textwidth,height=0.16\textwidth,clip]{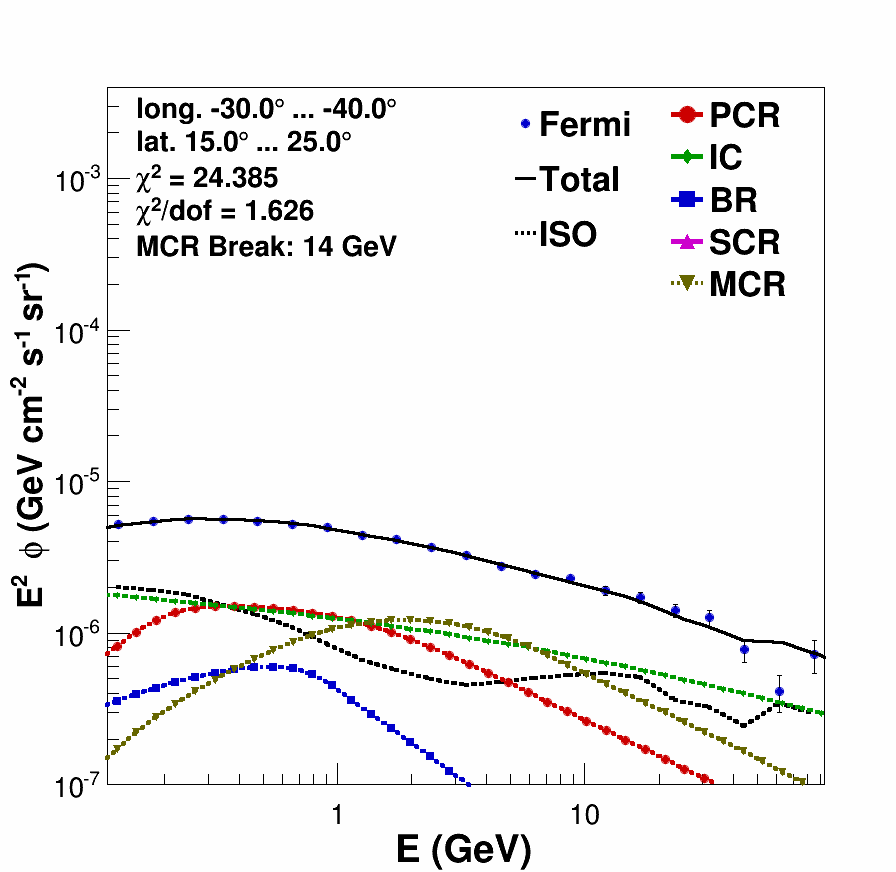}
\includegraphics[width=0.16\textwidth,height=0.16\textwidth,clip]{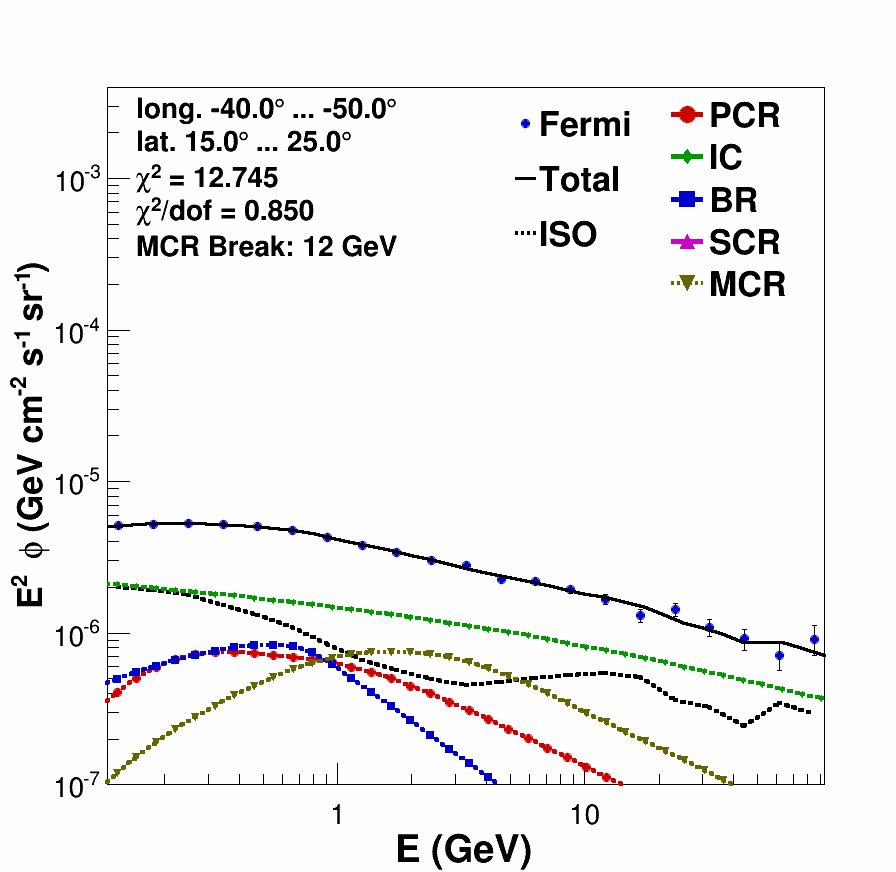}
\includegraphics[width=0.16\textwidth,height=0.16\textwidth,clip]{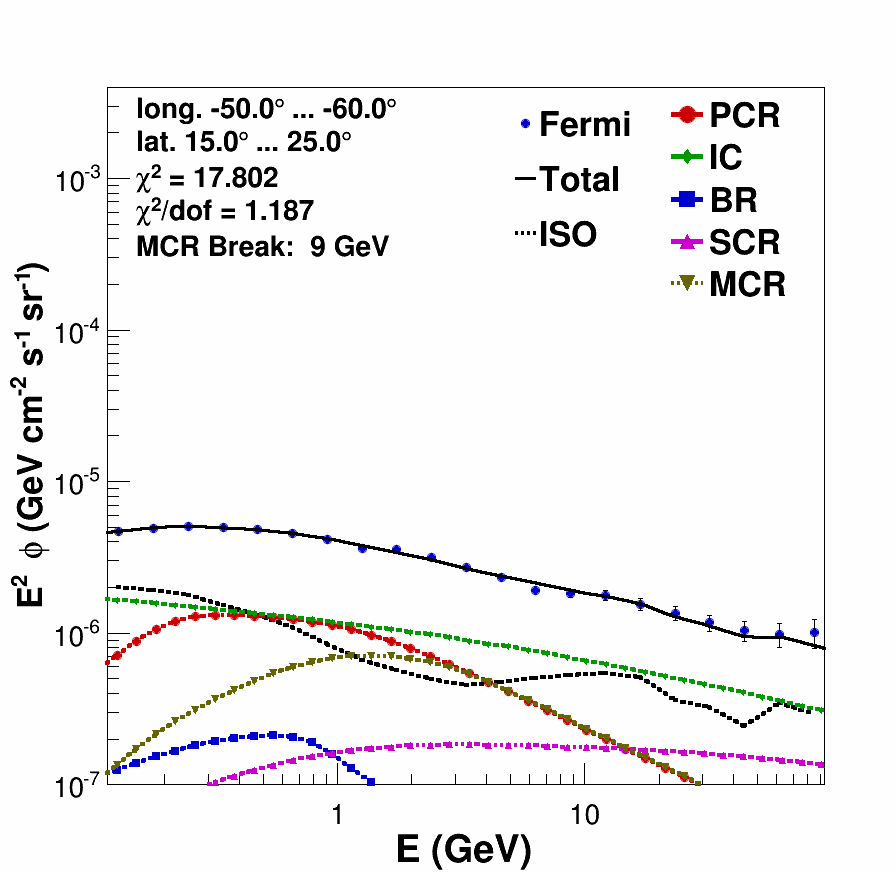}
\includegraphics[width=0.16\textwidth,height=0.16\textwidth,clip]{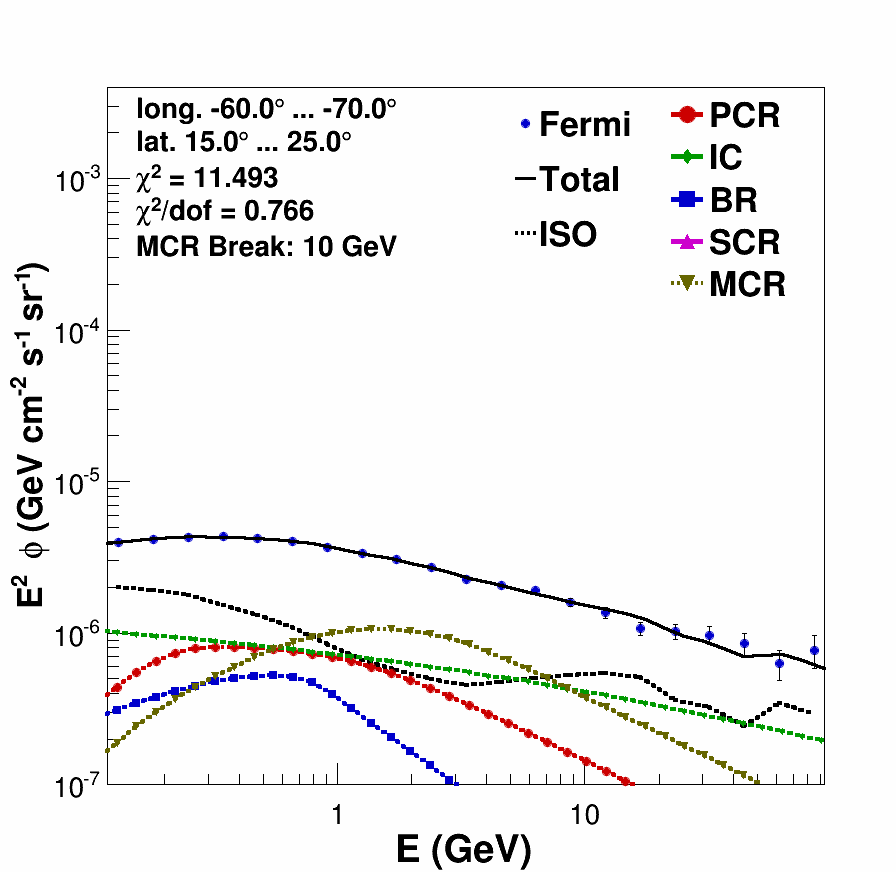}
\includegraphics[width=0.16\textwidth,height=0.16\textwidth,clip]{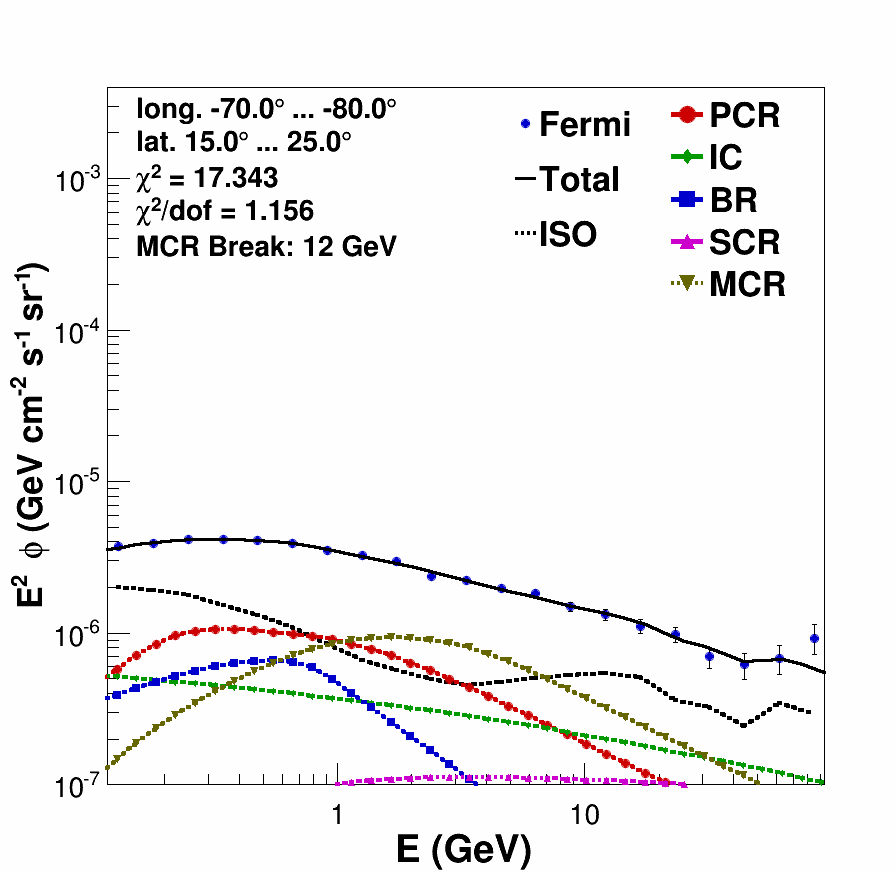}
\includegraphics[width=0.16\textwidth,height=0.16\textwidth,clip]{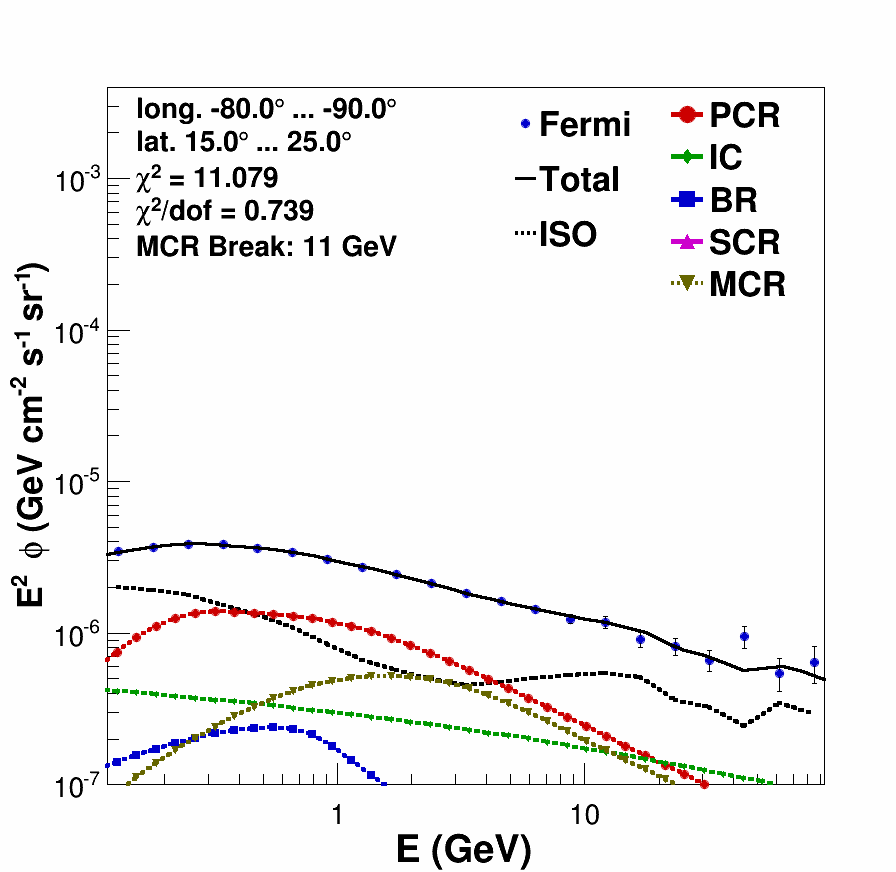}
\includegraphics[width=0.16\textwidth,height=0.16\textwidth,clip]{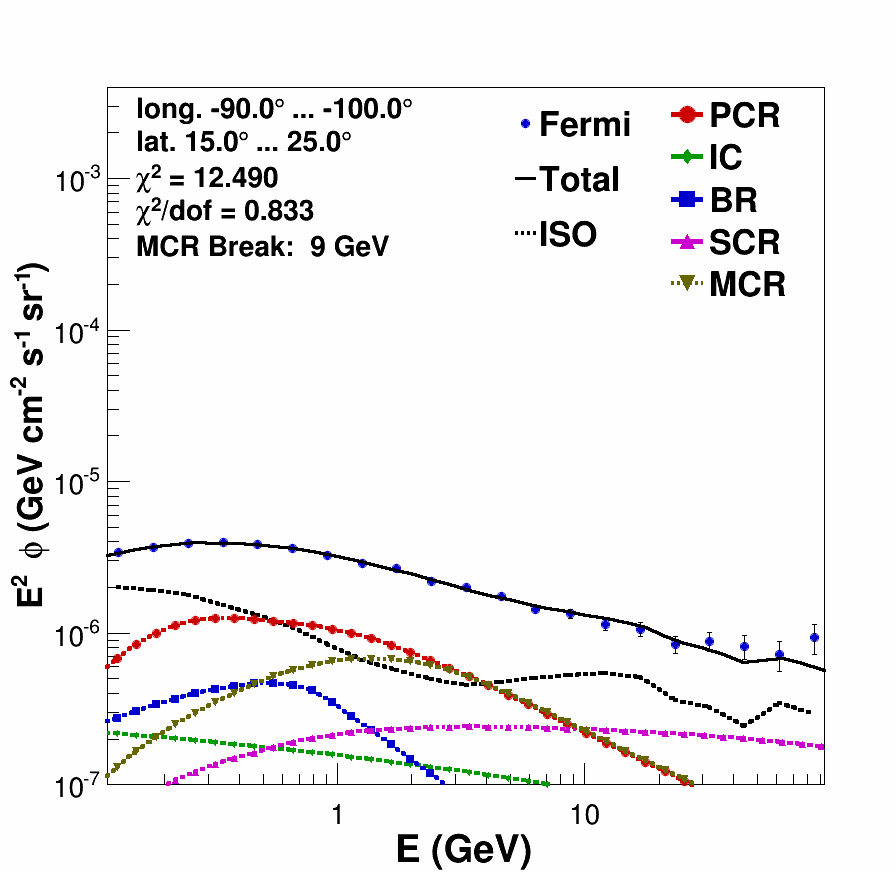}
\includegraphics[width=0.16\textwidth,height=0.16\textwidth,clip]{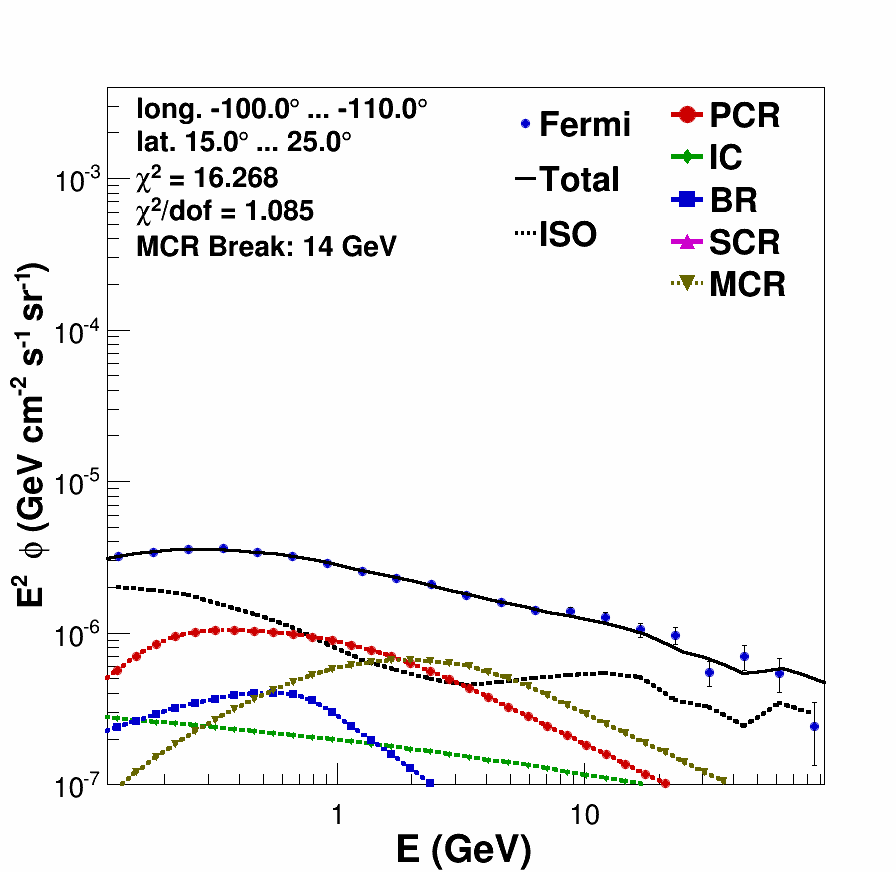}
\includegraphics[width=0.16\textwidth,height=0.16\textwidth,clip]{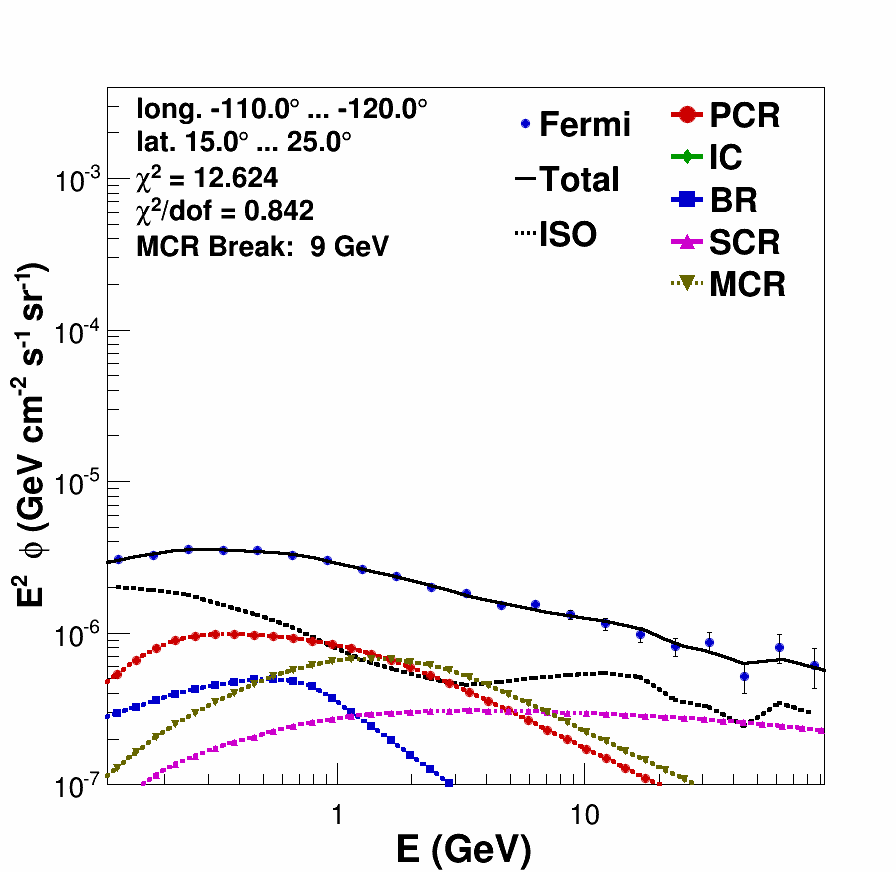}
\includegraphics[width=0.16\textwidth,height=0.16\textwidth,clip]{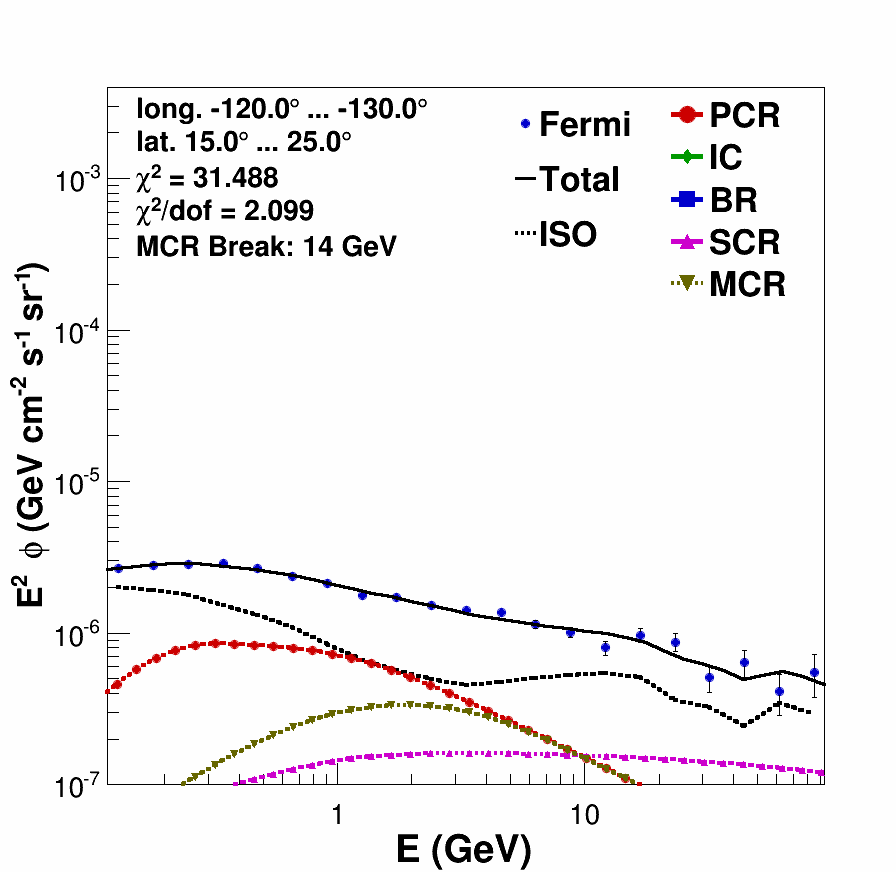}
\includegraphics[width=0.16\textwidth,height=0.16\textwidth,clip]{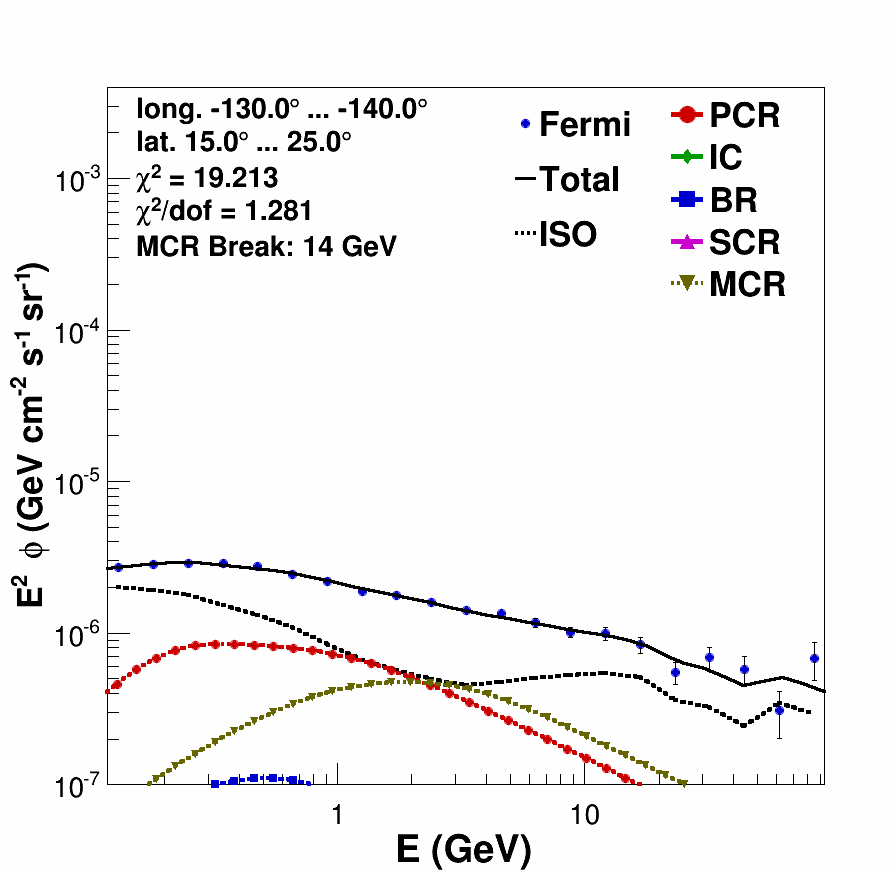}
\includegraphics[width=0.16\textwidth,height=0.16\textwidth,clip]{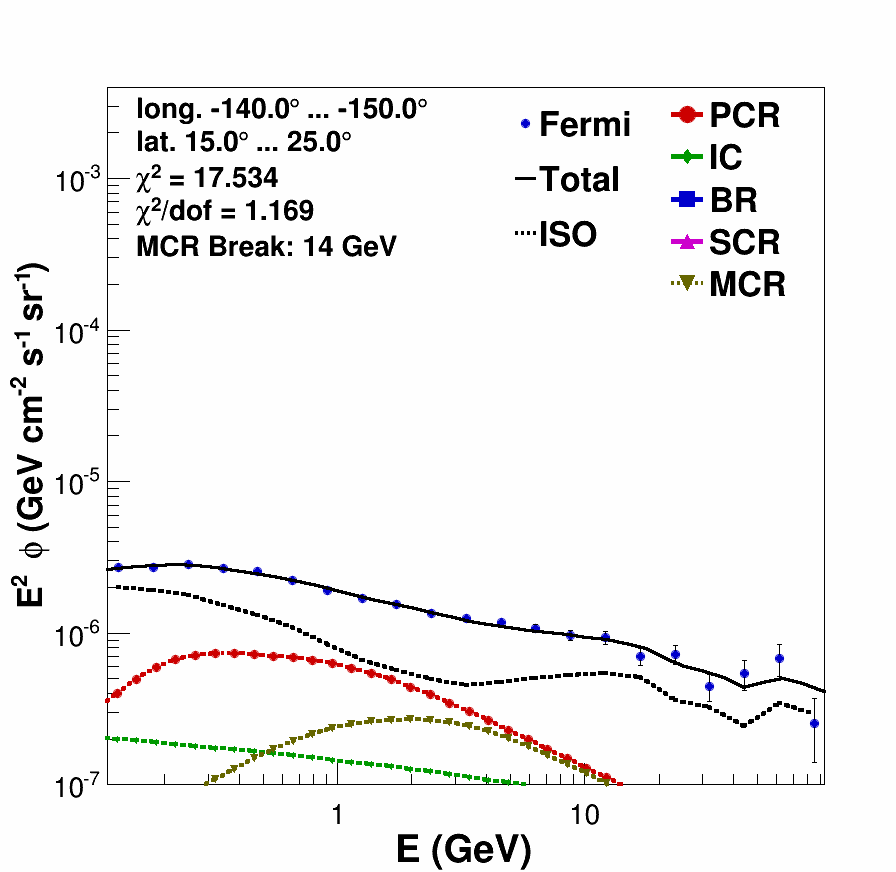}
\includegraphics[width=0.16\textwidth,height=0.16\textwidth,clip]{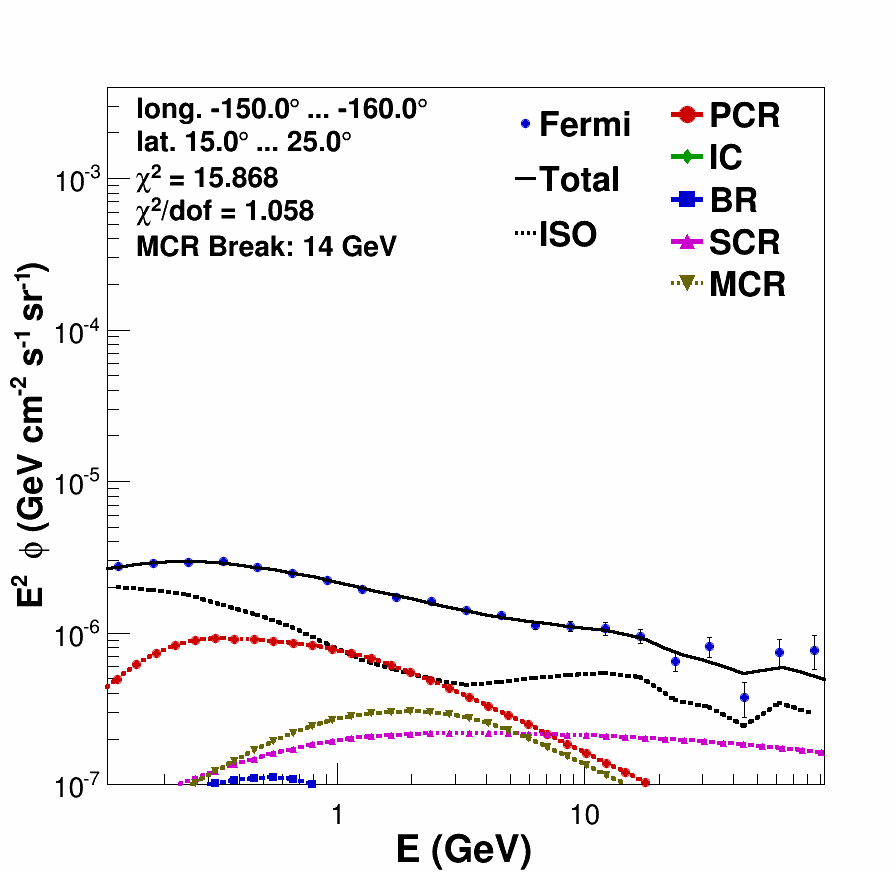}
\includegraphics[width=0.16\textwidth,height=0.16\textwidth,clip]{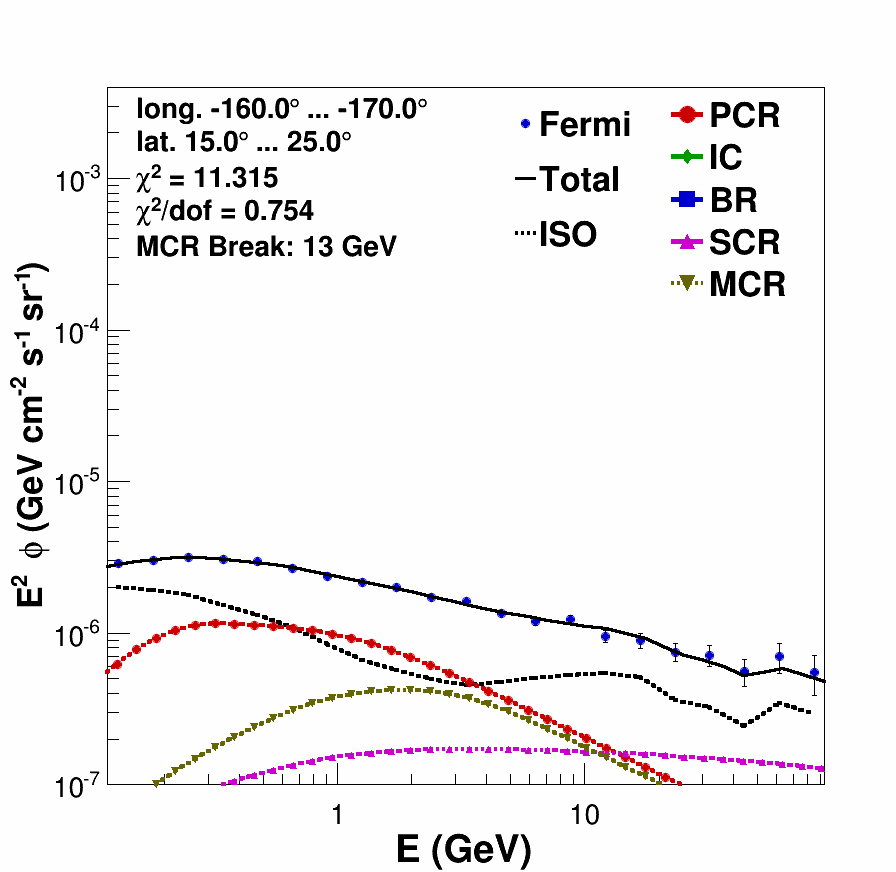}
\includegraphics[width=0.16\textwidth,height=0.16\textwidth,clip]{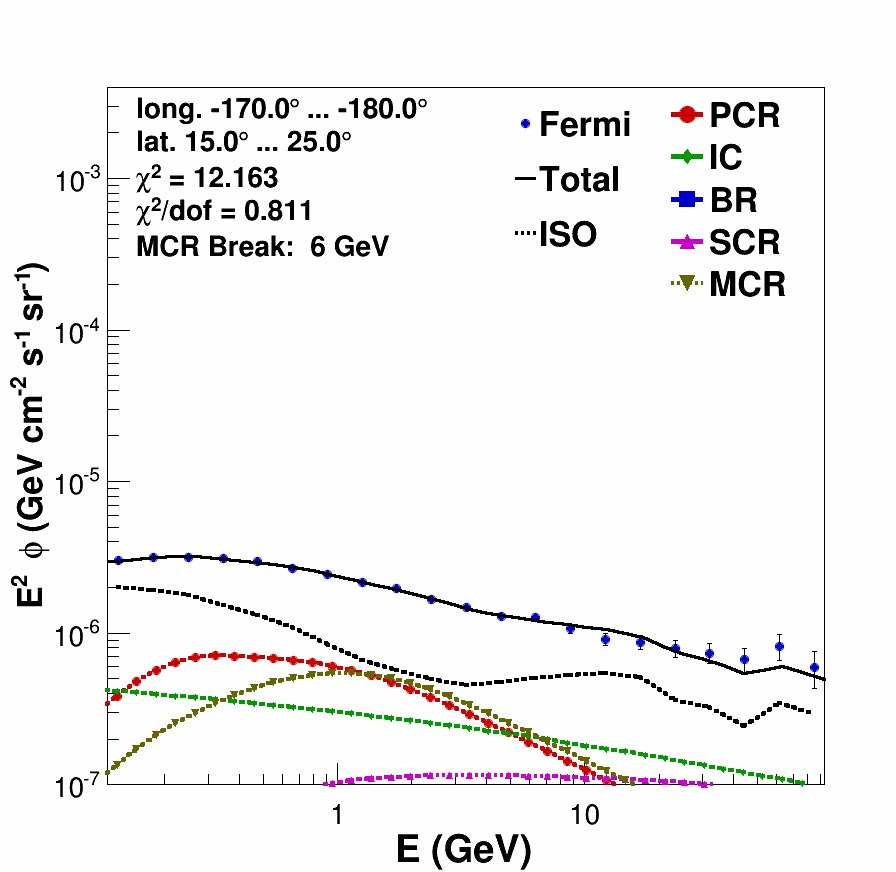}%%%%%%r6
\caption[]{Template fits for latitudes  with $15.0^\circ<b<25.0^\circ$ and longitudes decreasing from 180$^\circ$ to -180$^\circ$. \label{F16}
}
\end{figure}
\begin{figure}
\centering
\includegraphics[width=0.16\textwidth,height=0.16\textwidth,clip]{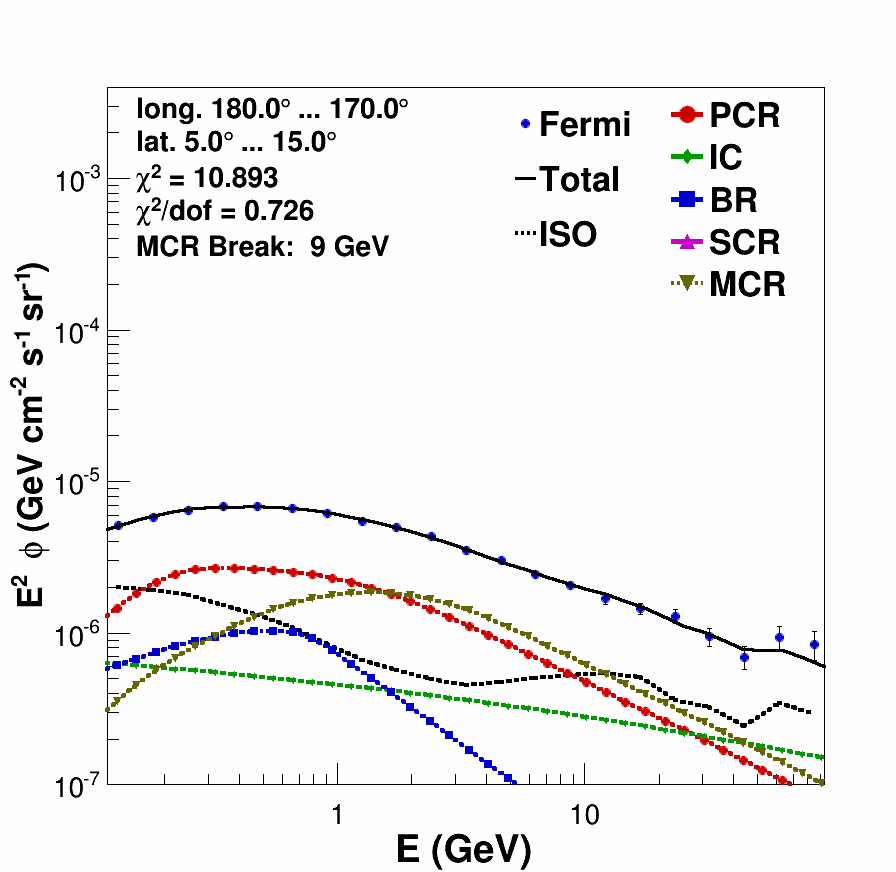}
\includegraphics[width=0.16\textwidth,height=0.16\textwidth,clip]{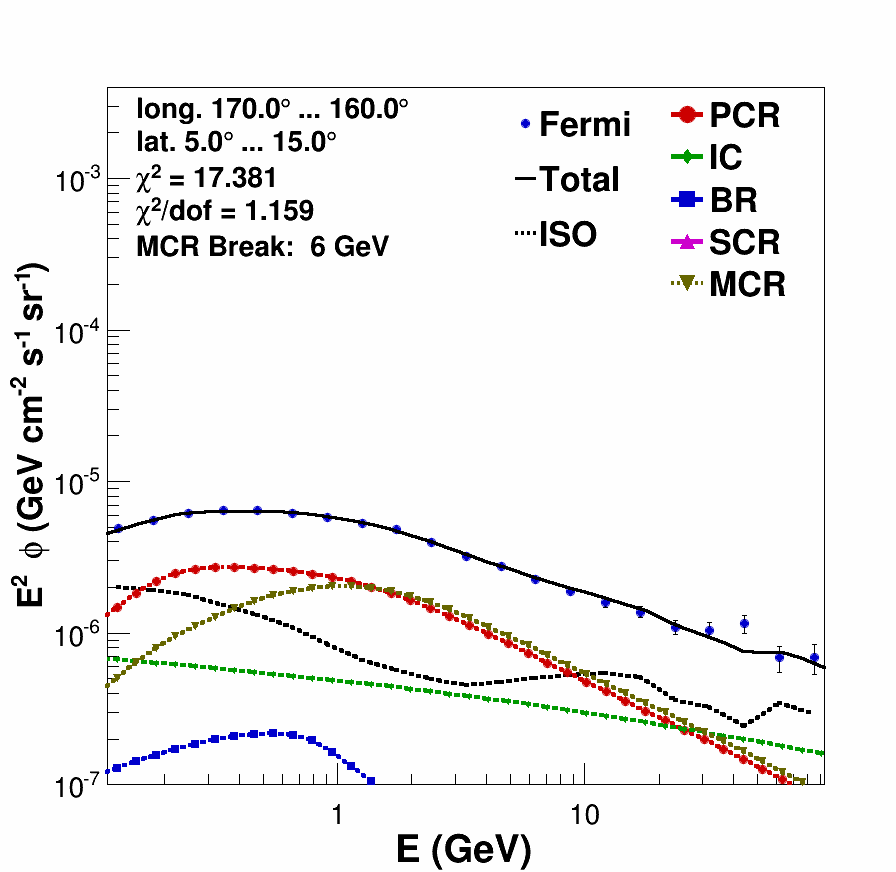}
\includegraphics[width=0.16\textwidth,height=0.16\textwidth,clip]{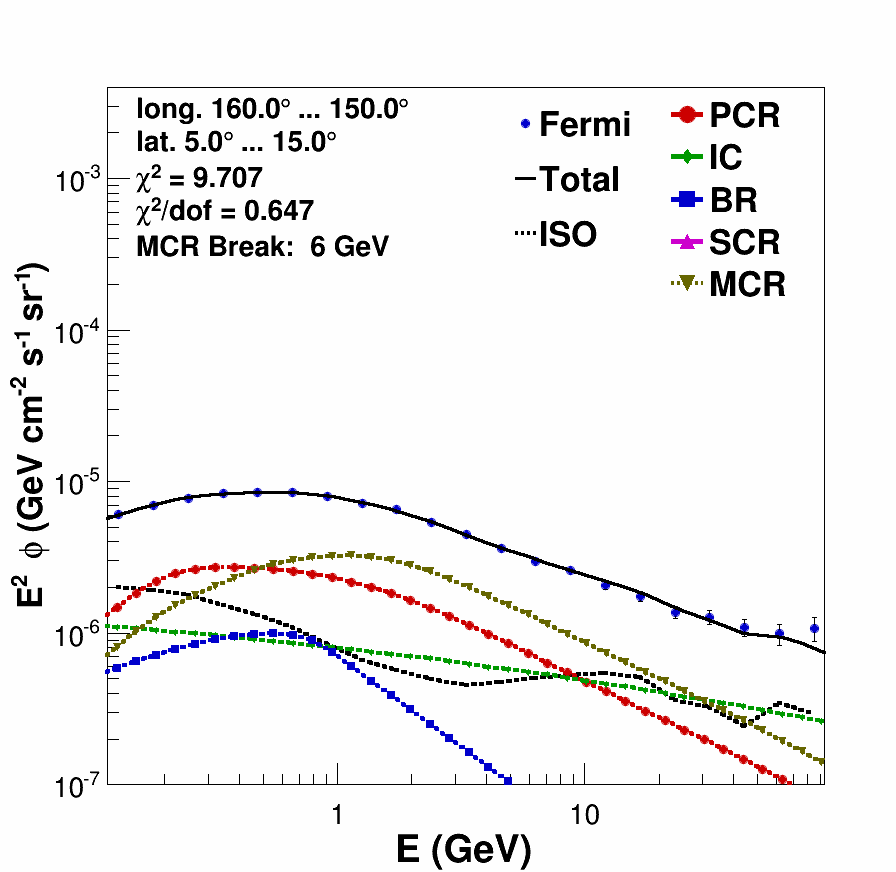}
\includegraphics[width=0.16\textwidth,height=0.16\textwidth,clip]{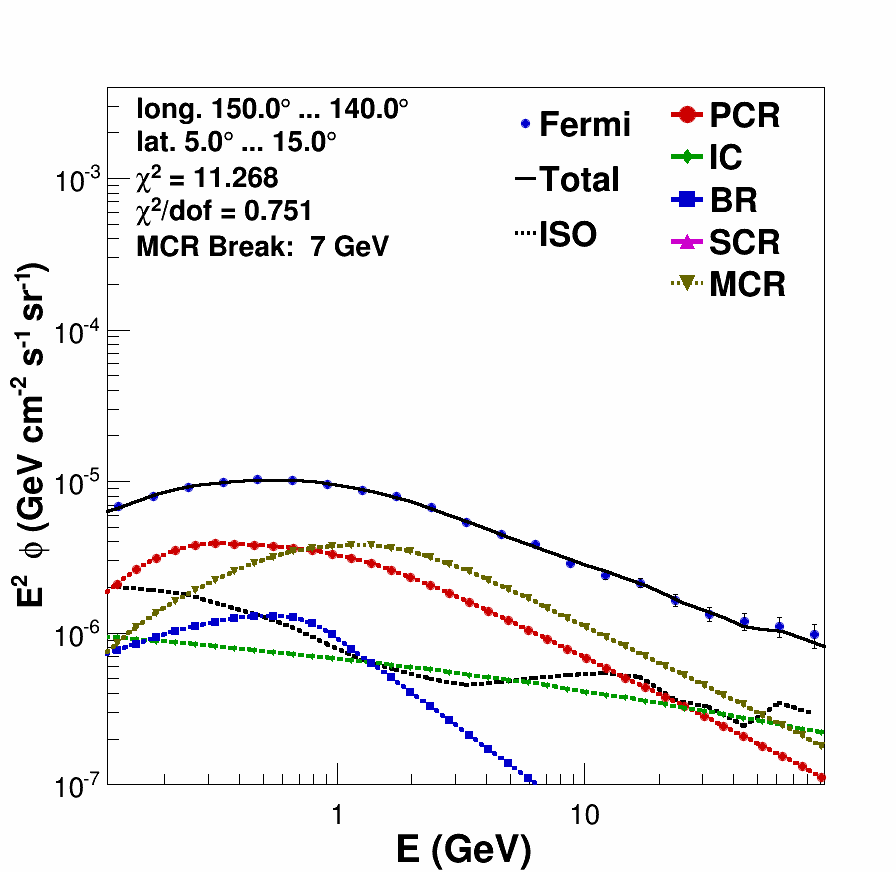}
\includegraphics[width=0.16\textwidth,height=0.16\textwidth,clip]{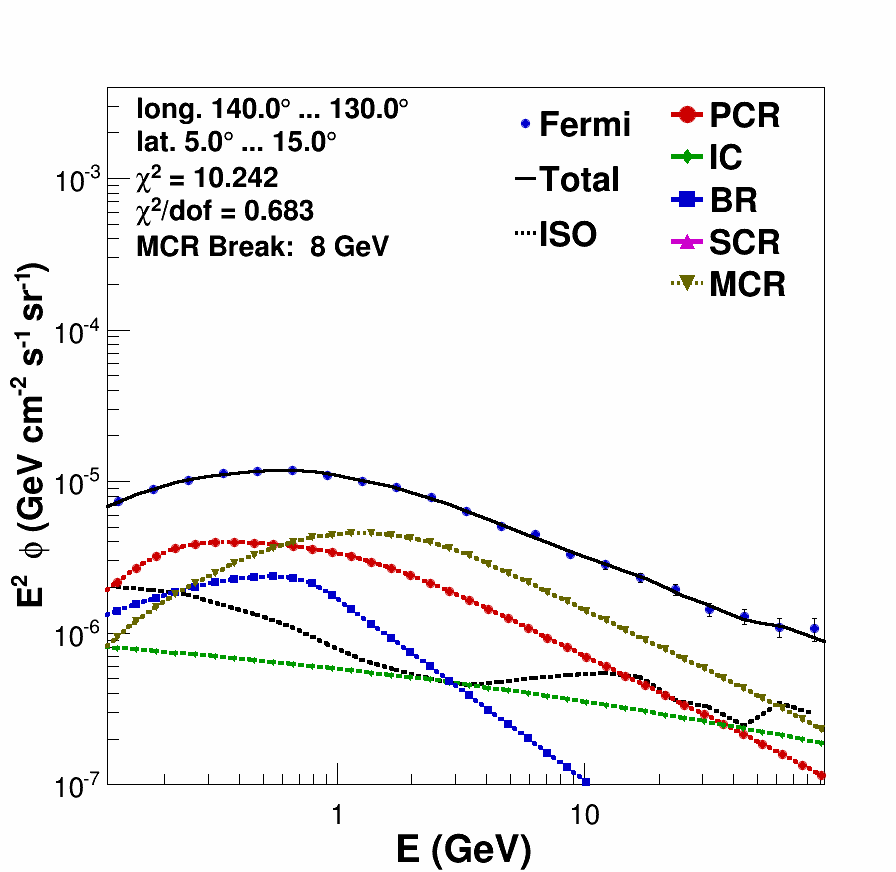}
\includegraphics[width=0.16\textwidth,height=0.16\textwidth,clip]{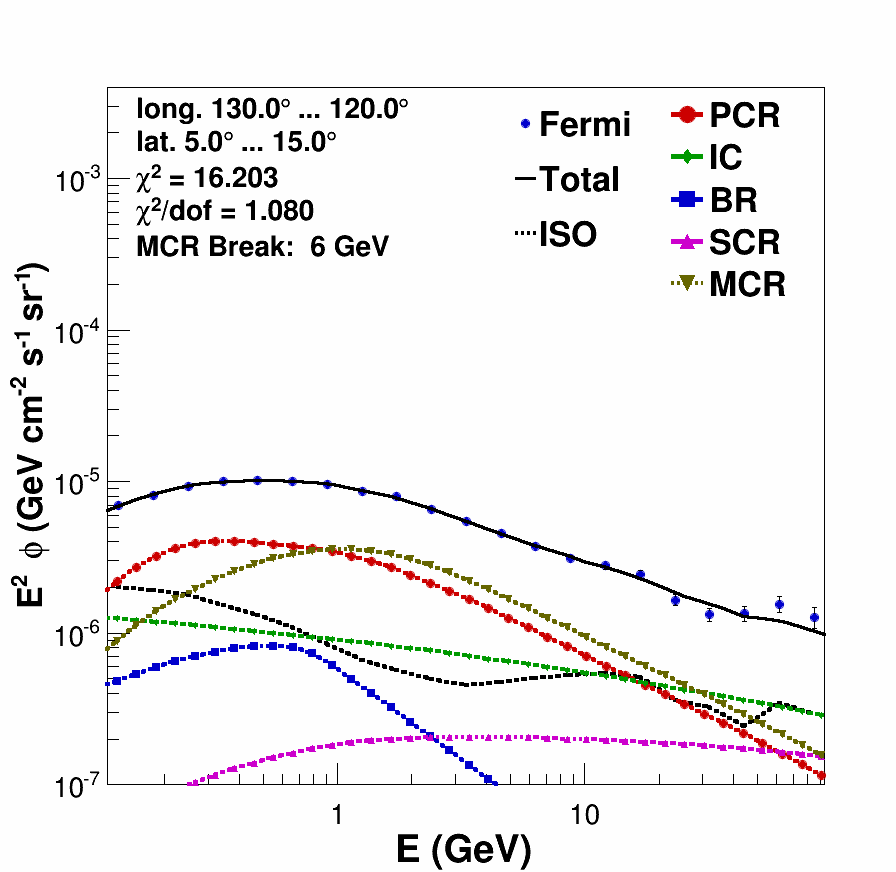}
\includegraphics[width=0.16\textwidth,height=0.16\textwidth,clip]{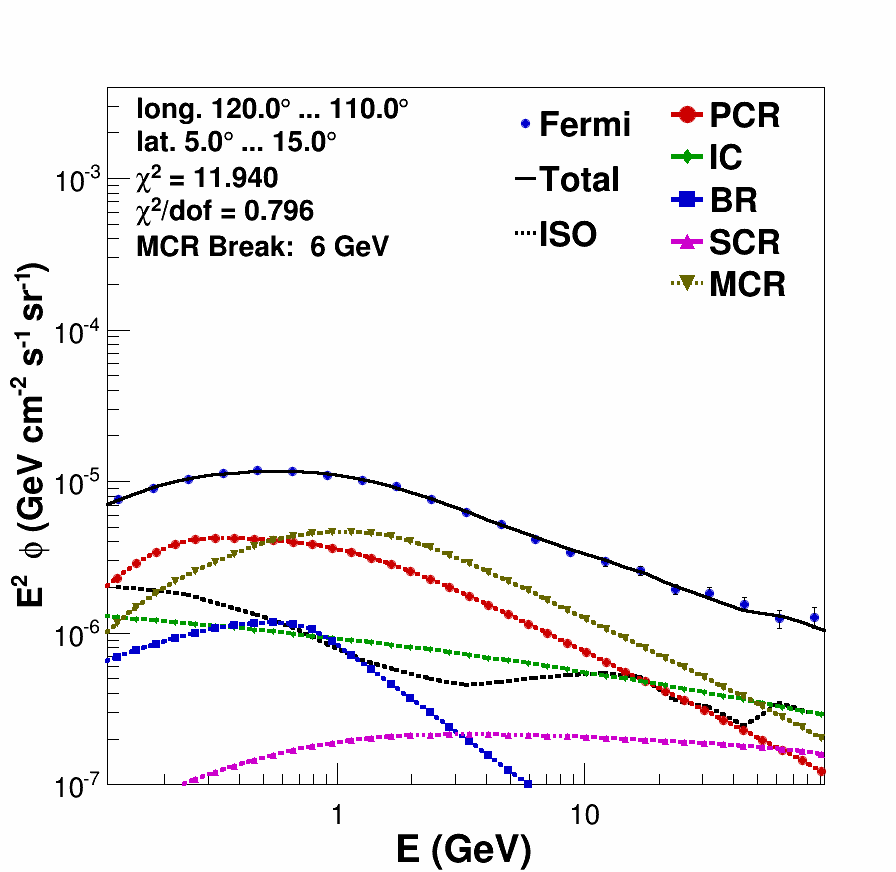}
\includegraphics[width=0.16\textwidth,height=0.16\textwidth,clip]{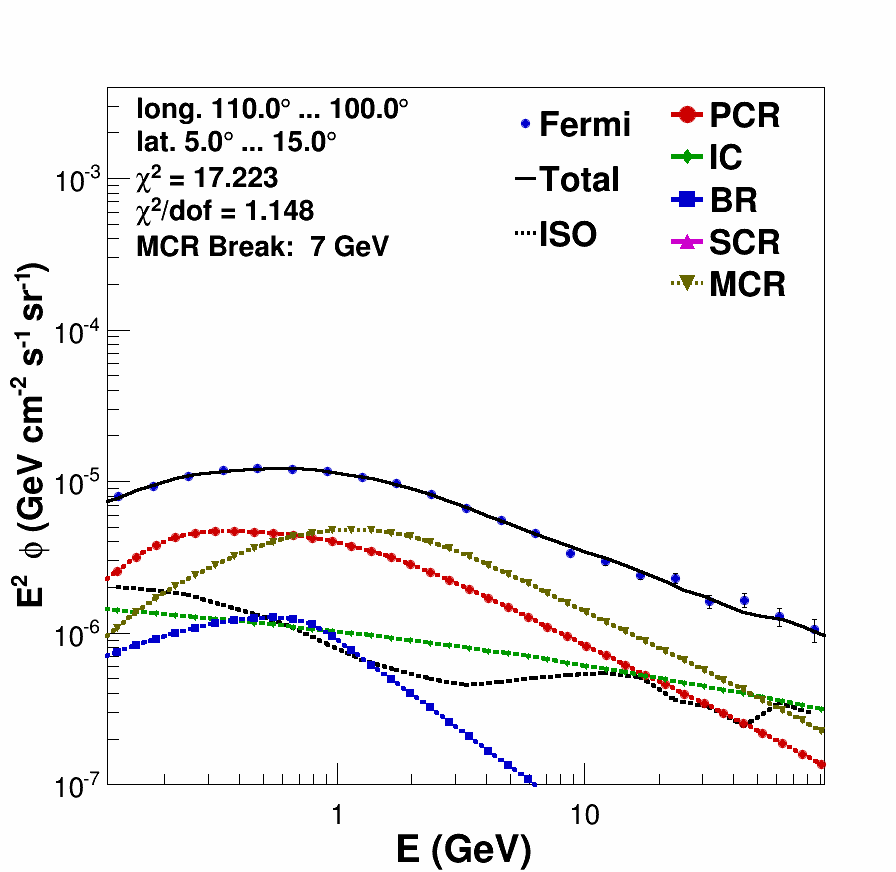}
\includegraphics[width=0.16\textwidth,height=0.16\textwidth,clip]{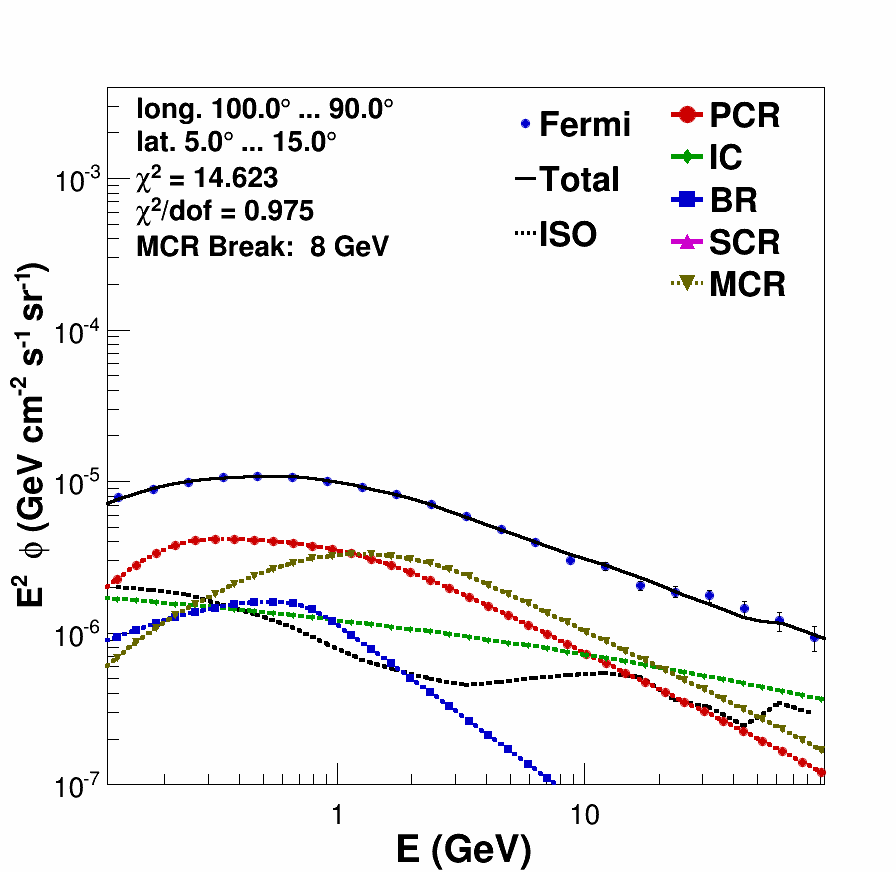}
\includegraphics[width=0.16\textwidth,height=0.16\textwidth,clip]{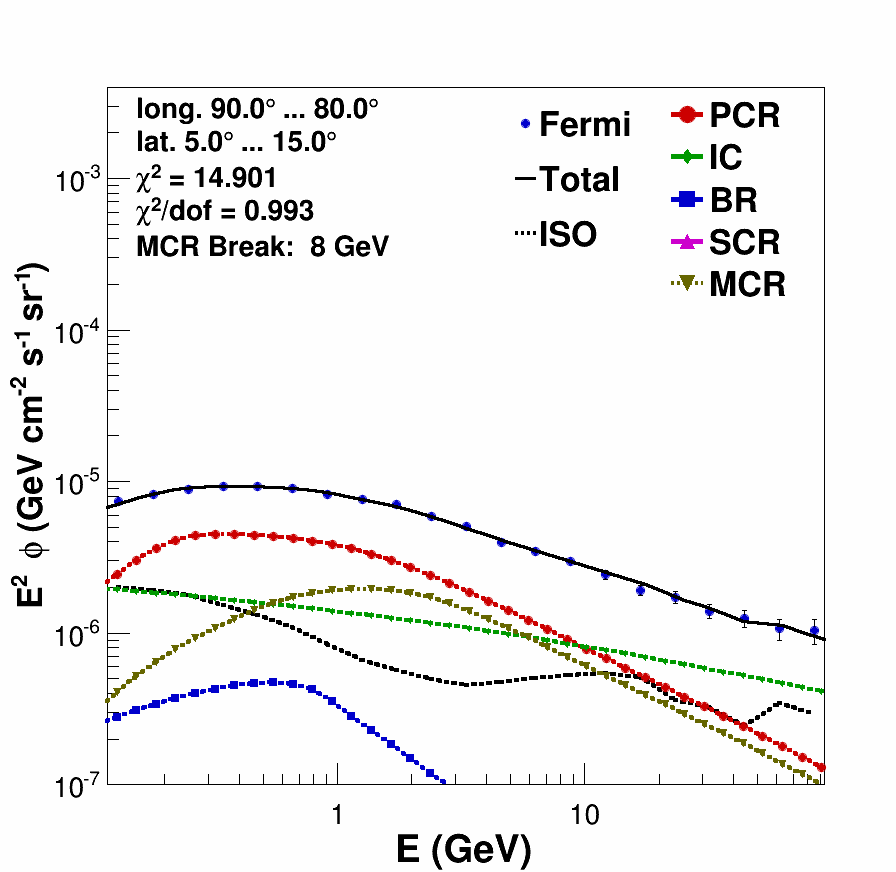}
\includegraphics[width=0.16\textwidth,height=0.16\textwidth,clip]{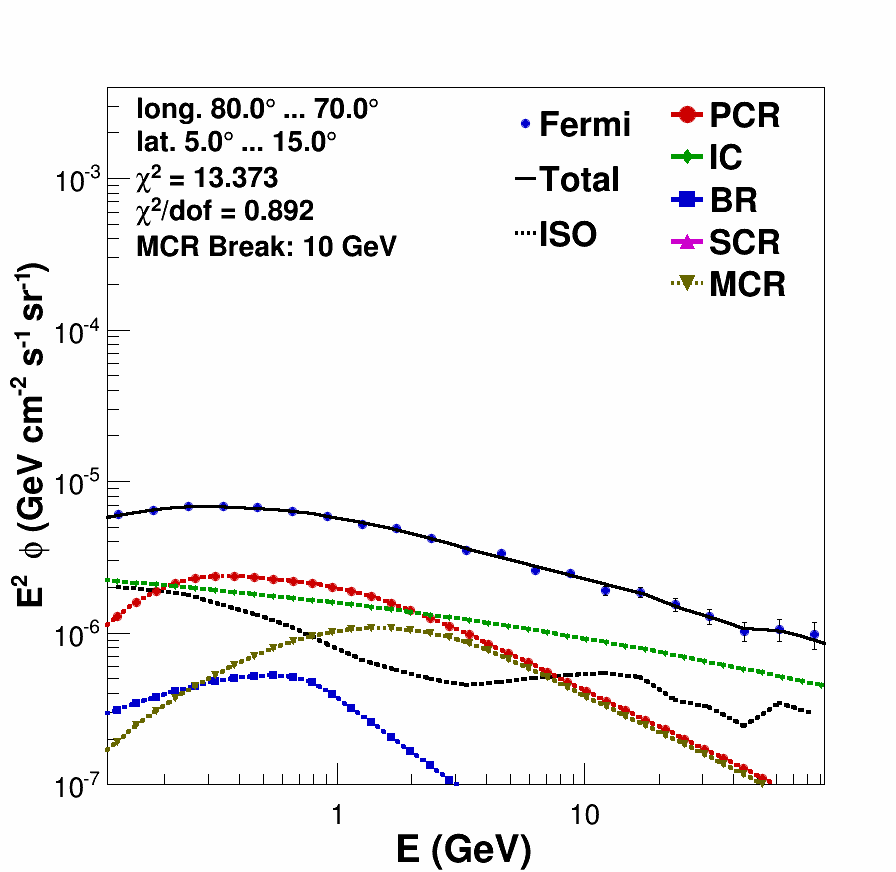}
\includegraphics[width=0.16\textwidth,height=0.16\textwidth,clip]{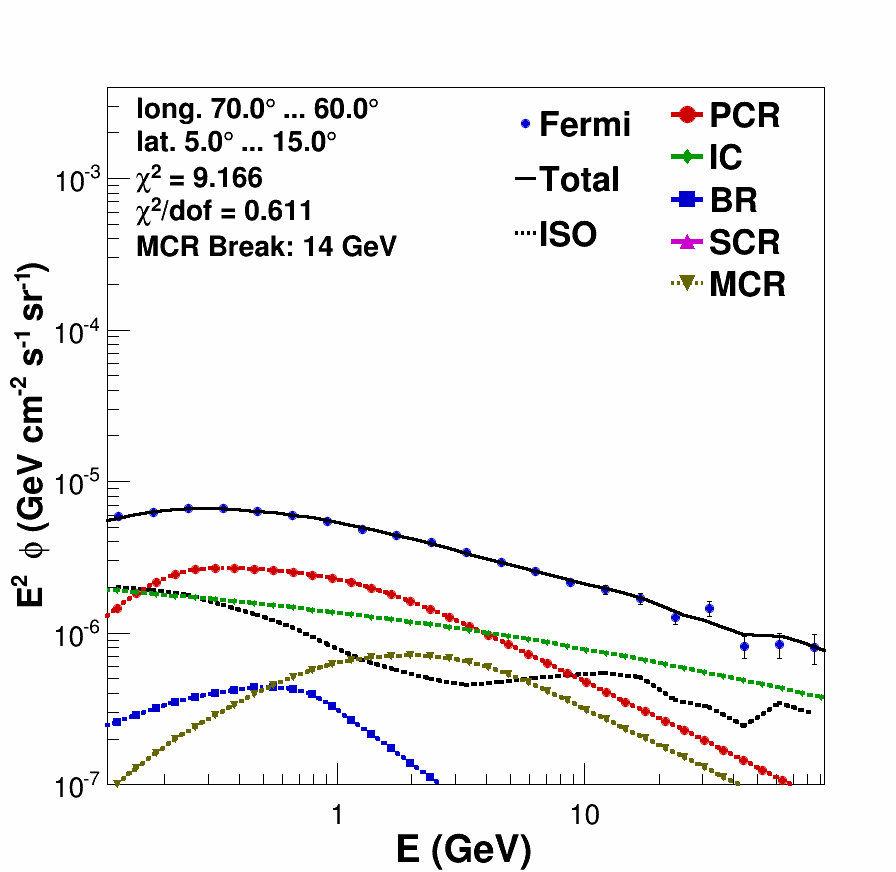}
\includegraphics[width=0.16\textwidth,height=0.16\textwidth,clip]{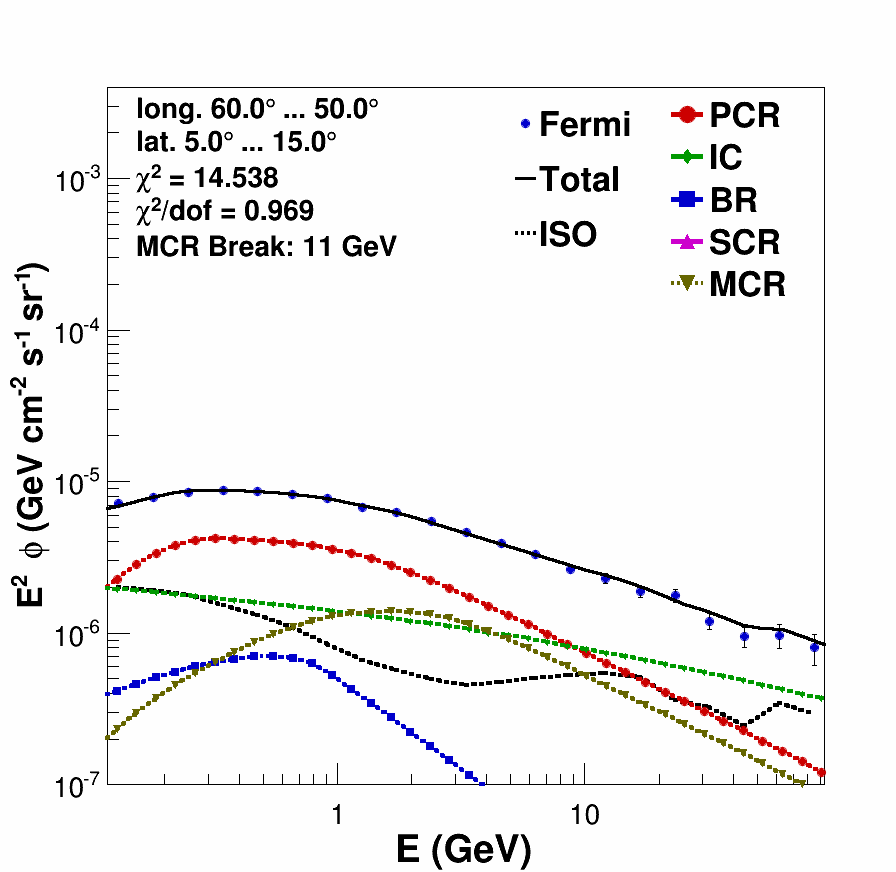}
\includegraphics[width=0.16\textwidth,height=0.16\textwidth,clip]{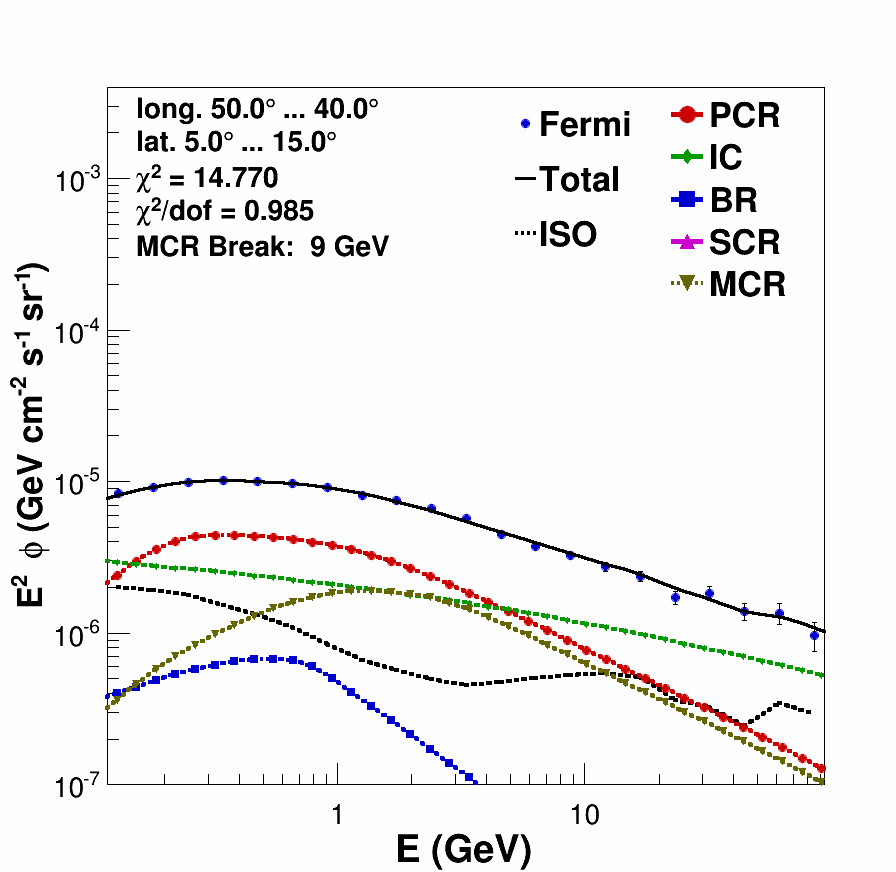}
\includegraphics[width=0.16\textwidth,height=0.16\textwidth,clip]{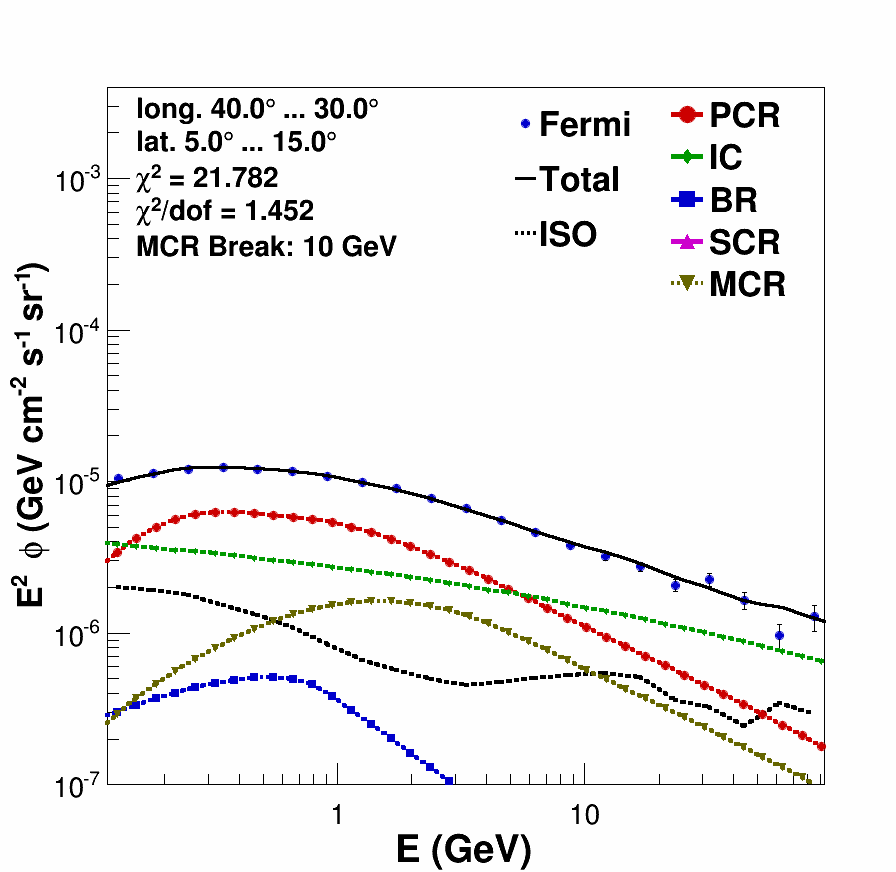}
\includegraphics[width=0.16\textwidth,height=0.16\textwidth,clip]{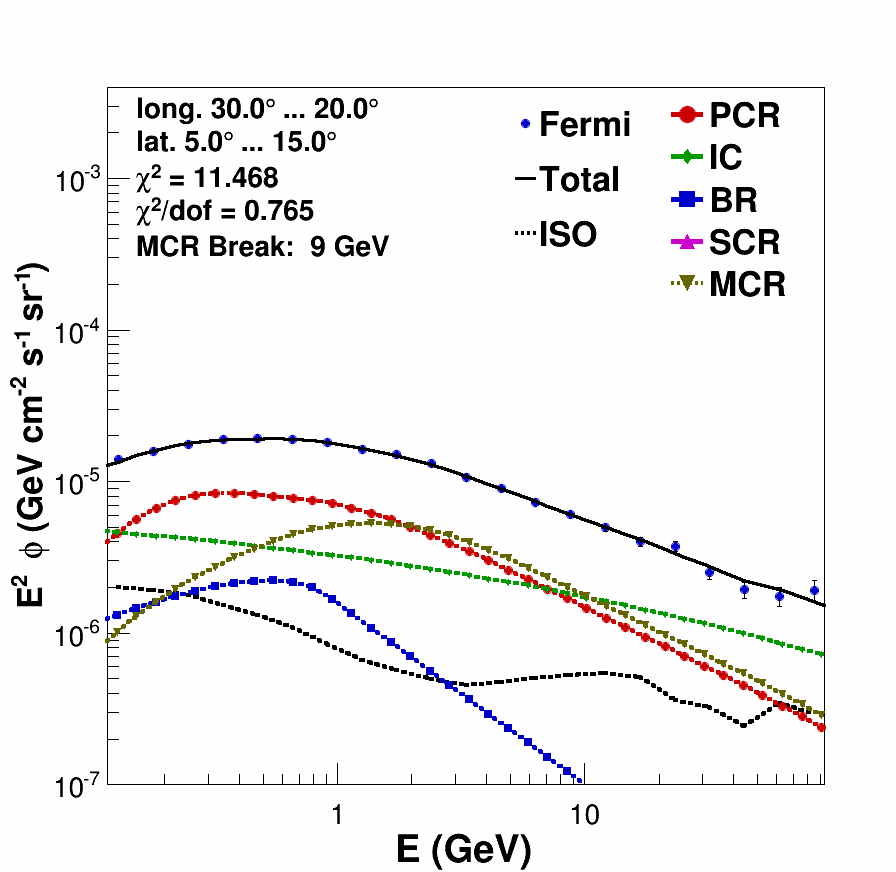}
\includegraphics[width=0.16\textwidth,height=0.16\textwidth,clip]{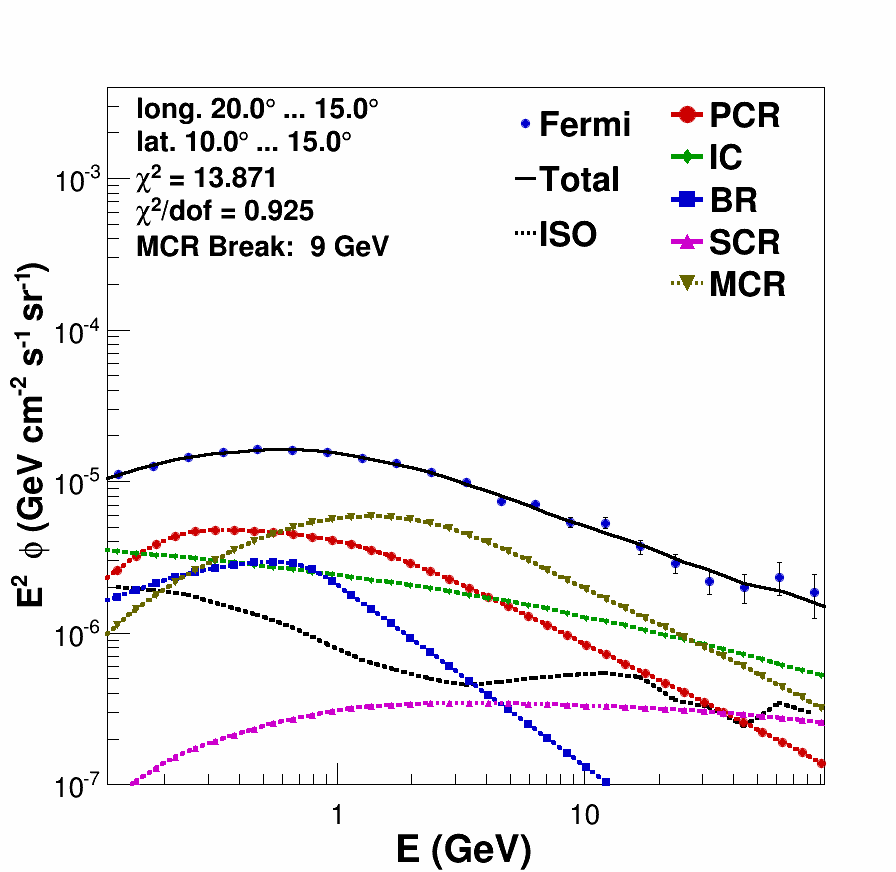}
\includegraphics[width=0.16\textwidth,height=0.16\textwidth,clip]{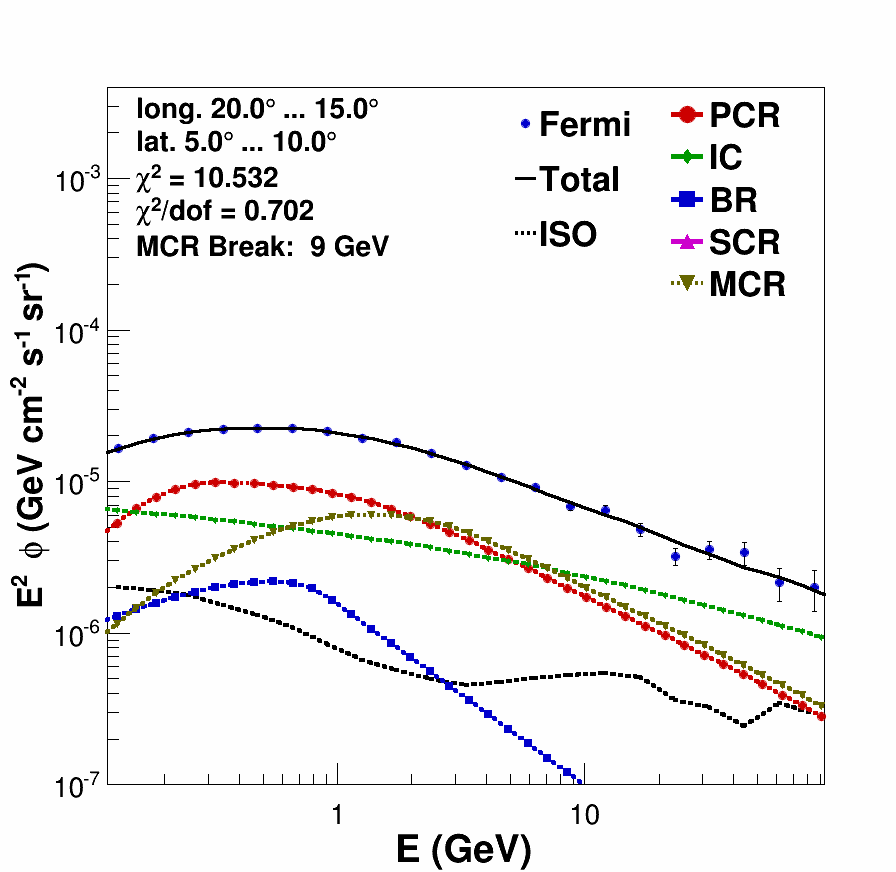}
\includegraphics[width=0.16\textwidth,height=0.16\textwidth,clip]{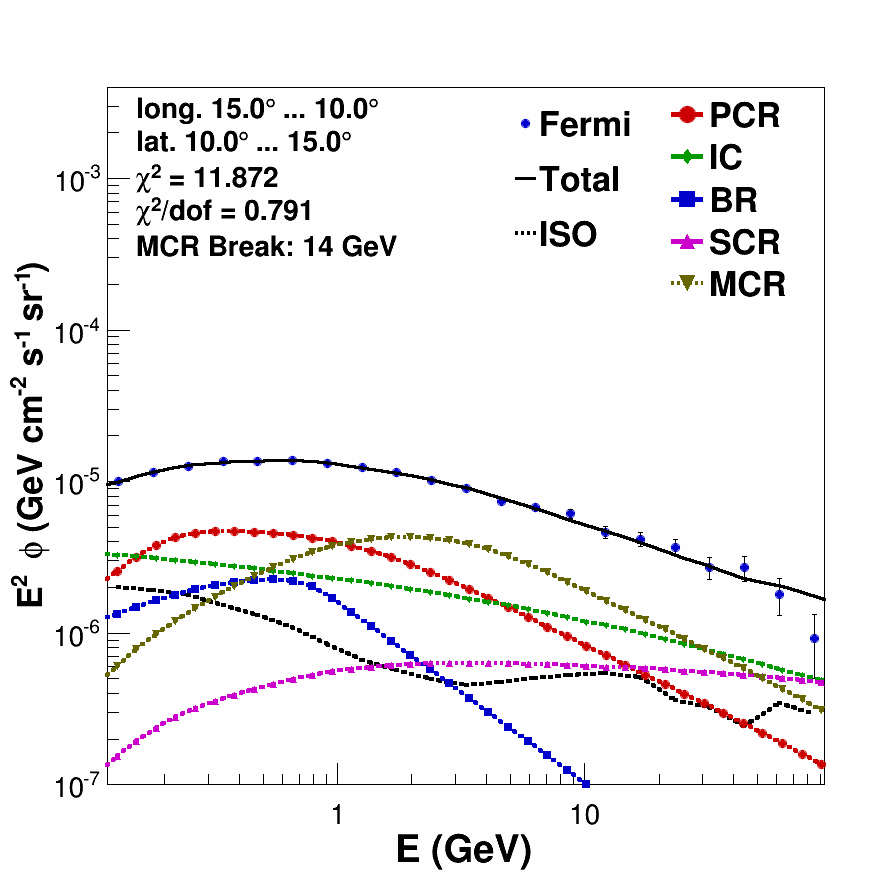}
\includegraphics[width=0.16\textwidth,height=0.16\textwidth,clip]{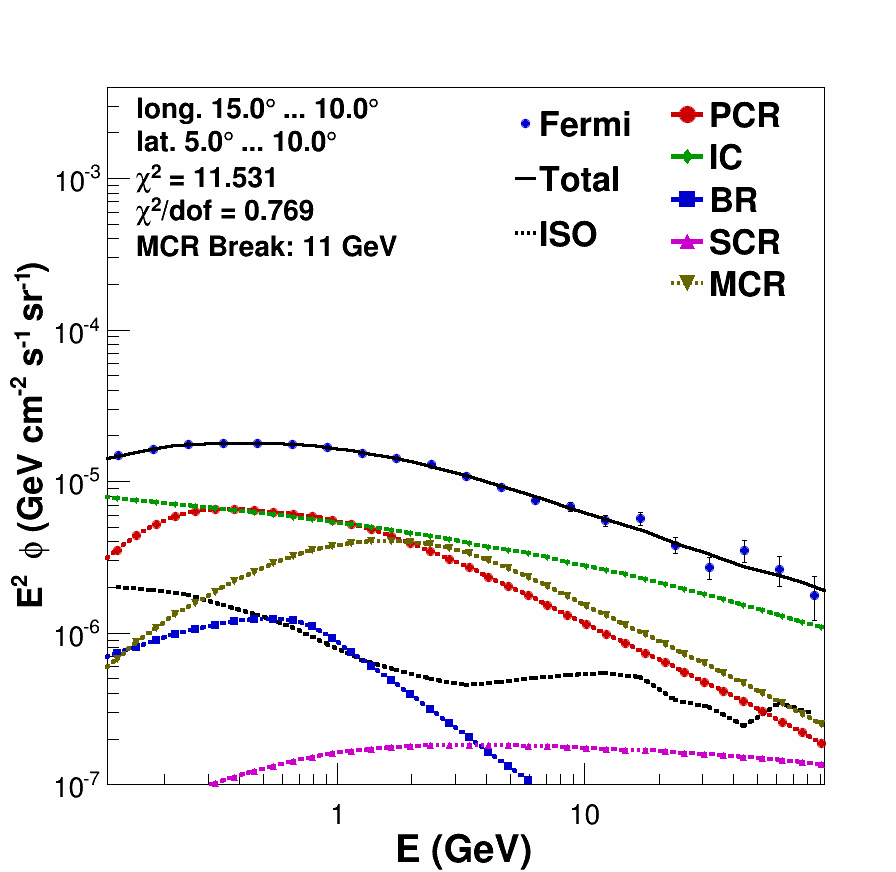}
\includegraphics[width=0.16\textwidth,height=0.16\textwidth,clip]{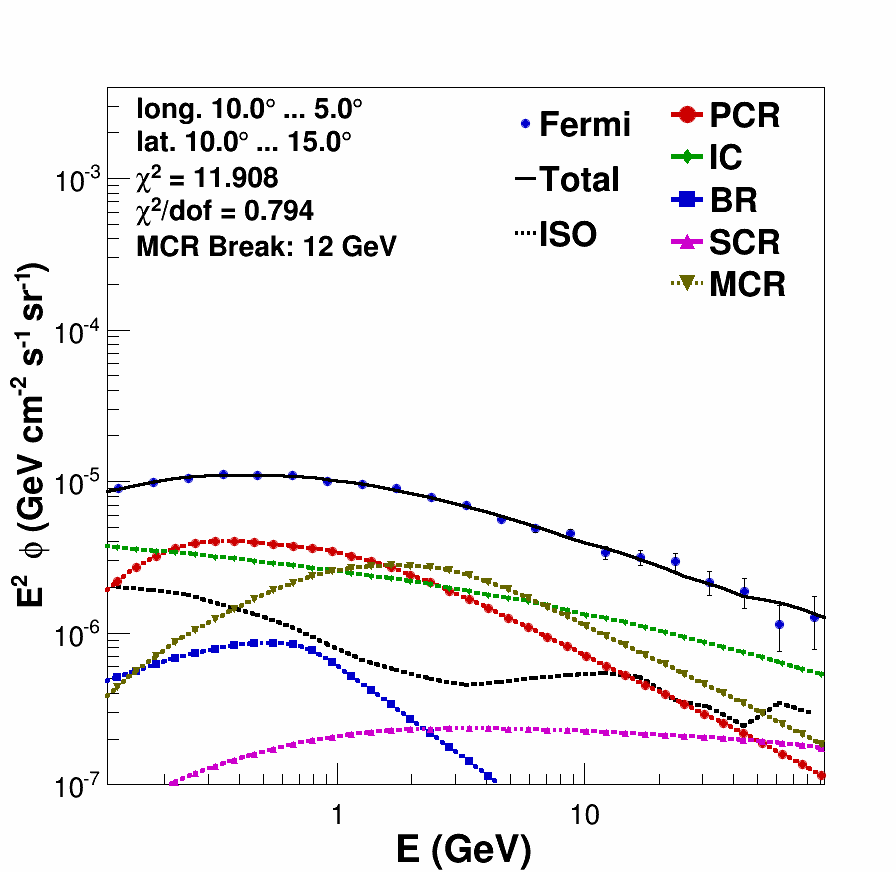}
\includegraphics[width=0.16\textwidth,height=0.16\textwidth,clip]{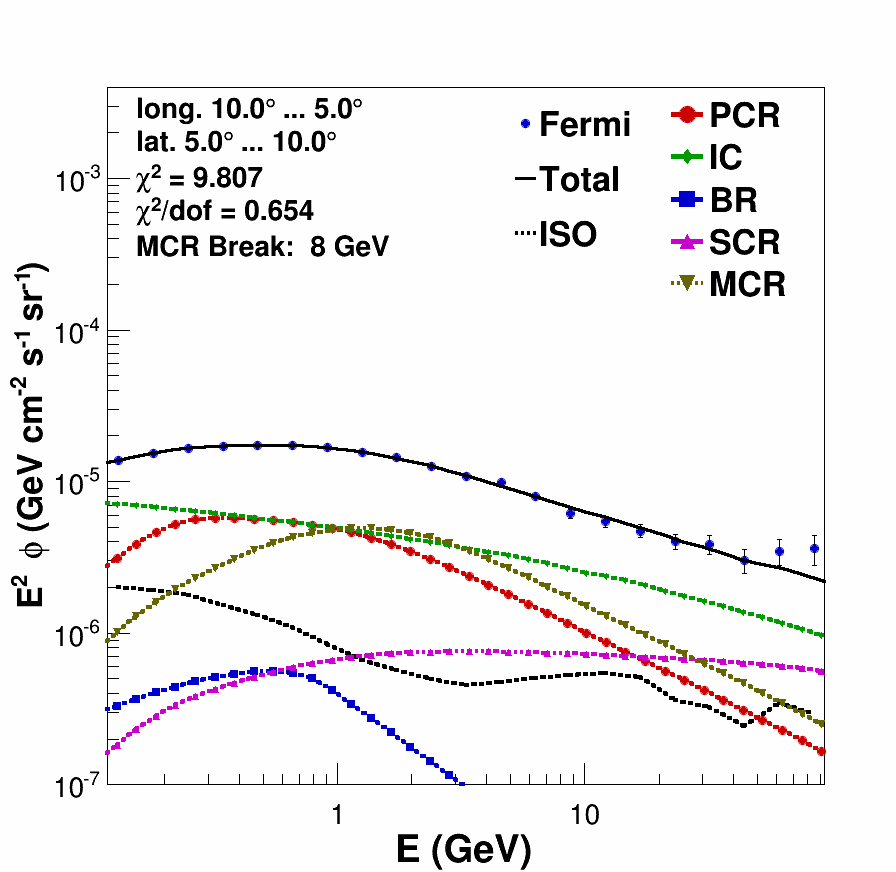}
\includegraphics[width=0.16\textwidth,height=0.16\textwidth,clip]{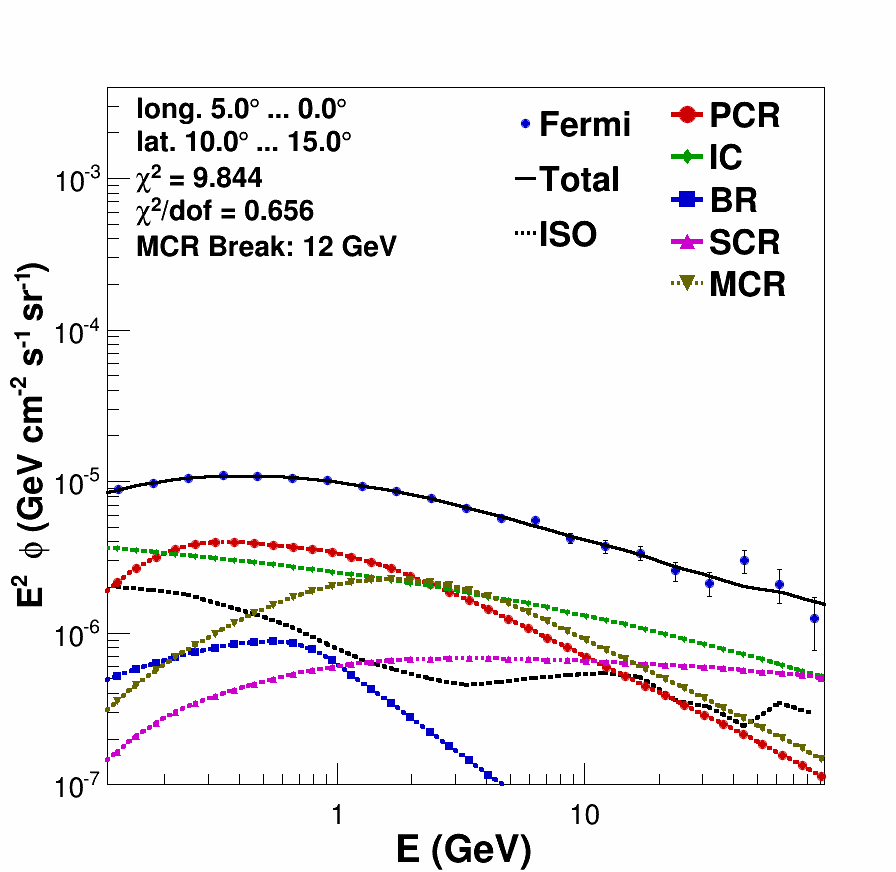}
\includegraphics[width=0.16\textwidth,height=0.16\textwidth,clip]{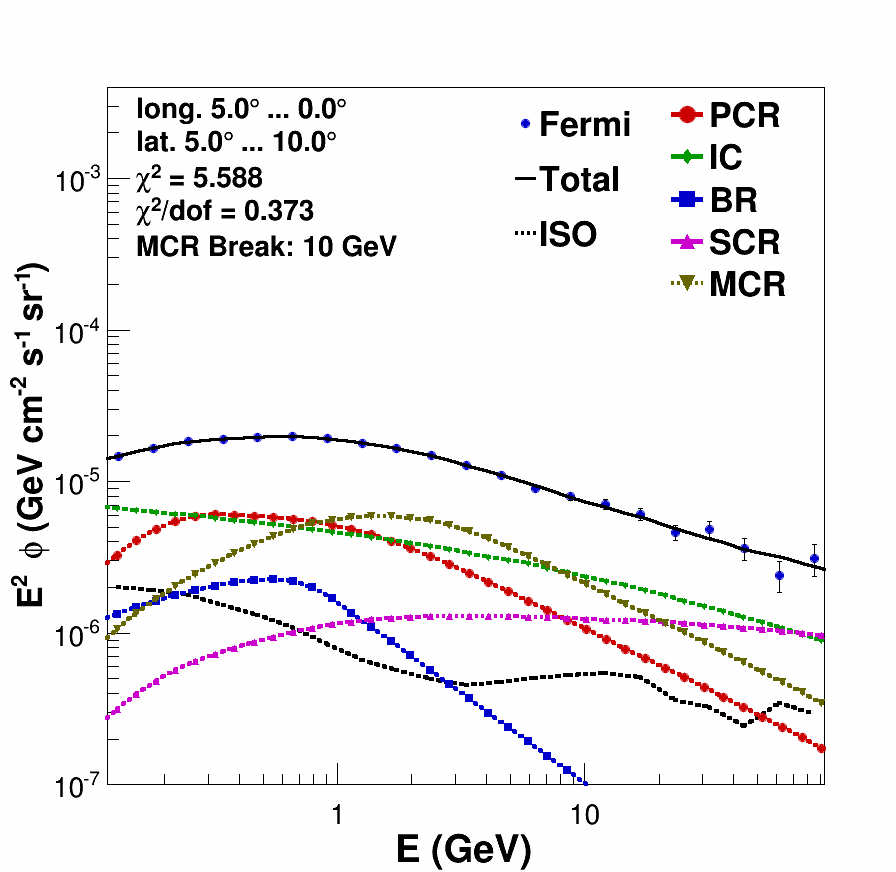}
\includegraphics[width=0.16\textwidth,height=0.16\textwidth,clip]{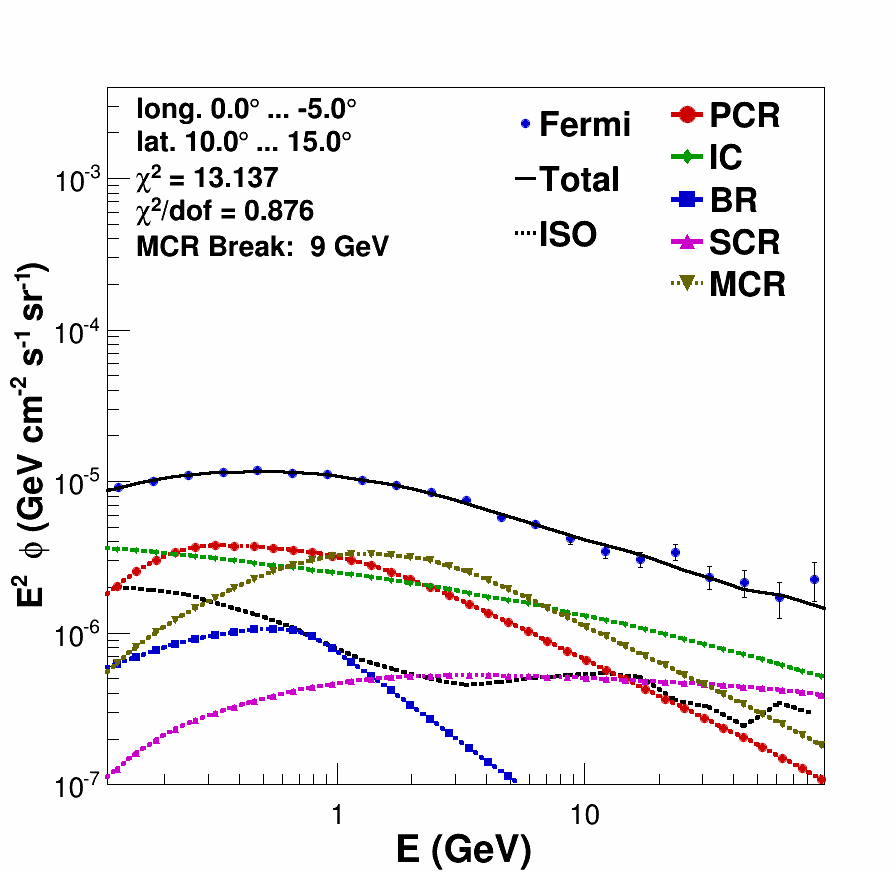}
\includegraphics[width=0.16\textwidth,height=0.16\textwidth,clip]{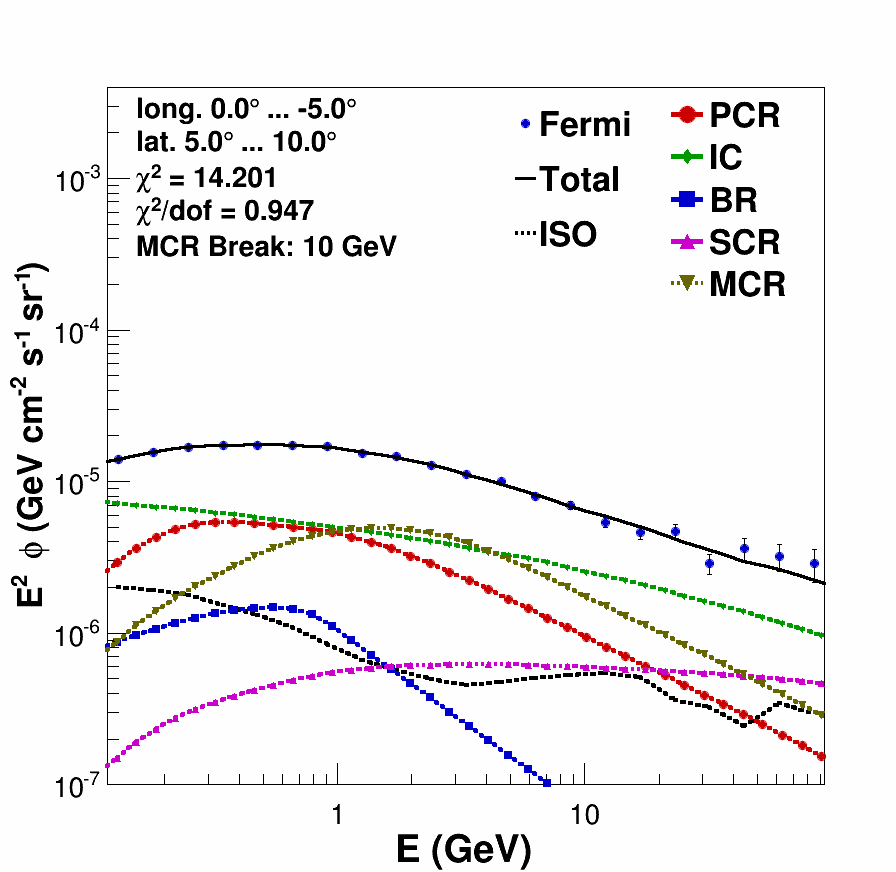}
\includegraphics[width=0.16\textwidth,height=0.16\textwidth,clip]{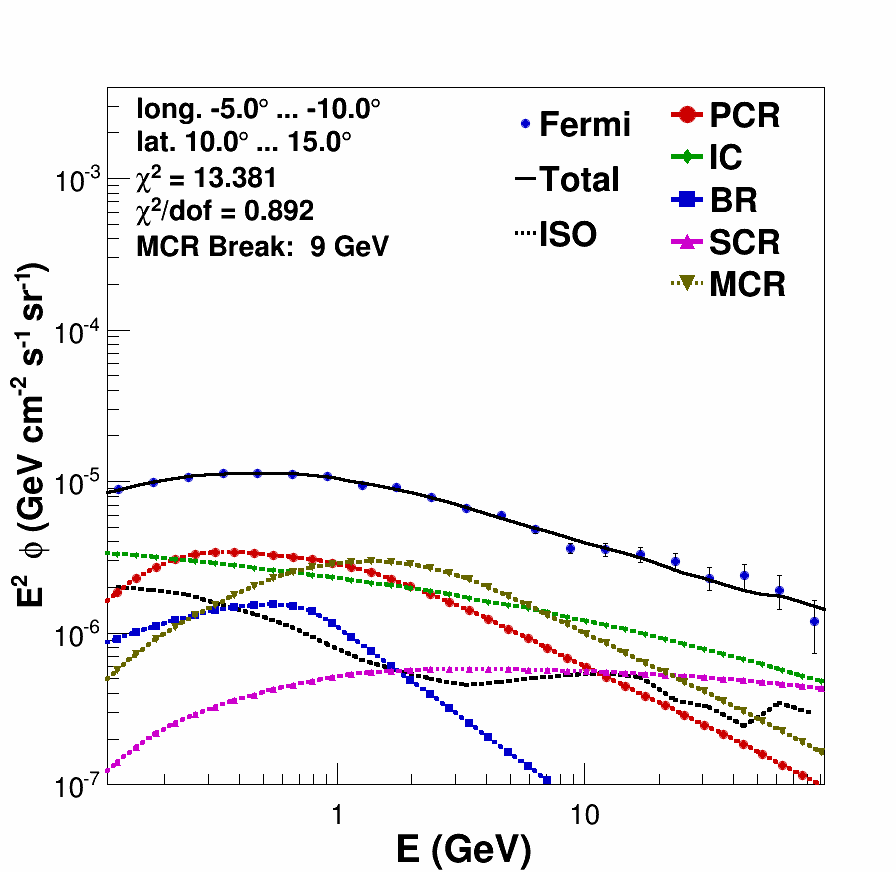}
\includegraphics[width=0.16\textwidth,height=0.16\textwidth,clip]{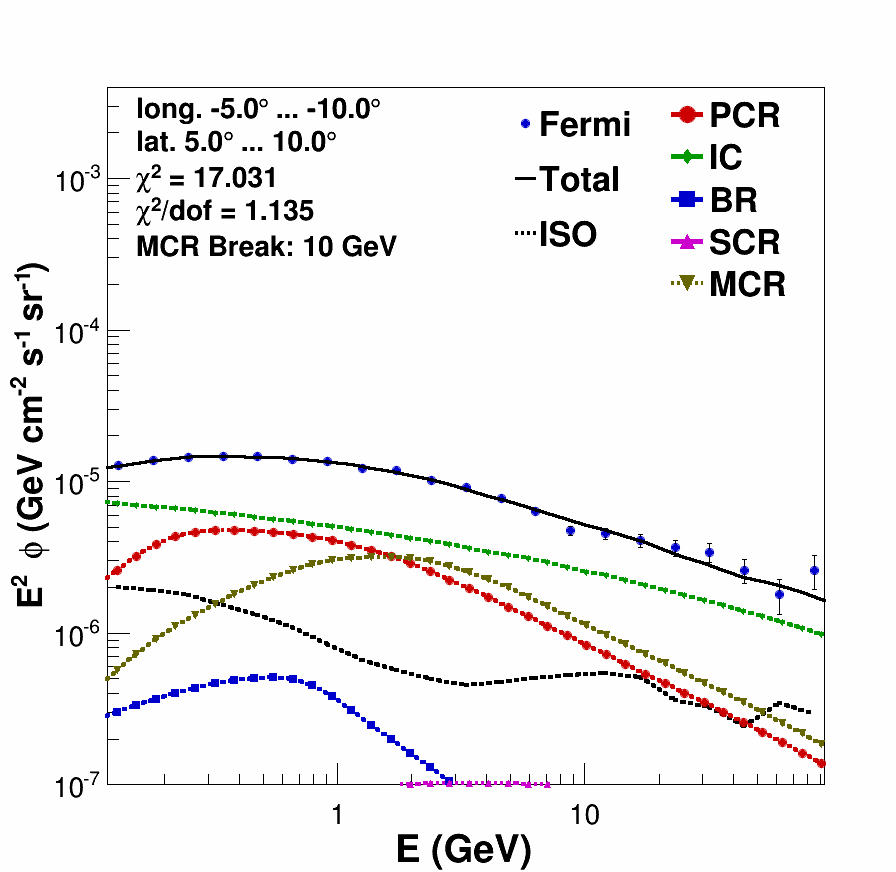}
\includegraphics[width=0.16\textwidth,height=0.16\textwidth,clip]{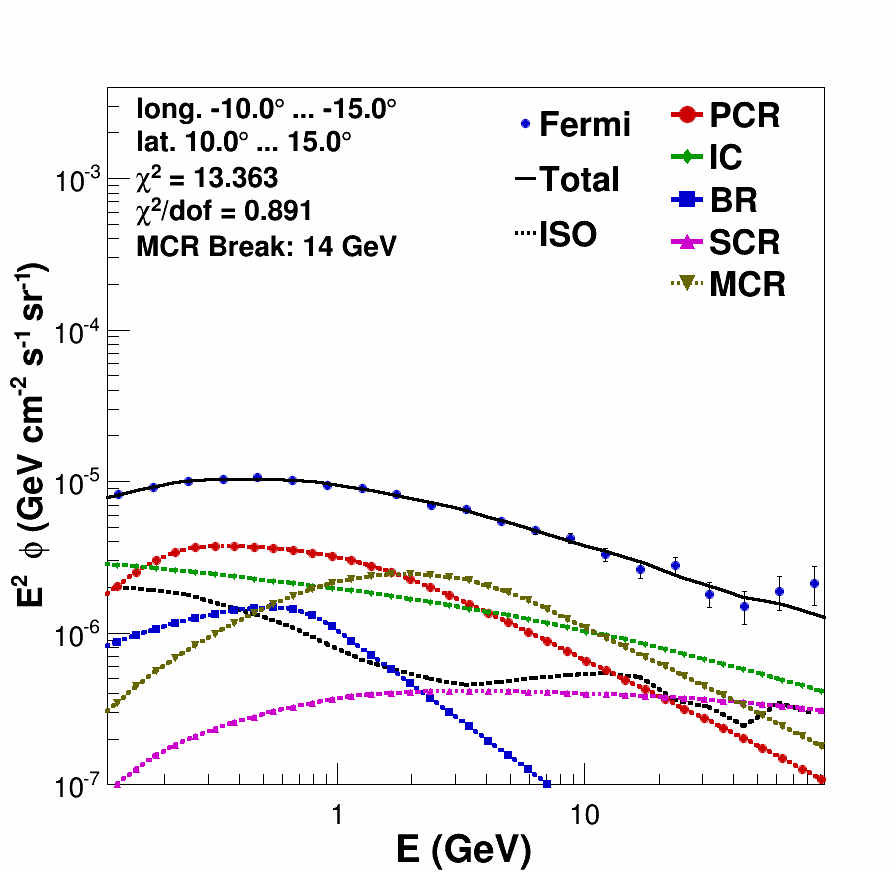}
\includegraphics[width=0.16\textwidth,height=0.16\textwidth,clip]{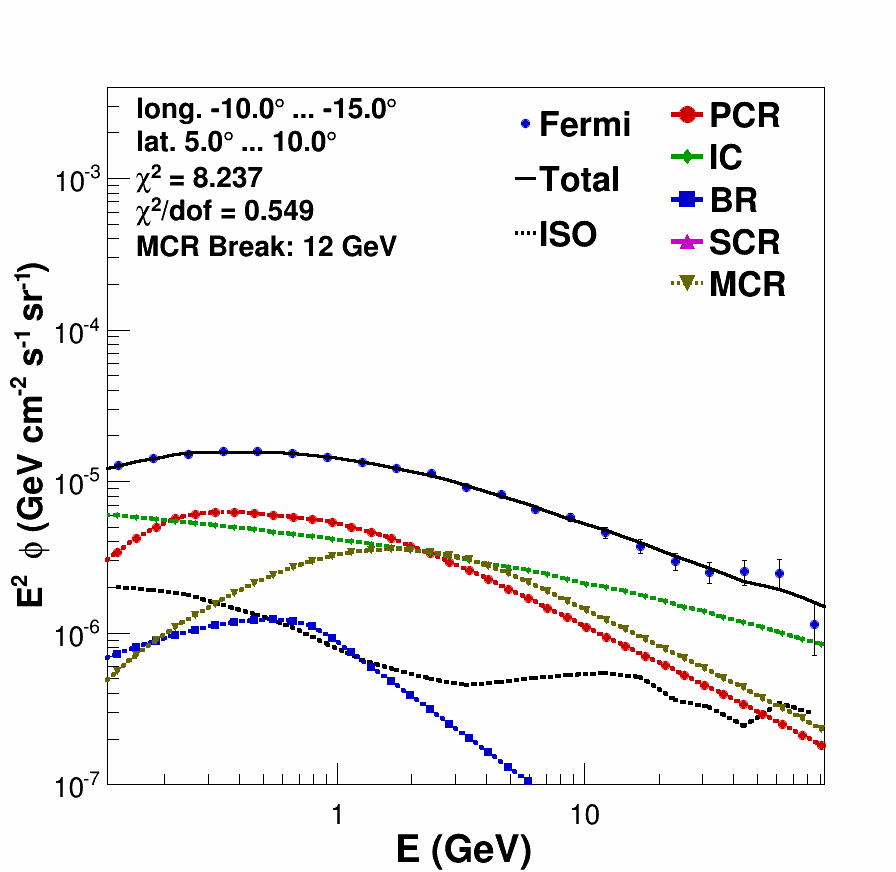}
\includegraphics[width=0.16\textwidth,height=0.16\textwidth,clip]{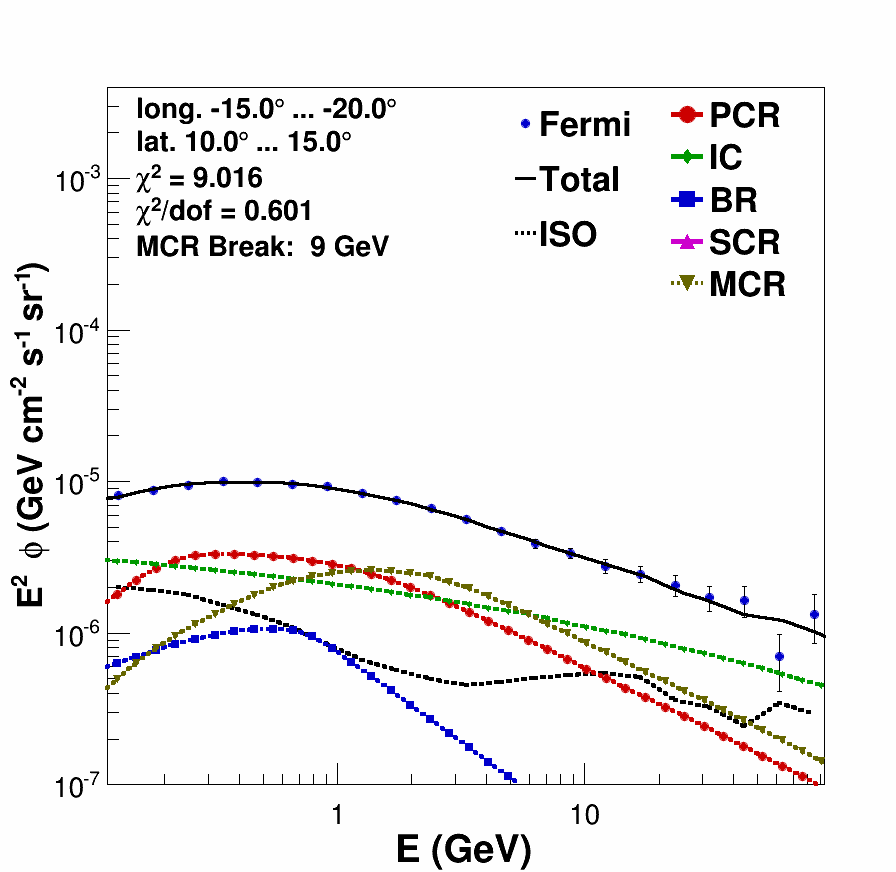}
\includegraphics[width=0.16\textwidth,height=0.16\textwidth,clip]{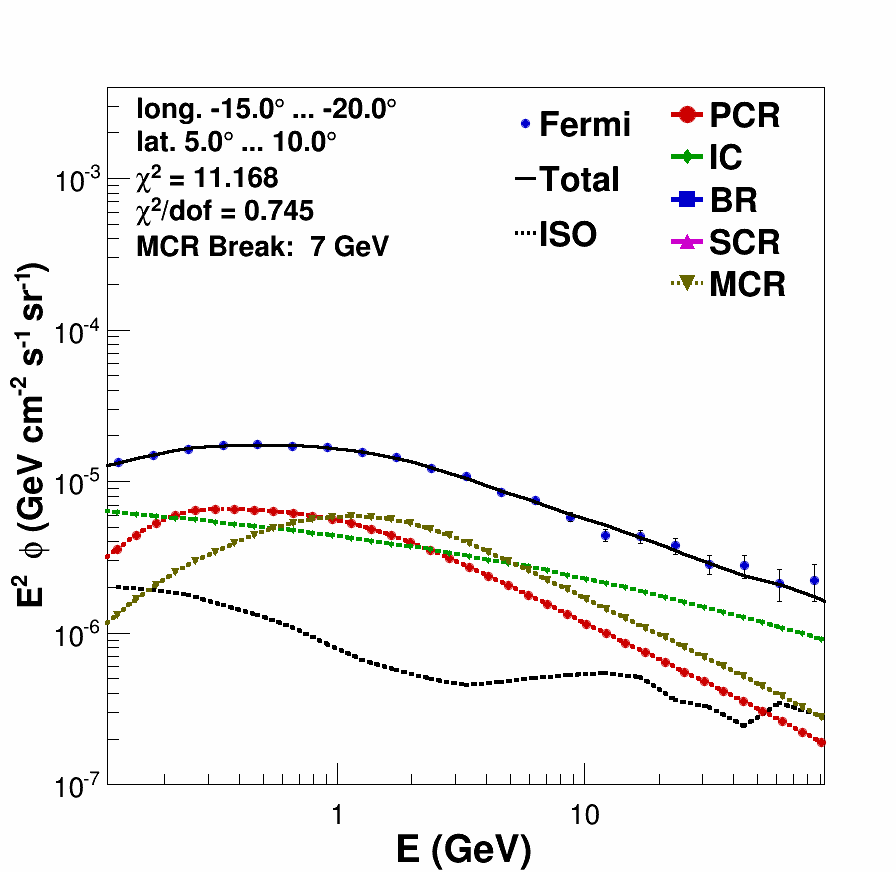}
\includegraphics[width=0.16\textwidth,height=0.16\textwidth,clip]{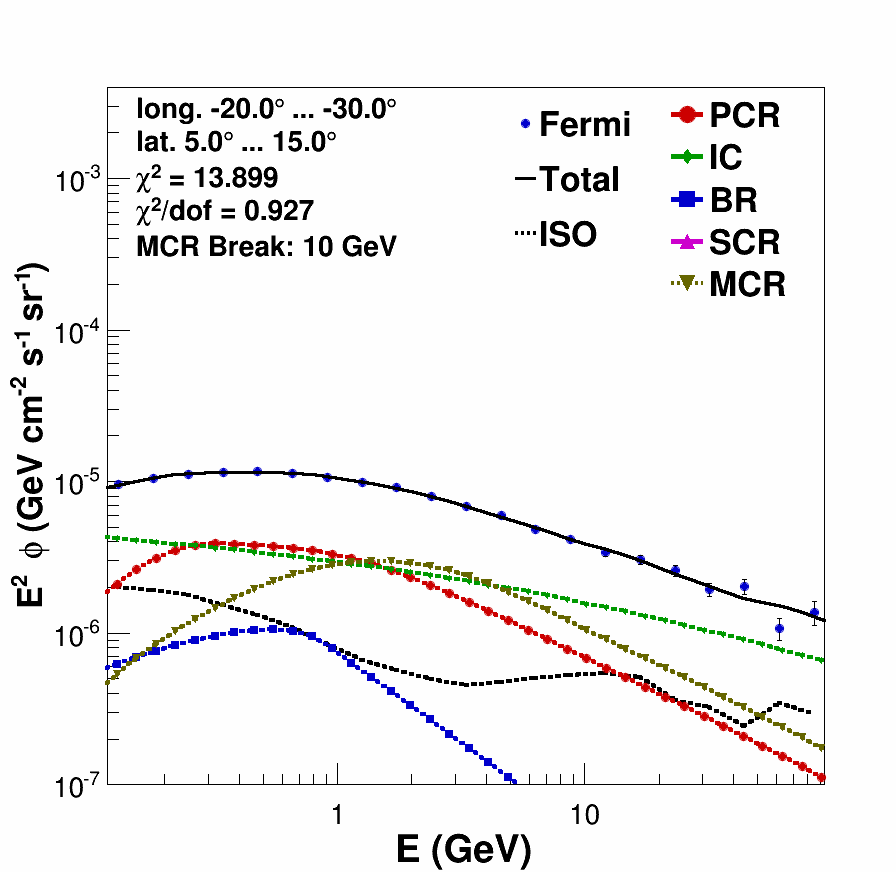}
\includegraphics[width=0.16\textwidth,height=0.16\textwidth,clip]{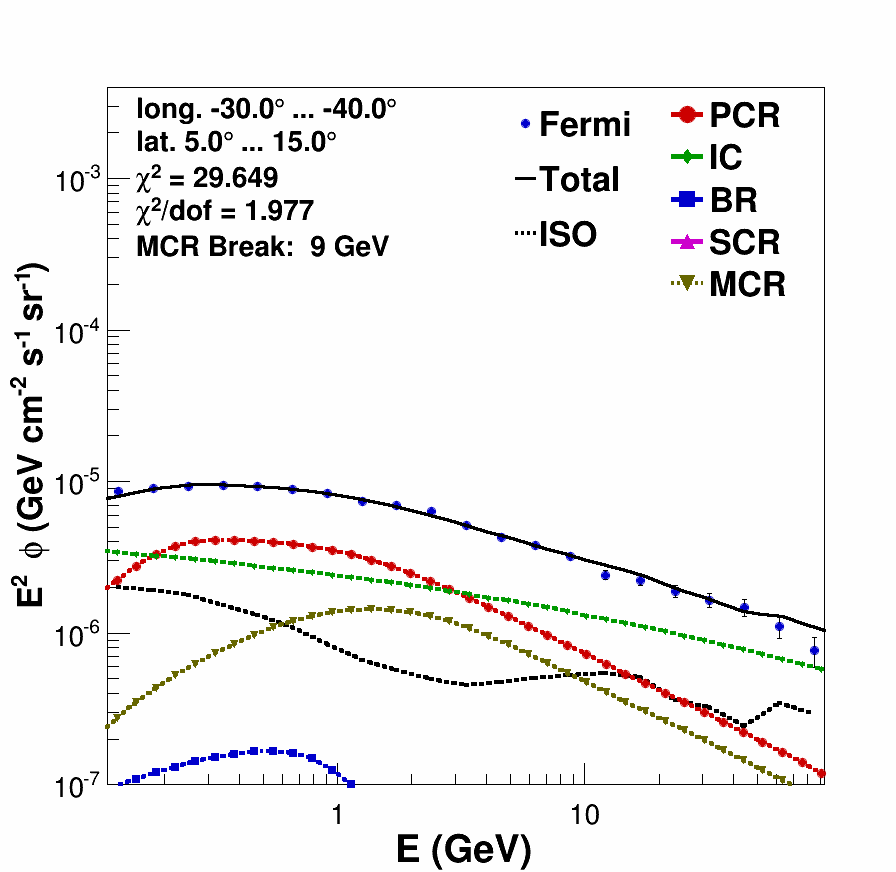}
\includegraphics[width=0.16\textwidth,height=0.16\textwidth,clip]{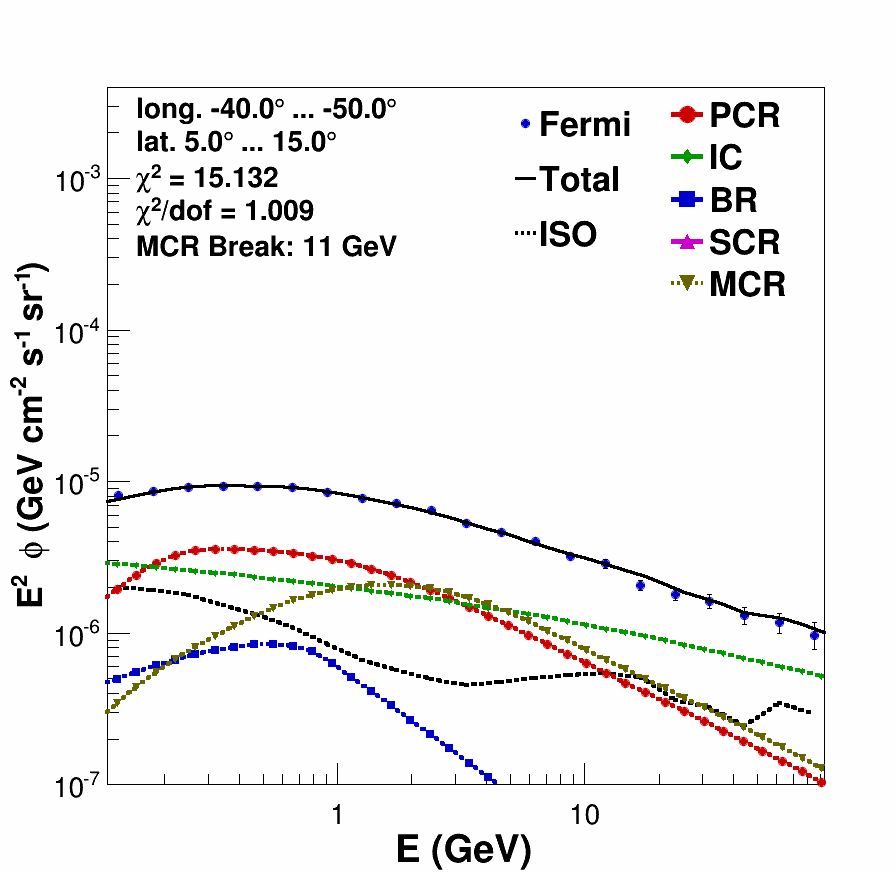}
\includegraphics[width=0.16\textwidth,height=0.16\textwidth,clip]{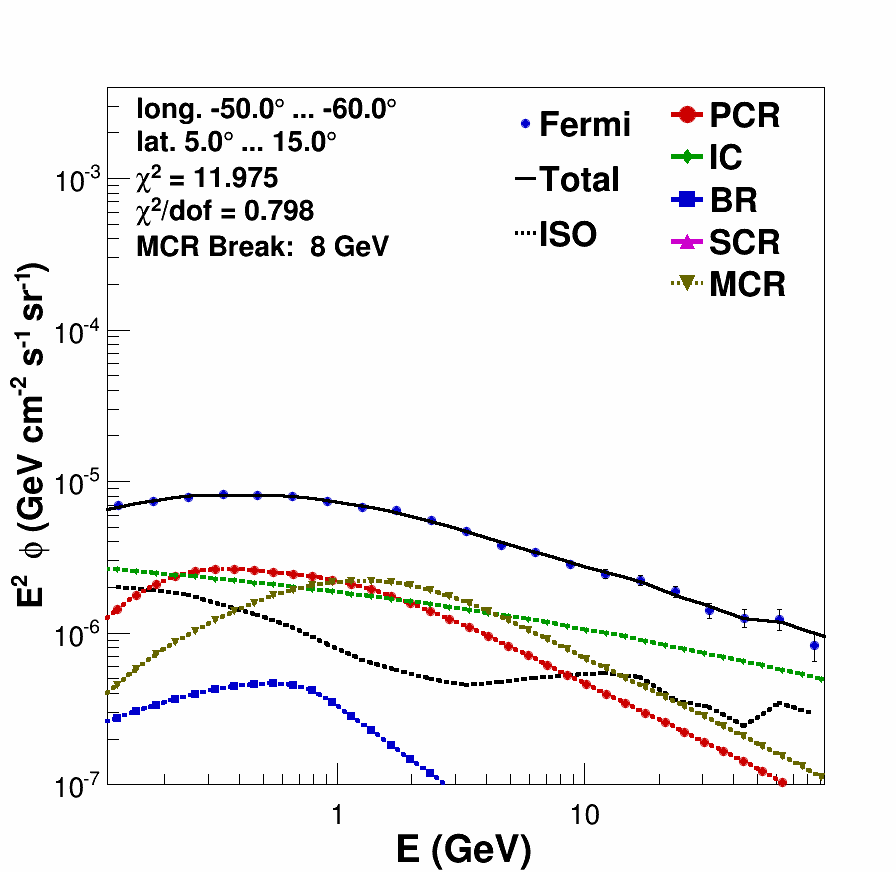}
\includegraphics[width=0.16\textwidth,height=0.16\textwidth,clip]{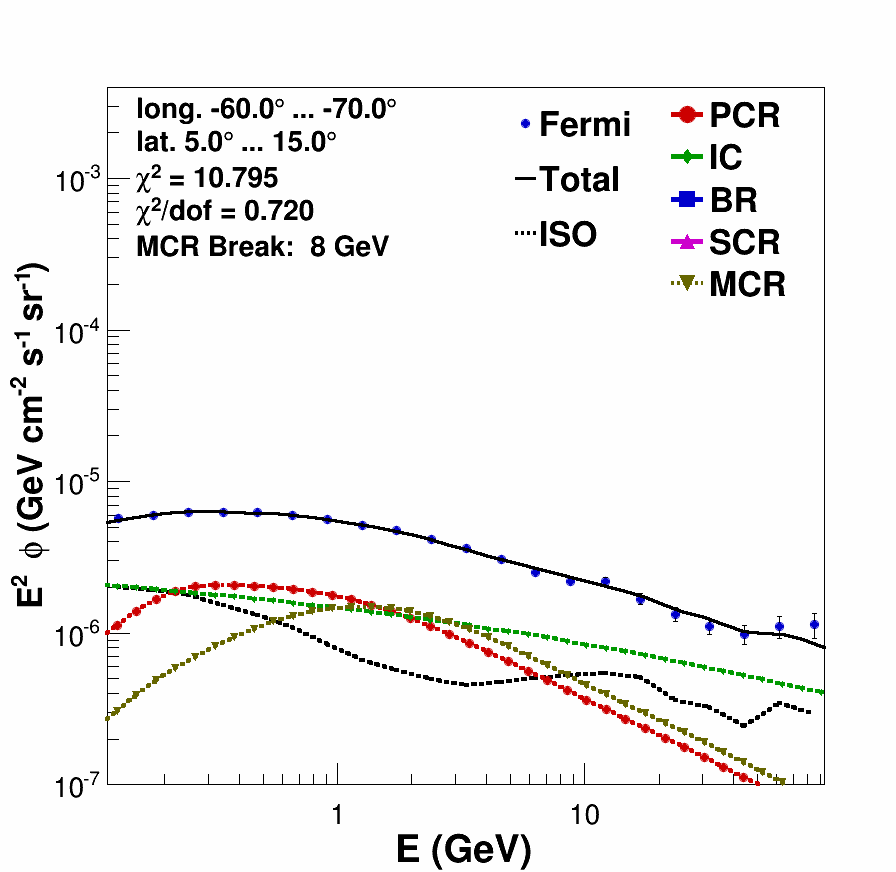}
\includegraphics[width=0.16\textwidth,height=0.16\textwidth,clip]{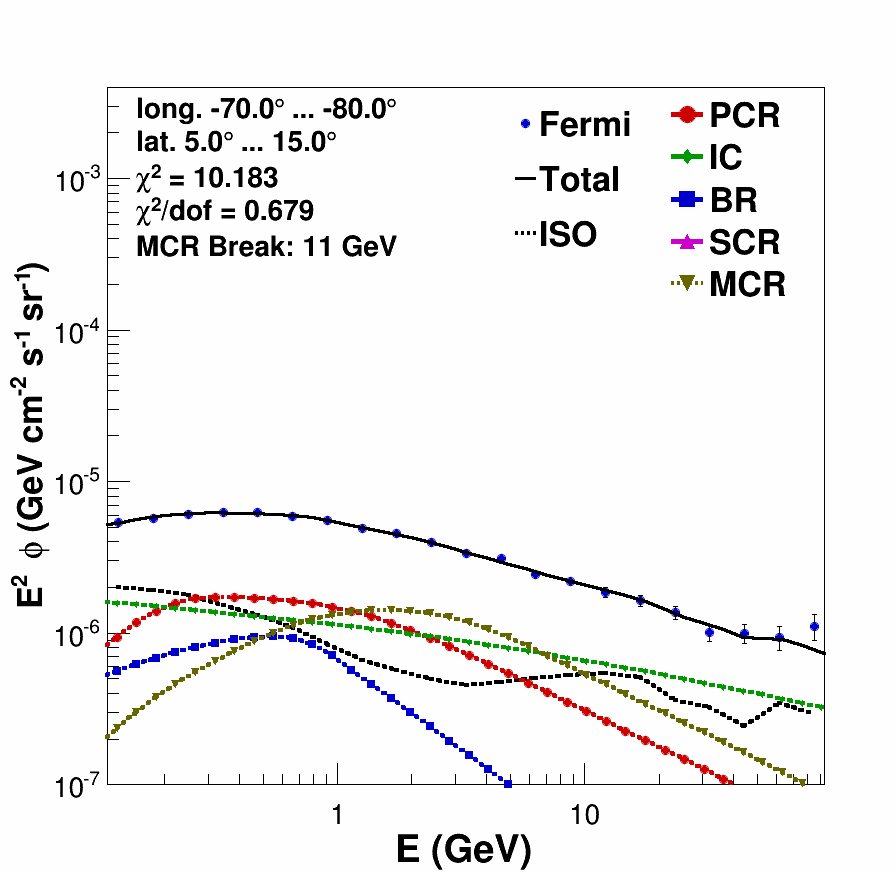}
\includegraphics[width=0.16\textwidth,height=0.16\textwidth,clip]{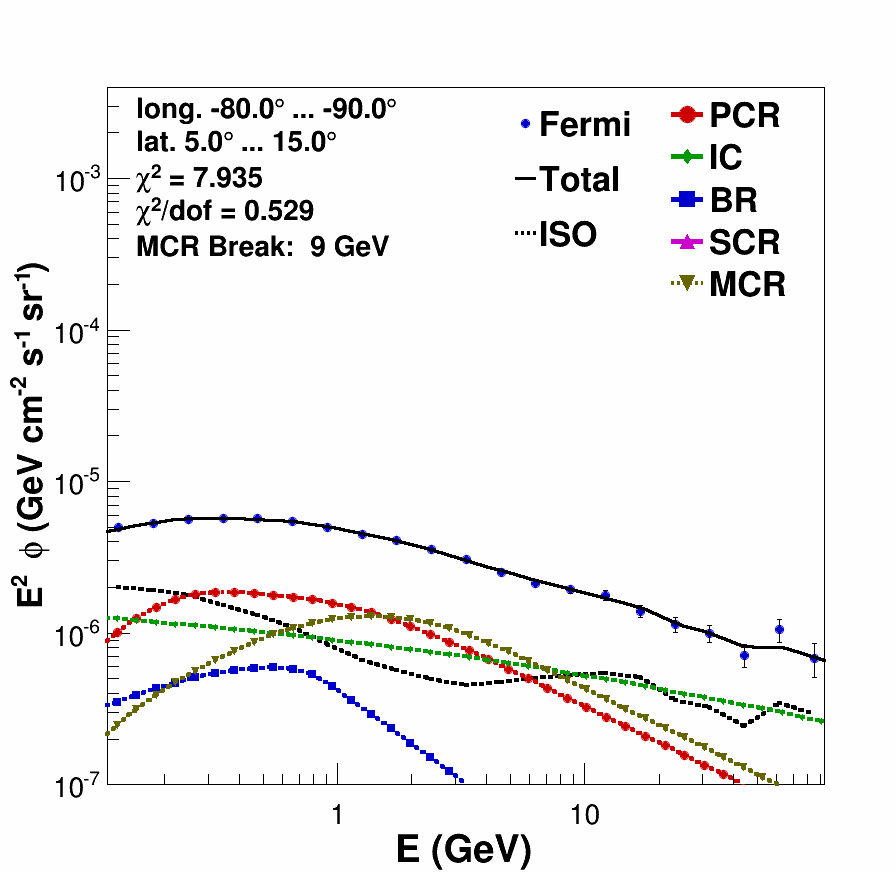}
\includegraphics[width=0.16\textwidth,height=0.16\textwidth,clip]{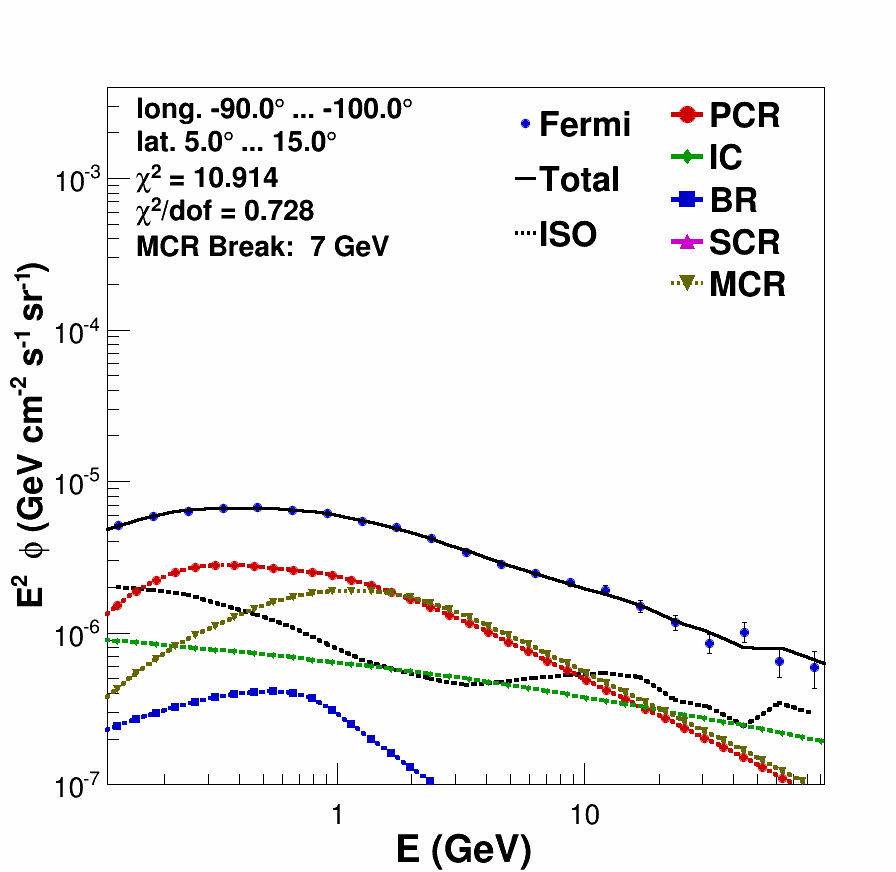}
\includegraphics[width=0.16\textwidth,height=0.16\textwidth,clip]{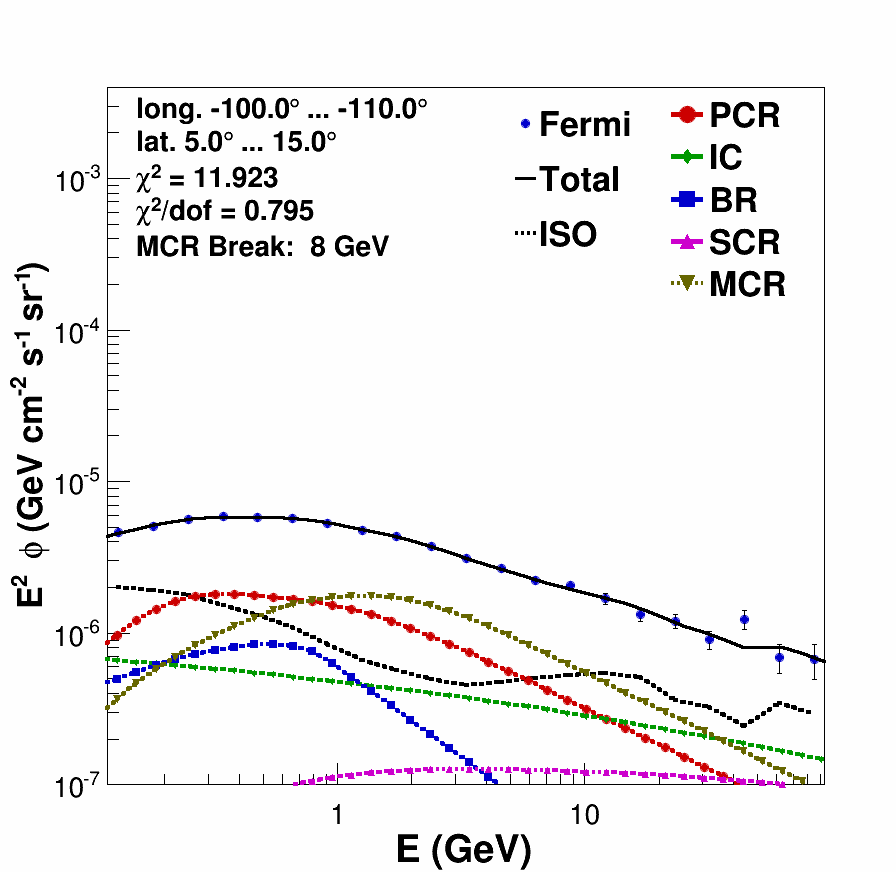}
\includegraphics[width=0.16\textwidth,height=0.16\textwidth,clip]{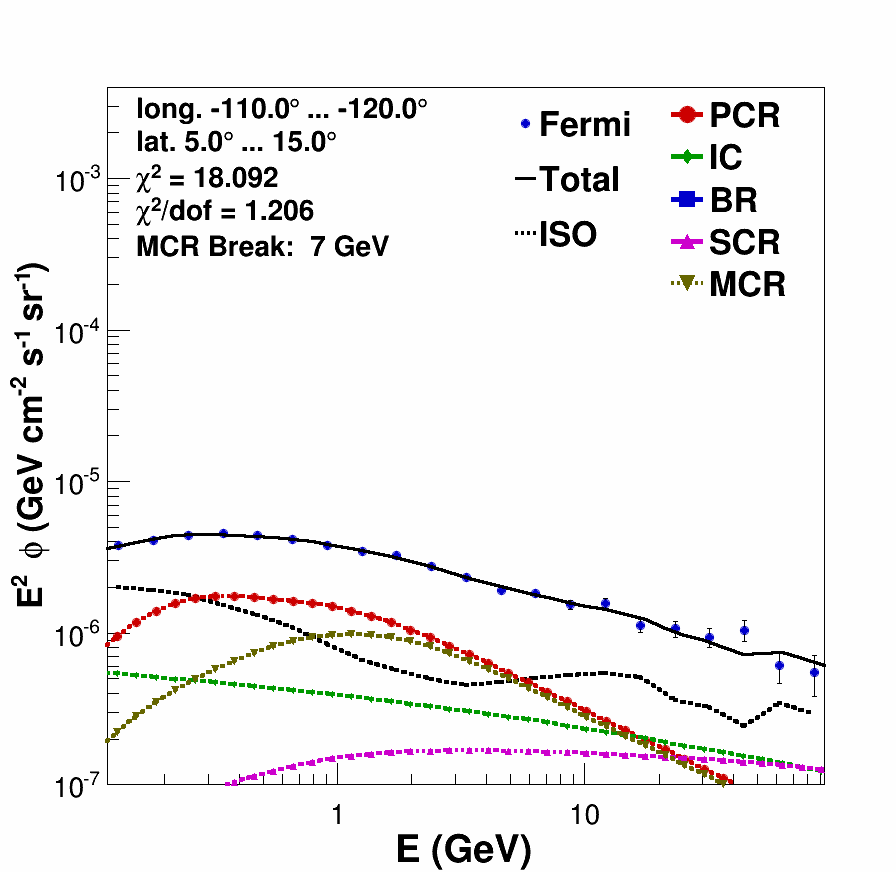}
\includegraphics[width=0.16\textwidth,height=0.16\textwidth,clip]{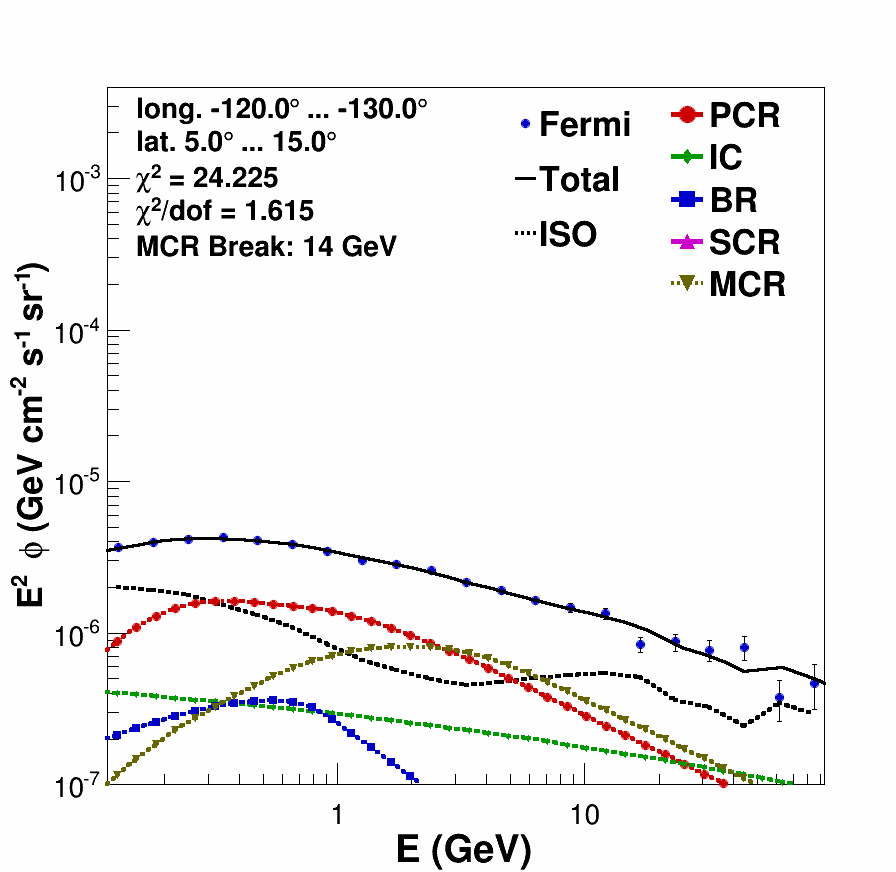}
\includegraphics[width=0.16\textwidth,height=0.16\textwidth,clip]{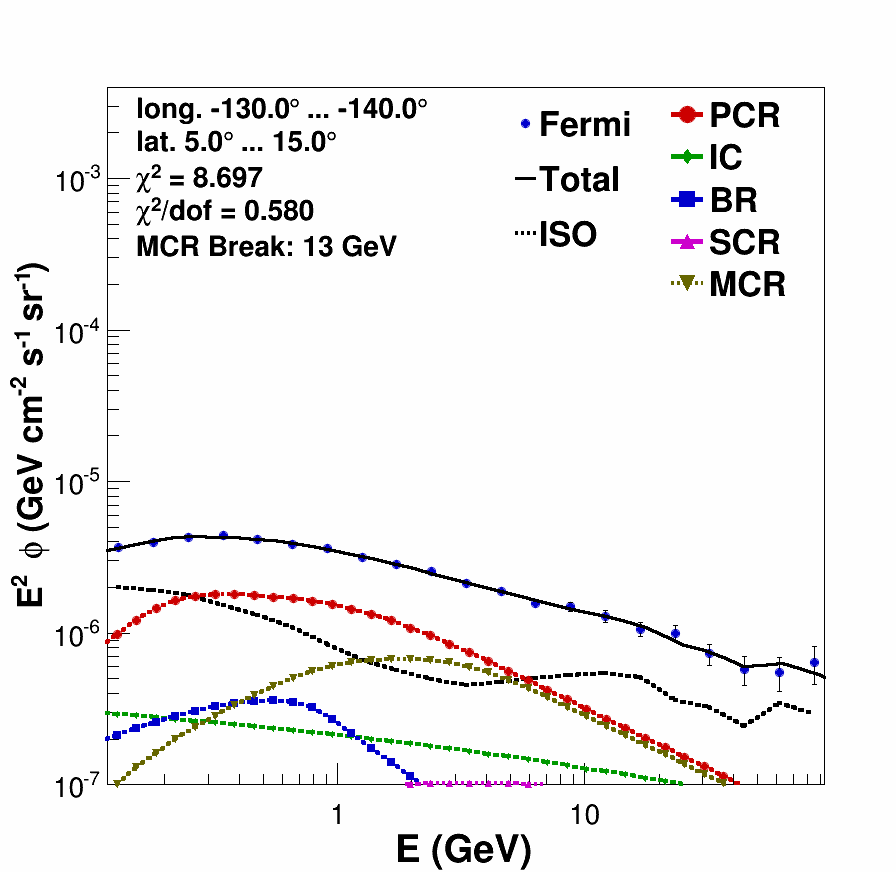}
\includegraphics[width=0.16\textwidth,height=0.16\textwidth,clip]{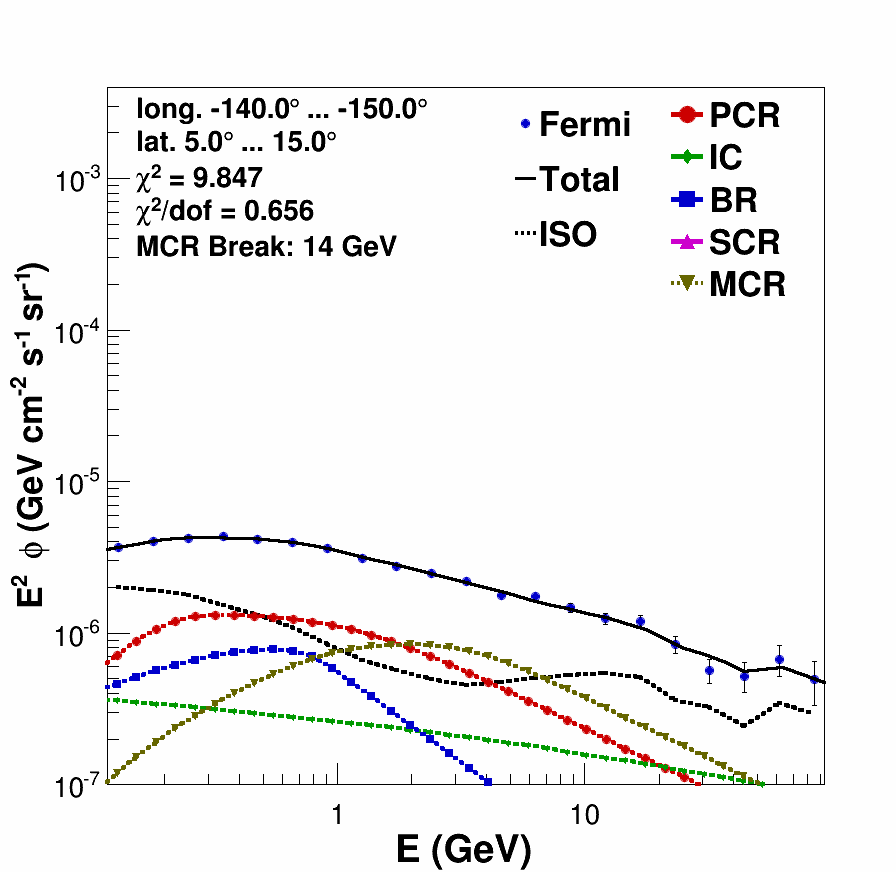}
\includegraphics[width=0.16\textwidth,height=0.16\textwidth,clip]{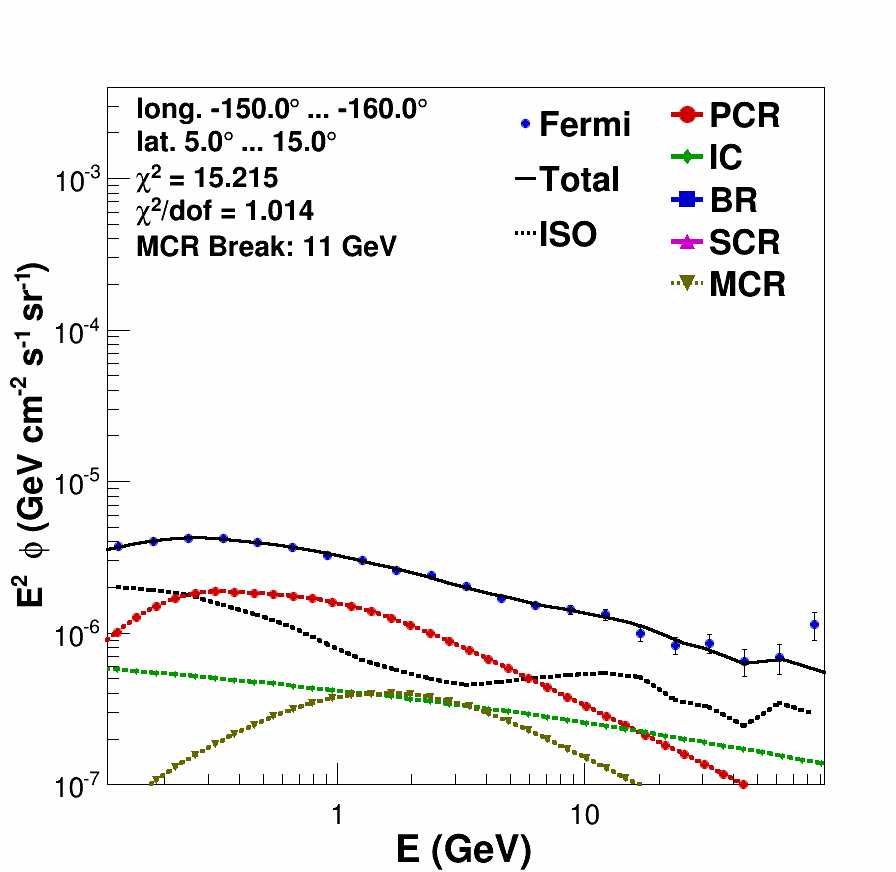}
\includegraphics[width=0.16\textwidth,height=0.16\textwidth,clip]{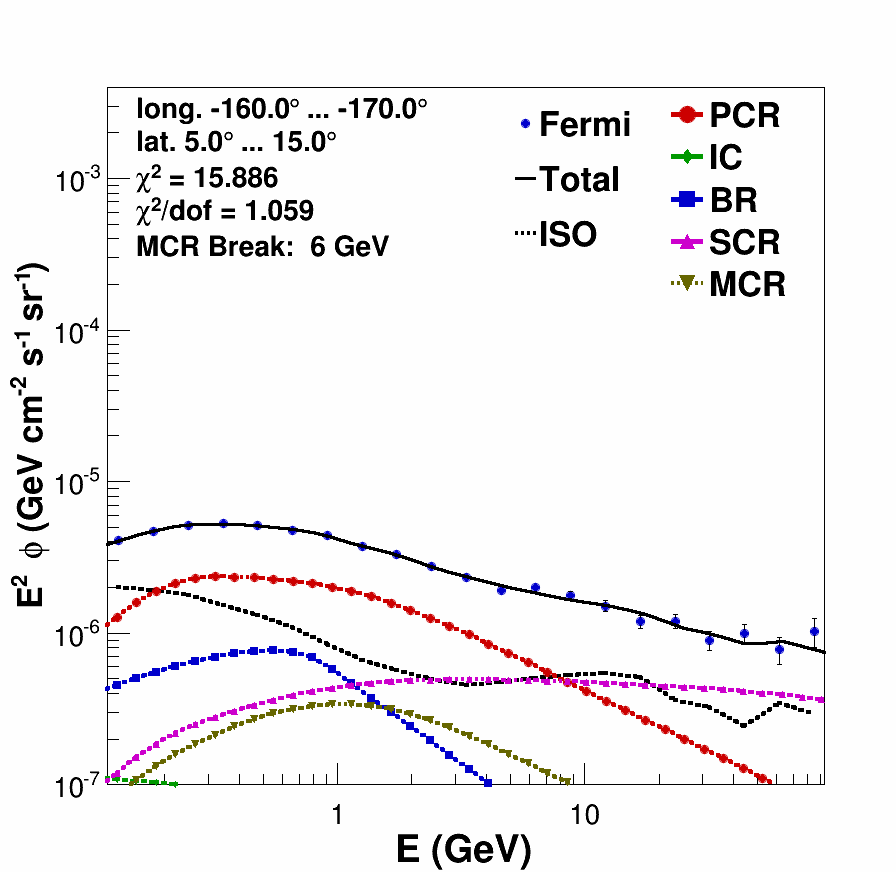}
\includegraphics[width=0.16\textwidth,height=0.16\textwidth,clip]{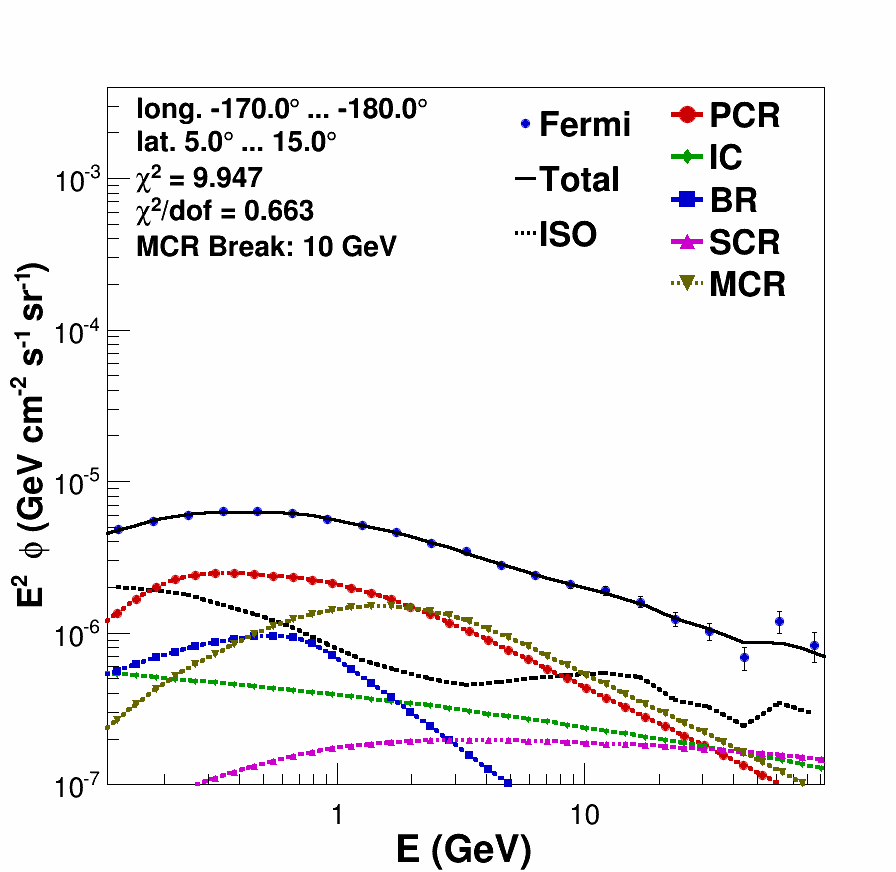}%%%%%r7
\caption[]{Template fits for latitudes  with $5.0^\circ<b<15.0^\circ$ and longitudes decreasing from 180$^\circ$ to -180$^\circ$. \label{F17}
}
\end{figure}
\begin{figure}
\centering
\includegraphics[width=0.16\textwidth,height=0.16\textwidth,clip]{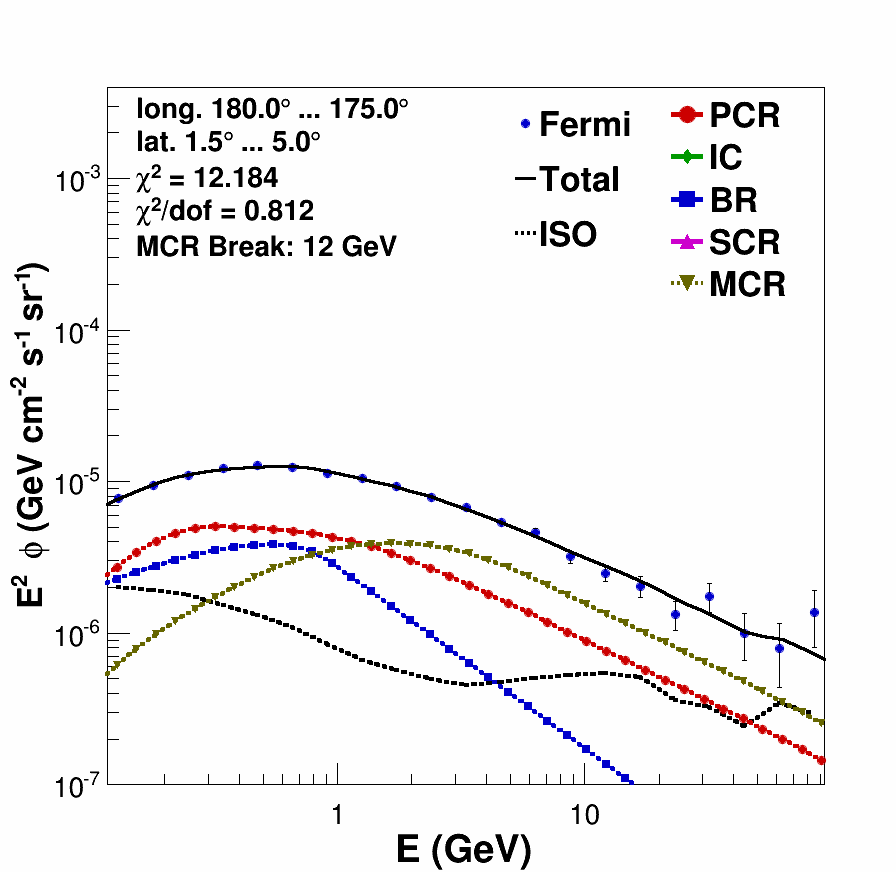}
\includegraphics[width=0.16\textwidth,height=0.16\textwidth,clip]{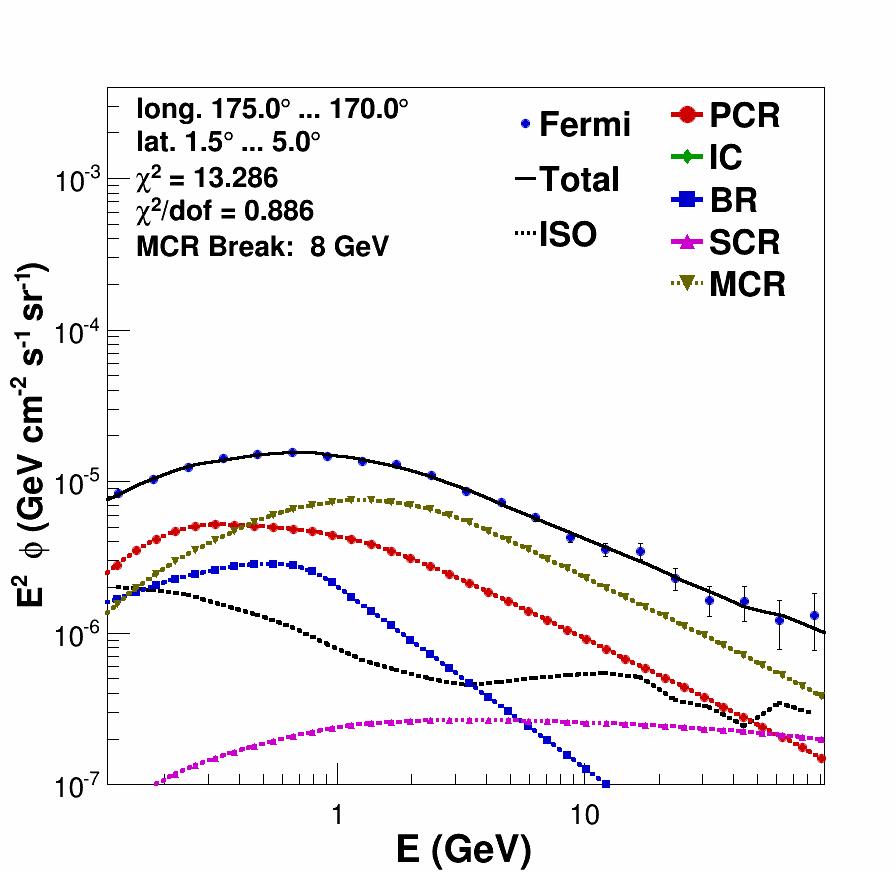}
\includegraphics[width=0.16\textwidth,height=0.16\textwidth,clip]{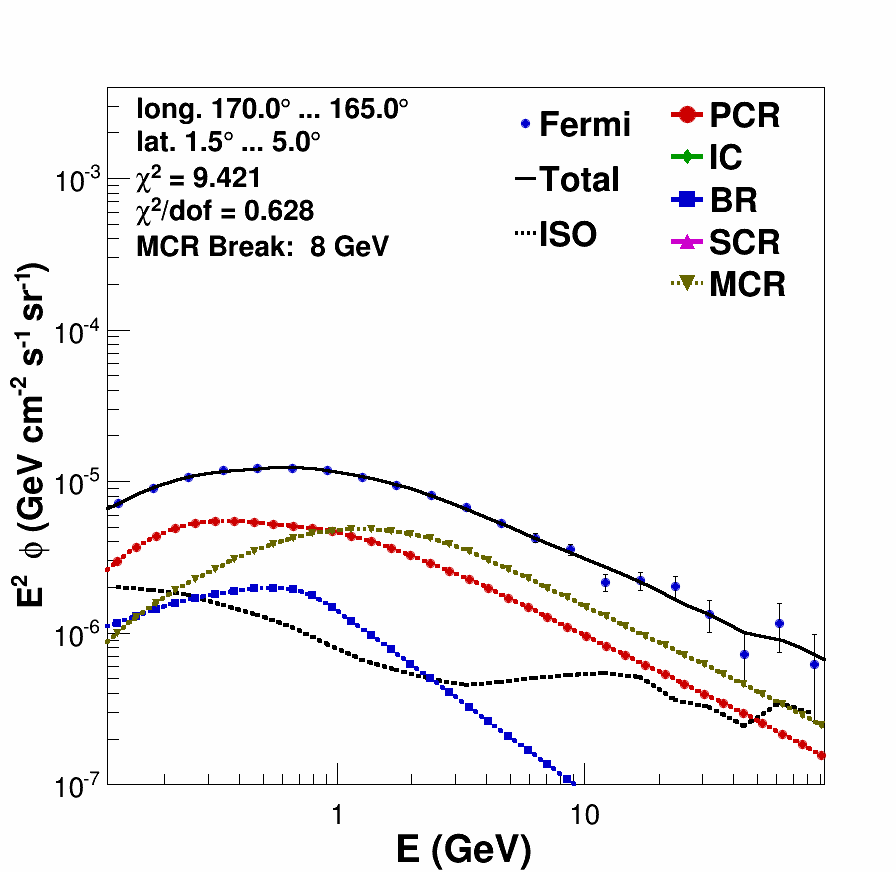}
\includegraphics[width=0.16\textwidth,height=0.16\textwidth,clip]{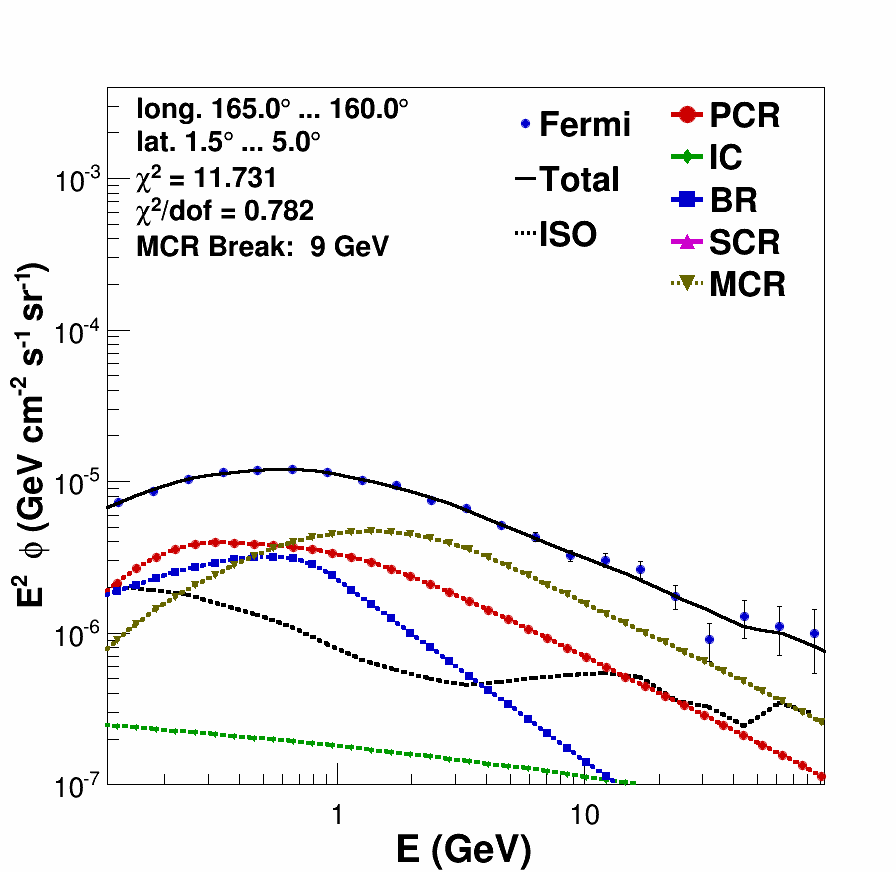}
\includegraphics[width=0.16\textwidth,height=0.16\textwidth,clip]{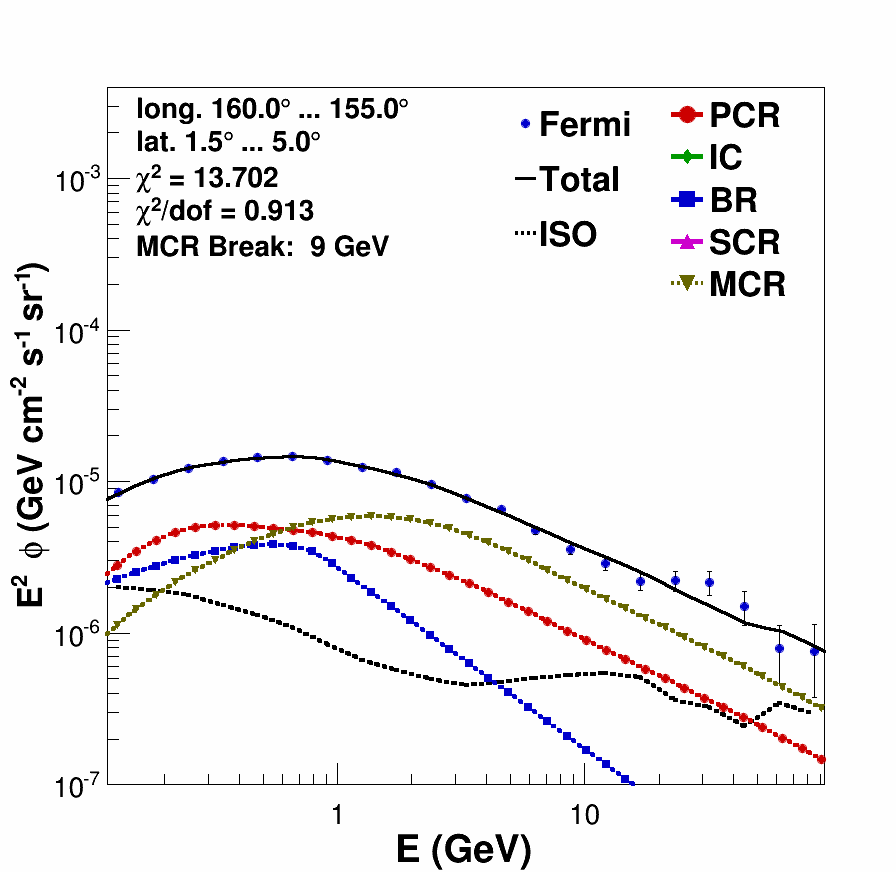}
\includegraphics[width=0.16\textwidth,height=0.16\textwidth,clip]{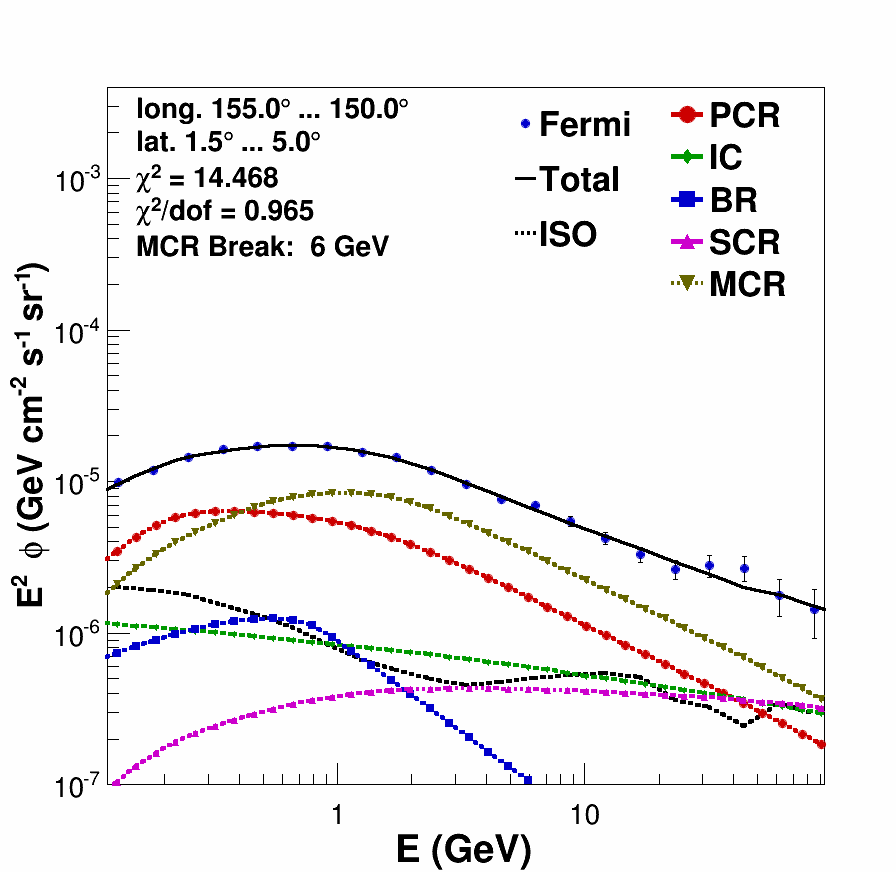}
\includegraphics[width=0.16\textwidth,height=0.16\textwidth,clip]{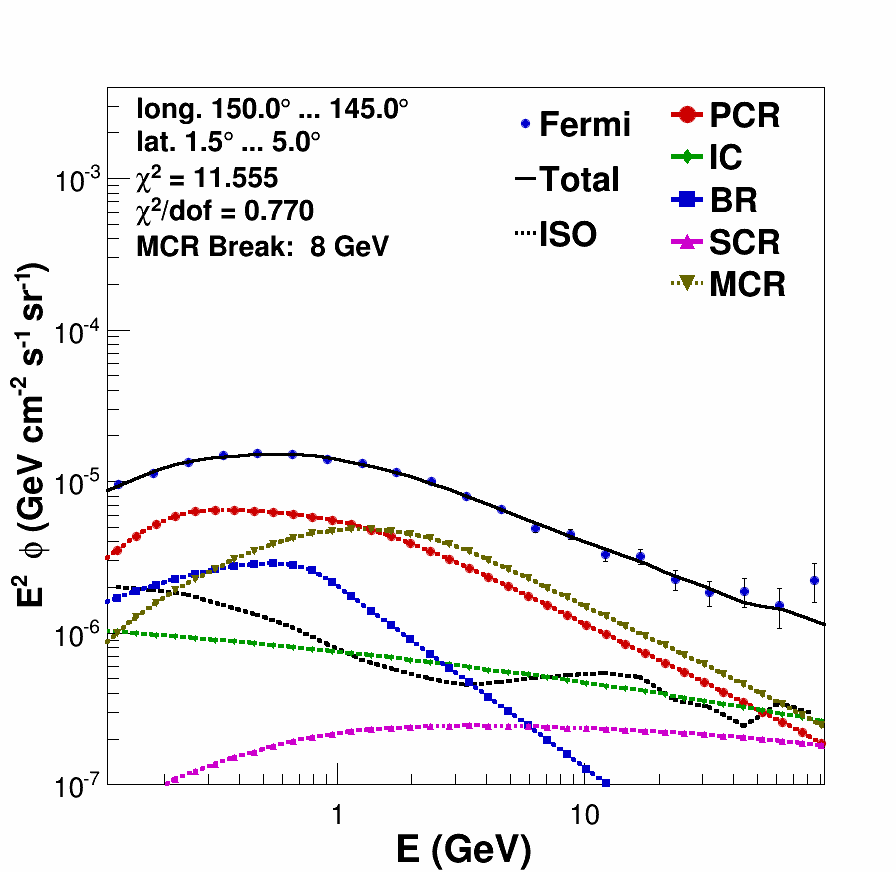}
\includegraphics[width=0.16\textwidth,height=0.16\textwidth,clip]{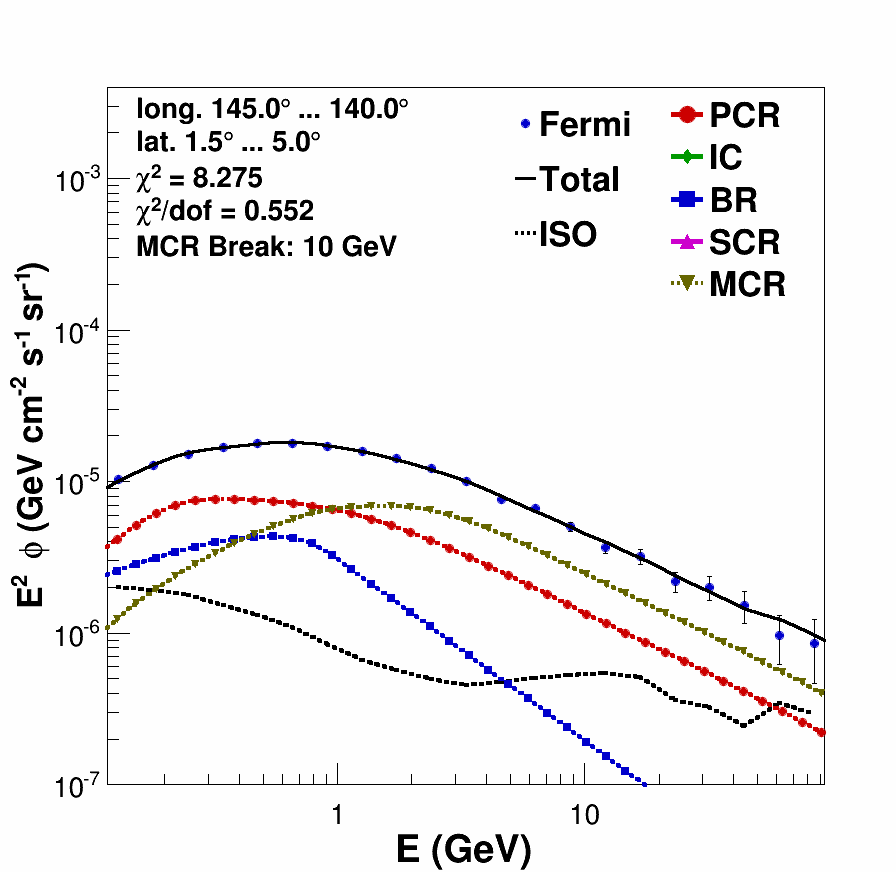}
\includegraphics[width=0.16\textwidth,height=0.16\textwidth,clip]{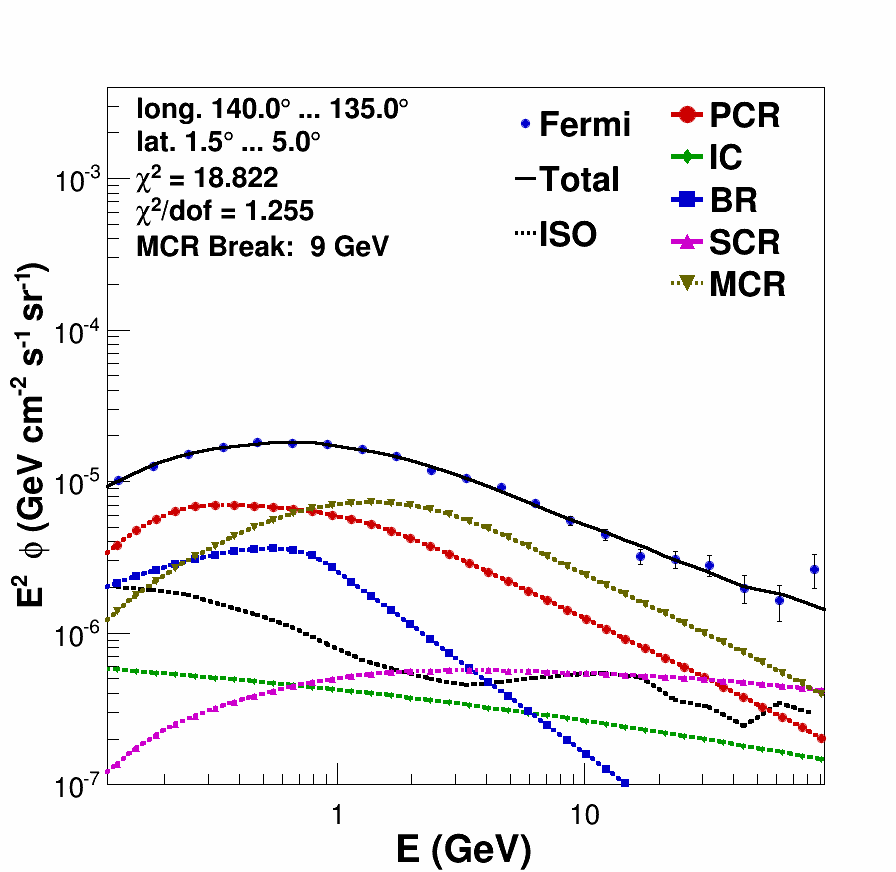}
\includegraphics[width=0.16\textwidth,height=0.16\textwidth,clip]{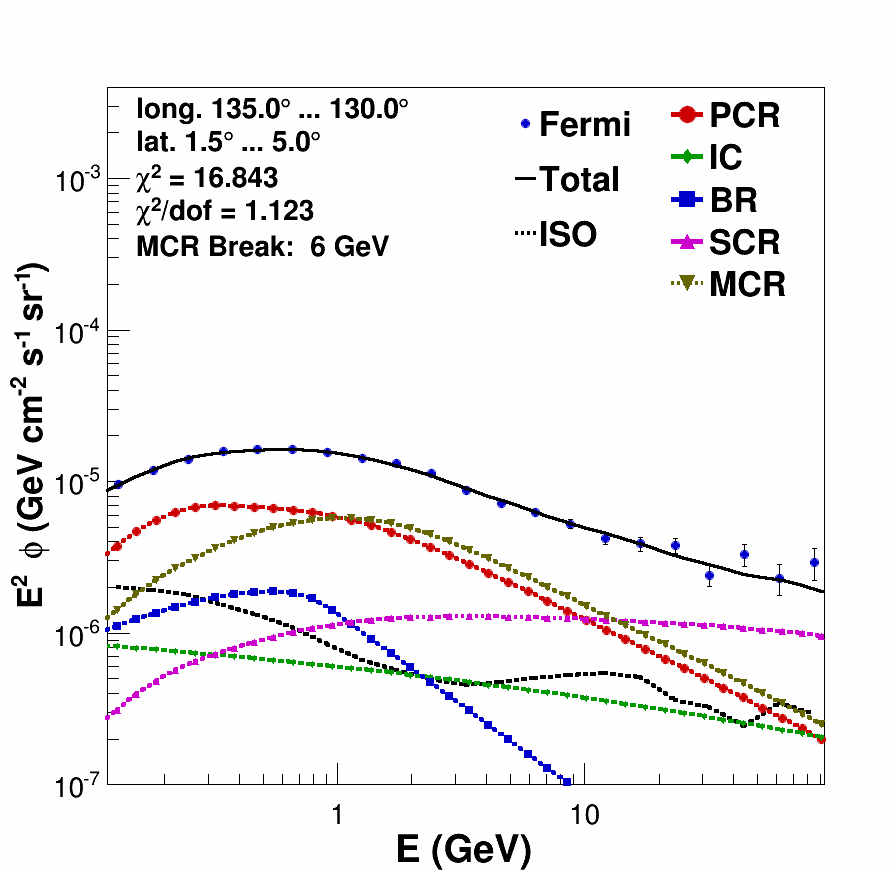}
\includegraphics[width=0.16\textwidth,height=0.16\textwidth,clip]{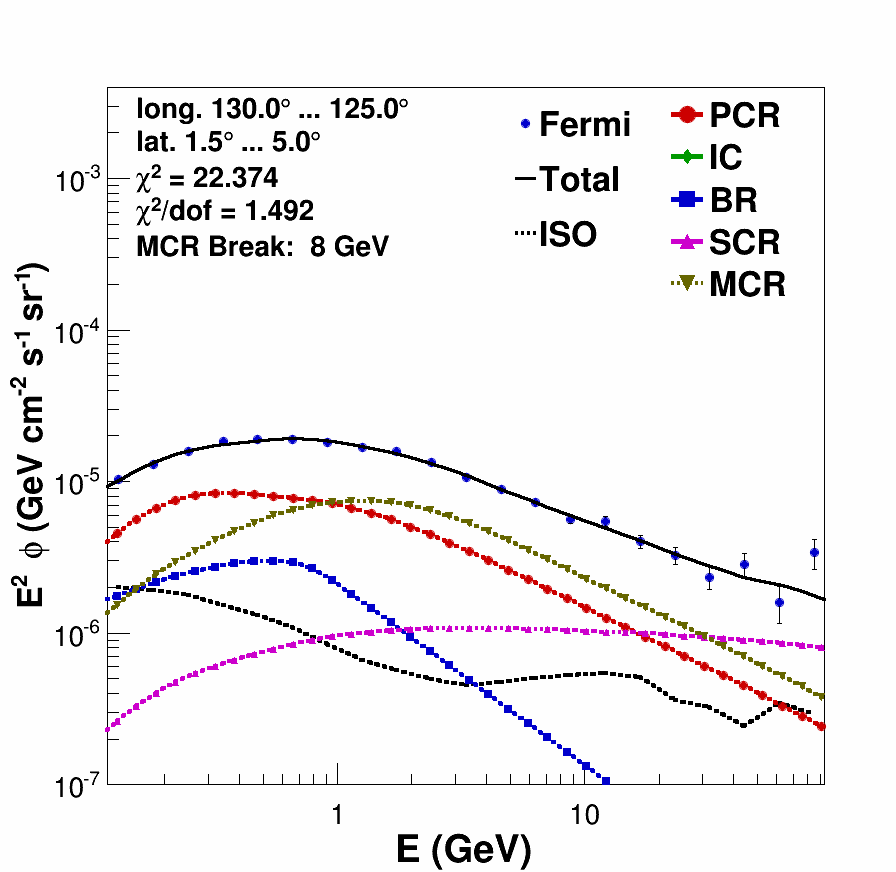}
\includegraphics[width=0.16\textwidth,height=0.16\textwidth,clip]{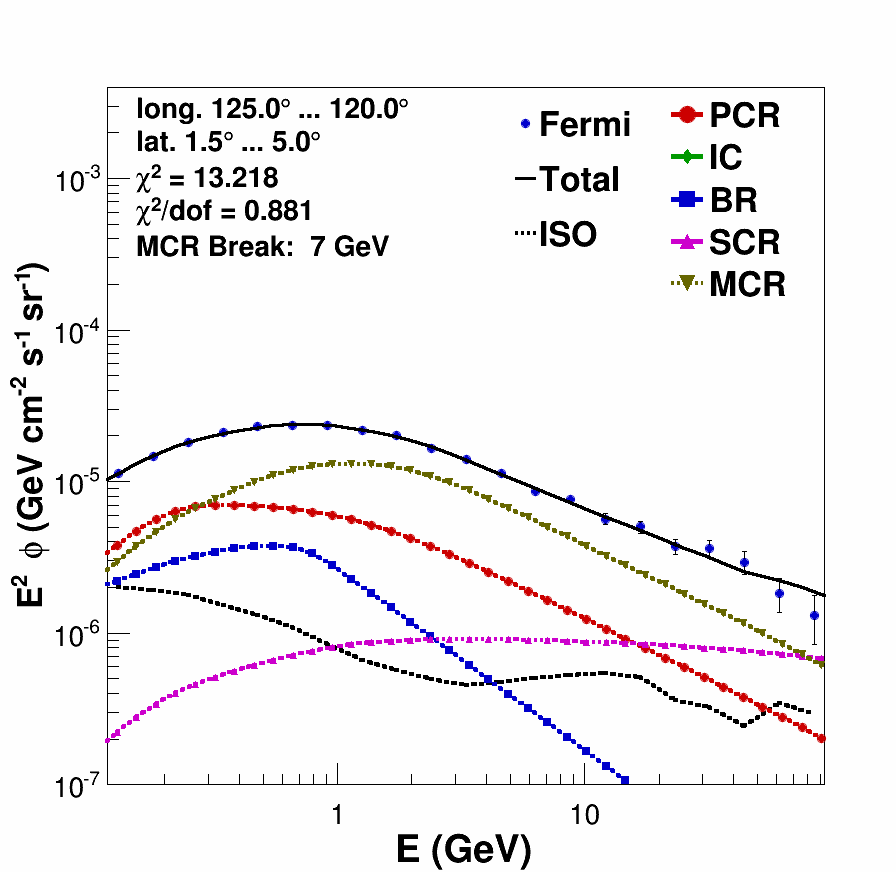}
\includegraphics[width=0.16\textwidth,height=0.16\textwidth,clip]{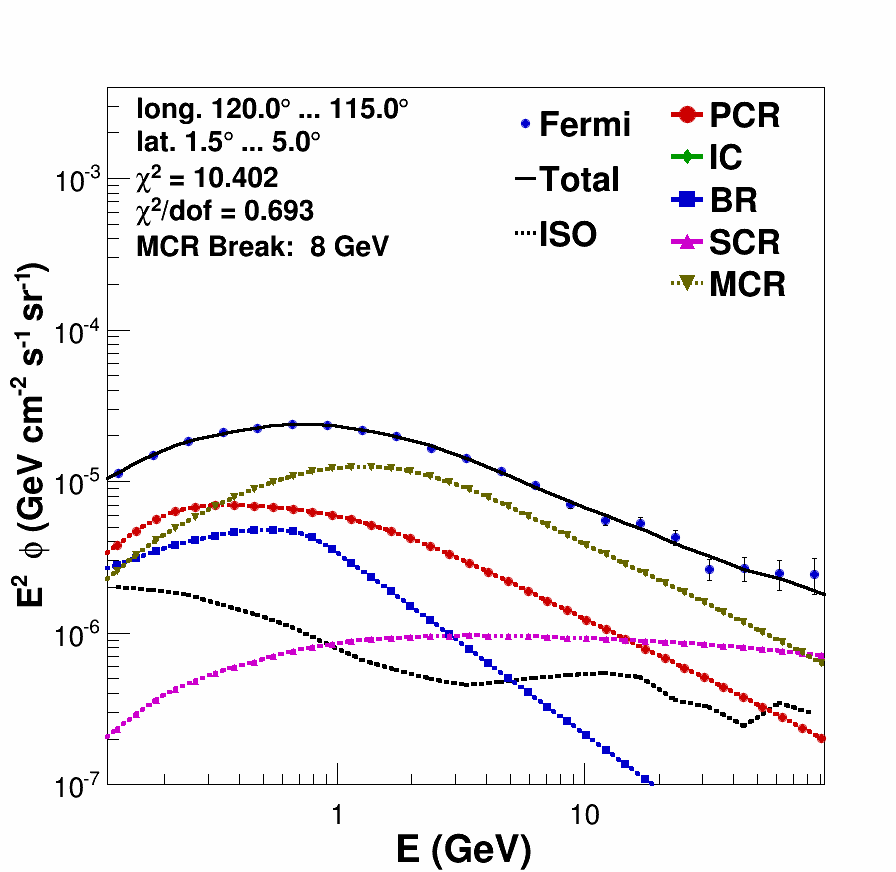}
\includegraphics[width=0.16\textwidth,height=0.16\textwidth,clip]{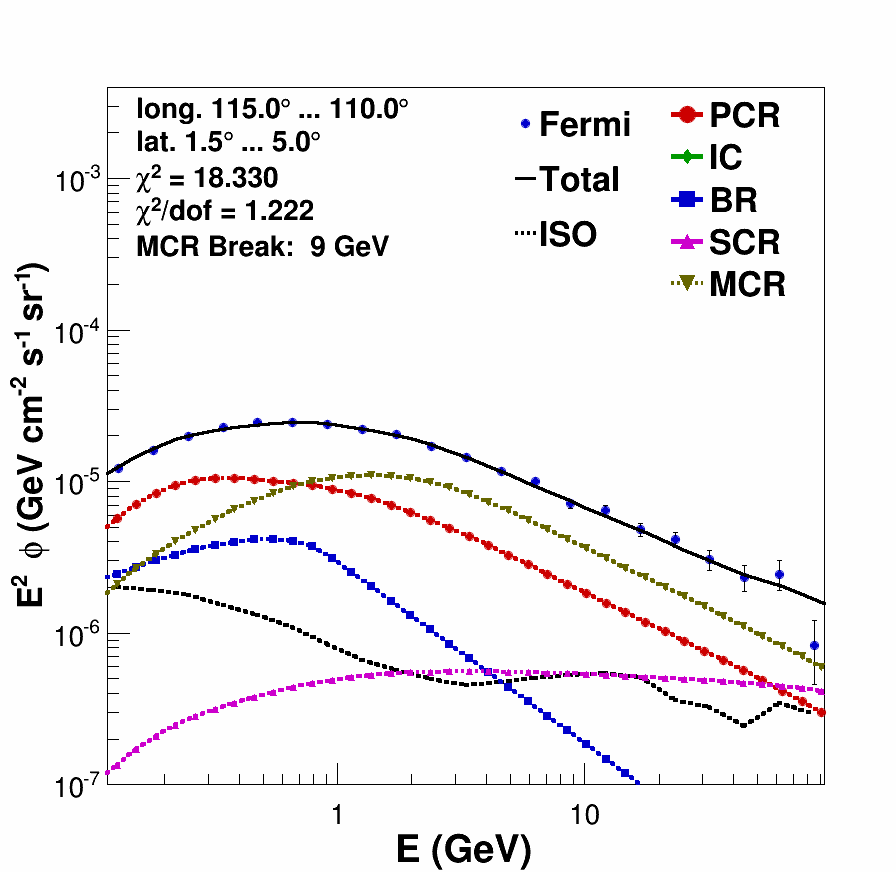}
\includegraphics[width=0.16\textwidth,height=0.16\textwidth,clip]{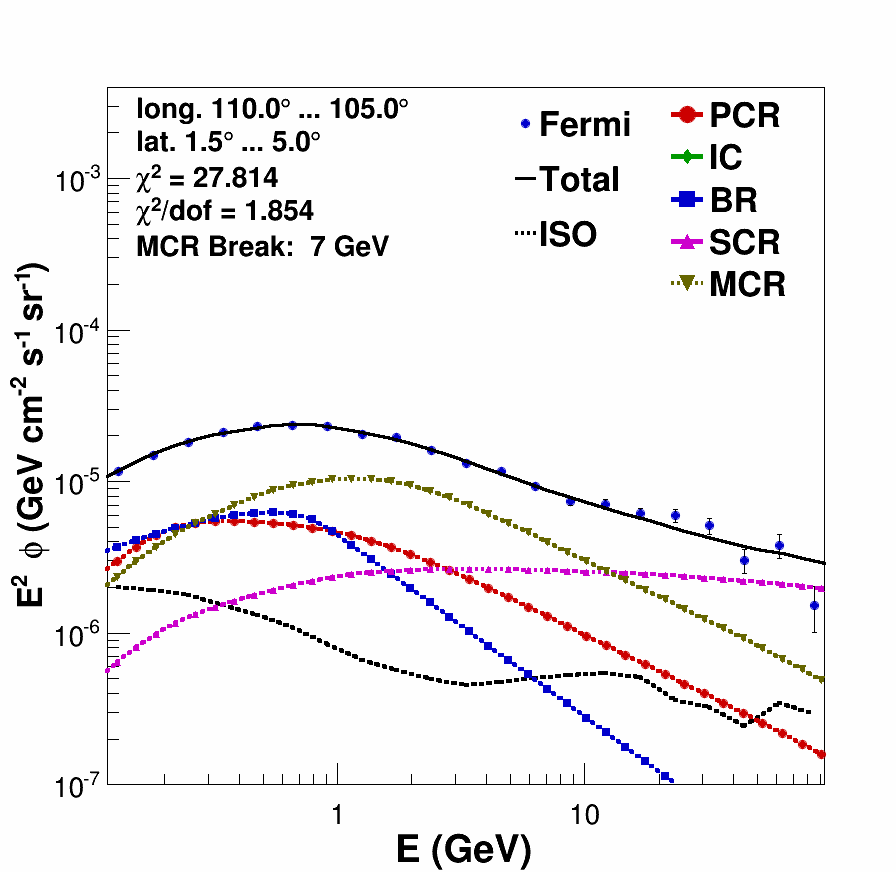}
\includegraphics[width=0.16\textwidth,height=0.16\textwidth,clip]{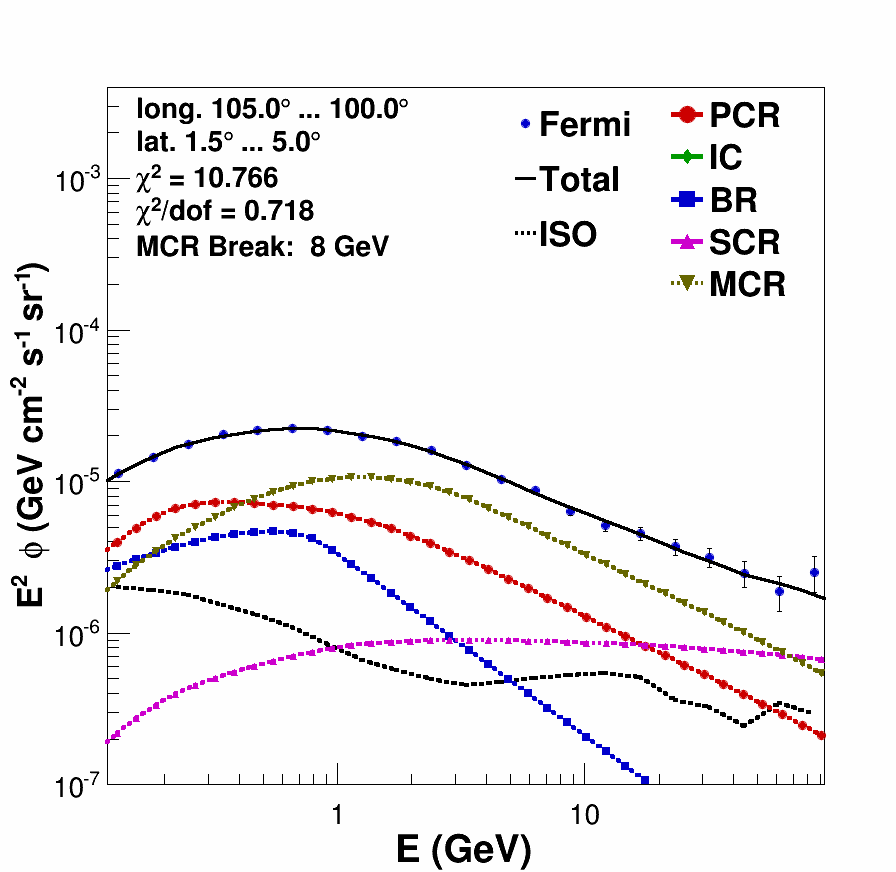}
\includegraphics[width=0.16\textwidth,height=0.16\textwidth,clip]{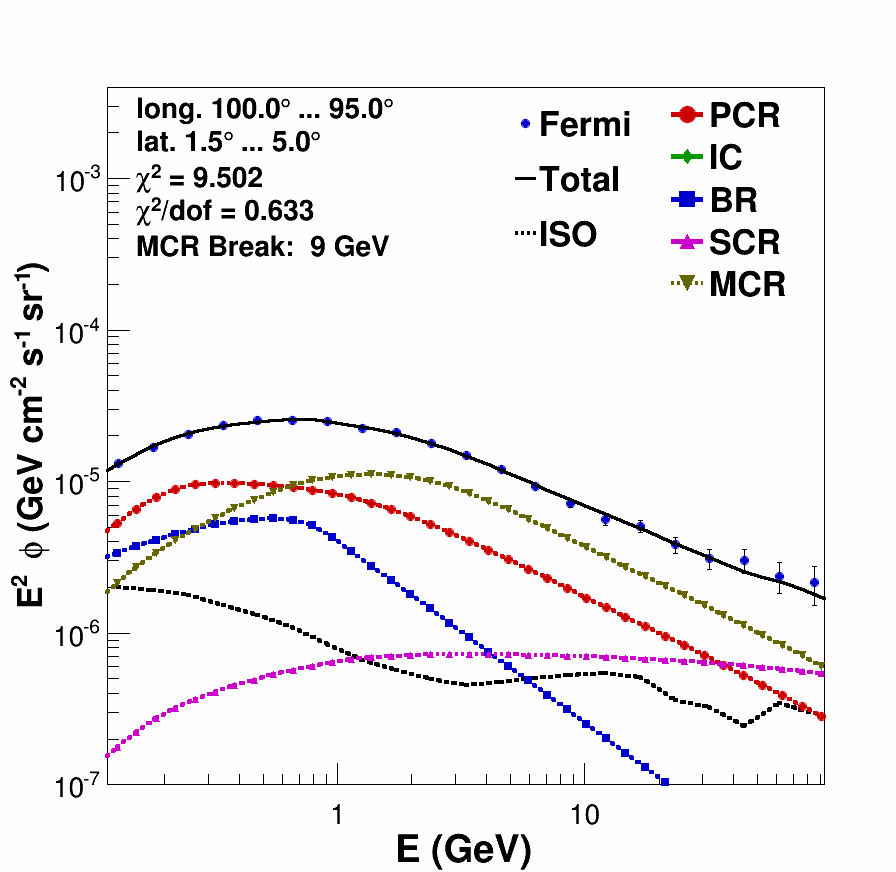}
\includegraphics[width=0.16\textwidth,height=0.16\textwidth,clip]{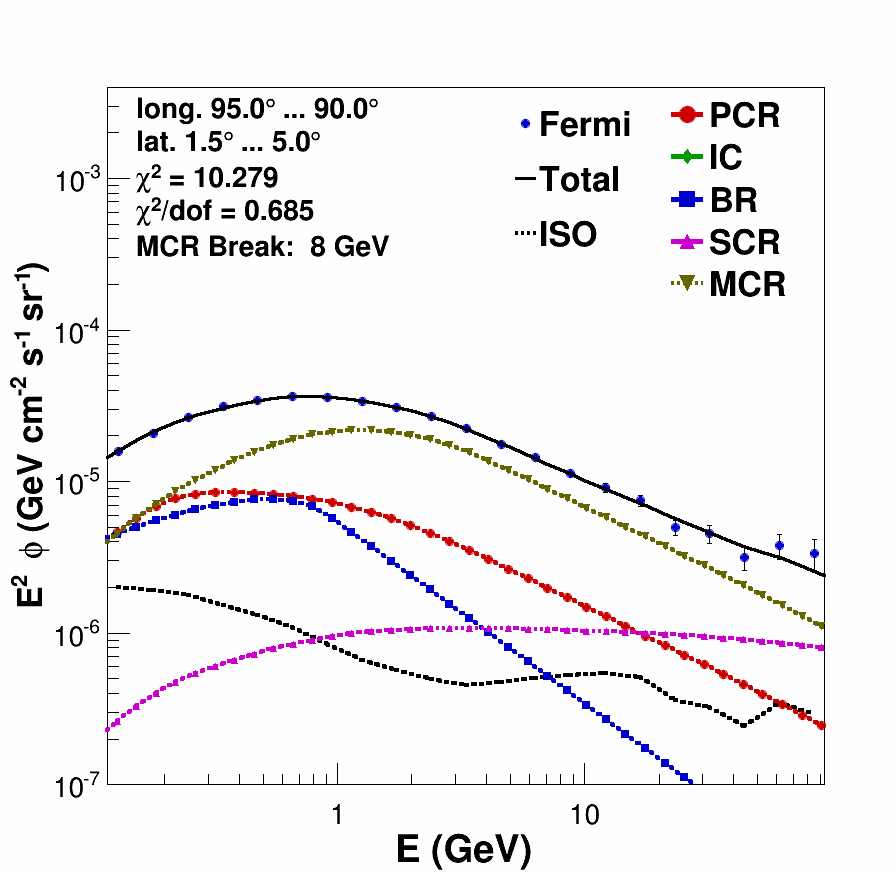}
\includegraphics[width=0.16\textwidth,height=0.16\textwidth,clip]{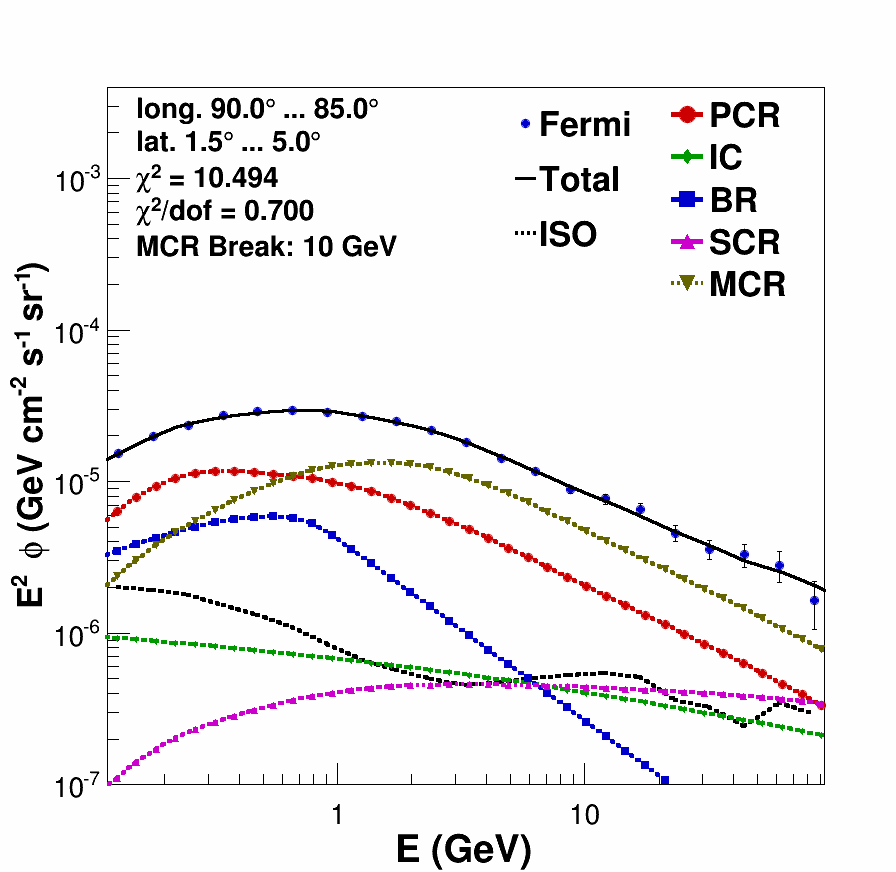}
\includegraphics[width=0.16\textwidth,height=0.16\textwidth,clip]{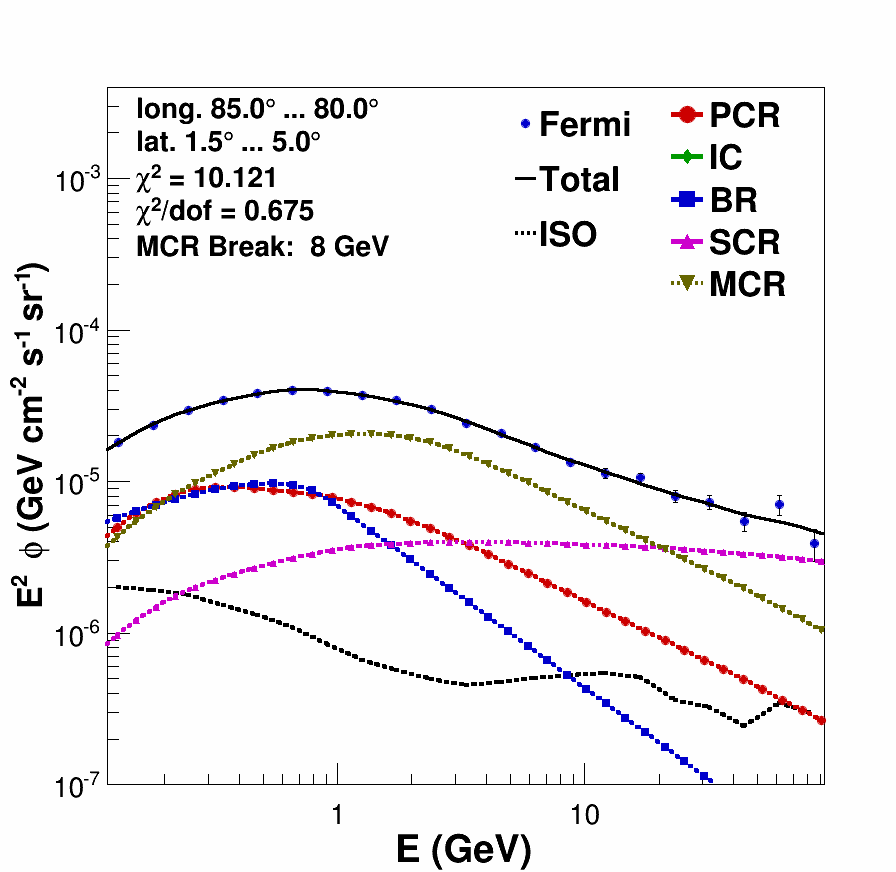}
\includegraphics[width=0.16\textwidth,height=0.16\textwidth,clip]{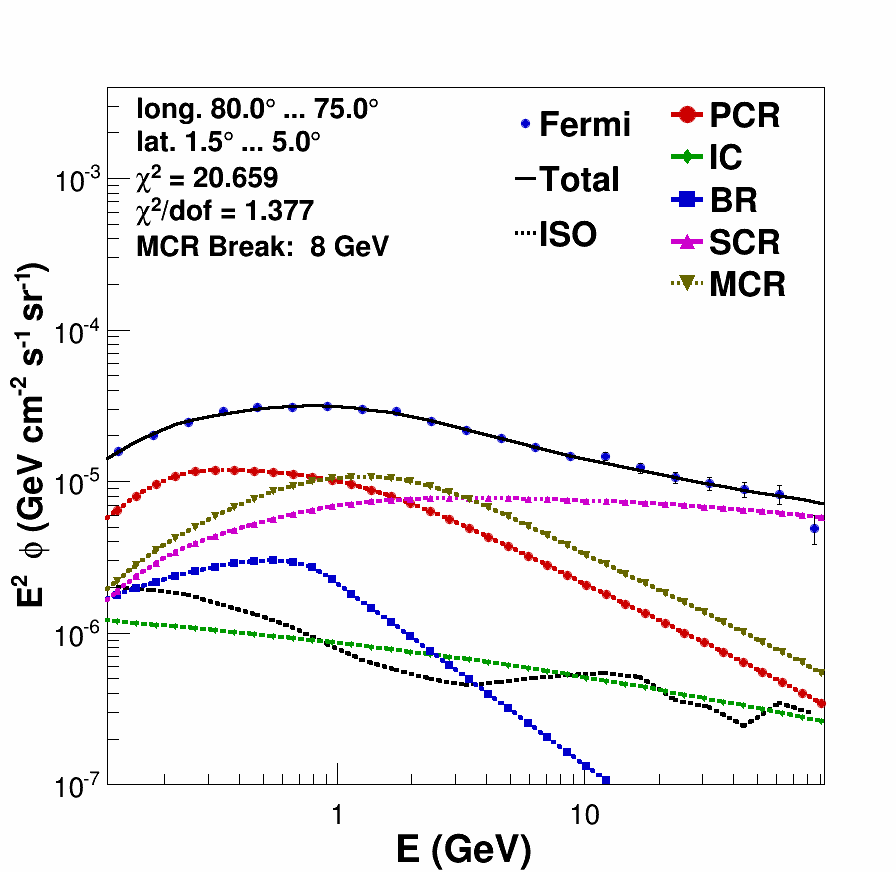}
\includegraphics[width=0.16\textwidth,height=0.16\textwidth,clip]{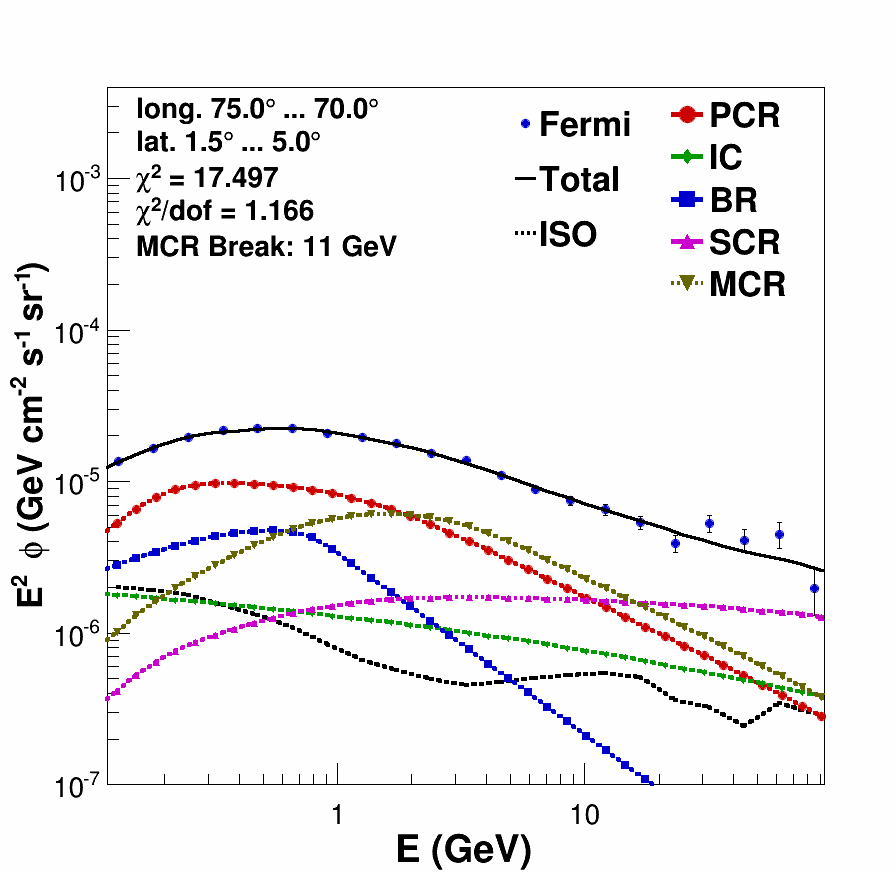}
\includegraphics[width=0.16\textwidth,height=0.16\textwidth,clip]{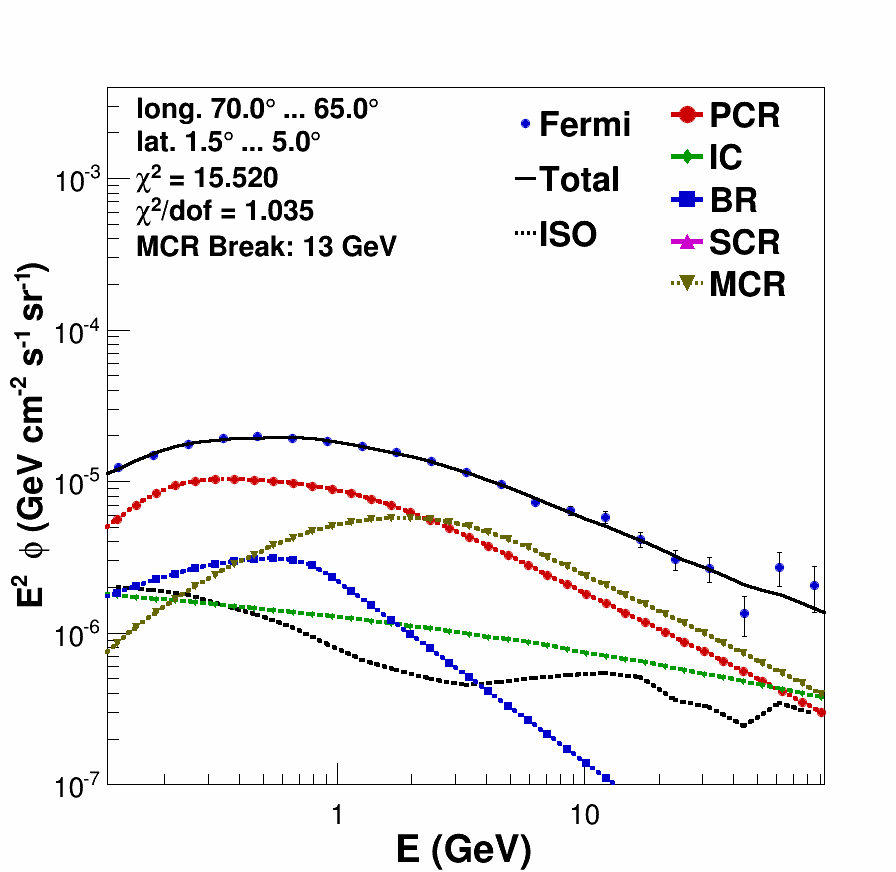}
\includegraphics[width=0.16\textwidth,height=0.16\textwidth,clip]{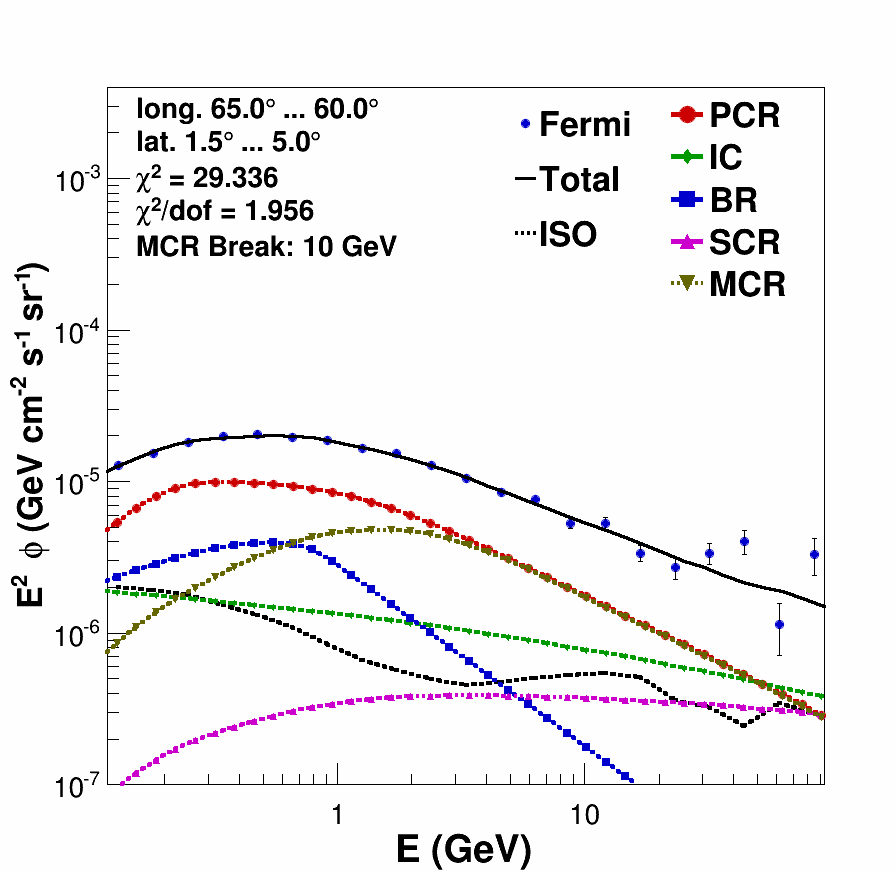}
\includegraphics[width=0.16\textwidth,height=0.16\textwidth,clip]{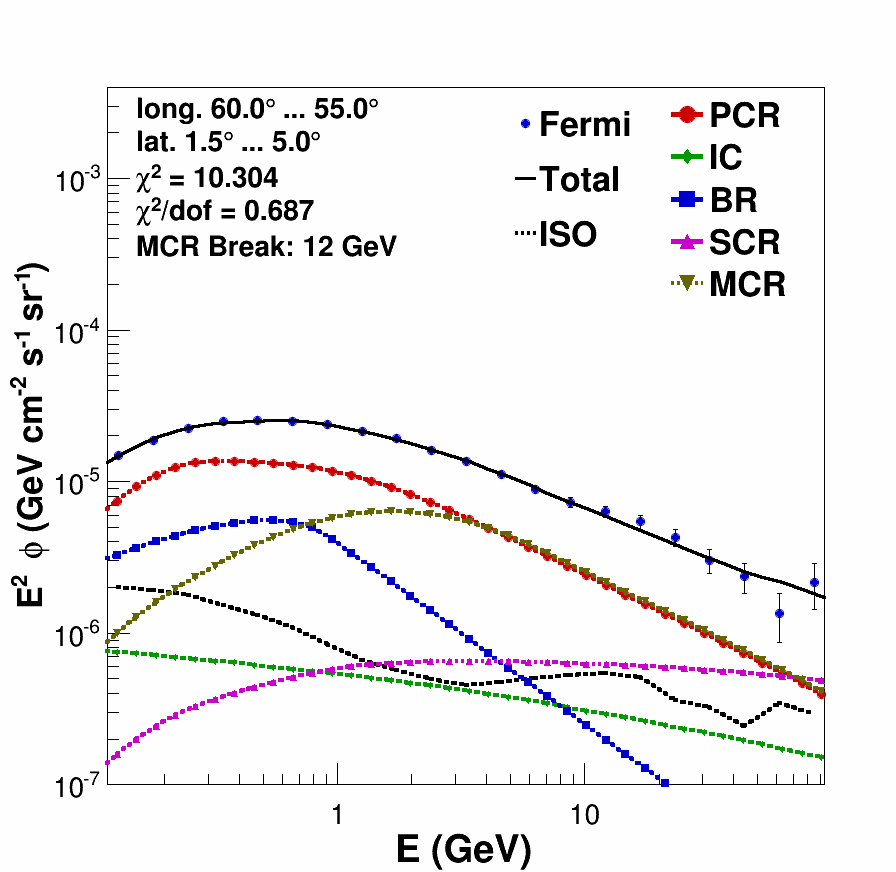}
\includegraphics[width=0.16\textwidth,height=0.16\textwidth,clip]{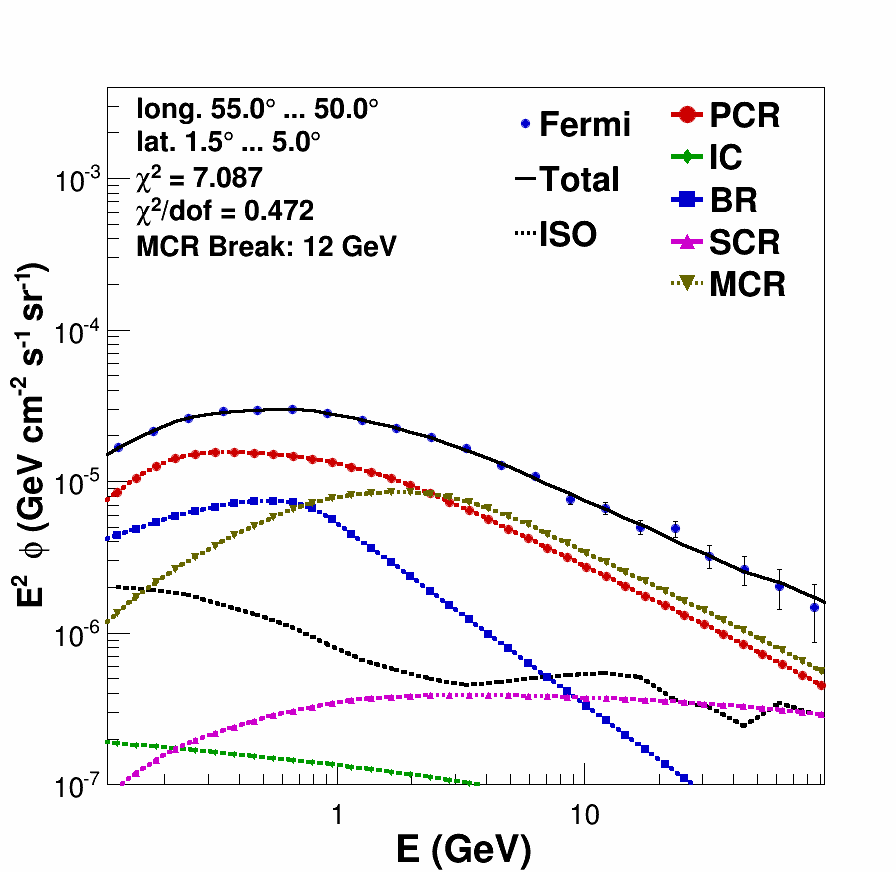}
\includegraphics[width=0.16\textwidth,height=0.16\textwidth,clip]{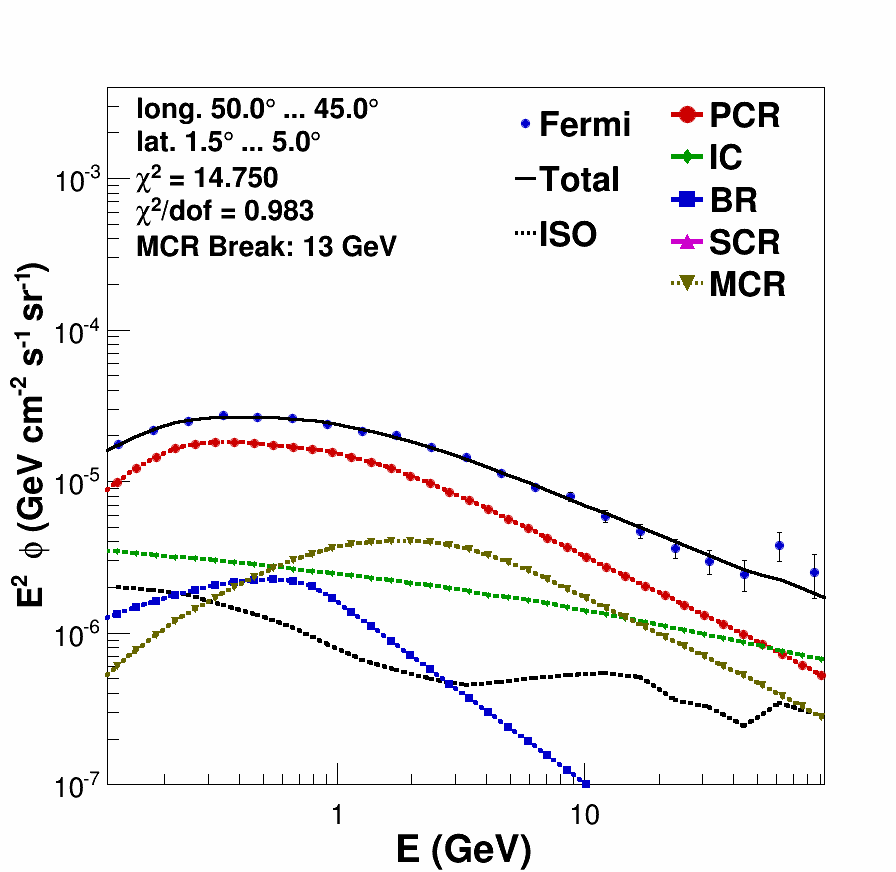}
\includegraphics[width=0.16\textwidth,height=0.16\textwidth,clip]{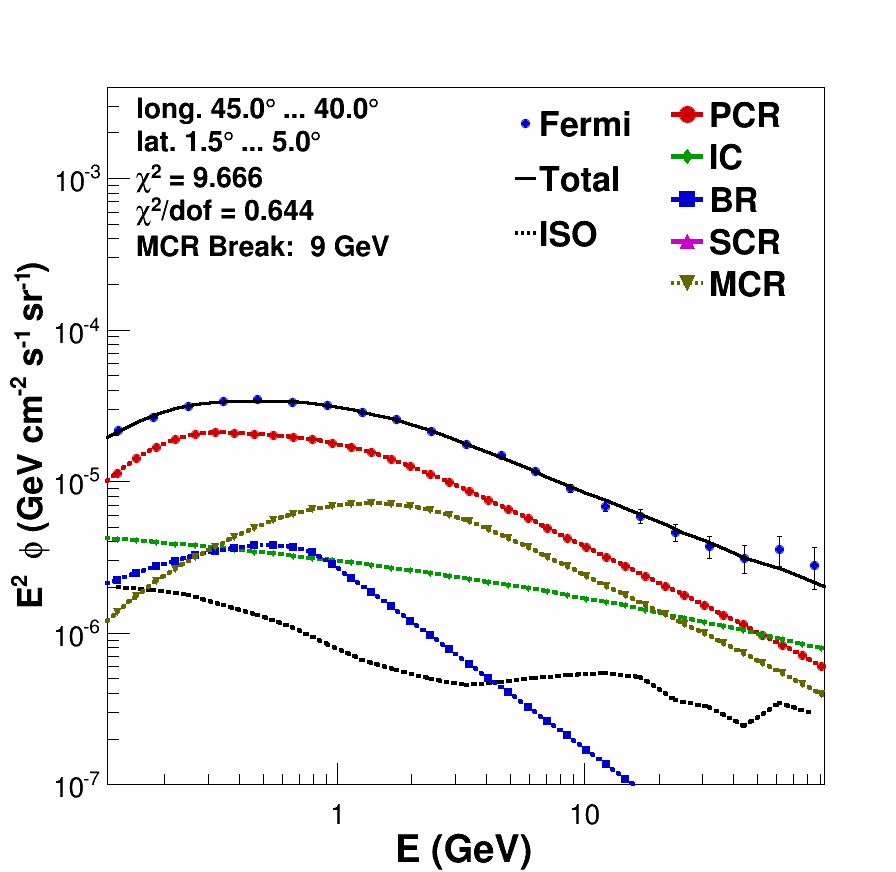}
\includegraphics[width=0.16\textwidth,height=0.16\textwidth,clip]{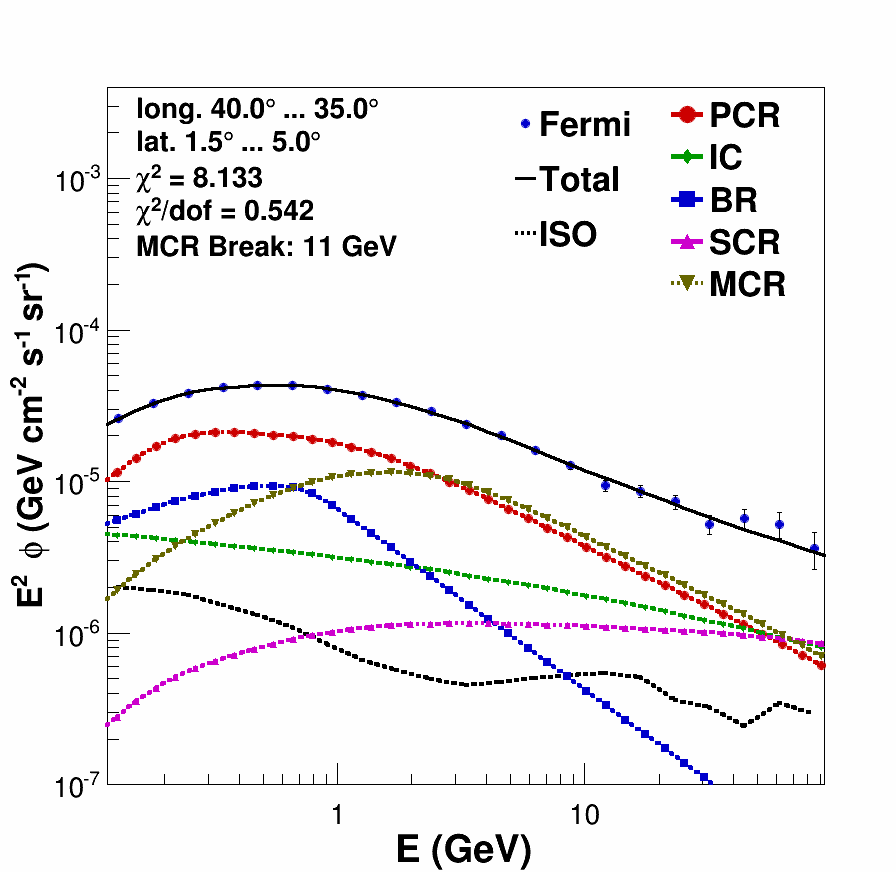}
\includegraphics[width=0.16\textwidth,height=0.16\textwidth,clip]{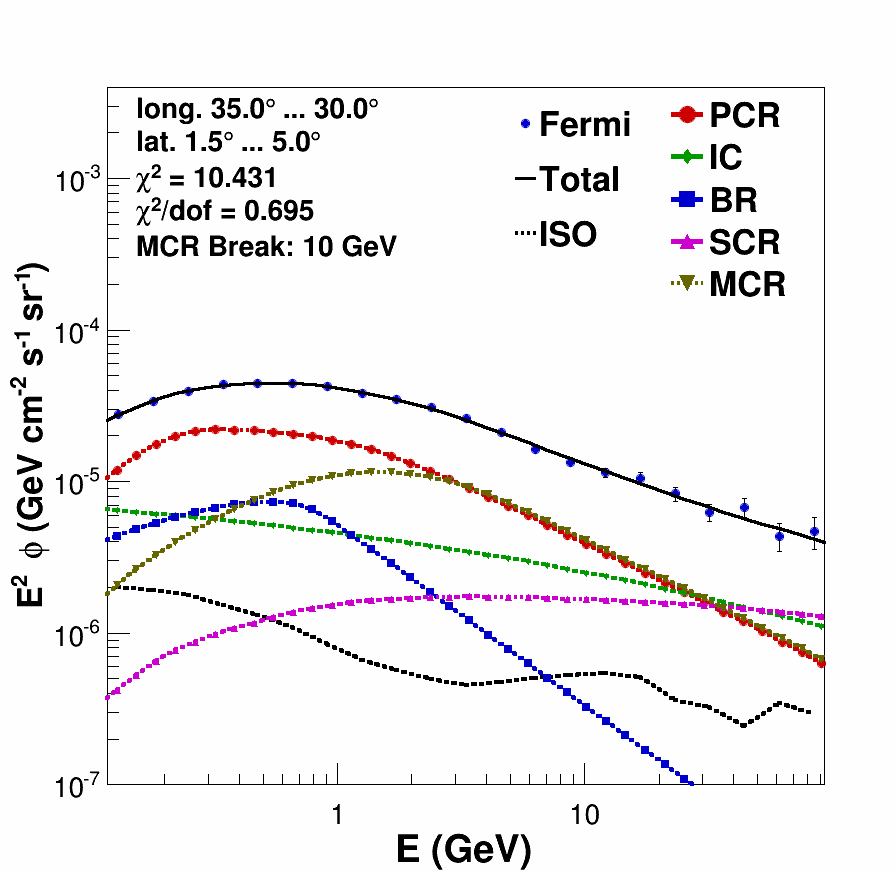}
\includegraphics[width=0.16\textwidth,height=0.16\textwidth,clip]{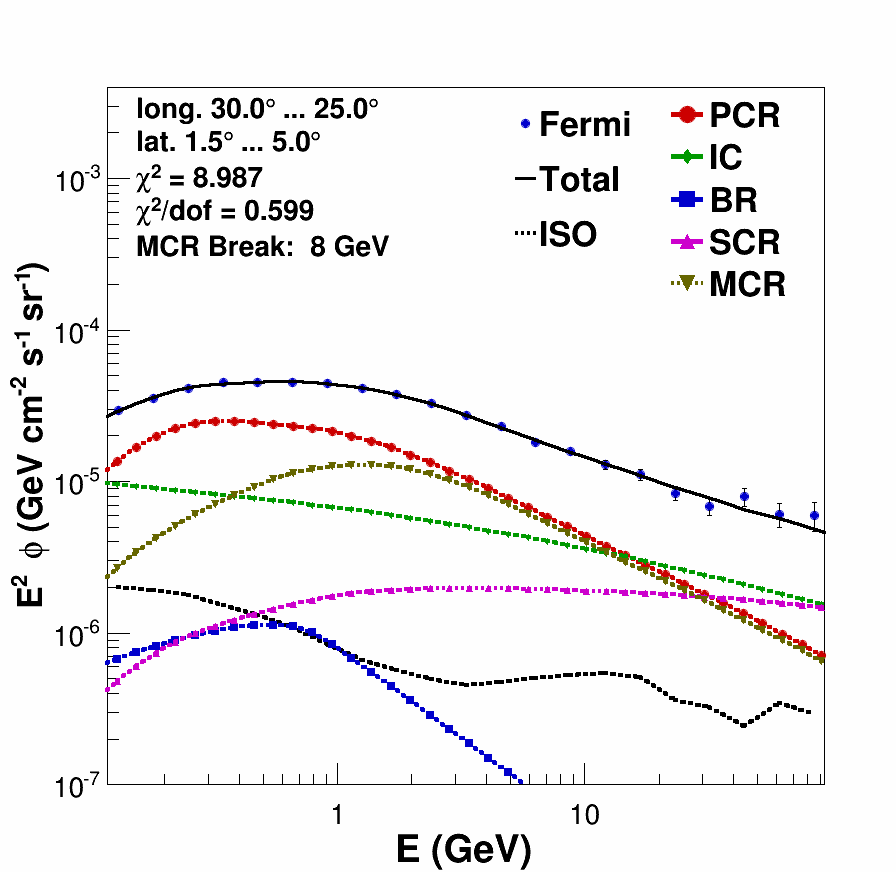}
\includegraphics[width=0.16\textwidth,height=0.16\textwidth,clip]{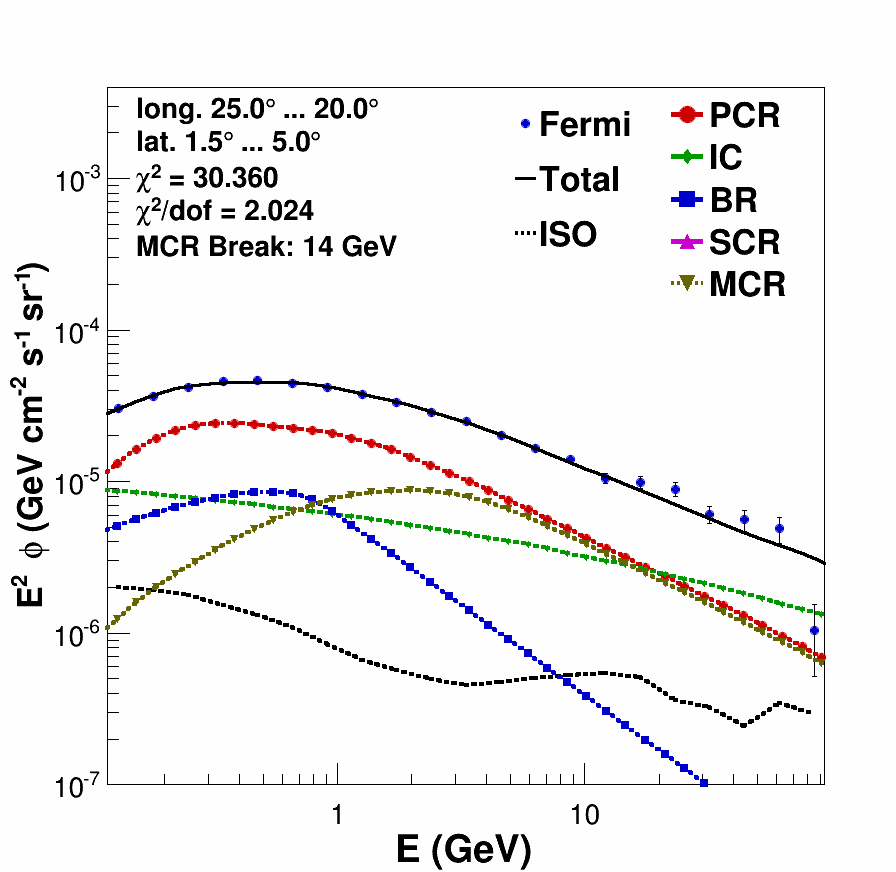}
\includegraphics[width=0.16\textwidth,height=0.16\textwidth,clip]{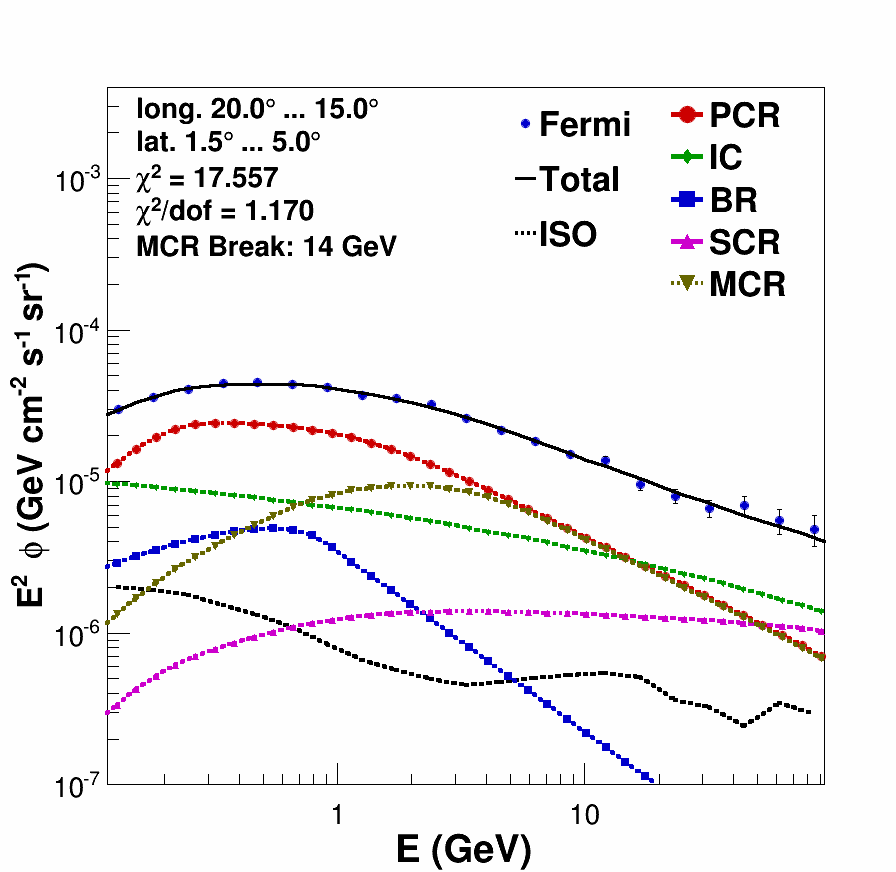}
\includegraphics[width=0.16\textwidth,height=0.16\textwidth,clip]{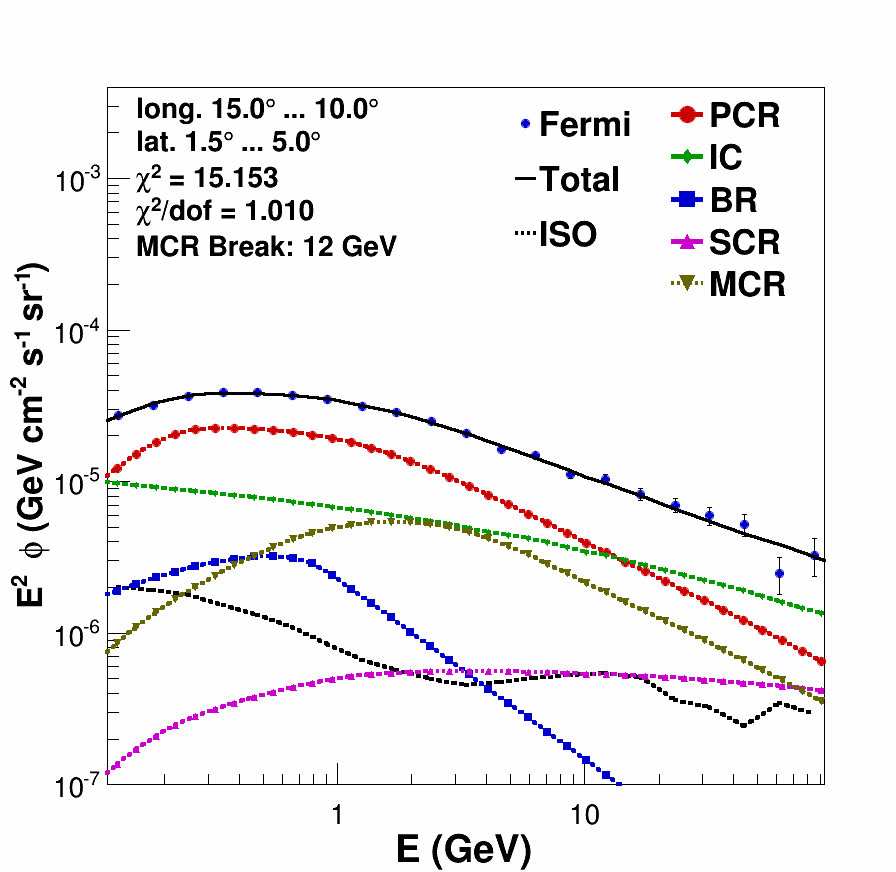}
\includegraphics[width=0.16\textwidth,height=0.16\textwidth,clip]{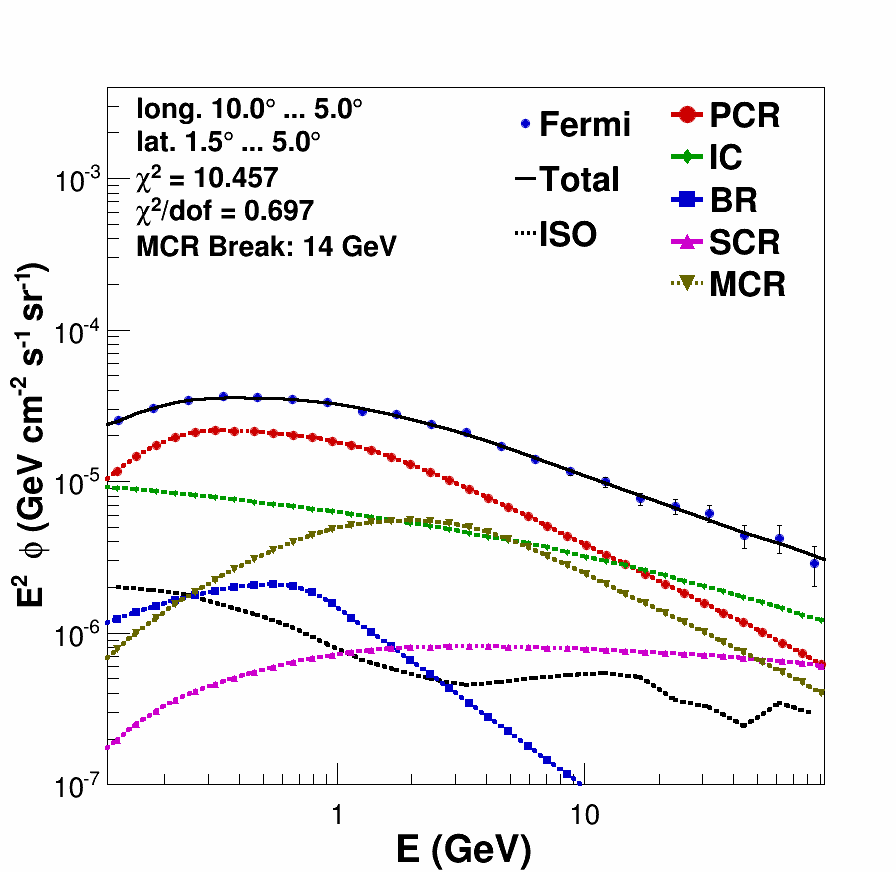}
\includegraphics[width=0.16\textwidth,height=0.16\textwidth,clip]{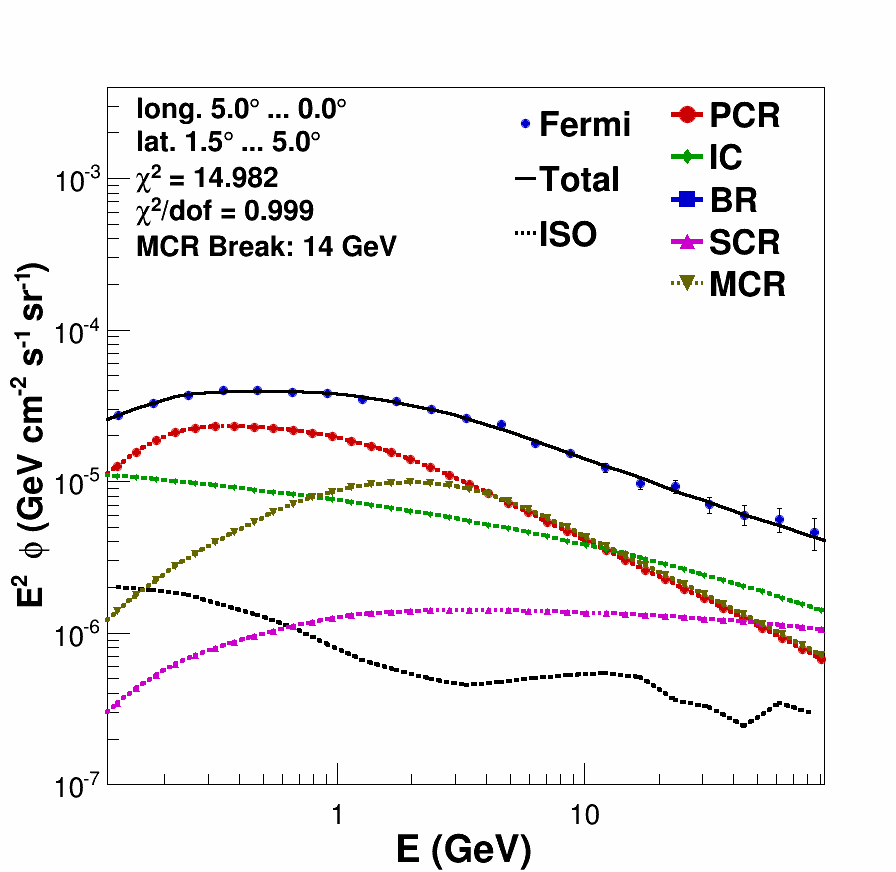}%%%%%r8a
\caption[]{Template fits for latitudes  with $1.5^\circ<b<5.0^\circ$ and longitudes decreasing from 180$^\circ$ to 0$^\circ$.} \label{F18}
\end{figure}
\begin{figure}
\centering
\includegraphics[width=0.16\textwidth,height=0.16\textwidth,clip]{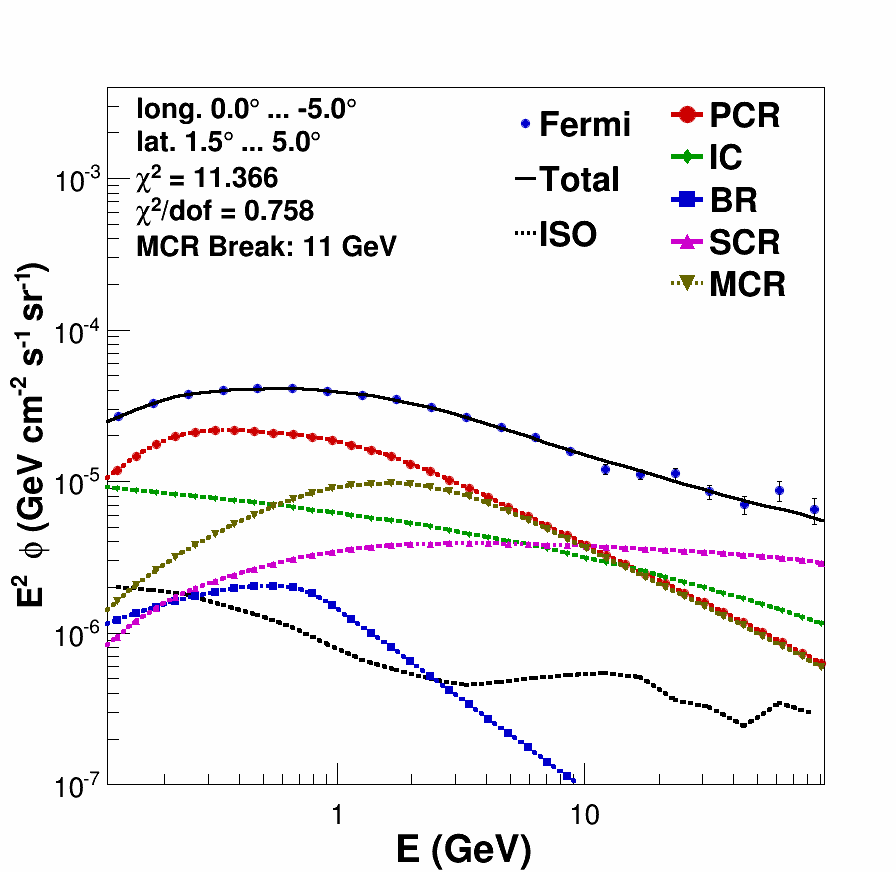}
\includegraphics[width=0.16\textwidth,height=0.16\textwidth,clip]{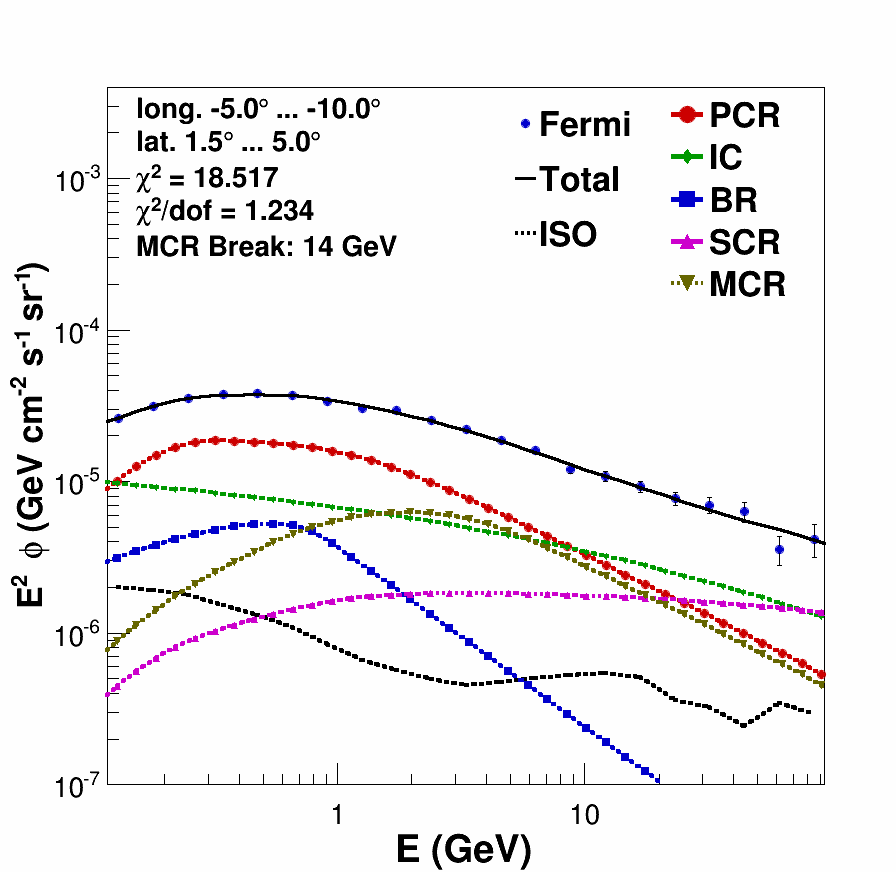}
\includegraphics[width=0.16\textwidth,height=0.16\textwidth,clip]{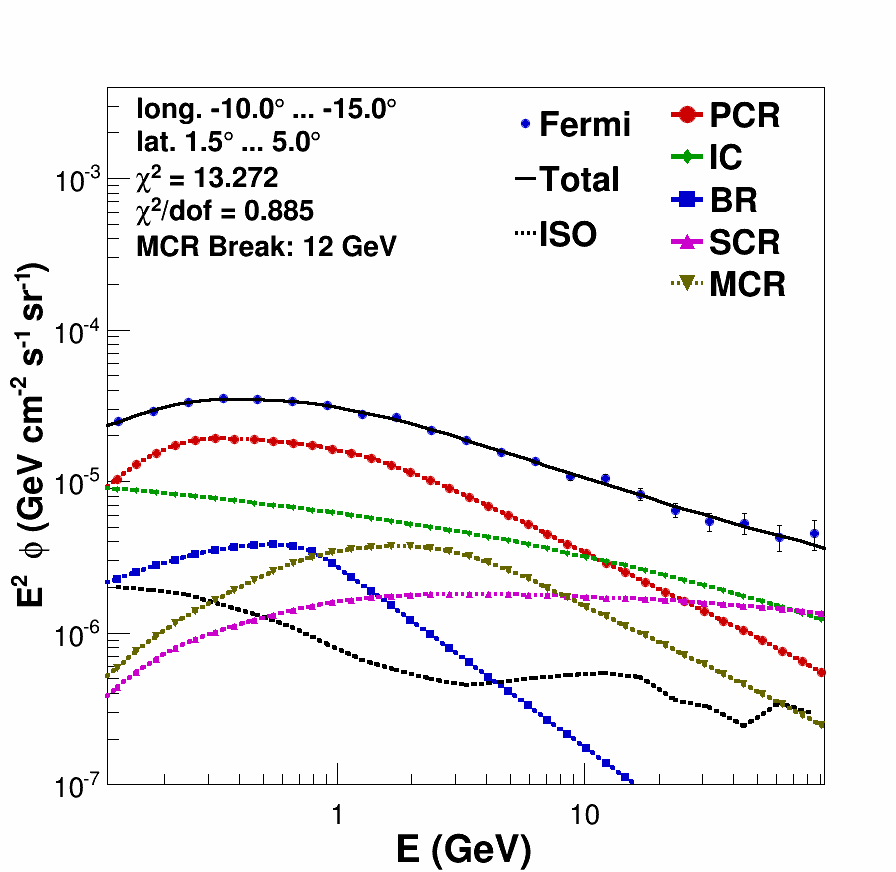}
\includegraphics[width=0.16\textwidth,height=0.16\textwidth,clip]{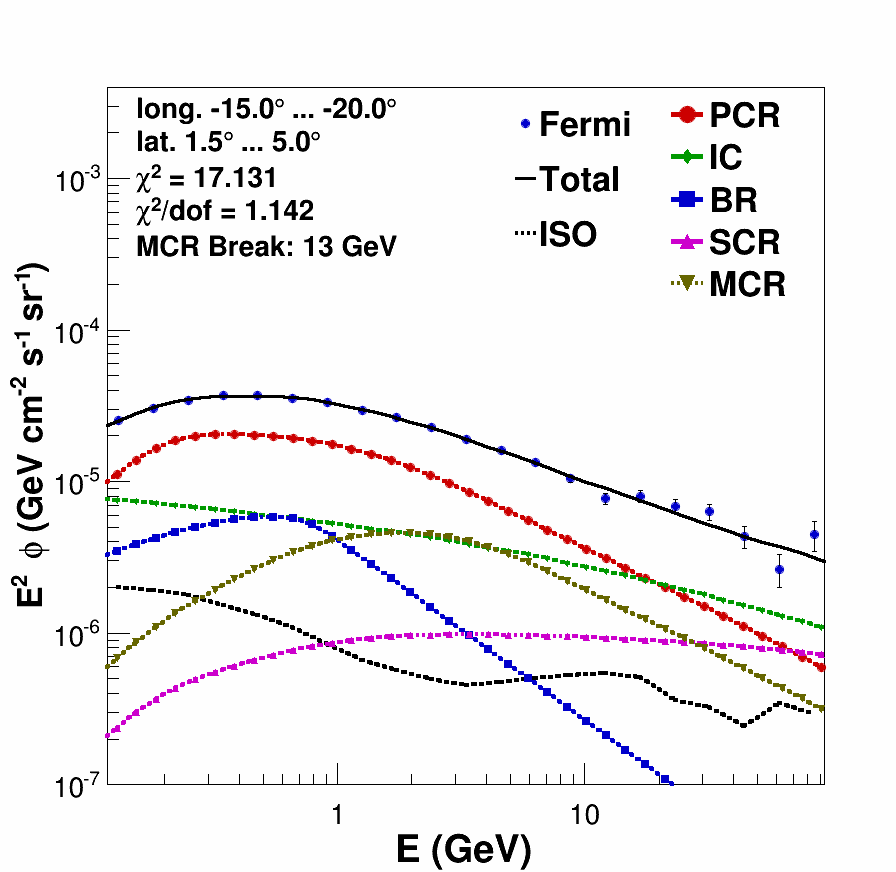}
\includegraphics[width=0.16\textwidth,height=0.16\textwidth,clip]{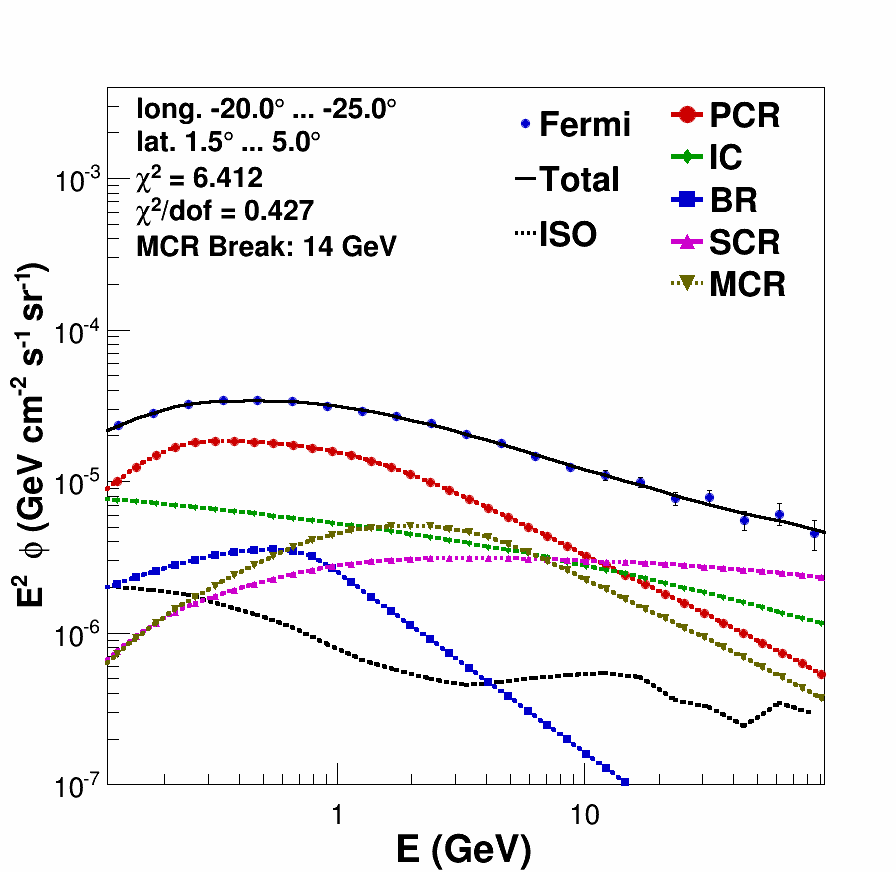}
\includegraphics[width=0.16\textwidth,height=0.16\textwidth,clip]{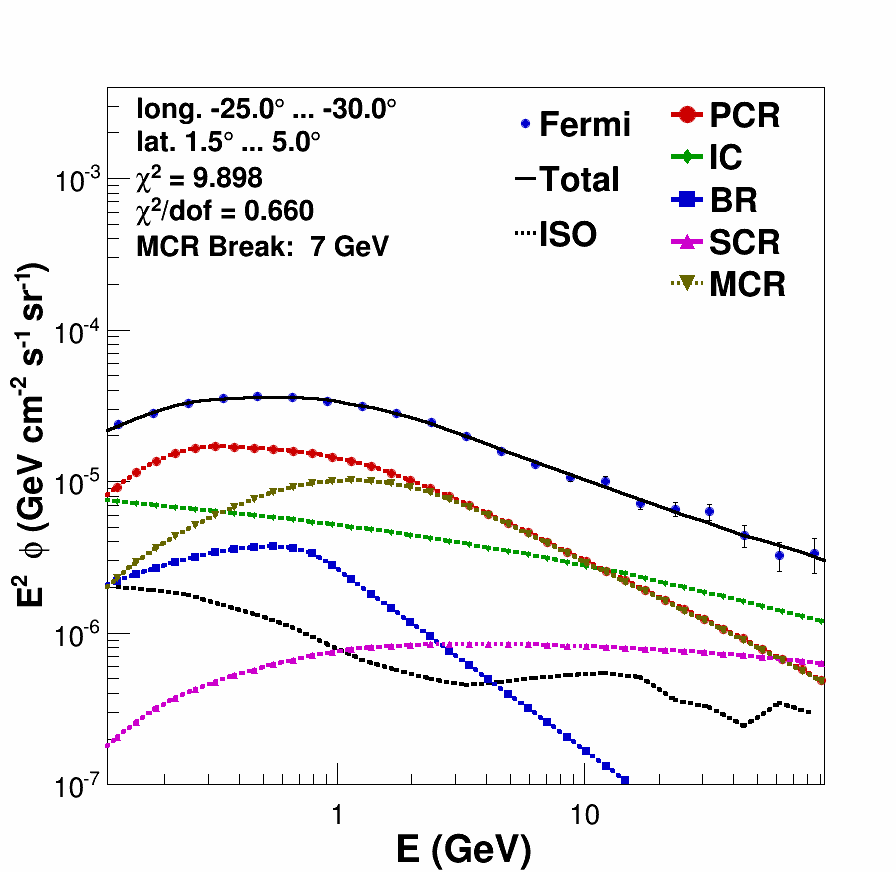}
\includegraphics[width=0.16\textwidth,height=0.16\textwidth,clip]{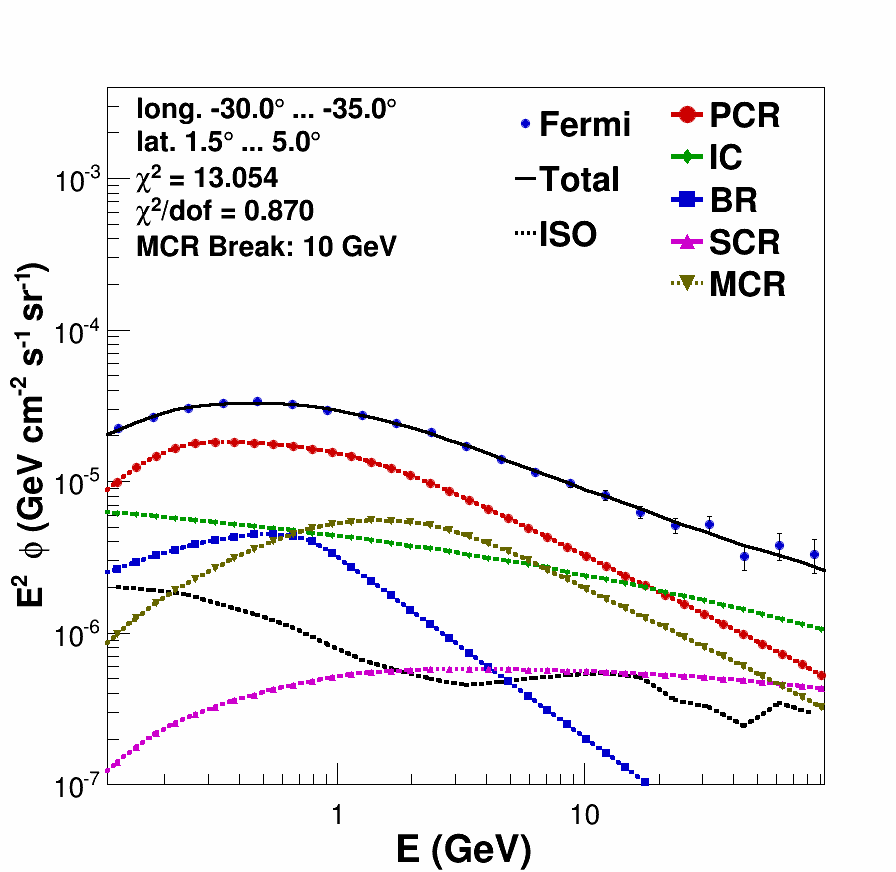}
\includegraphics[width=0.16\textwidth,height=0.16\textwidth,clip]{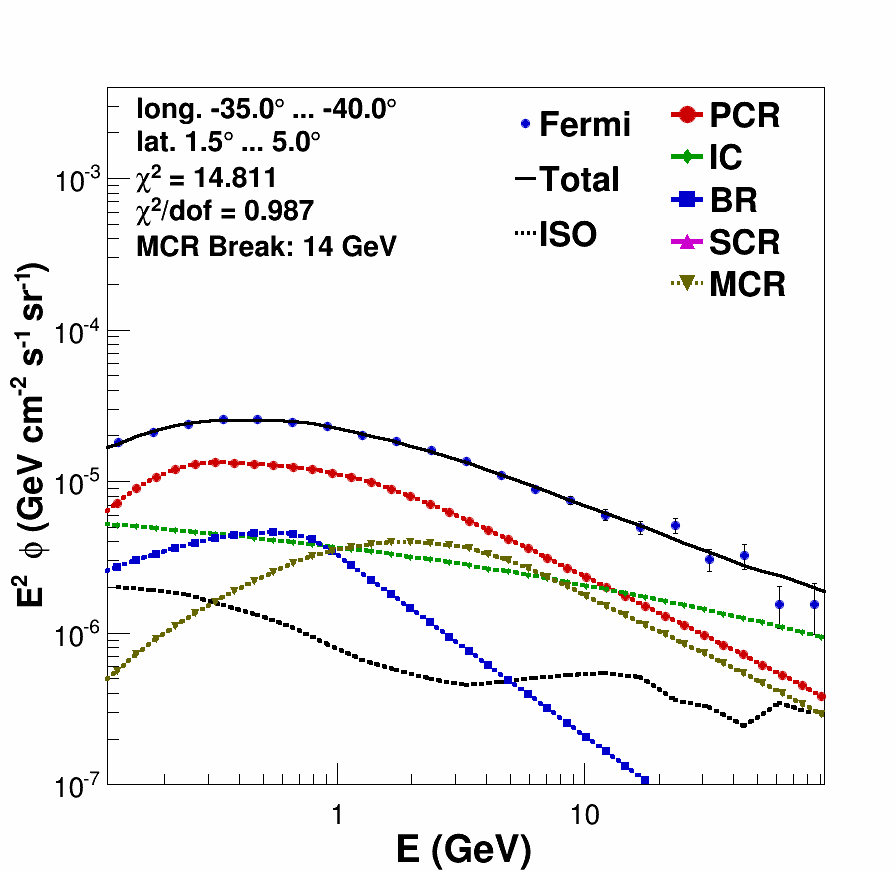}
\includegraphics[width=0.16\textwidth,height=0.16\textwidth,clip]{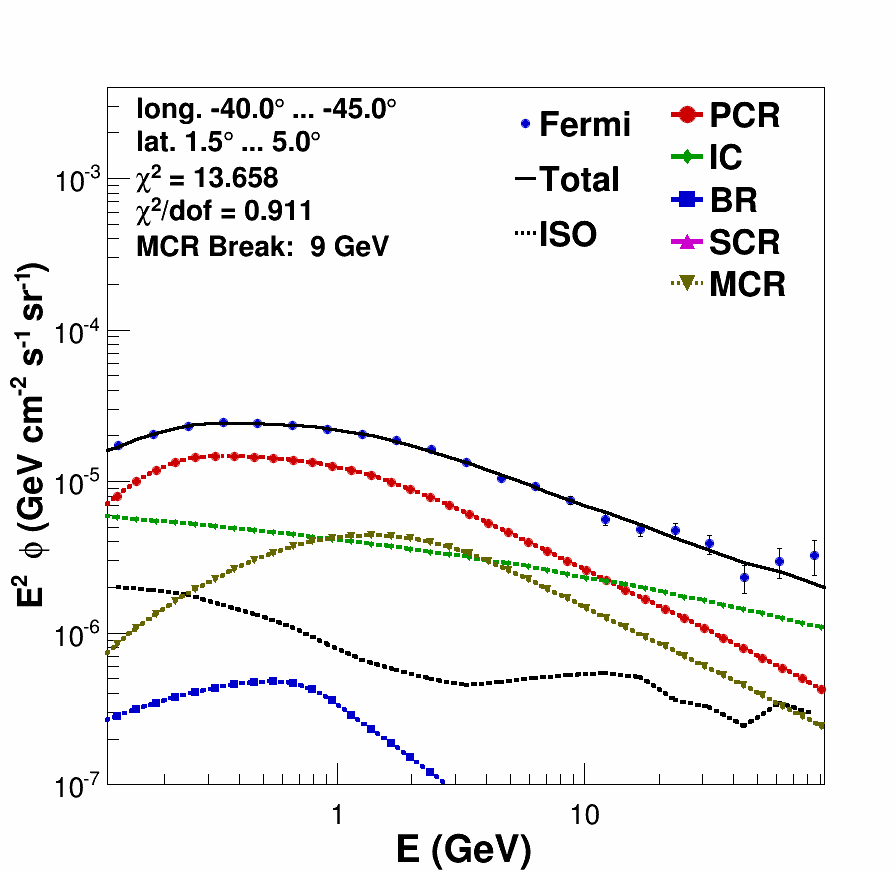}
\includegraphics[width=0.16\textwidth,height=0.16\textwidth,clip]{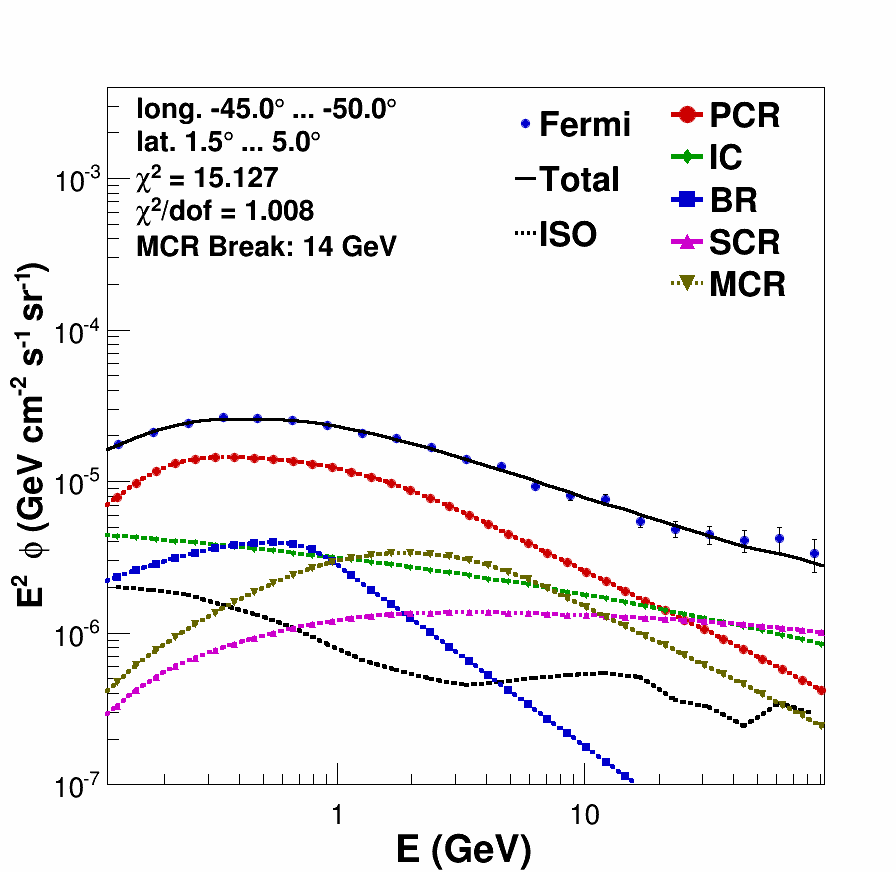}
\includegraphics[width=0.16\textwidth,height=0.16\textwidth,clip]{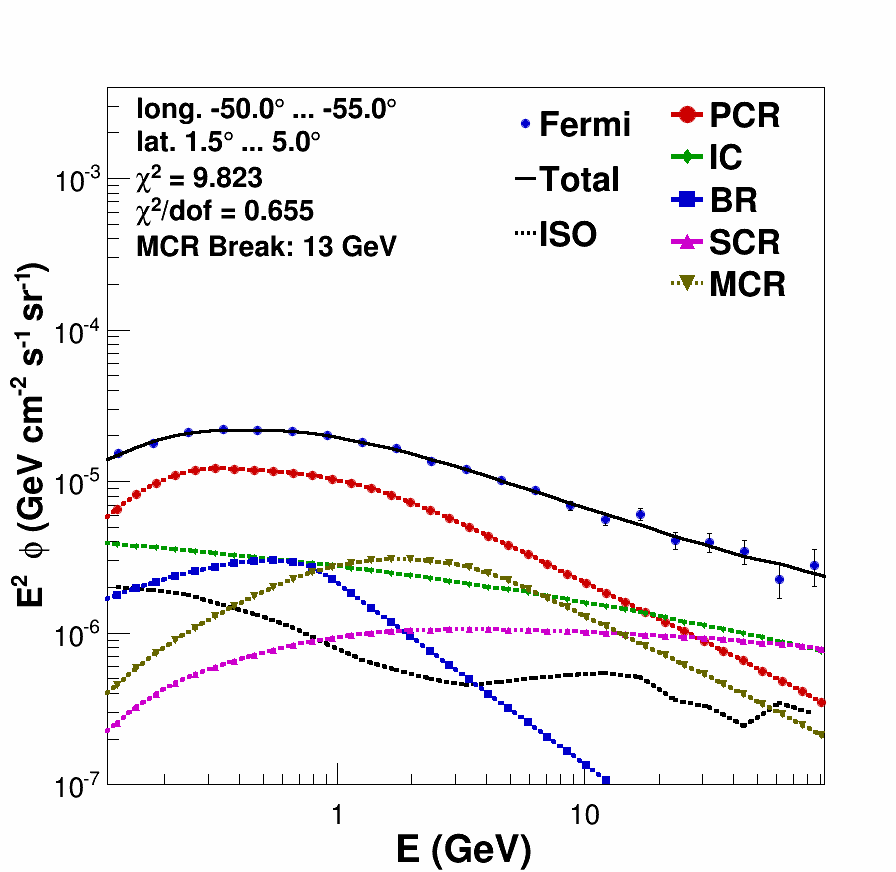}
\includegraphics[width=0.16\textwidth,height=0.16\textwidth,clip]{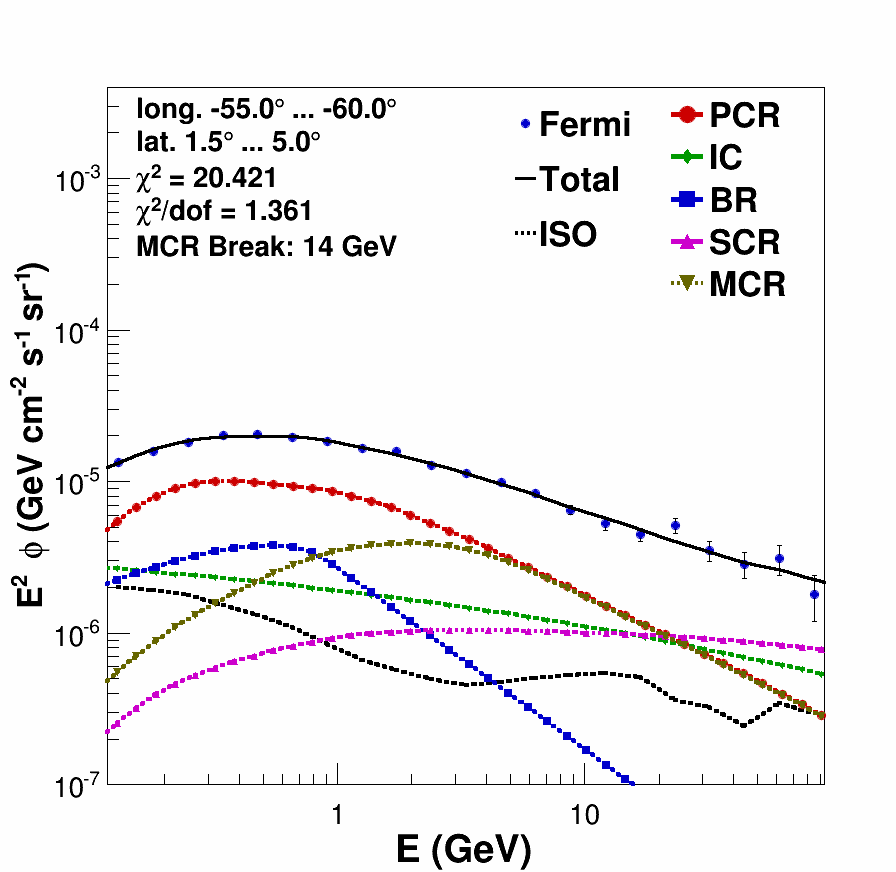}
\includegraphics[width=0.16\textwidth,height=0.16\textwidth,clip]{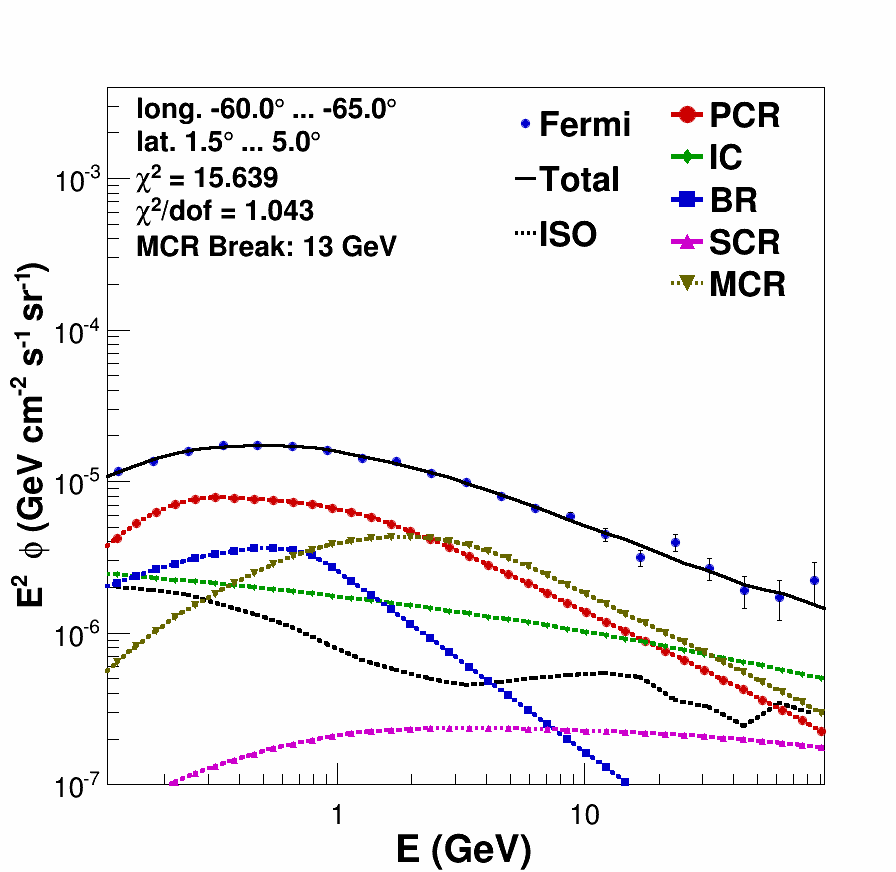}
\includegraphics[width=0.16\textwidth,height=0.16\textwidth,clip]{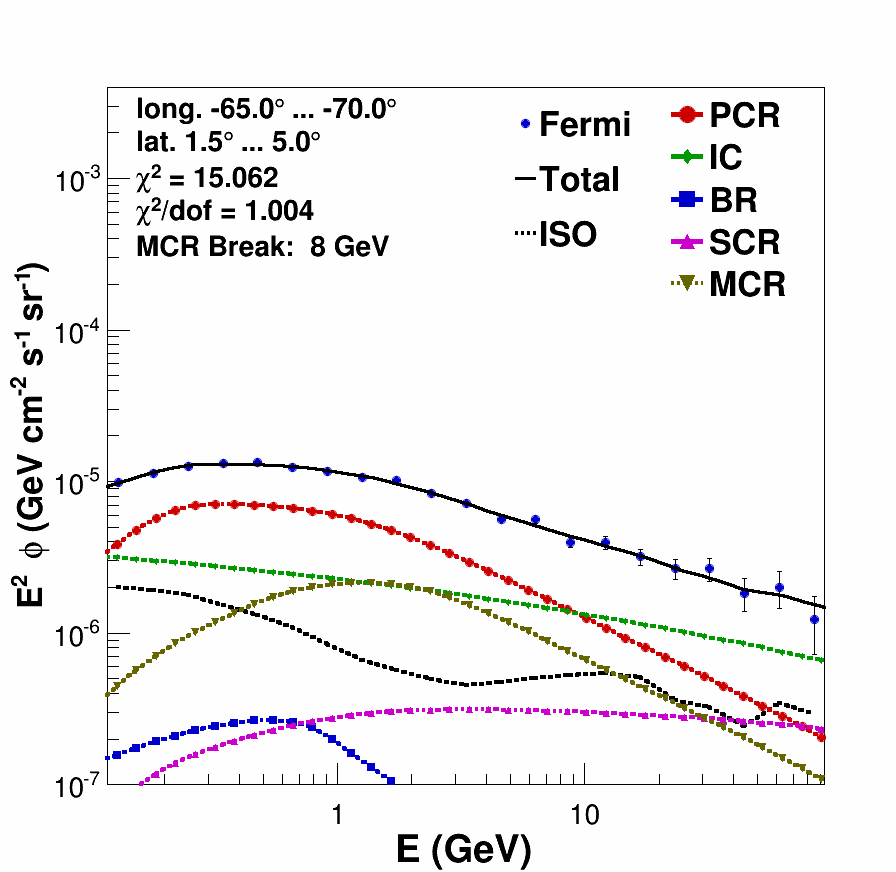}
\includegraphics[width=0.16\textwidth,height=0.16\textwidth,clip]{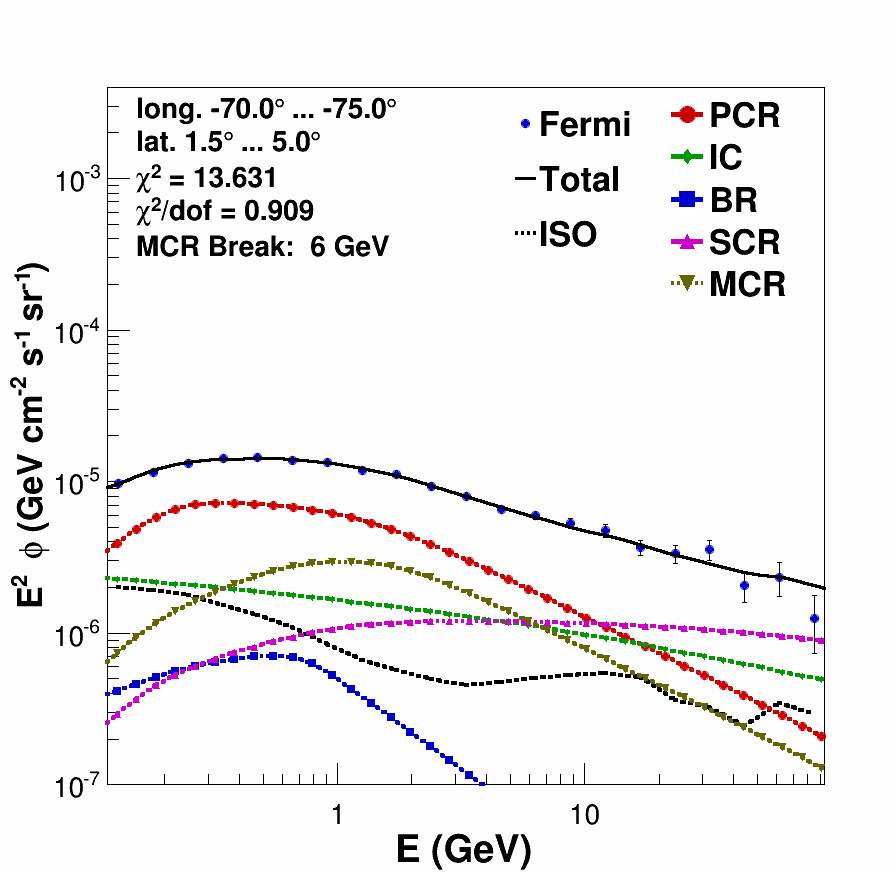}
\includegraphics[width=0.16\textwidth,height=0.16\textwidth,clip]{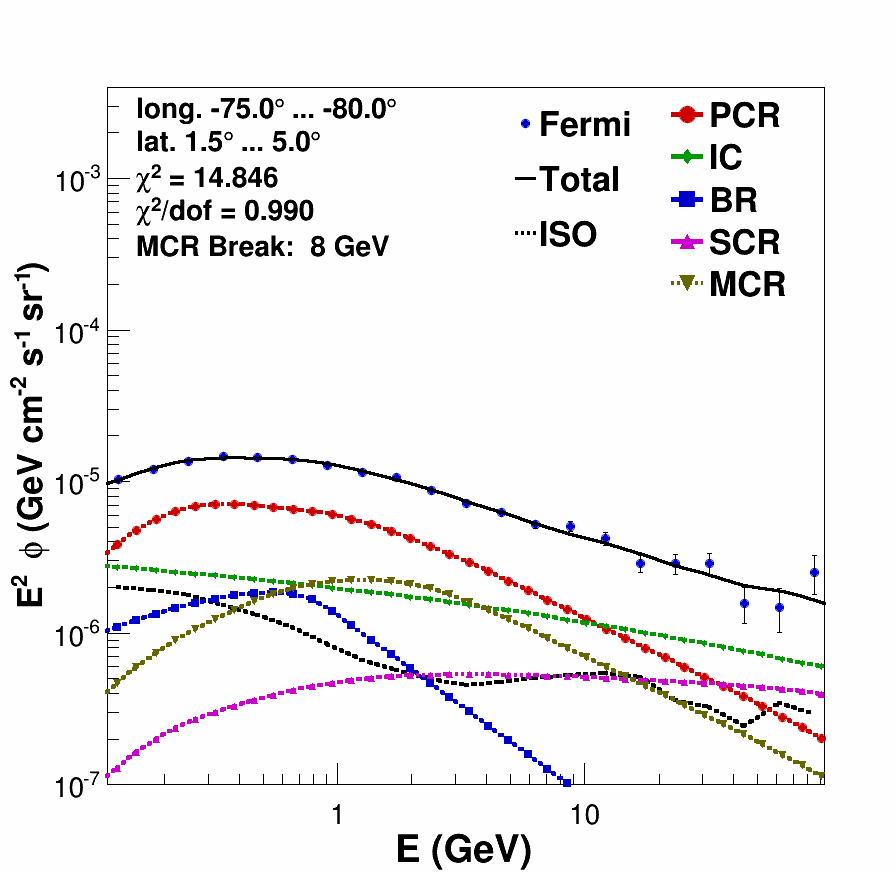}
\includegraphics[width=0.16\textwidth,height=0.16\textwidth,clip]{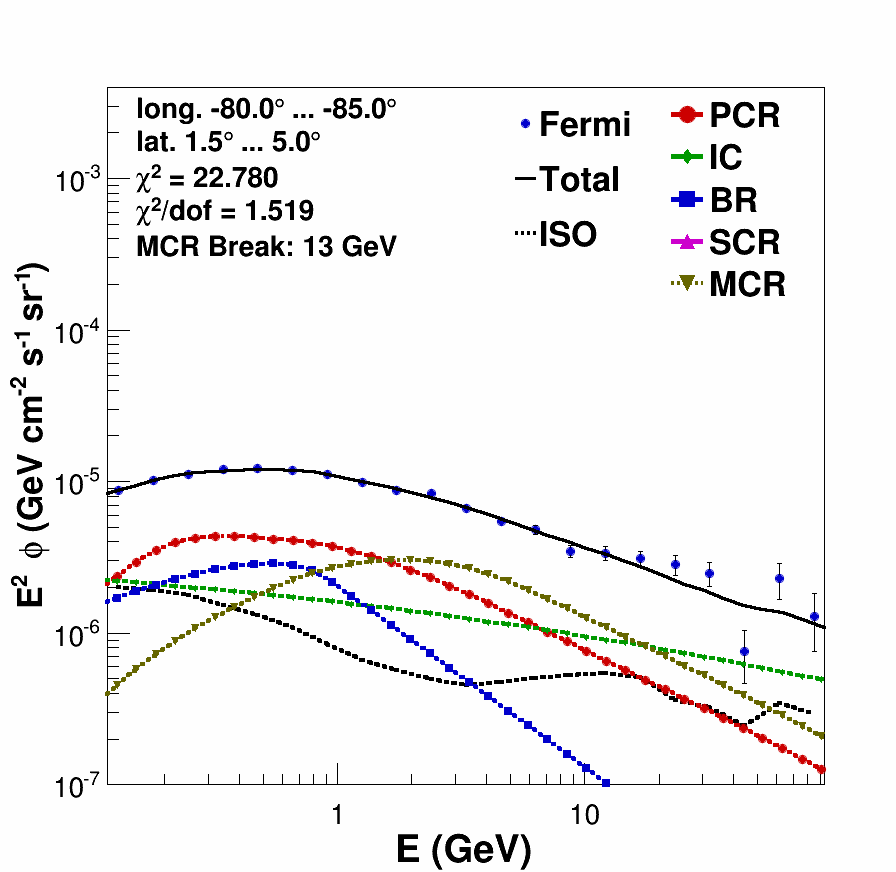}
\includegraphics[width=0.16\textwidth,height=0.16\textwidth,clip]{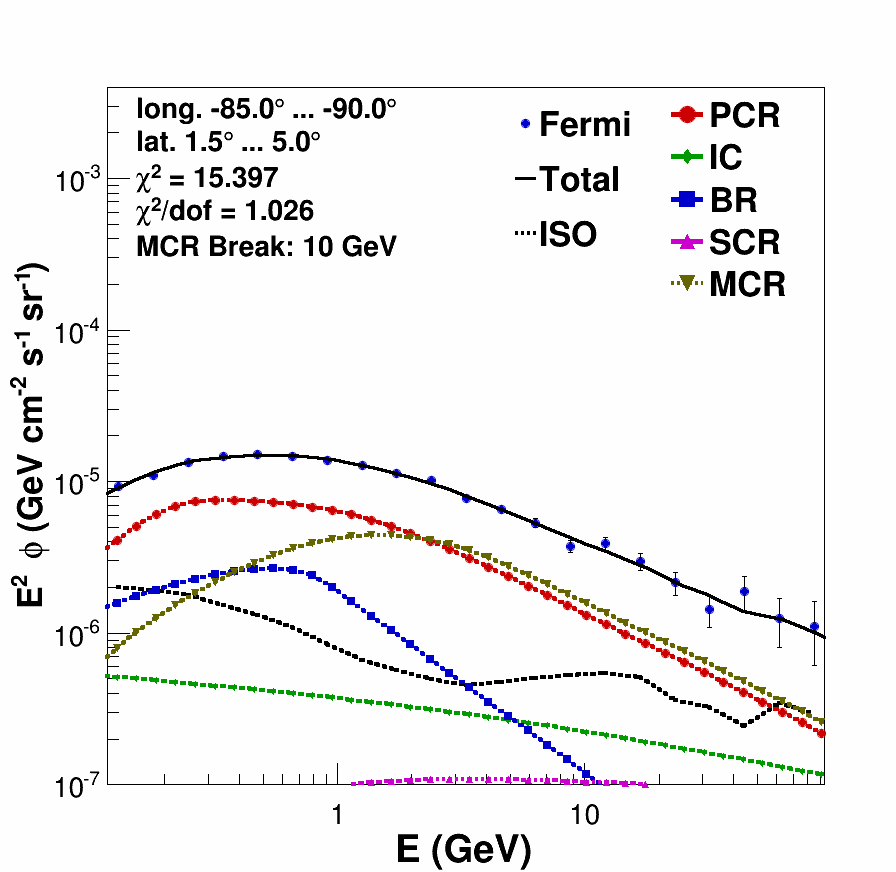}
\includegraphics[width=0.16\textwidth,height=0.16\textwidth,clip]{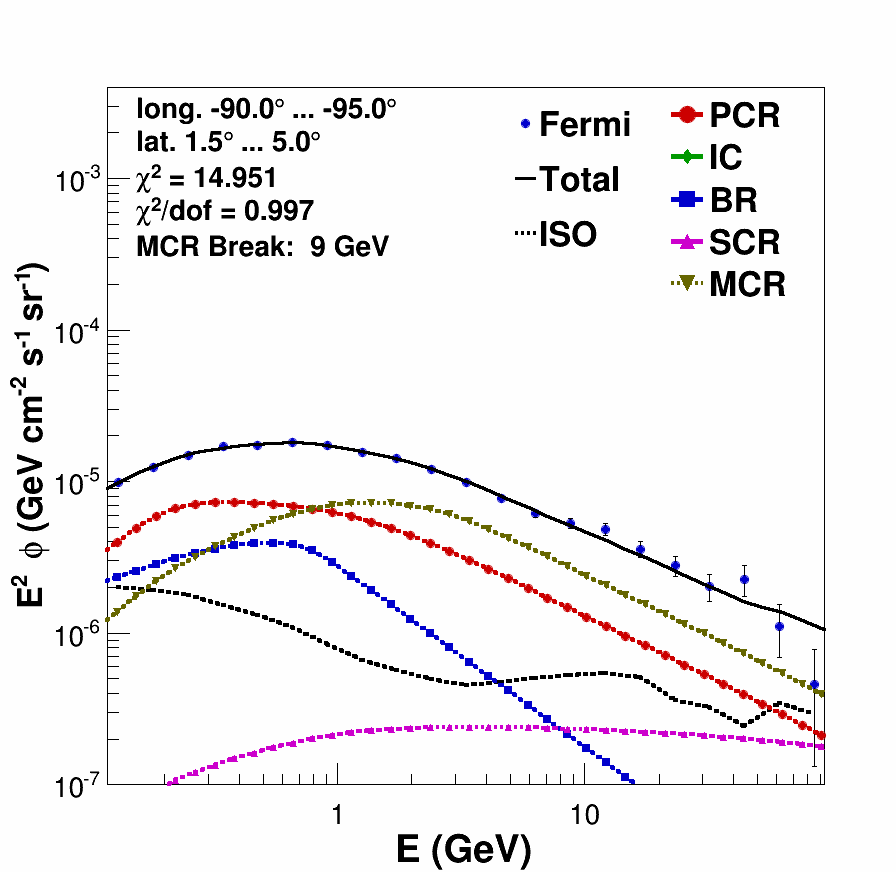}
\includegraphics[width=0.16\textwidth,height=0.16\textwidth,clip]{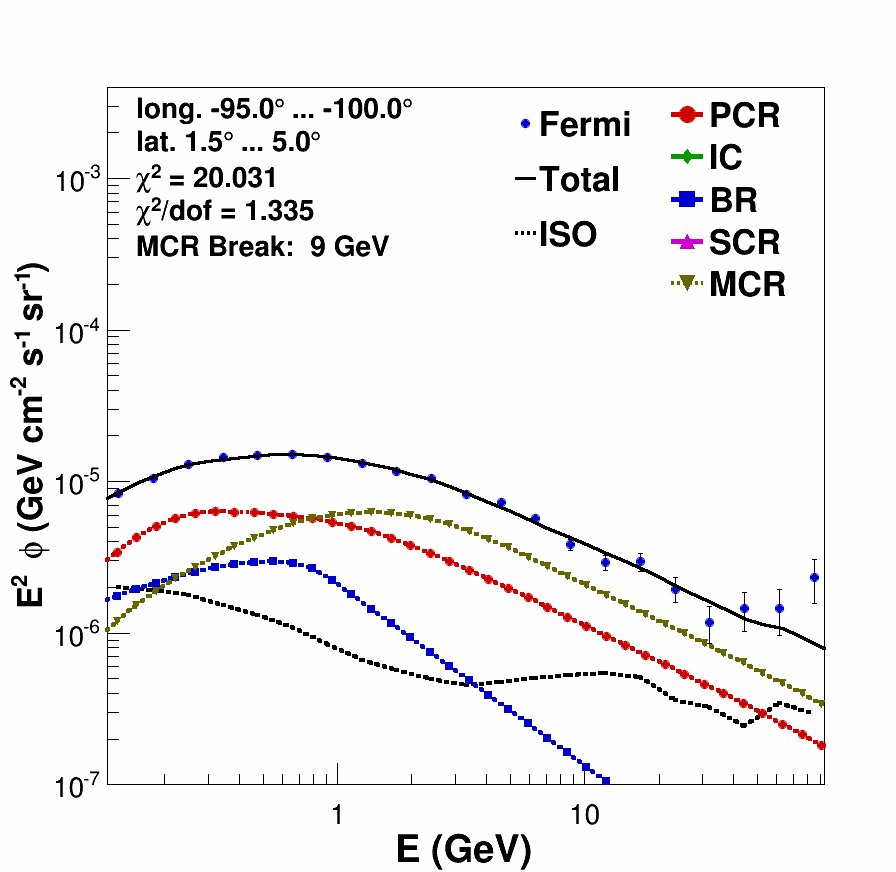}
\includegraphics[width=0.16\textwidth,height=0.16\textwidth,clip]{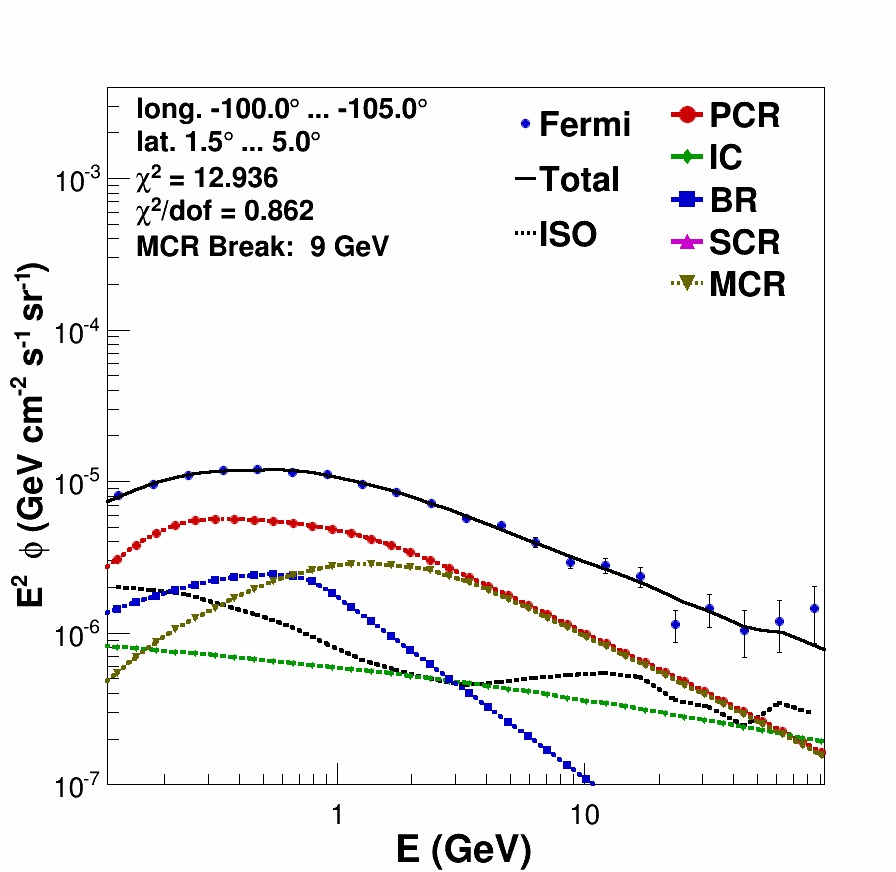}
\includegraphics[width=0.16\textwidth,height=0.16\textwidth,clip]{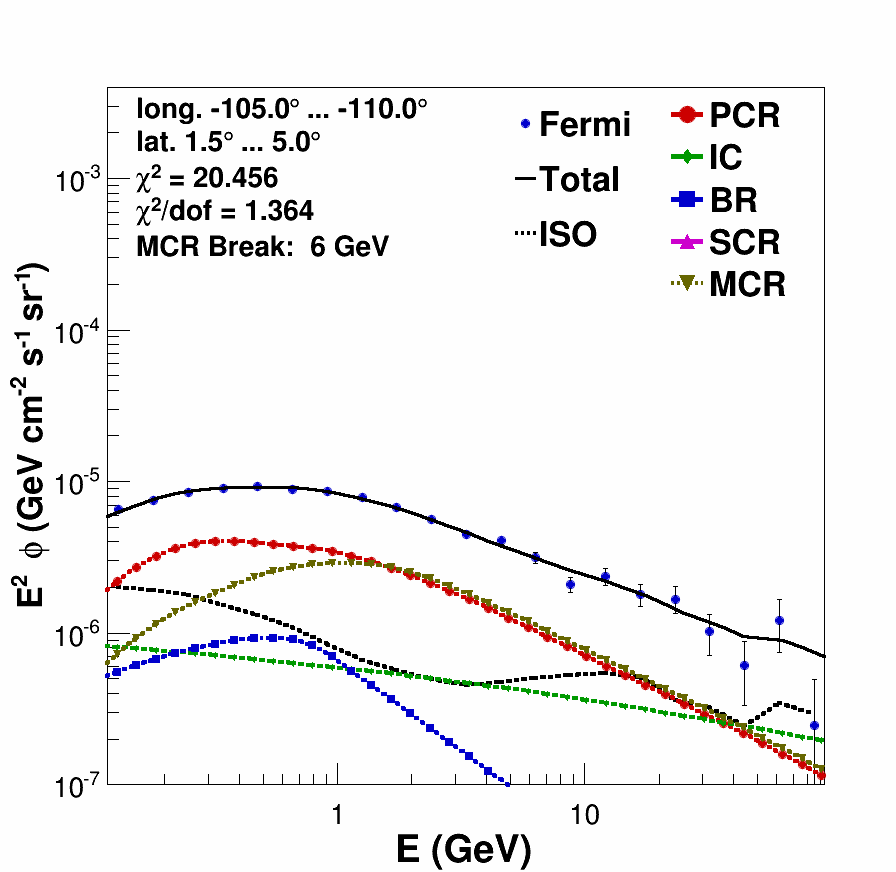}
\includegraphics[width=0.16\textwidth,height=0.16\textwidth,clip]{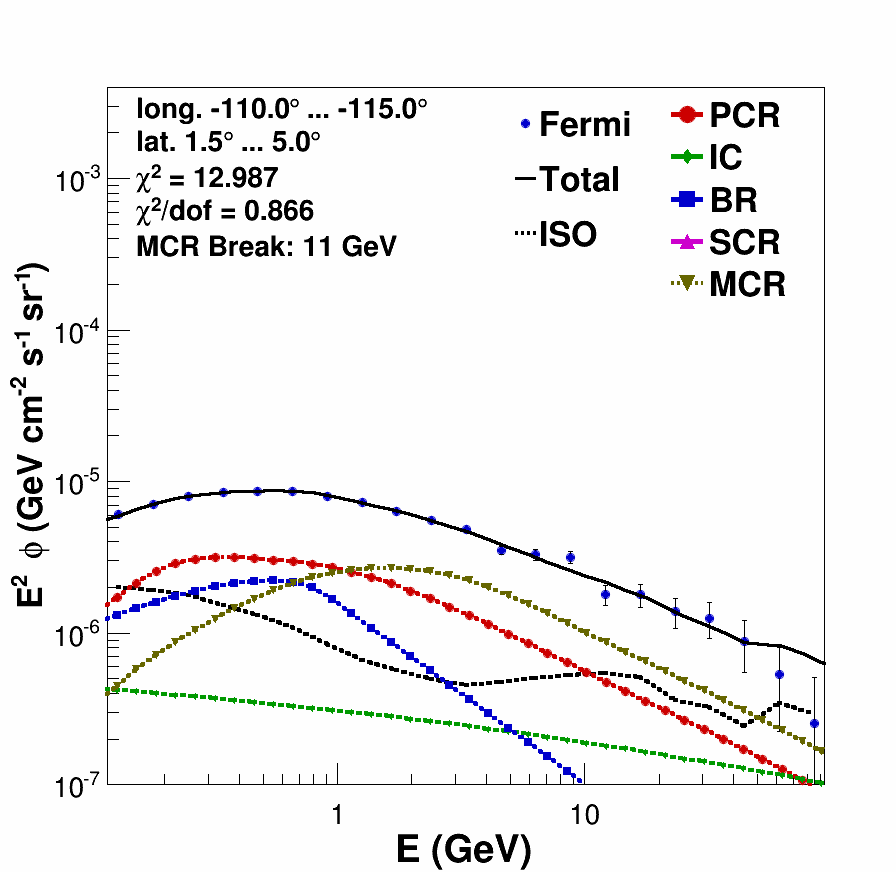}
\includegraphics[width=0.16\textwidth,height=0.16\textwidth,clip]{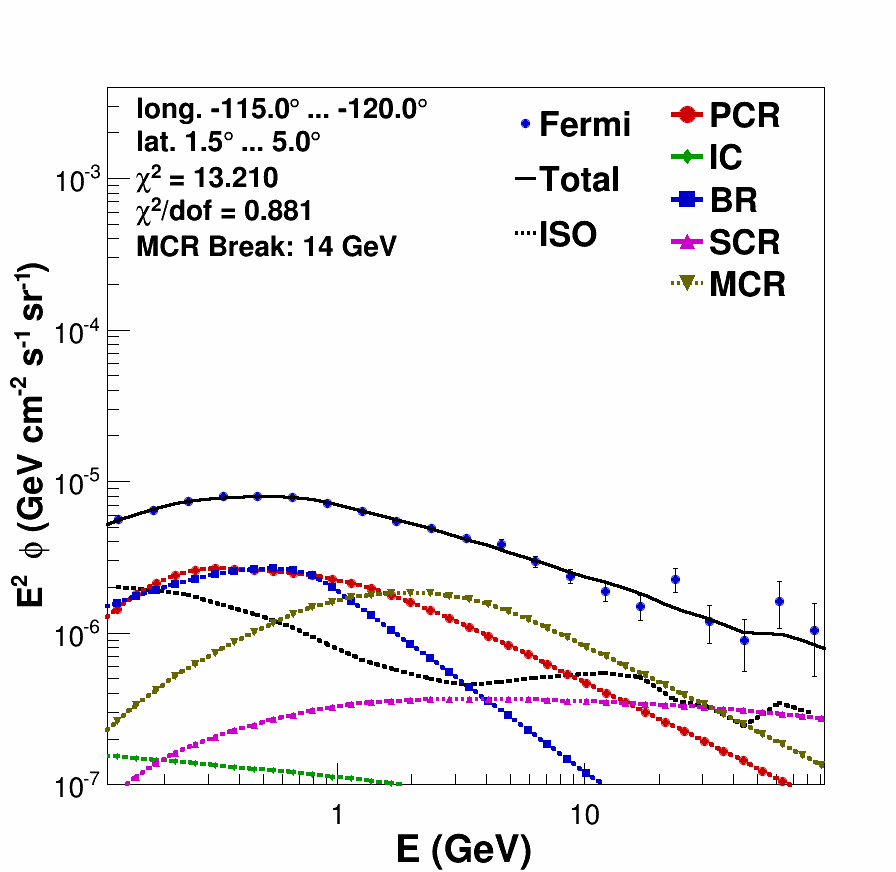}
\includegraphics[width=0.16\textwidth,height=0.16\textwidth,clip]{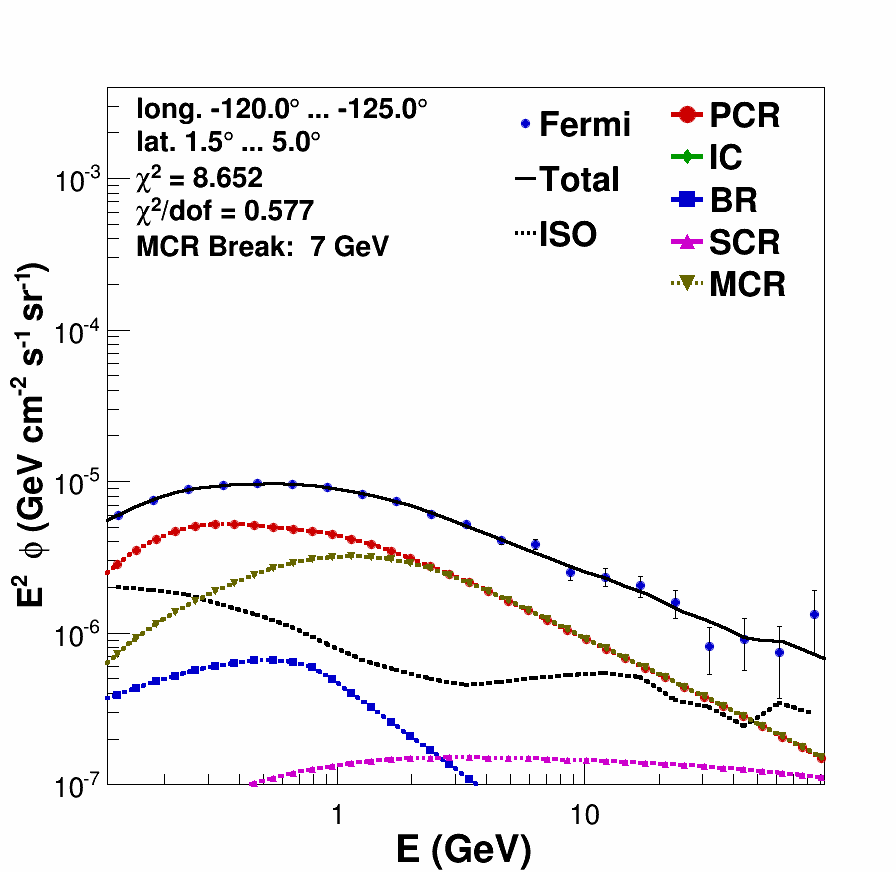}
\includegraphics[width=0.16\textwidth,height=0.16\textwidth,clip]{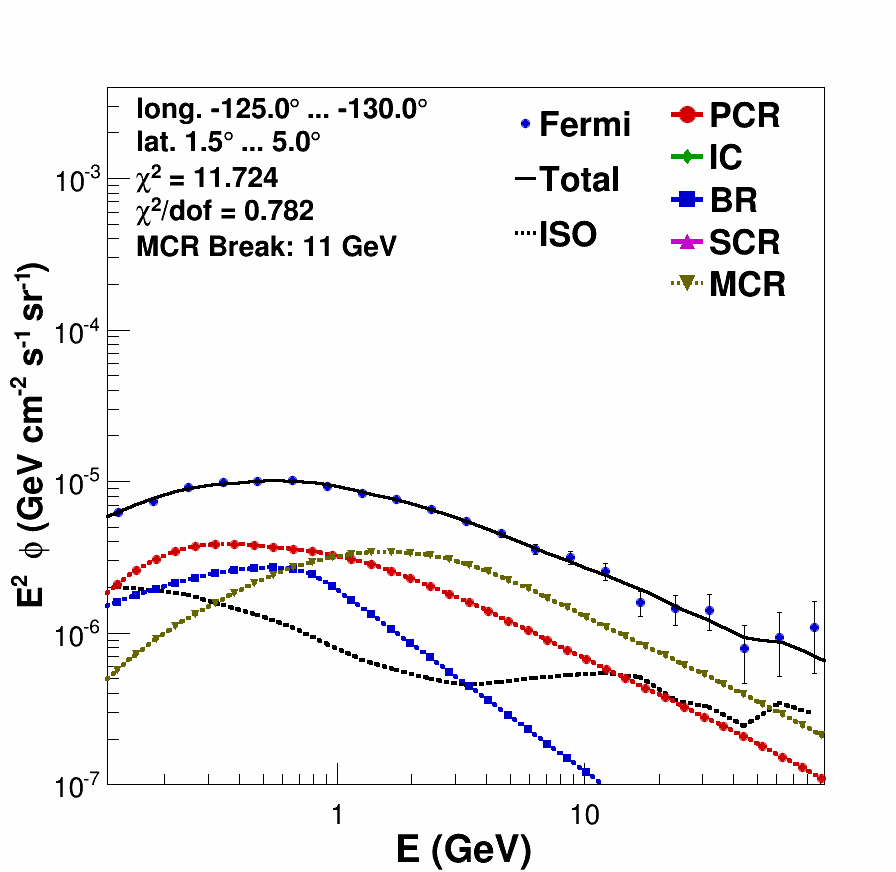}
\includegraphics[width=0.16\textwidth,height=0.16\textwidth,clip]{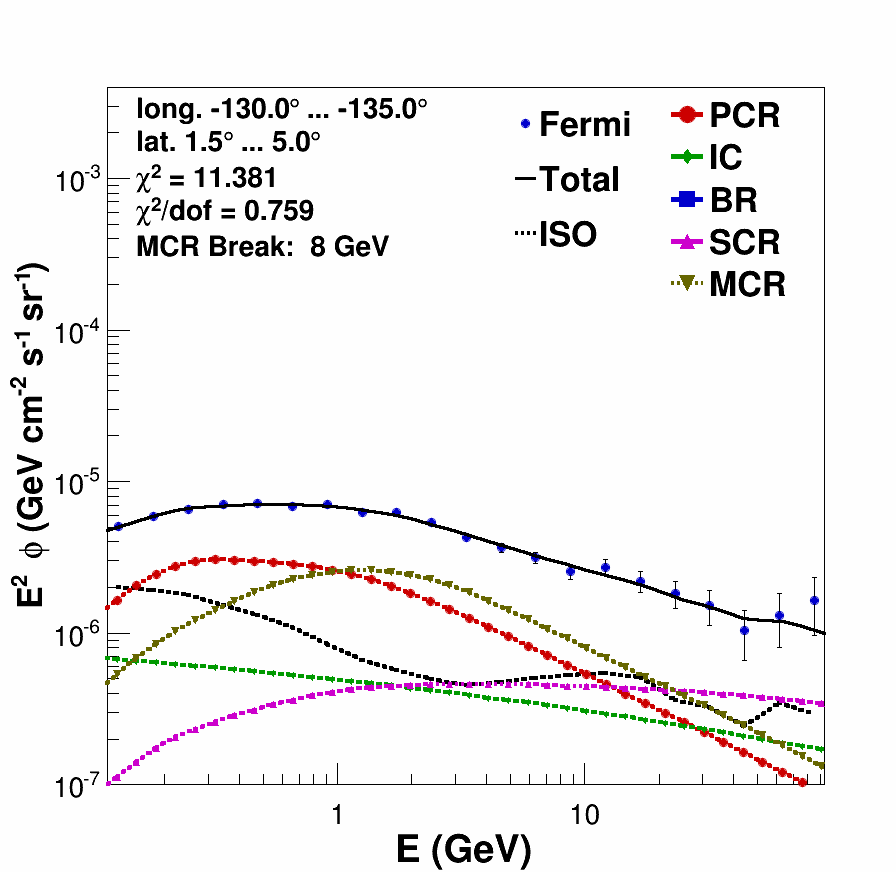}
\includegraphics[width=0.16\textwidth,height=0.16\textwidth,clip]{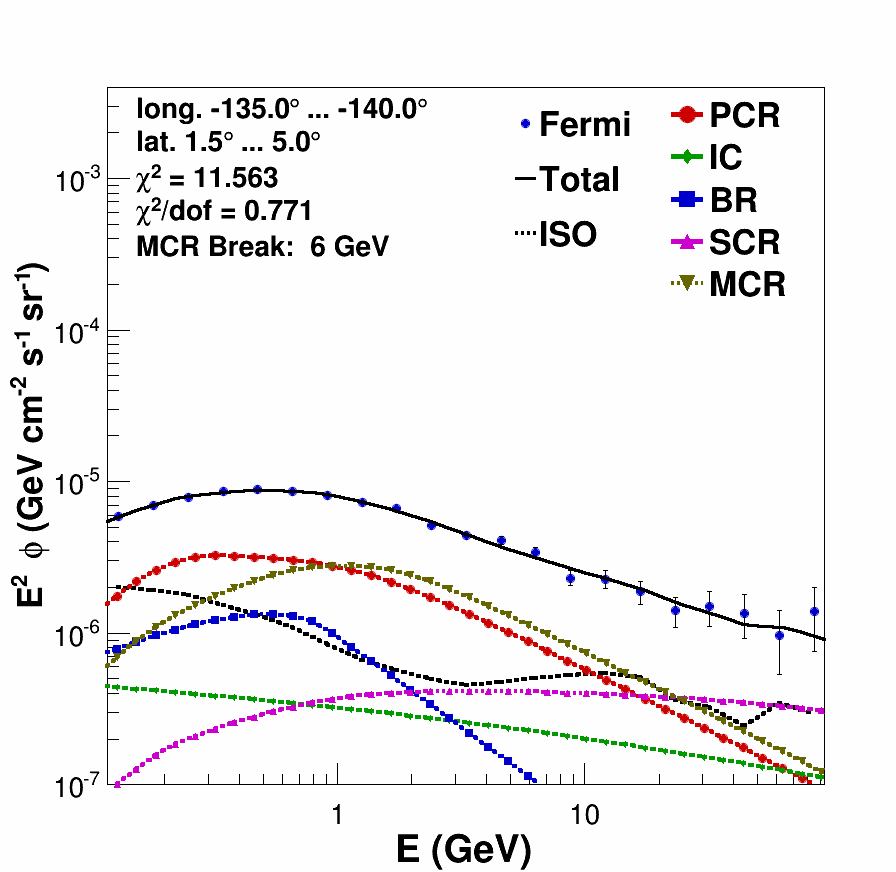}
\includegraphics[width=0.16\textwidth,height=0.16\textwidth,clip]{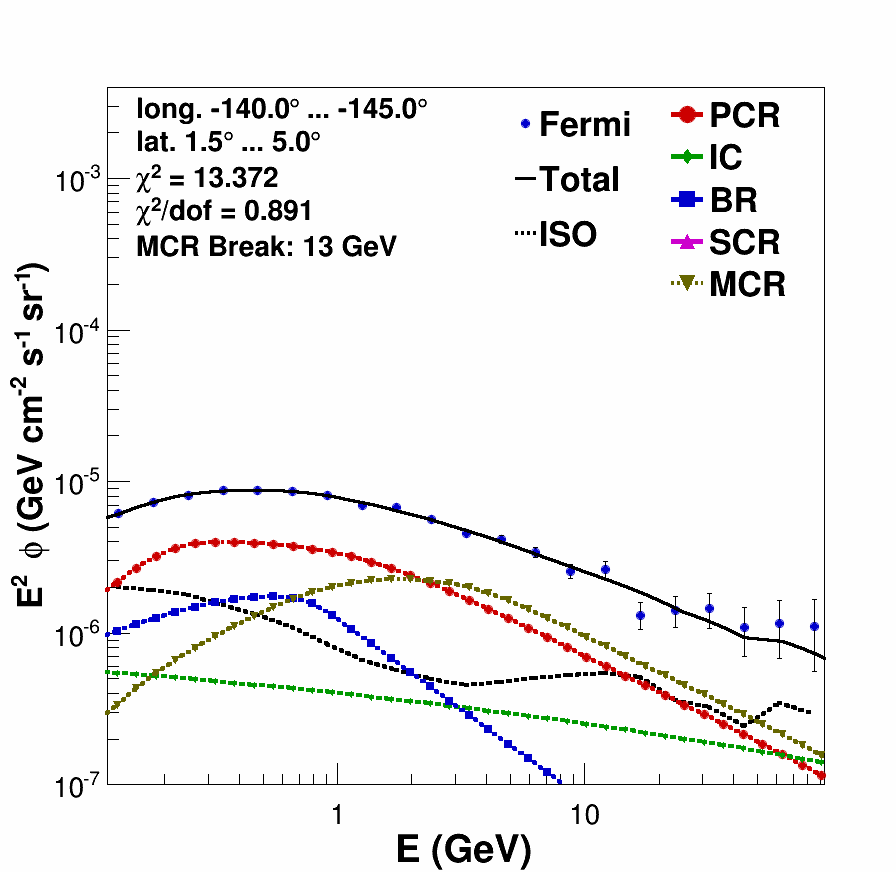}
\includegraphics[width=0.16\textwidth,height=0.16\textwidth,clip]{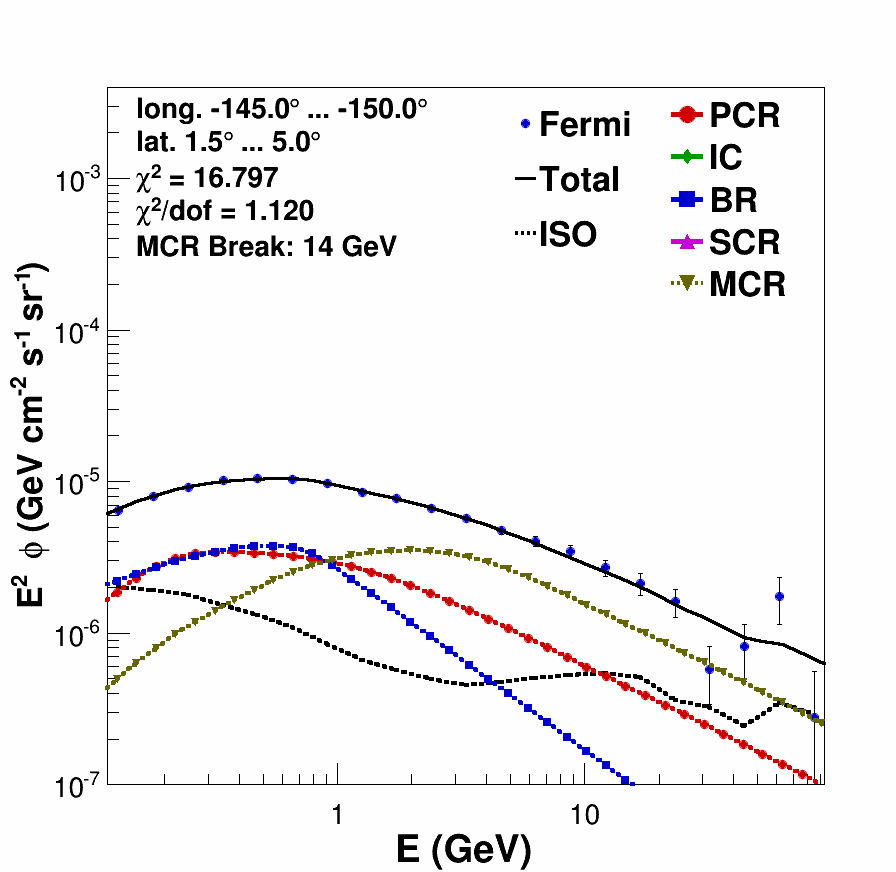}
\includegraphics[width=0.16\textwidth,height=0.16\textwidth,clip]{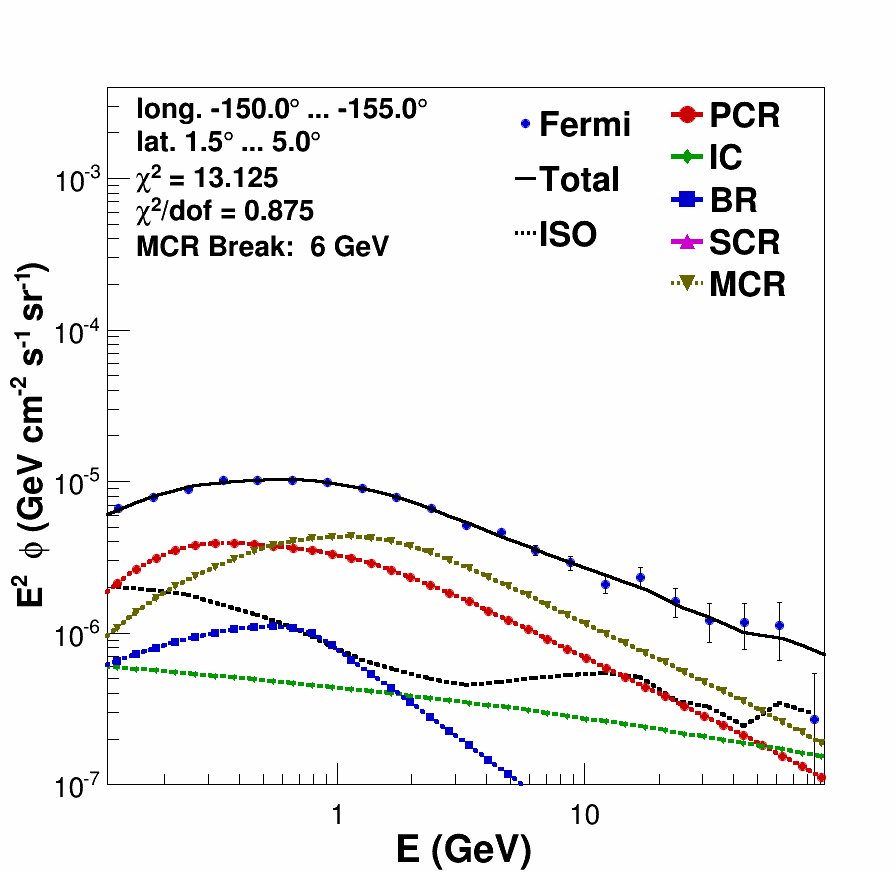}
\includegraphics[width=0.16\textwidth,height=0.16\textwidth,clip]{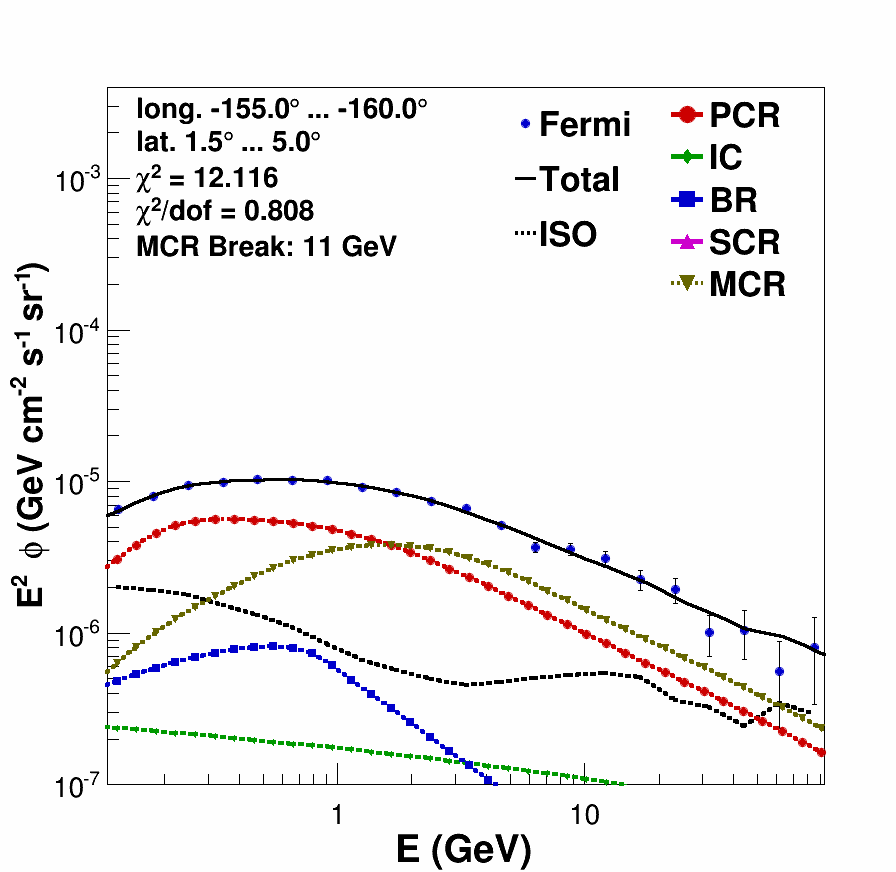}
\includegraphics[width=0.16\textwidth,height=0.16\textwidth,clip]{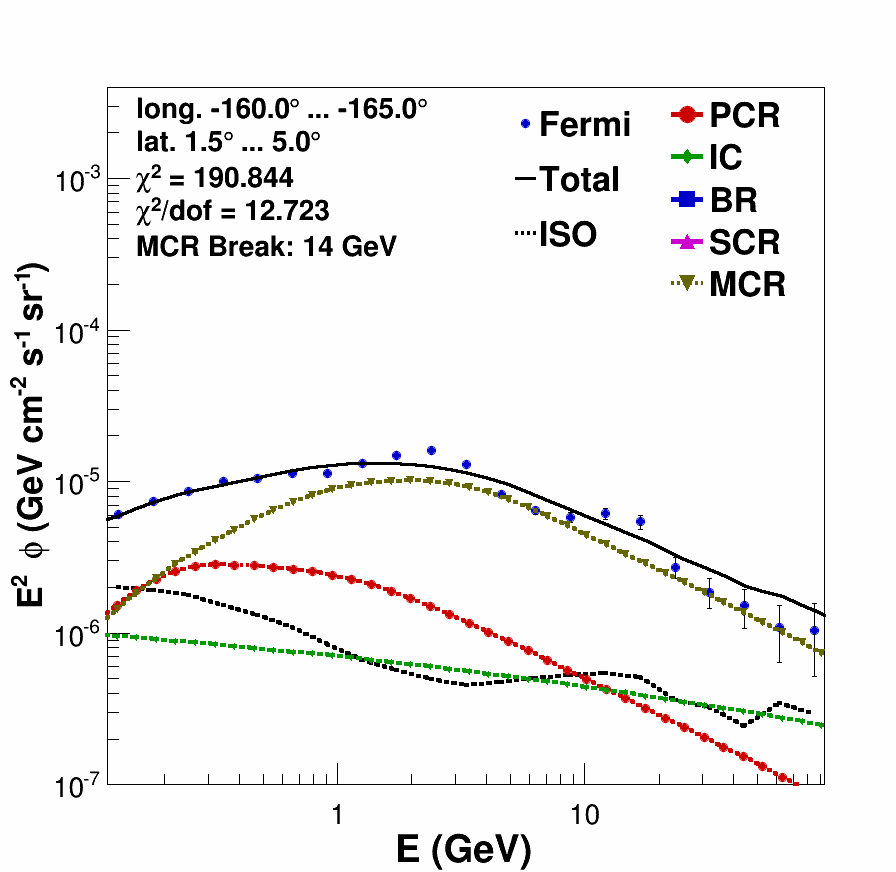}
\includegraphics[width=0.16\textwidth,height=0.16\textwidth,clip]{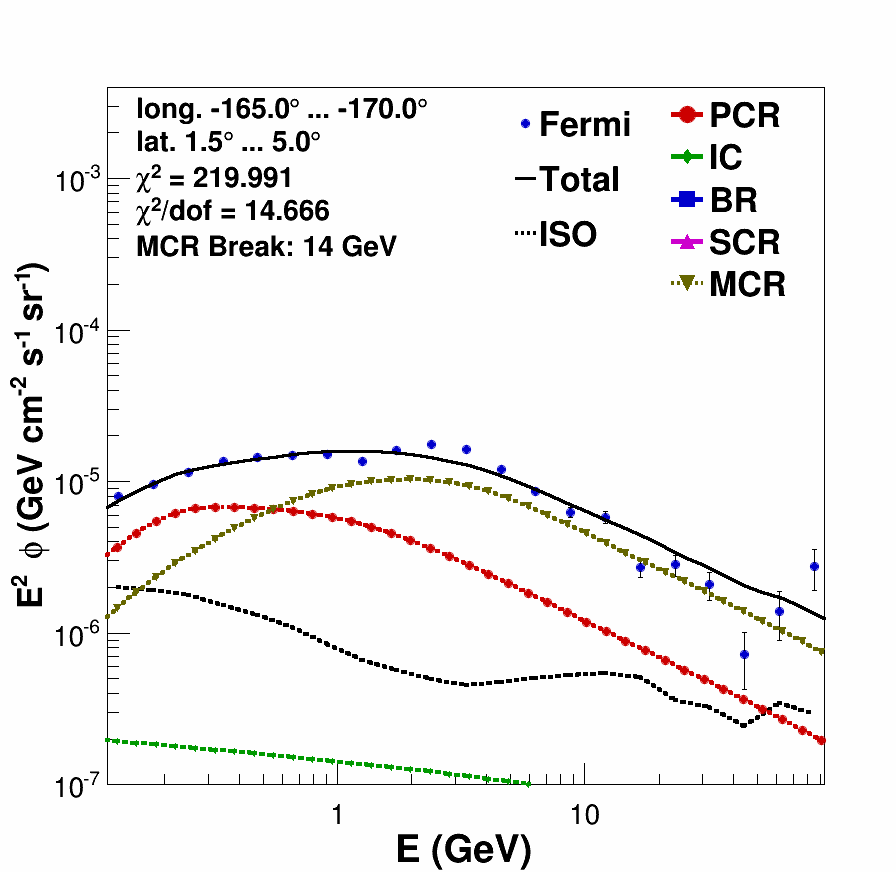}
\includegraphics[width=0.16\textwidth,height=0.16\textwidth,clip]{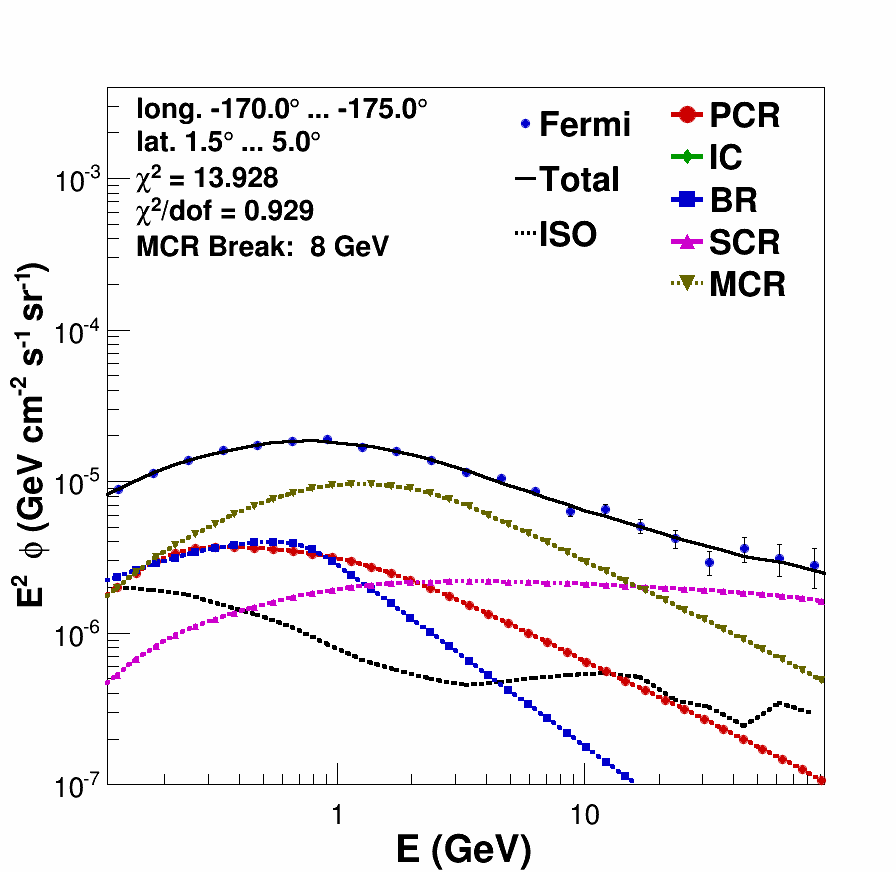}
\includegraphics[width=0.16\textwidth,height=0.16\textwidth,clip]{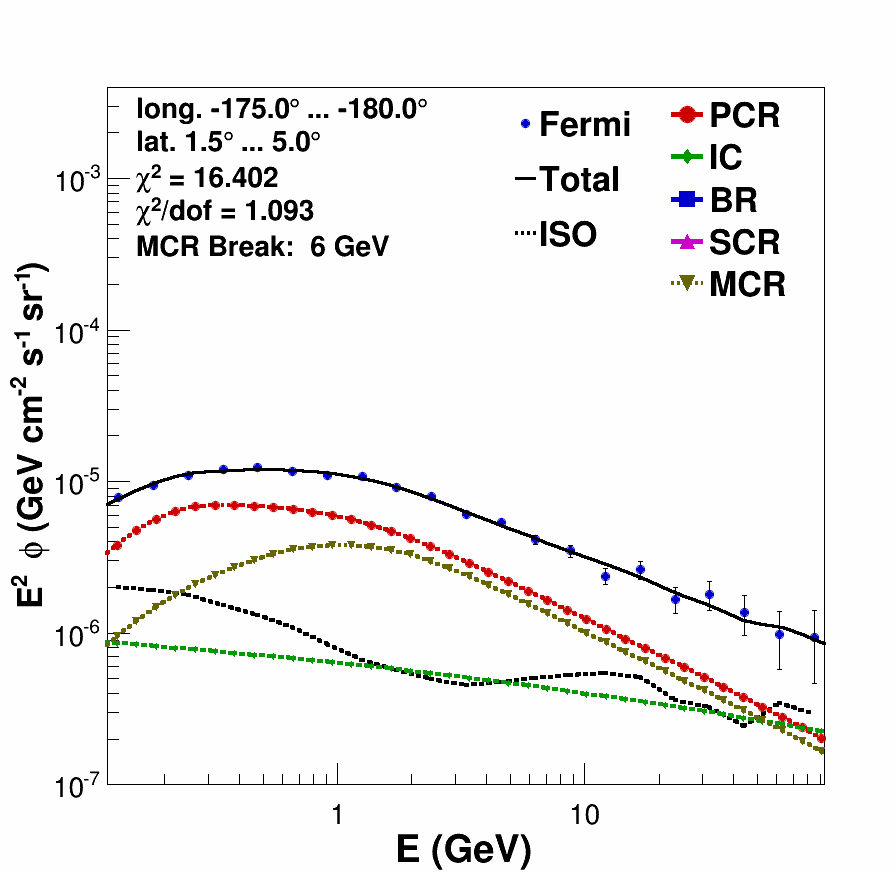}%%%%%%r8b
\caption[]{Template fits for latitudes  with $1.5^\circ<b<5.0^\circ$ and longitudes decreasing from 0$^\circ$ to -180$^\circ$.} \label{F19}
\end{figure}
\begin{figure}
\centering
\includegraphics[width=0.16\textwidth,height=0.16\textwidth,clip]{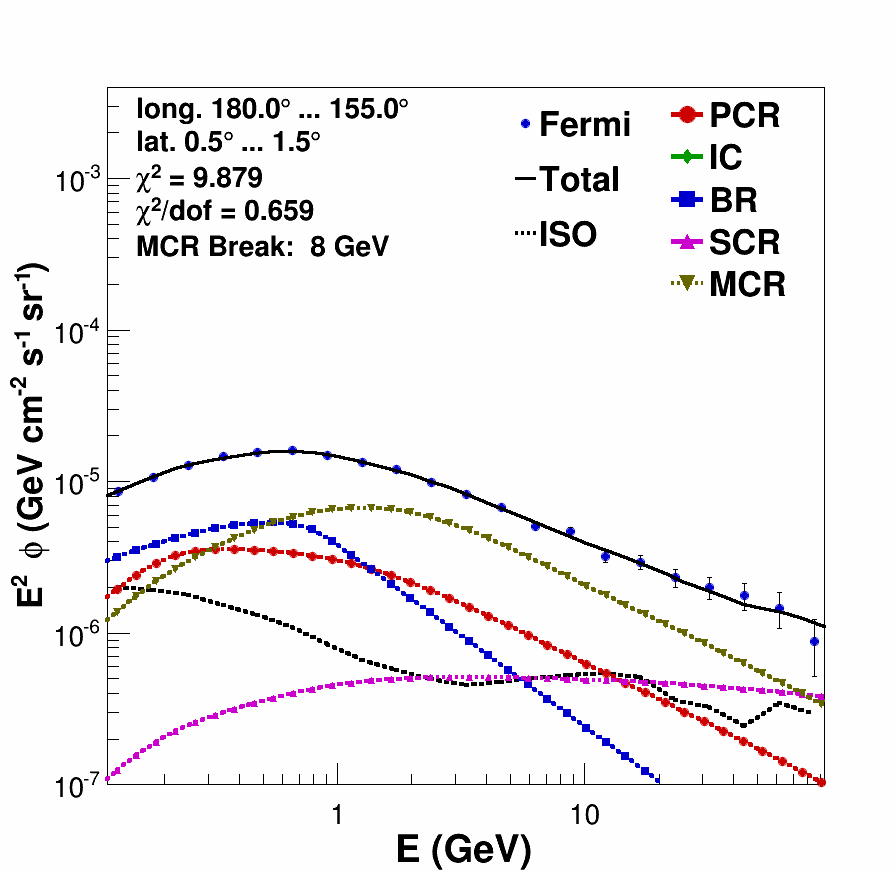}
\includegraphics[width=0.16\textwidth,height=0.16\textwidth,clip]{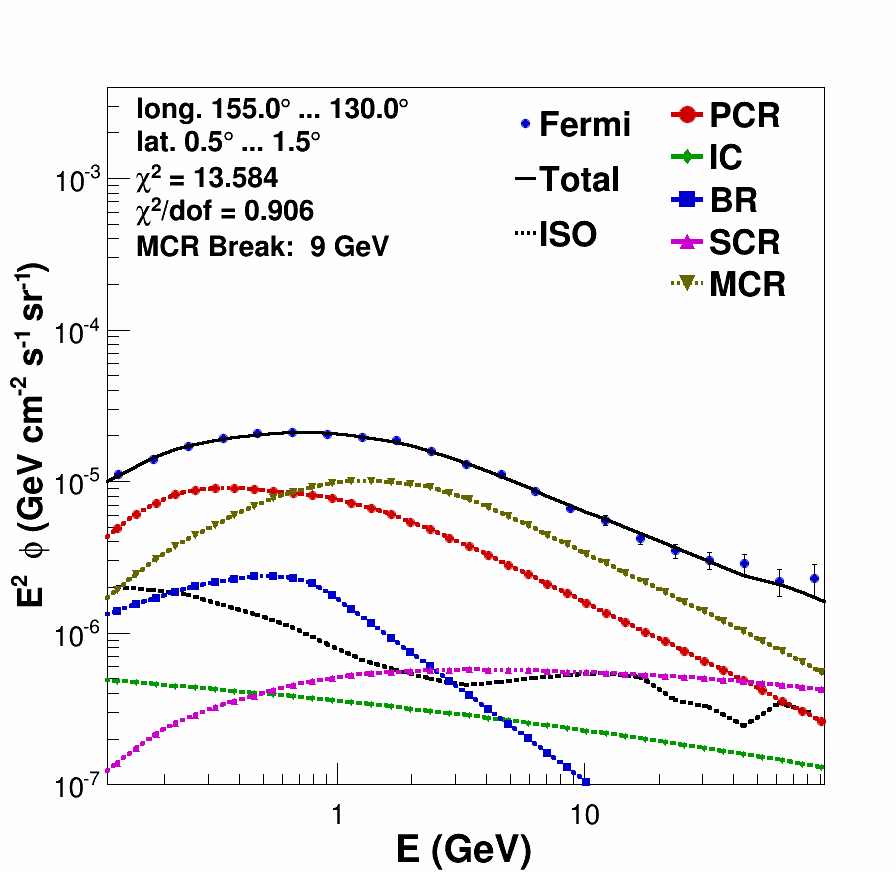}
\includegraphics[width=0.16\textwidth,height=0.16\textwidth,clip]{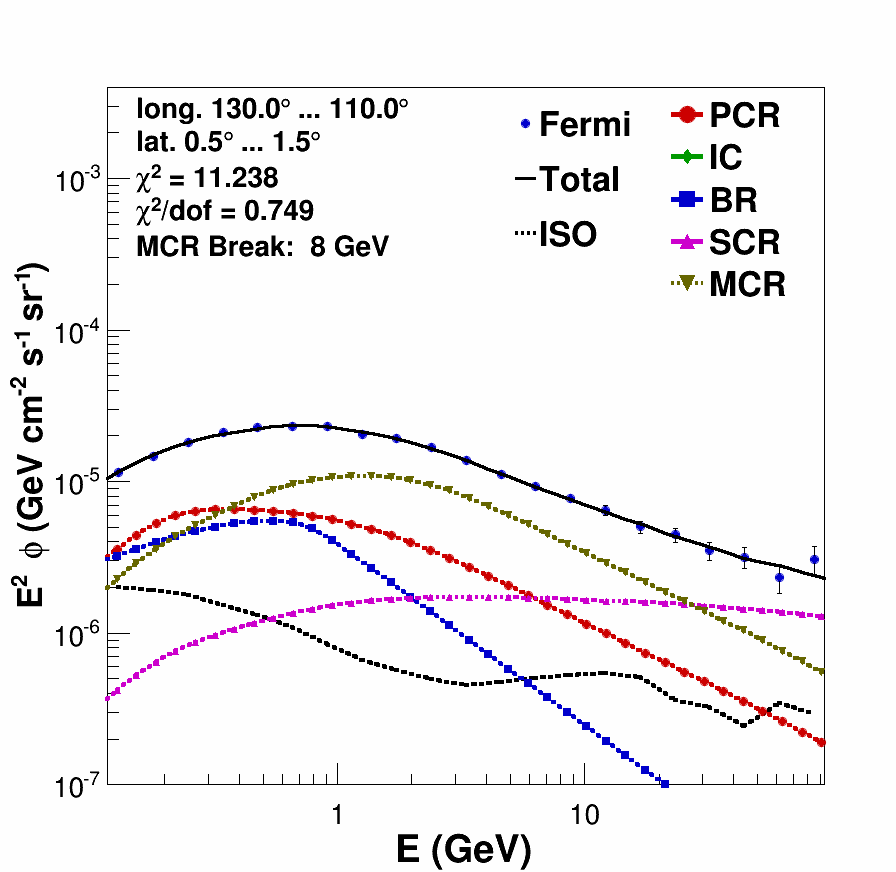}
\includegraphics[width=0.16\textwidth,height=0.16\textwidth,clip]{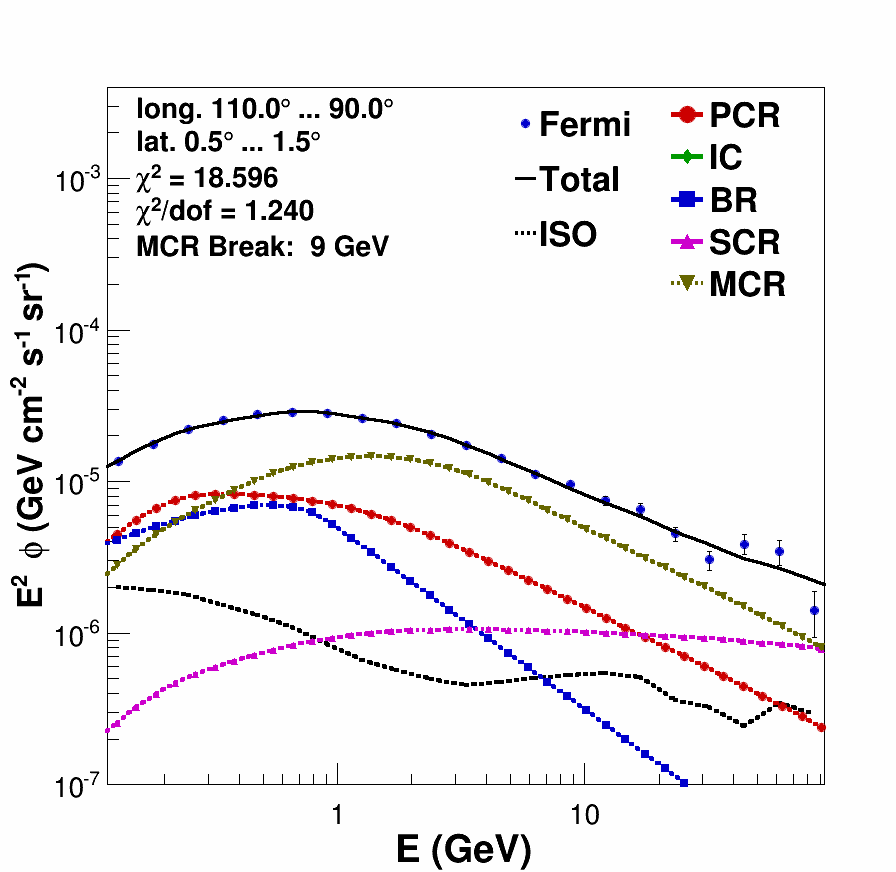}
\includegraphics[width=0.16\textwidth,height=0.16\textwidth,clip]{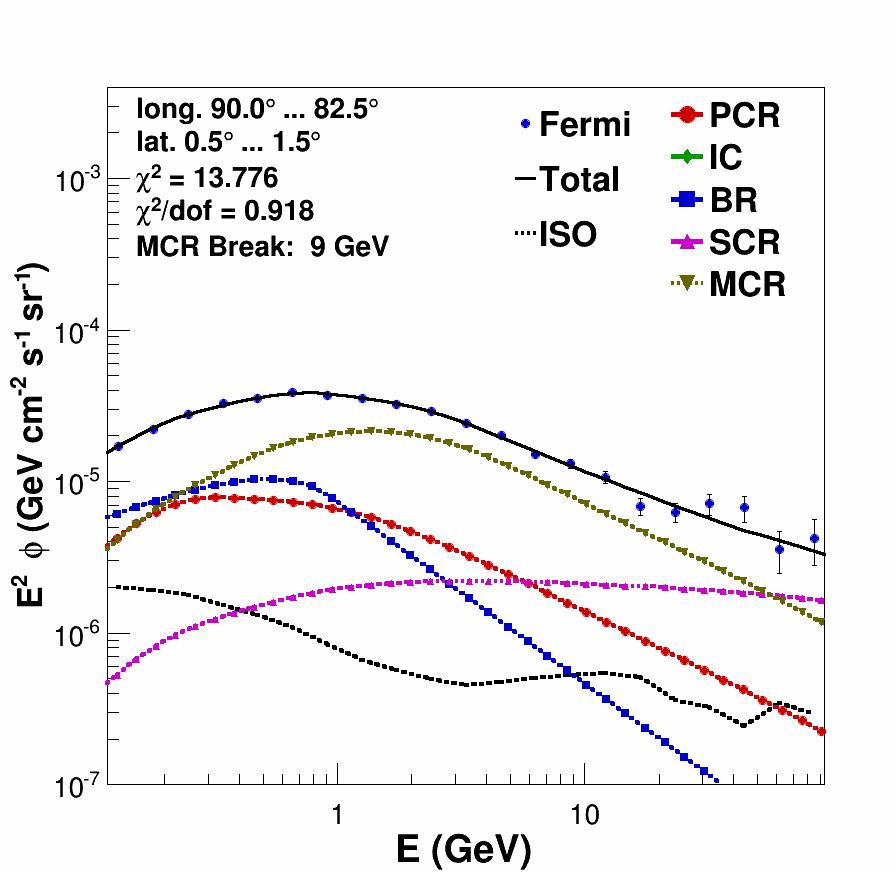}
\includegraphics[width=0.16\textwidth,height=0.16\textwidth,clip]{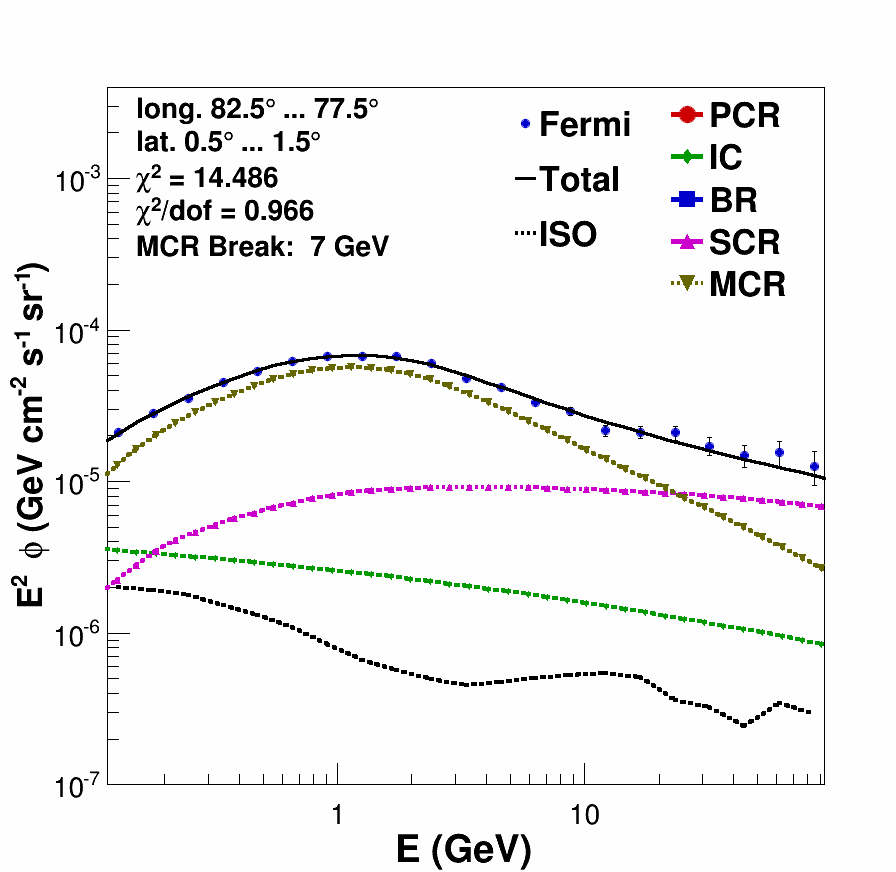}
\includegraphics[width=0.16\textwidth,height=0.16\textwidth,clip]{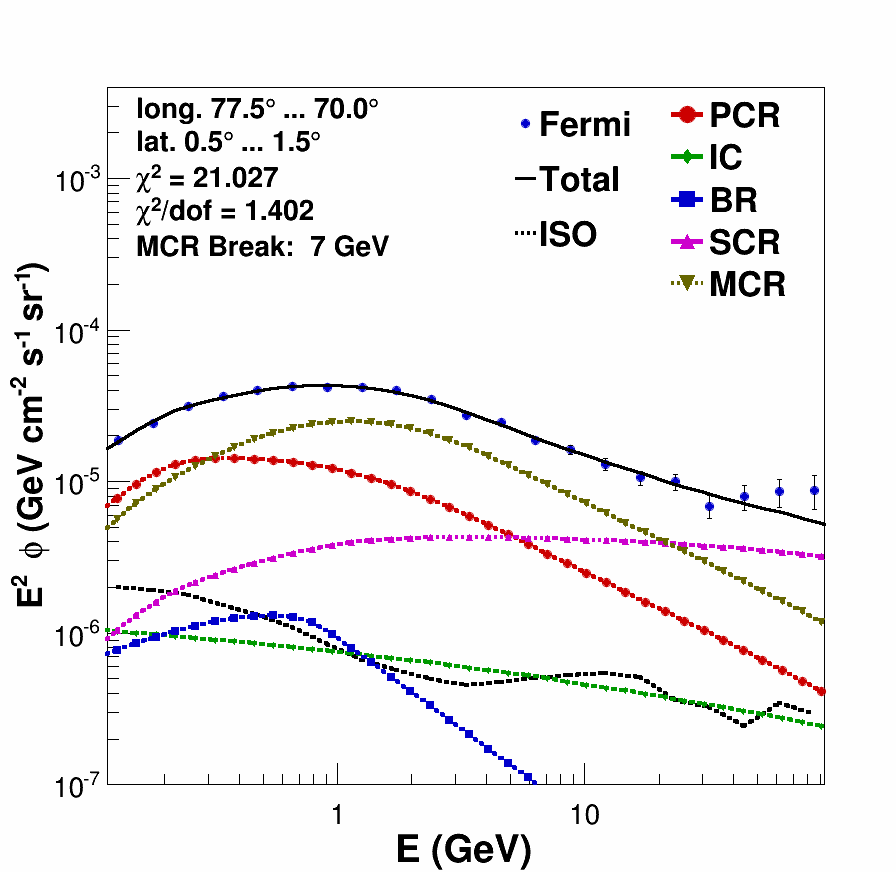}
\includegraphics[width=0.16\textwidth,height=0.16\textwidth,clip]{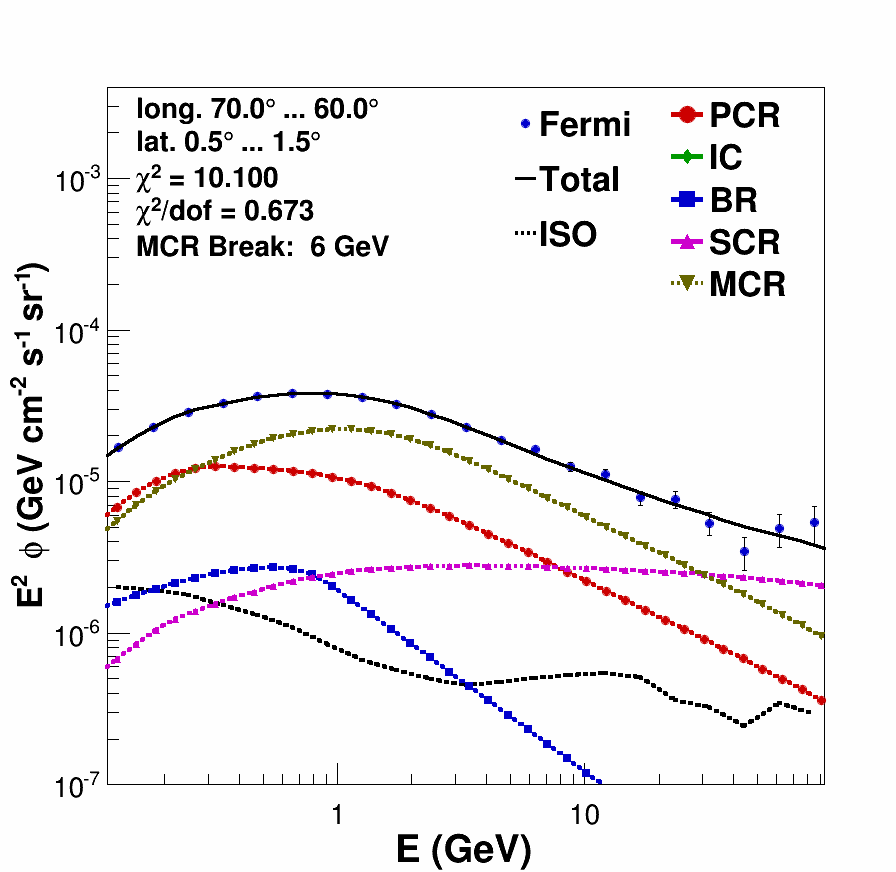}
\includegraphics[width=0.16\textwidth,height=0.16\textwidth,clip]{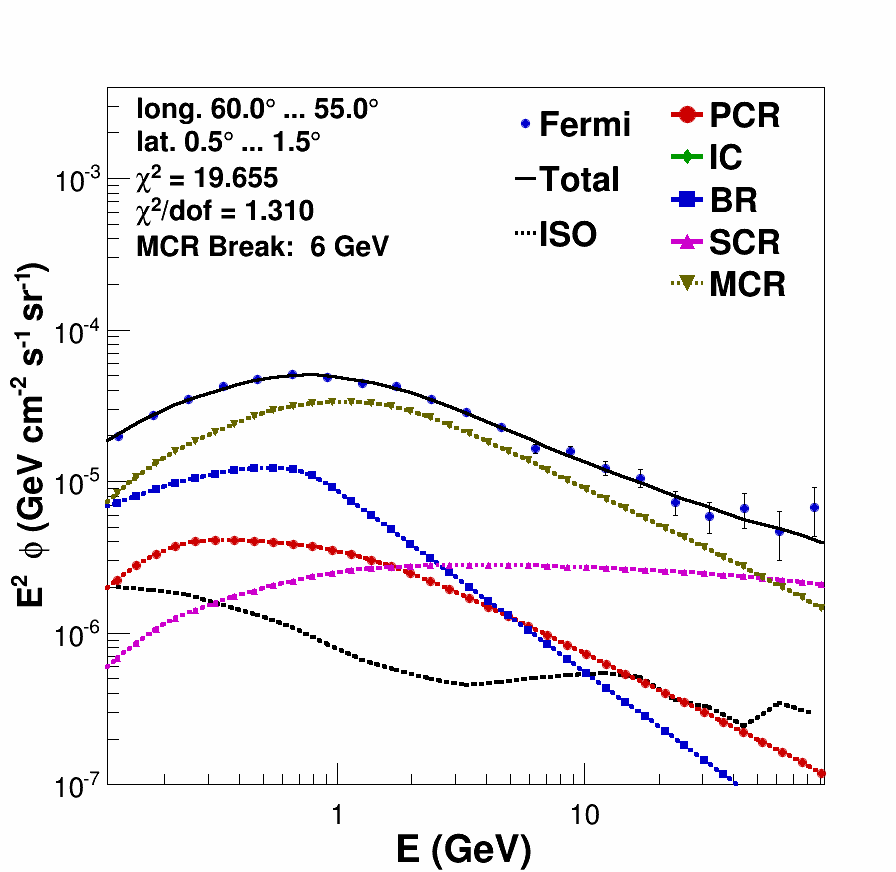}
\includegraphics[width=0.16\textwidth,height=0.16\textwidth,clip]{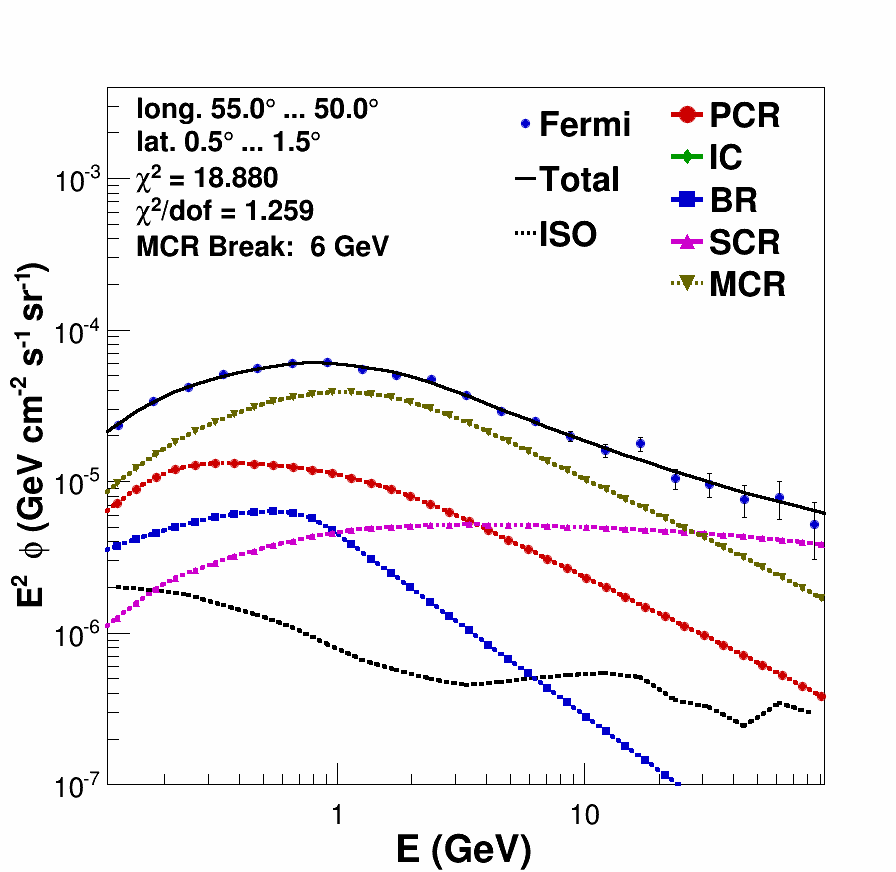}
\includegraphics[width=0.16\textwidth,height=0.16\textwidth,clip]{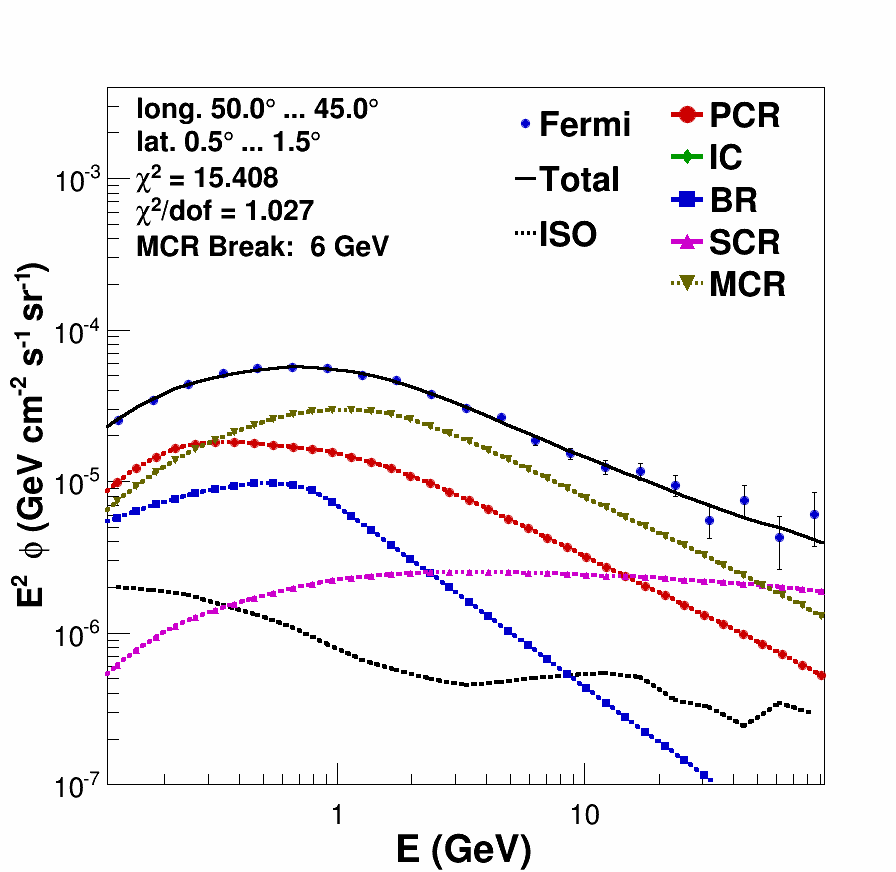}
\includegraphics[width=0.16\textwidth,height=0.16\textwidth,clip]{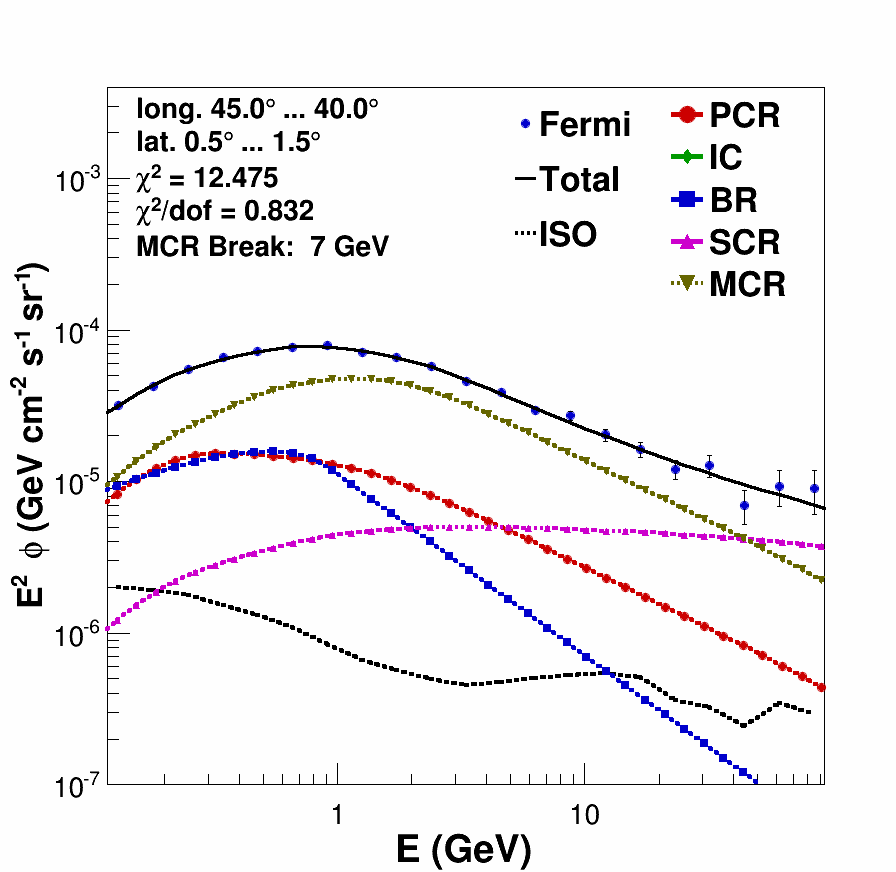}
\includegraphics[width=0.16\textwidth,height=0.16\textwidth,clip]{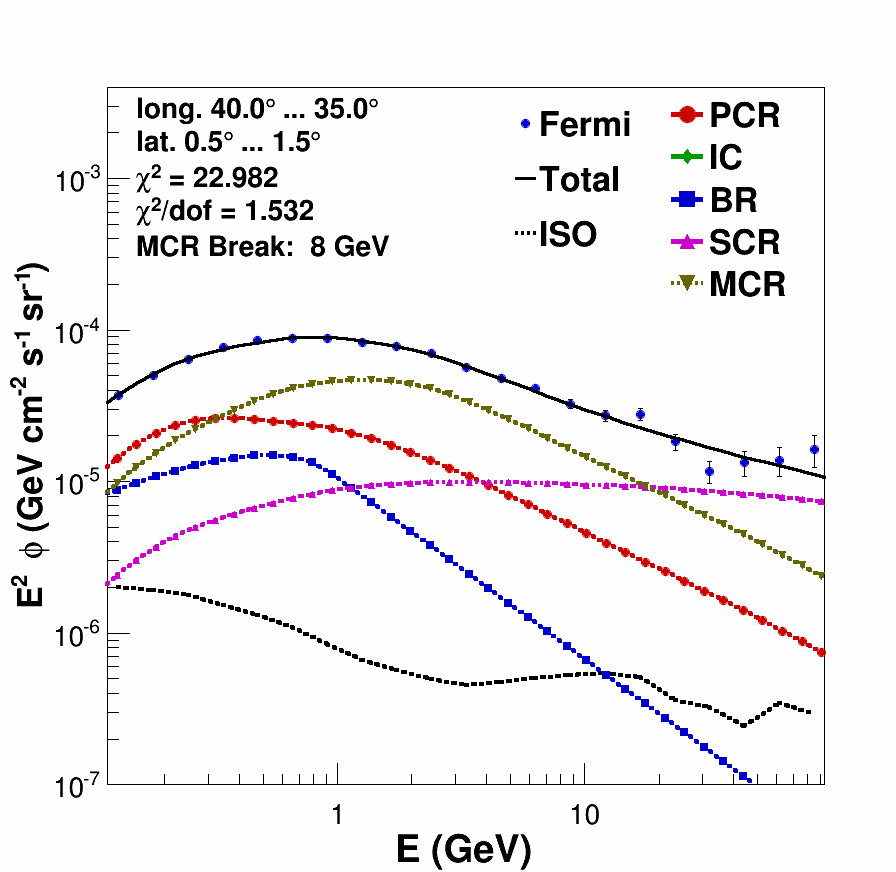}
\includegraphics[width=0.16\textwidth,height=0.16\textwidth,clip]{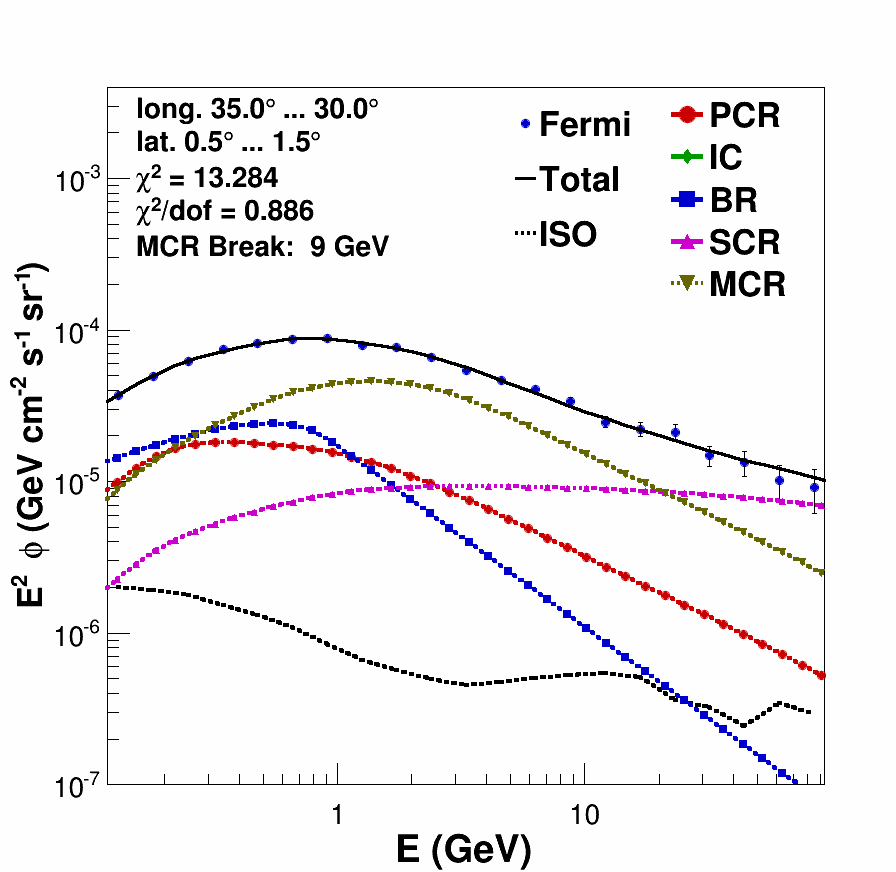}
\includegraphics[width=0.16\textwidth,height=0.16\textwidth,clip]{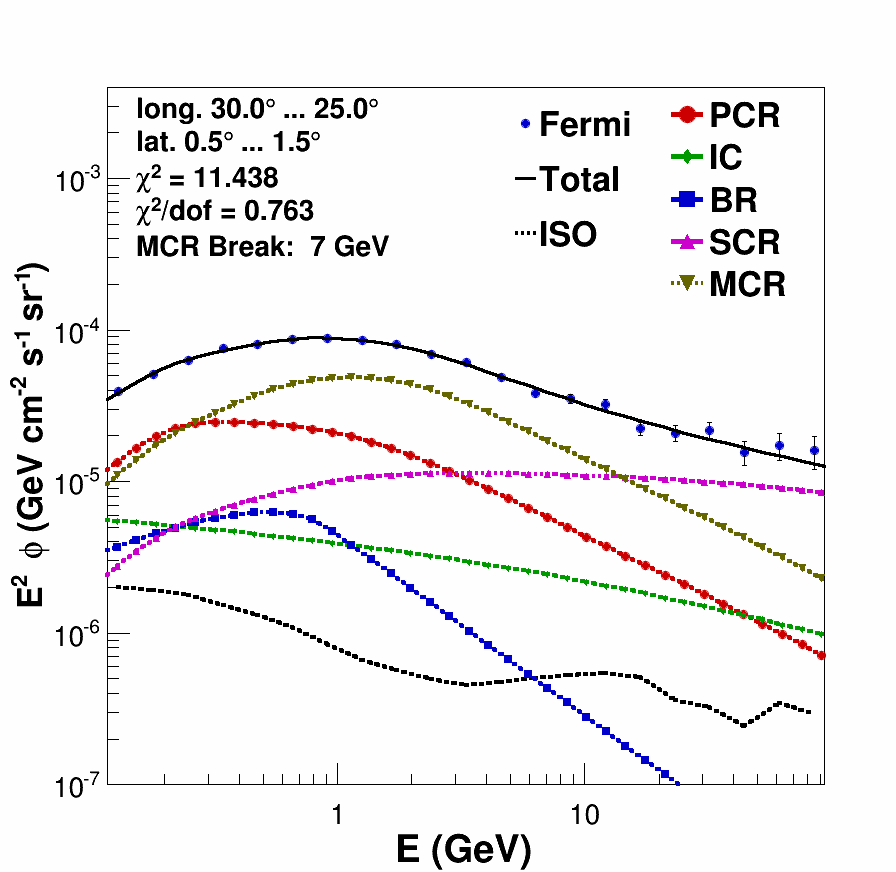}
\includegraphics[width=0.16\textwidth,height=0.16\textwidth,clip]{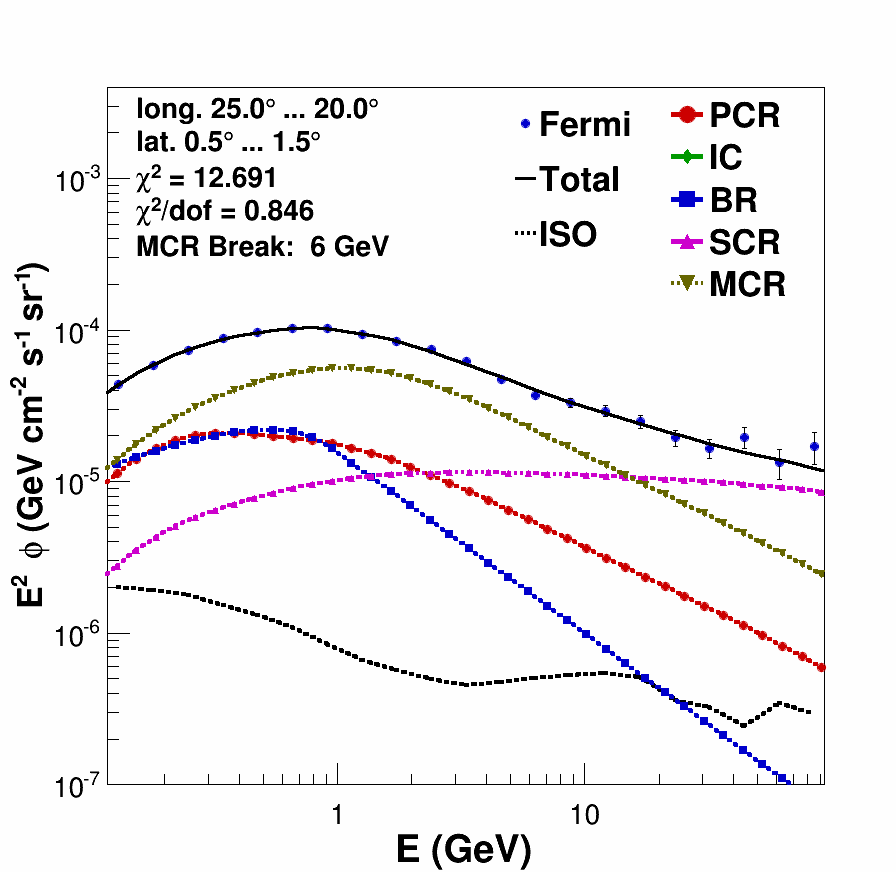}
\includegraphics[width=0.16\textwidth,height=0.16\textwidth,clip]{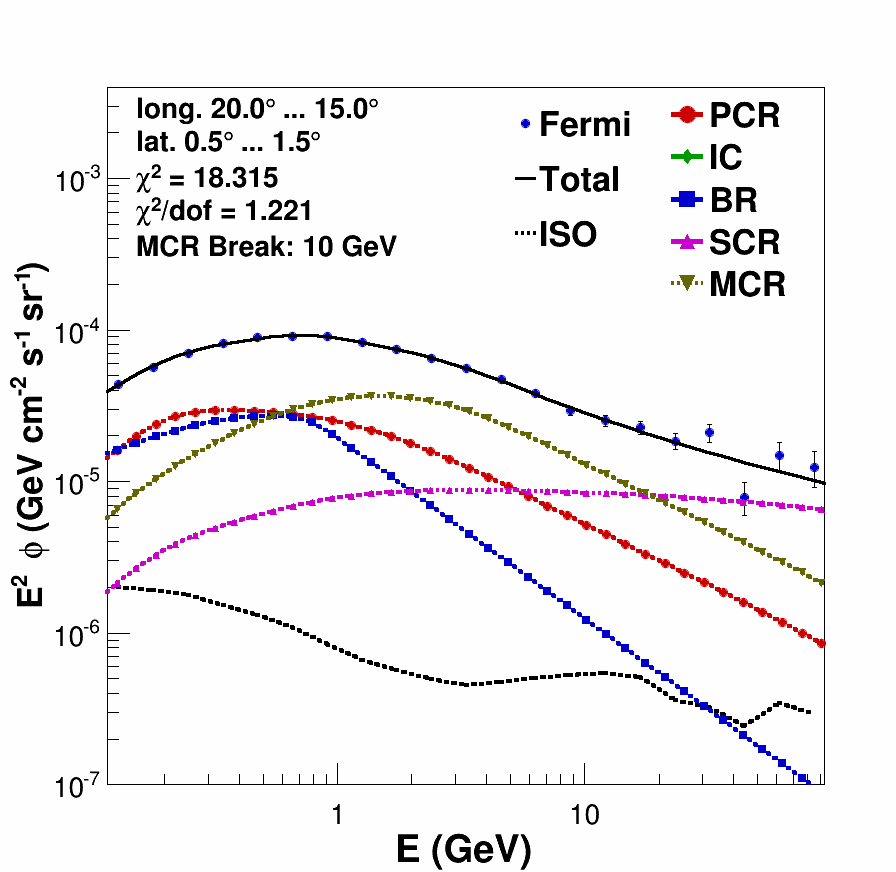}
\includegraphics[width=0.16\textwidth,height=0.16\textwidth,clip]{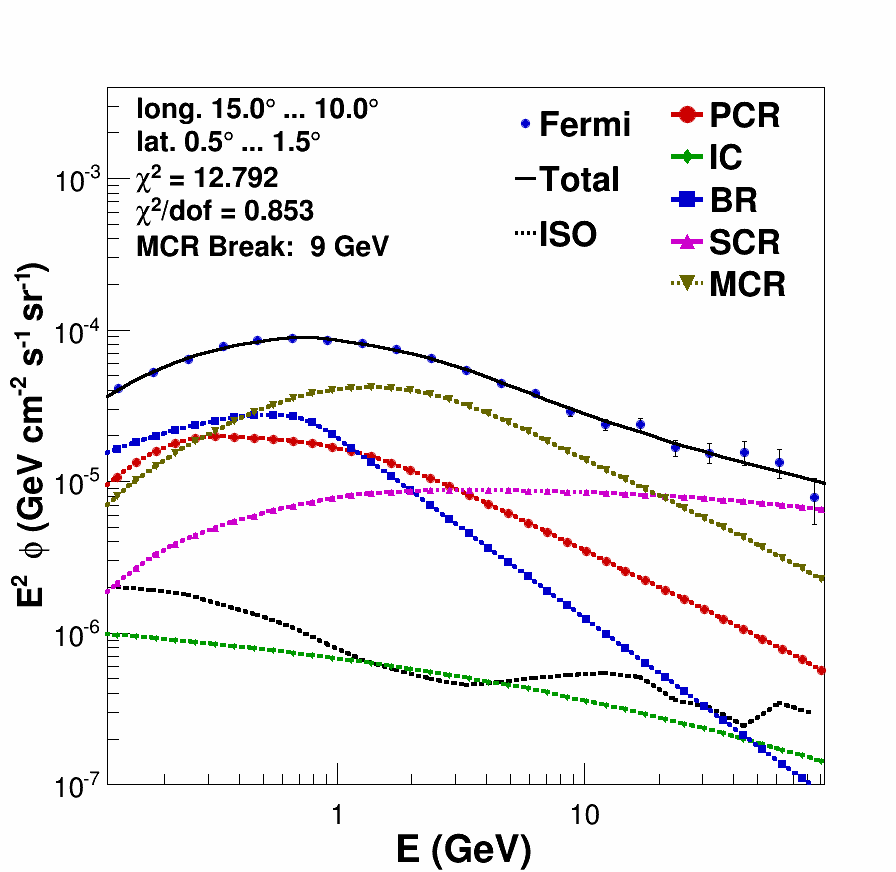}
\includegraphics[width=0.16\textwidth,height=0.16\textwidth,clip]{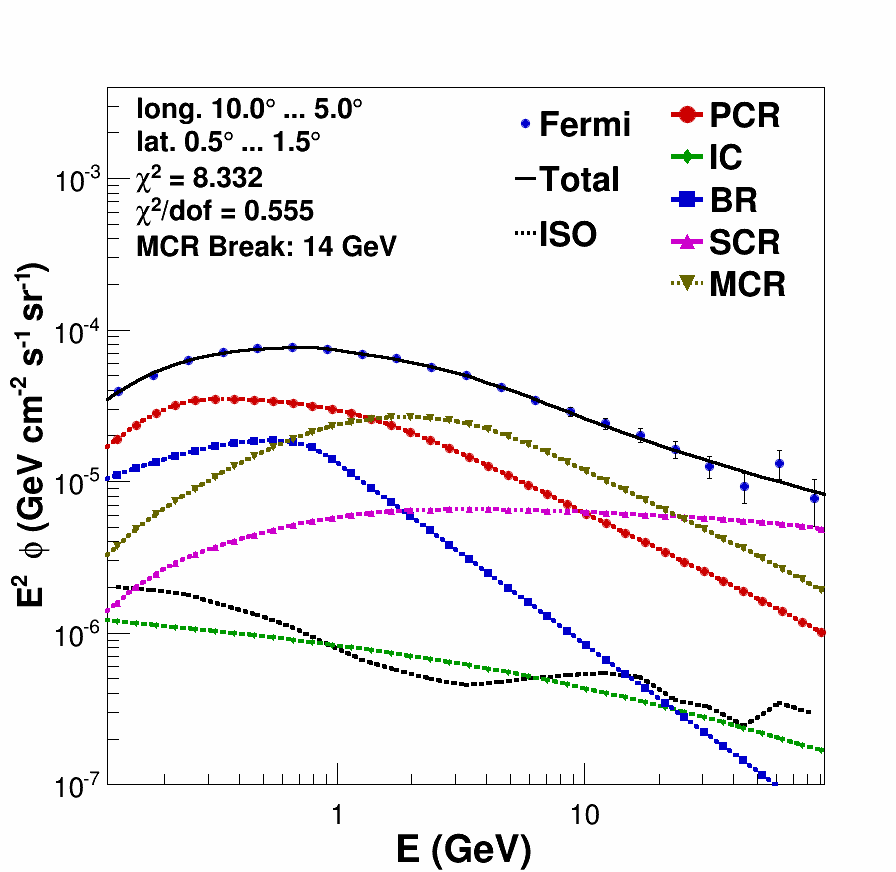}
\includegraphics[width=0.16\textwidth,height=0.16\textwidth,clip]{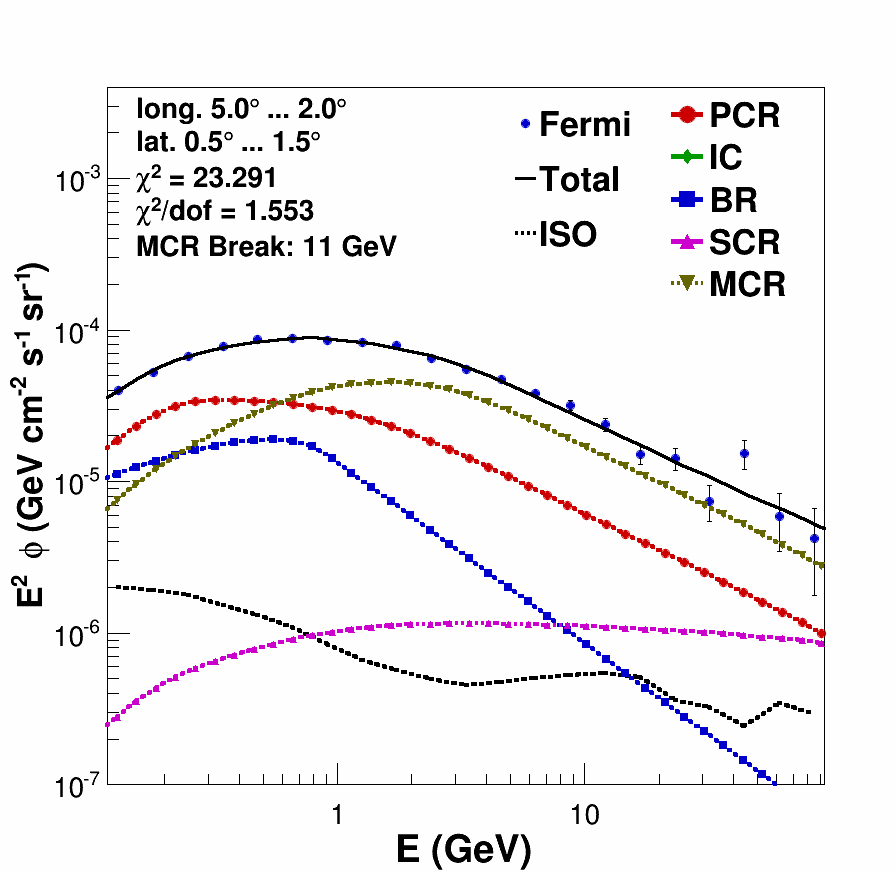}
\includegraphics[width=0.16\textwidth,height=0.16\textwidth,clip]{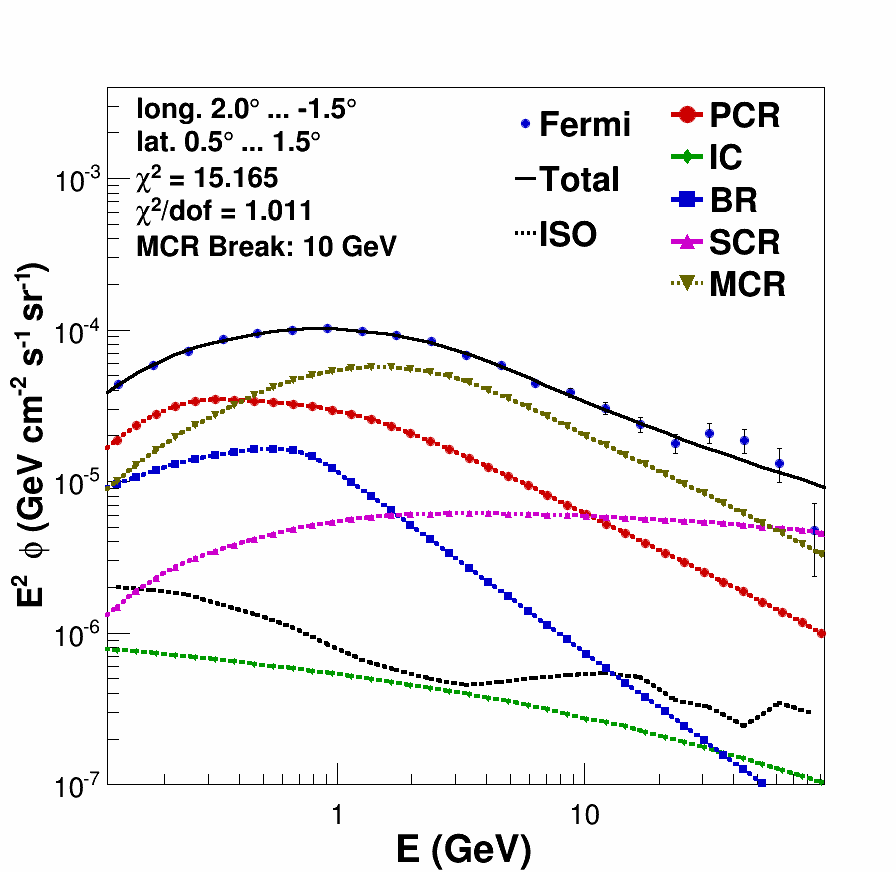}
\includegraphics[width=0.16\textwidth,height=0.16\textwidth,clip]{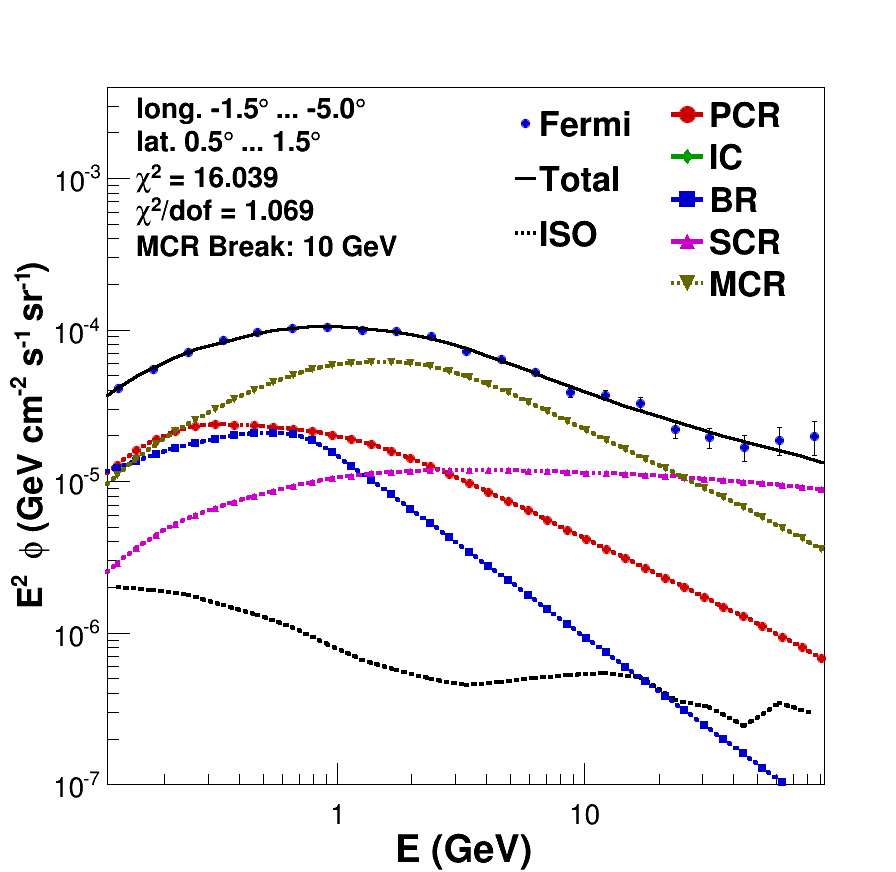}
\includegraphics[width=0.16\textwidth,height=0.16\textwidth,clip]{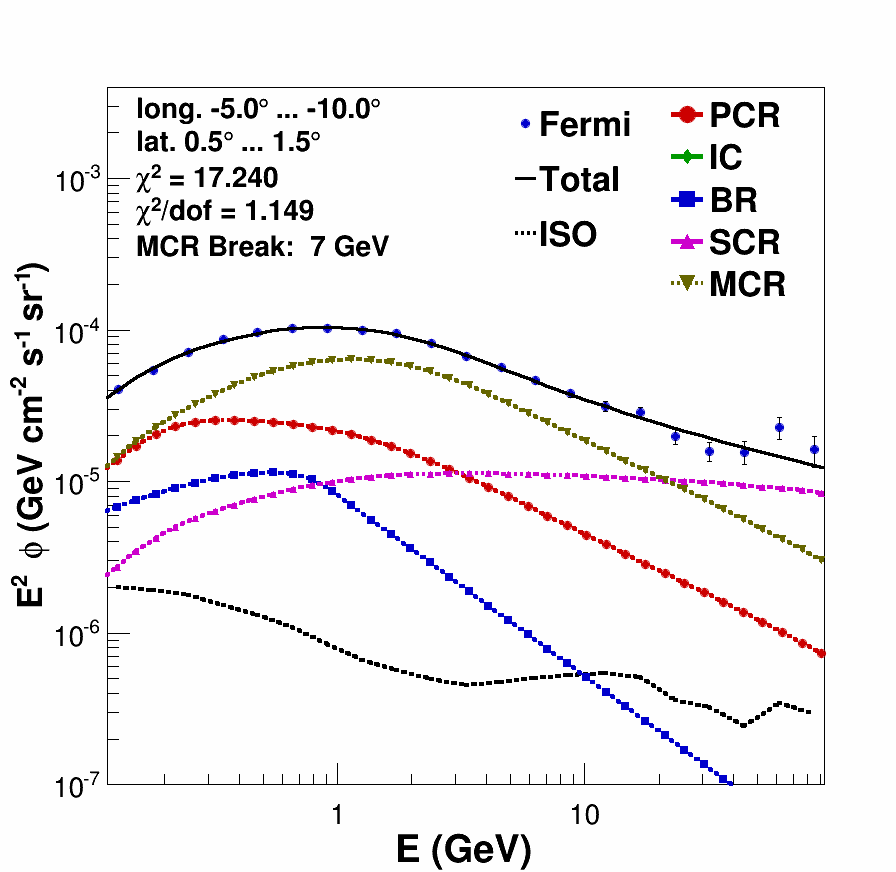}
\includegraphics[width=0.16\textwidth,height=0.16\textwidth,clip]{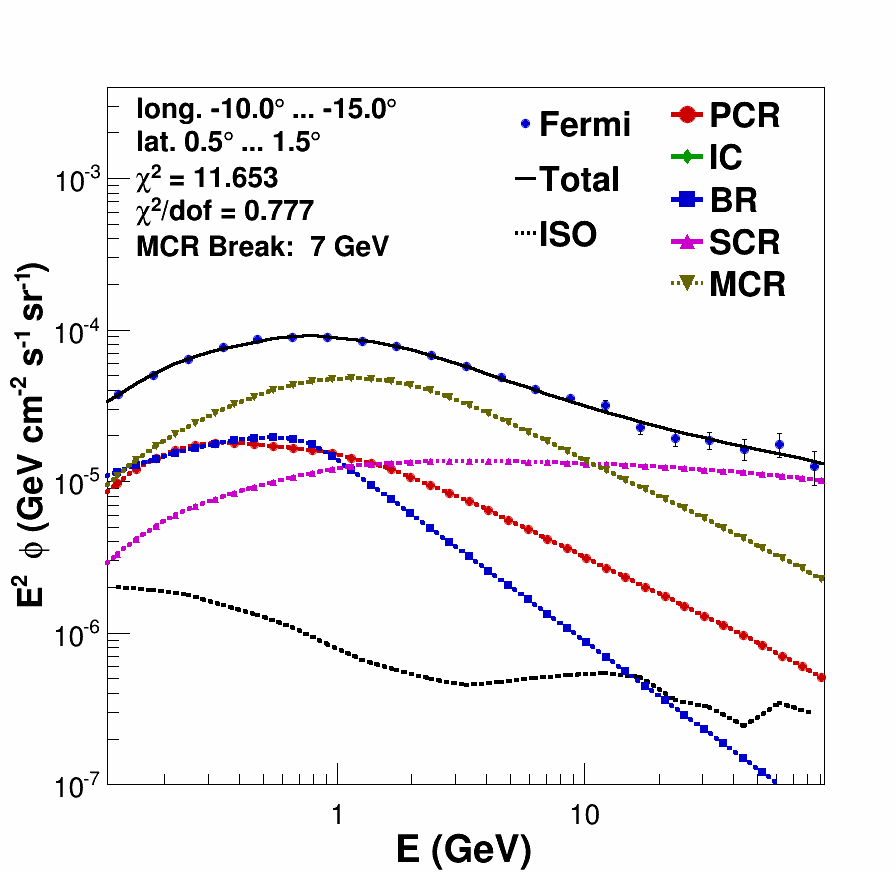}
\includegraphics[width=0.16\textwidth,height=0.16\textwidth,clip]{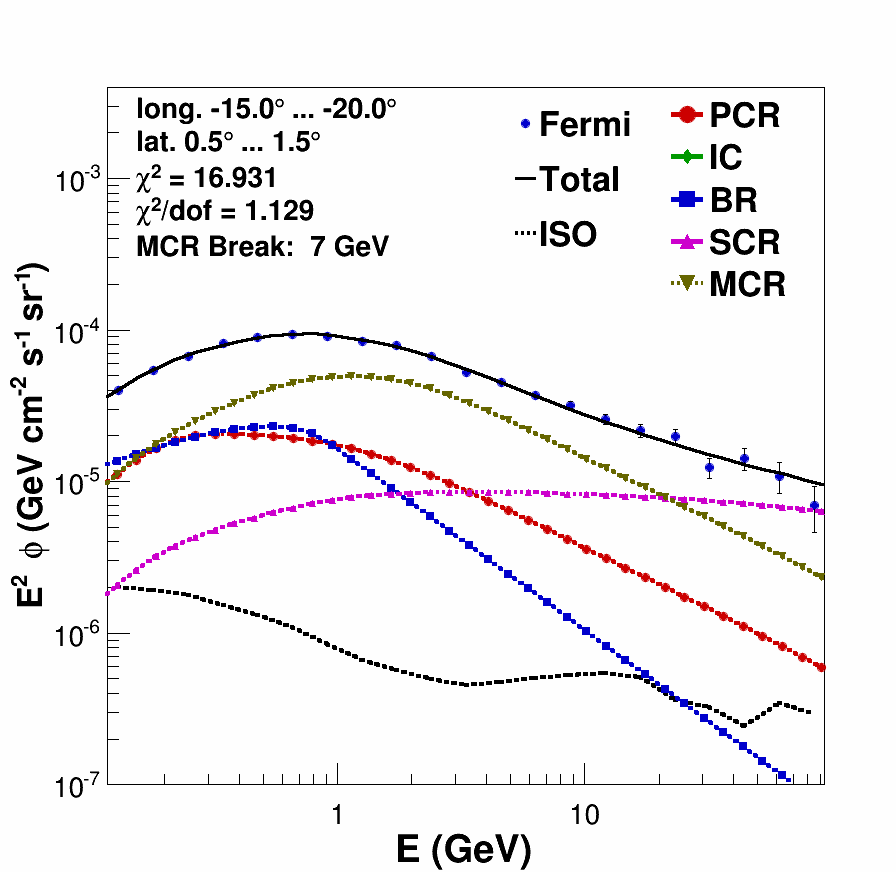}
\includegraphics[width=0.16\textwidth,height=0.16\textwidth,clip]{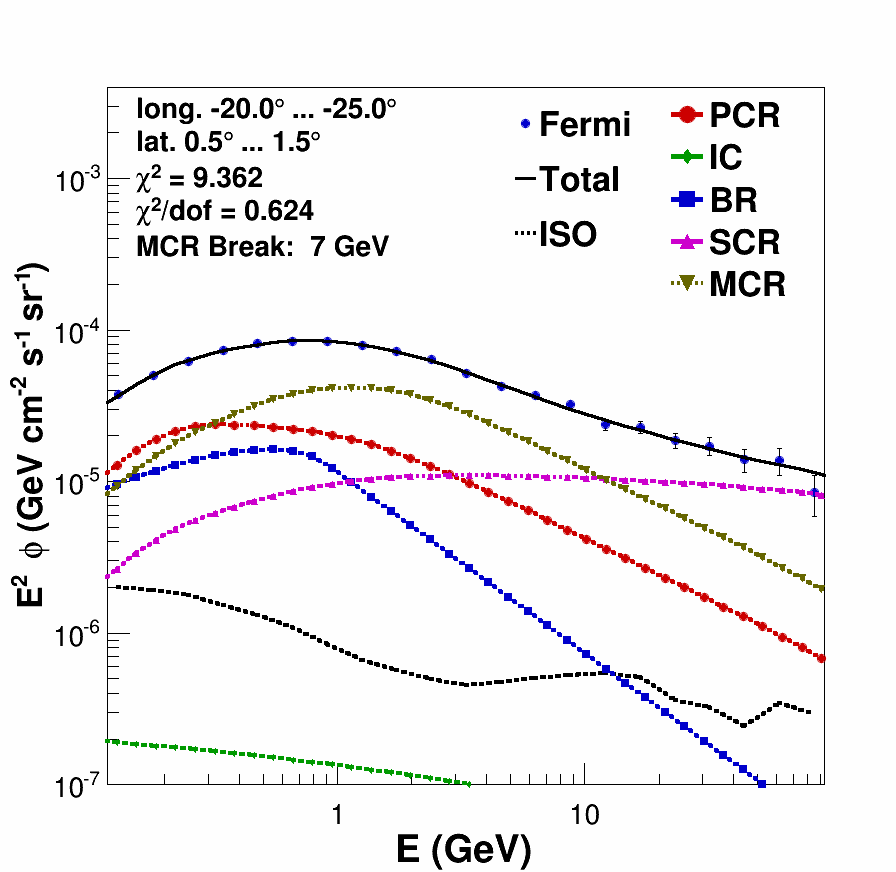}
\includegraphics[width=0.16\textwidth,height=0.16\textwidth,clip]{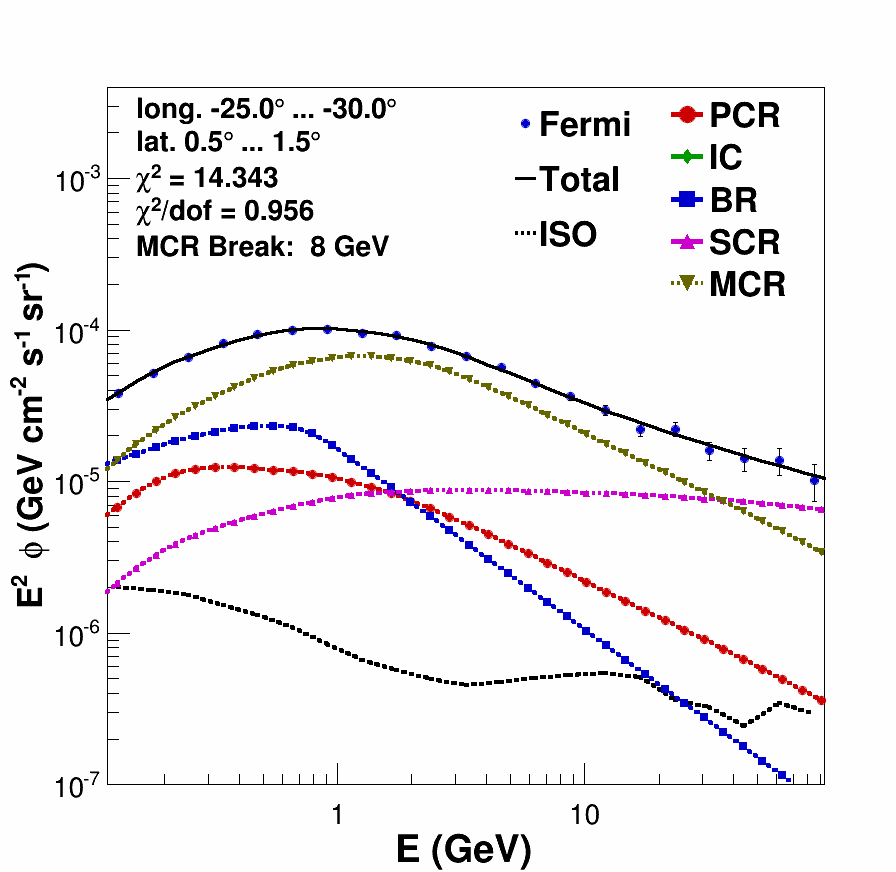}
\includegraphics[width=0.16\textwidth,height=0.16\textwidth,clip]{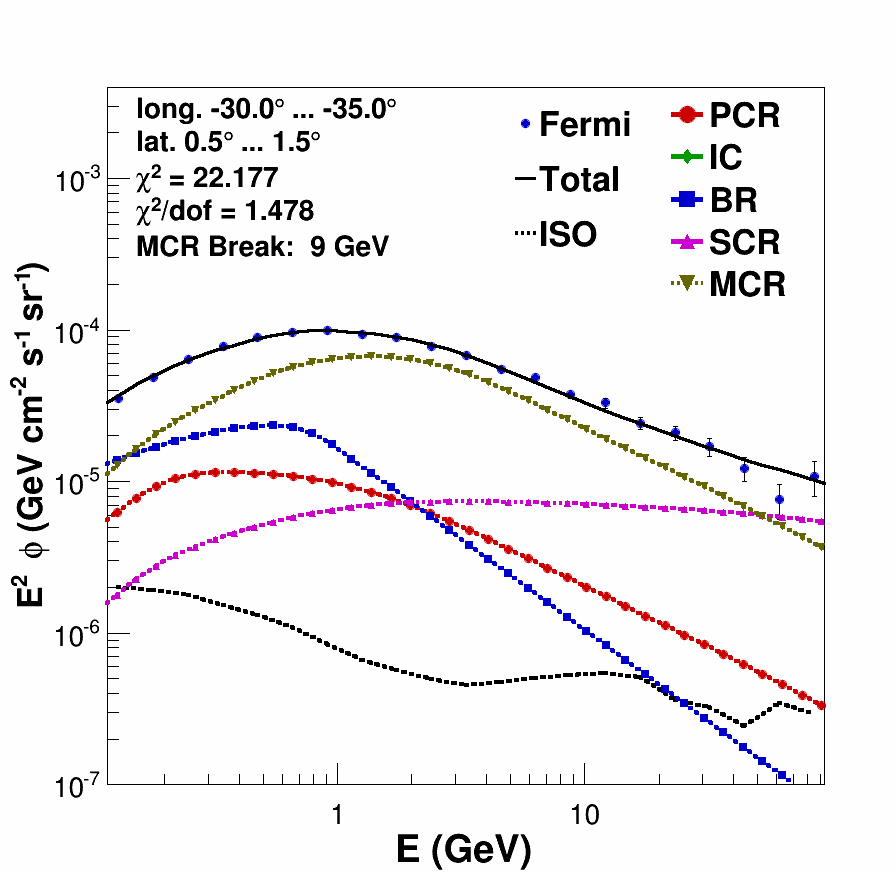}
\includegraphics[width=0.16\textwidth,height=0.16\textwidth,clip]{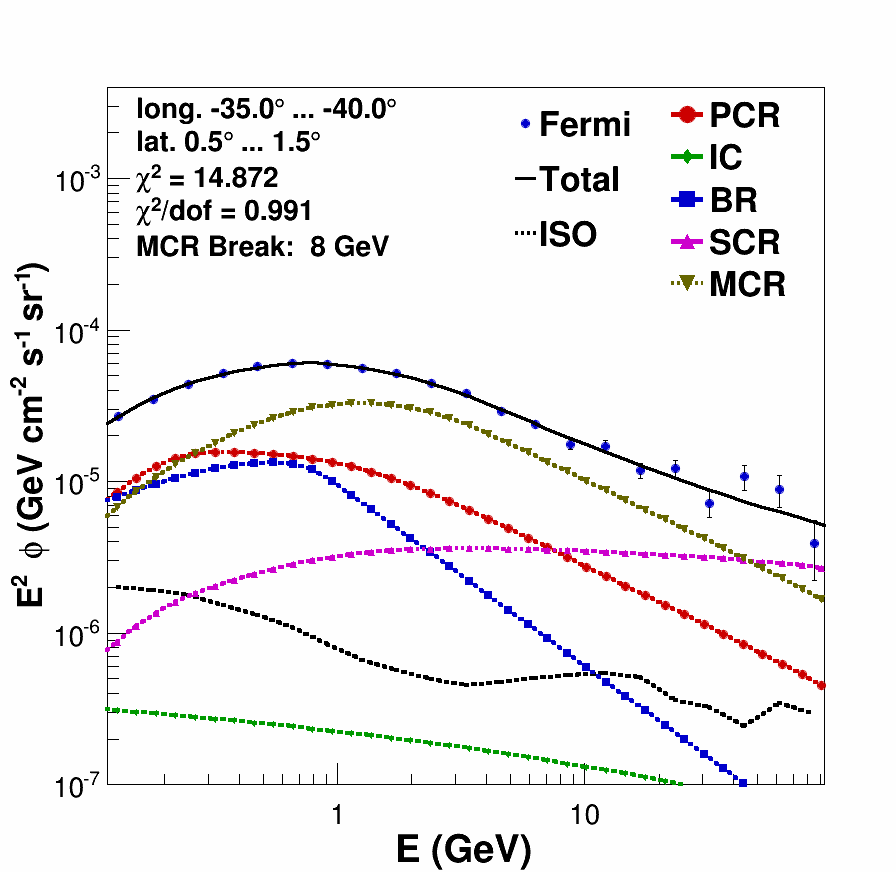}
\includegraphics[width=0.16\textwidth,height=0.16\textwidth,clip]{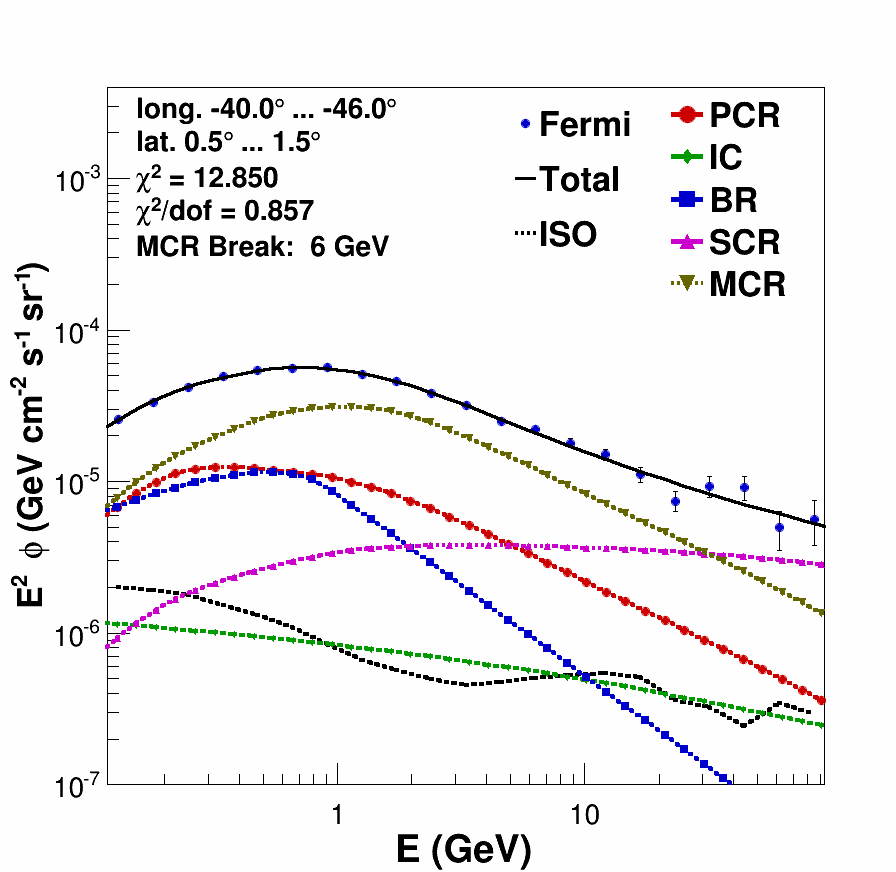}
\includegraphics[width=0.16\textwidth,height=0.16\textwidth,clip]{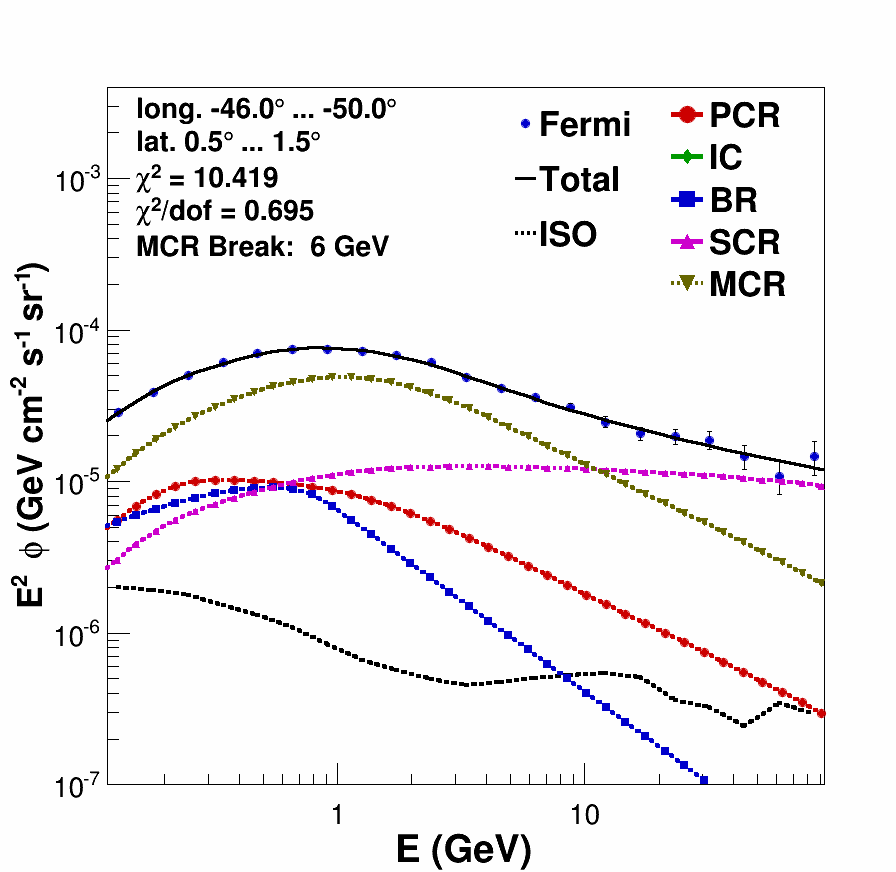}
\includegraphics[width=0.16\textwidth,height=0.16\textwidth,clip]{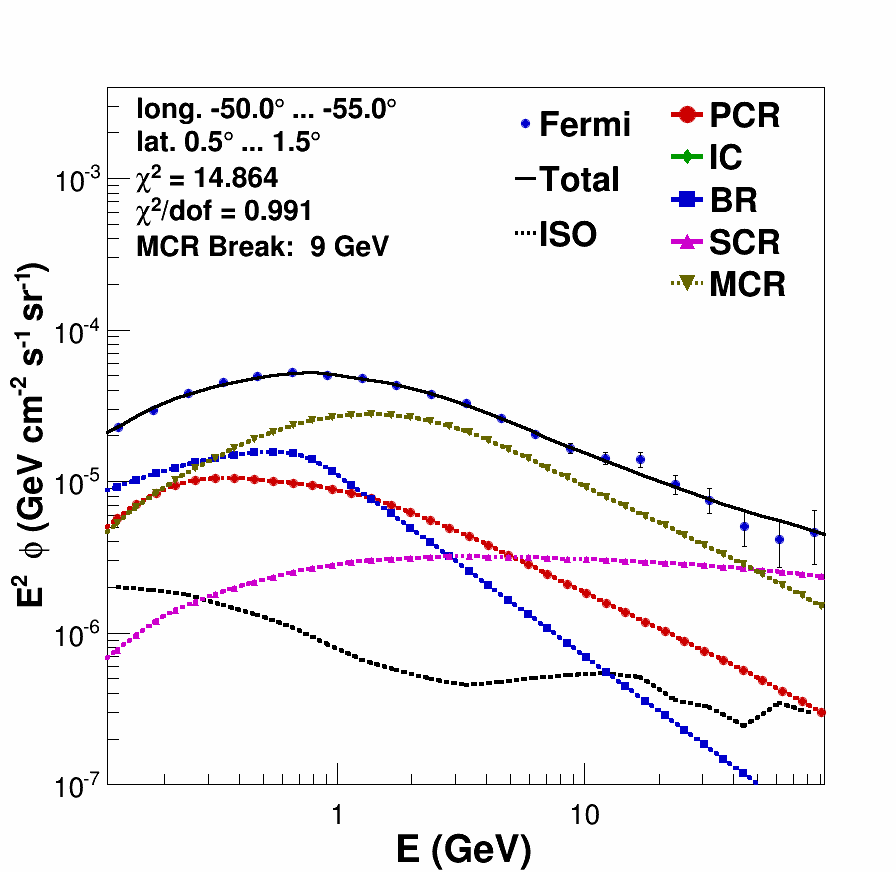}
\includegraphics[width=0.16\textwidth,height=0.16\textwidth,clip]{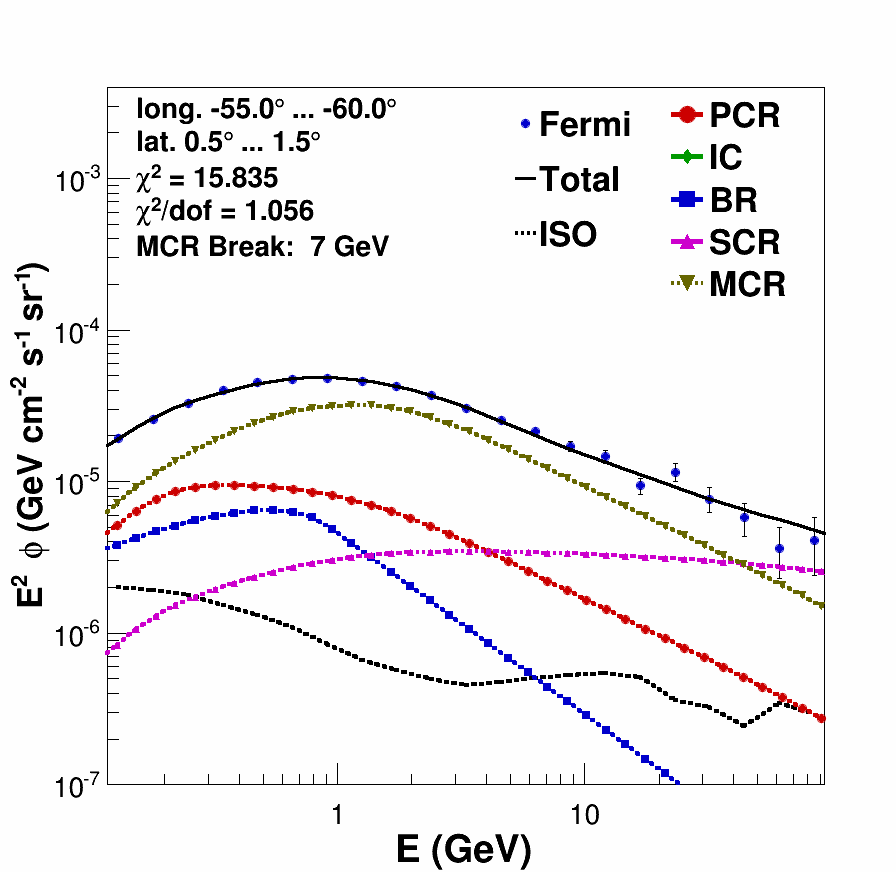}
\includegraphics[width=0.16\textwidth,height=0.16\textwidth,clip]{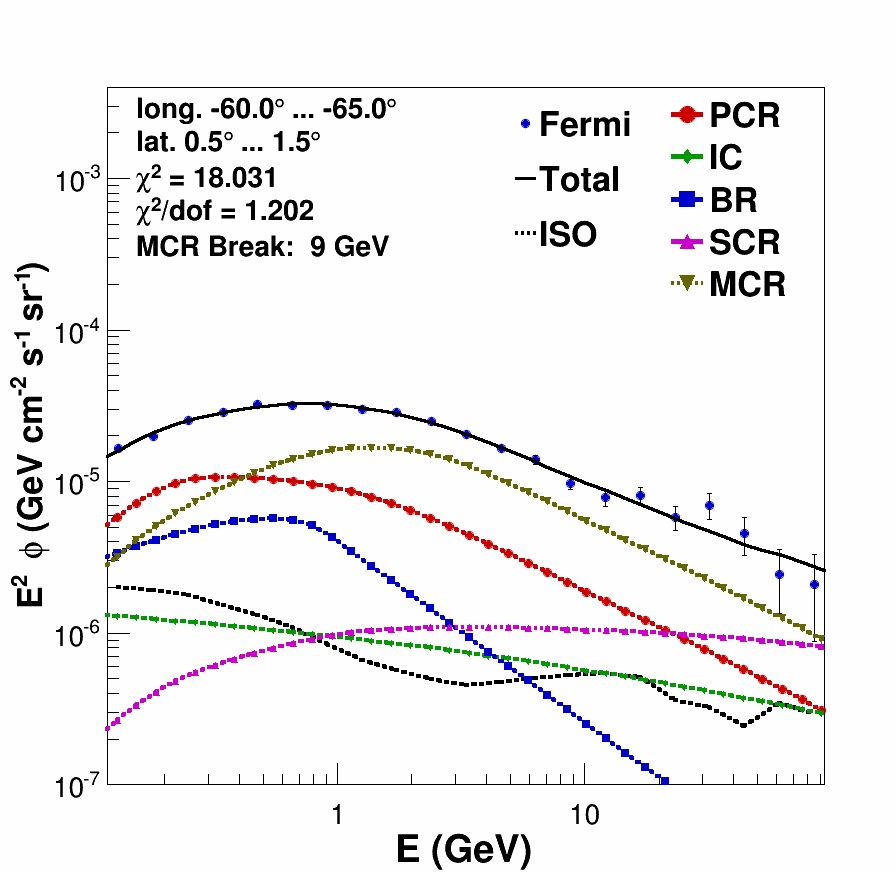}
\includegraphics[width=0.16\textwidth,height=0.16\textwidth,clip]{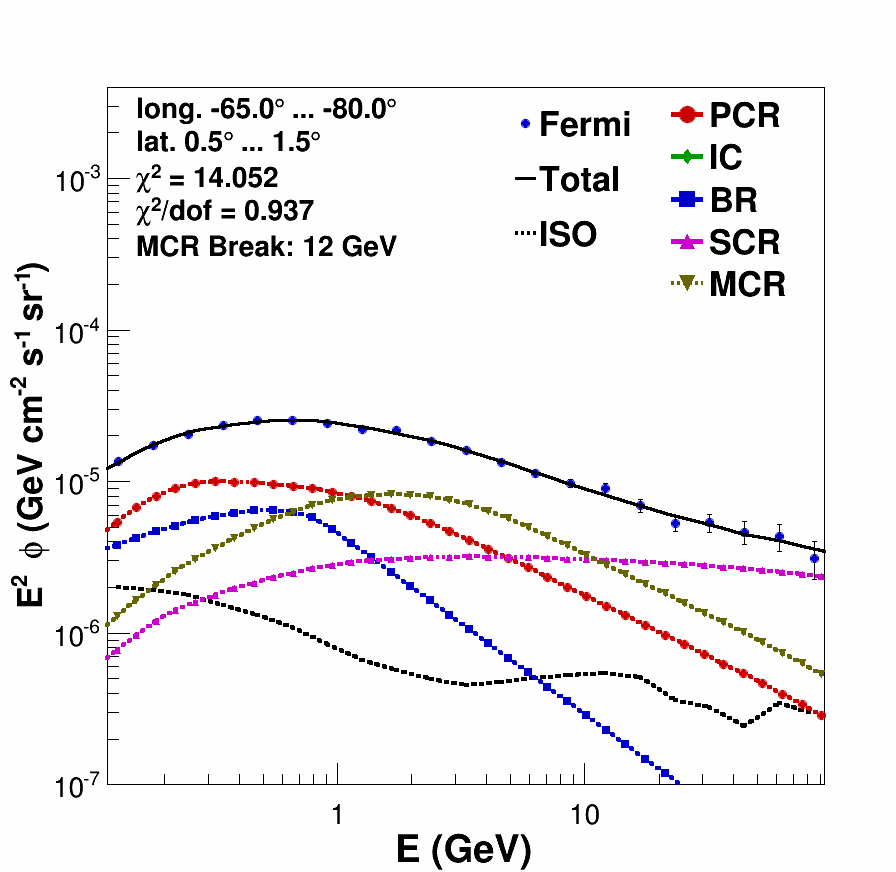}
\includegraphics[width=0.16\textwidth,height=0.16\textwidth,clip]{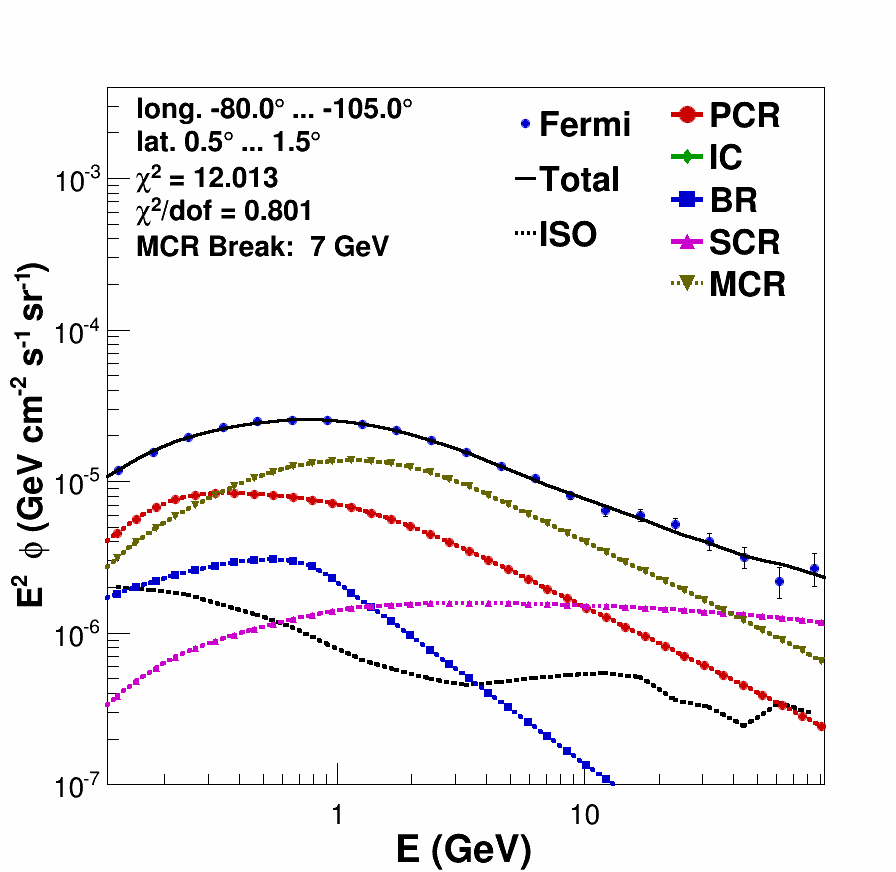}
\includegraphics[width=0.16\textwidth,height=0.16\textwidth,clip]{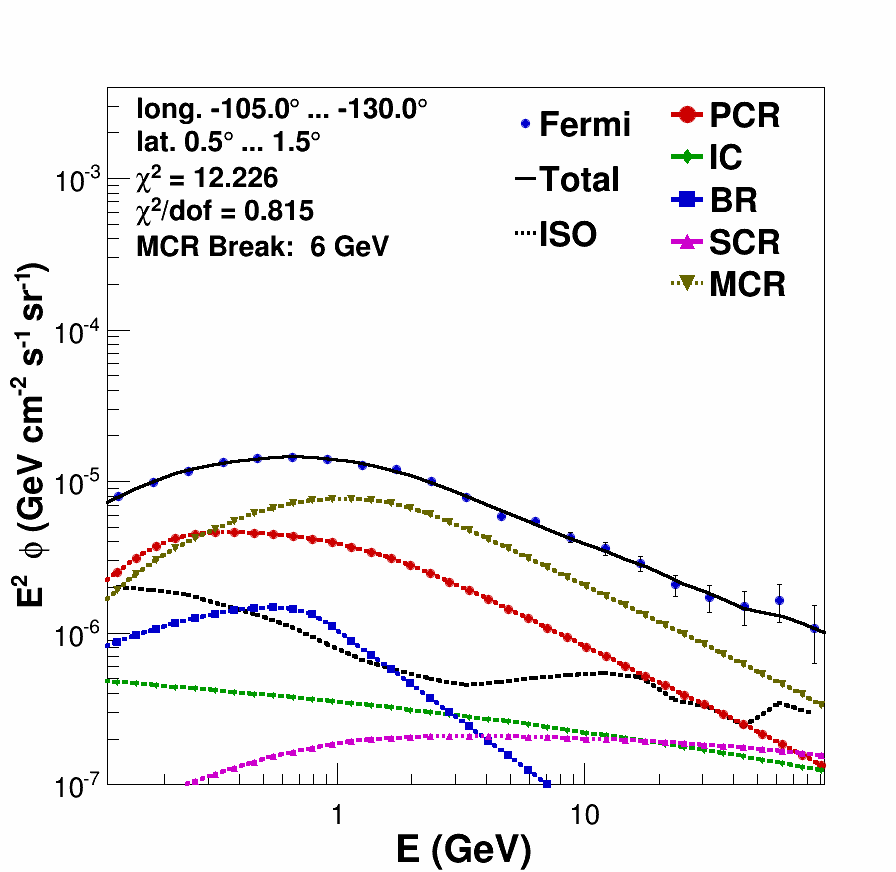}
\includegraphics[width=0.16\textwidth,height=0.16\textwidth,clip]{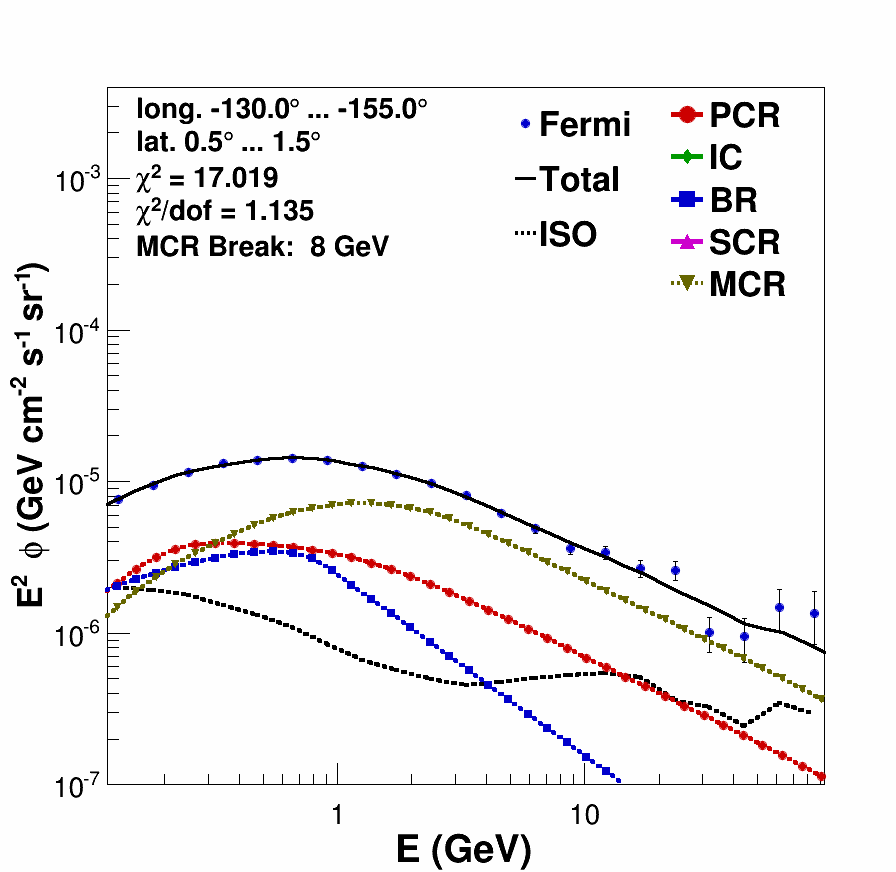}
\includegraphics[width=0.16\textwidth,height=0.16\textwidth,clip]{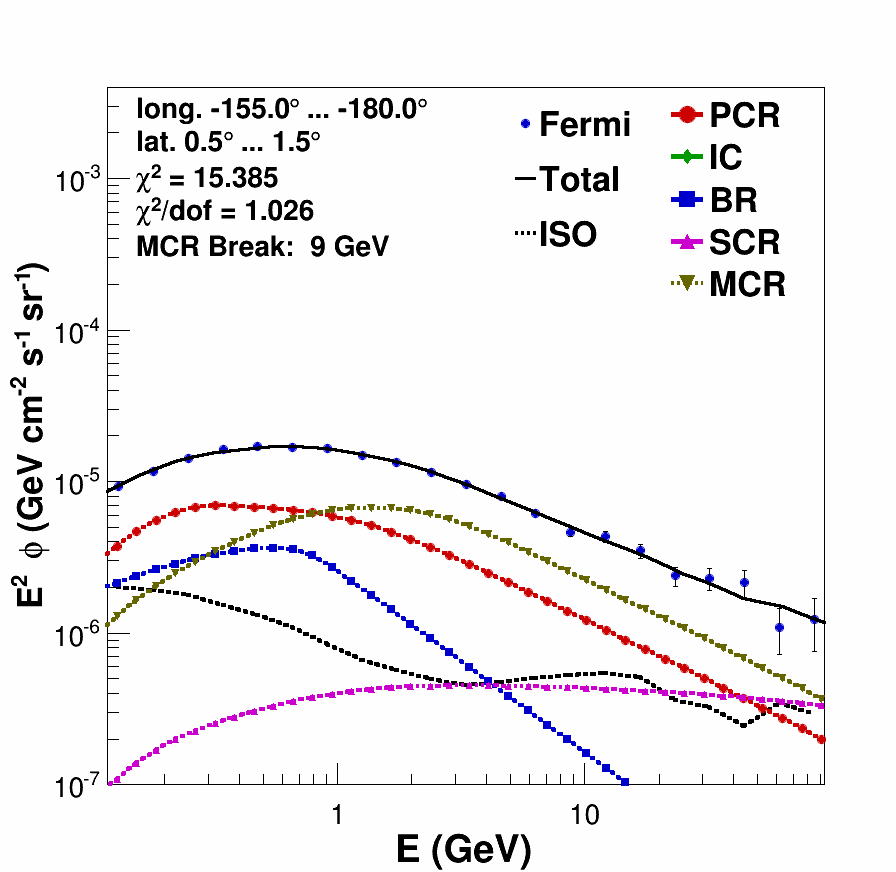}%%%%%%r9
\caption[]{Template fits for latitudes  with $0.5^\circ<b<1.5^\circ$ and longitudes decreasing from 180$^\circ$ to -180$^\circ$.} \label{F20}
\end{figure}
\clearpage
\begin{figure}
\centering
\includegraphics[width=0.16\textwidth,height=0.16\textwidth,clip]{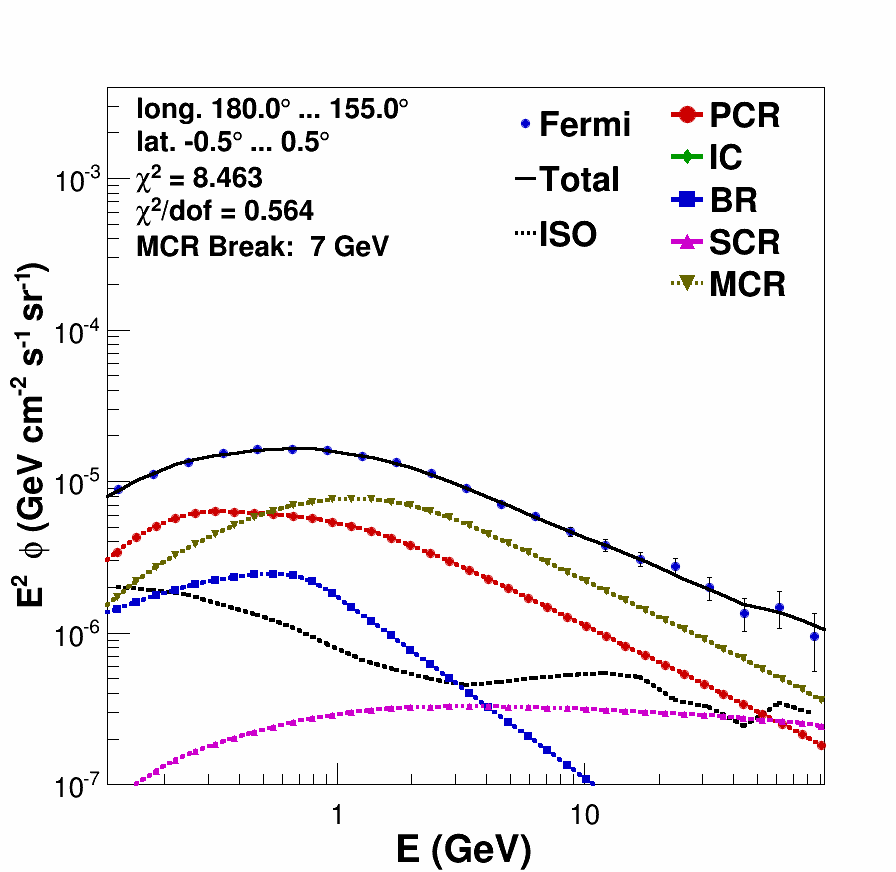}
\includegraphics[width=0.16\textwidth,height=0.16\textwidth,clip]{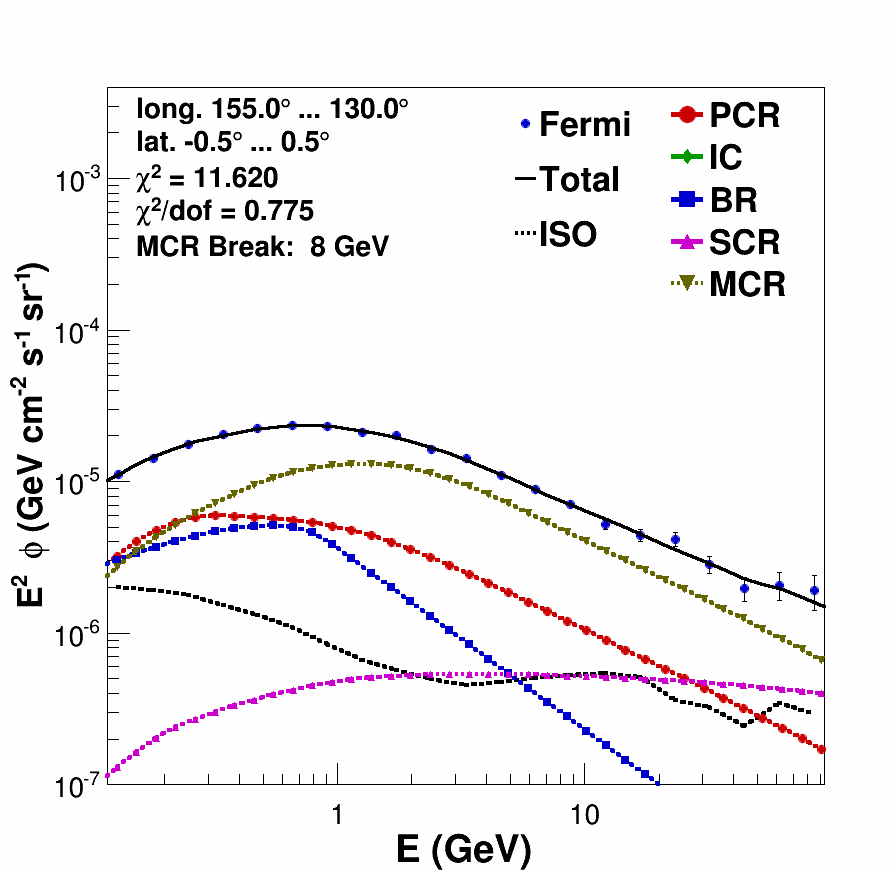}
\includegraphics[width=0.16\textwidth,height=0.16\textwidth,clip]{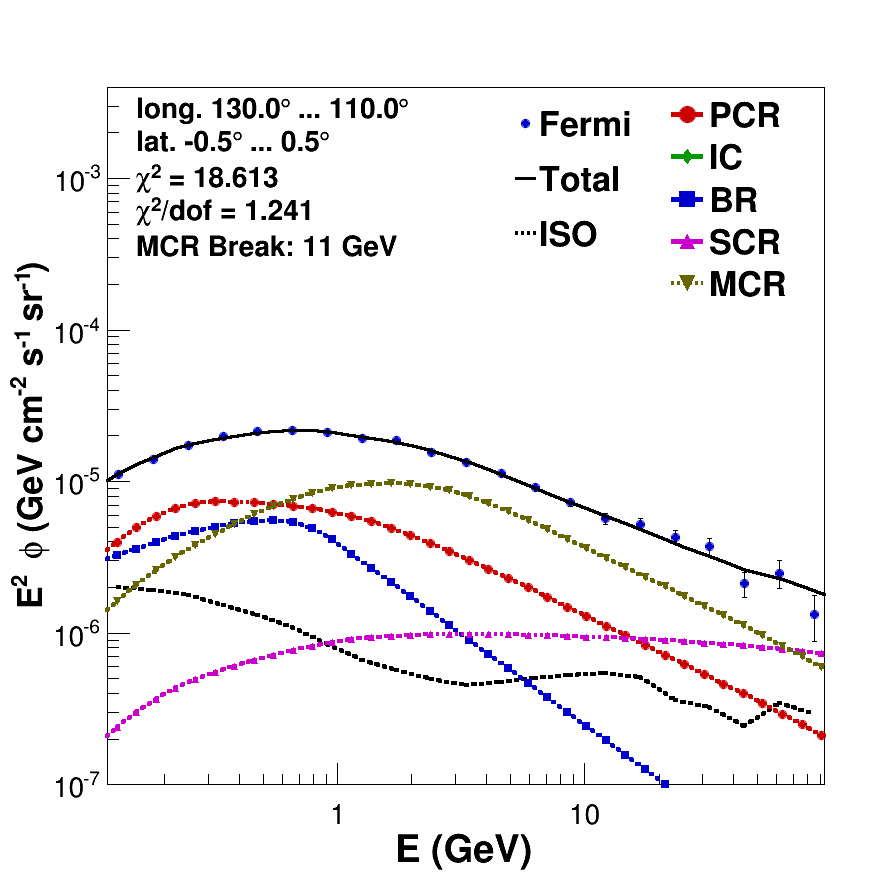}
\includegraphics[width=0.16\textwidth,height=0.16\textwidth,clip]{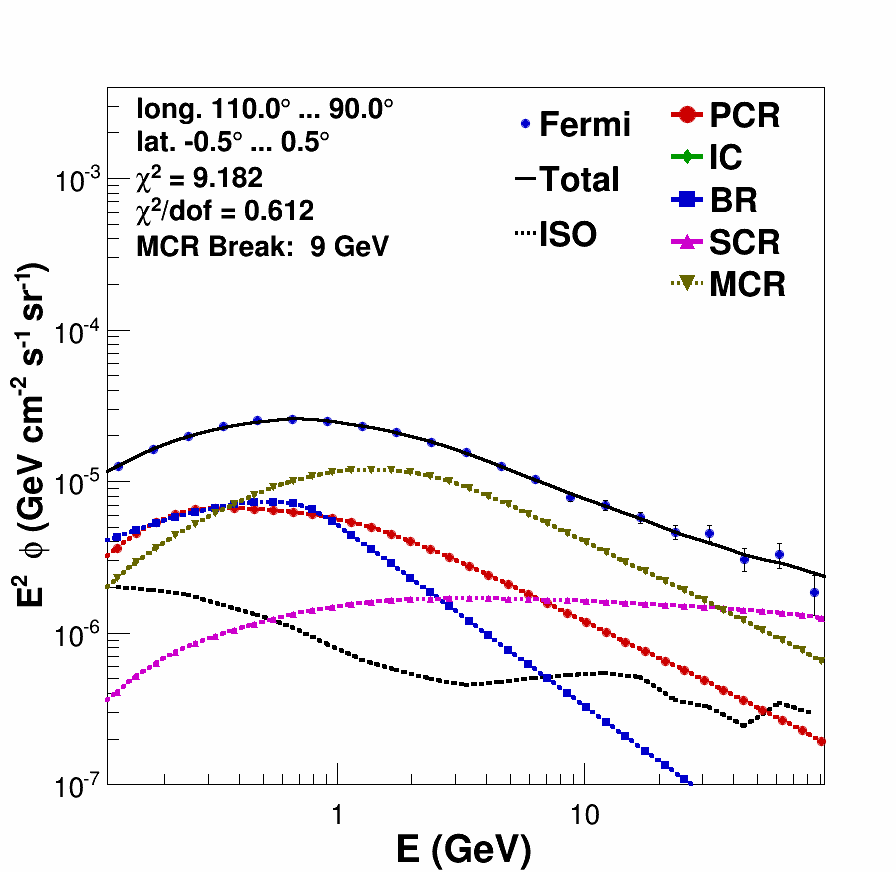}
\includegraphics[width=0.16\textwidth,height=0.16\textwidth,clip]{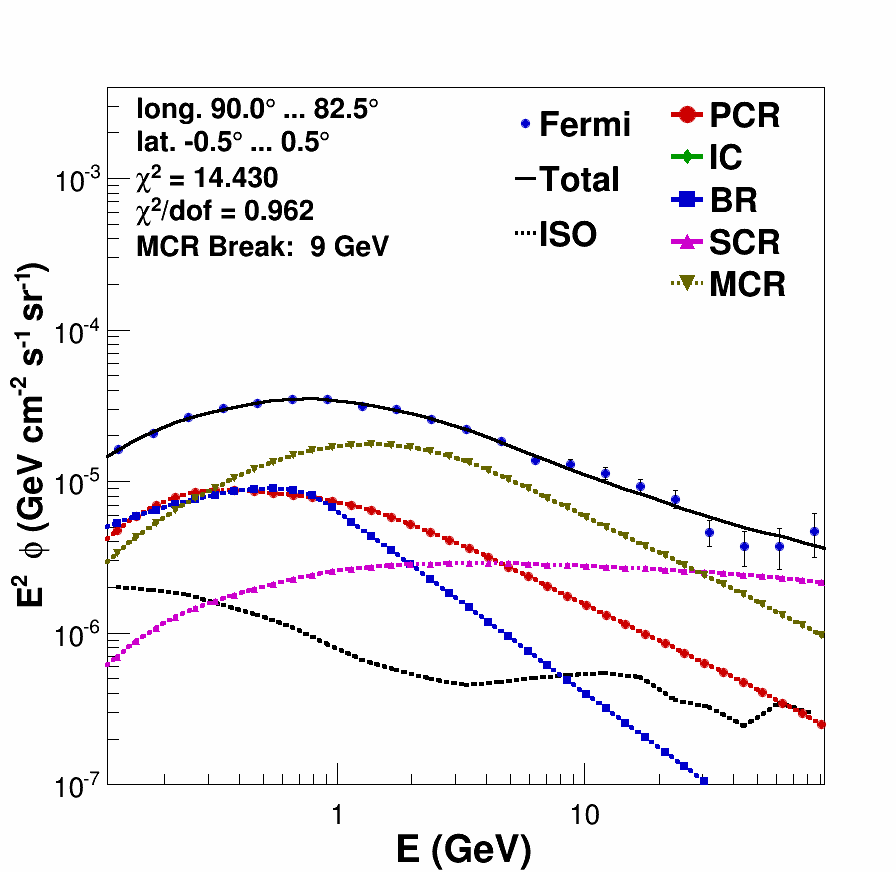}
\includegraphics[width=0.16\textwidth,height=0.16\textwidth,clip]{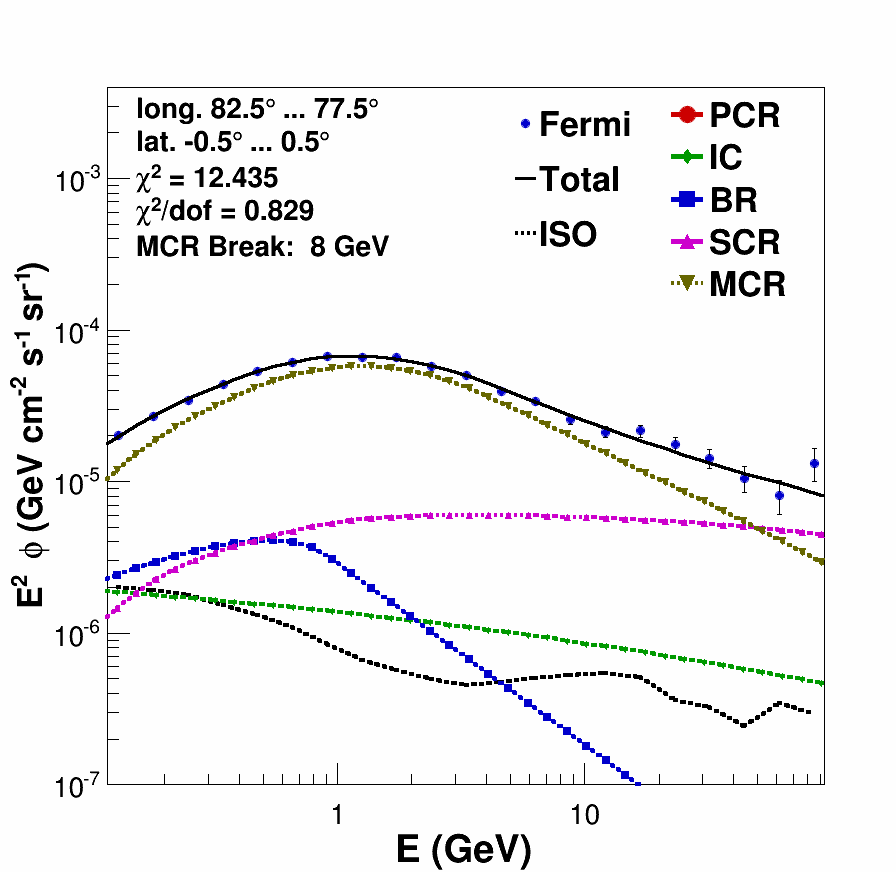}
\includegraphics[width=0.16\textwidth,height=0.16\textwidth,clip]{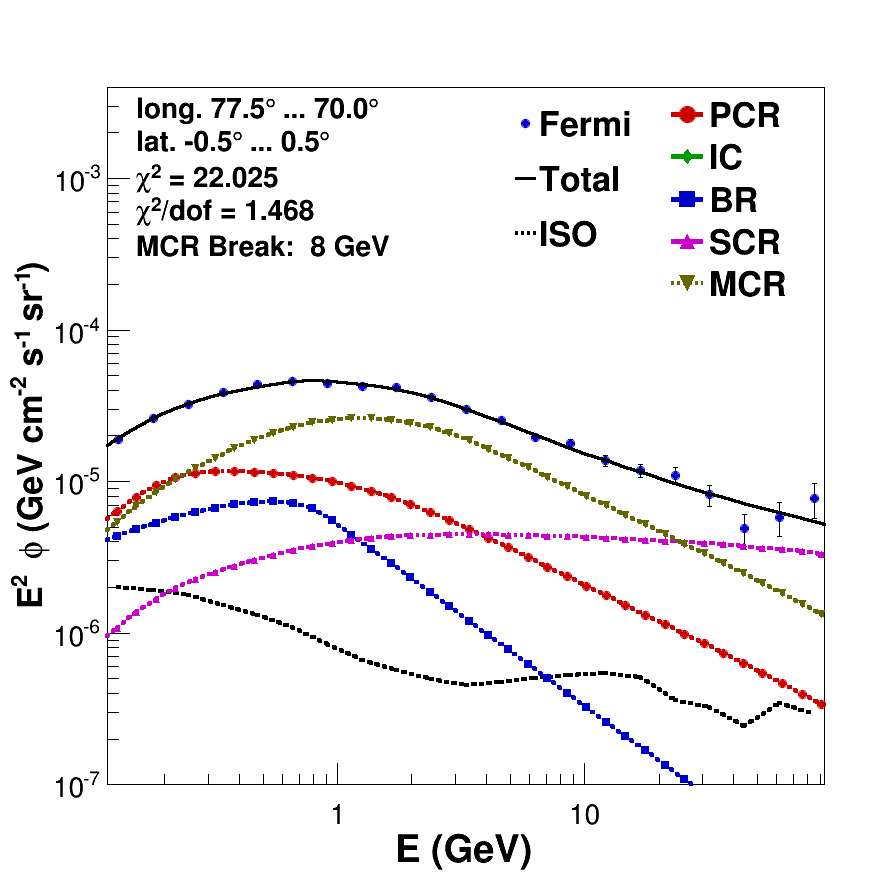}
\includegraphics[width=0.16\textwidth,height=0.16\textwidth,clip]{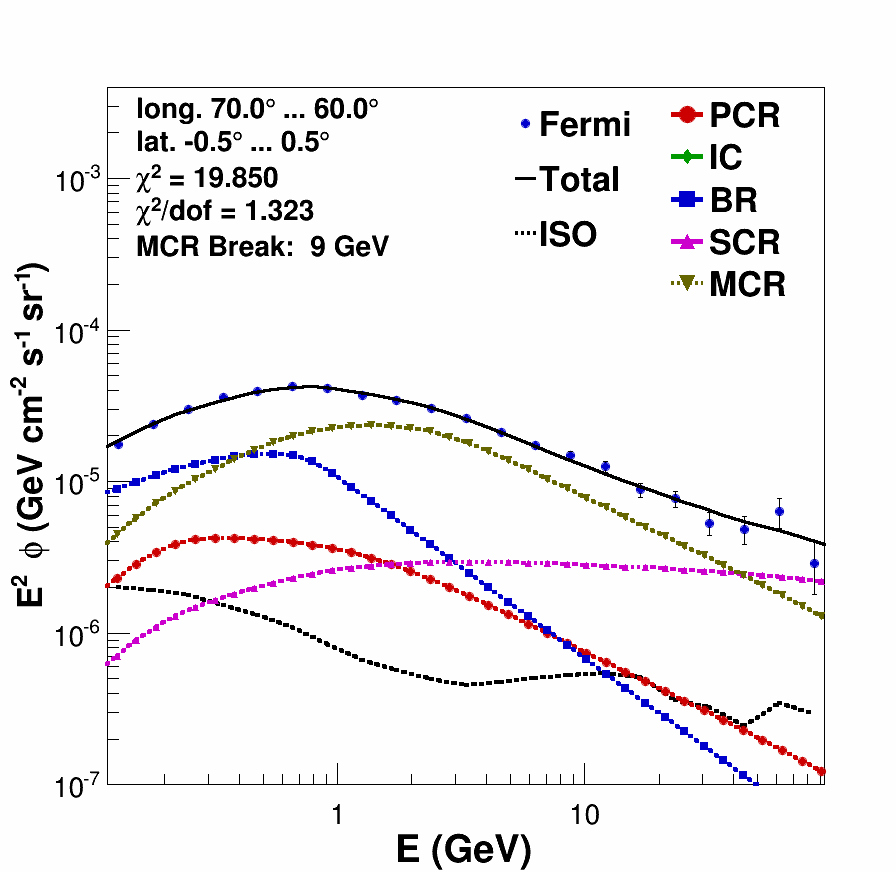}
\includegraphics[width=0.16\textwidth,height=0.16\textwidth,clip]{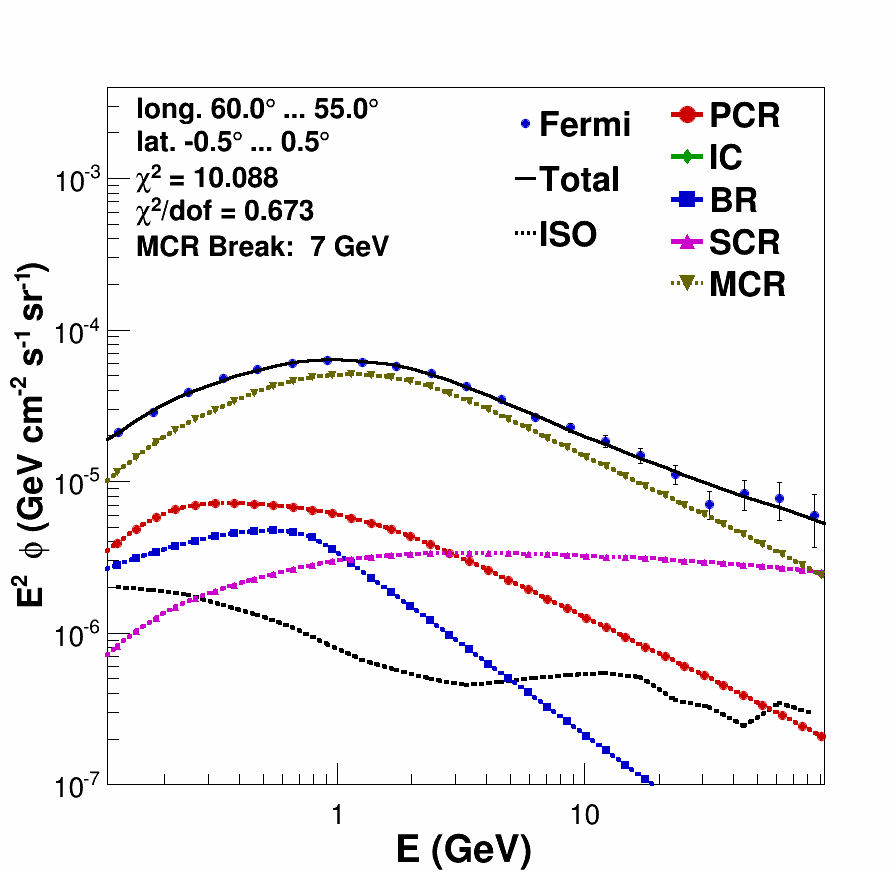}
\includegraphics[width=0.16\textwidth,height=0.16\textwidth,clip]{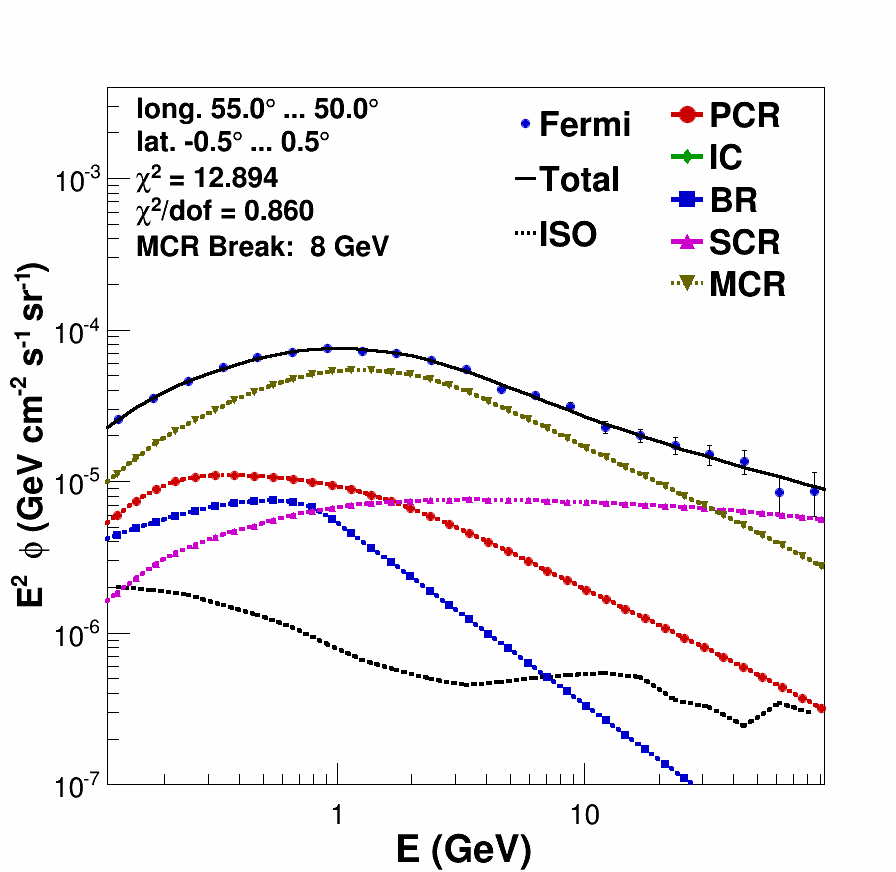}
\includegraphics[width=0.16\textwidth,height=0.16\textwidth,clip]{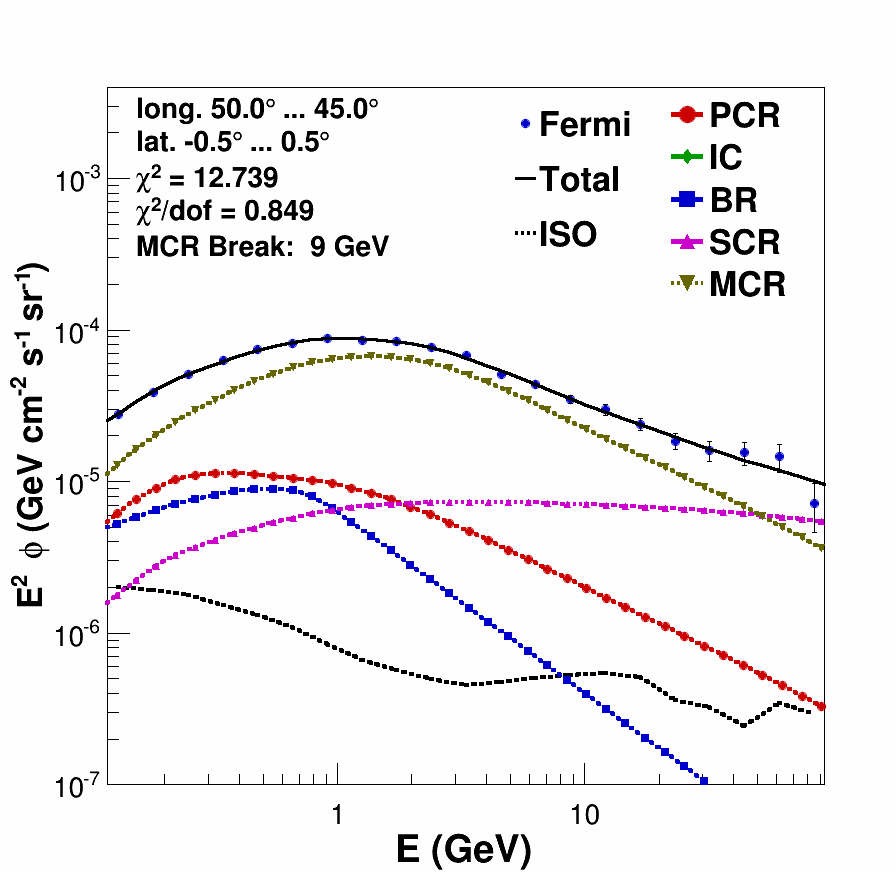}
\includegraphics[width=0.16\textwidth,height=0.16\textwidth,clip]{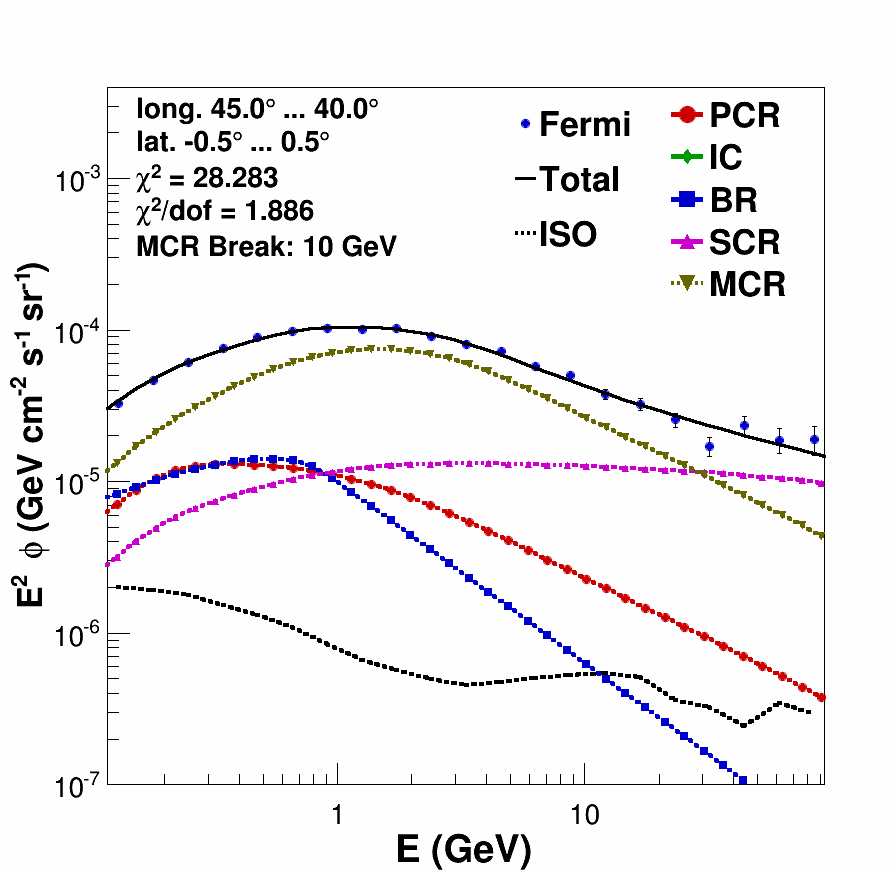}
\includegraphics[width=0.16\textwidth,height=0.16\textwidth,clip]{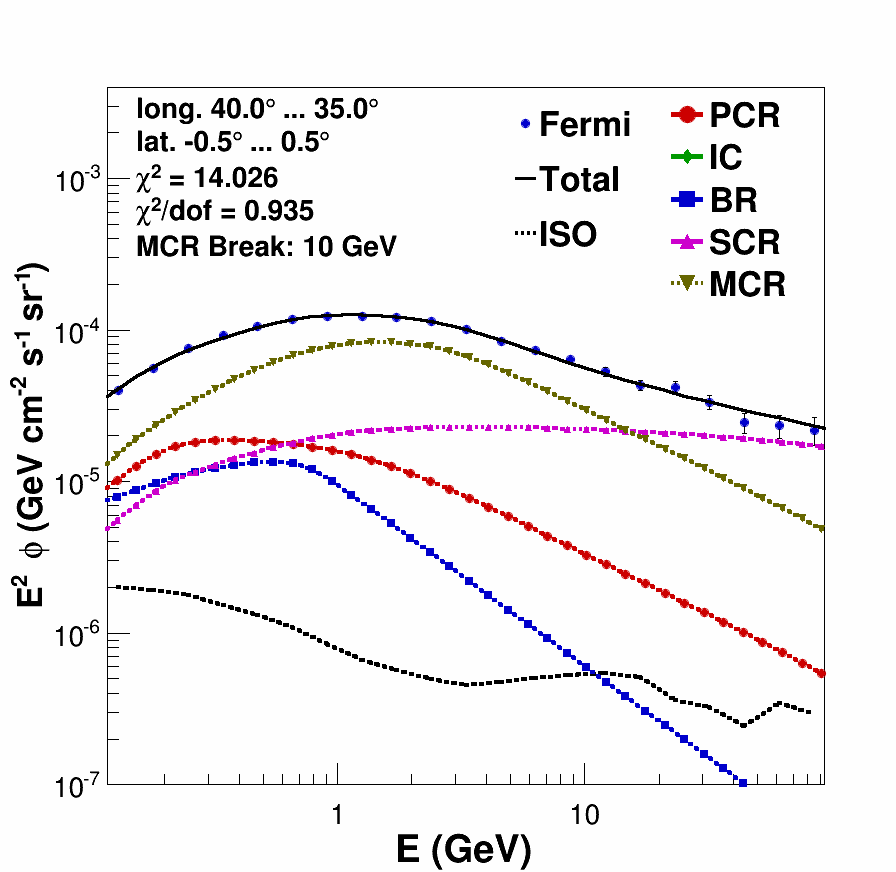}
\includegraphics[width=0.16\textwidth,height=0.16\textwidth,clip]{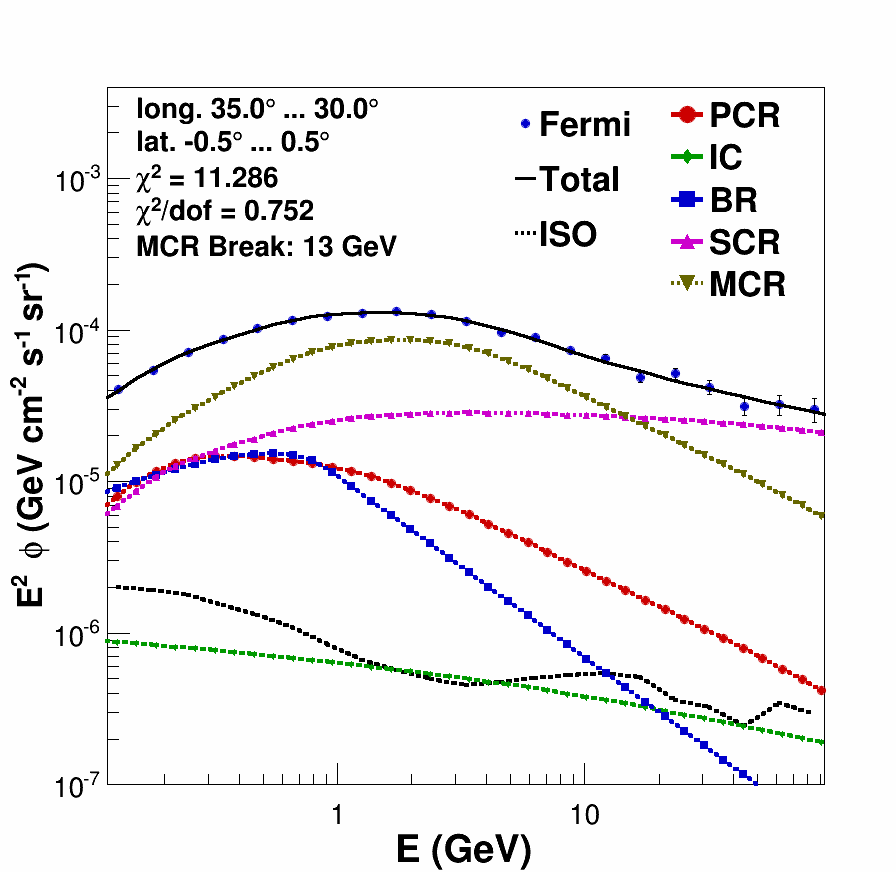}
\includegraphics[width=0.16\textwidth,height=0.16\textwidth,clip]{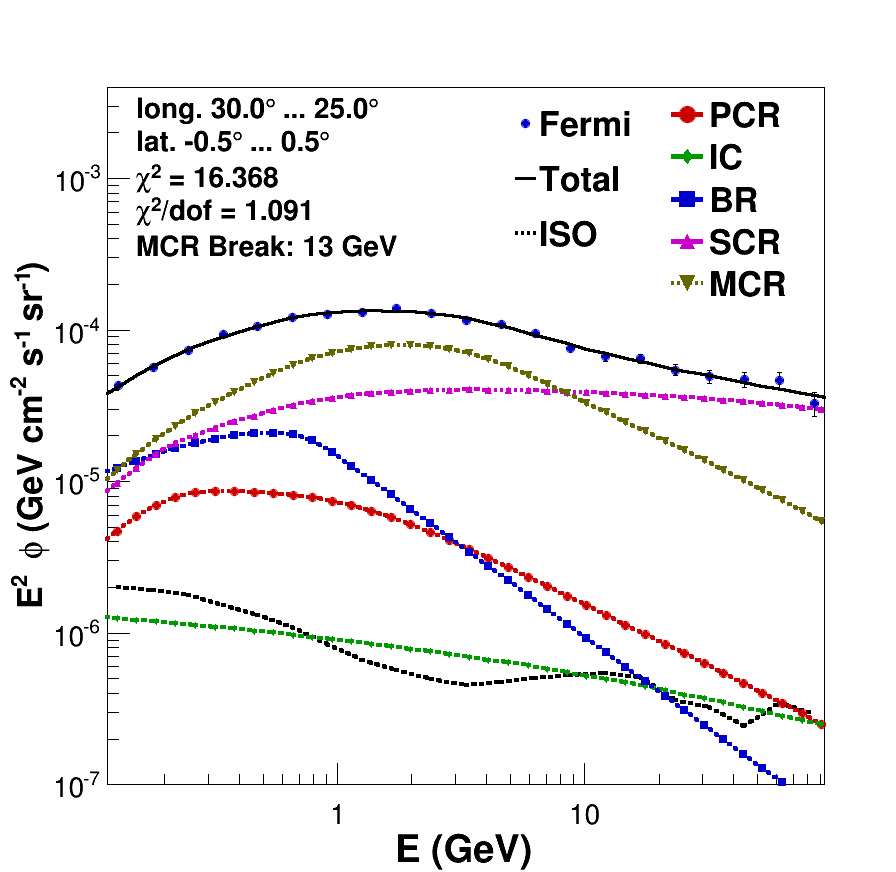}
\includegraphics[width=0.16\textwidth,height=0.16\textwidth,clip]{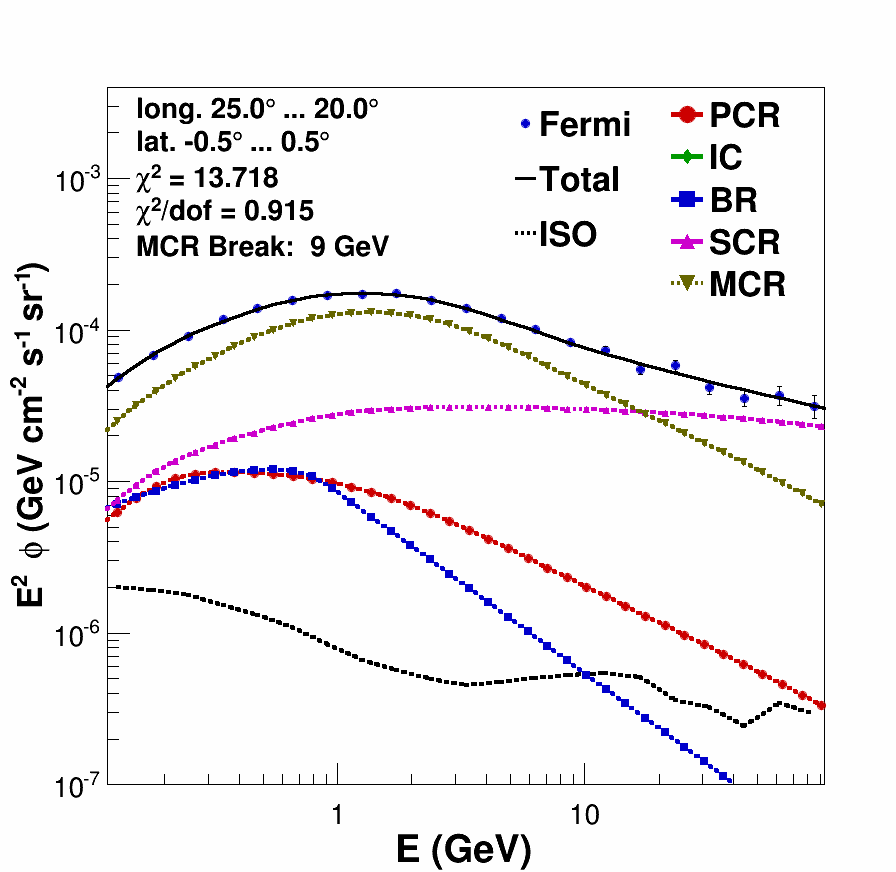}
\includegraphics[width=0.16\textwidth,height=0.16\textwidth,clip]{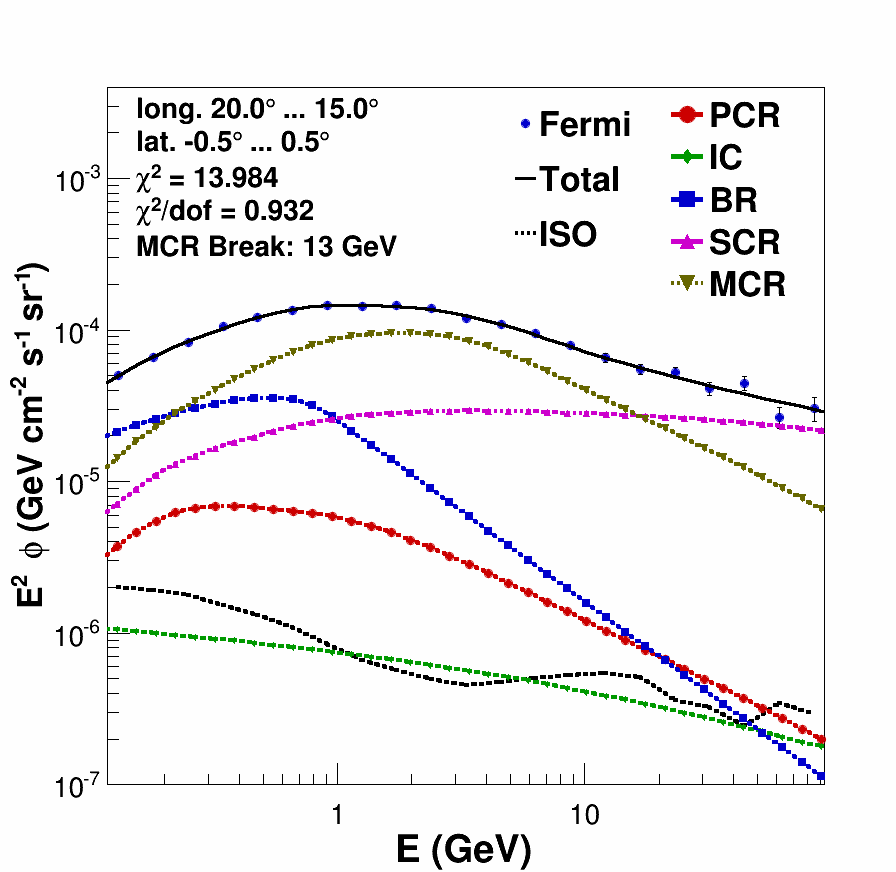}
\includegraphics[width=0.16\textwidth,height=0.16\textwidth,clip]{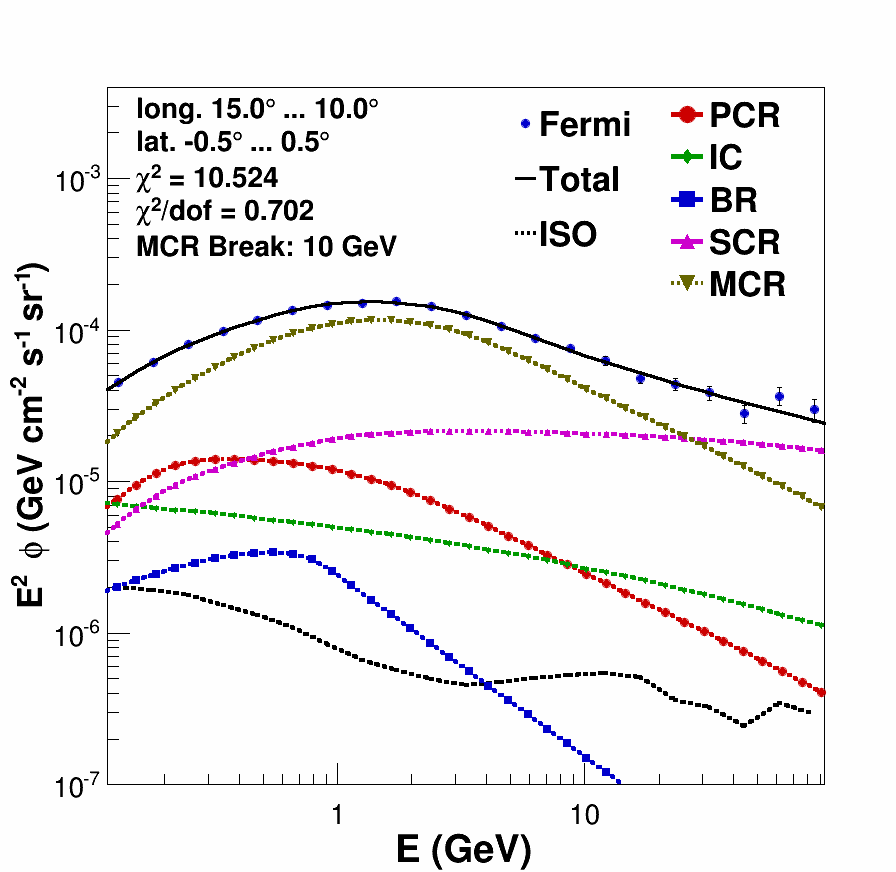}
\includegraphics[width=0.16\textwidth,height=0.16\textwidth,clip]{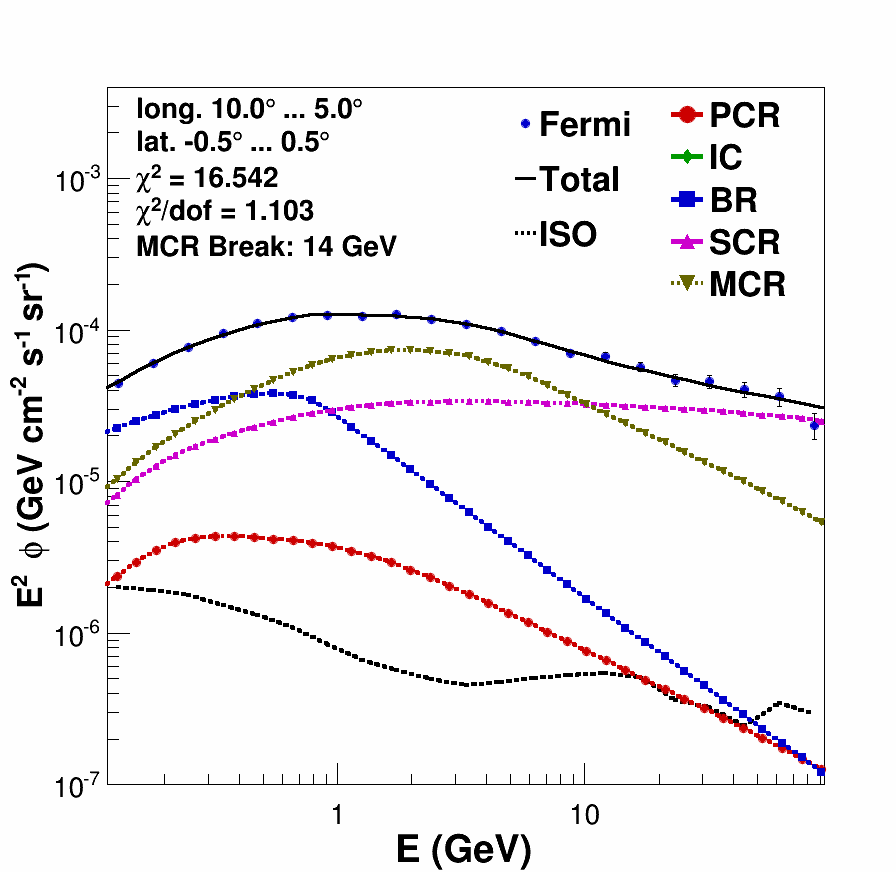}
\includegraphics[width=0.16\textwidth,height=0.16\textwidth,clip]{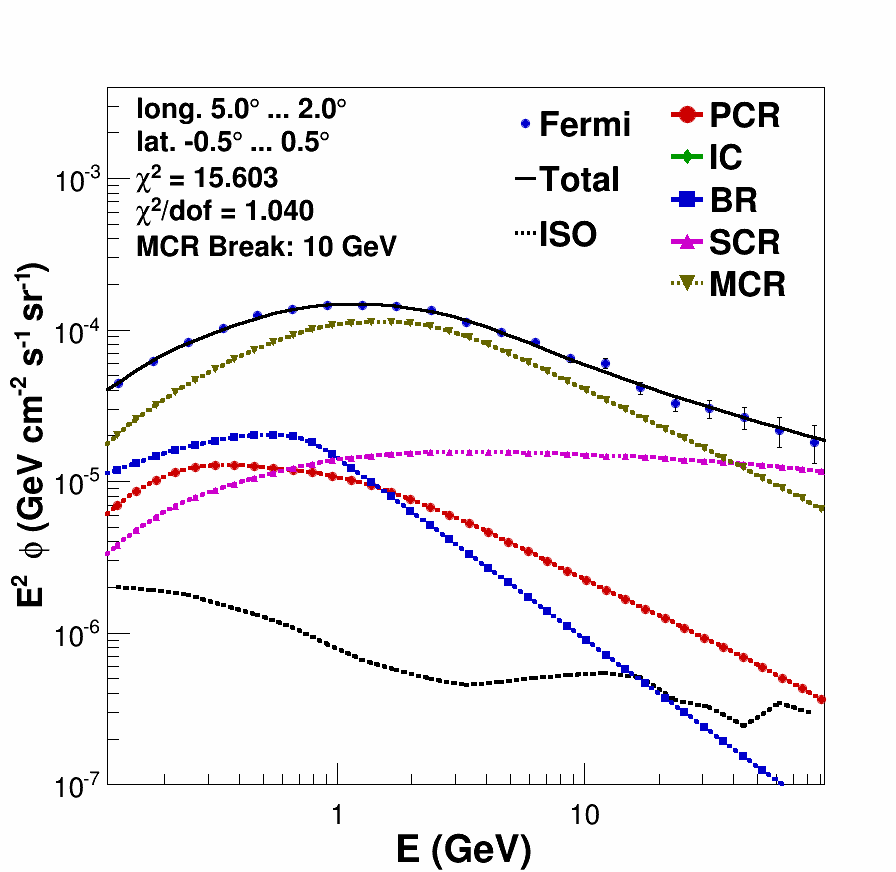}
\includegraphics[width=0.16\textwidth,height=0.16\textwidth,clip]{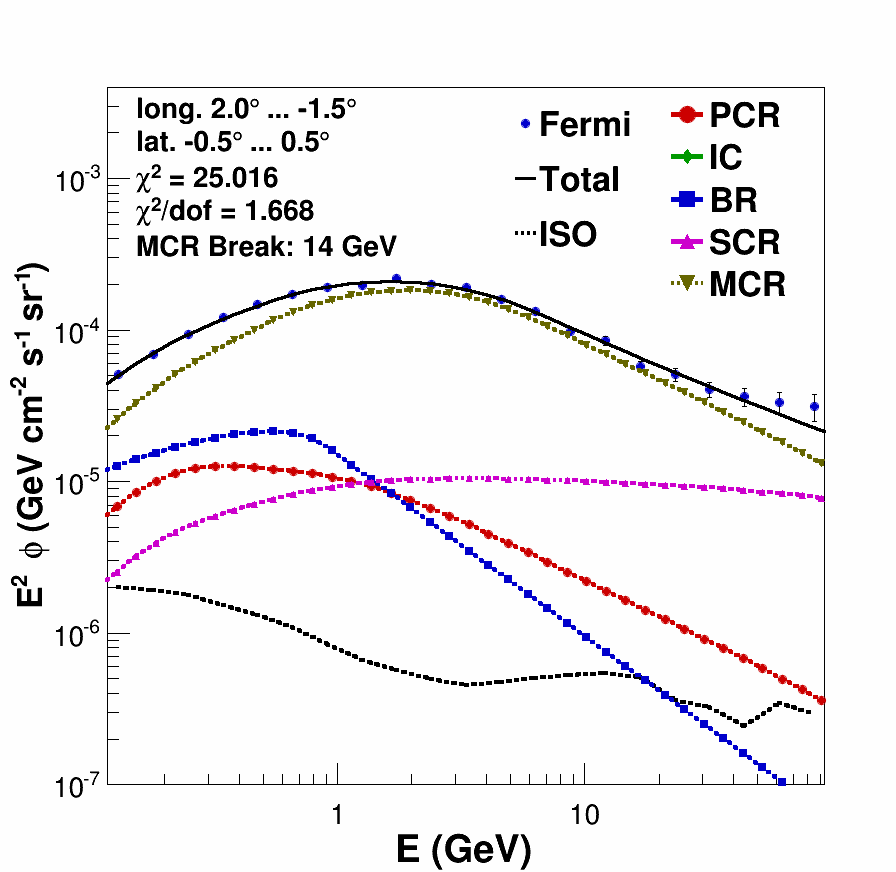}
\includegraphics[width=0.16\textwidth,height=0.16\textwidth,clip]{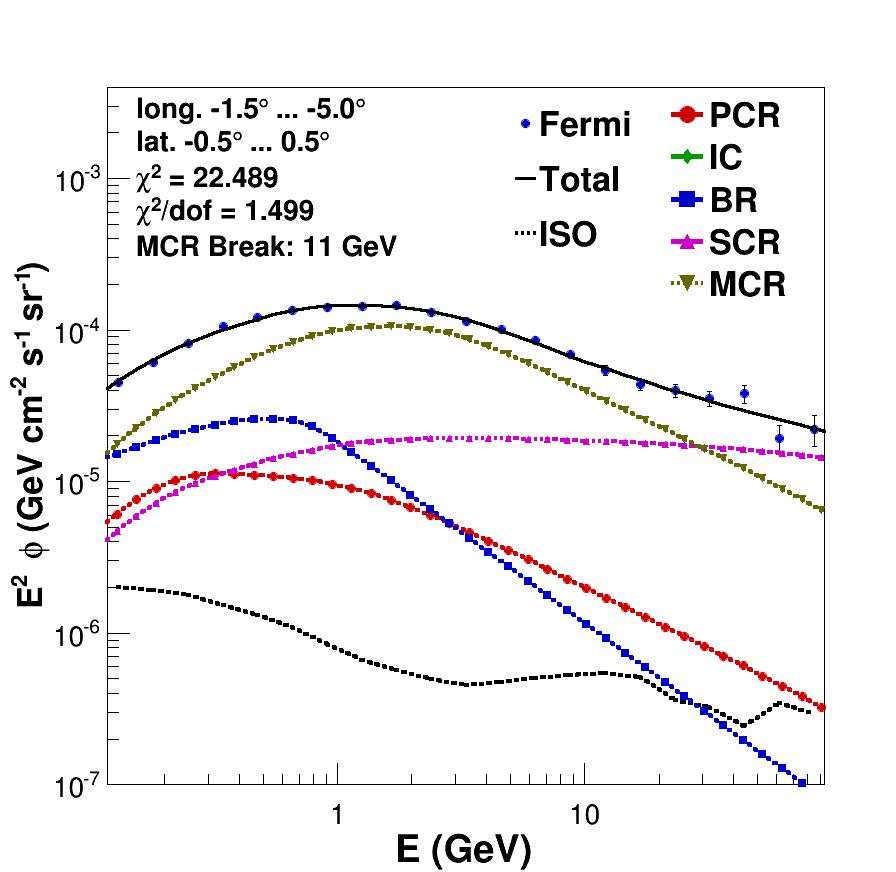}
\includegraphics[width=0.16\textwidth,height=0.16\textwidth,clip]{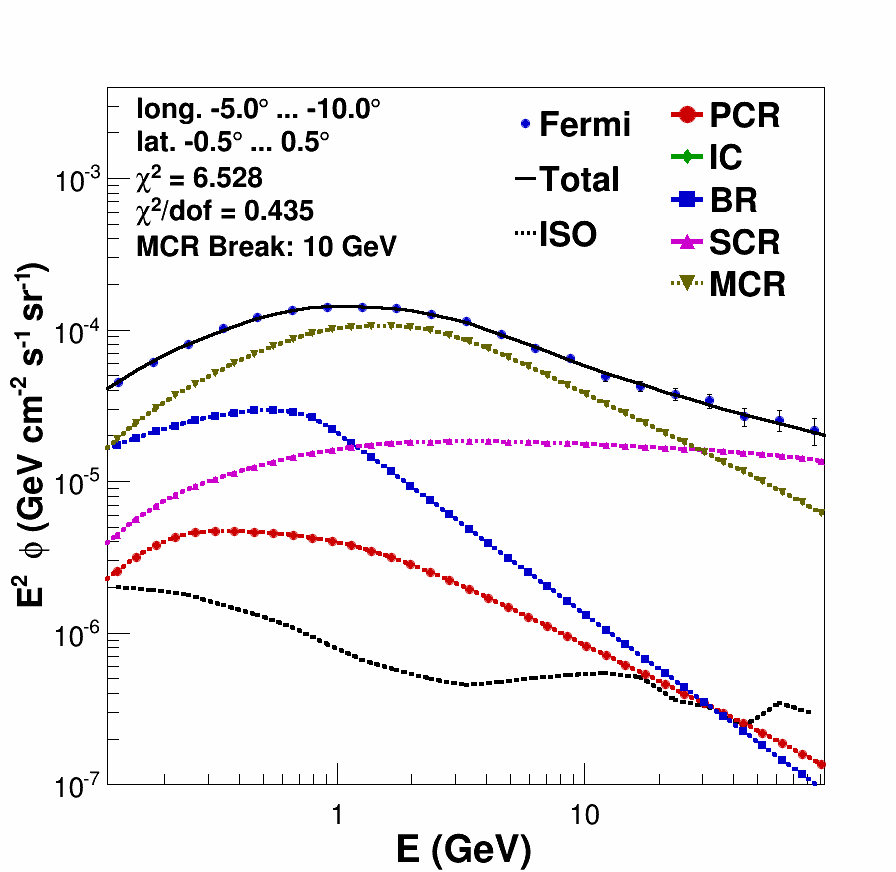}
\includegraphics[width=0.16\textwidth,height=0.16\textwidth,clip]{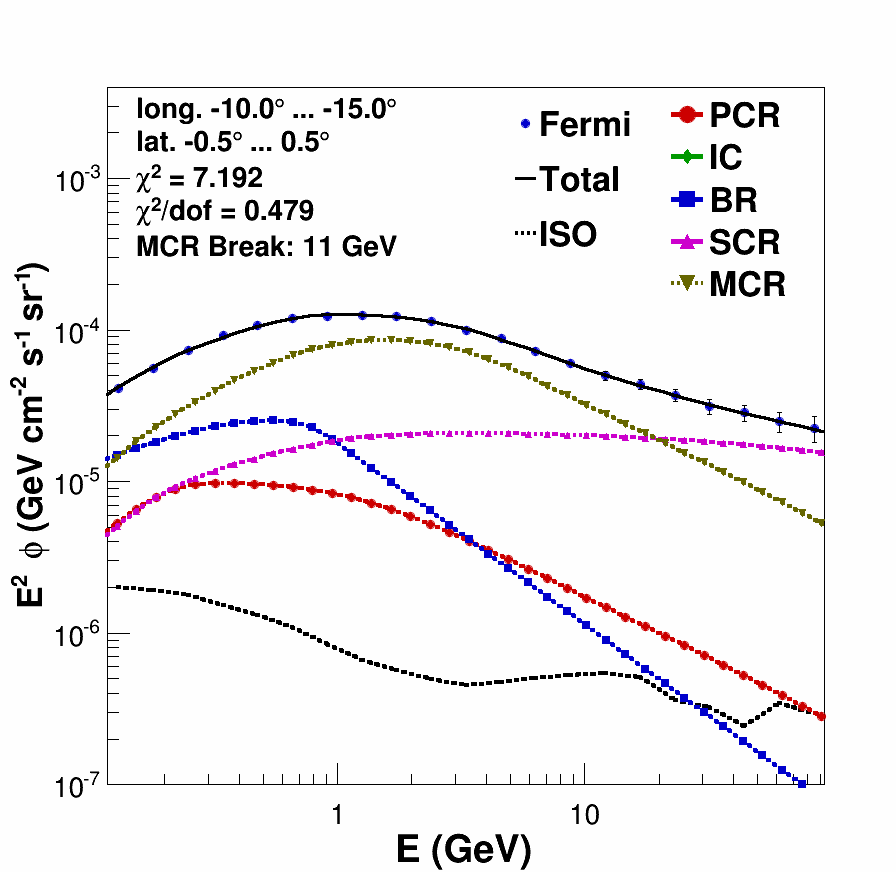}
\includegraphics[width=0.16\textwidth,height=0.16\textwidth,clip]{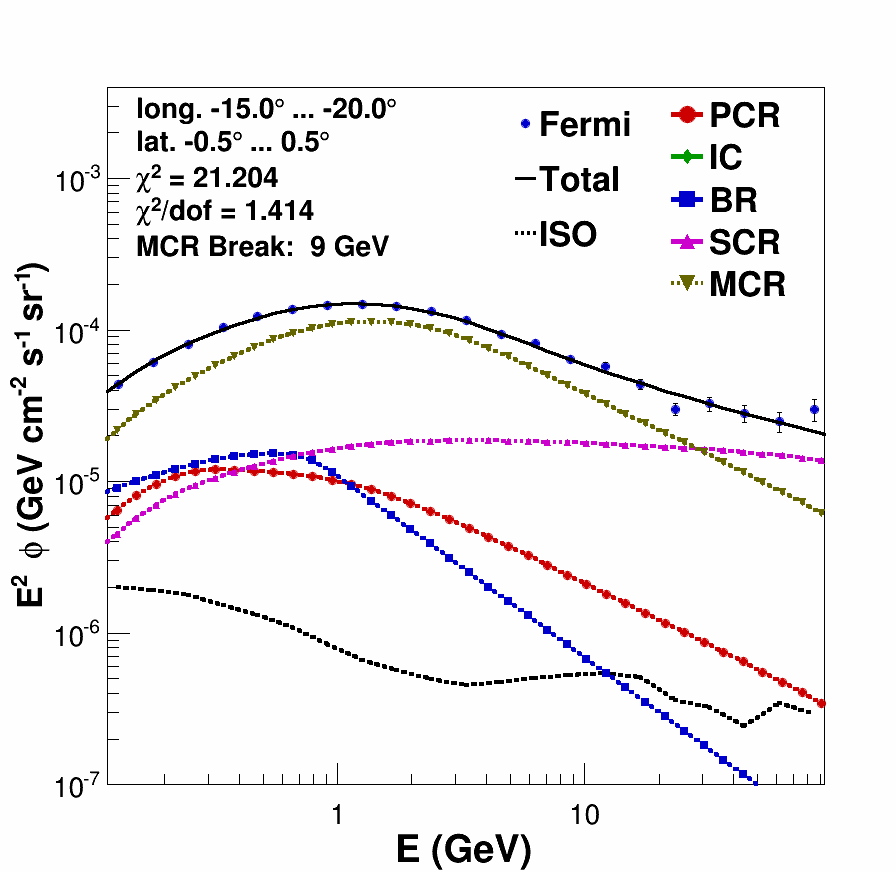}
\includegraphics[width=0.16\textwidth,height=0.16\textwidth,clip]{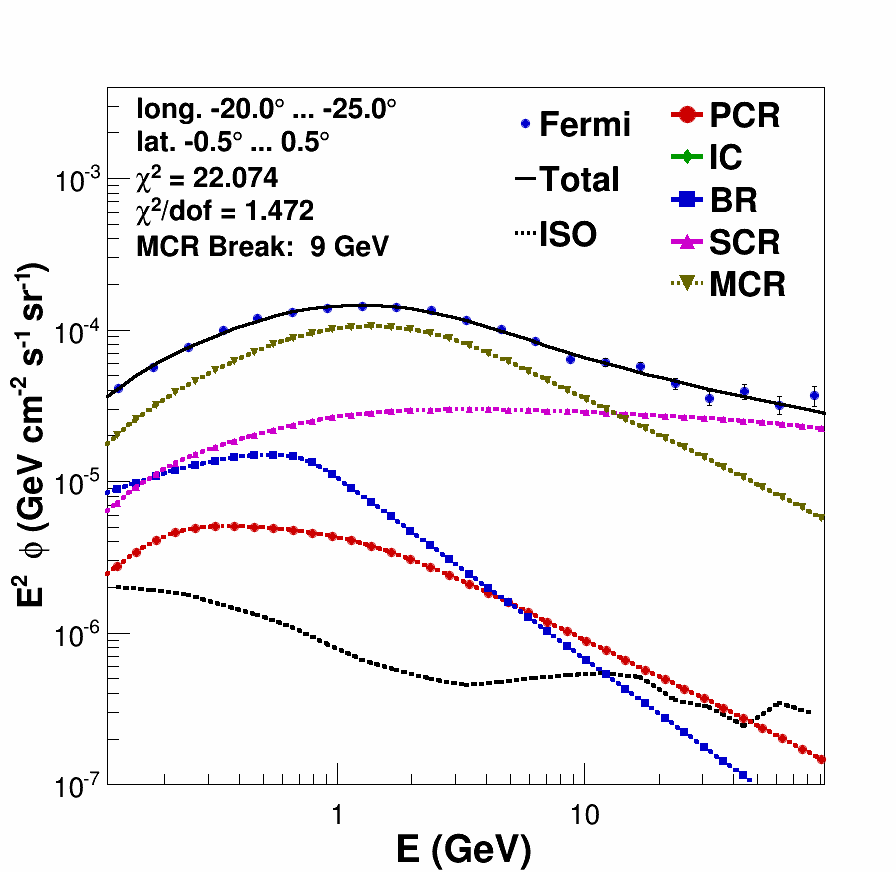}
\includegraphics[width=0.16\textwidth,height=0.16\textwidth,clip]{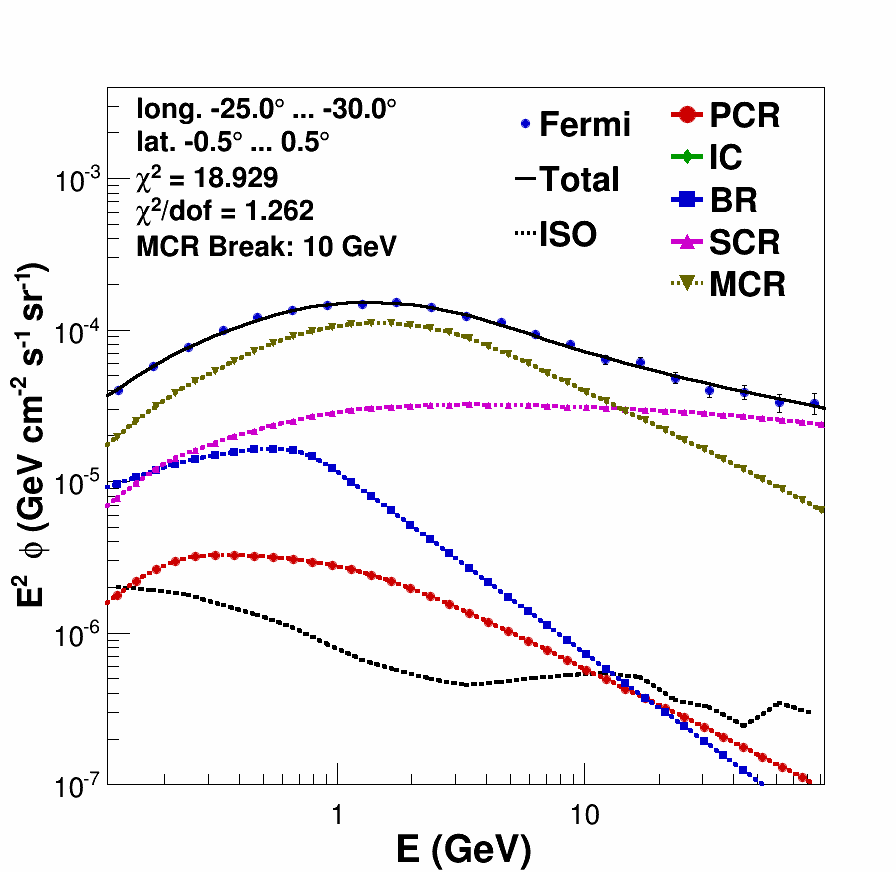}
\includegraphics[width=0.16\textwidth,height=0.16\textwidth,clip]{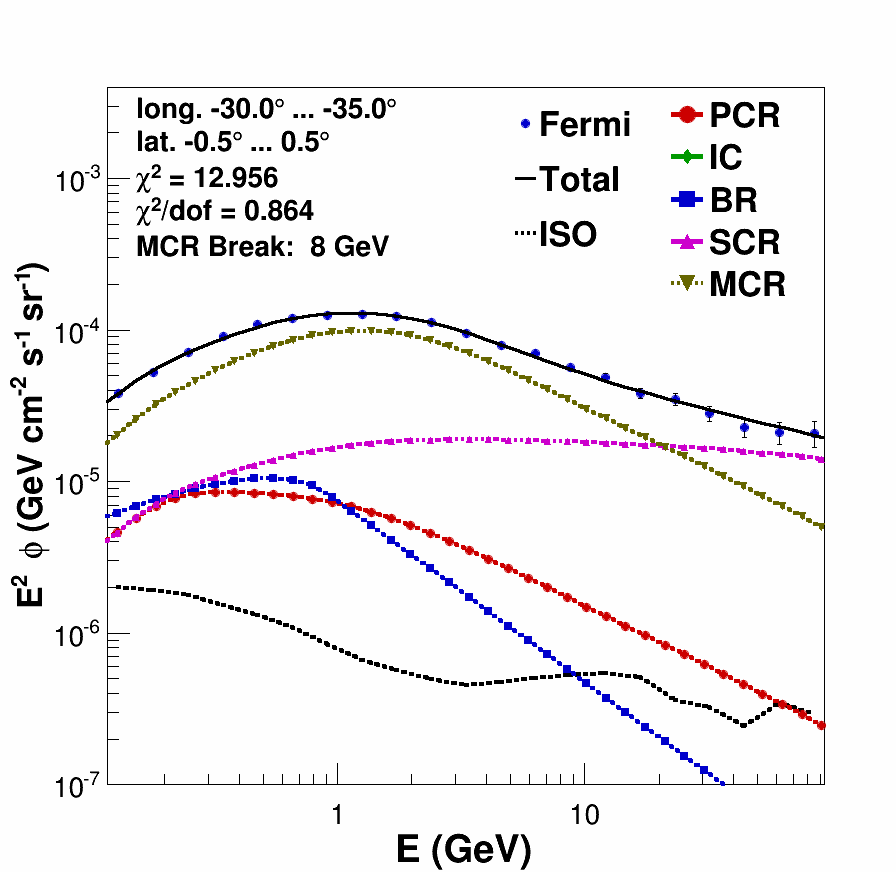}
\includegraphics[width=0.16\textwidth,height=0.16\textwidth,clip]{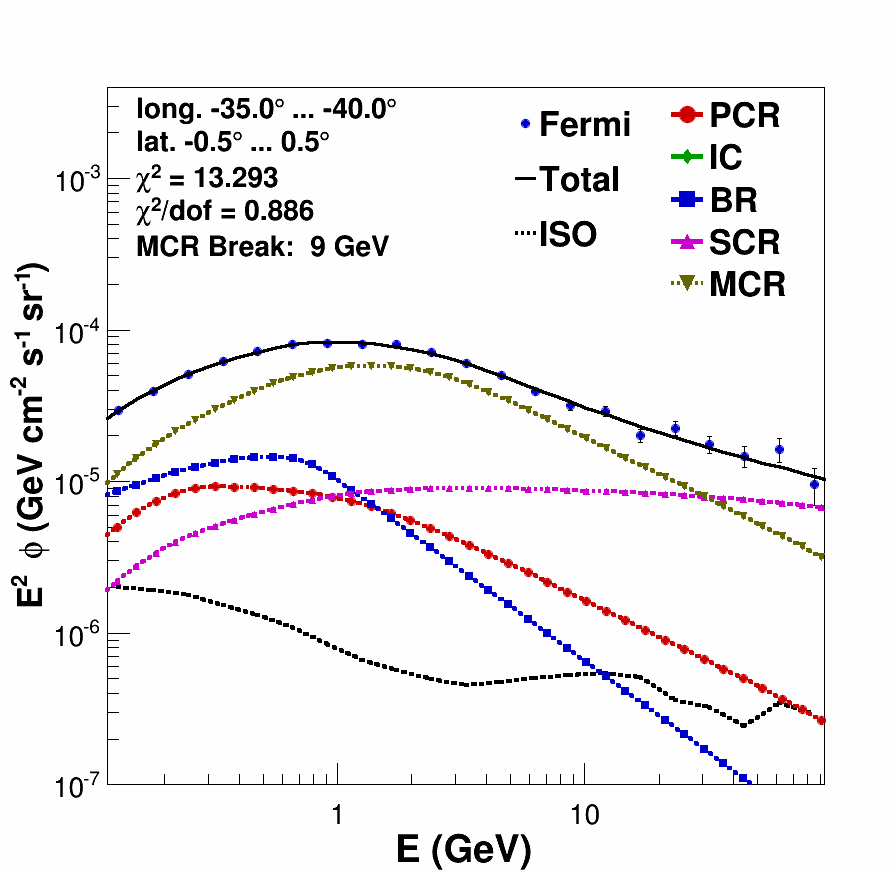}
\includegraphics[width=0.16\textwidth,height=0.16\textwidth,clip]{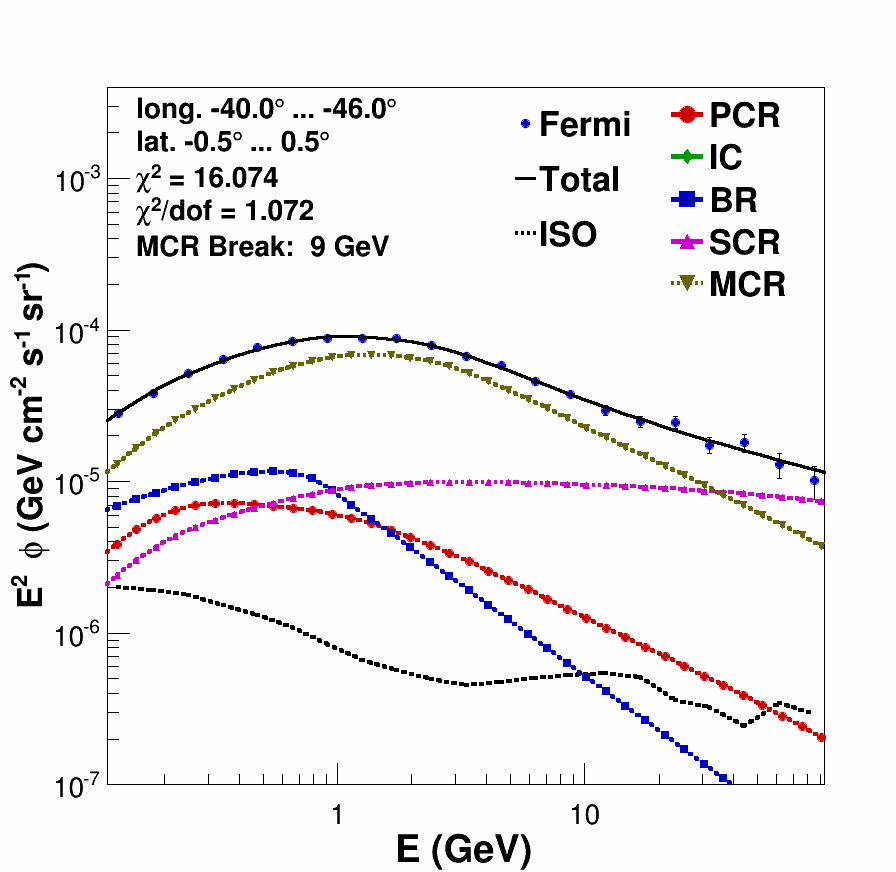}
\includegraphics[width=0.16\textwidth,height=0.16\textwidth,clip]{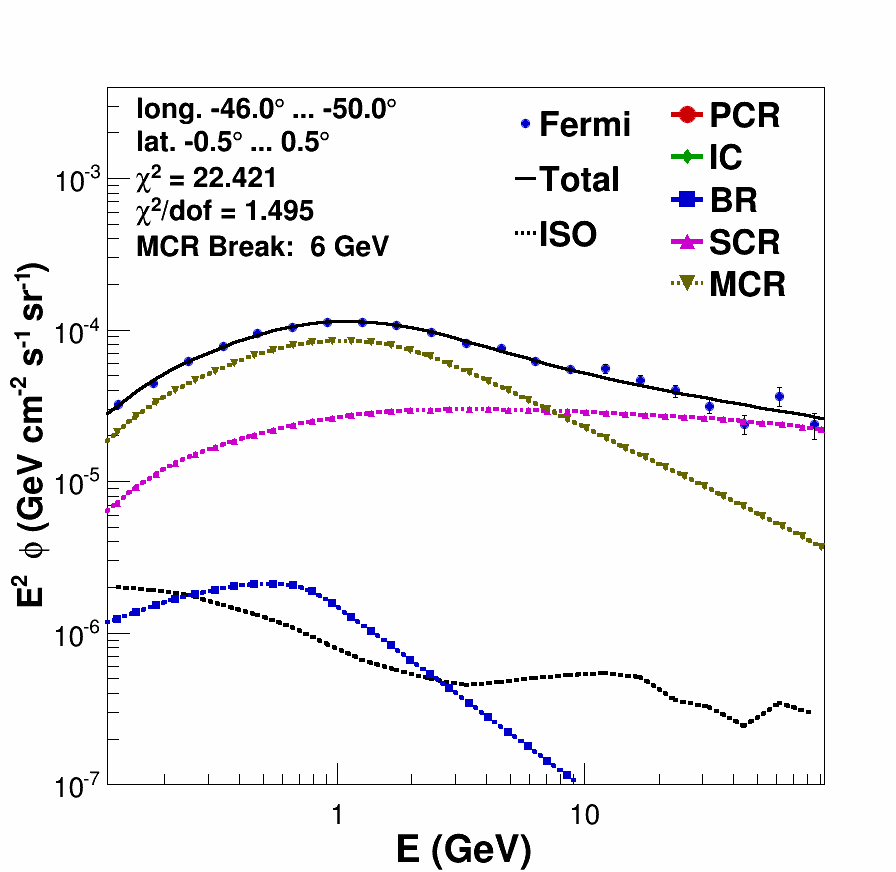}
\includegraphics[width=0.16\textwidth,height=0.16\textwidth,clip]{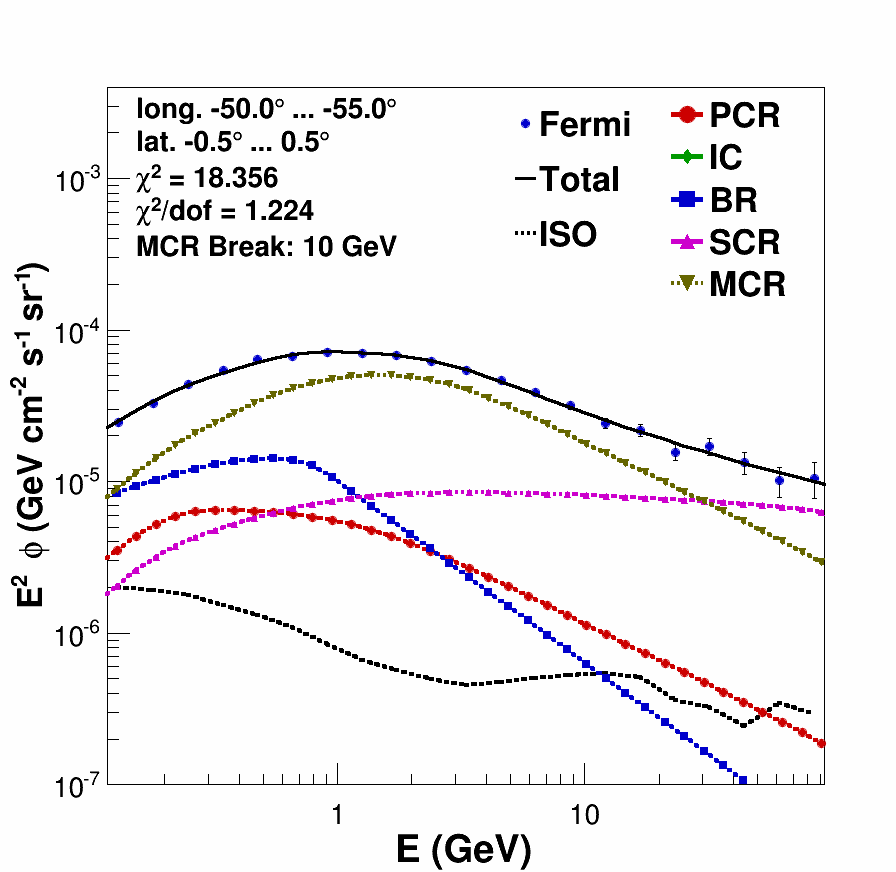}
\includegraphics[width=0.16\textwidth,height=0.16\textwidth,clip]{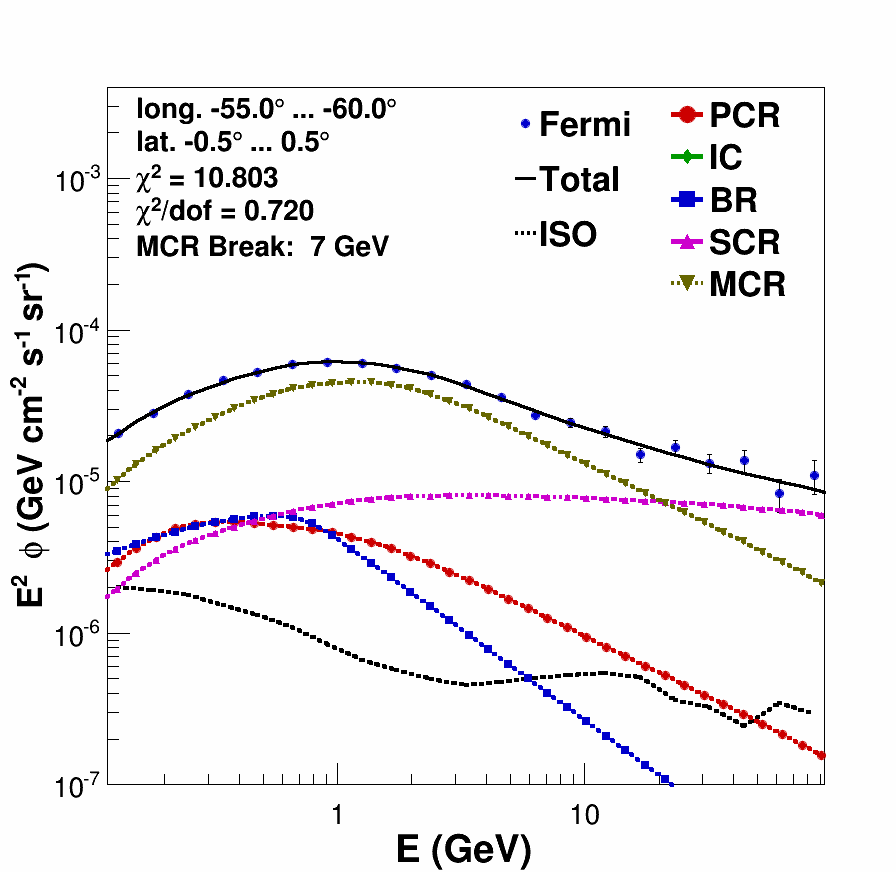}
\includegraphics[width=0.16\textwidth,height=0.16\textwidth,clip]{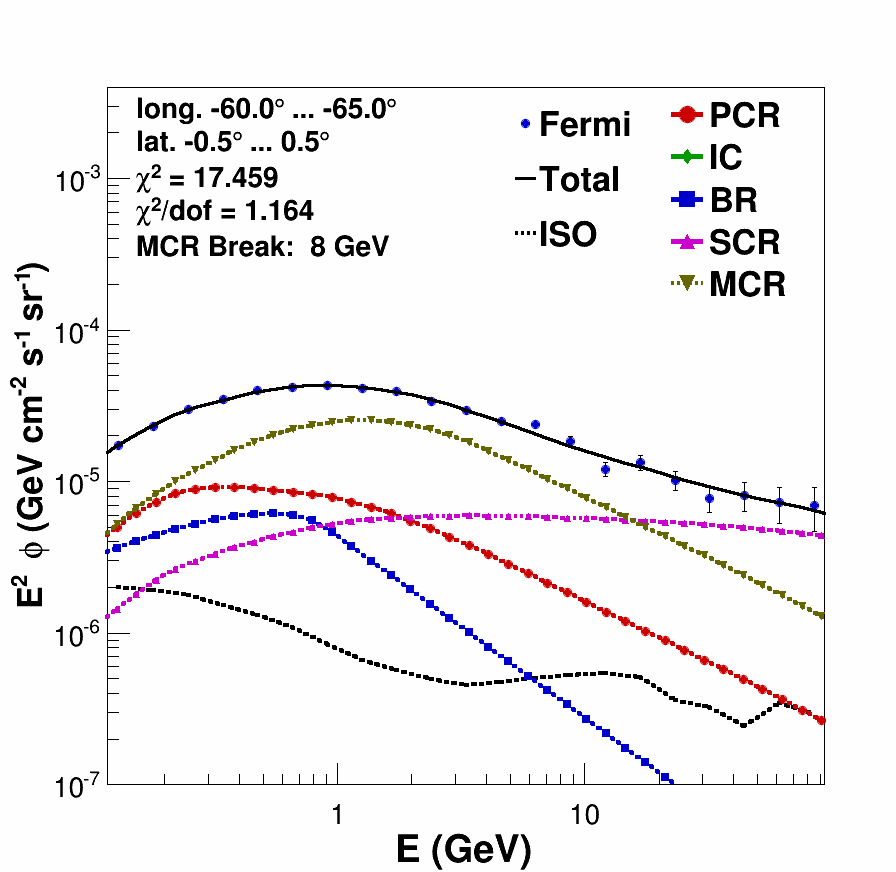}
\includegraphics[width=0.16\textwidth,height=0.16\textwidth,clip]{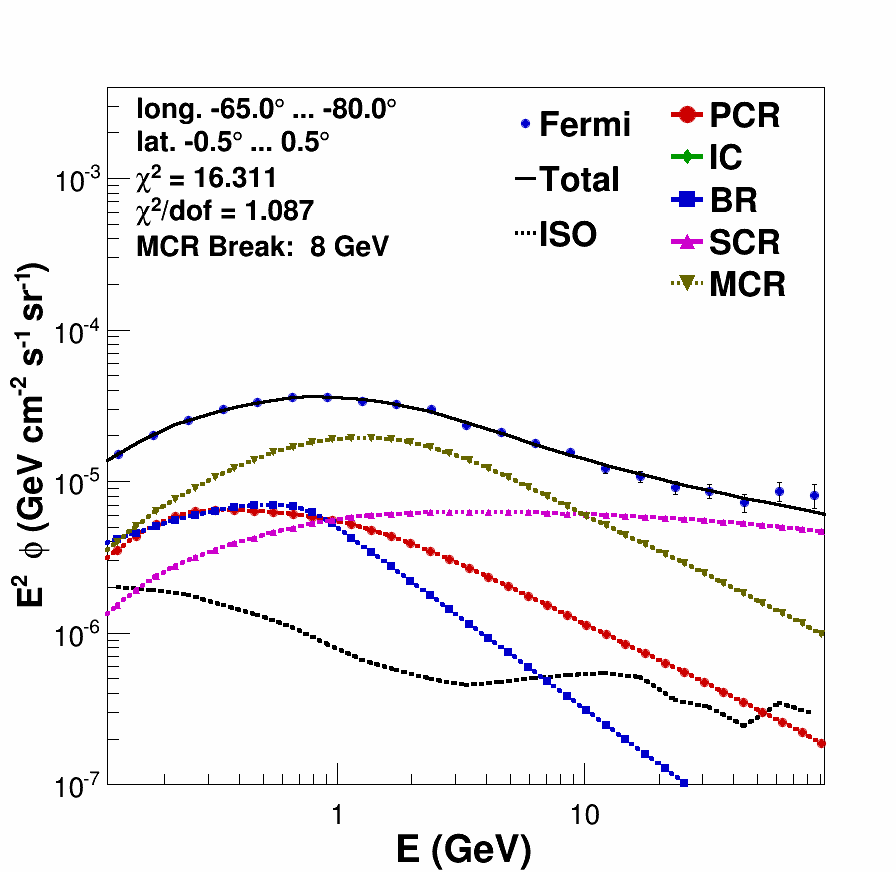}
\includegraphics[width=0.16\textwidth,height=0.16\textwidth,clip]{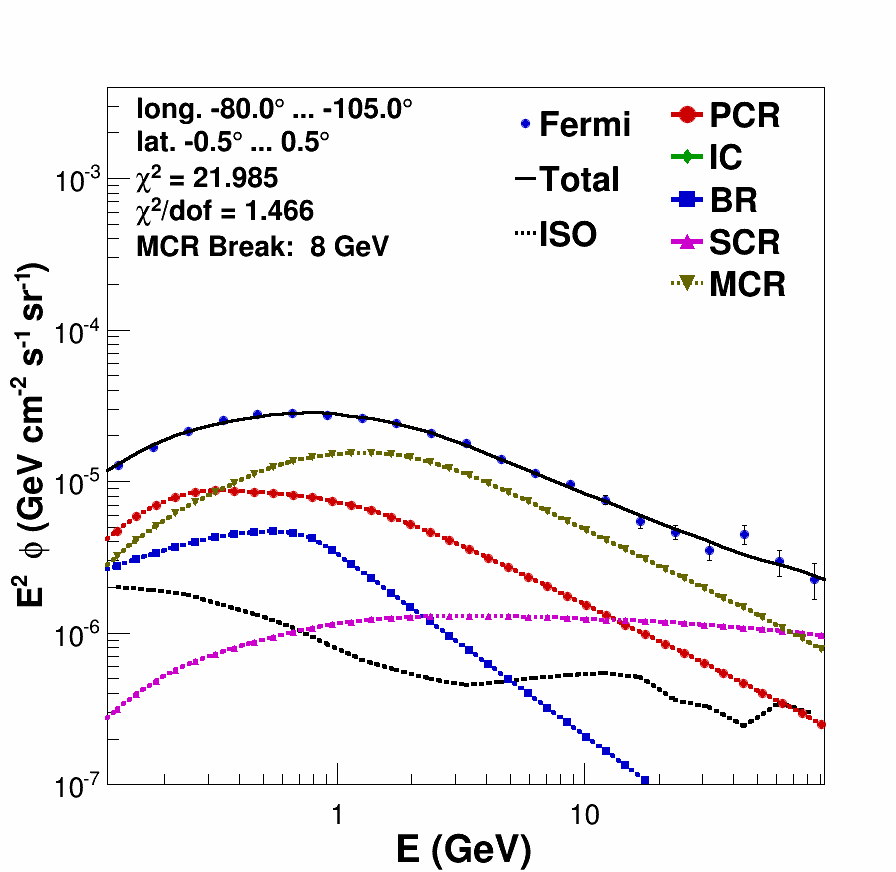}
\includegraphics[width=0.16\textwidth,height=0.16\textwidth,clip]{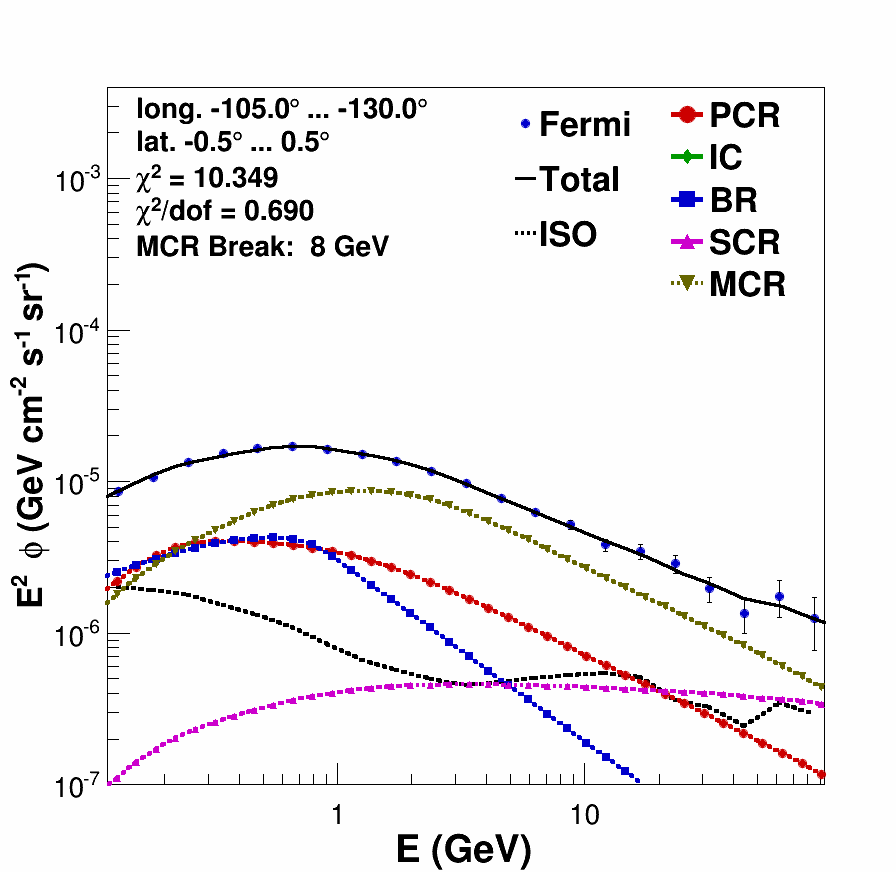}
\includegraphics[width=0.16\textwidth,height=0.16\textwidth,clip]{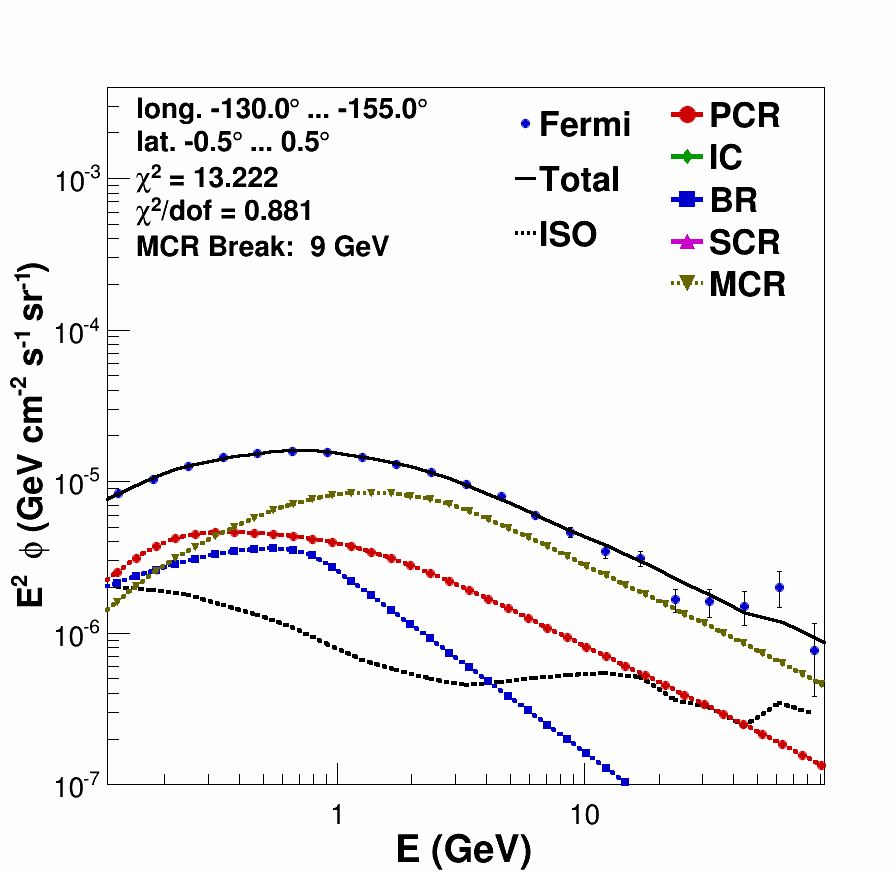}
\includegraphics[width=0.16\textwidth,height=0.16\textwidth,clip]{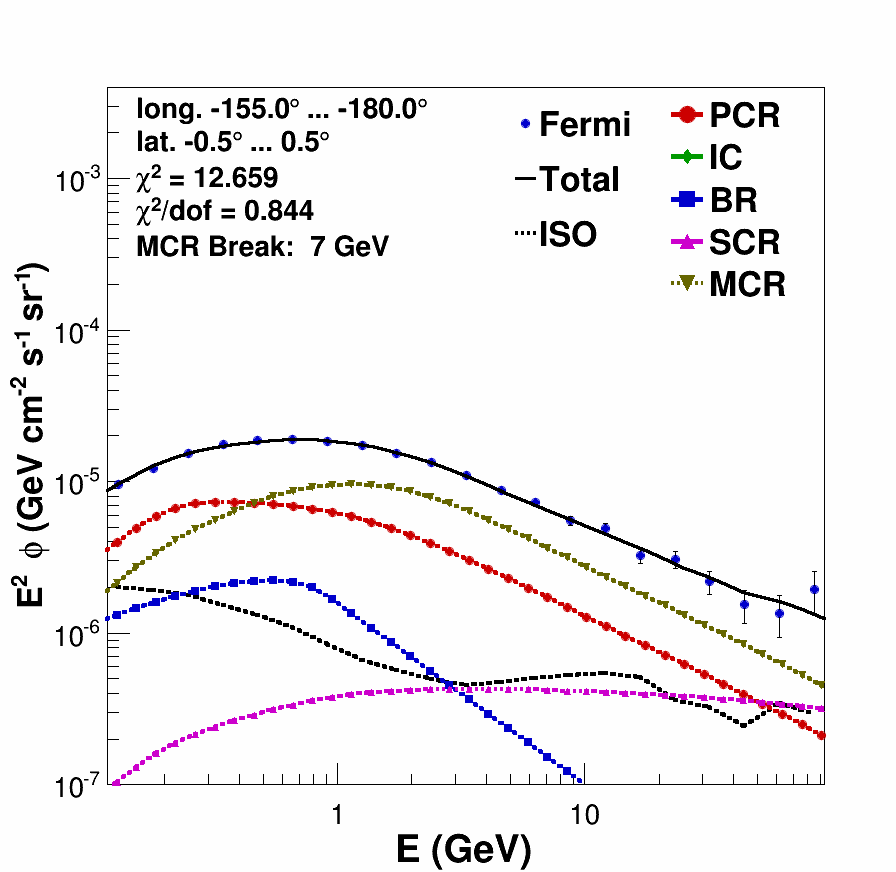}%%%%%%gd
\caption[]{Template fits for latitudes  with $-0.5^\circ<b<0.5^\circ$ and longitudes decreasing from 180$^\circ$ to -180$^\circ$.} \label{F21}
\end{figure}
\begin{figure}
\centering
	\includegraphics[width=0.16\textwidth,height=0.16\textwidth,clip]{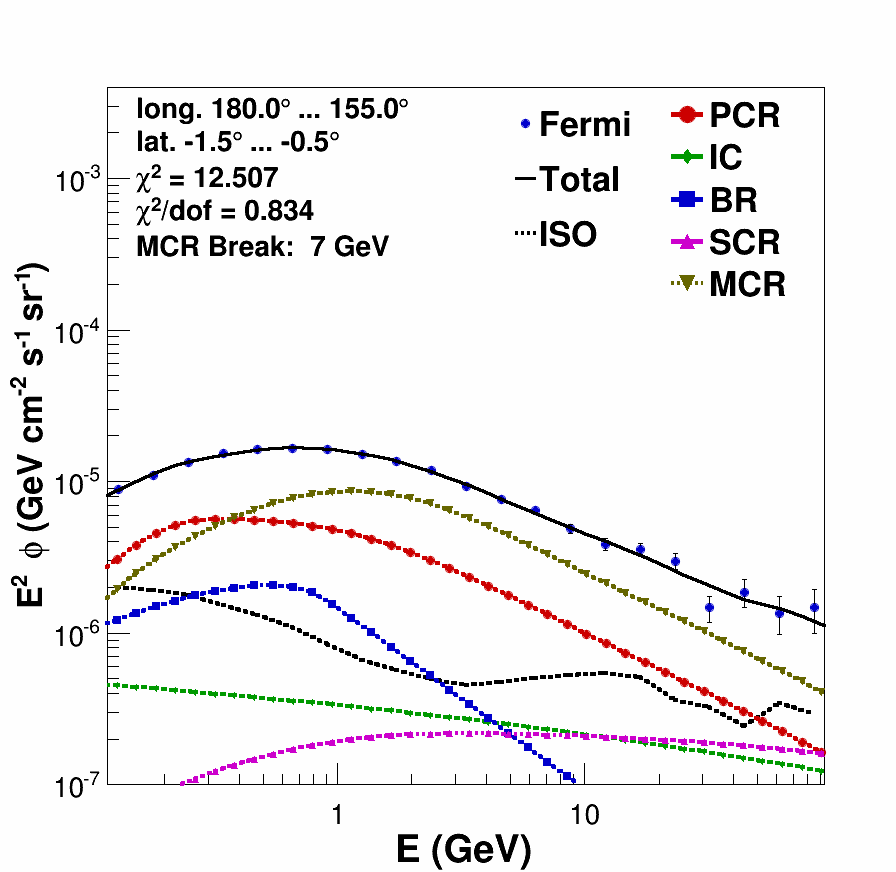}
	\includegraphics[width=0.16\textwidth,height=0.16\textwidth,clip]{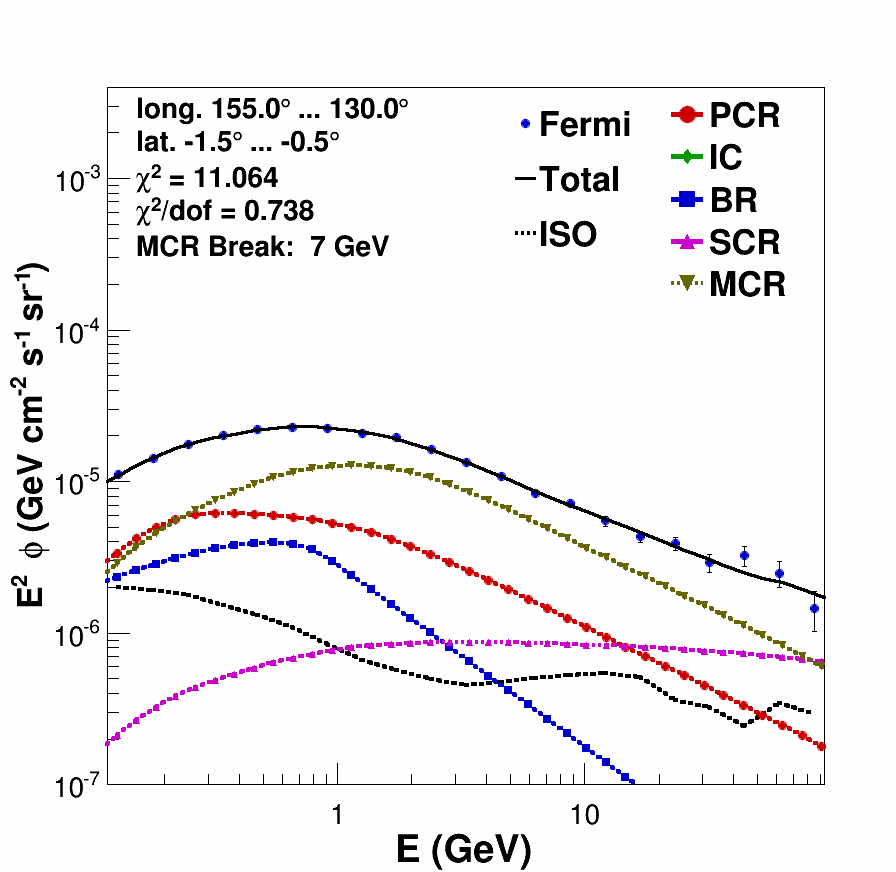}
	\includegraphics[width=0.16\textwidth,height=0.16\textwidth,clip]{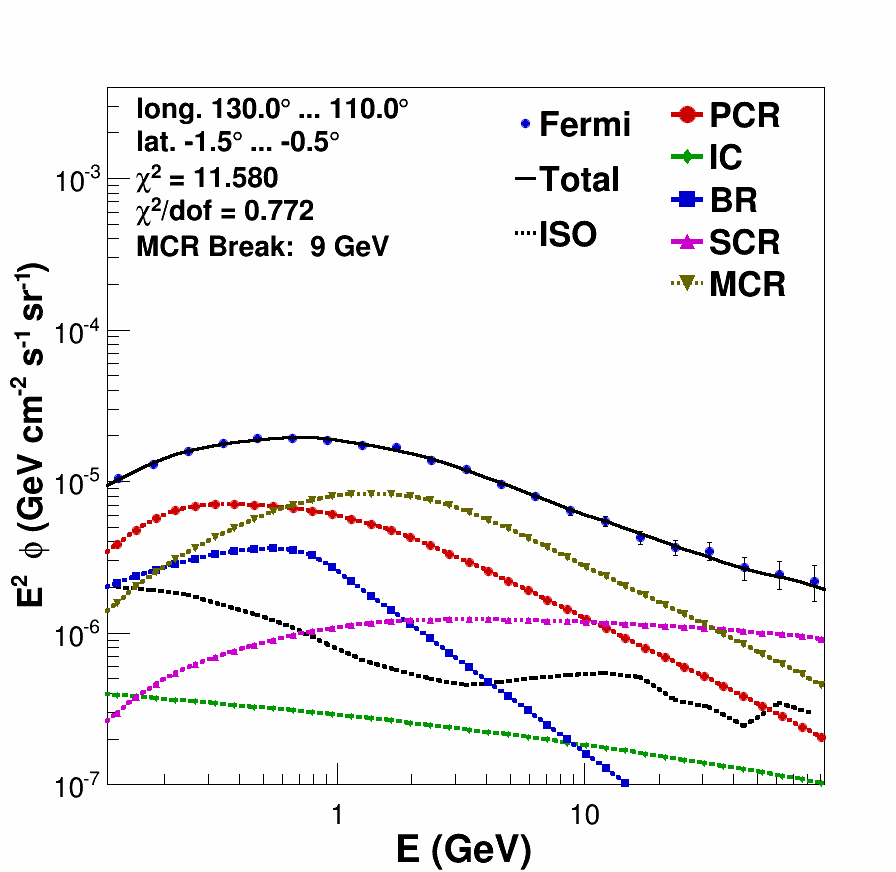}
	\includegraphics[width=0.16\textwidth,height=0.16\textwidth,clip]{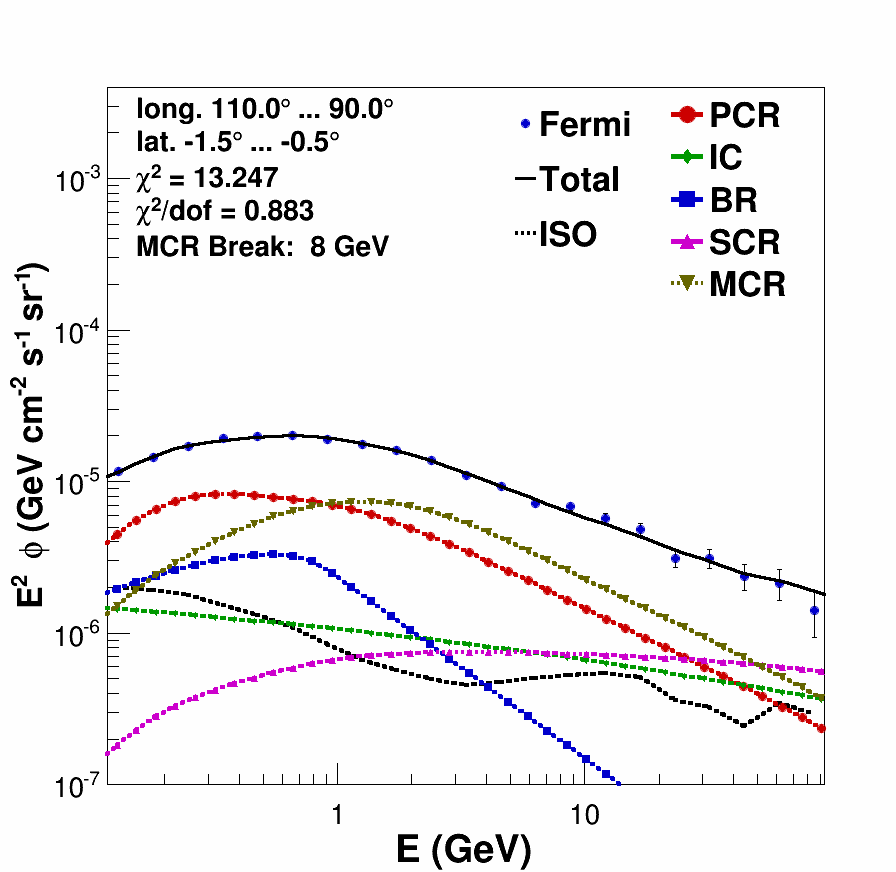}
	\includegraphics[width=0.16\textwidth,height=0.16\textwidth,clip]{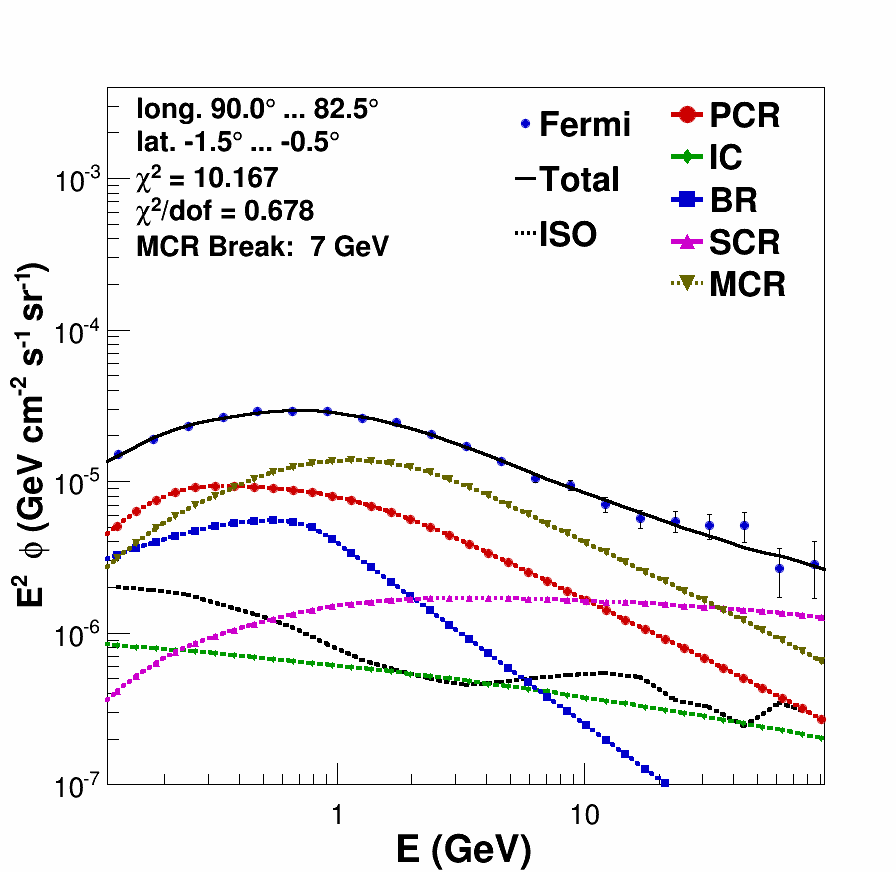}
	\includegraphics[width=0.16\textwidth,height=0.16\textwidth,clip]{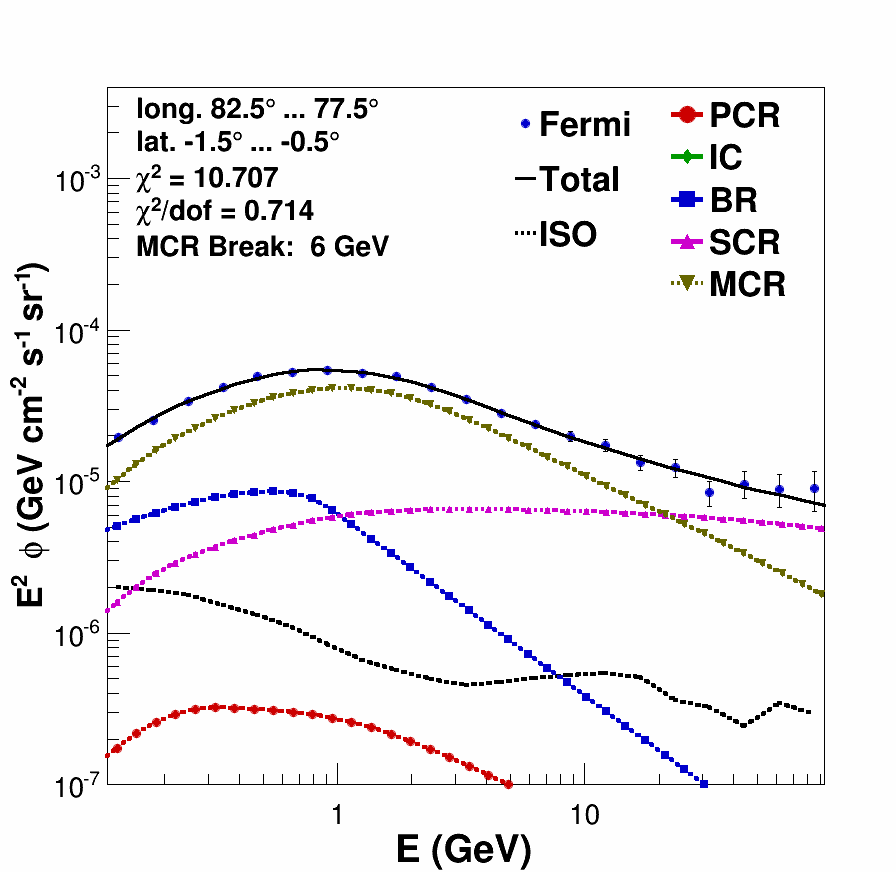}
	\includegraphics[width=0.16\textwidth,height=0.16\textwidth,clip]{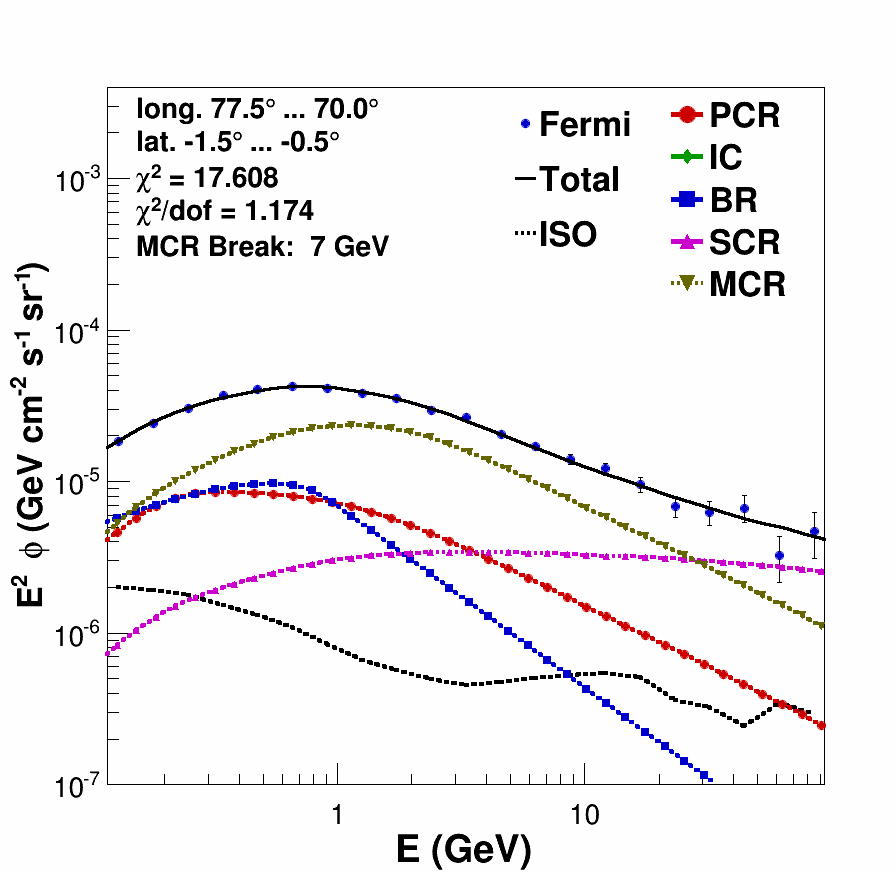}
	\includegraphics[width=0.16\textwidth,height=0.16\textwidth,clip]{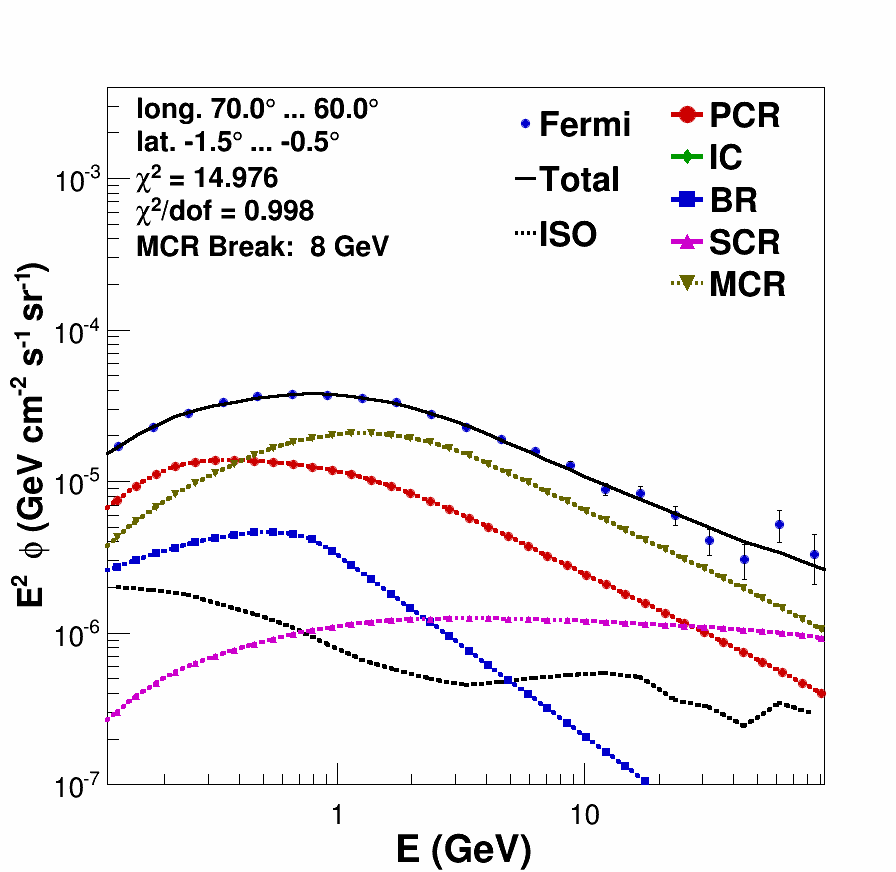}
	\includegraphics[width=0.16\textwidth,height=0.16\textwidth,clip]{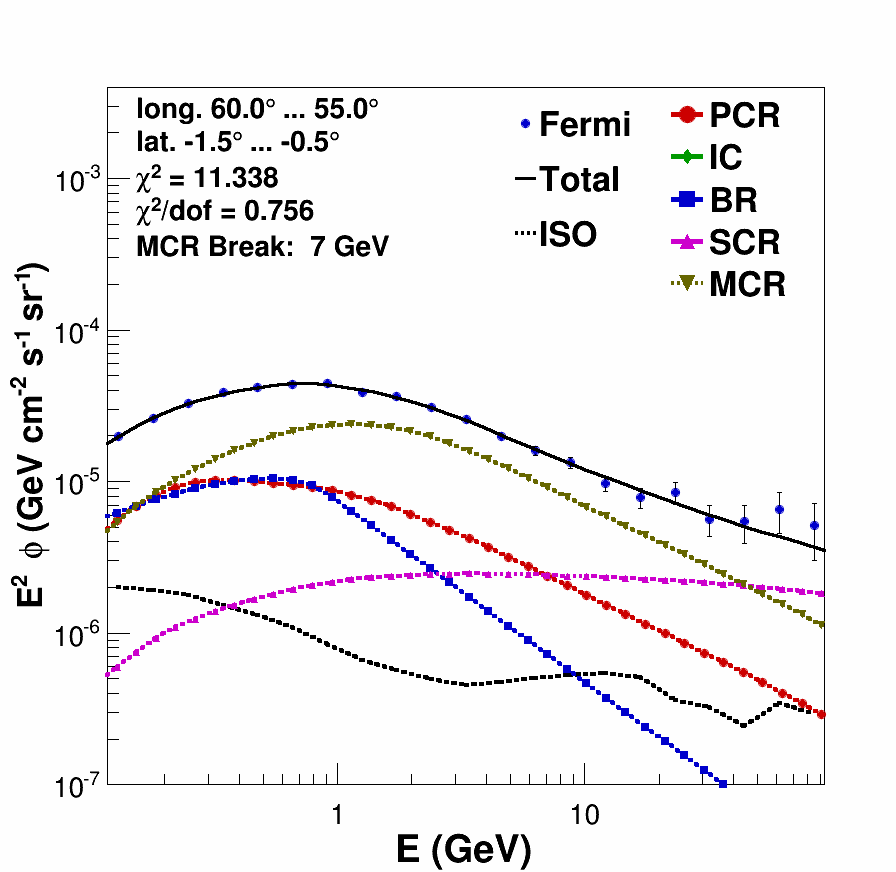}
	\includegraphics[width=0.16\textwidth,height=0.16\textwidth,clip]{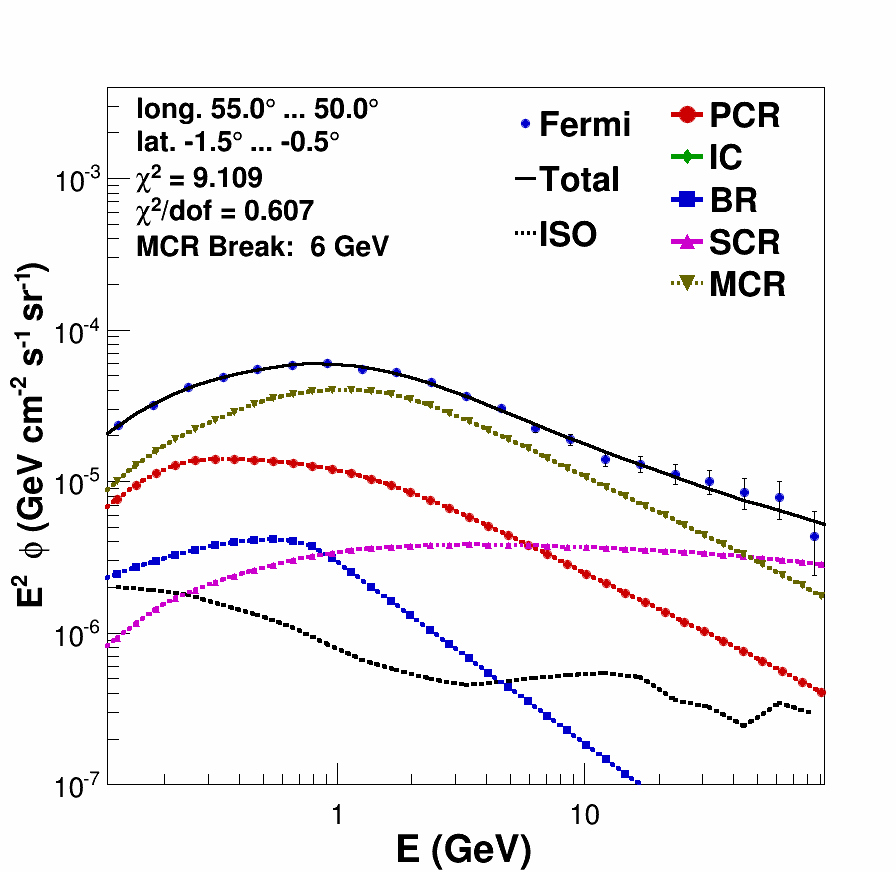}
	\includegraphics[width=0.16\textwidth,height=0.16\textwidth,clip]{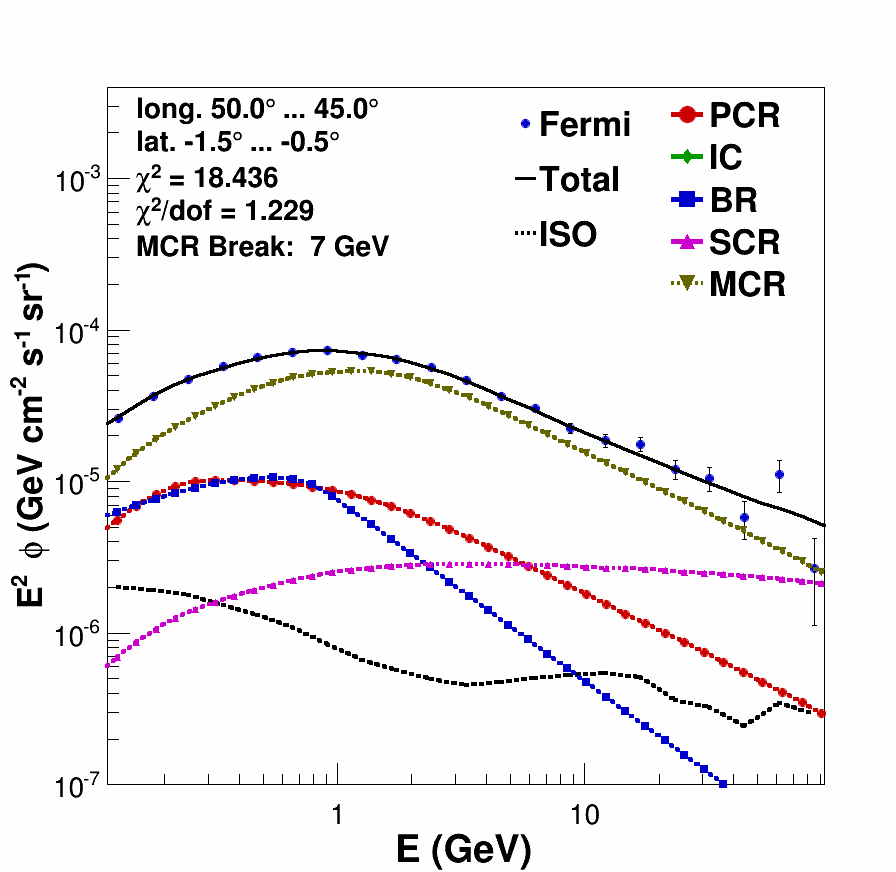}
	\includegraphics[width=0.16\textwidth,height=0.16\textwidth,clip]{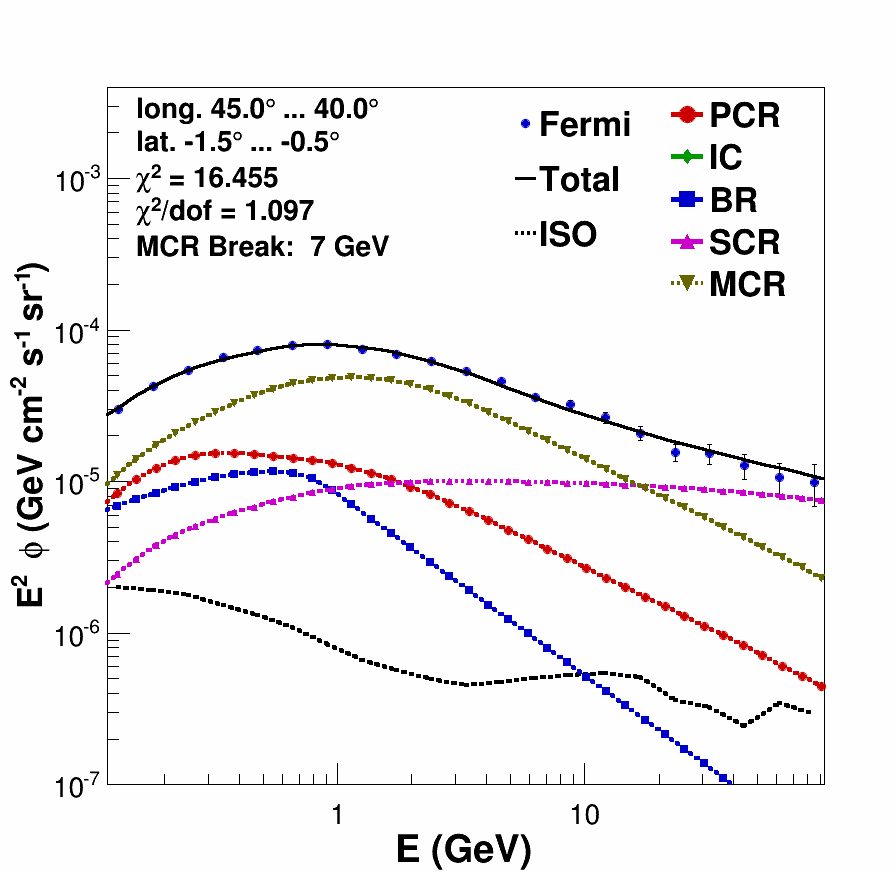}
	\includegraphics[width=0.16\textwidth,height=0.16\textwidth,clip]{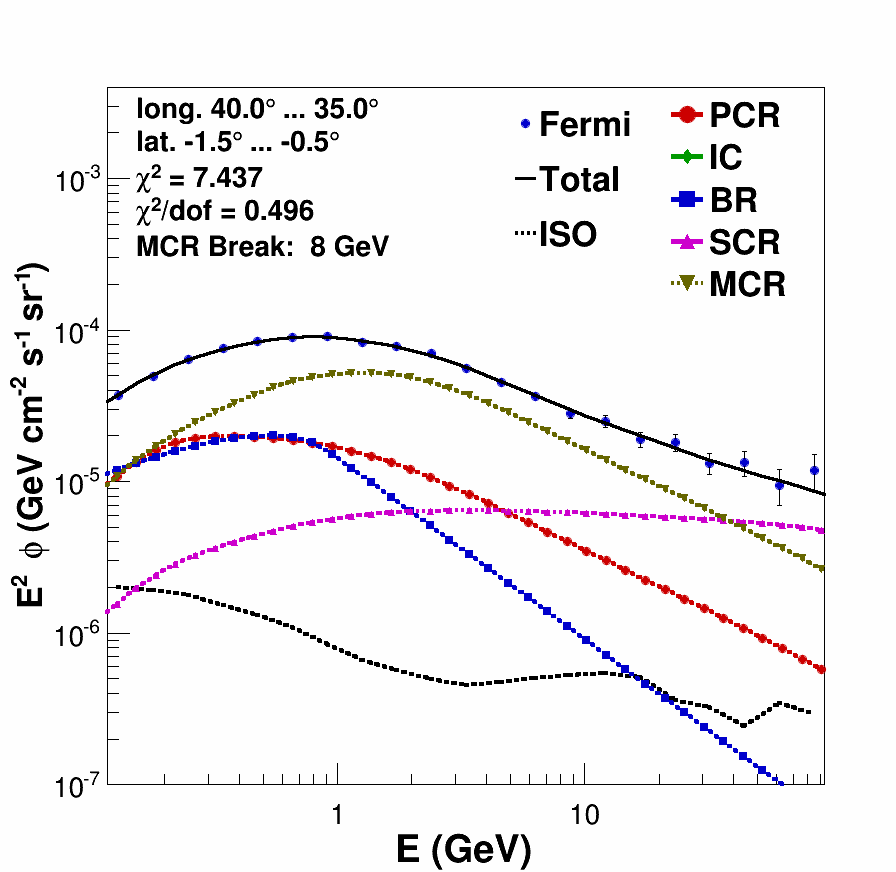}
	\includegraphics[width=0.16\textwidth,height=0.16\textwidth,clip]{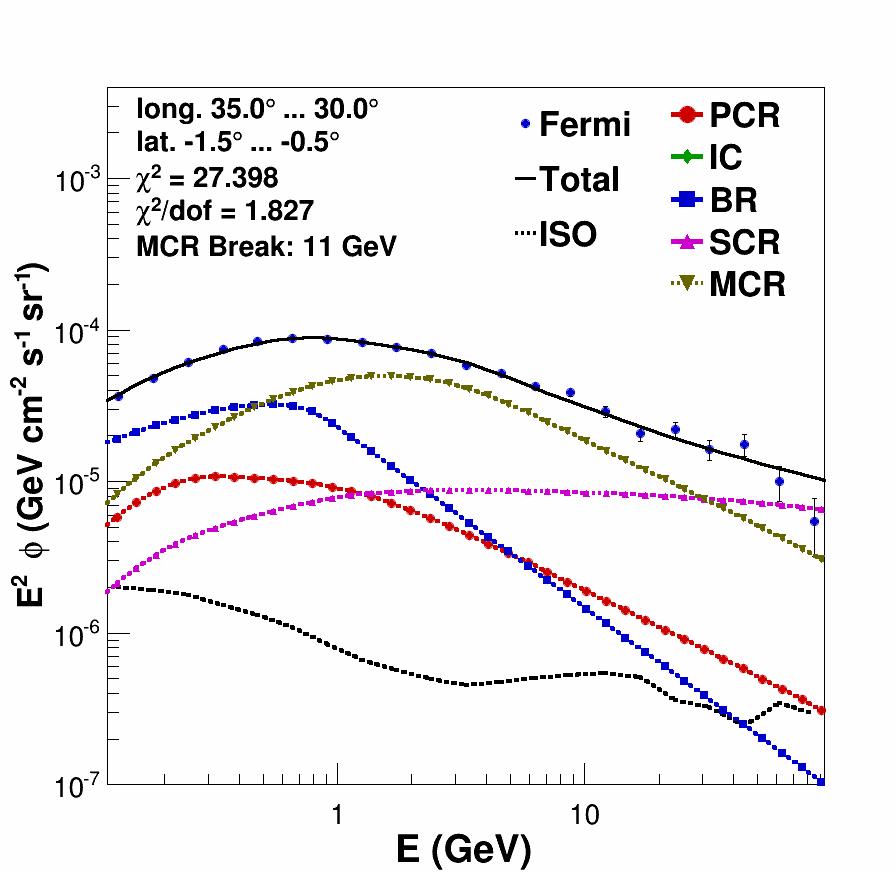}
	\includegraphics[width=0.16\textwidth,height=0.16\textwidth,clip]{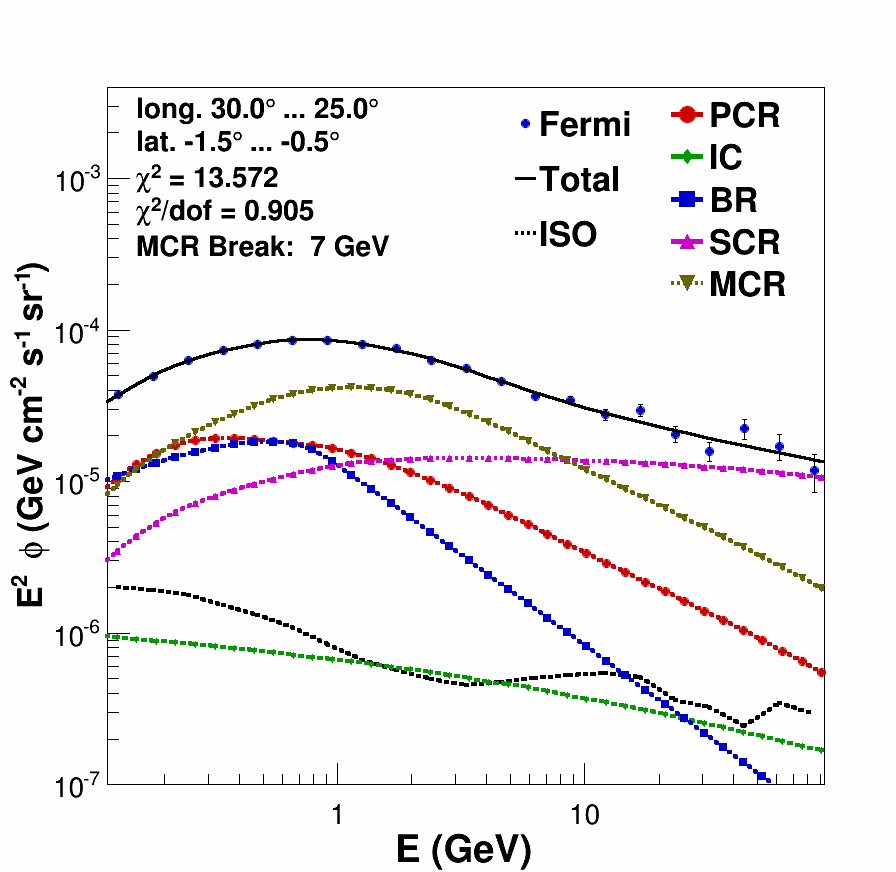}
	\includegraphics[width=0.16\textwidth,height=0.16\textwidth,clip]{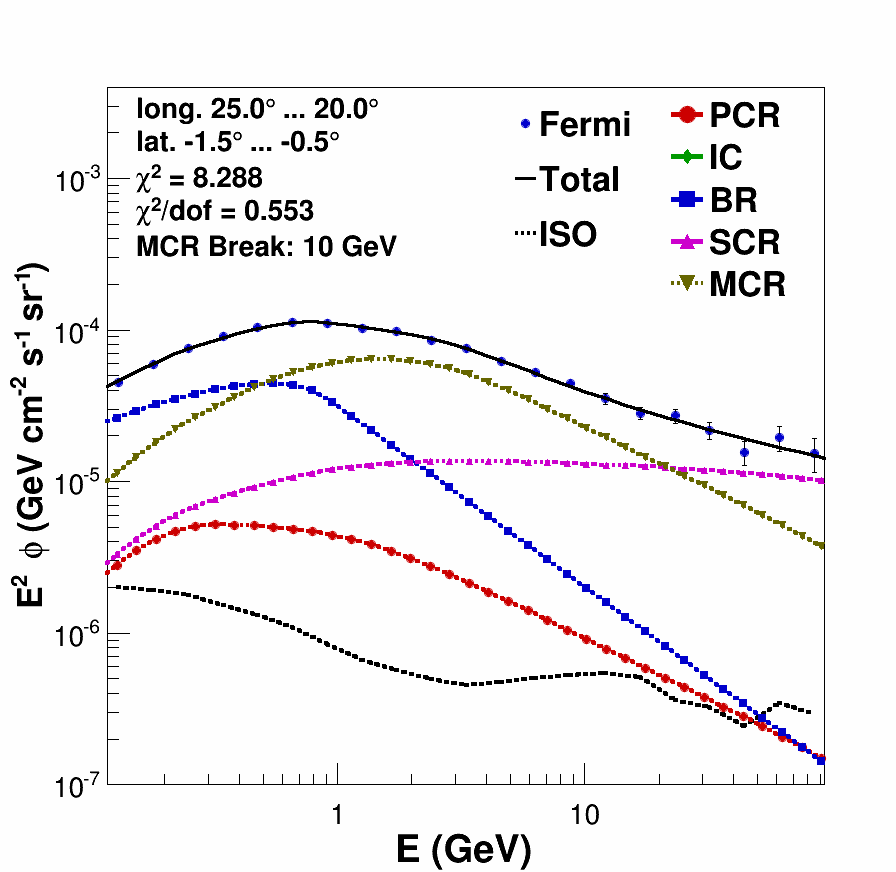}
	\includegraphics[width=0.16\textwidth,height=0.16\textwidth,clip]{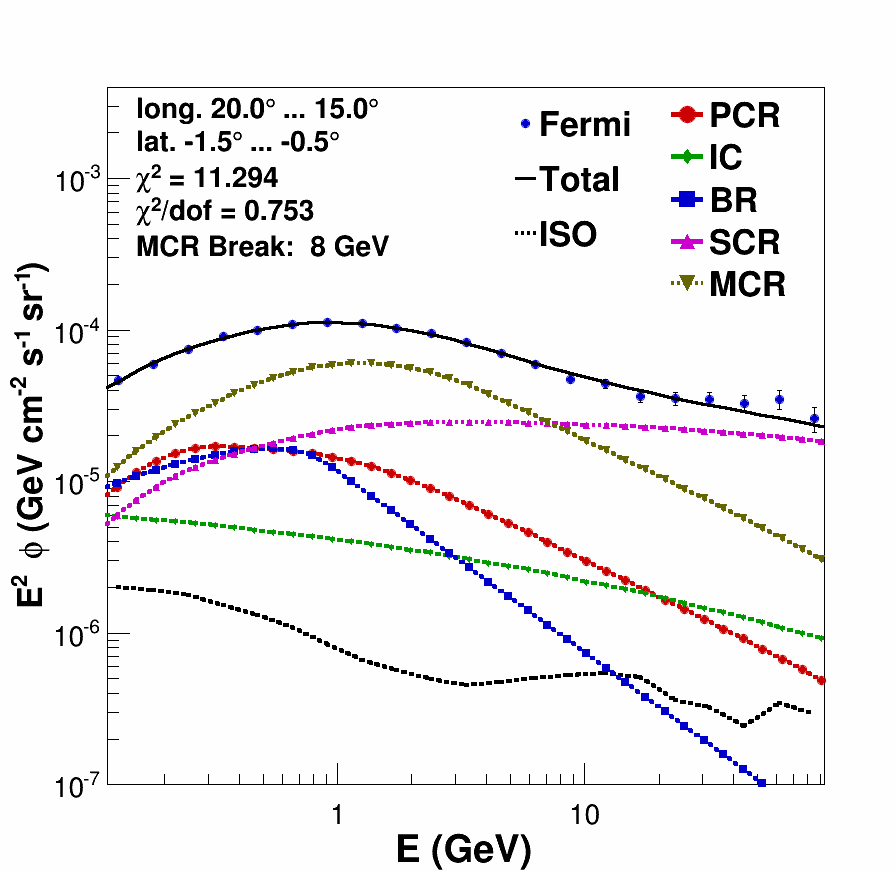}
	\includegraphics[width=0.16\textwidth,height=0.16\textwidth,clip]{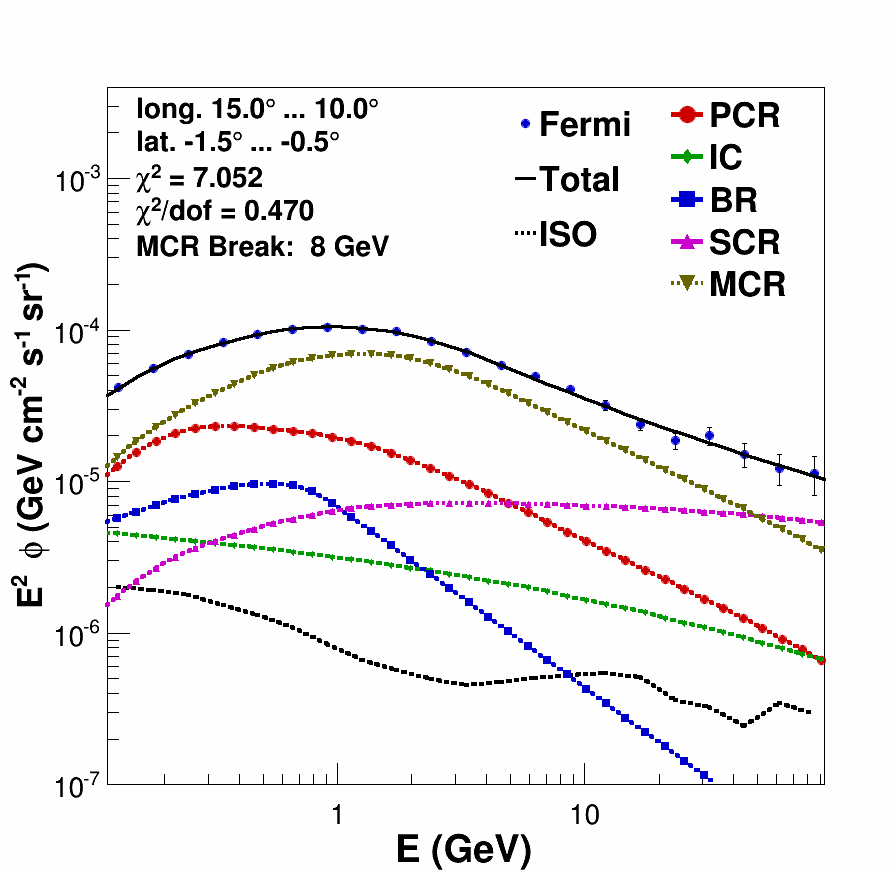}
	\includegraphics[width=0.16\textwidth,height=0.16\textwidth,clip]{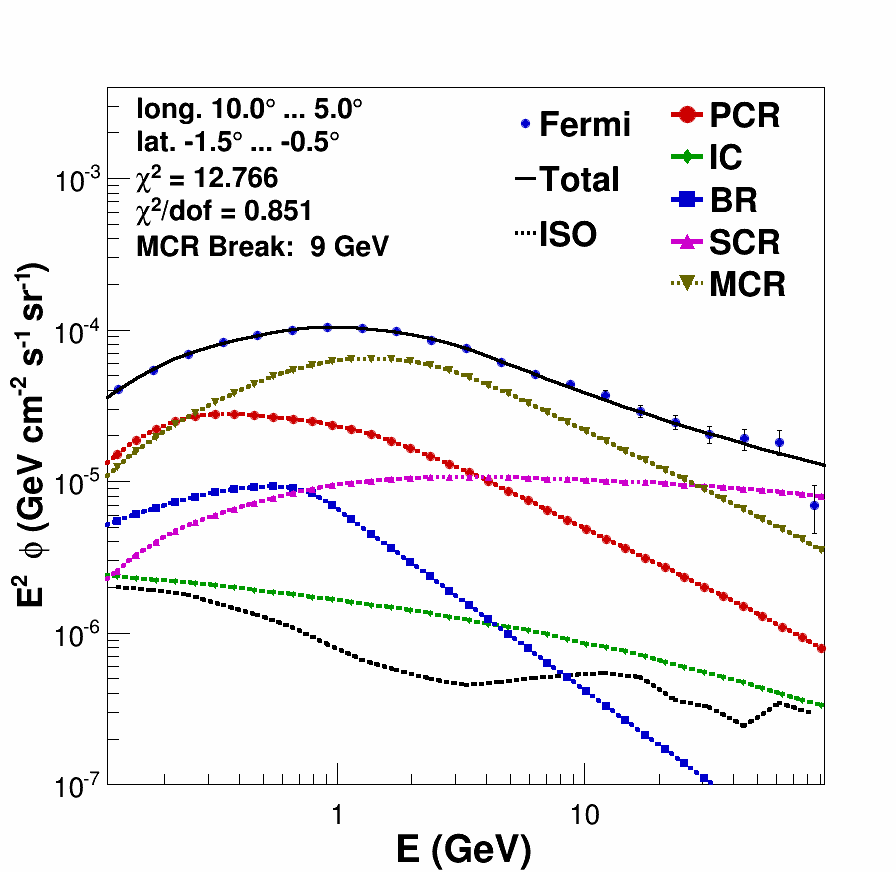}
	\includegraphics[width=0.16\textwidth,height=0.16\textwidth,clip]{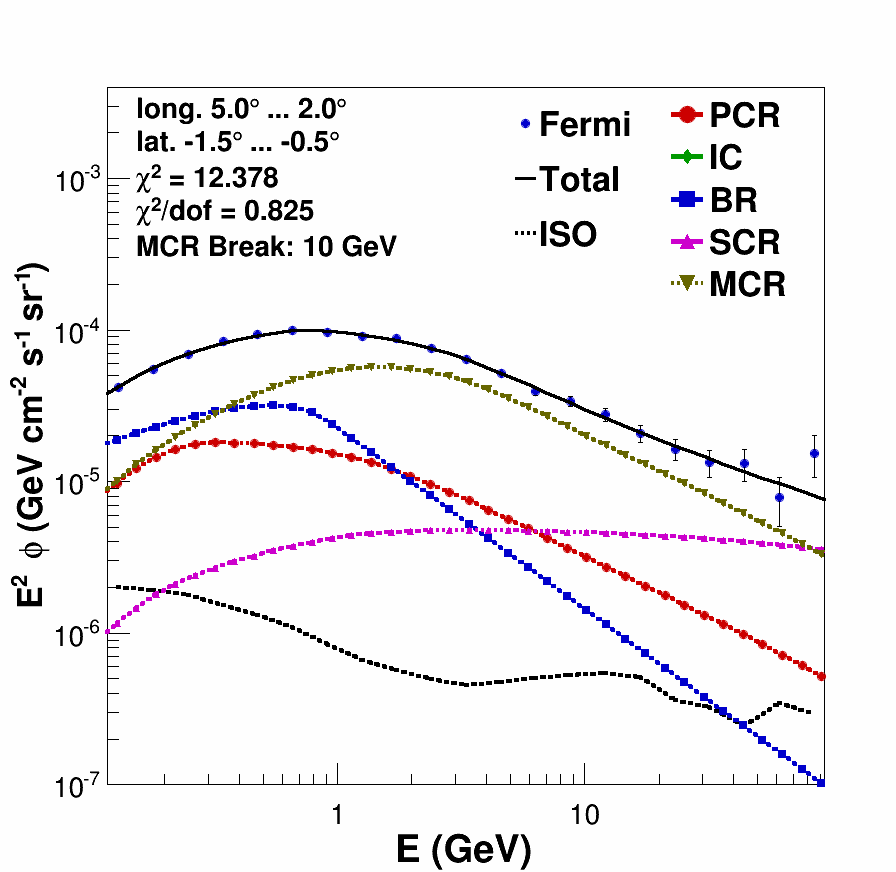}
	\includegraphics[width=0.16\textwidth,height=0.16\textwidth,clip]{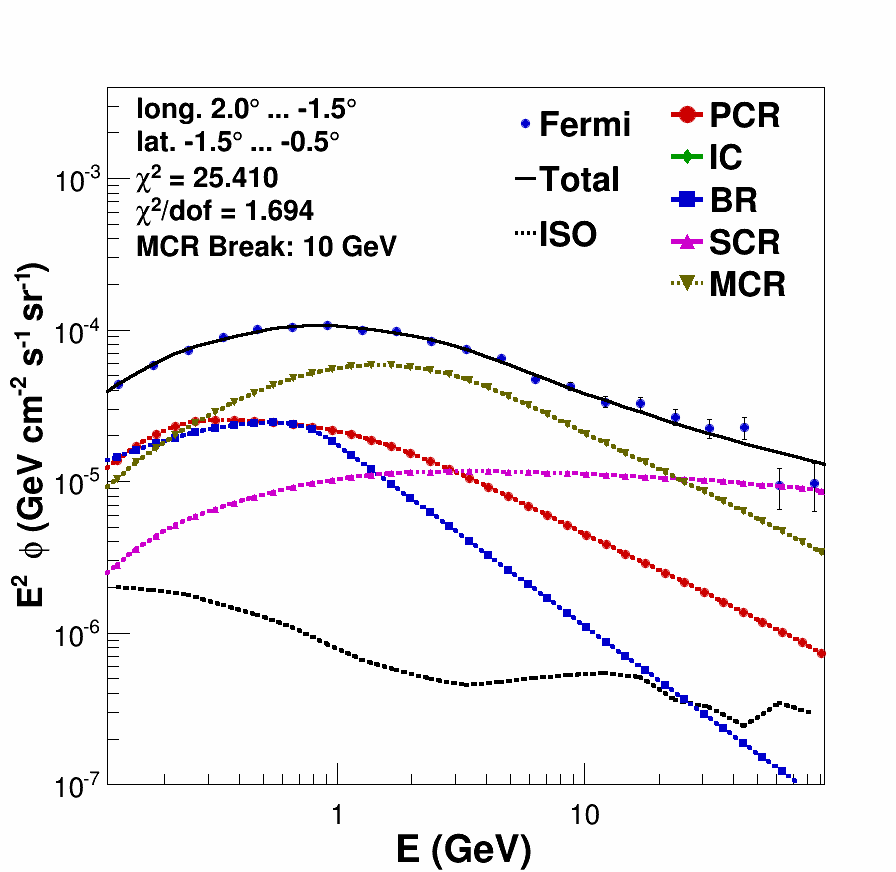}
	\includegraphics[width=0.16\textwidth,height=0.16\textwidth,clip]{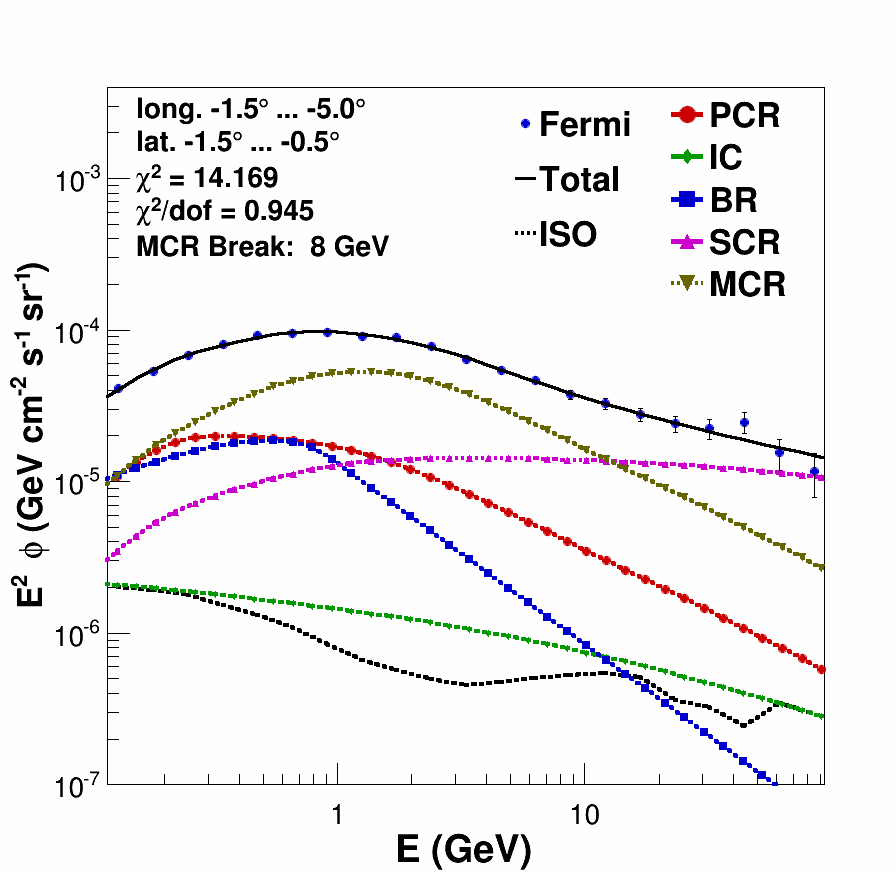}
	\includegraphics[width=0.16\textwidth,height=0.16\textwidth,clip]{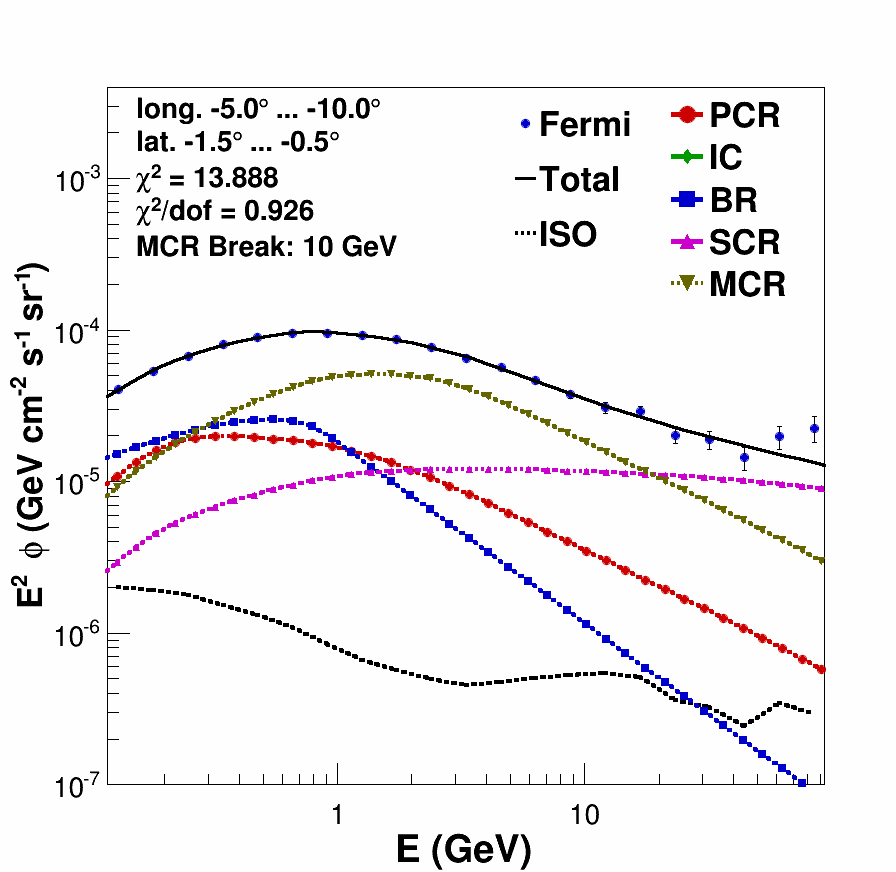}
	\includegraphics[width=0.16\textwidth,height=0.16\textwidth,clip]{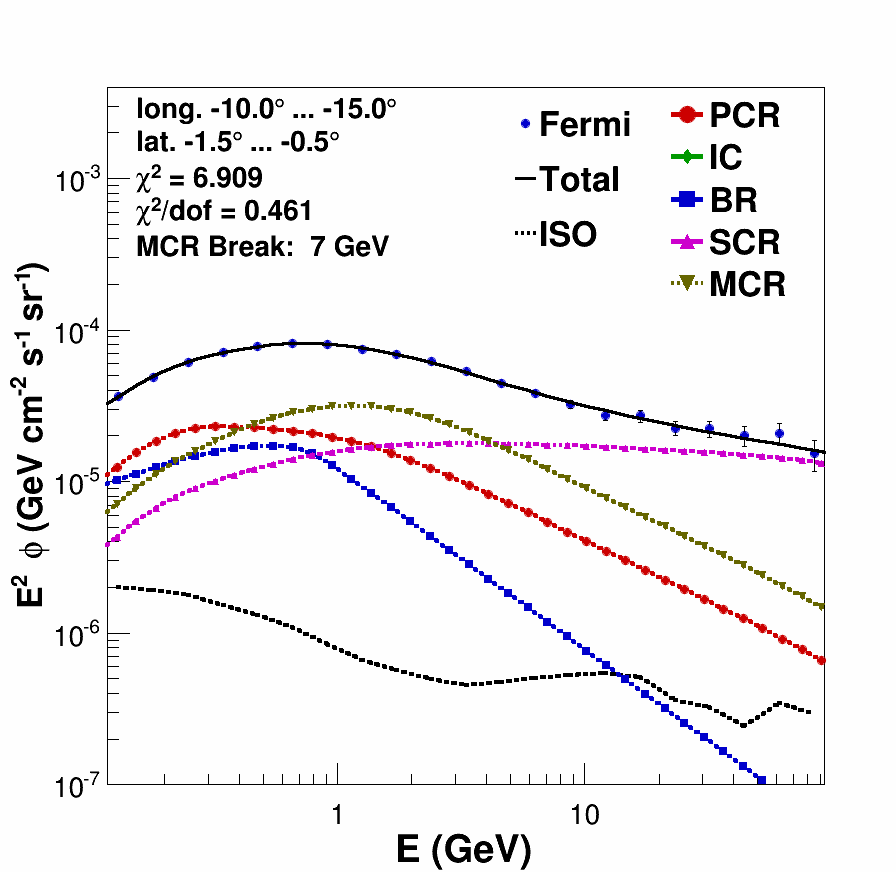}
	\includegraphics[width=0.16\textwidth,height=0.16\textwidth,clip]{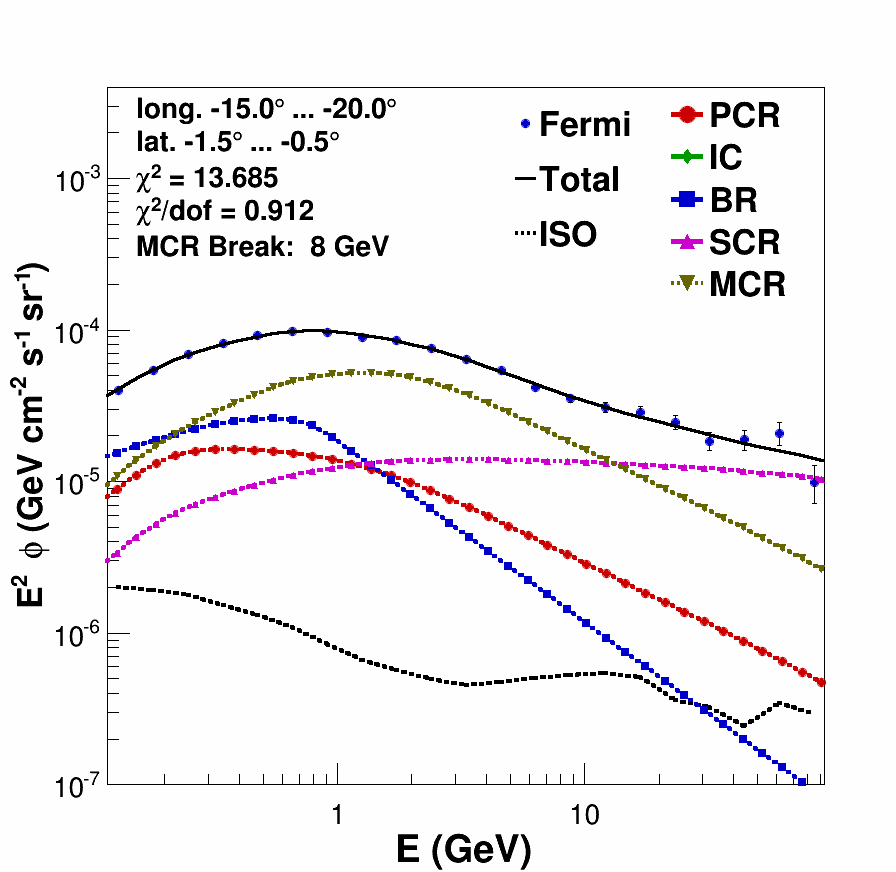}
	\includegraphics[width=0.16\textwidth,height=0.16\textwidth,clip]{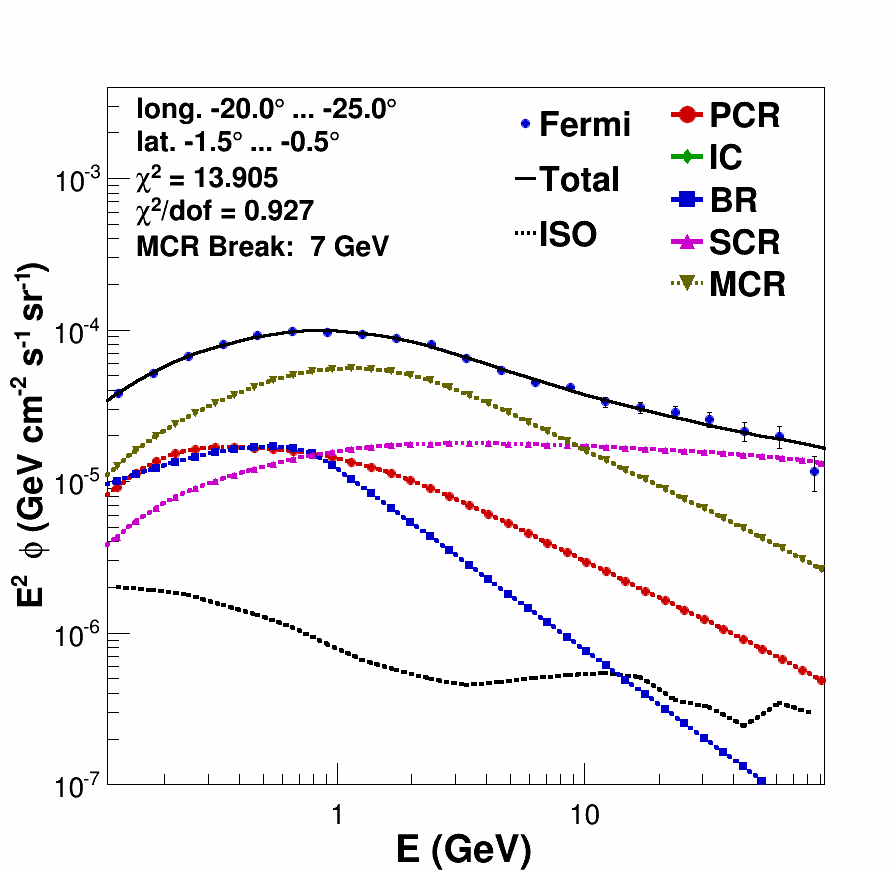}
	\includegraphics[width=0.16\textwidth,height=0.16\textwidth,clip]{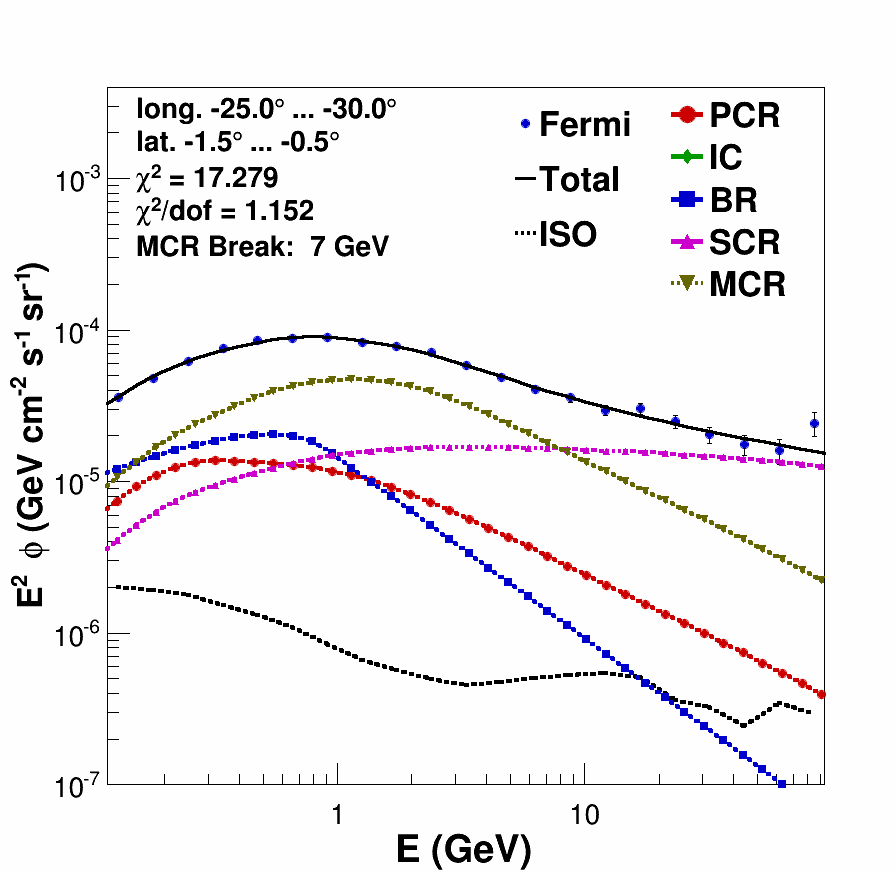}
	\includegraphics[width=0.16\textwidth,height=0.16\textwidth,clip]{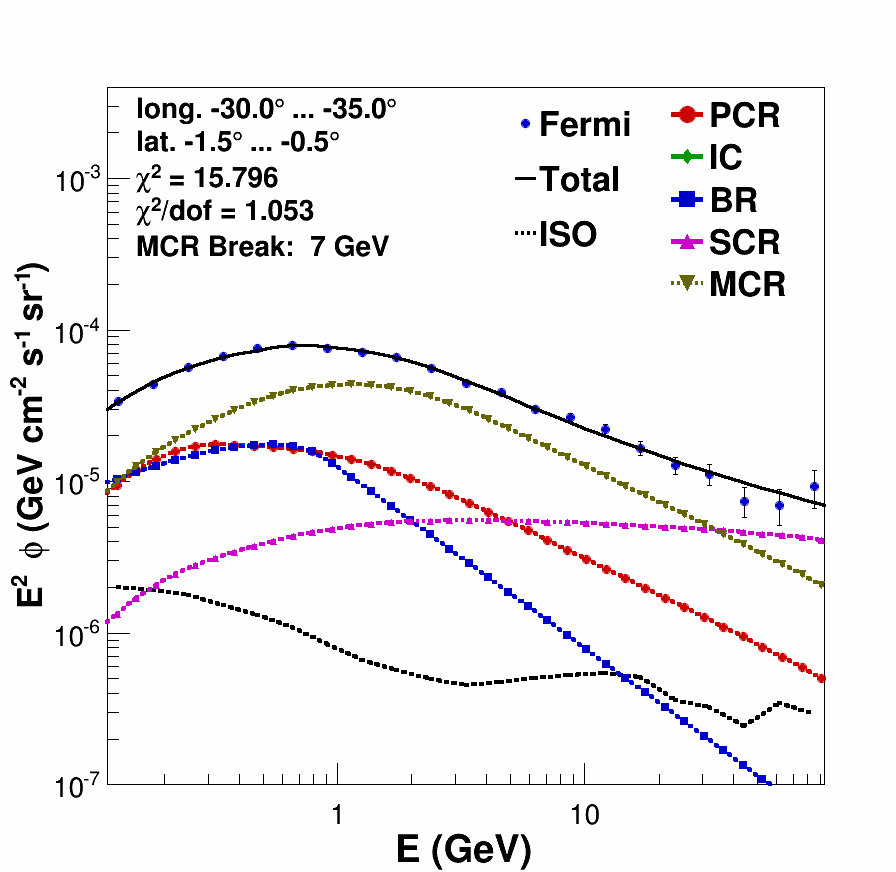}
	\includegraphics[width=0.16\textwidth,height=0.16\textwidth,clip]{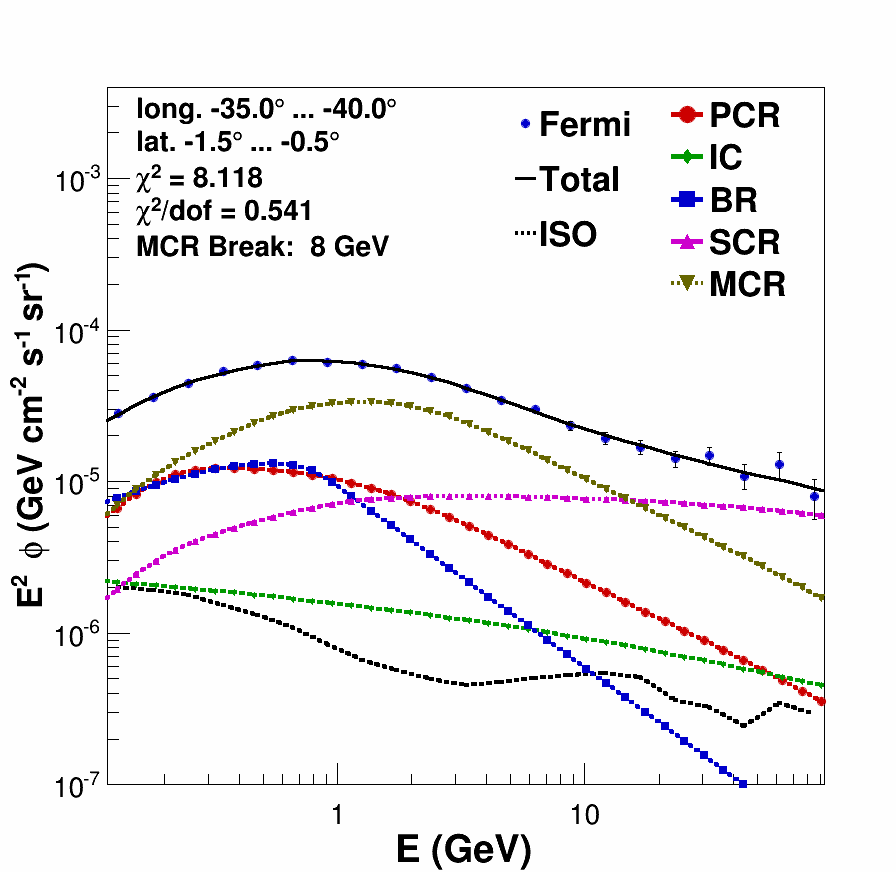}
	\includegraphics[width=0.16\textwidth,height=0.16\textwidth,clip]{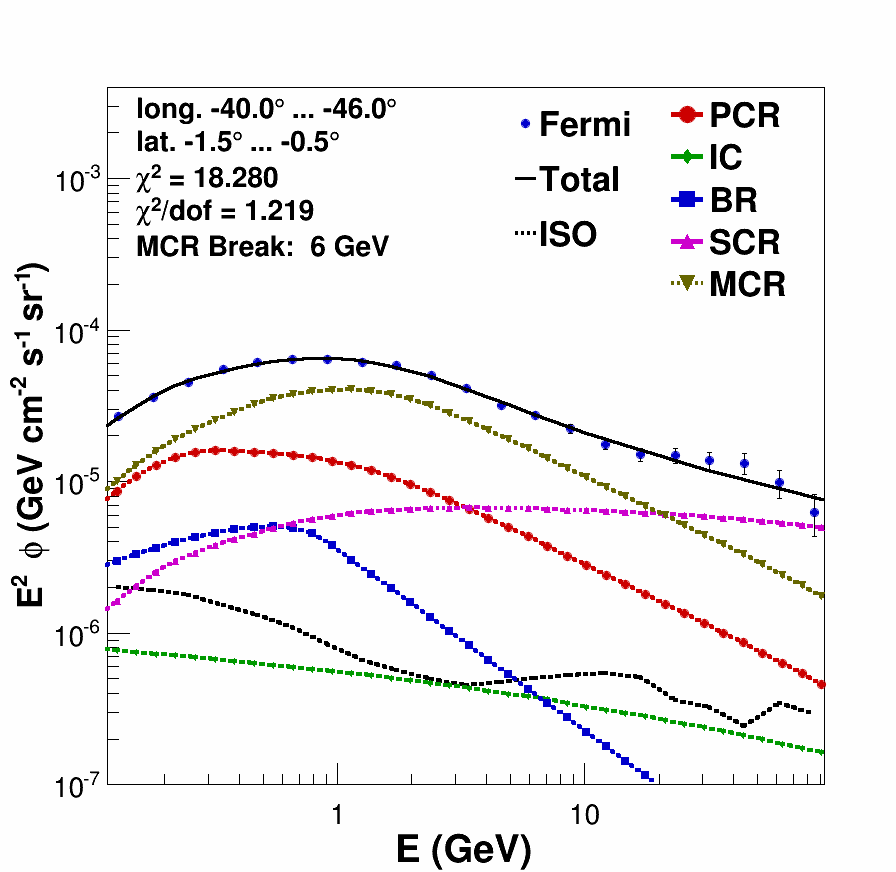}
	\includegraphics[width=0.16\textwidth,height=0.16\textwidth,clip]{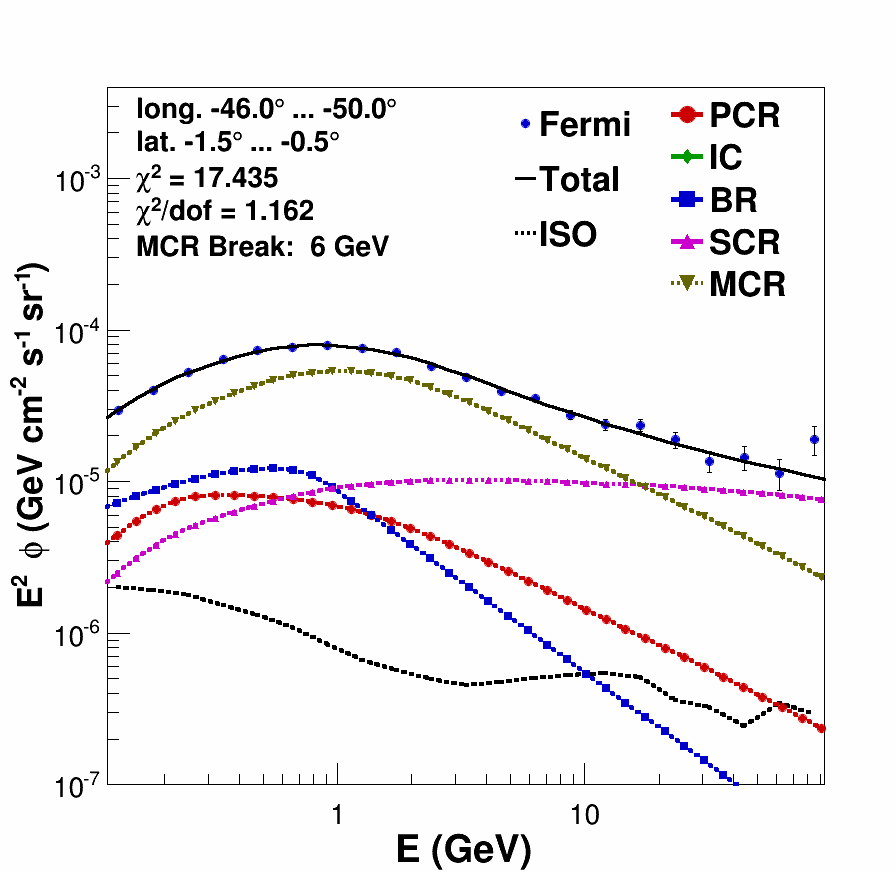}
	\includegraphics[width=0.16\textwidth,height=0.16\textwidth,clip]{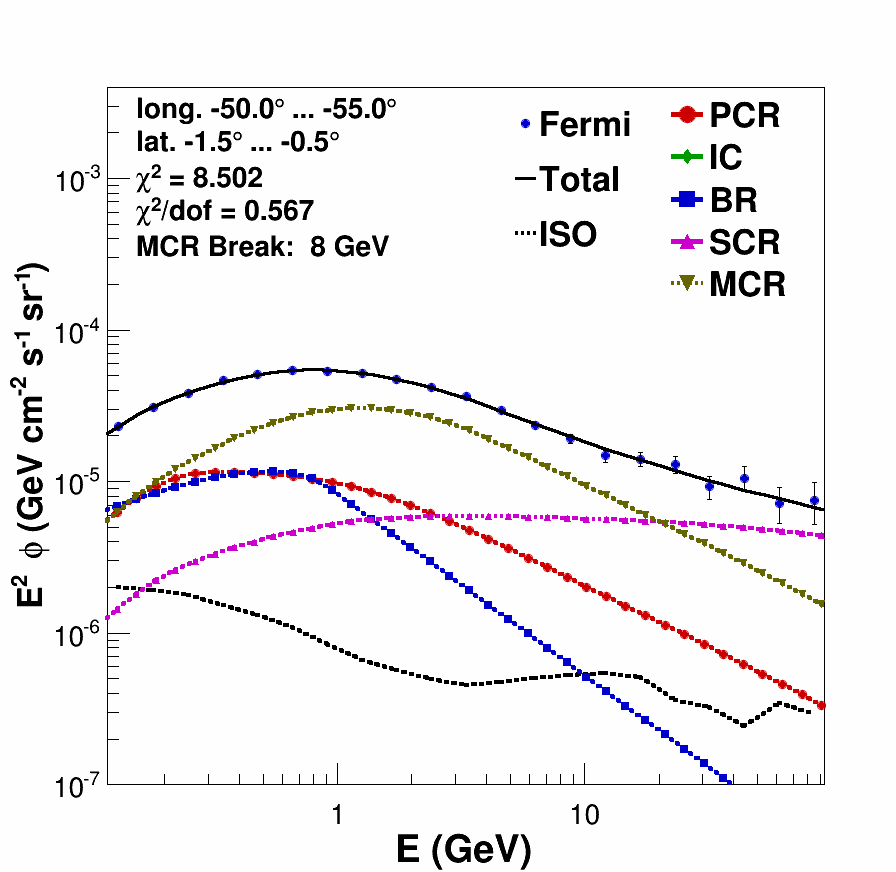}
	\includegraphics[width=0.16\textwidth,height=0.16\textwidth,clip]{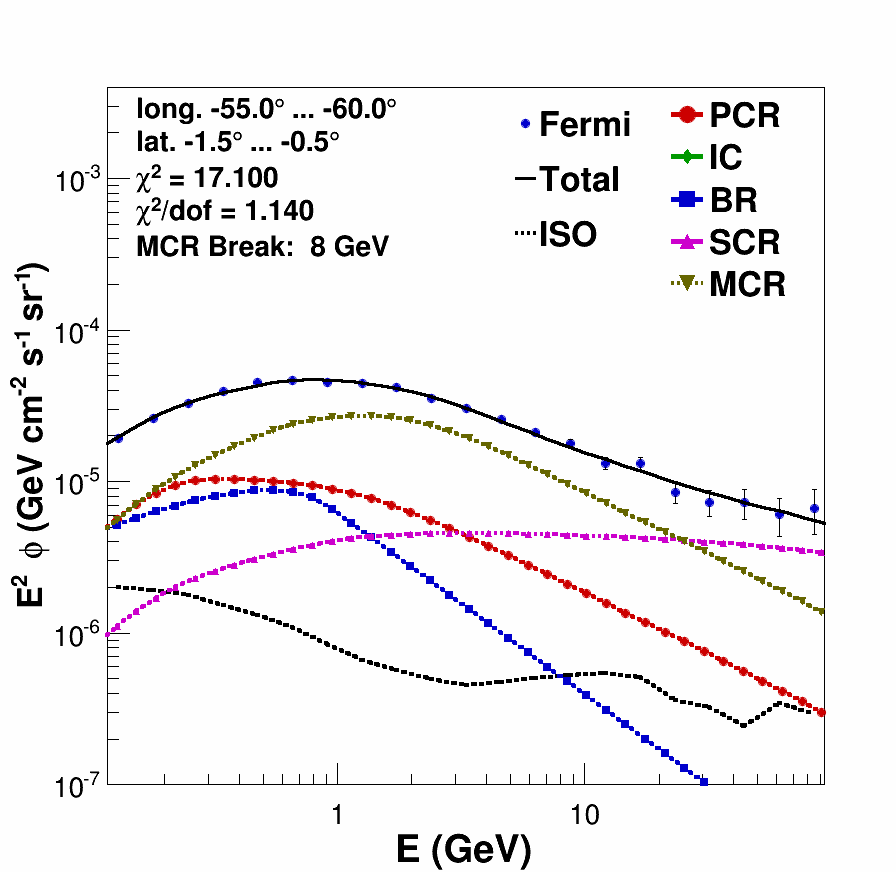}
	\includegraphics[width=0.16\textwidth,height=0.16\textwidth,clip]{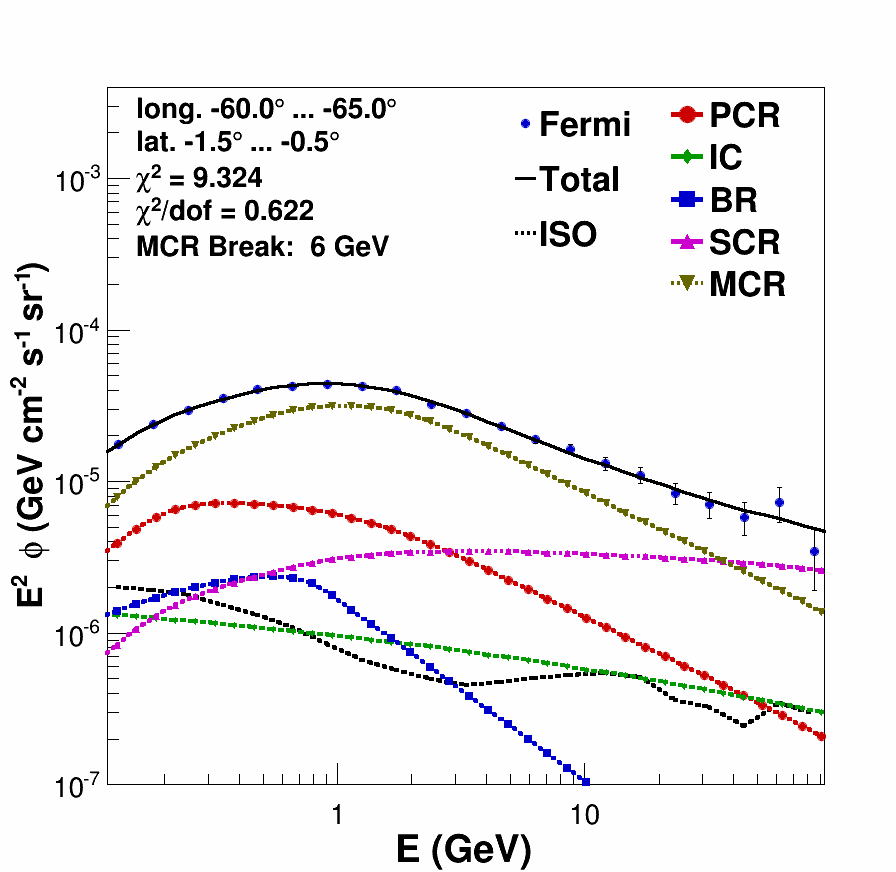}
	\includegraphics[width=0.16\textwidth,height=0.16\textwidth,clip]{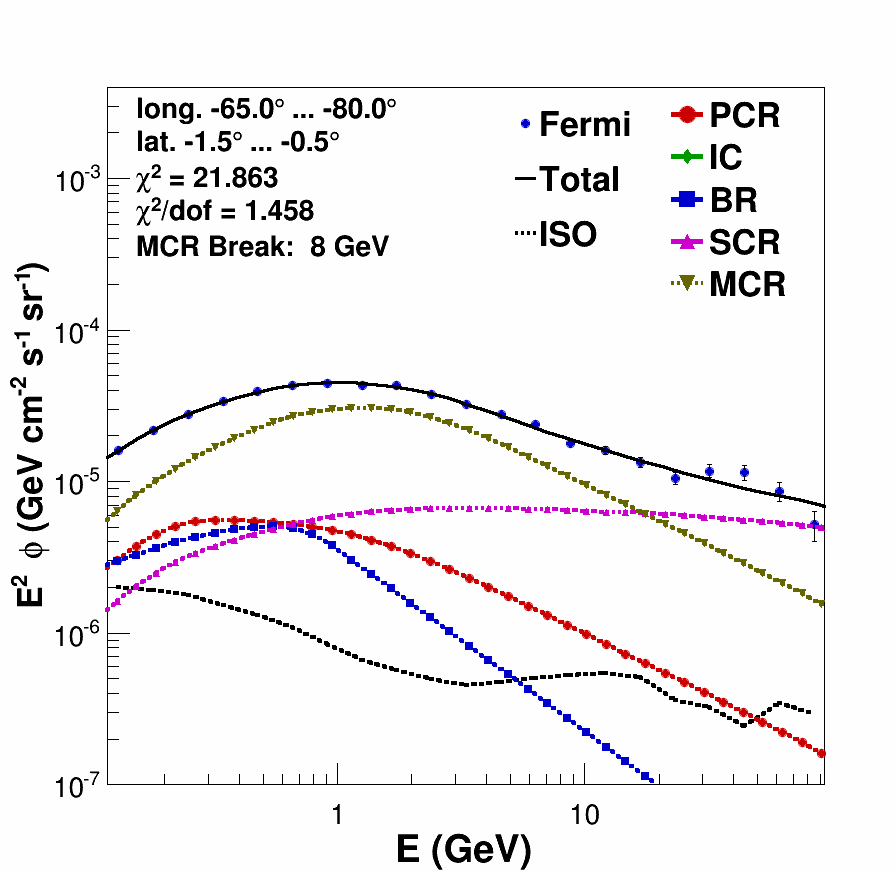}
	\includegraphics[width=0.16\textwidth,height=0.16\textwidth,clip]{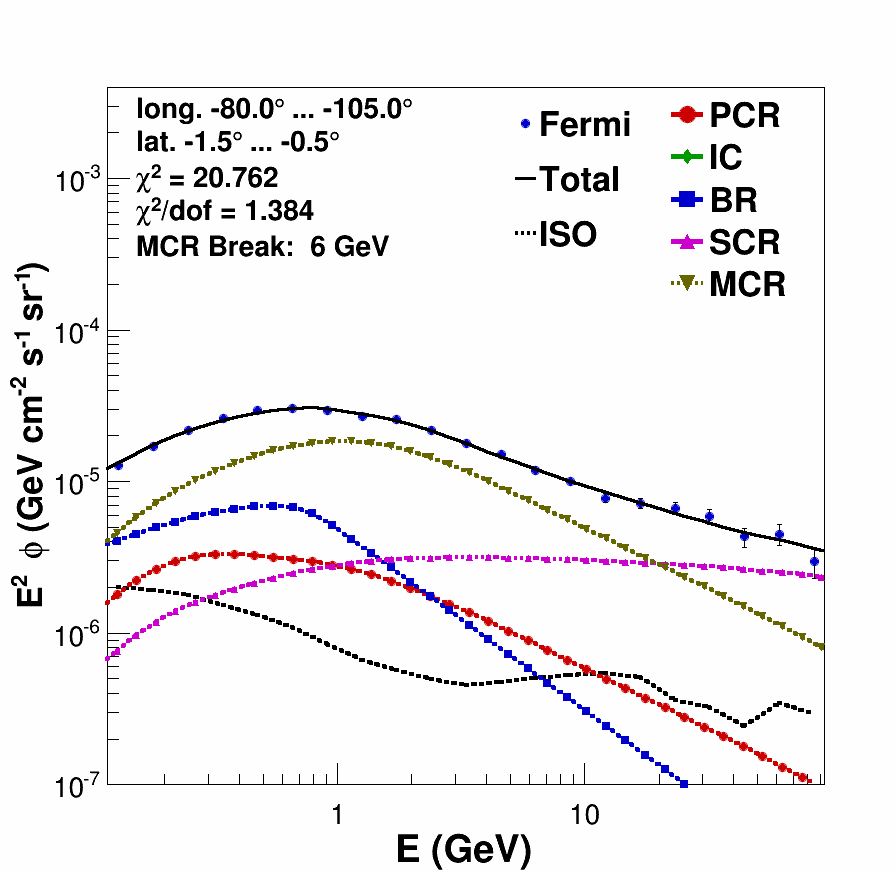}
	\includegraphics[width=0.16\textwidth,height=0.16\textwidth,clip]{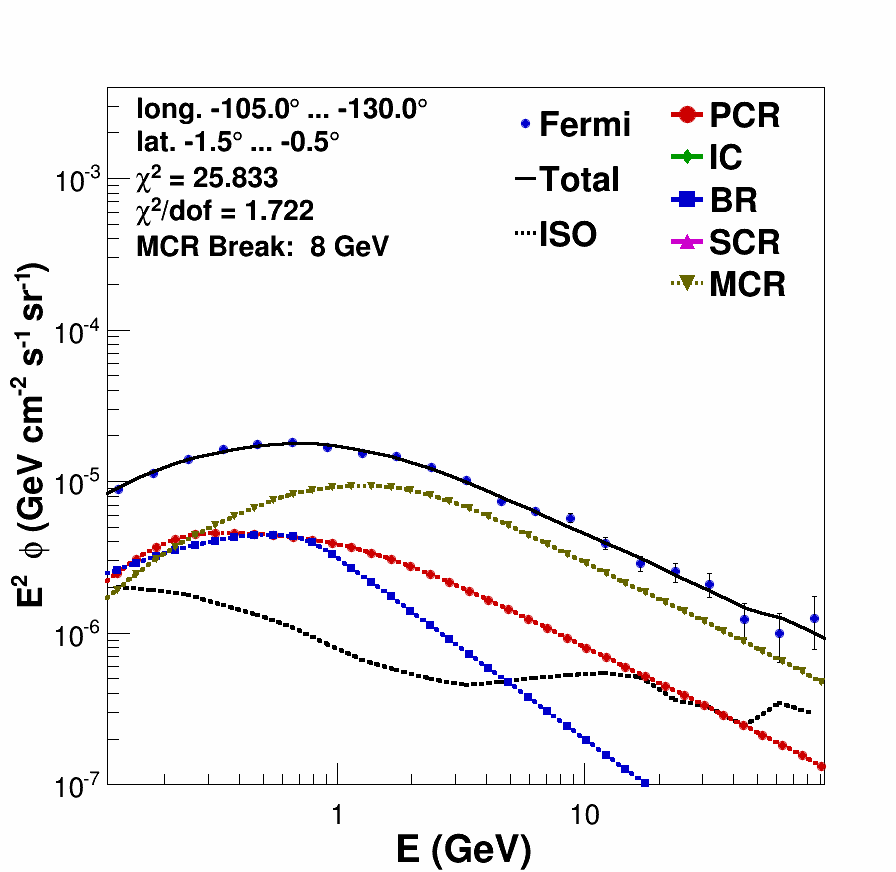}
38	\includegraphics[width=0.16\textwidth,height=0.16\textwidth,clip]{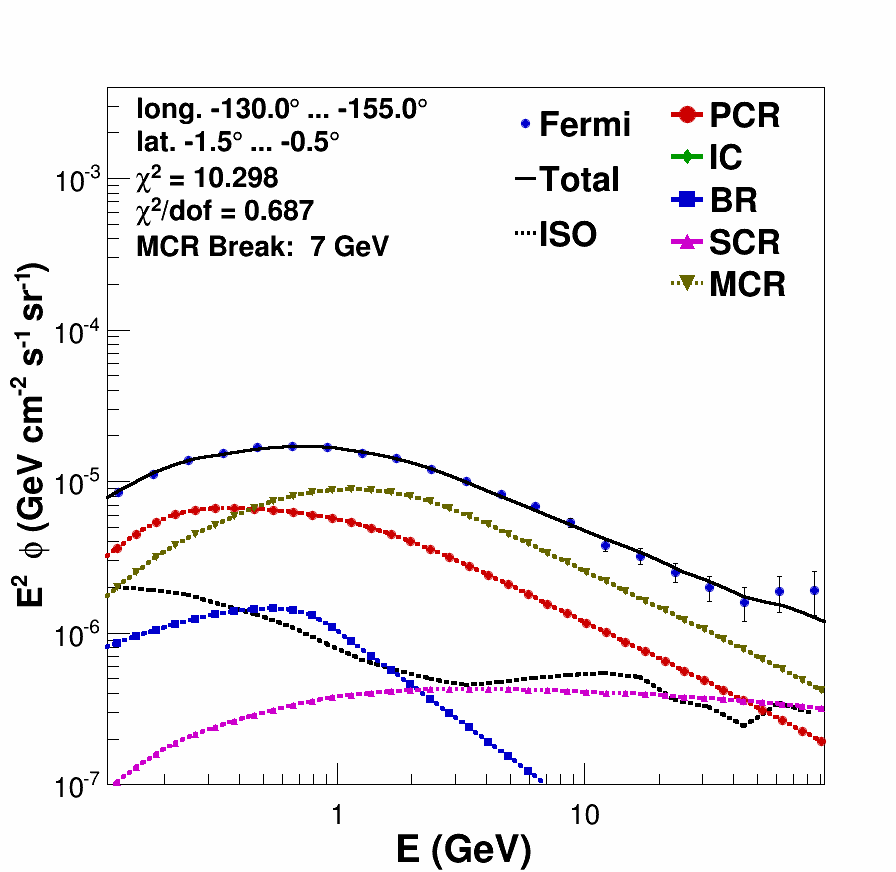}
39	\includegraphics[width=0.16\textwidth,height=0.16\textwidth,clip]{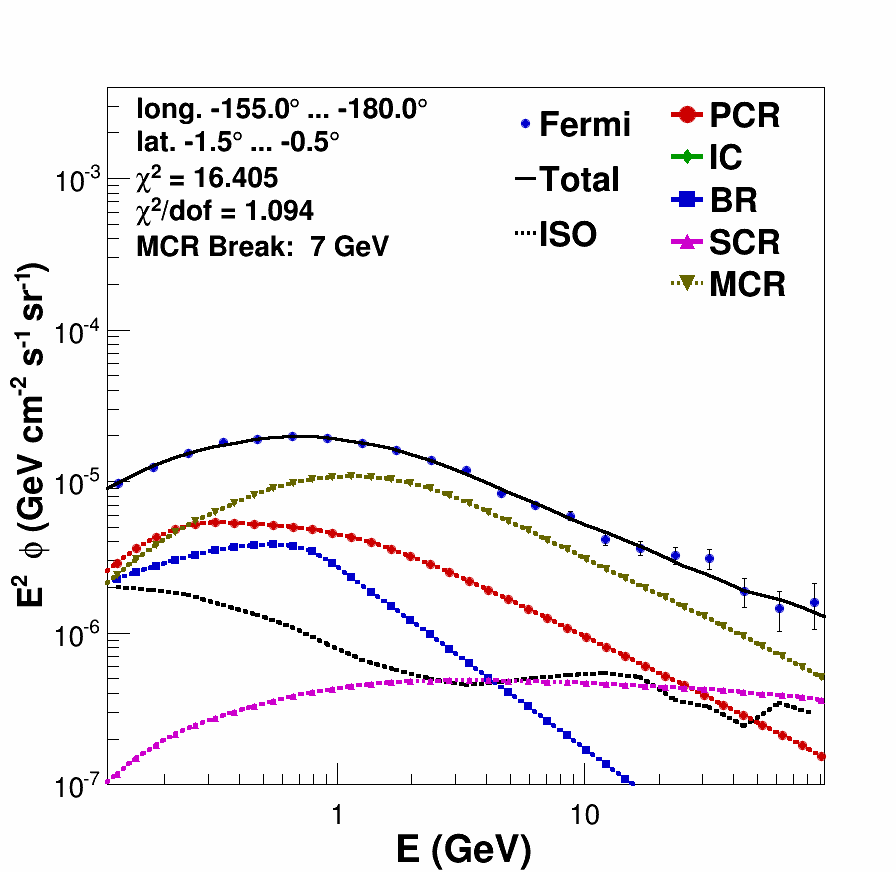}%%%%%r11
\caption[]{Template fits for latitudes  with $-1.5^\circ<b<-0.5^\circ$ and longitudes decreasing from 180$^\circ$ to -180$^\circ$.} \label{F22}
\end{figure}
\begin{figure}
\centering
\includegraphics[width=0.16\textwidth,height=0.16\textwidth,clip]{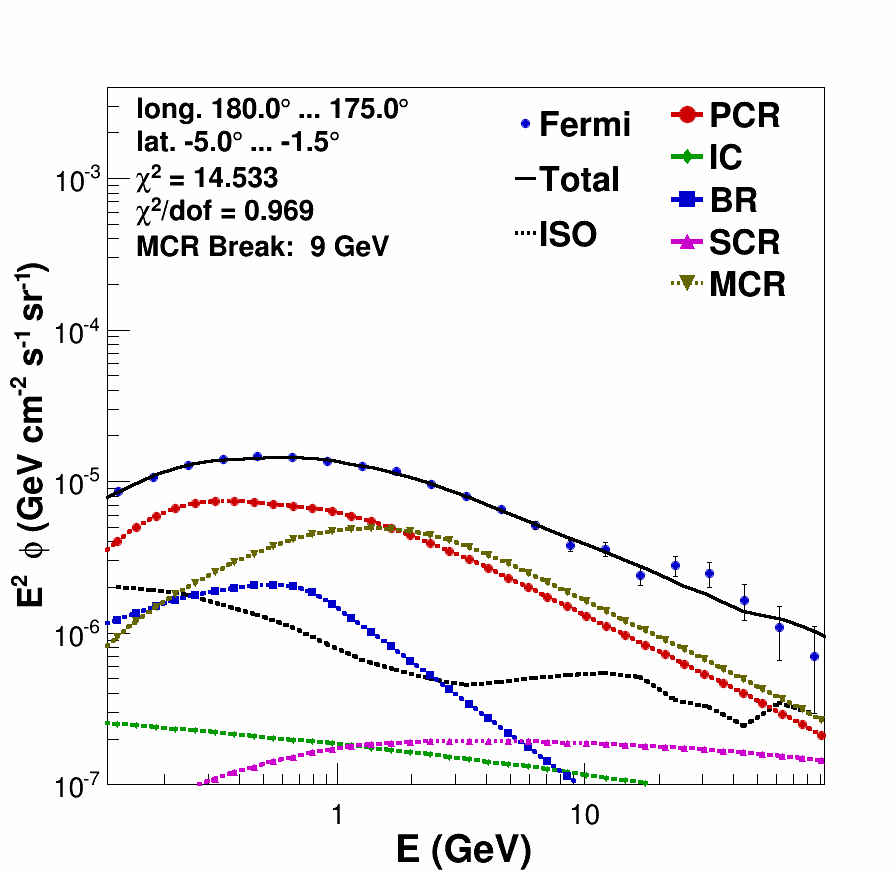}
\includegraphics[width=0.16\textwidth,height=0.16\textwidth,clip]{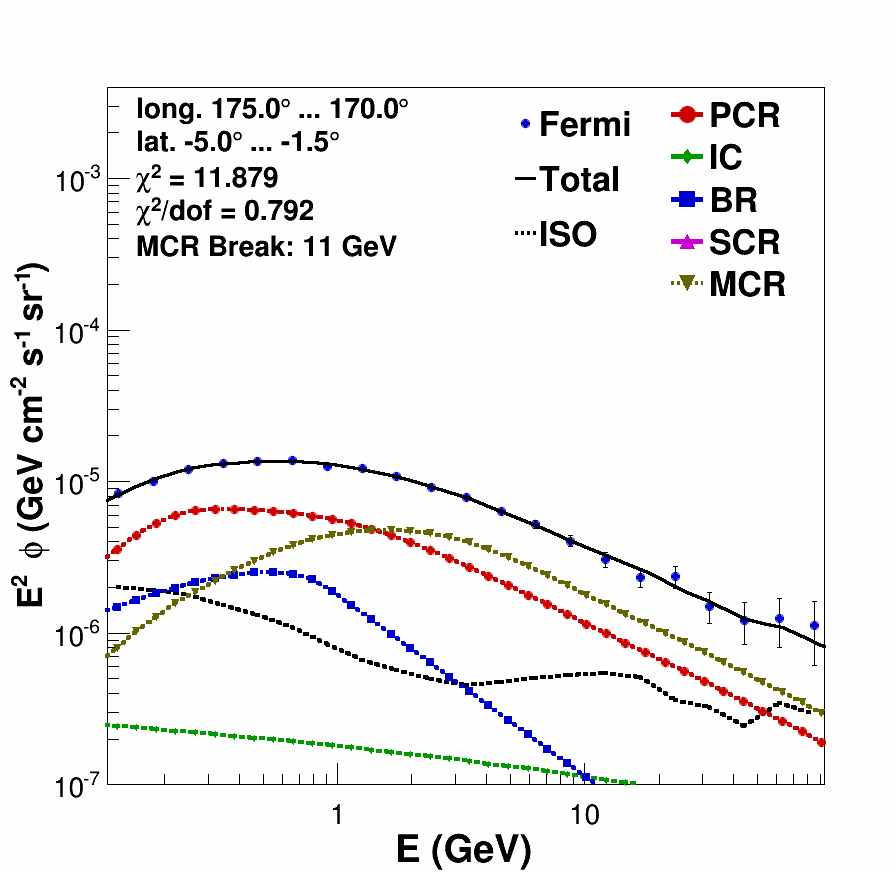}
\includegraphics[width=0.16\textwidth,height=0.16\textwidth,clip]{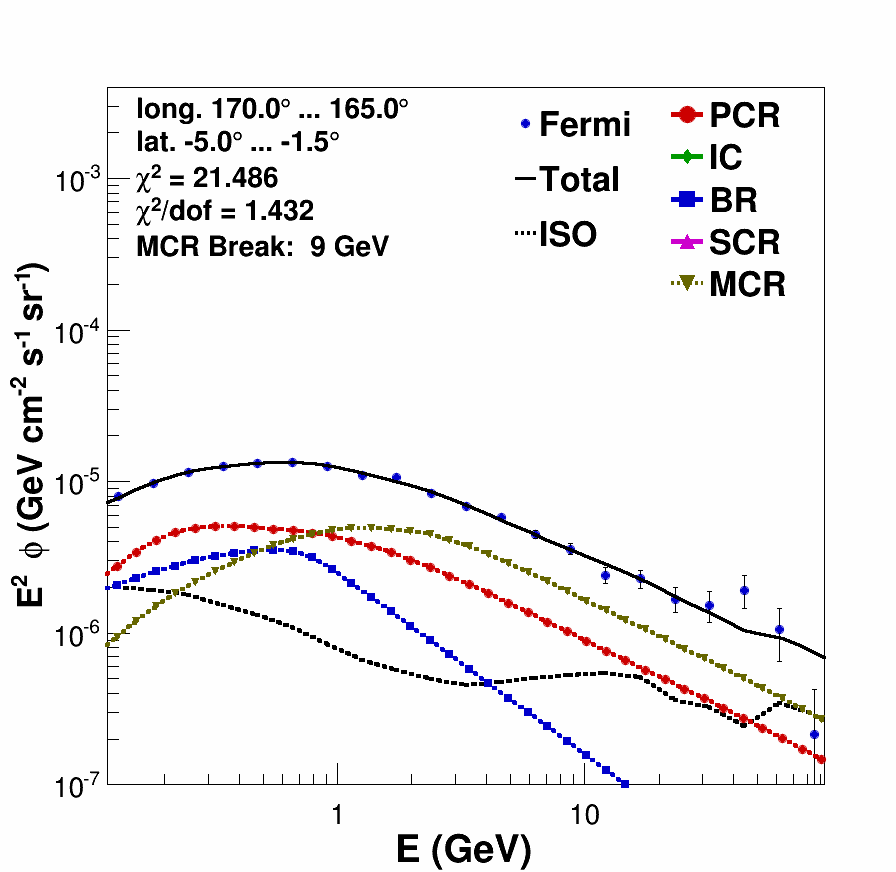}
\includegraphics[width=0.16\textwidth,height=0.16\textwidth,clip]{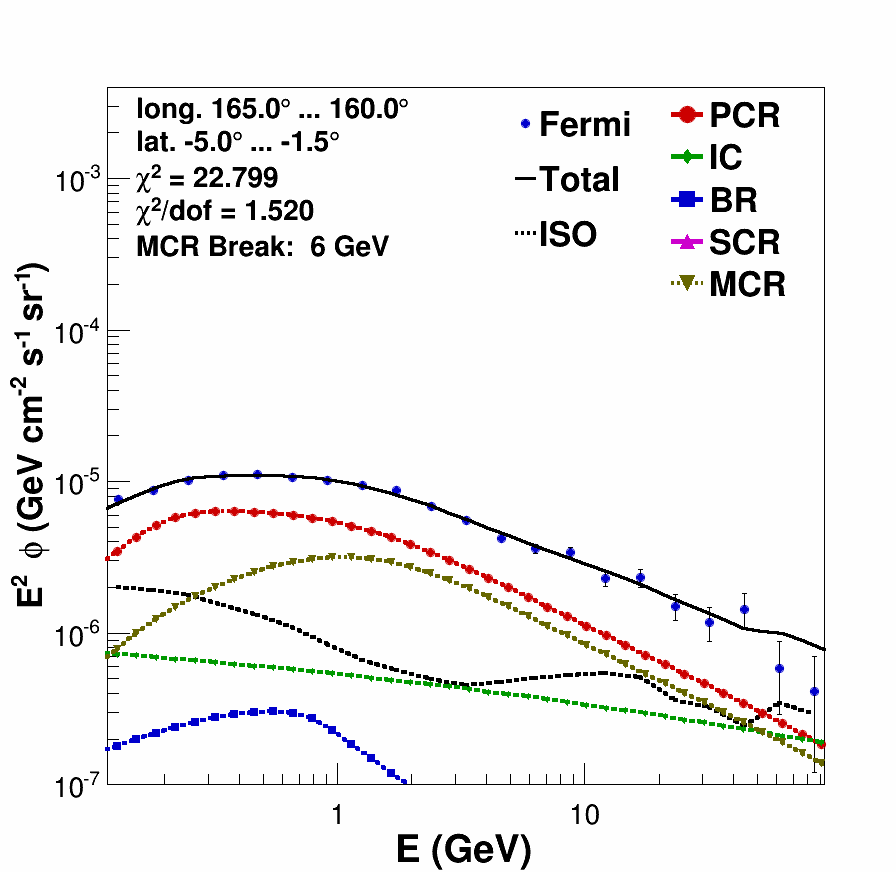}
\includegraphics[width=0.16\textwidth,height=0.16\textwidth,clip]{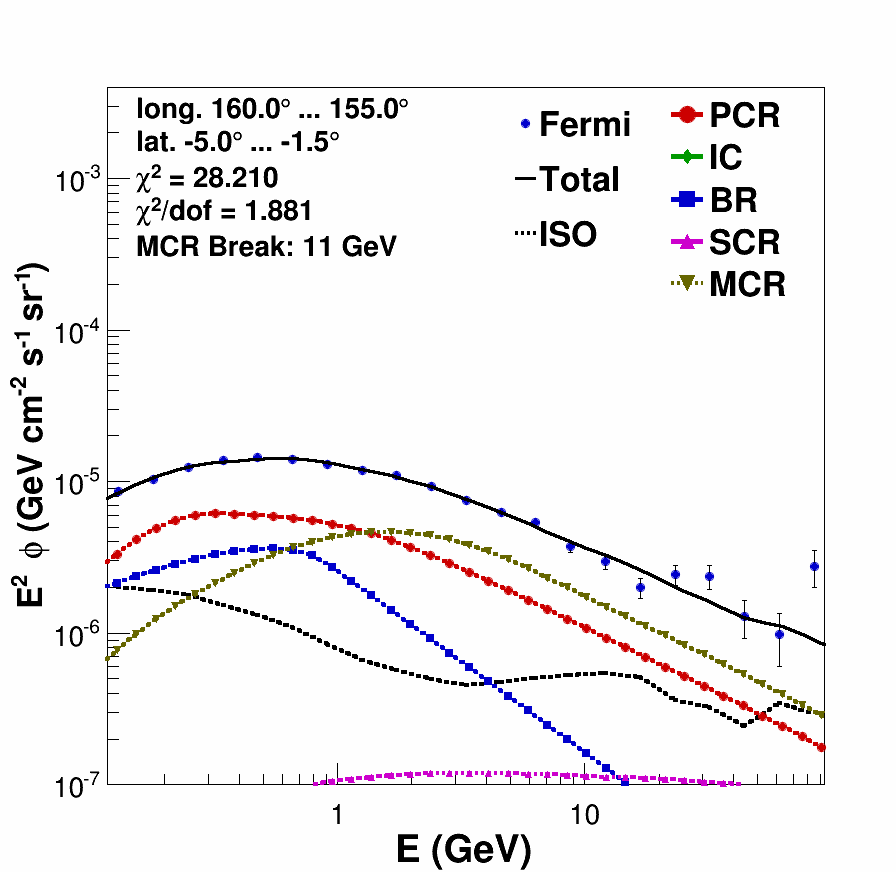}
\includegraphics[width=0.16\textwidth,height=0.16\textwidth,clip]{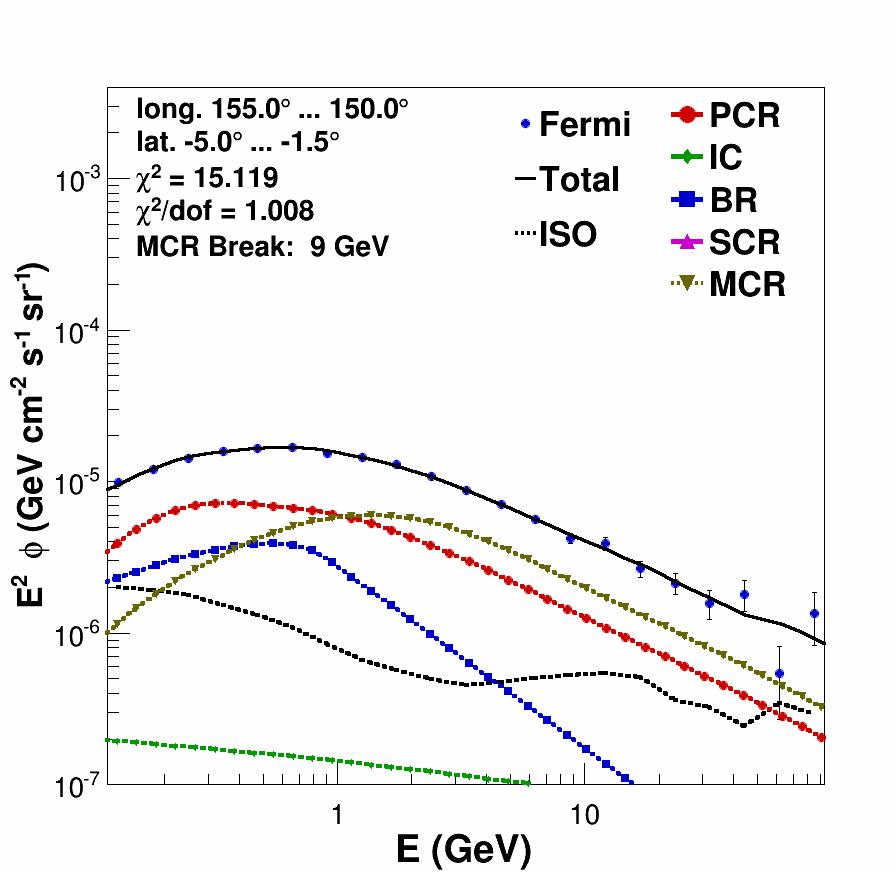}
\includegraphics[width=0.16\textwidth,height=0.16\textwidth,clip]{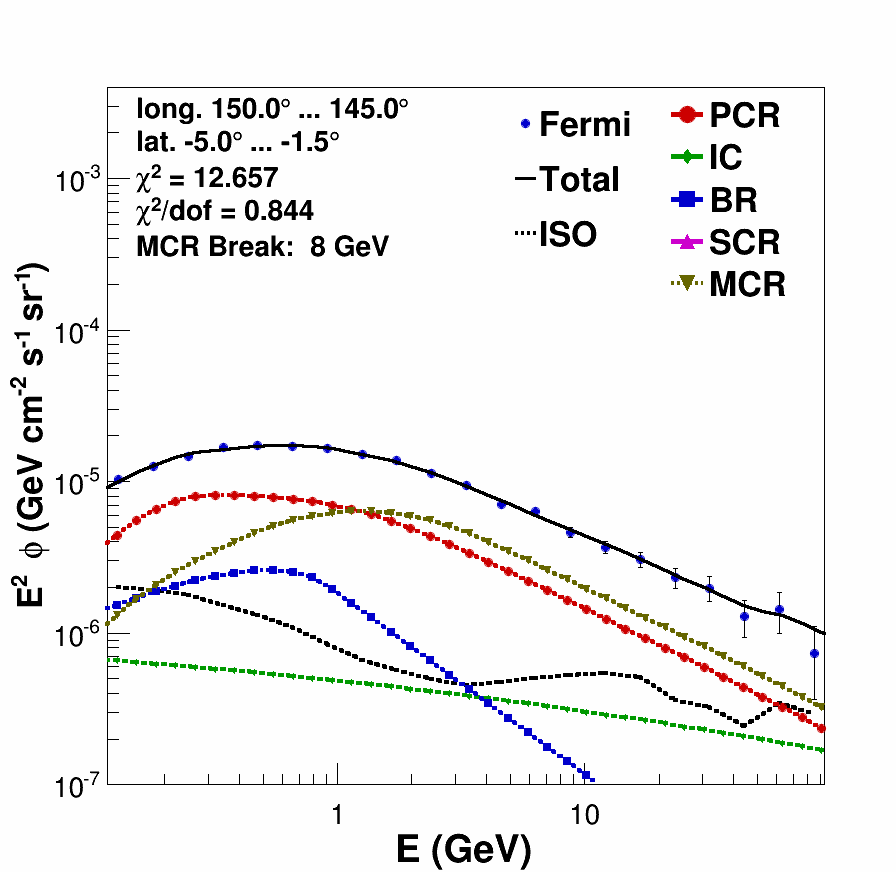}
\includegraphics[width=0.16\textwidth,height=0.16\textwidth,clip]{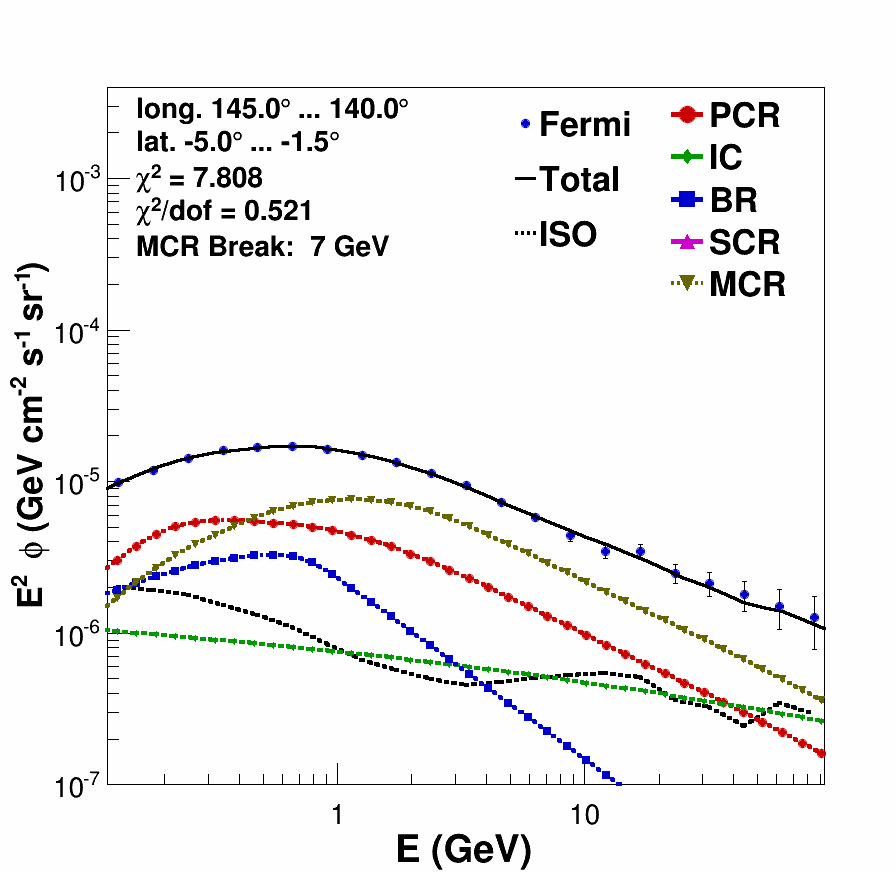}
\includegraphics[width=0.16\textwidth,height=0.16\textwidth,clip]{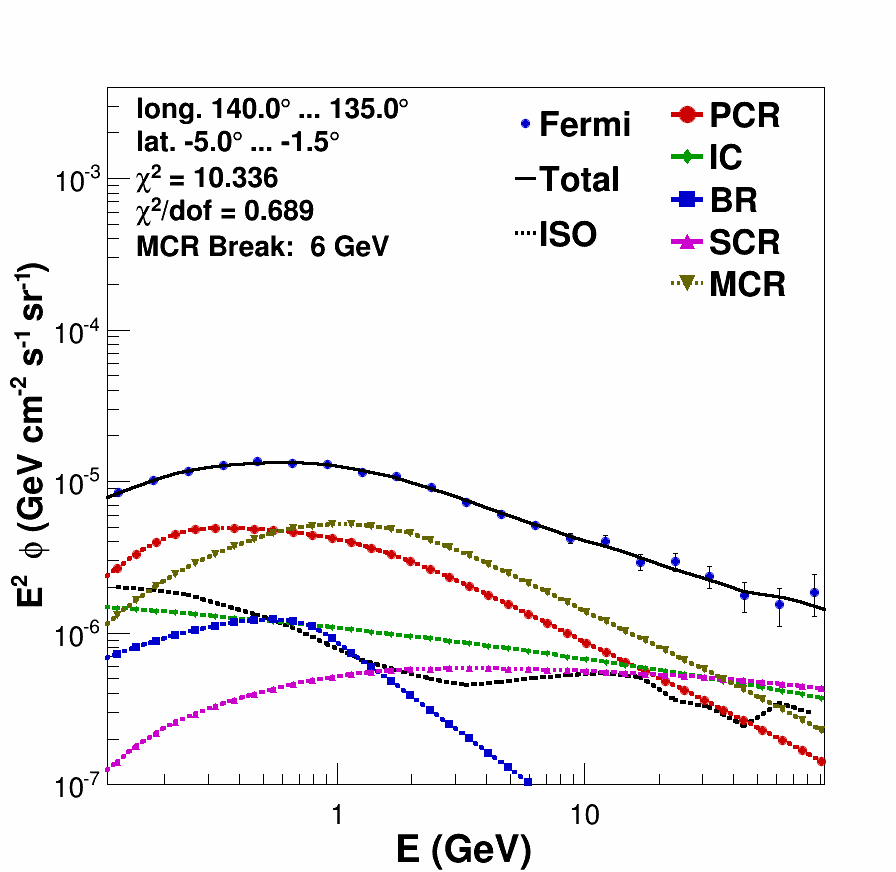}
\includegraphics[width=0.16\textwidth,height=0.16\textwidth,clip]{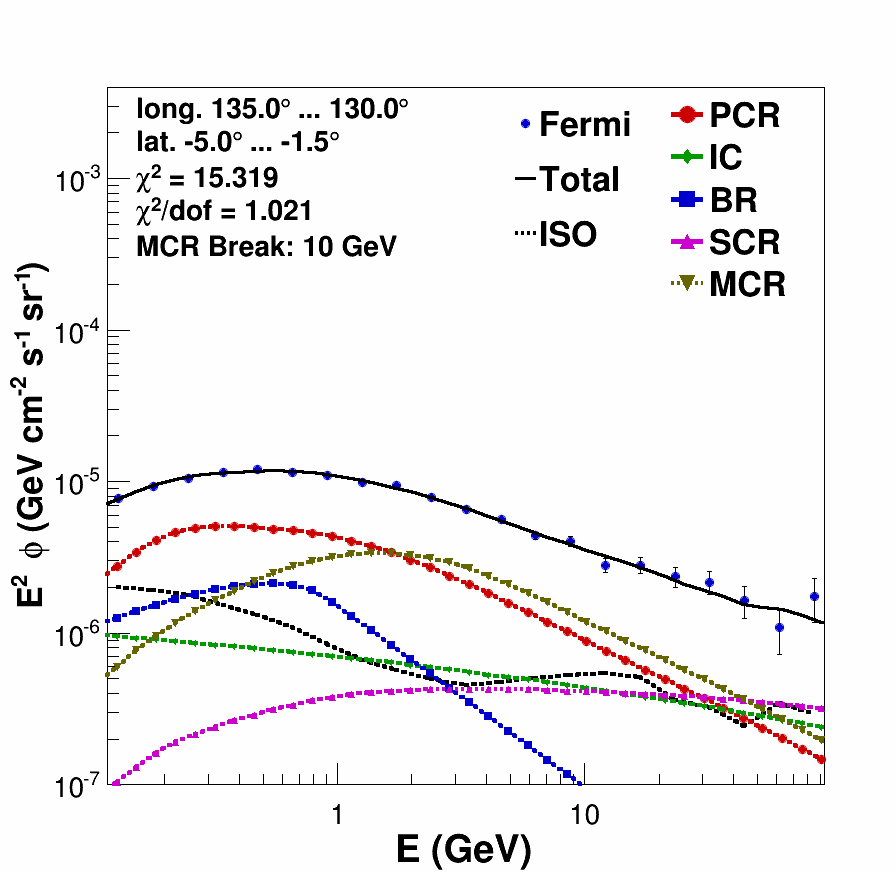}
\includegraphics[width=0.16\textwidth,height=0.16\textwidth,clip]{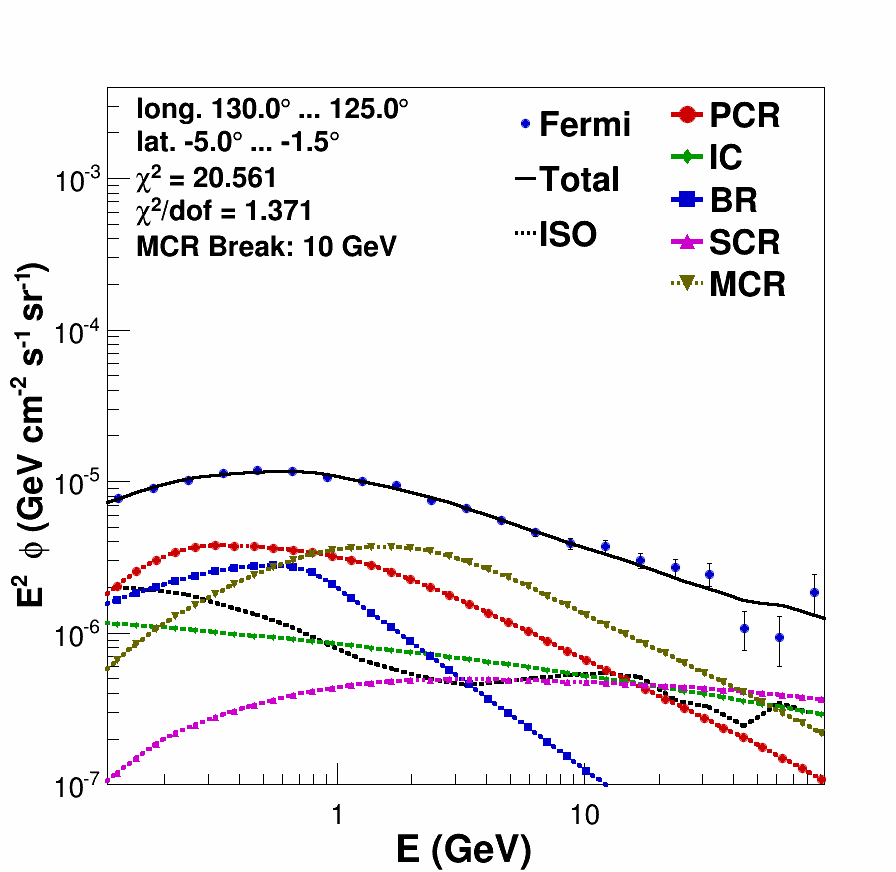}
\includegraphics[width=0.16\textwidth,height=0.16\textwidth,clip]{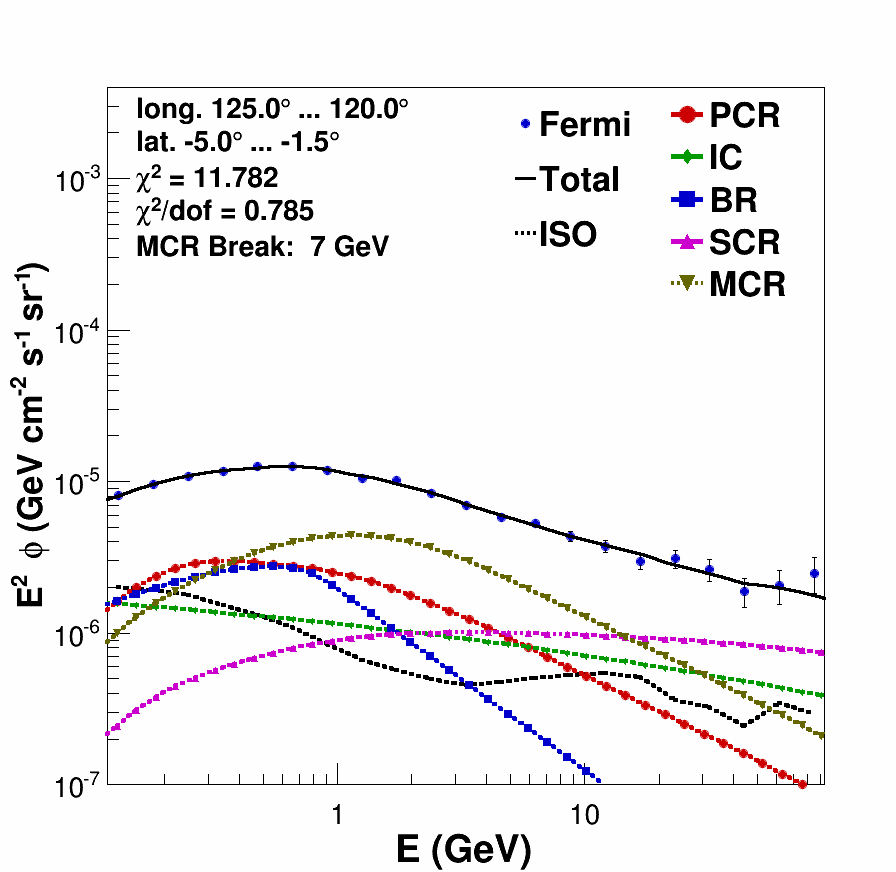}
\includegraphics[width=0.16\textwidth,height=0.16\textwidth,clip]{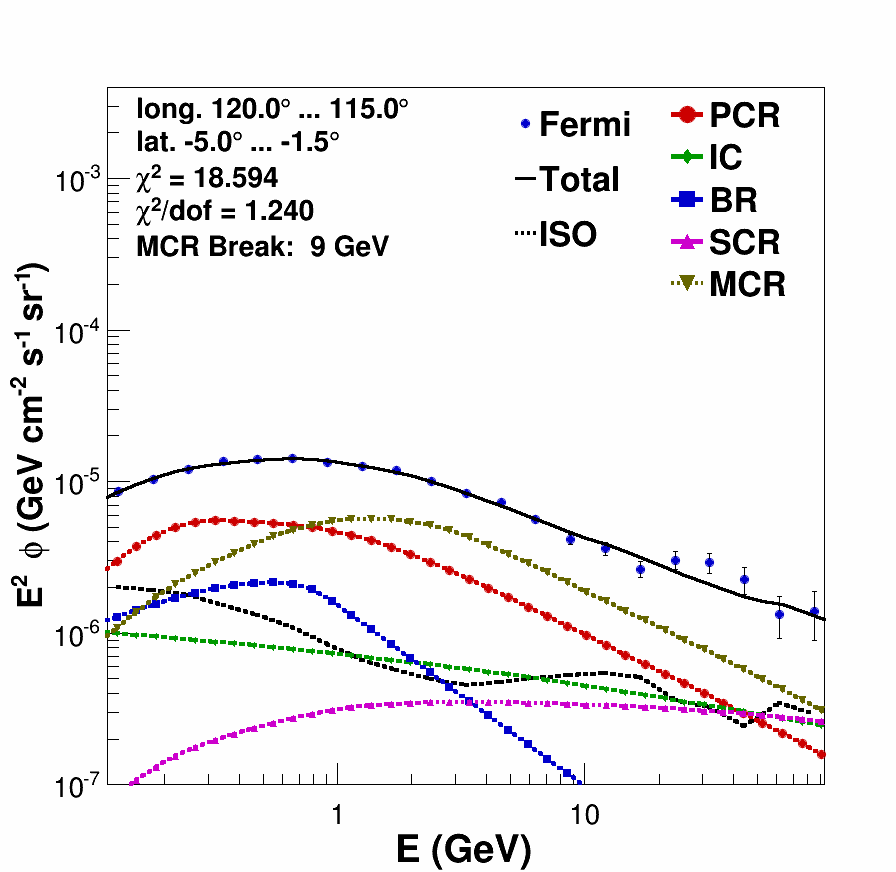}
\includegraphics[width=0.16\textwidth,height=0.16\textwidth,clip]{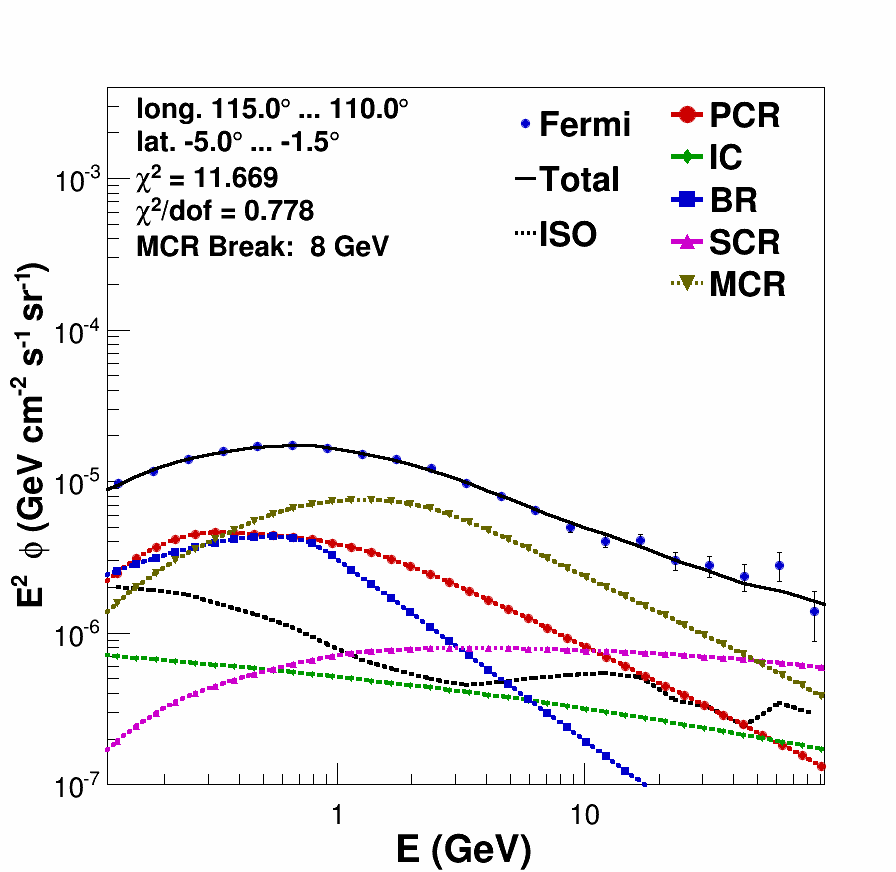}
\includegraphics[width=0.16\textwidth,height=0.16\textwidth,clip]{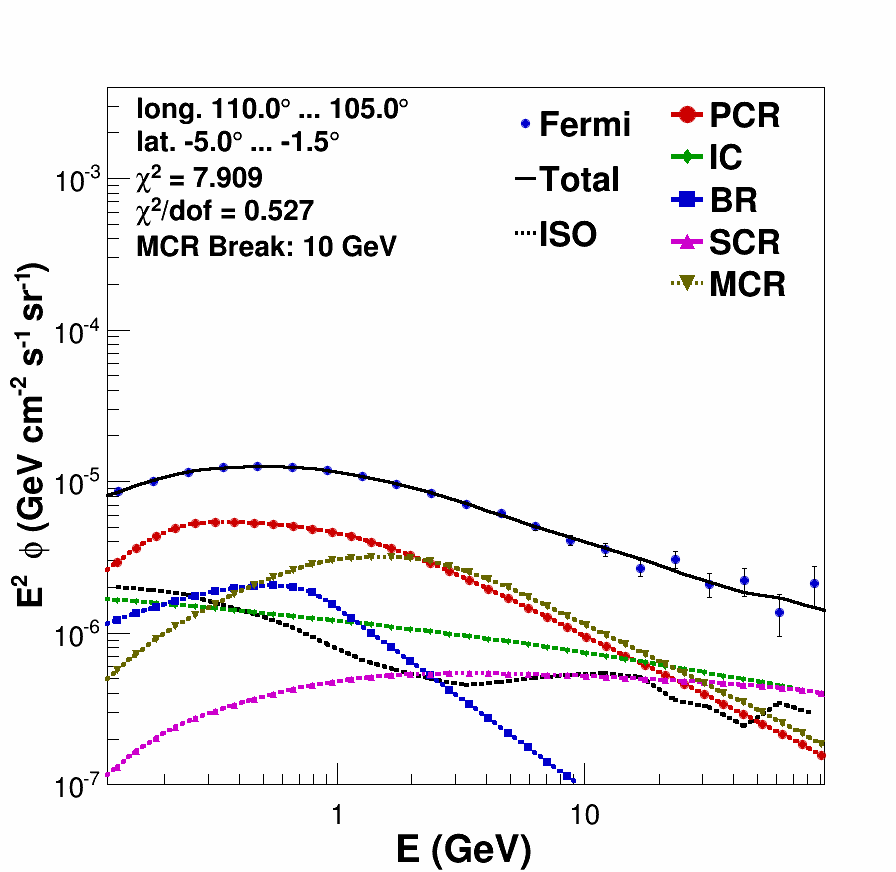}
\includegraphics[width=0.16\textwidth,height=0.16\textwidth,clip]{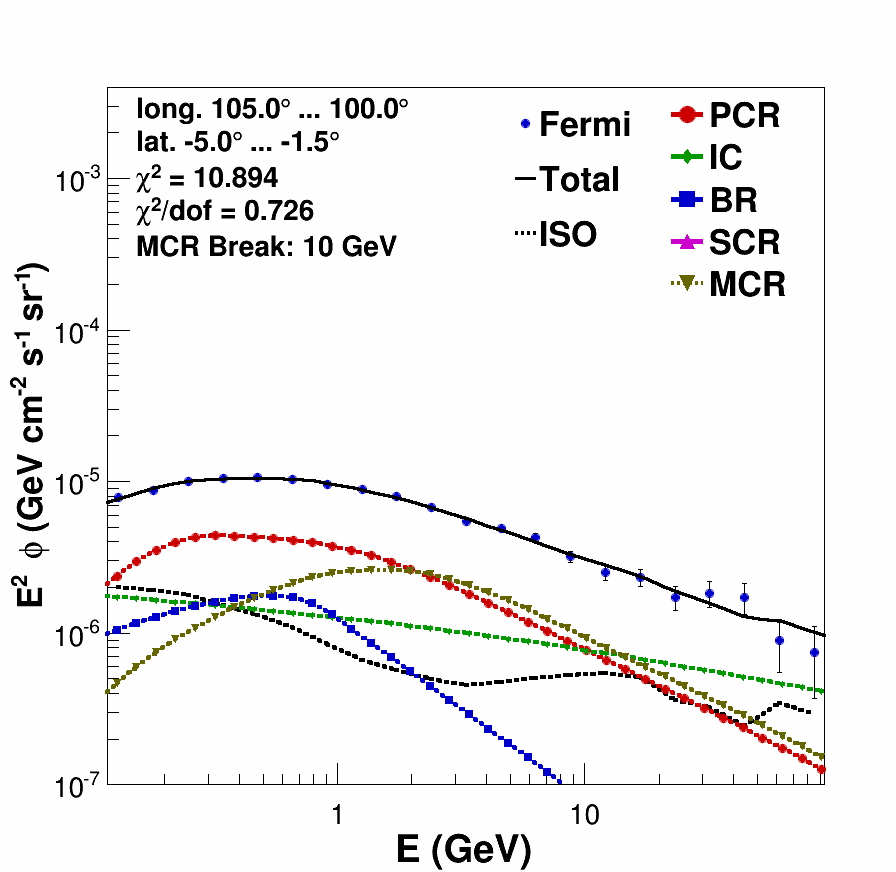}
\includegraphics[width=0.16\textwidth,height=0.16\textwidth,clip]{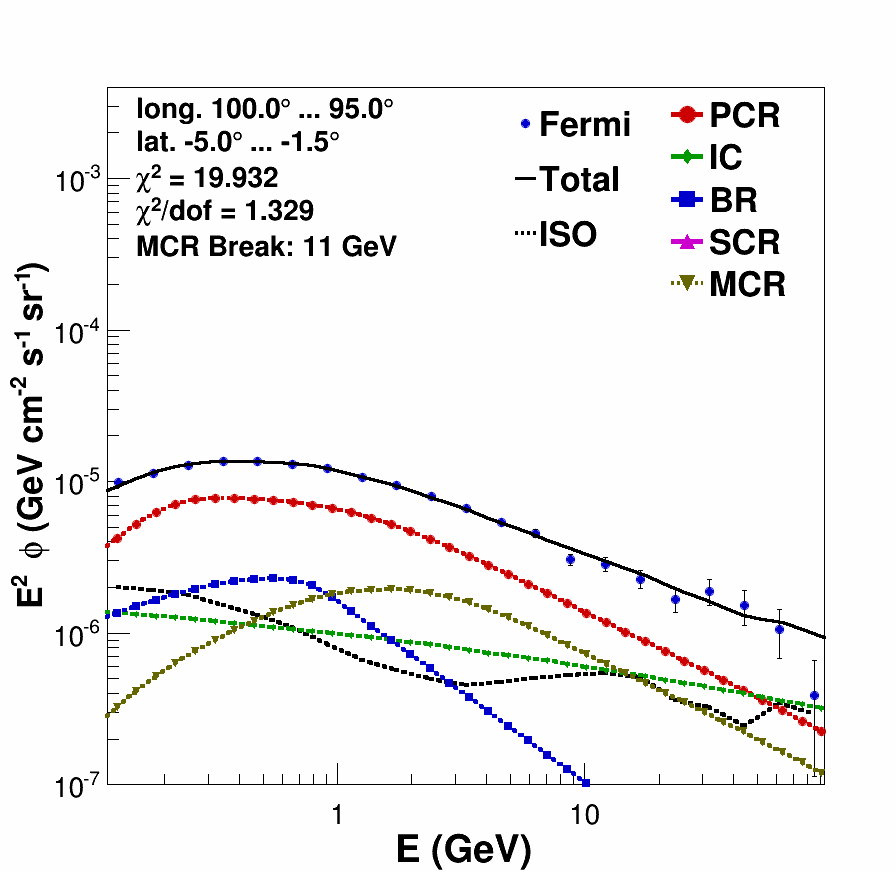}
\includegraphics[width=0.16\textwidth,height=0.16\textwidth,clip]{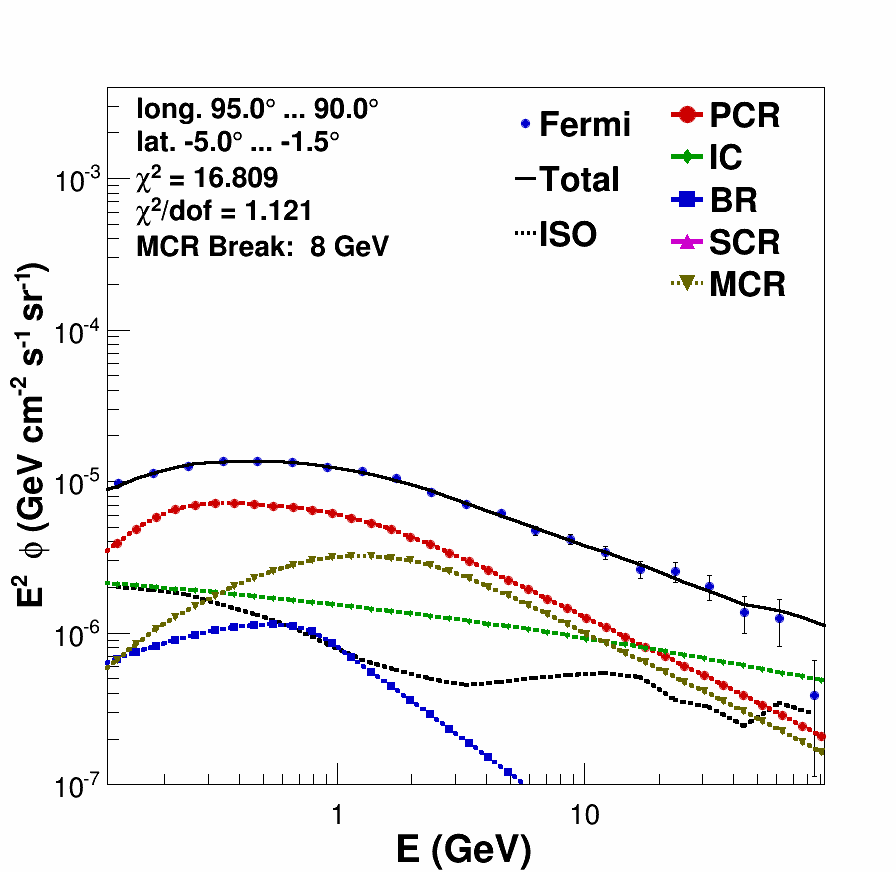}
\includegraphics[width=0.16\textwidth,height=0.16\textwidth,clip]{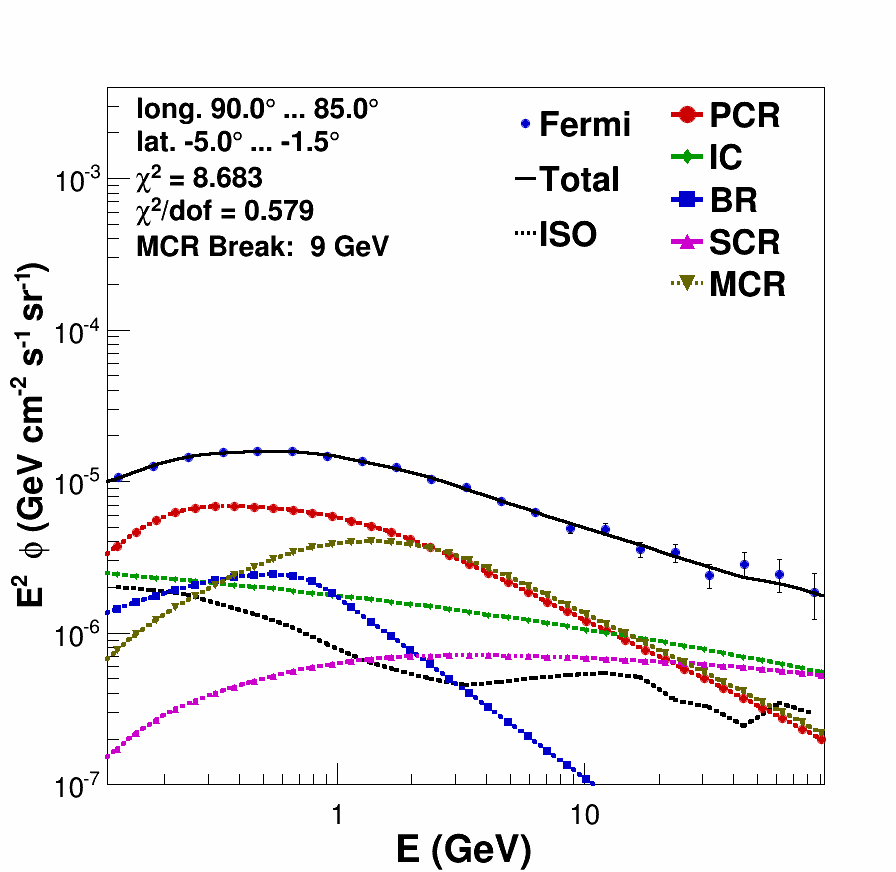}
\includegraphics[width=0.16\textwidth,height=0.16\textwidth,clip]{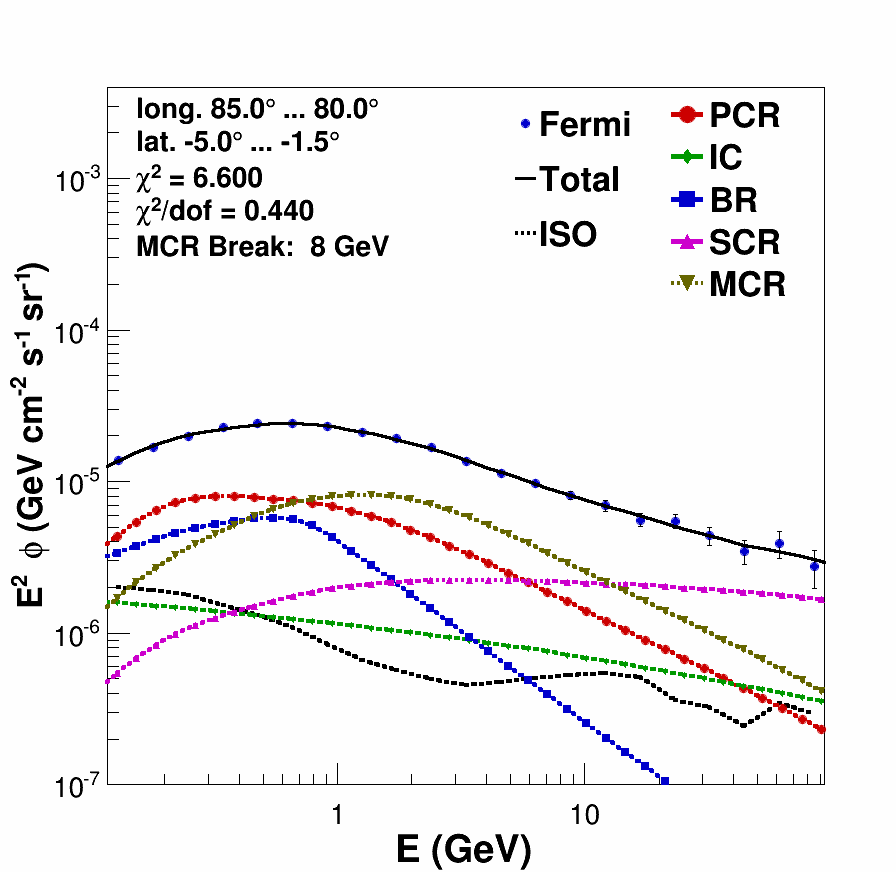}
\includegraphics[width=0.16\textwidth,height=0.16\textwidth,clip]{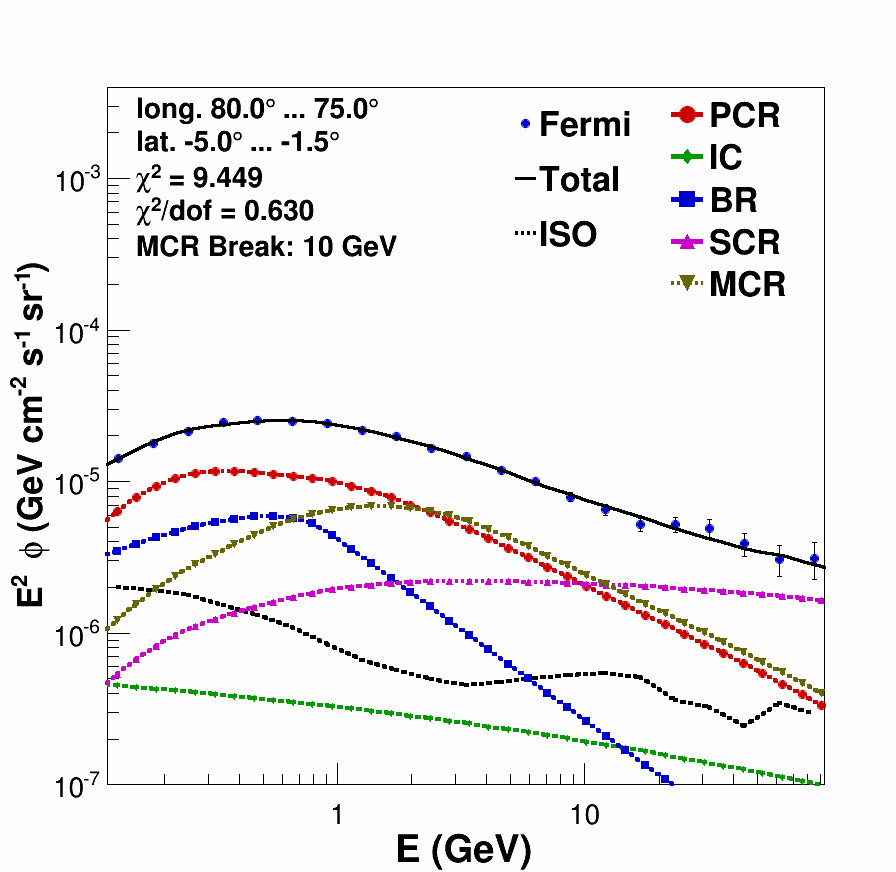}
\includegraphics[width=0.16\textwidth,height=0.16\textwidth,clip]{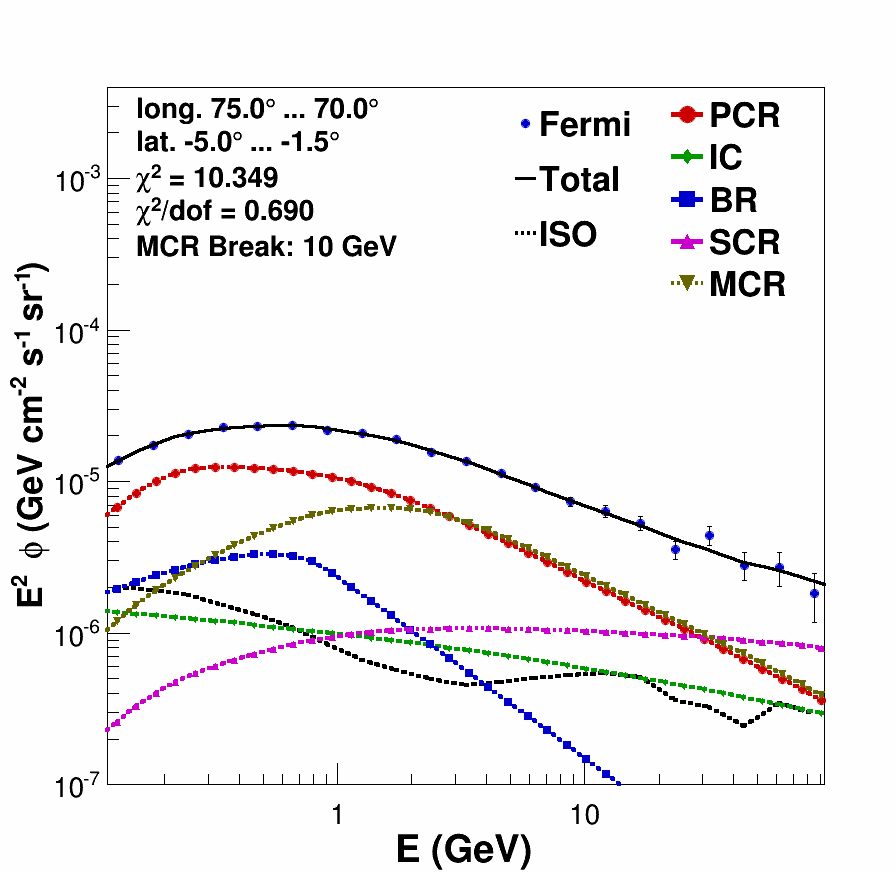}
\includegraphics[width=0.16\textwidth,height=0.16\textwidth,clip]{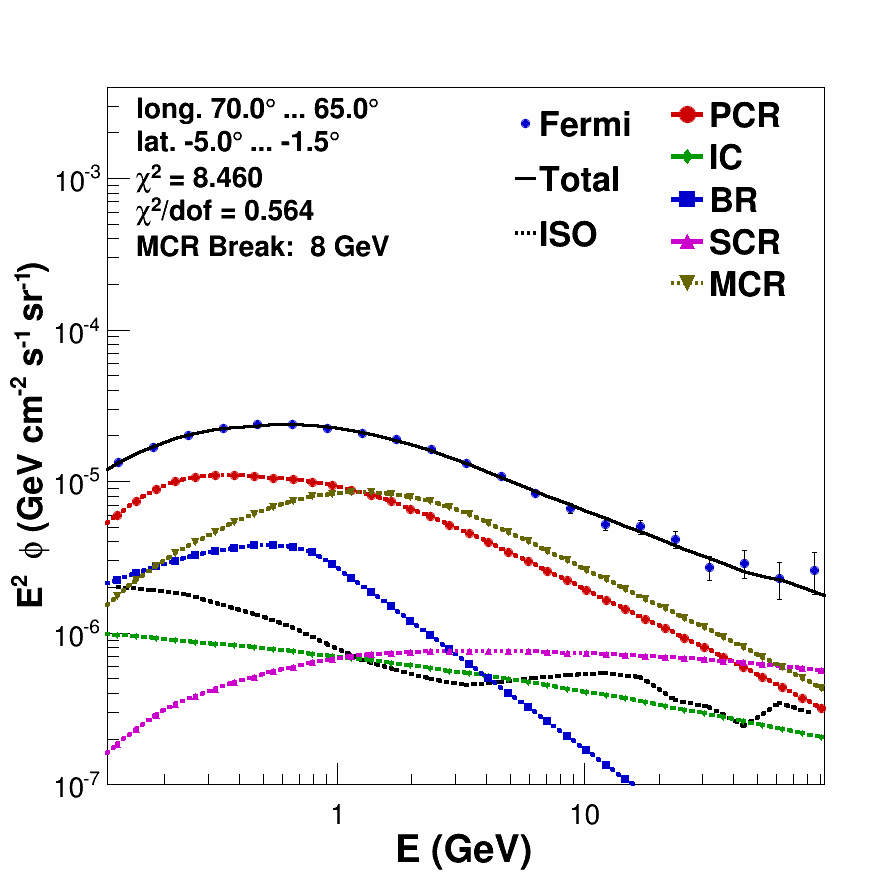}
\includegraphics[width=0.16\textwidth,height=0.16\textwidth,clip]{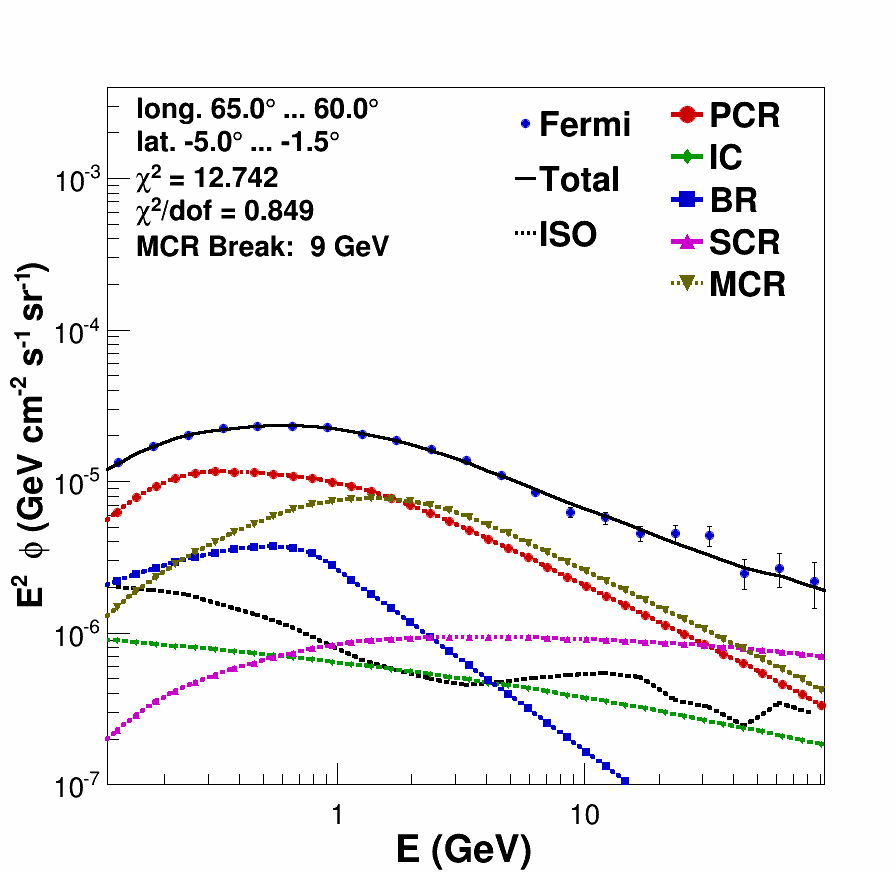}
\includegraphics[width=0.16\textwidth,height=0.16\textwidth,clip]{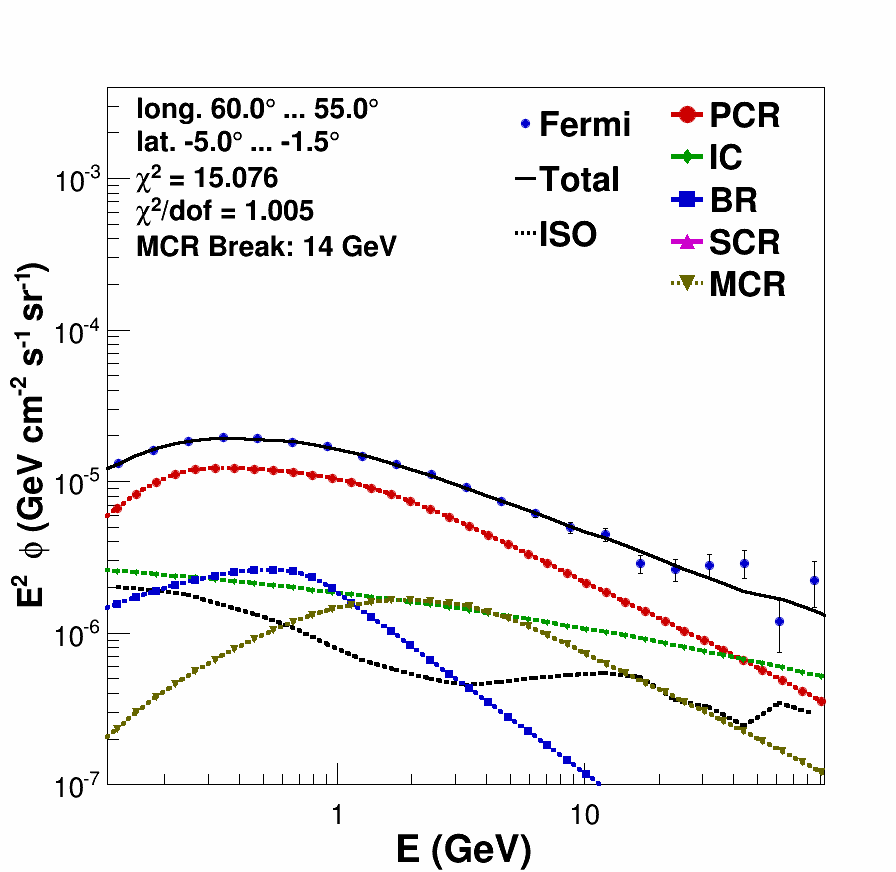}
\includegraphics[width=0.16\textwidth,height=0.16\textwidth,clip]{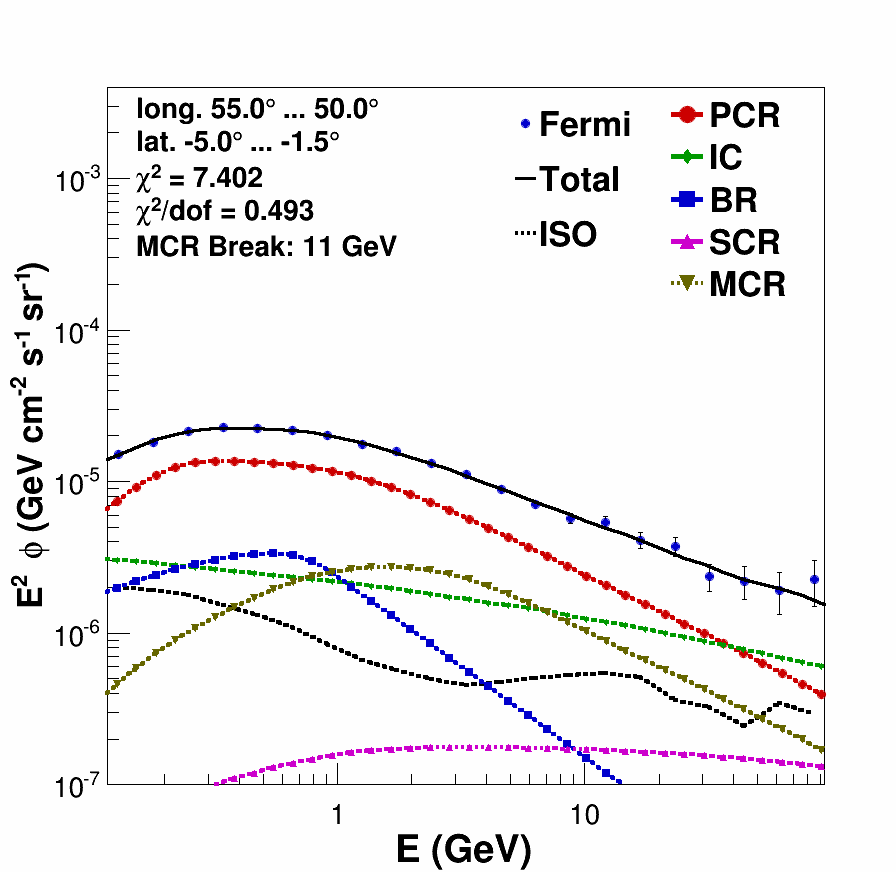}
\includegraphics[width=0.16\textwidth,height=0.16\textwidth,clip]{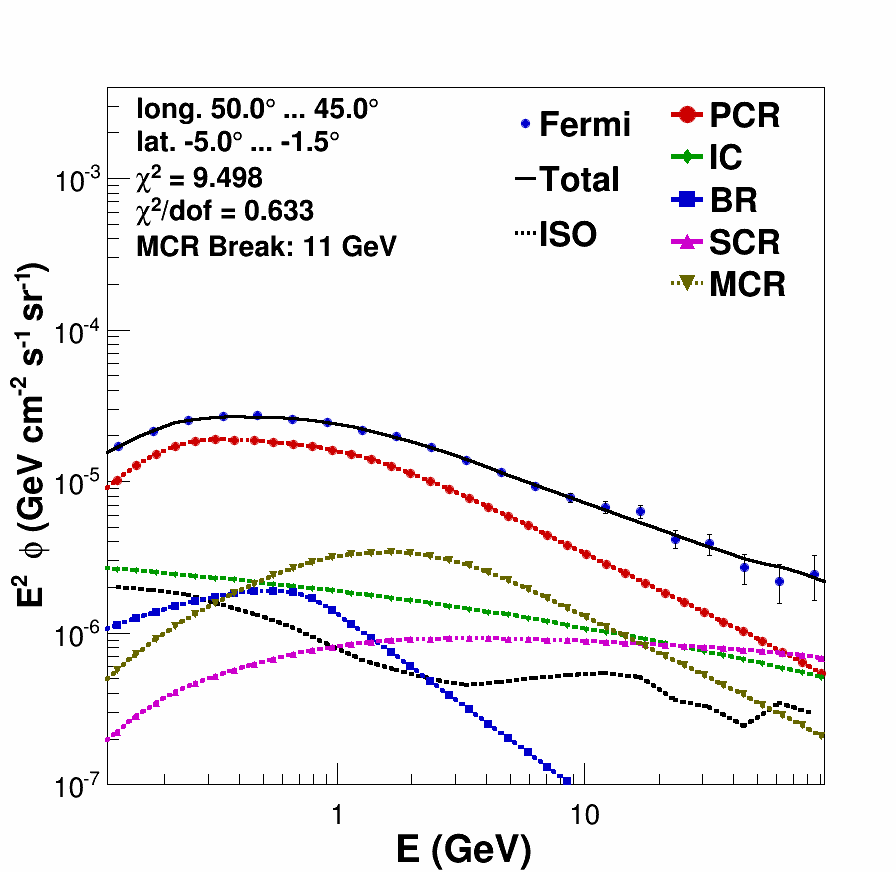}
\includegraphics[width=0.16\textwidth,height=0.16\textwidth,clip]{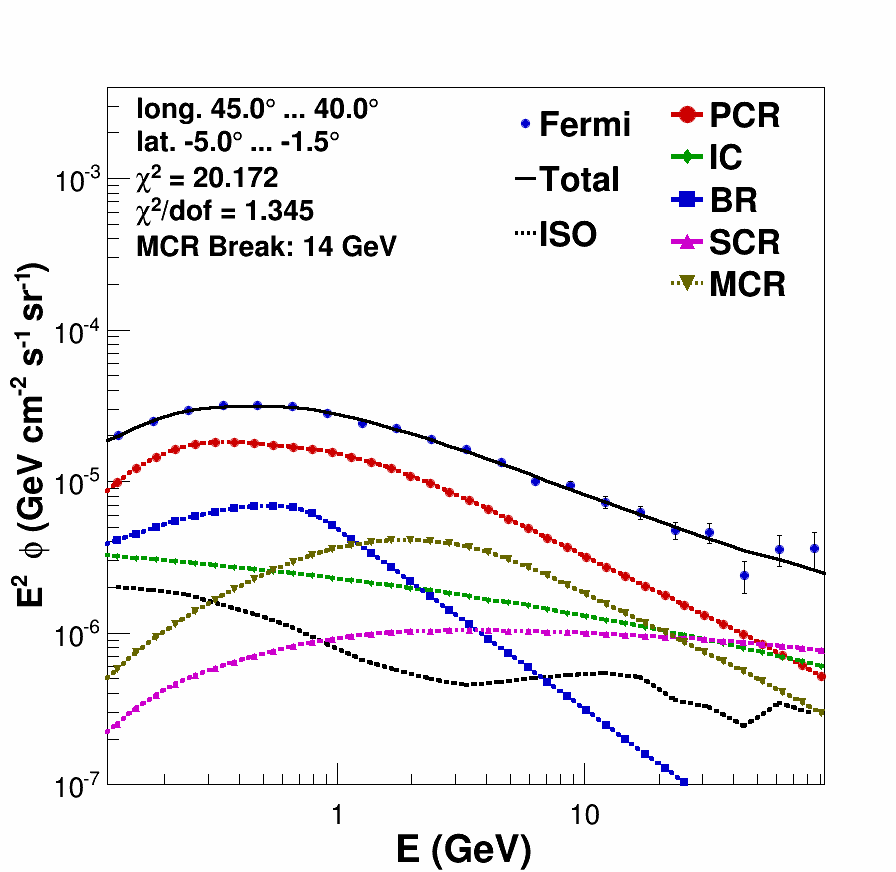}
\includegraphics[width=0.16\textwidth,height=0.16\textwidth,clip]{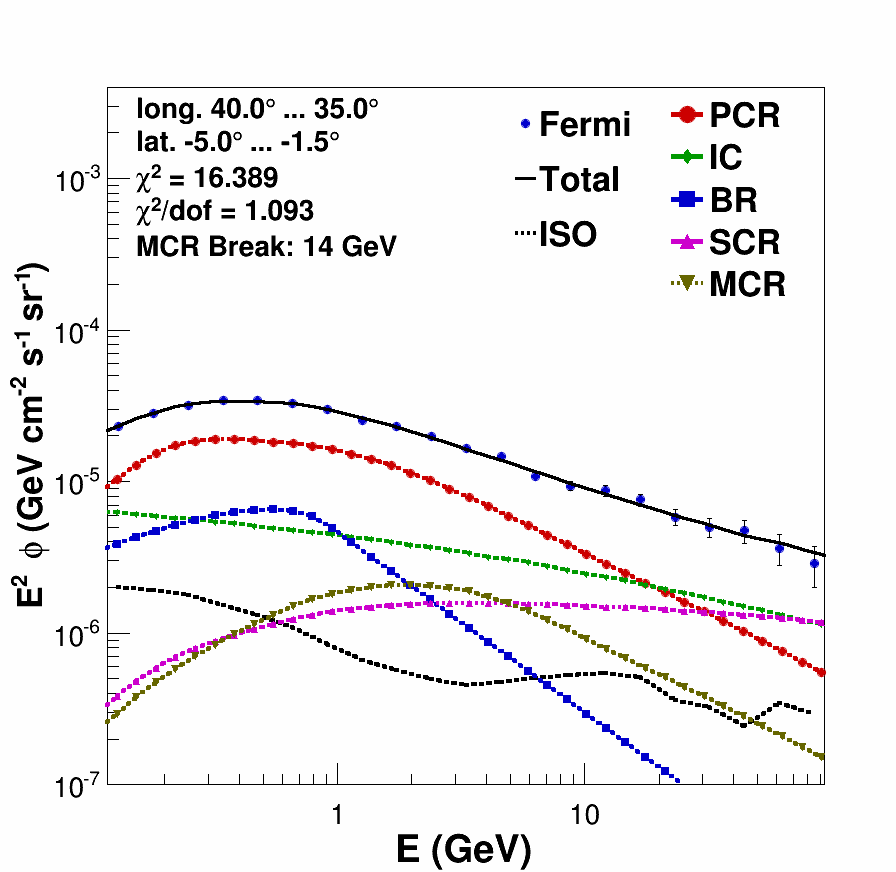}
\includegraphics[width=0.16\textwidth,height=0.16\textwidth,clip]{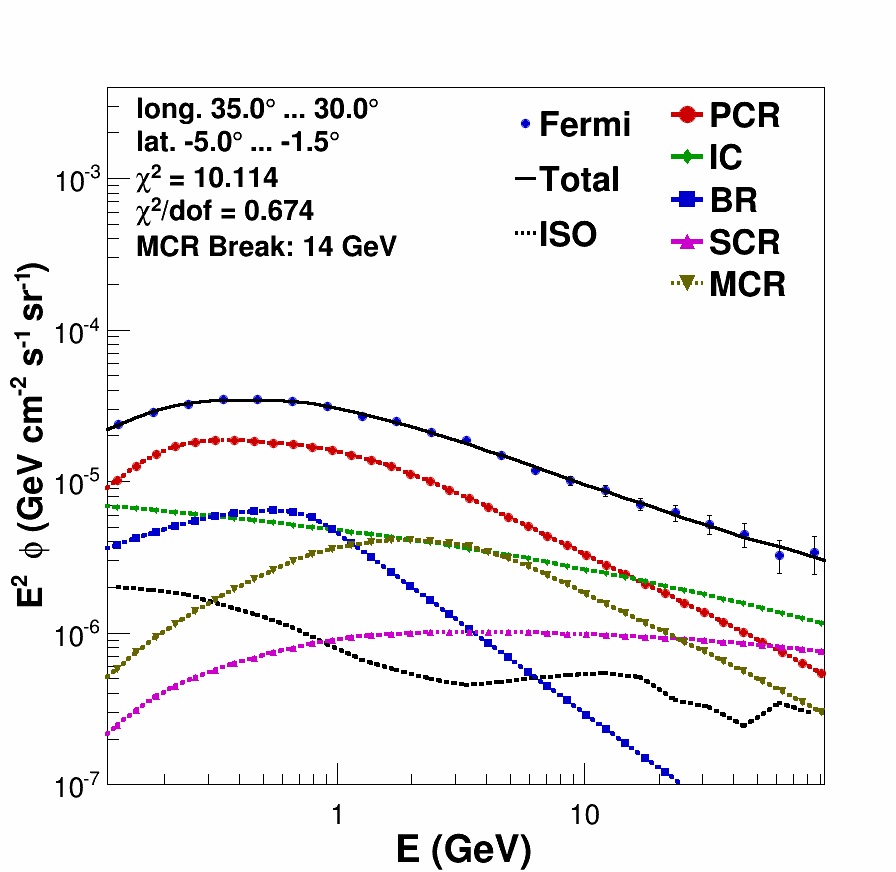}
\includegraphics[width=0.16\textwidth,height=0.16\textwidth,clip]{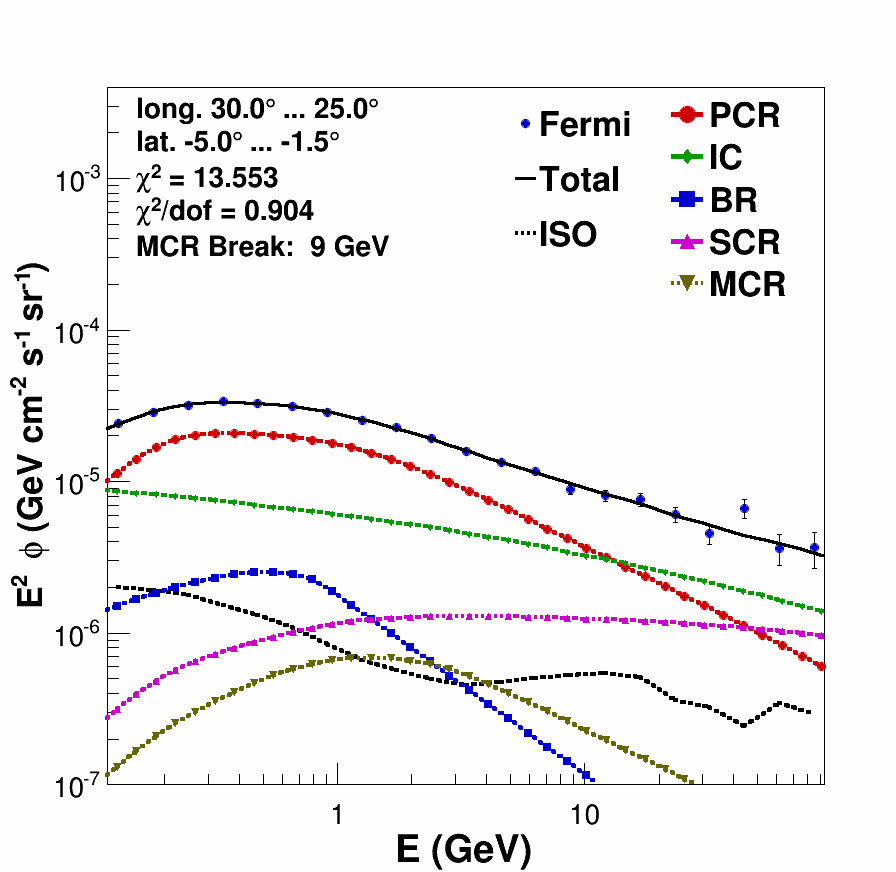}
\includegraphics[width=0.16\textwidth,height=0.16\textwidth,clip]{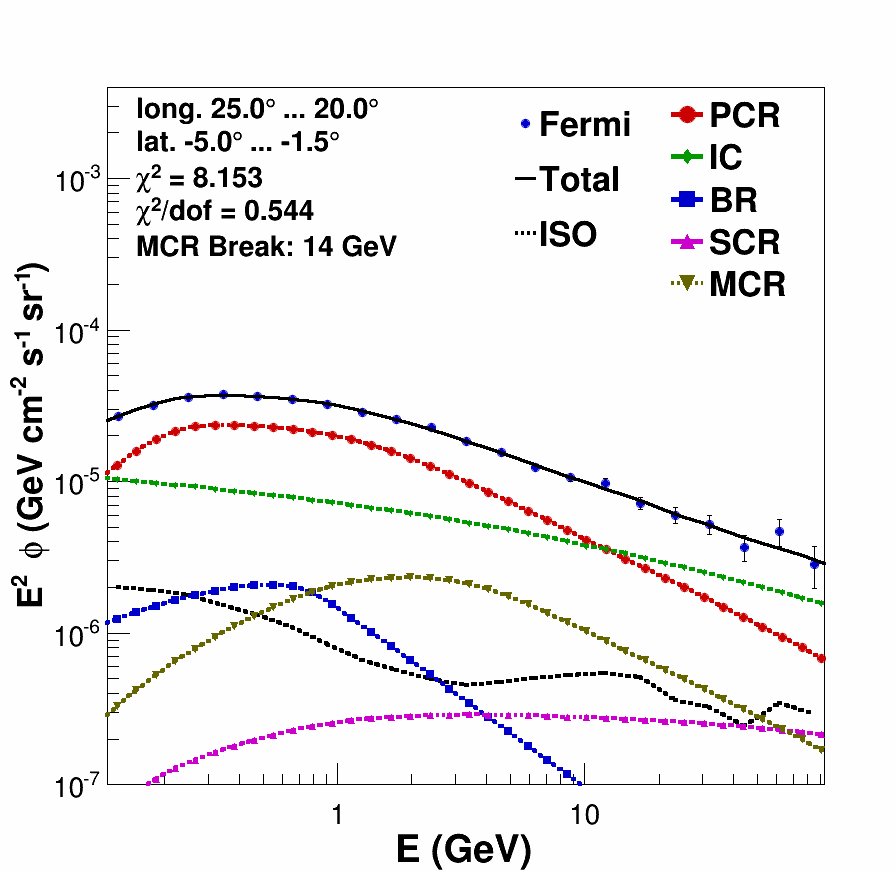}
\includegraphics[width=0.16\textwidth,height=0.16\textwidth,clip]{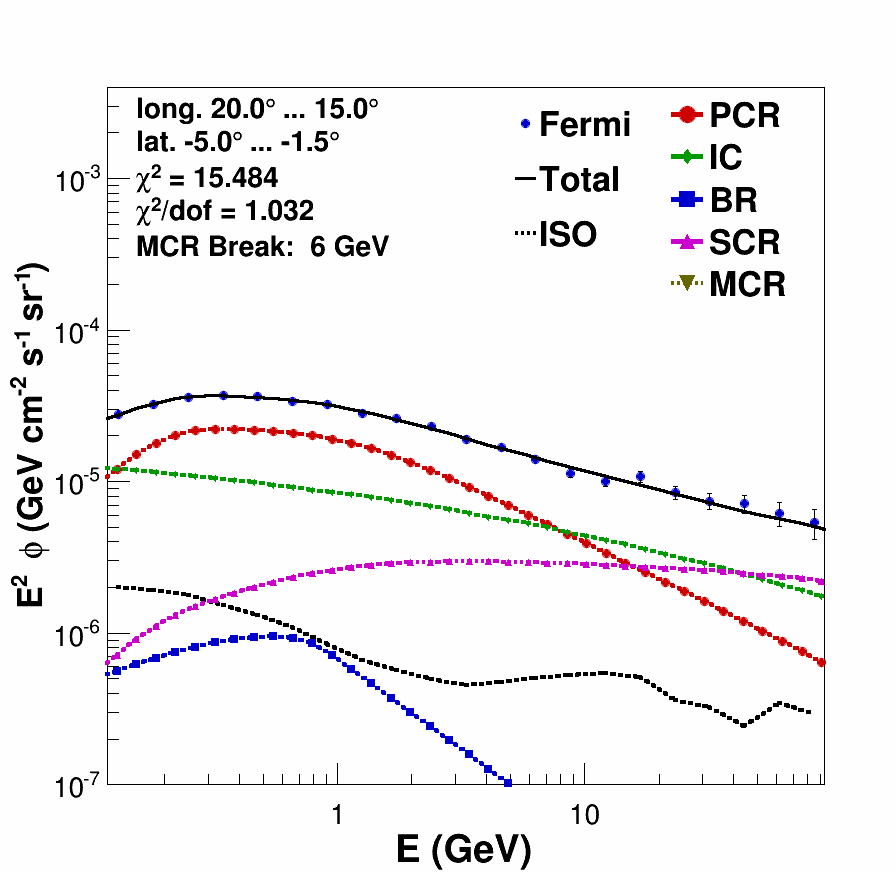}
\includegraphics[width=0.16\textwidth,height=0.16\textwidth,clip]{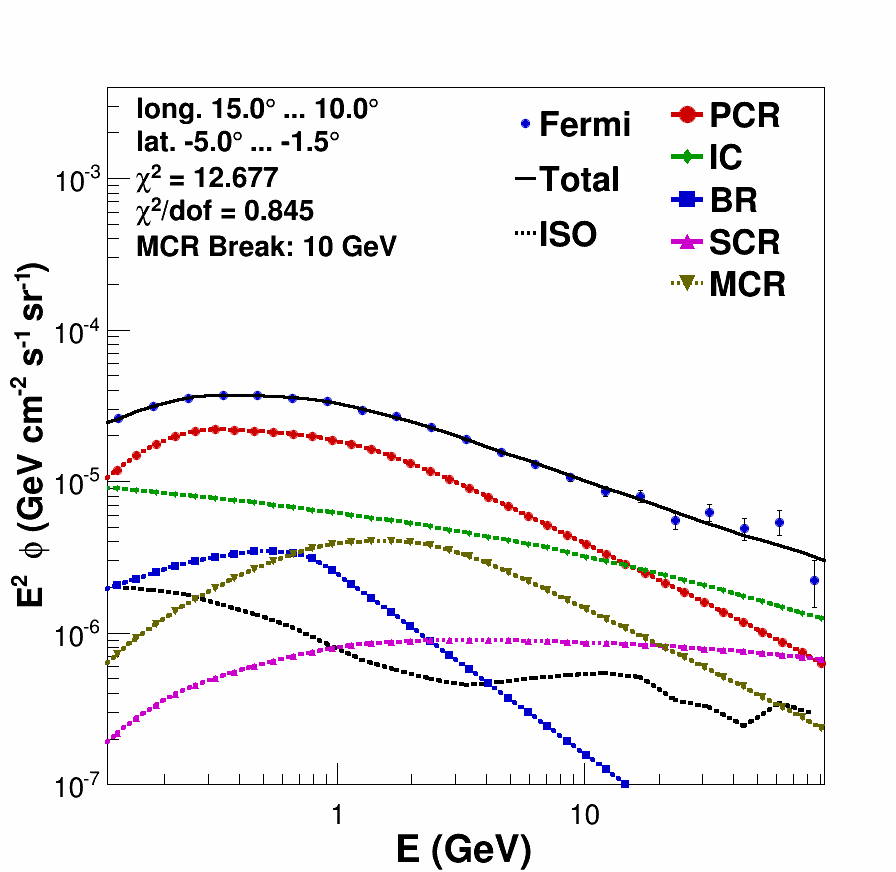}
\includegraphics[width=0.16\textwidth,height=0.16\textwidth,clip]{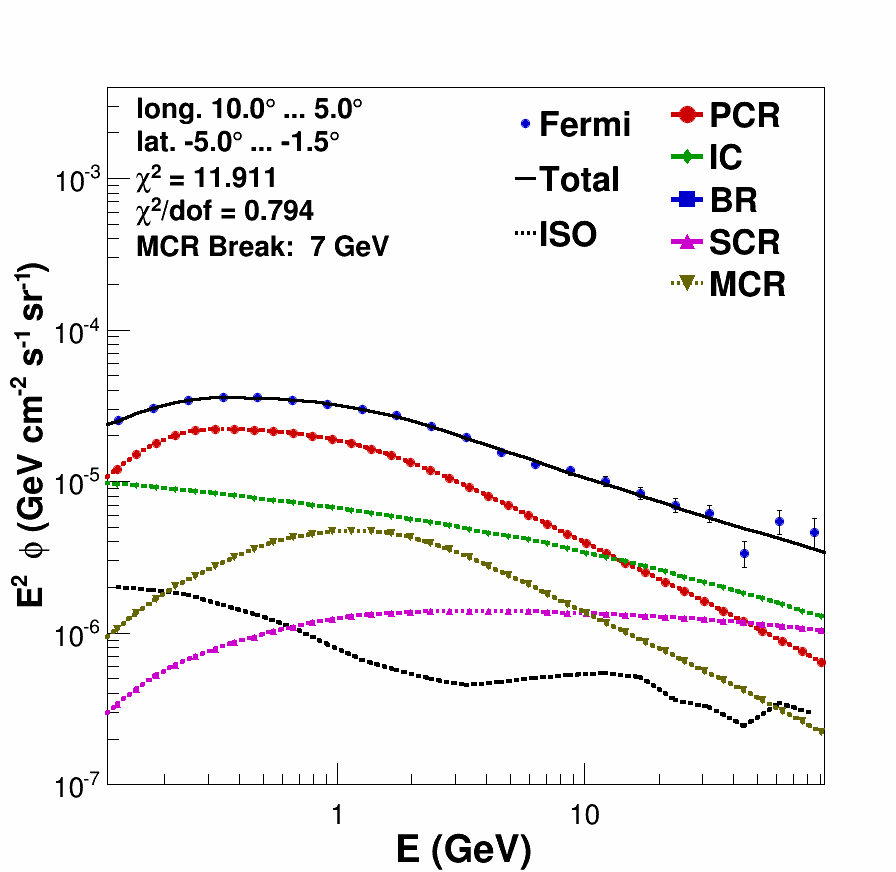}
\includegraphics[width=0.16\textwidth,height=0.16\textwidth,clip]{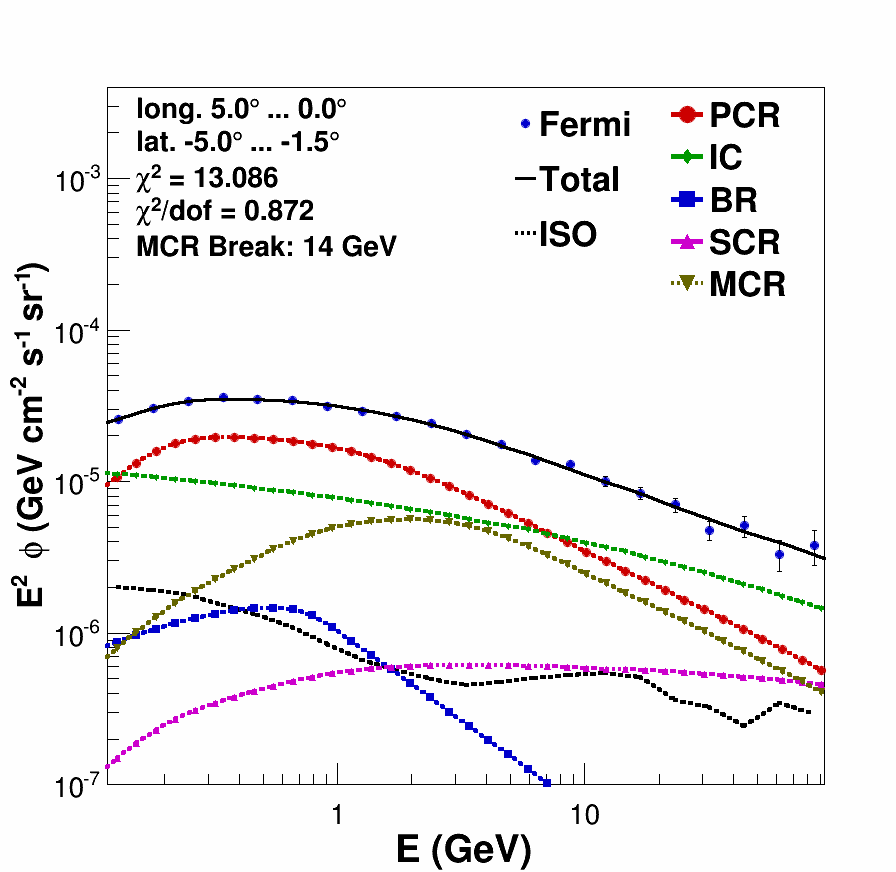}%%%%%%r12a
\caption[]{Template fits for latitudes  with $-5.0^\circ<b<-1.5^\circ$ and longitudes decreasing from 180$^\circ$ to 0$^\circ$.} \label{F23}
\end{figure}
\begin{figure}
\centering
\includegraphics[width=0.16\textwidth,height=0.16\textwidth,clip]{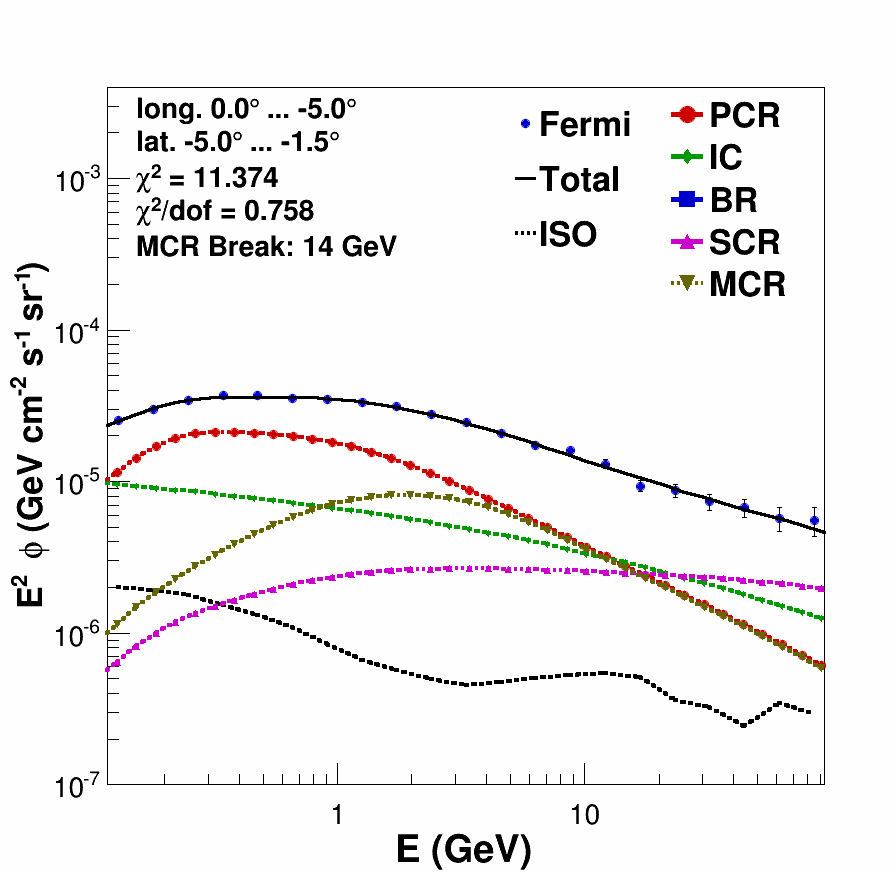}
\includegraphics[width=0.16\textwidth,height=0.16\textwidth,clip]{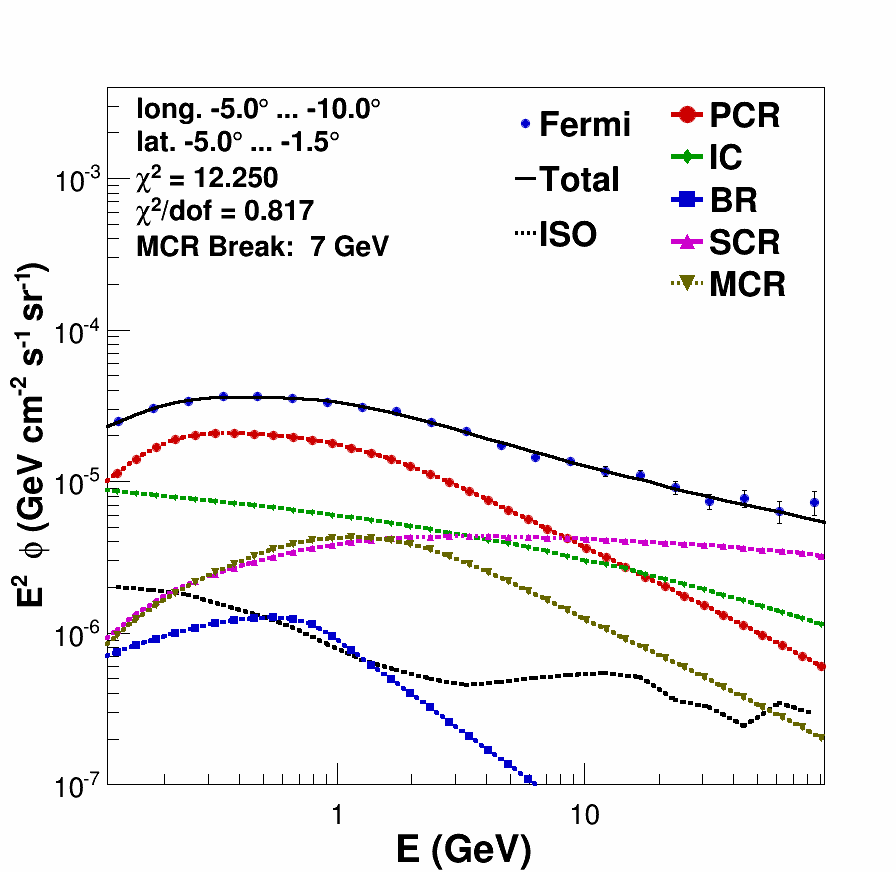}
\includegraphics[width=0.16\textwidth,height=0.16\textwidth,clip]{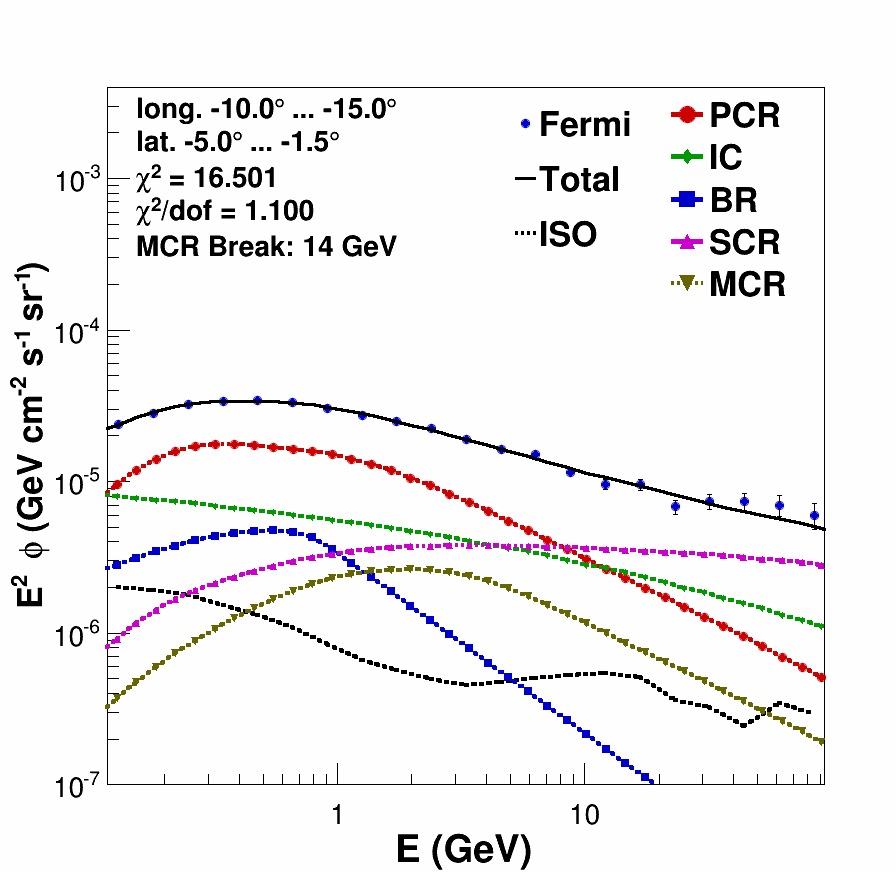}
\includegraphics[width=0.16\textwidth,height=0.16\textwidth,clip]{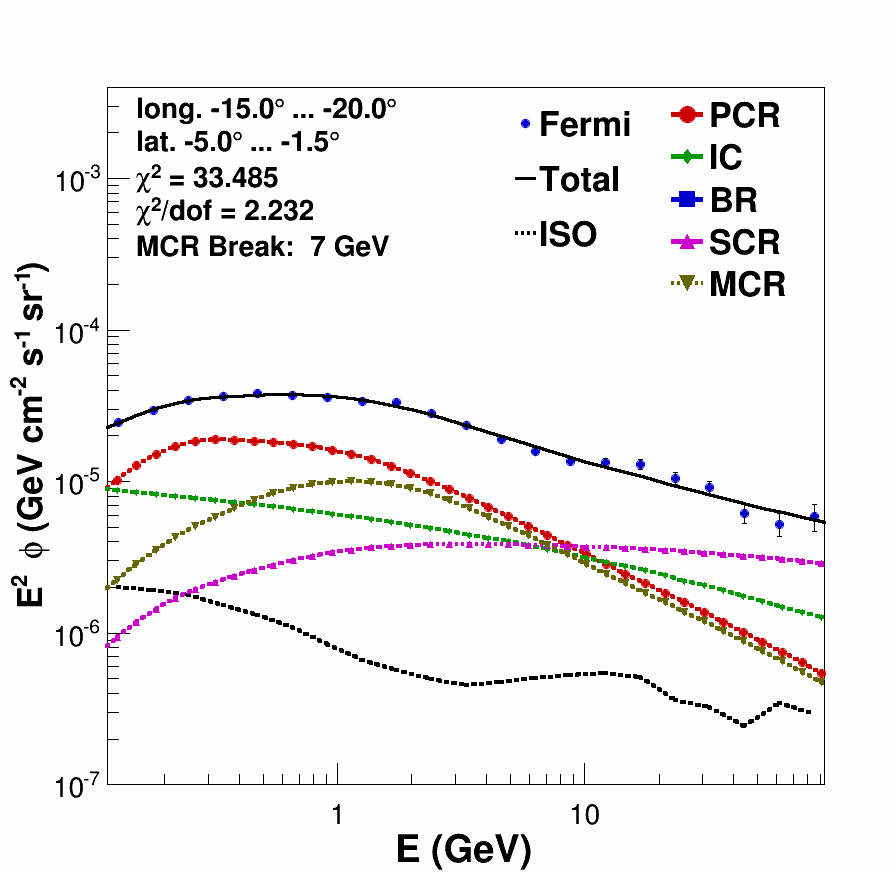}
\includegraphics[width=0.16\textwidth,height=0.16\textwidth,clip]{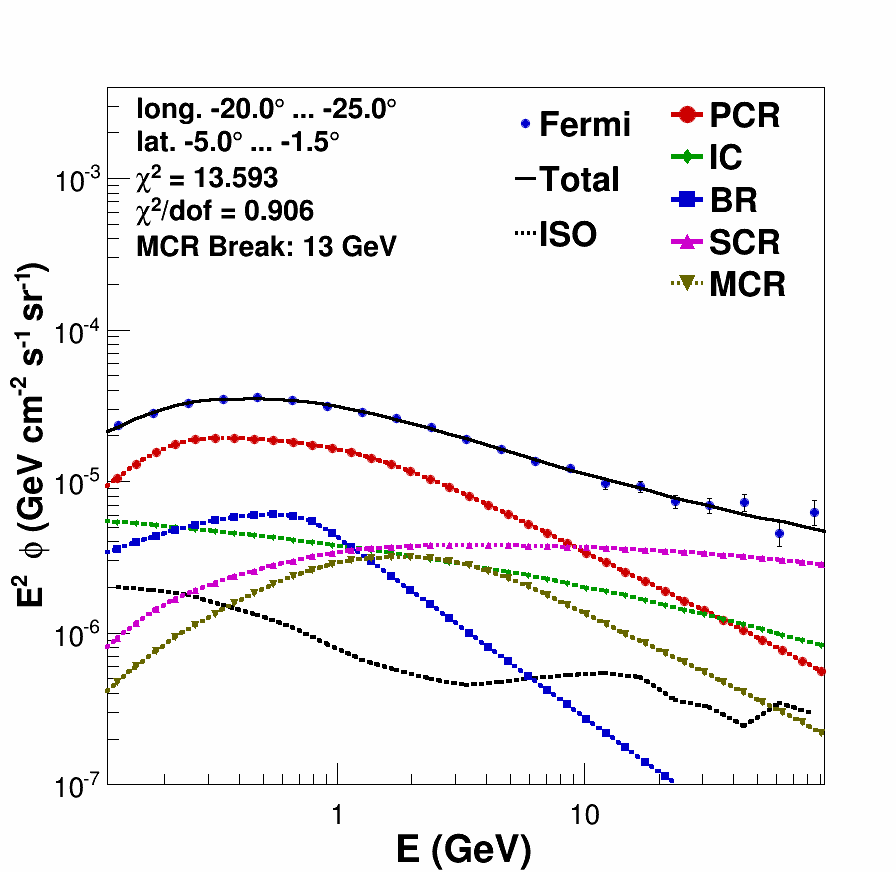}
\includegraphics[width=0.16\textwidth,height=0.16\textwidth,clip]{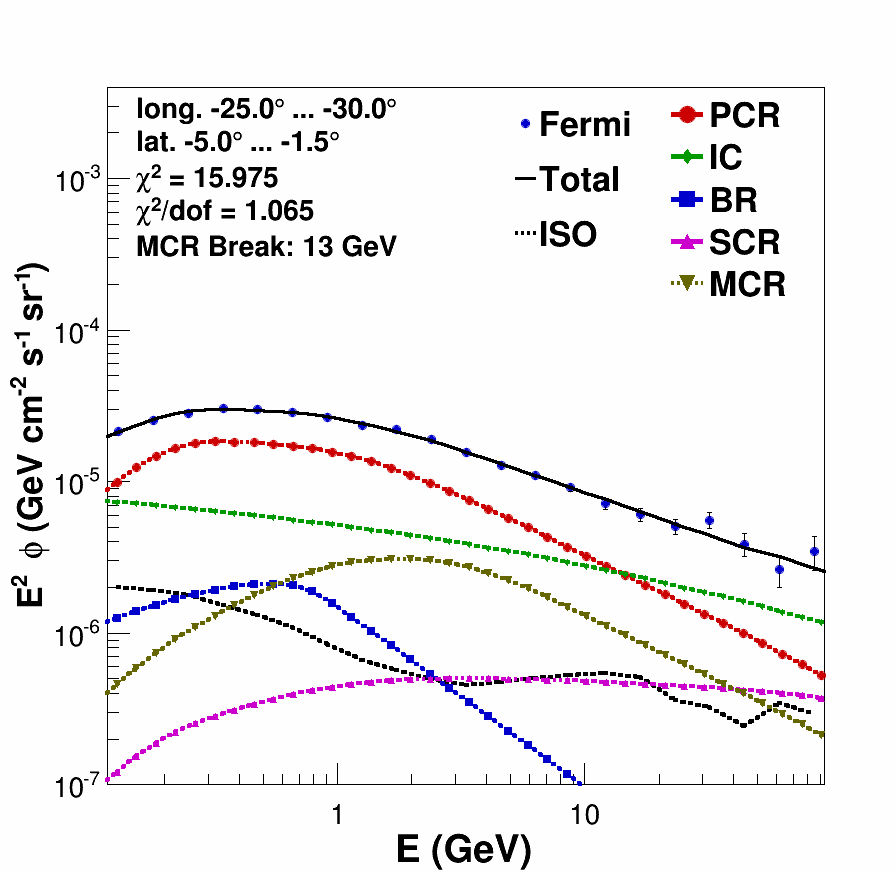}
\includegraphics[width=0.16\textwidth,height=0.16\textwidth,clip]{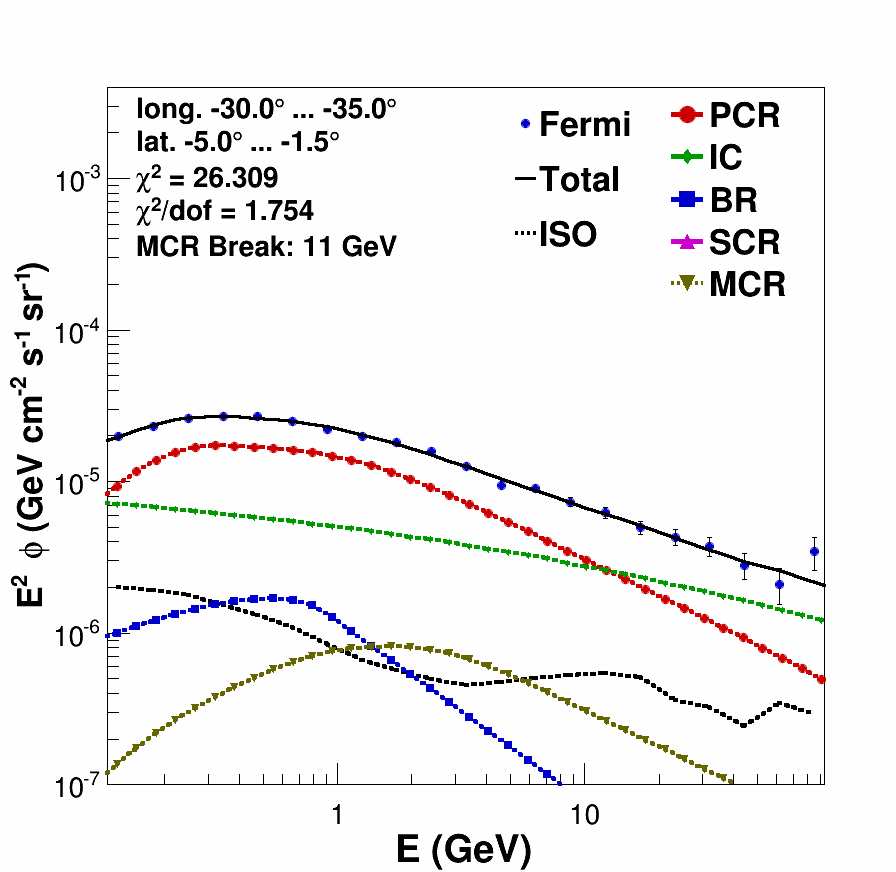}
\includegraphics[width=0.16\textwidth,height=0.16\textwidth,clip]{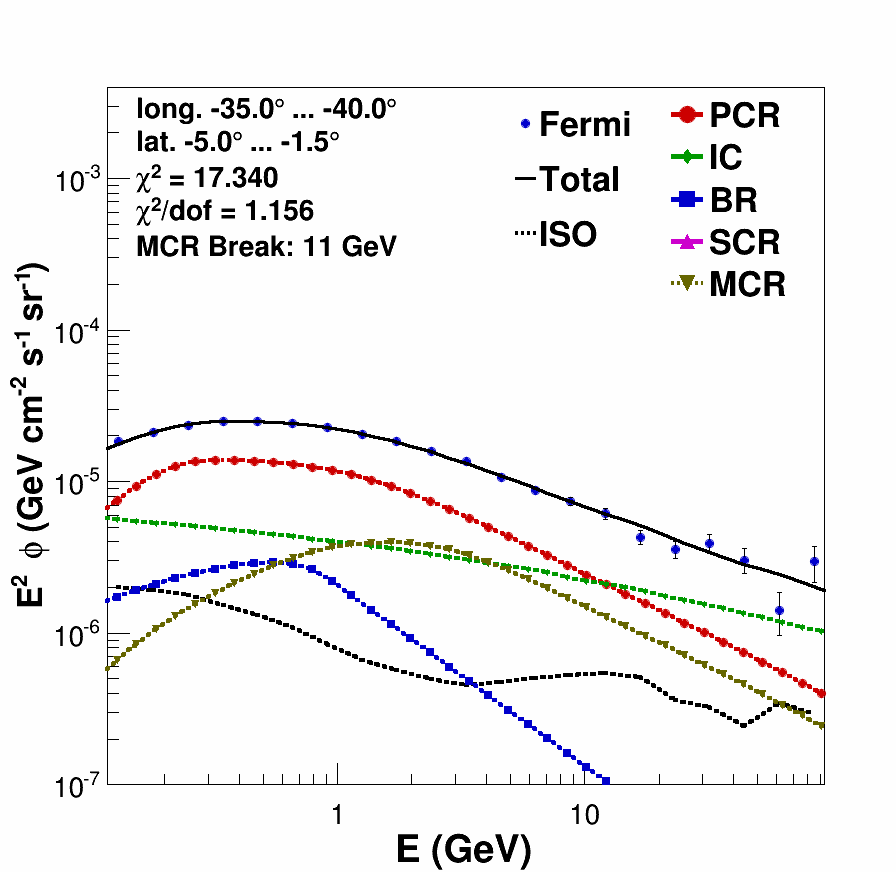}
\includegraphics[width=0.16\textwidth,height=0.16\textwidth,clip]{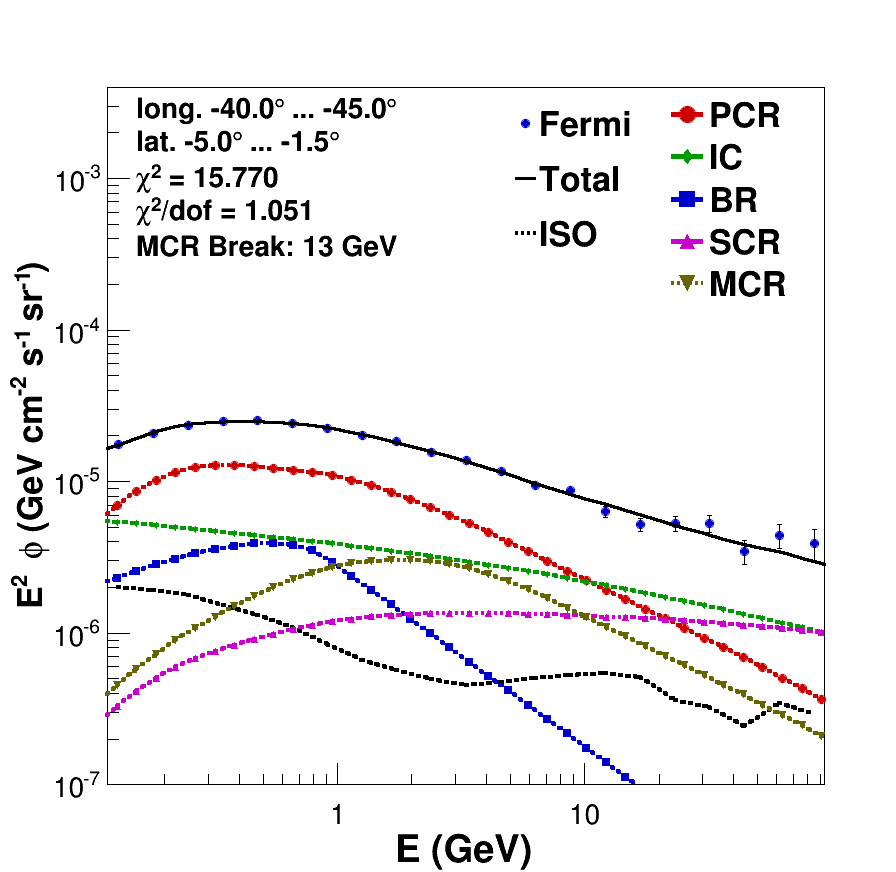}
\includegraphics[width=0.16\textwidth,height=0.16\textwidth,clip]{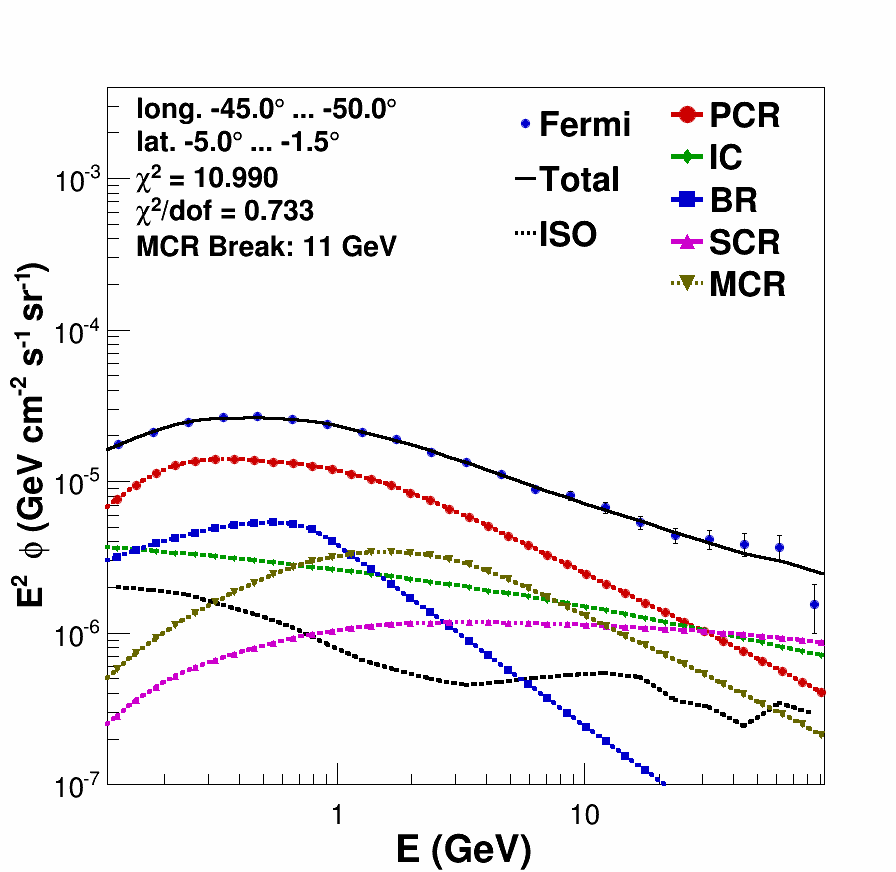}
\includegraphics[width=0.16\textwidth,height=0.16\textwidth,clip]{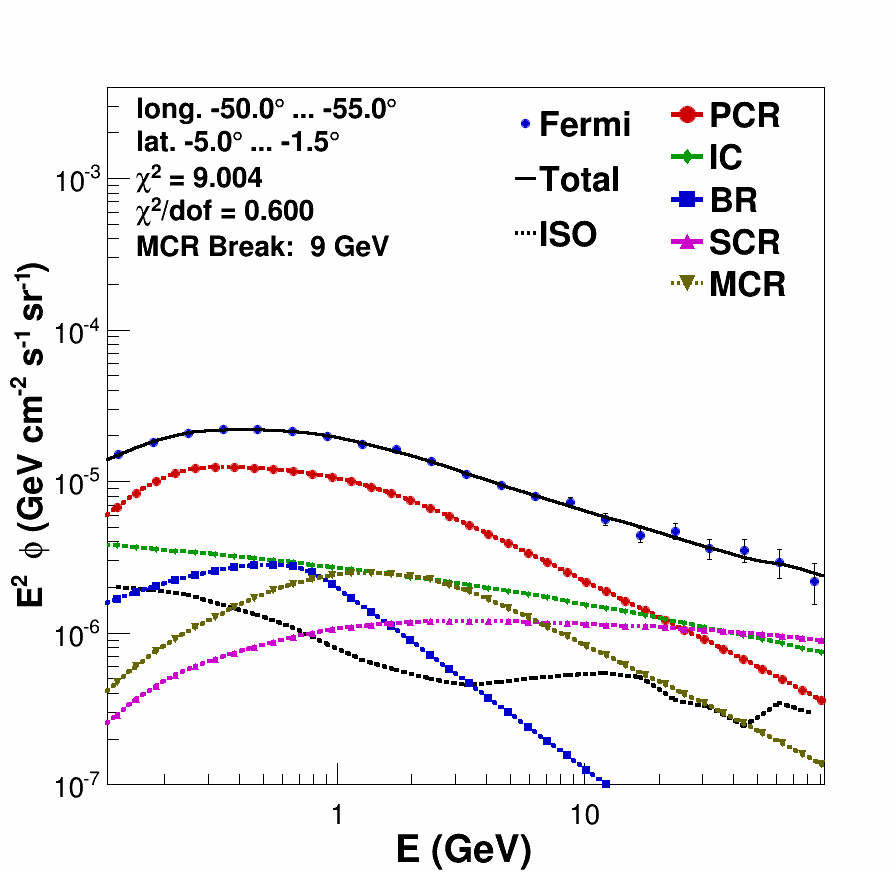}
\includegraphics[width=0.16\textwidth,height=0.16\textwidth,clip]{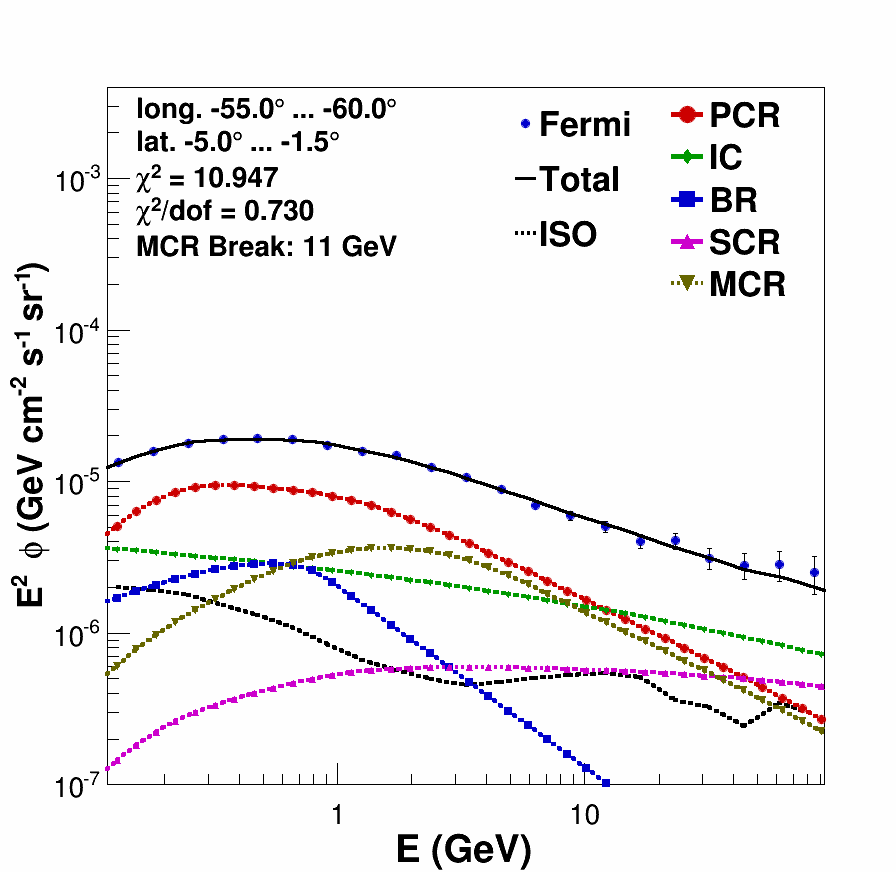}
\includegraphics[width=0.16\textwidth,height=0.16\textwidth,clip]{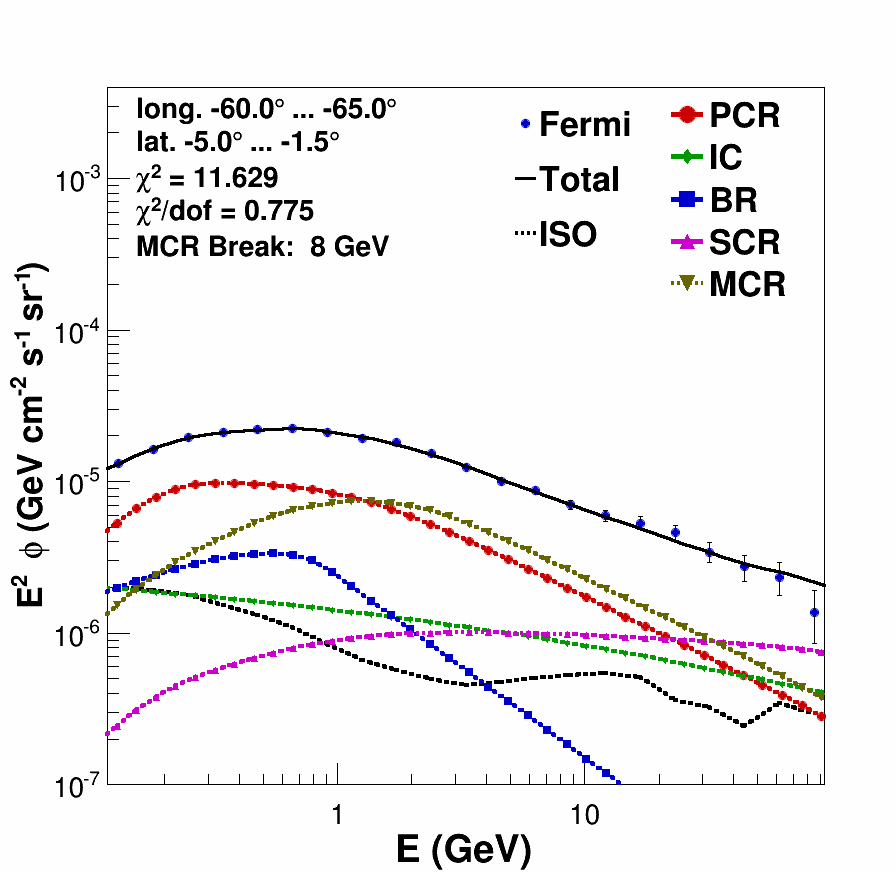}
\includegraphics[width=0.16\textwidth,height=0.16\textwidth,clip]{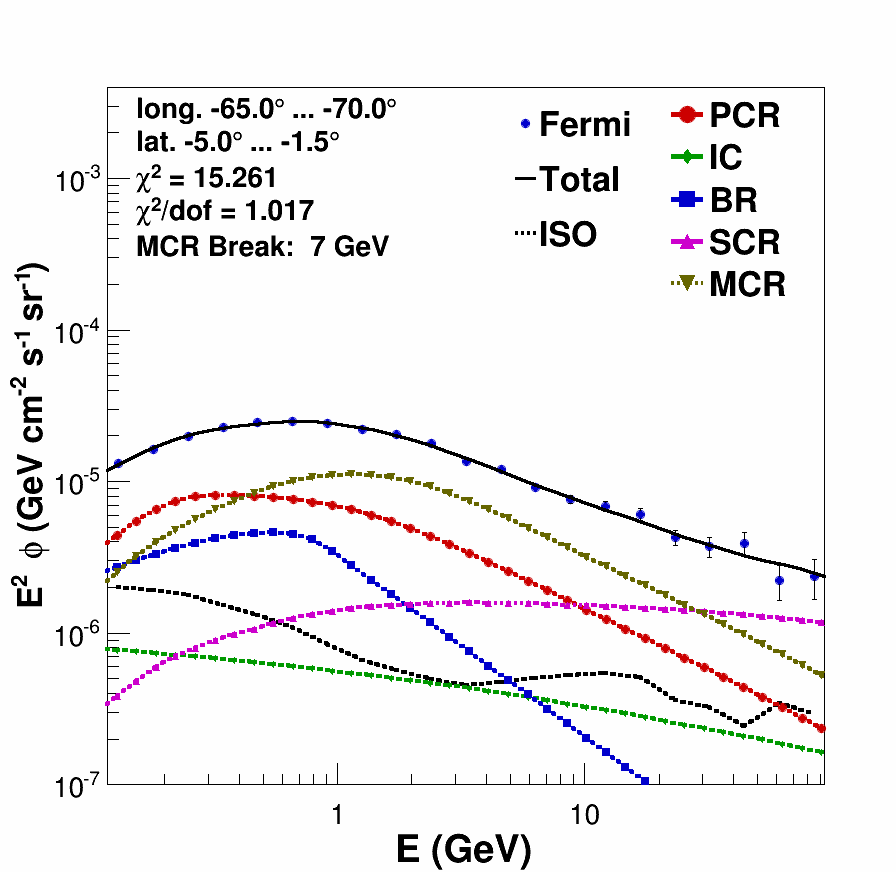}
\includegraphics[width=0.16\textwidth,height=0.16\textwidth,clip]{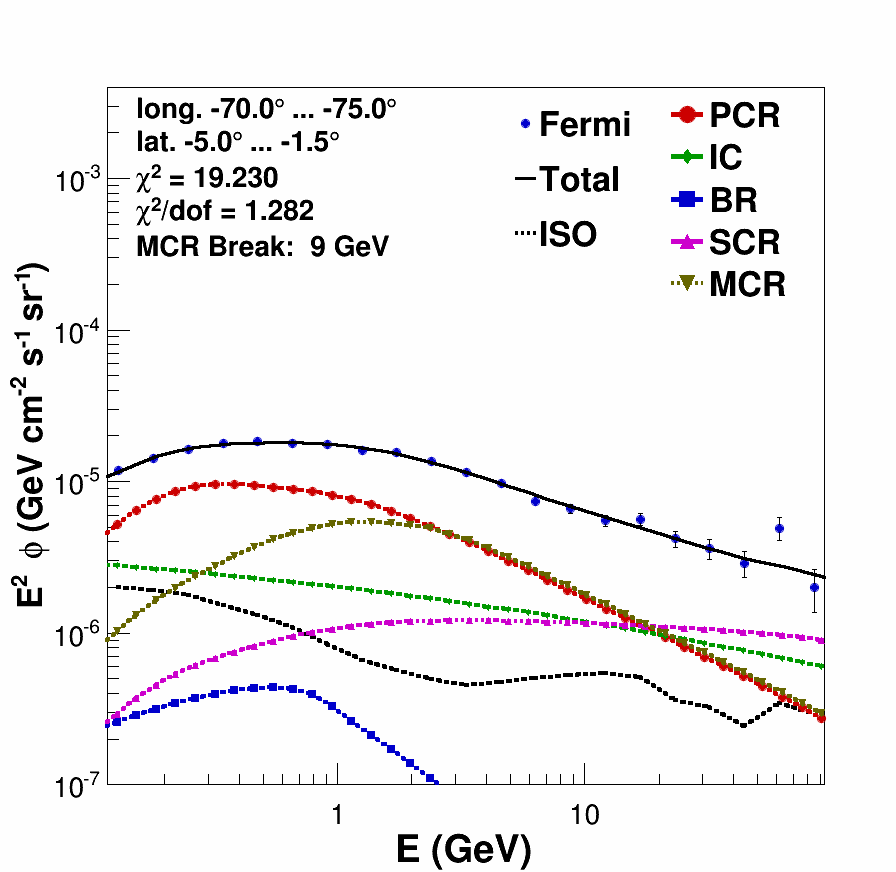}
\includegraphics[width=0.16\textwidth,height=0.16\textwidth,clip]{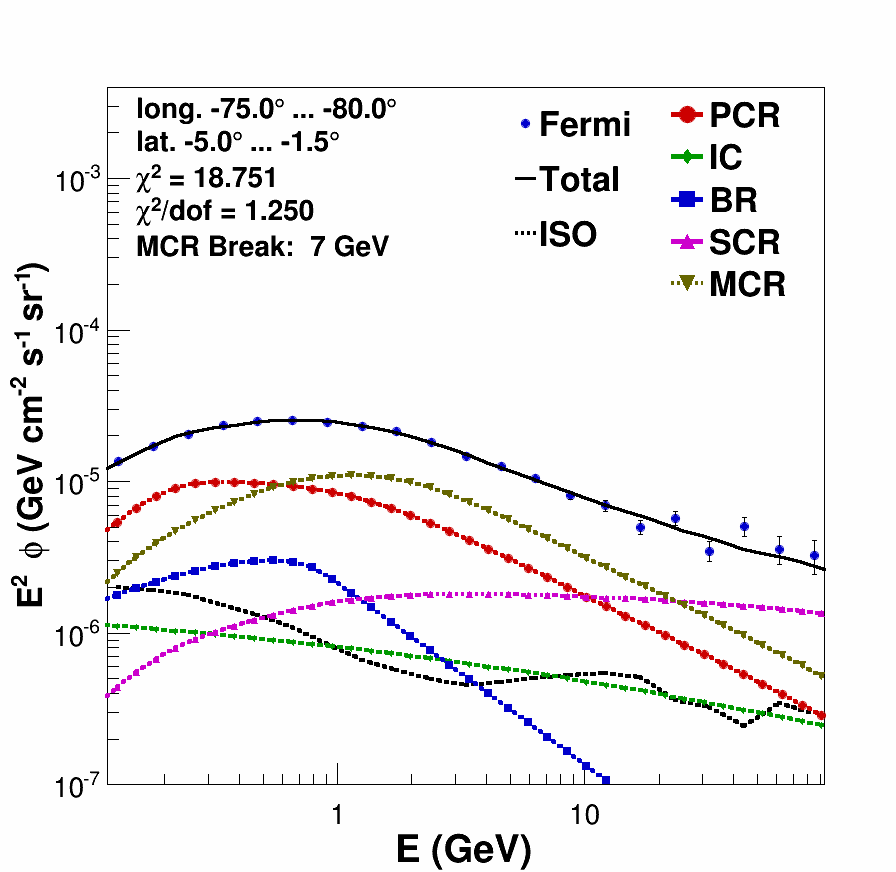}
\includegraphics[width=0.16\textwidth,height=0.16\textwidth,clip]{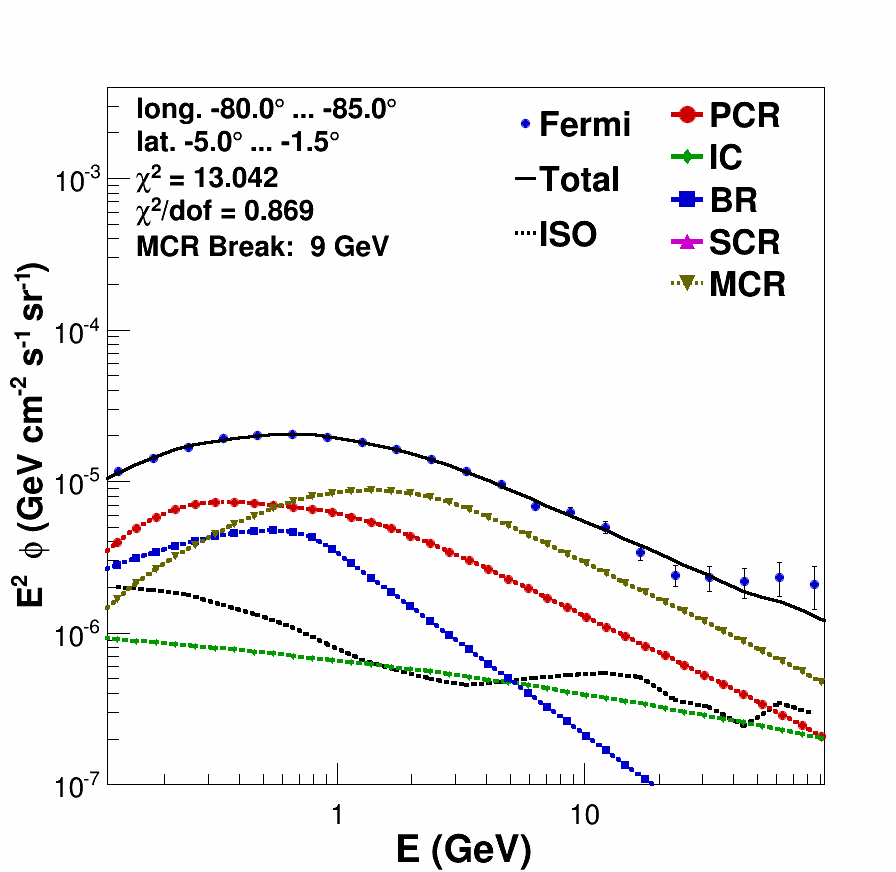}
\includegraphics[width=0.16\textwidth,height=0.16\textwidth,clip]{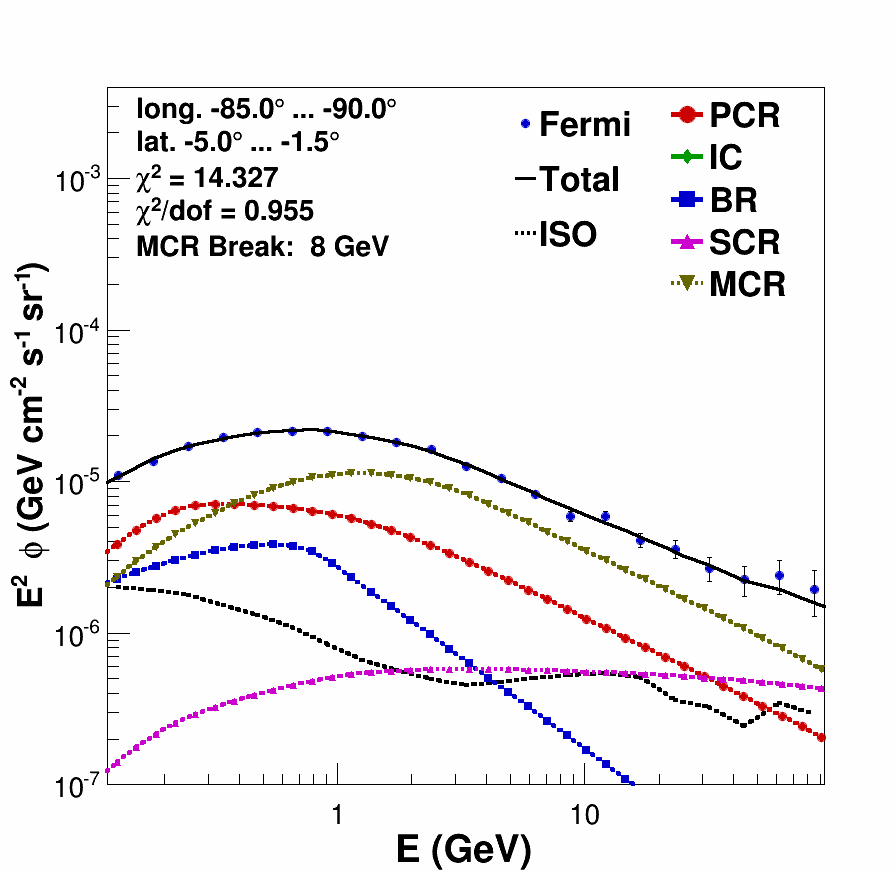}
\includegraphics[width=0.16\textwidth,height=0.16\textwidth,clip]{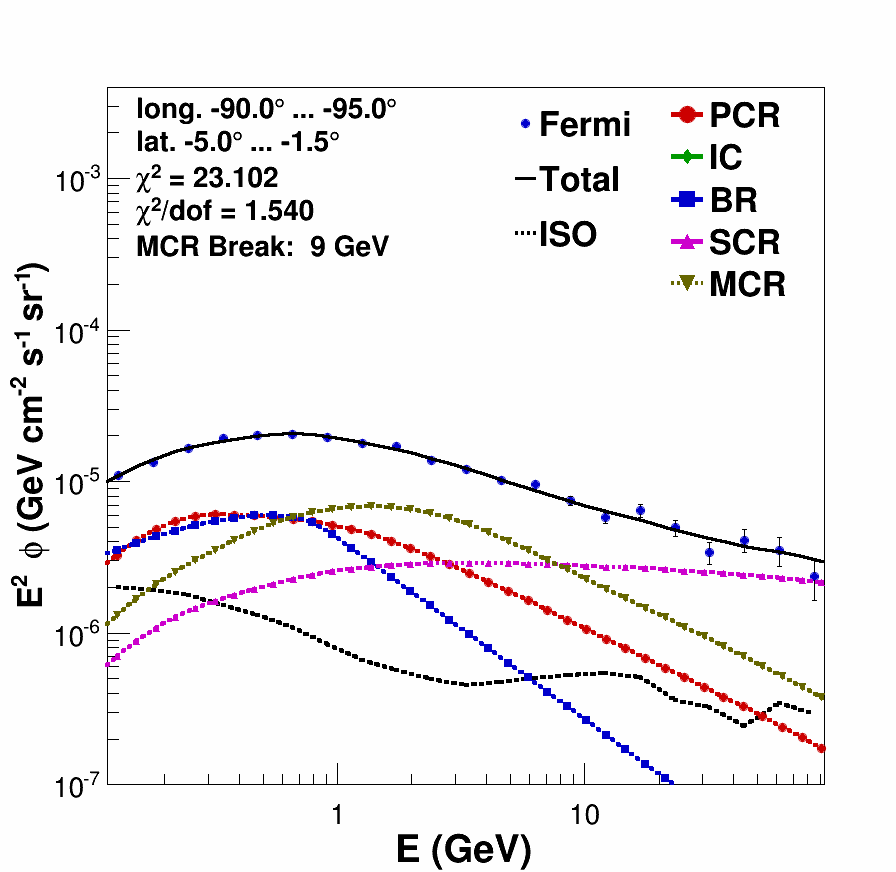}
\includegraphics[width=0.16\textwidth,height=0.16\textwidth,clip]{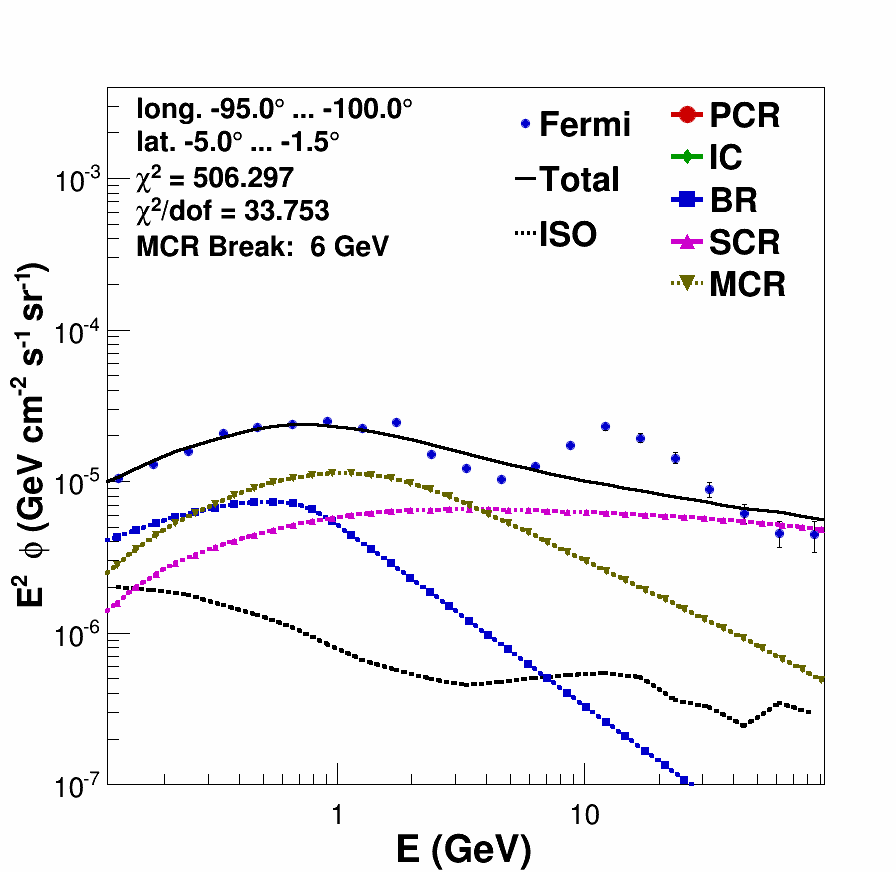}
\includegraphics[width=0.16\textwidth,height=0.16\textwidth,clip]{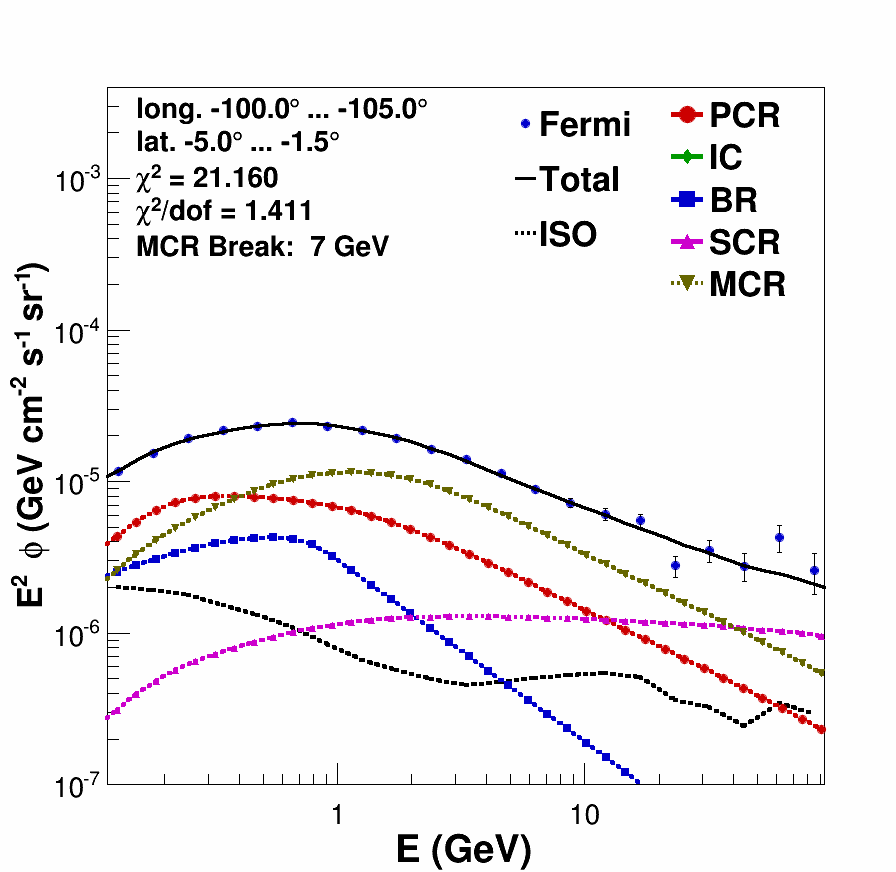}
\includegraphics[width=0.16\textwidth,height=0.16\textwidth,clip]{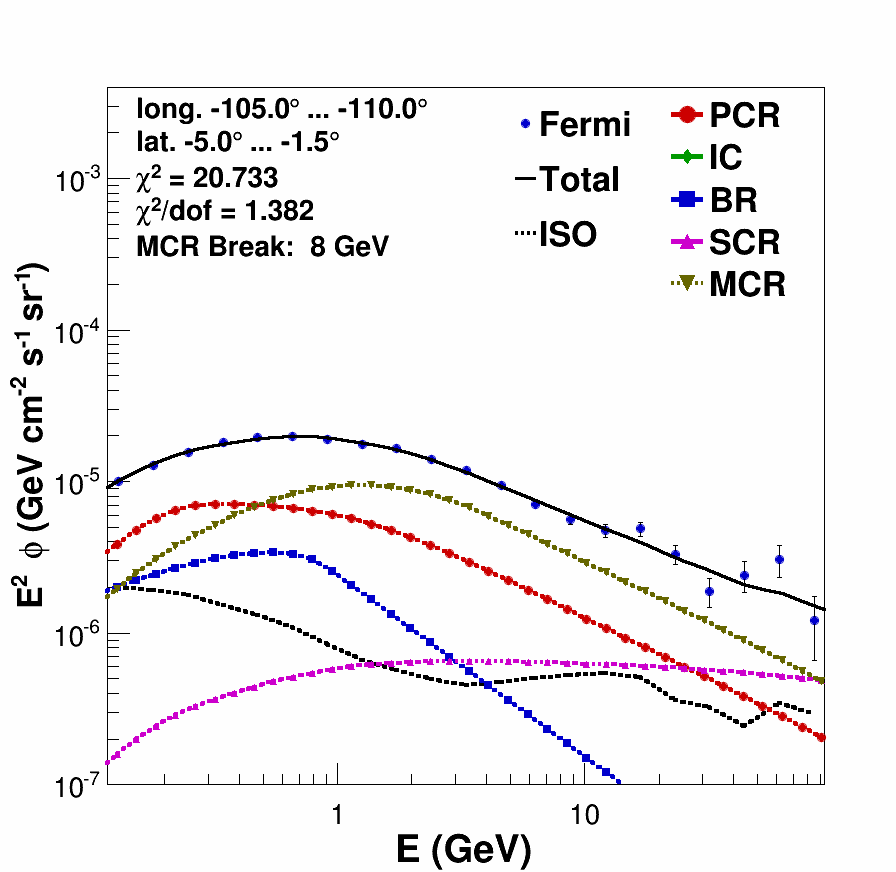}
\includegraphics[width=0.16\textwidth,height=0.16\textwidth,clip]{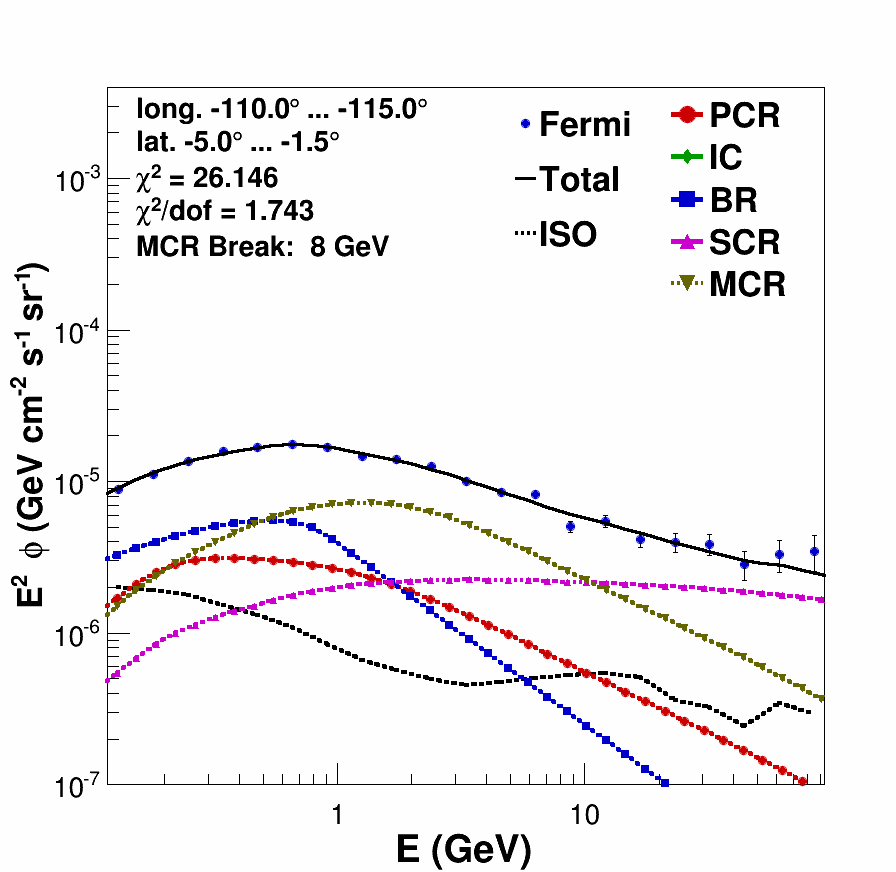}
\includegraphics[width=0.16\textwidth,height=0.16\textwidth,clip]{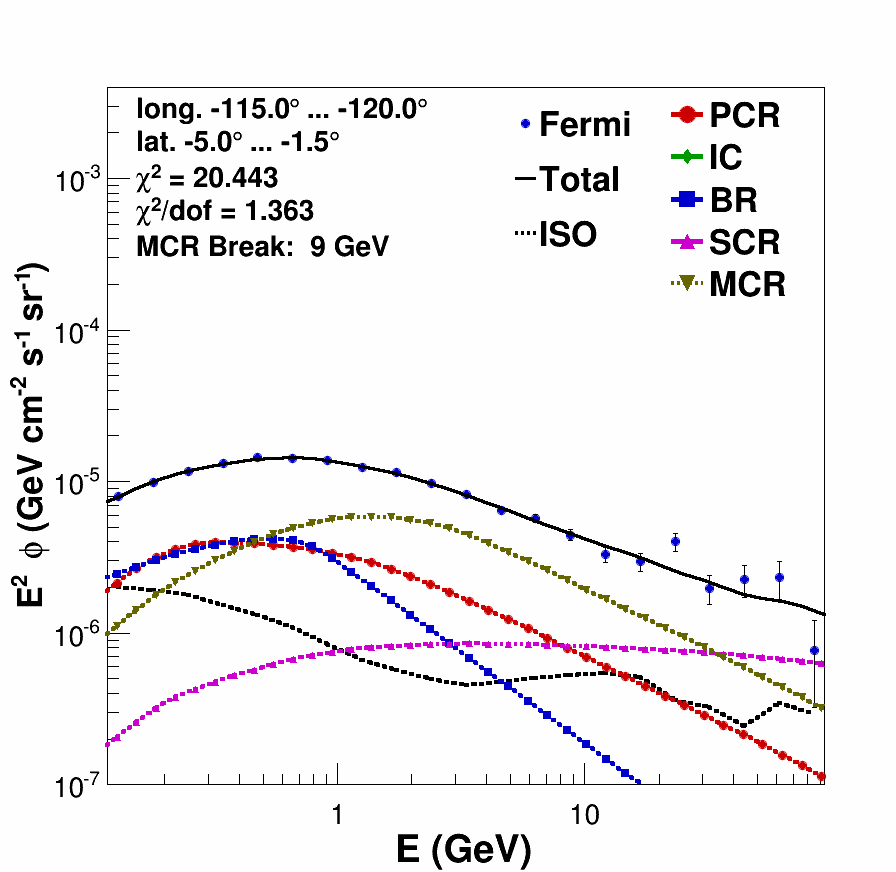}
\includegraphics[width=0.16\textwidth,height=0.16\textwidth,clip]{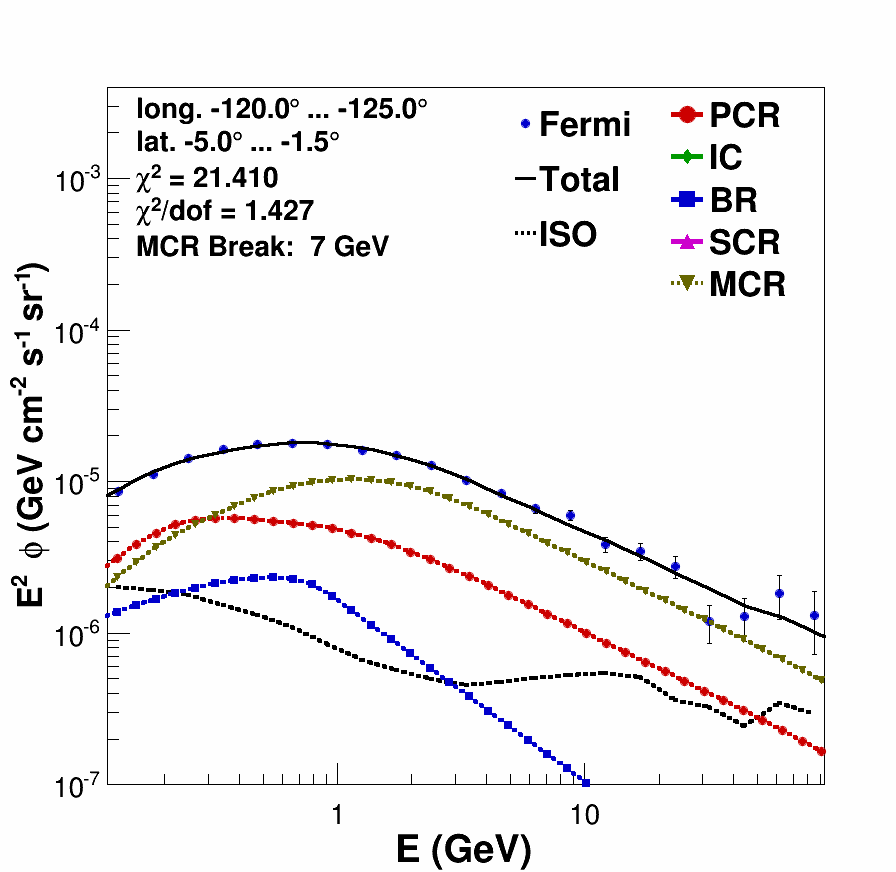}
\includegraphics[width=0.16\textwidth,height=0.16\textwidth,clip]{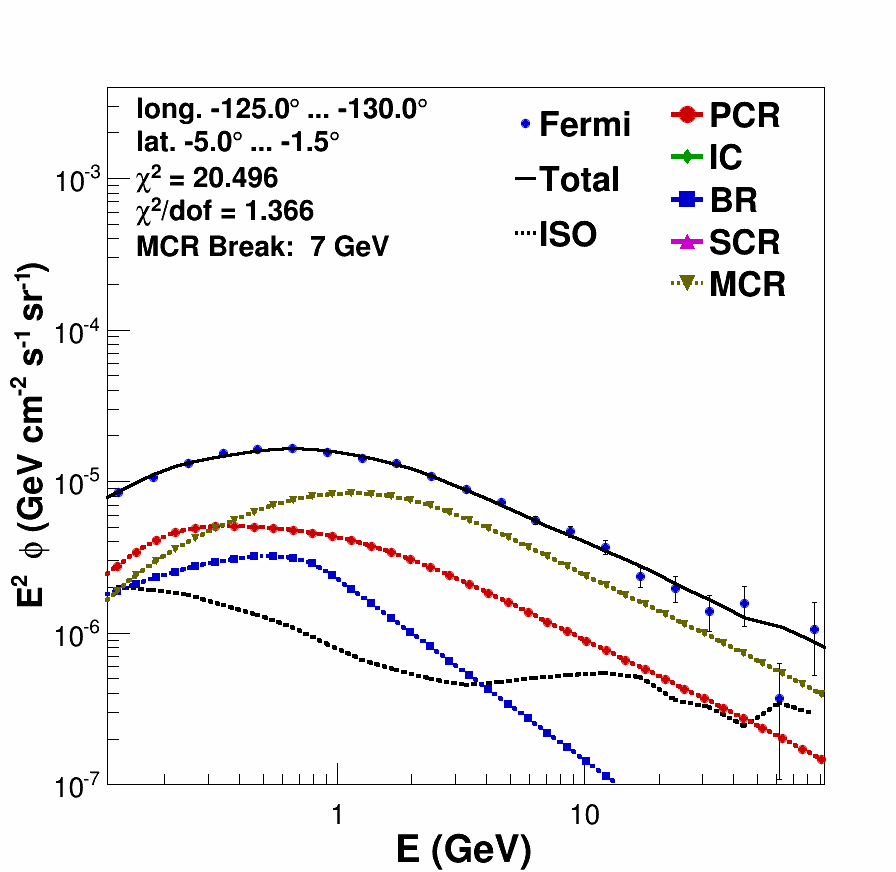}
\includegraphics[width=0.16\textwidth,height=0.16\textwidth,clip]{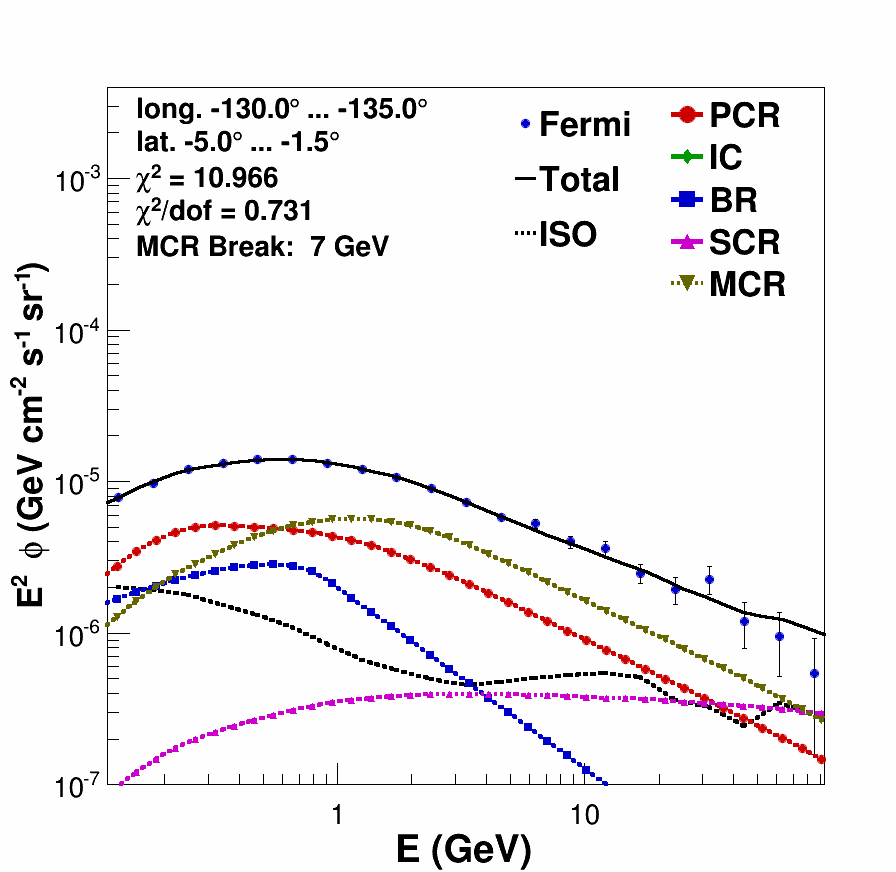}
\includegraphics[width=0.16\textwidth,height=0.16\textwidth,clip]{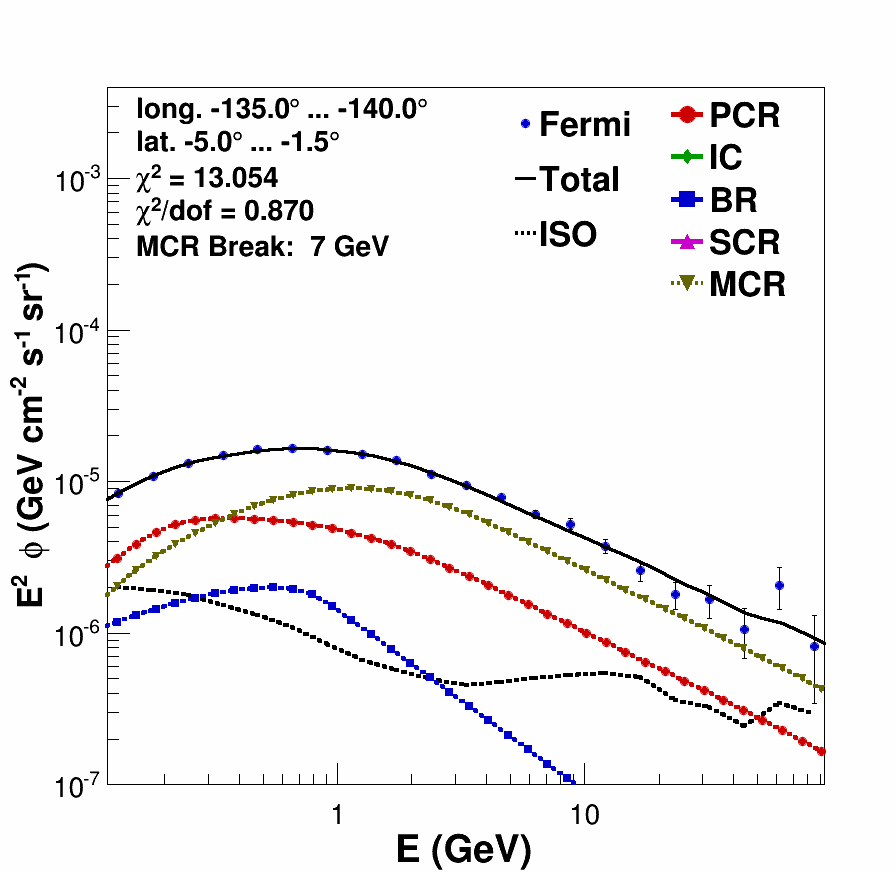}
\includegraphics[width=0.16\textwidth,height=0.16\textwidth,clip]{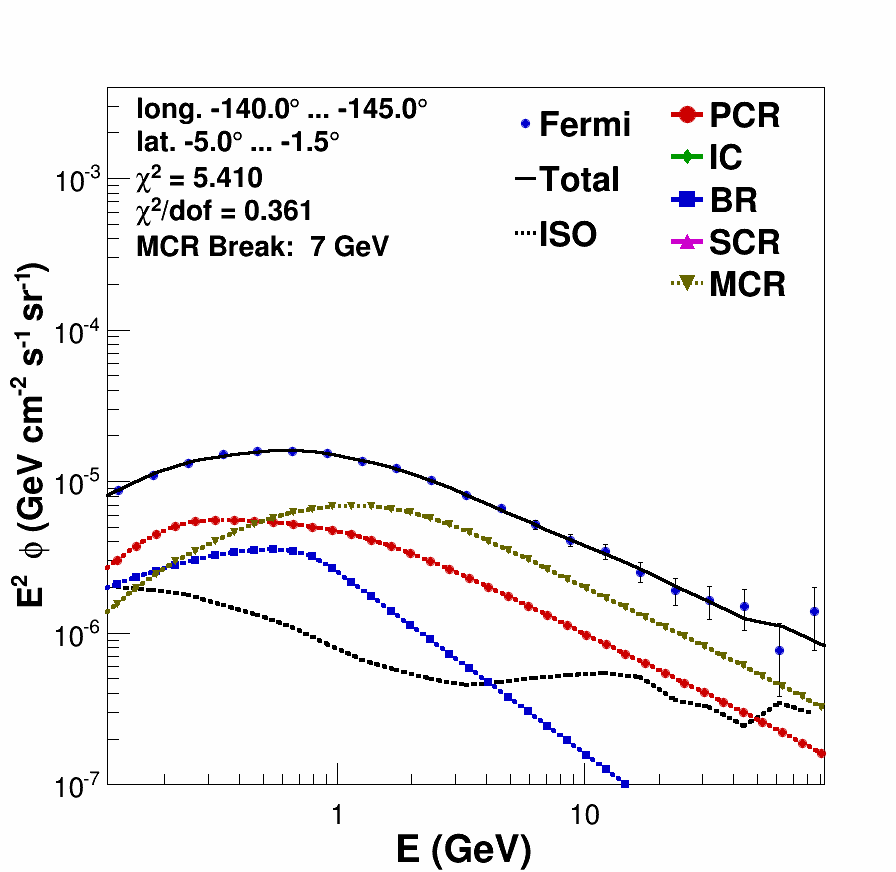}
\includegraphics[width=0.16\textwidth,height=0.16\textwidth,clip]{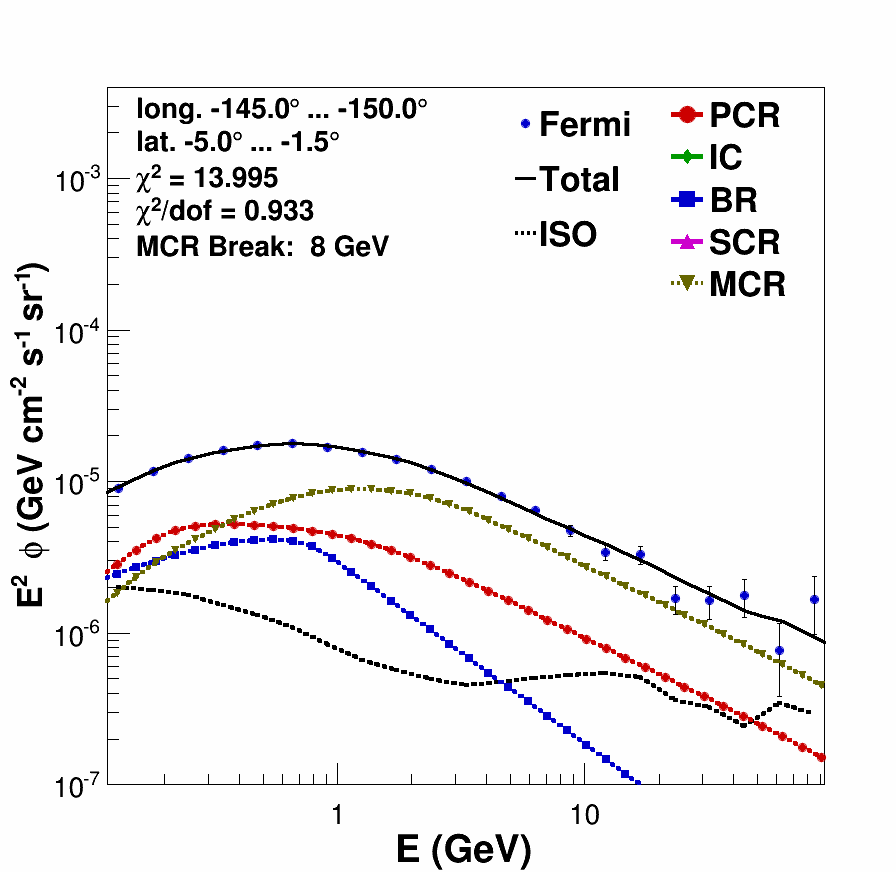}
\includegraphics[width=0.16\textwidth,height=0.16\textwidth,clip]{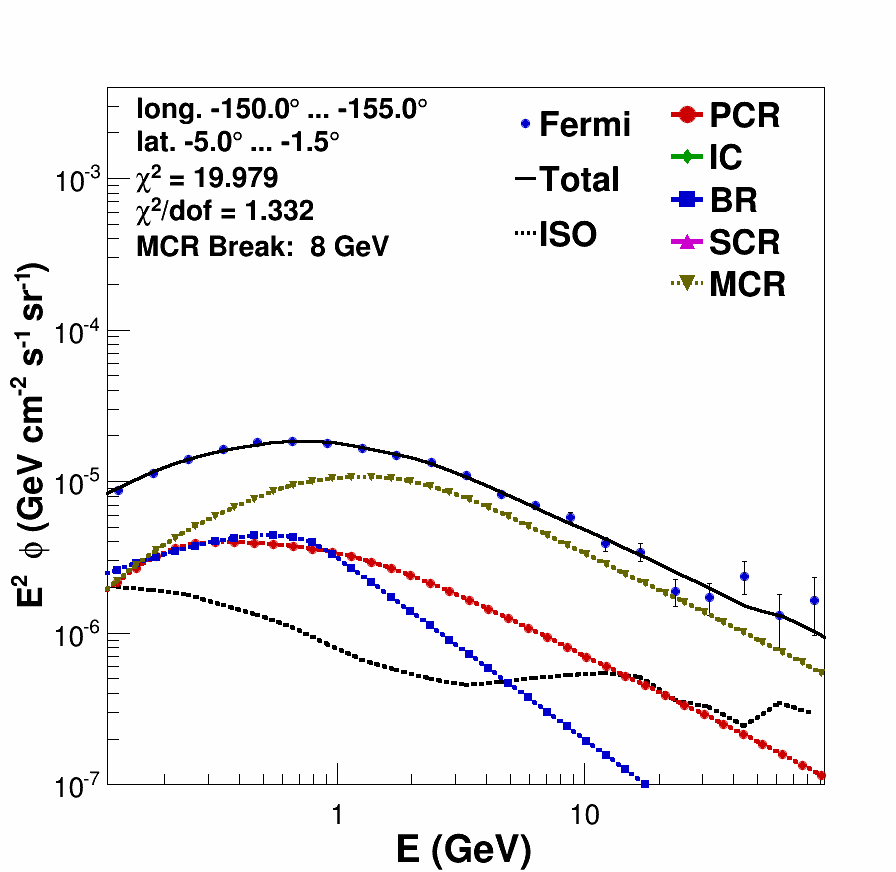}
\includegraphics[width=0.16\textwidth,height=0.16\textwidth,clip]{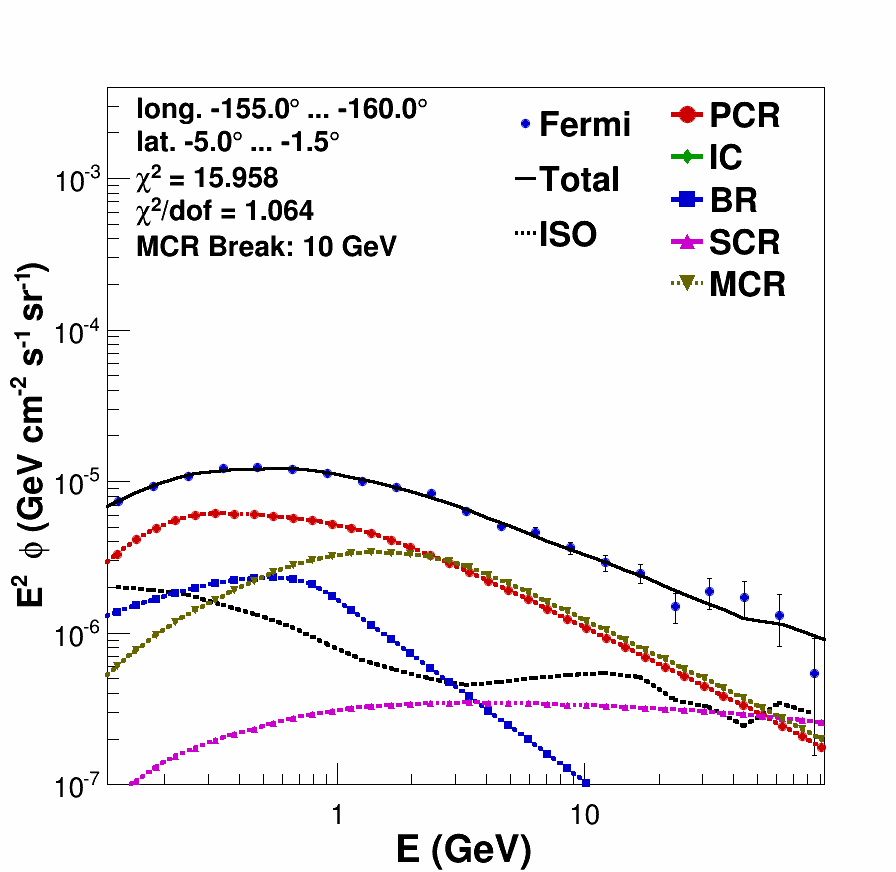}
\includegraphics[width=0.16\textwidth,height=0.16\textwidth,clip]{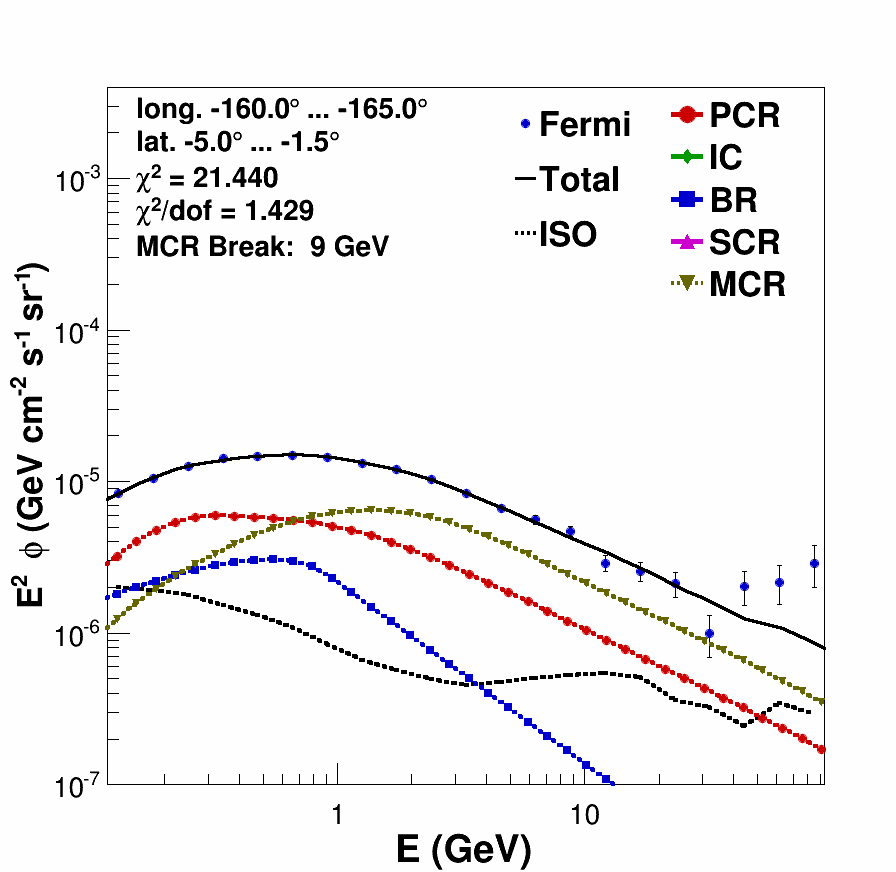}
\includegraphics[width=0.16\textwidth,height=0.16\textwidth,clip]{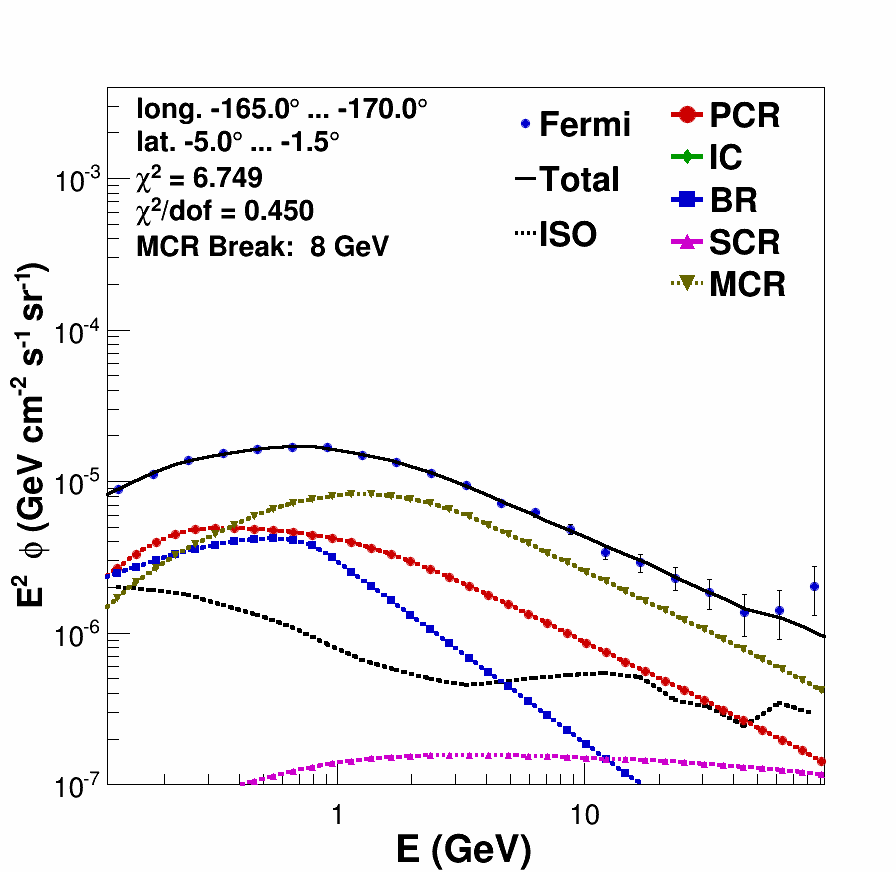}
\includegraphics[width=0.16\textwidth,height=0.16\textwidth,clip]{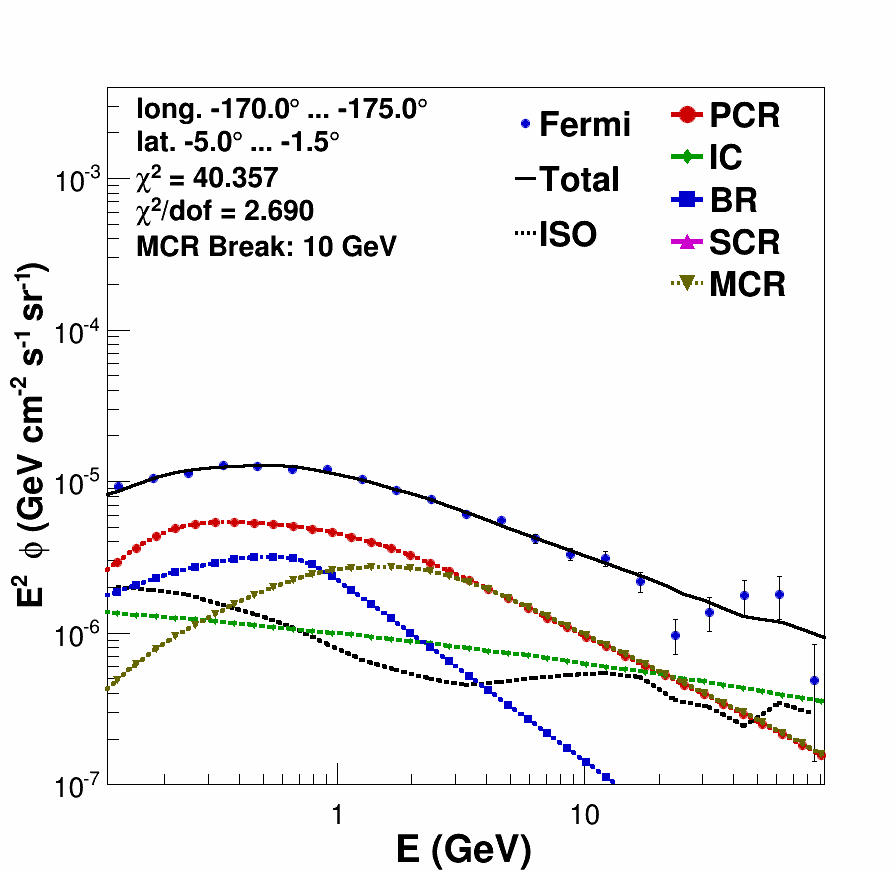}
\includegraphics[width=0.16\textwidth,height=0.16\textwidth,clip]{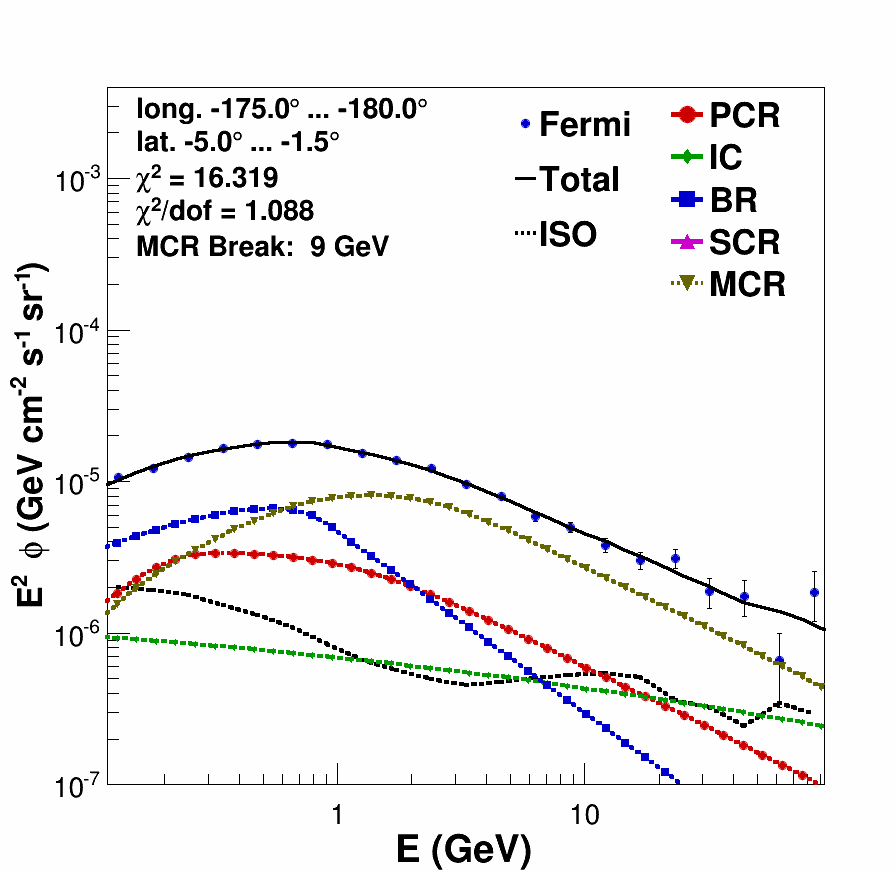}%%%%%r12b
\caption[]{Template fits for latitudes  with $-5.0^\circ<b<-1.5^\circ$ and longitudes decreasing from 0$^\circ$ to -180$^\circ$.} \label{F24}
\end{figure}
\begin{figure}
\centering
\includegraphics[width=0.16\textwidth,height=0.16\textwidth,clip]{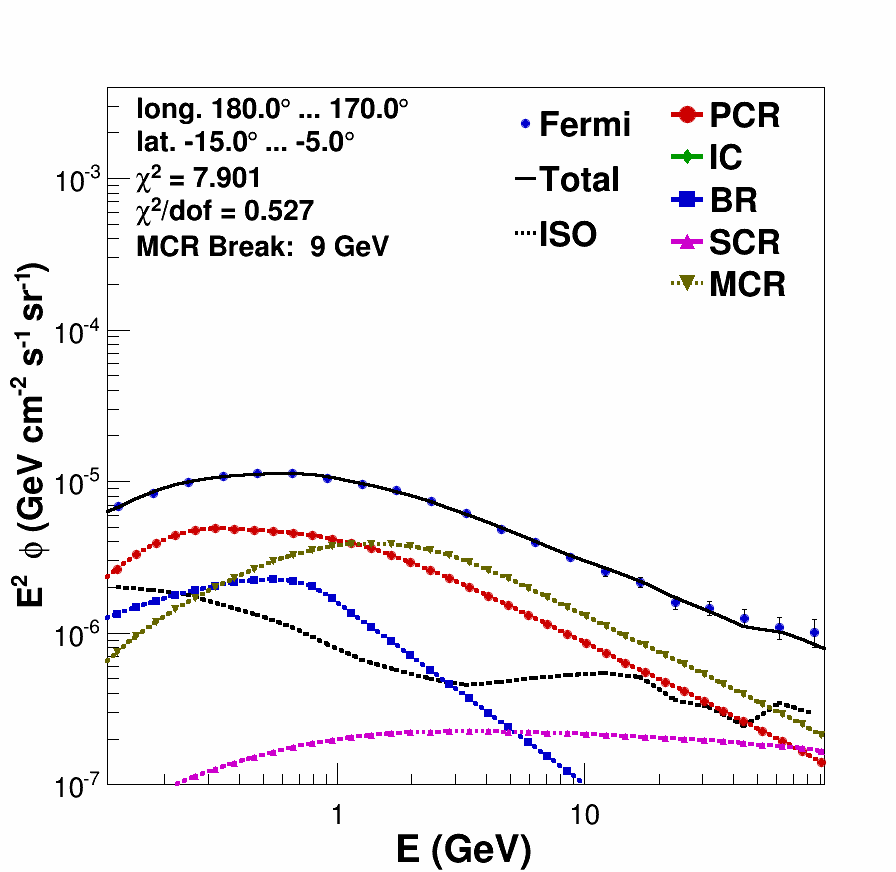}
\includegraphics[width=0.16\textwidth,height=0.16\textwidth,clip]{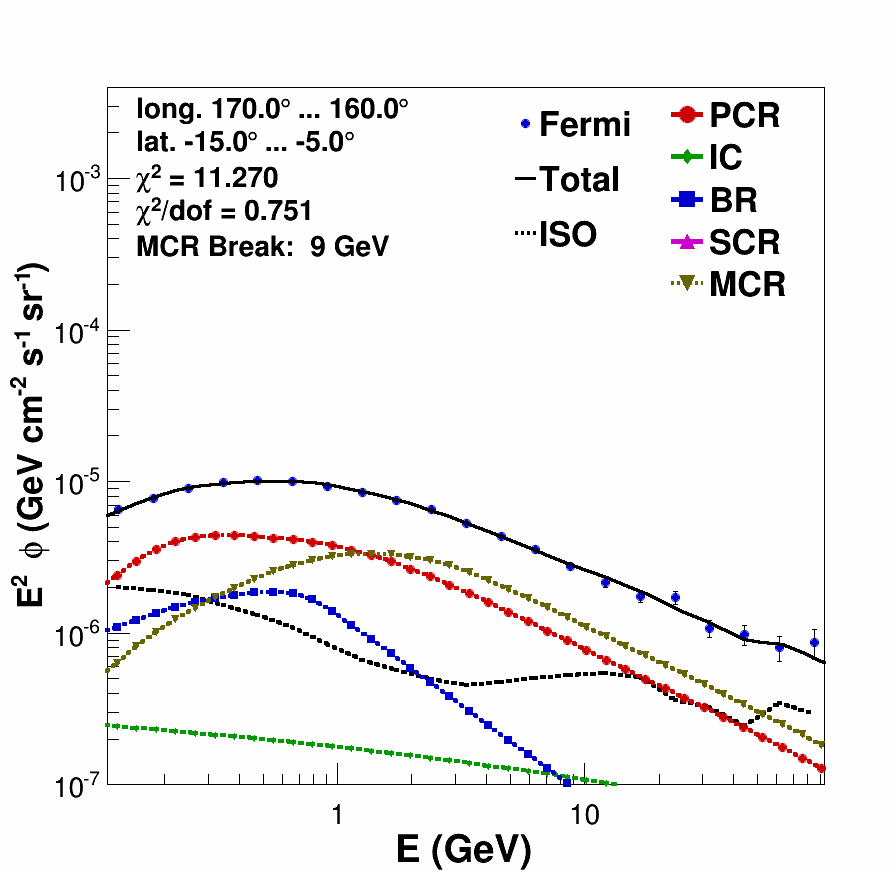}
\includegraphics[width=0.16\textwidth,height=0.16\textwidth,clip]{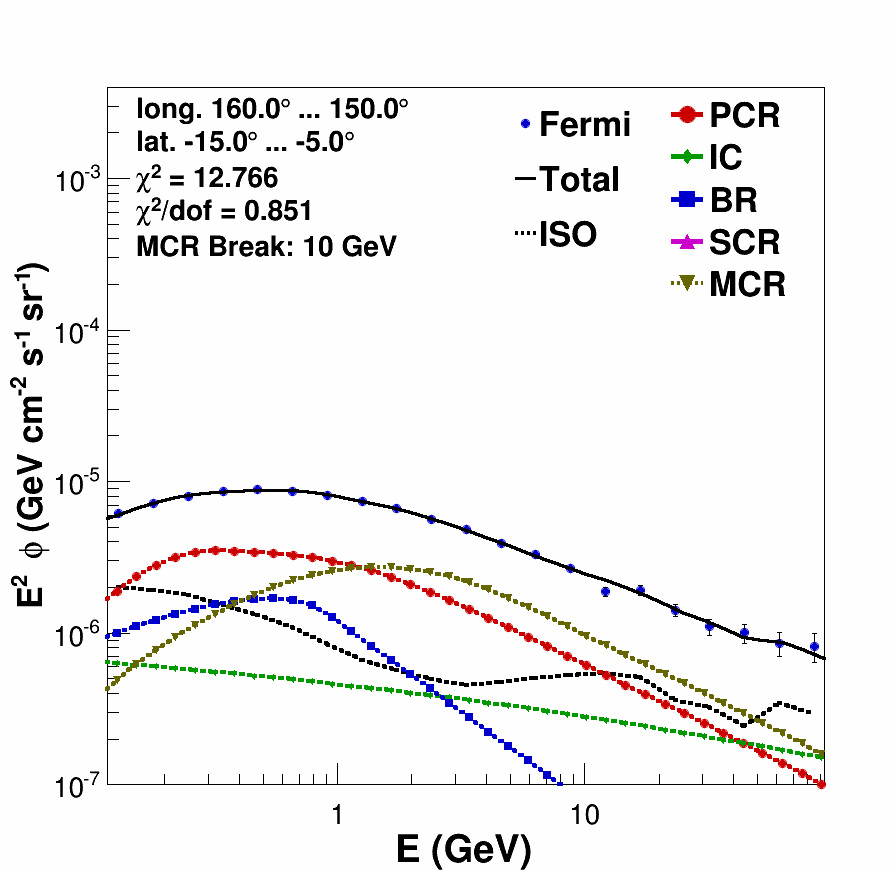}
\includegraphics[width=0.16\textwidth,height=0.16\textwidth,clip]{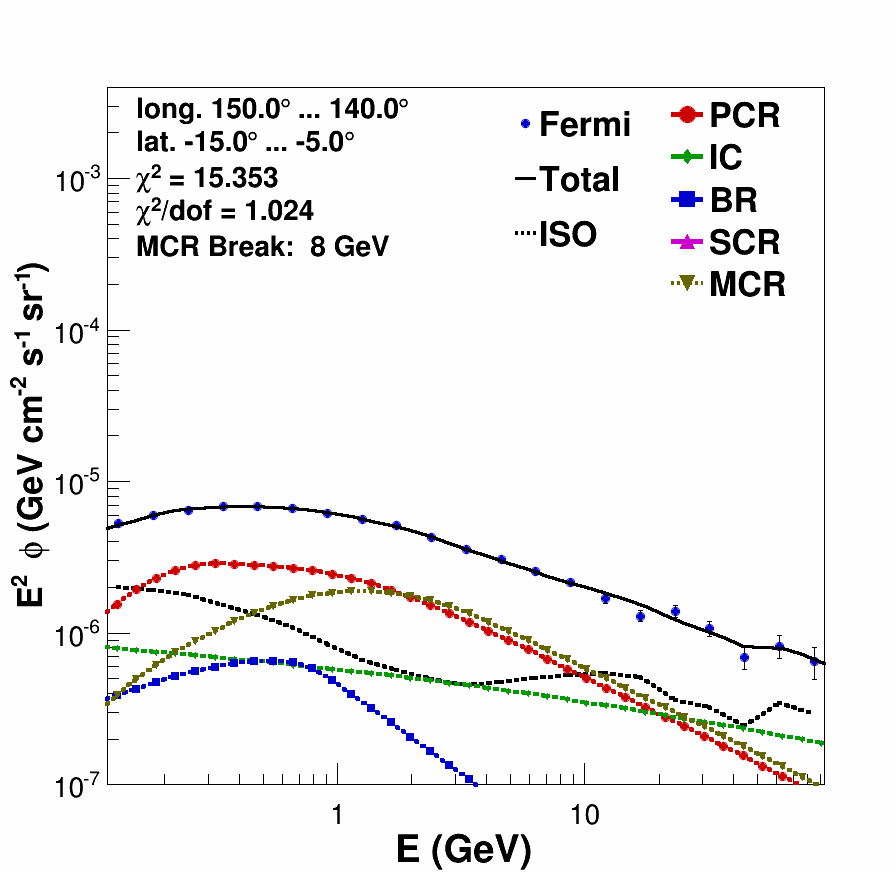}
\includegraphics[width=0.16\textwidth,height=0.16\textwidth,clip]{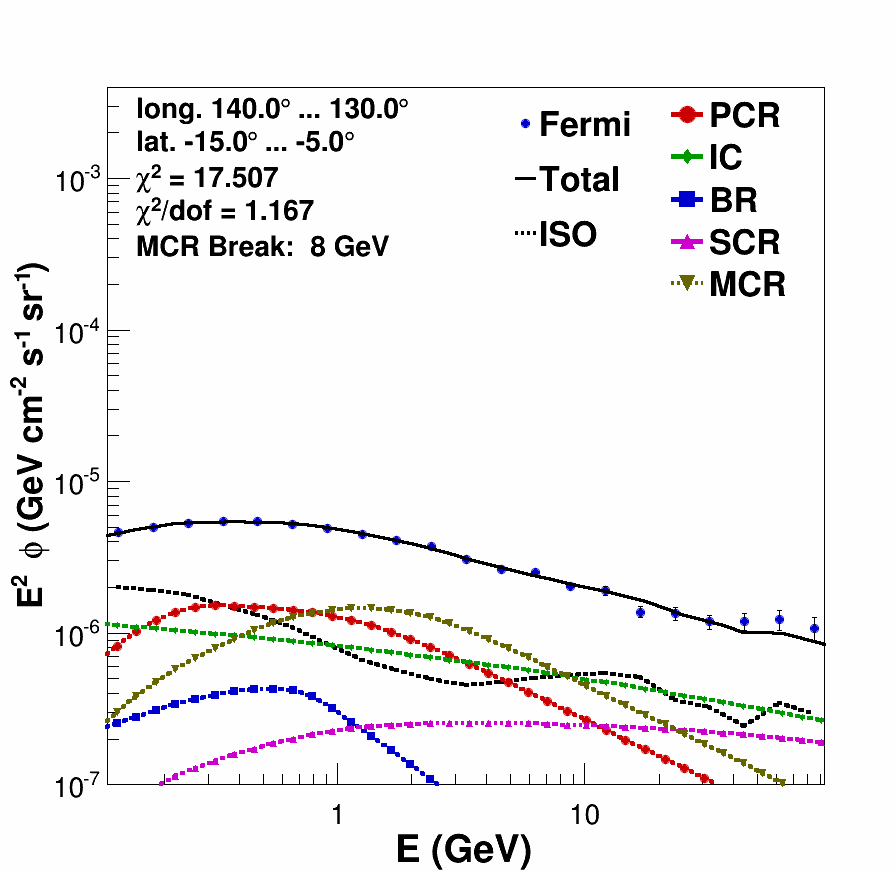}
\includegraphics[width=0.16\textwidth,height=0.16\textwidth,clip]{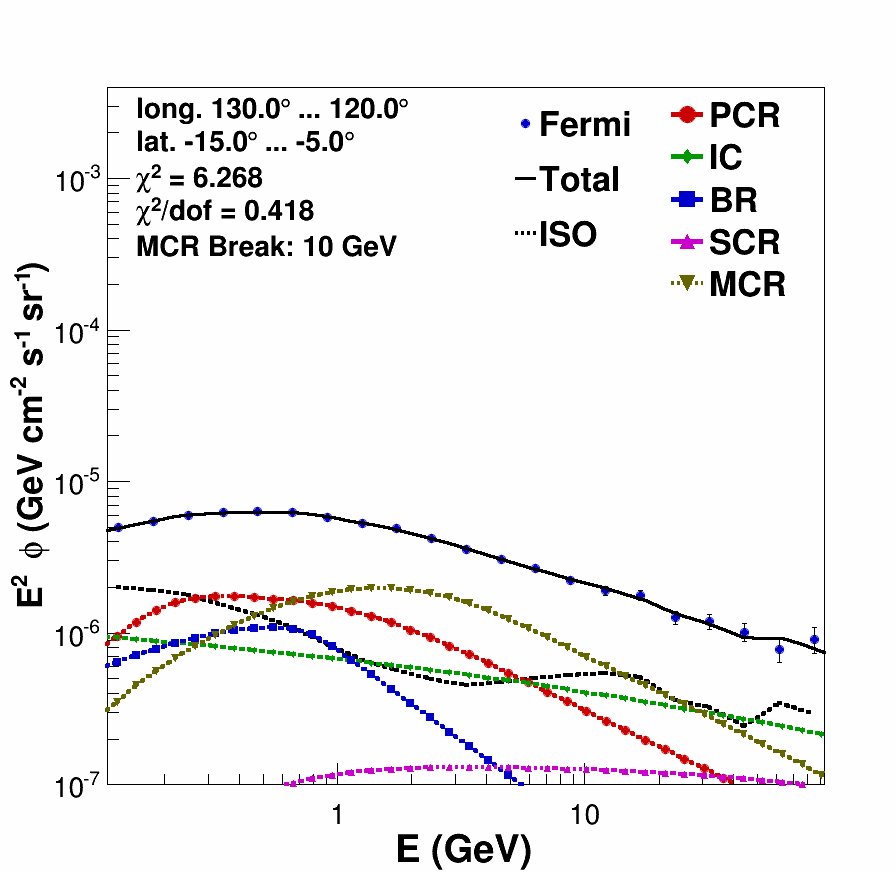}
\includegraphics[width=0.16\textwidth,height=0.16\textwidth,clip]{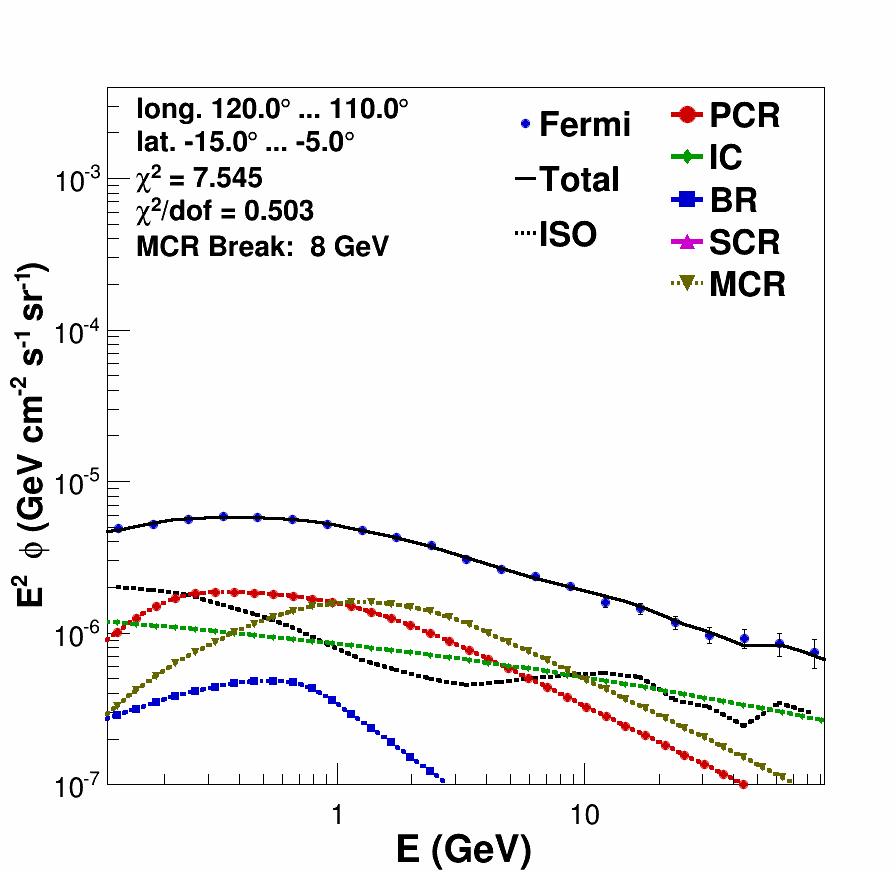}
\includegraphics[width=0.16\textwidth,height=0.16\textwidth,clip]{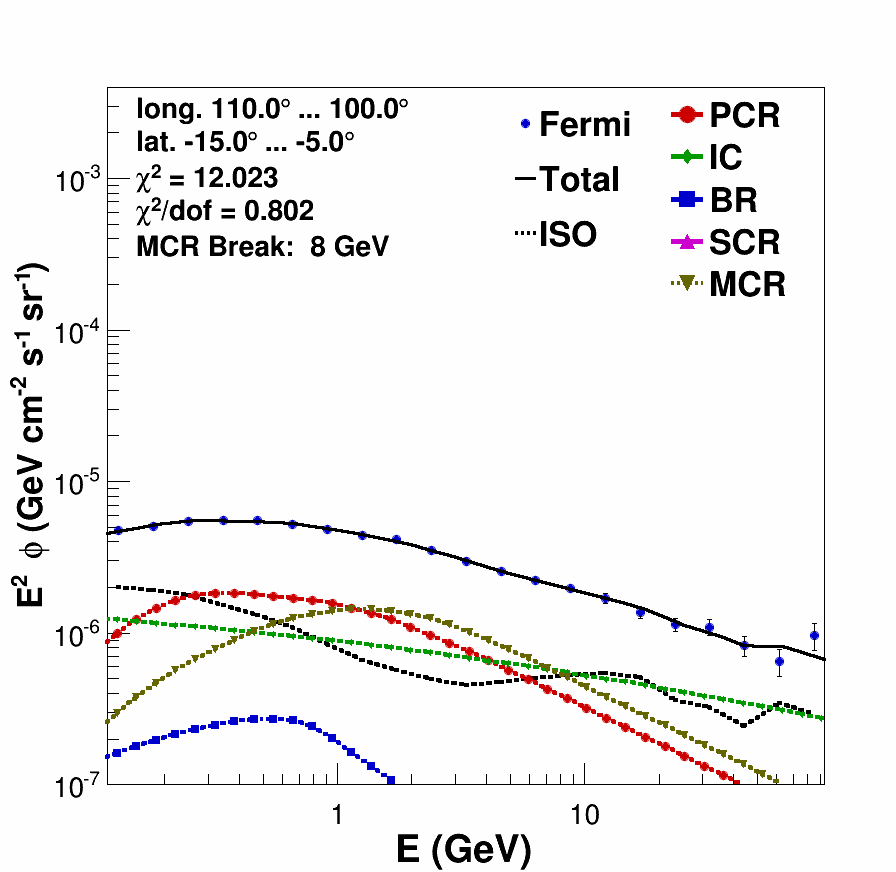}
\includegraphics[width=0.16\textwidth,height=0.16\textwidth,clip]{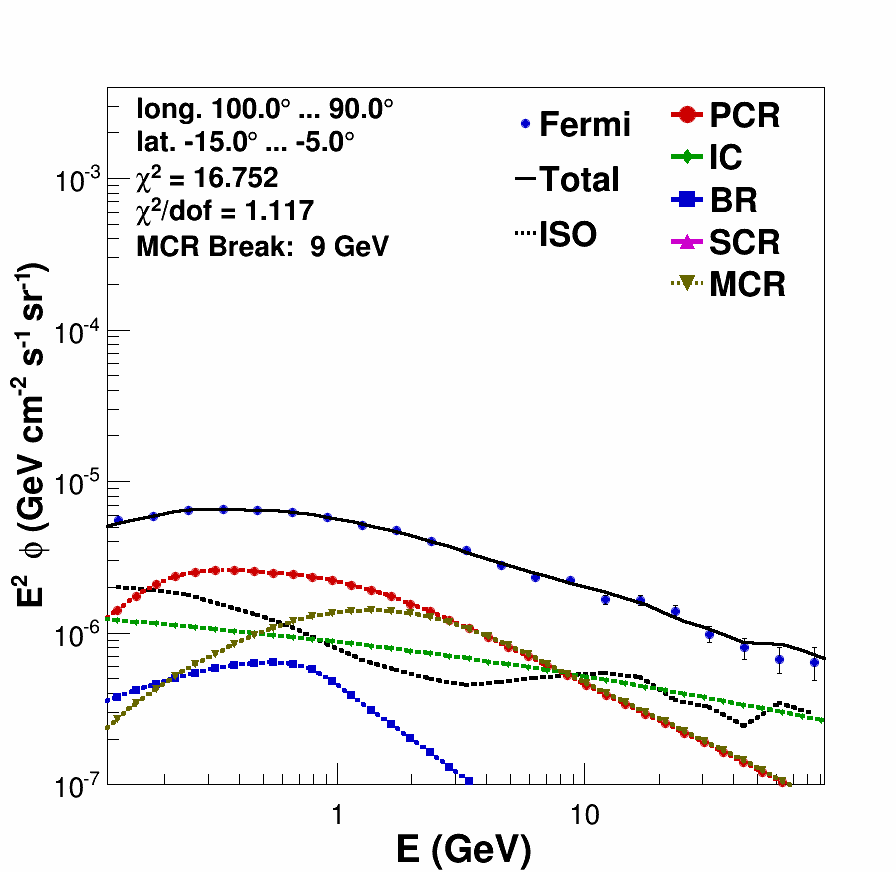}
\includegraphics[width=0.16\textwidth,height=0.16\textwidth,clip]{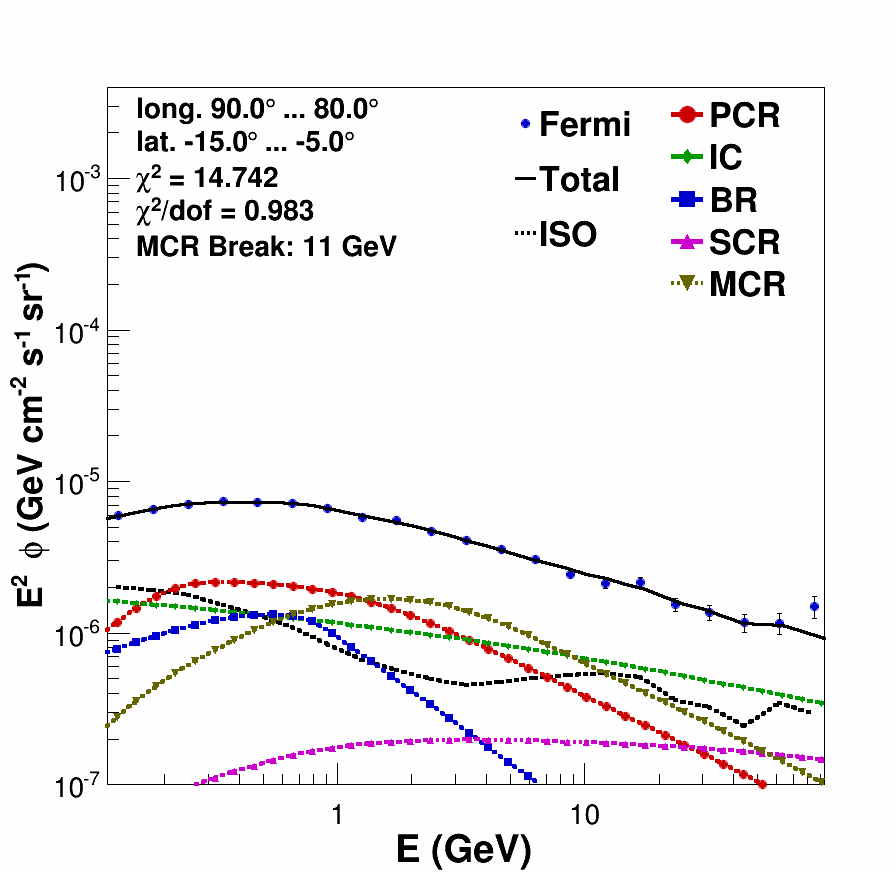}
\includegraphics[width=0.16\textwidth,height=0.16\textwidth,clip]{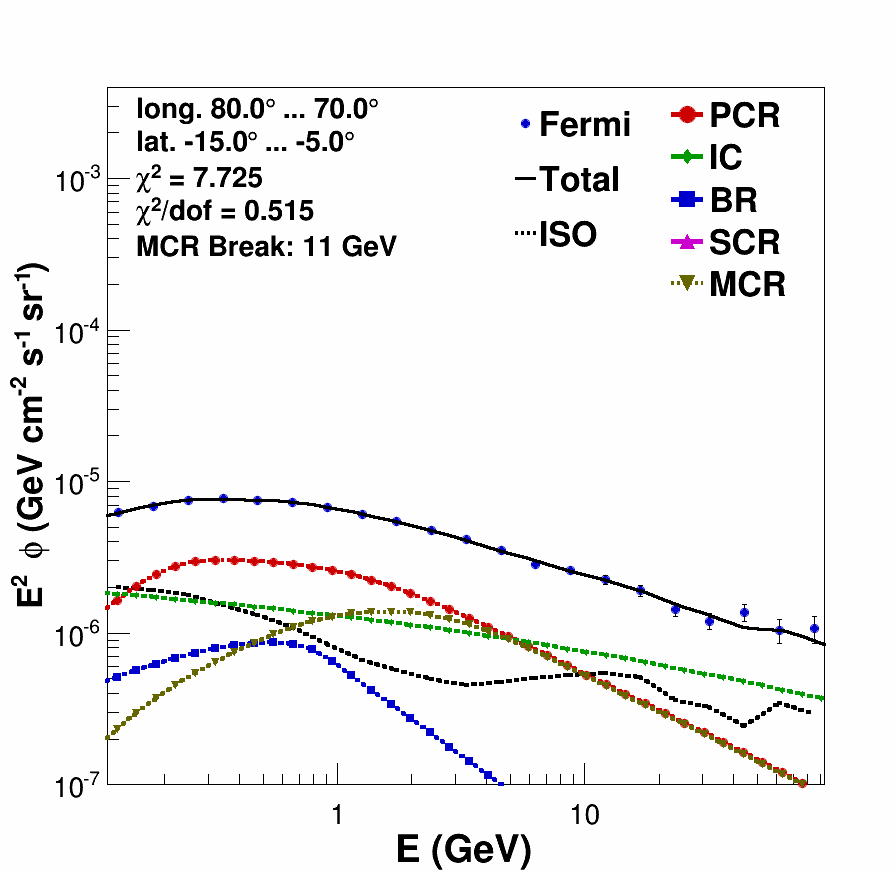}
\includegraphics[width=0.16\textwidth,height=0.16\textwidth,clip]{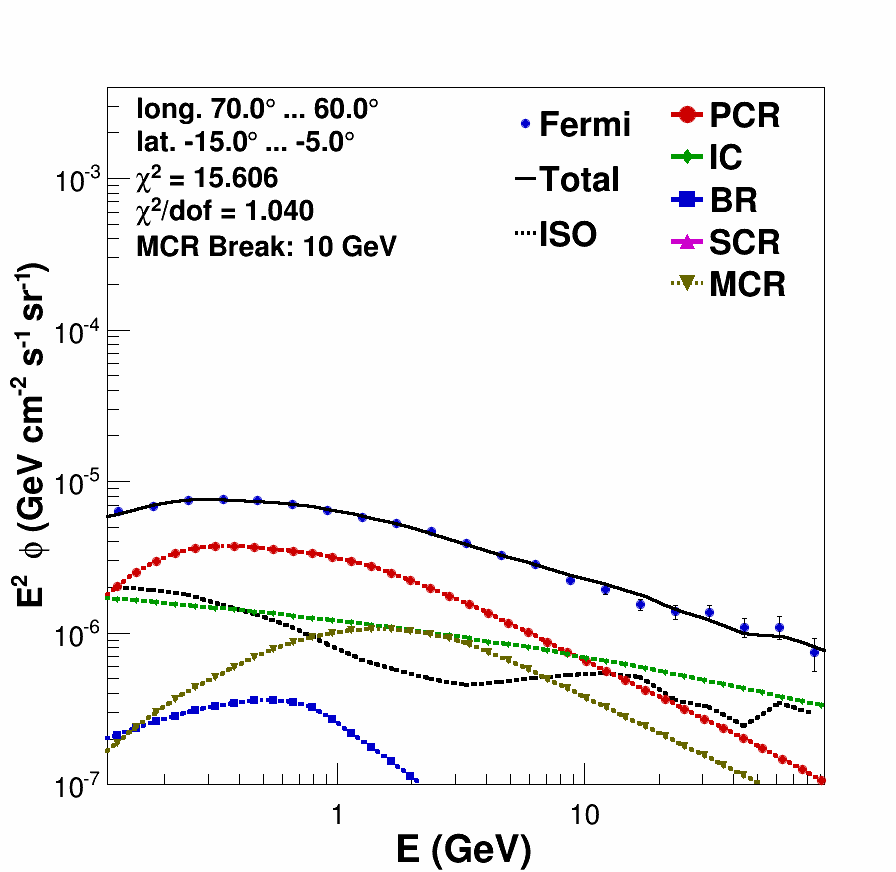}
\includegraphics[width=0.16\textwidth,height=0.16\textwidth,clip]{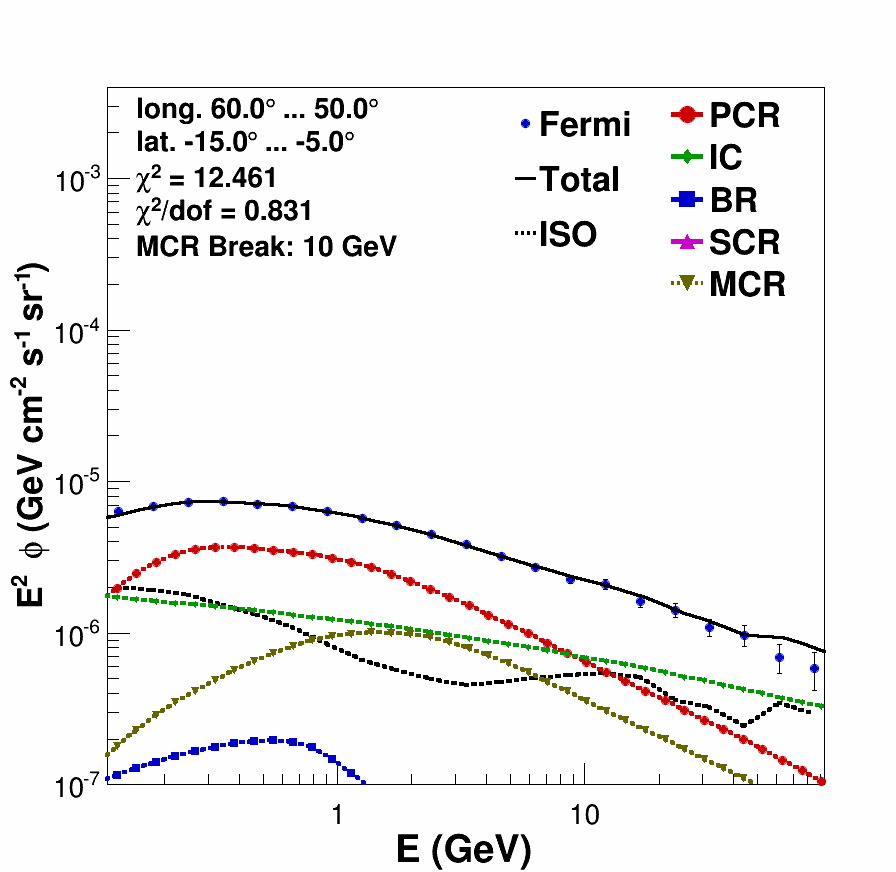}
\includegraphics[width=0.16\textwidth,height=0.16\textwidth,clip]{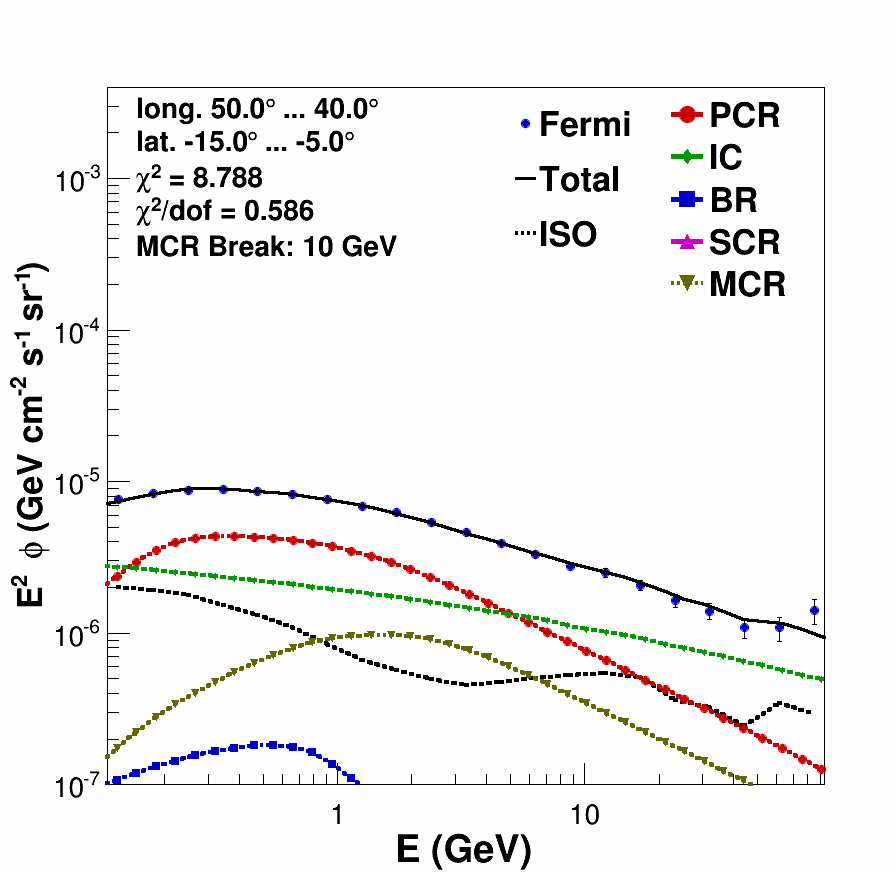}
\includegraphics[width=0.16\textwidth,height=0.16\textwidth,clip]{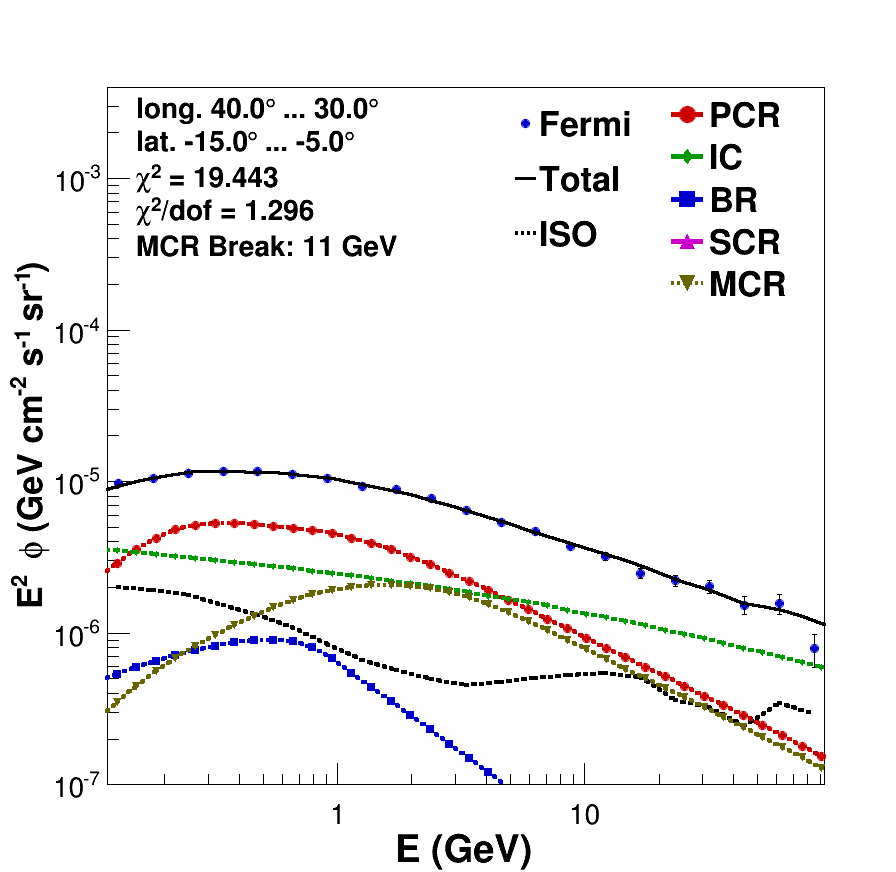}
\includegraphics[width=0.16\textwidth,height=0.16\textwidth,clip]{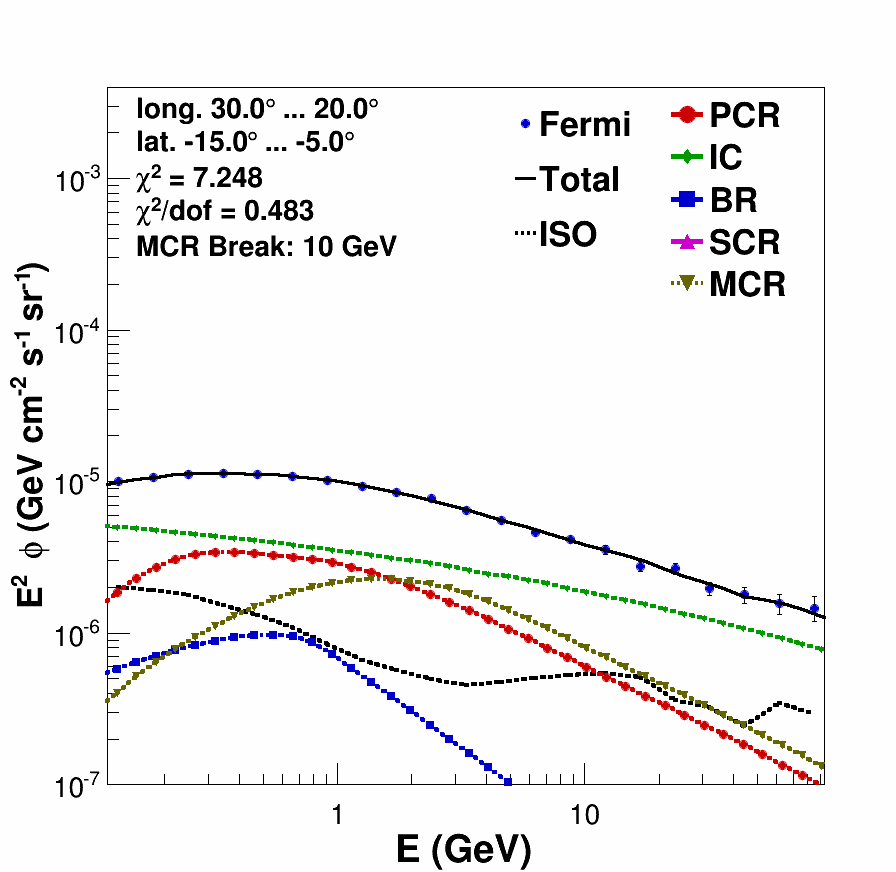}
\includegraphics[width=0.16\textwidth,height=0.16\textwidth,clip]{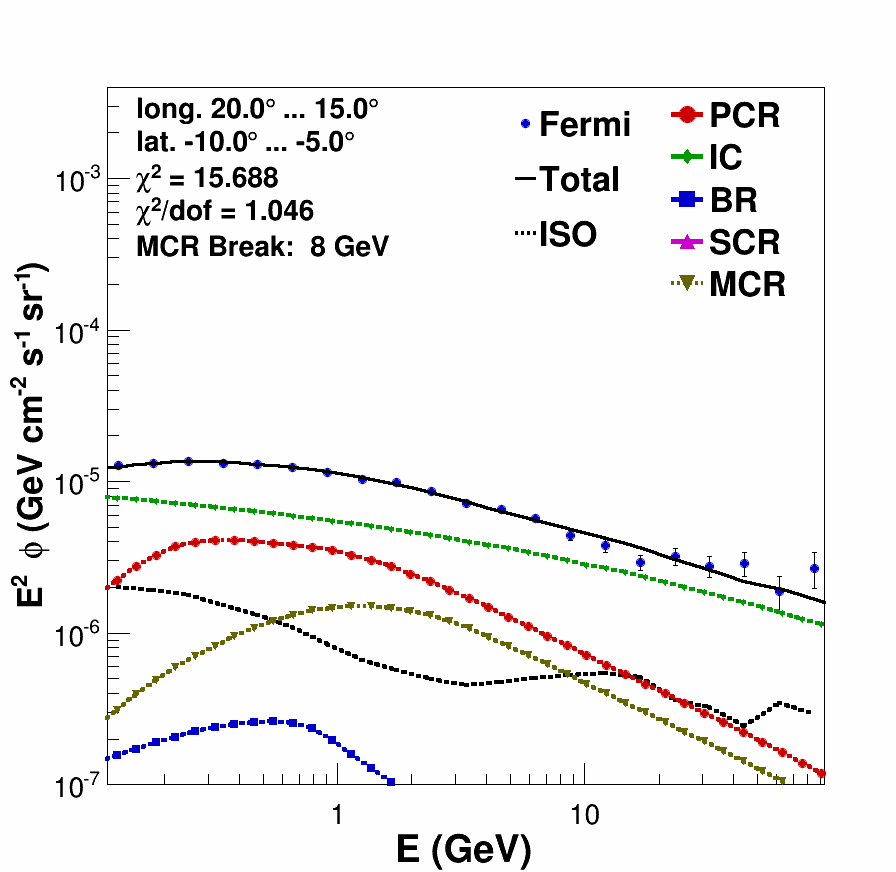}
\includegraphics[width=0.16\textwidth,height=0.16\textwidth,clip]{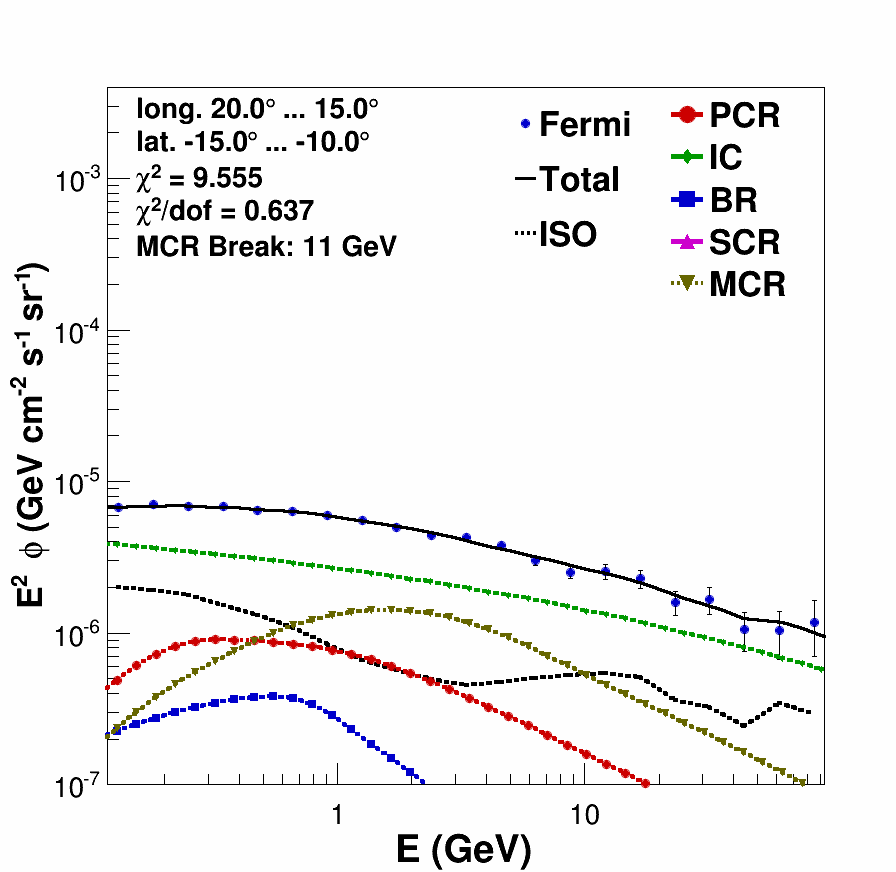}
\includegraphics[width=0.16\textwidth,height=0.16\textwidth,clip]{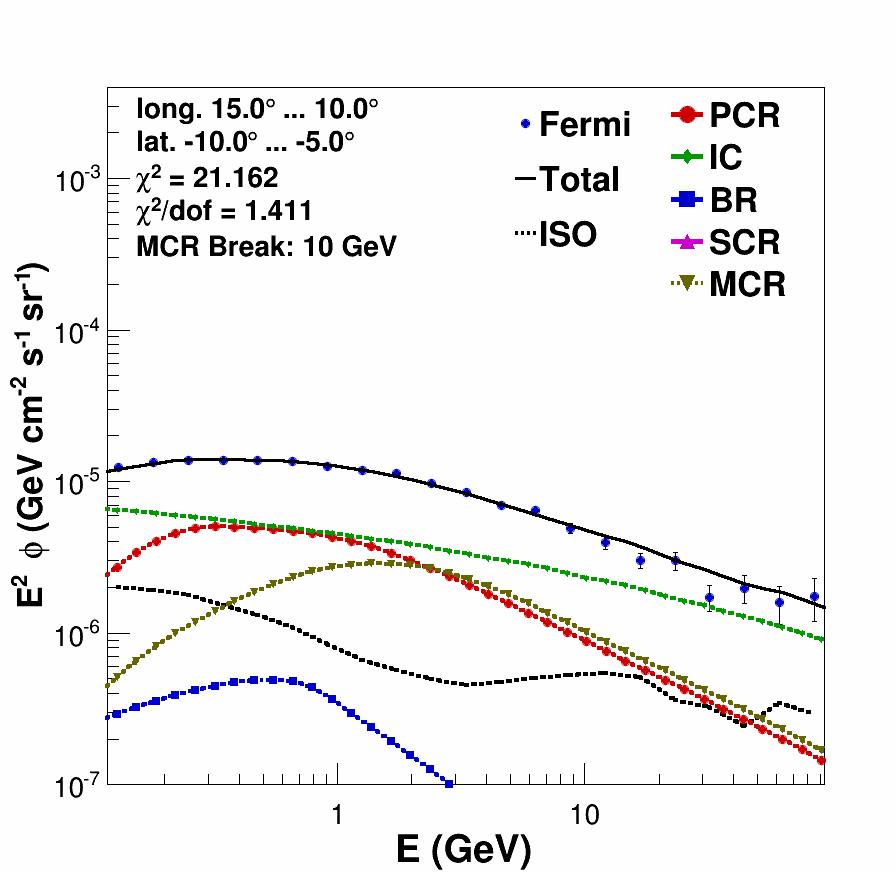}
\includegraphics[width=0.16\textwidth,height=0.16\textwidth,clip]{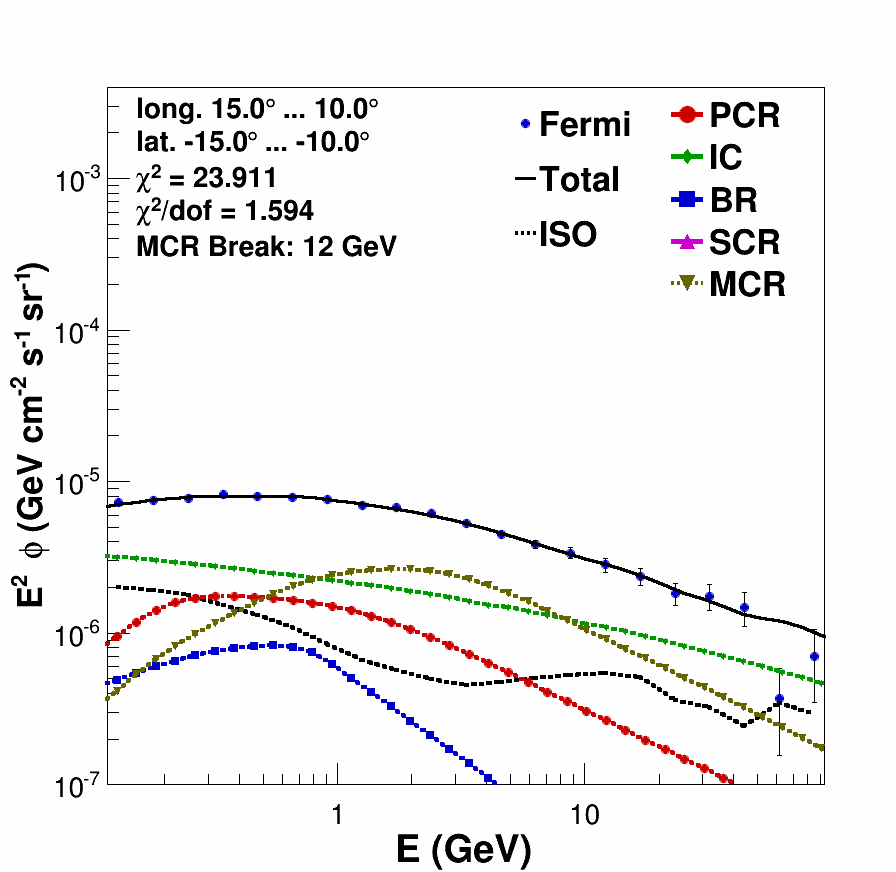}
\includegraphics[width=0.16\textwidth,height=0.16\textwidth,clip]{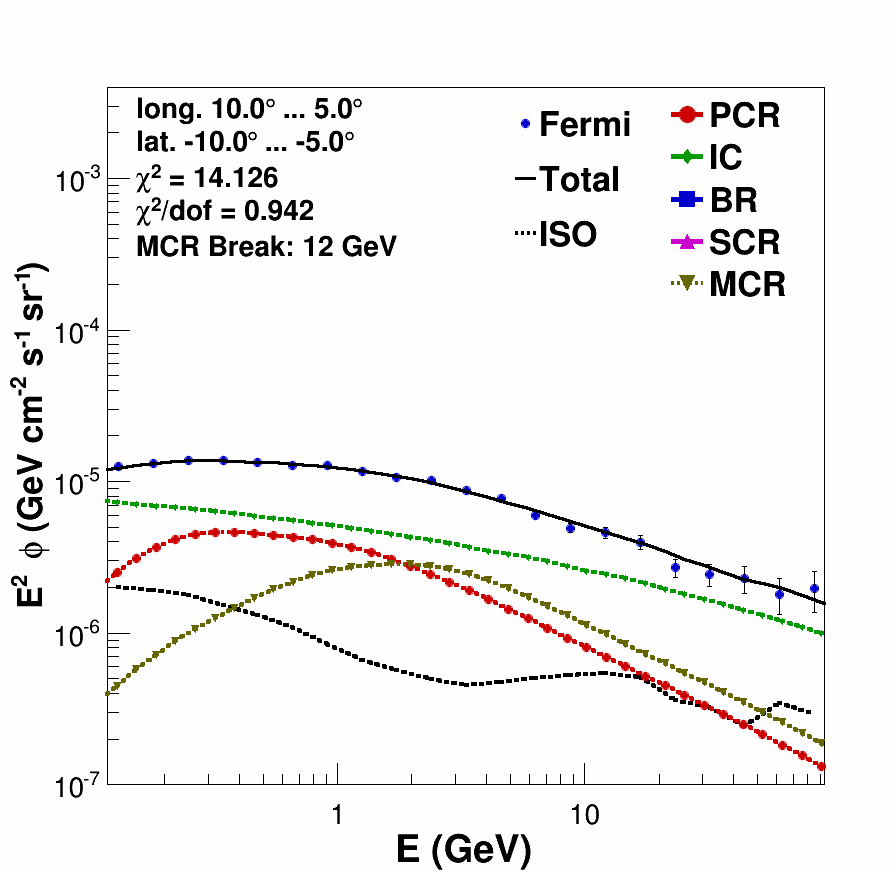}
\includegraphics[width=0.16\textwidth,height=0.16\textwidth,clip]{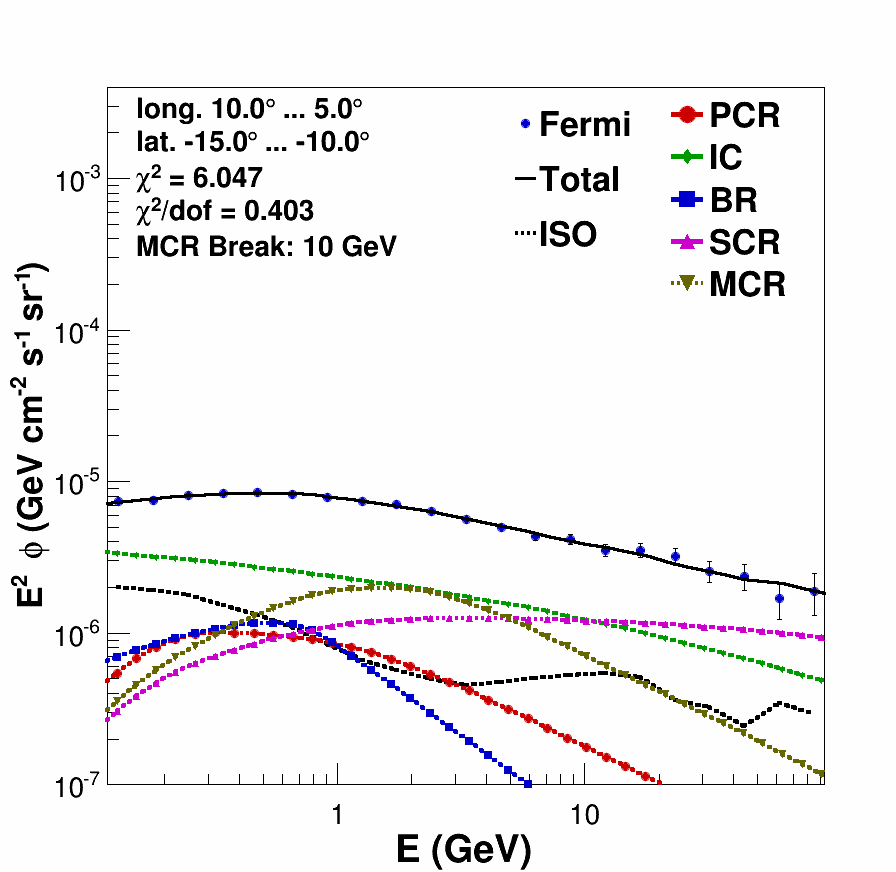}
\includegraphics[width=0.16\textwidth,height=0.16\textwidth,clip]{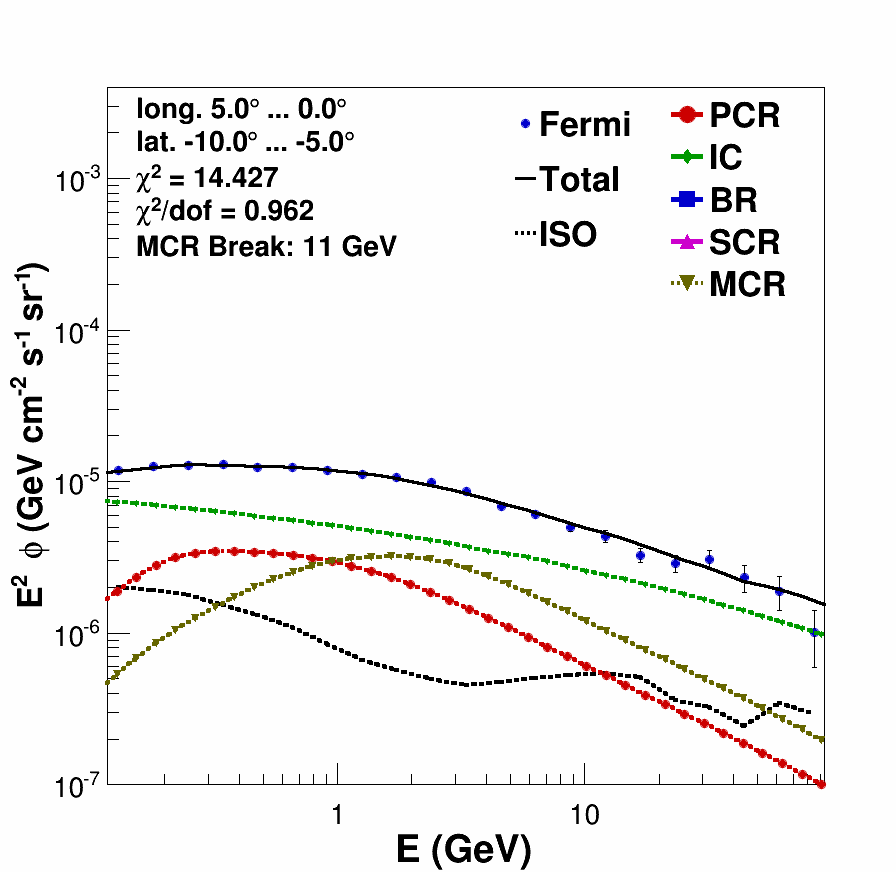}
\includegraphics[width=0.16\textwidth,height=0.16\textwidth,clip]{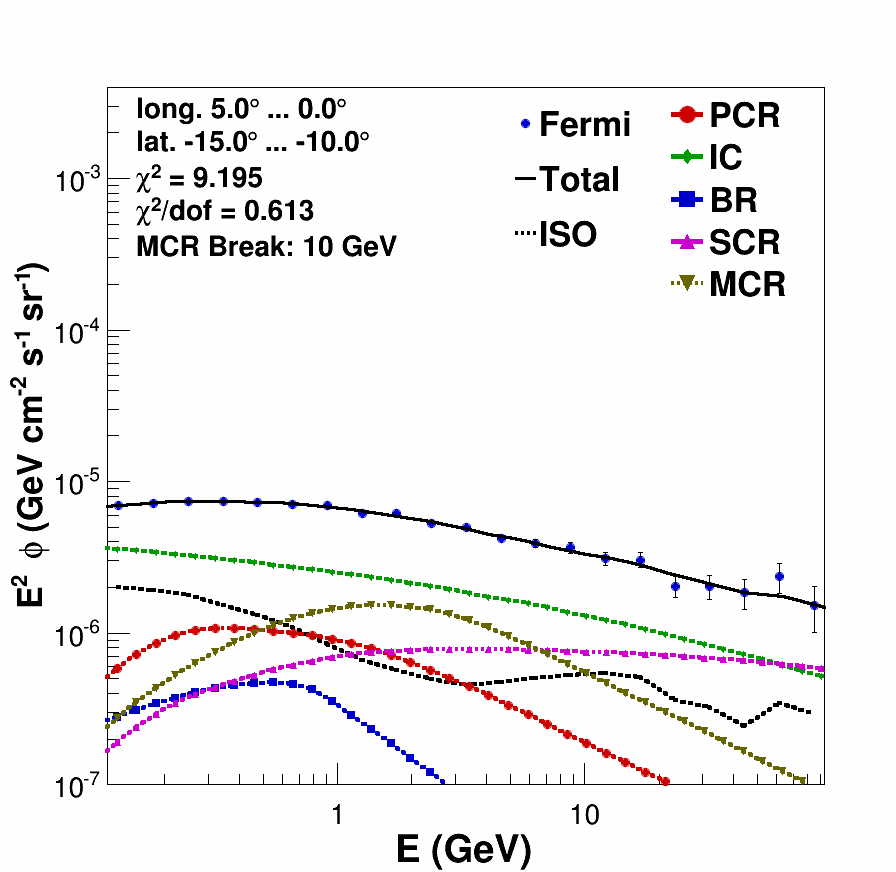}
\includegraphics[width=0.16\textwidth,height=0.16\textwidth,clip]{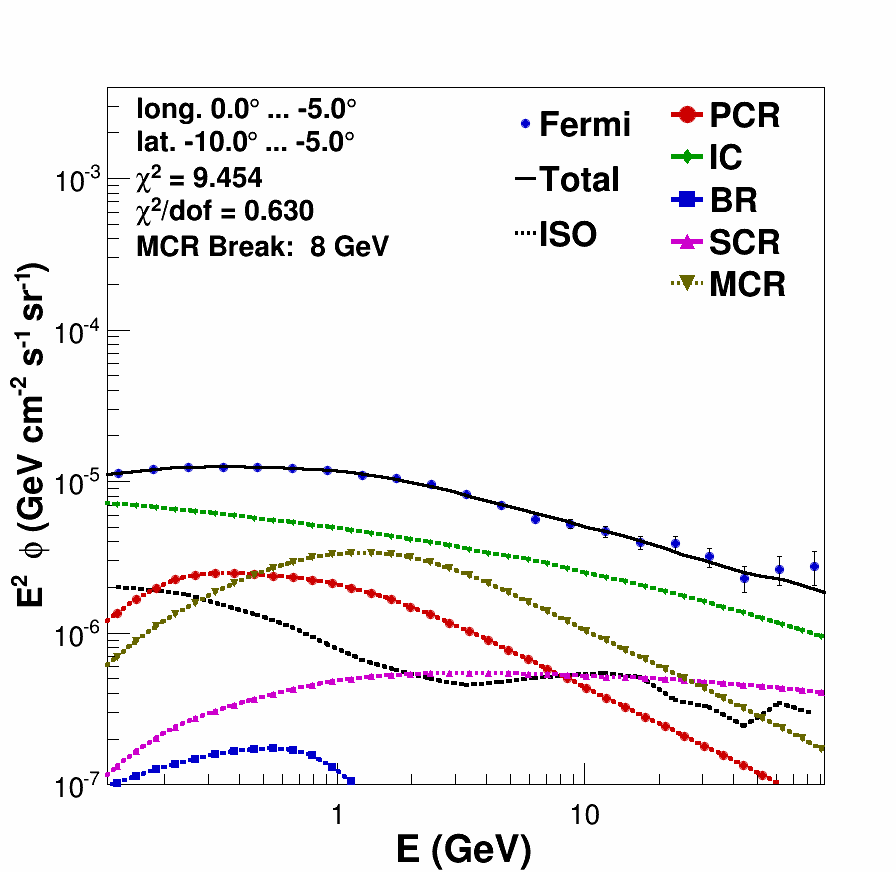}
\includegraphics[width=0.16\textwidth,height=0.16\textwidth,clip]{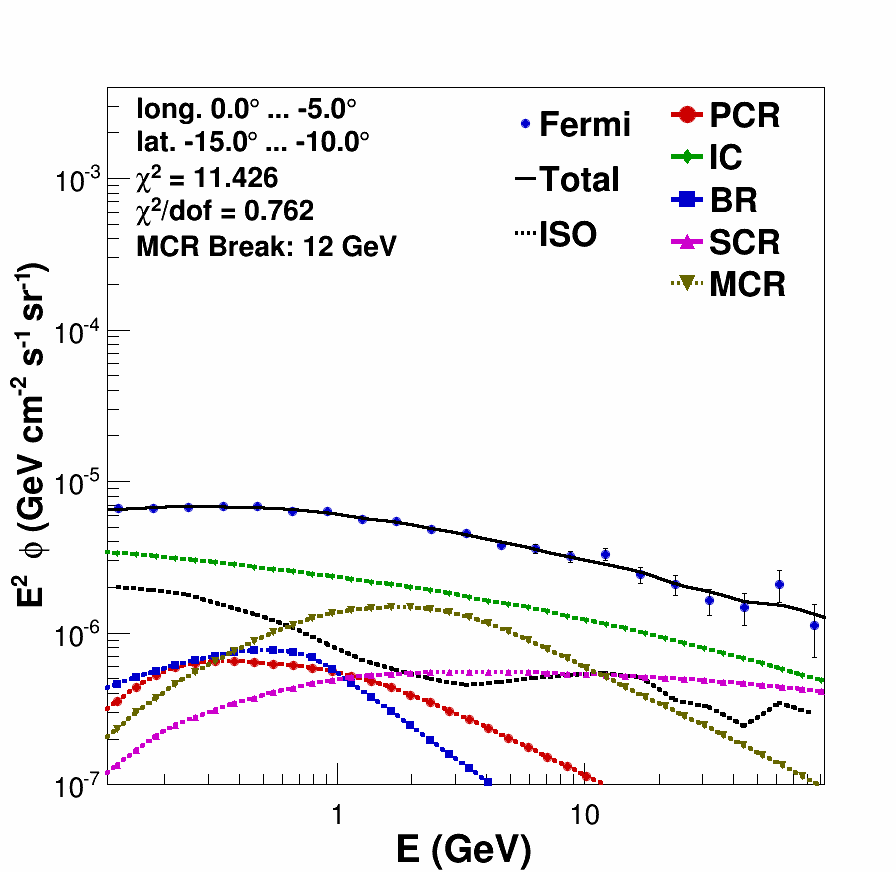}
\includegraphics[width=0.16\textwidth,height=0.16\textwidth,clip]{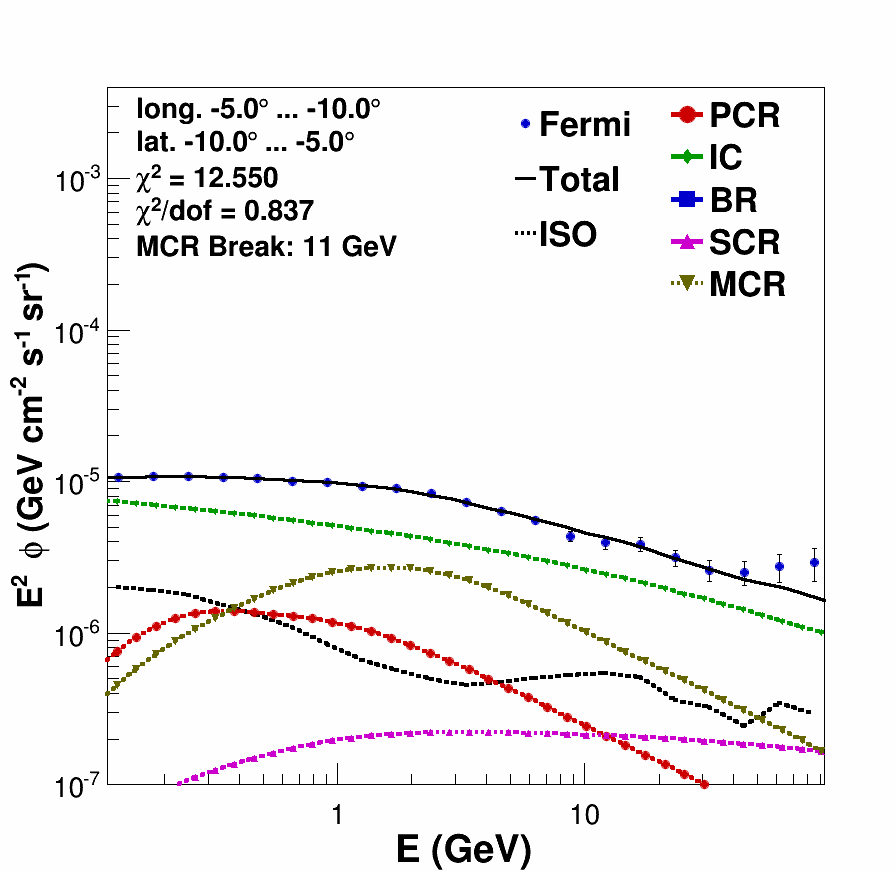}
\includegraphics[width=0.16\textwidth,height=0.16\textwidth,clip]{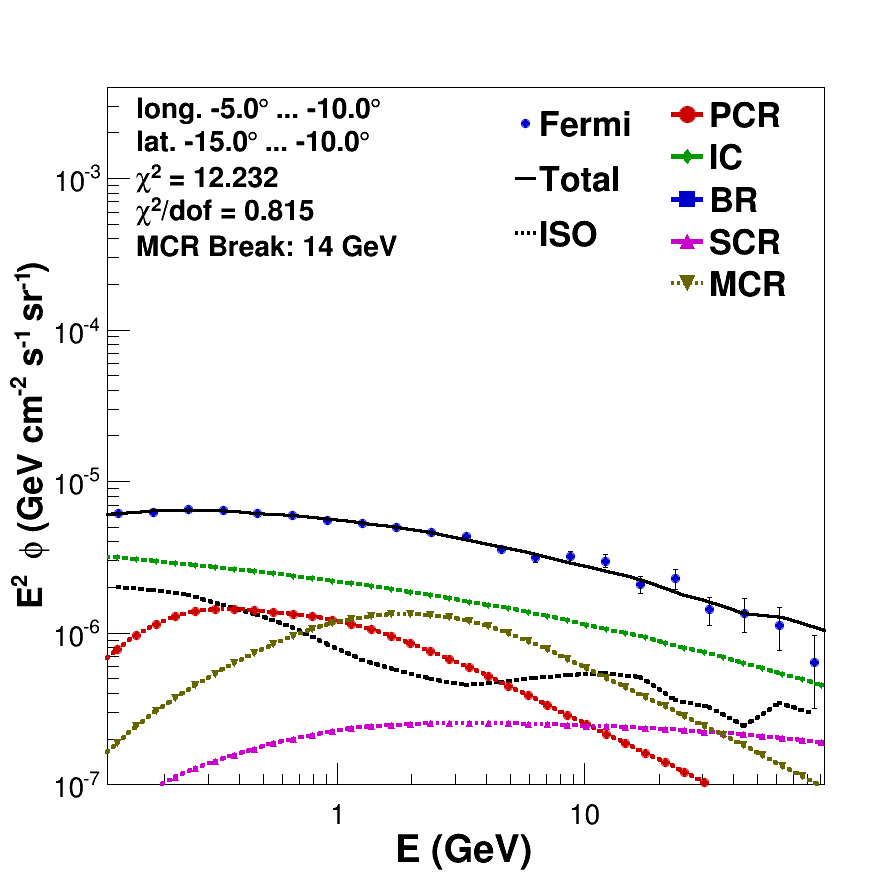}
\includegraphics[width=0.16\textwidth,height=0.16\textwidth,clip]{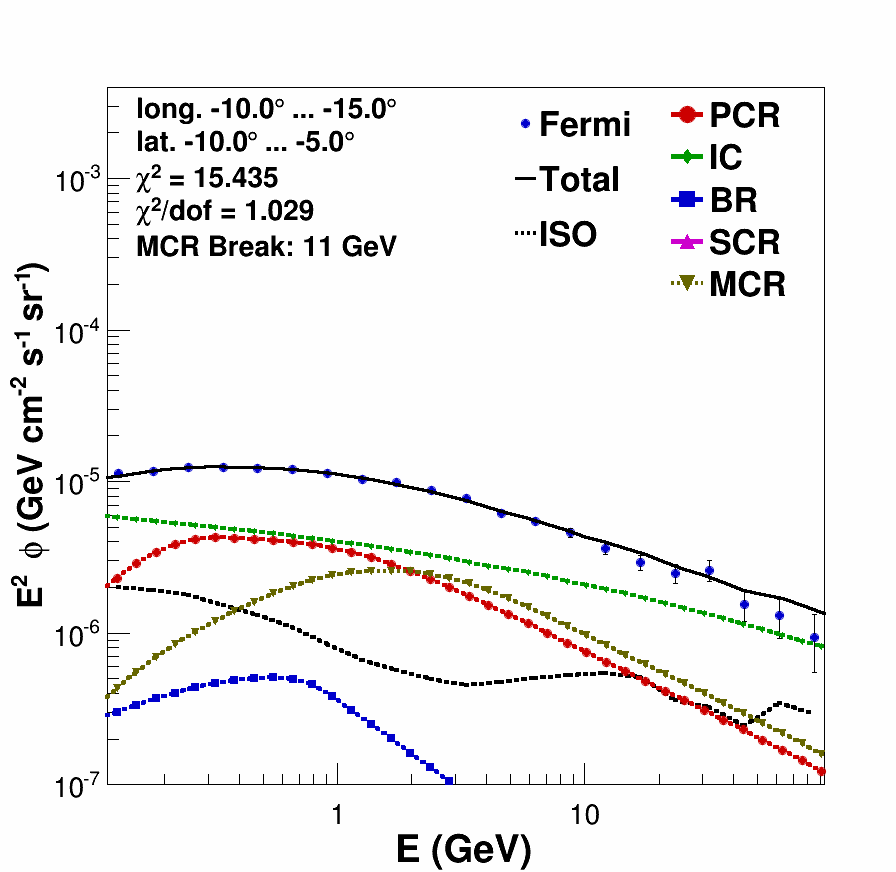}
\includegraphics[width=0.16\textwidth,height=0.16\textwidth,clip]{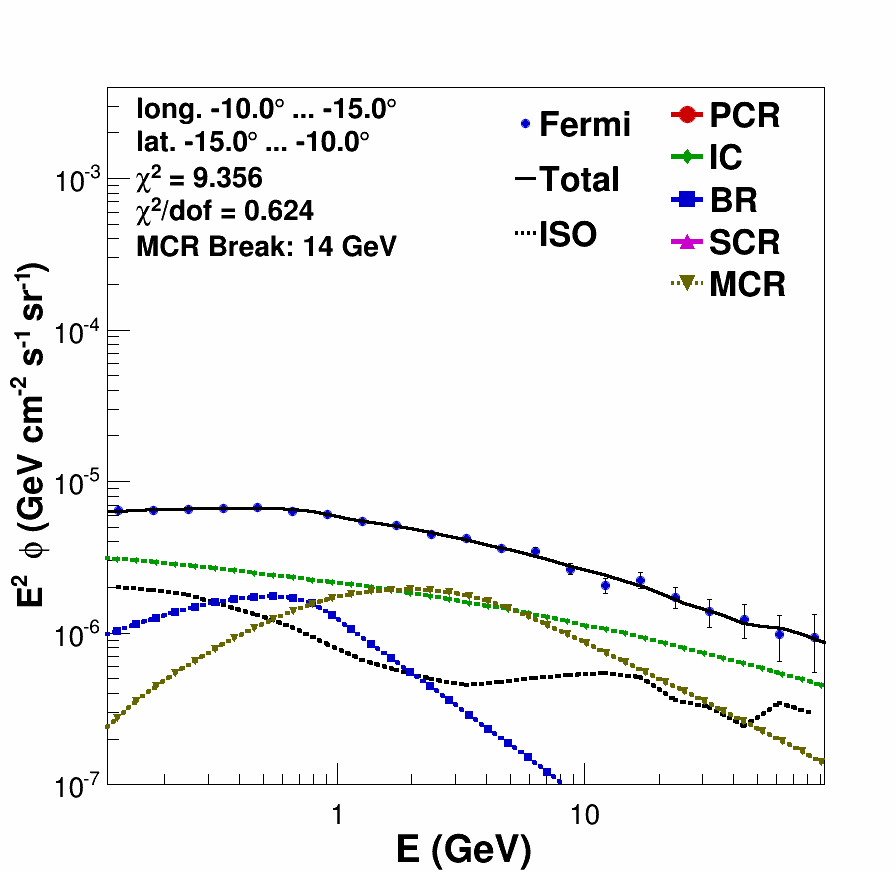}
\includegraphics[width=0.16\textwidth,height=0.16\textwidth,clip]{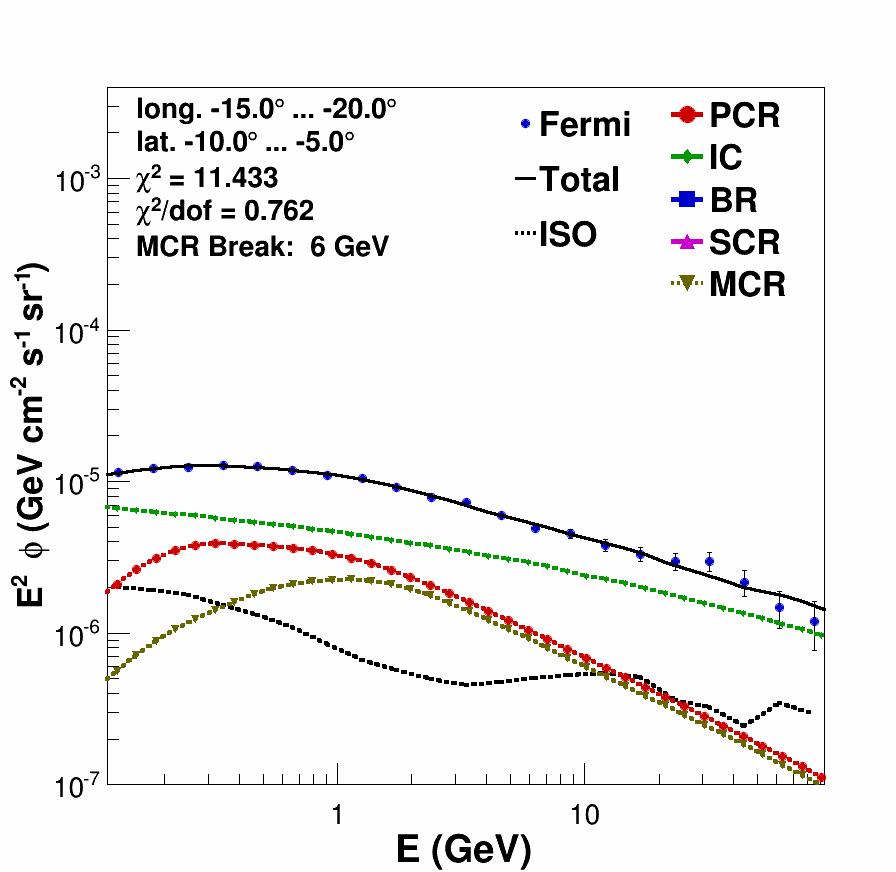}
\includegraphics[width=0.16\textwidth,height=0.16\textwidth,clip]{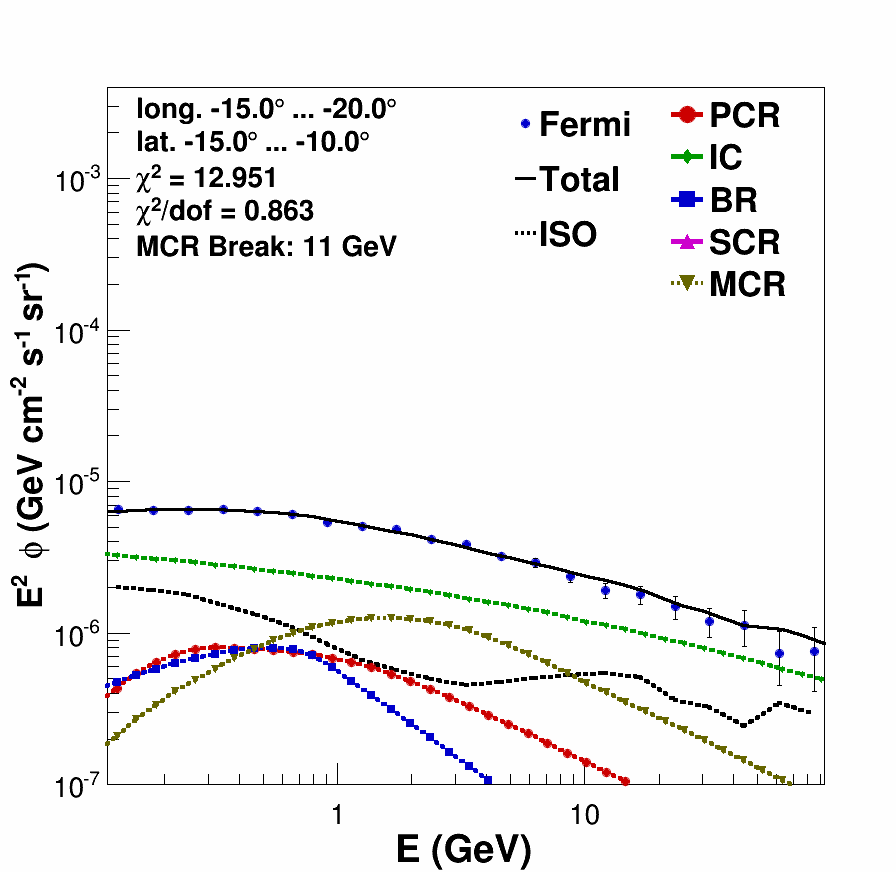}
\includegraphics[width=0.16\textwidth,height=0.16\textwidth,clip]{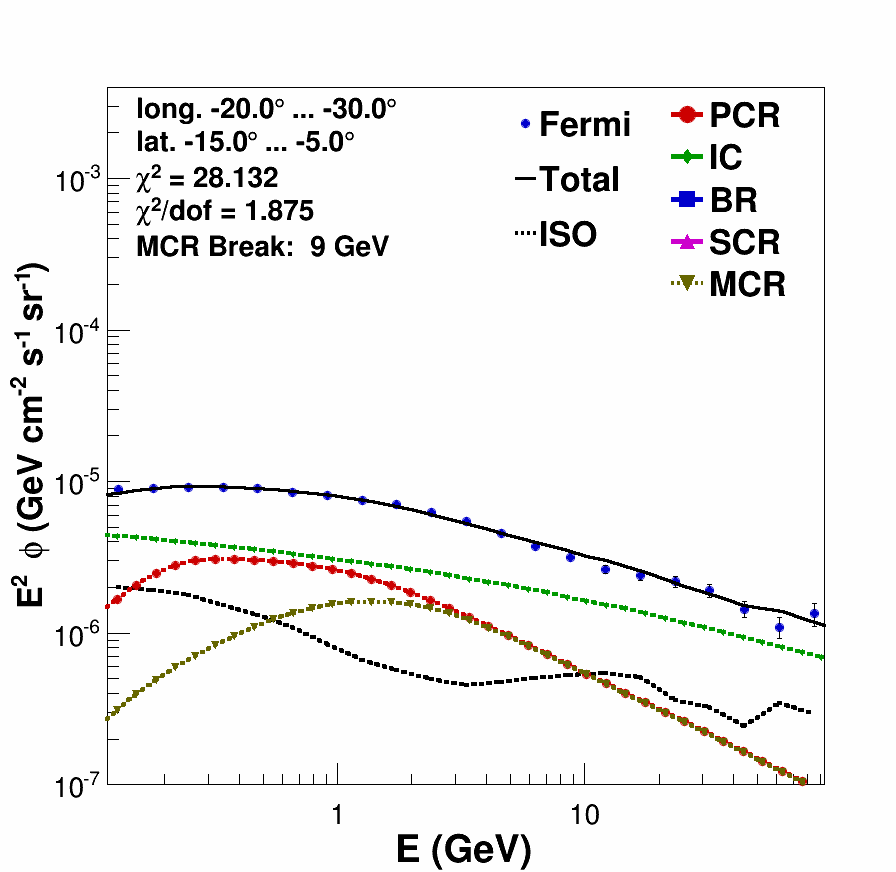}
\includegraphics[width=0.16\textwidth,height=0.16\textwidth,clip]{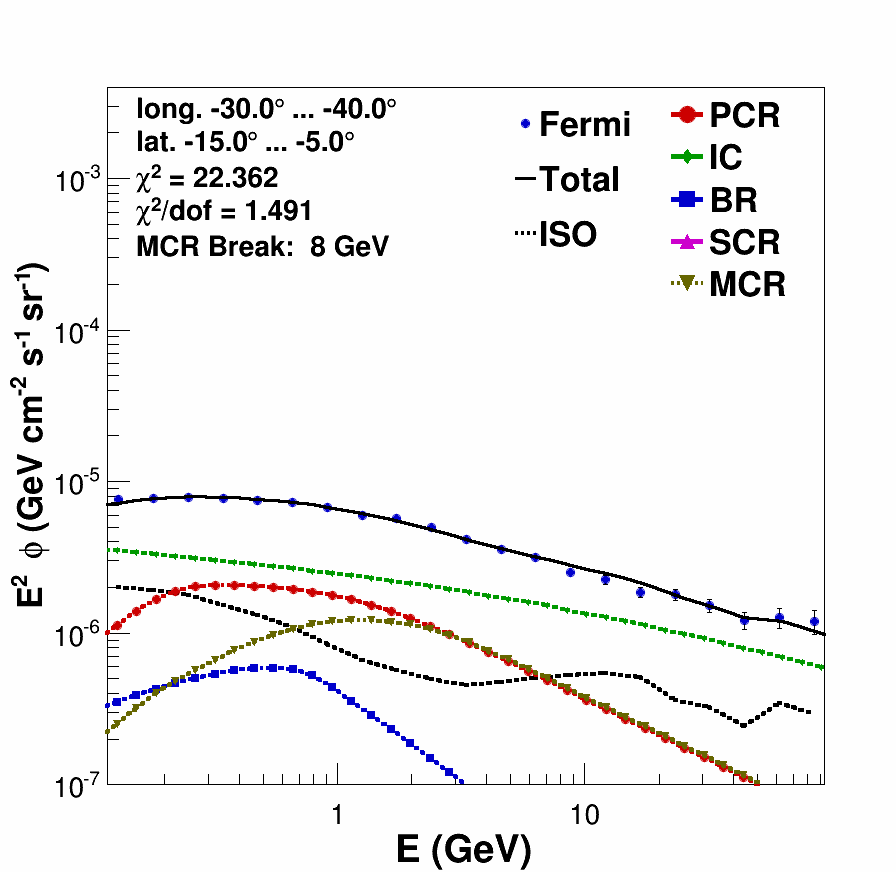}
\includegraphics[width=0.16\textwidth,height=0.16\textwidth,clip]{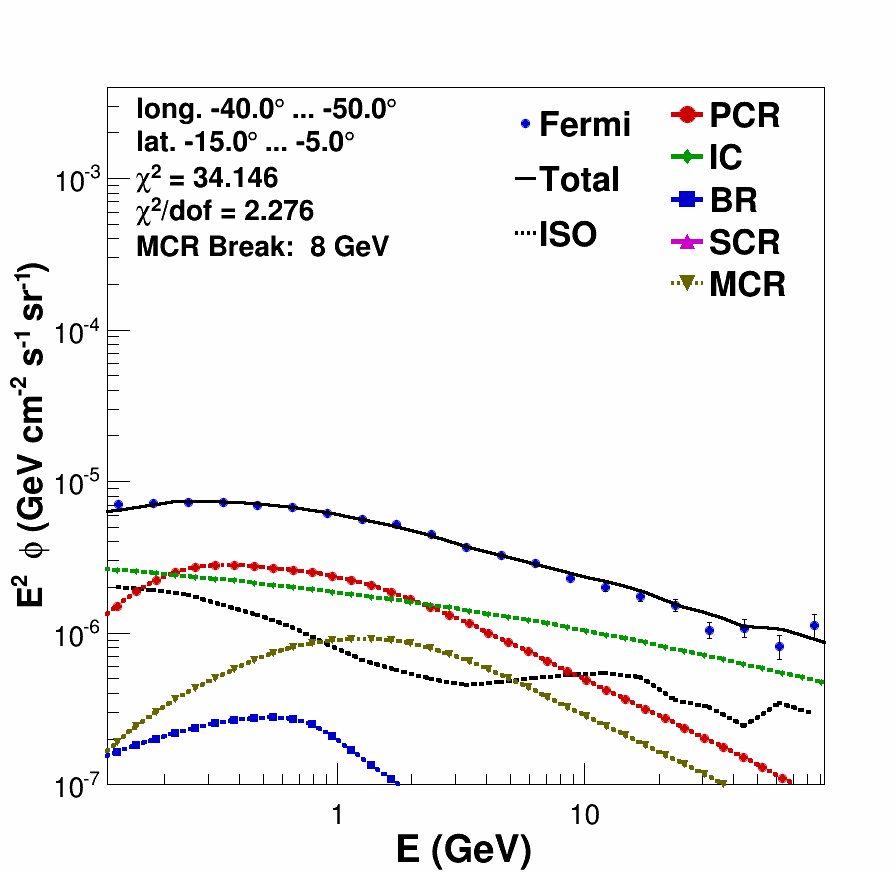}
\includegraphics[width=0.16\textwidth,height=0.16\textwidth,clip]{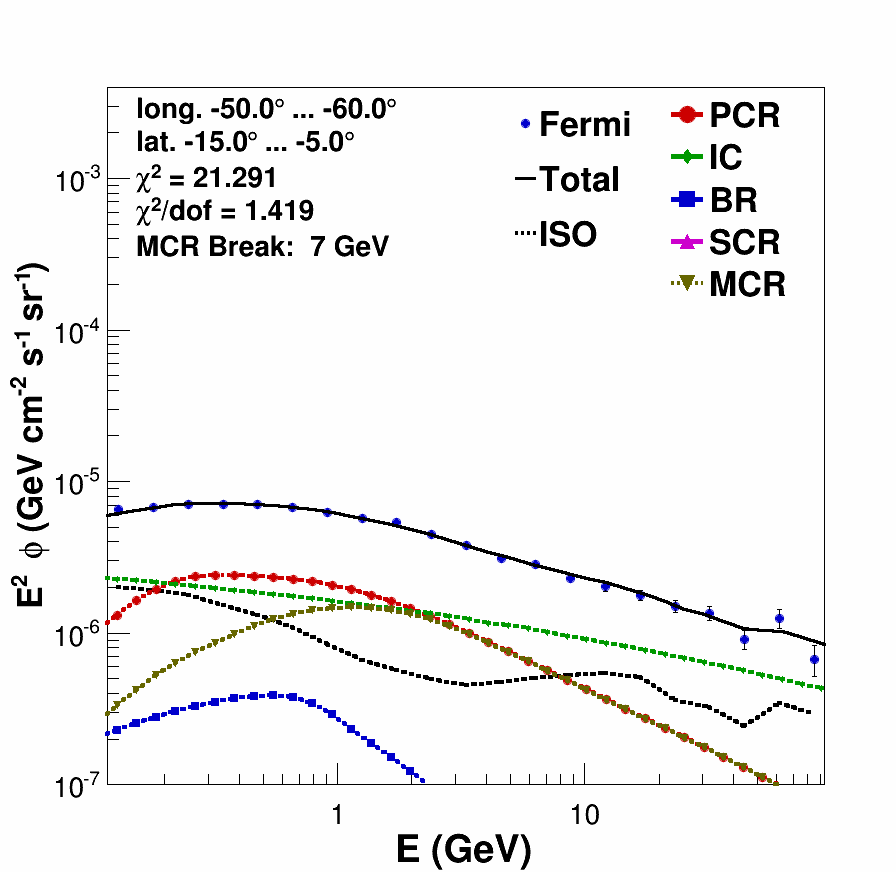}
\includegraphics[width=0.16\textwidth,height=0.16\textwidth,clip]{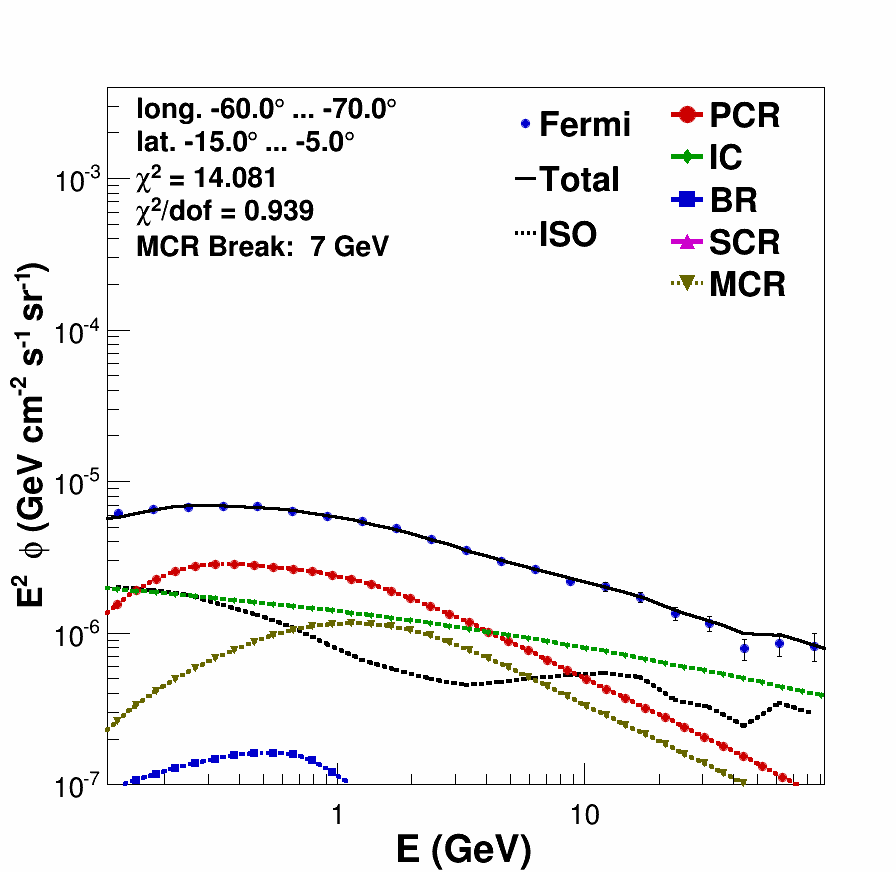}
\includegraphics[width=0.16\textwidth,height=0.16\textwidth,clip]{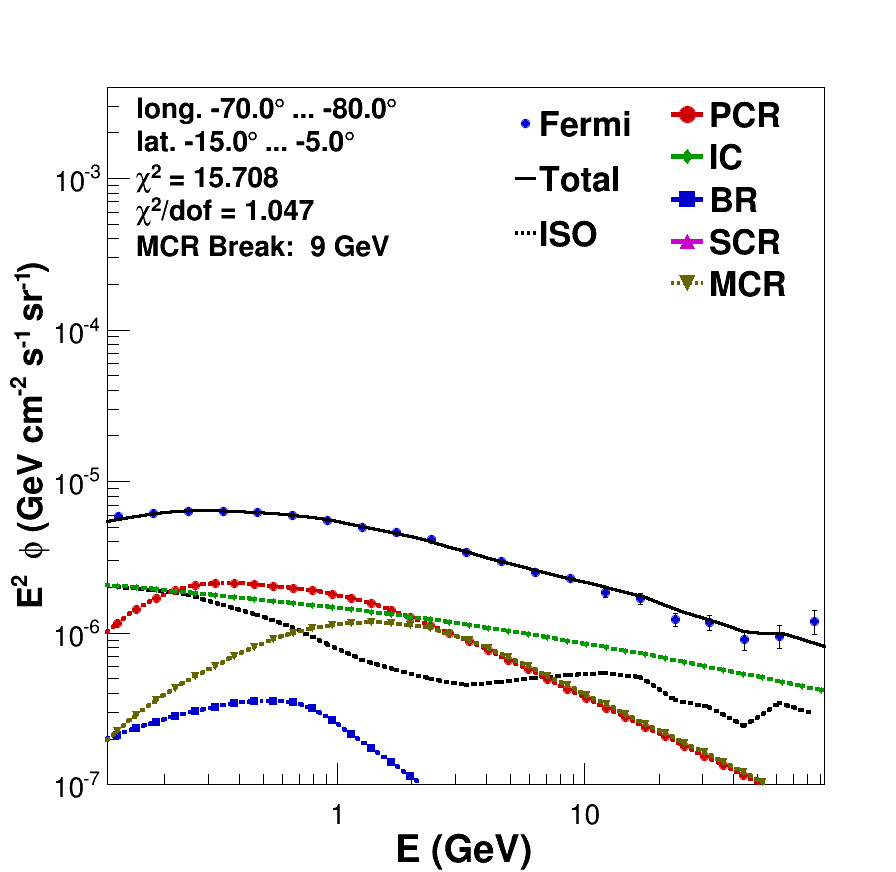}
\includegraphics[width=0.16\textwidth,height=0.16\textwidth,clip]{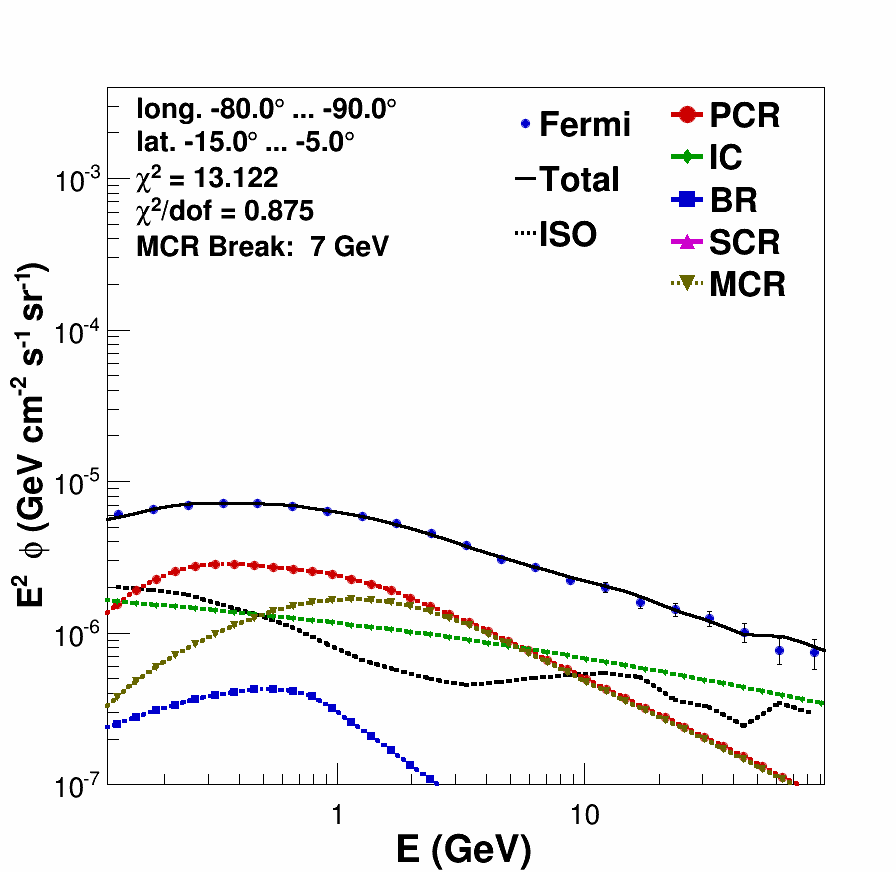}
\includegraphics[width=0.16\textwidth,height=0.16\textwidth,clip]{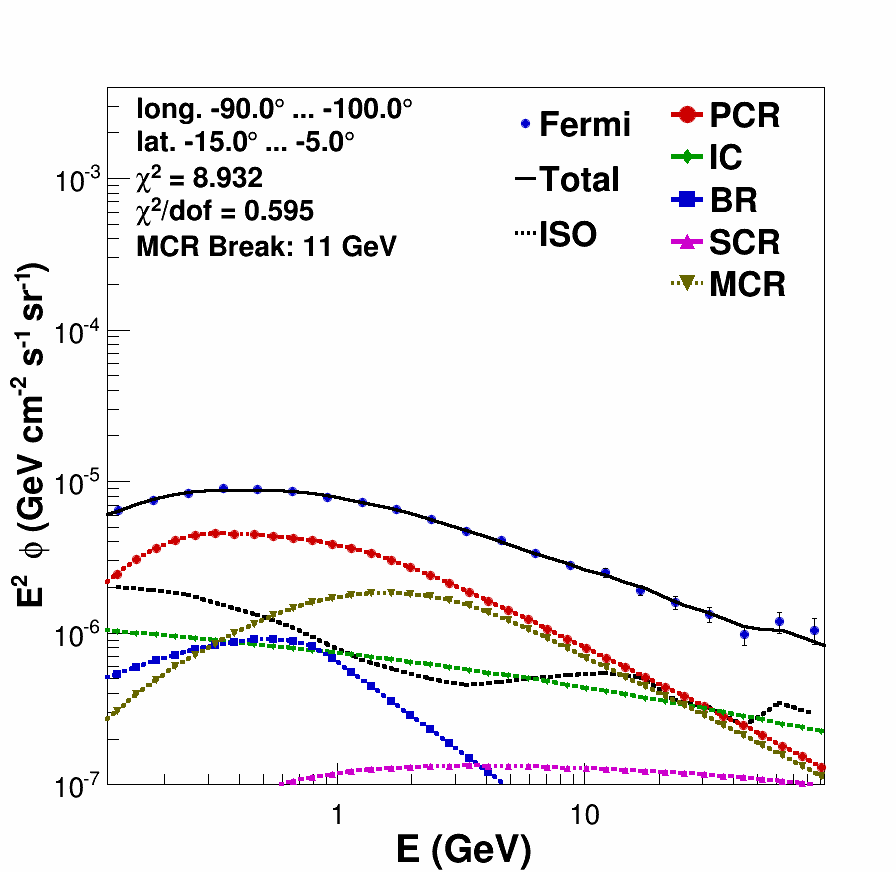}
\includegraphics[width=0.16\textwidth,height=0.16\textwidth,clip]{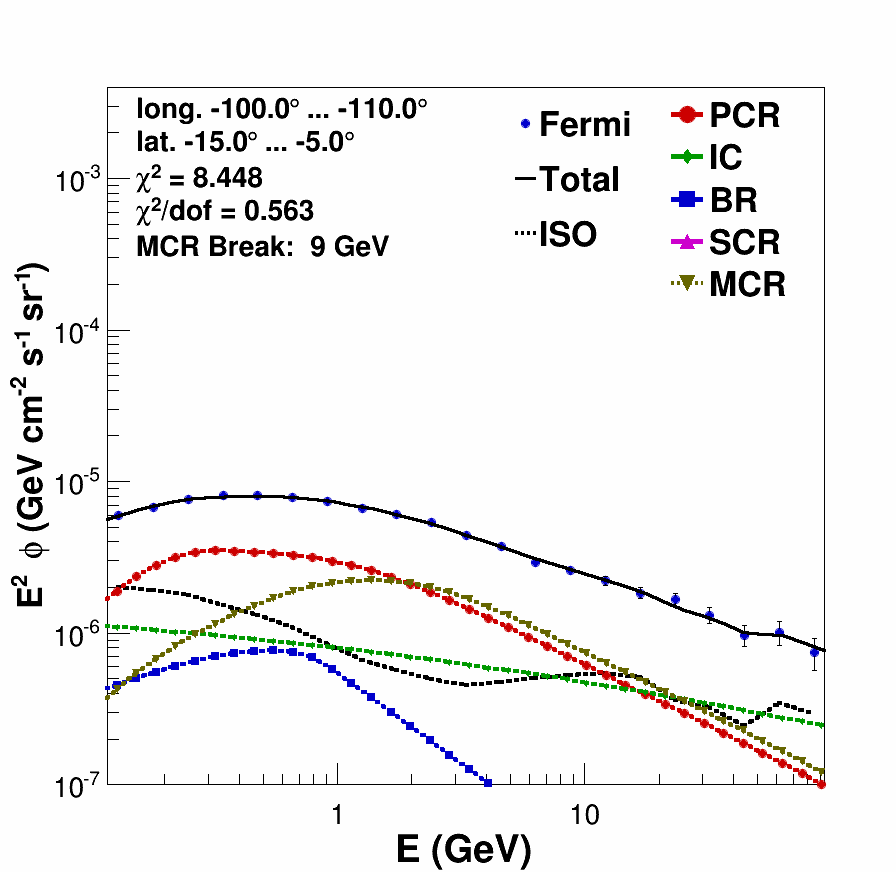}
\includegraphics[width=0.16\textwidth,height=0.16\textwidth,clip]{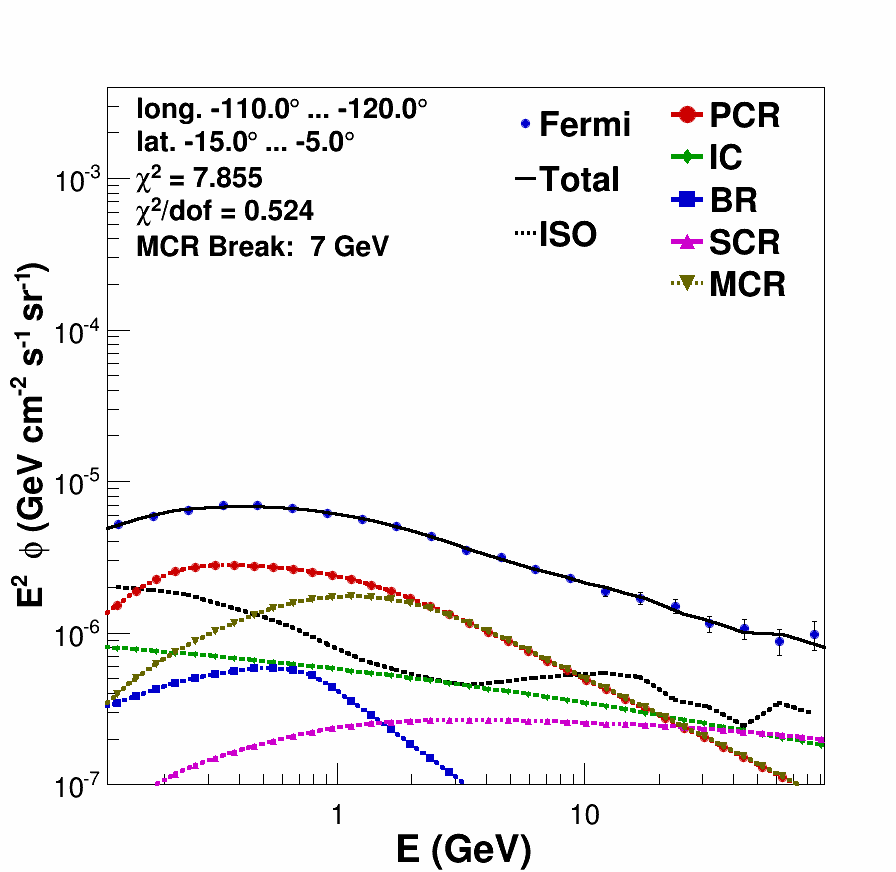}
\includegraphics[width=0.16\textwidth,height=0.16\textwidth,clip]{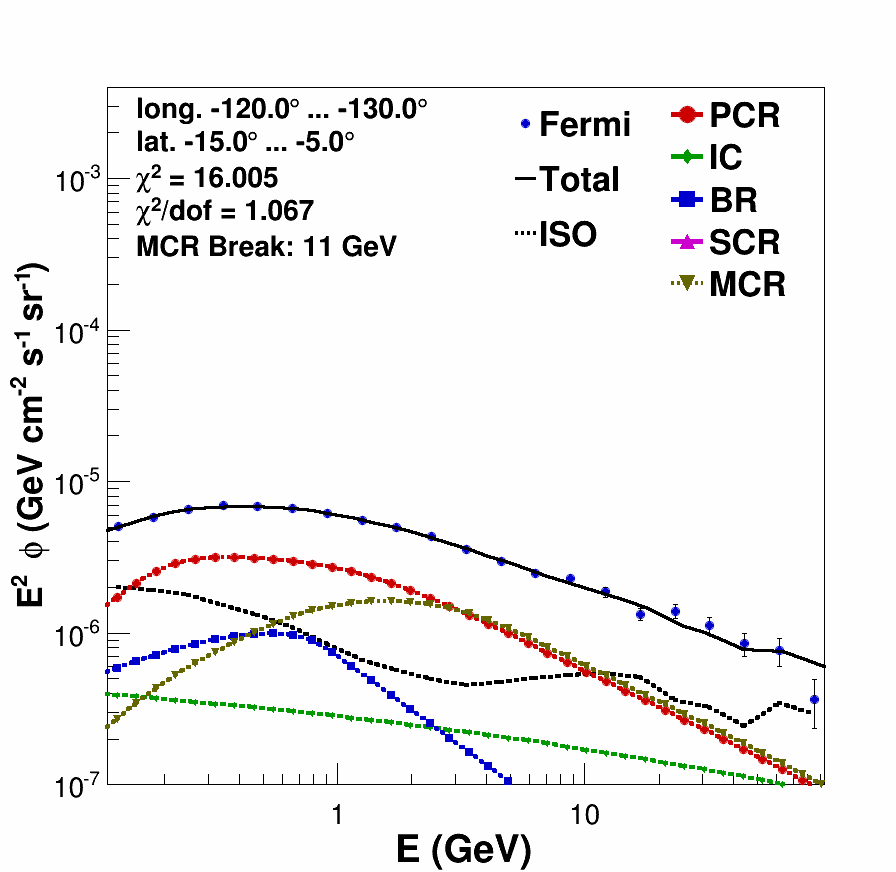}
\includegraphics[width=0.16\textwidth,height=0.16\textwidth,clip]{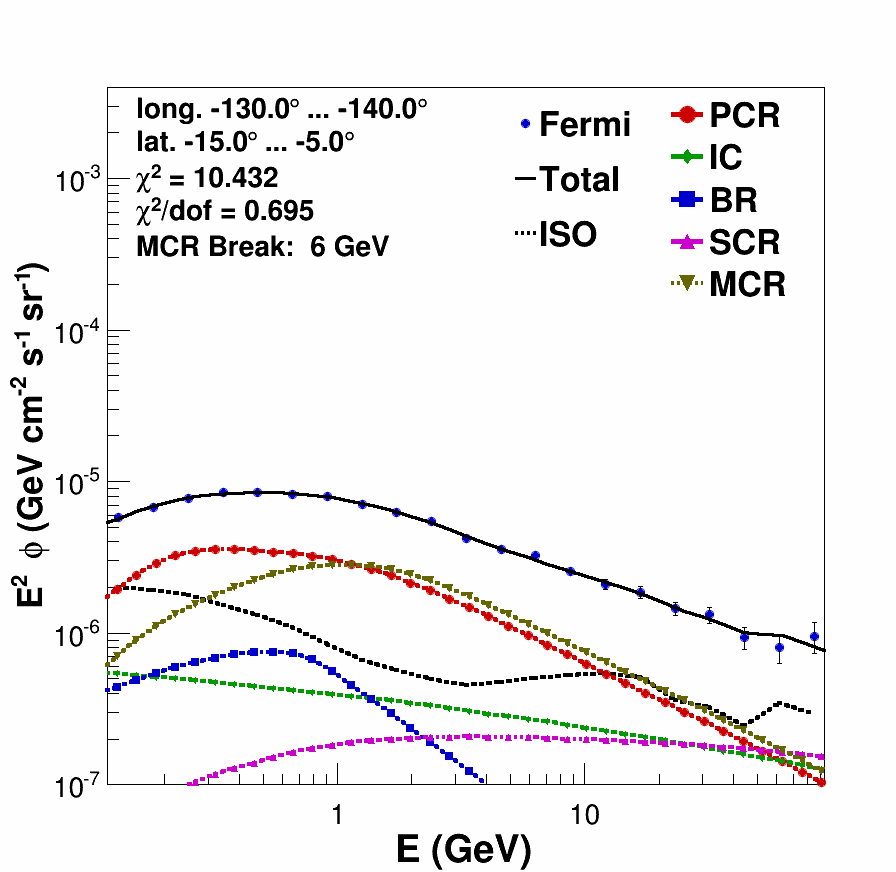}
\includegraphics[width=0.16\textwidth,height=0.16\textwidth,clip]{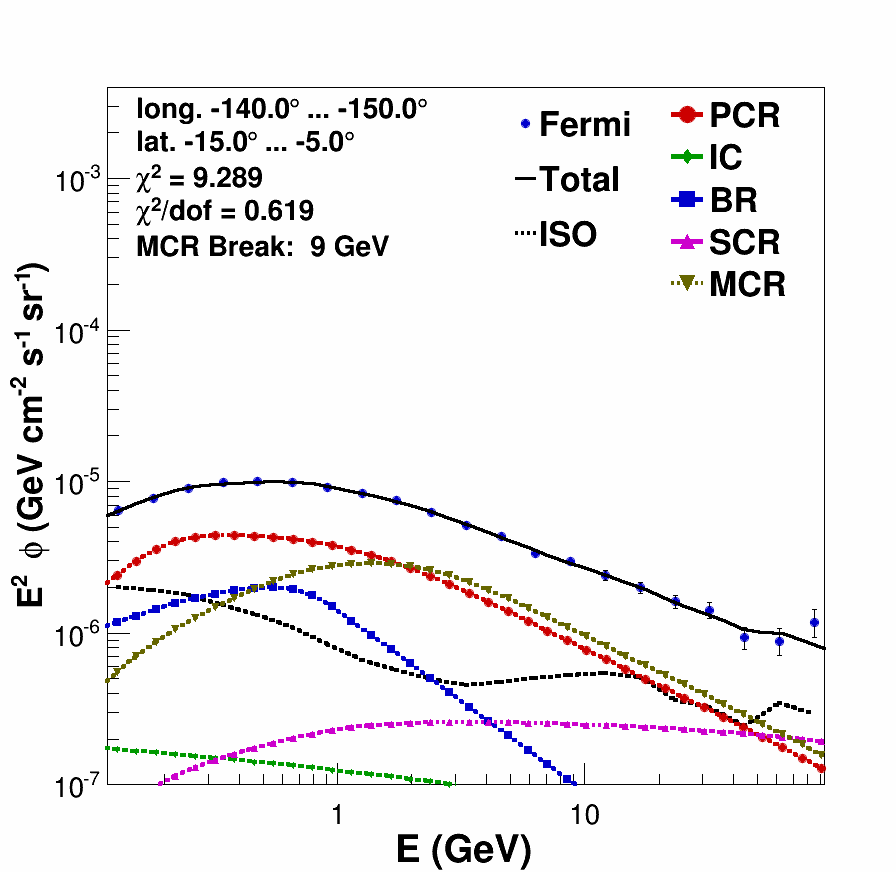}
\includegraphics[width=0.16\textwidth,height=0.16\textwidth,clip]{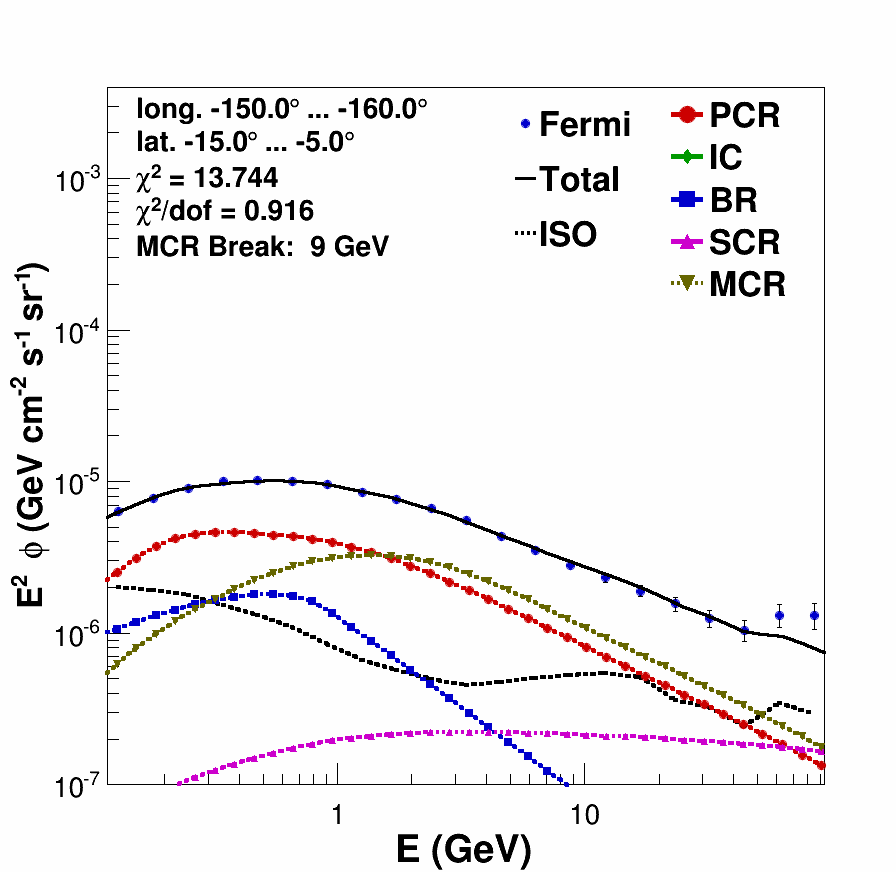}
\includegraphics[width=0.16\textwidth,height=0.16\textwidth,clip]{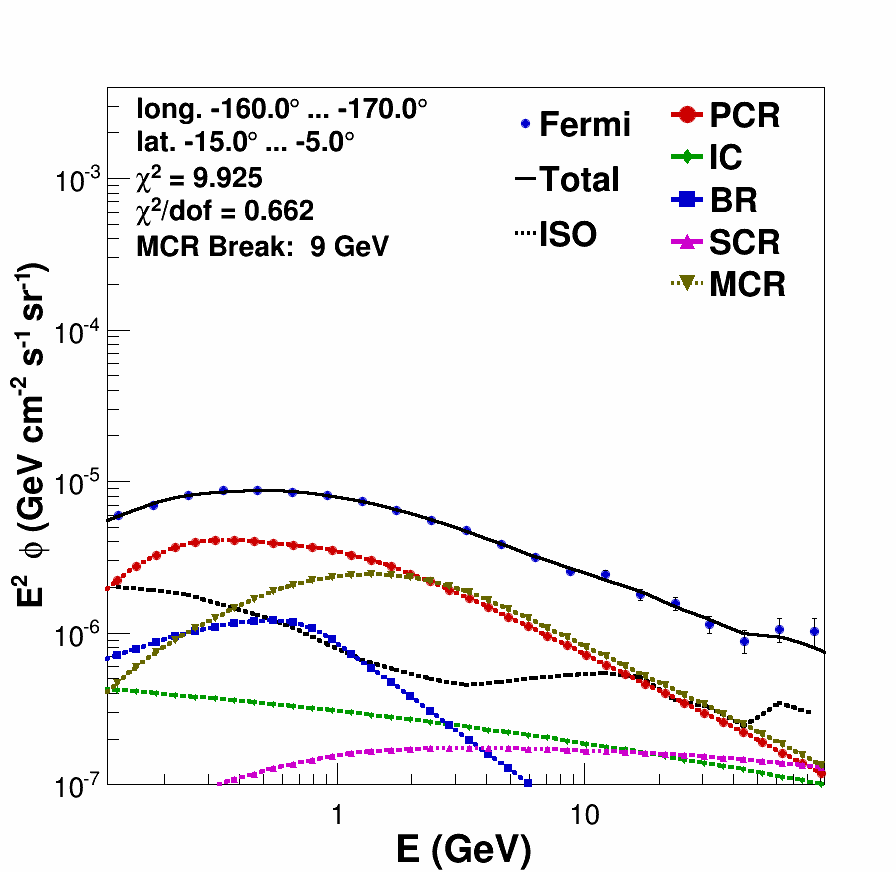}
\includegraphics[width=0.16\textwidth,height=0.16\textwidth,clip]{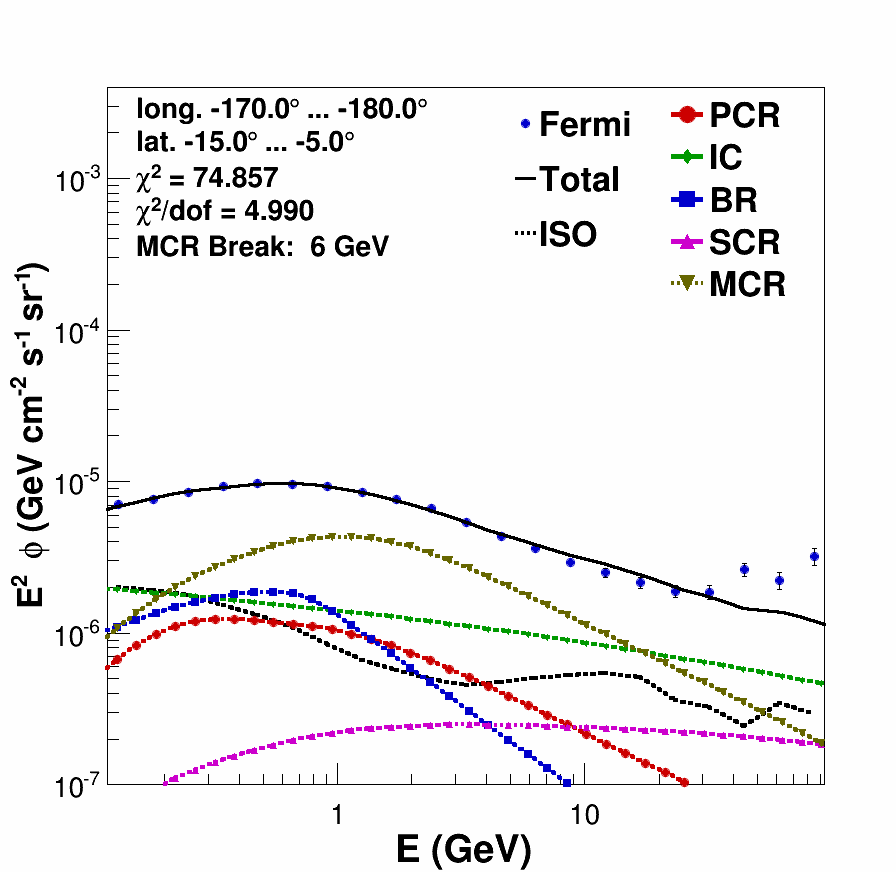}%%%%%%r13
\caption[]{Template fits for latitudes  with $-15.0^\circ<b<-5.0^\circ$ and longitudes decreasing from 180$^\circ$ to -180$^\circ$.} \label{F25}
\end{figure}
\begin{figure}
\centering
\includegraphics[width=0.16\textwidth,height=0.16\textwidth,clip]{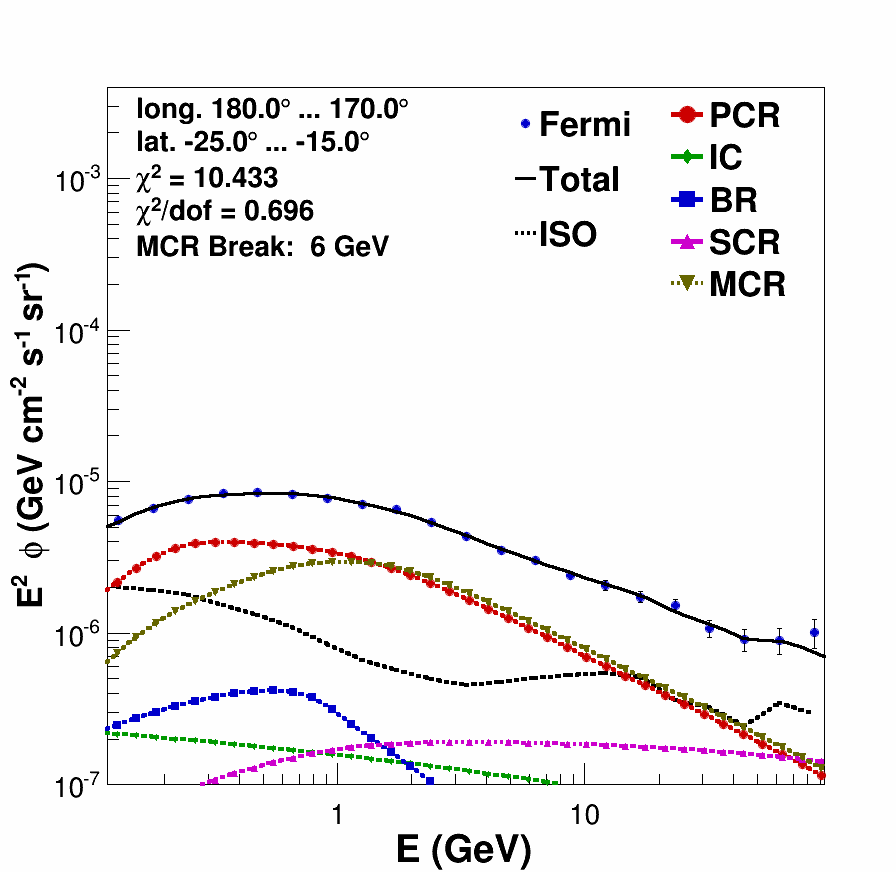}
\includegraphics[width=0.16\textwidth,height=0.16\textwidth,clip]{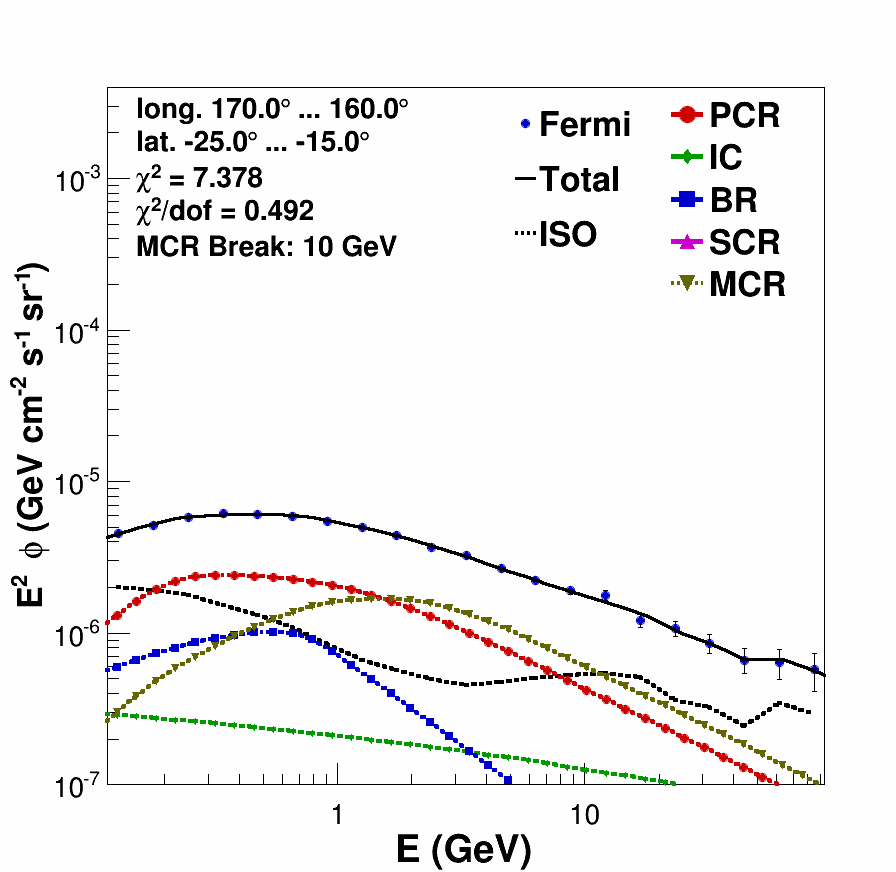}
\includegraphics[width=0.16\textwidth,height=0.16\textwidth,clip]{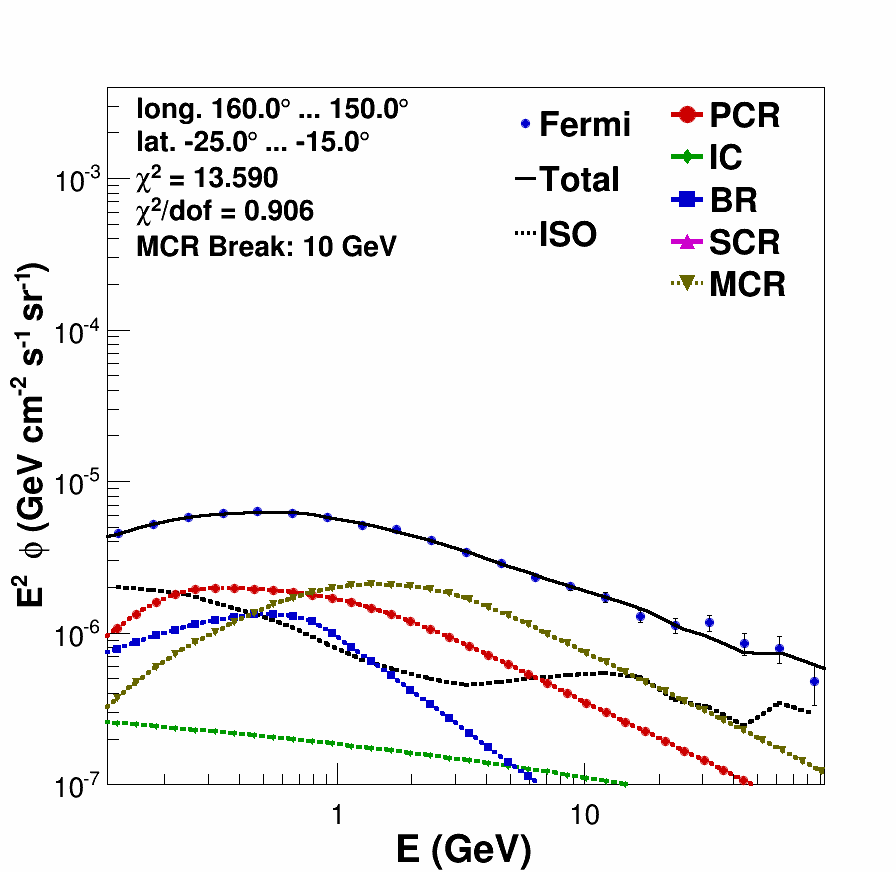}
\includegraphics[width=0.16\textwidth,height=0.16\textwidth,clip]{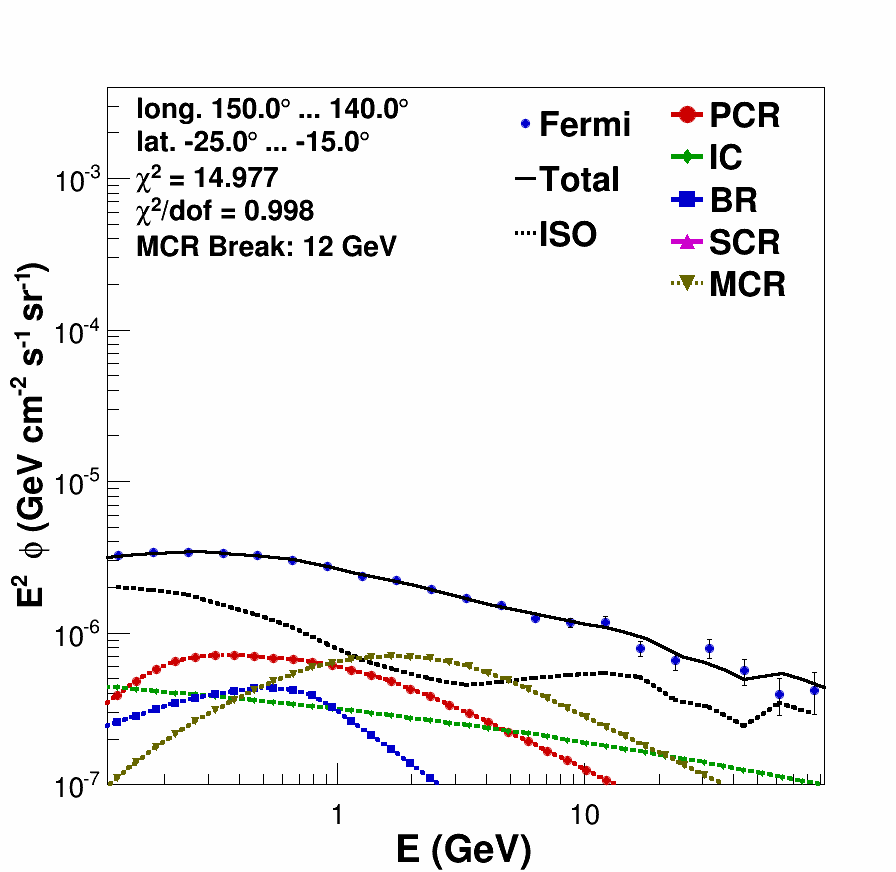}
\includegraphics[width=0.16\textwidth,height=0.16\textwidth,clip]{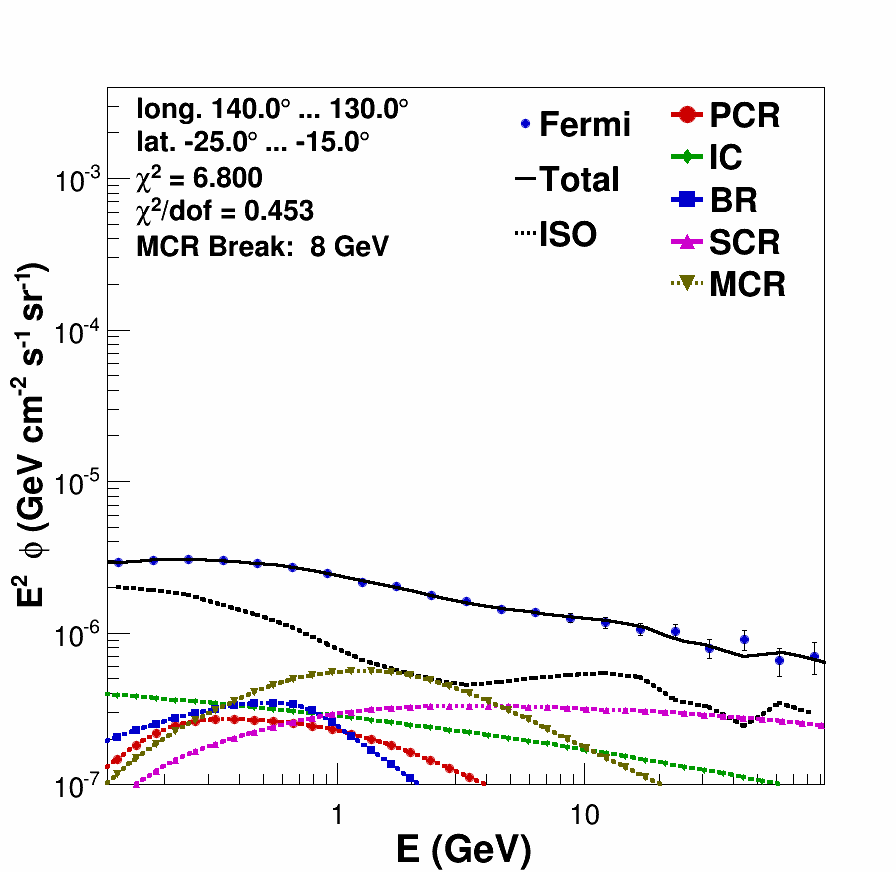}
\includegraphics[width=0.16\textwidth,height=0.16\textwidth,clip]{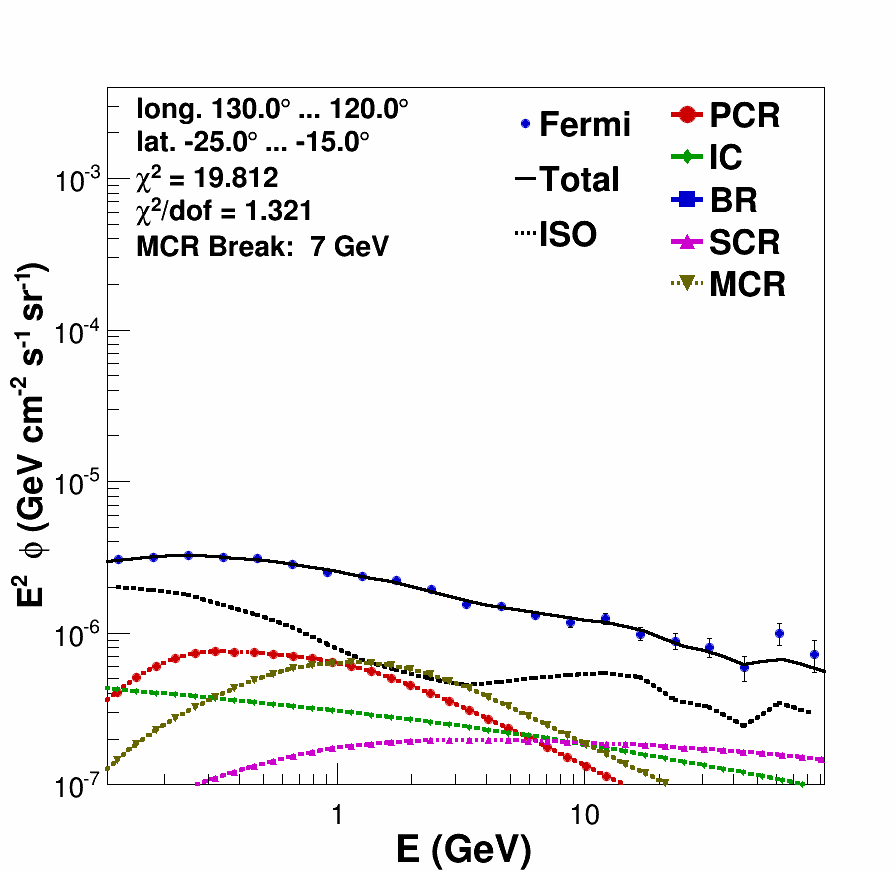}
\includegraphics[width=0.16\textwidth,height=0.16\textwidth,clip]{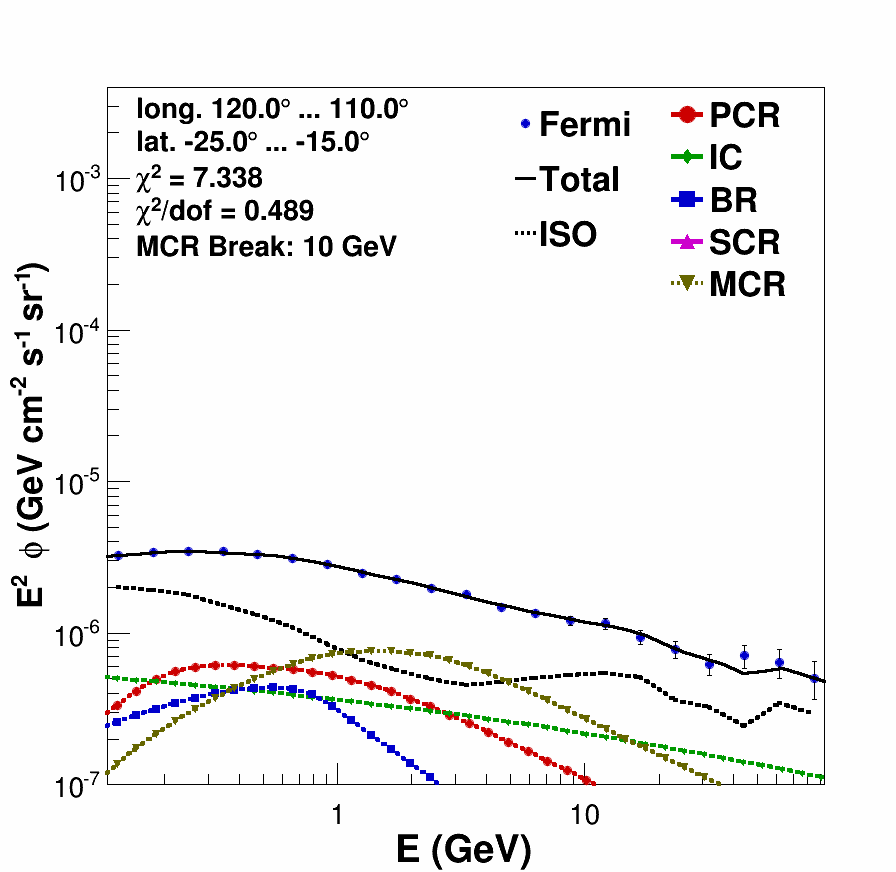}
\includegraphics[width=0.16\textwidth,height=0.16\textwidth,clip]{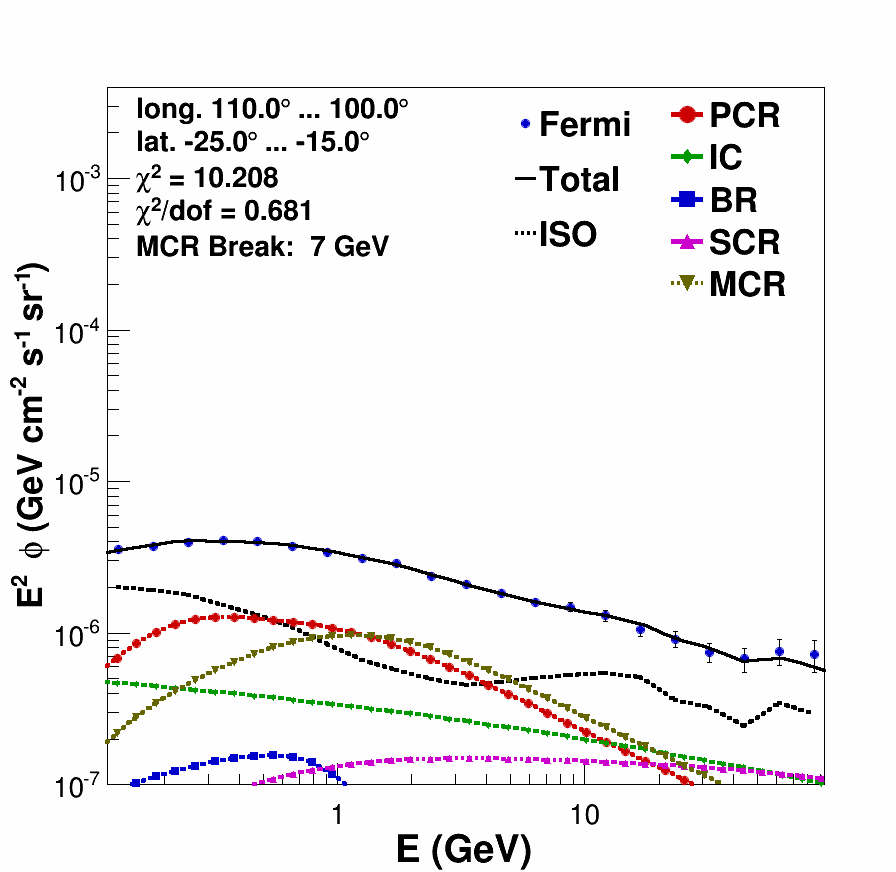}
\includegraphics[width=0.16\textwidth,height=0.16\textwidth,clip]{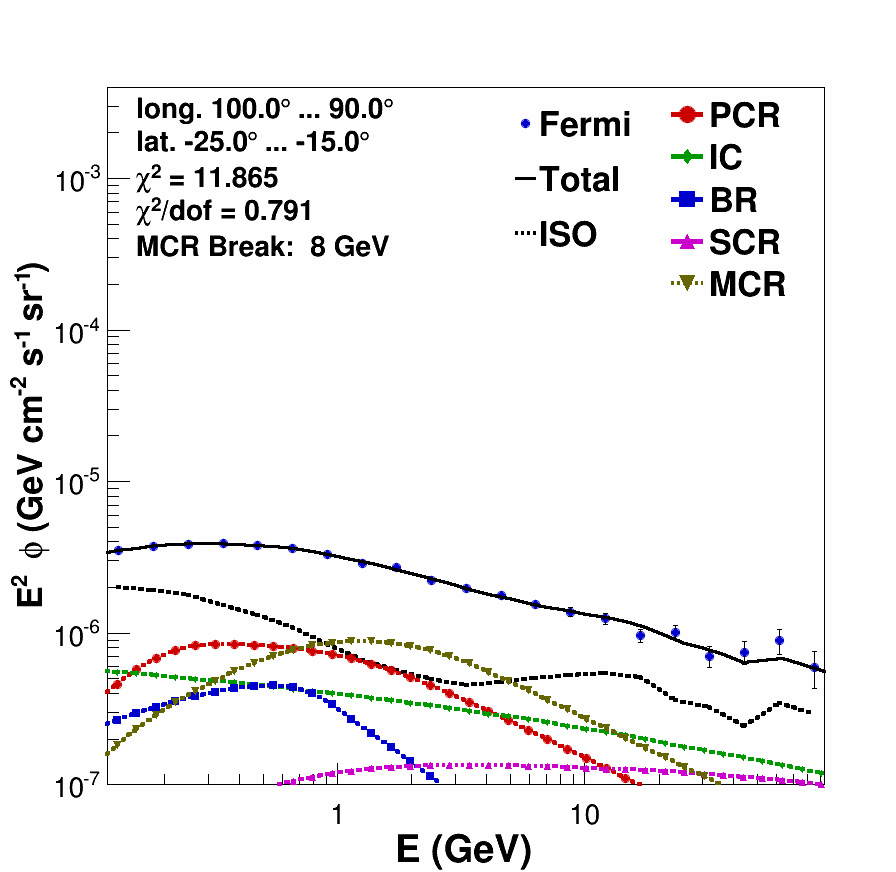}
\includegraphics[width=0.16\textwidth,height=0.16\textwidth,clip]{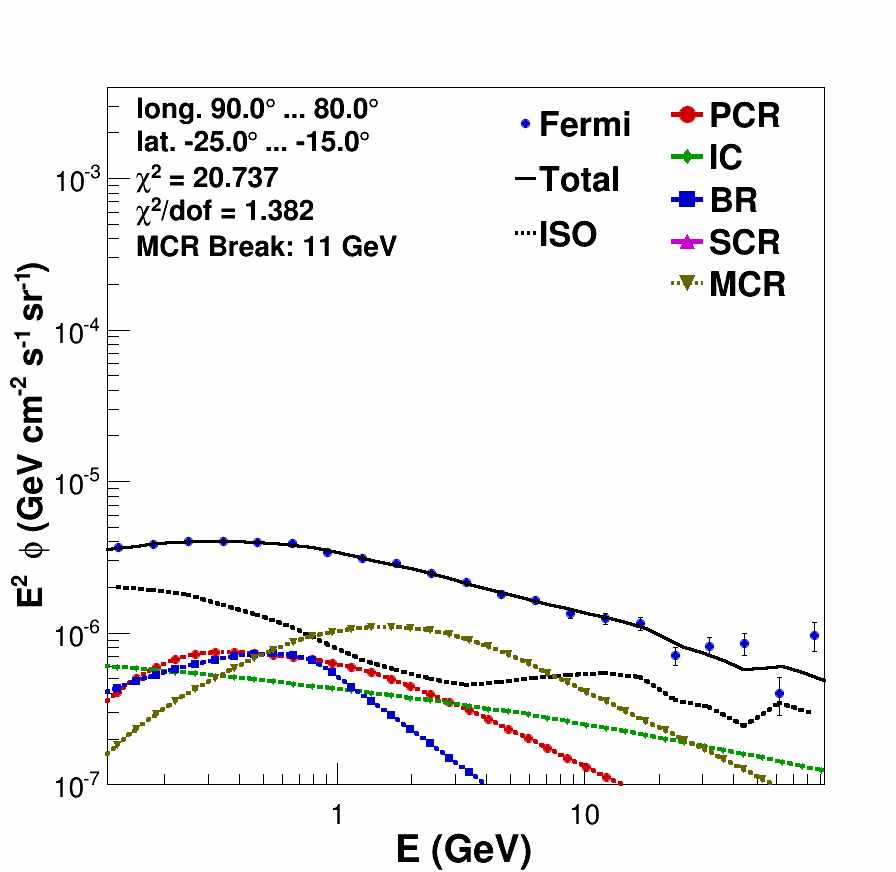}
\includegraphics[width=0.16\textwidth,height=0.16\textwidth,clip]{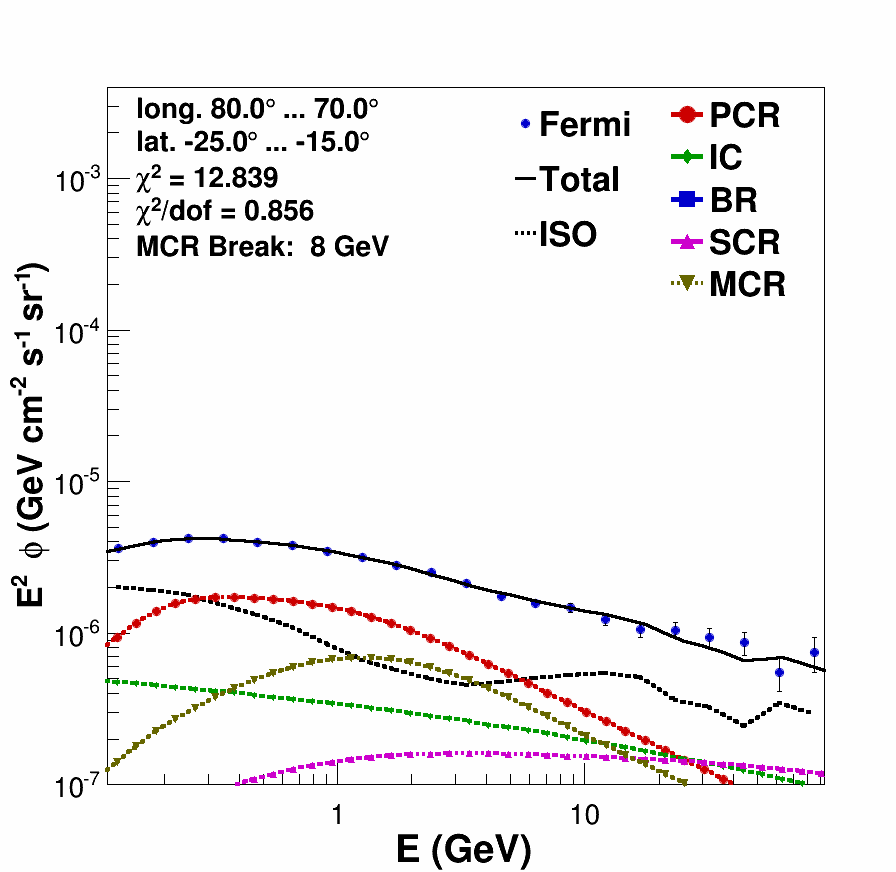}
\includegraphics[width=0.16\textwidth,height=0.16\textwidth,clip]{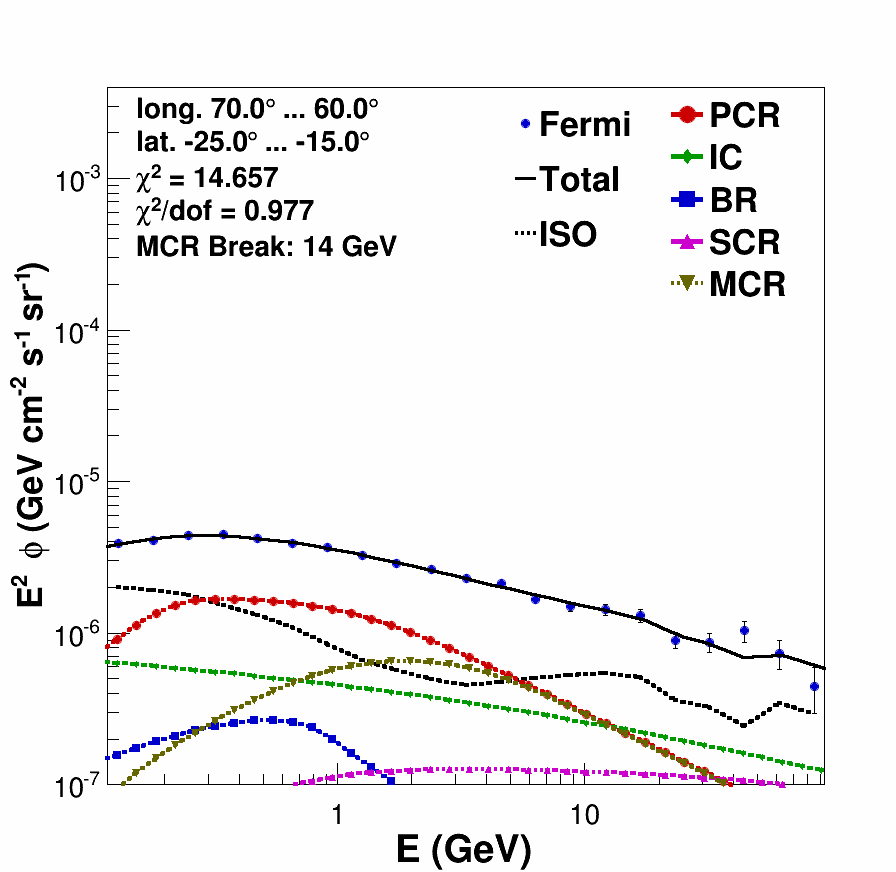}
\includegraphics[width=0.16\textwidth,height=0.16\textwidth,clip]{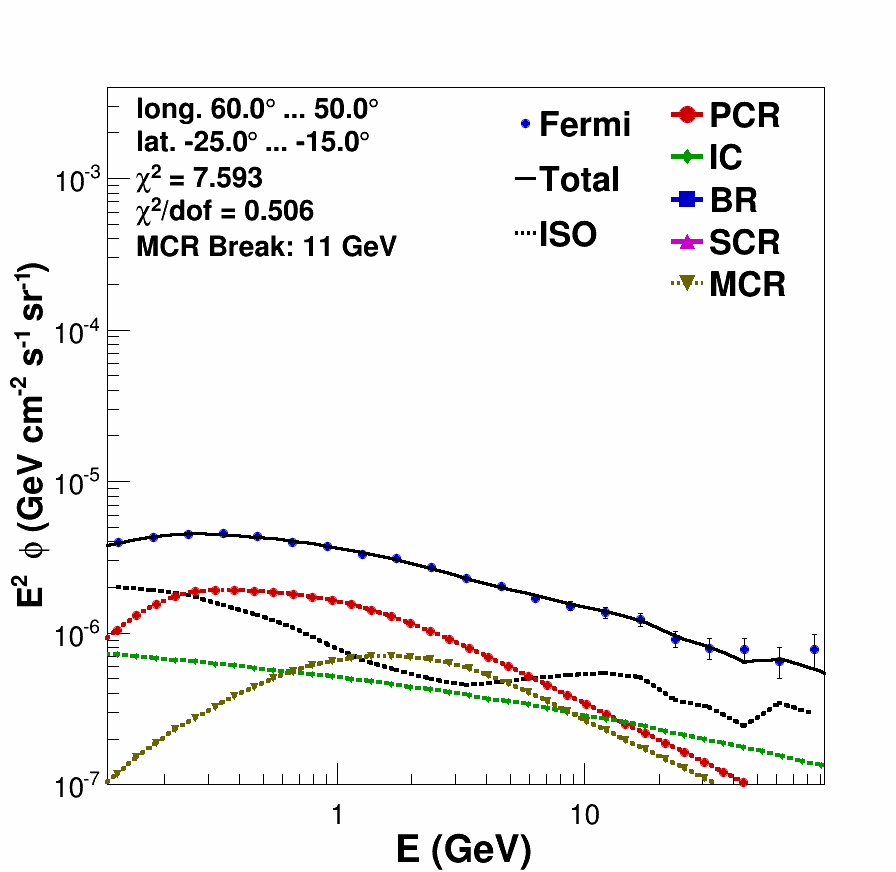}
\includegraphics[width=0.16\textwidth,height=0.16\textwidth,clip]{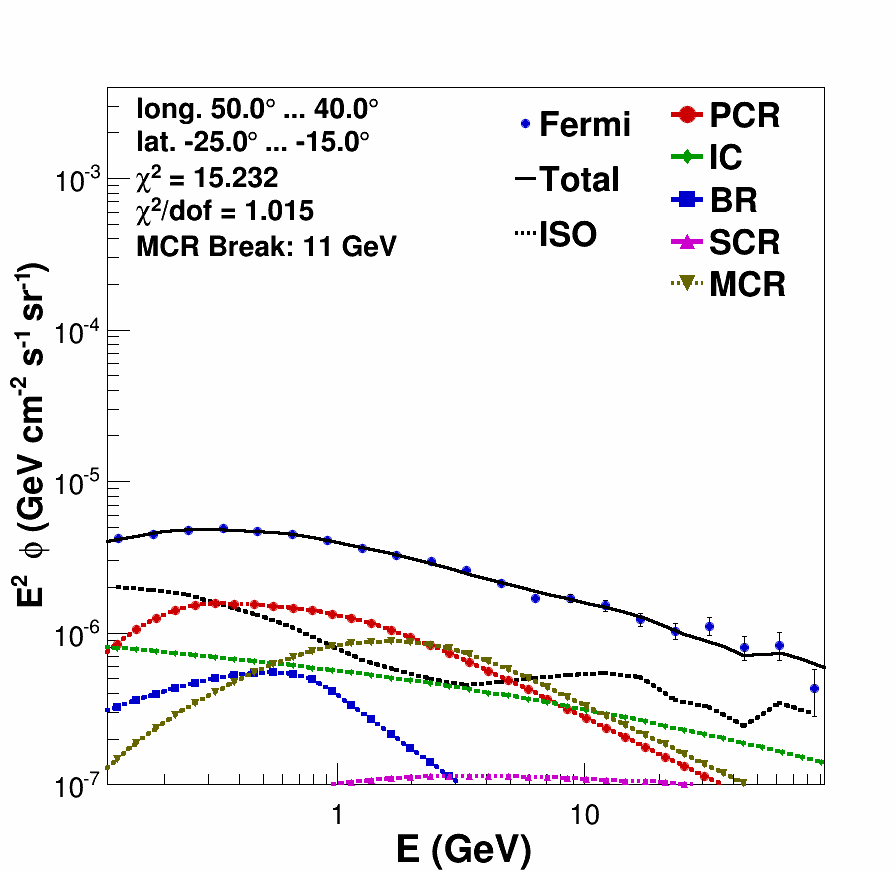}
\includegraphics[width=0.16\textwidth,height=0.16\textwidth,clip]{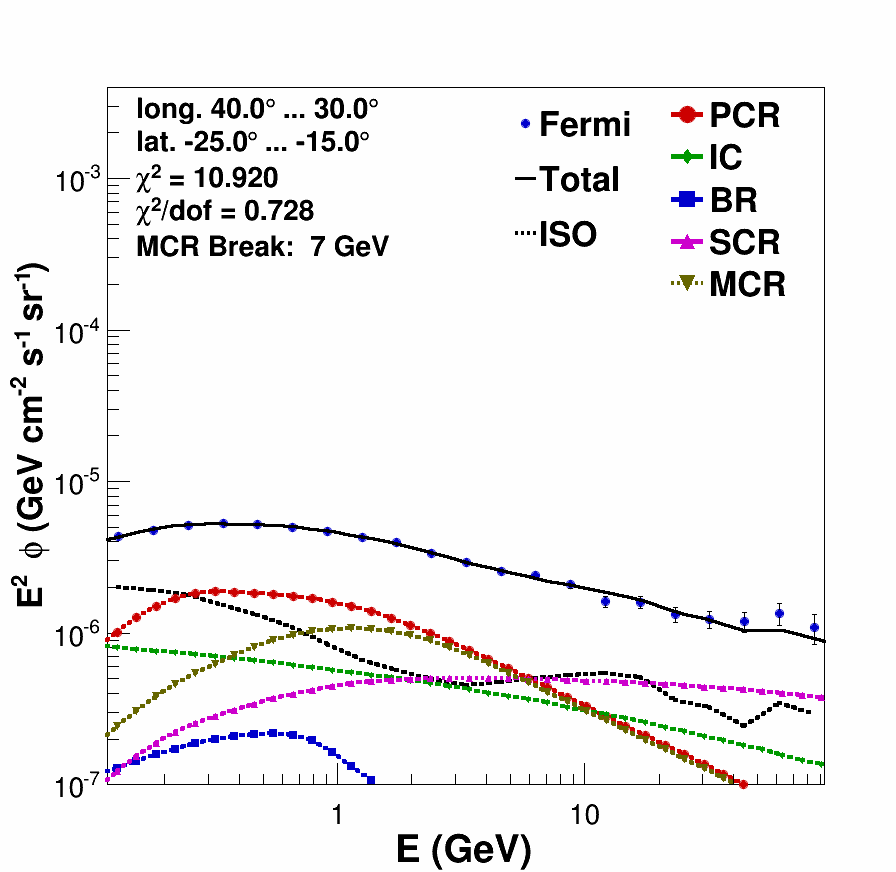}
\includegraphics[width=0.16\textwidth,height=0.16\textwidth,clip]{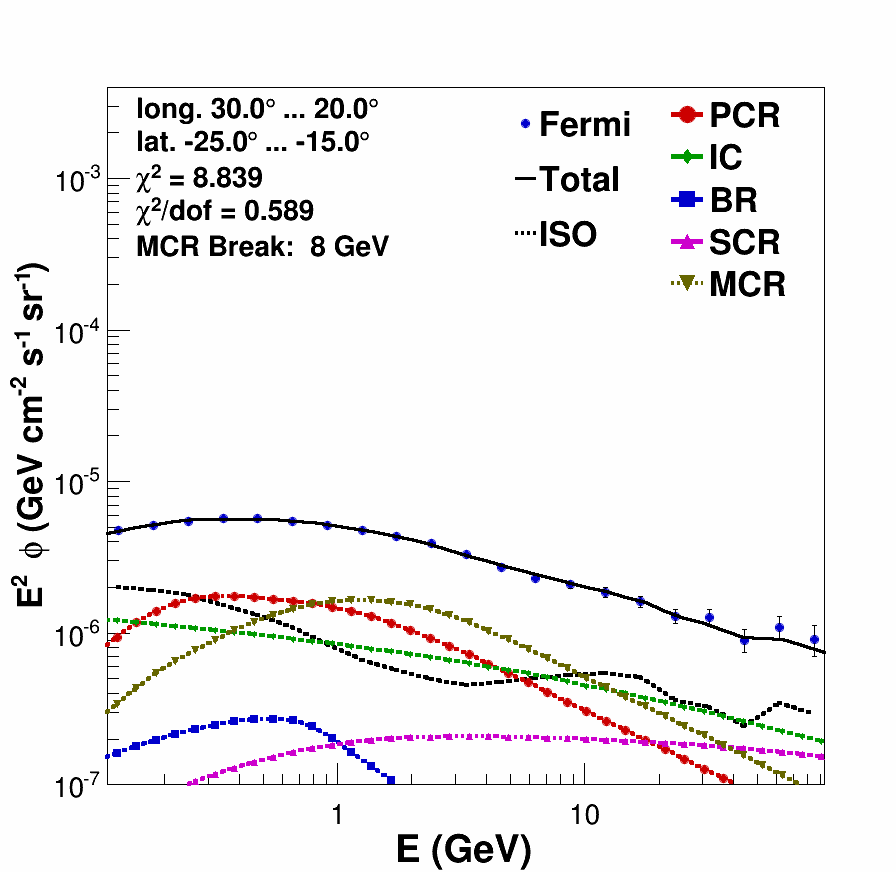}
\includegraphics[width=0.16\textwidth,height=0.16\textwidth,clip]{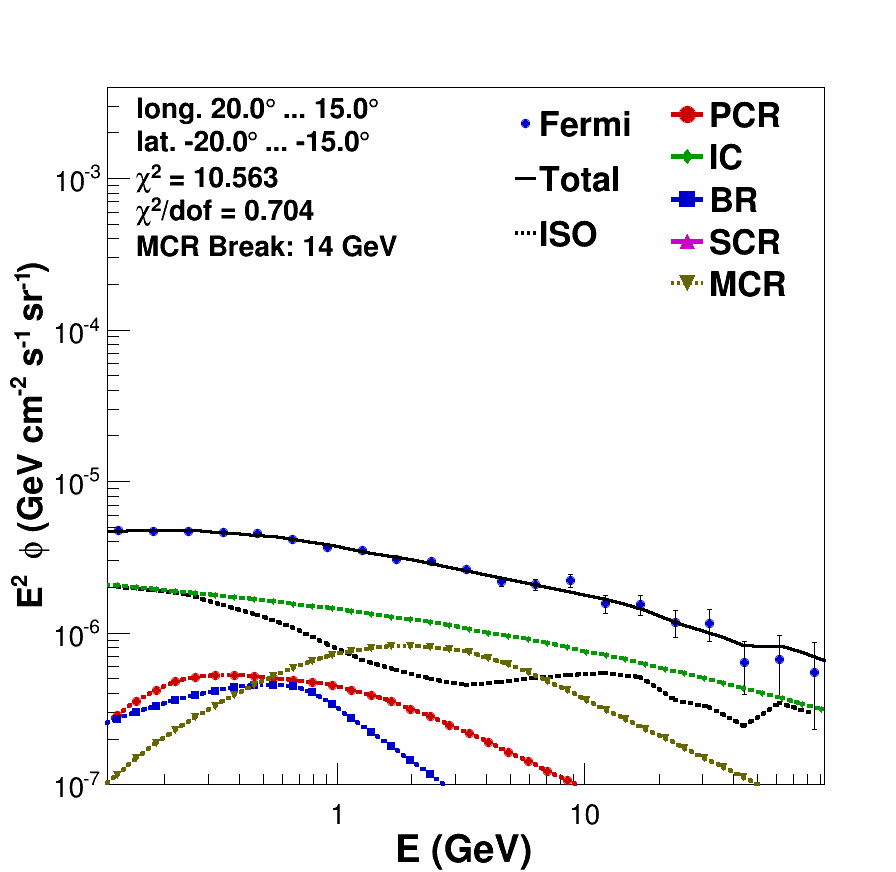}
\includegraphics[width=0.16\textwidth,height=0.16\textwidth,clip]{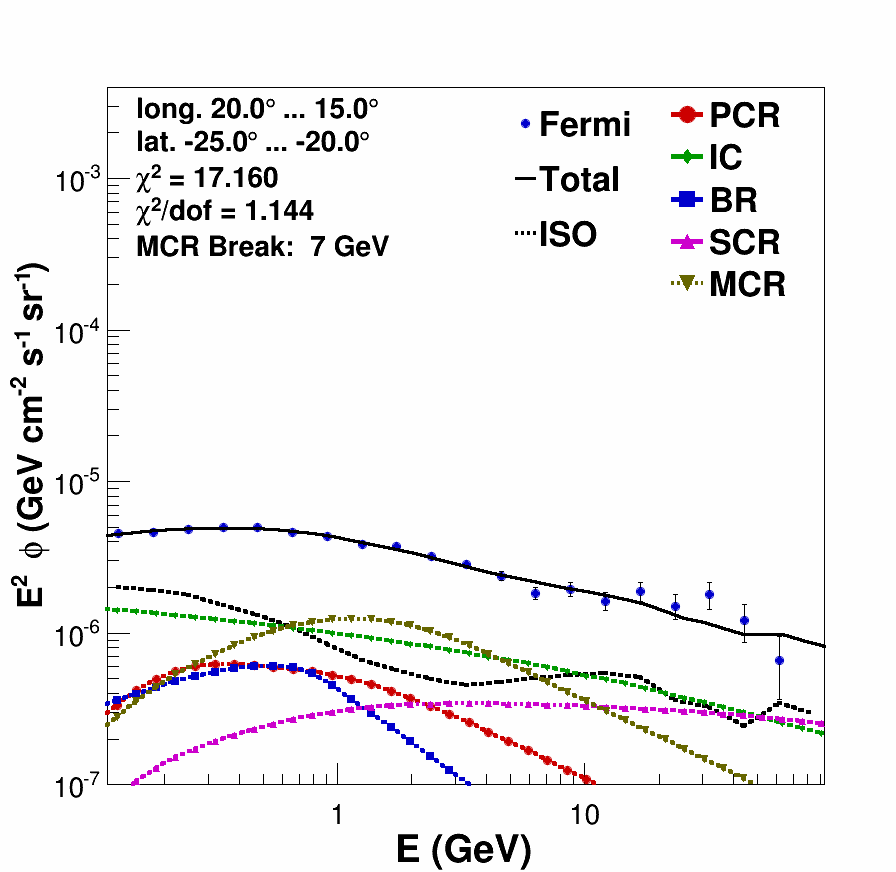}
\includegraphics[width=0.16\textwidth,height=0.16\textwidth,clip]{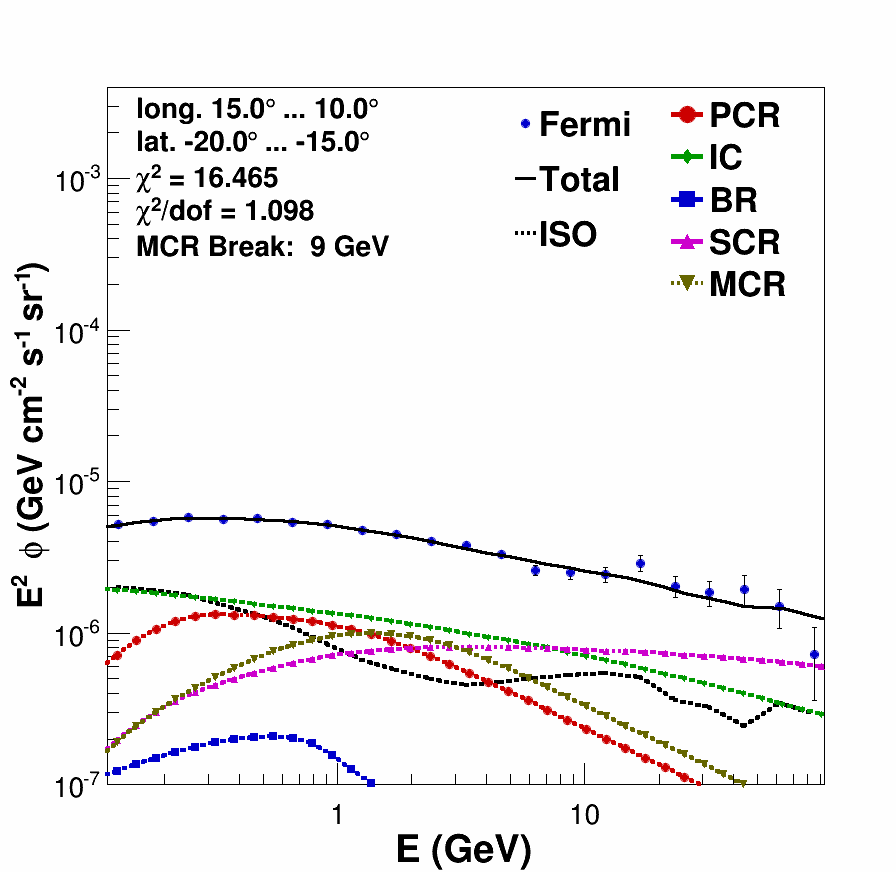}
\includegraphics[width=0.16\textwidth,height=0.16\textwidth,clip]{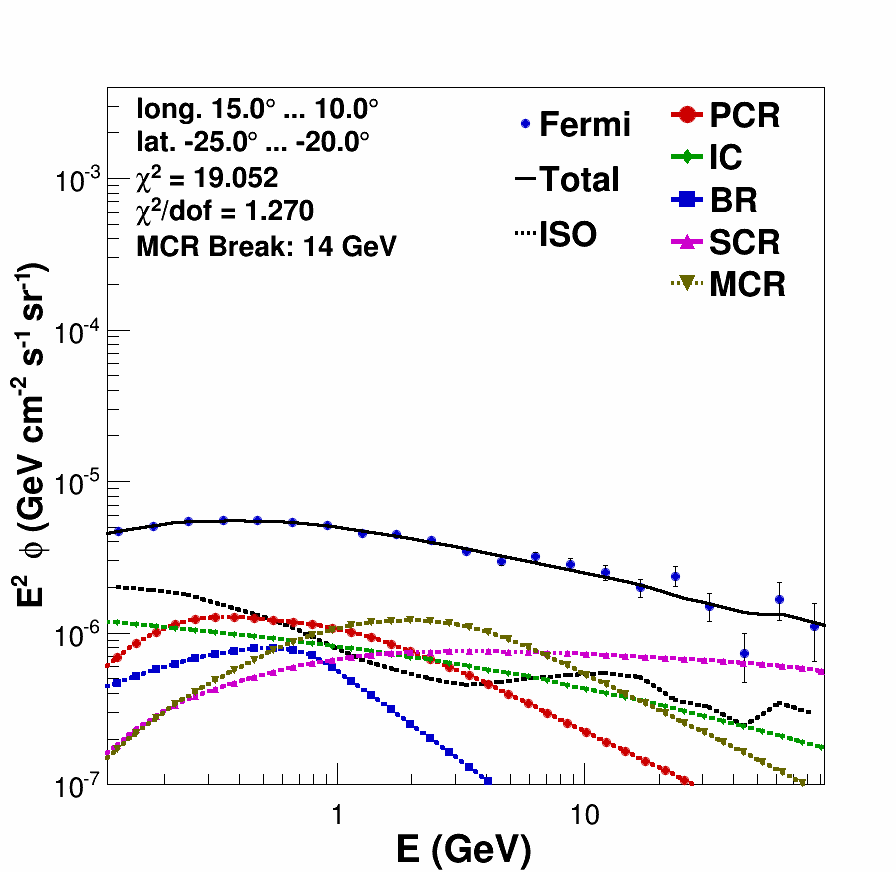}
\includegraphics[width=0.16\textwidth,height=0.16\textwidth,clip]{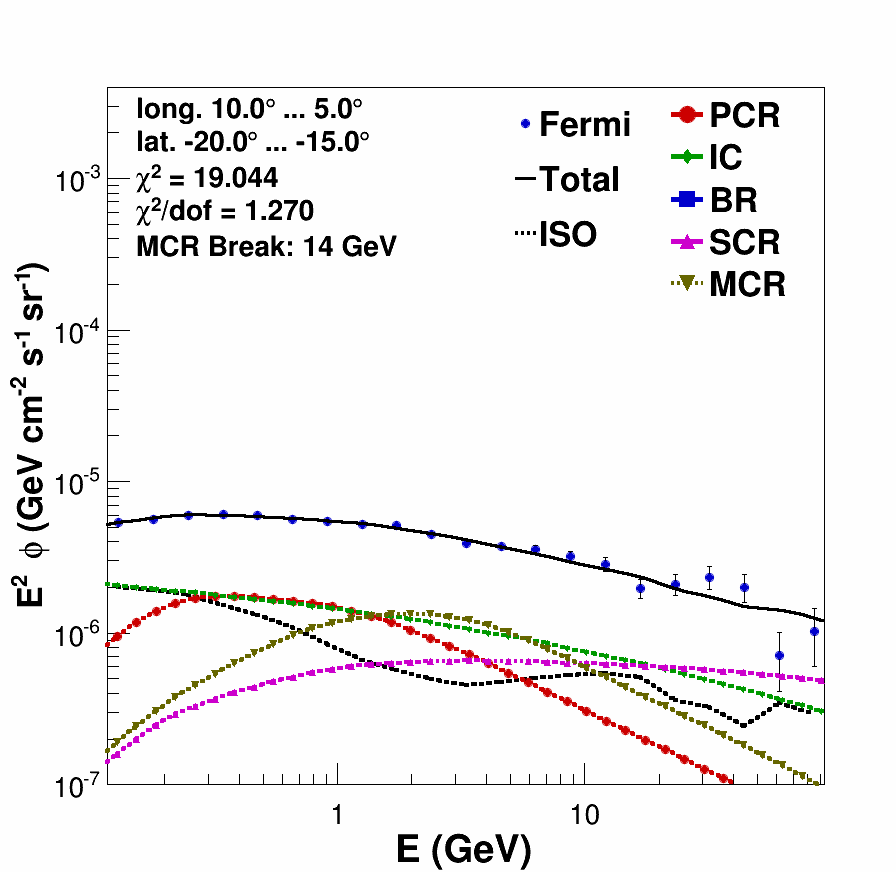}
\includegraphics[width=0.16\textwidth,height=0.16\textwidth,clip]{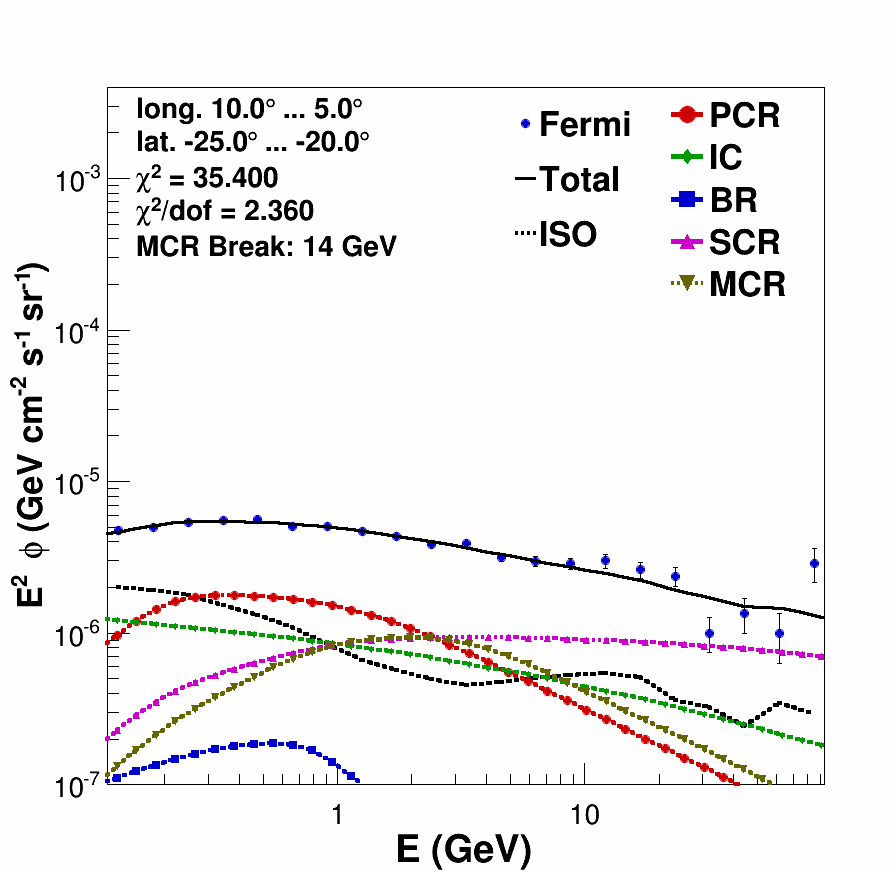}
\includegraphics[width=0.16\textwidth,height=0.16\textwidth,clip]{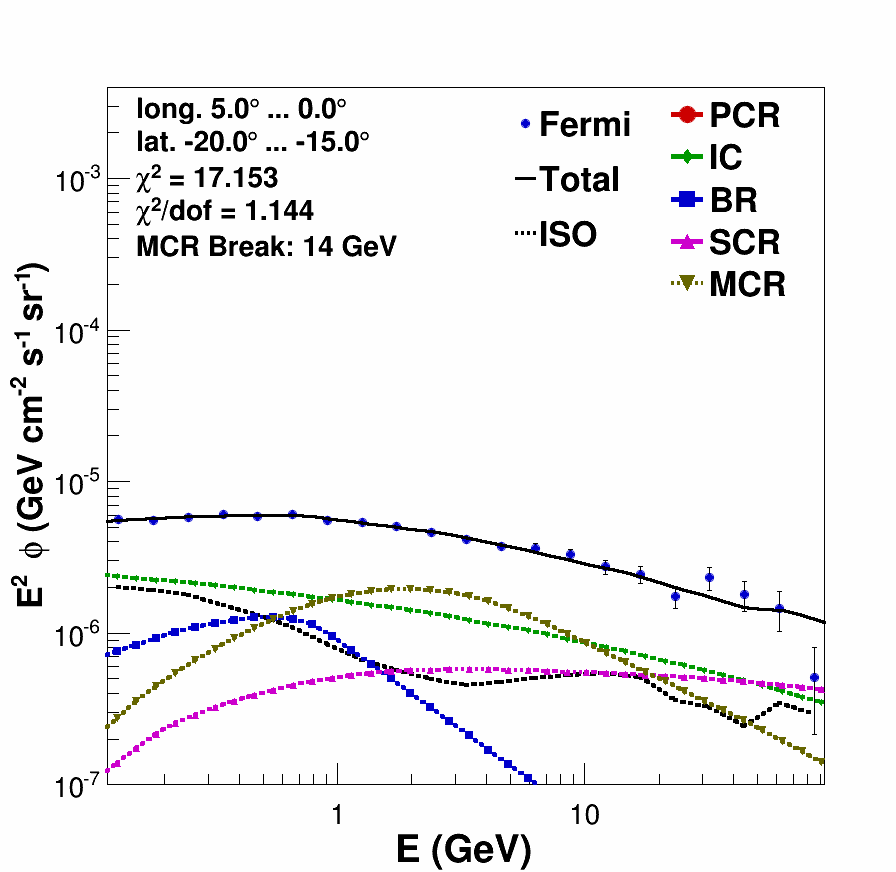}
\includegraphics[width=0.16\textwidth,height=0.16\textwidth,clip]{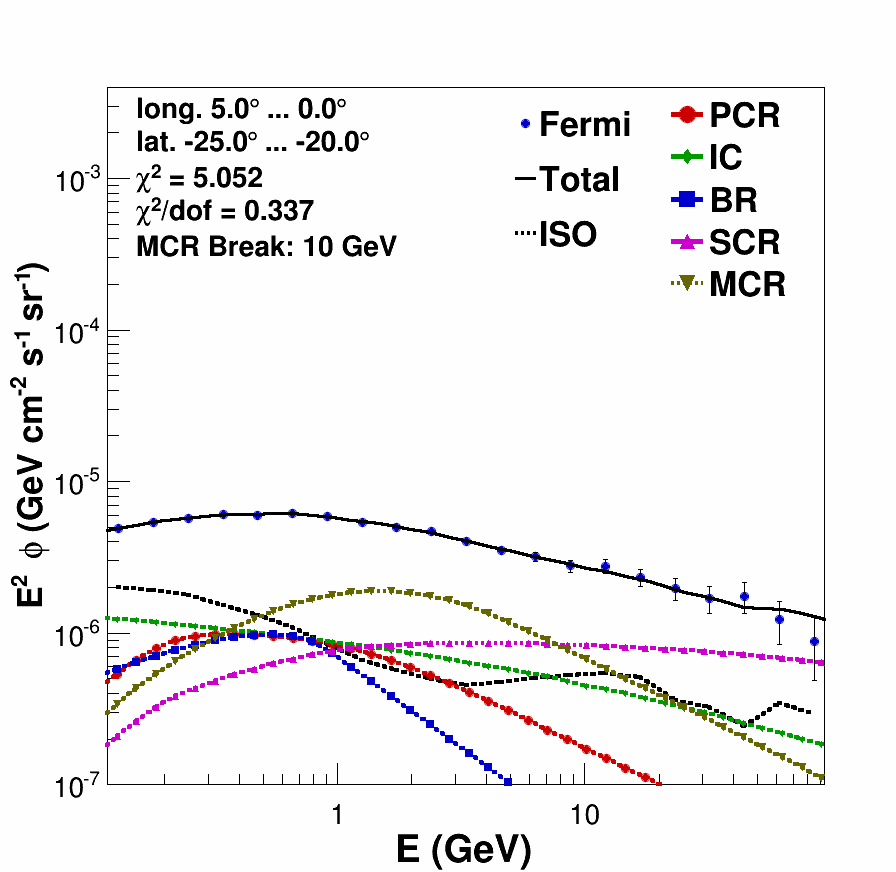}
\includegraphics[width=0.16\textwidth,height=0.16\textwidth,clip]{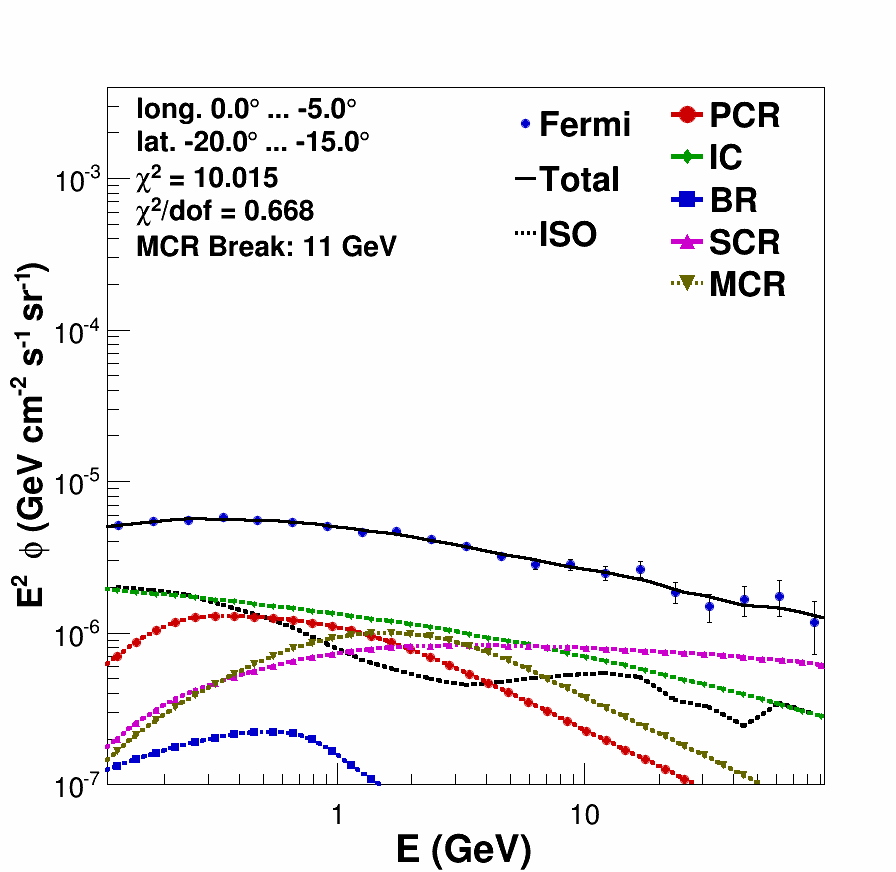}
\includegraphics[width=0.16\textwidth,height=0.16\textwidth,clip]{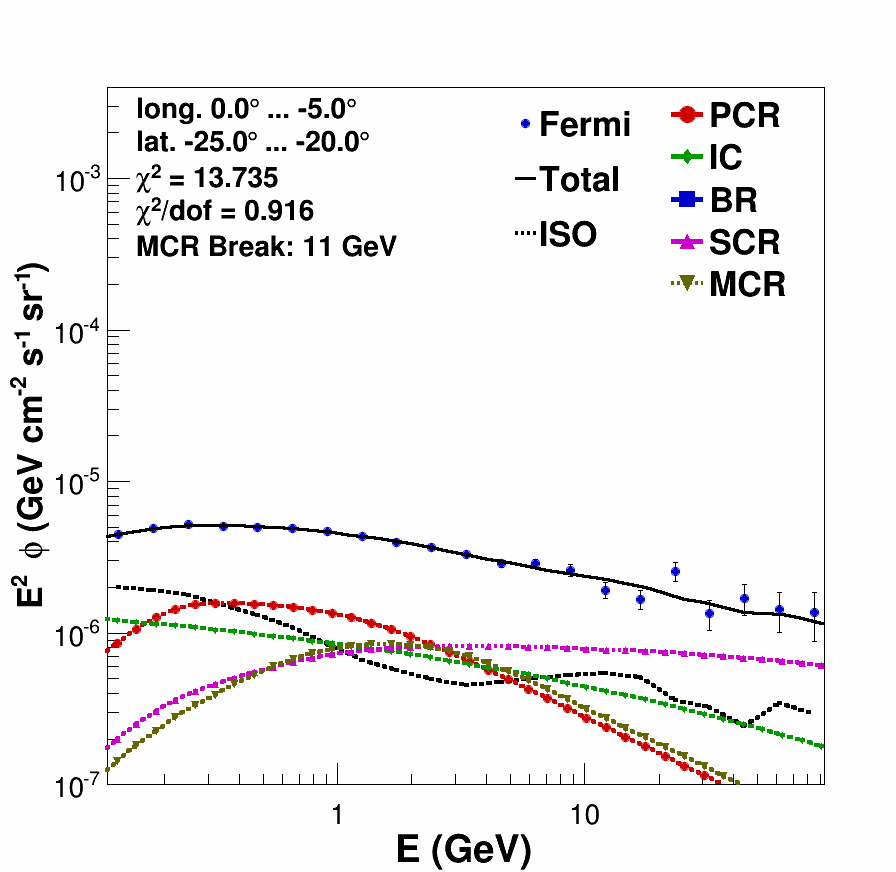}
\includegraphics[width=0.16\textwidth,height=0.16\textwidth,clip]{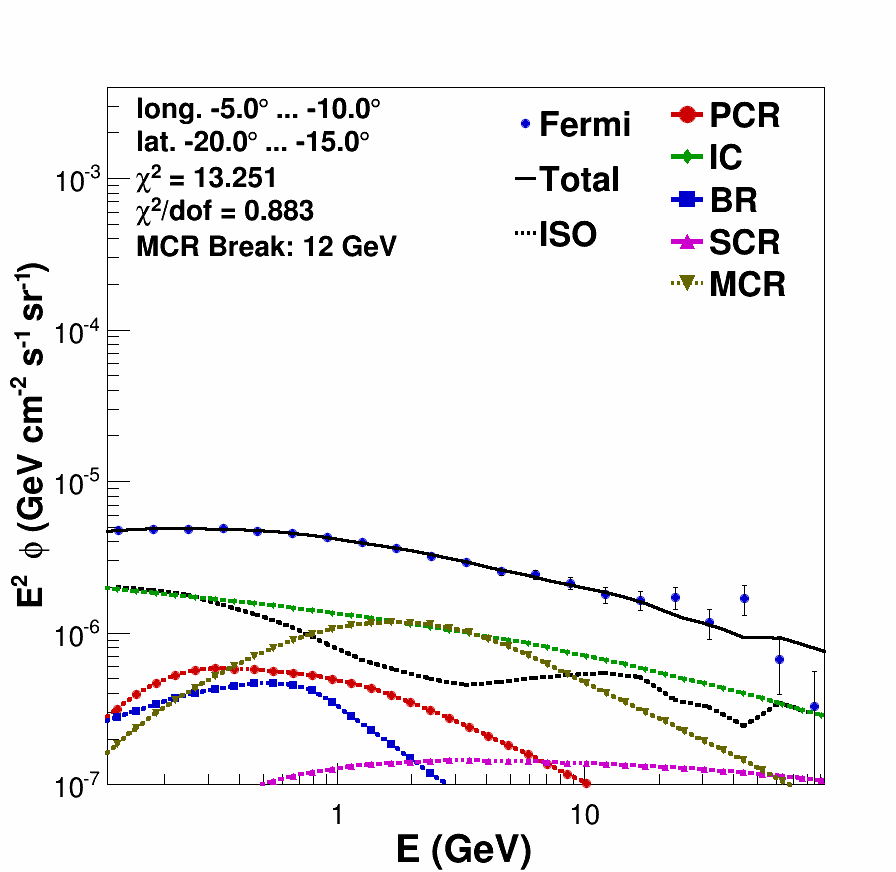}
\includegraphics[width=0.16\textwidth,height=0.16\textwidth,clip]{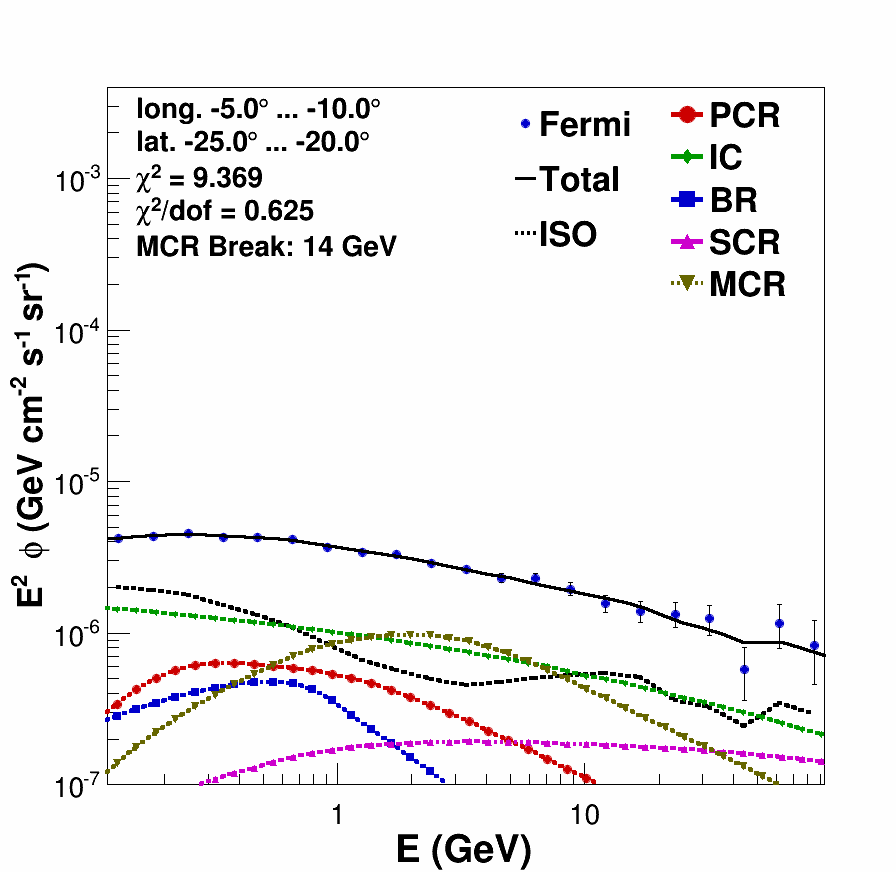}
\includegraphics[width=0.16\textwidth,height=0.16\textwidth,clip]{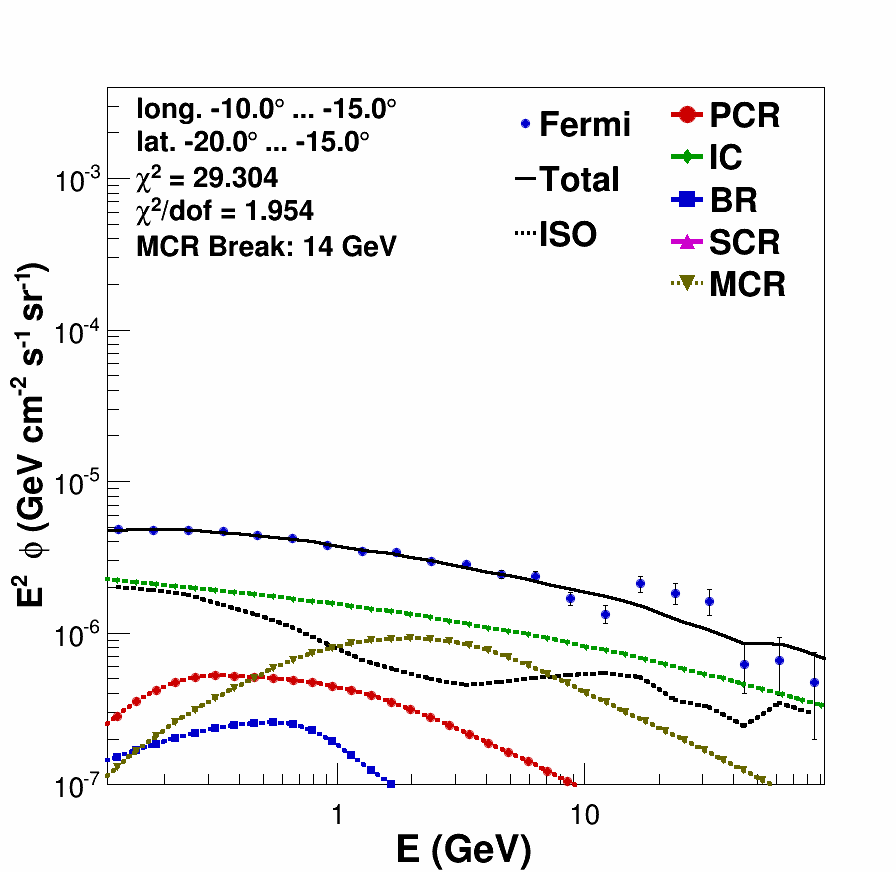}
\includegraphics[width=0.16\textwidth,height=0.16\textwidth,clip]{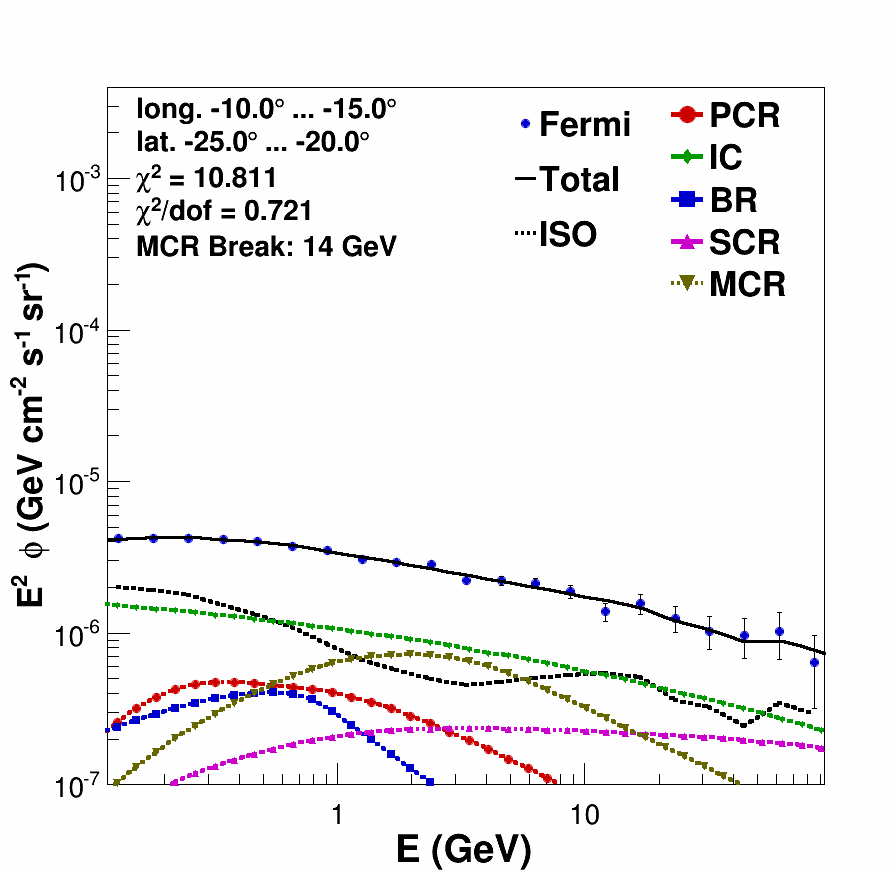}
\includegraphics[width=0.16\textwidth,height=0.16\textwidth,clip]{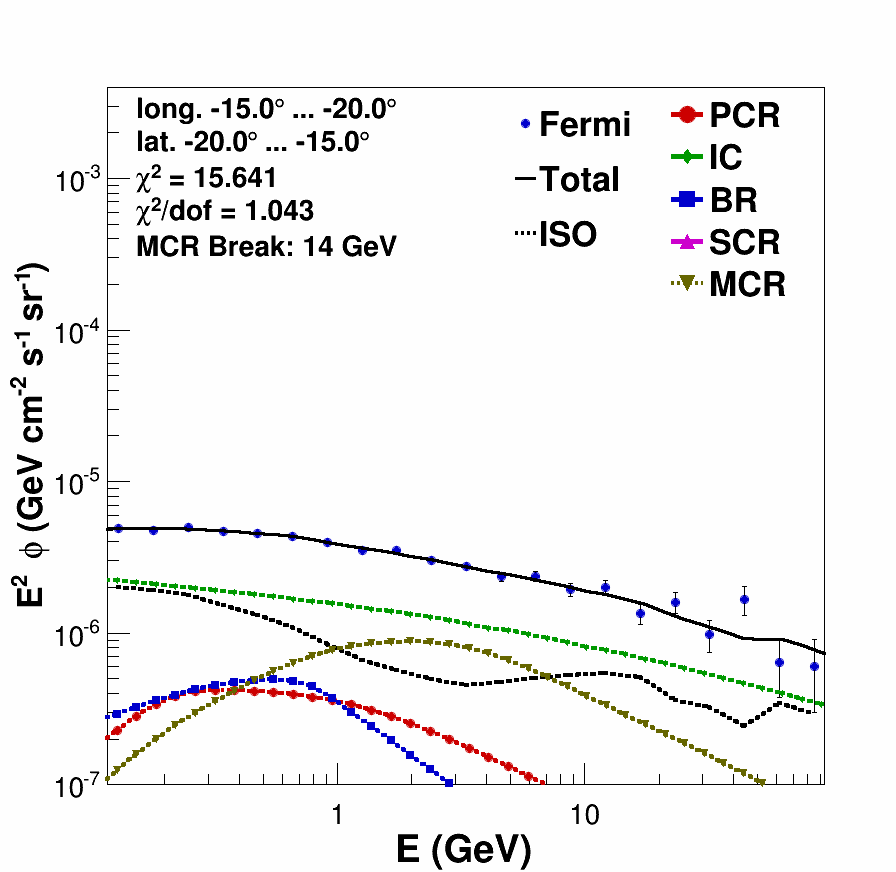}
\includegraphics[width=0.16\textwidth,height=0.16\textwidth,clip]{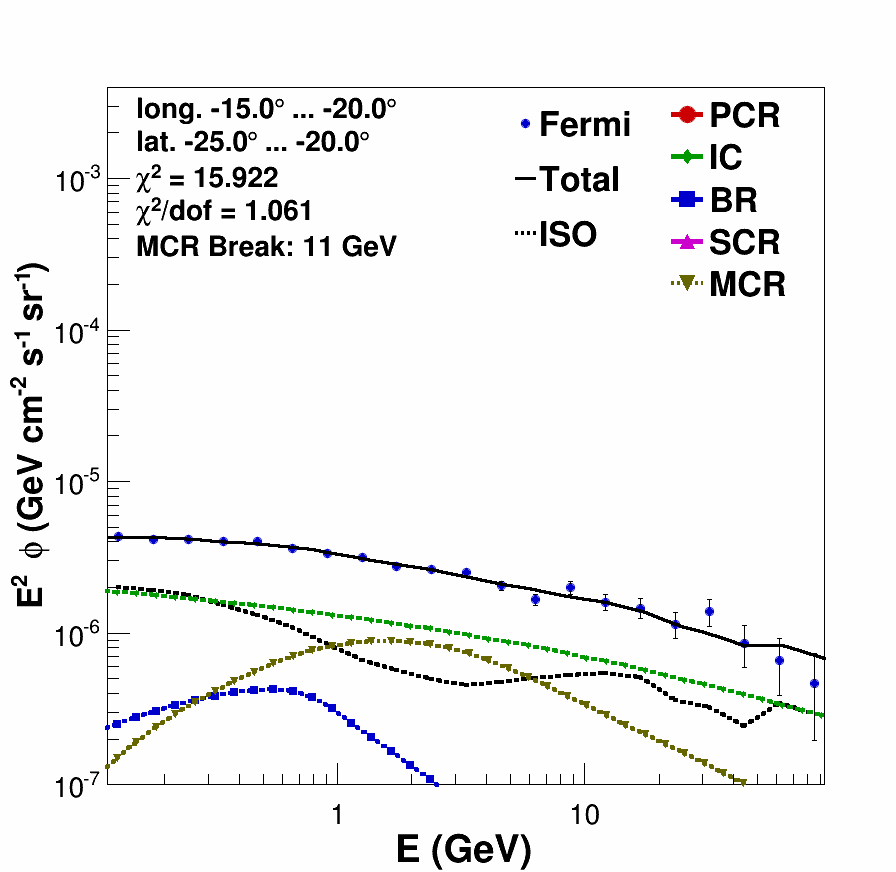}
\includegraphics[width=0.16\textwidth,height=0.16\textwidth,clip]{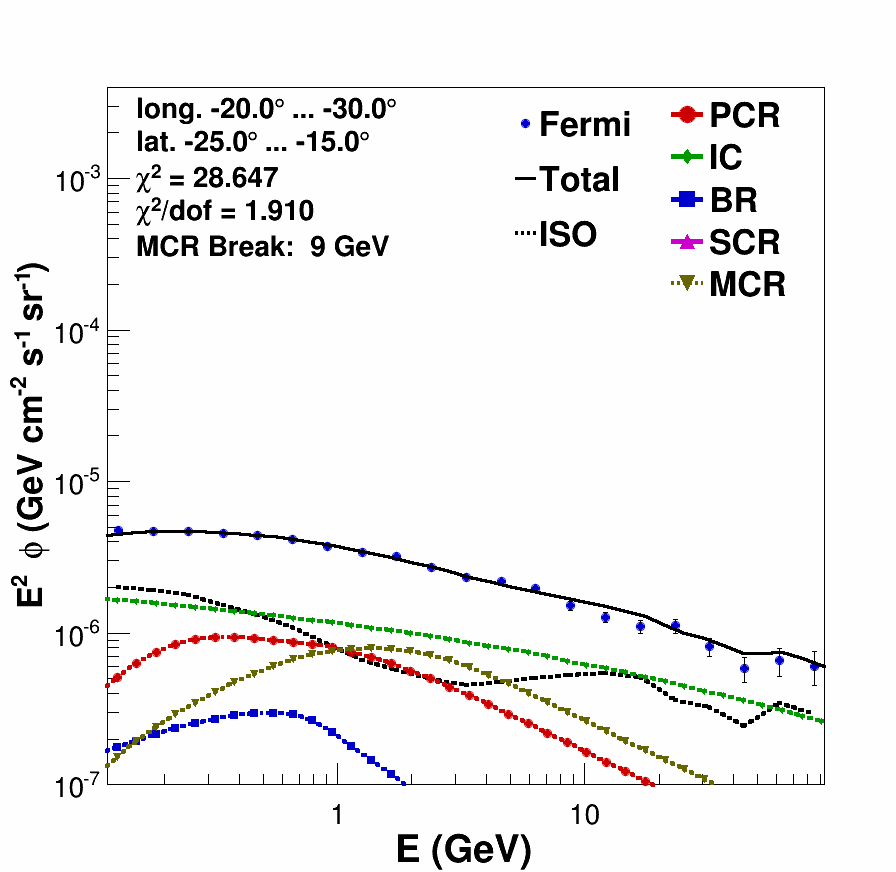}
\includegraphics[width=0.16\textwidth,height=0.16\textwidth,clip]{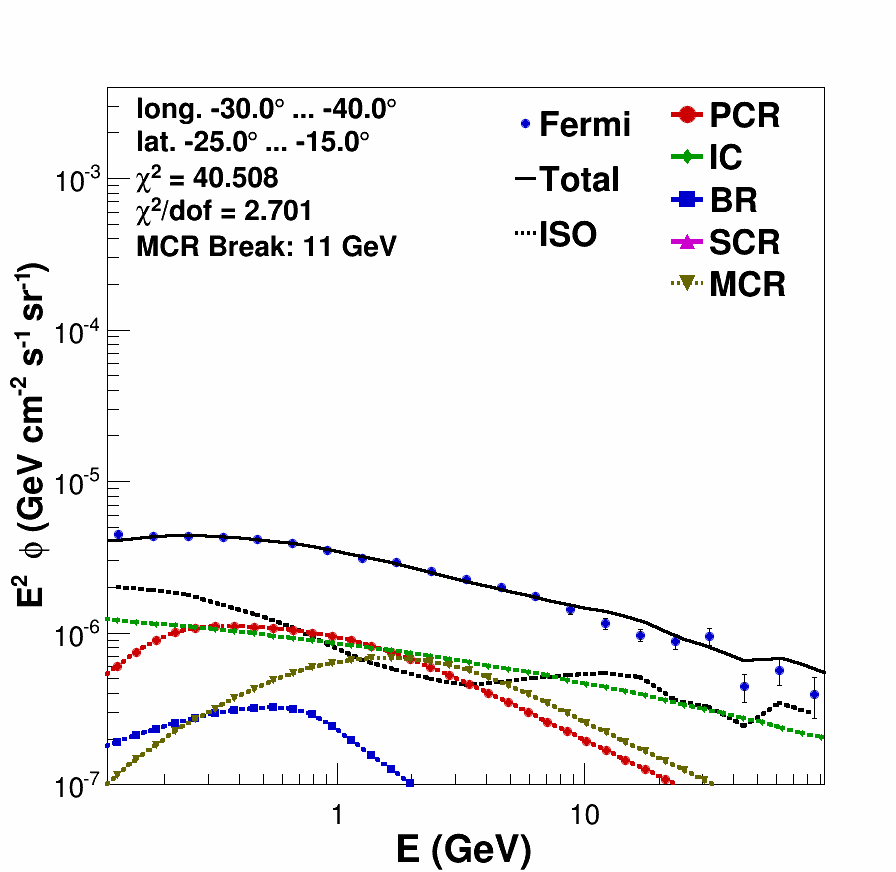}
\includegraphics[width=0.16\textwidth,height=0.16\textwidth,clip]{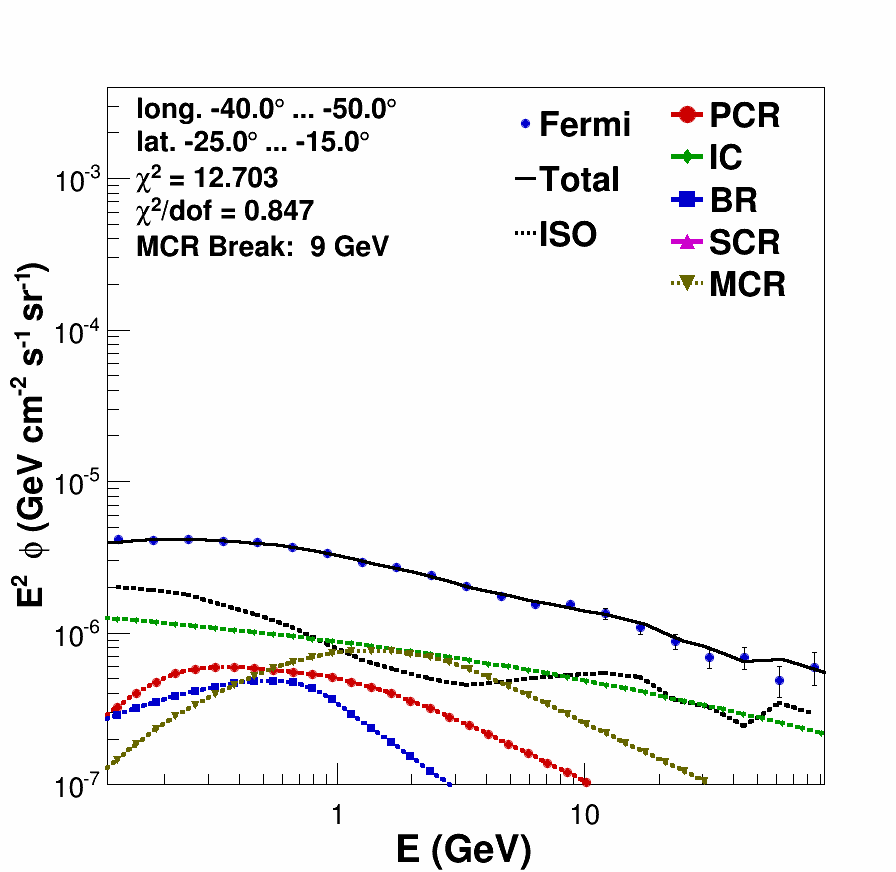}
\includegraphics[width=0.16\textwidth,height=0.16\textwidth,clip]{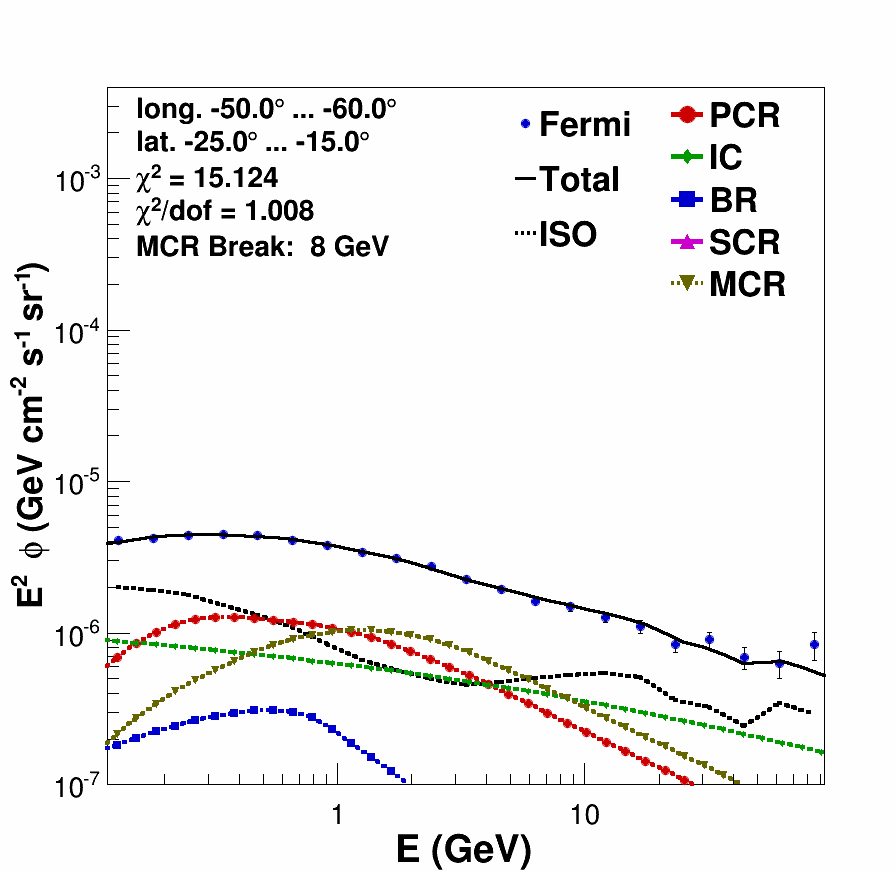}
\includegraphics[width=0.16\textwidth,height=0.16\textwidth,clip]{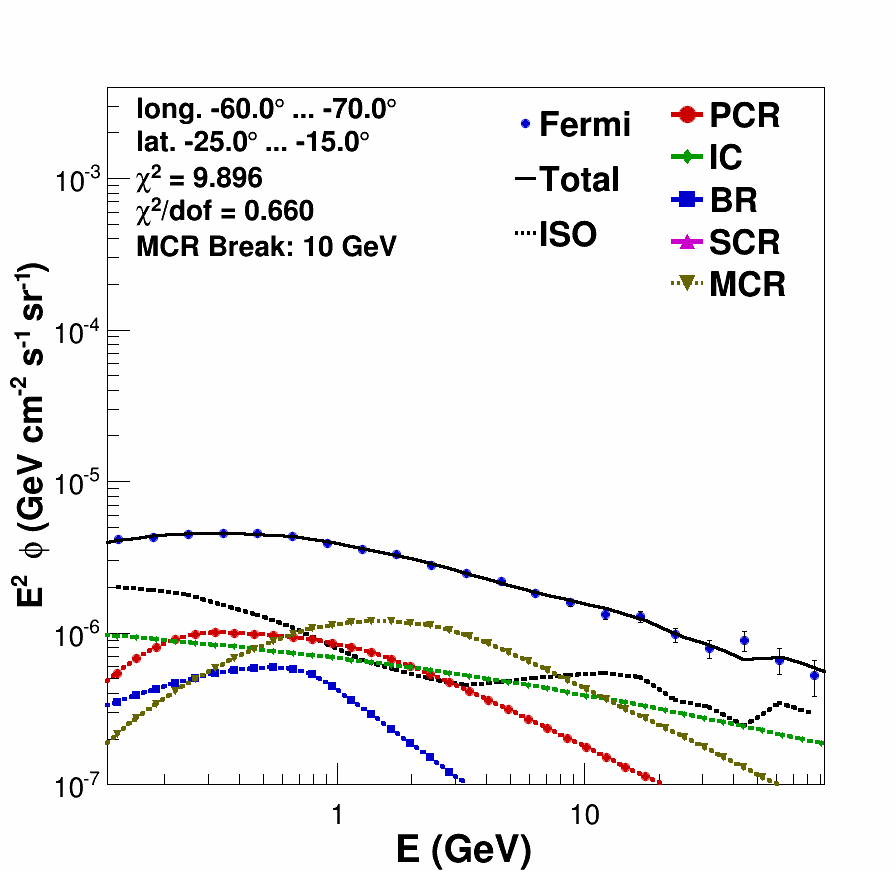}
\includegraphics[width=0.16\textwidth,height=0.16\textwidth,clip]{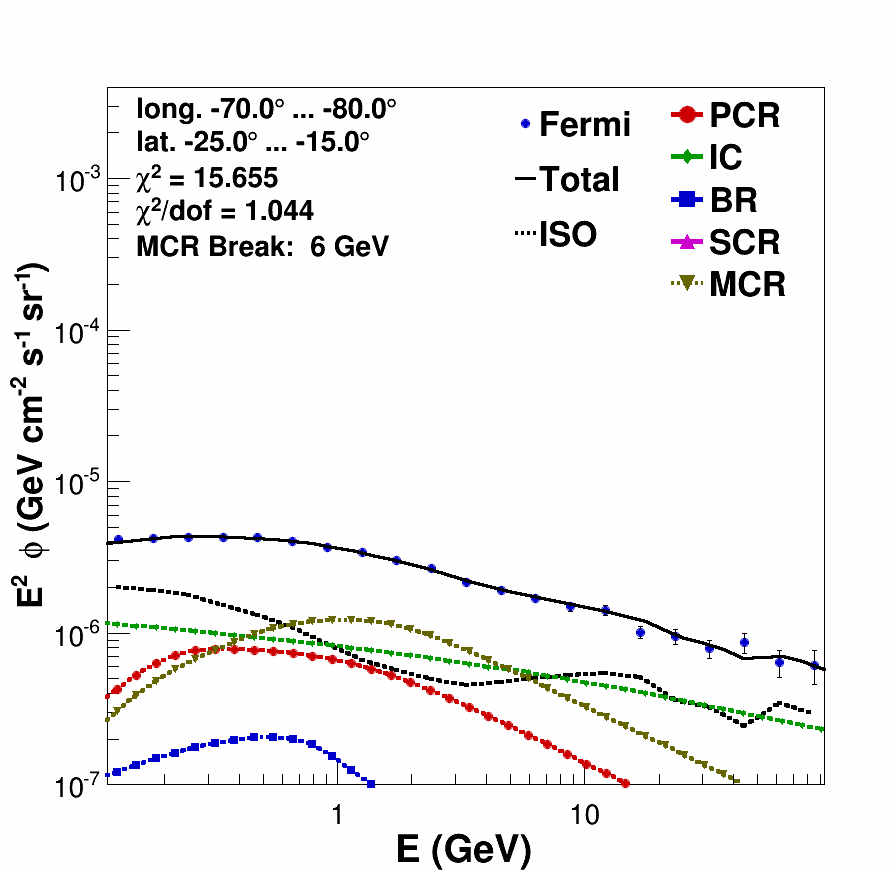}
\includegraphics[width=0.16\textwidth,height=0.16\textwidth,clip]{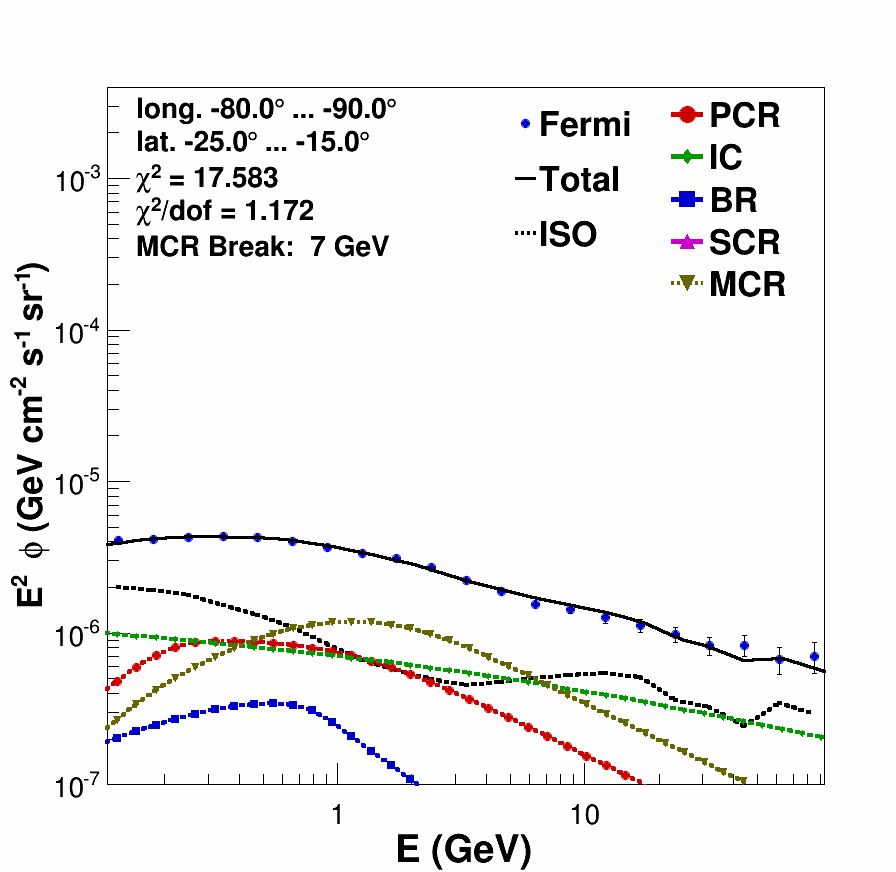}
\includegraphics[width=0.16\textwidth,height=0.16\textwidth,clip]{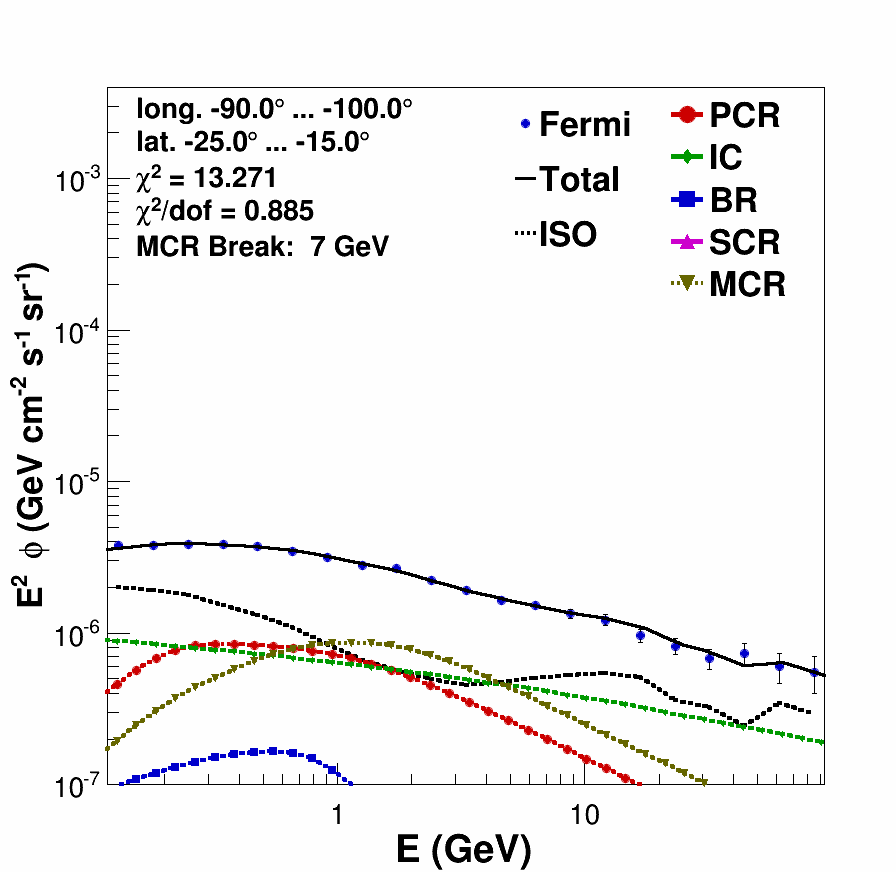}
\includegraphics[width=0.16\textwidth,height=0.16\textwidth,clip]{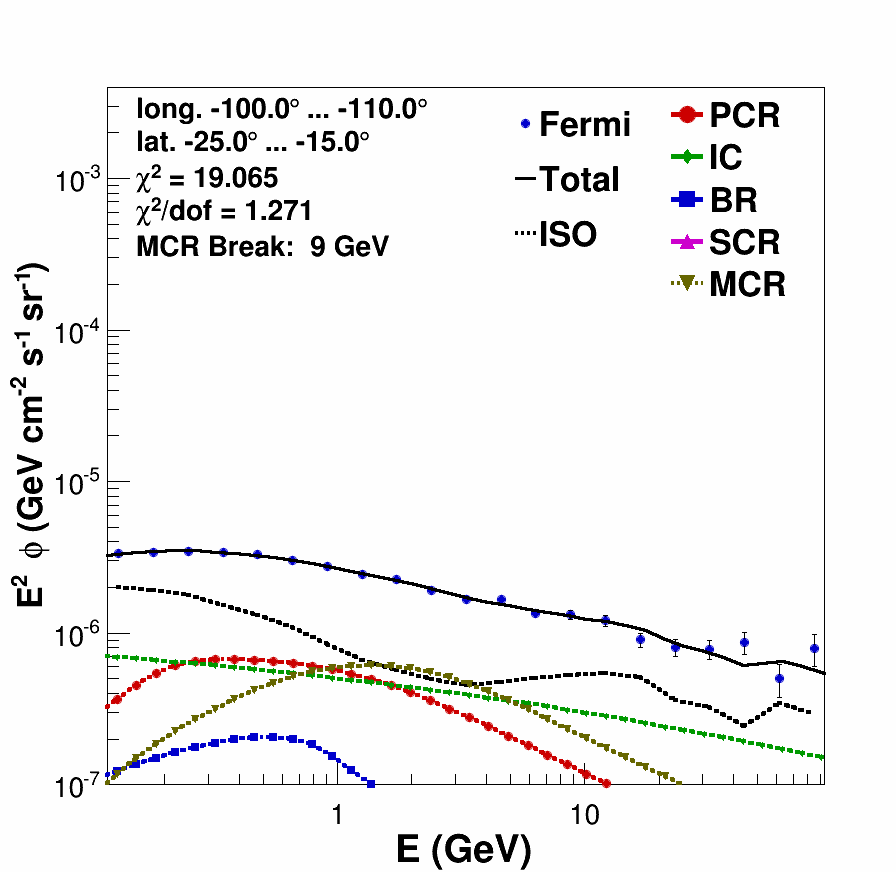}
\includegraphics[width=0.16\textwidth,height=0.16\textwidth,clip]{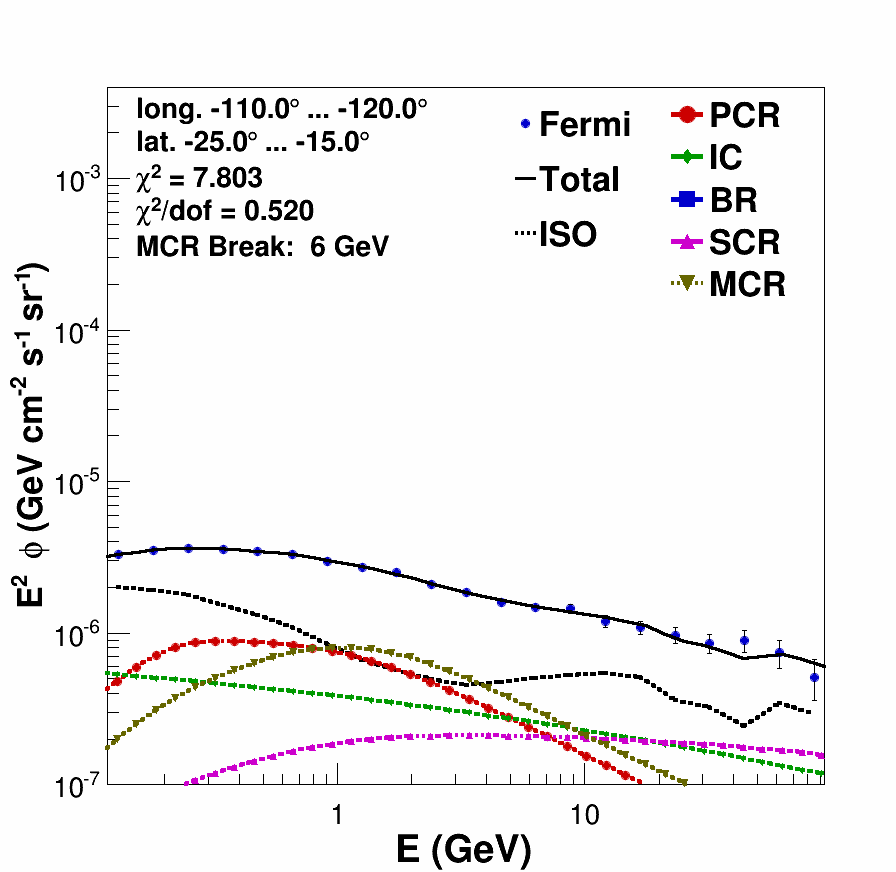}
\includegraphics[width=0.16\textwidth,height=0.16\textwidth,clip]{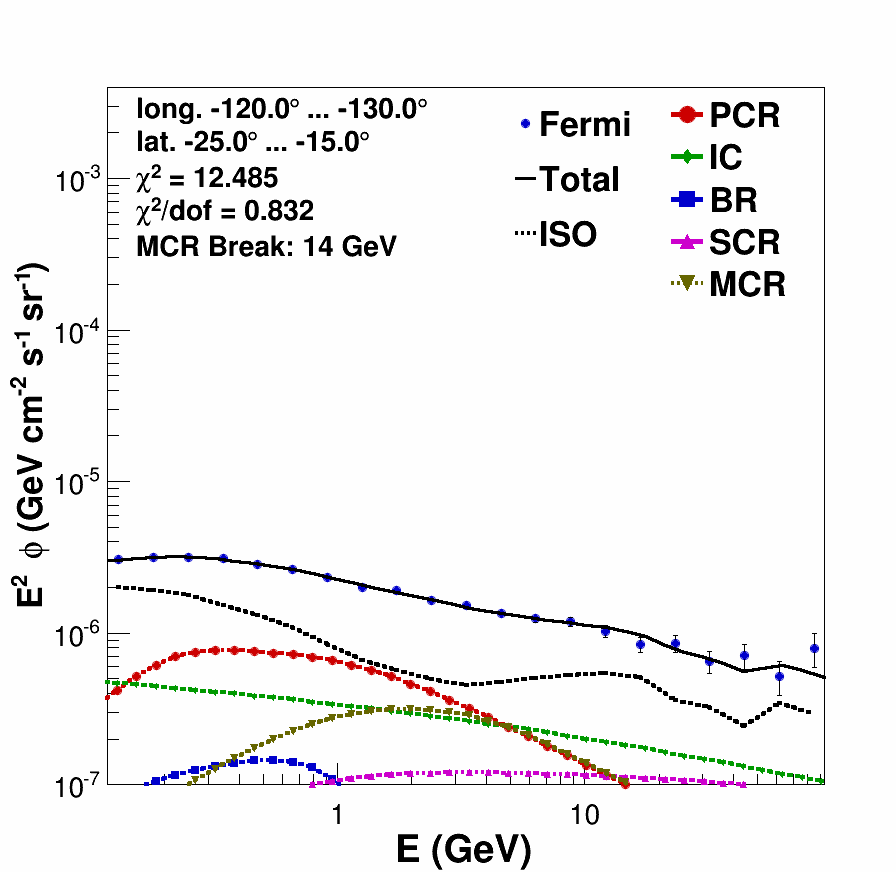}
\includegraphics[width=0.16\textwidth,height=0.16\textwidth,clip]{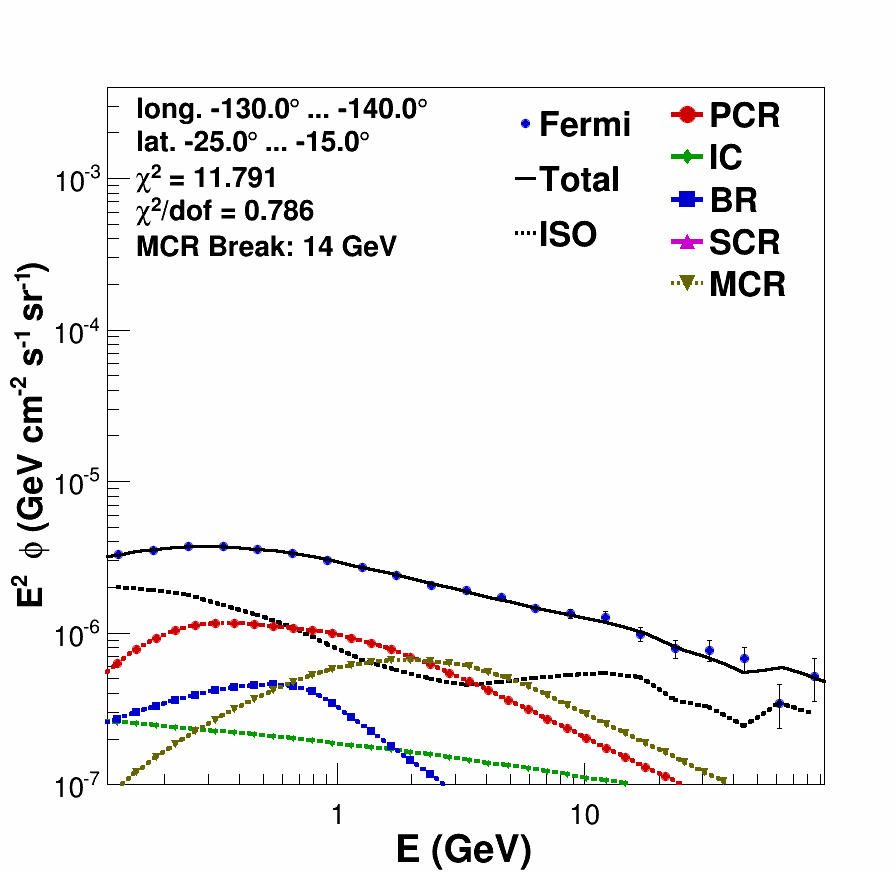}
\includegraphics[width=0.16\textwidth,height=0.16\textwidth,clip]{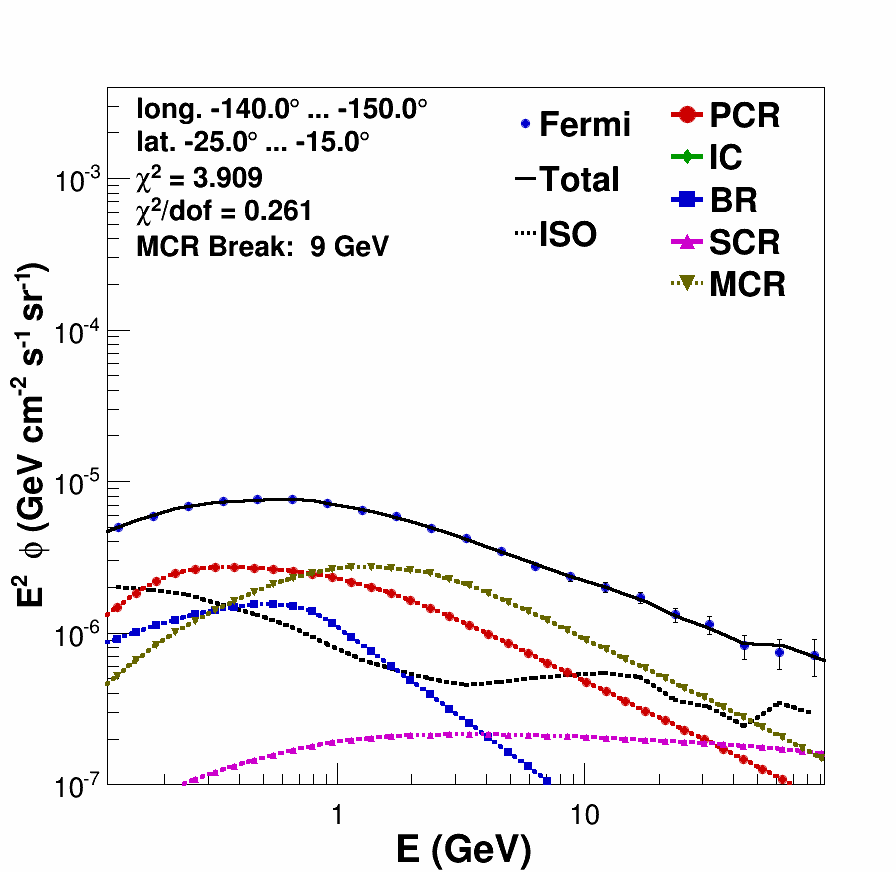}
\includegraphics[width=0.16\textwidth,height=0.16\textwidth,clip]{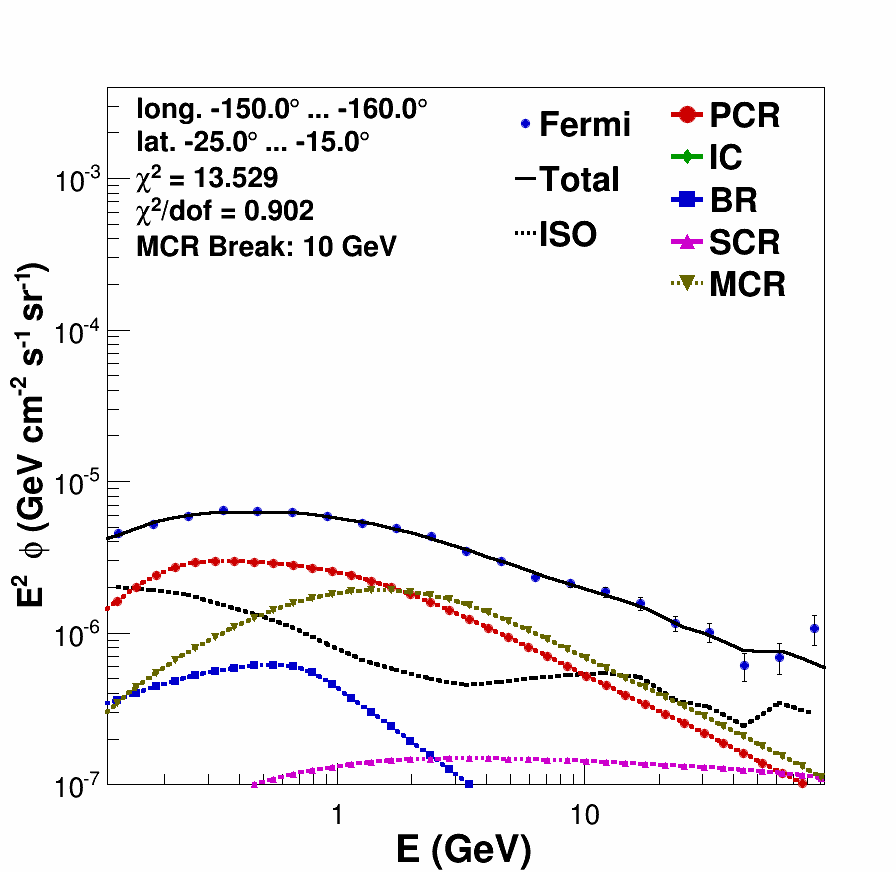}
\includegraphics[width=0.16\textwidth,height=0.16\textwidth,clip]{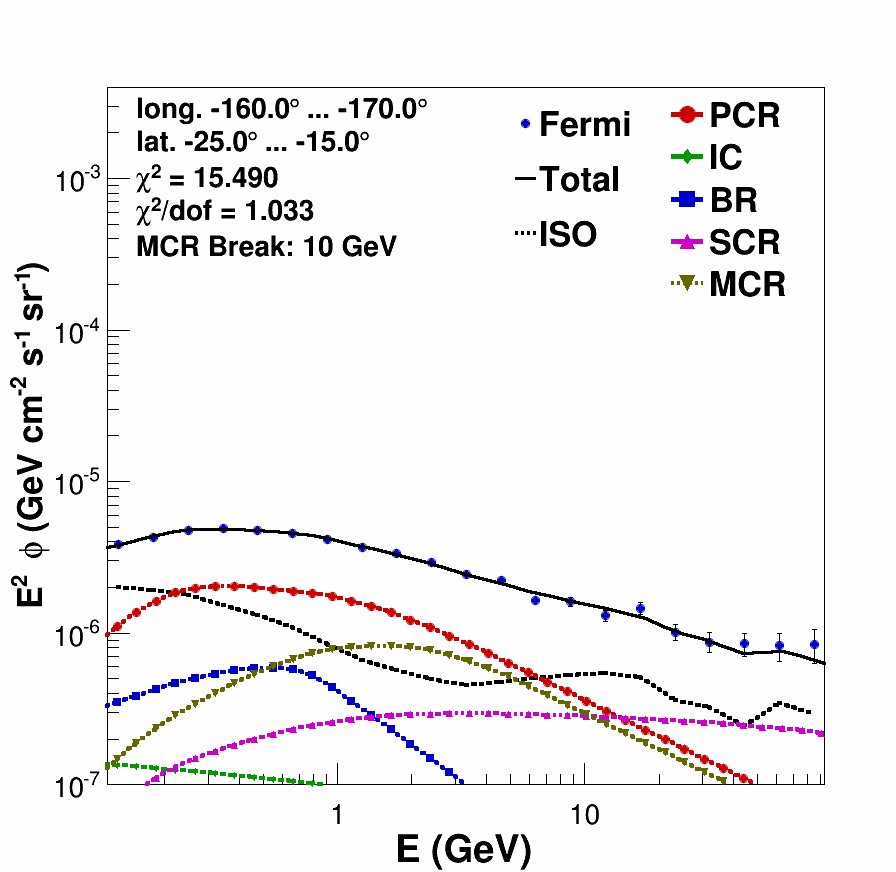}
\includegraphics[width=0.16\textwidth,height=0.16\textwidth,clip]{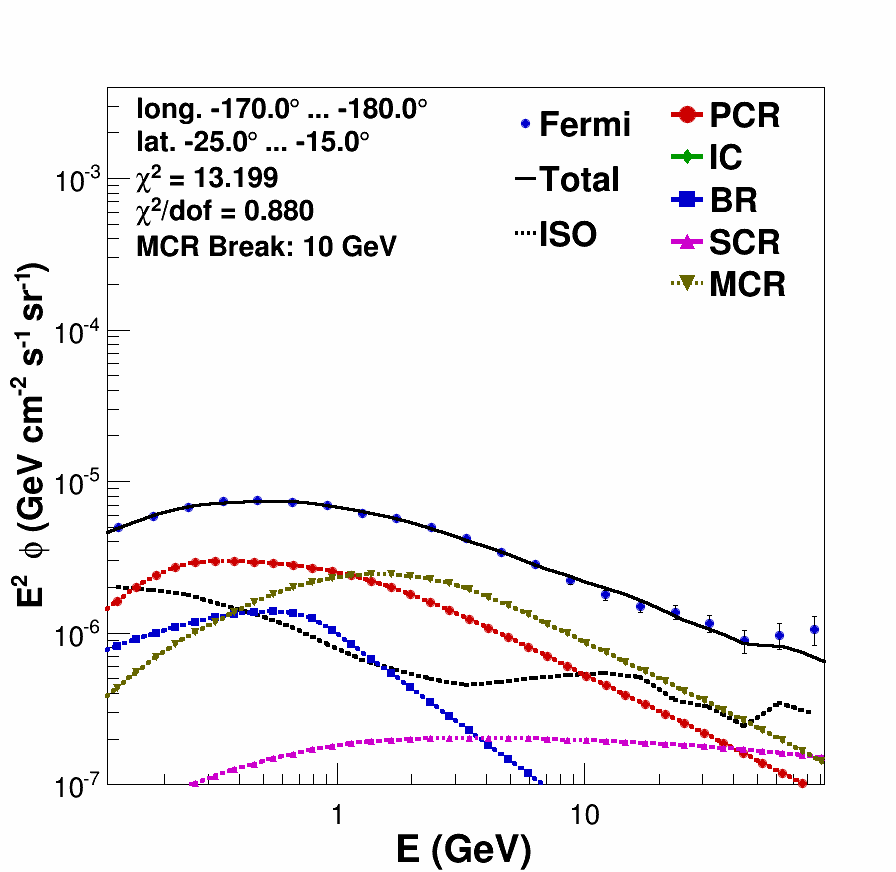}%%%%%%r14
\caption[]{Template fits for latitudes  with $-25.0^\circ<b<-15.0^\circ$ and longitudes decreasing from 180$^\circ$ to -180$^\circ$.} \label{F26}
\end{figure}
\clearpage
\begin{figure}
\centering
\includegraphics[width=0.16\textwidth,height=0.16\textwidth,clip]{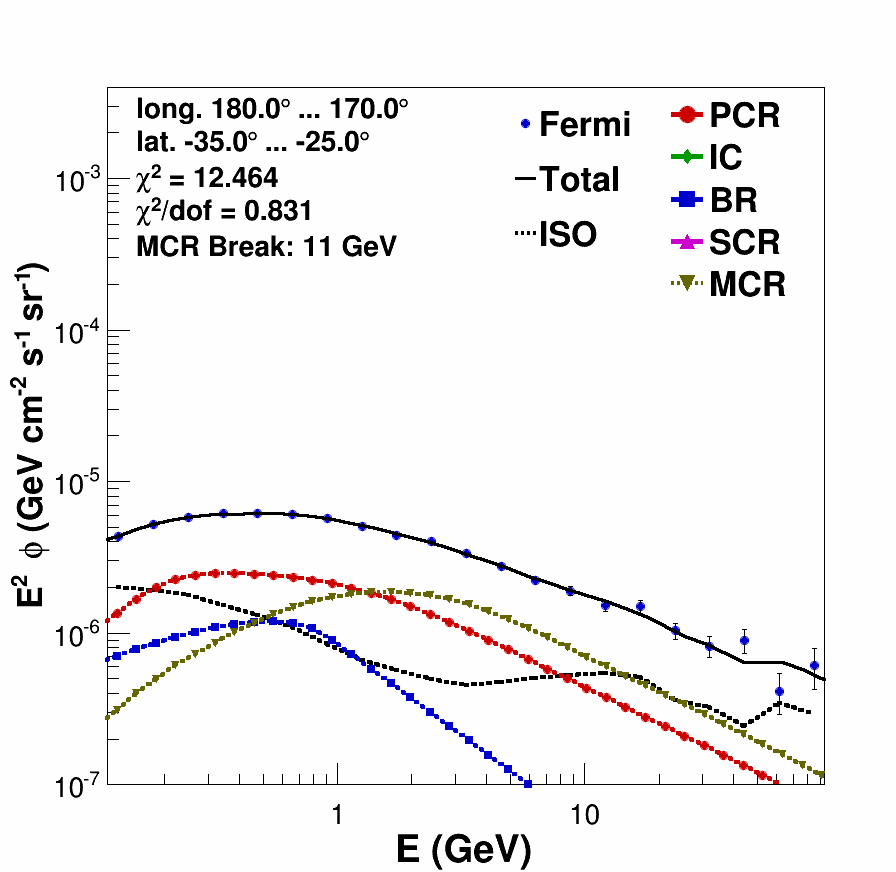}
\includegraphics[width=0.16\textwidth,height=0.16\textwidth,clip]{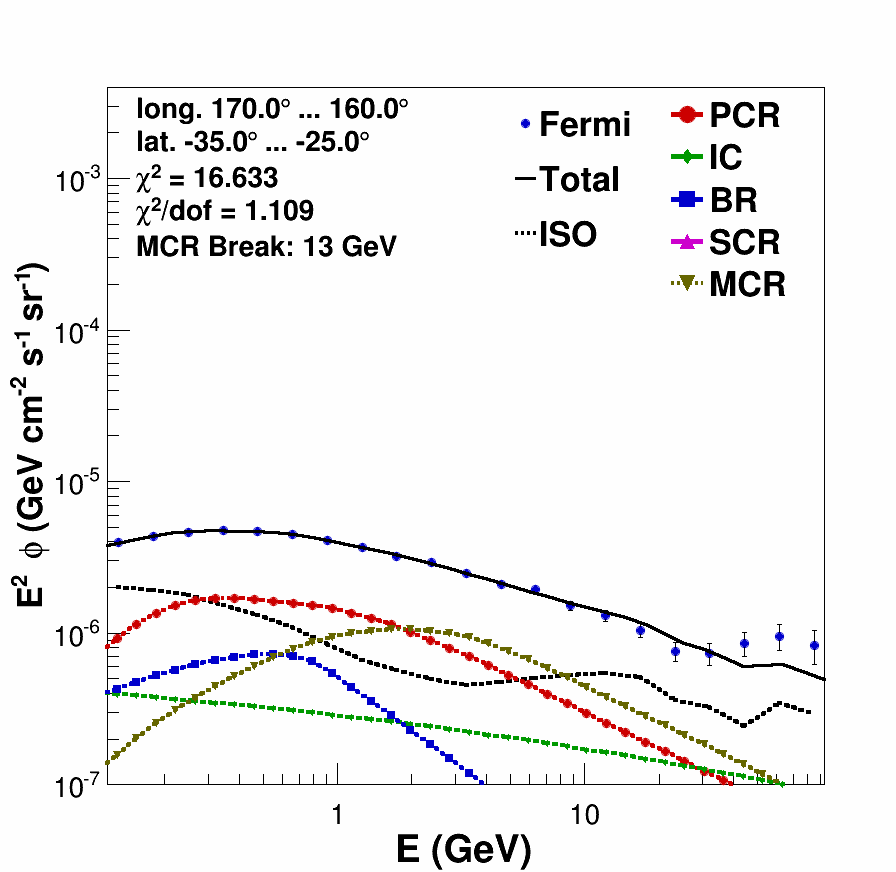}
\includegraphics[width=0.16\textwidth,height=0.16\textwidth,clip]{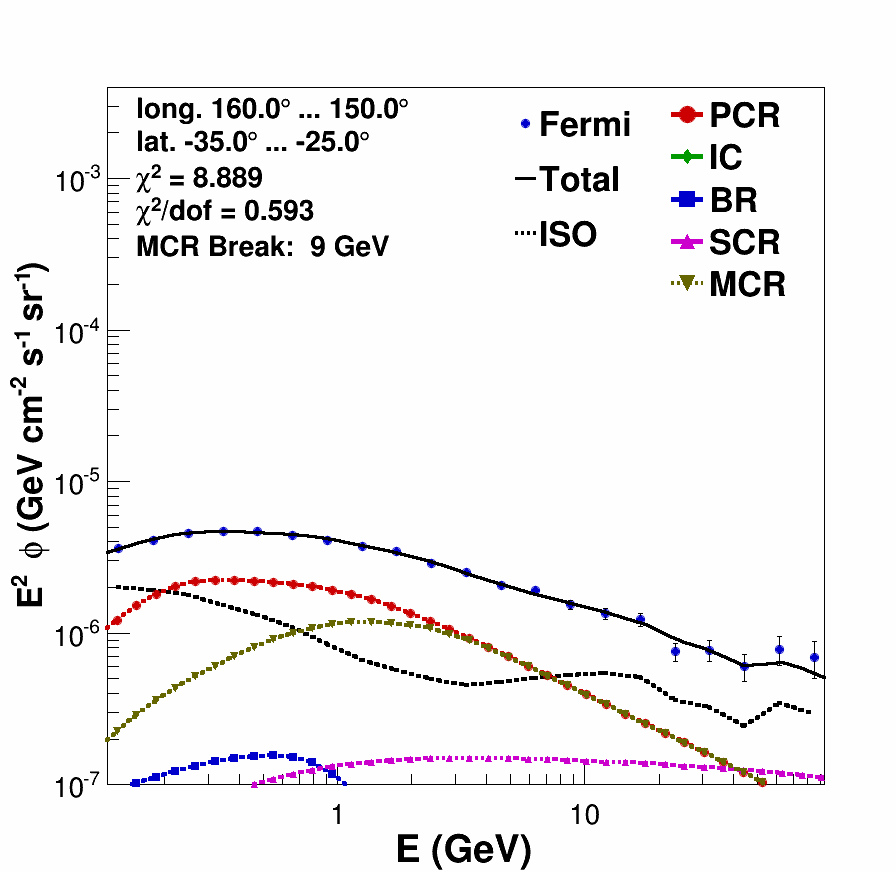}
\includegraphics[width=0.16\textwidth,height=0.16\textwidth,clip]{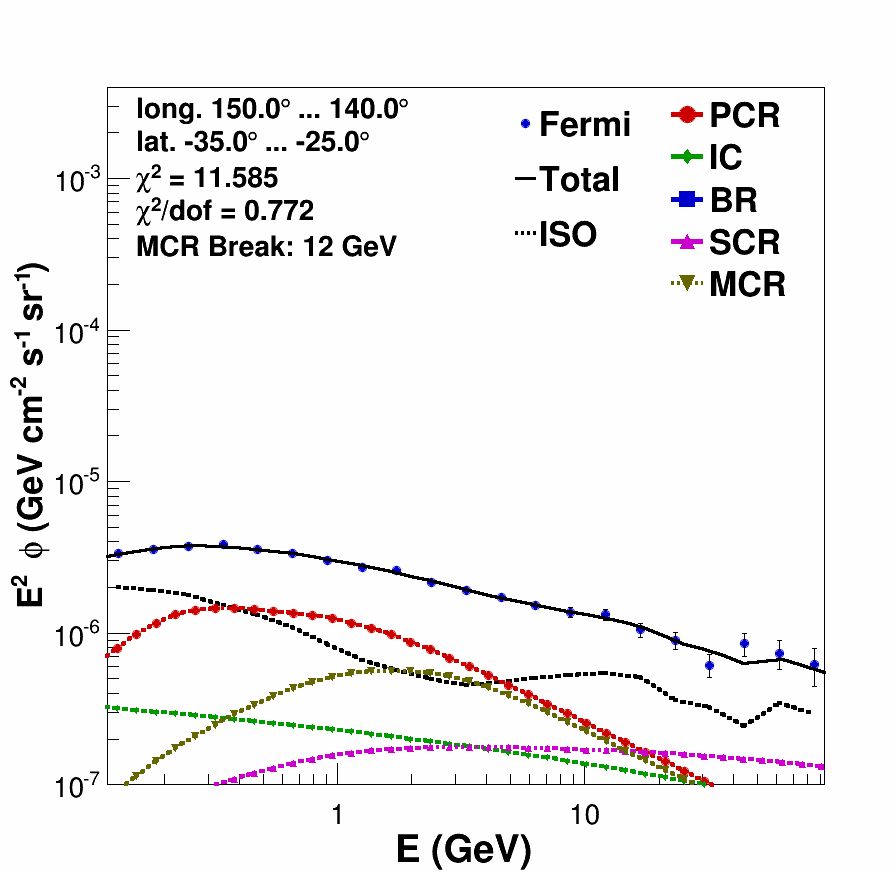}
\includegraphics[width=0.16\textwidth,height=0.16\textwidth,clip]{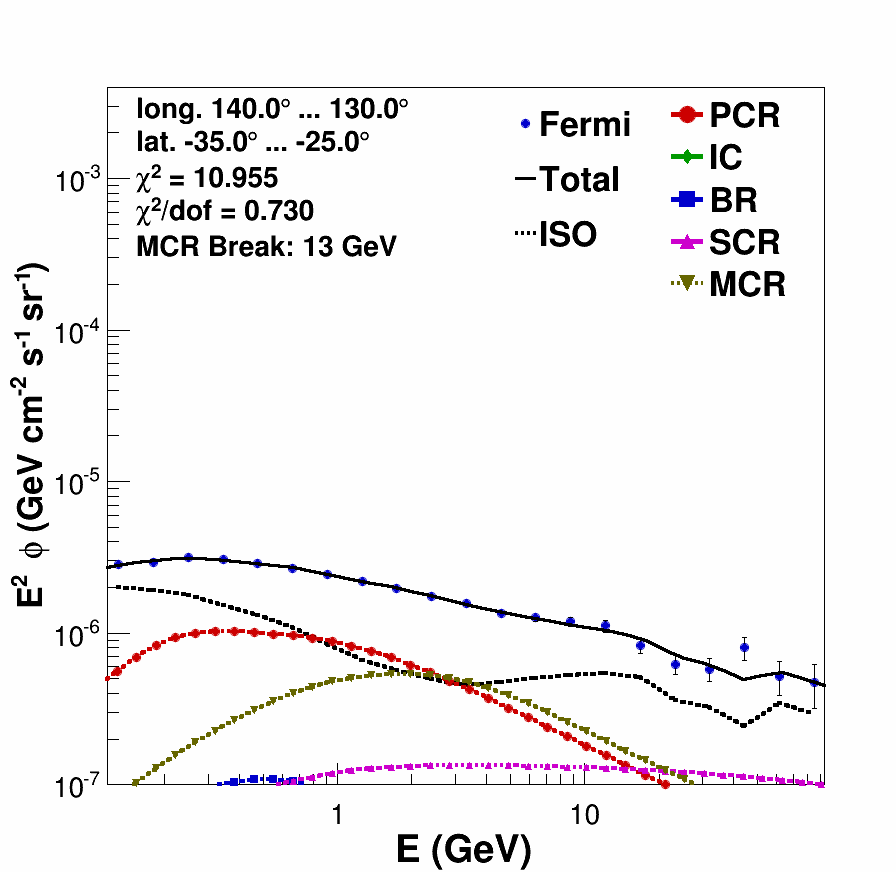}
\includegraphics[width=0.16\textwidth,height=0.16\textwidth,clip]{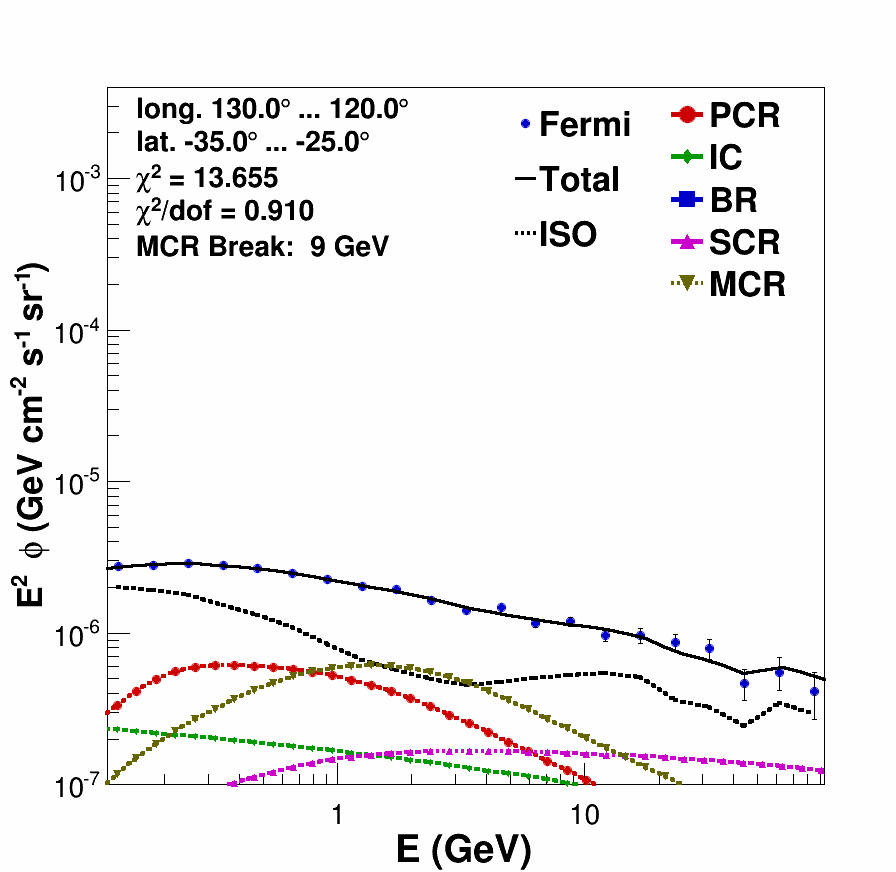}
\includegraphics[width=0.16\textwidth,height=0.16\textwidth,clip]{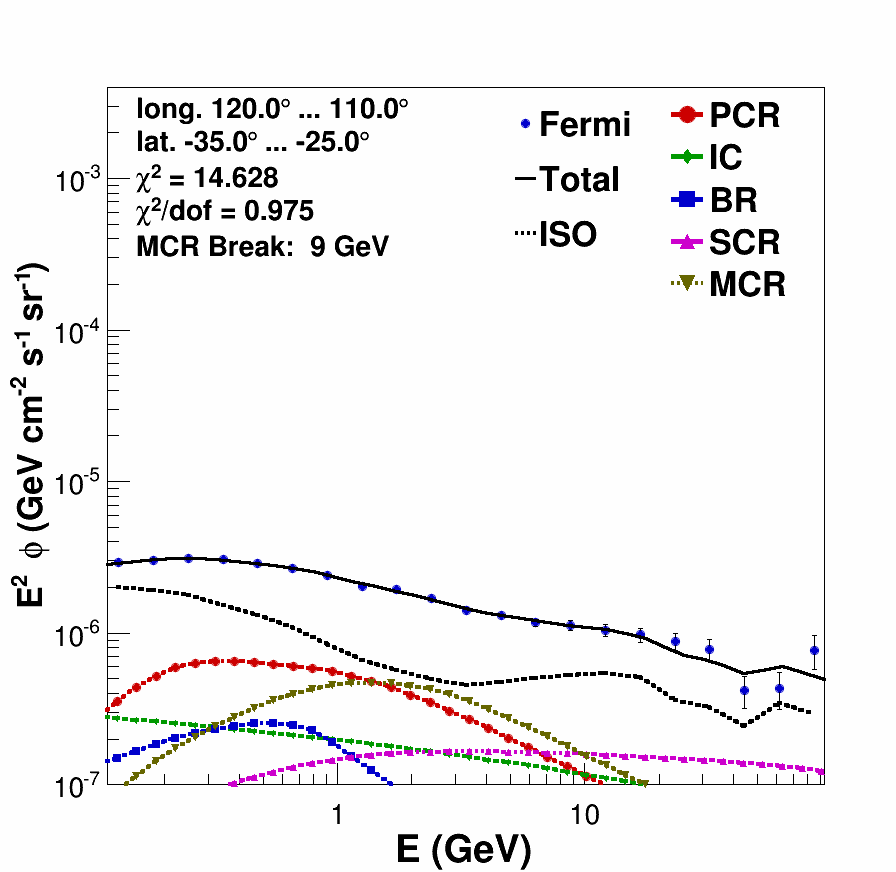}
\includegraphics[width=0.16\textwidth,height=0.16\textwidth,clip]{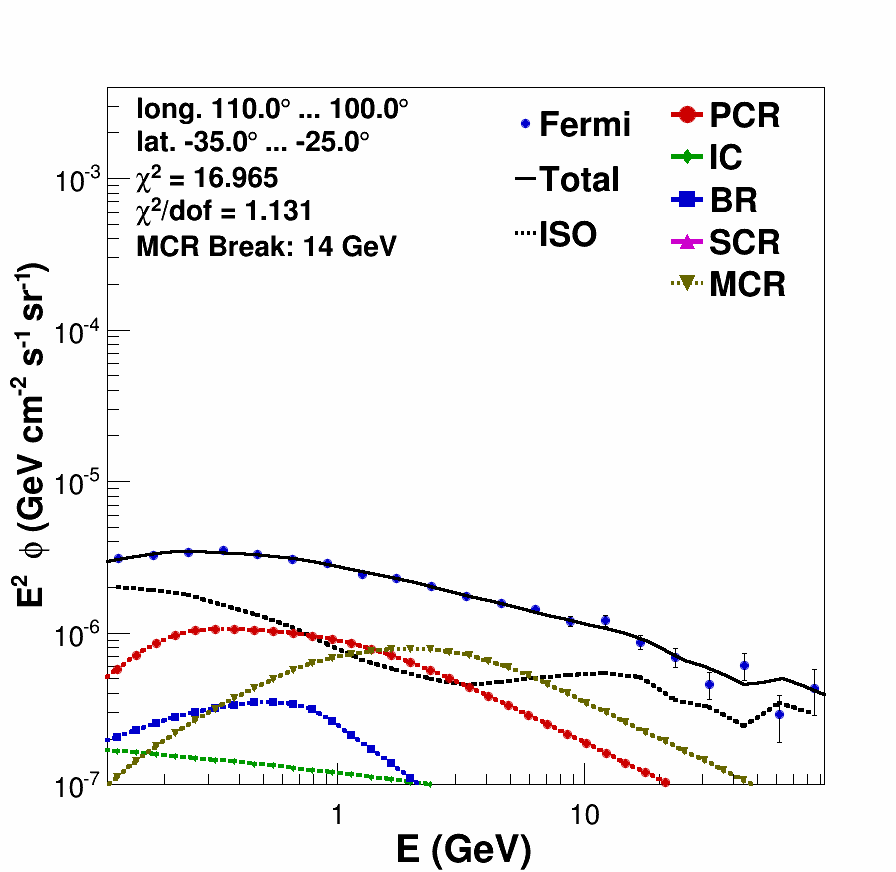}
\includegraphics[width=0.16\textwidth,height=0.16\textwidth,clip]{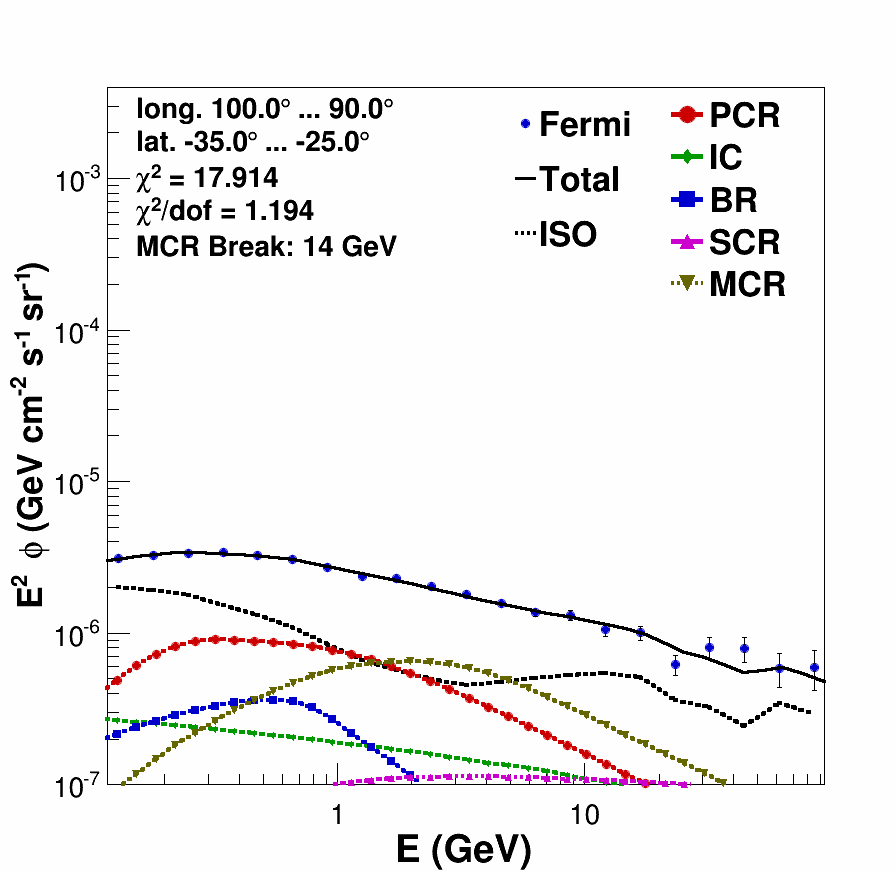}
\includegraphics[width=0.16\textwidth,height=0.16\textwidth,clip]{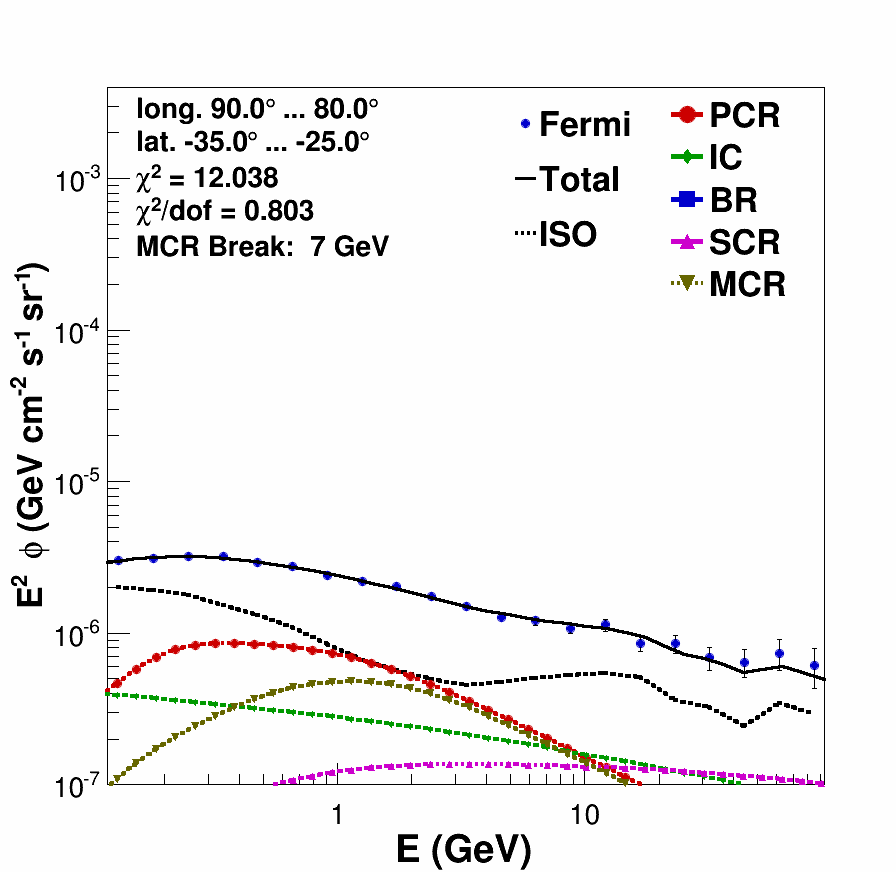}
\includegraphics[width=0.16\textwidth,height=0.16\textwidth,clip]{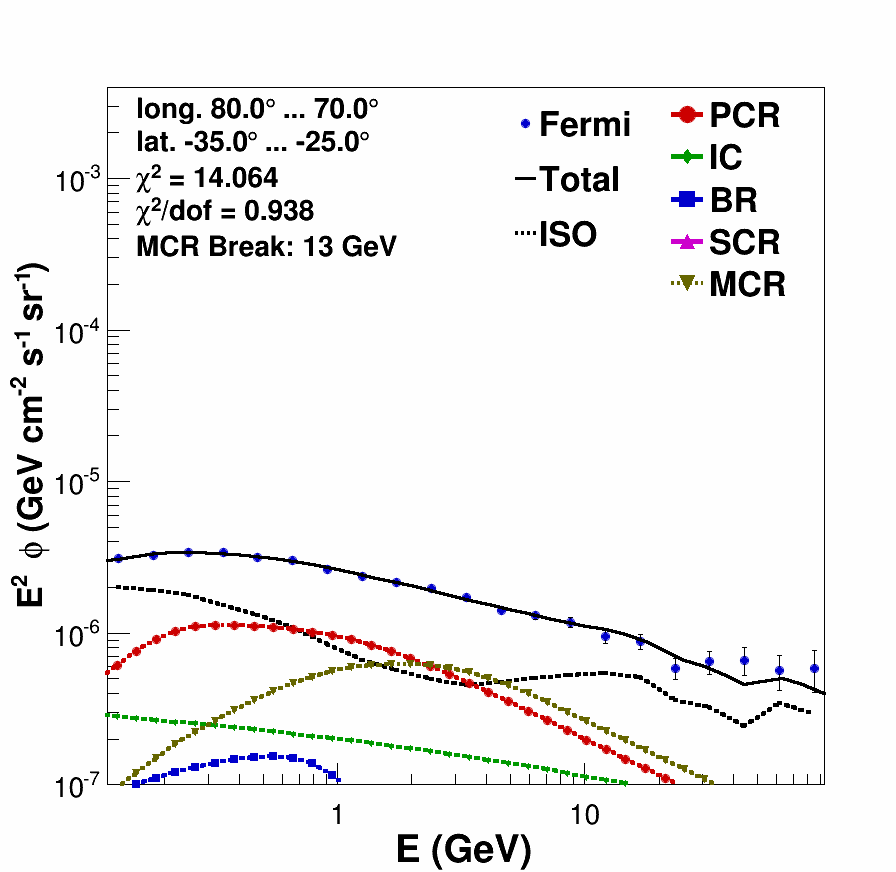}
\includegraphics[width=0.16\textwidth,height=0.16\textwidth,clip]{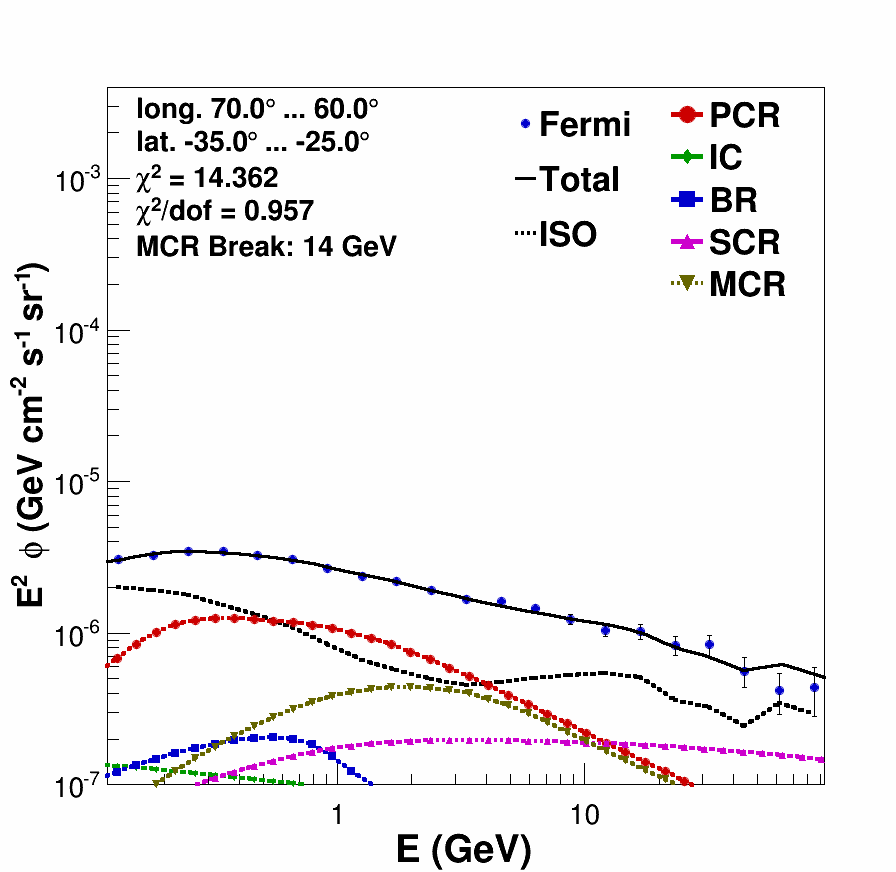}
\includegraphics[width=0.16\textwidth,height=0.16\textwidth,clip]{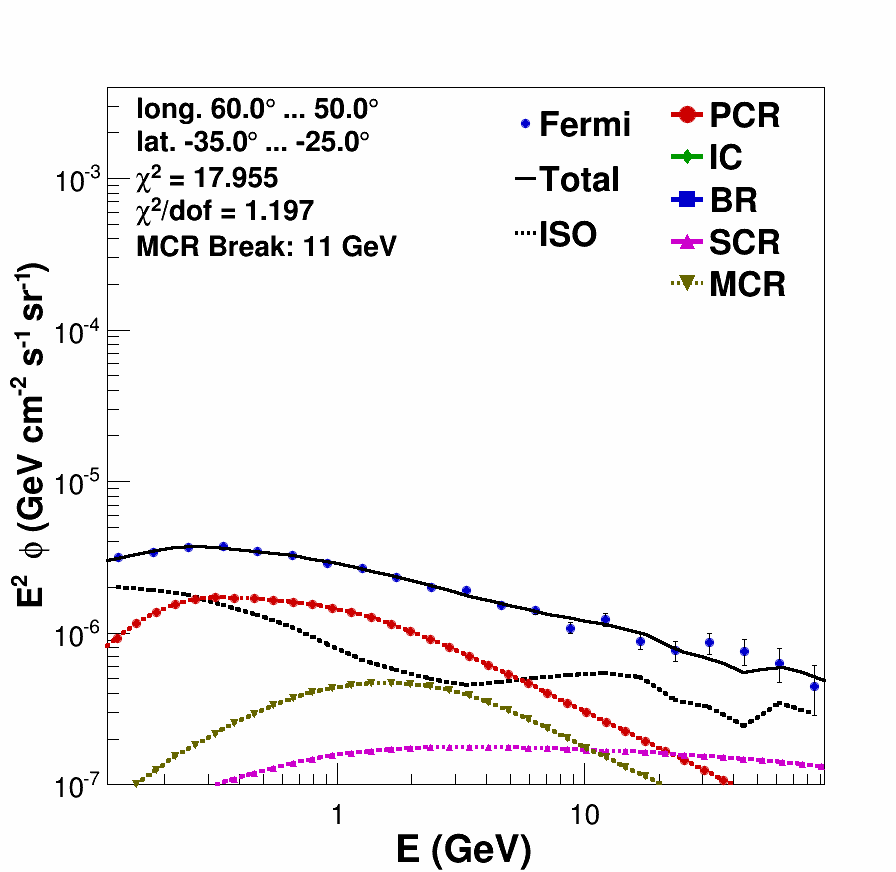}
\includegraphics[width=0.16\textwidth,height=0.16\textwidth,clip]{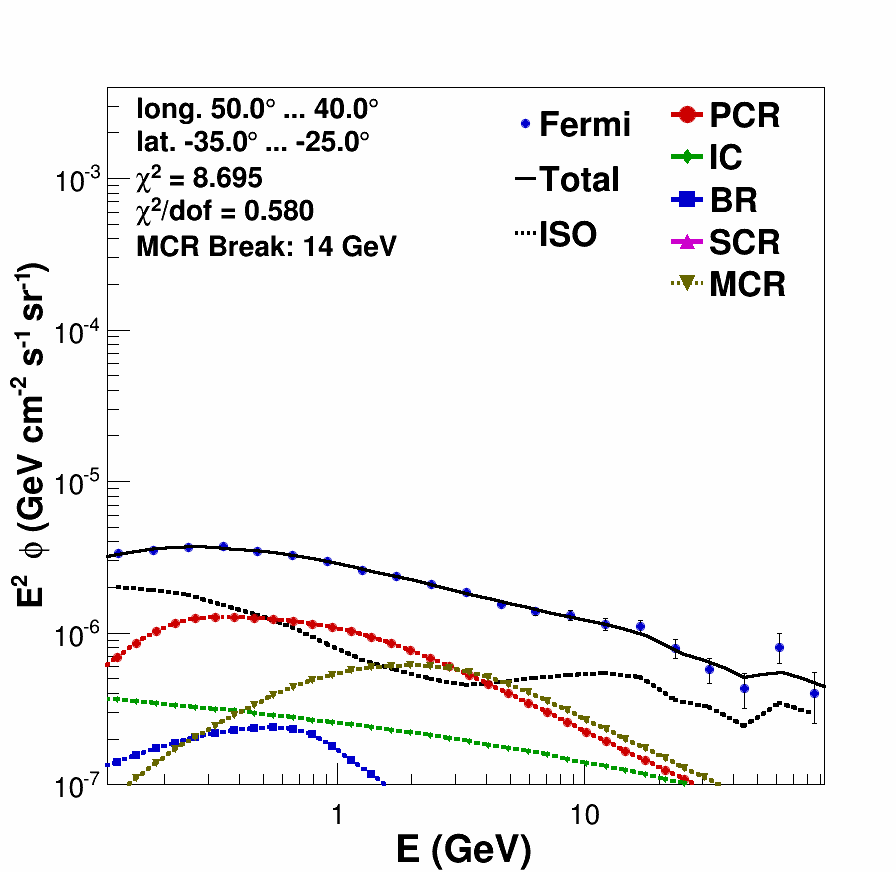}
\includegraphics[width=0.16\textwidth,height=0.16\textwidth,clip]{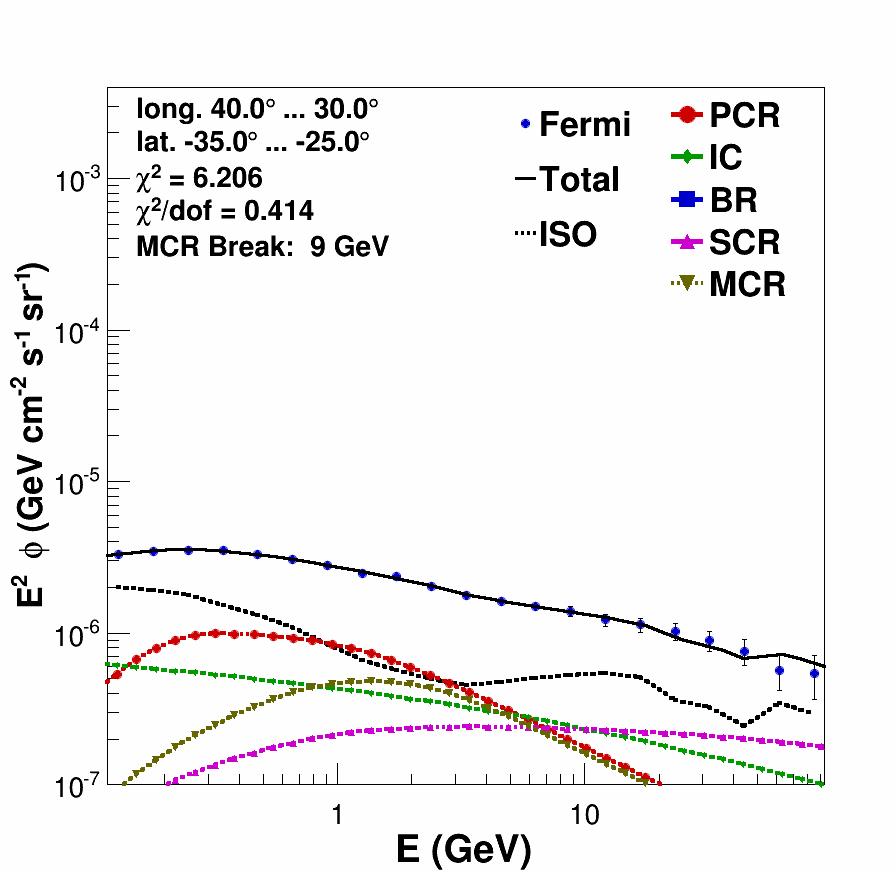}
\includegraphics[width=0.16\textwidth,height=0.16\textwidth,clip]{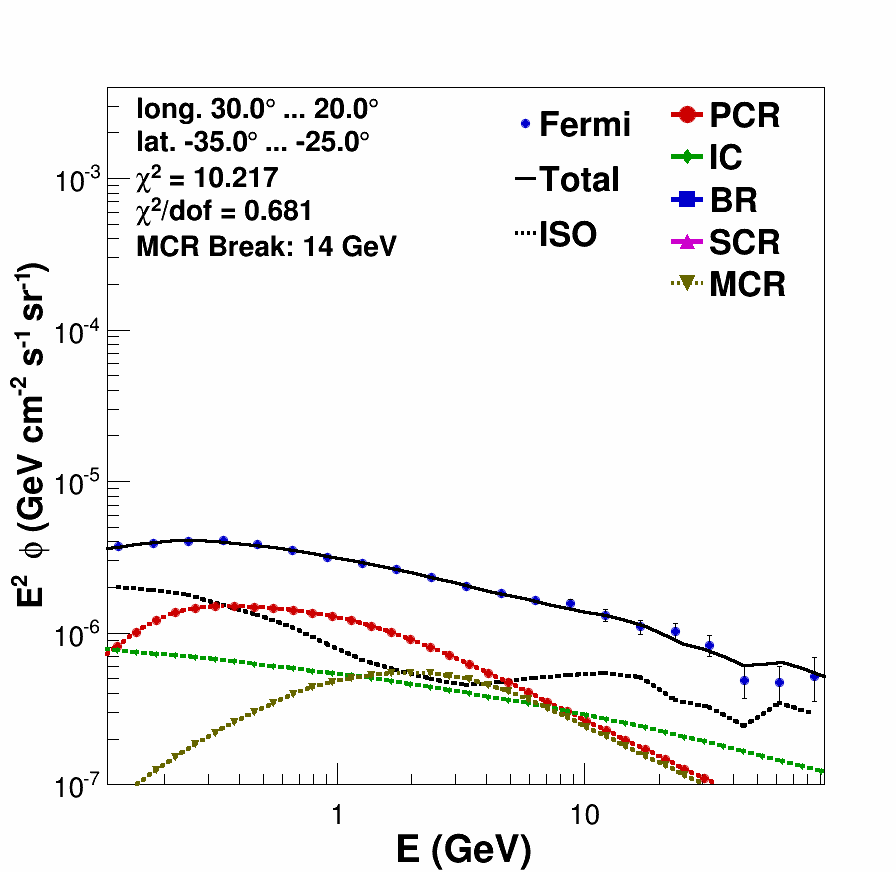}
\includegraphics[width=0.16\textwidth,height=0.16\textwidth,clip]{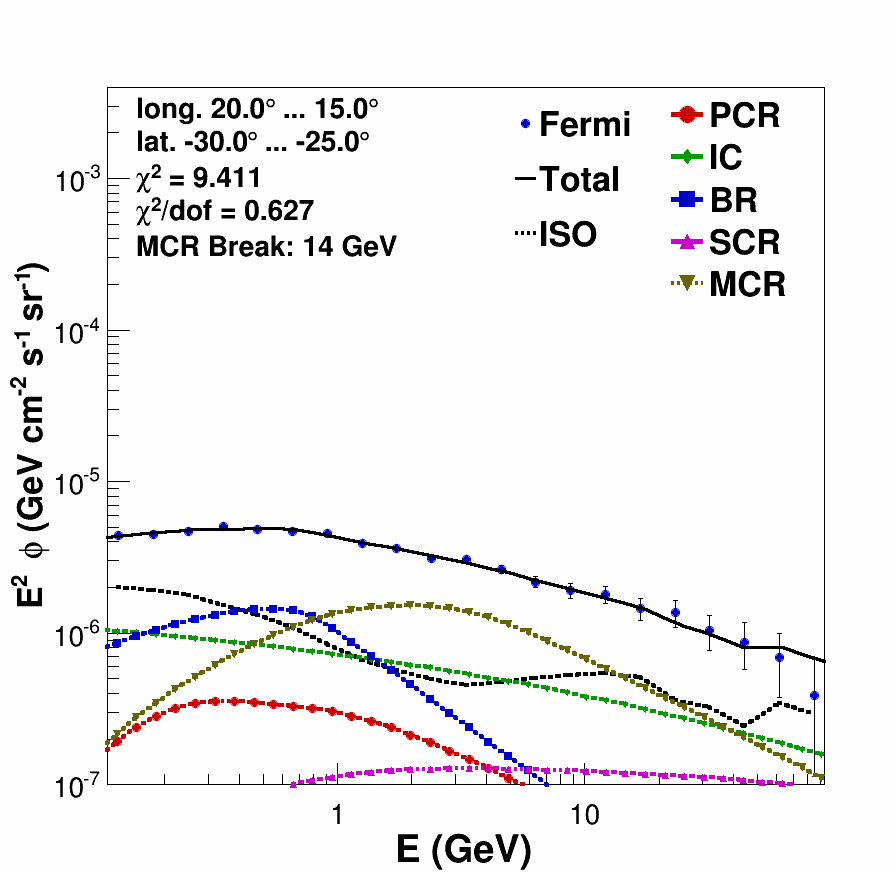}
\includegraphics[width=0.16\textwidth,height=0.16\textwidth,clip]{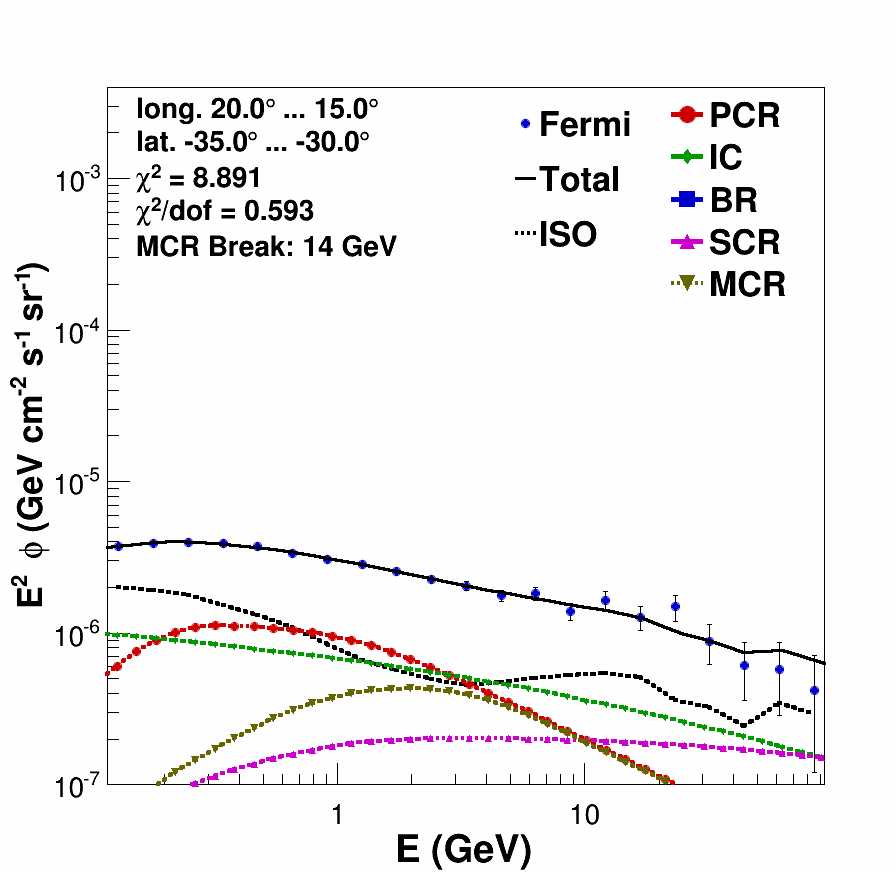}
\includegraphics[width=0.16\textwidth,height=0.16\textwidth,clip]{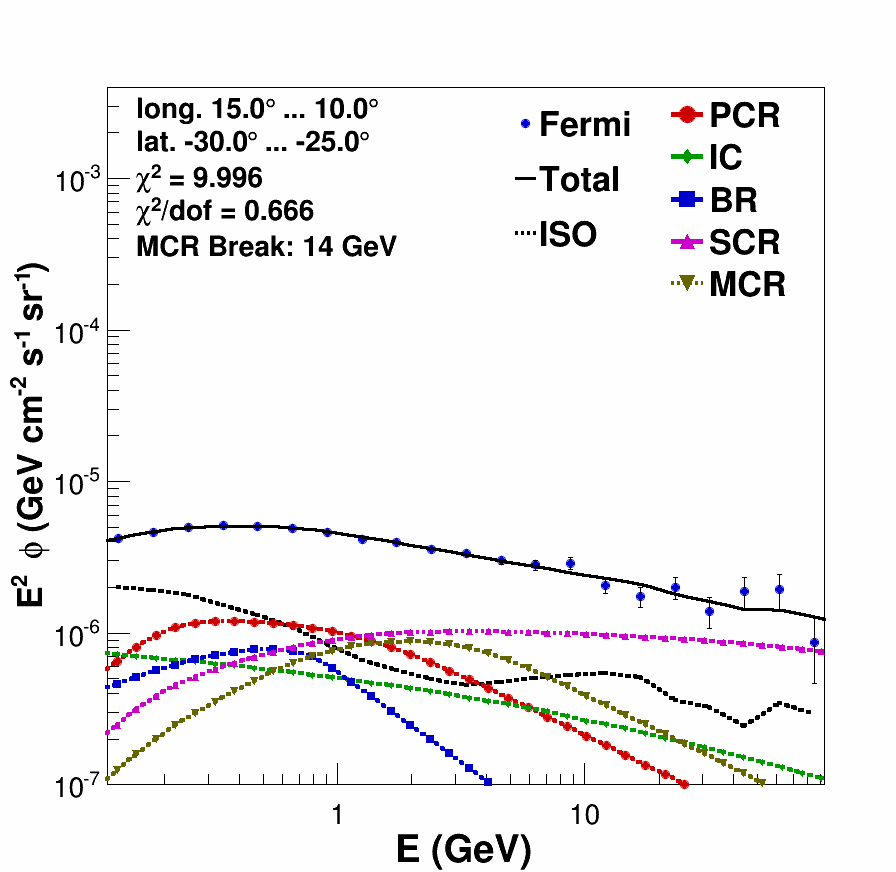}
\includegraphics[width=0.16\textwidth,height=0.16\textwidth,clip]{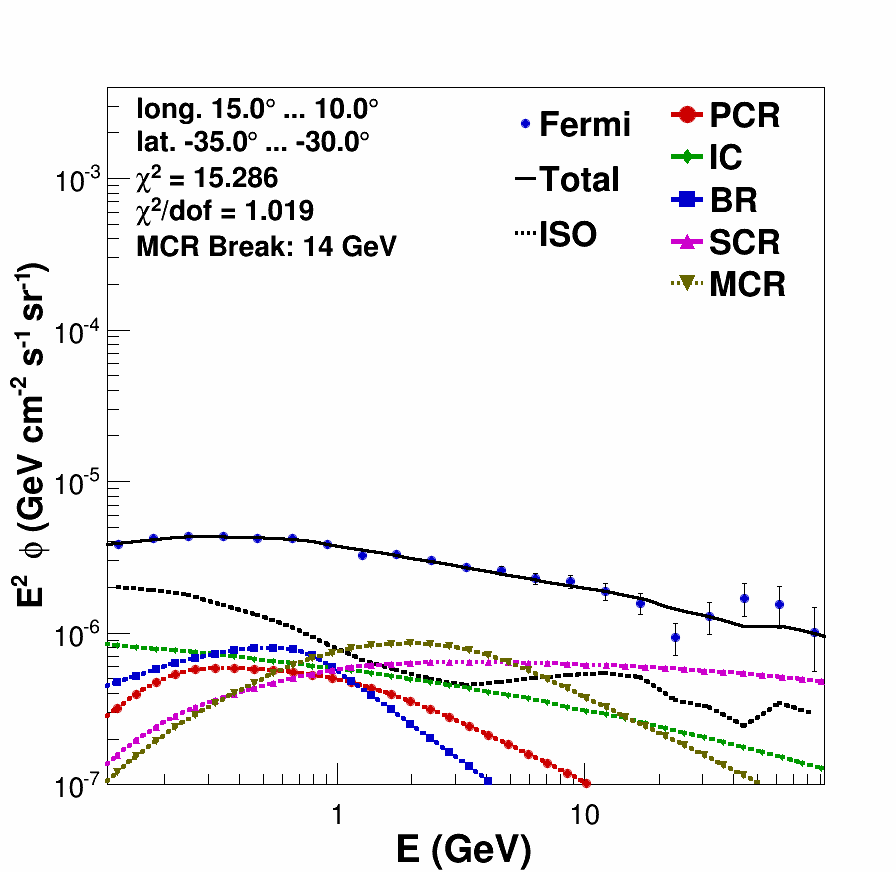}
\includegraphics[width=0.16\textwidth,height=0.16\textwidth,clip]{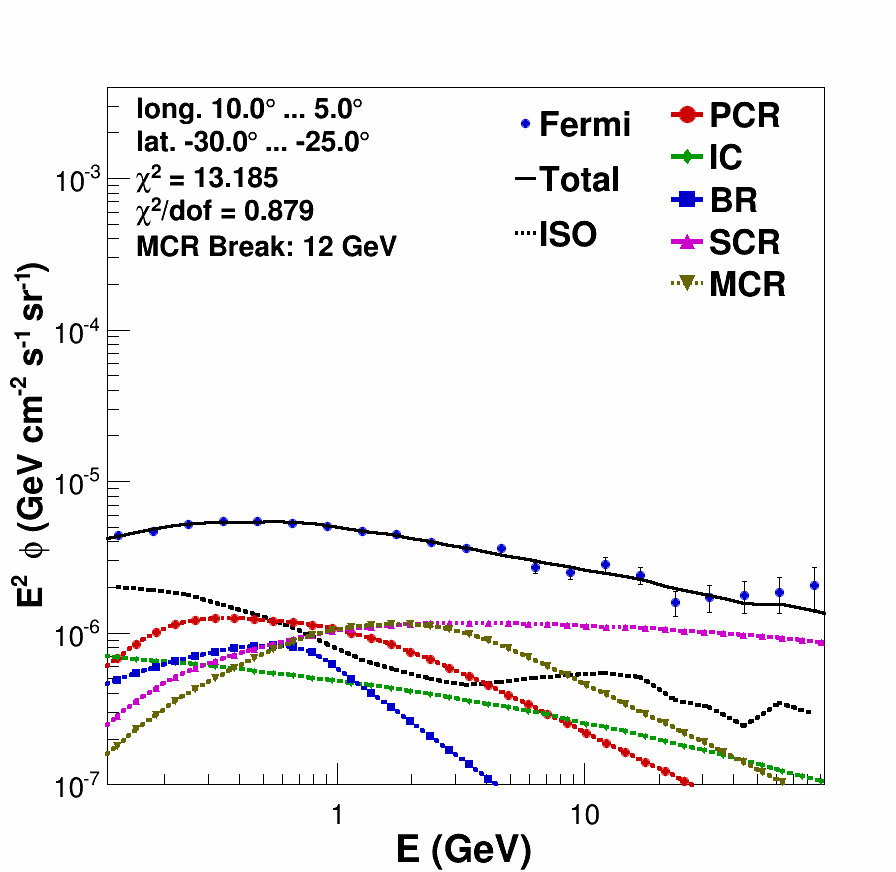}
\includegraphics[width=0.16\textwidth,height=0.16\textwidth,clip]{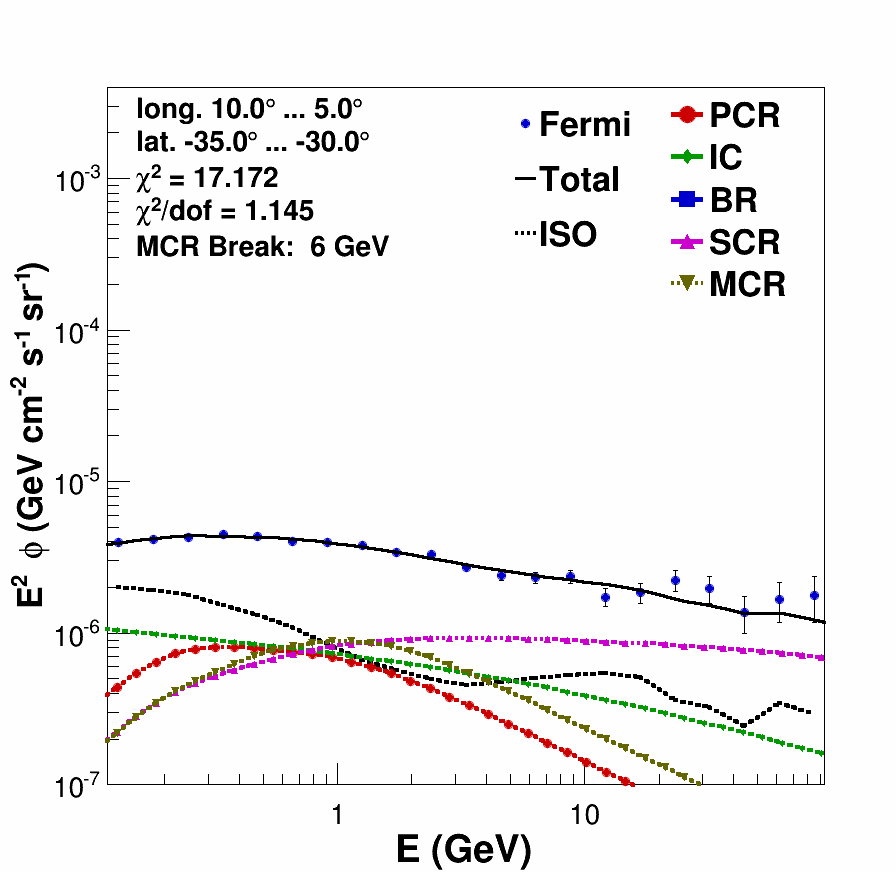}
\includegraphics[width=0.16\textwidth,height=0.16\textwidth,clip]{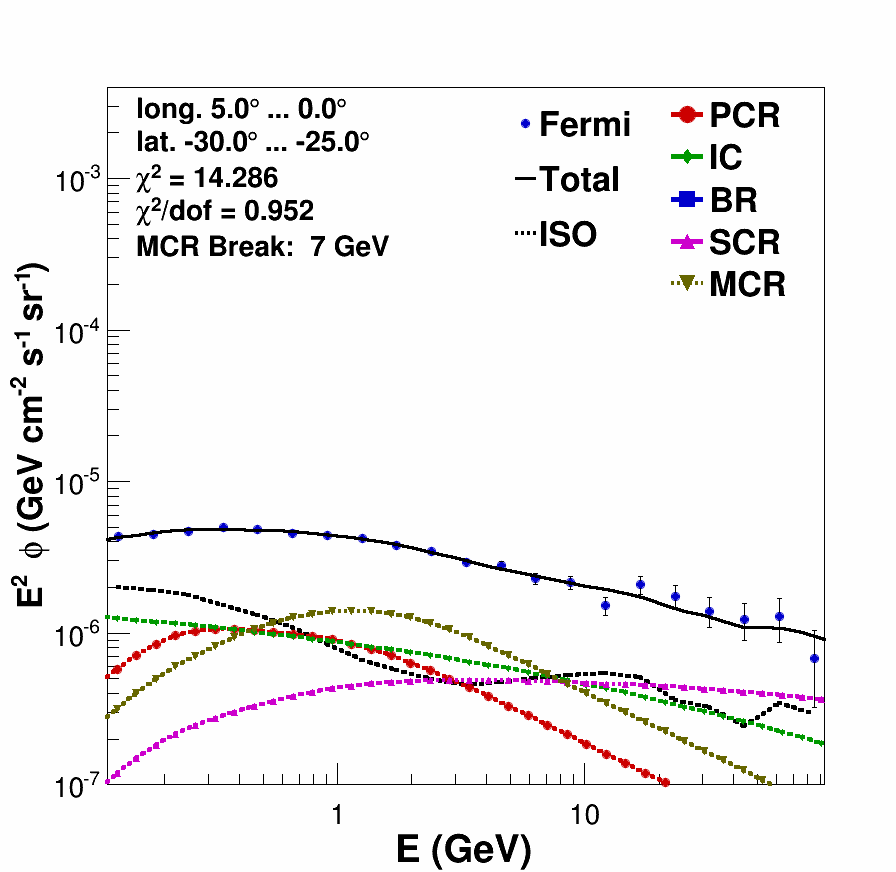}
\includegraphics[width=0.16\textwidth,height=0.16\textwidth,clip]{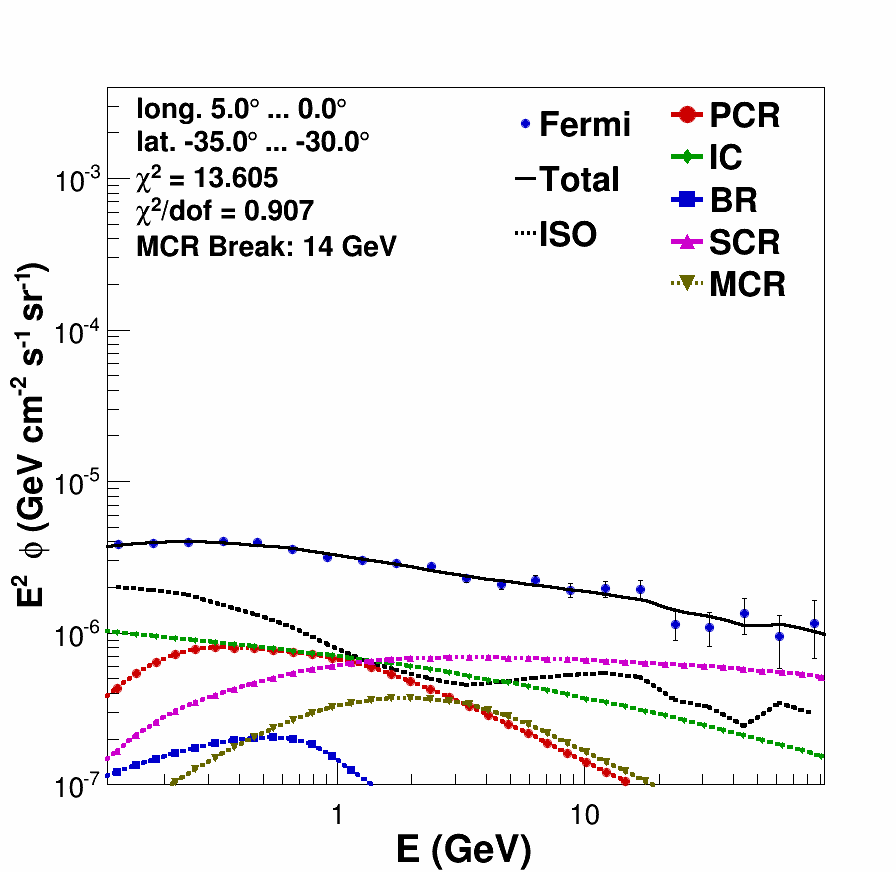}
\includegraphics[width=0.16\textwidth,height=0.16\textwidth,clip]{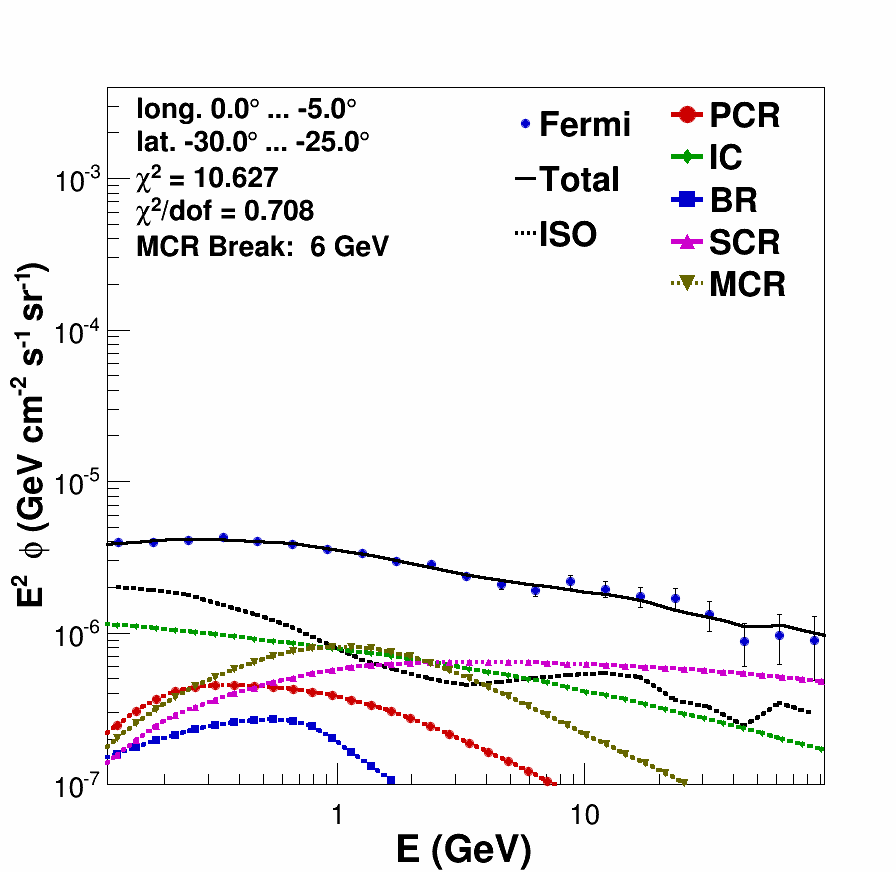}
\includegraphics[width=0.16\textwidth,height=0.16\textwidth,clip]{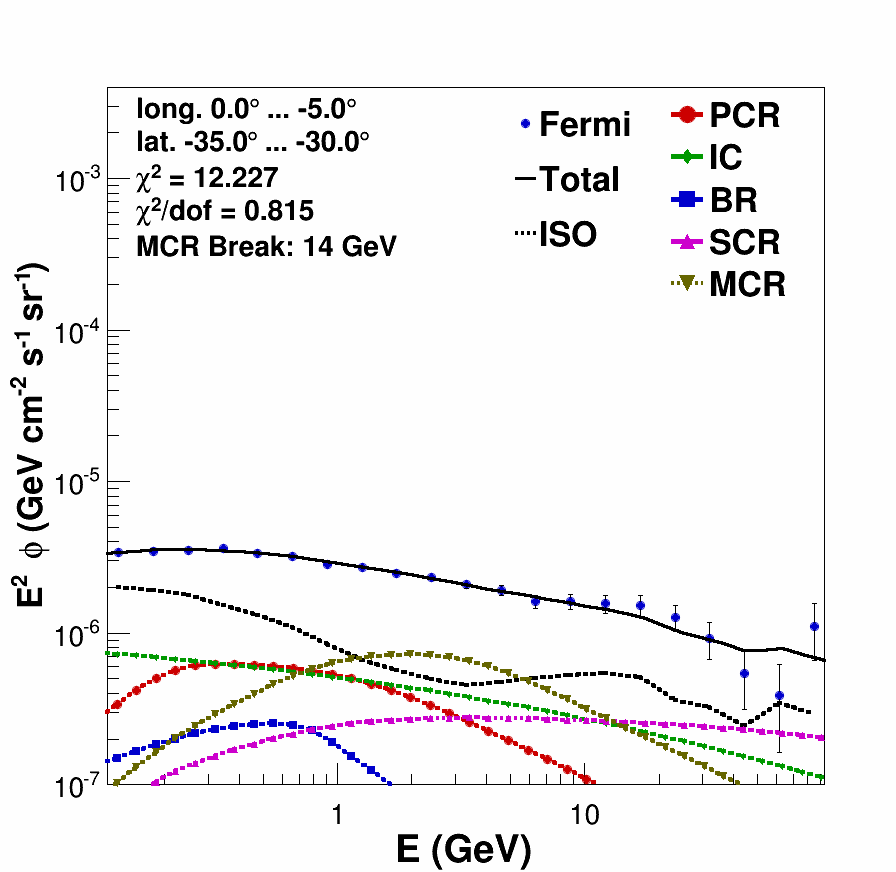}
\includegraphics[width=0.16\textwidth,height=0.16\textwidth,clip]{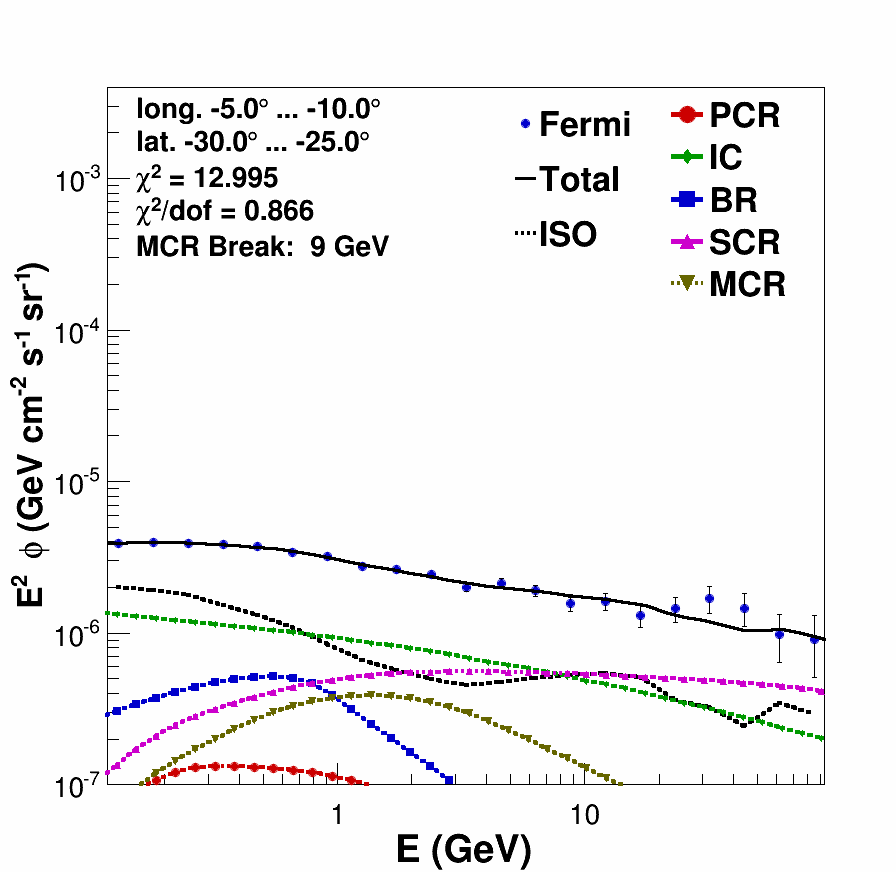}
\includegraphics[width=0.16\textwidth,height=0.16\textwidth,clip]{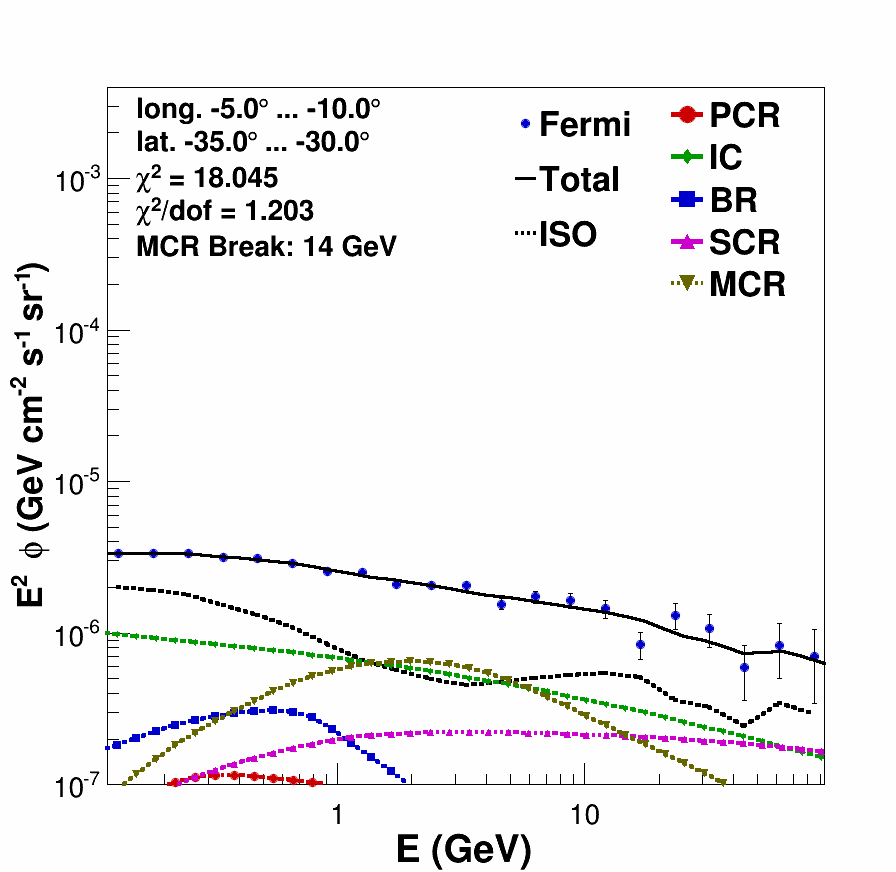}
\includegraphics[width=0.16\textwidth,height=0.16\textwidth,clip]{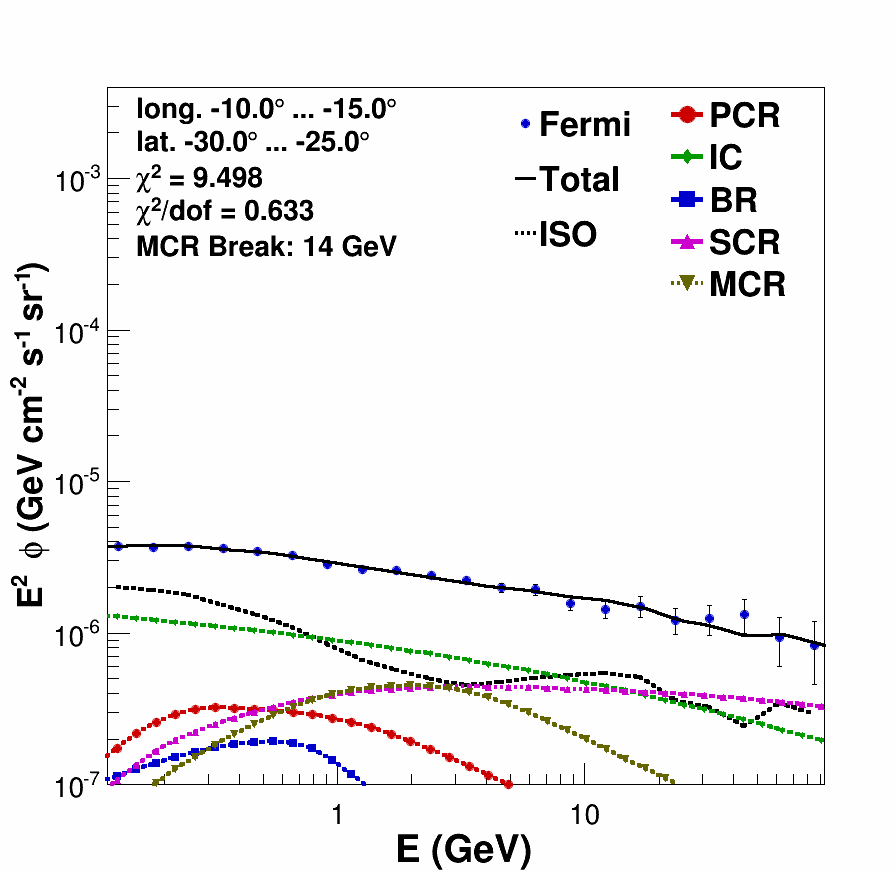}
\includegraphics[width=0.16\textwidth,height=0.16\textwidth,clip]{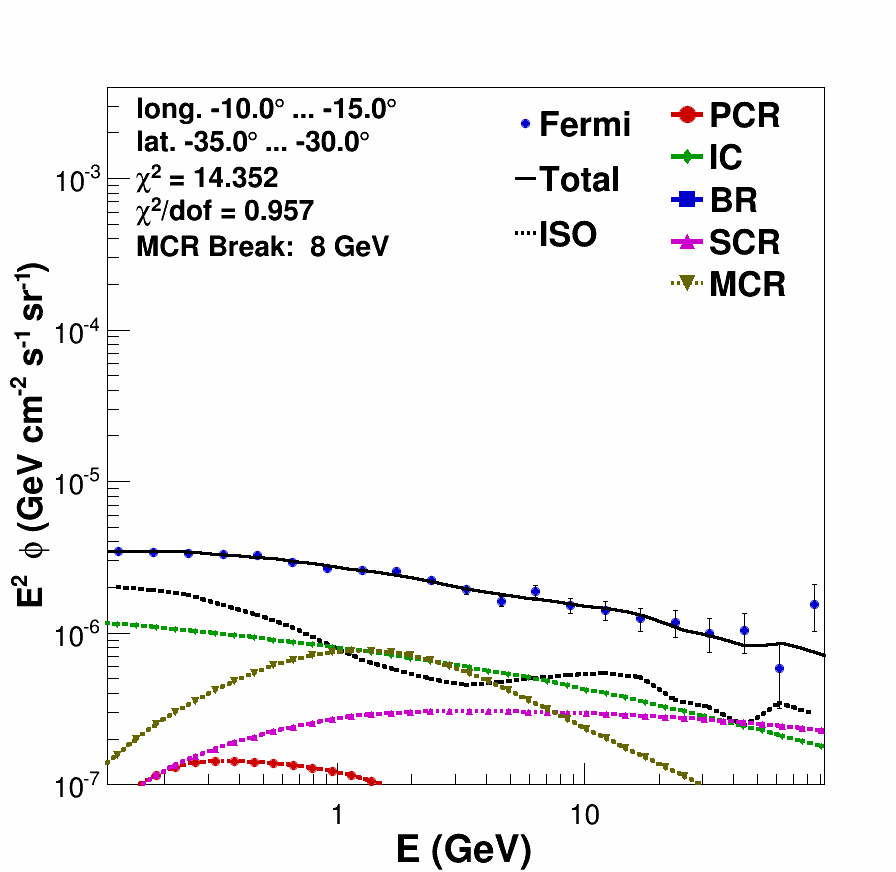}
\includegraphics[width=0.16\textwidth,height=0.16\textwidth,clip]{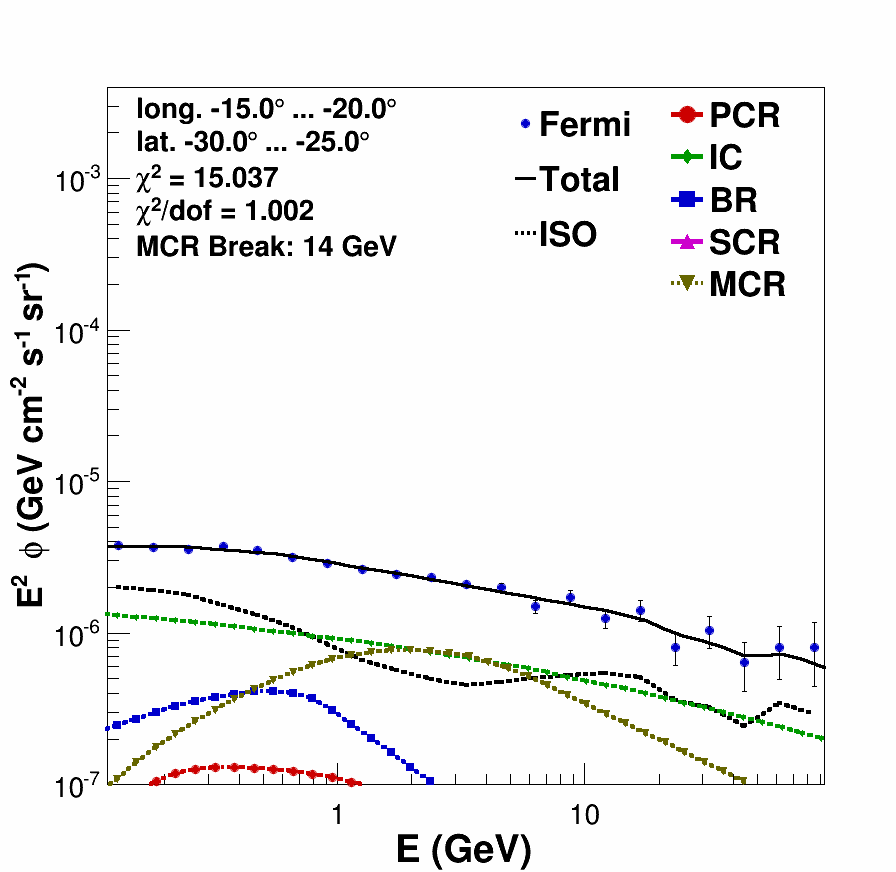}
\includegraphics[width=0.16\textwidth,height=0.16\textwidth,clip]{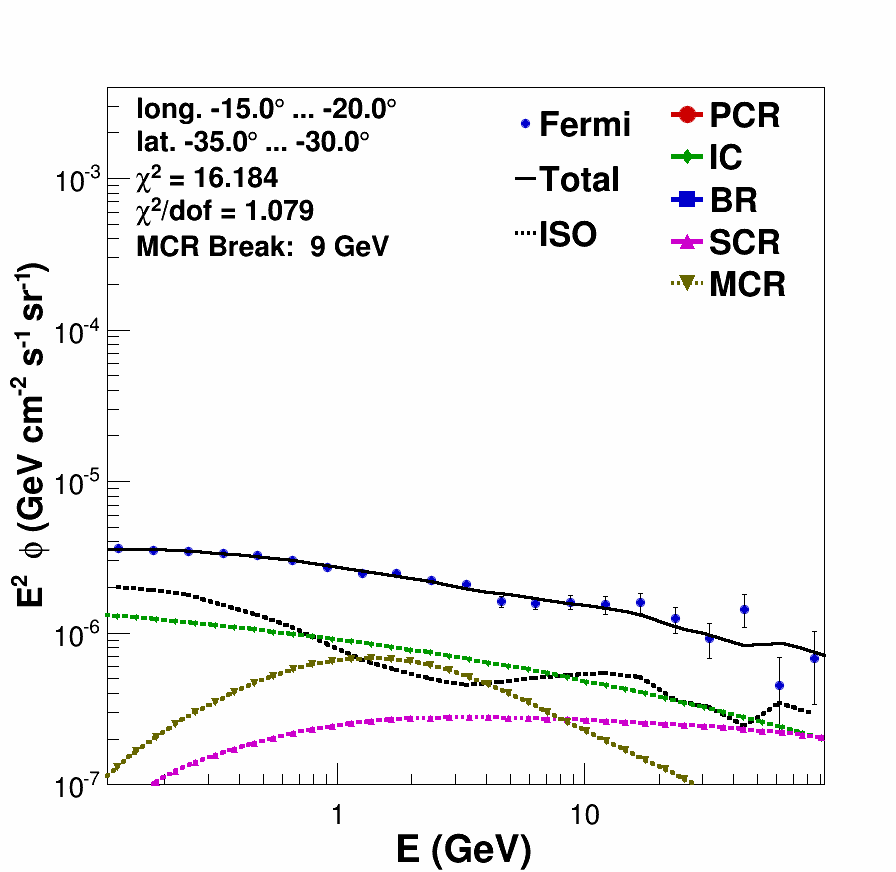}
\includegraphics[width=0.16\textwidth,height=0.16\textwidth,clip]{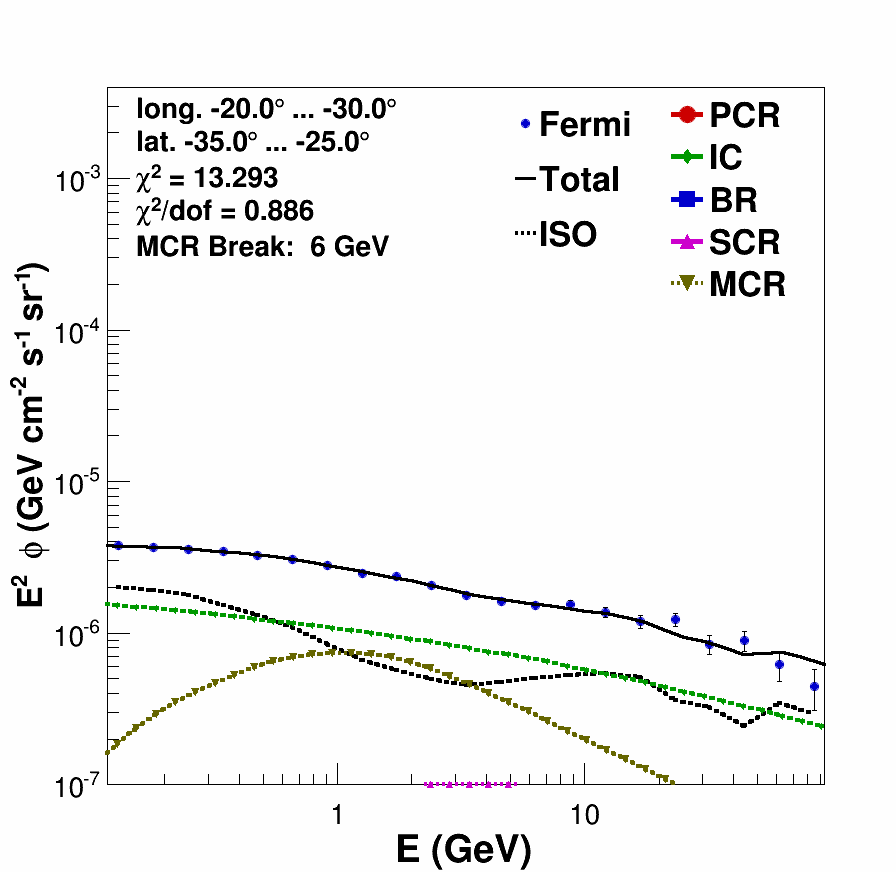}
\includegraphics[width=0.16\textwidth,height=0.16\textwidth,clip]{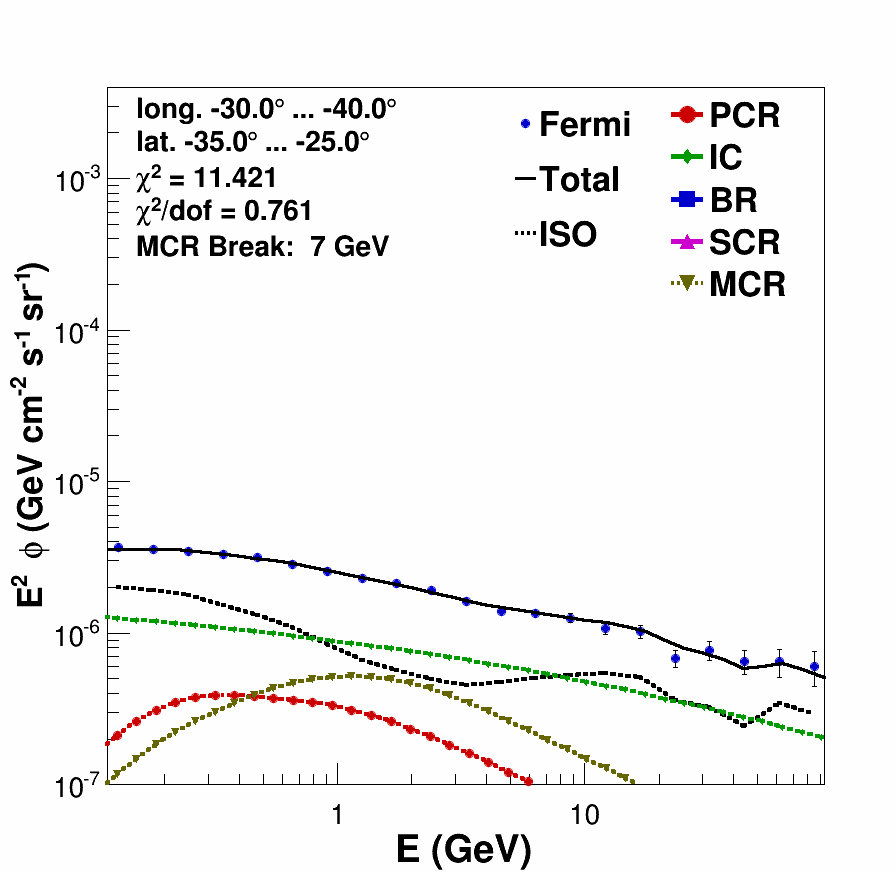}
\includegraphics[width=0.16\textwidth,height=0.16\textwidth,clip]{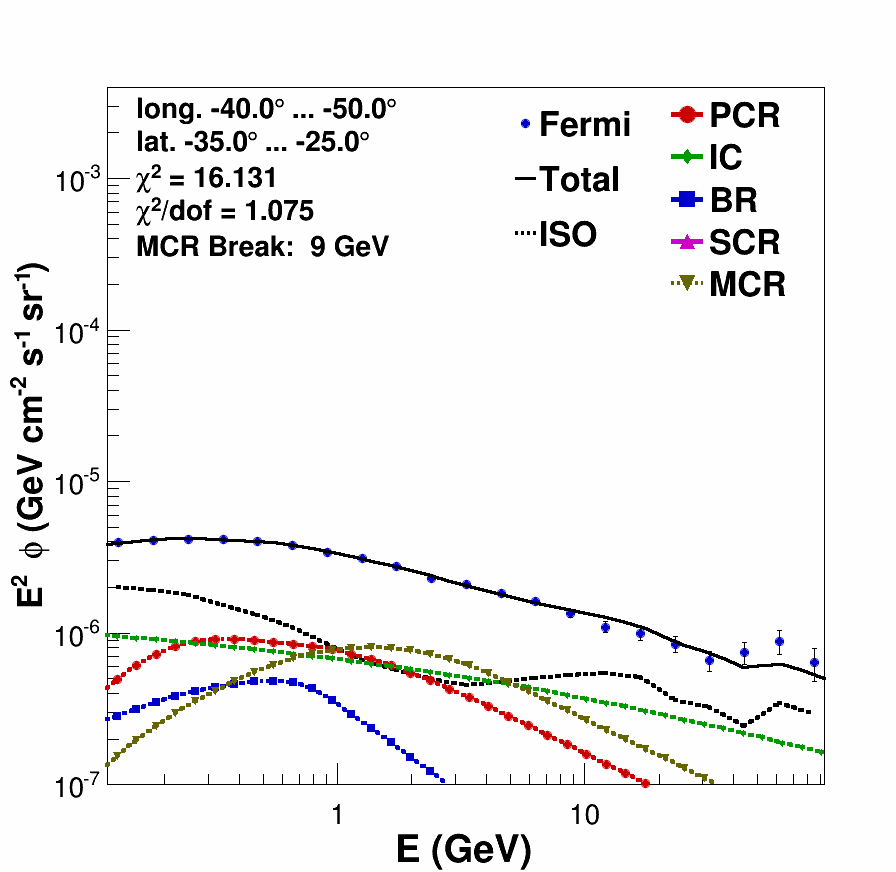}
\includegraphics[width=0.16\textwidth,height=0.16\textwidth,clip]{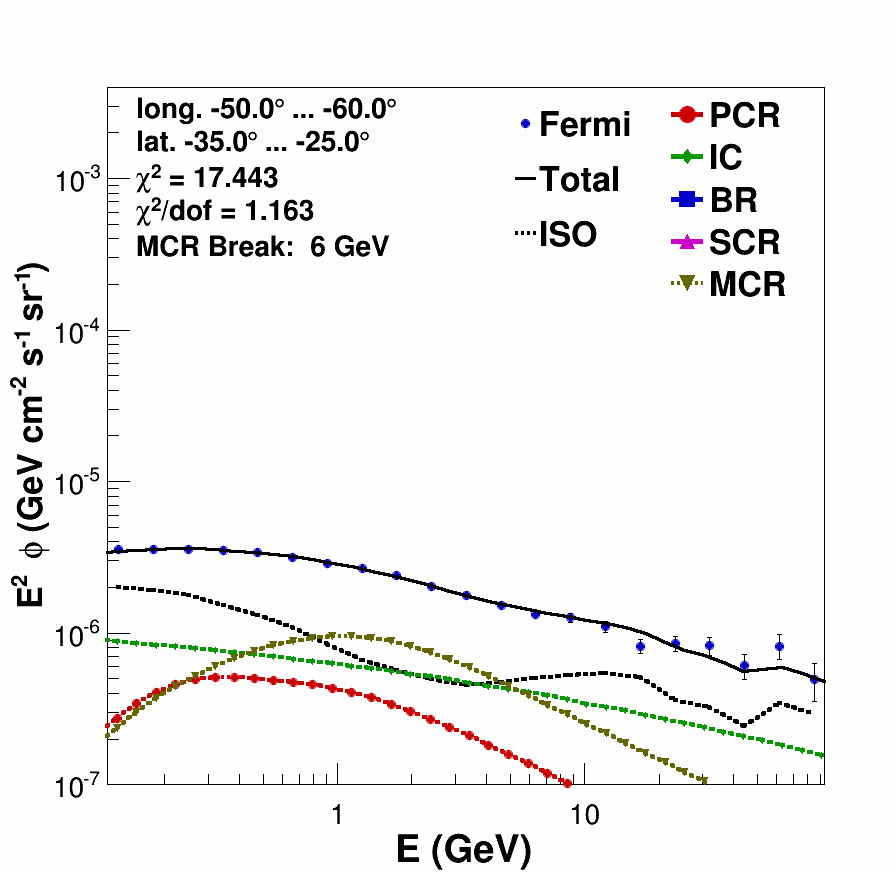}
\includegraphics[width=0.16\textwidth,height=0.16\textwidth,clip]{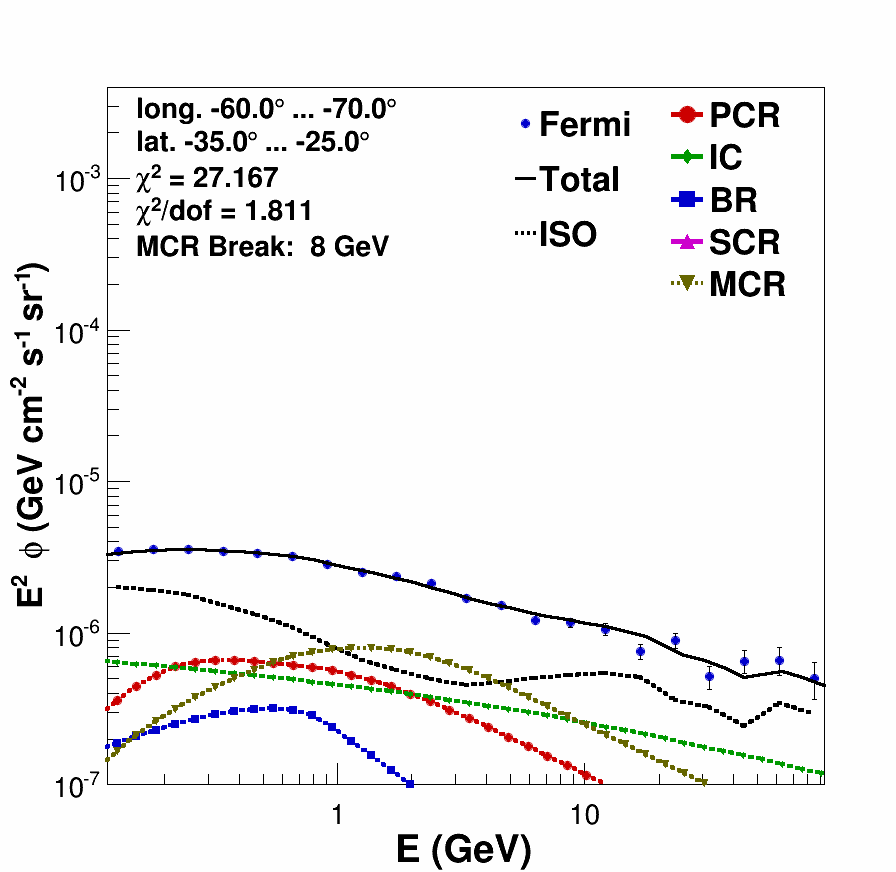}
\includegraphics[width=0.16\textwidth,height=0.16\textwidth,clip]{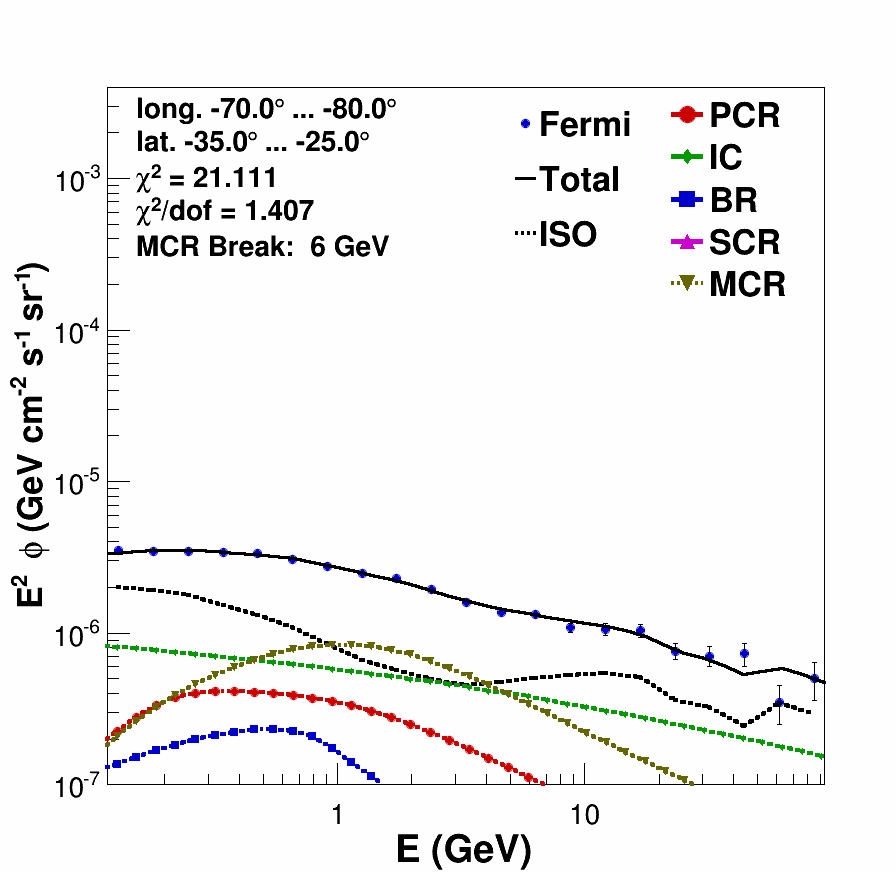}
\includegraphics[width=0.16\textwidth,height=0.16\textwidth,clip]{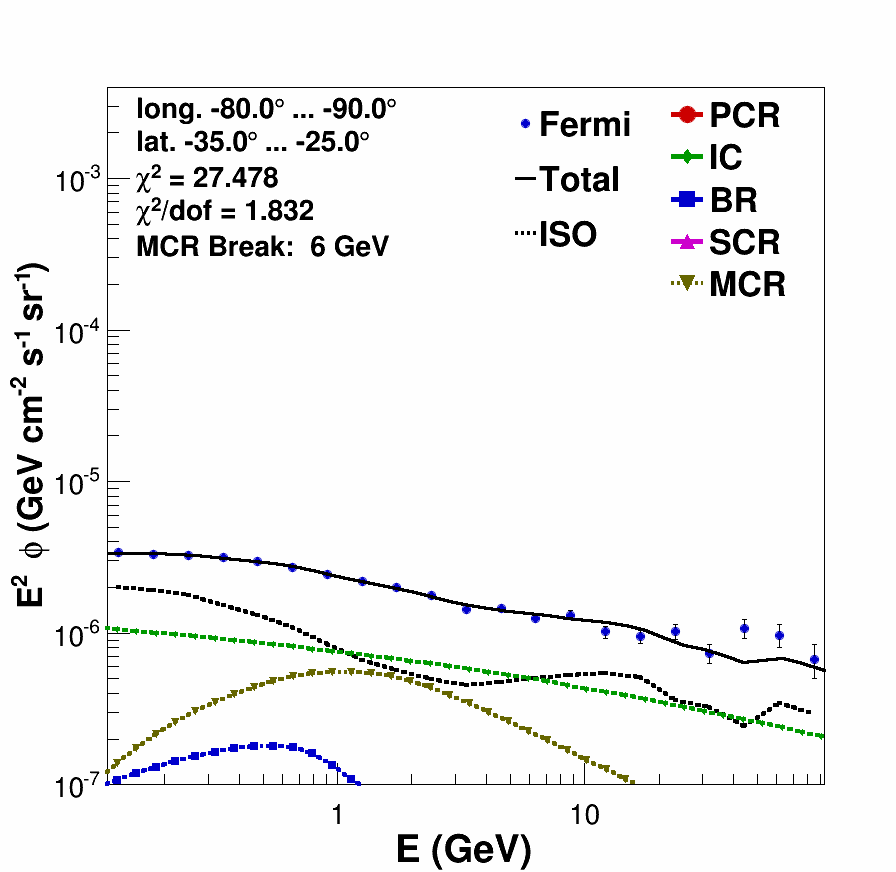}
\includegraphics[width=0.16\textwidth,height=0.16\textwidth,clip]{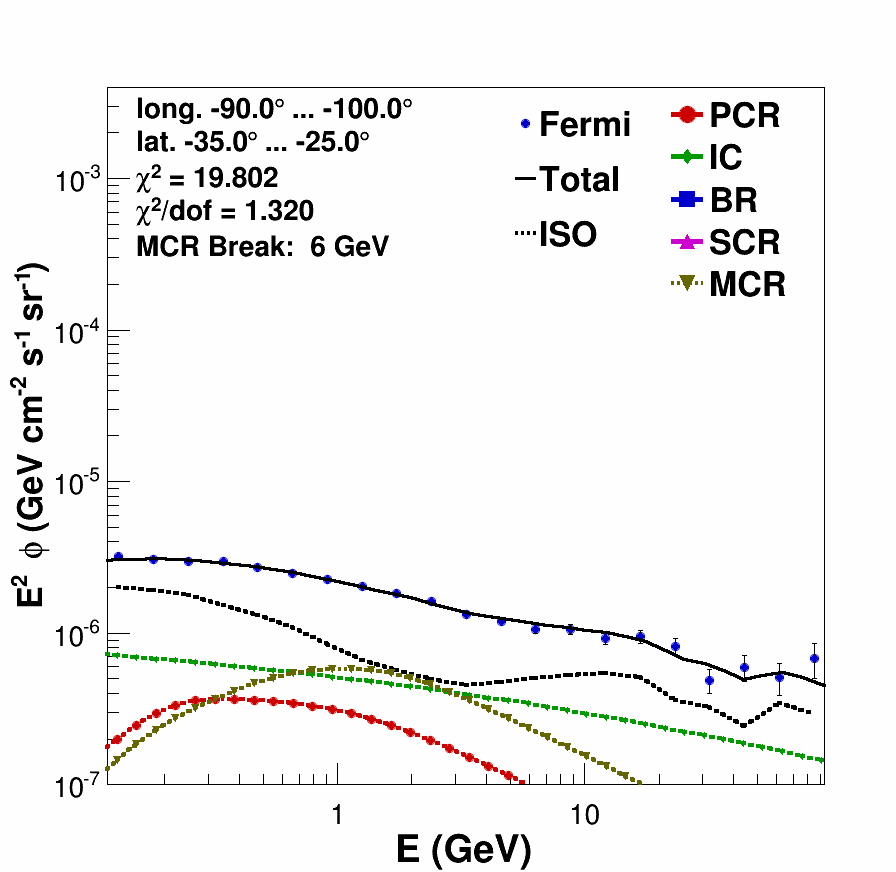}
\includegraphics[width=0.16\textwidth,height=0.16\textwidth,clip]{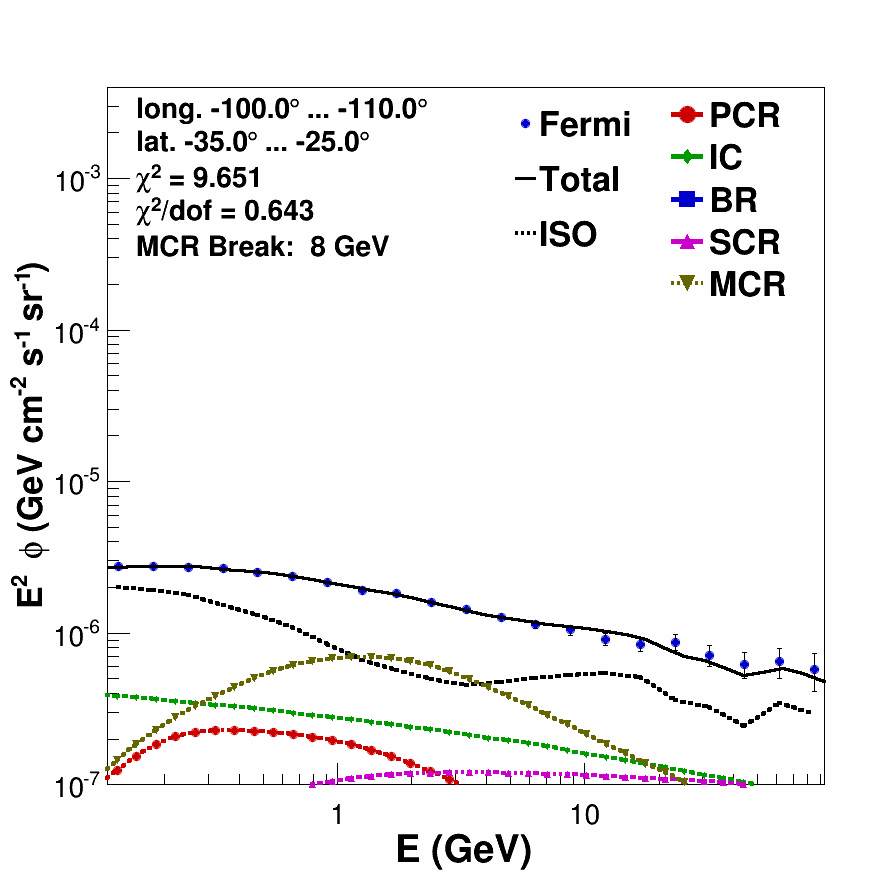}
\includegraphics[width=0.16\textwidth,height=0.16\textwidth,clip]{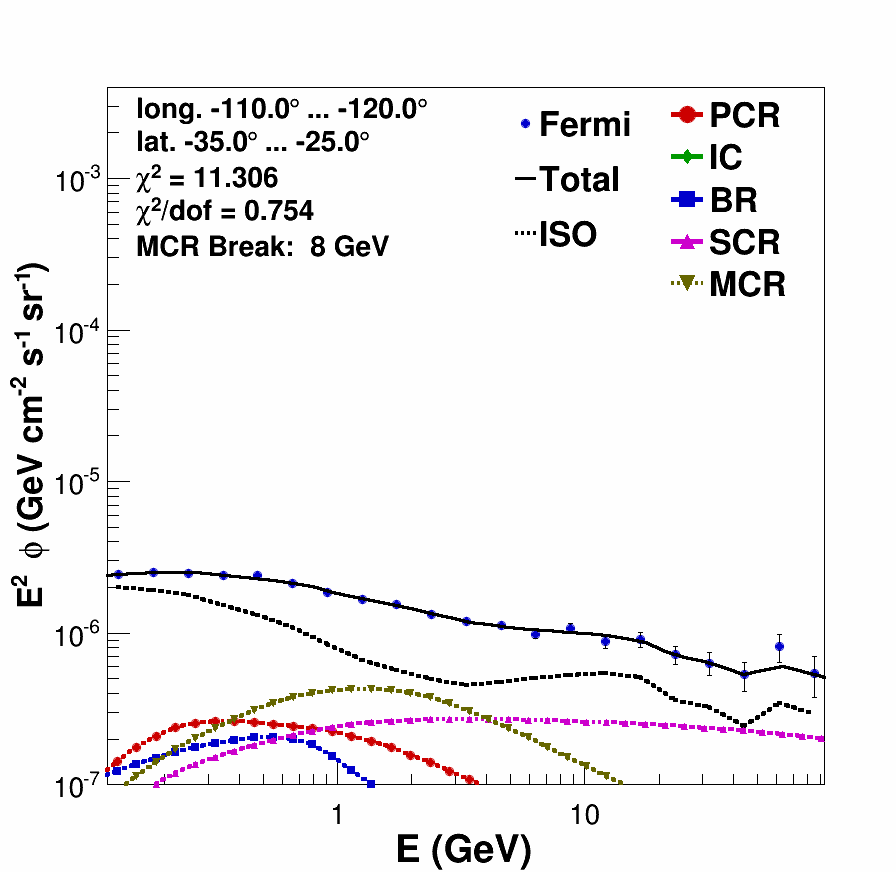}
\includegraphics[width=0.16\textwidth,height=0.16\textwidth,clip]{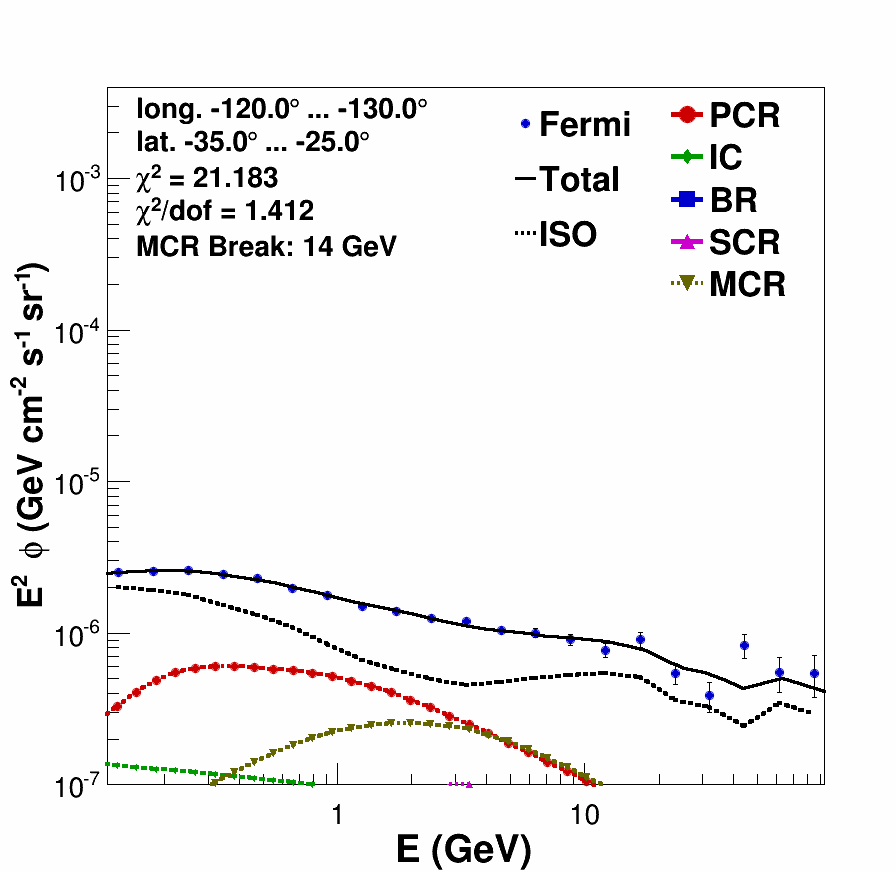}
\includegraphics[width=0.16\textwidth,height=0.16\textwidth,clip]{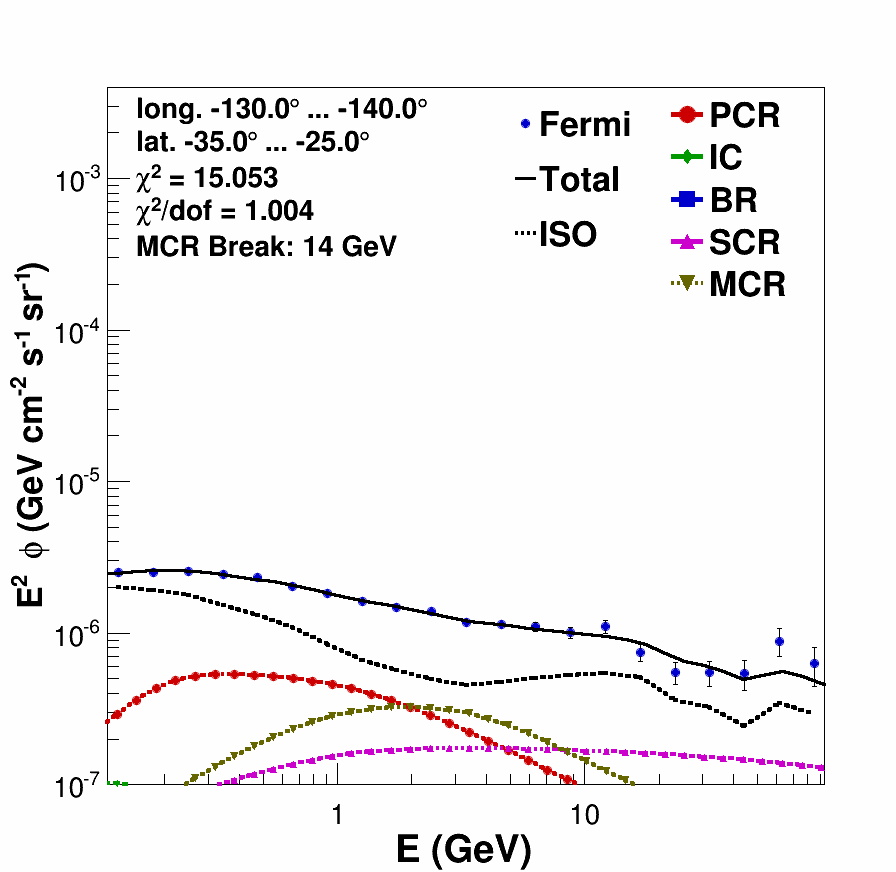}
\includegraphics[width=0.16\textwidth,height=0.16\textwidth,clip]{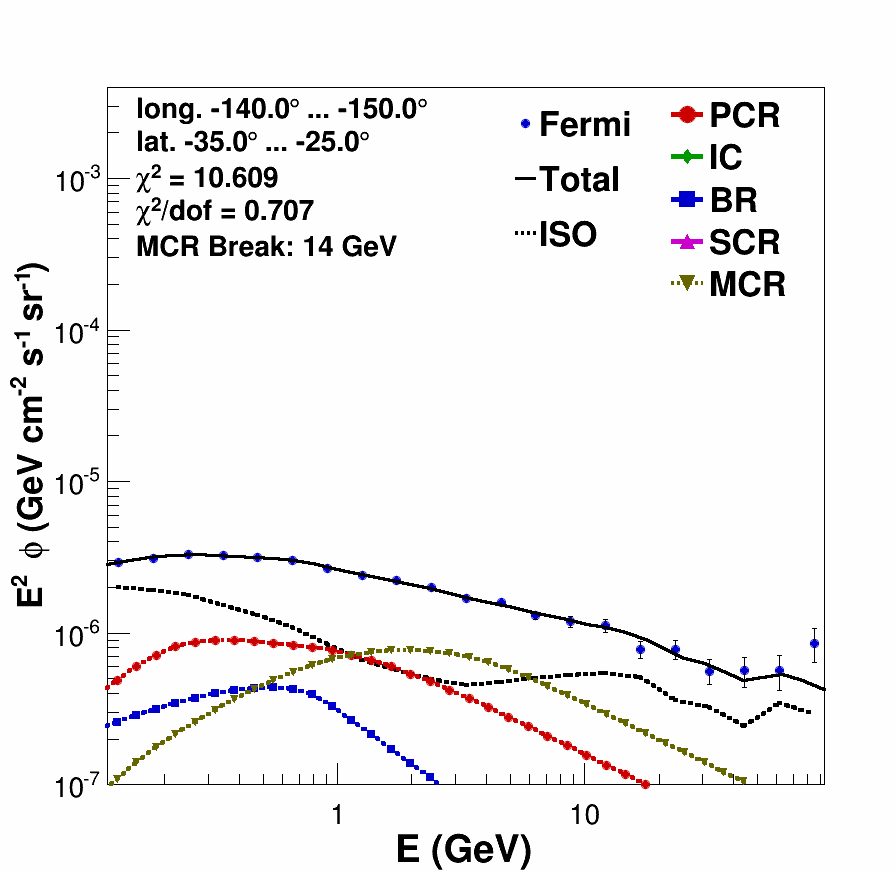}
\includegraphics[width=0.16\textwidth,height=0.16\textwidth,clip]{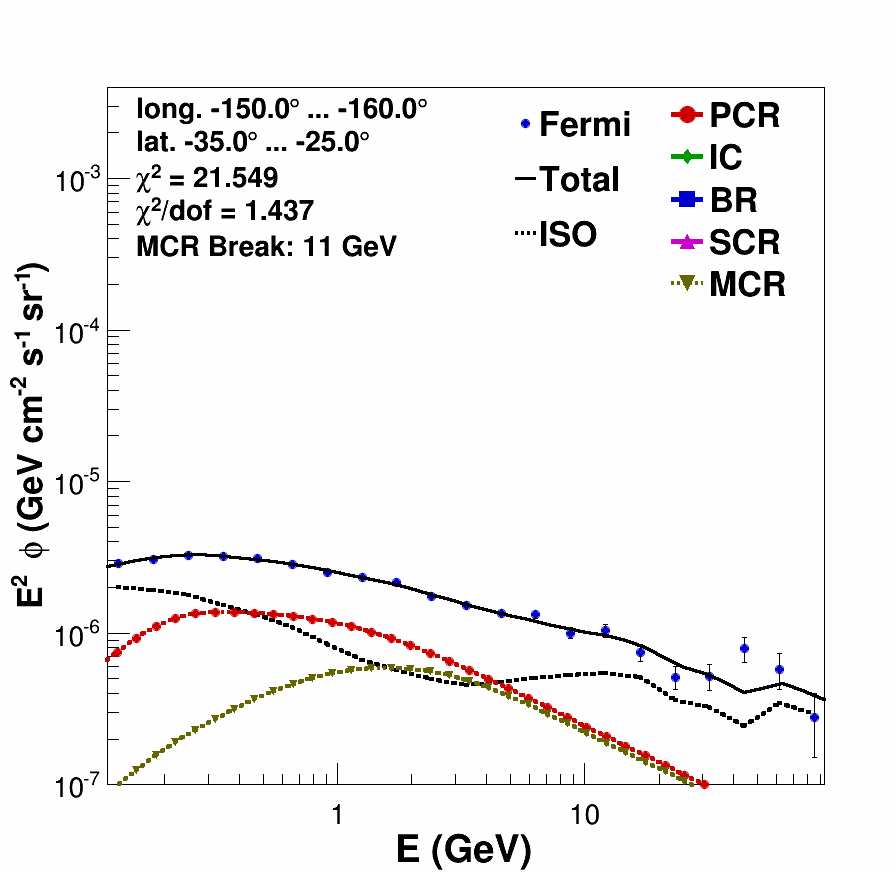}
\includegraphics[width=0.16\textwidth,height=0.16\textwidth,clip]{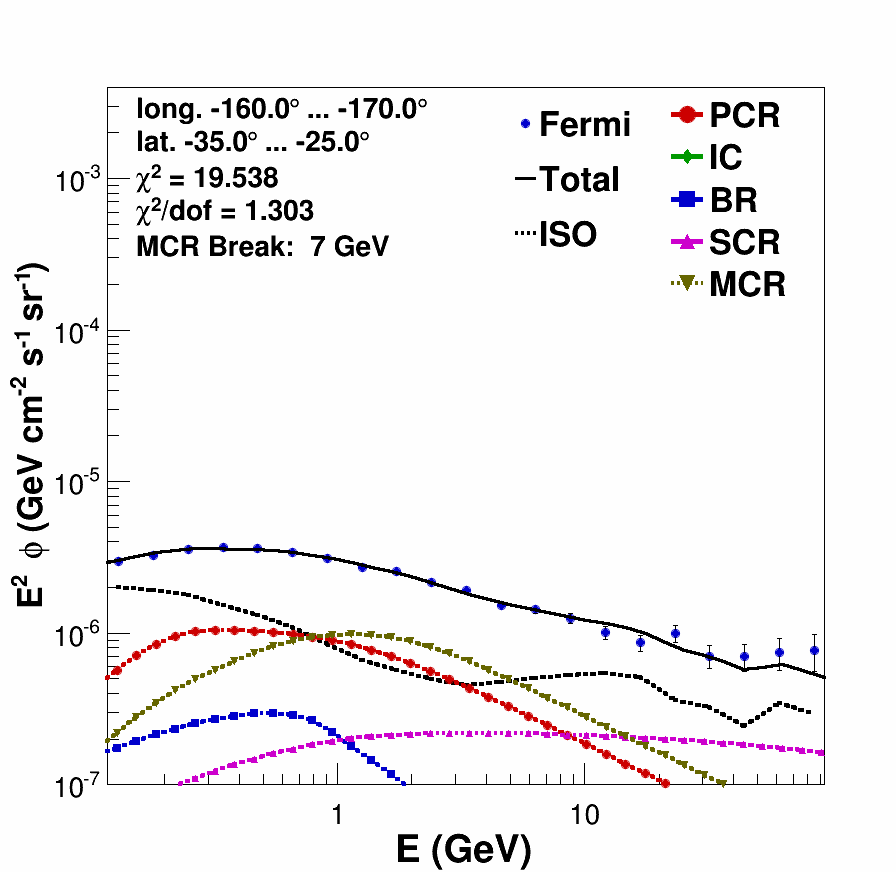}
\includegraphics[width=0.16\textwidth,height=0.16\textwidth,clip]{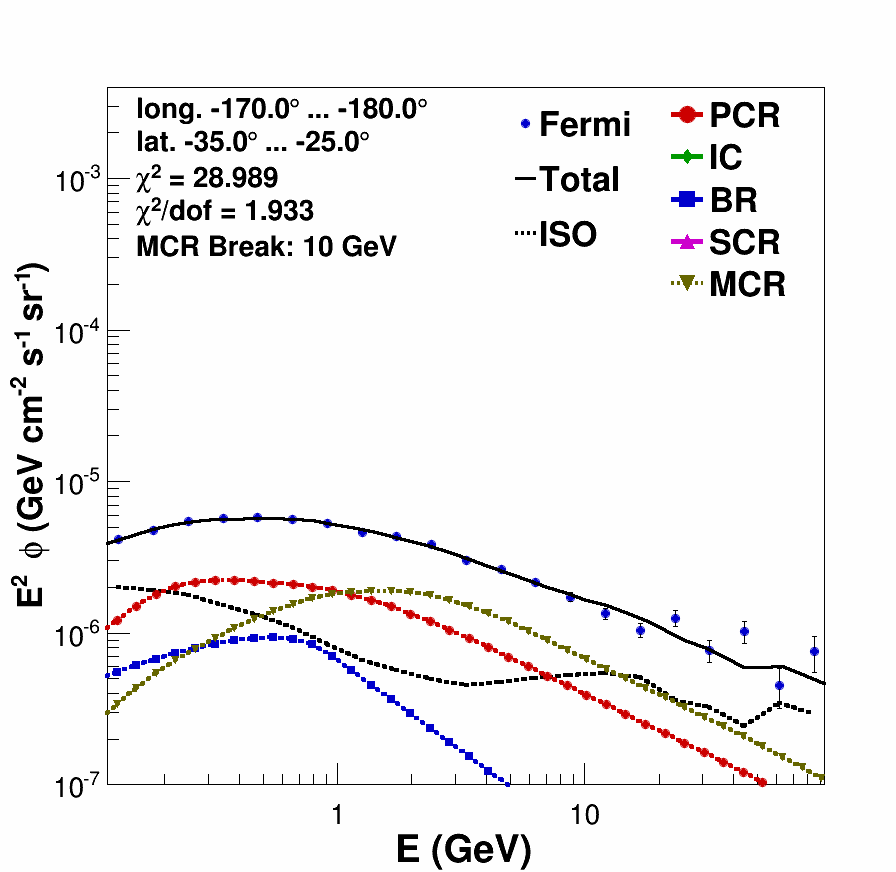}%%%%%r15
\caption[]{Template fits for latitudes  with $-35.0^\circ<b<-25.0^\circ$ and longitudes decreasing from 180$^\circ$ to -180$^\circ$.} \label{F27}
\end{figure}
\begin{figure}
\centering
\includegraphics[width=0.16\textwidth,height=0.16\textwidth,clip]{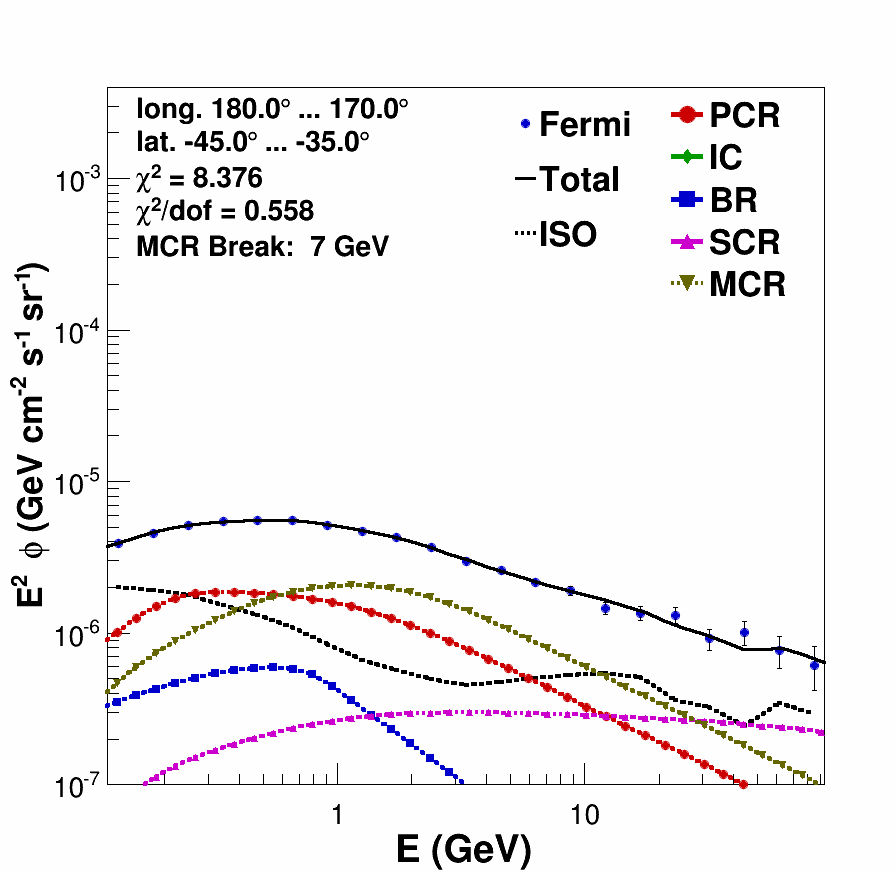}
\includegraphics[width=0.16\textwidth,height=0.16\textwidth,clip]{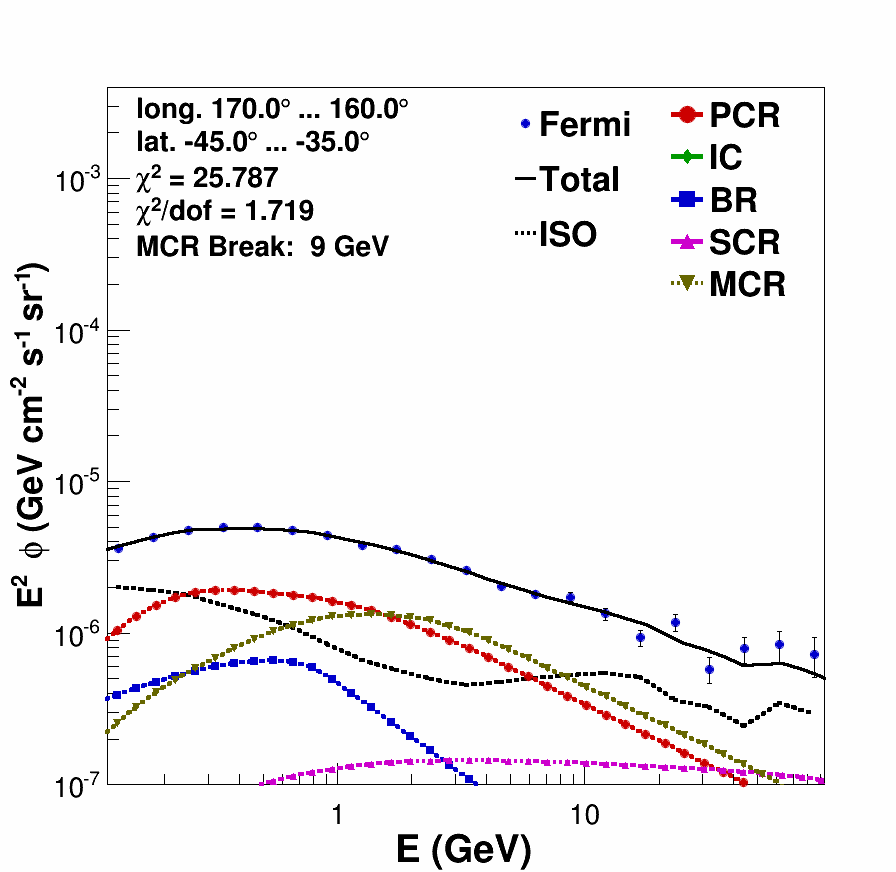}
\includegraphics[width=0.16\textwidth,height=0.16\textwidth,clip]{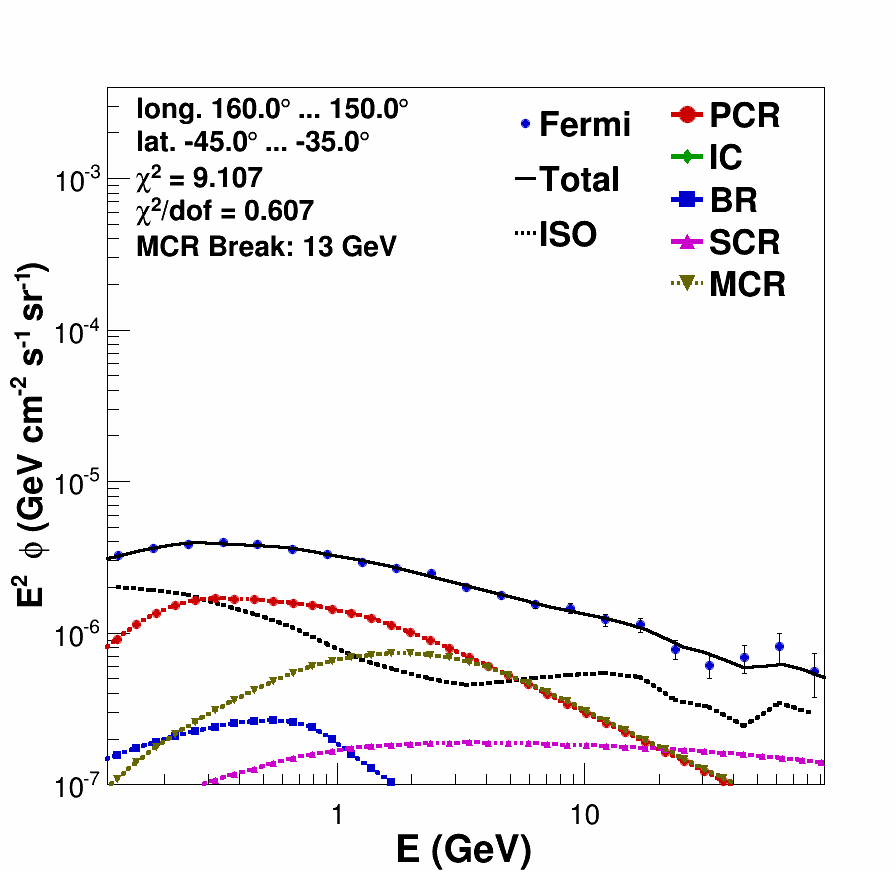}
\includegraphics[width=0.16\textwidth,height=0.16\textwidth,clip]{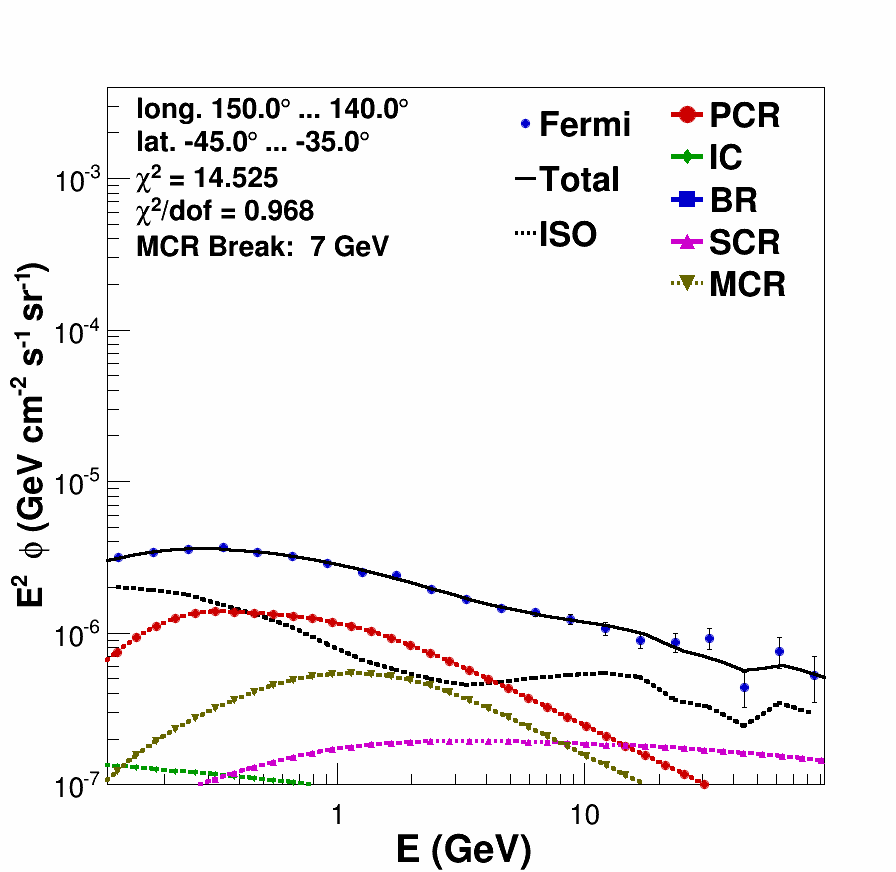}
\includegraphics[width=0.16\textwidth,height=0.16\textwidth,clip]{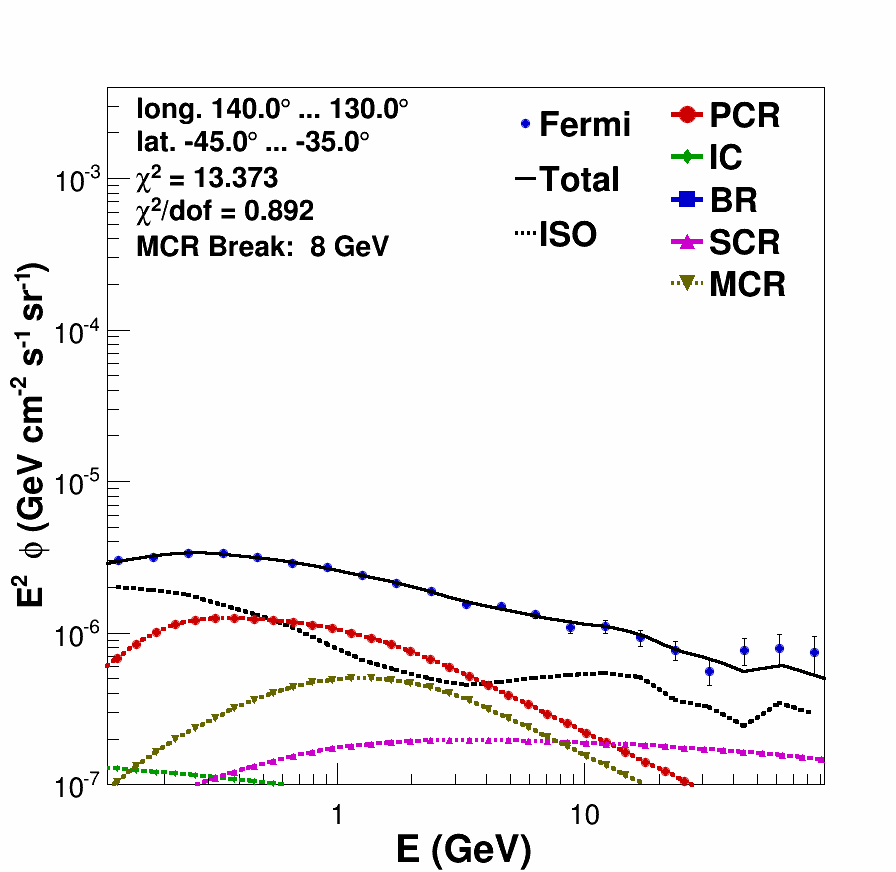}
\includegraphics[width=0.16\textwidth,height=0.16\textwidth,clip]{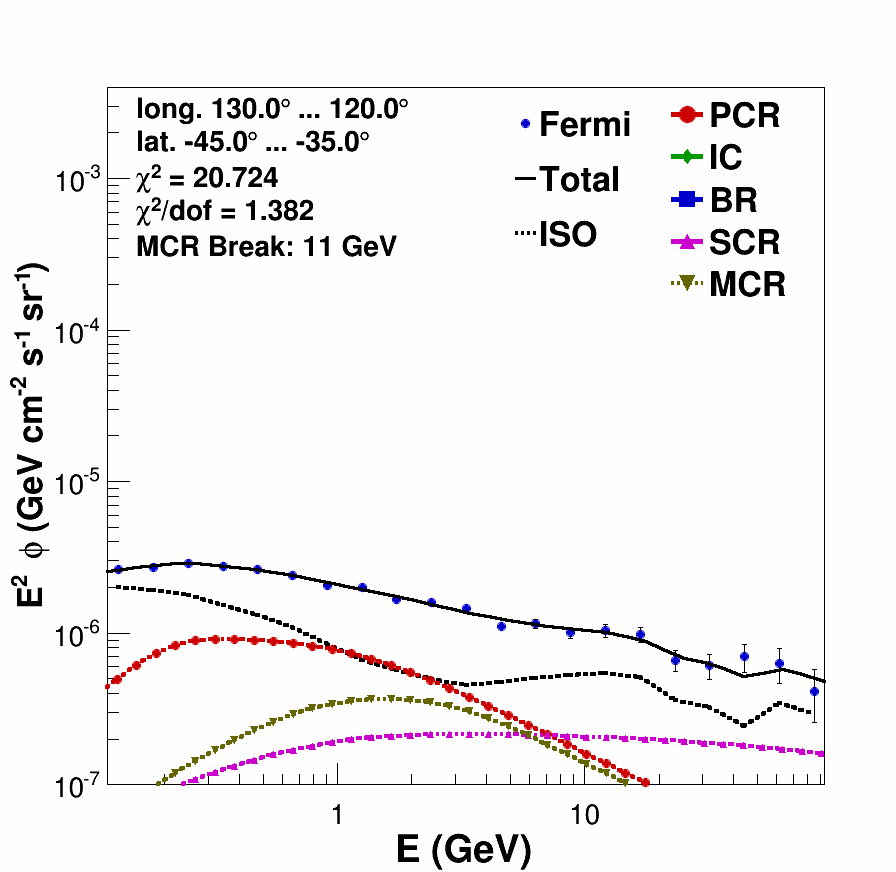}
\includegraphics[width=0.16\textwidth,height=0.16\textwidth,clip]{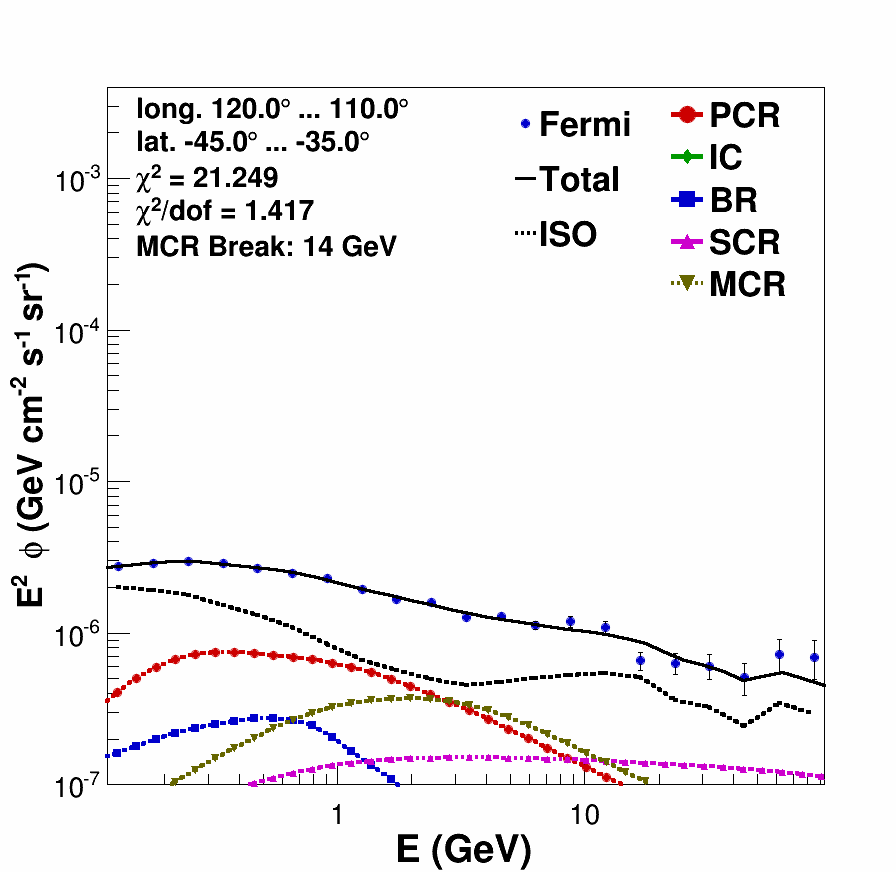}
\includegraphics[width=0.16\textwidth,height=0.16\textwidth,clip]{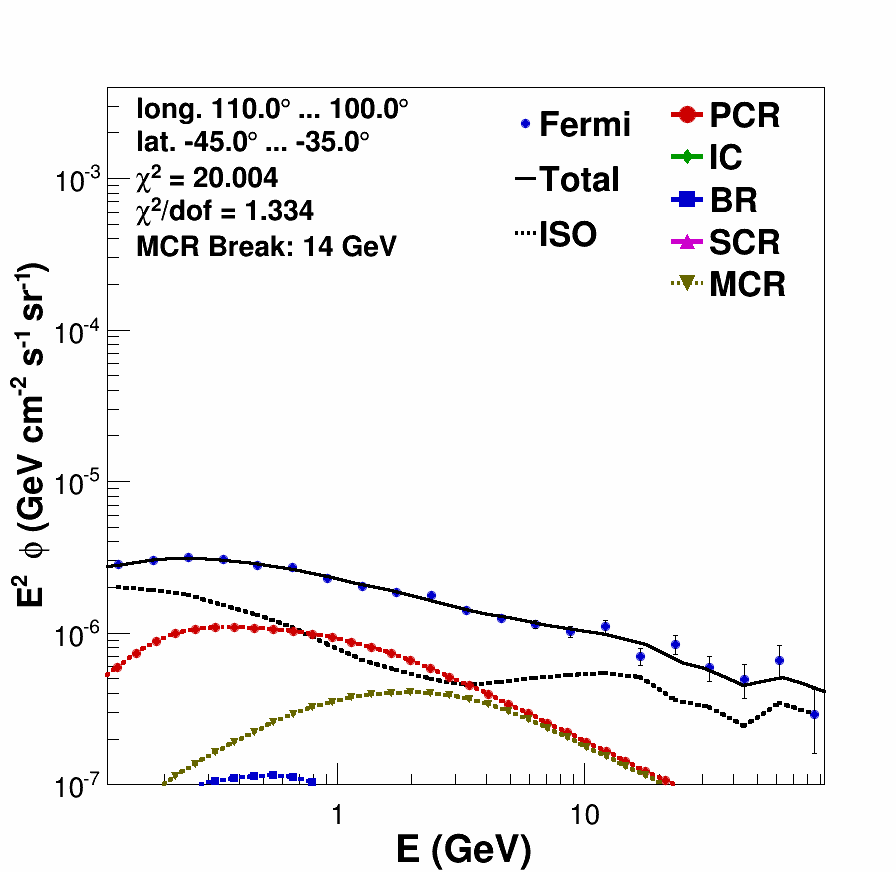}
\includegraphics[width=0.16\textwidth,height=0.16\textwidth,clip]{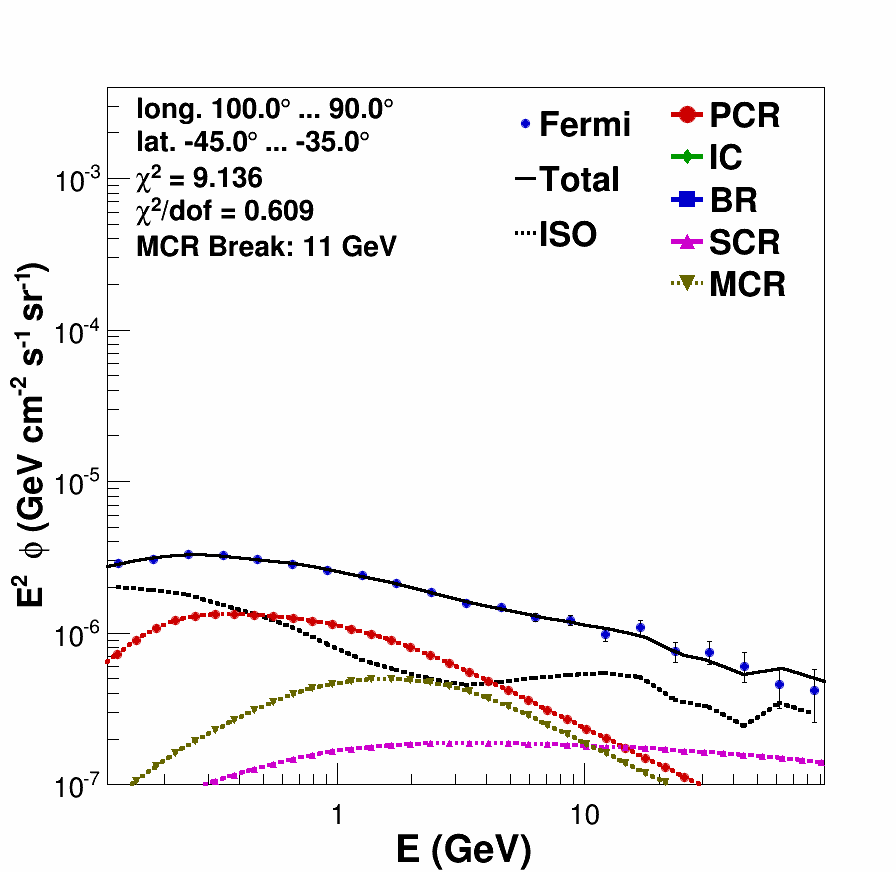}
\includegraphics[width=0.16\textwidth,height=0.16\textwidth,clip]{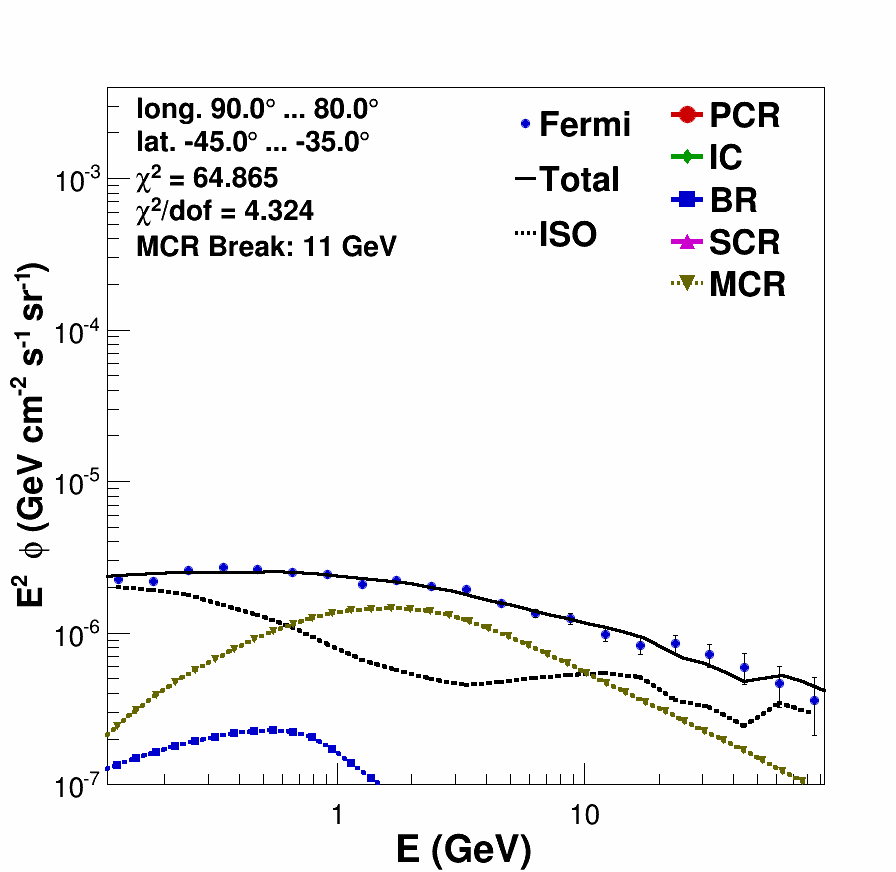}
\includegraphics[width=0.16\textwidth,height=0.16\textwidth,clip]{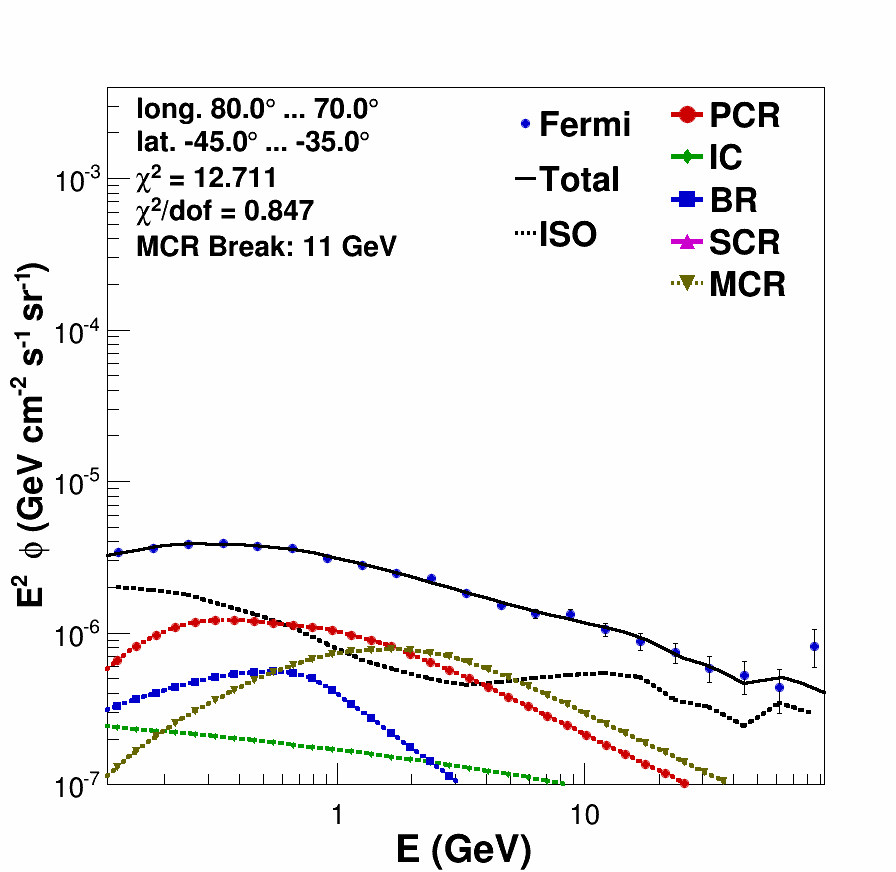}
\includegraphics[width=0.16\textwidth,height=0.16\textwidth,clip]{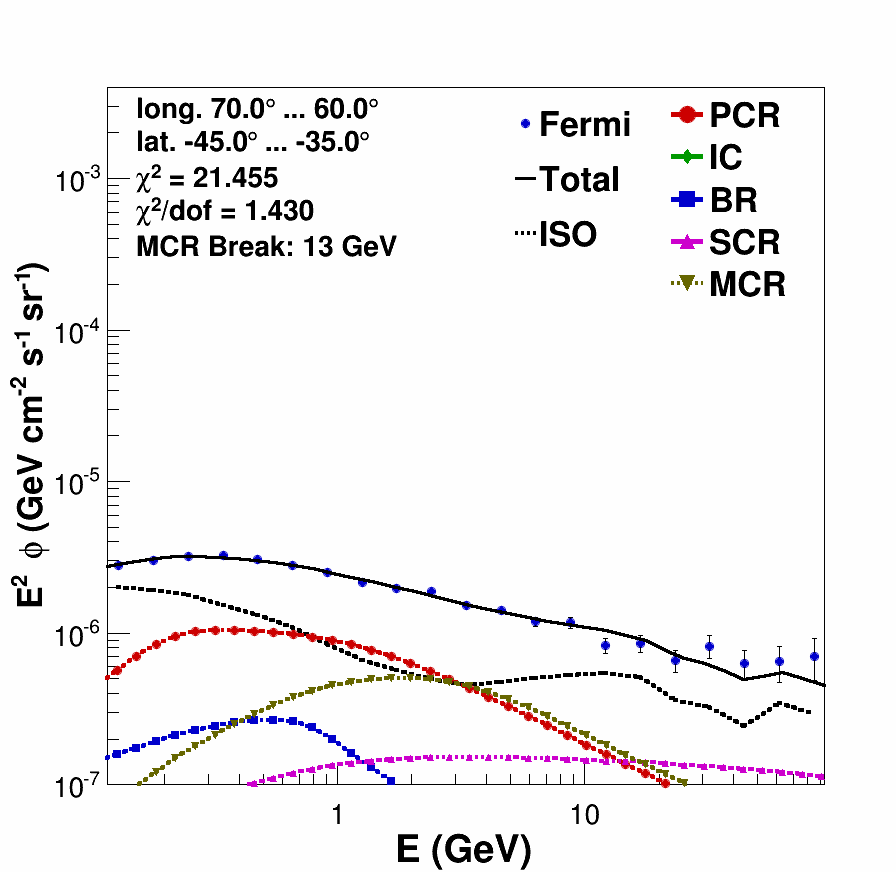}
\includegraphics[width=0.16\textwidth,height=0.16\textwidth,clip]{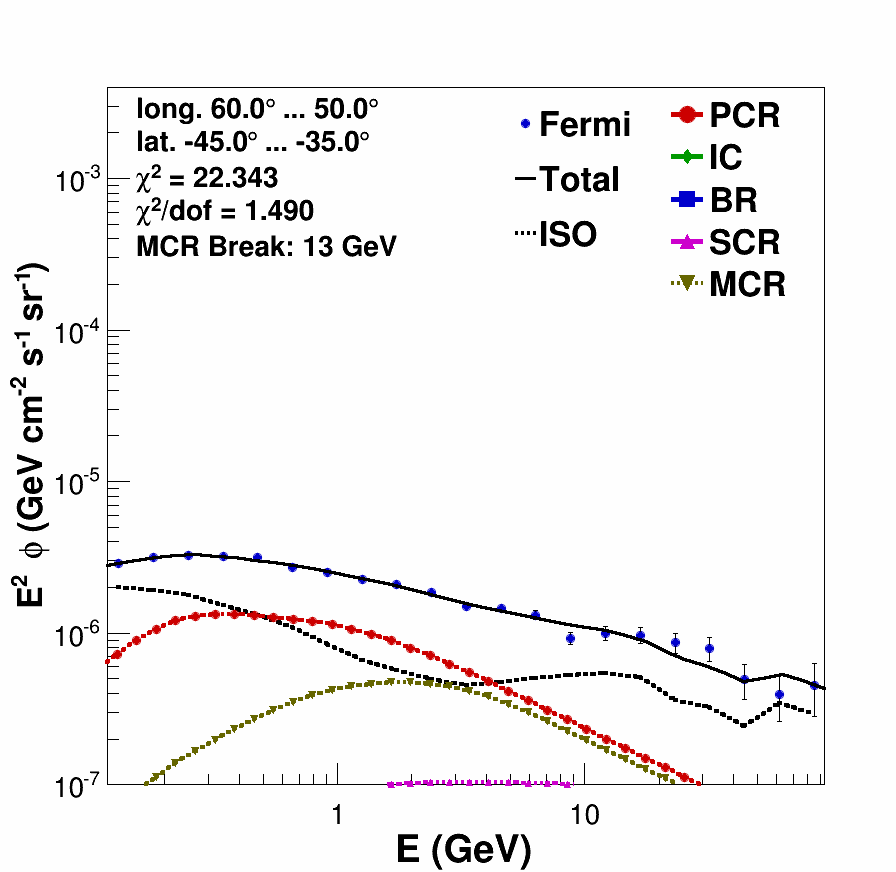}
\includegraphics[width=0.16\textwidth,height=0.16\textwidth,clip]{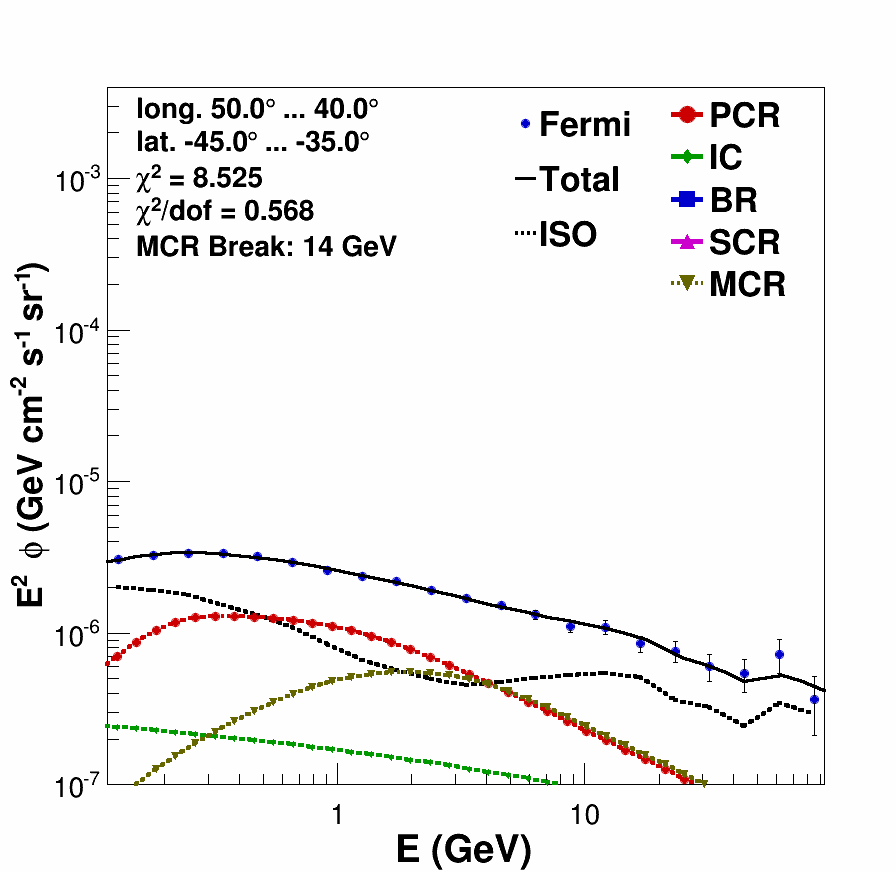}
\includegraphics[width=0.16\textwidth,height=0.16\textwidth,clip]{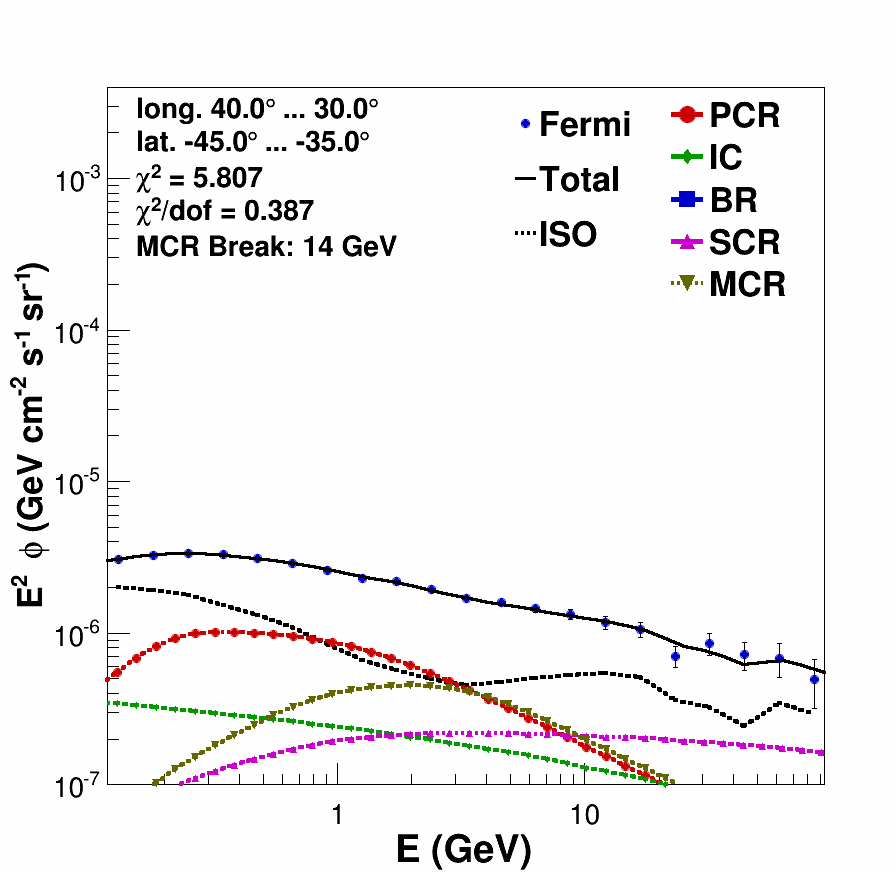}
\includegraphics[width=0.16\textwidth,height=0.16\textwidth,clip]{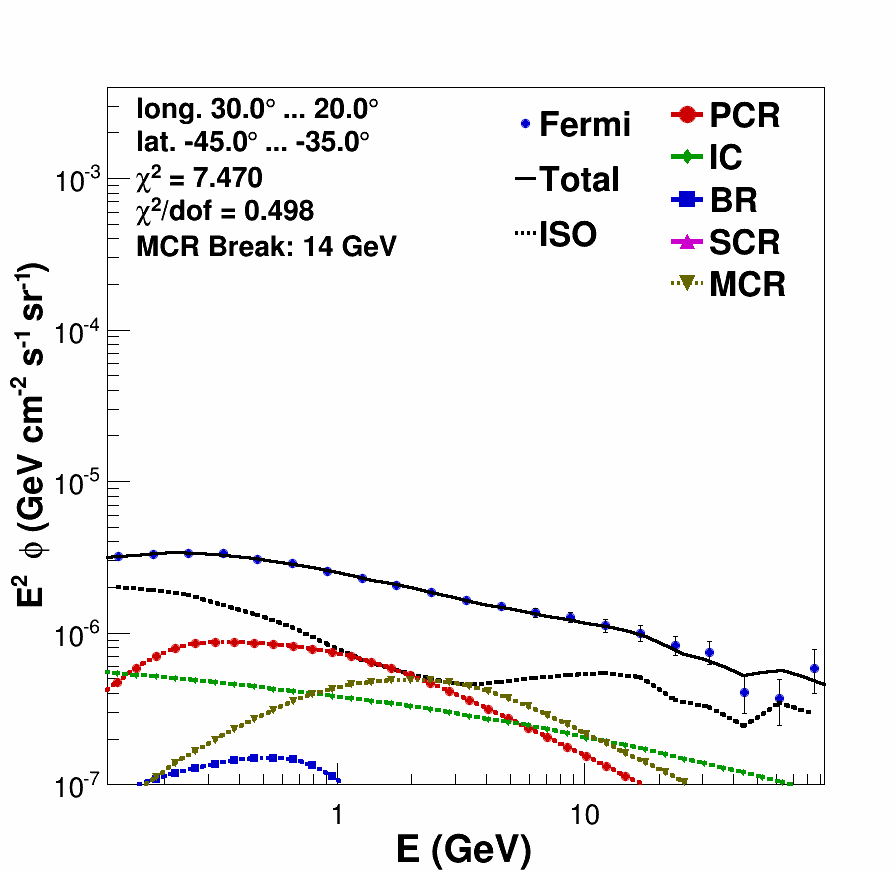}
\includegraphics[width=0.16\textwidth,height=0.16\textwidth,clip]{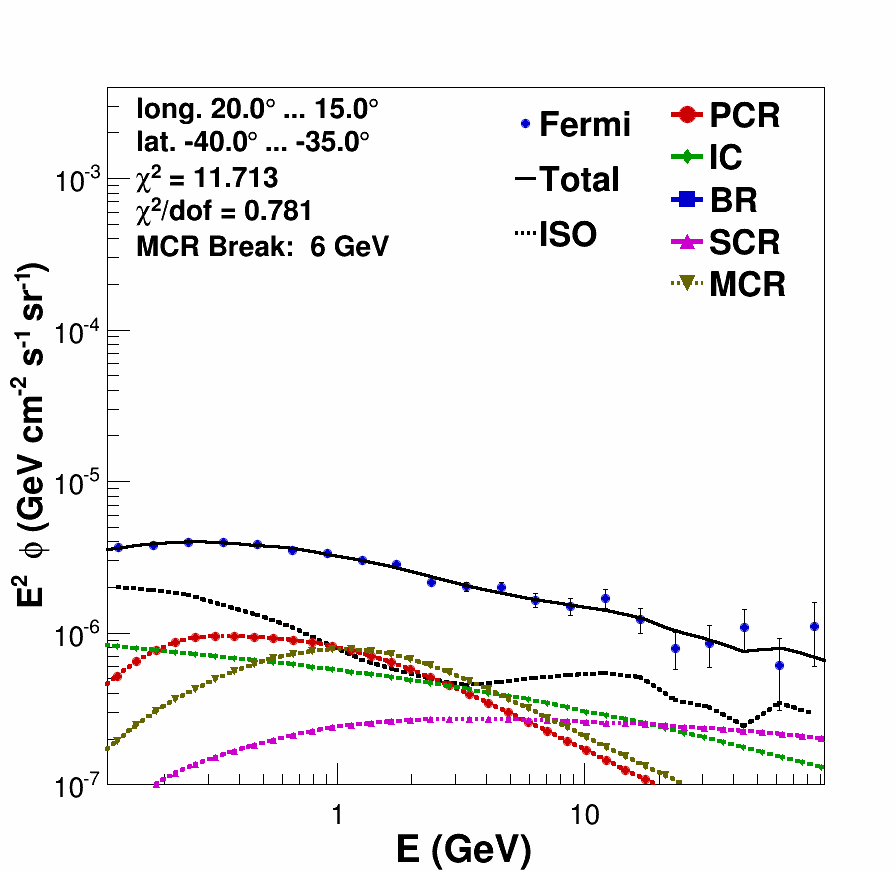}
\includegraphics[width=0.16\textwidth,height=0.16\textwidth,clip]{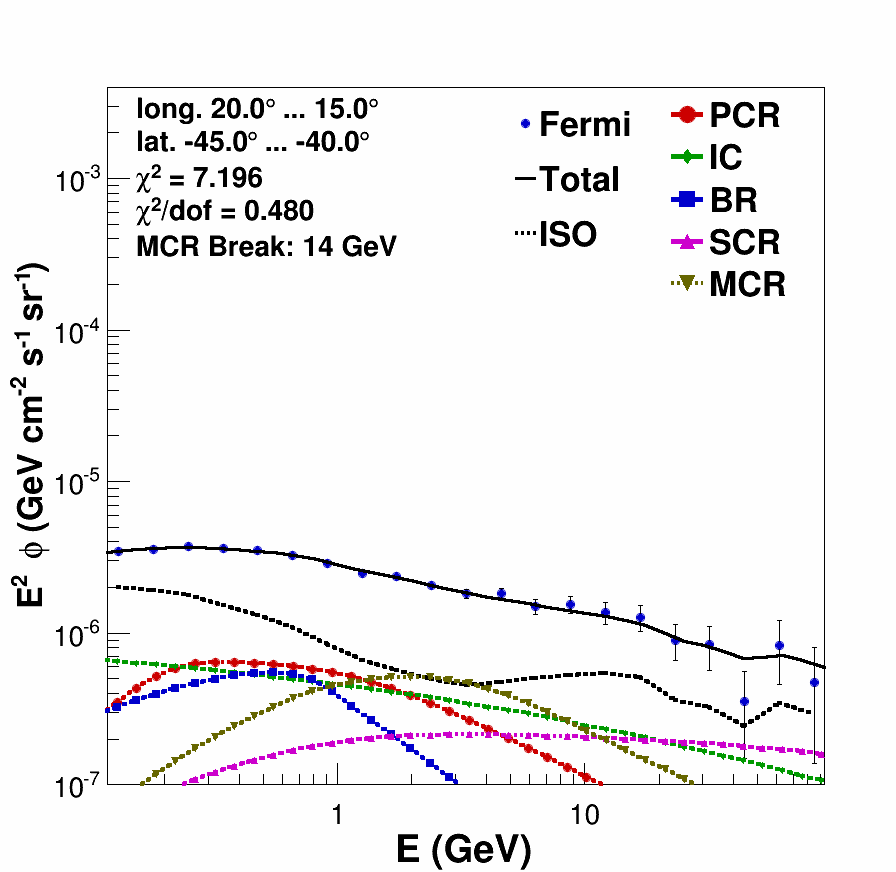}
\includegraphics[width=0.16\textwidth,height=0.16\textwidth,clip]{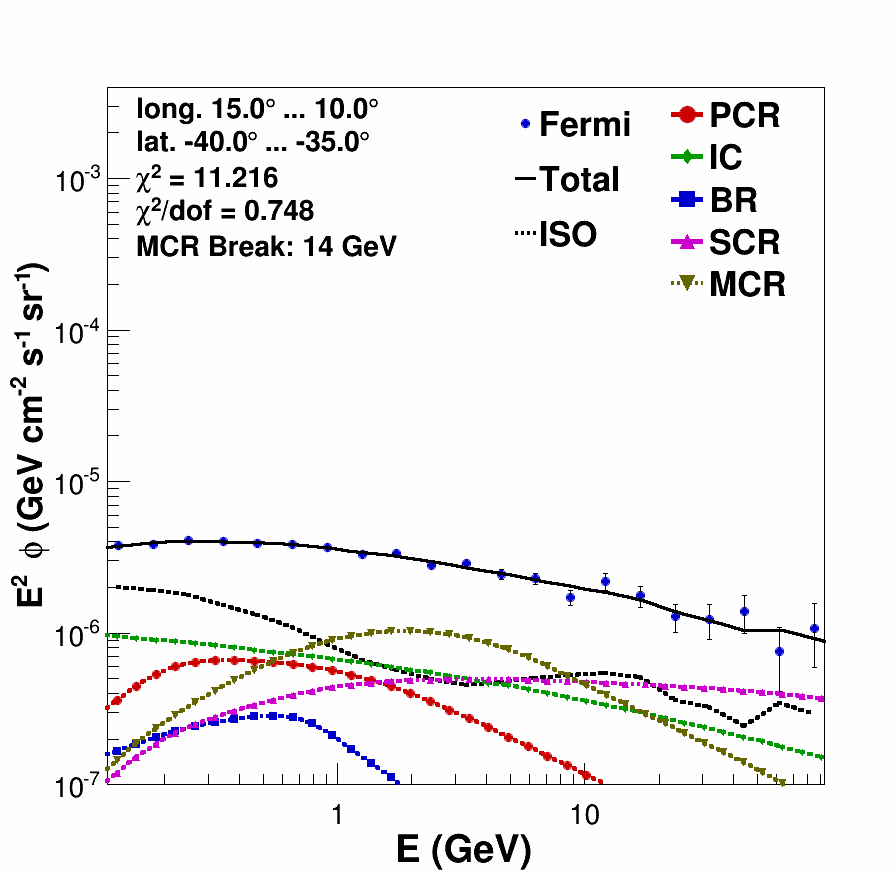}
\includegraphics[width=0.16\textwidth,height=0.16\textwidth,clip]{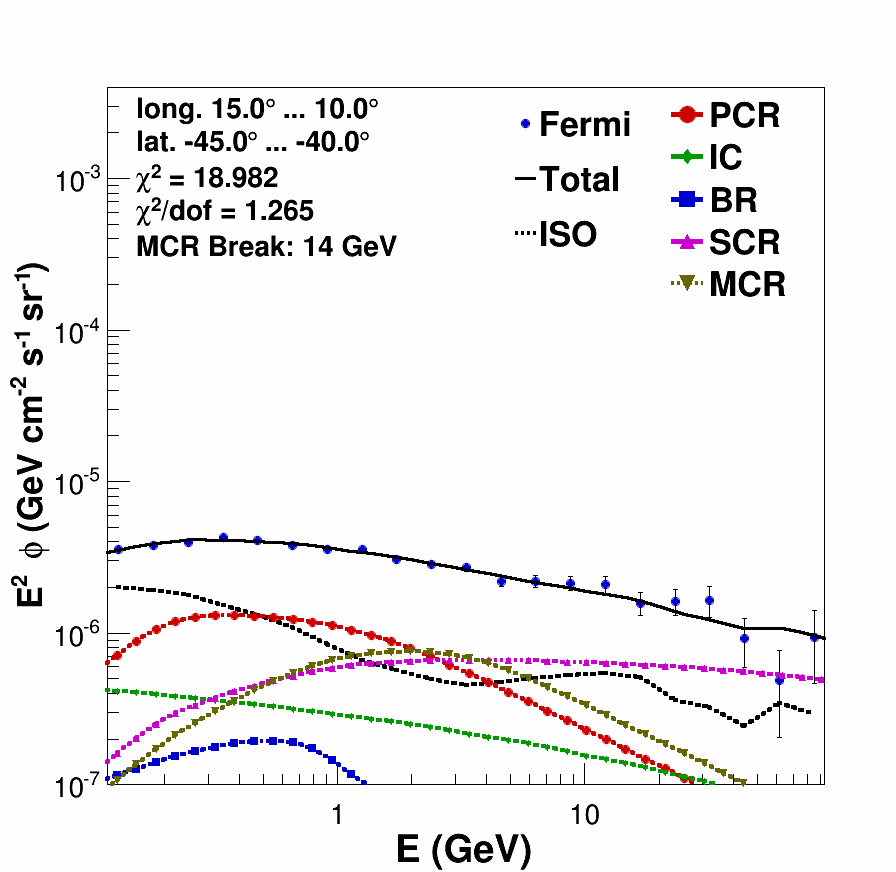}
\includegraphics[width=0.16\textwidth,height=0.16\textwidth,clip]{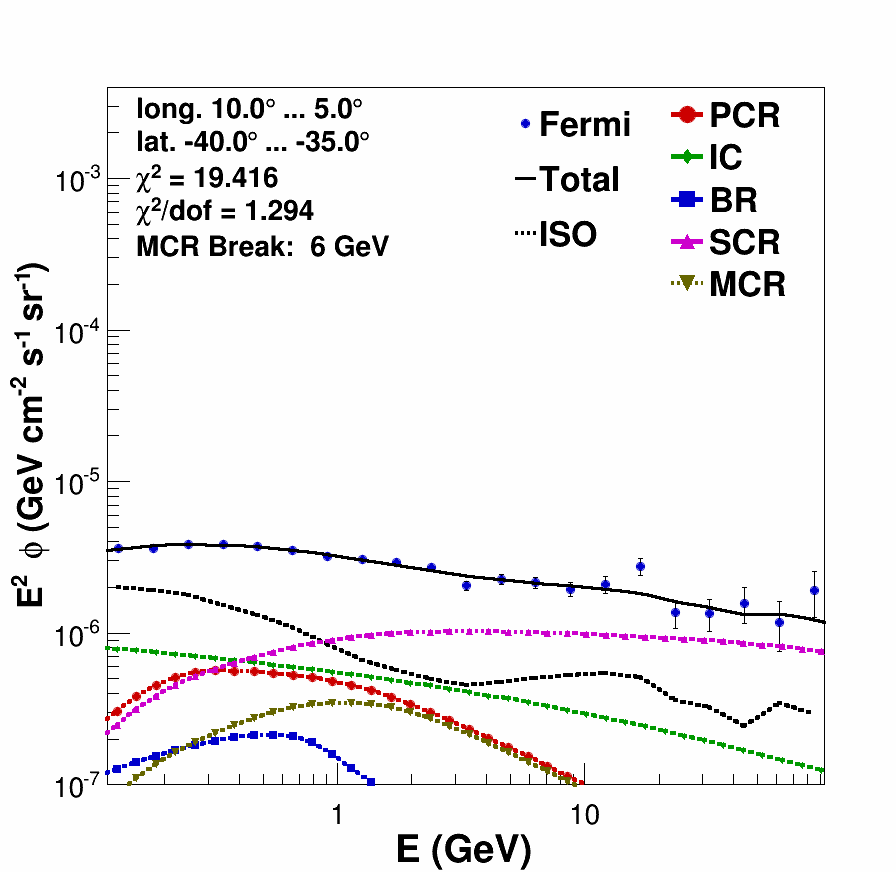}
\includegraphics[width=0.16\textwidth,height=0.16\textwidth,clip]{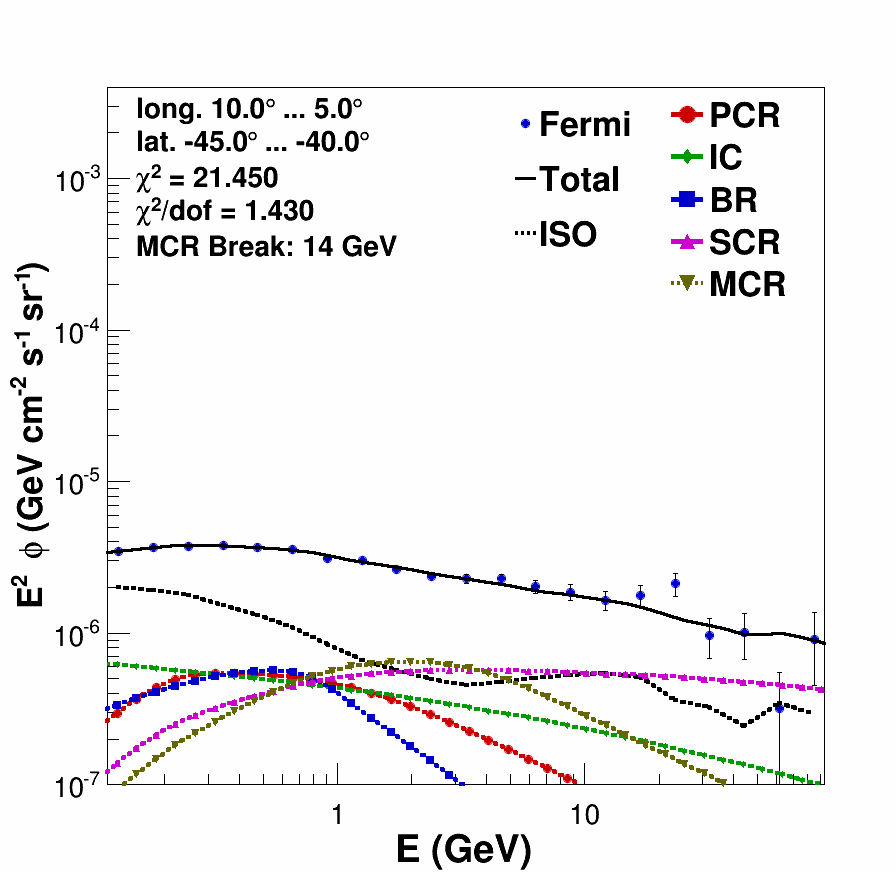}
\includegraphics[width=0.16\textwidth,height=0.16\textwidth,clip]{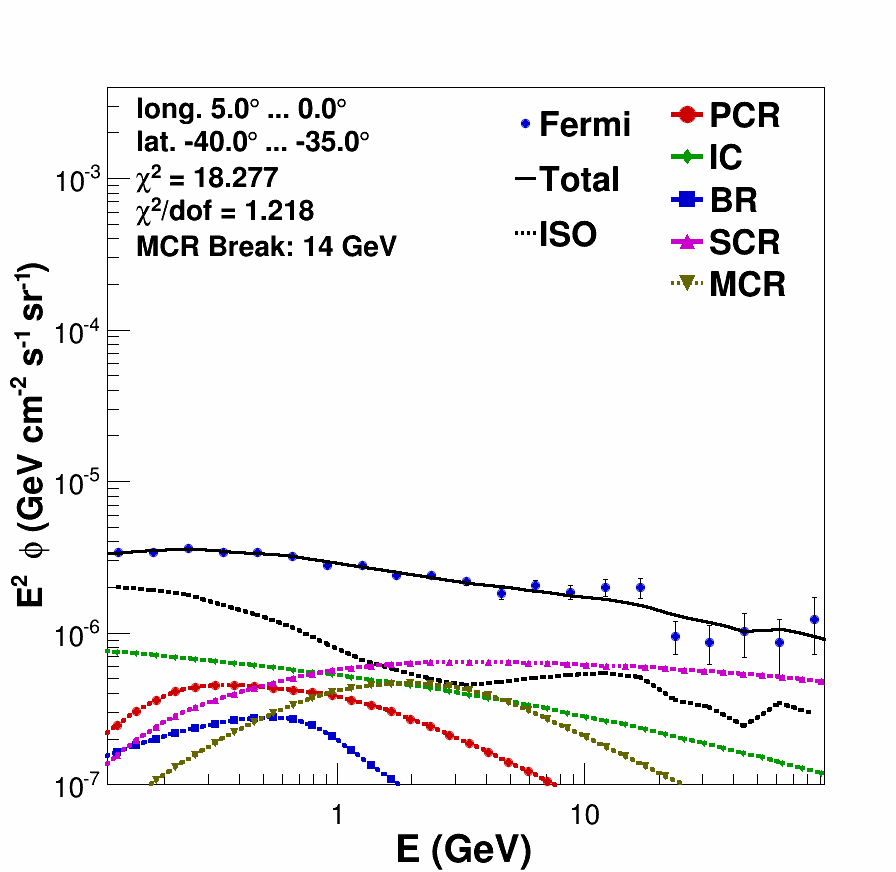}
\includegraphics[width=0.16\textwidth,height=0.16\textwidth,clip]{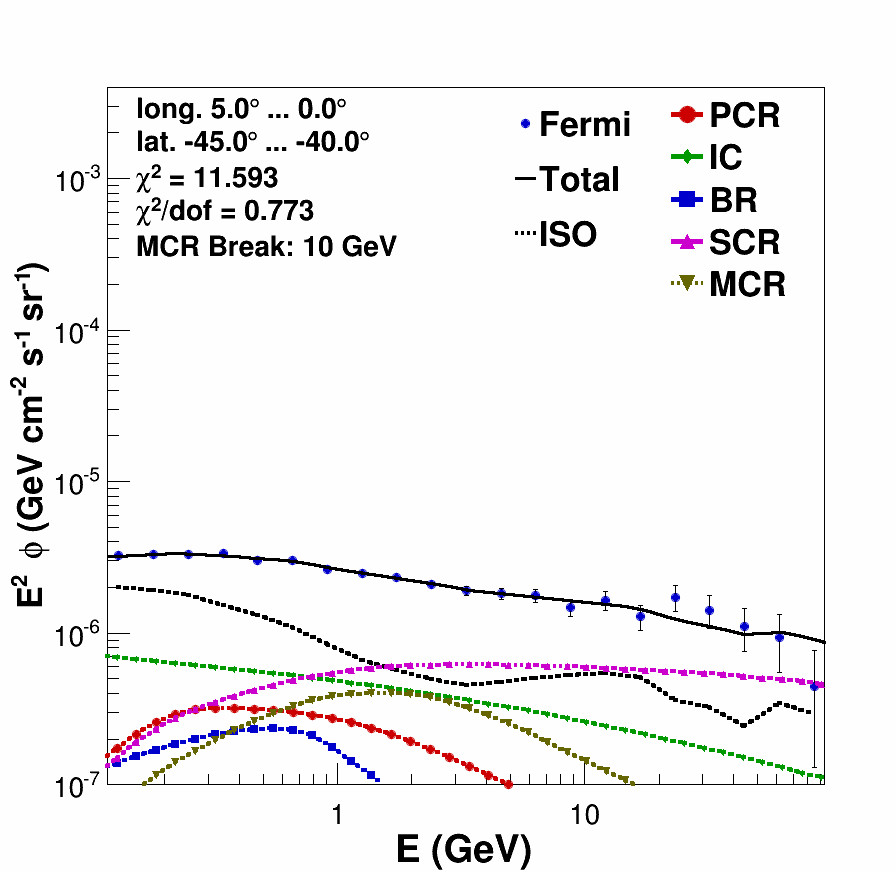}
\includegraphics[width=0.16\textwidth,height=0.16\textwidth,clip]{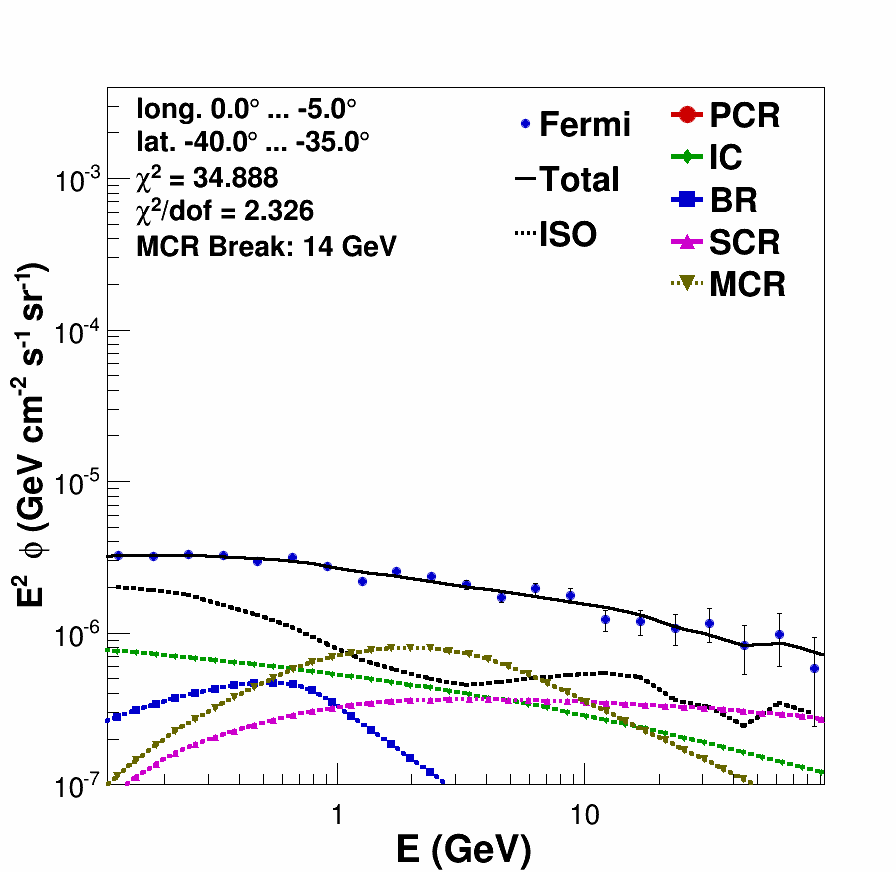}
\includegraphics[width=0.16\textwidth,height=0.16\textwidth,clip]{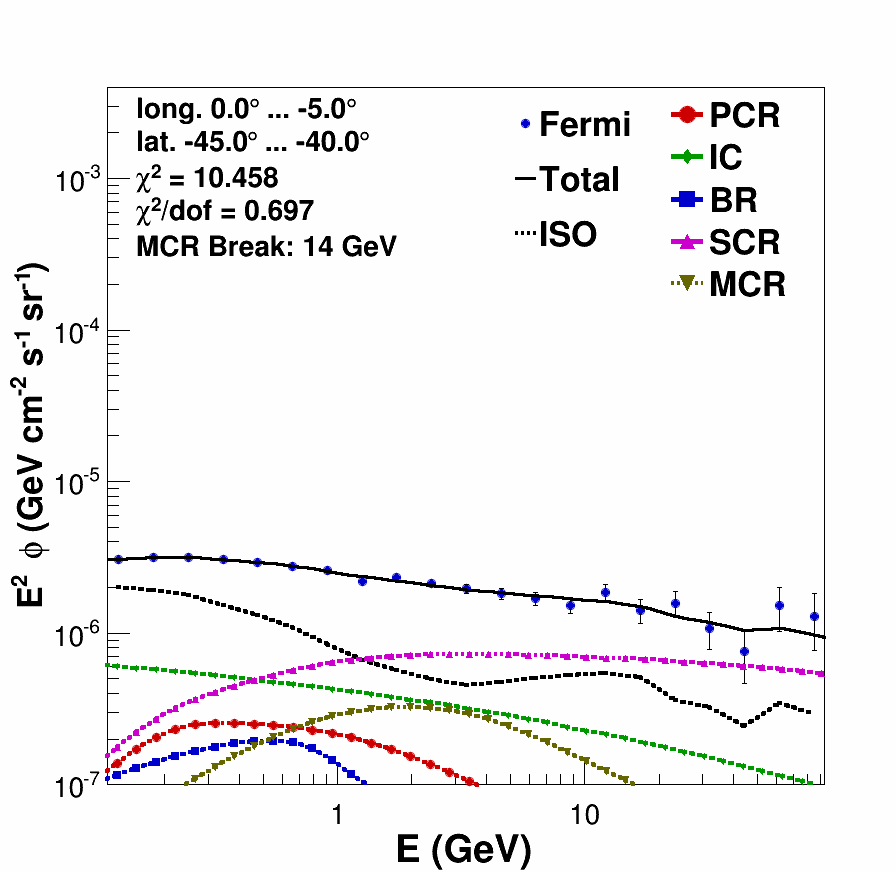}
\includegraphics[width=0.16\textwidth,height=0.16\textwidth,clip]{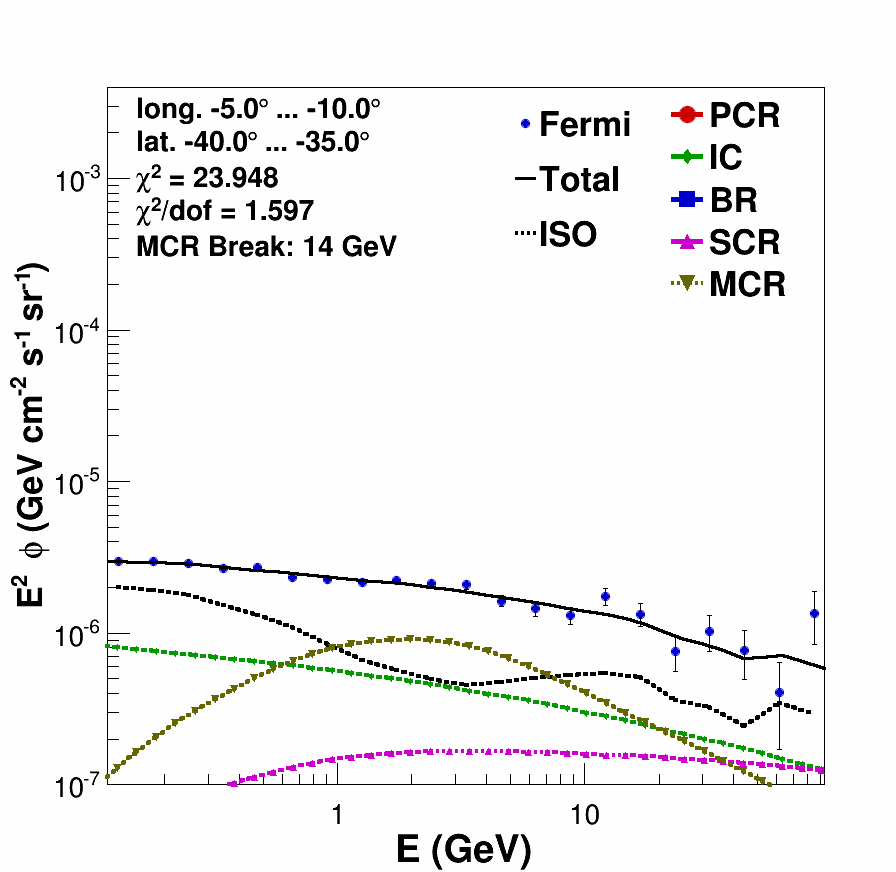}
\includegraphics[width=0.16\textwidth,height=0.16\textwidth,clip]{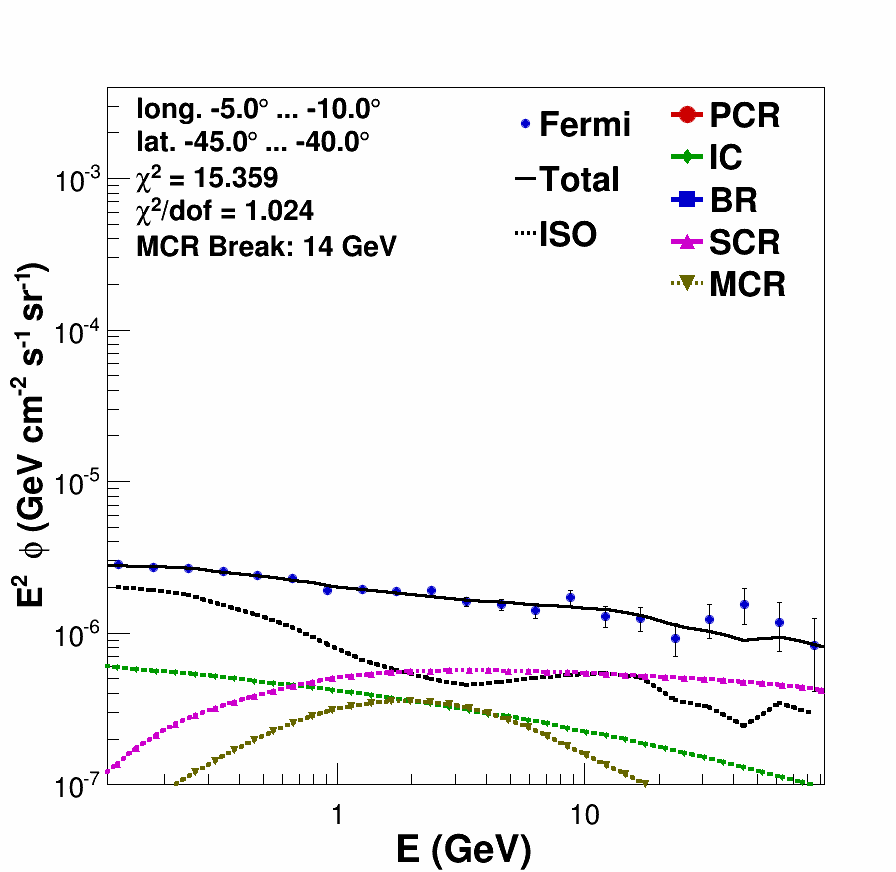}
\includegraphics[width=0.16\textwidth,height=0.16\textwidth,clip]{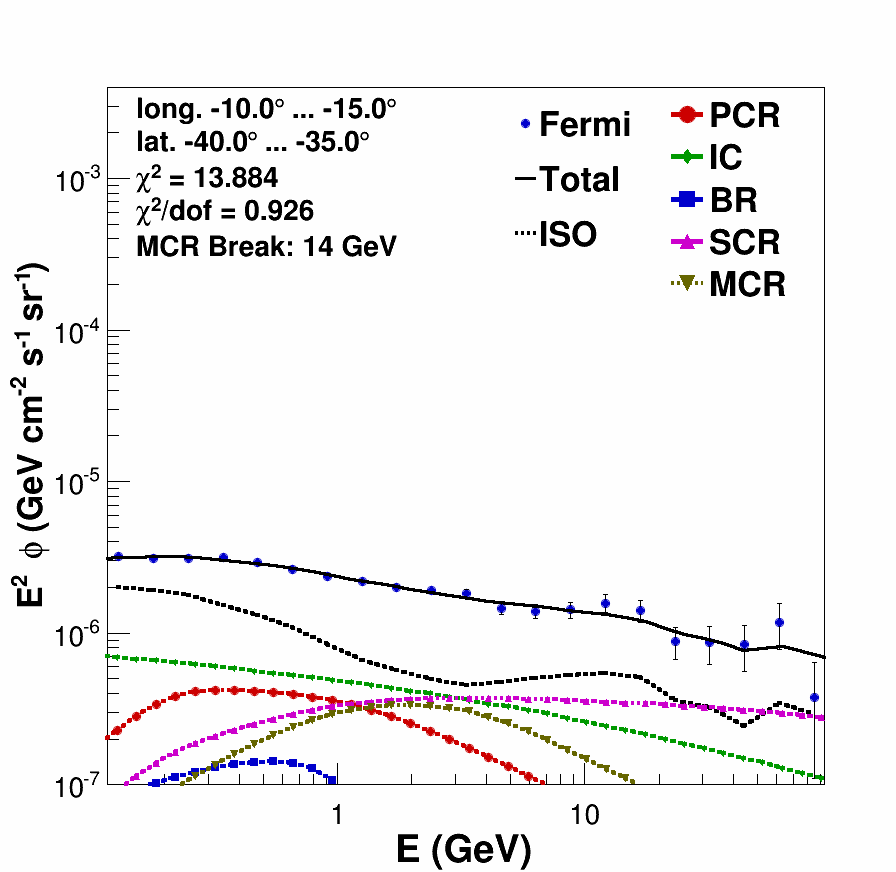}
\includegraphics[width=0.16\textwidth,height=0.16\textwidth,clip]{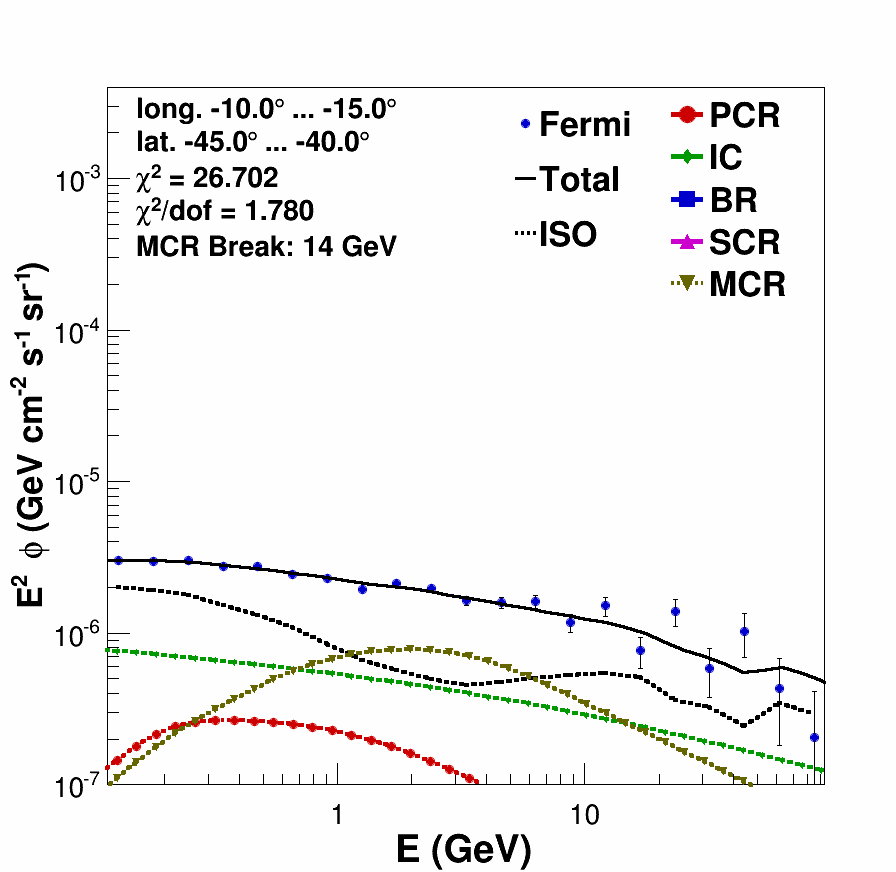}
\includegraphics[width=0.16\textwidth,height=0.16\textwidth,clip]{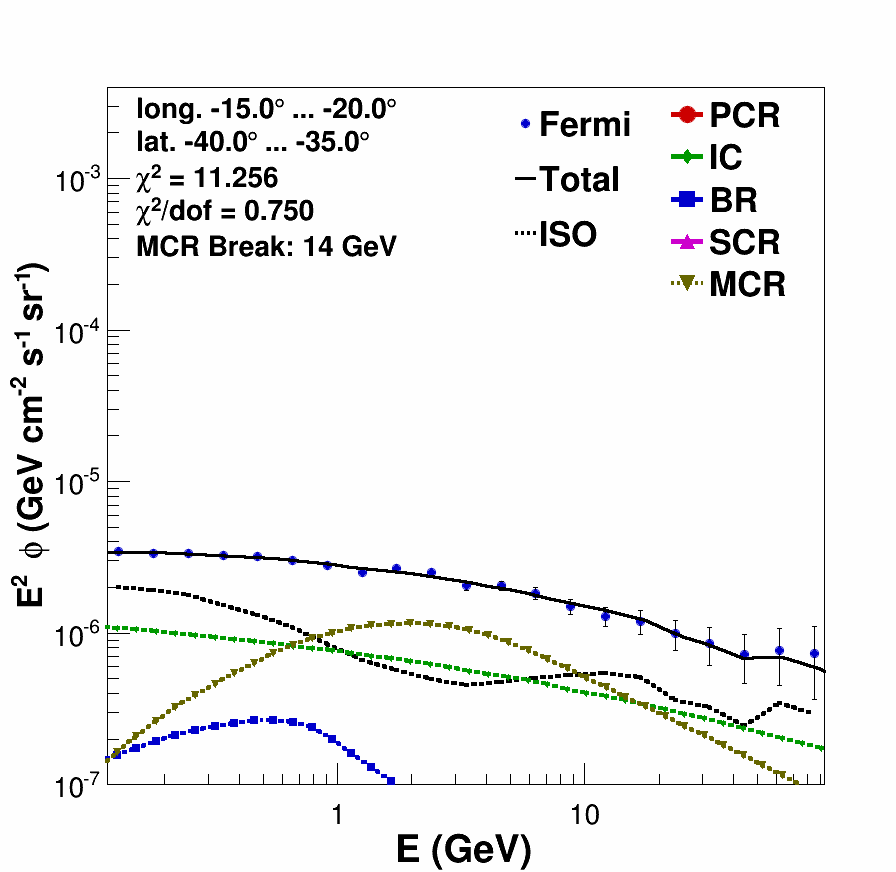}
\includegraphics[width=0.16\textwidth,height=0.16\textwidth,clip]{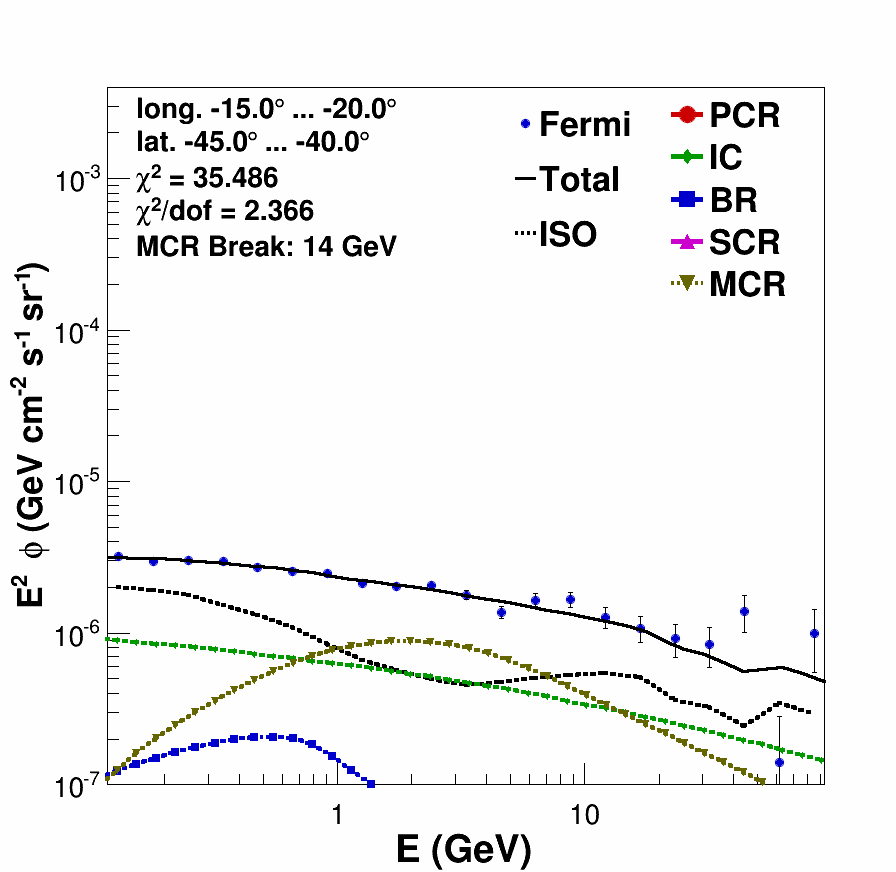}
\includegraphics[width=0.16\textwidth,height=0.16\textwidth,clip]{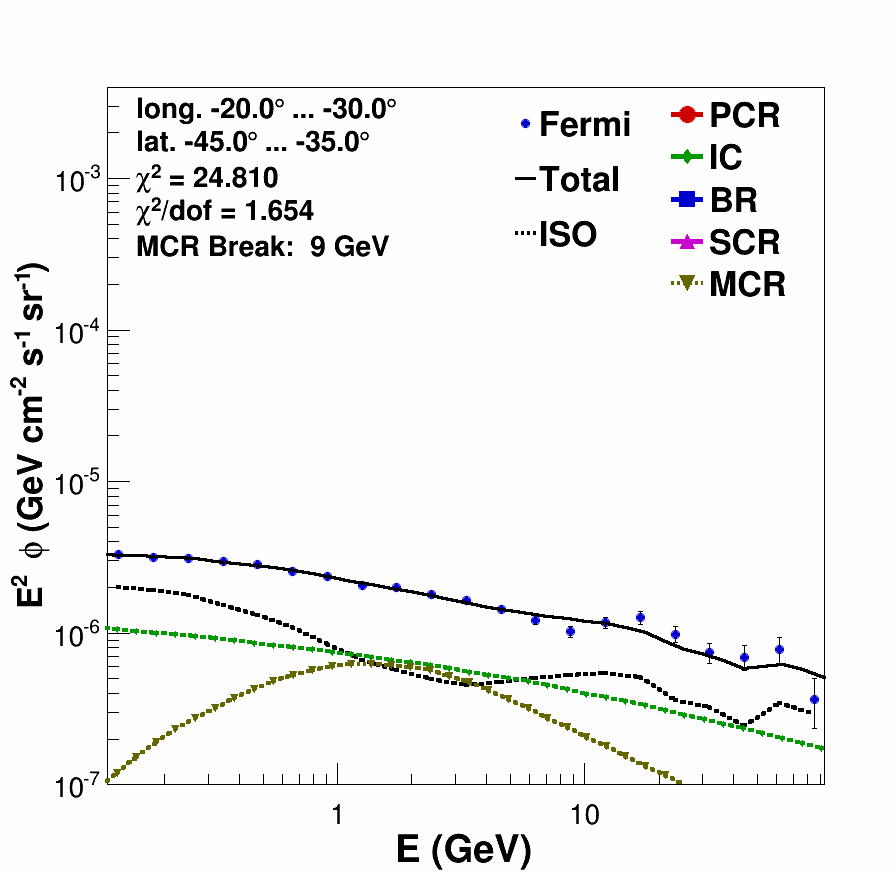}
\includegraphics[width=0.16\textwidth,height=0.16\textwidth,clip]{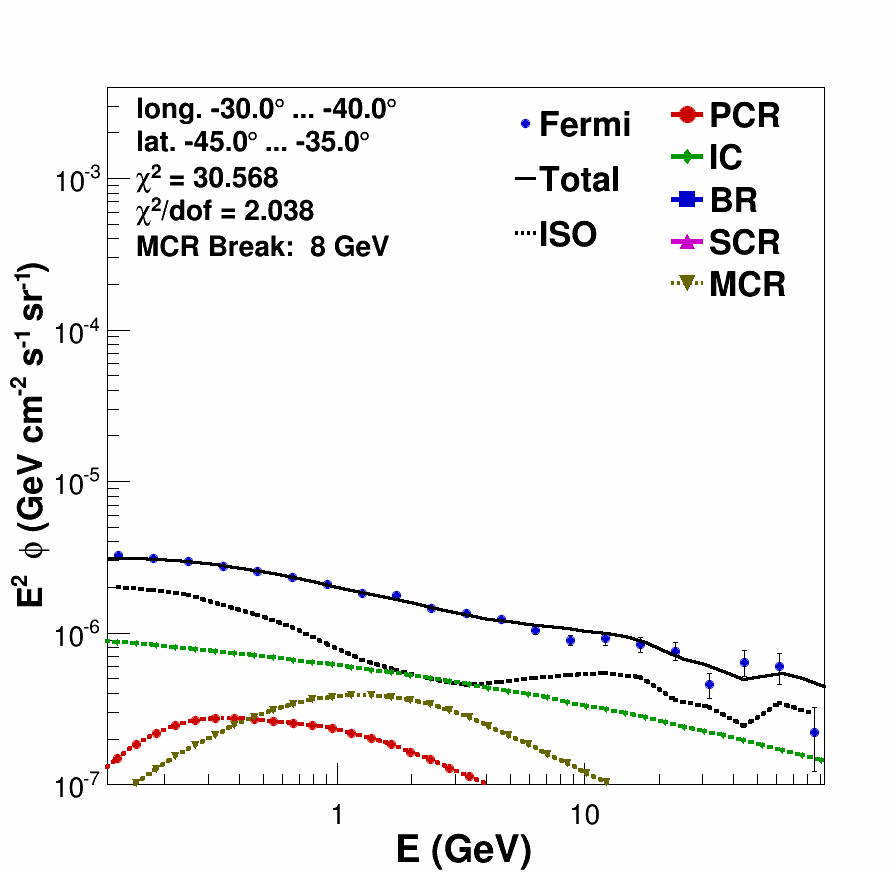}
\includegraphics[width=0.16\textwidth,height=0.16\textwidth,clip]{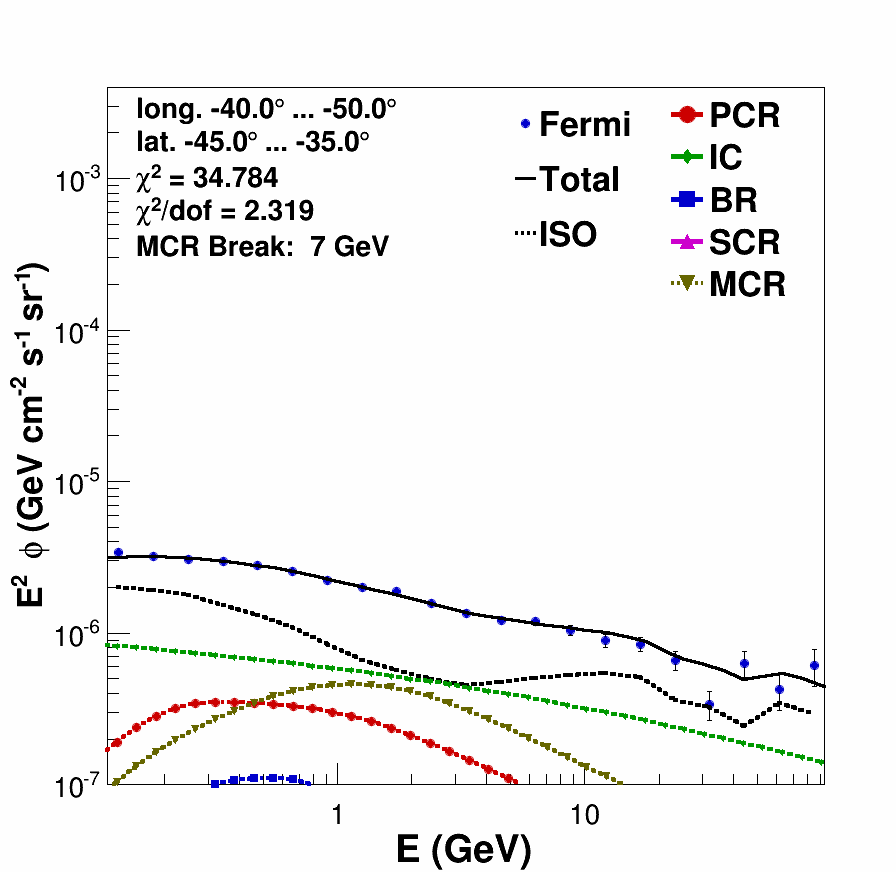}
\includegraphics[width=0.16\textwidth,height=0.16\textwidth,clip]{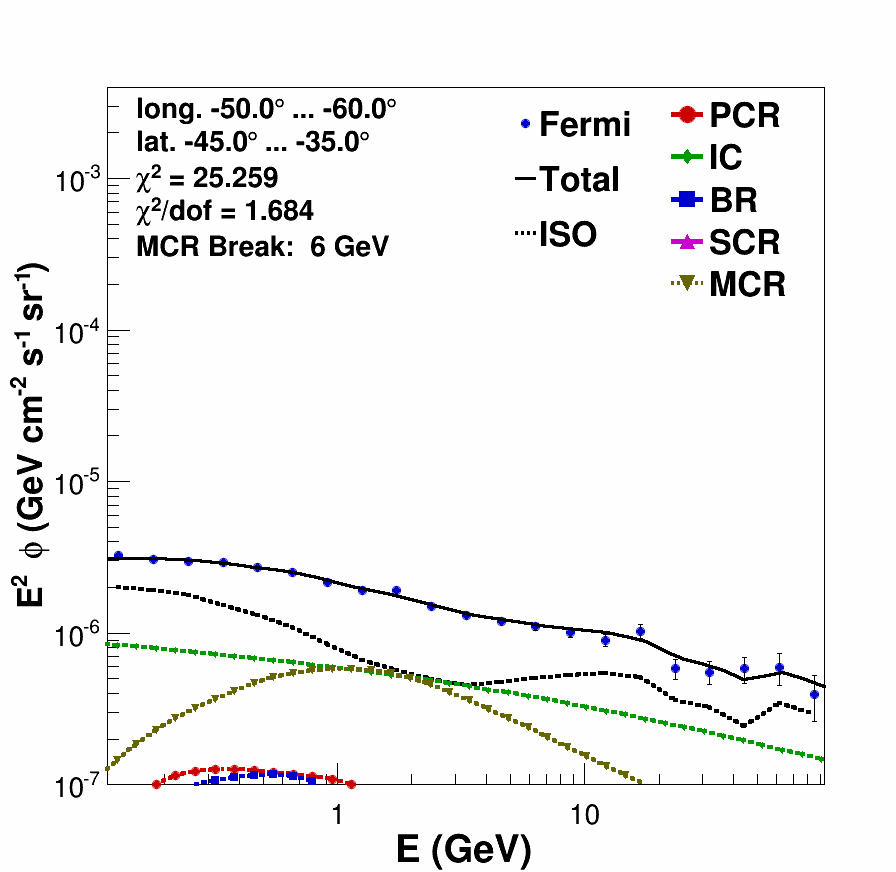}
\includegraphics[width=0.16\textwidth,height=0.16\textwidth,clip]{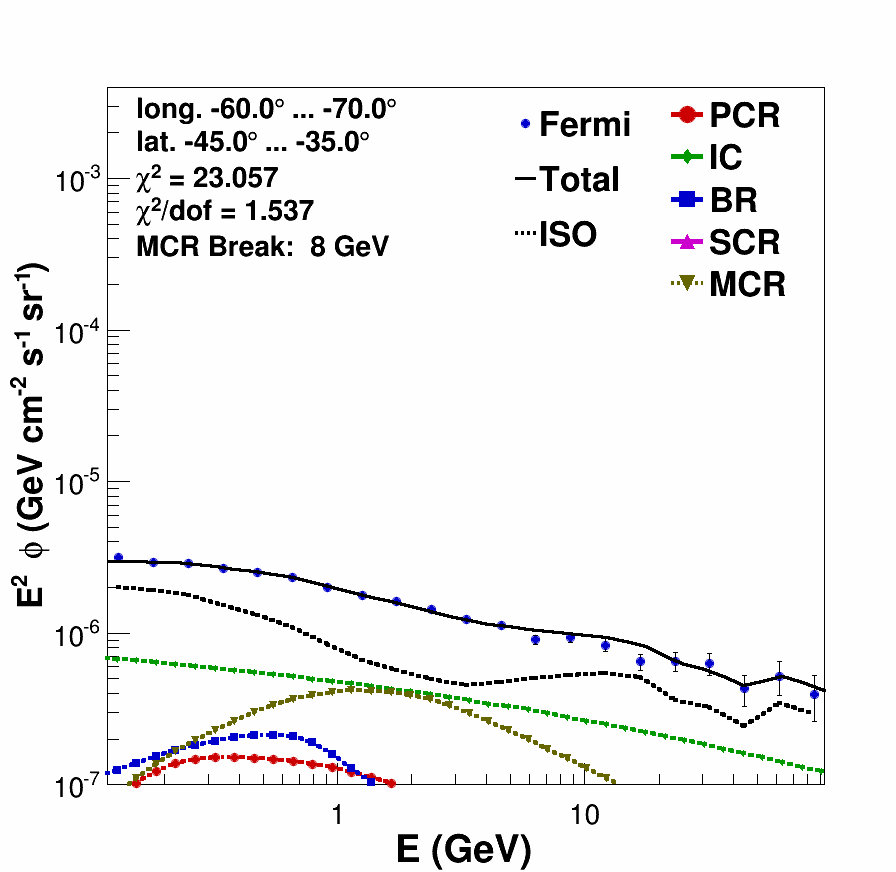}
\includegraphics[width=0.16\textwidth,height=0.16\textwidth,clip]{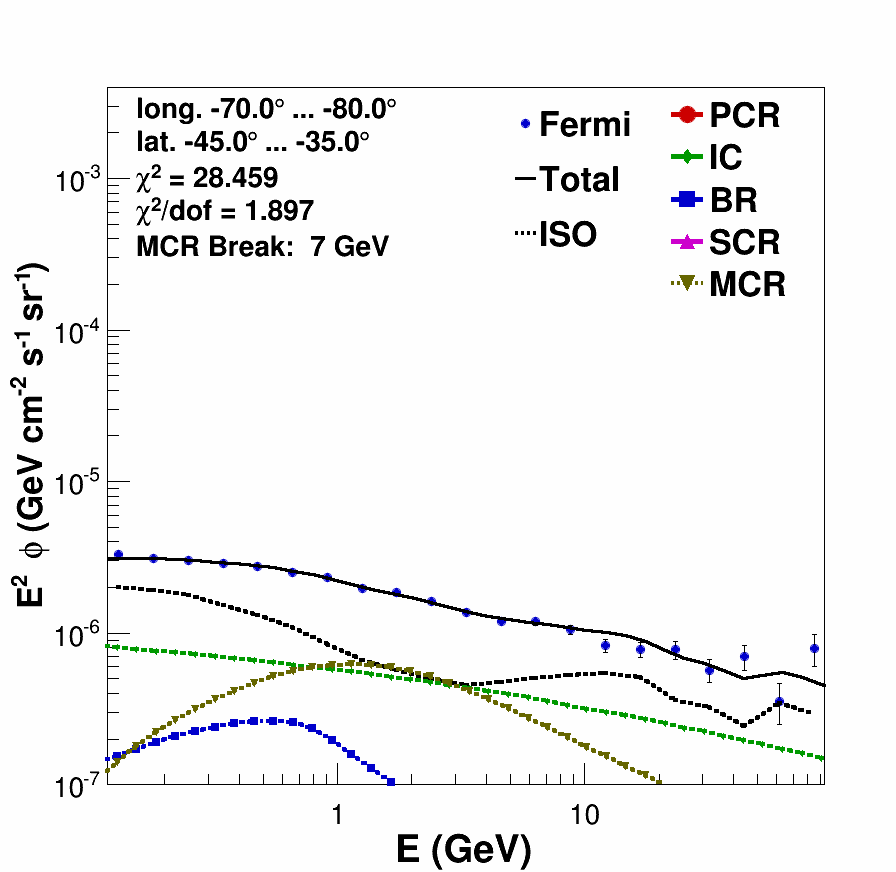}
\includegraphics[width=0.16\textwidth,height=0.16\textwidth,clip]{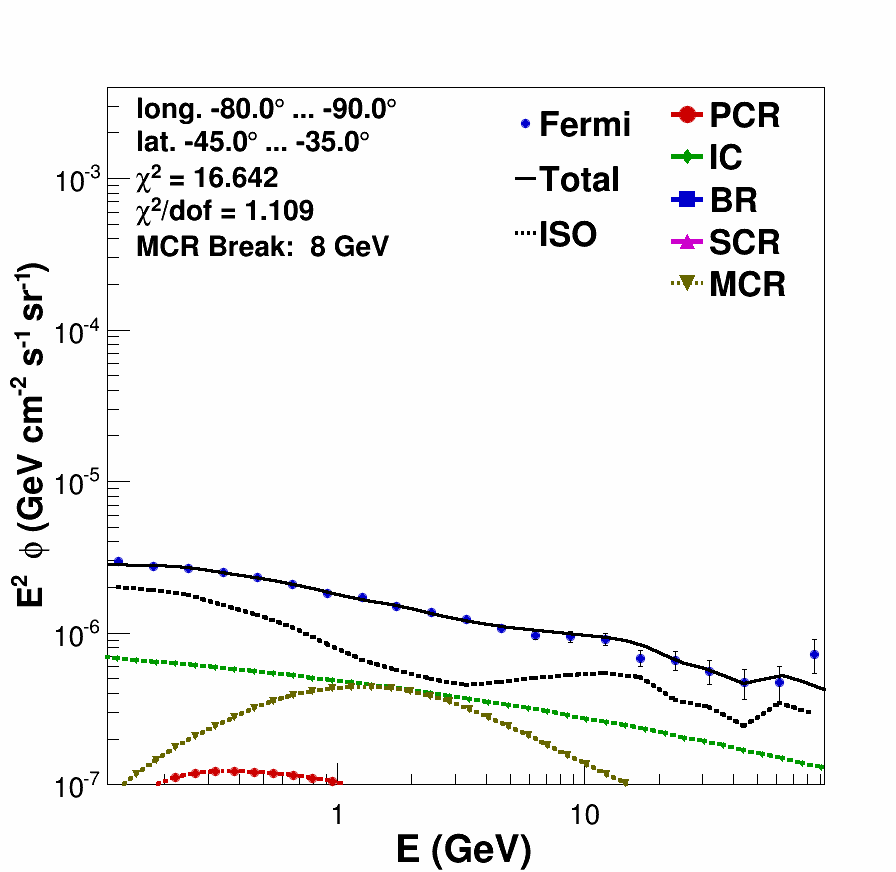}
\includegraphics[width=0.16\textwidth,height=0.16\textwidth,clip]{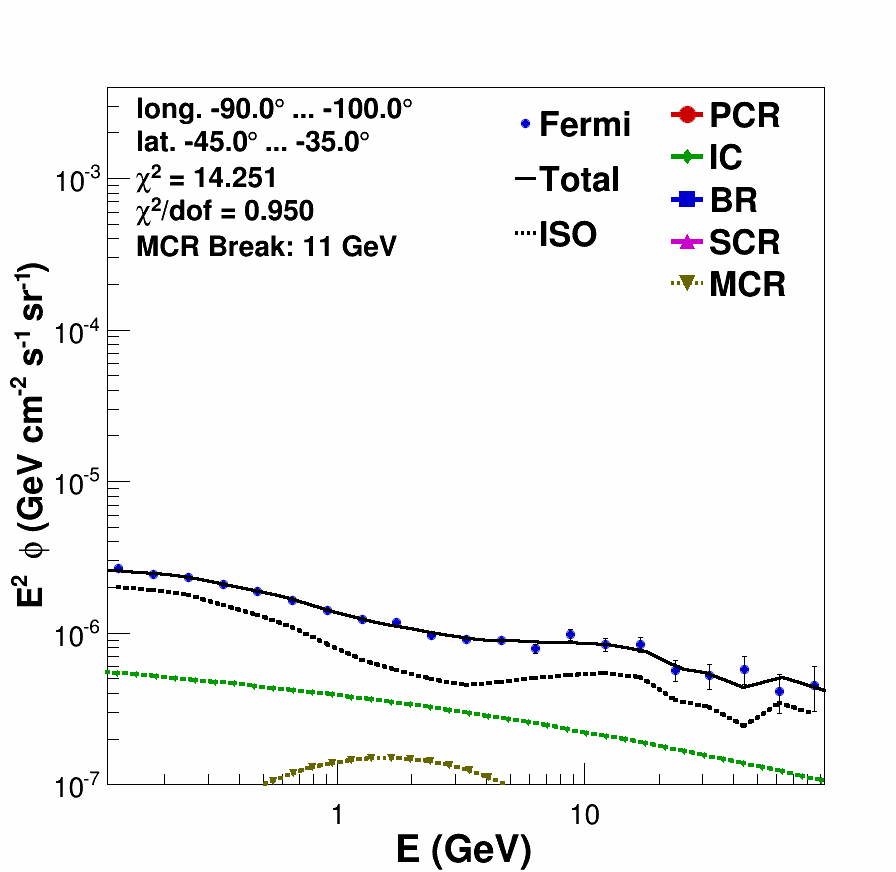}
\includegraphics[width=0.16\textwidth,height=0.16\textwidth,clip]{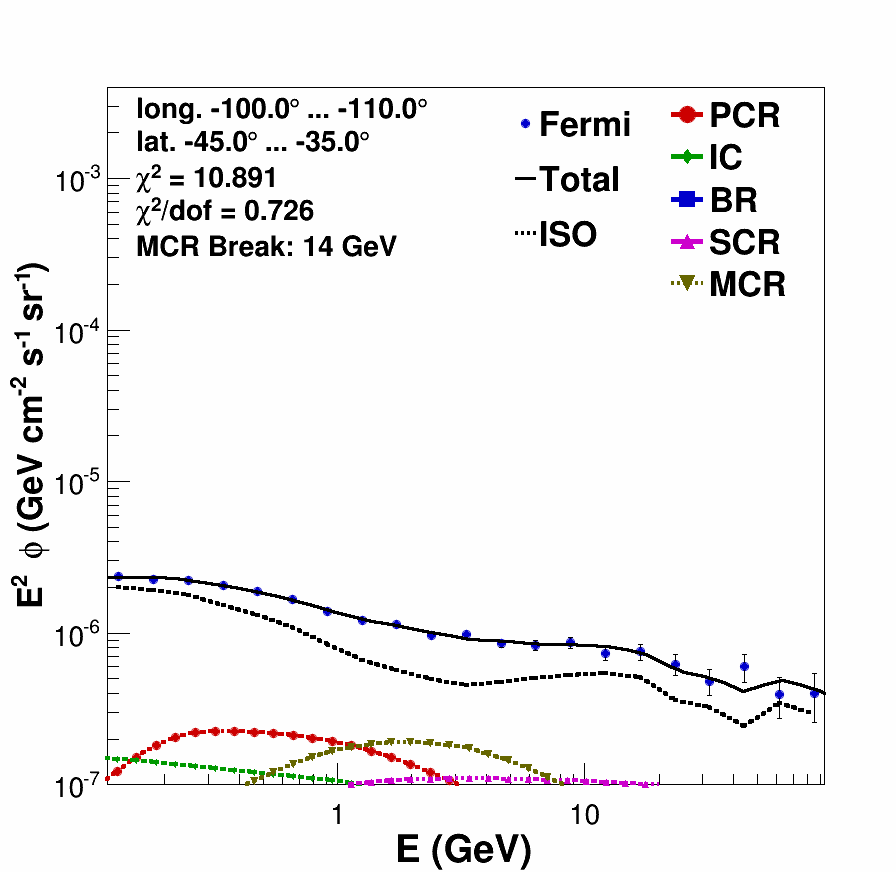}
\includegraphics[width=0.16\textwidth,height=0.16\textwidth,clip]{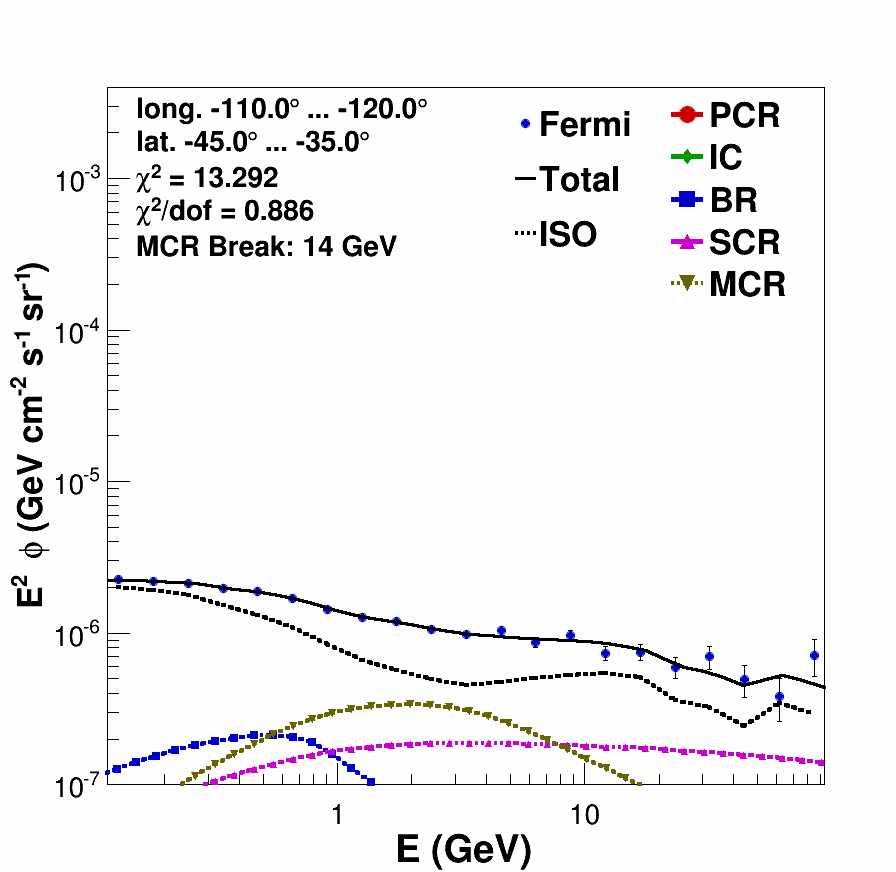}
\includegraphics[width=0.16\textwidth,height=0.16\textwidth,clip]{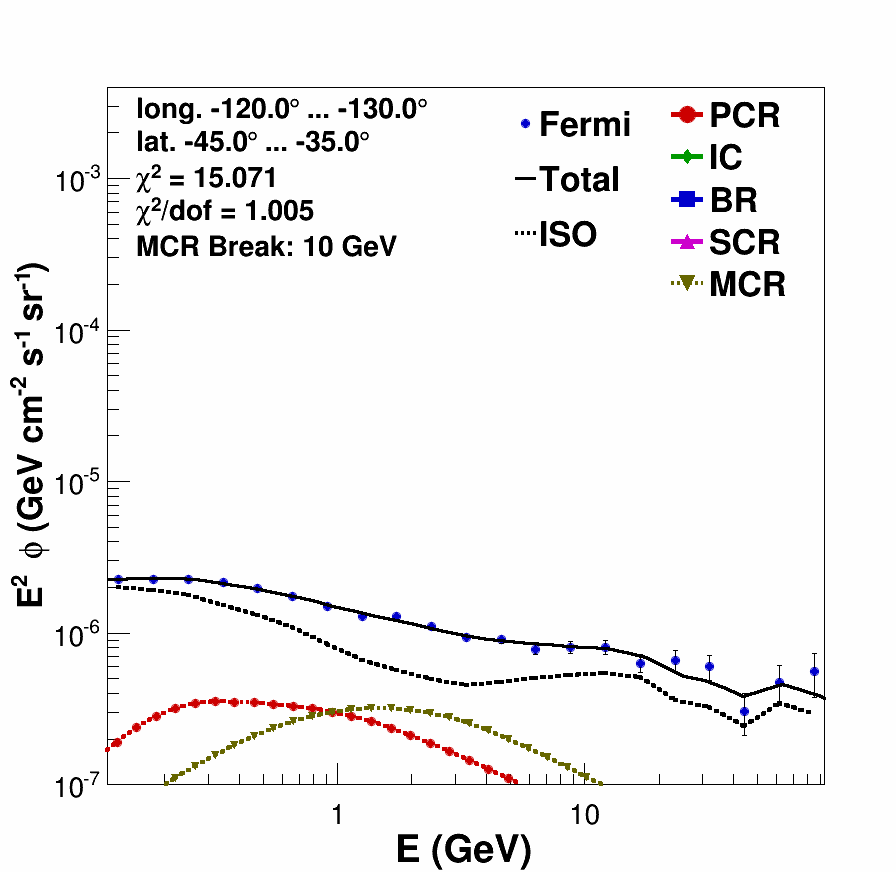}
\includegraphics[width=0.16\textwidth,height=0.16\textwidth,clip]{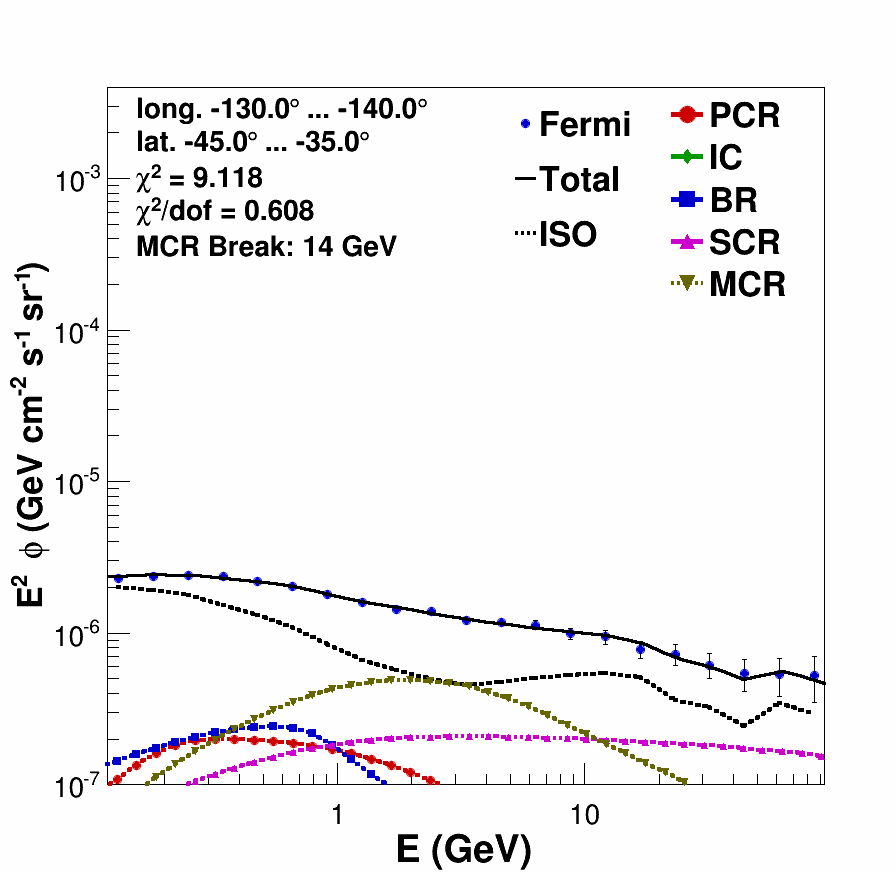}
\includegraphics[width=0.16\textwidth,height=0.16\textwidth,clip]{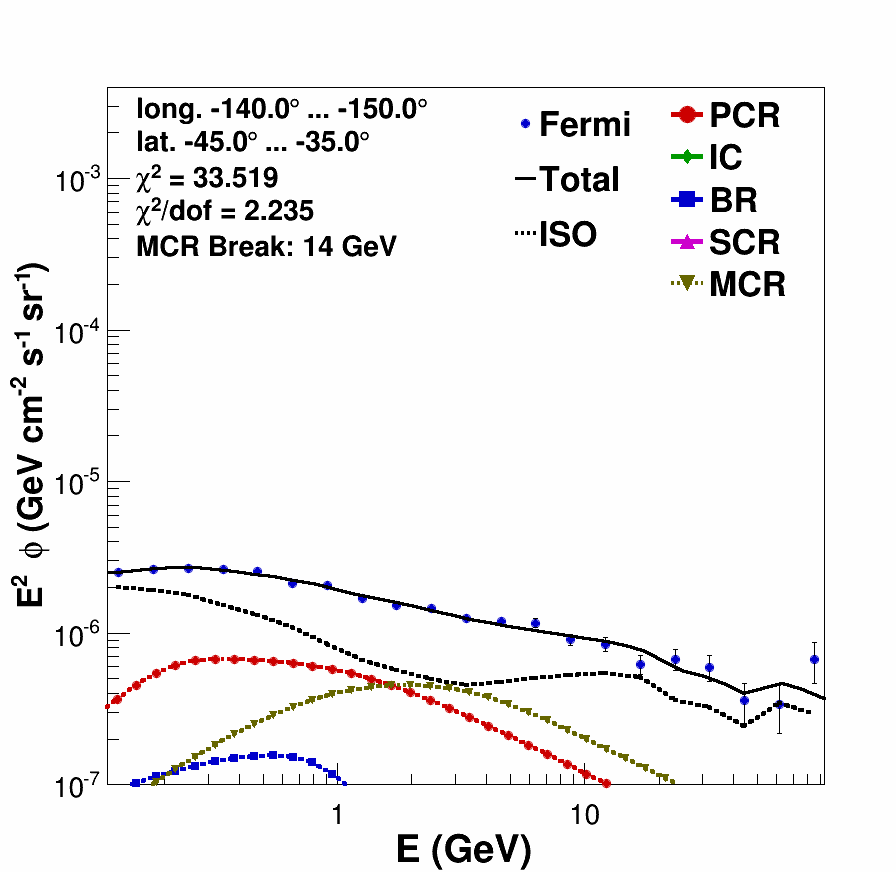}
\includegraphics[width=0.16\textwidth,height=0.16\textwidth,clip]{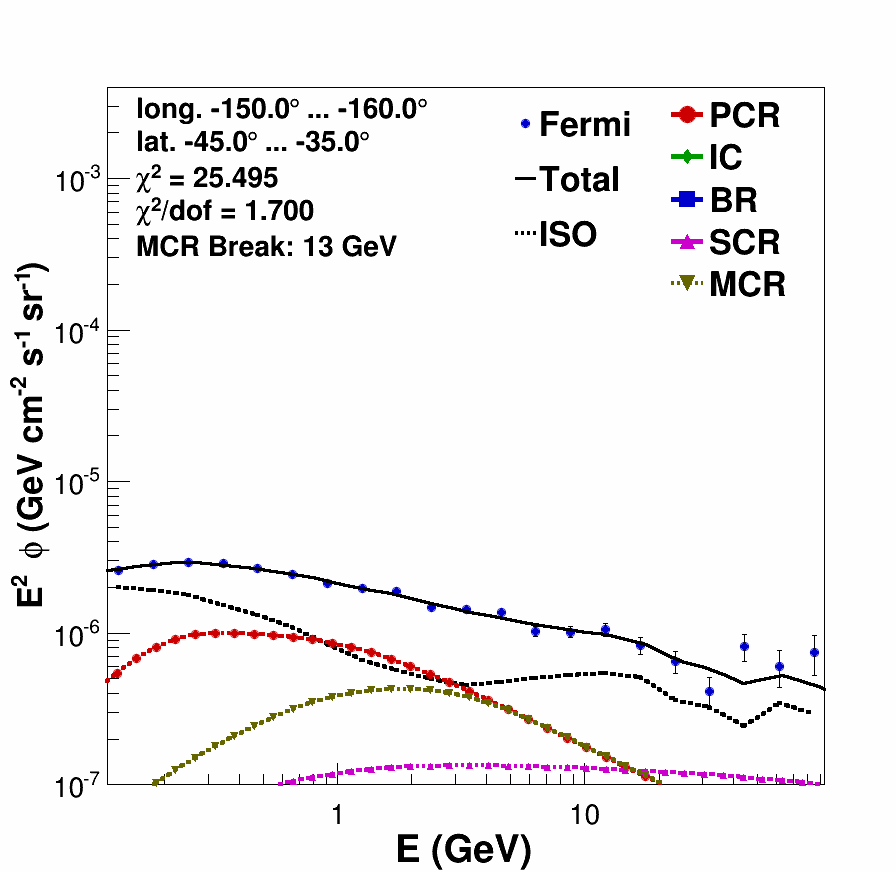}
\includegraphics[width=0.16\textwidth,height=0.16\textwidth,clip]{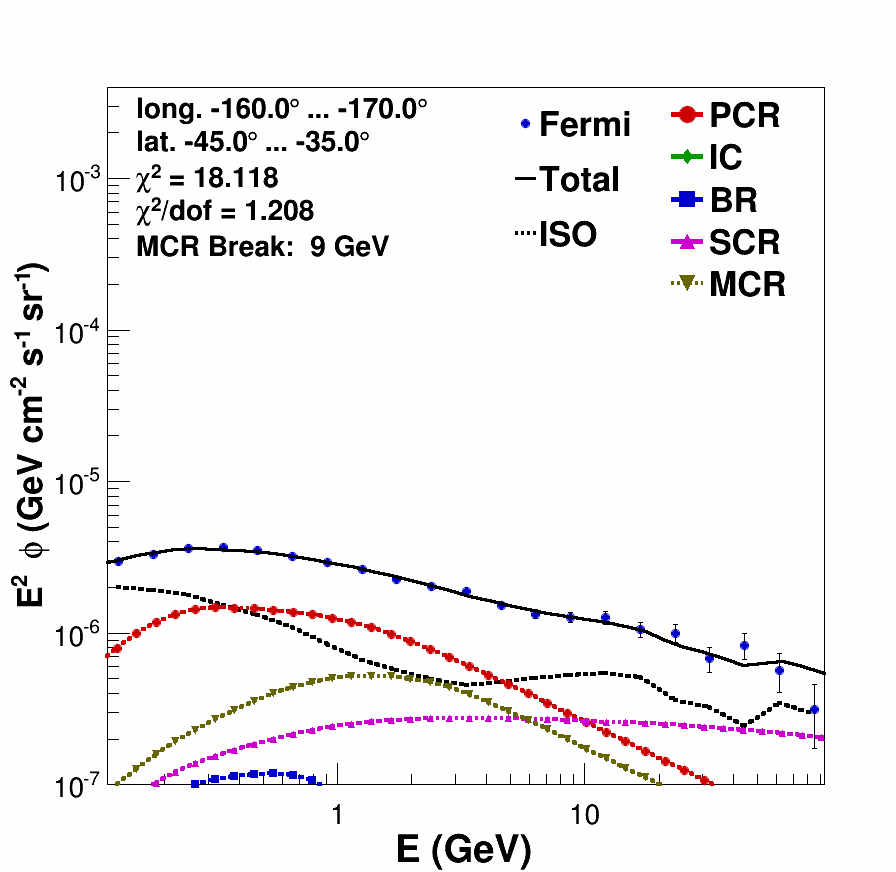}
\includegraphics[width=0.16\textwidth,height=0.16\textwidth,clip]{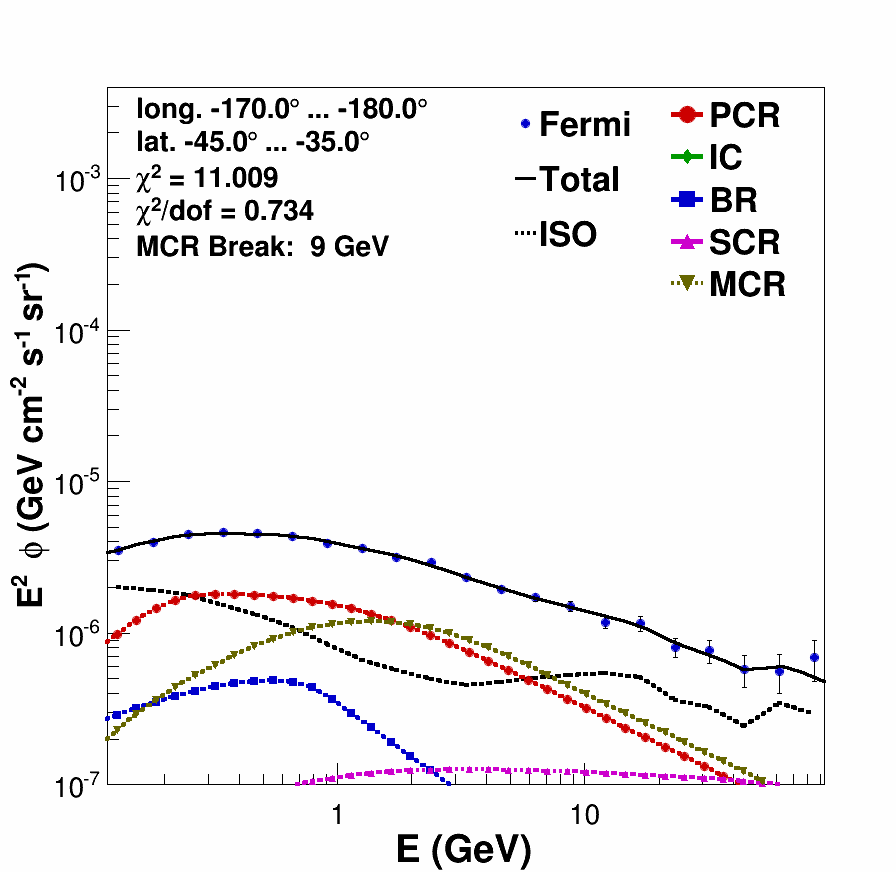}%%%%r16
\caption[]{Template fits for latitudes  with $-45.0^\circ<b<-35.0^\circ$ and longitudes decreasing from 180$^\circ$ to -180$^\circ$.} \label{F28}
\end{figure}
\begin{figure}
\centering
\includegraphics[width=0.16\textwidth,height=0.16\textwidth,clip]{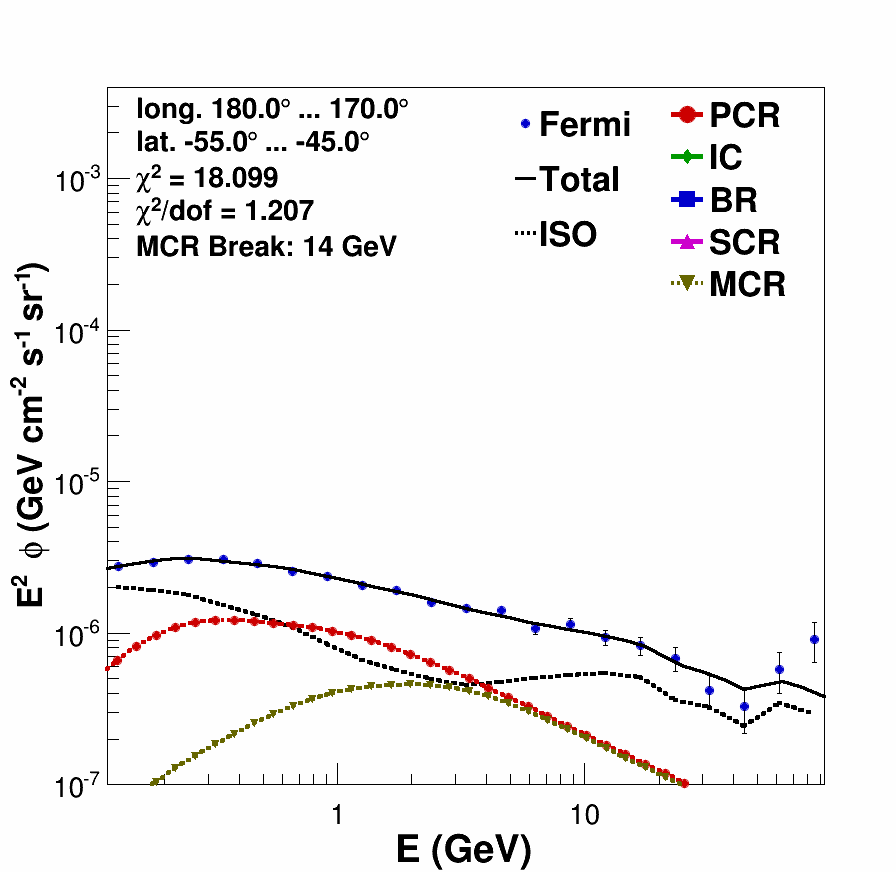}
\includegraphics[width=0.16\textwidth,height=0.16\textwidth,clip]{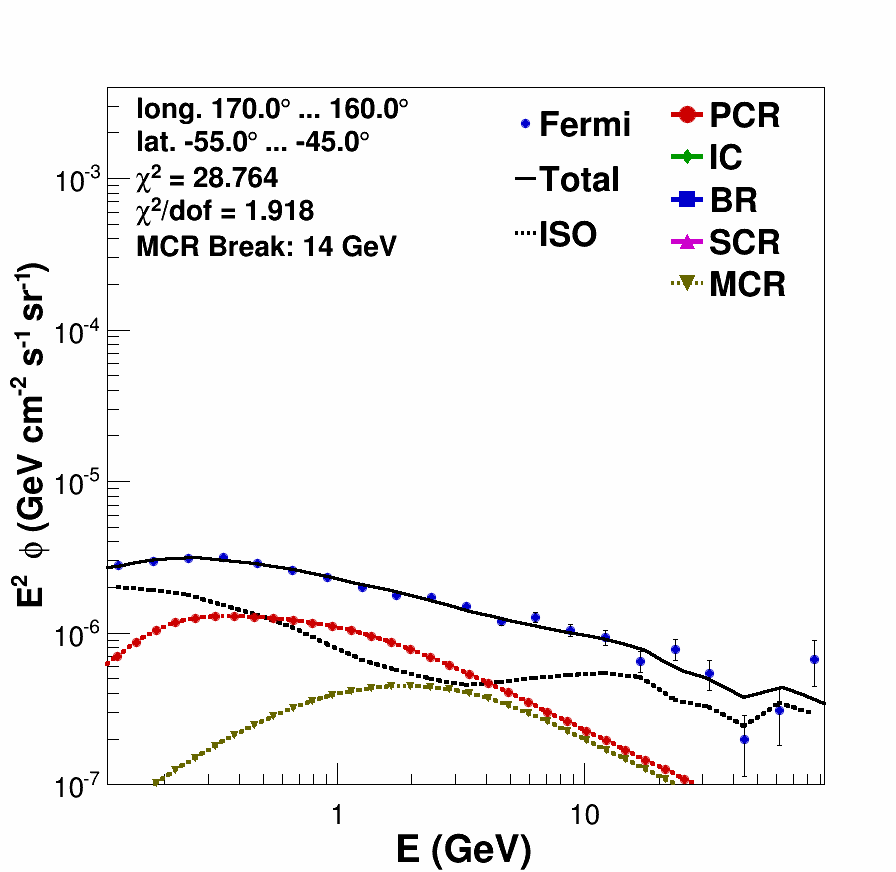}
\includegraphics[width=0.16\textwidth,height=0.16\textwidth,clip]{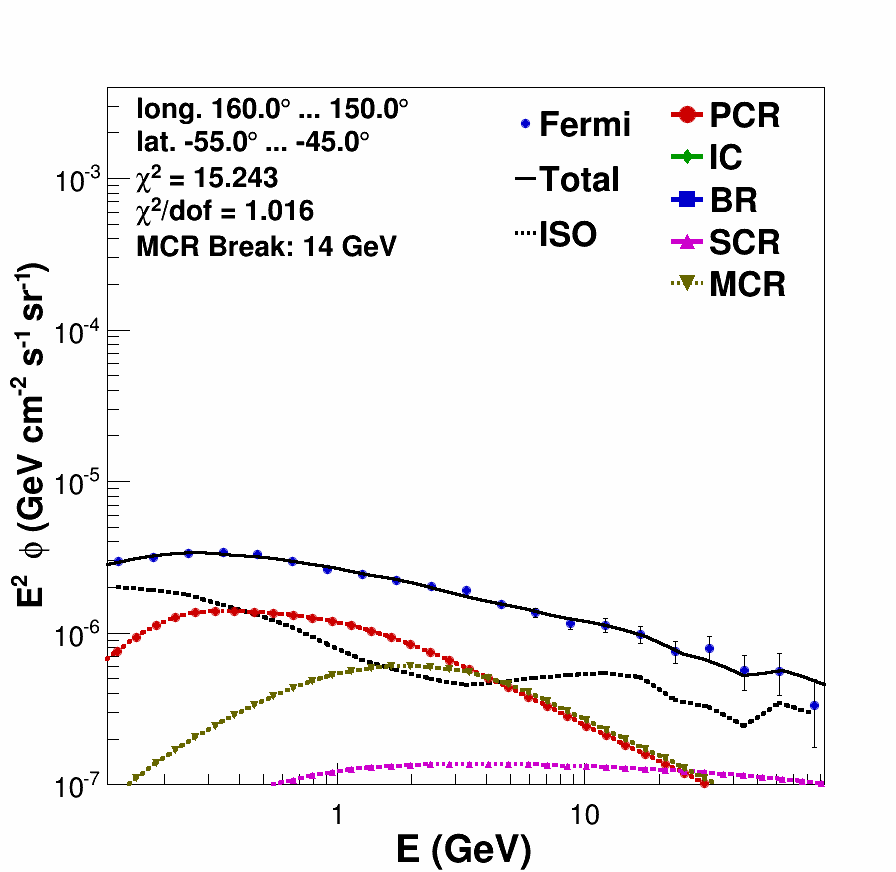}
\includegraphics[width=0.16\textwidth,height=0.16\textwidth,clip]{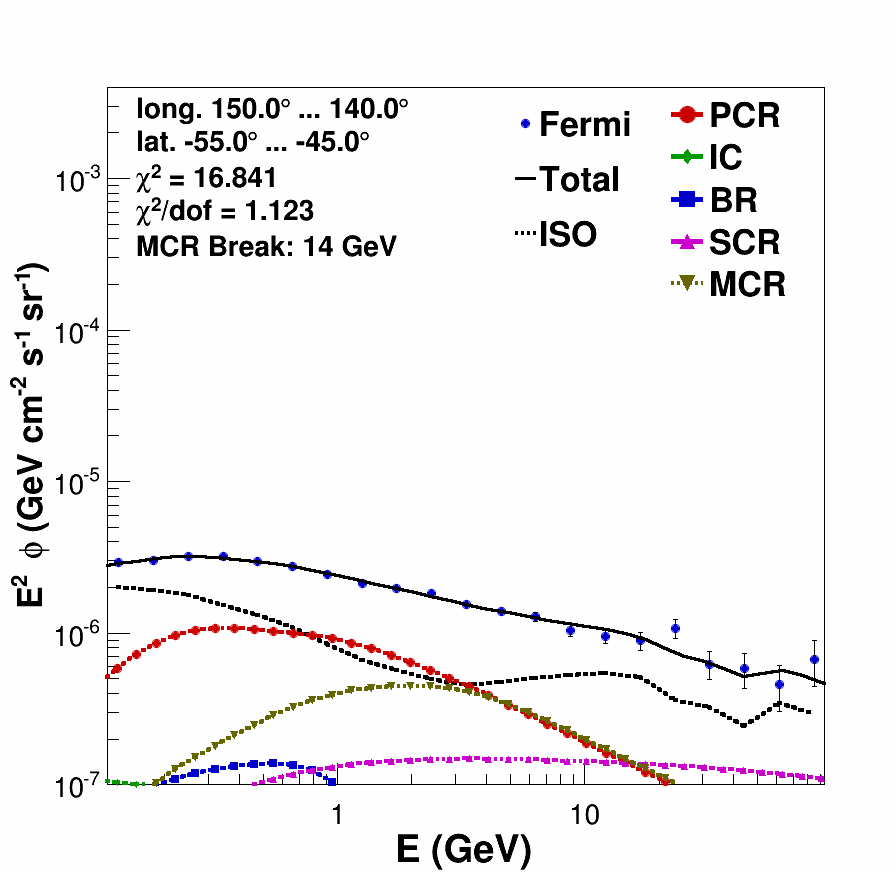}
\includegraphics[width=0.16\textwidth,height=0.16\textwidth,clip]{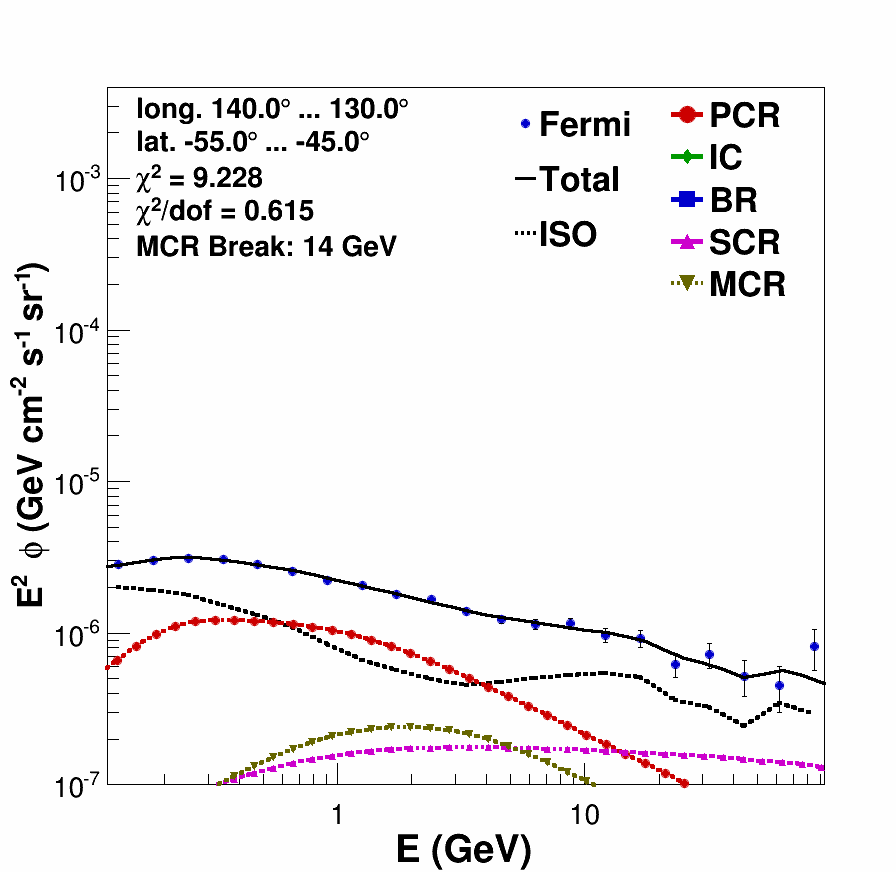}
\includegraphics[width=0.16\textwidth,height=0.16\textwidth,clip]{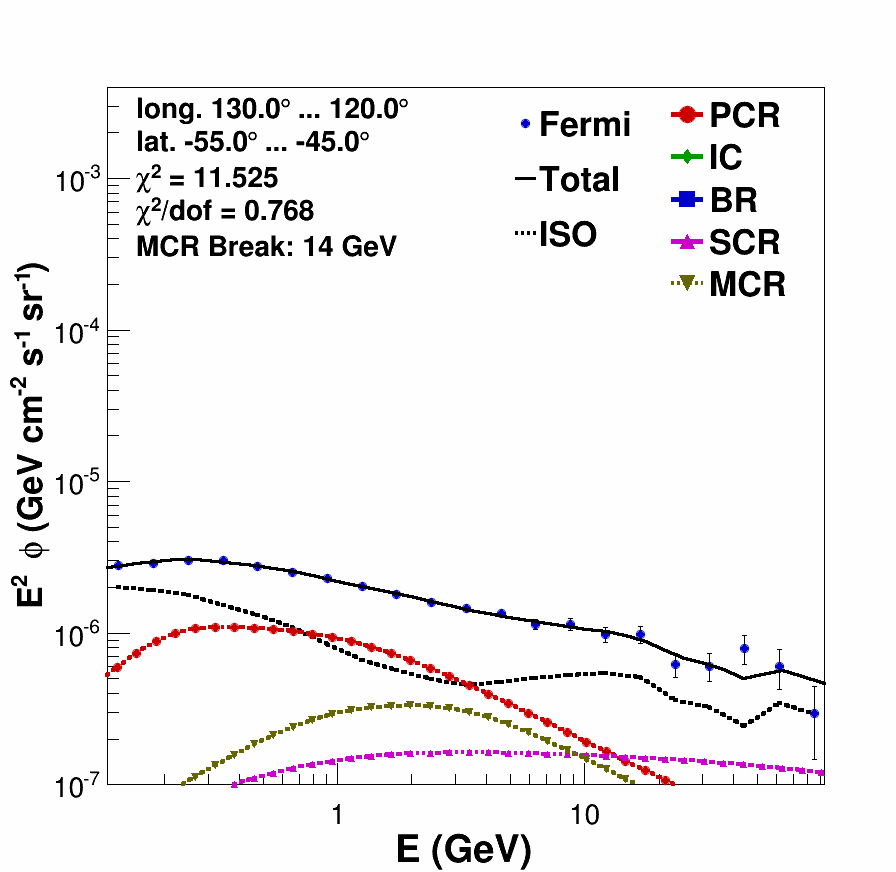}
\includegraphics[width=0.16\textwidth,height=0.16\textwidth,clip]{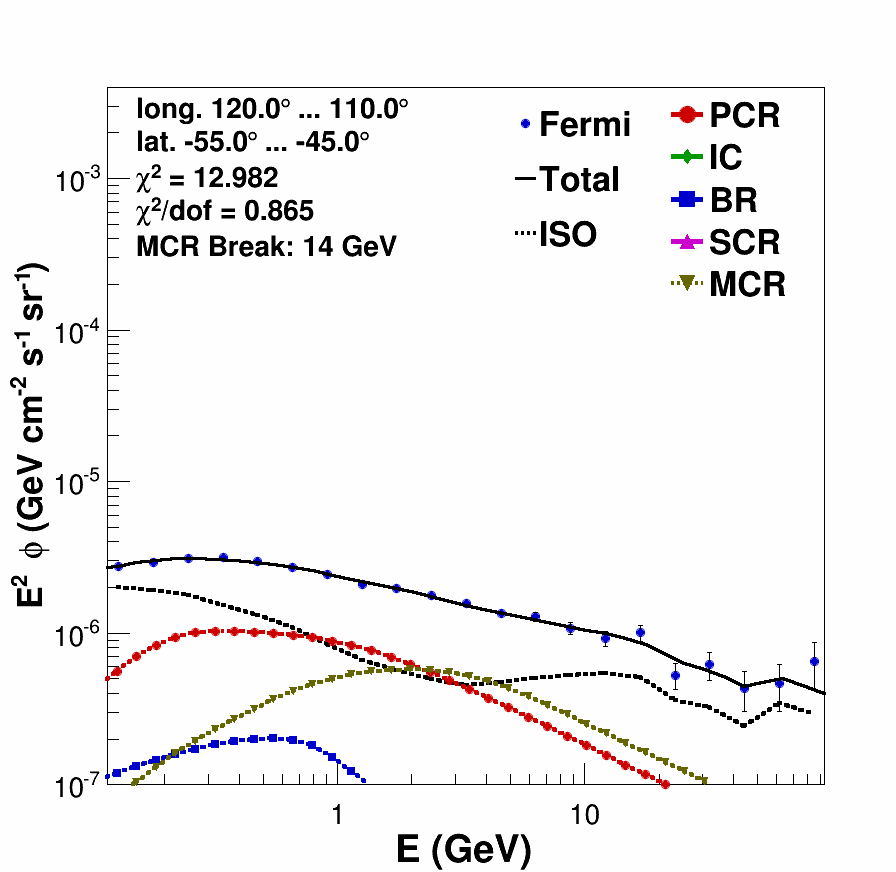}
\includegraphics[width=0.16\textwidth,height=0.16\textwidth,clip]{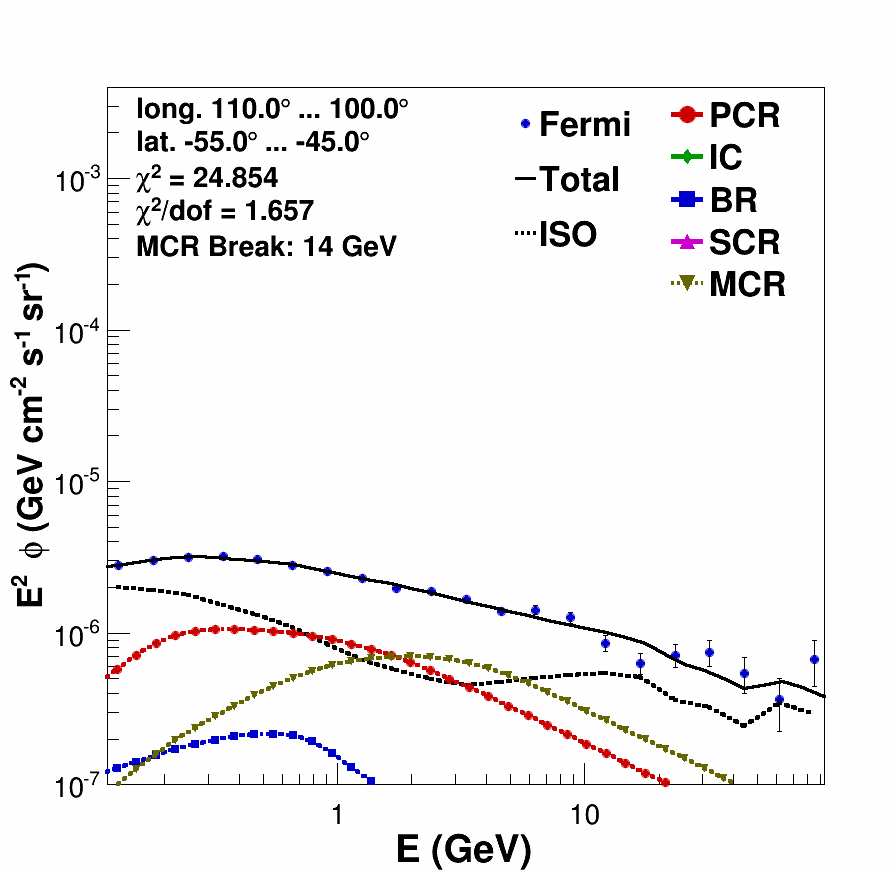}
\includegraphics[width=0.16\textwidth,height=0.16\textwidth,clip]{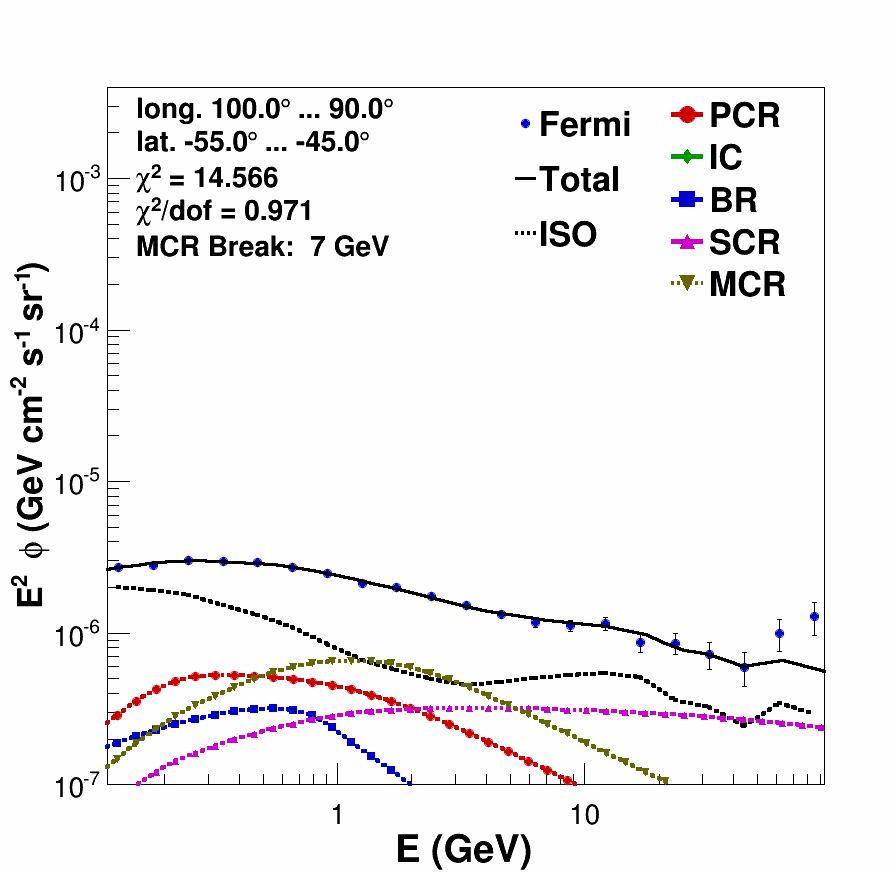}
\includegraphics[width=0.16\textwidth,height=0.16\textwidth,clip]{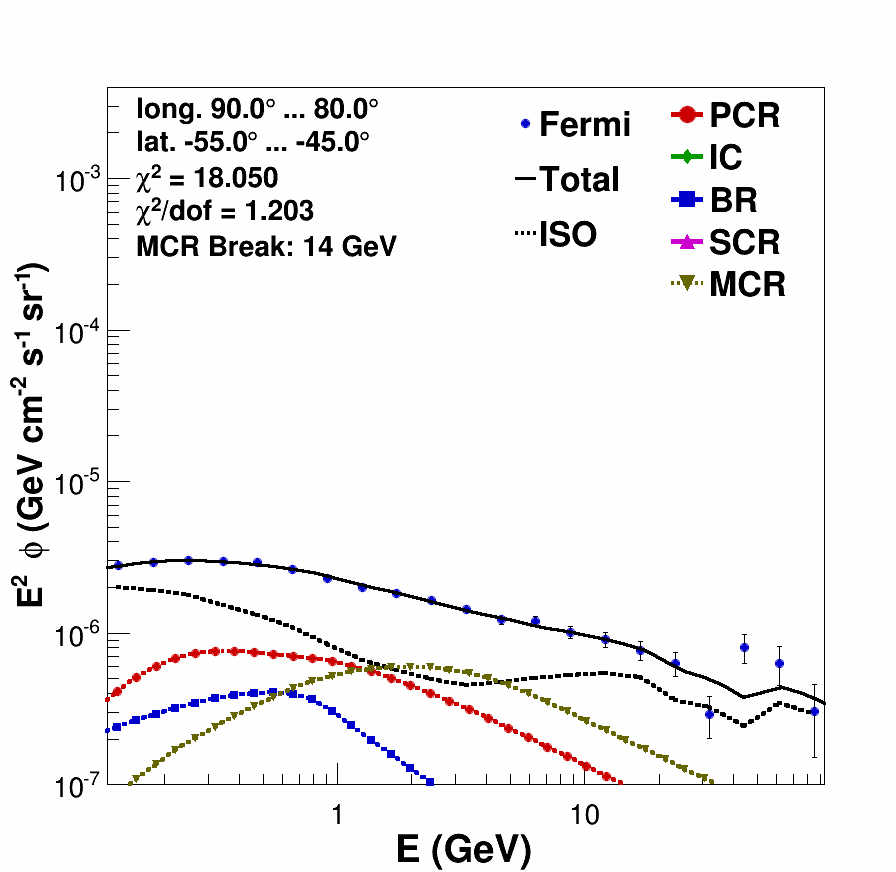}
\includegraphics[width=0.16\textwidth,height=0.16\textwidth,clip]{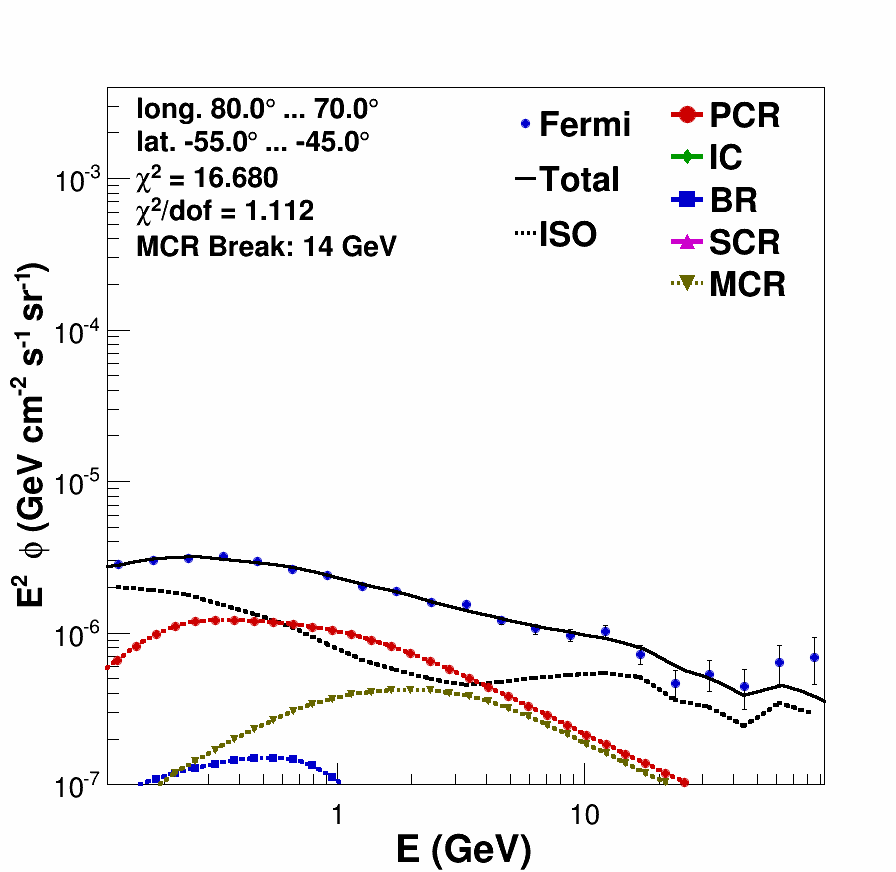}
\includegraphics[width=0.16\textwidth,height=0.16\textwidth,clip]{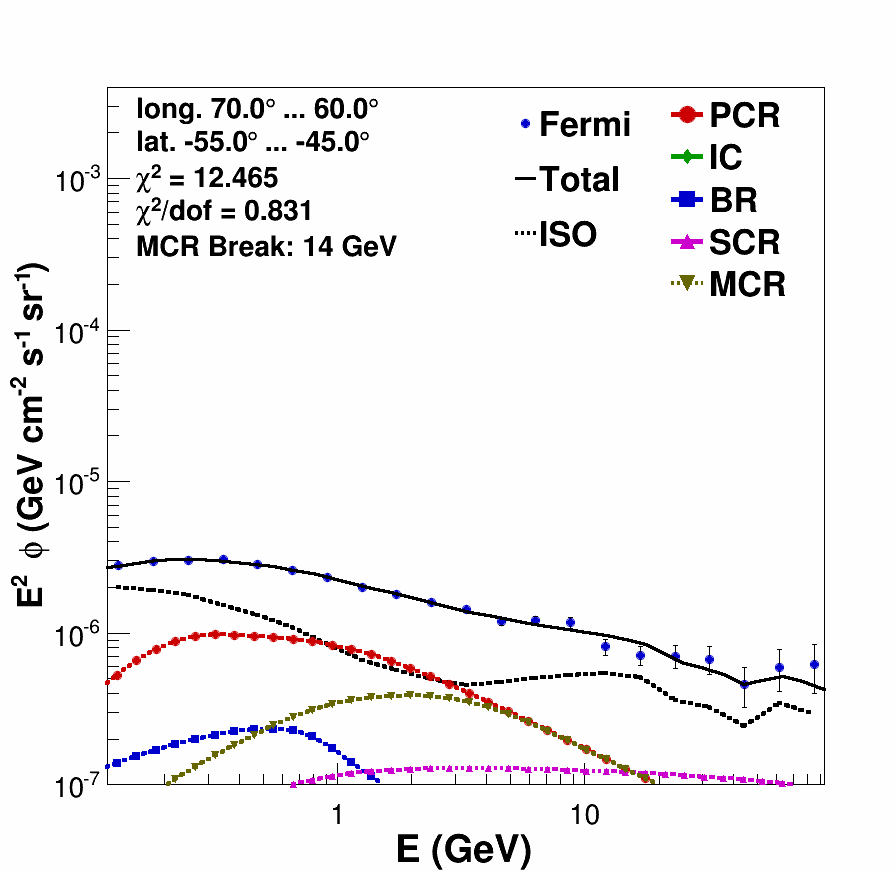}
\includegraphics[width=0.16\textwidth,height=0.16\textwidth,clip]{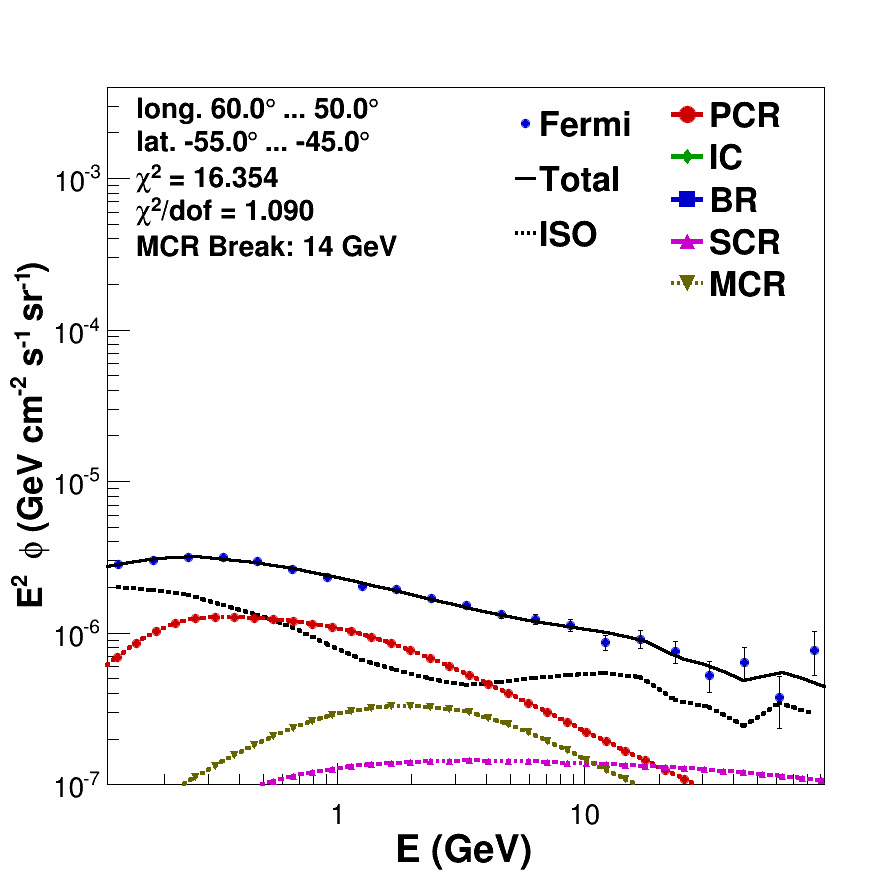}
\includegraphics[width=0.16\textwidth,height=0.16\textwidth,clip]{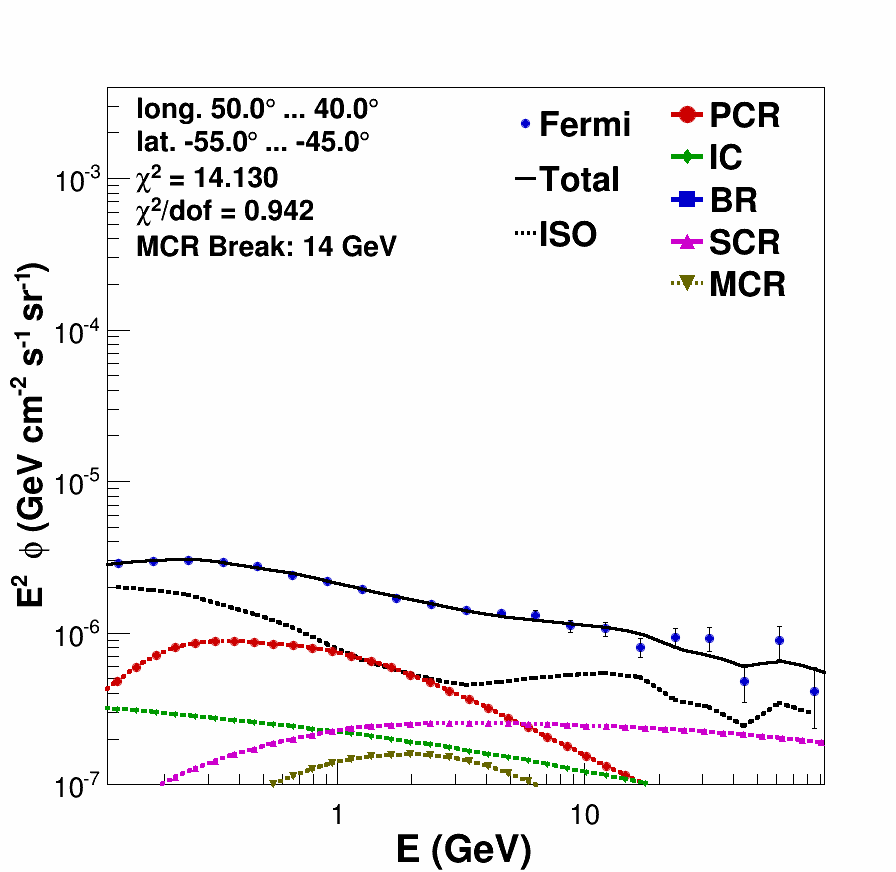}
\includegraphics[width=0.16\textwidth,height=0.16\textwidth,clip]{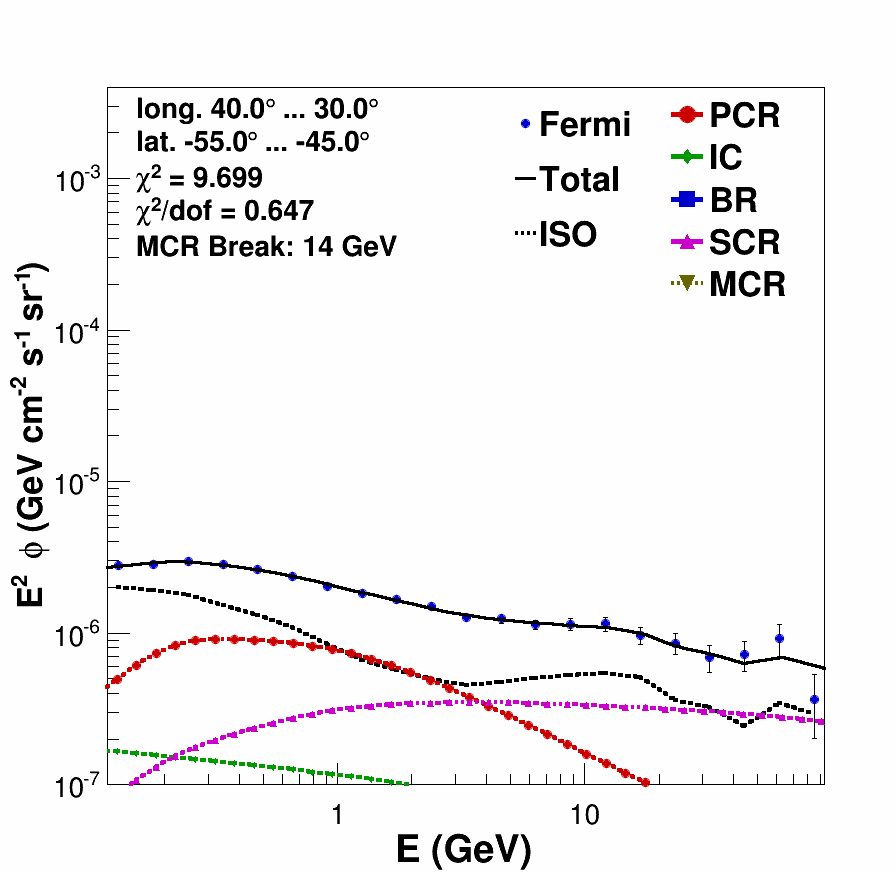}
\includegraphics[width=0.16\textwidth,height=0.16\textwidth,clip]{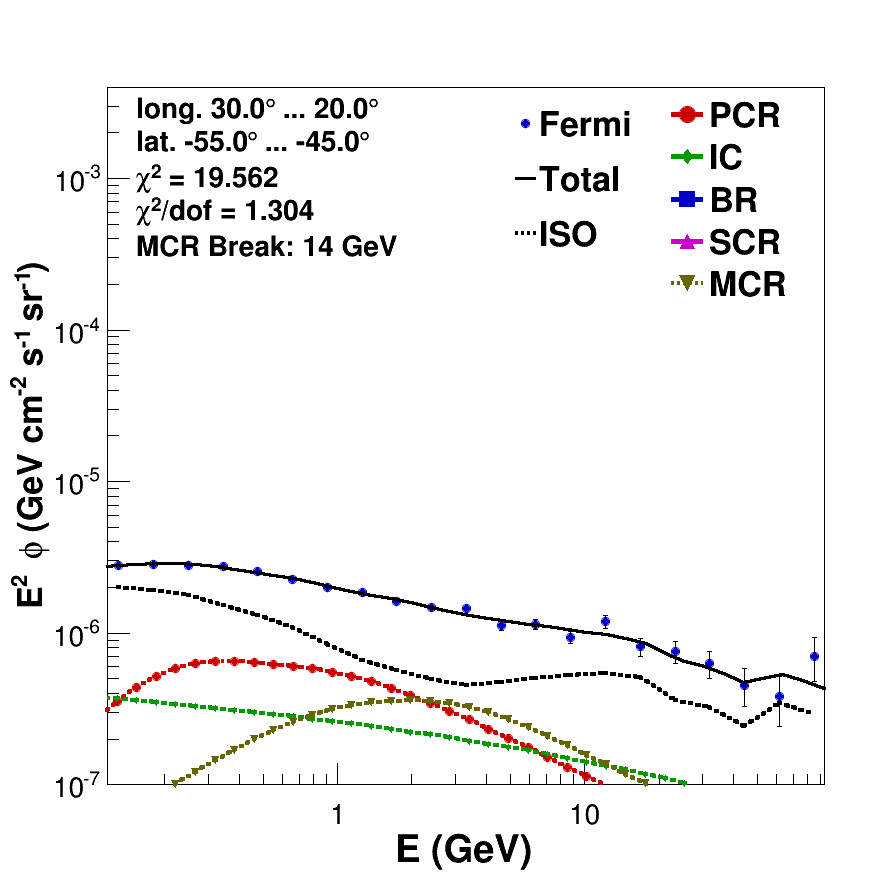}
\includegraphics[width=0.16\textwidth,height=0.16\textwidth,clip]{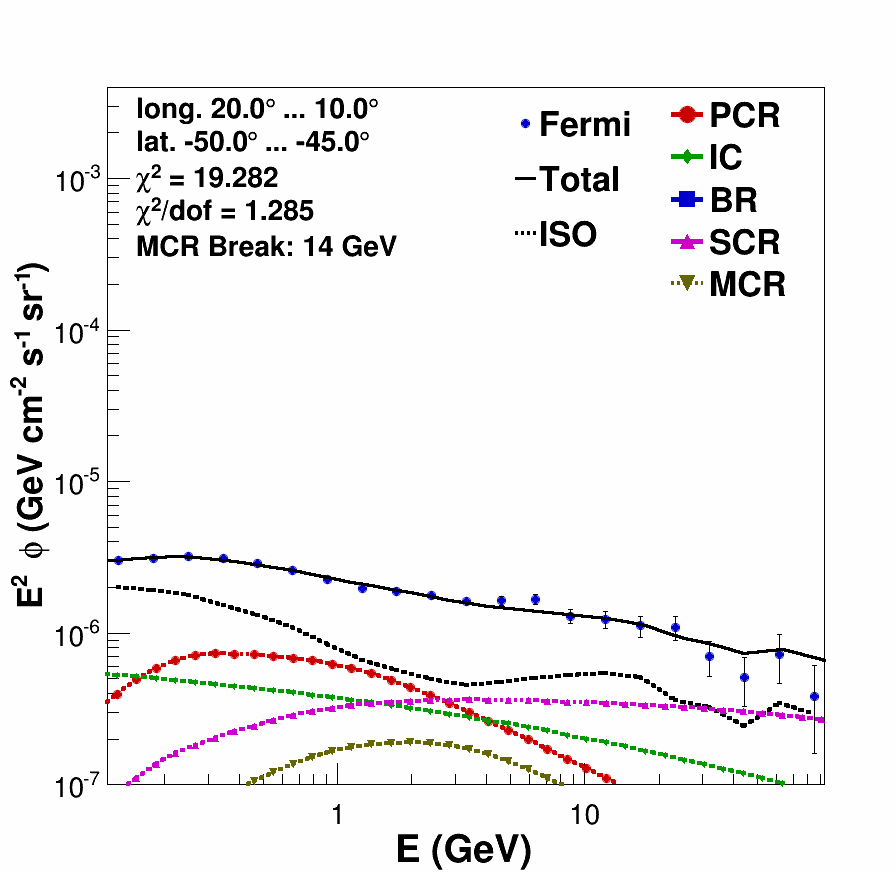}
\includegraphics[width=0.16\textwidth,height=0.16\textwidth,clip]{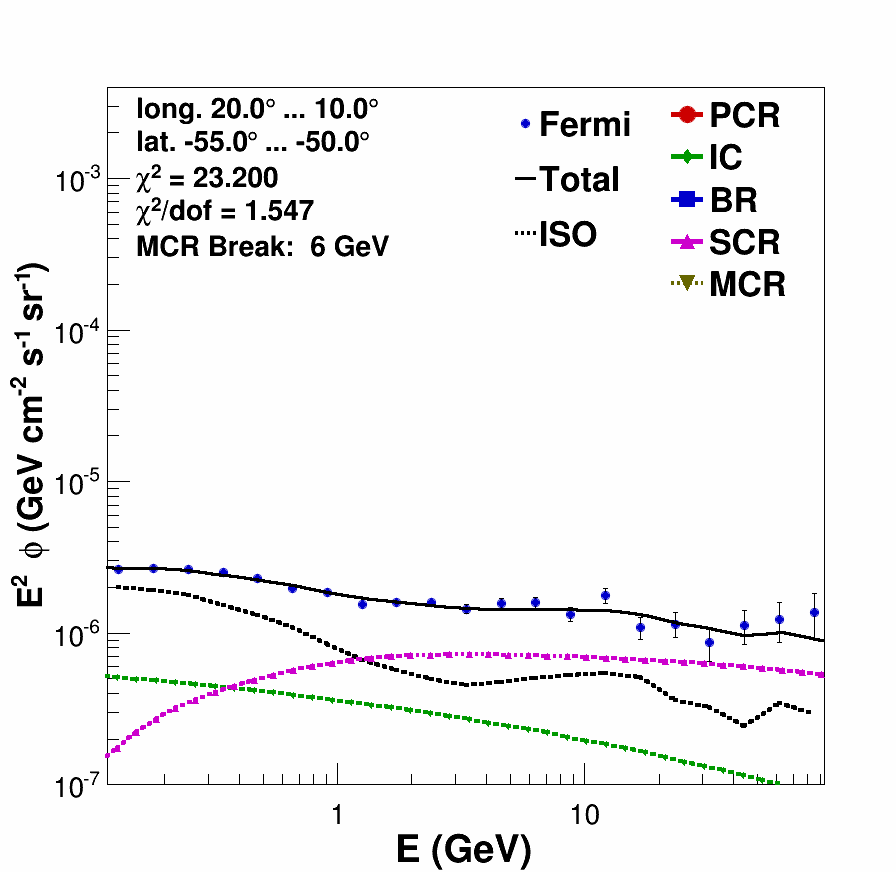}
\includegraphics[width=0.16\textwidth,height=0.16\textwidth,clip]{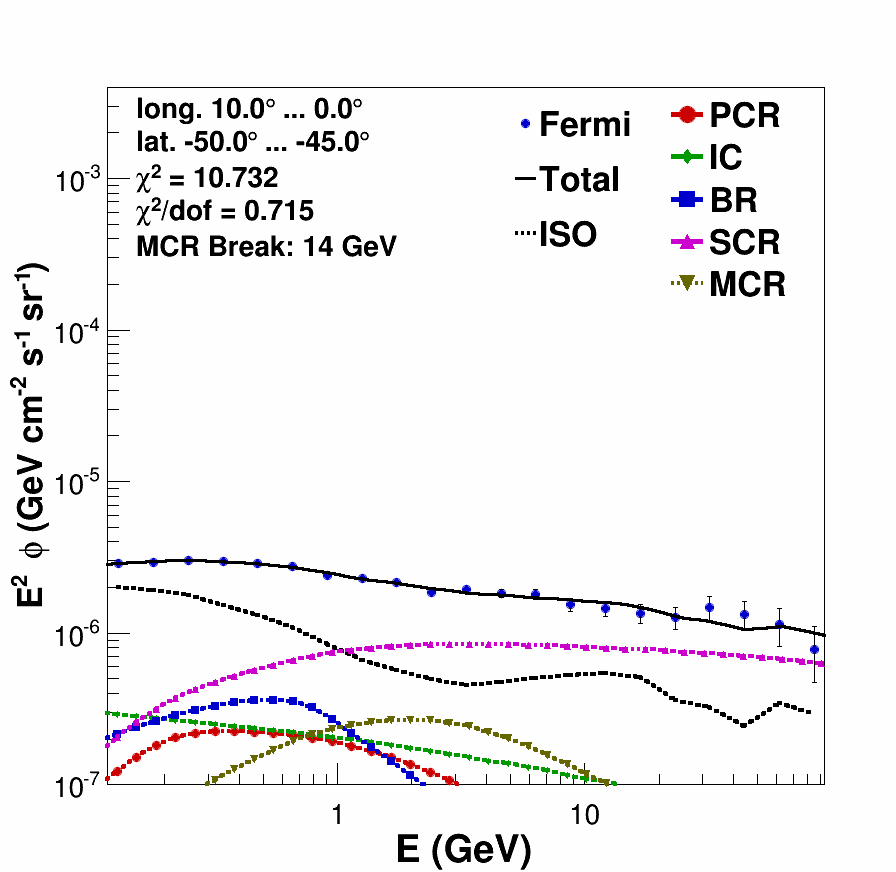}
\includegraphics[width=0.16\textwidth,height=0.16\textwidth,clip]{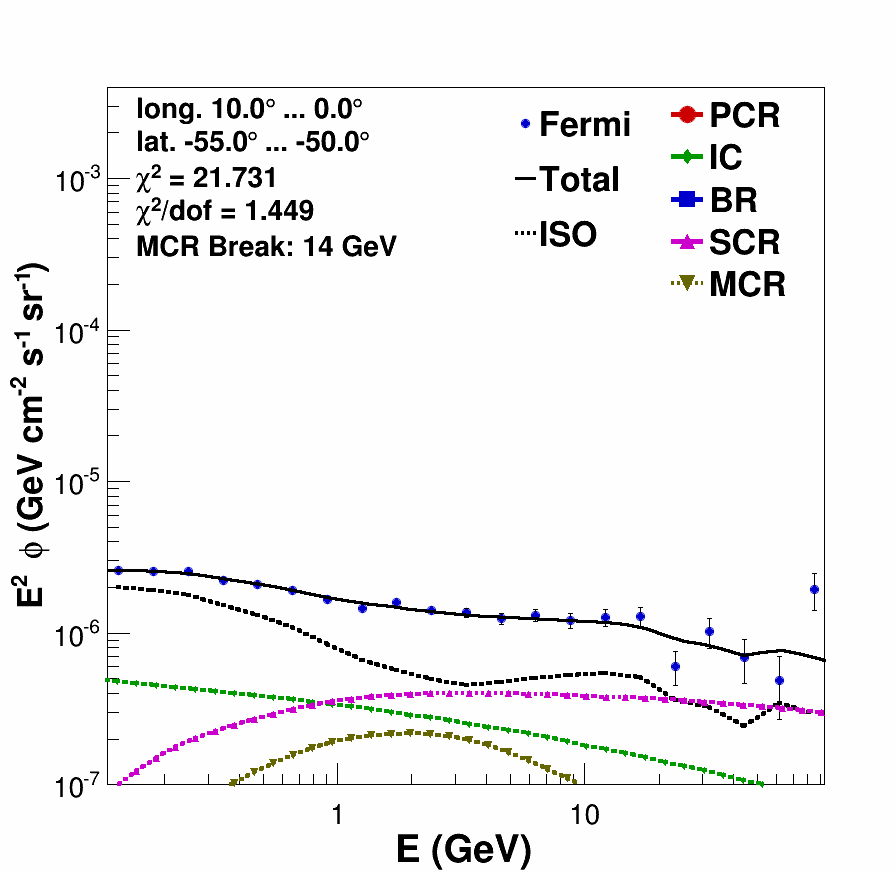}
\includegraphics[width=0.16\textwidth,height=0.16\textwidth,clip]{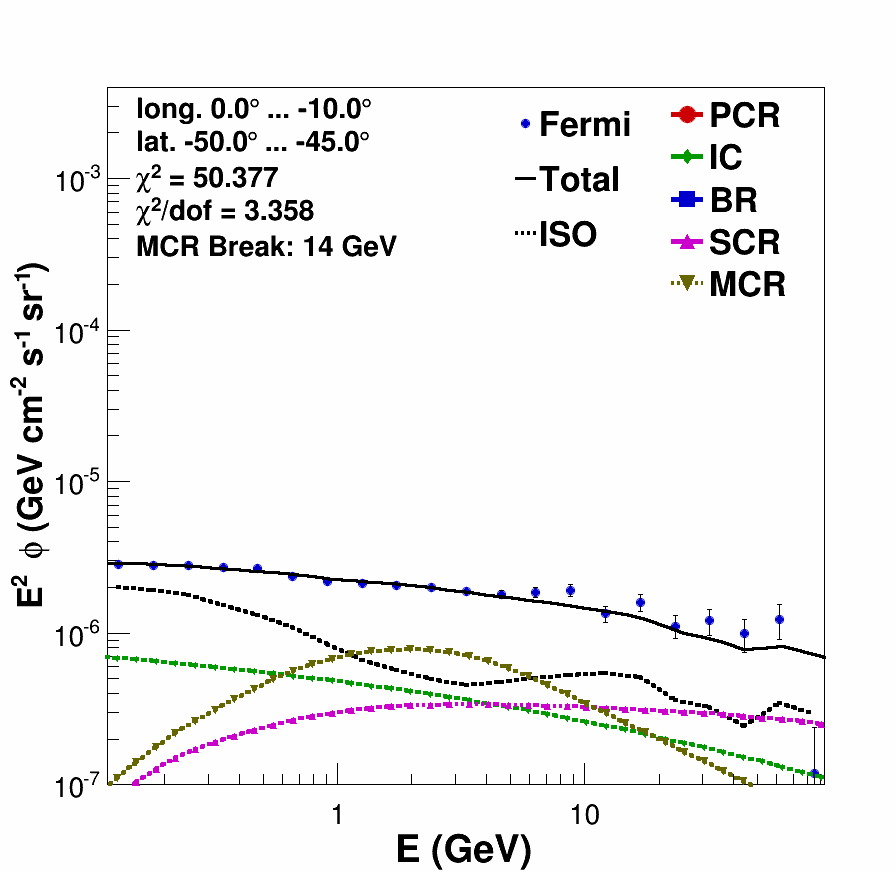}
\includegraphics[width=0.16\textwidth,height=0.16\textwidth,clip]{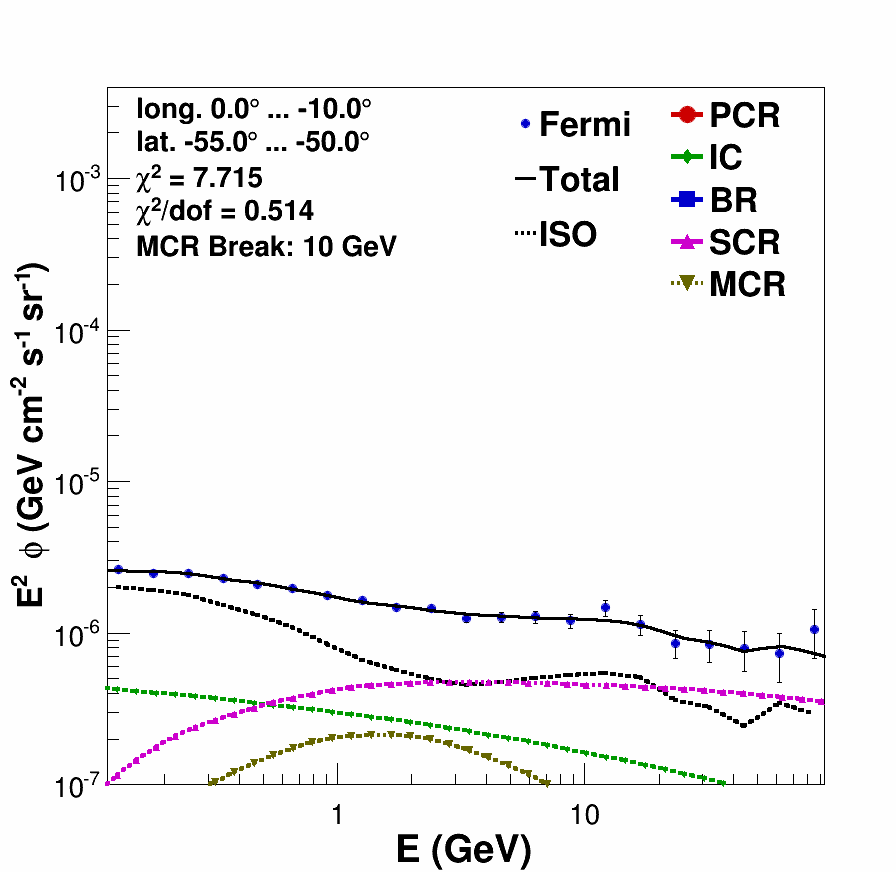}
\includegraphics[width=0.16\textwidth,height=0.16\textwidth,clip]{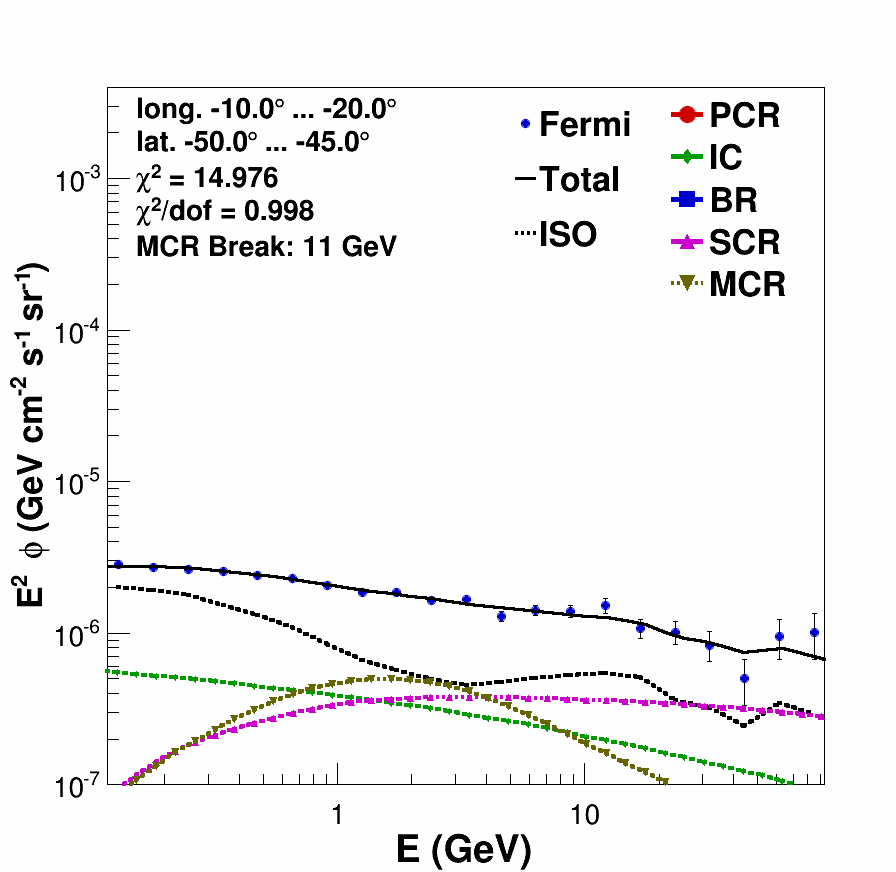}
\includegraphics[width=0.16\textwidth,height=0.16\textwidth,clip]{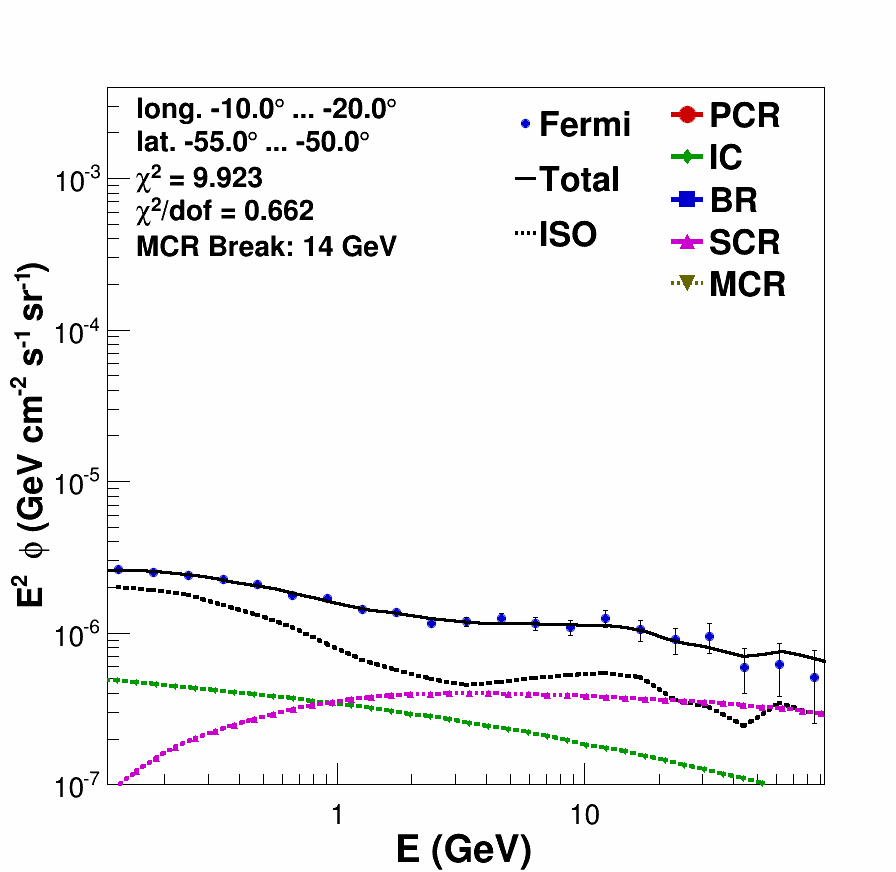}
\includegraphics[width=0.16\textwidth,height=0.16\textwidth,clip]{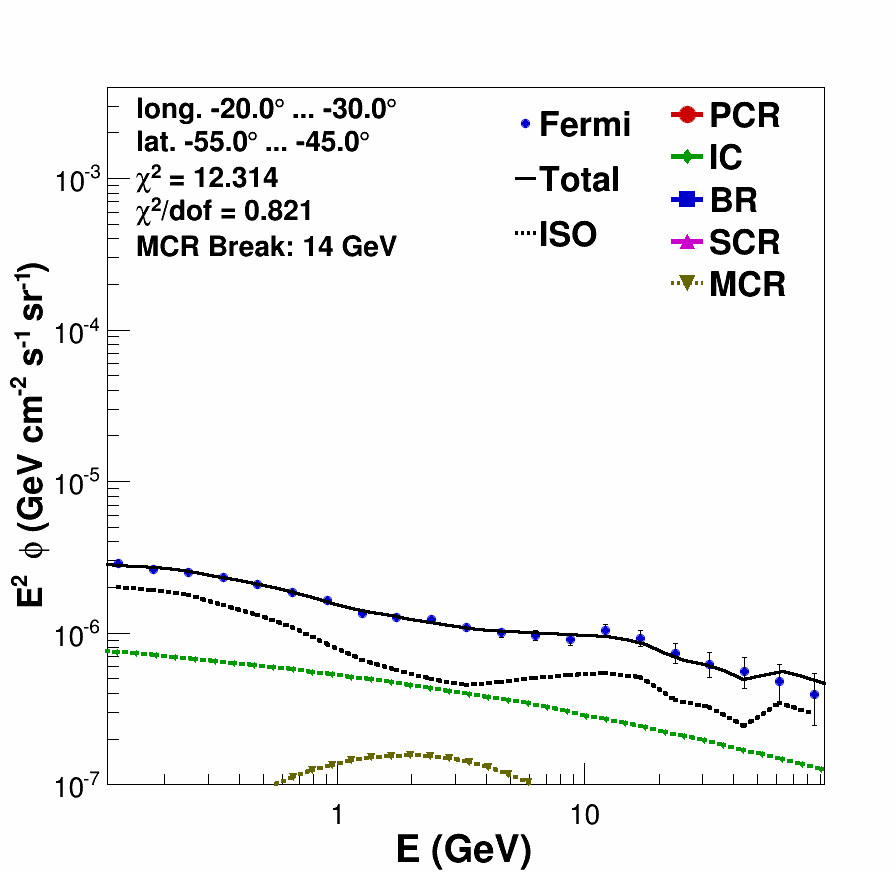}
\includegraphics[width=0.16\textwidth,height=0.16\textwidth,clip]{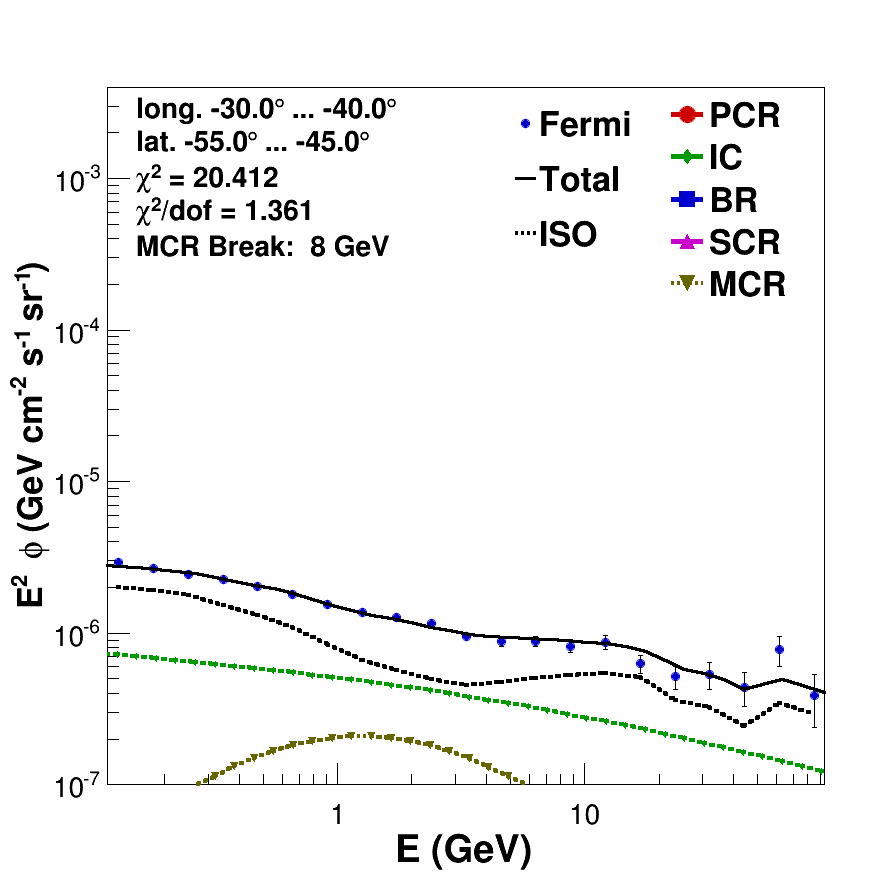}
\includegraphics[width=0.16\textwidth,height=0.16\textwidth,clip]{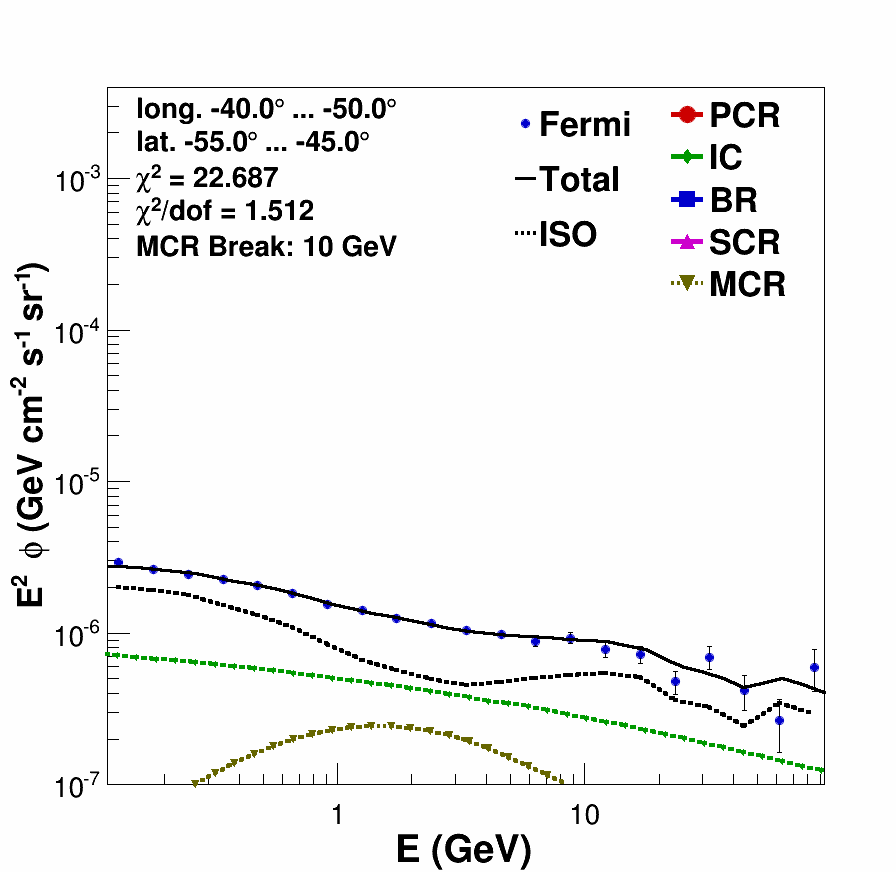}
\includegraphics[width=0.16\textwidth,height=0.16\textwidth,clip]{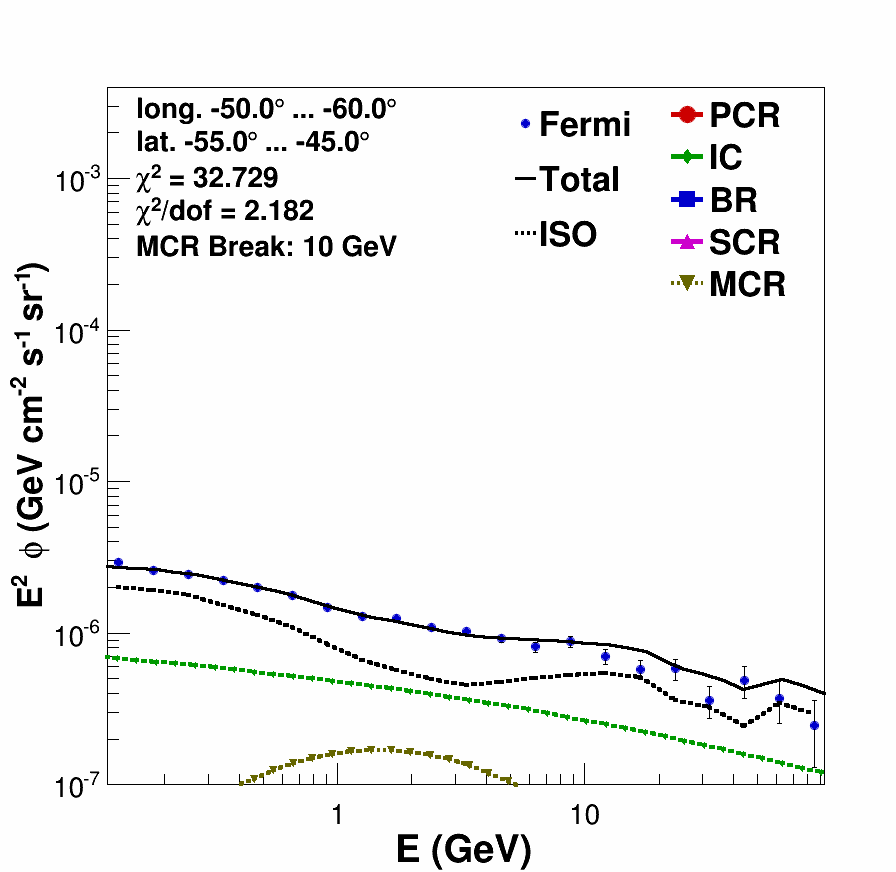}
\includegraphics[width=0.16\textwidth,height=0.16\textwidth,clip]{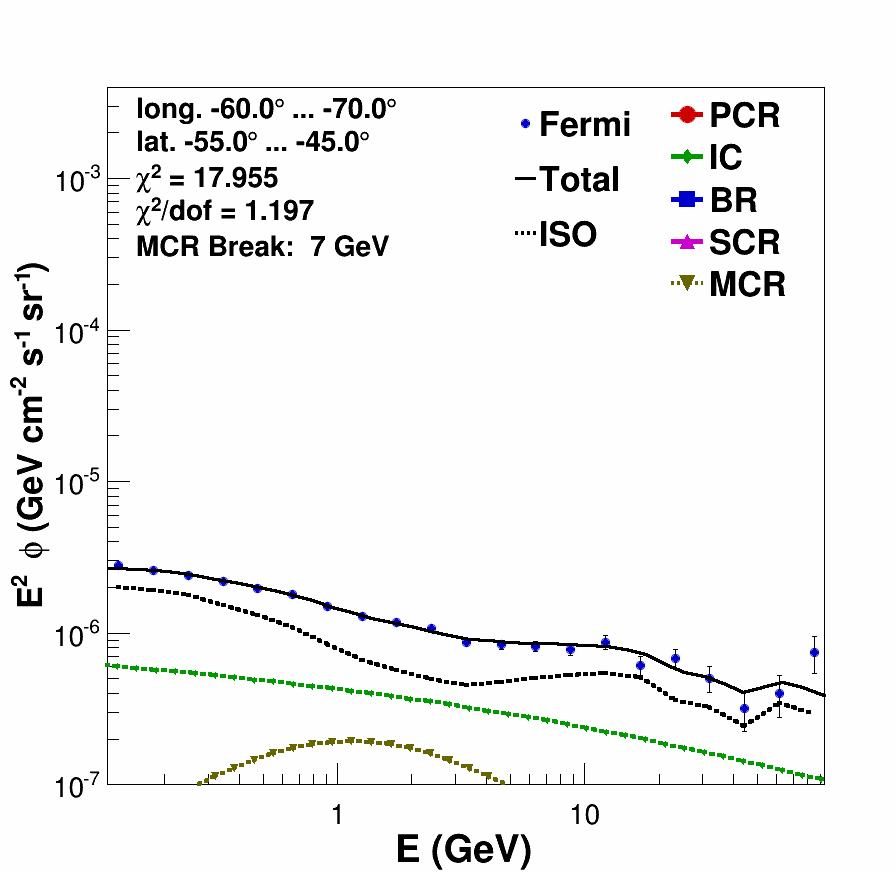}
\includegraphics[width=0.16\textwidth,height=0.16\textwidth,clip]{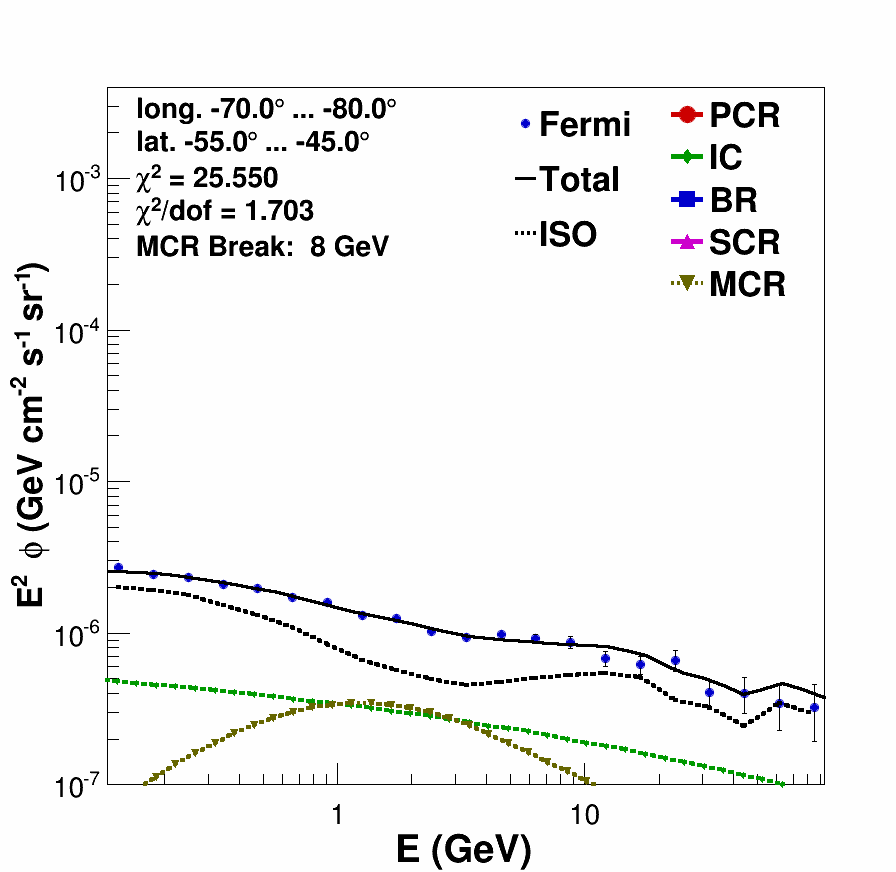}
\includegraphics[width=0.16\textwidth,height=0.16\textwidth,clip]{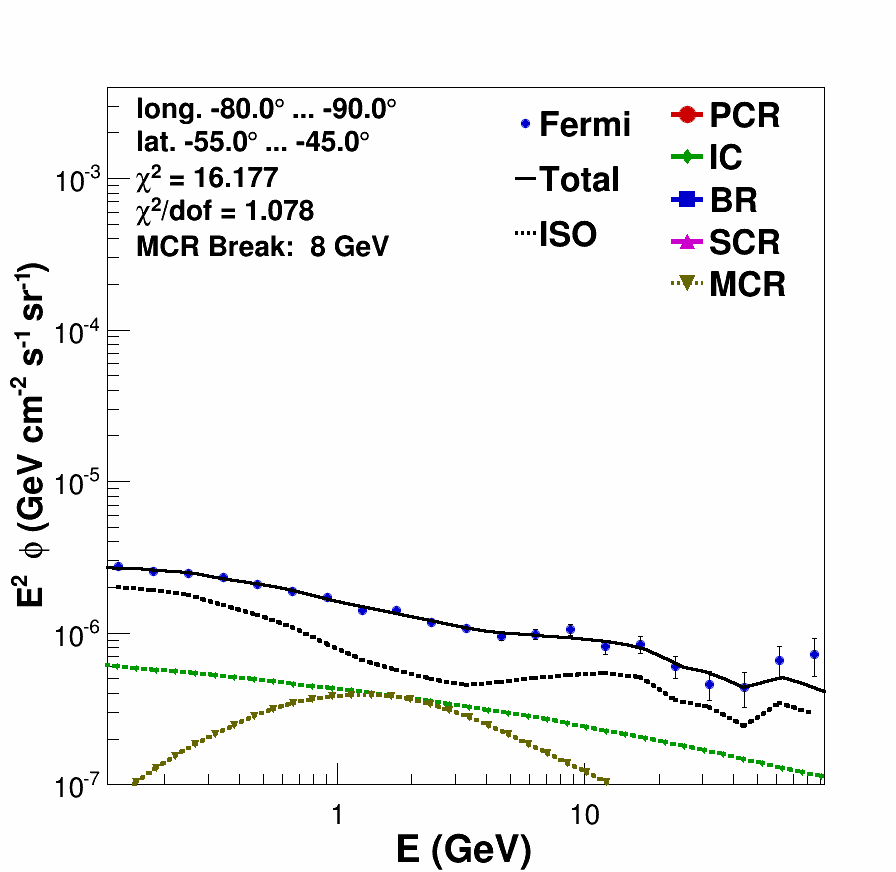}
\includegraphics[width=0.16\textwidth,height=0.16\textwidth,clip]{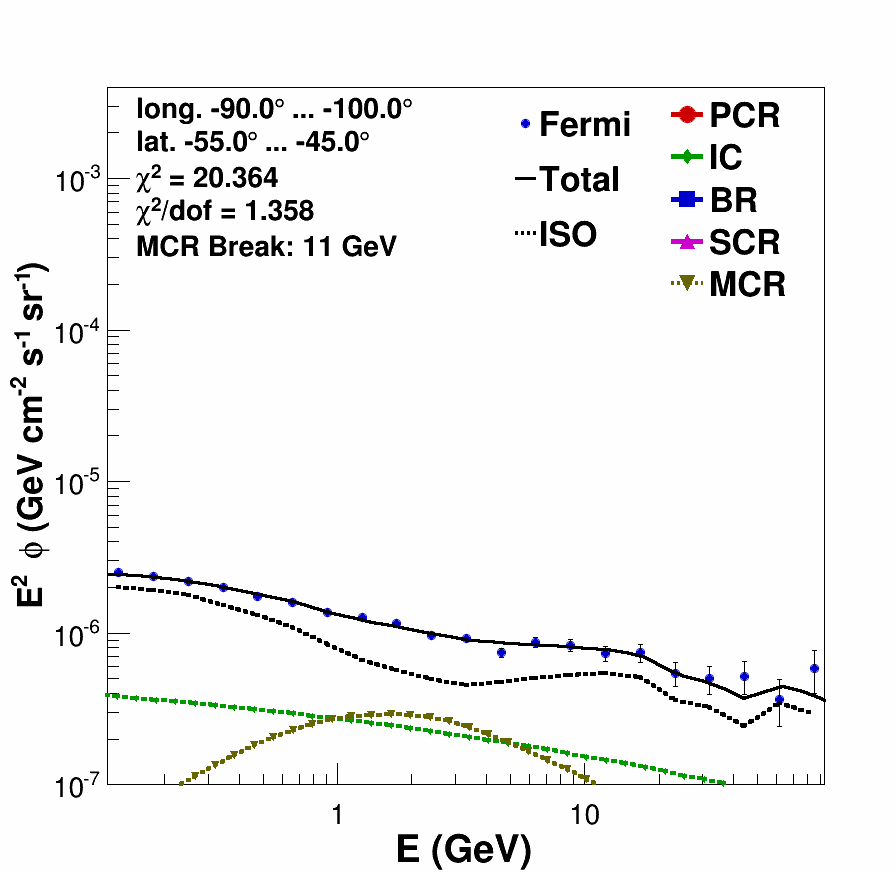}
\includegraphics[width=0.16\textwidth,height=0.16\textwidth,clip]{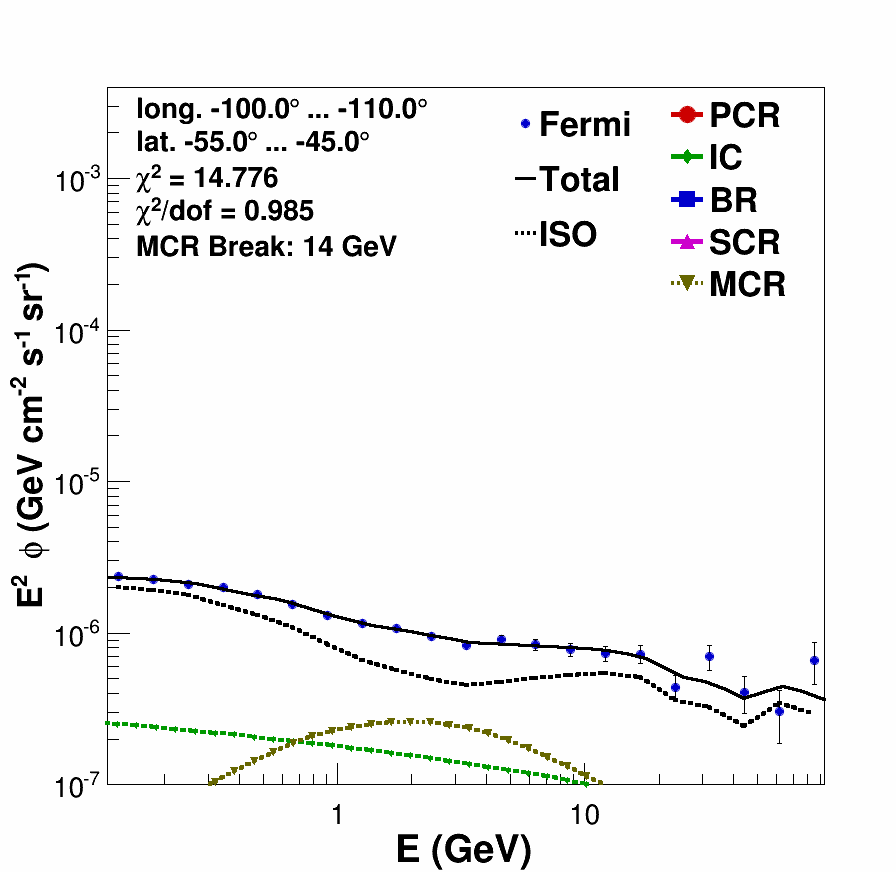}
\includegraphics[width=0.16\textwidth,height=0.16\textwidth,clip]{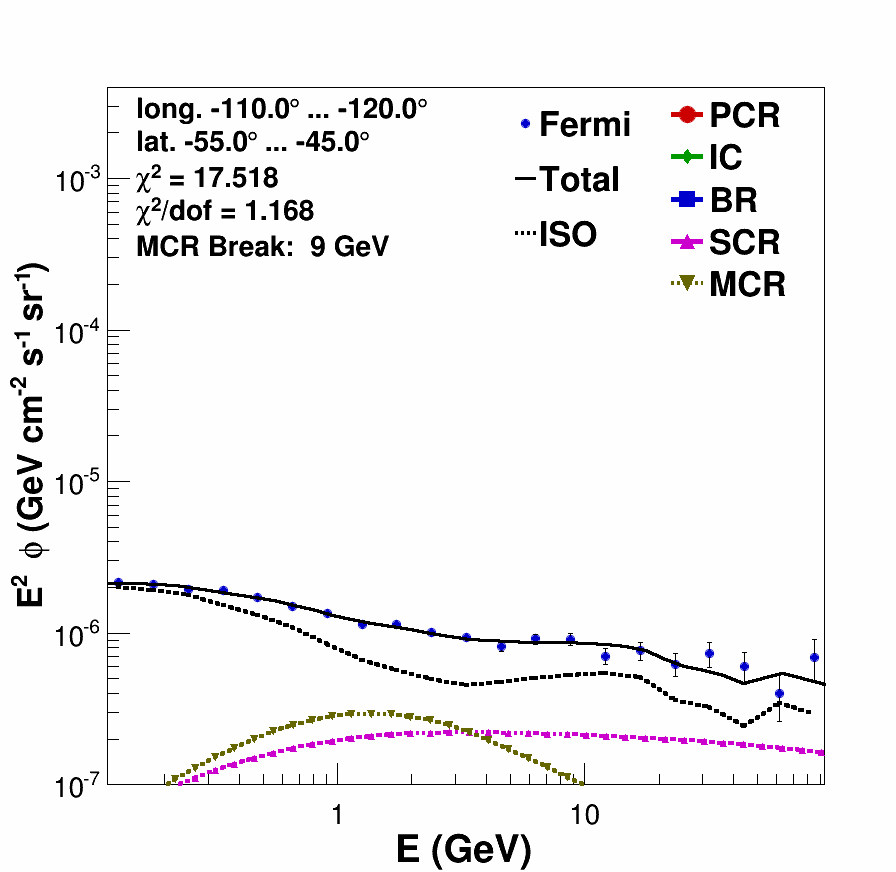}
\includegraphics[width=0.16\textwidth,height=0.16\textwidth,clip]{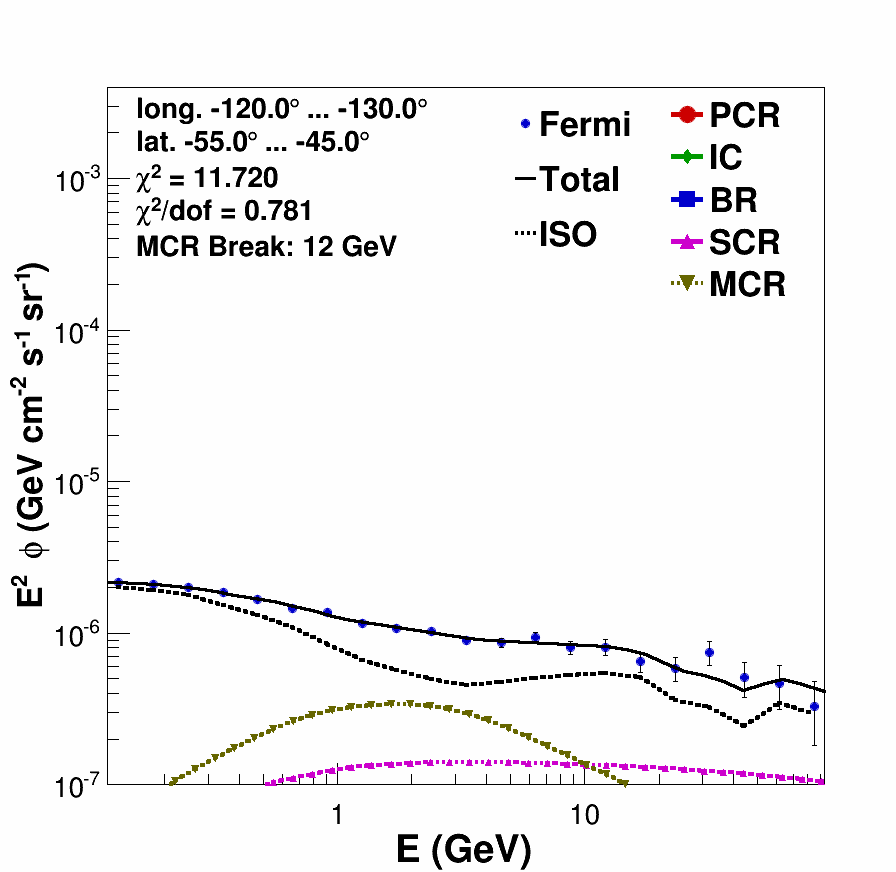}
\includegraphics[width=0.16\textwidth,height=0.16\textwidth,clip]{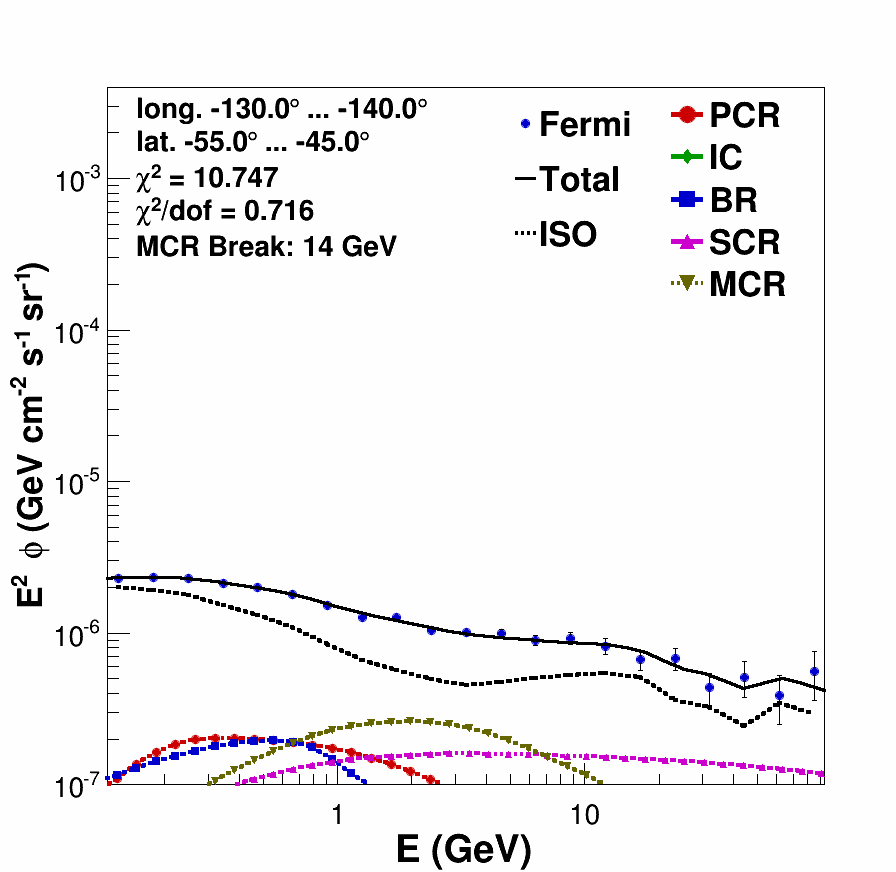}
\includegraphics[width=0.16\textwidth,height=0.16\textwidth,clip]{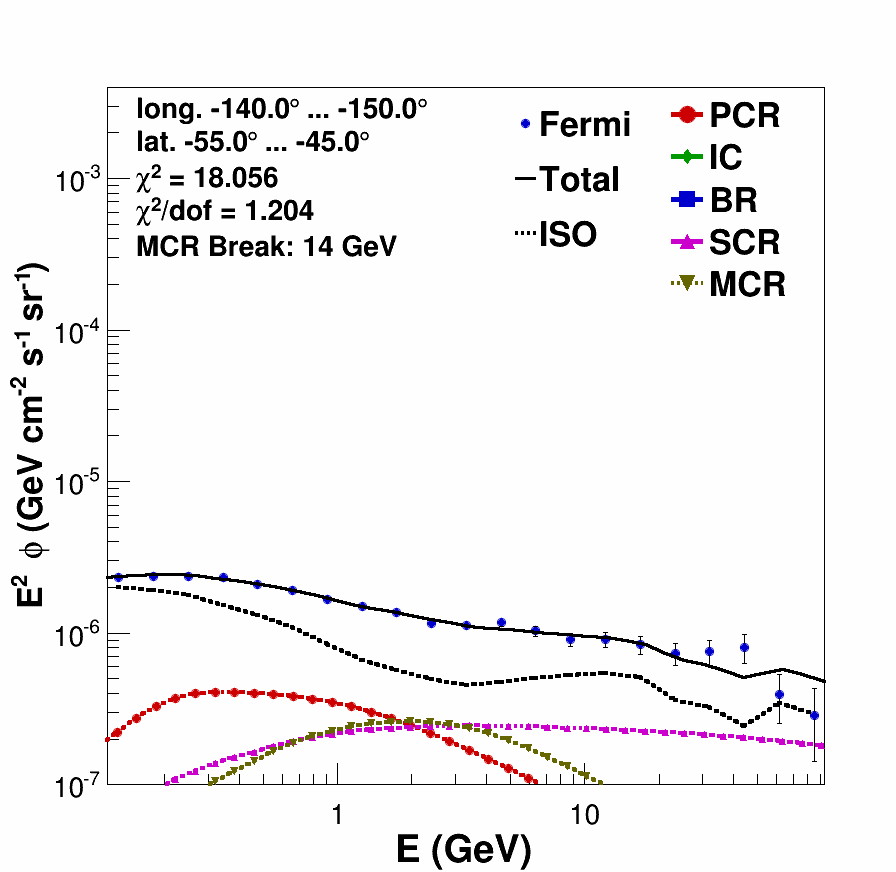}
\includegraphics[width=0.16\textwidth,height=0.16\textwidth,clip]{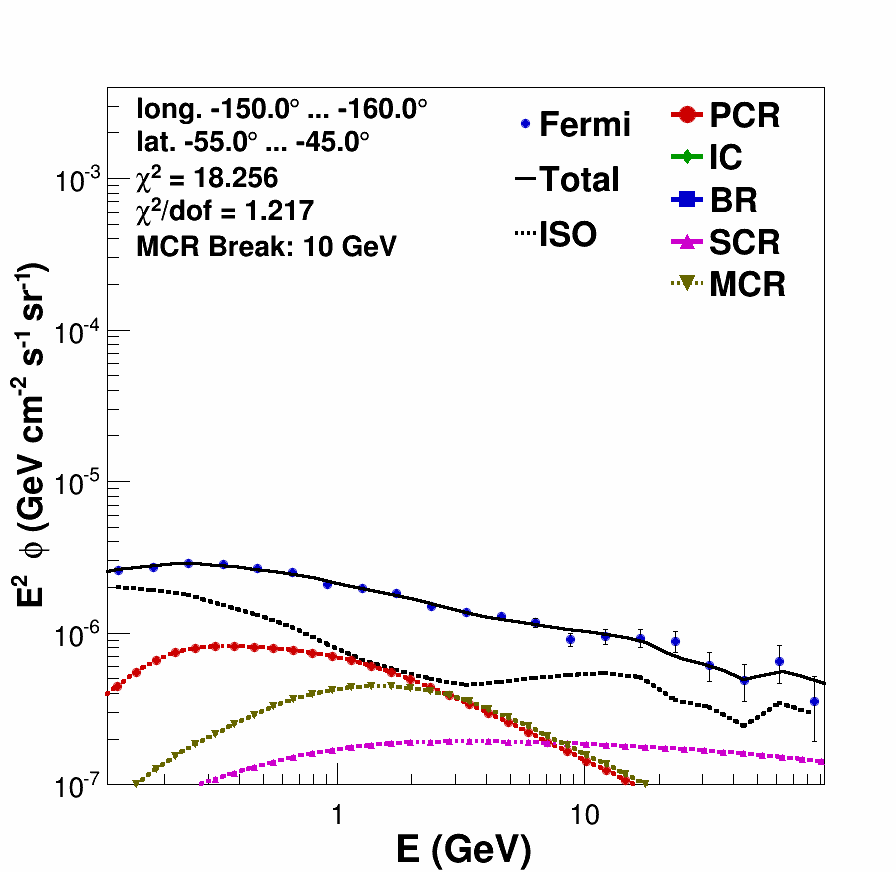}
\includegraphics[width=0.16\textwidth,height=0.16\textwidth,clip]{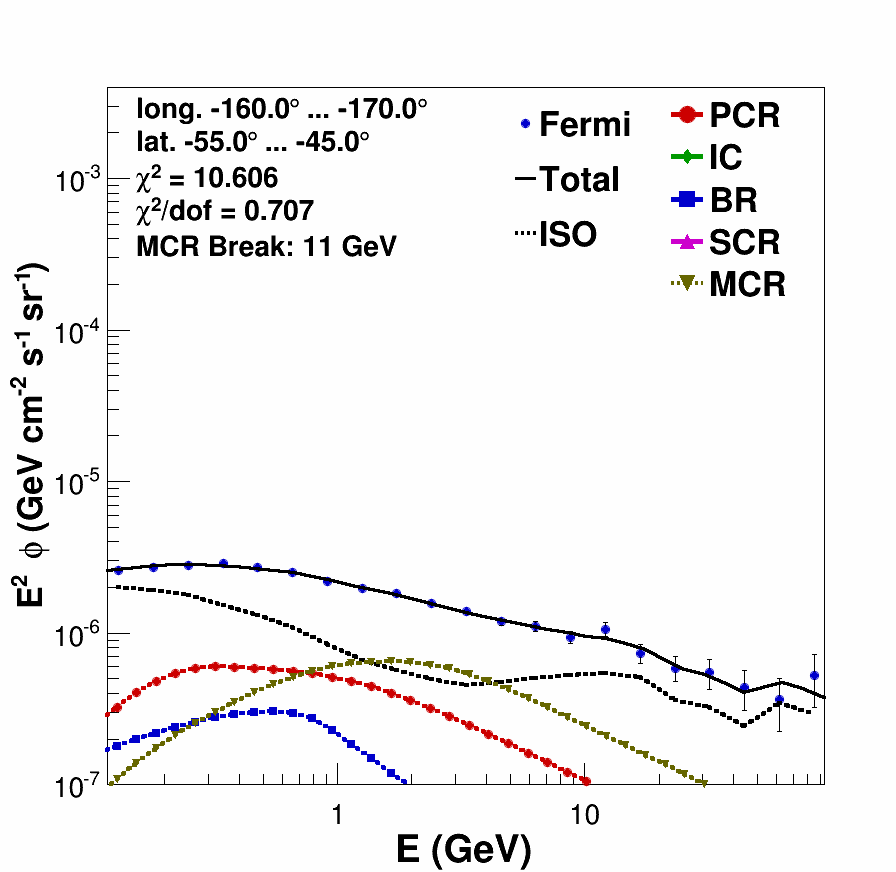}
\includegraphics[width=0.16\textwidth,height=0.16\textwidth,clip]{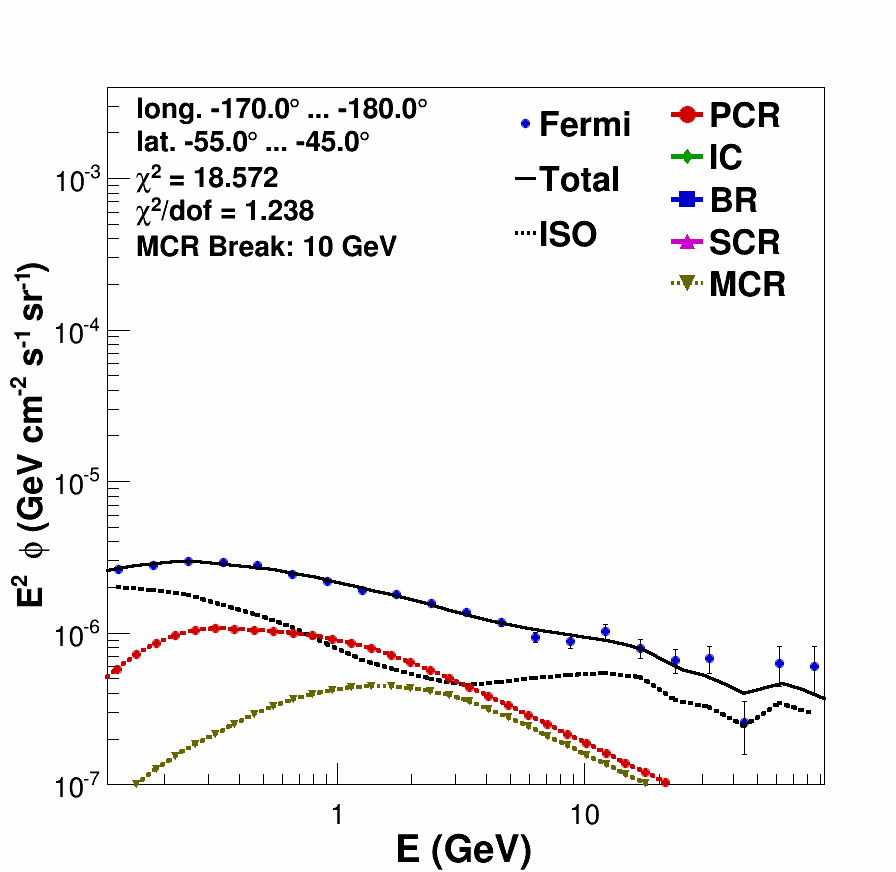}%%%%%r17
\caption[]{Template fits for latitudes  with $-55.0^\circ<b<-45.0^\circ$ and longitudes decreasing from 180$^\circ$ to -180$^\circ$.} \label{F29}
\end{figure}
\begin{figure} 
\centering
\includegraphics[width=0.16\textwidth,height=0.16\textwidth,clip]{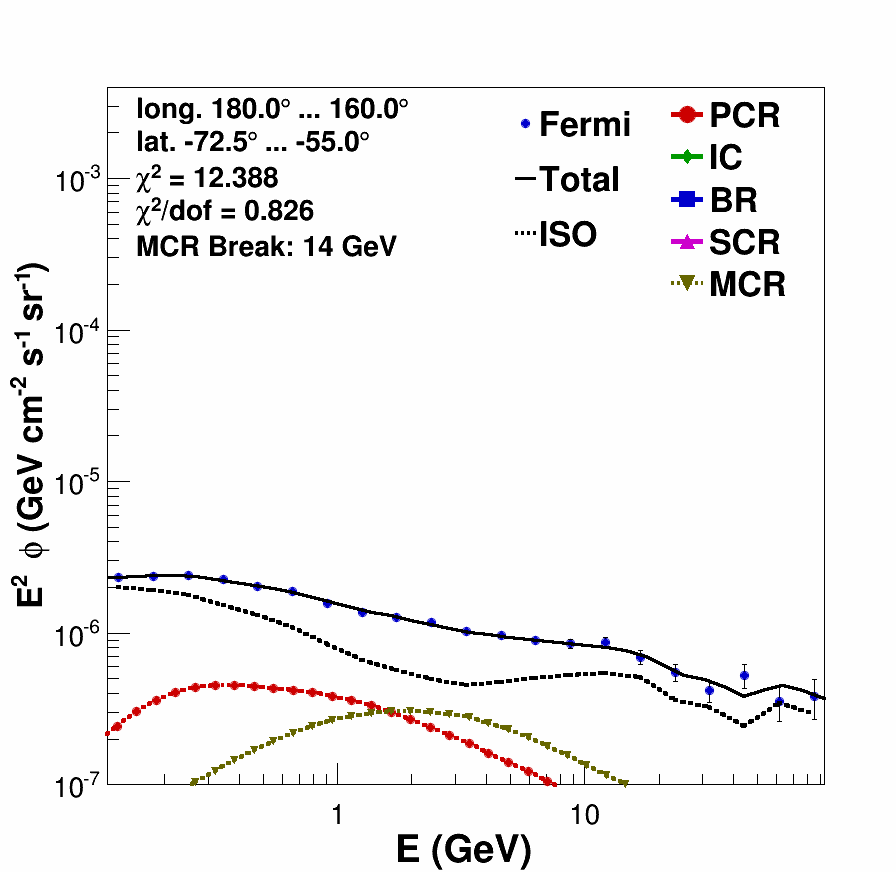}
\includegraphics[width=0.16\textwidth,height=0.16\textwidth,clip]{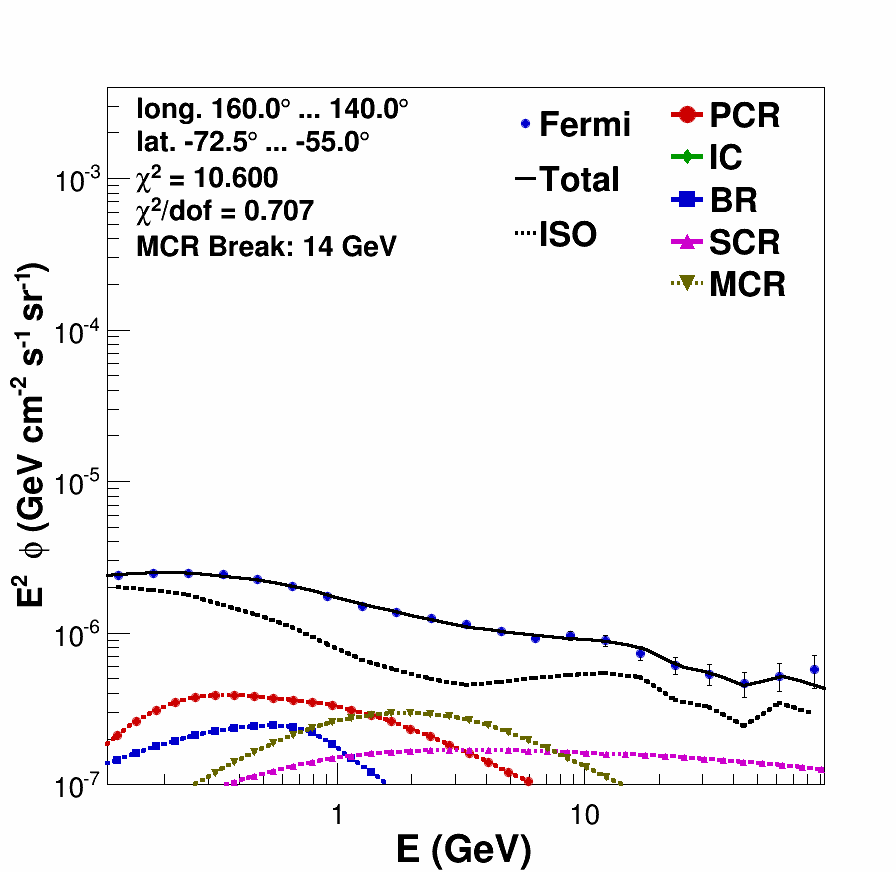}
\includegraphics[width=0.16\textwidth,height=0.16\textwidth,clip]{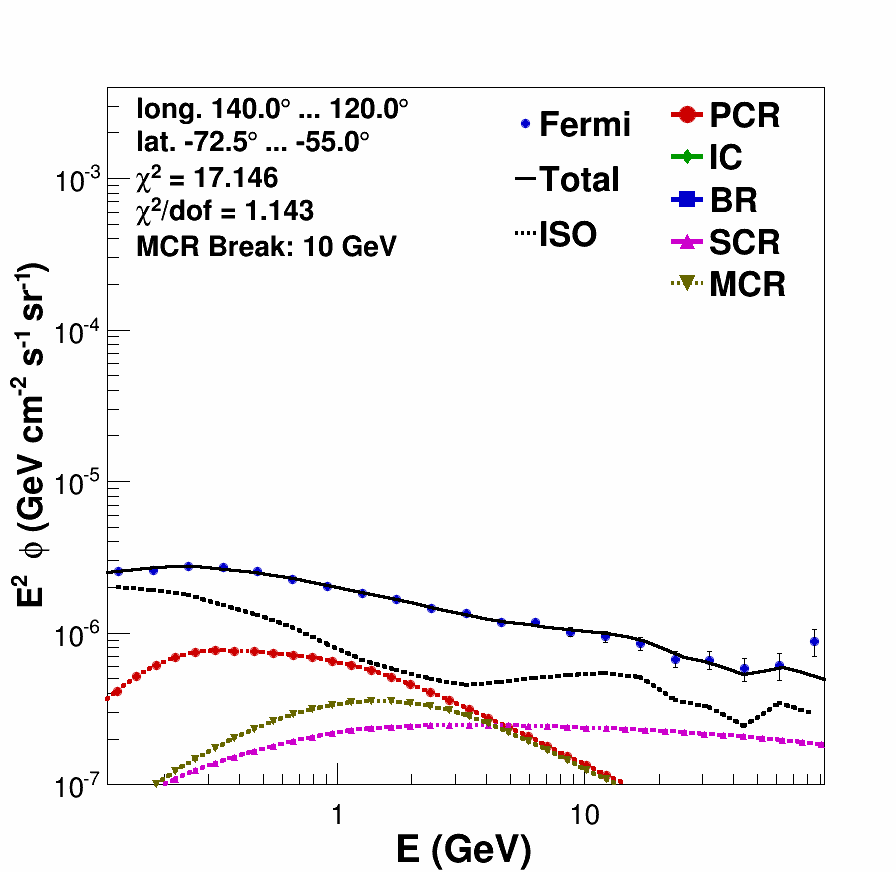}
\includegraphics[width=0.16\textwidth,height=0.16\textwidth,clip]{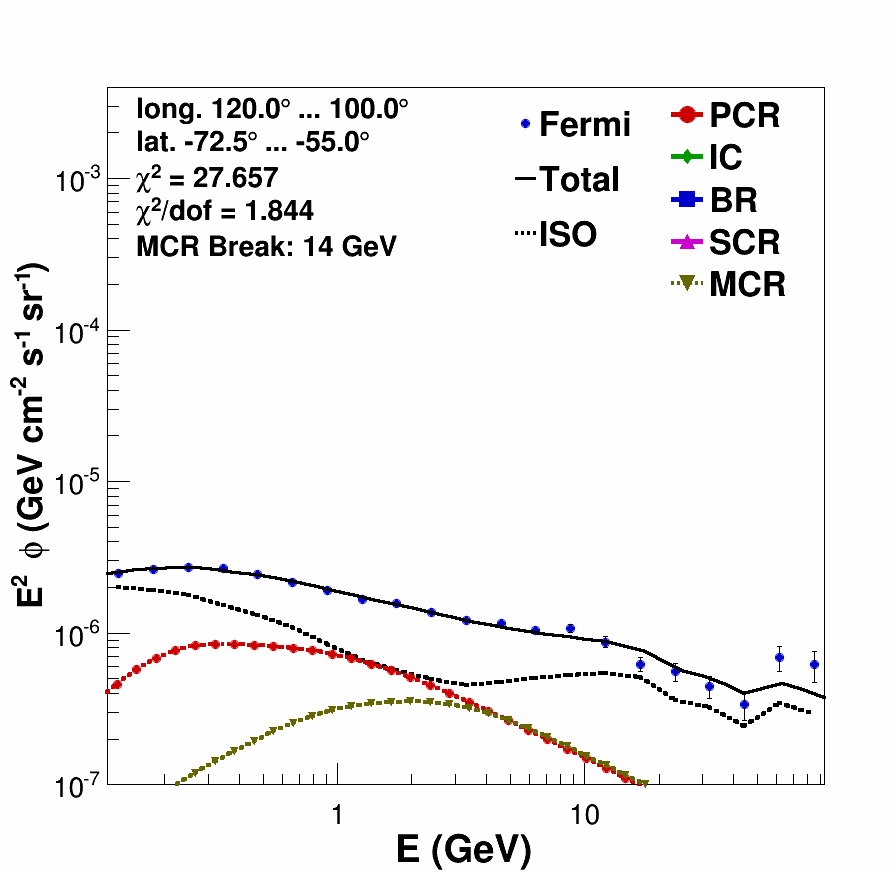}
\includegraphics[width=0.16\textwidth,height=0.16\textwidth,clip]{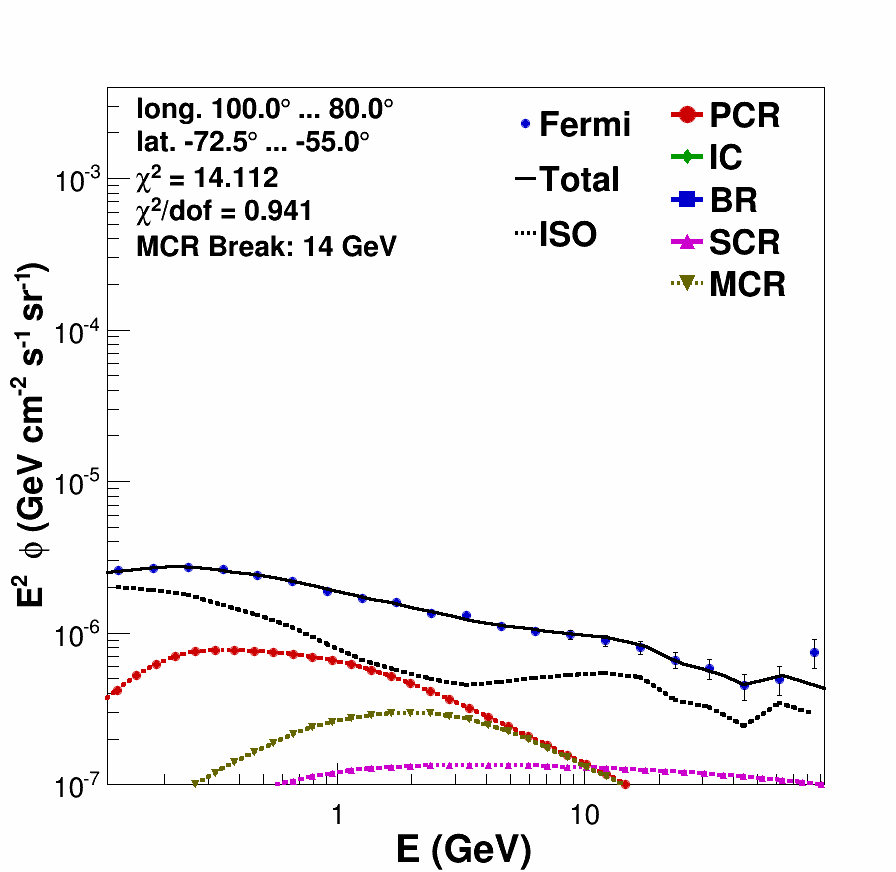}
\includegraphics[width=0.16\textwidth,height=0.16\textwidth,clip]{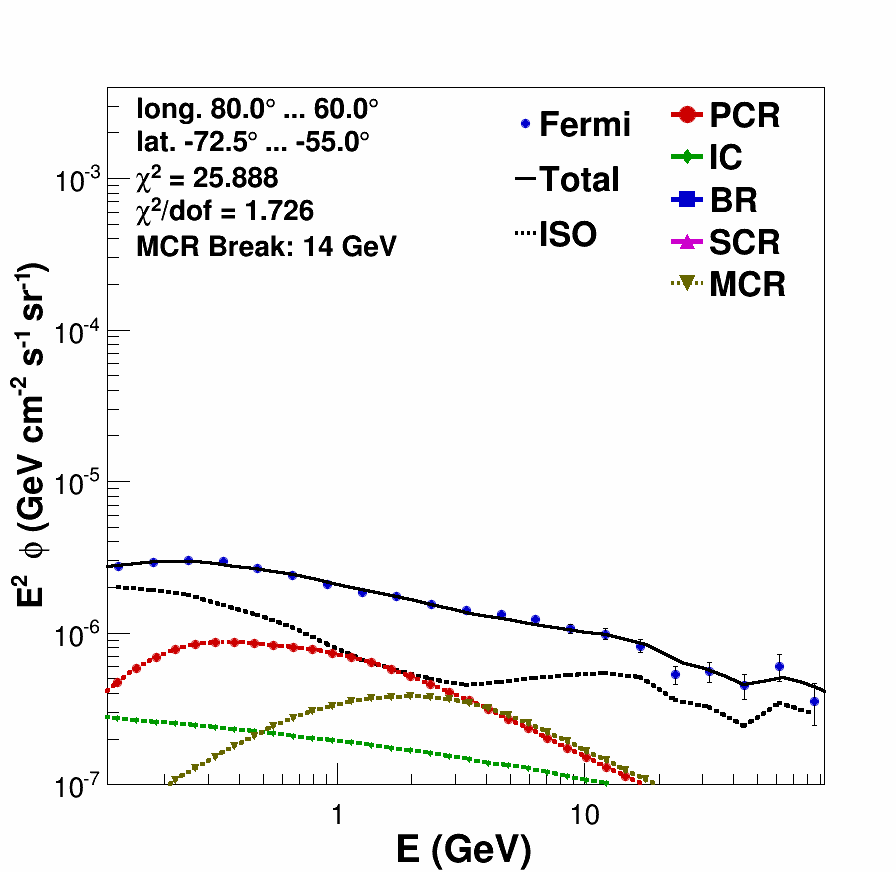}
\includegraphics[width=0.16\textwidth,height=0.16\textwidth,clip]{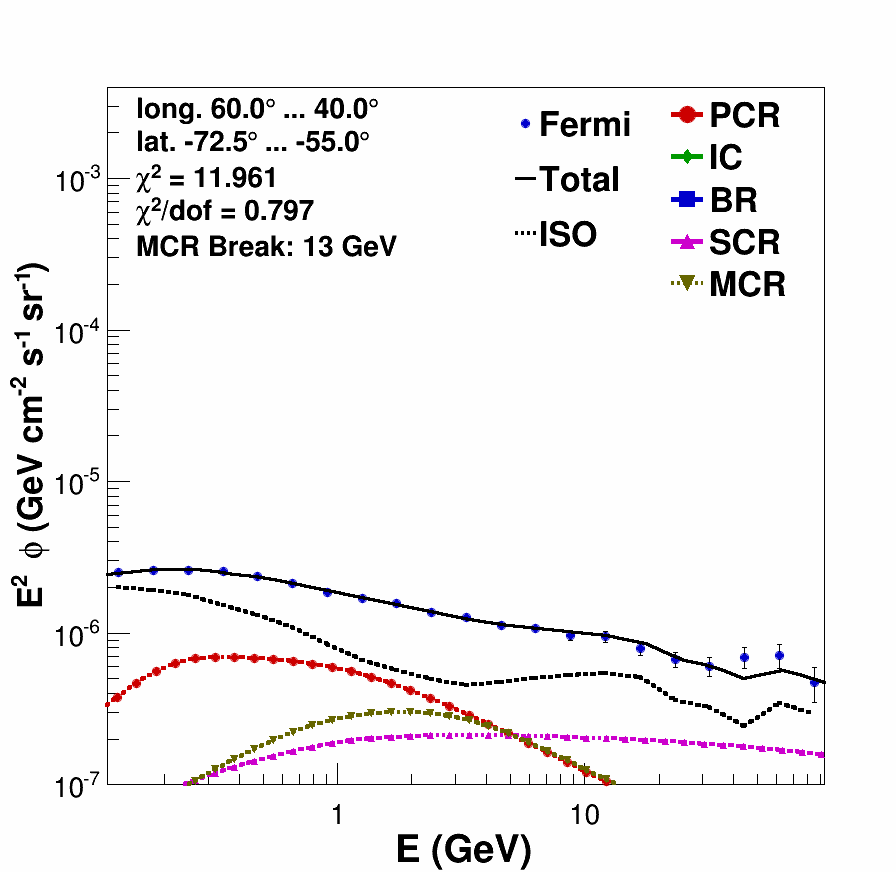}
\includegraphics[width=0.16\textwidth,height=0.16\textwidth,clip]{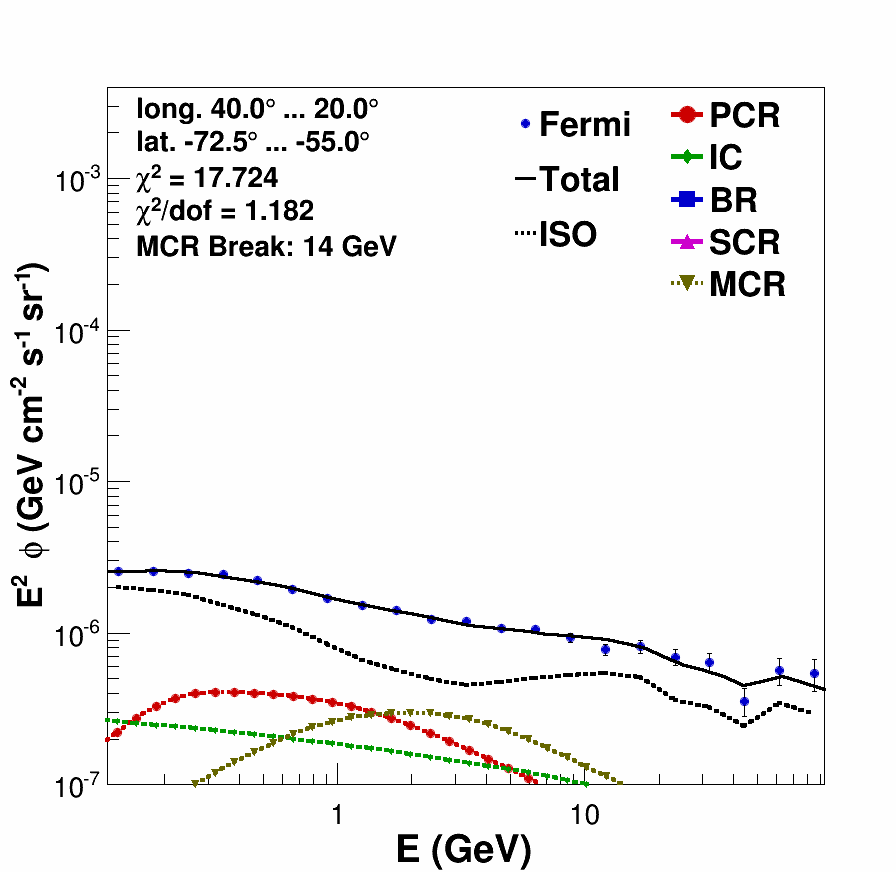}
\includegraphics[width=0.16\textwidth,height=0.16\textwidth,clip]{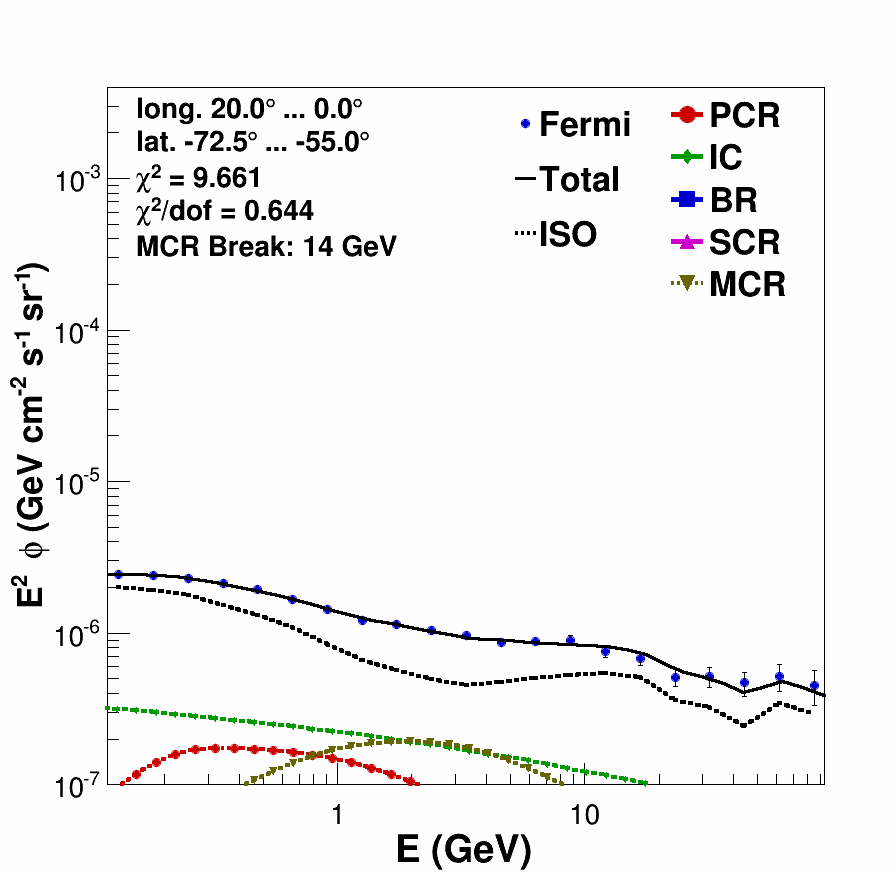}
\includegraphics[width=0.16\textwidth,height=0.16\textwidth,clip]{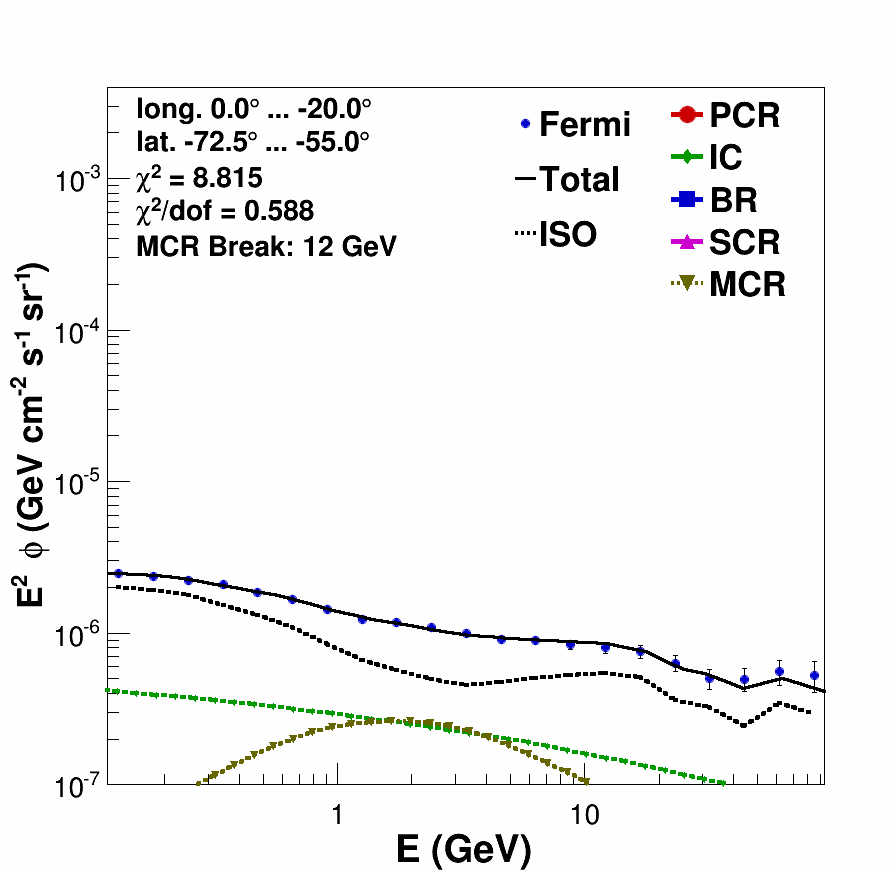}
\includegraphics[width=0.16\textwidth,height=0.16\textwidth,clip]{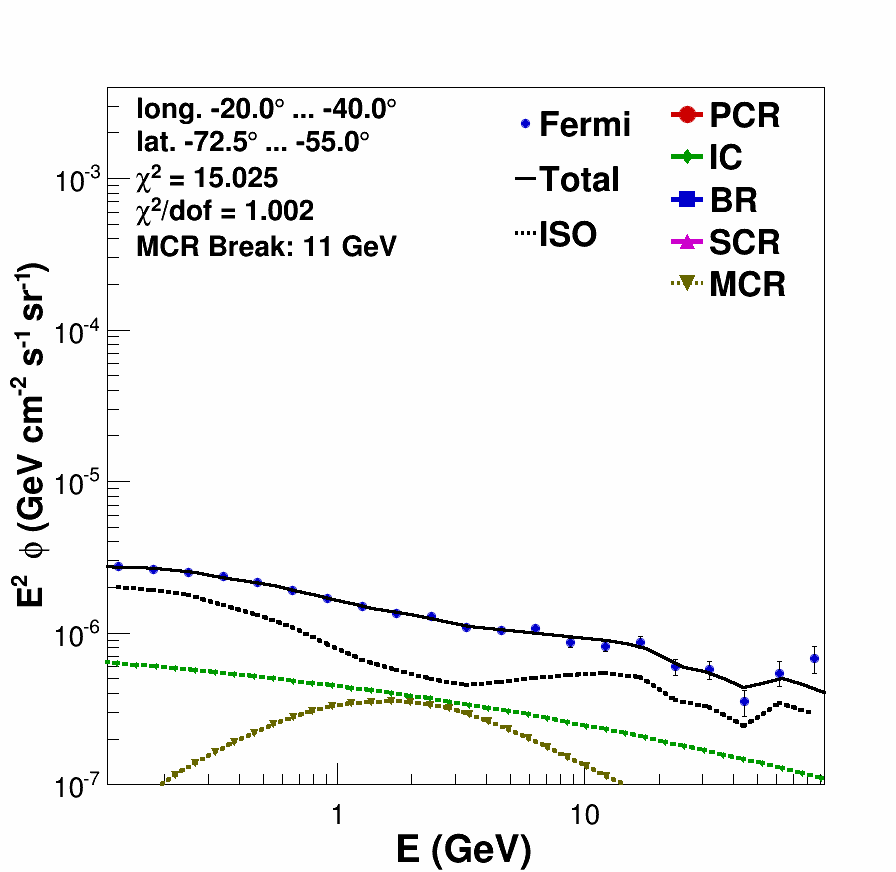}
\includegraphics[width=0.16\textwidth,height=0.16\textwidth,clip]{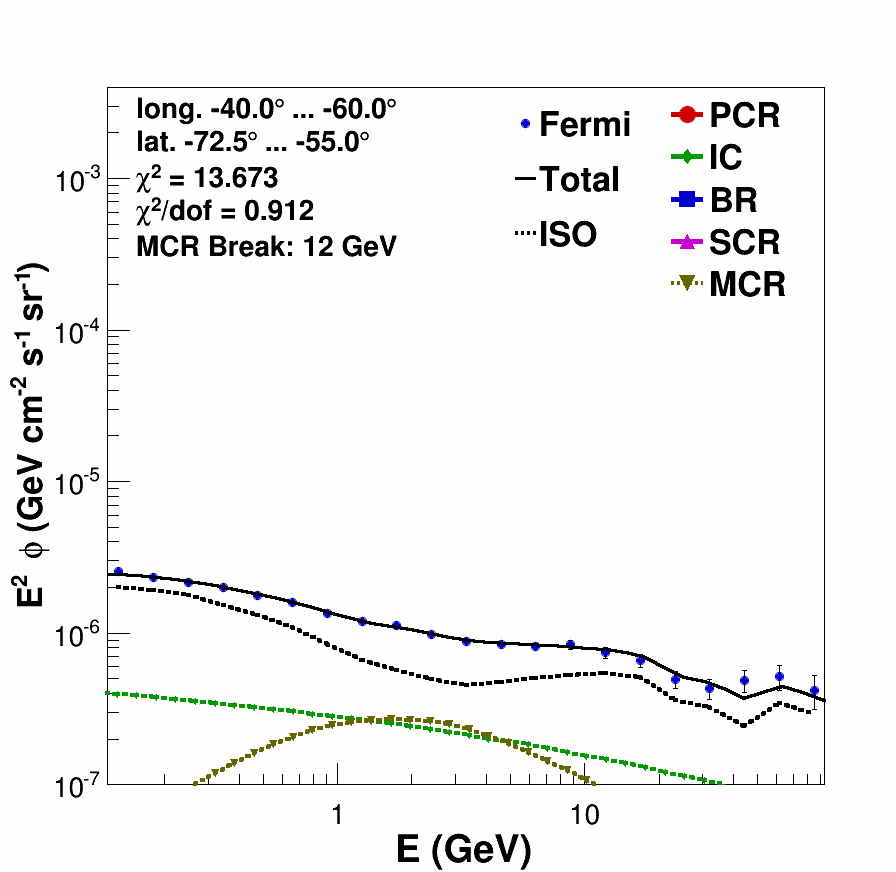}
\includegraphics[width=0.16\textwidth,height=0.16\textwidth,clip]{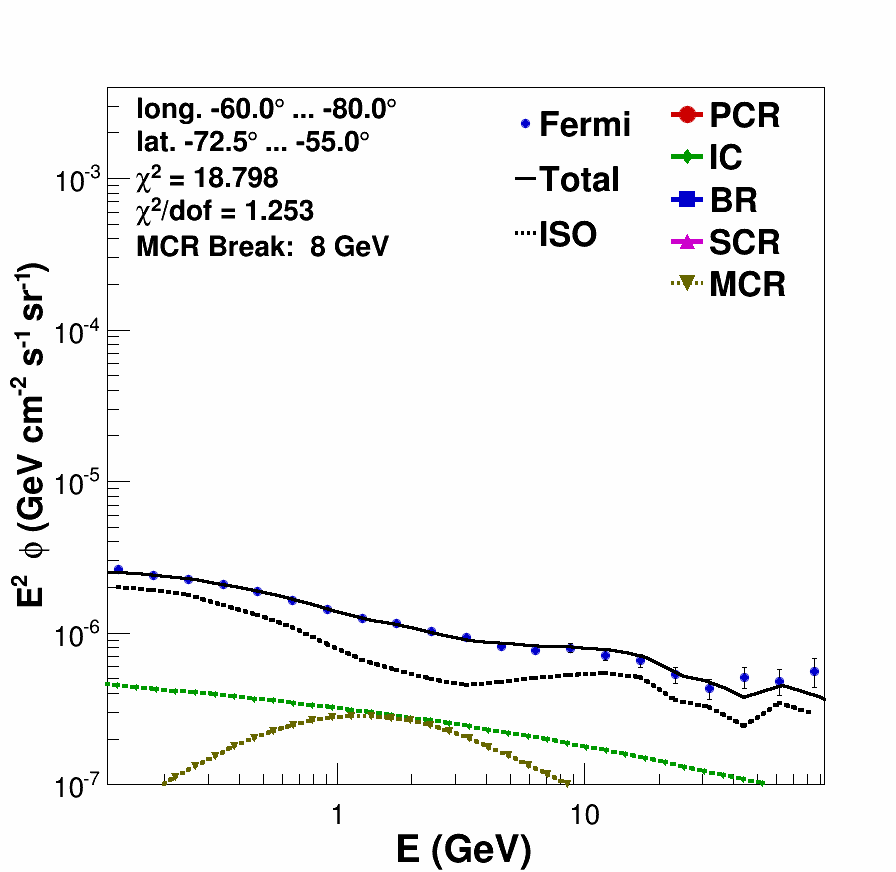}
\includegraphics[width=0.16\textwidth,height=0.16\textwidth,clip]{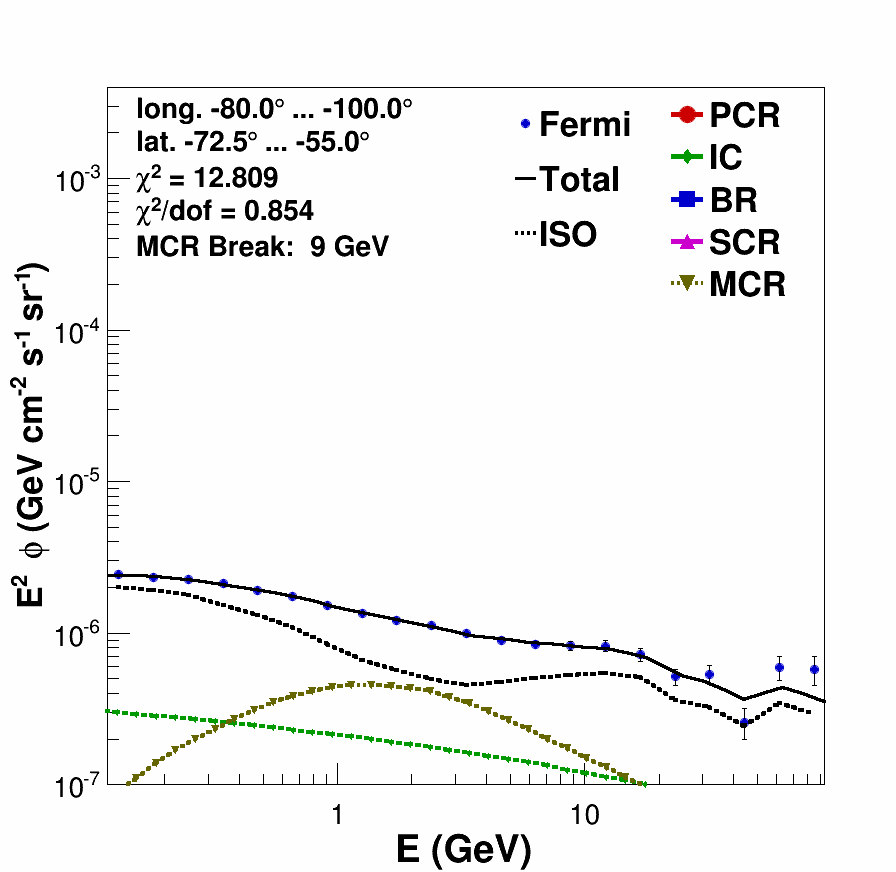}
\includegraphics[width=0.16\textwidth,height=0.16\textwidth,clip]{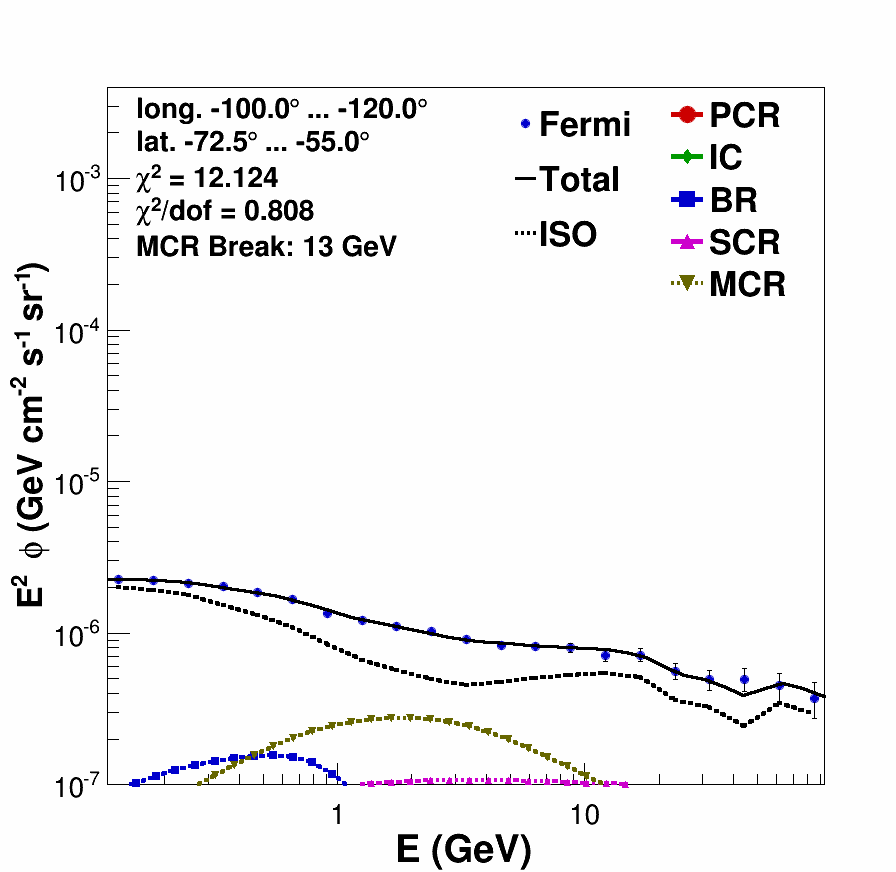}
\includegraphics[width=0.16\textwidth,height=0.16\textwidth,clip]{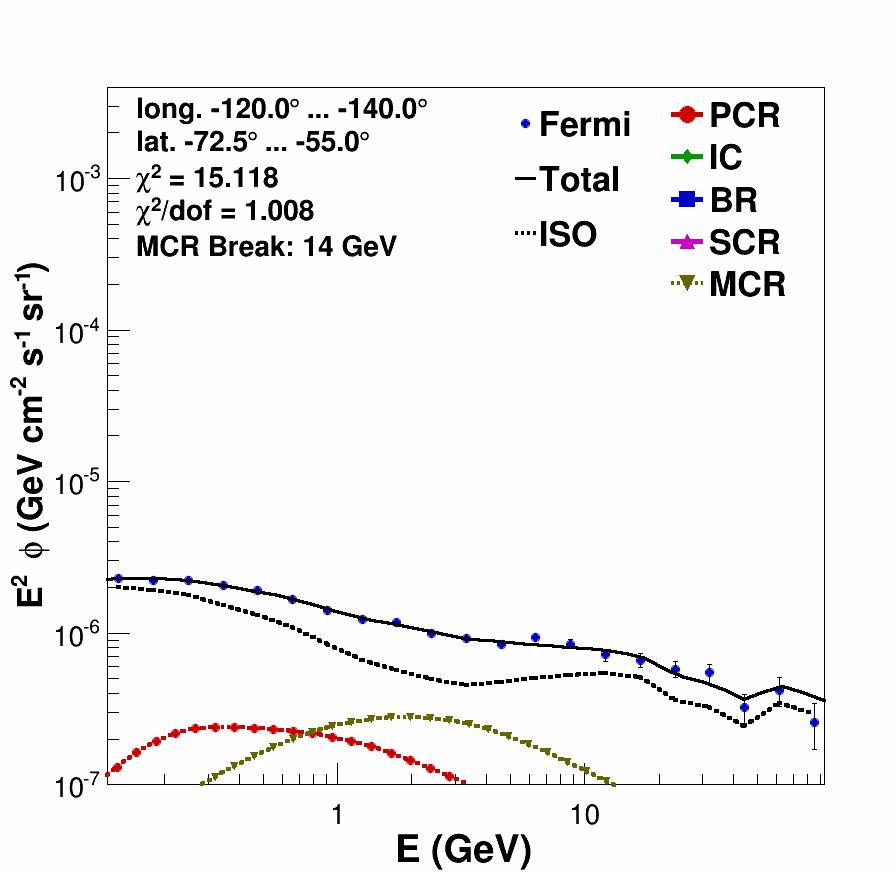}
\includegraphics[width=0.16\textwidth,height=0.16\textwidth,clip]{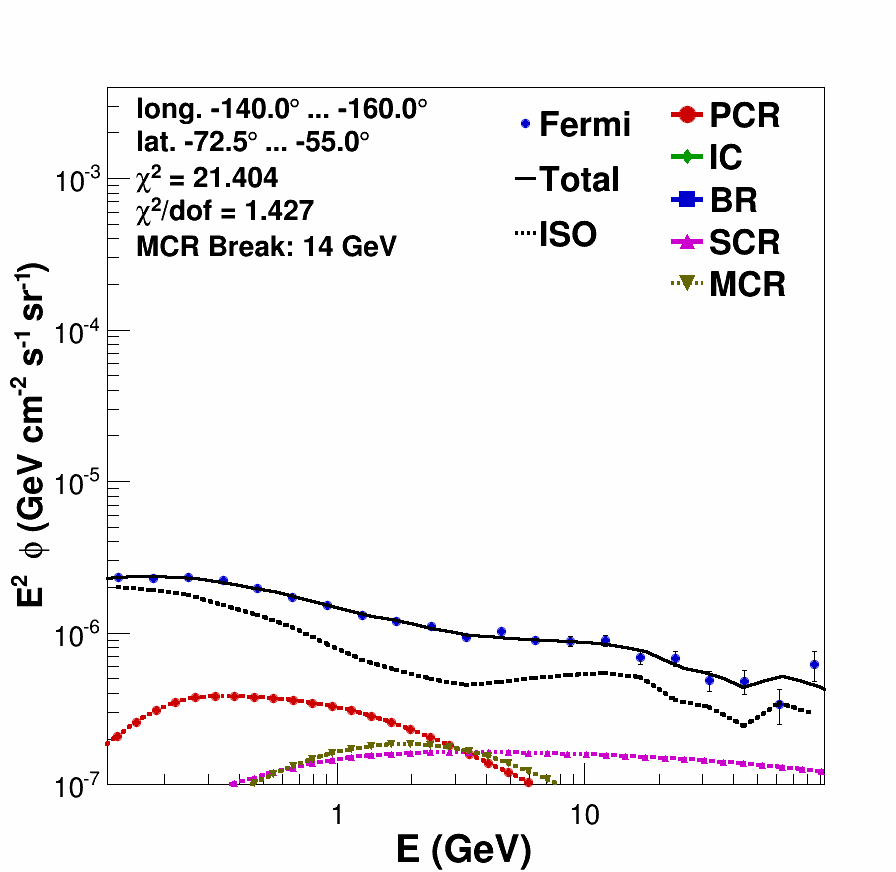}
\includegraphics[width=0.16\textwidth,height=0.16\textwidth,clip]{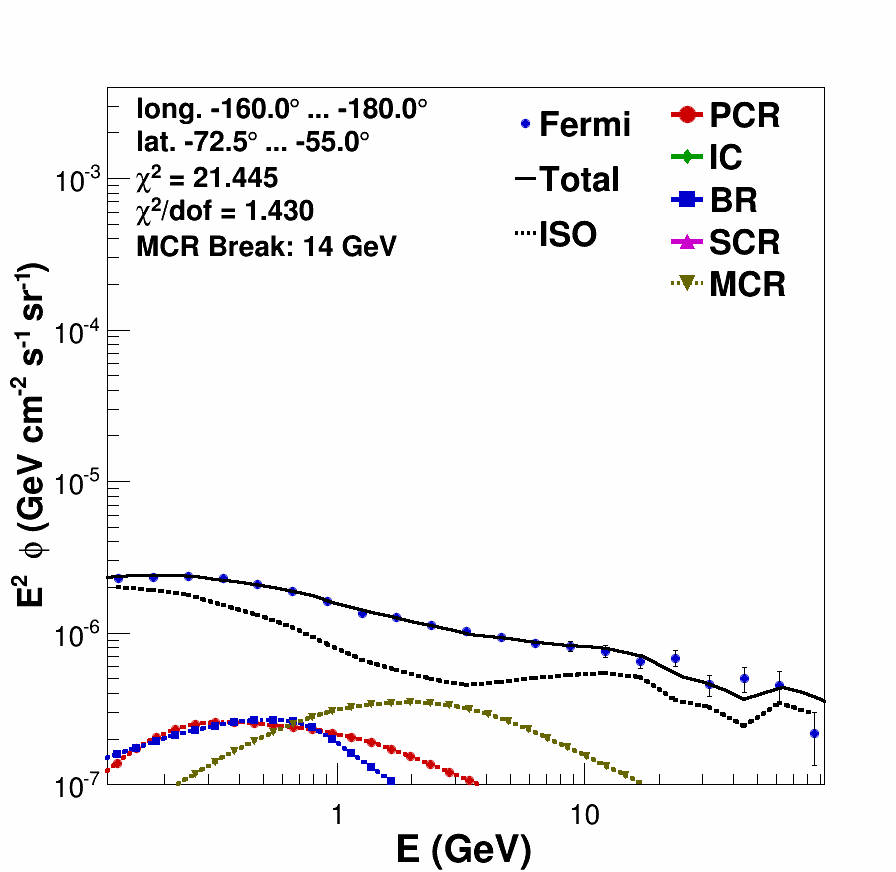}%%%r18
\caption[]{Template fits for latitudes  with $-72.5^\circ<b<-55.0^\circ$ and longitudes decreasing from 180$^\circ$ to -180$^\circ$.} \label{F30}
\end{figure}
\begin{figure}
\includegraphics[width=0.16\textwidth,height=0.16\textwidth,clip]{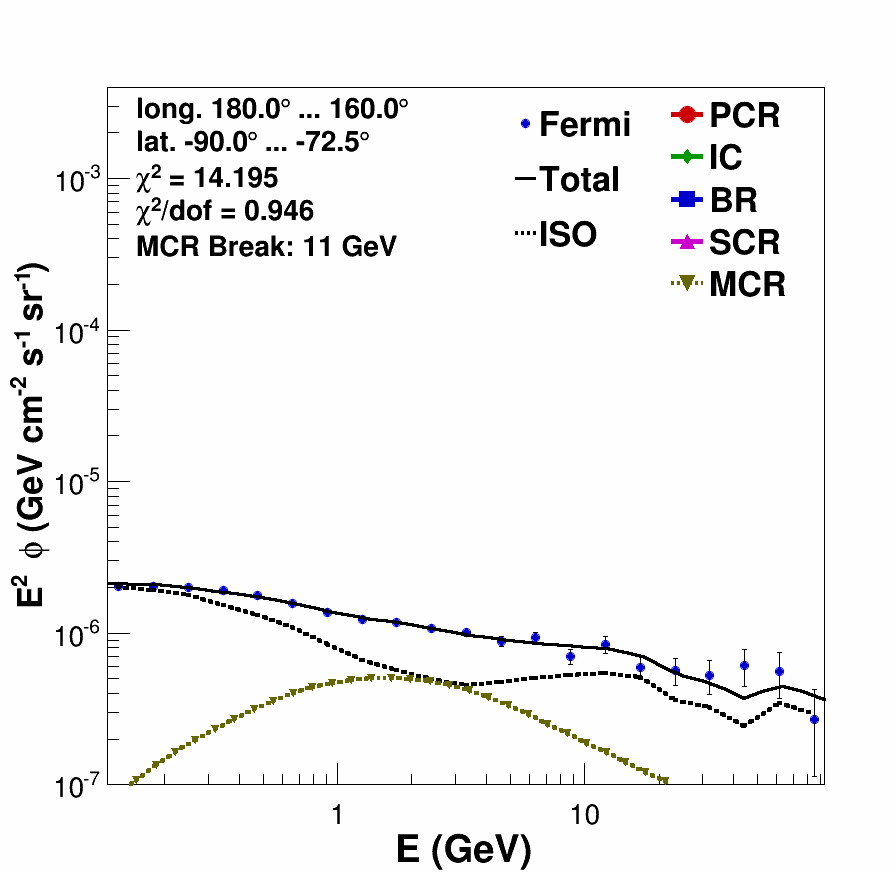}
\includegraphics[width=0.16\textwidth,height=0.16\textwidth,clip]{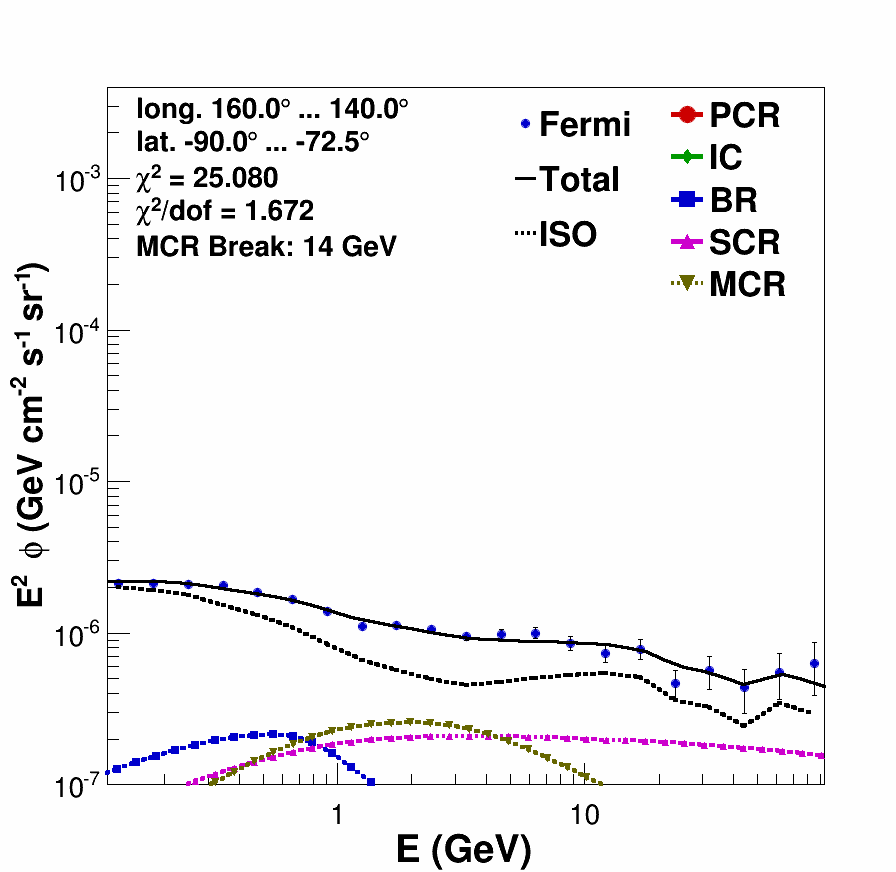}
\includegraphics[width=0.16\textwidth,height=0.16\textwidth,clip]{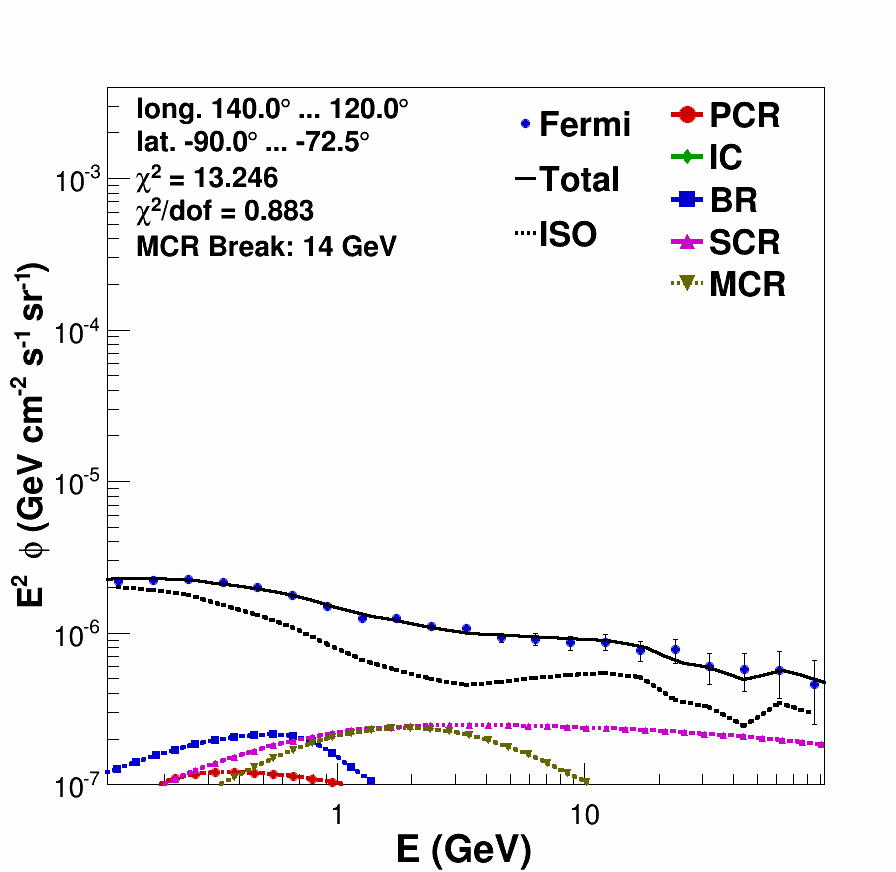}
\includegraphics[width=0.16\textwidth,height=0.16\textwidth,clip]{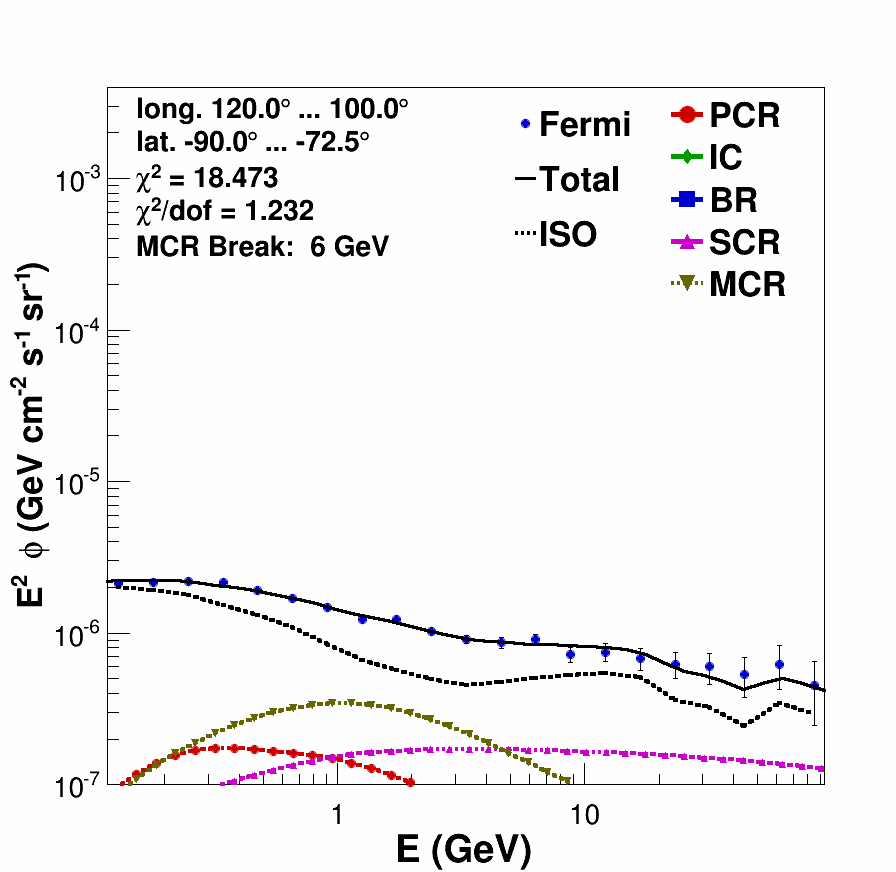}
\includegraphics[width=0.16\textwidth,height=0.16\textwidth,clip]{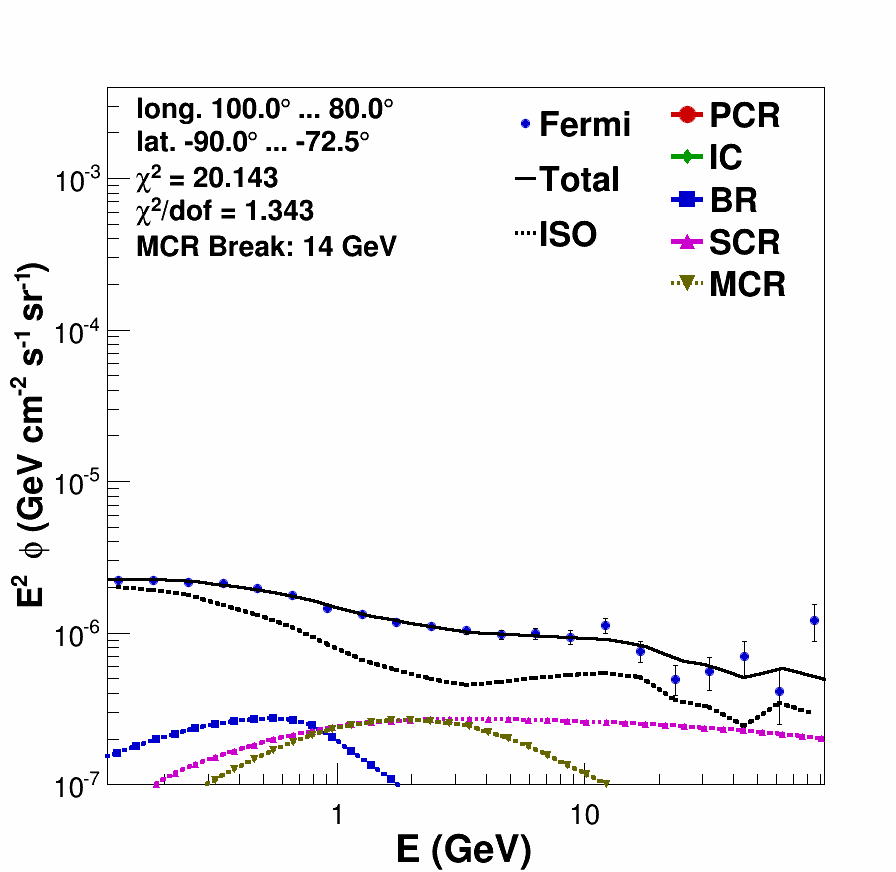}
\includegraphics[width=0.16\textwidth,height=0.16\textwidth,clip]{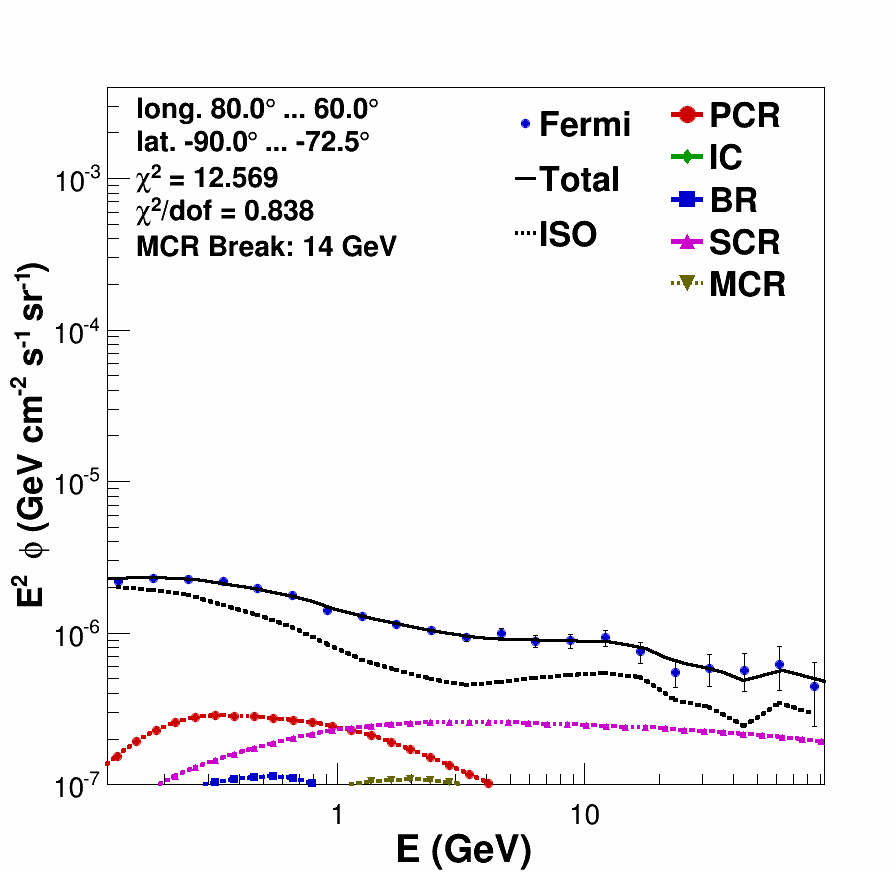}
\includegraphics[width=0.16\textwidth,height=0.16\textwidth,clip]{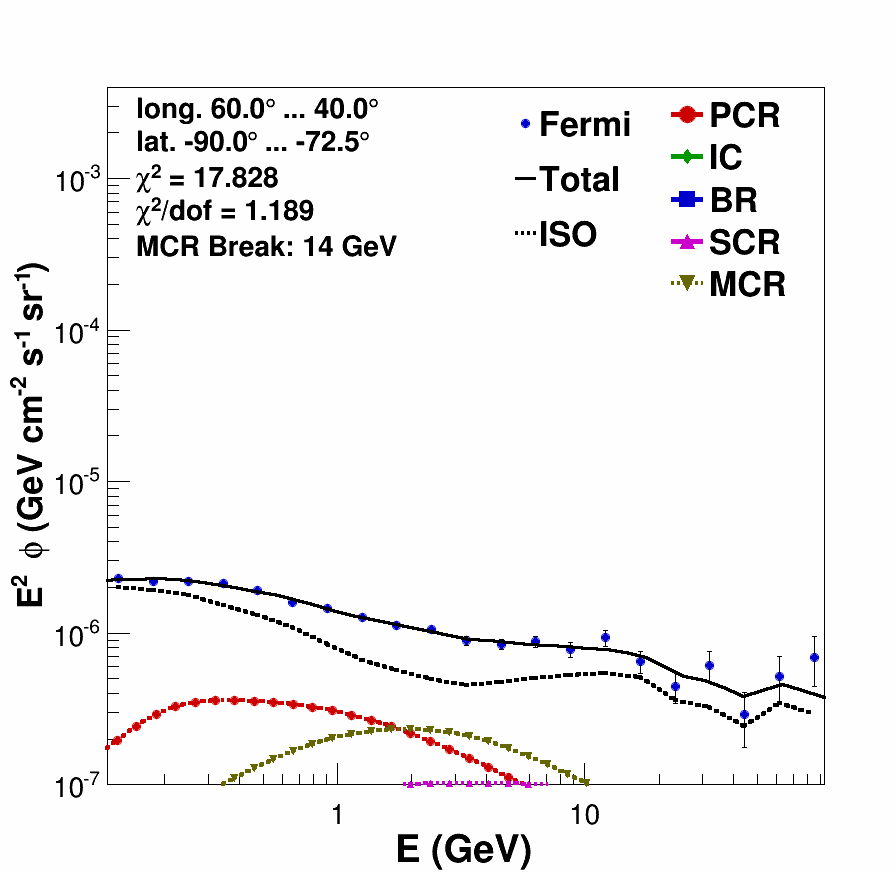}
\includegraphics[width=0.16\textwidth,height=0.16\textwidth,clip]{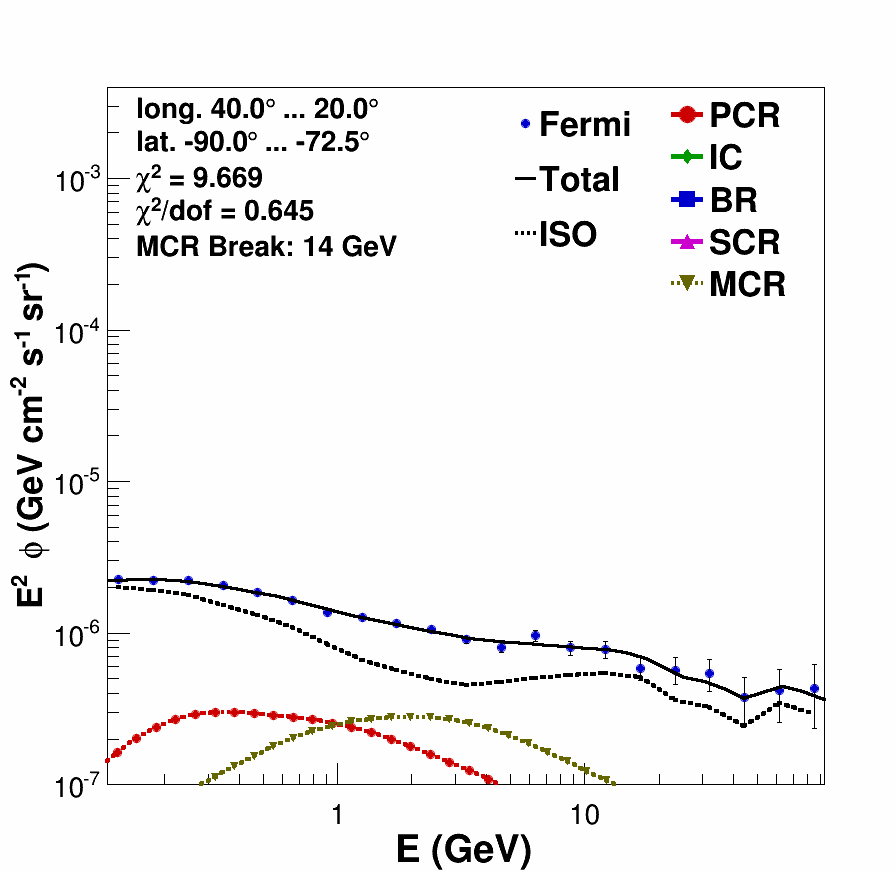}
\includegraphics[width=0.16\textwidth,height=0.16\textwidth,clip]{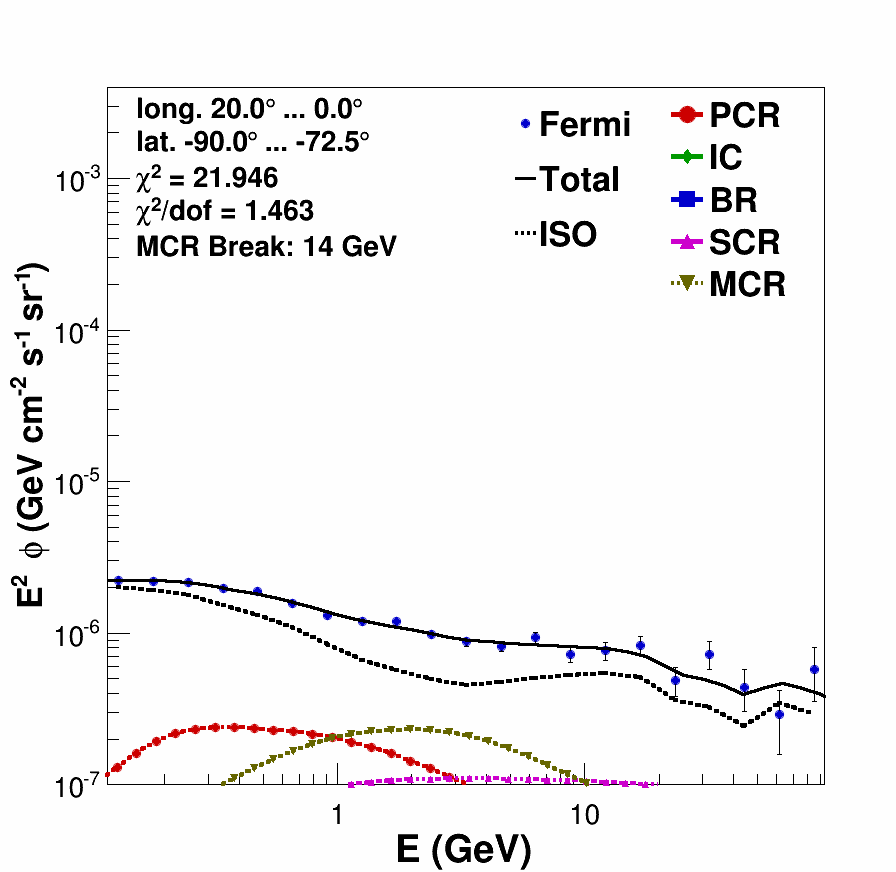}
\includegraphics[width=0.16\textwidth,height=0.16\textwidth,clip]{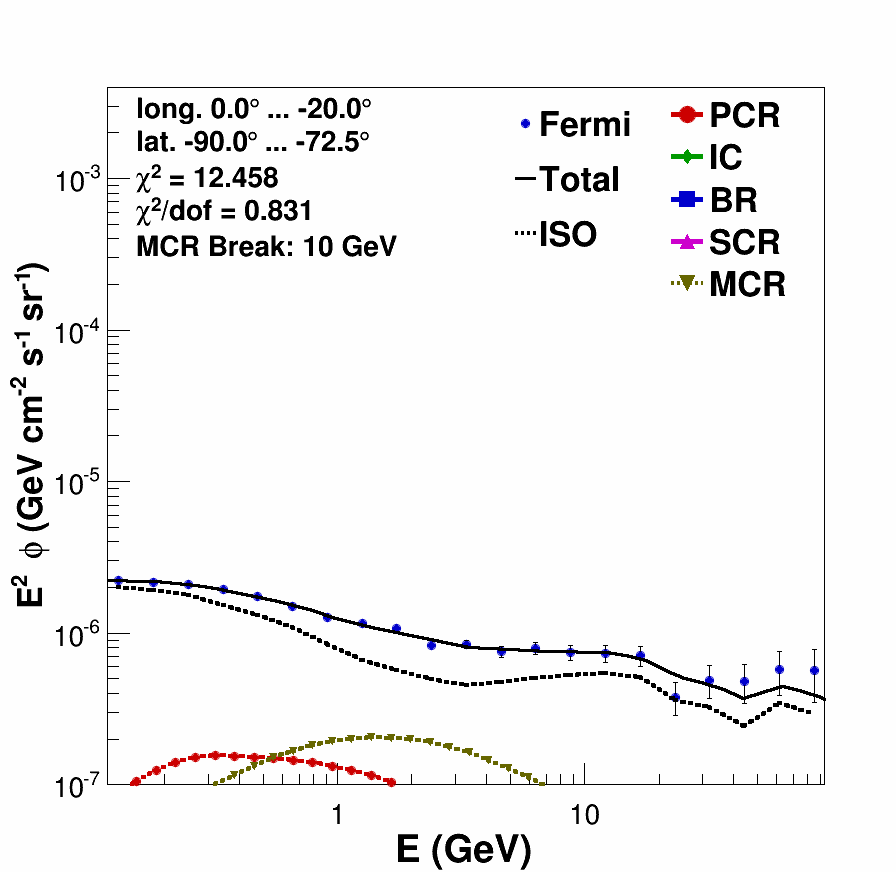}
\includegraphics[width=0.16\textwidth,height=0.16\textwidth,clip]{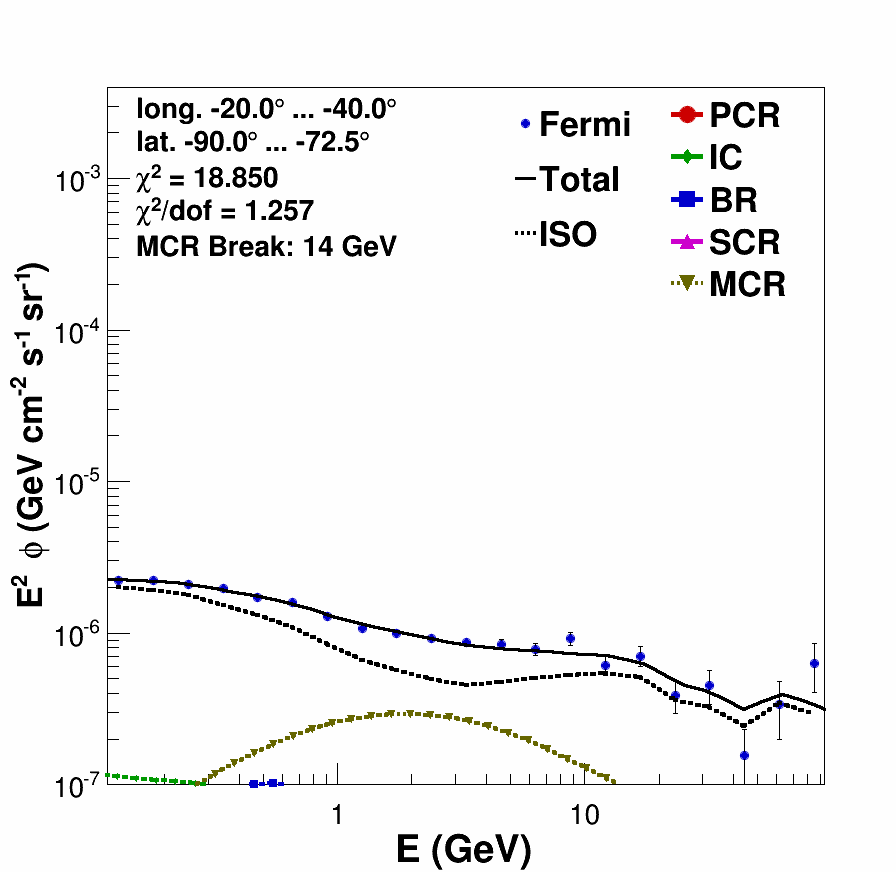}
\includegraphics[width=0.16\textwidth,height=0.16\textwidth,clip]{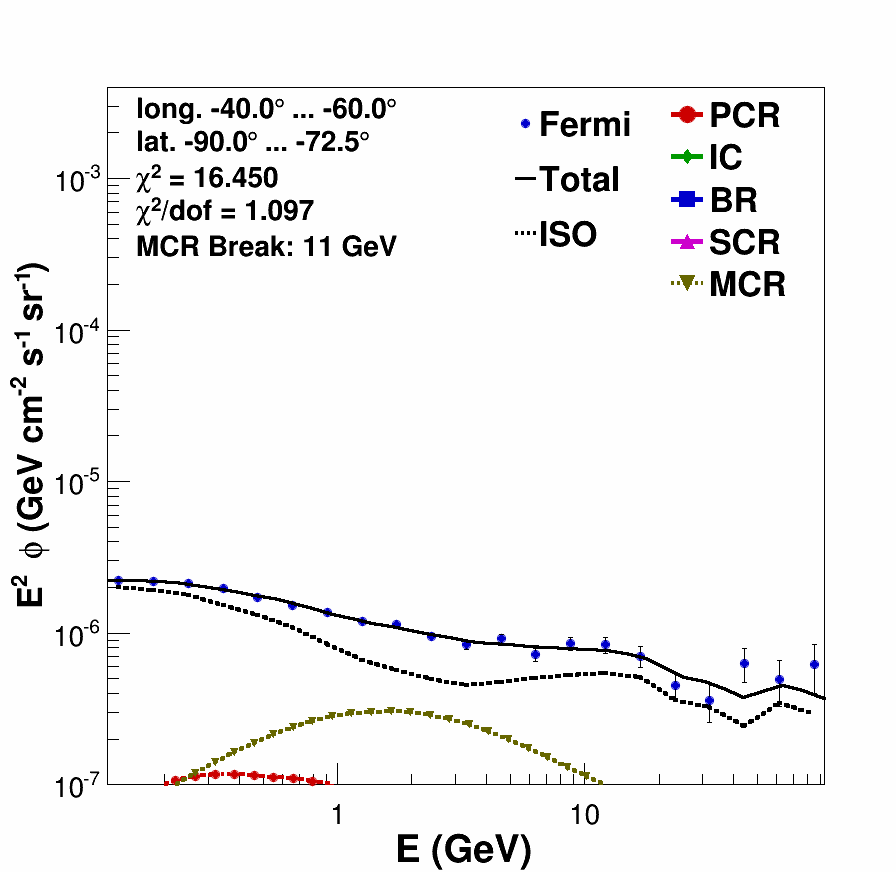}
\includegraphics[width=0.16\textwidth,height=0.16\textwidth,clip]{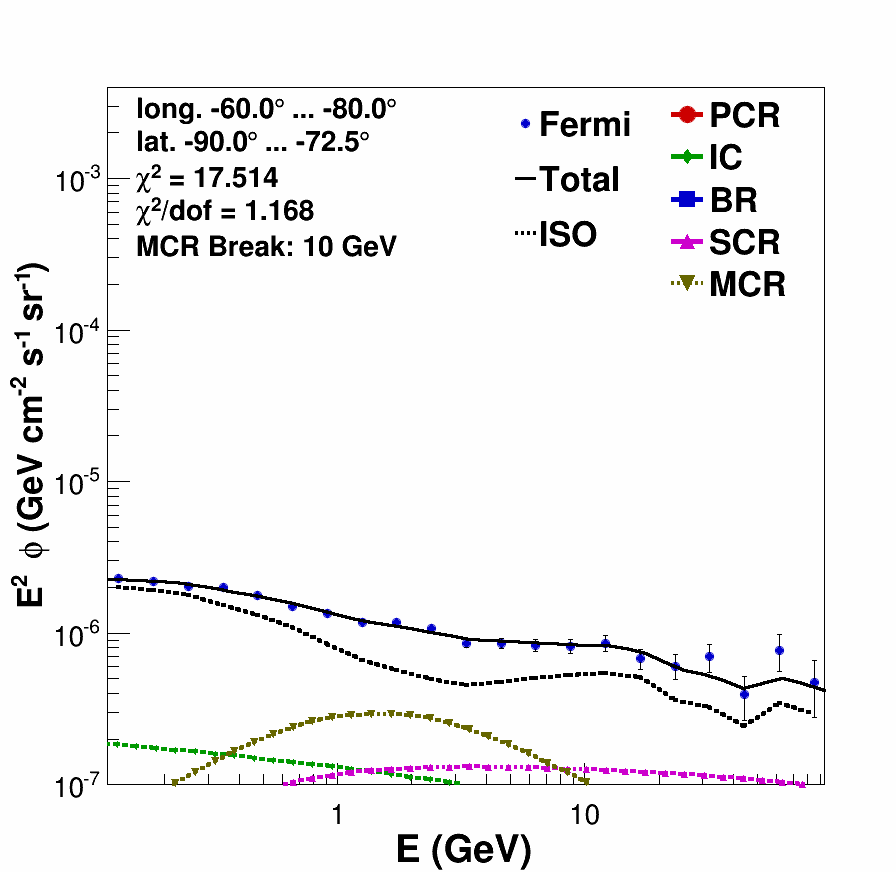}
\includegraphics[width=0.16\textwidth,height=0.16\textwidth,clip]{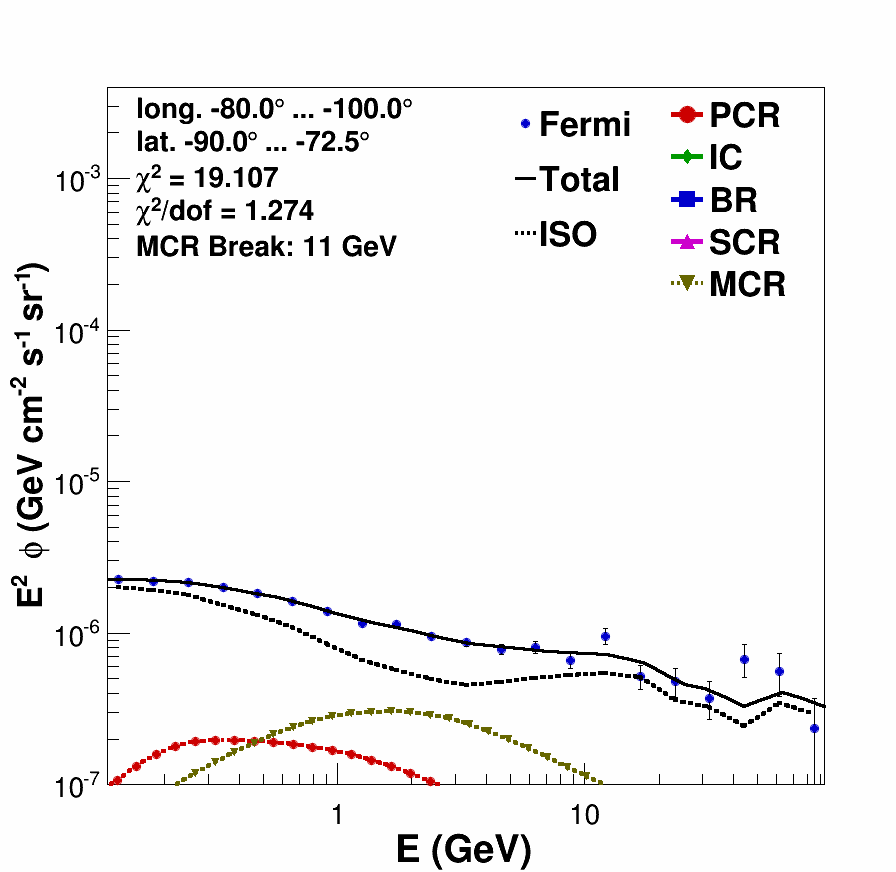}
\includegraphics[width=0.16\textwidth,height=0.16\textwidth,clip]{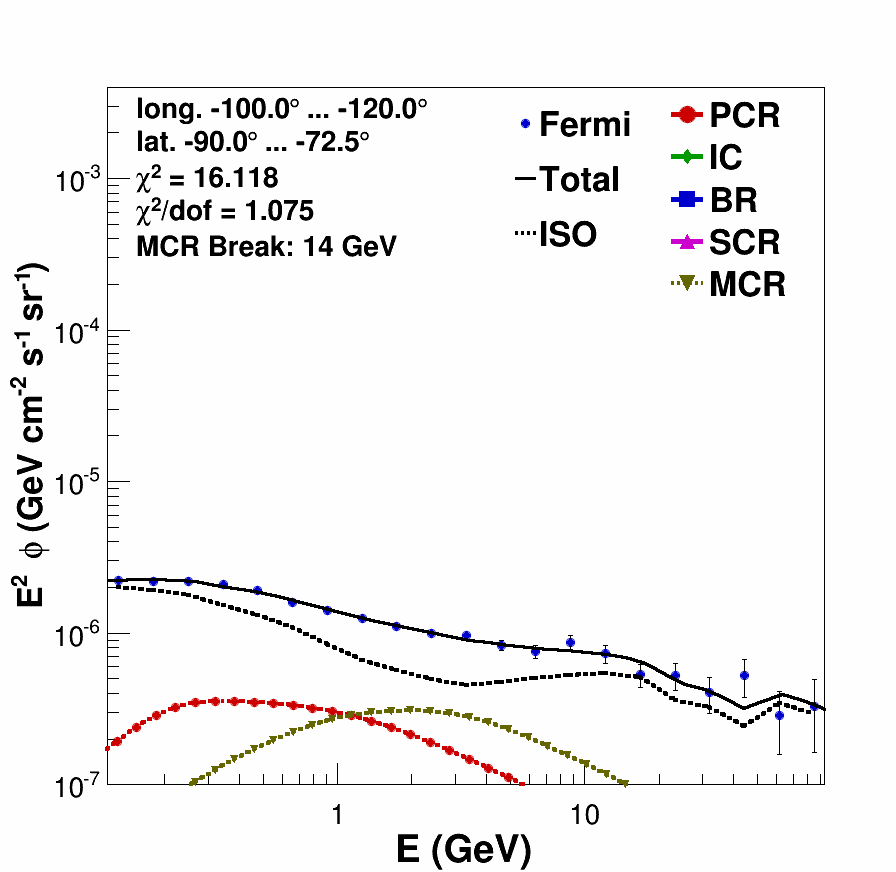}
\includegraphics[width=0.16\textwidth,height=0.16\textwidth,clip]{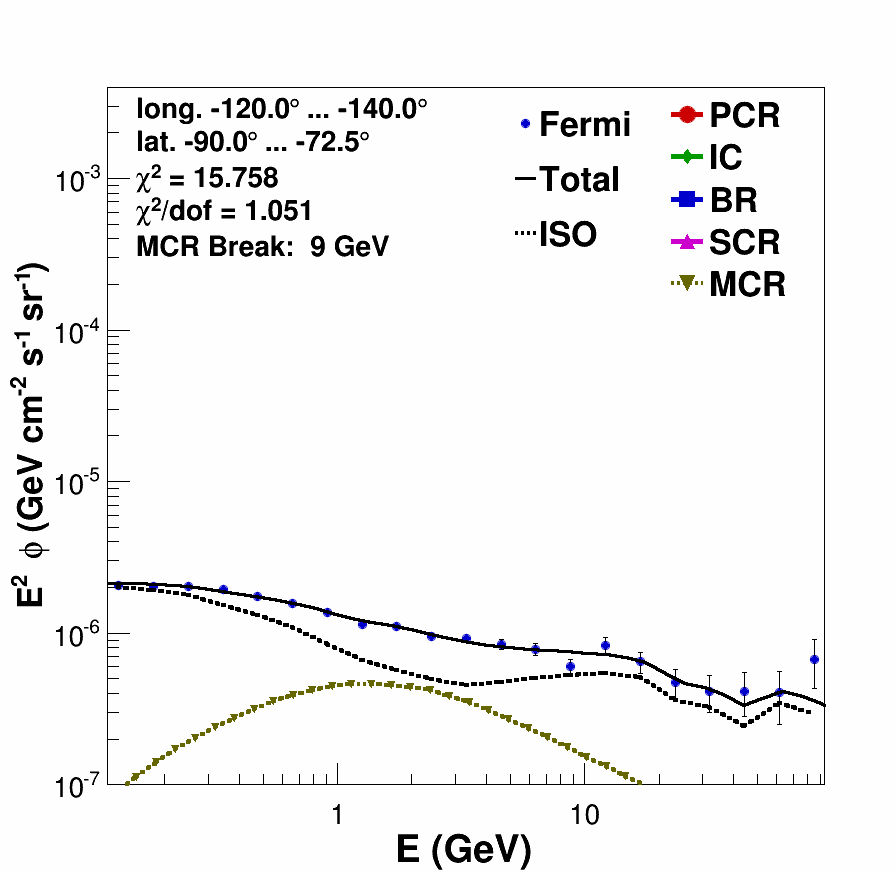}
\includegraphics[width=0.16\textwidth,height=0.16\textwidth,clip]{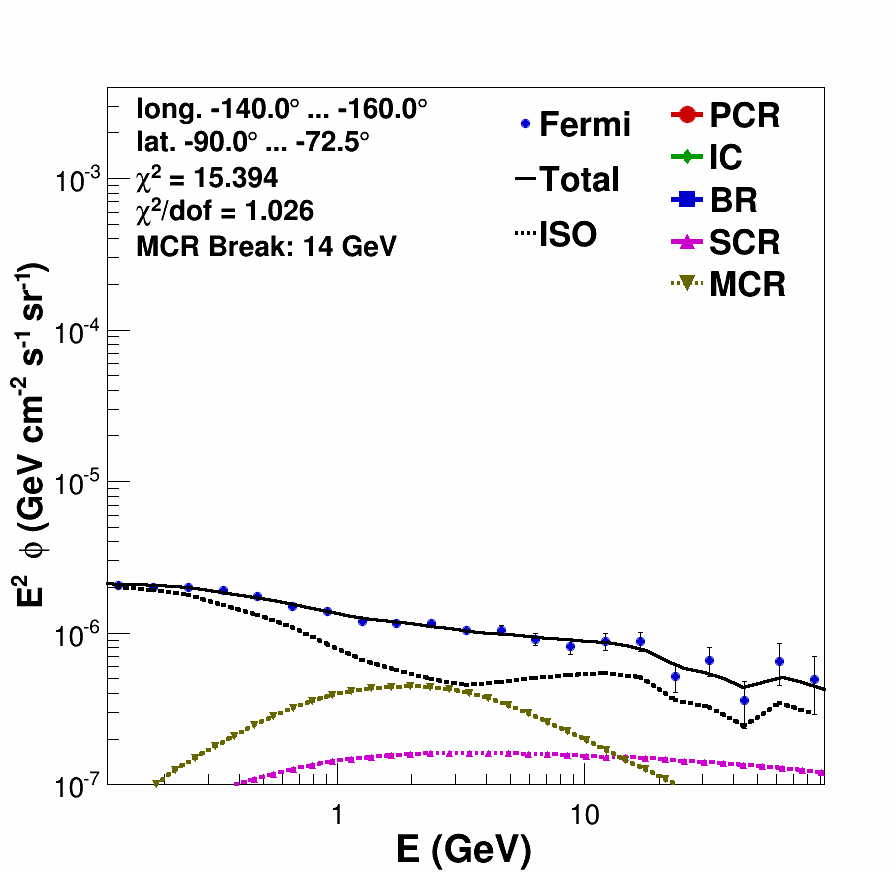}
\includegraphics[width=0.16\textwidth,height=0.16\textwidth,clip]{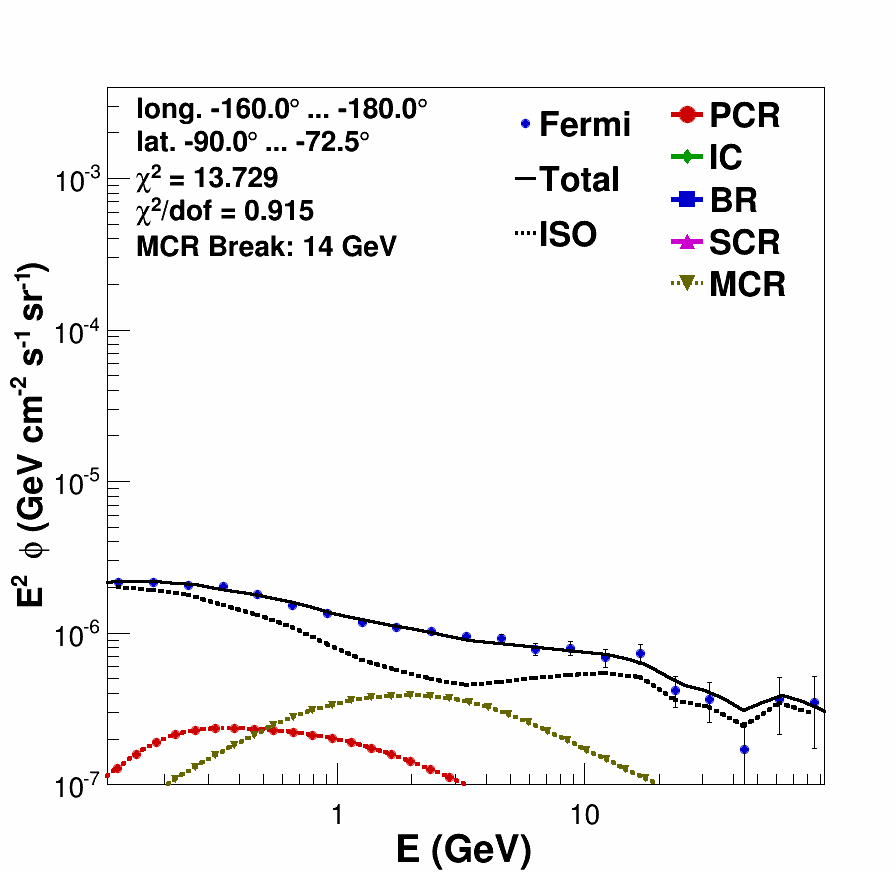}%%%%r19
\caption[]{Template fits for latitudes  with $-90.0^\circ<b<-72.5^\circ$ and longitudes decreasing from 180$^\circ$ to -180$^\circ$.} \label{F31}
\end{figure}
\clearpage
%%%%%%%%%%%%%%%%%%%%%%%%%%%%%%%%%%%%%%%%%%%%%%%%%%%%%%%%%%%%%% now DM
\begin{figure}
\centering
{\bf Appendix B: Figs. \ref{F32}-\ref{F52} show the template fits in each of the 797 cones using the DM template to describe the ``GeV-excess''.  The figures start with the highest latitudes and in each figure the longitude varies for a given stripe in latitude, as indicated in the legends.}\vspace*{3mm}\\
\includegraphics[width=0.16\textwidth,height=0.16\textwidth,clip]{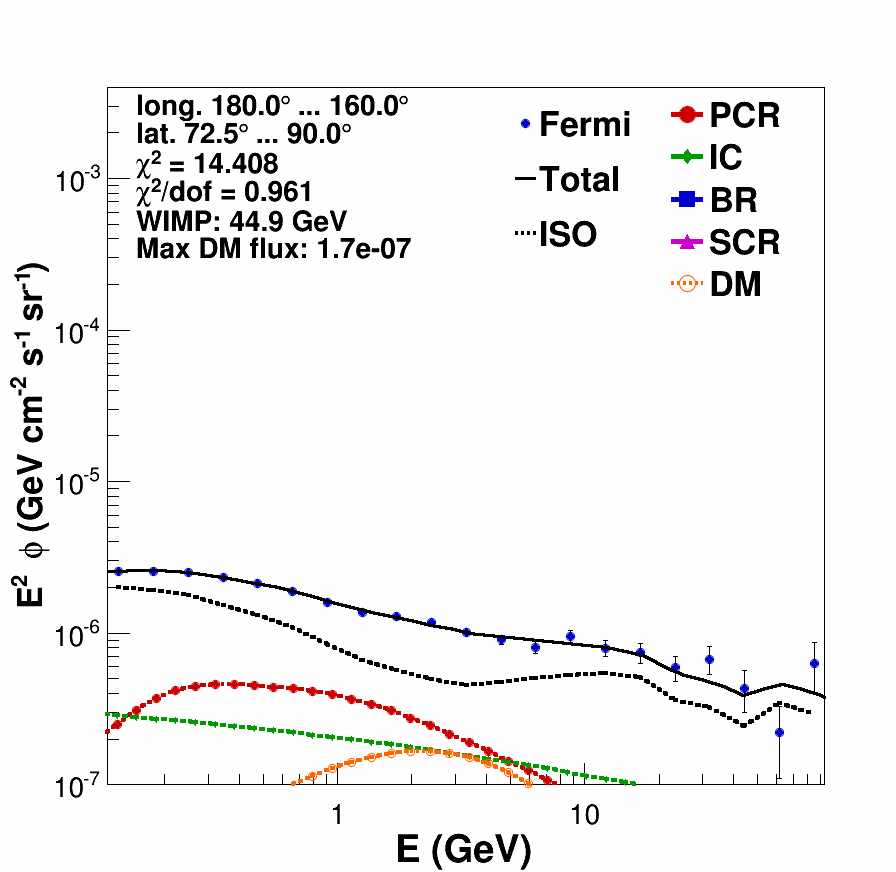}
\includegraphics[width=0.16\textwidth,height=0.16\textwidth,clip]{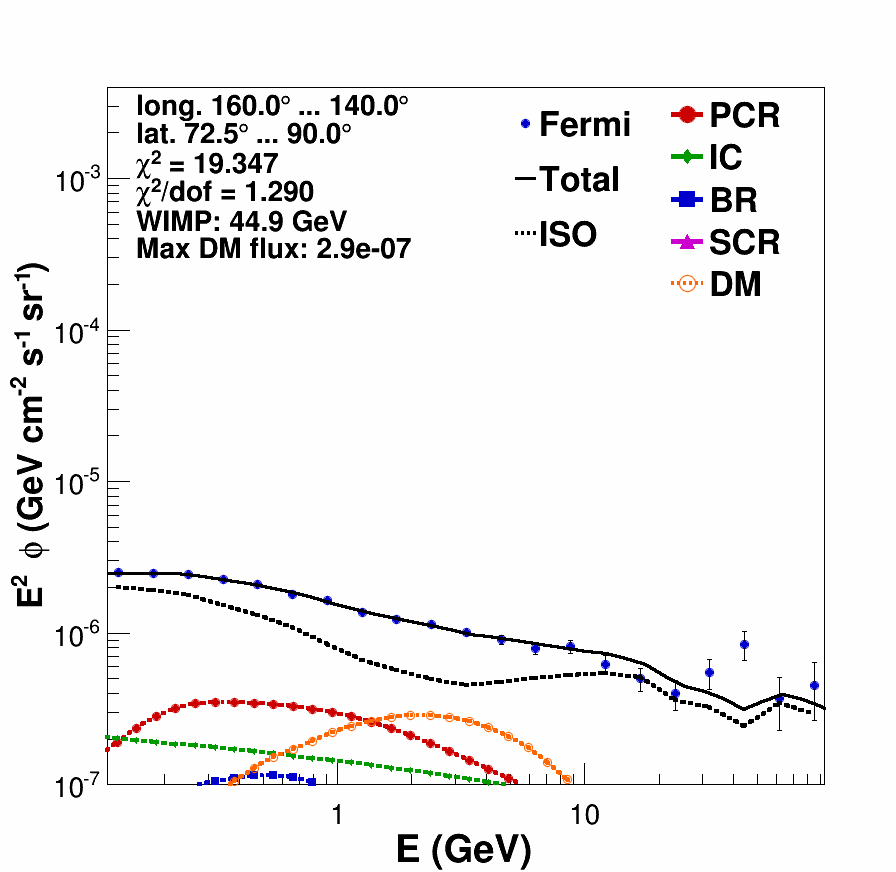}
\includegraphics[width=0.16\textwidth,height=0.16\textwidth,clip]{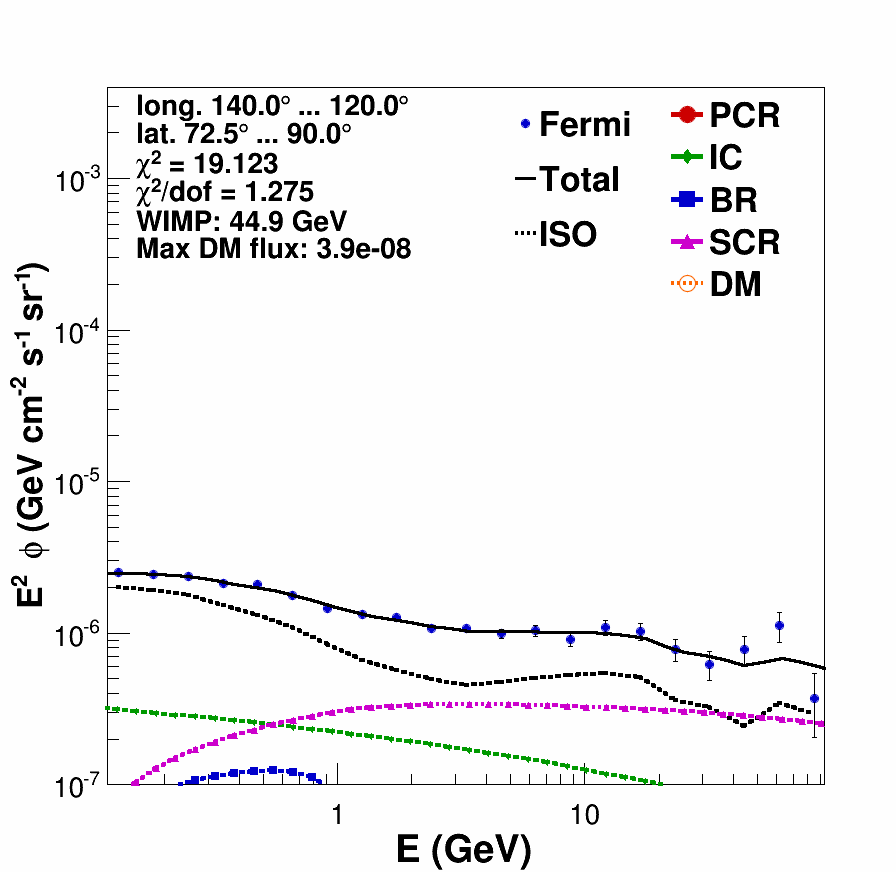}
\includegraphics[width=0.16\textwidth,height=0.16\textwidth,clip]{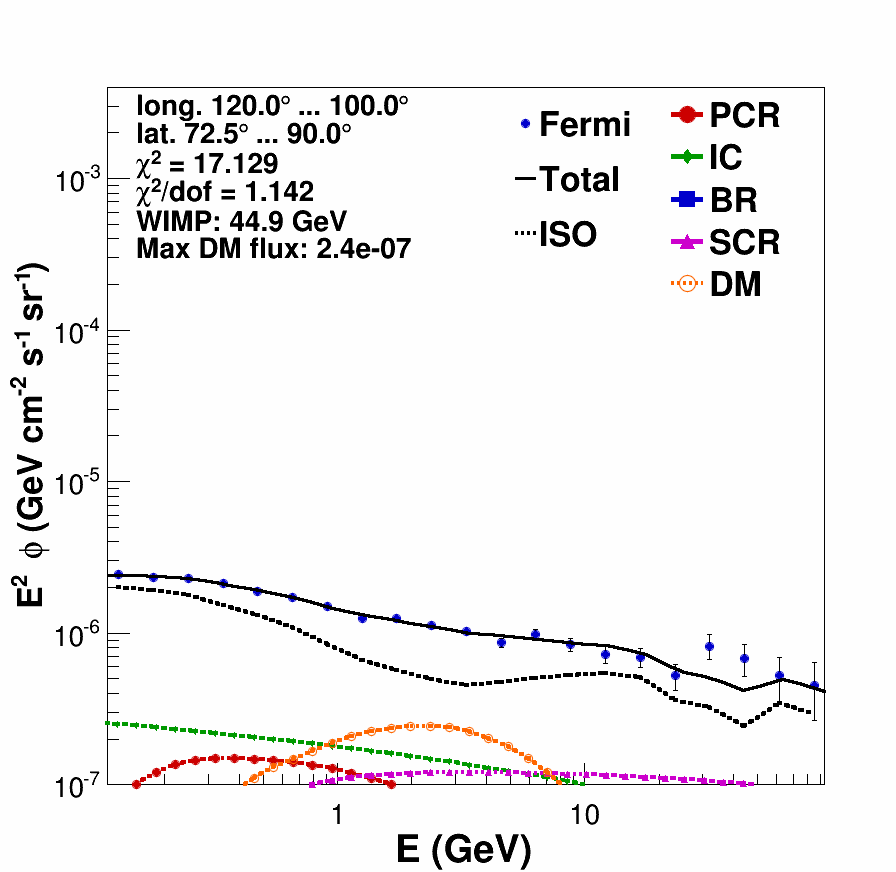}
\includegraphics[width=0.16\textwidth,height=0.16\textwidth,clip]{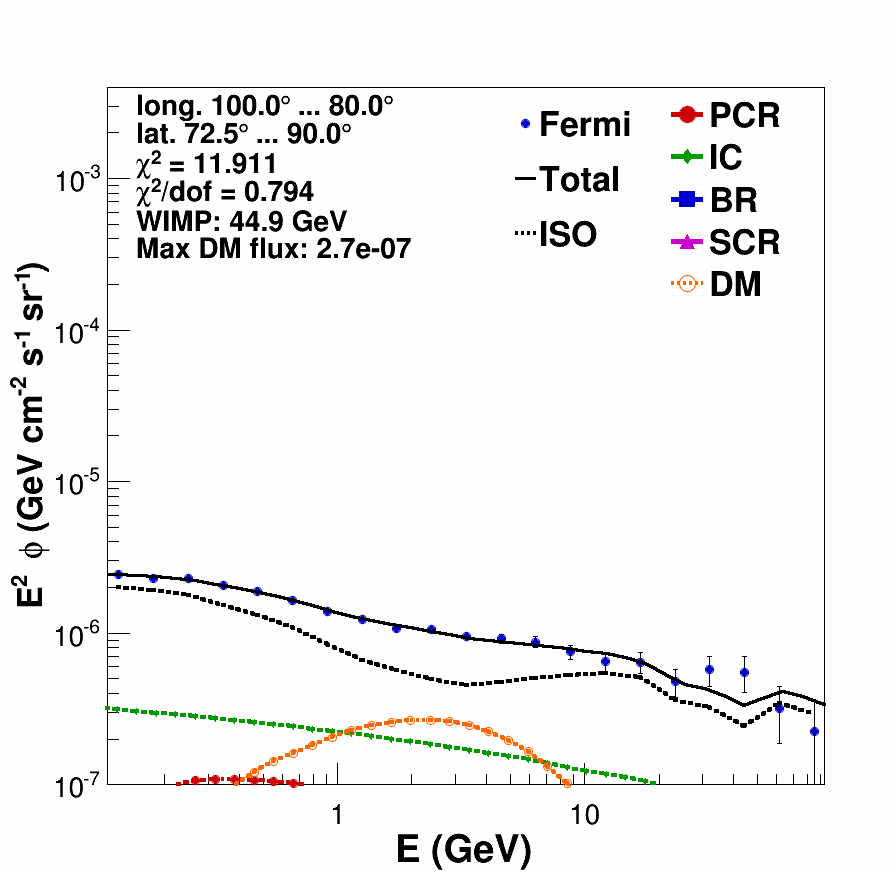}
\includegraphics[width=0.16\textwidth,height=0.16\textwidth,clip]{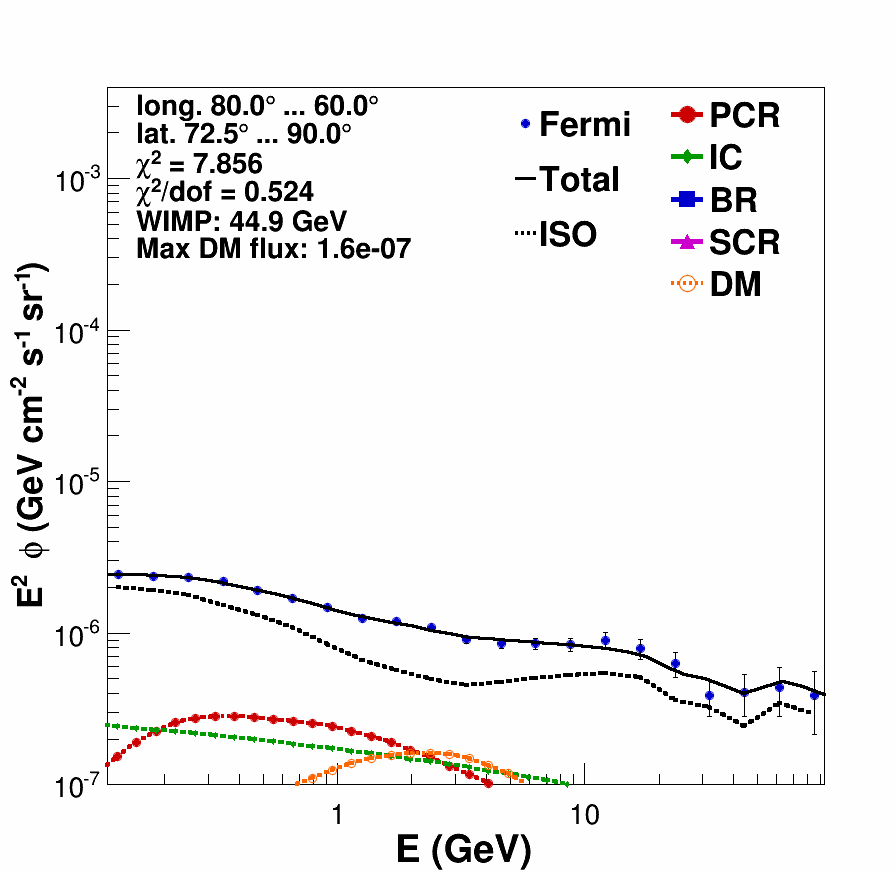}
\includegraphics[width=0.16\textwidth,height=0.16\textwidth,clip]{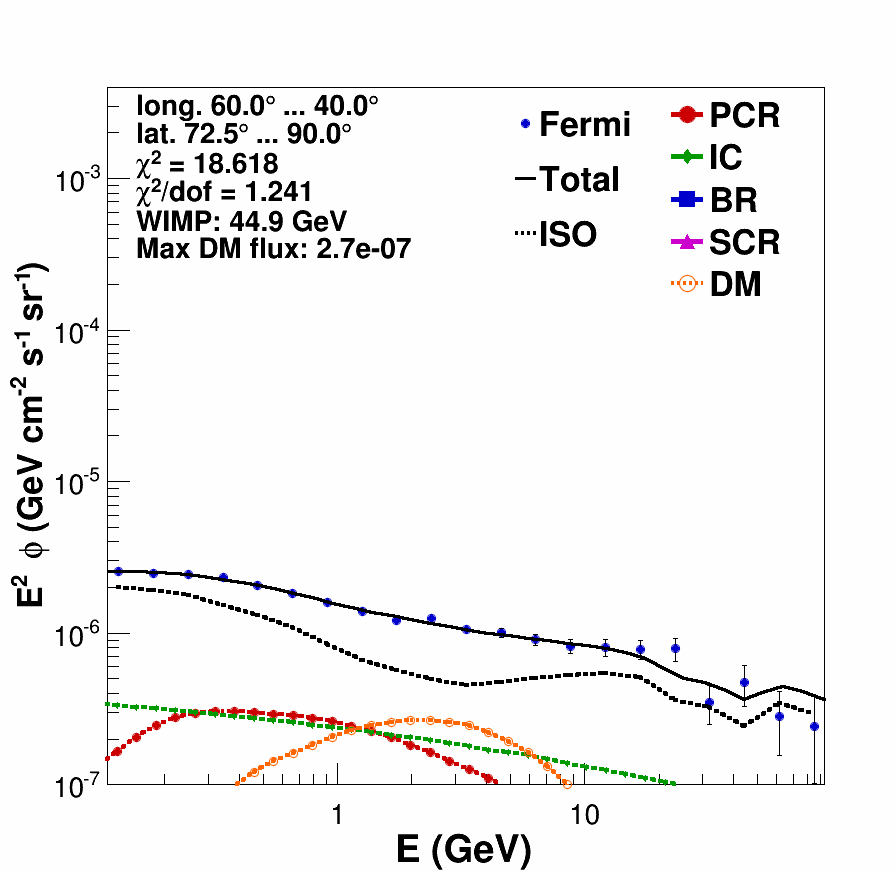}
\includegraphics[width=0.16\textwidth,height=0.16\textwidth,clip]{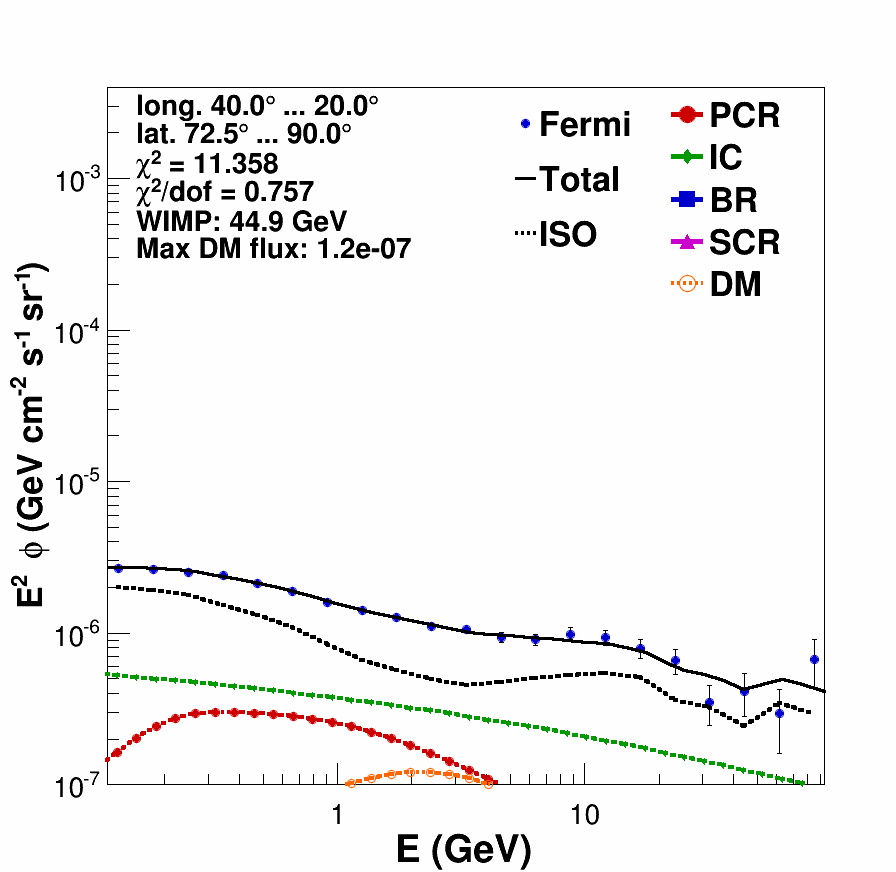}
\includegraphics[width=0.16\textwidth,height=0.16\textwidth,clip]{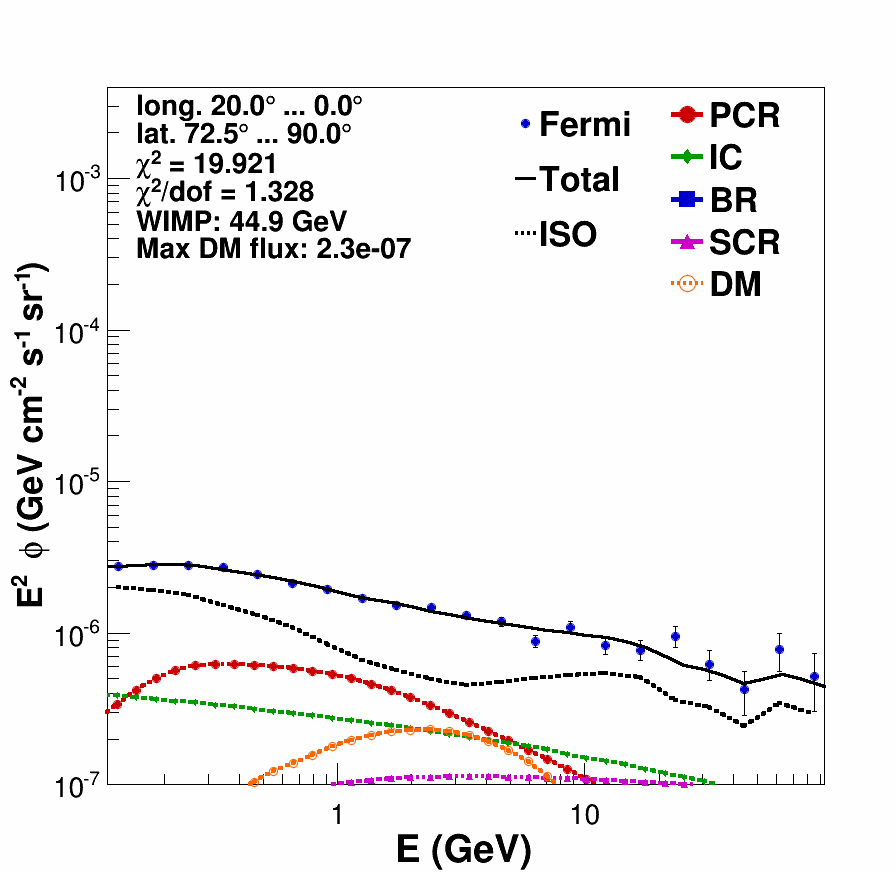}
\includegraphics[width=0.16\textwidth,height=0.16\textwidth,clip]{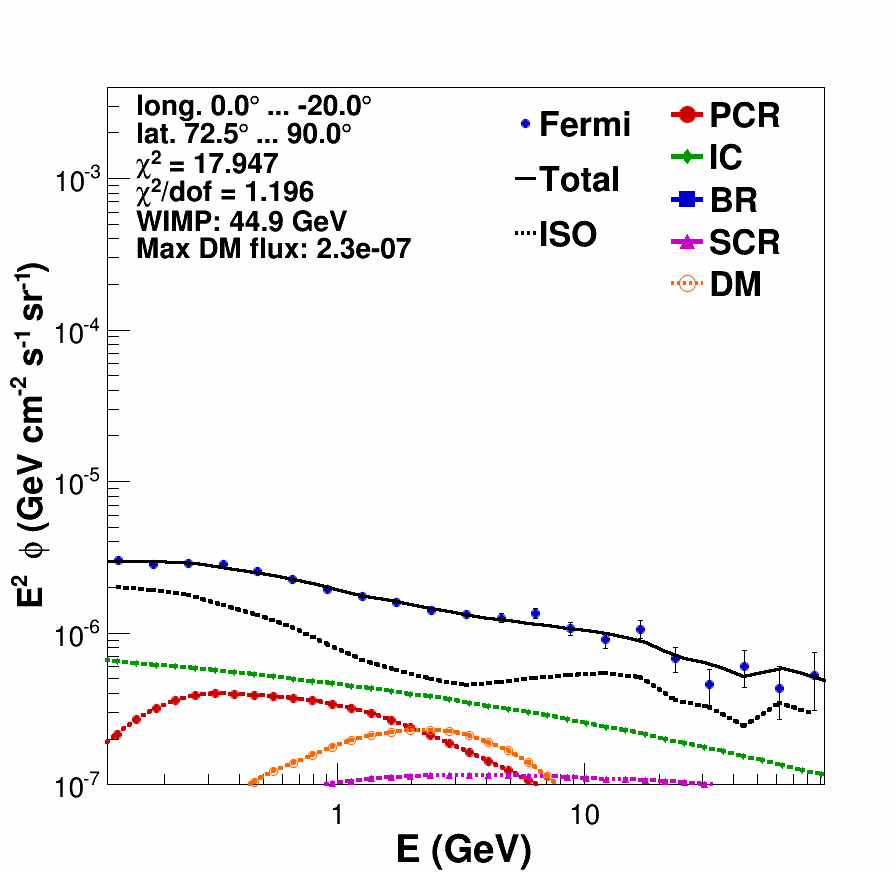}
\includegraphics[width=0.16\textwidth,height=0.16\textwidth,clip]{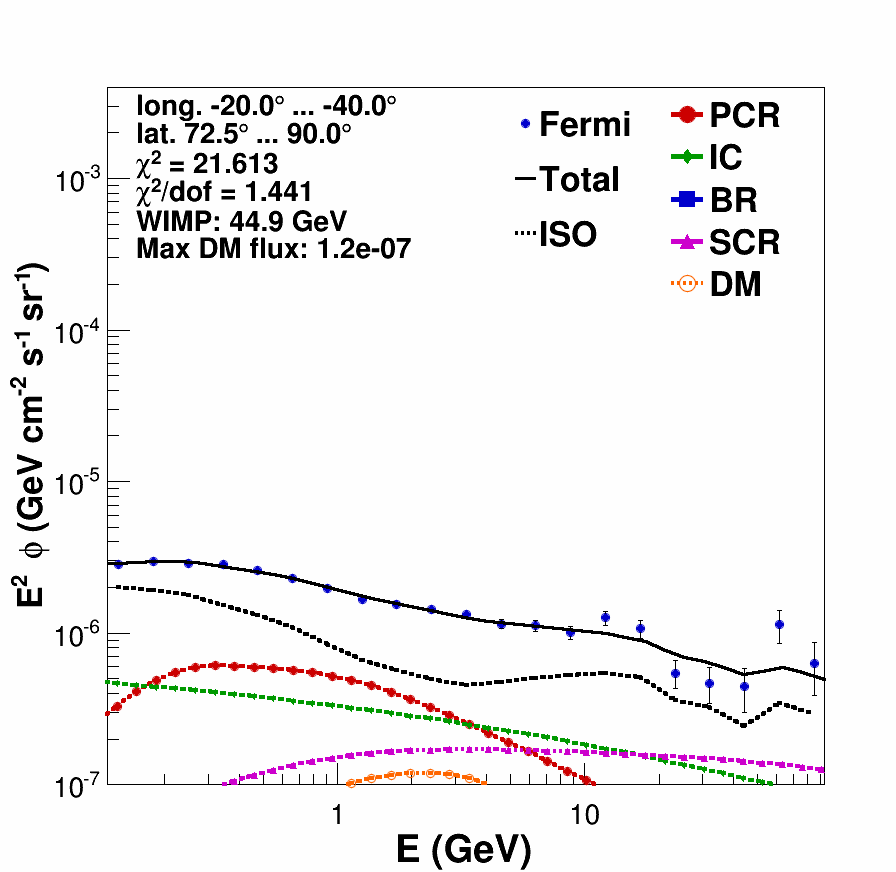}
\includegraphics[width=0.16\textwidth,height=0.16\textwidth,clip]{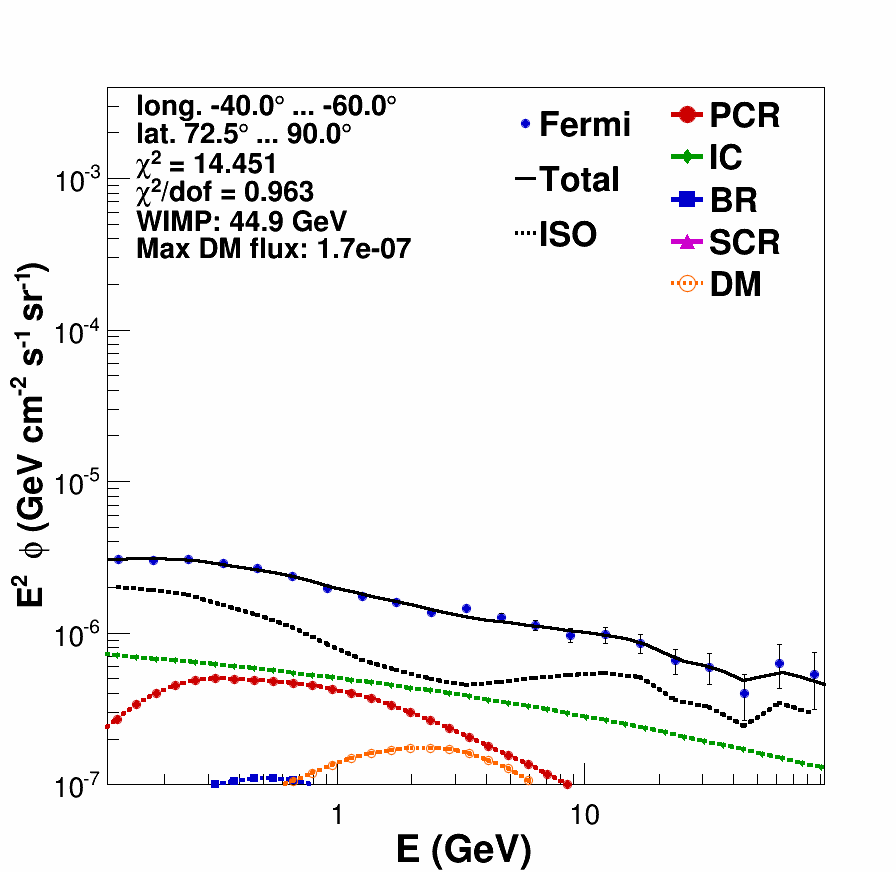}
\includegraphics[width=0.16\textwidth,height=0.16\textwidth,clip]{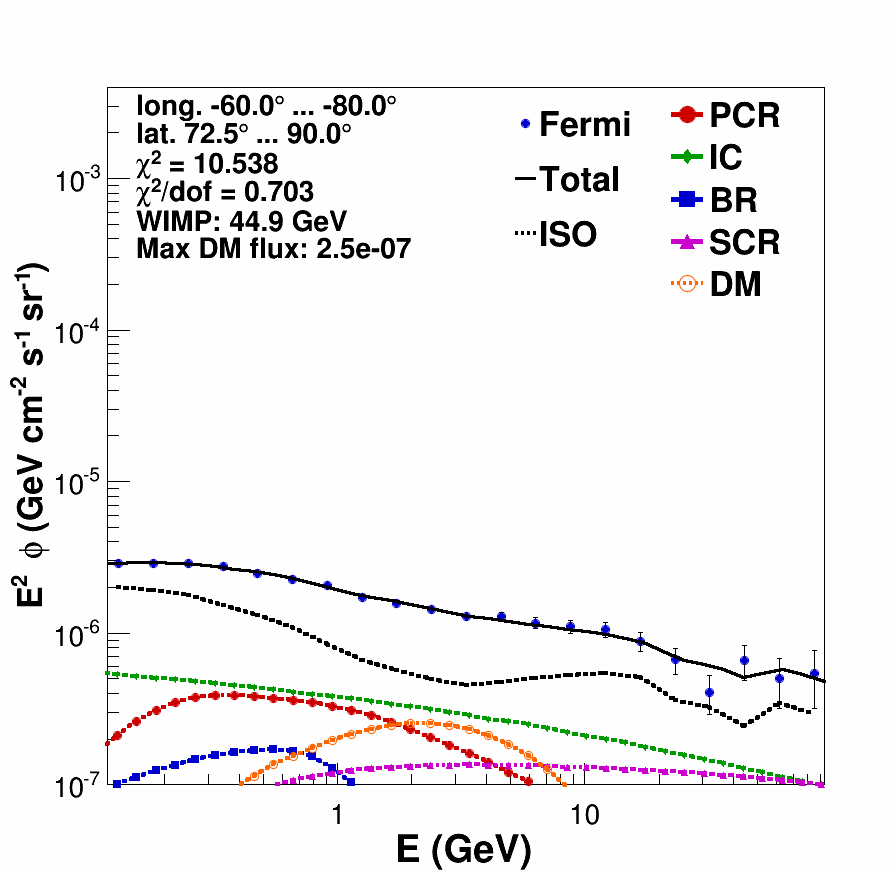}
\includegraphics[width=0.16\textwidth,height=0.16\textwidth,clip]{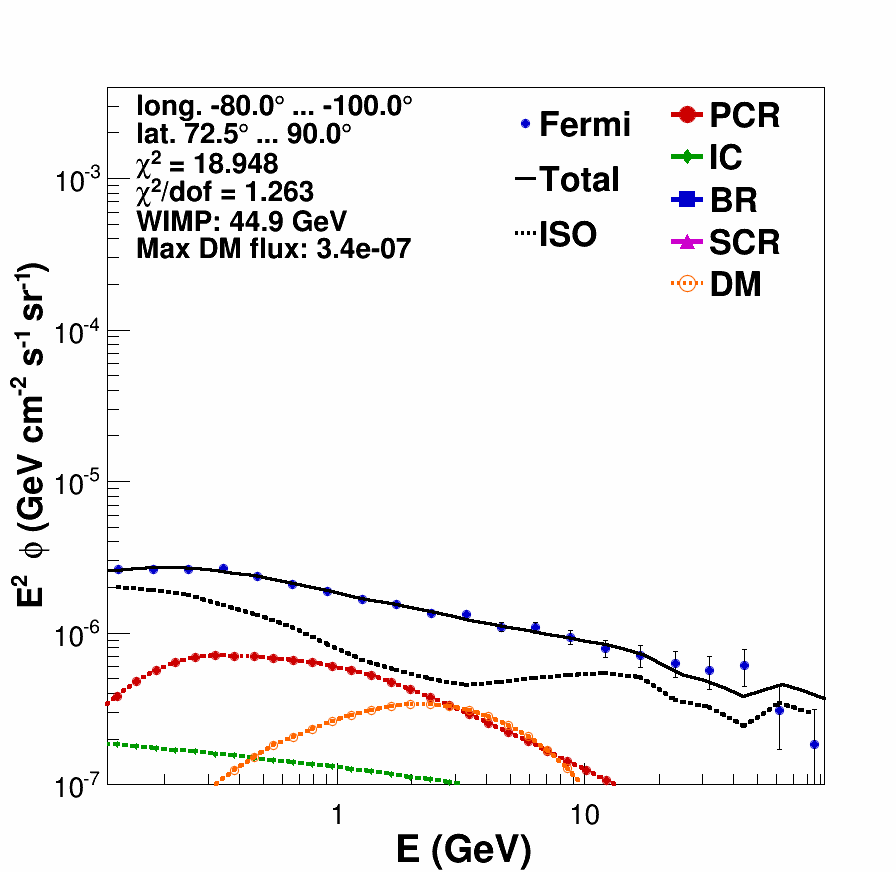}
\includegraphics[width=0.16\textwidth,height=0.16\textwidth,clip]{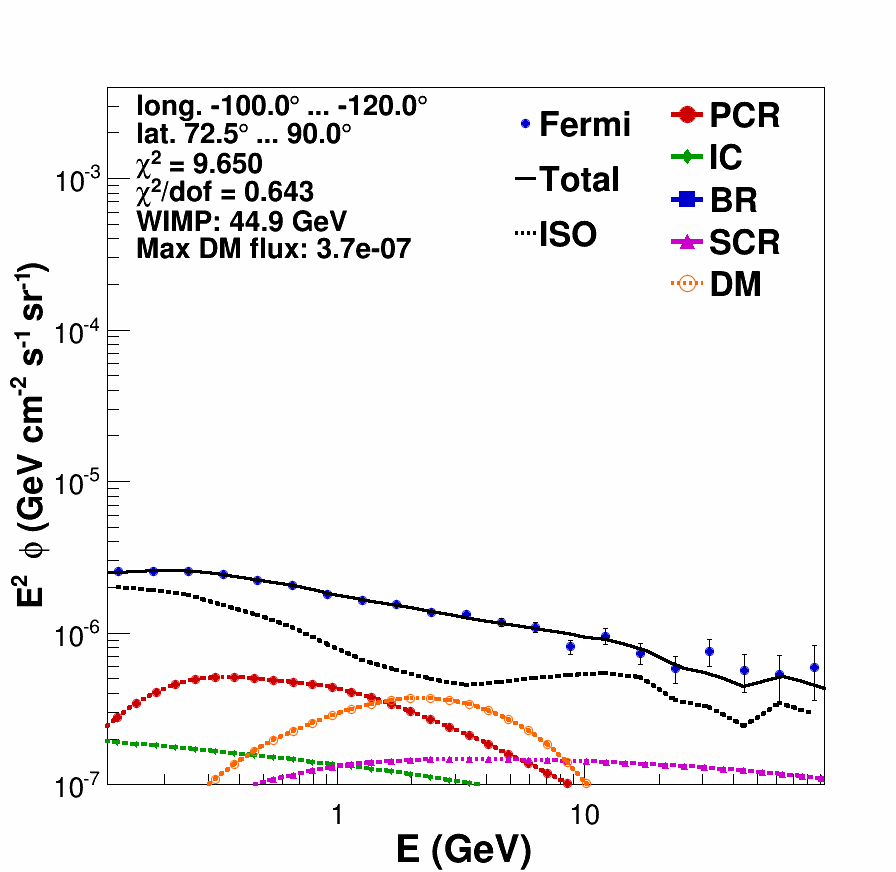}
\includegraphics[width=0.16\textwidth,height=0.16\textwidth,clip]{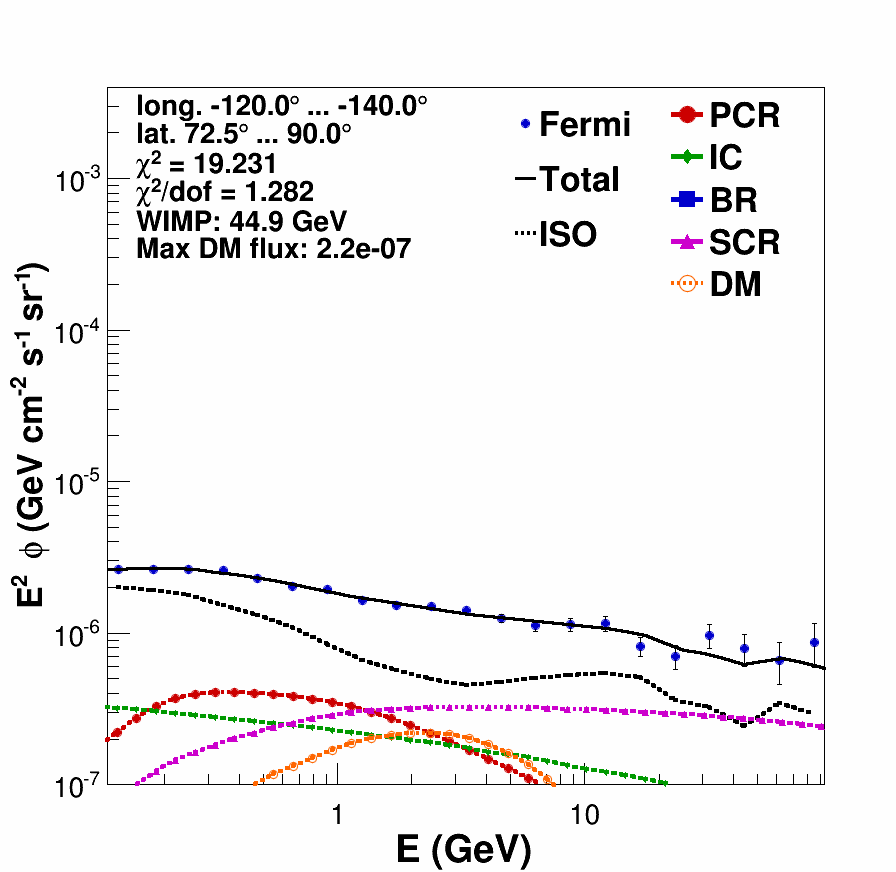}
\includegraphics[width=0.16\textwidth,height=0.16\textwidth,clip]{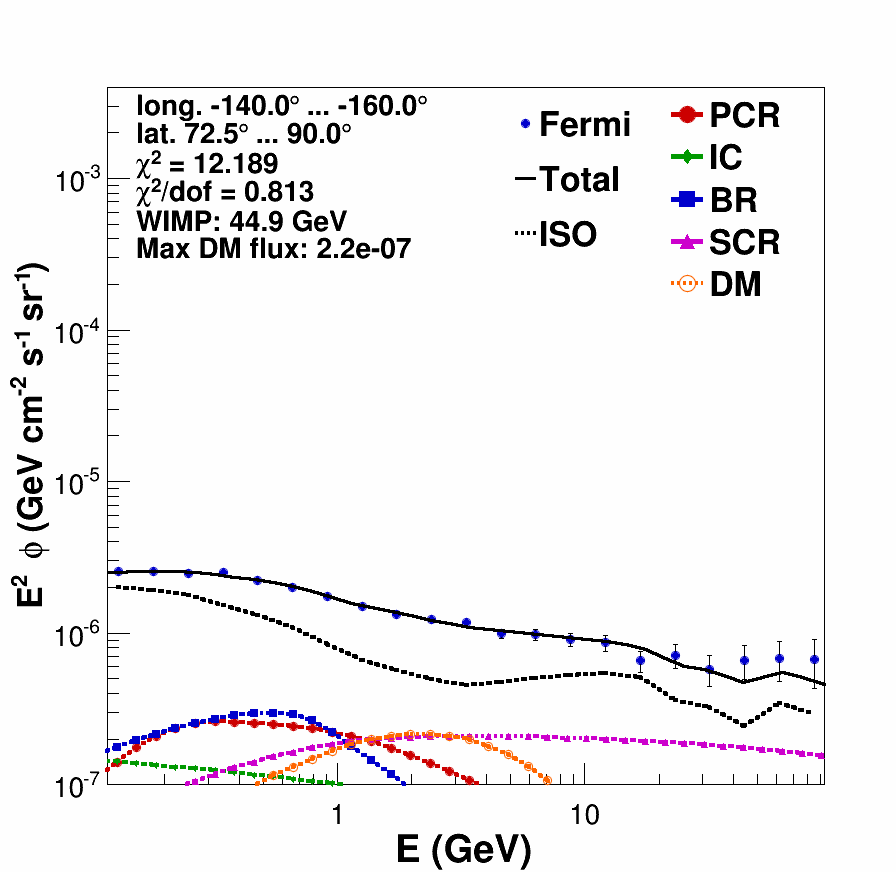}
\includegraphics[width=0.16\textwidth,height=0.16\textwidth,clip]{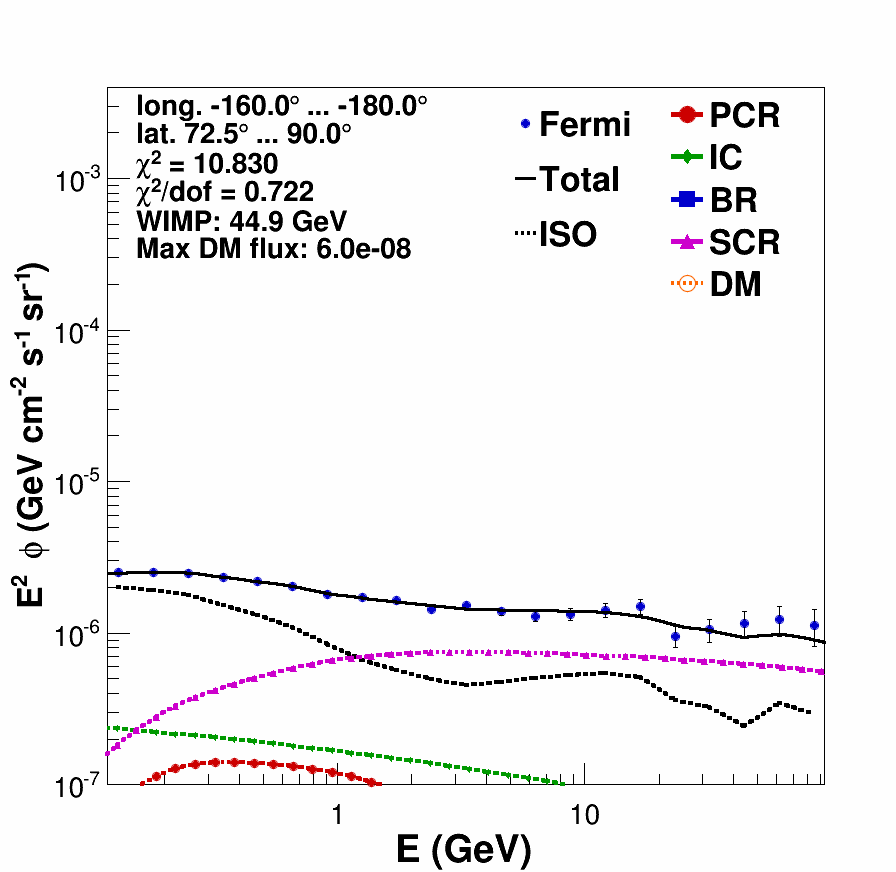}    %\\[2cm]%%%%%% first row
\caption[]{Template fits for latitudes  with $72.5^\circ<b<90.0^\circ$ and longitudes decreasing from 180$^\circ$ to -180$^\circ$.} 
\label{F32}
\end{figure}
\begin{figure}
\includegraphics[width=0.16\textwidth,height=0.16\textwidth,clip]{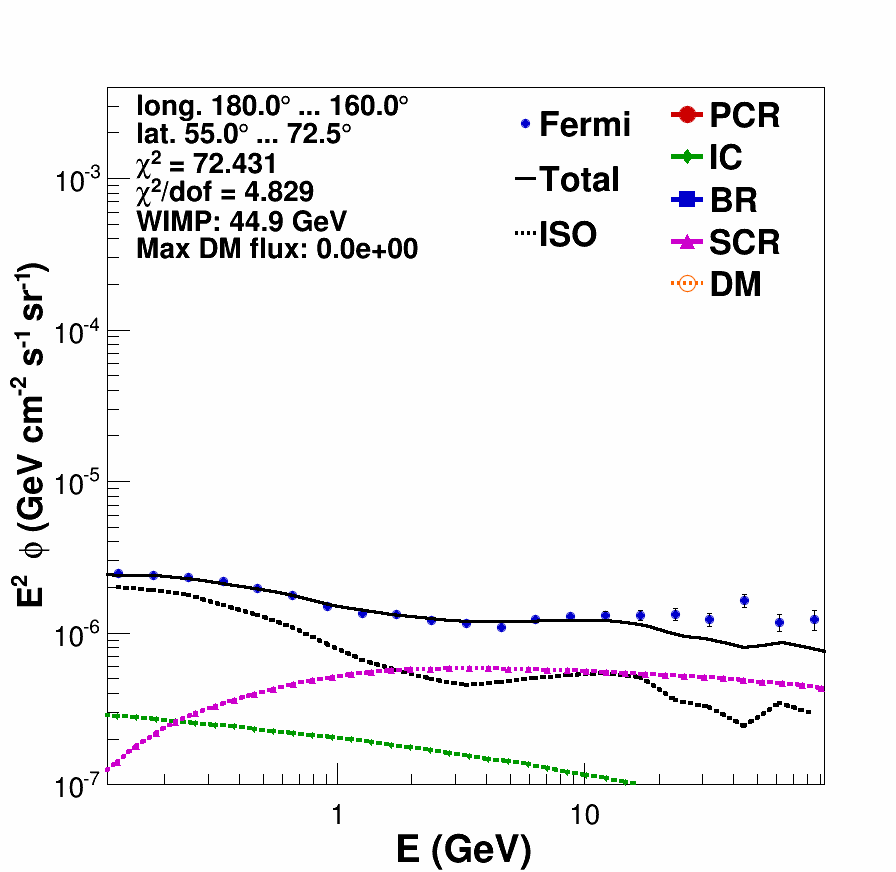}
\includegraphics[width=0.16\textwidth,height=0.16\textwidth,clip]{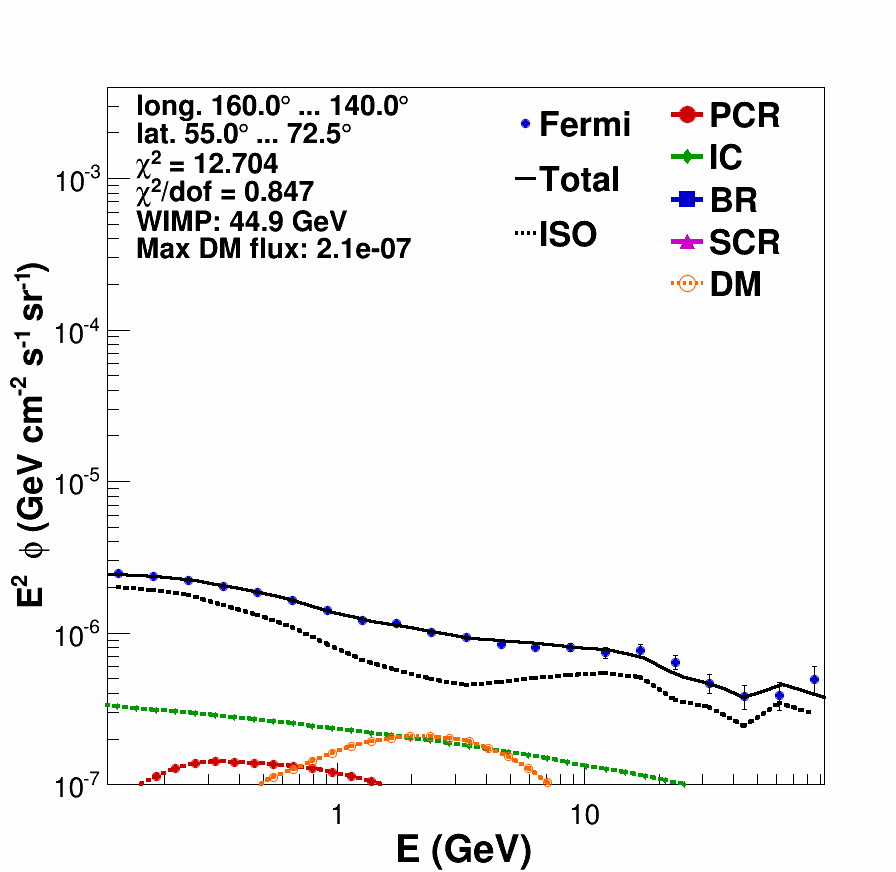}
\includegraphics[width=0.16\textwidth,height=0.16\textwidth,clip]{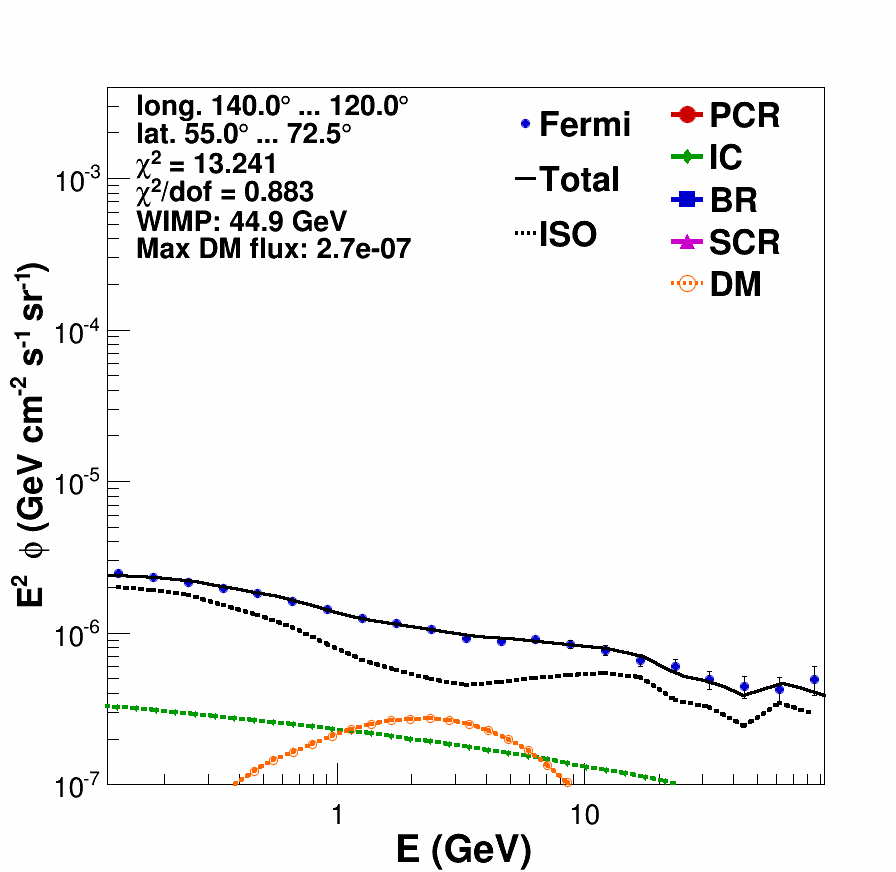}
\includegraphics[width=0.16\textwidth,height=0.16\textwidth,clip]{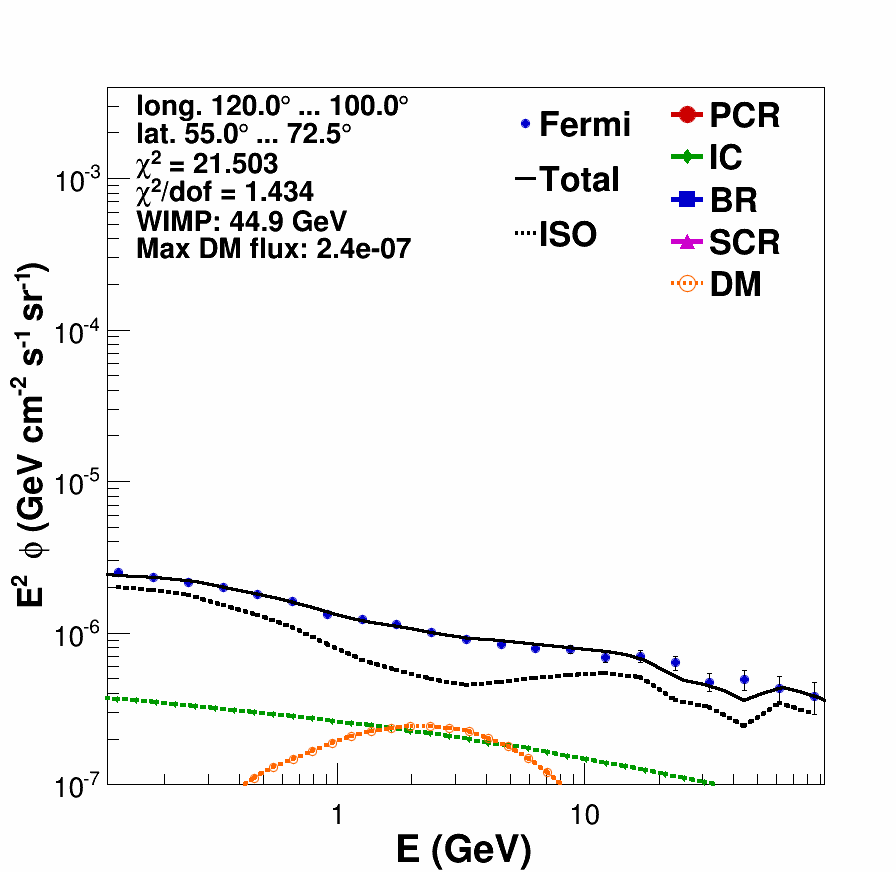}
\includegraphics[width=0.16\textwidth,height=0.16\textwidth,clip]{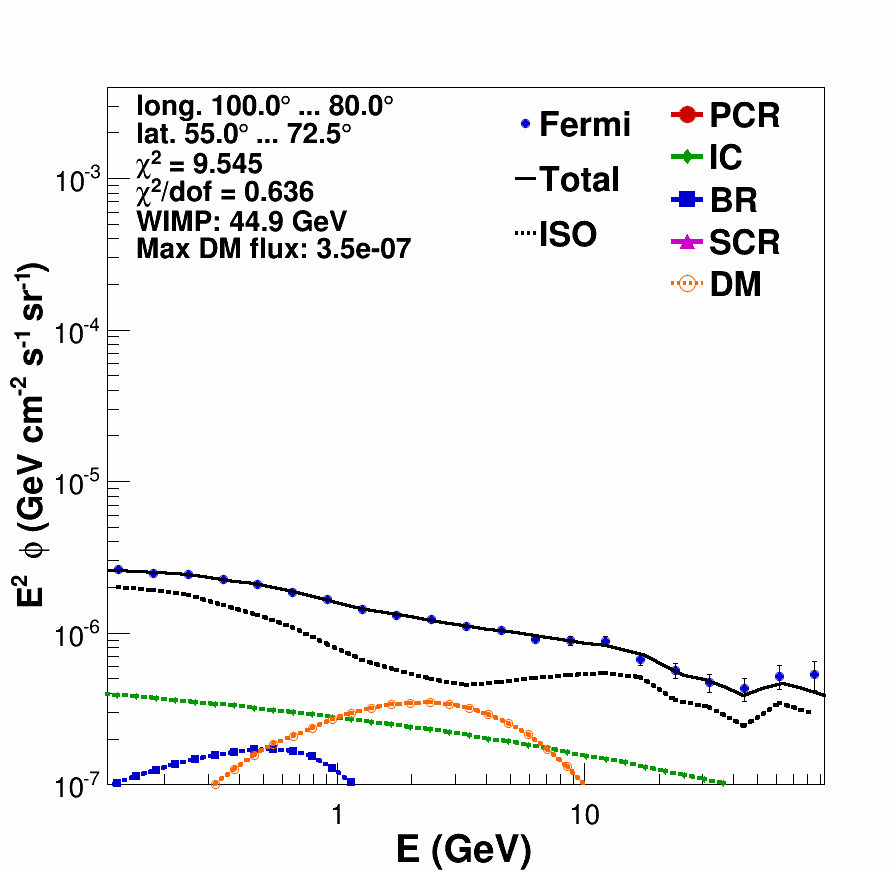}
\includegraphics[width=0.16\textwidth,height=0.16\textwidth,clip]{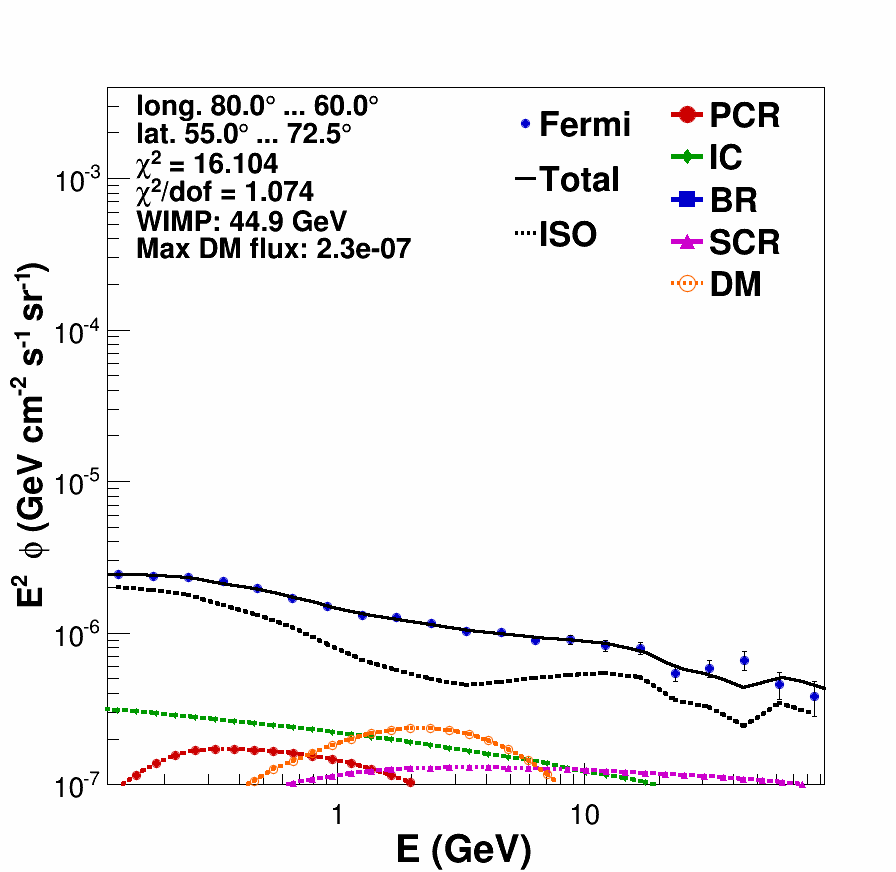}
\includegraphics[width=0.16\textwidth,height=0.16\textwidth,clip]{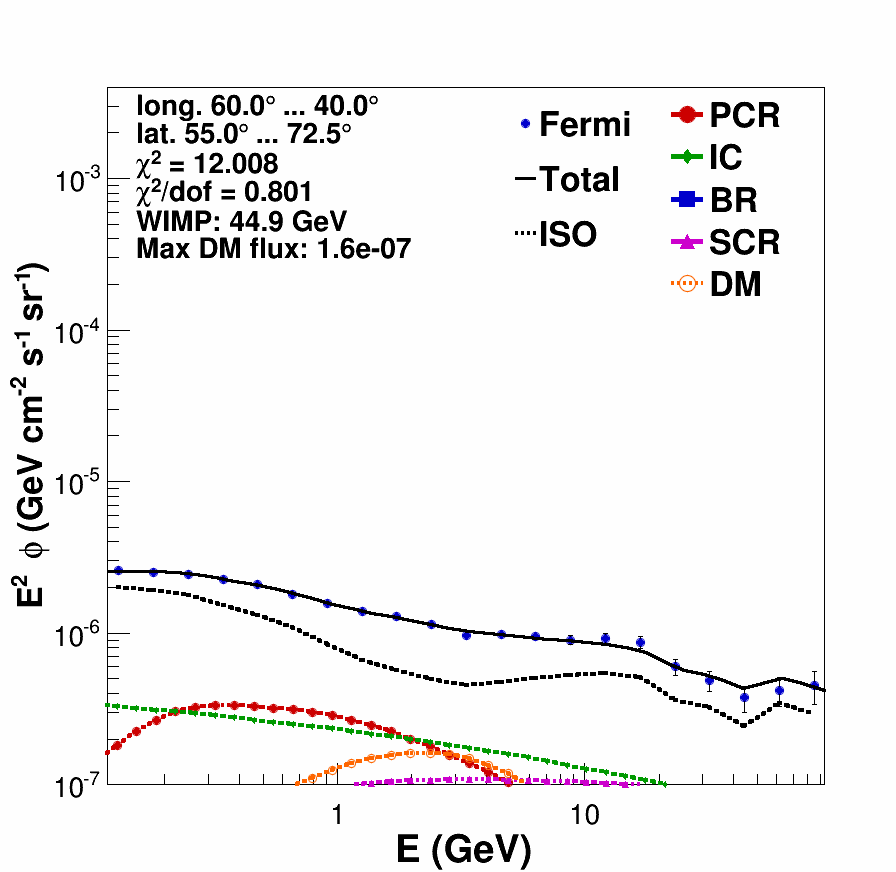}
\includegraphics[width=0.16\textwidth,height=0.16\textwidth,clip]{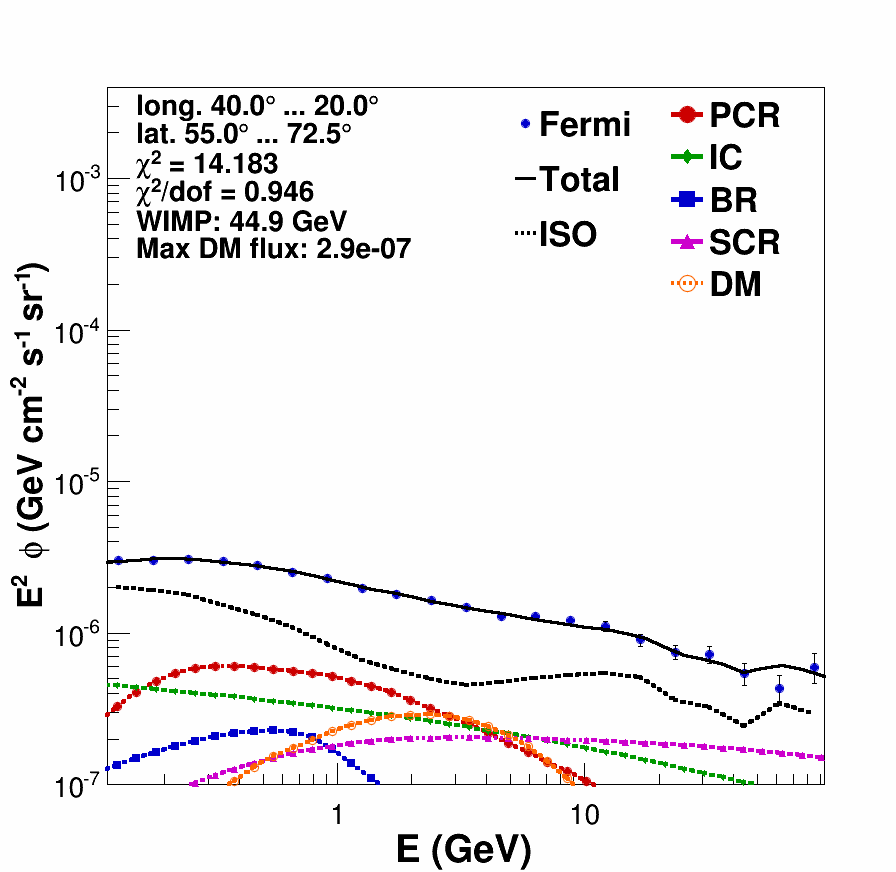}
\includegraphics[width=0.16\textwidth,height=0.16\textwidth,clip]{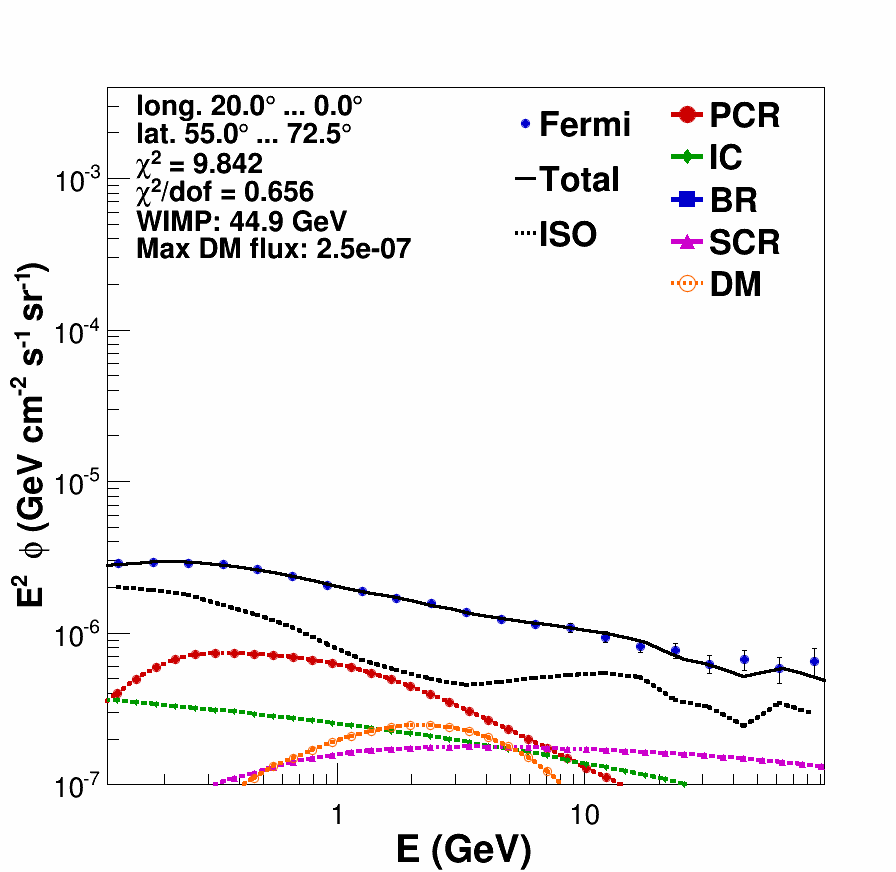}
\includegraphics[width=0.16\textwidth,height=0.16\textwidth,clip]{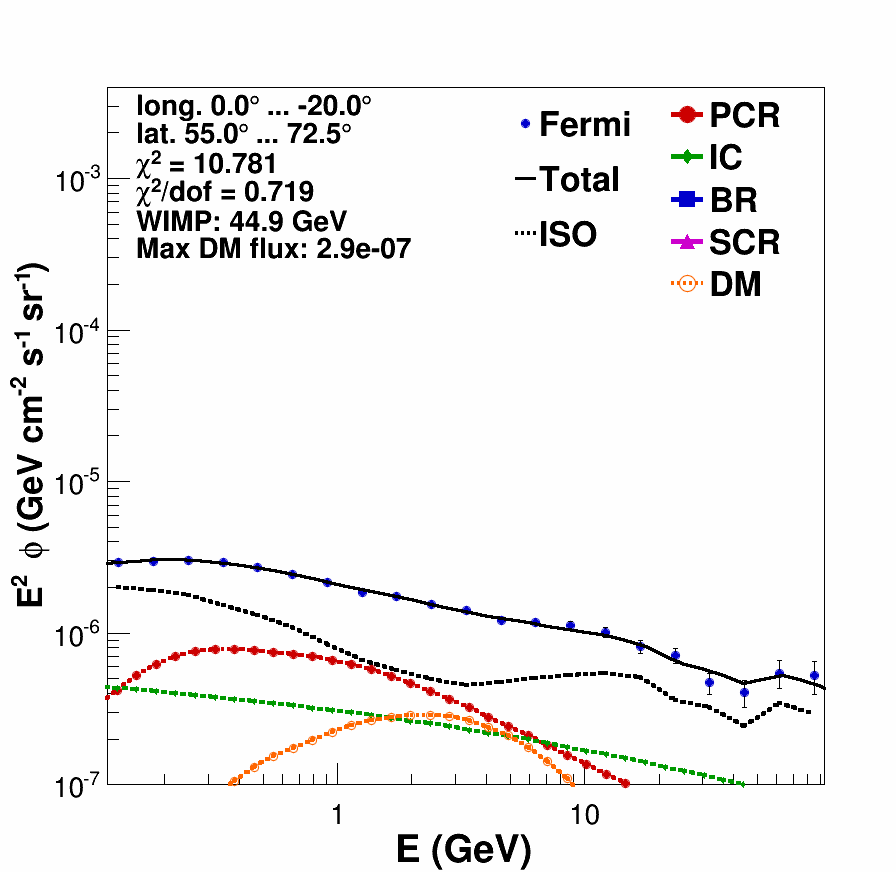}
\includegraphics[width=0.16\textwidth,height=0.16\textwidth,clip]{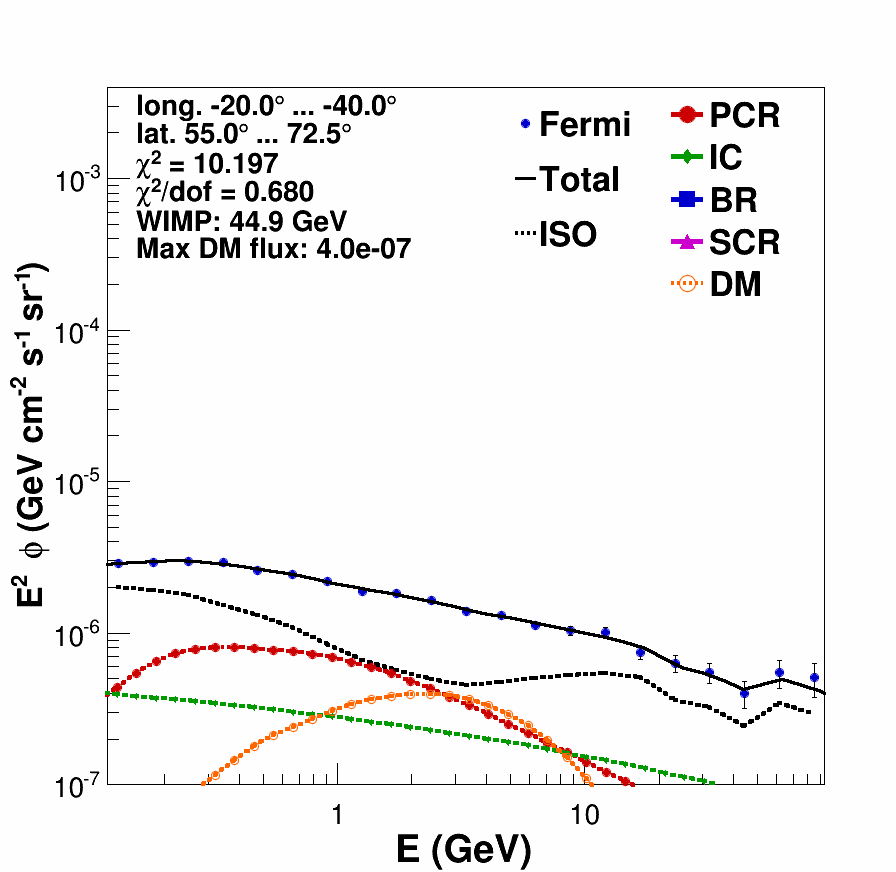}
\includegraphics[width=0.16\textwidth,height=0.16\textwidth,clip]{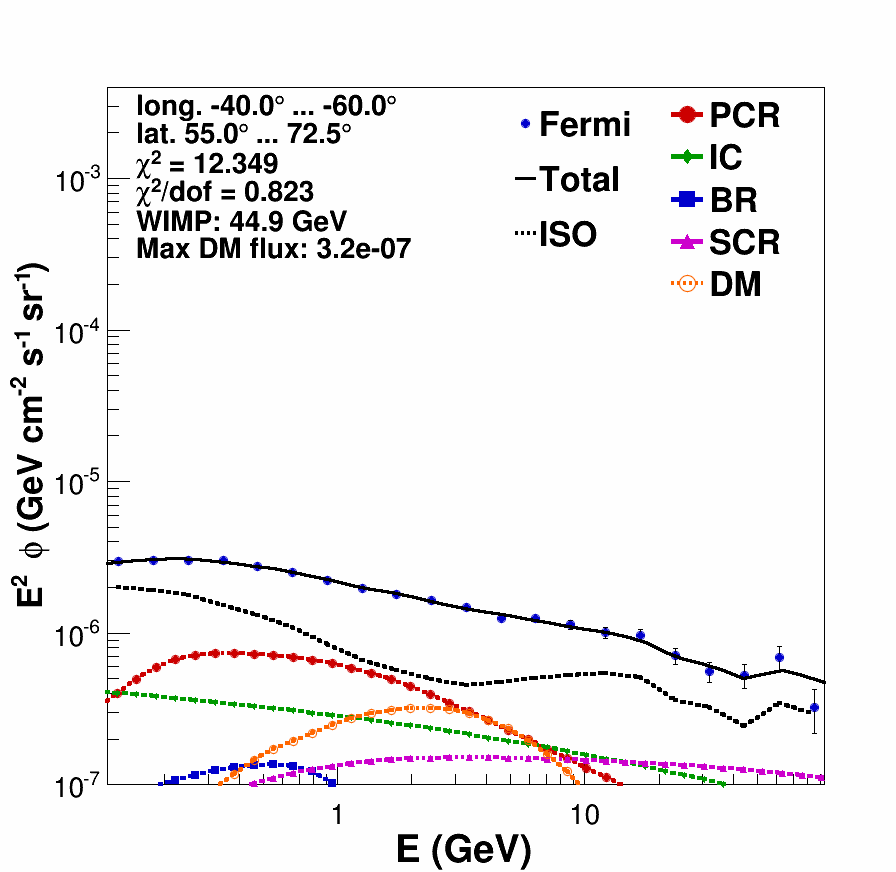}
\includegraphics[width=0.16\textwidth,height=0.16\textwidth,clip]{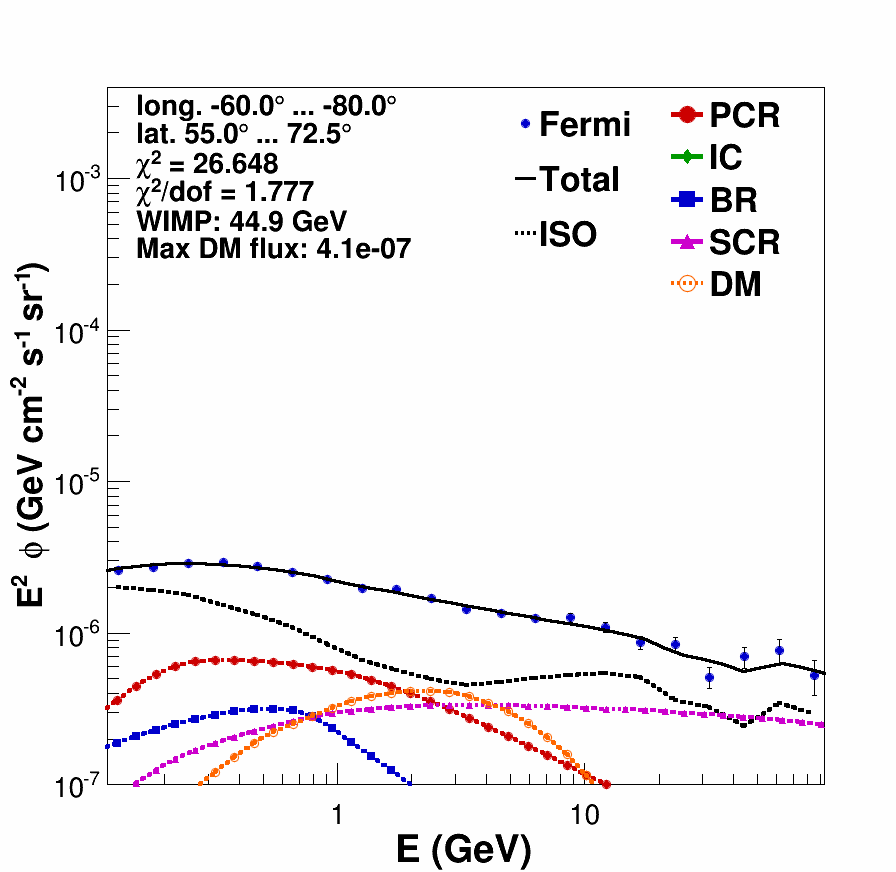}
\includegraphics[width=0.16\textwidth,height=0.16\textwidth,clip]{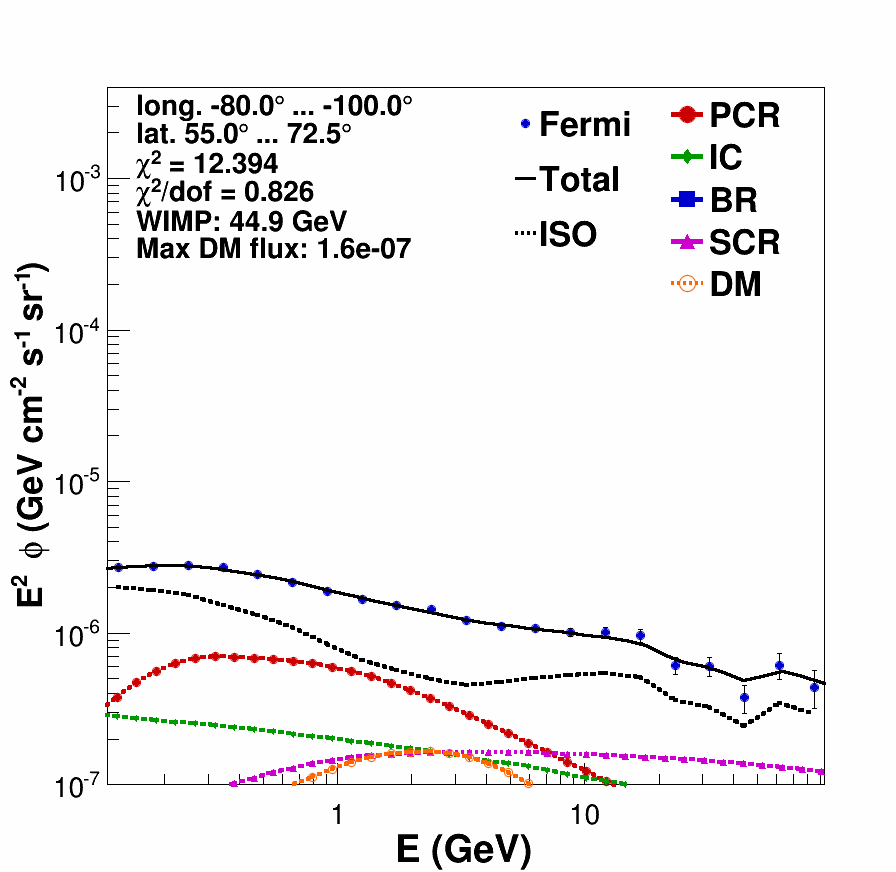}
\includegraphics[width=0.16\textwidth,height=0.16\textwidth,clip]{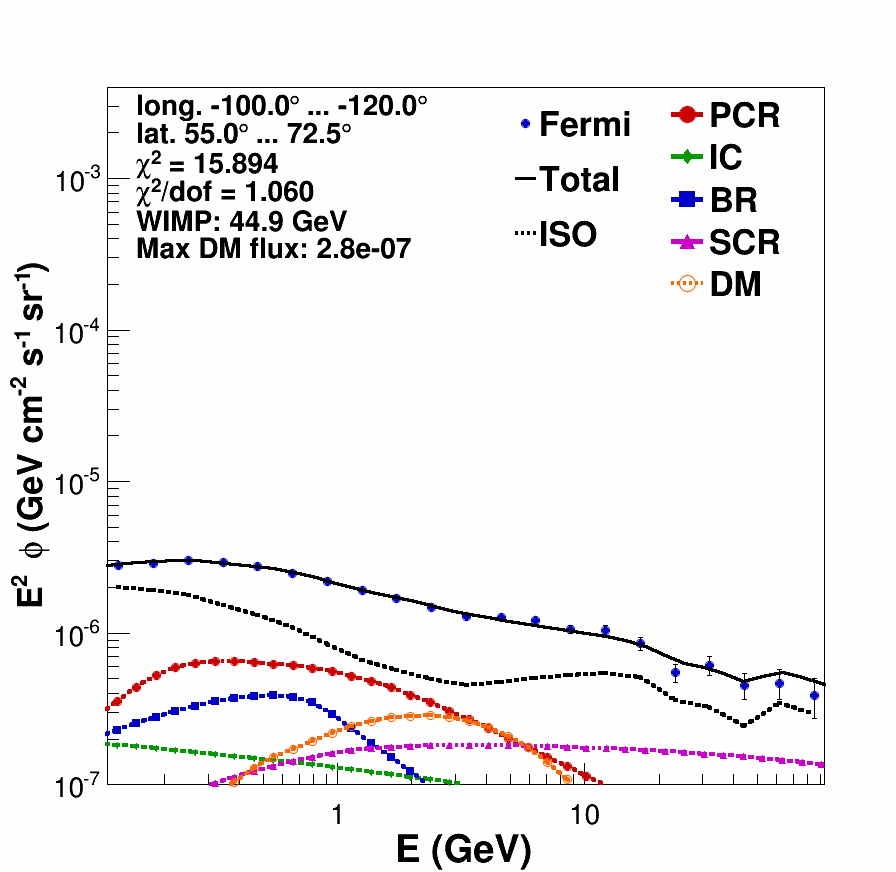}
\includegraphics[width=0.16\textwidth,height=0.16\textwidth,clip]{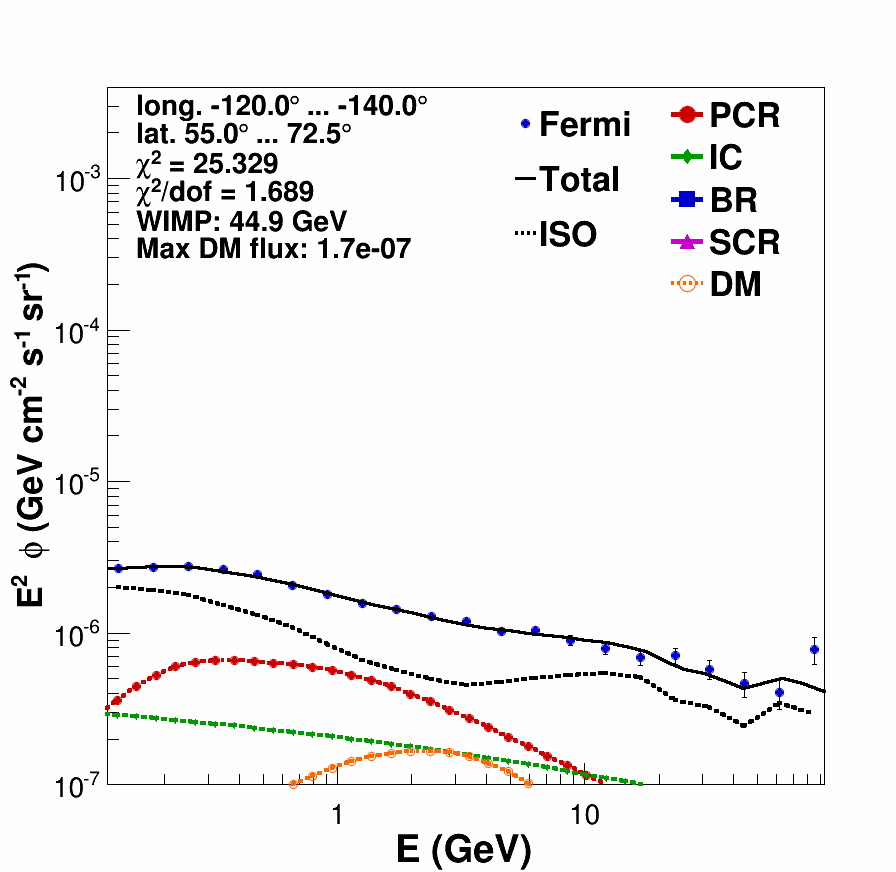}
\includegraphics[width=0.16\textwidth,height=0.16\textwidth,clip]{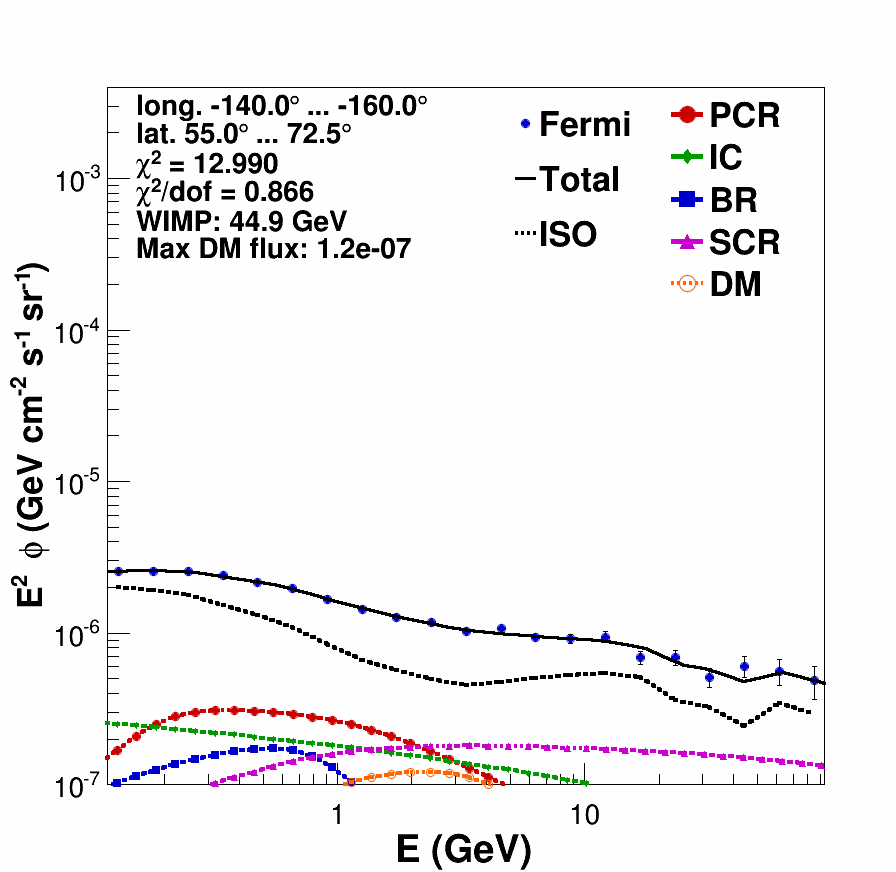}
\includegraphics[width=0.16\textwidth,height=0.16\textwidth,clip]{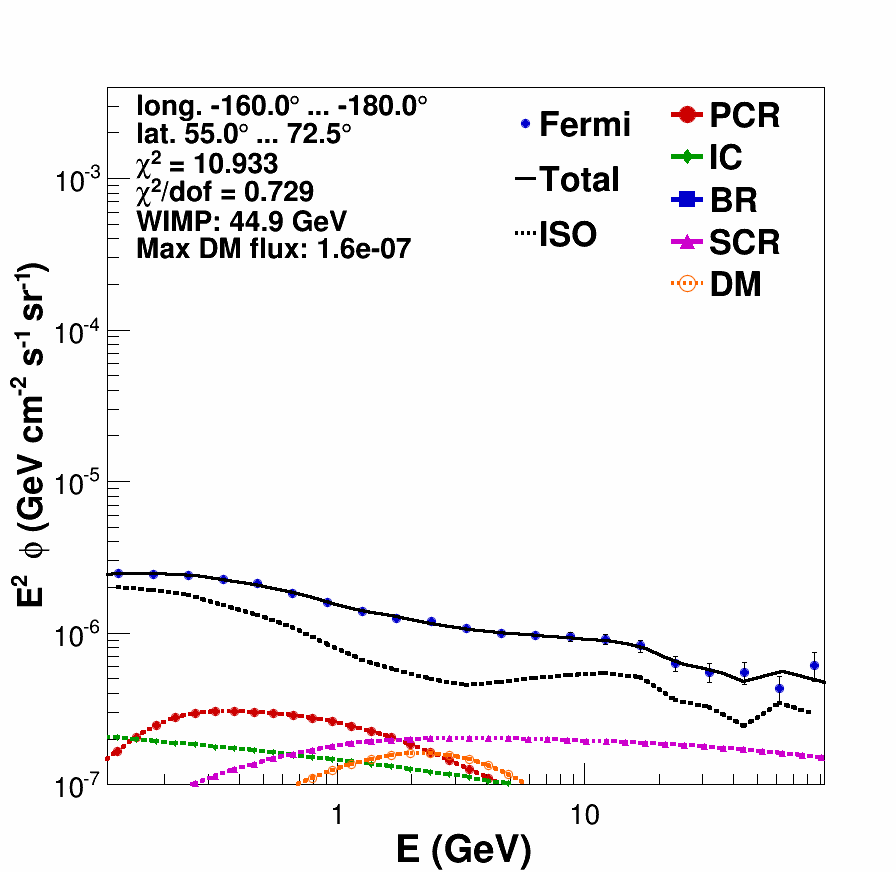}%\\%%%%%r2
\caption[]{Template fits for latitudes  with $55.0^\circ<b<72.5^\circ$ and longitudes decreasing from 180$^\circ$ to -180$^\circ$.} 
\label{F33}
\end{figure}
\begin{figure}
\centering
\includegraphics[width=0.16\textwidth,height=0.16\textwidth,clip]{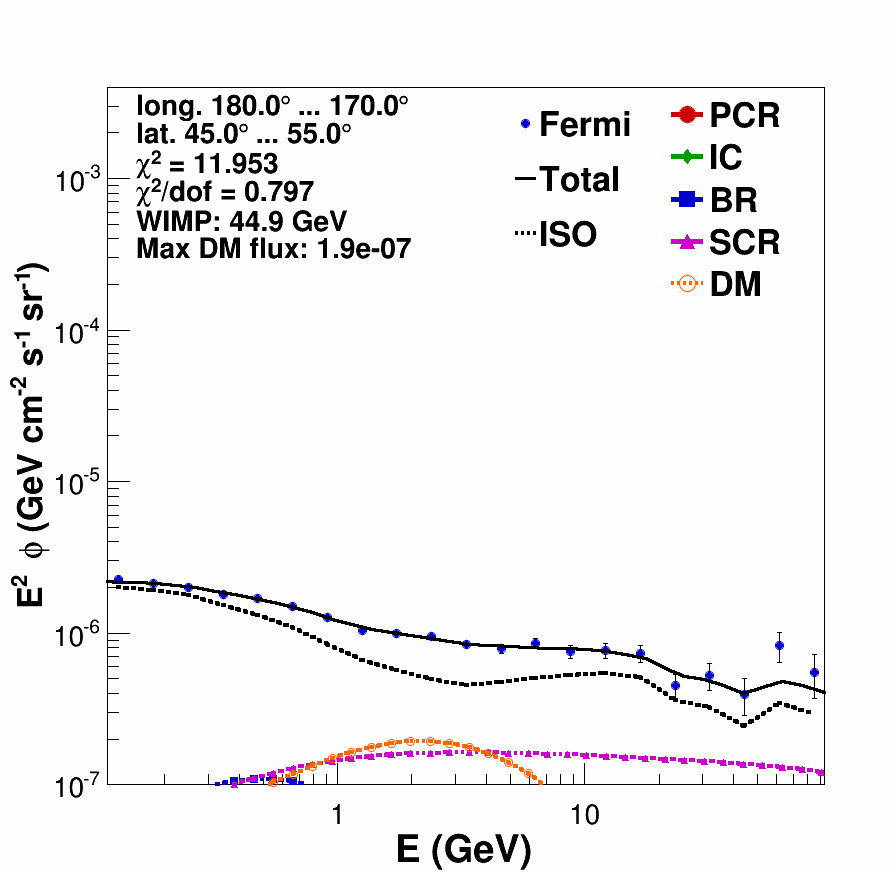}
\includegraphics[width=0.16\textwidth,height=0.16\textwidth,clip]{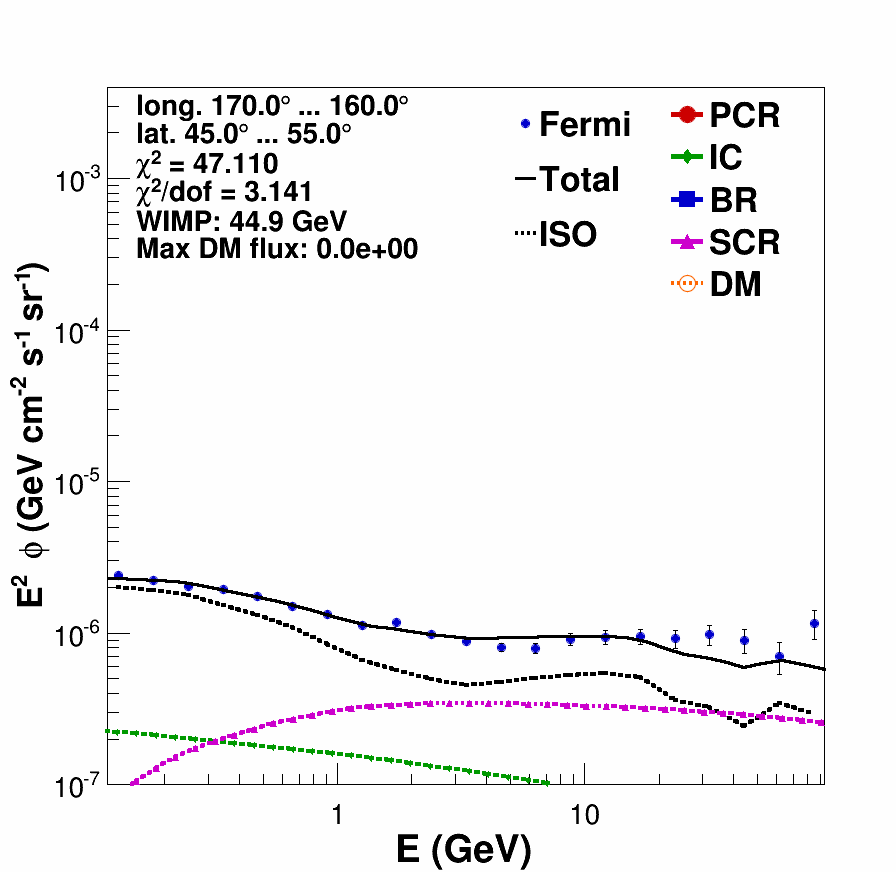}
\includegraphics[width=0.16\textwidth,height=0.16\textwidth,clip]{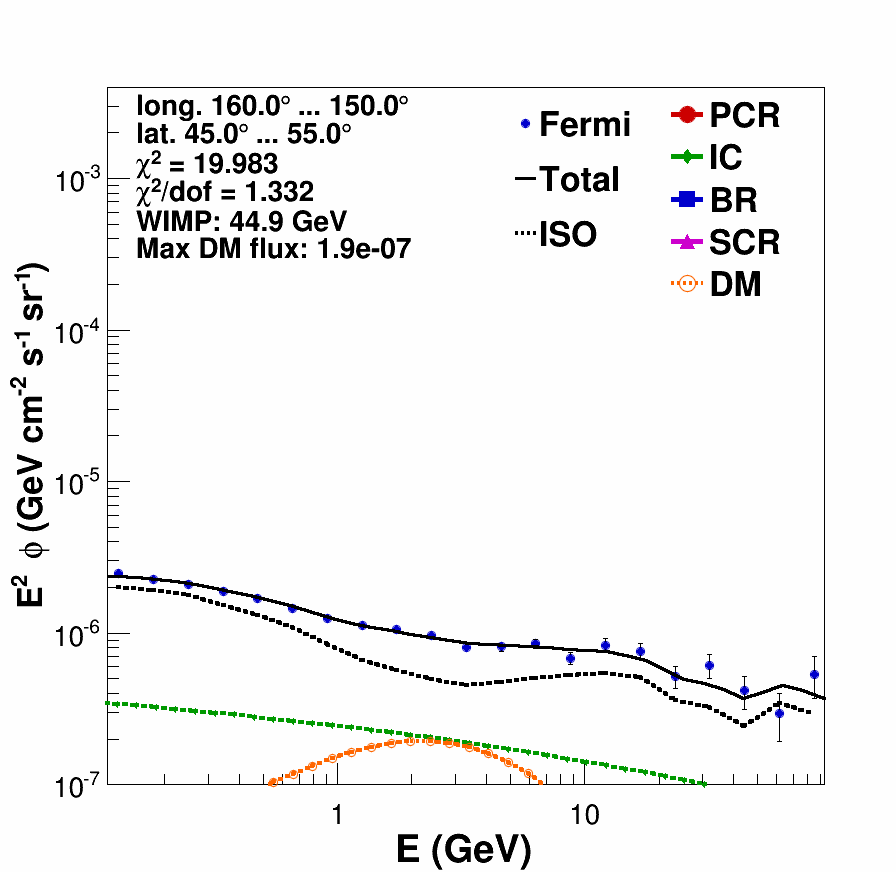}
\includegraphics[width=0.16\textwidth,height=0.16\textwidth,clip]{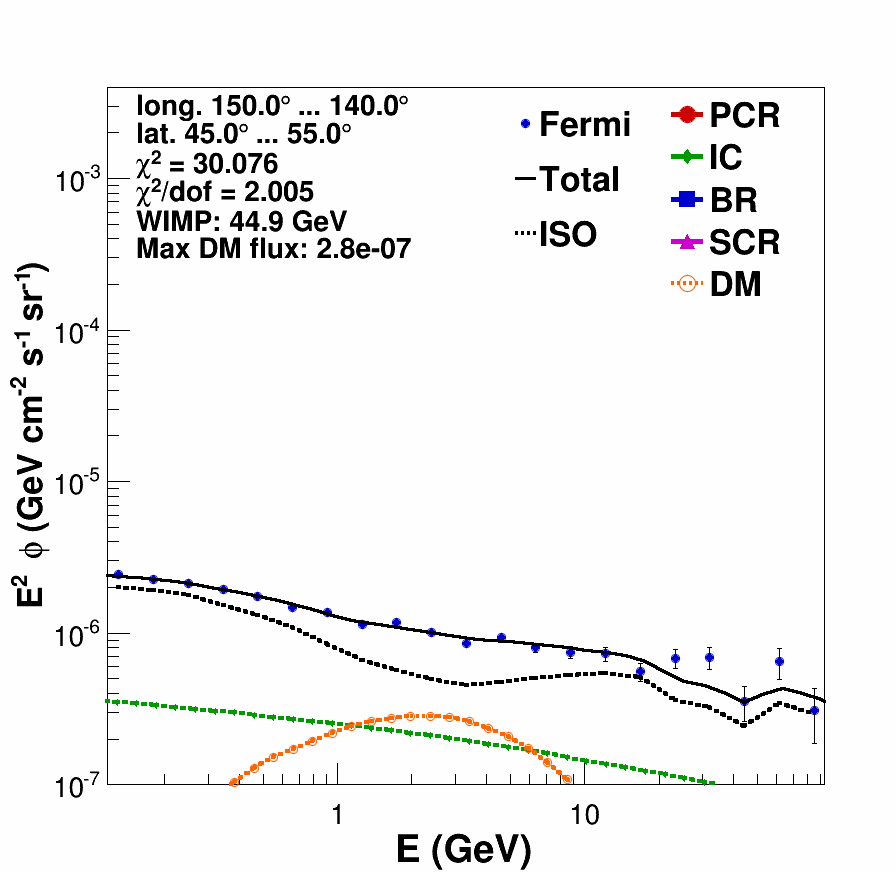}
\includegraphics[width=0.16\textwidth,height=0.16\textwidth,clip]{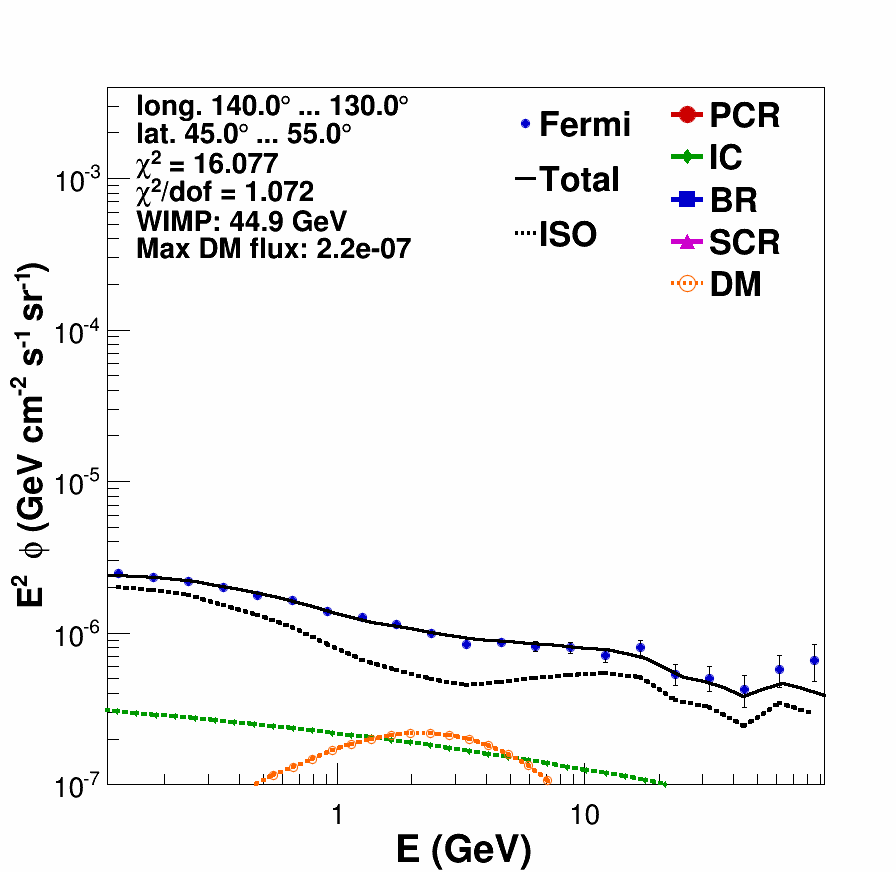}
\includegraphics[width=0.16\textwidth,height=0.16\textwidth,clip]{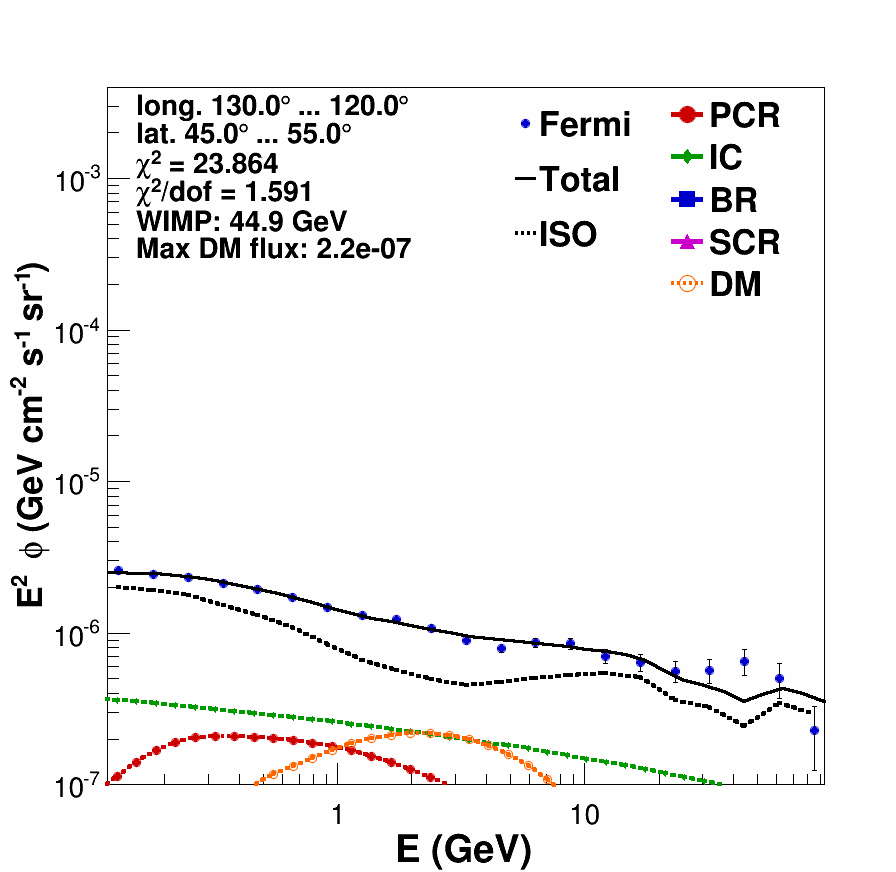}
\includegraphics[width=0.16\textwidth,height=0.16\textwidth,clip]{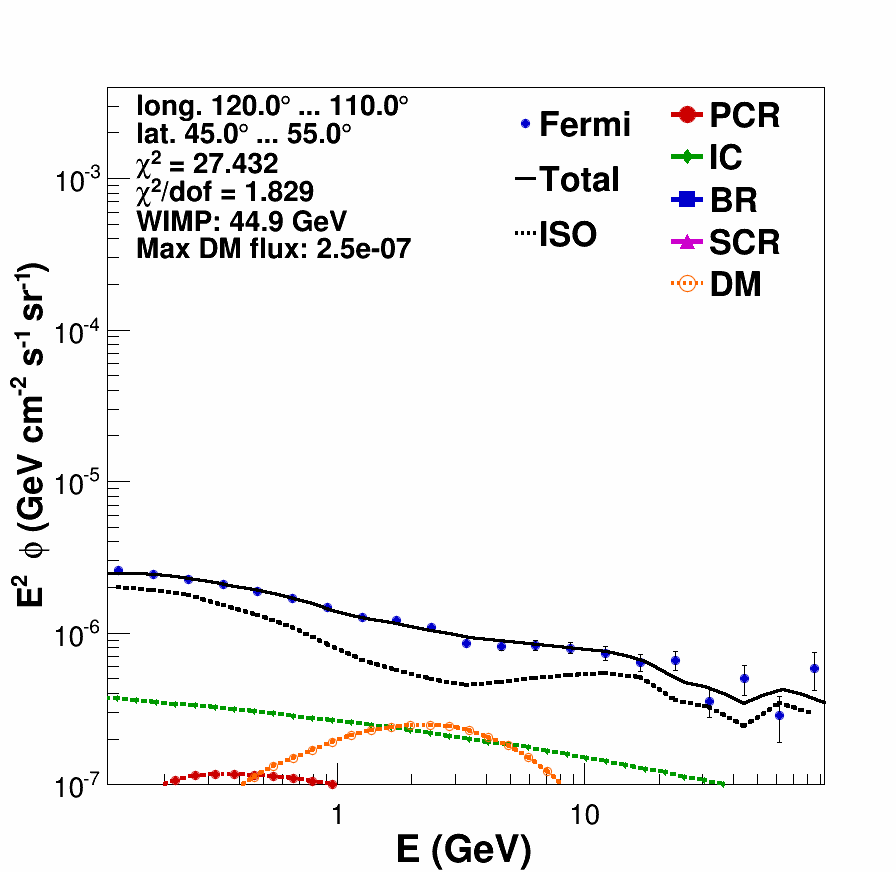}
\includegraphics[width=0.16\textwidth,height=0.16\textwidth,clip]{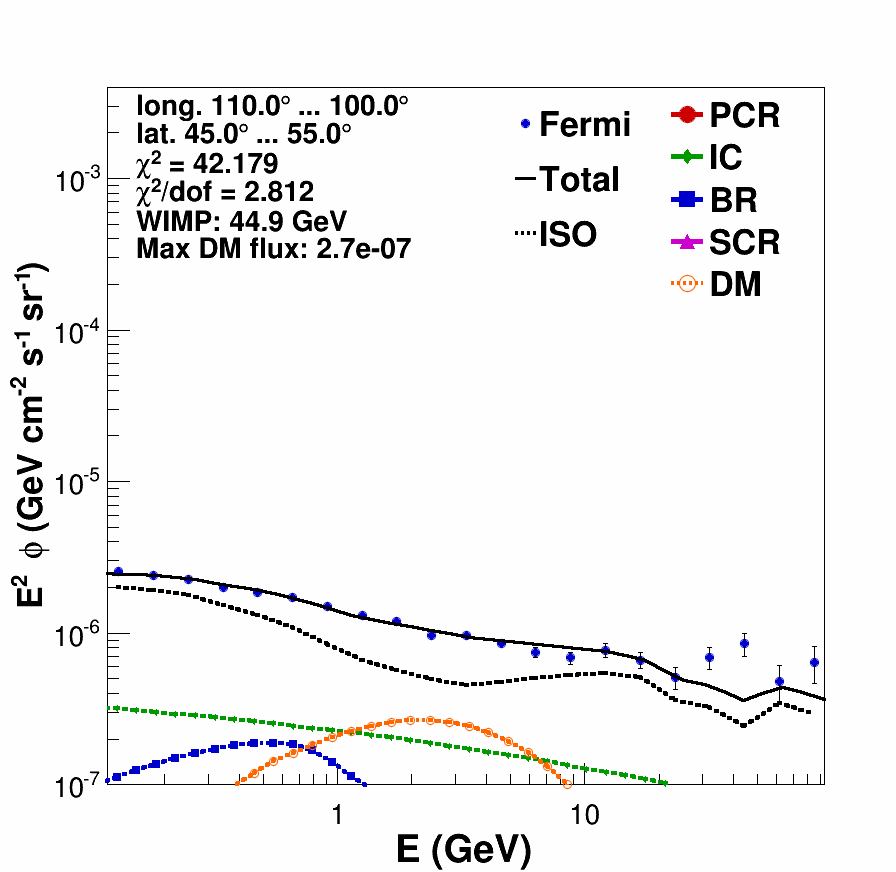}
\includegraphics[width=0.16\textwidth,height=0.16\textwidth,clip]{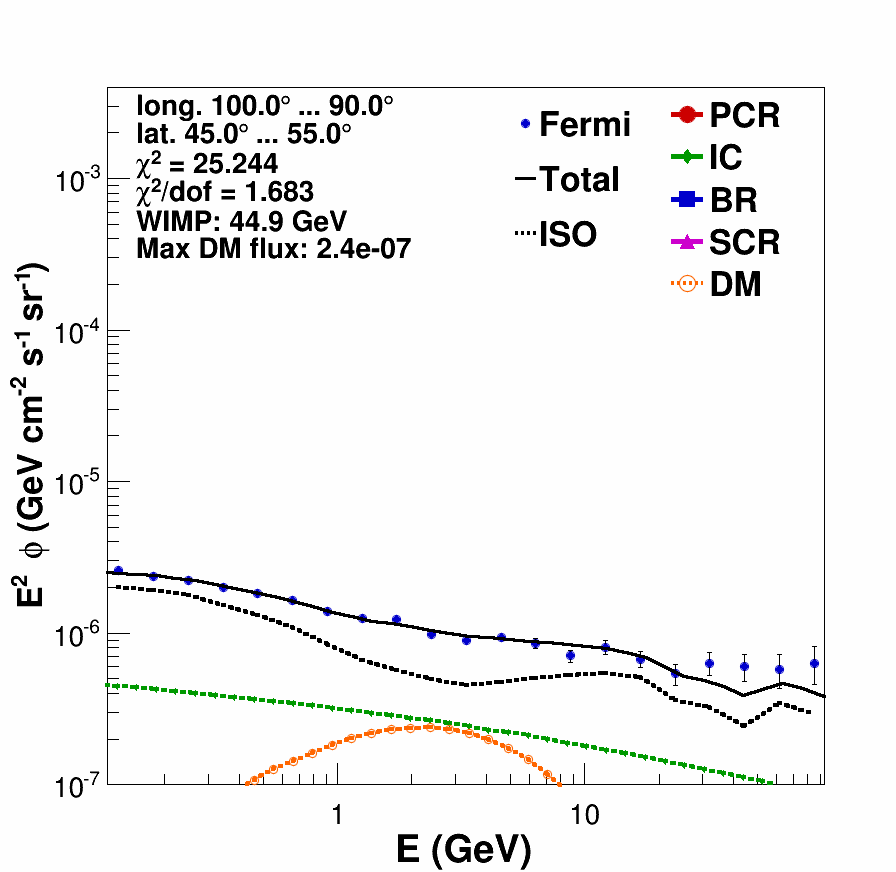}
\includegraphics[width=0.16\textwidth,height=0.16\textwidth,clip]{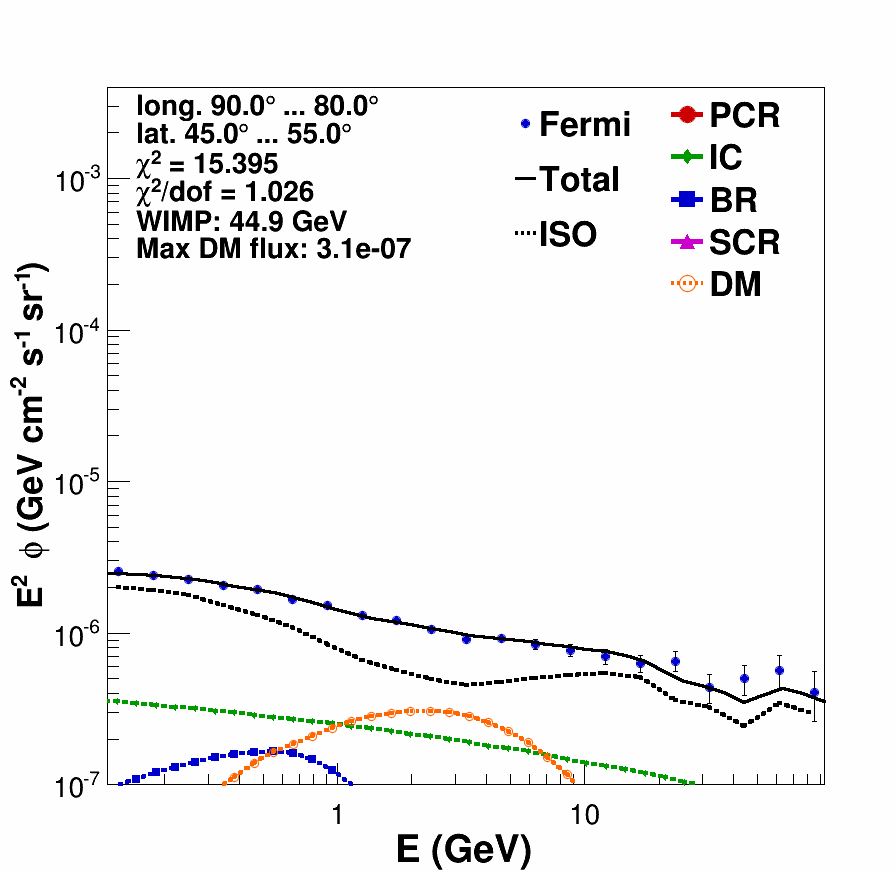}
\includegraphics[width=0.16\textwidth,height=0.16\textwidth,clip]{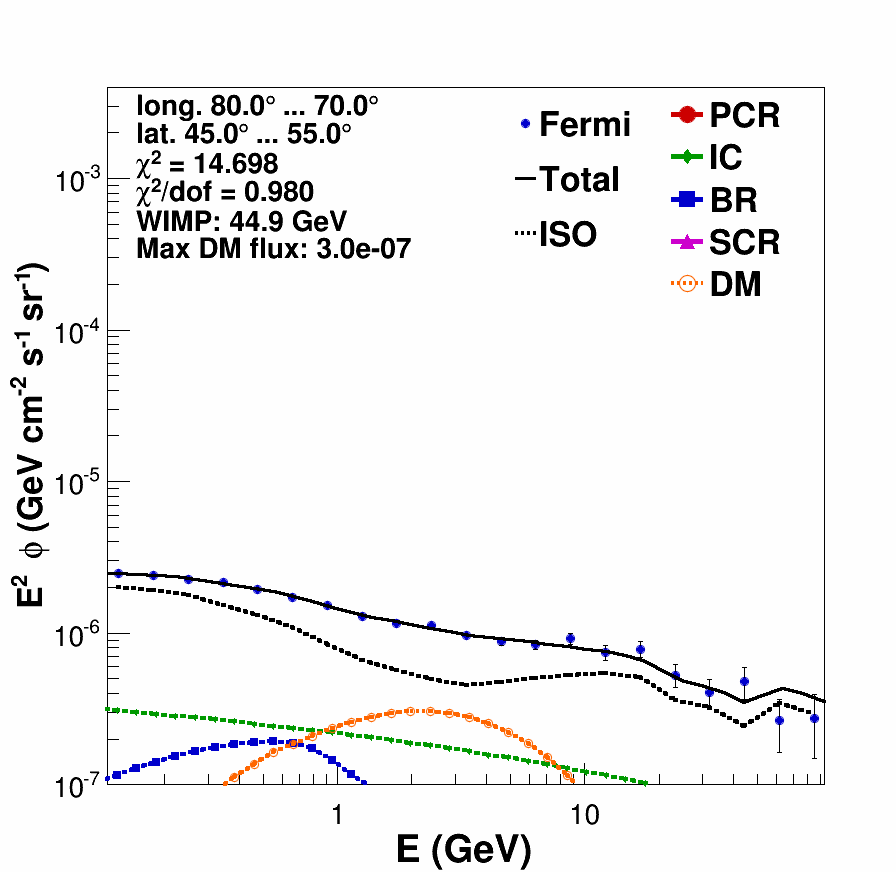}
\includegraphics[width=0.16\textwidth,height=0.16\textwidth,clip]{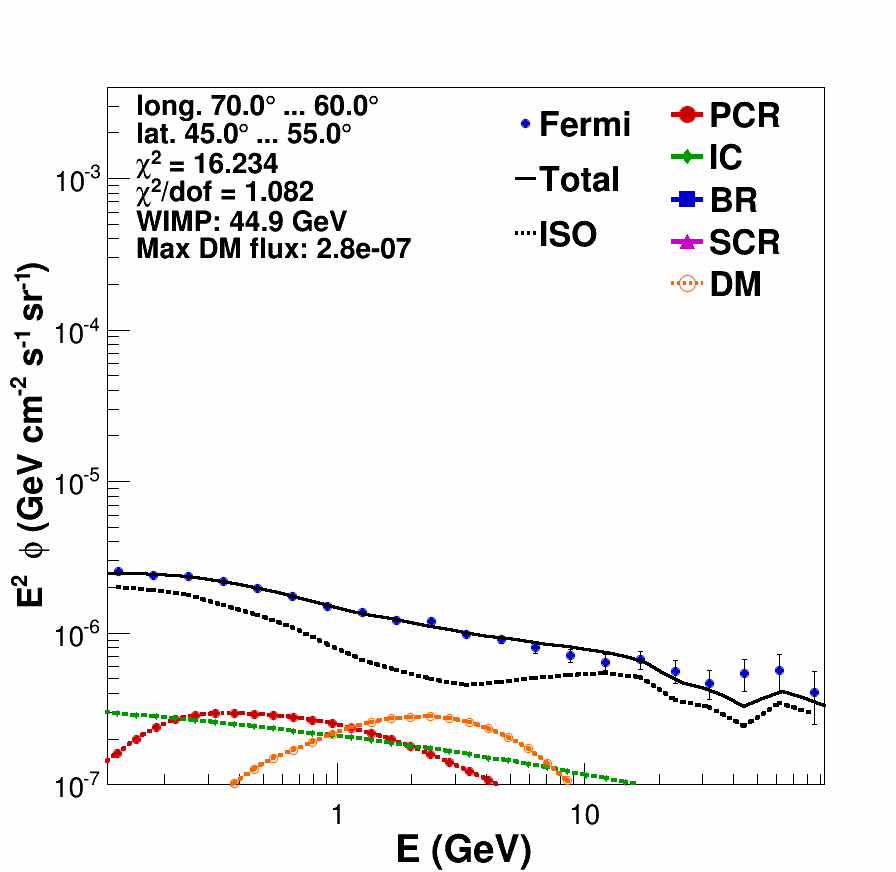}
\includegraphics[width=0.16\textwidth,height=0.16\textwidth,clip]{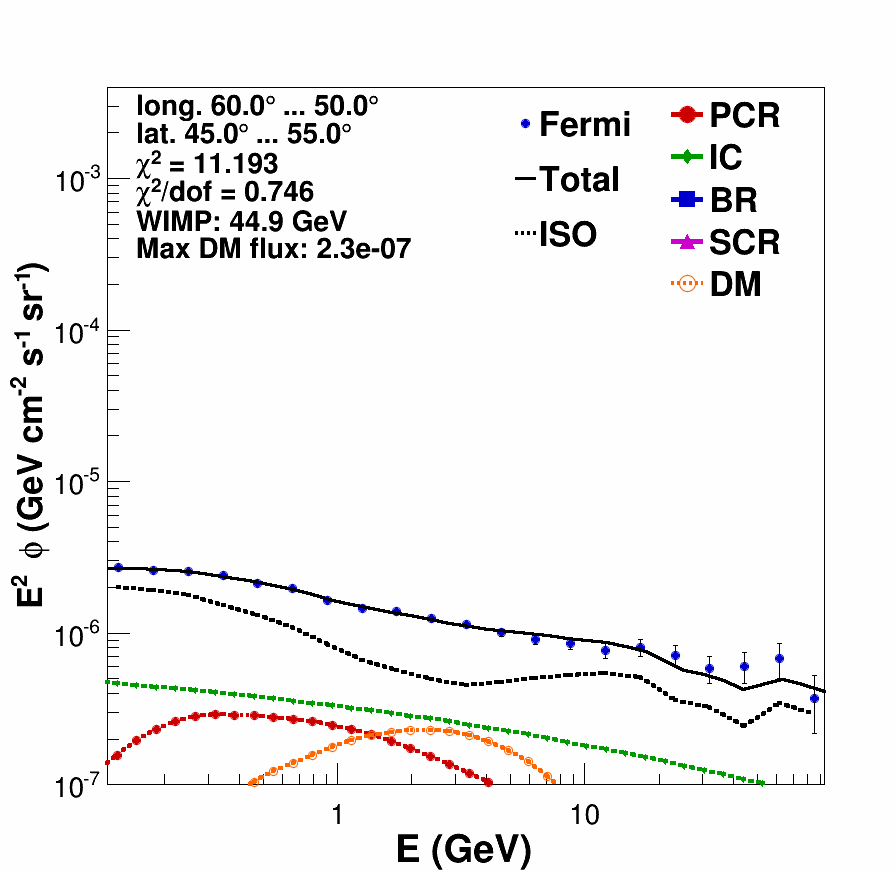}
\includegraphics[width=0.16\textwidth,height=0.16\textwidth,clip]{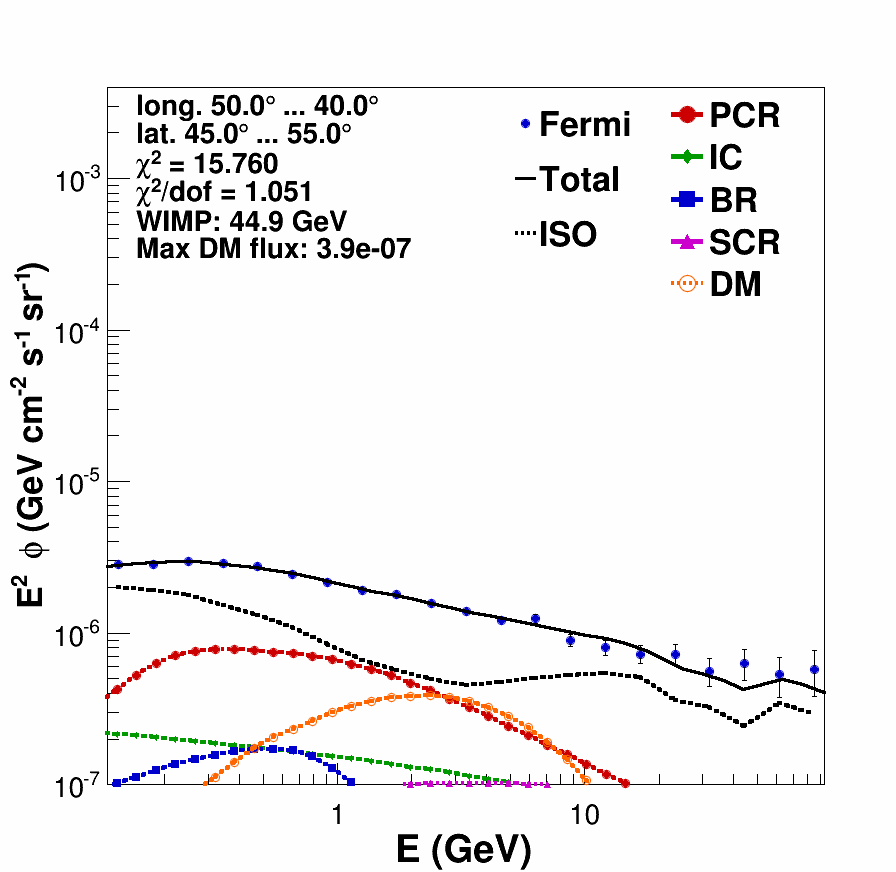}
\includegraphics[width=0.16\textwidth,height=0.16\textwidth,clip]{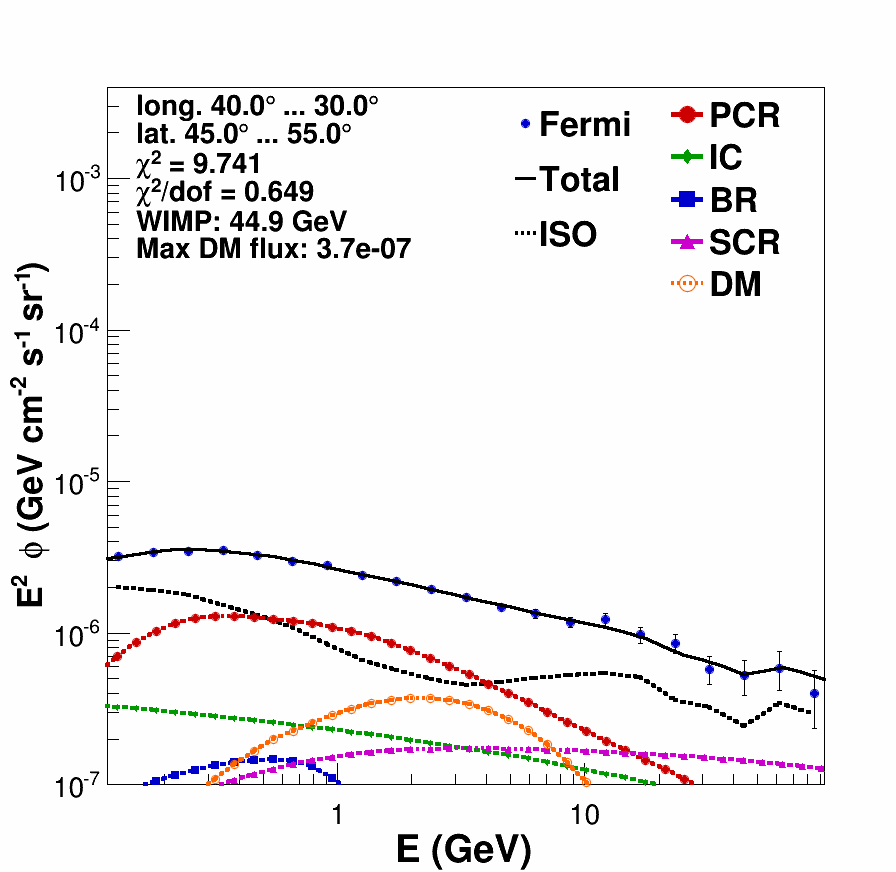}
\includegraphics[width=0.16\textwidth,height=0.16\textwidth,clip]{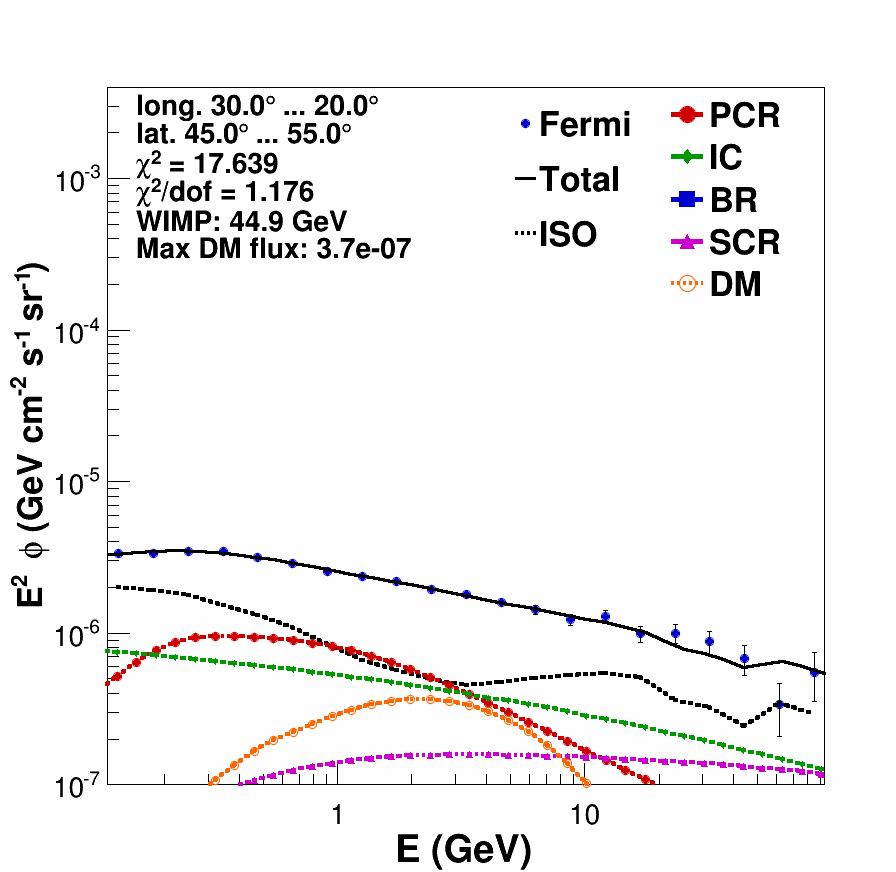}
\includegraphics[width=0.16\textwidth,height=0.16\textwidth,clip]{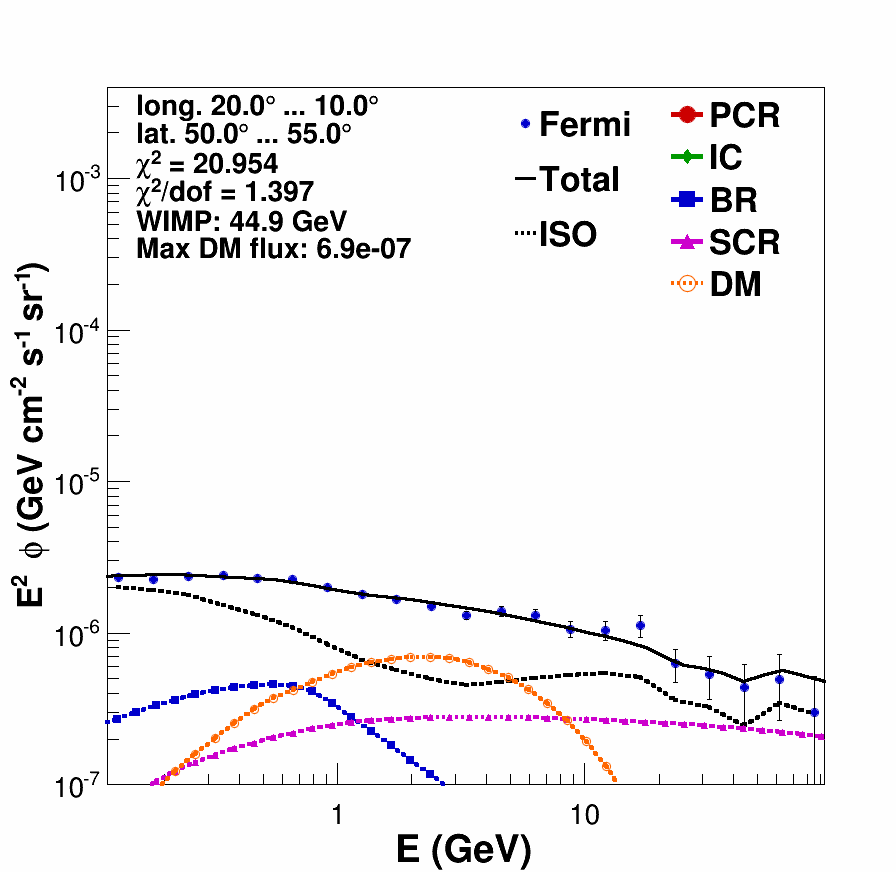}
\includegraphics[width=0.16\textwidth,height=0.16\textwidth,clip]{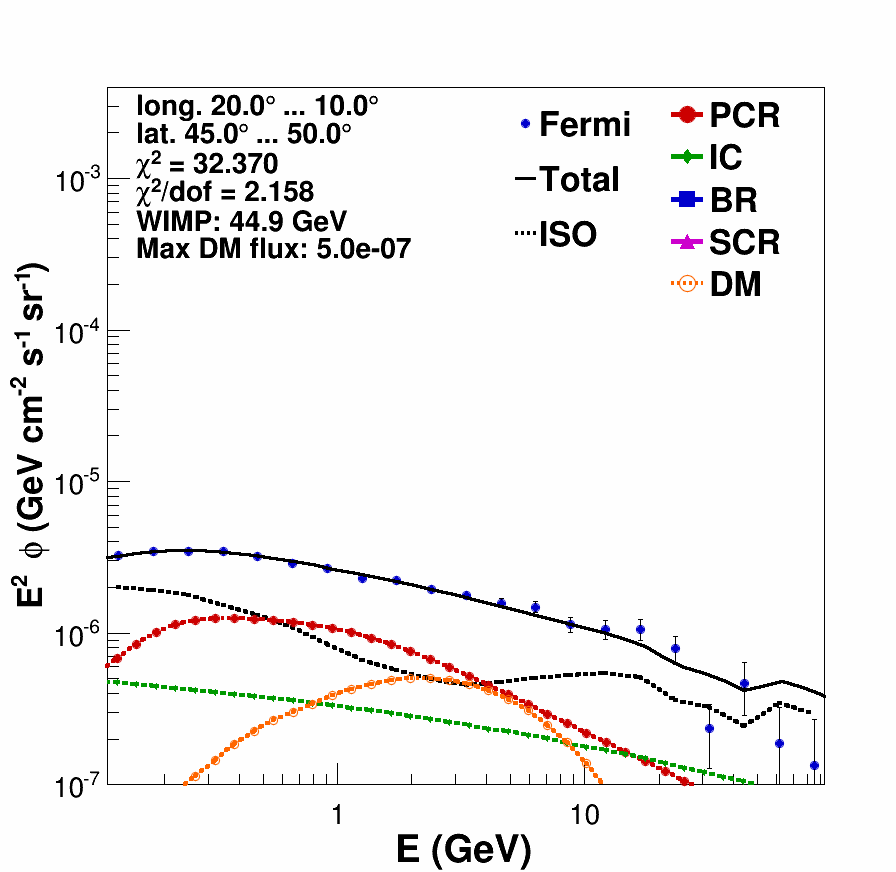}
\includegraphics[width=0.16\textwidth,height=0.16\textwidth,clip]{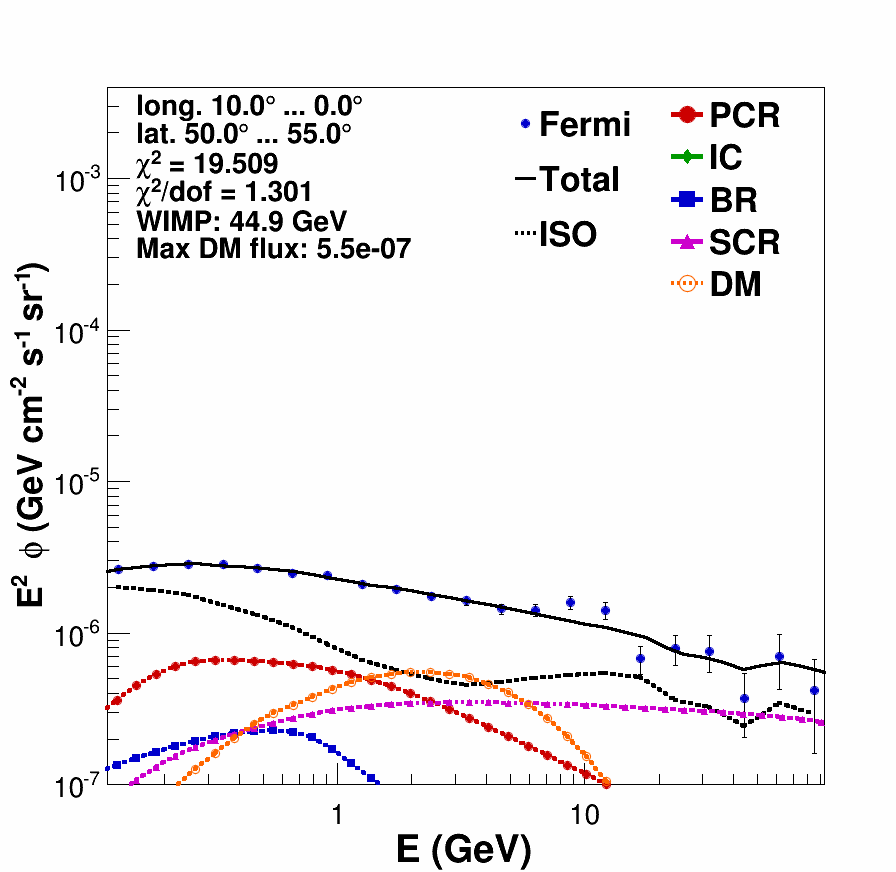}
\includegraphics[width=0.16\textwidth,height=0.16\textwidth,clip]{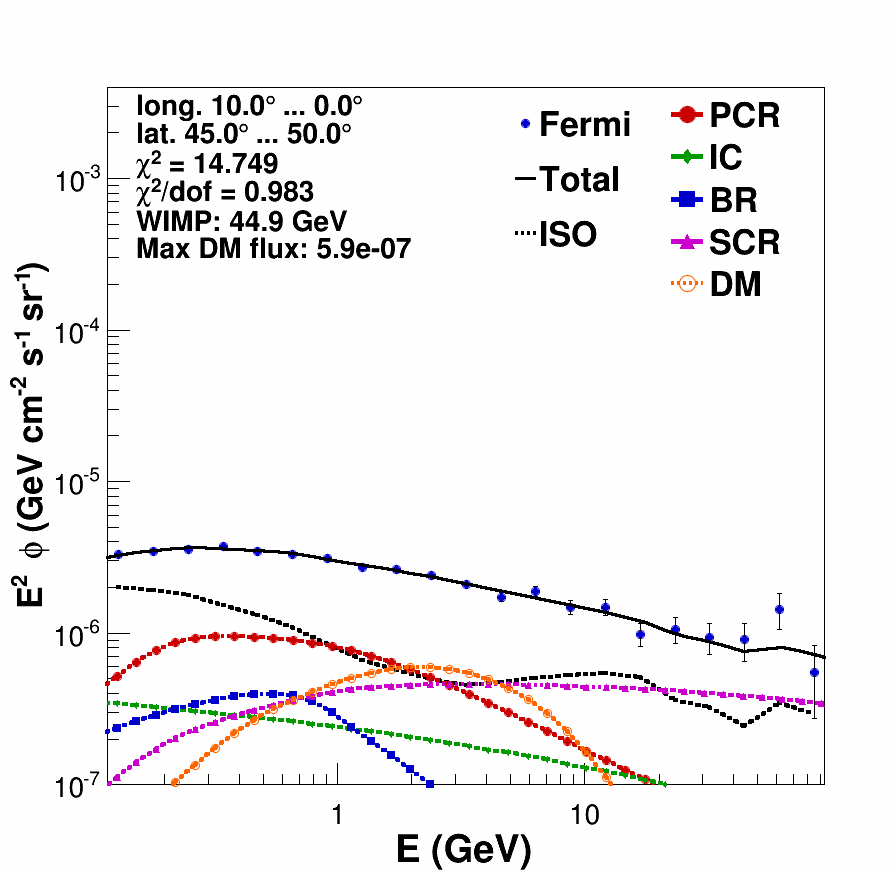}
\includegraphics[width=0.16\textwidth,height=0.16\textwidth,clip]{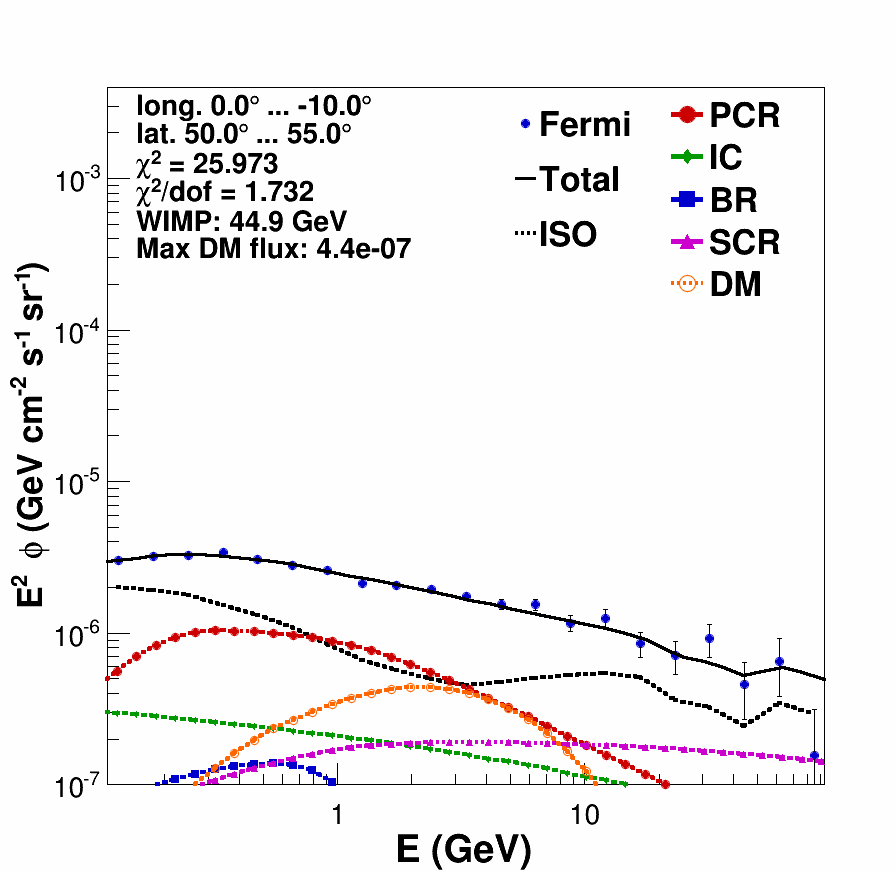}
\includegraphics[width=0.16\textwidth,height=0.16\textwidth,clip]{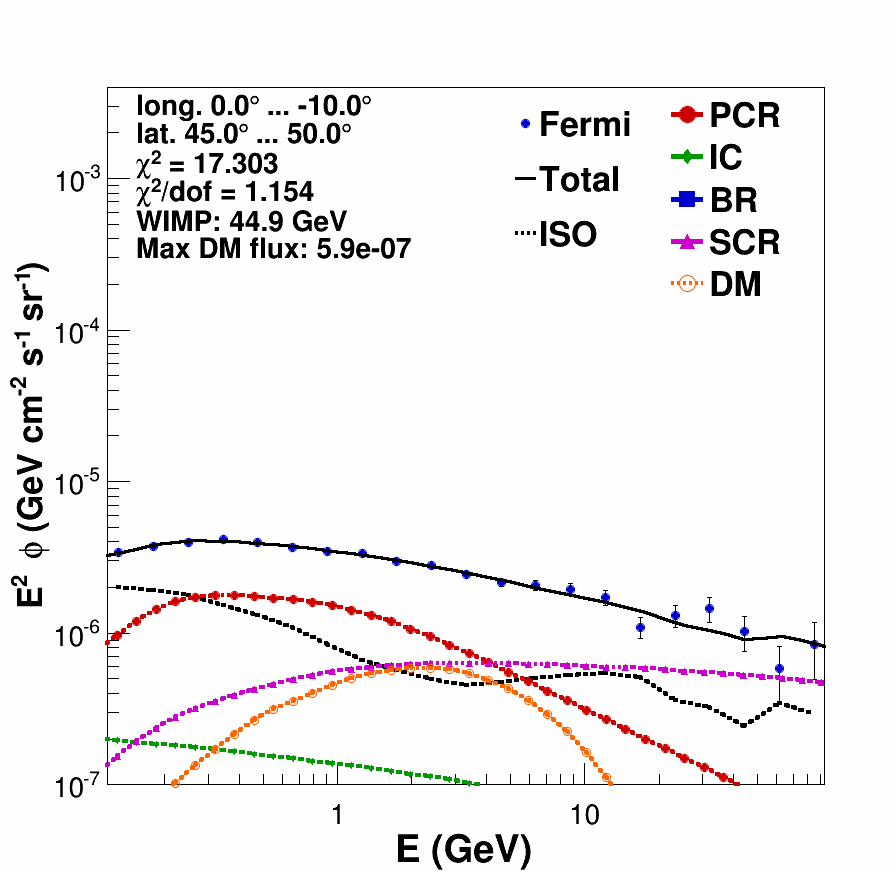}
\includegraphics[width=0.16\textwidth,height=0.16\textwidth,clip]{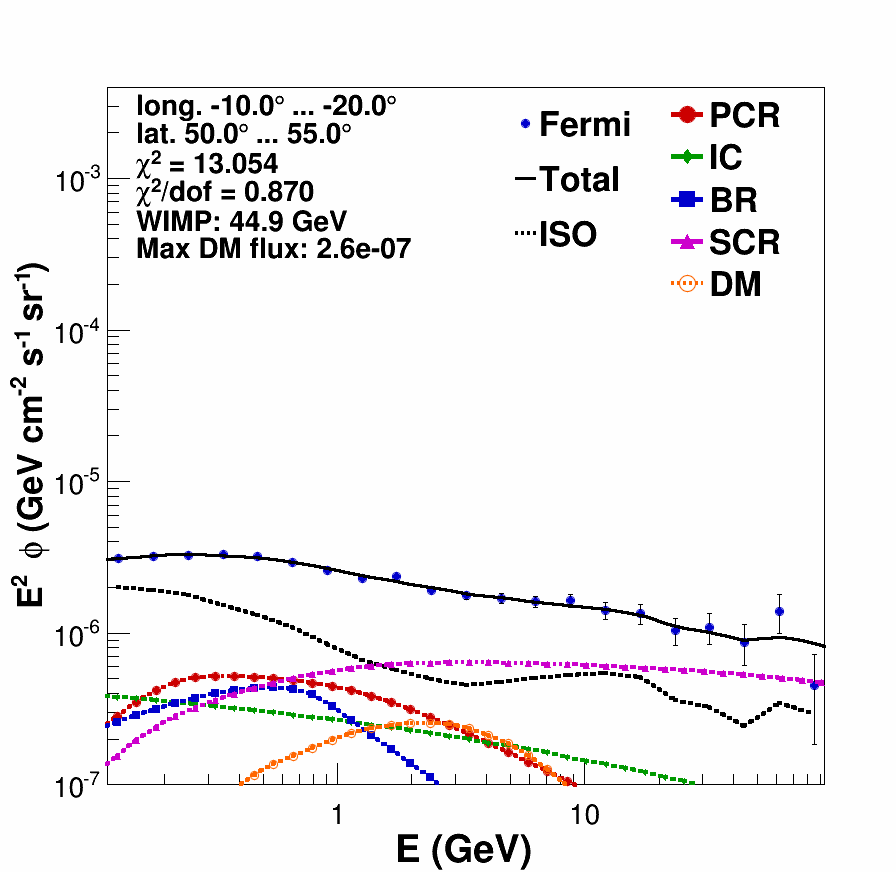}
\includegraphics[width=0.16\textwidth,height=0.16\textwidth,clip]{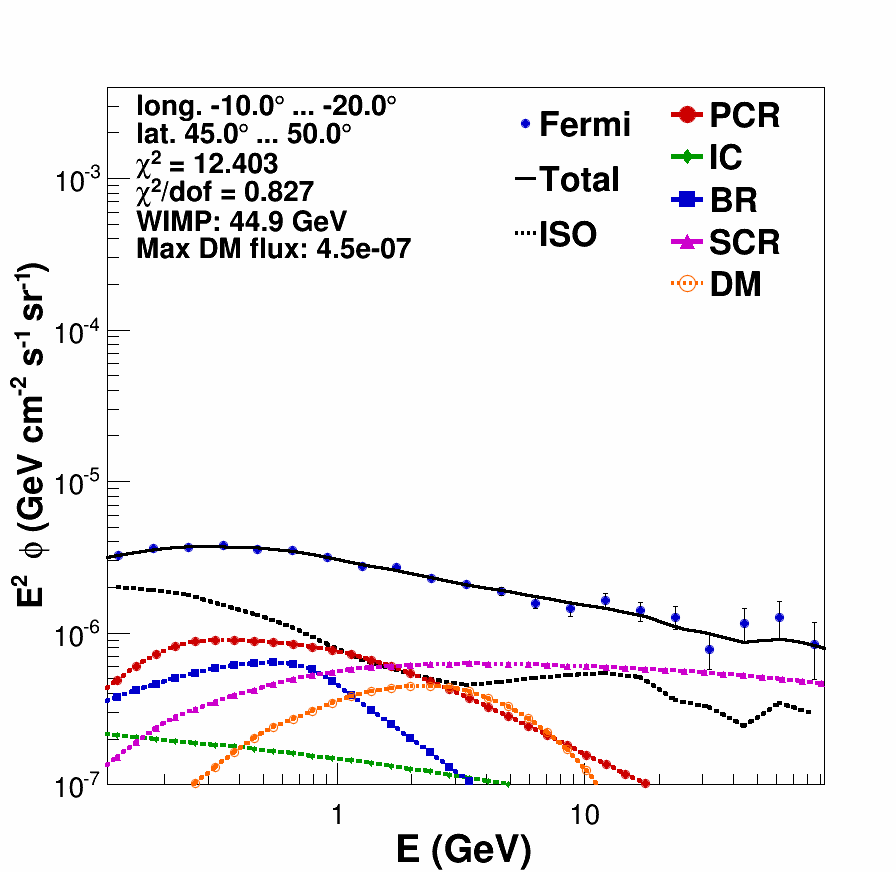}
\includegraphics[width=0.16\textwidth,height=0.16\textwidth,clip]{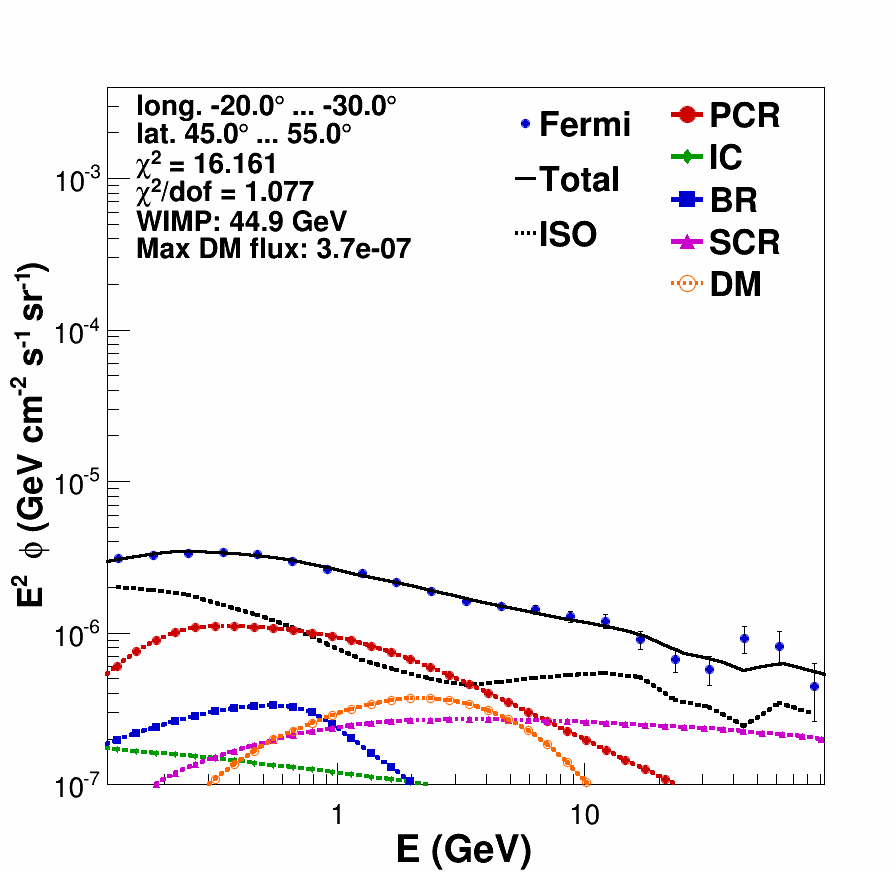}
\includegraphics[width=0.16\textwidth,height=0.16\textwidth,clip]{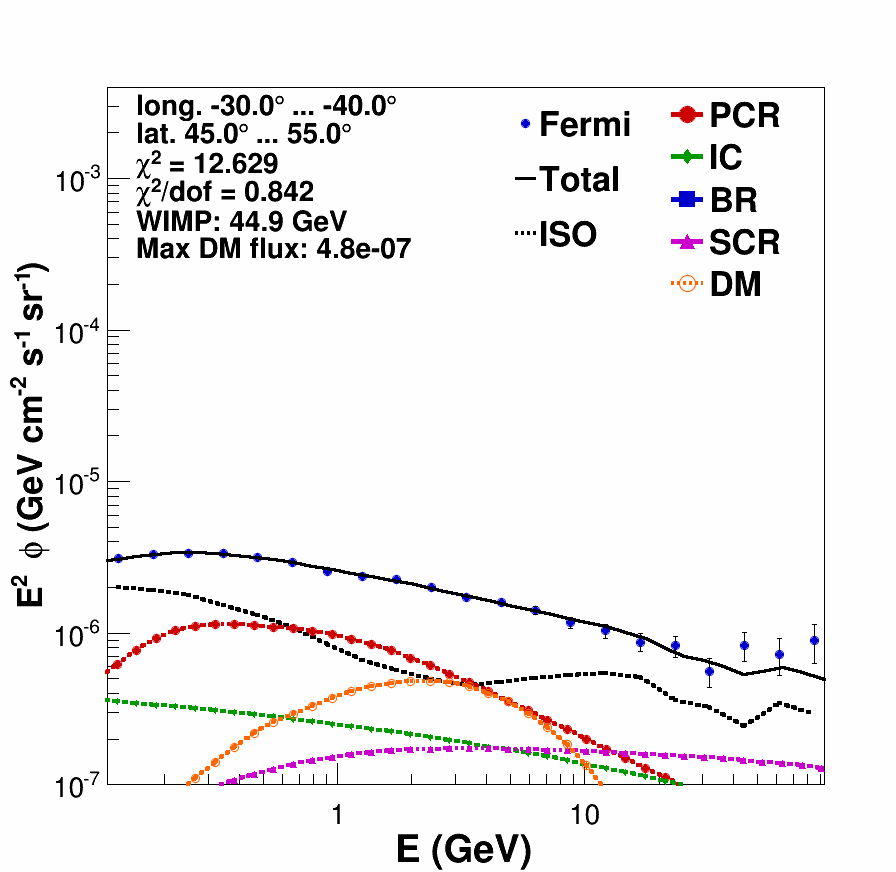}
\includegraphics[width=0.16\textwidth,height=0.16\textwidth,clip]{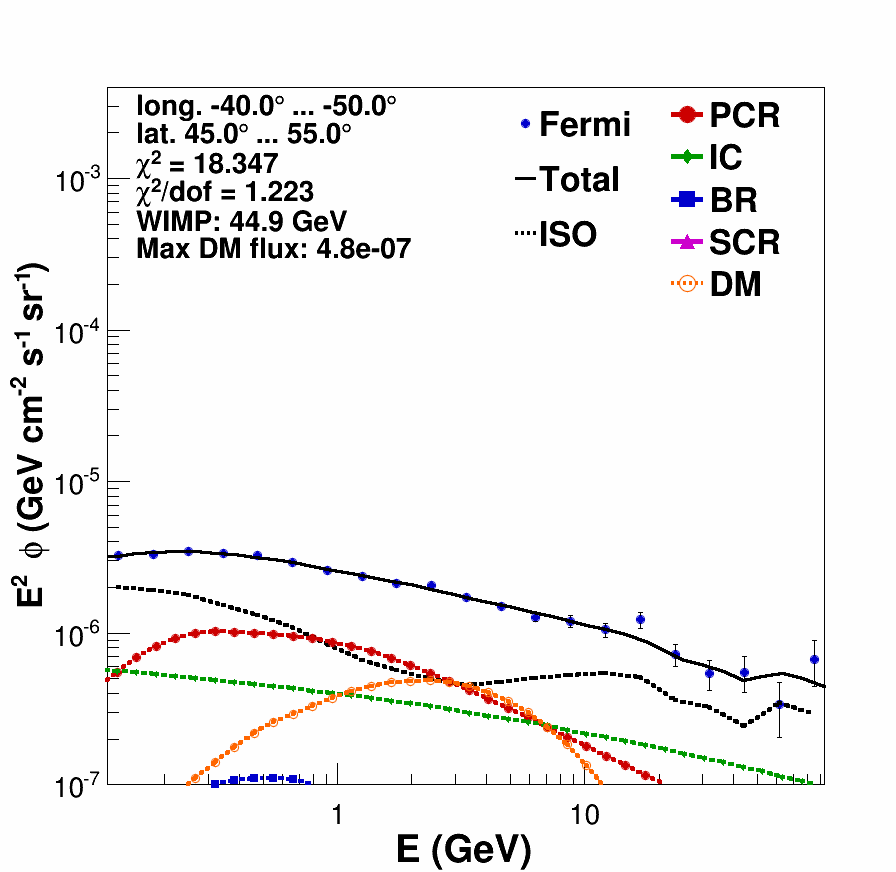}
\includegraphics[width=0.16\textwidth,height=0.16\textwidth,clip]{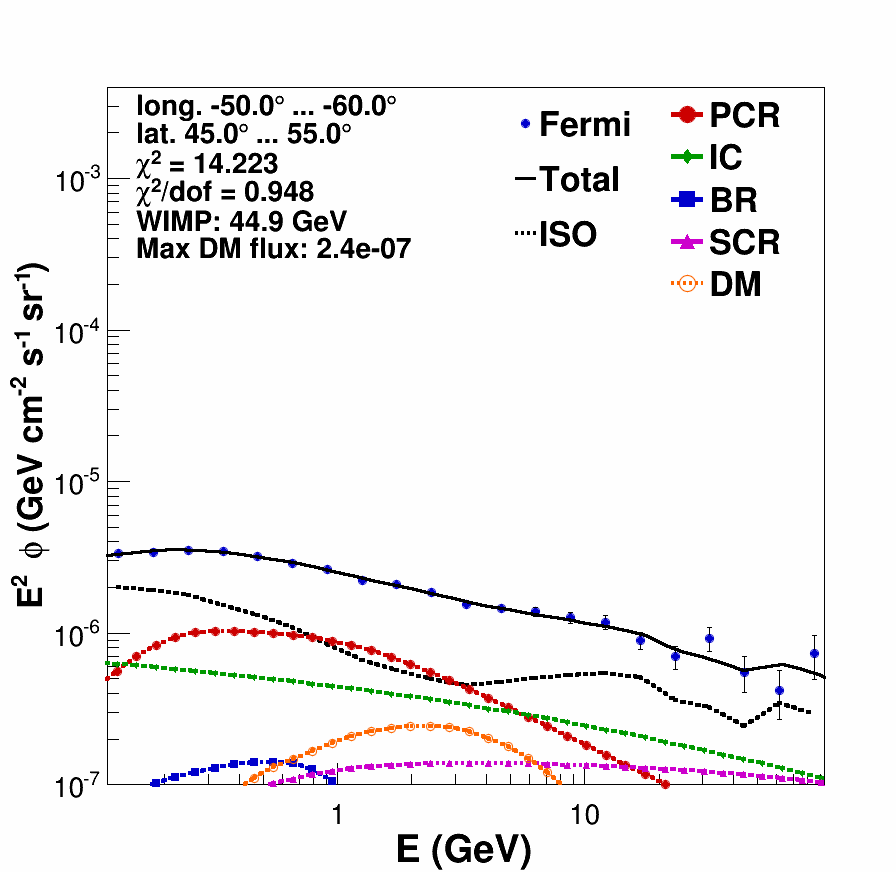}
\includegraphics[width=0.16\textwidth,height=0.16\textwidth,clip]{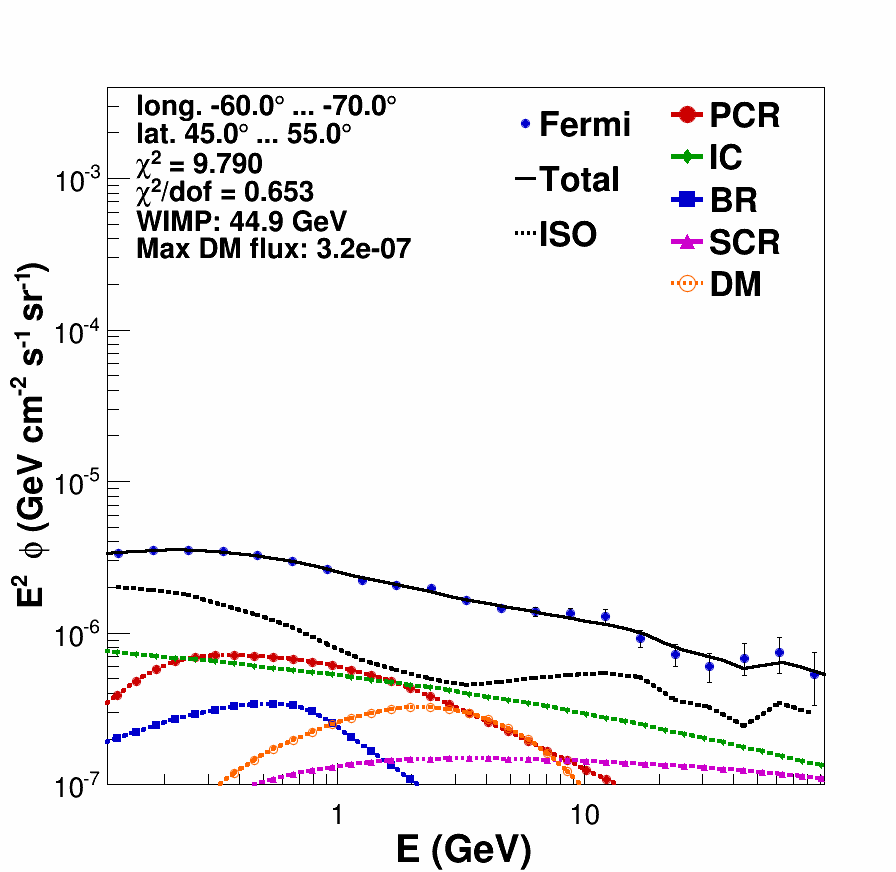}
\includegraphics[width=0.16\textwidth,height=0.16\textwidth,clip]{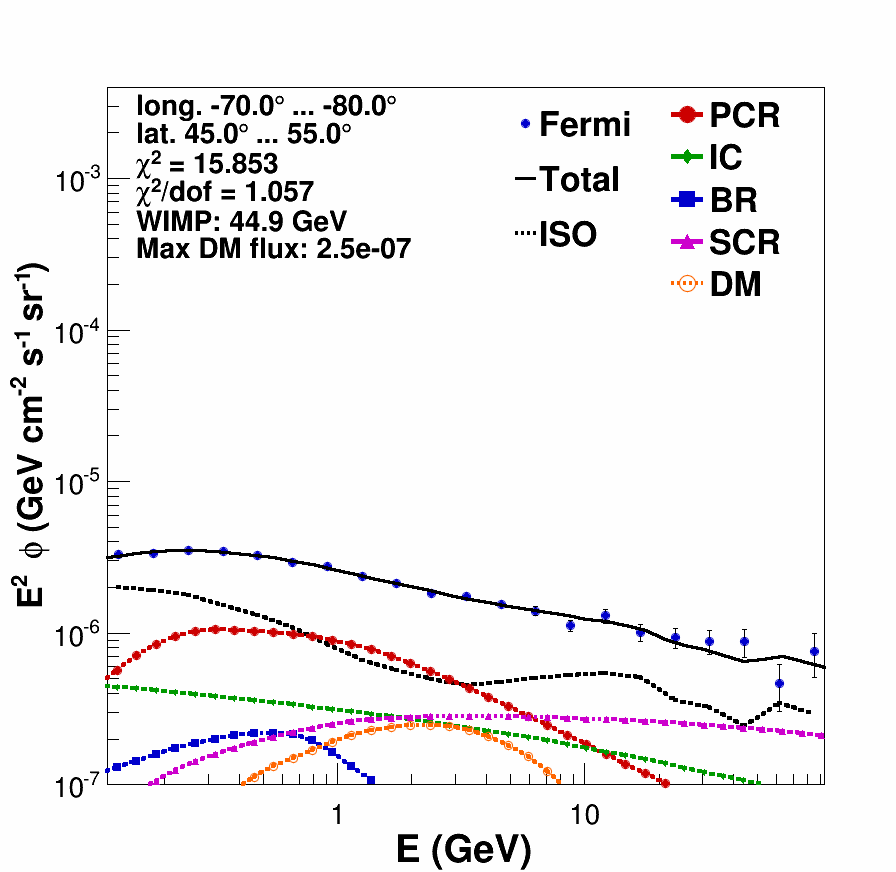}
\includegraphics[width=0.16\textwidth,height=0.16\textwidth,clip]{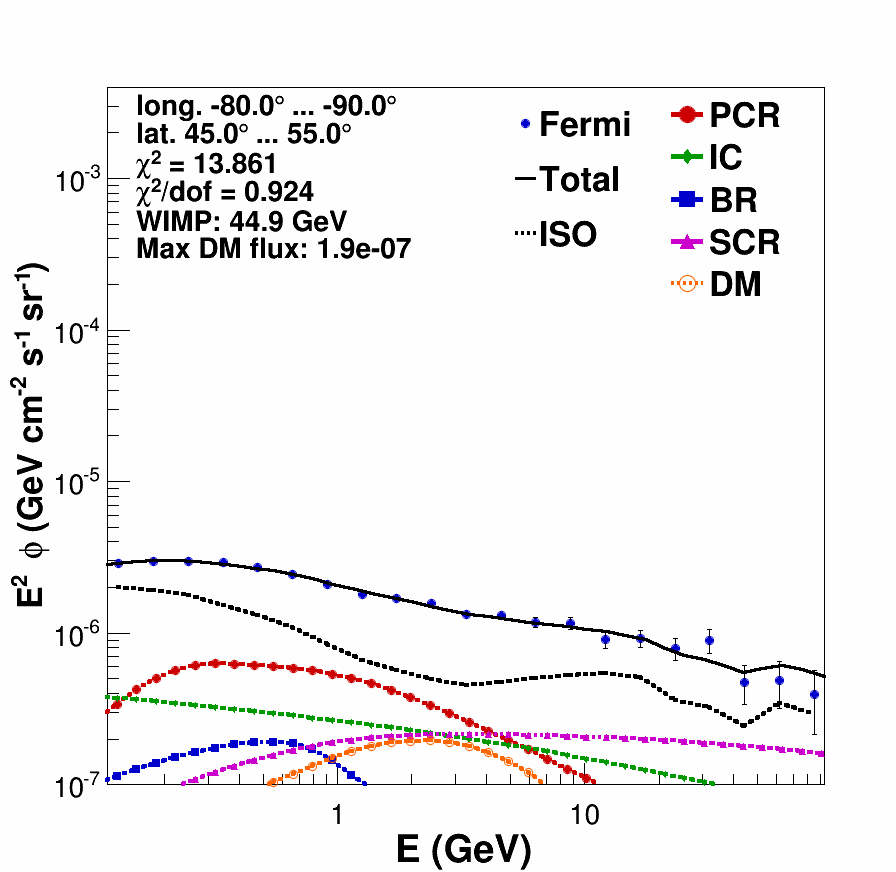}
\includegraphics[width=0.16\textwidth,height=0.16\textwidth,clip]{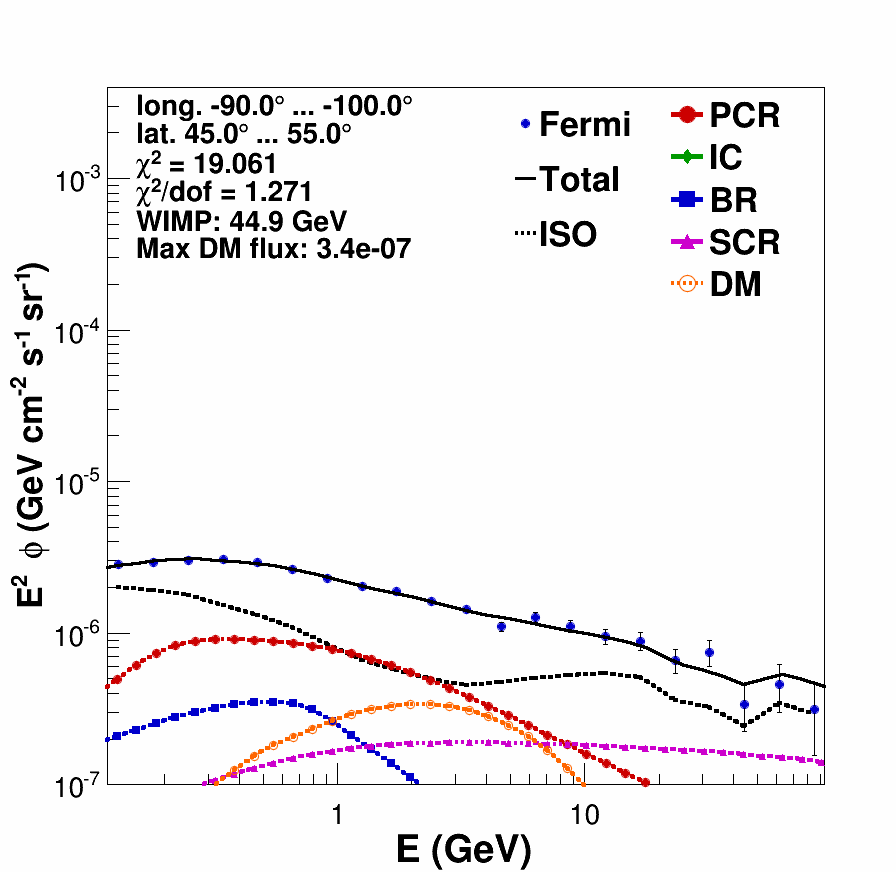}
\includegraphics[width=0.16\textwidth,height=0.16\textwidth,clip]{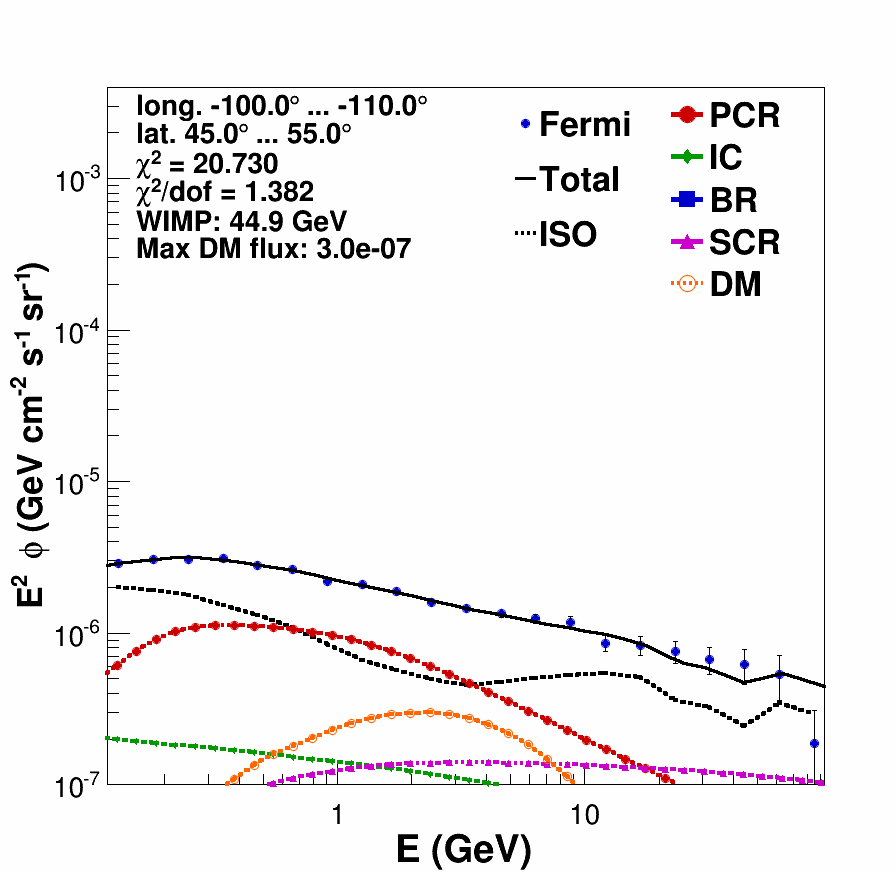}
\includegraphics[width=0.16\textwidth,height=0.16\textwidth,clip]{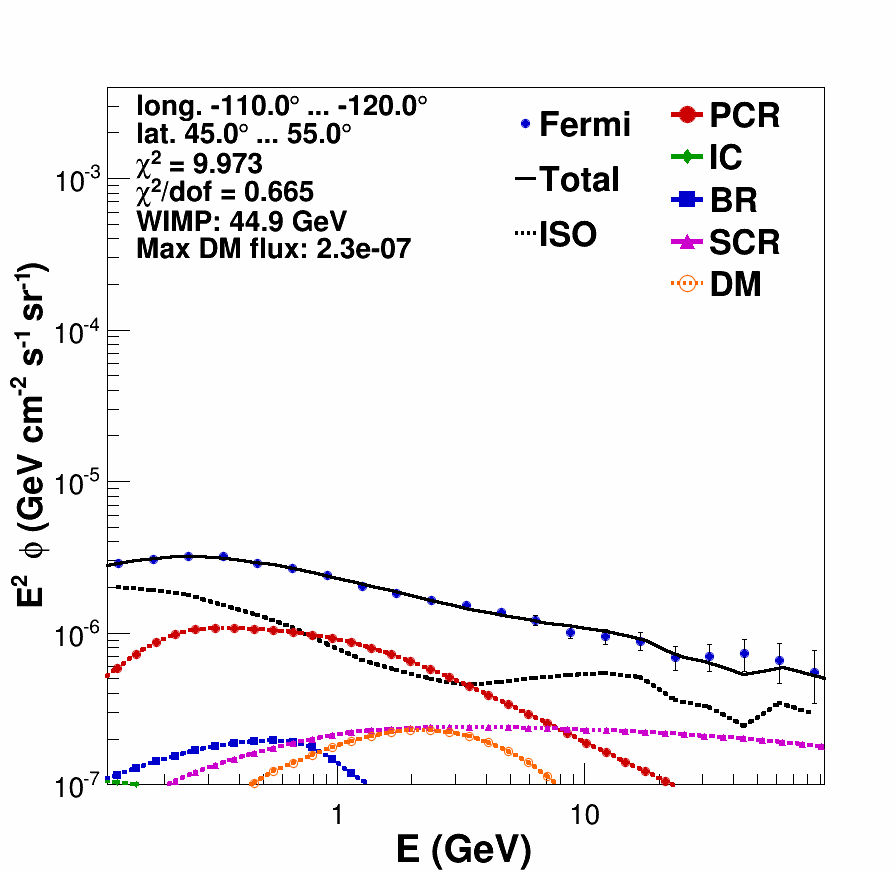}
\includegraphics[width=0.16\textwidth,height=0.16\textwidth,clip]{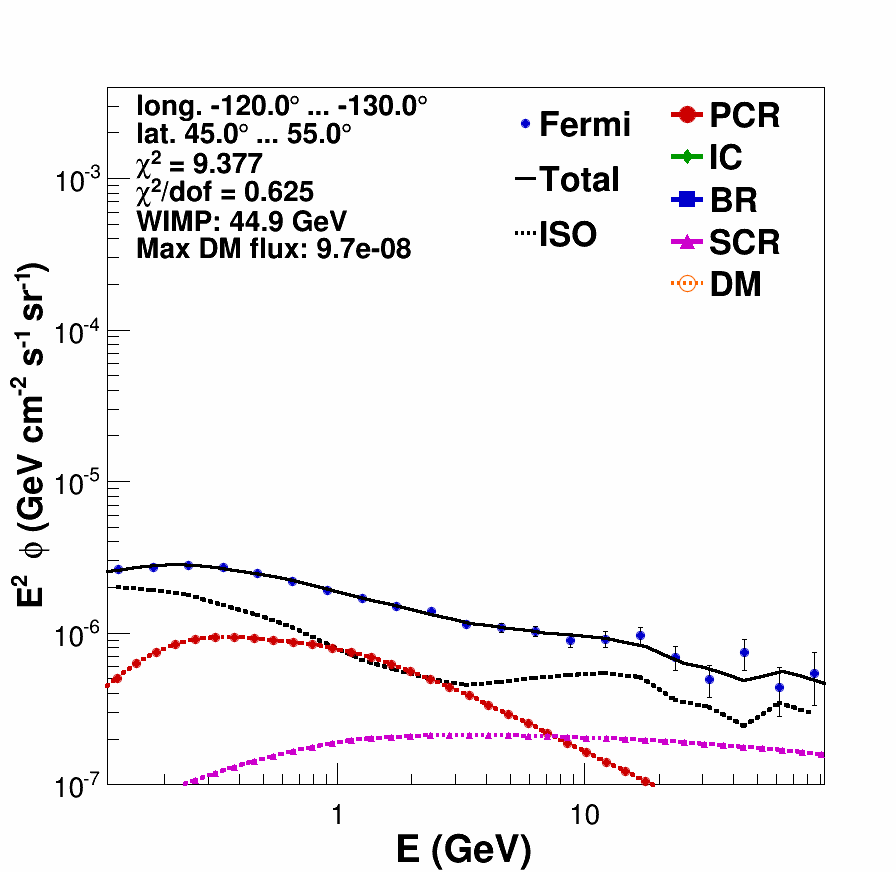}
\includegraphics[width=0.16\textwidth,height=0.16\textwidth,clip]{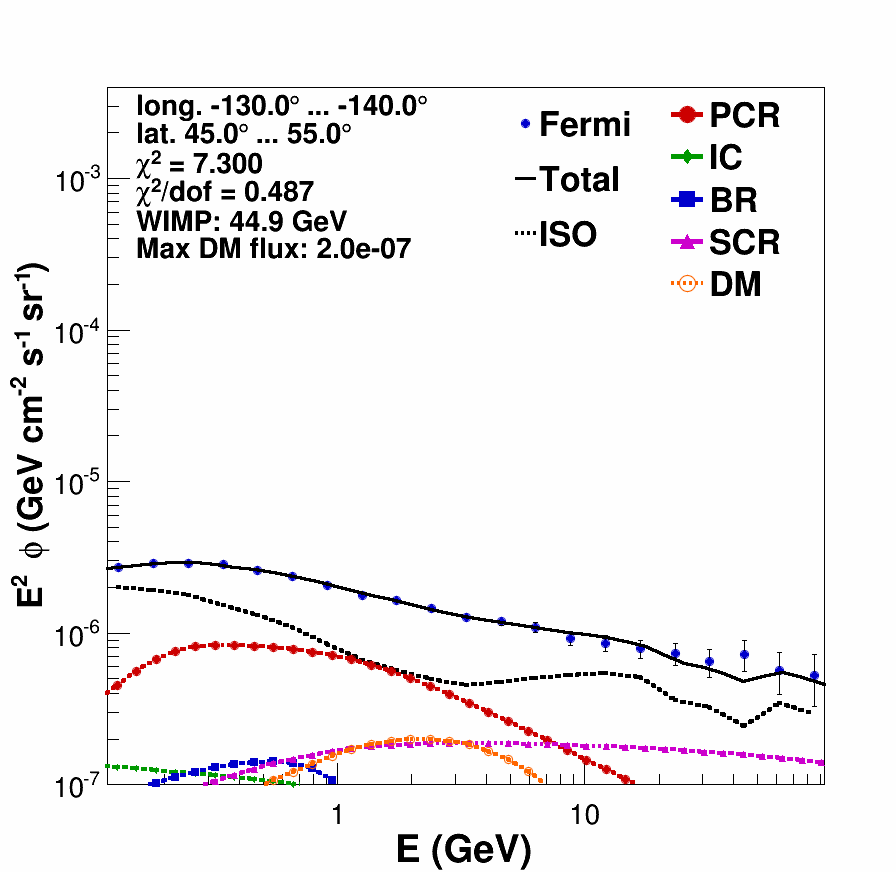}
\includegraphics[width=0.16\textwidth,height=0.16\textwidth,clip]{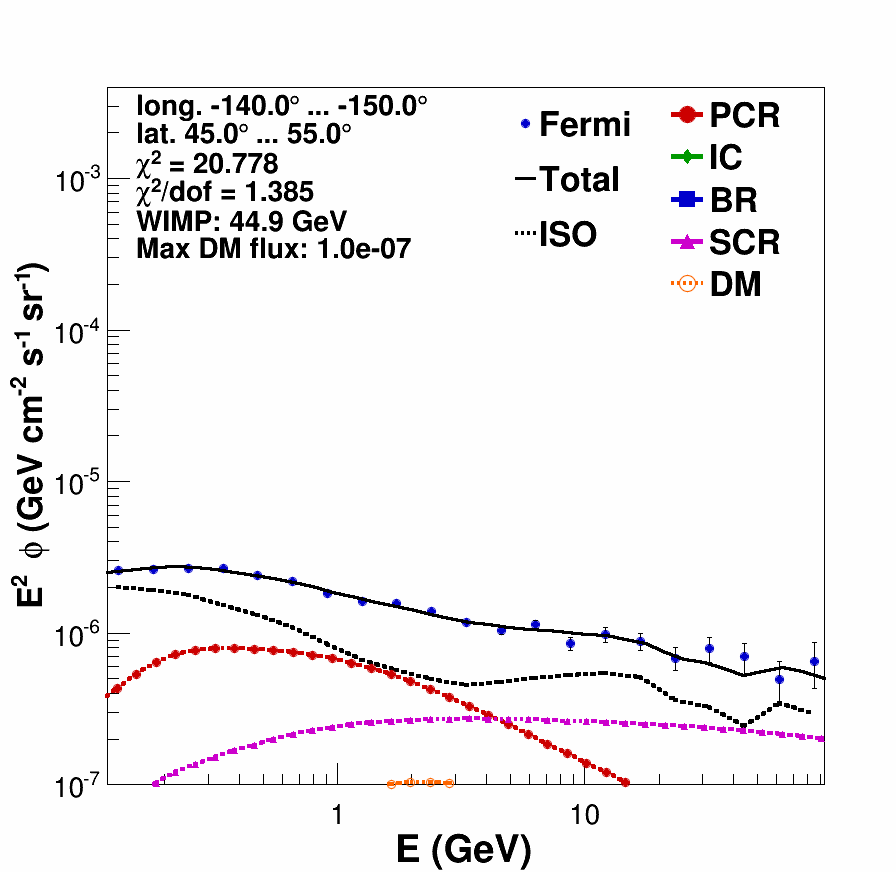}
\includegraphics[width=0.16\textwidth,height=0.16\textwidth,clip]{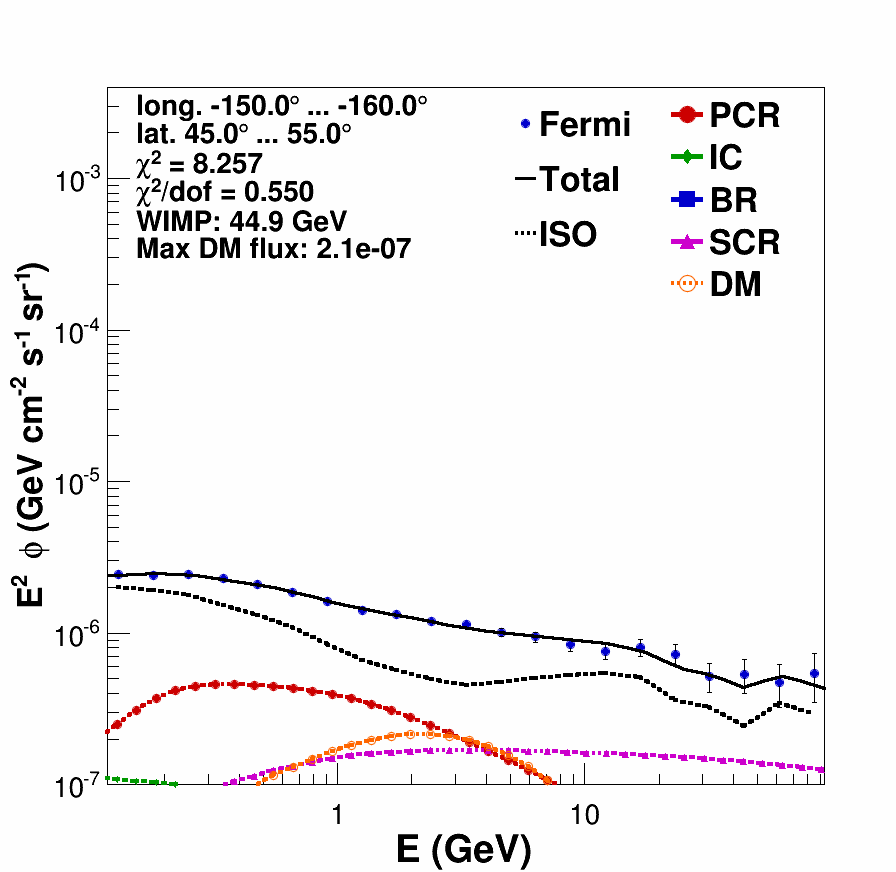}
\includegraphics[width=0.16\textwidth,height=0.16\textwidth,clip]{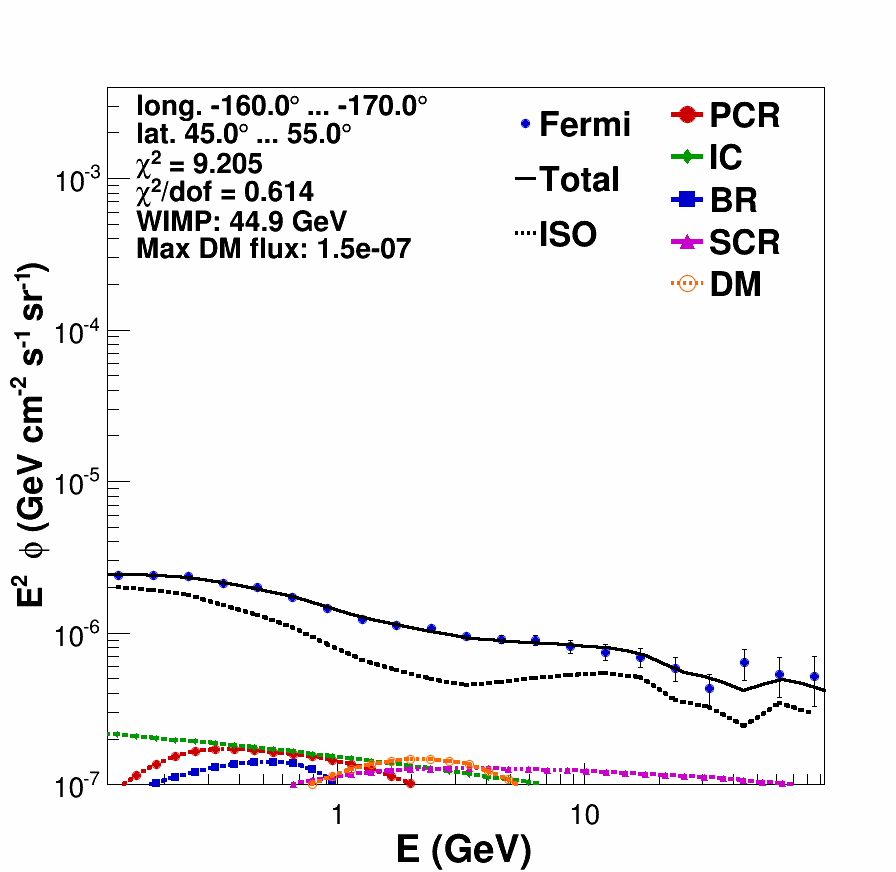}
\includegraphics[width=0.16\textwidth,height=0.16\textwidth,clip]{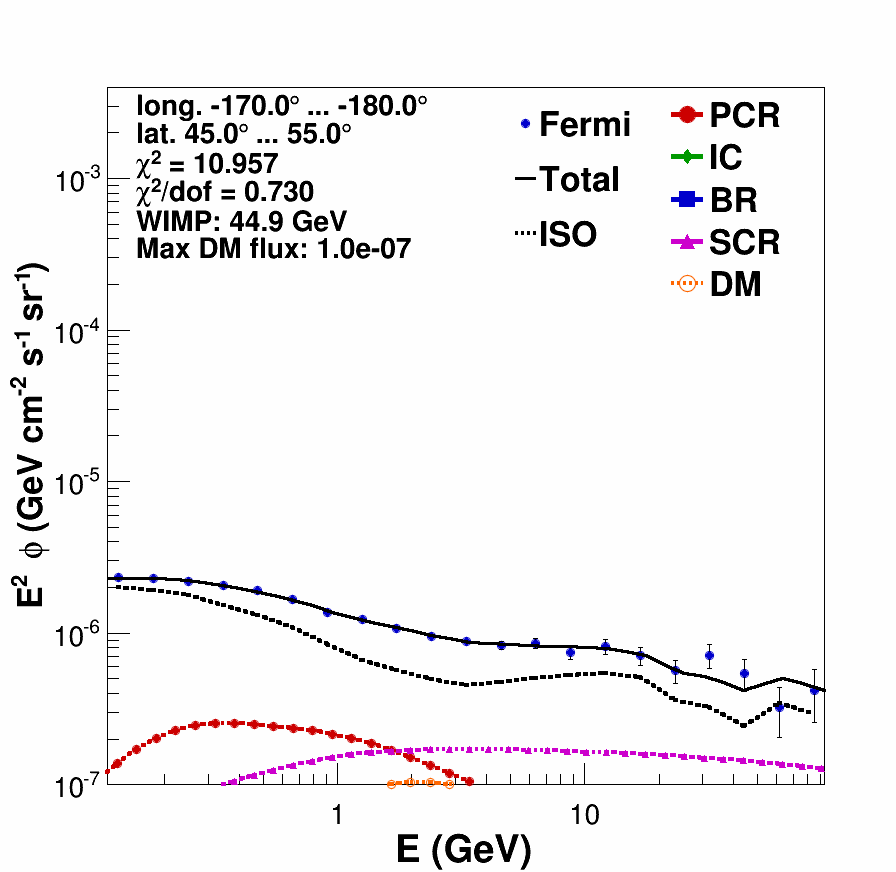}%%%%%%%r3
\caption[]{Template fits for latitudes  with $45.0^\circ<b<55.0^\circ$ and longitudes decreasing from 180$^\circ$ to -180$^\circ$. \label{F34}
}
\end{figure}
\begin{figure}
\centering
\includegraphics[width=0.16\textwidth,height=0.16\textwidth,clip]{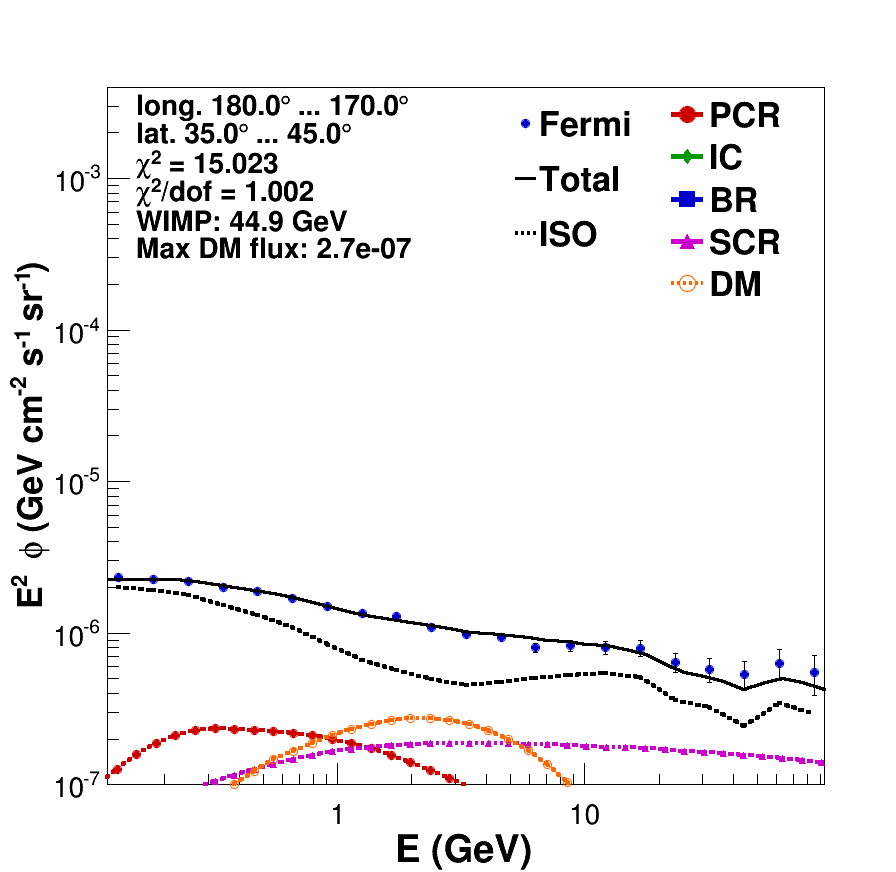}
\includegraphics[width=0.16\textwidth,height=0.16\textwidth,clip]{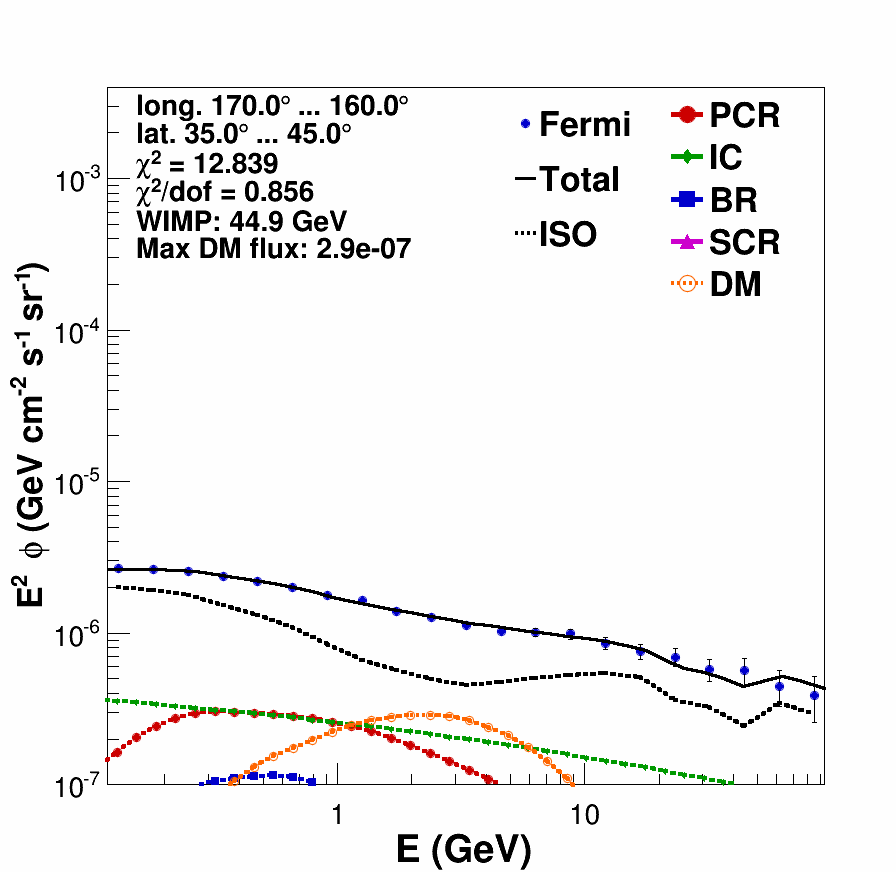}
\includegraphics[width=0.16\textwidth,height=0.16\textwidth,clip]{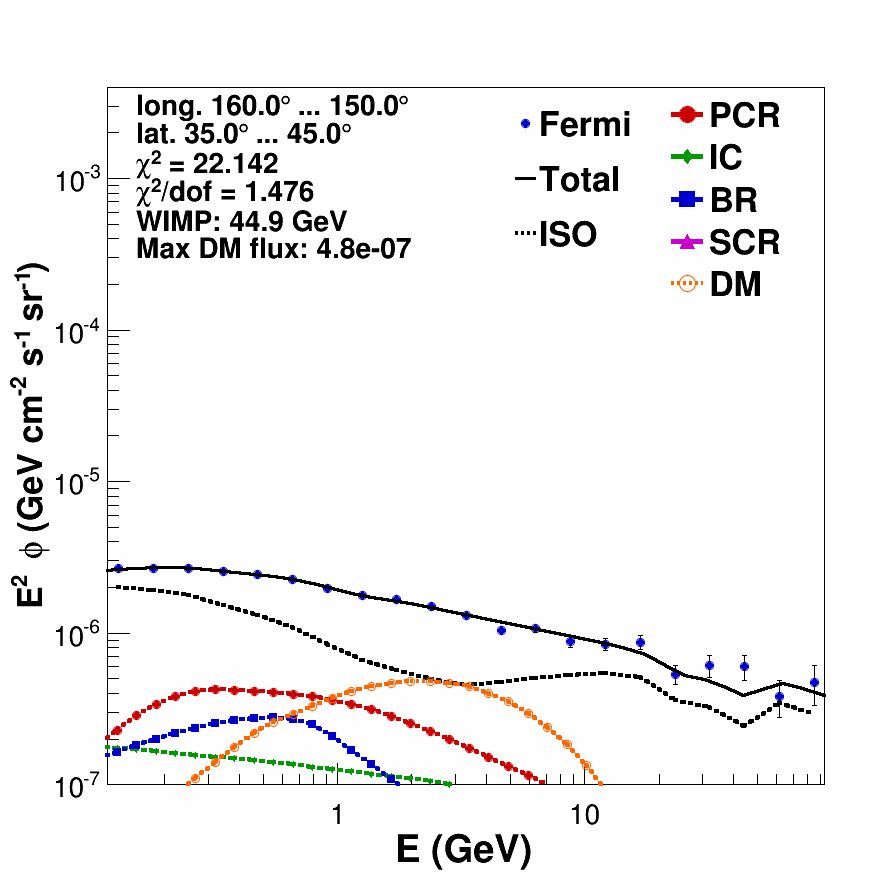}
\includegraphics[width=0.16\textwidth,height=0.16\textwidth,clip]{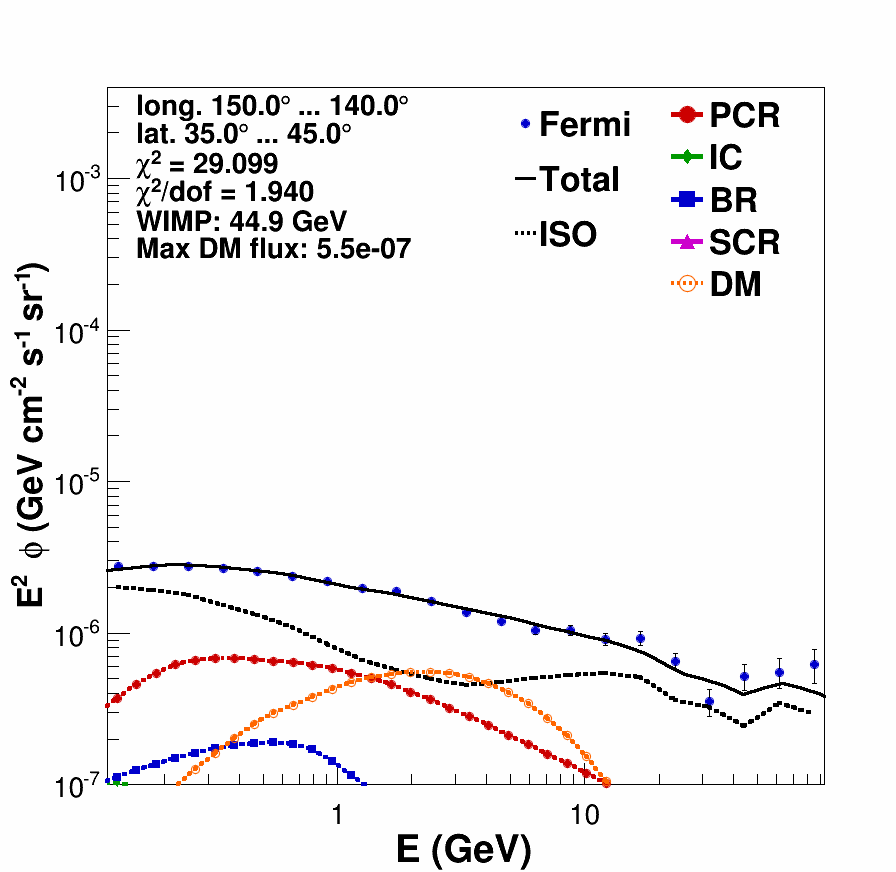}
\includegraphics[width=0.16\textwidth,height=0.16\textwidth,clip]{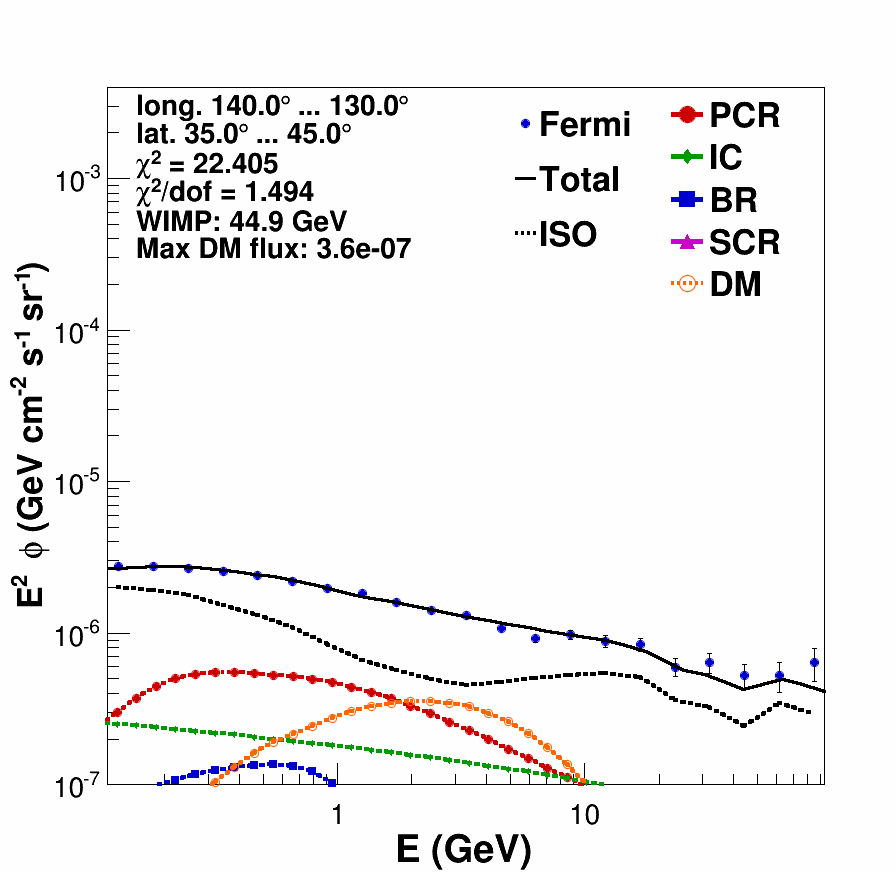}
\includegraphics[width=0.16\textwidth,height=0.16\textwidth,clip]{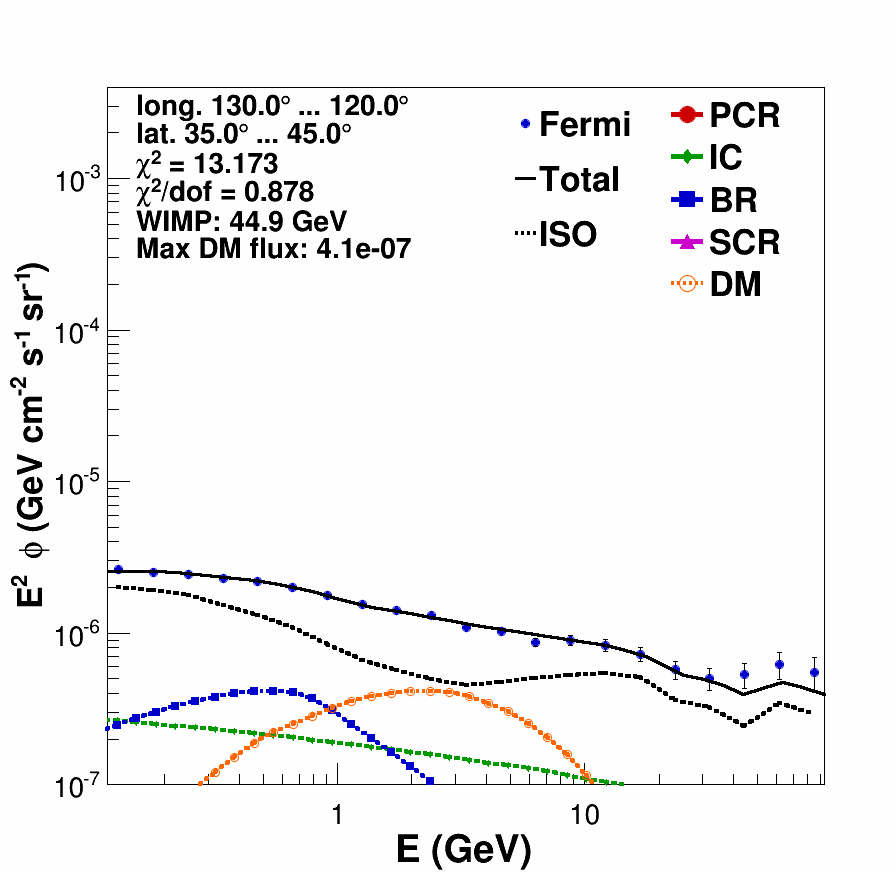}
\includegraphics[width=0.16\textwidth,height=0.16\textwidth,clip]{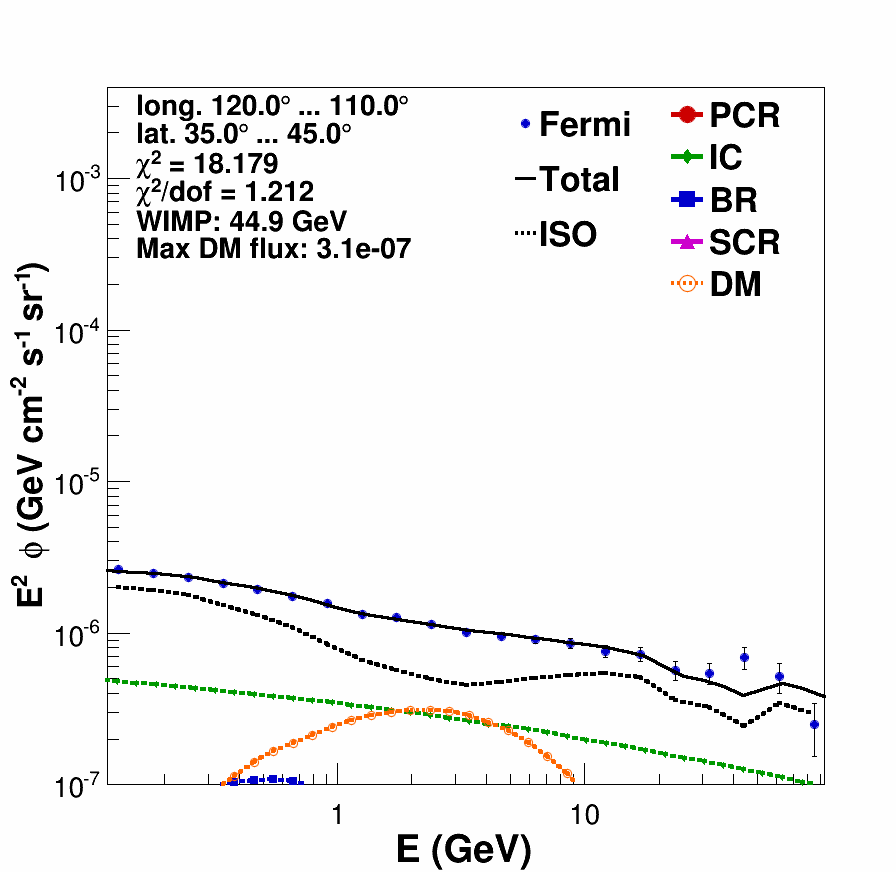}
\includegraphics[width=0.16\textwidth,height=0.16\textwidth,clip]{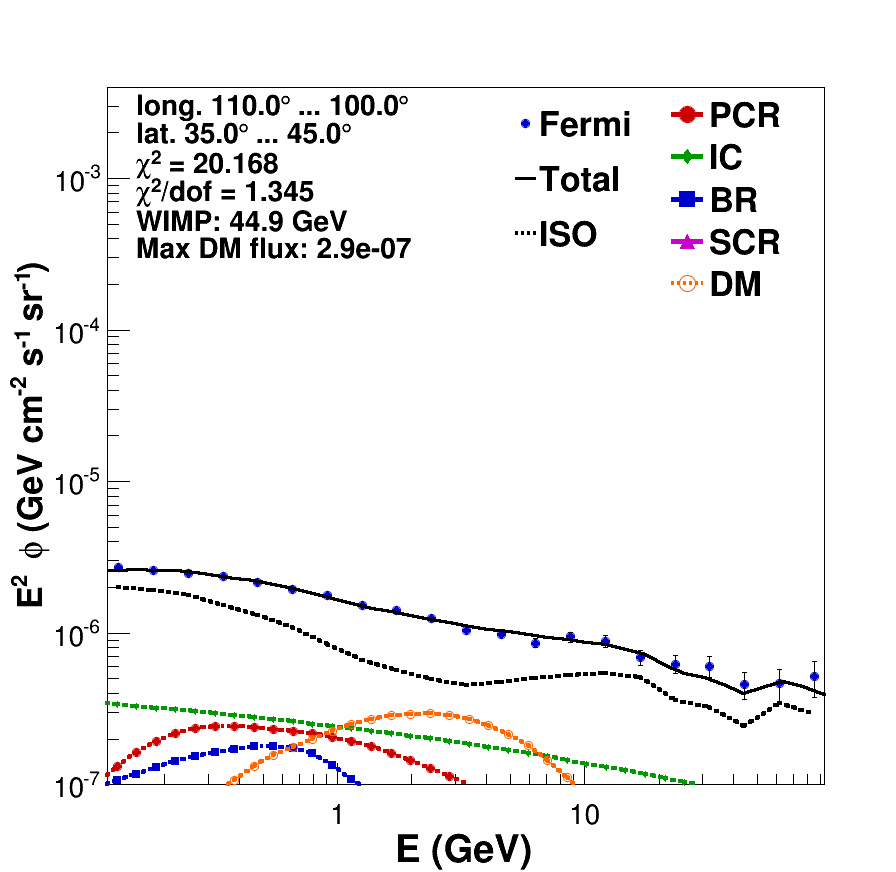}
\includegraphics[width=0.16\textwidth,height=0.16\textwidth,clip]{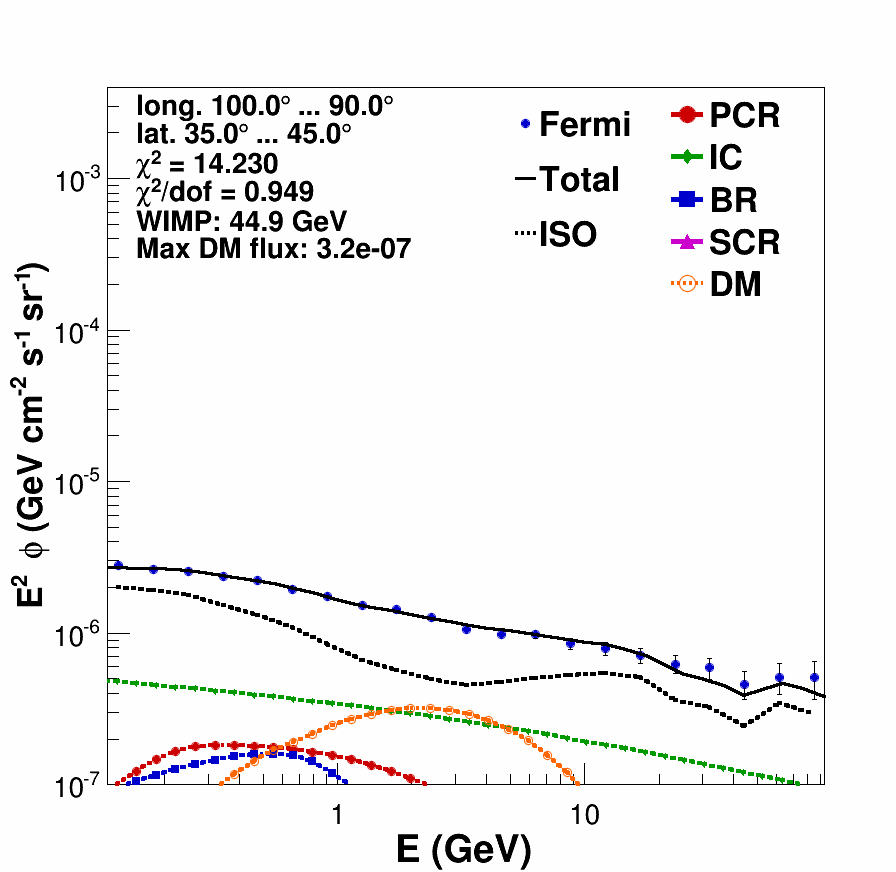}
\includegraphics[width=0.16\textwidth,height=0.16\textwidth,clip]{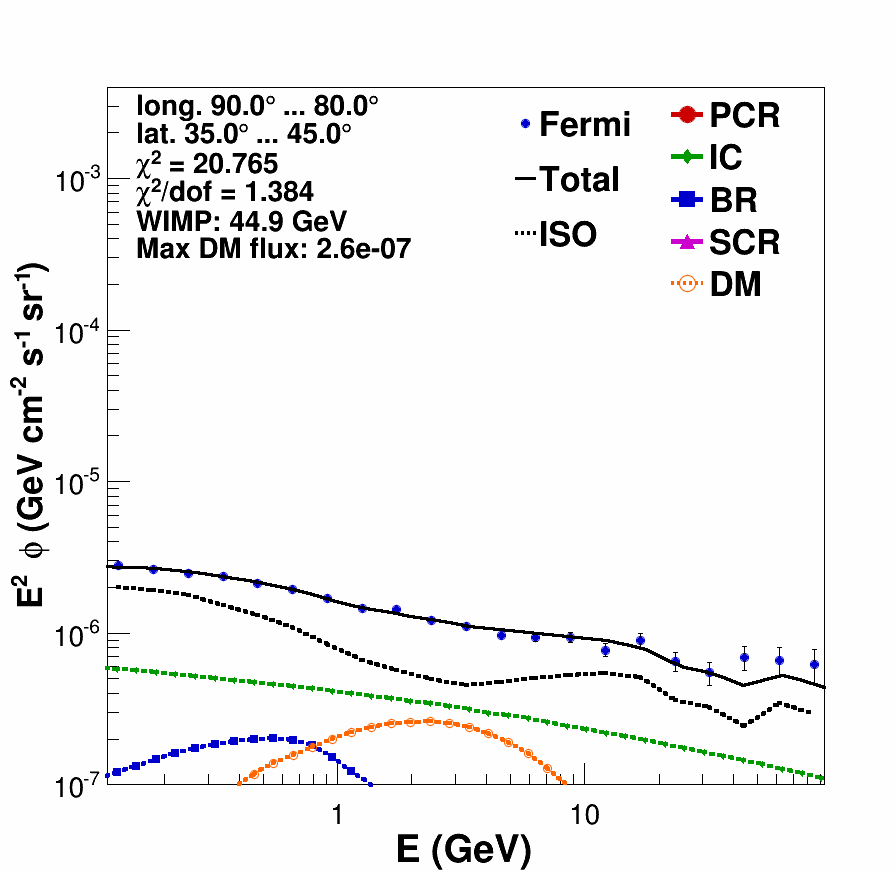}
\includegraphics[width=0.16\textwidth,height=0.16\textwidth,clip]{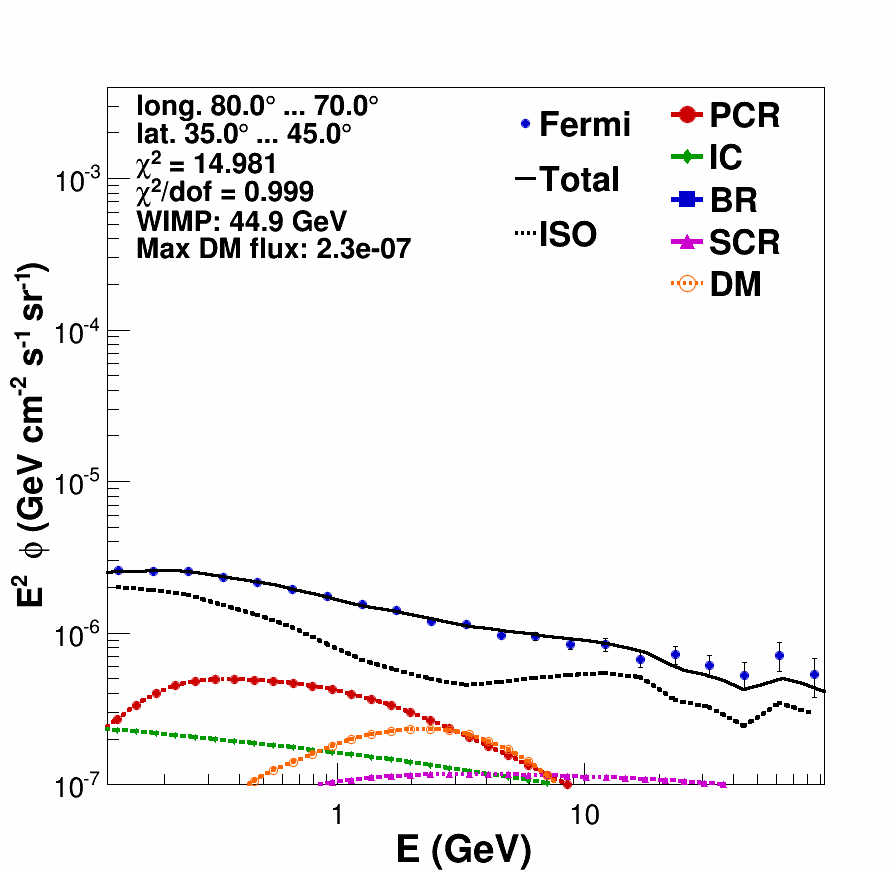}
\includegraphics[width=0.16\textwidth,height=0.16\textwidth,clip]{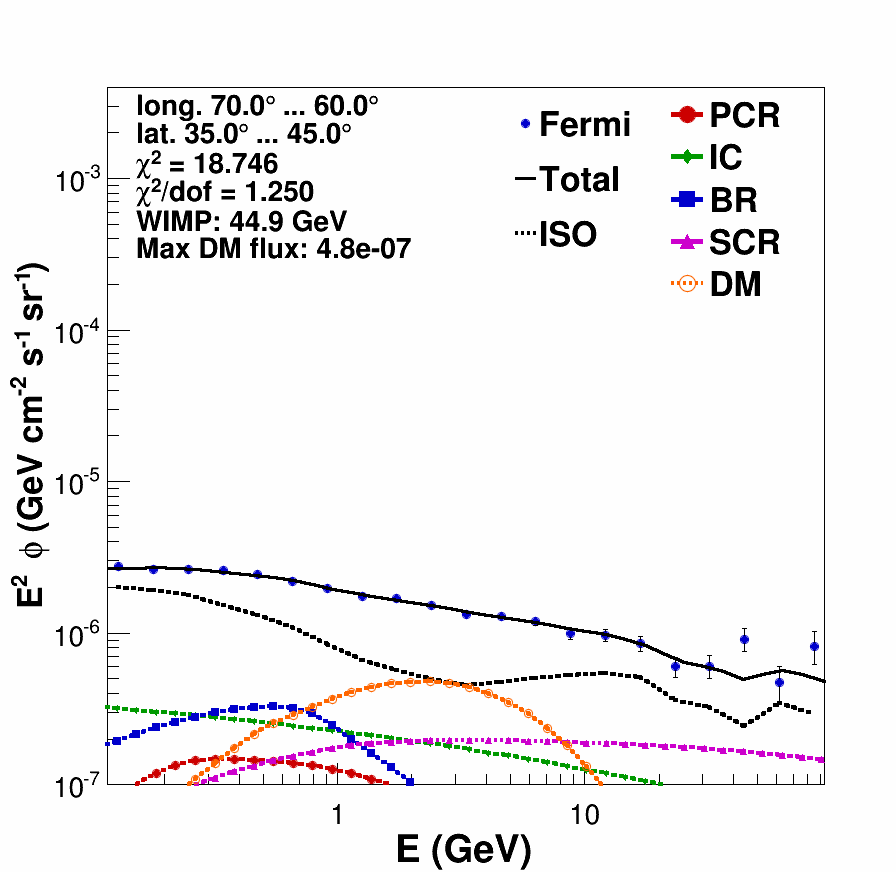}
\includegraphics[width=0.16\textwidth,height=0.16\textwidth,clip]{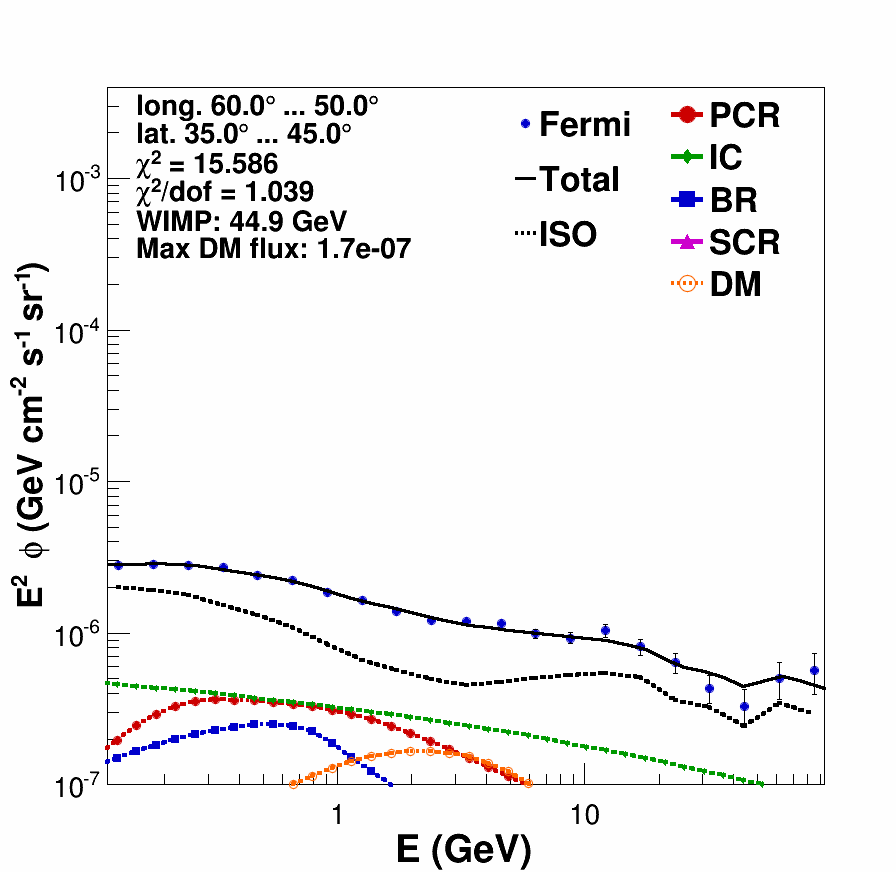}
\includegraphics[width=0.16\textwidth,height=0.16\textwidth,clip]{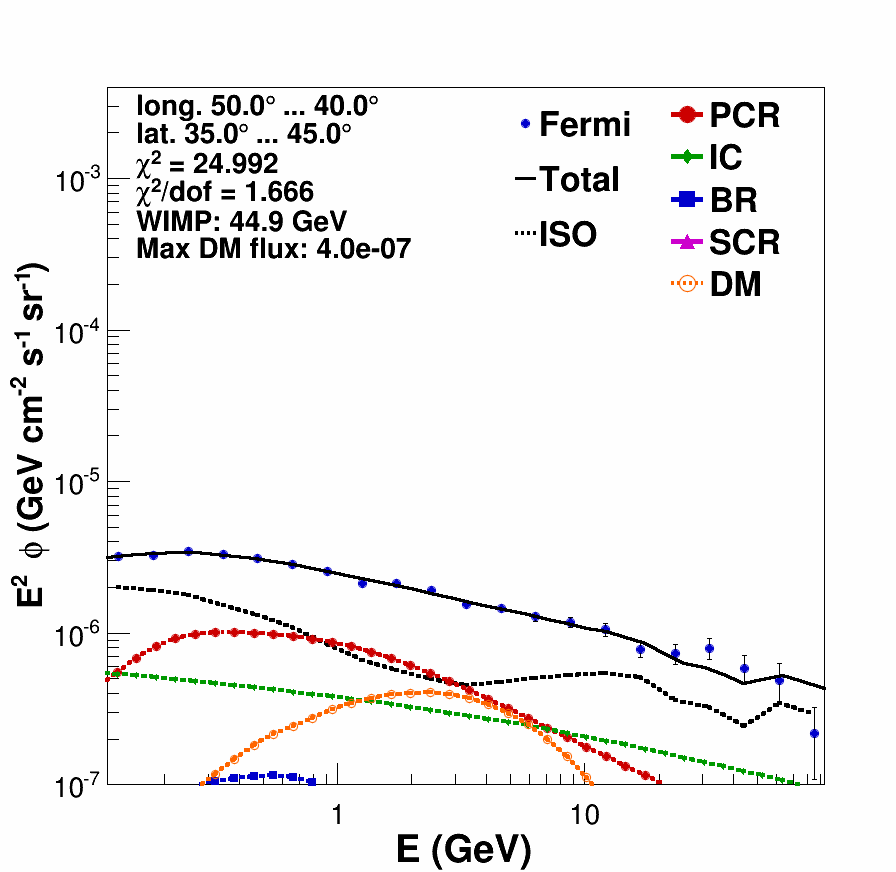}
\includegraphics[width=0.16\textwidth,height=0.16\textwidth,clip]{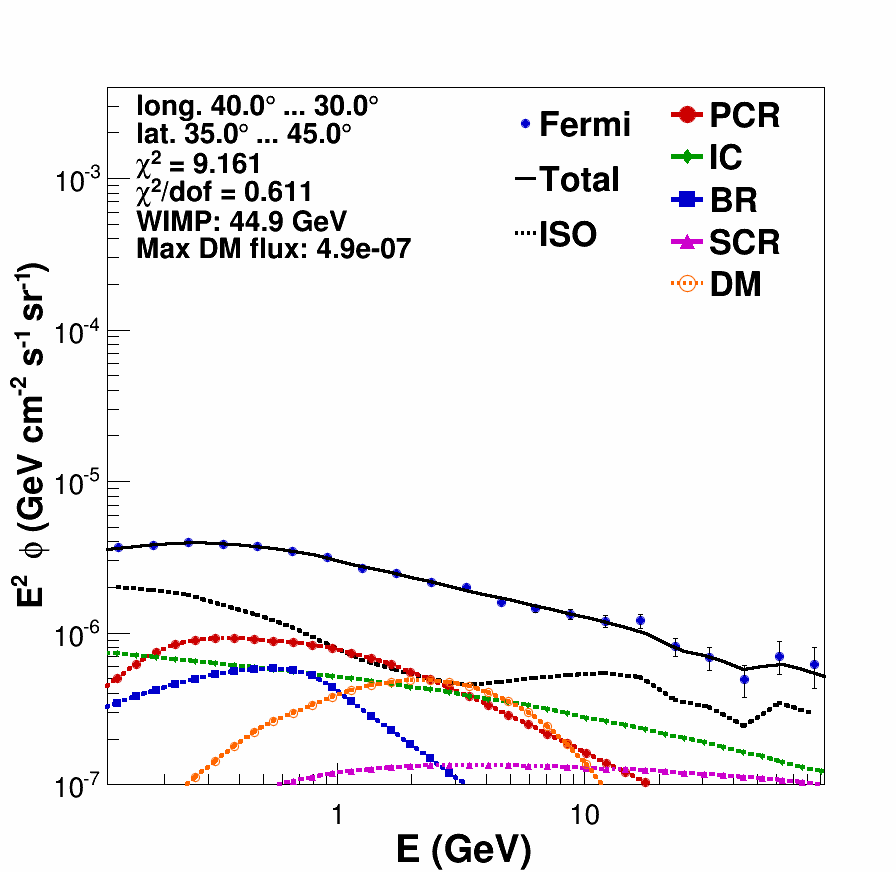}
\includegraphics[width=0.16\textwidth,height=0.16\textwidth,clip]{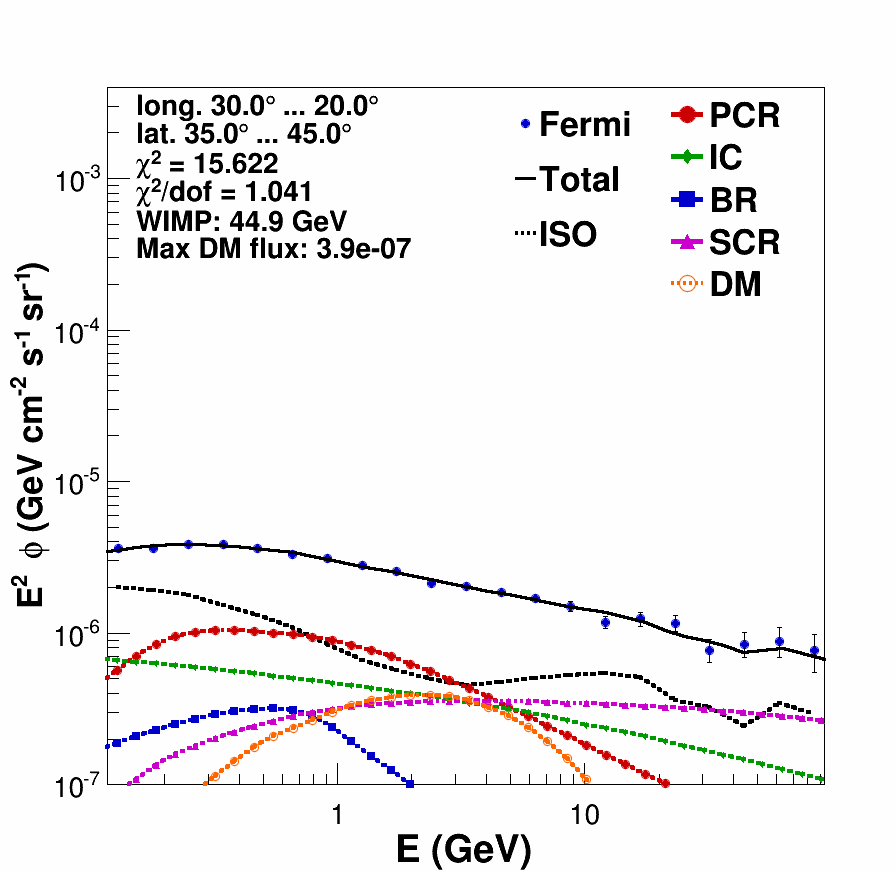}
\includegraphics[width=0.16\textwidth,height=0.16\textwidth,clip]{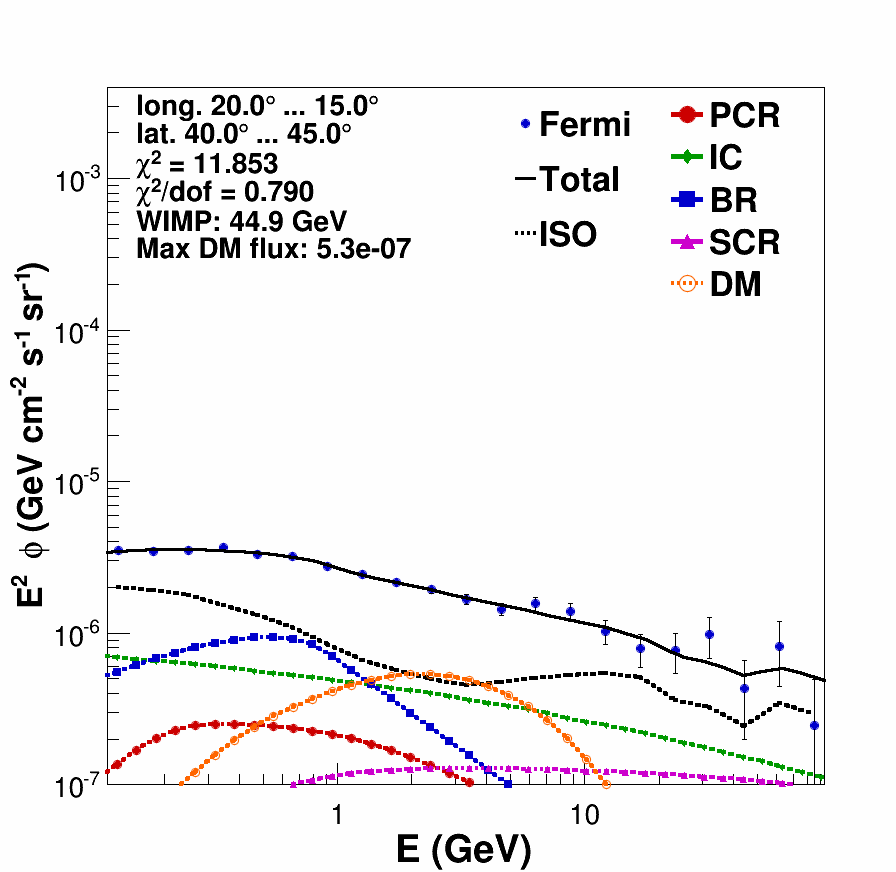}
\includegraphics[width=0.16\textwidth,height=0.16\textwidth,clip]{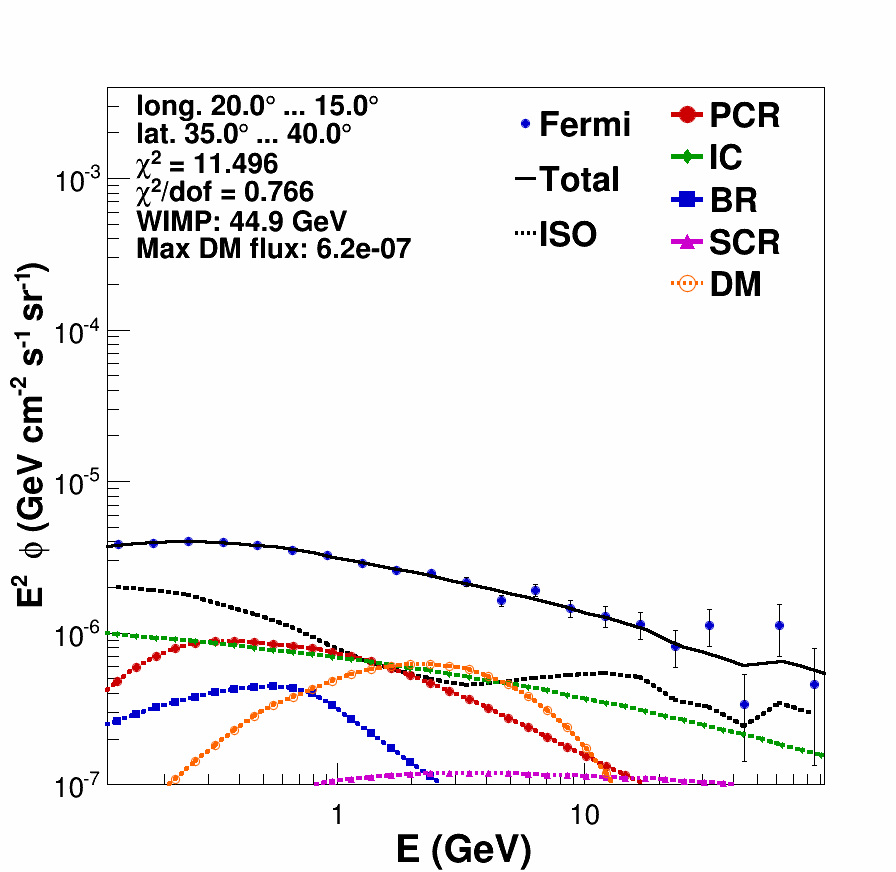}
\includegraphics[width=0.16\textwidth,height=0.16\textwidth,clip]{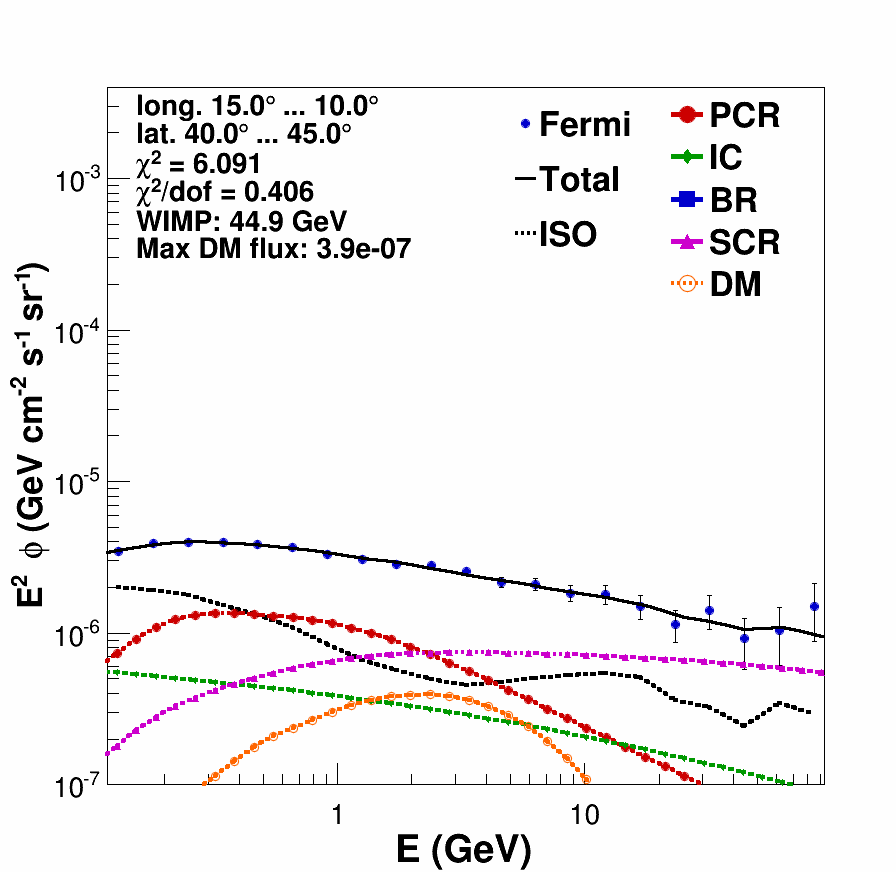}
\includegraphics[width=0.16\textwidth,height=0.16\textwidth,clip]{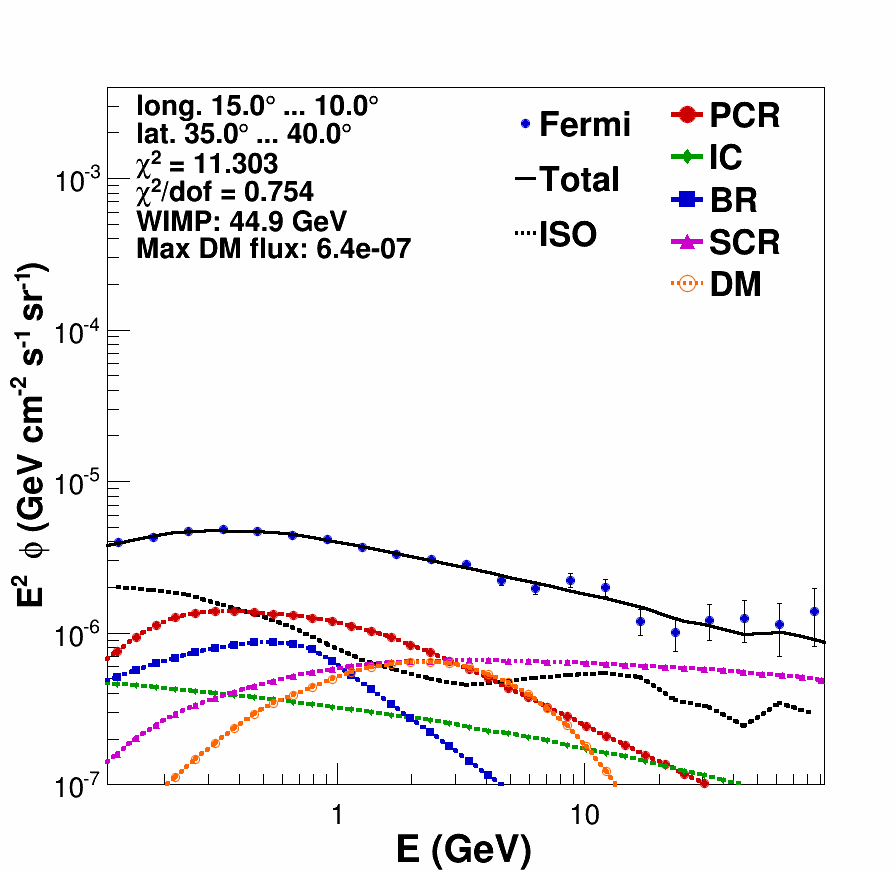}
\includegraphics[width=0.16\textwidth,height=0.16\textwidth,clip]{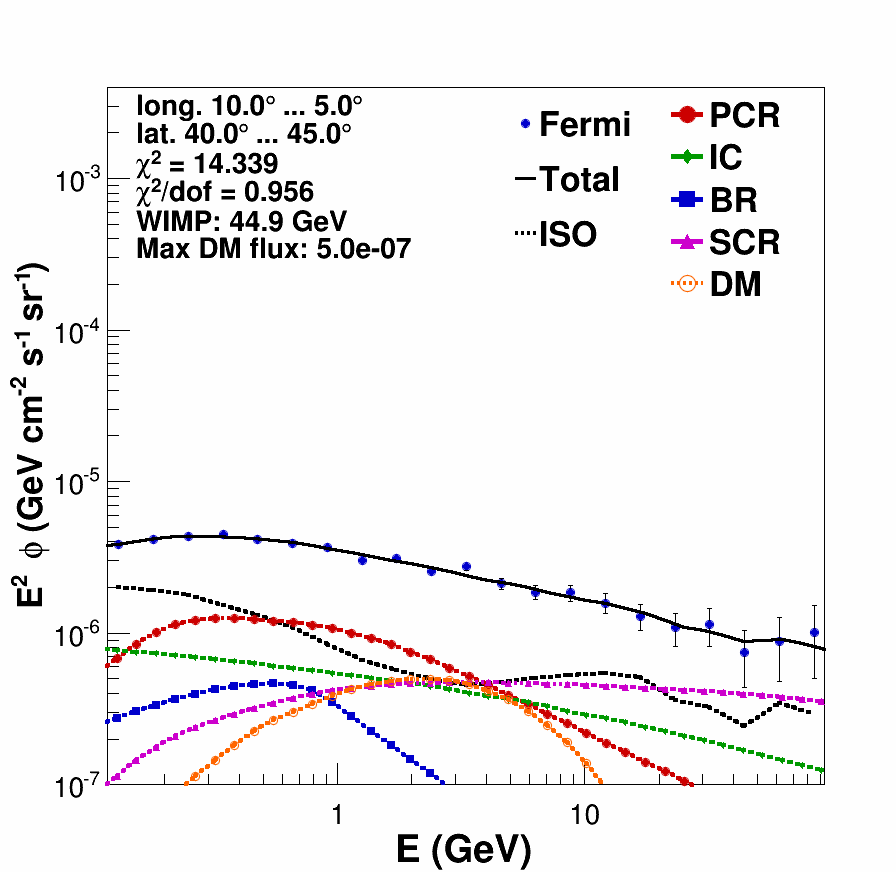}
\includegraphics[width=0.16\textwidth,height=0.16\textwidth,clip]{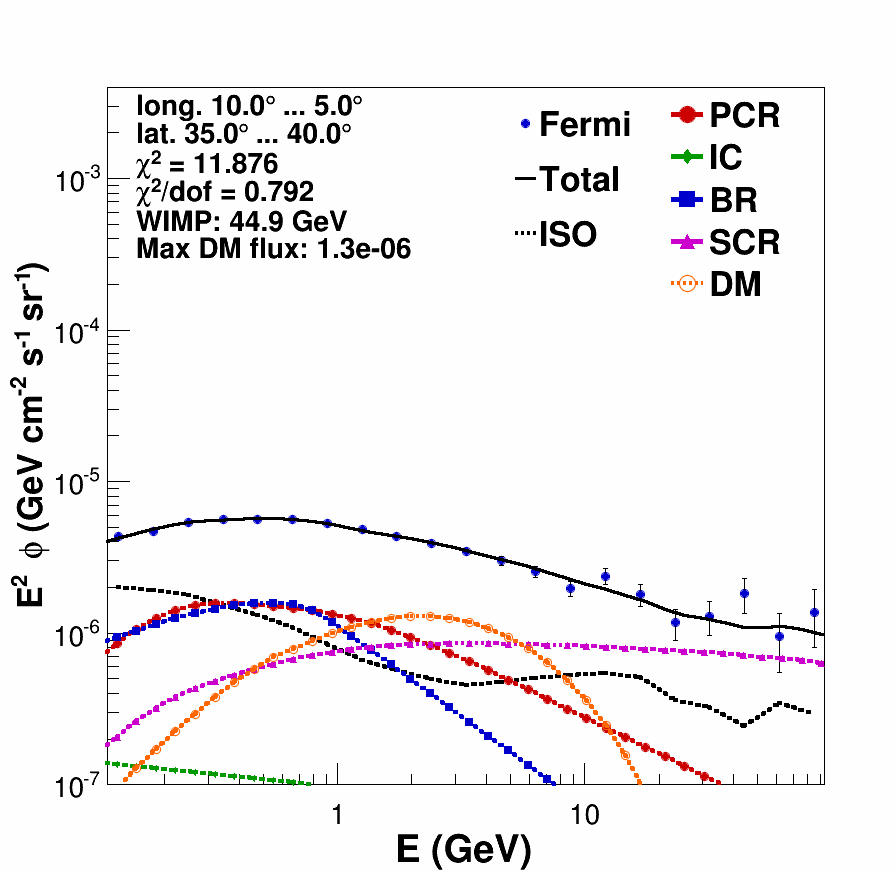}
\includegraphics[width=0.16\textwidth,height=0.16\textwidth,clip]{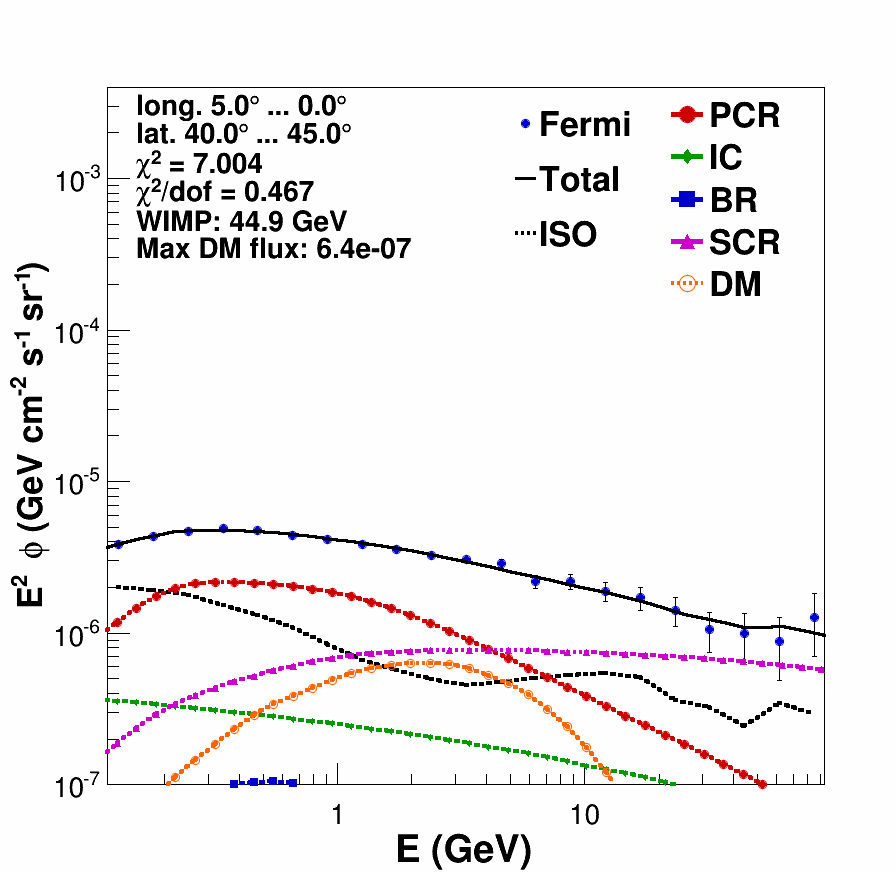}
\includegraphics[width=0.16\textwidth,height=0.16\textwidth,clip]{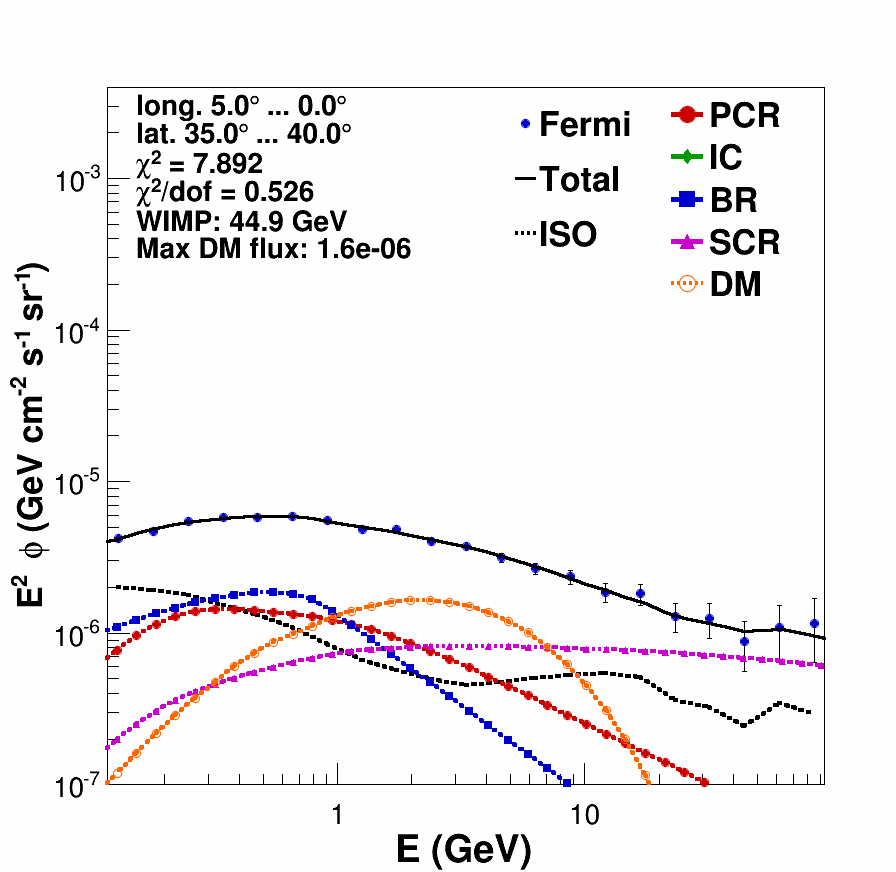}
\includegraphics[width=0.16\textwidth,height=0.16\textwidth,clip]{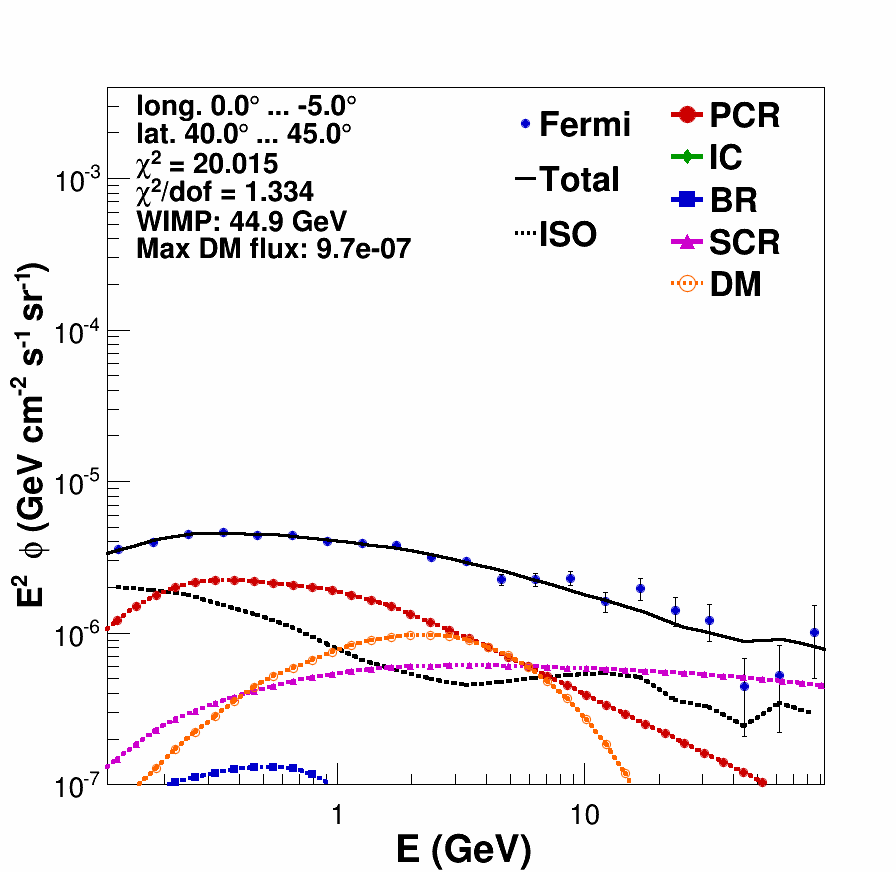}
\includegraphics[width=0.16\textwidth,height=0.16\textwidth,clip]{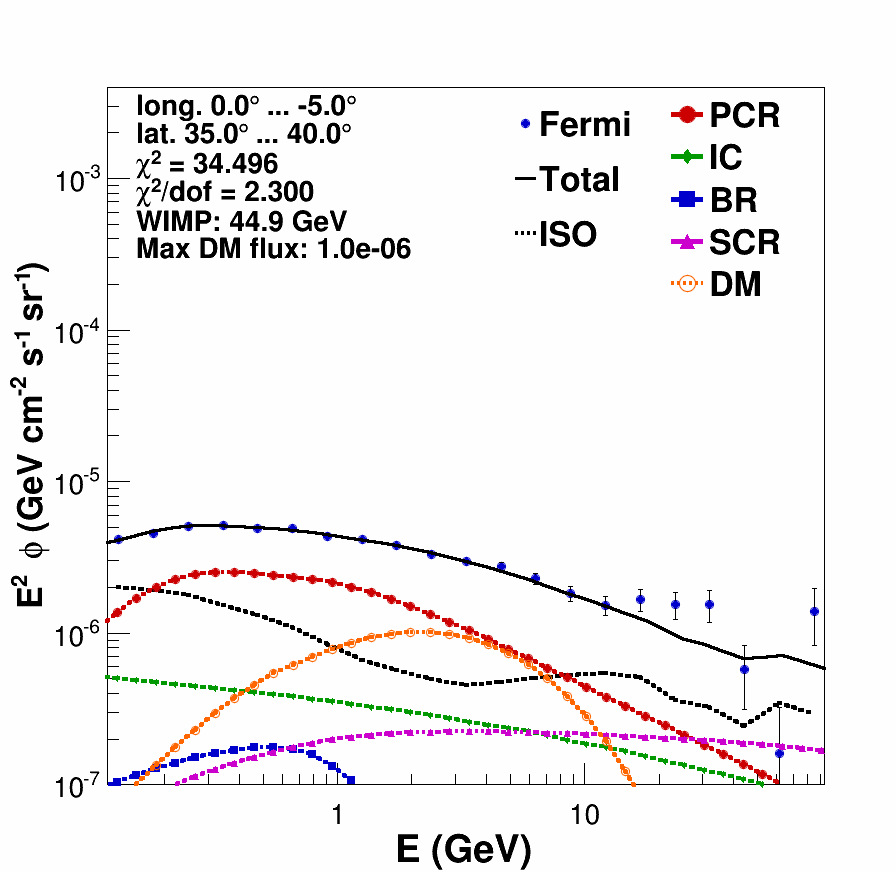}
\includegraphics[width=0.16\textwidth,height=0.16\textwidth,clip]{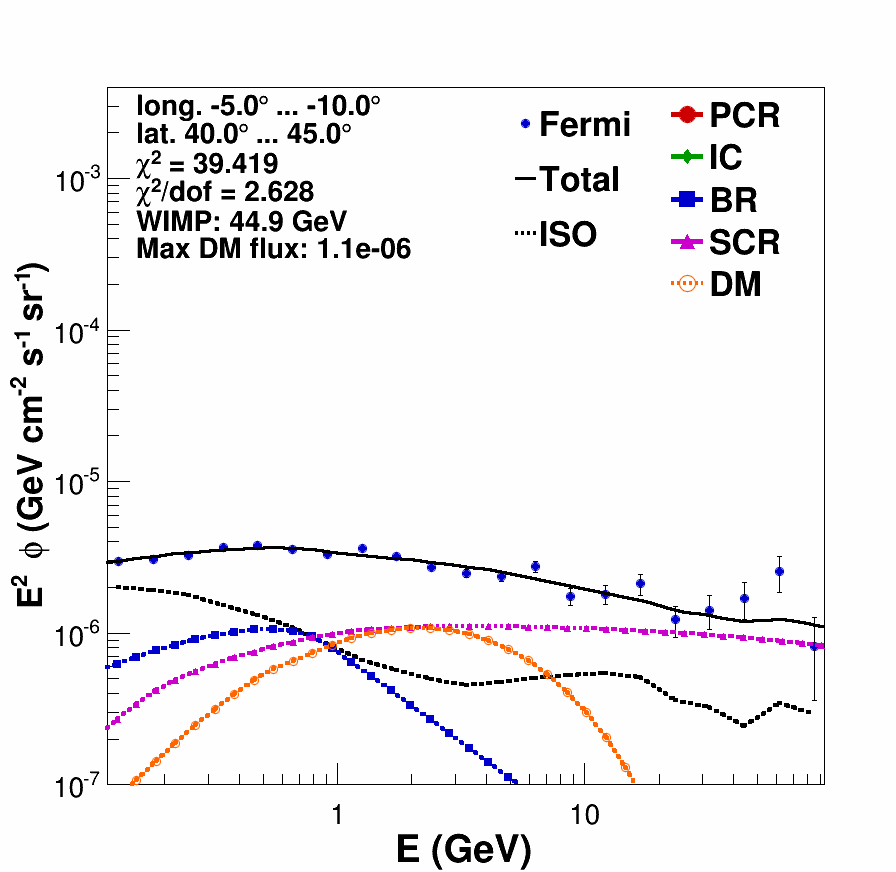}
\includegraphics[width=0.16\textwidth,height=0.16\textwidth,clip]{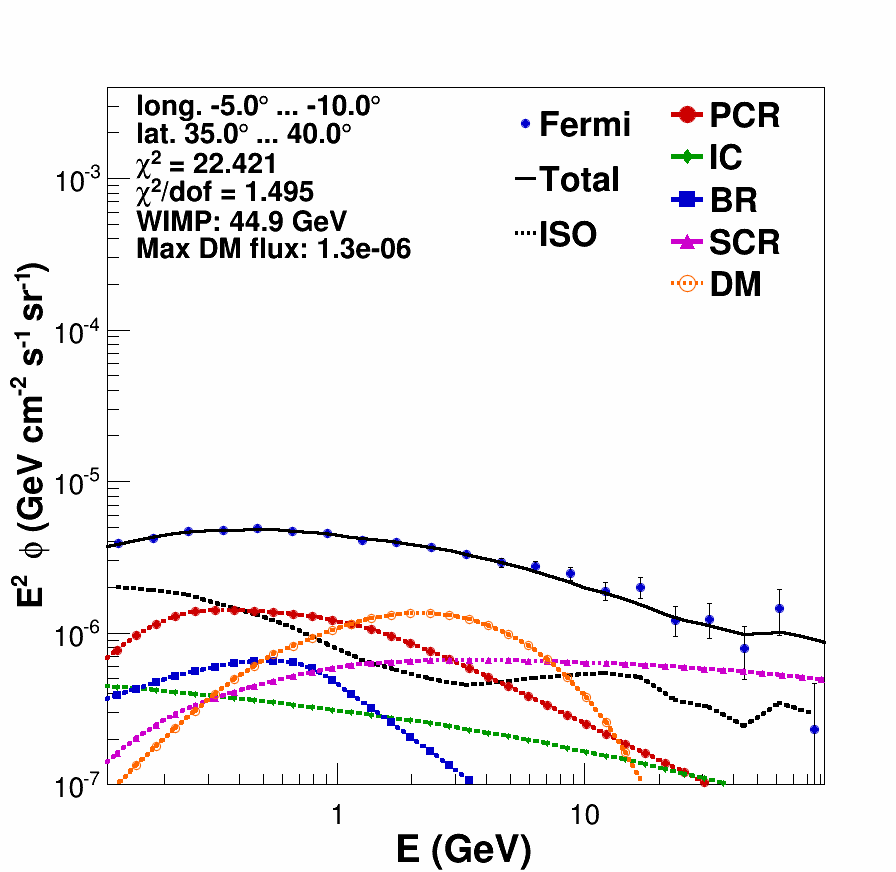}
\includegraphics[width=0.16\textwidth,height=0.16\textwidth,clip]{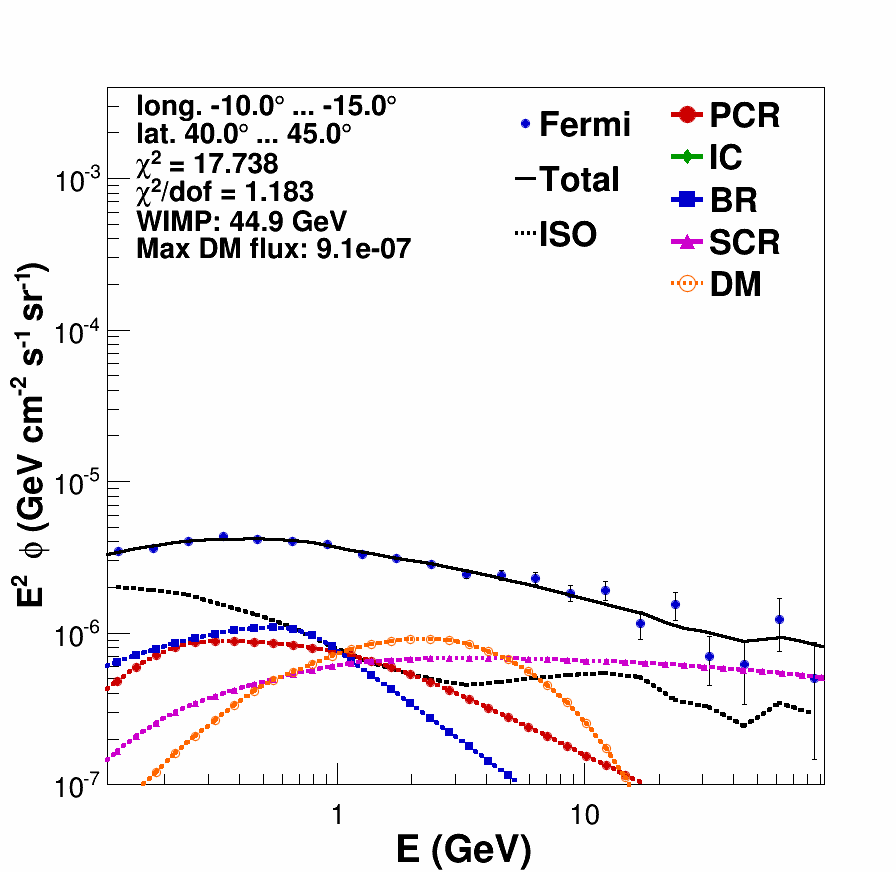}
\includegraphics[width=0.16\textwidth,height=0.16\textwidth,clip]{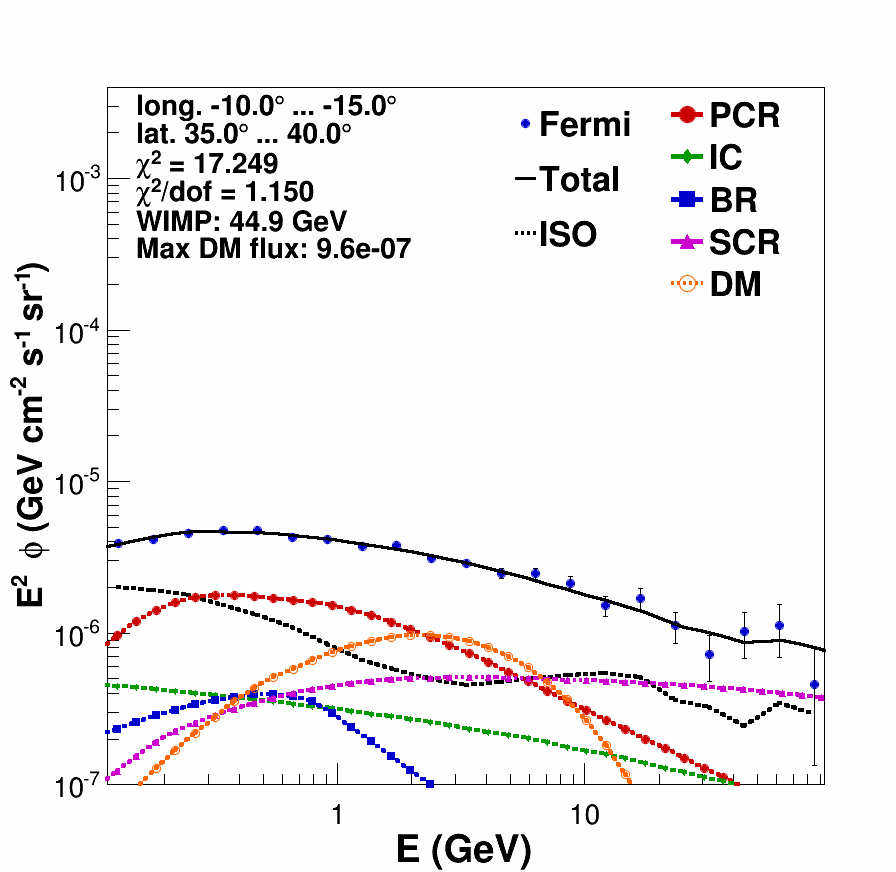}
\includegraphics[width=0.16\textwidth,height=0.16\textwidth,clip]{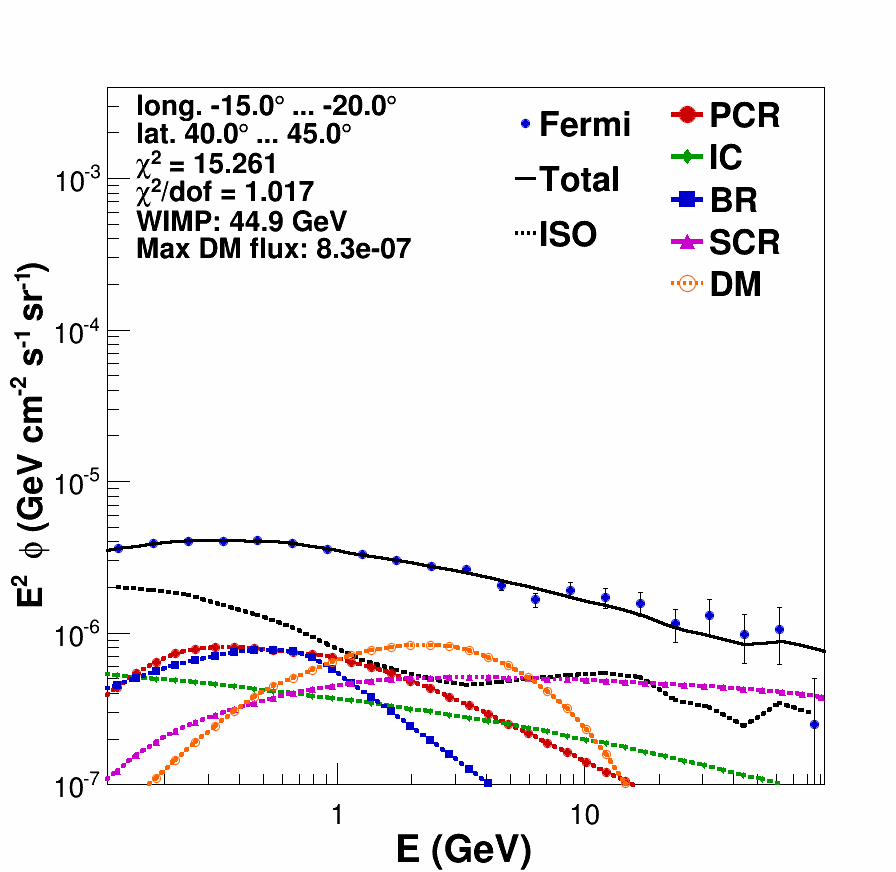}
\includegraphics[width=0.16\textwidth,height=0.16\textwidth,clip]{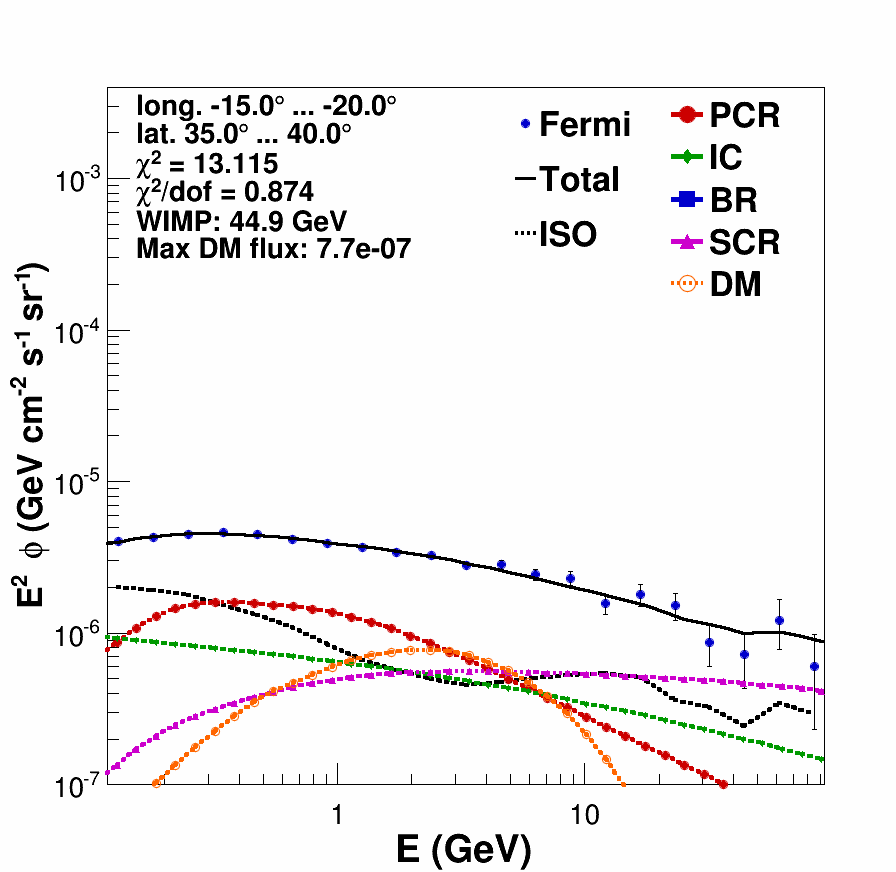}
\includegraphics[width=0.16\textwidth,height=0.16\textwidth,clip]{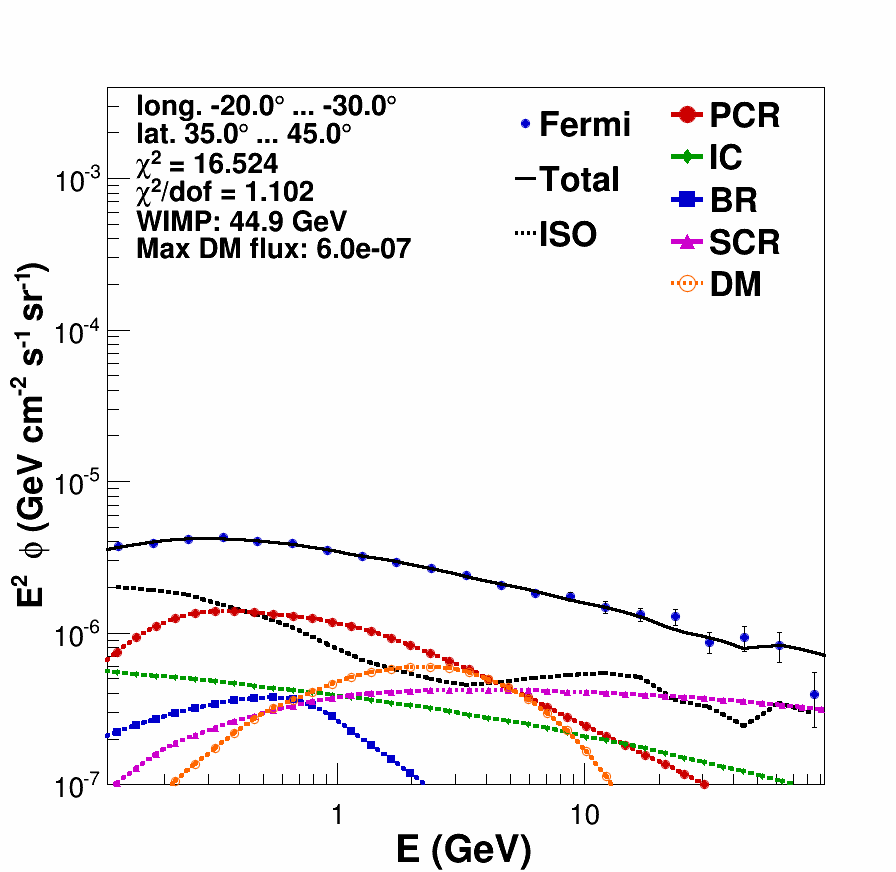}
\includegraphics[width=0.16\textwidth,height=0.16\textwidth,clip]{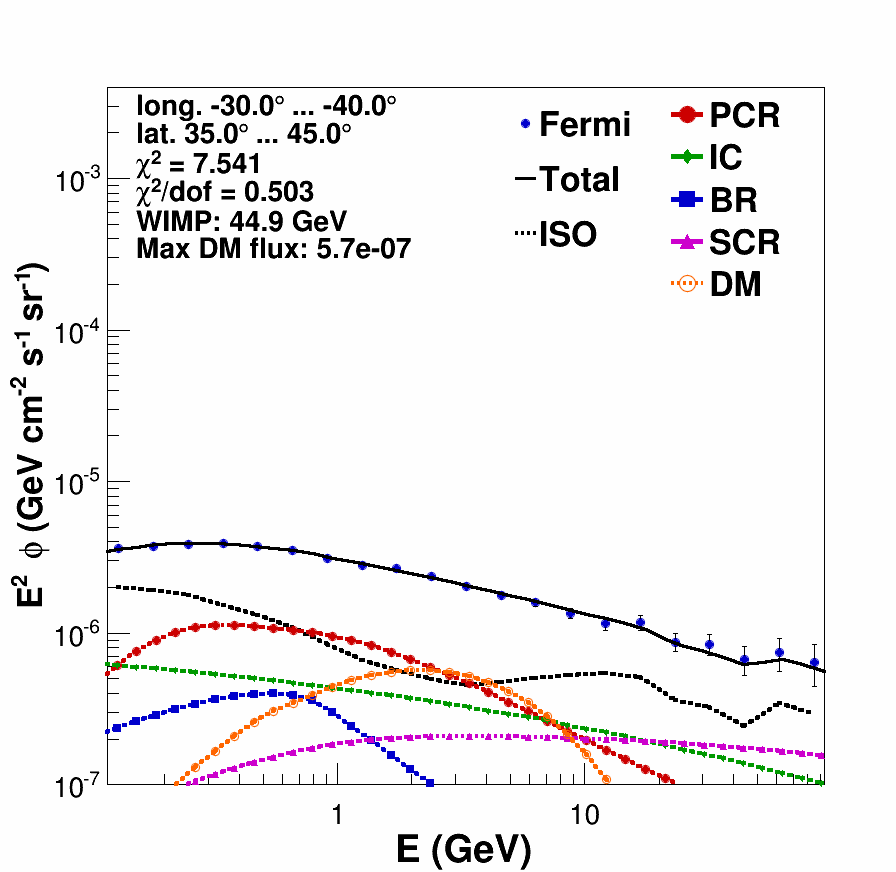}
\includegraphics[width=0.16\textwidth,height=0.16\textwidth,clip]{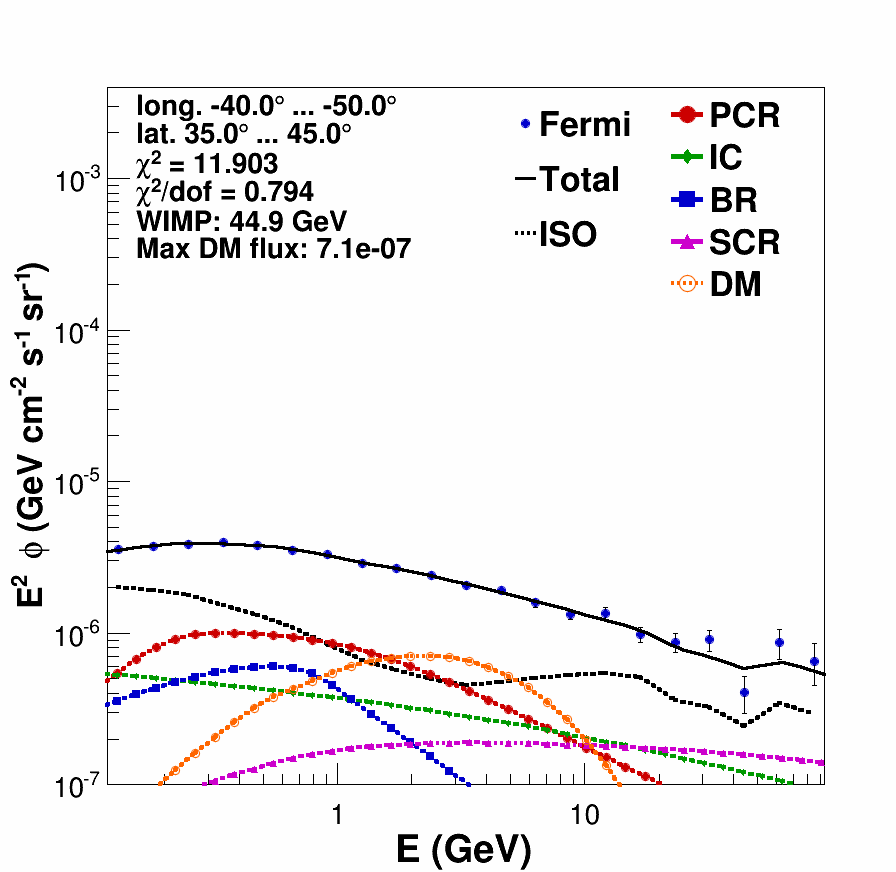}
\includegraphics[width=0.16\textwidth,height=0.16\textwidth,clip]{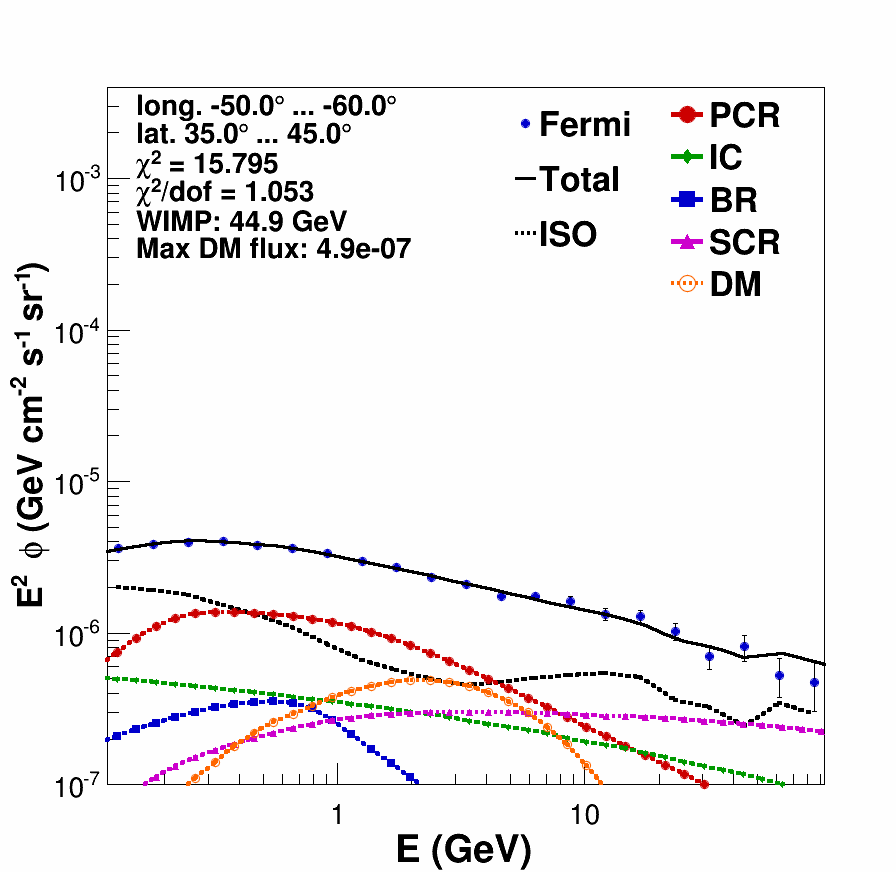}
\includegraphics[width=0.16\textwidth,height=0.16\textwidth,clip]{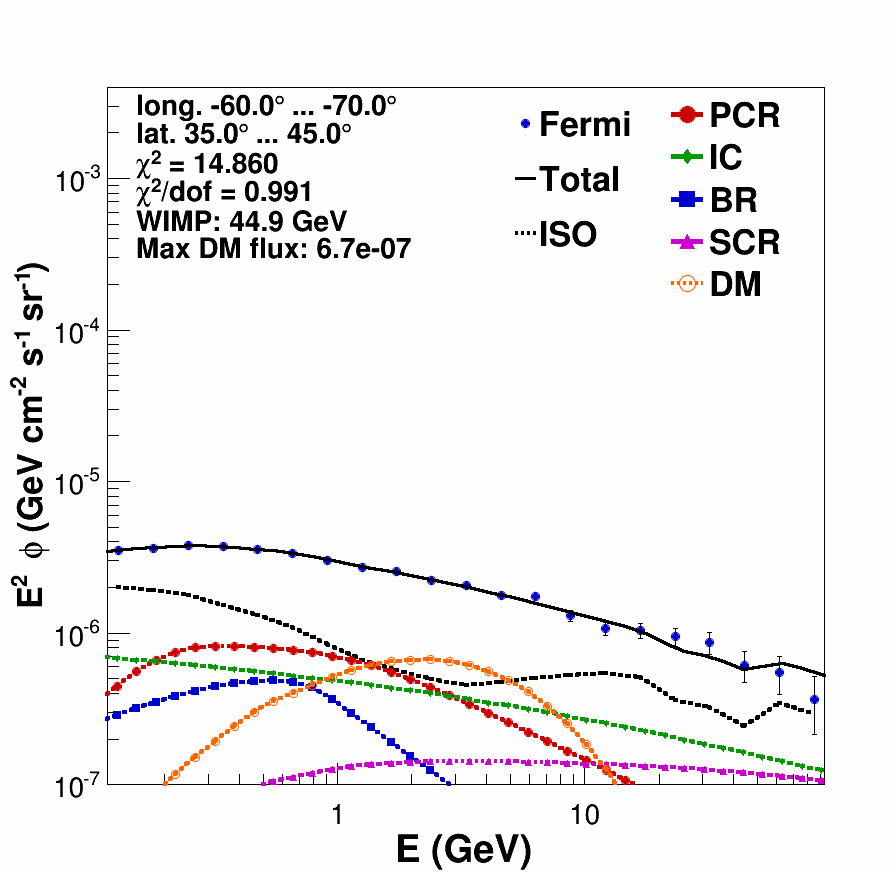}
\includegraphics[width=0.16\textwidth,height=0.16\textwidth,clip]{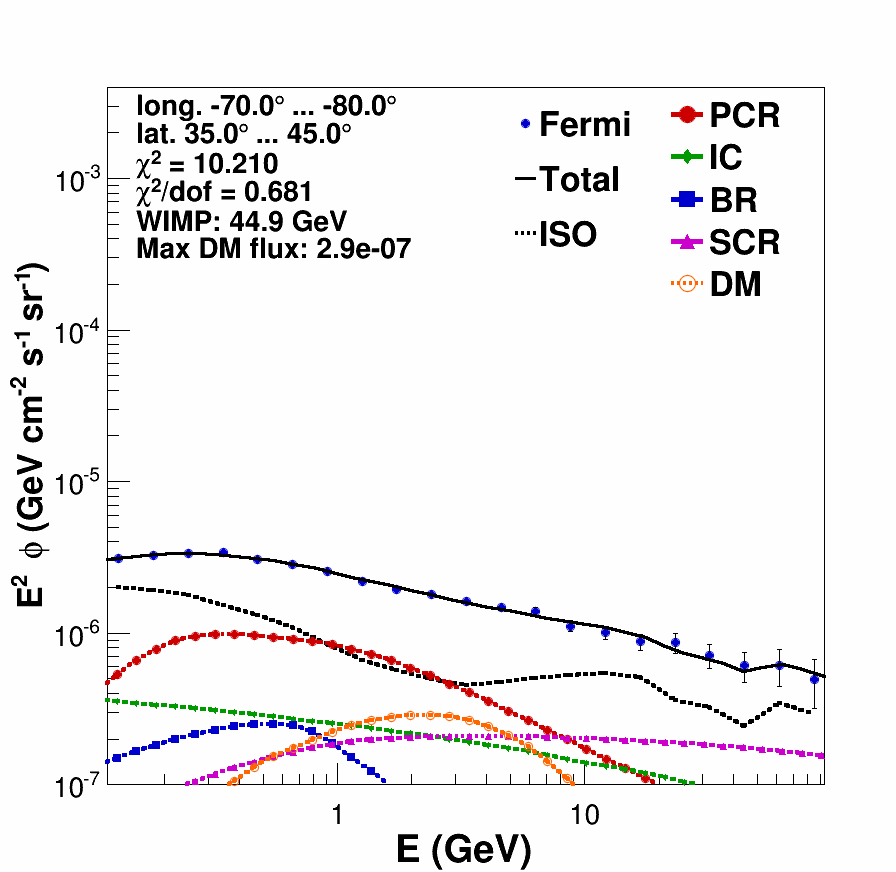}
\includegraphics[width=0.16\textwidth,height=0.16\textwidth,clip]{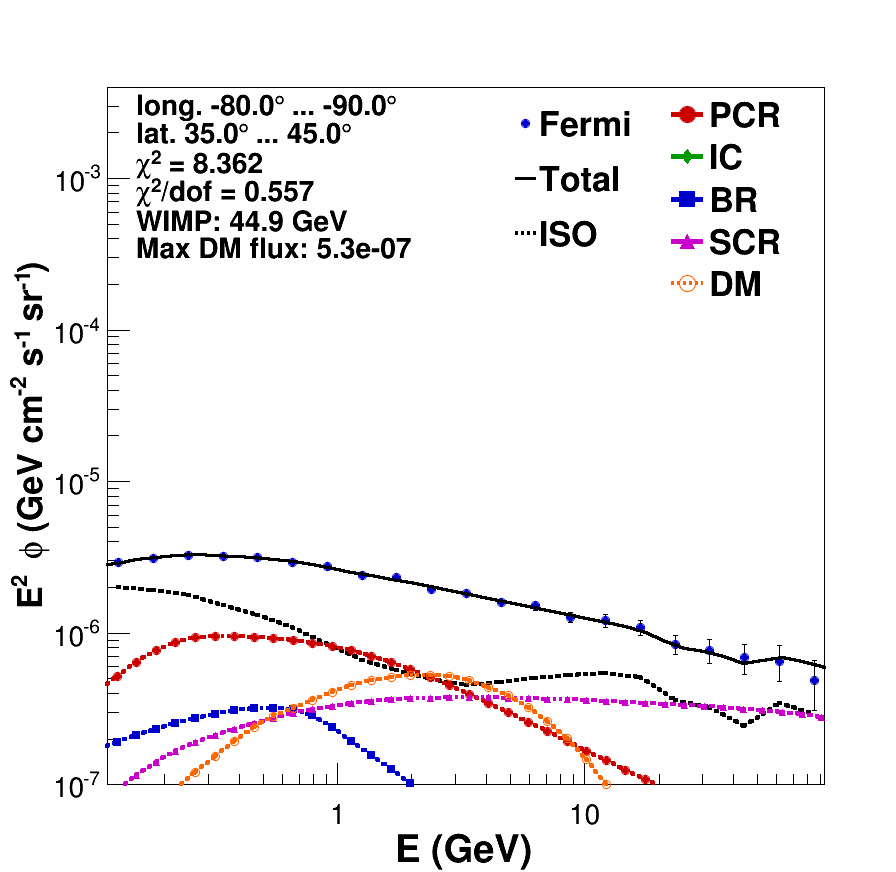}
\includegraphics[width=0.16\textwidth,height=0.16\textwidth,clip]{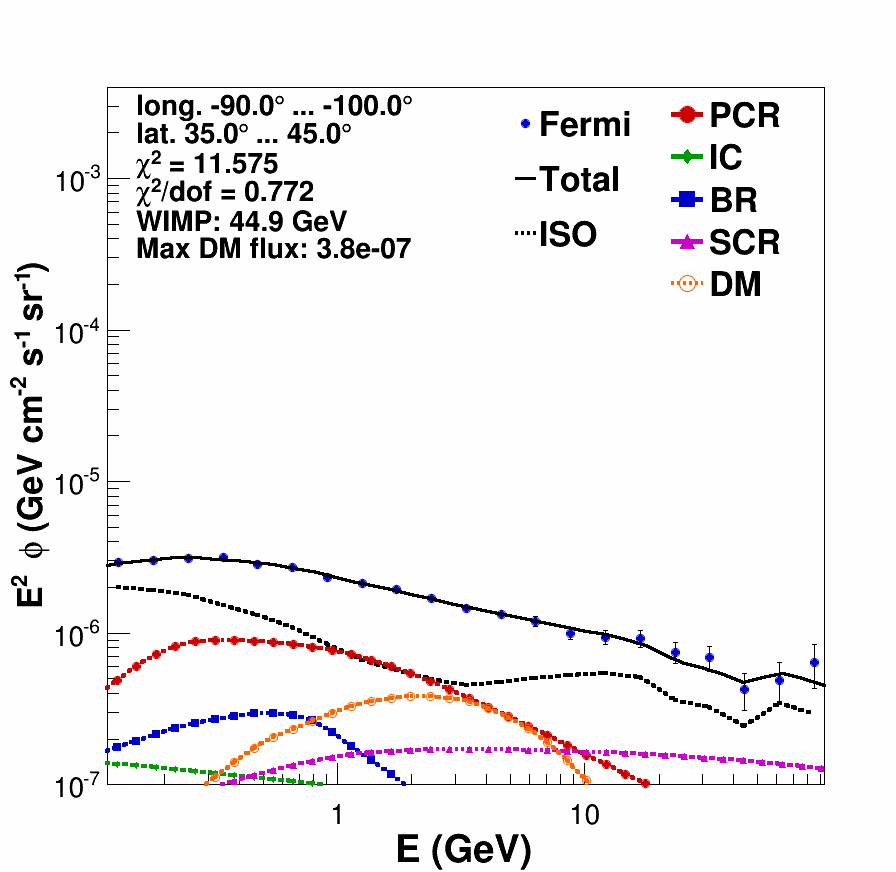}
\includegraphics[width=0.16\textwidth,height=0.16\textwidth,clip]{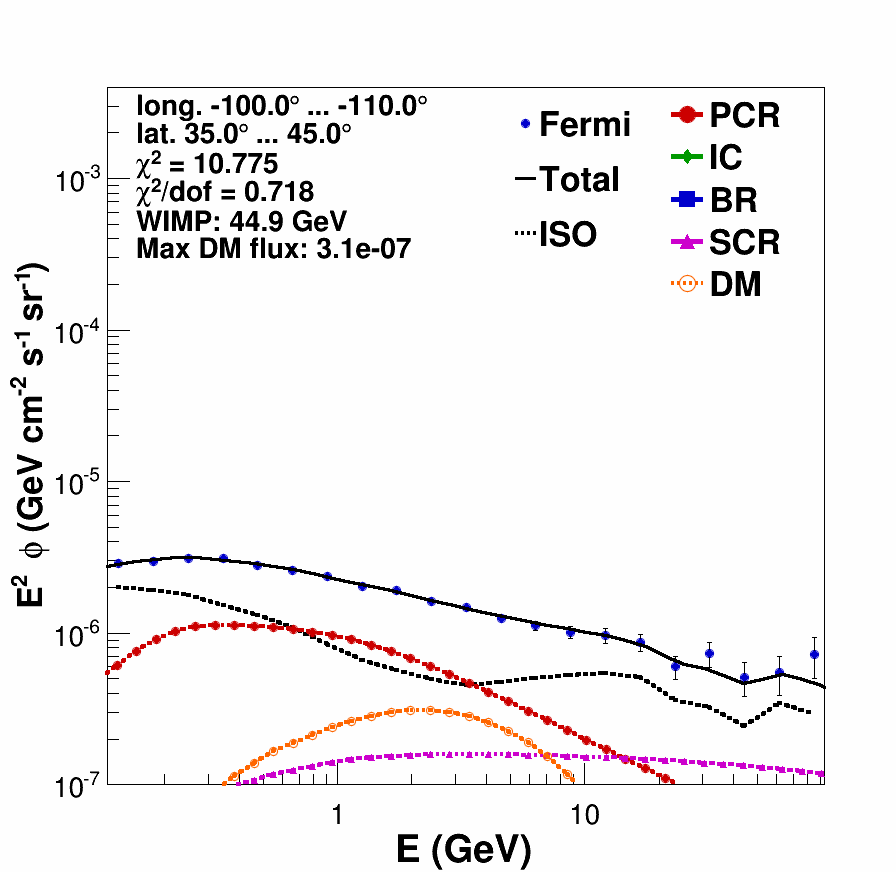}
\includegraphics[width=0.16\textwidth,height=0.16\textwidth,clip]{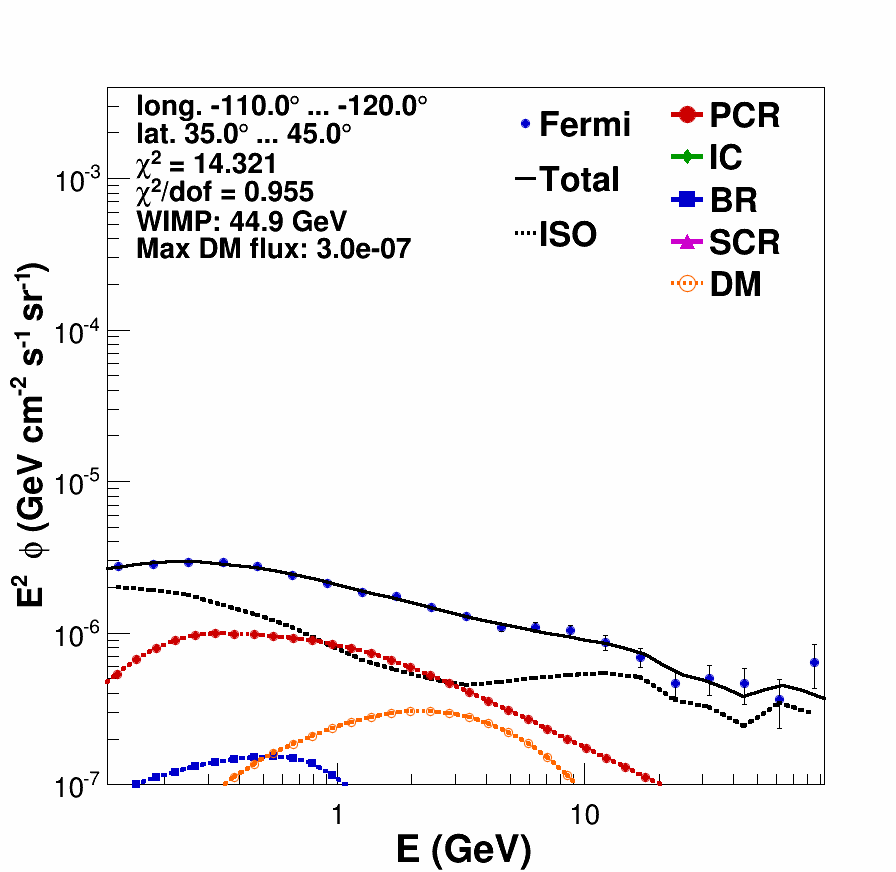}
\includegraphics[width=0.16\textwidth,height=0.16\textwidth,clip]{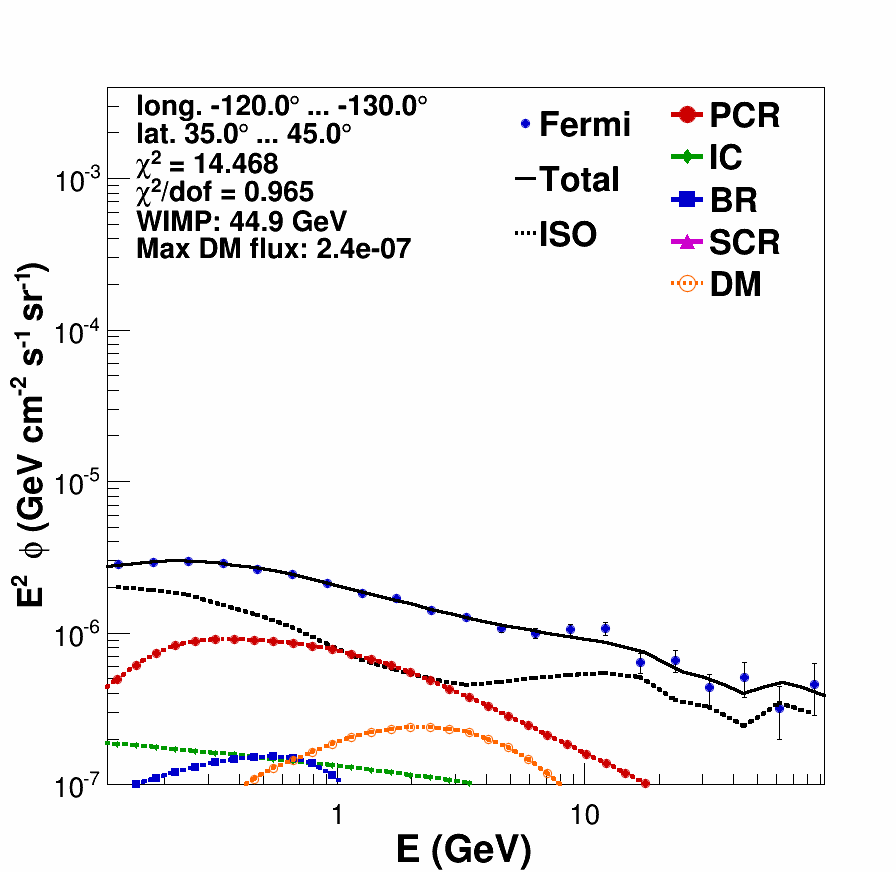}
\includegraphics[width=0.16\textwidth,height=0.16\textwidth,clip]{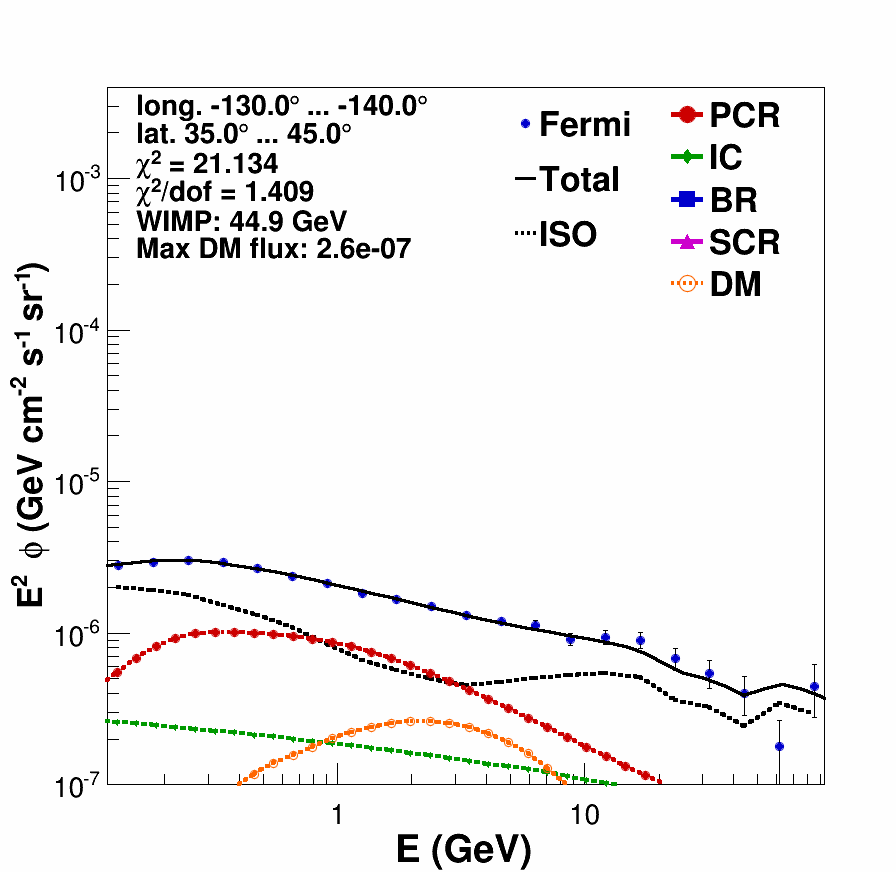}
\includegraphics[width=0.16\textwidth,height=0.16\textwidth,clip]{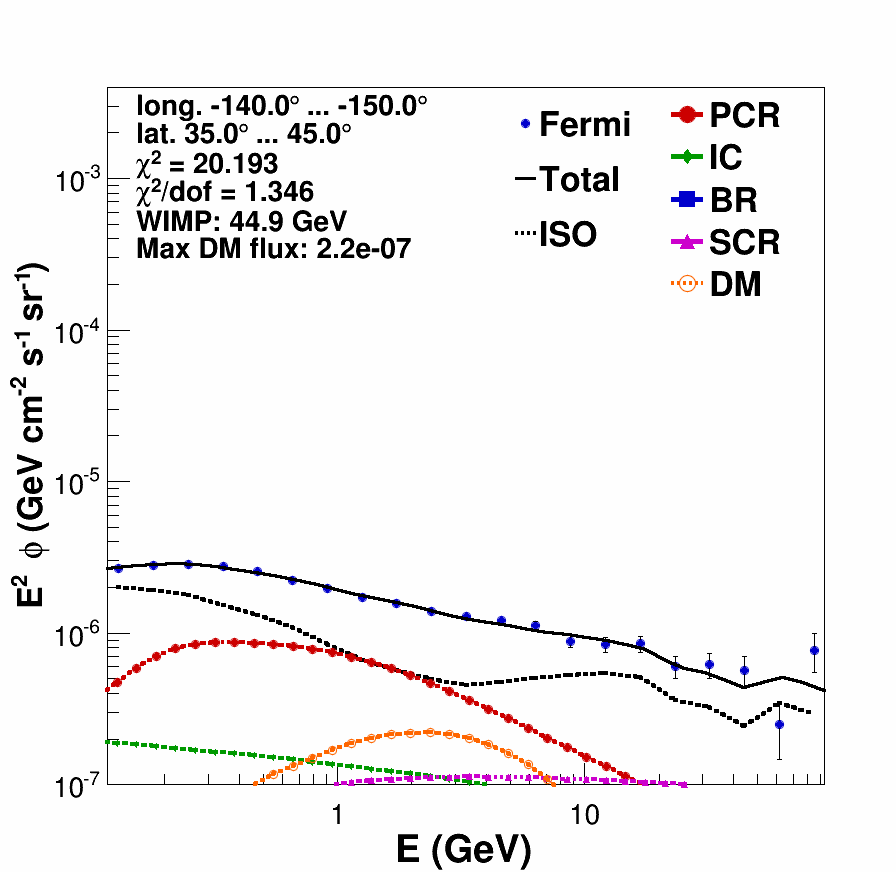}
\includegraphics[width=0.16\textwidth,height=0.16\textwidth,clip]{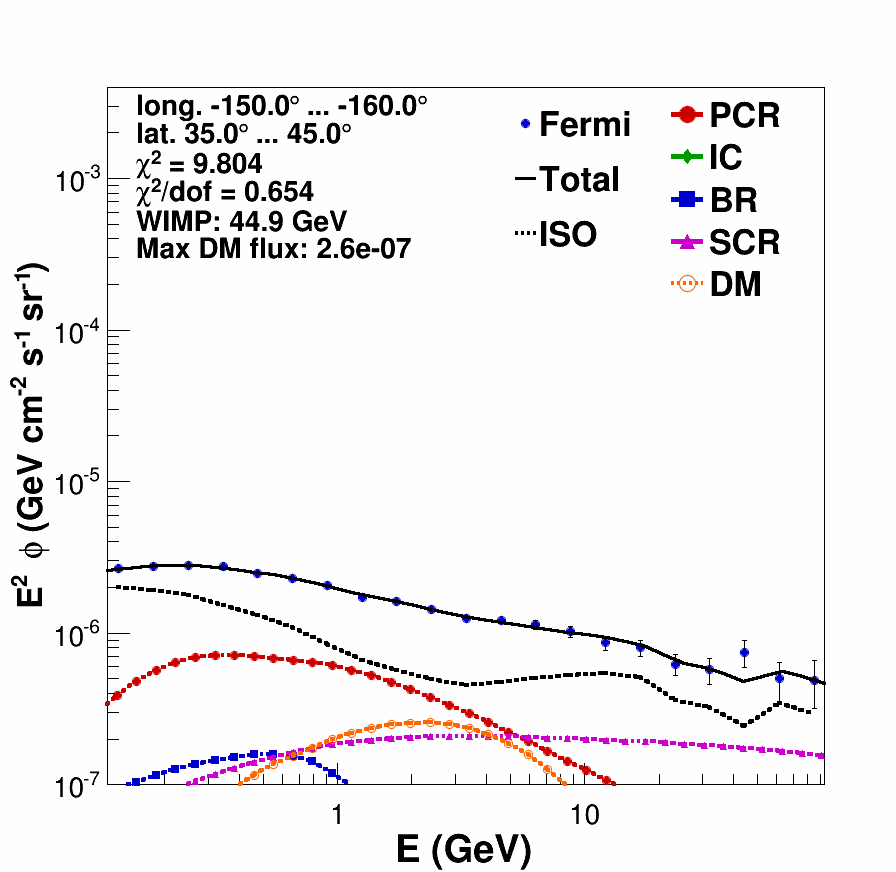}
\includegraphics[width=0.16\textwidth,height=0.16\textwidth,clip]{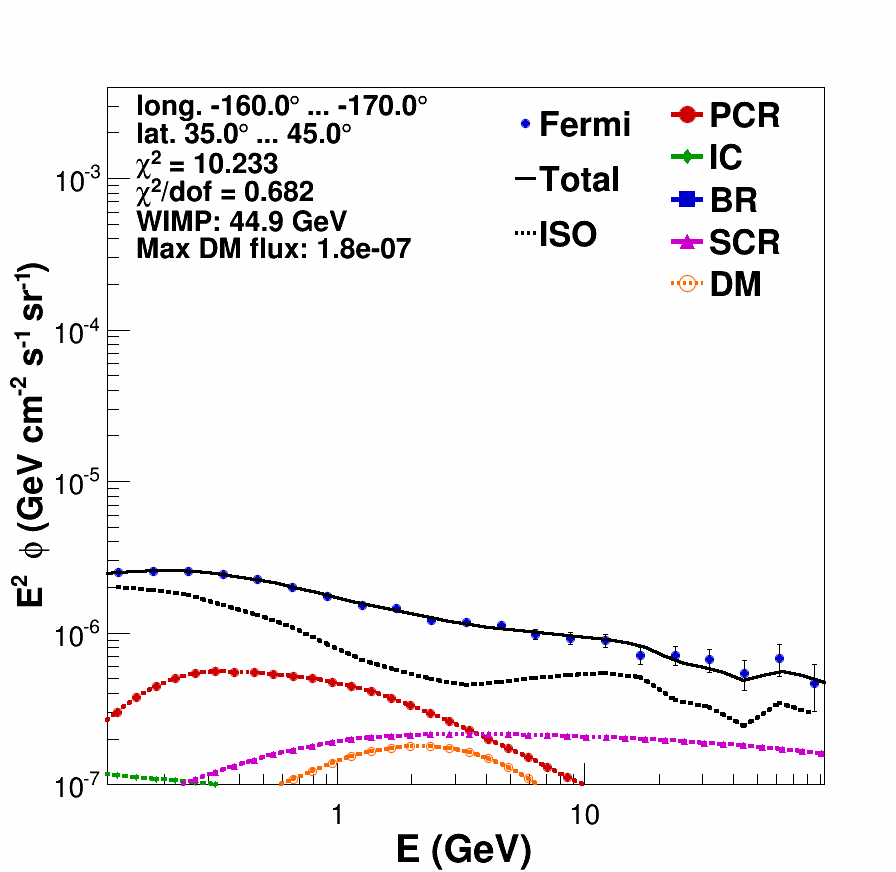}
\includegraphics[width=0.16\textwidth,height=0.16\textwidth,clip]{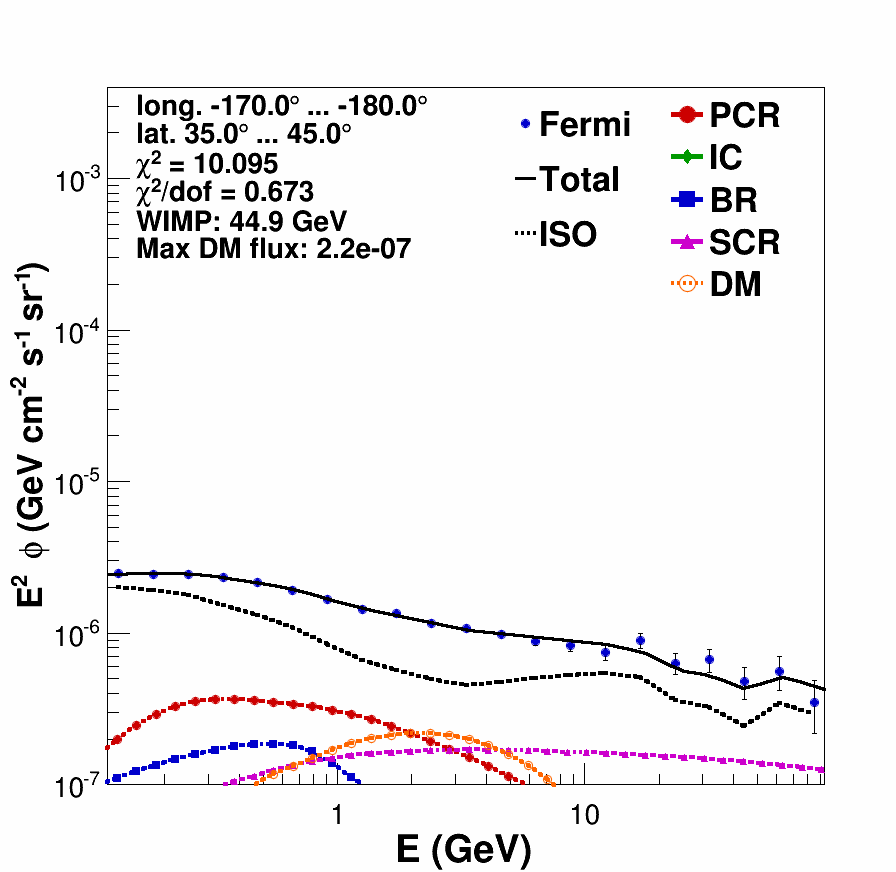}%%%%%r4
\caption[]{Template fits for latitudes  with $35.0^\circ<b<45.0^\circ$ and longitudes decreasing from 180$^\circ$ to -180$^\circ$. \label{F35}
}
\end{figure}
\clearpage
\begin{figure}
\centering
\includegraphics[width=0.16\textwidth,height=0.16\textwidth,clip]{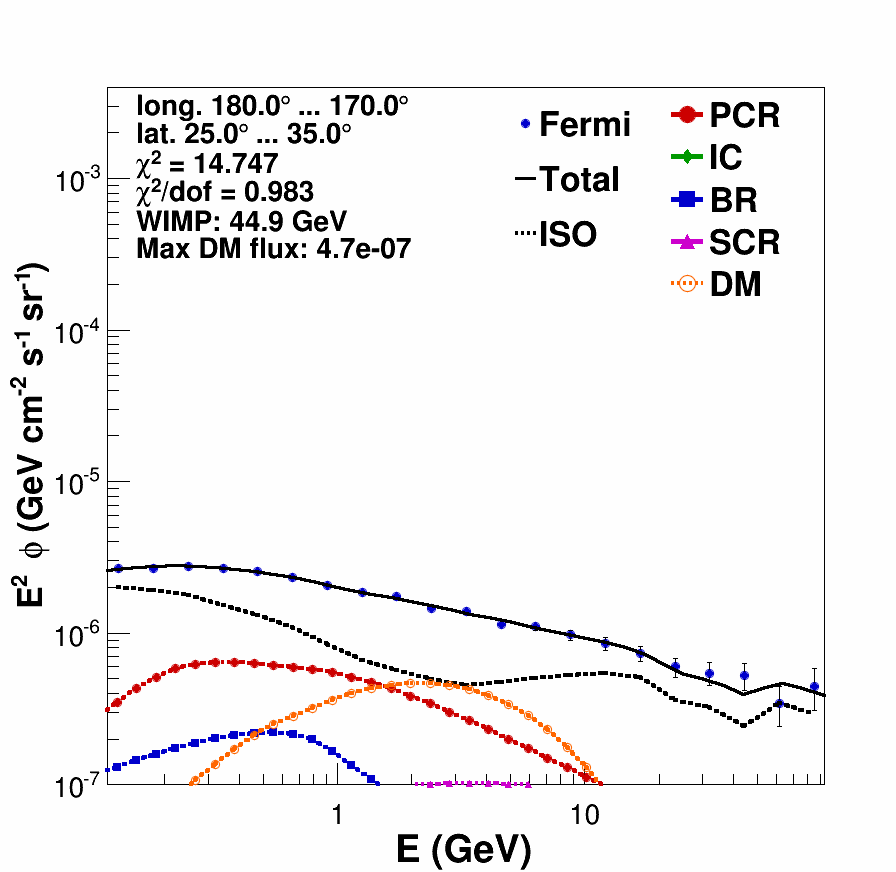}
\includegraphics[width=0.16\textwidth,height=0.16\textwidth,clip]{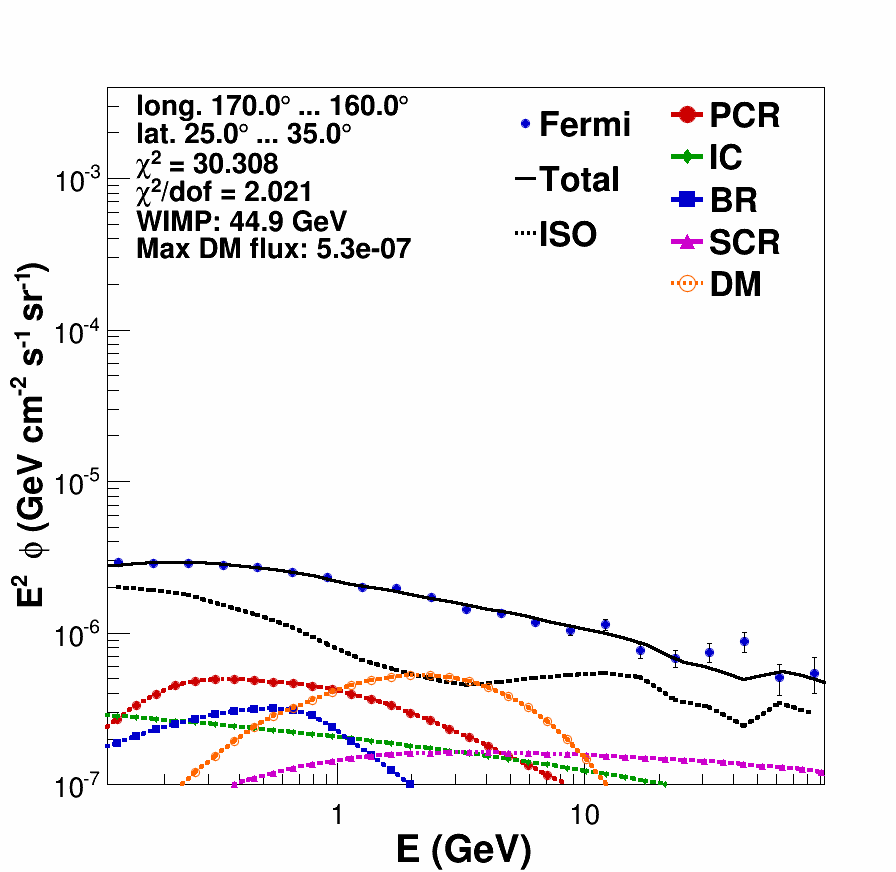}
\includegraphics[width=0.16\textwidth,height=0.16\textwidth,clip]{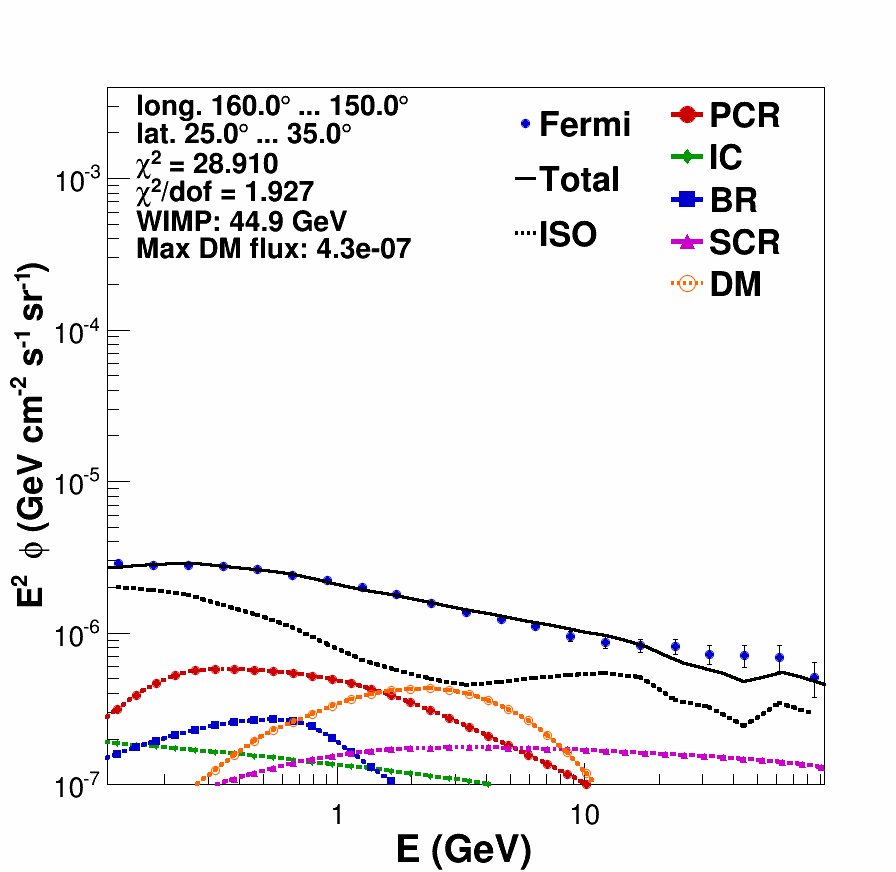}
\includegraphics[width=0.16\textwidth,height=0.16\textwidth,clip]{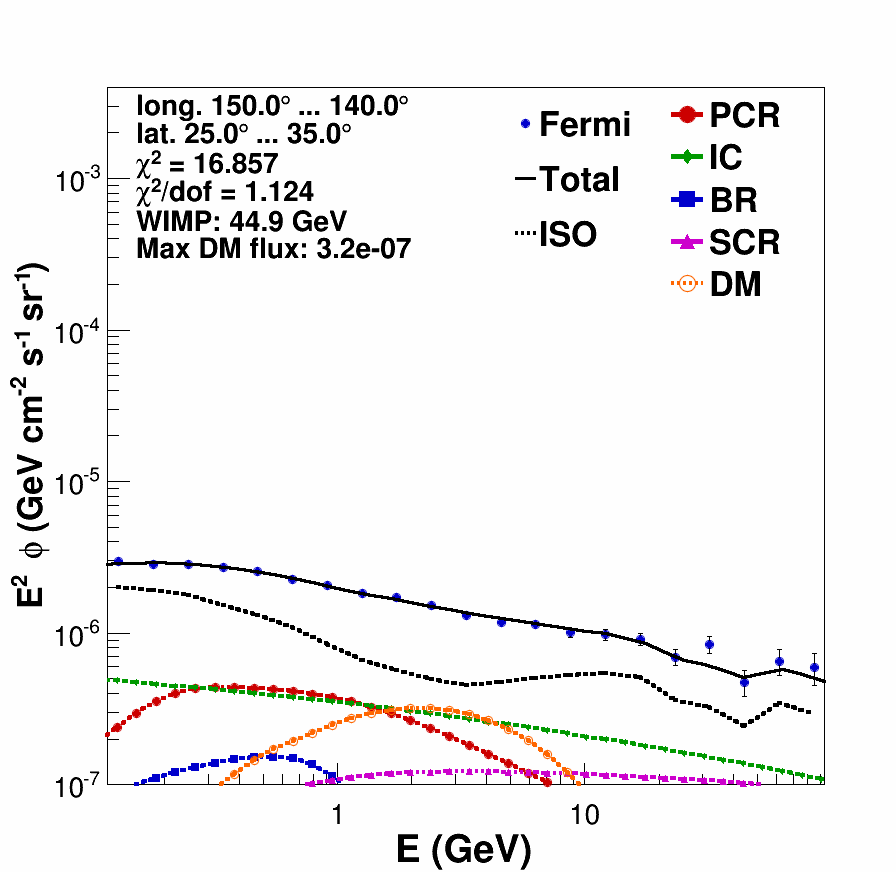}
\includegraphics[width=0.16\textwidth,height=0.16\textwidth,clip]{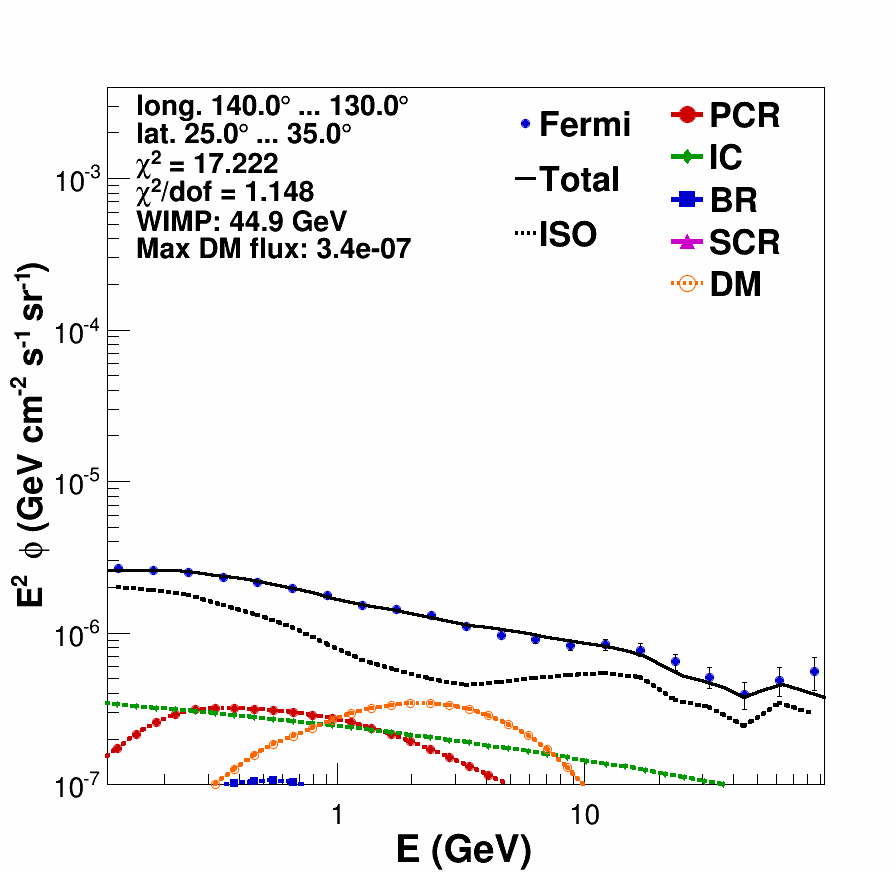}
\includegraphics[width=0.16\textwidth,height=0.16\textwidth,clip]{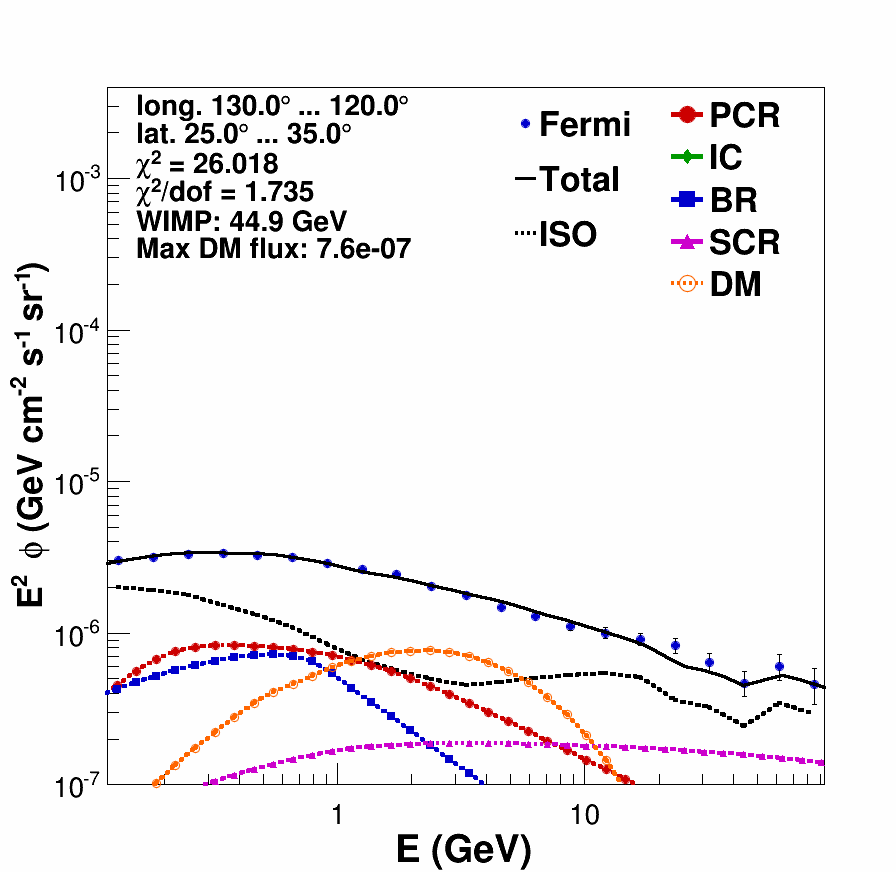}
\includegraphics[width=0.16\textwidth,height=0.16\textwidth,clip]{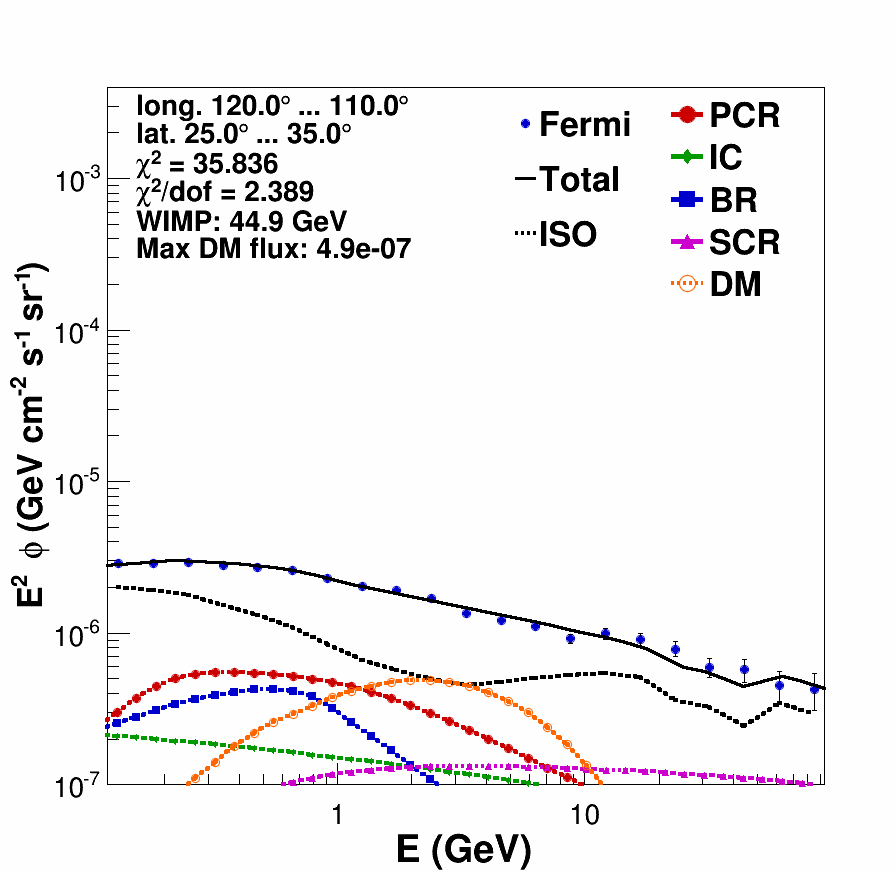}
\includegraphics[width=0.16\textwidth,height=0.16\textwidth,clip]{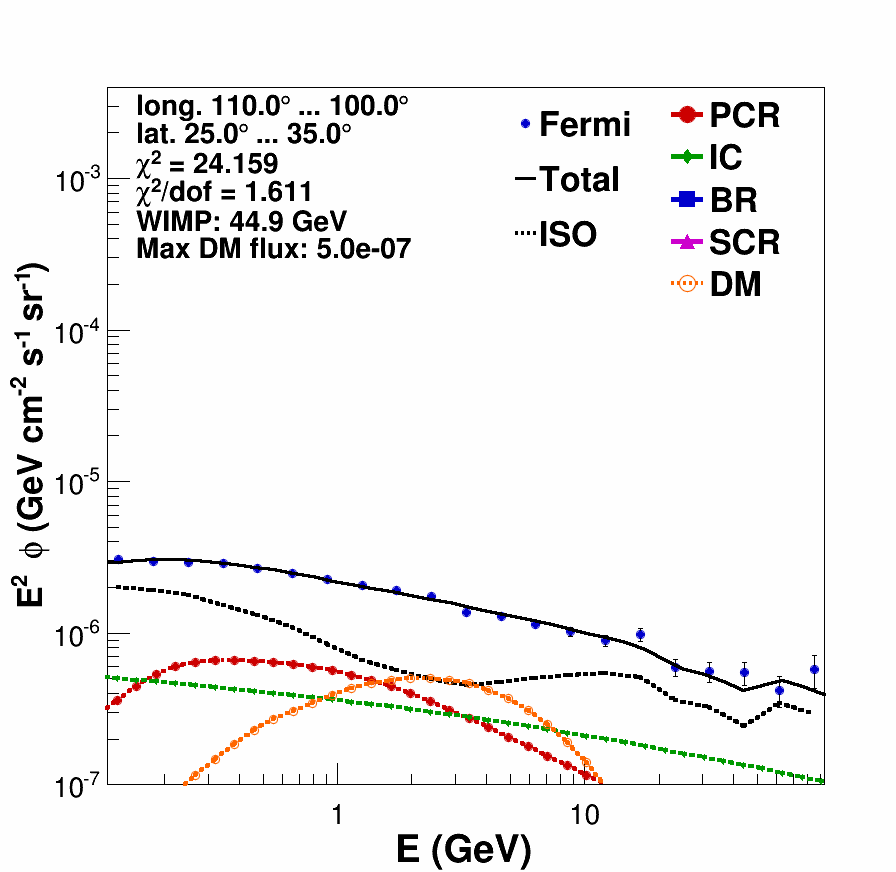}
\includegraphics[width=0.16\textwidth,height=0.16\textwidth,clip]{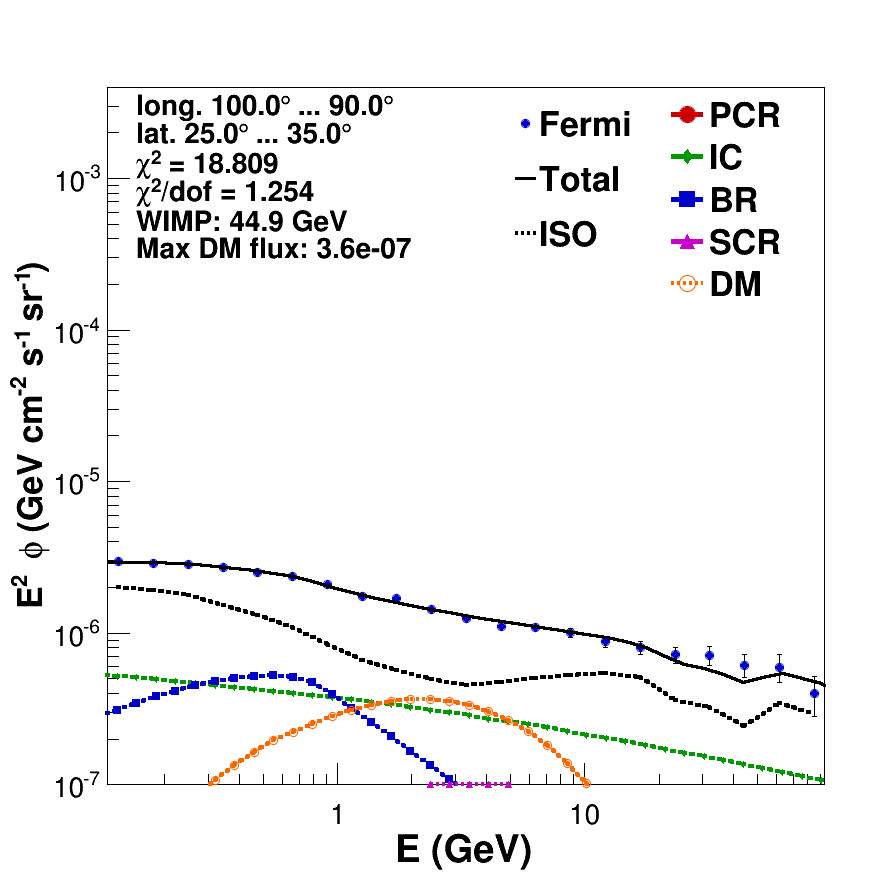}
\includegraphics[width=0.16\textwidth,height=0.16\textwidth,clip]{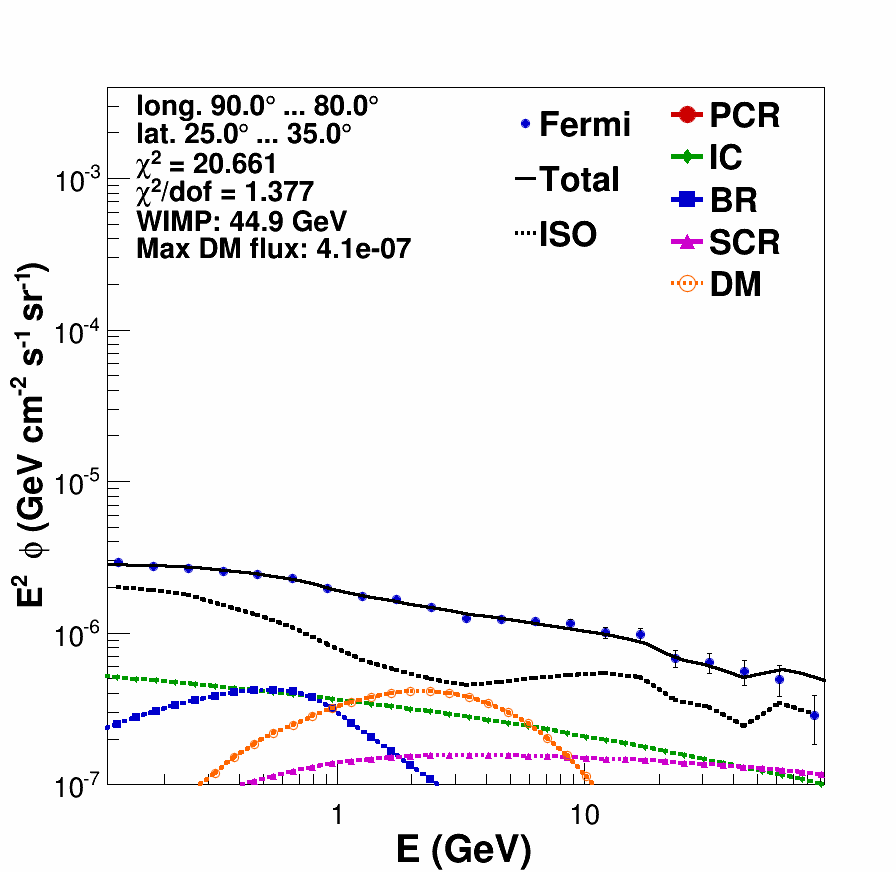}
\includegraphics[width=0.16\textwidth,height=0.16\textwidth,clip]{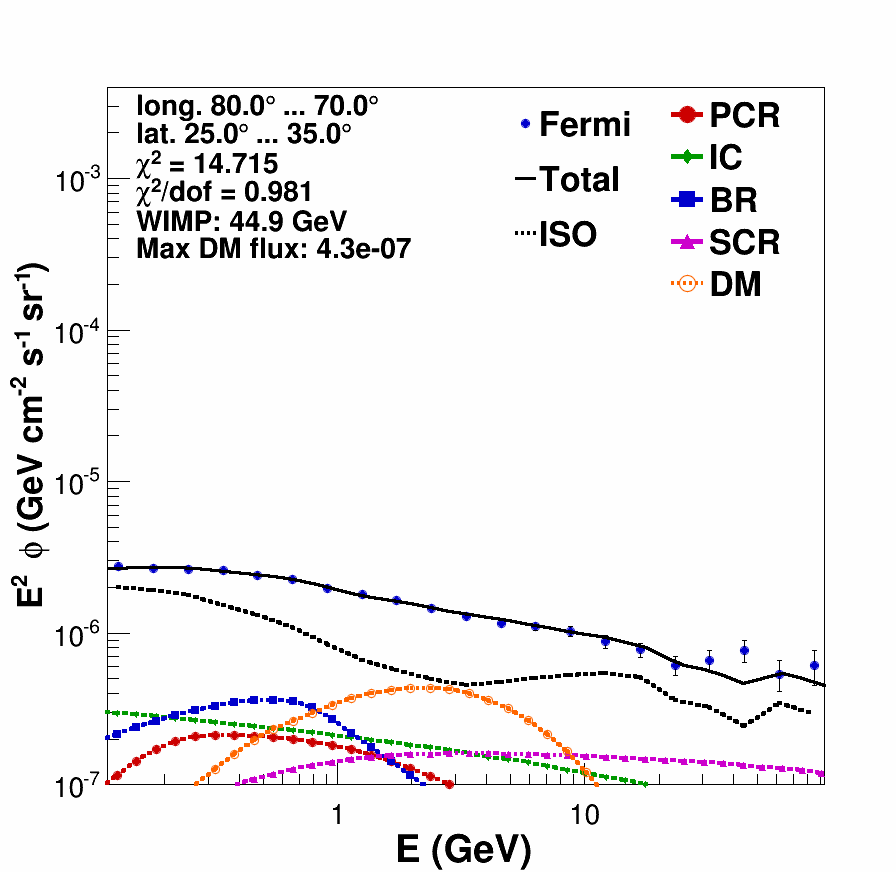}
\includegraphics[width=0.16\textwidth,height=0.16\textwidth,clip]{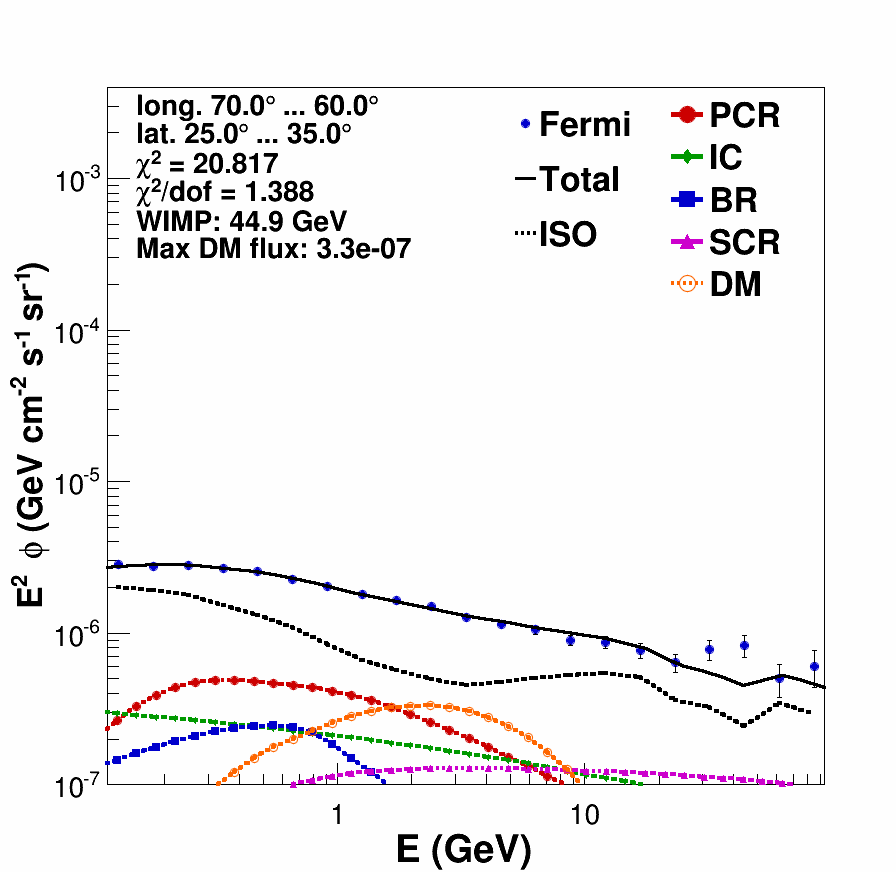}
\includegraphics[width=0.16\textwidth,height=0.16\textwidth,clip]{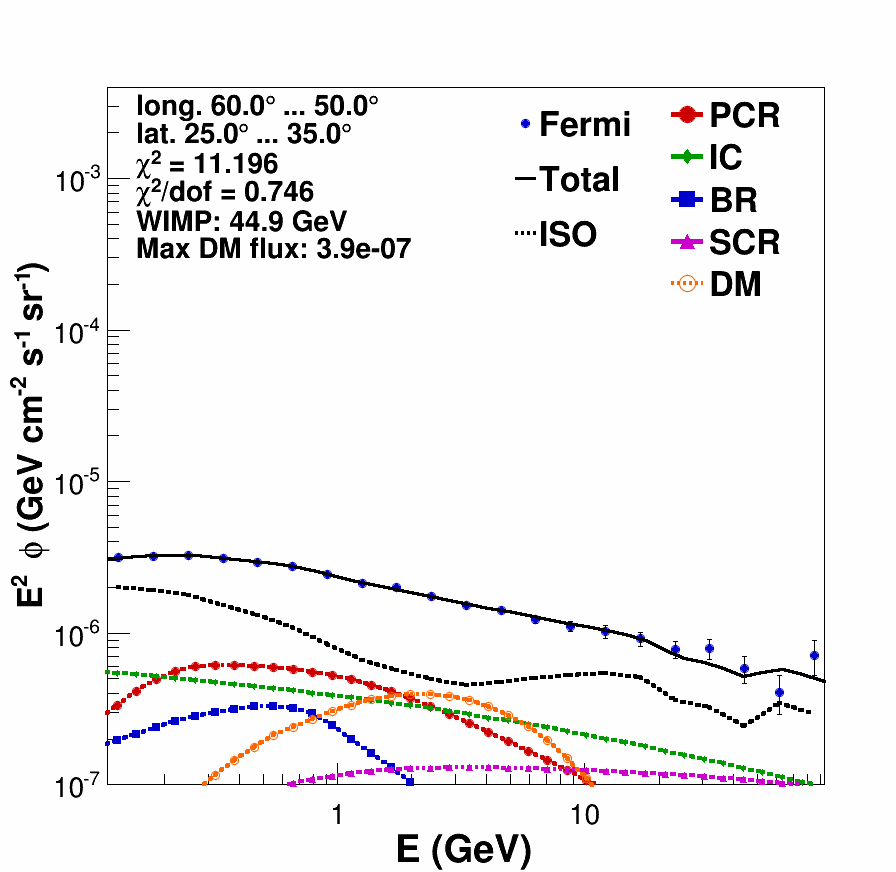}
\includegraphics[width=0.16\textwidth,height=0.16\textwidth,clip]{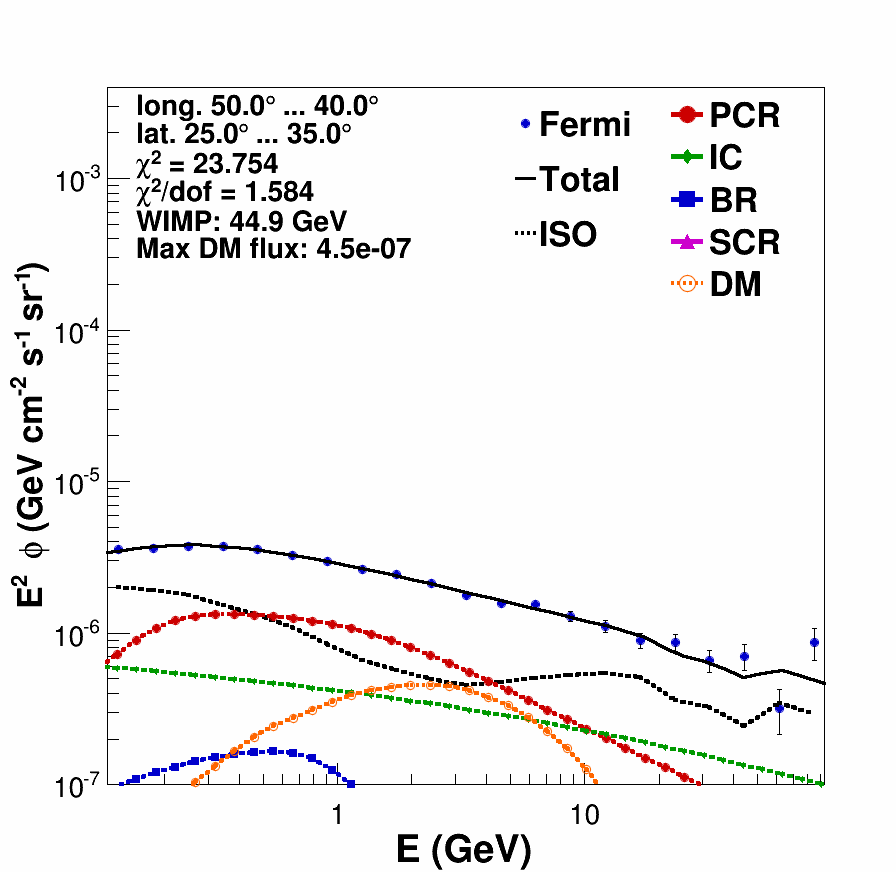}
\includegraphics[width=0.16\textwidth,height=0.16\textwidth,clip]{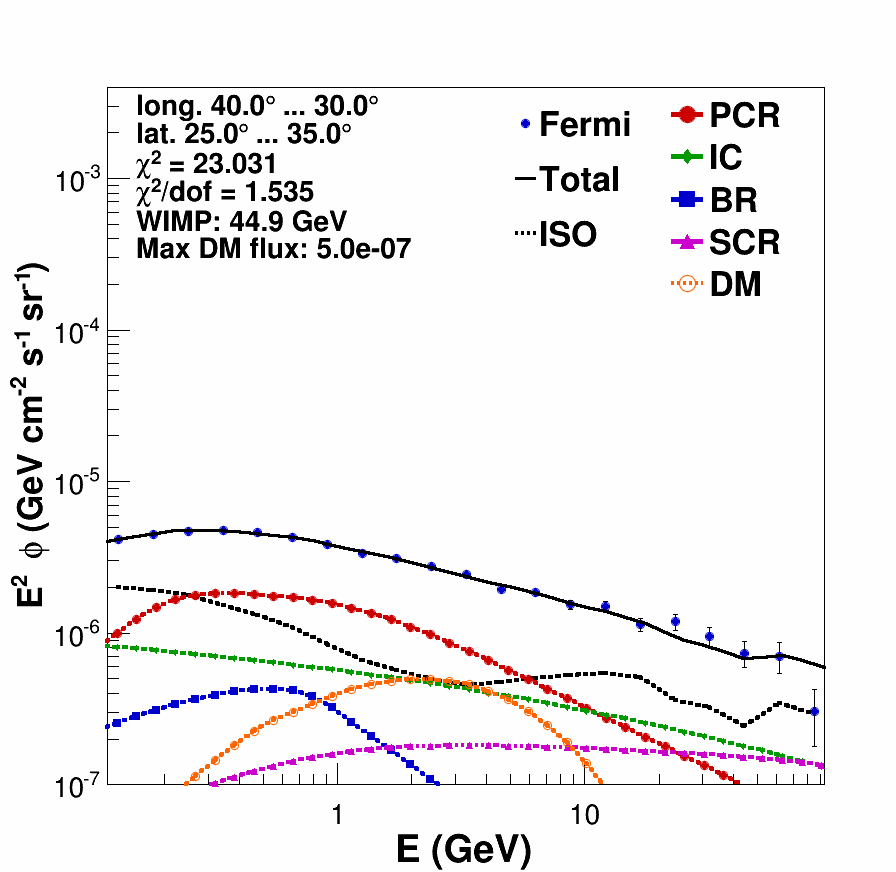}
\includegraphics[width=0.16\textwidth,height=0.16\textwidth,clip]{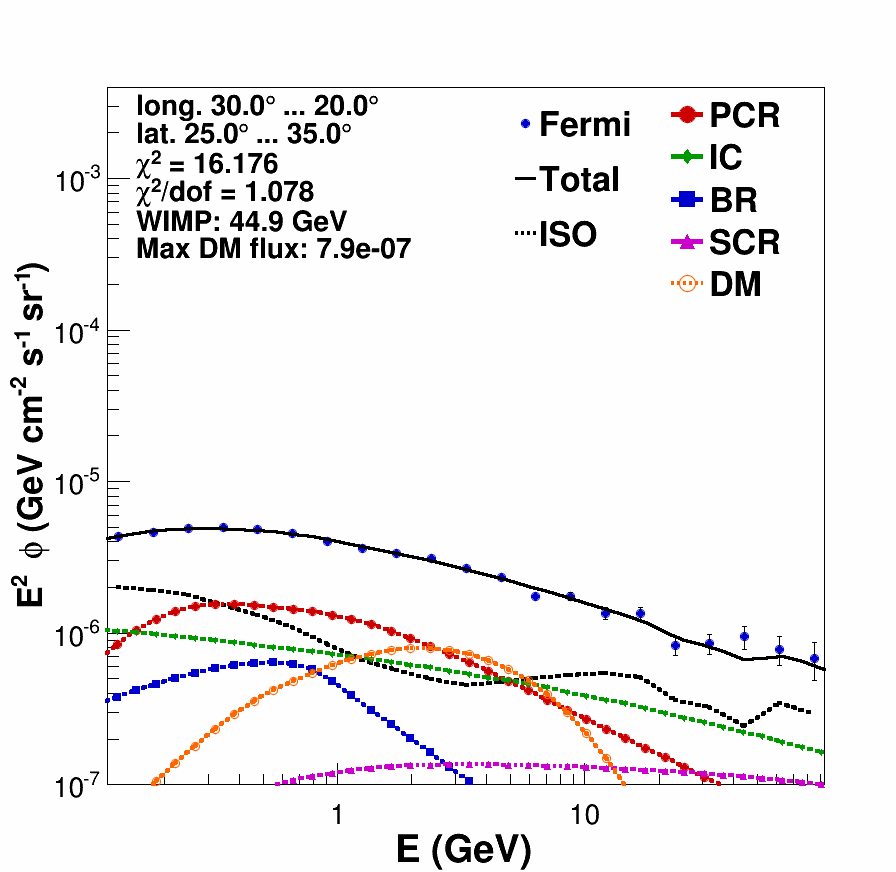}
\includegraphics[width=0.16\textwidth,height=0.16\textwidth,clip]{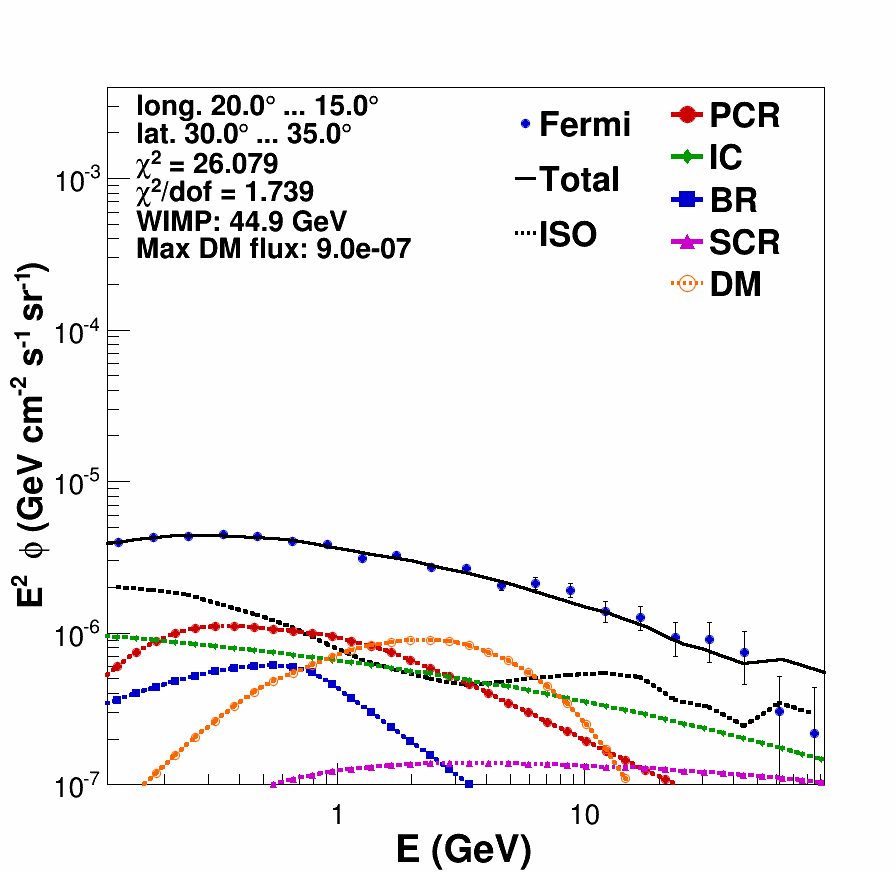}
\includegraphics[width=0.16\textwidth,height=0.16\textwidth,clip]{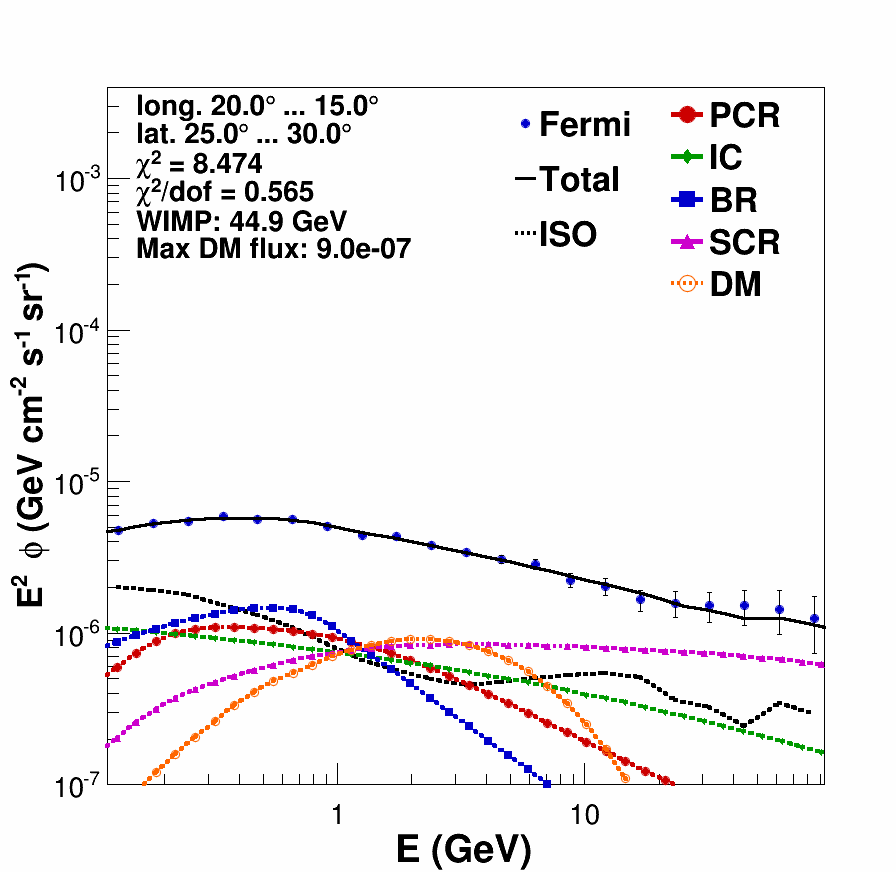}
\includegraphics[width=0.16\textwidth,height=0.16\textwidth,clip]{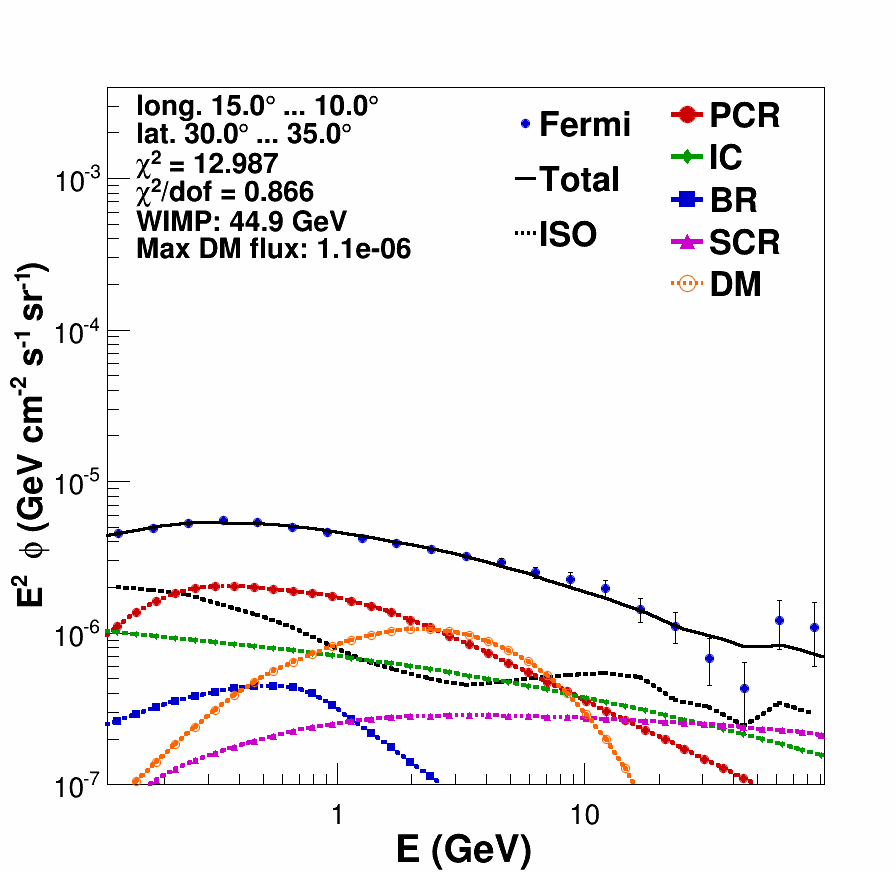}
\includegraphics[width=0.16\textwidth,height=0.16\textwidth,clip]{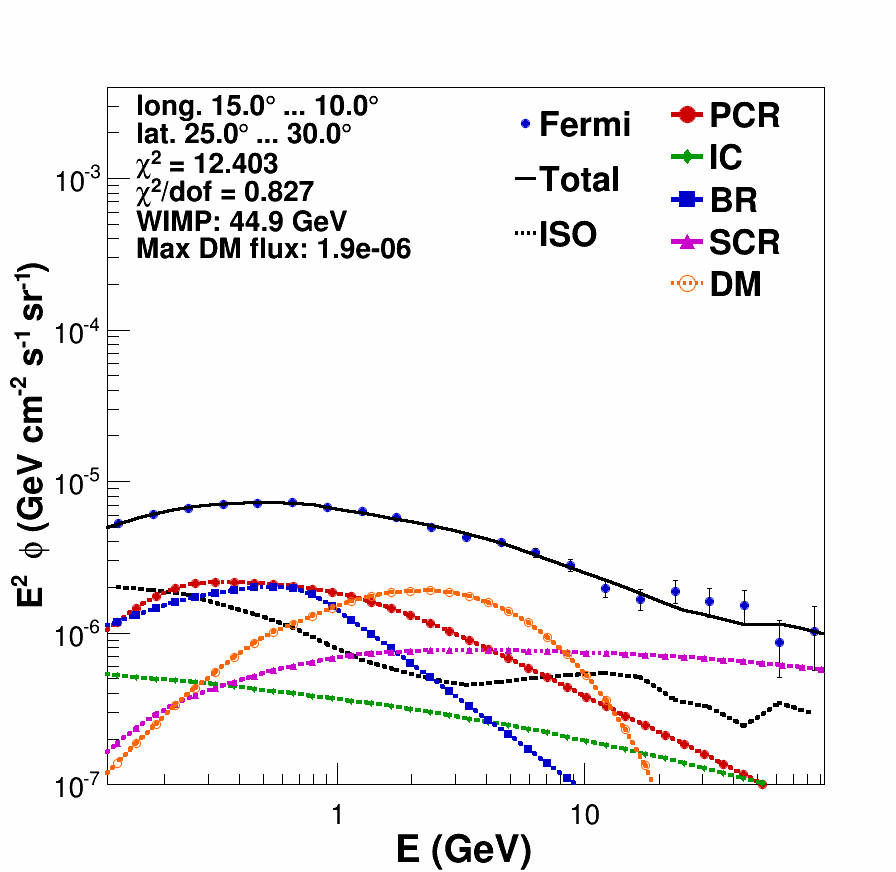}
\includegraphics[width=0.16\textwidth,height=0.16\textwidth,clip]{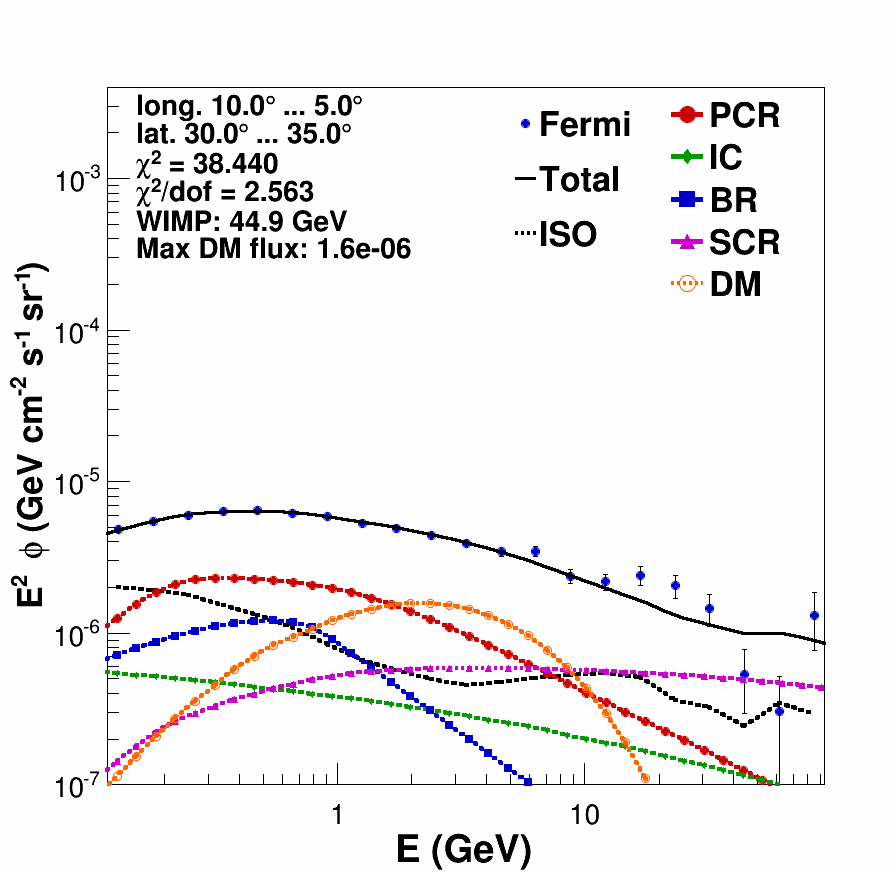}
\includegraphics[width=0.16\textwidth,height=0.16\textwidth,clip]{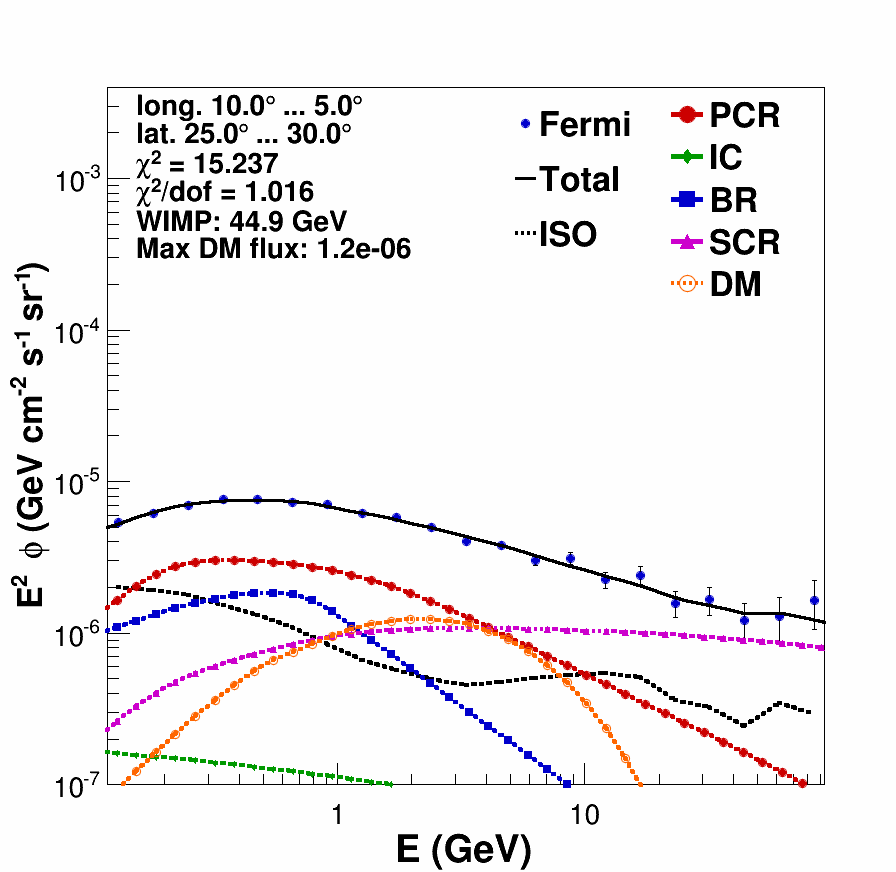}
\includegraphics[width=0.16\textwidth,height=0.16\textwidth,clip]{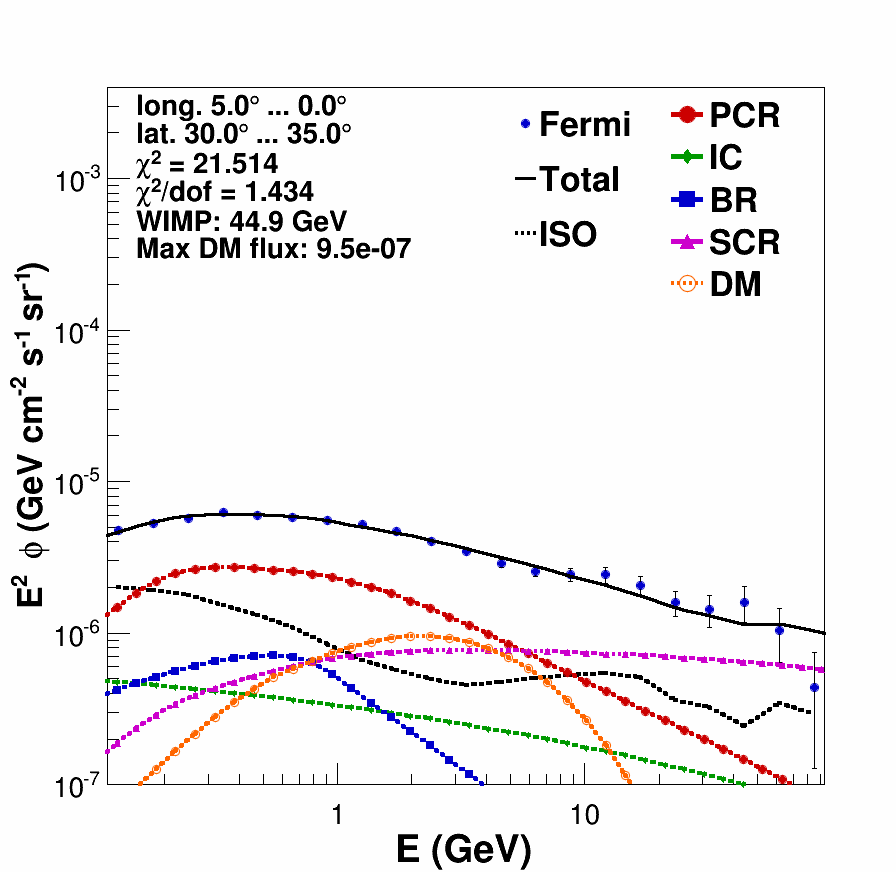}
\includegraphics[width=0.16\textwidth,height=0.16\textwidth,clip]{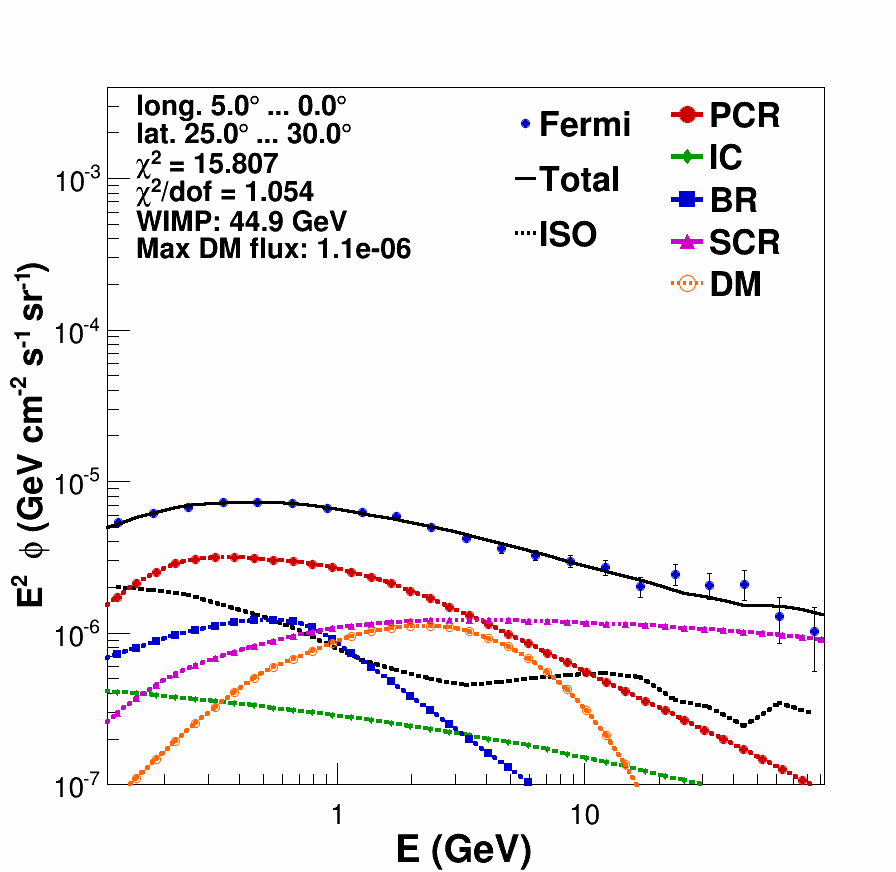}
\includegraphics[width=0.16\textwidth,height=0.16\textwidth,clip]{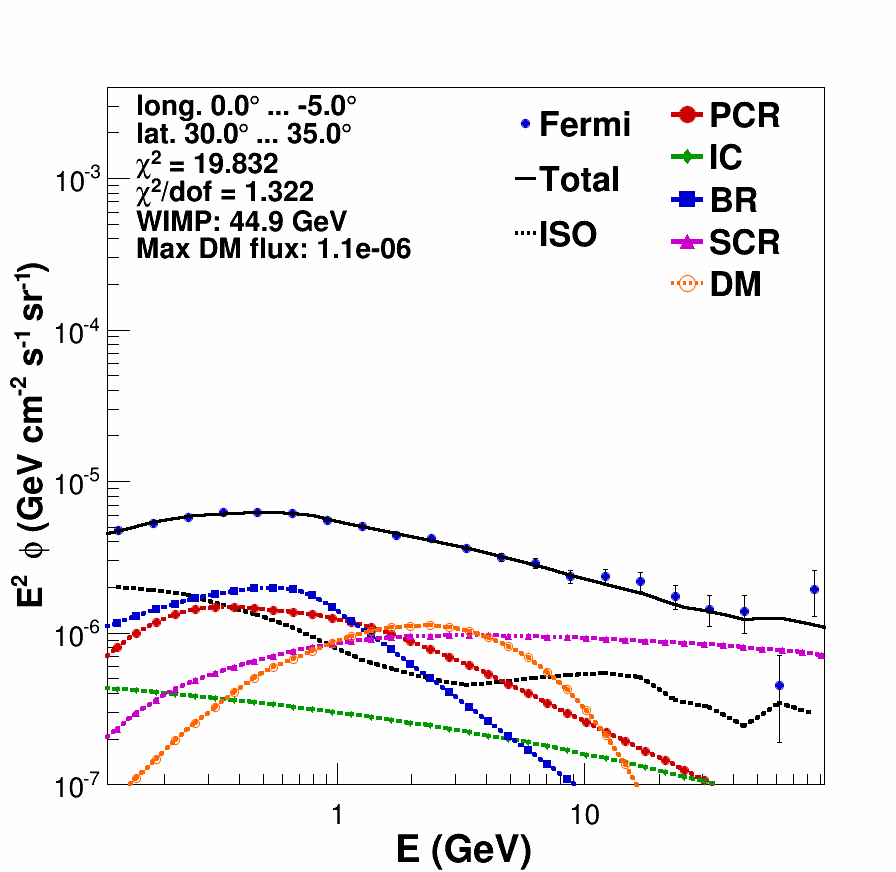}
\includegraphics[width=0.16\textwidth,height=0.16\textwidth,clip]{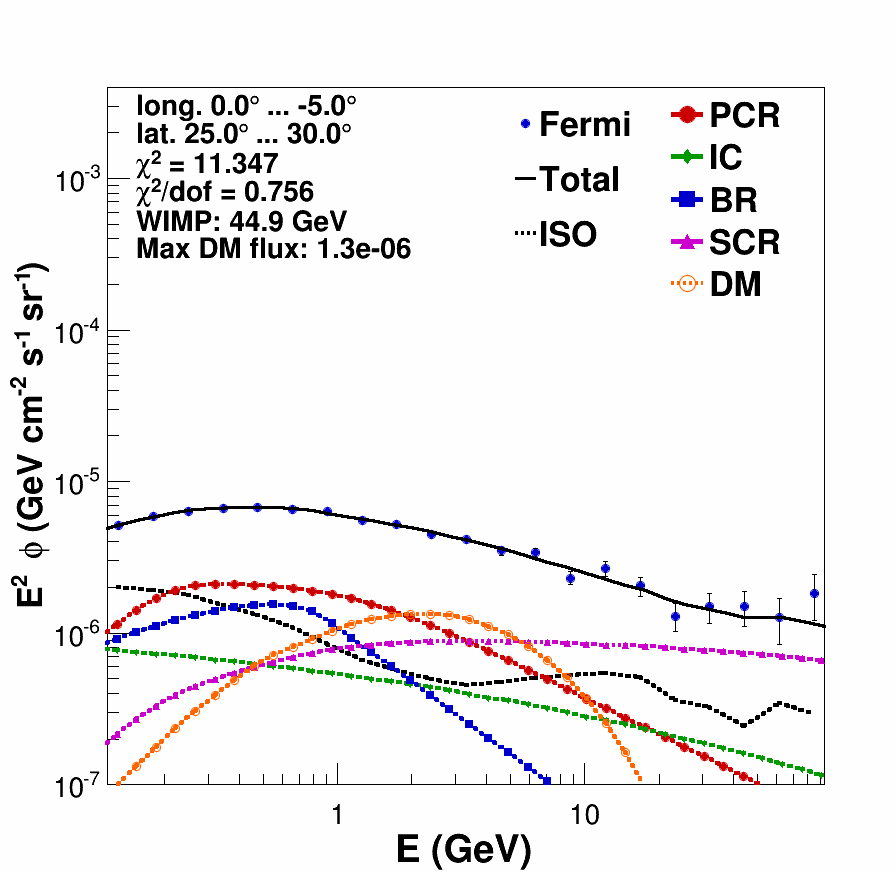}
\includegraphics[width=0.16\textwidth,height=0.16\textwidth,clip]{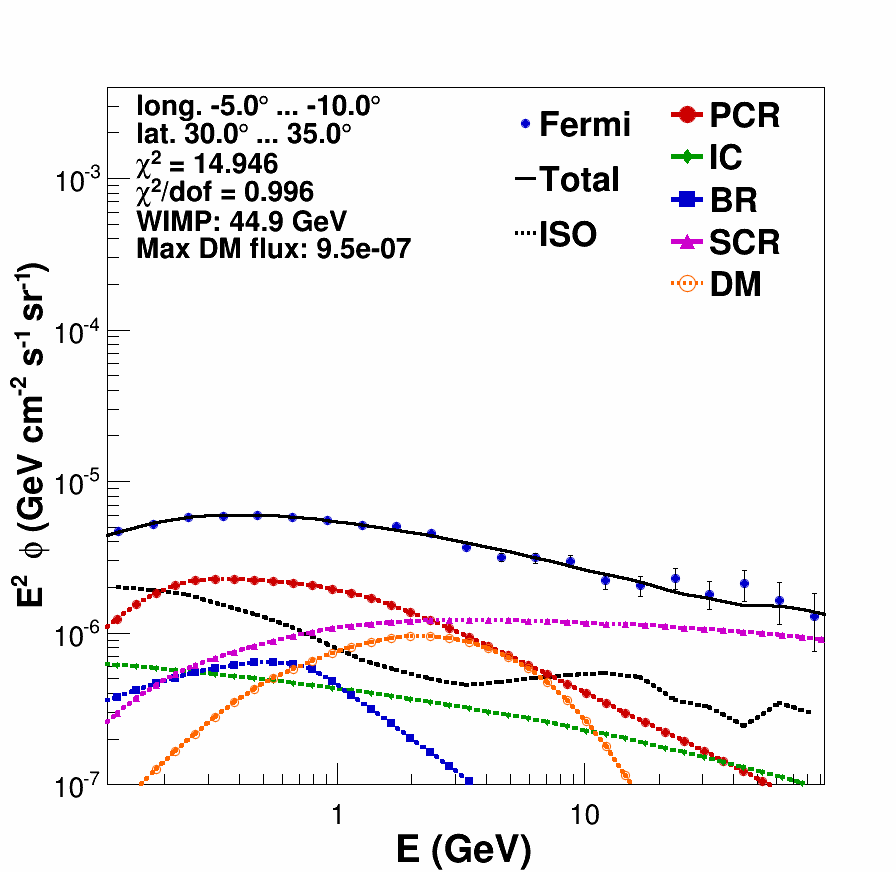}
\includegraphics[width=0.16\textwidth,height=0.16\textwidth,clip]{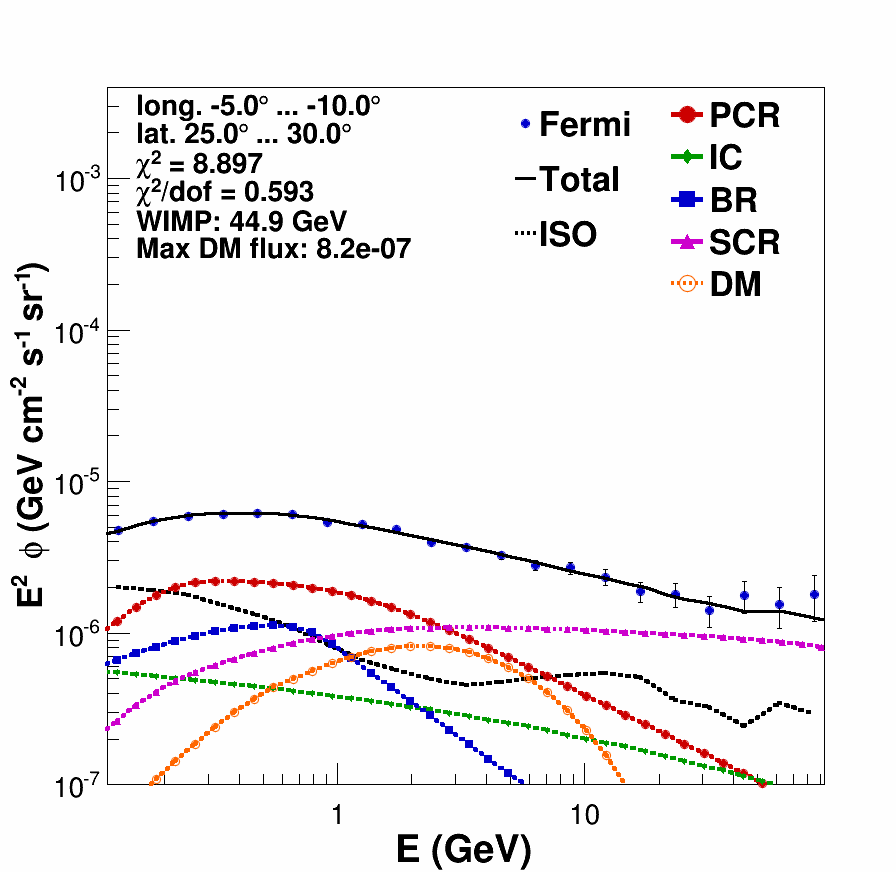}
\includegraphics[width=0.16\textwidth,height=0.16\textwidth,clip]{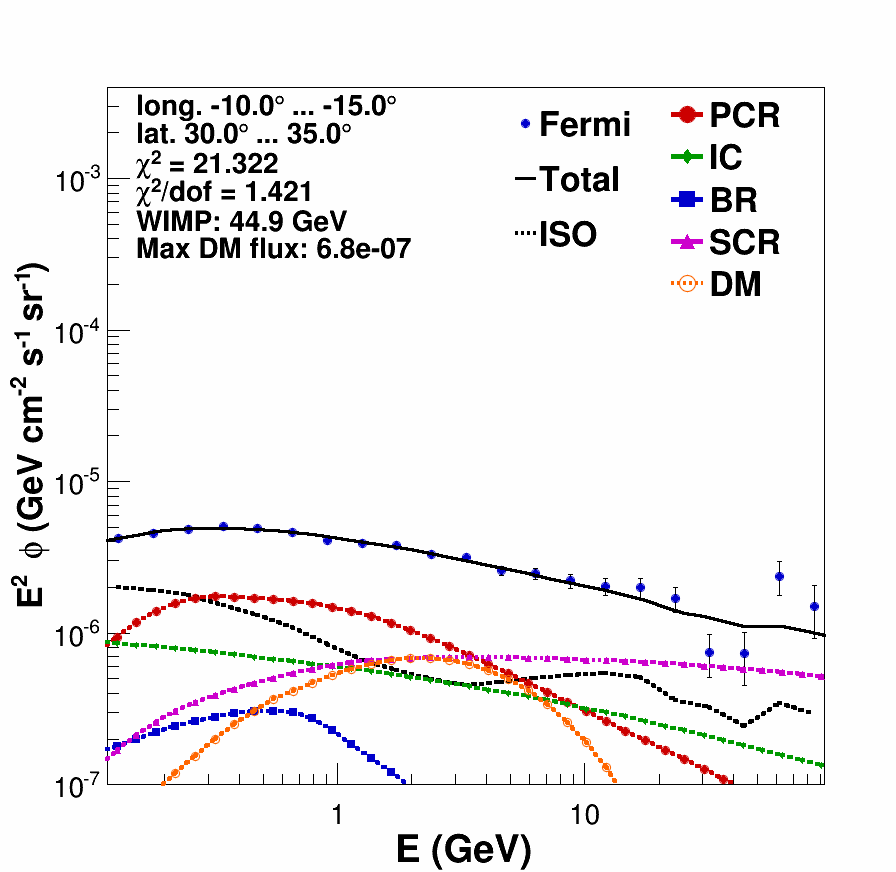}
\includegraphics[width=0.16\textwidth,height=0.16\textwidth,clip]{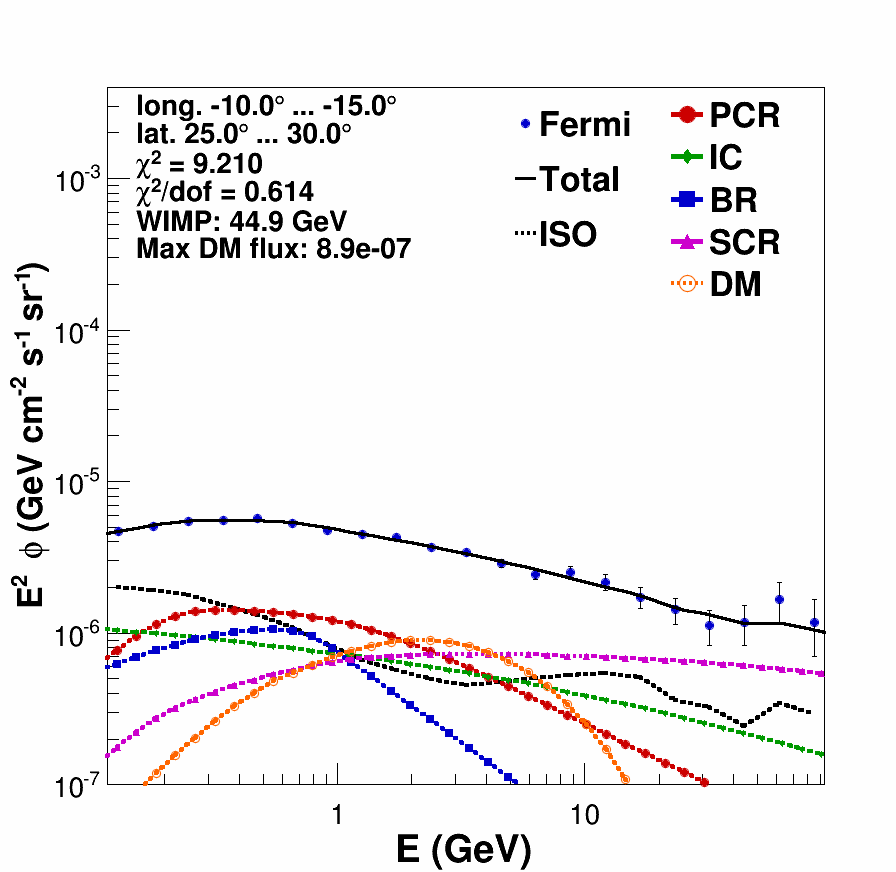}
\includegraphics[width=0.16\textwidth,height=0.16\textwidth,clip]{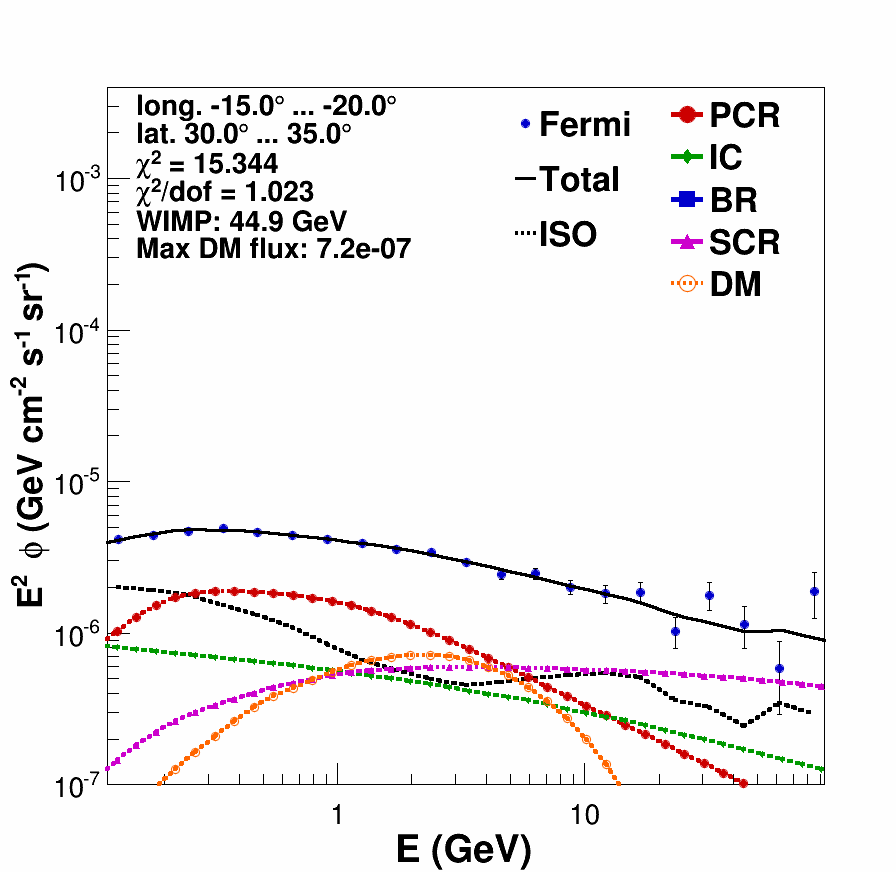}
\includegraphics[width=0.16\textwidth,height=0.16\textwidth,clip]{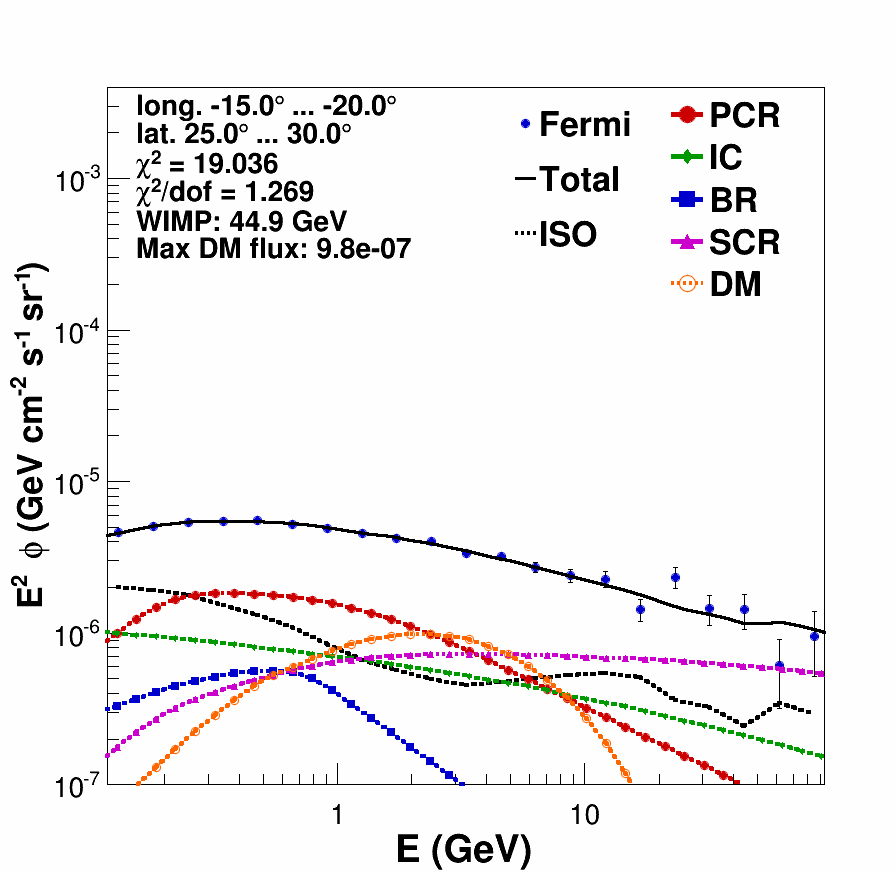}
\includegraphics[width=0.16\textwidth,height=0.16\textwidth,clip]{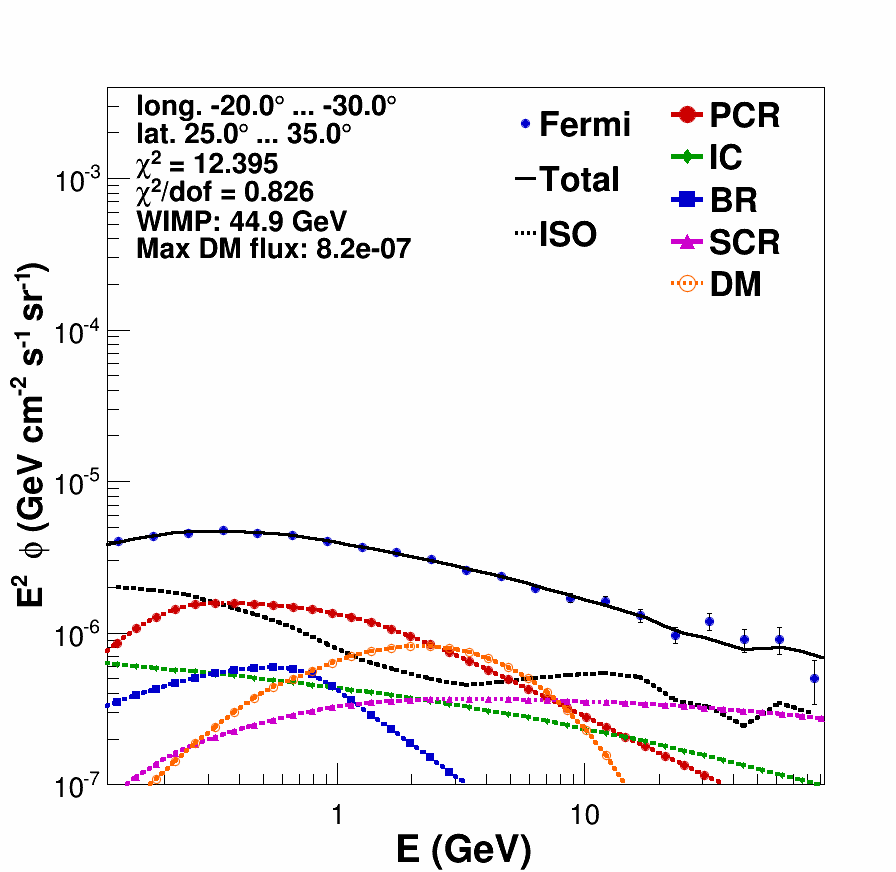}
\includegraphics[width=0.16\textwidth,height=0.16\textwidth,clip]{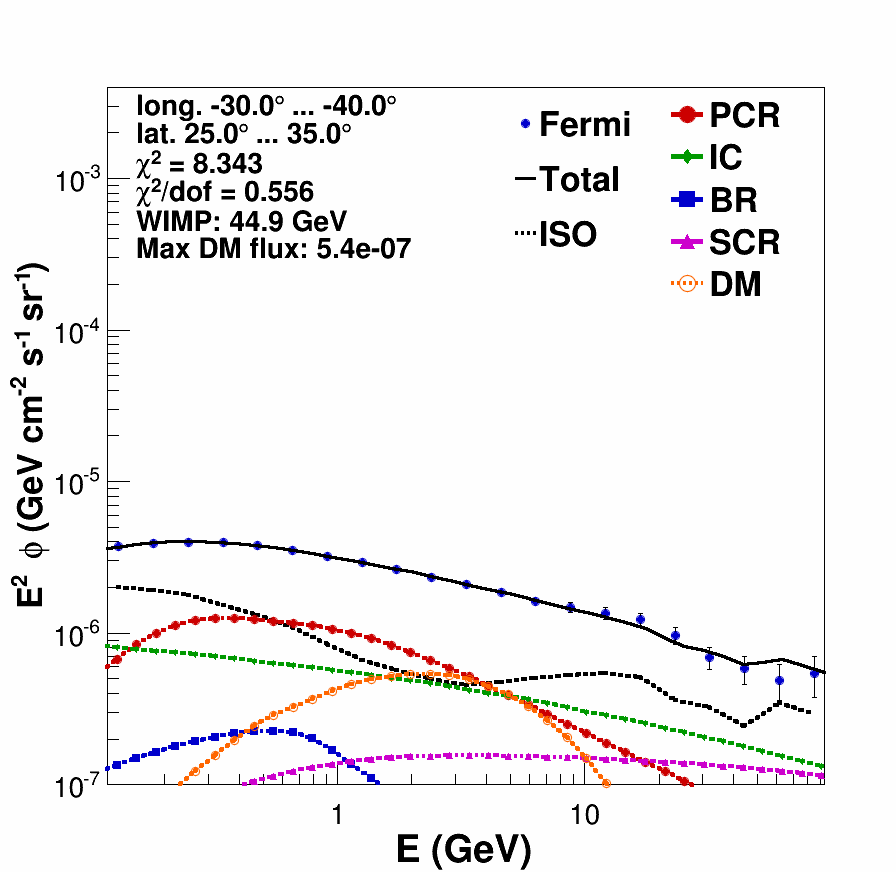}
\includegraphics[width=0.16\textwidth,height=0.16\textwidth,clip]{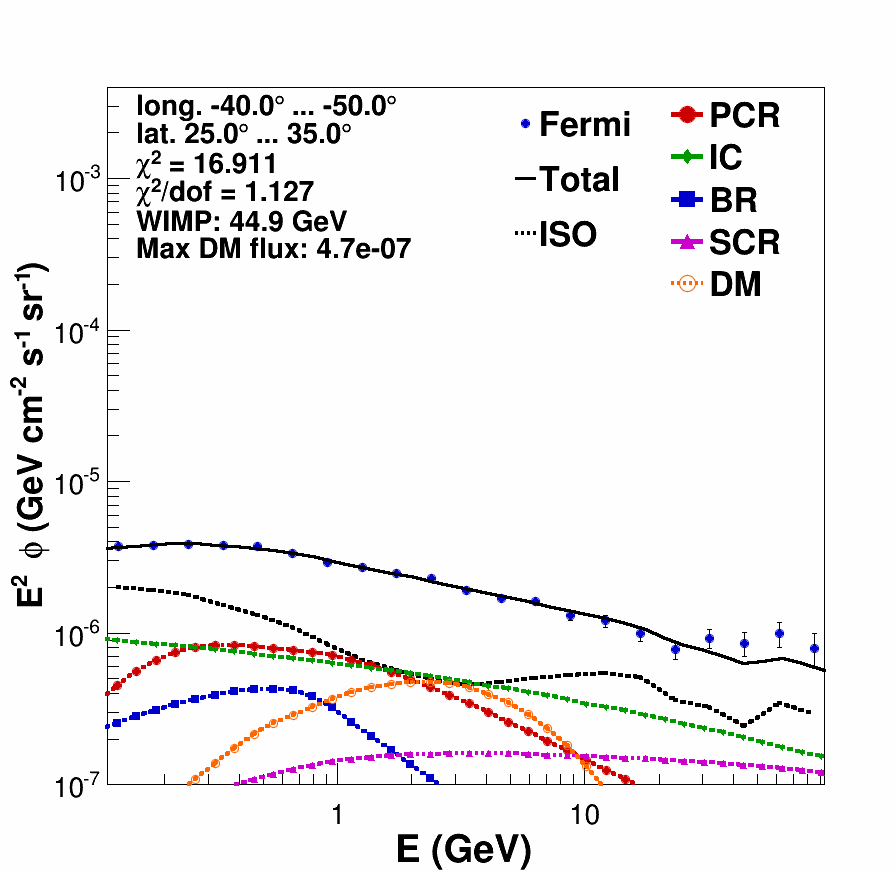}
\includegraphics[width=0.16\textwidth,height=0.16\textwidth,clip]{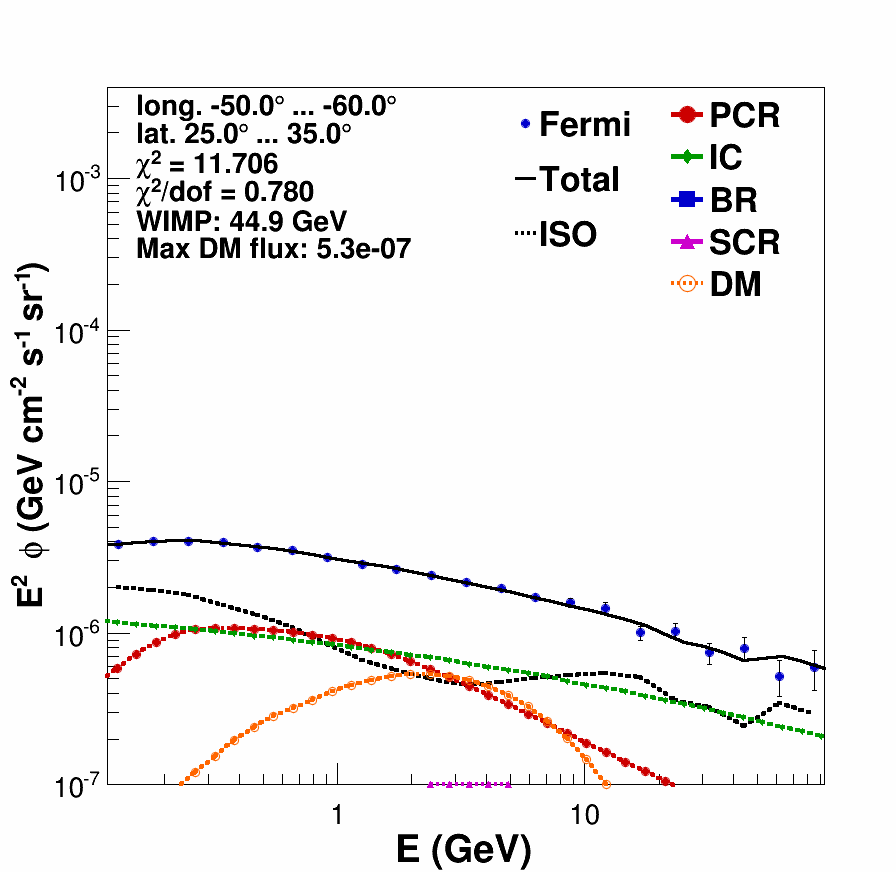}
\includegraphics[width=0.16\textwidth,height=0.16\textwidth,clip]{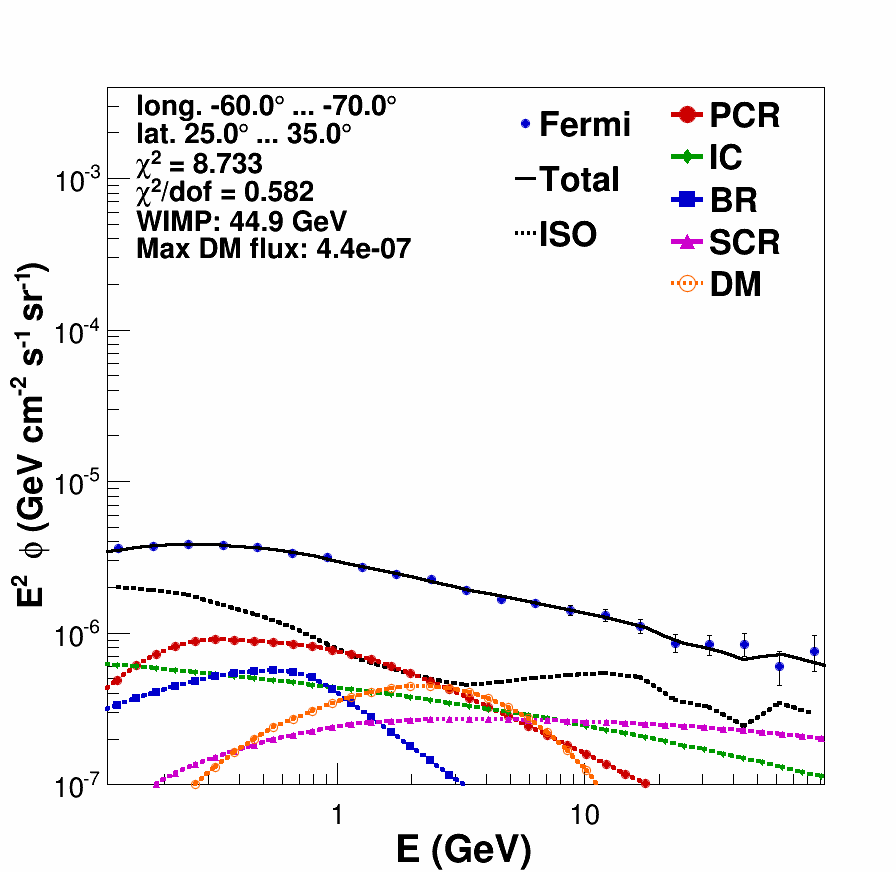}
\includegraphics[width=0.16\textwidth,height=0.16\textwidth,clip]{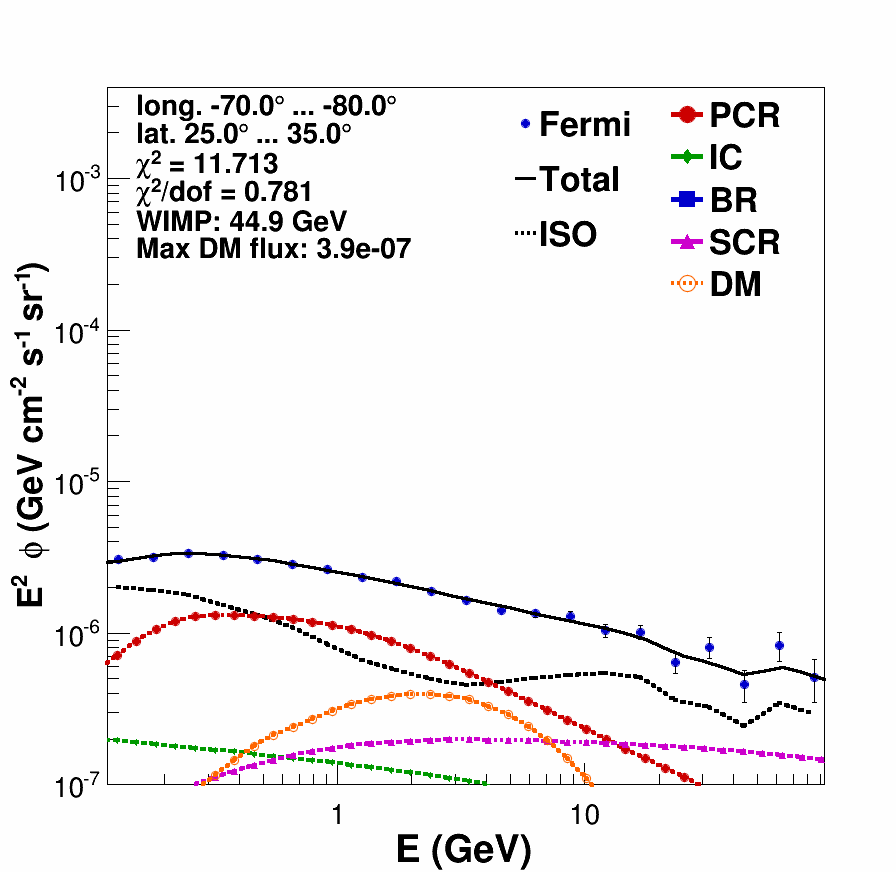}
\includegraphics[width=0.16\textwidth,height=0.16\textwidth,clip]{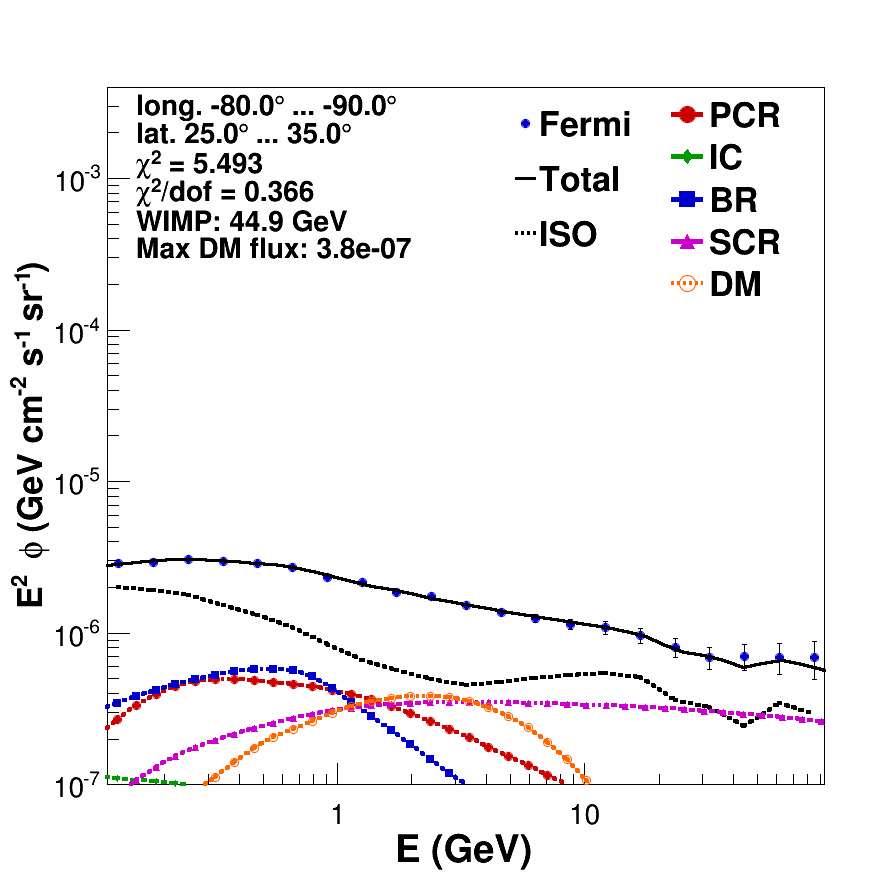}
\includegraphics[width=0.16\textwidth,height=0.16\textwidth,clip]{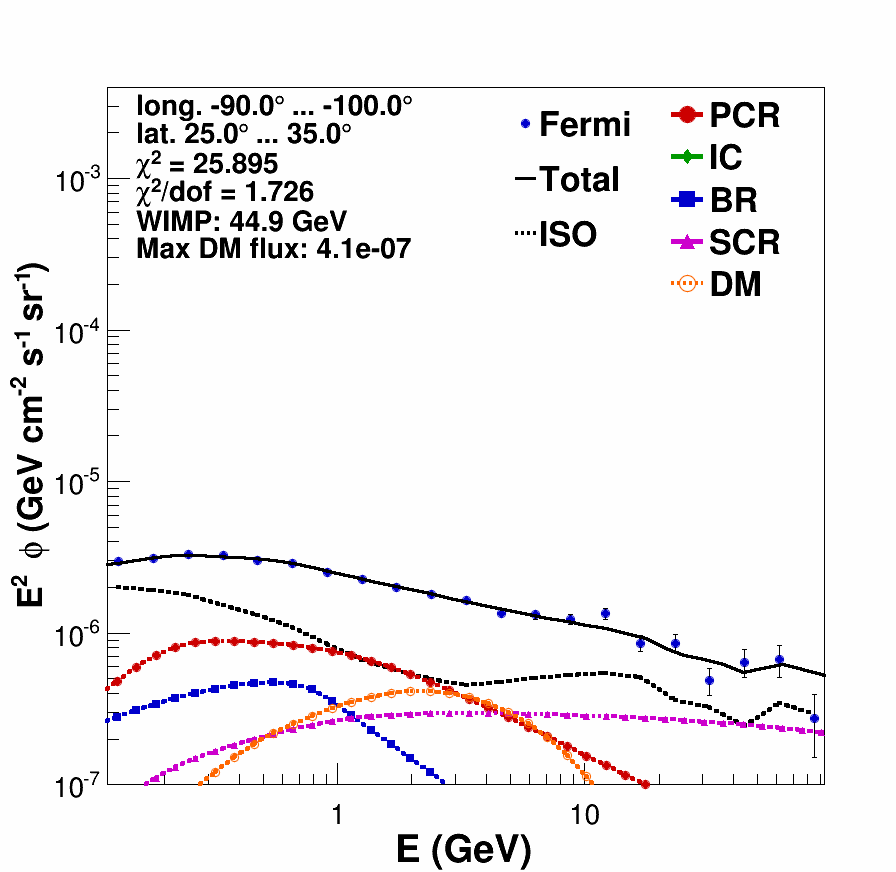}
\includegraphics[width=0.16\textwidth,height=0.16\textwidth,clip]{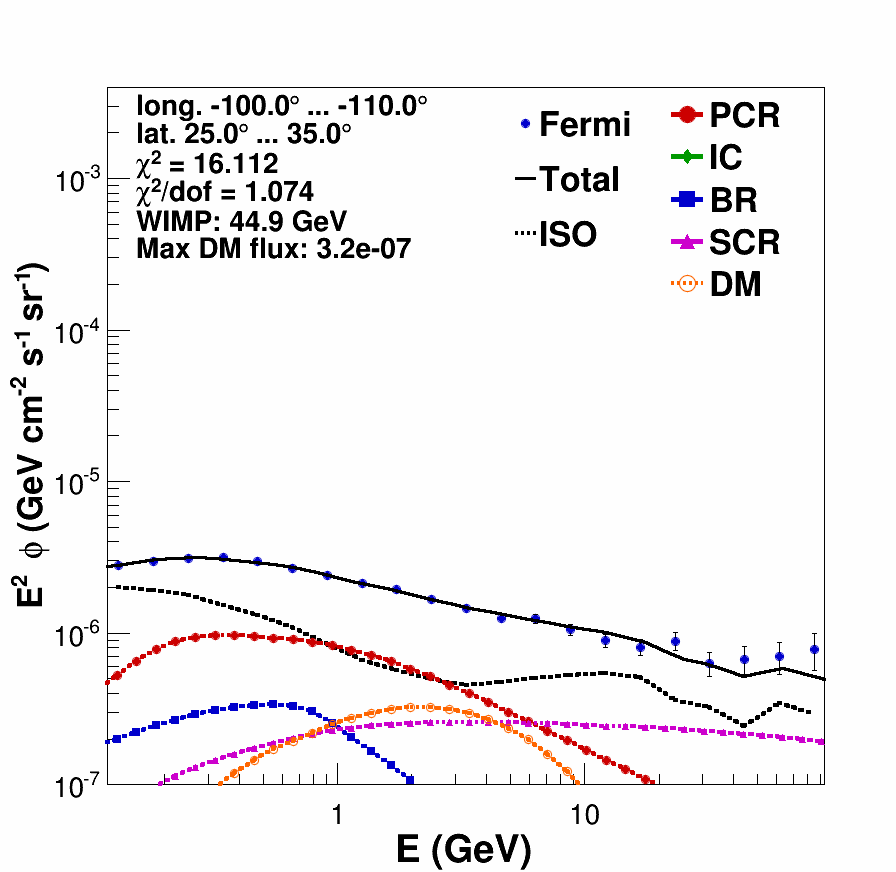}
\includegraphics[width=0.16\textwidth,height=0.16\textwidth,clip]{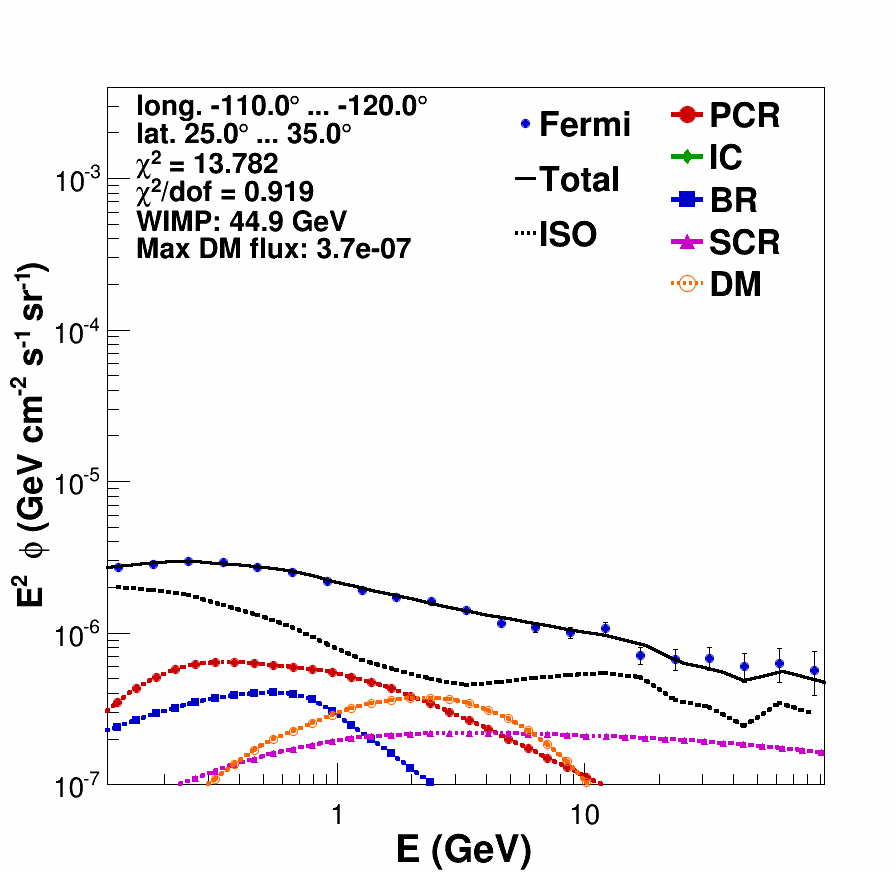}
\includegraphics[width=0.16\textwidth,height=0.16\textwidth,clip]{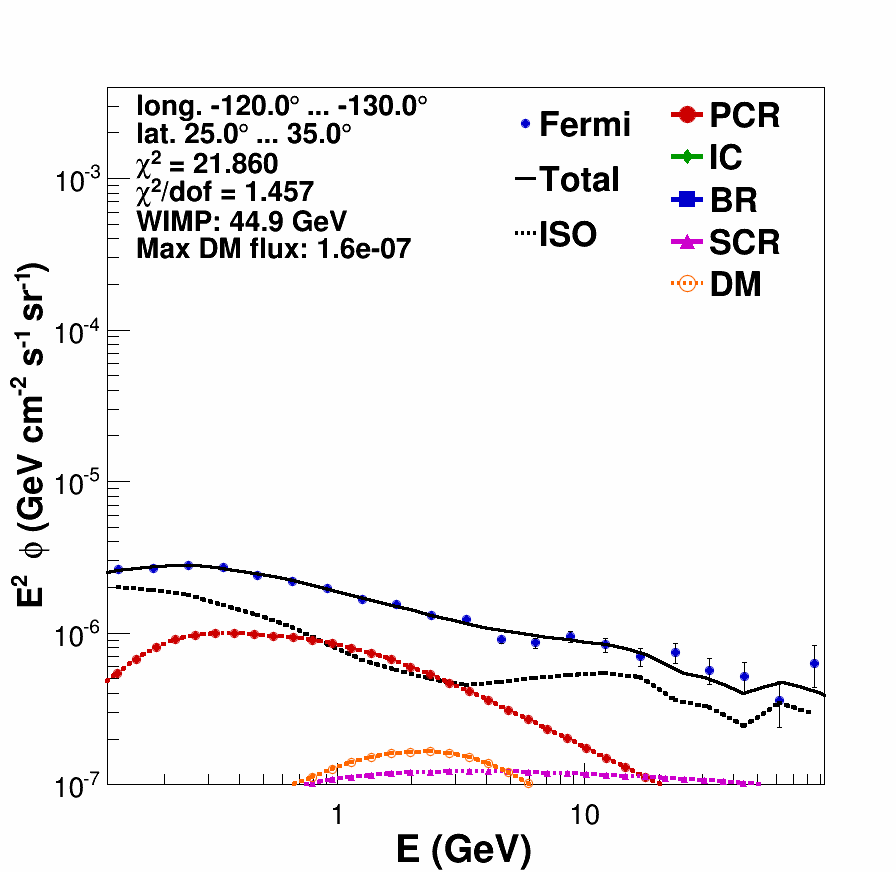}
\includegraphics[width=0.16\textwidth,height=0.16\textwidth,clip]{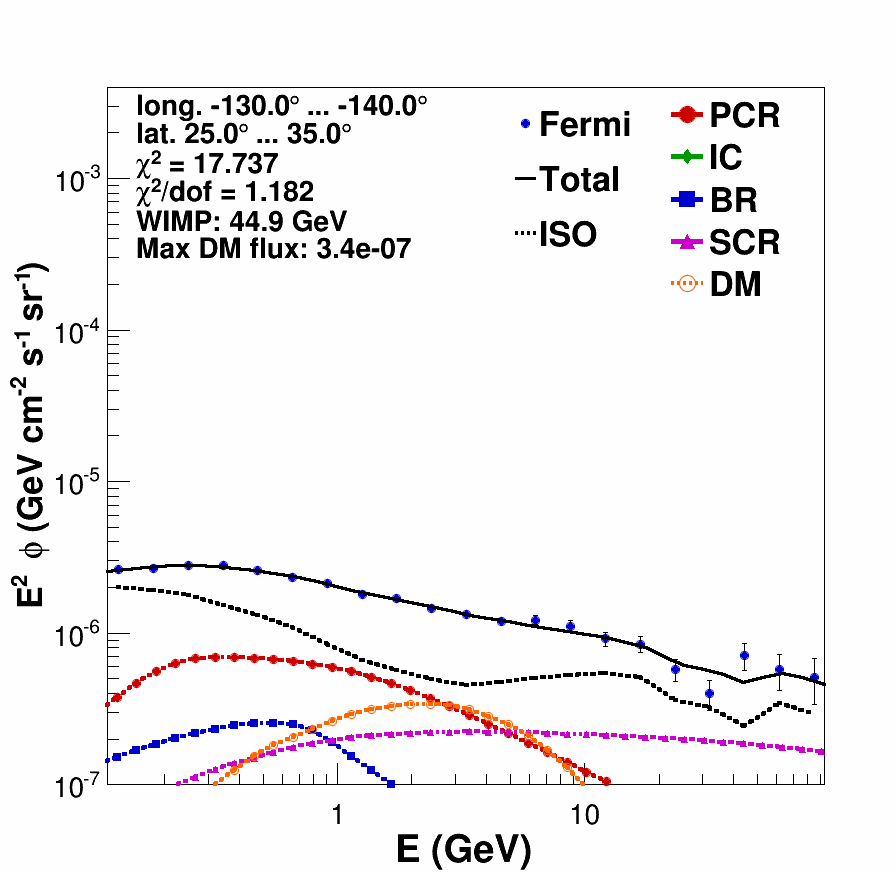}
\includegraphics[width=0.16\textwidth,height=0.16\textwidth,clip]{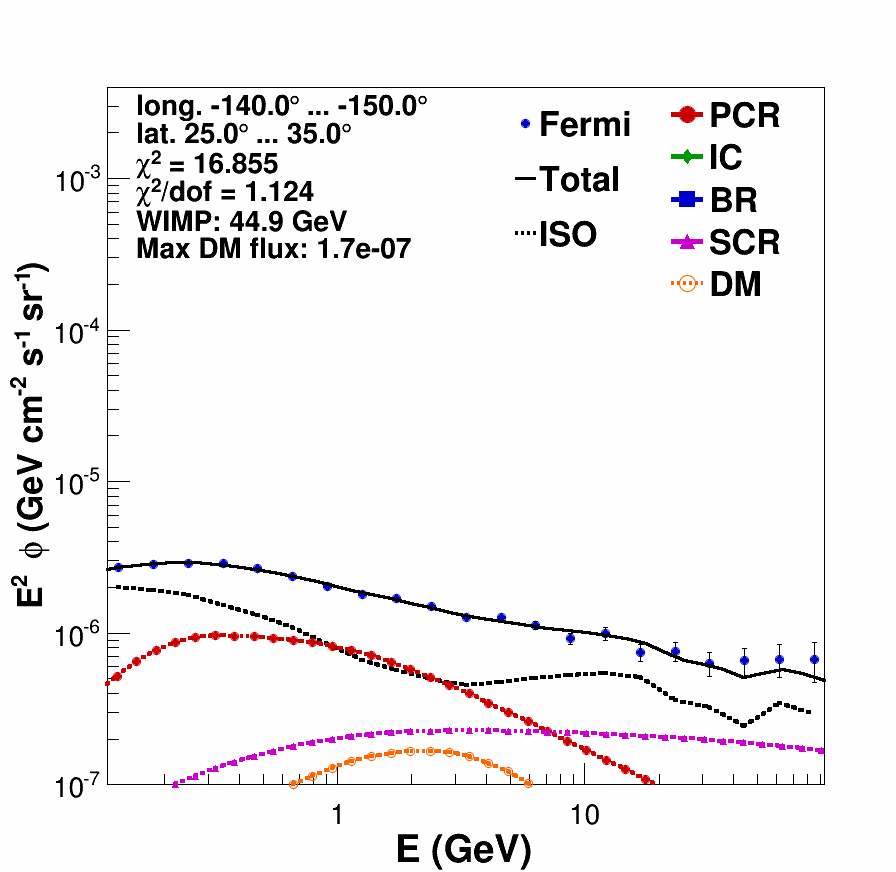}
\includegraphics[width=0.16\textwidth,height=0.16\textwidth,clip]{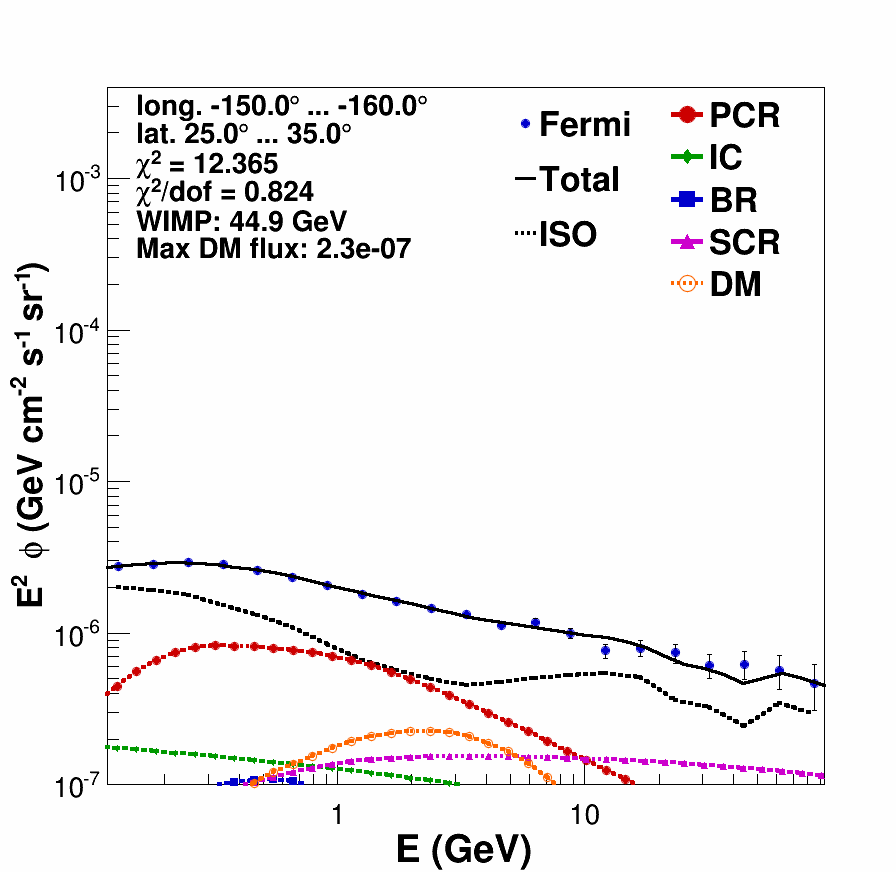}
\includegraphics[width=0.16\textwidth,height=0.16\textwidth,clip]{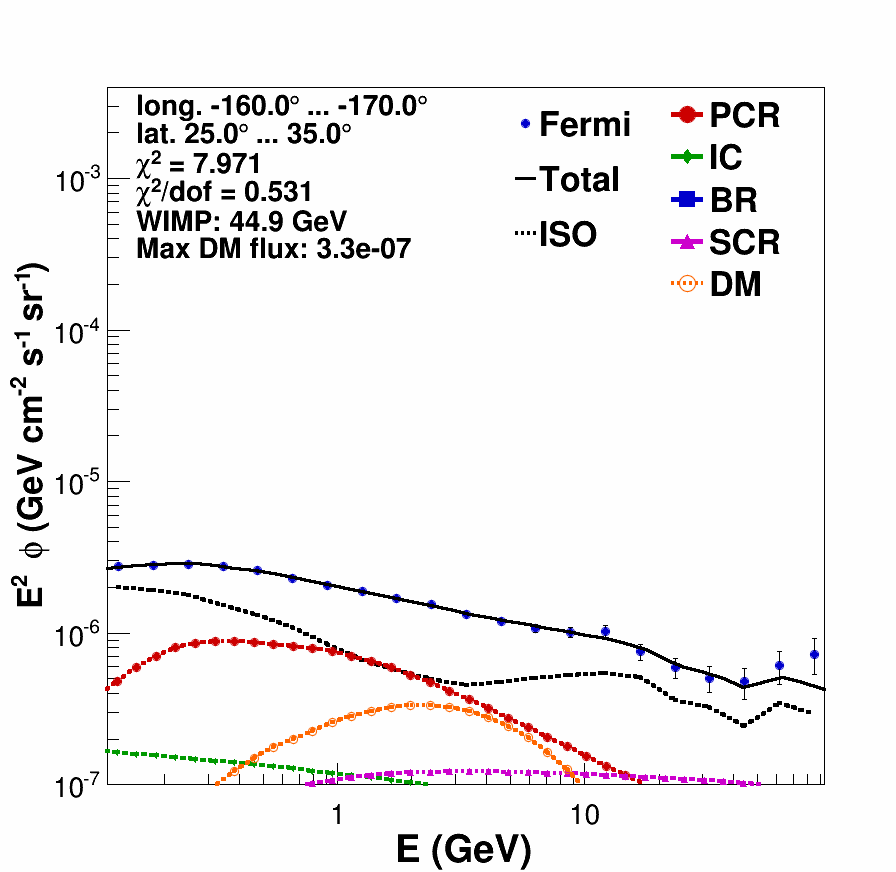}
\includegraphics[width=0.16\textwidth,height=0.16\textwidth,clip]{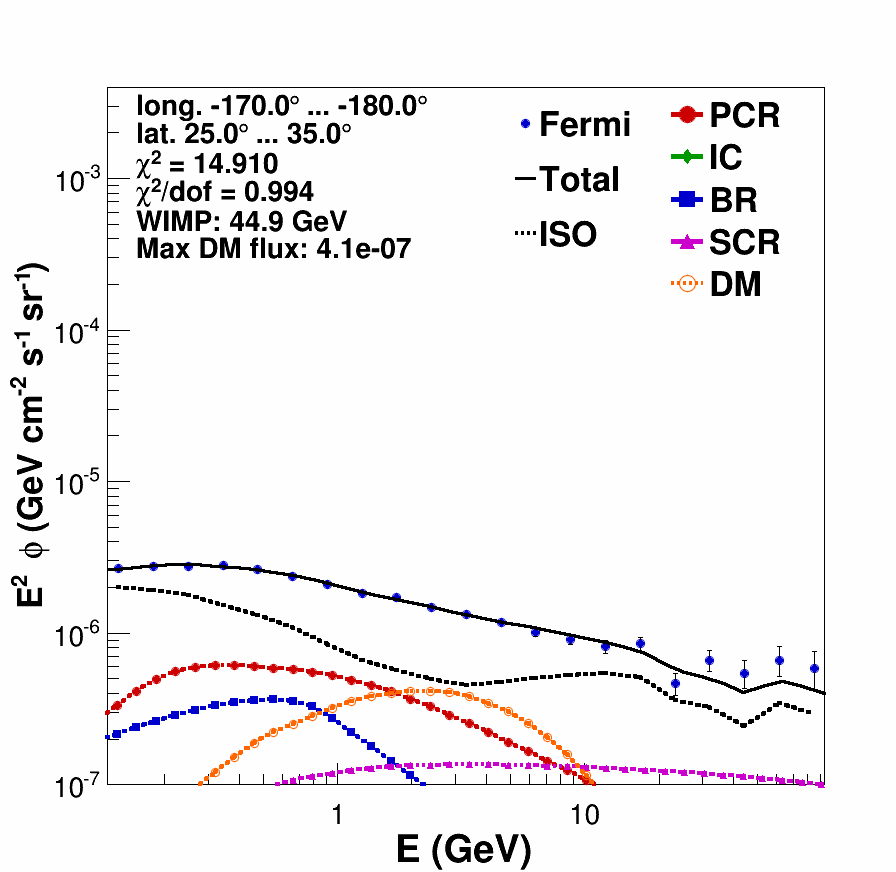}%%%%%r5
\caption[]{Template fits for latitudes  with $25.0^\circ<b<35.0^\circ$ and longitudes decreasing from 180$^\circ$ to -180$^\circ$. \label{F36}
}
\end{figure}
\begin{figure}
\centering
\includegraphics[width=0.16\textwidth,height=0.16\textwidth,clip]{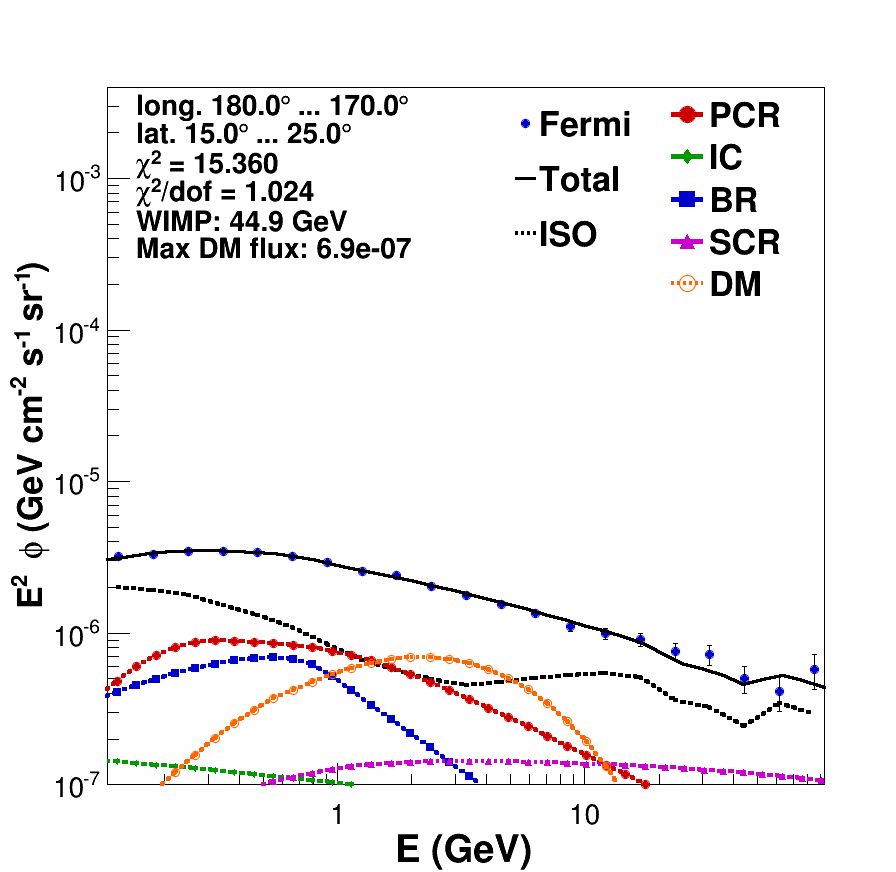}
\includegraphics[width=0.16\textwidth,height=0.16\textwidth,clip]{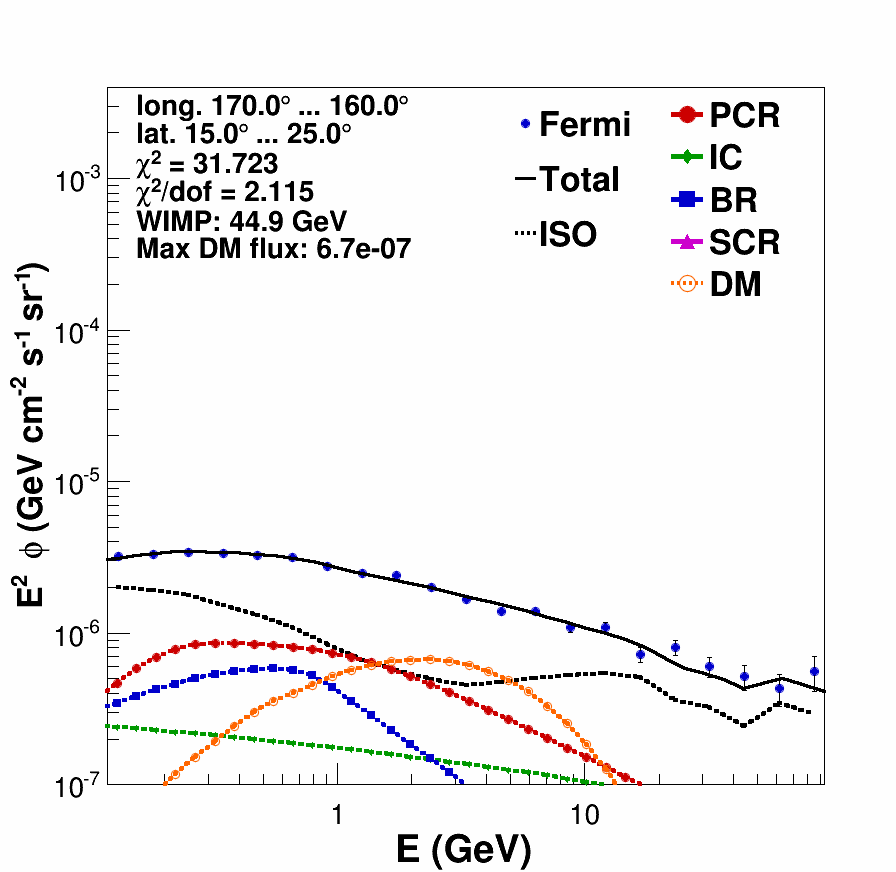}
\includegraphics[width=0.16\textwidth,height=0.16\textwidth,clip]{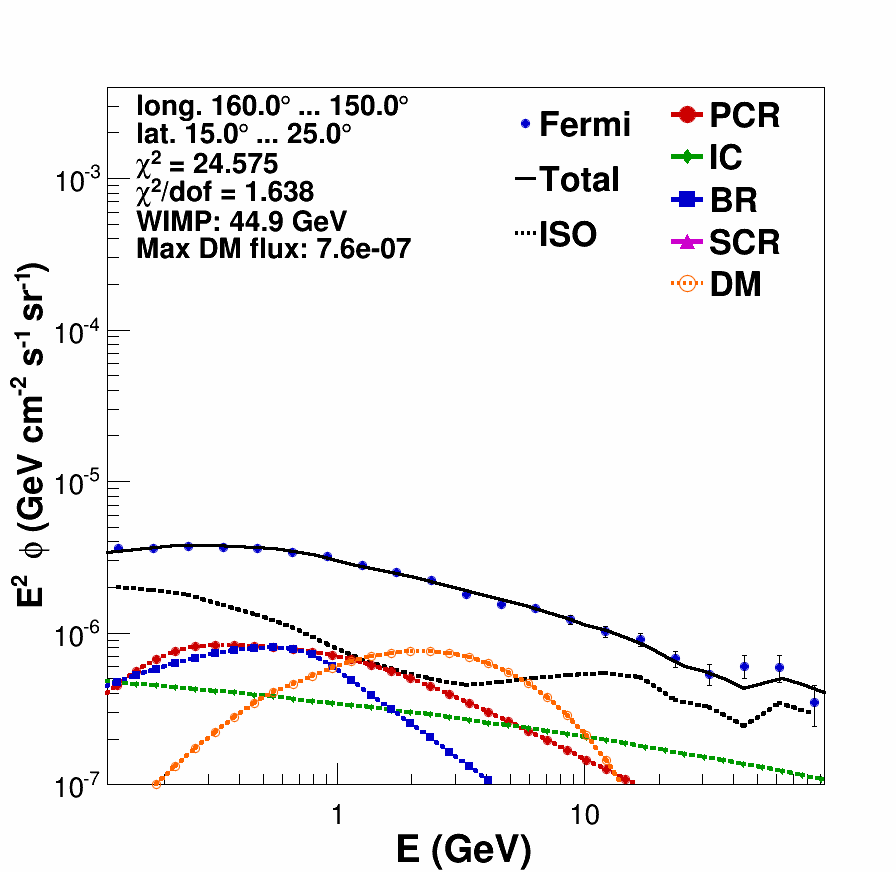}
\includegraphics[width=0.16\textwidth,height=0.16\textwidth,clip]{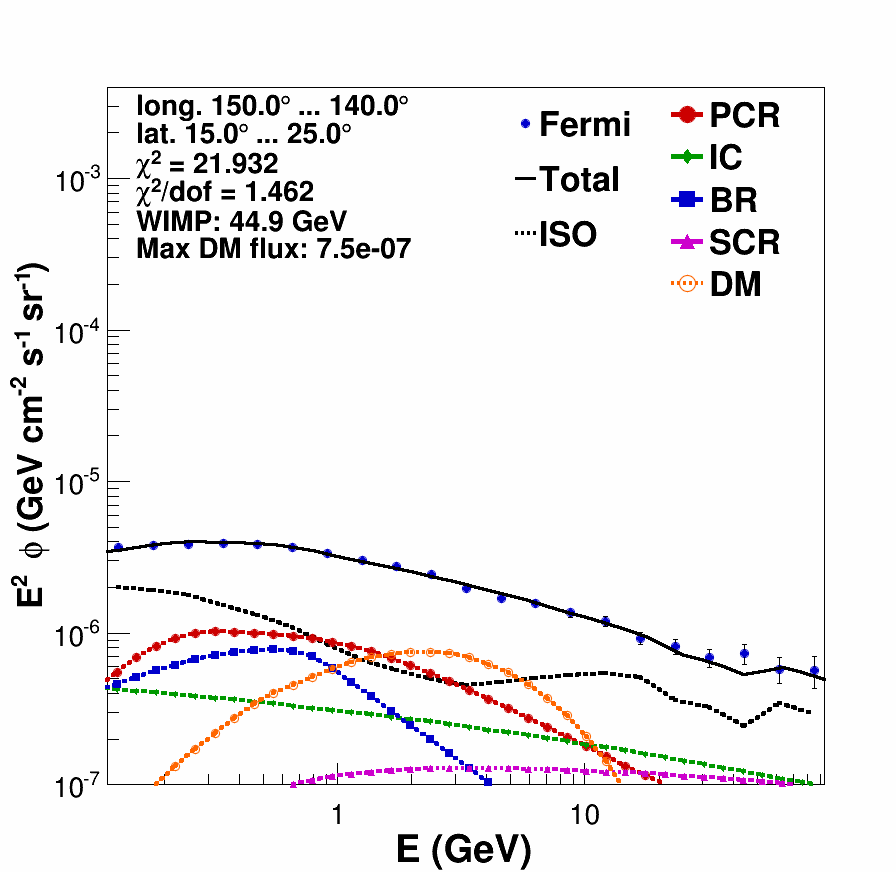}
\includegraphics[width=0.16\textwidth,height=0.16\textwidth,clip]{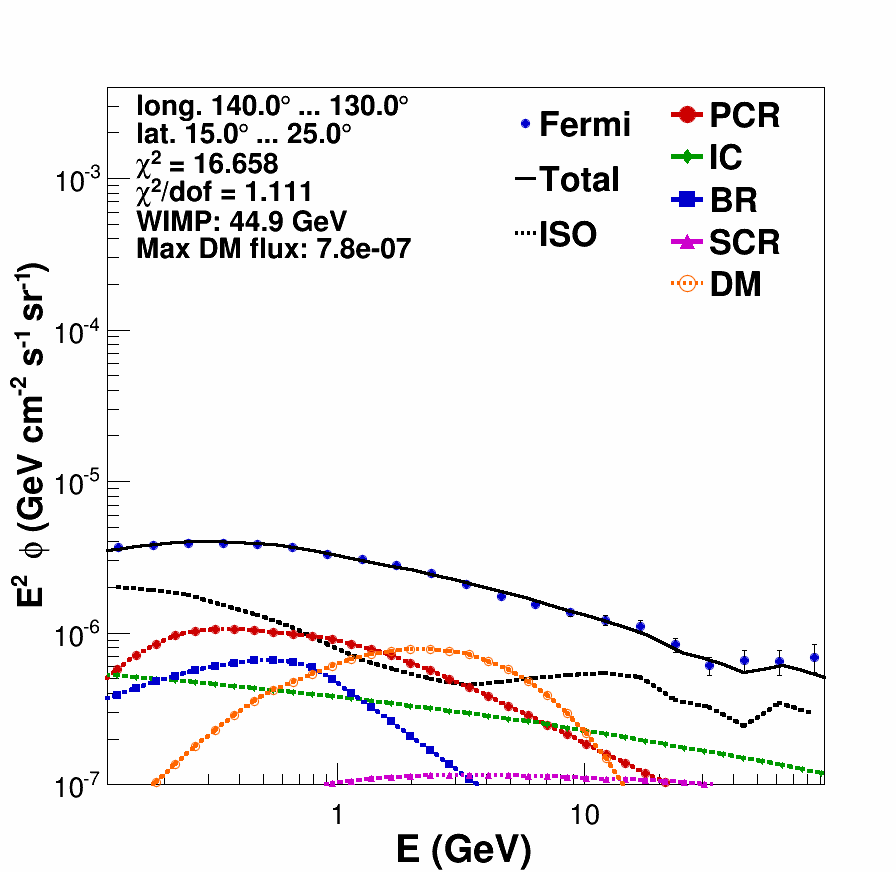}
\includegraphics[width=0.16\textwidth,height=0.16\textwidth,clip]{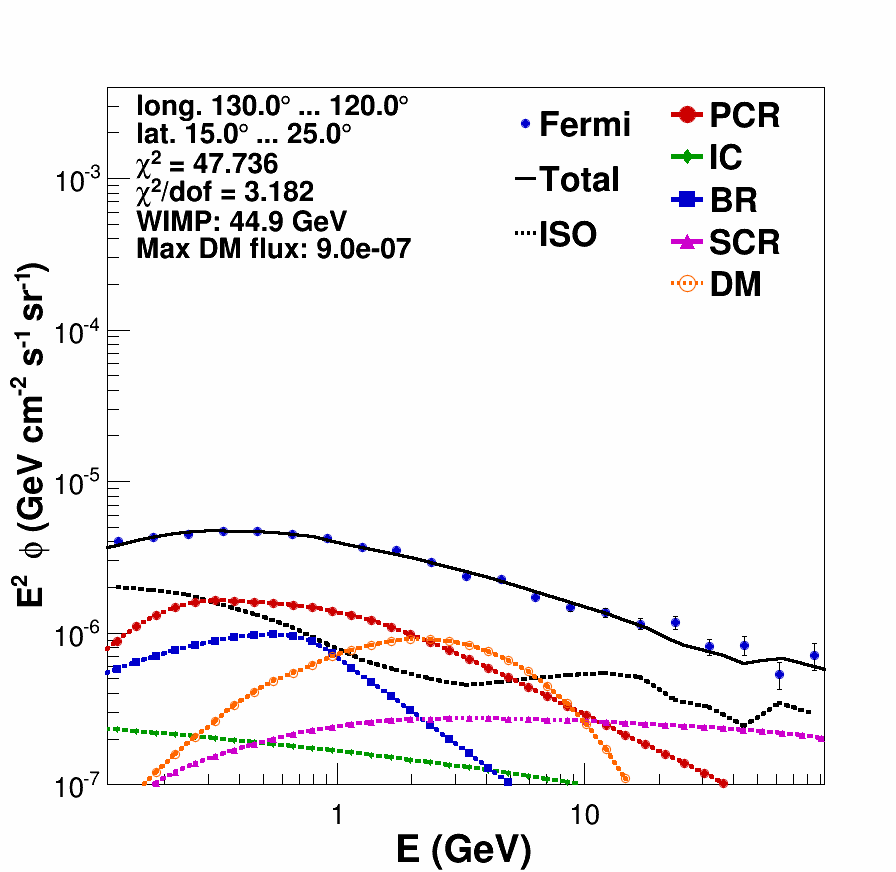}
\includegraphics[width=0.16\textwidth,height=0.16\textwidth,clip]{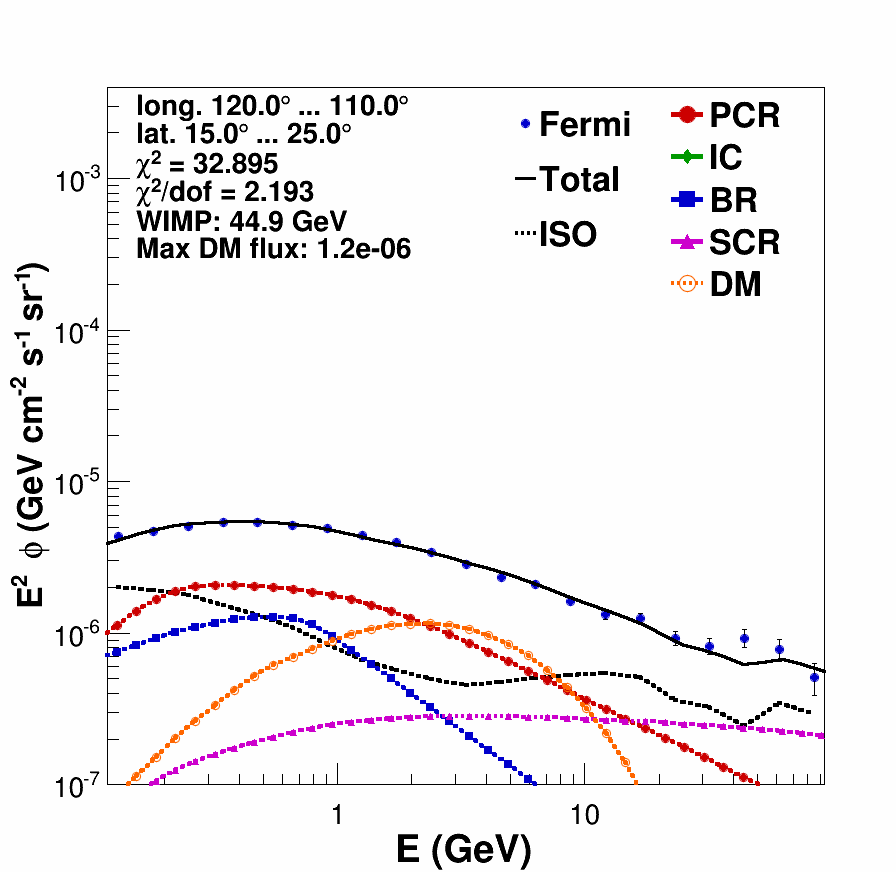}
\includegraphics[width=0.16\textwidth,height=0.16\textwidth,clip]{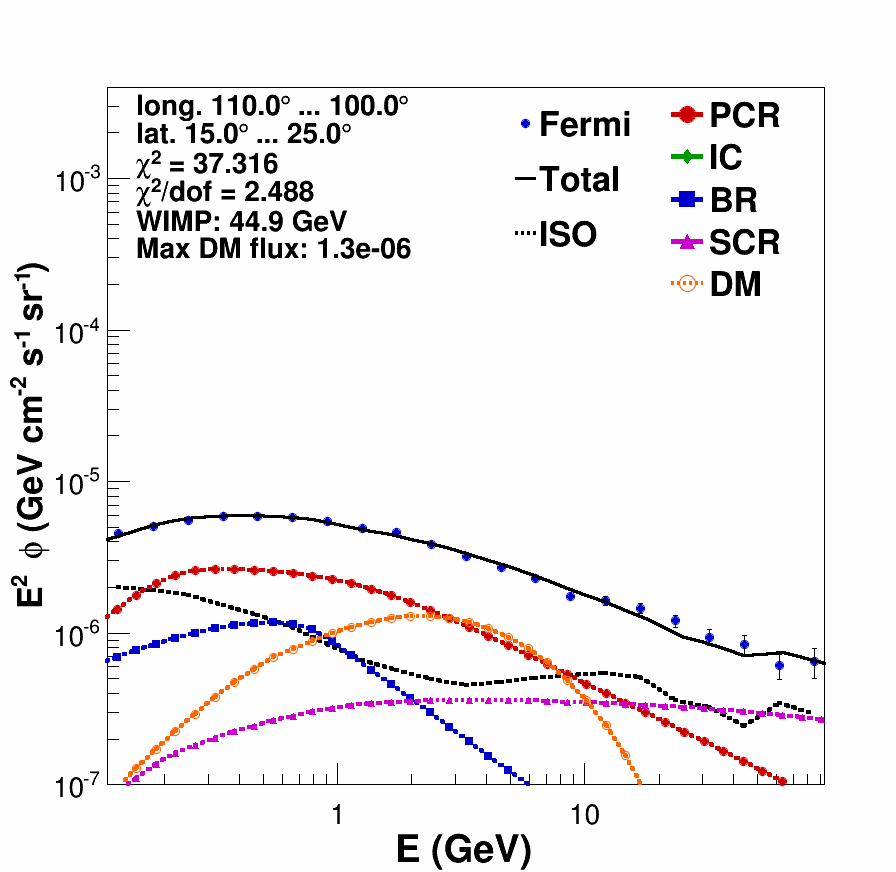}
\includegraphics[width=0.16\textwidth,height=0.16\textwidth,clip]{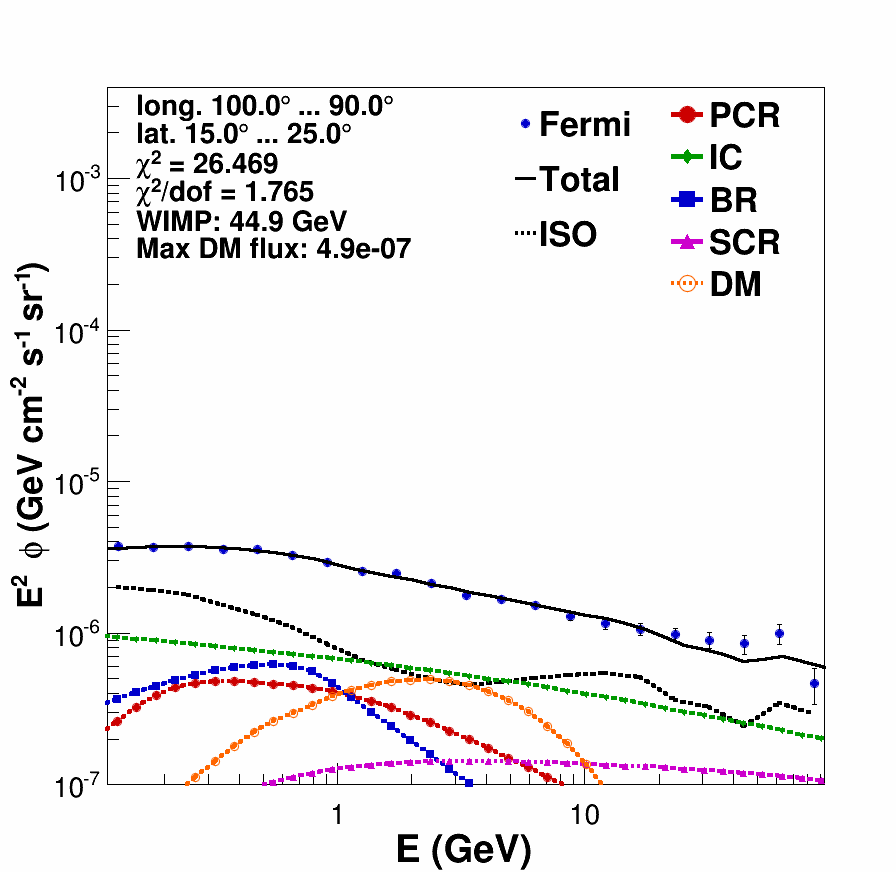}
\includegraphics[width=0.16\textwidth,height=0.16\textwidth,clip]{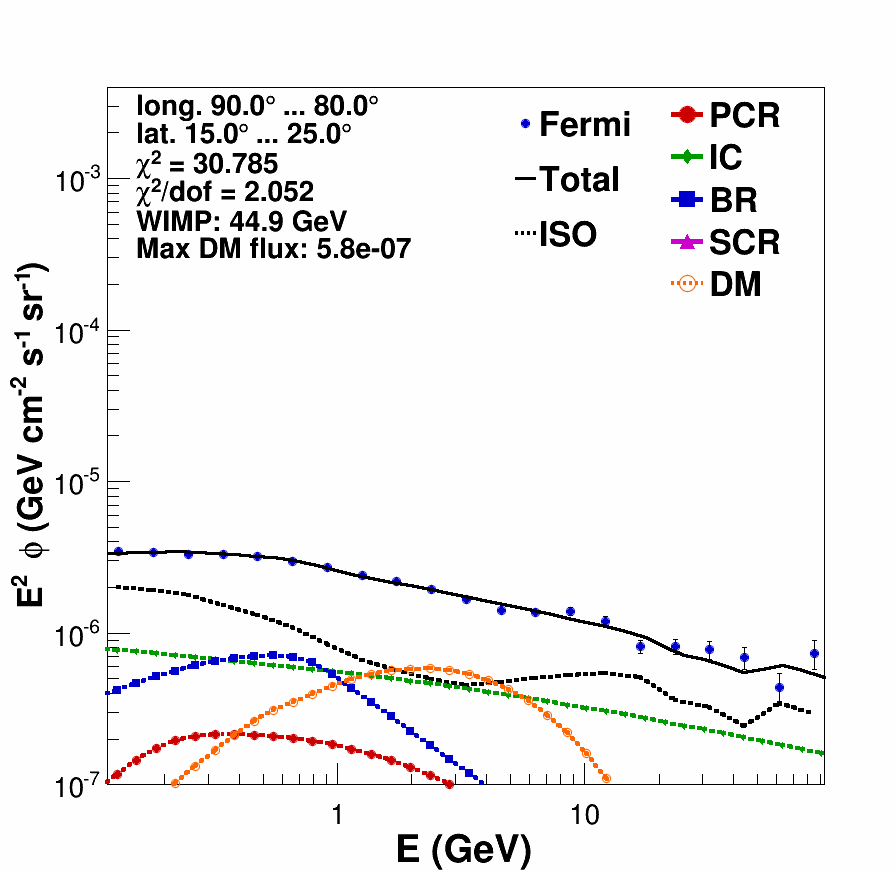}
\includegraphics[width=0.16\textwidth,height=0.16\textwidth,clip]{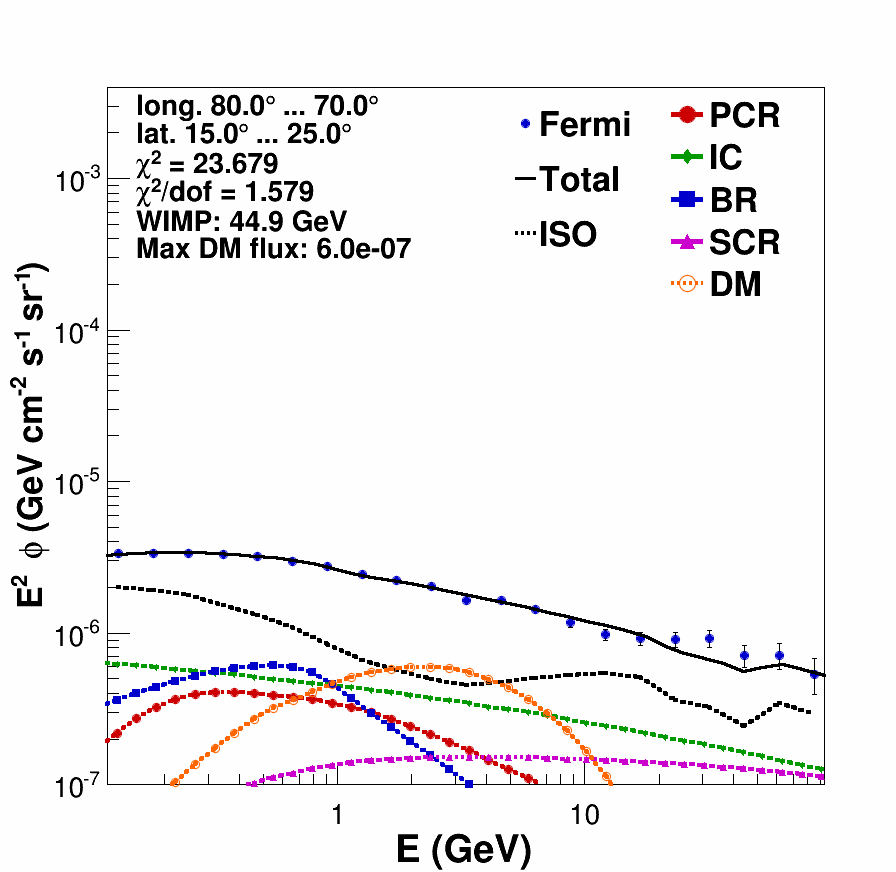}
\includegraphics[width=0.16\textwidth,height=0.16\textwidth,clip]{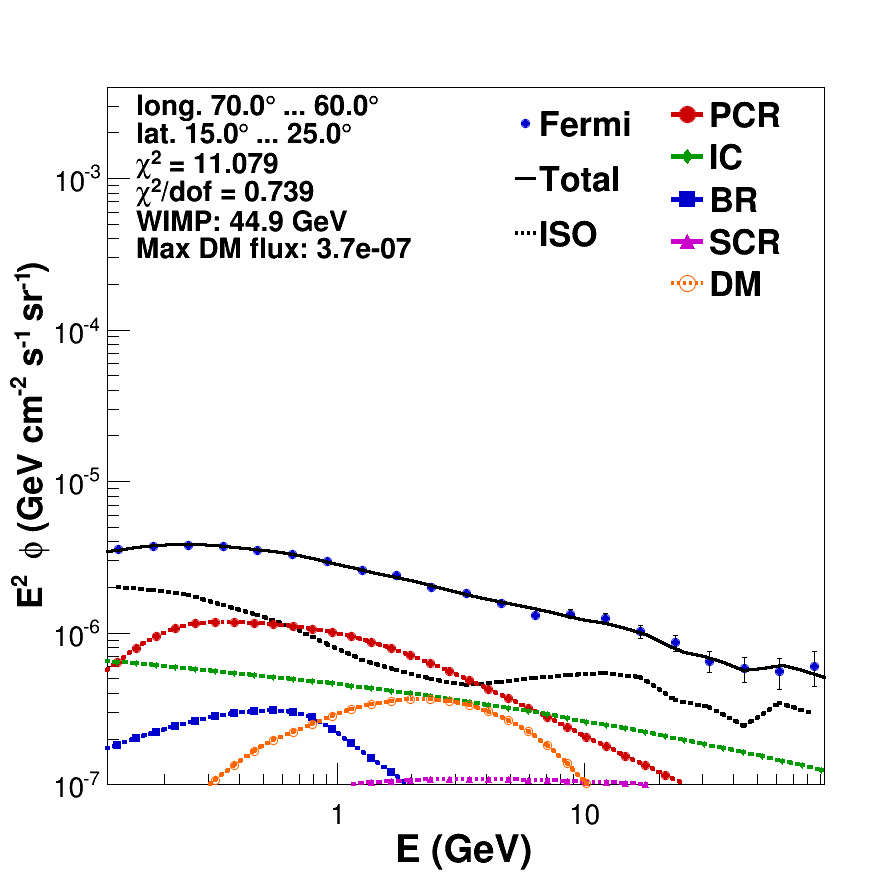}
\includegraphics[width=0.16\textwidth,height=0.16\textwidth,clip]{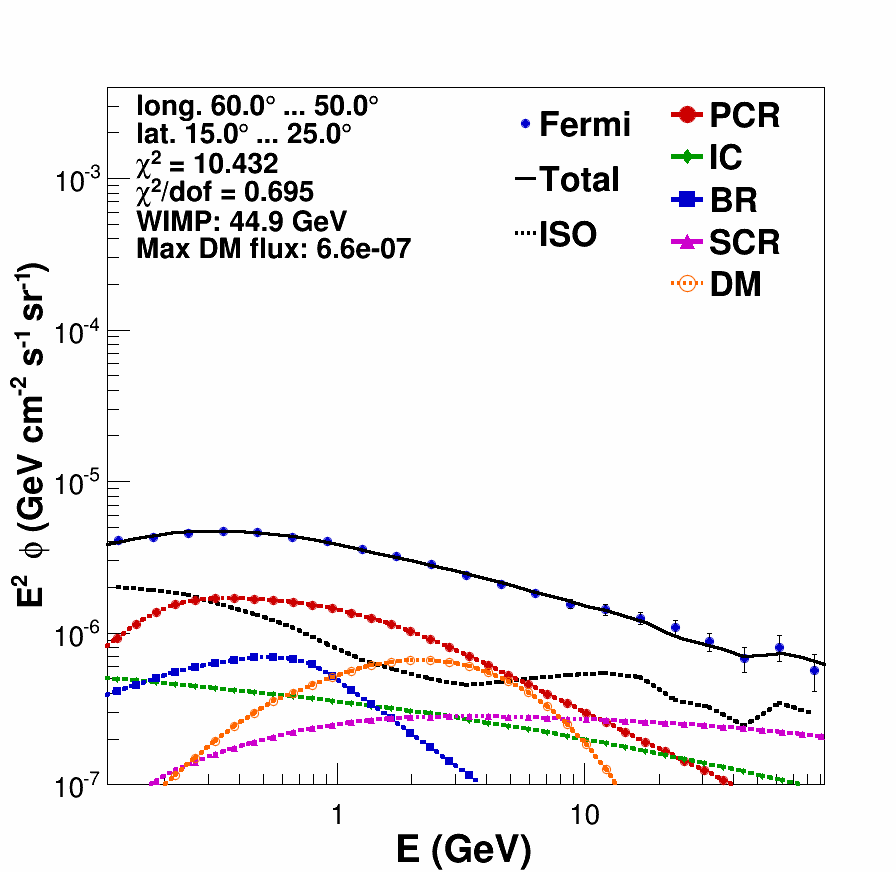}
\includegraphics[width=0.16\textwidth,height=0.16\textwidth,clip]{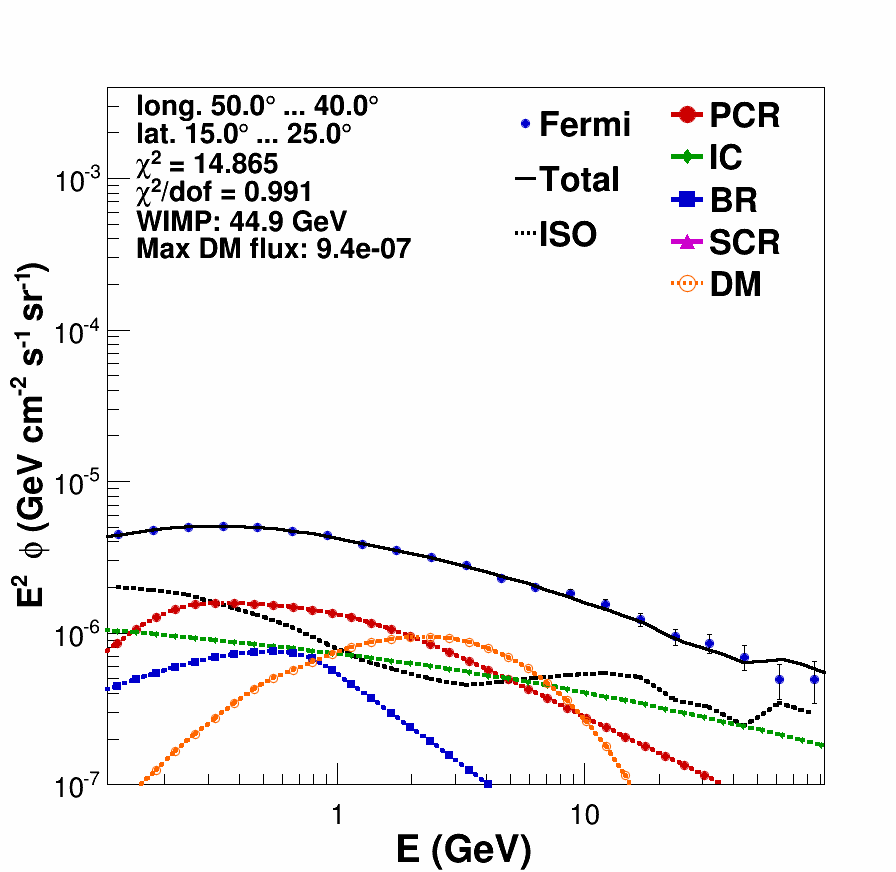}
\includegraphics[width=0.16\textwidth,height=0.16\textwidth,clip]{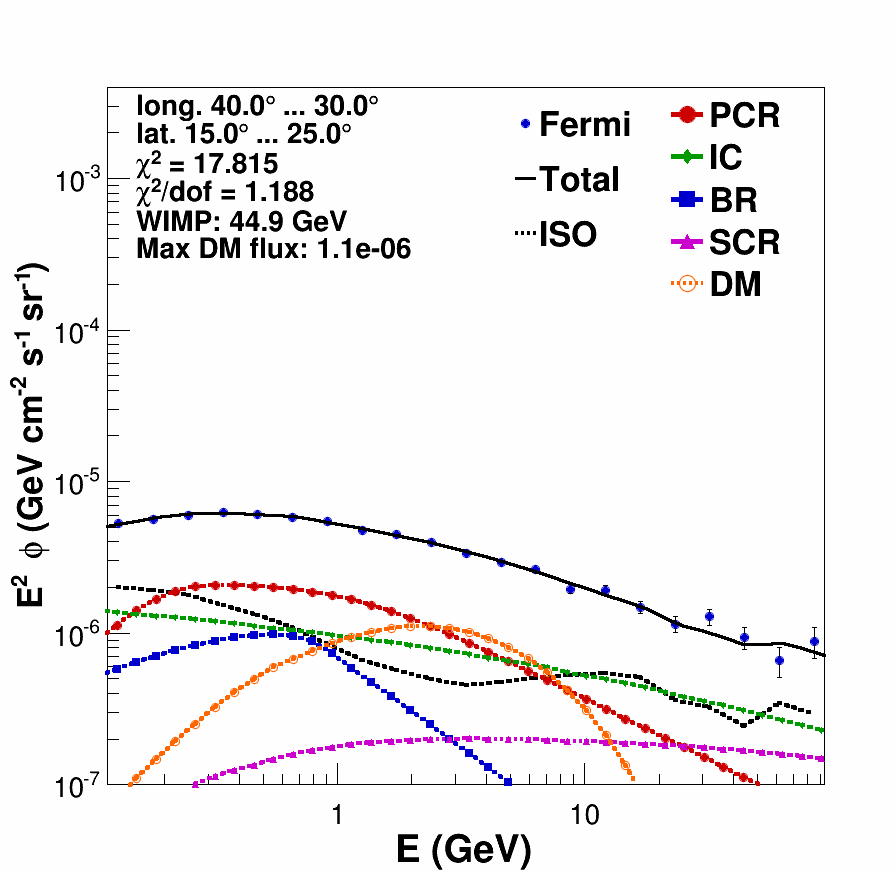}
\includegraphics[width=0.16\textwidth,height=0.16\textwidth,clip]{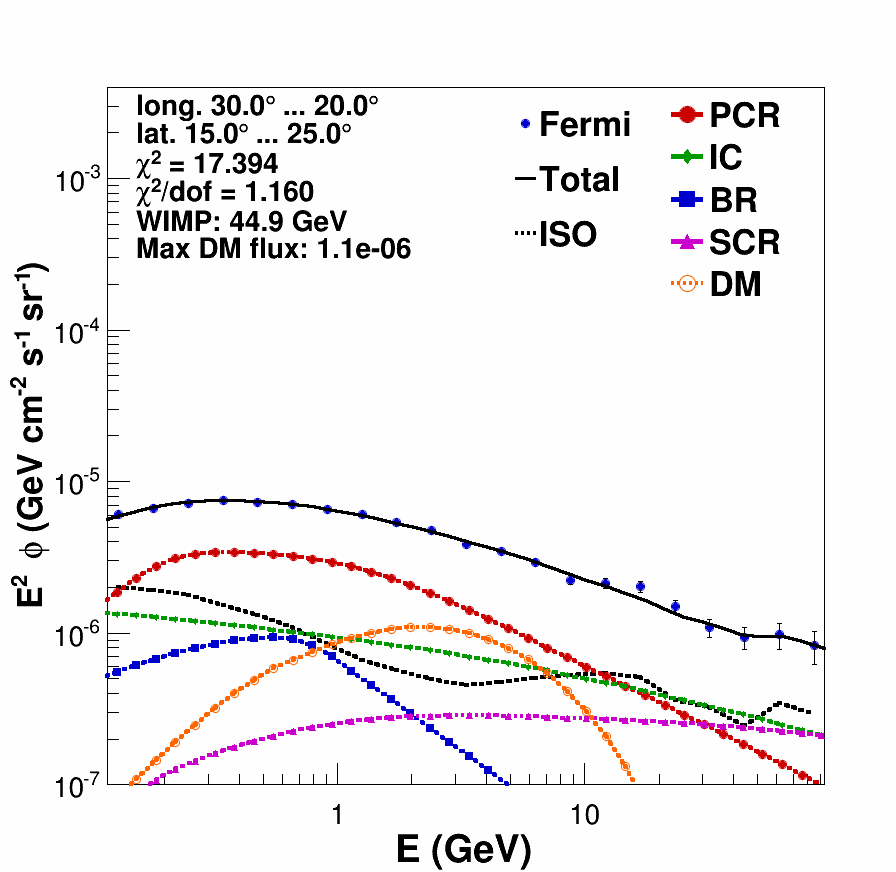}
\includegraphics[width=0.16\textwidth,height=0.16\textwidth,clip]{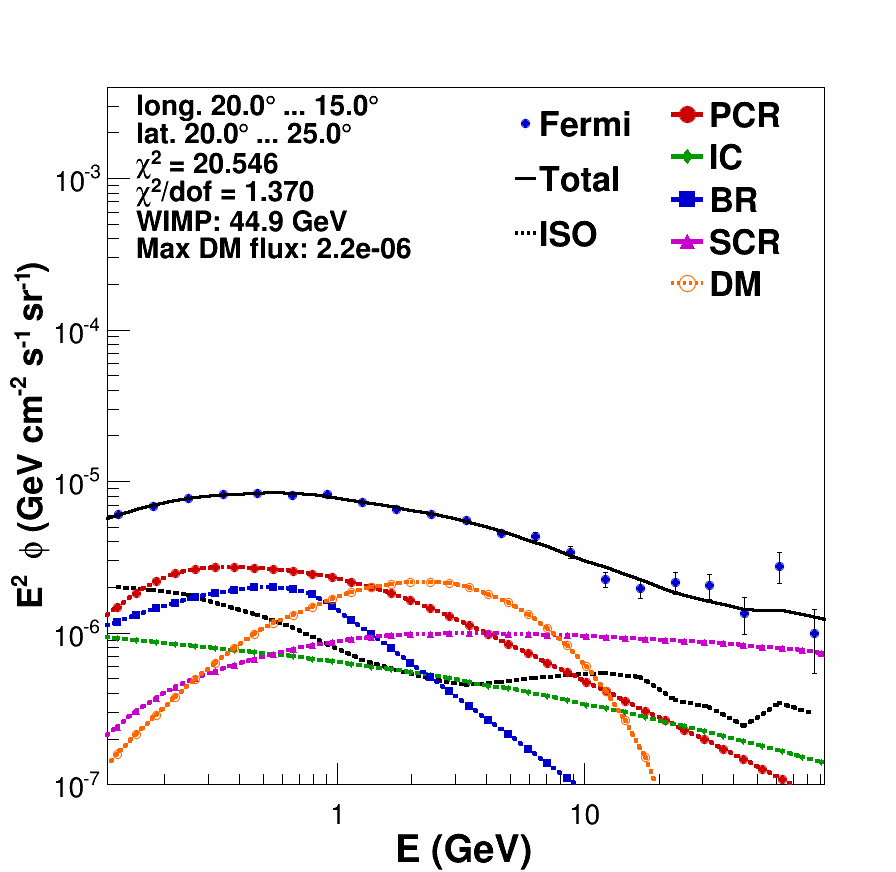}
\includegraphics[width=0.16\textwidth,height=0.16\textwidth,clip]{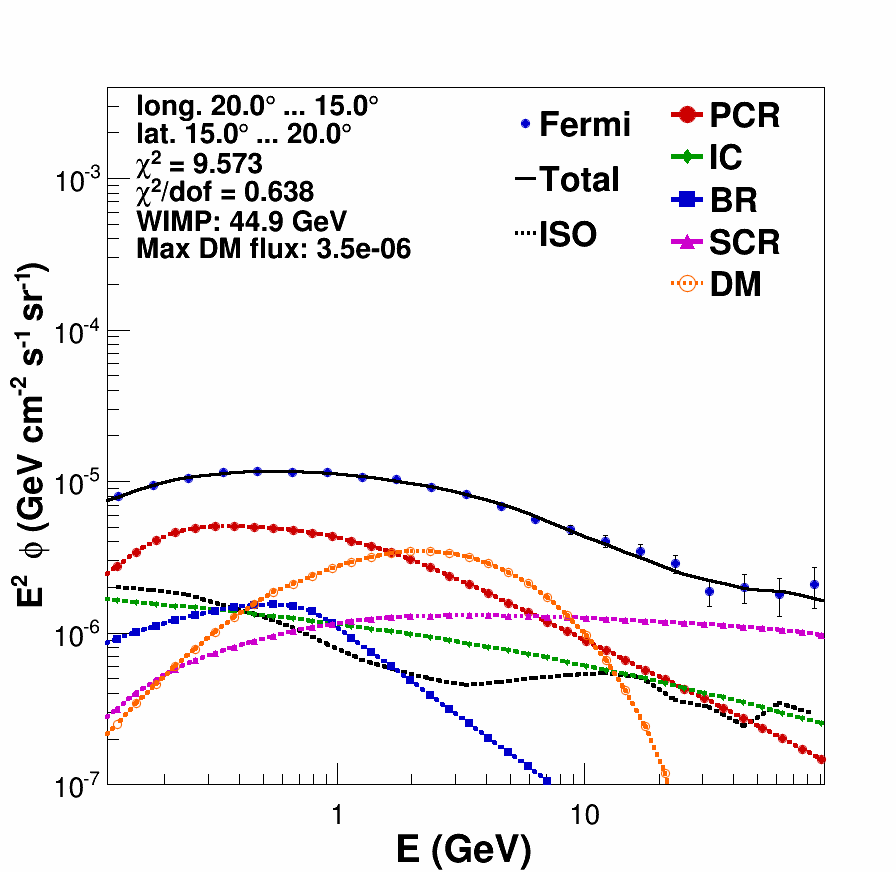}
\includegraphics[width=0.16\textwidth,height=0.16\textwidth,clip]{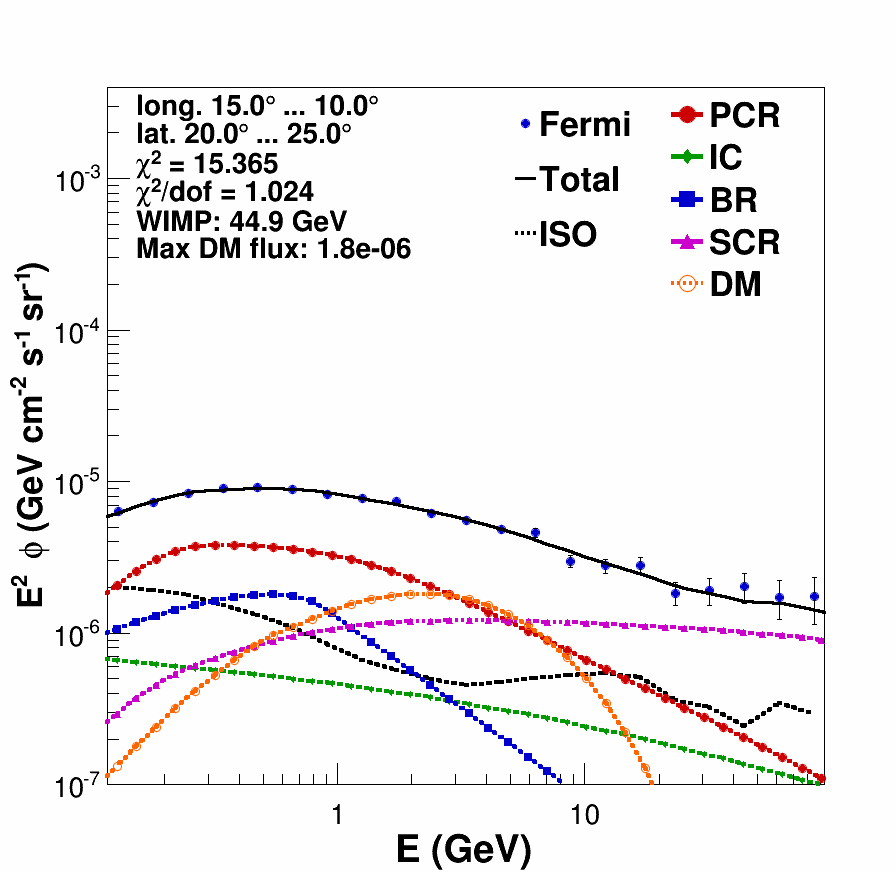}
\includegraphics[width=0.16\textwidth,height=0.16\textwidth,clip]{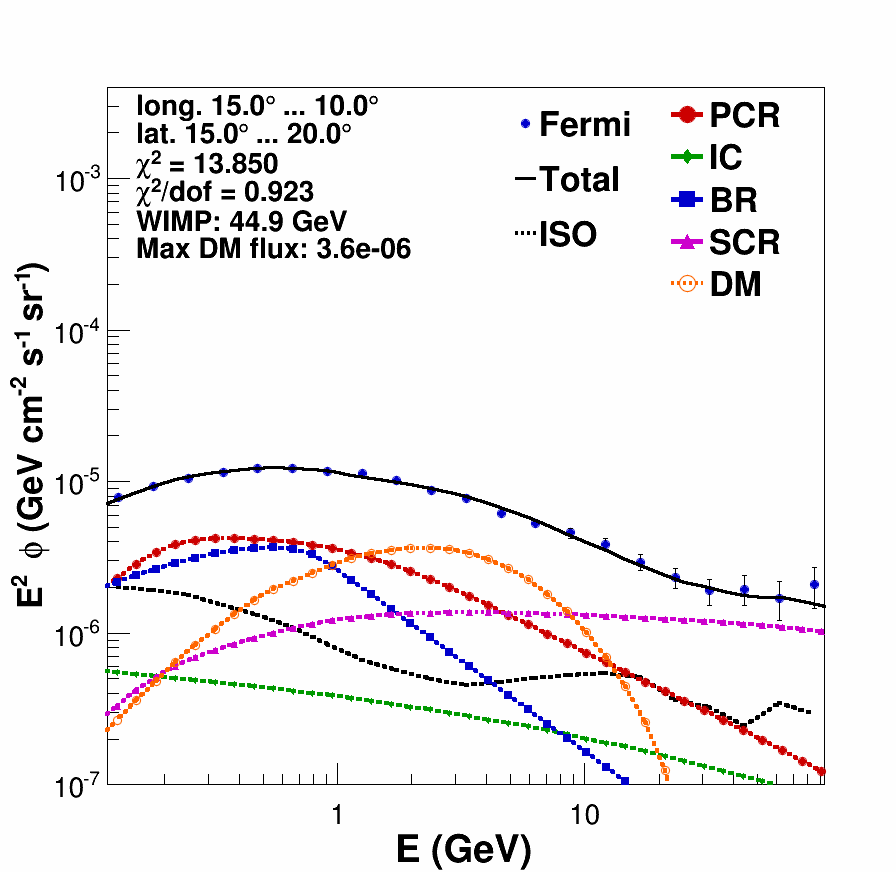}
\includegraphics[width=0.16\textwidth,height=0.16\textwidth,clip]{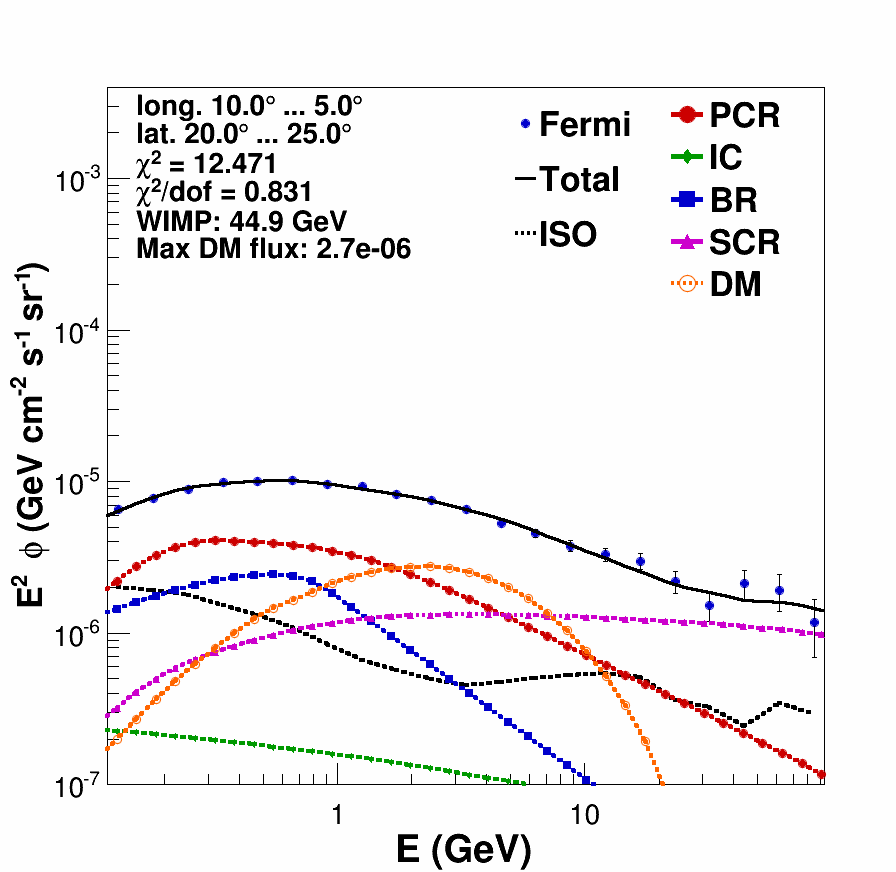}
\includegraphics[width=0.16\textwidth,height=0.16\textwidth,clip]{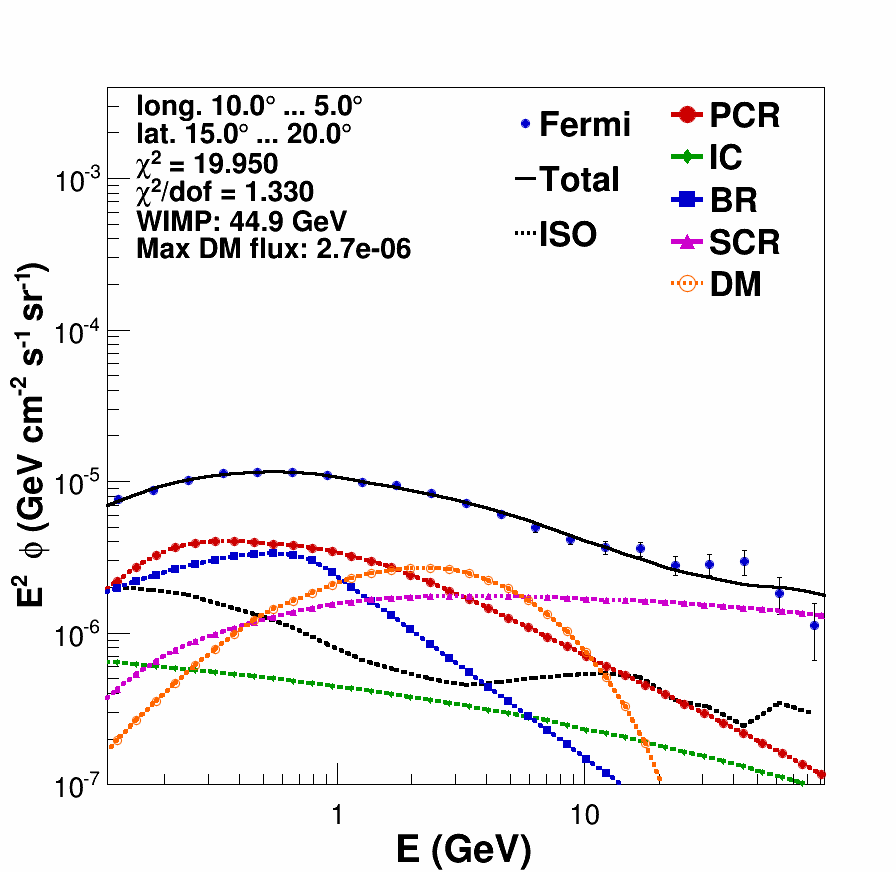}
\includegraphics[width=0.16\textwidth,height=0.16\textwidth,clip]{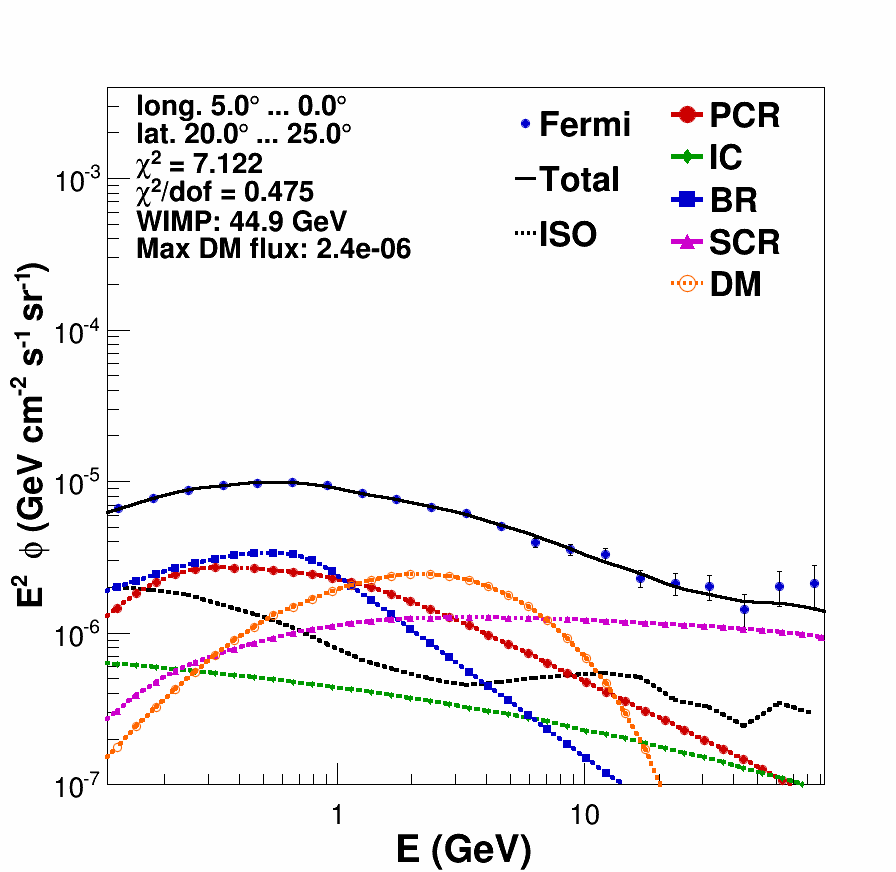}
\includegraphics[width=0.16\textwidth,height=0.16\textwidth,clip]{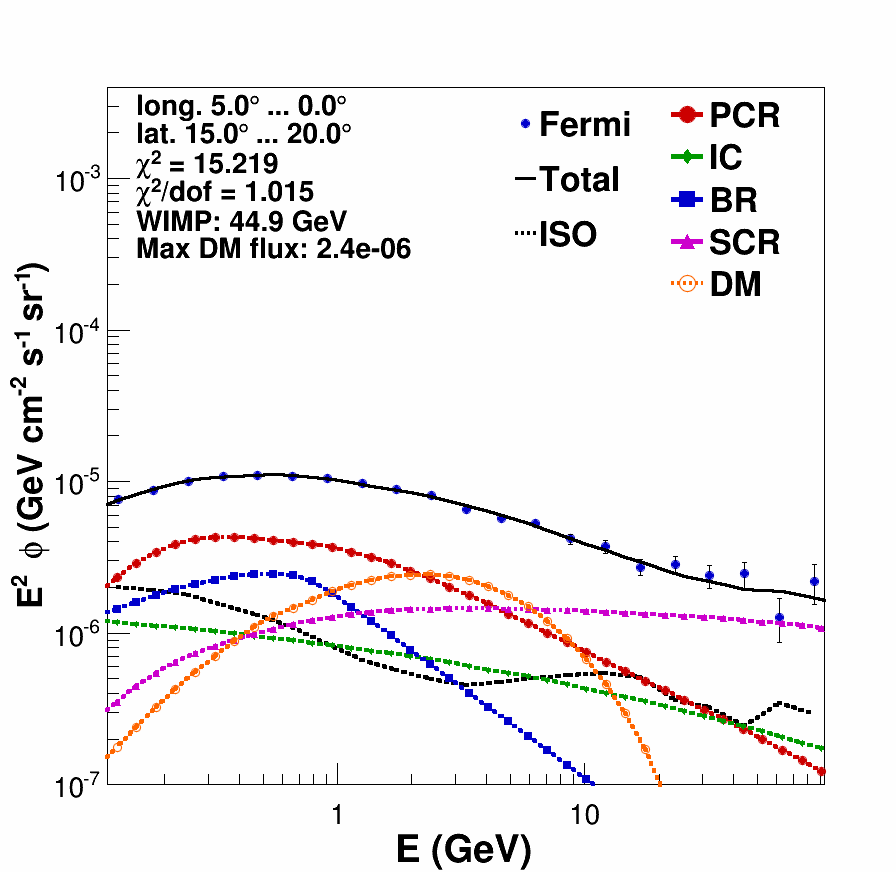}
\includegraphics[width=0.16\textwidth,height=0.16\textwidth,clip]{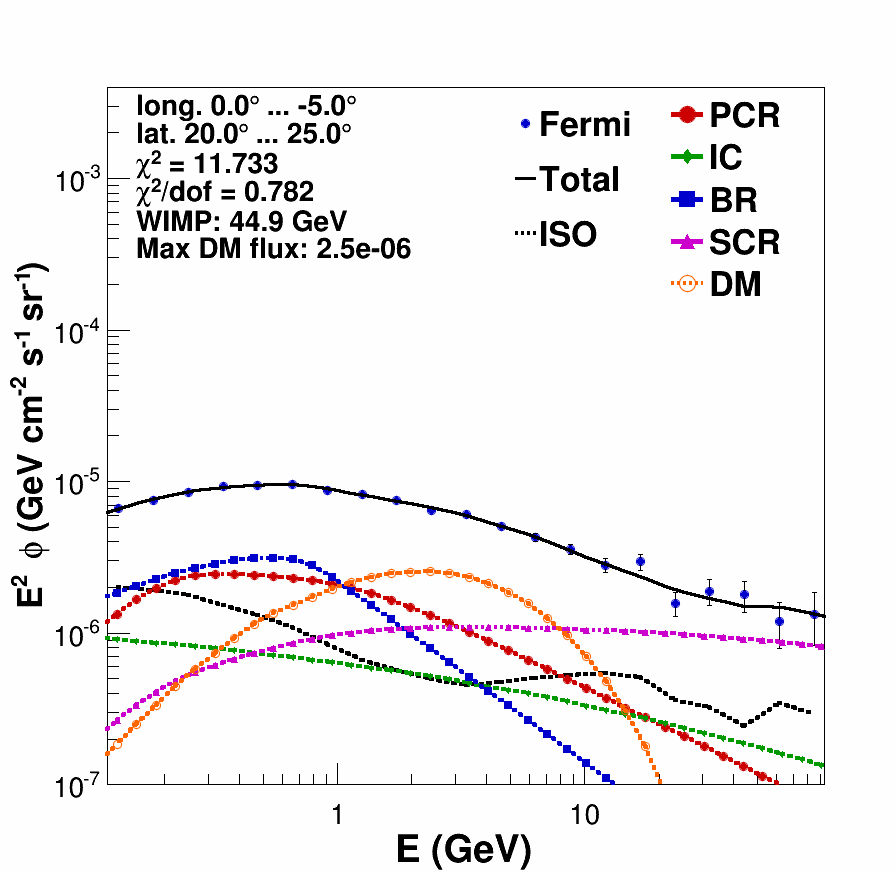}
\includegraphics[width=0.16\textwidth,height=0.16\textwidth,clip]{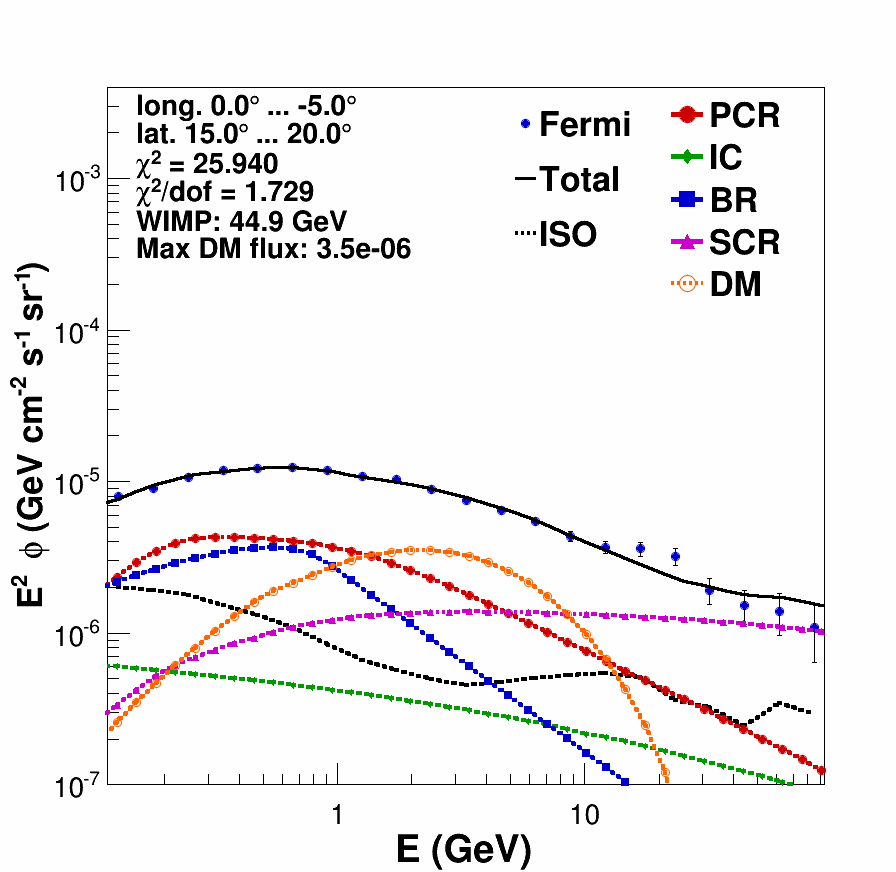}
\includegraphics[width=0.16\textwidth,height=0.16\textwidth,clip]{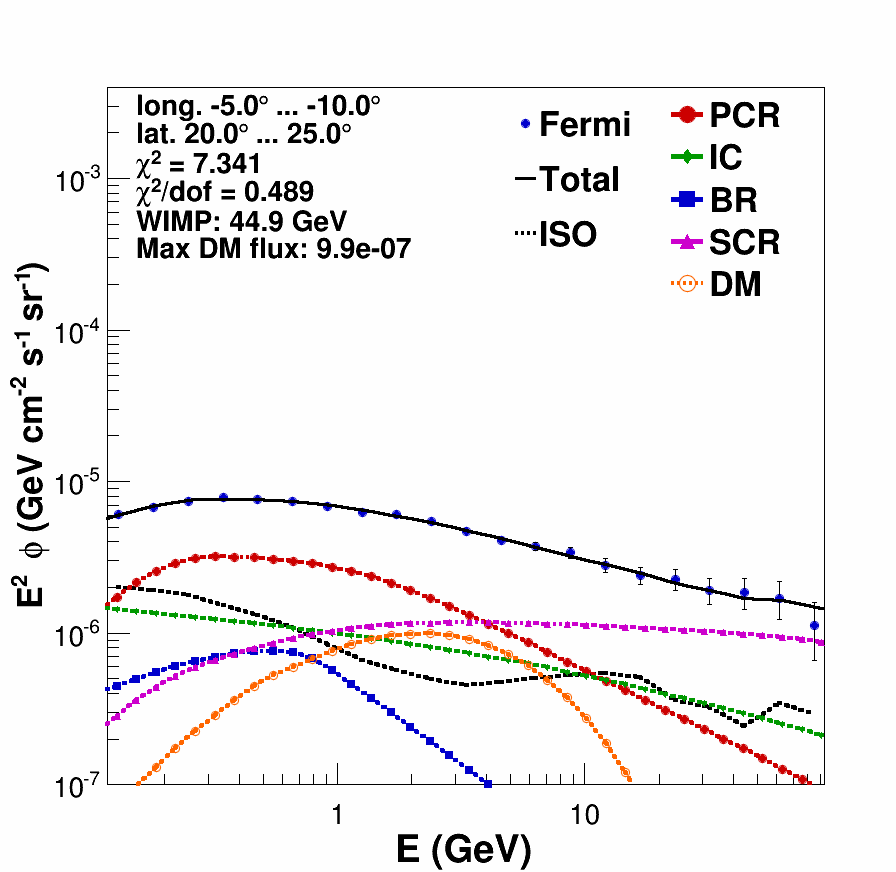}
\includegraphics[width=0.16\textwidth,height=0.16\textwidth,clip]{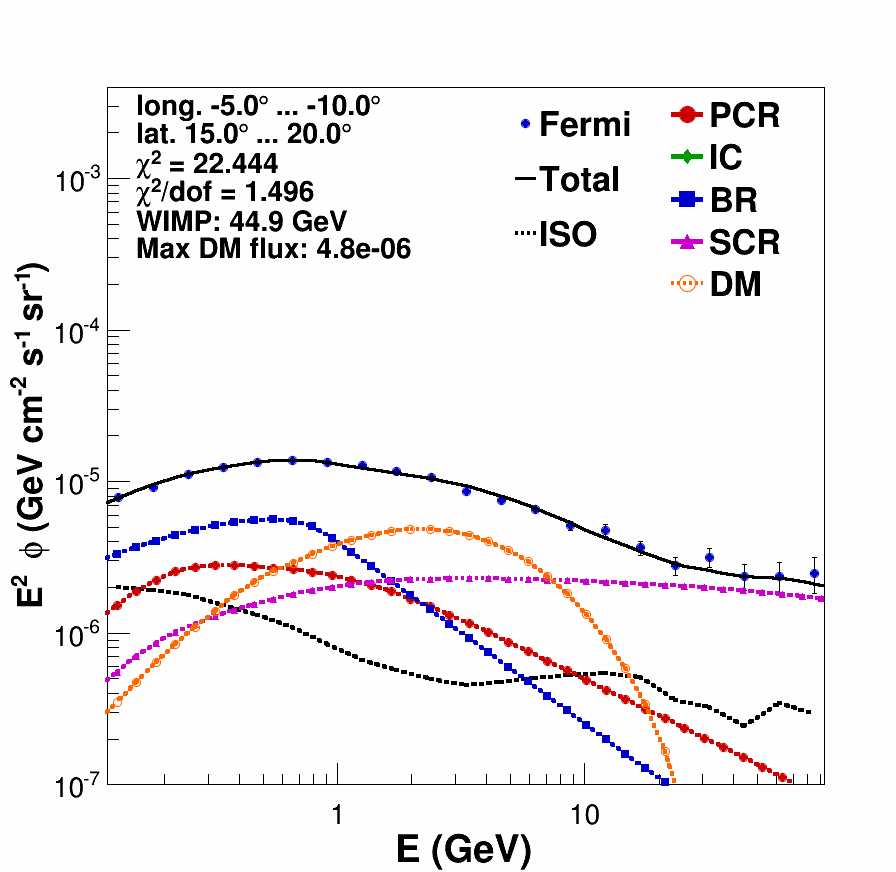}
\includegraphics[width=0.16\textwidth,height=0.16\textwidth,clip]{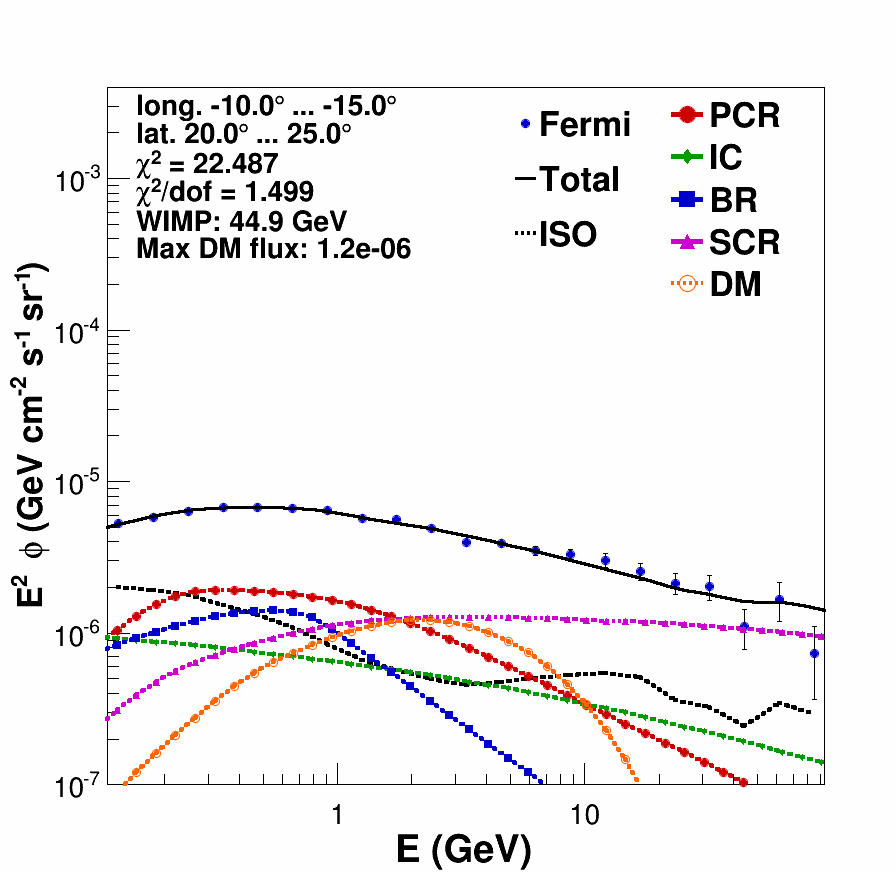}
\includegraphics[width=0.16\textwidth,height=0.16\textwidth,clip]{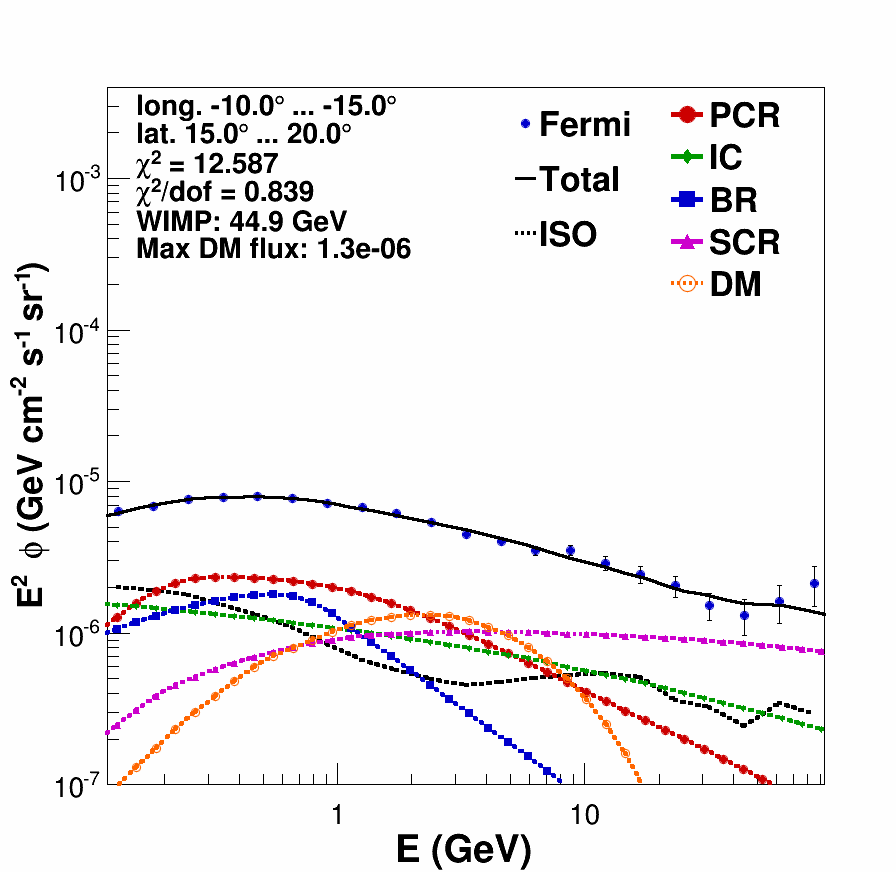}
\includegraphics[width=0.16\textwidth,height=0.16\textwidth,clip]{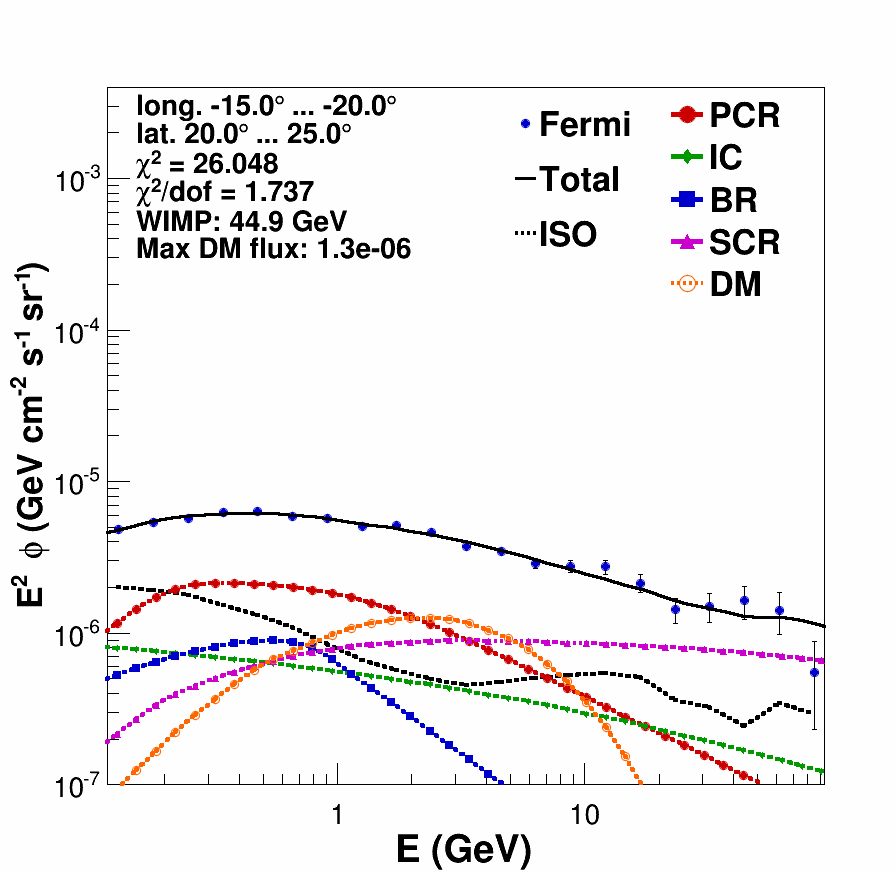}
\includegraphics[width=0.16\textwidth,height=0.16\textwidth,clip]{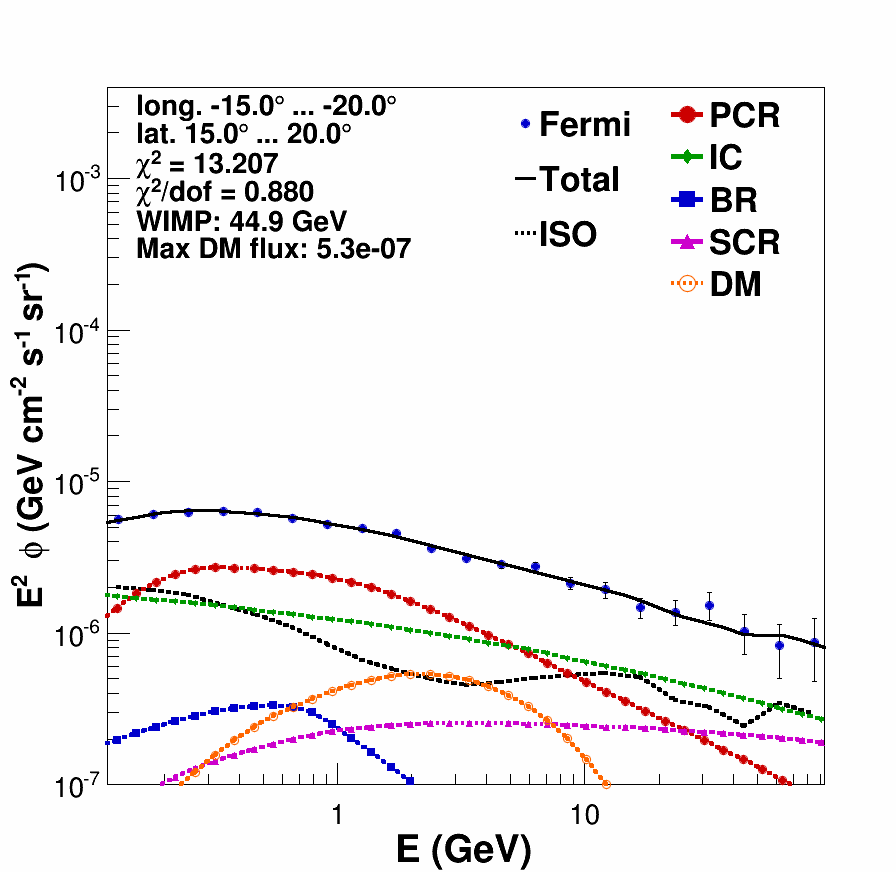}
\includegraphics[width=0.16\textwidth,height=0.16\textwidth,clip]{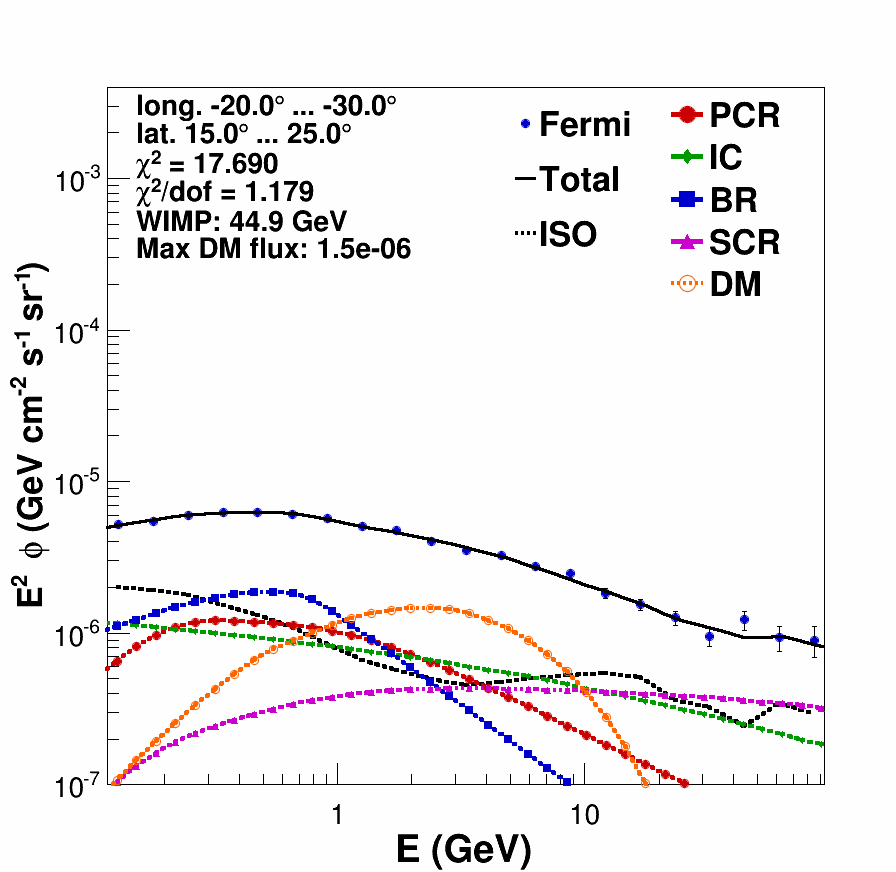}
\includegraphics[width=0.16\textwidth,height=0.16\textwidth,clip]{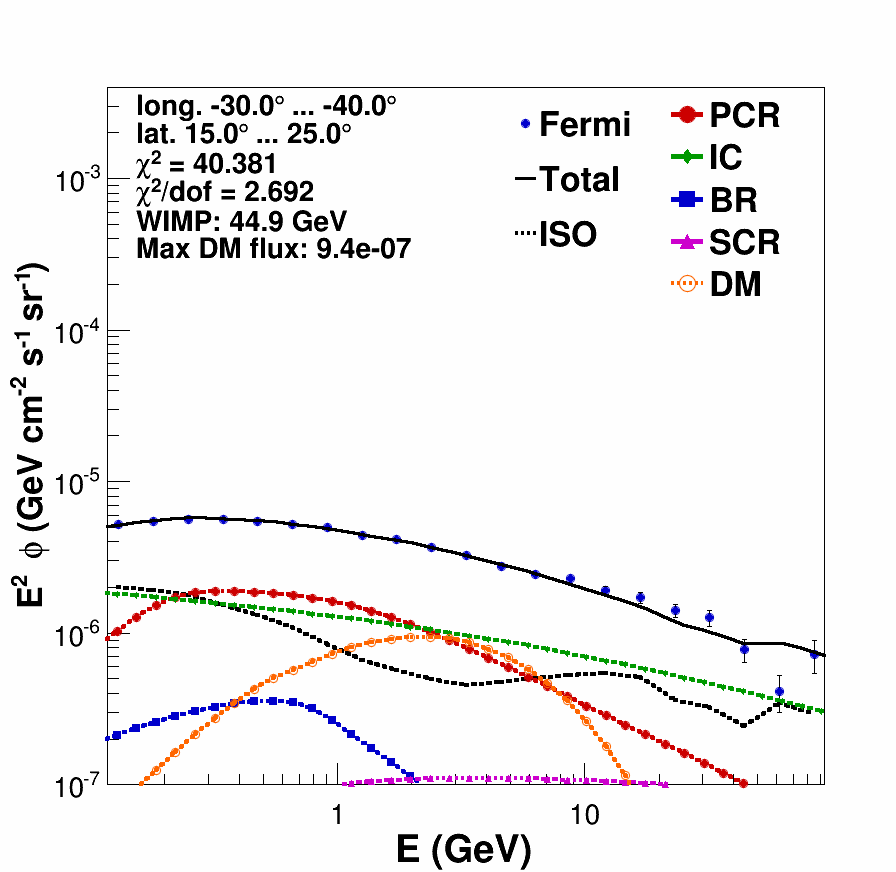}
\includegraphics[width=0.16\textwidth,height=0.16\textwidth,clip]{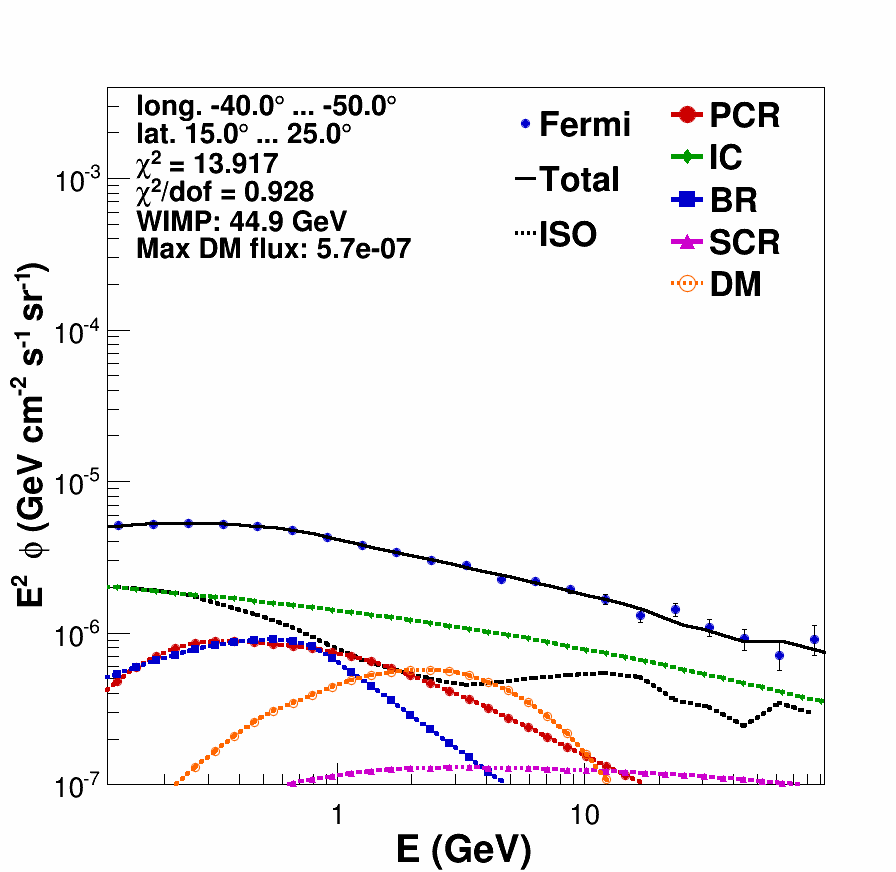}
\includegraphics[width=0.16\textwidth,height=0.16\textwidth,clip]{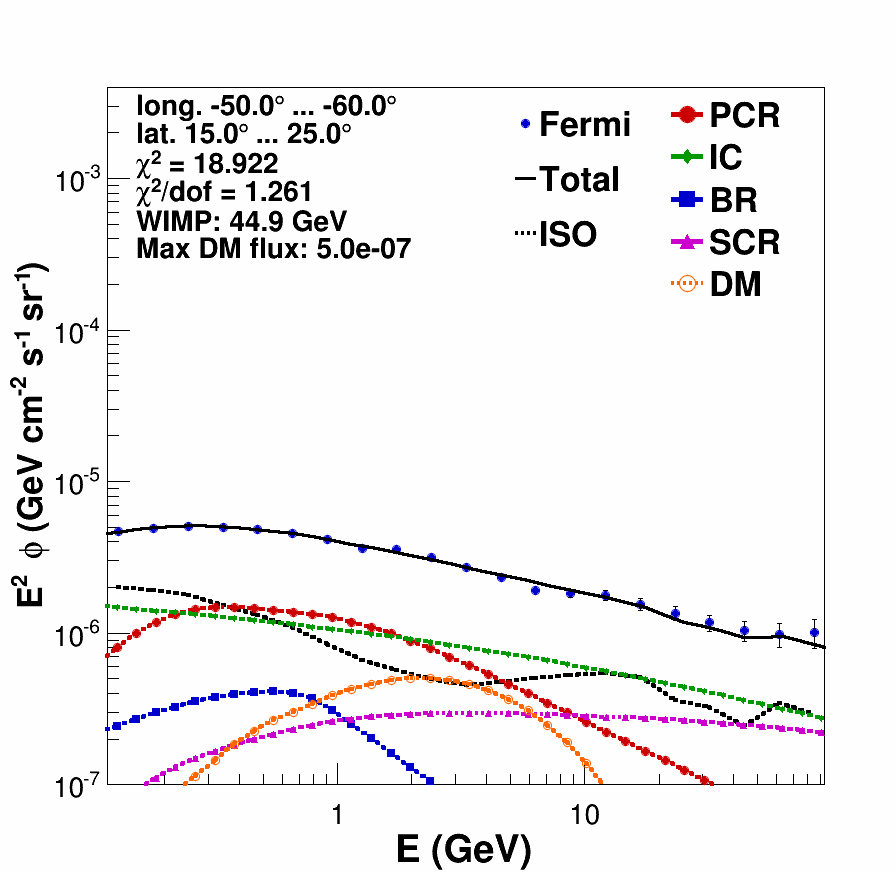}
\includegraphics[width=0.16\textwidth,height=0.16\textwidth,clip]{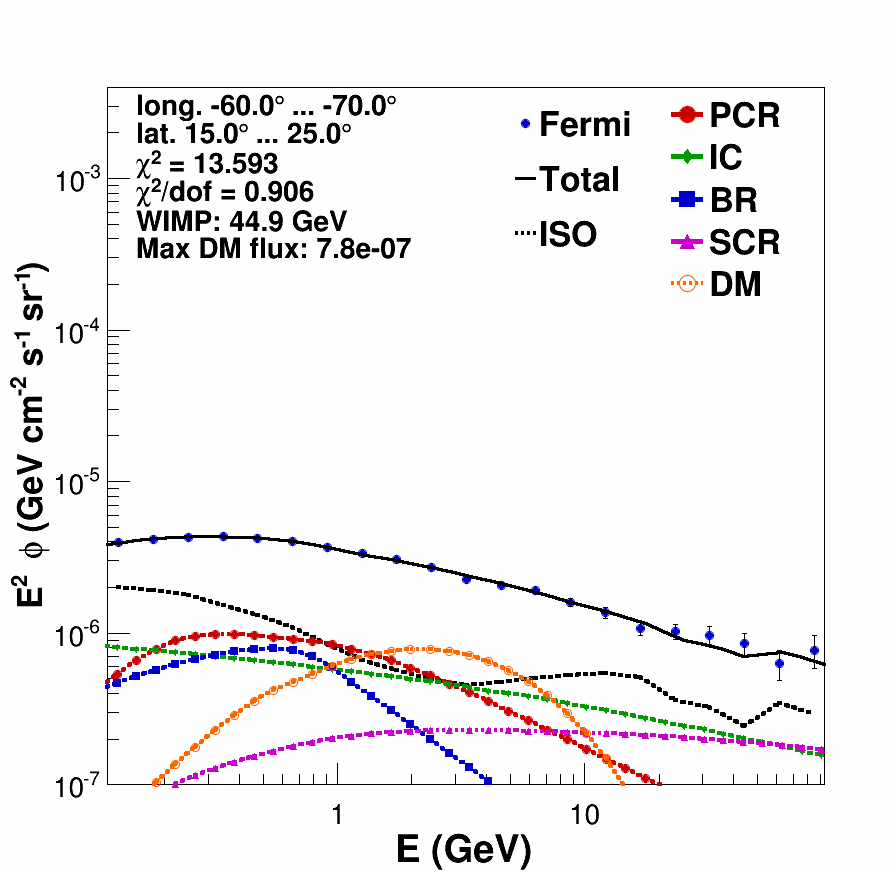}
\includegraphics[width=0.16\textwidth,height=0.16\textwidth,clip]{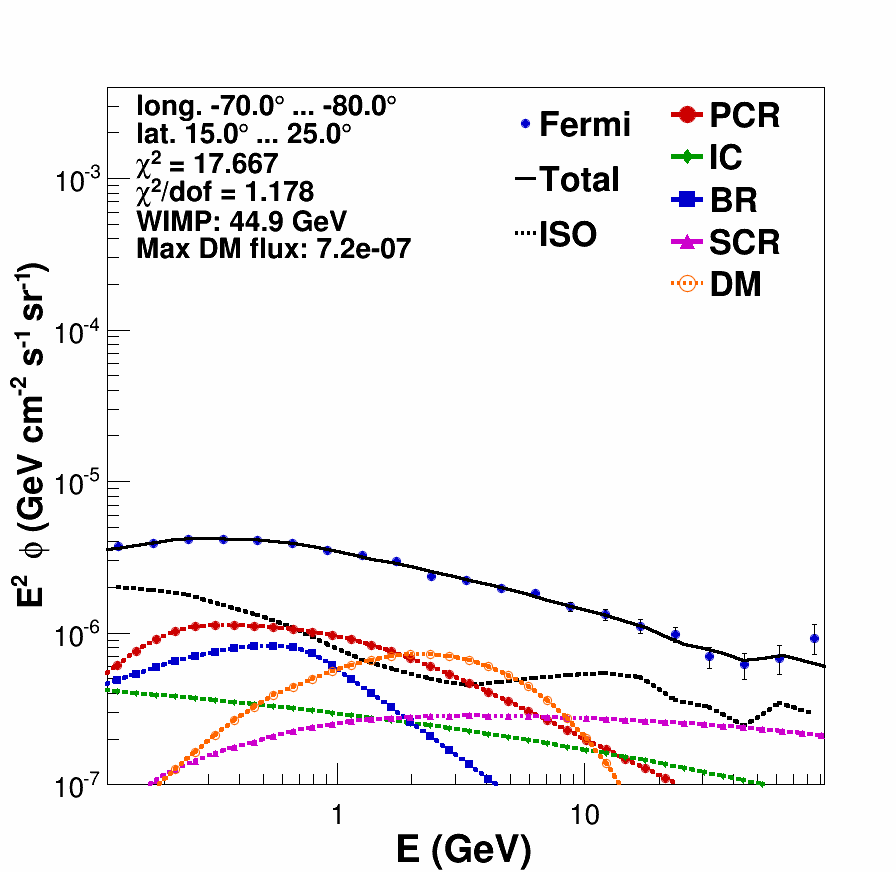}
\includegraphics[width=0.16\textwidth,height=0.16\textwidth,clip]{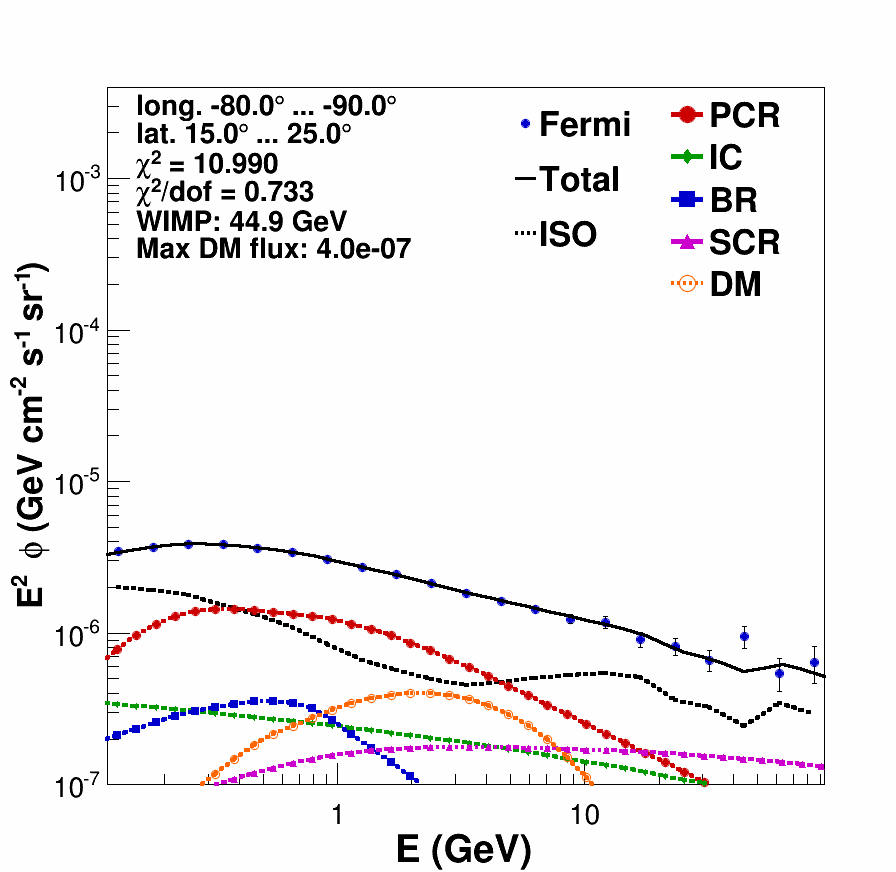}
\includegraphics[width=0.16\textwidth,height=0.16\textwidth,clip]{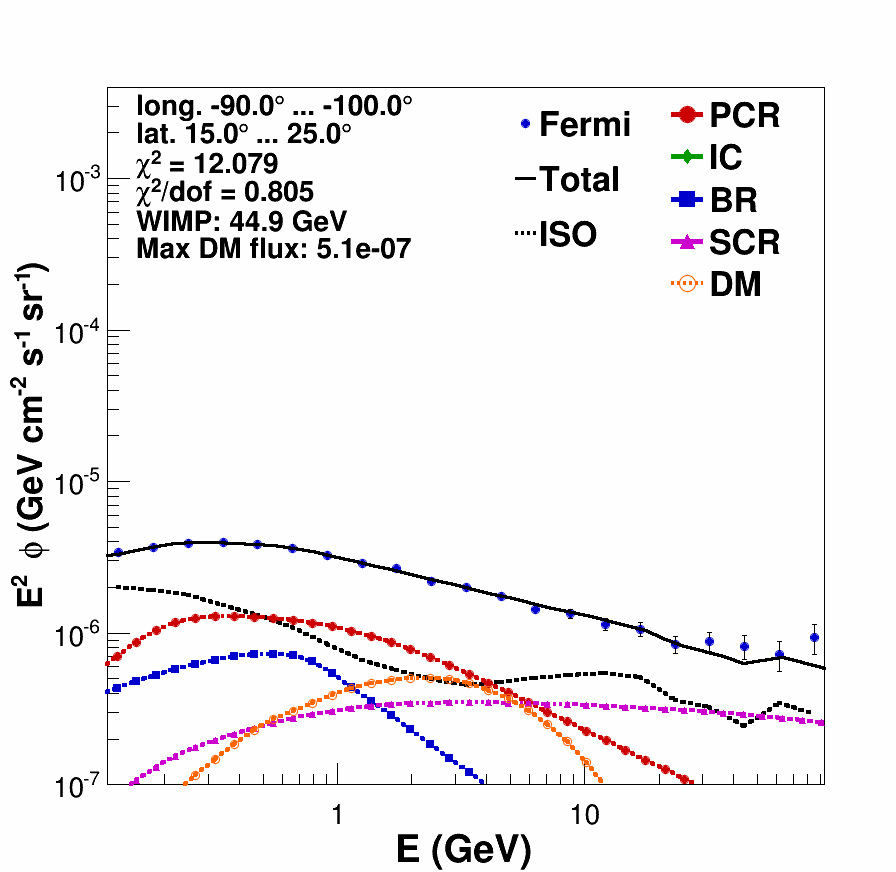}
\includegraphics[width=0.16\textwidth,height=0.16\textwidth,clip]{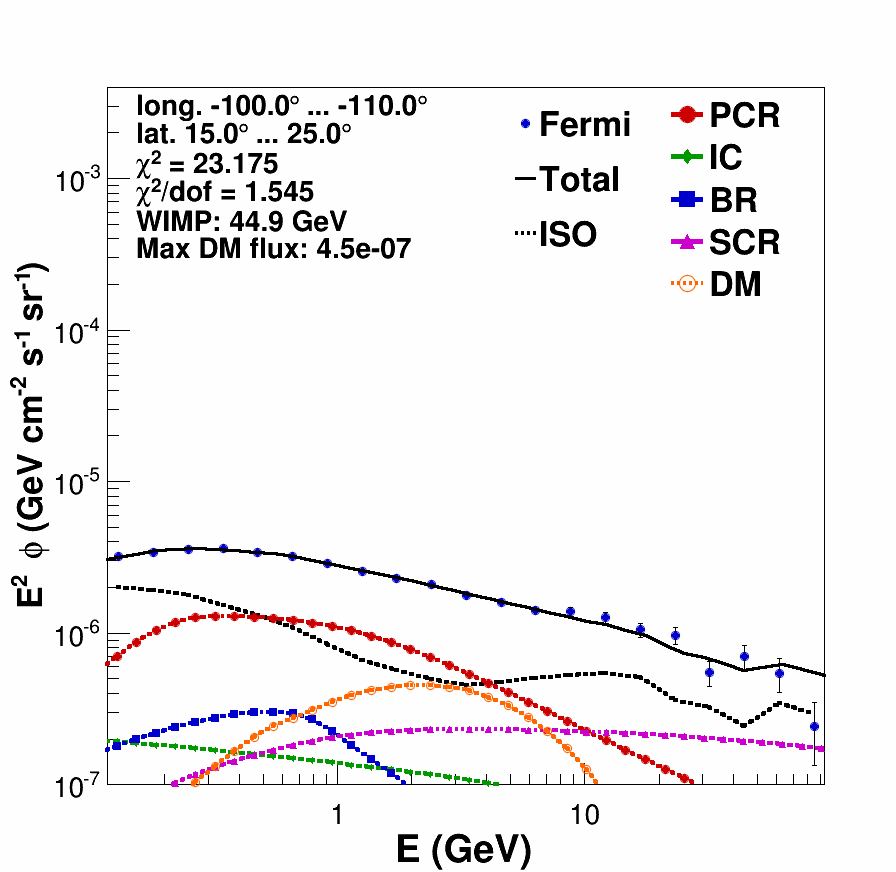}
\includegraphics[width=0.16\textwidth,height=0.16\textwidth,clip]{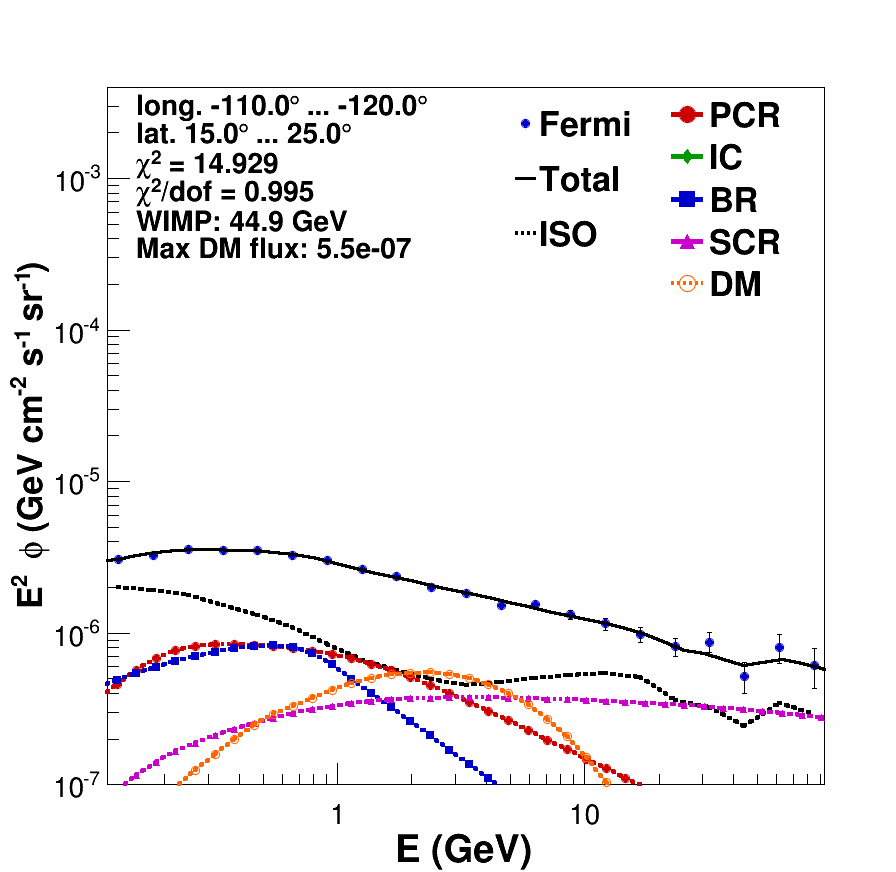}
\includegraphics[width=0.16\textwidth,height=0.16\textwidth,clip]{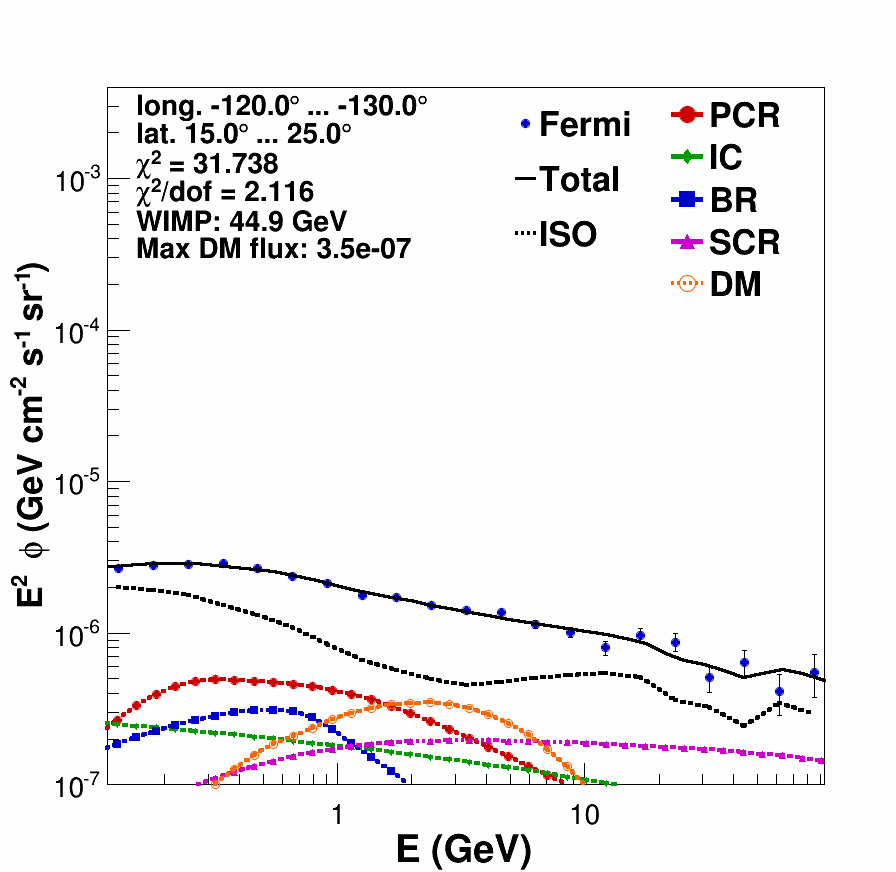}
\includegraphics[width=0.16\textwidth,height=0.16\textwidth,clip]{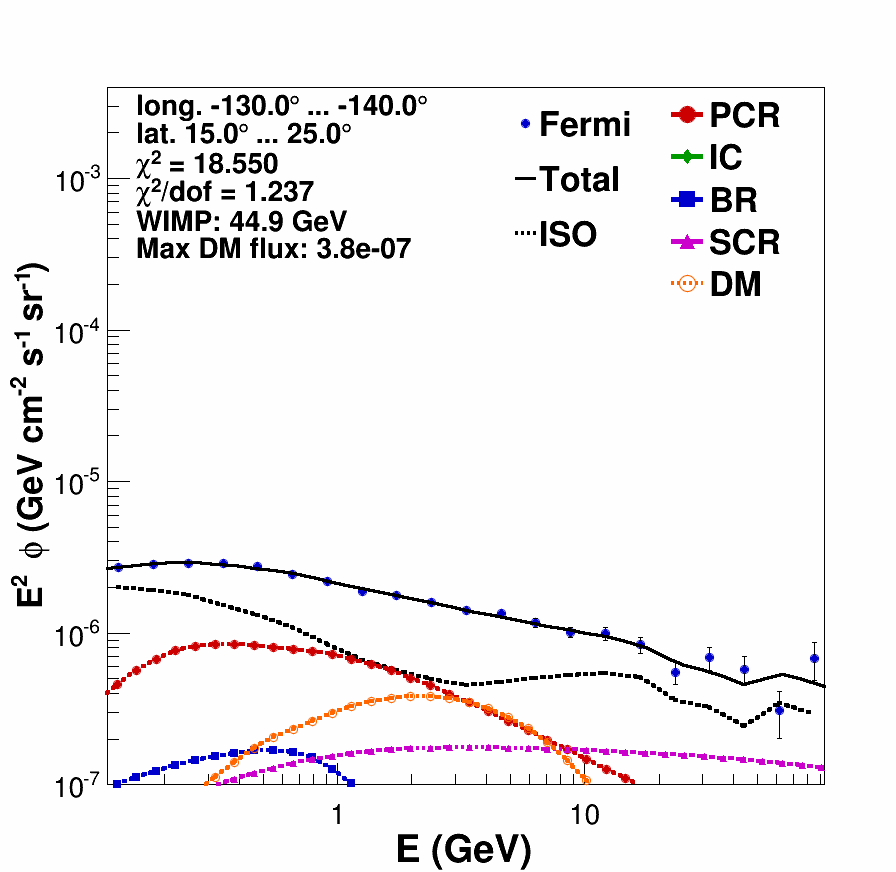}
\includegraphics[width=0.16\textwidth,height=0.16\textwidth,clip]{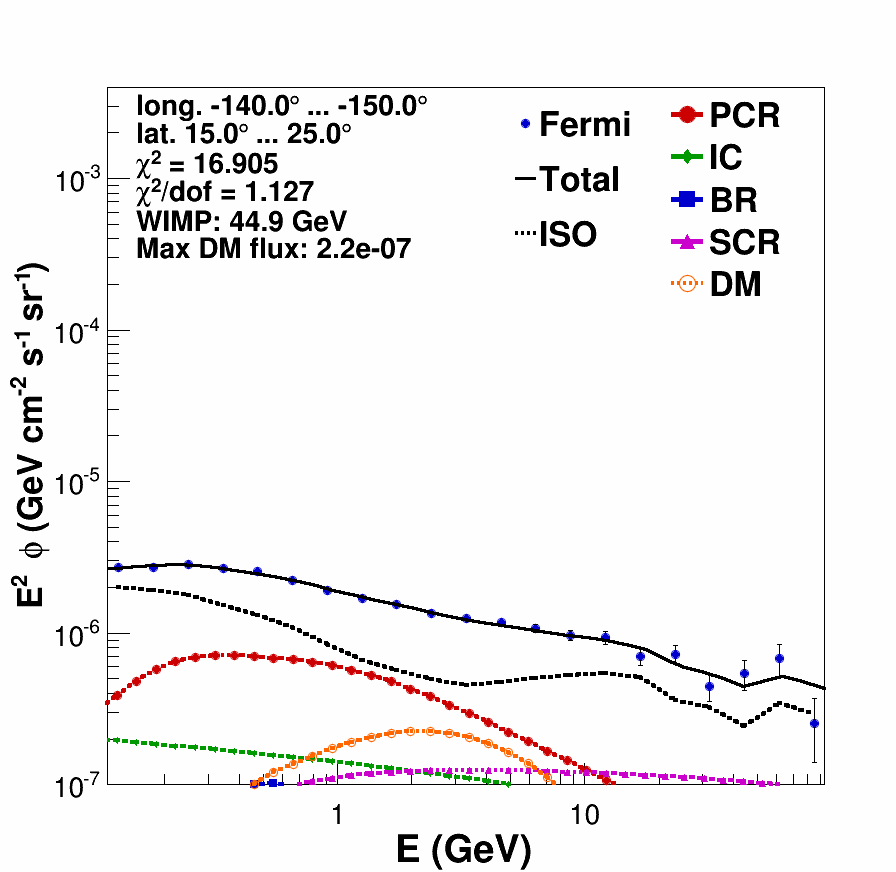}
\includegraphics[width=0.16\textwidth,height=0.16\textwidth,clip]{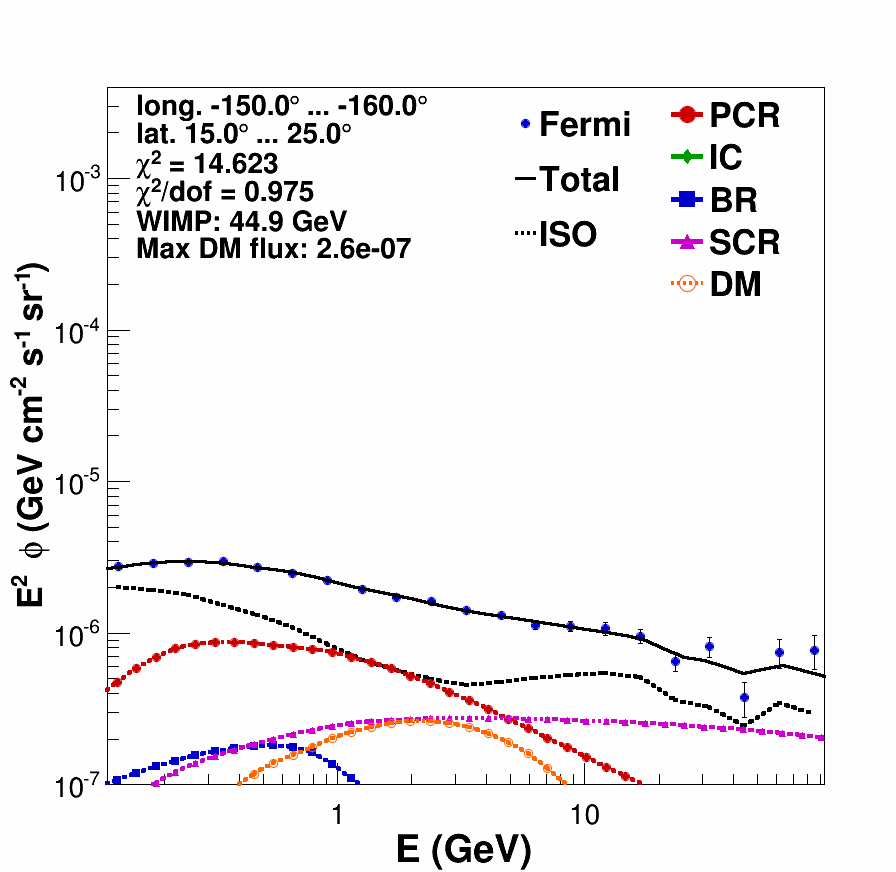}
\includegraphics[width=0.16\textwidth,height=0.16\textwidth,clip]{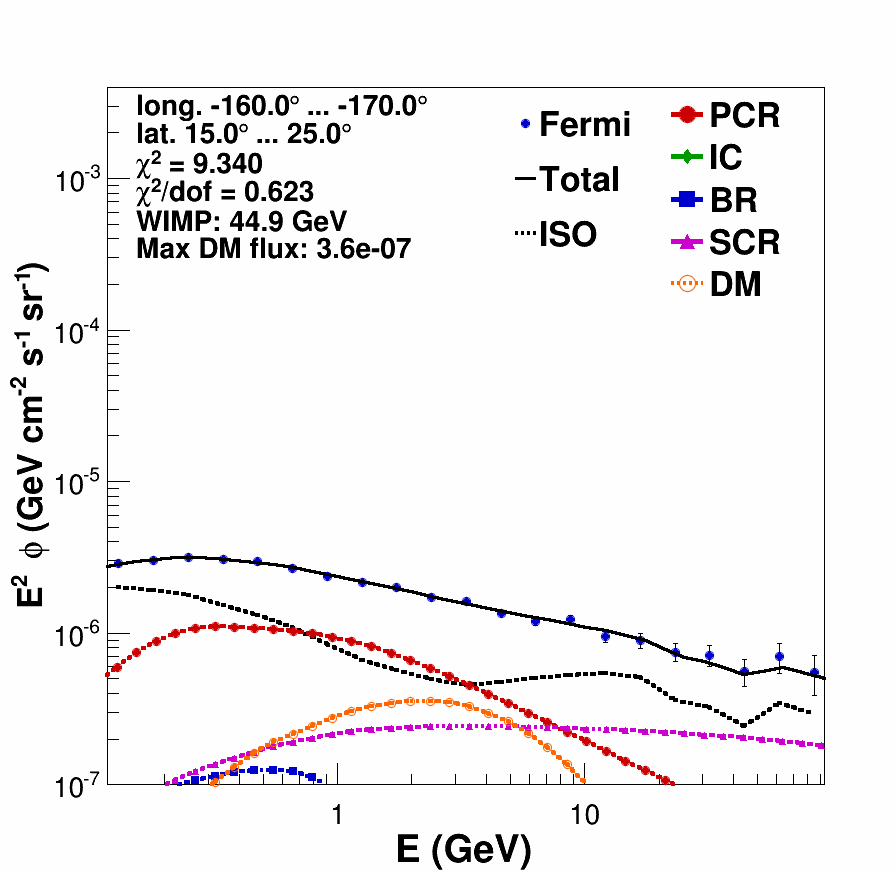}
\includegraphics[width=0.16\textwidth,height=0.16\textwidth,clip]{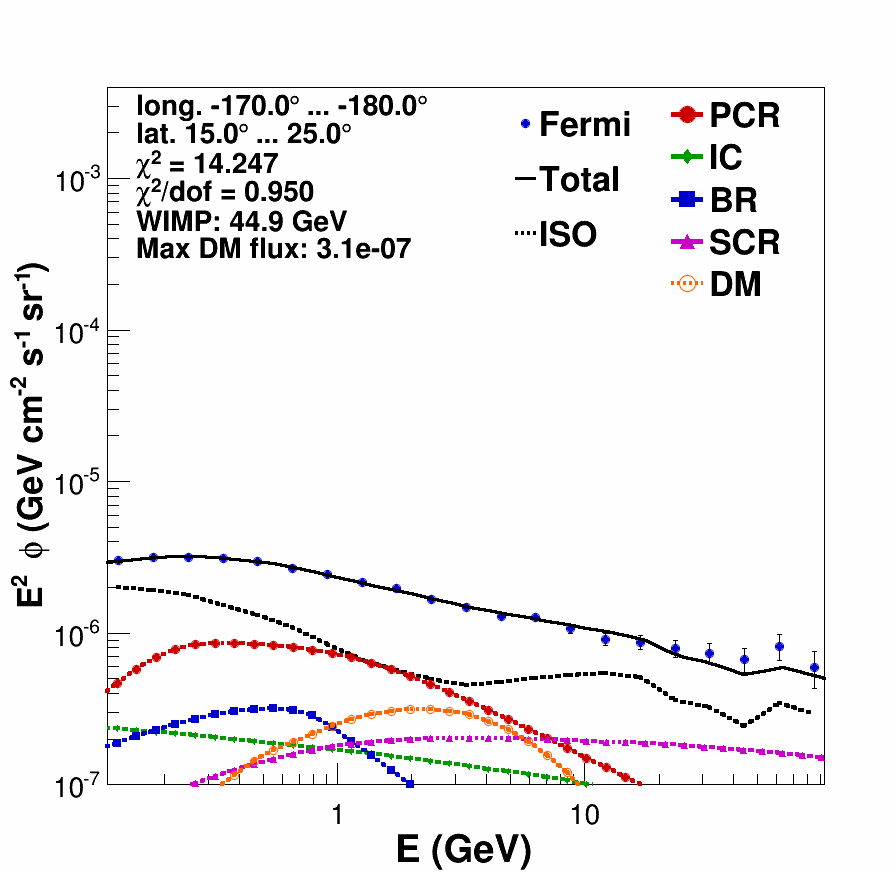}%%%%%%r6
\caption[]{Template fits for latitudes  with $15.0^\circ<b<25.0^\circ$ and longitudes decreasing from 180$^\circ$ to -180$^\circ$. \label{F37}
}
\end{figure}
\begin{figure}
\centering
\includegraphics[width=0.16\textwidth,height=0.16\textwidth,clip]{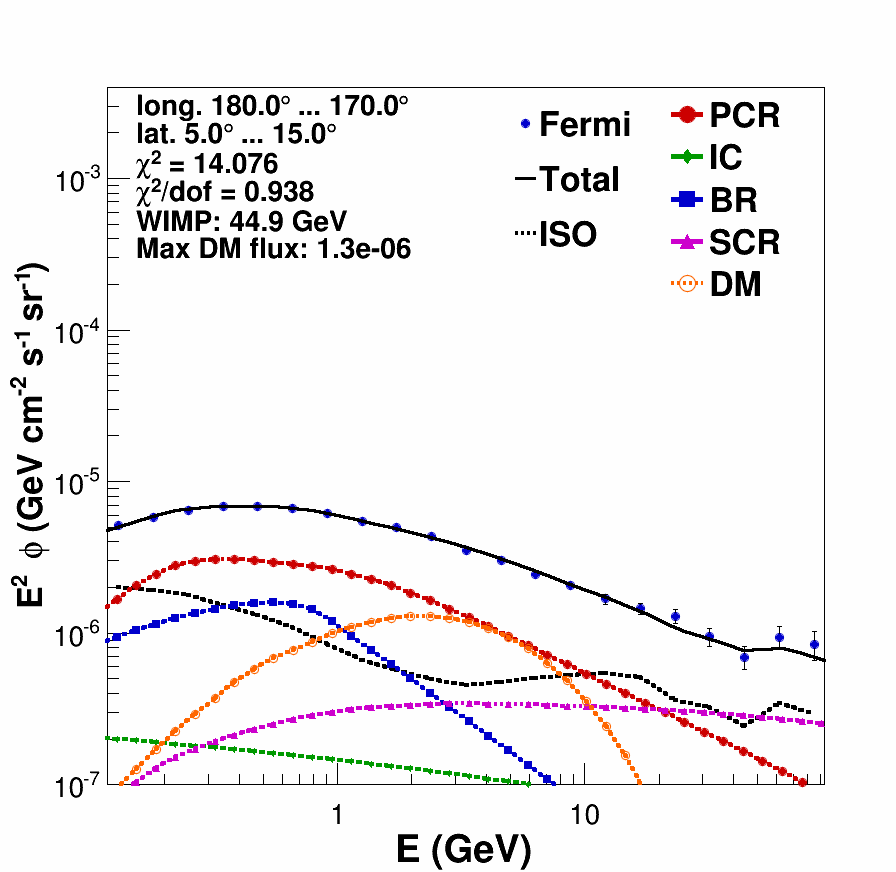}
\includegraphics[width=0.16\textwidth,height=0.16\textwidth,clip]{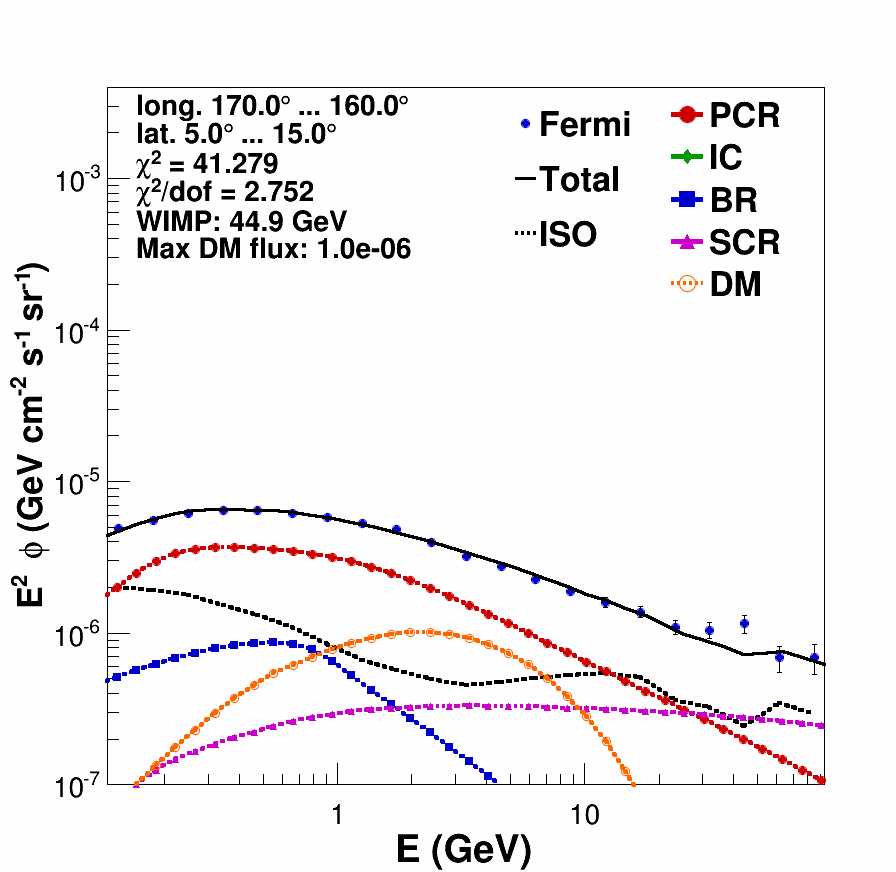}
\includegraphics[width=0.16\textwidth,height=0.16\textwidth,clip]{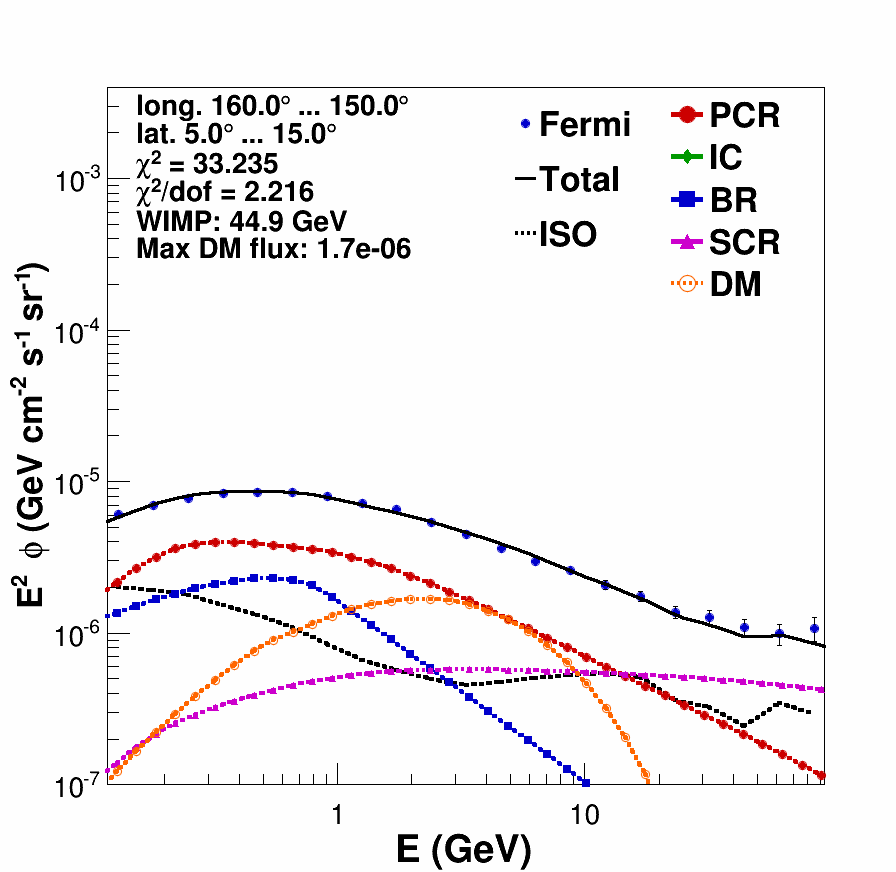}
\includegraphics[width=0.16\textwidth,height=0.16\textwidth,clip]{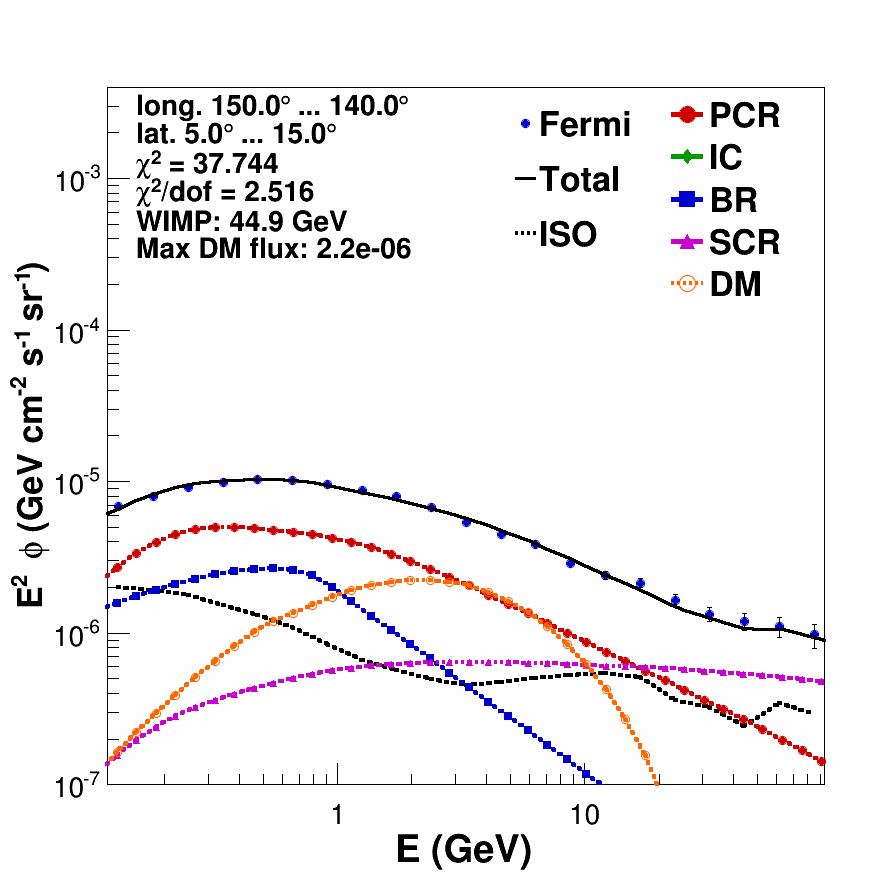}
\includegraphics[width=0.16\textwidth,height=0.16\textwidth,clip]{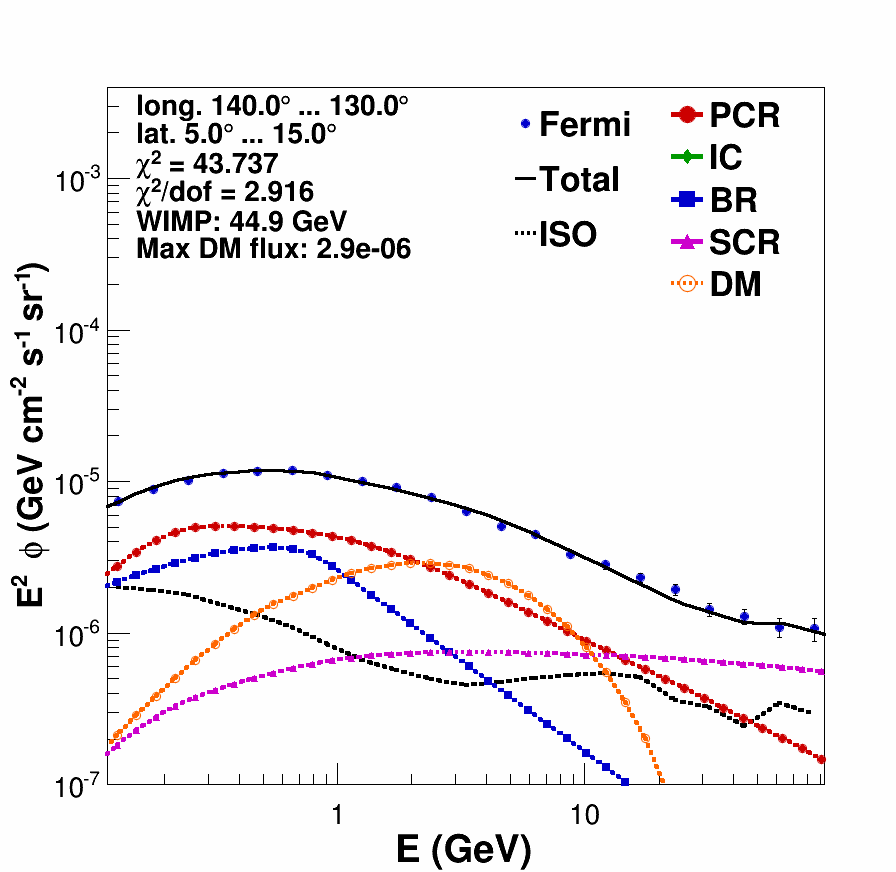}
\includegraphics[width=0.16\textwidth,height=0.16\textwidth,clip]{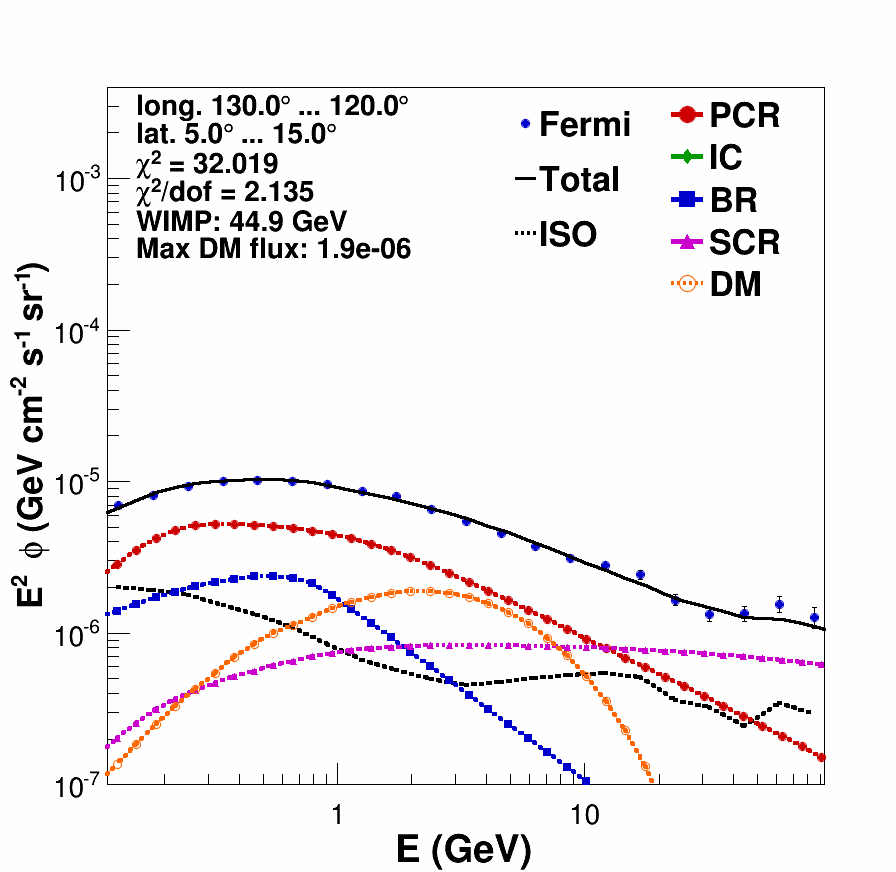}
\includegraphics[width=0.16\textwidth,height=0.16\textwidth,clip]{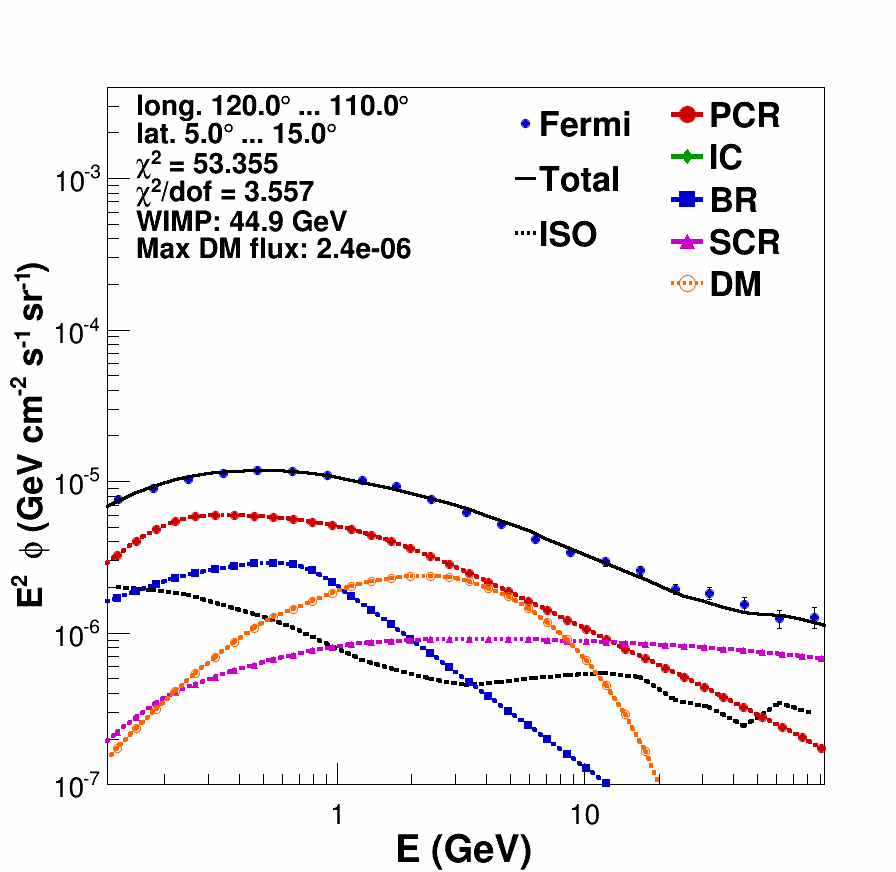}
\includegraphics[width=0.16\textwidth,height=0.16\textwidth,clip]{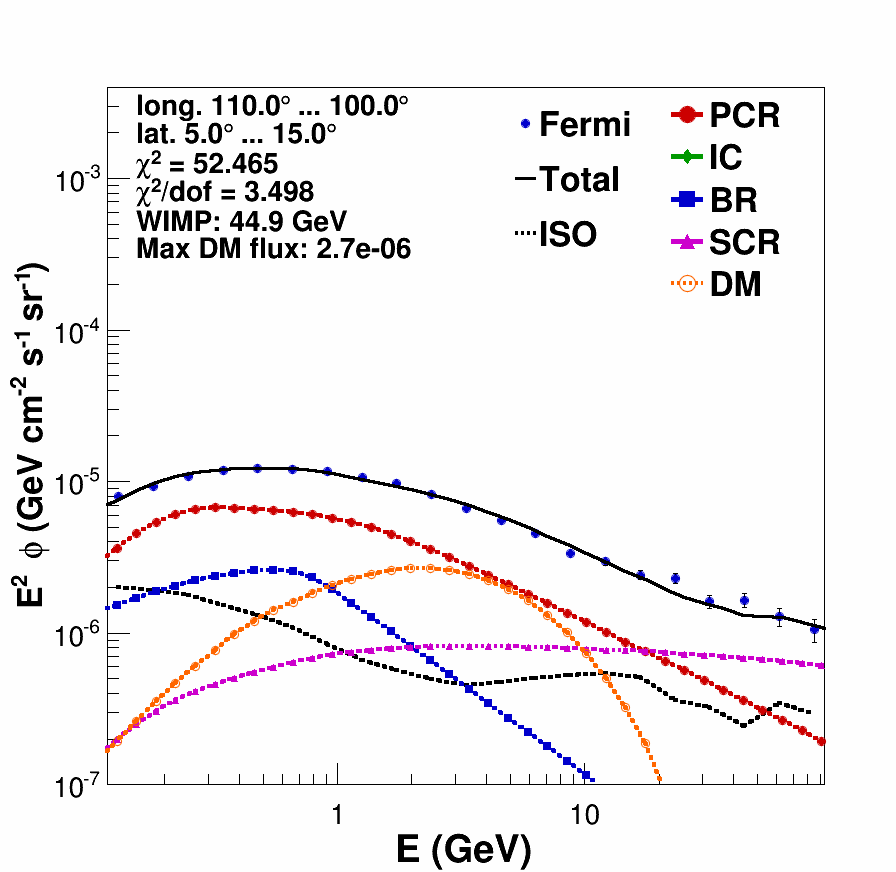}
\includegraphics[width=0.16\textwidth,height=0.16\textwidth,clip]{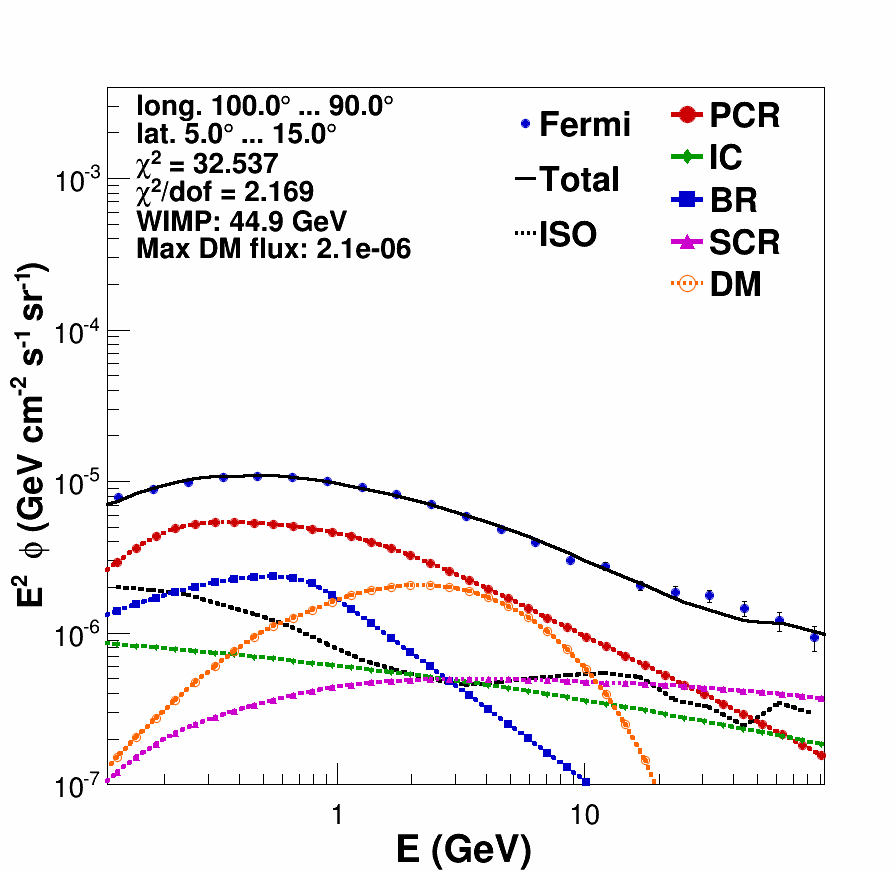}
\includegraphics[width=0.16\textwidth,height=0.16\textwidth,clip]{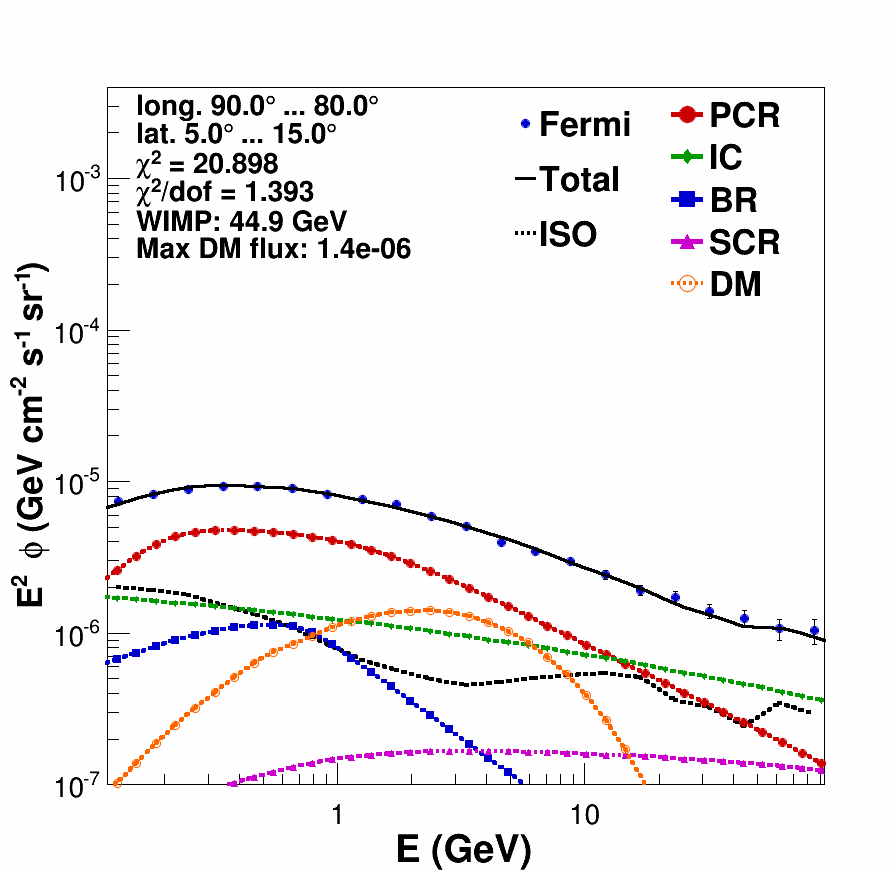}
\includegraphics[width=0.16\textwidth,height=0.16\textwidth,clip]{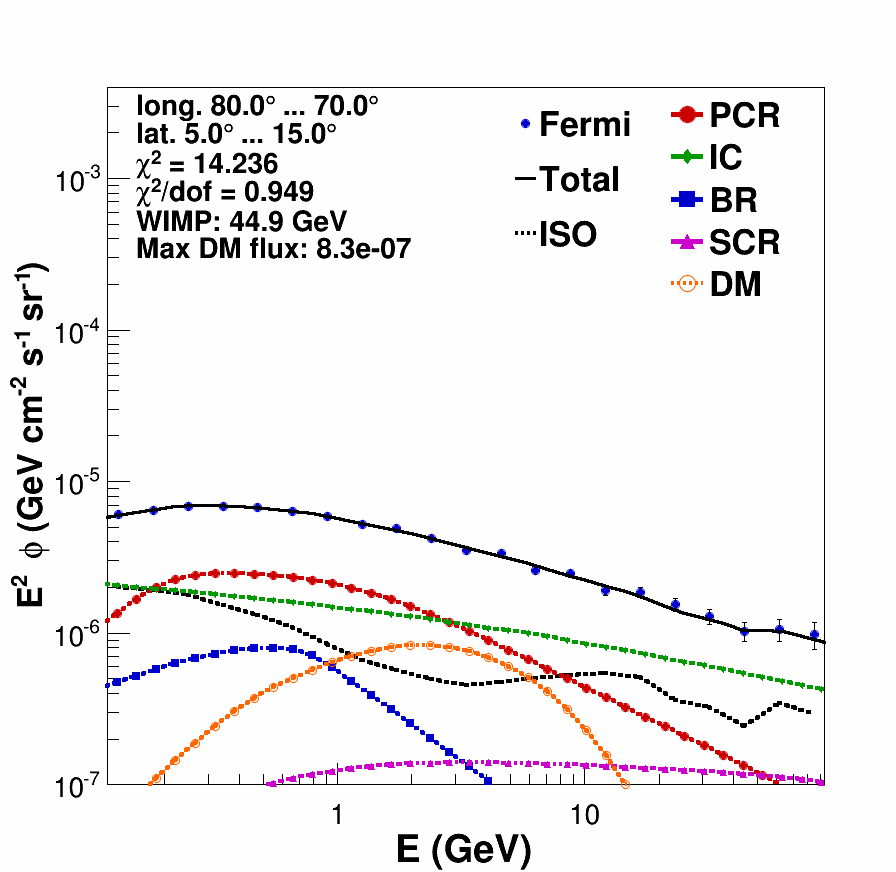}
\includegraphics[width=0.16\textwidth,height=0.16\textwidth,clip]{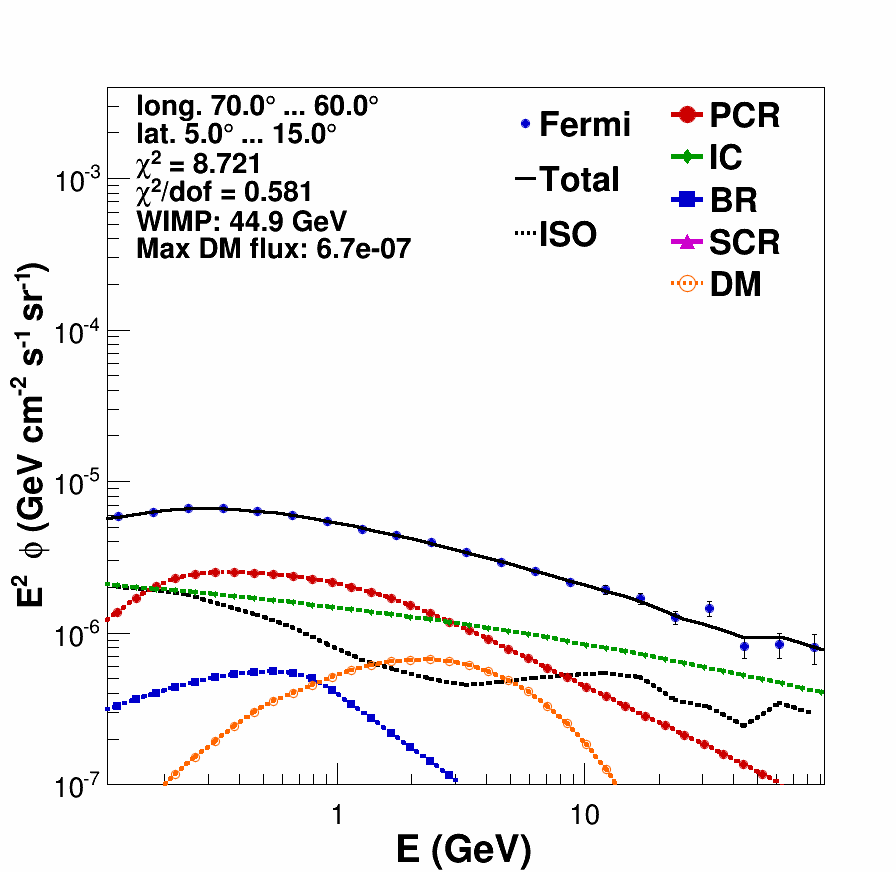}
\includegraphics[width=0.16\textwidth,height=0.16\textwidth,clip]{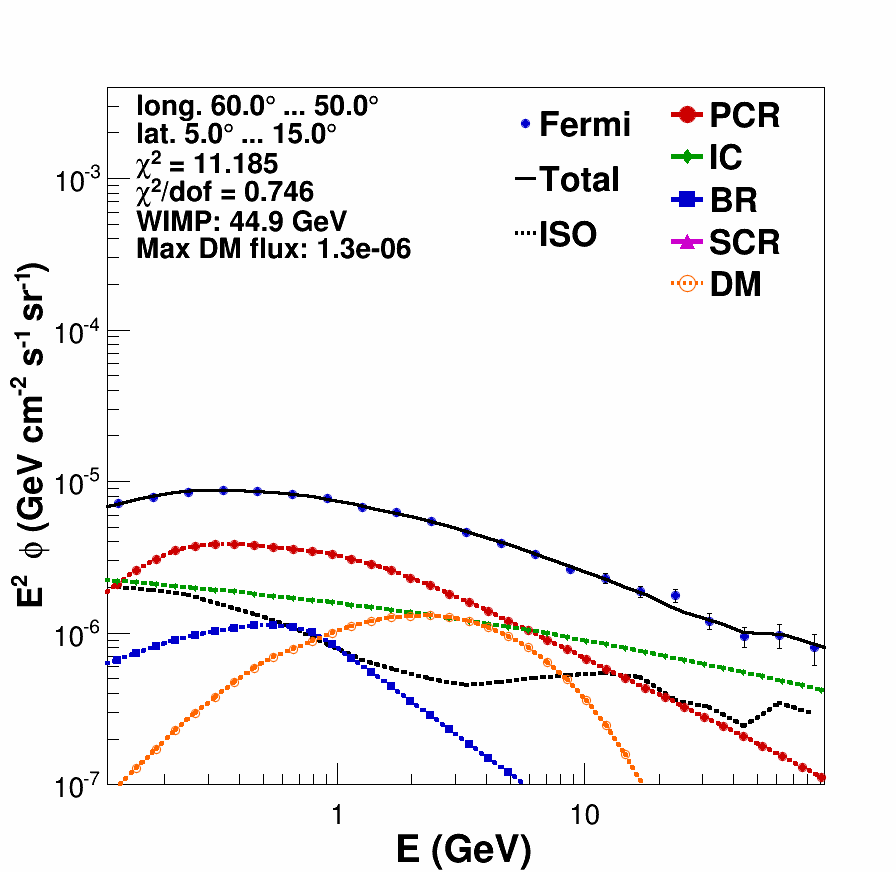}
\includegraphics[width=0.16\textwidth,height=0.16\textwidth,clip]{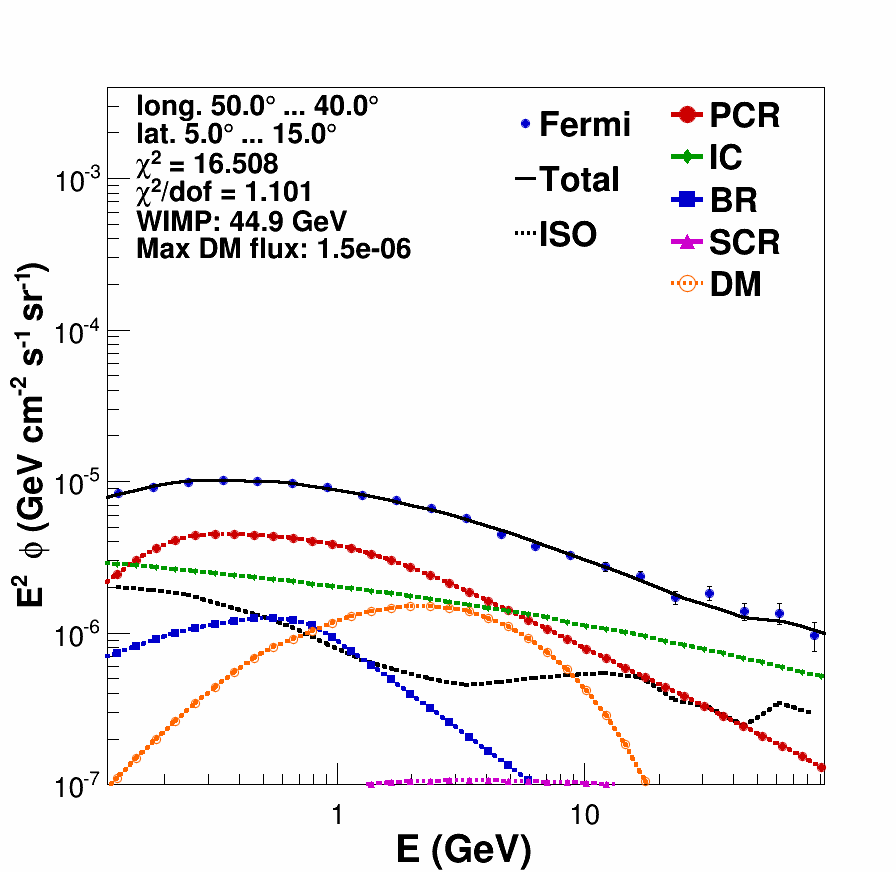}
\includegraphics[width=0.16\textwidth,height=0.16\textwidth,clip]{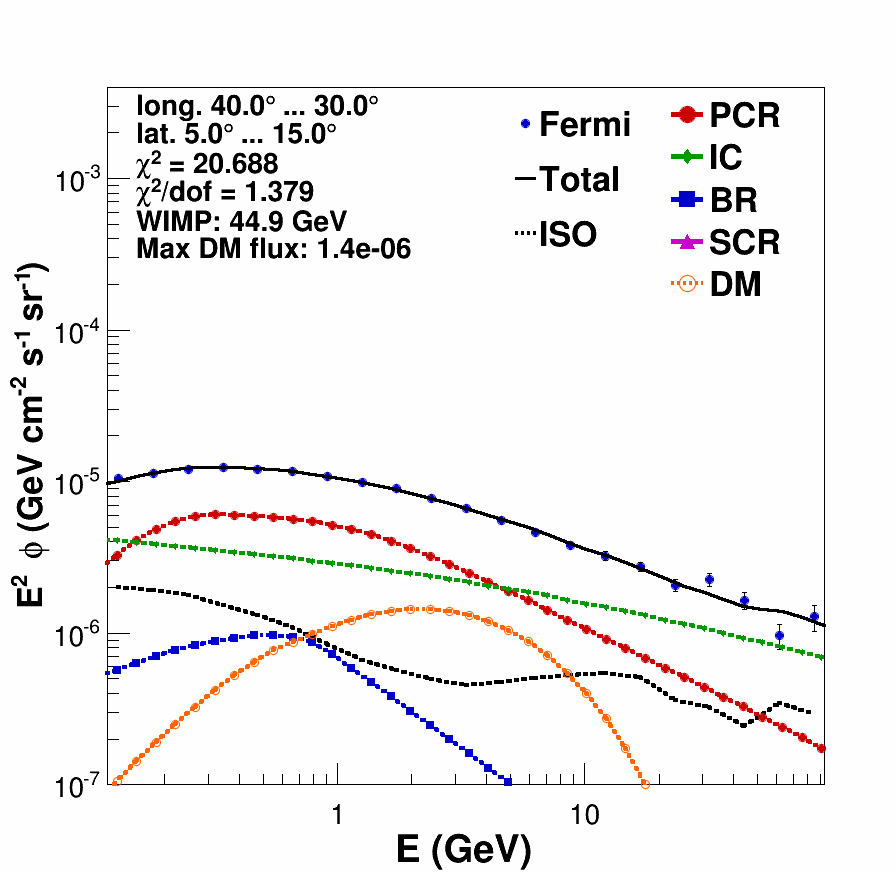}
\includegraphics[width=0.16\textwidth,height=0.16\textwidth,clip]{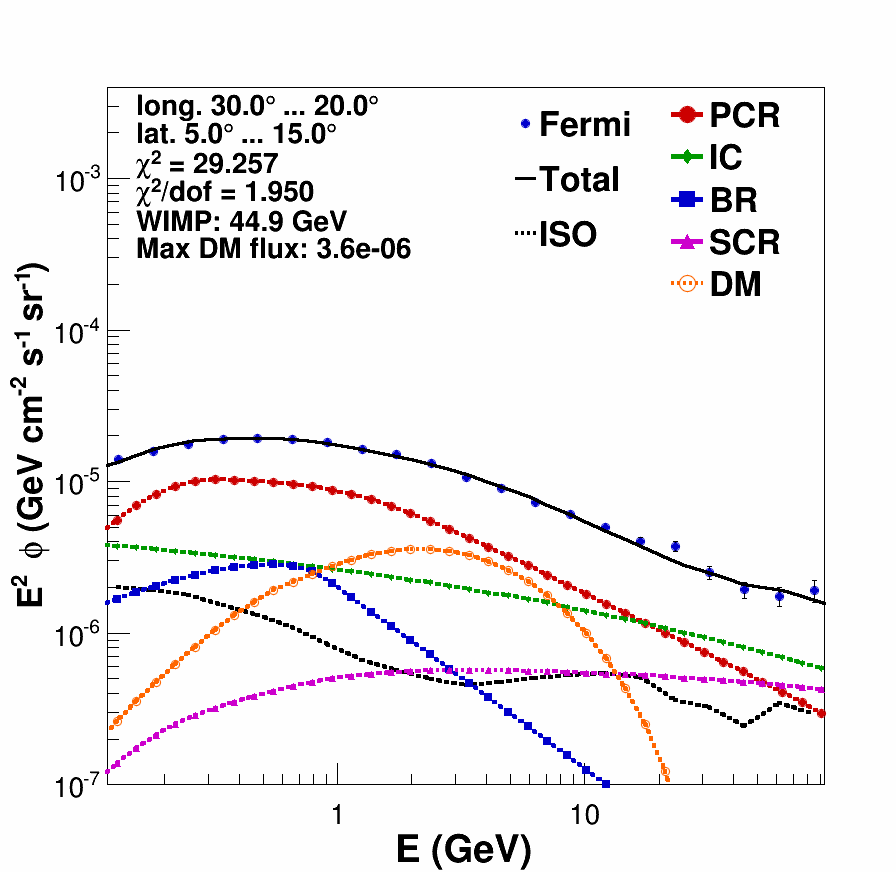}
\includegraphics[width=0.16\textwidth,height=0.16\textwidth,clip]{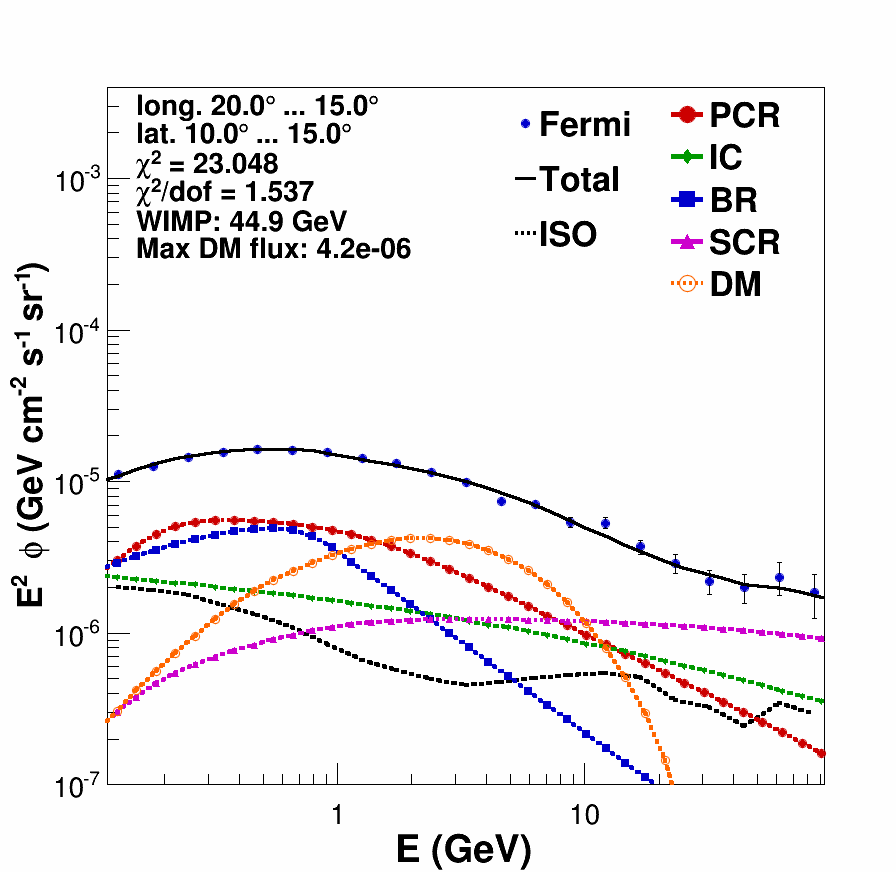}
\includegraphics[width=0.16\textwidth,height=0.16\textwidth,clip]{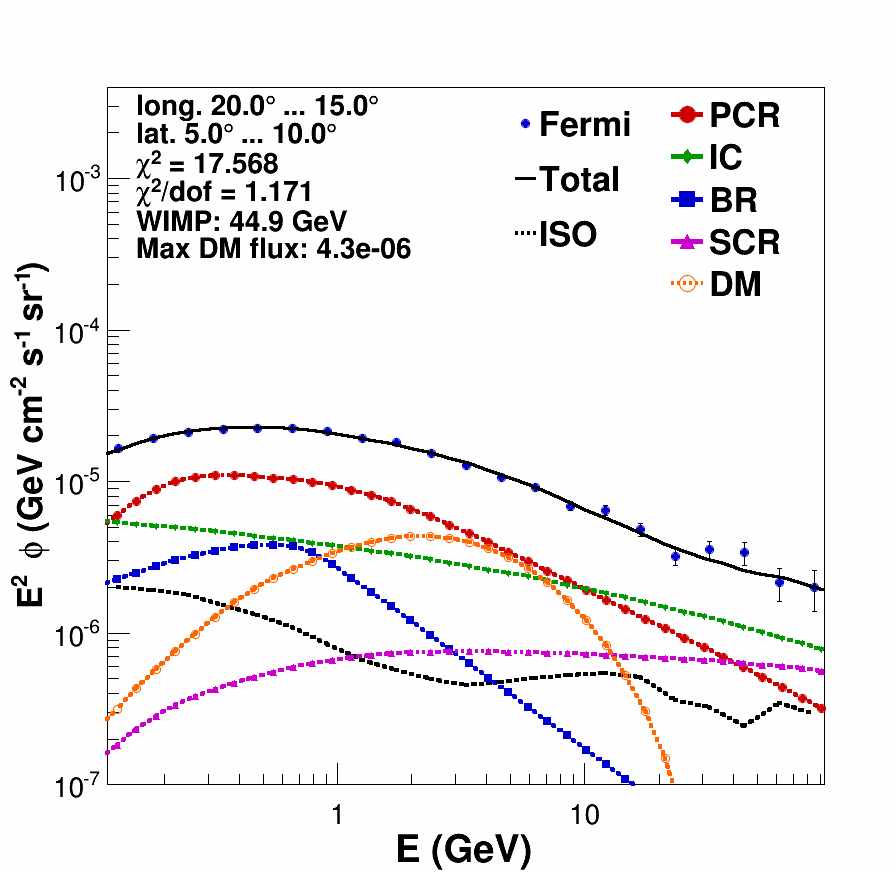}
\includegraphics[width=0.16\textwidth,height=0.16\textwidth,clip]{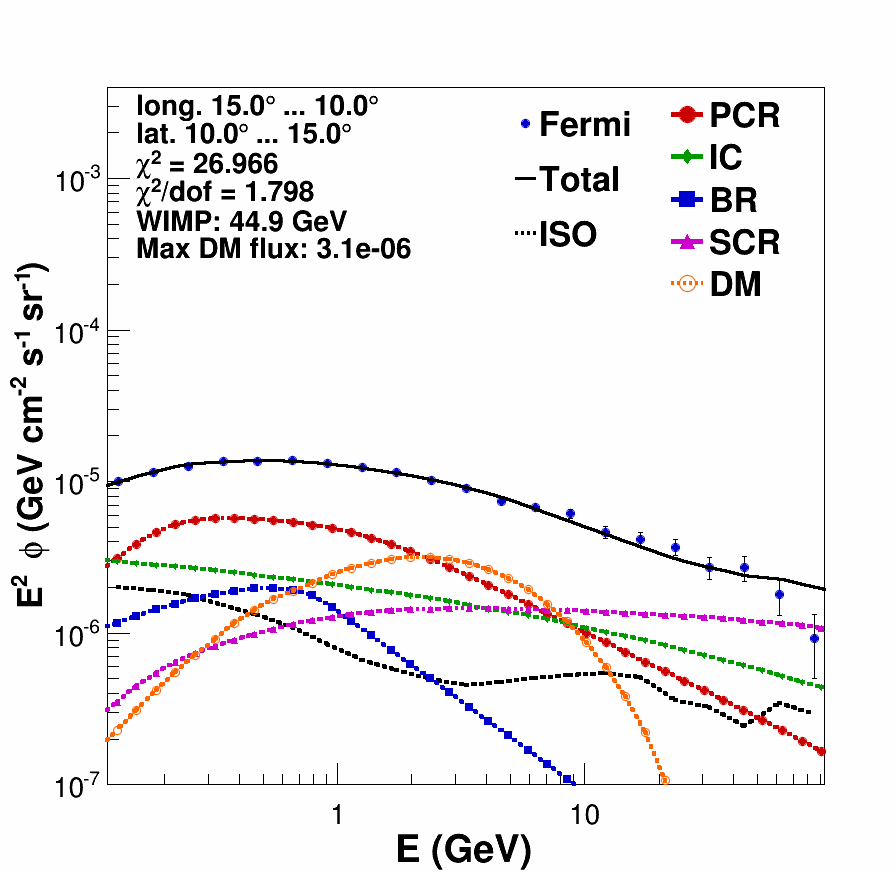}
\includegraphics[width=0.16\textwidth,height=0.16\textwidth,clip]{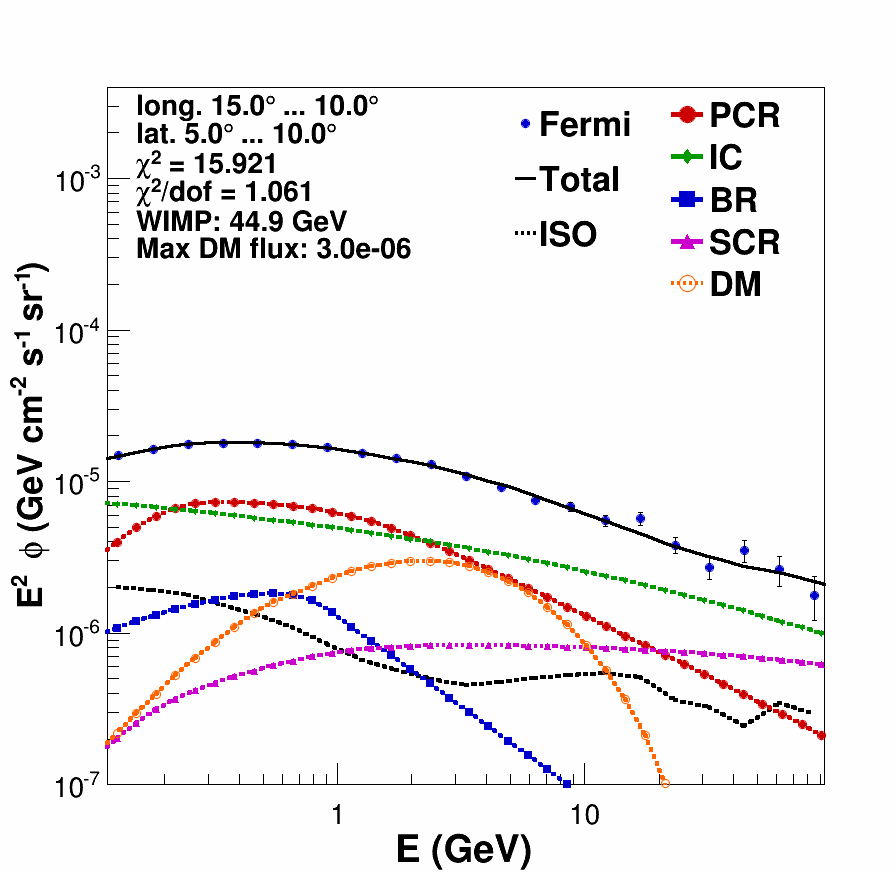}
\includegraphics[width=0.16\textwidth,height=0.16\textwidth,clip]{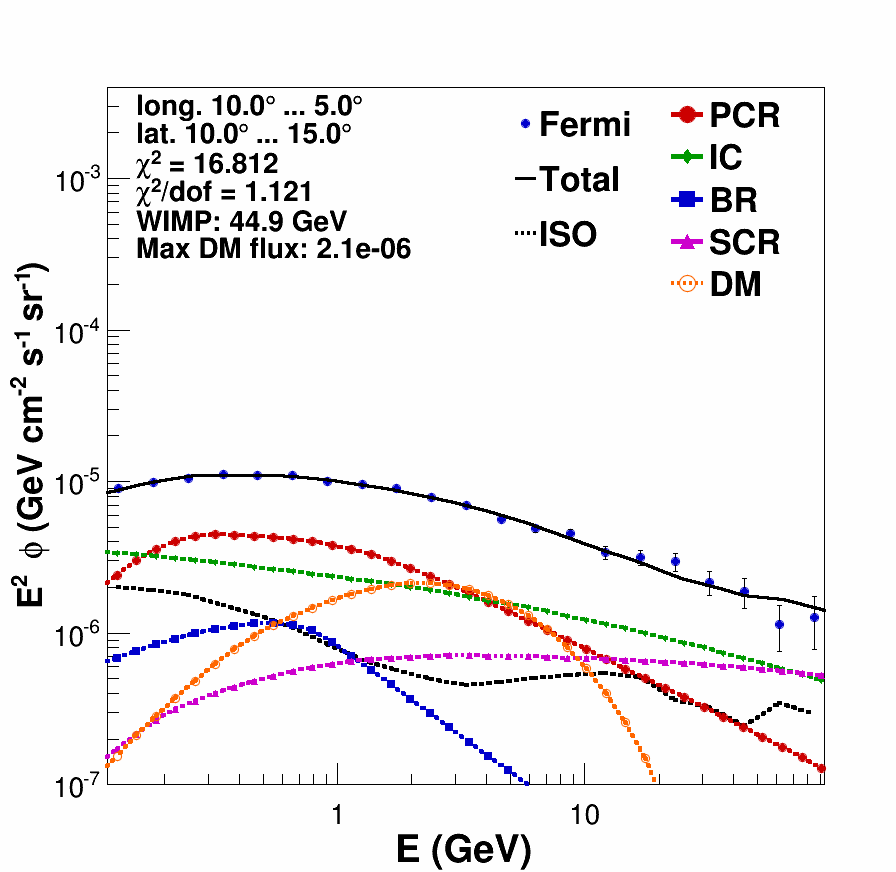}
\includegraphics[width=0.16\textwidth,height=0.16\textwidth,clip]{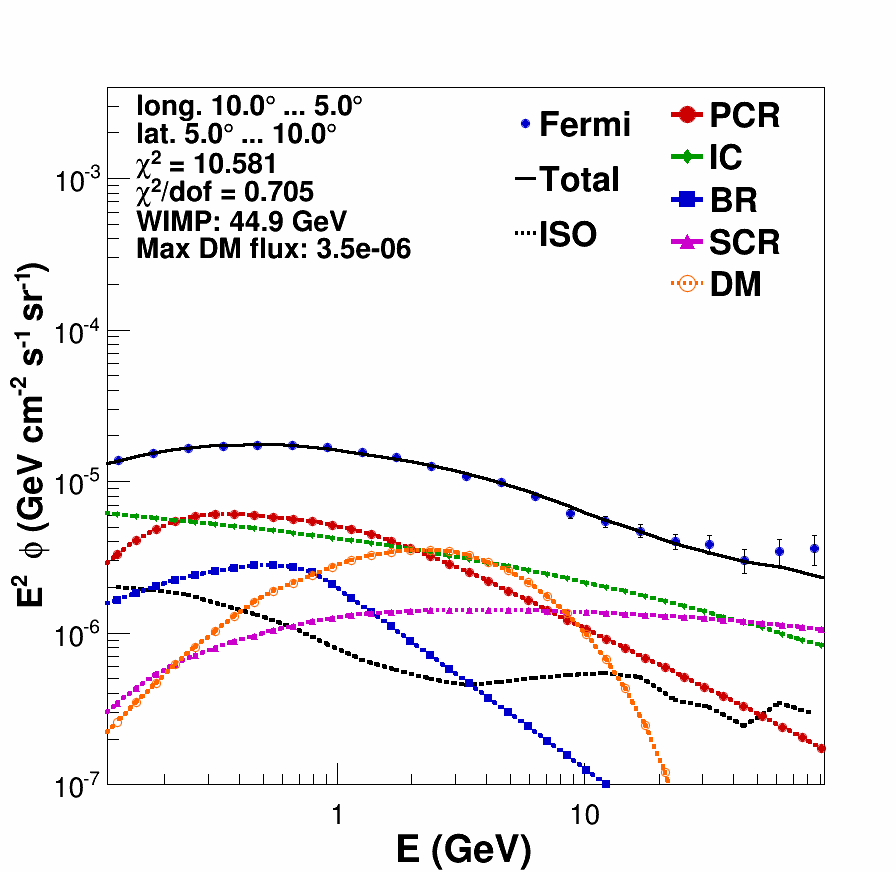}
\includegraphics[width=0.16\textwidth,height=0.16\textwidth,clip]{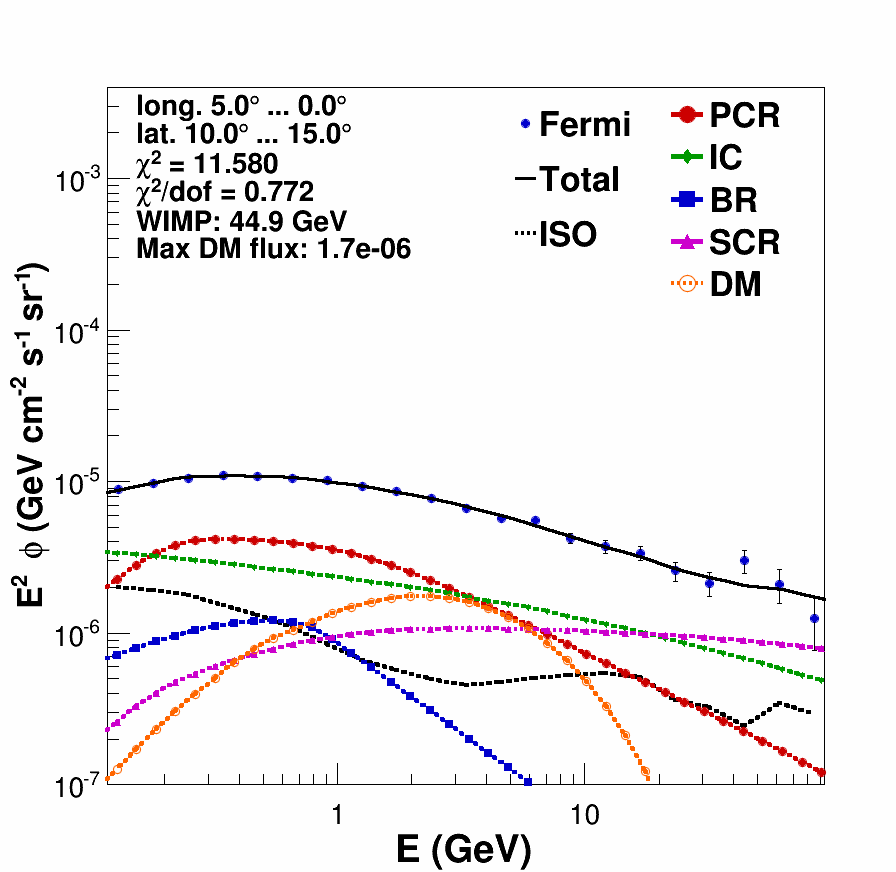}
\includegraphics[width=0.16\textwidth,height=0.16\textwidth,clip]{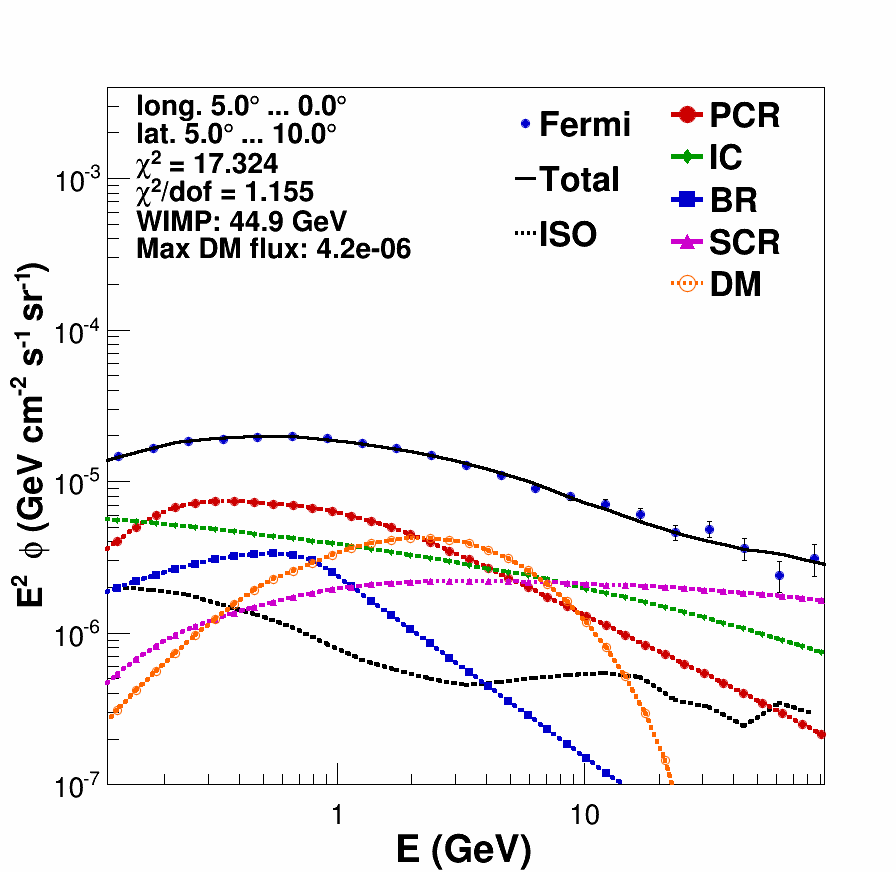}
\includegraphics[width=0.16\textwidth,height=0.16\textwidth,clip]{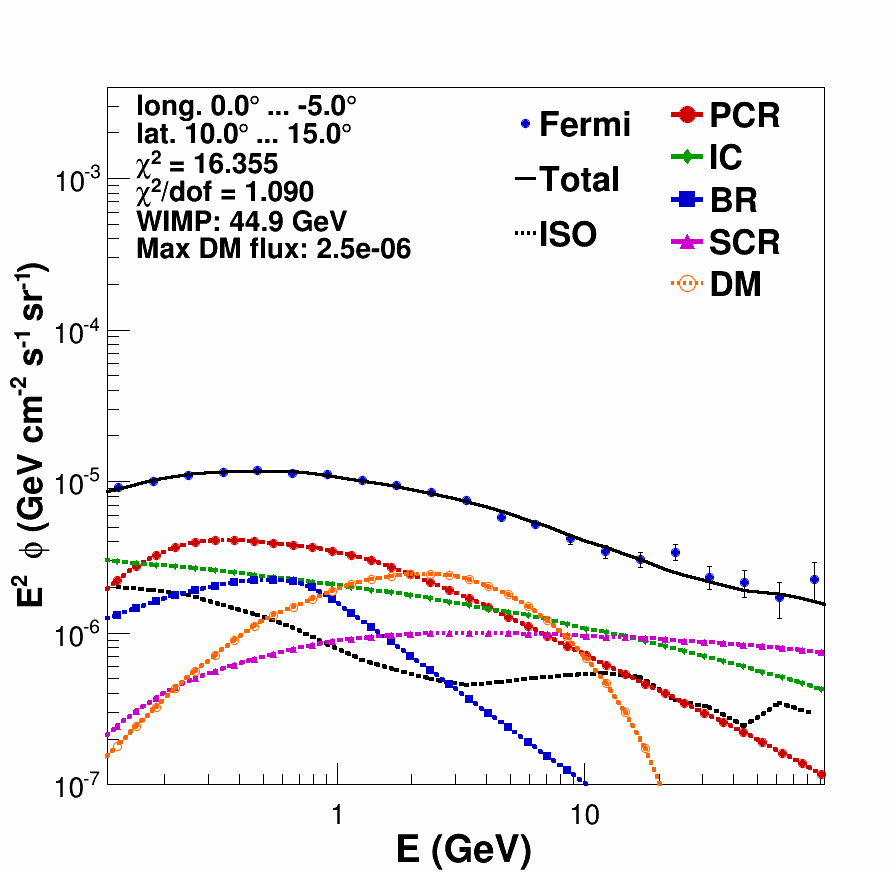}
\includegraphics[width=0.16\textwidth,height=0.16\textwidth,clip]{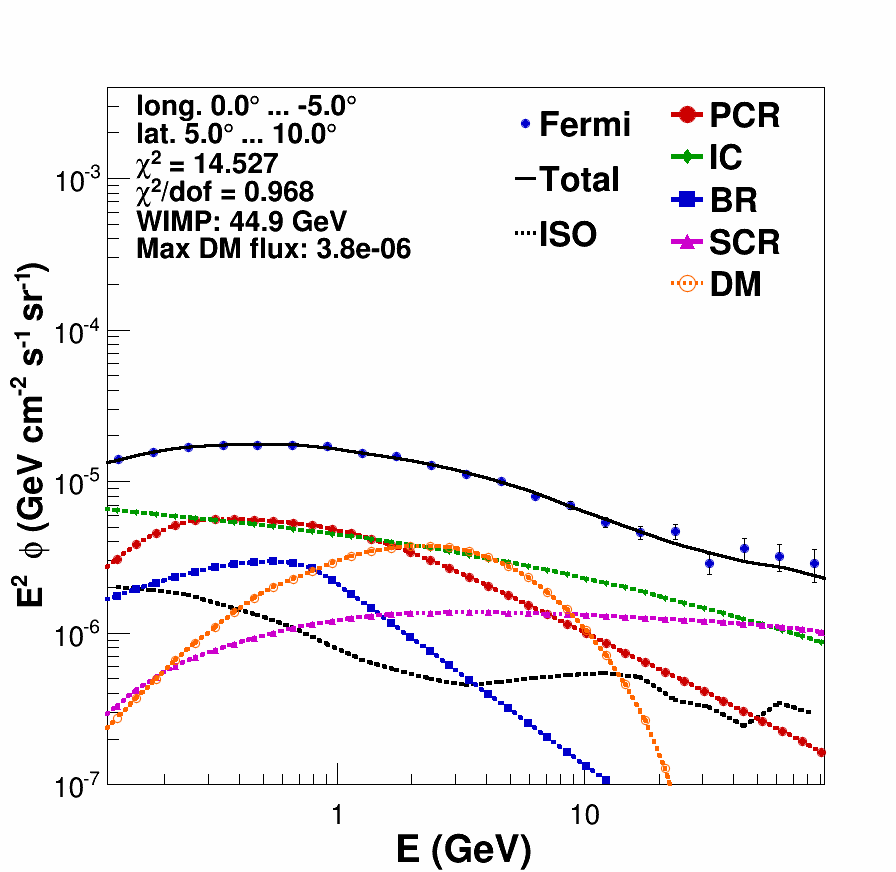}
\includegraphics[width=0.16\textwidth,height=0.16\textwidth,clip]{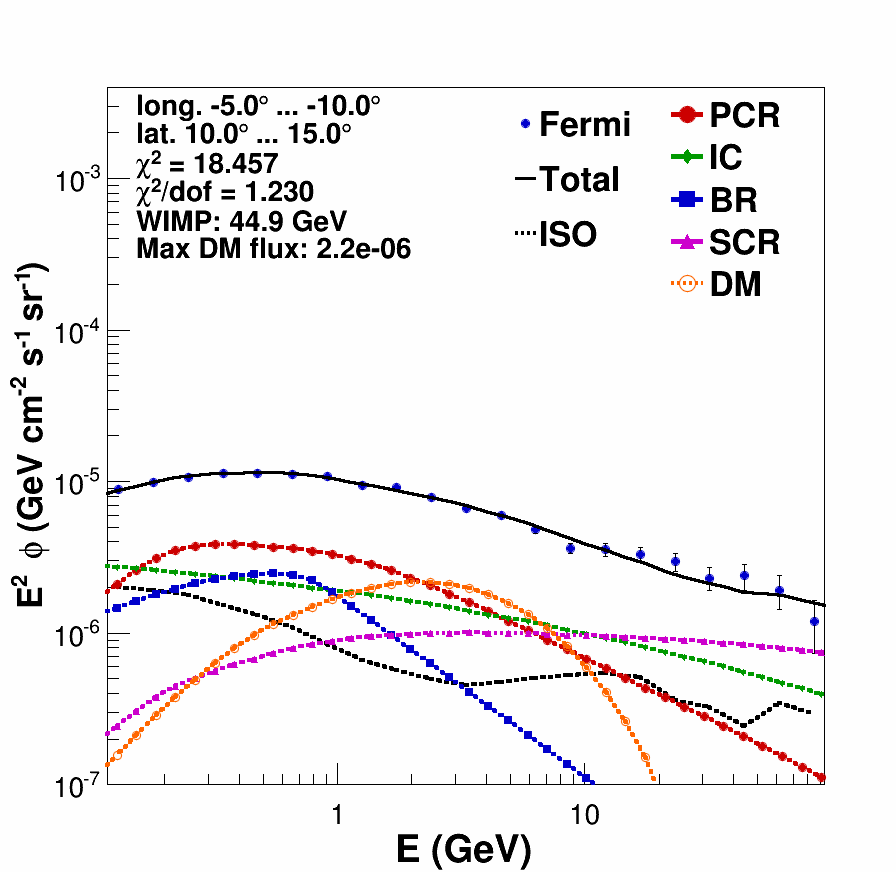}
\includegraphics[width=0.16\textwidth,height=0.16\textwidth,clip]{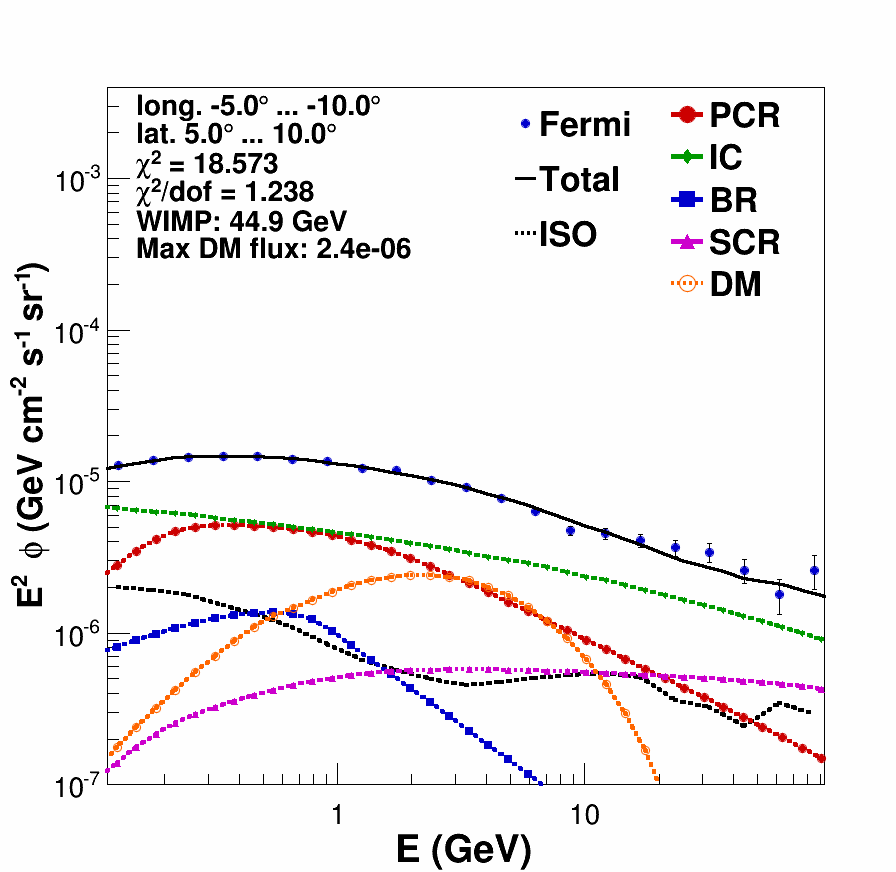}
\includegraphics[width=0.16\textwidth,height=0.16\textwidth,clip]{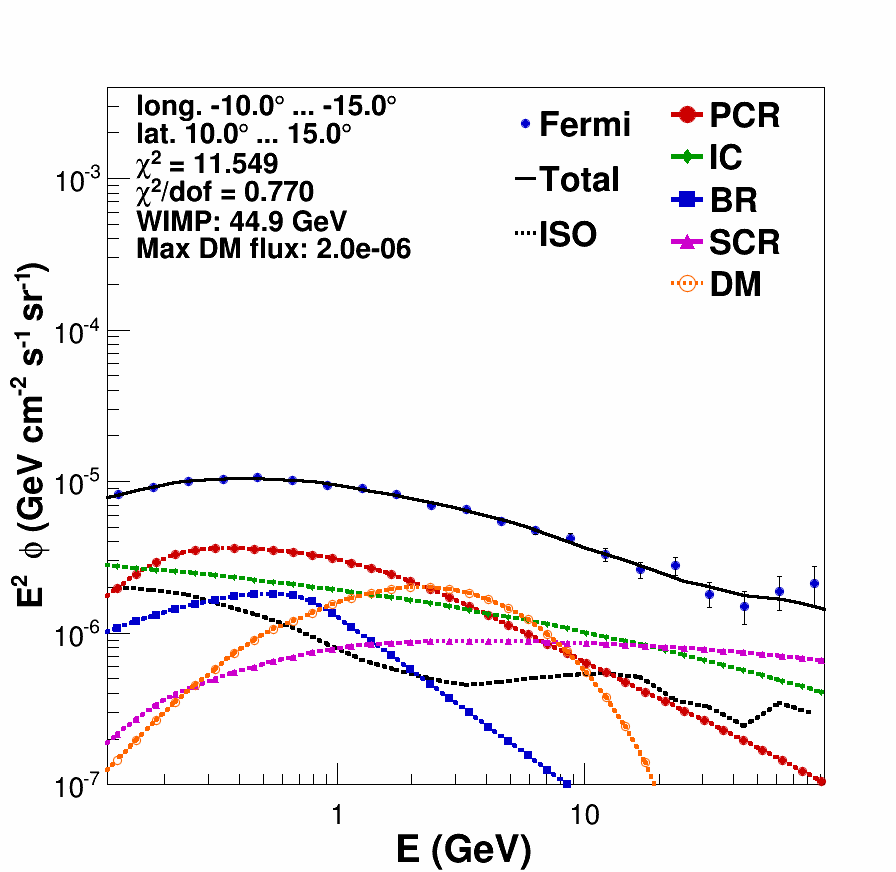}
\includegraphics[width=0.16\textwidth,height=0.16\textwidth,clip]{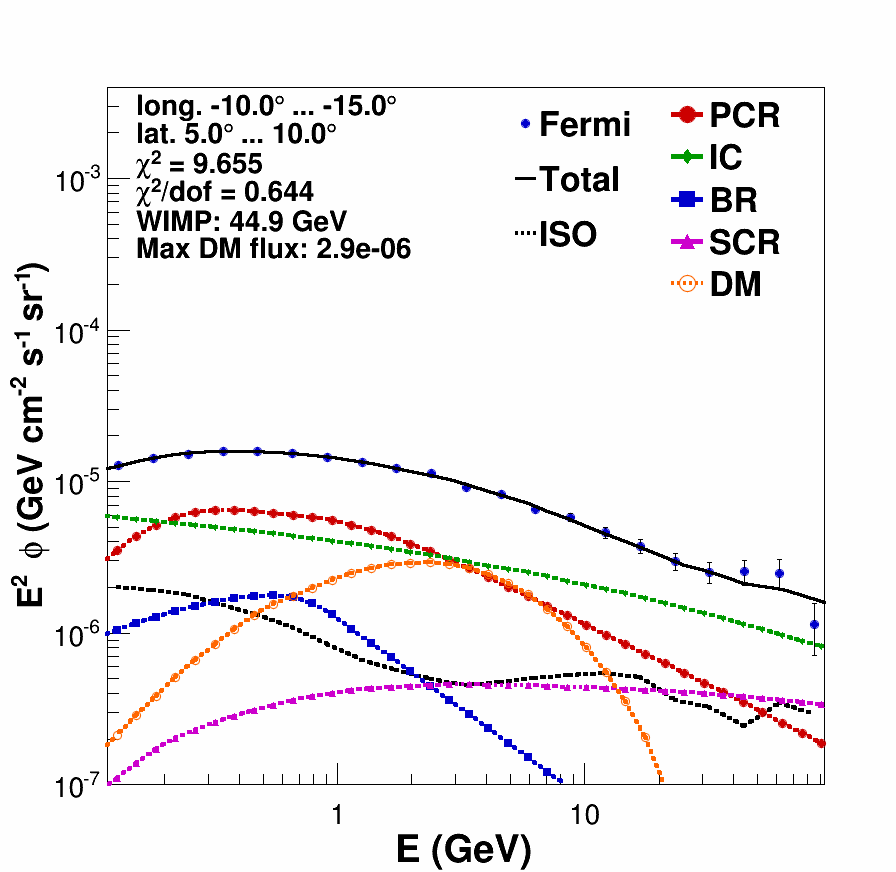}
\includegraphics[width=0.16\textwidth,height=0.16\textwidth,clip]{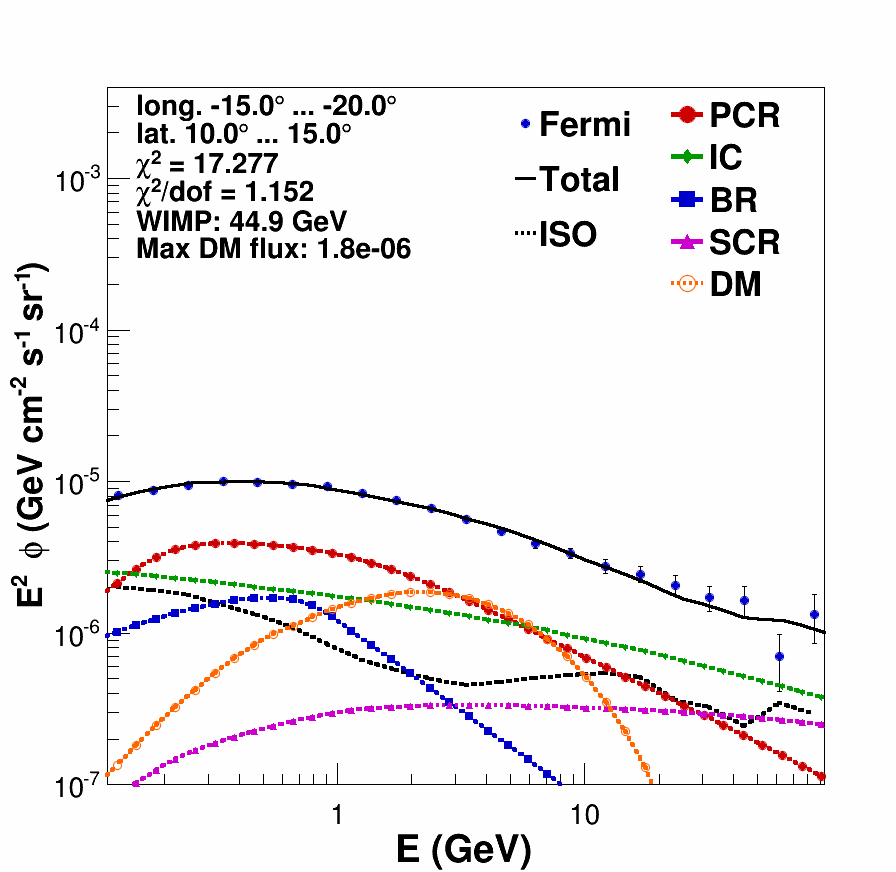}
\includegraphics[width=0.16\textwidth,height=0.16\textwidth,clip]{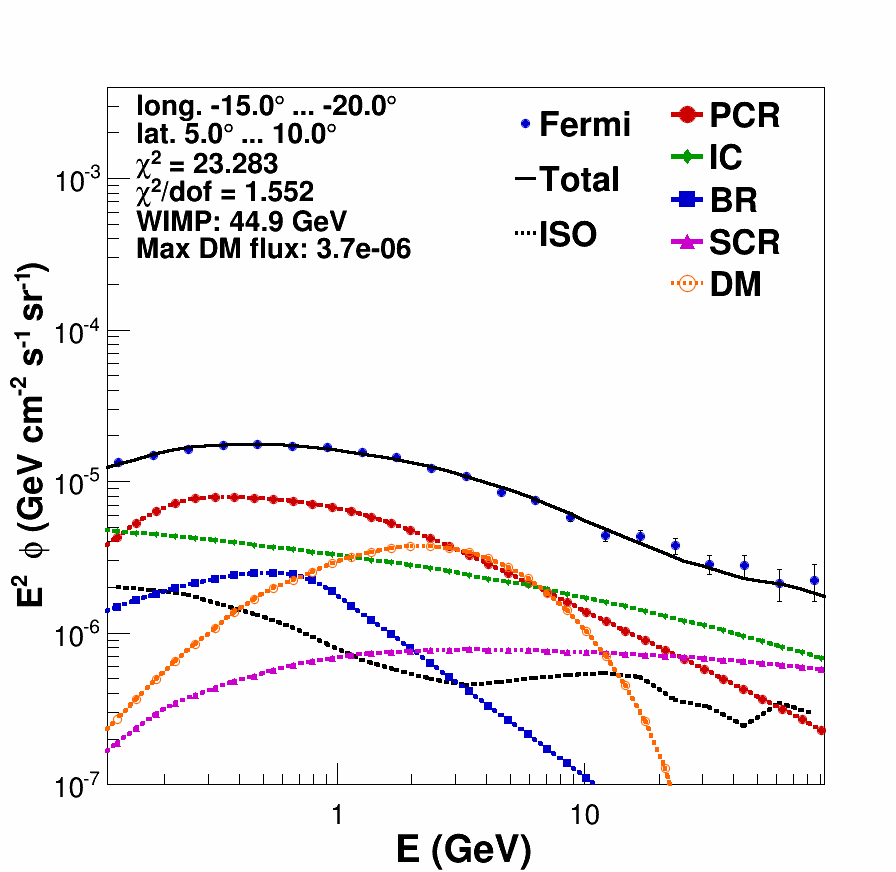}
\includegraphics[width=0.16\textwidth,height=0.16\textwidth,clip]{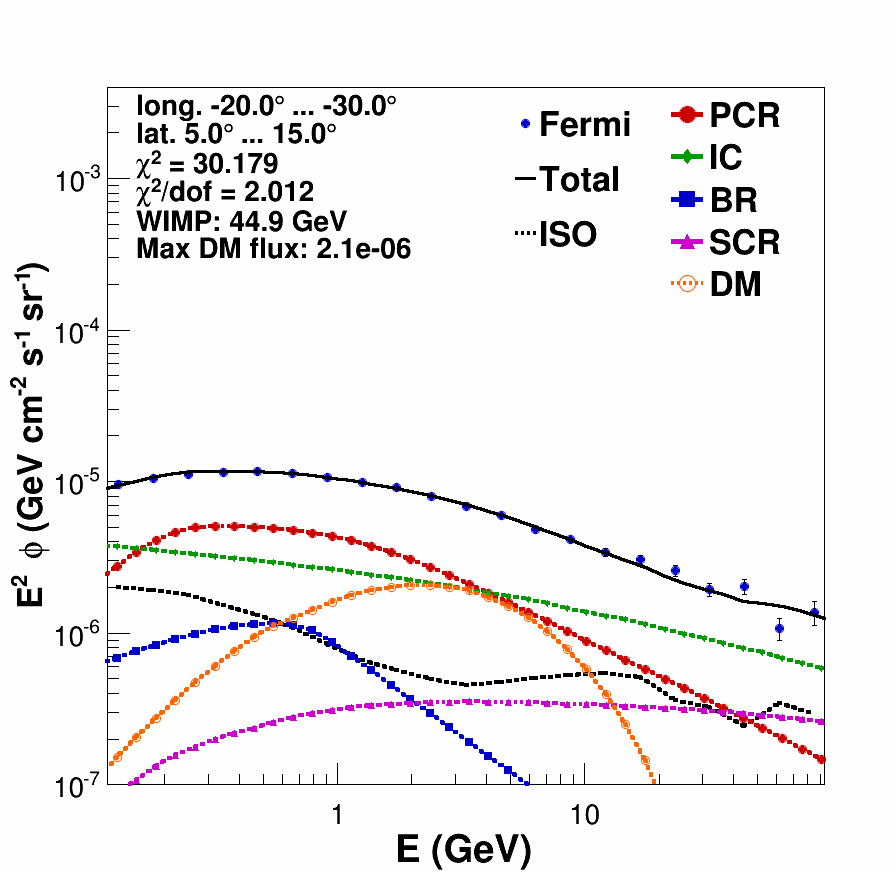}
\includegraphics[width=0.16\textwidth,height=0.16\textwidth,clip]{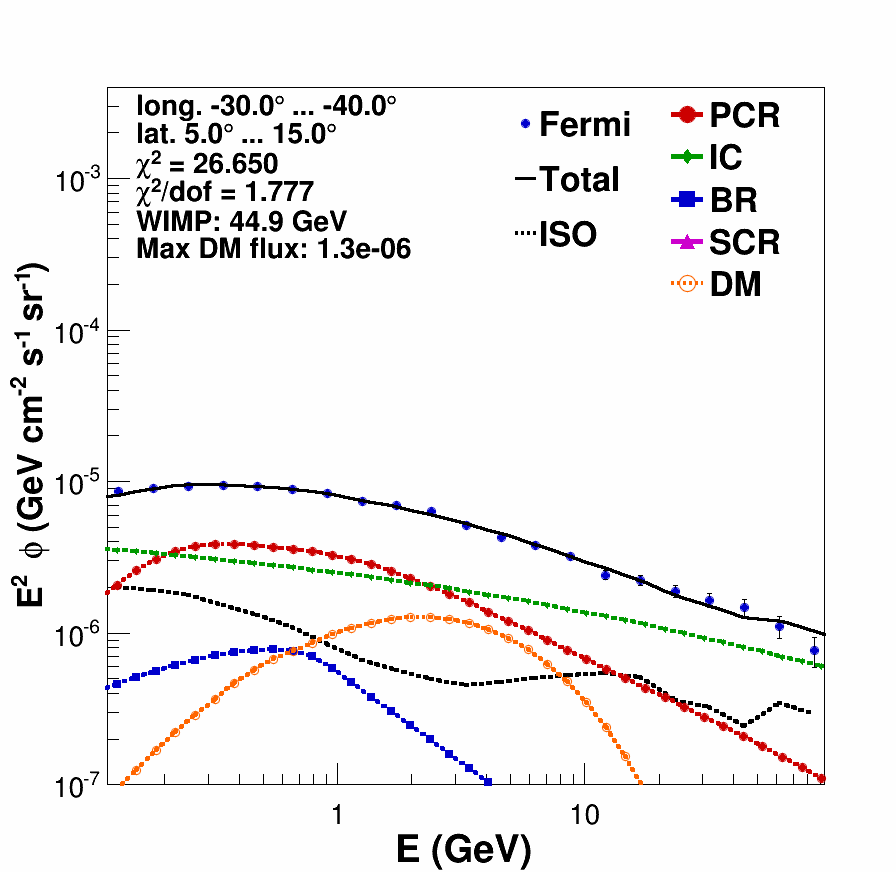}
\includegraphics[width=0.16\textwidth,height=0.16\textwidth,clip]{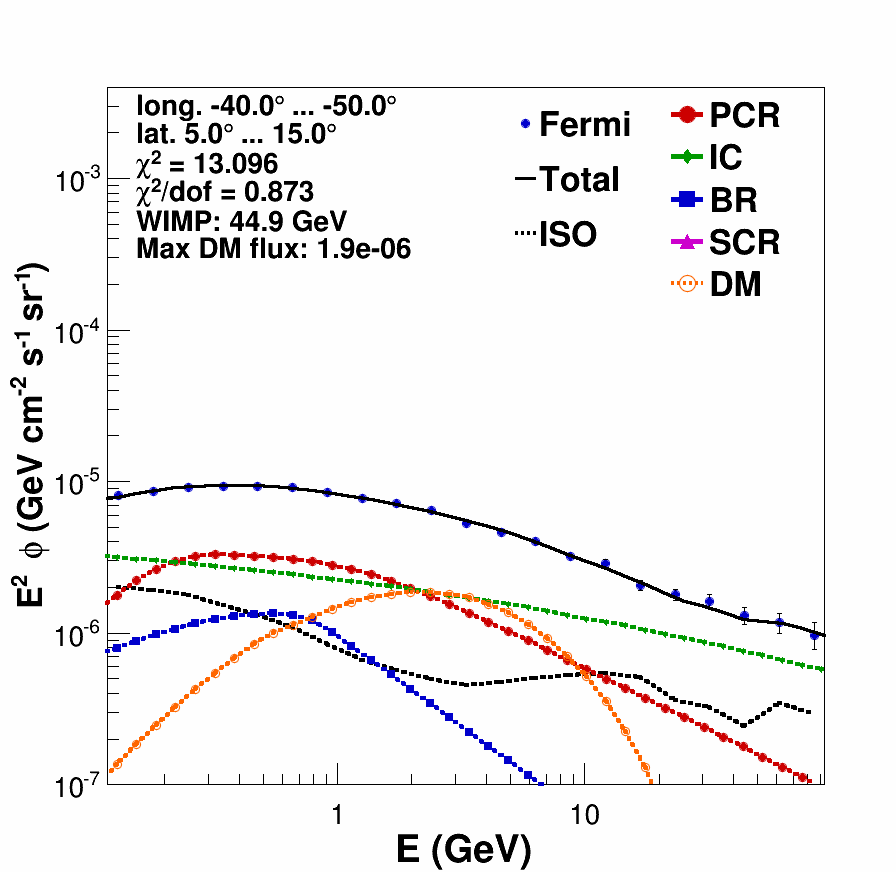}
\includegraphics[width=0.16\textwidth,height=0.16\textwidth,clip]{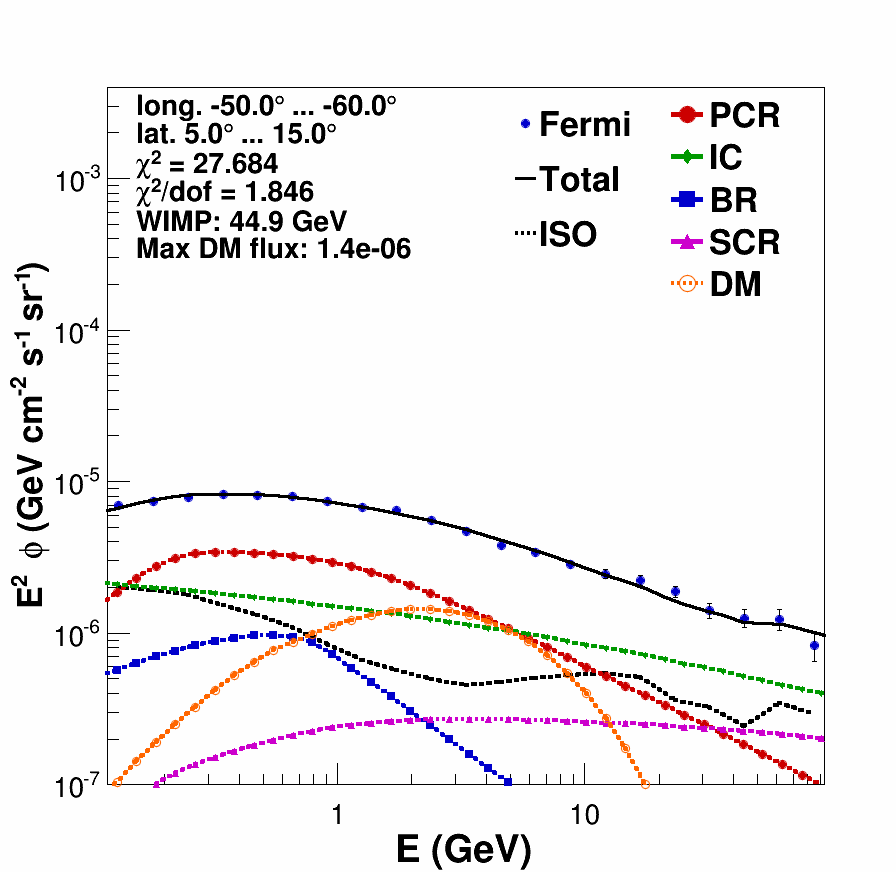}
\includegraphics[width=0.16\textwidth,height=0.16\textwidth,clip]{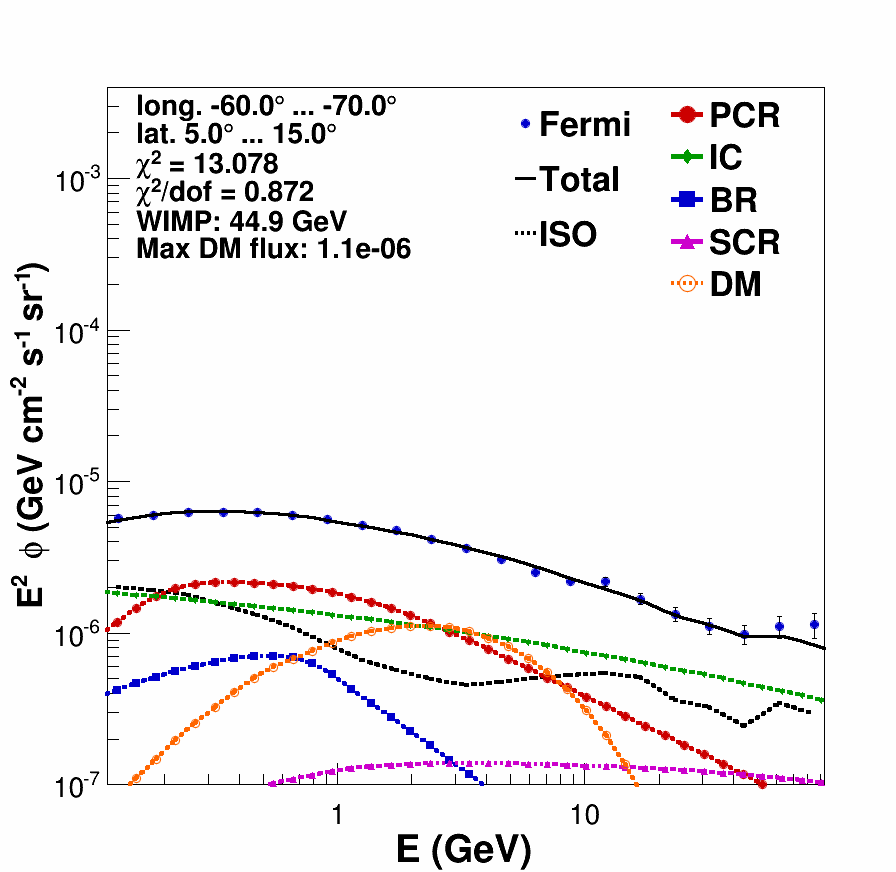}
\includegraphics[width=0.16\textwidth,height=0.16\textwidth,clip]{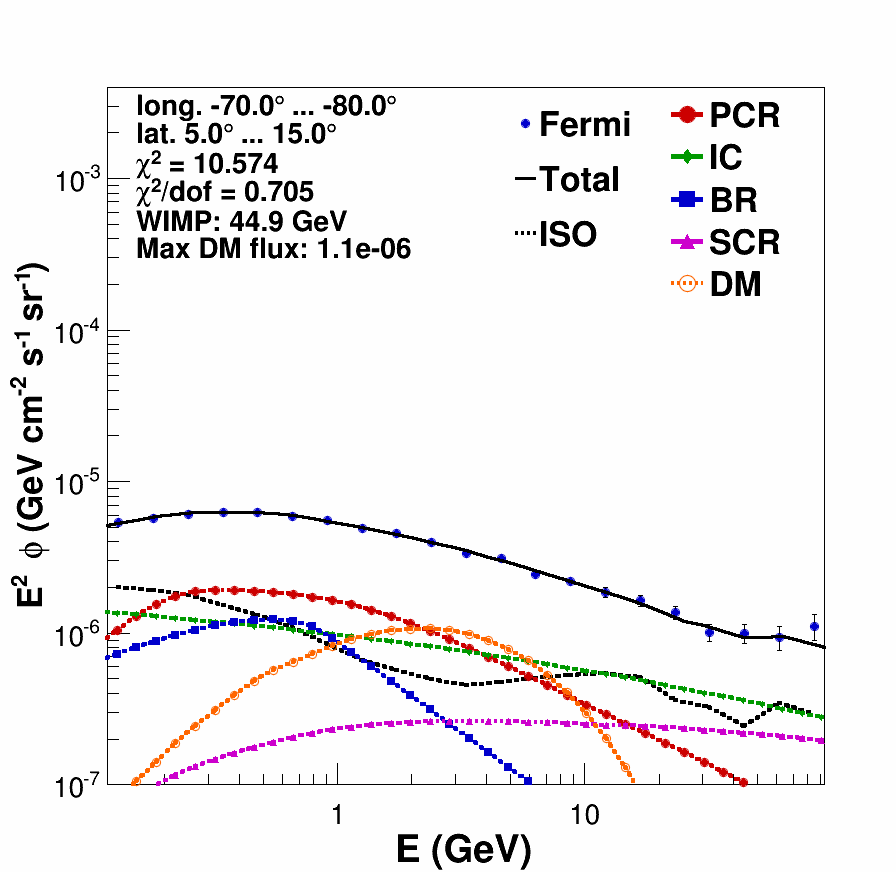}
\includegraphics[width=0.16\textwidth,height=0.16\textwidth,clip]{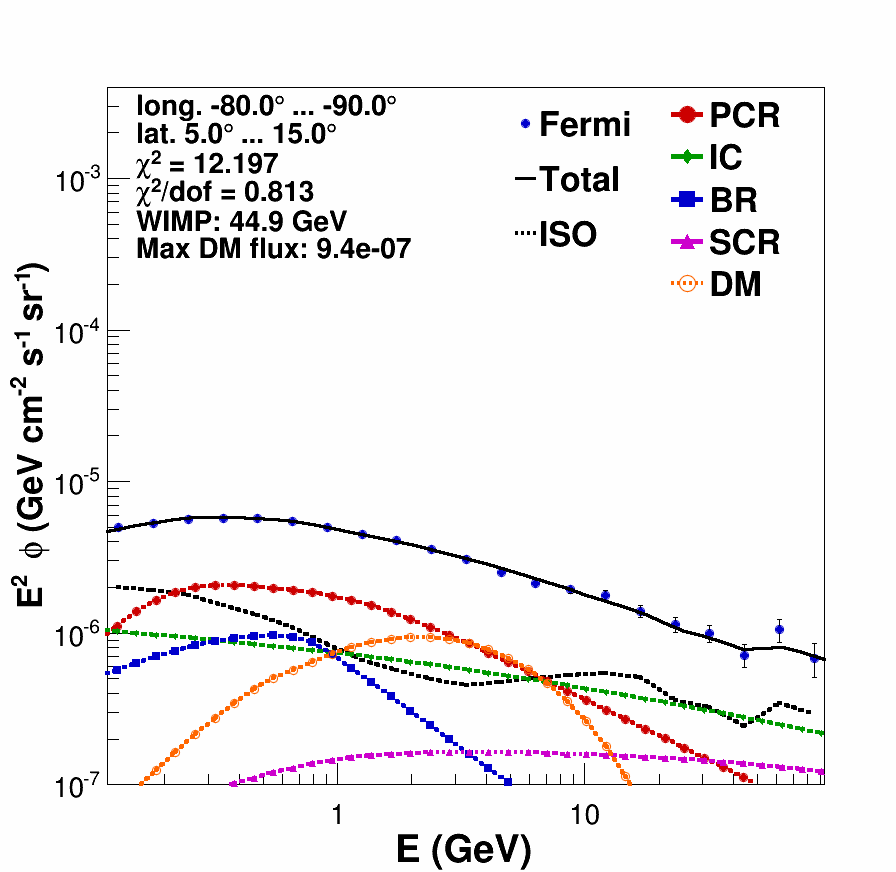}
\includegraphics[width=0.16\textwidth,height=0.16\textwidth,clip]{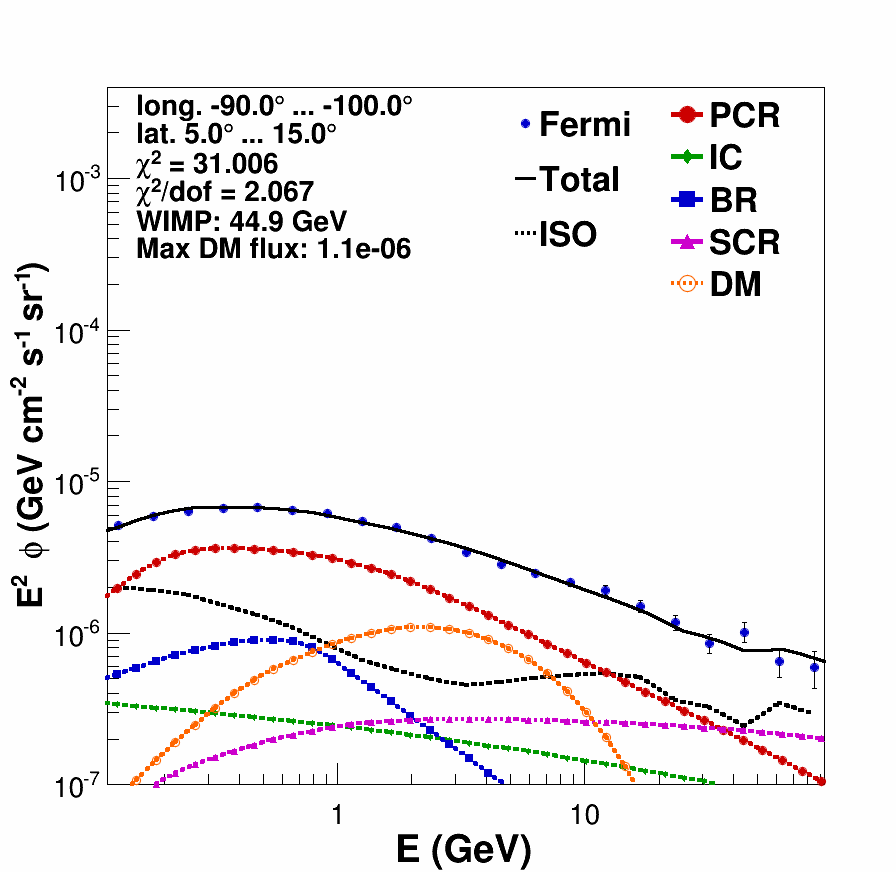}
\includegraphics[width=0.16\textwidth,height=0.16\textwidth,clip]{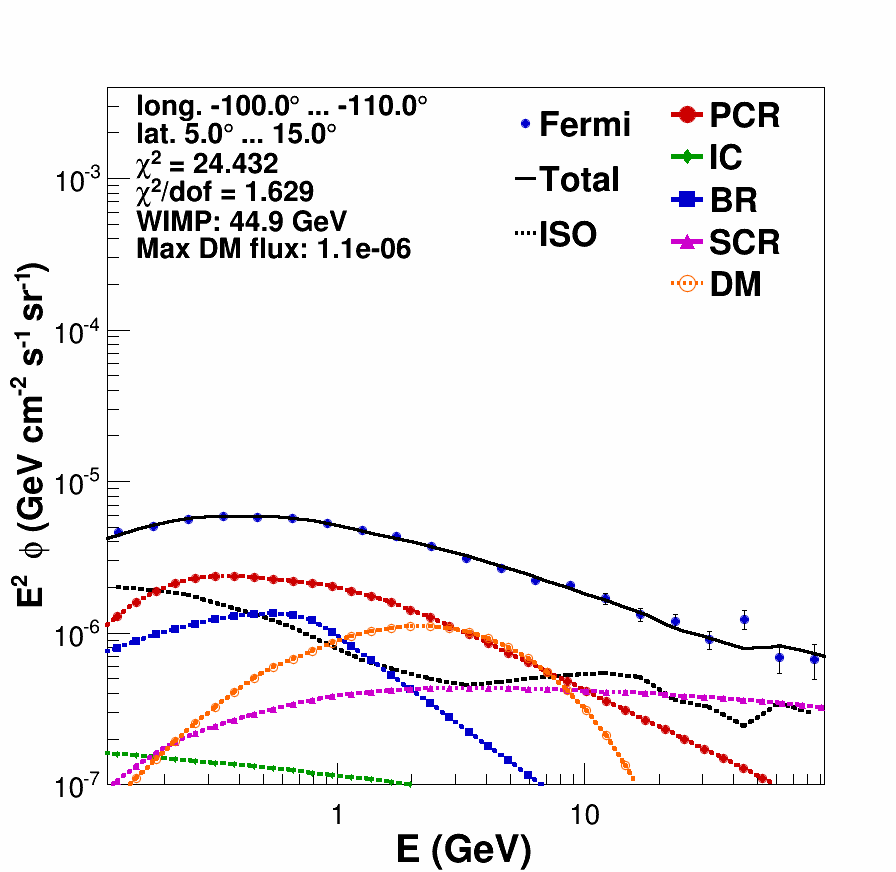}
\includegraphics[width=0.16\textwidth,height=0.16\textwidth,clip]{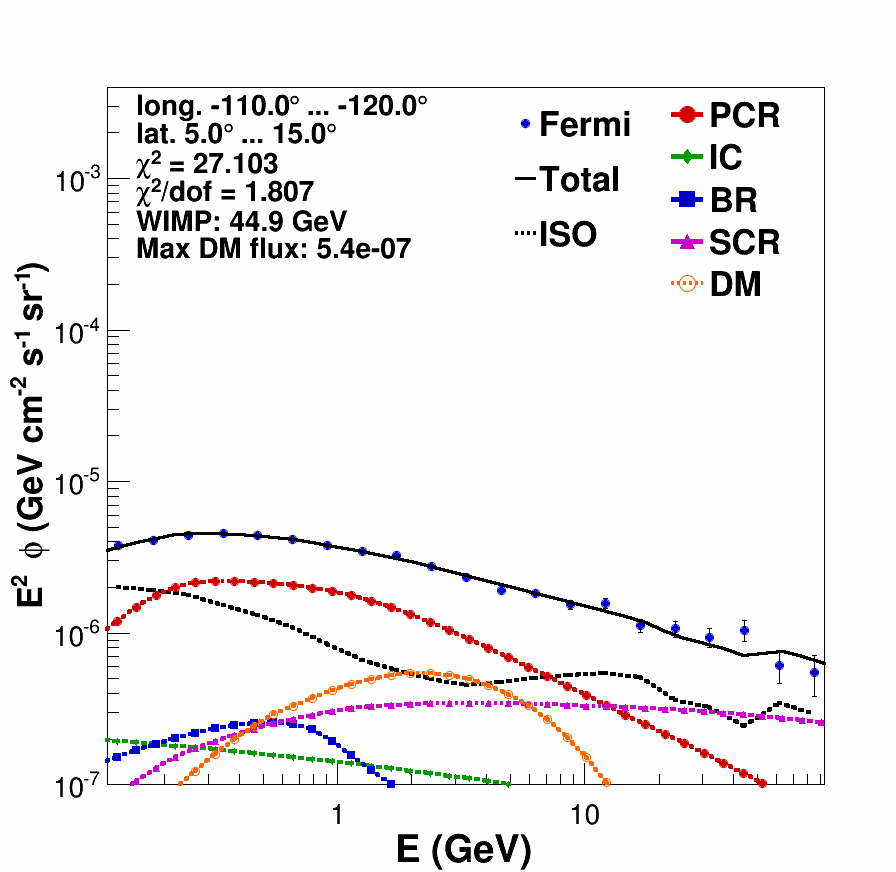}
\includegraphics[width=0.16\textwidth,height=0.16\textwidth,clip]{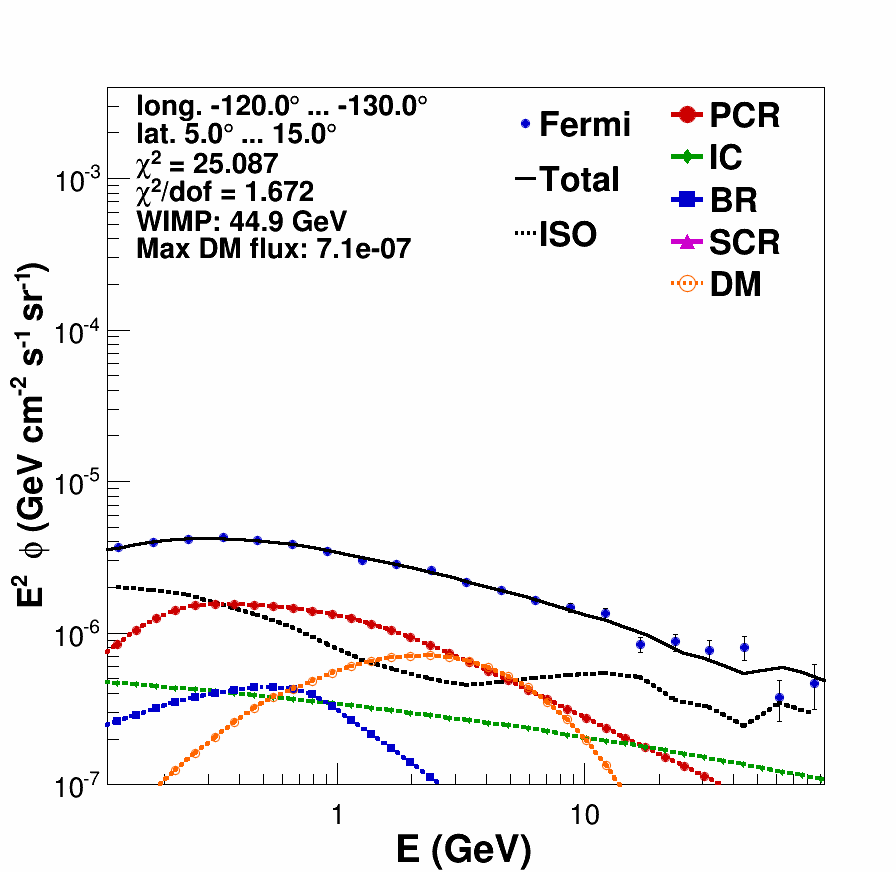}
\includegraphics[width=0.16\textwidth,height=0.16\textwidth,clip]{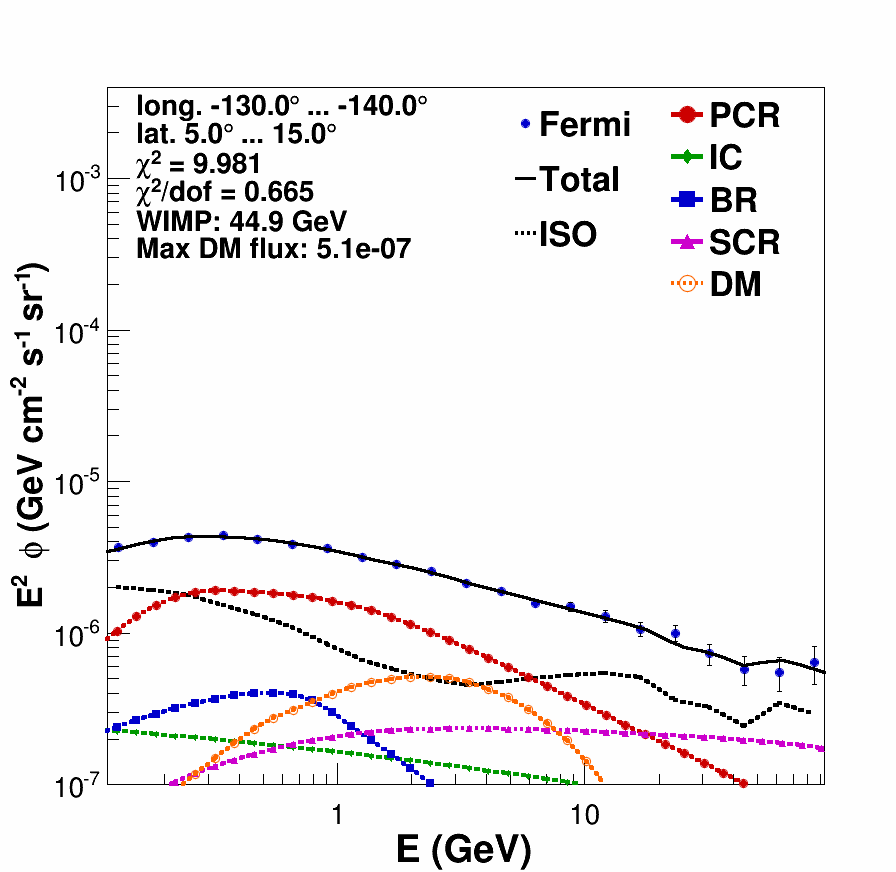}
\includegraphics[width=0.16\textwidth,height=0.16\textwidth,clip]{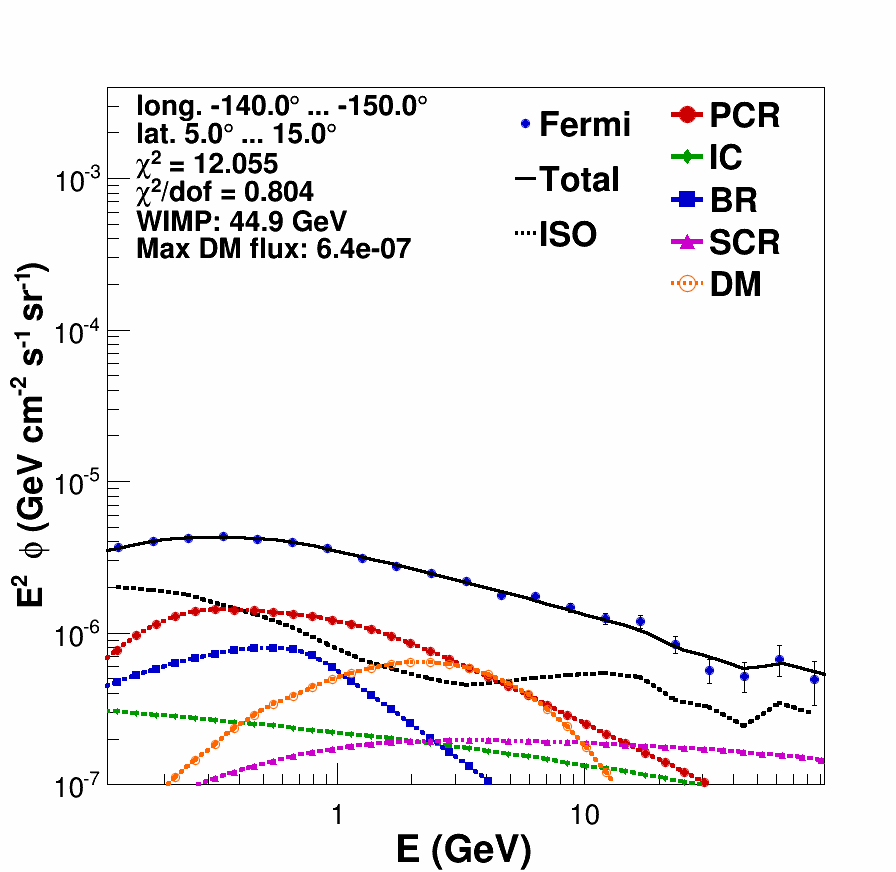}
\includegraphics[width=0.16\textwidth,height=0.16\textwidth,clip]{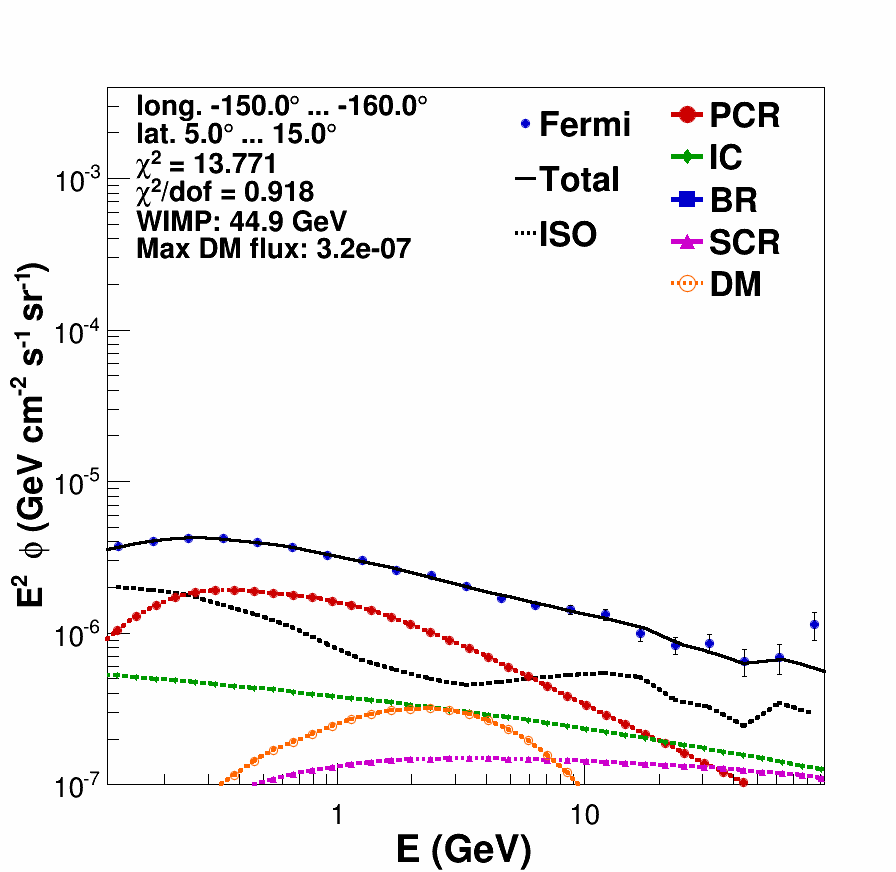}
\includegraphics[width=0.16\textwidth,height=0.16\textwidth,clip]{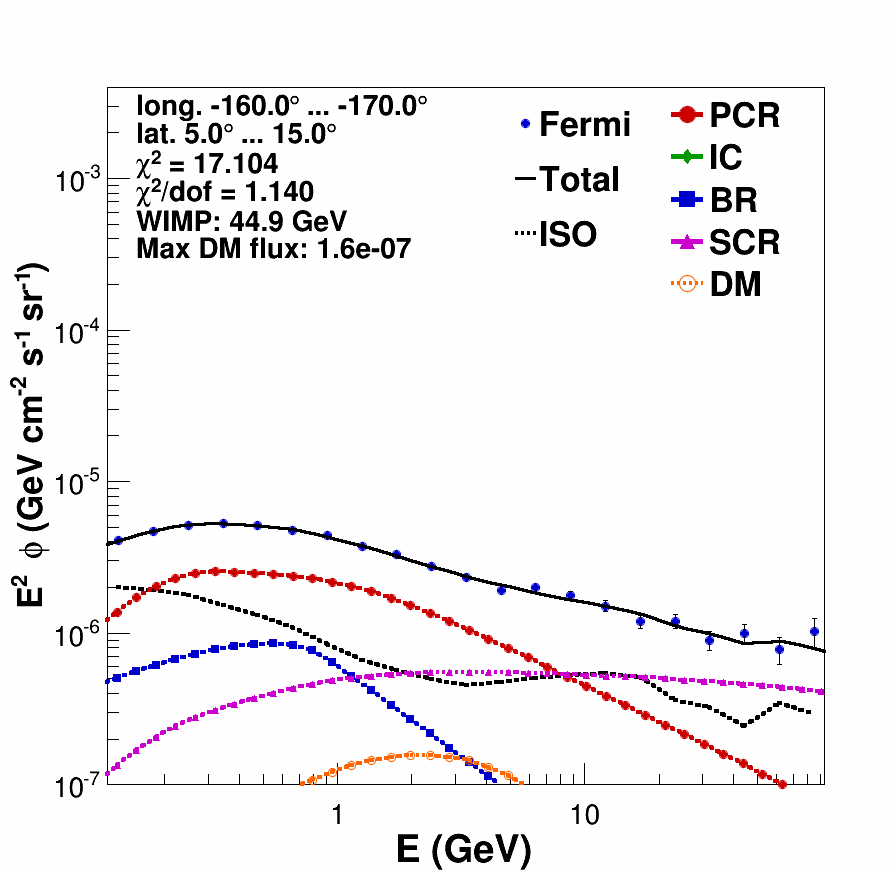}
\includegraphics[width=0.16\textwidth,height=0.16\textwidth,clip]{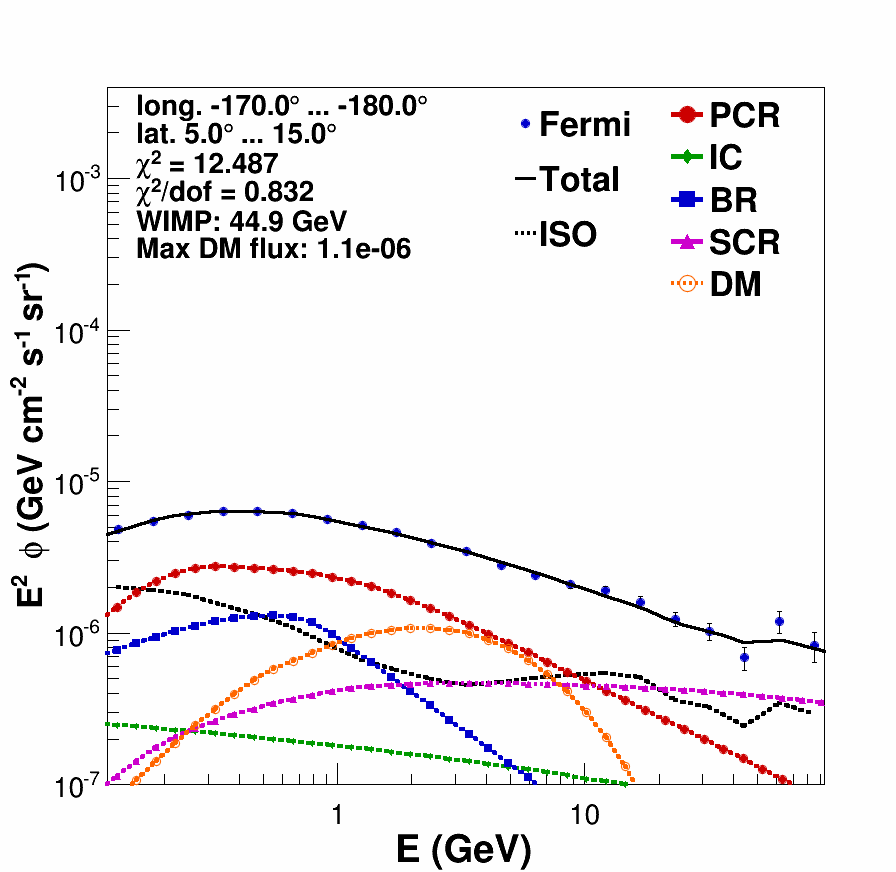}%%%%%r7
\caption[]{Template fits for latitudes  with $5.0^\circ<b<15.0^\circ$ and longitudes decreasing from 180$^\circ$ to -180$^\circ$. \label{F38}
}
\end{figure}
\begin{figure}
\centering
\includegraphics[width=0.16\textwidth,height=0.16\textwidth,clip]{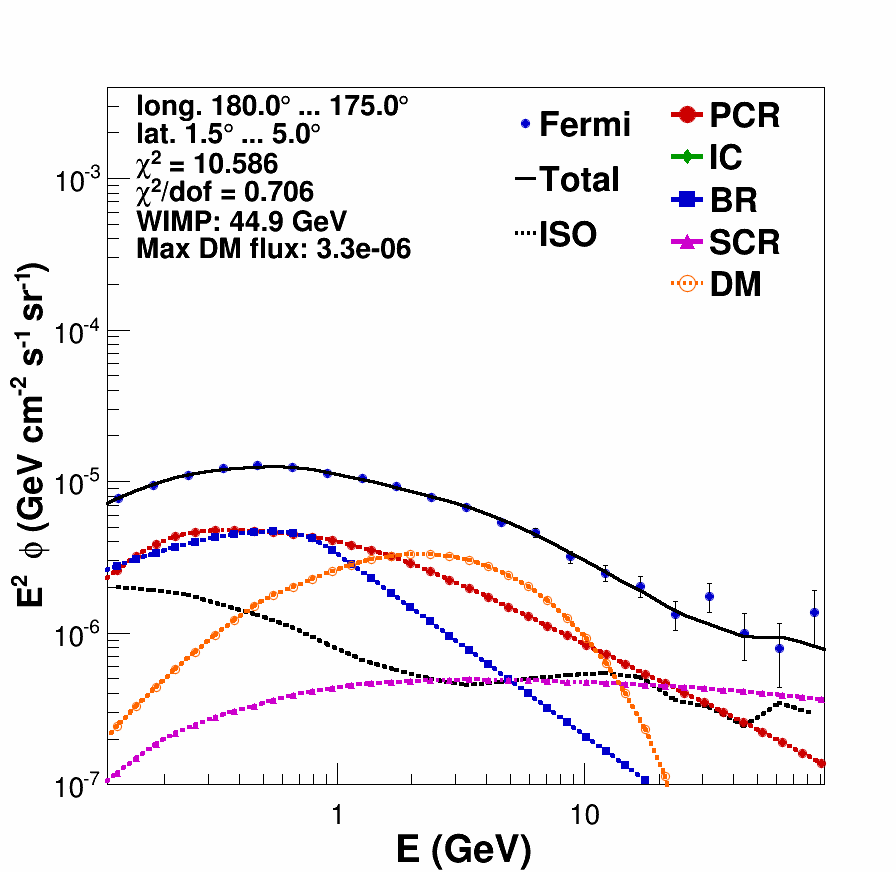}
\includegraphics[width=0.16\textwidth,height=0.16\textwidth,clip]{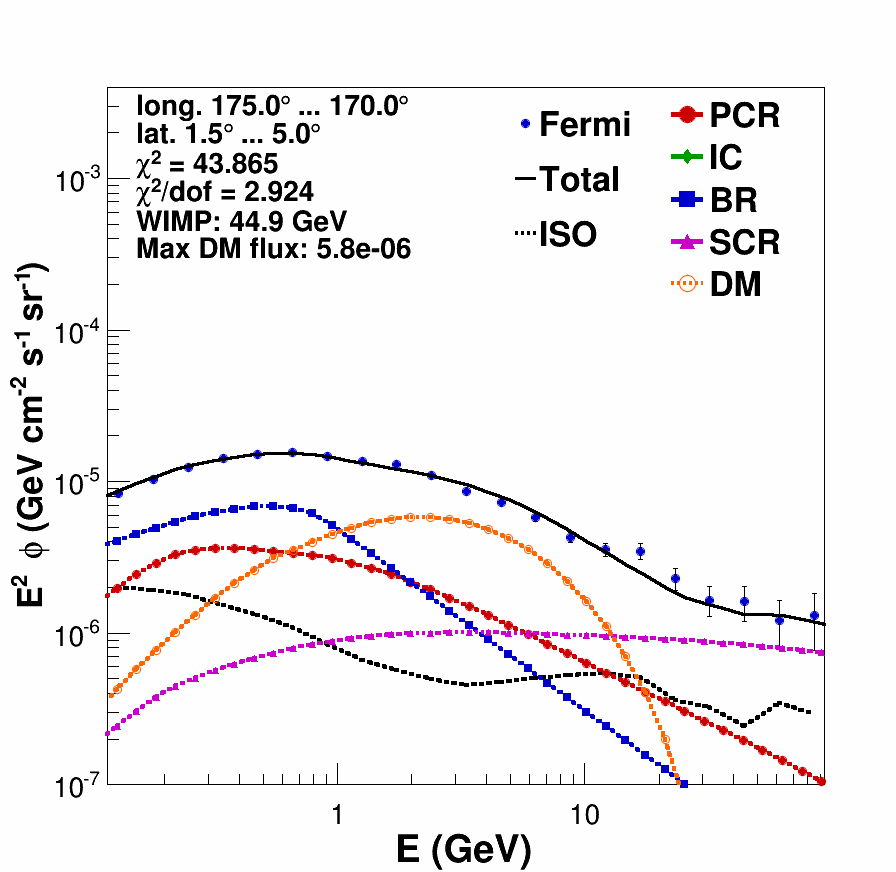}
\includegraphics[width=0.16\textwidth,height=0.16\textwidth,clip]{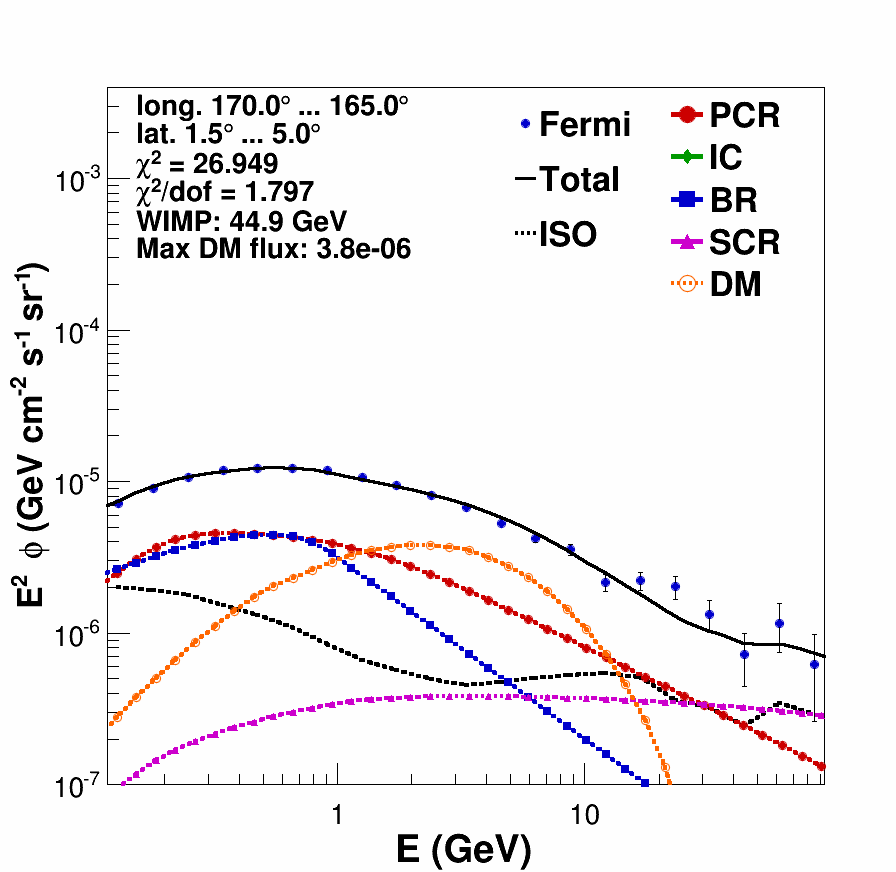}
\includegraphics[width=0.16\textwidth,height=0.16\textwidth,clip]{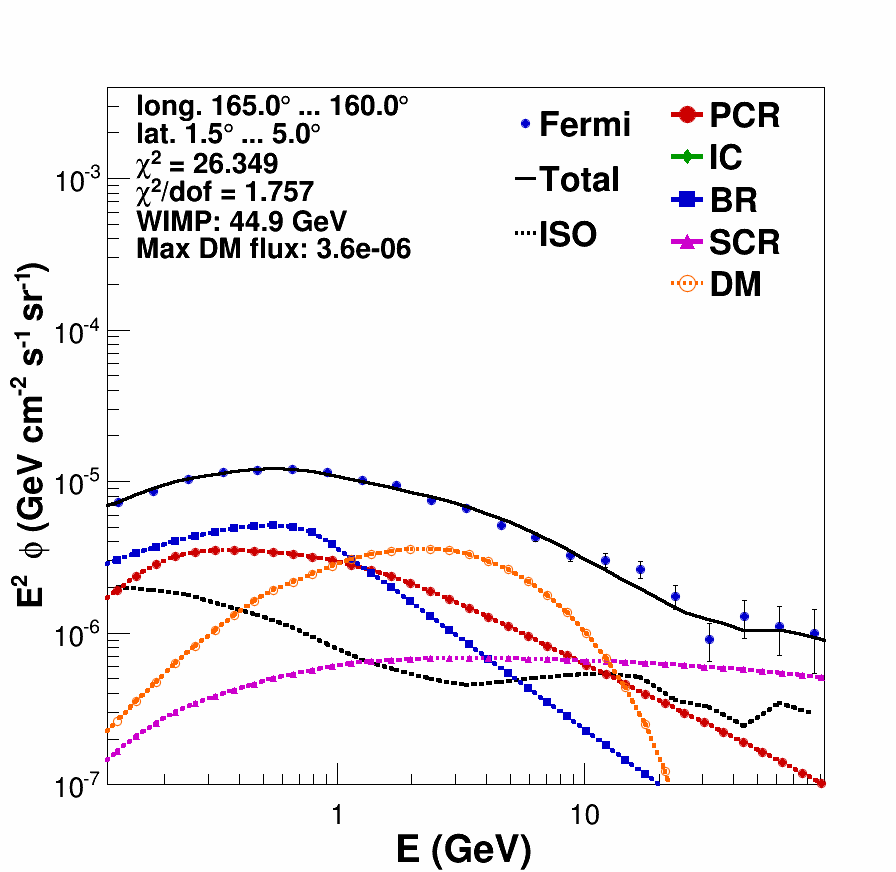}
\includegraphics[width=0.16\textwidth,height=0.16\textwidth,clip]{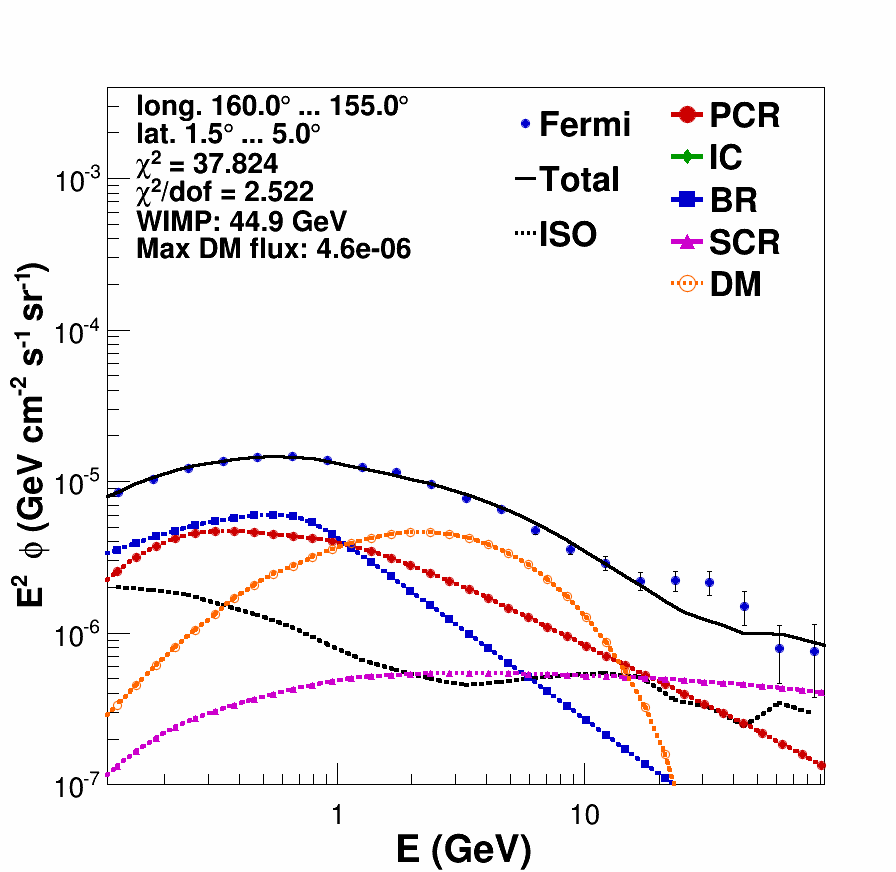}
\includegraphics[width=0.16\textwidth,height=0.16\textwidth,clip]{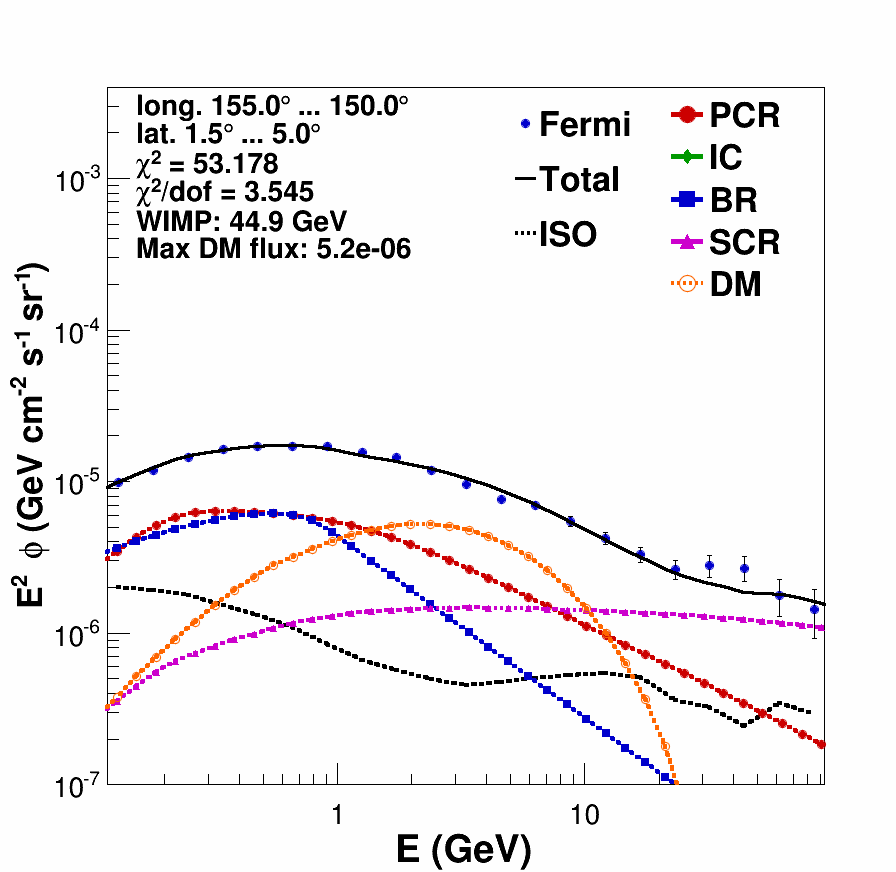}
\includegraphics[width=0.16\textwidth,height=0.16\textwidth,clip]{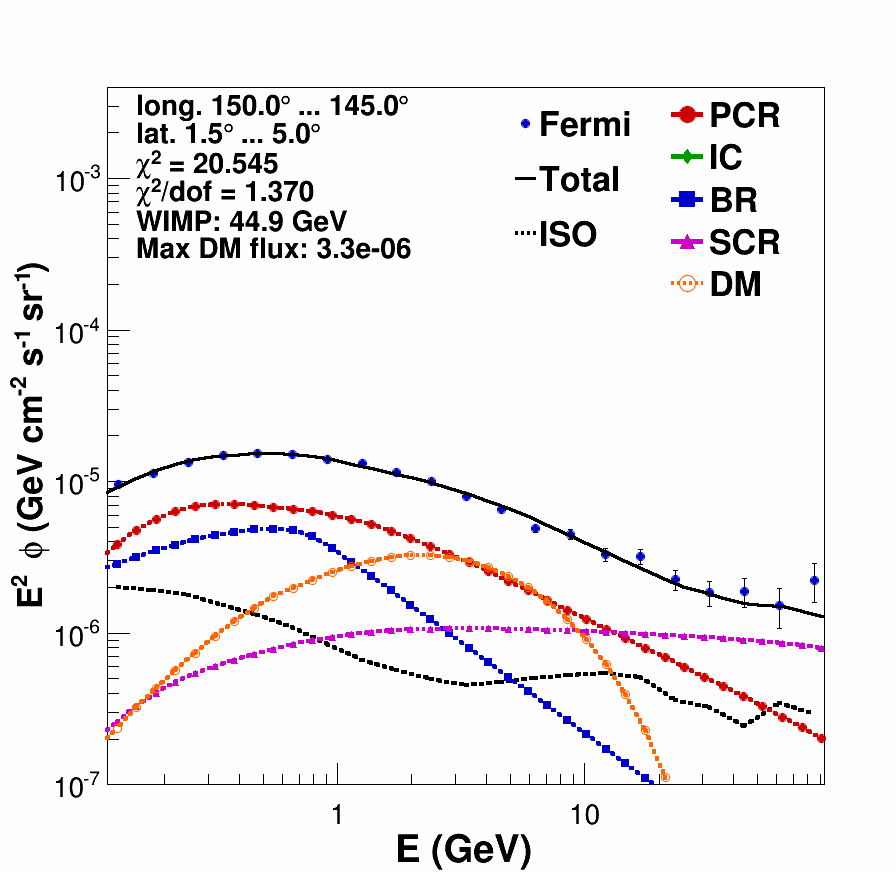}
\includegraphics[width=0.16\textwidth,height=0.16\textwidth,clip]{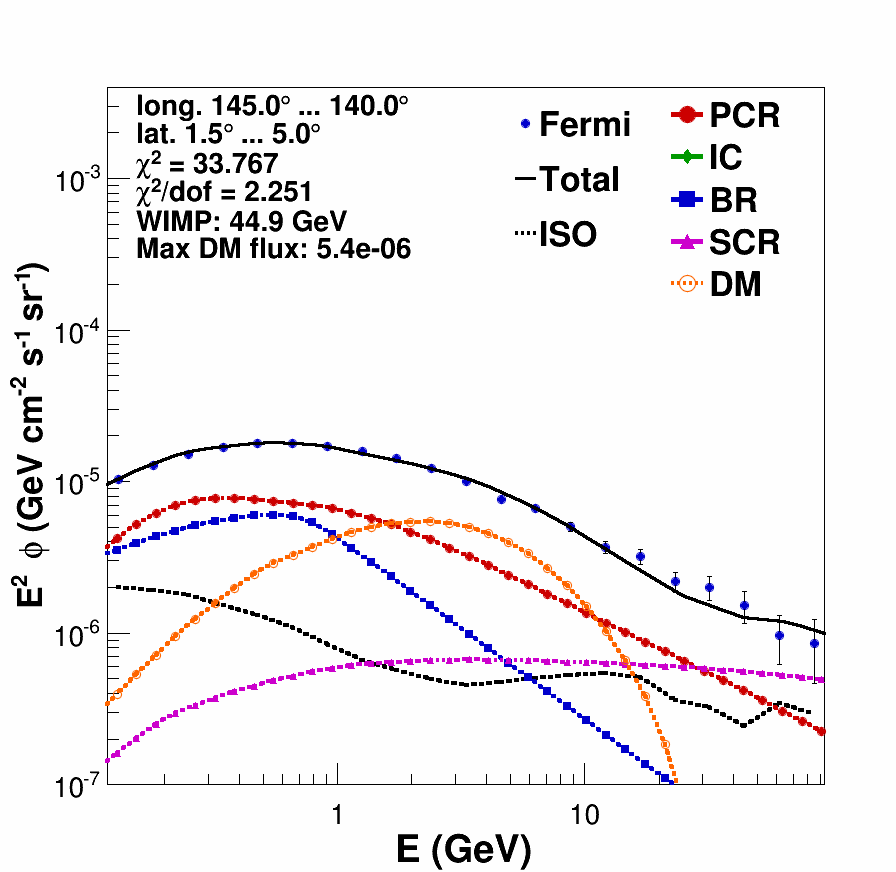}
\includegraphics[width=0.16\textwidth,height=0.16\textwidth,clip]{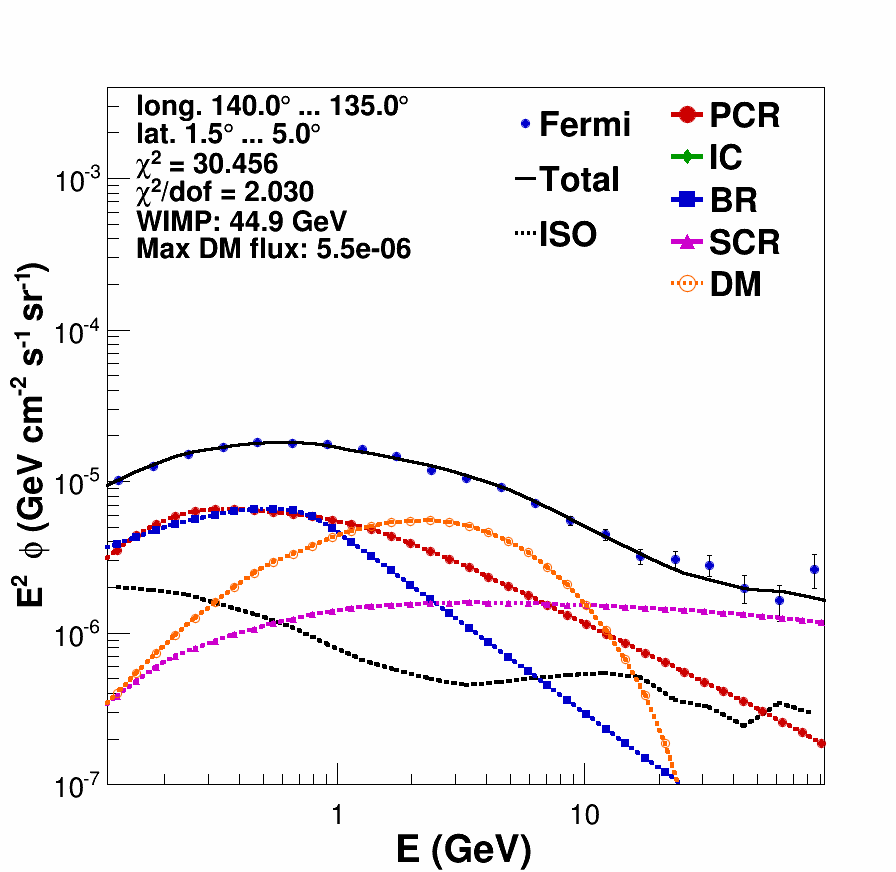}
\includegraphics[width=0.16\textwidth,height=0.16\textwidth,clip]{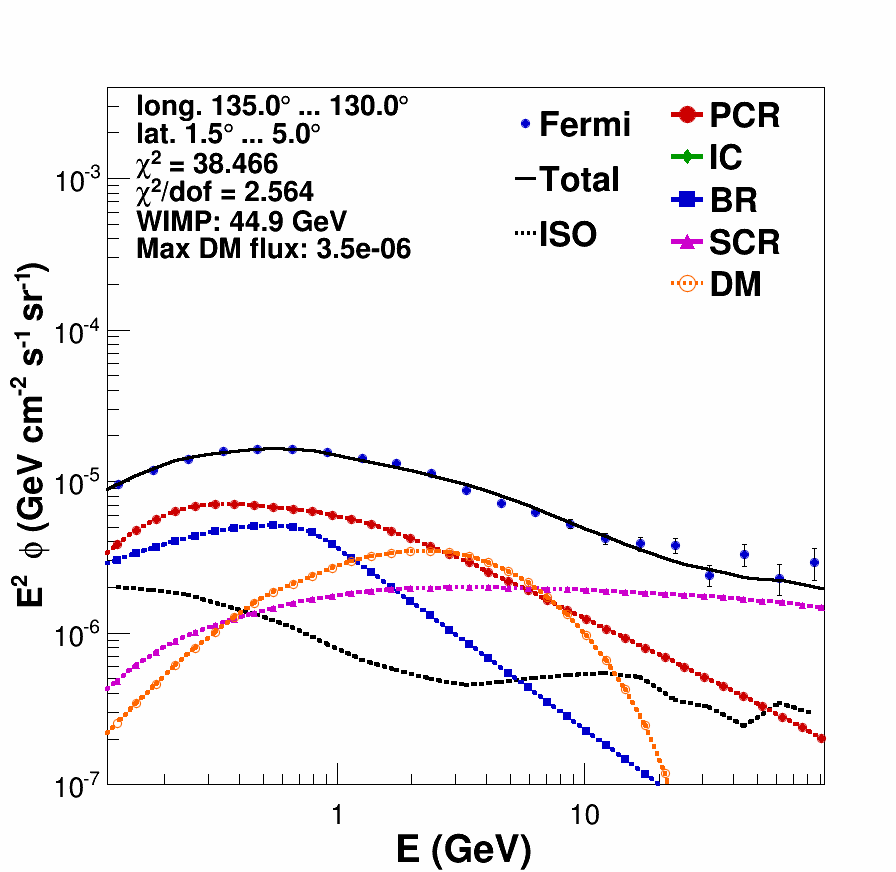}
\includegraphics[width=0.16\textwidth,height=0.16\textwidth,clip]{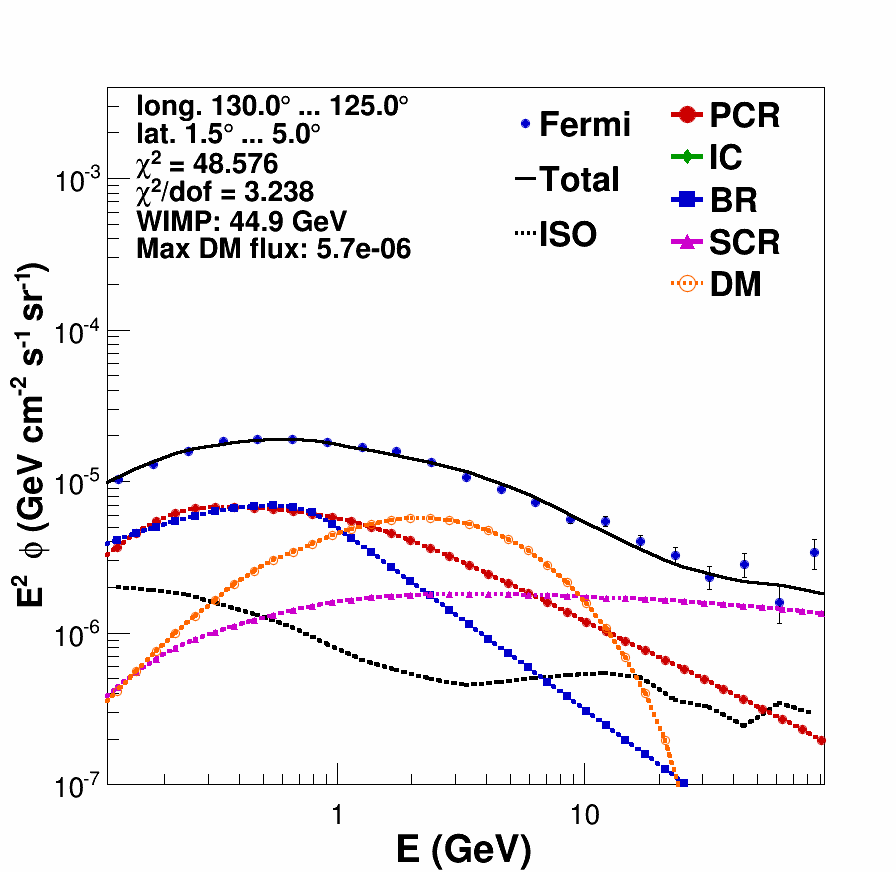}
\includegraphics[width=0.16\textwidth,height=0.16\textwidth,clip]{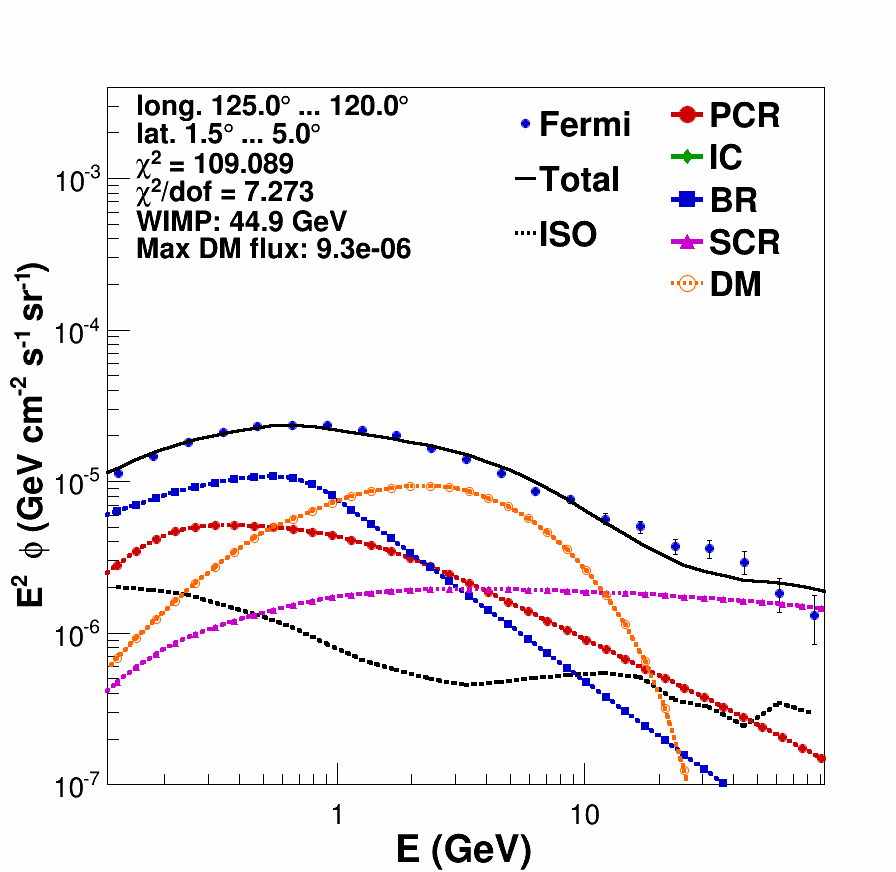}
\includegraphics[width=0.16\textwidth,height=0.16\textwidth,clip]{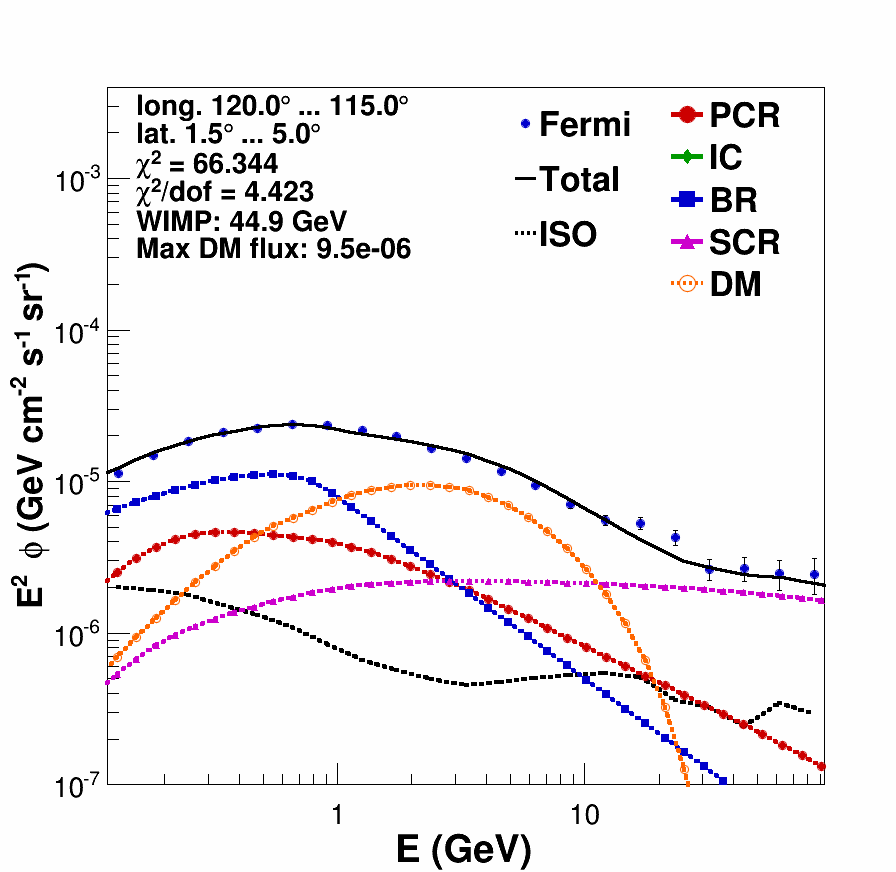}
\includegraphics[width=0.16\textwidth,height=0.16\textwidth,clip]{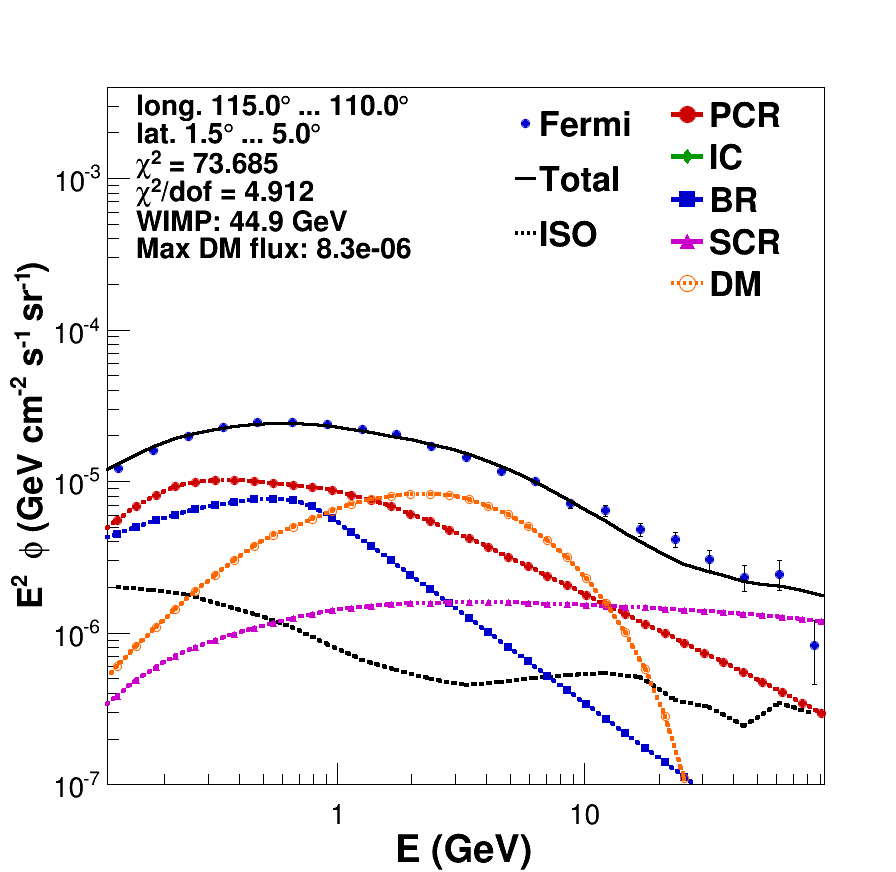}
\includegraphics[width=0.16\textwidth,height=0.16\textwidth,clip]{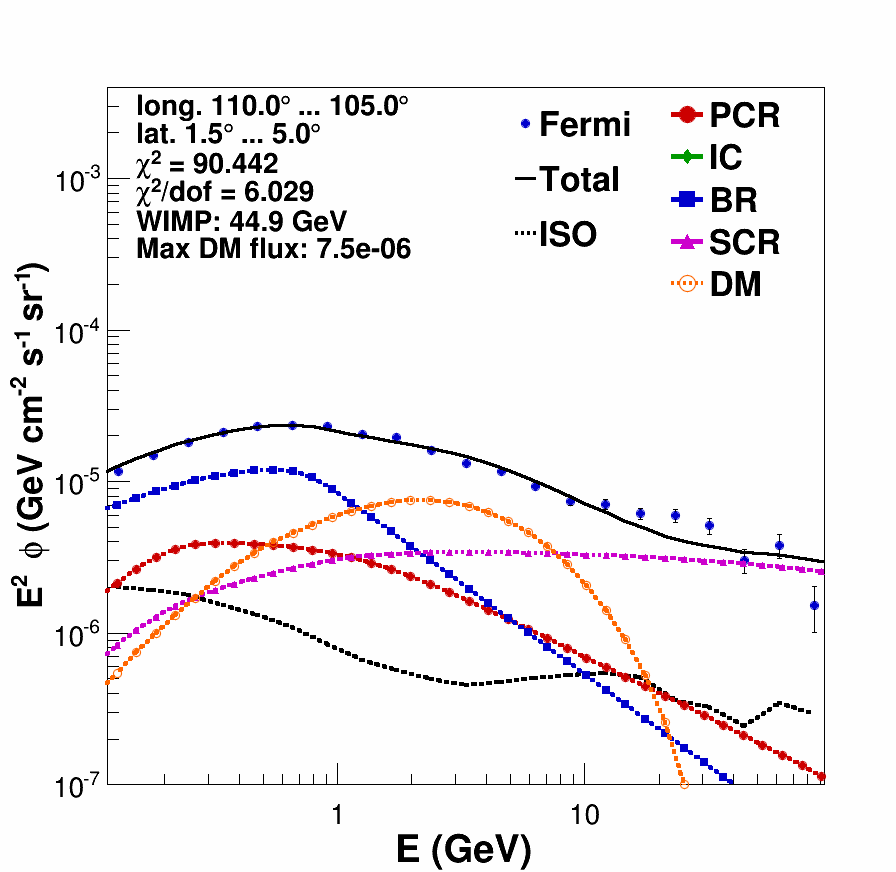}
\includegraphics[width=0.16\textwidth,height=0.16\textwidth,clip]{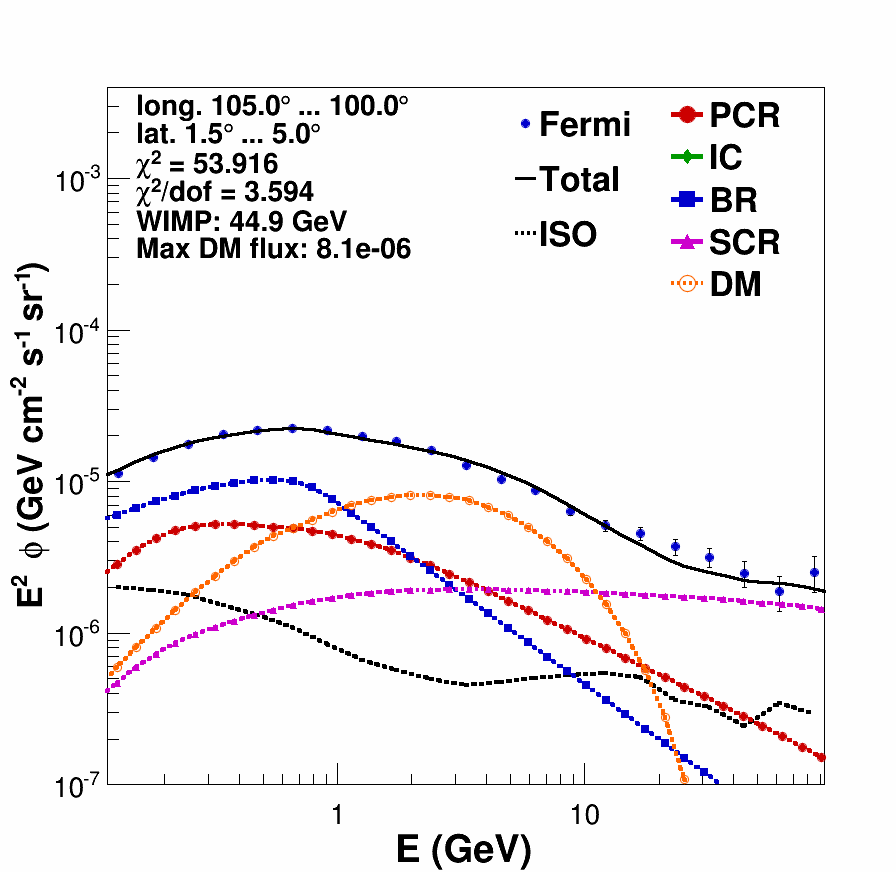}
\includegraphics[width=0.16\textwidth,height=0.16\textwidth,clip]{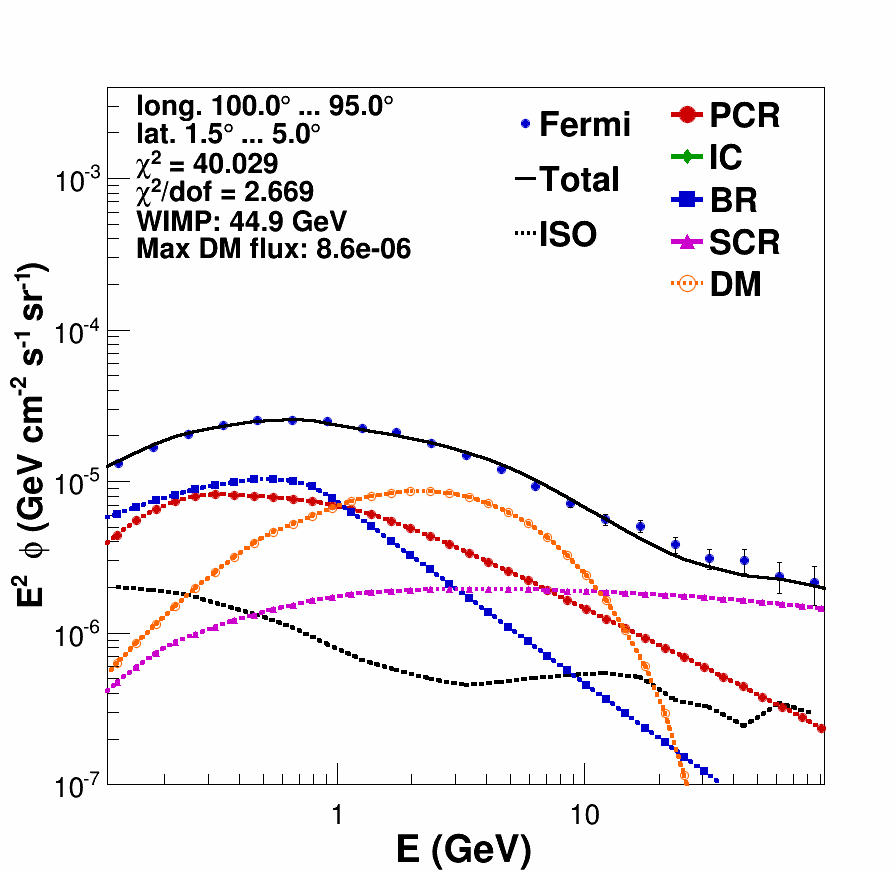}
\includegraphics[width=0.16\textwidth,height=0.16\textwidth,clip]{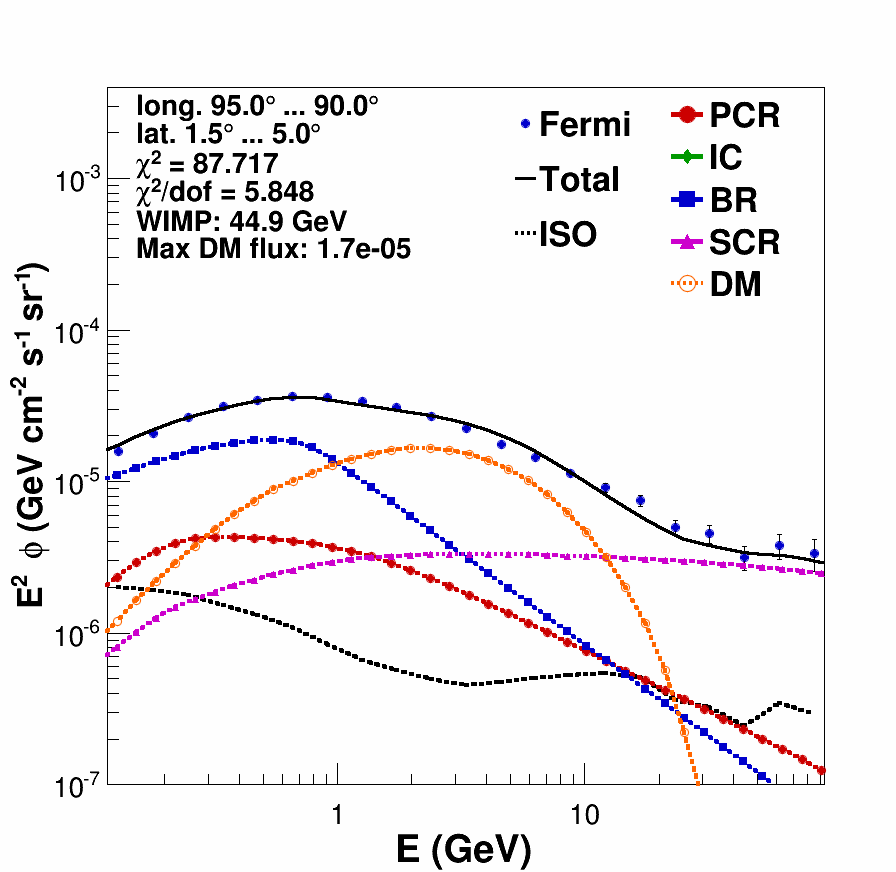}
\includegraphics[width=0.16\textwidth,height=0.16\textwidth,clip]{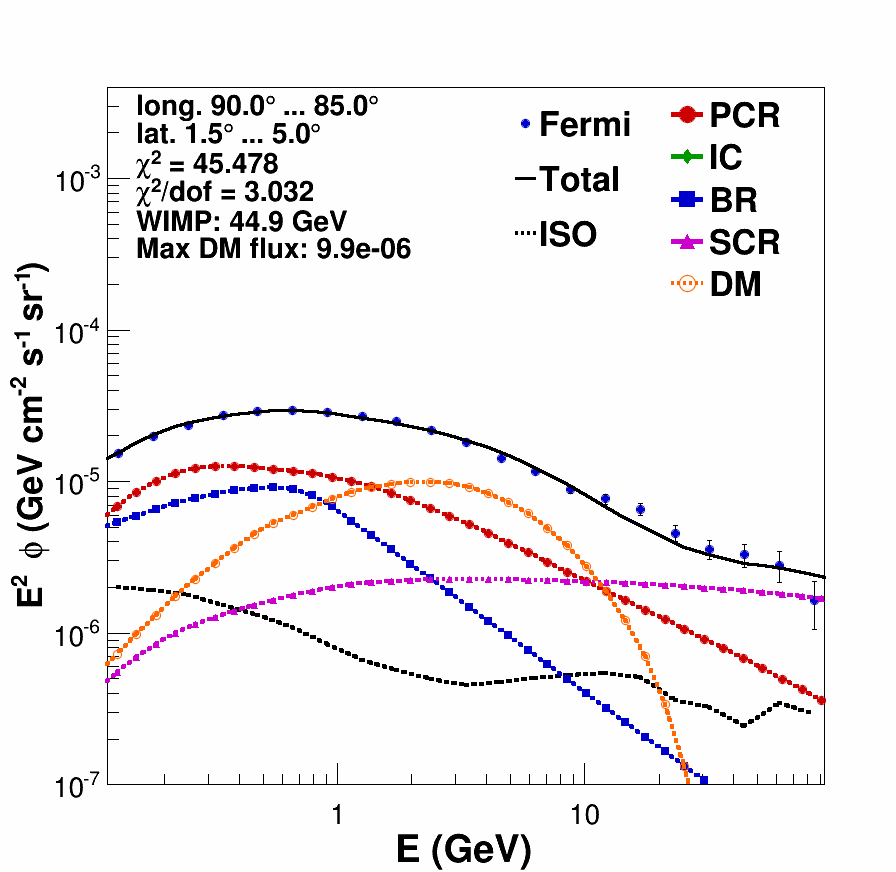}
\includegraphics[width=0.16\textwidth,height=0.16\textwidth,clip]{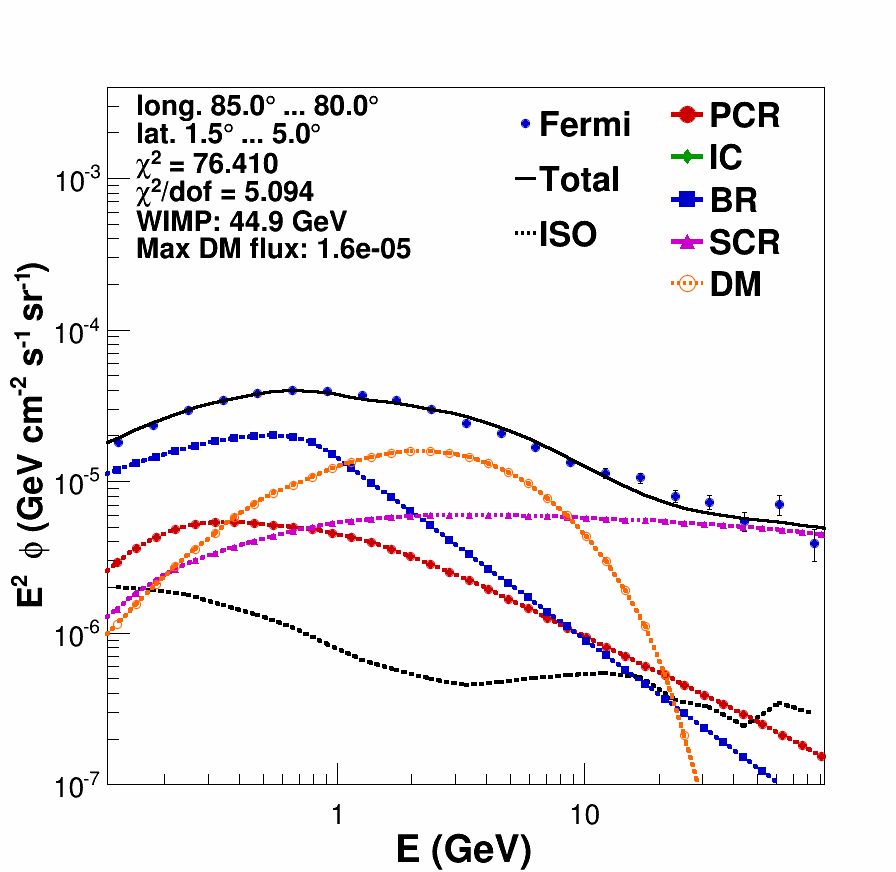}
\includegraphics[width=0.16\textwidth,height=0.16\textwidth,clip]{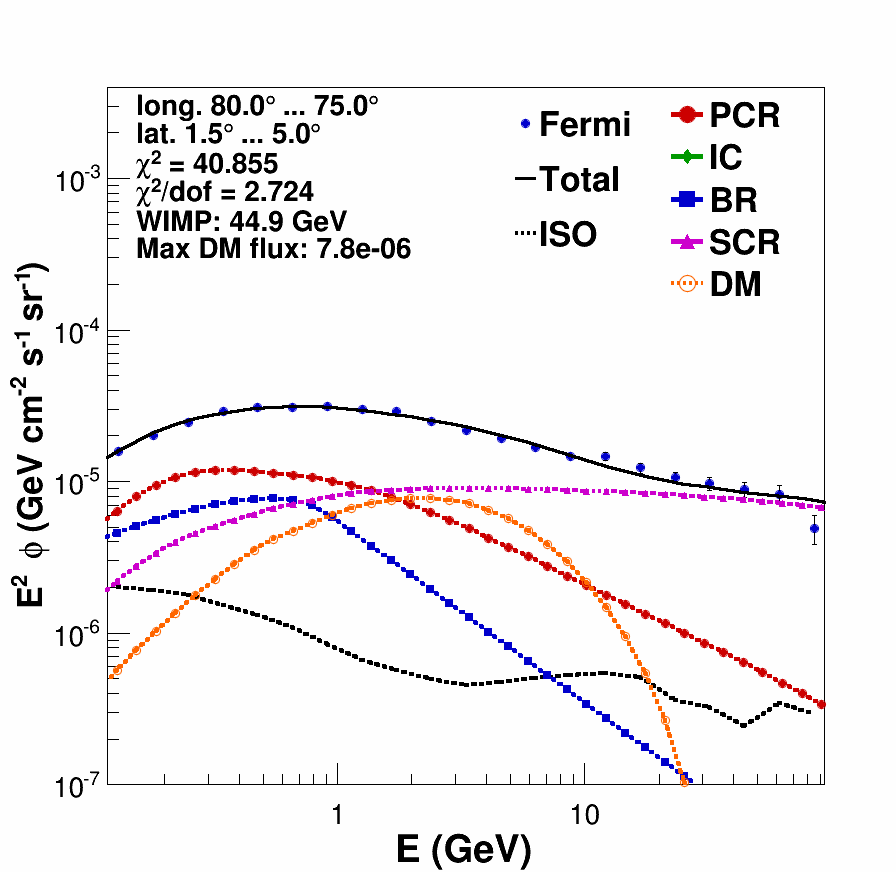}
\includegraphics[width=0.16\textwidth,height=0.16\textwidth,clip]{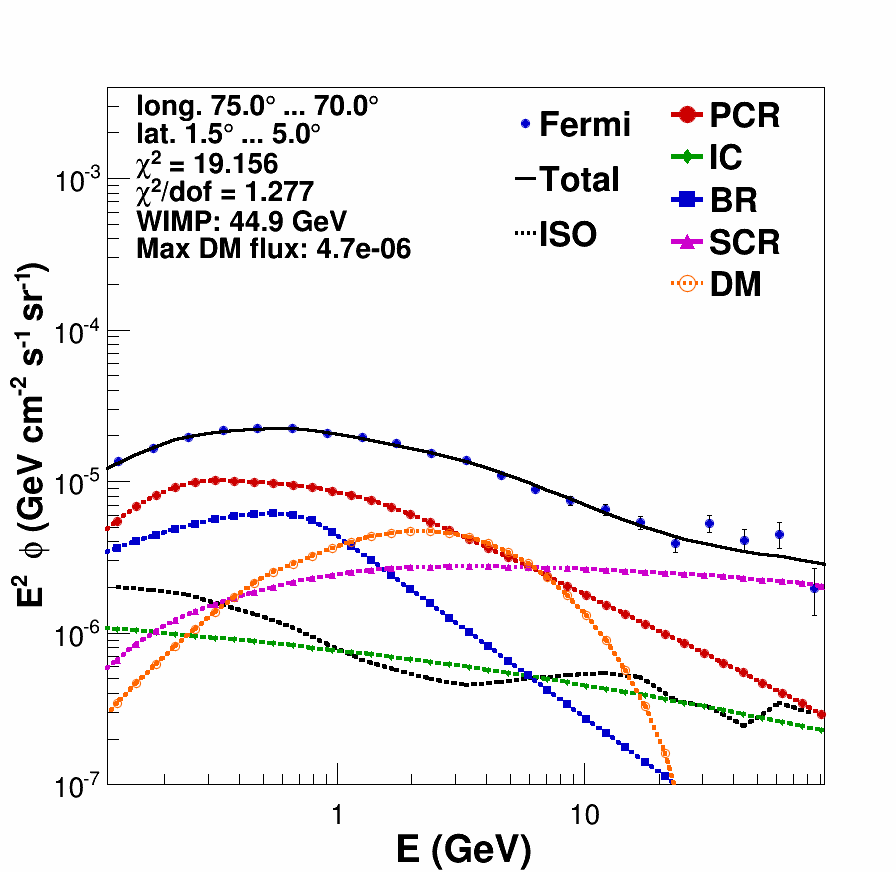}
\includegraphics[width=0.16\textwidth,height=0.16\textwidth,clip]{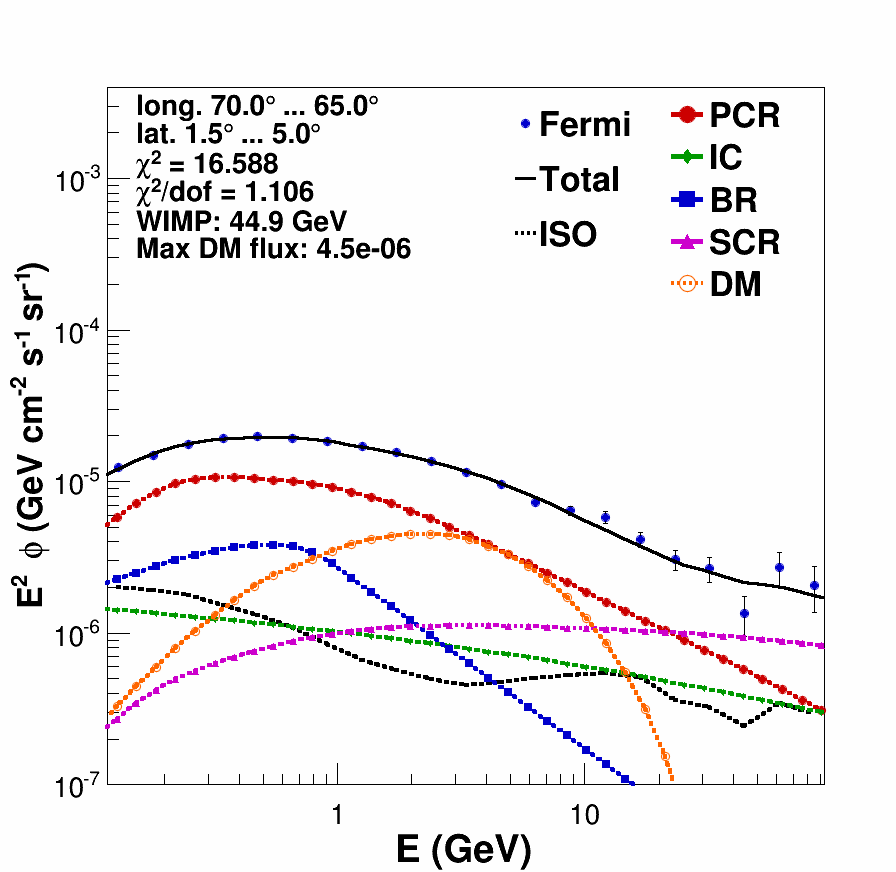}
\includegraphics[width=0.16\textwidth,height=0.16\textwidth,clip]{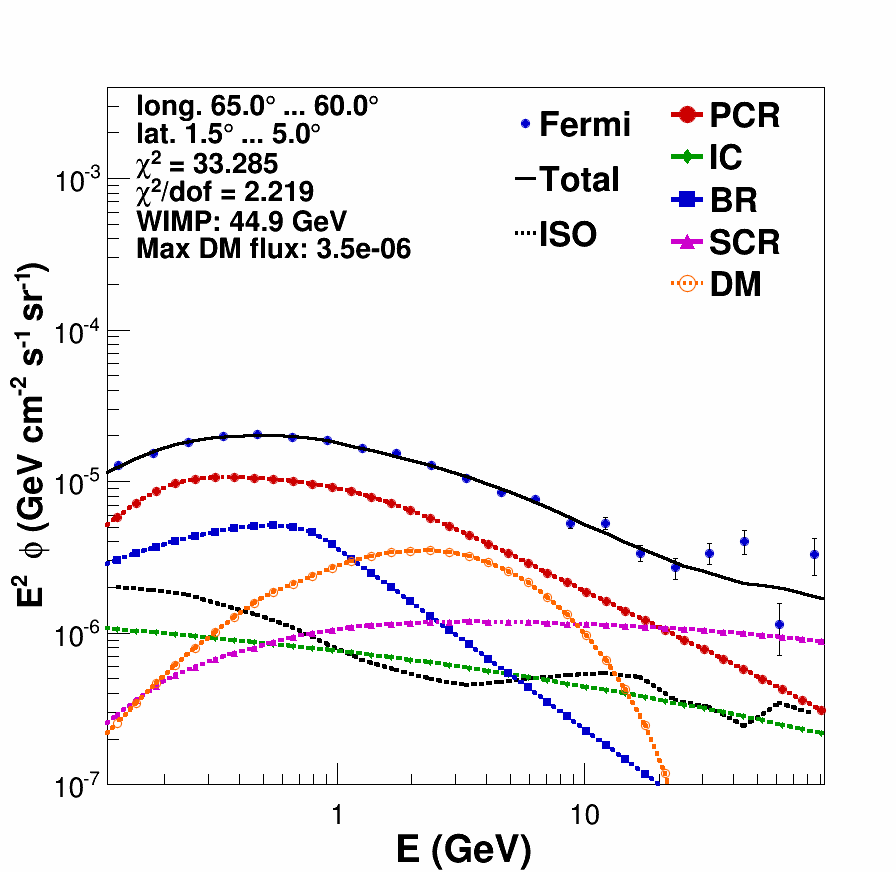}
\includegraphics[width=0.16\textwidth,height=0.16\textwidth,clip]{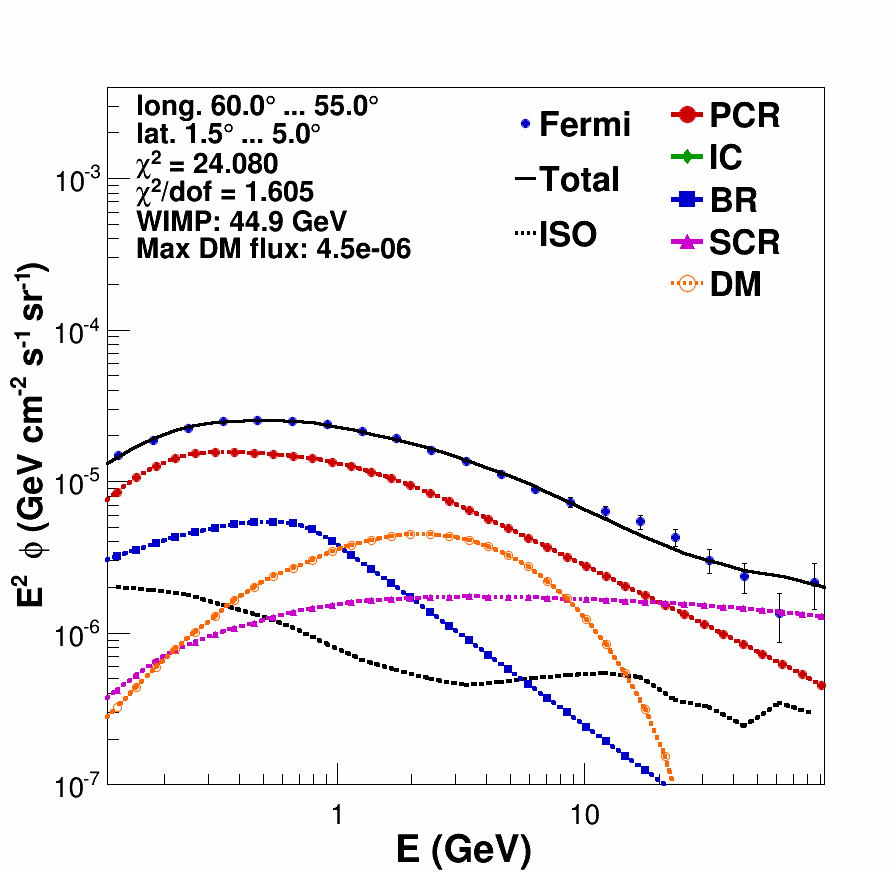}
\includegraphics[width=0.16\textwidth,height=0.16\textwidth,clip]{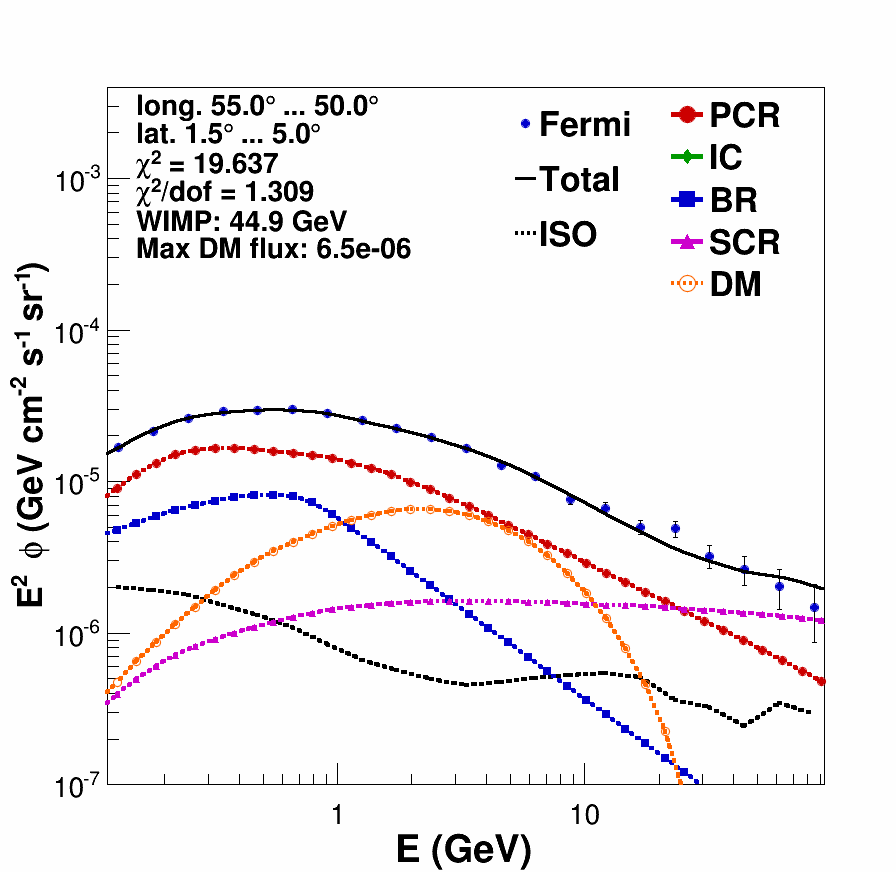}
\includegraphics[width=0.16\textwidth,height=0.16\textwidth,clip]{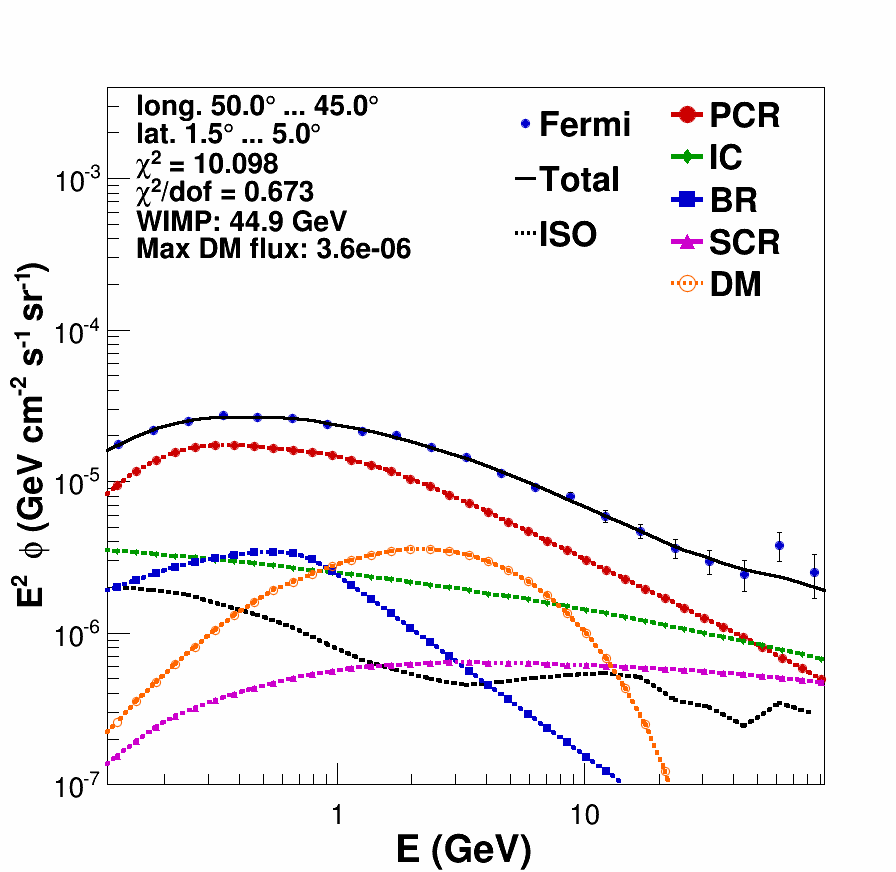}
\includegraphics[width=0.16\textwidth,height=0.16\textwidth,clip]{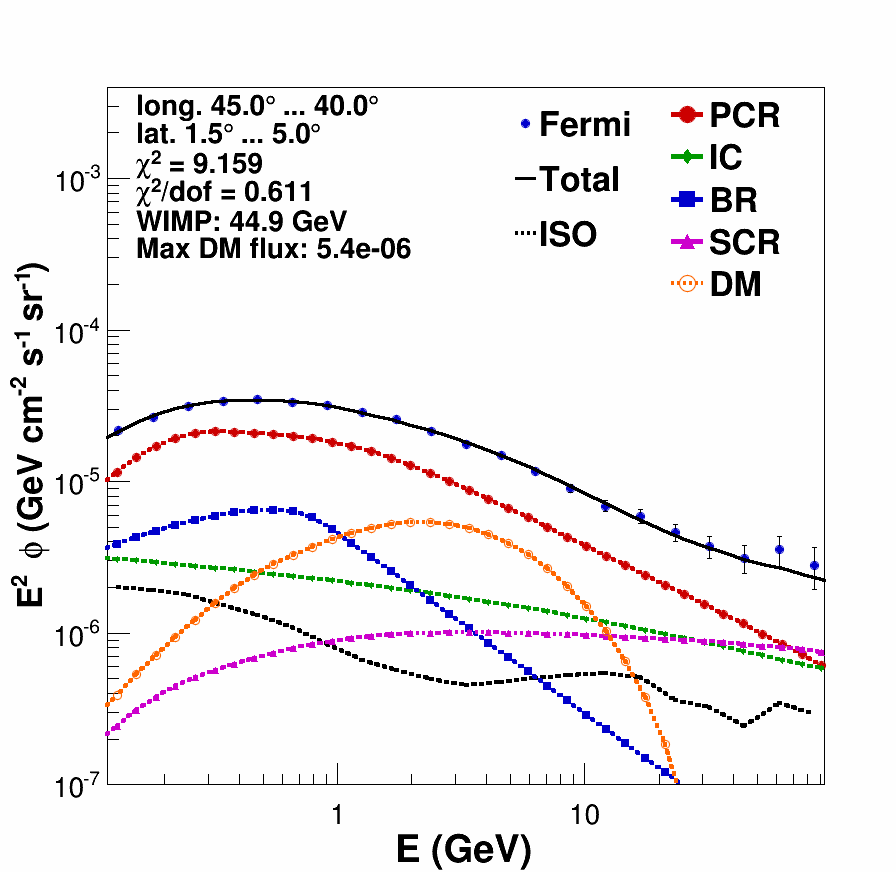}
\includegraphics[width=0.16\textwidth,height=0.16\textwidth,clip]{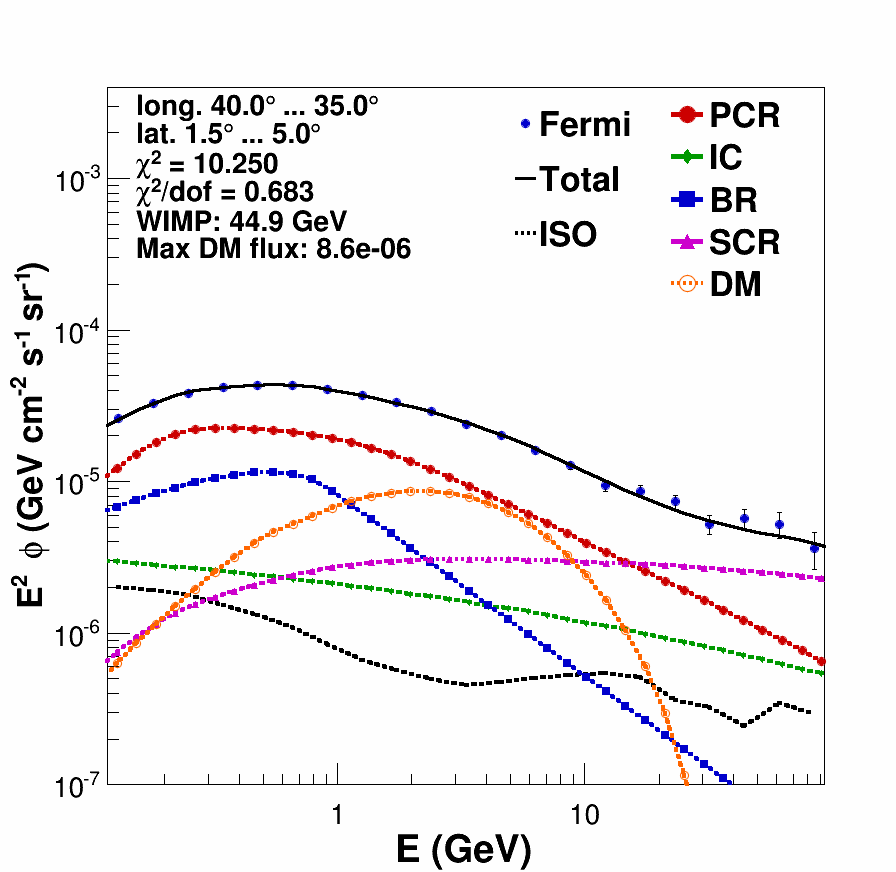}
\includegraphics[width=0.16\textwidth,height=0.16\textwidth,clip]{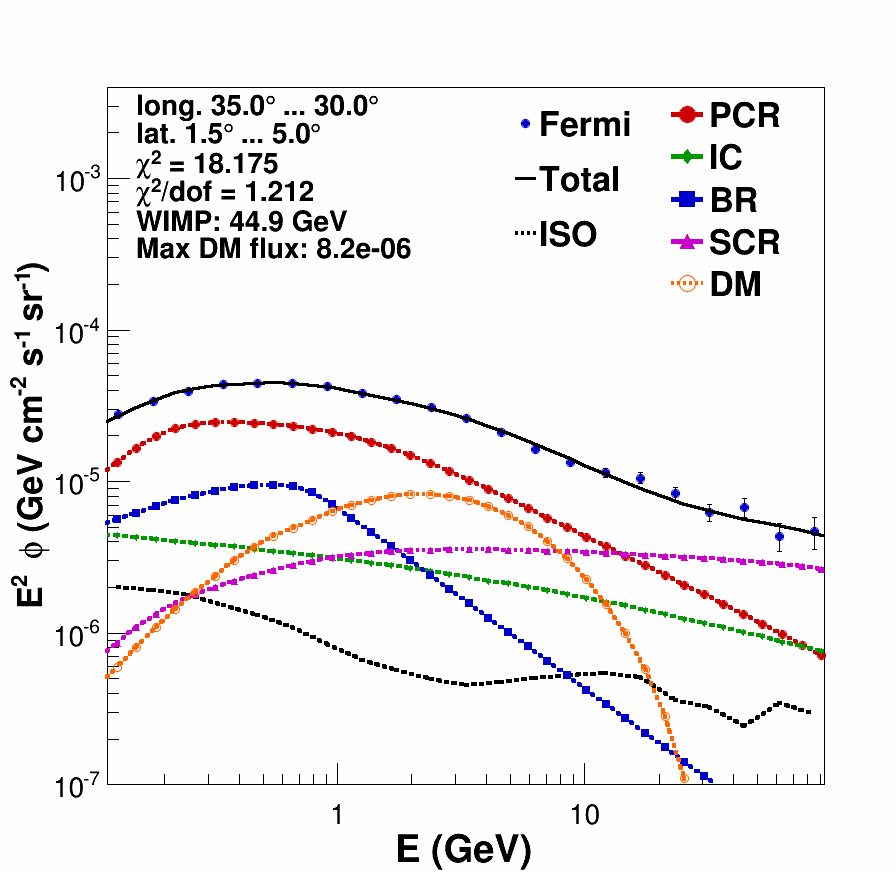}
\includegraphics[width=0.16\textwidth,height=0.16\textwidth,clip]{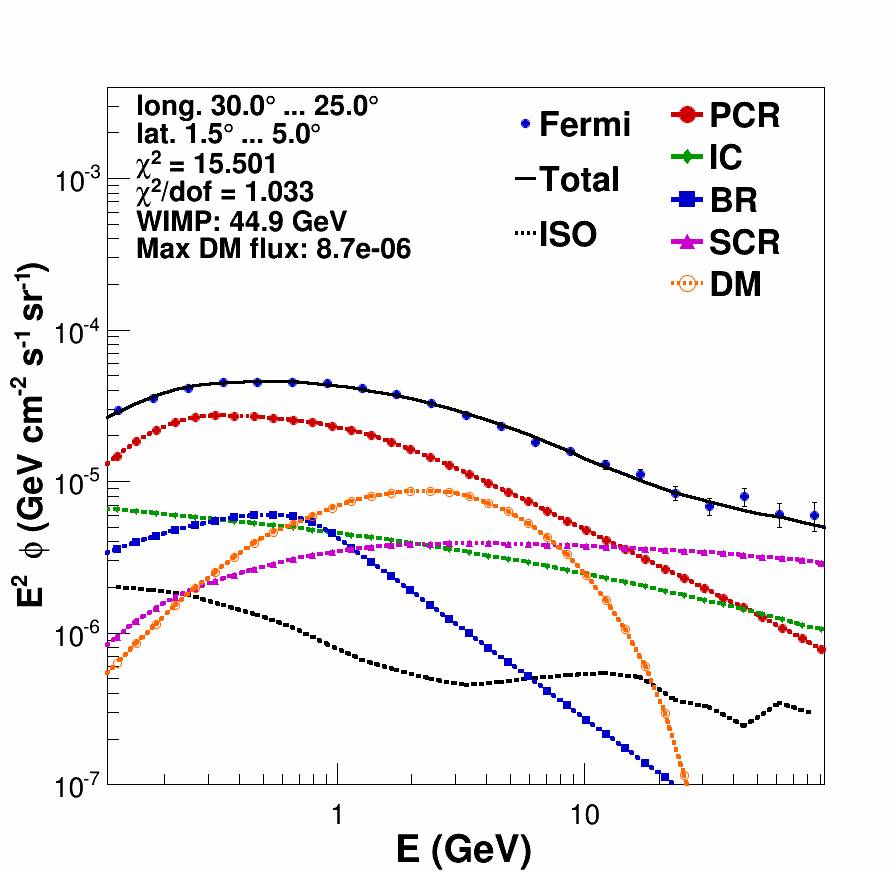}
\includegraphics[width=0.16\textwidth,height=0.16\textwidth,clip]{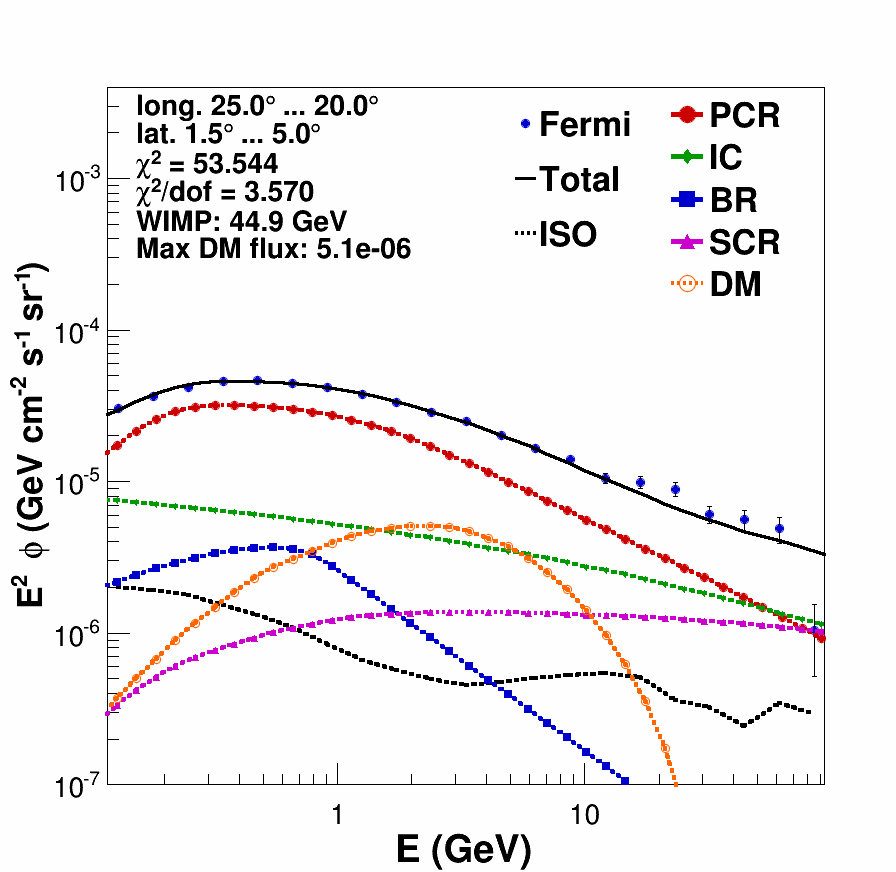}
\includegraphics[width=0.16\textwidth,height=0.16\textwidth,clip]{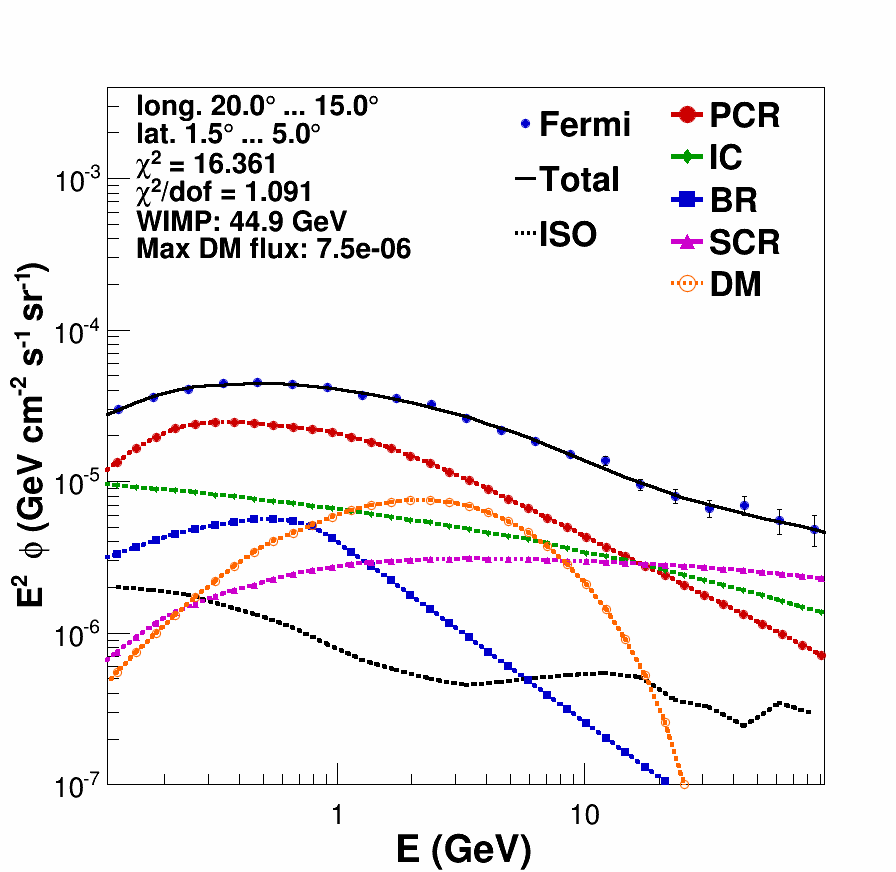}
\includegraphics[width=0.16\textwidth,height=0.16\textwidth,clip]{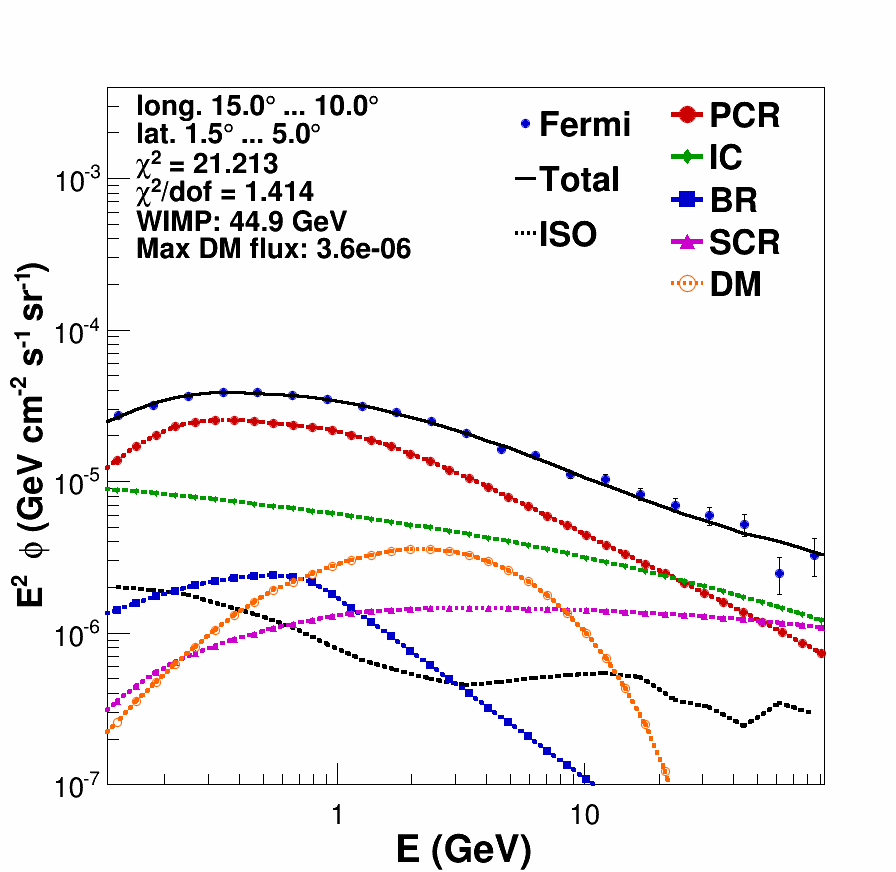}
\includegraphics[width=0.16\textwidth,height=0.16\textwidth,clip]{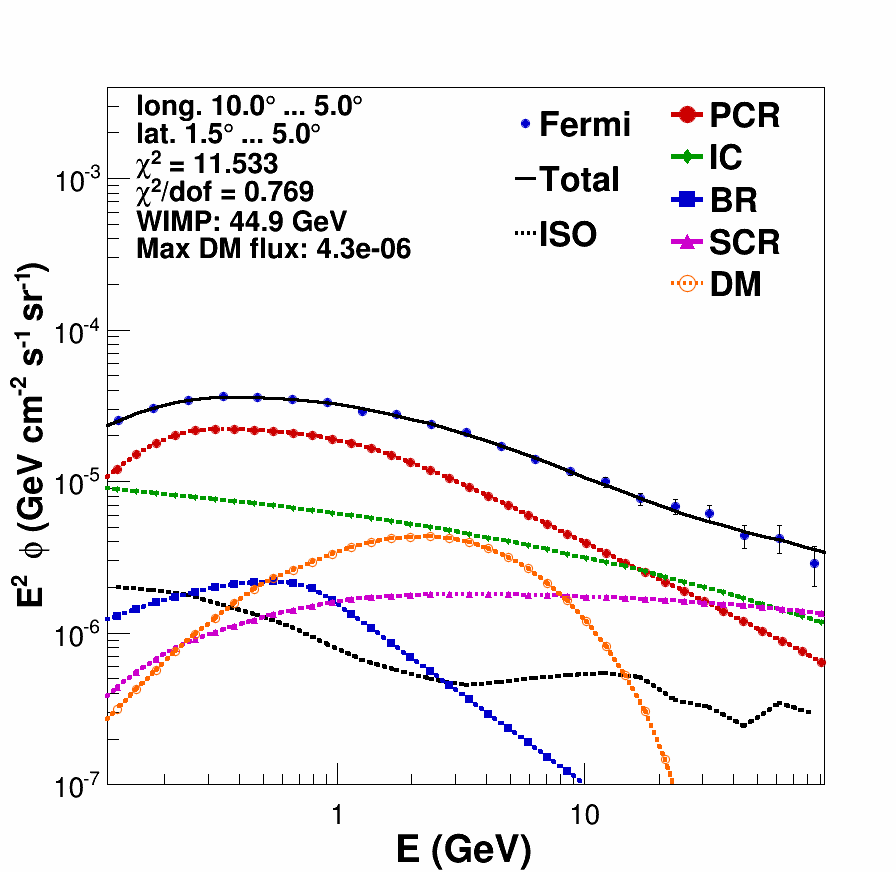}
\includegraphics[width=0.16\textwidth,height=0.16\textwidth,clip]{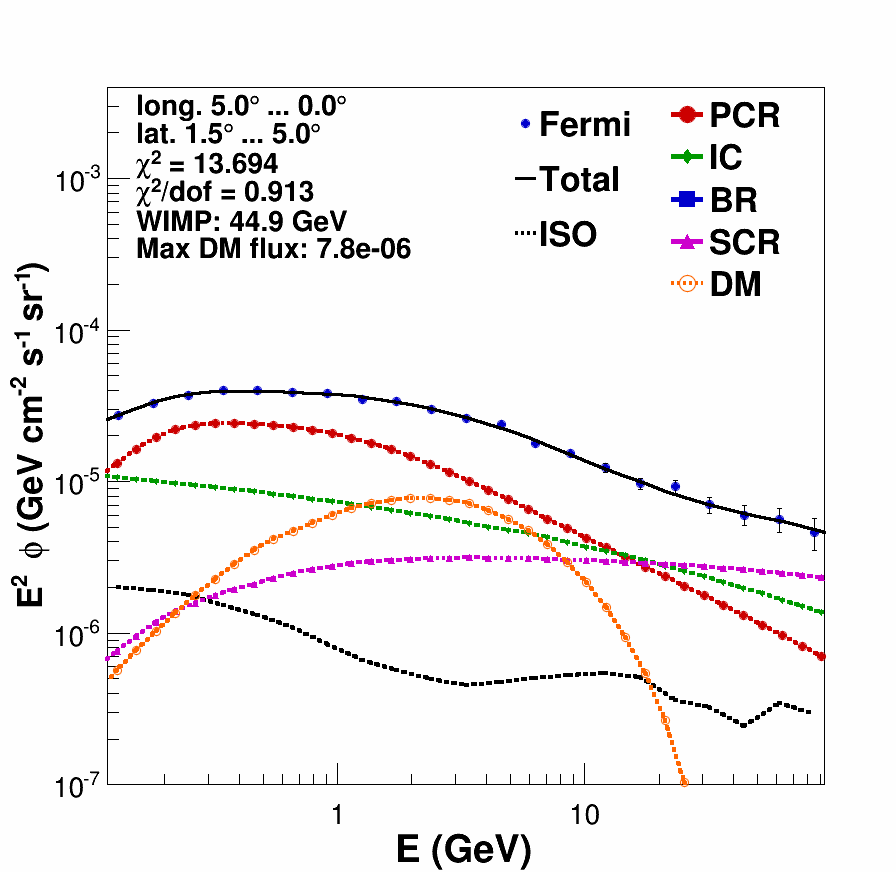}%%%%%r8a
\caption[]{Template fits for latitudes  with $1.5^\circ<b<5.0^\circ$ and longitudes decreasing from 180$^\circ$ to 0$^\circ$.} \label{F39}
\end{figure}
\begin{figure}
\centering
\includegraphics[width=0.16\textwidth,height=0.16\textwidth,clip]{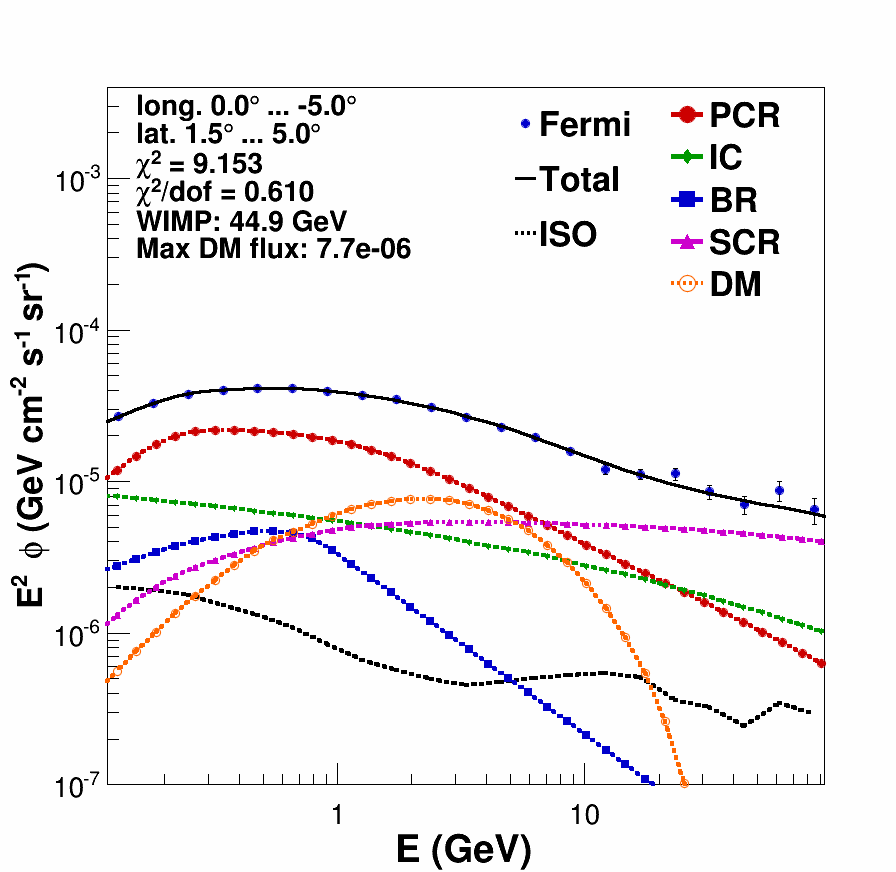}
\includegraphics[width=0.16\textwidth,height=0.16\textwidth,clip]{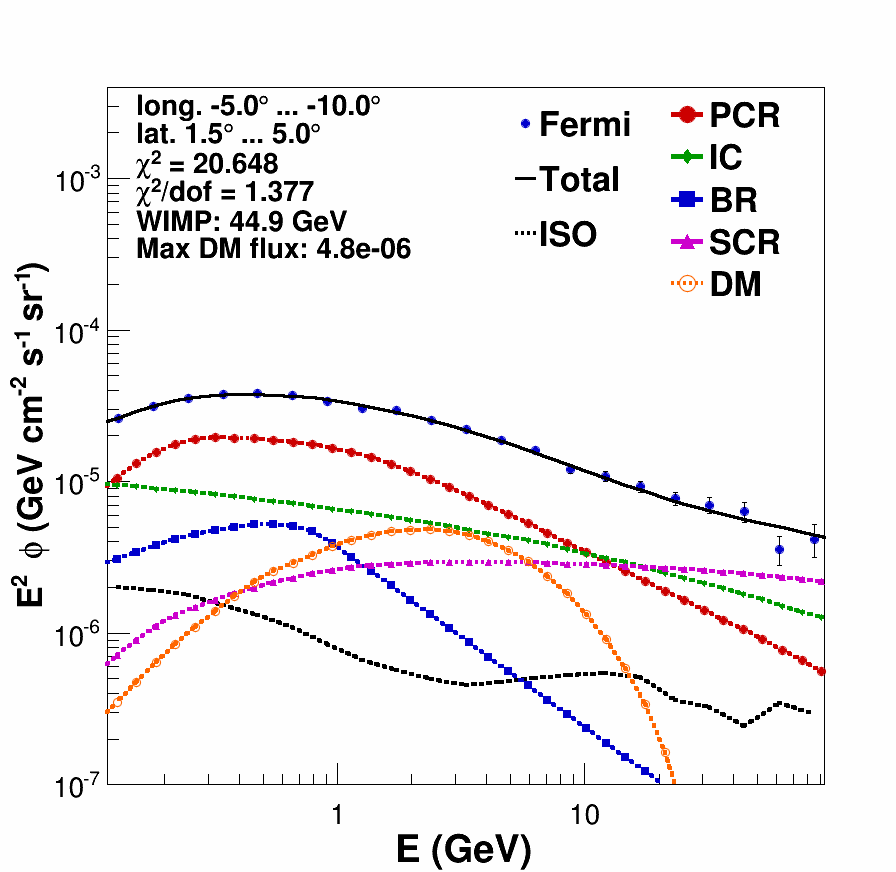}
\includegraphics[width=0.16\textwidth,height=0.16\textwidth,clip]{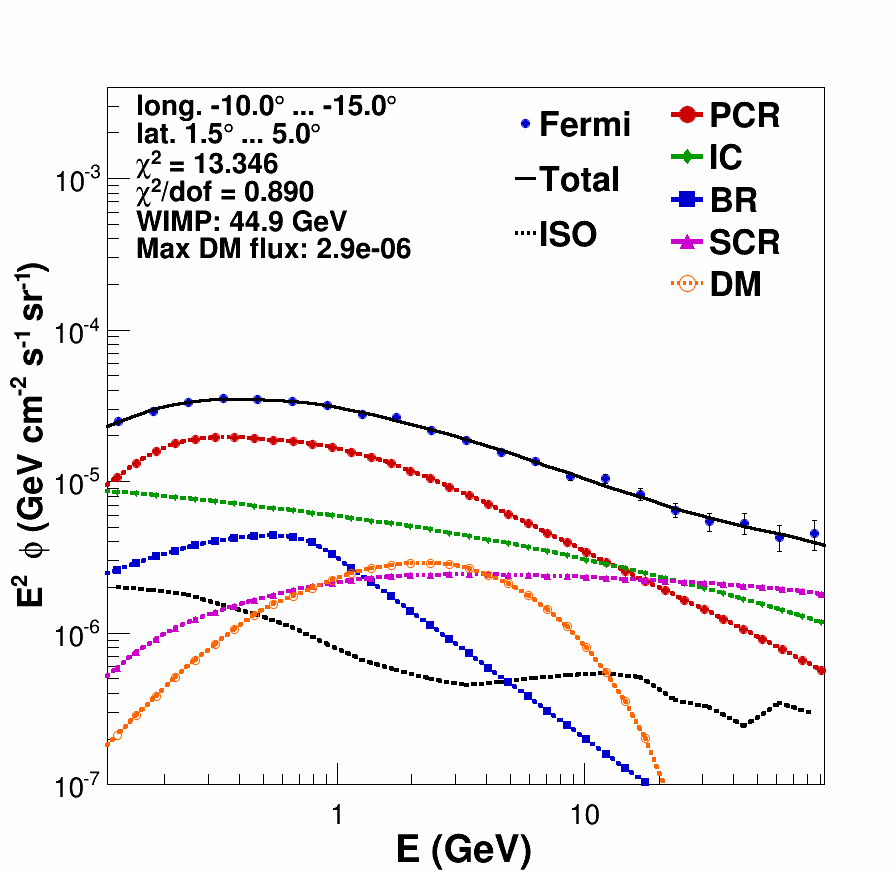}
\includegraphics[width=0.16\textwidth,height=0.16\textwidth,clip]{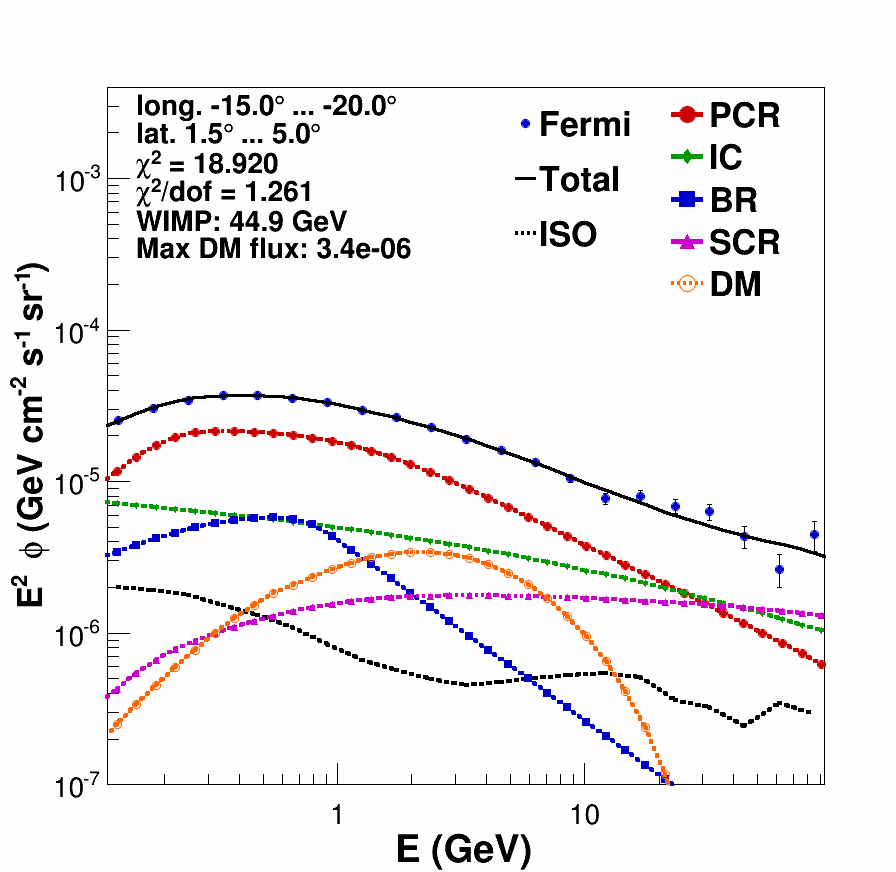}
\includegraphics[width=0.16\textwidth,height=0.16\textwidth,clip]{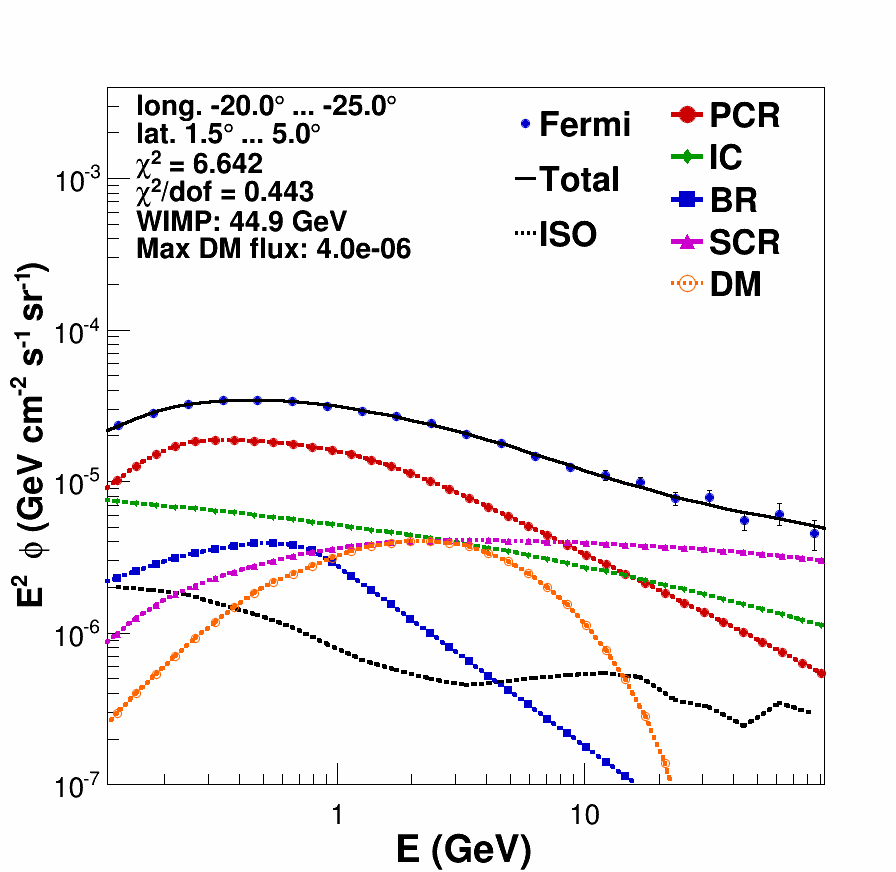}
\includegraphics[width=0.16\textwidth,height=0.16\textwidth,clip]{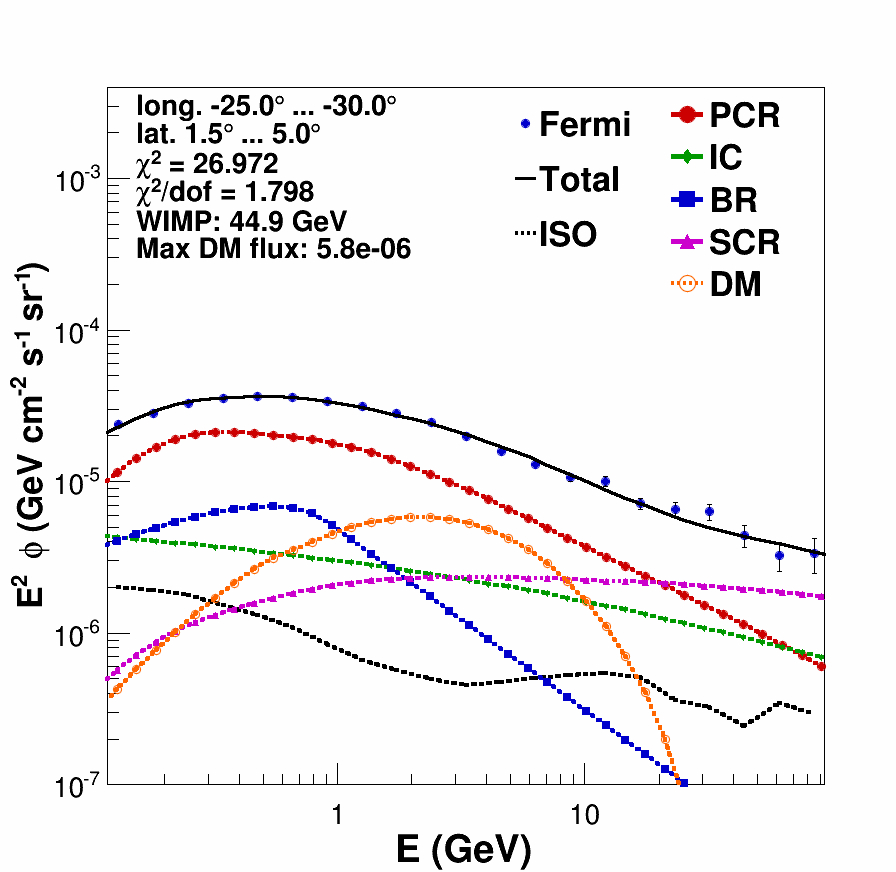}
\includegraphics[width=0.16\textwidth,height=0.16\textwidth,clip]{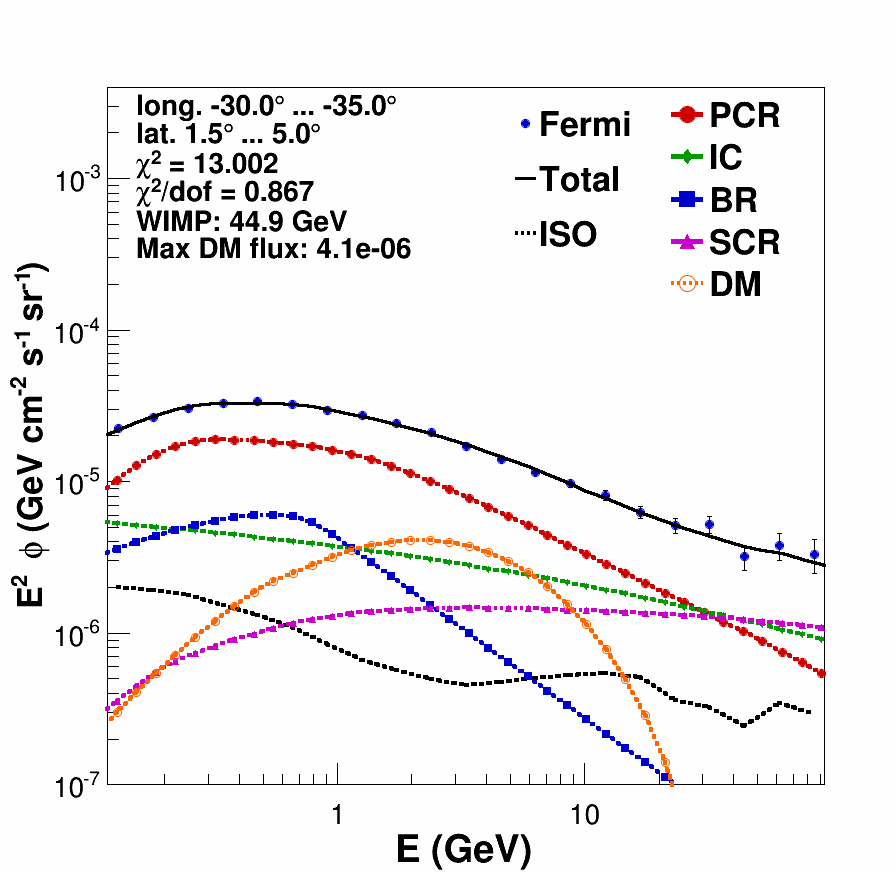}
\includegraphics[width=0.16\textwidth,height=0.16\textwidth,clip]{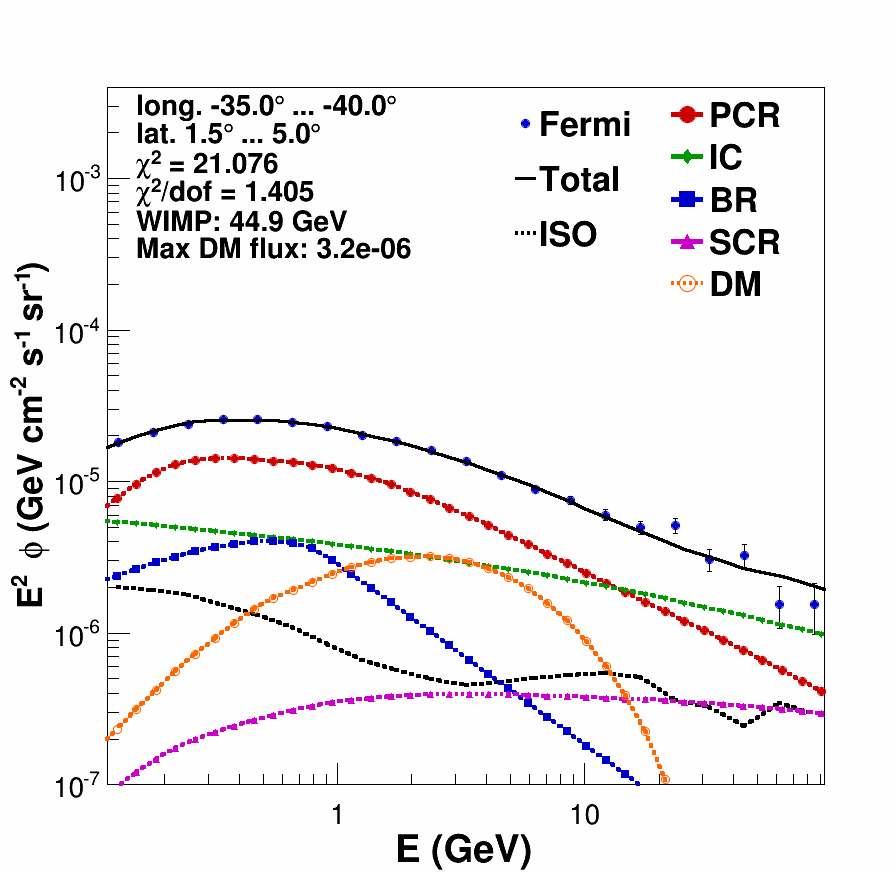}
\includegraphics[width=0.16\textwidth,height=0.16\textwidth,clip]{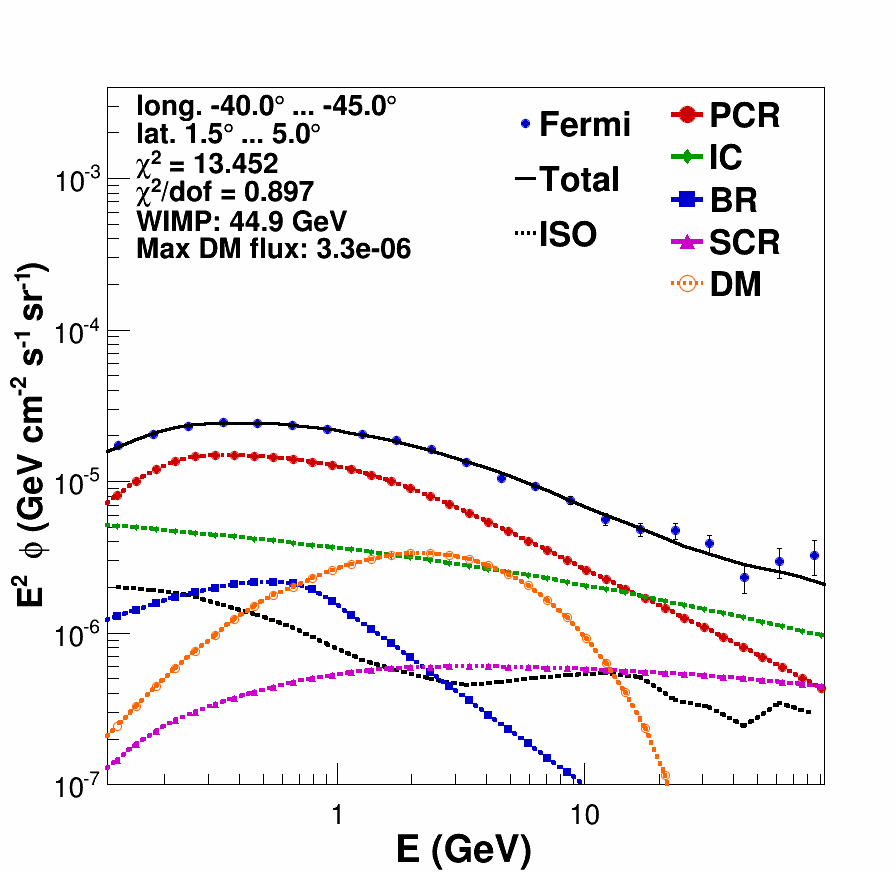}
\includegraphics[width=0.16\textwidth,height=0.16\textwidth,clip]{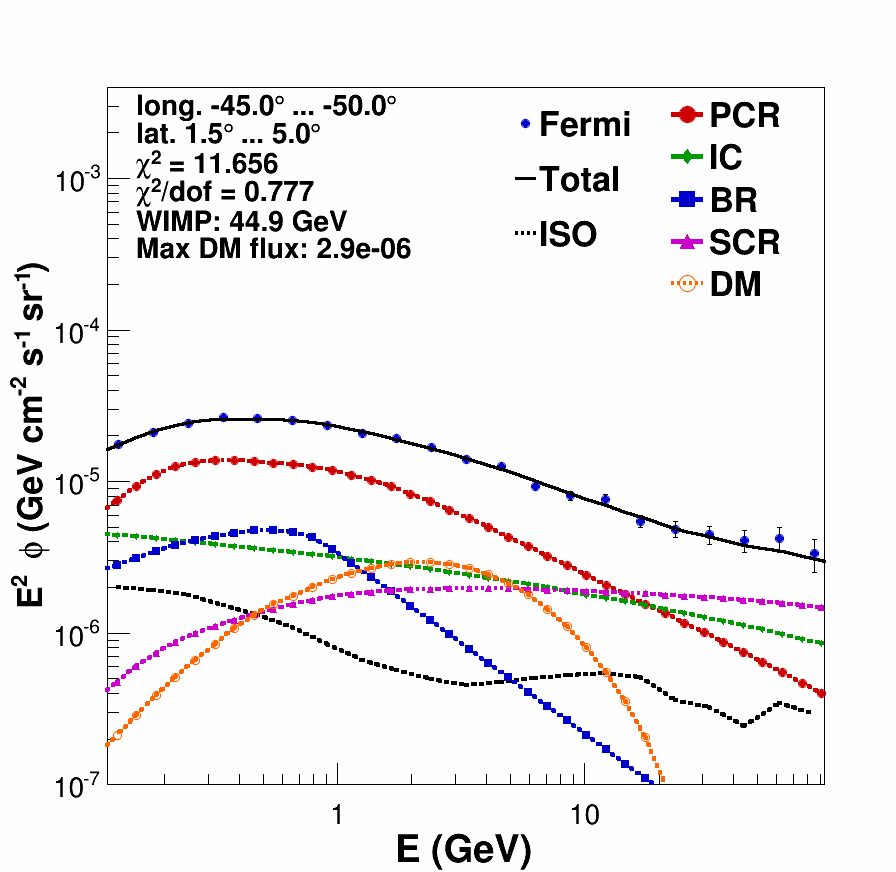}
\includegraphics[width=0.16\textwidth,height=0.16\textwidth,clip]{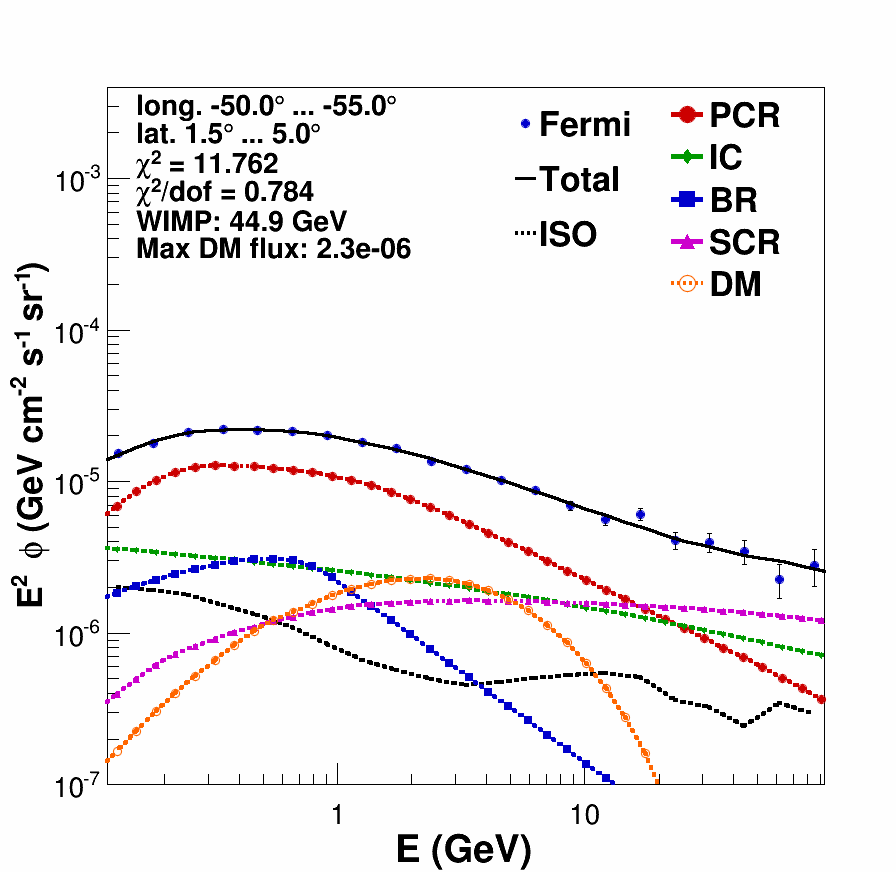}
\includegraphics[width=0.16\textwidth,height=0.16\textwidth,clip]{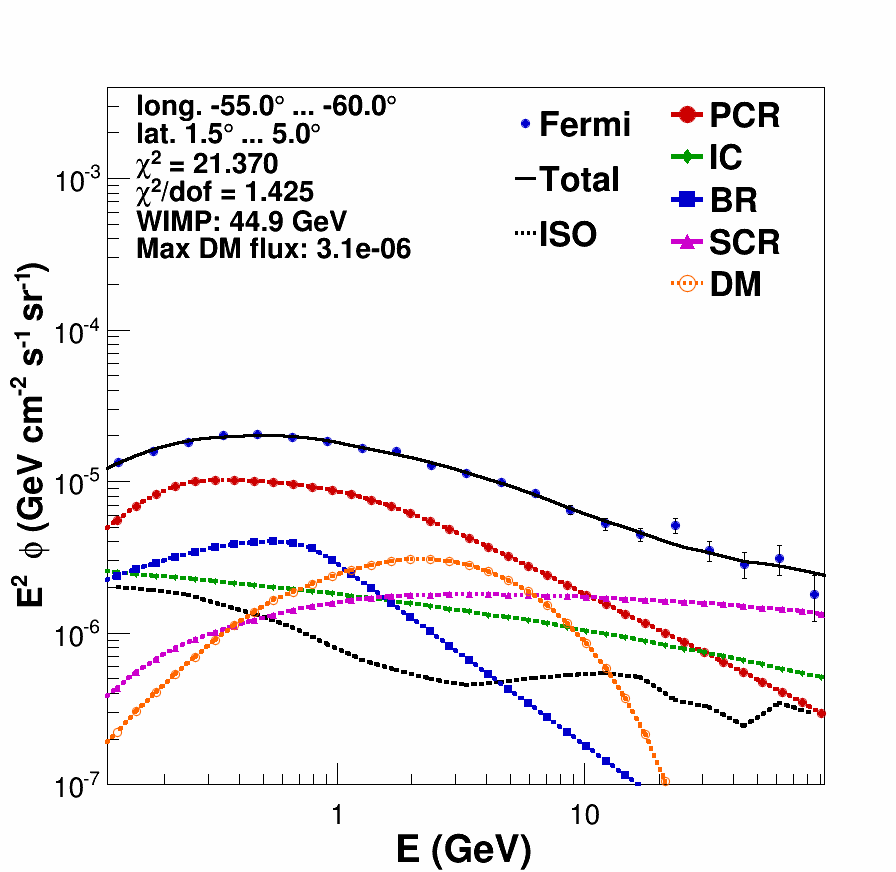}
\includegraphics[width=0.16\textwidth,height=0.16\textwidth,clip]{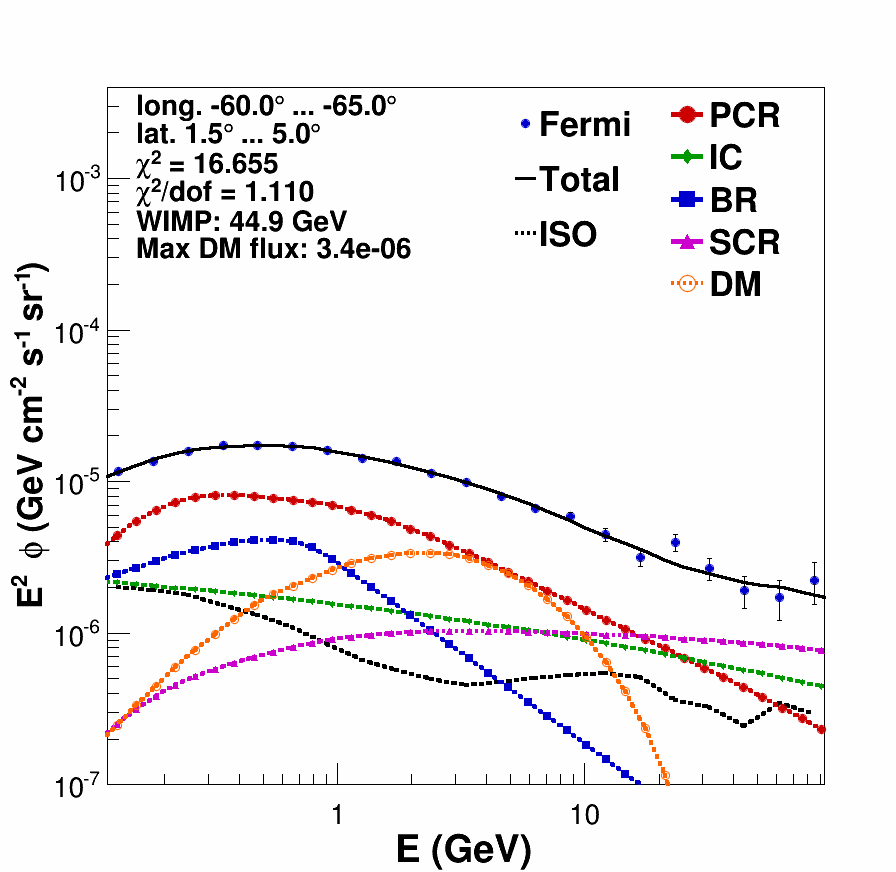}
\includegraphics[width=0.16\textwidth,height=0.16\textwidth,clip]{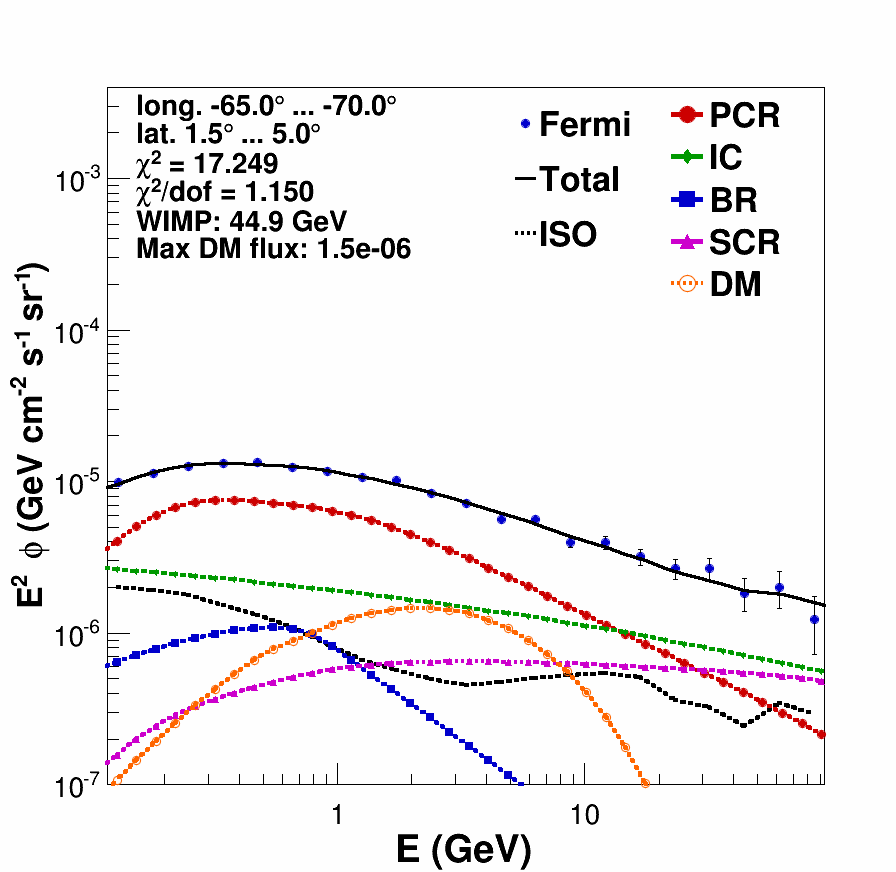}
\includegraphics[width=0.16\textwidth,height=0.16\textwidth,clip]{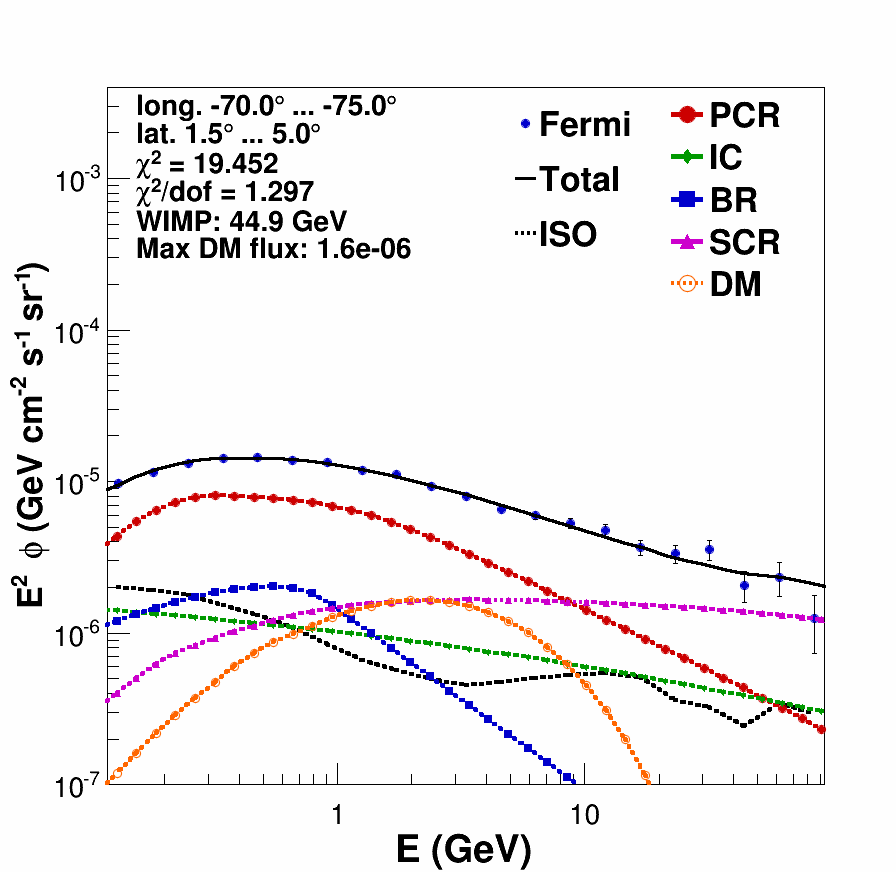}
\includegraphics[width=0.16\textwidth,height=0.16\textwidth,clip]{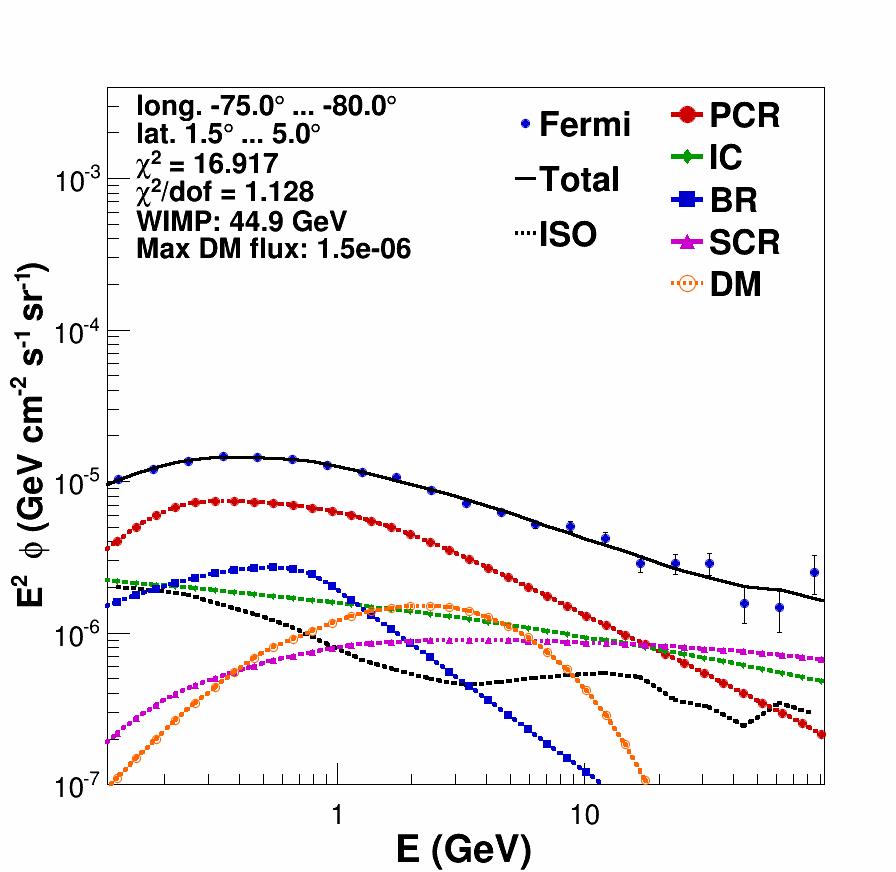}
\includegraphics[width=0.16\textwidth,height=0.16\textwidth,clip]{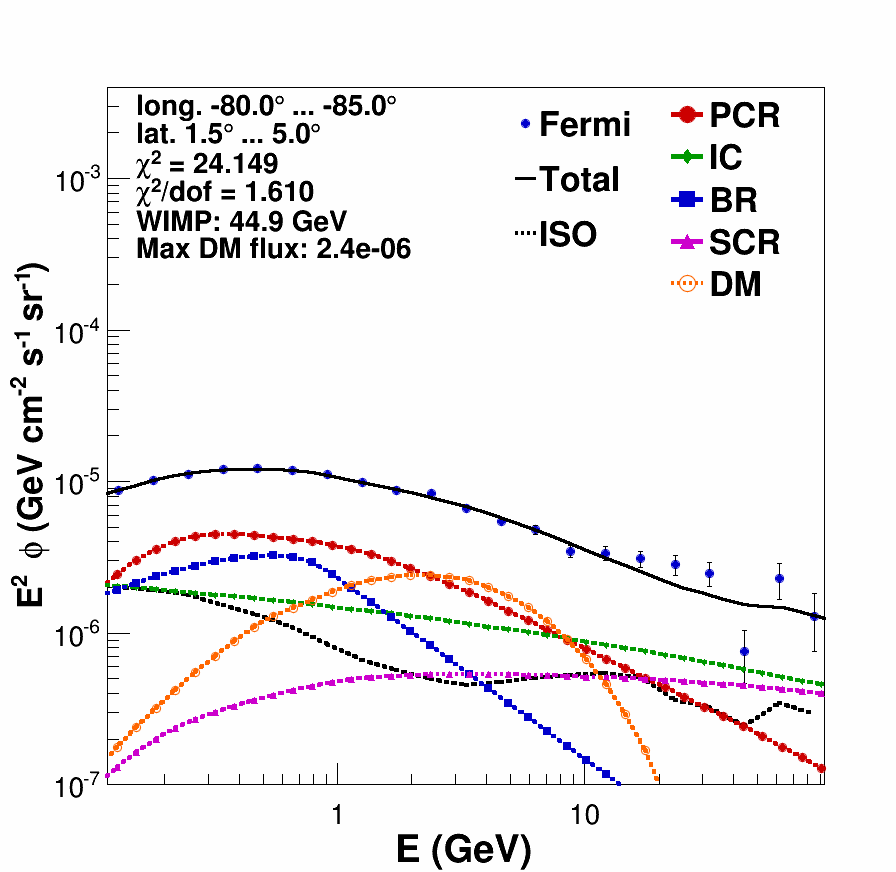}
\includegraphics[width=0.16\textwidth,height=0.16\textwidth,clip]{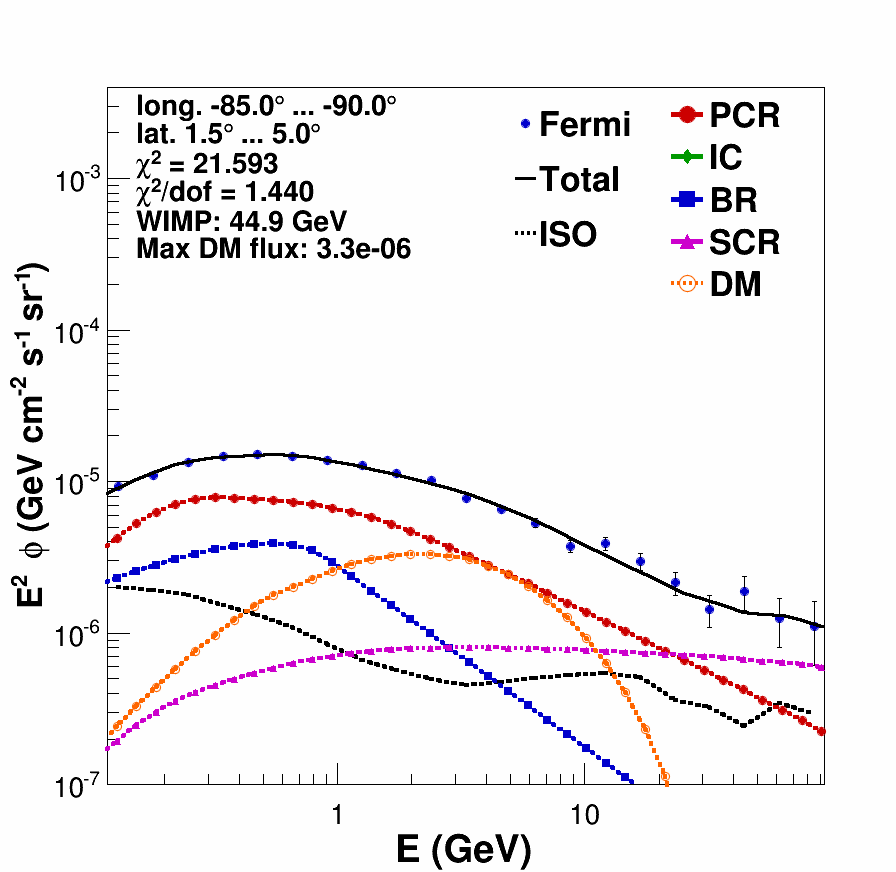}
\includegraphics[width=0.16\textwidth,height=0.16\textwidth,clip]{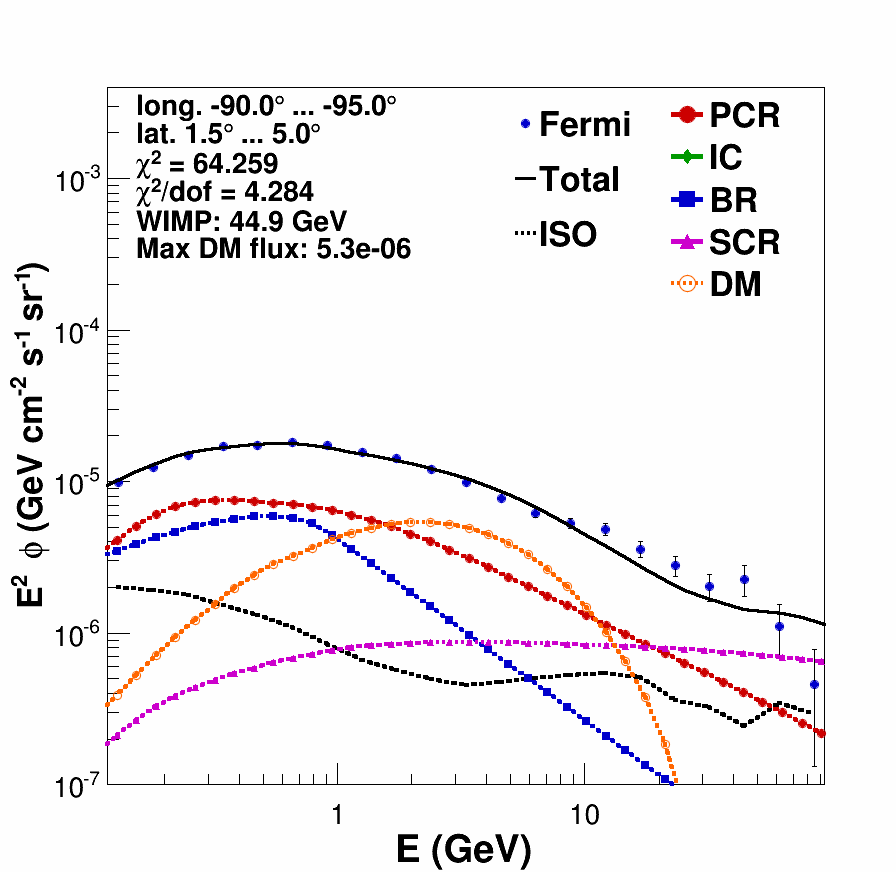}
\includegraphics[width=0.16\textwidth,height=0.16\textwidth,clip]{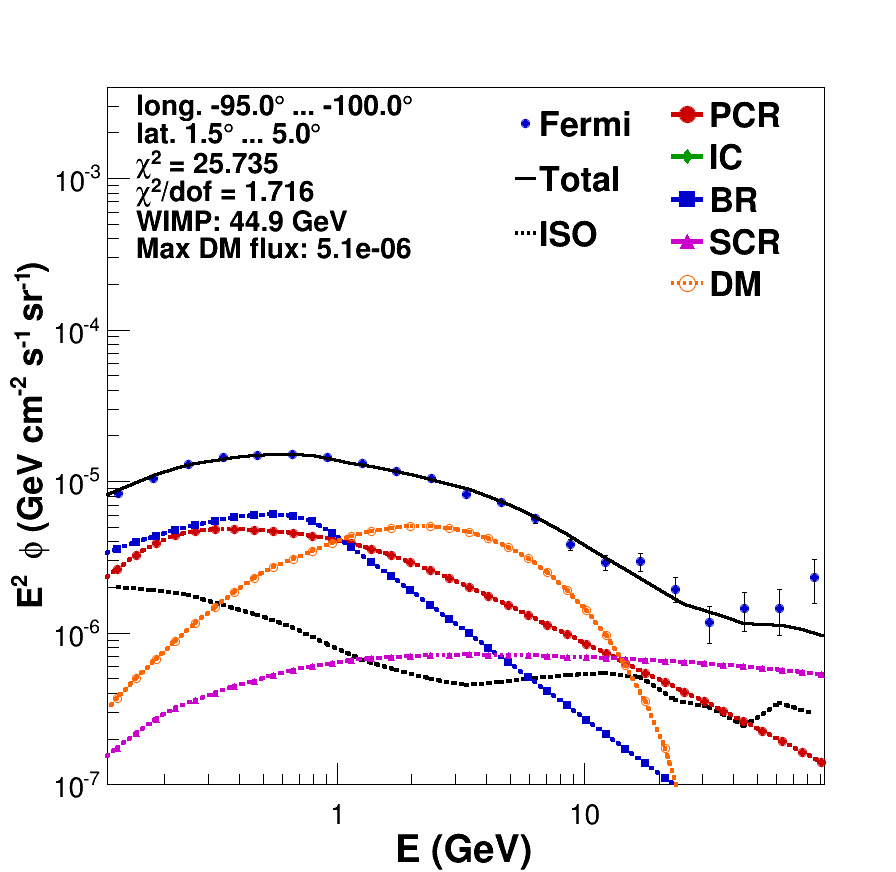}
\includegraphics[width=0.16\textwidth,height=0.16\textwidth,clip]{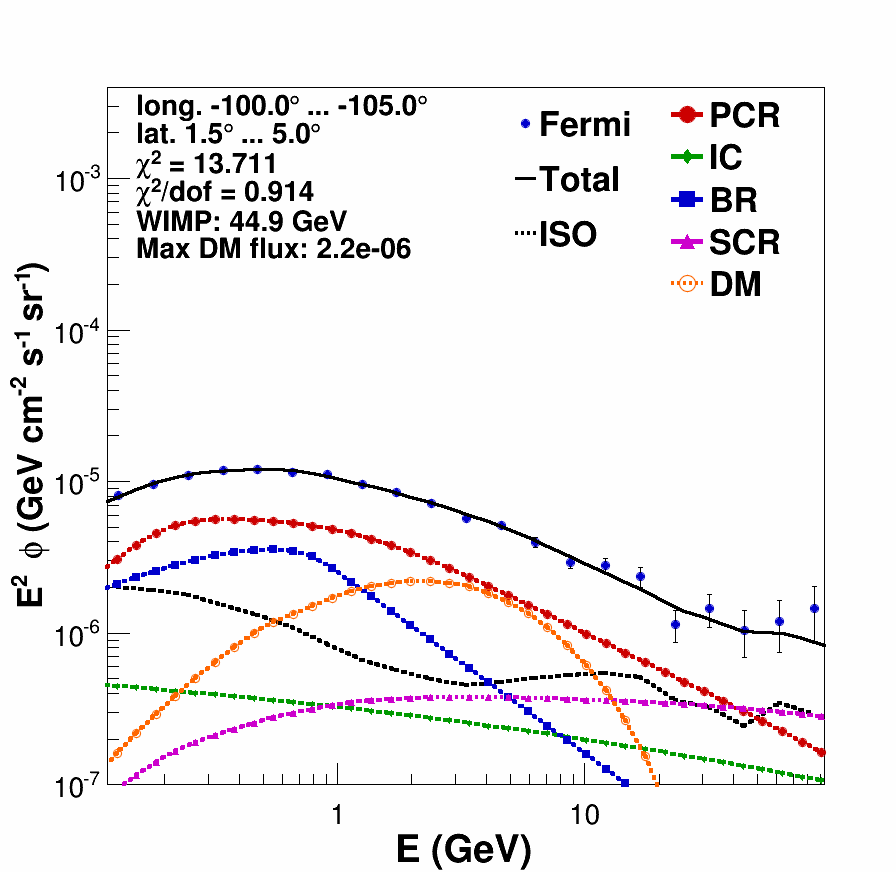}
\includegraphics[width=0.16\textwidth,height=0.16\textwidth,clip]{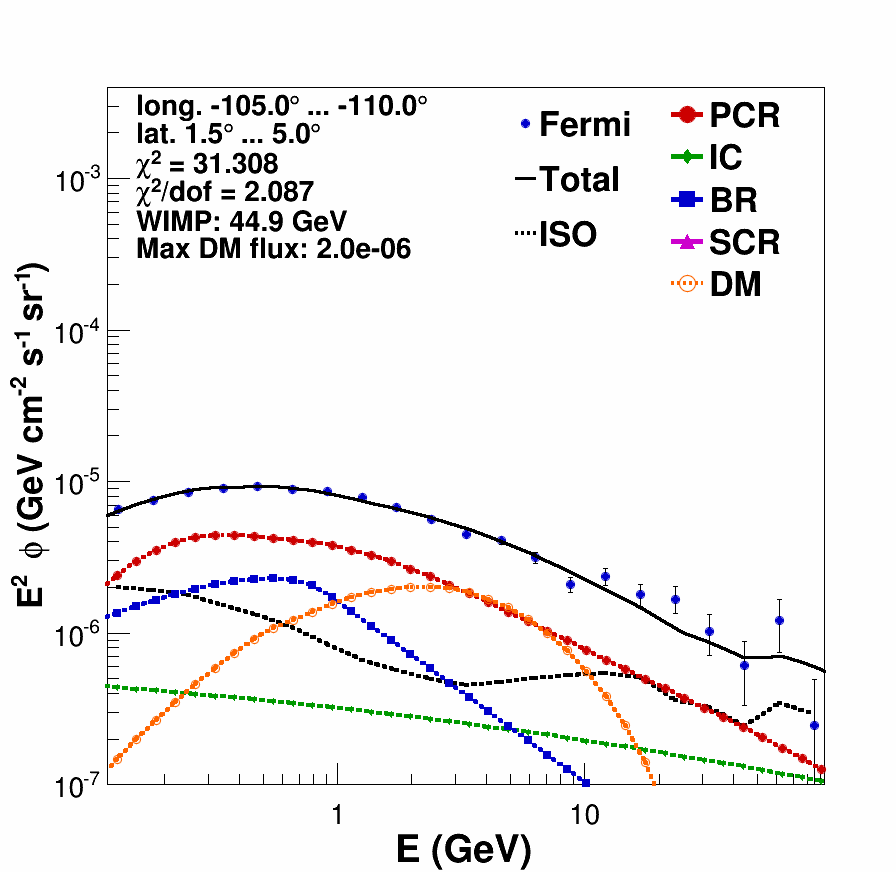}
\includegraphics[width=0.16\textwidth,height=0.16\textwidth,clip]{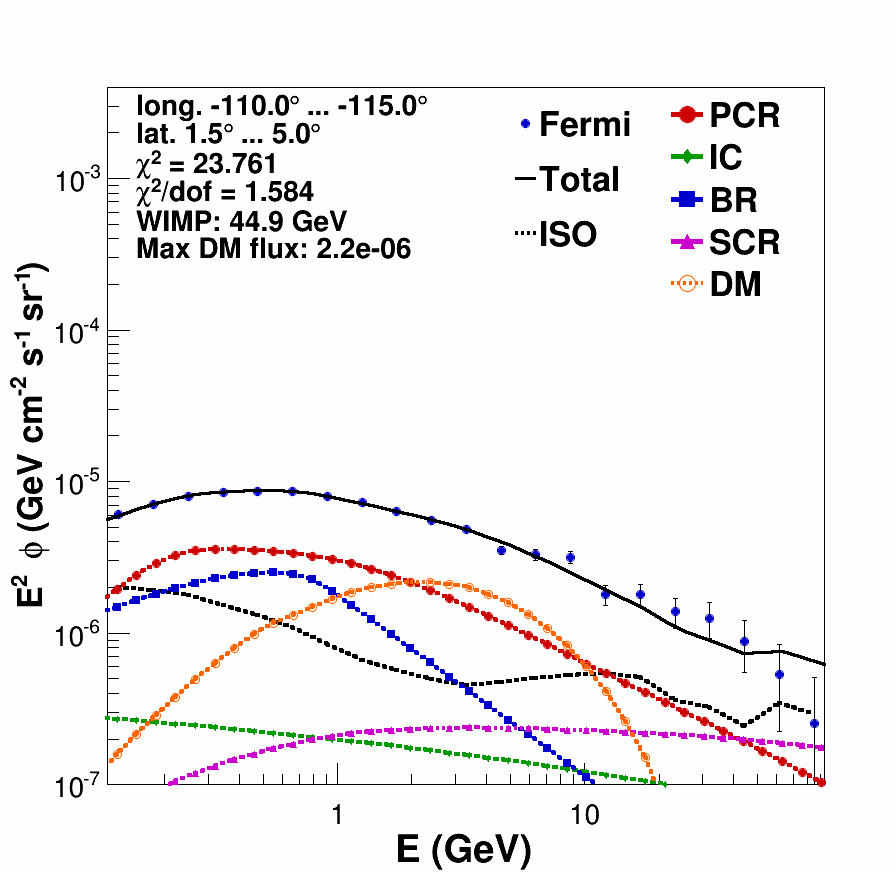}
\includegraphics[width=0.16\textwidth,height=0.16\textwidth,clip]{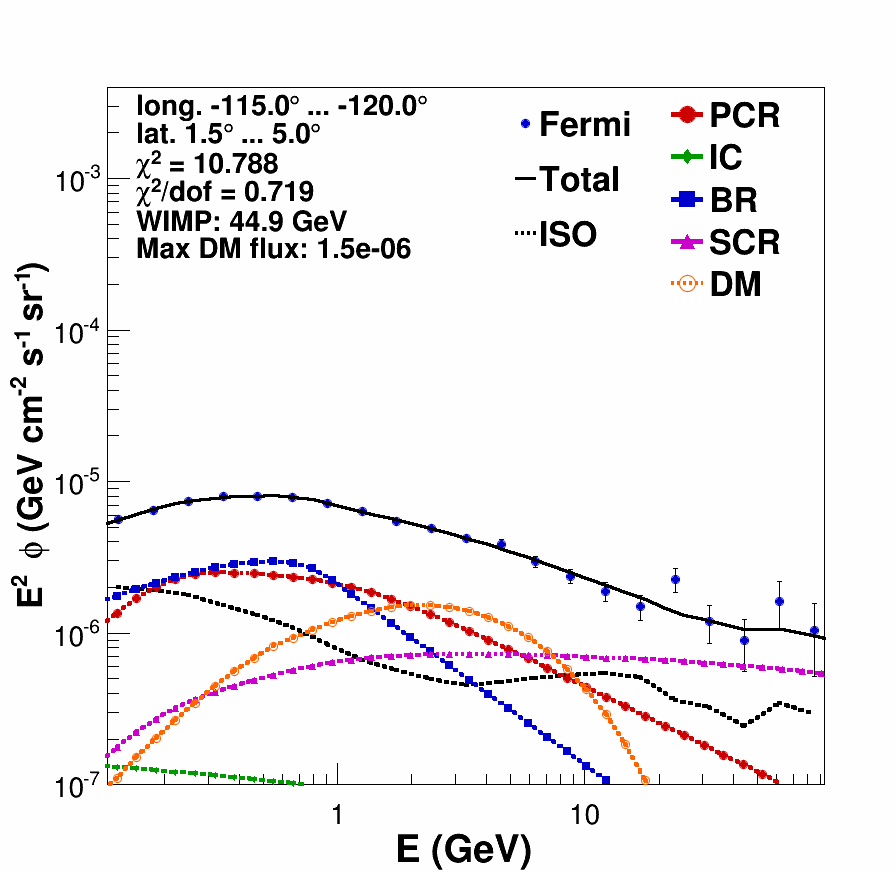}
\includegraphics[width=0.16\textwidth,height=0.16\textwidth,clip]{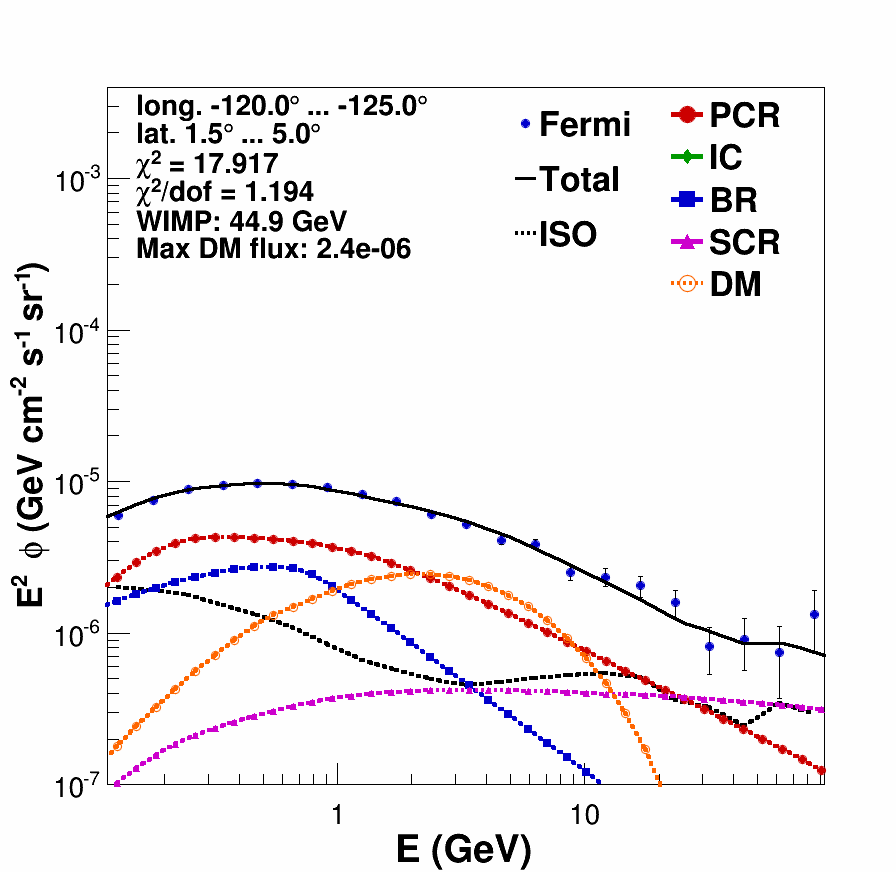}
\includegraphics[width=0.16\textwidth,height=0.16\textwidth,clip]{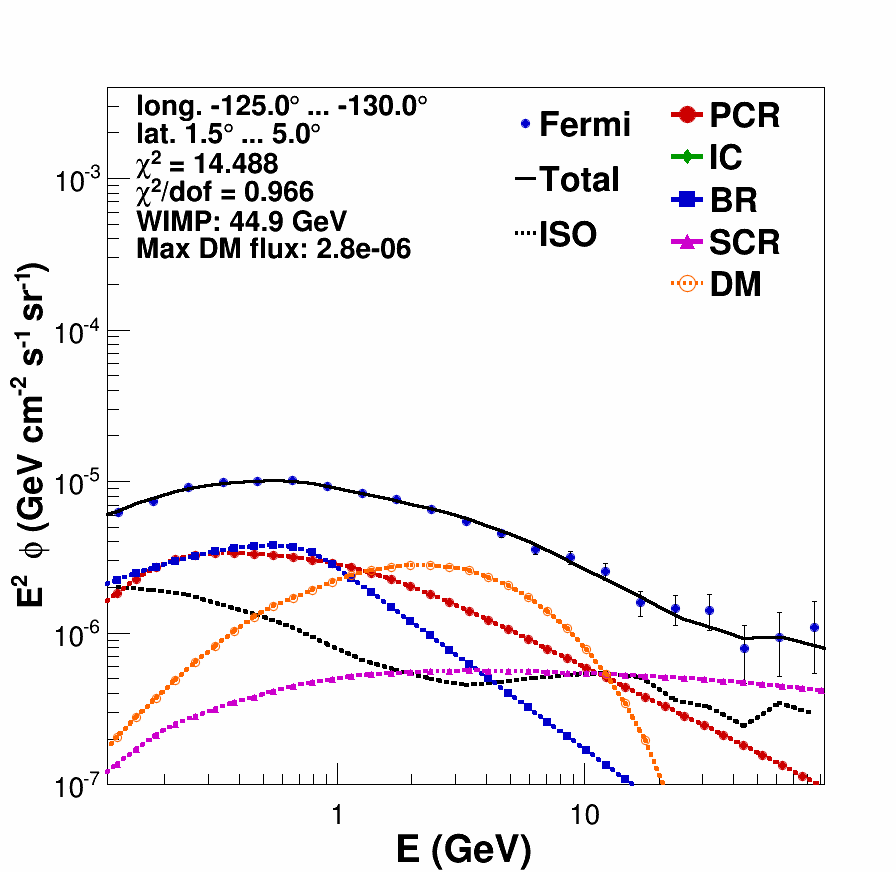}
\includegraphics[width=0.16\textwidth,height=0.16\textwidth,clip]{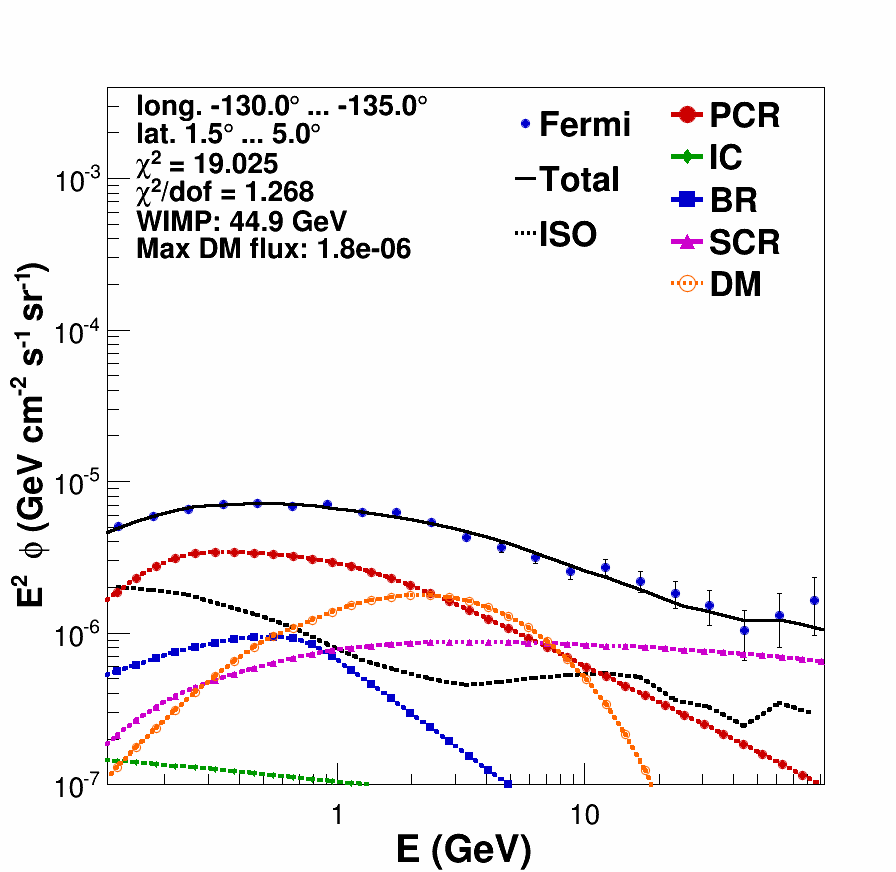}
\includegraphics[width=0.16\textwidth,height=0.16\textwidth,clip]{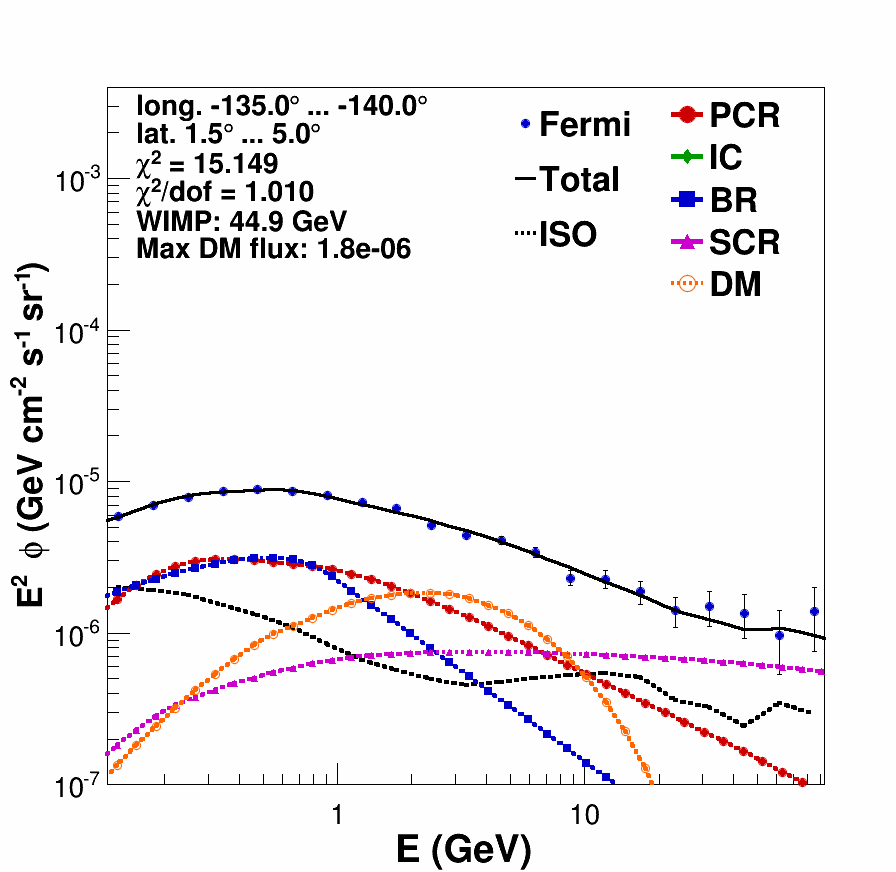}
\includegraphics[width=0.16\textwidth,height=0.16\textwidth,clip]{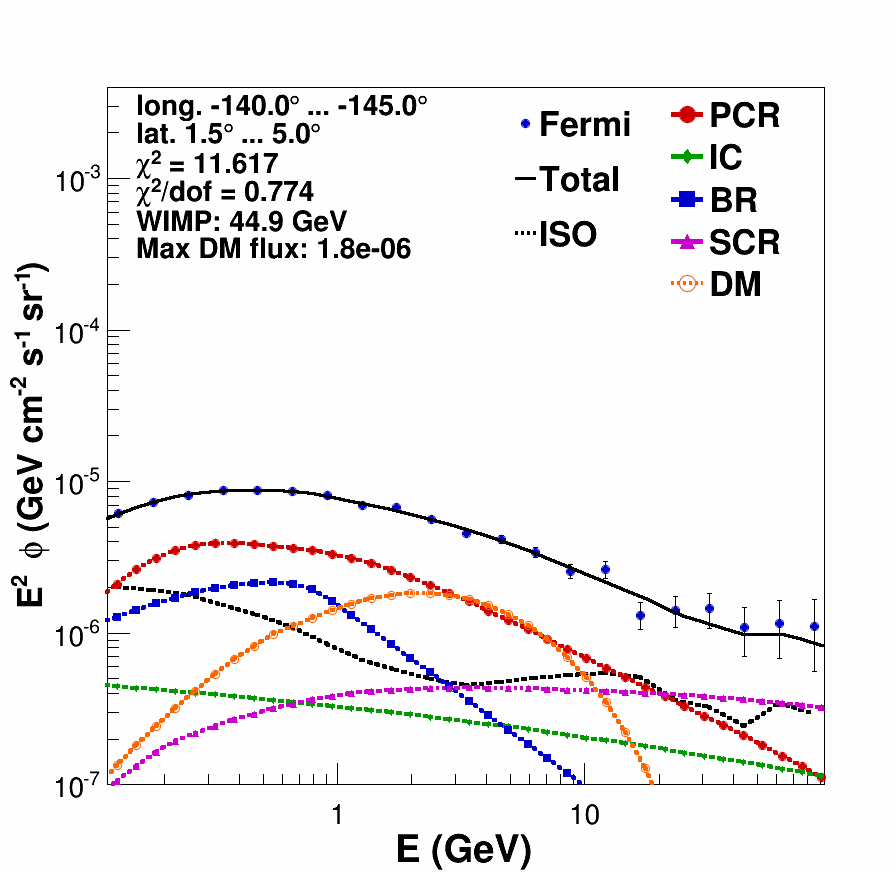}
\includegraphics[width=0.16\textwidth,height=0.16\textwidth,clip]{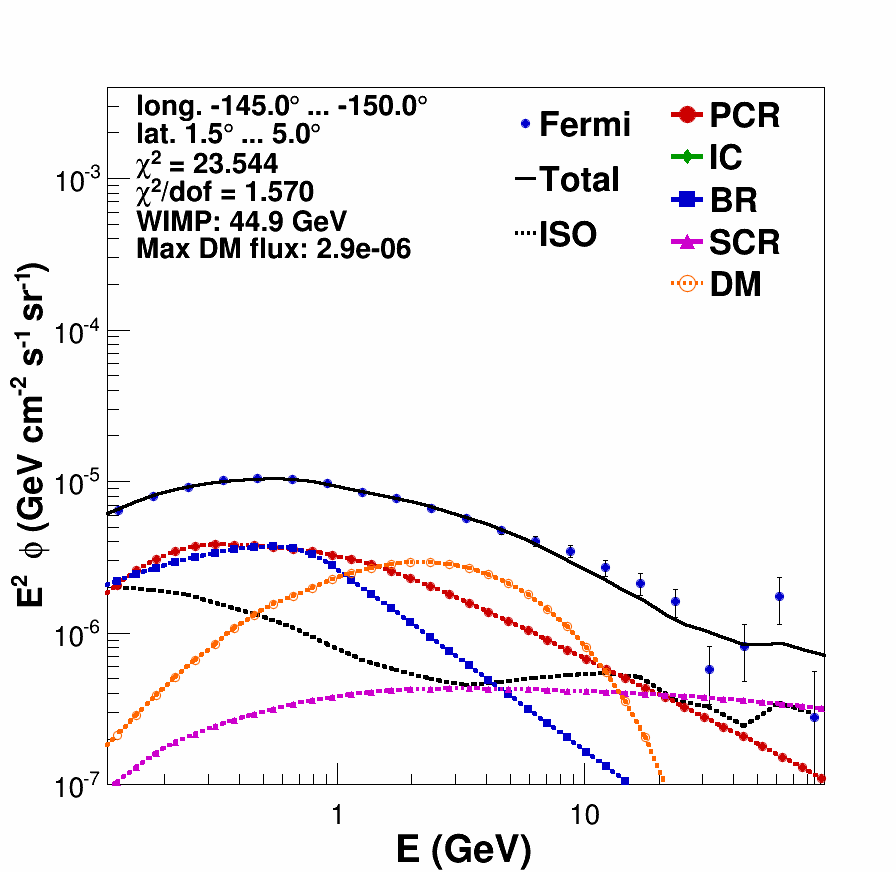}
\includegraphics[width=0.16\textwidth,height=0.16\textwidth,clip]{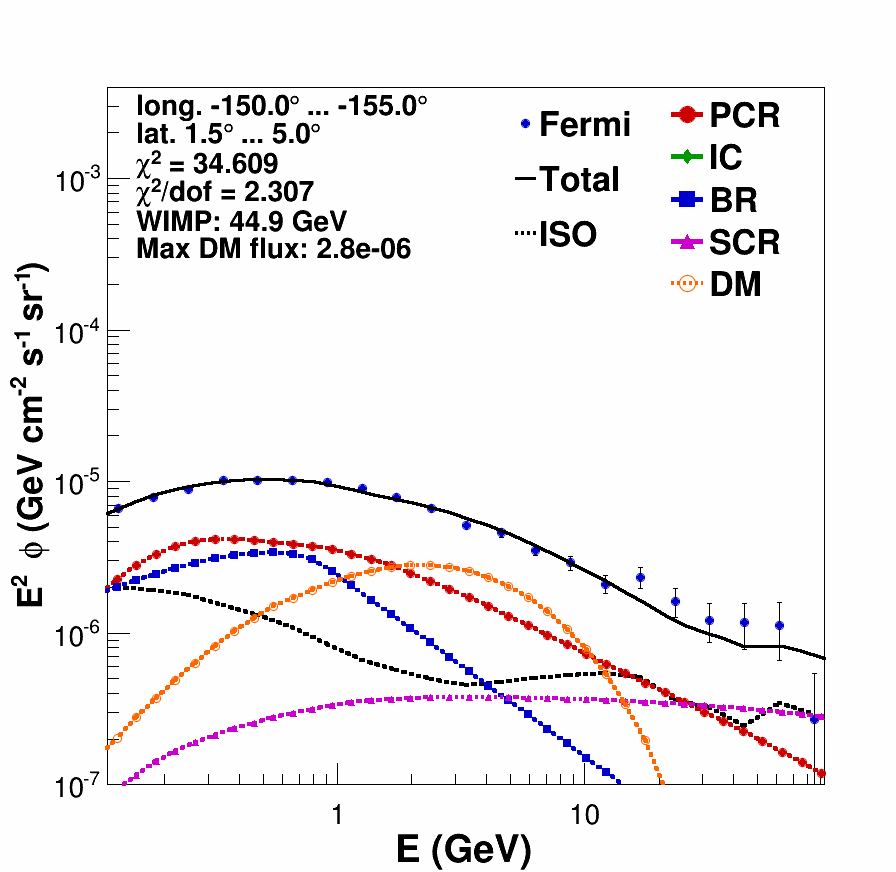}
\includegraphics[width=0.16\textwidth,height=0.16\textwidth,clip]{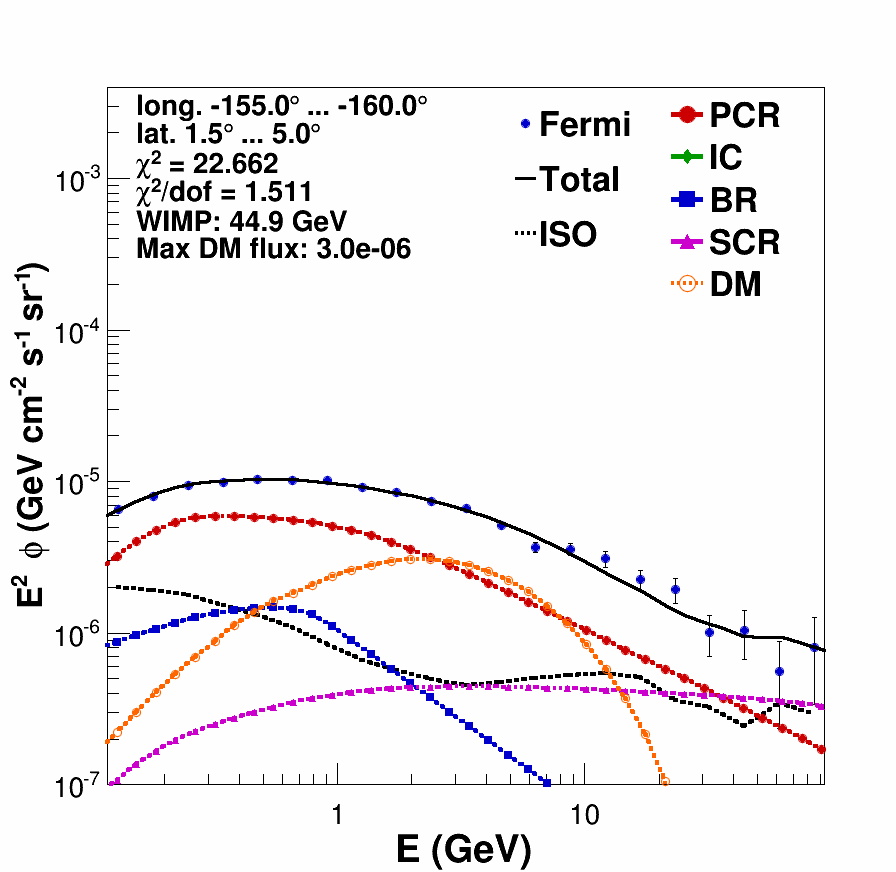}
\includegraphics[width=0.16\textwidth,height=0.16\textwidth,clip]{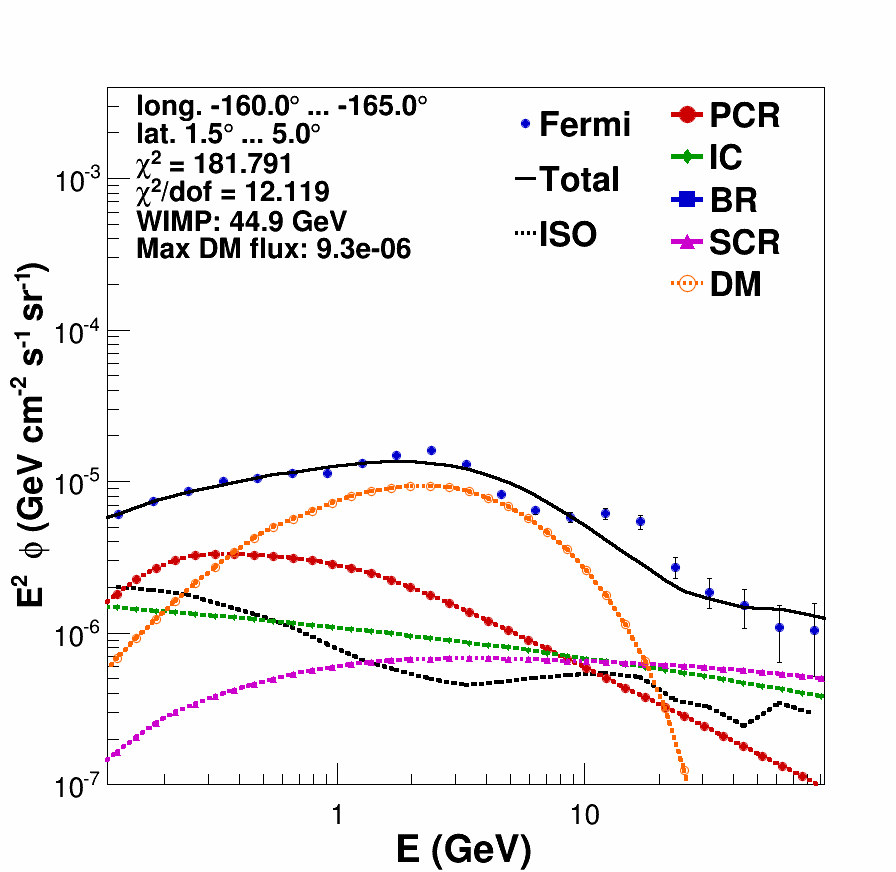}
\includegraphics[width=0.16\textwidth,height=0.16\textwidth,clip]{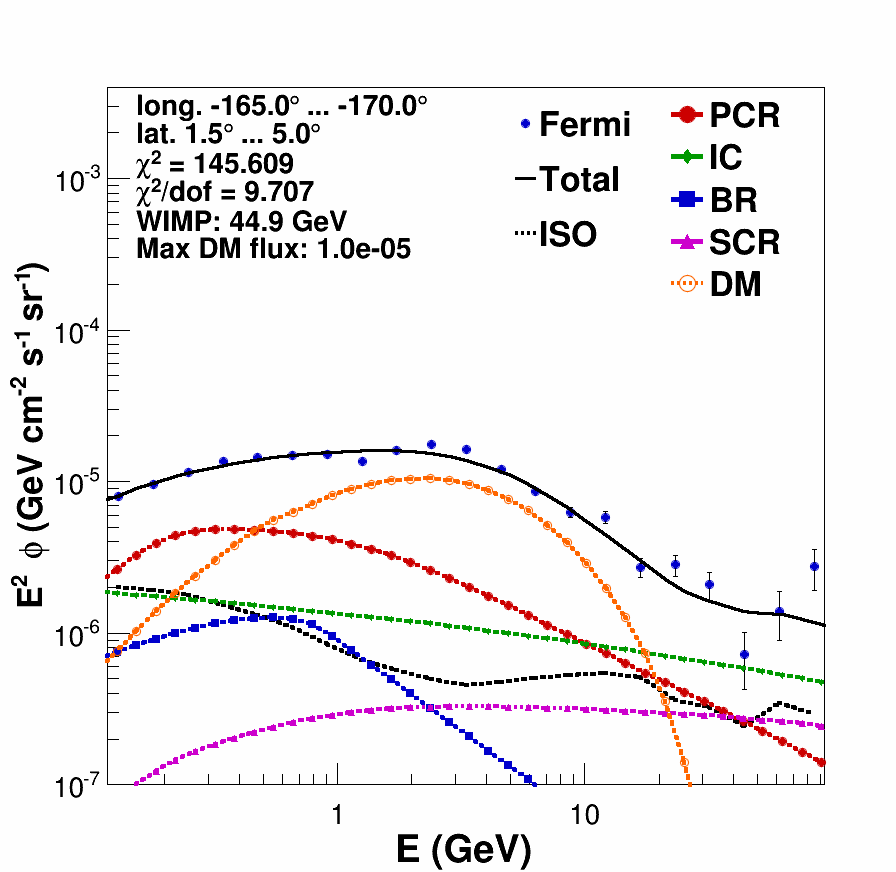}
\includegraphics[width=0.16\textwidth,height=0.16\textwidth,clip]{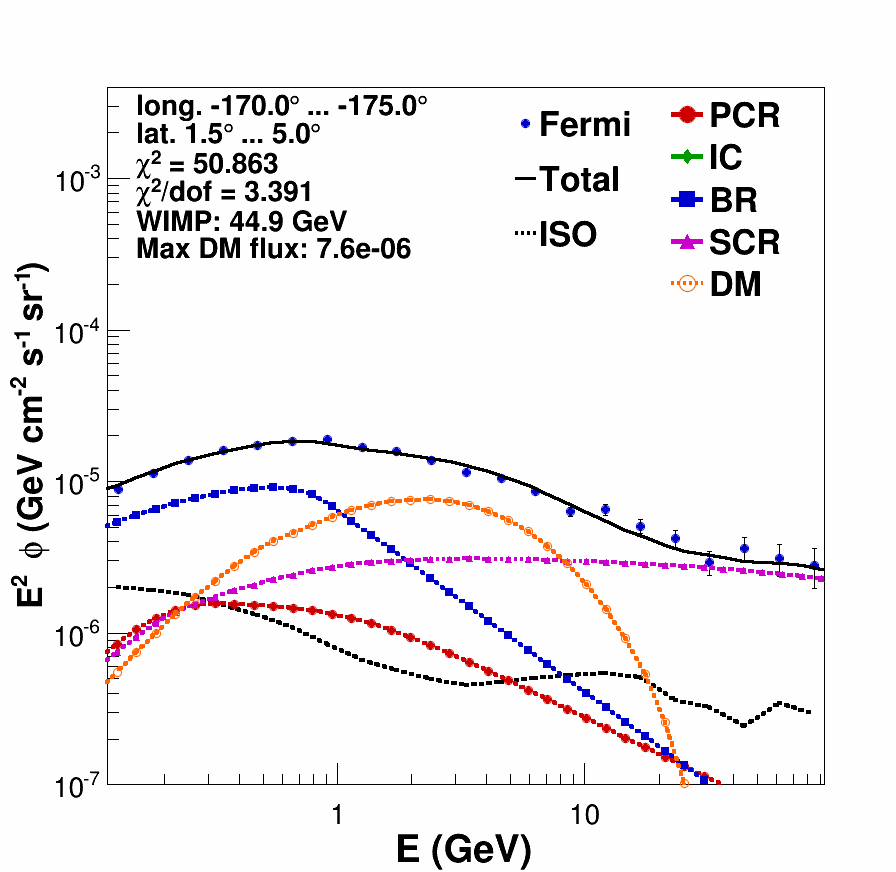}
\includegraphics[width=0.16\textwidth,height=0.16\textwidth,clip]{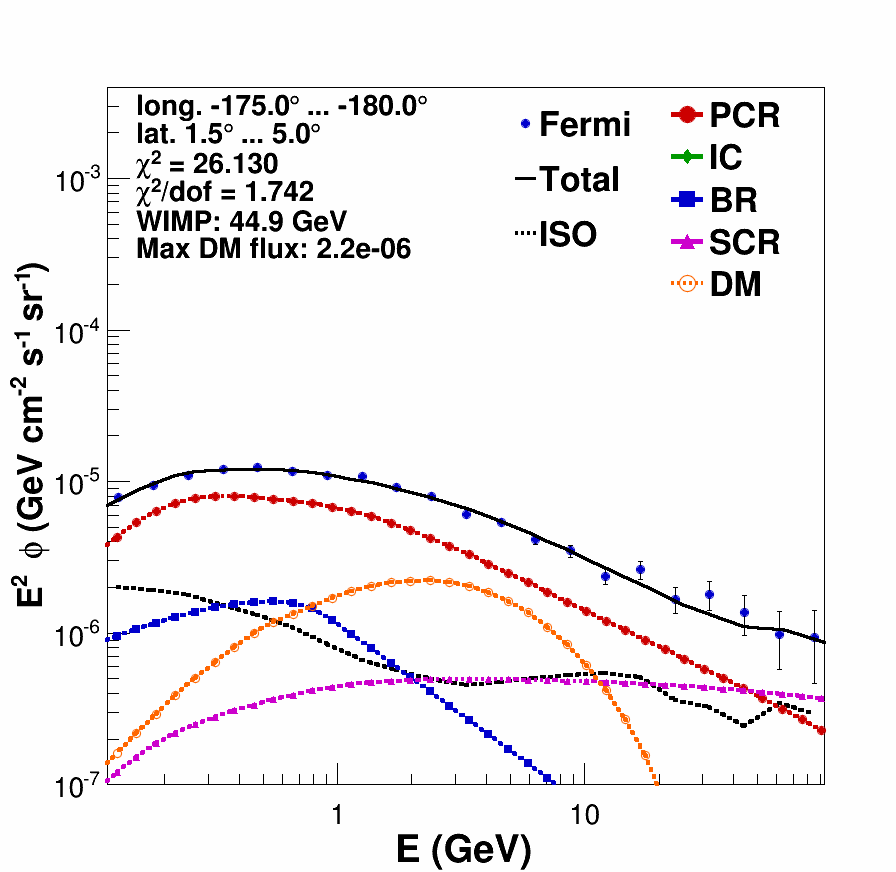}%%%%%%r8b
\caption[]{Template fits for latitudes  with $1.5^\circ<b<5.0^\circ$ and longitudes decreasing from 0$^\circ$ to -180$^\circ$.} \label{F40}
\end{figure}
\begin{figure}
\centering
\includegraphics[width=0.16\textwidth,height=0.16\textwidth,clip]{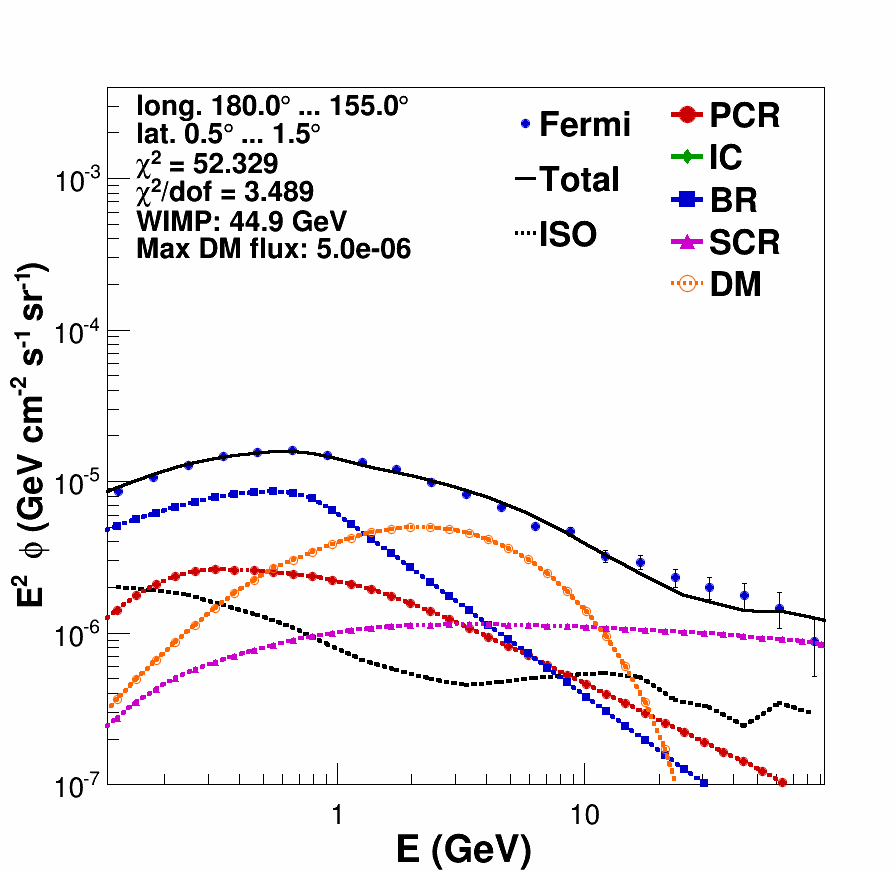}
\includegraphics[width=0.16\textwidth,height=0.16\textwidth,clip]{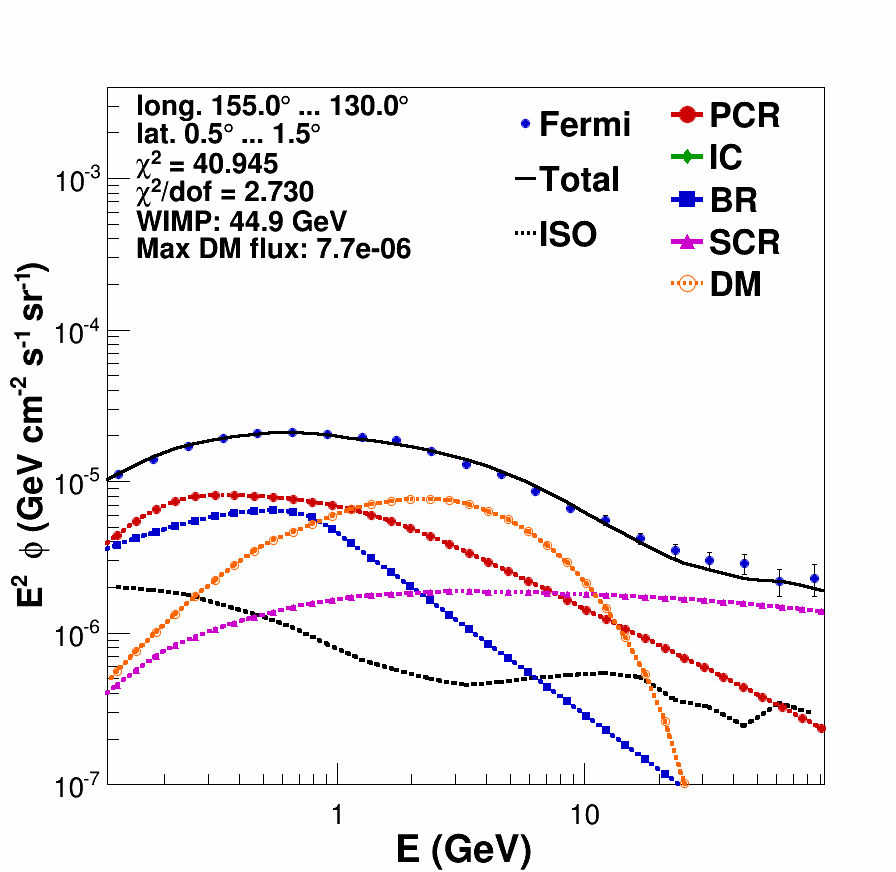}
\includegraphics[width=0.16\textwidth,height=0.16\textwidth,clip]{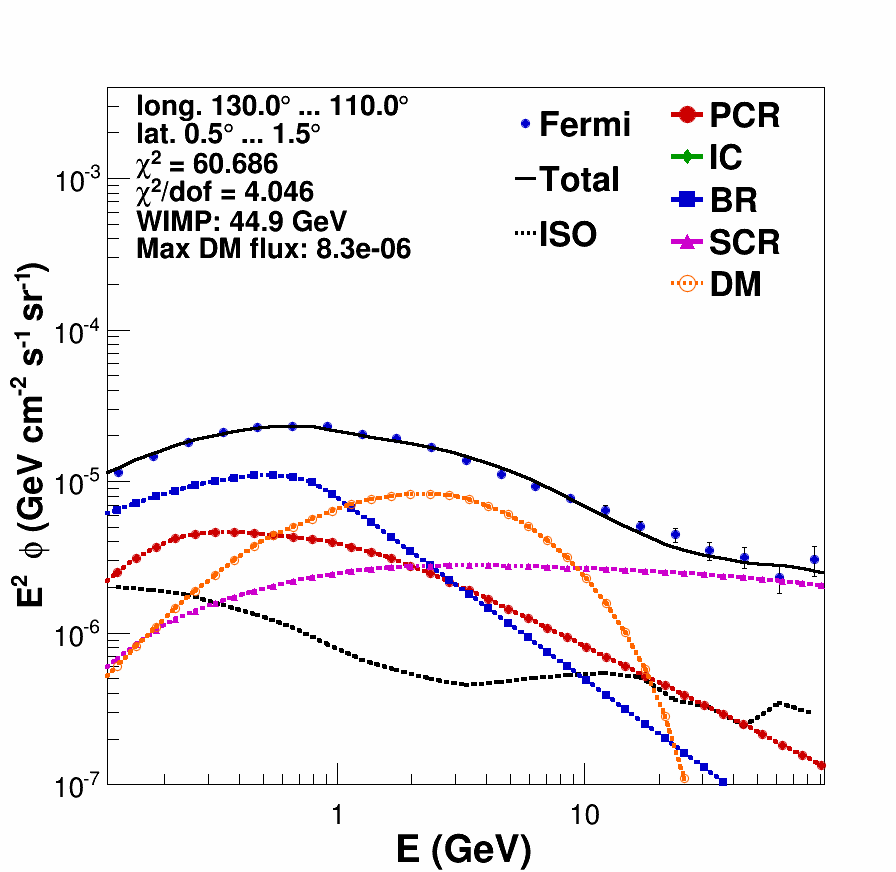}
\includegraphics[width=0.16\textwidth,height=0.16\textwidth,clip]{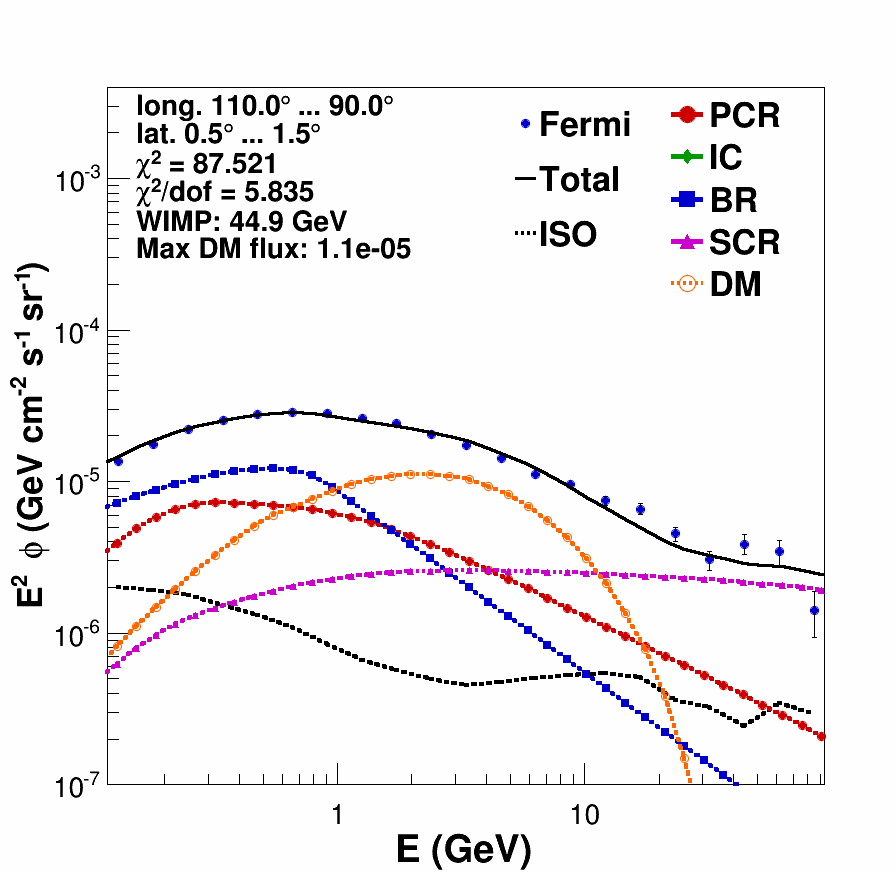}
\includegraphics[width=0.16\textwidth,height=0.16\textwidth,clip]{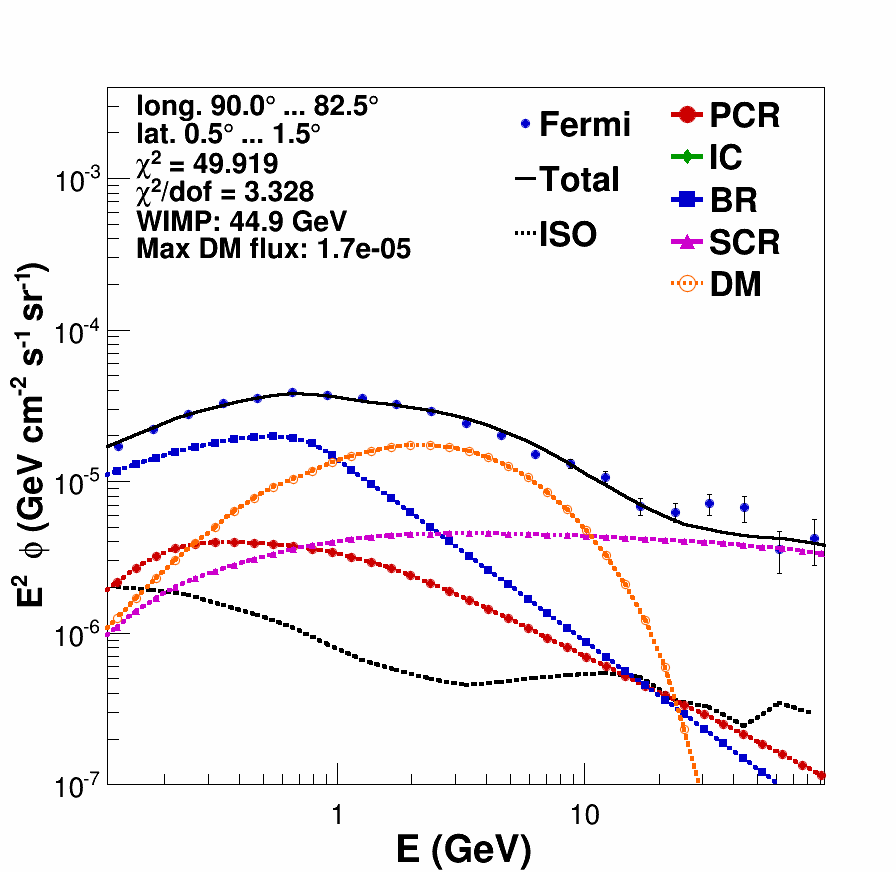}
\includegraphics[width=0.16\textwidth,height=0.16\textwidth,clip]{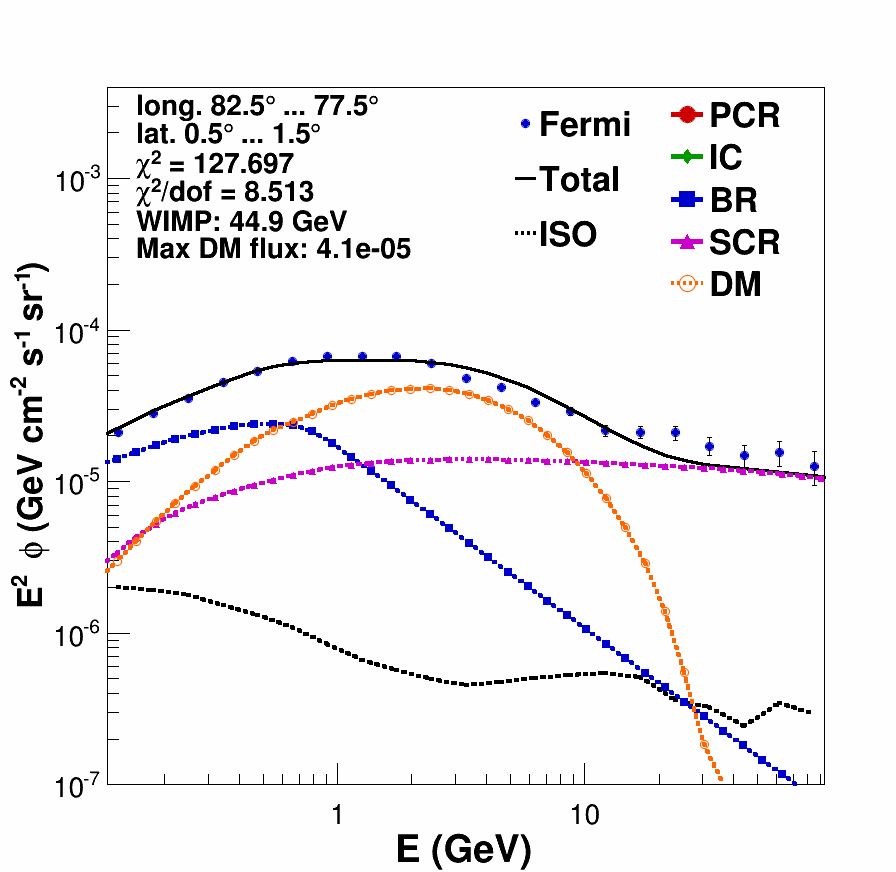}
\includegraphics[width=0.16\textwidth,height=0.16\textwidth,clip]{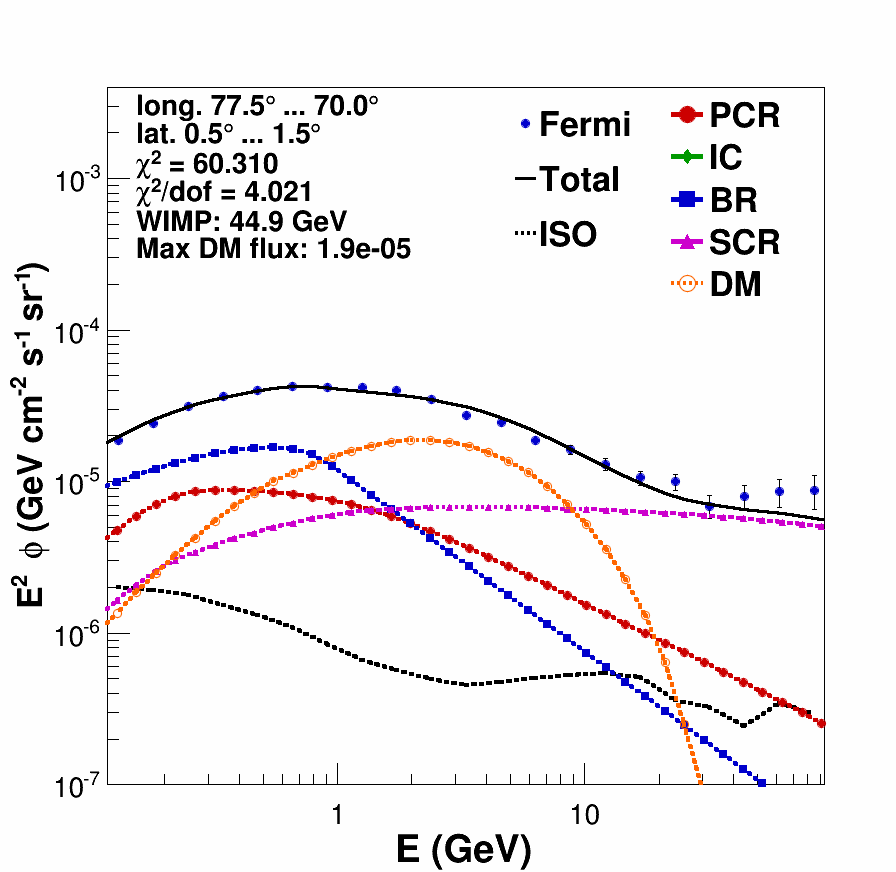}
\includegraphics[width=0.16\textwidth,height=0.16\textwidth,clip]{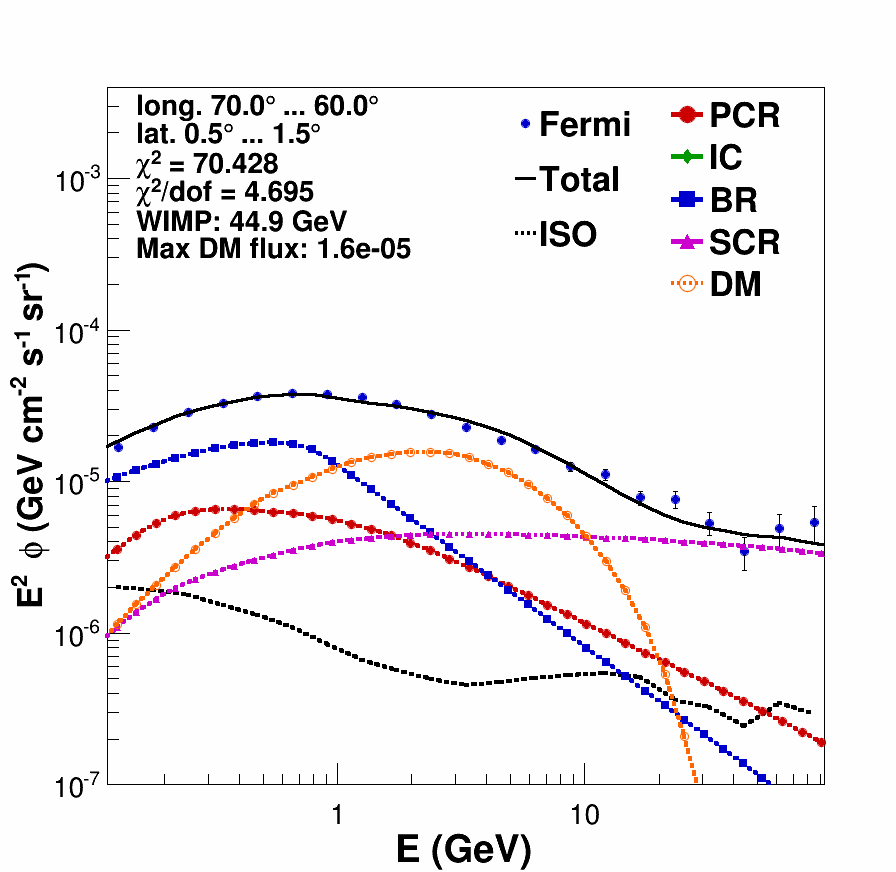}
\includegraphics[width=0.16\textwidth,height=0.16\textwidth,clip]{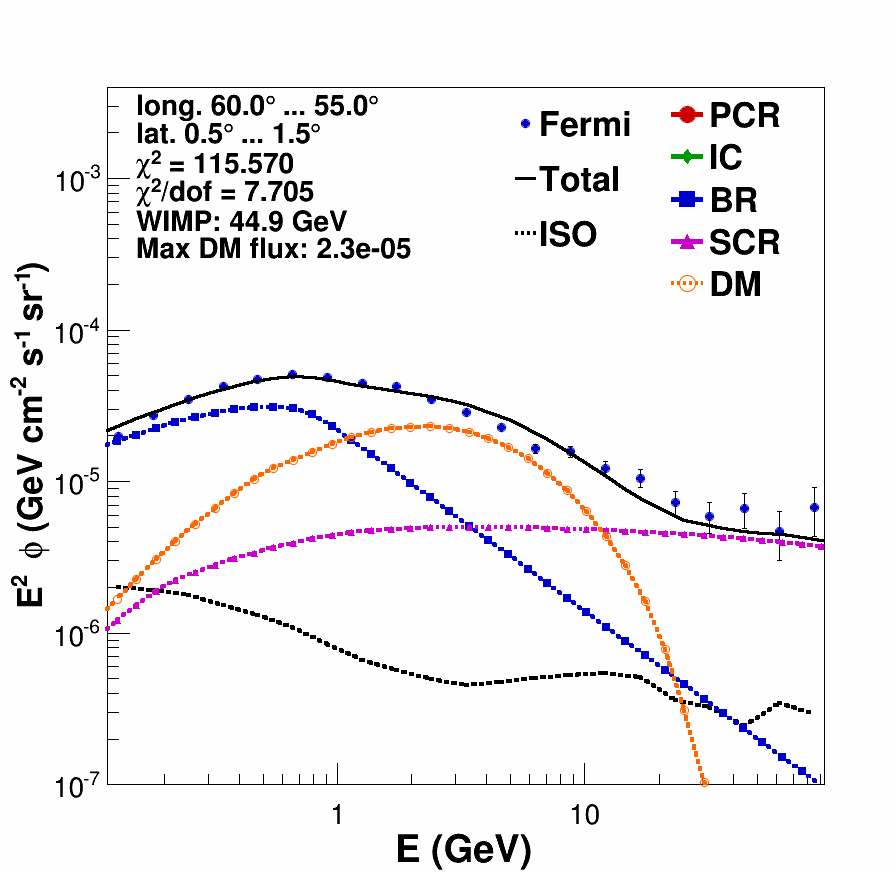}
\includegraphics[width=0.16\textwidth,height=0.16\textwidth,clip]{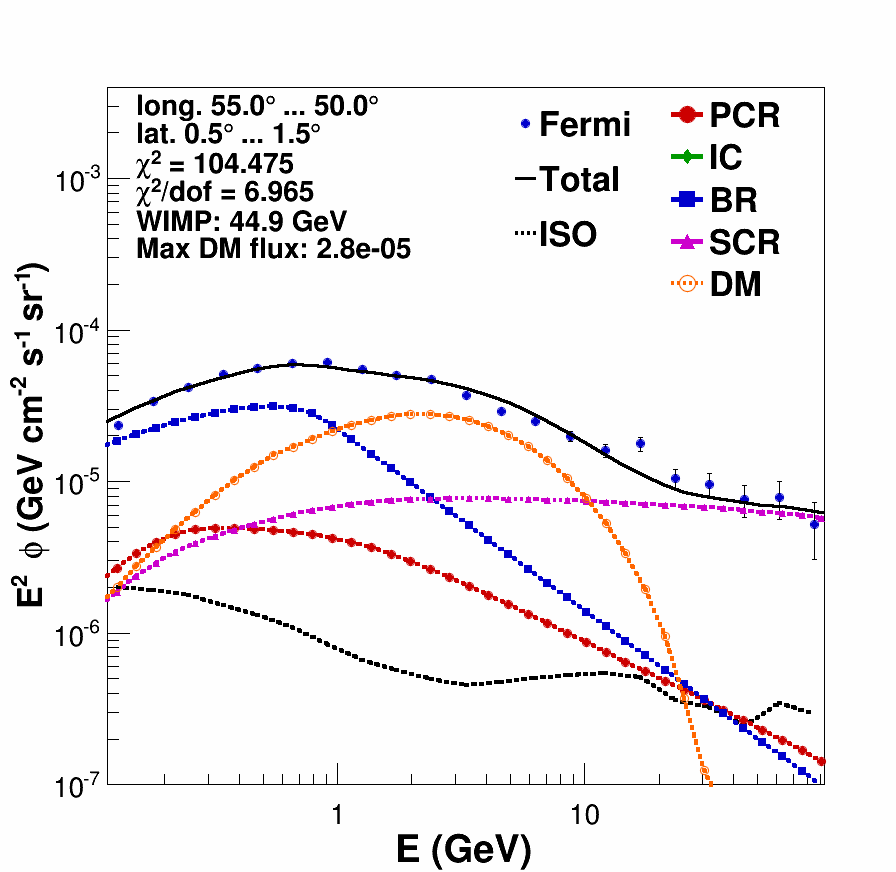}
\includegraphics[width=0.16\textwidth,height=0.16\textwidth,clip]{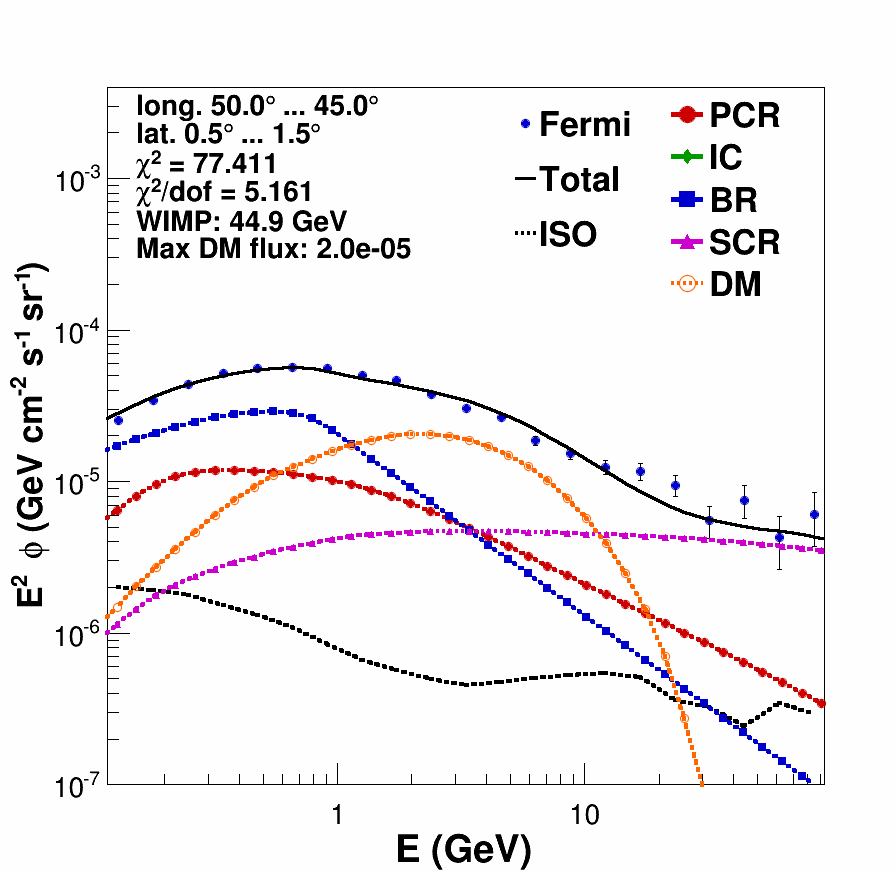}
\includegraphics[width=0.16\textwidth,height=0.16\textwidth,clip]{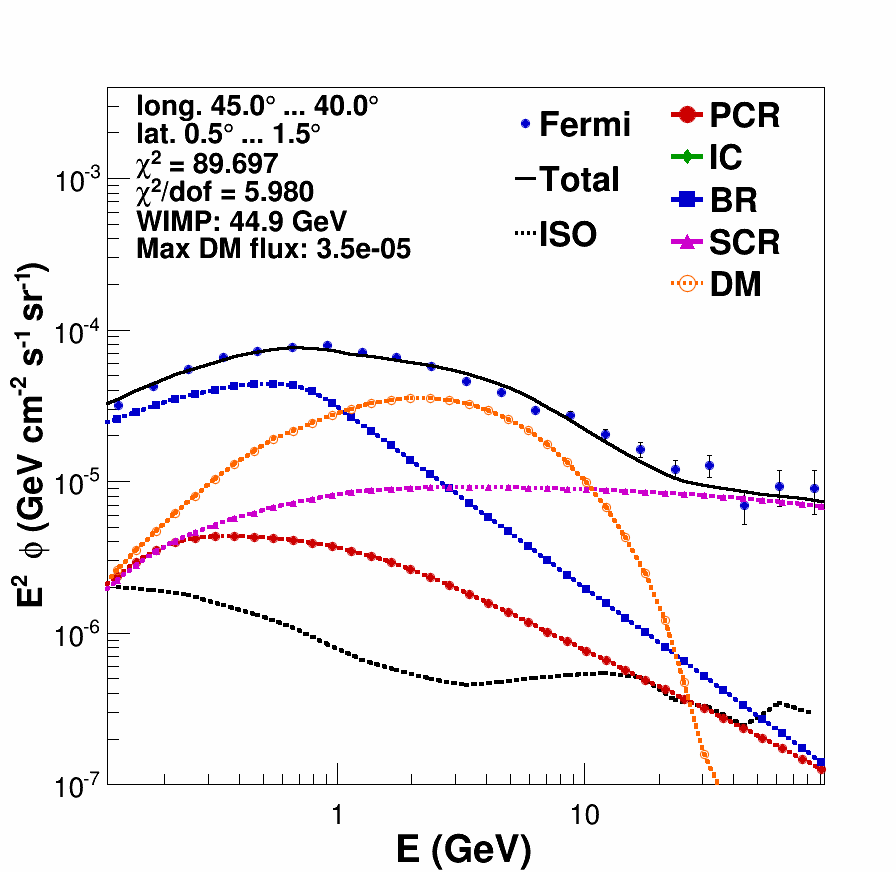}
\includegraphics[width=0.16\textwidth,height=0.16\textwidth,clip]{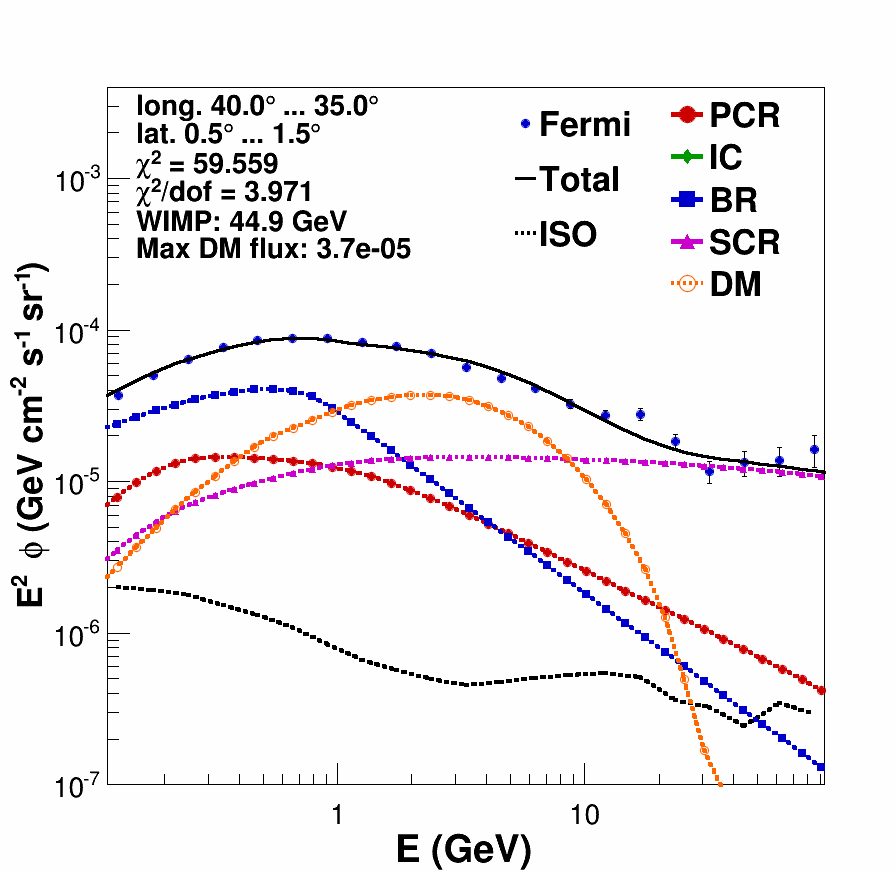}
\includegraphics[width=0.16\textwidth,height=0.16\textwidth,clip]{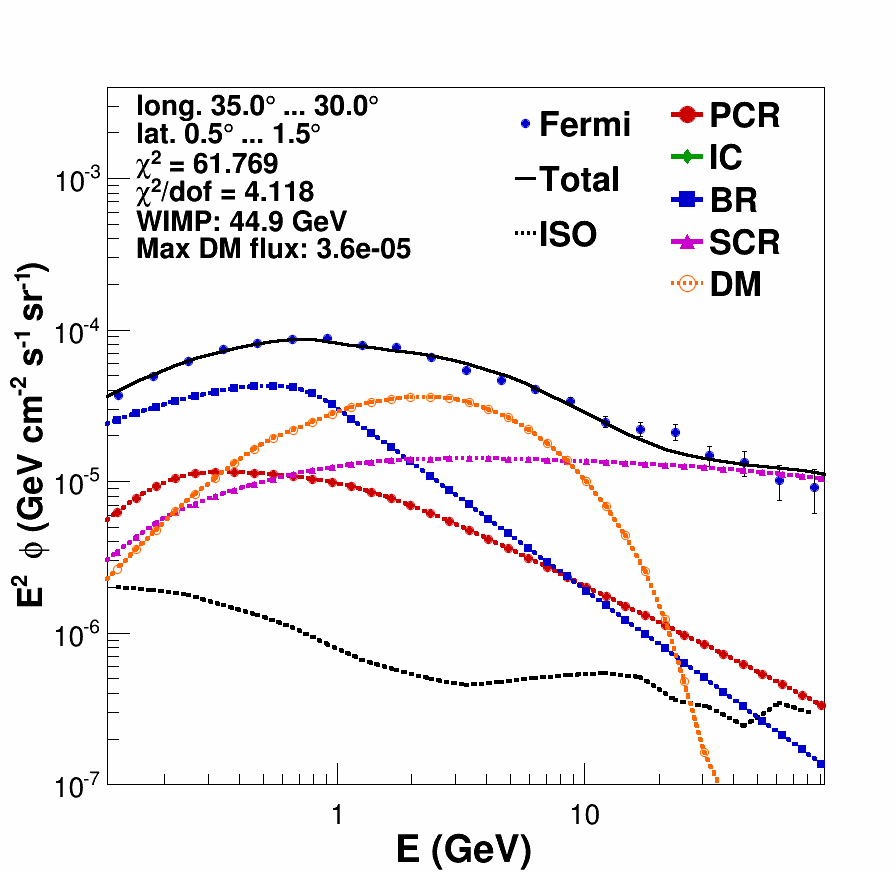}
\includegraphics[width=0.16\textwidth,height=0.16\textwidth,clip]{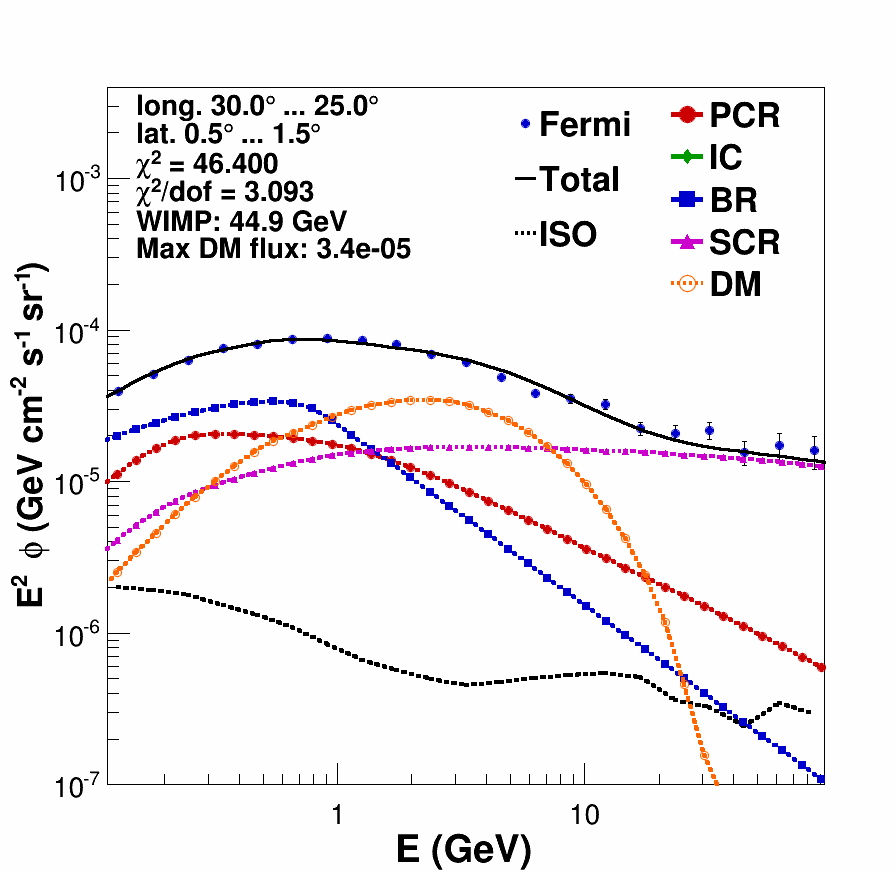}
\includegraphics[width=0.16\textwidth,height=0.16\textwidth,clip]{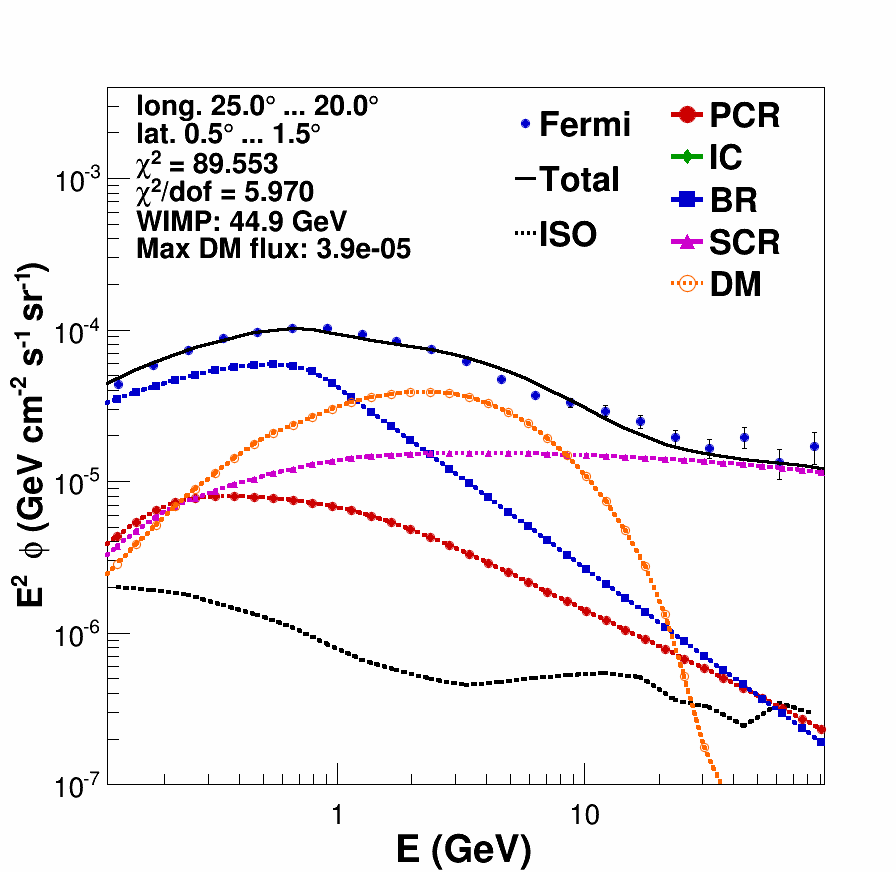}
\includegraphics[width=0.16\textwidth,height=0.16\textwidth,clip]{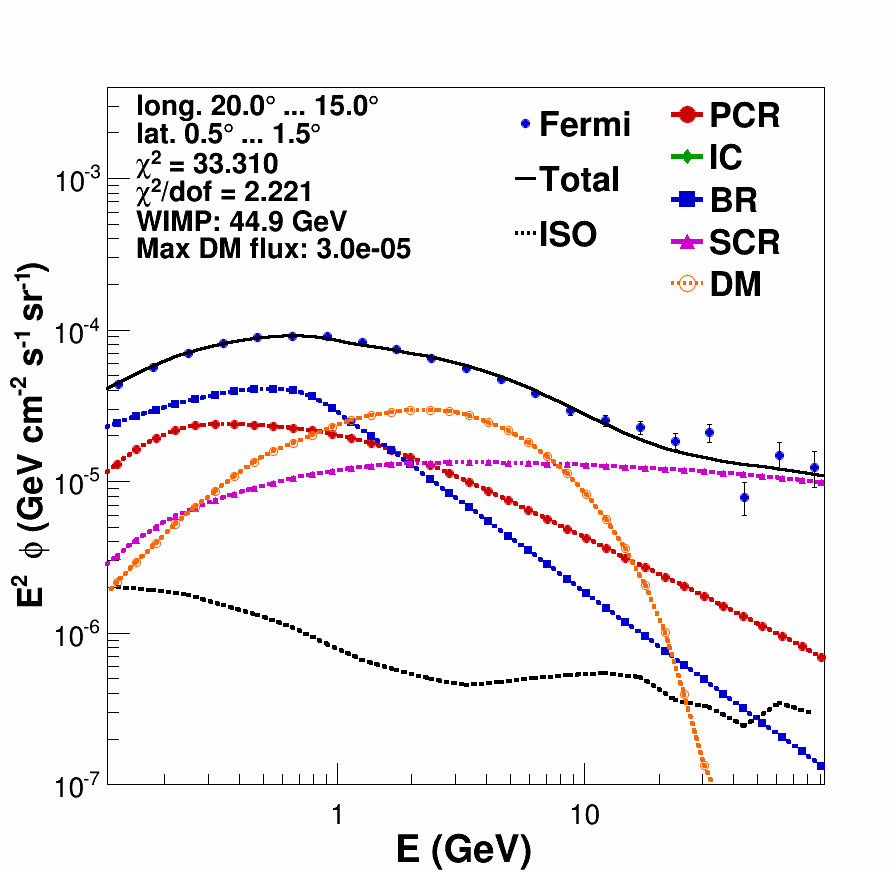}
\includegraphics[width=0.16\textwidth,height=0.16\textwidth,clip]{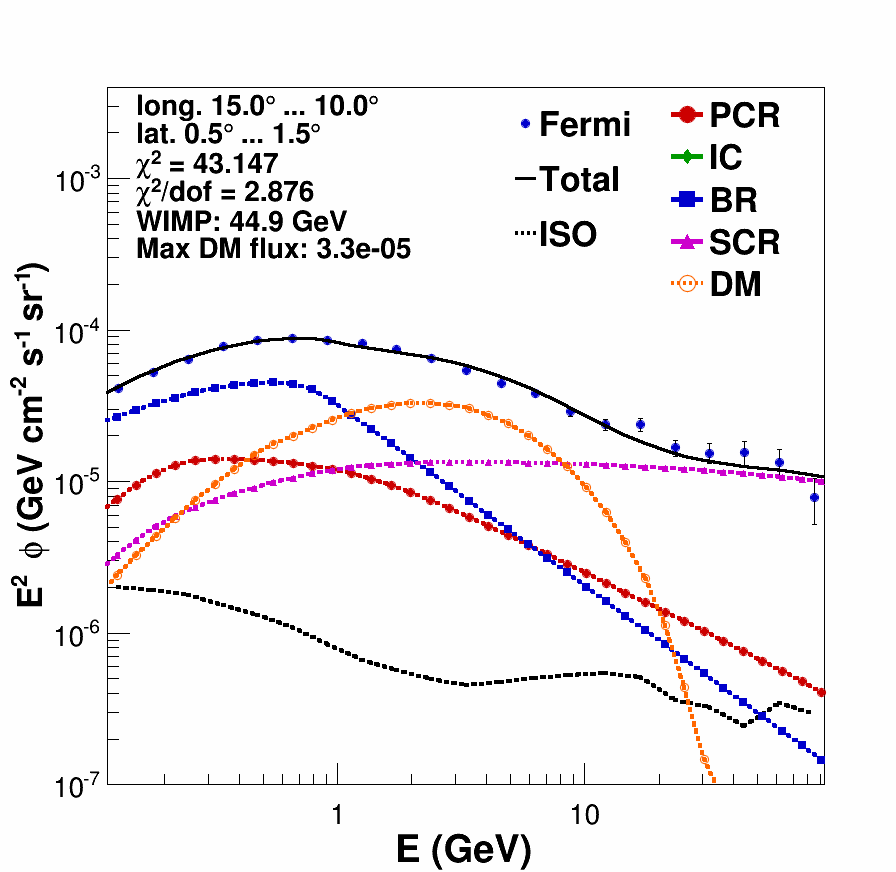}
\includegraphics[width=0.16\textwidth,height=0.16\textwidth,clip]{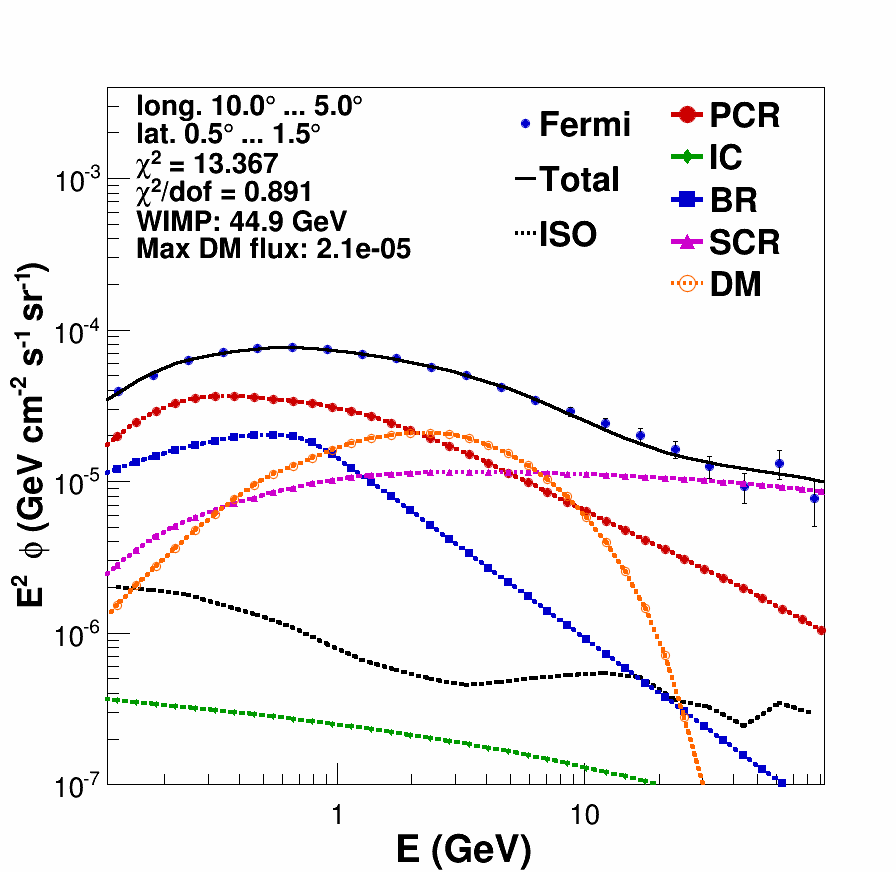}
\includegraphics[width=0.16\textwidth,height=0.16\textwidth,clip]{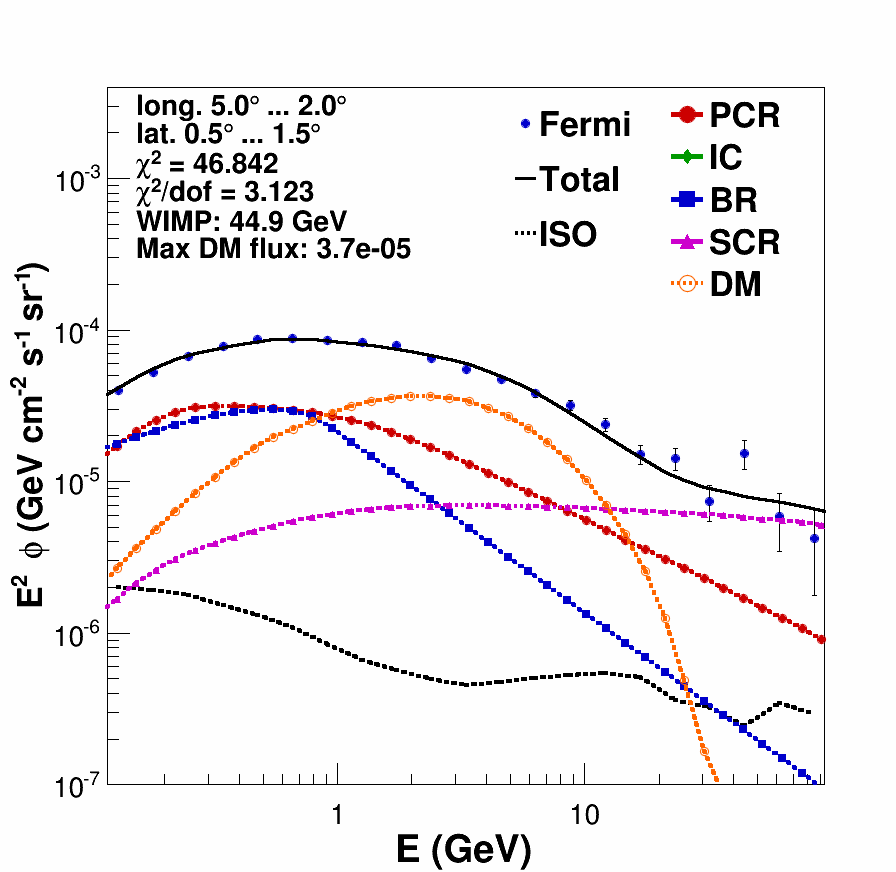}
\includegraphics[width=0.16\textwidth,height=0.16\textwidth,clip]{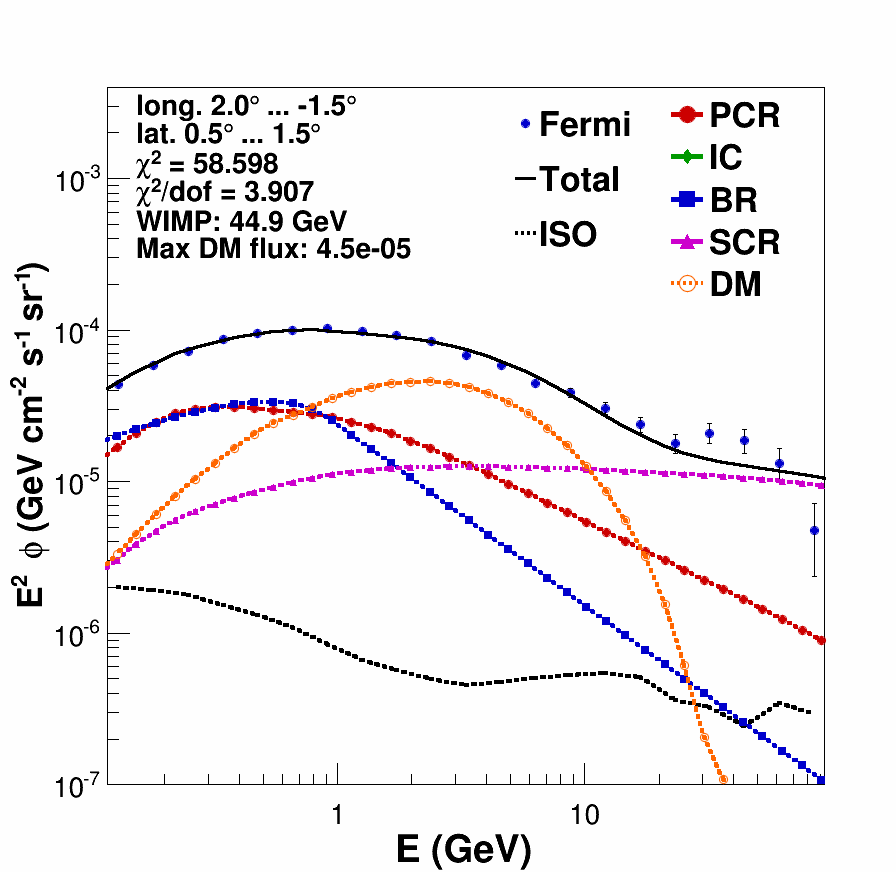}
\includegraphics[width=0.16\textwidth,height=0.16\textwidth,clip]{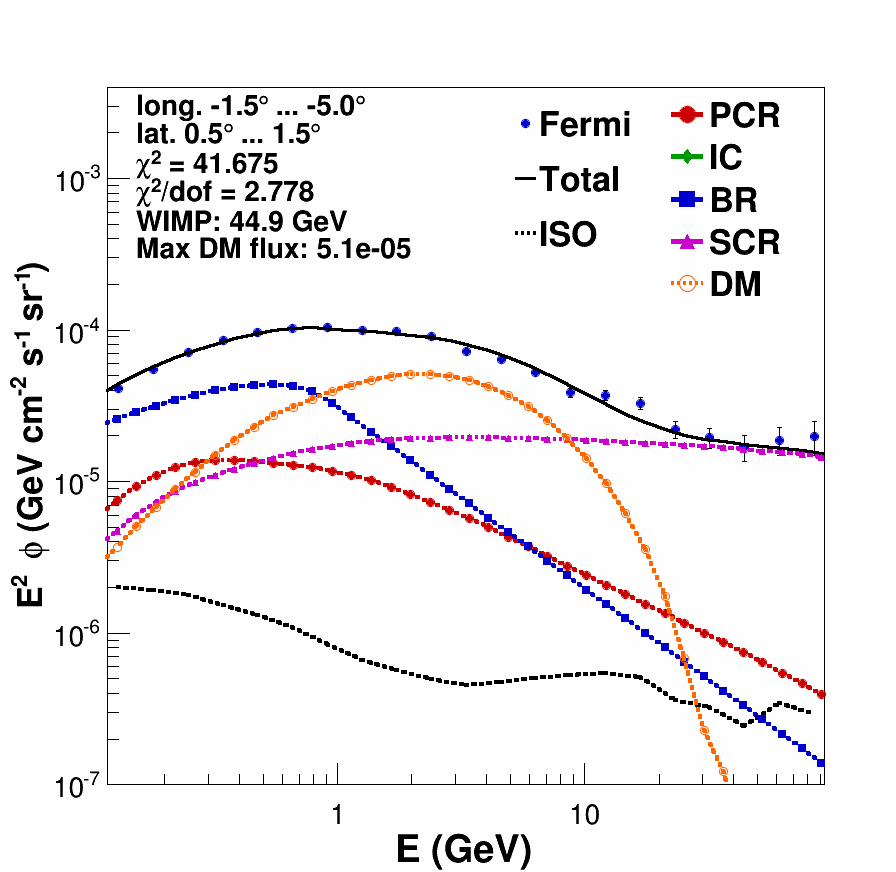}
\includegraphics[width=0.16\textwidth,height=0.16\textwidth,clip]{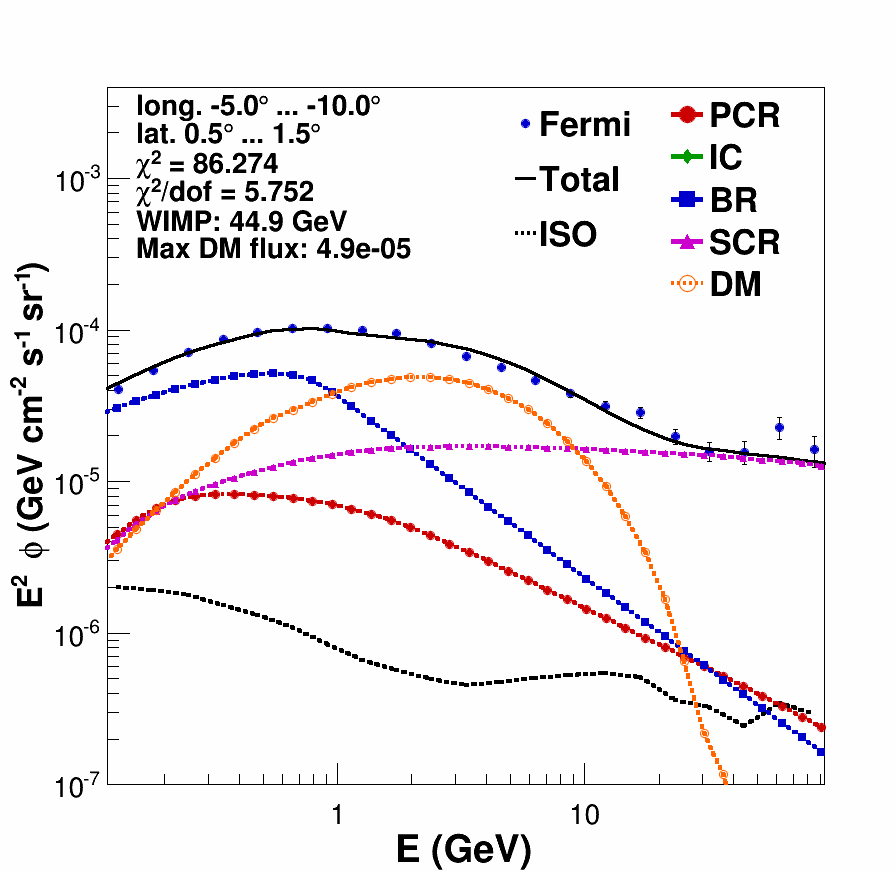}
\includegraphics[width=0.16\textwidth,height=0.16\textwidth,clip]{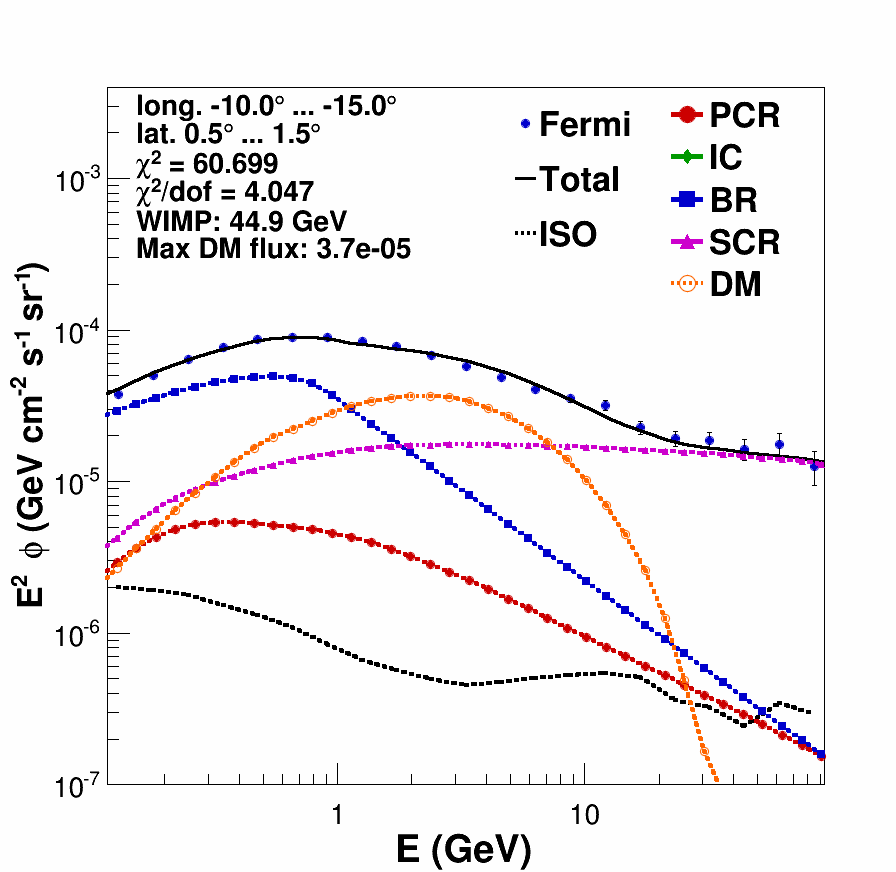}
\includegraphics[width=0.16\textwidth,height=0.16\textwidth,clip]{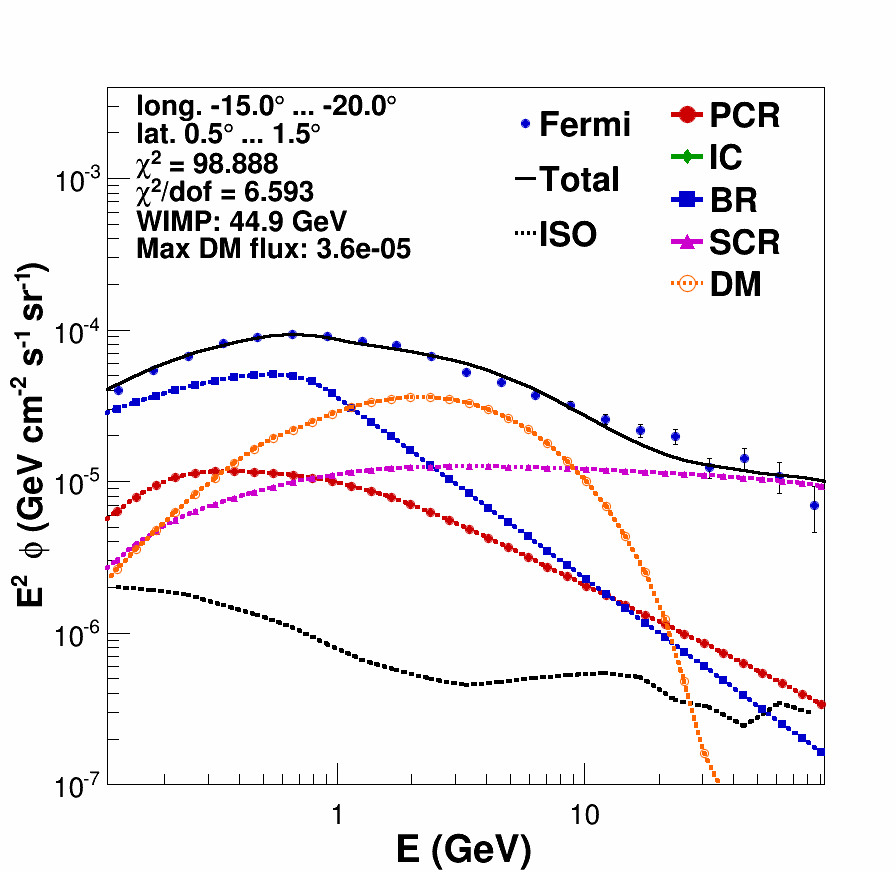}
\includegraphics[width=0.16\textwidth,height=0.16\textwidth,clip]{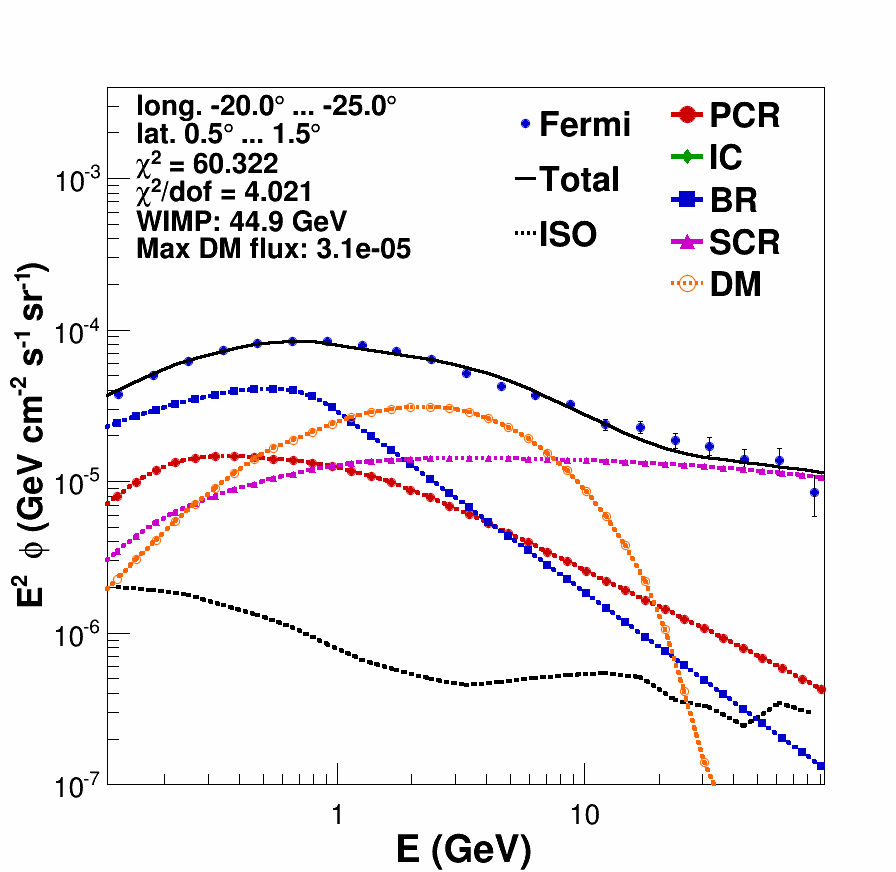}
\includegraphics[width=0.16\textwidth,height=0.16\textwidth,clip]{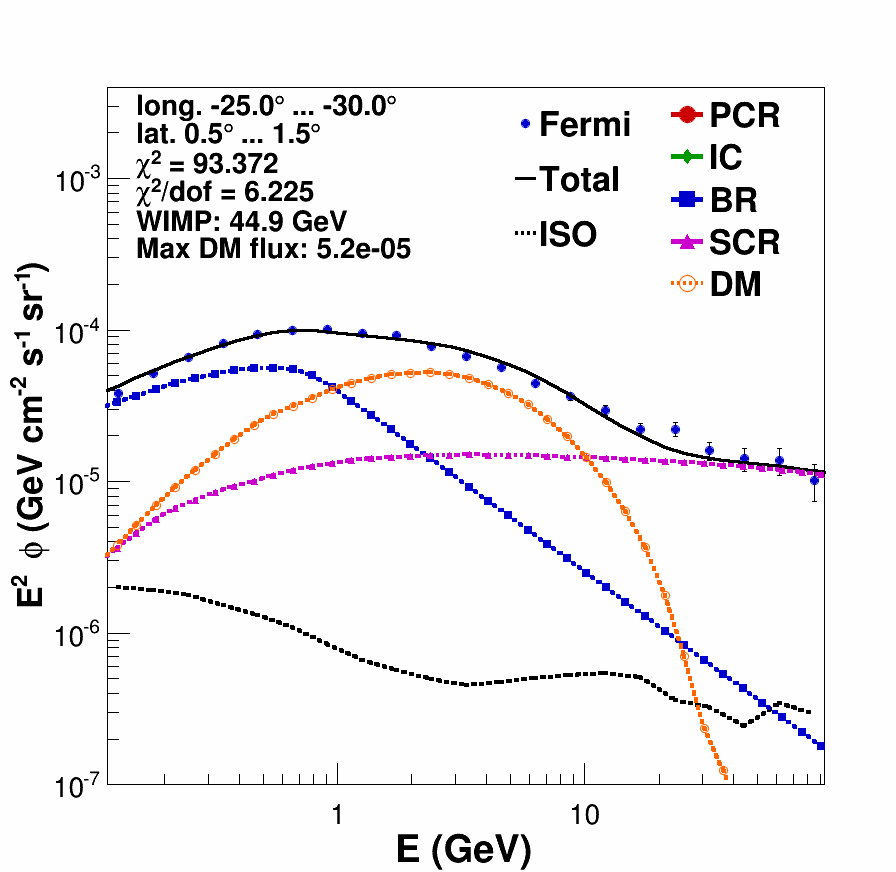}
\includegraphics[width=0.16\textwidth,height=0.16\textwidth,clip]{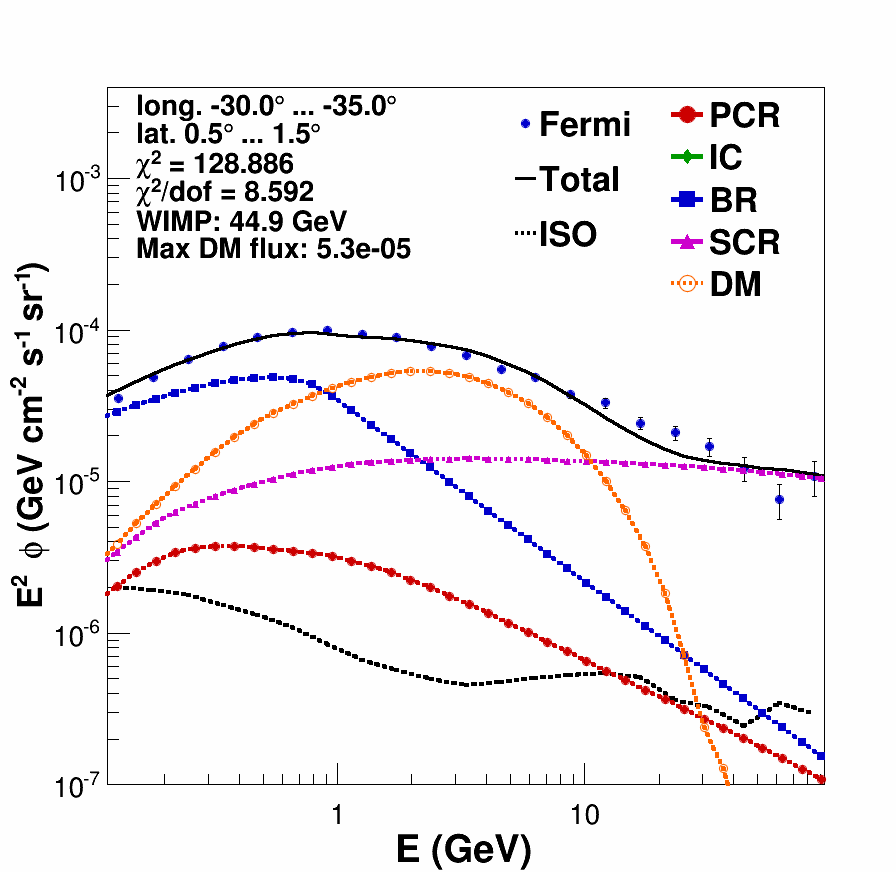}
\includegraphics[width=0.16\textwidth,height=0.16\textwidth,clip]{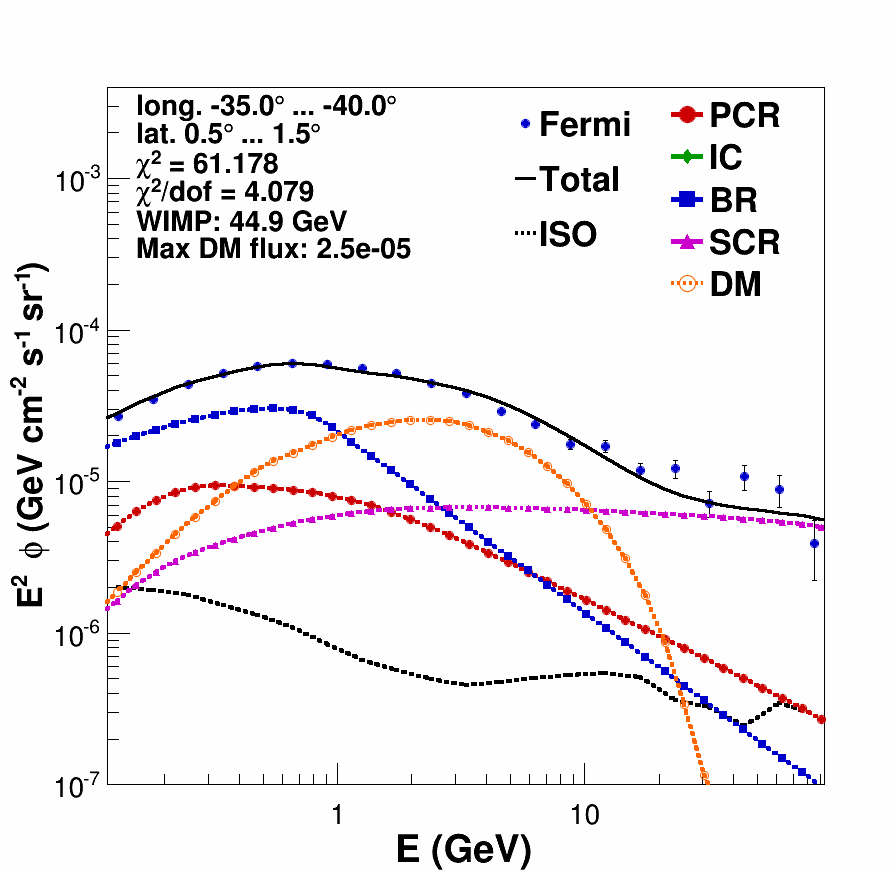}
\includegraphics[width=0.16\textwidth,height=0.16\textwidth,clip]{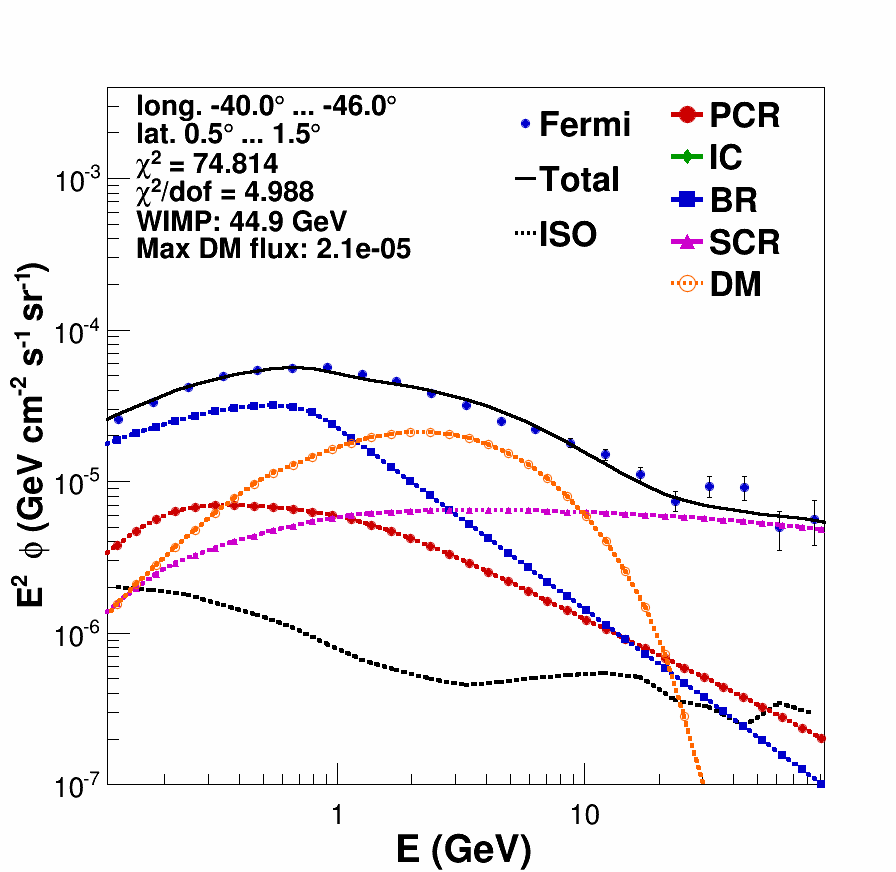}
\includegraphics[width=0.16\textwidth,height=0.16\textwidth,clip]{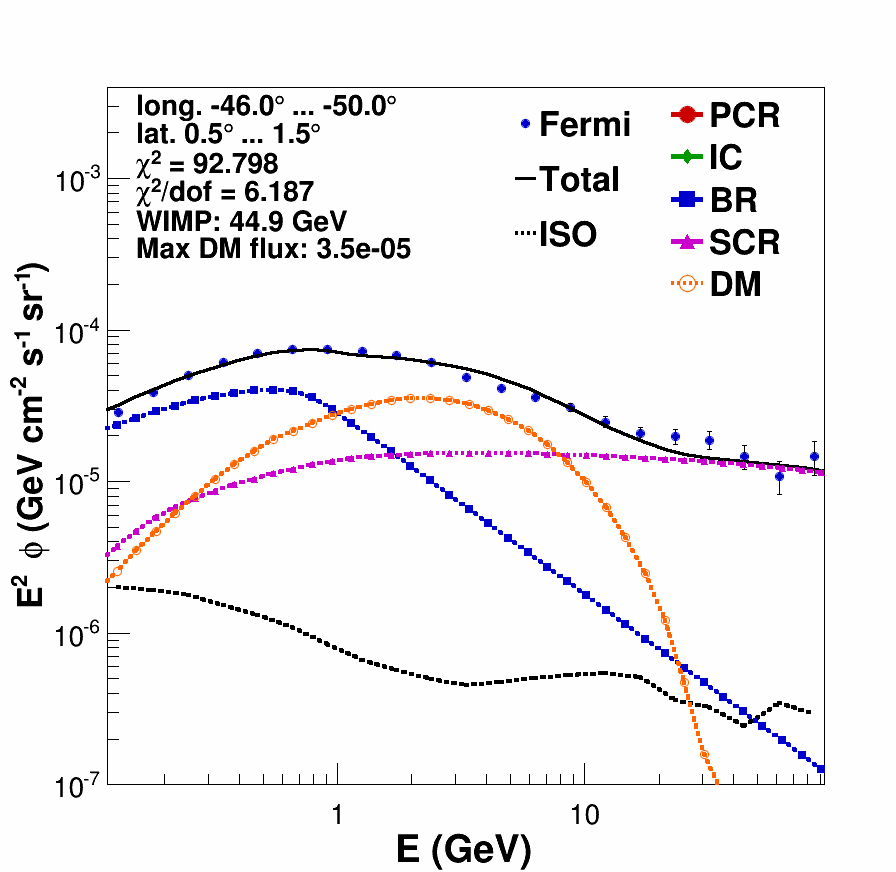}
\includegraphics[width=0.16\textwidth,height=0.16\textwidth,clip]{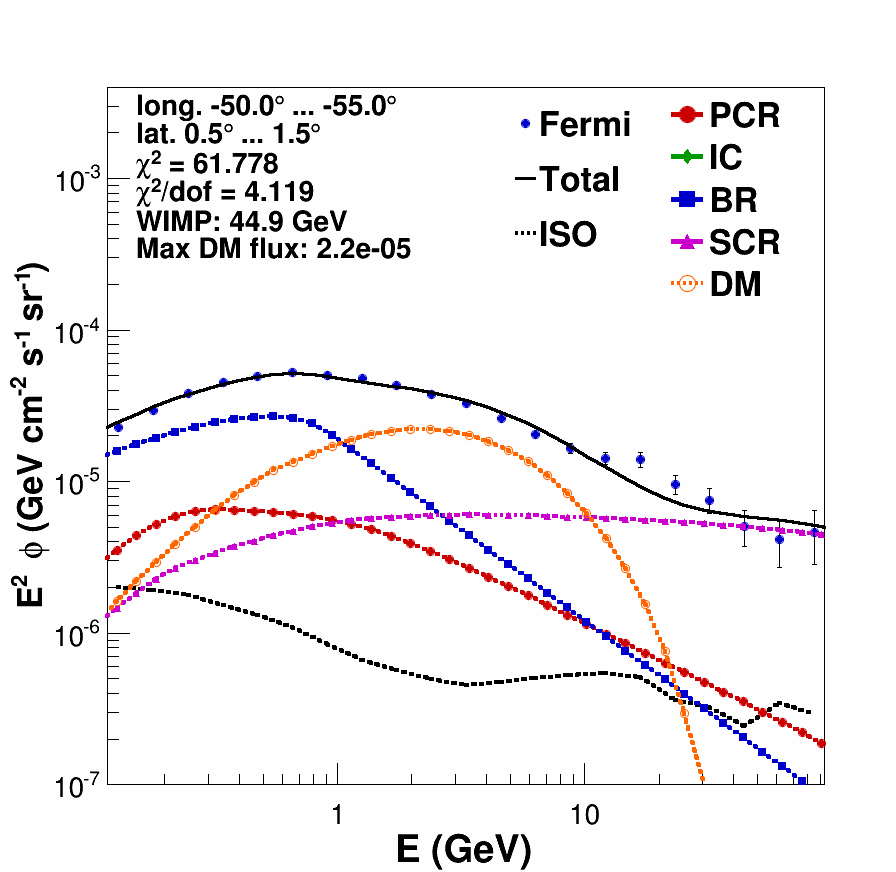}
\includegraphics[width=0.16\textwidth,height=0.16\textwidth,clip]{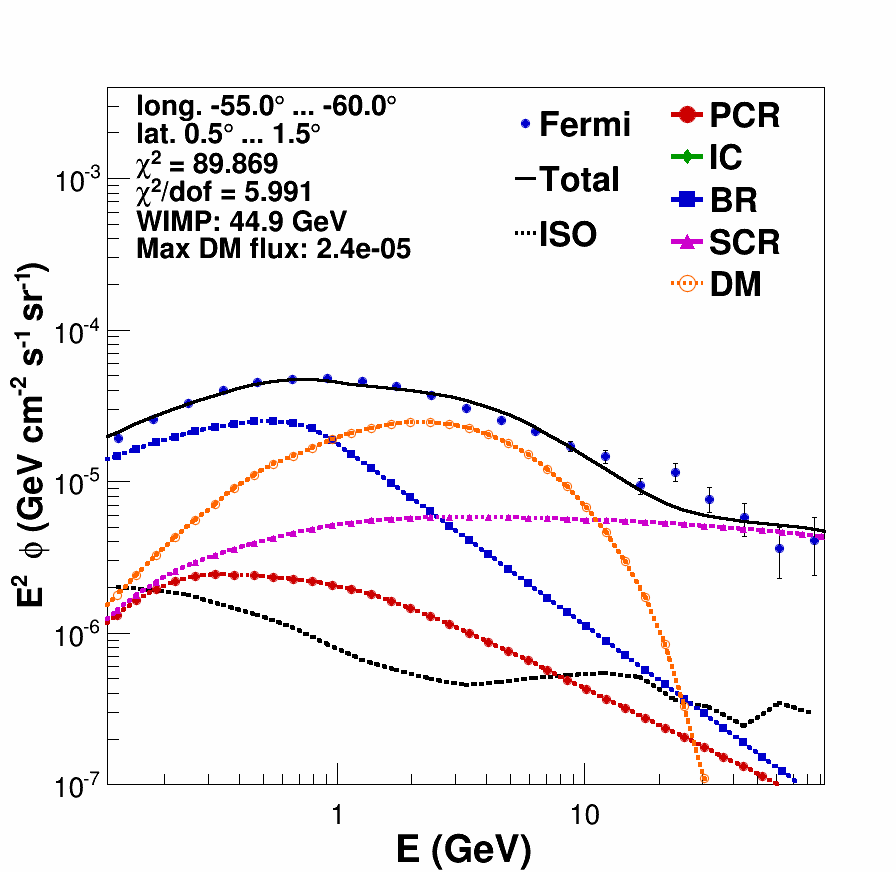}
\includegraphics[width=0.16\textwidth,height=0.16\textwidth,clip]{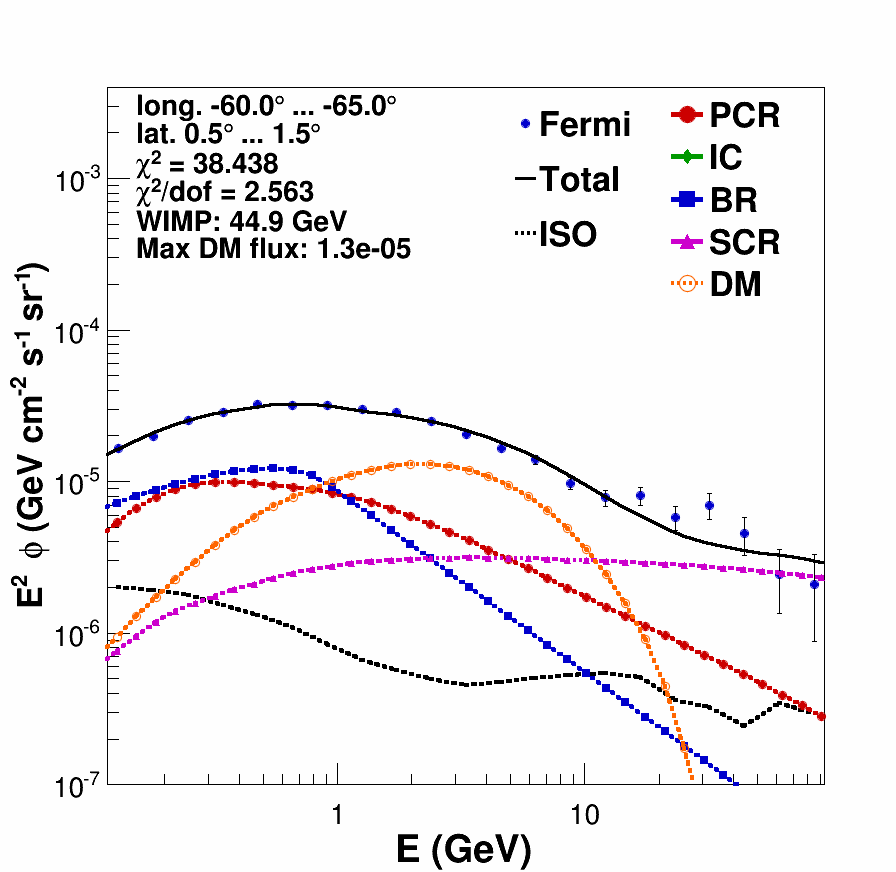}
\includegraphics[width=0.16\textwidth,height=0.16\textwidth,clip]{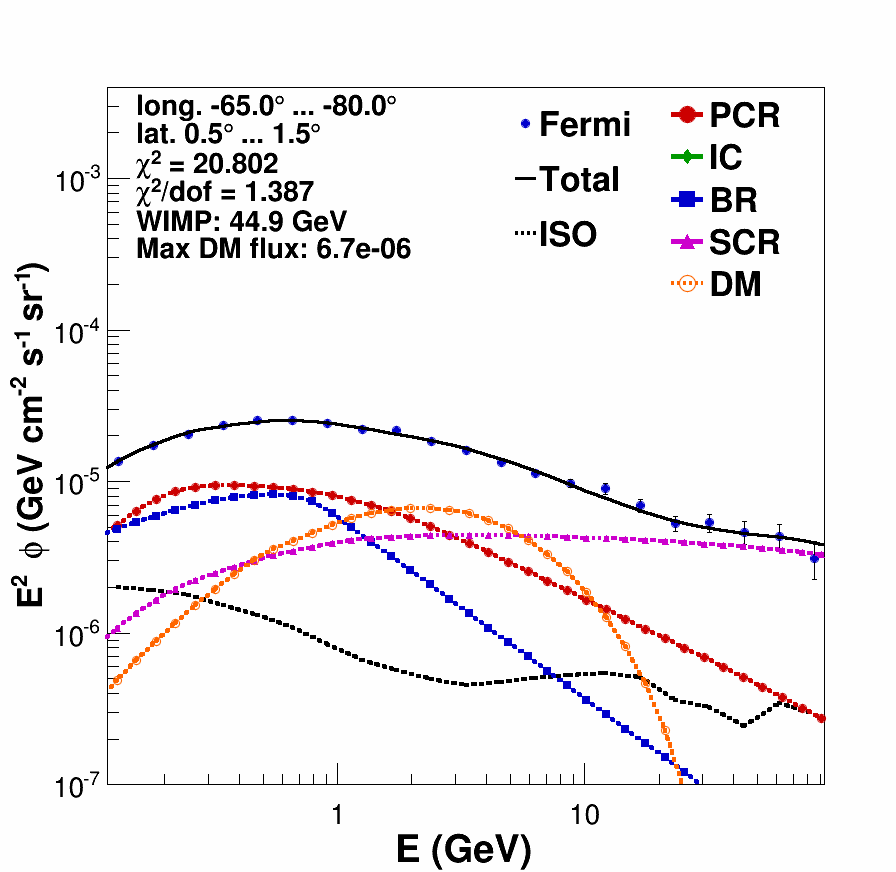}
\includegraphics[width=0.16\textwidth,height=0.16\textwidth,clip]{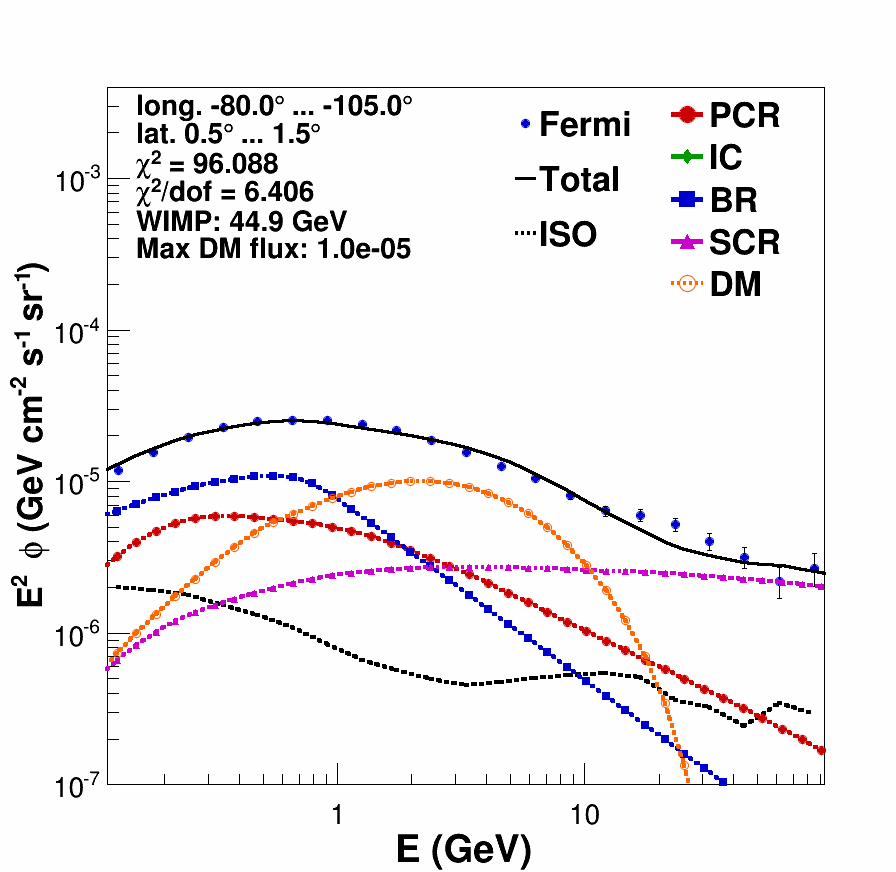}
\includegraphics[width=0.16\textwidth,height=0.16\textwidth,clip]{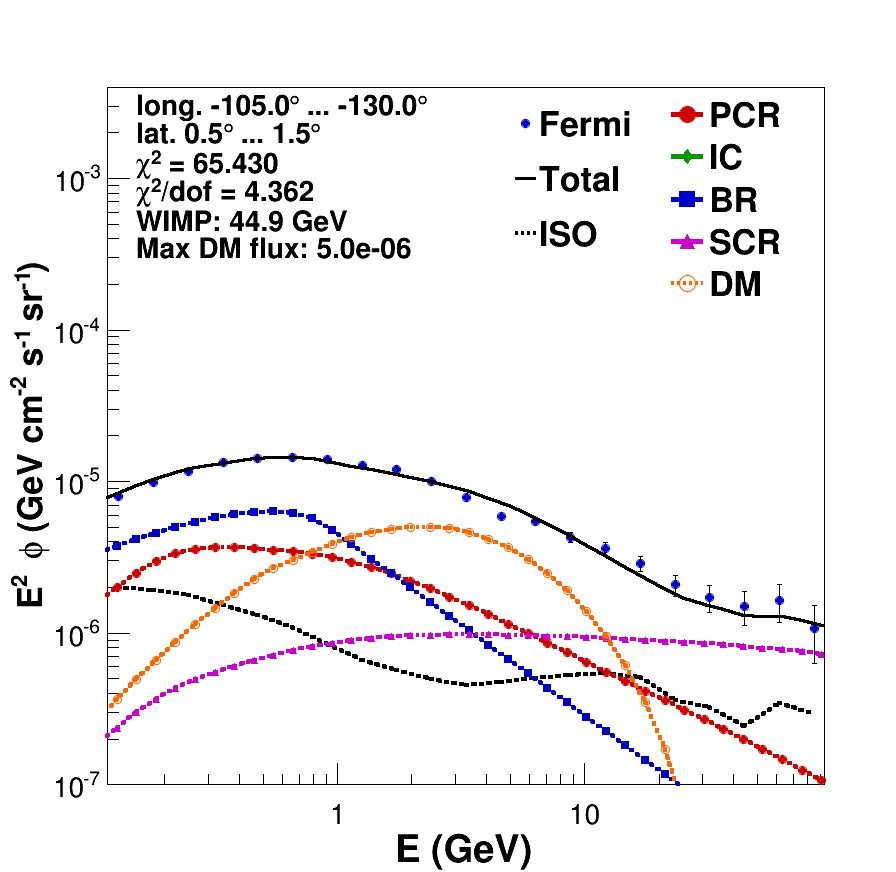}
\includegraphics[width=0.16\textwidth,height=0.16\textwidth,clip]{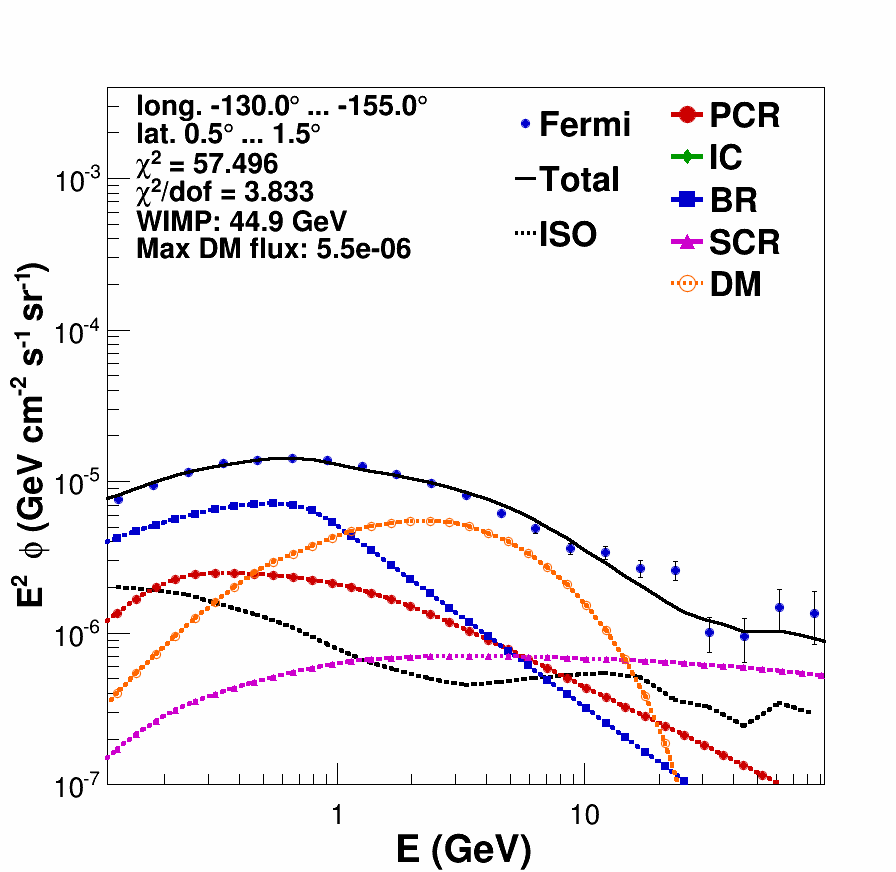}
\includegraphics[width=0.16\textwidth,height=0.16\textwidth,clip]{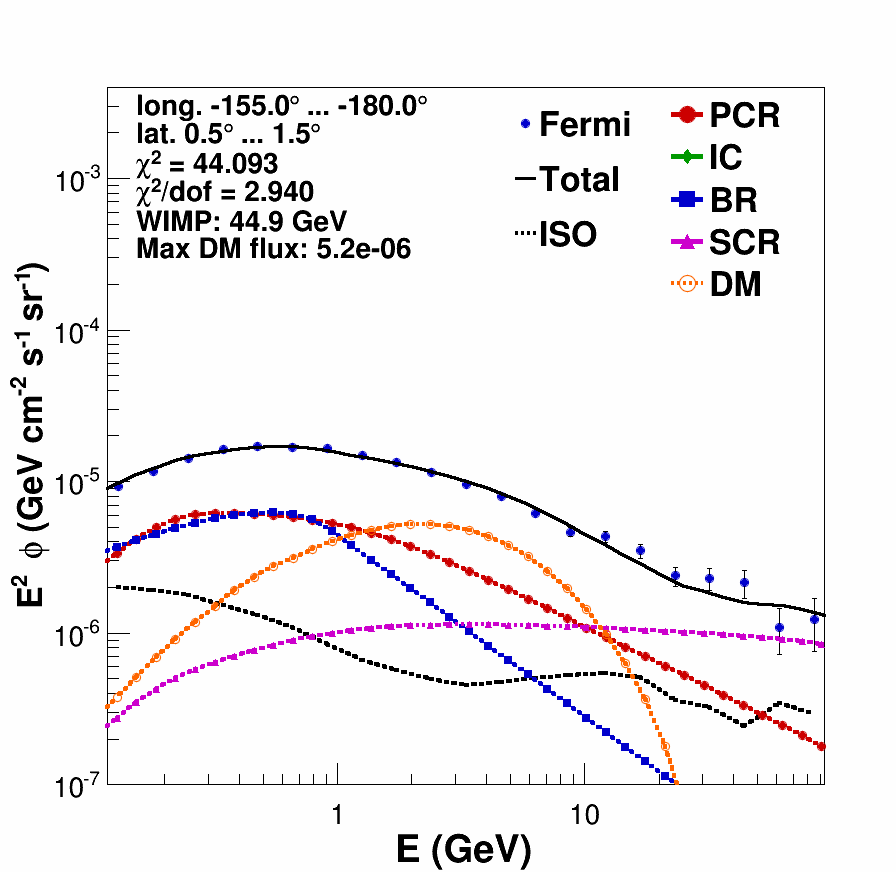}%%%%%%r9
\caption[]{Template fits for latitudes  with $0.5^\circ<b<1.5^\circ$ and longitudes decreasing from 180$^\circ$ to -180$^\circ$.} \label{F41}
\end{figure}
\clearpage
\begin{figure}
\centering
\includegraphics[width=0.16\textwidth,height=0.16\textwidth,clip]{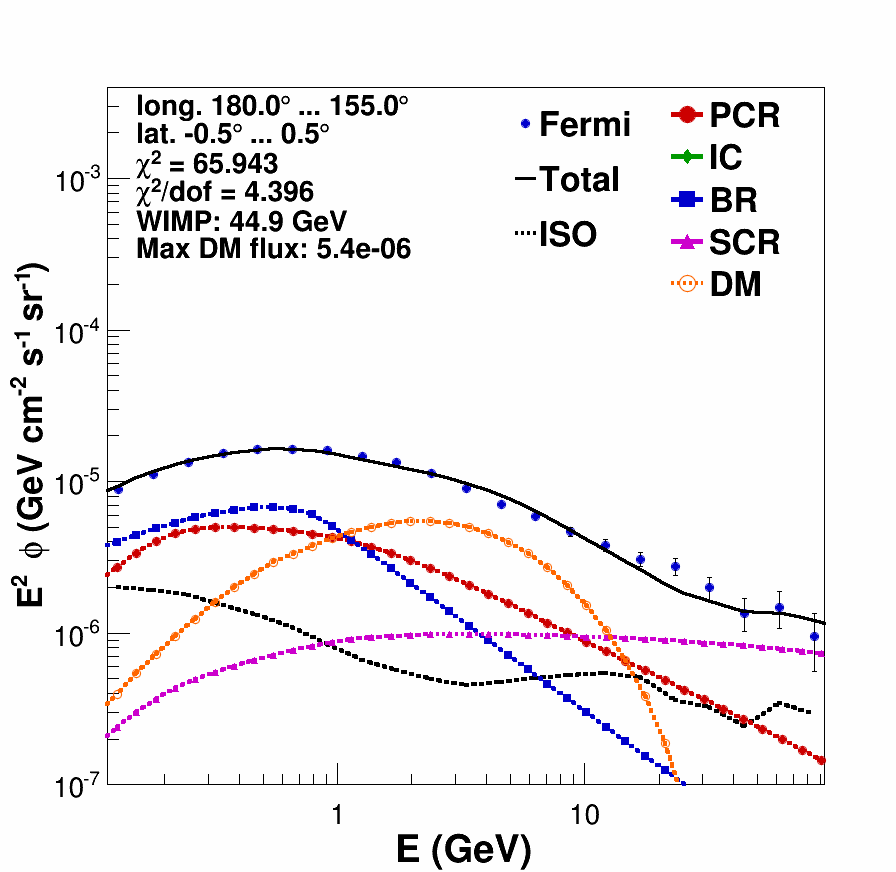}
\includegraphics[width=0.16\textwidth,height=0.16\textwidth,clip]{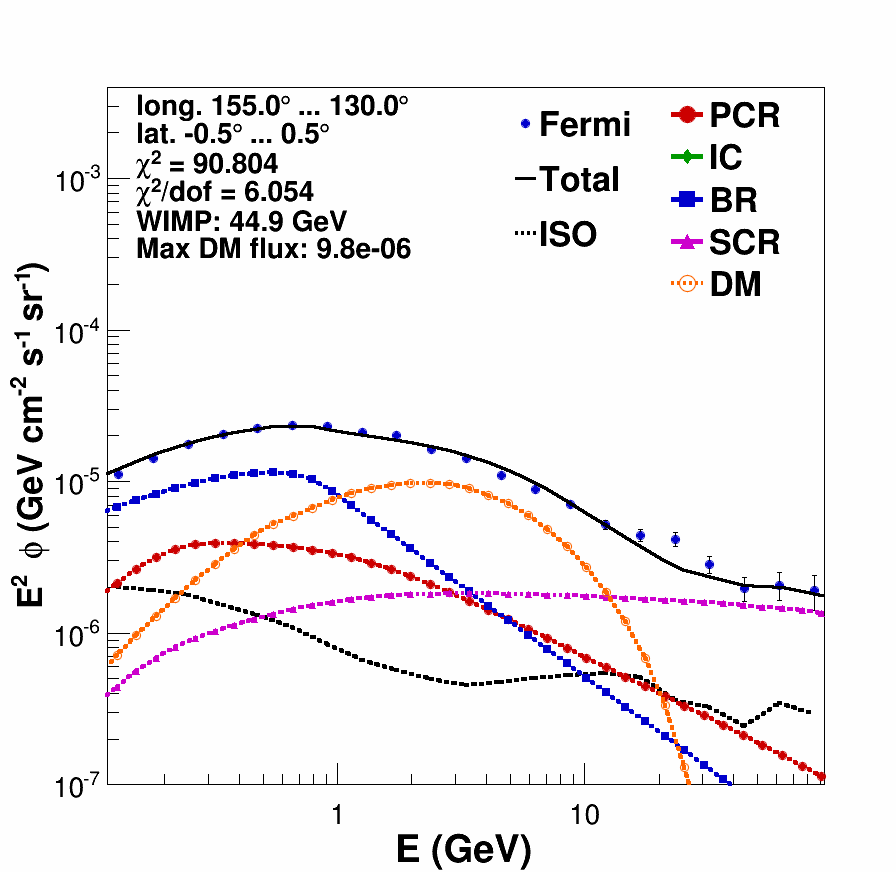}
\includegraphics[width=0.16\textwidth,height=0.16\textwidth,clip]{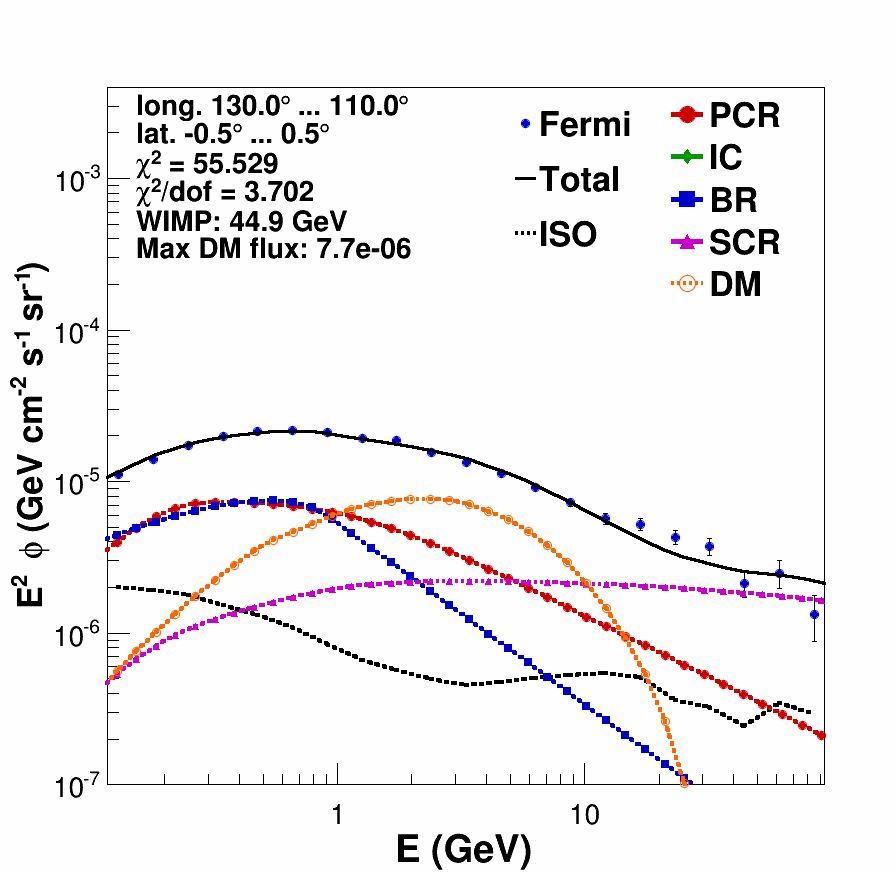}
\includegraphics[width=0.16\textwidth,height=0.16\textwidth,clip]{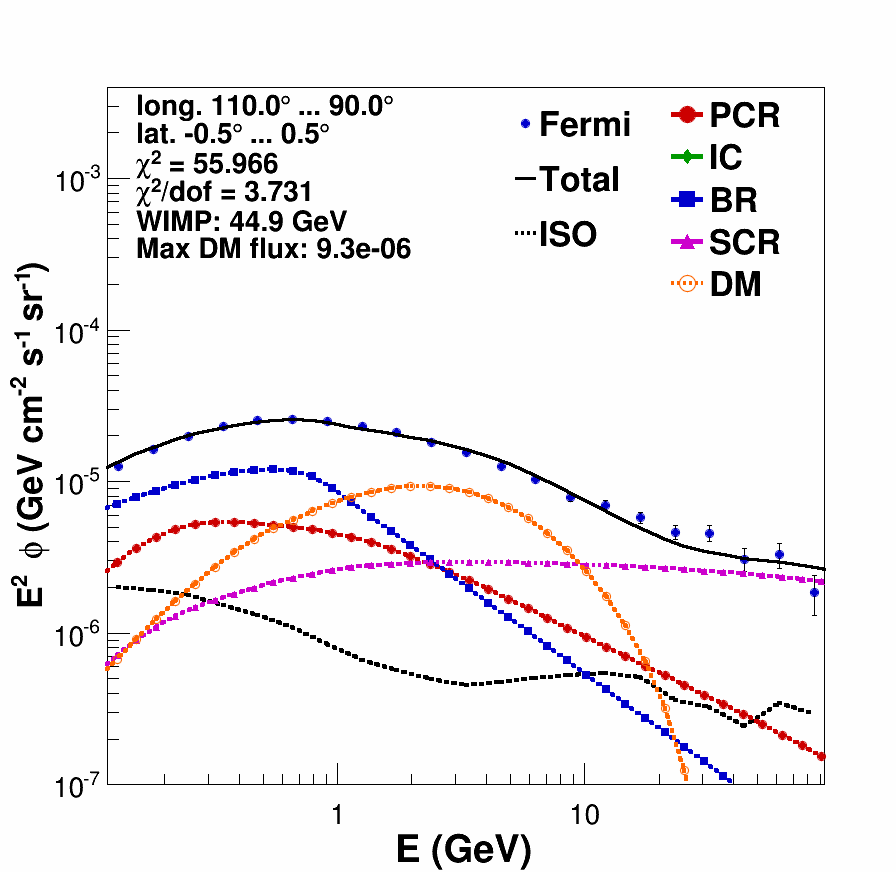}
\includegraphics[width=0.16\textwidth,height=0.16\textwidth,clip]{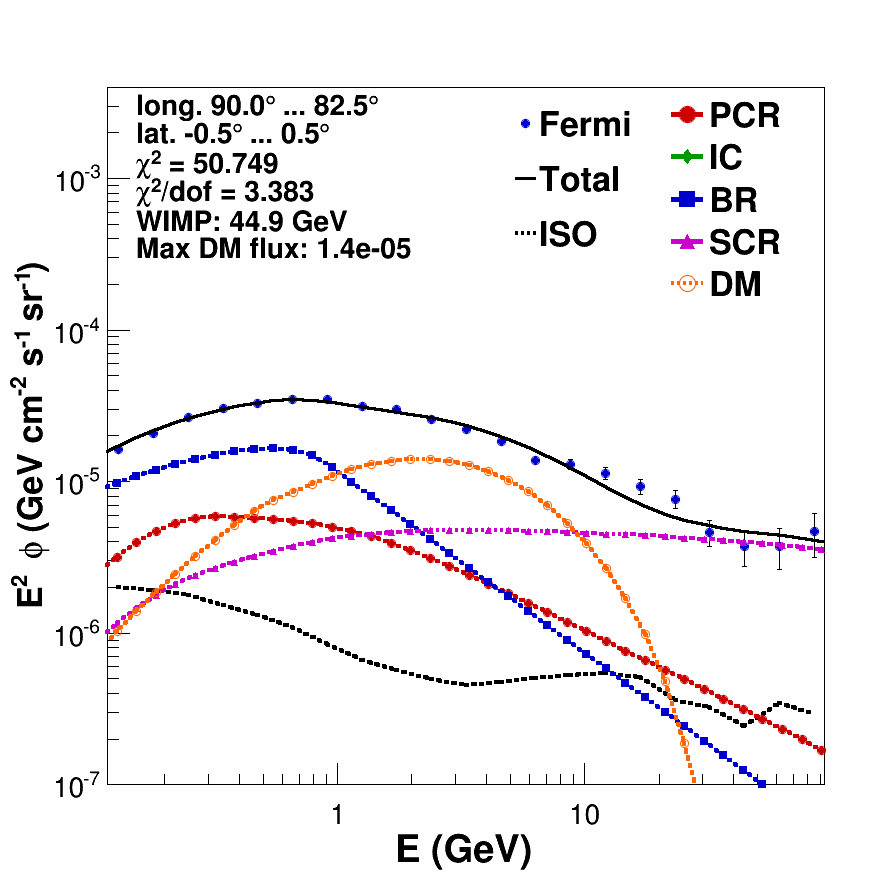}
\includegraphics[width=0.16\textwidth,height=0.16\textwidth,clip]{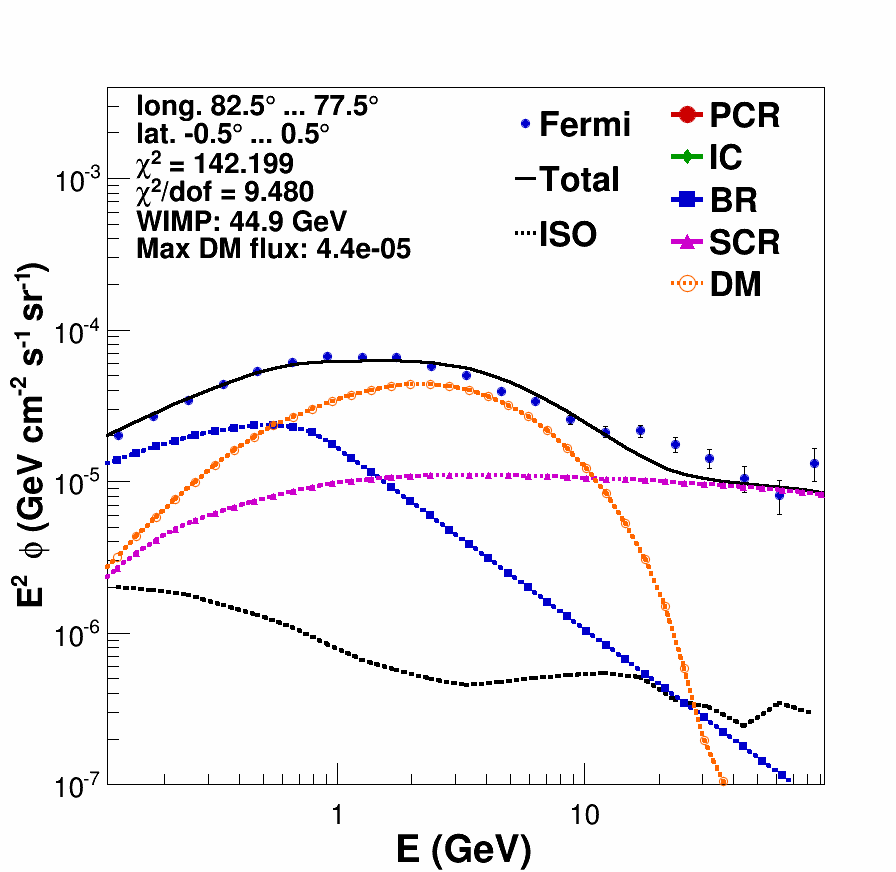}
\includegraphics[width=0.16\textwidth,height=0.16\textwidth,clip]{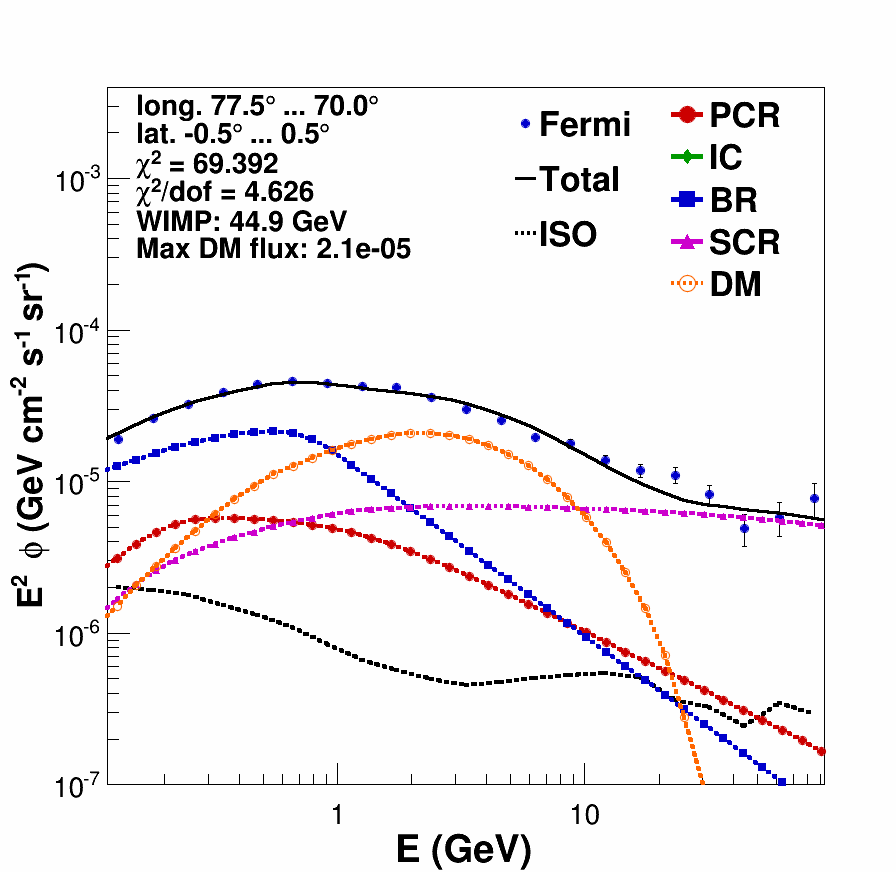}
\includegraphics[width=0.16\textwidth,height=0.16\textwidth,clip]{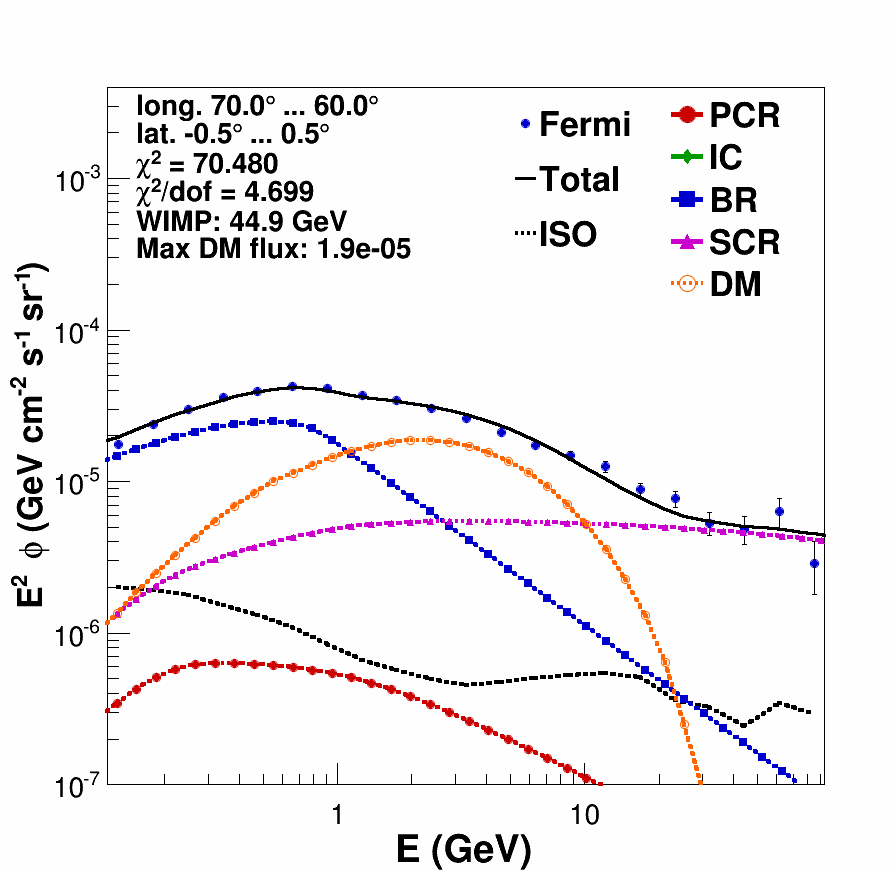}
\includegraphics[width=0.16\textwidth,height=0.16\textwidth,clip]{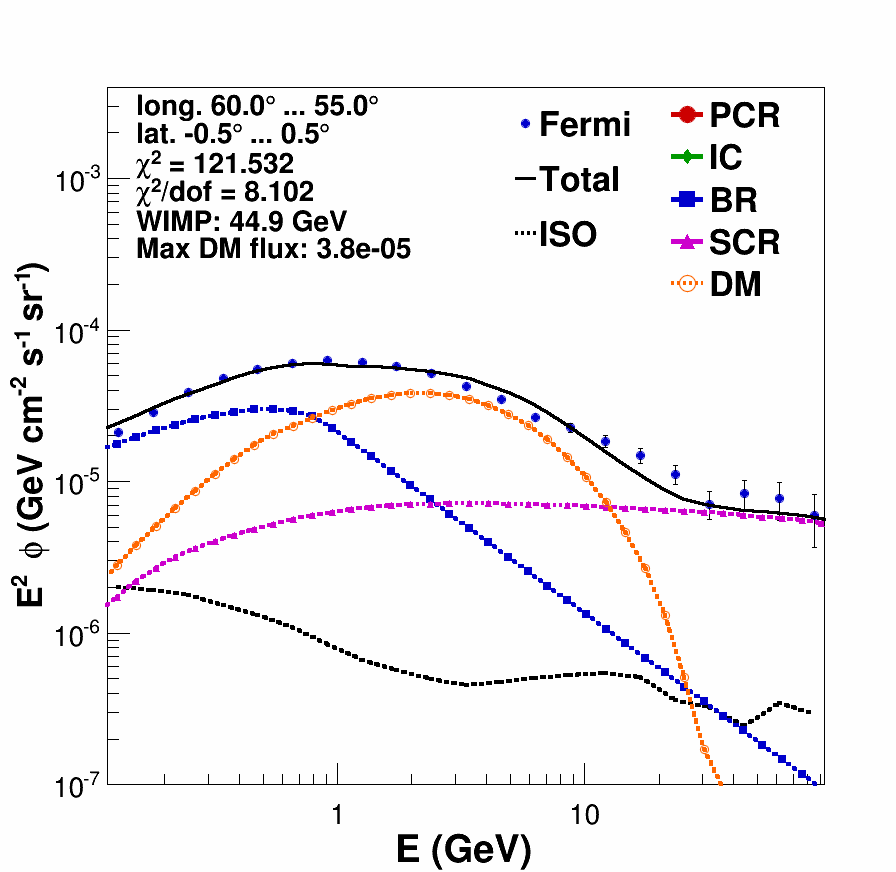}
\includegraphics[width=0.16\textwidth,height=0.16\textwidth,clip]{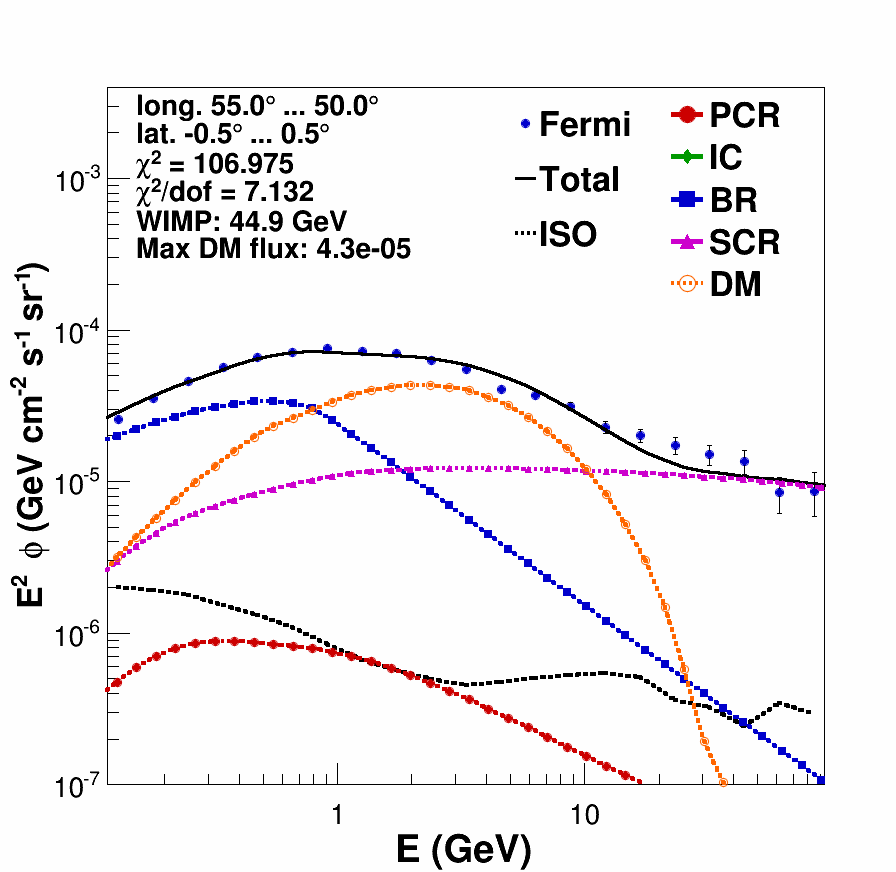}
\includegraphics[width=0.16\textwidth,height=0.16\textwidth,clip]{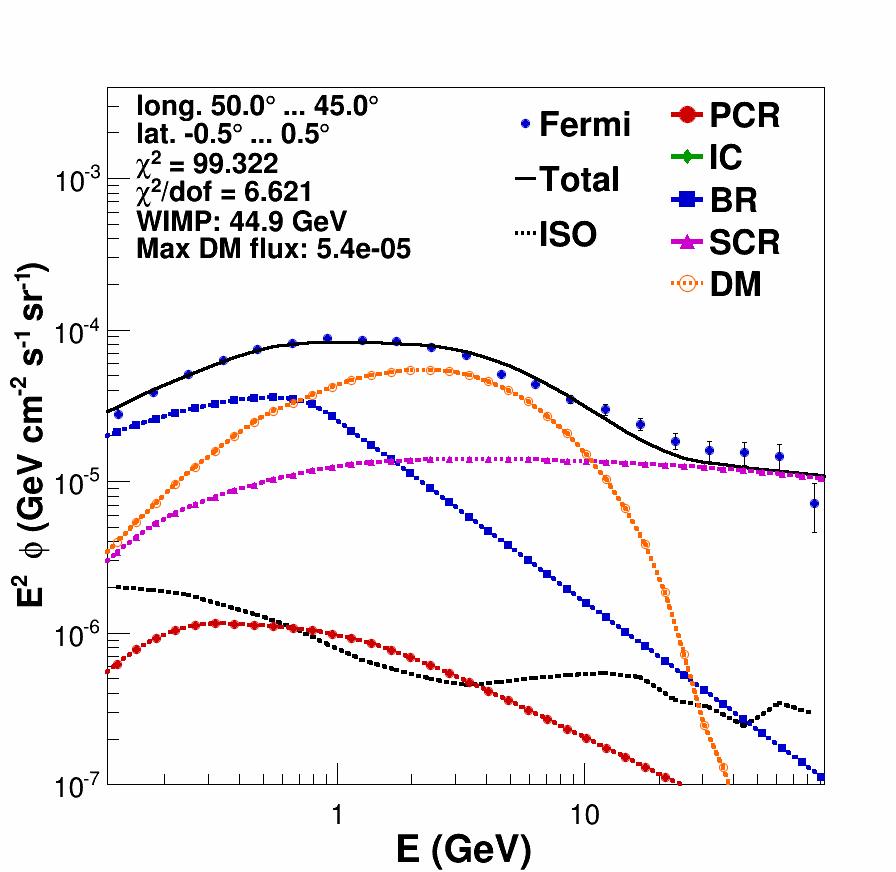}
\includegraphics[width=0.16\textwidth,height=0.16\textwidth,clip]{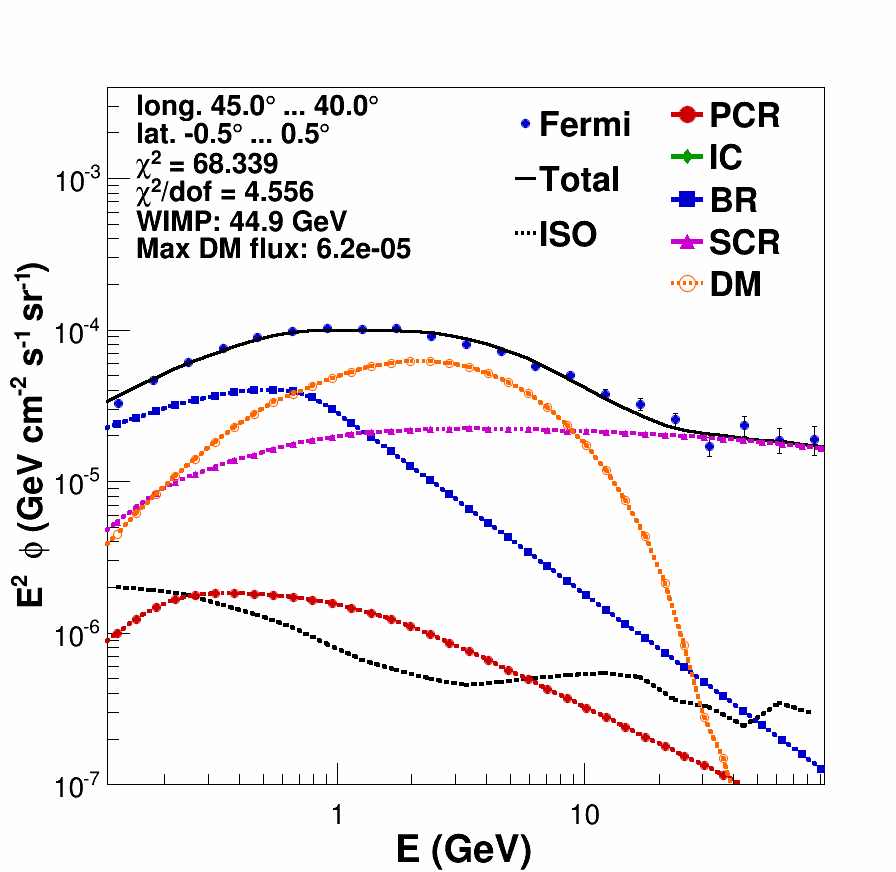}
\includegraphics[width=0.16\textwidth,height=0.16\textwidth,clip]{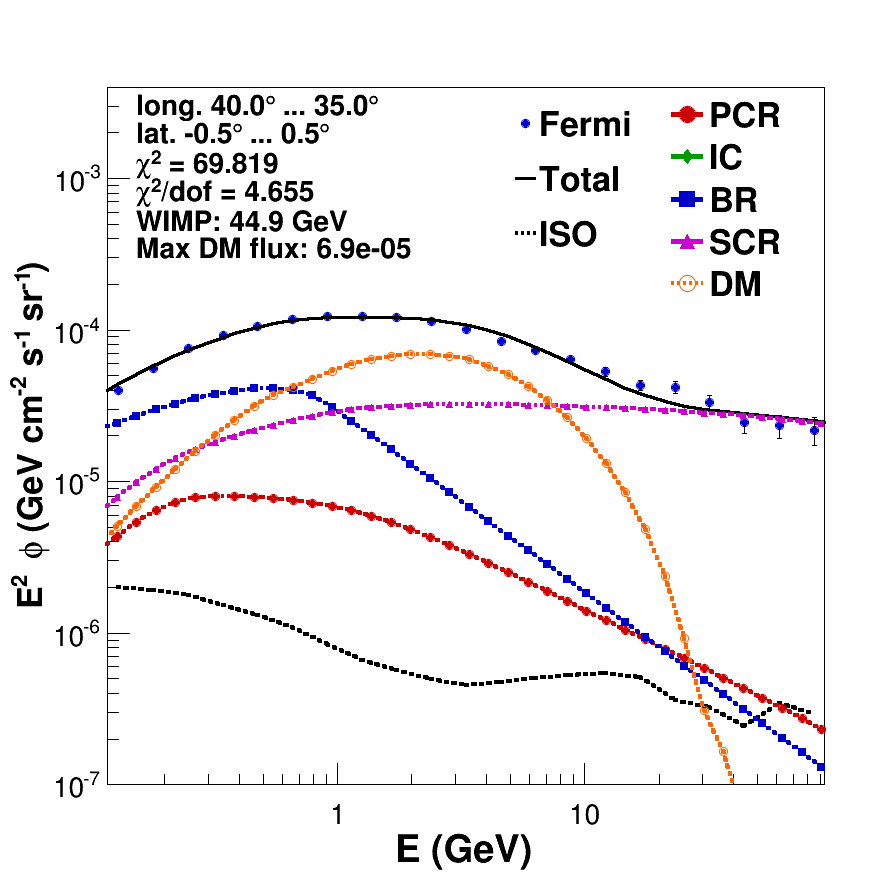}
\includegraphics[width=0.16\textwidth,height=0.16\textwidth,clip]{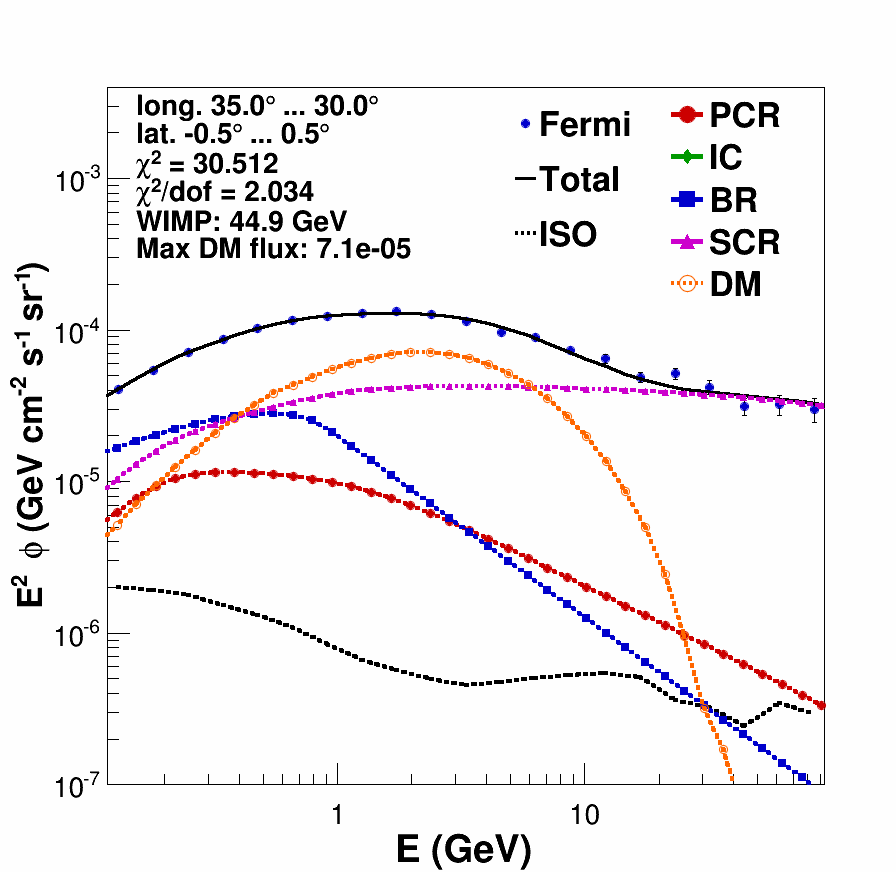}
\includegraphics[width=0.16\textwidth,height=0.16\textwidth,clip]{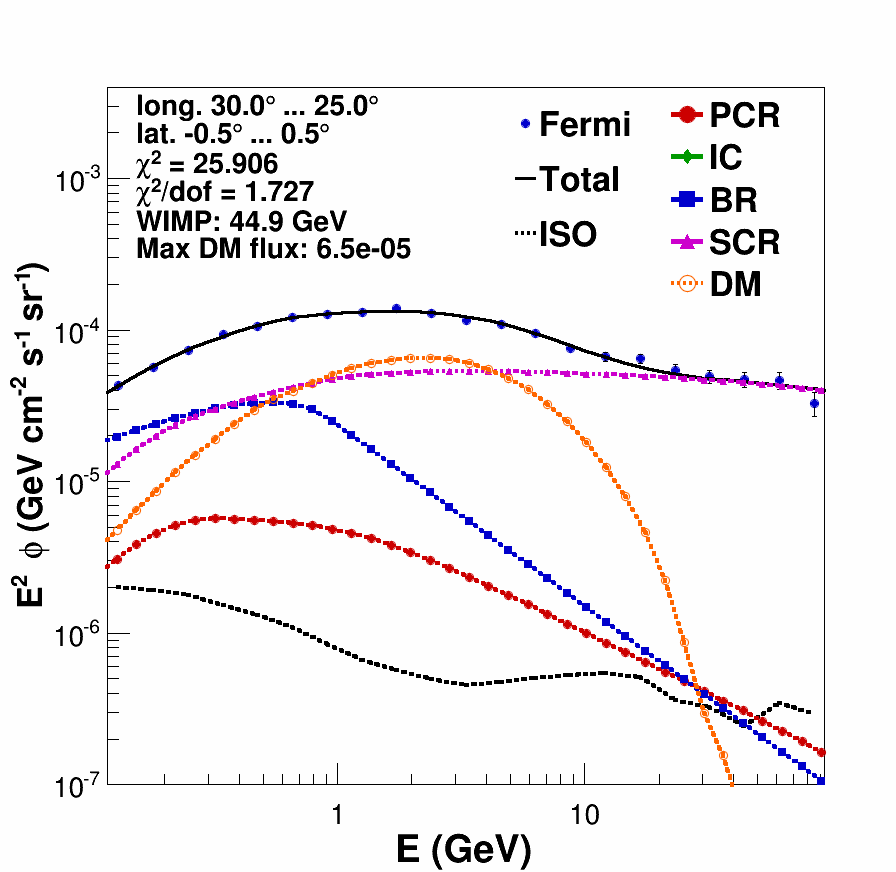}
\includegraphics[width=0.16\textwidth,height=0.16\textwidth,clip]{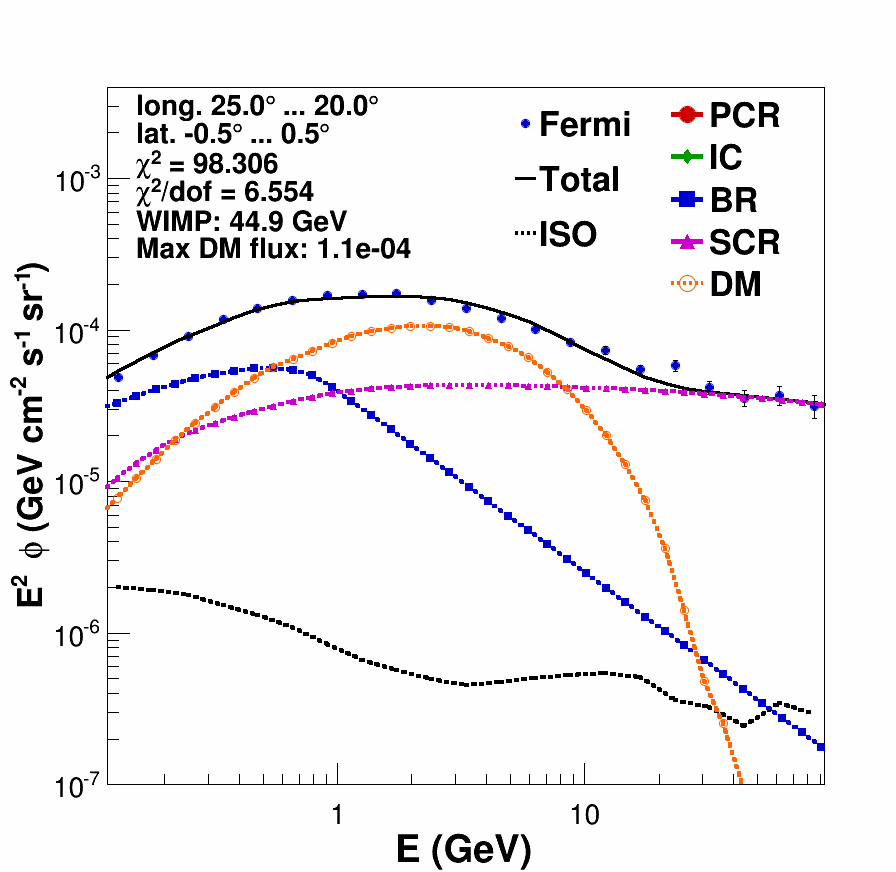}
\includegraphics[width=0.16\textwidth,height=0.16\textwidth,clip]{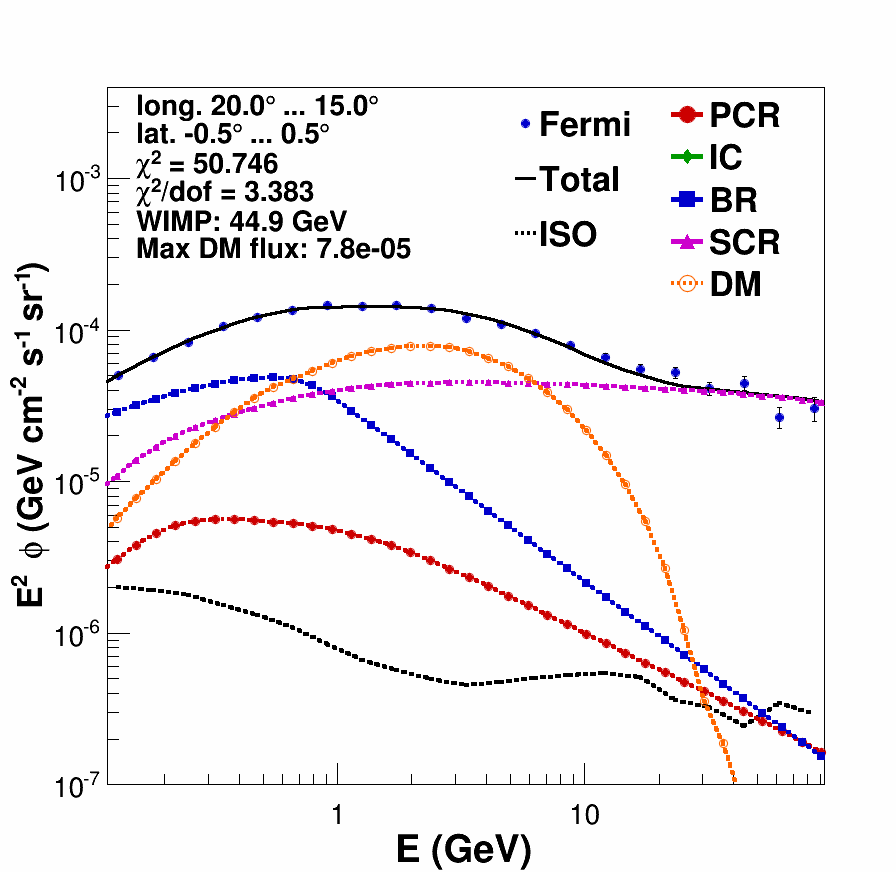}
\includegraphics[width=0.16\textwidth,height=0.16\textwidth,clip]{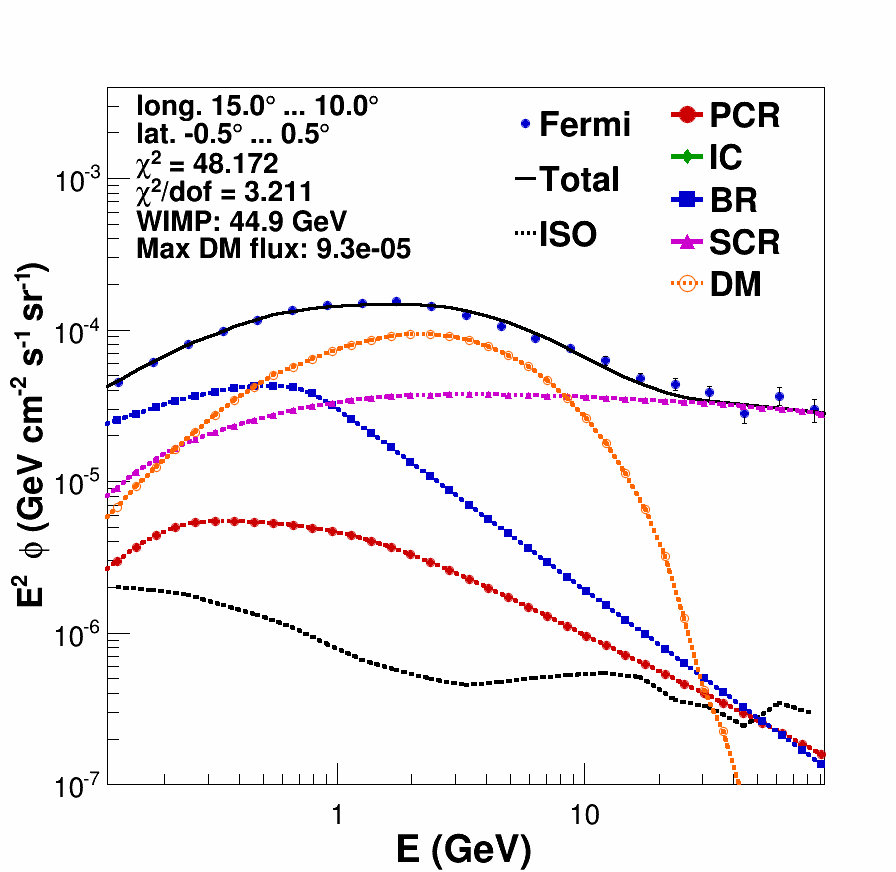}
\includegraphics[width=0.16\textwidth,height=0.16\textwidth,clip]{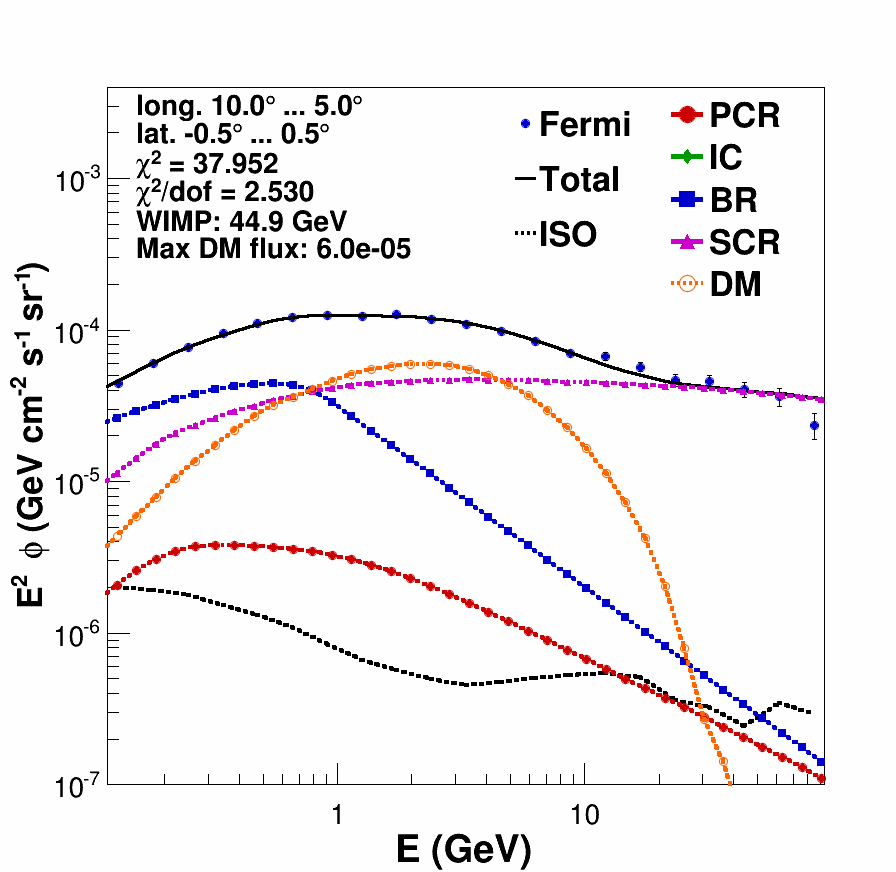}
\includegraphics[width=0.16\textwidth,height=0.16\textwidth,clip]{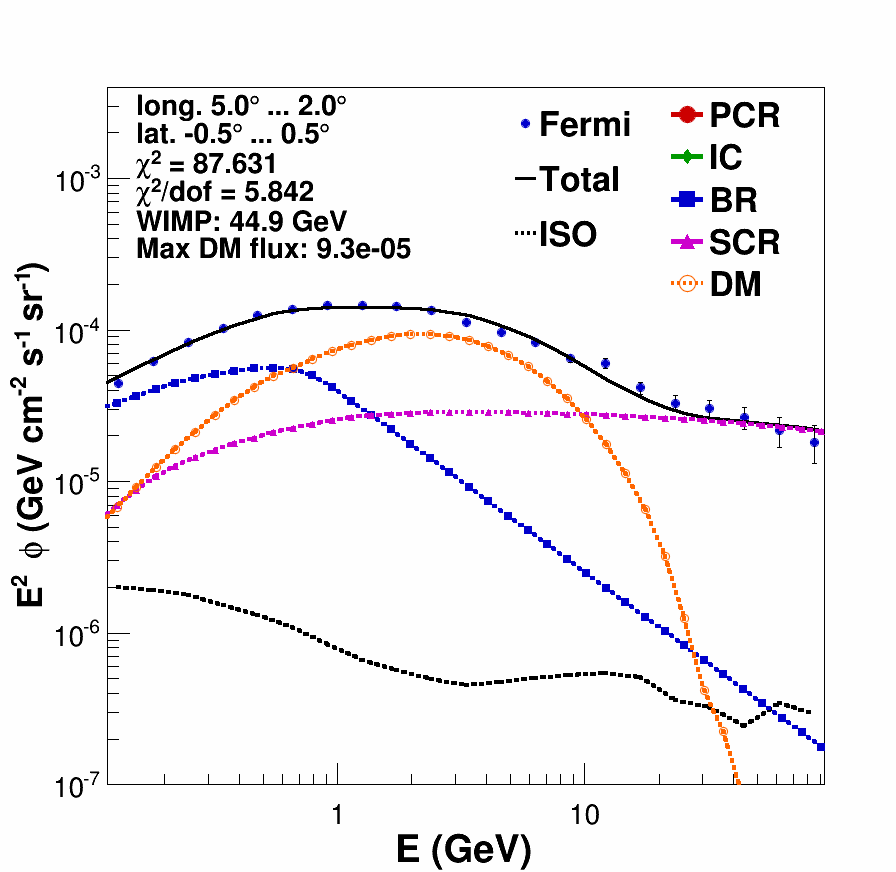}
\includegraphics[width=0.16\textwidth,height=0.16\textwidth,clip]{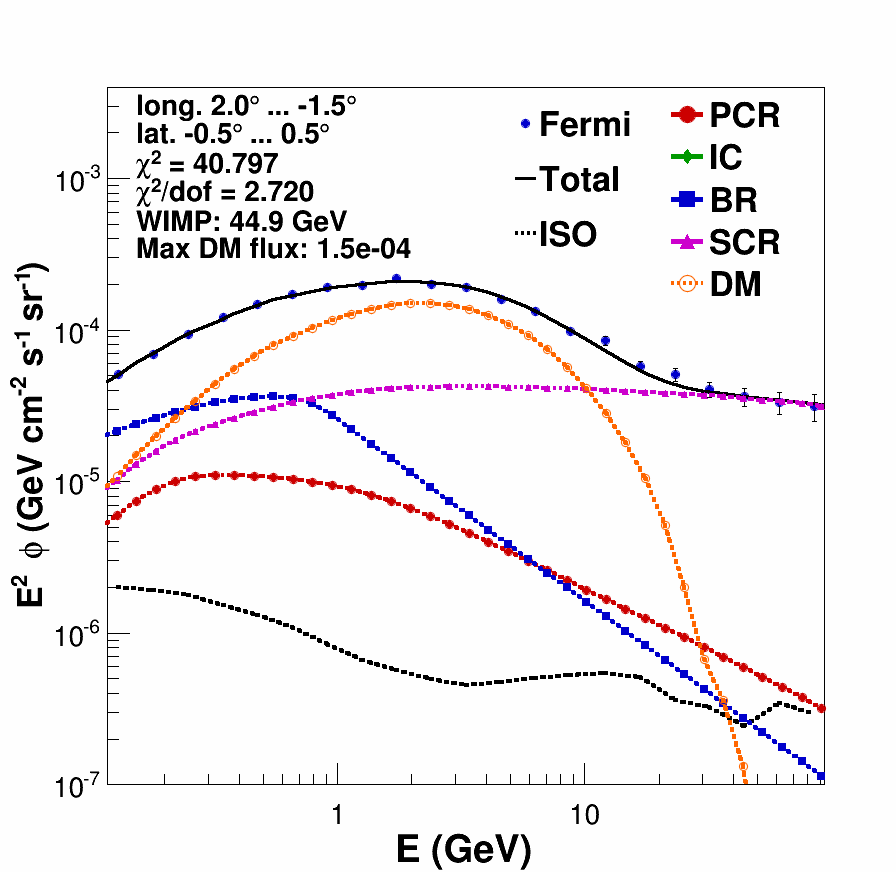}
\includegraphics[width=0.16\textwidth,height=0.16\textwidth,clip]{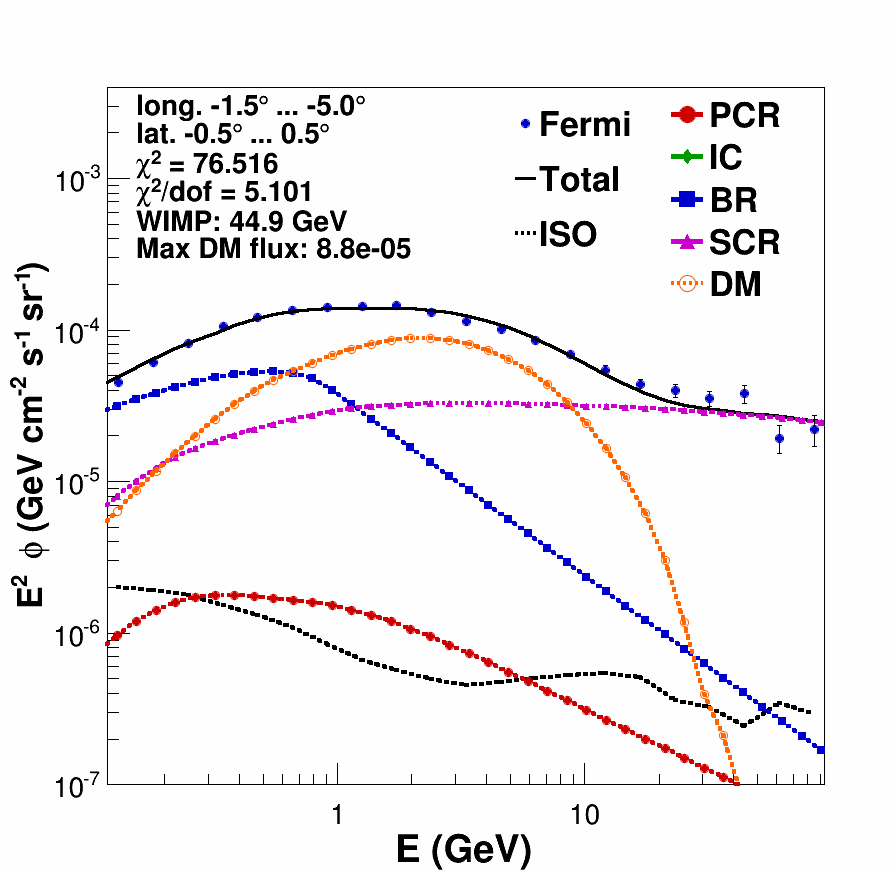}
\includegraphics[width=0.16\textwidth,height=0.16\textwidth,clip]{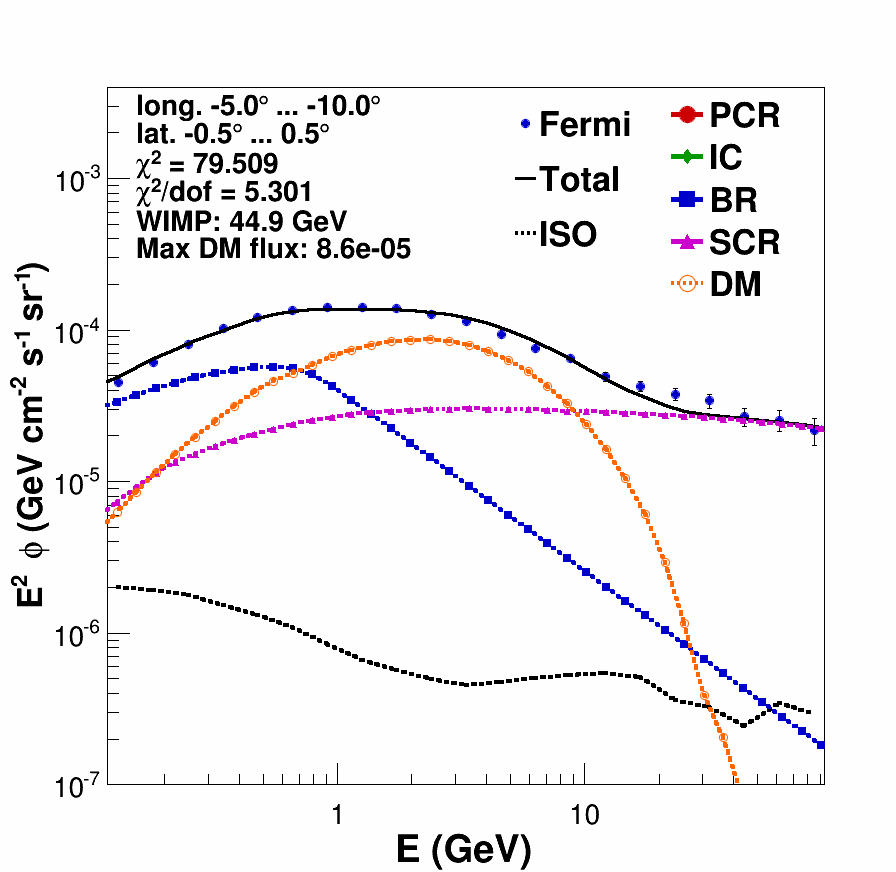}
\includegraphics[width=0.16\textwidth,height=0.16\textwidth,clip]{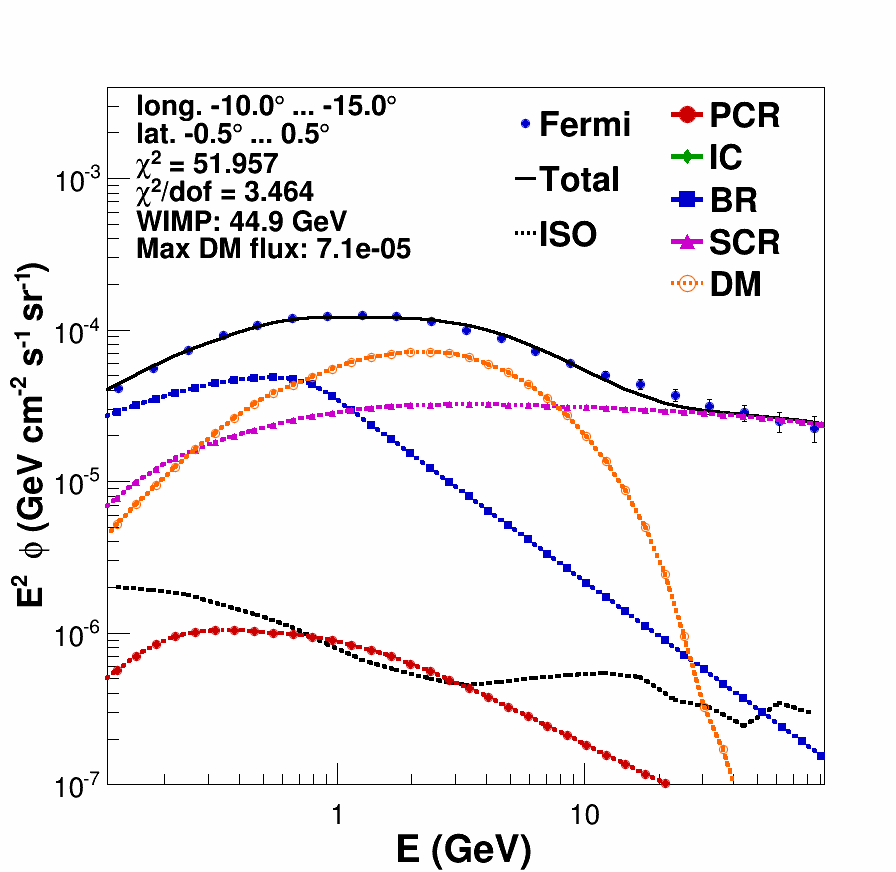}
\includegraphics[width=0.16\textwidth,height=0.16\textwidth,clip]{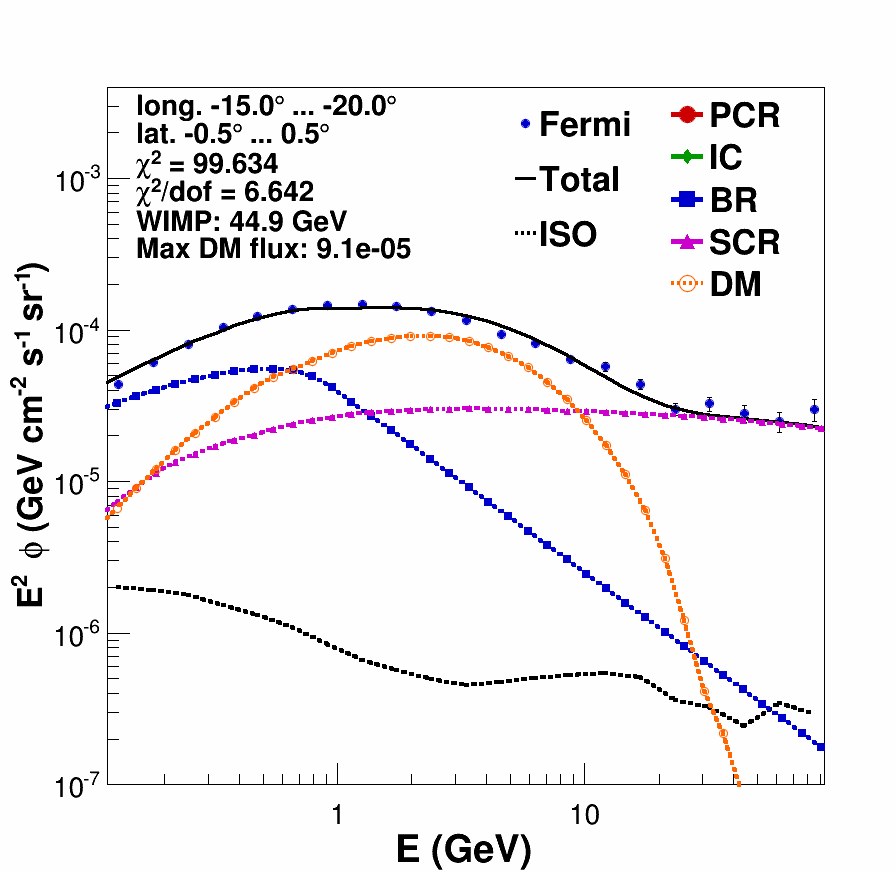}
\includegraphics[width=0.16\textwidth,height=0.16\textwidth,clip]{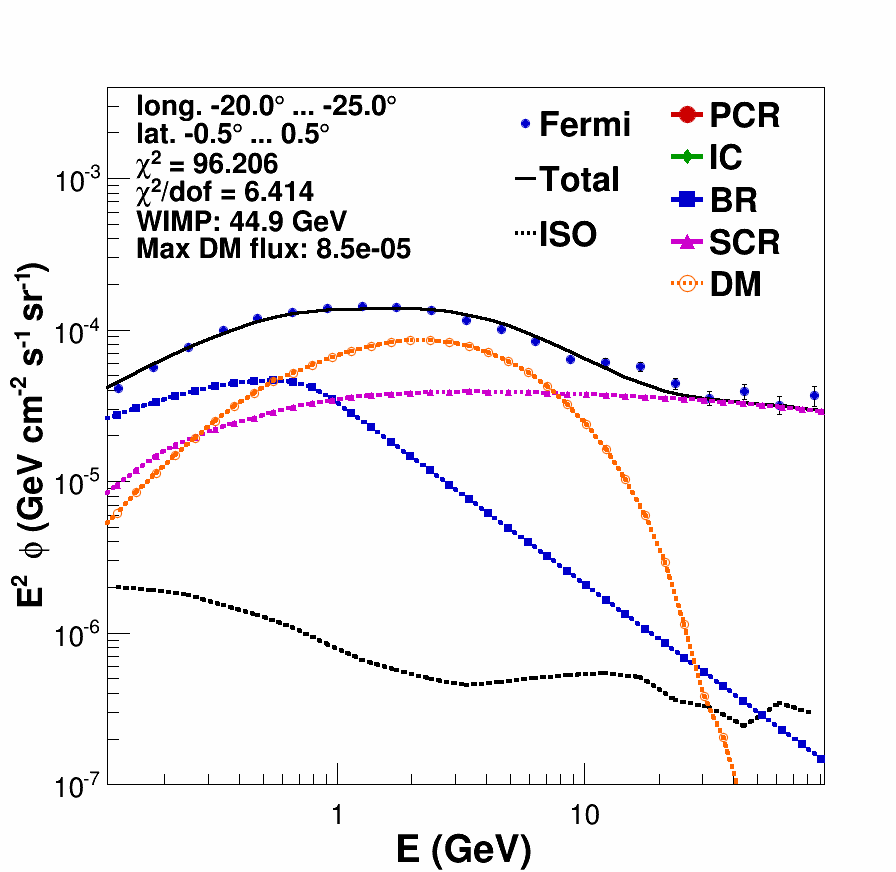}
\includegraphics[width=0.16\textwidth,height=0.16\textwidth,clip]{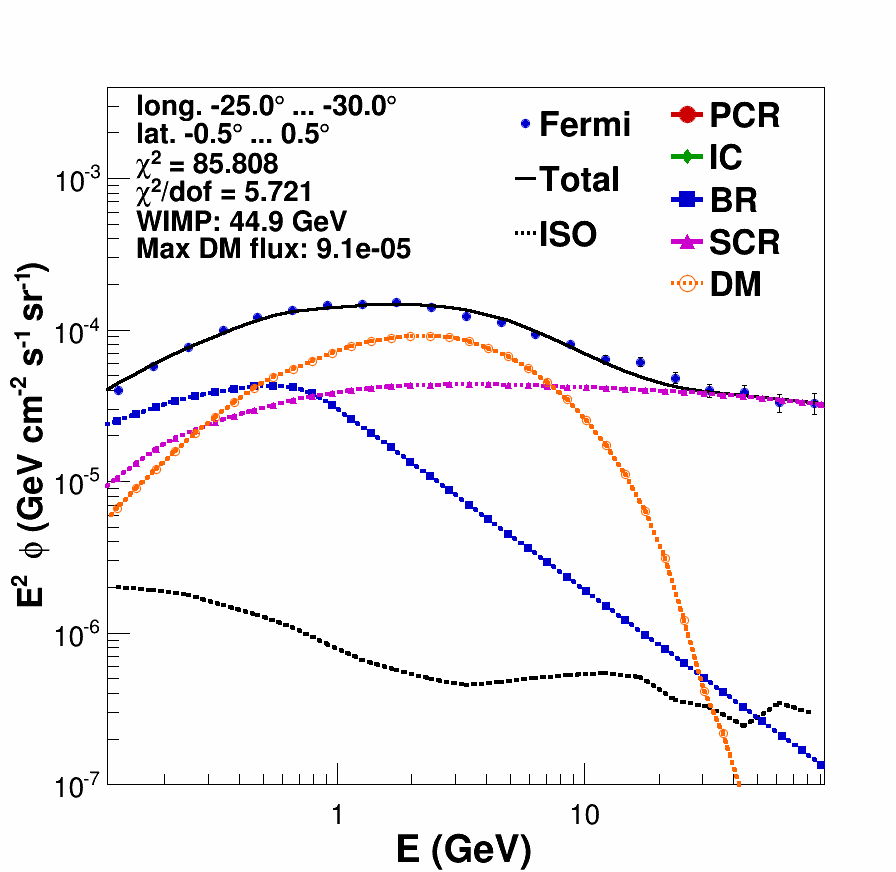}
\includegraphics[width=0.16\textwidth,height=0.16\textwidth,clip]{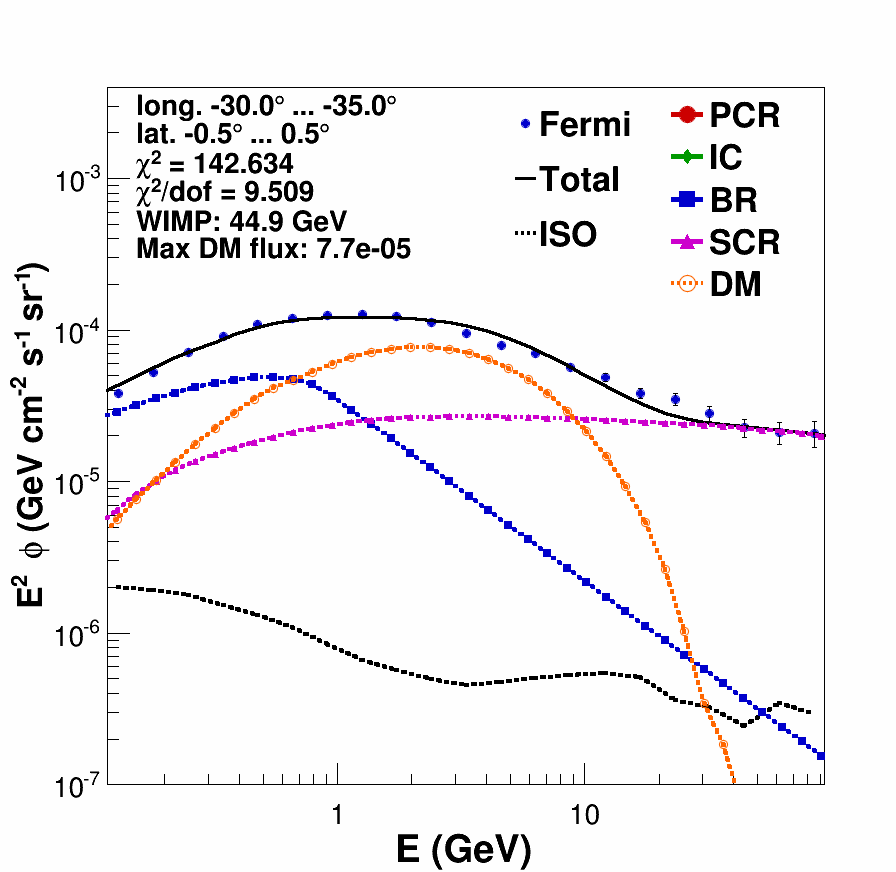}
\includegraphics[width=0.16\textwidth,height=0.16\textwidth,clip]{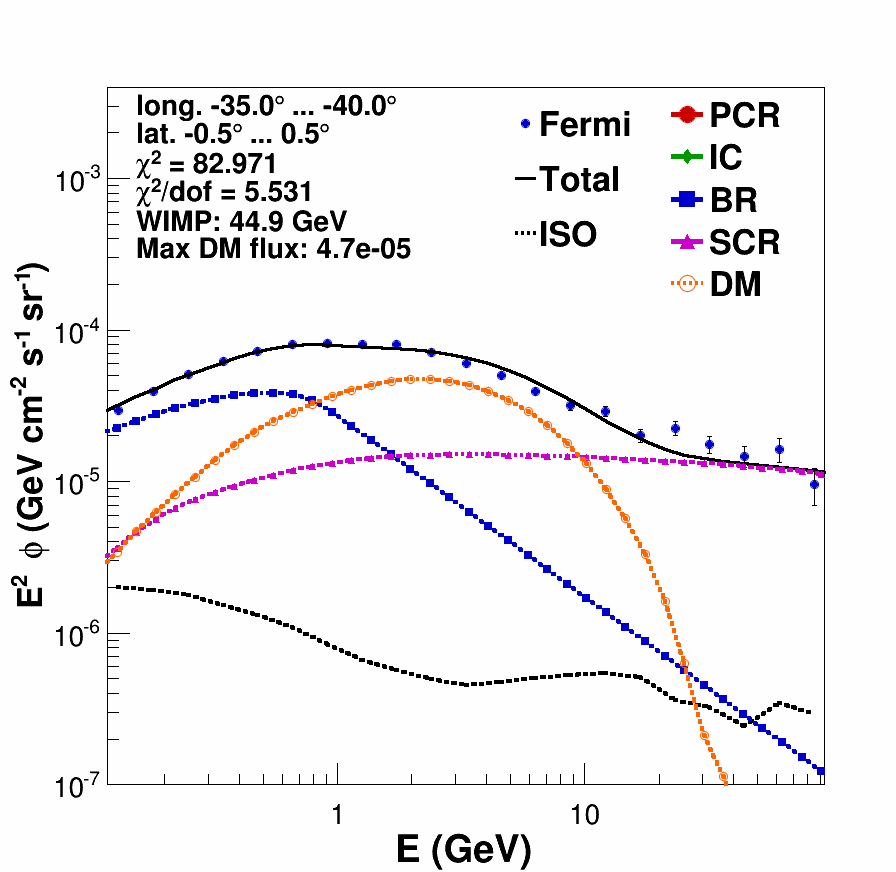}
\includegraphics[width=0.16\textwidth,height=0.16\textwidth,clip]{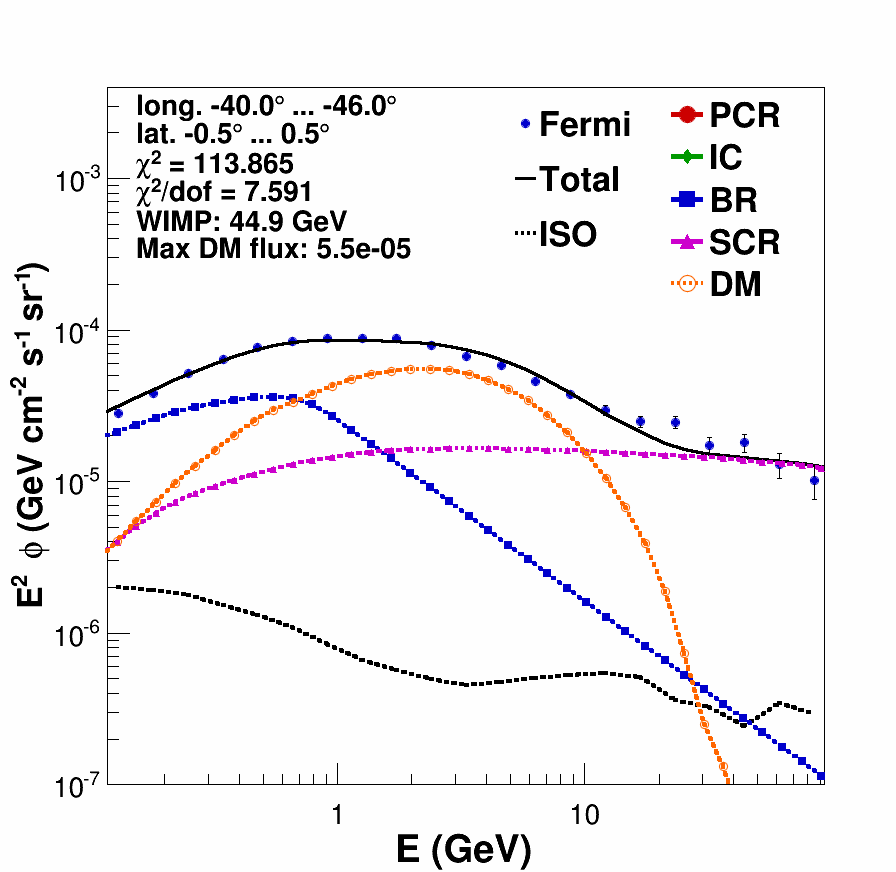}
\includegraphics[width=0.16\textwidth,height=0.16\textwidth,clip]{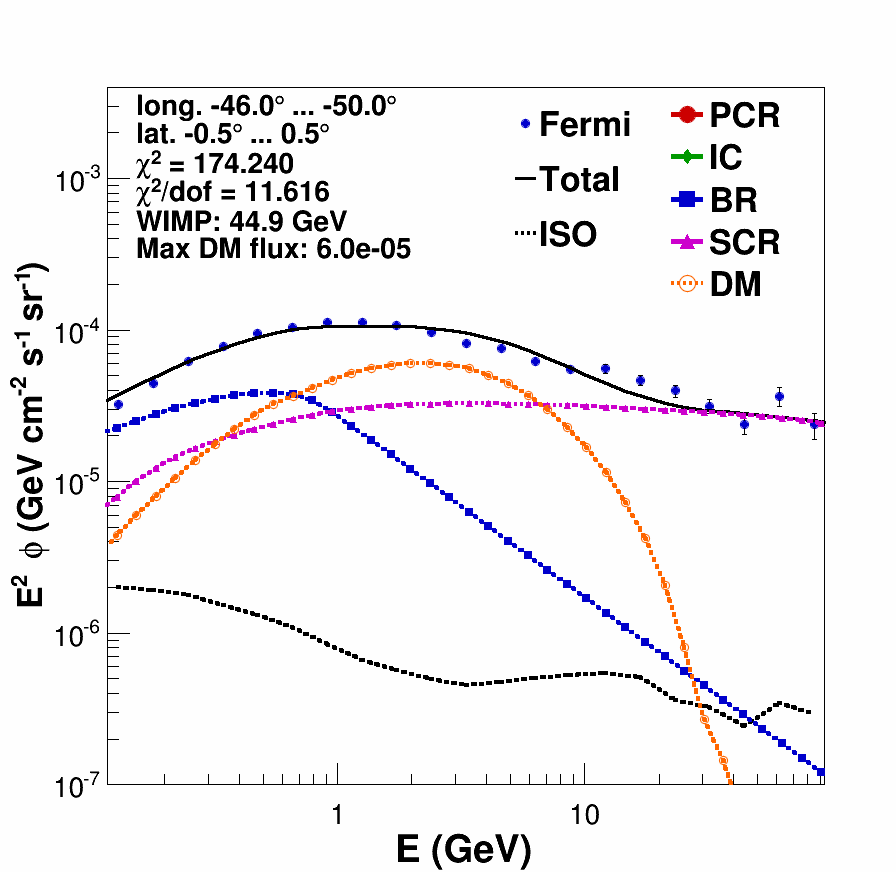}
\includegraphics[width=0.16\textwidth,height=0.16\textwidth,clip]{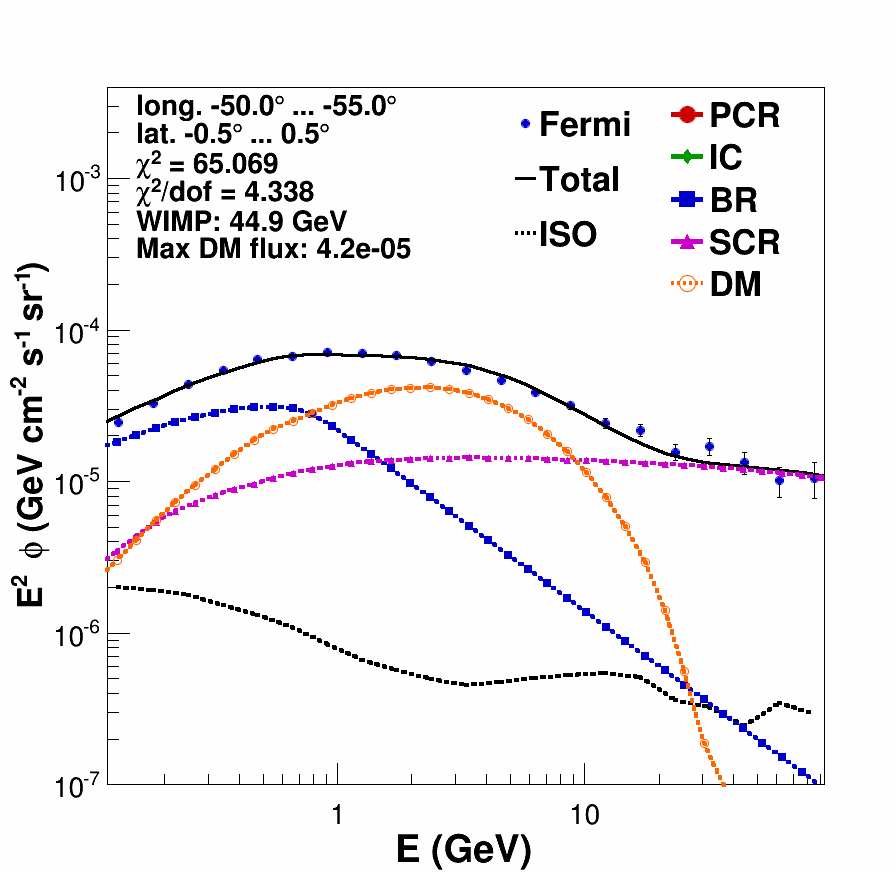}
\includegraphics[width=0.16\textwidth,height=0.16\textwidth,clip]{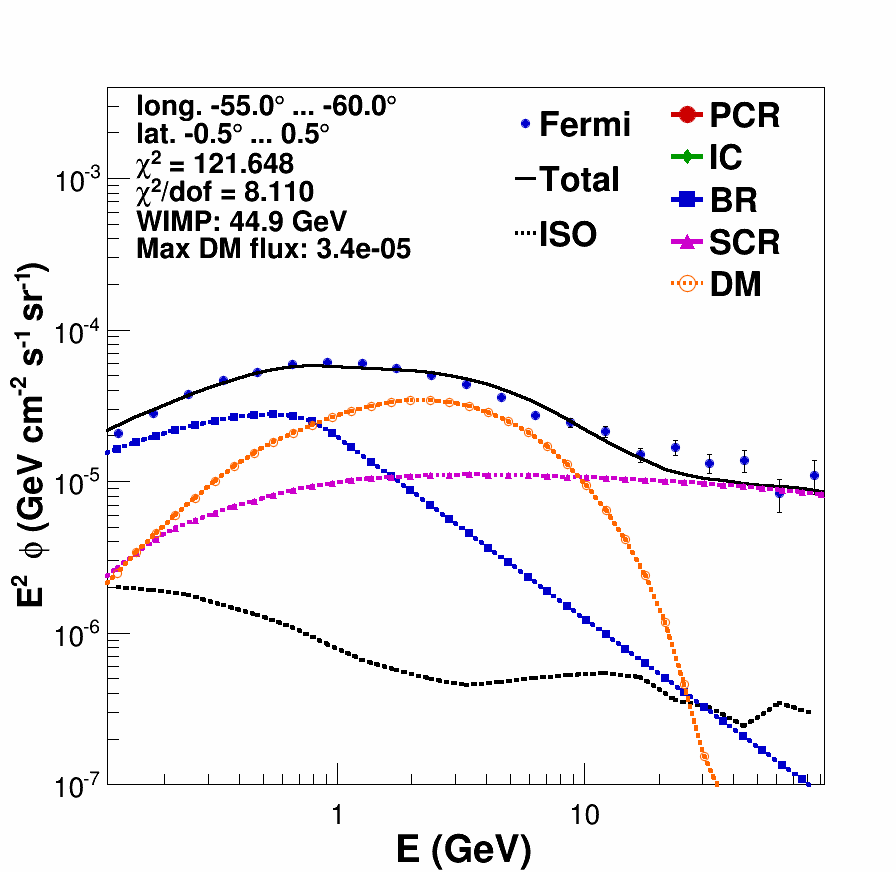}
\includegraphics[width=0.16\textwidth,height=0.16\textwidth,clip]{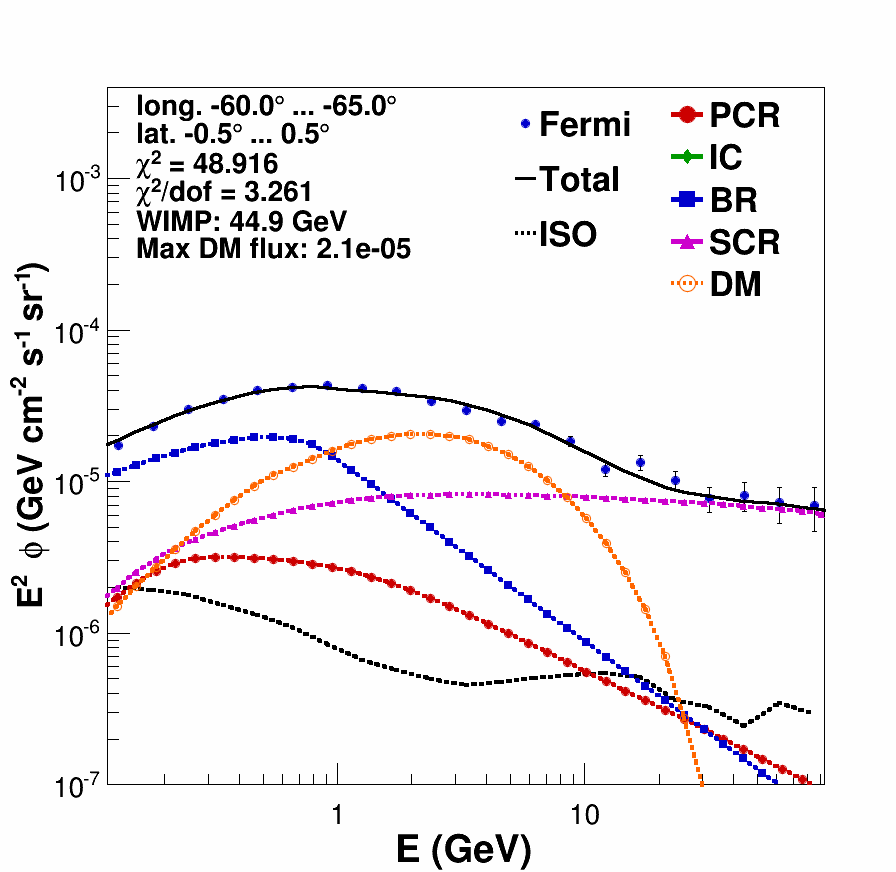}
\includegraphics[width=0.16\textwidth,height=0.16\textwidth,clip]{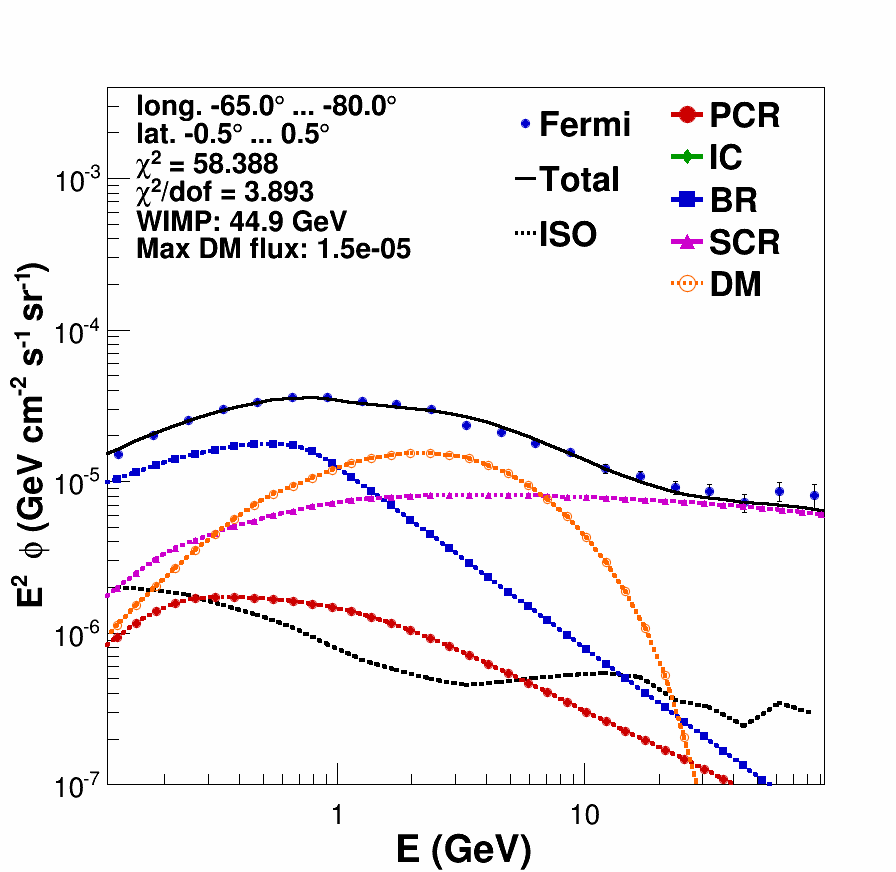}
\includegraphics[width=0.16\textwidth,height=0.16\textwidth,clip]{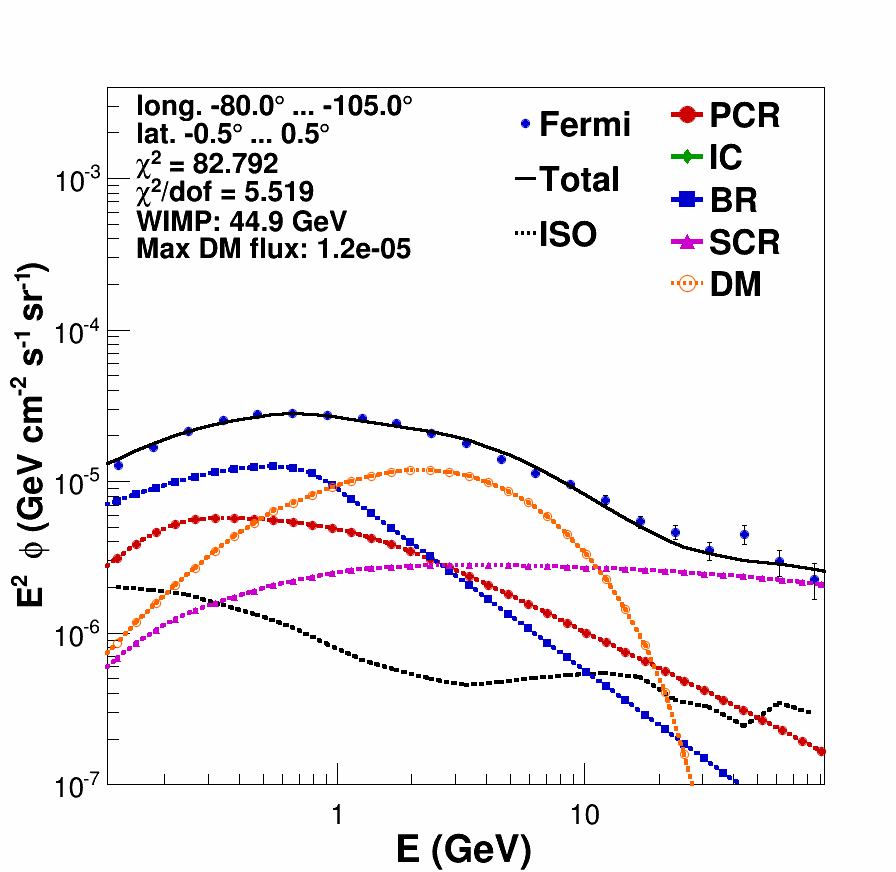}
\includegraphics[width=0.16\textwidth,height=0.16\textwidth,clip]{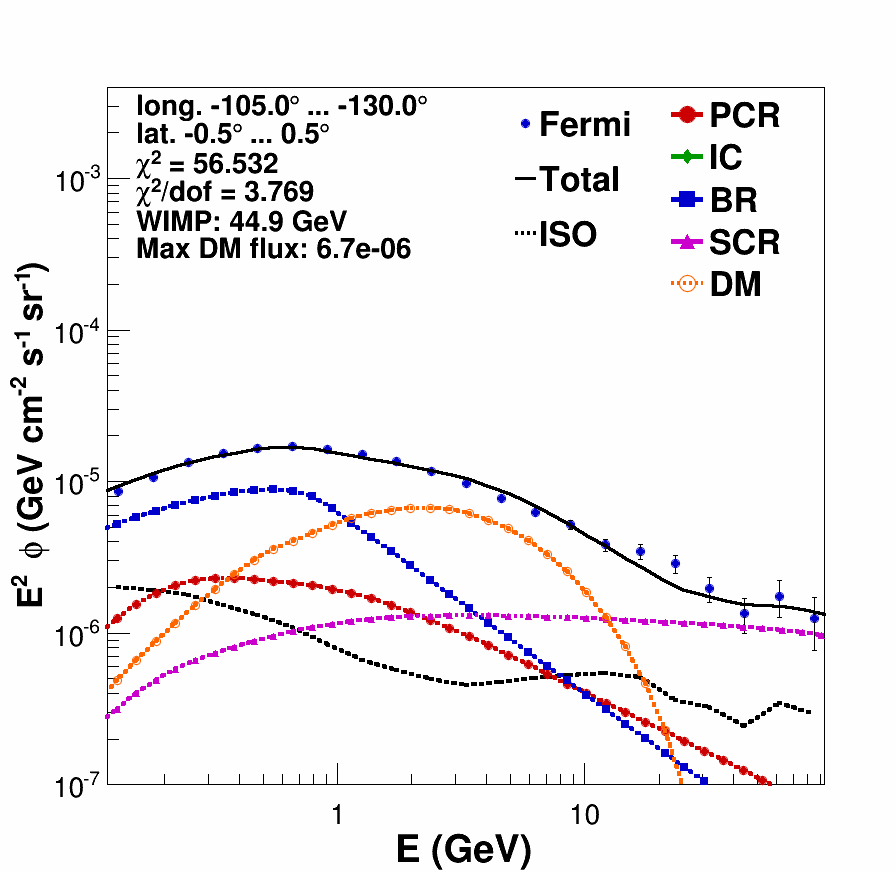}
\includegraphics[width=0.16\textwidth,height=0.16\textwidth,clip]{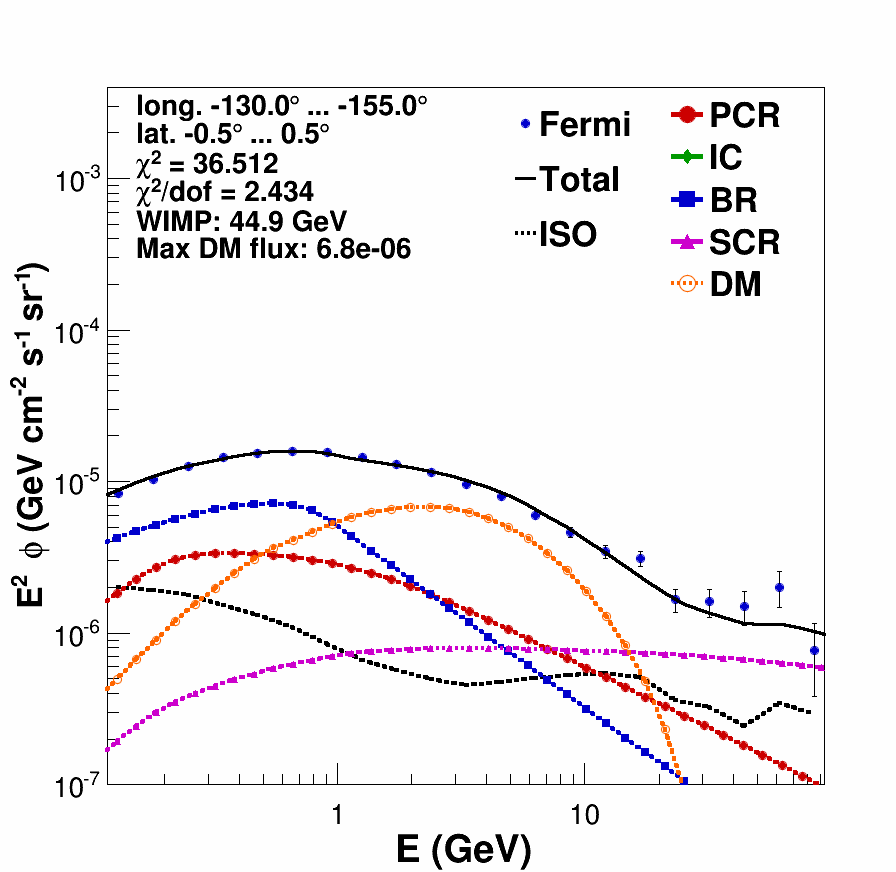}
\includegraphics[width=0.16\textwidth,height=0.16\textwidth,clip]{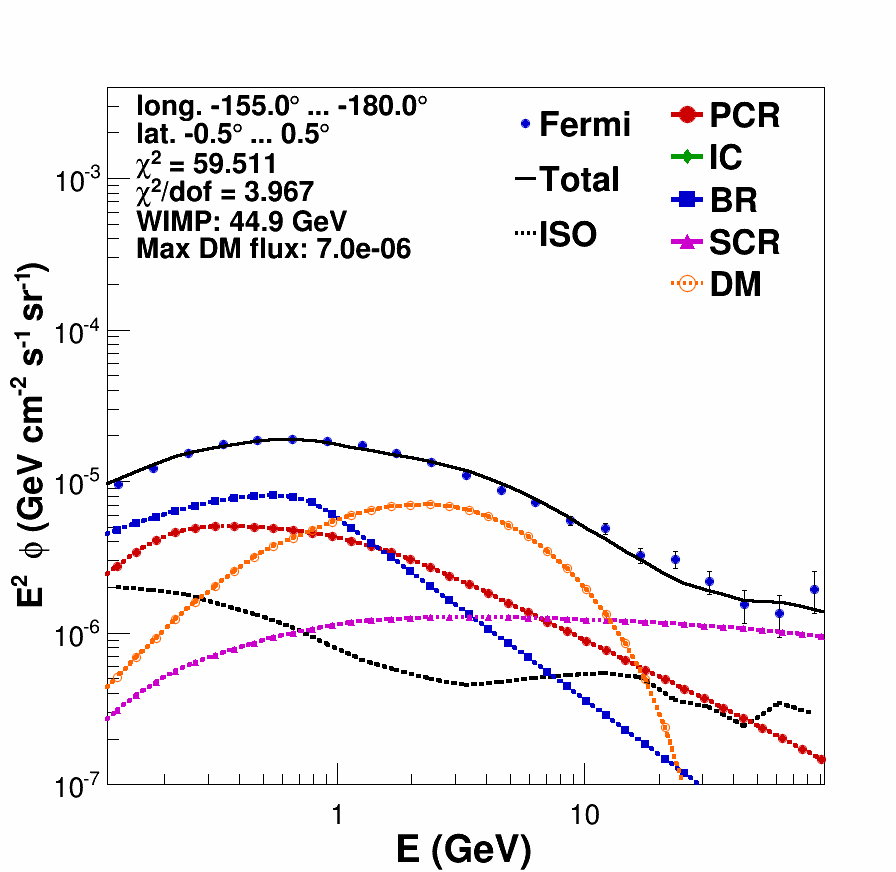}%%%%%%gd
\caption[]{Template fits for latitudes  with $-0.5^\circ<b<0.5^\circ$ and longitudes decreasing from 180$^\circ$ to -180$^\circ$.} \label{F42}
\end{figure}
\begin{figure}
\centering
	\includegraphics[width=0.16\textwidth,height=0.16\textwidth,clip]{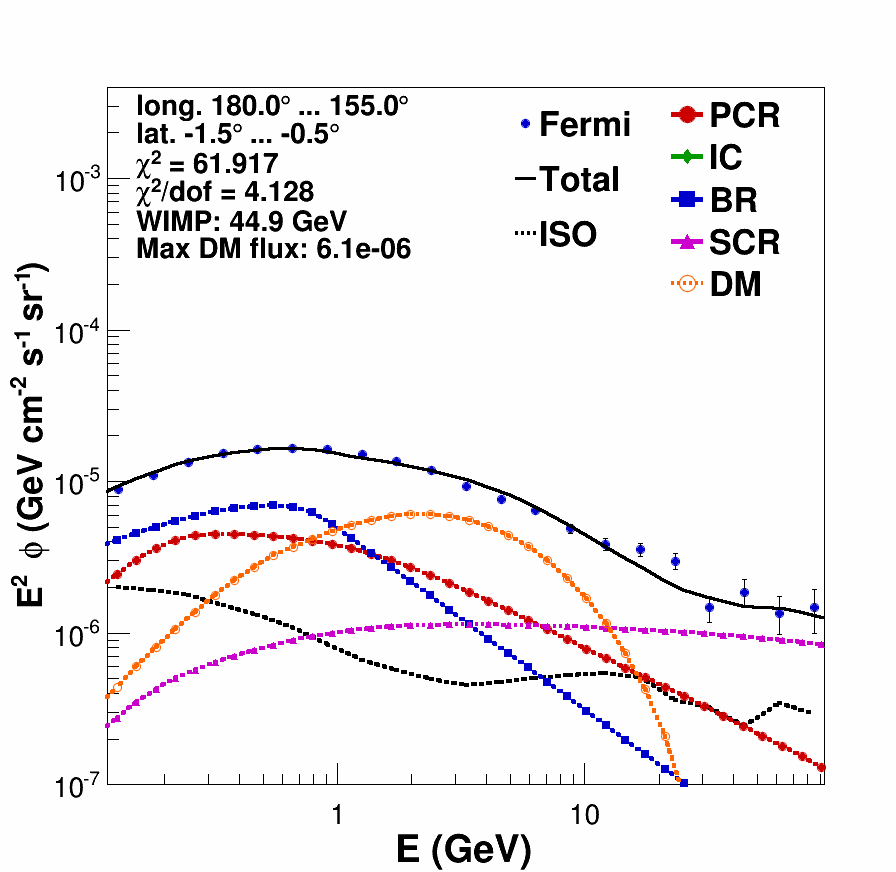}
	\includegraphics[width=0.16\textwidth,height=0.16\textwidth,clip]{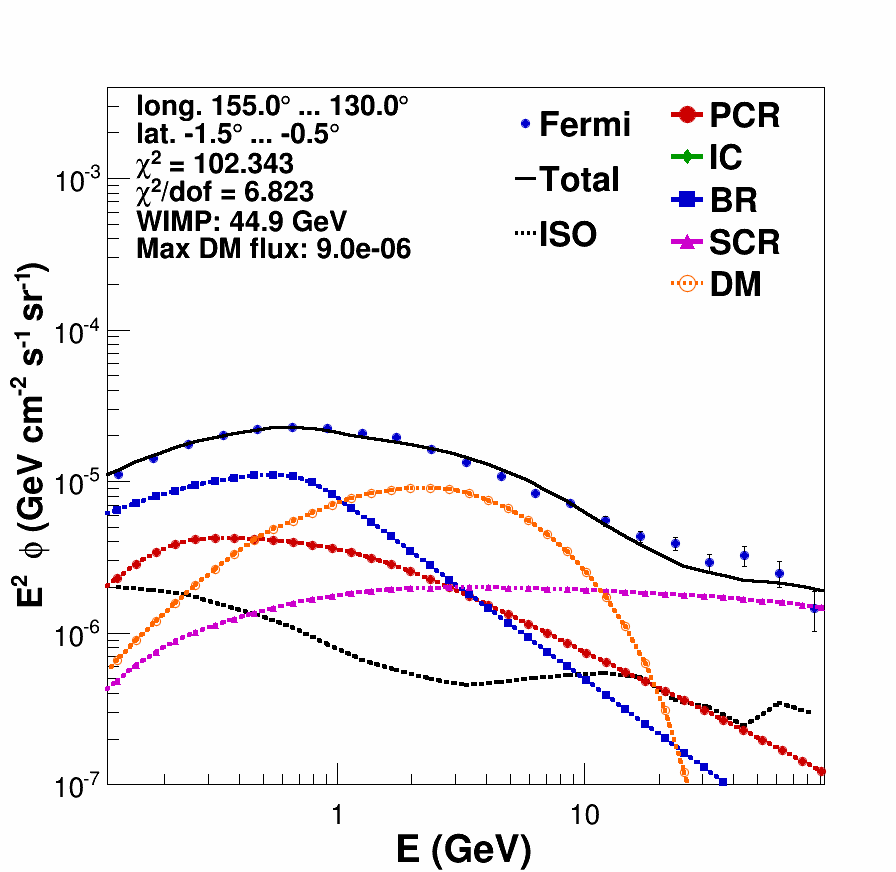}
	\includegraphics[width=0.16\textwidth,height=0.16\textwidth,clip]{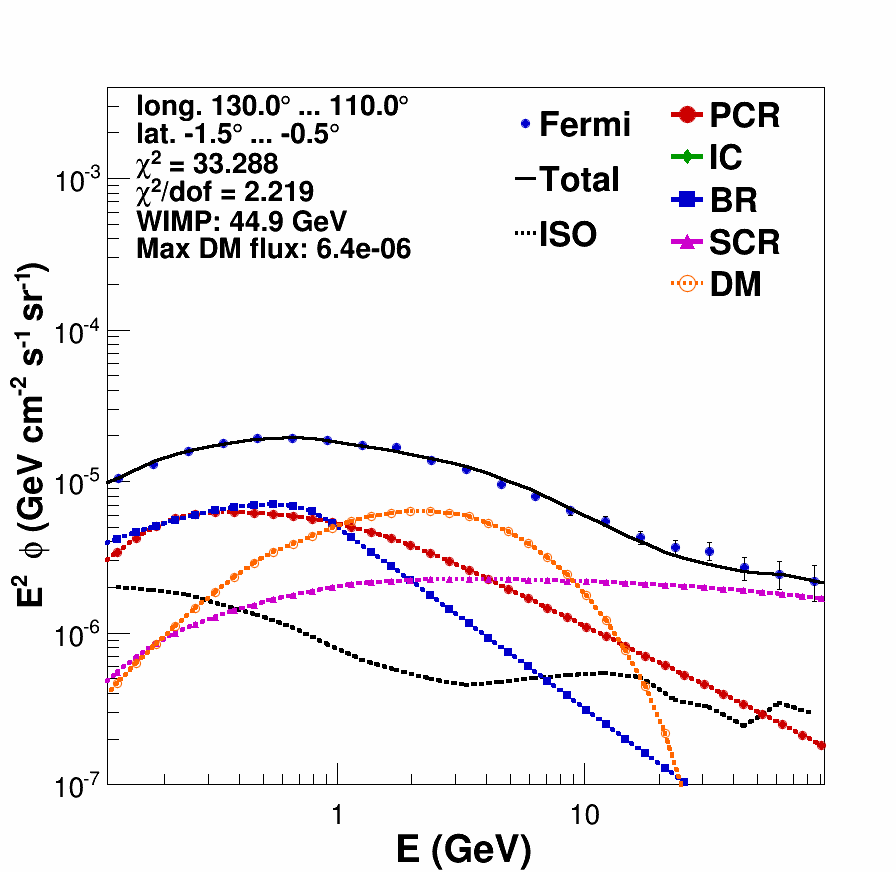}
	\includegraphics[width=0.16\textwidth,height=0.16\textwidth,clip]{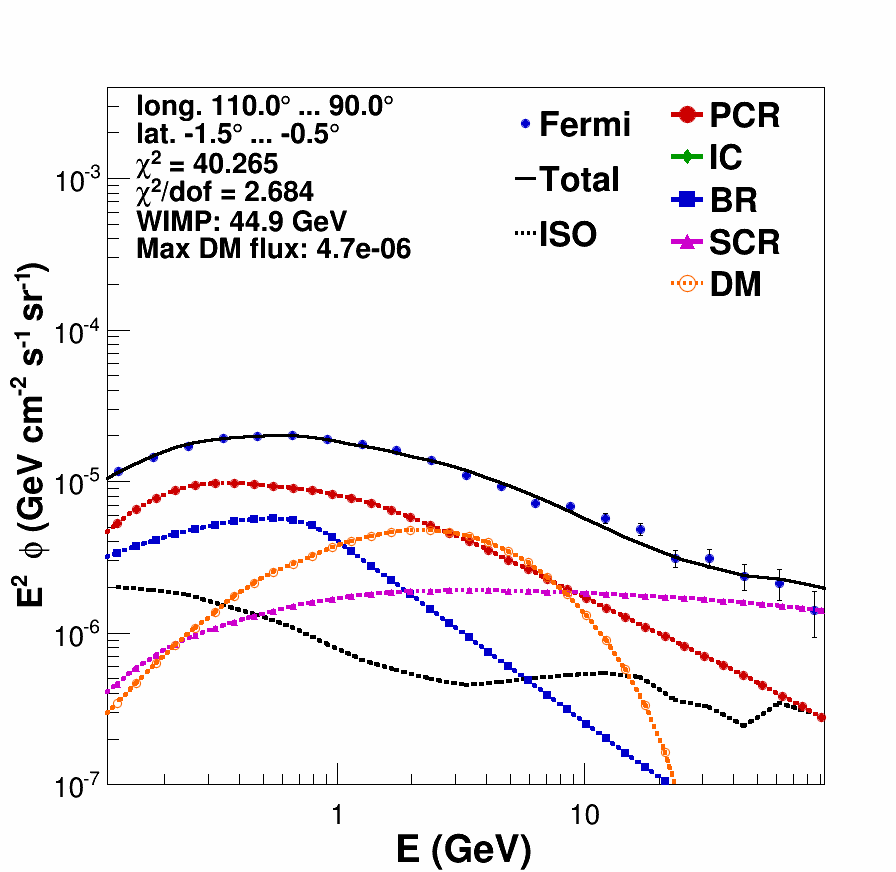}
	\includegraphics[width=0.16\textwidth,height=0.16\textwidth,clip]{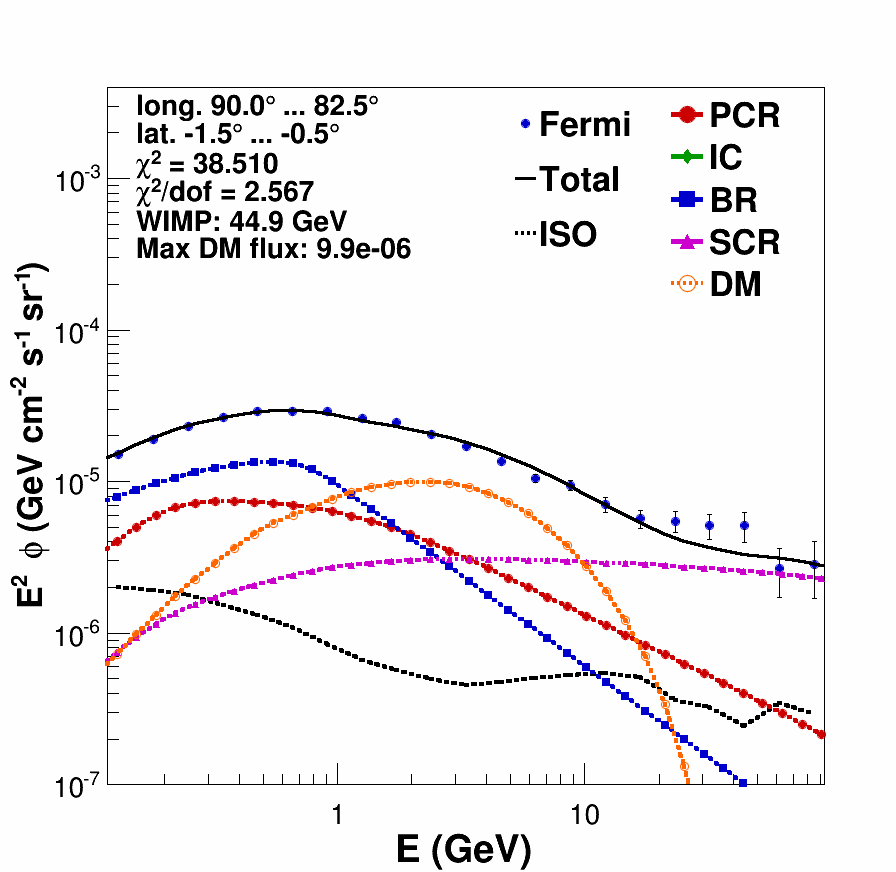}
	\includegraphics[width=0.16\textwidth,height=0.16\textwidth,clip]{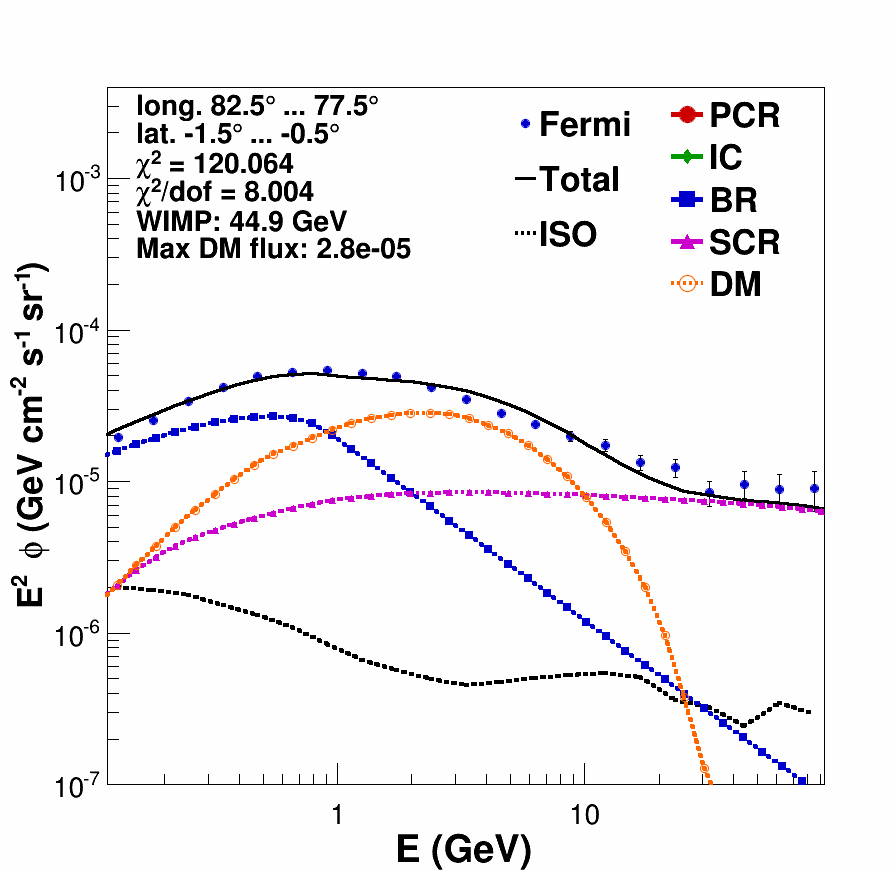}
	\includegraphics[width=0.16\textwidth,height=0.16\textwidth,clip]{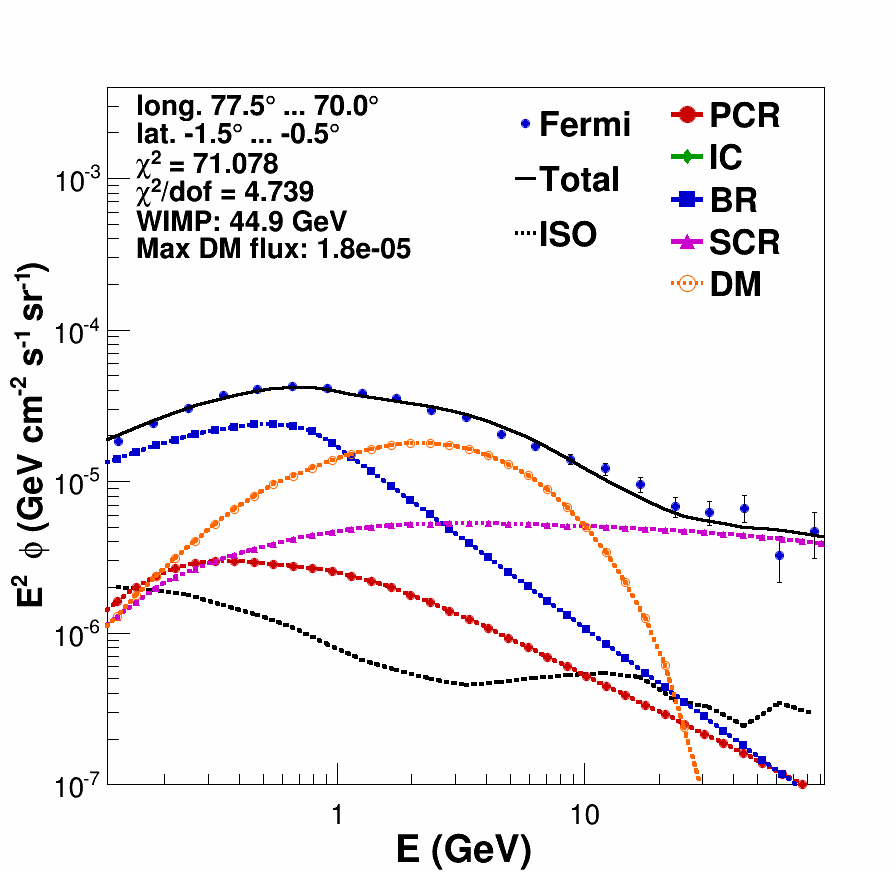}
	\includegraphics[width=0.16\textwidth,height=0.16\textwidth,clip]{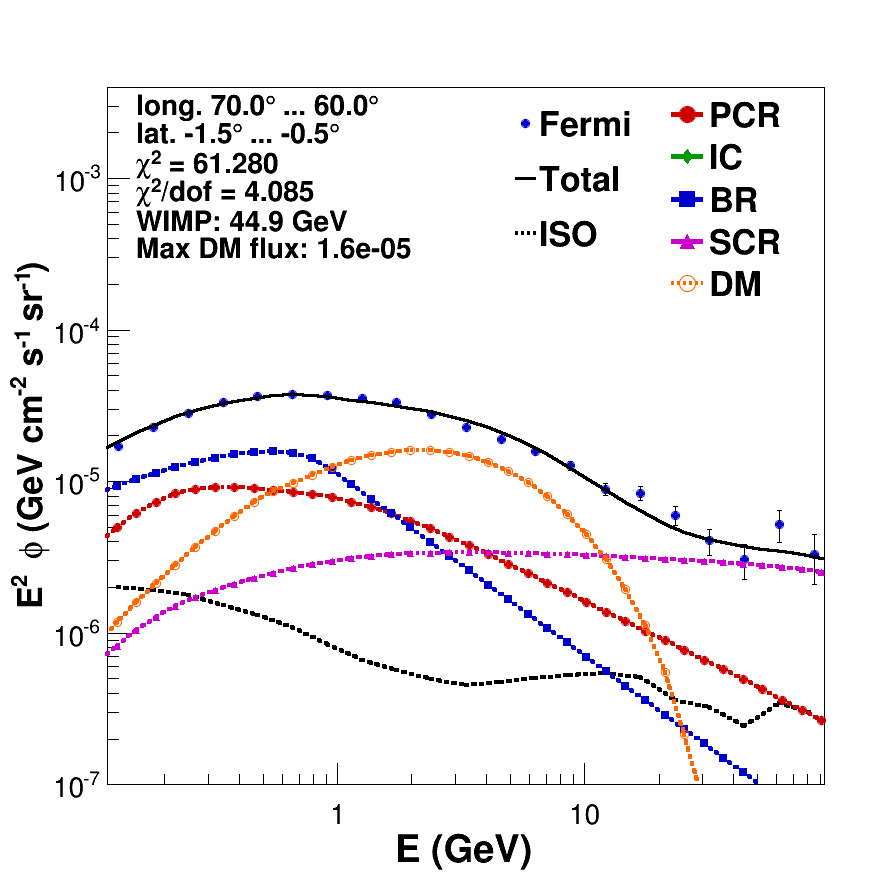}
	\includegraphics[width=0.16\textwidth,height=0.16\textwidth,clip]{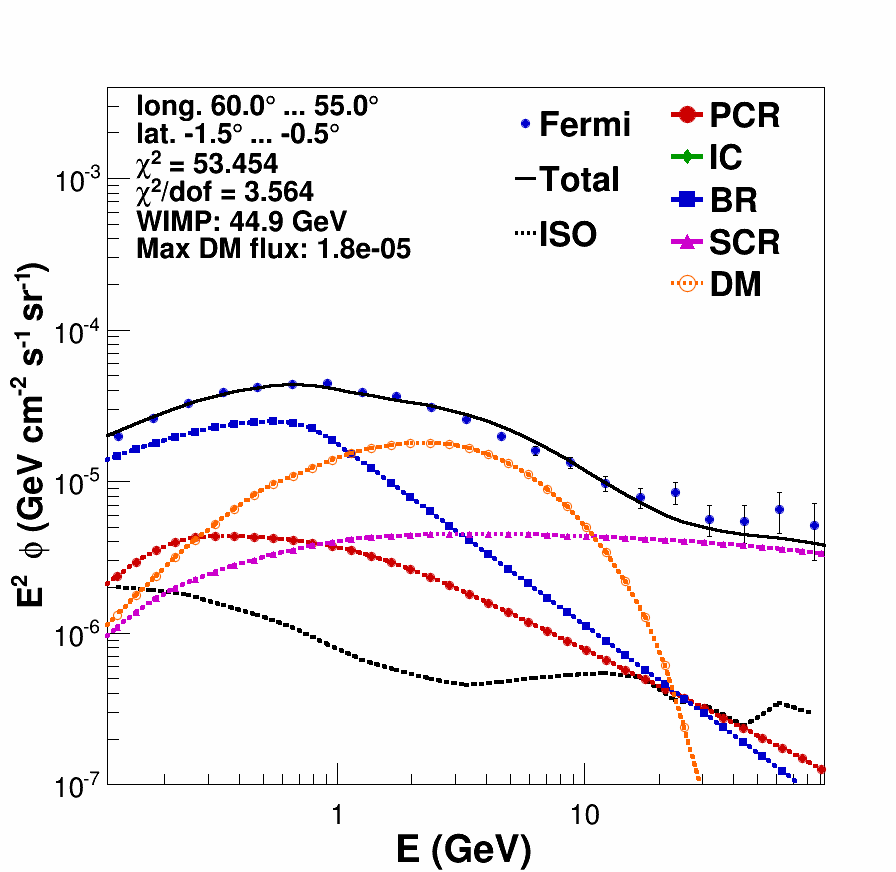}
	\includegraphics[width=0.16\textwidth,height=0.16\textwidth,clip]{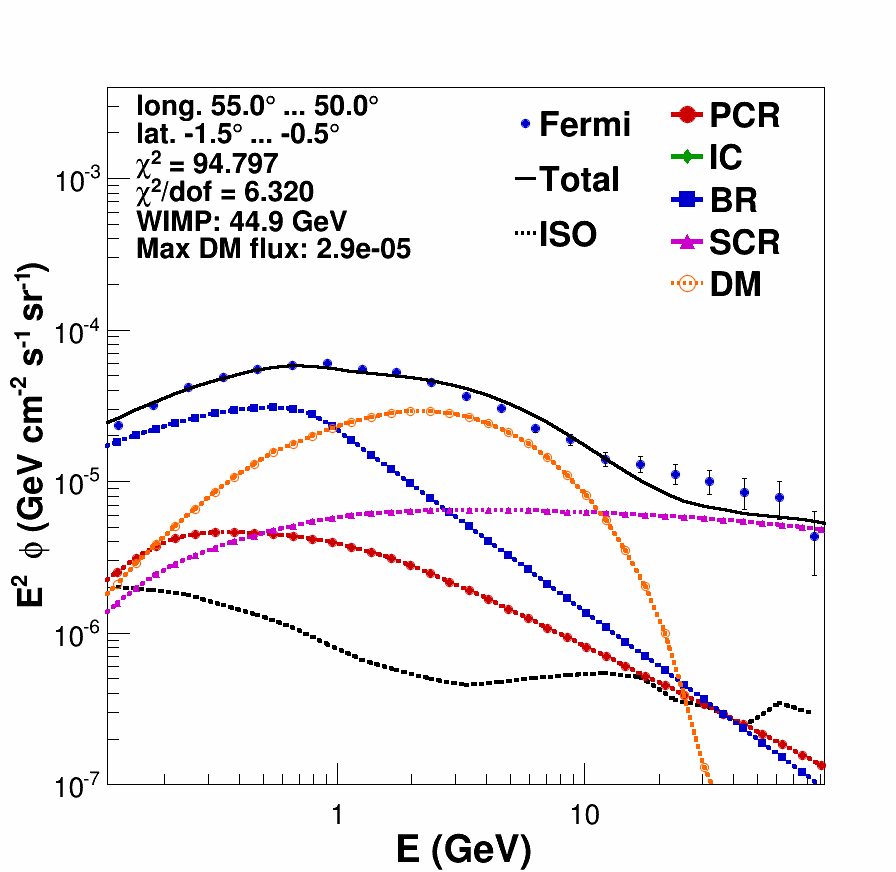}
	\includegraphics[width=0.16\textwidth,height=0.16\textwidth,clip]{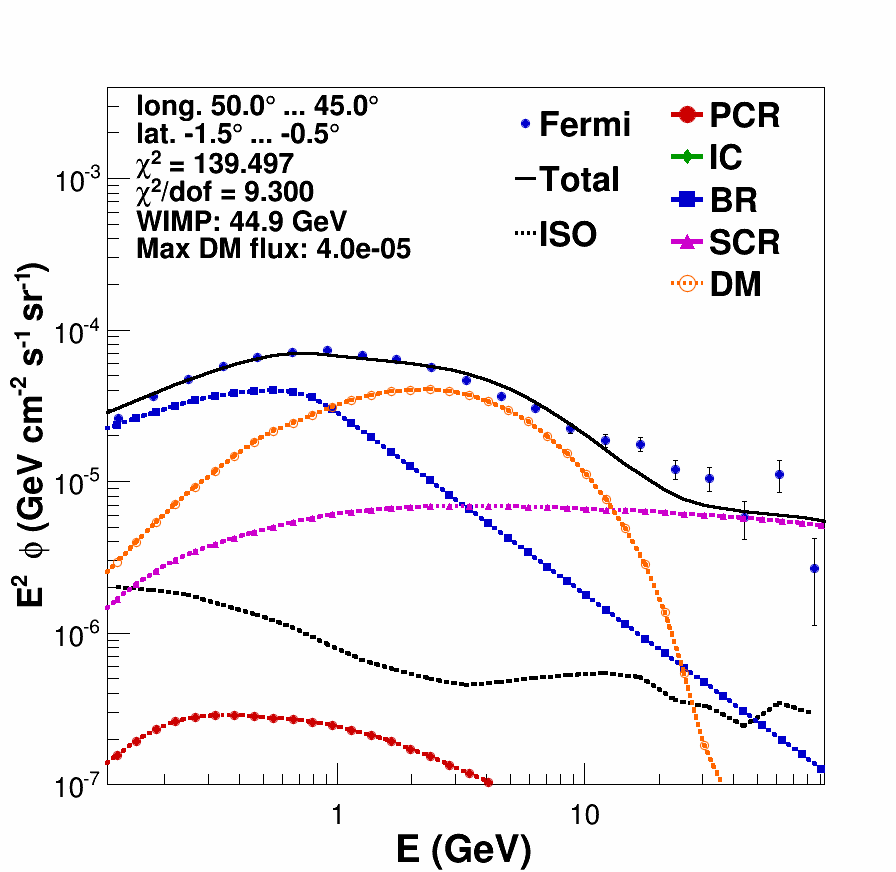}
	\includegraphics[width=0.16\textwidth,height=0.16\textwidth,clip]{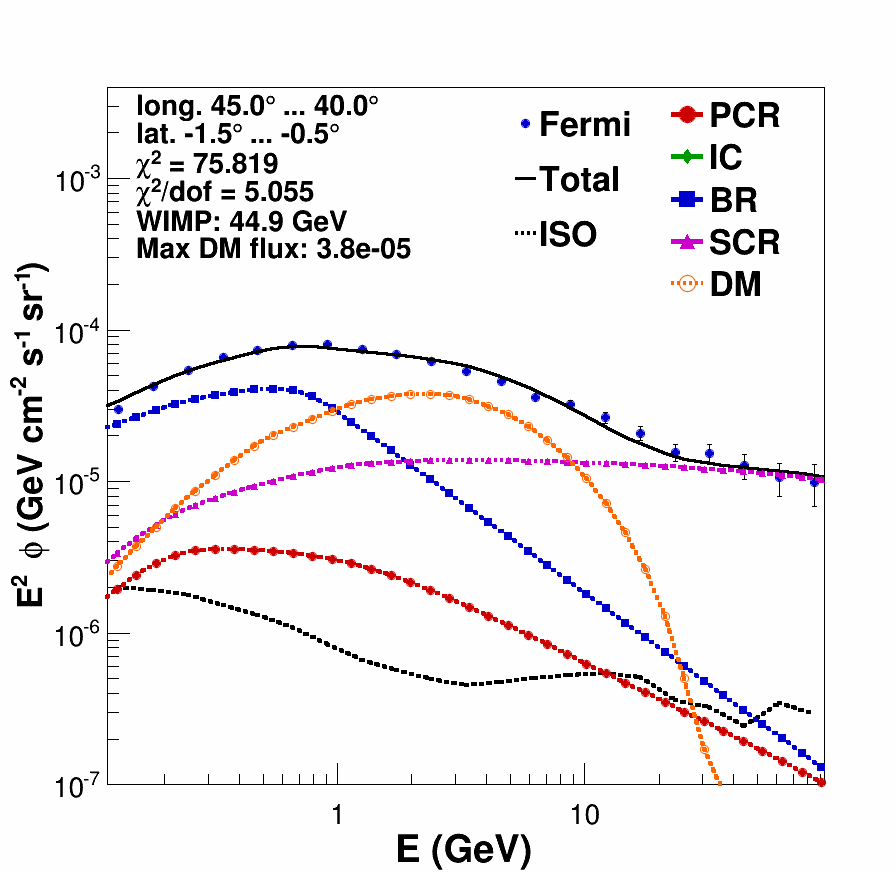}
	\includegraphics[width=0.16\textwidth,height=0.16\textwidth,clip]{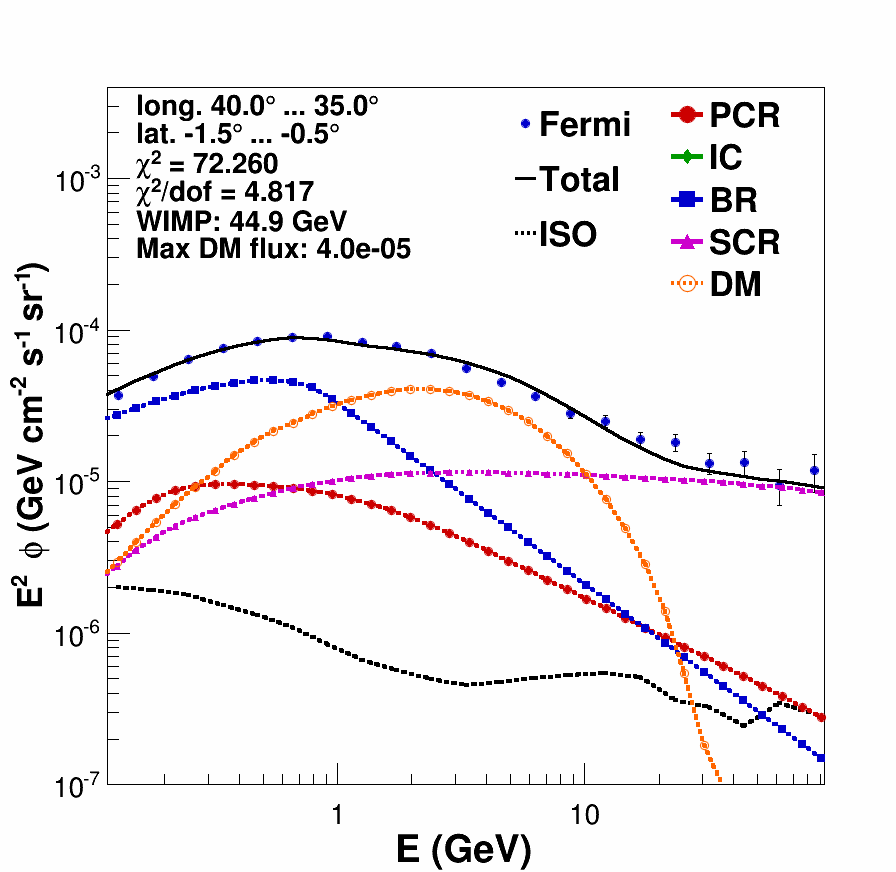}
	\includegraphics[width=0.16\textwidth,height=0.16\textwidth,clip]{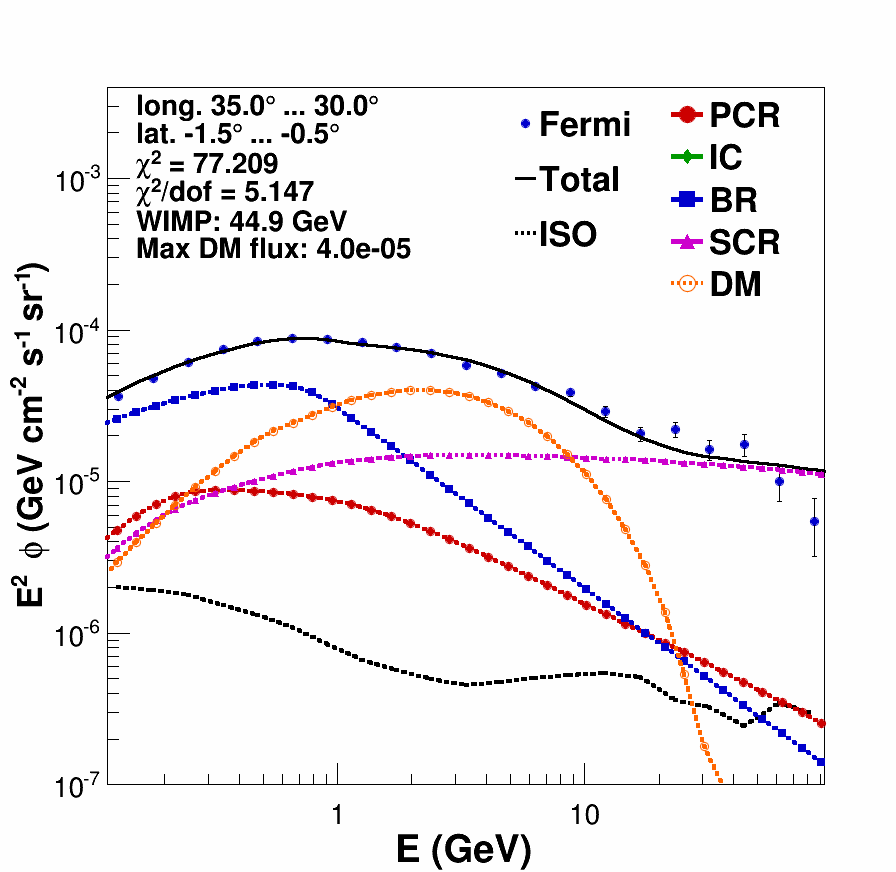}
	\includegraphics[width=0.16\textwidth,height=0.16\textwidth,clip]{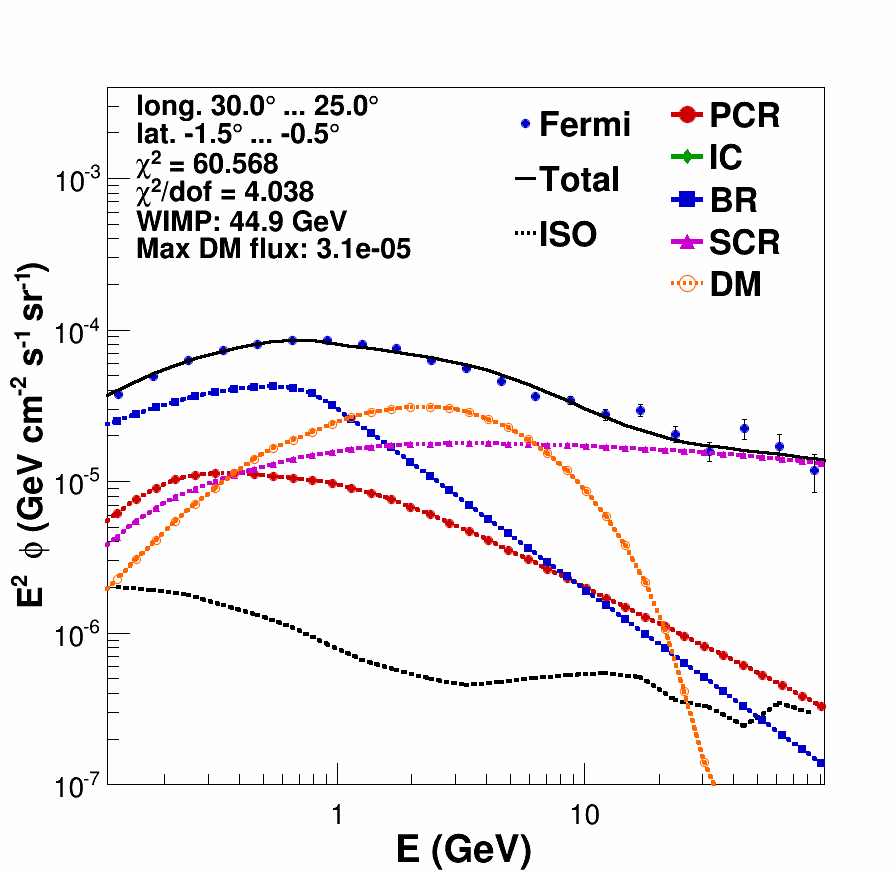}
	\includegraphics[width=0.16\textwidth,height=0.16\textwidth,clip]{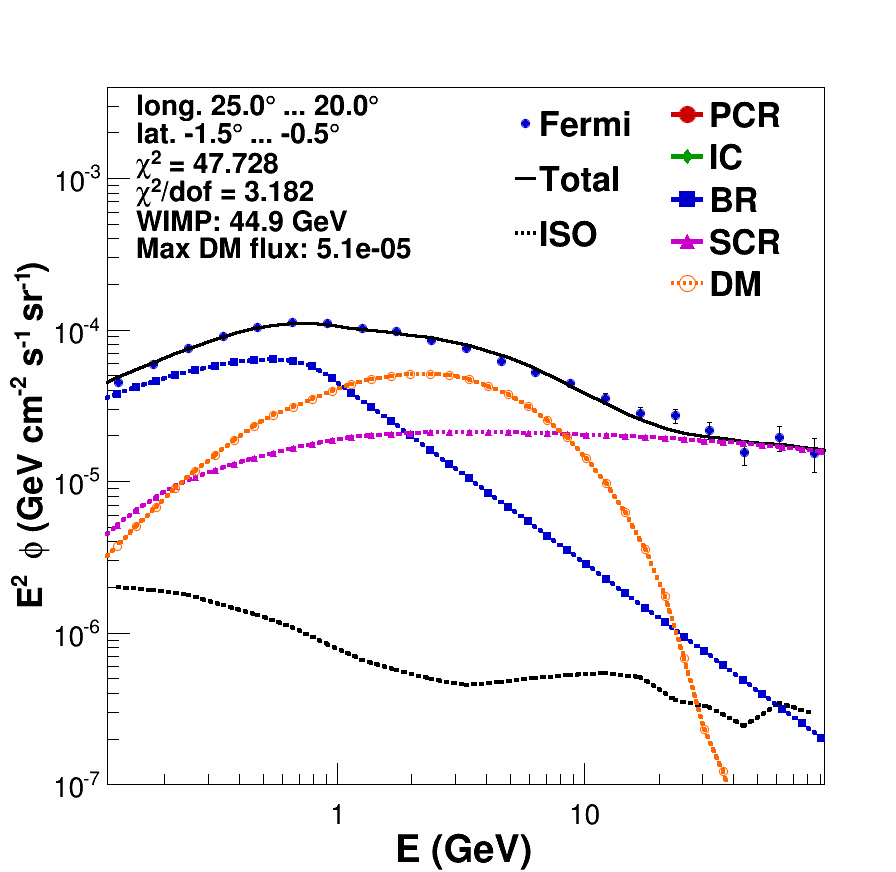}
	\includegraphics[width=0.16\textwidth,height=0.16\textwidth,clip]{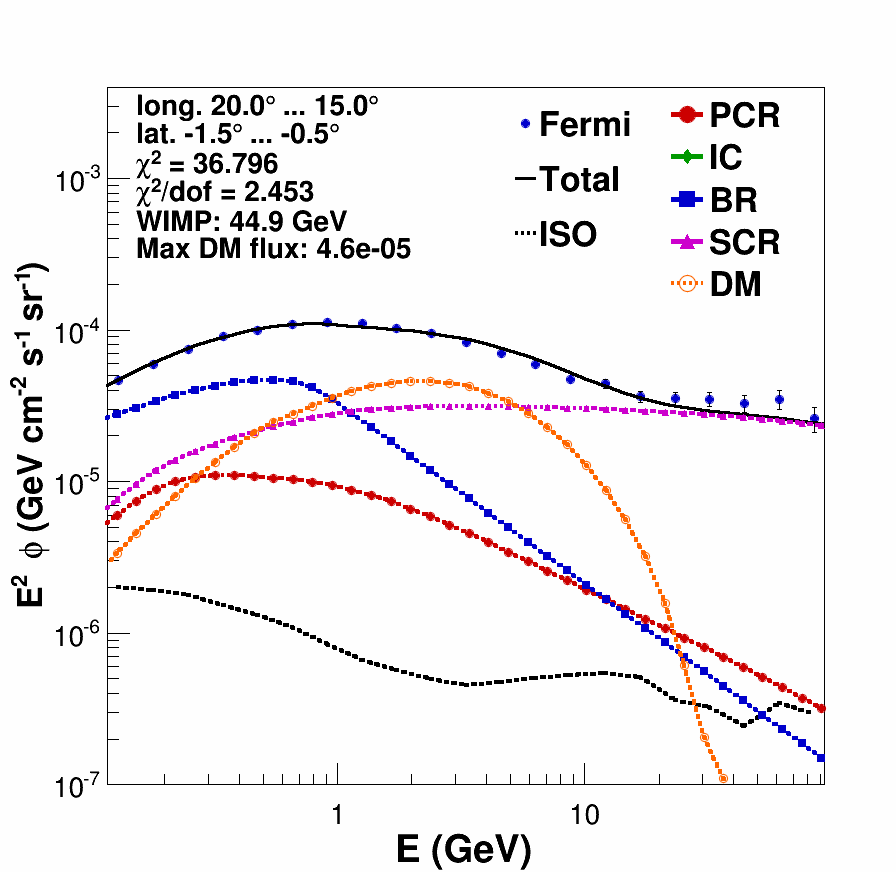}
	\includegraphics[width=0.16\textwidth,height=0.16\textwidth,clip]{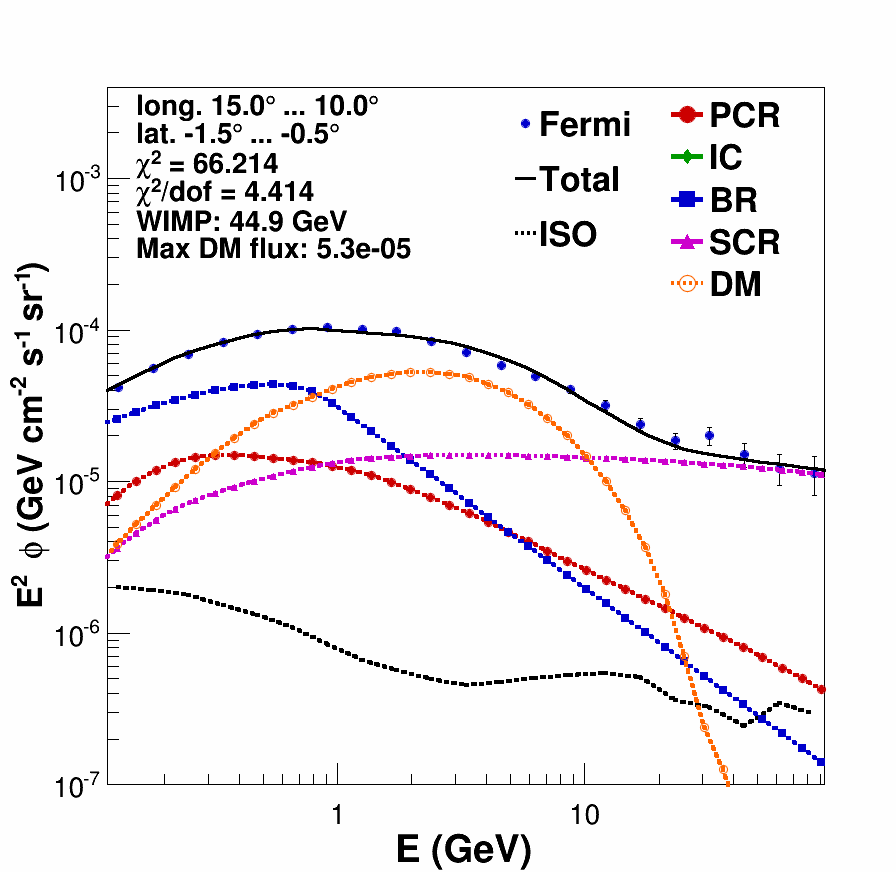}
	\includegraphics[width=0.16\textwidth,height=0.16\textwidth,clip]{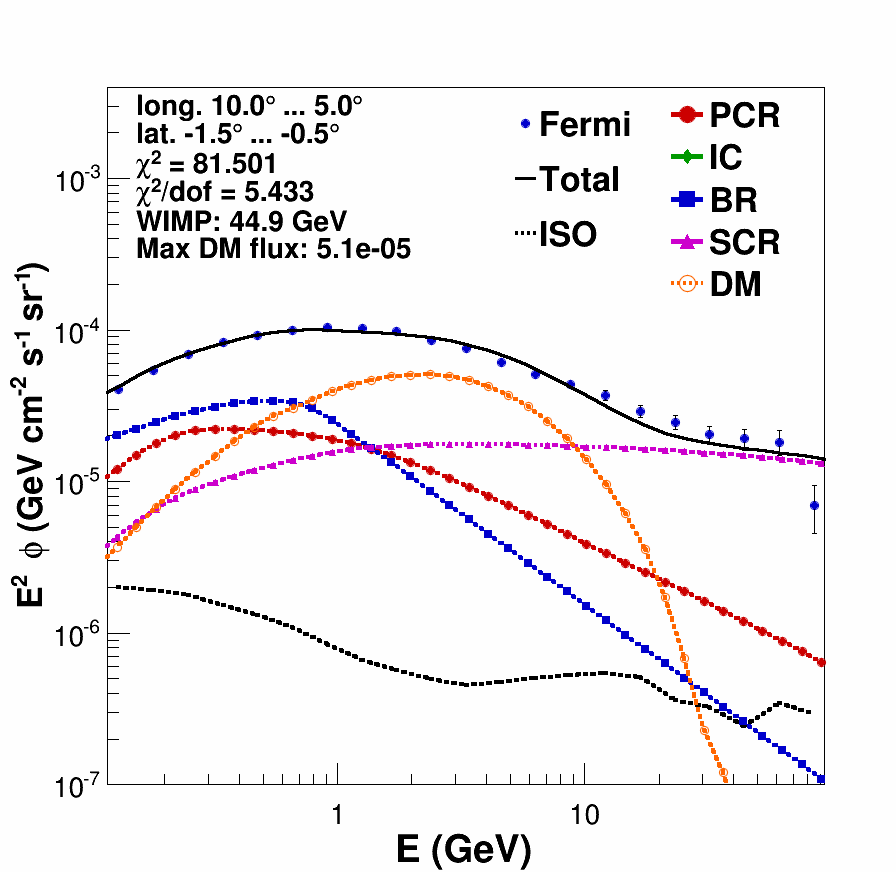}
	\includegraphics[width=0.16\textwidth,height=0.16\textwidth,clip]{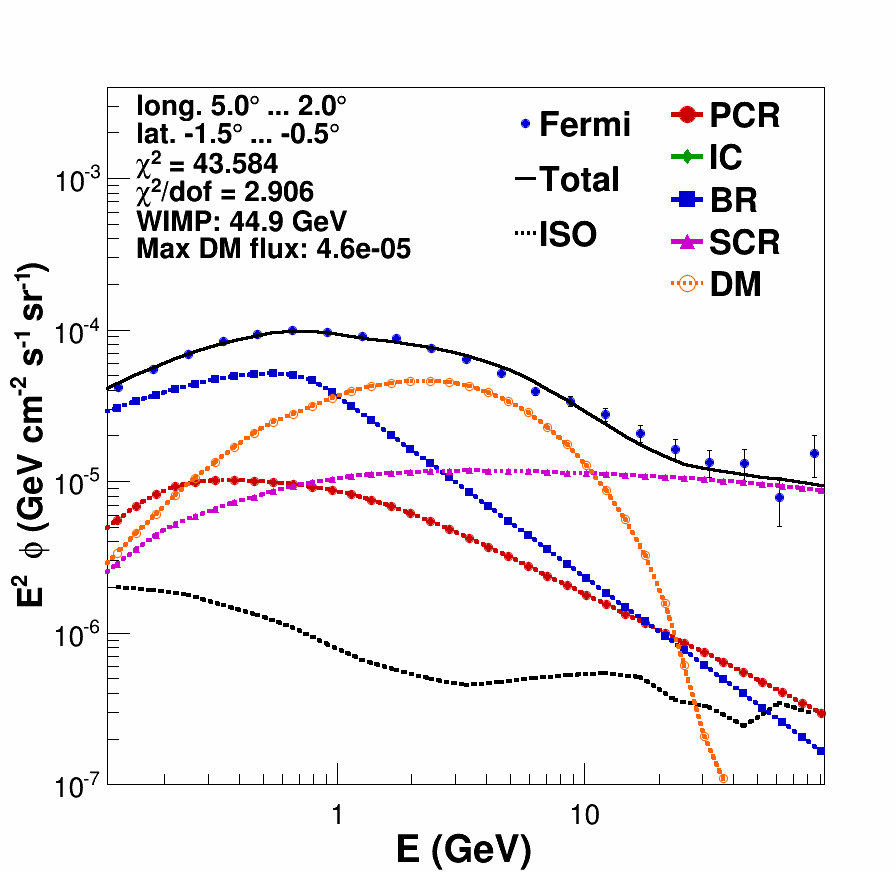}
	\includegraphics[width=0.16\textwidth,height=0.16\textwidth,clip]{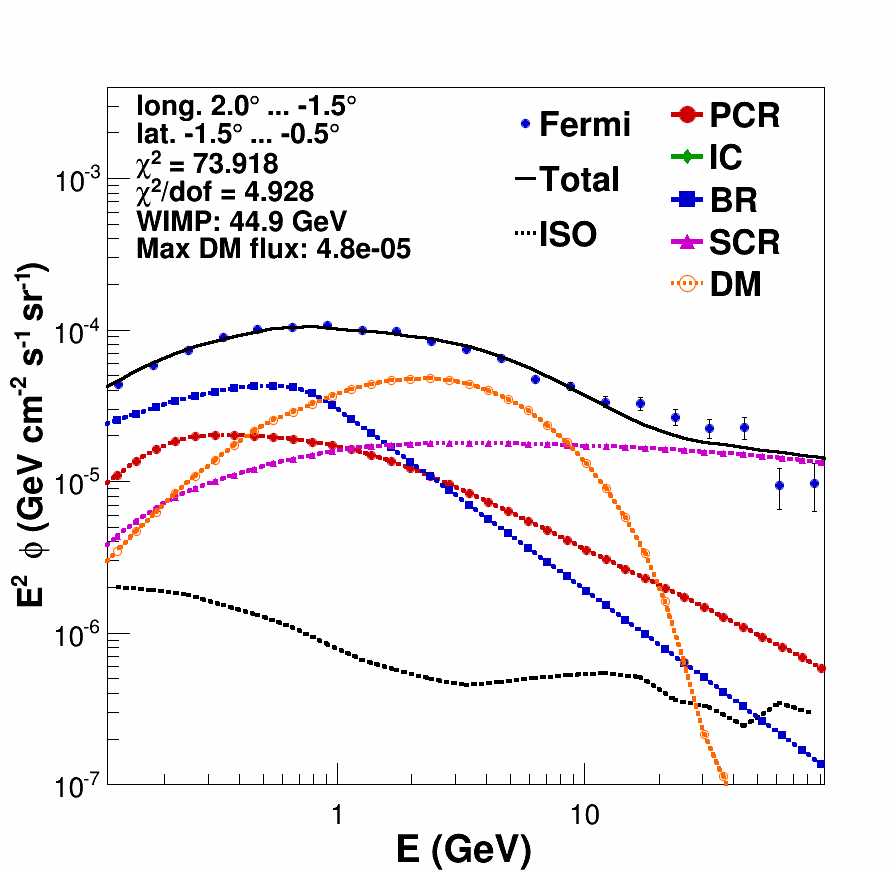}
	\includegraphics[width=0.16\textwidth,height=0.16\textwidth,clip]{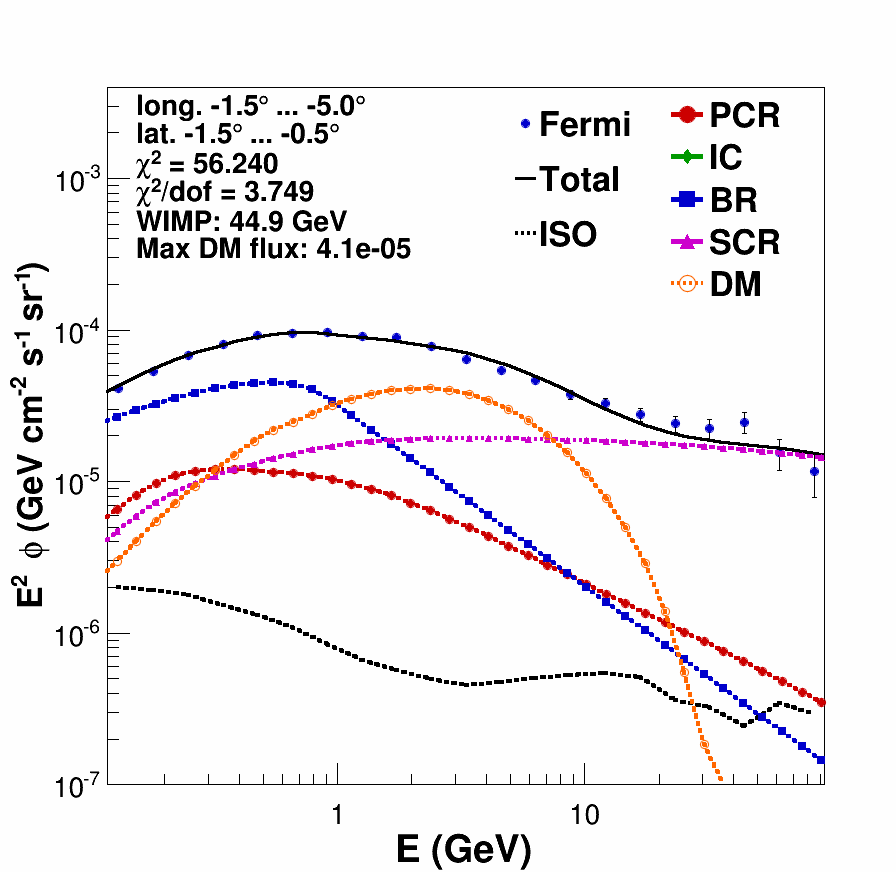}
	\includegraphics[width=0.16\textwidth,height=0.16\textwidth,clip]{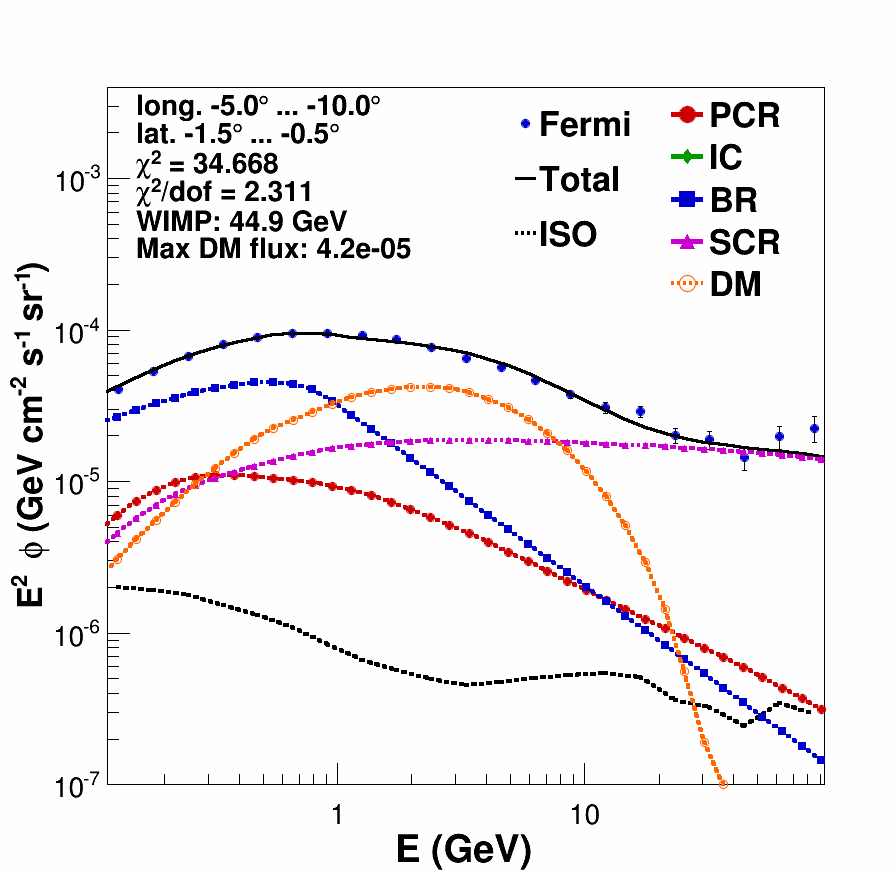}
	\includegraphics[width=0.16\textwidth,height=0.16\textwidth,clip]{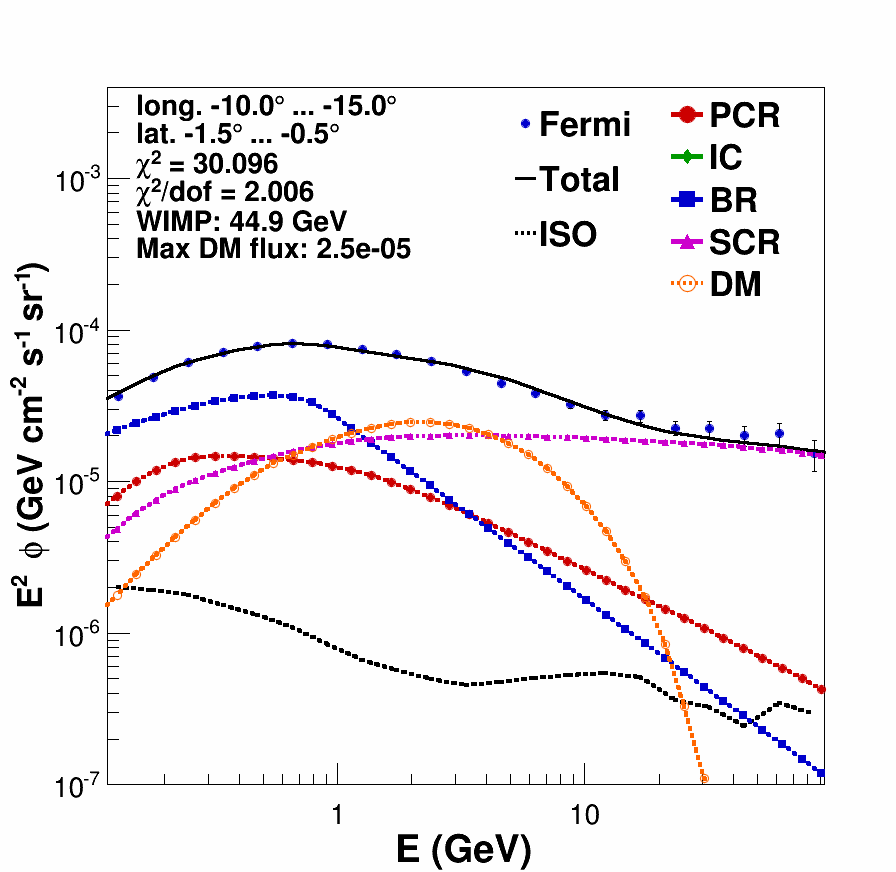}
	\includegraphics[width=0.16\textwidth,height=0.16\textwidth,clip]{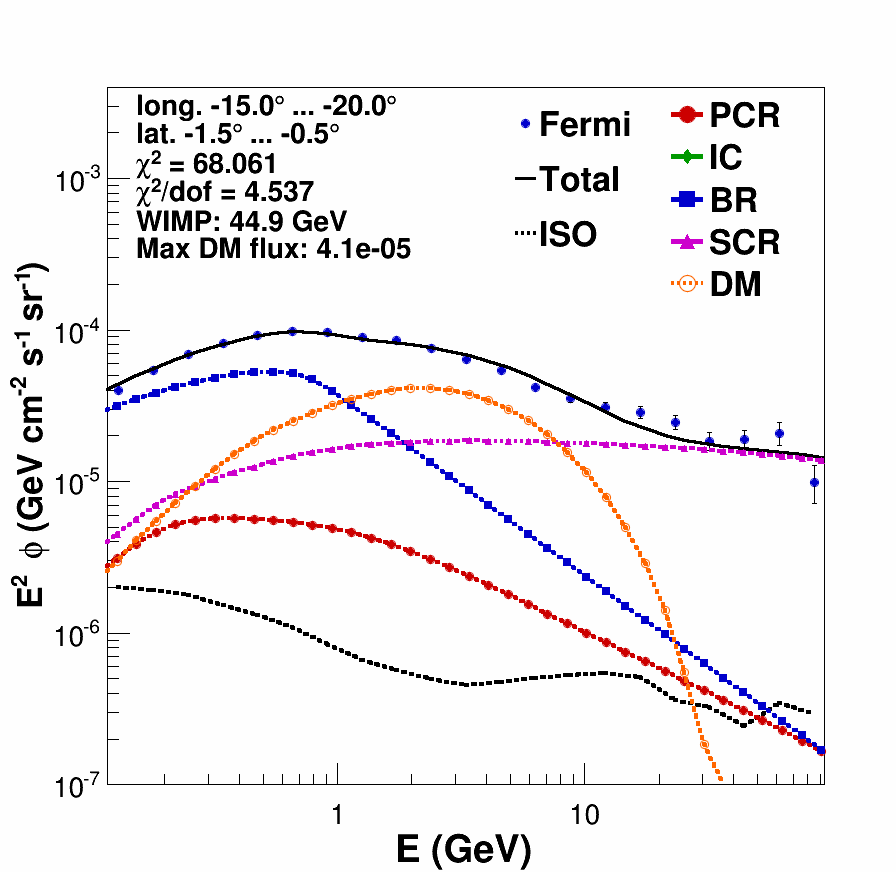}
	\includegraphics[width=0.16\textwidth,height=0.16\textwidth,clip]{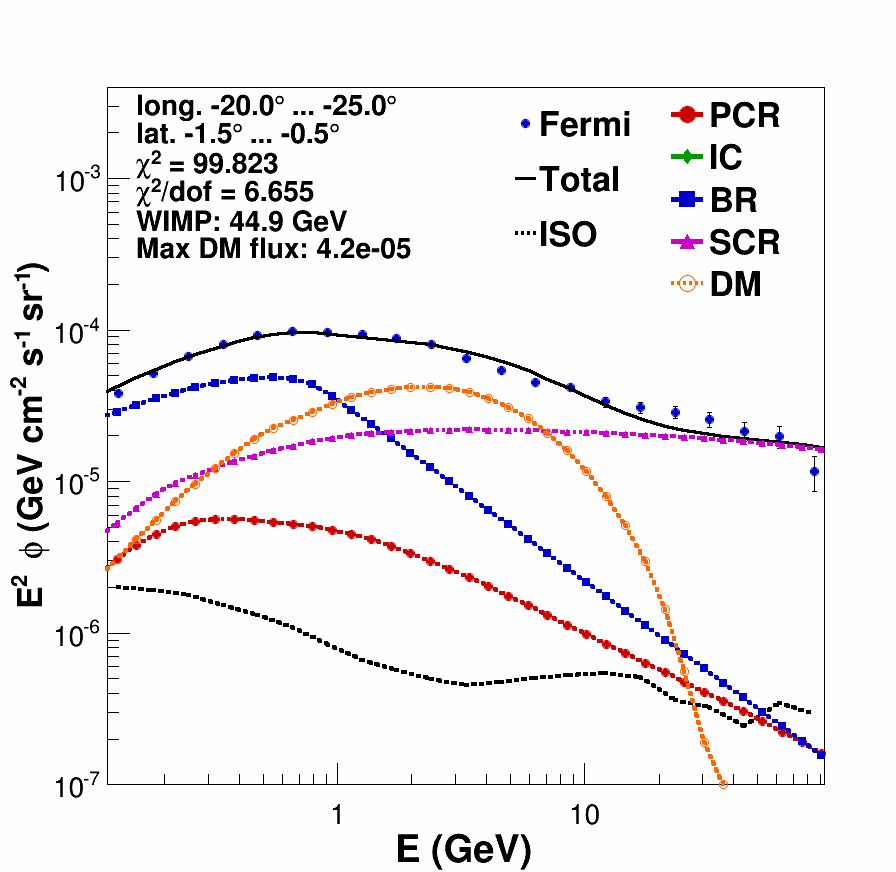}
	\includegraphics[width=0.16\textwidth,height=0.16\textwidth,clip]{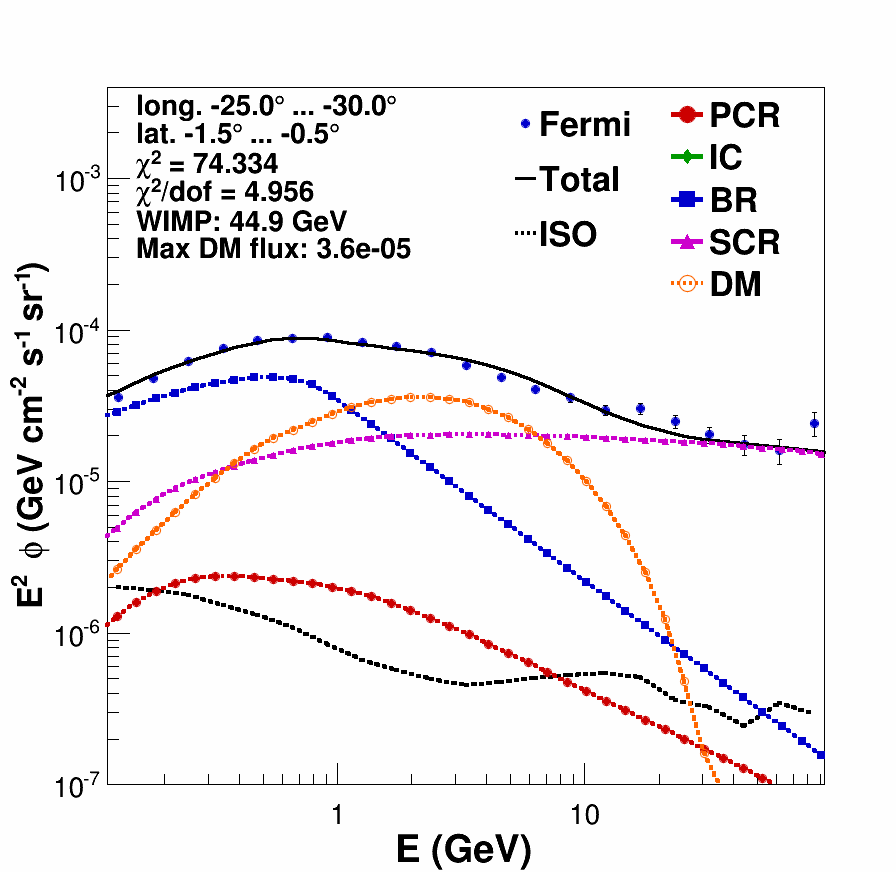}
	\includegraphics[width=0.16\textwidth,height=0.16\textwidth,clip]{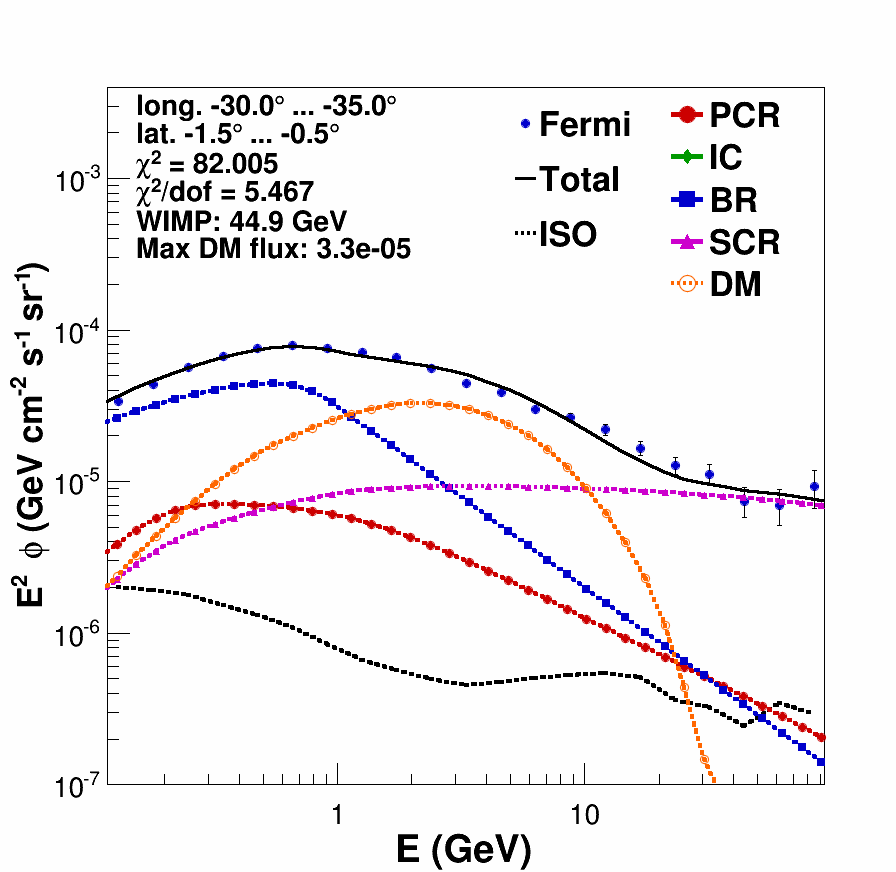}
	\includegraphics[width=0.16\textwidth,height=0.16\textwidth,clip]{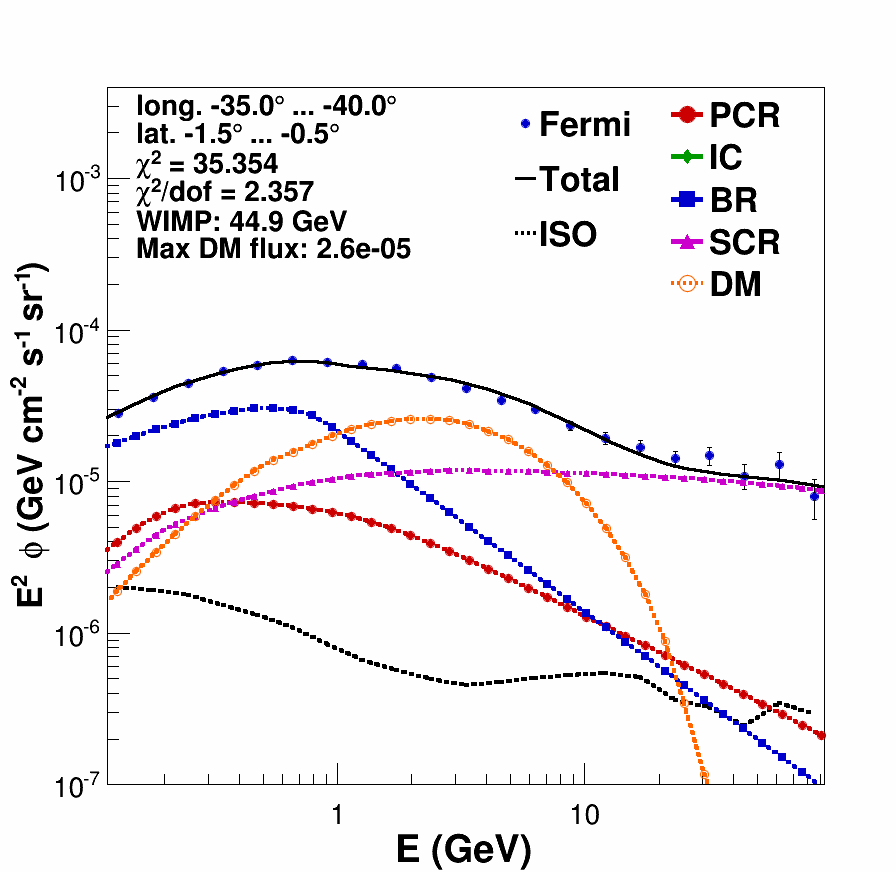}
	\includegraphics[width=0.16\textwidth,height=0.16\textwidth,clip]{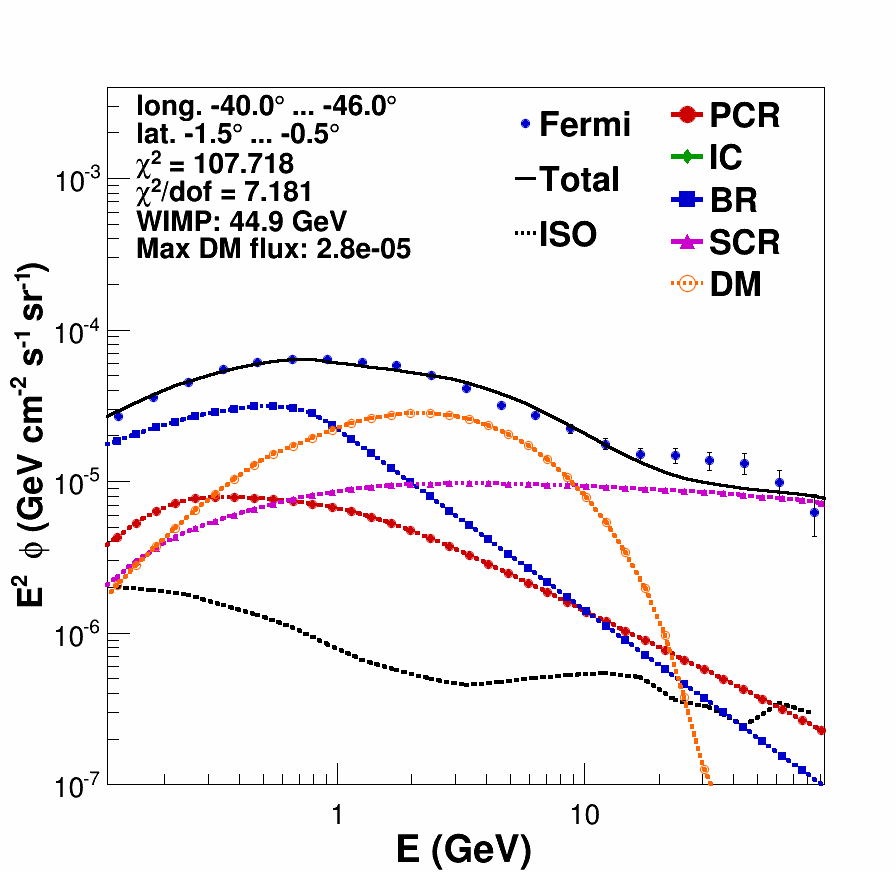}
	\includegraphics[width=0.16\textwidth,height=0.16\textwidth,clip]{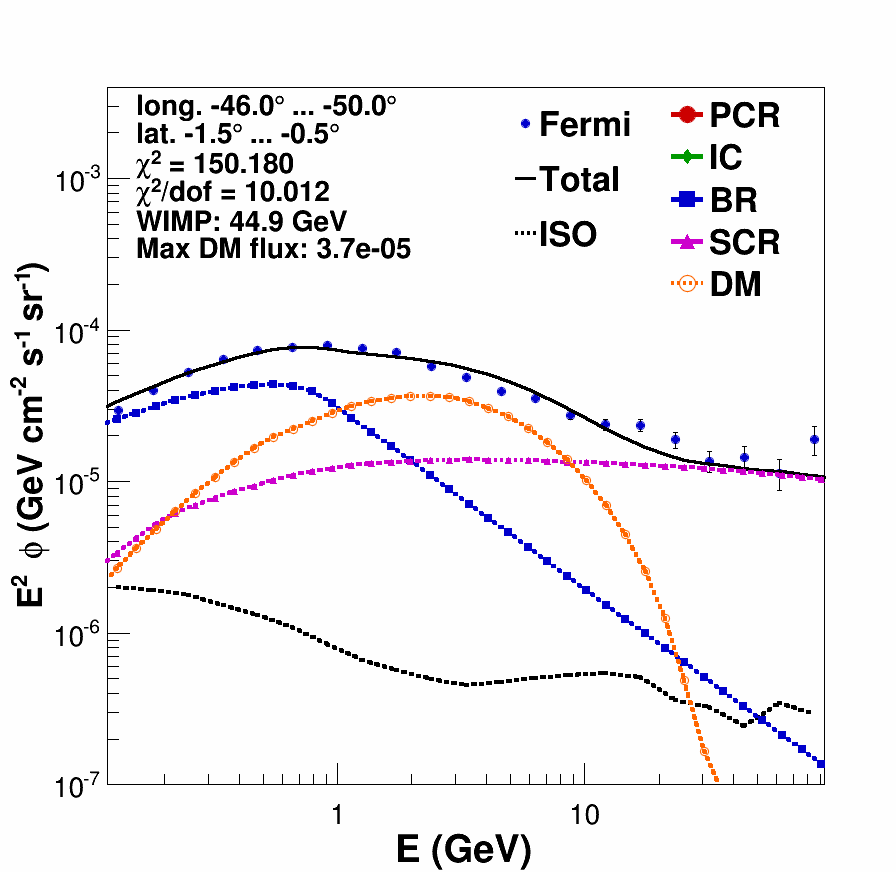}
	\includegraphics[width=0.16\textwidth,height=0.16\textwidth,clip]{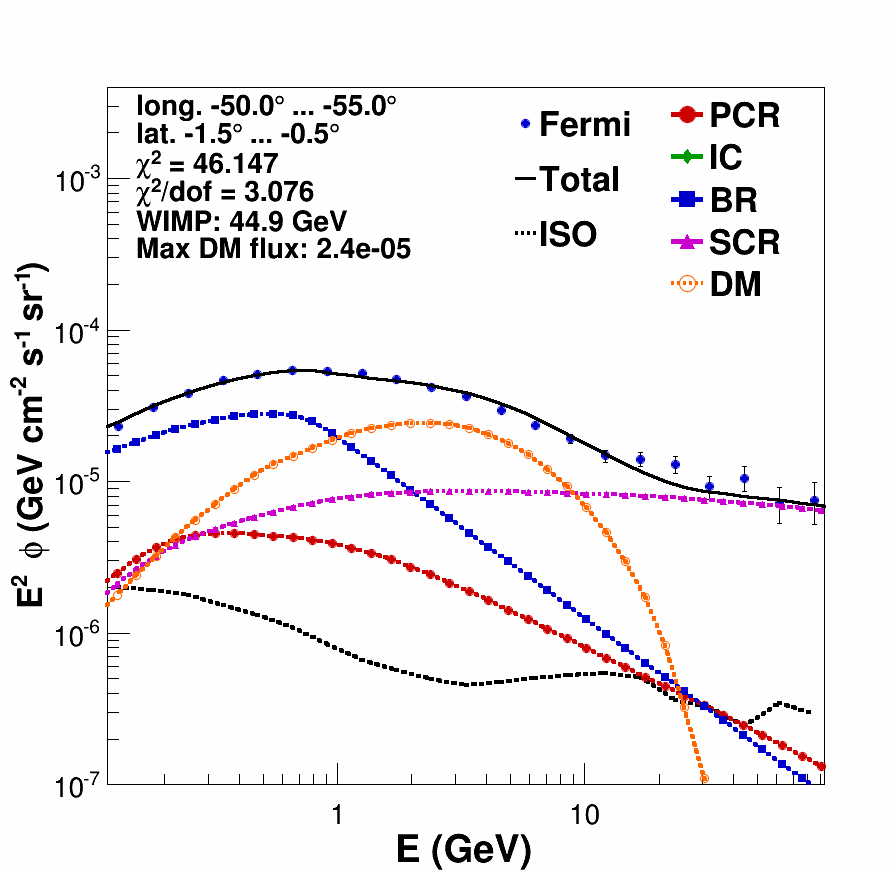}
	\includegraphics[width=0.16\textwidth,height=0.16\textwidth,clip]{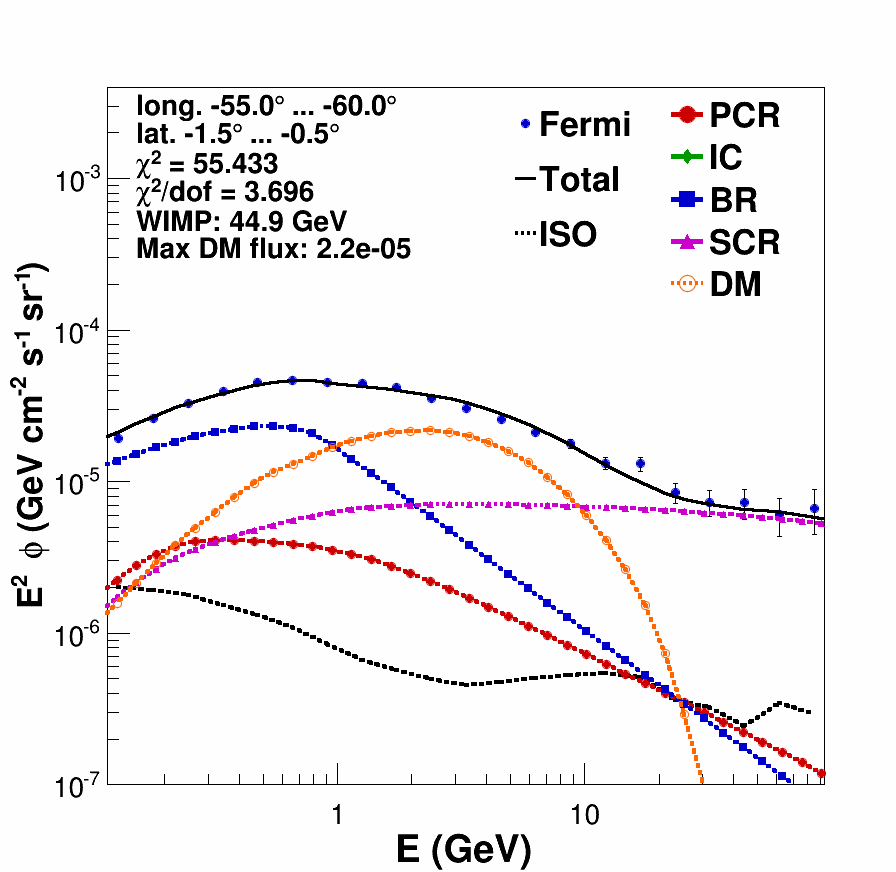}
	\includegraphics[width=0.16\textwidth,height=0.16\textwidth,clip]{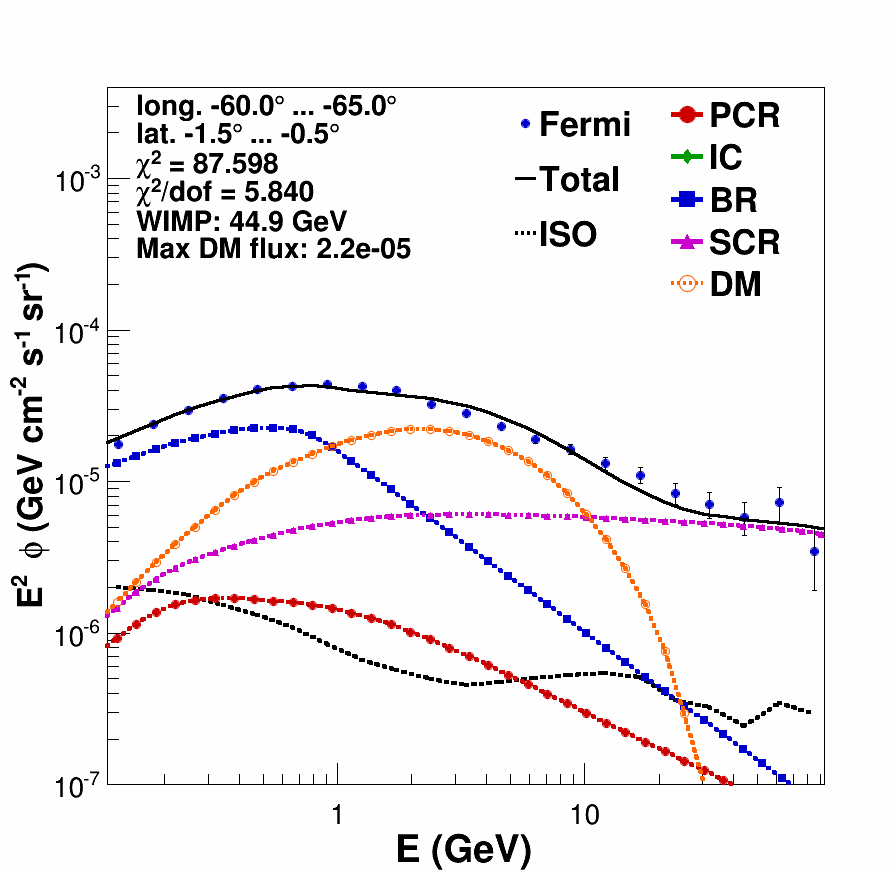}
	\includegraphics[width=0.16\textwidth,height=0.16\textwidth,clip]{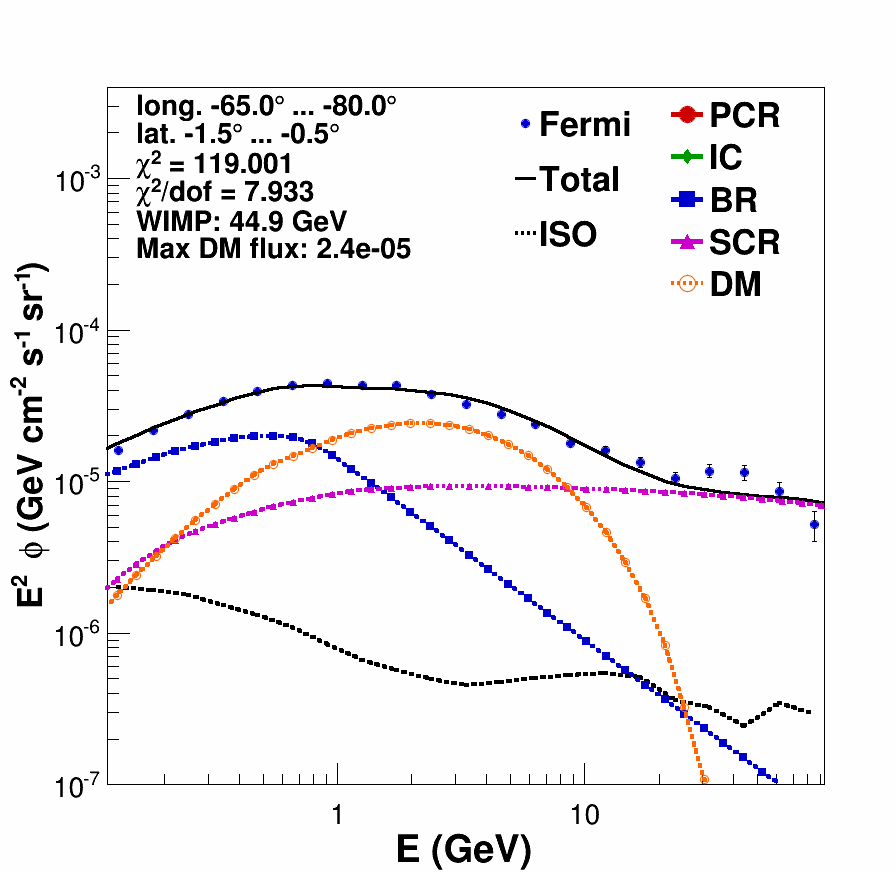}
	\includegraphics[width=0.16\textwidth,height=0.16\textwidth,clip]{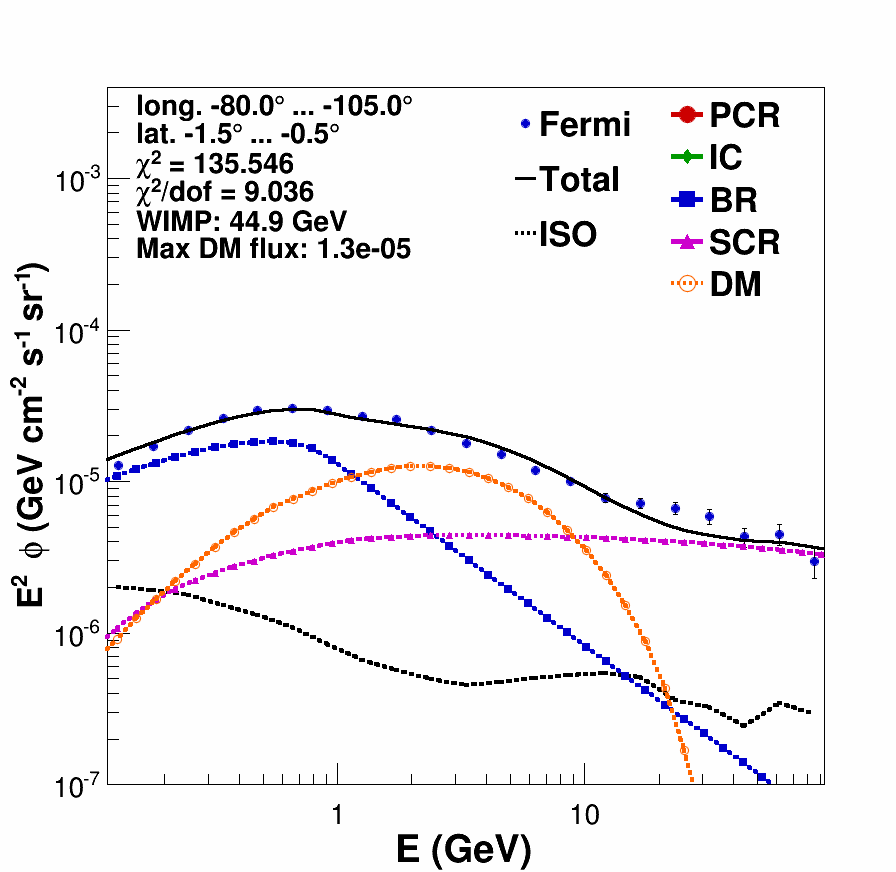}
	\includegraphics[width=0.16\textwidth,height=0.16\textwidth,clip]{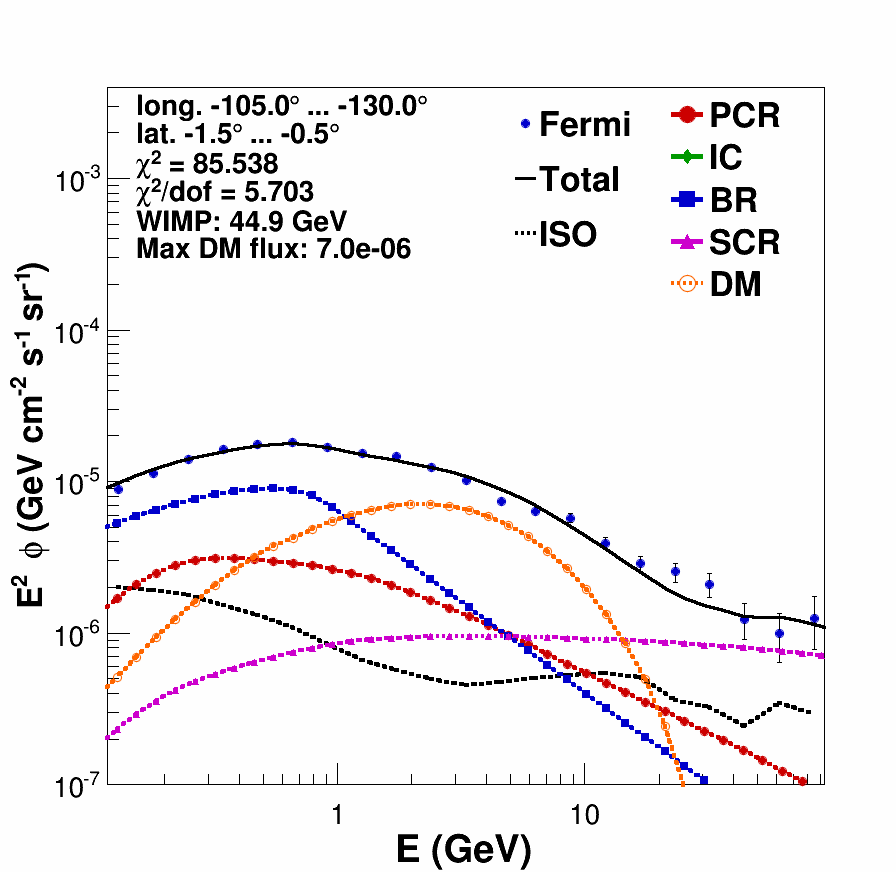}
38	\includegraphics[width=0.16\textwidth,height=0.16\textwidth,clip]{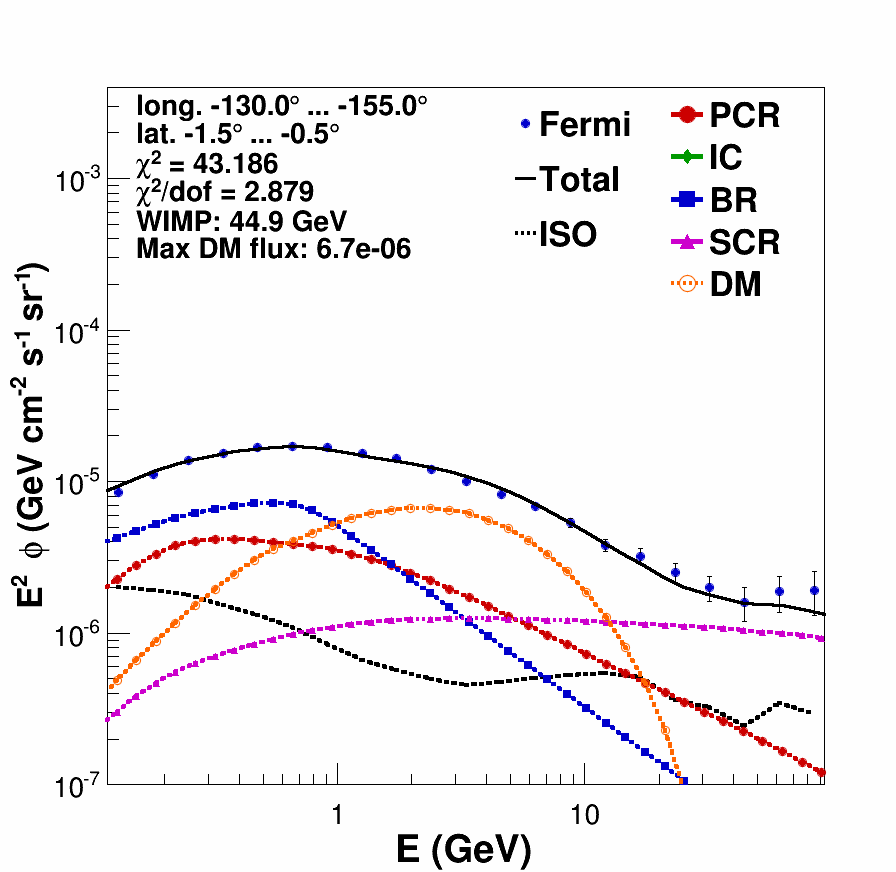}
39	\includegraphics[width=0.16\textwidth,height=0.16\textwidth,clip]{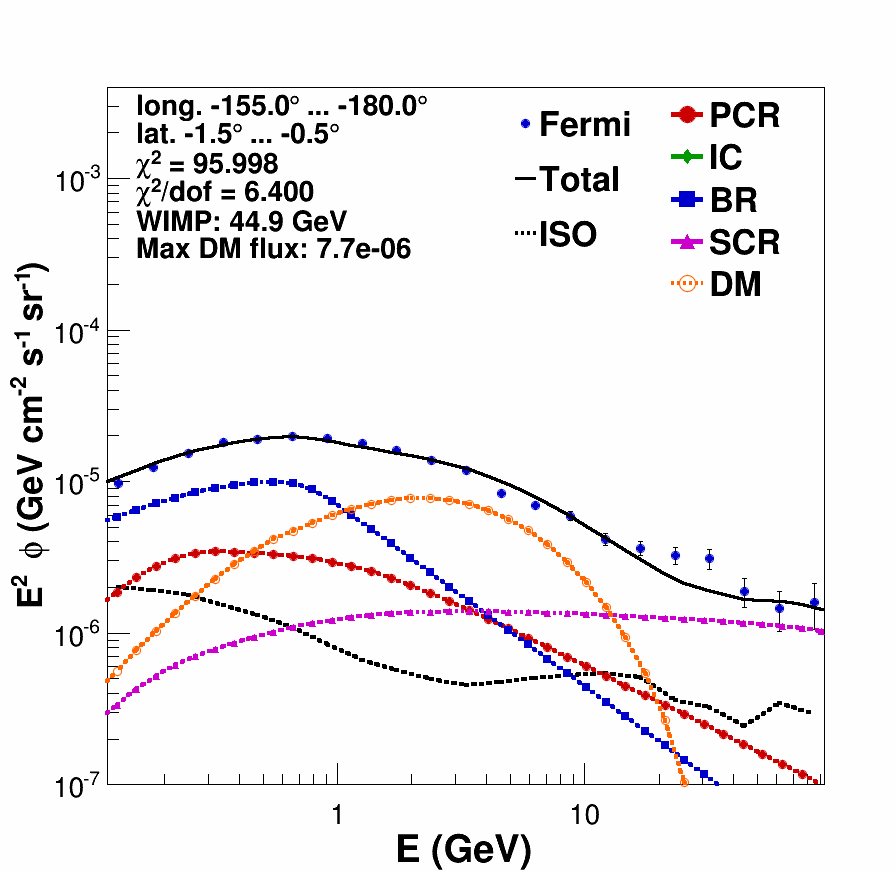}%%%%%r11
\caption[]{Template fits for latitudes  with $-1.5^\circ<b<-0.5^\circ$ and longitudes decreasing from 180$^\circ$ to -180$^\circ$.} \label{F43}
\end{figure}
\begin{figure}
\centering
\includegraphics[width=0.16\textwidth,height=0.16\textwidth,clip]{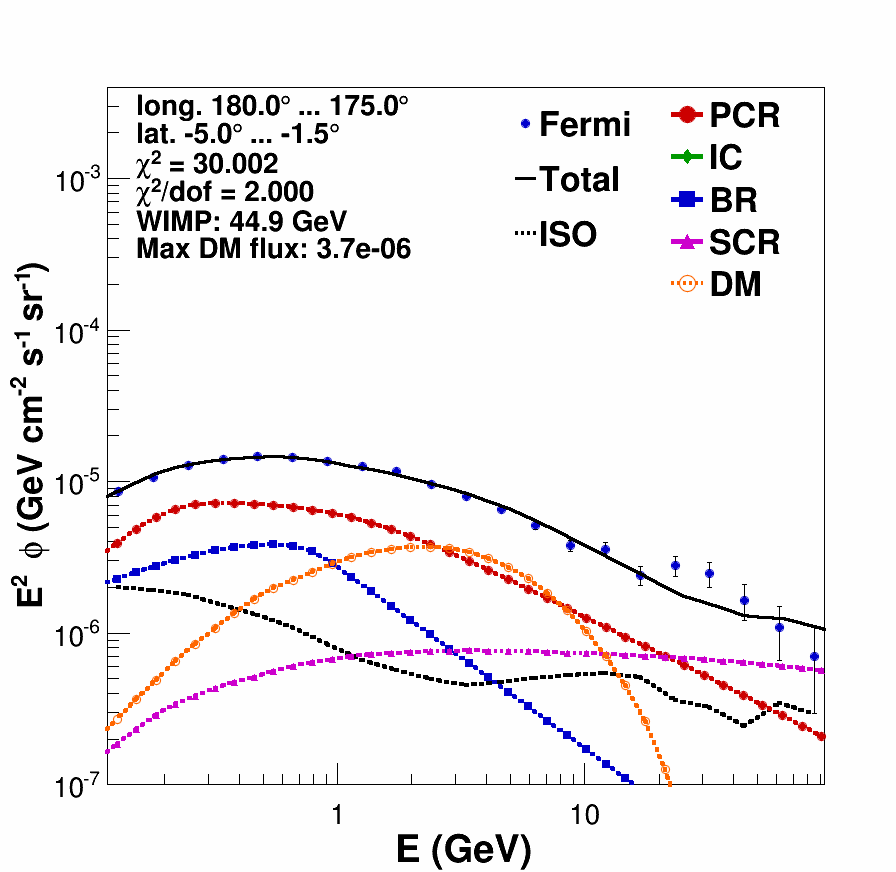}
\includegraphics[width=0.16\textwidth,height=0.16\textwidth,clip]{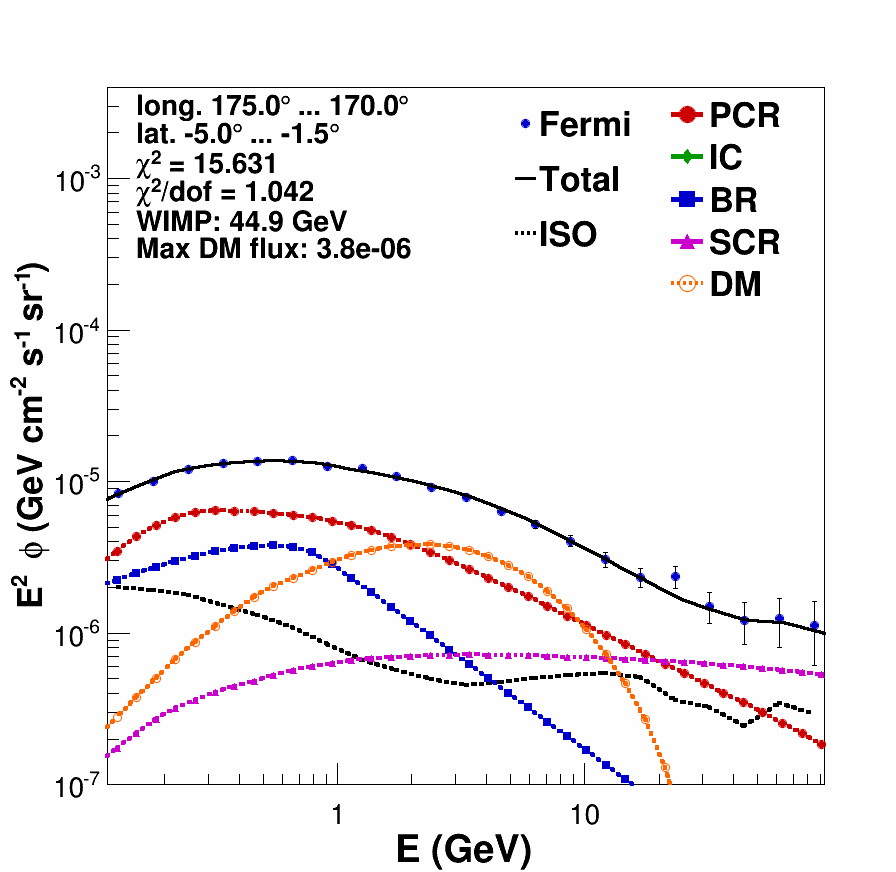}
\includegraphics[width=0.16\textwidth,height=0.16\textwidth,clip]{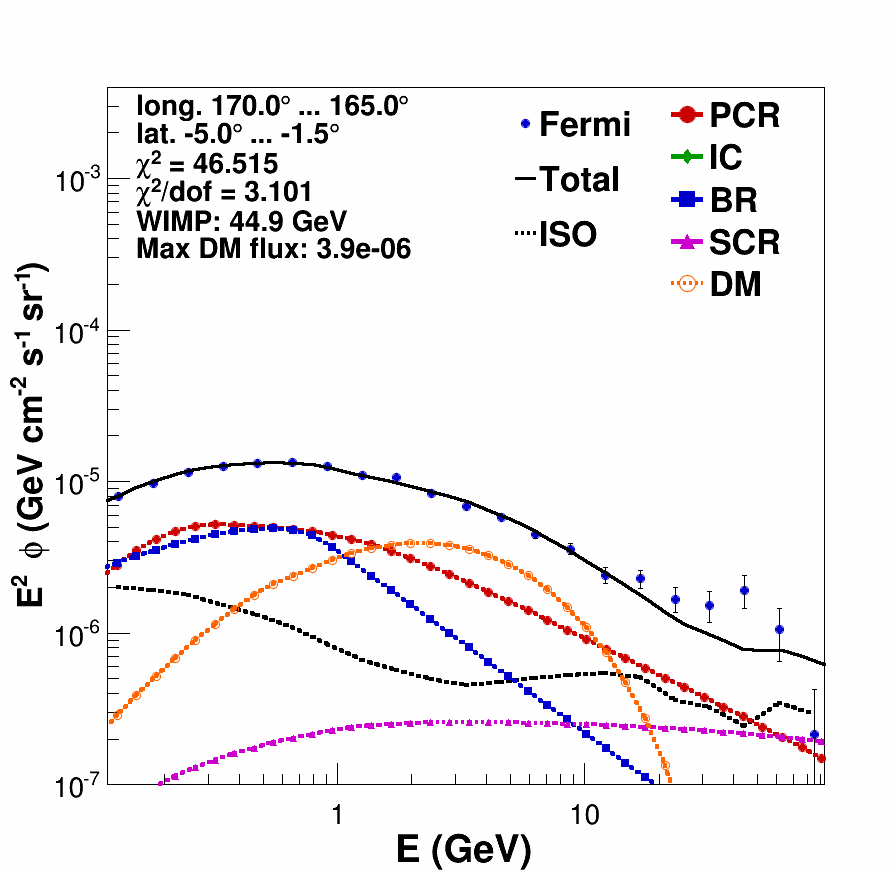}
\includegraphics[width=0.16\textwidth,height=0.16\textwidth,clip]{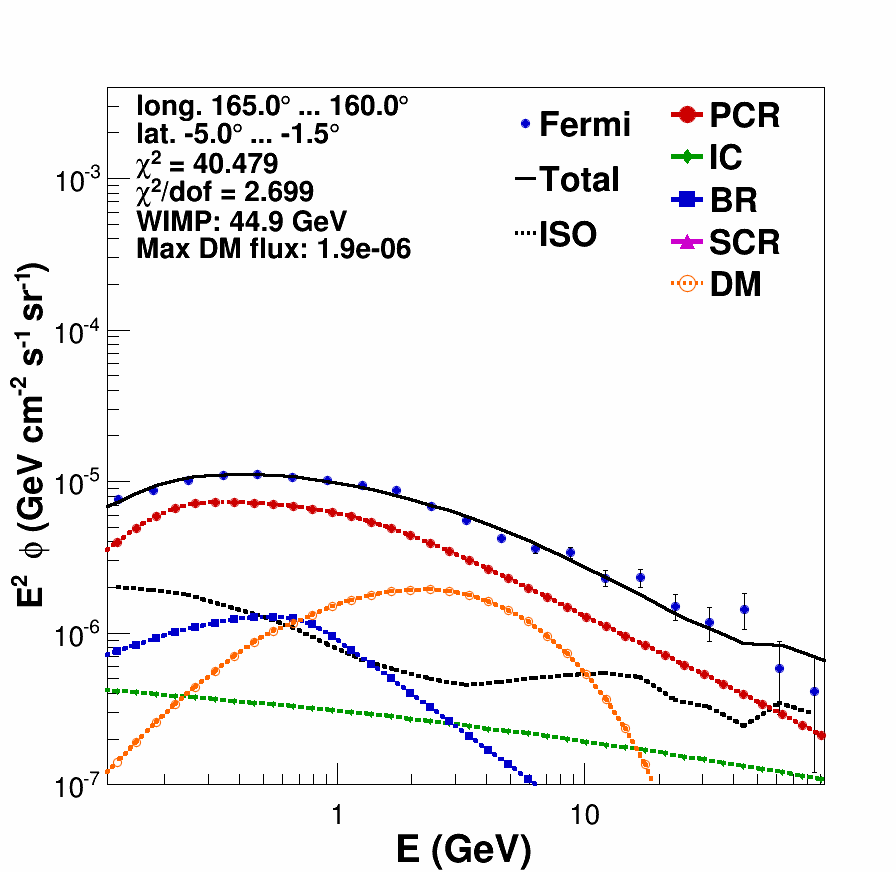}
\includegraphics[width=0.16\textwidth,height=0.16\textwidth,clip]{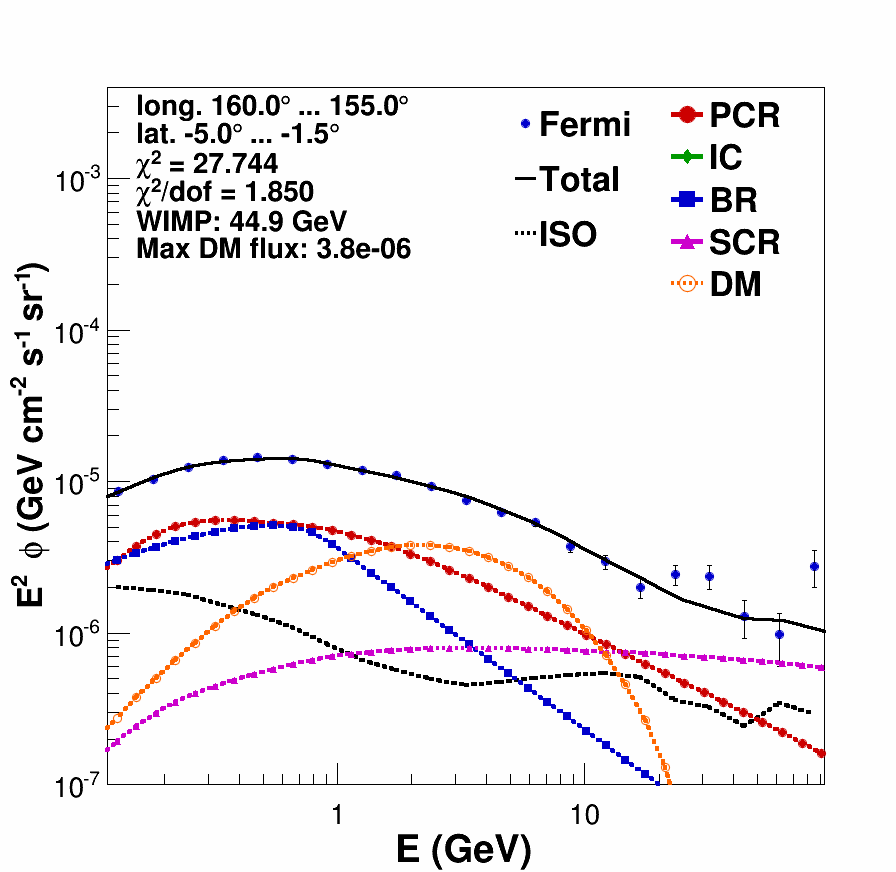}
\includegraphics[width=0.16\textwidth,height=0.16\textwidth,clip]{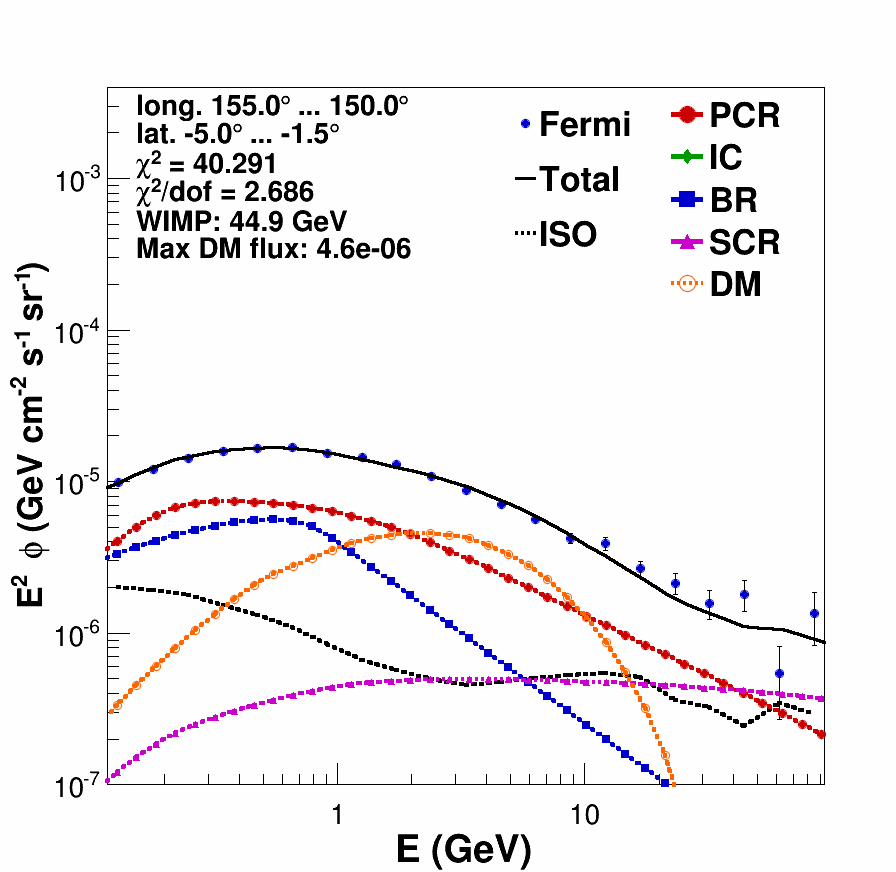}
\includegraphics[width=0.16\textwidth,height=0.16\textwidth,clip]{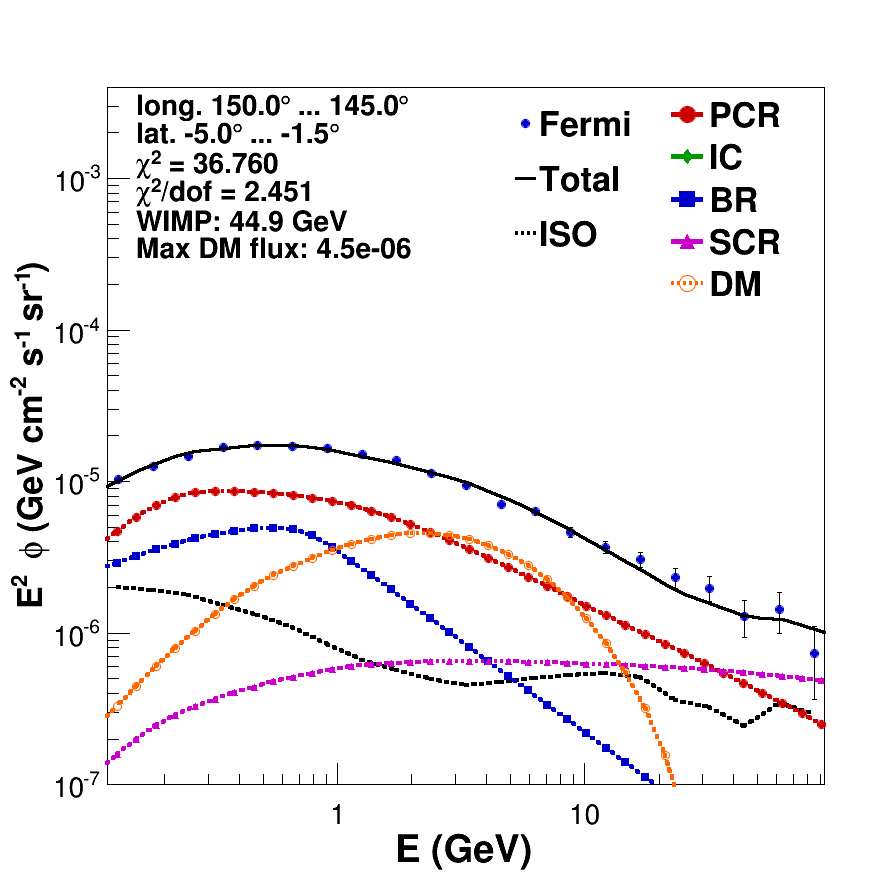}
\includegraphics[width=0.16\textwidth,height=0.16\textwidth,clip]{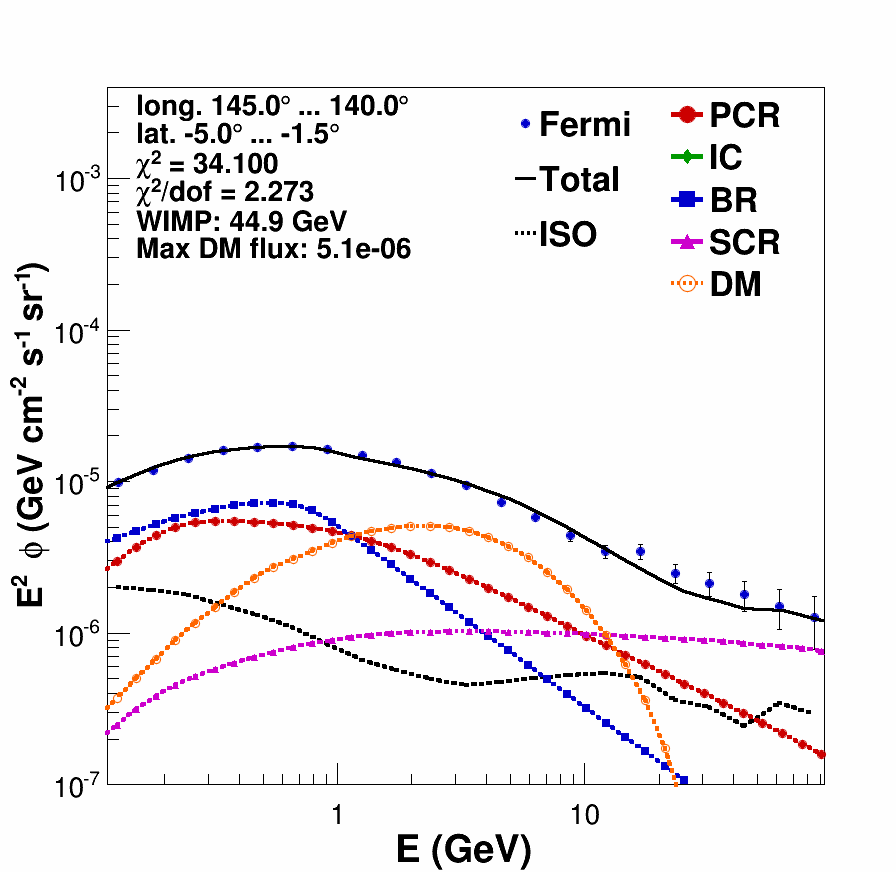}
\includegraphics[width=0.16\textwidth,height=0.16\textwidth,clip]{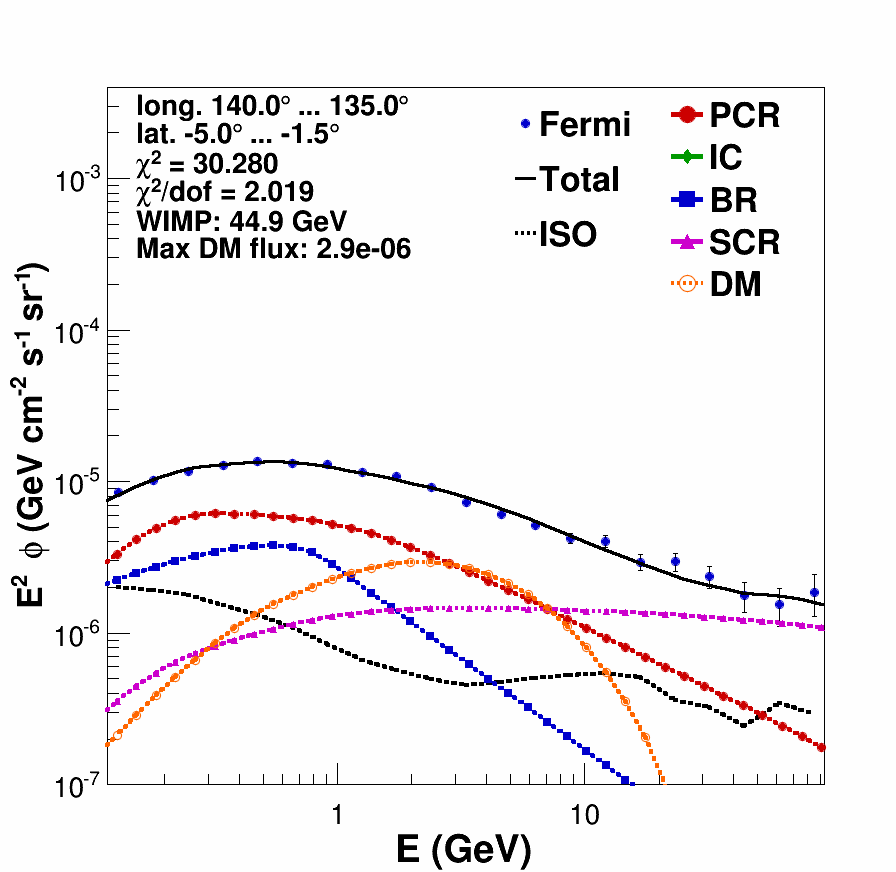}
\includegraphics[width=0.16\textwidth,height=0.16\textwidth,clip]{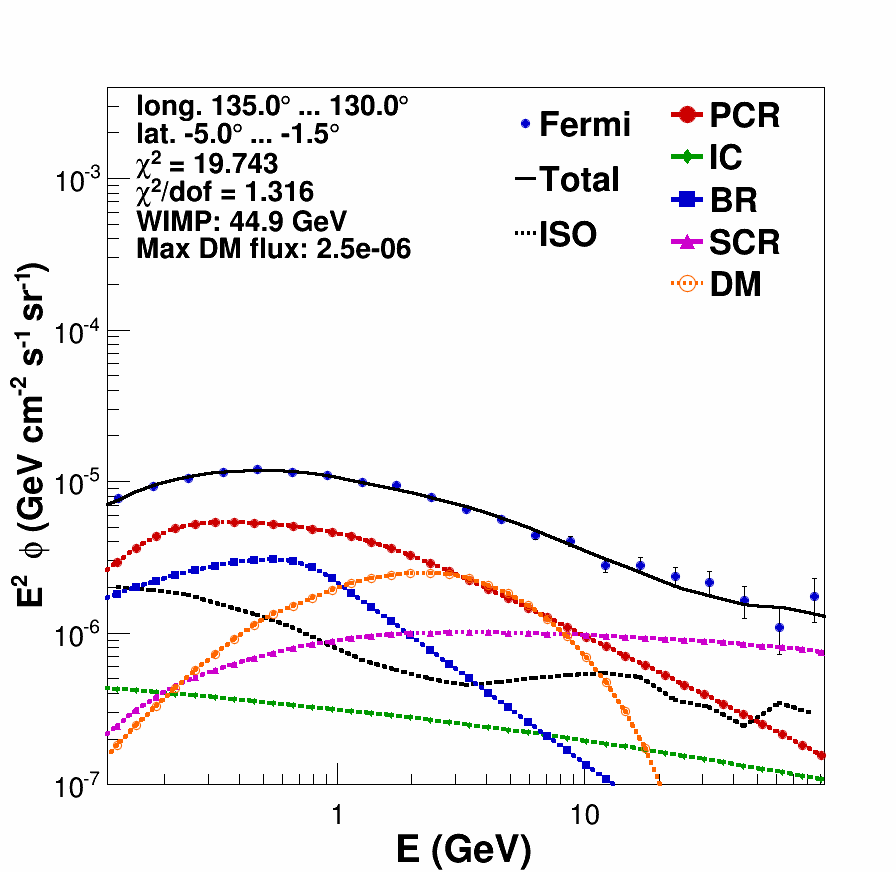}
\includegraphics[width=0.16\textwidth,height=0.16\textwidth,clip]{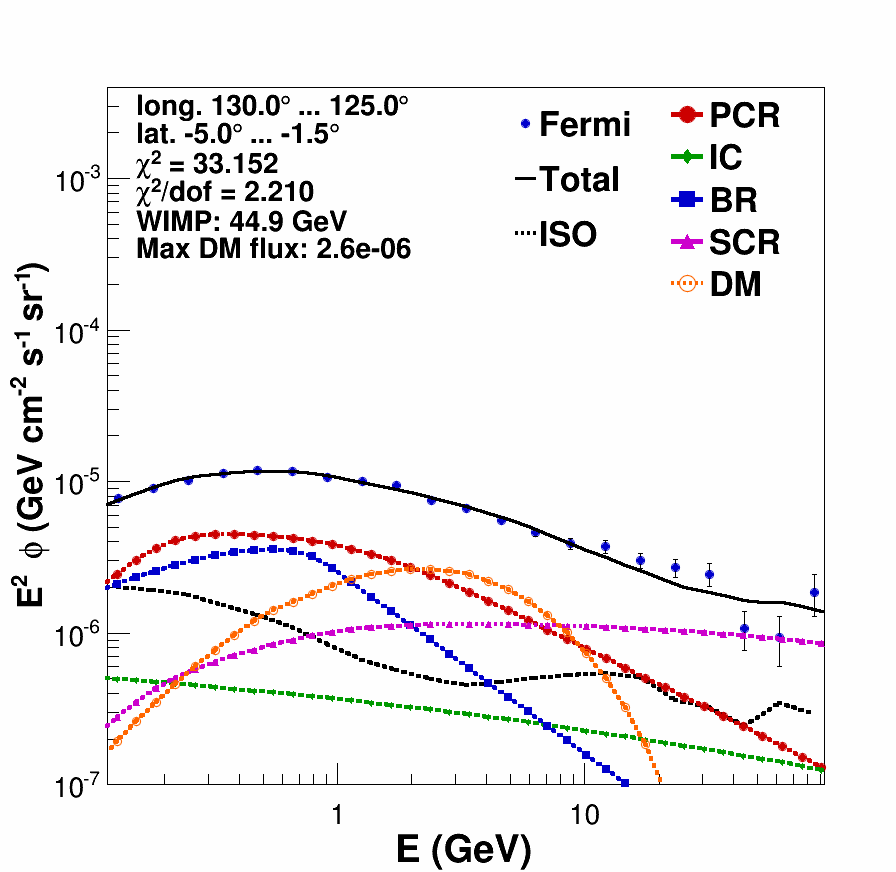}
\includegraphics[width=0.16\textwidth,height=0.16\textwidth,clip]{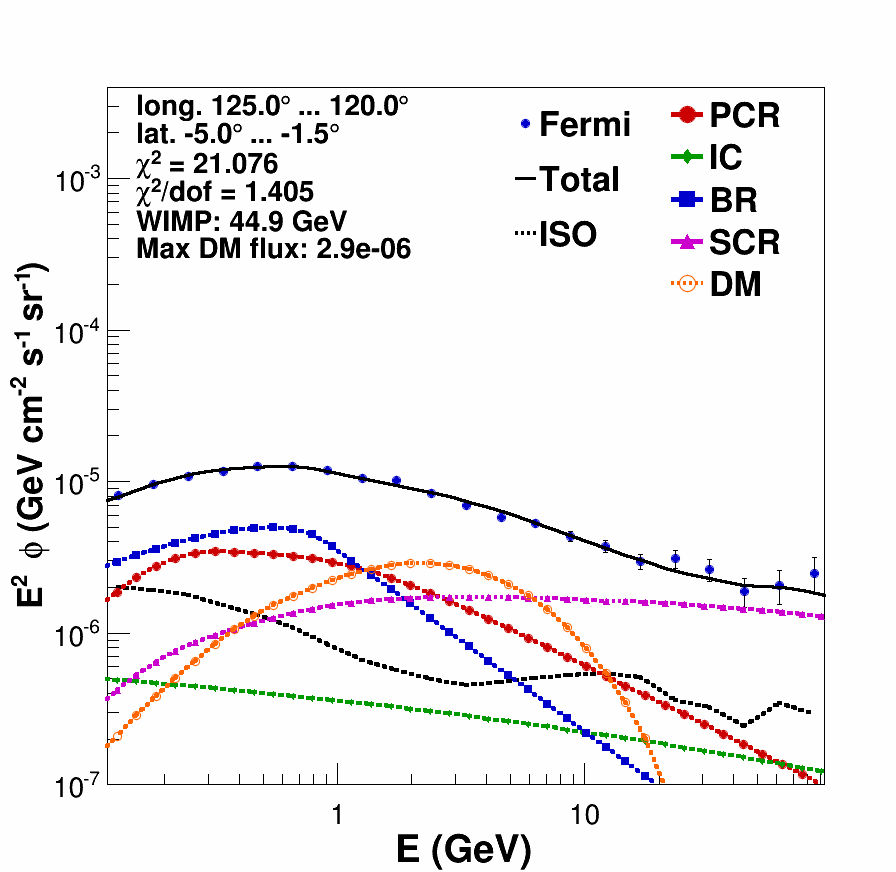}
\includegraphics[width=0.16\textwidth,height=0.16\textwidth,clip]{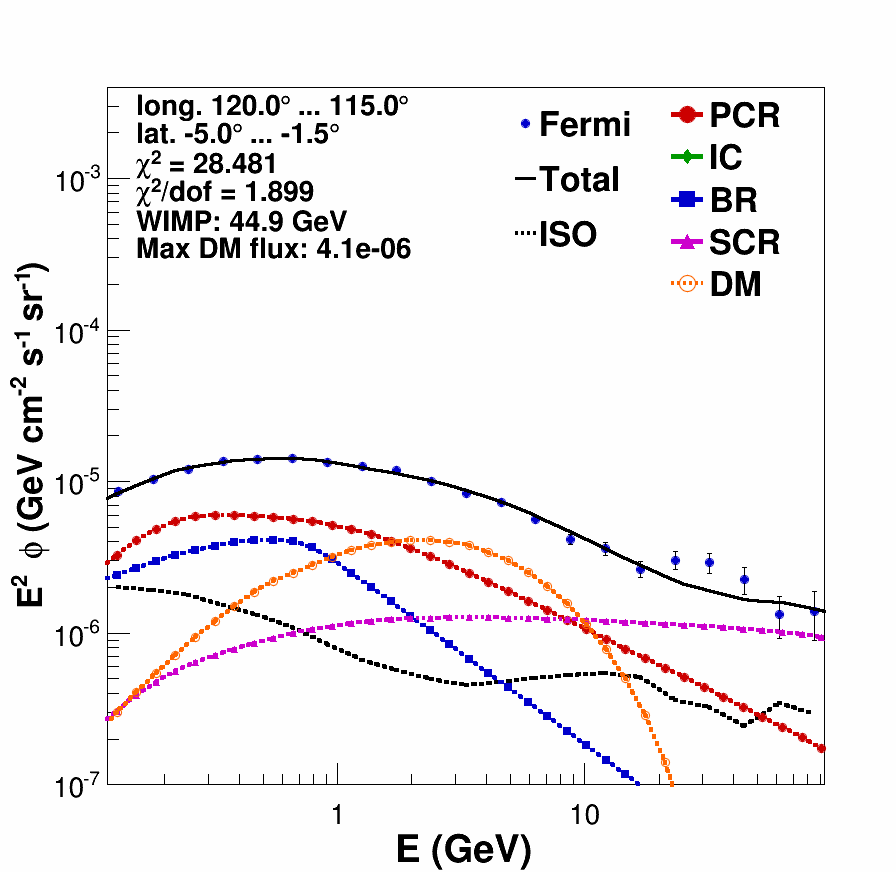}
\includegraphics[width=0.16\textwidth,height=0.16\textwidth,clip]{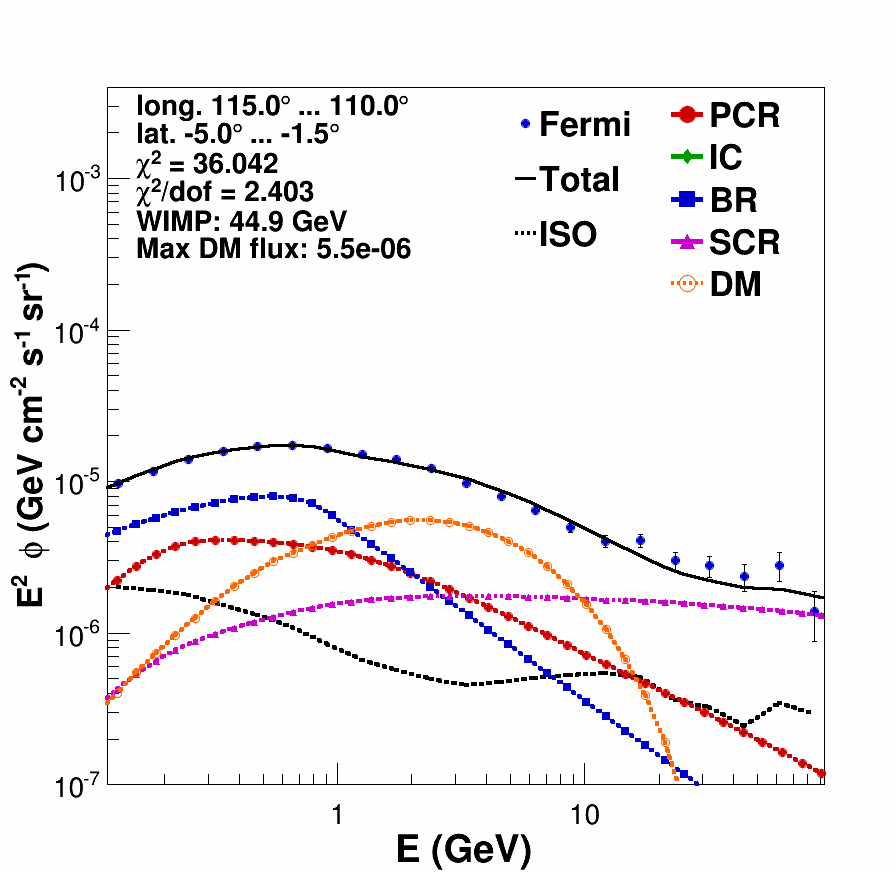}
\includegraphics[width=0.16\textwidth,height=0.16\textwidth,clip]{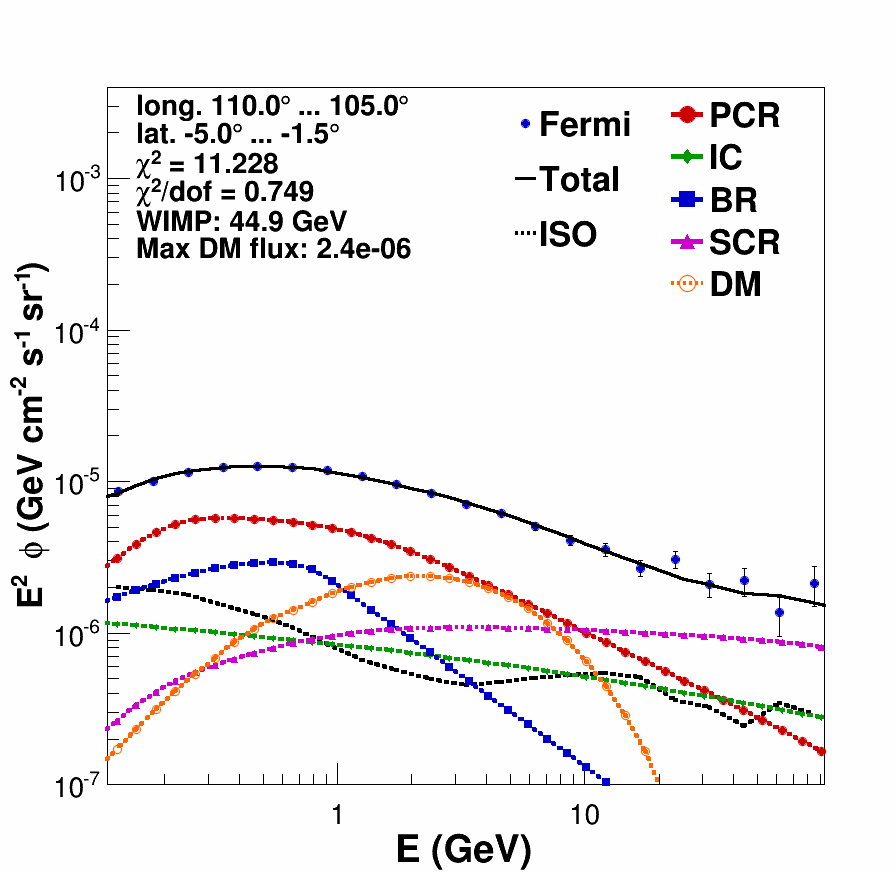}
\includegraphics[width=0.16\textwidth,height=0.16\textwidth,clip]{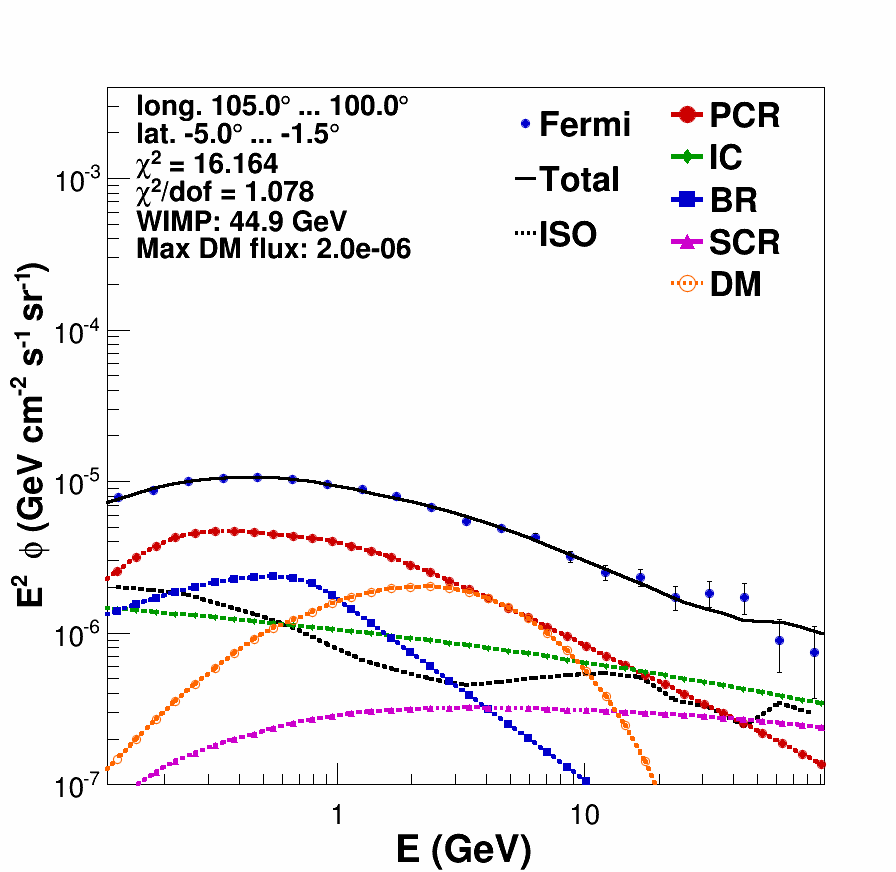}
\includegraphics[width=0.16\textwidth,height=0.16\textwidth,clip]{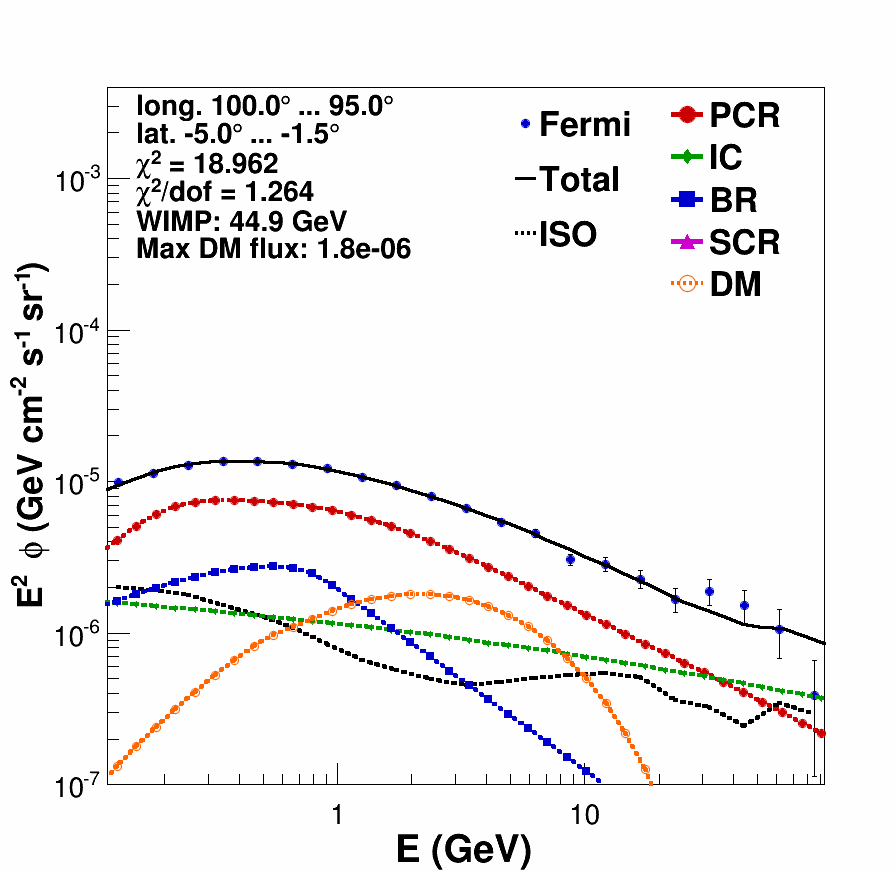}
\includegraphics[width=0.16\textwidth,height=0.16\textwidth,clip]{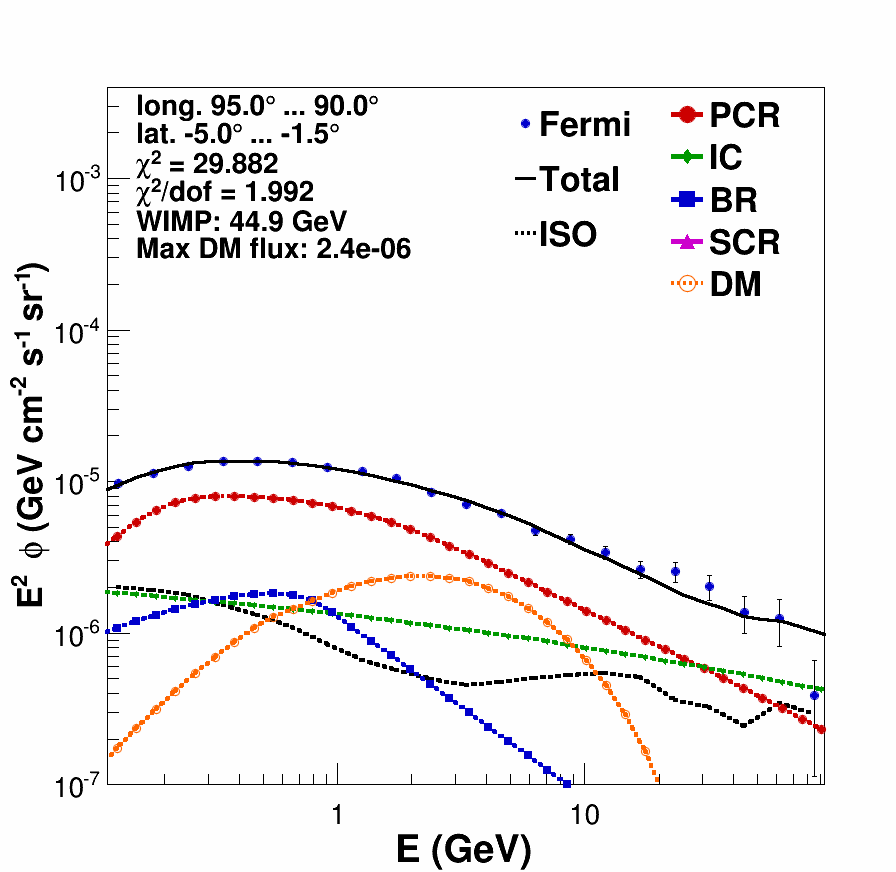}
\includegraphics[width=0.16\textwidth,height=0.16\textwidth,clip]{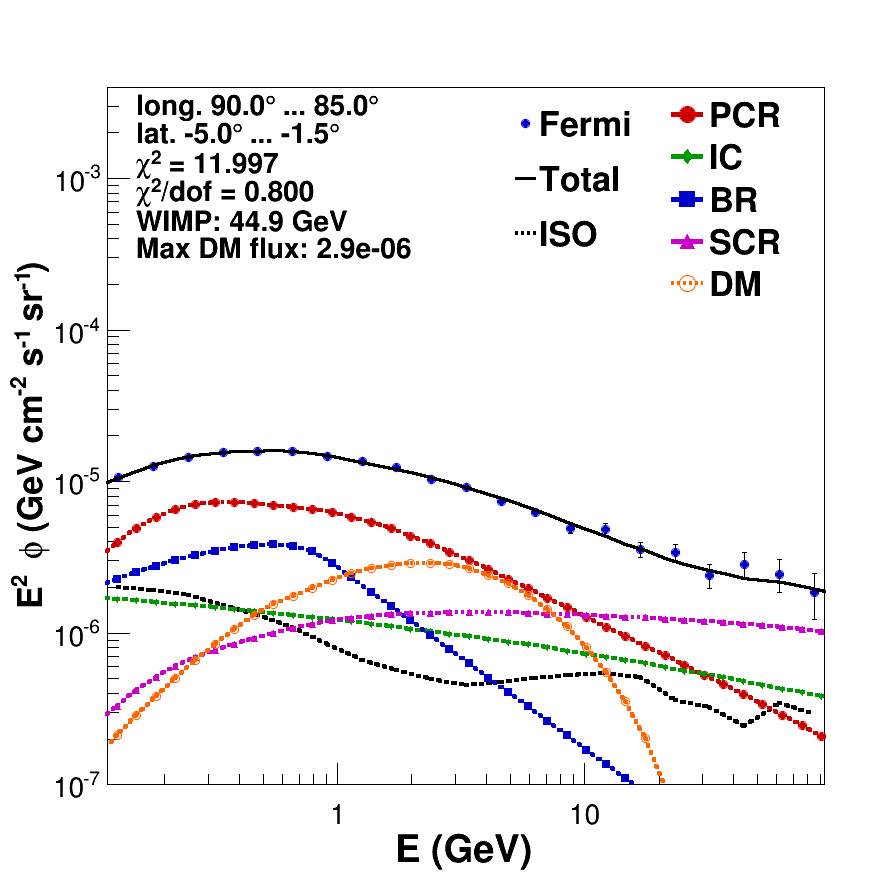}
\includegraphics[width=0.16\textwidth,height=0.16\textwidth,clip]{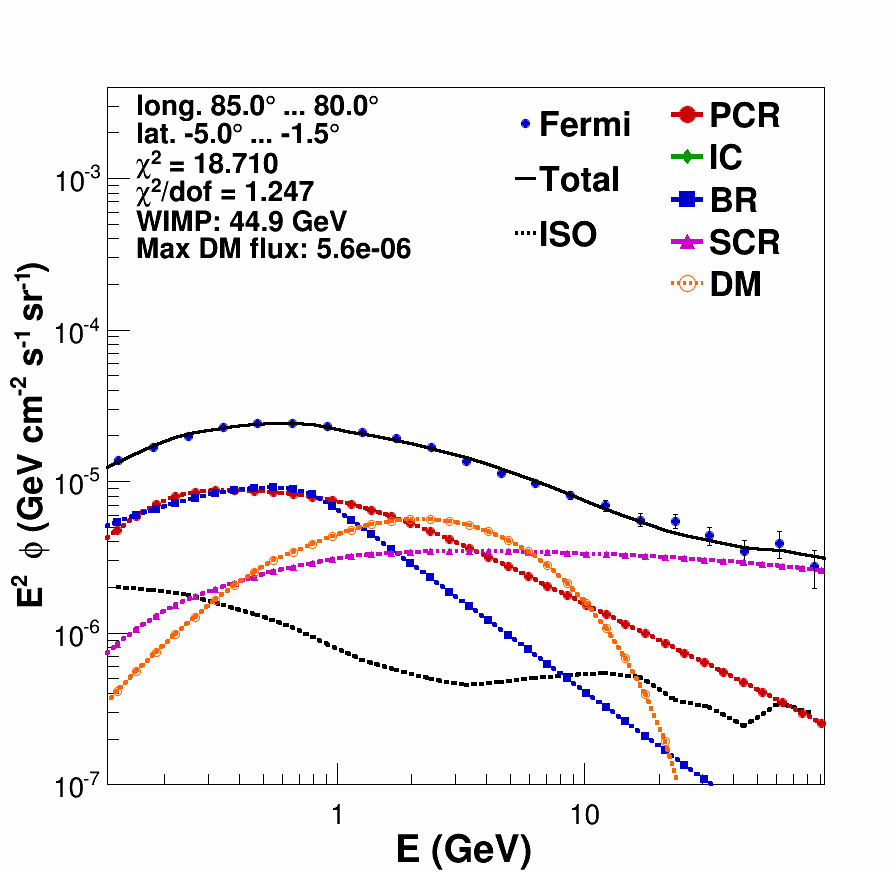}
\includegraphics[width=0.16\textwidth,height=0.16\textwidth,clip]{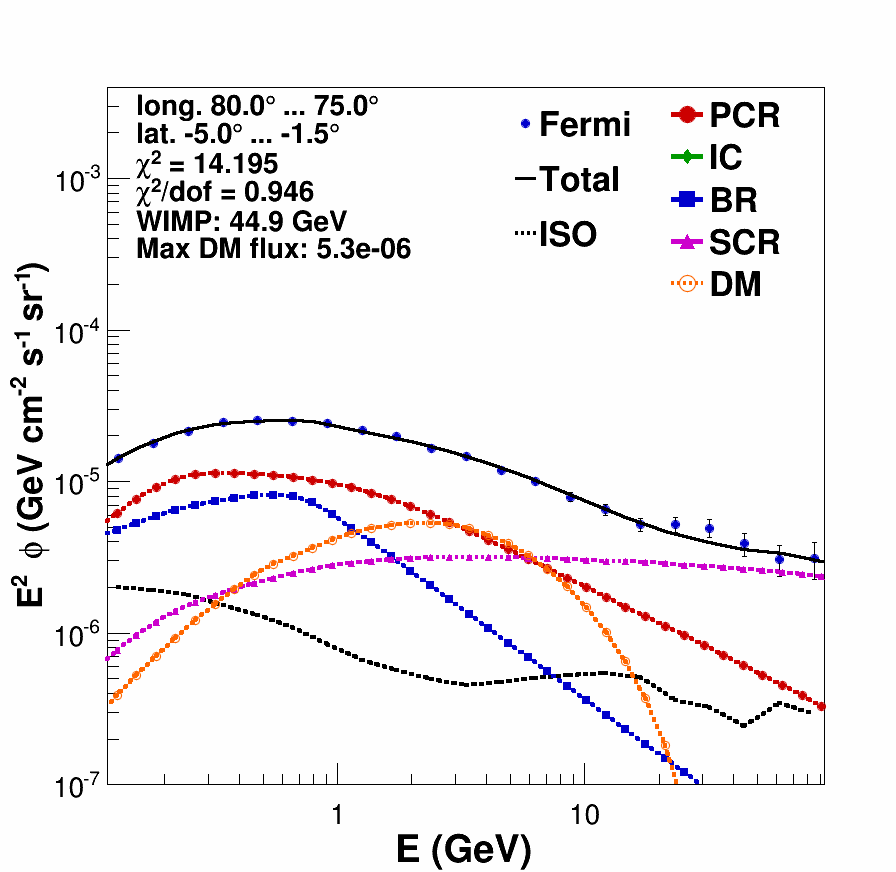}
\includegraphics[width=0.16\textwidth,height=0.16\textwidth,clip]{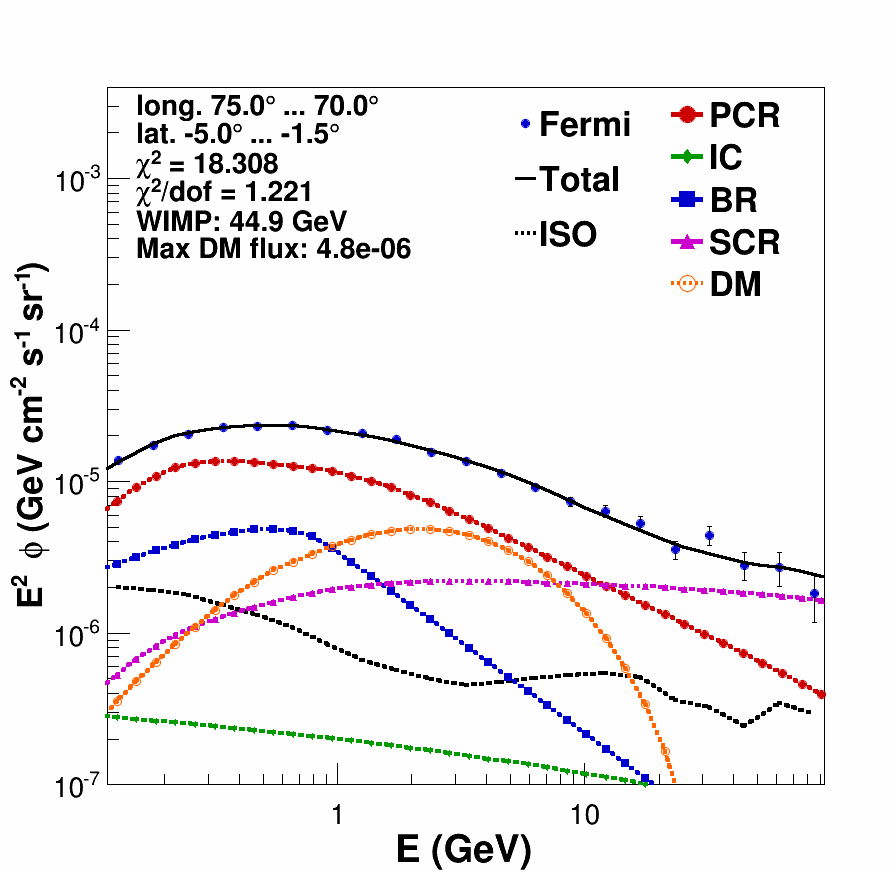}
\includegraphics[width=0.16\textwidth,height=0.16\textwidth,clip]{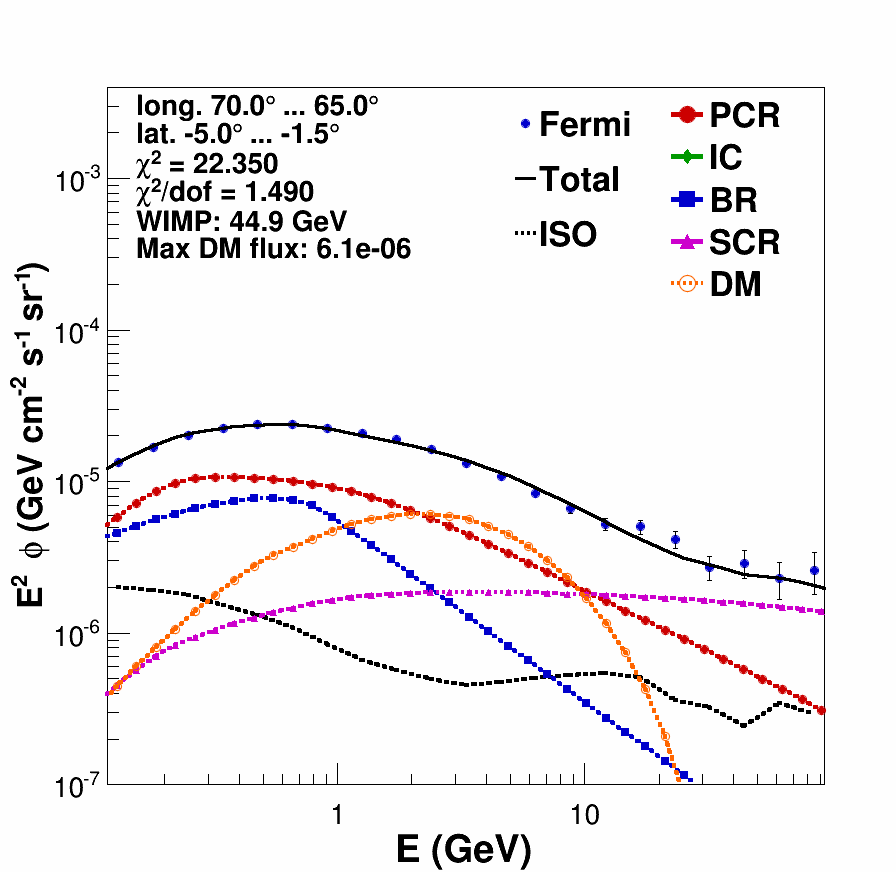}
\includegraphics[width=0.16\textwidth,height=0.16\textwidth,clip]{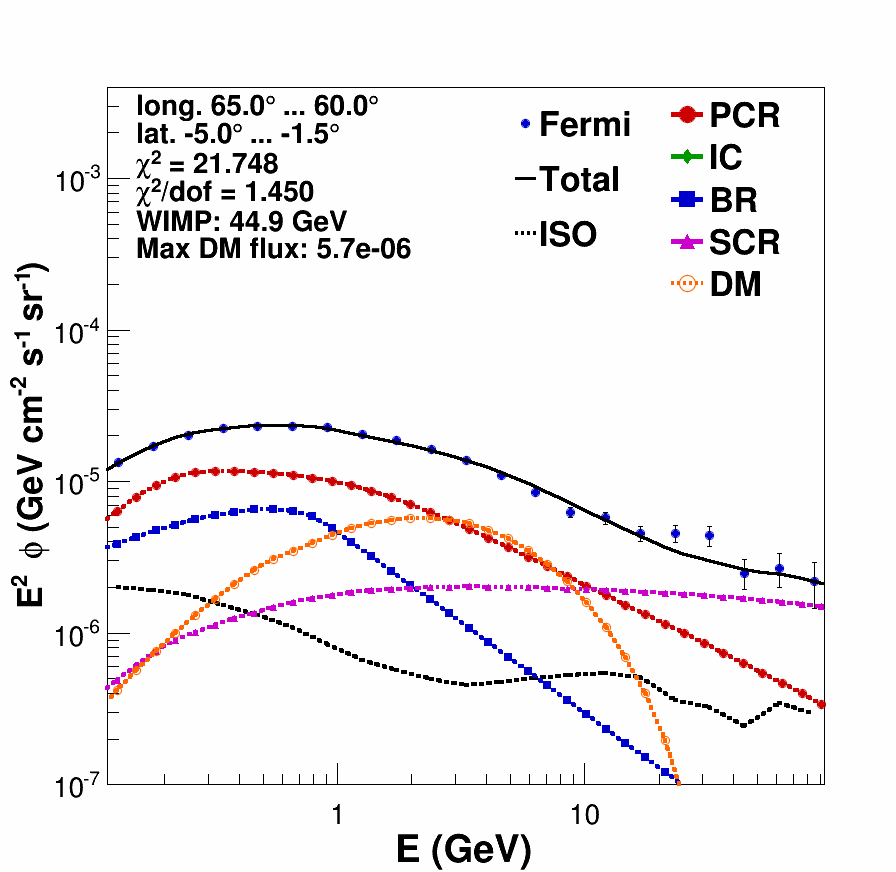}
\includegraphics[width=0.16\textwidth,height=0.16\textwidth,clip]{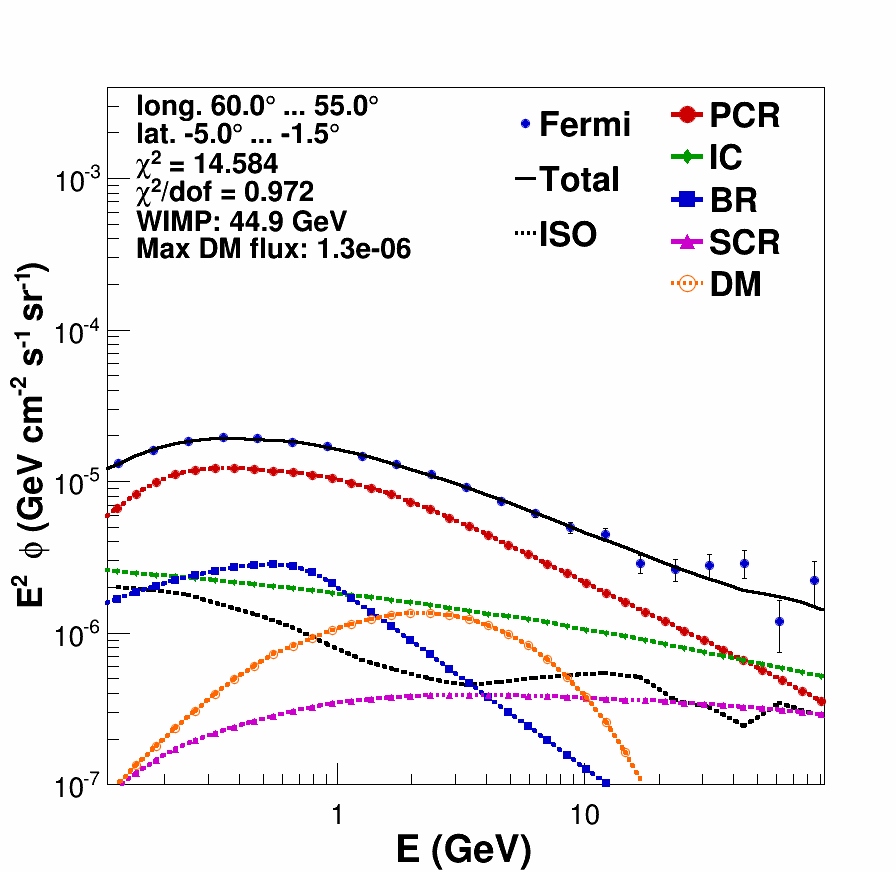}
\includegraphics[width=0.16\textwidth,height=0.16\textwidth,clip]{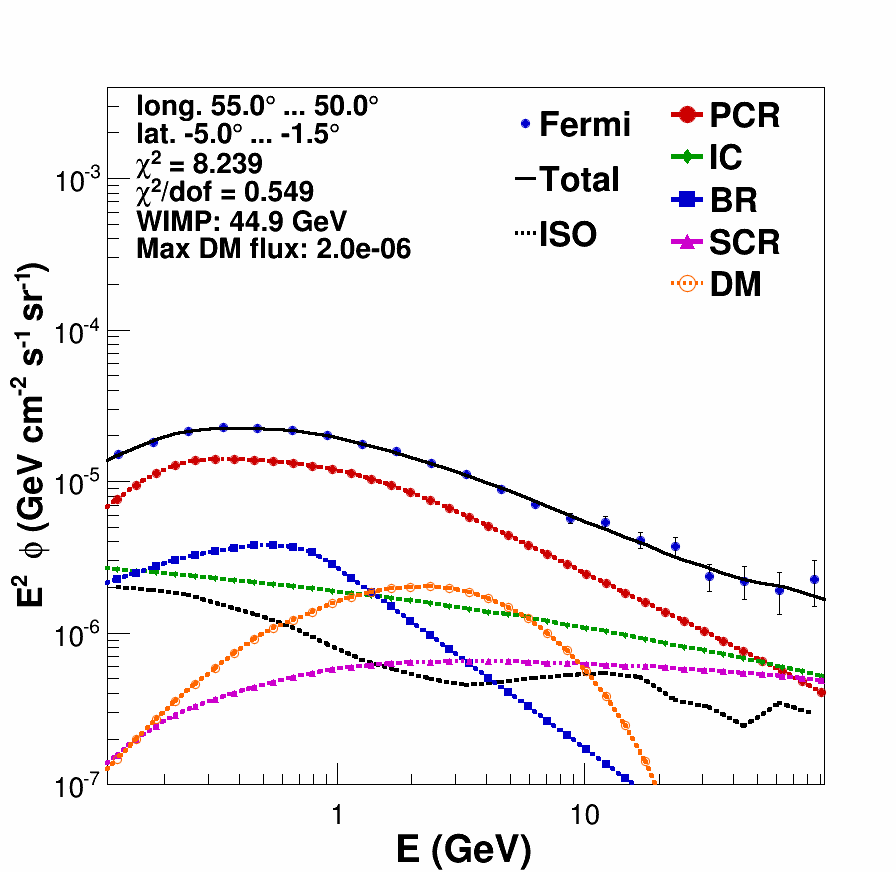}
\includegraphics[width=0.16\textwidth,height=0.16\textwidth,clip]{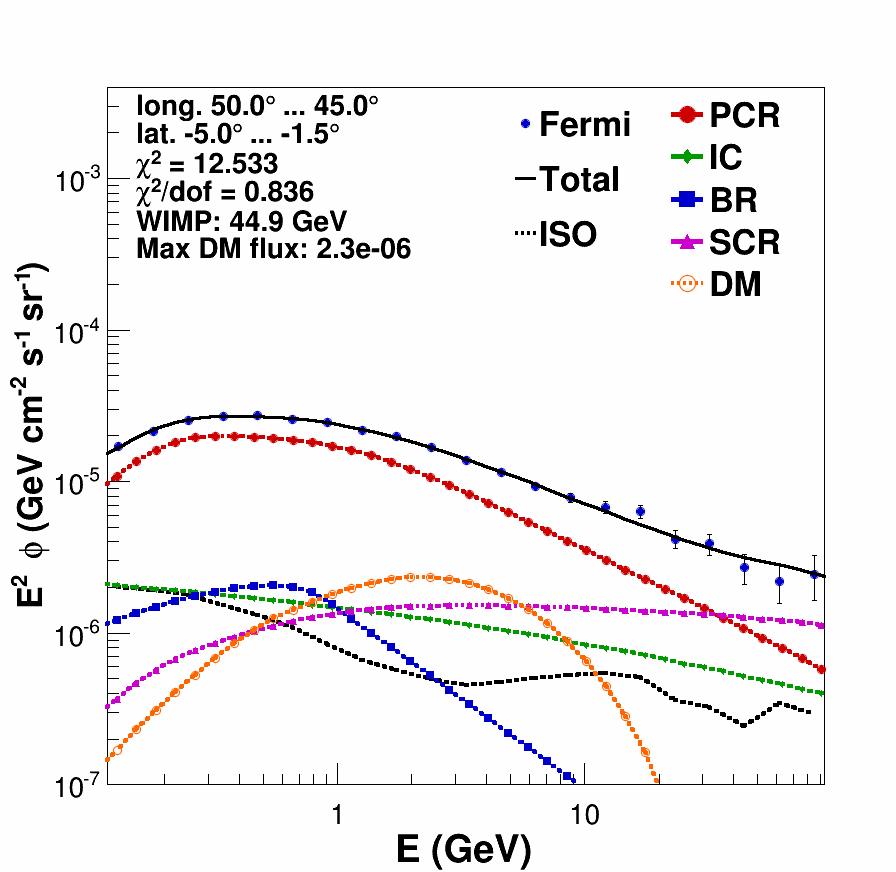}
\includegraphics[width=0.16\textwidth,height=0.16\textwidth,clip]{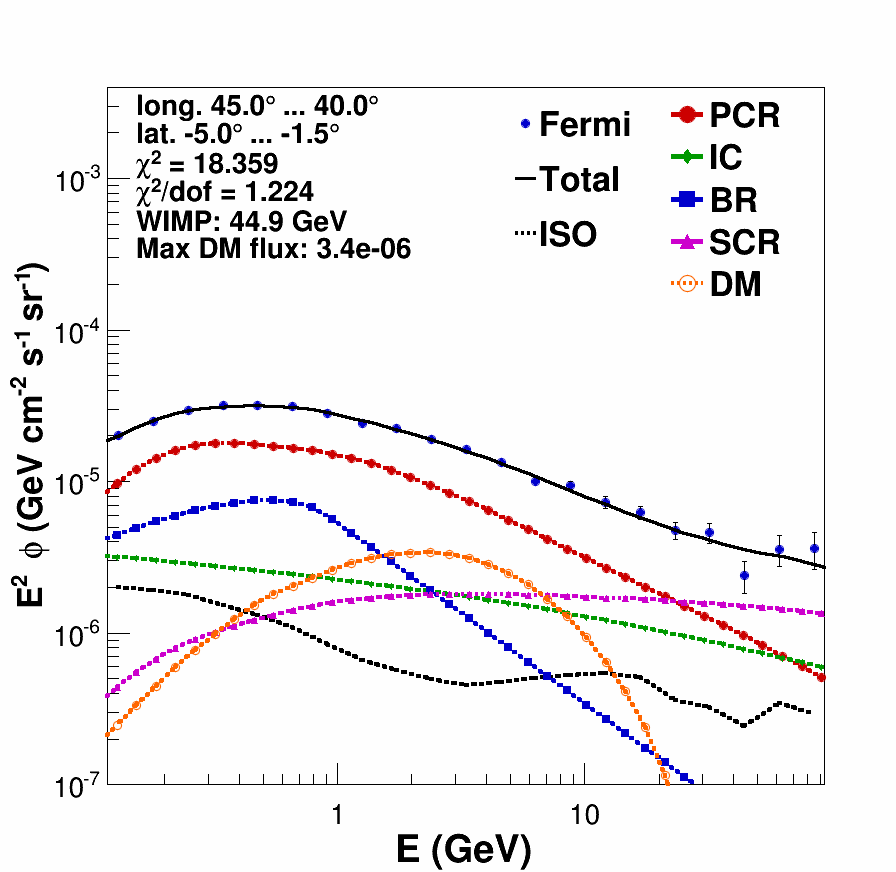}
\includegraphics[width=0.16\textwidth,height=0.16\textwidth,clip]{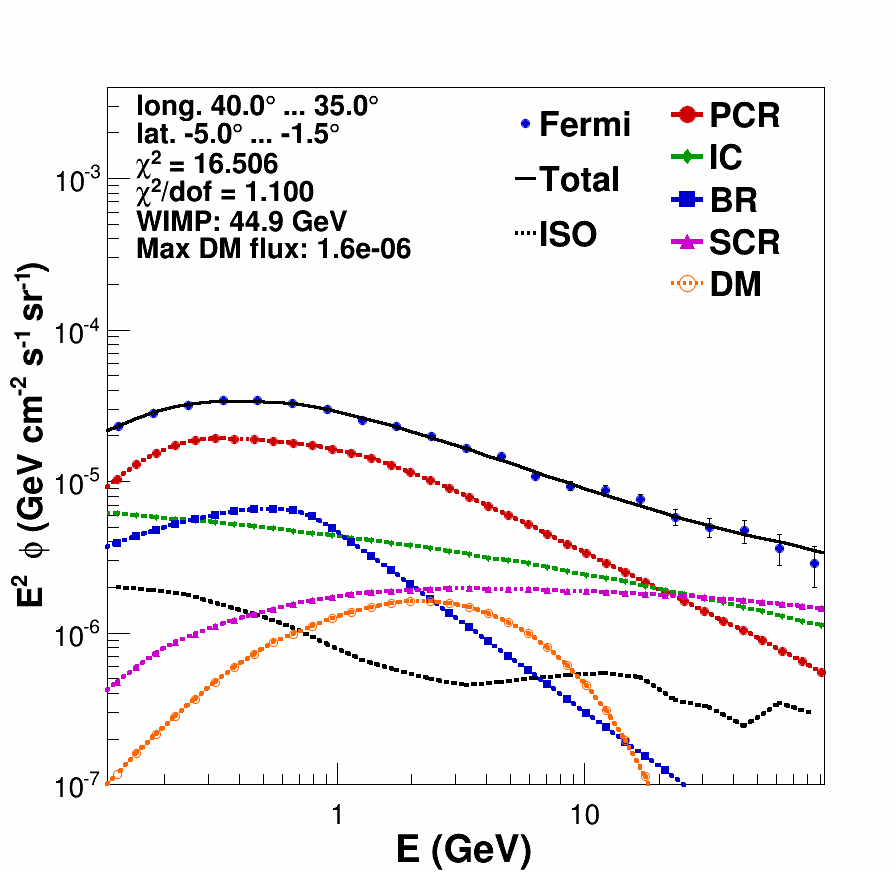}
\includegraphics[width=0.16\textwidth,height=0.16\textwidth,clip]{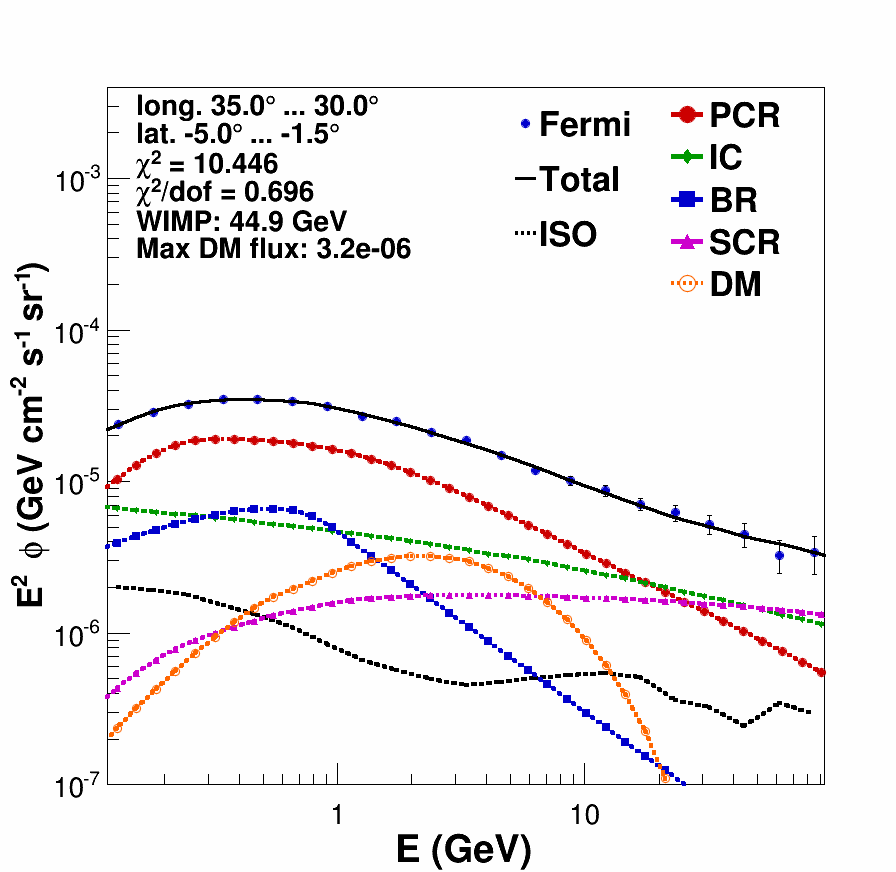}
\includegraphics[width=0.16\textwidth,height=0.16\textwidth,clip]{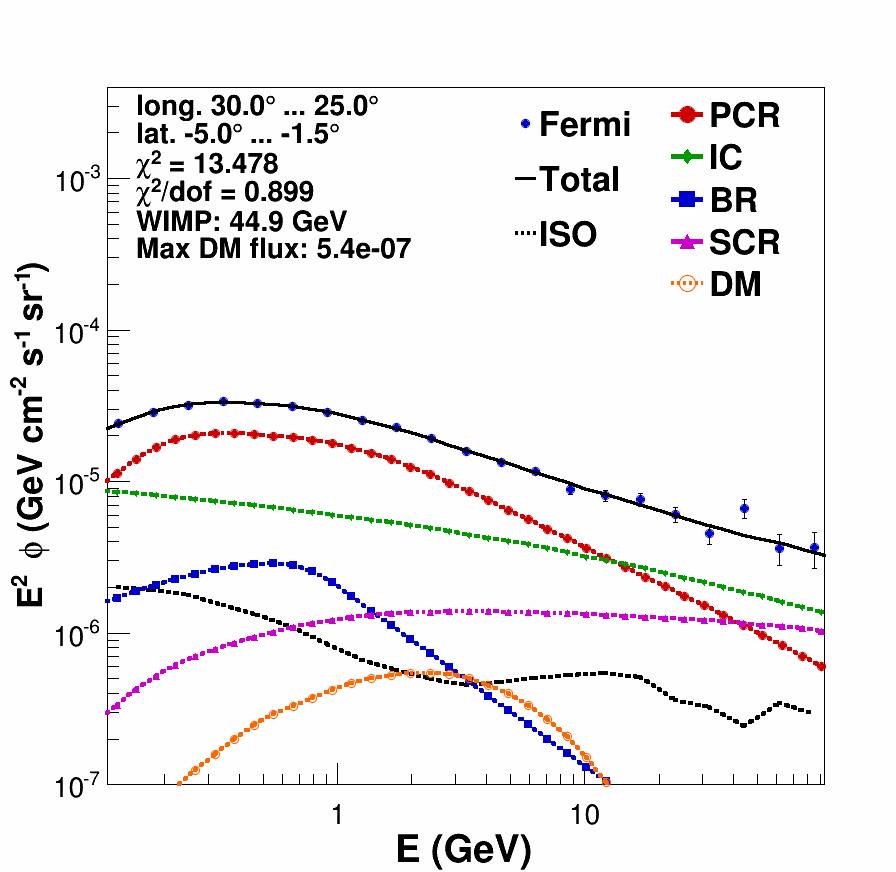}
\includegraphics[width=0.16\textwidth,height=0.16\textwidth,clip]{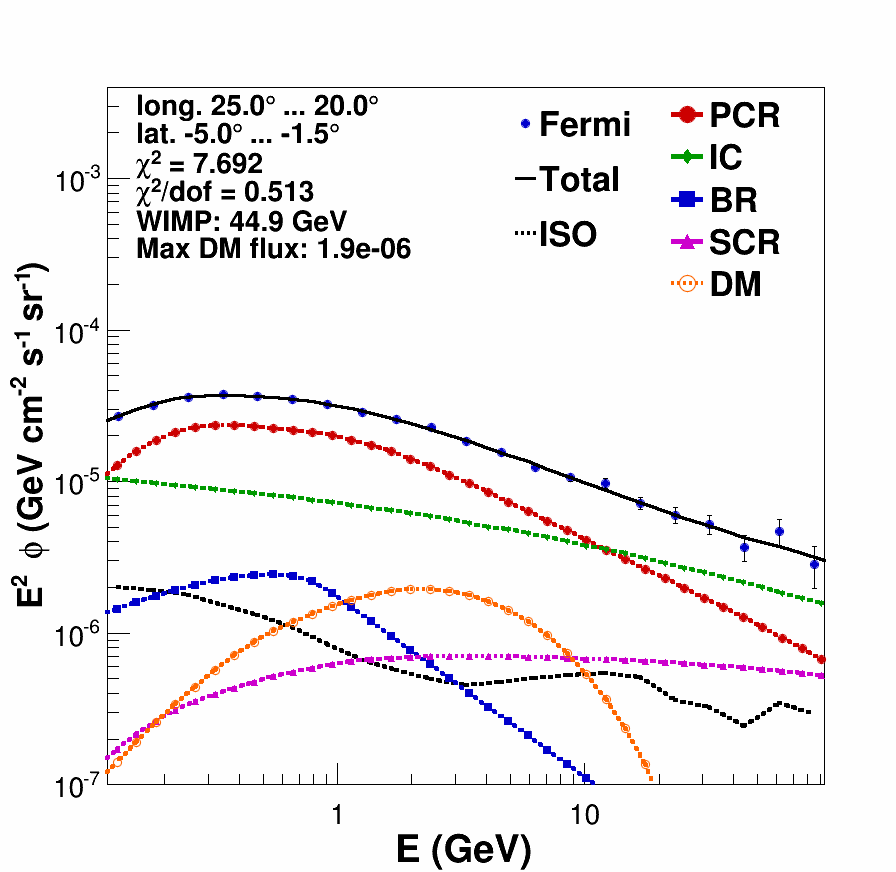}
\includegraphics[width=0.16\textwidth,height=0.16\textwidth,clip]{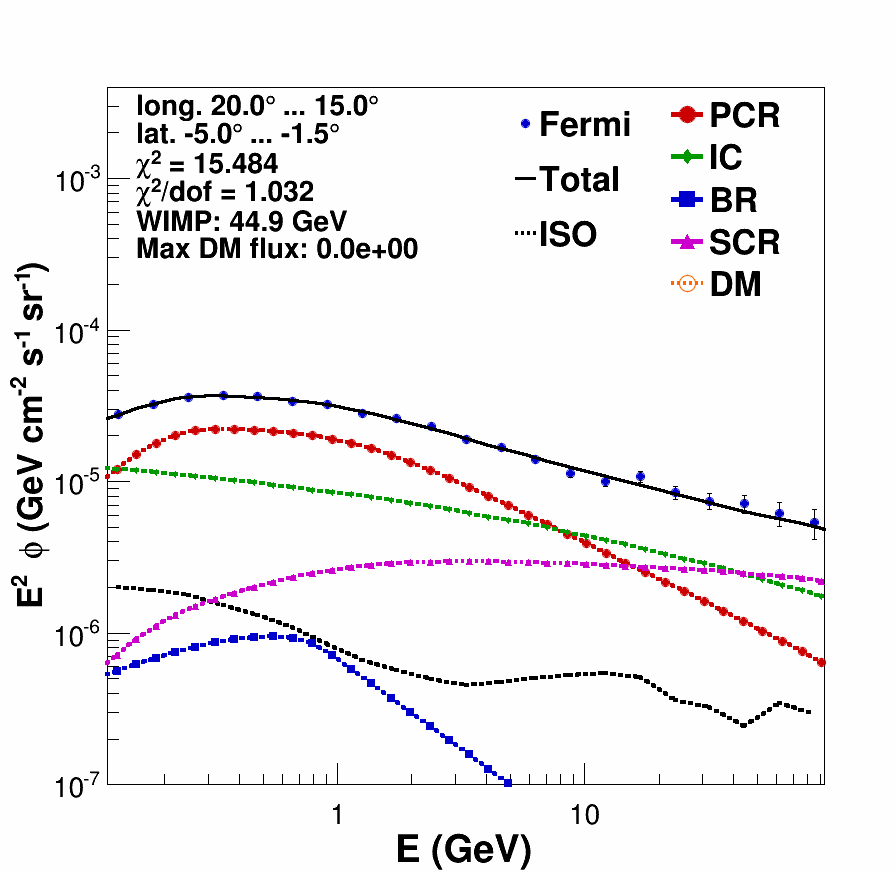}
\includegraphics[width=0.16\textwidth,height=0.16\textwidth,clip]{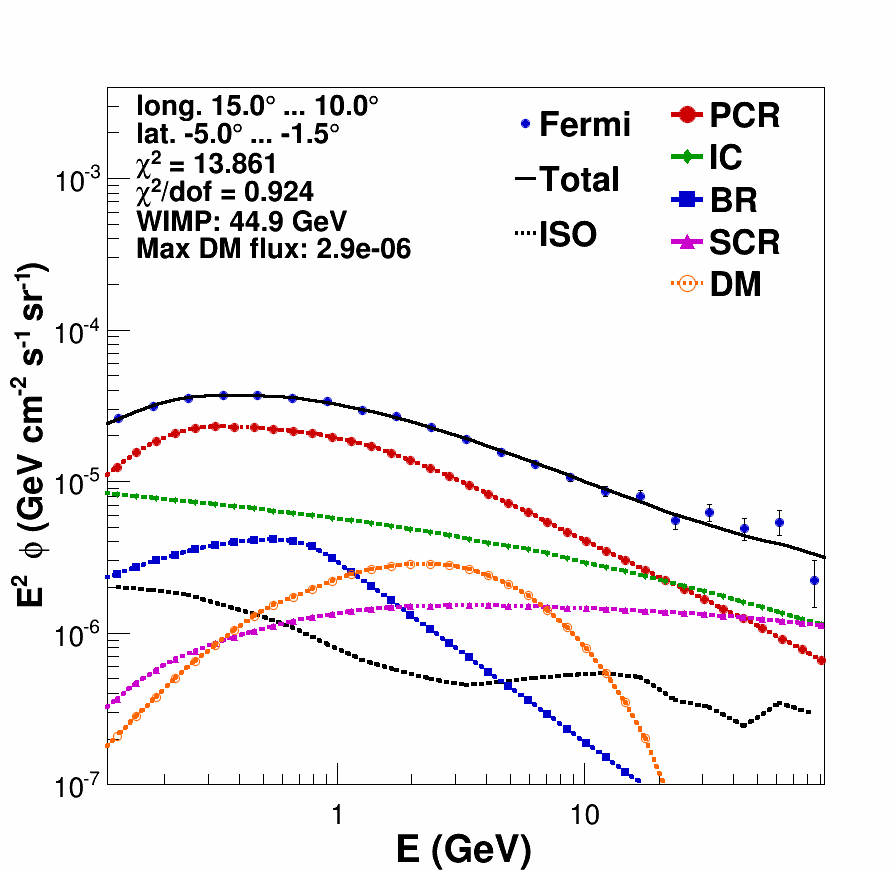}
\includegraphics[width=0.16\textwidth,height=0.16\textwidth,clip]{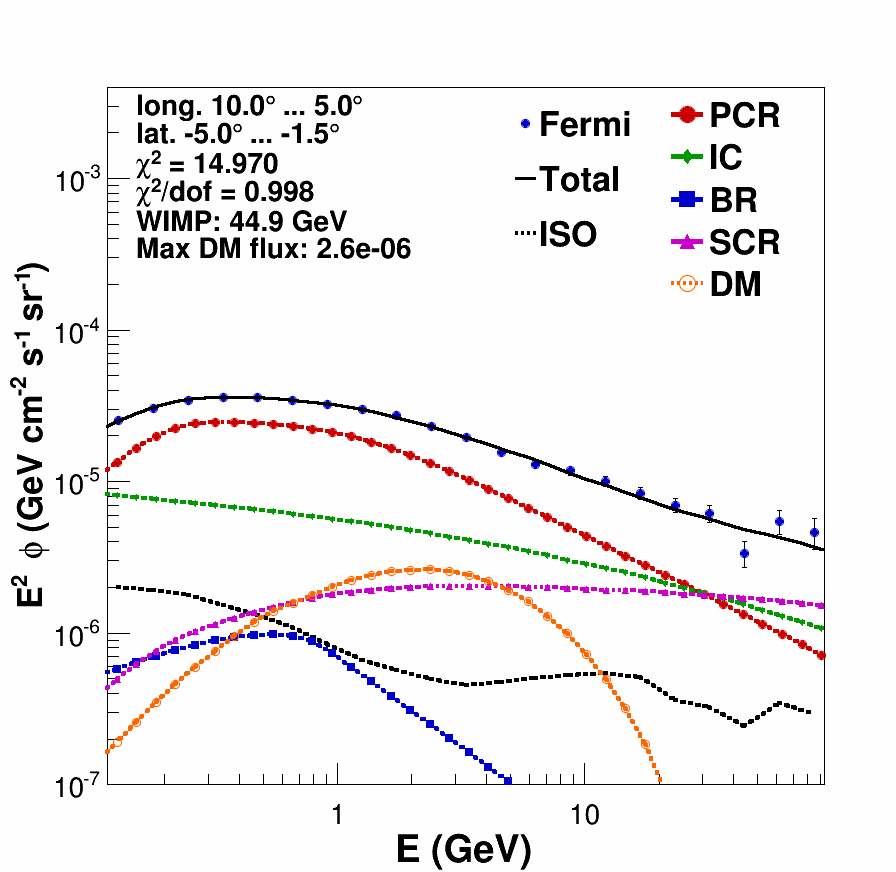}
\includegraphics[width=0.16\textwidth,height=0.16\textwidth,clip]{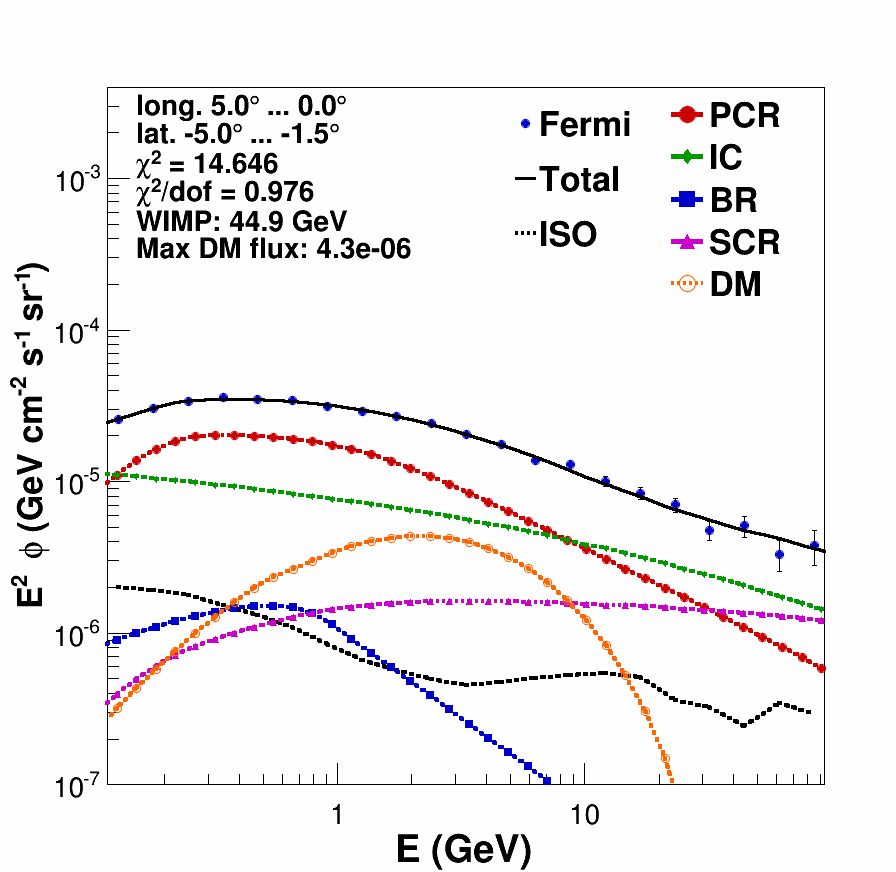}%%%%%%r12a
\caption[]{Template fits for latitudes  with $-5.0^\circ<b<-1.5^\circ$ and longitudes decreasing from 180$^\circ$ to 0$^\circ$.} \label{F44}
\end{figure}
\begin{figure}
\centering
\includegraphics[width=0.16\textwidth,height=0.16\textwidth,clip]{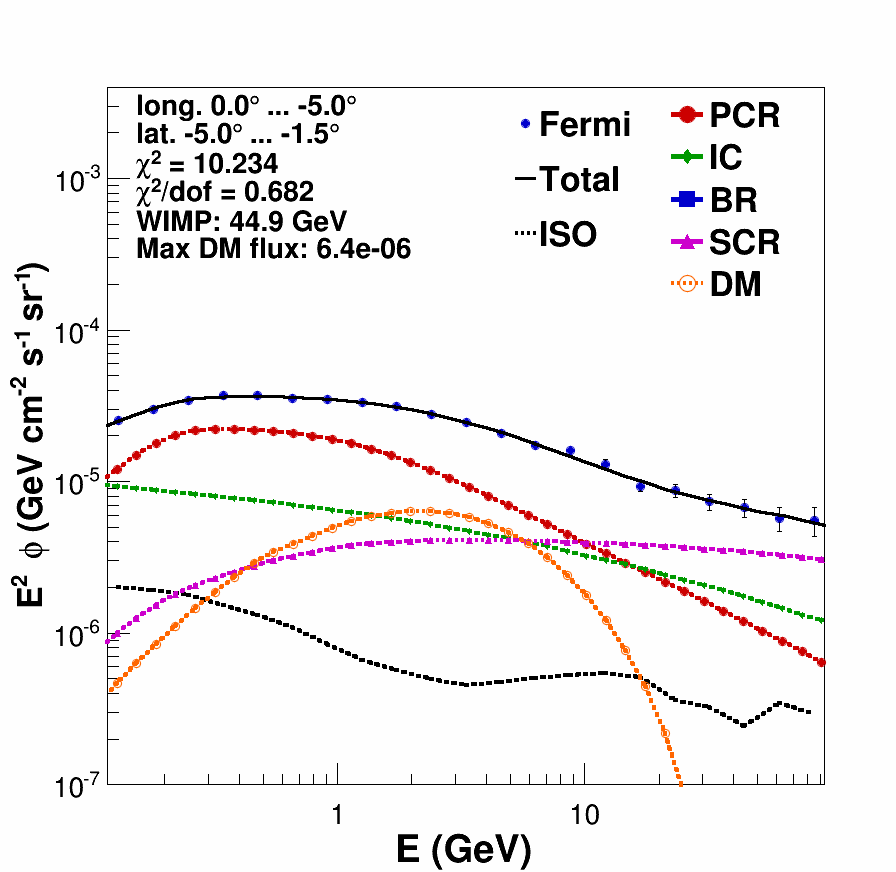}
\includegraphics[width=0.16\textwidth,height=0.16\textwidth,clip]{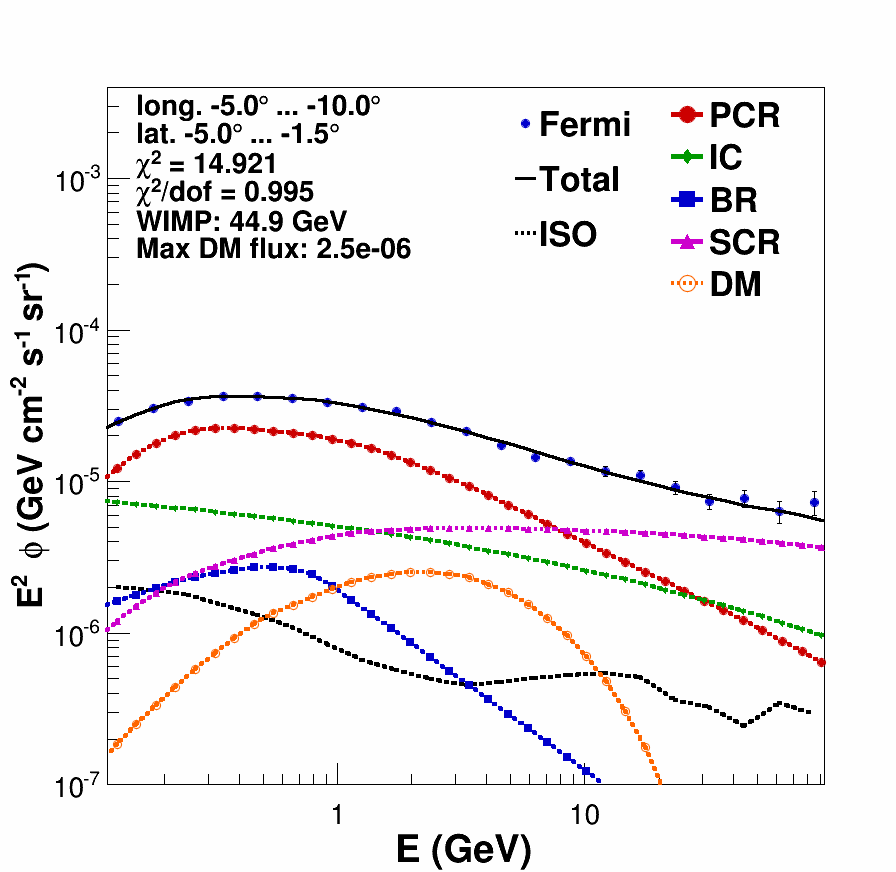}
\includegraphics[width=0.16\textwidth,height=0.16\textwidth,clip]{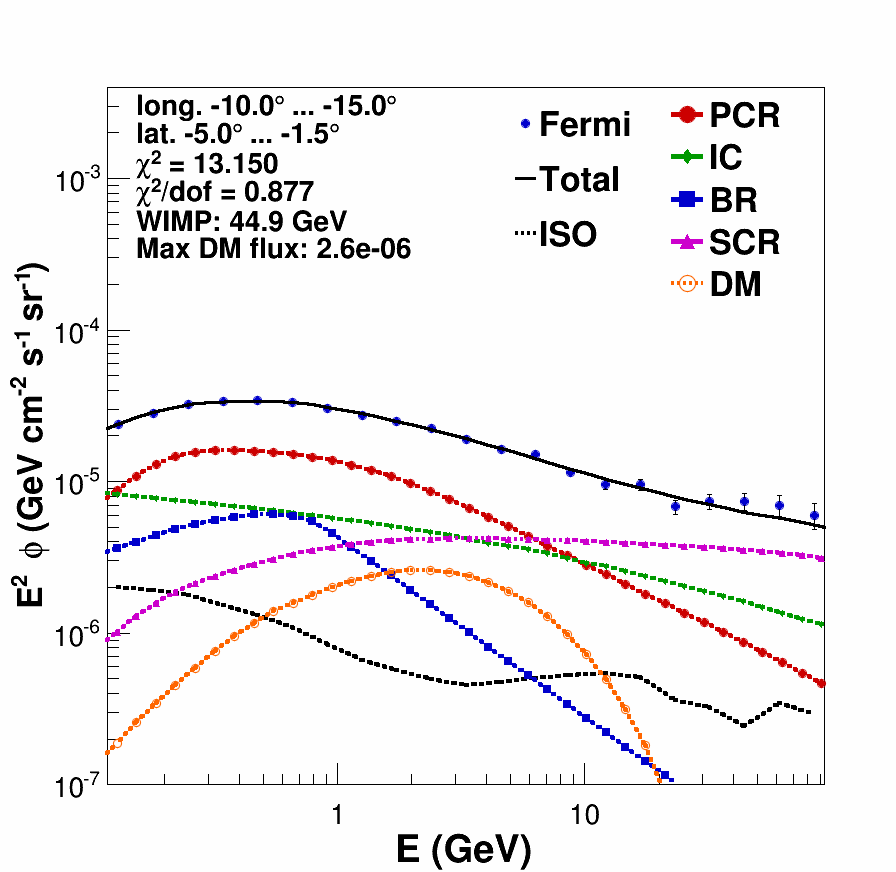}
\includegraphics[width=0.16\textwidth,height=0.16\textwidth,clip]{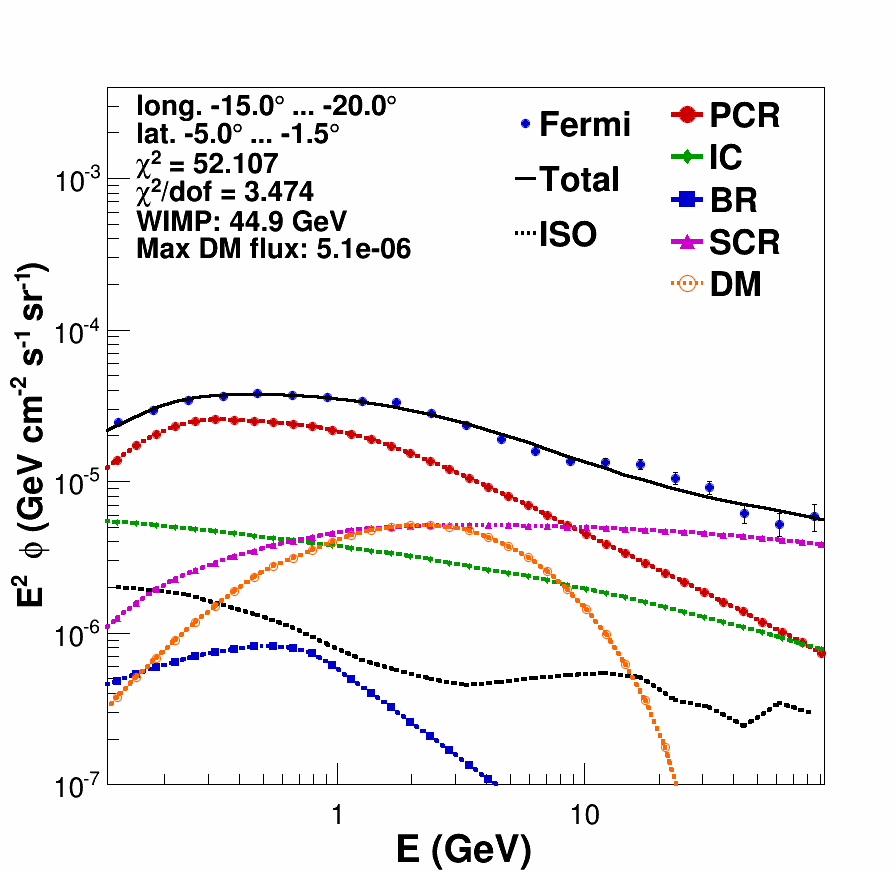}
\includegraphics[width=0.16\textwidth,height=0.16\textwidth,clip]{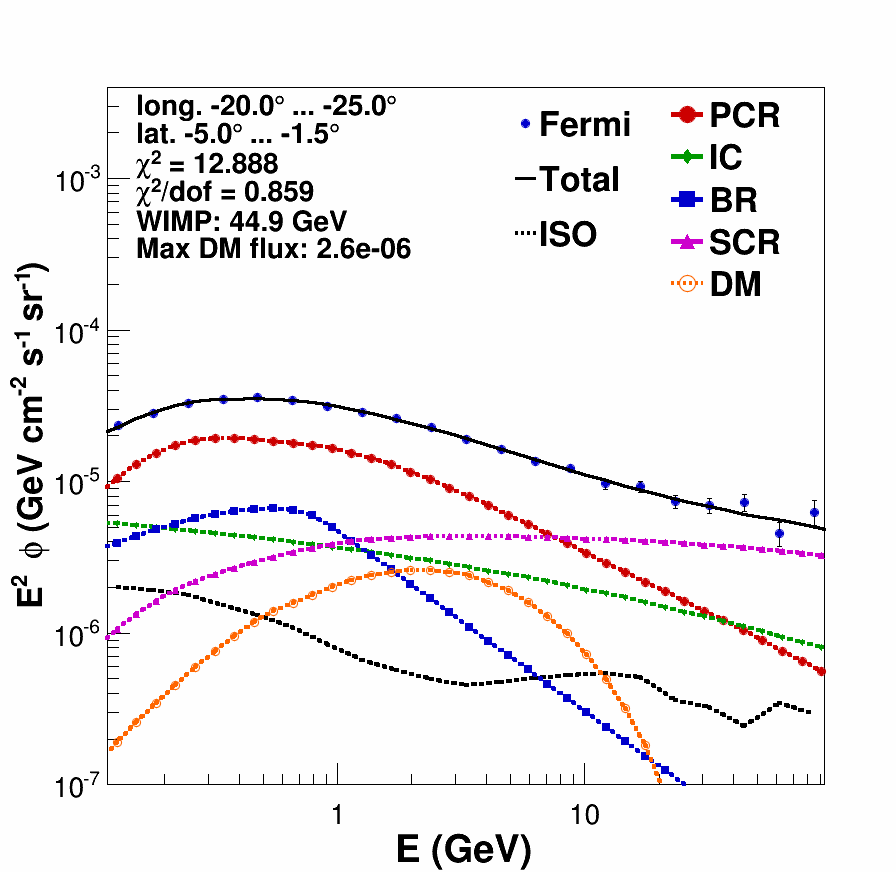}
\includegraphics[width=0.16\textwidth,height=0.16\textwidth,clip]{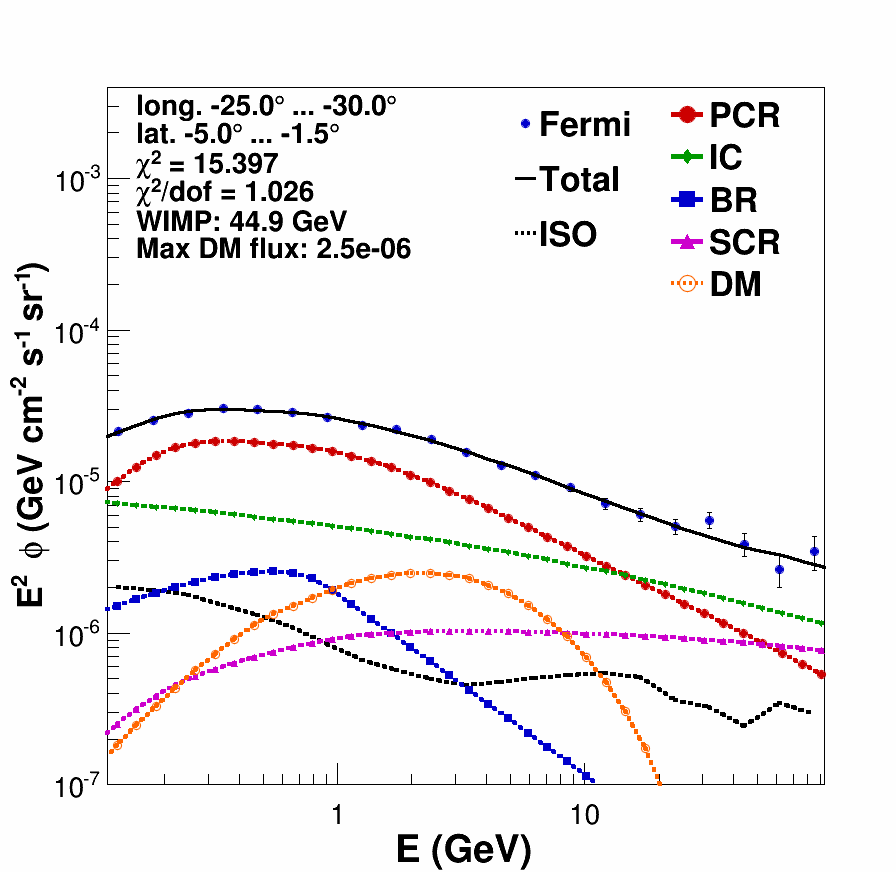}
\includegraphics[width=0.16\textwidth,height=0.16\textwidth,clip]{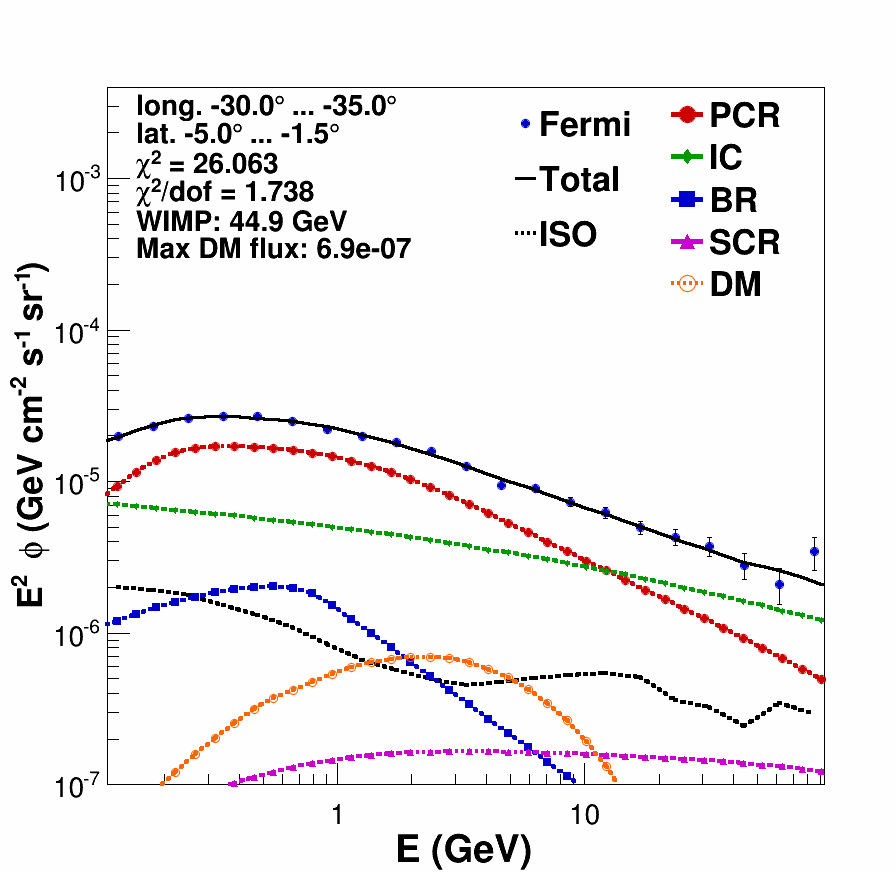}
\includegraphics[width=0.16\textwidth,height=0.16\textwidth,clip]{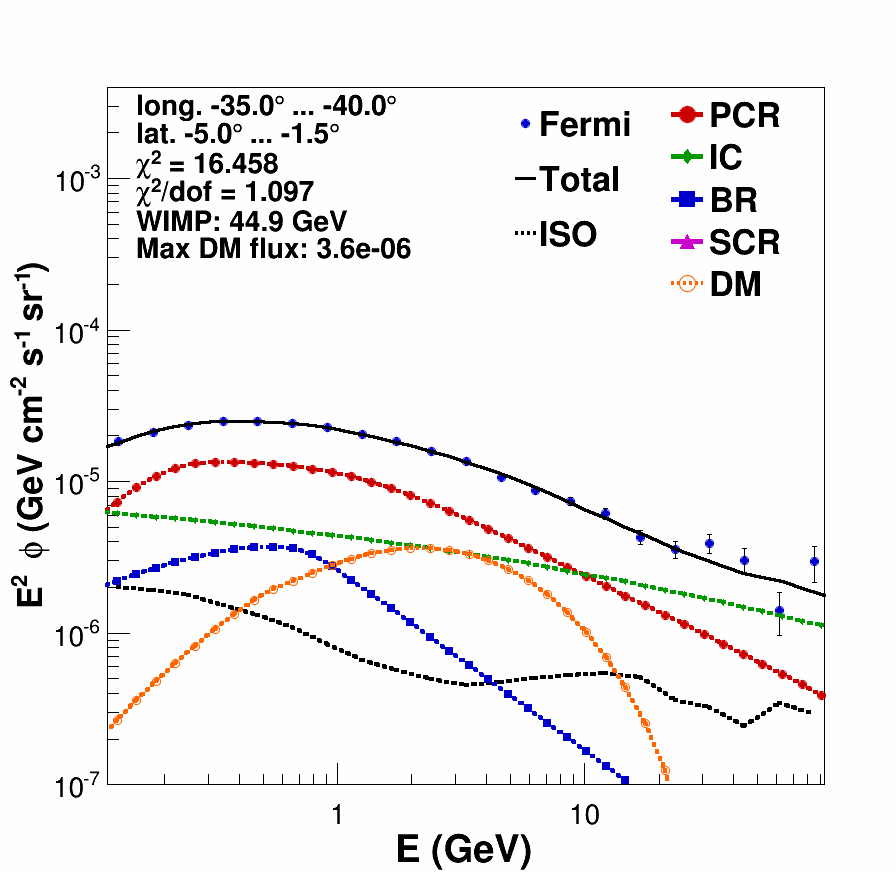}
\includegraphics[width=0.16\textwidth,height=0.16\textwidth,clip]{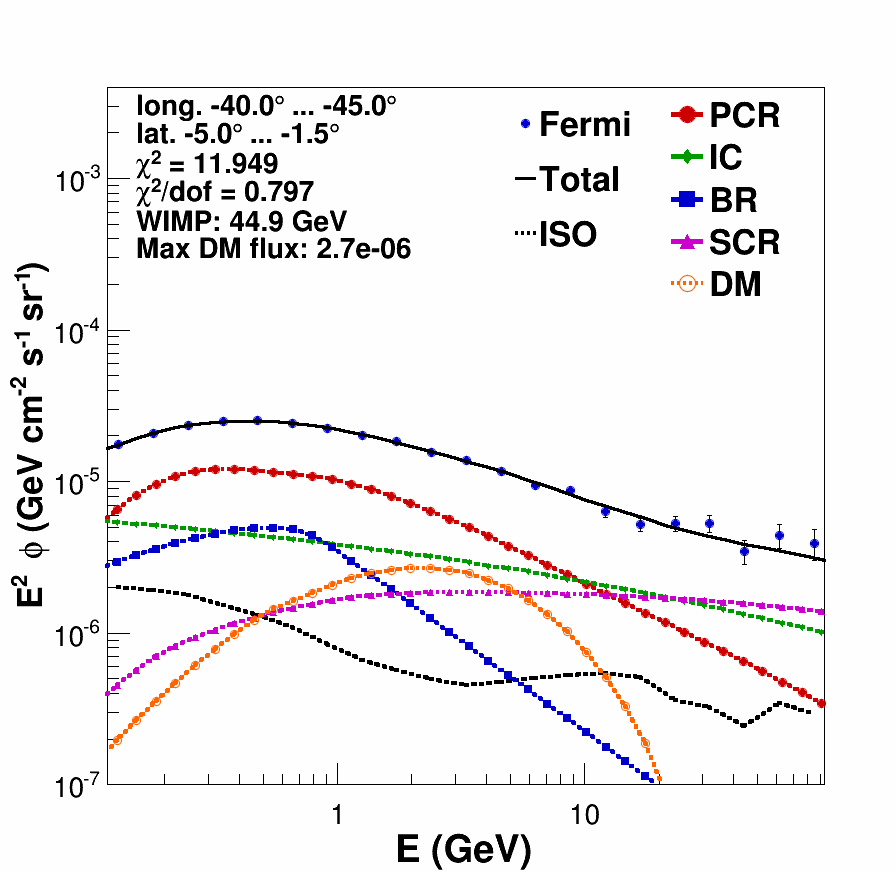}
\includegraphics[width=0.16\textwidth,height=0.16\textwidth,clip]{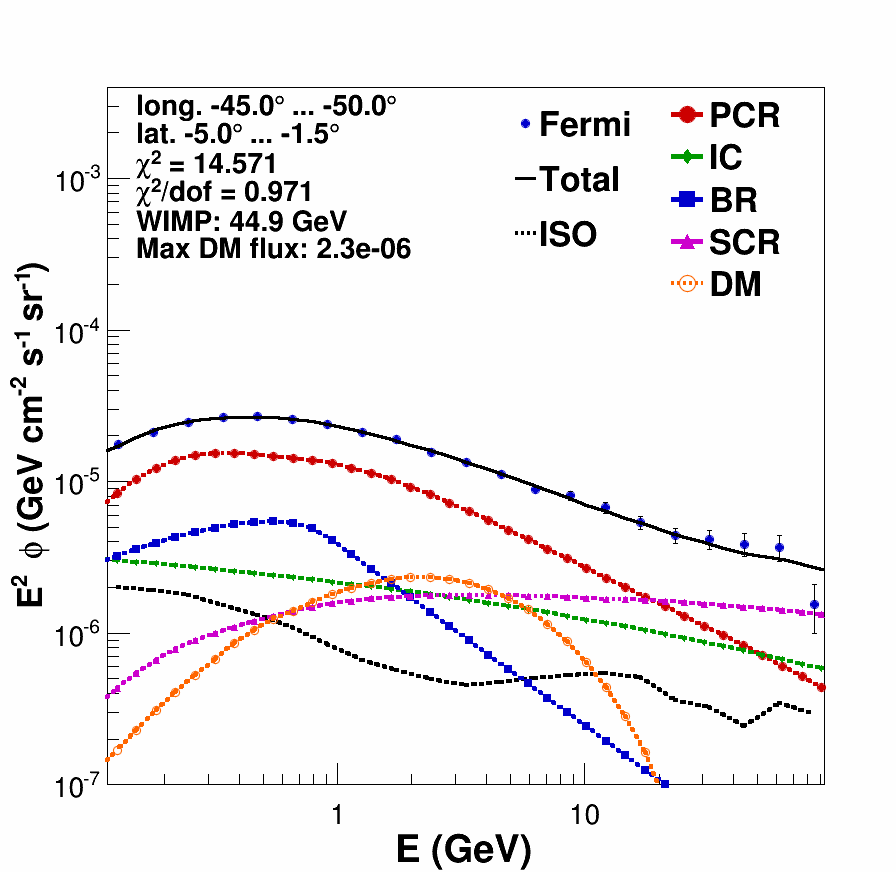}
\includegraphics[width=0.16\textwidth,height=0.16\textwidth,clip]{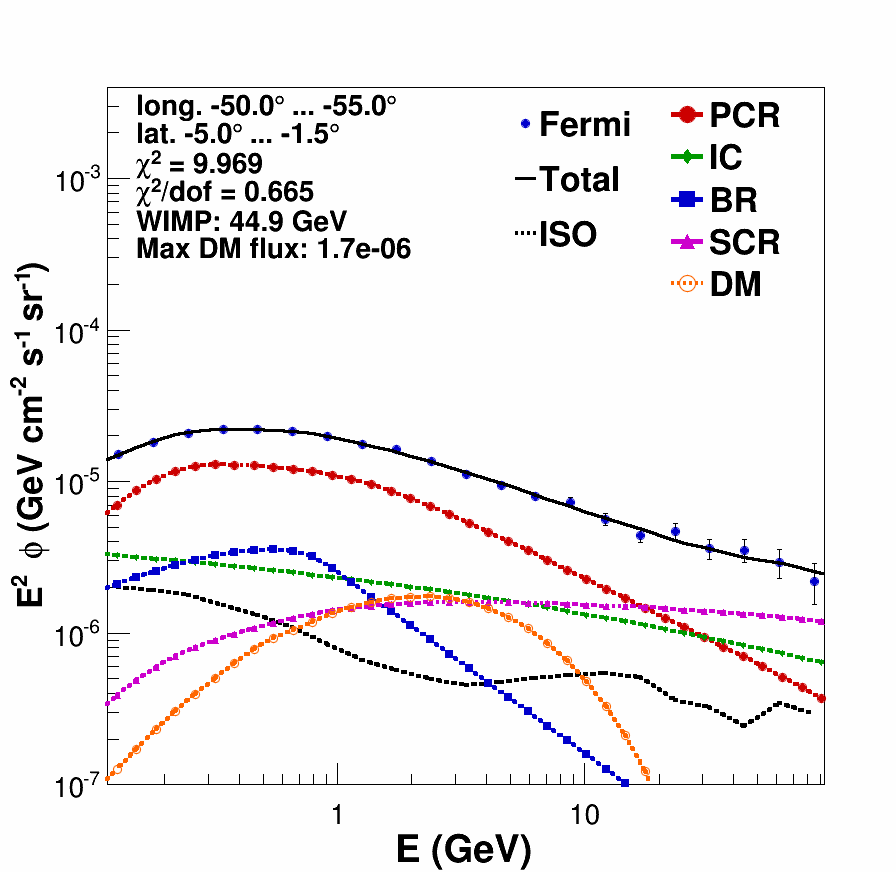}
\includegraphics[width=0.16\textwidth,height=0.16\textwidth,clip]{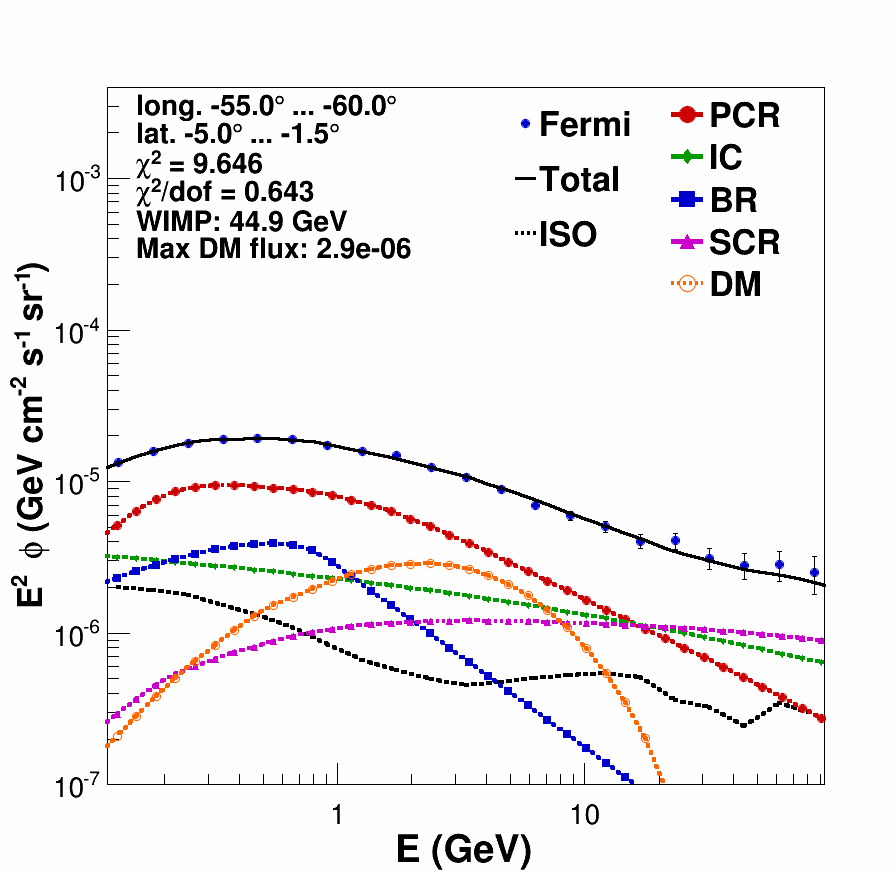}
\includegraphics[width=0.16\textwidth,height=0.16\textwidth,clip]{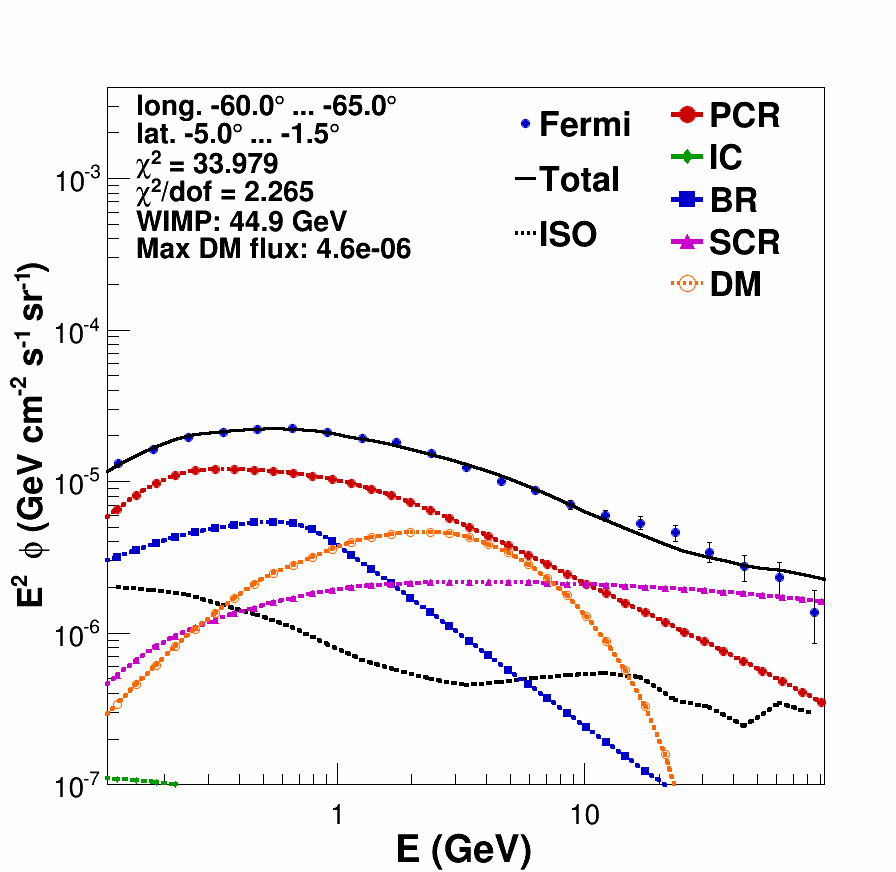}
\includegraphics[width=0.16\textwidth,height=0.16\textwidth,clip]{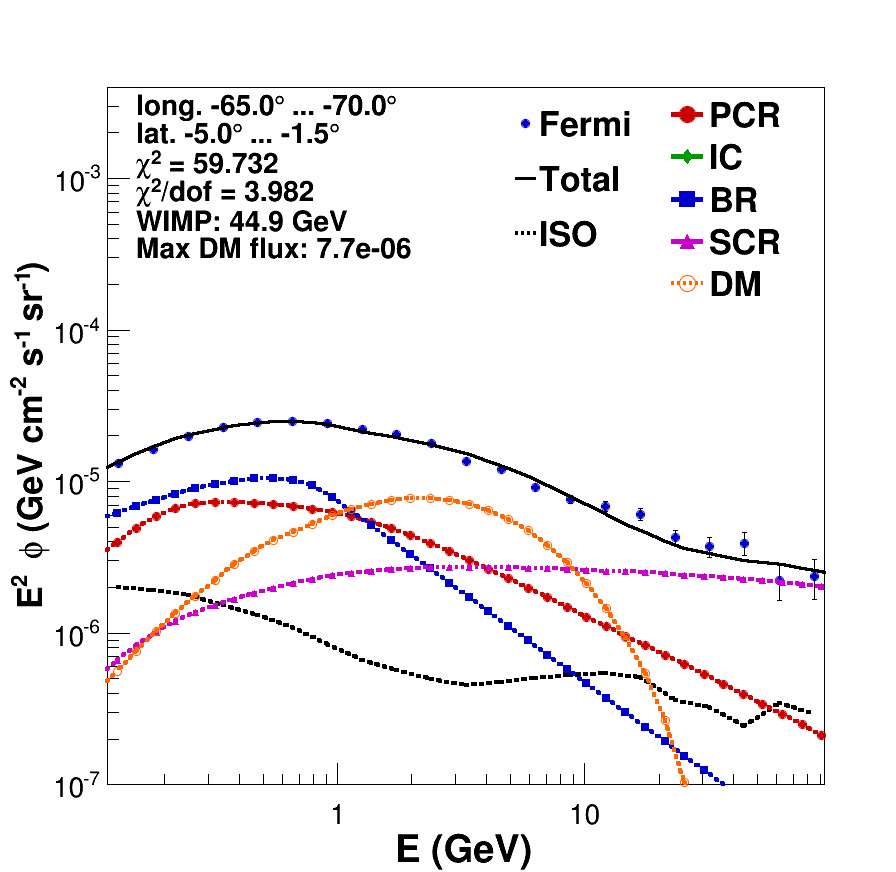}
\includegraphics[width=0.16\textwidth,height=0.16\textwidth,clip]{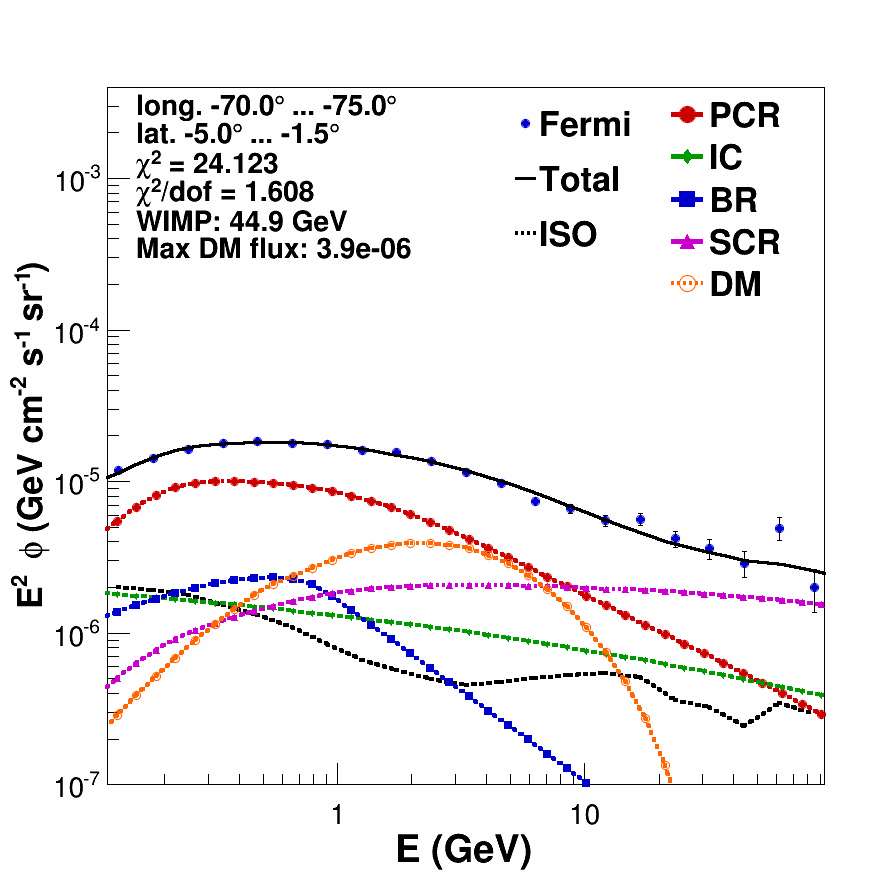}
\includegraphics[width=0.16\textwidth,height=0.16\textwidth,clip]{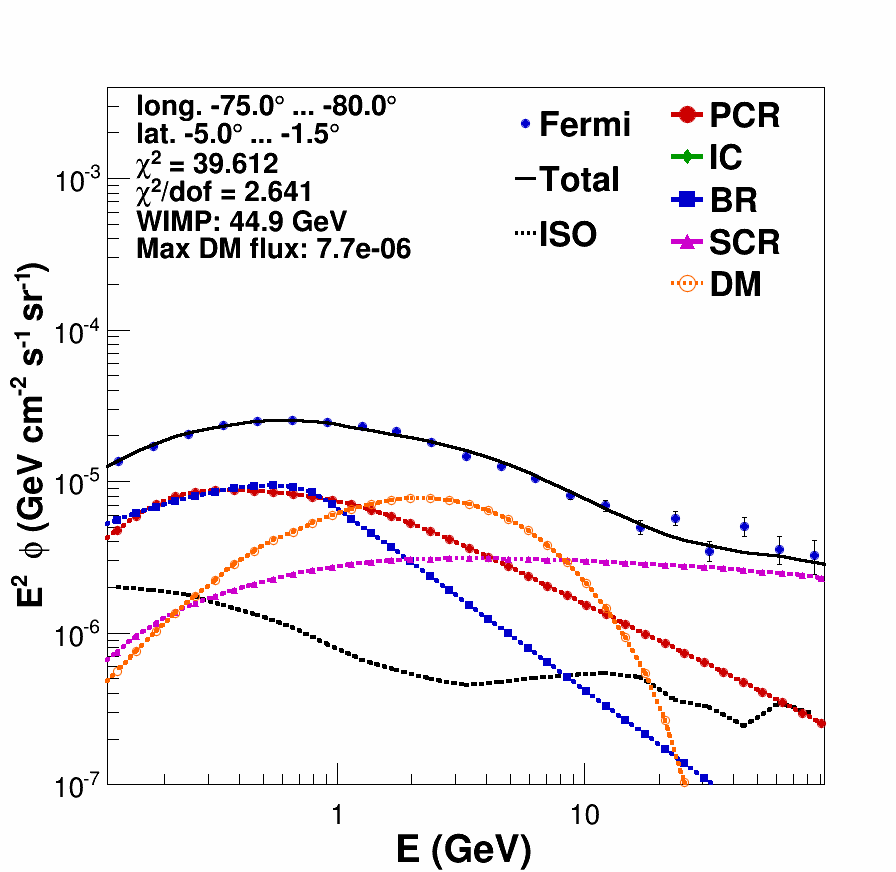}
\includegraphics[width=0.16\textwidth,height=0.16\textwidth,clip]{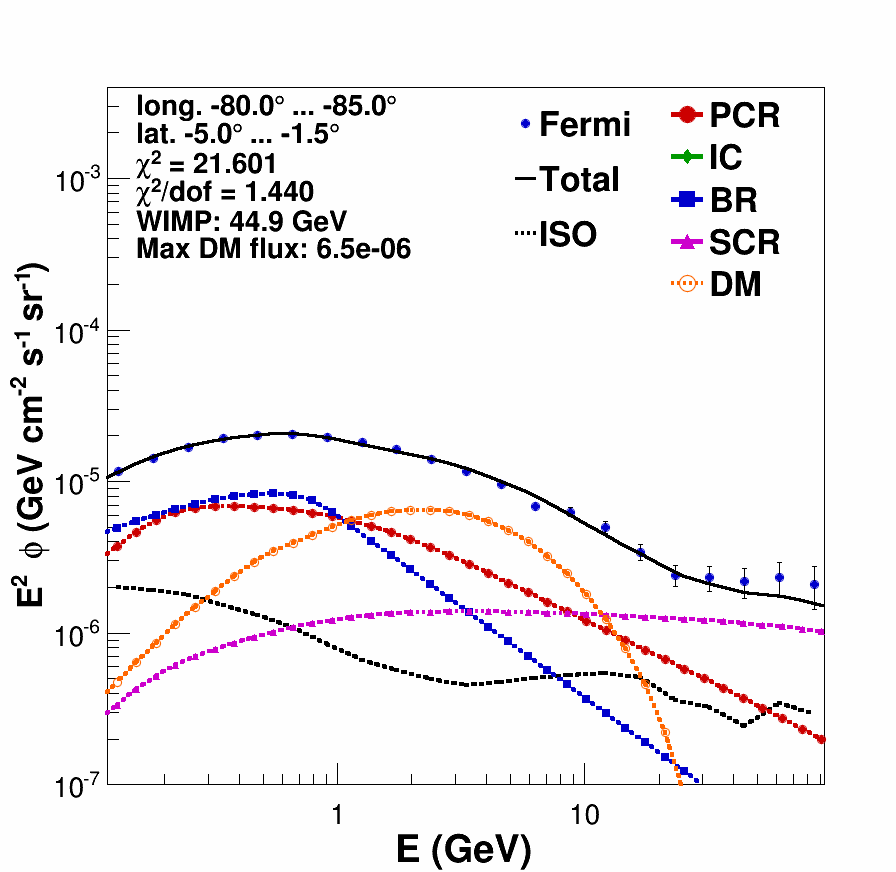}
\includegraphics[width=0.16\textwidth,height=0.16\textwidth,clip]{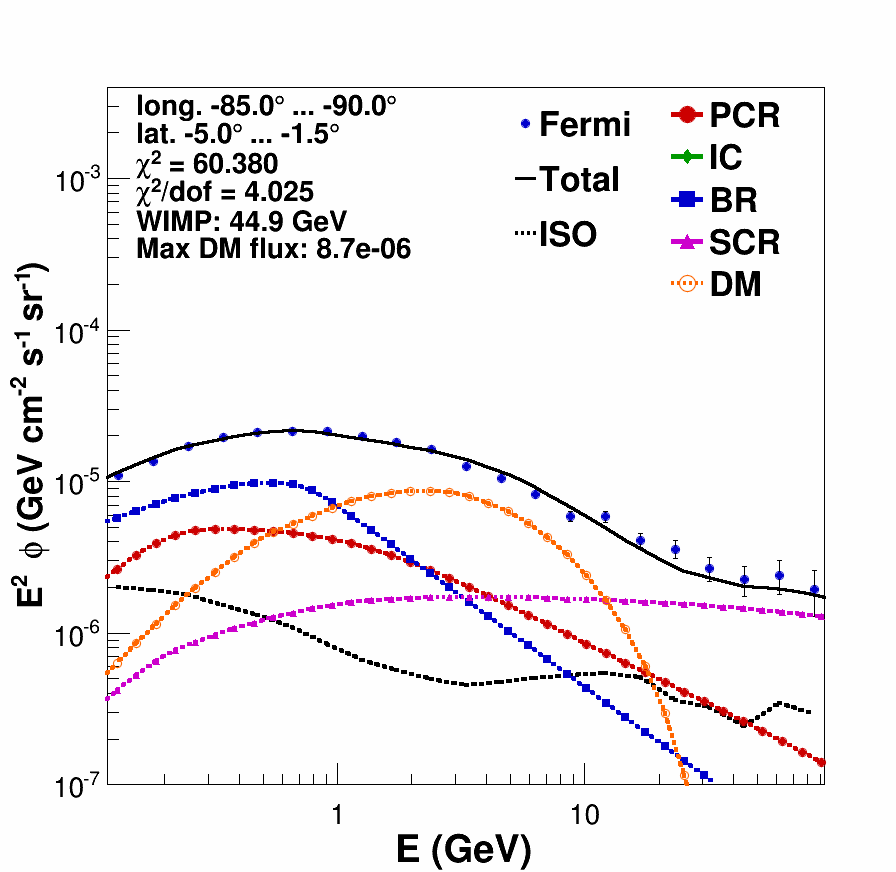}
\includegraphics[width=0.16\textwidth,height=0.16\textwidth,clip]{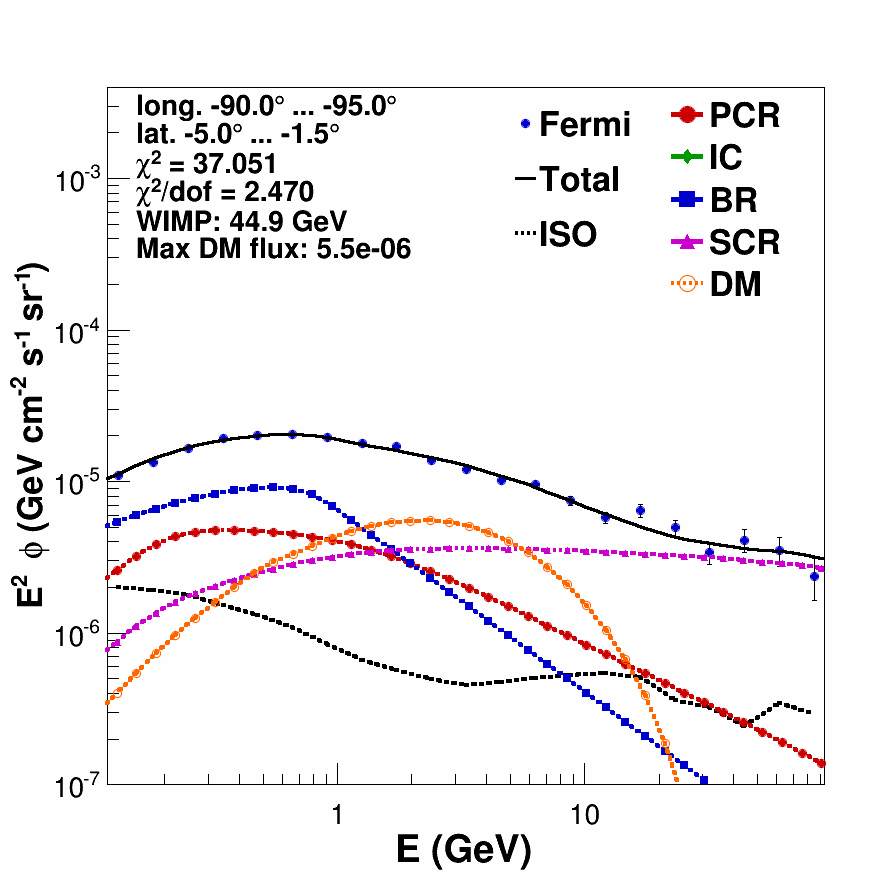}
\includegraphics[width=0.16\textwidth,height=0.16\textwidth,clip]{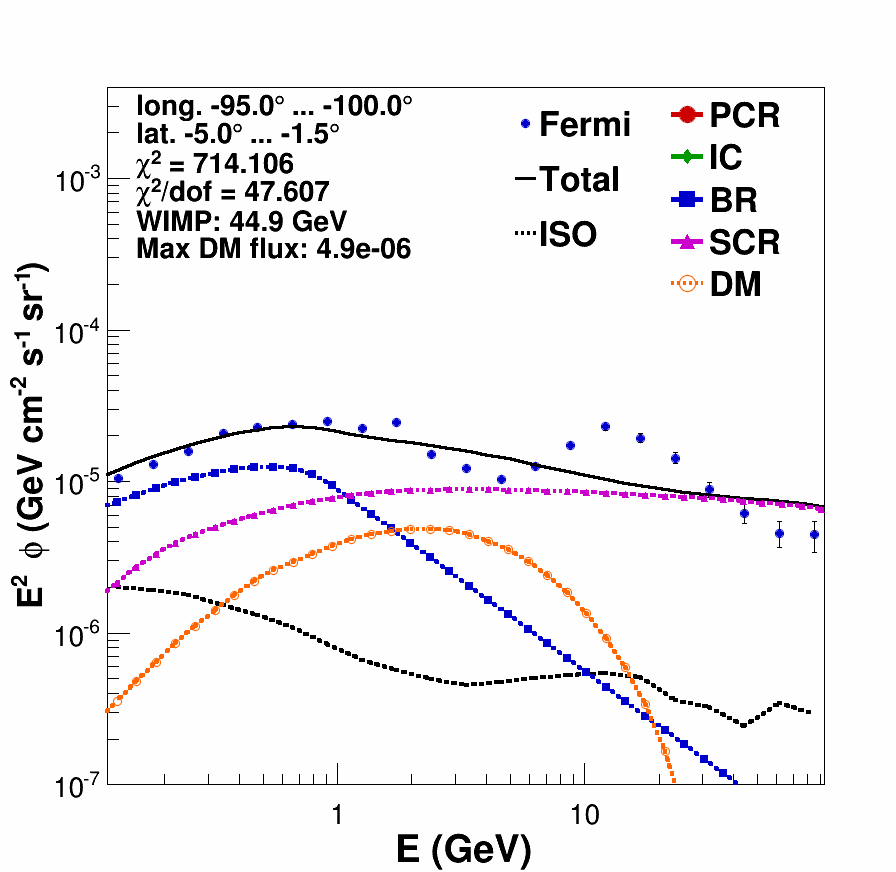}
\includegraphics[width=0.16\textwidth,height=0.16\textwidth,clip]{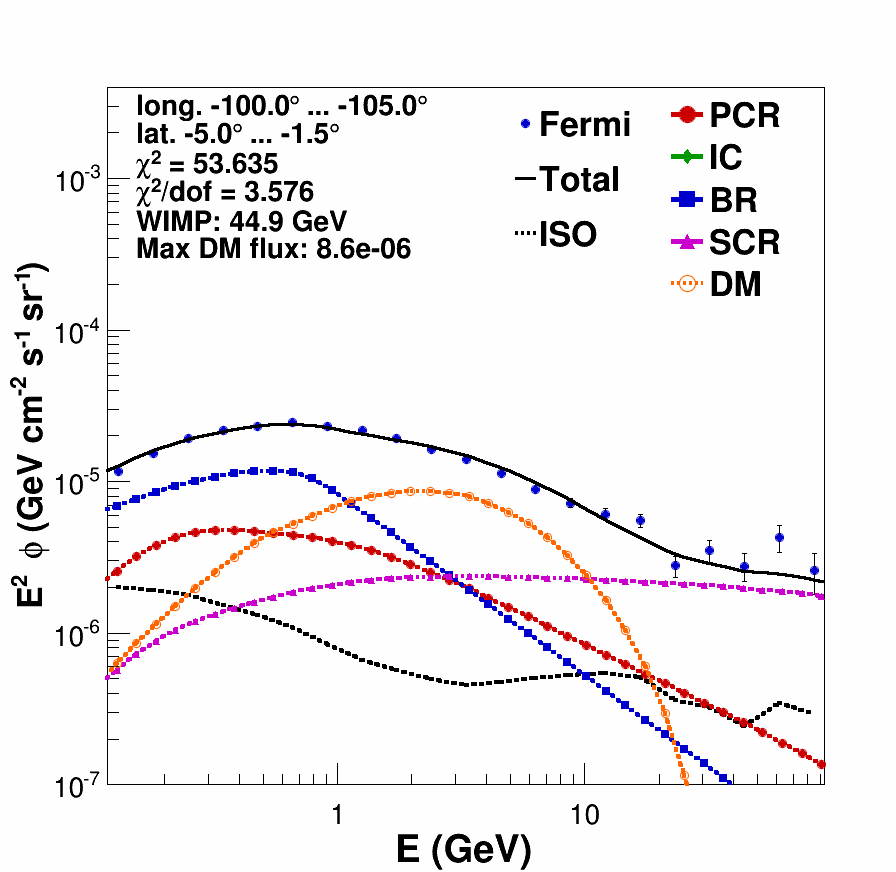}
\includegraphics[width=0.16\textwidth,height=0.16\textwidth,clip]{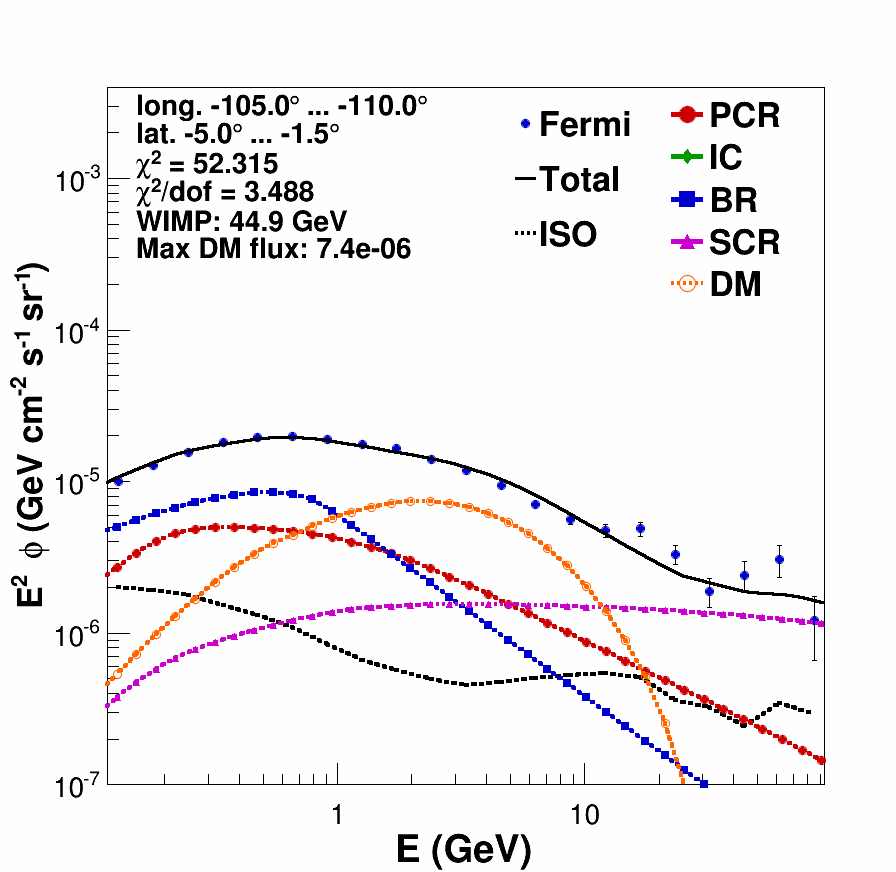}
\includegraphics[width=0.16\textwidth,height=0.16\textwidth,clip]{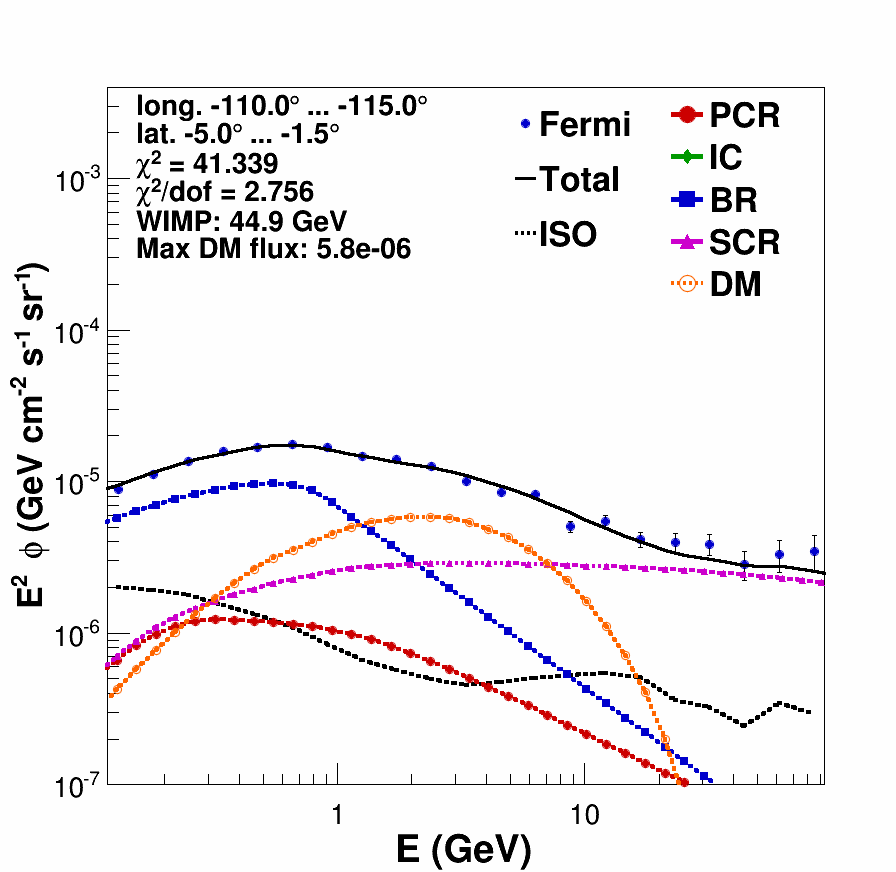}
\includegraphics[width=0.16\textwidth,height=0.16\textwidth,clip]{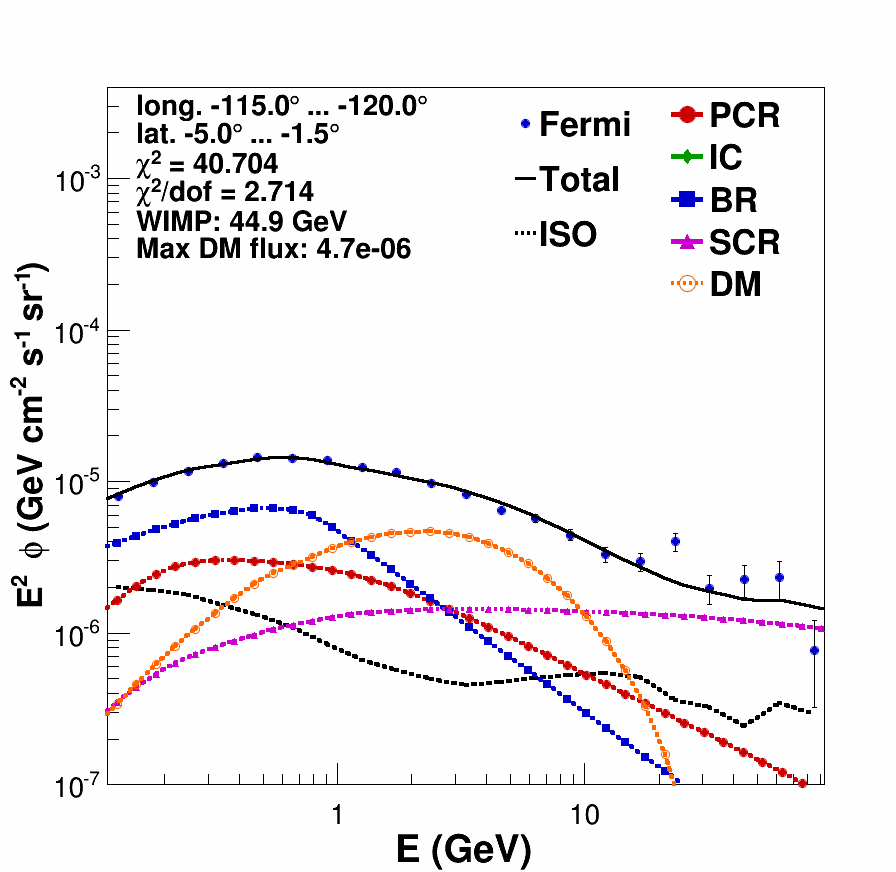}
\includegraphics[width=0.16\textwidth,height=0.16\textwidth,clip]{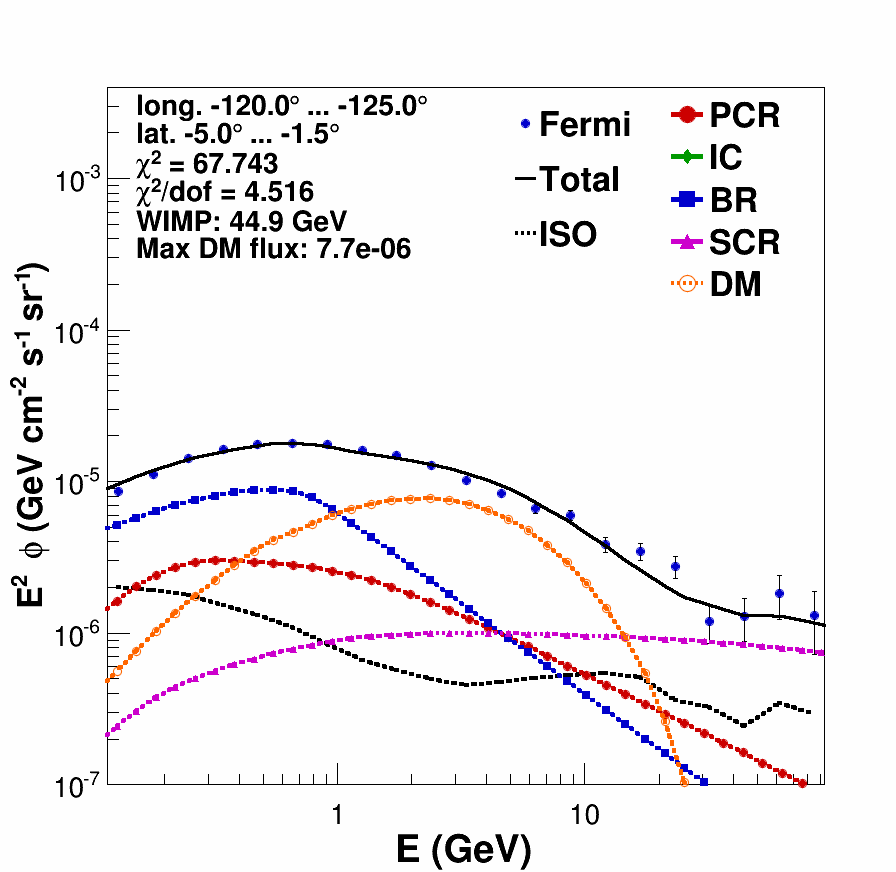}
\includegraphics[width=0.16\textwidth,height=0.16\textwidth,clip]{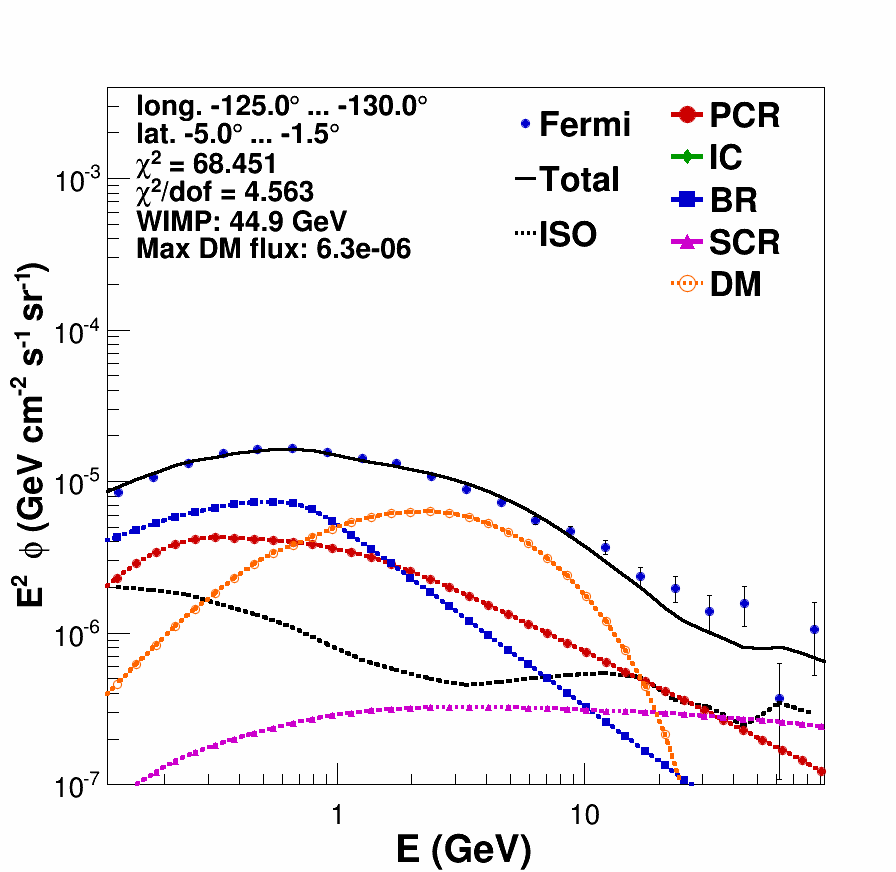}
\includegraphics[width=0.16\textwidth,height=0.16\textwidth,clip]{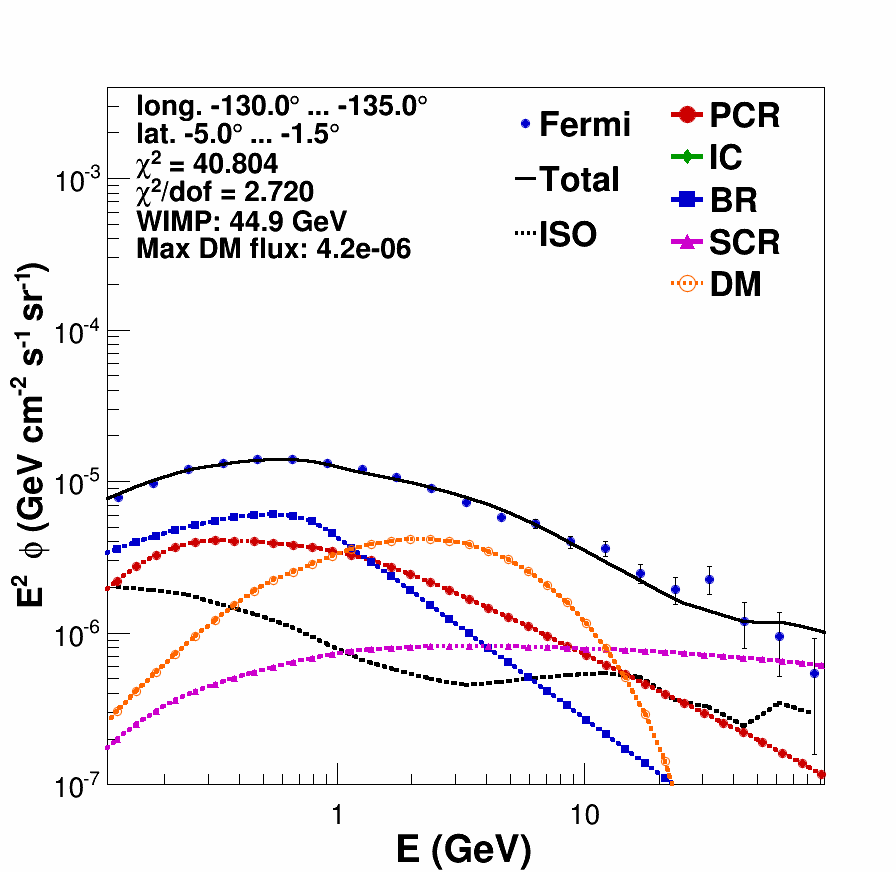}
\includegraphics[width=0.16\textwidth,height=0.16\textwidth,clip]{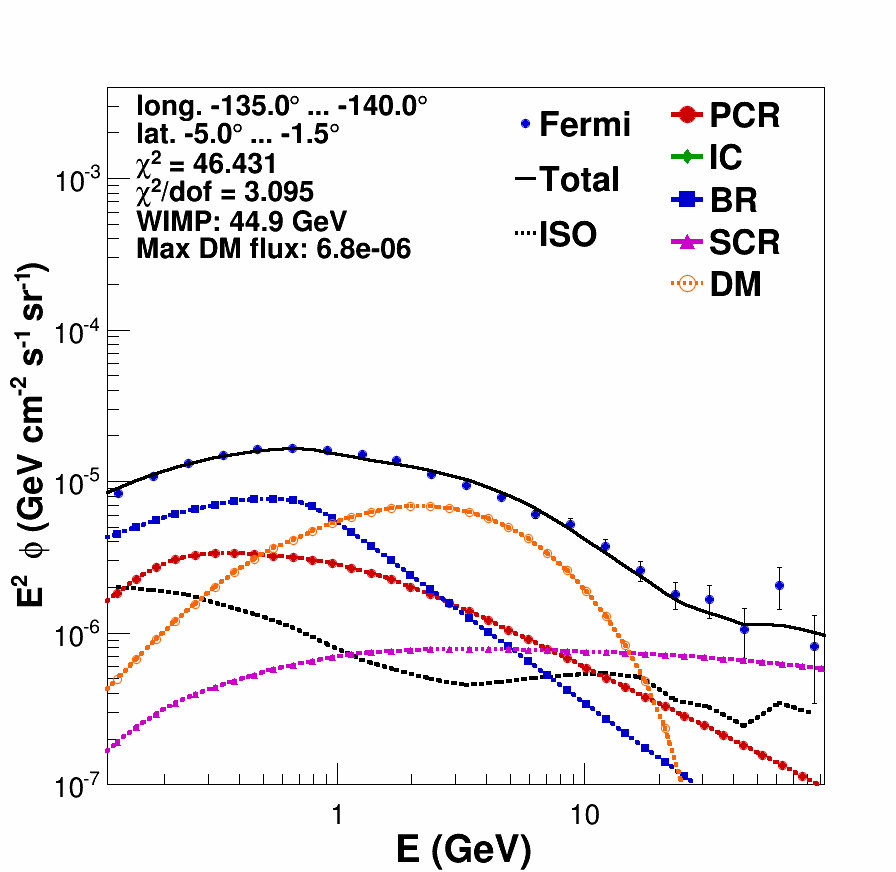}
\includegraphics[width=0.16\textwidth,height=0.16\textwidth,clip]{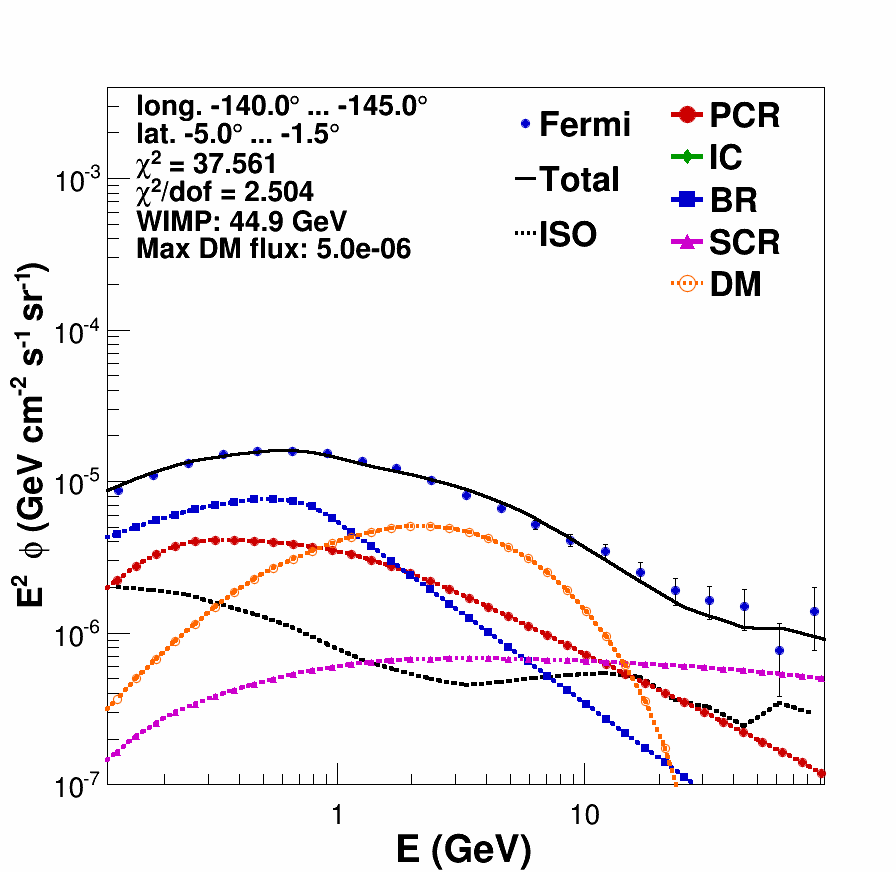}
\includegraphics[width=0.16\textwidth,height=0.16\textwidth,clip]{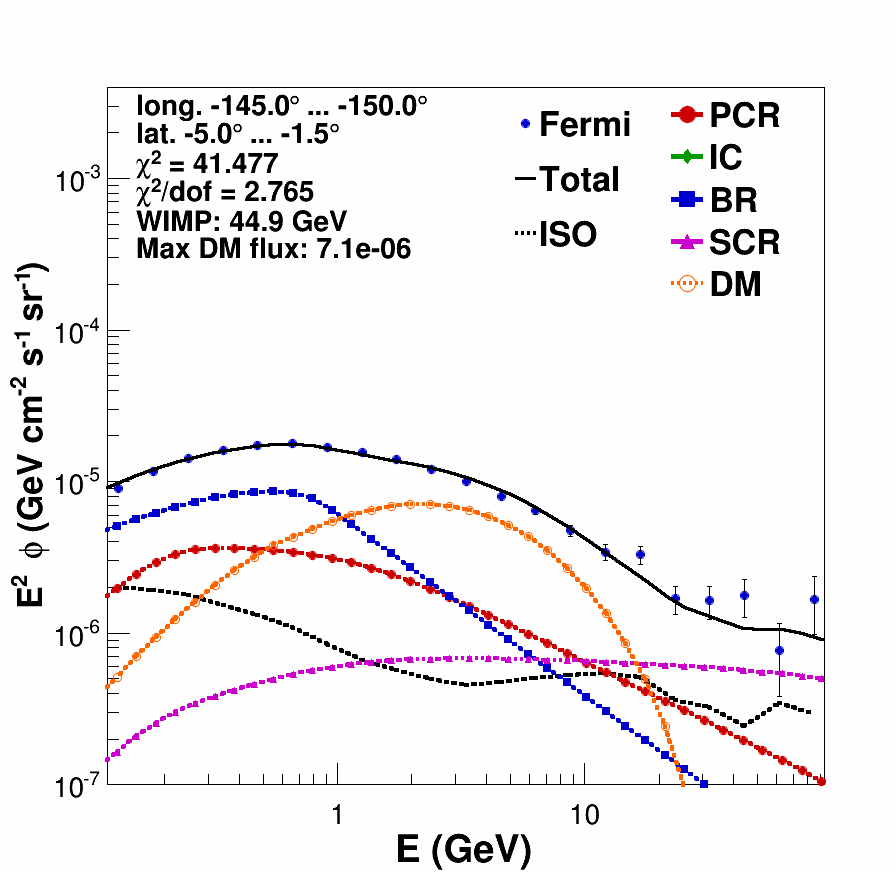}
\includegraphics[width=0.16\textwidth,height=0.16\textwidth,clip]{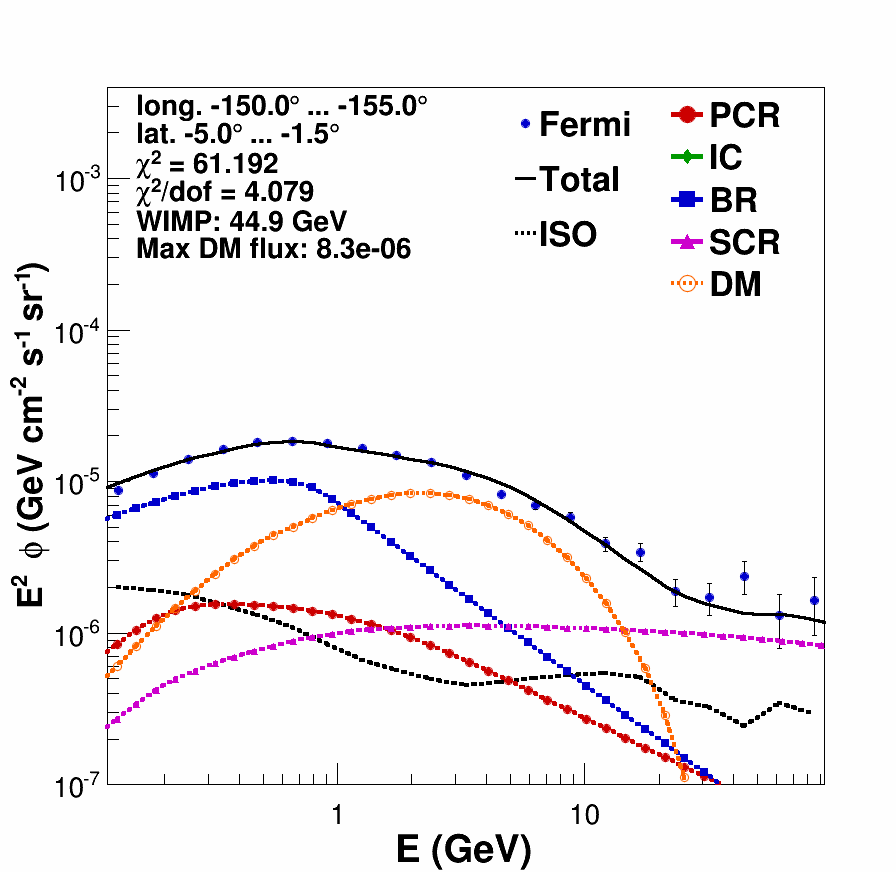}
\includegraphics[width=0.16\textwidth,height=0.16\textwidth,clip]{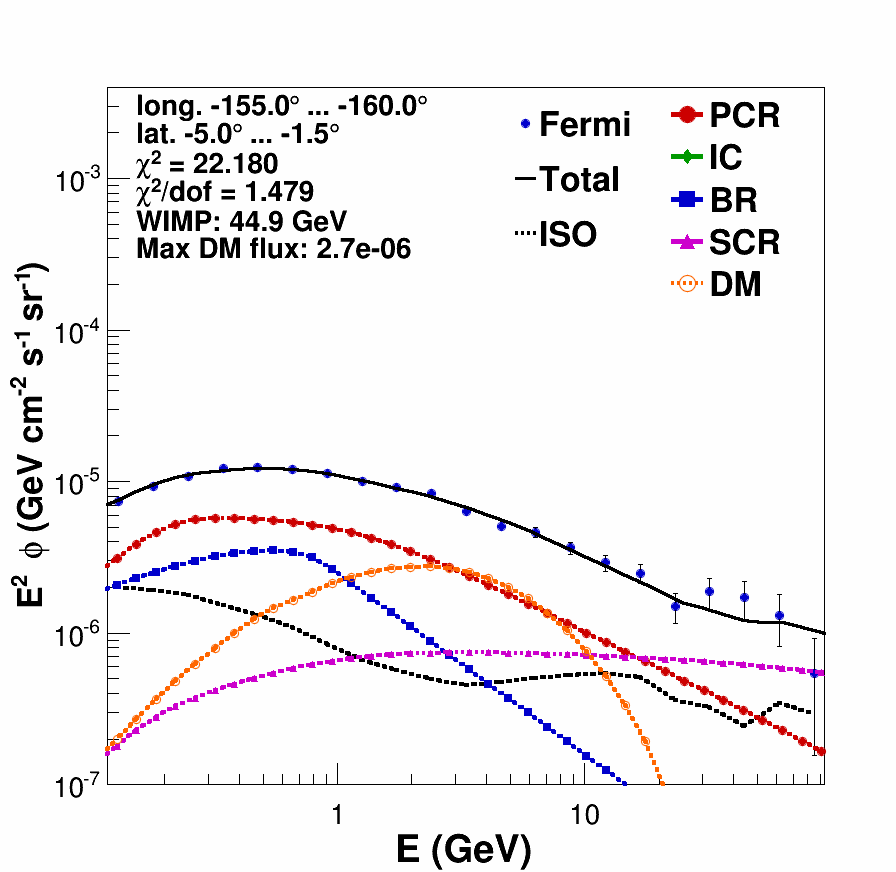}
\includegraphics[width=0.16\textwidth,height=0.16\textwidth,clip]{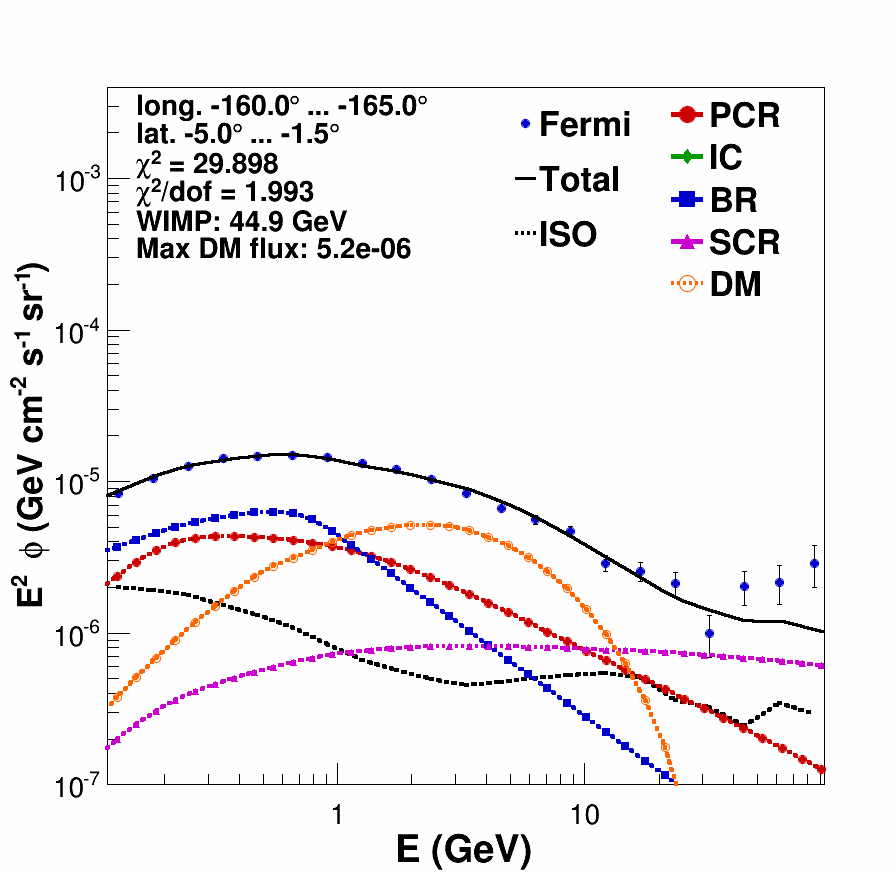}
\includegraphics[width=0.16\textwidth,height=0.16\textwidth,clip]{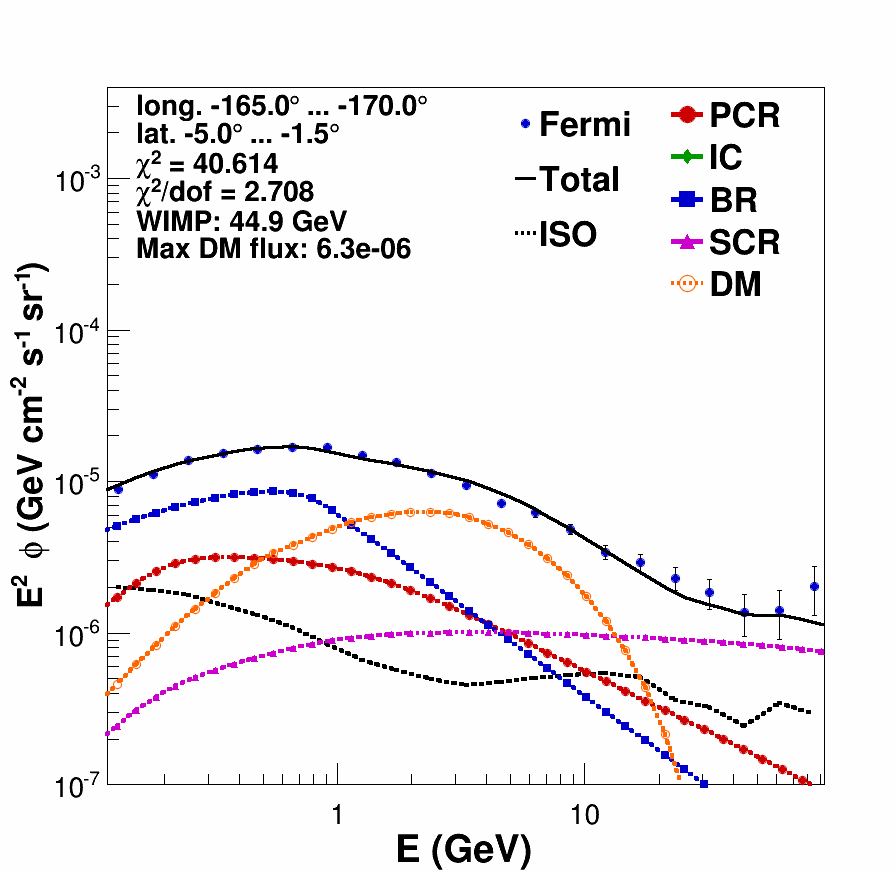}
\includegraphics[width=0.16\textwidth,height=0.16\textwidth,clip]{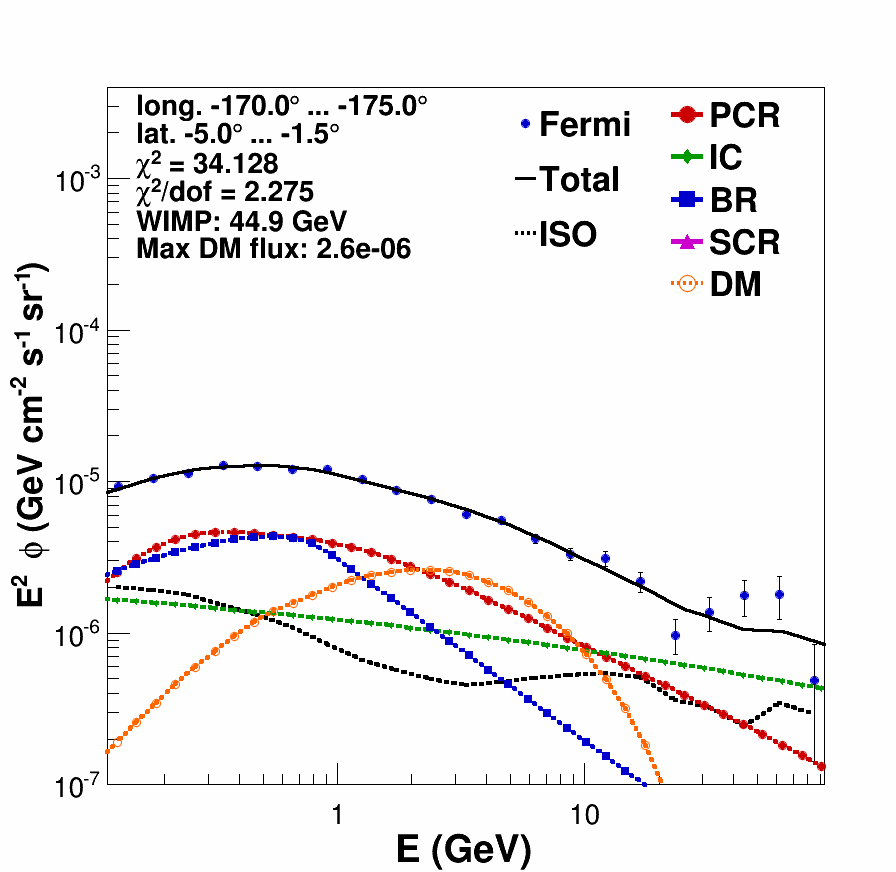}
\includegraphics[width=0.16\textwidth,height=0.16\textwidth,clip]{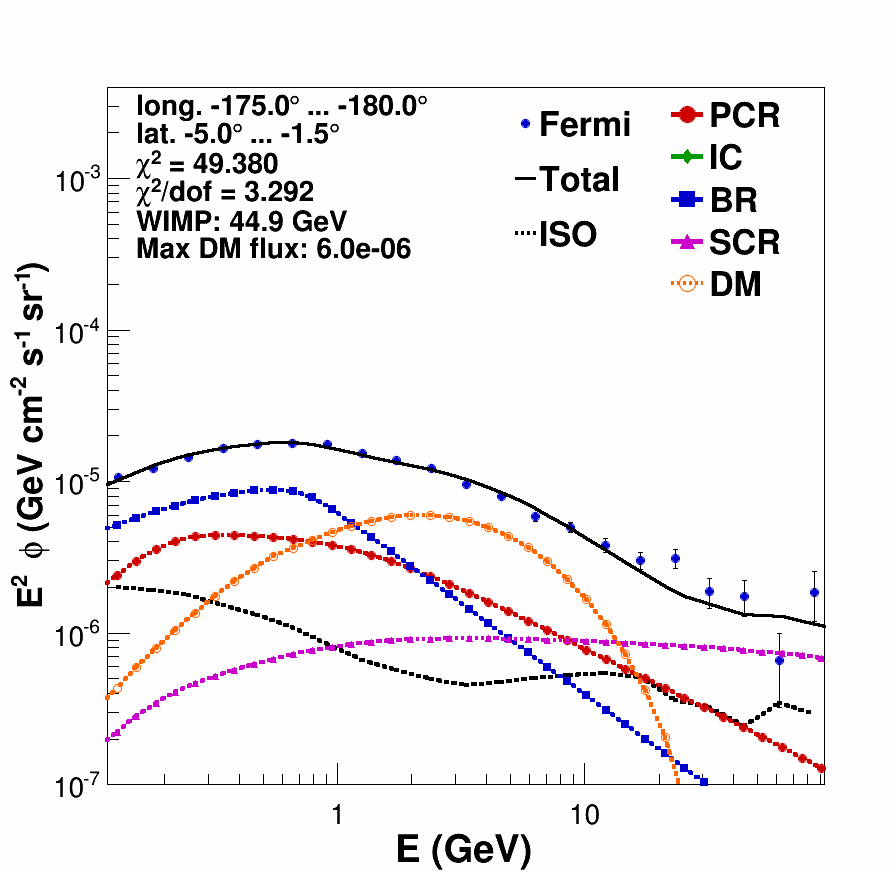}%%%%%r12b
\caption[]{Template fits for latitudes  with $-5.0^\circ<b<-1.5^\circ$ and longitudes decreasing from 0$^\circ$ to -180$^\circ$.} \label{F45}
\end{figure}
\begin{figure}
\centering
\includegraphics[width=0.16\textwidth,height=0.16\textwidth,clip]{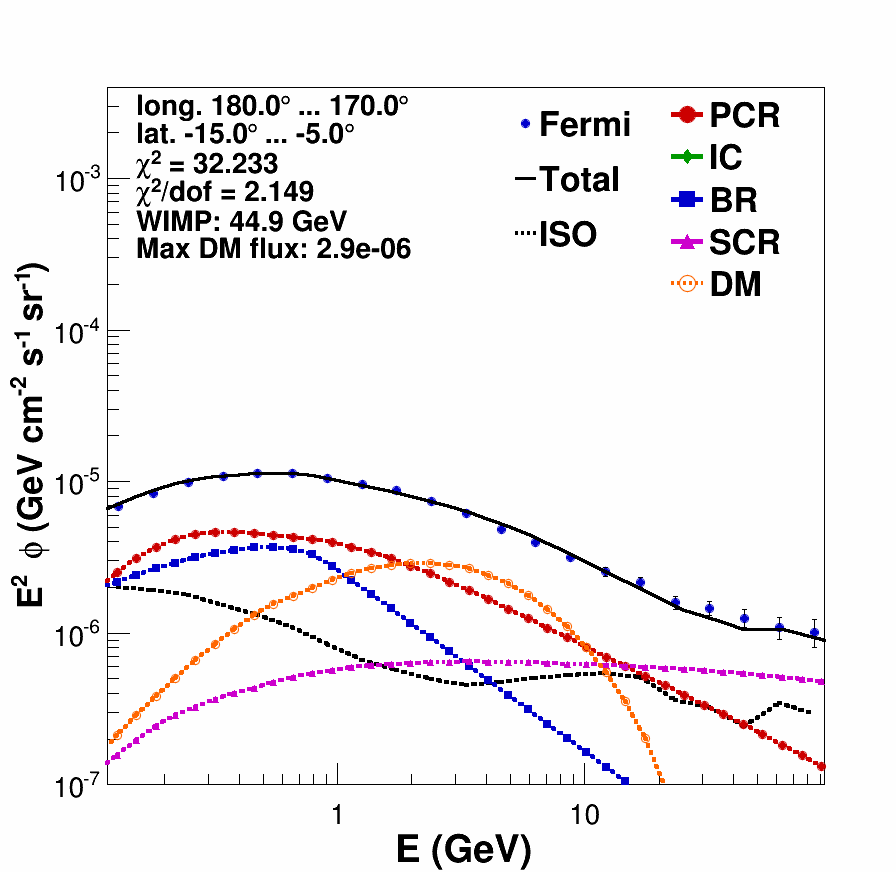}
\includegraphics[width=0.16\textwidth,height=0.16\textwidth,clip]{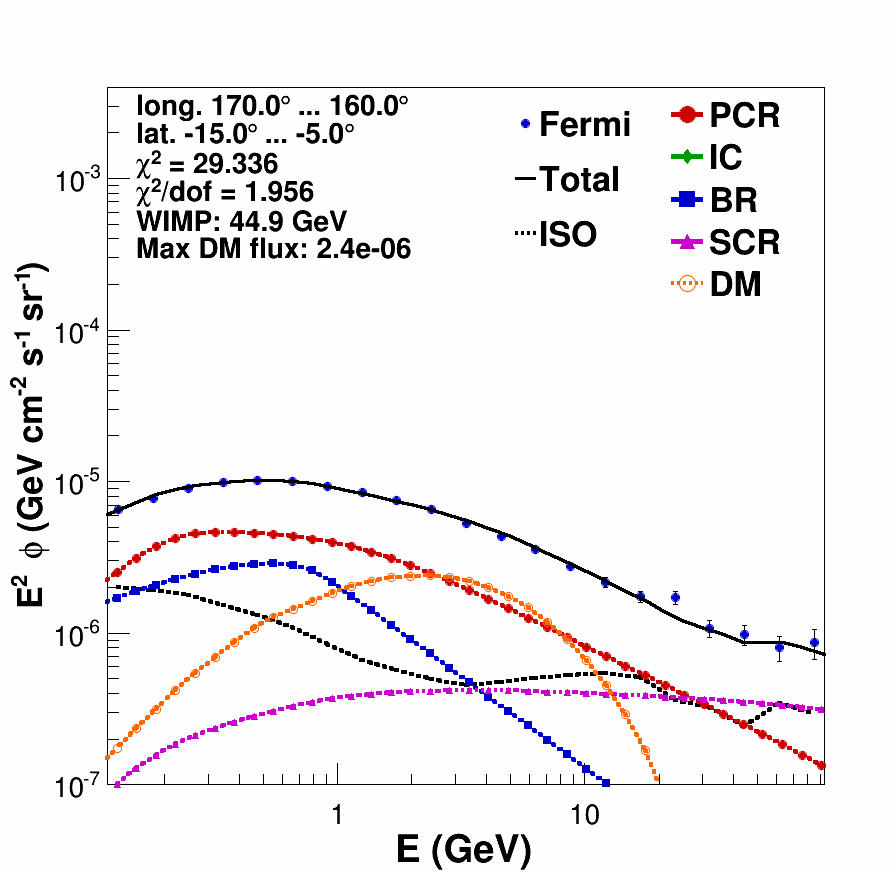}
\includegraphics[width=0.16\textwidth,height=0.16\textwidth,clip]{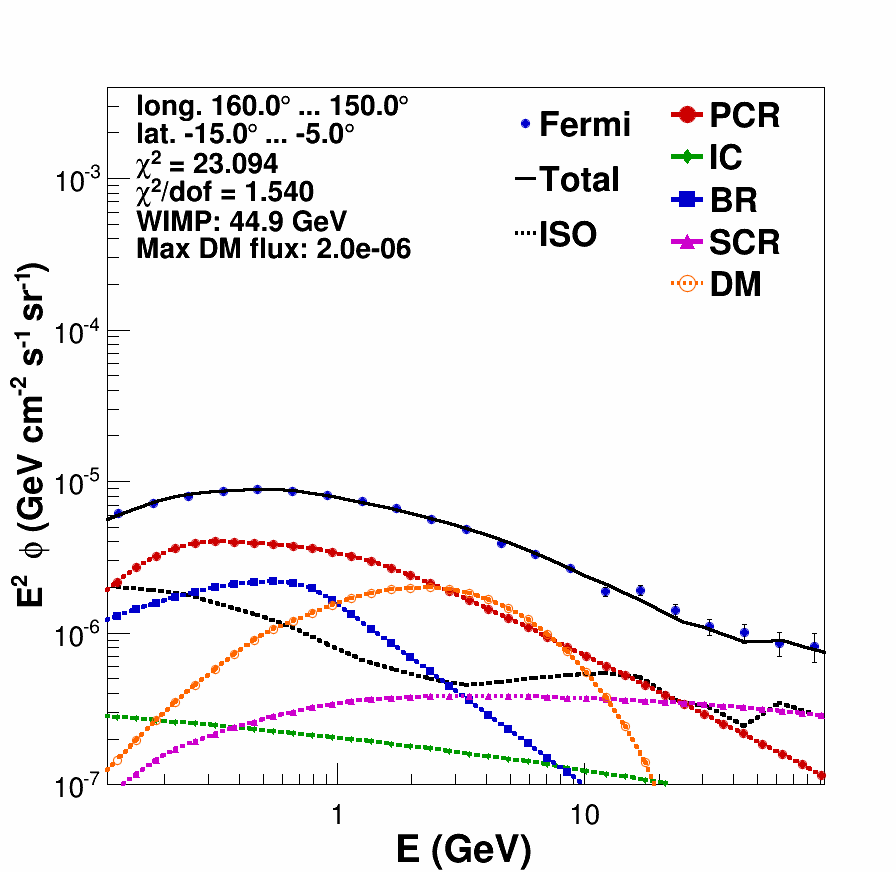}
\includegraphics[width=0.16\textwidth,height=0.16\textwidth,clip]{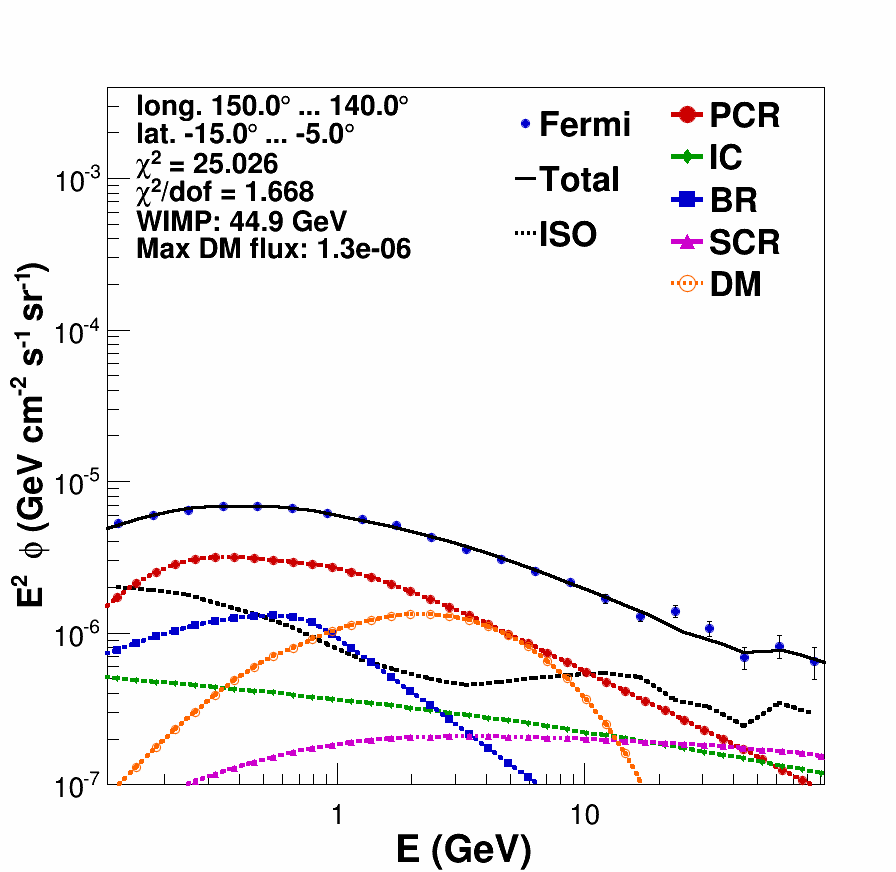}
\includegraphics[width=0.16\textwidth,height=0.16\textwidth,clip]{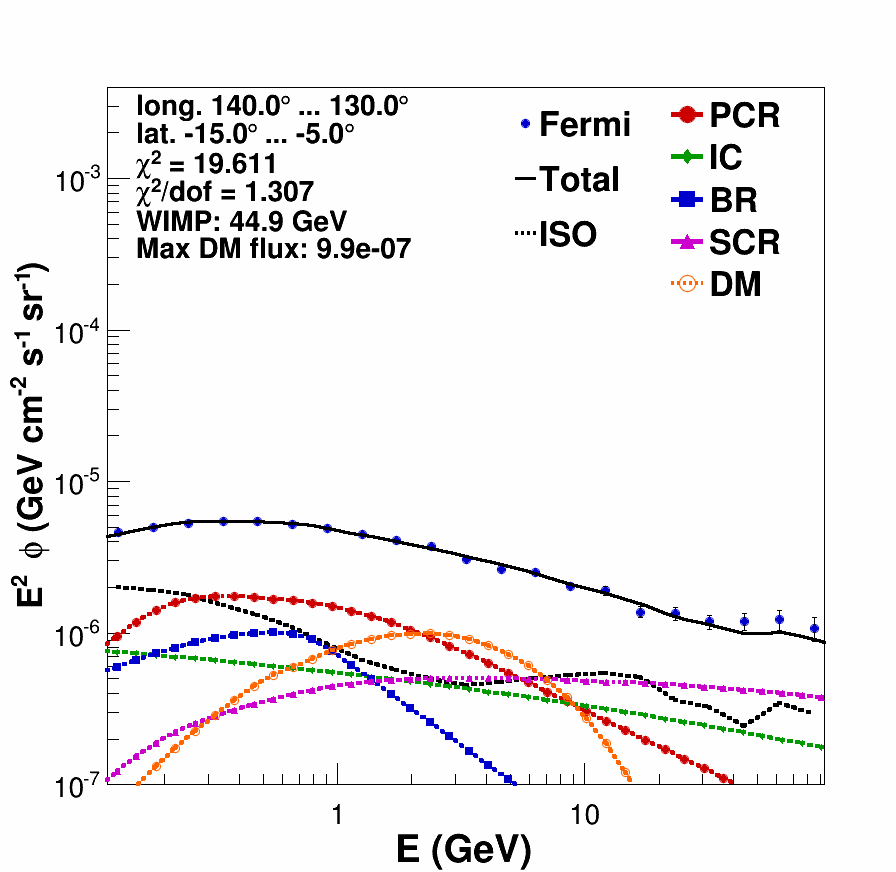}
\includegraphics[width=0.16\textwidth,height=0.16\textwidth,clip]{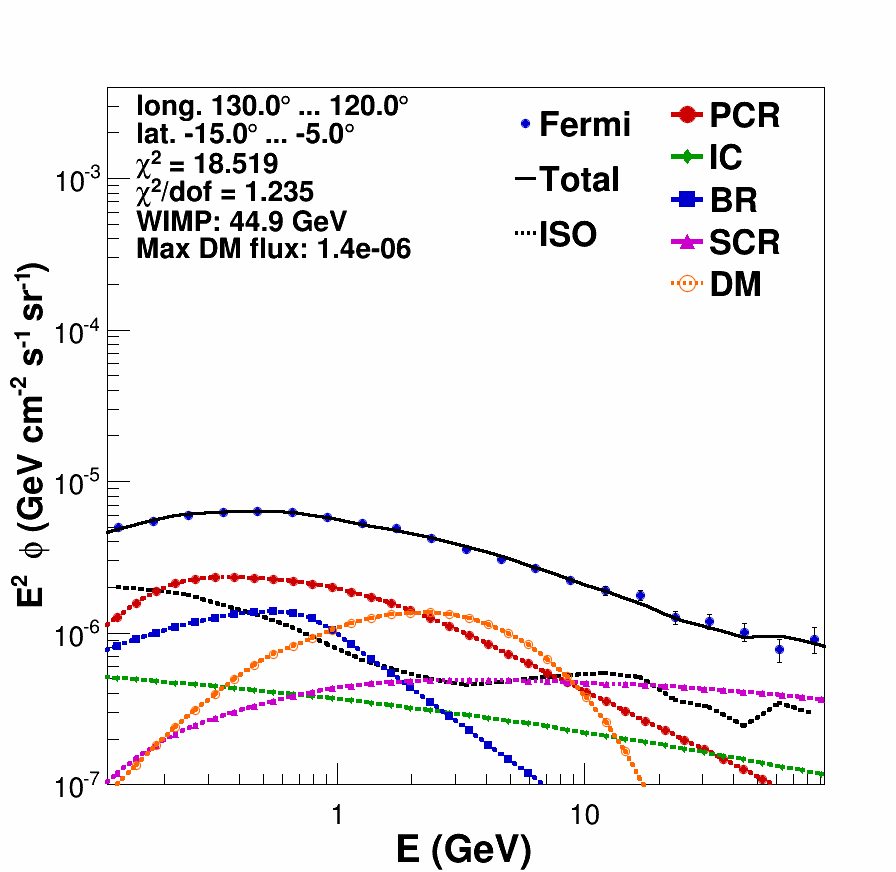}
\includegraphics[width=0.16\textwidth,height=0.16\textwidth,clip]{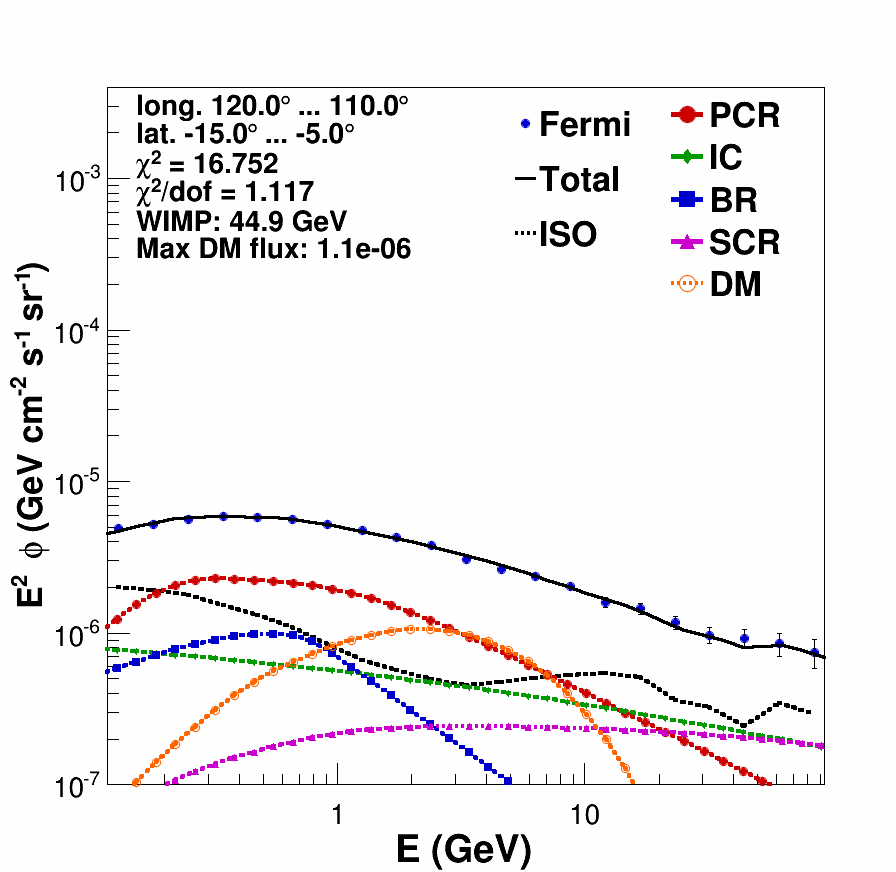}
\includegraphics[width=0.16\textwidth,height=0.16\textwidth,clip]{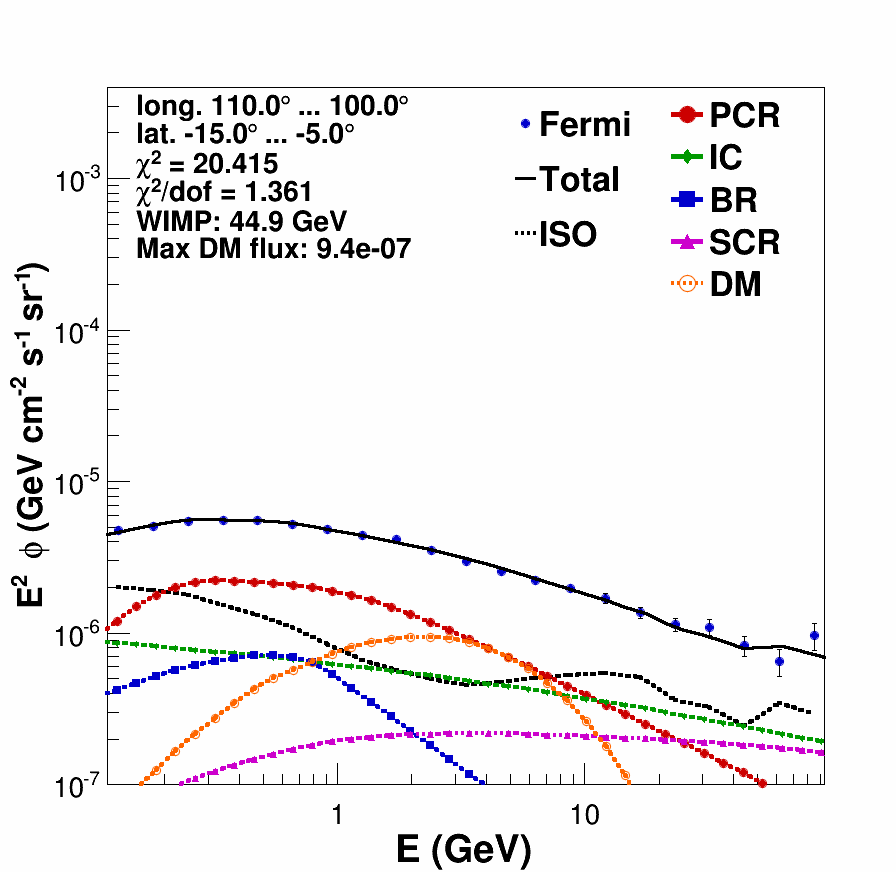}
\includegraphics[width=0.16\textwidth,height=0.16\textwidth,clip]{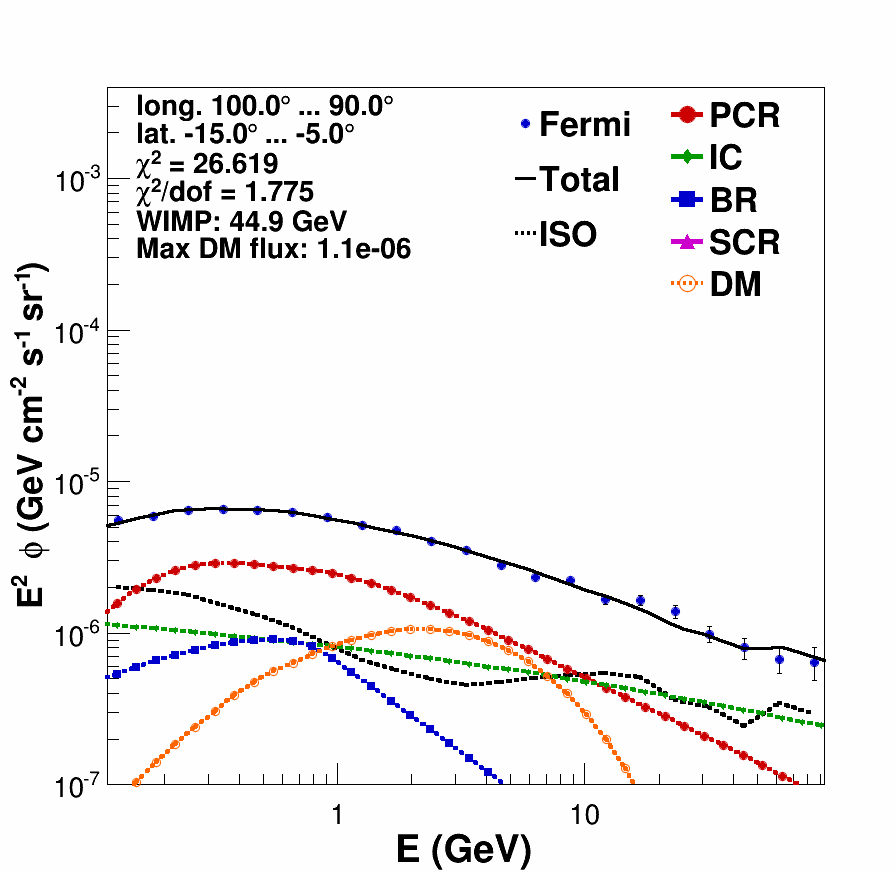}
\includegraphics[width=0.16\textwidth,height=0.16\textwidth,clip]{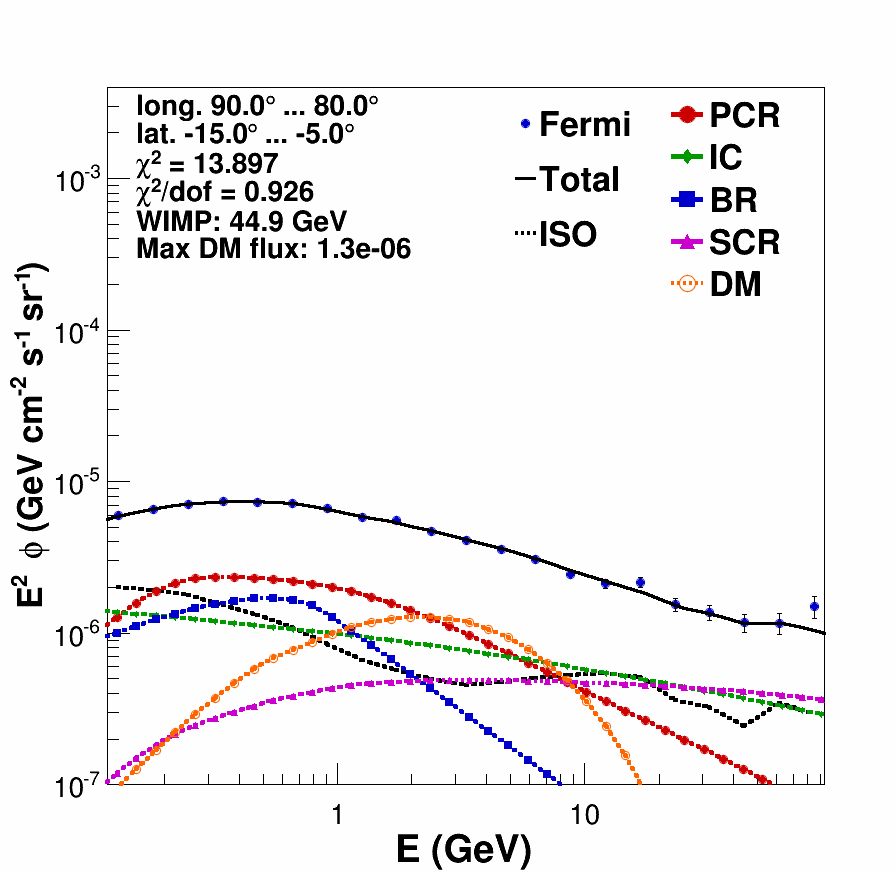}
\includegraphics[width=0.16\textwidth,height=0.16\textwidth,clip]{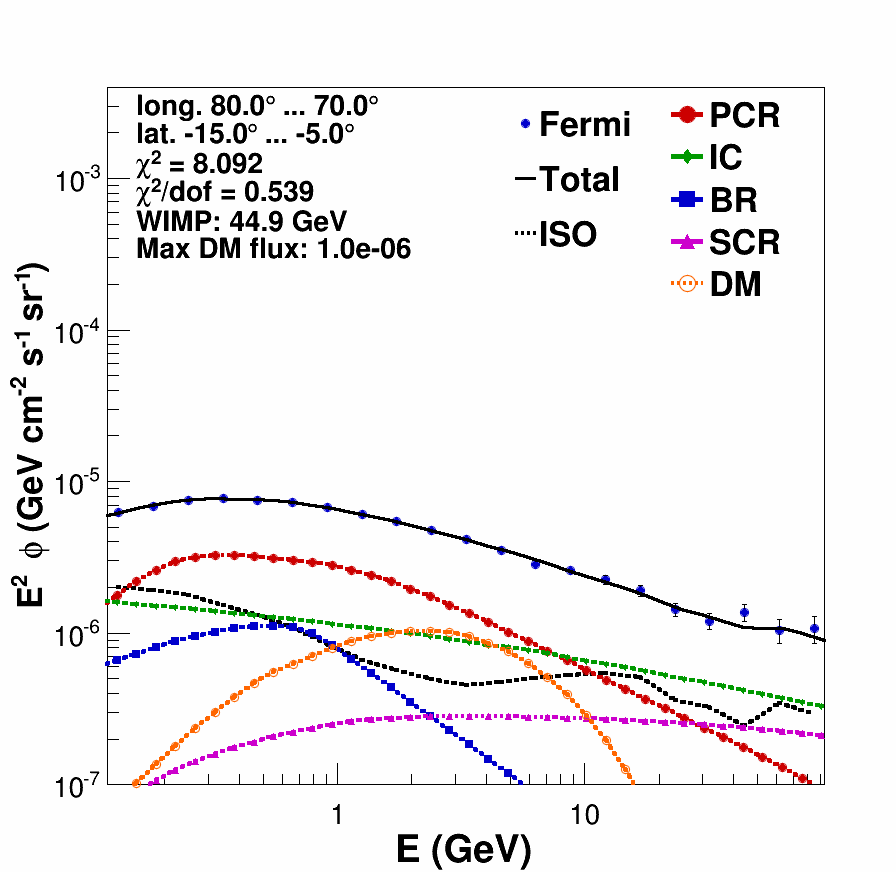}
\includegraphics[width=0.16\textwidth,height=0.16\textwidth,clip]{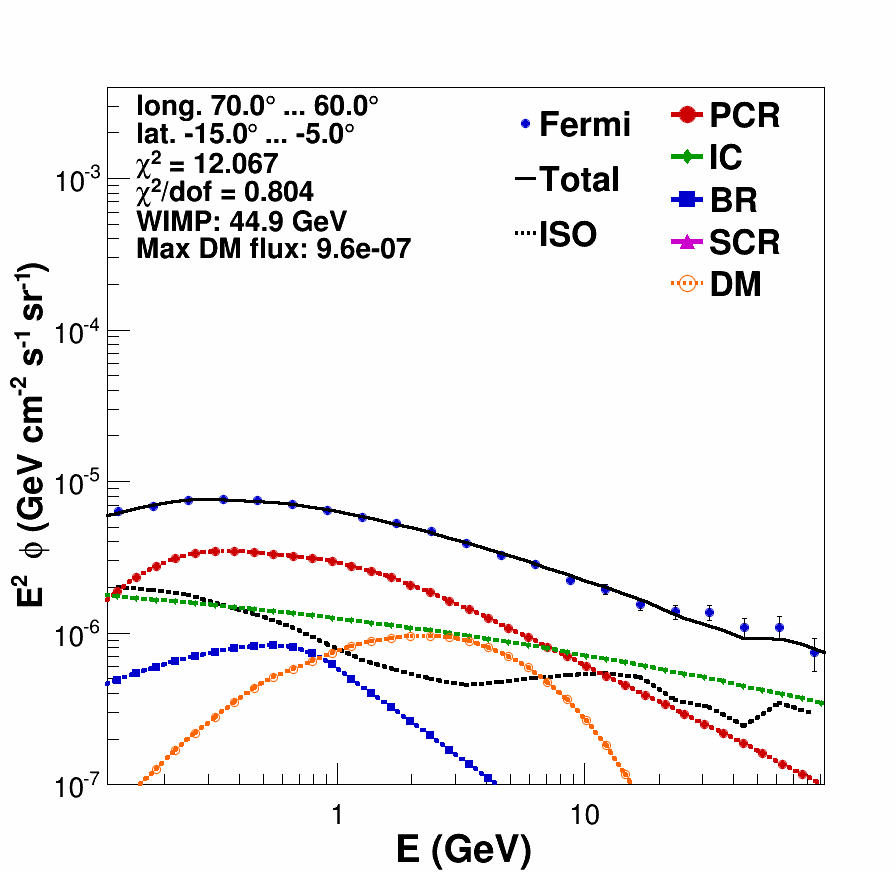}
\includegraphics[width=0.16\textwidth,height=0.16\textwidth,clip]{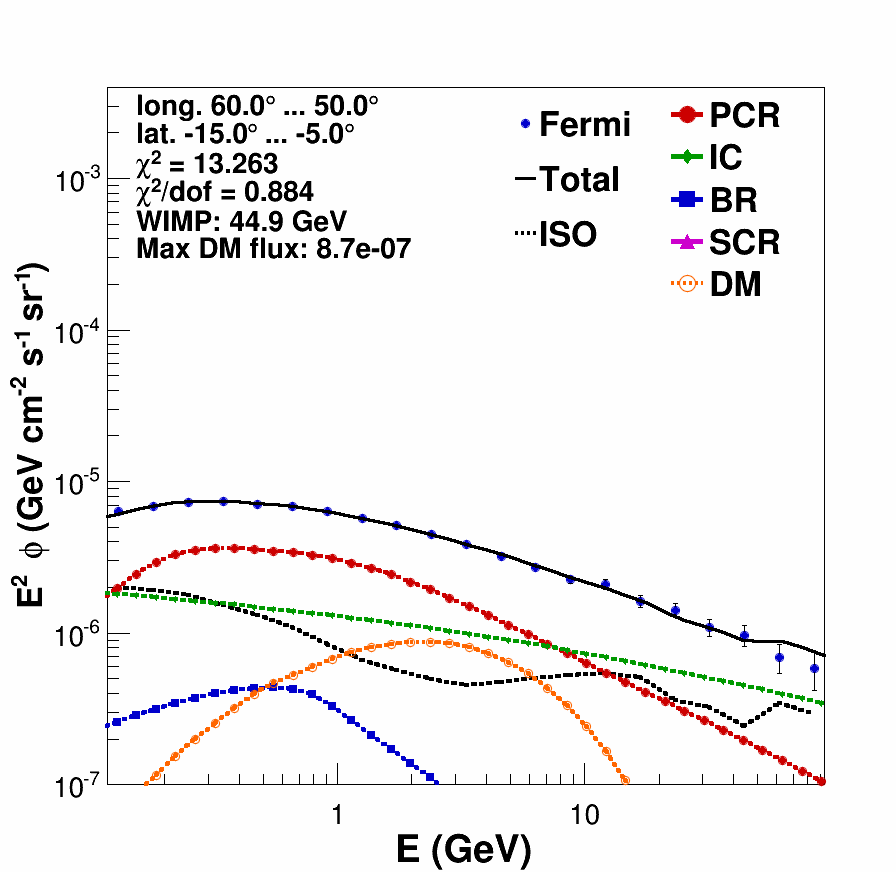}
\includegraphics[width=0.16\textwidth,height=0.16\textwidth,clip]{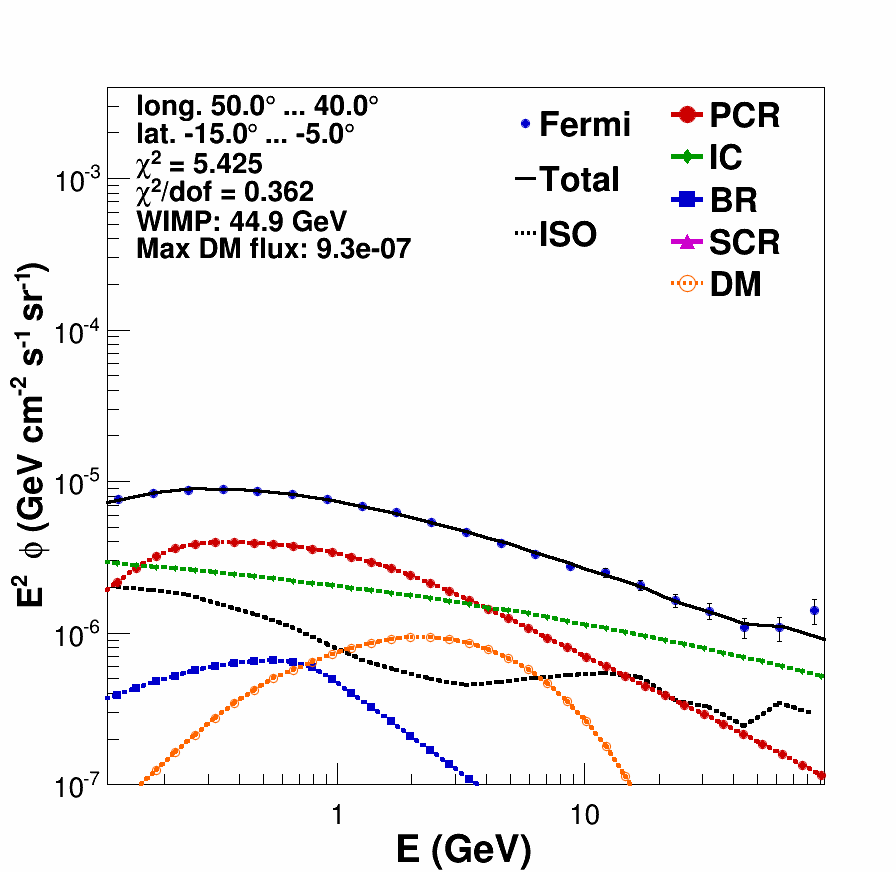}
\includegraphics[width=0.16\textwidth,height=0.16\textwidth,clip]{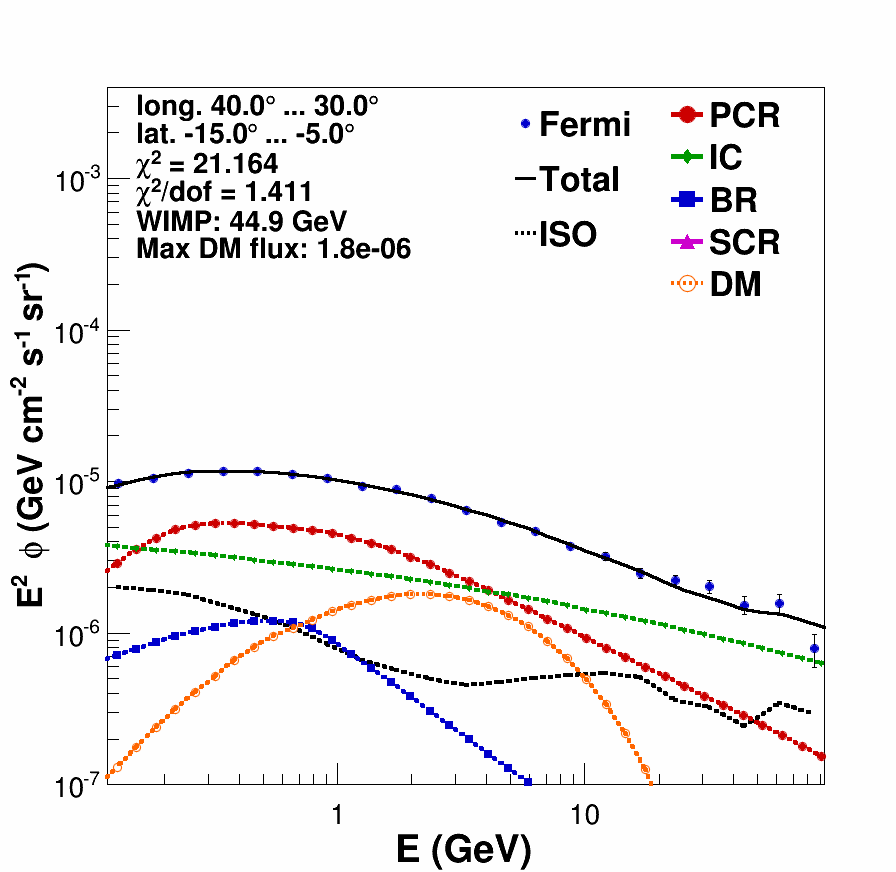}
\includegraphics[width=0.16\textwidth,height=0.16\textwidth,clip]{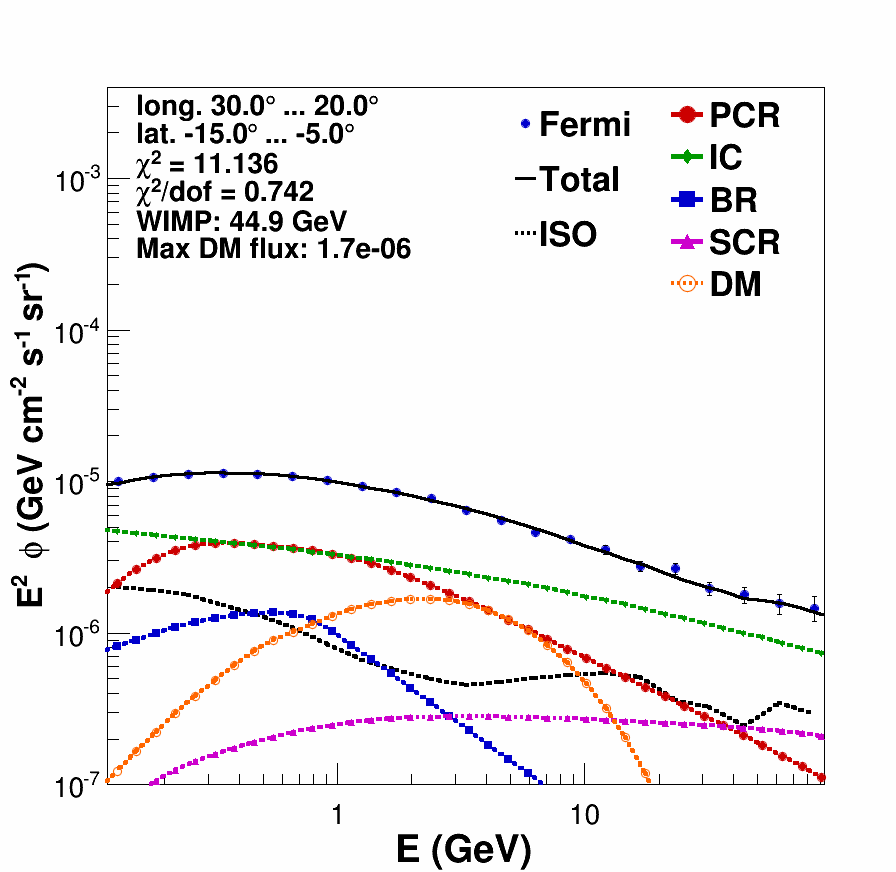}
\includegraphics[width=0.16\textwidth,height=0.16\textwidth,clip]{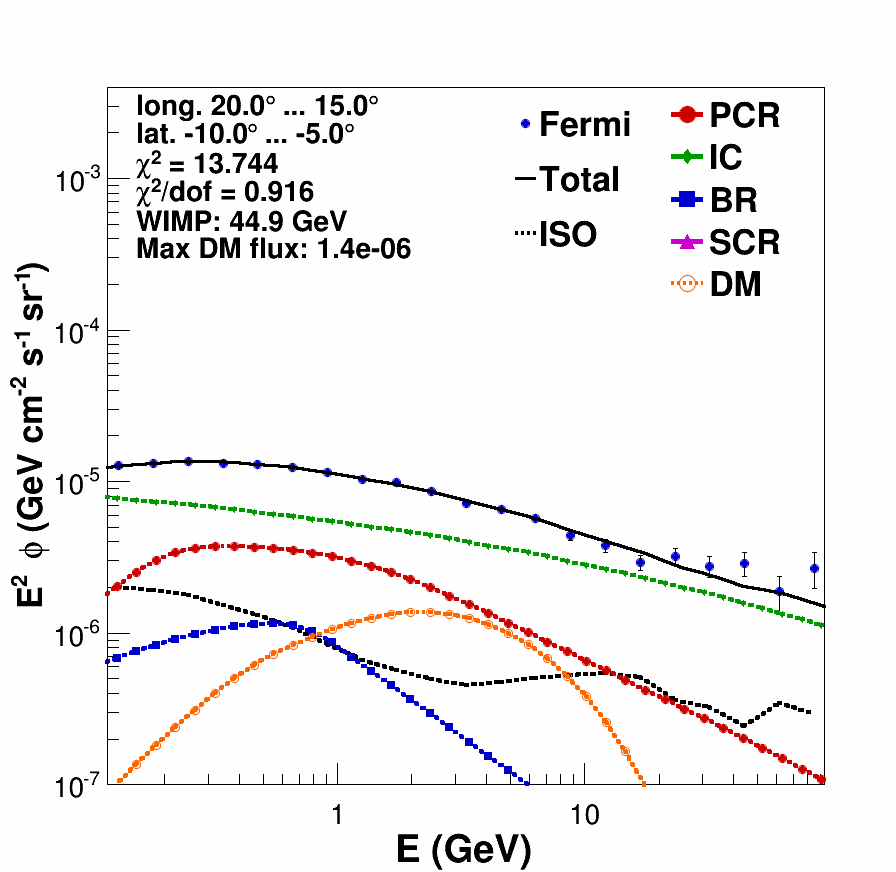}
\includegraphics[width=0.16\textwidth,height=0.16\textwidth,clip]{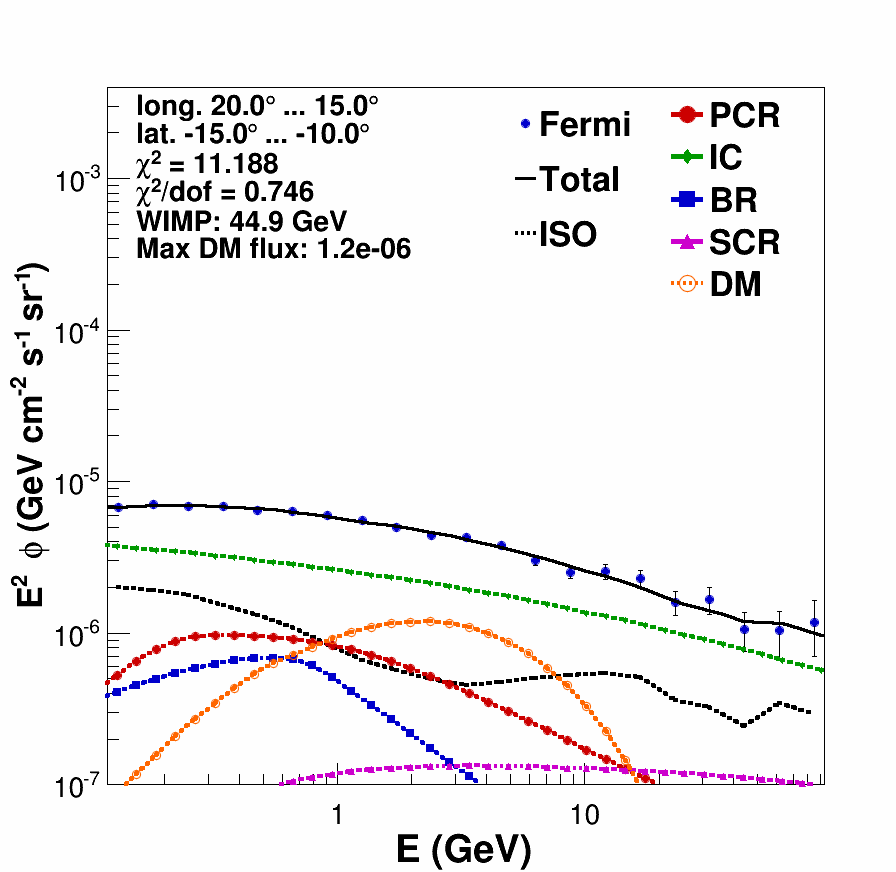}
\includegraphics[width=0.16\textwidth,height=0.16\textwidth,clip]{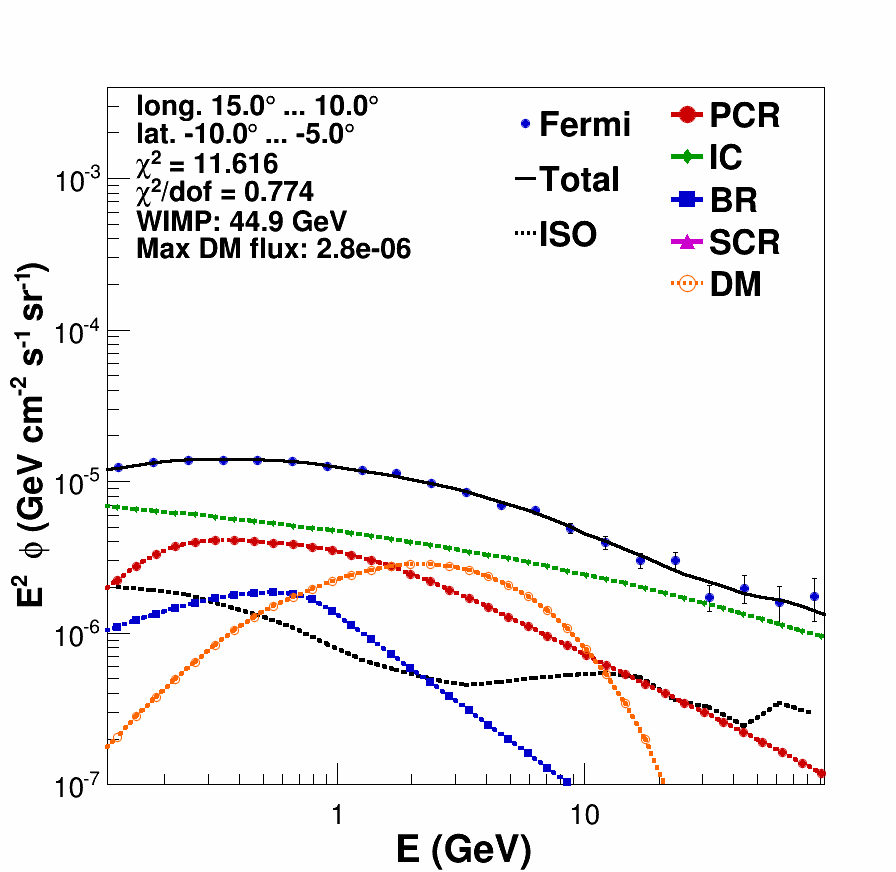}
\includegraphics[width=0.16\textwidth,height=0.16\textwidth,clip]{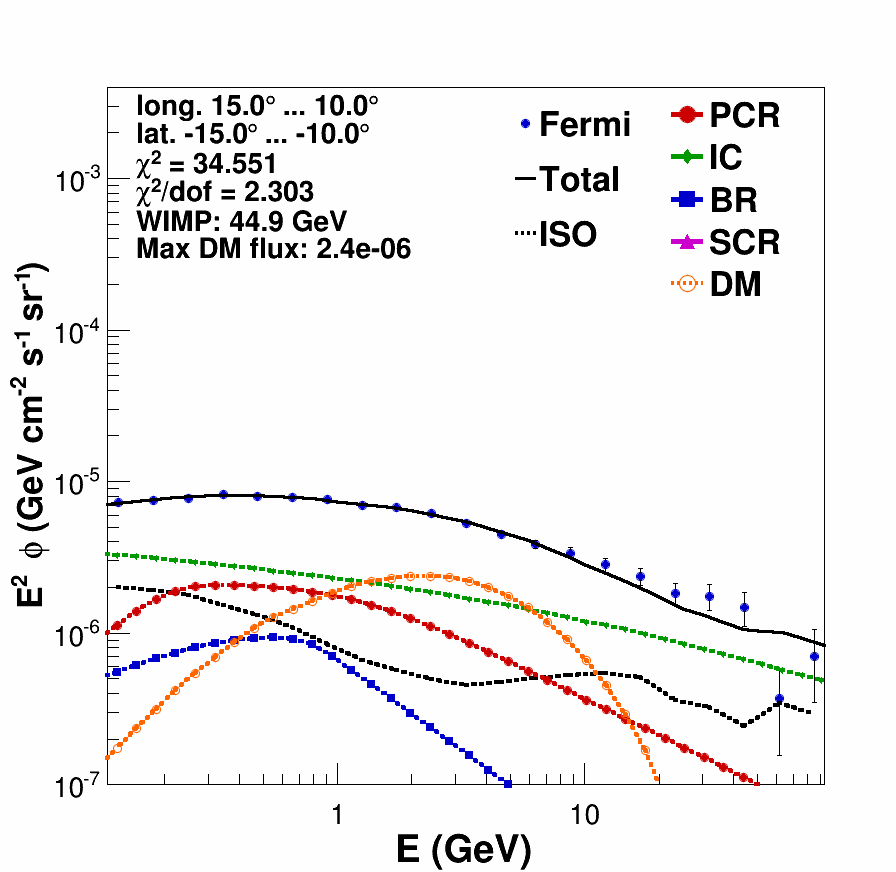}
\includegraphics[width=0.16\textwidth,height=0.16\textwidth,clip]{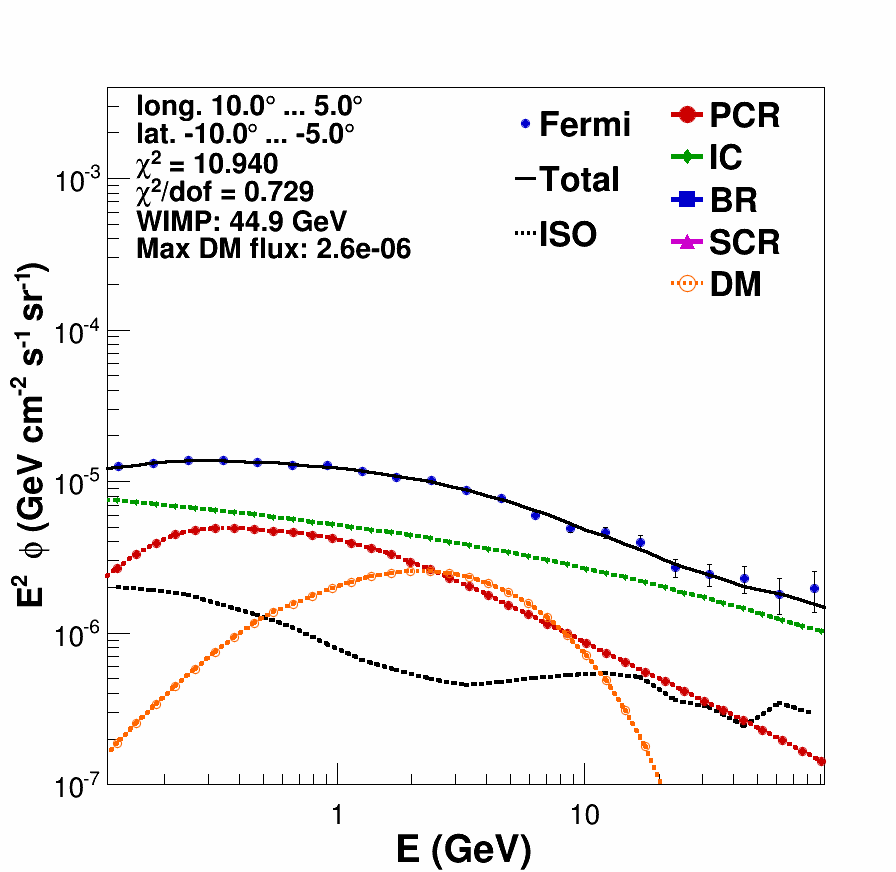}
\includegraphics[width=0.16\textwidth,height=0.16\textwidth,clip]{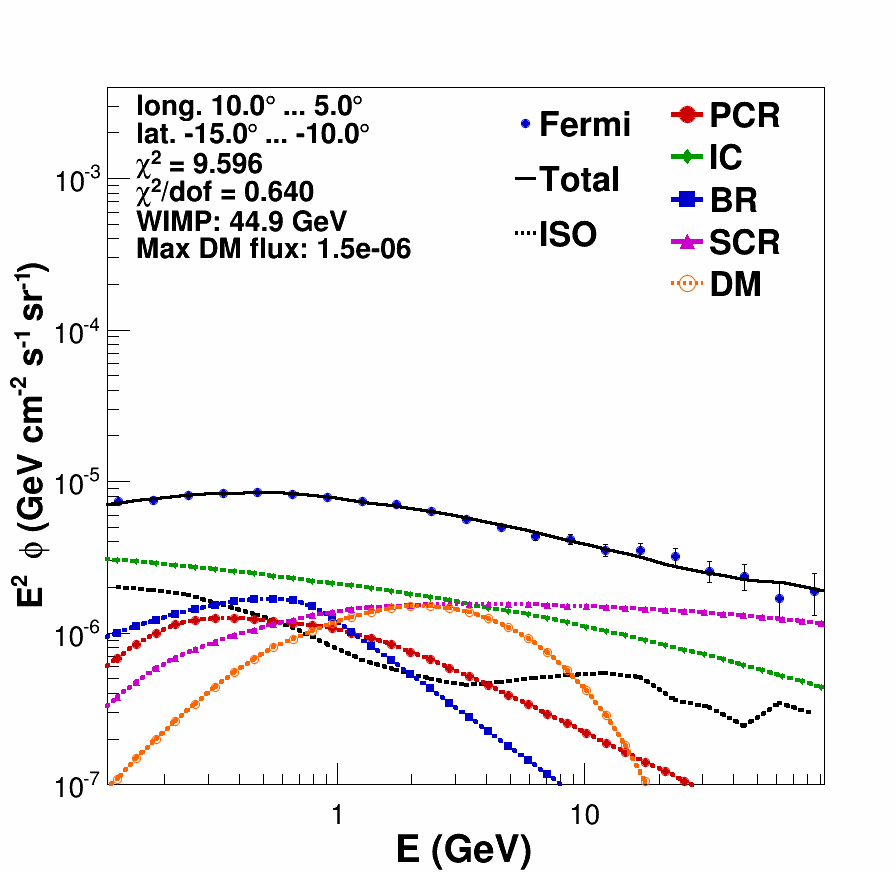}
\includegraphics[width=0.16\textwidth,height=0.16\textwidth,clip]{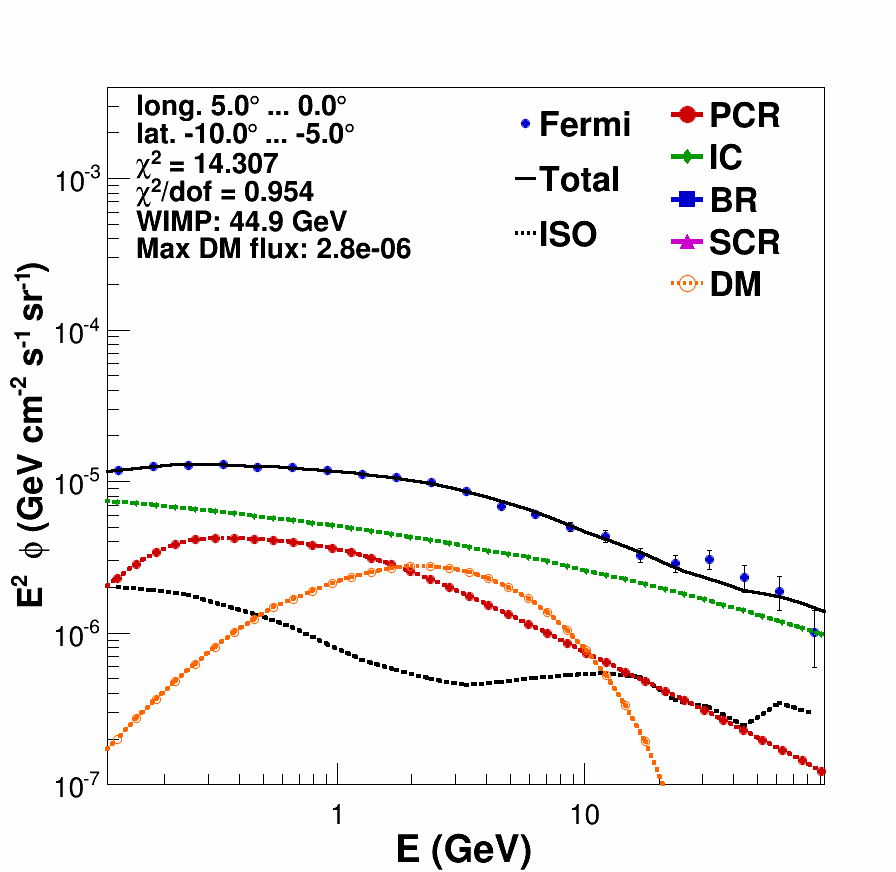}
\includegraphics[width=0.16\textwidth,height=0.16\textwidth,clip]{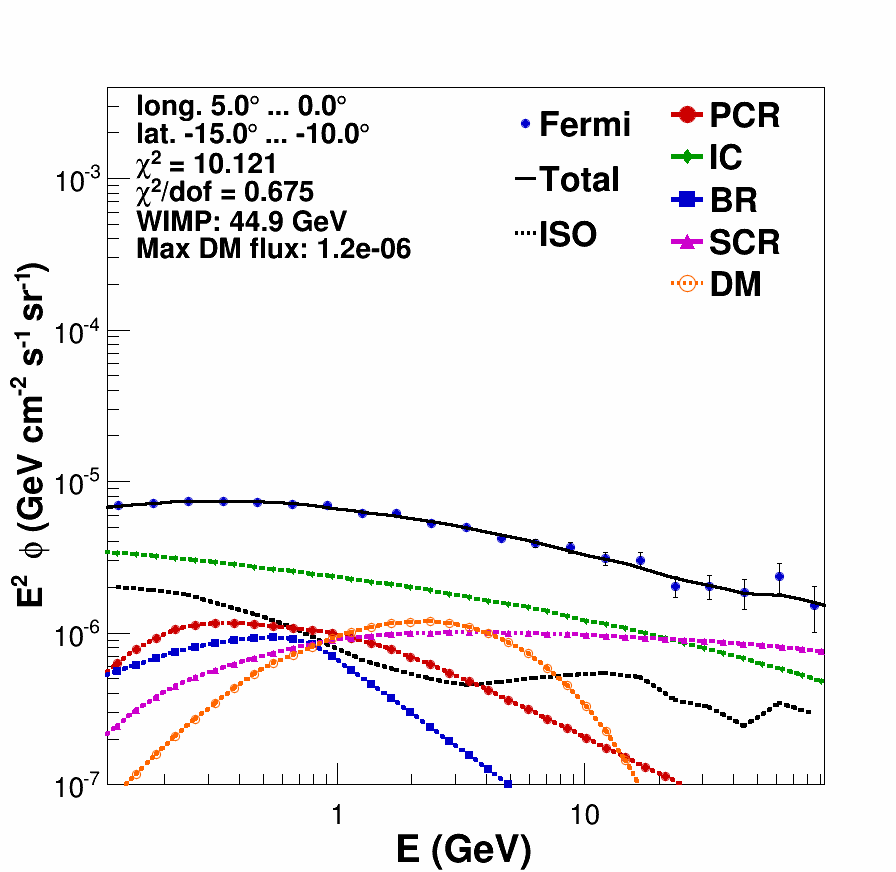}
\includegraphics[width=0.16\textwidth,height=0.16\textwidth,clip]{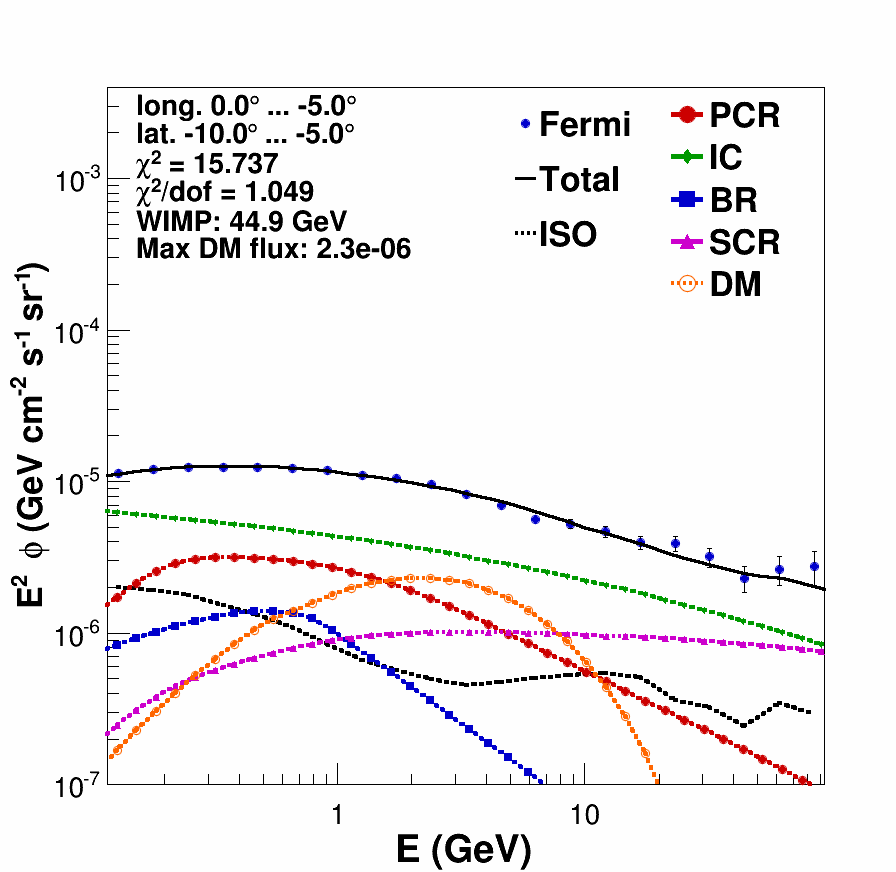}
\includegraphics[width=0.16\textwidth,height=0.16\textwidth,clip]{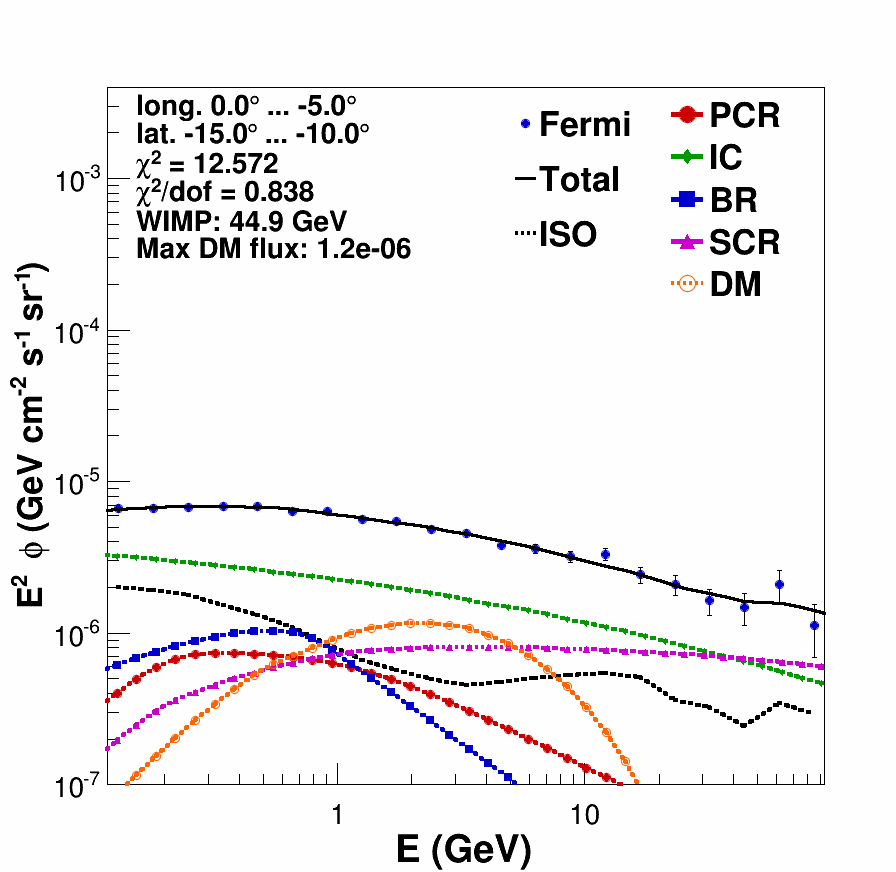}
\includegraphics[width=0.16\textwidth,height=0.16\textwidth,clip]{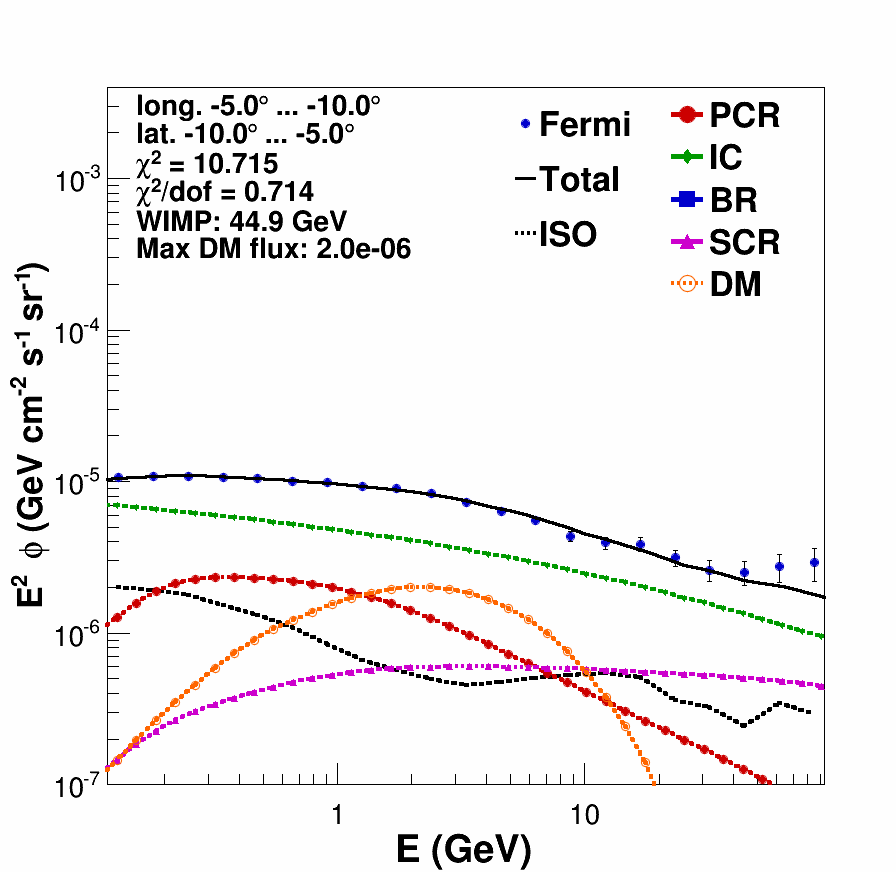}
\includegraphics[width=0.16\textwidth,height=0.16\textwidth,clip]{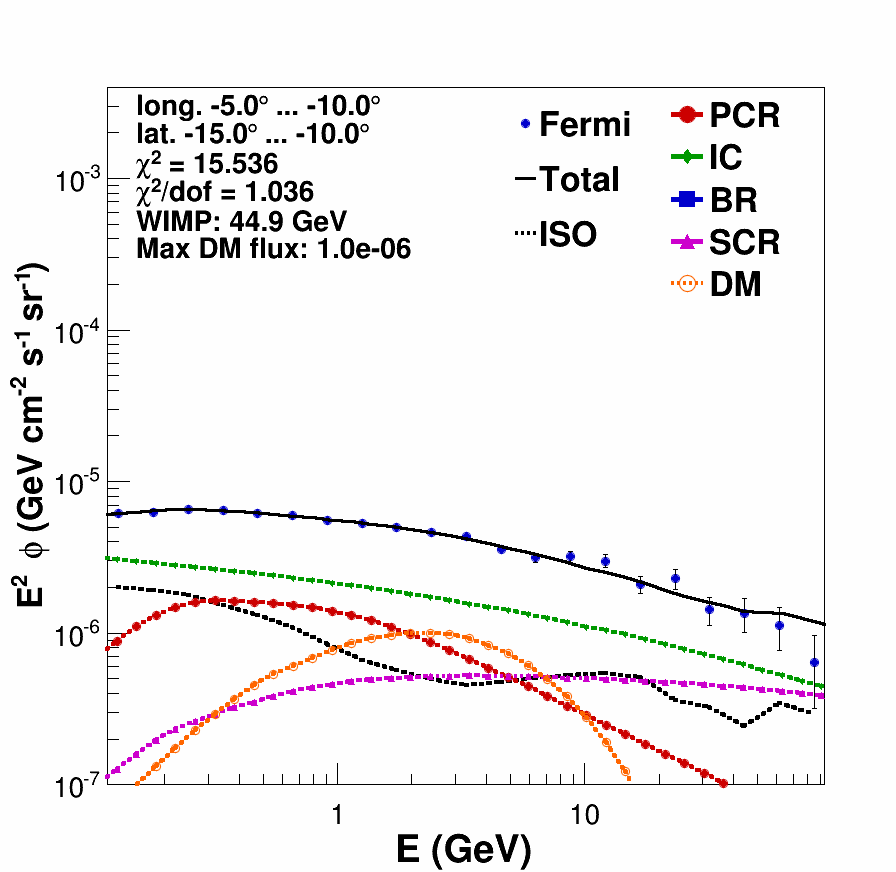}
\includegraphics[width=0.16\textwidth,height=0.16\textwidth,clip]{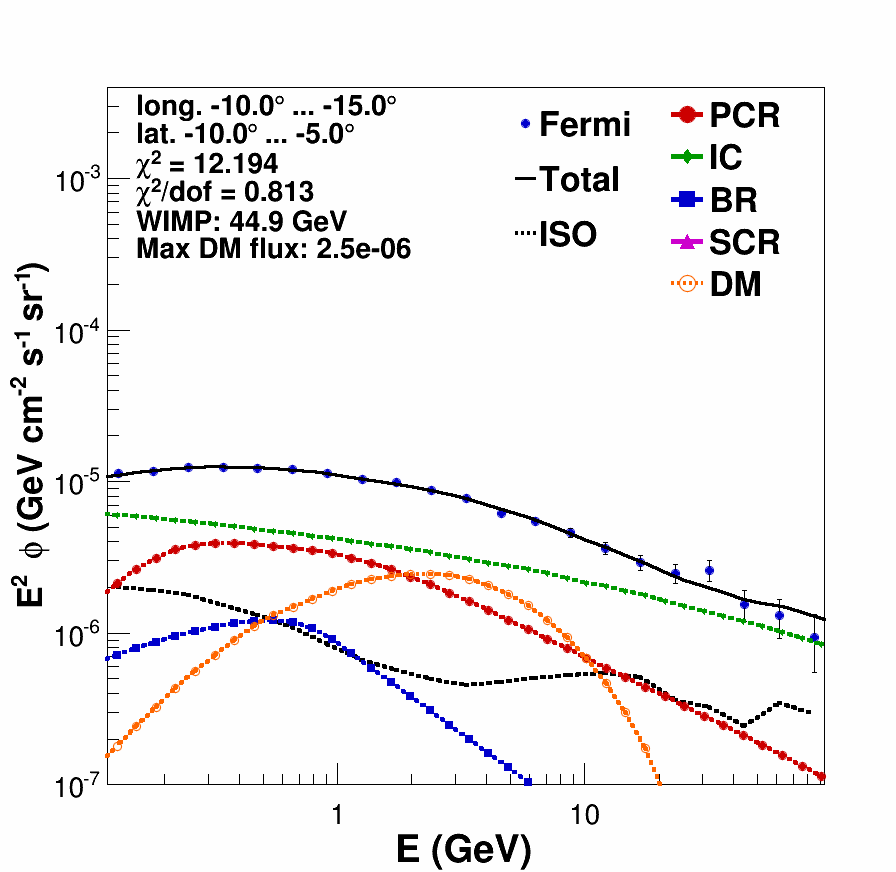}
\includegraphics[width=0.16\textwidth,height=0.16\textwidth,clip]{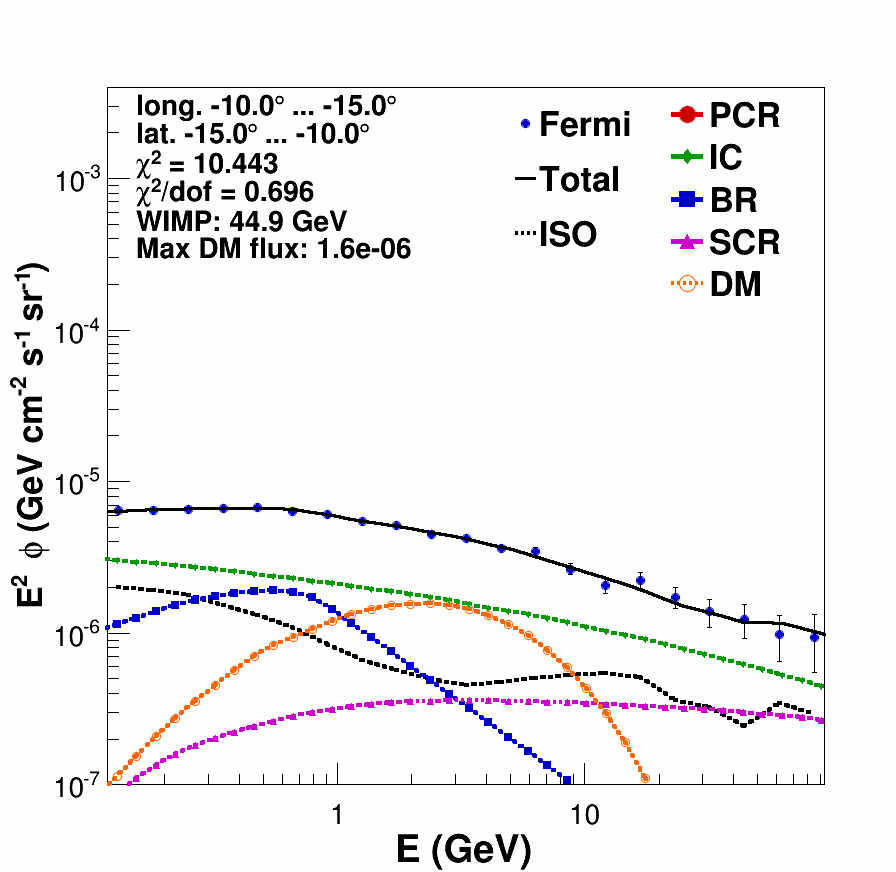}
\includegraphics[width=0.16\textwidth,height=0.16\textwidth,clip]{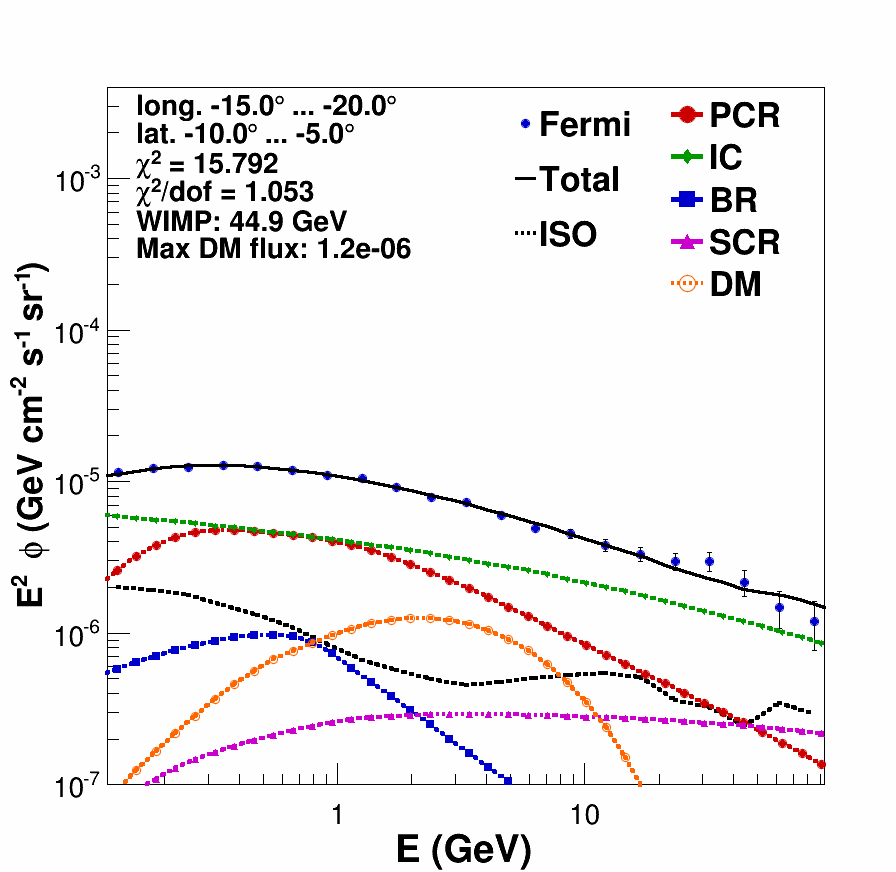}
\includegraphics[width=0.16\textwidth,height=0.16\textwidth,clip]{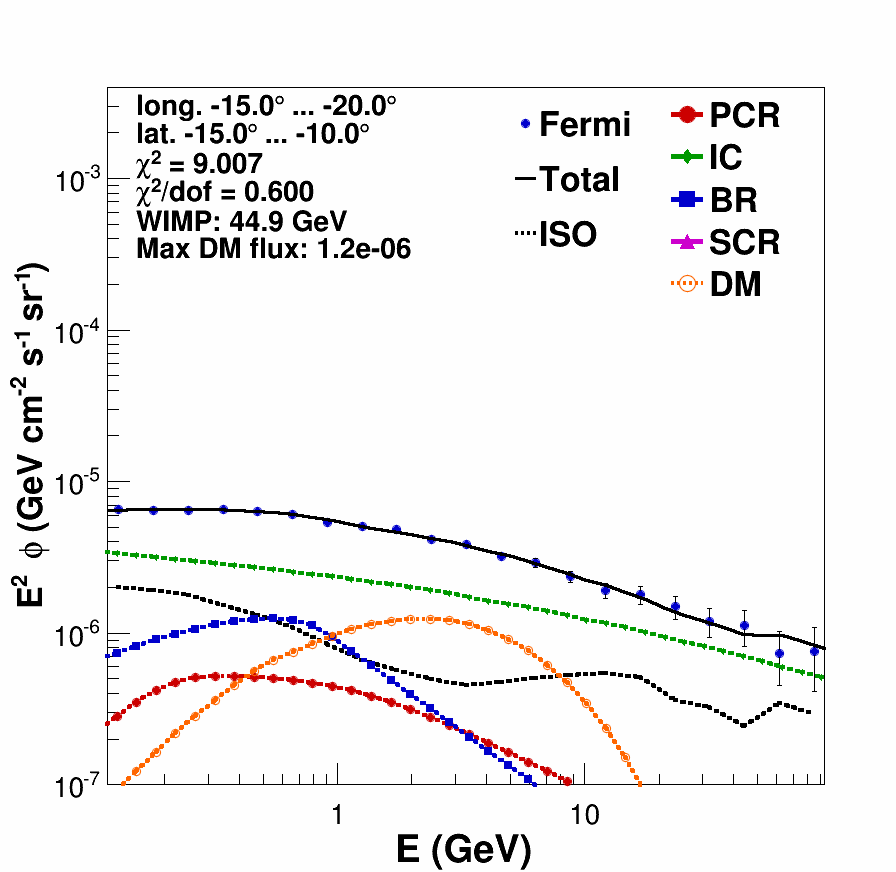}
\includegraphics[width=0.16\textwidth,height=0.16\textwidth,clip]{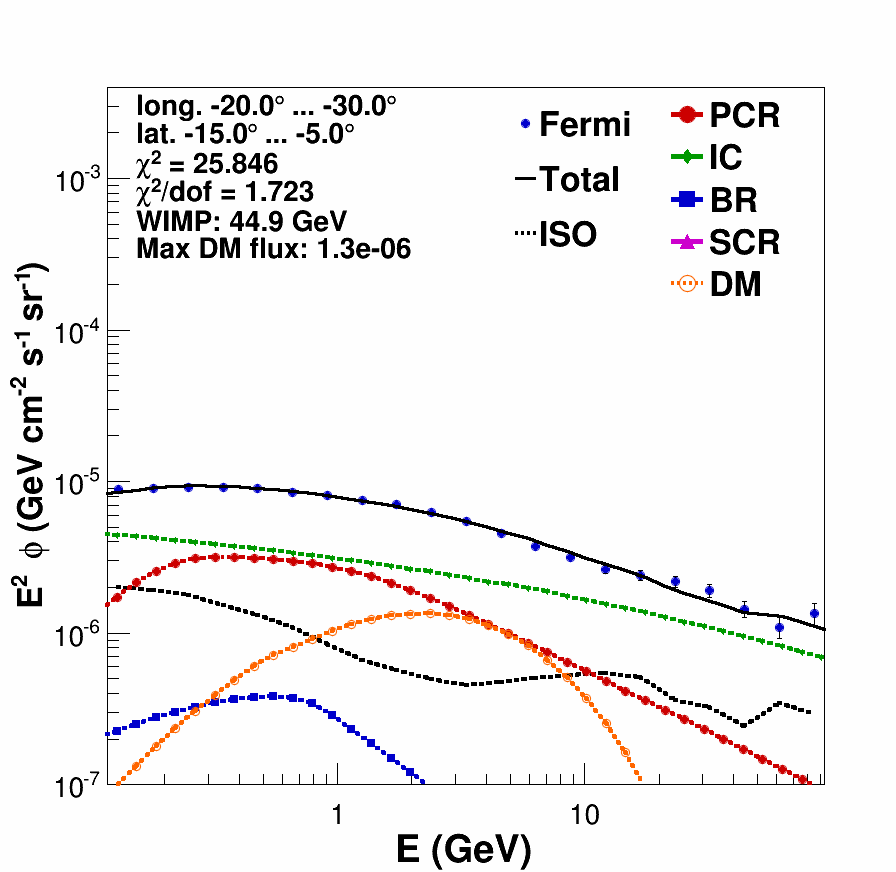}
\includegraphics[width=0.16\textwidth,height=0.16\textwidth,clip]{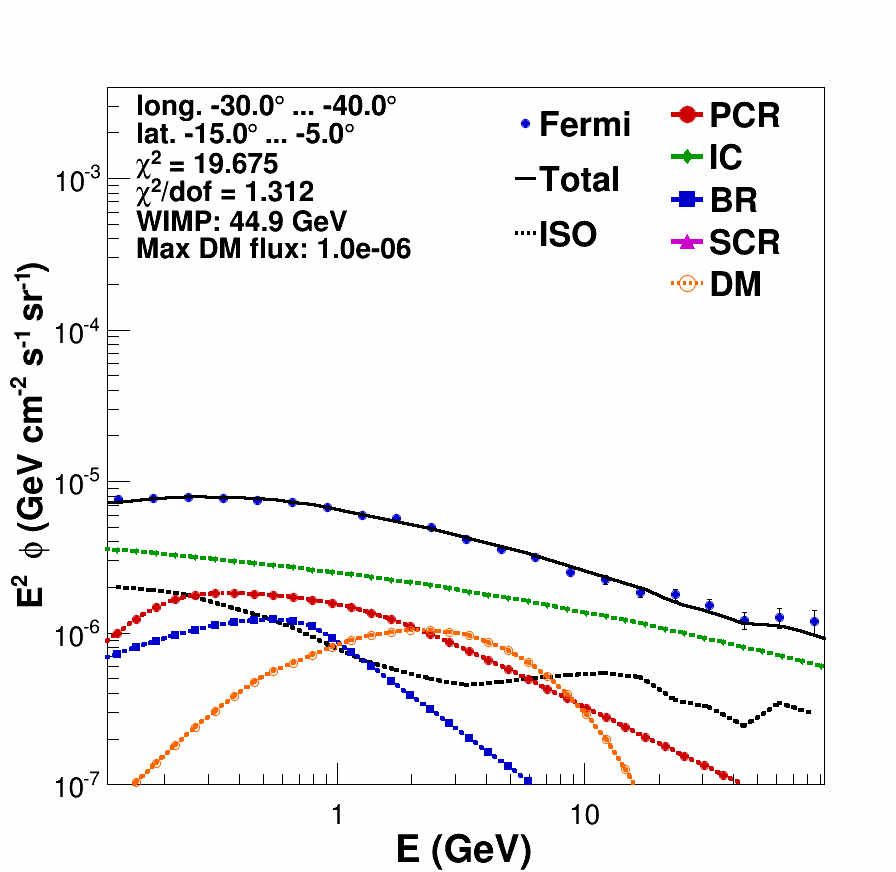}
\includegraphics[width=0.16\textwidth,height=0.16\textwidth,clip]{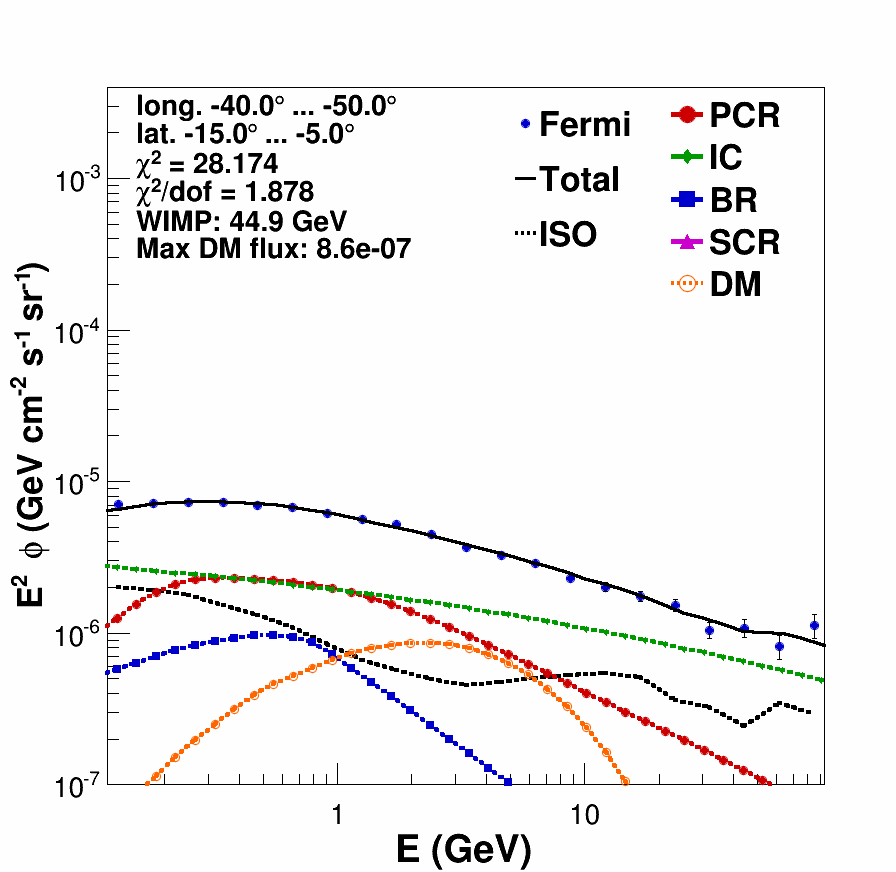}
\includegraphics[width=0.16\textwidth,height=0.16\textwidth,clip]{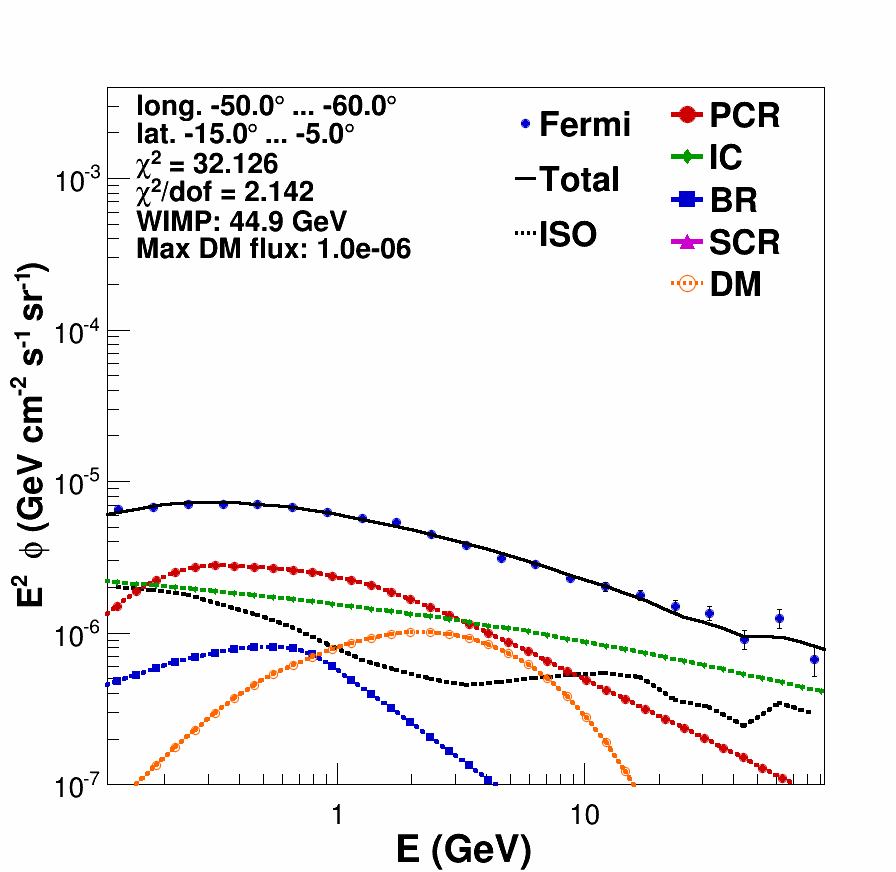}
\includegraphics[width=0.16\textwidth,height=0.16\textwidth,clip]{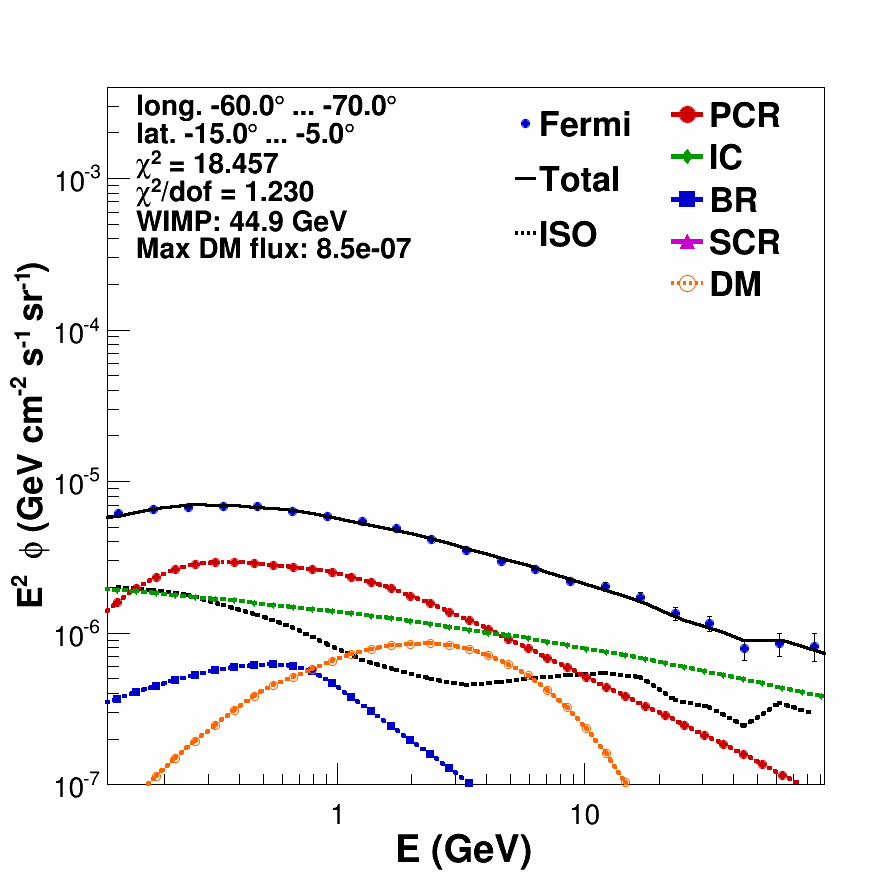}
\includegraphics[width=0.16\textwidth,height=0.16\textwidth,clip]{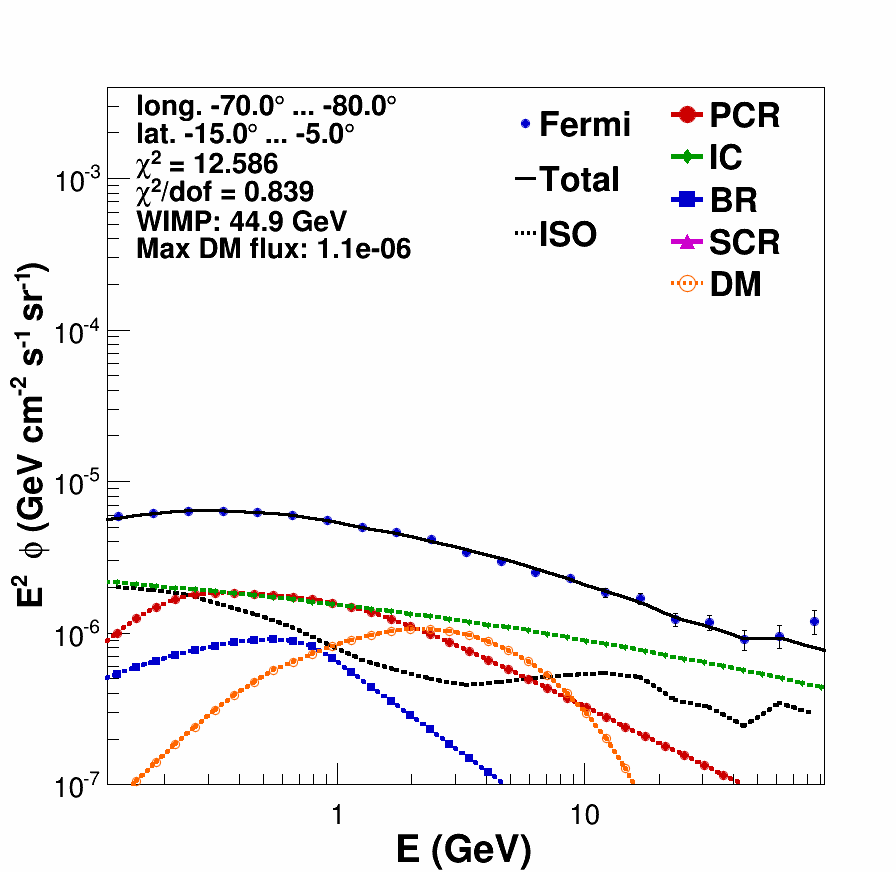}
\includegraphics[width=0.16\textwidth,height=0.16\textwidth,clip]{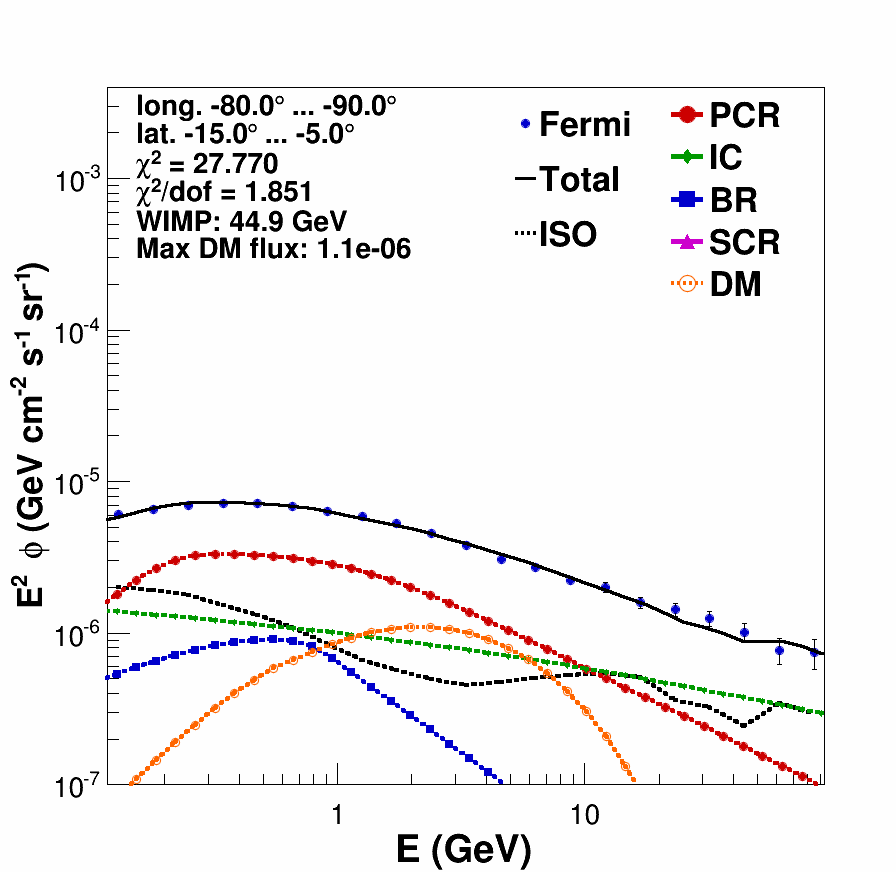}
\includegraphics[width=0.16\textwidth,height=0.16\textwidth,clip]{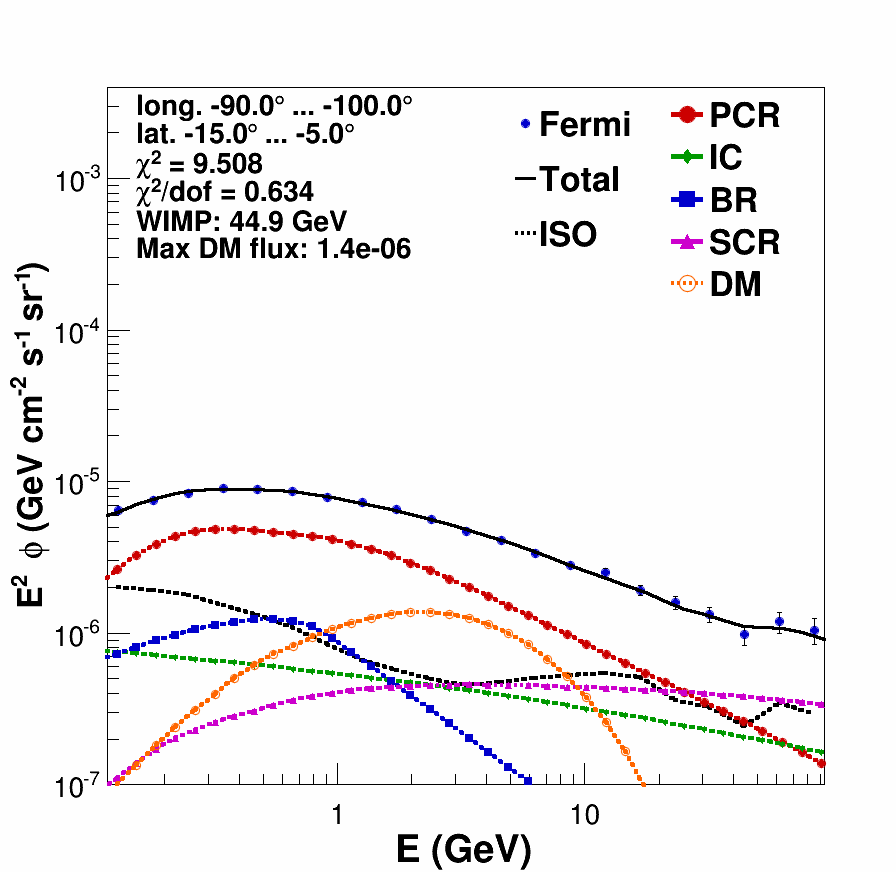}
\includegraphics[width=0.16\textwidth,height=0.16\textwidth,clip]{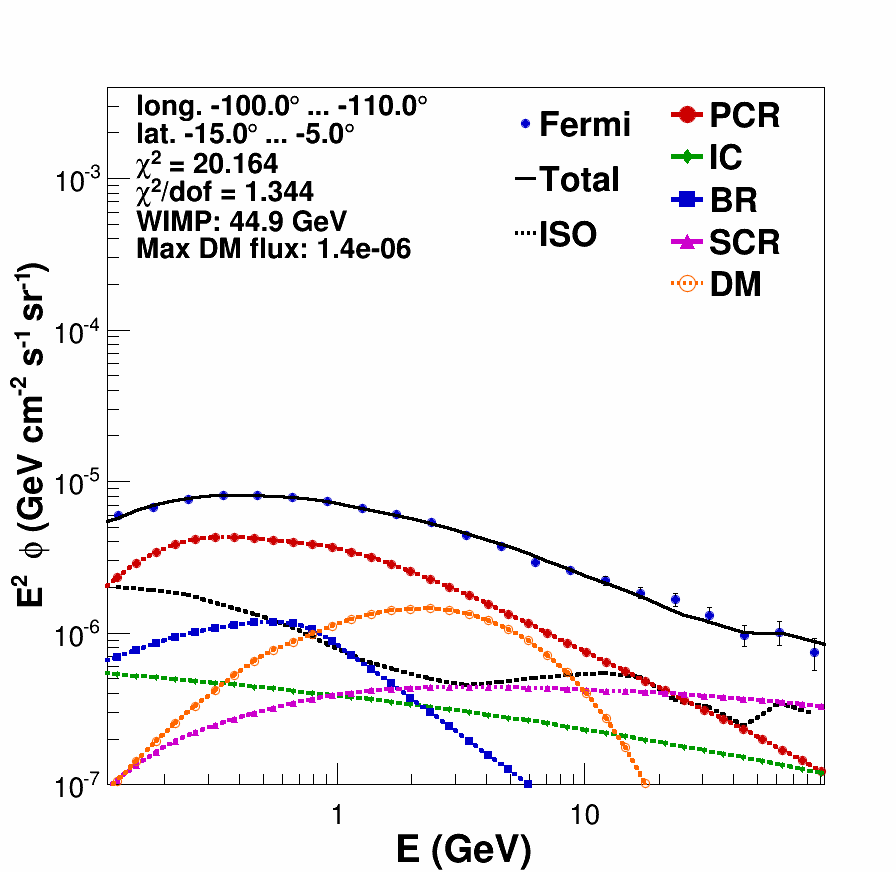}
\includegraphics[width=0.16\textwidth,height=0.16\textwidth,clip]{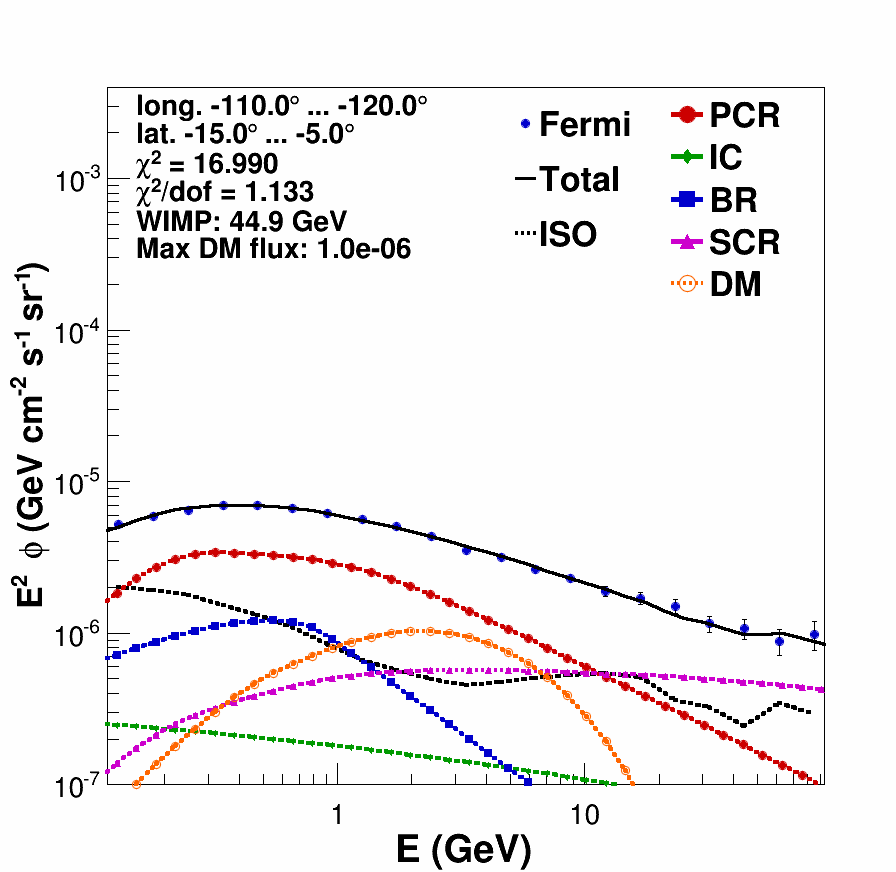}
\includegraphics[width=0.16\textwidth,height=0.16\textwidth,clip]{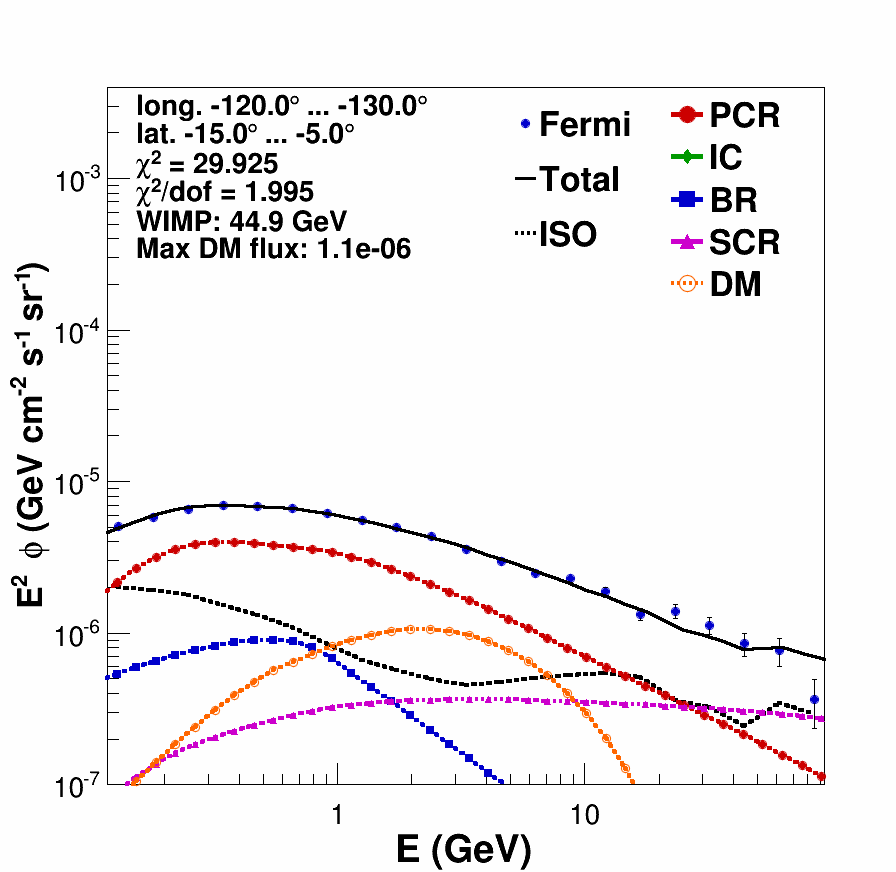}
\includegraphics[width=0.16\textwidth,height=0.16\textwidth,clip]{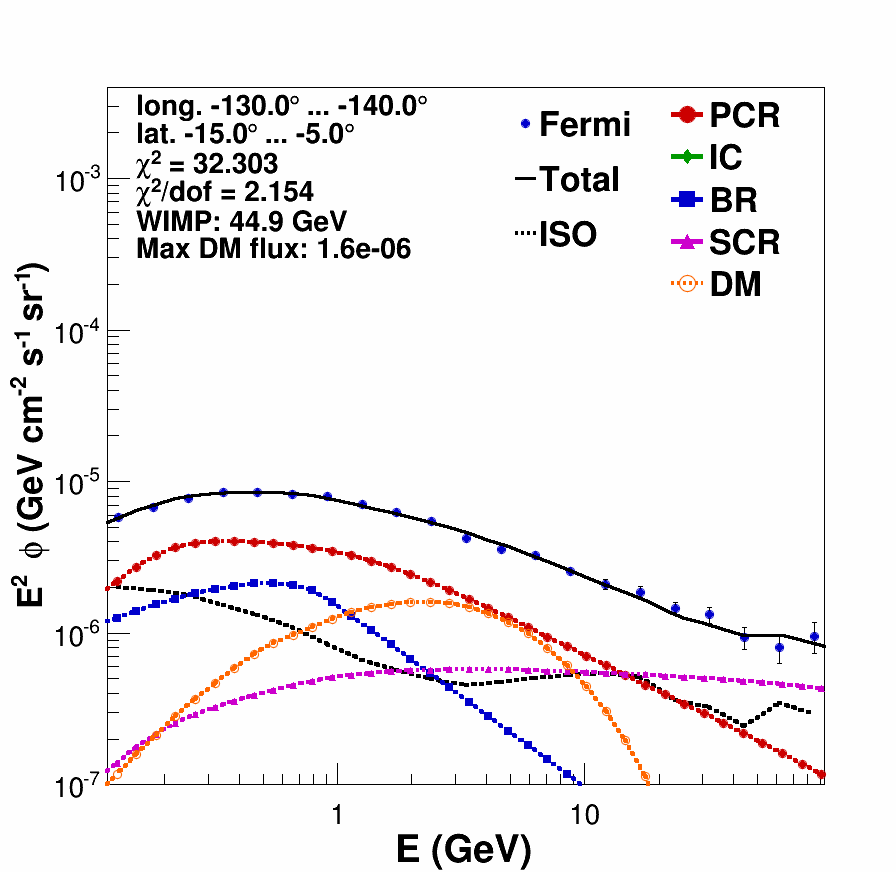}
\includegraphics[width=0.16\textwidth,height=0.16\textwidth,clip]{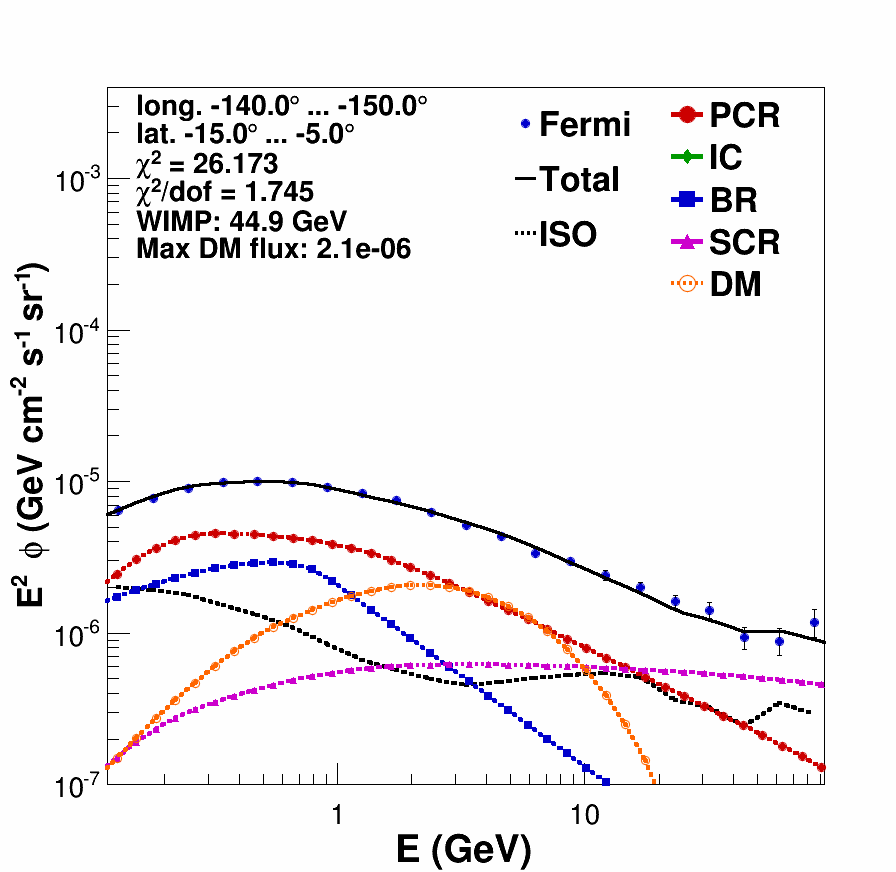}
\includegraphics[width=0.16\textwidth,height=0.16\textwidth,clip]{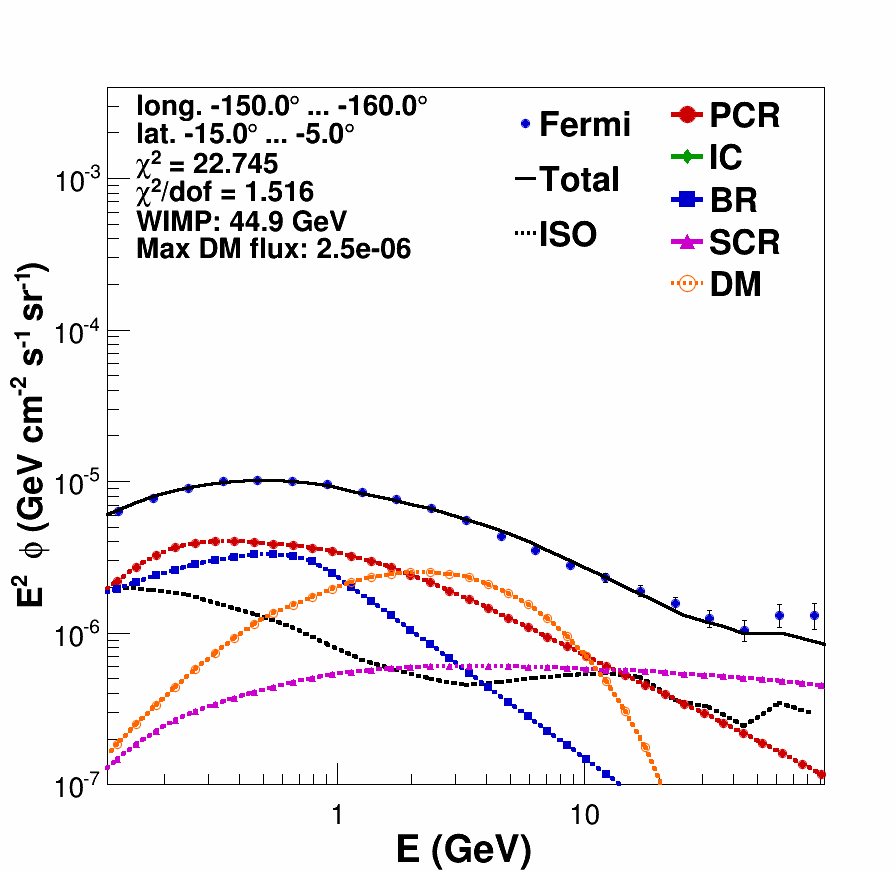}
\includegraphics[width=0.16\textwidth,height=0.16\textwidth,clip]{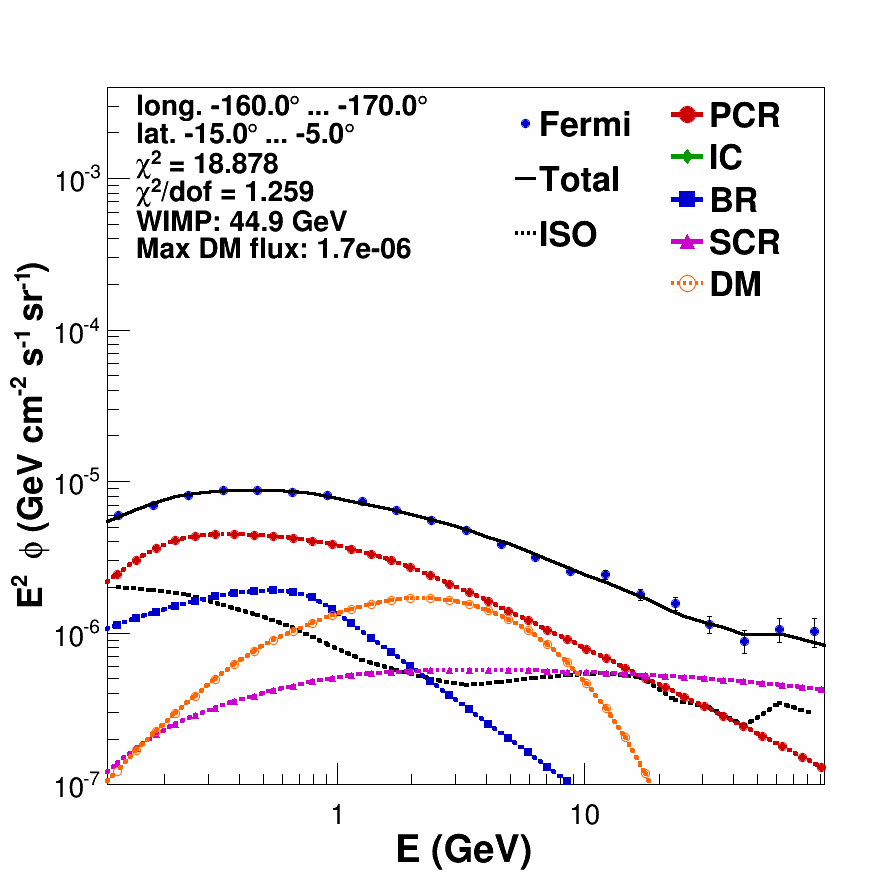}
\includegraphics[width=0.16\textwidth,height=0.16\textwidth,clip]{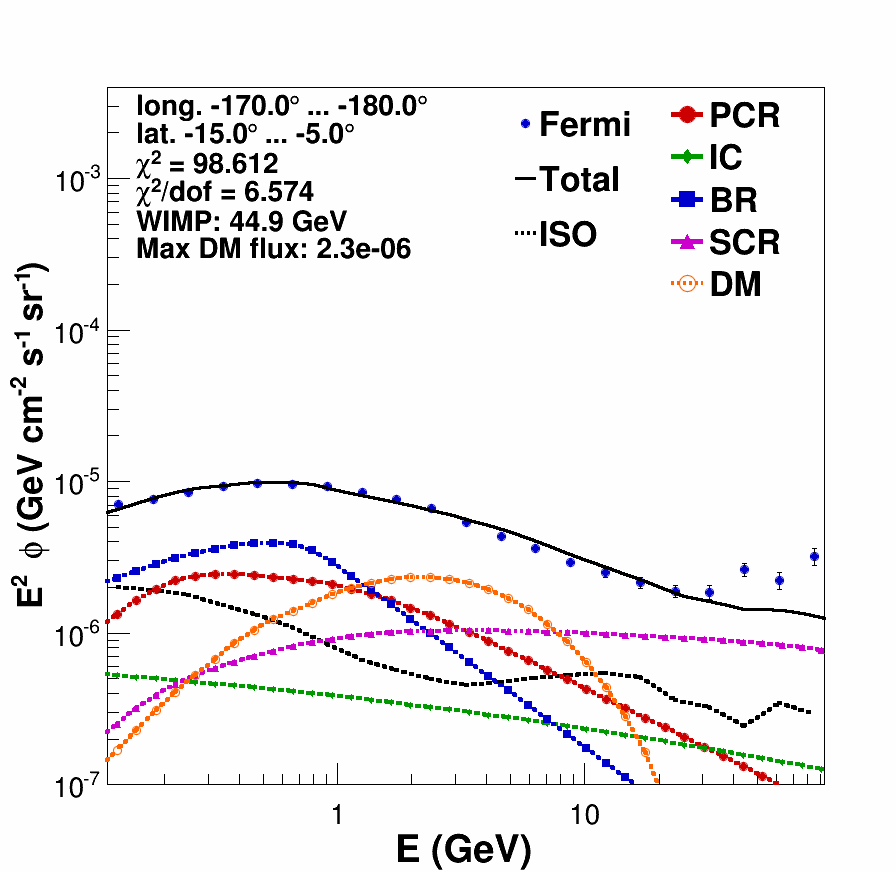}%%%%%%r13
\caption[]{Template fits for latitudes  with $-15.0^\circ<b<-5.0^\circ$ and longitudes decreasing from 180$^\circ$ to -180$^\circ$.} \label{F46}
\end{figure}
\begin{figure}
\centering
\includegraphics[width=0.16\textwidth,height=0.16\textwidth,clip]{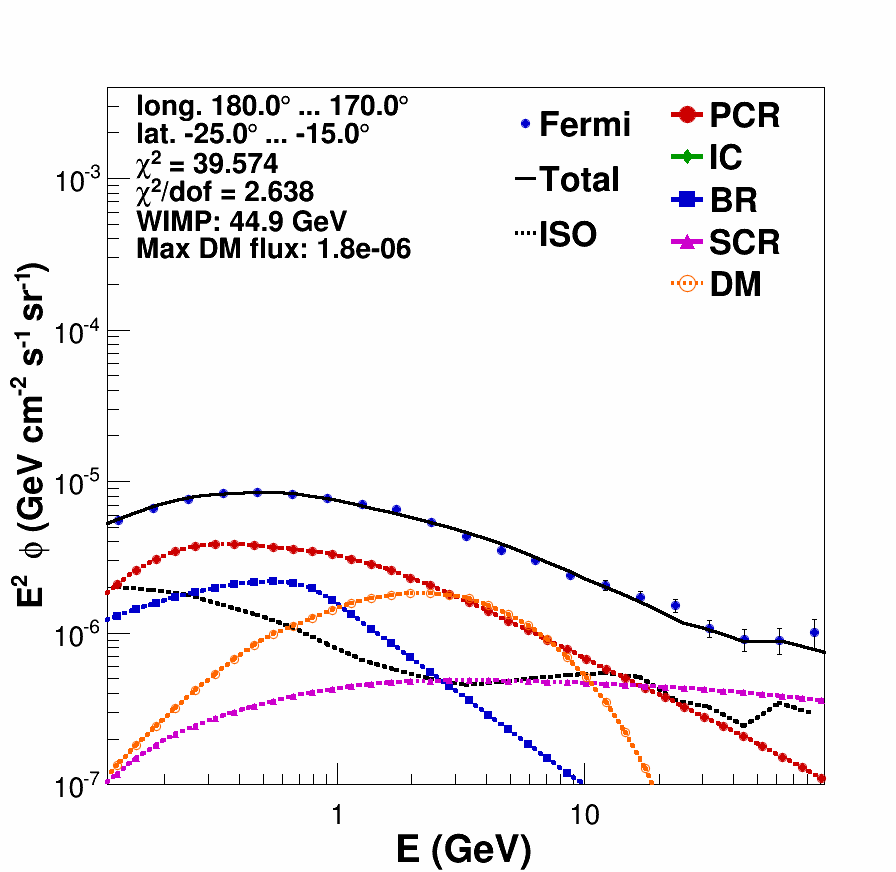}
\includegraphics[width=0.16\textwidth,height=0.16\textwidth,clip]{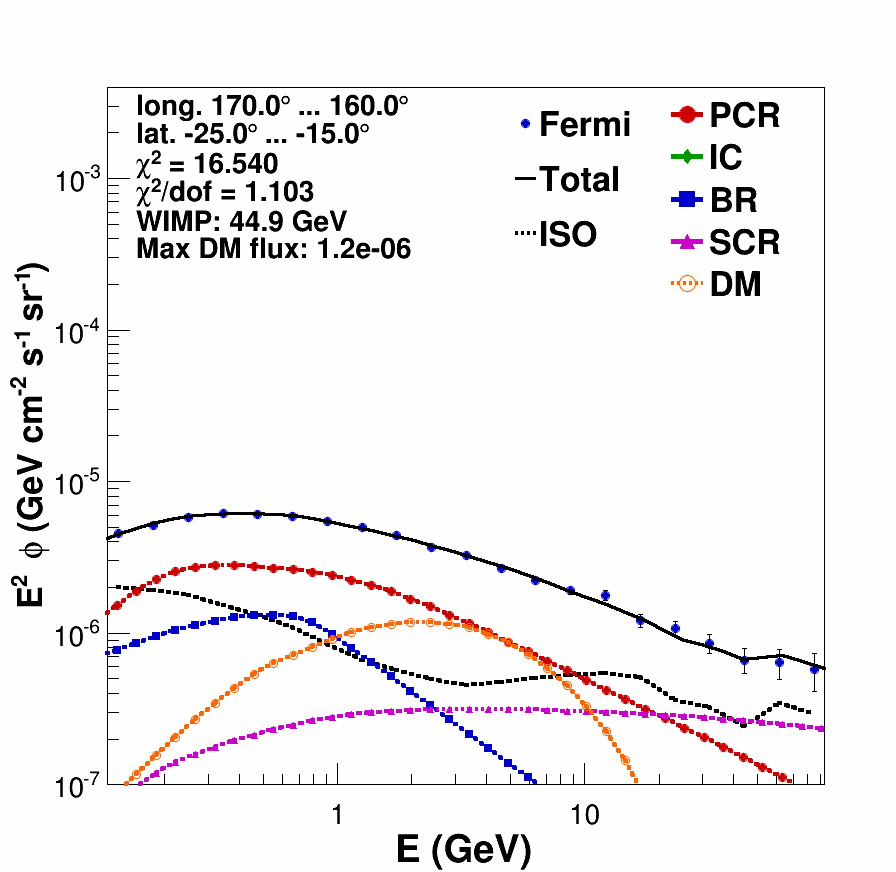}
\includegraphics[width=0.16\textwidth,height=0.16\textwidth,clip]{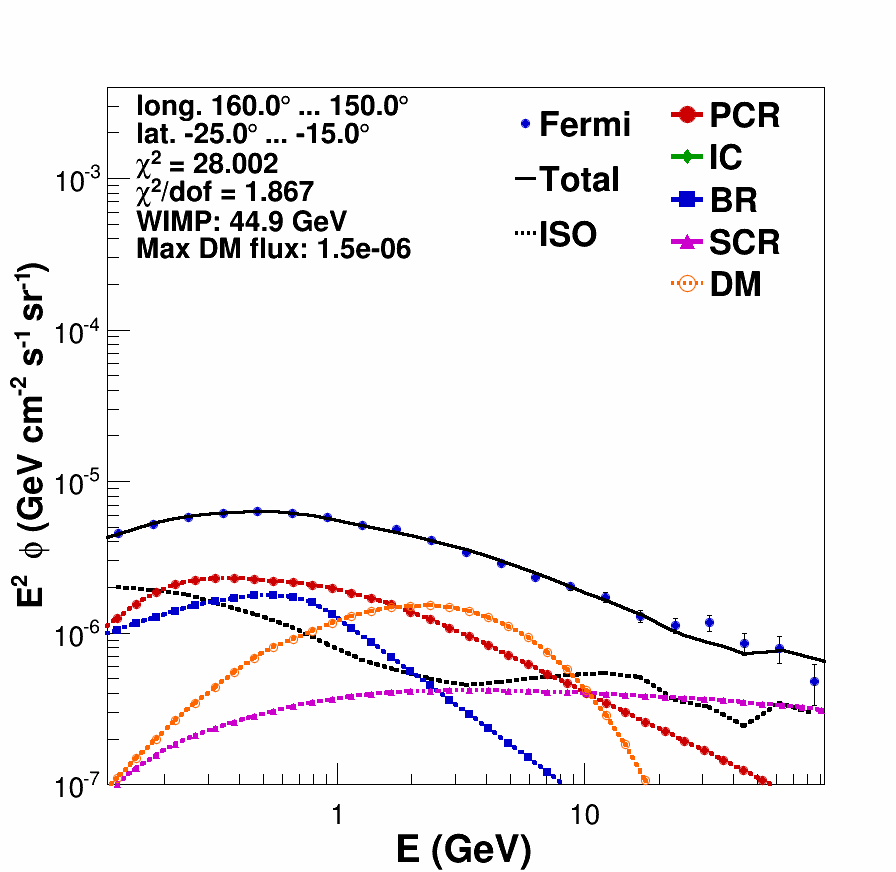}
\includegraphics[width=0.16\textwidth,height=0.16\textwidth,clip]{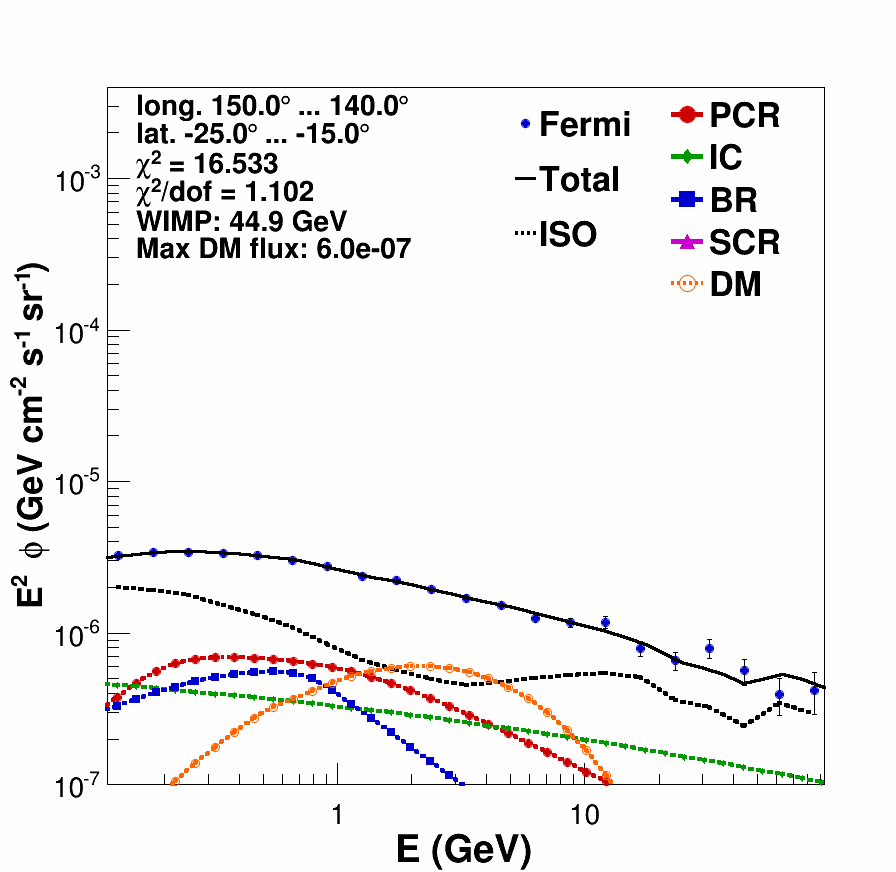}
\includegraphics[width=0.16\textwidth,height=0.16\textwidth,clip]{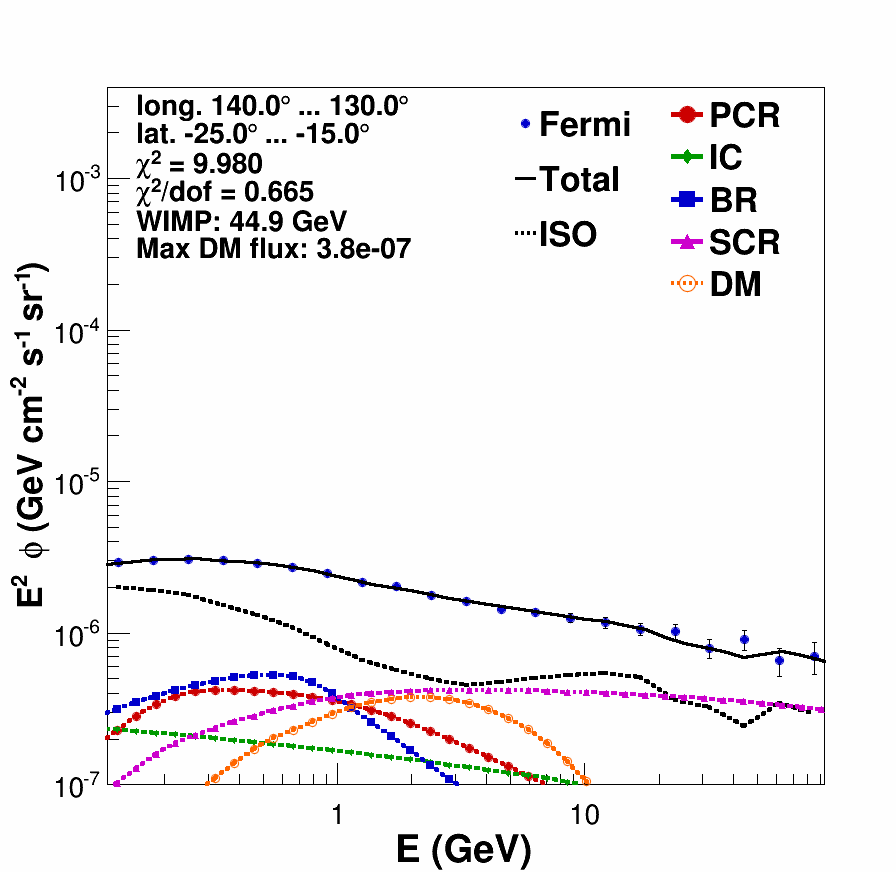}
\includegraphics[width=0.16\textwidth,height=0.16\textwidth,clip]{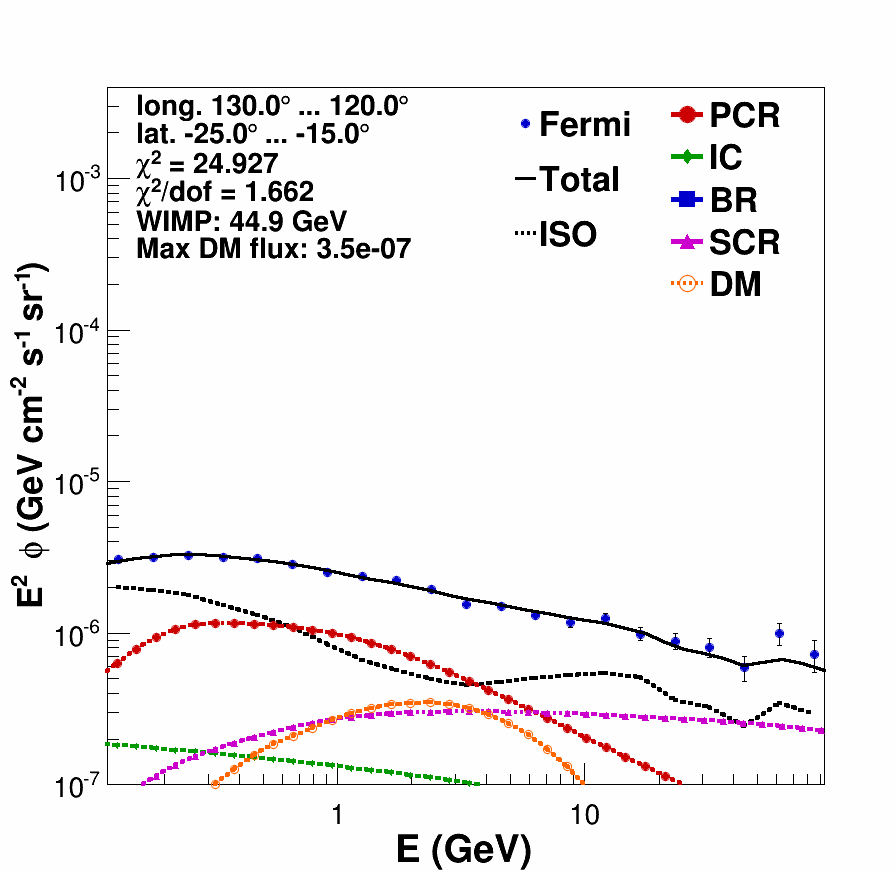}
\includegraphics[width=0.16\textwidth,height=0.16\textwidth,clip]{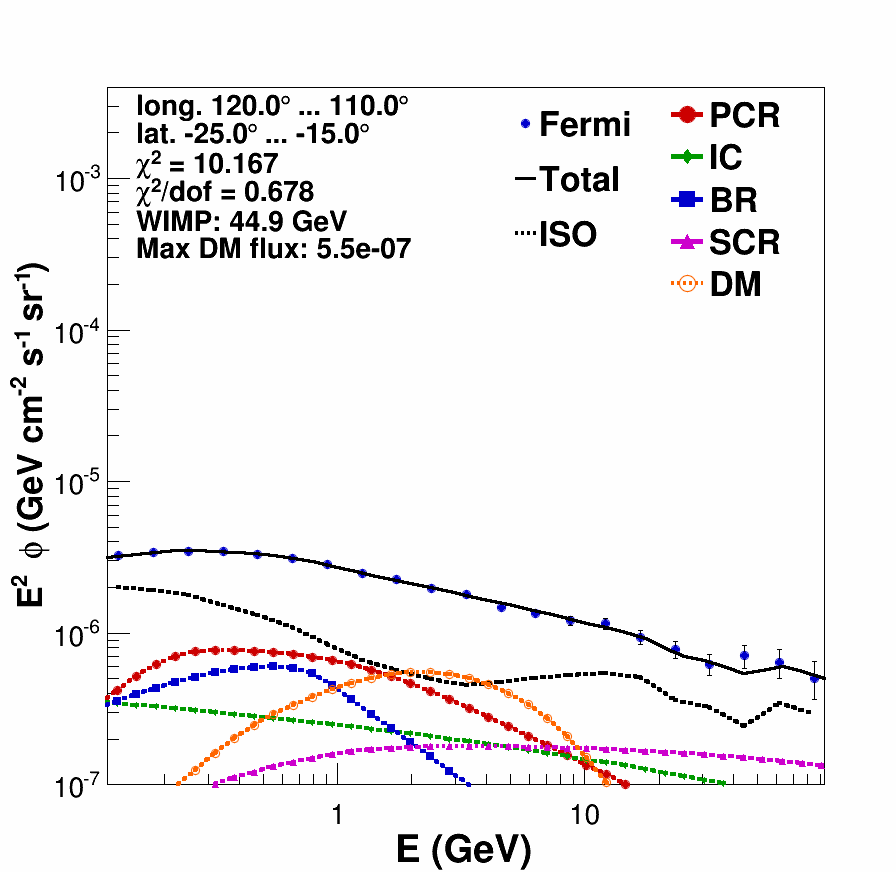}
\includegraphics[width=0.16\textwidth,height=0.16\textwidth,clip]{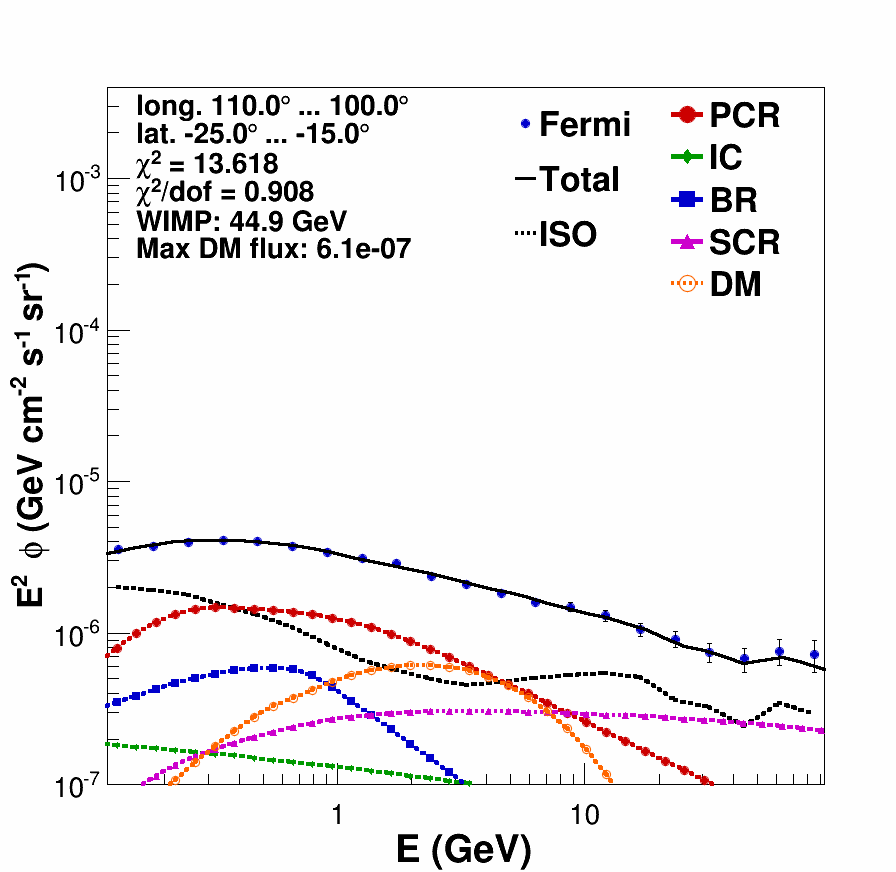}
\includegraphics[width=0.16\textwidth,height=0.16\textwidth,clip]{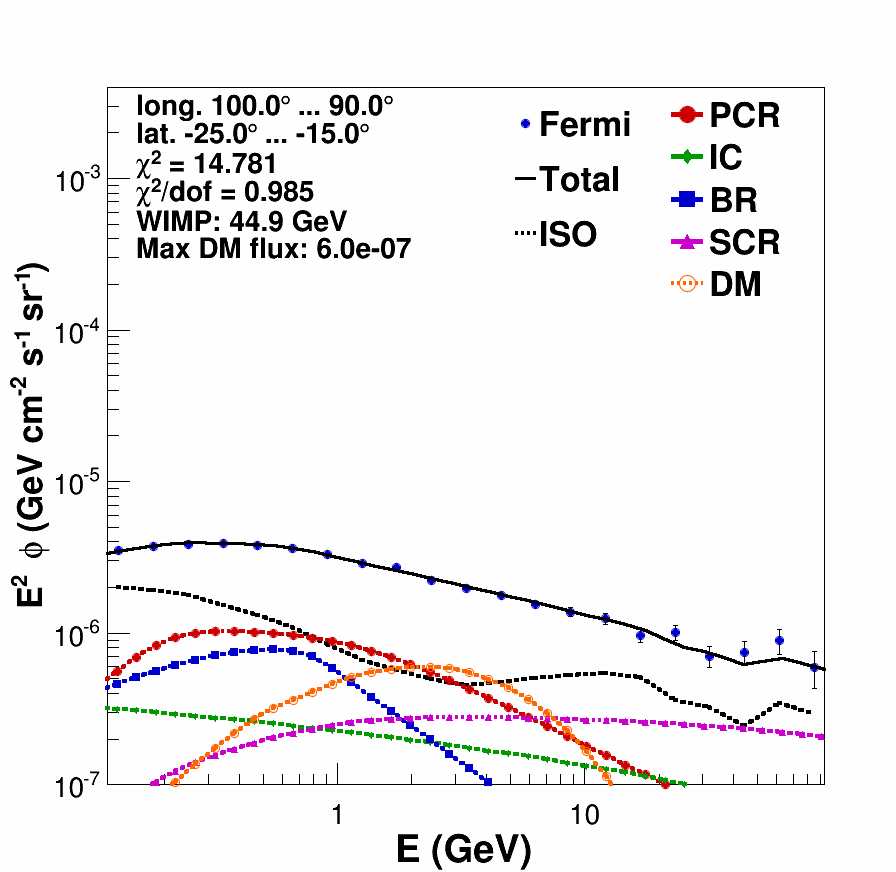}
\includegraphics[width=0.16\textwidth,height=0.16\textwidth,clip]{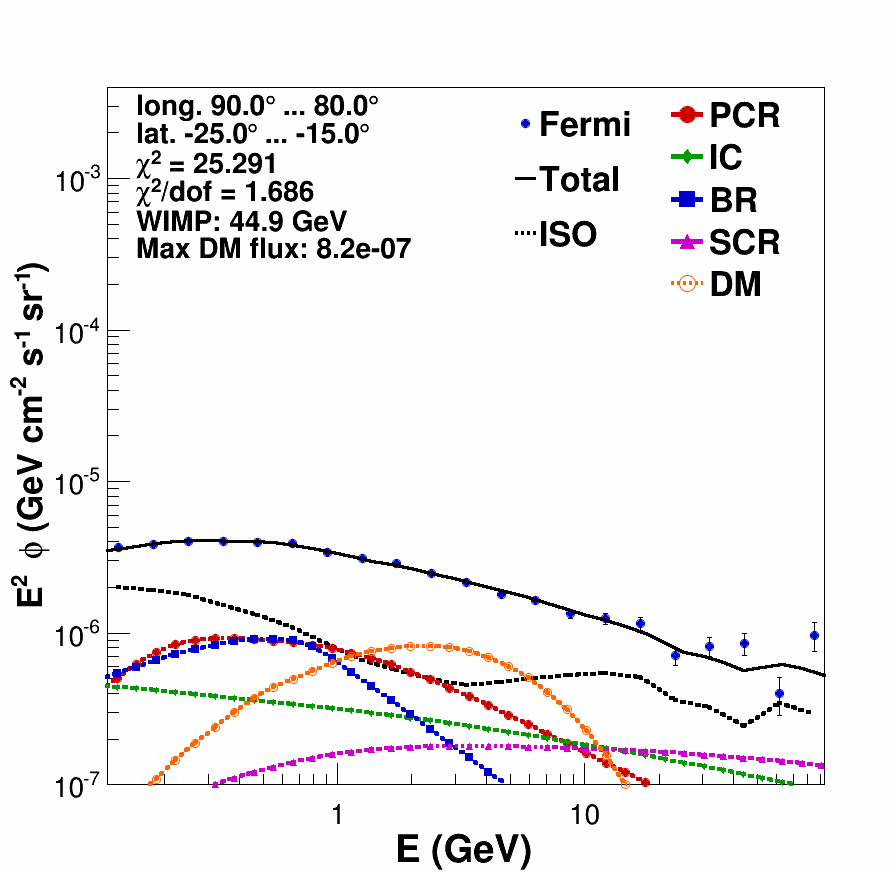}
\includegraphics[width=0.16\textwidth,height=0.16\textwidth,clip]{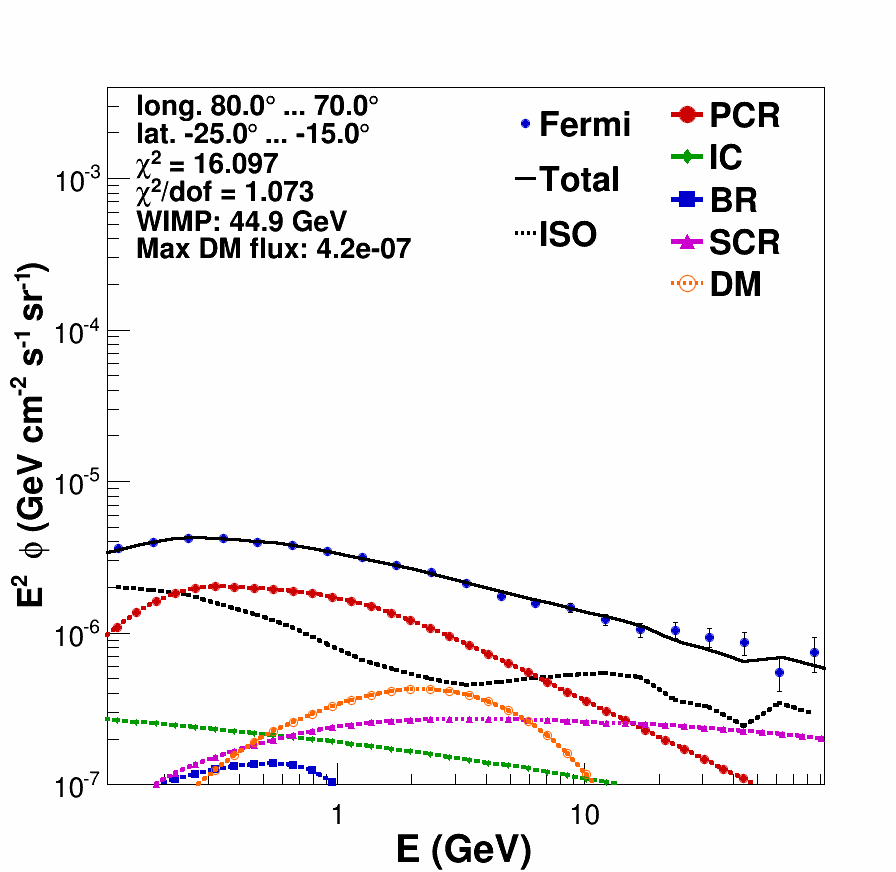}
\includegraphics[width=0.16\textwidth,height=0.16\textwidth,clip]{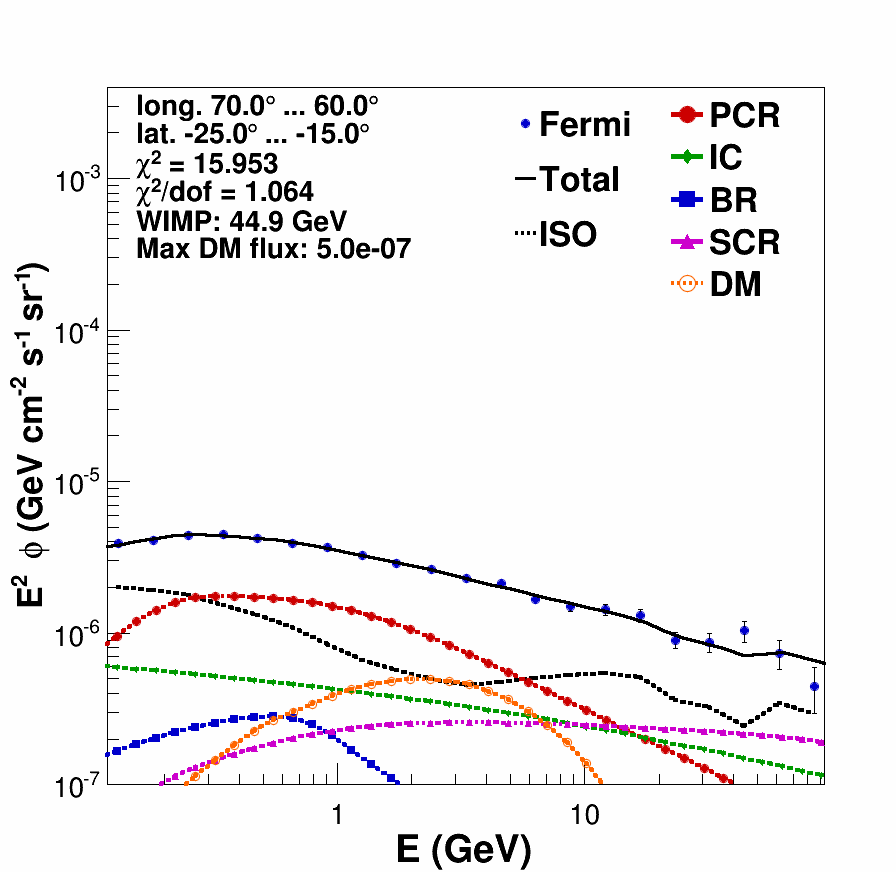}
\includegraphics[width=0.16\textwidth,height=0.16\textwidth,clip]{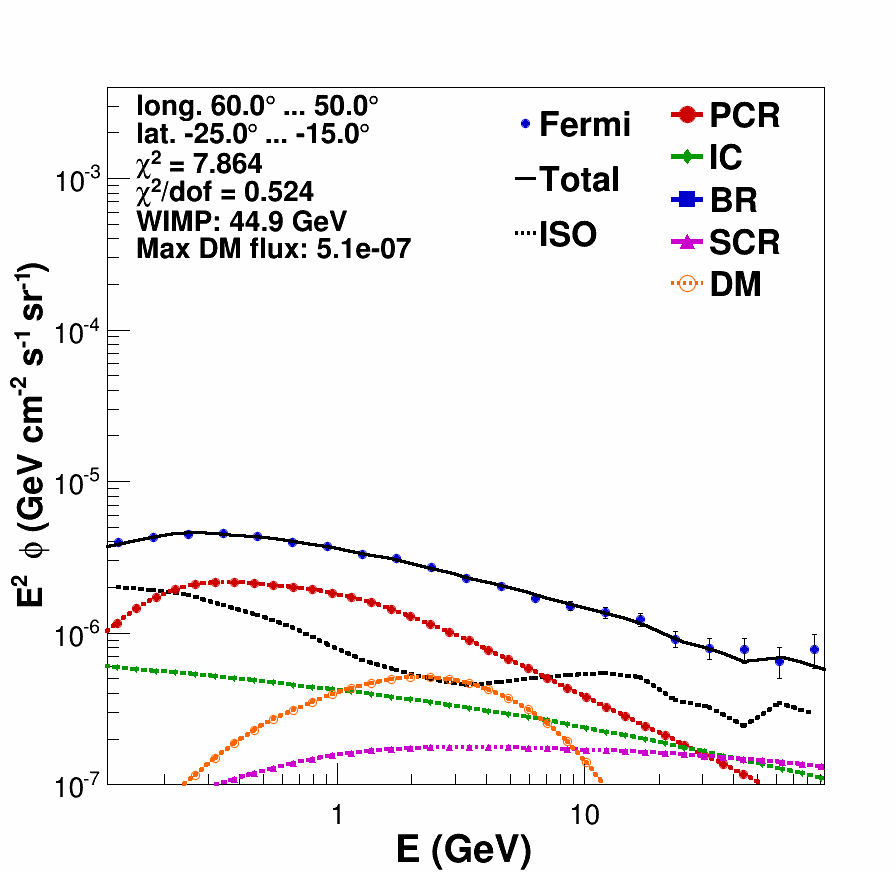}
\includegraphics[width=0.16\textwidth,height=0.16\textwidth,clip]{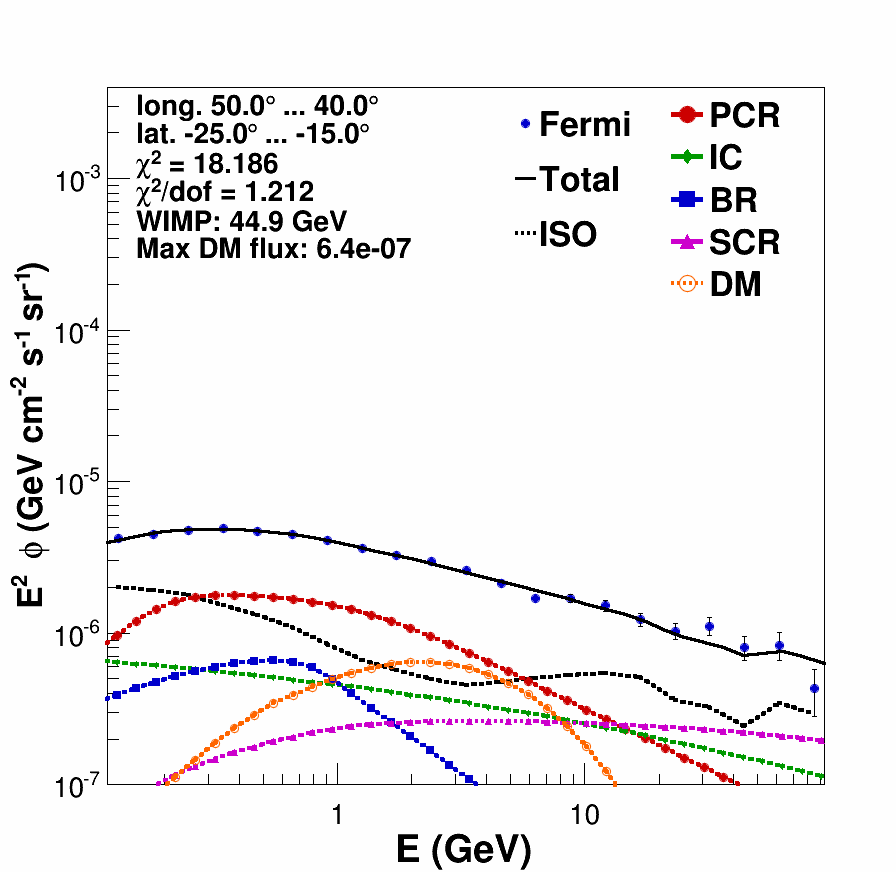}
\includegraphics[width=0.16\textwidth,height=0.16\textwidth,clip]{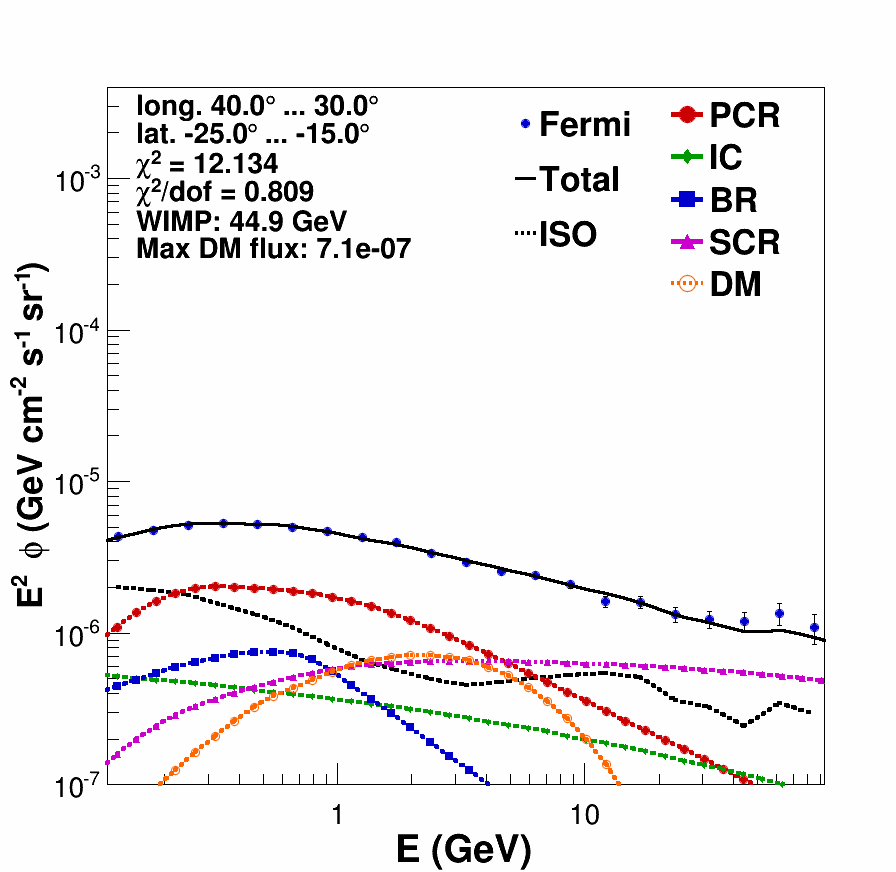}
\includegraphics[width=0.16\textwidth,height=0.16\textwidth,clip]{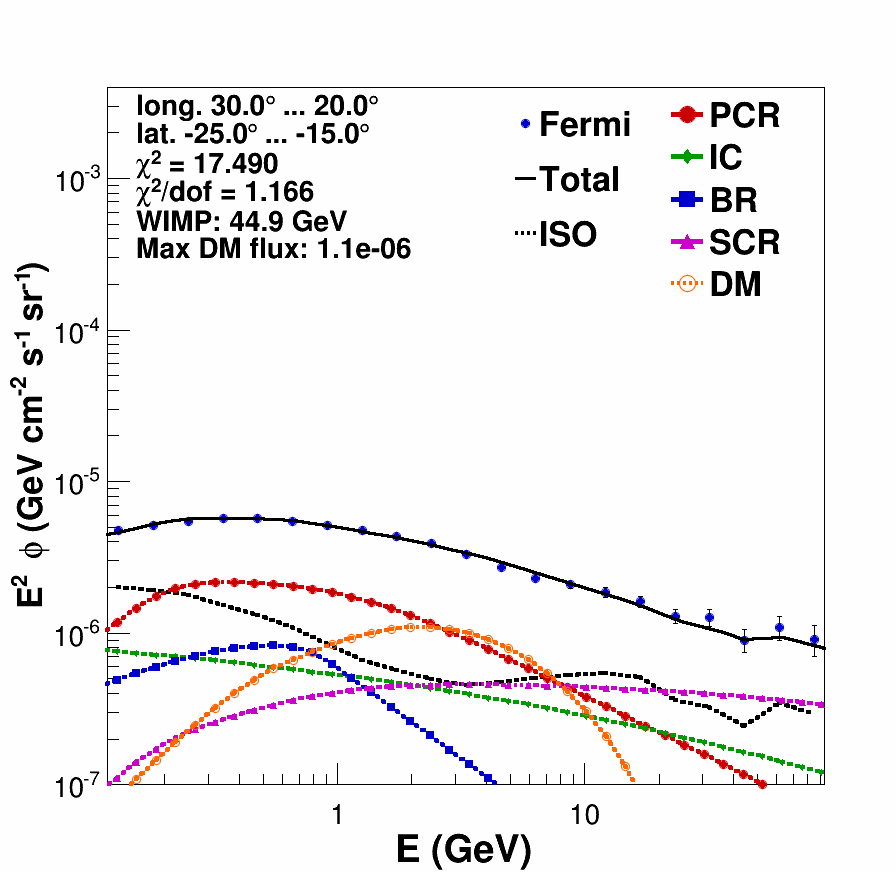}
\includegraphics[width=0.16\textwidth,height=0.16\textwidth,clip]{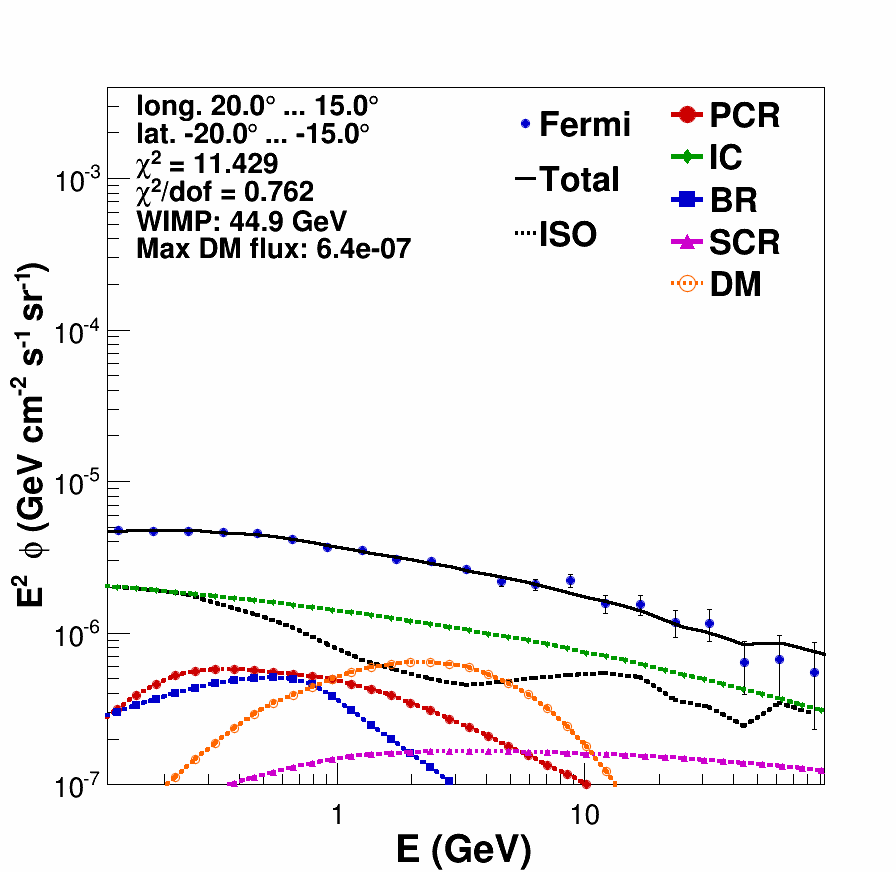}
\includegraphics[width=0.16\textwidth,height=0.16\textwidth,clip]{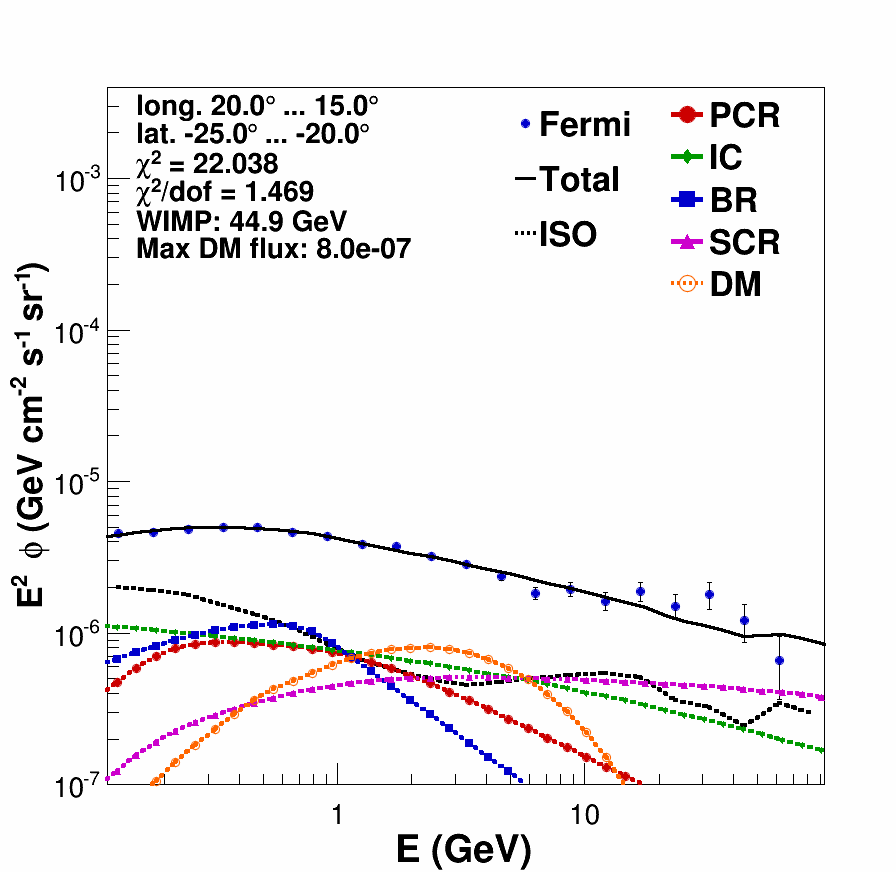}
\includegraphics[width=0.16\textwidth,height=0.16\textwidth,clip]{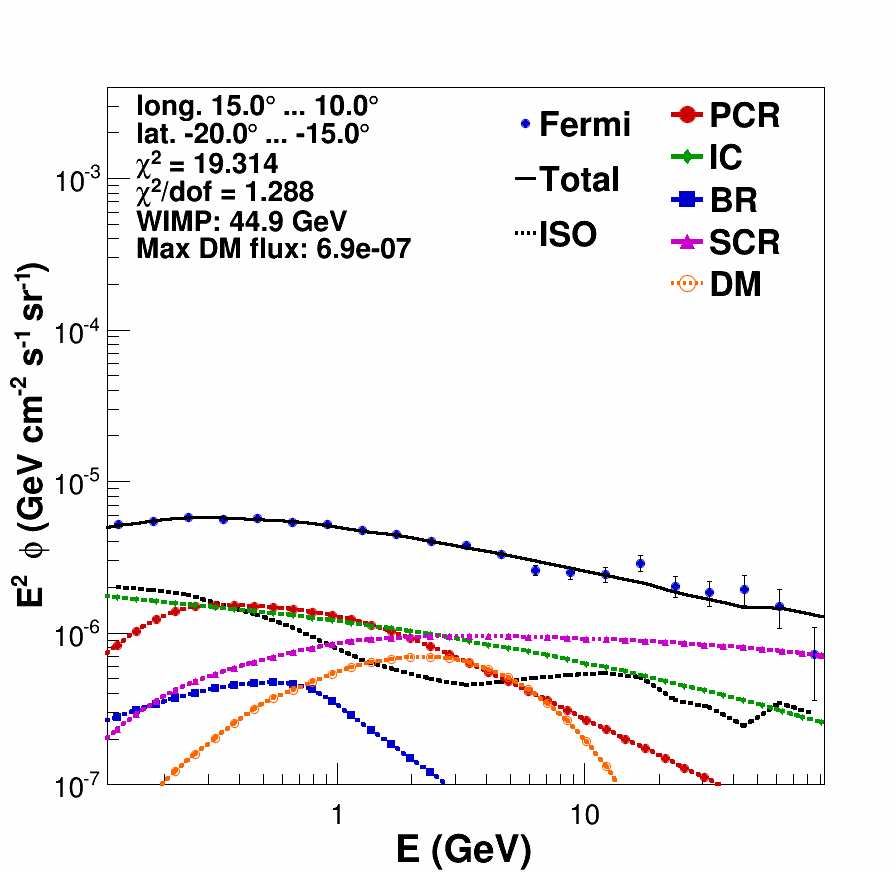}
\includegraphics[width=0.16\textwidth,height=0.16\textwidth,clip]{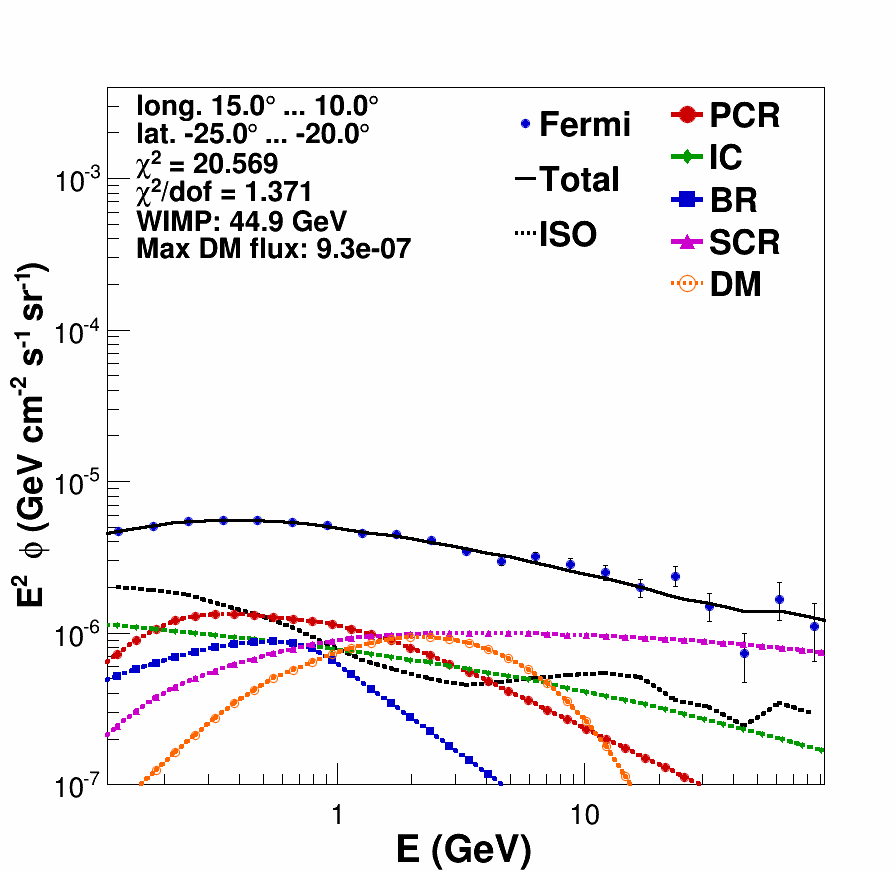}
\includegraphics[width=0.16\textwidth,height=0.16\textwidth,clip]{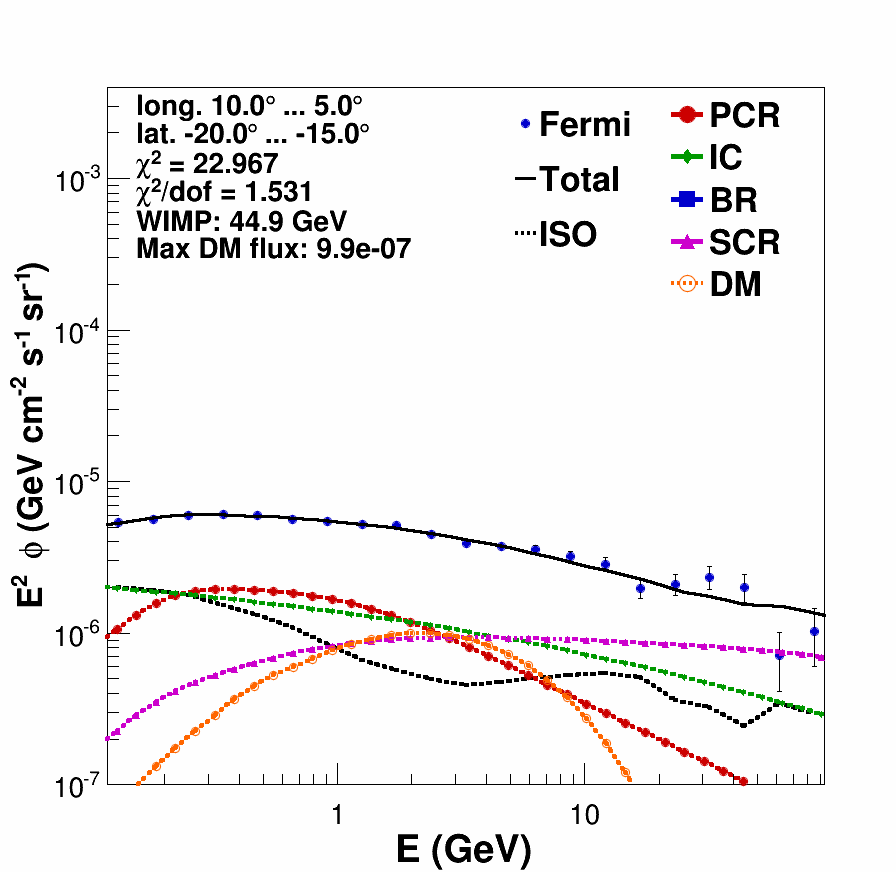}
\includegraphics[width=0.16\textwidth,height=0.16\textwidth,clip]{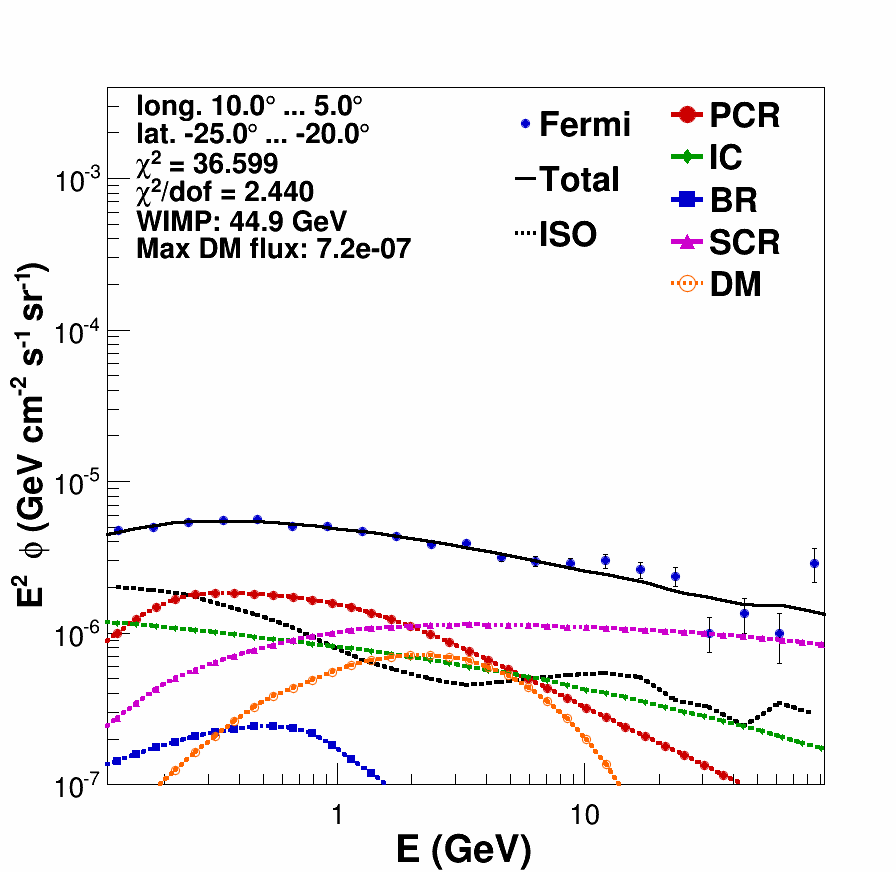}
\includegraphics[width=0.16\textwidth,height=0.16\textwidth,clip]{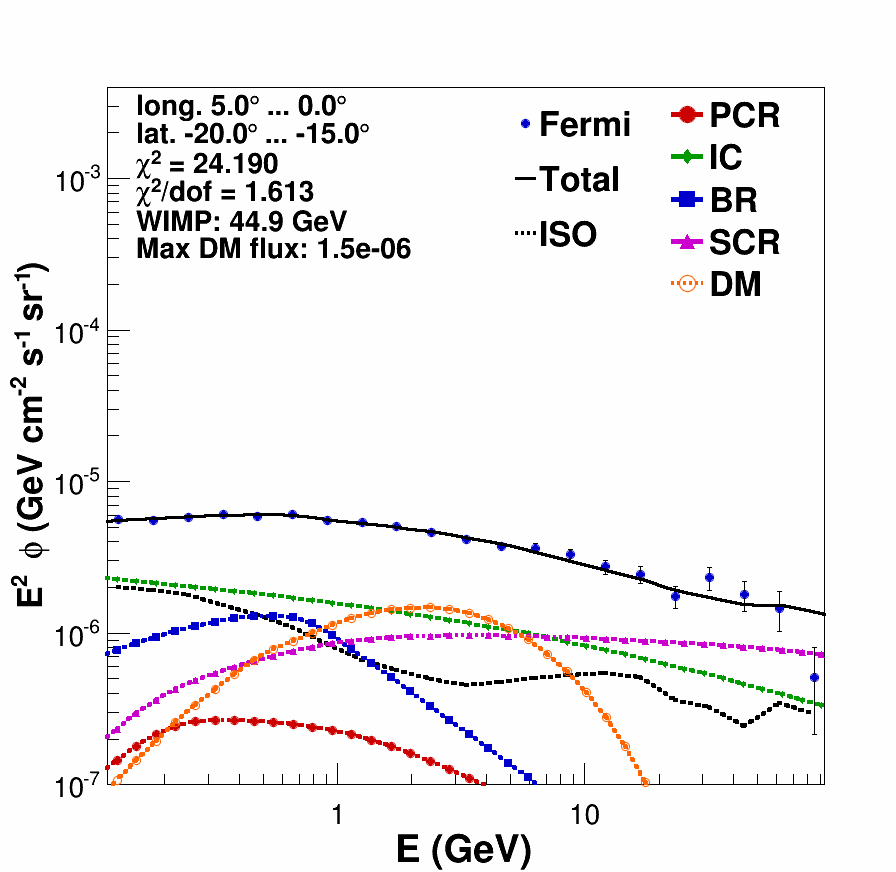}
\includegraphics[width=0.16\textwidth,height=0.16\textwidth,clip]{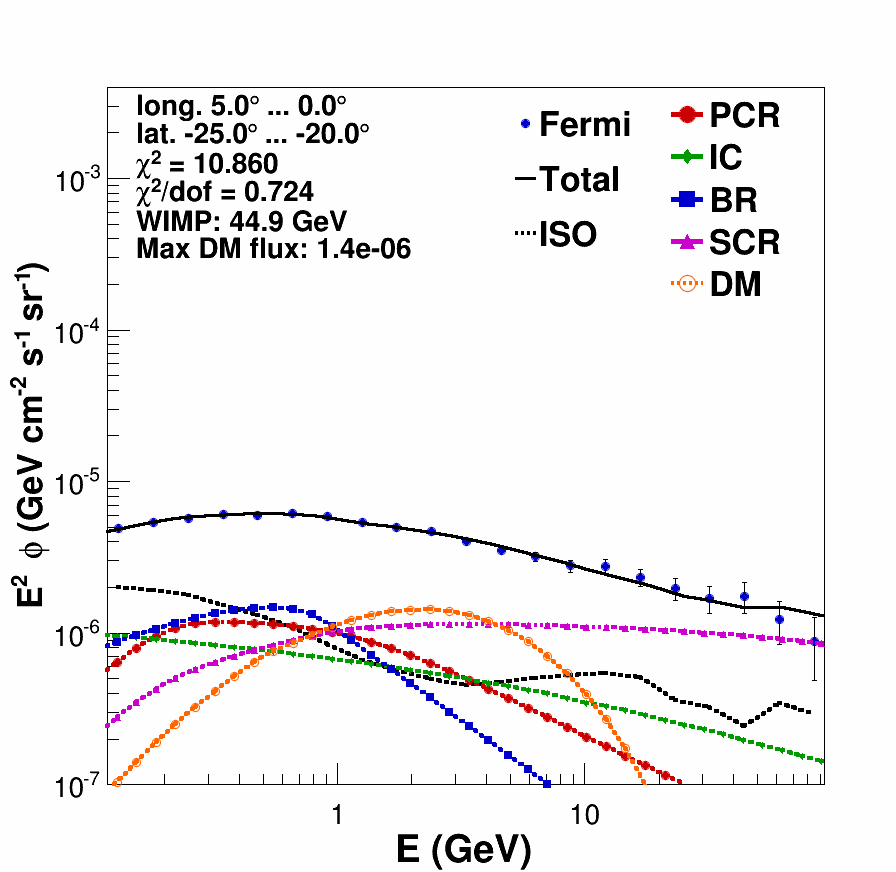}
\includegraphics[width=0.16\textwidth,height=0.16\textwidth,clip]{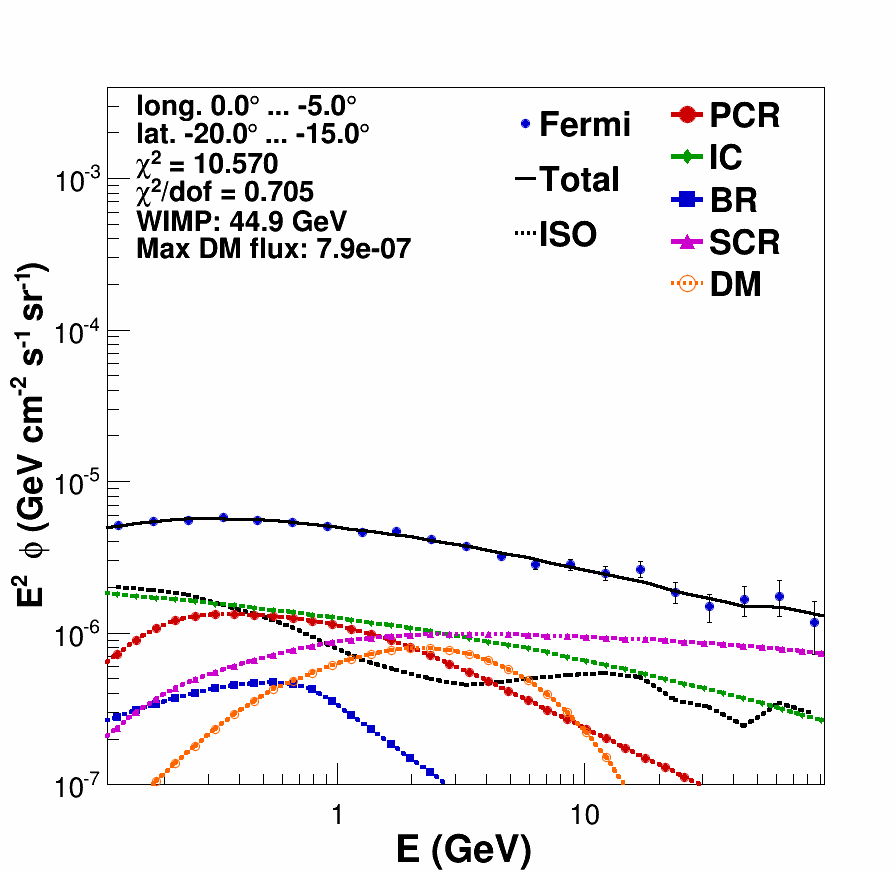}
\includegraphics[width=0.16\textwidth,height=0.16\textwidth,clip]{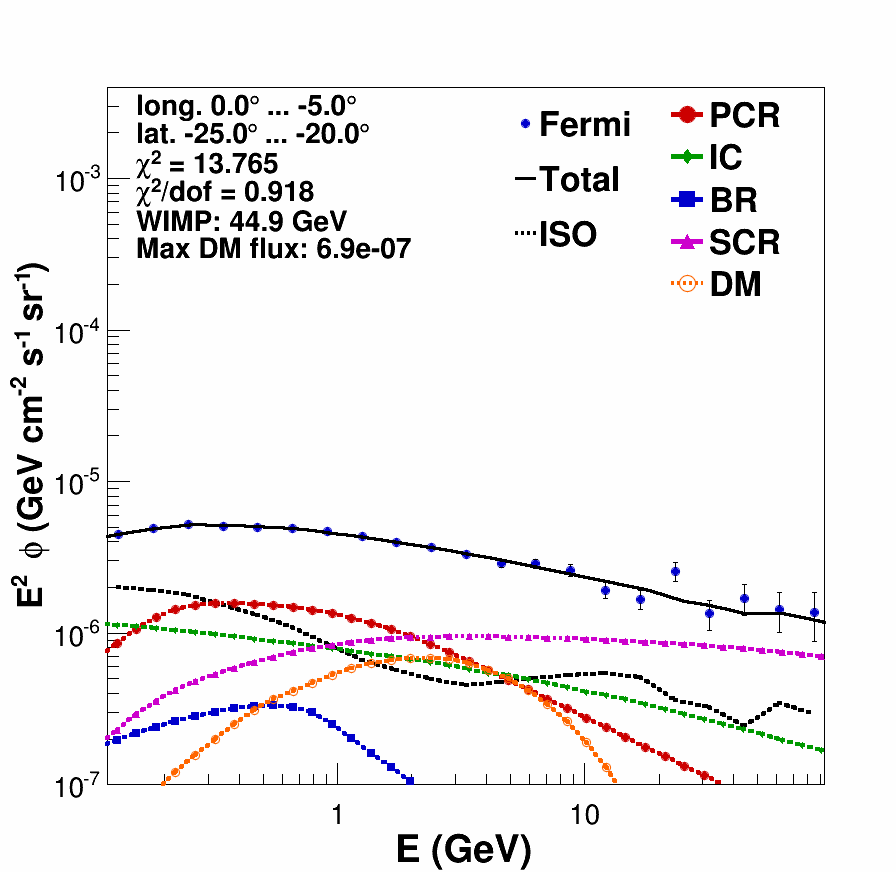}
\includegraphics[width=0.16\textwidth,height=0.16\textwidth,clip]{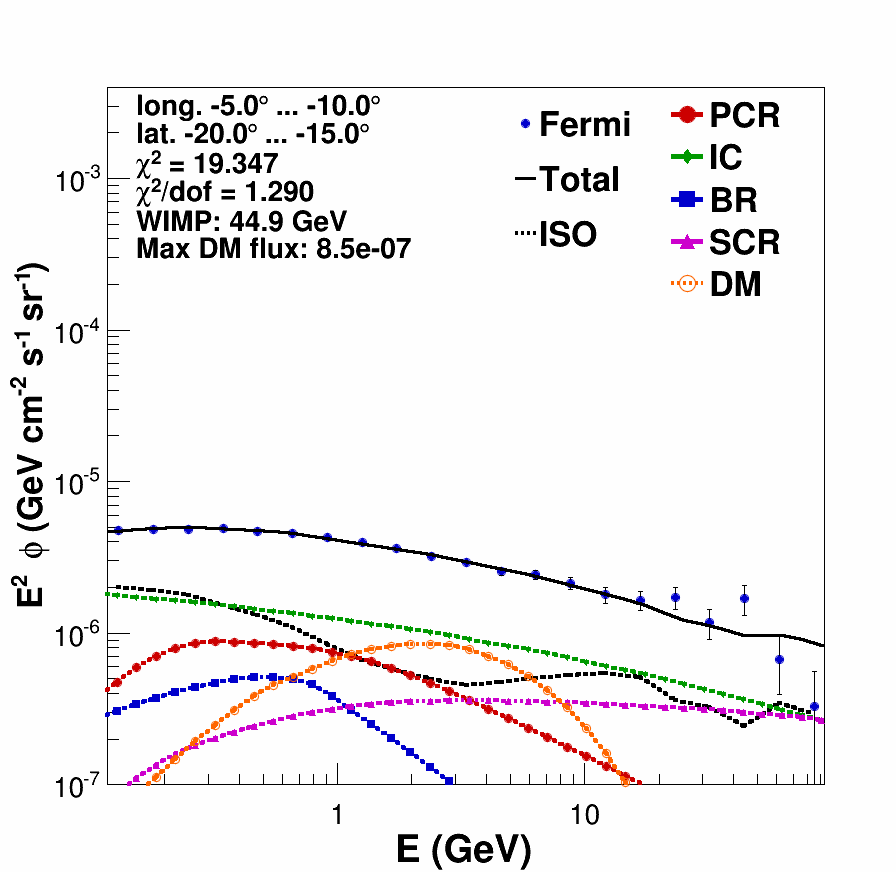}
\includegraphics[width=0.16\textwidth,height=0.16\textwidth,clip]{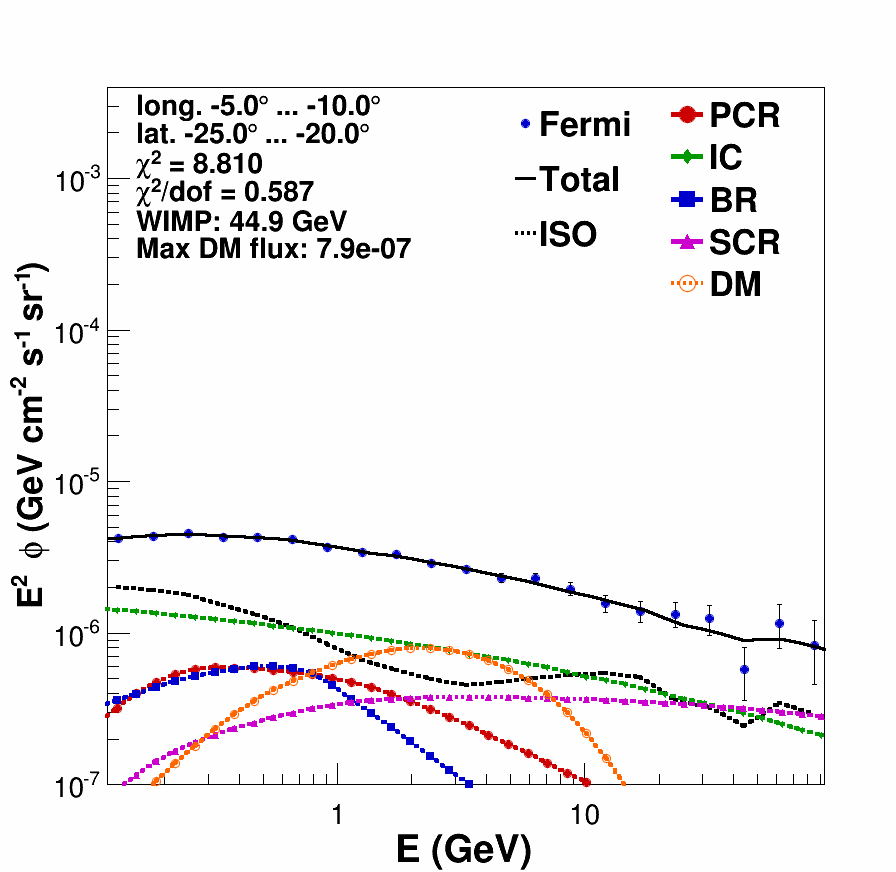}
\includegraphics[width=0.16\textwidth,height=0.16\textwidth,clip]{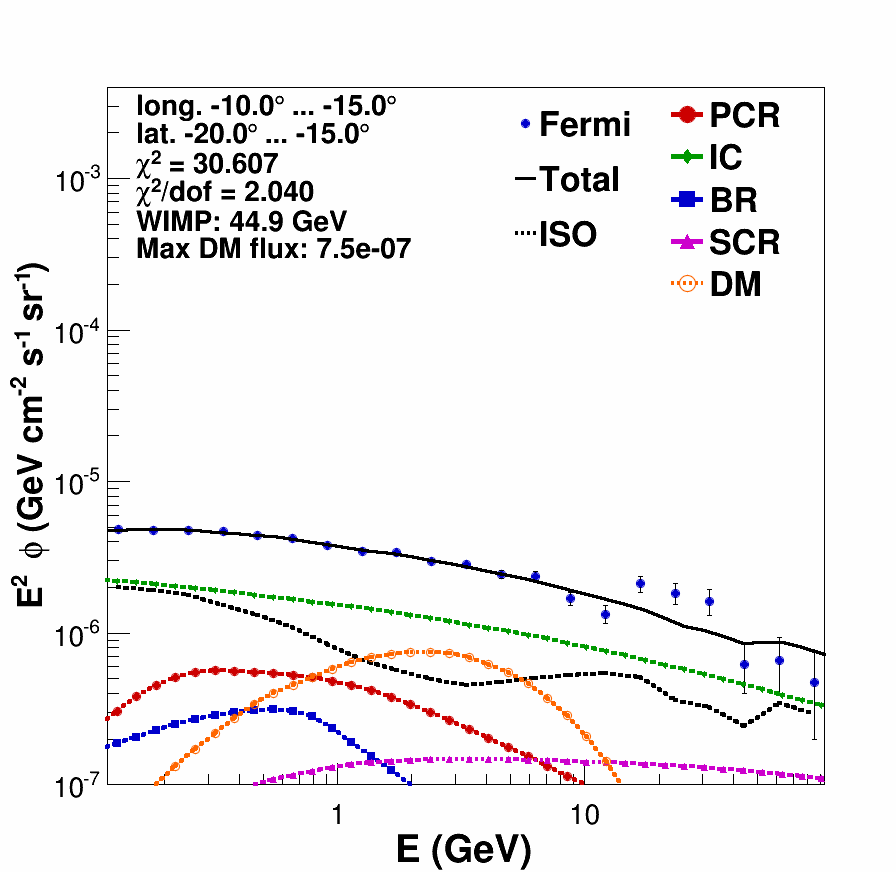}
\includegraphics[width=0.16\textwidth,height=0.16\textwidth,clip]{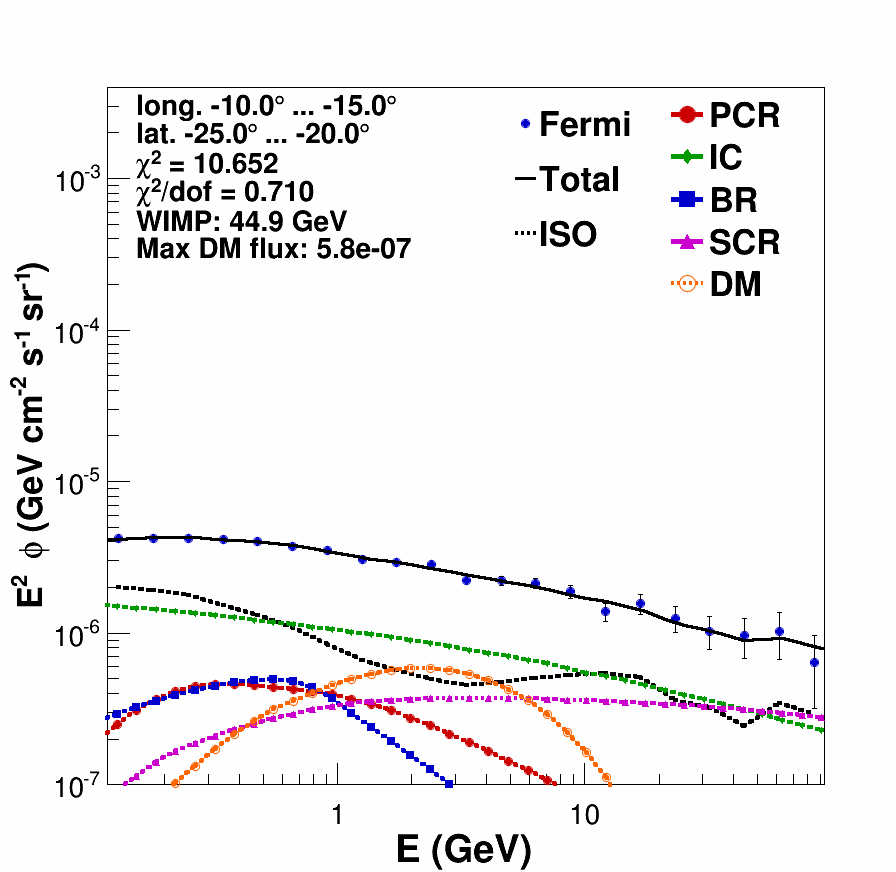}
\includegraphics[width=0.16\textwidth,height=0.16\textwidth,clip]{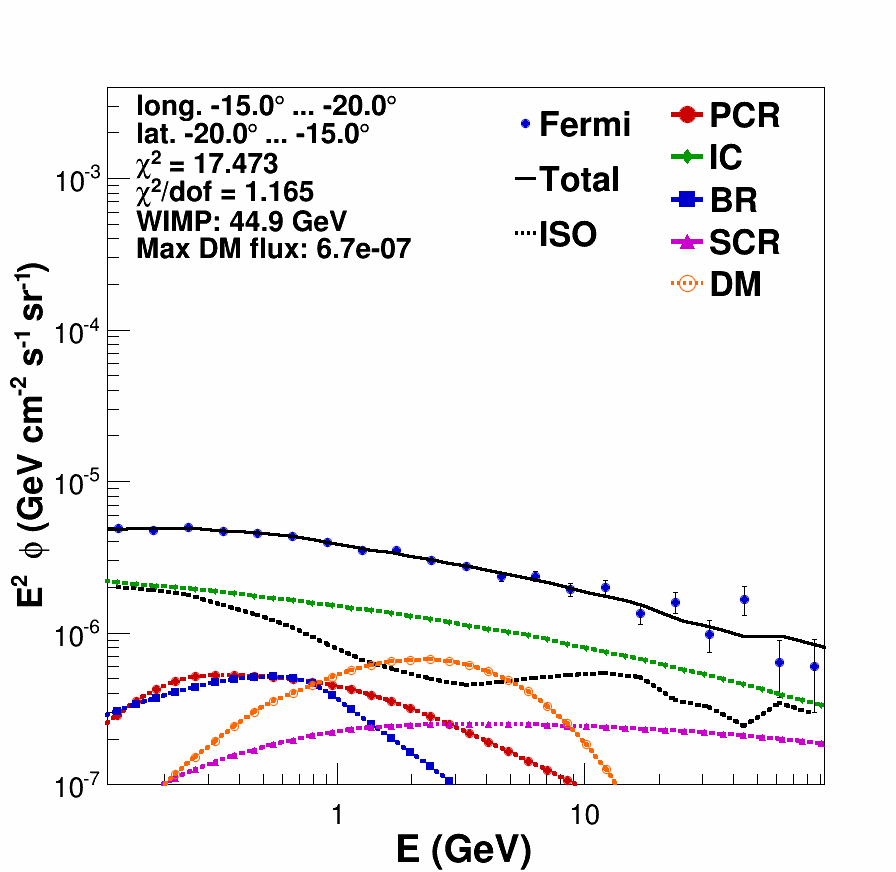}
\includegraphics[width=0.16\textwidth,height=0.16\textwidth,clip]{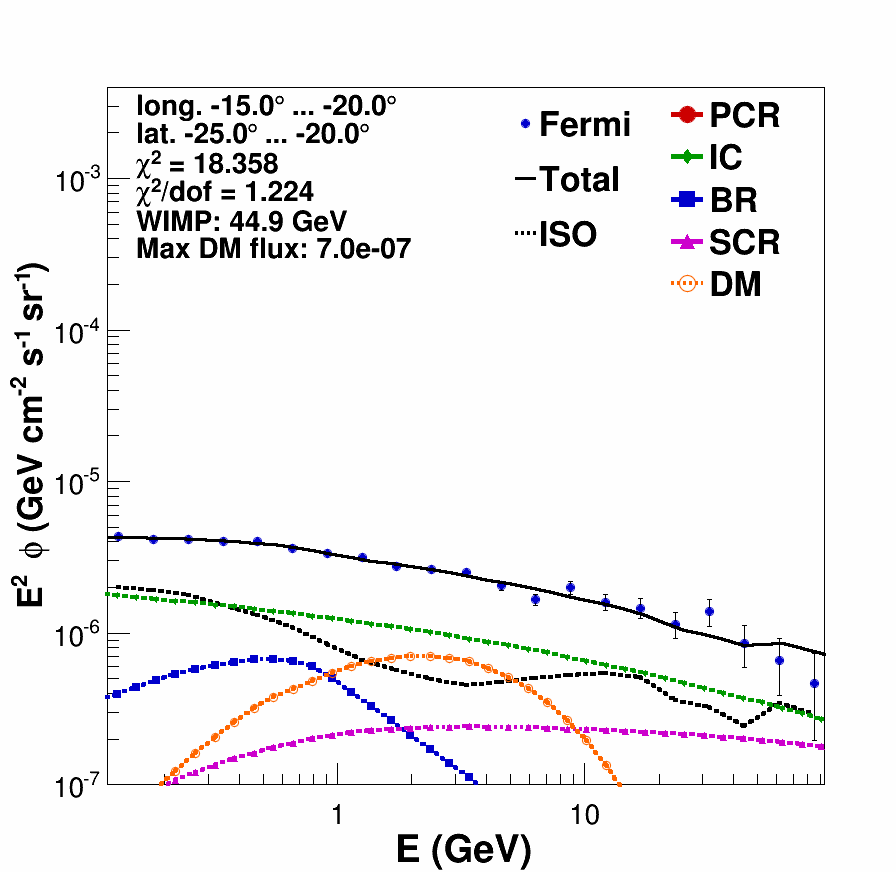}
\includegraphics[width=0.16\textwidth,height=0.16\textwidth,clip]{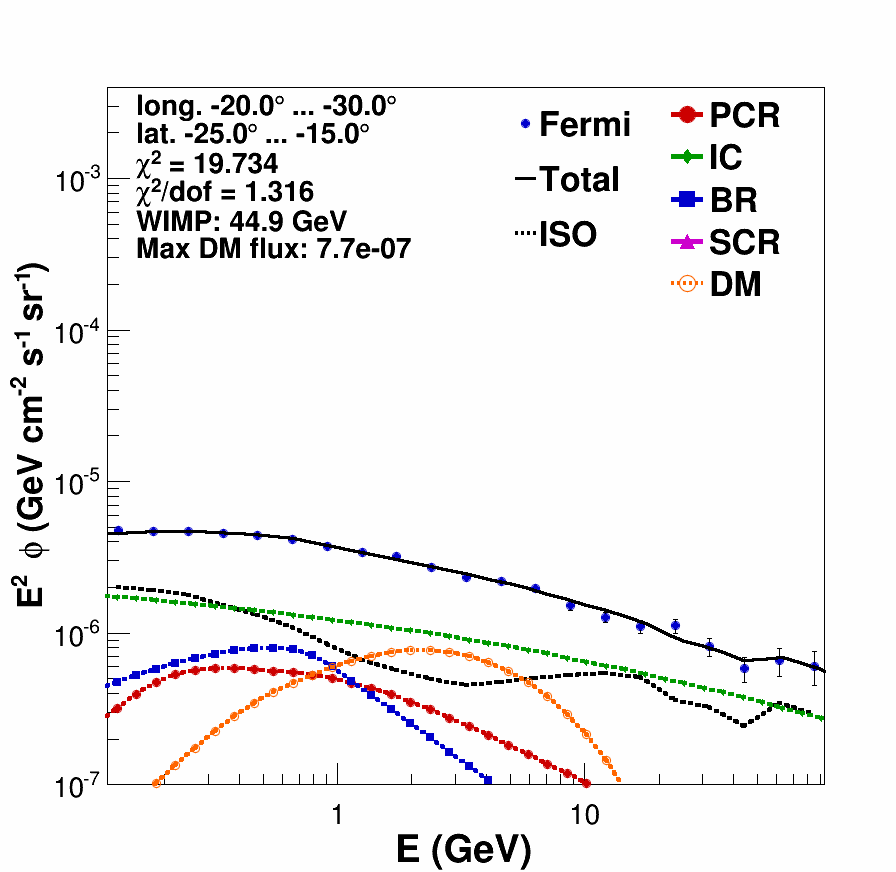}
\includegraphics[width=0.16\textwidth,height=0.16\textwidth,clip]{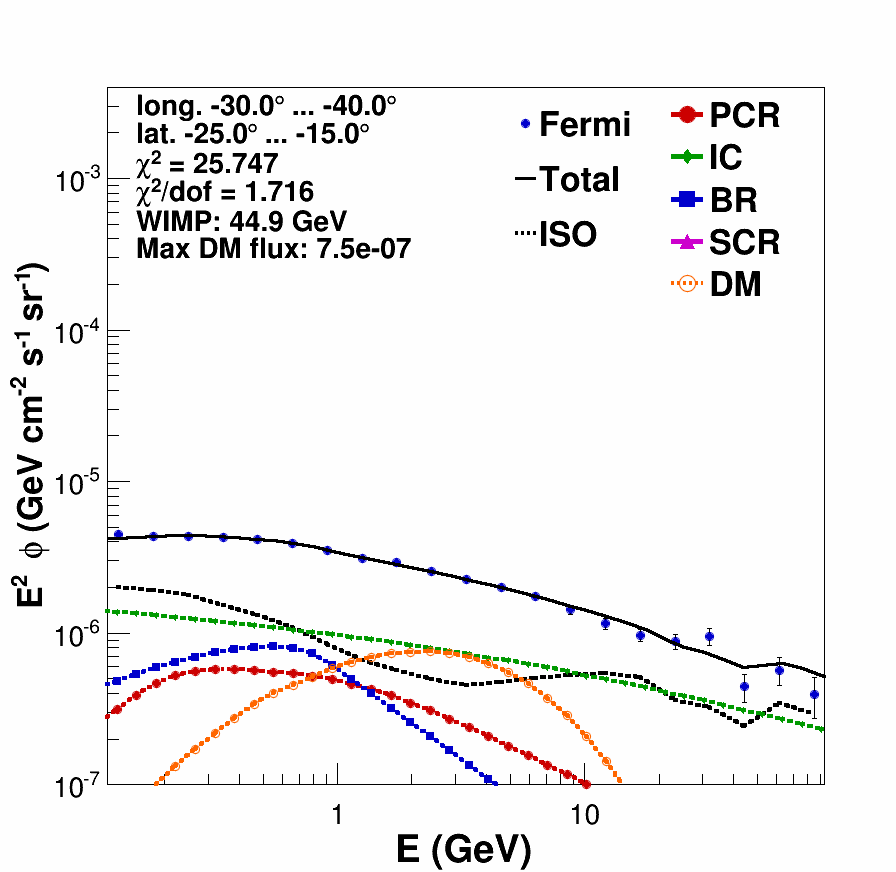}
\includegraphics[width=0.16\textwidth,height=0.16\textwidth,clip]{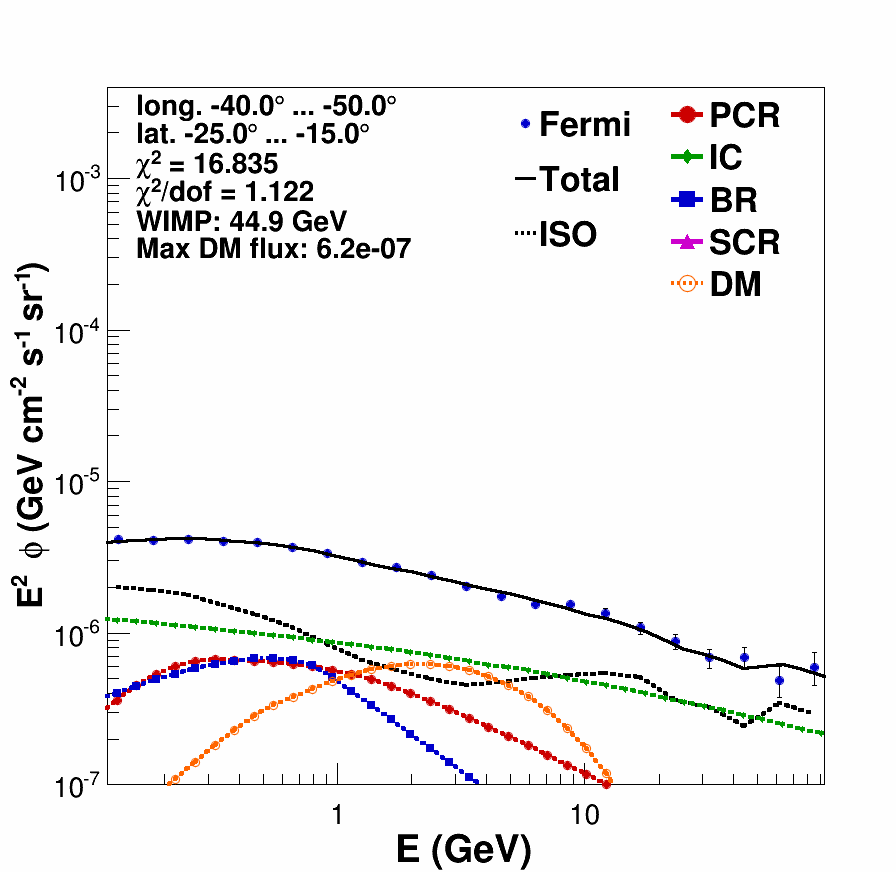}
\includegraphics[width=0.16\textwidth,height=0.16\textwidth,clip]{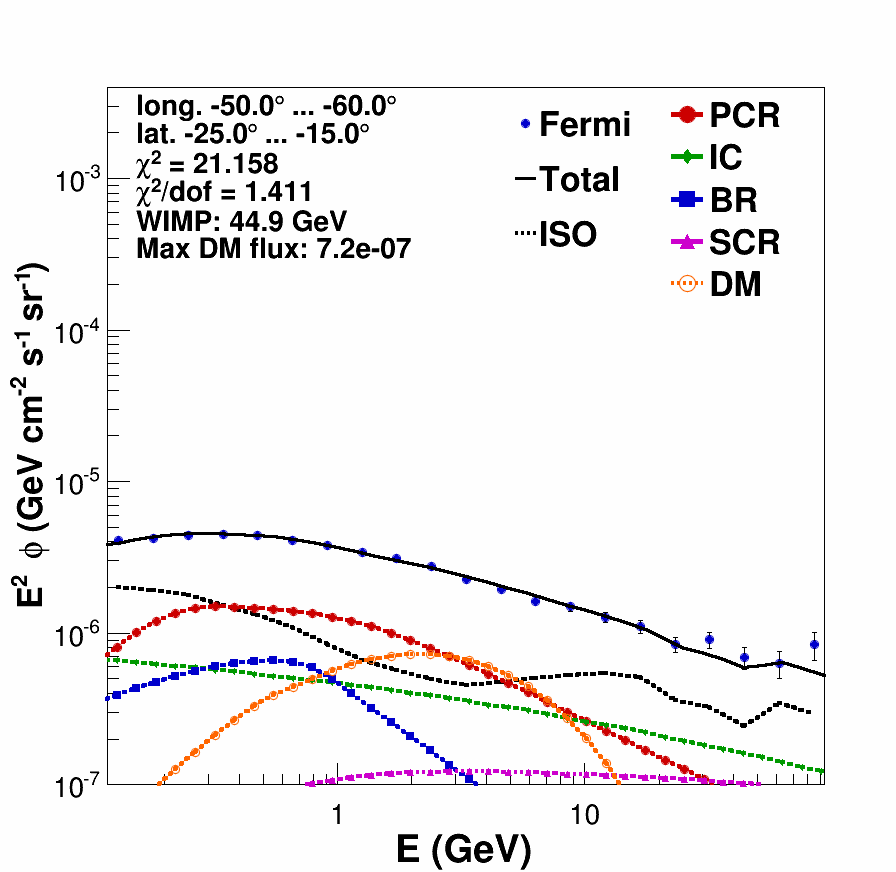}
\includegraphics[width=0.16\textwidth,height=0.16\textwidth,clip]{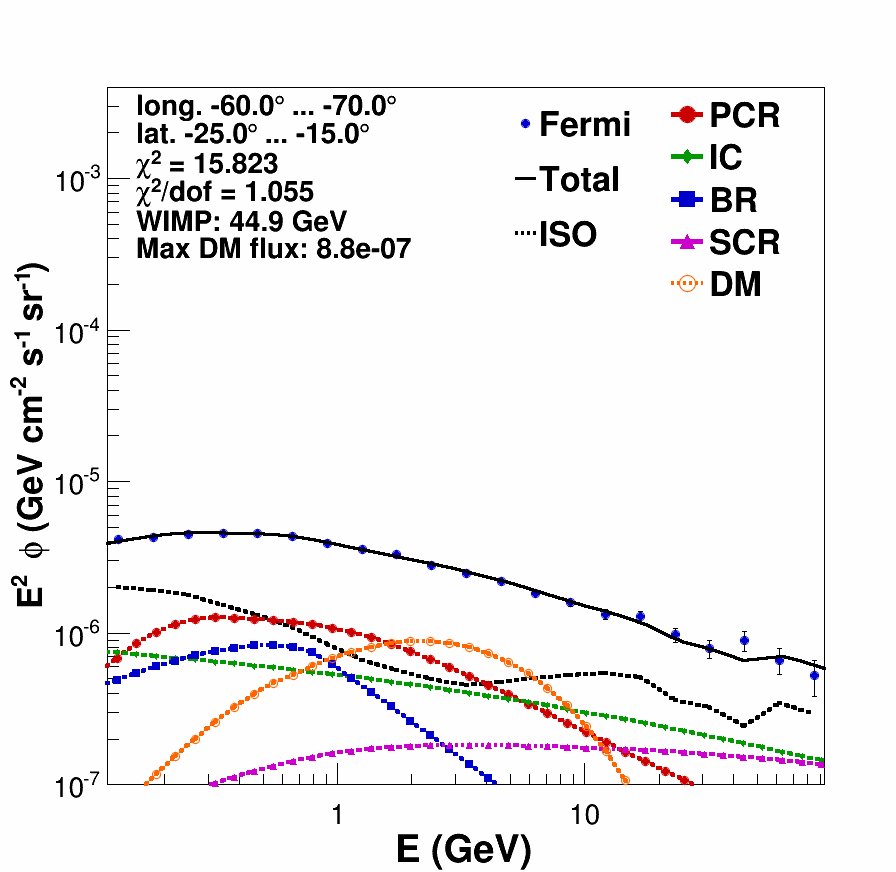}
\includegraphics[width=0.16\textwidth,height=0.16\textwidth,clip]{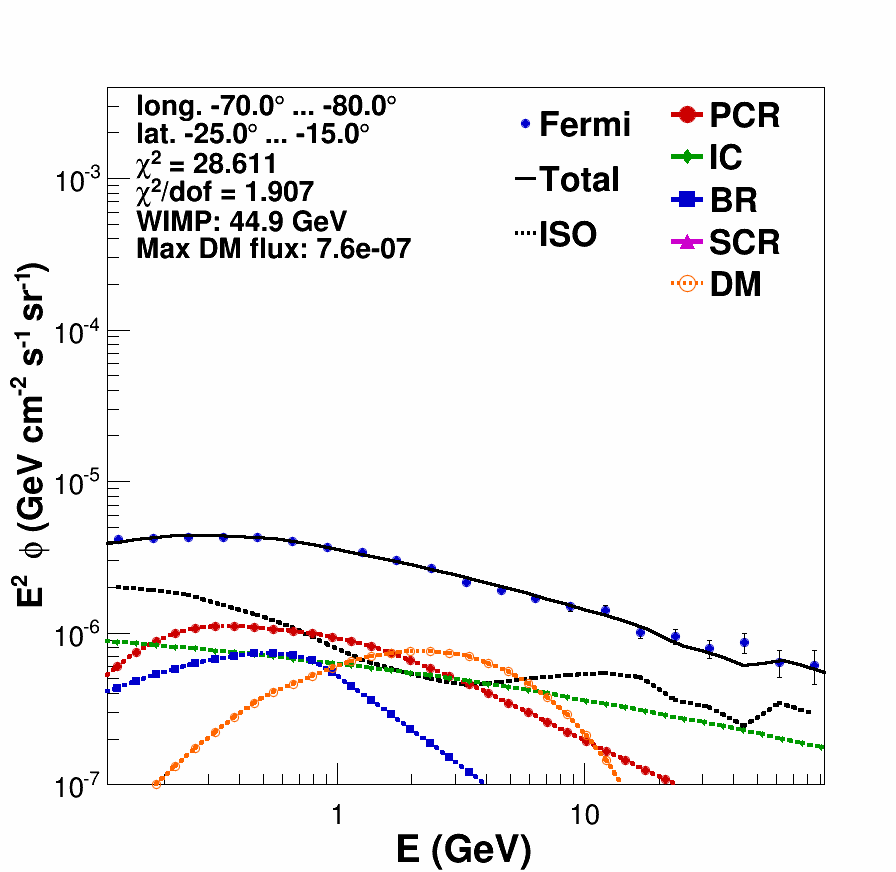}
\includegraphics[width=0.16\textwidth,height=0.16\textwidth,clip]{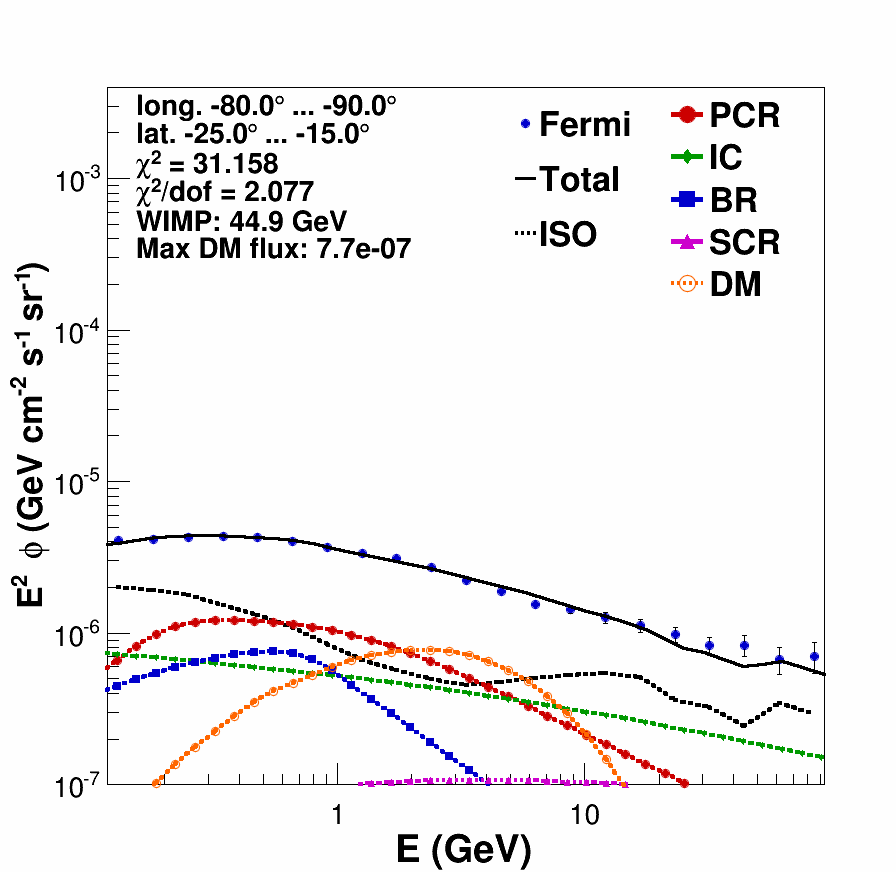}
\includegraphics[width=0.16\textwidth,height=0.16\textwidth,clip]{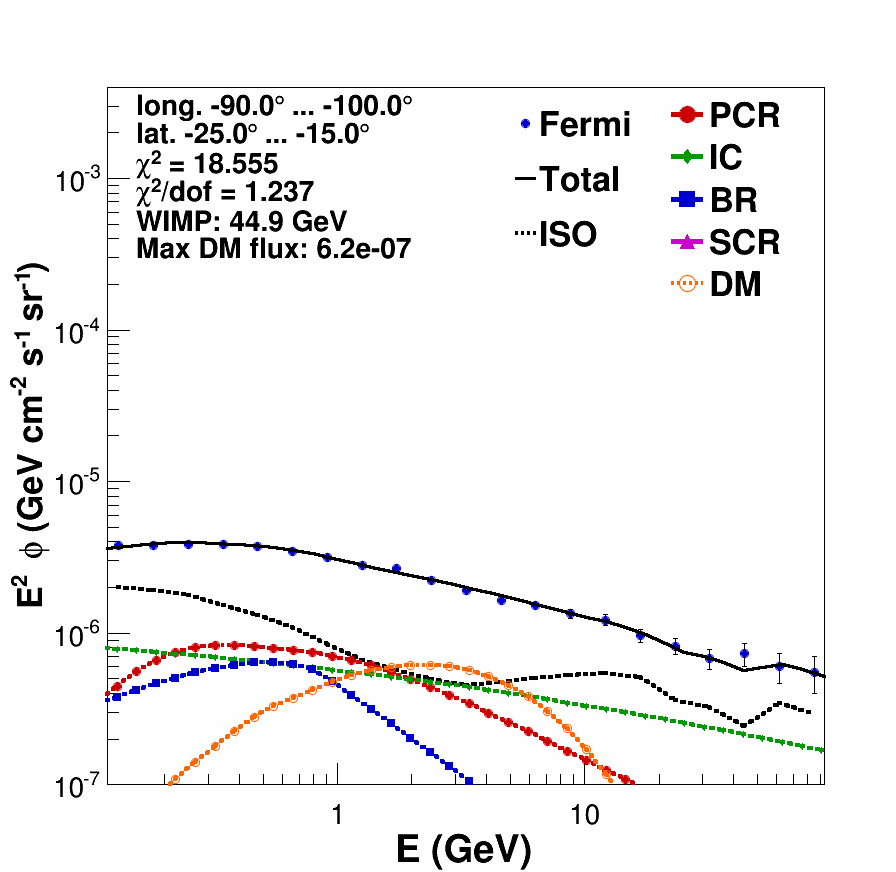}
\includegraphics[width=0.16\textwidth,height=0.16\textwidth,clip]{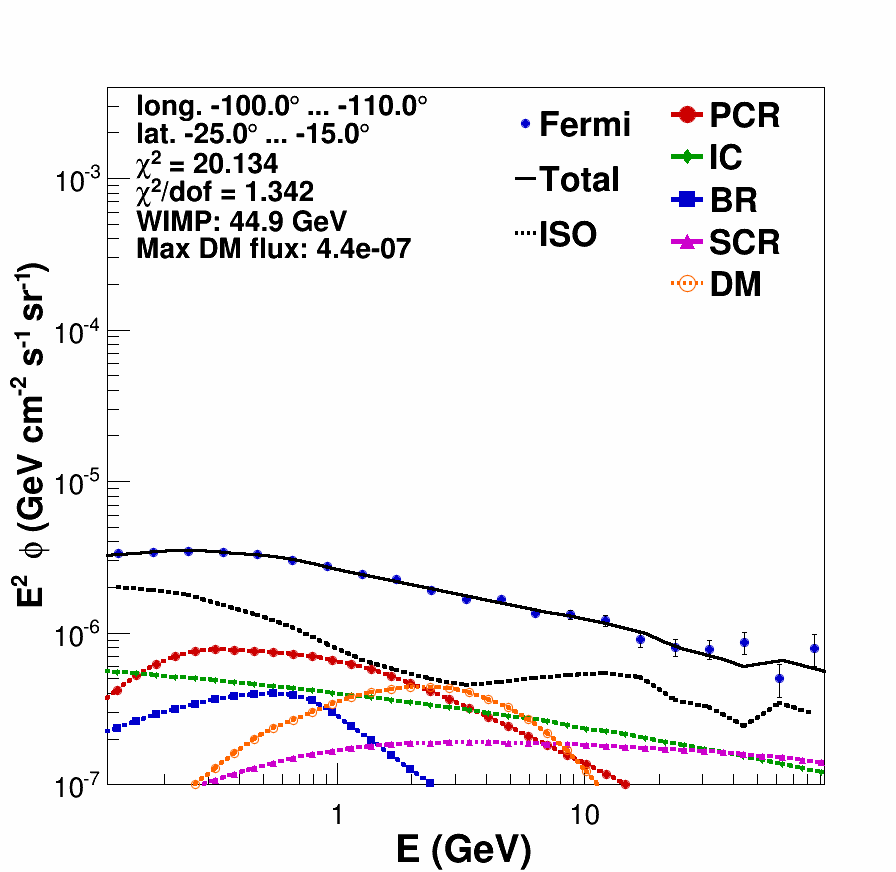}
\includegraphics[width=0.16\textwidth,height=0.16\textwidth,clip]{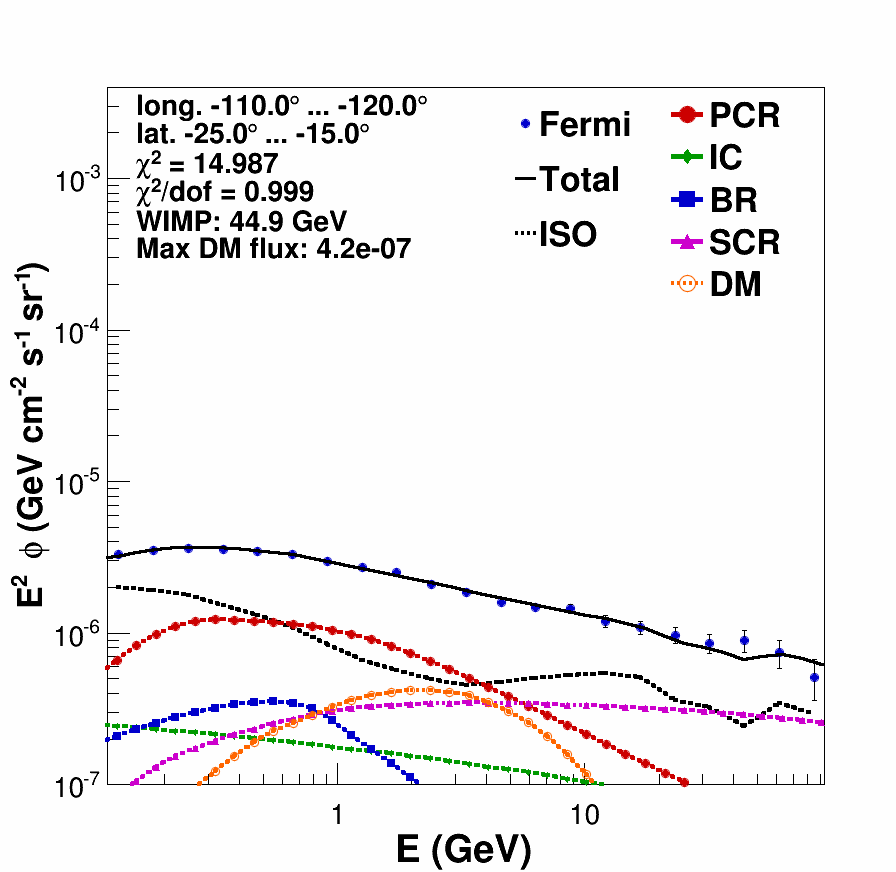}
\includegraphics[width=0.16\textwidth,height=0.16\textwidth,clip]{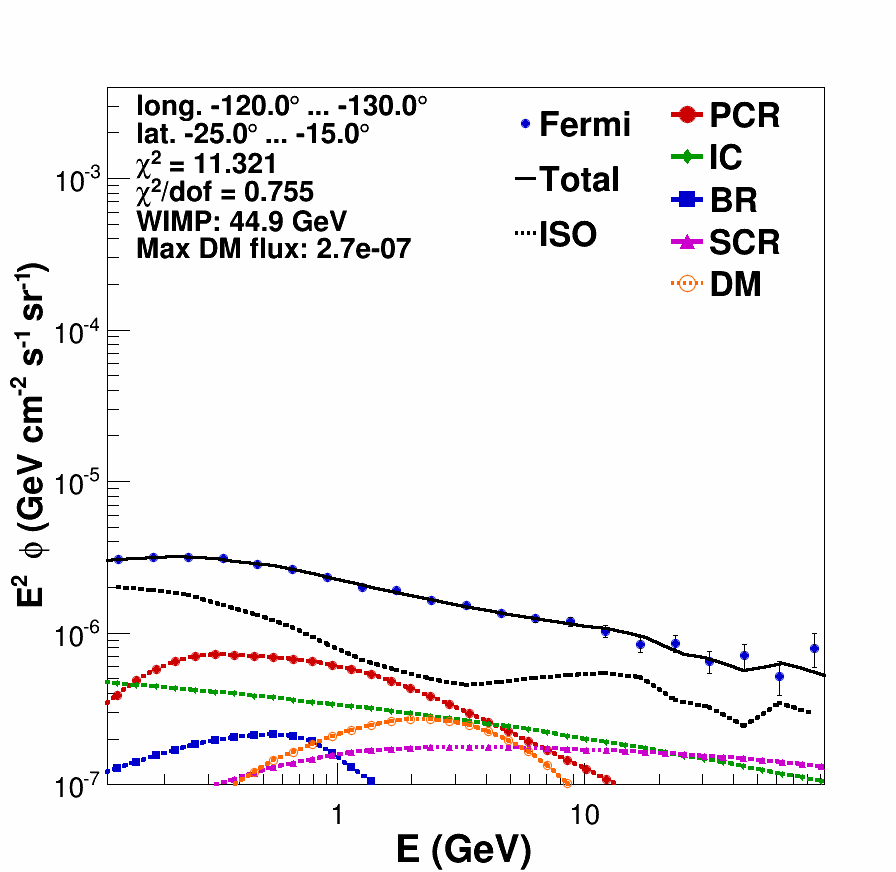}
\includegraphics[width=0.16\textwidth,height=0.16\textwidth,clip]{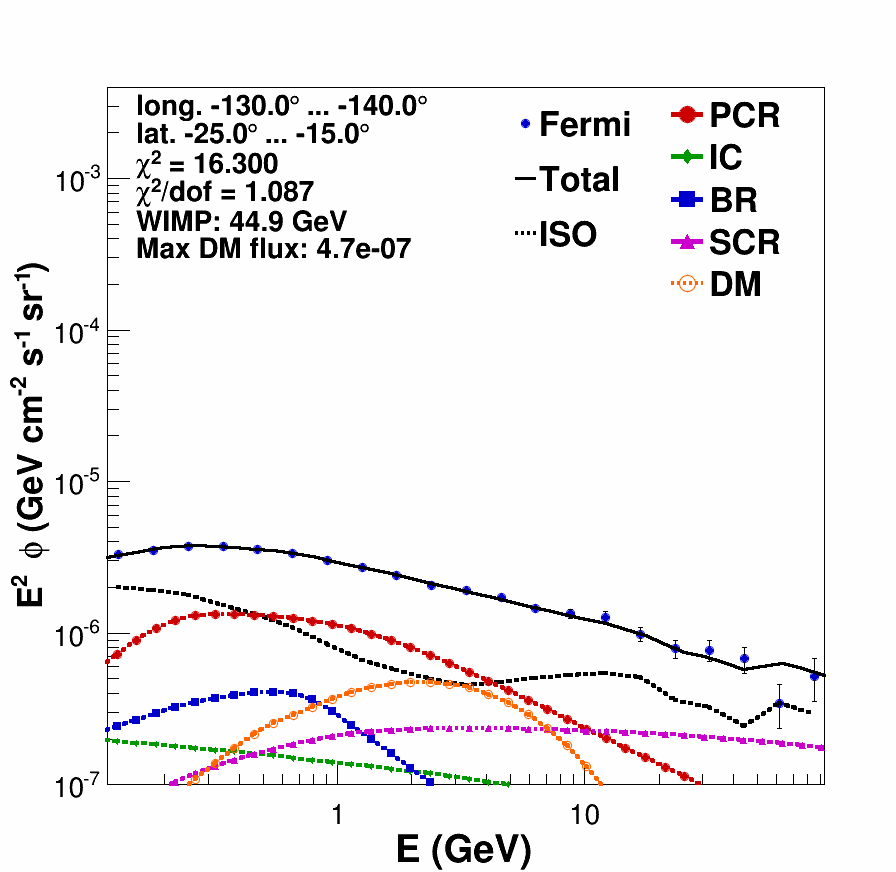}
\includegraphics[width=0.16\textwidth,height=0.16\textwidth,clip]{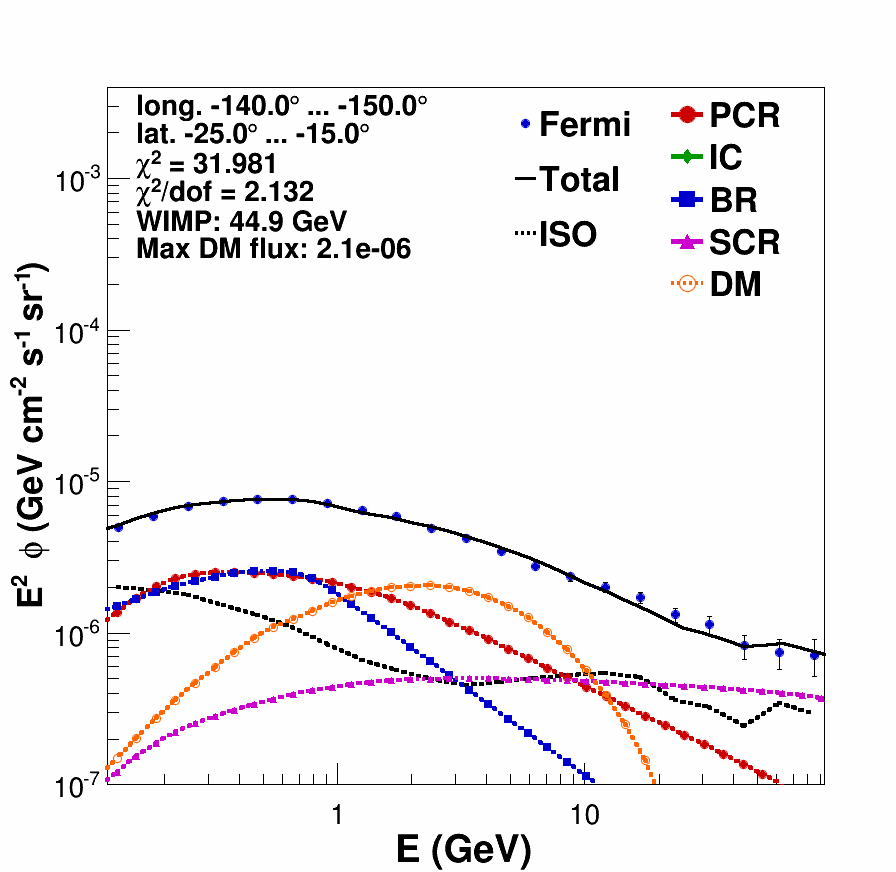}
\includegraphics[width=0.16\textwidth,height=0.16\textwidth,clip]{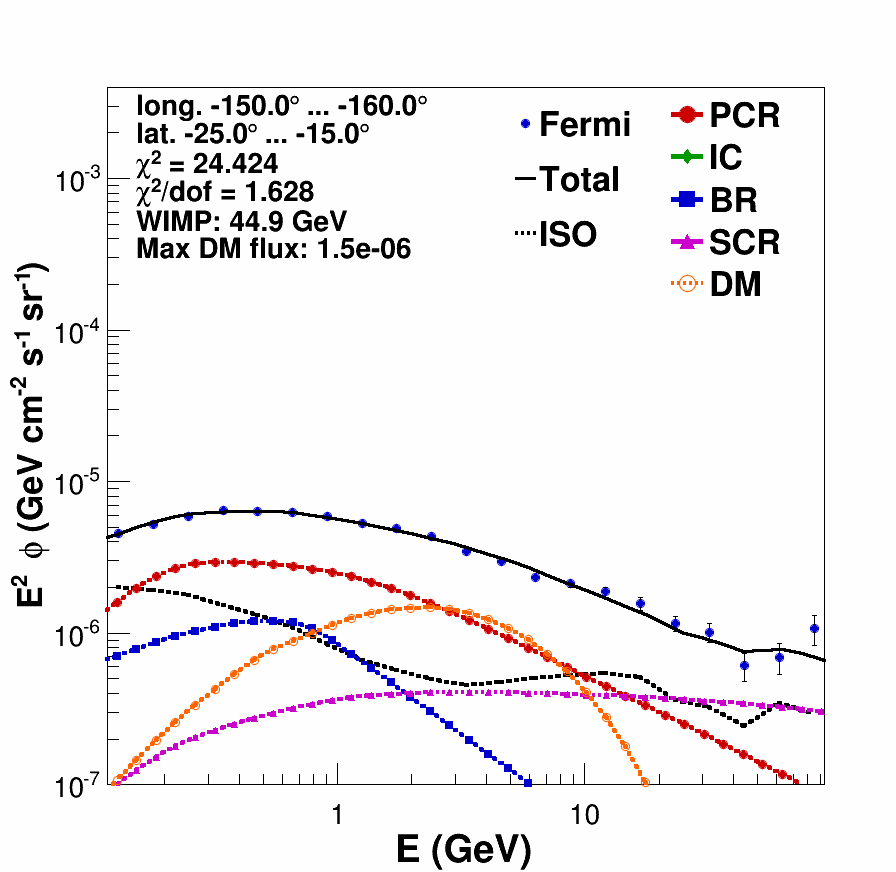}
\includegraphics[width=0.16\textwidth,height=0.16\textwidth,clip]{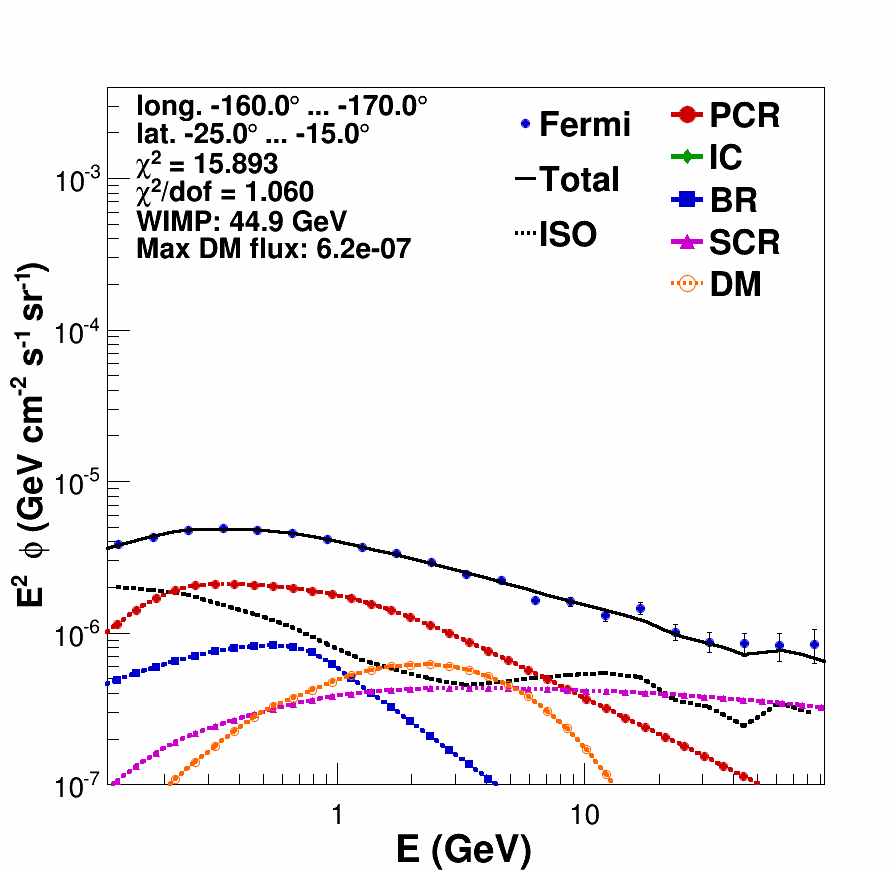}
\includegraphics[width=0.16\textwidth,height=0.16\textwidth,clip]{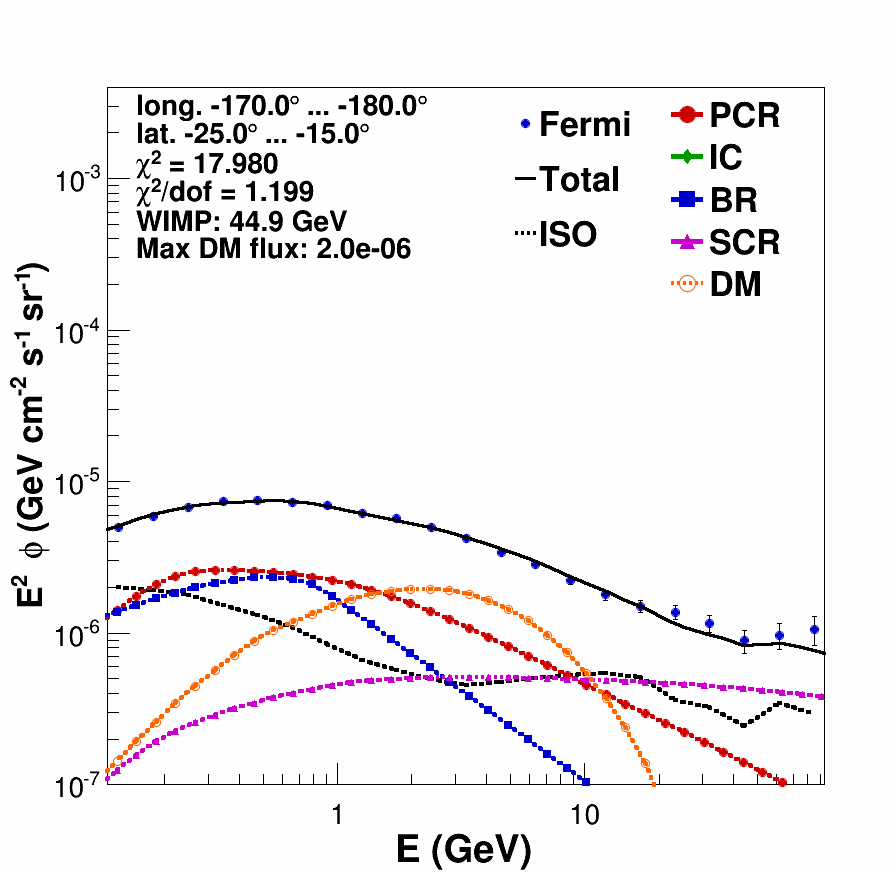}%%%%%%r14
\caption[]{Template fits for latitudes  with $-25.0^\circ<b<-15.0^\circ$ and longitudes decreasing from 180$^\circ$ to -180$^\circ$.} \label{F47}
\end{figure}
\clearpage
\begin{figure}
\centering
\includegraphics[width=0.16\textwidth,height=0.16\textwidth,clip]{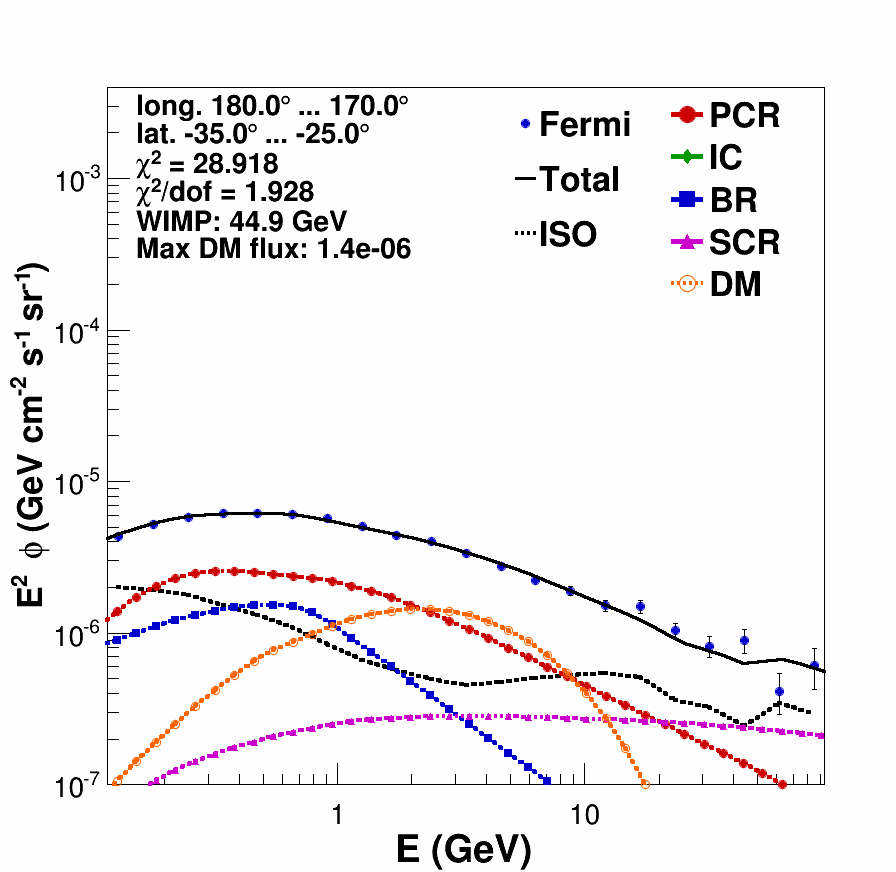}
\includegraphics[width=0.16\textwidth,height=0.16\textwidth,clip]{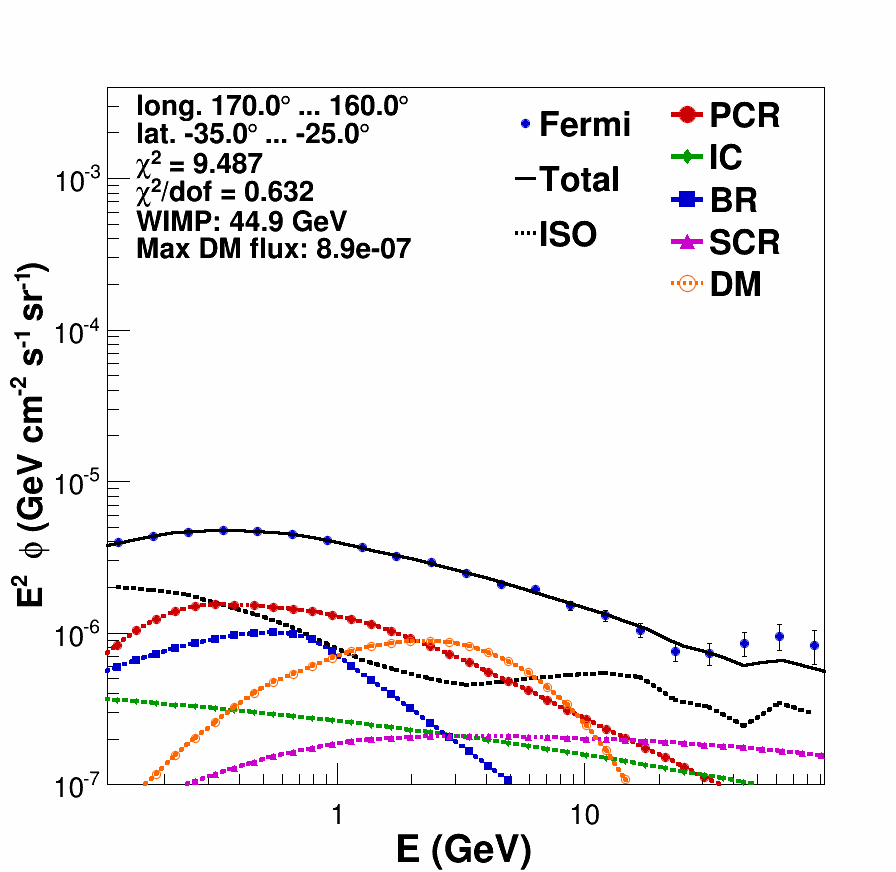}
\includegraphics[width=0.16\textwidth,height=0.16\textwidth,clip]{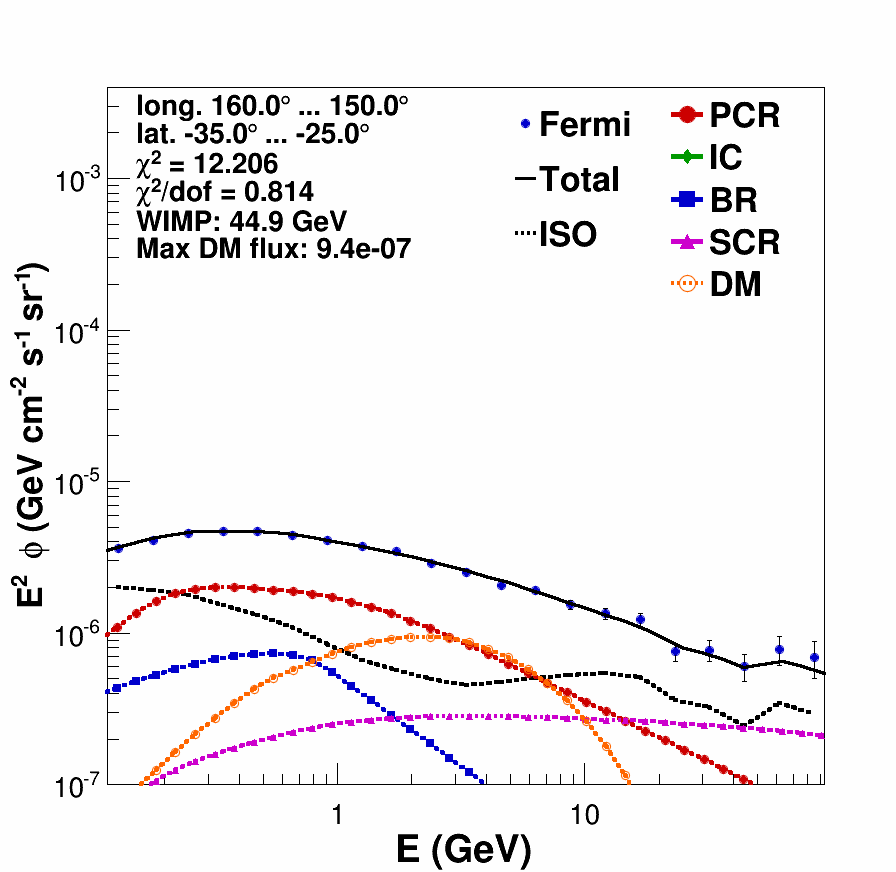}
\includegraphics[width=0.16\textwidth,height=0.16\textwidth,clip]{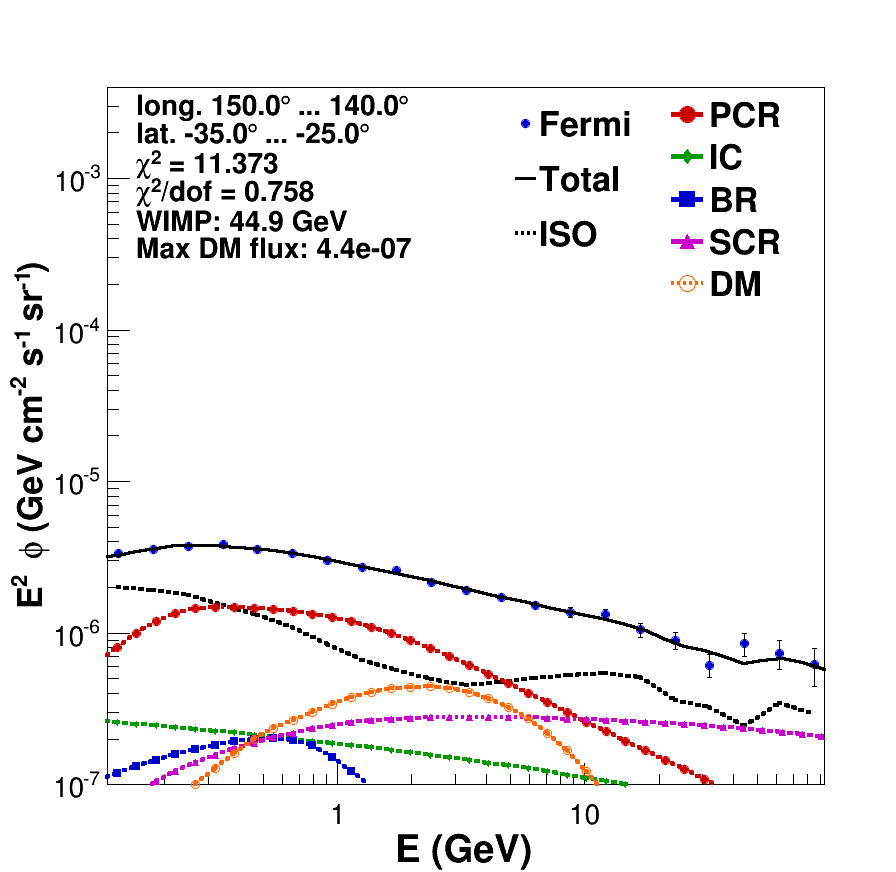}
\includegraphics[width=0.16\textwidth,height=0.16\textwidth,clip]{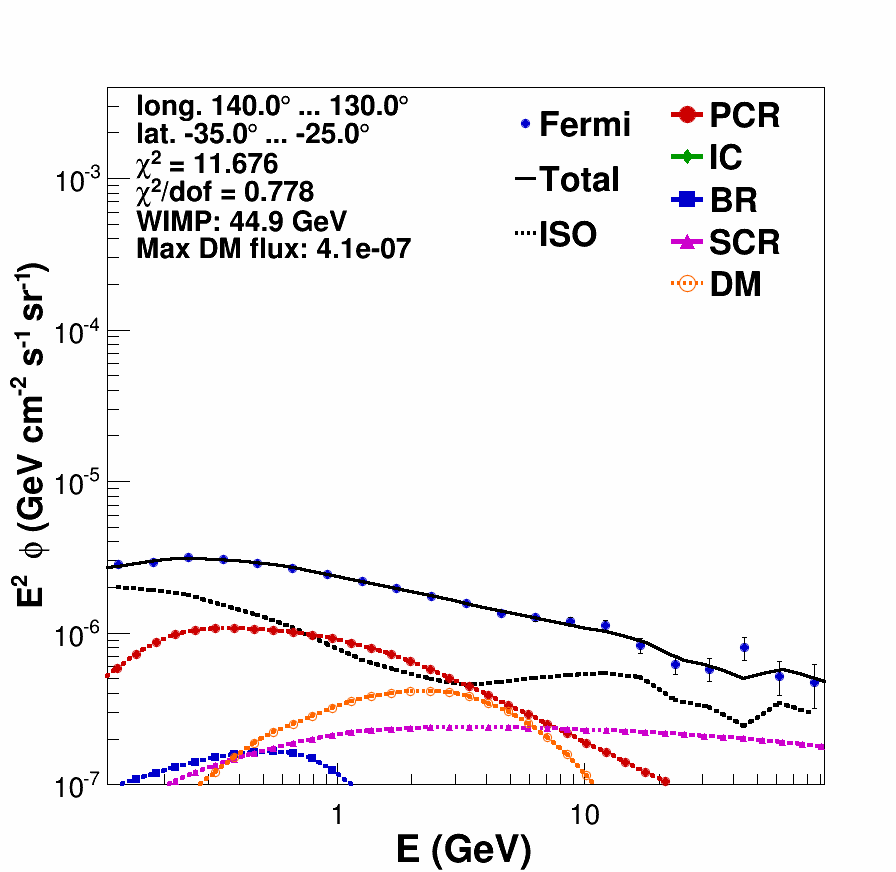}
\includegraphics[width=0.16\textwidth,height=0.16\textwidth,clip]{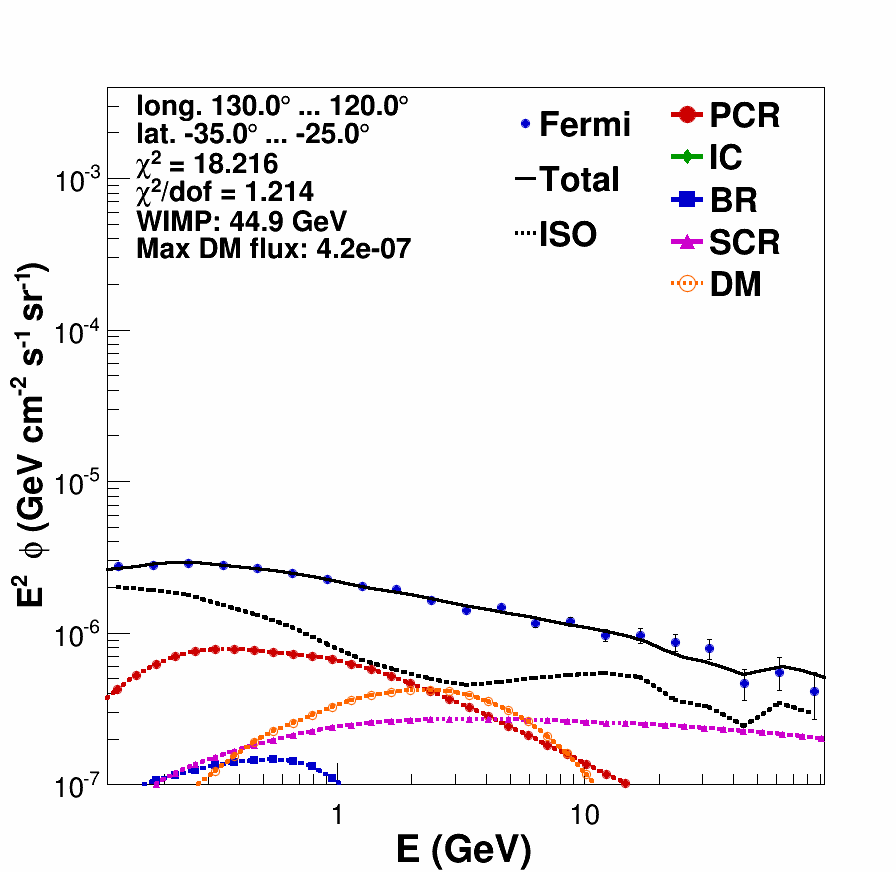}
\includegraphics[width=0.16\textwidth,height=0.16\textwidth,clip]{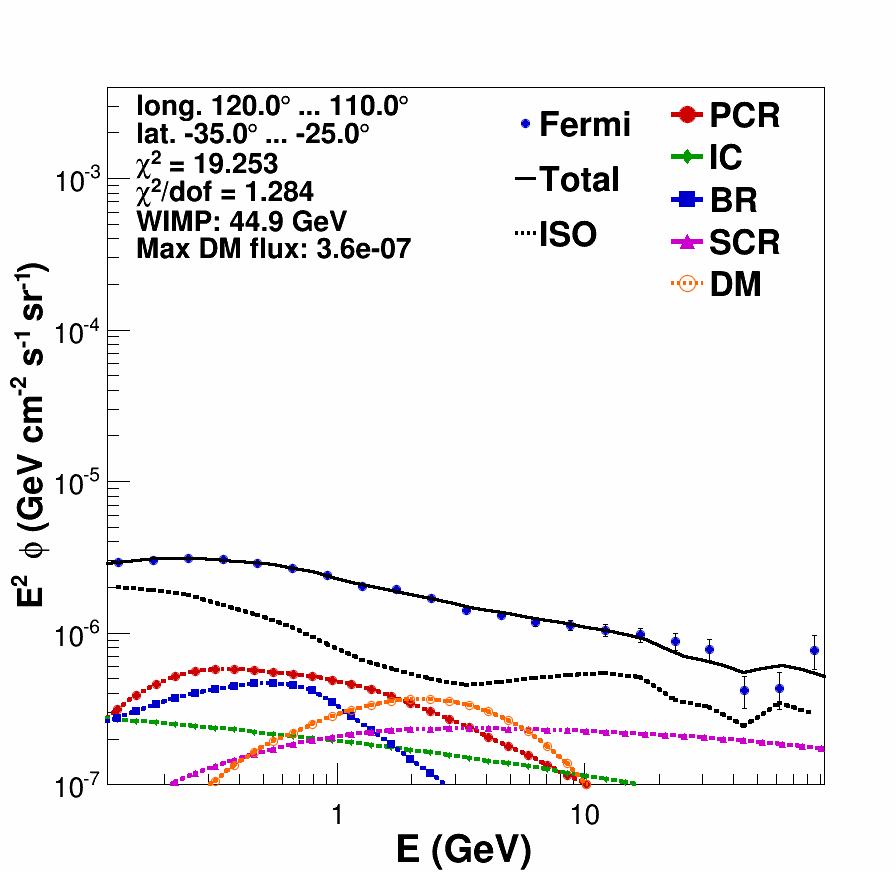}
\includegraphics[width=0.16\textwidth,height=0.16\textwidth,clip]{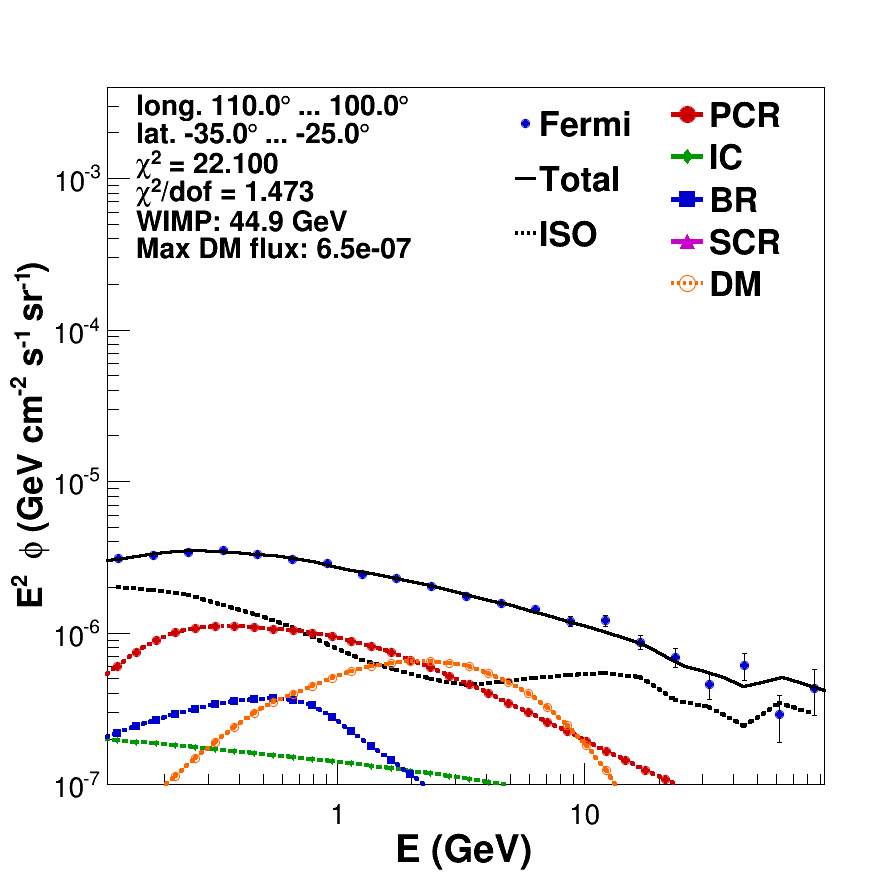}
\includegraphics[width=0.16\textwidth,height=0.16\textwidth,clip]{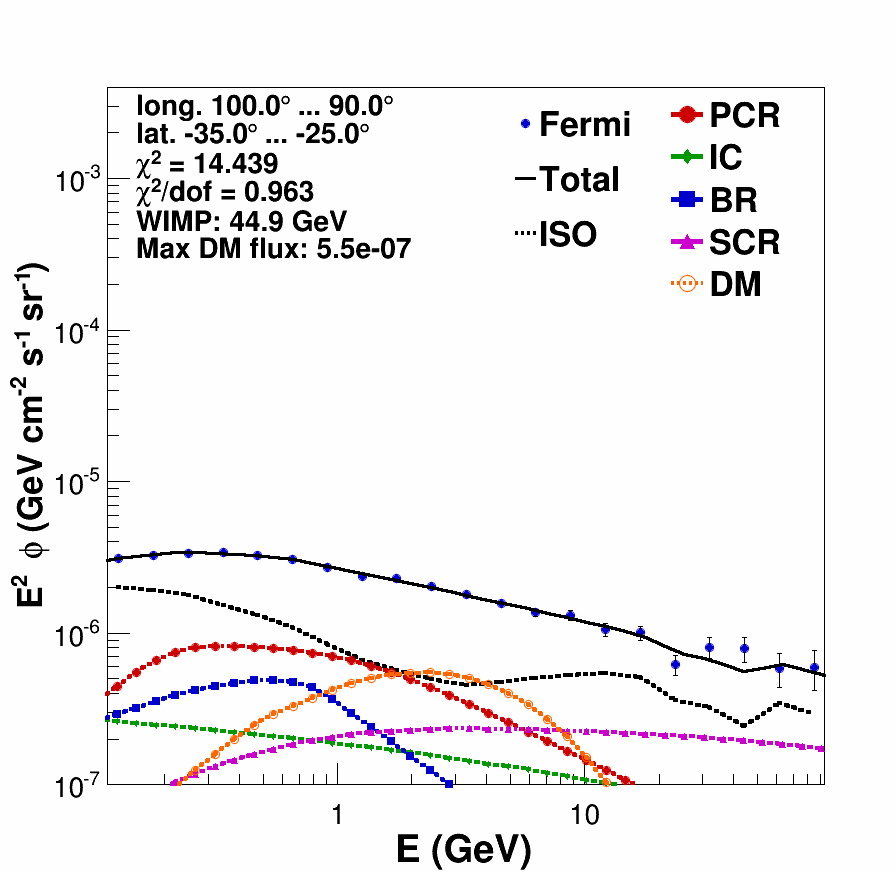}
\includegraphics[width=0.16\textwidth,height=0.16\textwidth,clip]{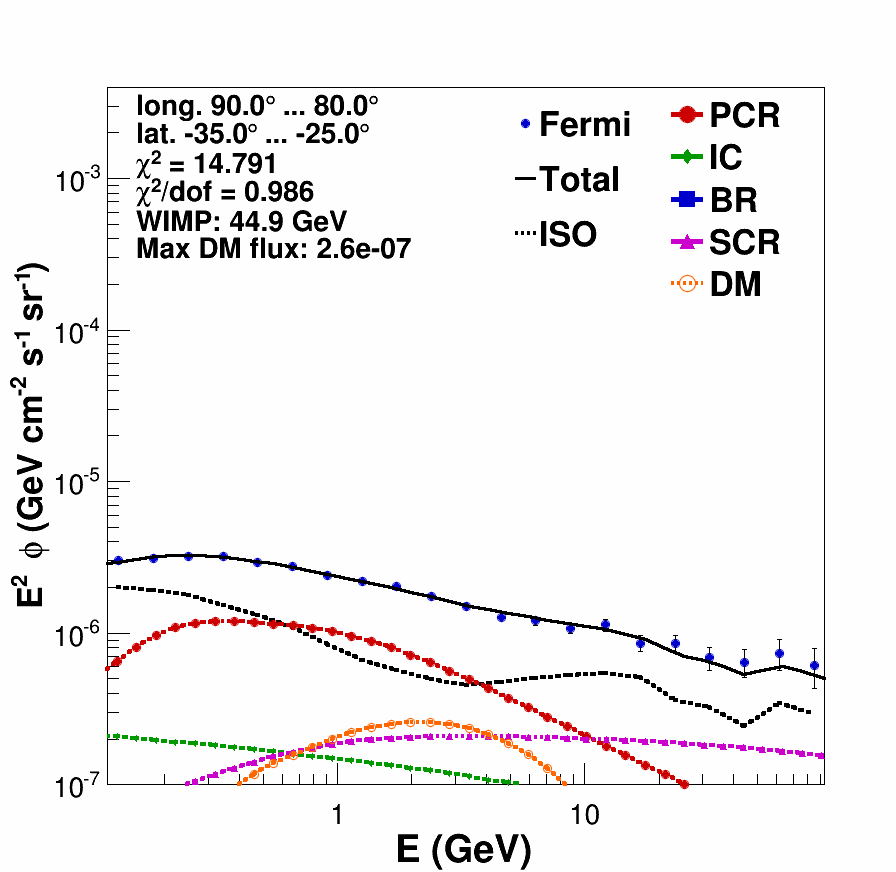}
\includegraphics[width=0.16\textwidth,height=0.16\textwidth,clip]{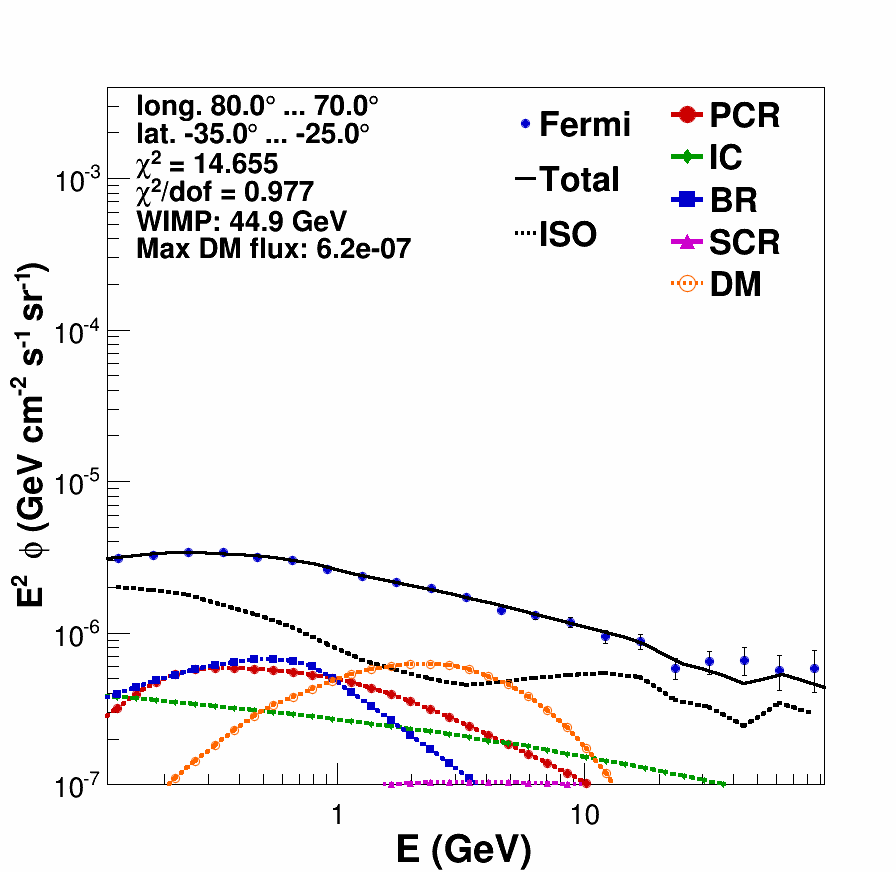}
\includegraphics[width=0.16\textwidth,height=0.16\textwidth,clip]{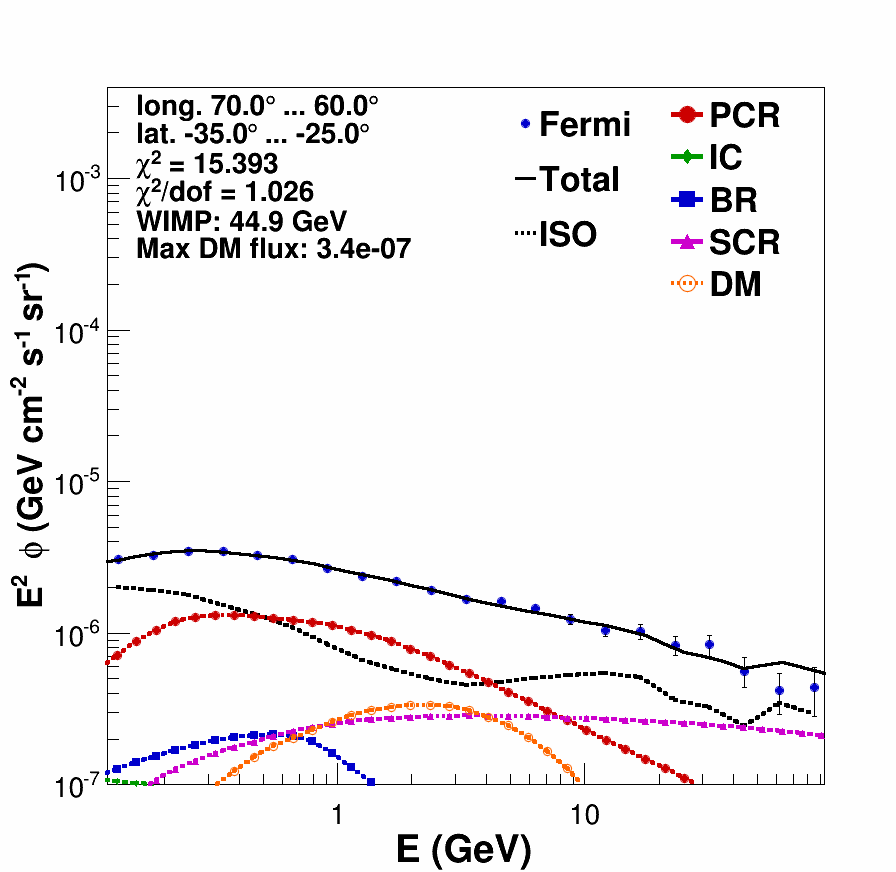}
\includegraphics[width=0.16\textwidth,height=0.16\textwidth,clip]{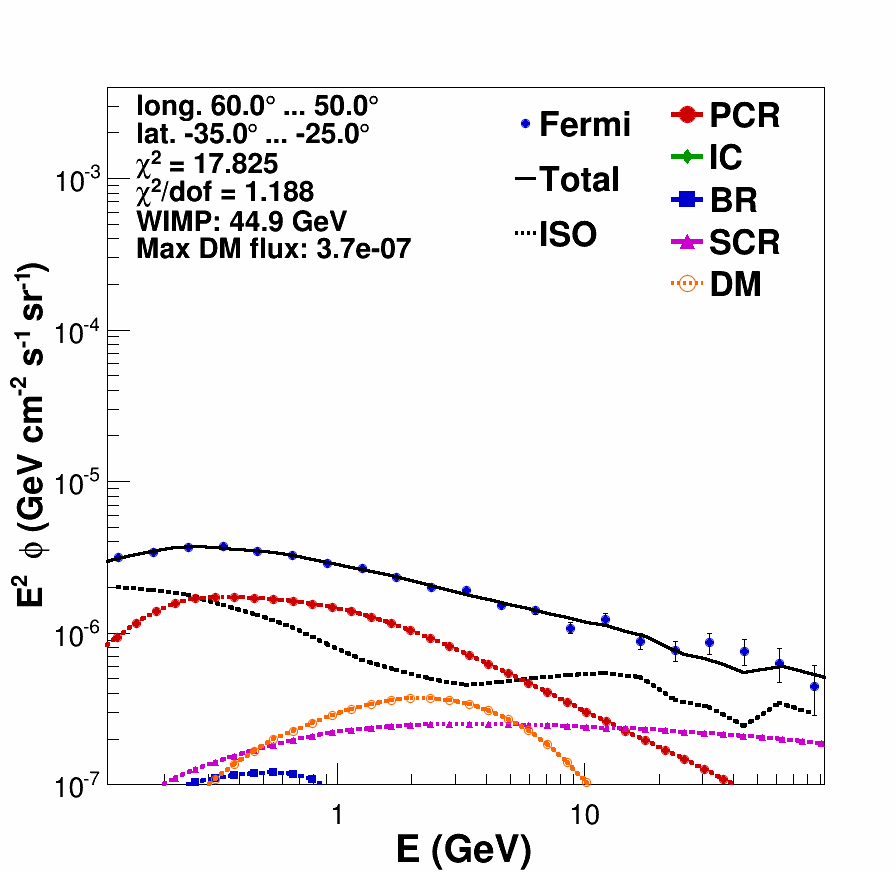}
\includegraphics[width=0.16\textwidth,height=0.16\textwidth,clip]{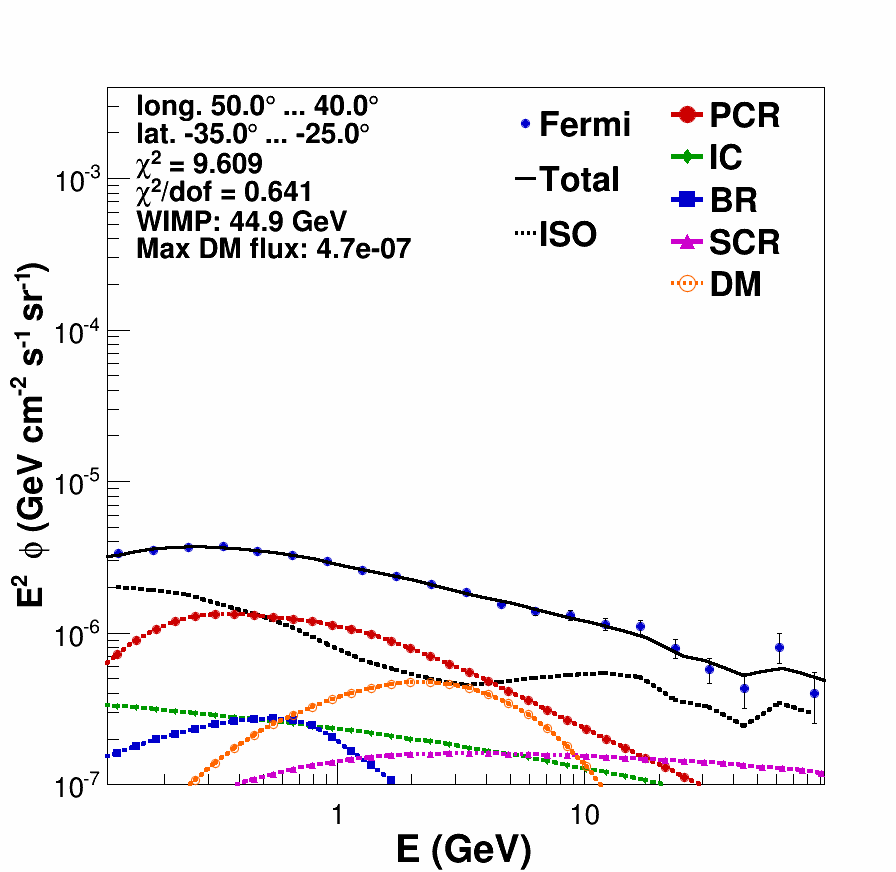}
\includegraphics[width=0.16\textwidth,height=0.16\textwidth,clip]{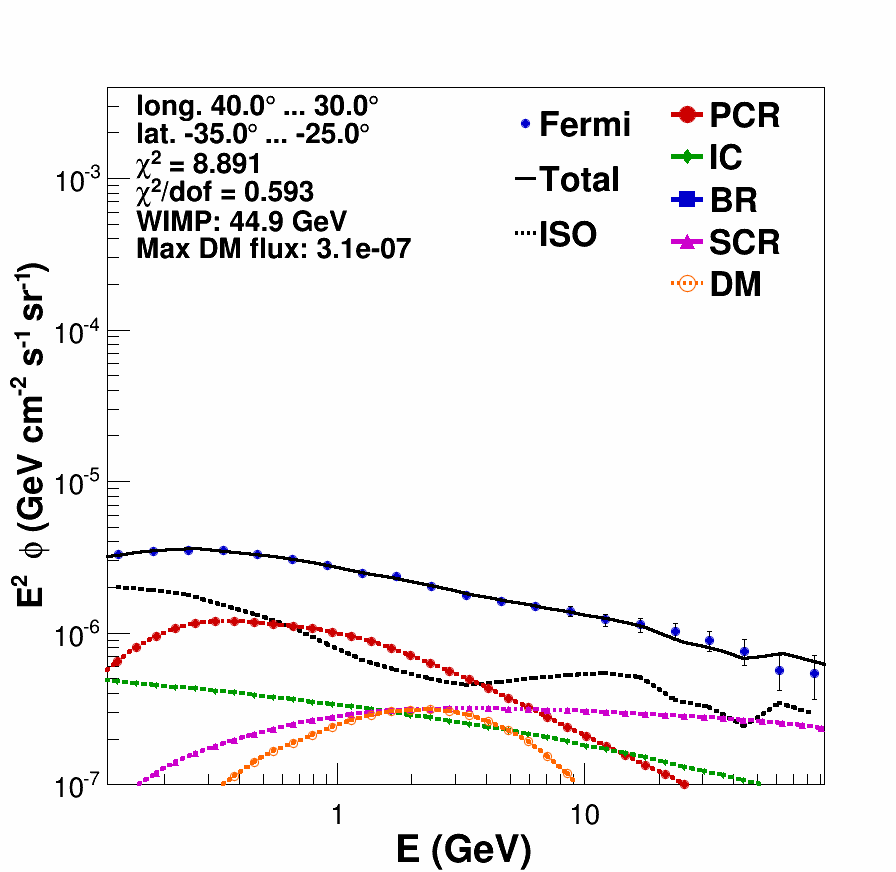}
\includegraphics[width=0.16\textwidth,height=0.16\textwidth,clip]{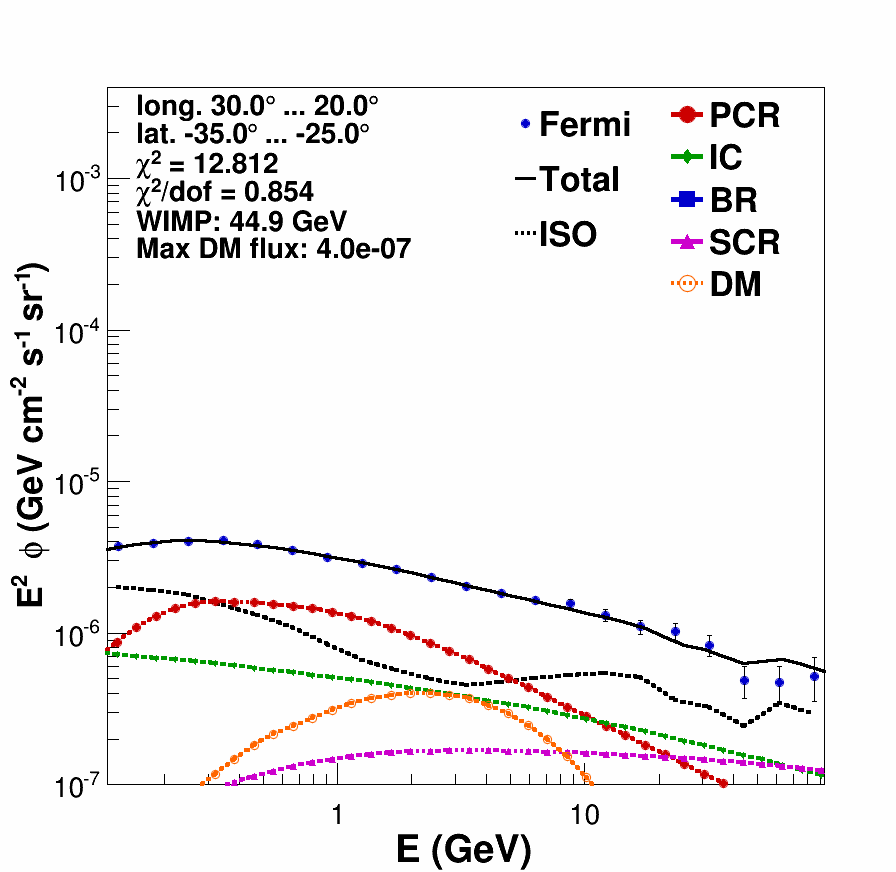}
\includegraphics[width=0.16\textwidth,height=0.16\textwidth,clip]{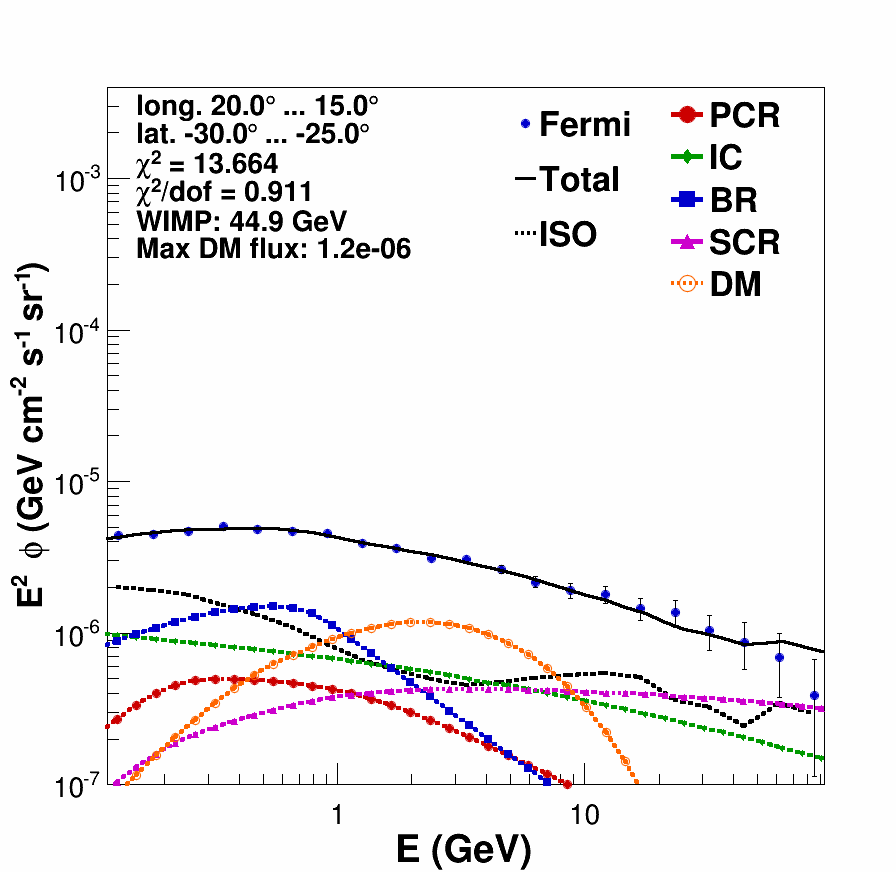}
\includegraphics[width=0.16\textwidth,height=0.16\textwidth,clip]{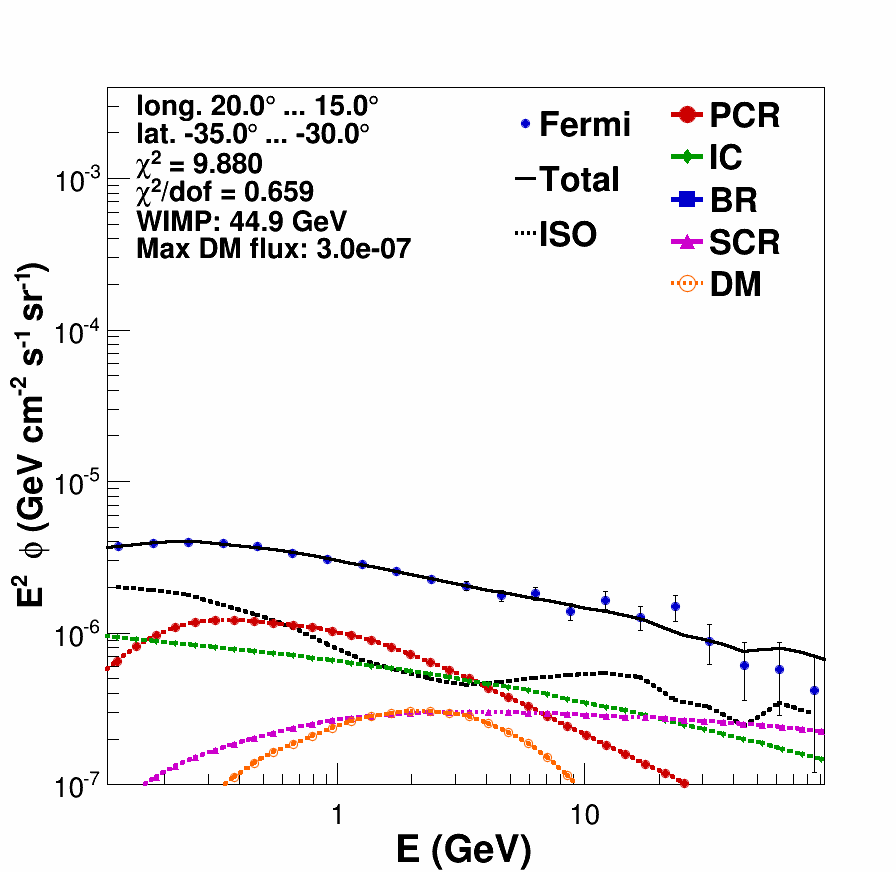}
\includegraphics[width=0.16\textwidth,height=0.16\textwidth,clip]{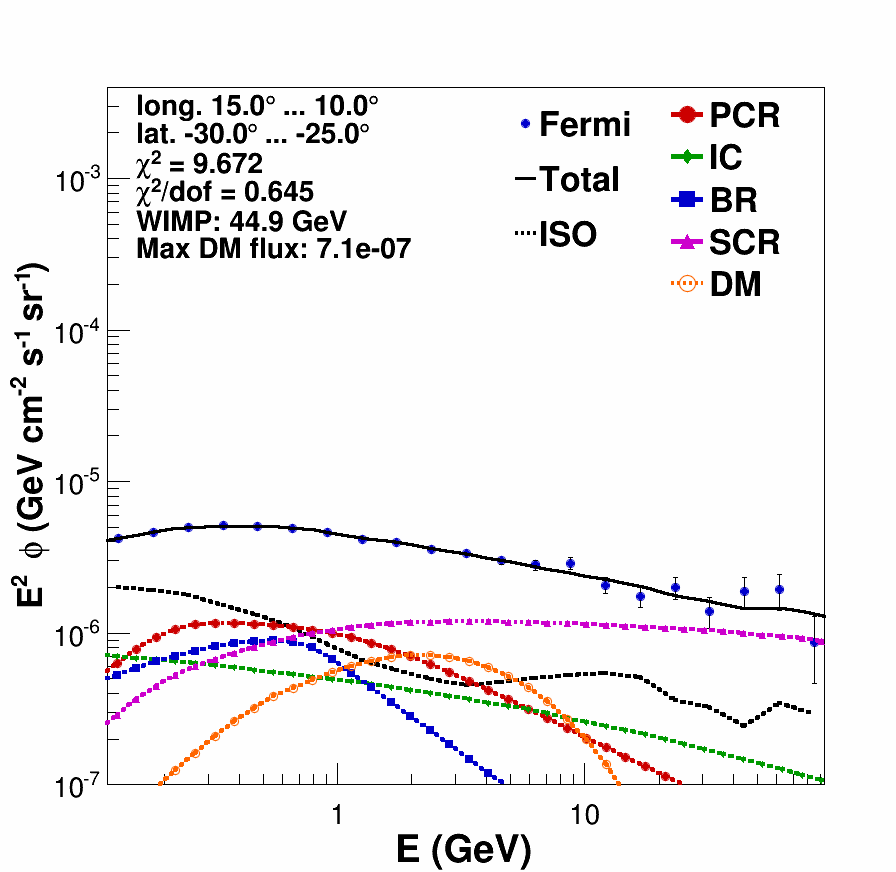}
\includegraphics[width=0.16\textwidth,height=0.16\textwidth,clip]{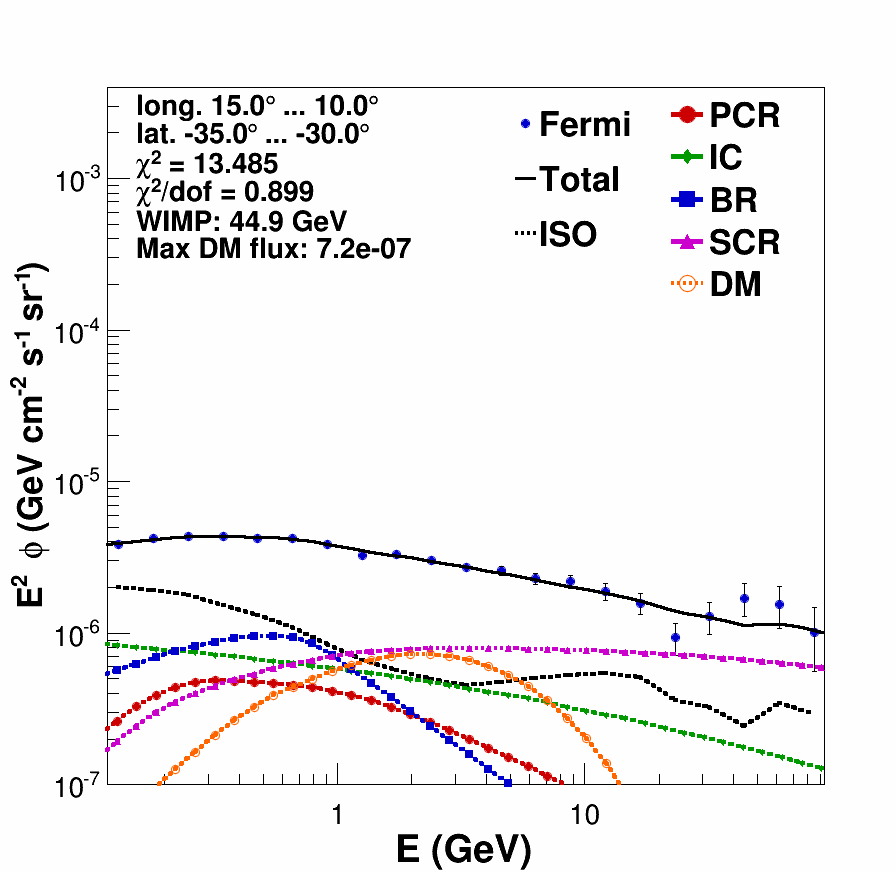}
\includegraphics[width=0.16\textwidth,height=0.16\textwidth,clip]{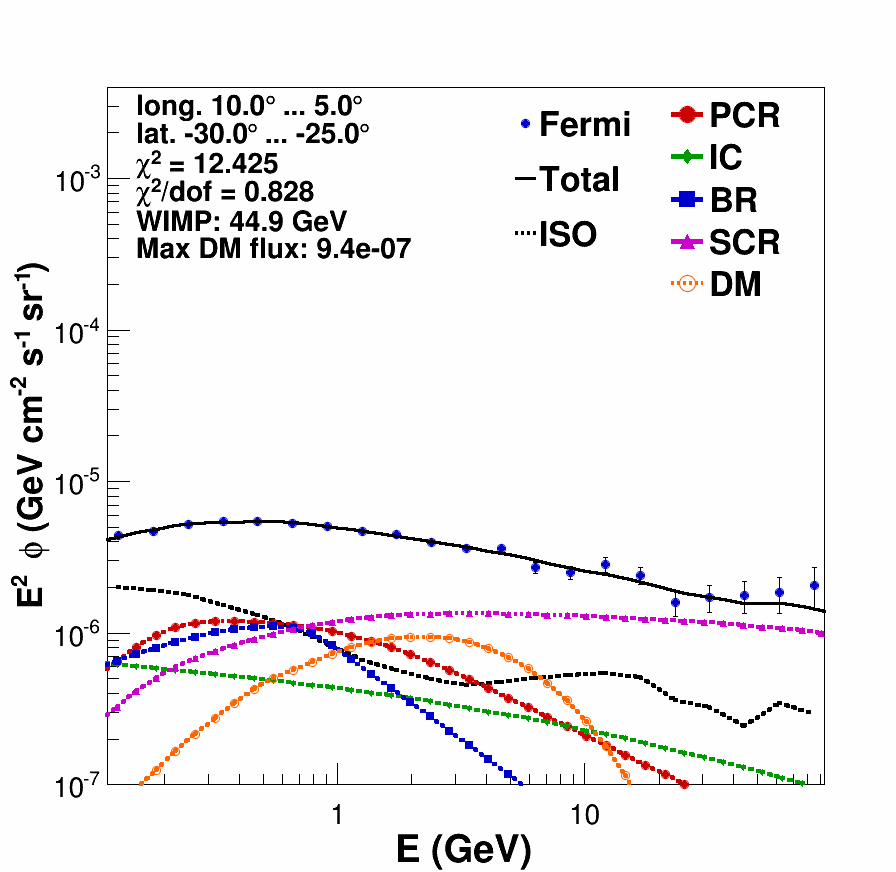}
\includegraphics[width=0.16\textwidth,height=0.16\textwidth,clip]{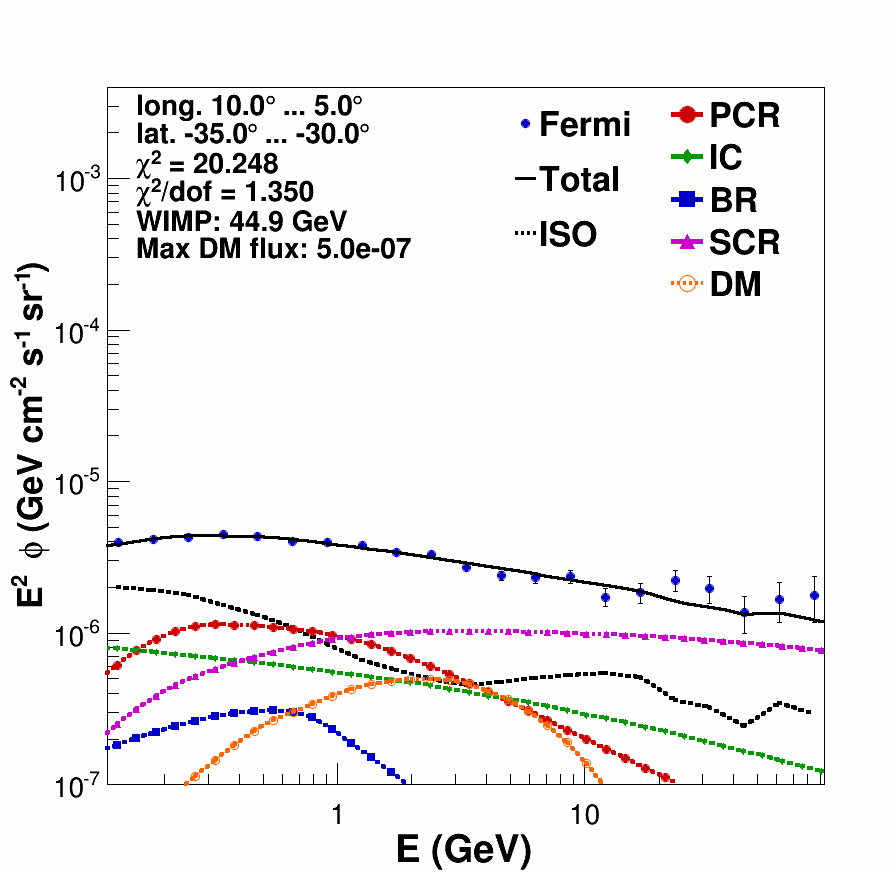}
\includegraphics[width=0.16\textwidth,height=0.16\textwidth,clip]{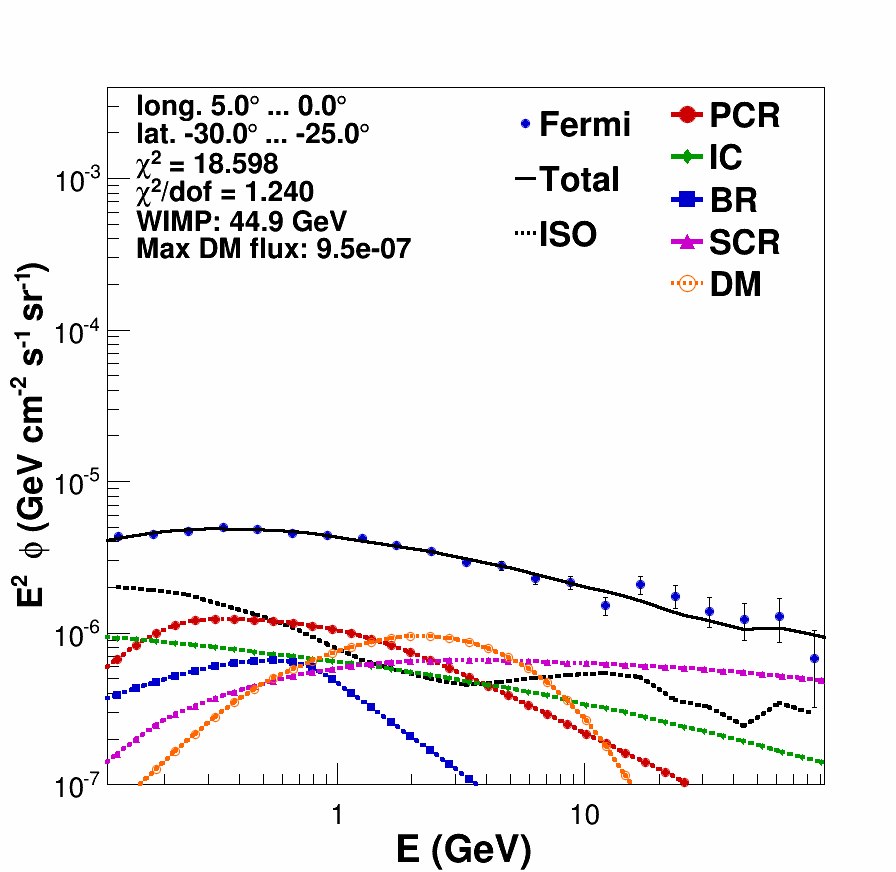}
\includegraphics[width=0.16\textwidth,height=0.16\textwidth,clip]{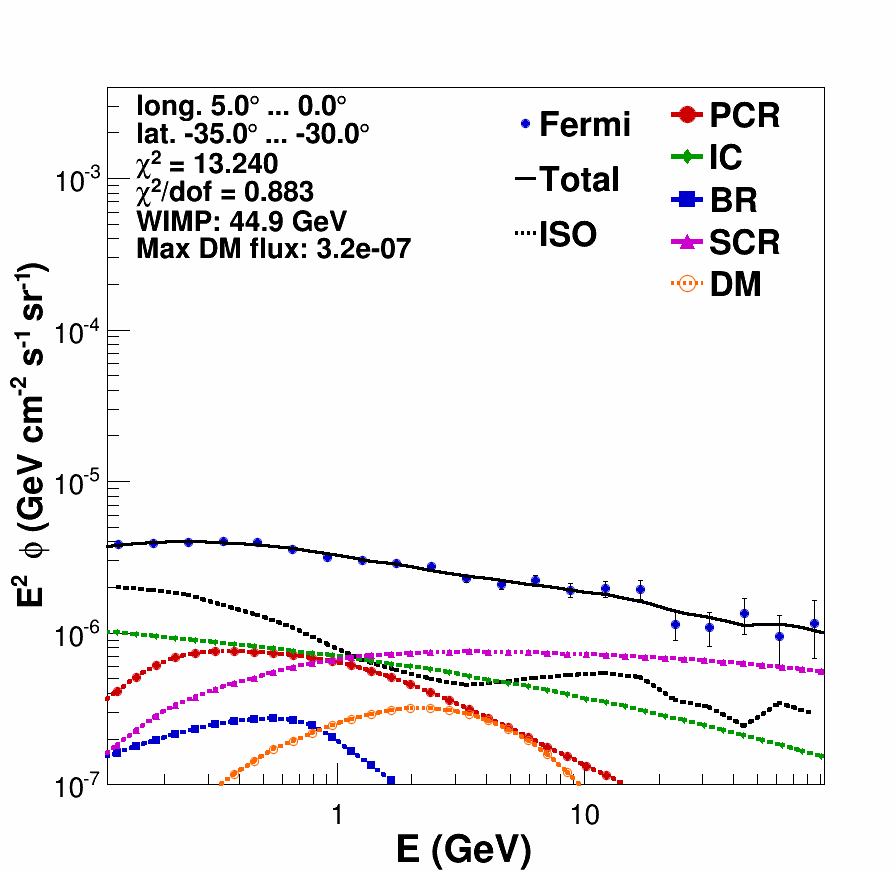}
\includegraphics[width=0.16\textwidth,height=0.16\textwidth,clip]{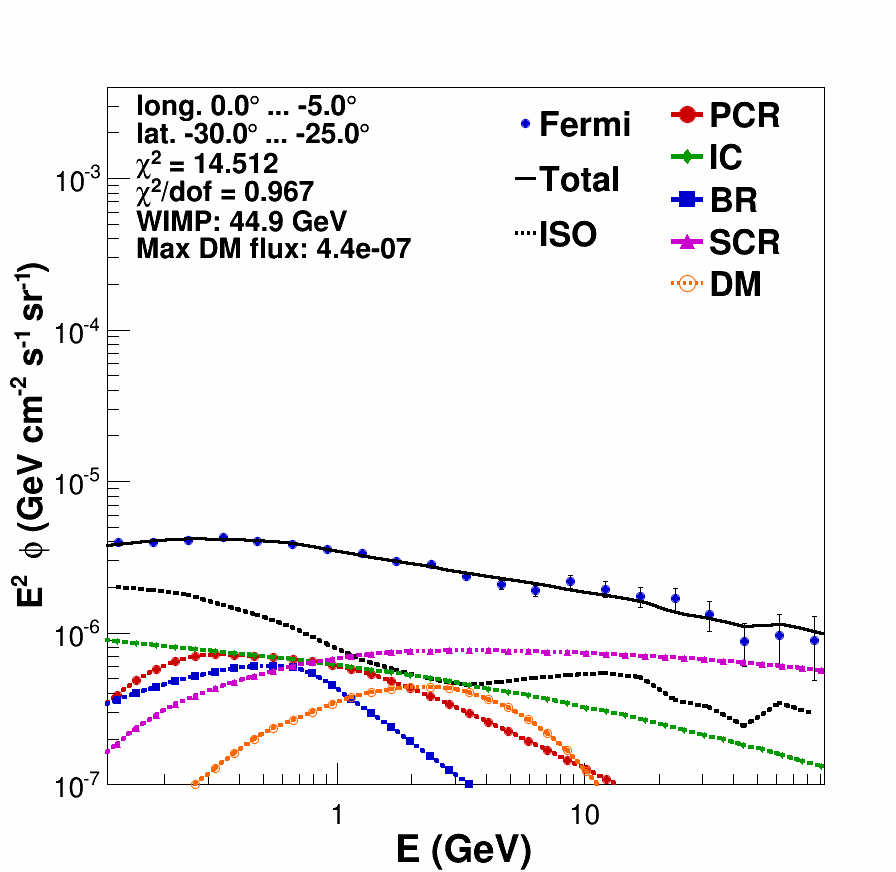}
\includegraphics[width=0.16\textwidth,height=0.16\textwidth,clip]{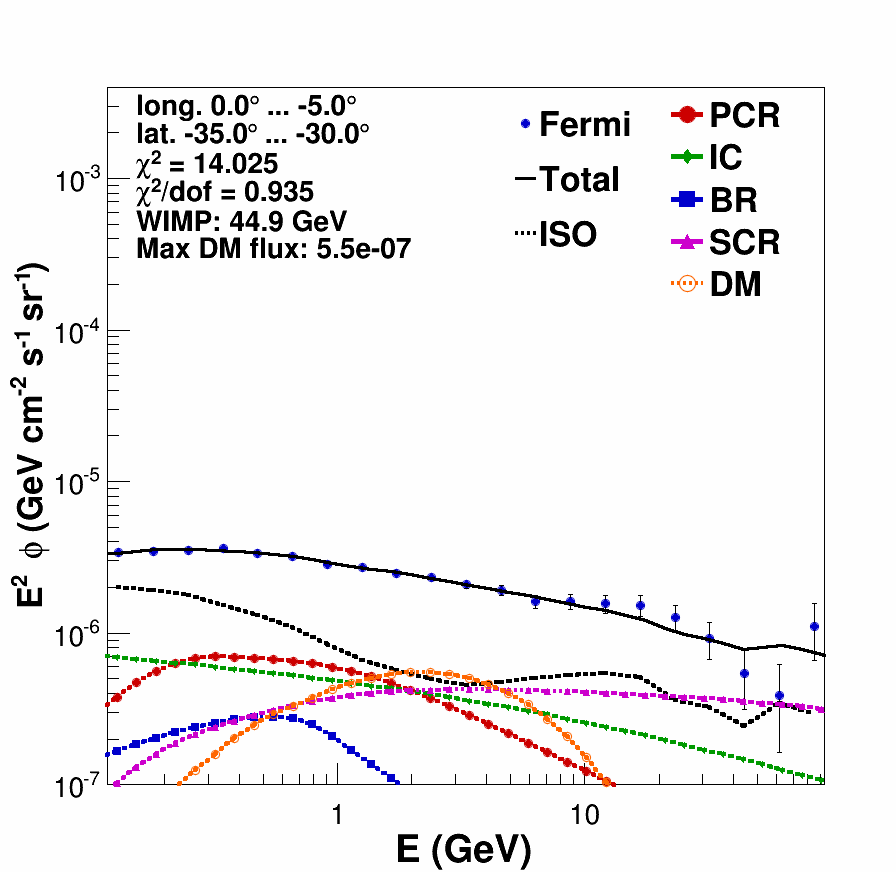}
\includegraphics[width=0.16\textwidth,height=0.16\textwidth,clip]{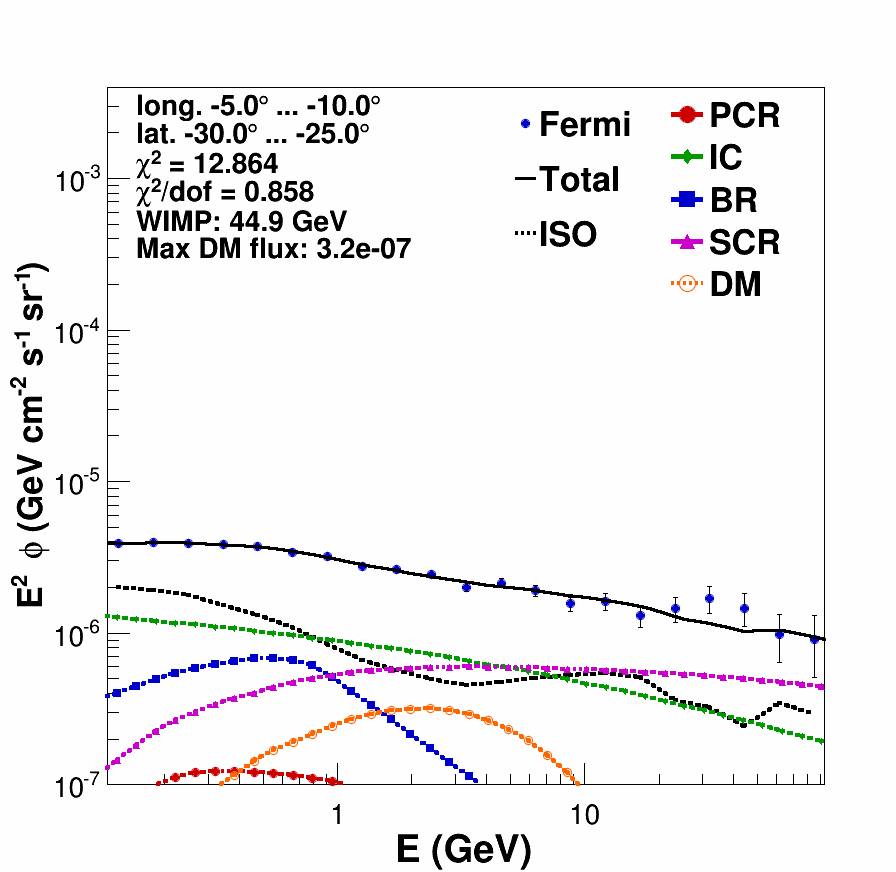}
\includegraphics[width=0.16\textwidth,height=0.16\textwidth,clip]{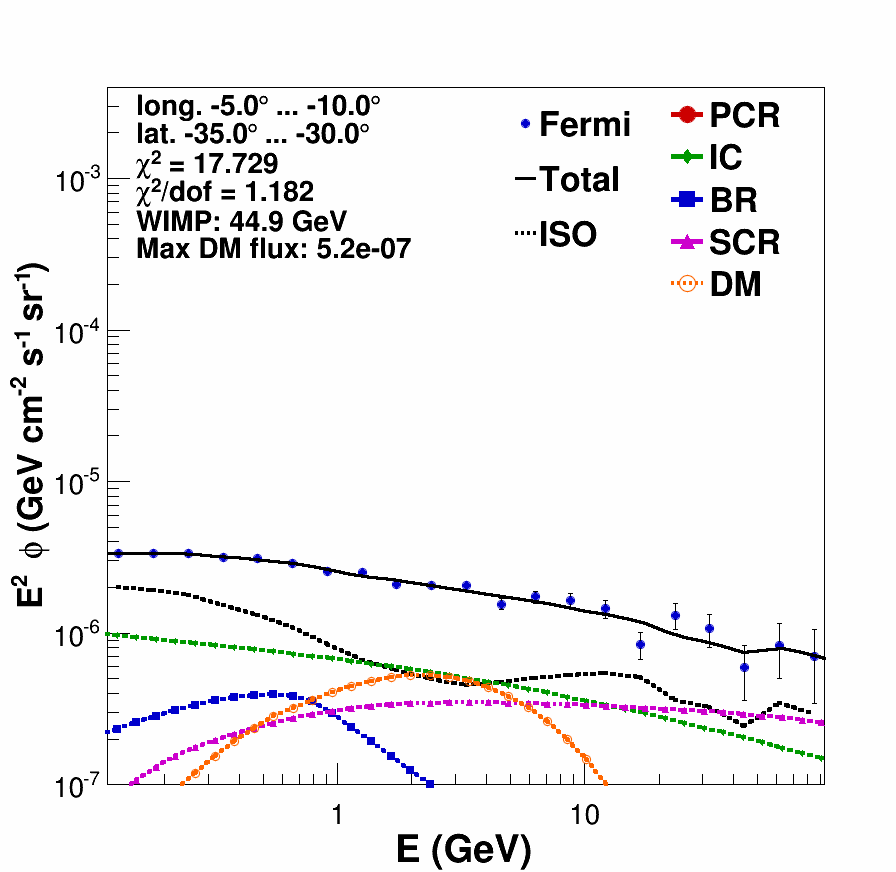}
\includegraphics[width=0.16\textwidth,height=0.16\textwidth,clip]{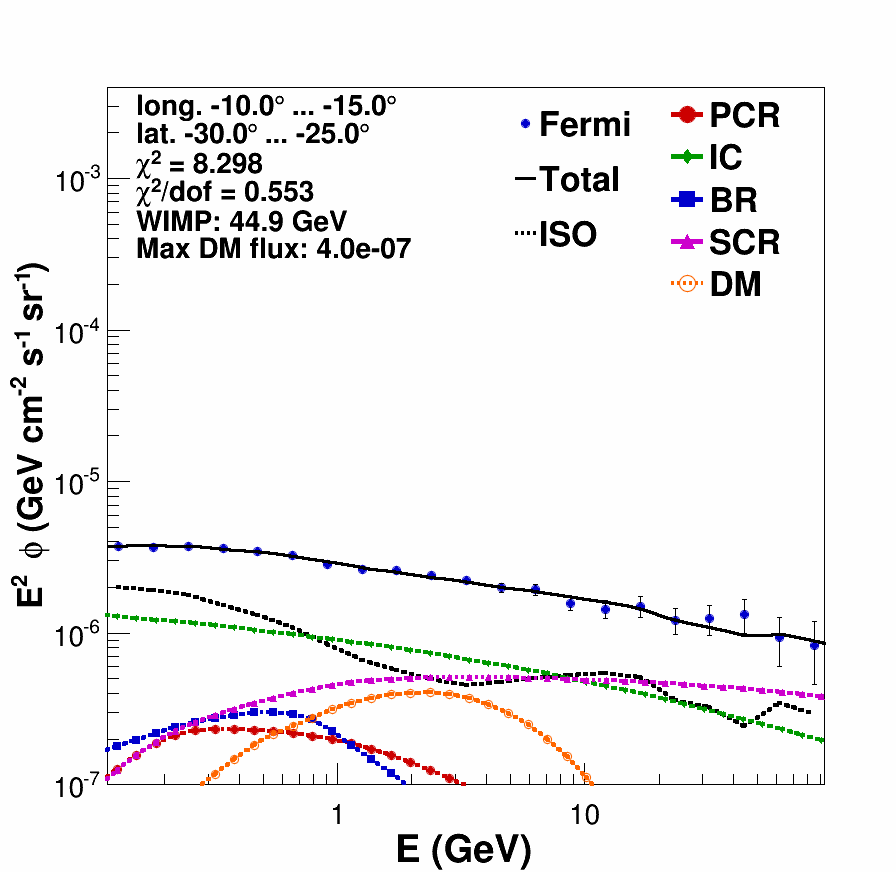}
\includegraphics[width=0.16\textwidth,height=0.16\textwidth,clip]{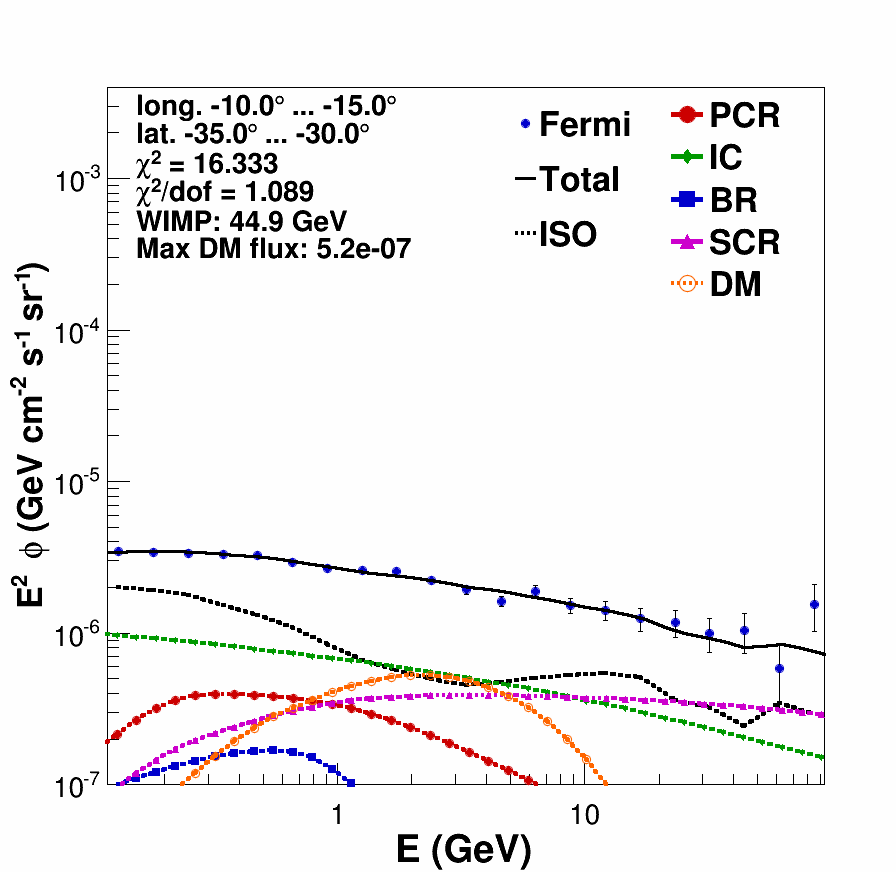}
\includegraphics[width=0.16\textwidth,height=0.16\textwidth,clip]{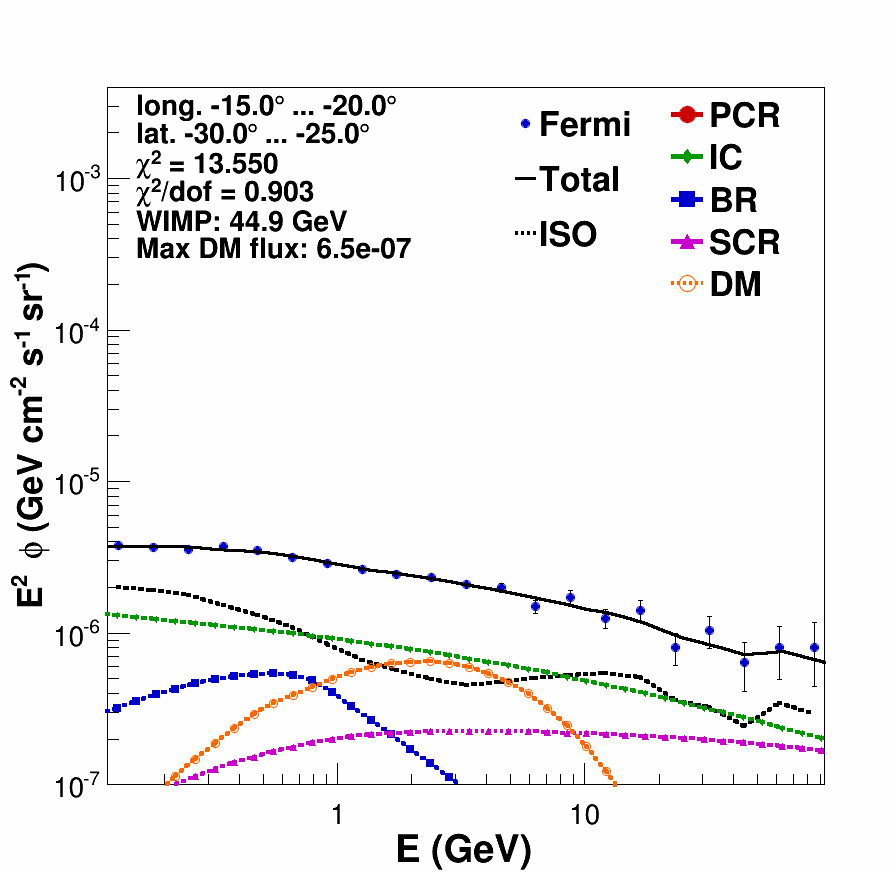}
\includegraphics[width=0.16\textwidth,height=0.16\textwidth,clip]{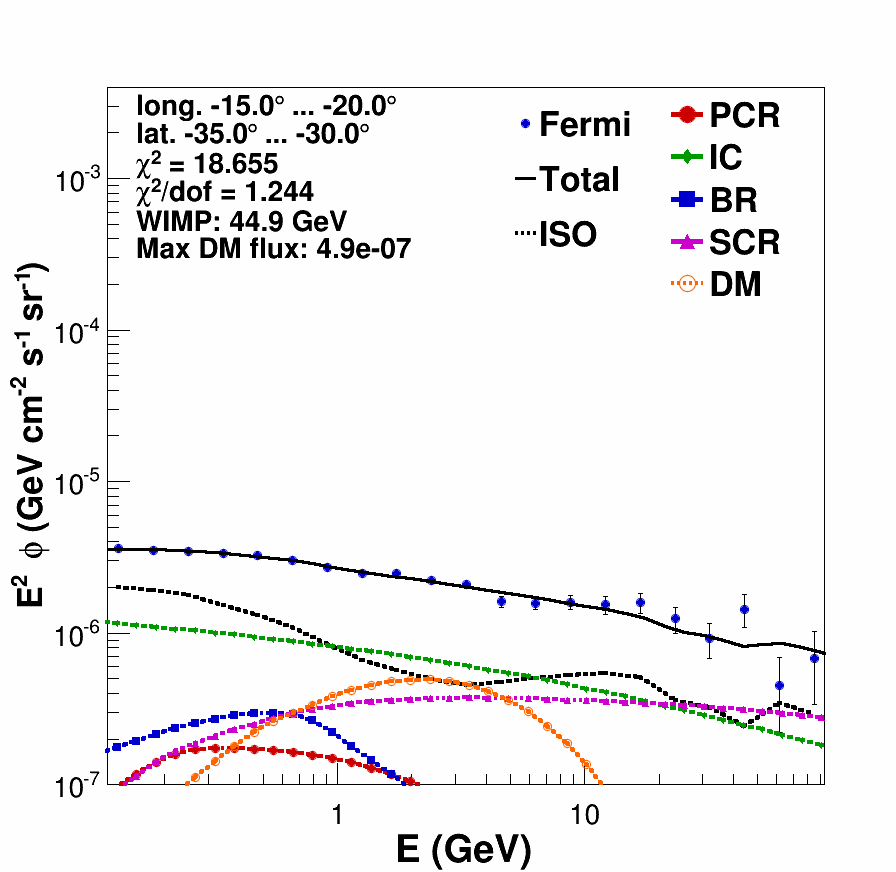}
\includegraphics[width=0.16\textwidth,height=0.16\textwidth,clip]{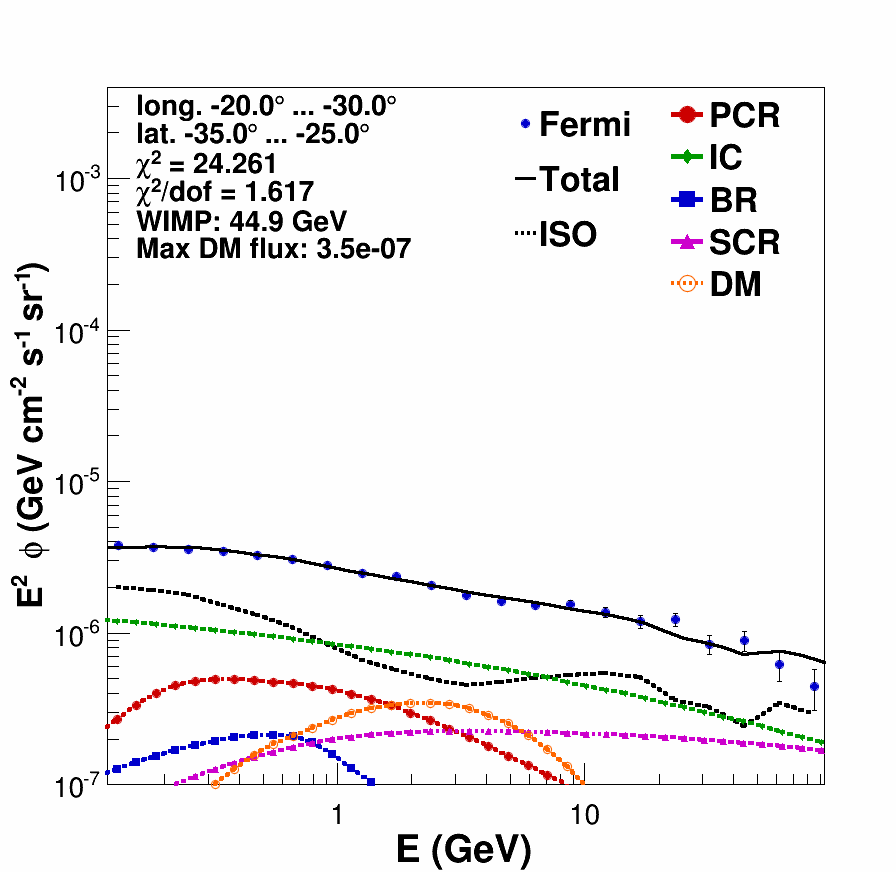}
\includegraphics[width=0.16\textwidth,height=0.16\textwidth,clip]{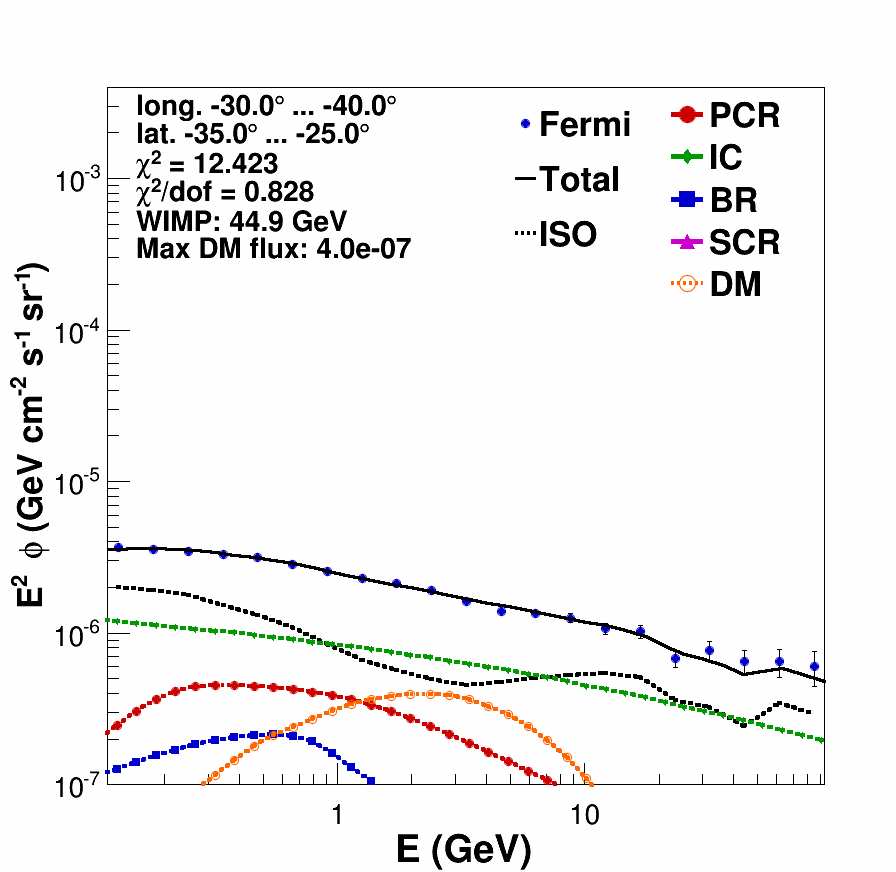}
\includegraphics[width=0.16\textwidth,height=0.16\textwidth,clip]{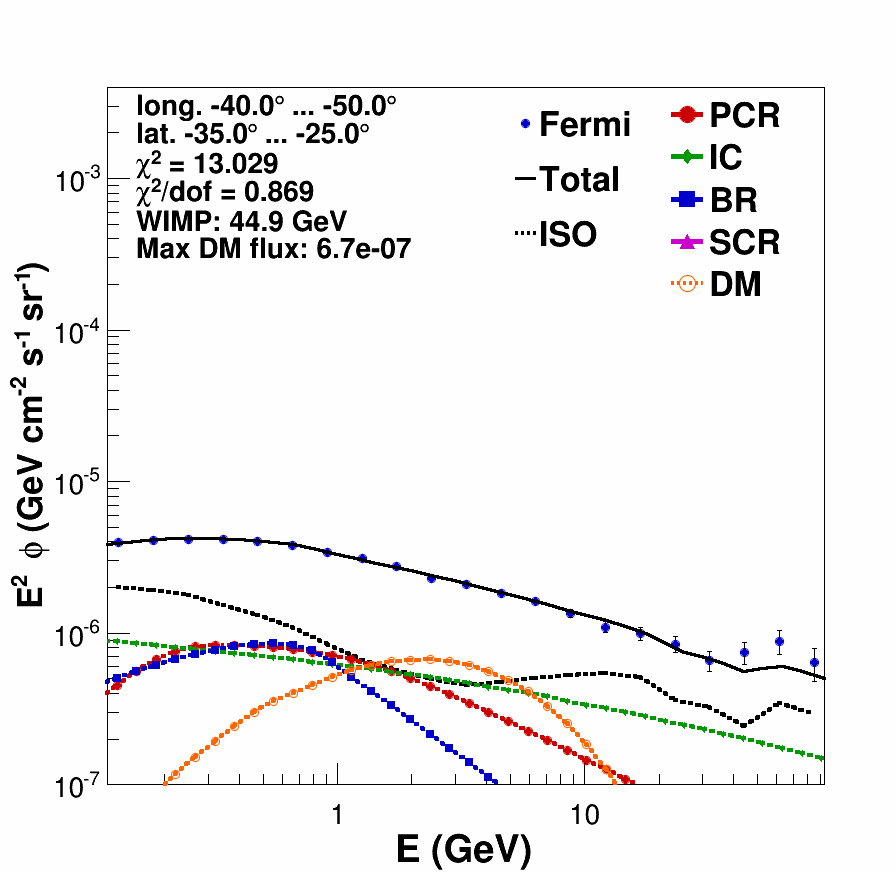}
\includegraphics[width=0.16\textwidth,height=0.16\textwidth,clip]{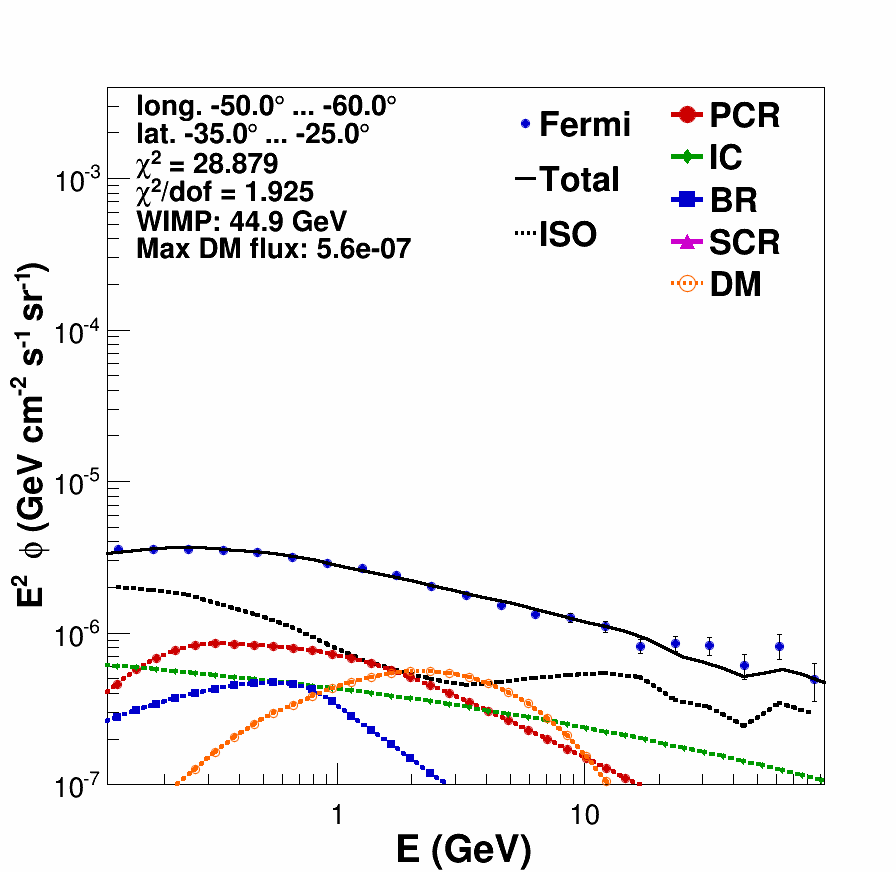}
\includegraphics[width=0.16\textwidth,height=0.16\textwidth,clip]{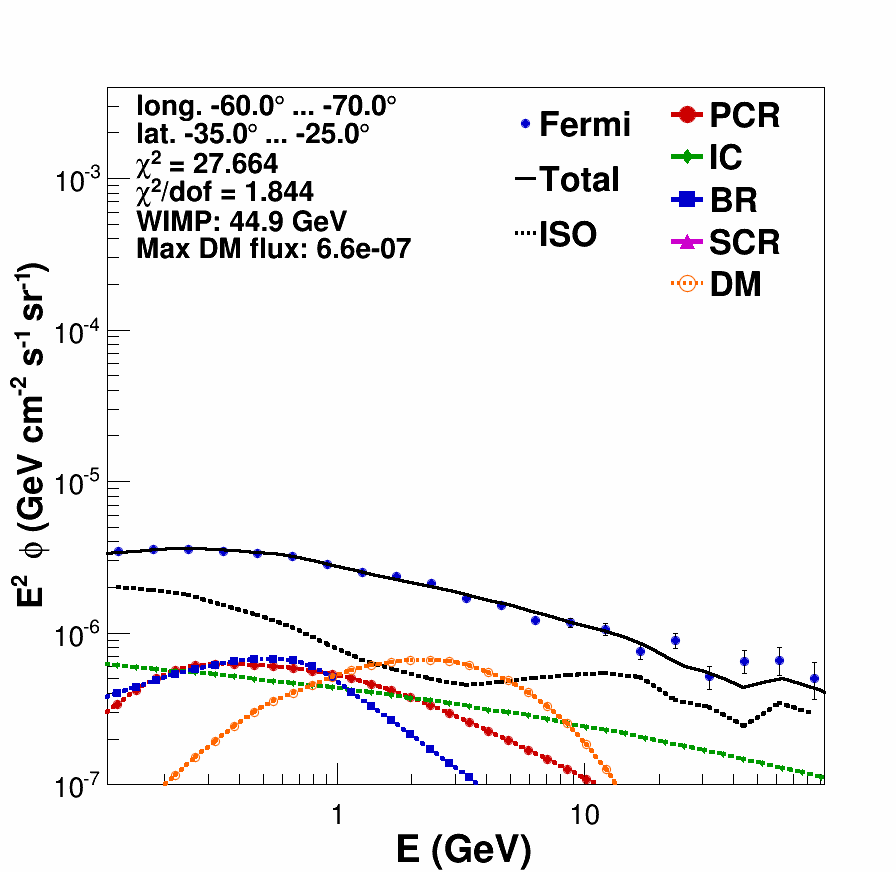}
\includegraphics[width=0.16\textwidth,height=0.16\textwidth,clip]{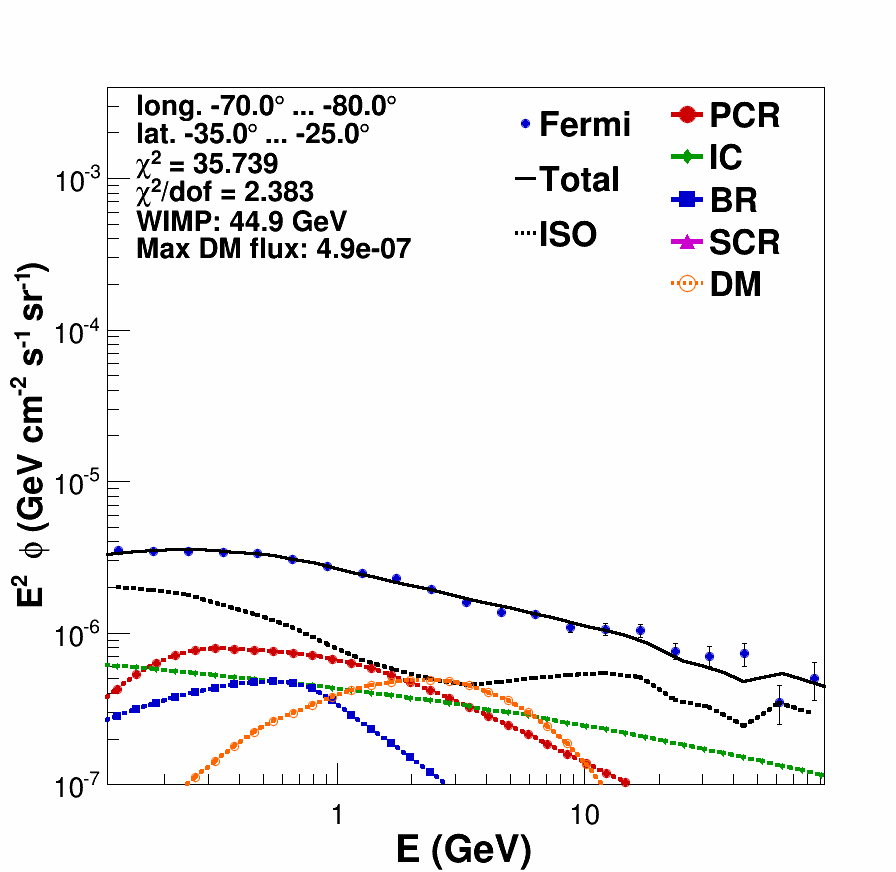}
\includegraphics[width=0.16\textwidth,height=0.16\textwidth,clip]{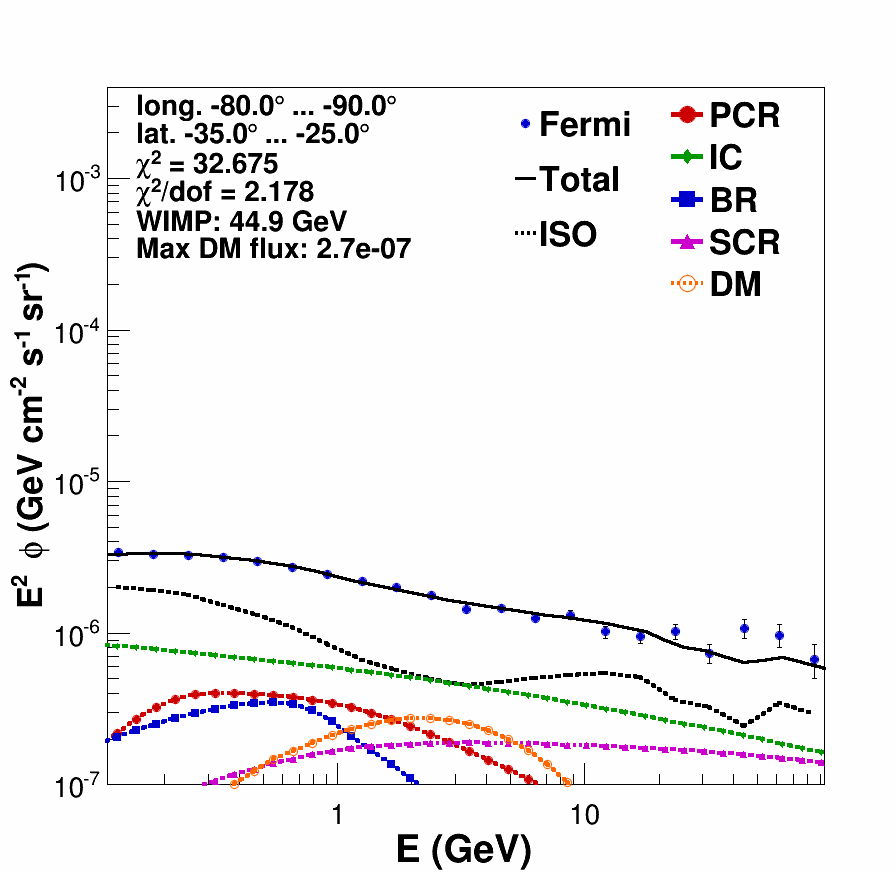}
\includegraphics[width=0.16\textwidth,height=0.16\textwidth,clip]{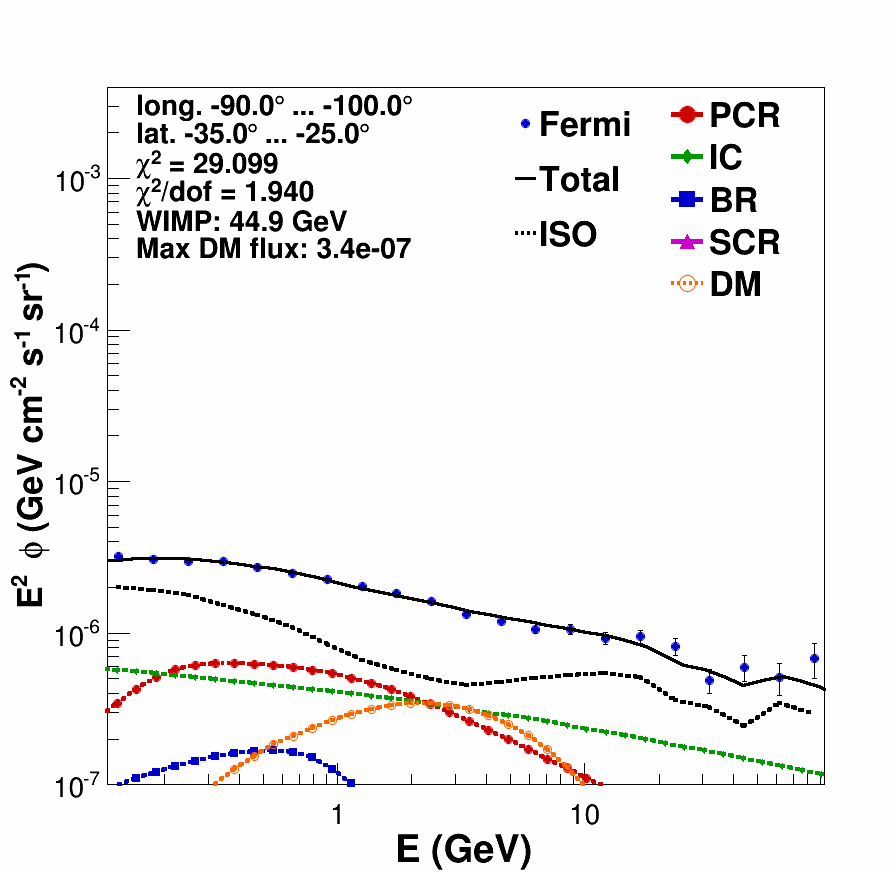}
\includegraphics[width=0.16\textwidth,height=0.16\textwidth,clip]{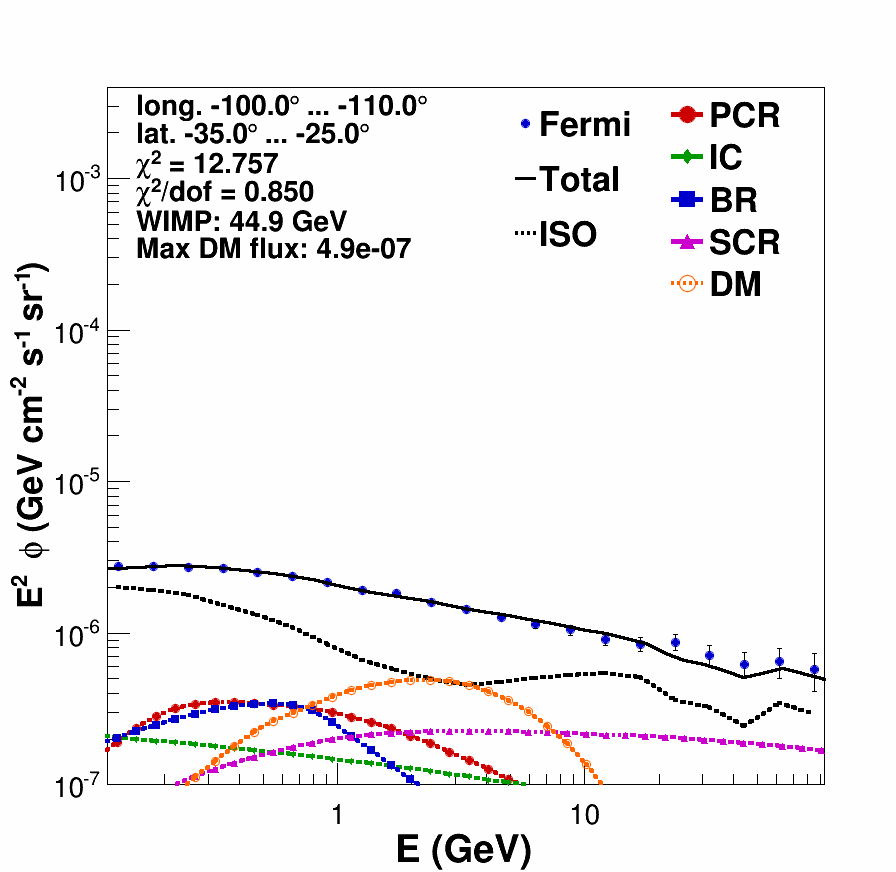}
\includegraphics[width=0.16\textwidth,height=0.16\textwidth,clip]{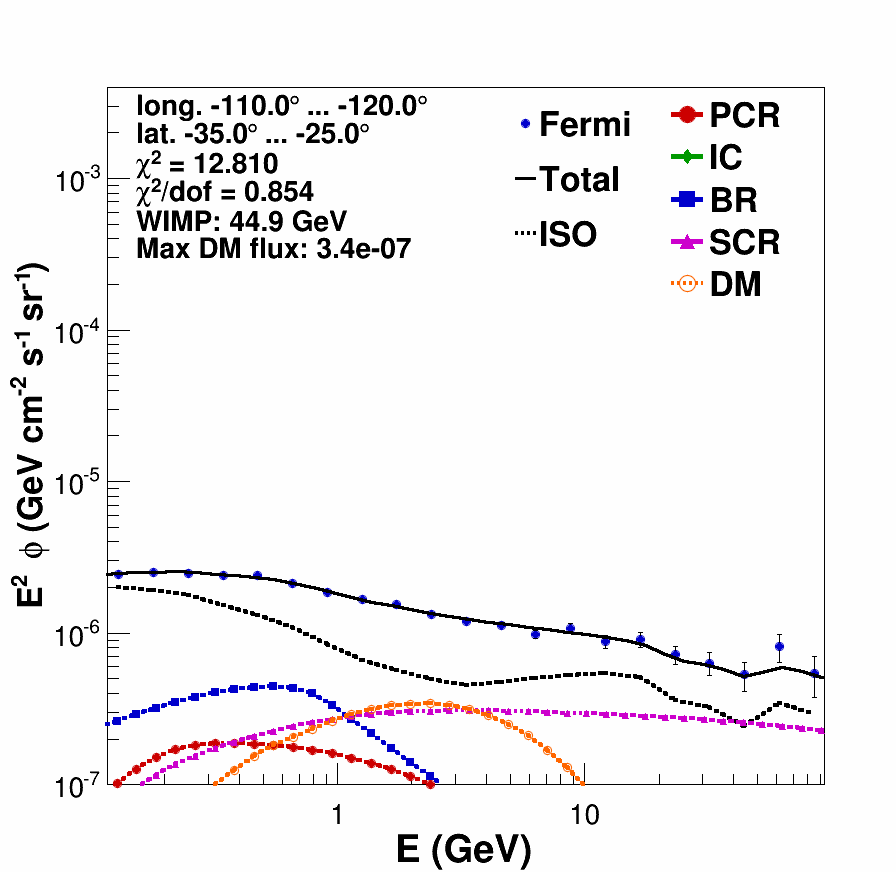}
\includegraphics[width=0.16\textwidth,height=0.16\textwidth,clip]{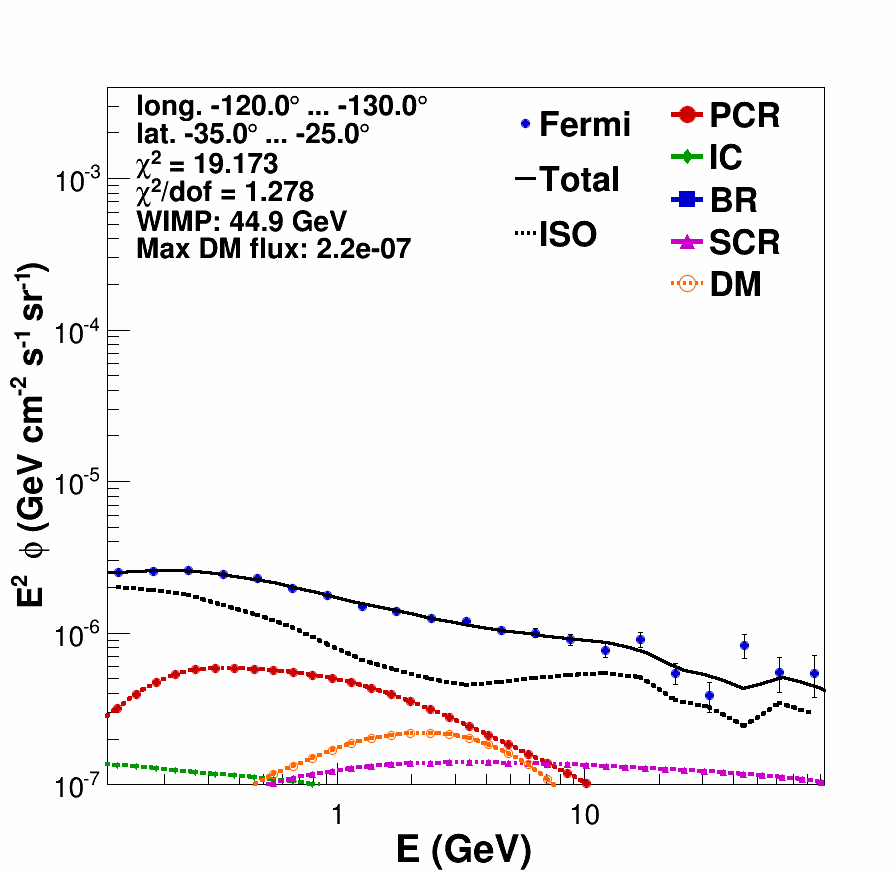}
\includegraphics[width=0.16\textwidth,height=0.16\textwidth,clip]{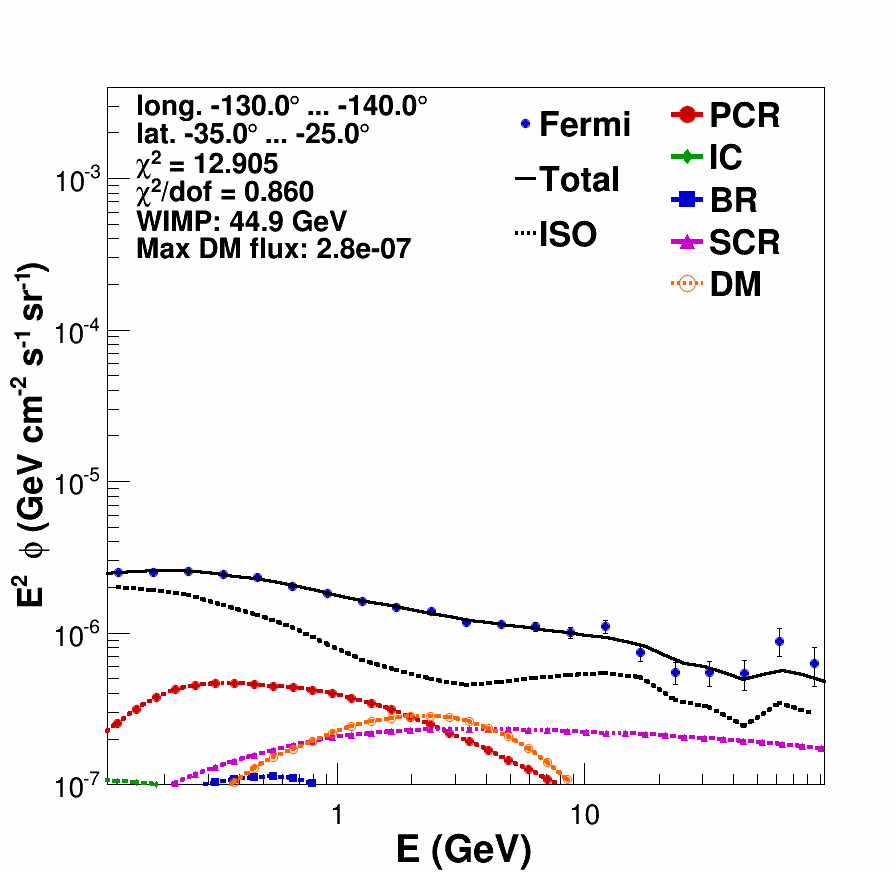}
\includegraphics[width=0.16\textwidth,height=0.16\textwidth,clip]{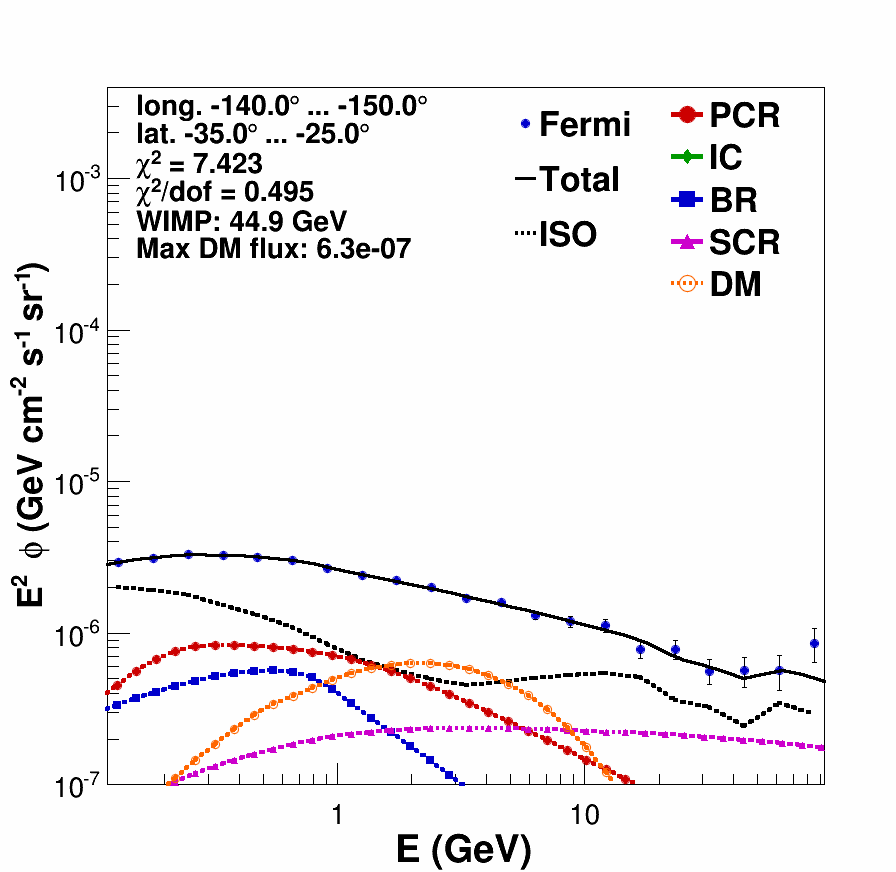}
\includegraphics[width=0.16\textwidth,height=0.16\textwidth,clip]{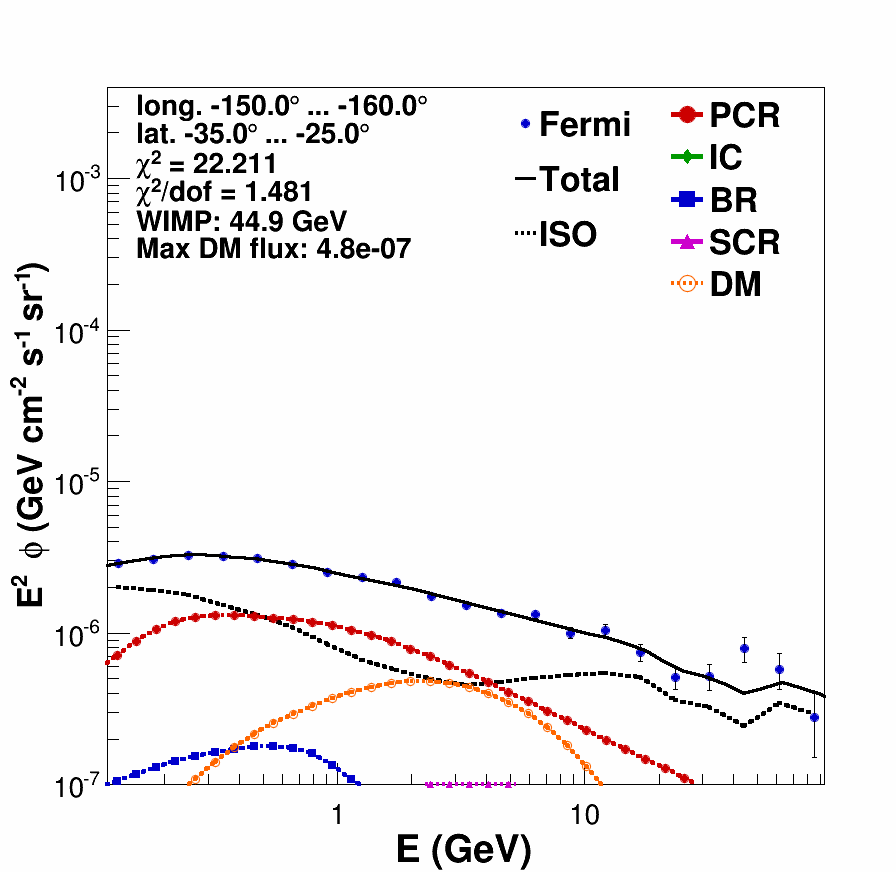}
\includegraphics[width=0.16\textwidth,height=0.16\textwidth,clip]{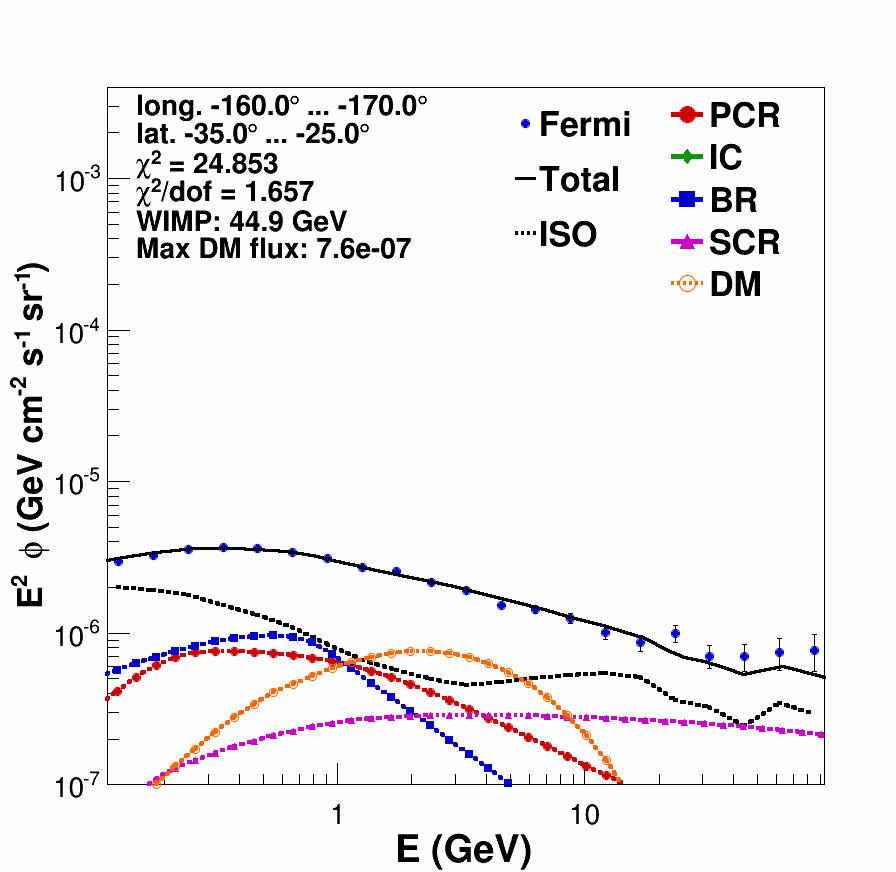}
\includegraphics[width=0.16\textwidth,height=0.16\textwidth,clip]{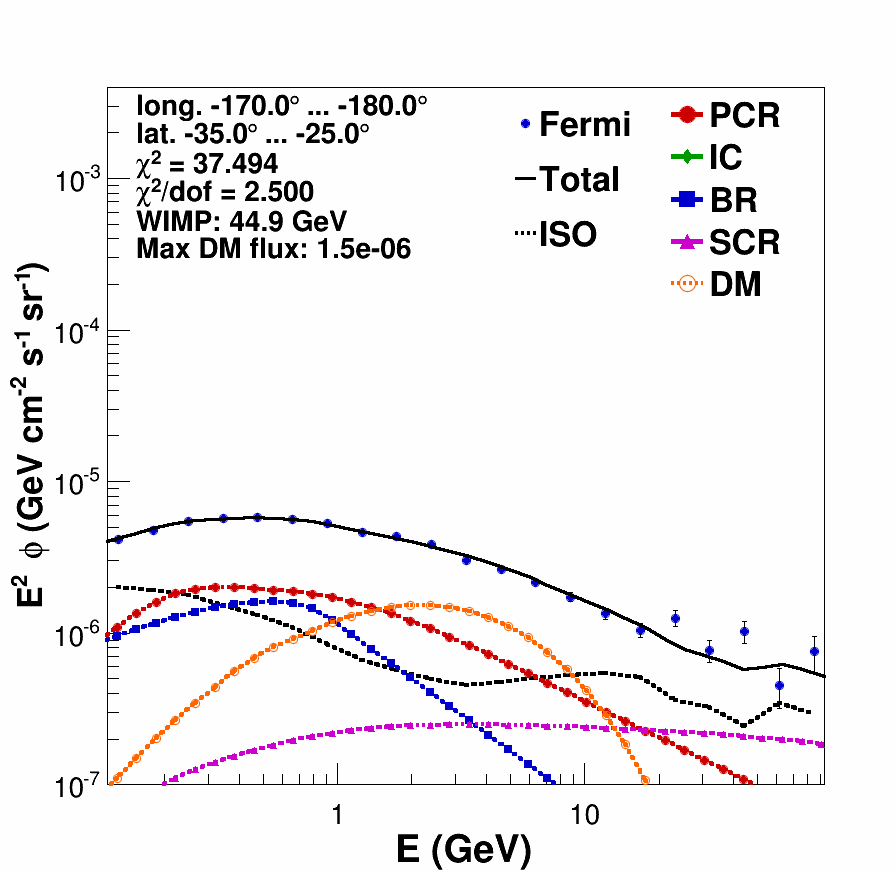}%%%%%r15
\caption[]{Template fits for latitudes  with $-35.0^\circ<b<-25.0^\circ$ and longitudes decreasing from 180$^\circ$ to -180$^\circ$.} \label{F48}
\end{figure}
\begin{figure}
\centering
\includegraphics[width=0.16\textwidth,height=0.16\textwidth,clip]{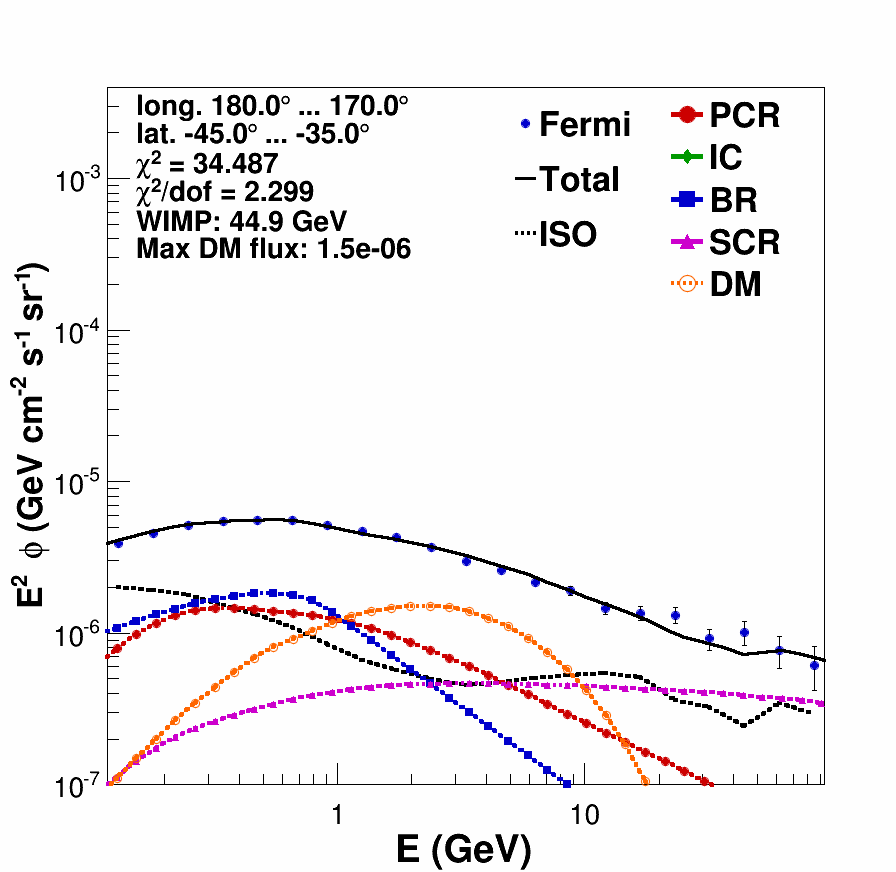}
\includegraphics[width=0.16\textwidth,height=0.16\textwidth,clip]{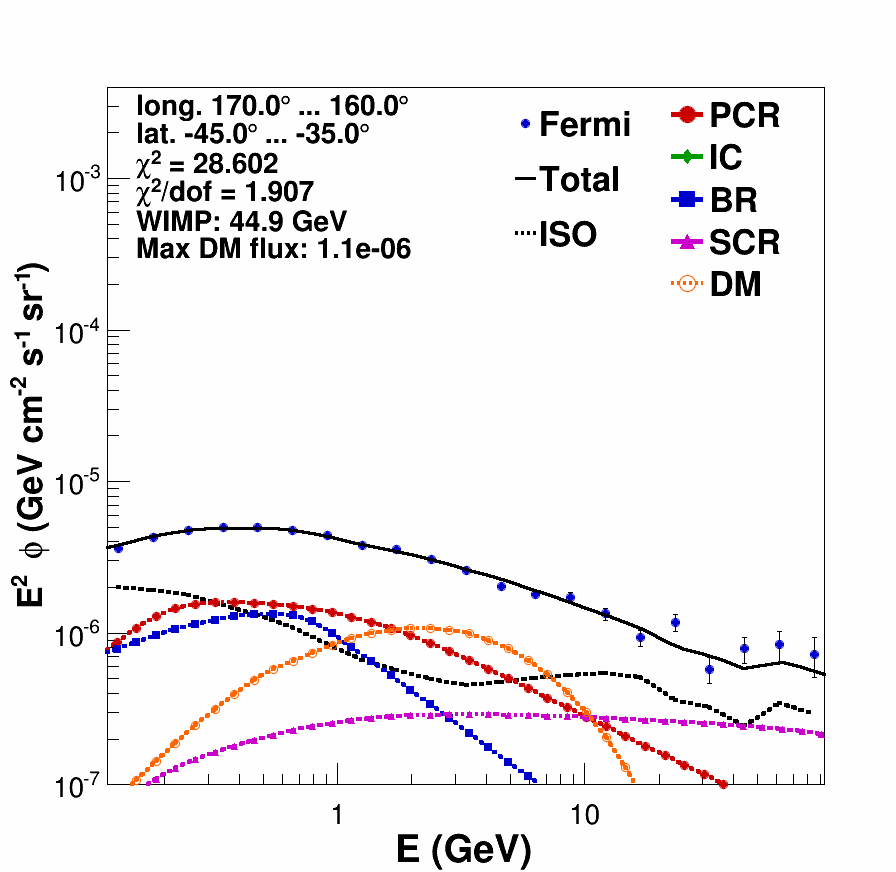}
\includegraphics[width=0.16\textwidth,height=0.16\textwidth,clip]{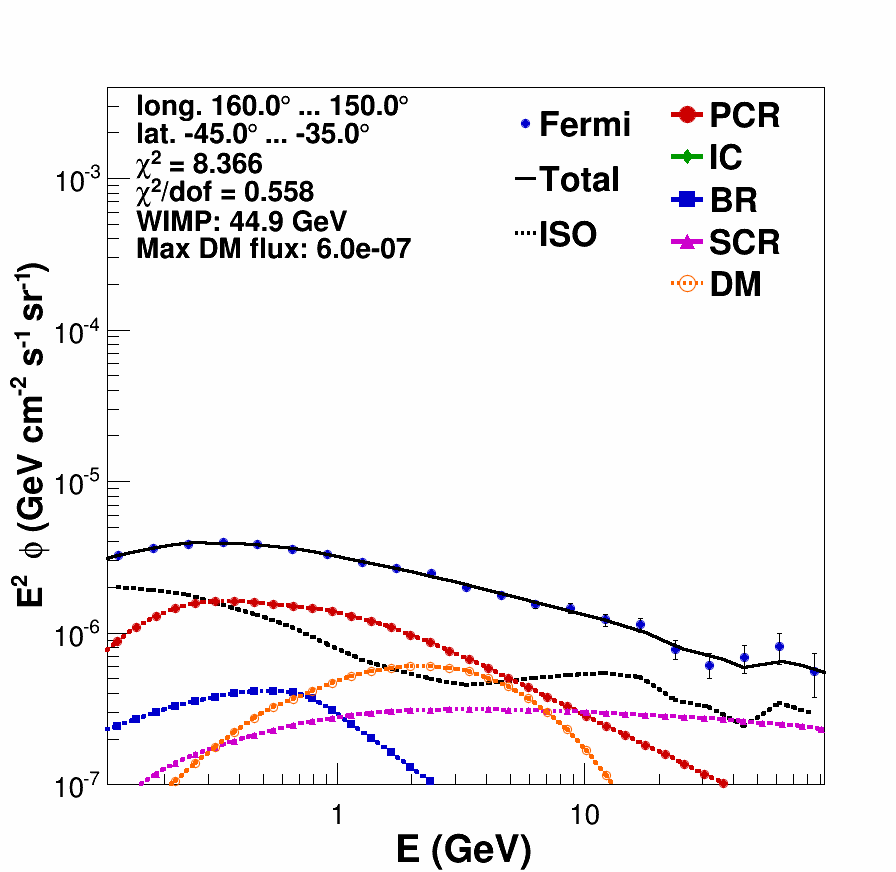}
\includegraphics[width=0.16\textwidth,height=0.16\textwidth,clip]{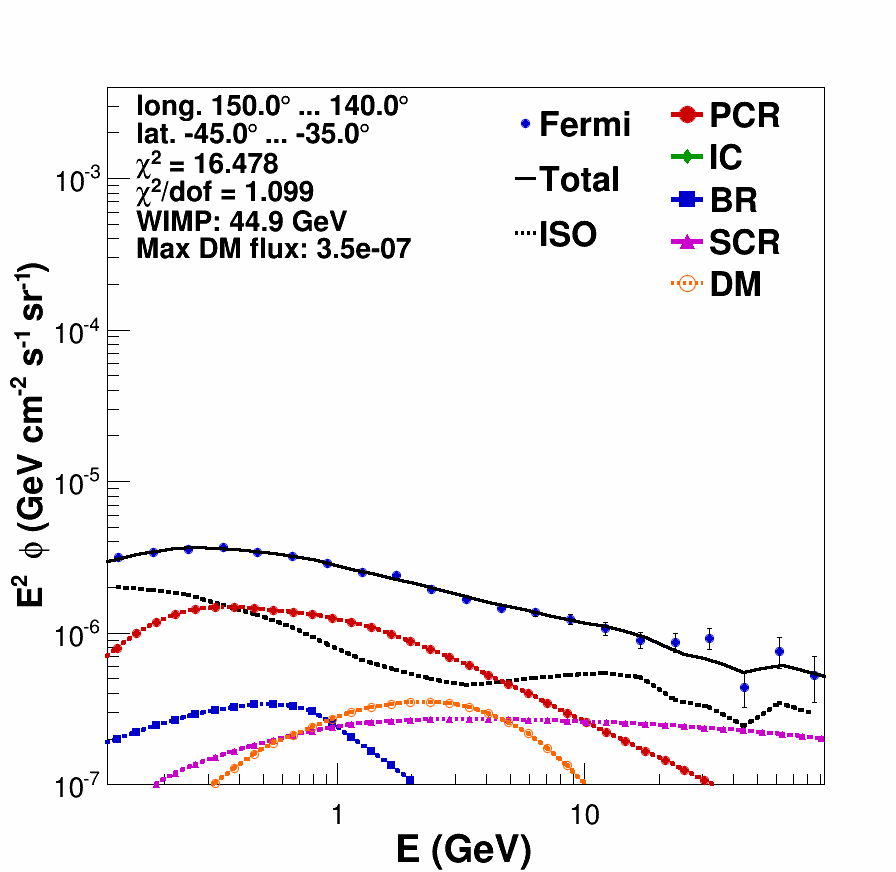}
\includegraphics[width=0.16\textwidth,height=0.16\textwidth,clip]{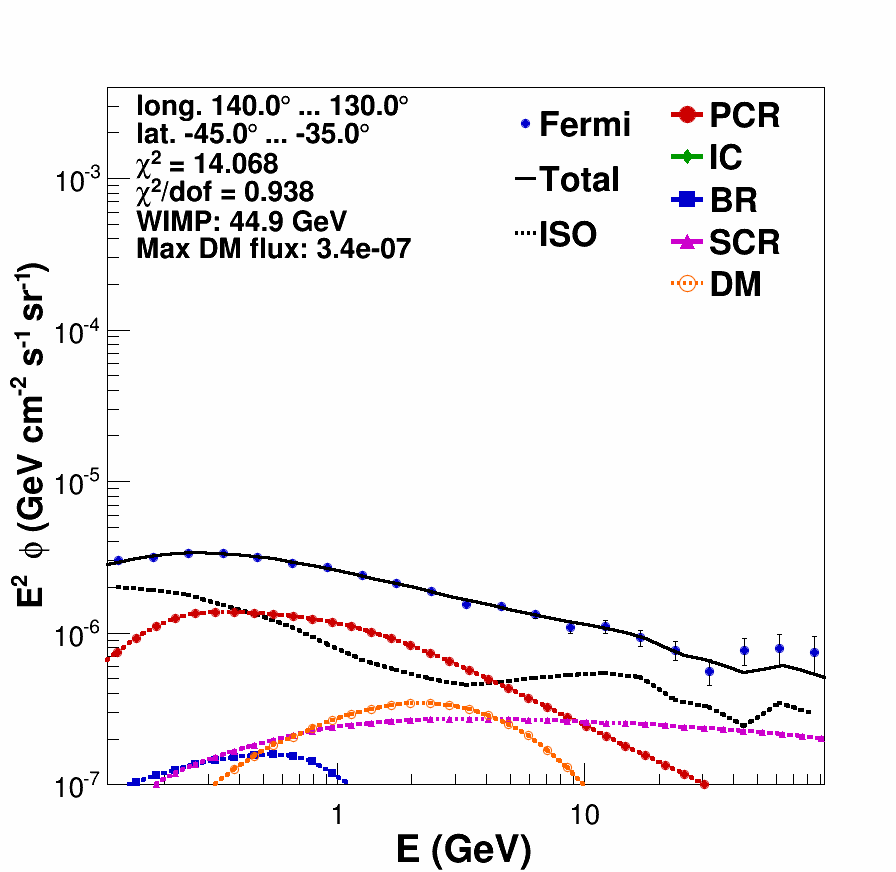}
\includegraphics[width=0.16\textwidth,height=0.16\textwidth,clip]{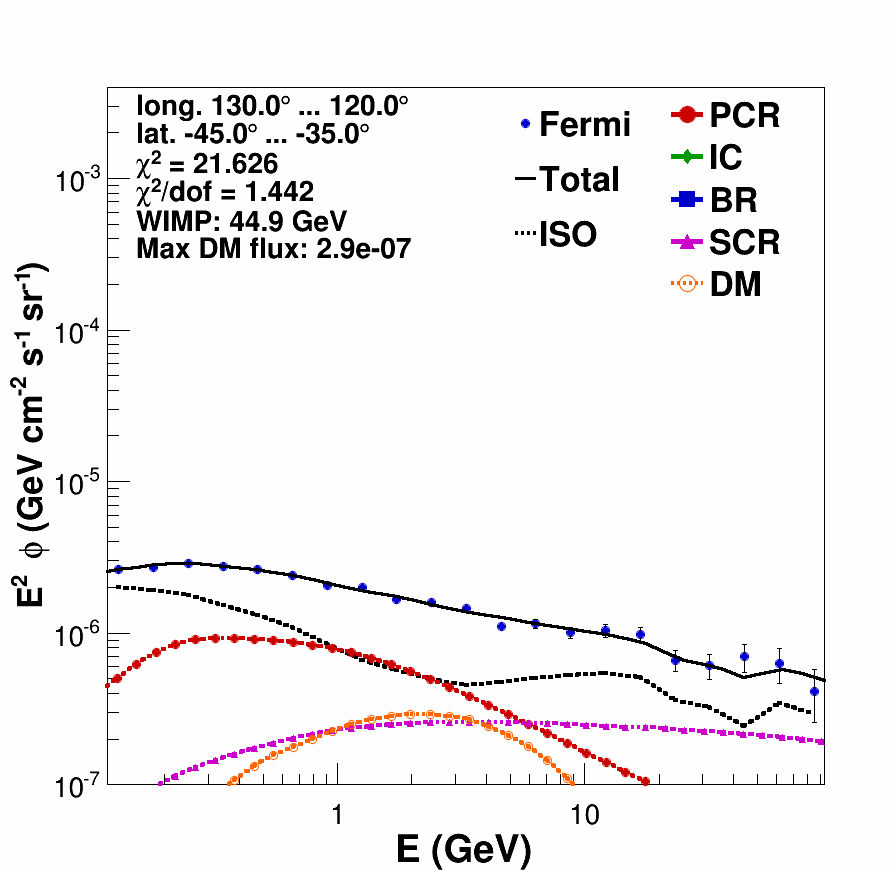}
\includegraphics[width=0.16\textwidth,height=0.16\textwidth,clip]{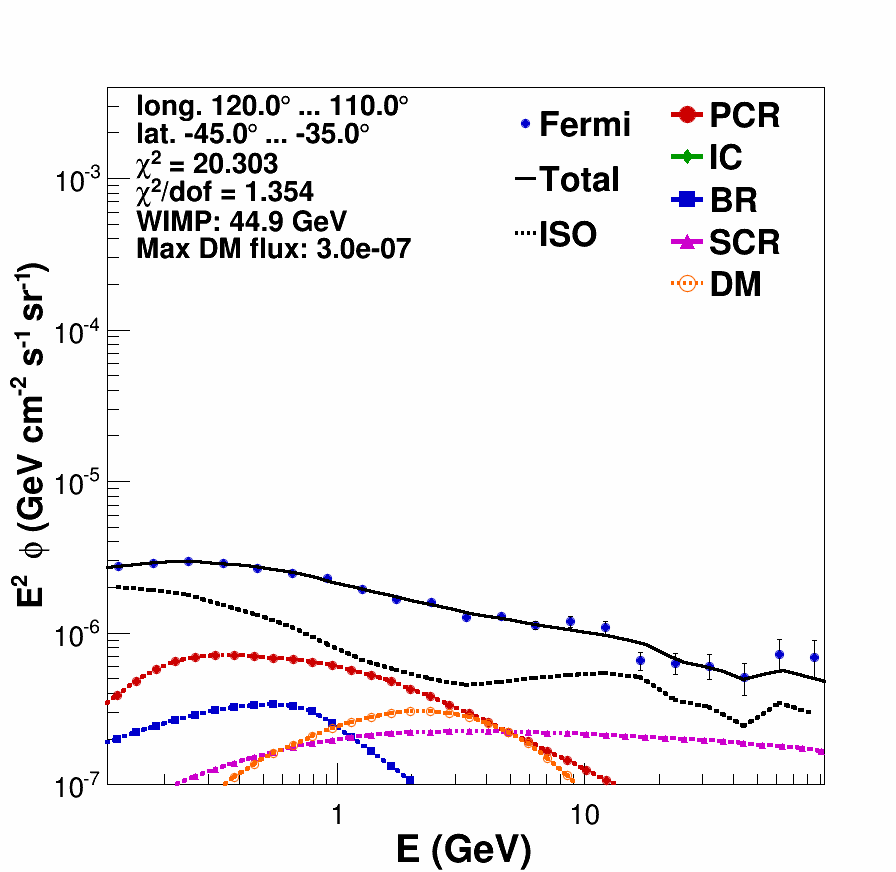}
\includegraphics[width=0.16\textwidth,height=0.16\textwidth,clip]{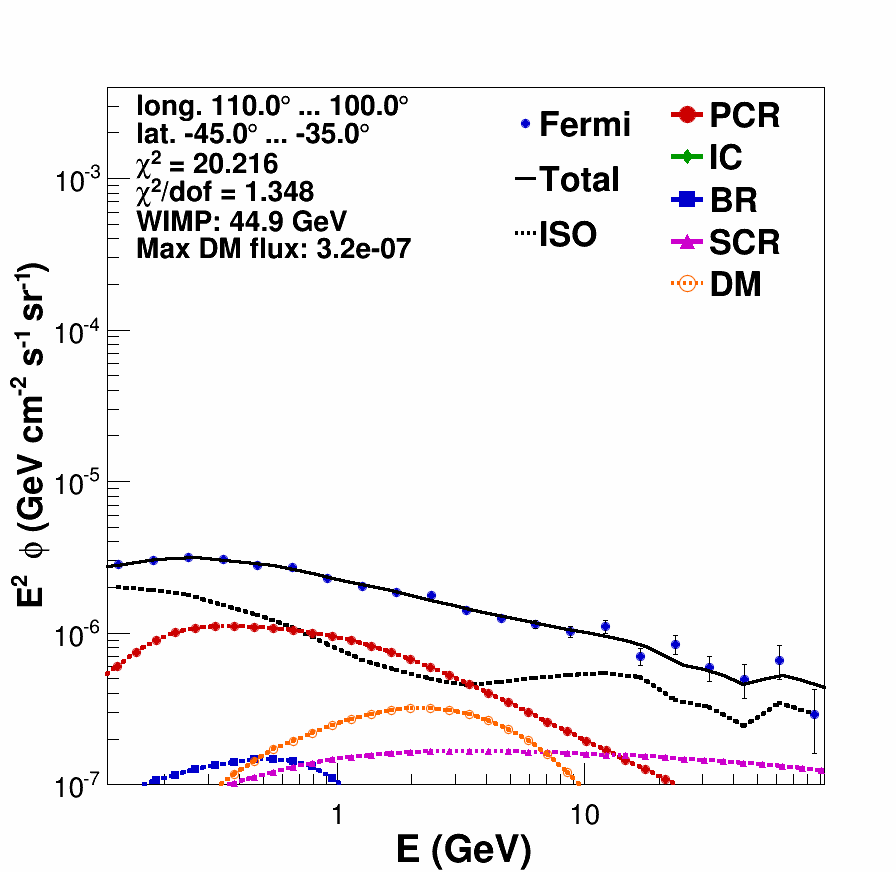}
\includegraphics[width=0.16\textwidth,height=0.16\textwidth,clip]{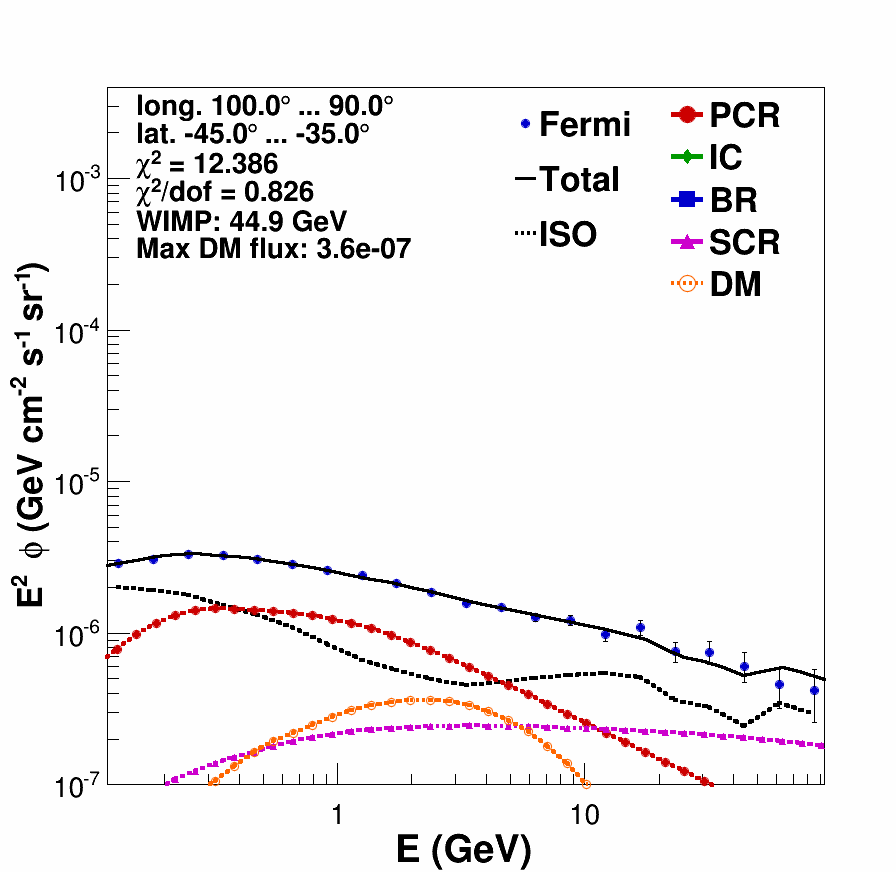}
\includegraphics[width=0.16\textwidth,height=0.16\textwidth,clip]{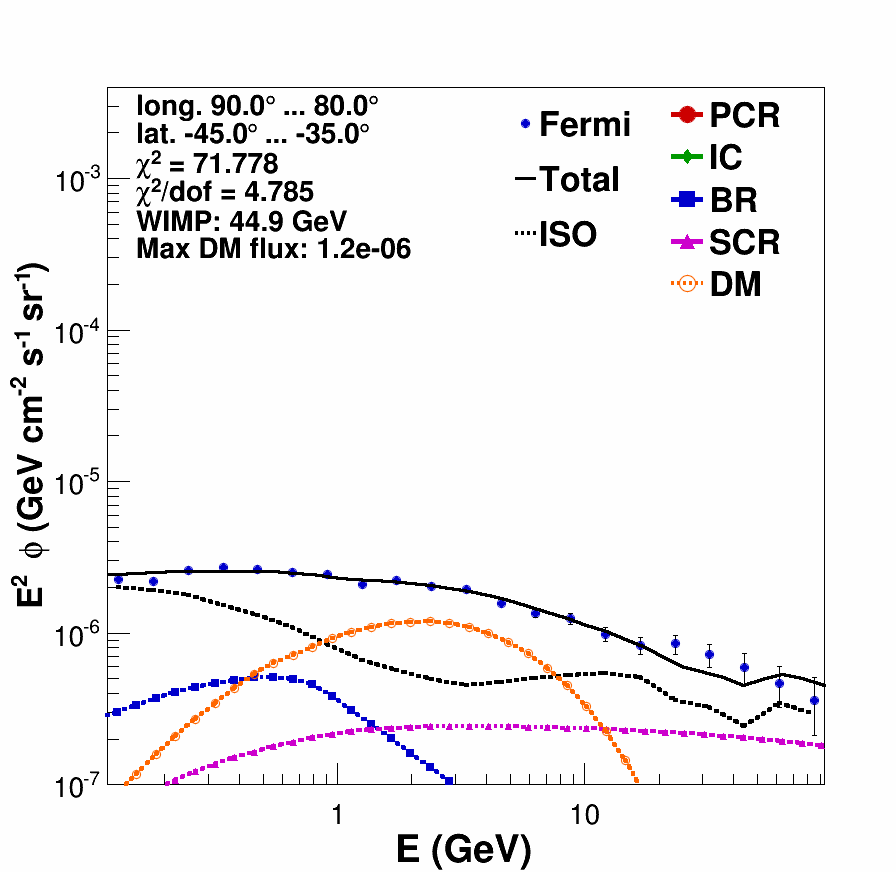}
\includegraphics[width=0.16\textwidth,height=0.16\textwidth,clip]{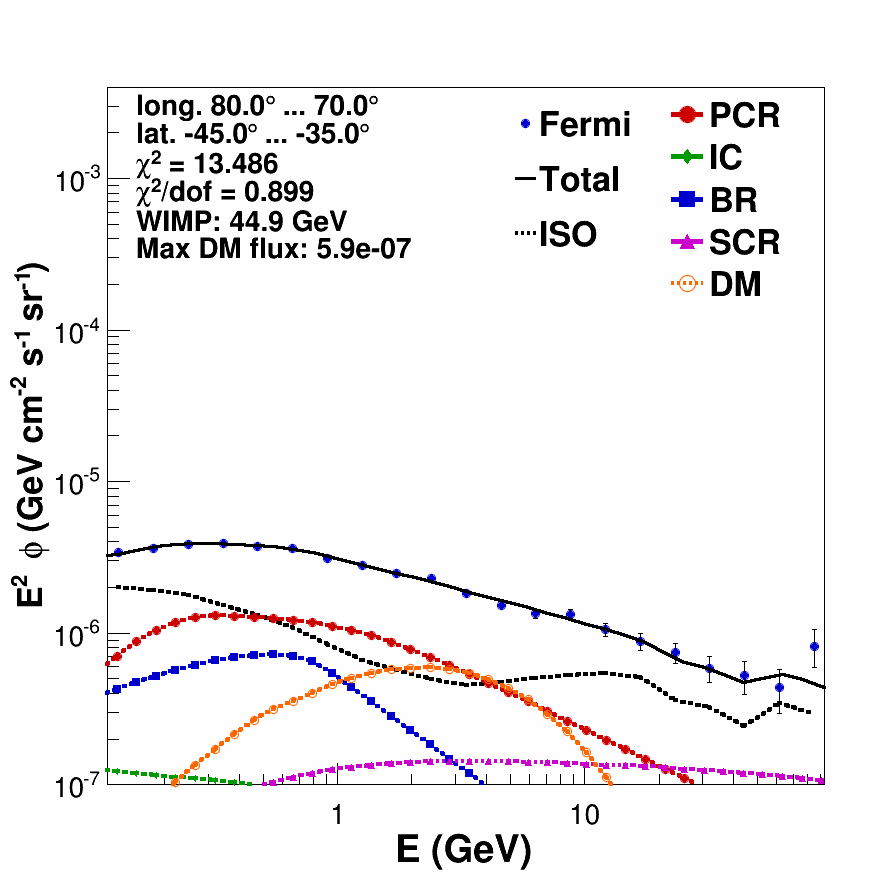}
\includegraphics[width=0.16\textwidth,height=0.16\textwidth,clip]{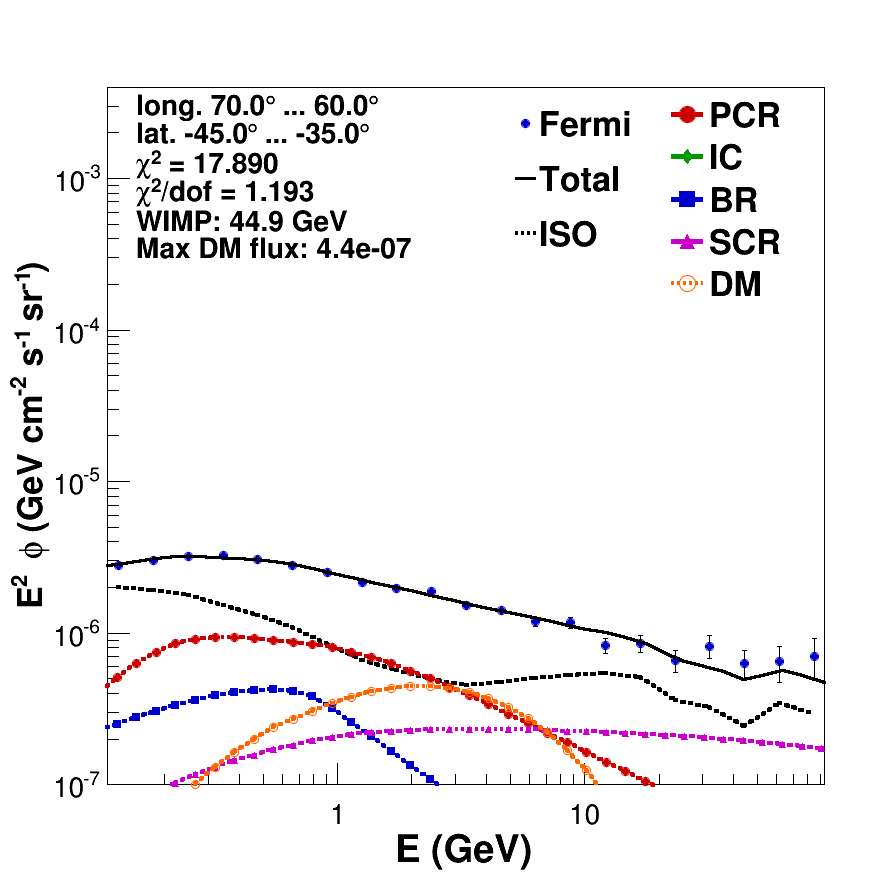}
\includegraphics[width=0.16\textwidth,height=0.16\textwidth,clip]{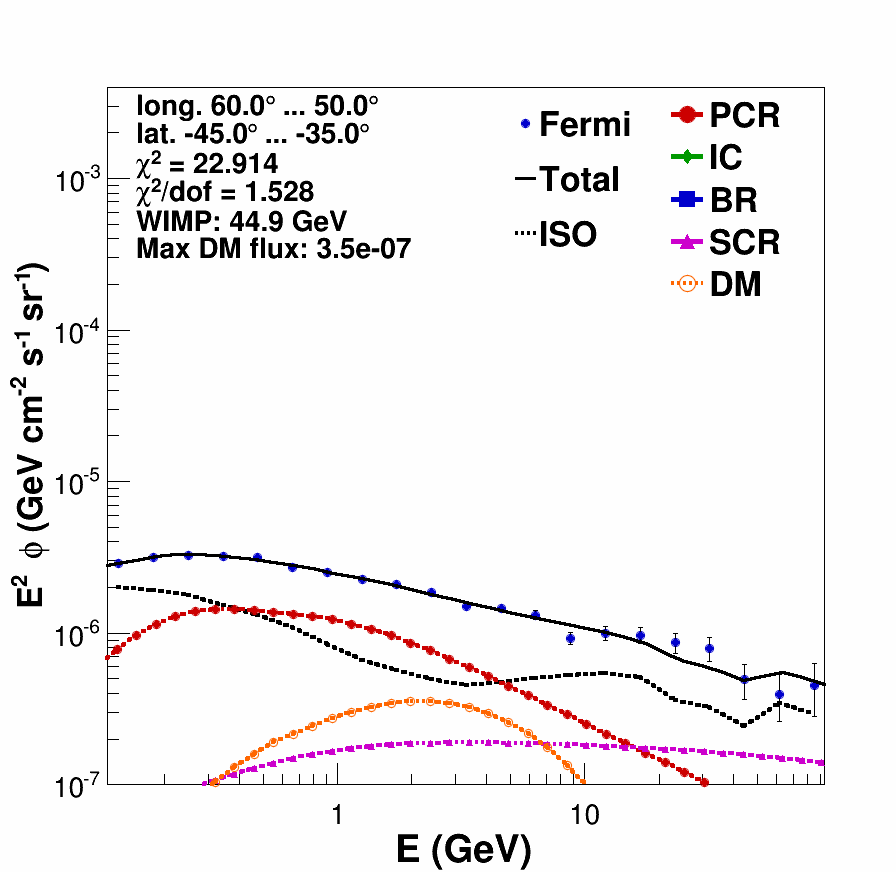}
\includegraphics[width=0.16\textwidth,height=0.16\textwidth,clip]{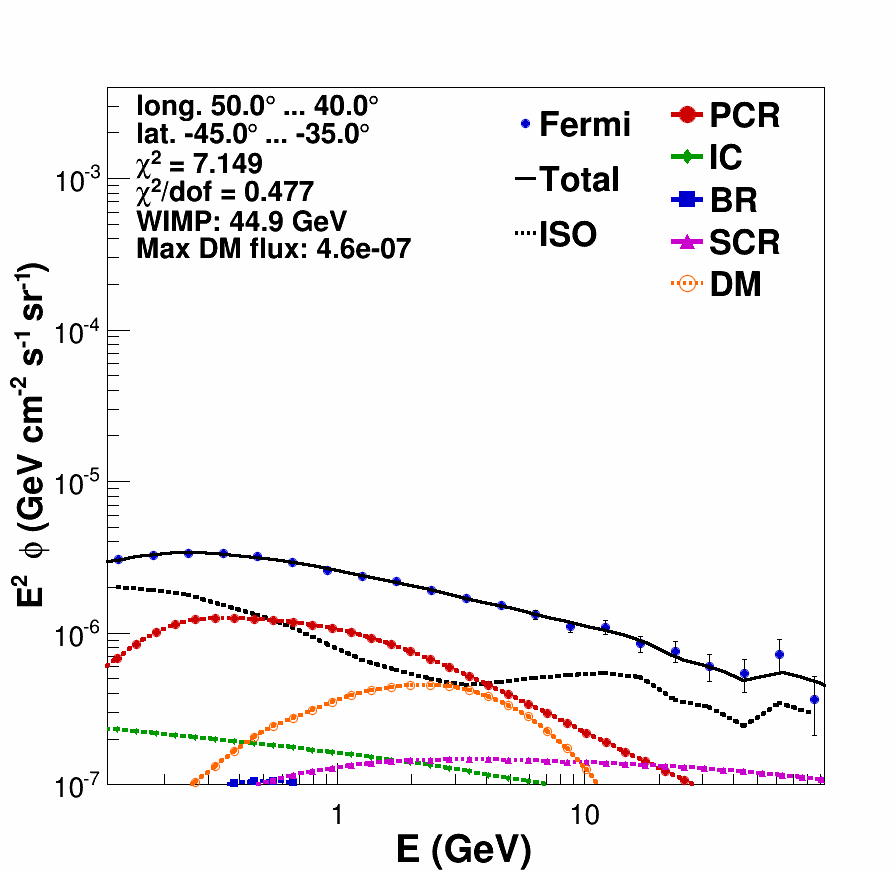}
\includegraphics[width=0.16\textwidth,height=0.16\textwidth,clip]{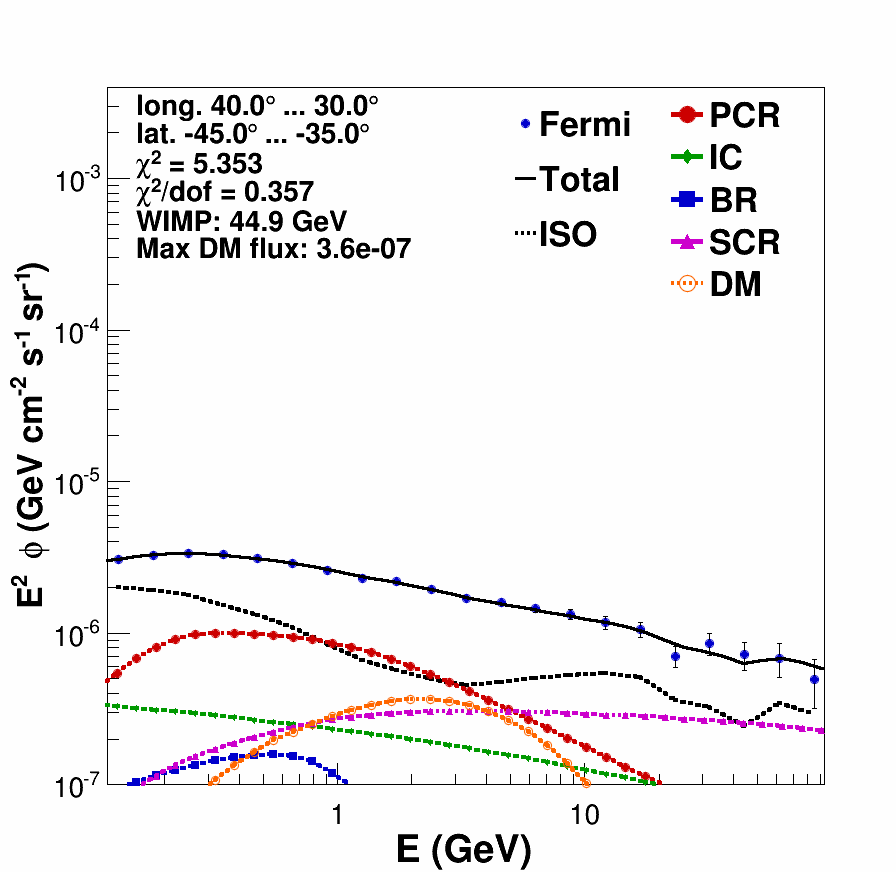}
\includegraphics[width=0.16\textwidth,height=0.16\textwidth,clip]{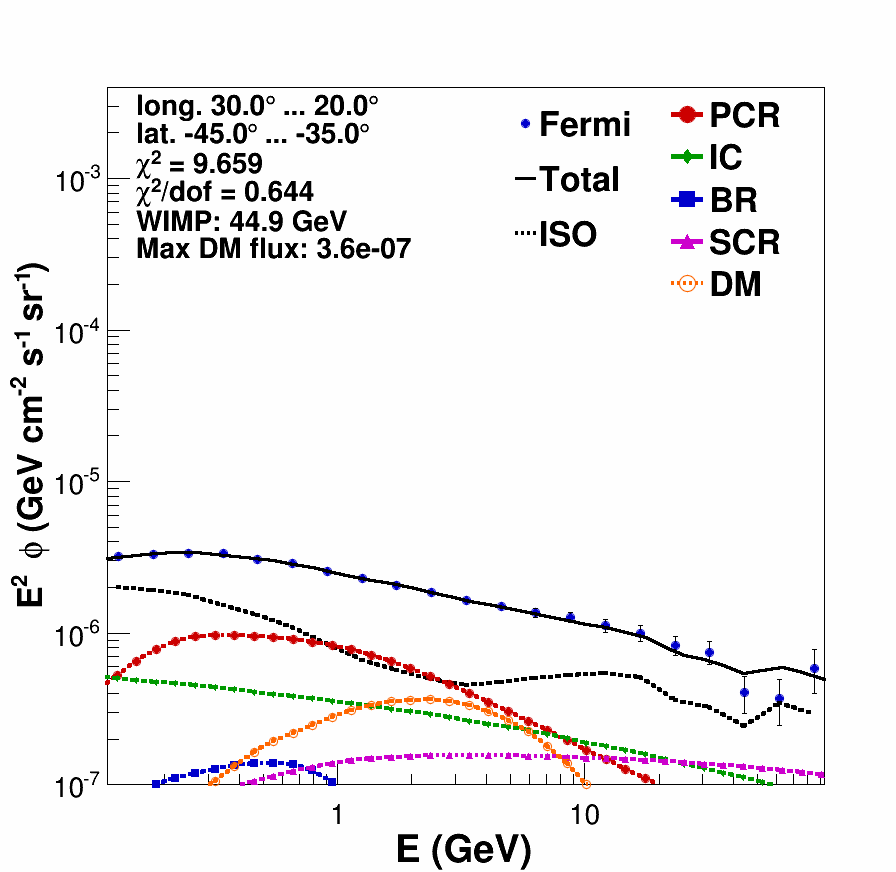}
\includegraphics[width=0.16\textwidth,height=0.16\textwidth,clip]{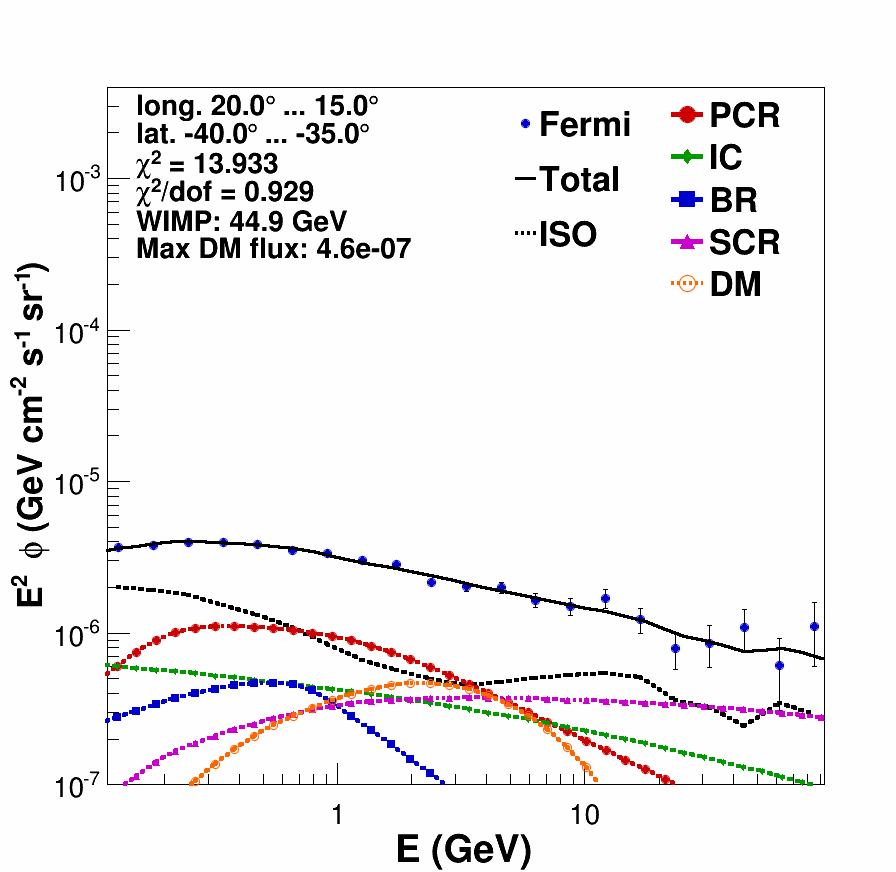}
\includegraphics[width=0.16\textwidth,height=0.16\textwidth,clip]{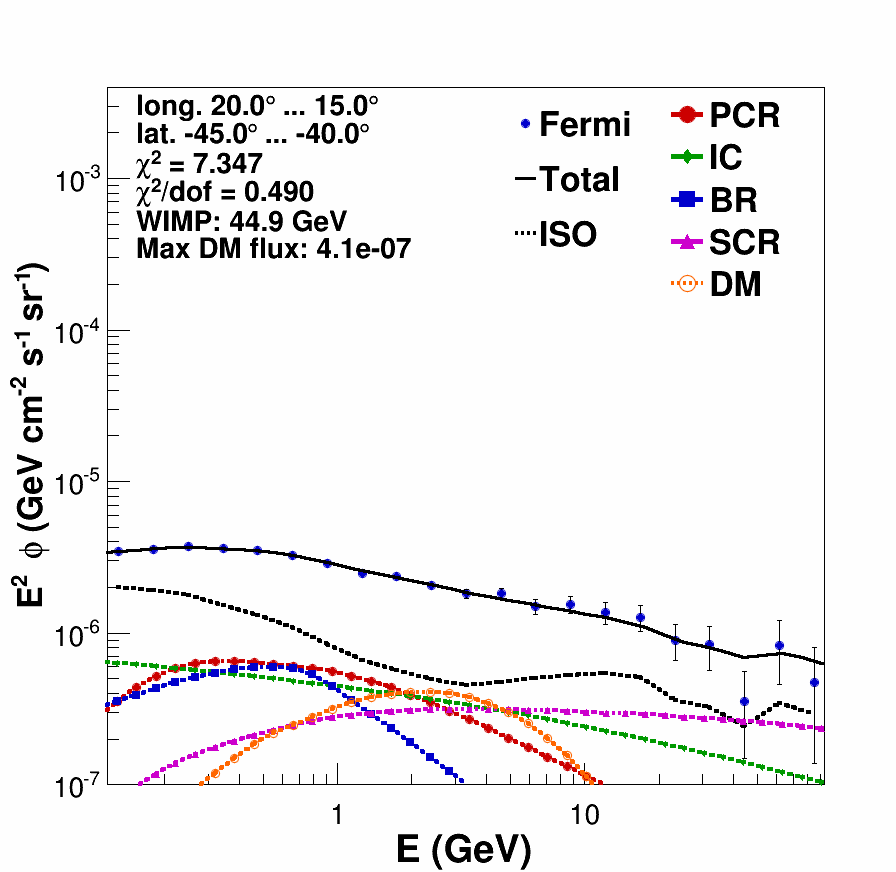}
\includegraphics[width=0.16\textwidth,height=0.16\textwidth,clip]{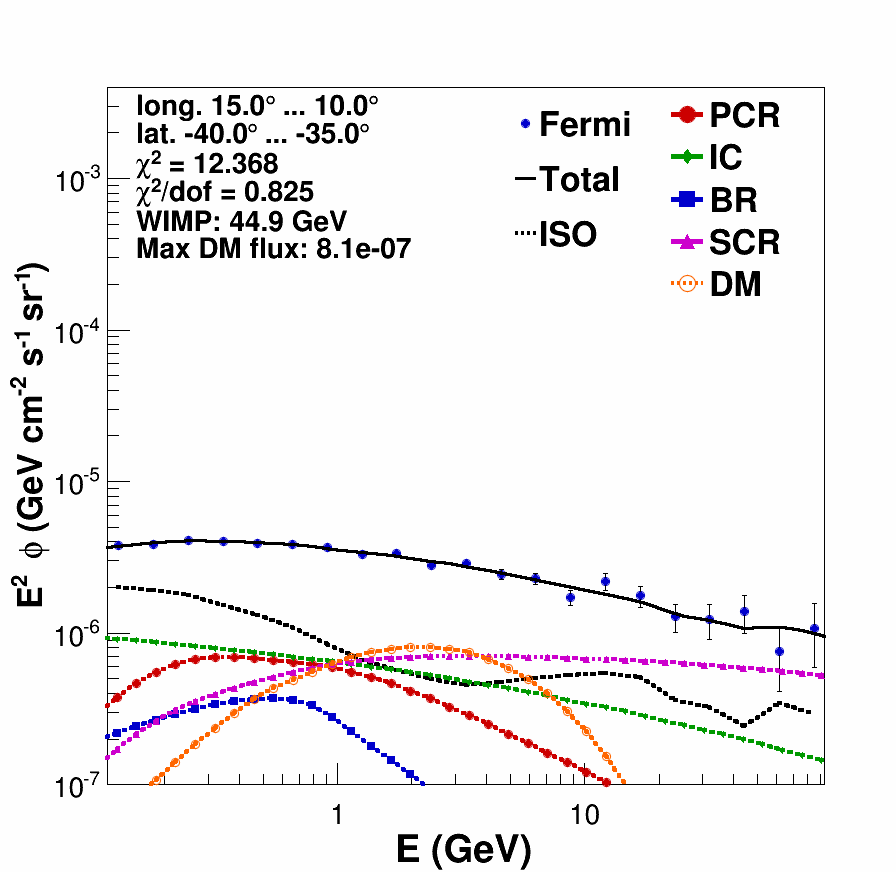}
\includegraphics[width=0.16\textwidth,height=0.16\textwidth,clip]{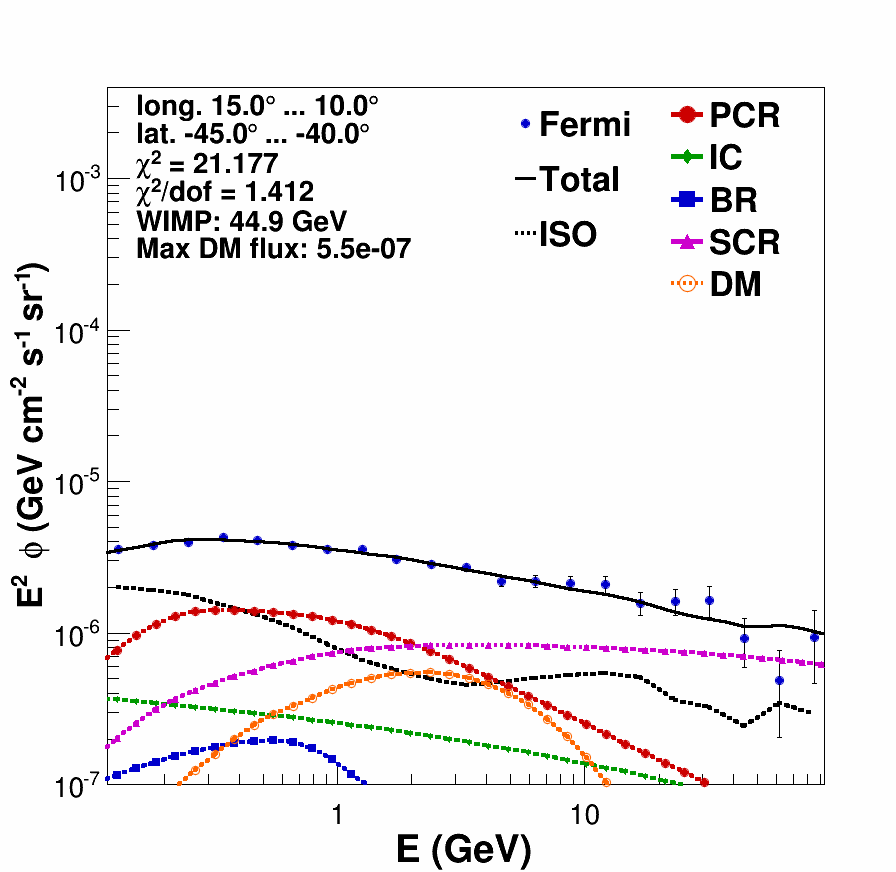}
\includegraphics[width=0.16\textwidth,height=0.16\textwidth,clip]{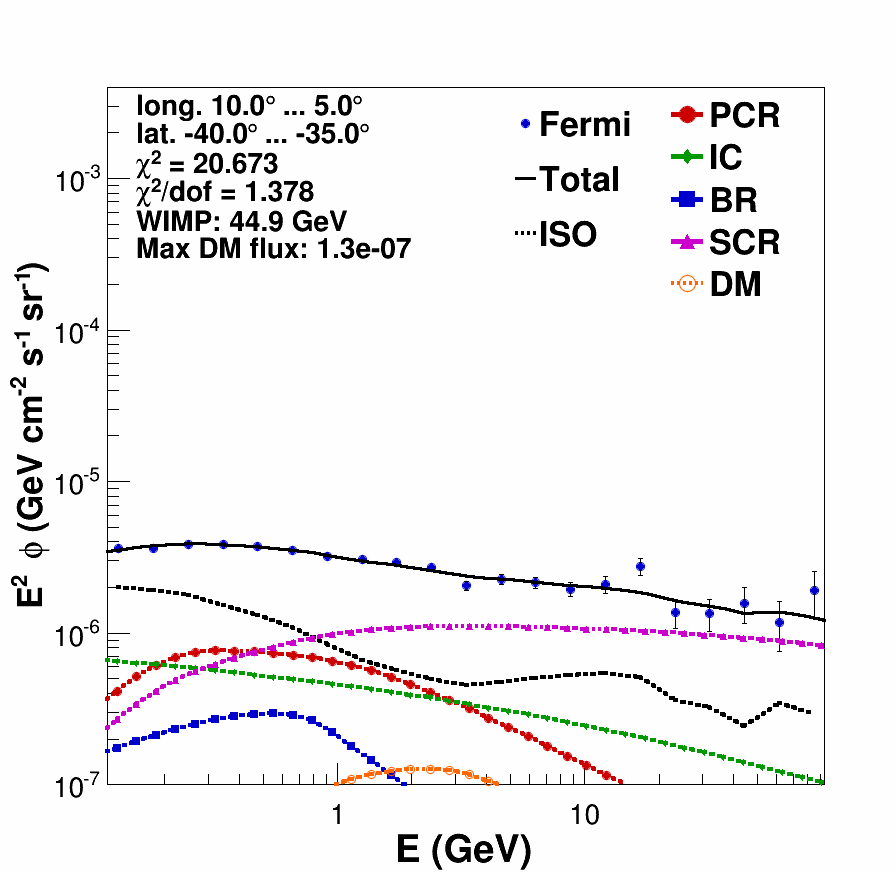}
\includegraphics[width=0.16\textwidth,height=0.16\textwidth,clip]{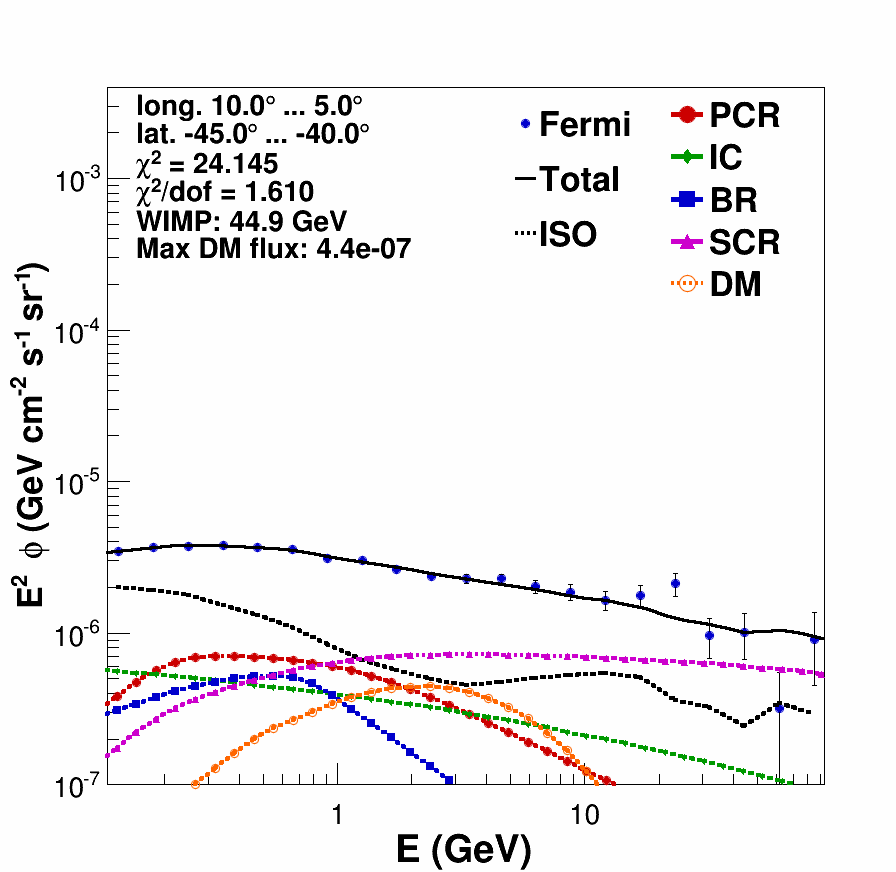}
\includegraphics[width=0.16\textwidth,height=0.16\textwidth,clip]{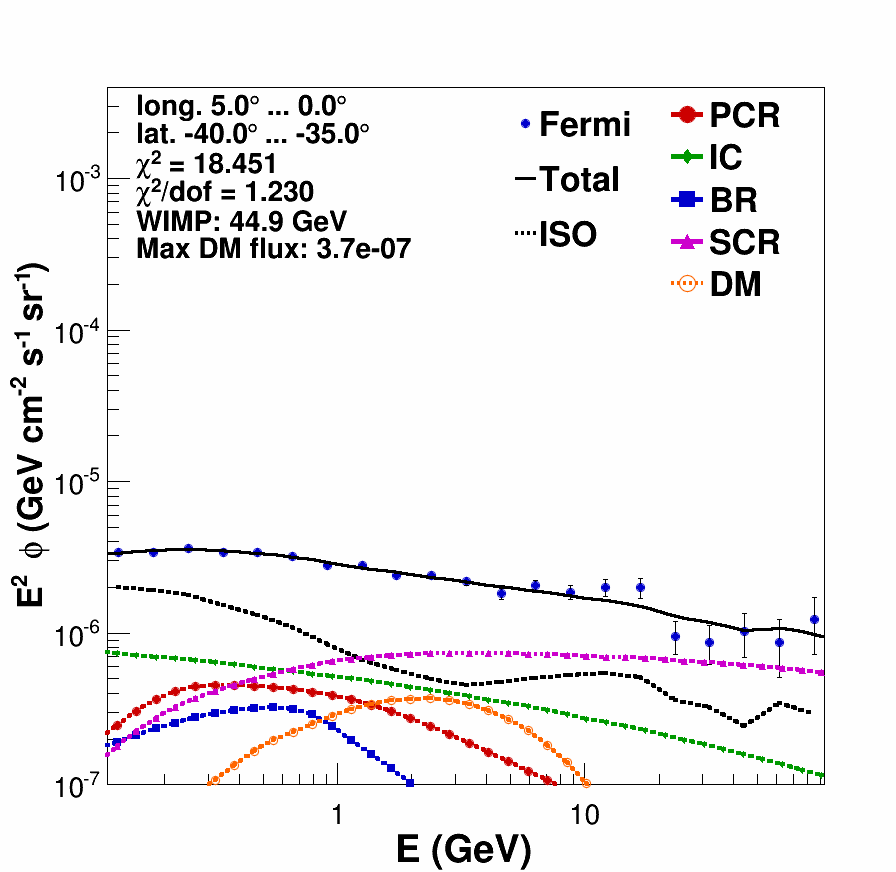}
\includegraphics[width=0.16\textwidth,height=0.16\textwidth,clip]{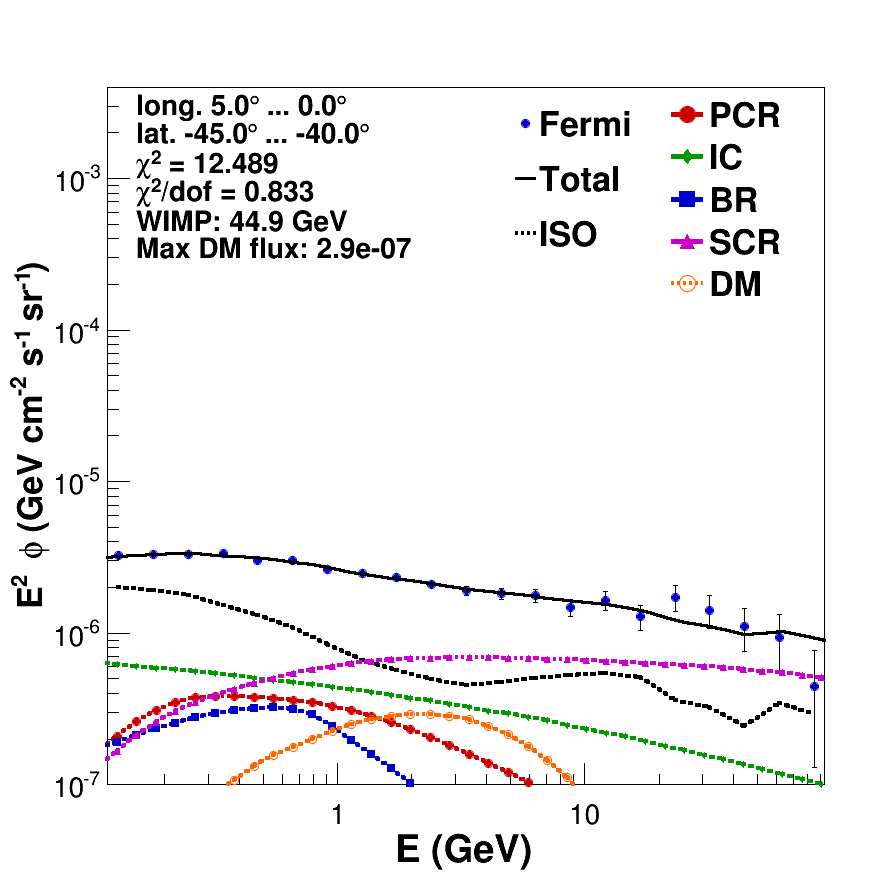}
\includegraphics[width=0.16\textwidth,height=0.16\textwidth,clip]{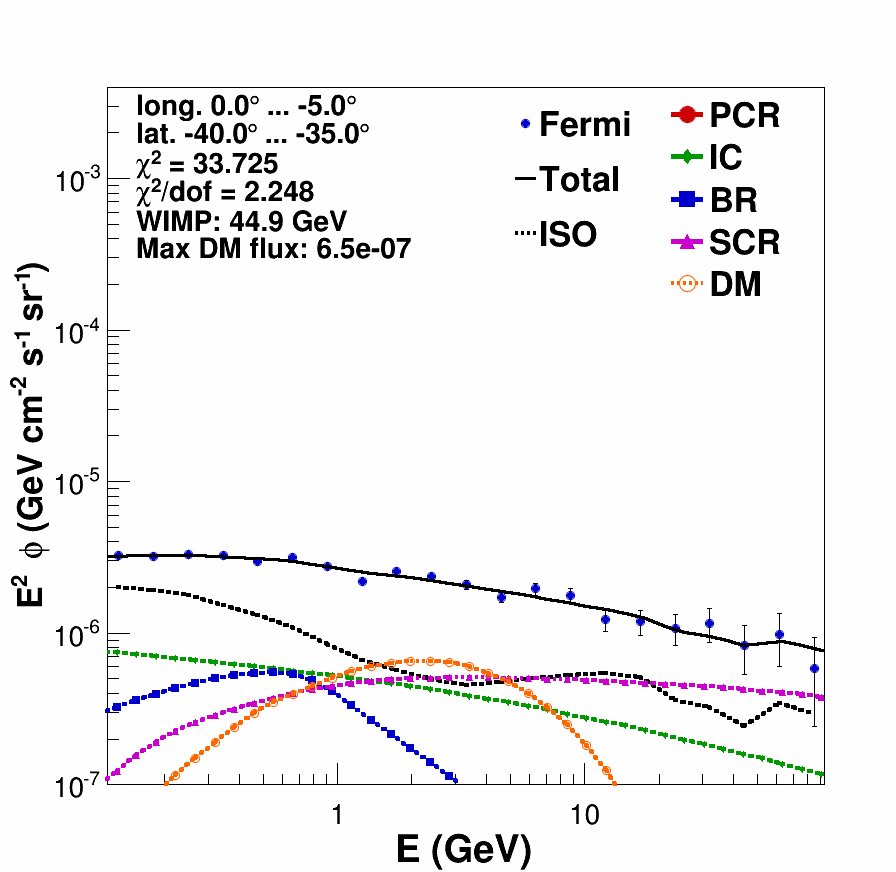}
\includegraphics[width=0.16\textwidth,height=0.16\textwidth,clip]{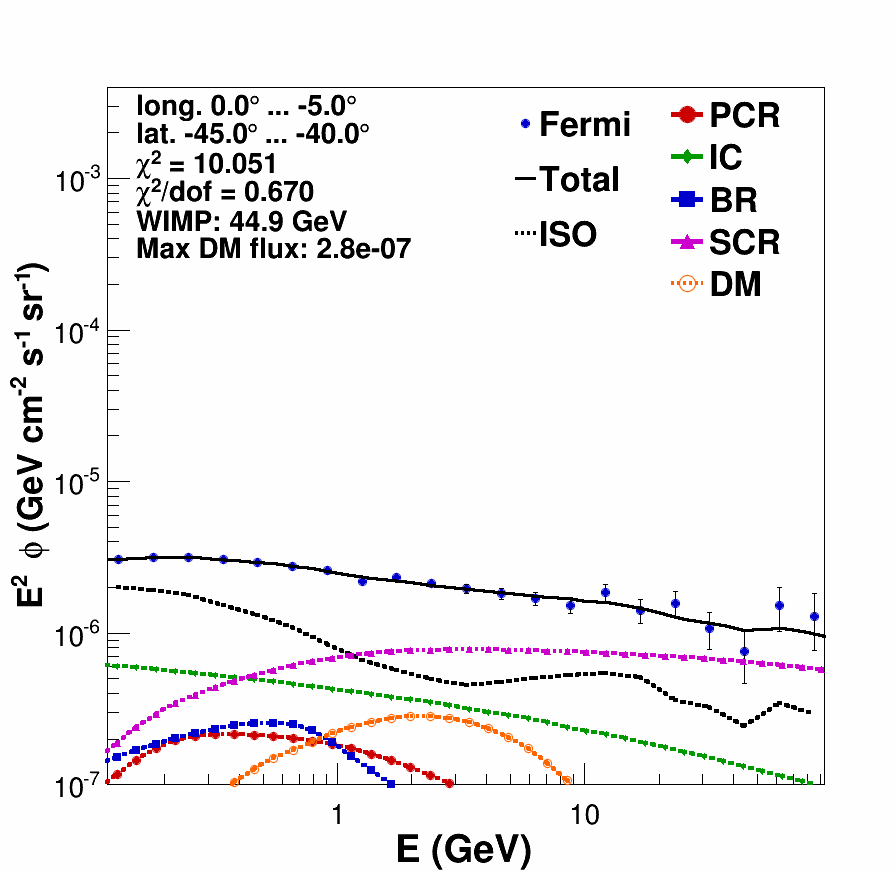}
\includegraphics[width=0.16\textwidth,height=0.16\textwidth,clip]{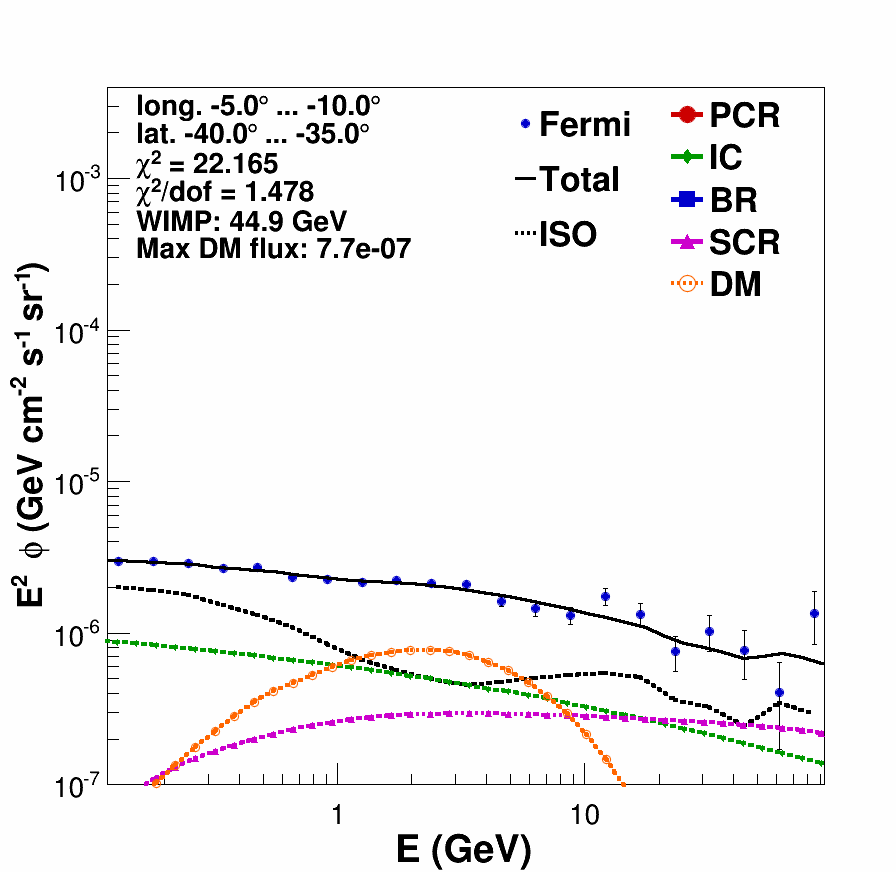}
\includegraphics[width=0.16\textwidth,height=0.16\textwidth,clip]{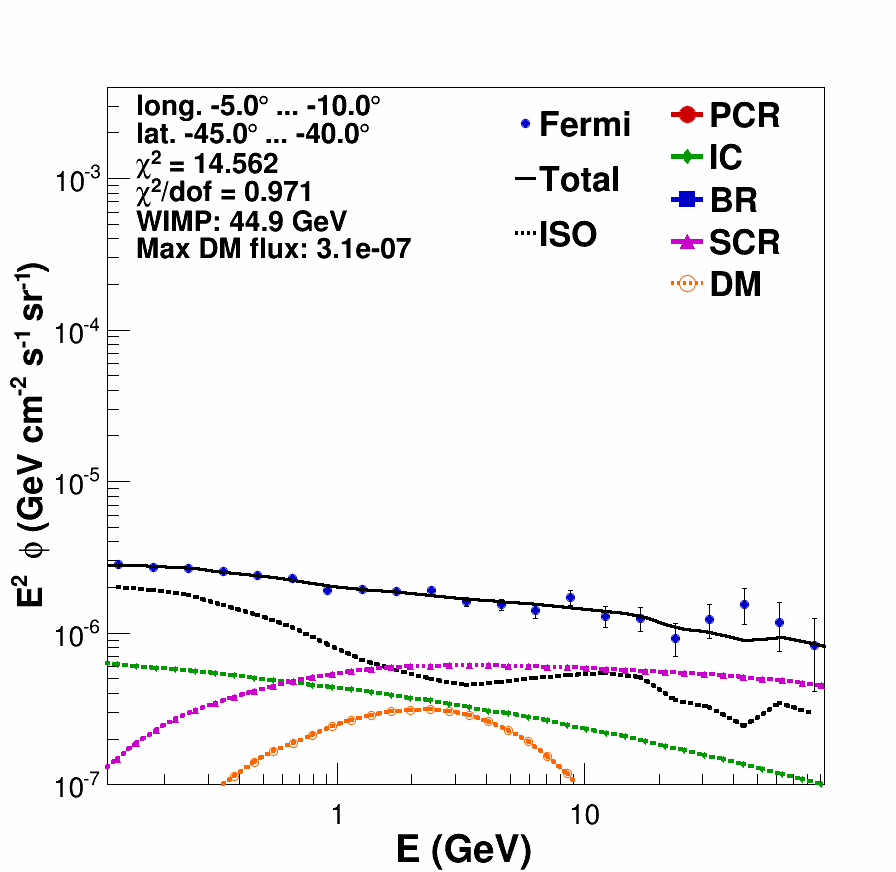}
\includegraphics[width=0.16\textwidth,height=0.16\textwidth,clip]{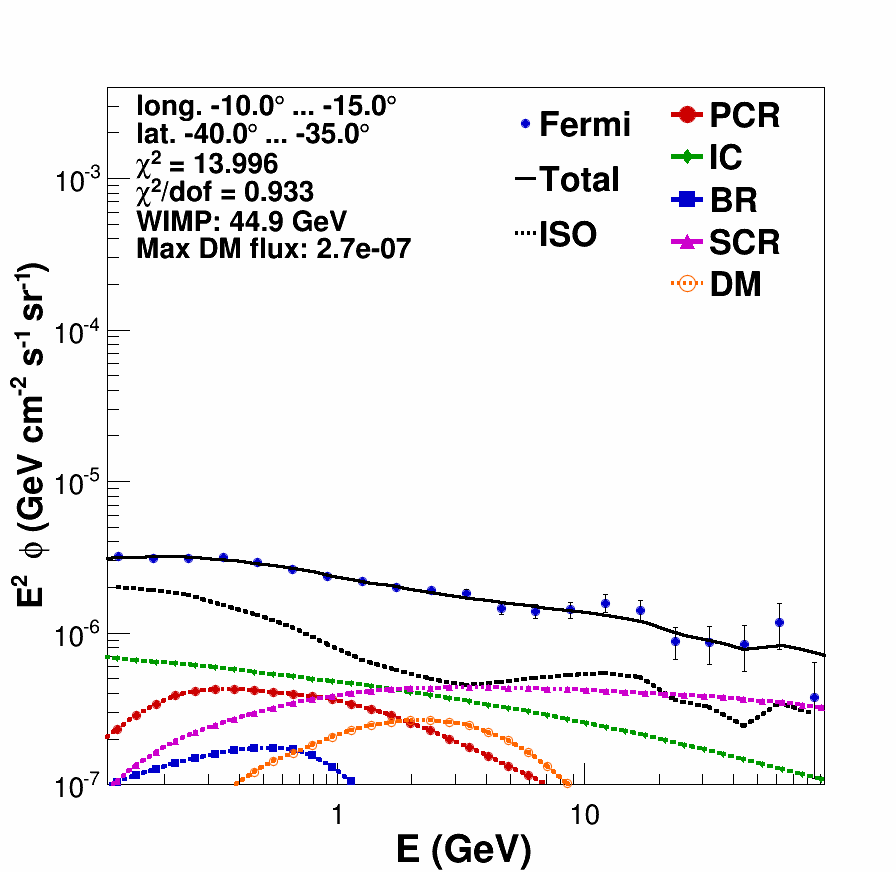}
\includegraphics[width=0.16\textwidth,height=0.16\textwidth,clip]{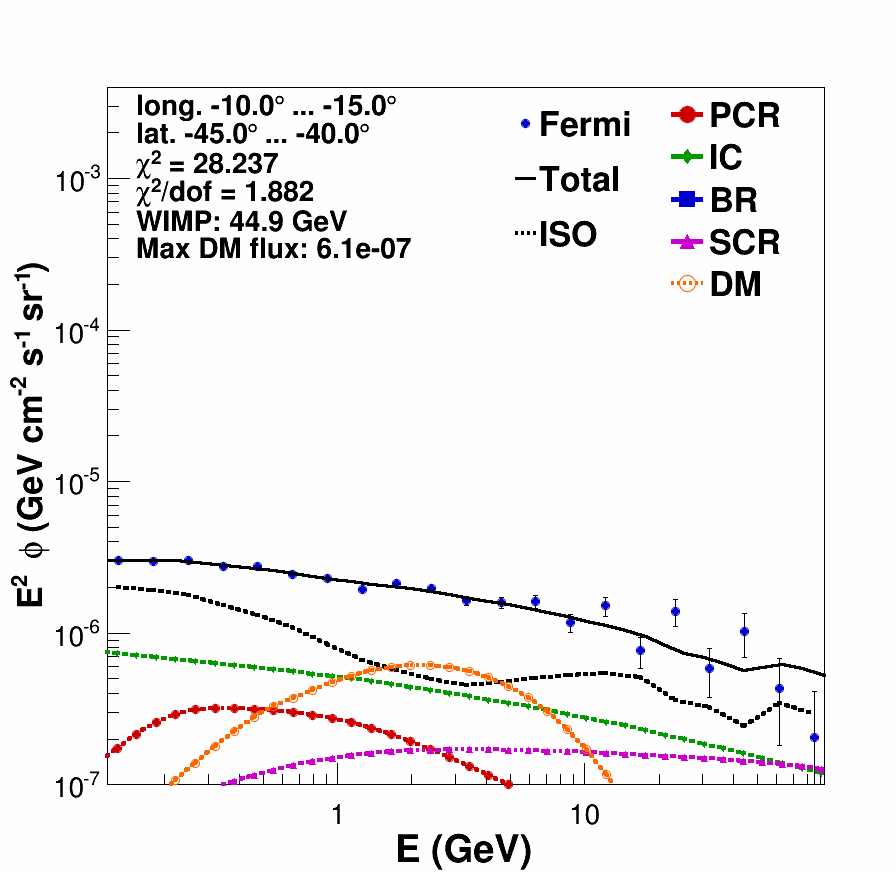}
\includegraphics[width=0.16\textwidth,height=0.16\textwidth,clip]{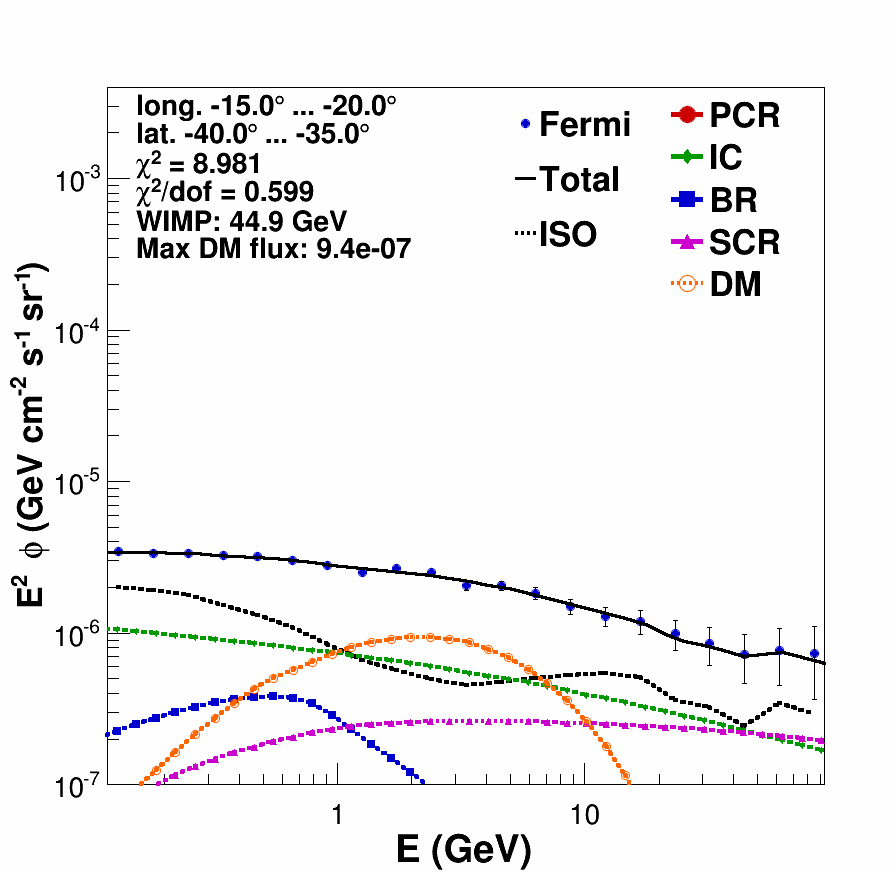}
\includegraphics[width=0.16\textwidth,height=0.16\textwidth,clip]{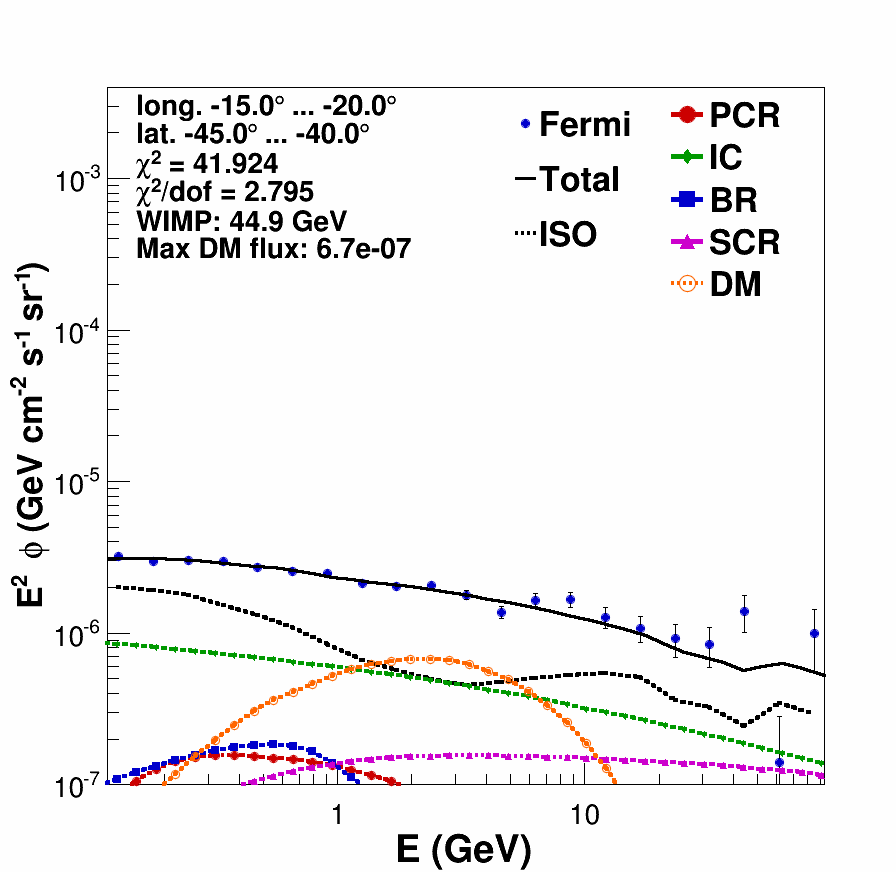}
\includegraphics[width=0.16\textwidth,height=0.16\textwidth,clip]{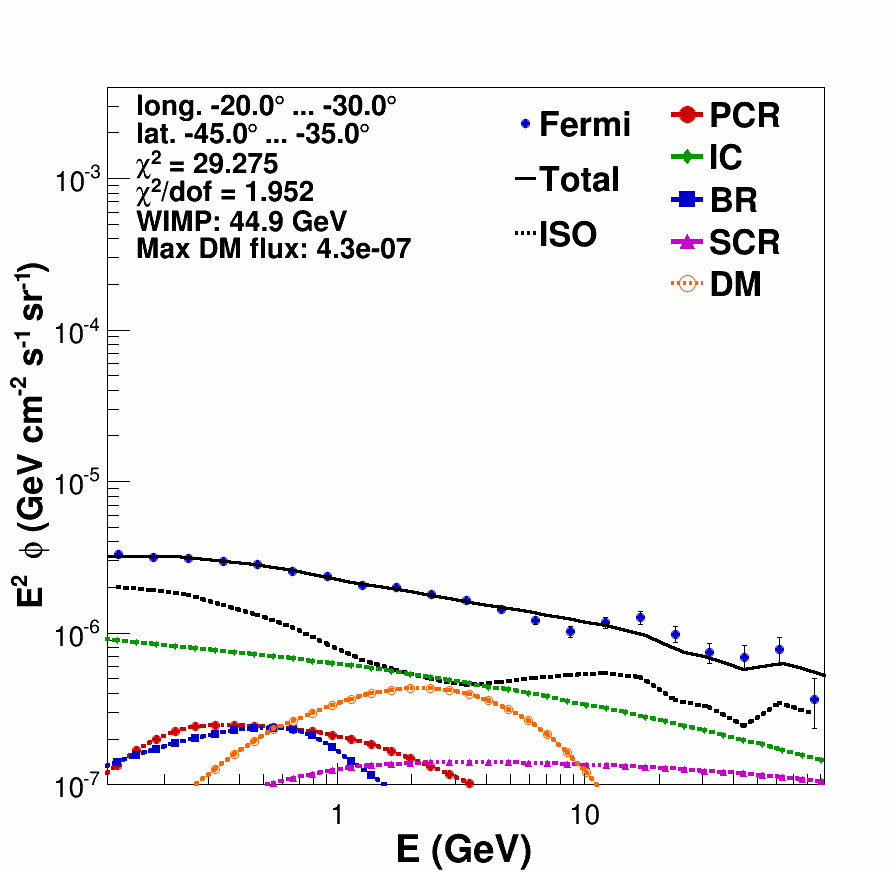}
\includegraphics[width=0.16\textwidth,height=0.16\textwidth,clip]{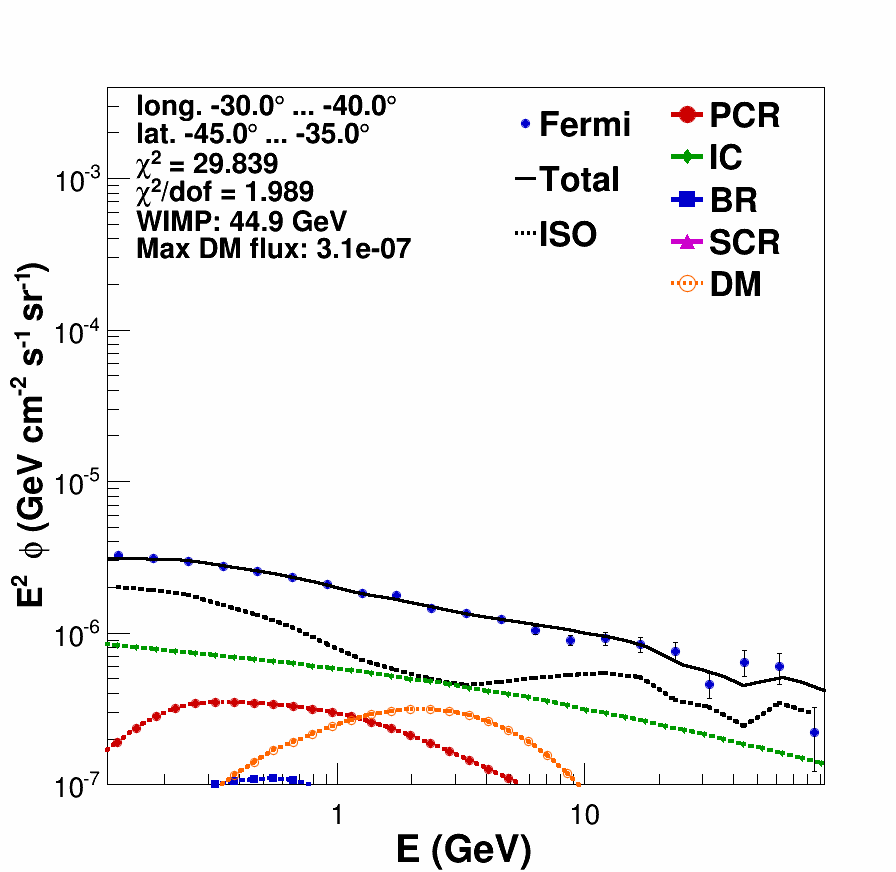}
\includegraphics[width=0.16\textwidth,height=0.16\textwidth,clip]{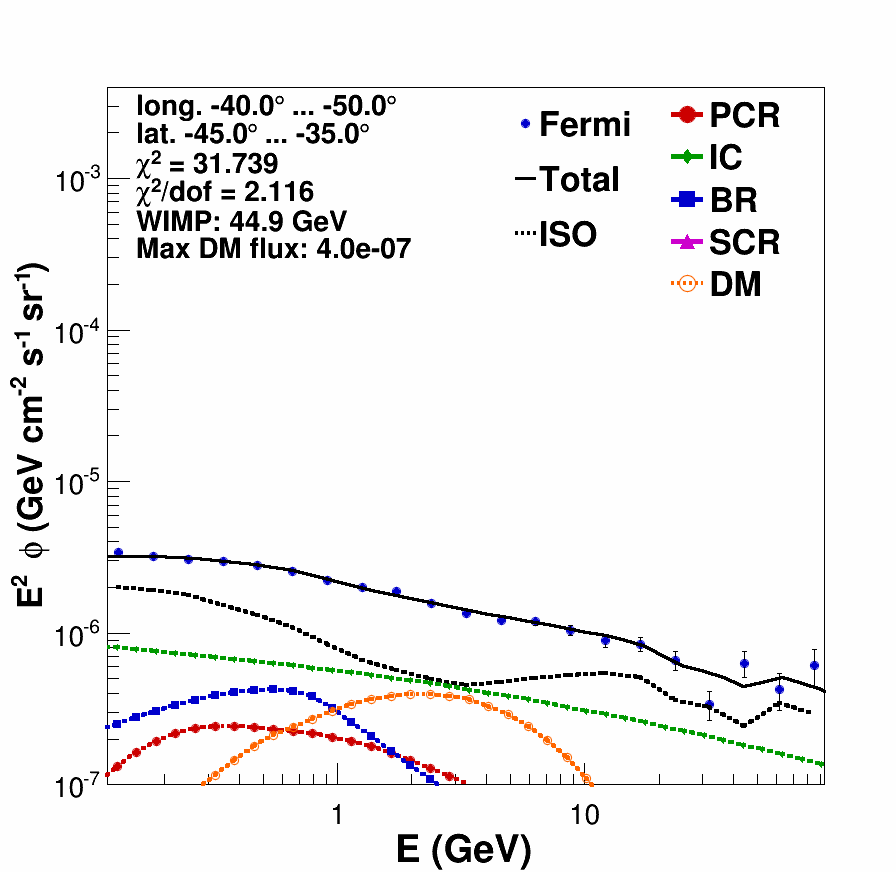}
\includegraphics[width=0.16\textwidth,height=0.16\textwidth,clip]{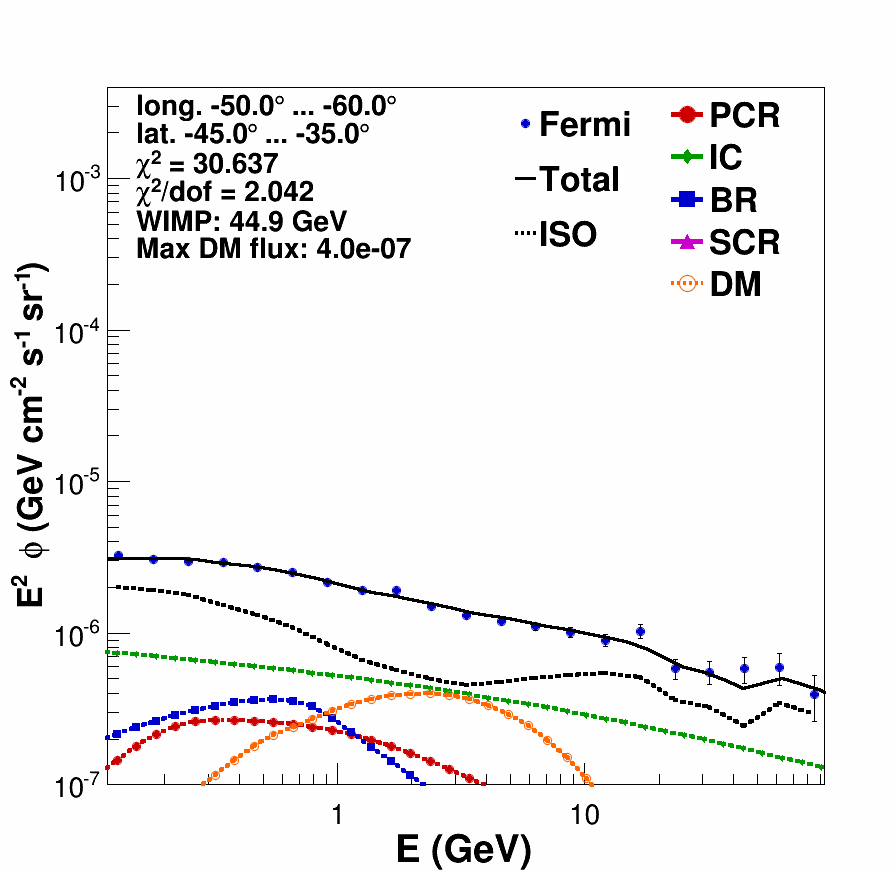}
\includegraphics[width=0.16\textwidth,height=0.16\textwidth,clip]{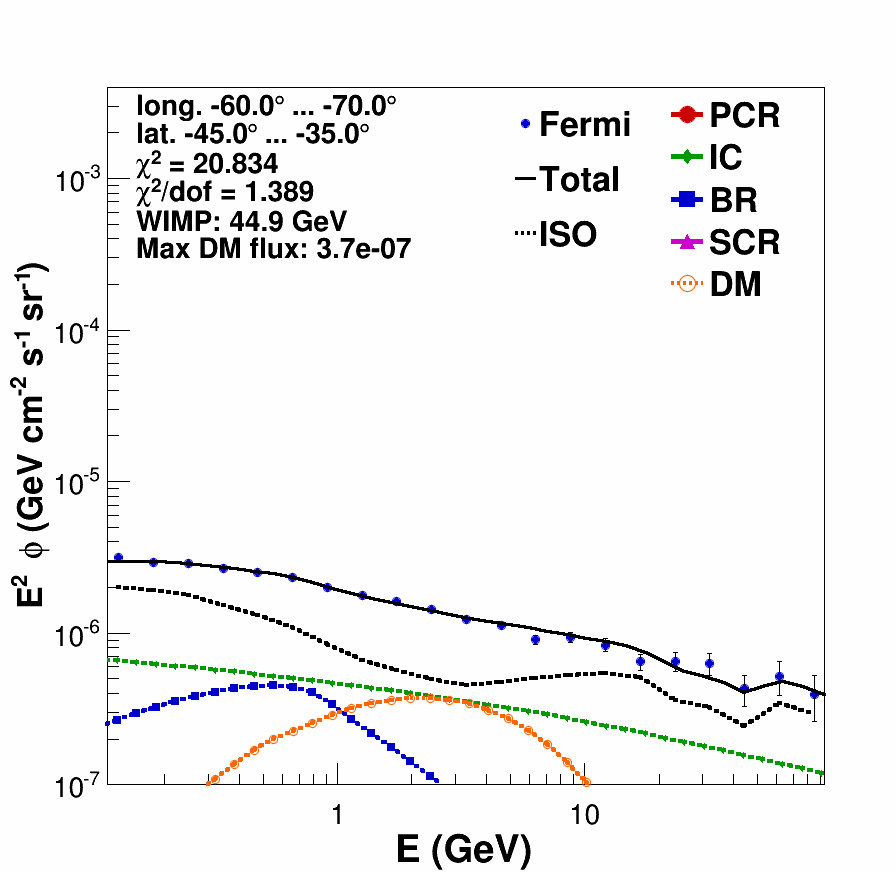}
\includegraphics[width=0.16\textwidth,height=0.16\textwidth,clip]{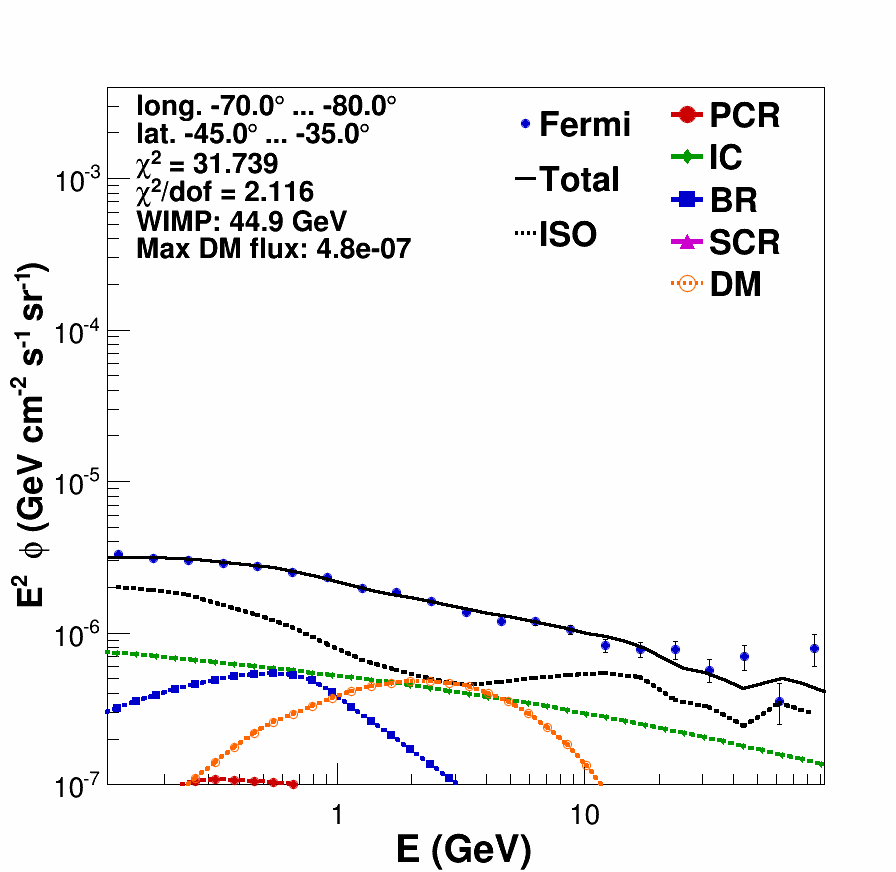}
\includegraphics[width=0.16\textwidth,height=0.16\textwidth,clip]{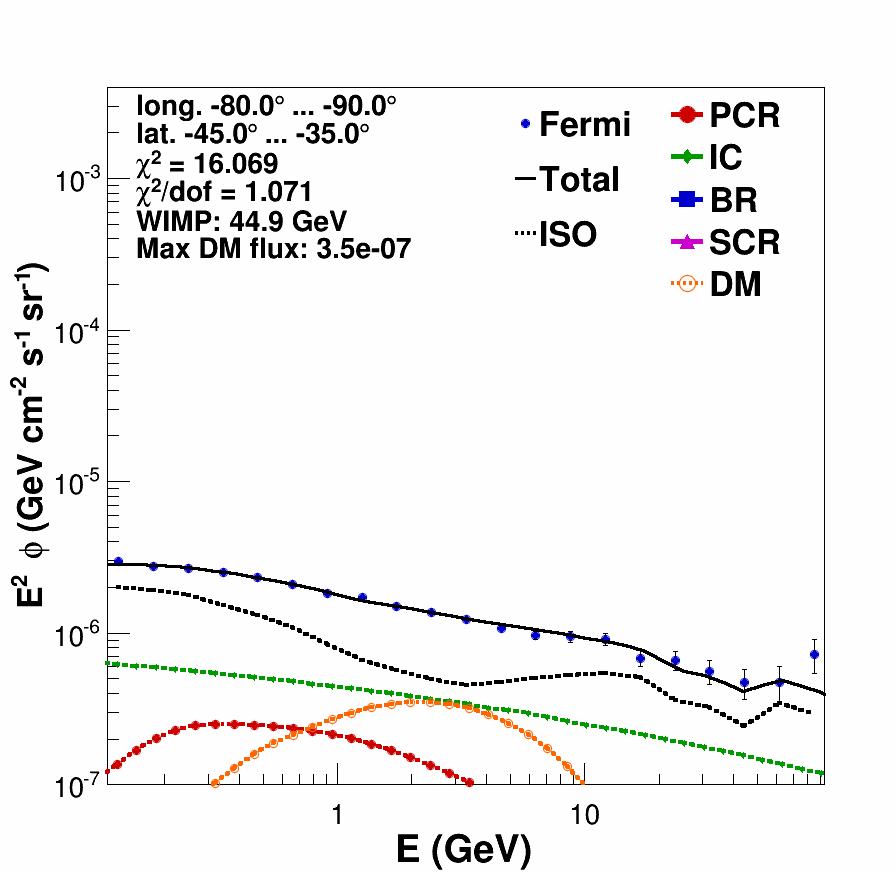}
\includegraphics[width=0.16\textwidth,height=0.16\textwidth,clip]{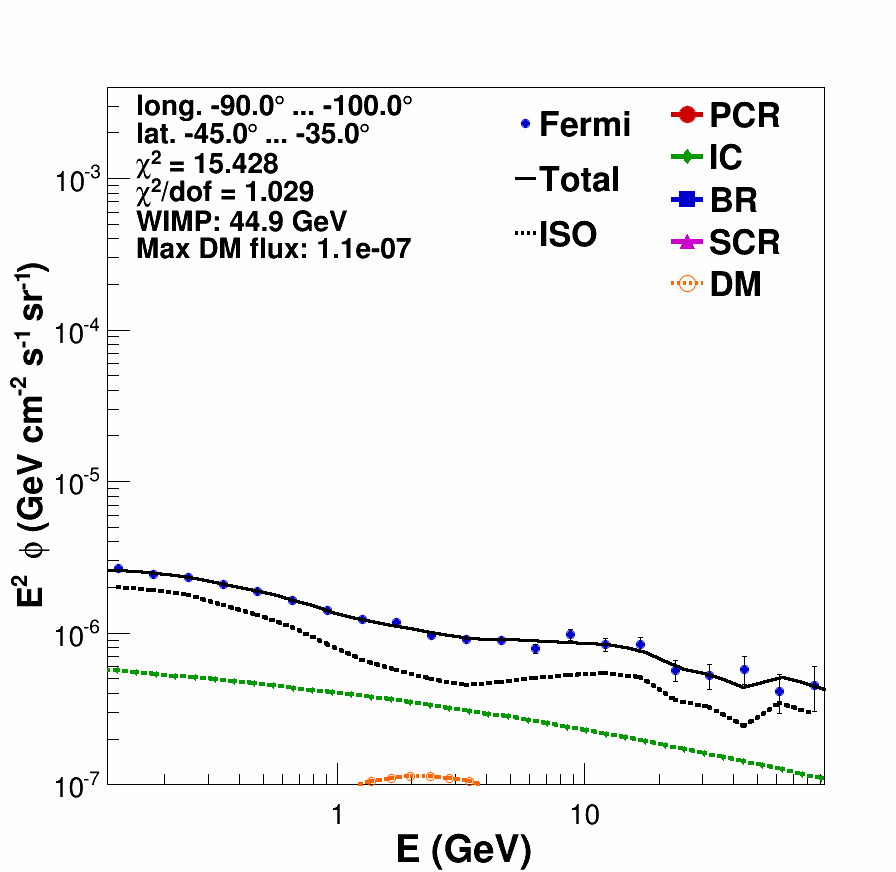}
\includegraphics[width=0.16\textwidth,height=0.16\textwidth,clip]{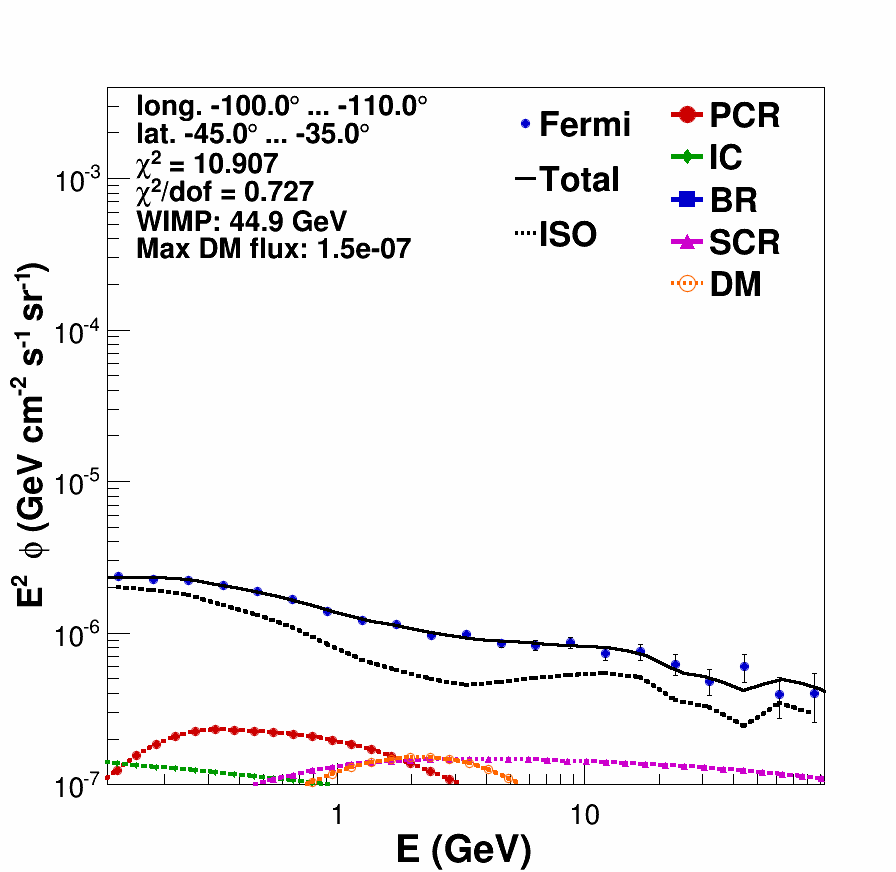}
\includegraphics[width=0.16\textwidth,height=0.16\textwidth,clip]{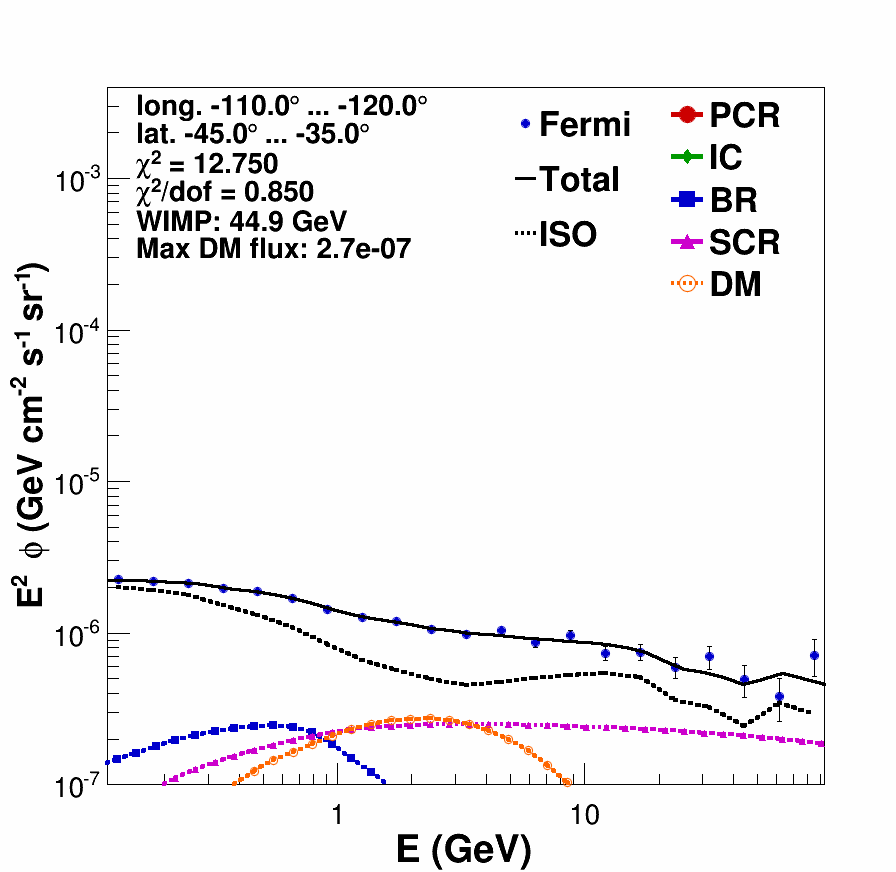}
\includegraphics[width=0.16\textwidth,height=0.16\textwidth,clip]{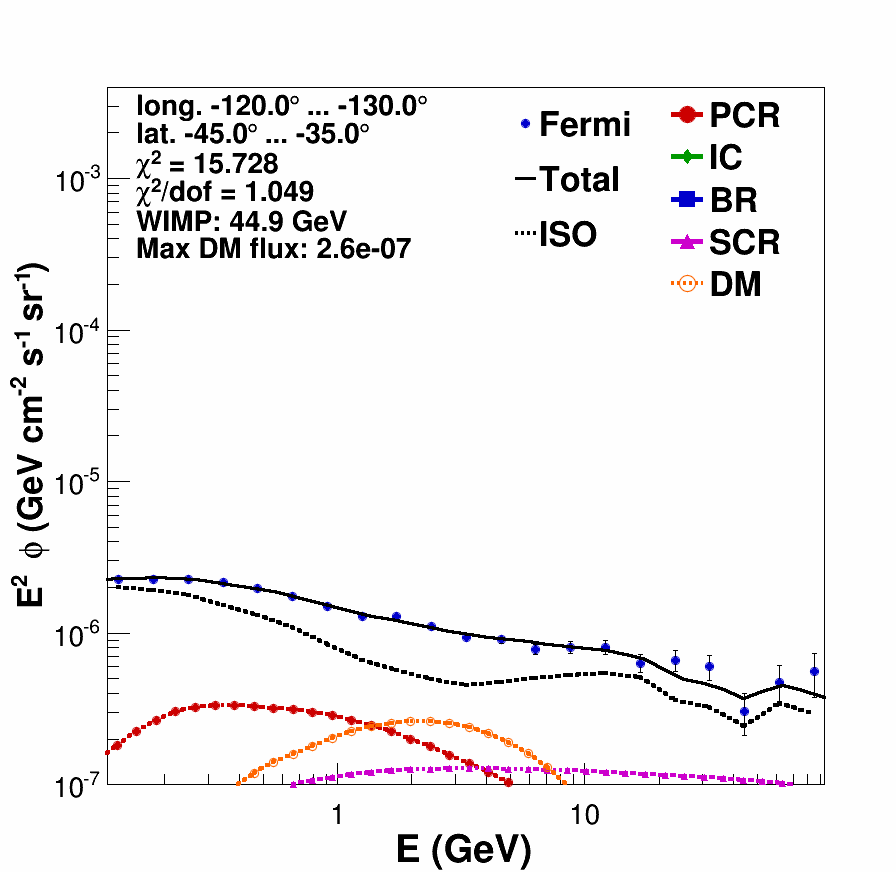}
\includegraphics[width=0.16\textwidth,height=0.16\textwidth,clip]{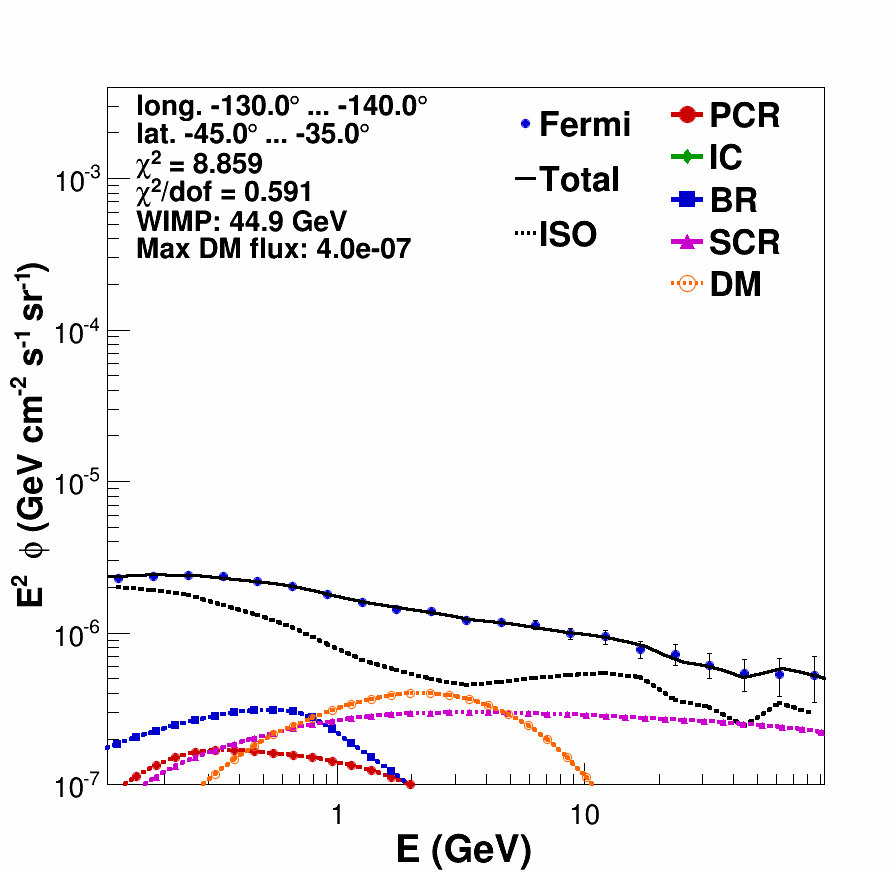}
\includegraphics[width=0.16\textwidth,height=0.16\textwidth,clip]{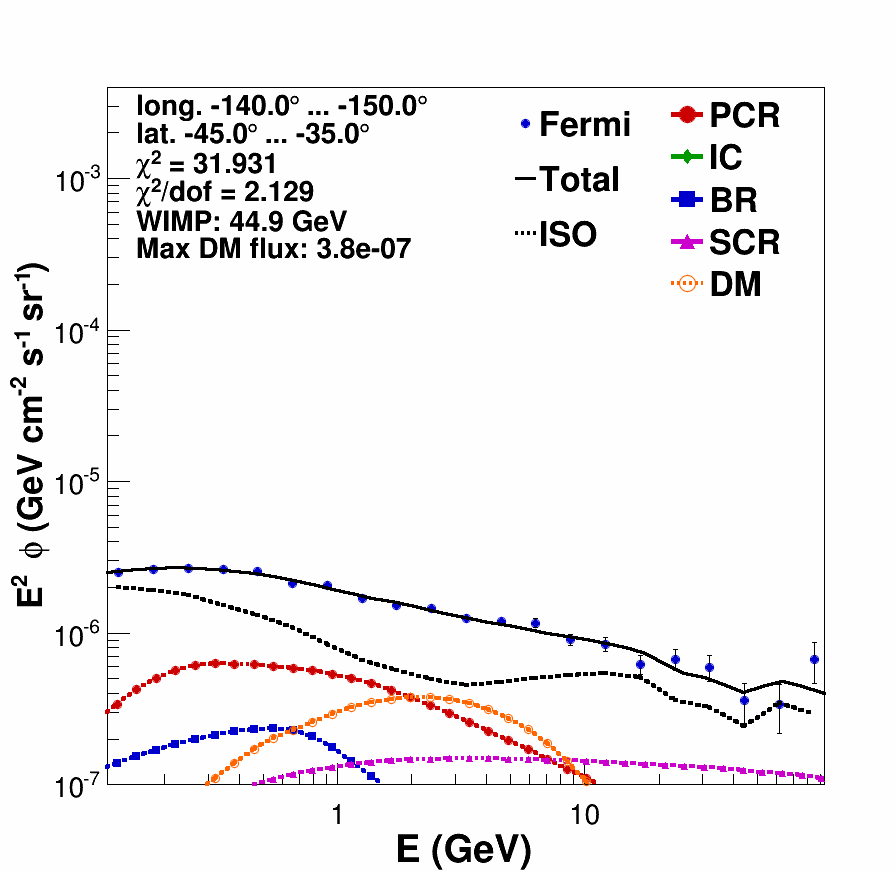}
\includegraphics[width=0.16\textwidth,height=0.16\textwidth,clip]{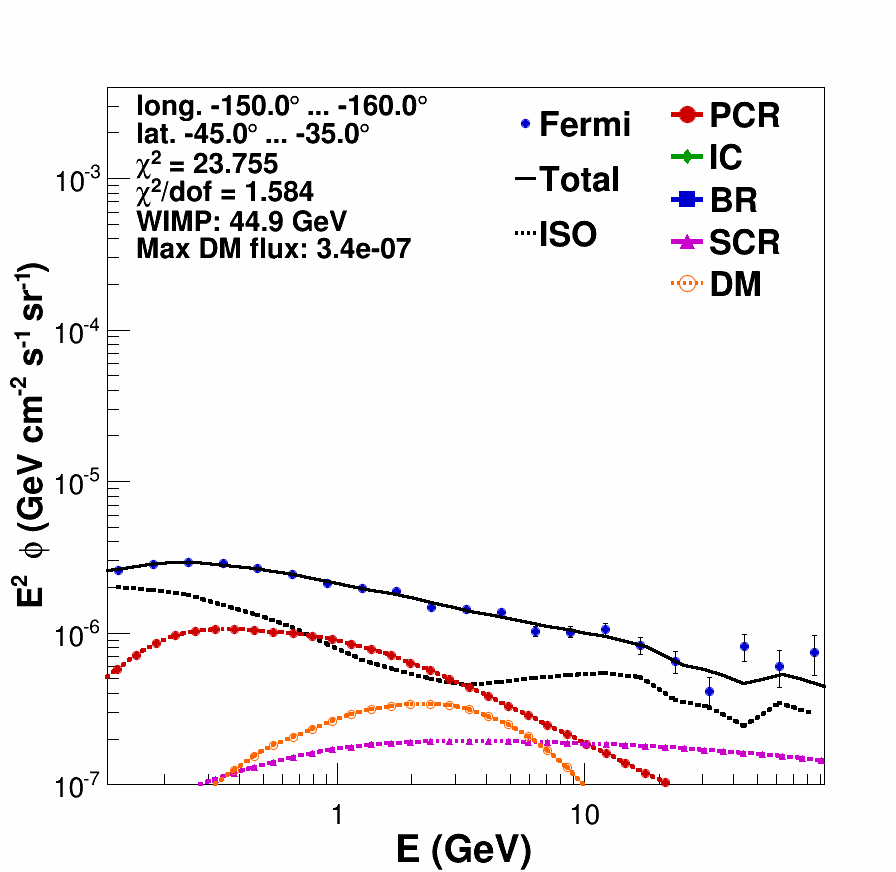}
\includegraphics[width=0.16\textwidth,height=0.16\textwidth,clip]{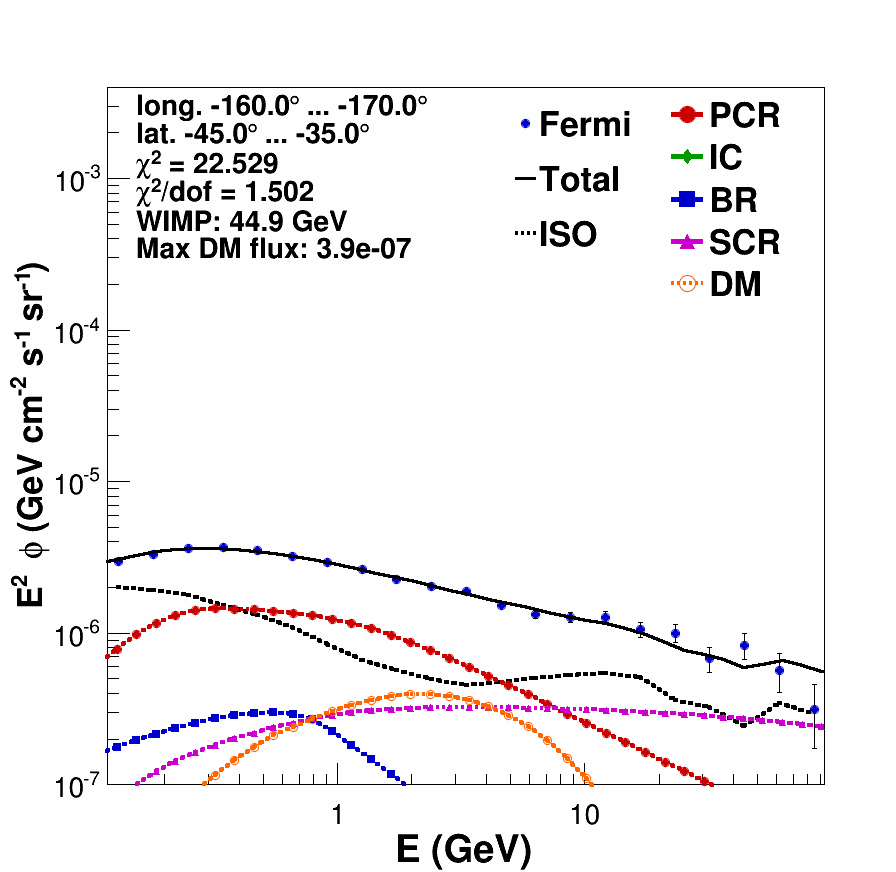}
\includegraphics[width=0.16\textwidth,height=0.16\textwidth,clip]{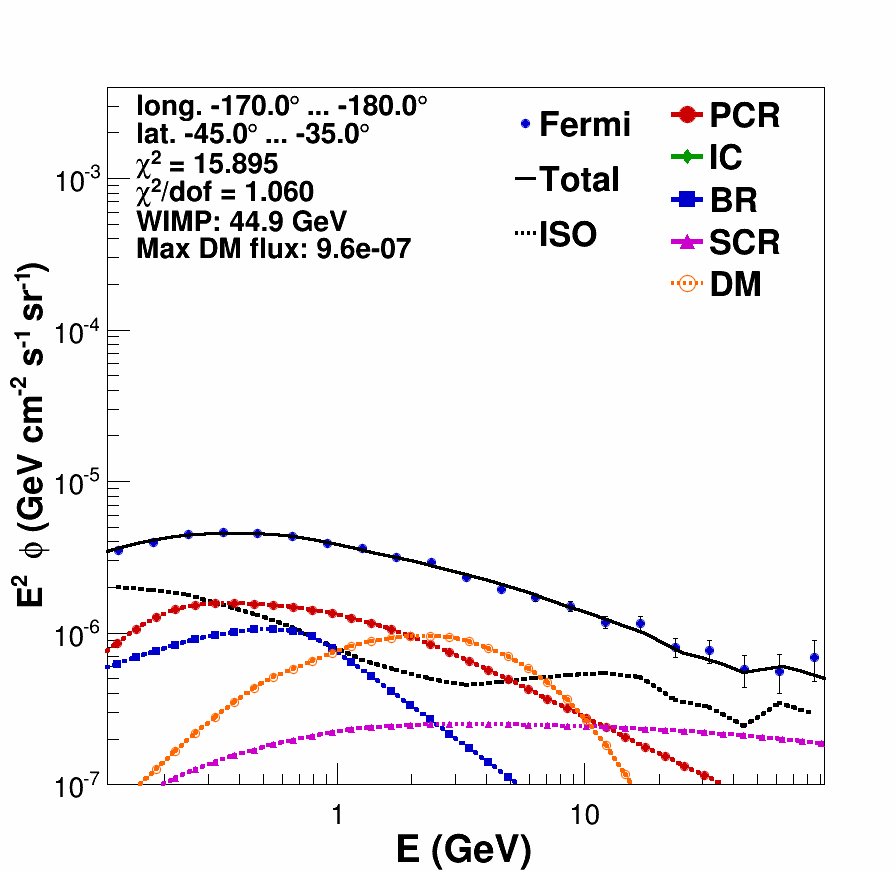}%%%%r16
\caption[]{Template fits for latitudes  with $-45.0^\circ<b<-35.0^\circ$ and longitudes decreasing from 180$^\circ$ to -180$^\circ$.} \label{F49}
\end{figure}
\begin{figure}
\centering
\includegraphics[width=0.16\textwidth,height=0.16\textwidth,clip]{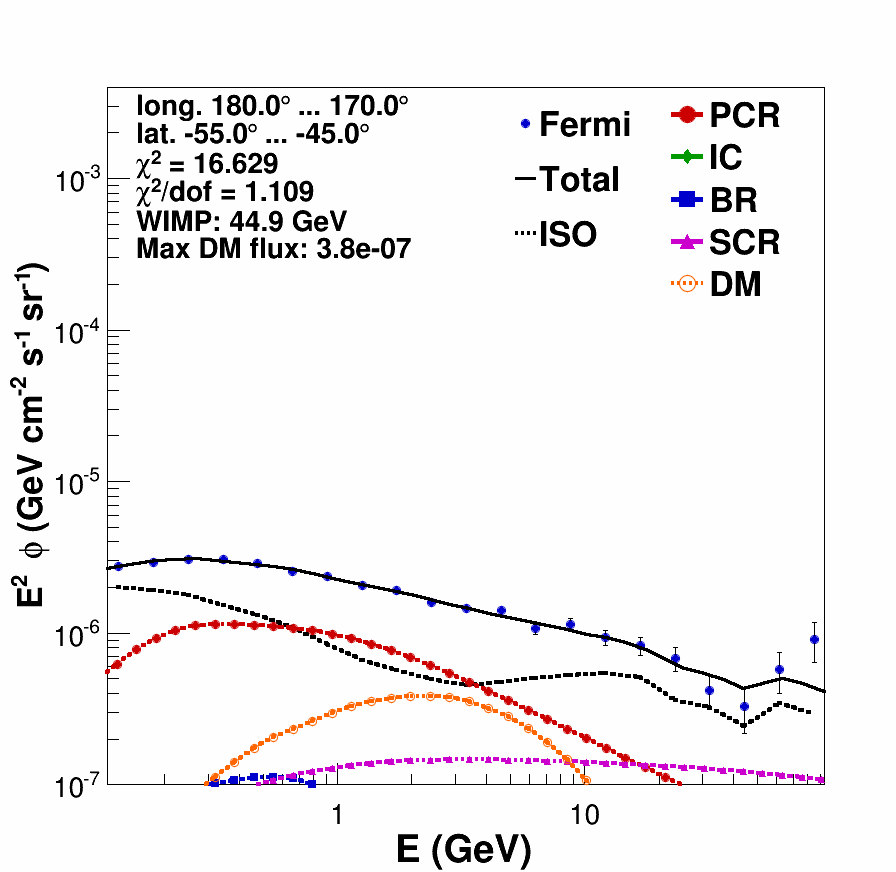}
\includegraphics[width=0.16\textwidth,height=0.16\textwidth,clip]{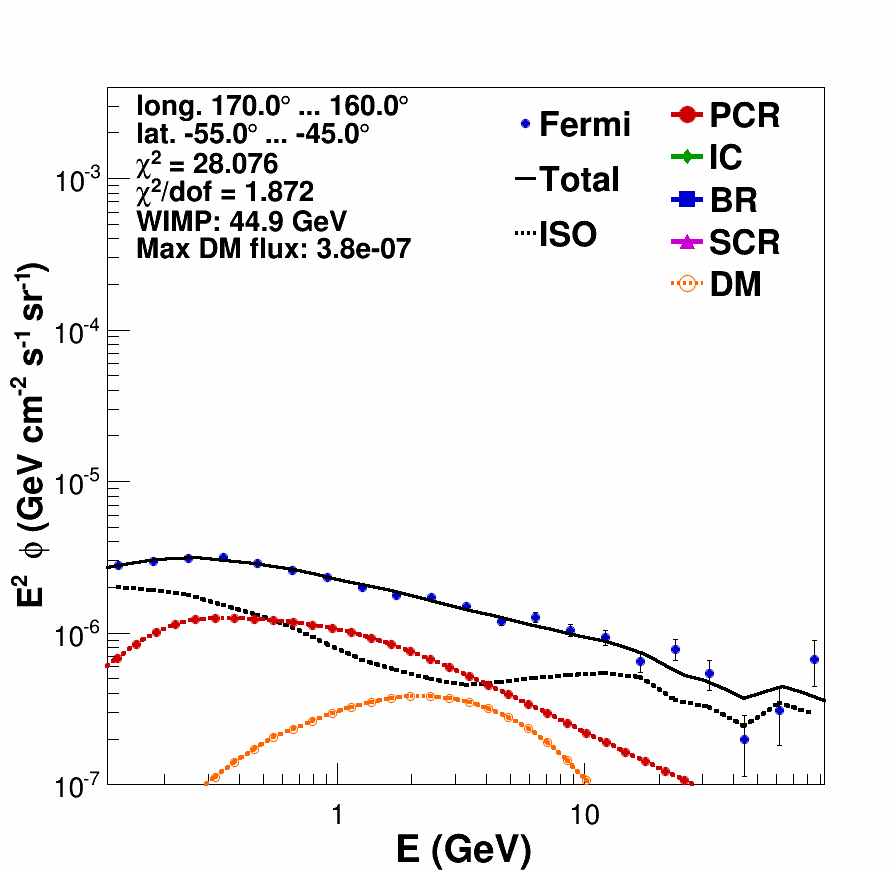}
\includegraphics[width=0.16\textwidth,height=0.16\textwidth,clip]{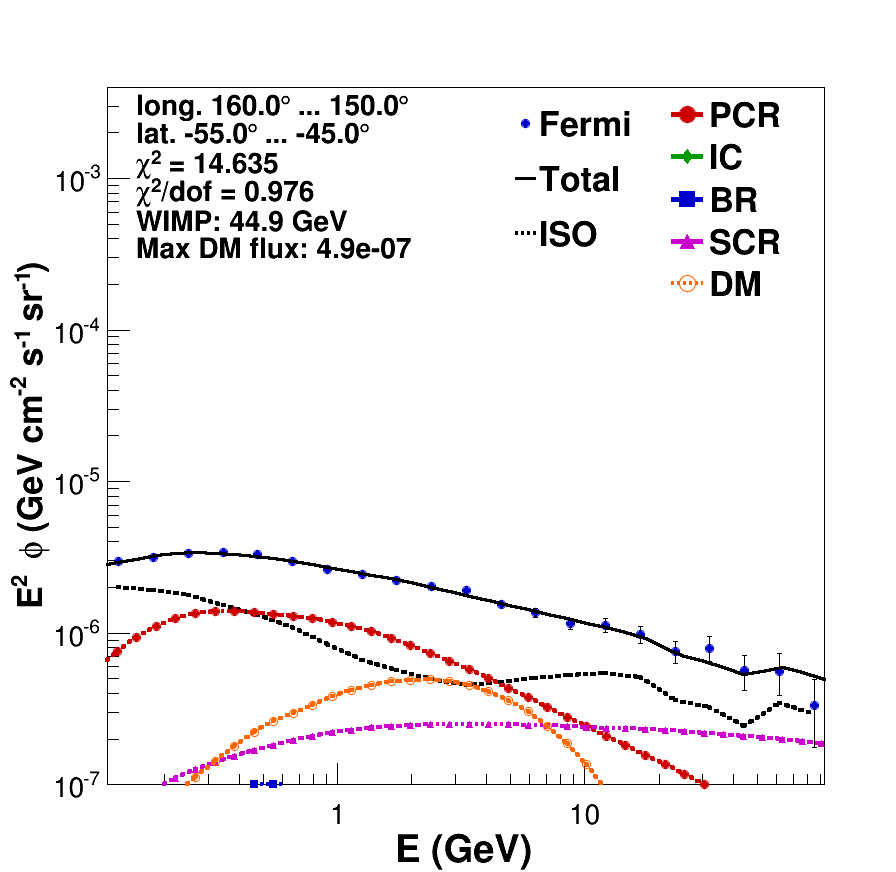}
\includegraphics[width=0.16\textwidth,height=0.16\textwidth,clip]{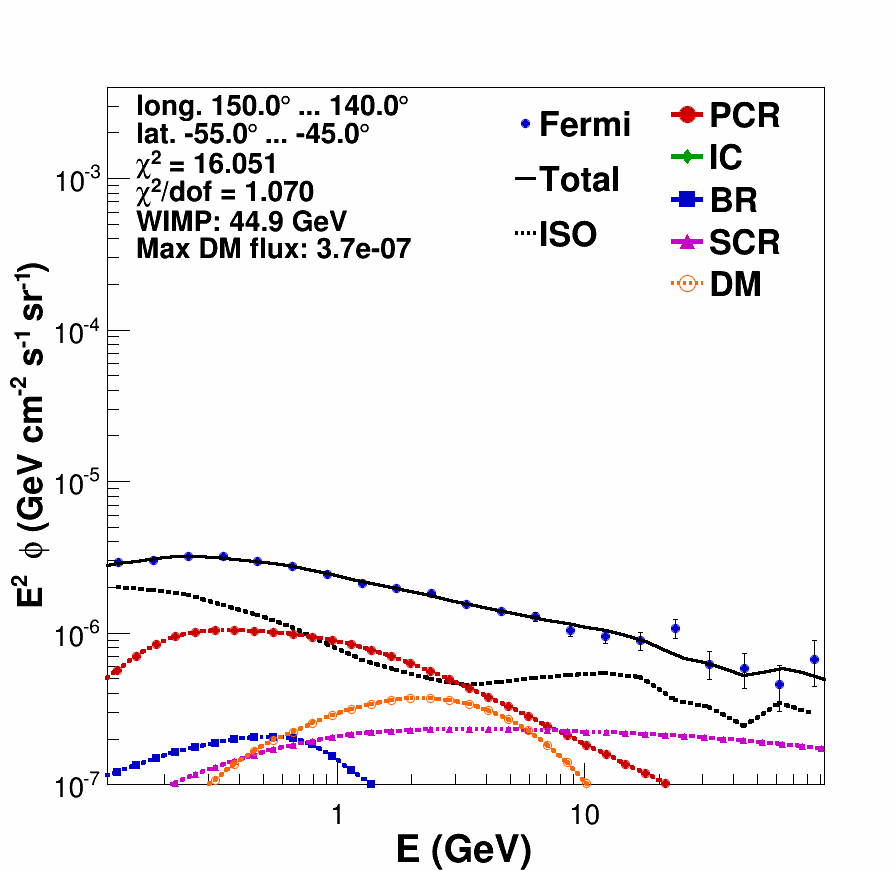}
\includegraphics[width=0.16\textwidth,height=0.16\textwidth,clip]{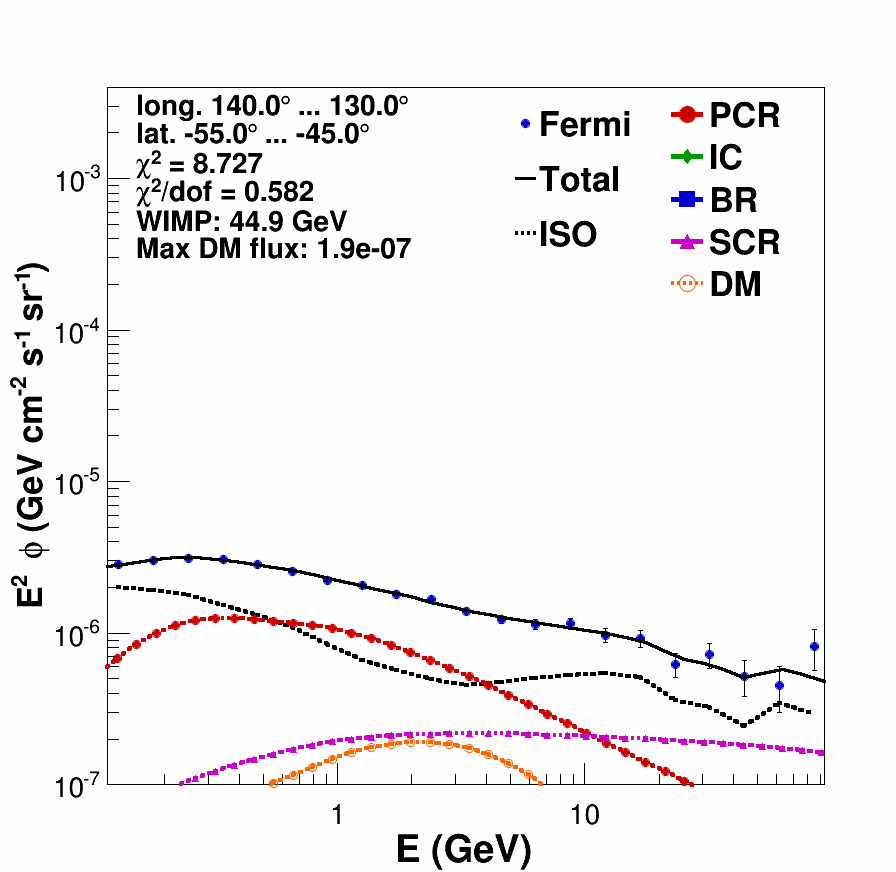}
\includegraphics[width=0.16\textwidth,height=0.16\textwidth,clip]{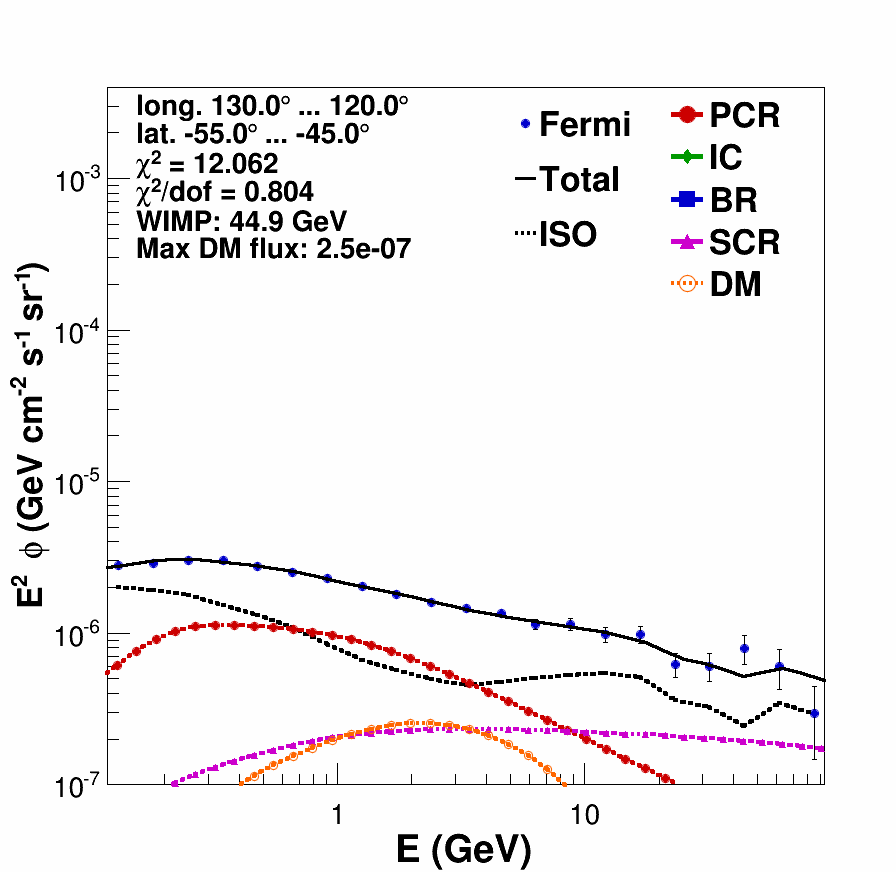}
\includegraphics[width=0.16\textwidth,height=0.16\textwidth,clip]{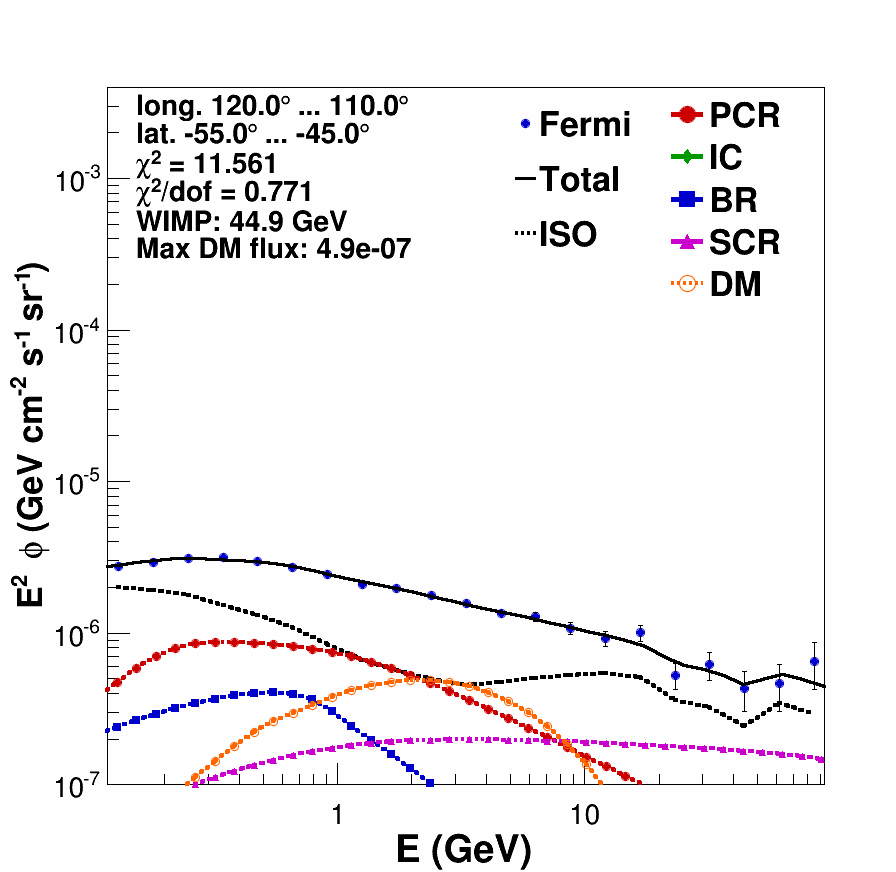}
\includegraphics[width=0.16\textwidth,height=0.16\textwidth,clip]{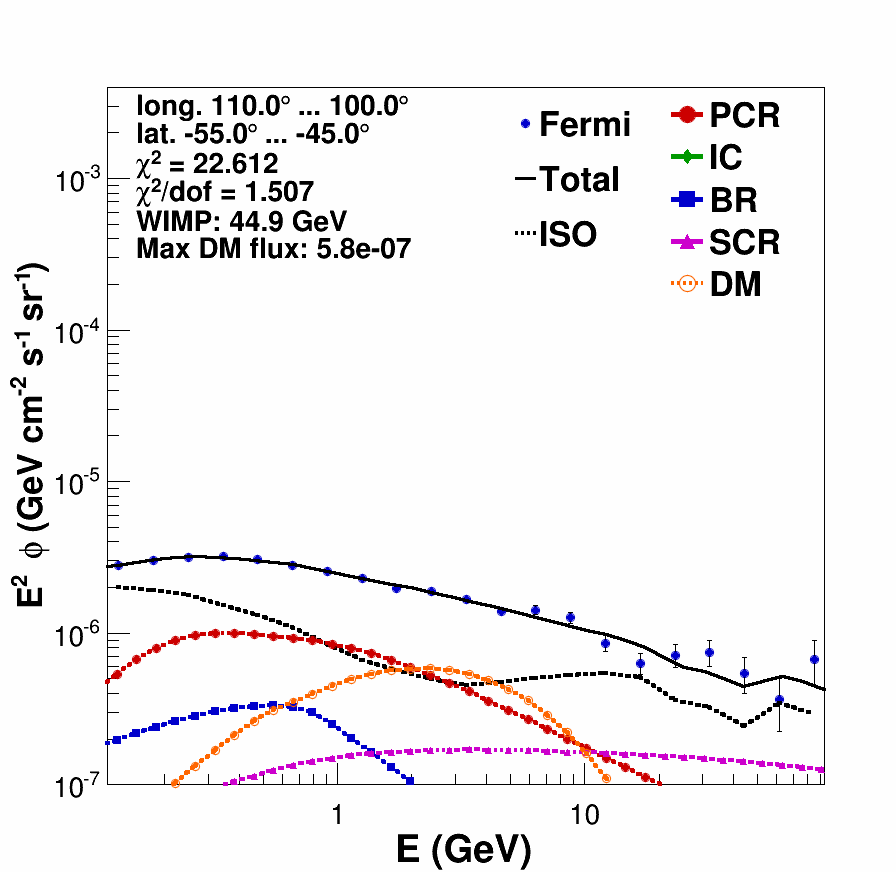}
\includegraphics[width=0.16\textwidth,height=0.16\textwidth,clip]{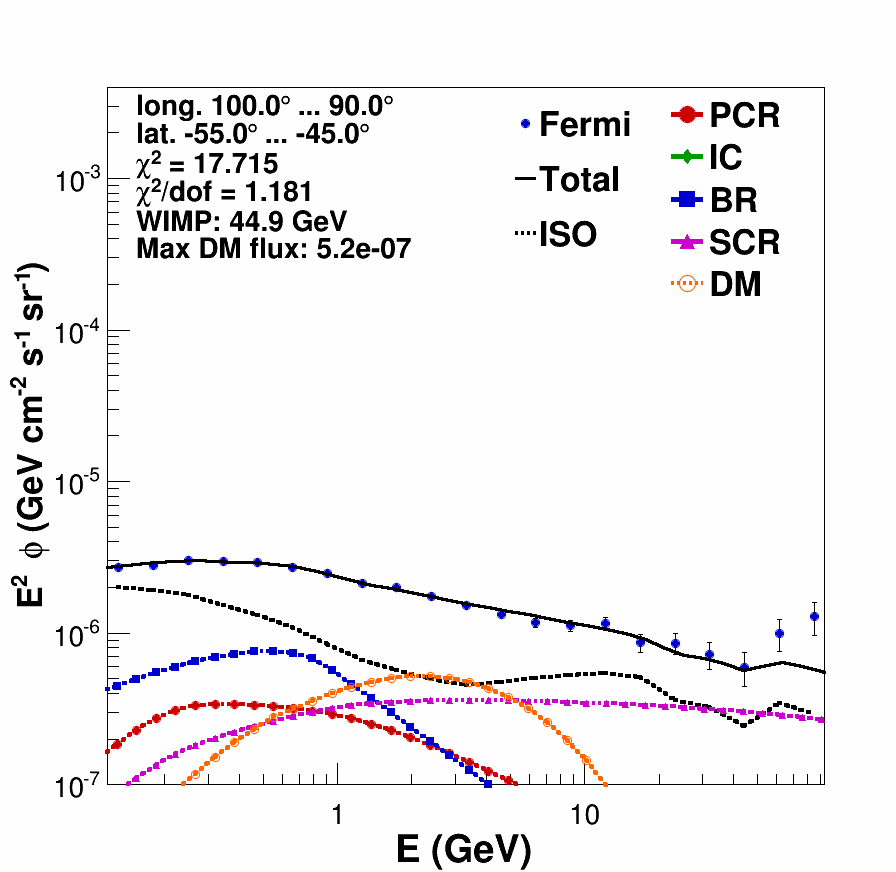}
\includegraphics[width=0.16\textwidth,height=0.16\textwidth,clip]{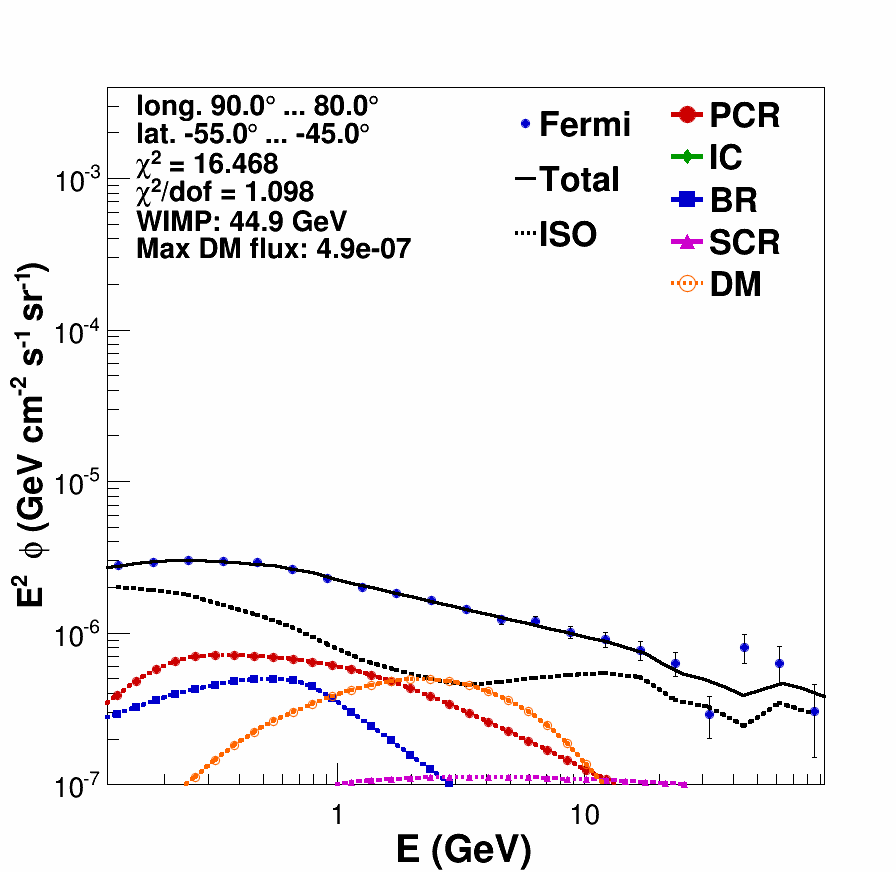}
\includegraphics[width=0.16\textwidth,height=0.16\textwidth,clip]{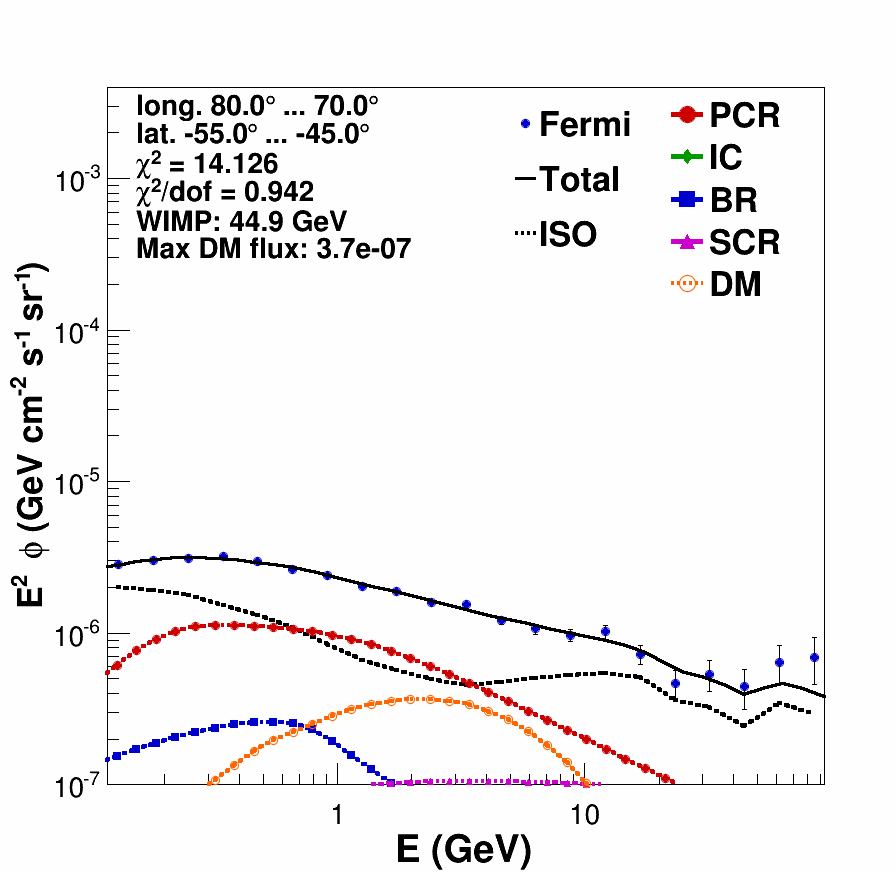}
\includegraphics[width=0.16\textwidth,height=0.16\textwidth,clip]{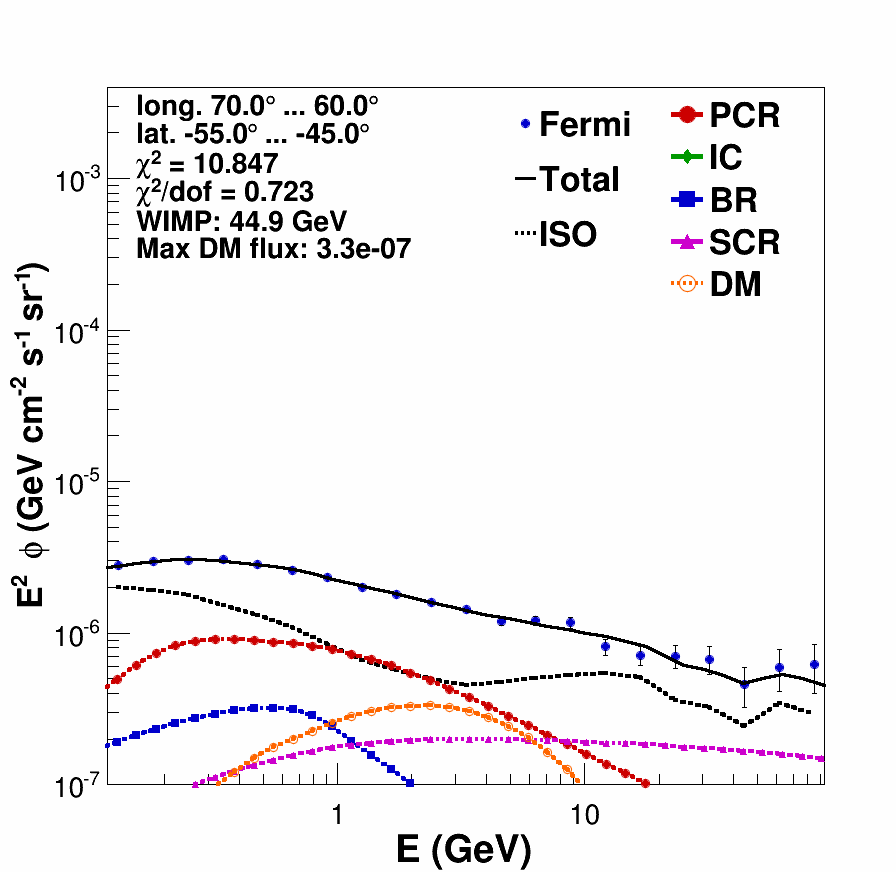}
\includegraphics[width=0.16\textwidth,height=0.16\textwidth,clip]{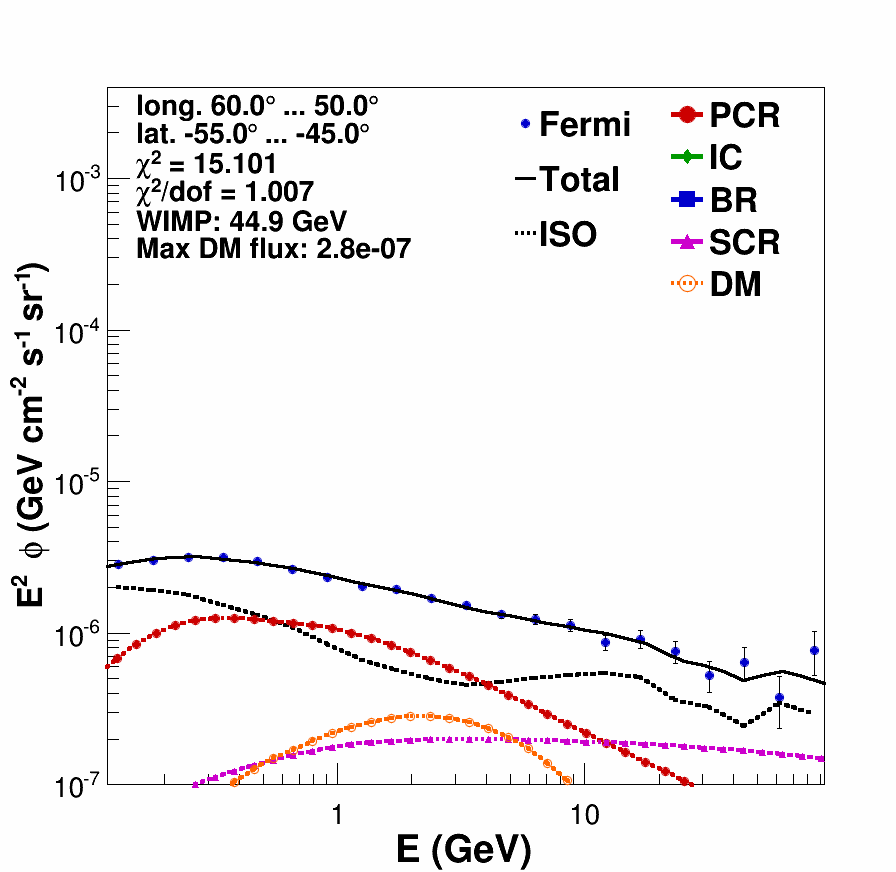}
\includegraphics[width=0.16\textwidth,height=0.16\textwidth,clip]{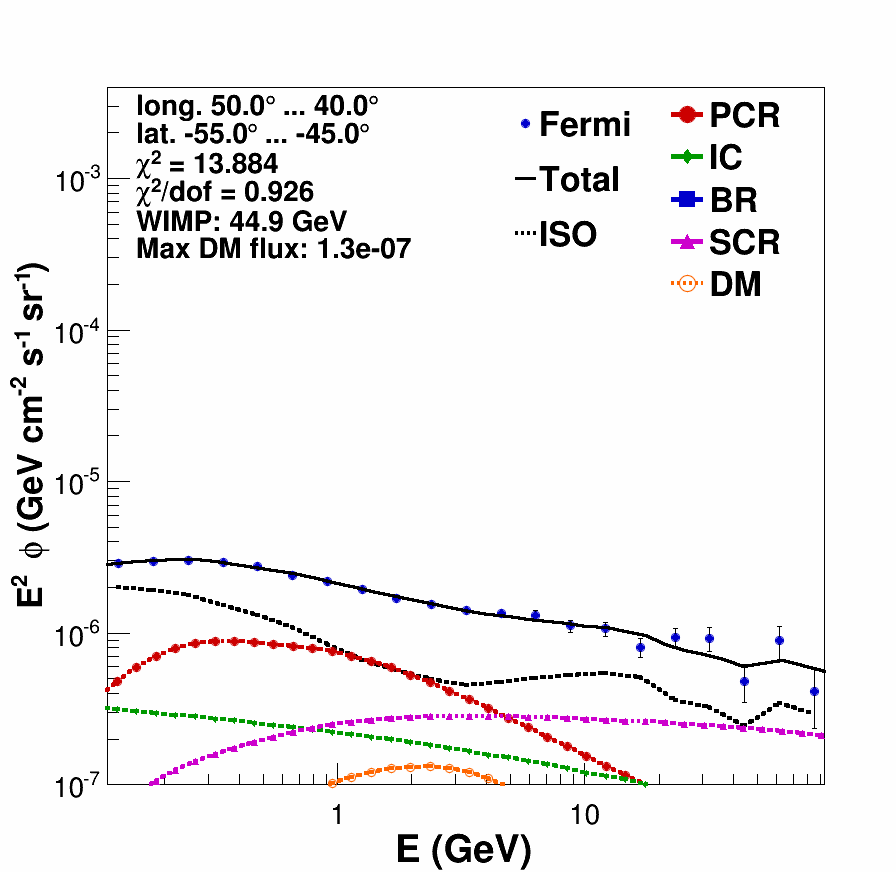}
\includegraphics[width=0.16\textwidth,height=0.16\textwidth,clip]{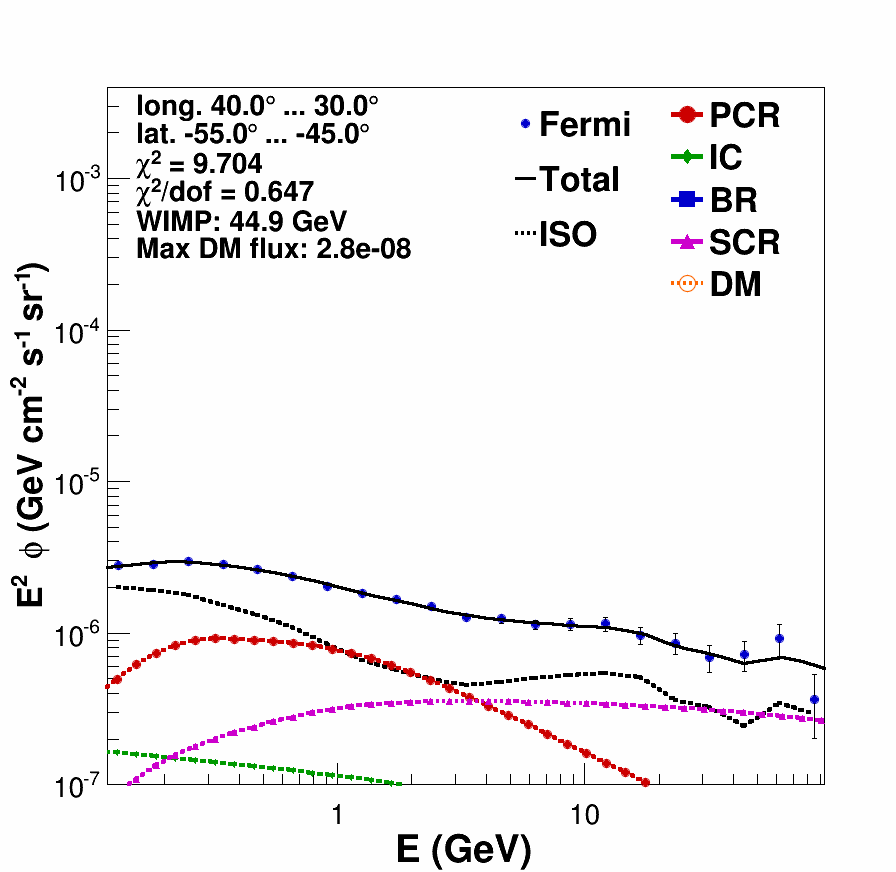}
\includegraphics[width=0.16\textwidth,height=0.16\textwidth,clip]{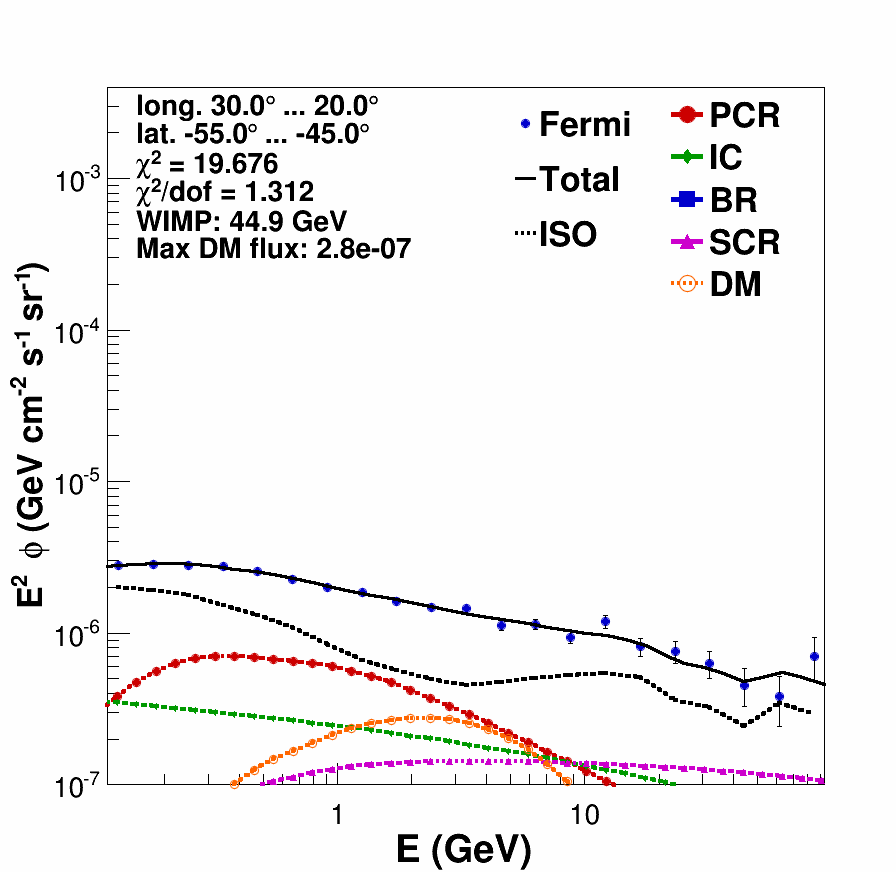}
\includegraphics[width=0.16\textwidth,height=0.16\textwidth,clip]{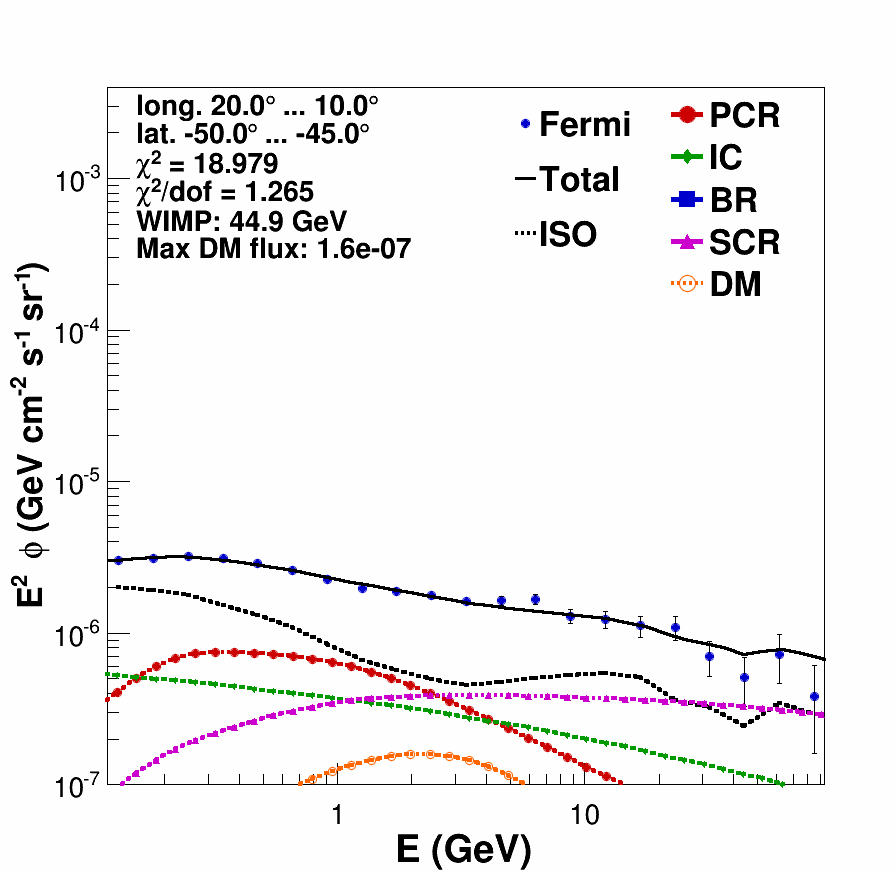}
\includegraphics[width=0.16\textwidth,height=0.16\textwidth,clip]{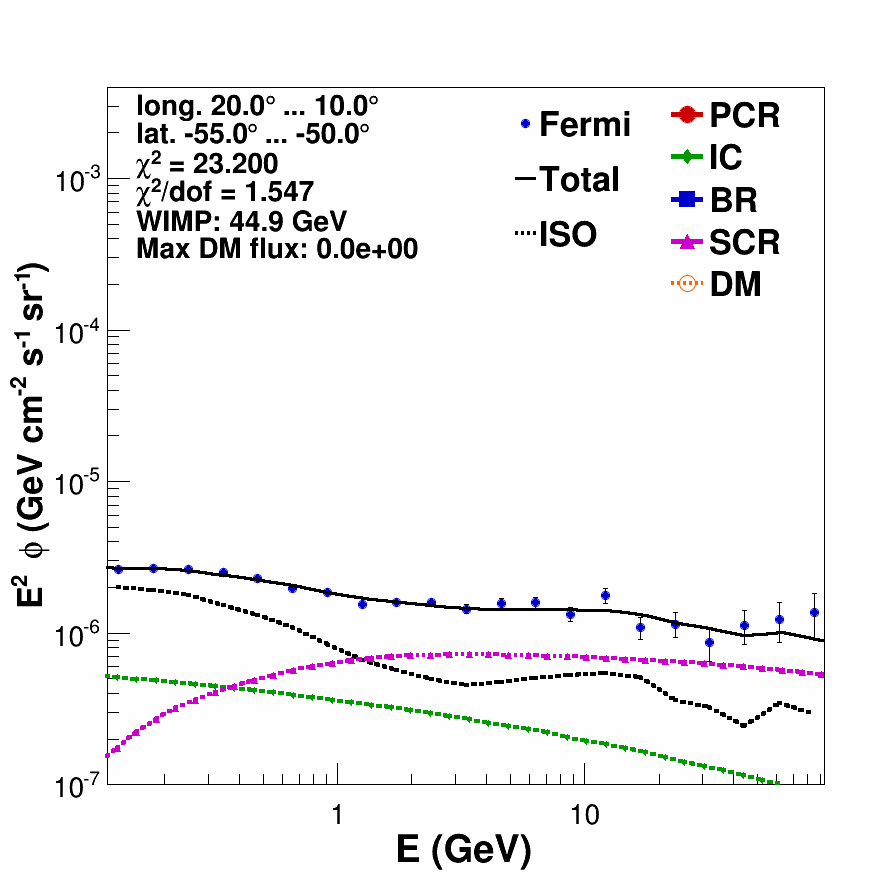}
\includegraphics[width=0.16\textwidth,height=0.16\textwidth,clip]{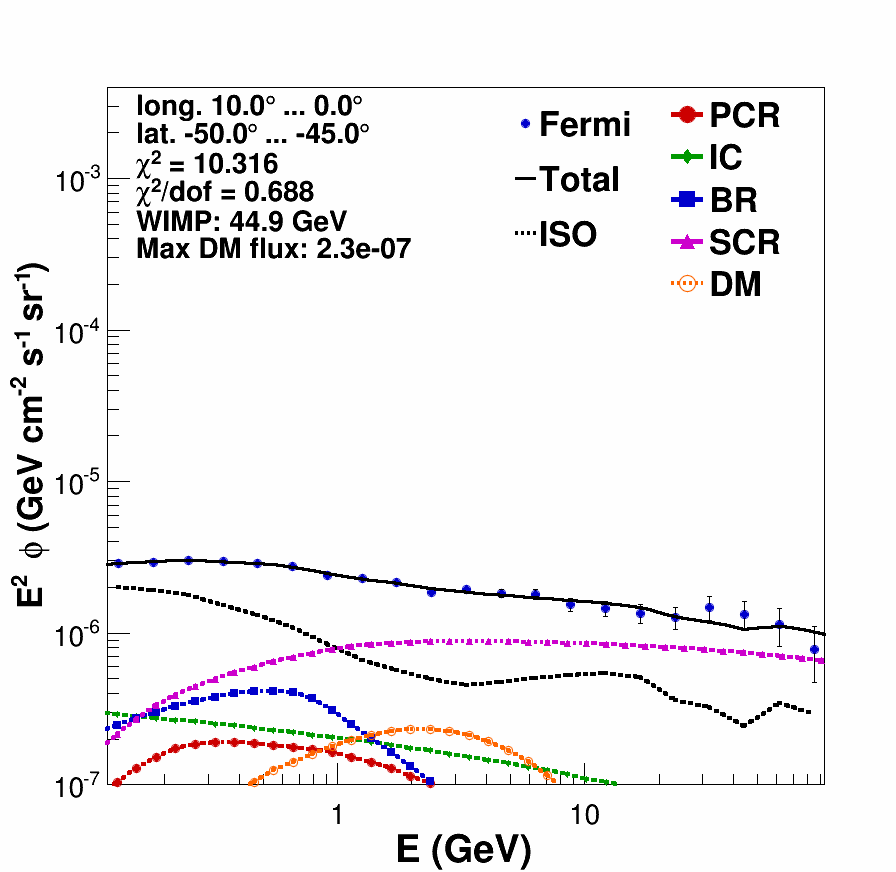}
\includegraphics[width=0.16\textwidth,height=0.16\textwidth,clip]{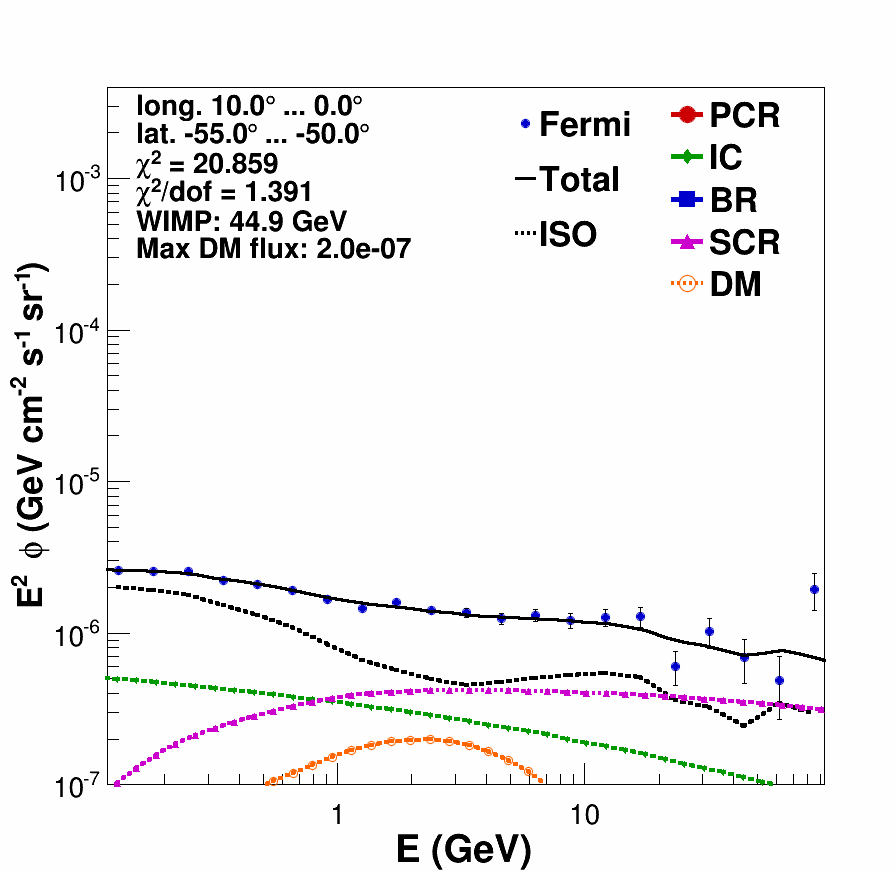}
\includegraphics[width=0.16\textwidth,height=0.16\textwidth,clip]{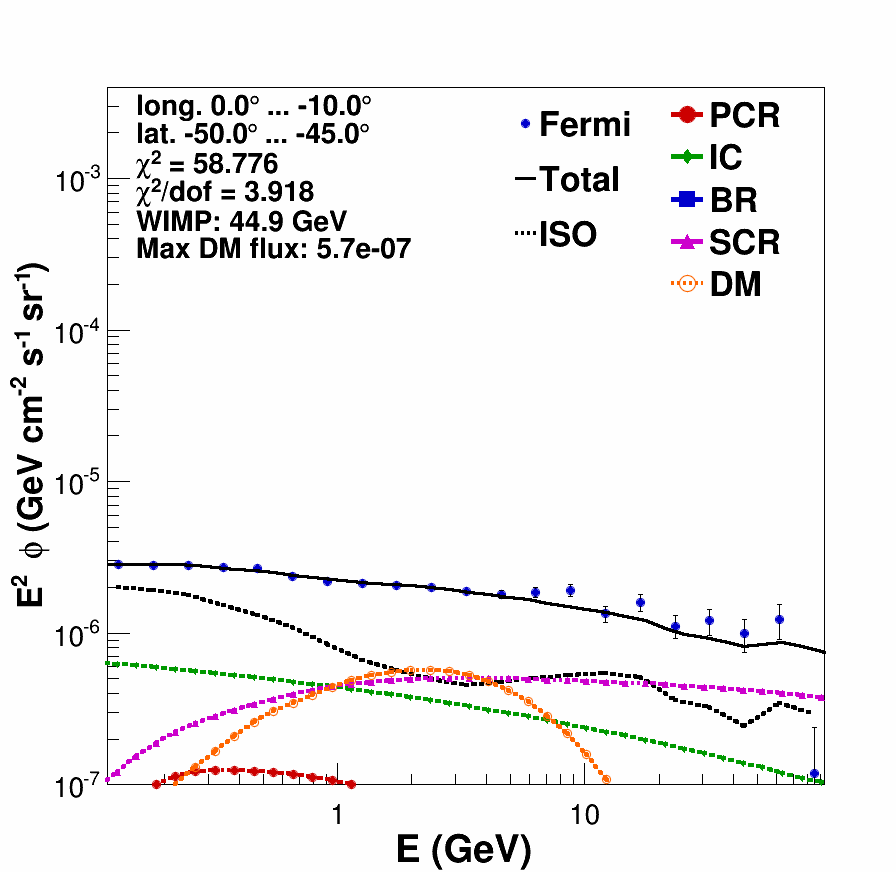}
\includegraphics[width=0.16\textwidth,height=0.16\textwidth,clip]{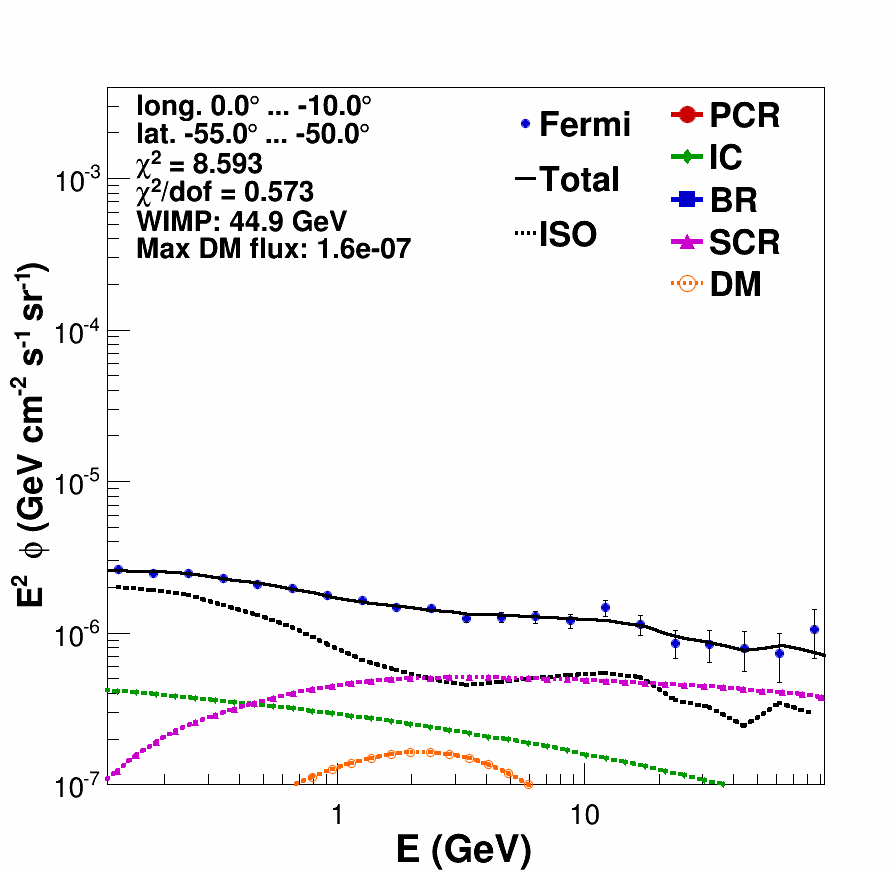}
\includegraphics[width=0.16\textwidth,height=0.16\textwidth,clip]{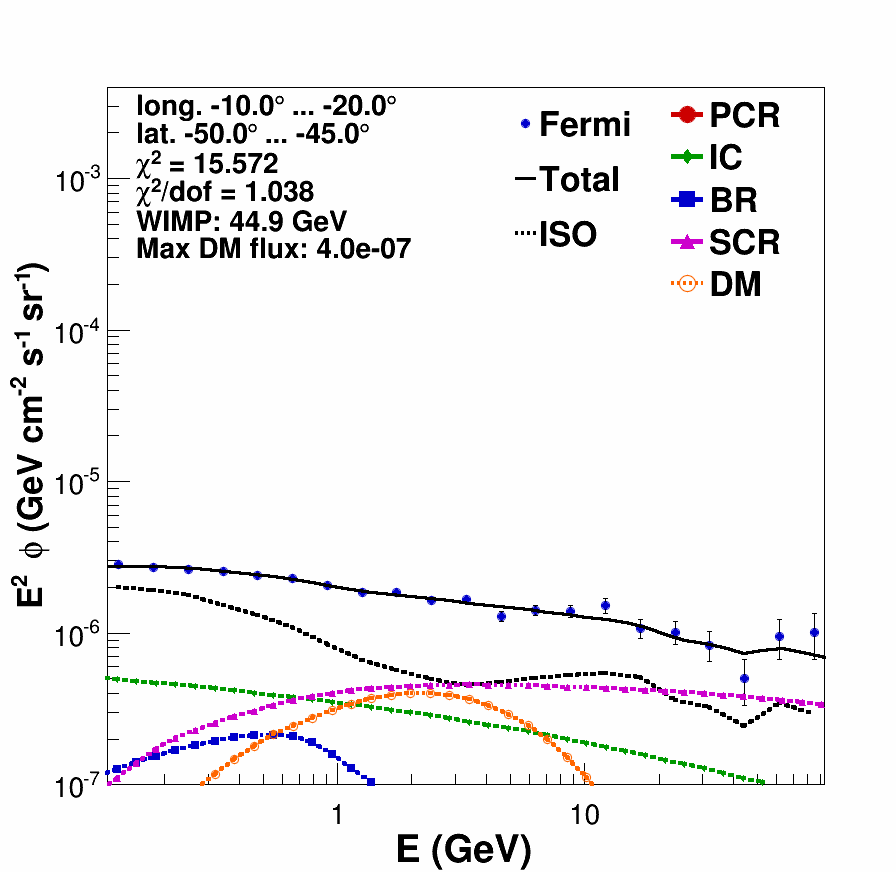}
\includegraphics[width=0.16\textwidth,height=0.16\textwidth,clip]{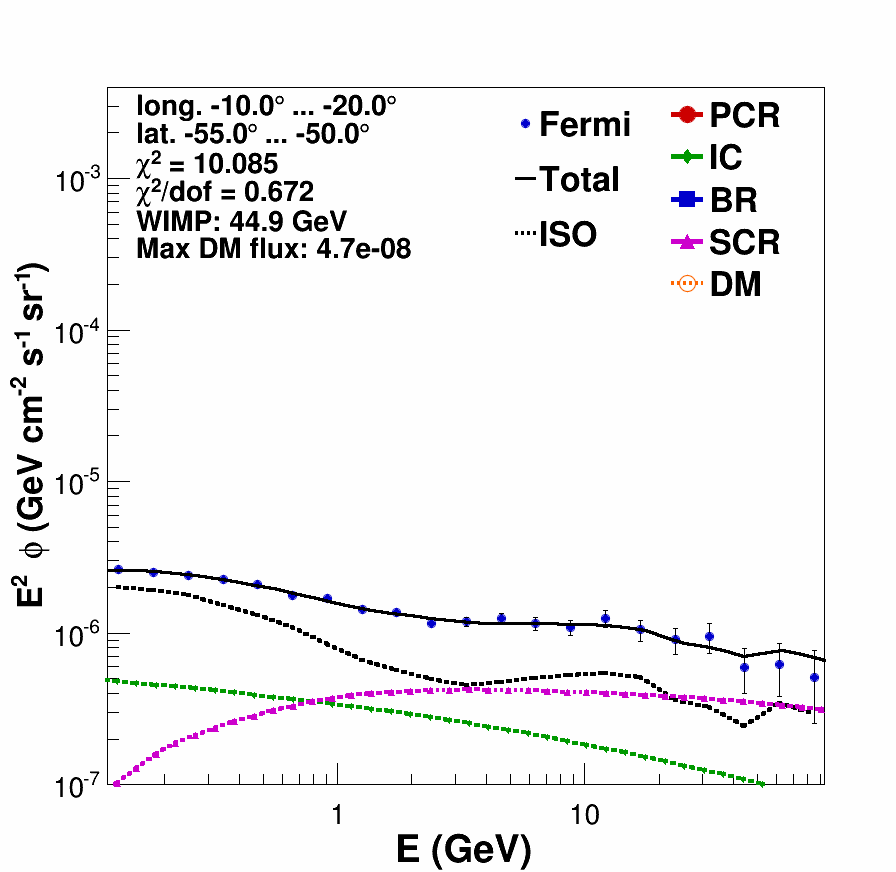}
\includegraphics[width=0.16\textwidth,height=0.16\textwidth,clip]{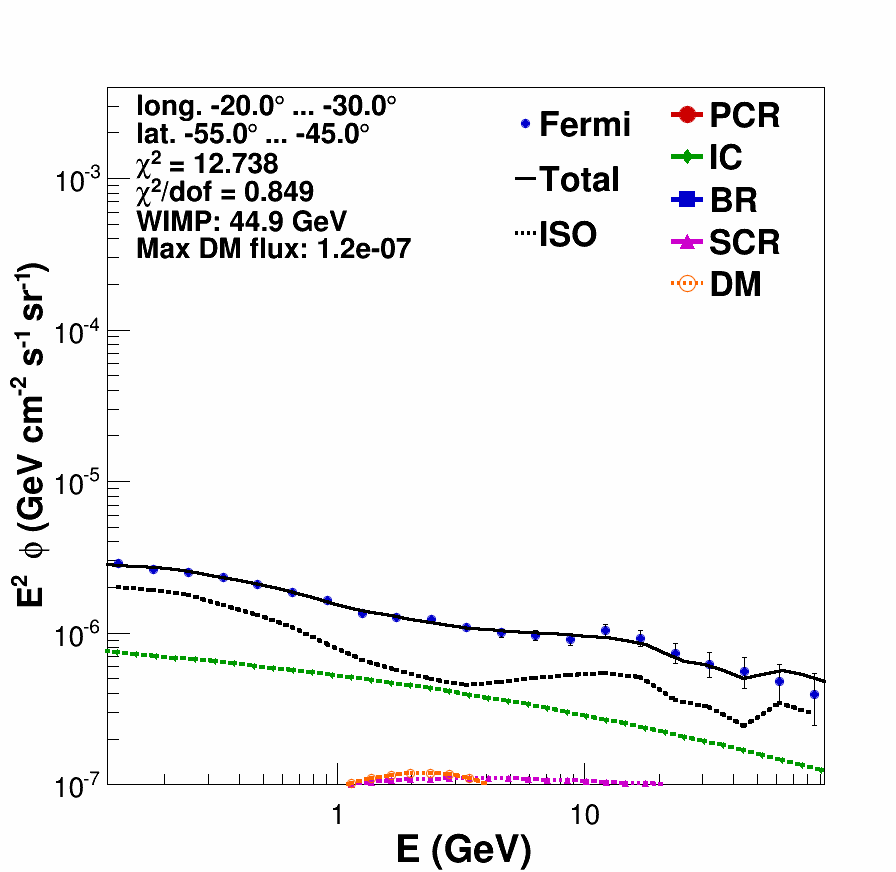}
\includegraphics[width=0.16\textwidth,height=0.16\textwidth,clip]{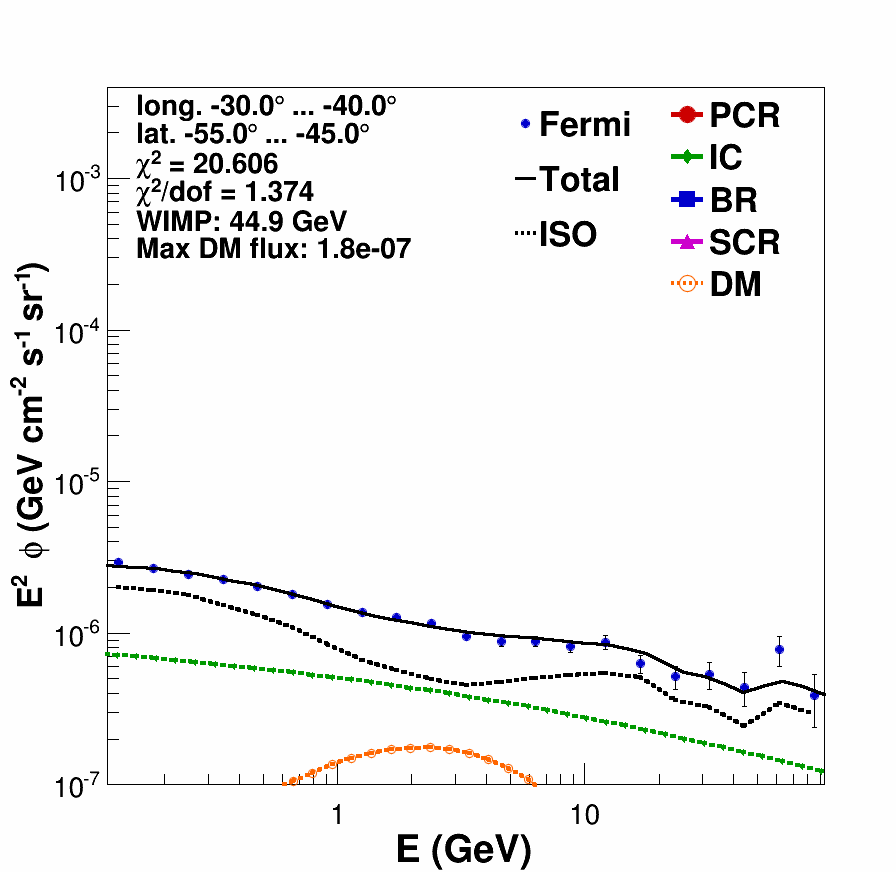}
\includegraphics[width=0.16\textwidth,height=0.16\textwidth,clip]{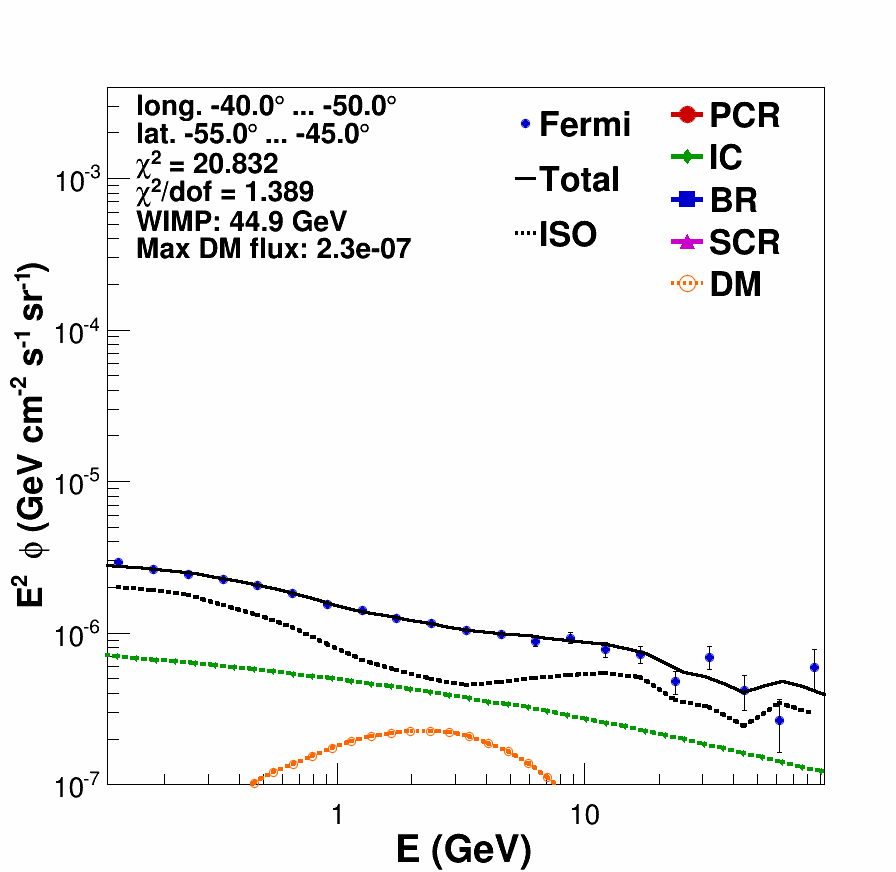}
\includegraphics[width=0.16\textwidth,height=0.16\textwidth,clip]{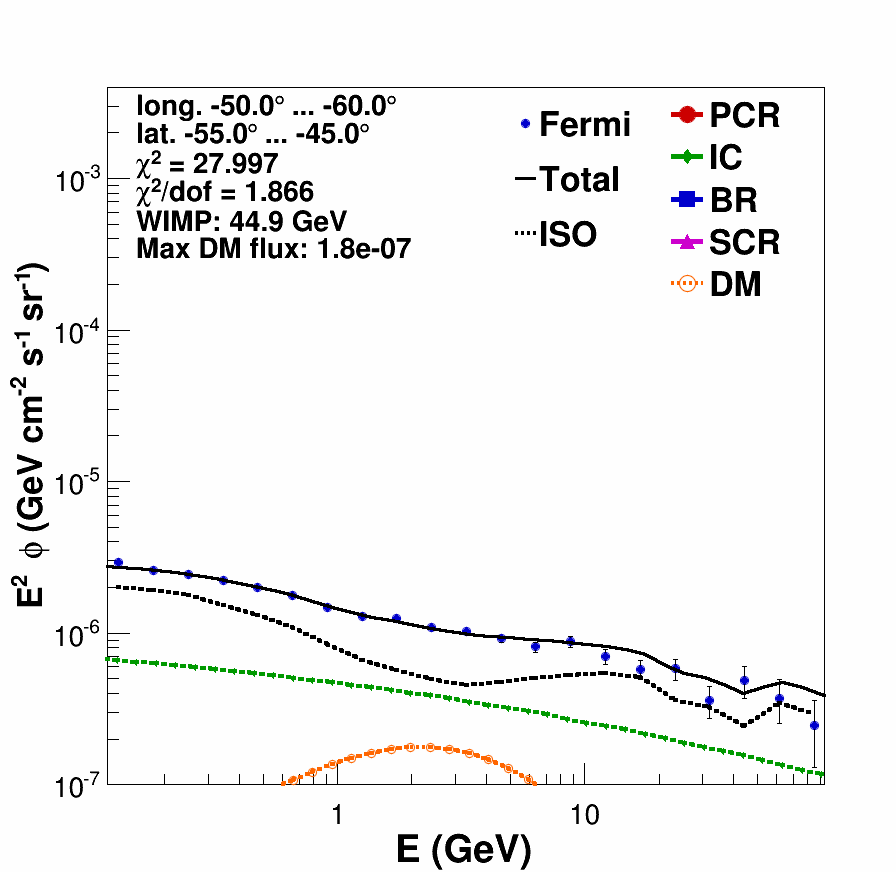}
\includegraphics[width=0.16\textwidth,height=0.16\textwidth,clip]{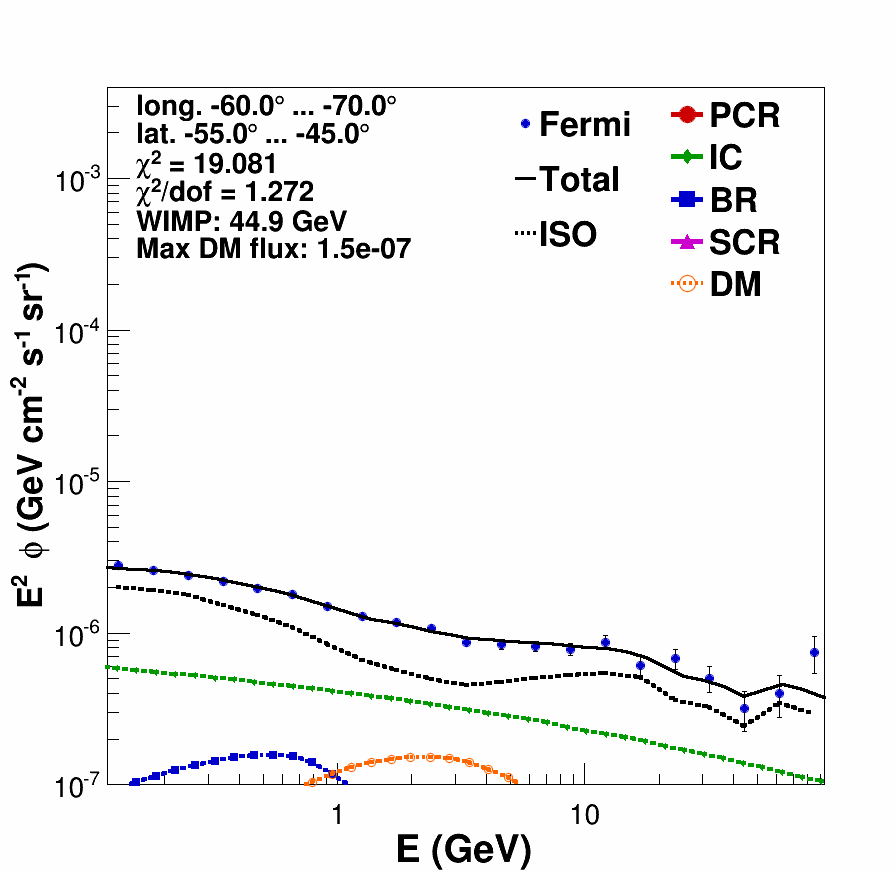}
\includegraphics[width=0.16\textwidth,height=0.16\textwidth,clip]{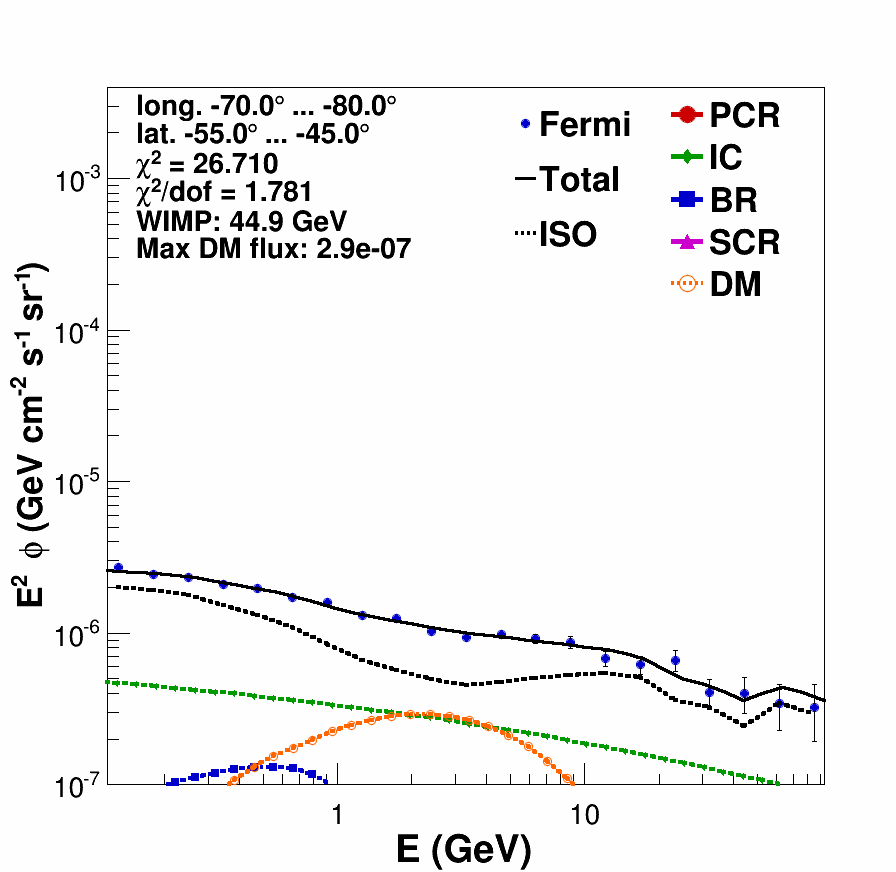}
\includegraphics[width=0.16\textwidth,height=0.16\textwidth,clip]{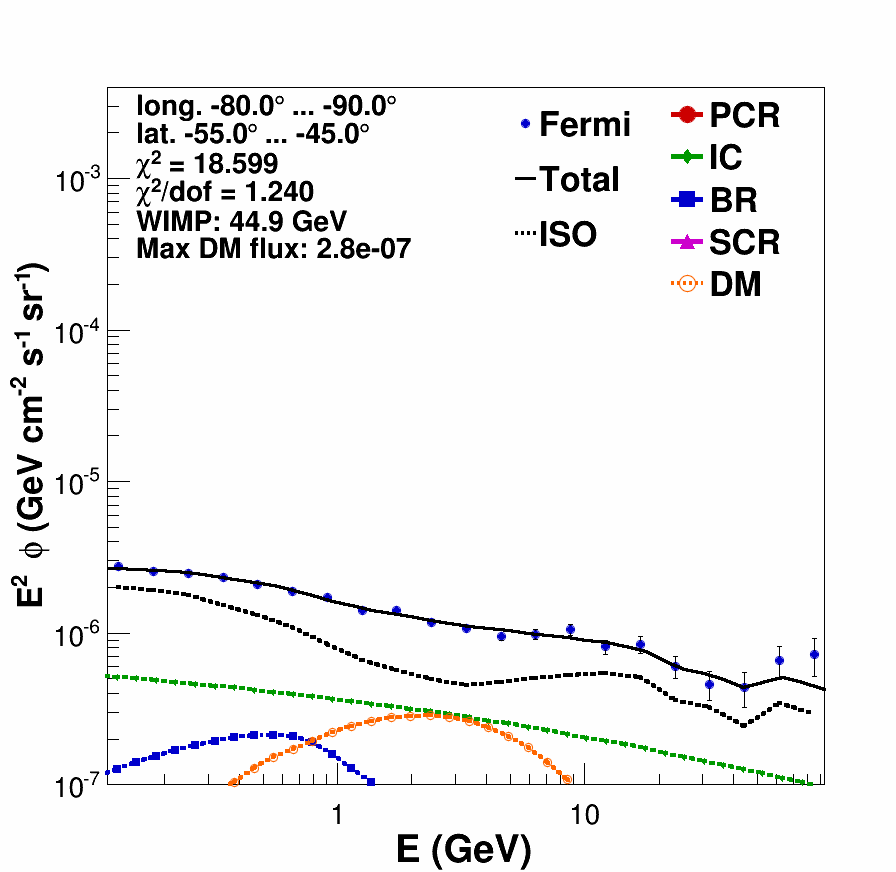}
\includegraphics[width=0.16\textwidth,height=0.16\textwidth,clip]{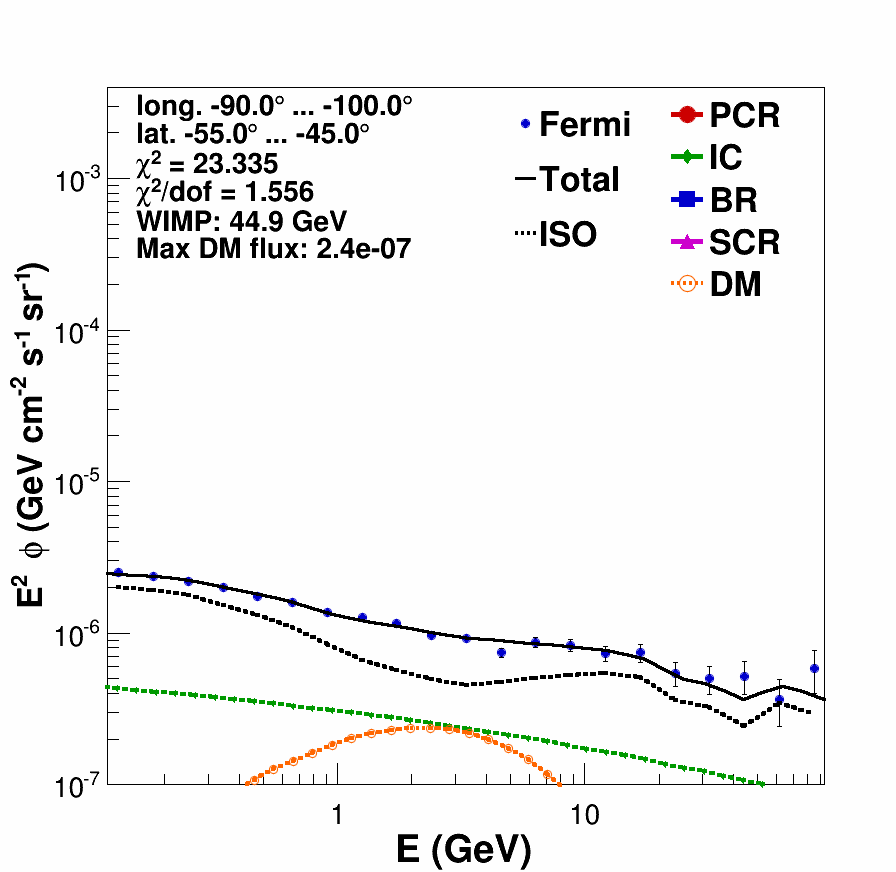}
\includegraphics[width=0.16\textwidth,height=0.16\textwidth,clip]{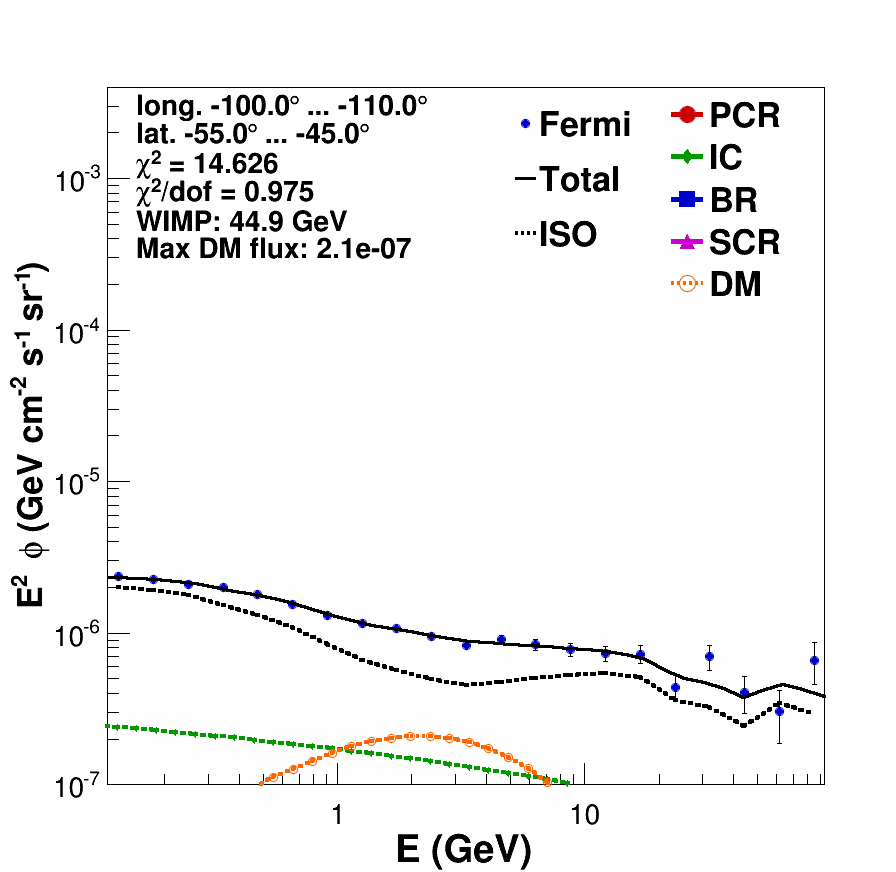}
\includegraphics[width=0.16\textwidth,height=0.16\textwidth,clip]{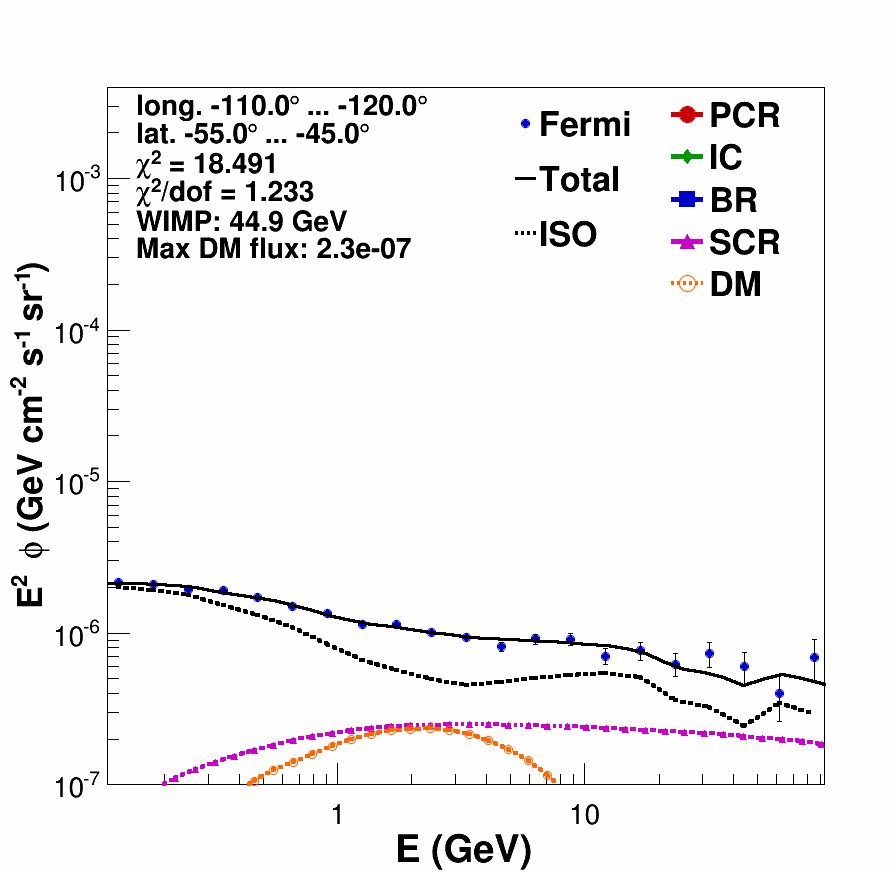}
\includegraphics[width=0.16\textwidth,height=0.16\textwidth,clip]{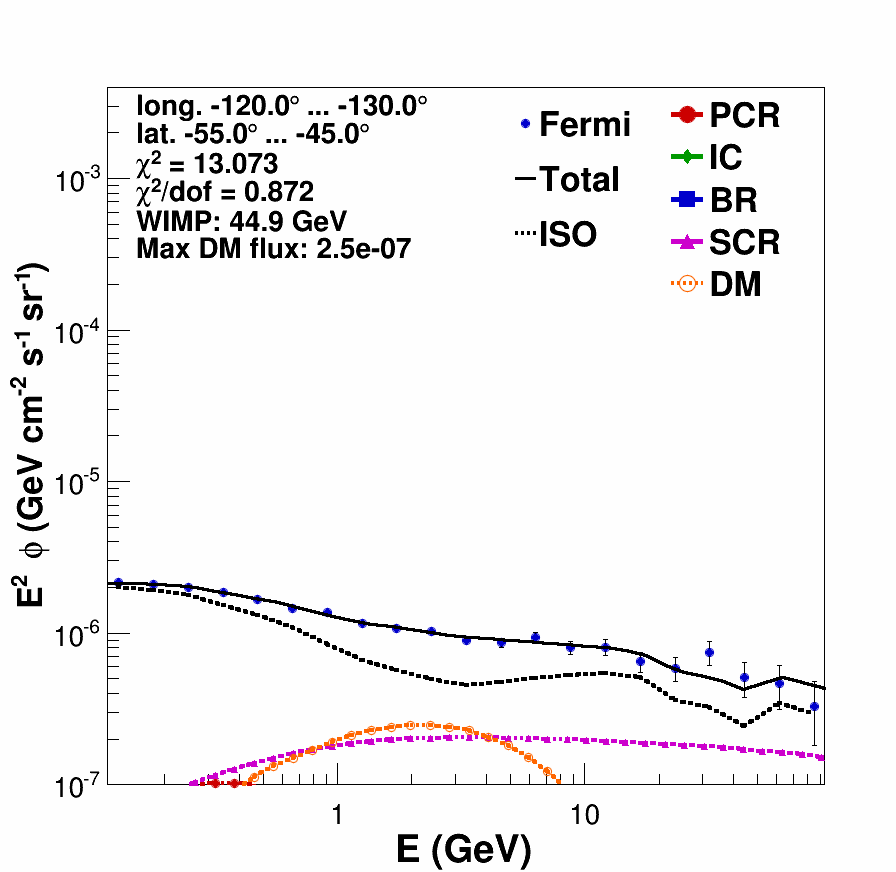}
\includegraphics[width=0.16\textwidth,height=0.16\textwidth,clip]{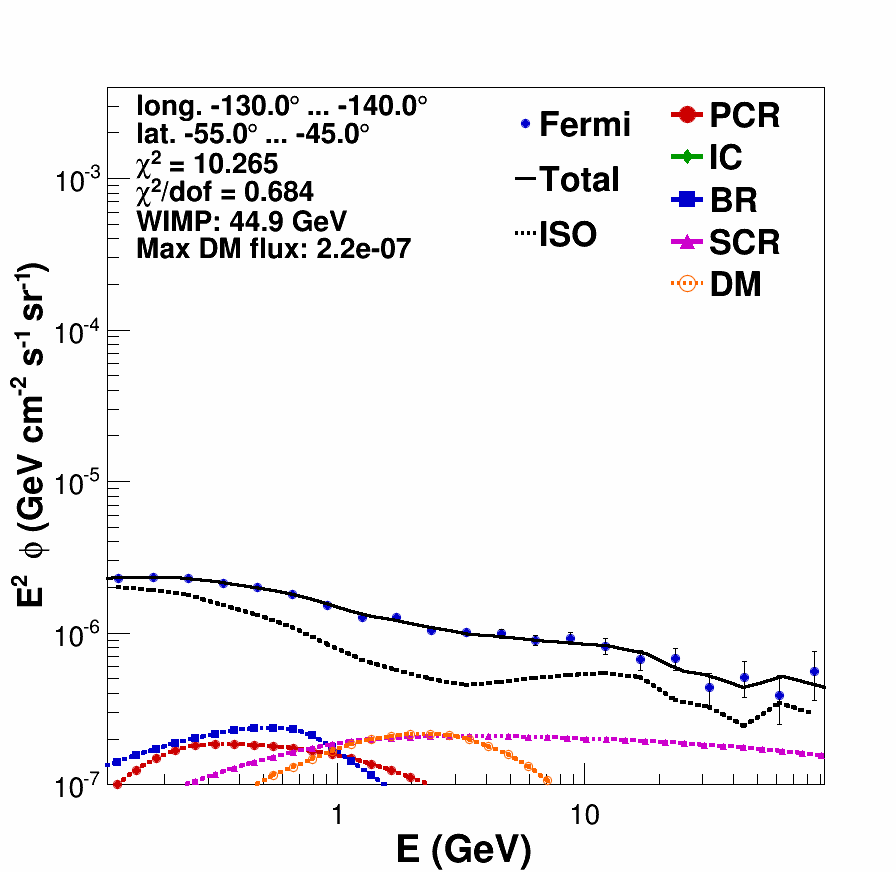}
\includegraphics[width=0.16\textwidth,height=0.16\textwidth,clip]{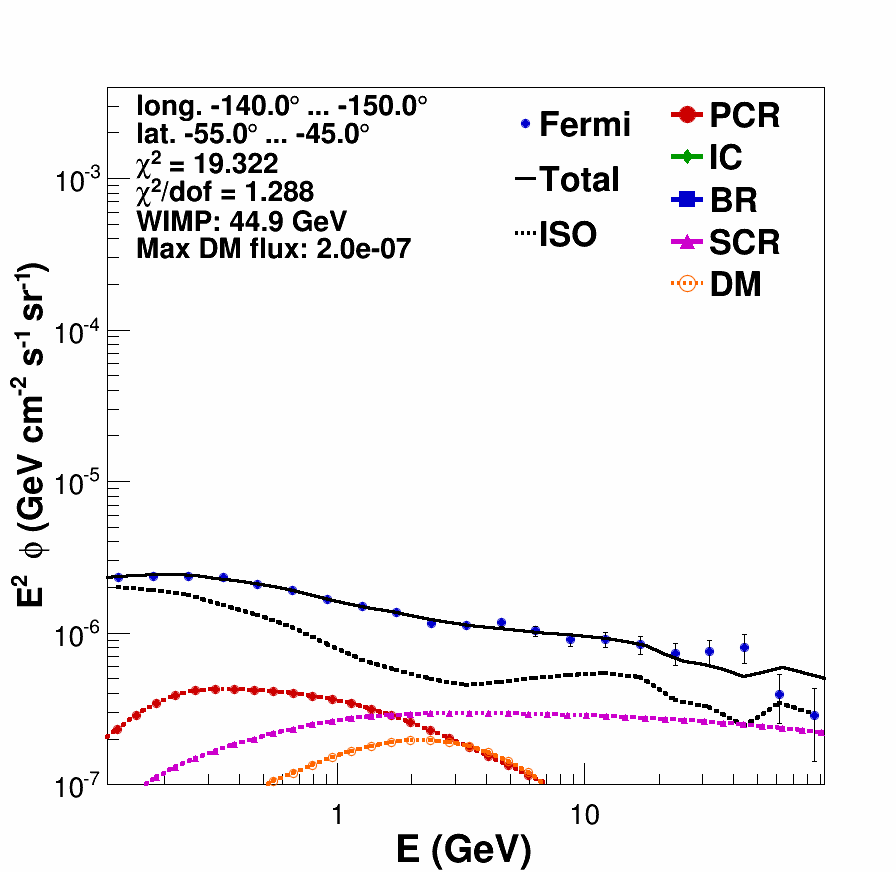}
\includegraphics[width=0.16\textwidth,height=0.16\textwidth,clip]{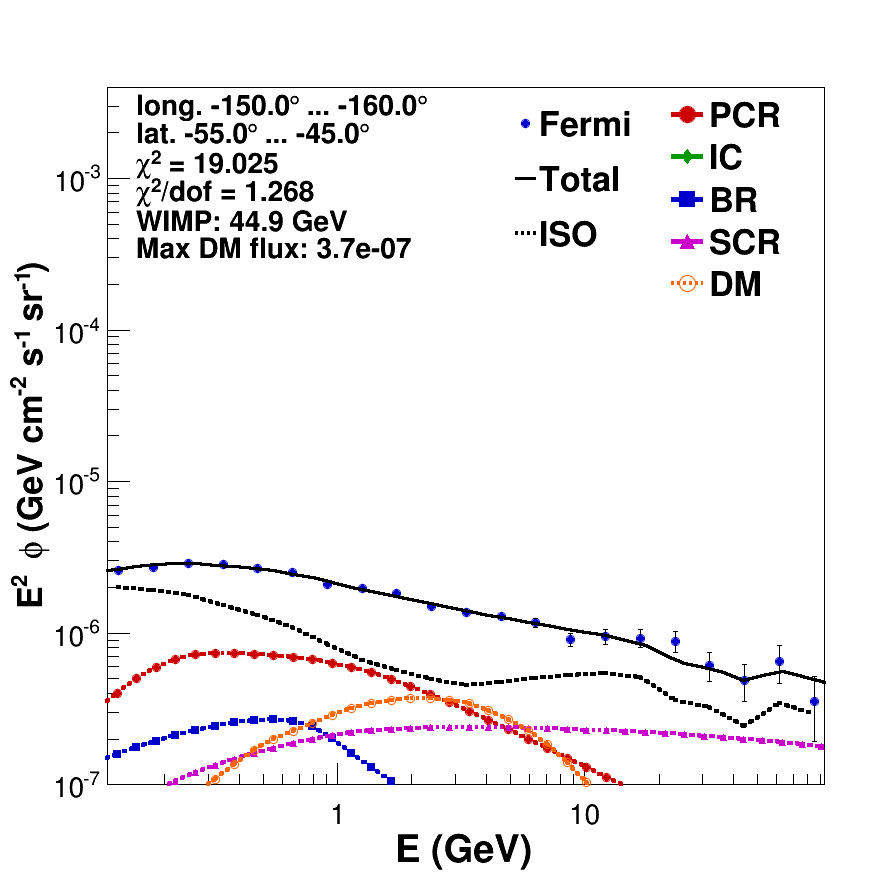}
\includegraphics[width=0.16\textwidth,height=0.16\textwidth,clip]{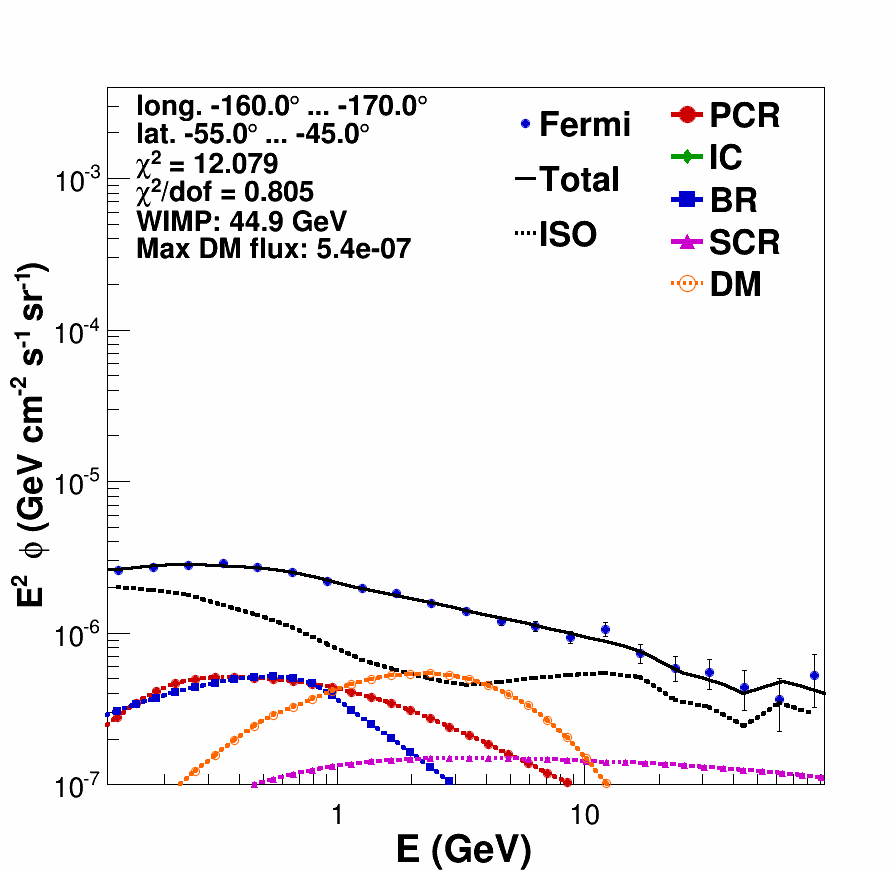}
\includegraphics[width=0.16\textwidth,height=0.16\textwidth,clip]{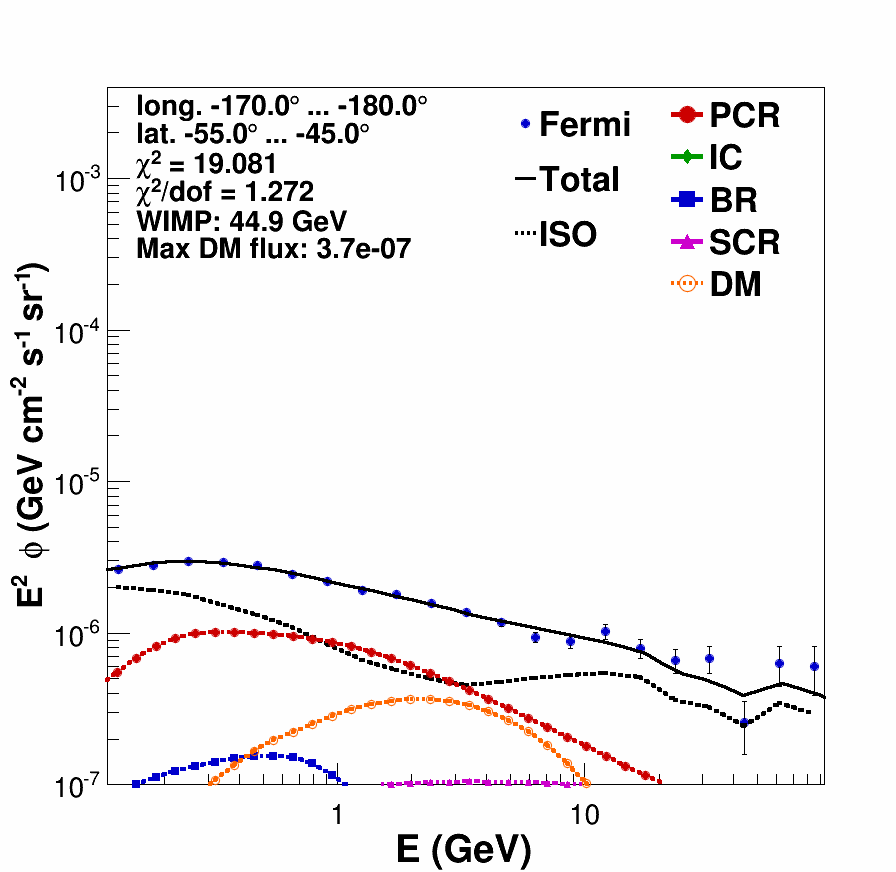}%%%%%r17
\caption[]{Template fits for latitudes  with $-55.0^\circ<b<-45.0^\circ$ and longitudes decreasing from 180$^\circ$ to -180$^\circ$.} \label{F50}
\end{figure}
\begin{figure} 
\centering
\includegraphics[width=0.16\textwidth,height=0.16\textwidth,clip]{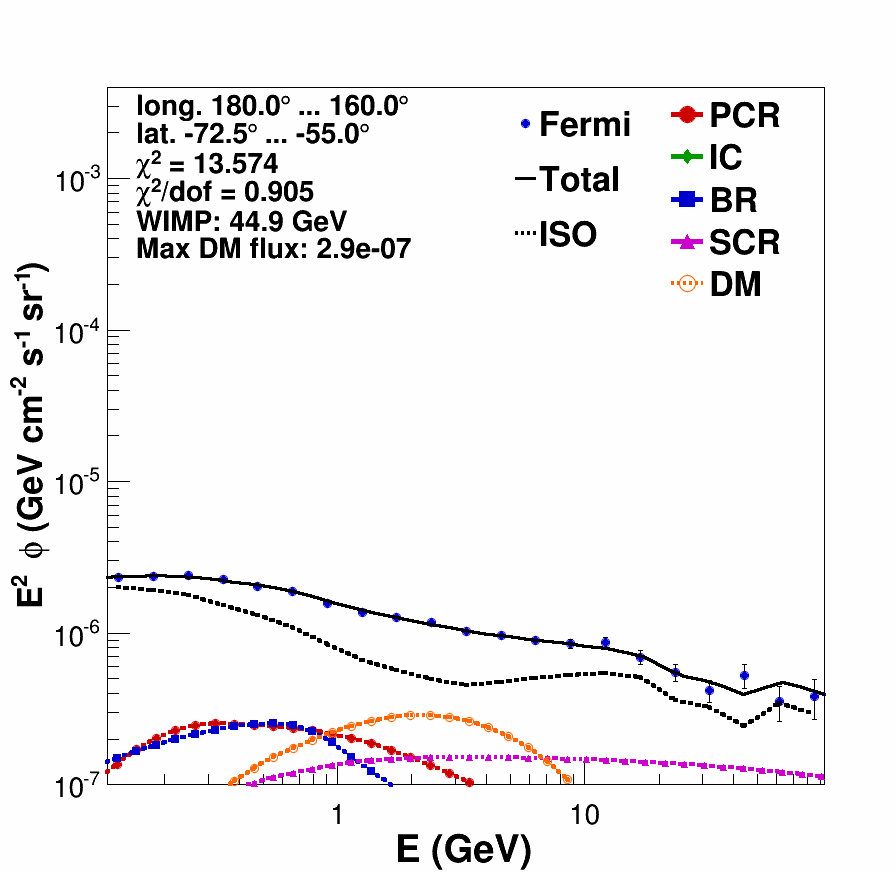}
\includegraphics[width=0.16\textwidth,height=0.16\textwidth,clip]{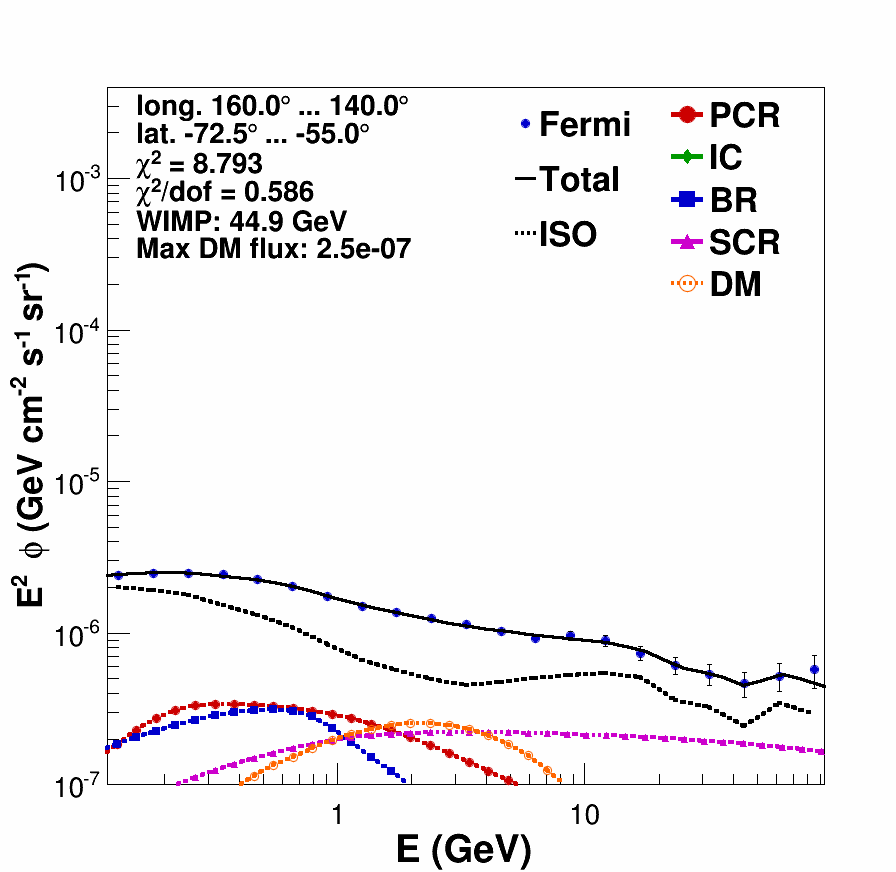}
\includegraphics[width=0.16\textwidth,height=0.16\textwidth,clip]{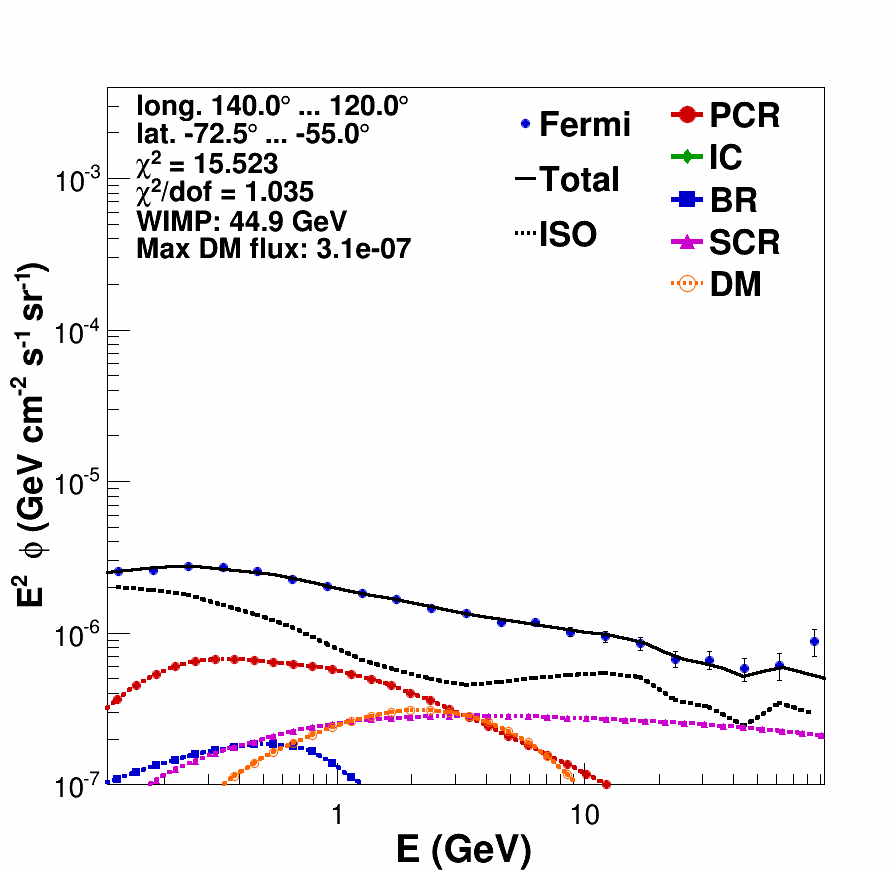}
\includegraphics[width=0.16\textwidth,height=0.16\textwidth,clip]{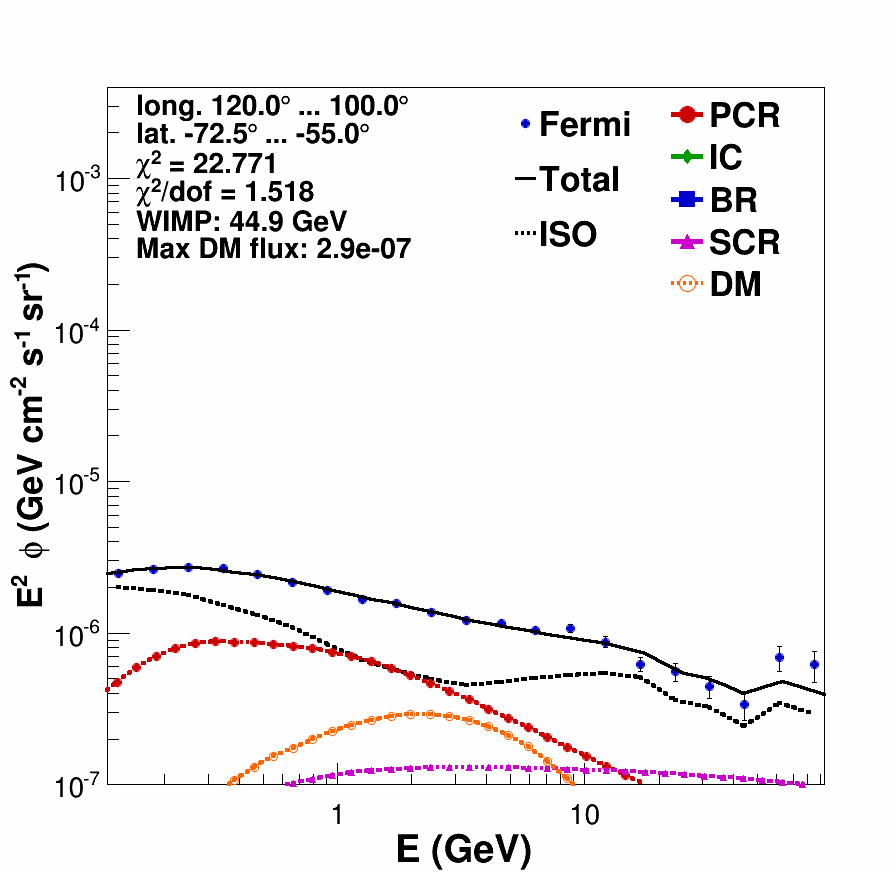}
\includegraphics[width=0.16\textwidth,height=0.16\textwidth,clip]{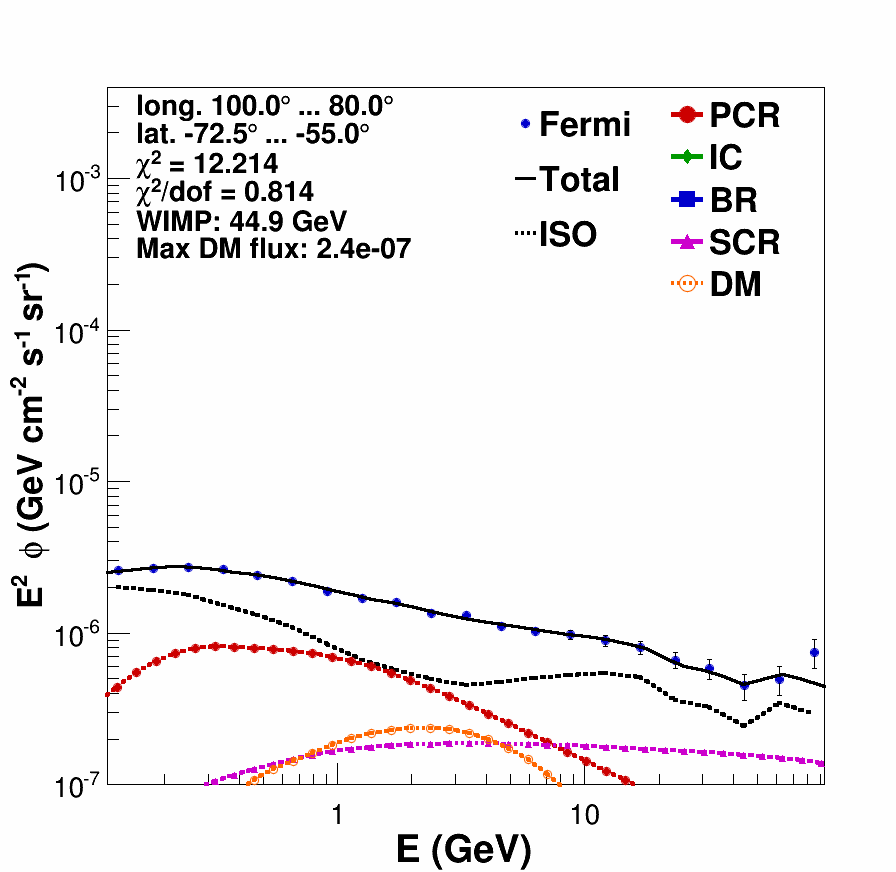}
\includegraphics[width=0.16\textwidth,height=0.16\textwidth,clip]{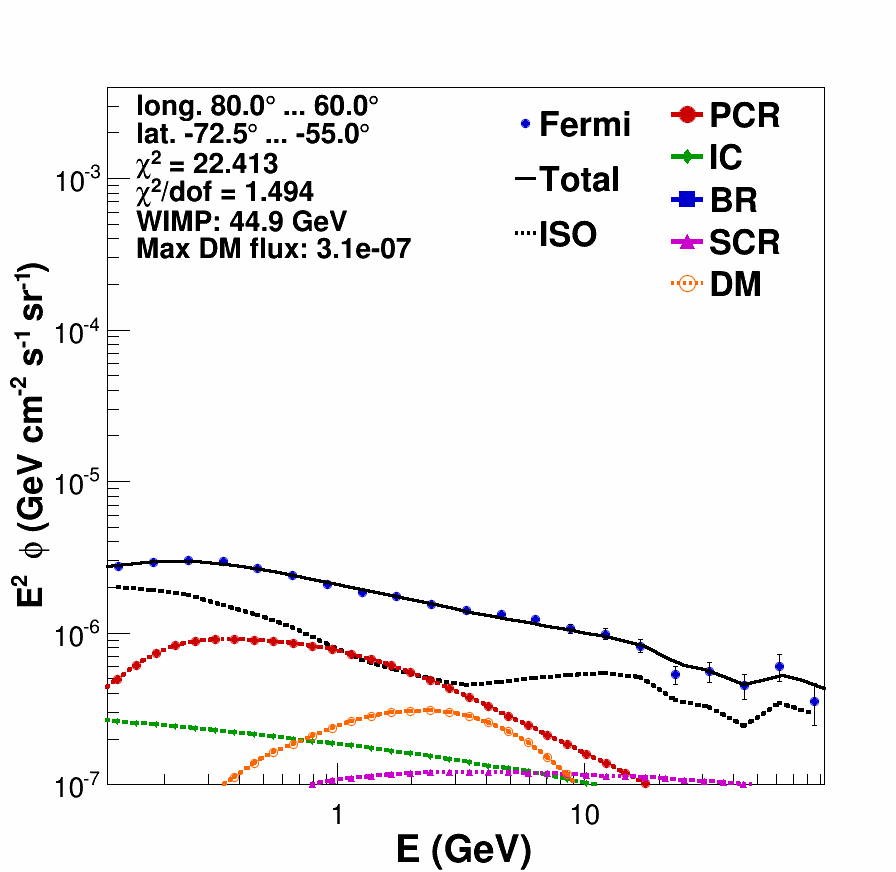}
\includegraphics[width=0.16\textwidth,height=0.16\textwidth,clip]{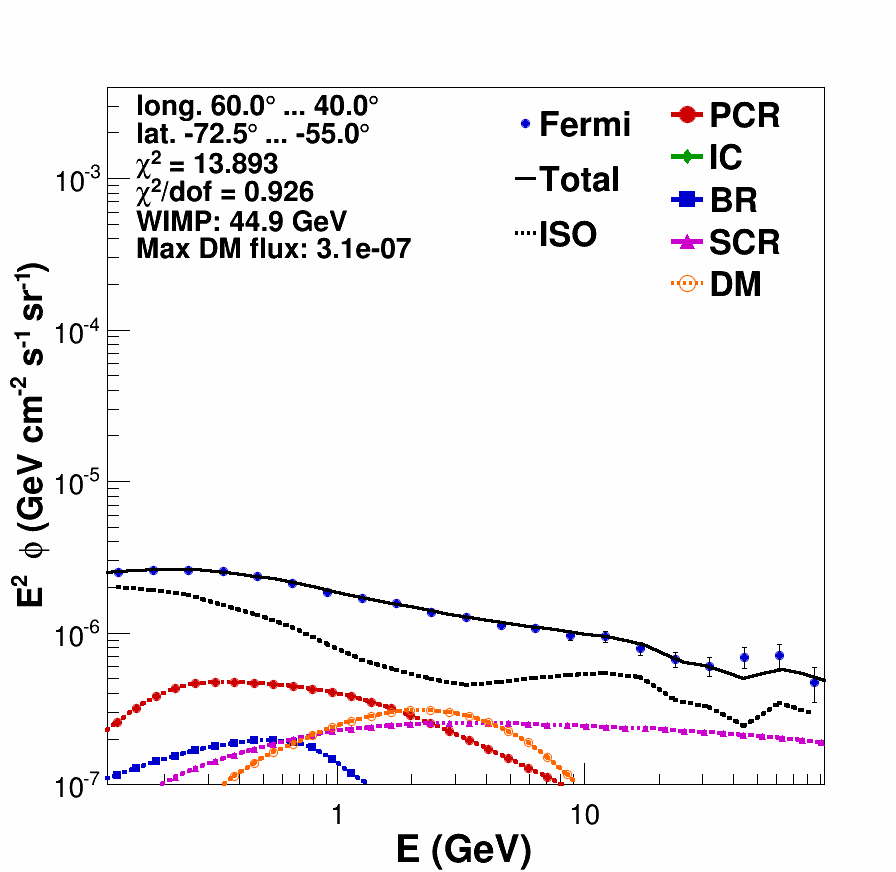}
\includegraphics[width=0.16\textwidth,height=0.16\textwidth,clip]{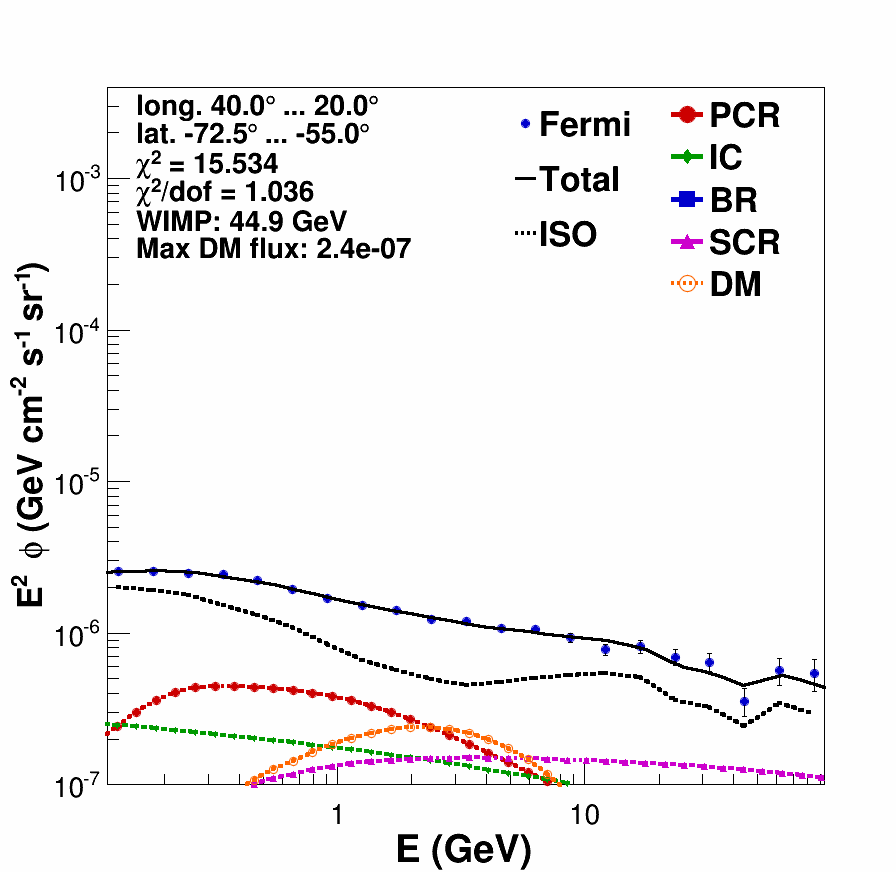}
\includegraphics[width=0.16\textwidth,height=0.16\textwidth,clip]{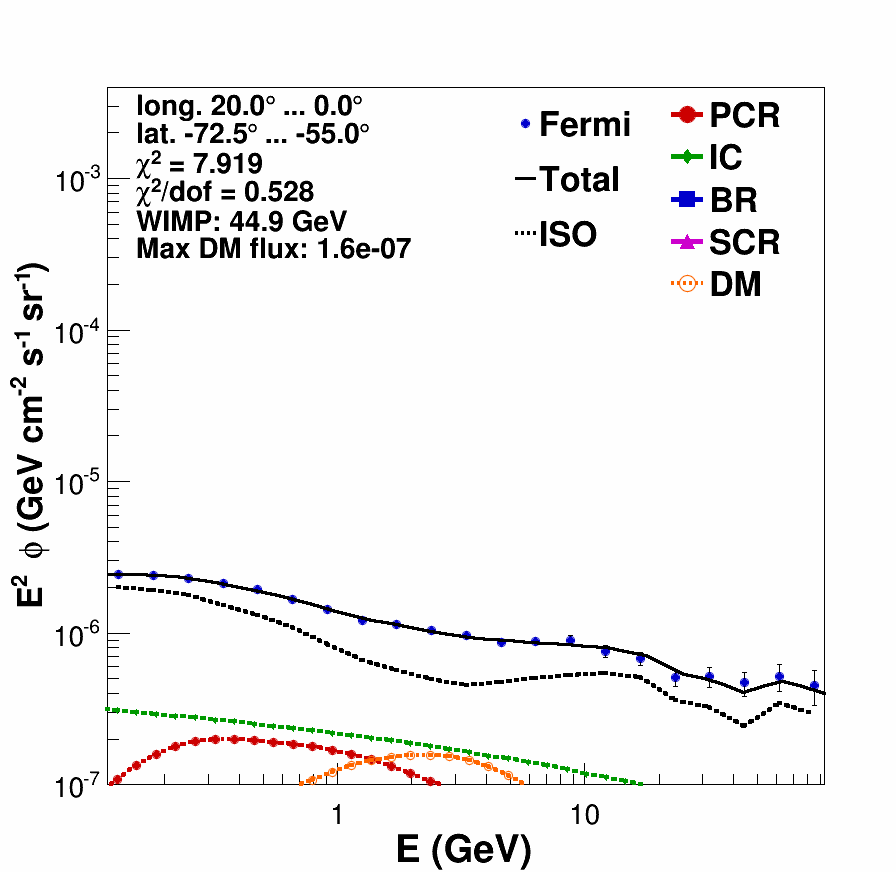}
\includegraphics[width=0.16\textwidth,height=0.16\textwidth,clip]{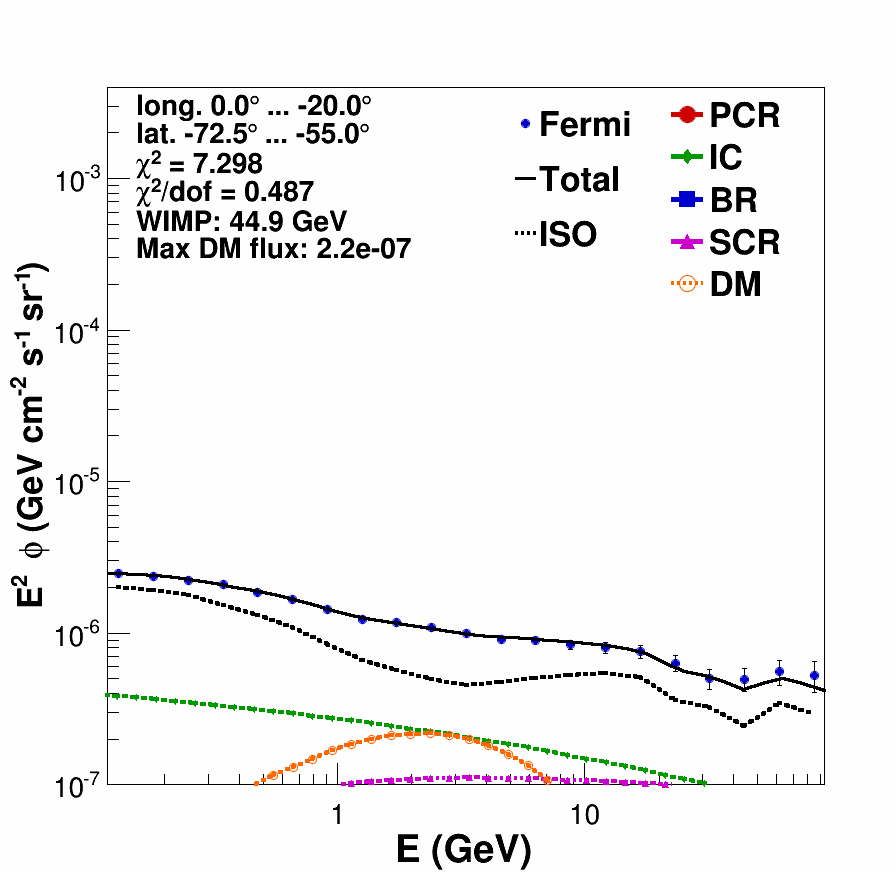}
\includegraphics[width=0.16\textwidth,height=0.16\textwidth,clip]{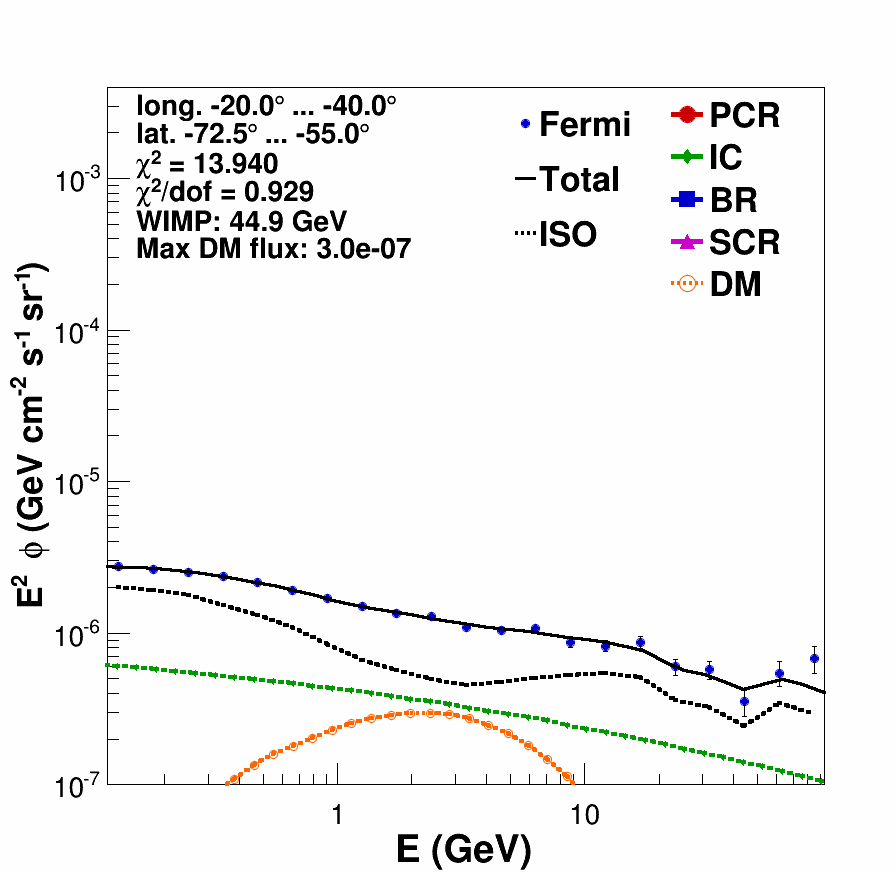}
\includegraphics[width=0.16\textwidth,height=0.16\textwidth,clip]{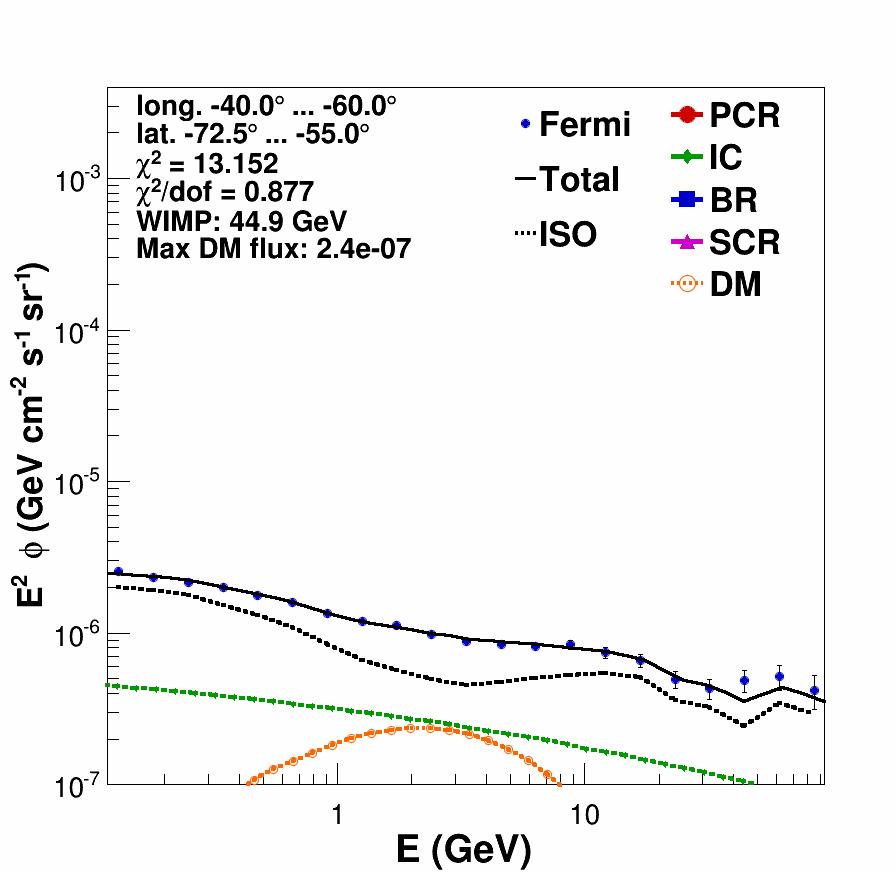}
\includegraphics[width=0.16\textwidth,height=0.16\textwidth,clip]{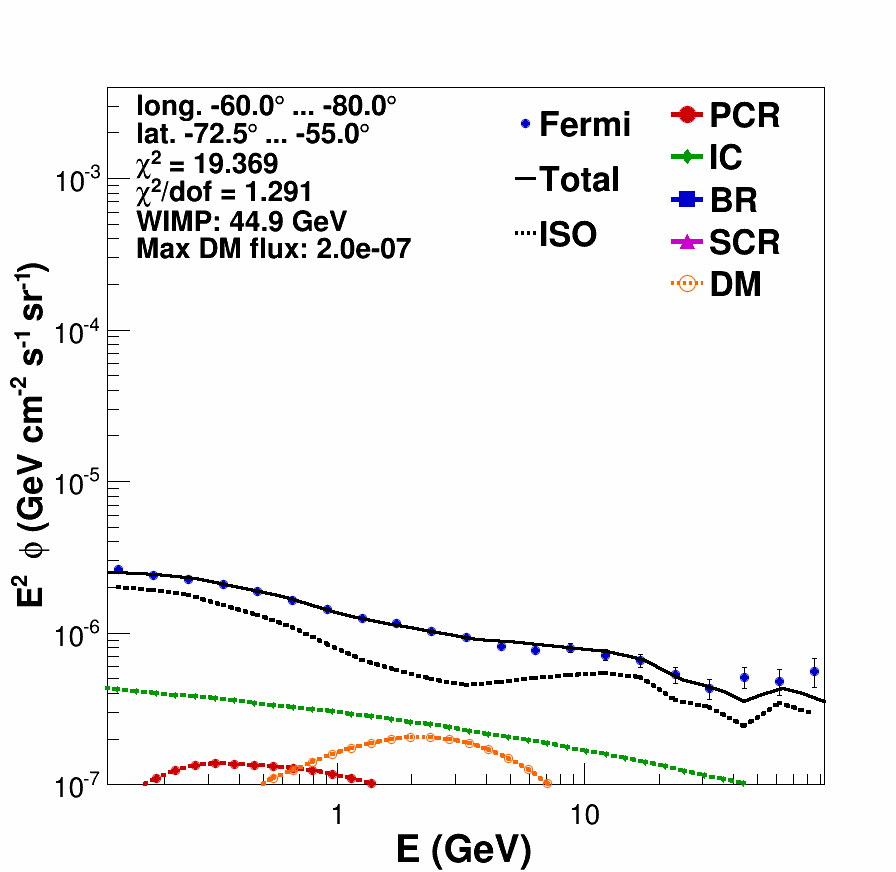}
\includegraphics[width=0.16\textwidth,height=0.16\textwidth,clip]{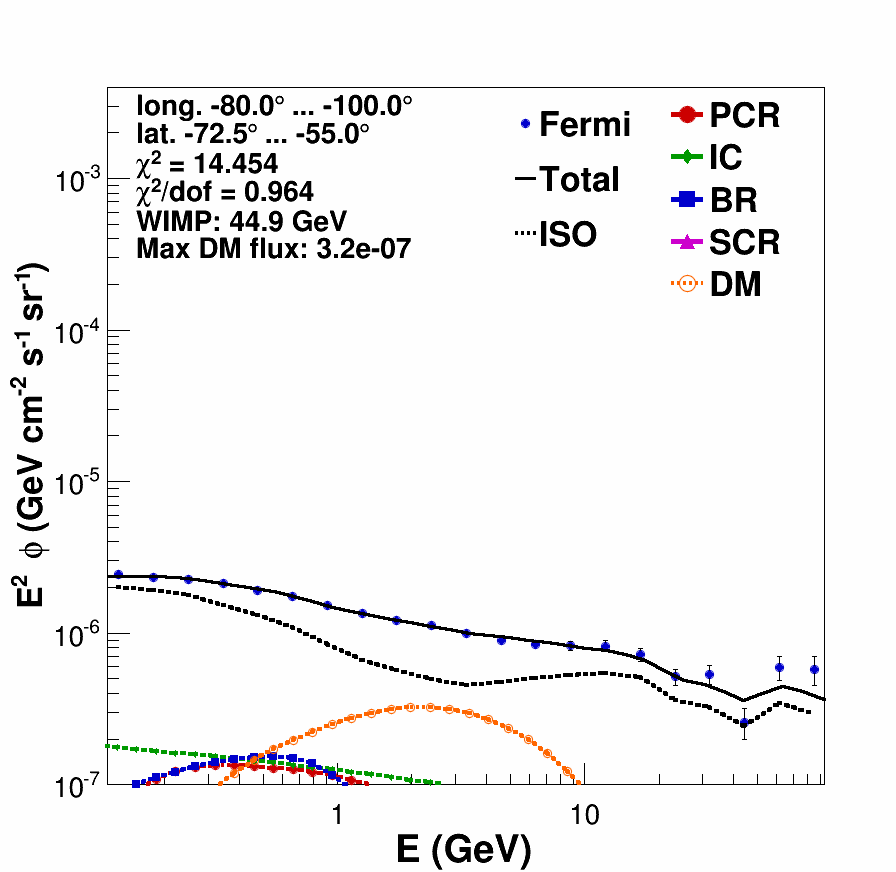}
\includegraphics[width=0.16\textwidth,height=0.16\textwidth,clip]{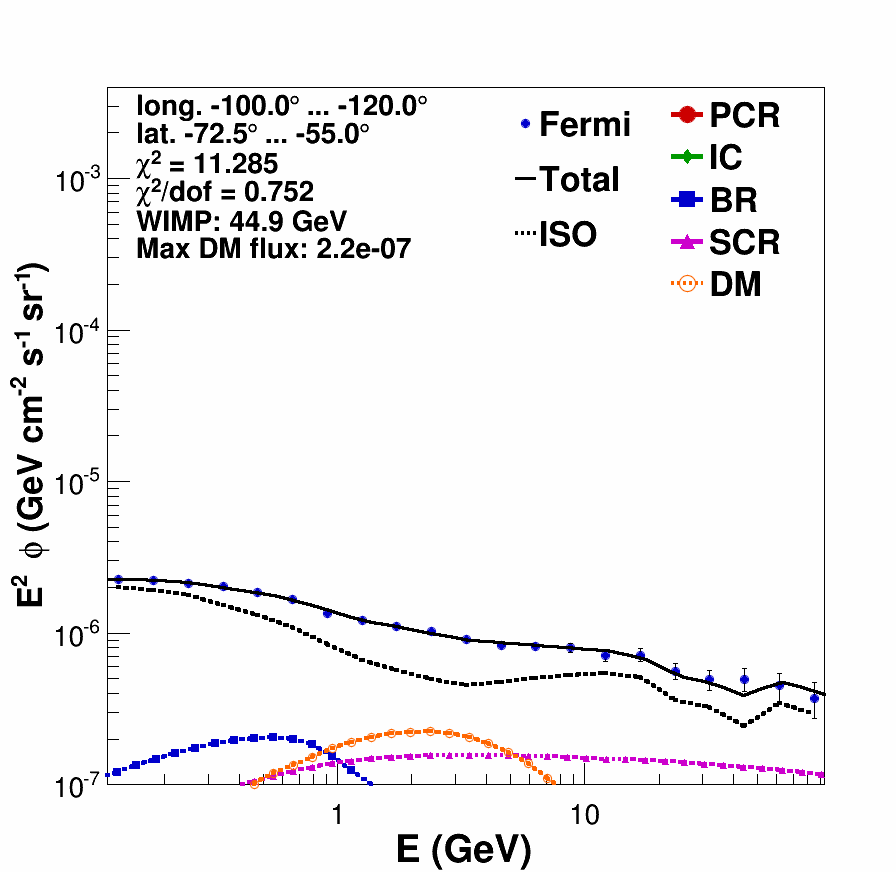}
\includegraphics[width=0.16\textwidth,height=0.16\textwidth,clip]{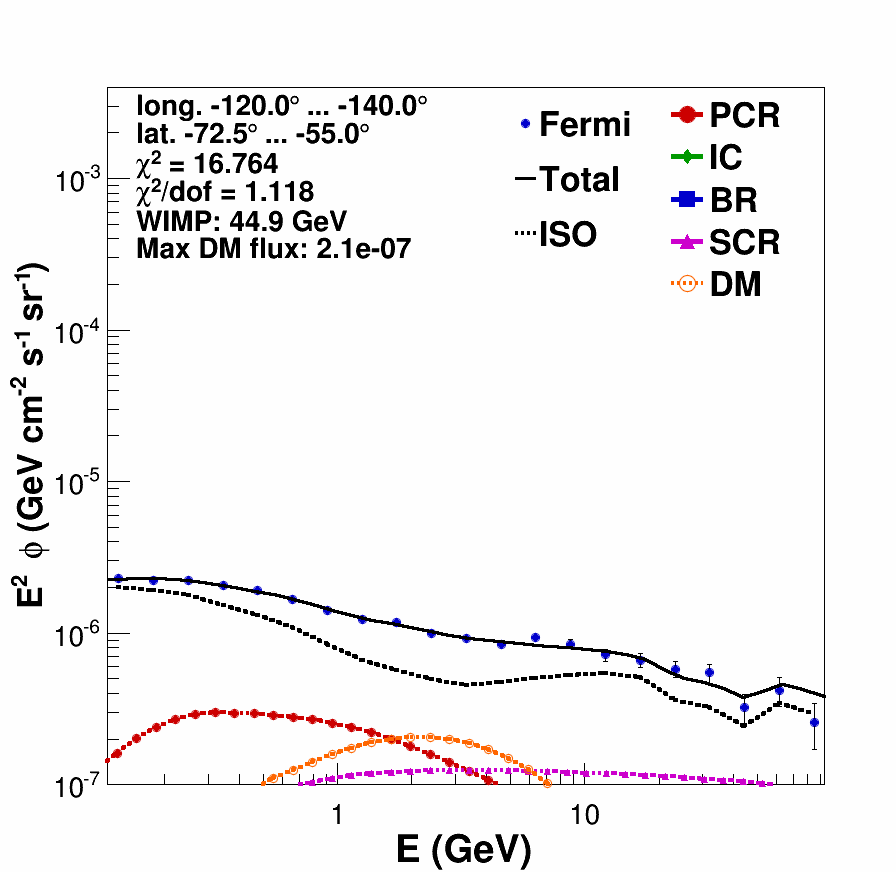}
\includegraphics[width=0.16\textwidth,height=0.16\textwidth,clip]{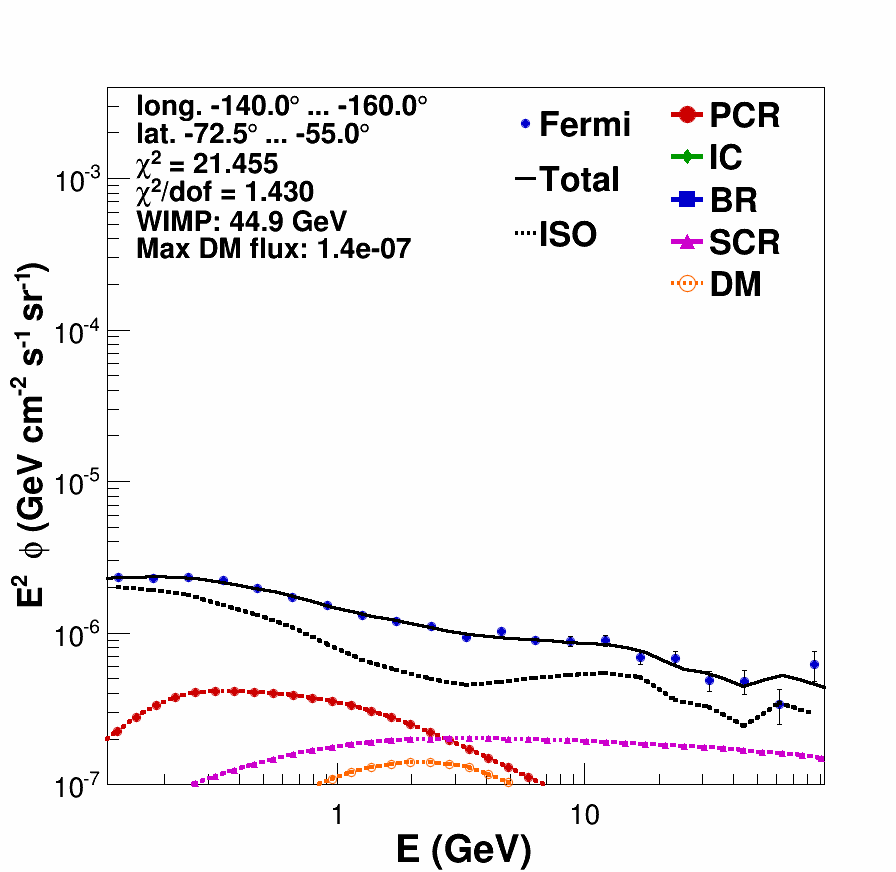}
\includegraphics[width=0.16\textwidth,height=0.16\textwidth,clip]{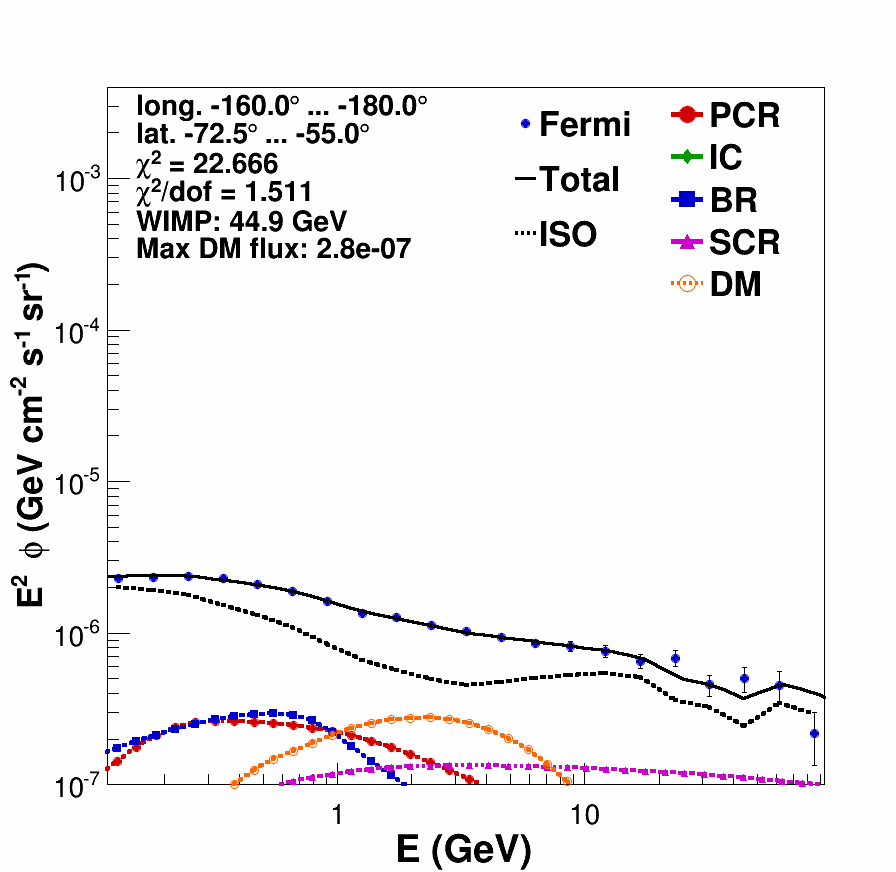}%%%r18
\caption[]{Template fits for latitudes  with $-72.5^\circ<b<-55.0^\circ$ and longitudes decreasing from 180$^\circ$ to -180$^\circ$.} \label{F51}
\end{figure}
\begin{figure}
\includegraphics[width=0.16\textwidth,height=0.16\textwidth,clip]{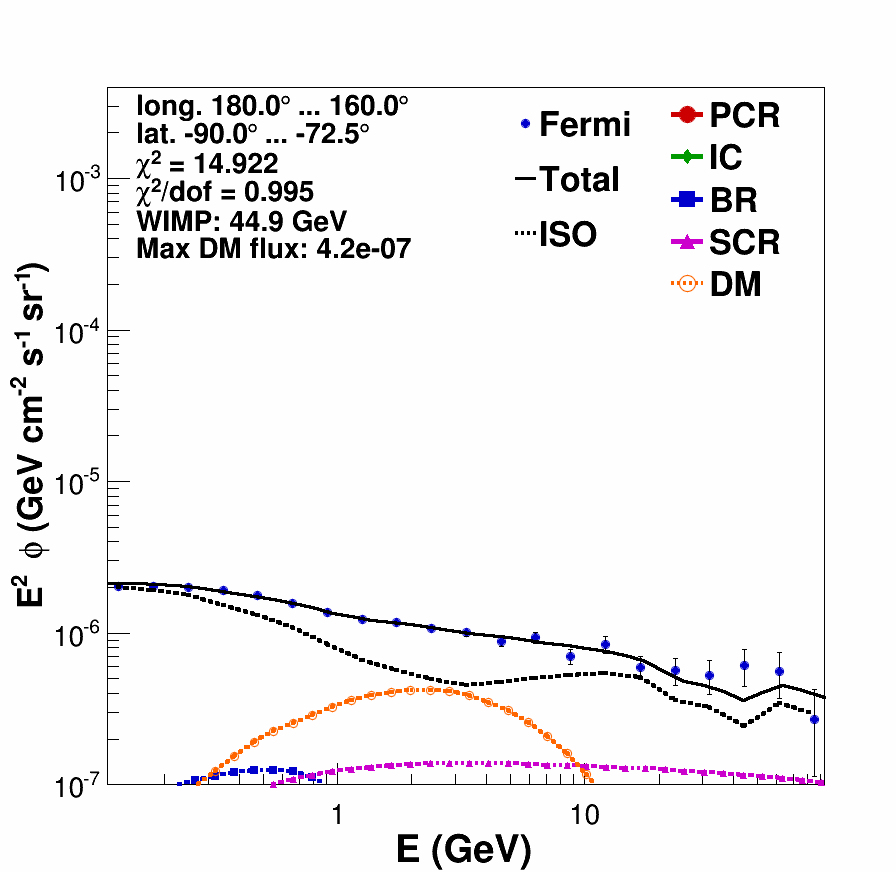}
\includegraphics[width=0.16\textwidth,height=0.16\textwidth,clip]{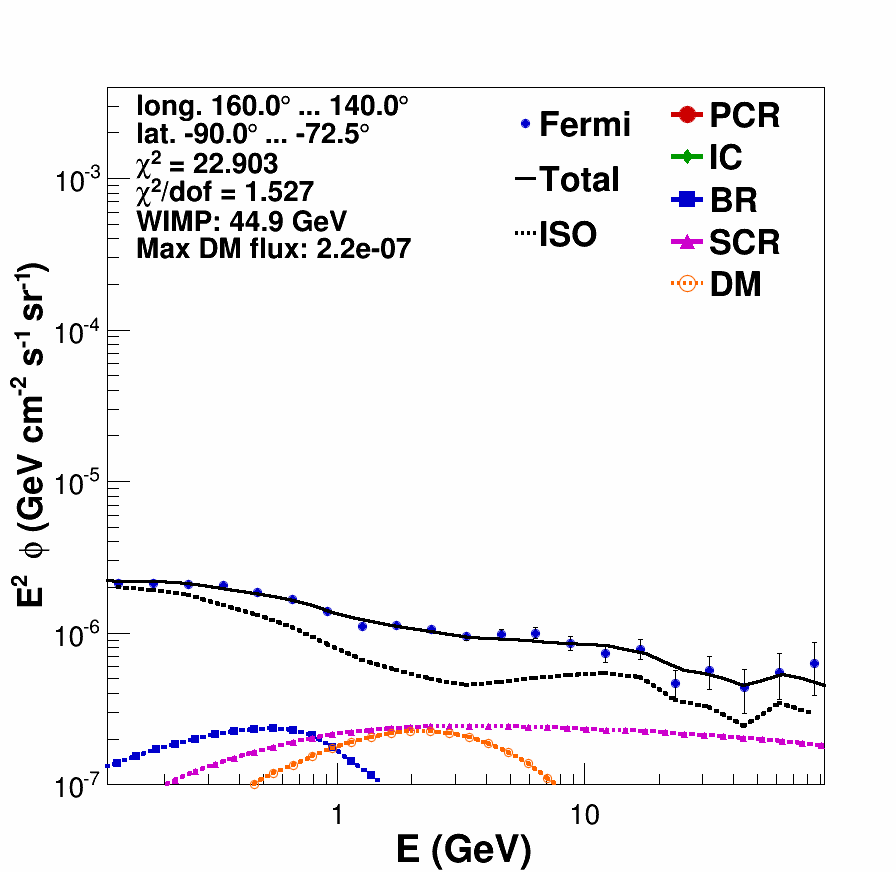}
\includegraphics[width=0.16\textwidth,height=0.16\textwidth,clip]{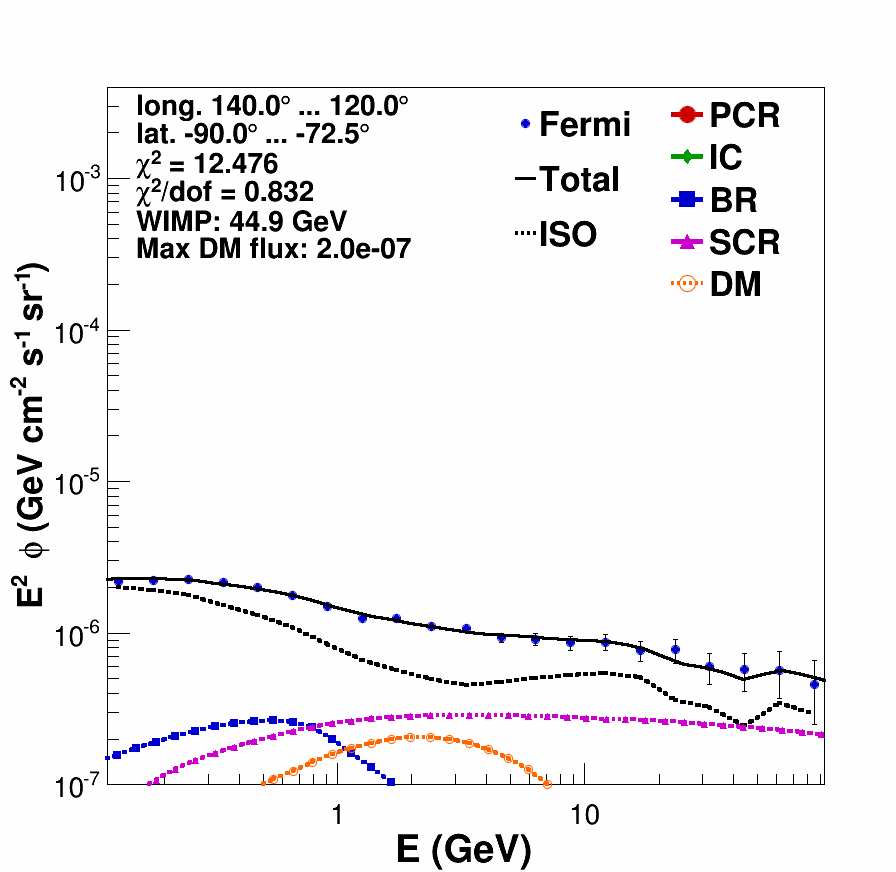}
\includegraphics[width=0.16\textwidth,height=0.16\textwidth,clip]{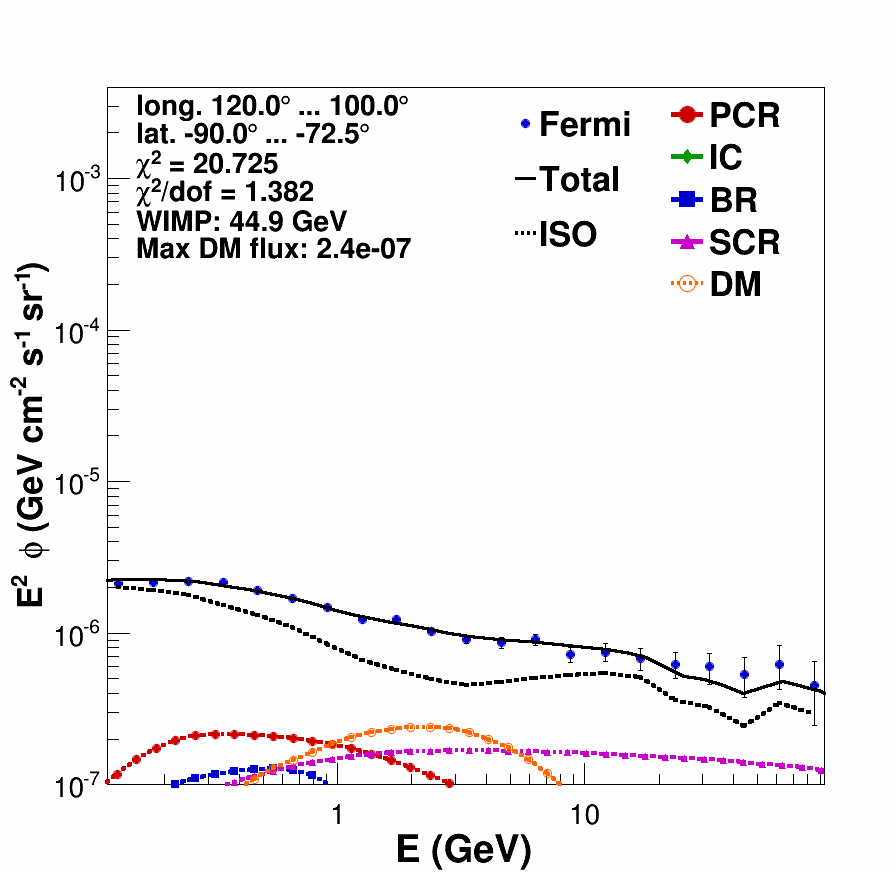}
\includegraphics[width=0.16\textwidth,height=0.16\textwidth,clip]{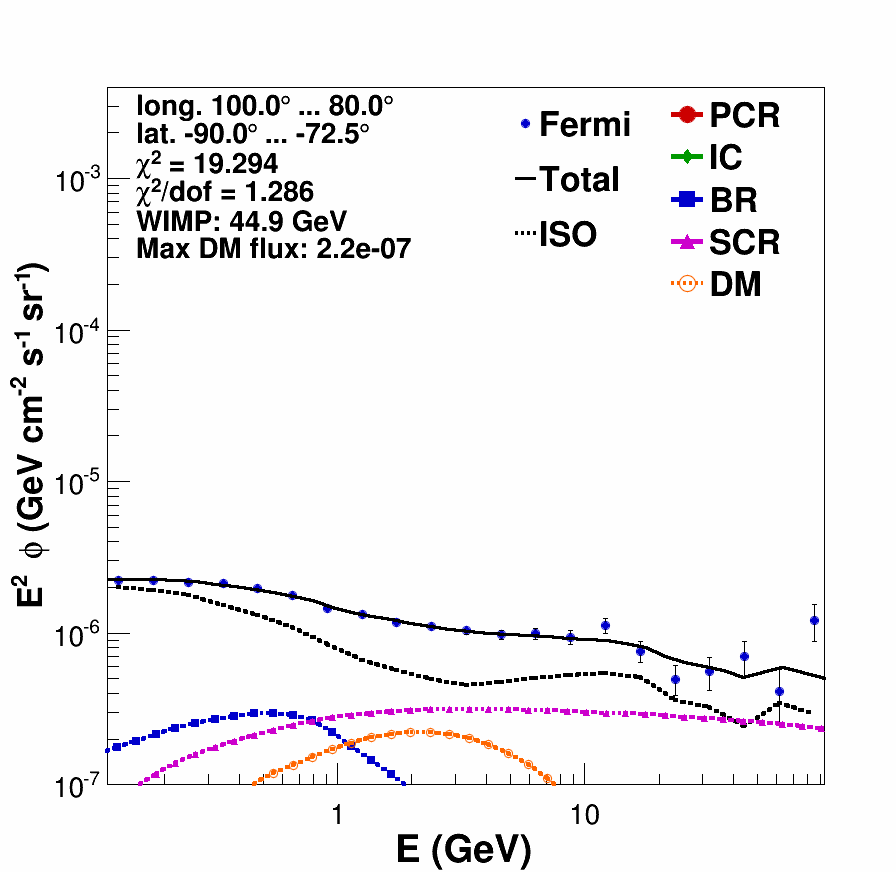}
\includegraphics[width=0.16\textwidth,height=0.16\textwidth,clip]{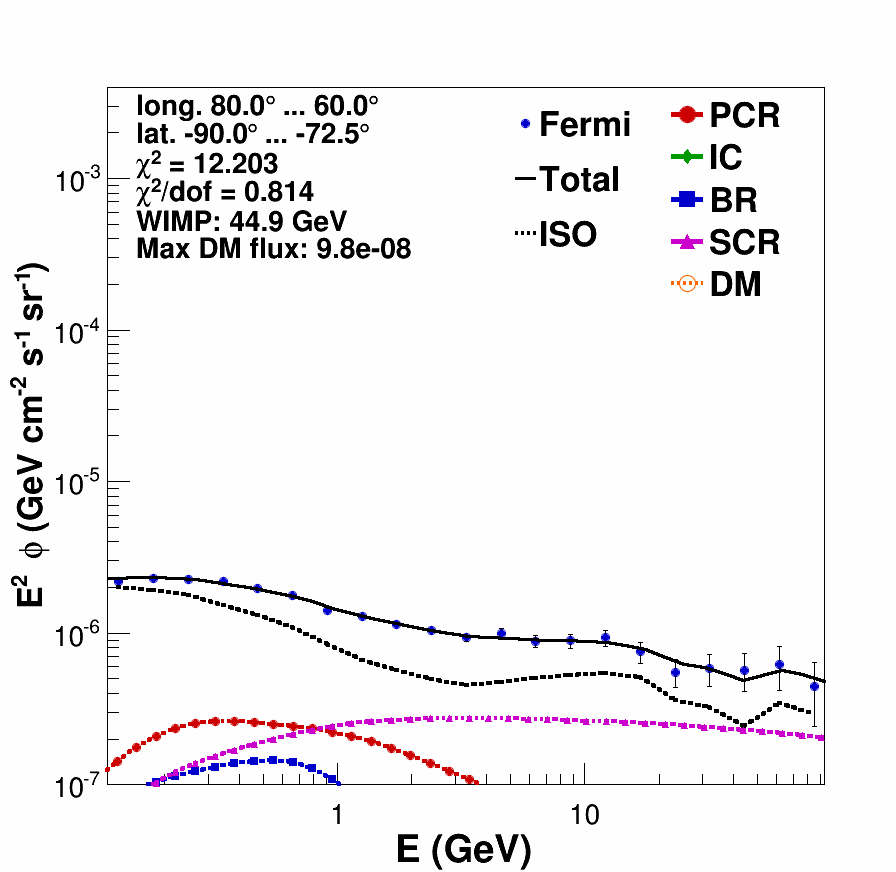}
\includegraphics[width=0.16\textwidth,height=0.16\textwidth,clip]{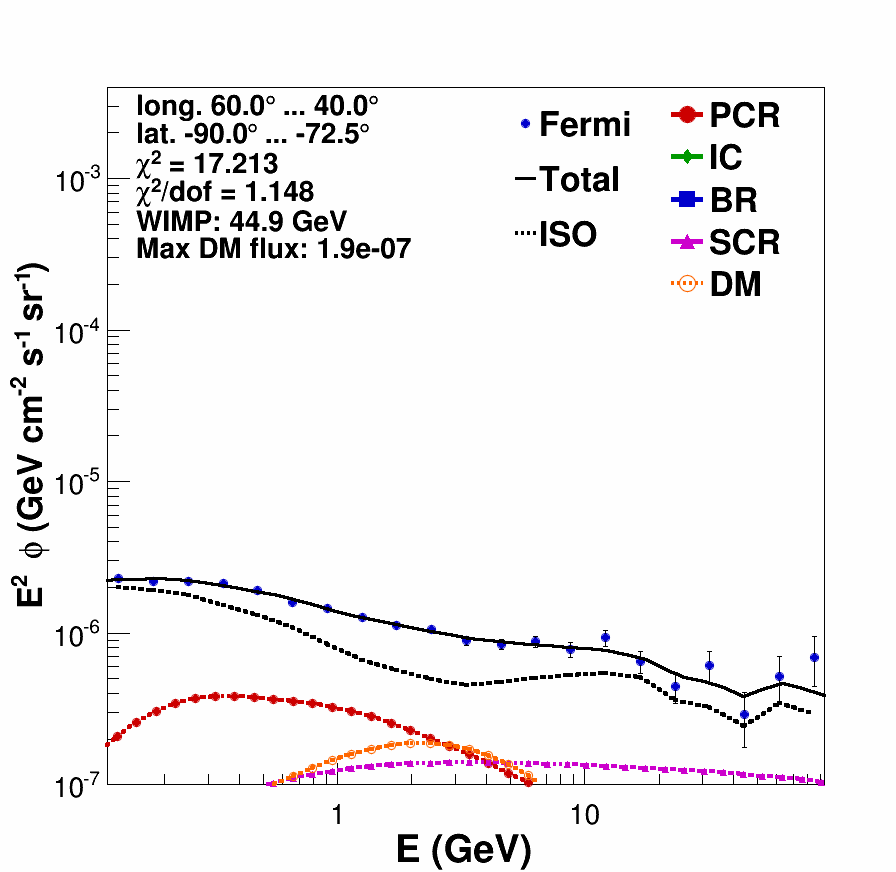}
\includegraphics[width=0.16\textwidth,height=0.16\textwidth,clip]{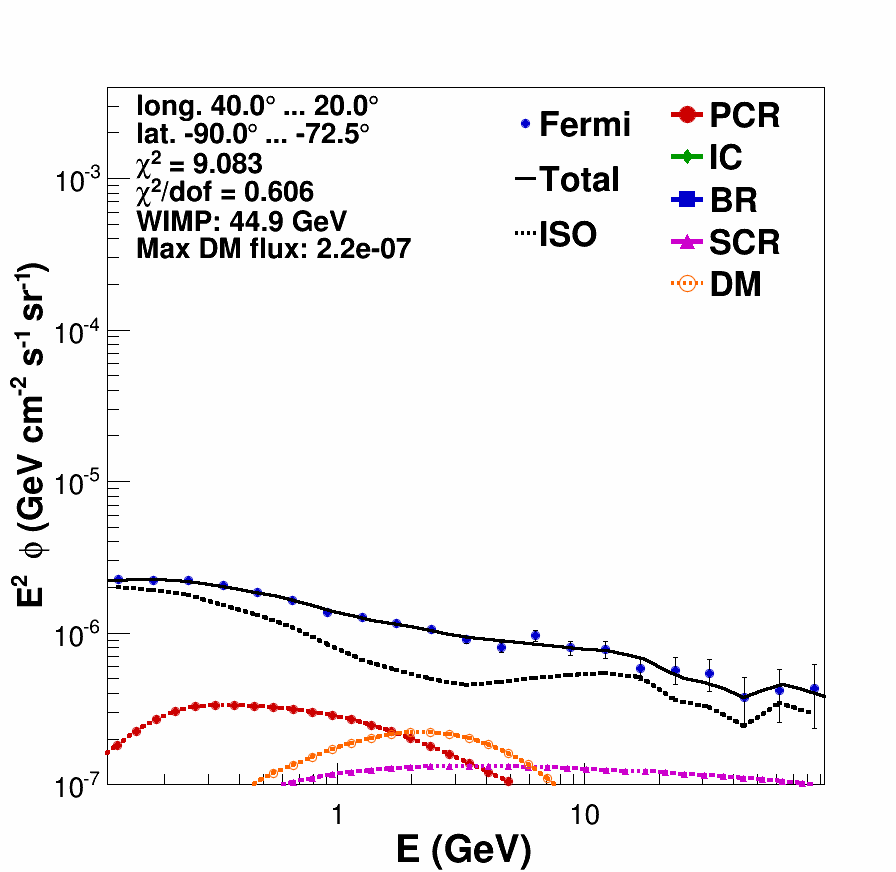}
\includegraphics[width=0.16\textwidth,height=0.16\textwidth,clip]{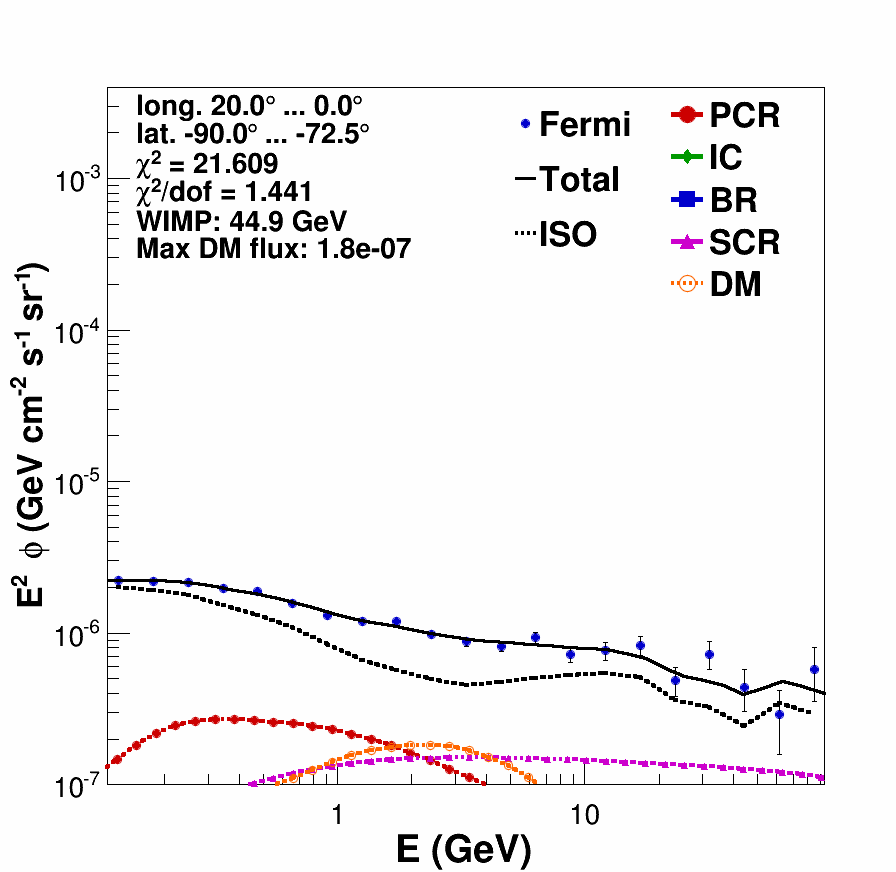}
\includegraphics[width=0.16\textwidth,height=0.16\textwidth,clip]{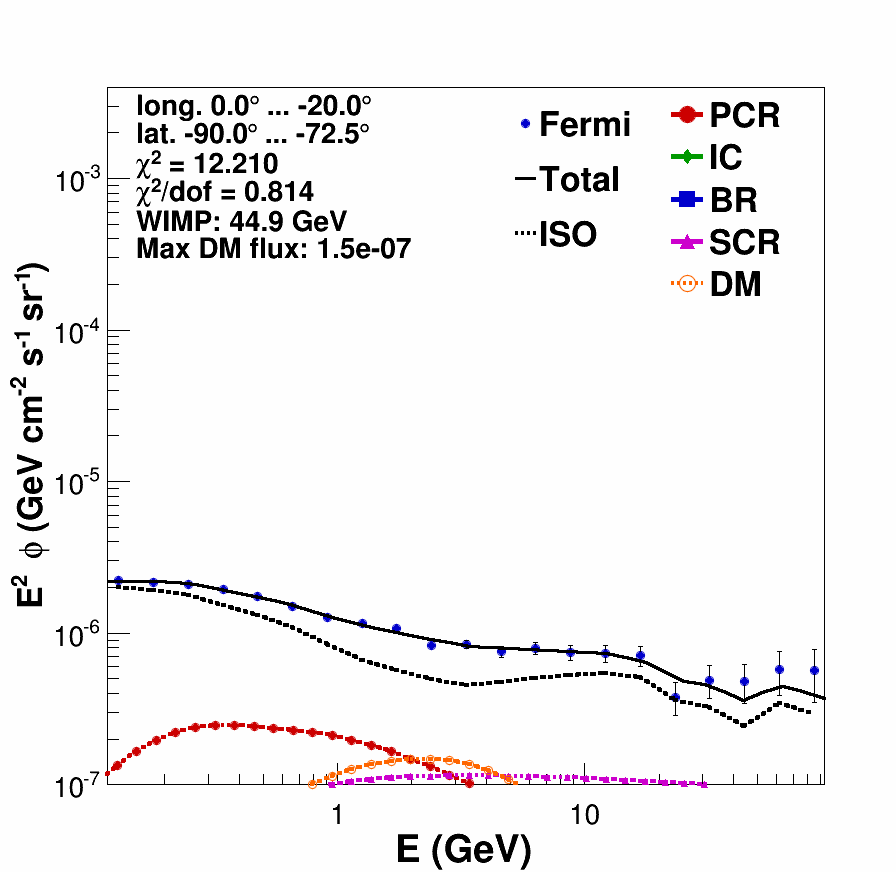}
\includegraphics[width=0.16\textwidth,height=0.16\textwidth,clip]{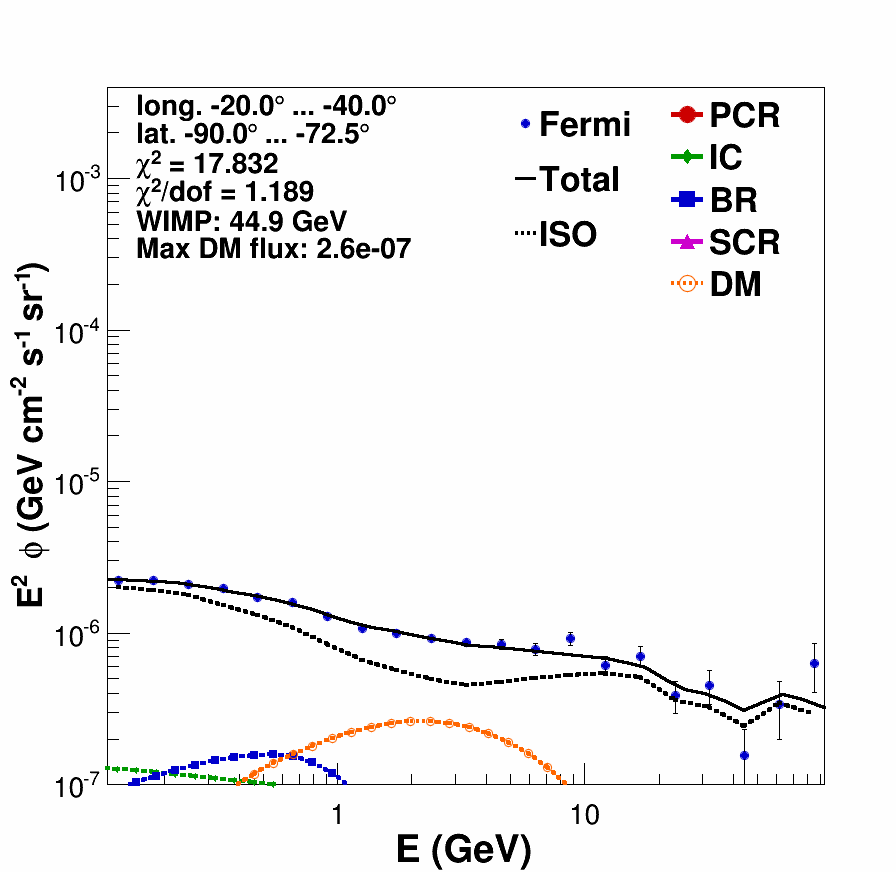}
\includegraphics[width=0.16\textwidth,height=0.16\textwidth,clip]{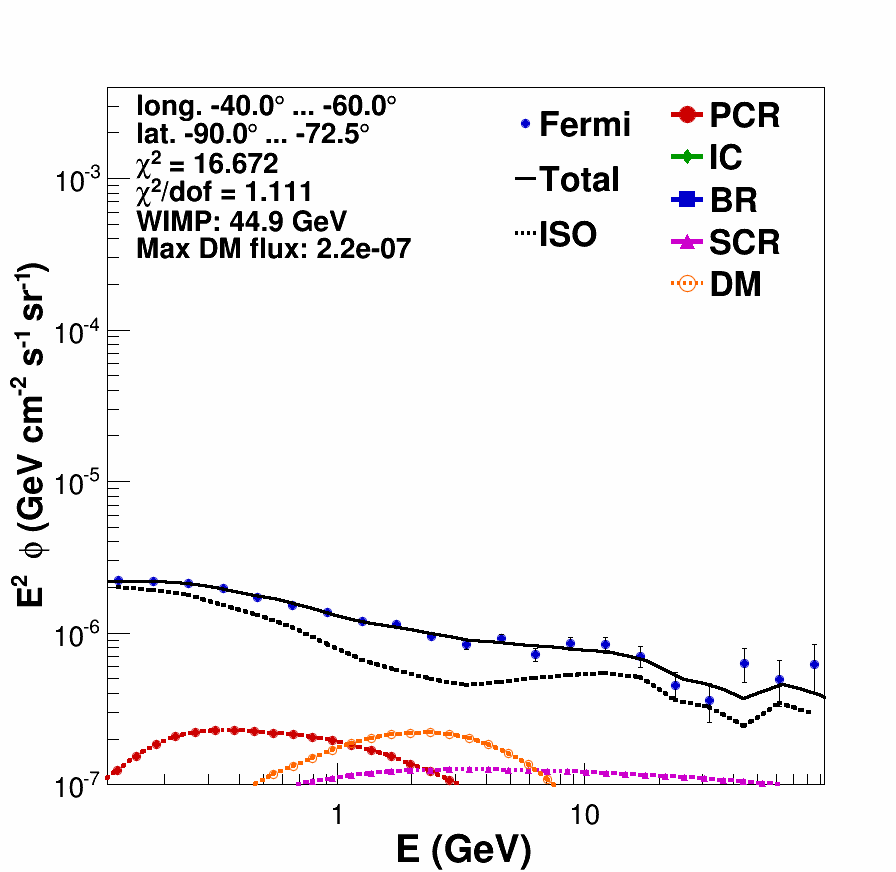}
\includegraphics[width=0.16\textwidth,height=0.16\textwidth,clip]{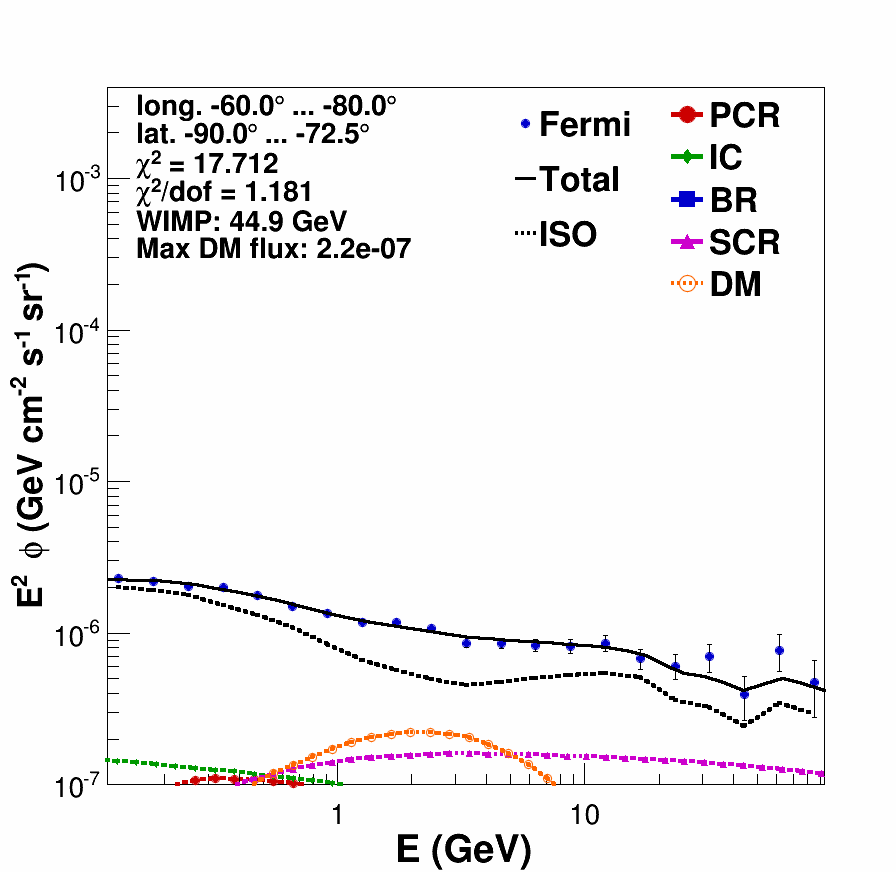}
\includegraphics[width=0.16\textwidth,height=0.16\textwidth,clip]{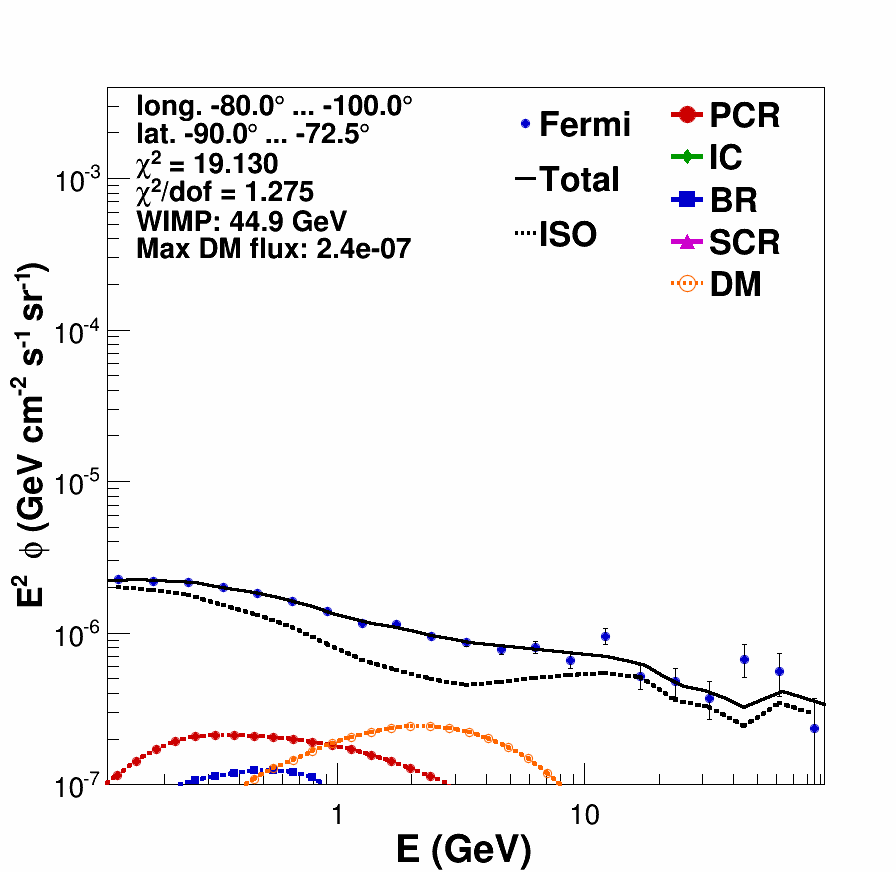}
\includegraphics[width=0.16\textwidth,height=0.16\textwidth,clip]{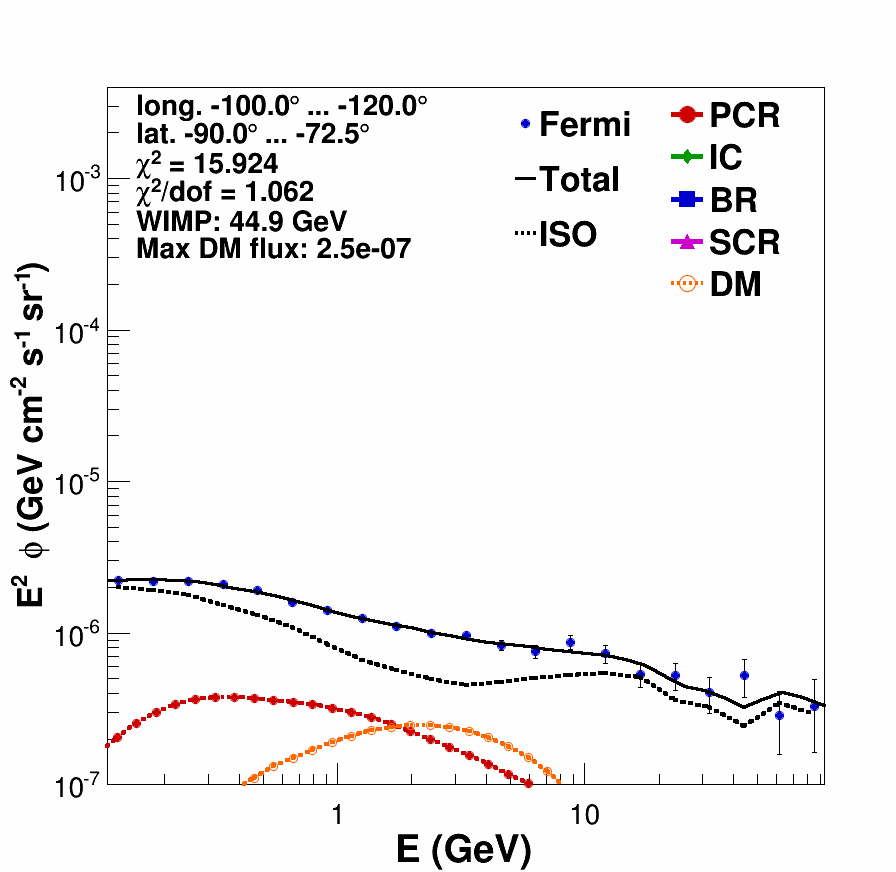}
\includegraphics[width=0.16\textwidth,height=0.16\textwidth,clip]{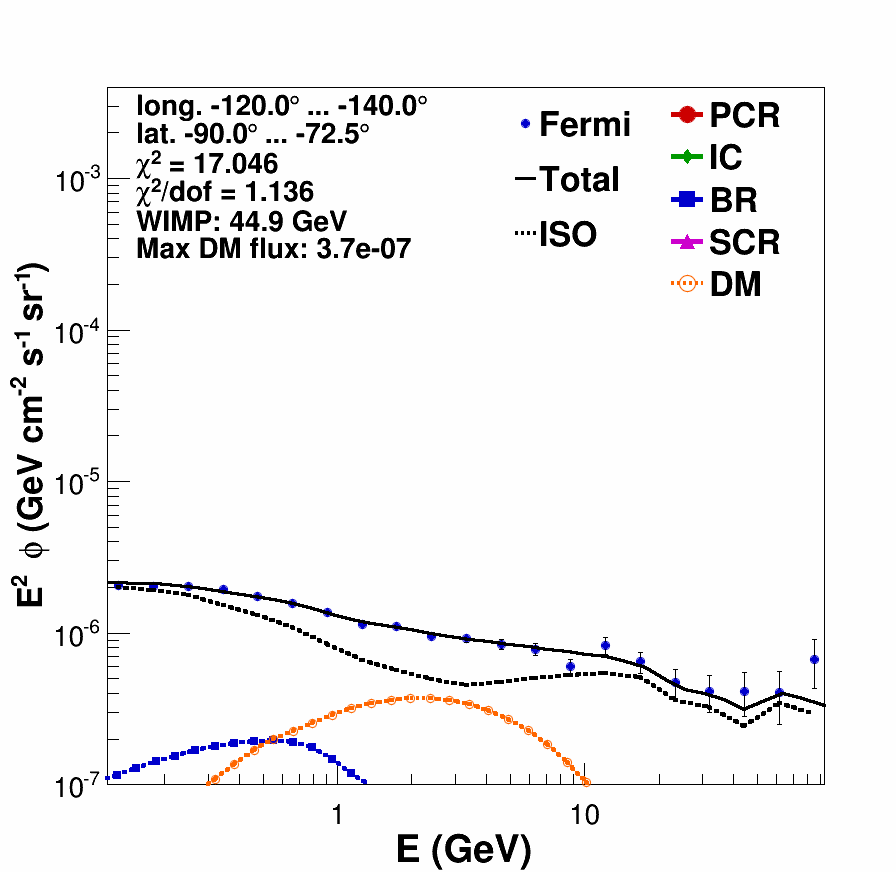}
\includegraphics[width=0.16\textwidth,height=0.16\textwidth,clip]{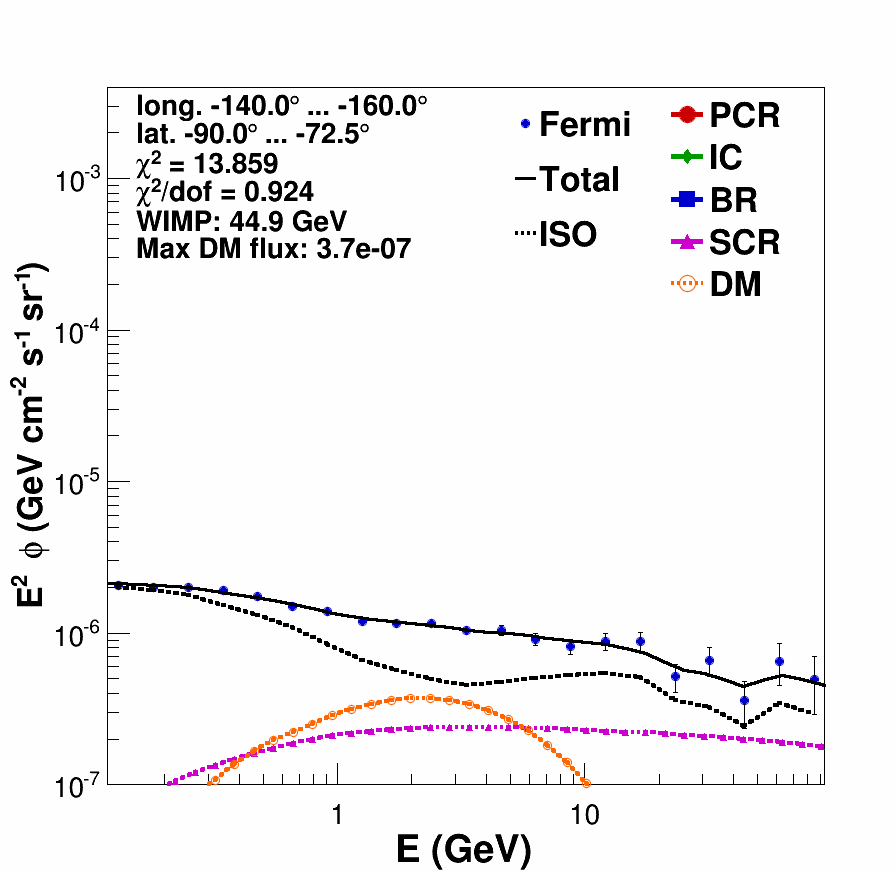}
\includegraphics[width=0.16\textwidth,height=0.16\textwidth,clip]{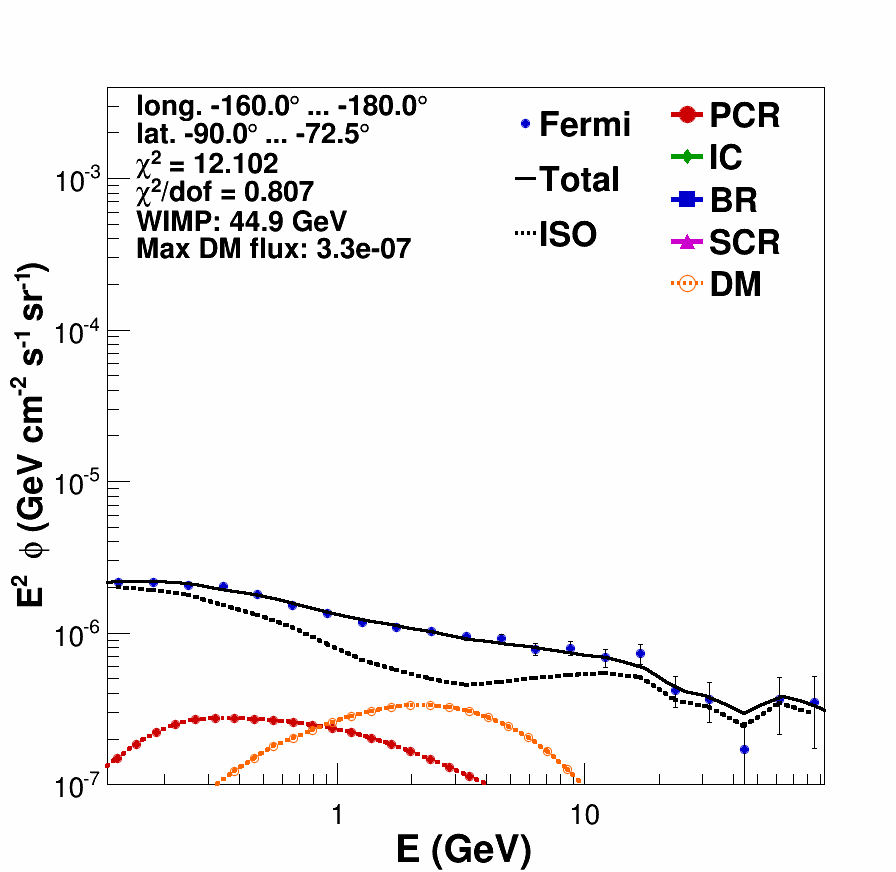}%%%%r19
\caption[]{Template fits for latitudes  with $-90.0^\circ<b<-72.5^\circ$ and longitudes decreasing from 180$^\circ$ to -180$^\circ$.} \label{F52}
\end{figure}